\documentclass{article}

\usepackage{amssymb}                                    
\usepackage{mathrsfs}                    
\usepackage{amsmath}                    
\usepackage{amsfonts}                   
\usepackage{amsfonts}                   
\usepackage{amsthm}                     
\usepackage{makeidx}
\usepackage{graphics}
\usepackage{hyperref}

\usepackage{shorttoc}

\usepackage{fullpage}

\usepackage[all]{xy}

\makeindex

\theoremstyle{plain}
\newtheorem{theorem}{Theorem}[subsection]
\newtheorem{lemma}[theorem]{Lemma}
\newtheorem{fact}[theorem]{Fact}
\newtheorem{proposition}[theorem]{Proposition}
\newtheorem{corollary}[theorem]{Corollary}
\newtheorem{claim}[theorem]{Claim}

\theoremstyle{definition}
\newtheorem{definition}[theorem]{Definition}
\newtheorem{example}[theorem]{Example}
\newtheorem{examples}[theorem]{Examples}
\newtheorem{warning}[theorem]{Warning}
\newtheorem{remark}[theorem]{Remark}
\newtheorem{observation}[theorem]{Observation}
\newtheorem{assumption}[theorem]{Assumption}

\numberwithin{equation}{section}

\renewcommand{\a}{\alpha}
\renewcommand{\b}{\beta}
\renewcommand{\c}{\gamma}
\newcommand{\cG}{\mathcal{G}}
\newcommand{\maps}{\colon}

\def\gg {\mathfrak{g}}

\renewcommand{\(}{\begin{equation}}
\renewcommand{\)}{\end{equation}}
\newcommand{\bea}{\begin{eqnarray}}
\newcommand{\eea}{\end{eqnarray}}
\newcommand{\R}{{\mathbb R}}
\newcommand{\C}{{\mathbb C}}
\newcommand{\Z}{{\mathbb Z}}

\newcommand{\Wbar}{\overline{W}}

\def\nnu {\boldsymbol{\nu}}

\def\proof {{Proof.}\hspace{7pt}}
\def\proofoftheorem #1 {{Proof of theorem \ref{#1}.}\hspace{7pt}}
\def\endofproof {\hfill{$\Box$}\\ \vspace{4pt}}

\def\hpull{\mathbf{\rfloor\!\!\rfloor}}

\begin{document}

\author{Urs Schreiber}

\title{Differential cohomology in a cohesive $\infty$-topos}
\date{21st century}

\maketitle

\begin{abstract}
  We formulate differential cohomology and Chern-Weil theory -- the theory of connections 
  on fiber bundles and of gauge fields -- abstractly in the context of a certain class 
  of higher toposes that we call 
  \emph{cohesive}. Cocycles in this differential cohomology classify higher principal bundles 
equipped with \emph{cohesive structure} 
(topological, smooth, synthetic differential, supergeometric, etc.) and equipped 
with \emph{connections}, hence \emph{higher gauge fields}.
We discuss various models of the axioms and 
applications to fundamental notions and constructions in 
quantum field theory and string theory.  
In particular we show that the cohesive and differential refinement of 
universal characteristic cocycles constitutes a higher Chern-Weil homomorphism 
refined from secondary caracteristic classes to morphisms of higher moduli stacks of higher gauge fields,
and at the same time constitutes extended geometric prequantization
-- in the sense of extended/multi-tiered quantum field theory --
of hierarchies of higher dimensional Chern-Simons-type field theories, 
their higher Wess-Zumino-Witten-type boundary field theories and all further higher codimension
defect field theories. We close with an outlook on the cohomological quantization 
of such higher boundary prequantum field theories by a kind of cohesive motives.

\end{abstract}

\vfill


\begin{center}
  This document is kept online with accompanying material at
  \\
  \medskip
  \href{http://ncatlab.org/schreiber/show/differential+cohomology+in+a+cohesive+topos}{ncatlab.org/schreiber/show/differential+cohomology+in+a+cohesive+topos}
\end{center}

\newpage

\noindent{\bf General Abstract.}
  We formulate differential cohomology (e.g. \cite{BunkeDifferentialCohomology})
  and Chern-Weil theory (e.g. \cite{BottTu})
   -- 
  the theory of connections on fiber bundles and of gauge fields -- 
   abstractly in the context of a certain class of $\infty$-toposes 
   (\cite{Lurie}) that we call 
\emph{cohesive}. Cocycles in this differential cohomology classify principal $\infty$-bundles 
equipped with \emph{cohesive structure} 
(topological, smooth, synthetic differential, supergeometric etc.) and equipped 
with \emph{$\infty$-connections}, hence \emph{higher gauge fields} (e.g. \cite{Freed}).

We construct the cohesive $\infty$-topos of 
smooth $\infty$-groupoids and 
$\infty$-Lie algebroids and show that in this concrete context the general abstract theory 
reproduces ordinary differential cohomology (Deligne cohomology/differential characters), 
ordinary Chern-Weil theory, the traditional notions of smooth 
principal bundles with connection, abelian and nonabelian gerbes/bundle gerbes with connection, 
principal 2-bundles with 2-connection, connections on 3-bundles, etc. and generalizes these to 
higher degree and to base spaces that are orbifolds and generally smooth $\infty$-groupoids, 
such as smooth realizations of classifying spaces/moduli stacks for principal $\infty$-bundles 
and configuration spaces of gauge theories.

We exhibit a general abstract \emph{$\infty$-Chern-Weil homomorphism} and observe that 
it generalizes the Lagrangian of Chern-Simons theory to \emph{$\infty$-Chern-Simons theory}. 
For every invariant polynomial on an $\infty$-Lie algebroid it 
sends principal $\infty$-connections to 
\emph{Chern-Simons circle $(n+1)$-bundles} ($n$-gerbes) with connection, whose 
higher parallel transport is the corresponding higher Chern-Simons Lagrangian. 
There is a general abstract formulation of the higher holonomy of this parallel transport and
this provides the action functional of $\infty$-Chern-Simons theory as a morphism on its 
cohesive configuration $\infty$-groupoid. 
Moreover, to each of these higher Chern-Simons Lagrangian is 
canonically associated a differentially twisted looping, which we identify
with the corresponding \emph{higher Wess-Zumino-Witten Lagrangian}. 

We show that, when interpreted in smooth $\infty$-groupoids and their variants, 
these intrinsic constructions reproduce the ordinary Chern-Weil homomorphism, 
hence ordinary Chern-Simons functionals and ordinary Wess-Zumino-Witten functionals,
provide their geometric prequantization in higher codimension
(localized down to the point)
and generalize this to a fairly extensive list of action functionals of 
quantum field theories and string theories, some of them new. All of these
appear in their refinement from functionals on local differential form data
to global functionals defined on the full moduli $\infty$-stacks 
of field configurations/$\infty$-connections, where they represent higher
prequantum line bundles. 
We show that these moduli $\infty$-stacks naturally 
encode fermionic $\sigma$-model anomaly cancellation conditions, such as given by
higher analogs of $\mathrm{Spin}$-structures and of $\mathrm{Spin}^c$-structures.

We moreover show that \emph{higher symplectic geometry} is naturally subsumed in 
higher Chern-Weil theory, such that the passage from the unrefined to the 
refined Chern-Weil homomorphism induced from higher symplectic forms
implements \emph{geometric prequantization}
of the above higher Chern-Simons and higher Wess-Zumino-Witten functionals.
We study the resulting formulation of \emph{local prequantum field theory},
show how it subsumes traditional classical field theory and how it illuminates
the boundary and defect structure of higher Chern-Simons-type field theories,
their higher Wess-Zumino-Witten type theories, etc.

We close with an outlook on the ``motivic quantization'' 
of such local prequantum field theory of higher moduli stacks of fields 
to genuine local quantum field theory with 
boundaries and defects, by pull-push in twisted generalized cohomology of higher
stacks and conclude that cohesive $\infty$-toposes provide a ``synthetic''
axiomatization of local quantum gauge field theories obtained from geometric Lagrangian data
\cite{sQFT}.

\medskip 

We think of these results as providing a further 
ingredient of the recent identification of the
mathematical foundations of quantum field and 
perturbative string theory \cite{SatiSchreiber}:  while the cobordism theorem
\cite{LurieTFT} identifies topological quantum field theories and their
boundary and defect theories
with a universal construction in higher category theory 
(representations of free symmetric monoidal $(\infty,n)$-categories with full duals), our results 
indicate that the geometric structures that these arise
from under geometric motivic quantization originate in a universal construction in 
higher topos theory: \emph{cohesion}.

\newpage

\noindent{\bf Acknowledgements.}
The program discussed here 
was initiated around the writing of
\cite{SSSIII}, following an unpublished precursor set of notes \cite{nactwist},
presented at \cite{OberwolfachTalk}, which was motivated in parts by the desire
to put the explicit constructions of 
\cite{SSW, SWI, SWII, SWIII, BCSS, RobertsSchreiber} on a broader
conceptual basis.
The present text has grown out of and subsumes these and the series of publications
\cite{OberwolfachTalk, SchreiberSkoda, FSS, FSScfield,
FiorenzaSatiSchreiberI, FiorenzaSatiSchreiberIV,  NSSa, NSSb, 
ScSh, FiorenzaSatiSchreiberCS,  hgp, LocalObservables, InfinityWZW, Nuiten}.
Notes from a lecture series introducing some of the central ideas
with emphasis on applications to string theory are available as
\cite{TwistedStructuresLecture}. 
(The basic idea of considering differential cohomology in the 
$\infty$-topos over smooth manifolds has then also been voiced in 
\cite{Hopkins}, 
together with the statement that this is the context in which 
the seminal article \cite{HopkinsSinger} on differential cohomology
was eventually meant to be considered, then done in \cite{BunkeNikolausVoelkl},
see \ref{DiagramsOfCohesiveGroupoids} below.) 
A survey of the general project of
``Synthetic quantum field theory'' in cohesive $\infty$-toposes is in \cite{sQFT}.
The following text aims to provide a comprehensive
theory and account of these developments. In as far as it uses paragraphs  
taken from the above joint publications, these paragraphs have been primarily authored 
by the present author.

\medskip

I cordially thank all my coauthors for their input and work and for discussion.
Especially I thank Domenico Fiorenza for joining in on working out details
of cohesive higher geometry and for improving on my expositions; and
Hisham Sati for providing a steady stream of crucial examples and motivations
from string theory and M-theory.
I am grateful to Richard Williamson for an extra derived left adjoint, to
David Carchedi for an extra derived right adjoint and to a talk by Peter Johnstone for  
making me recognize their 1-categorical shadow in Lawvere's writing, all at an early stage of this
work. I am indebted to Mike Shulman for plenty of discussion of and input on
higher topos theory and homotopy type theory in general. I had and have 
plenty of inspiring conversations with David Corfield on conceptual questions, a
good bit of which has influenced the developments presented here. 
Then I thank Igor Khavkine for discussion of covariant field theory.
I thank Uli Bunke, Thomas Nikolaus, and Michael V{\"o}lkl for 
discussion of stable cohesion and for finding the ``differential cohomology diagram''
from just the axioms of stable cohesion. I also thank the participants of the 
\emph{third String Geometry Network meeting} for discussion of supergeometric
cohesion and supergeometric higher WZW models; special thanks to Bas Janssen
for being really careful with super Lie algebra cohomology.
I am grateful to Charles Rezk for telling me about cohesion in global equivariant homotopy theory
and for providing pointers related to stable cohesion.
Last but not least I am thankful to Ieke Moerdijk.
David Corfield and Geoffrey Cruttwell kindly commented on the text
at some stage or through some stages of its development.  
Further typos were spotted by
Igor Khavkine, Fosco Loregian, Zhen Lin Low, David Roberts.

\newpage

\shorttableofcontents{Main contents}{2}

\newpage

\noindent
{\bf In \ref{Introduction} } we motivate our discussion, give an informal 
introduction to the main
concepts involved and survey various of our constructions and applications
in a more concrete, more traditional and more expository way than in the sections to follow.
This may be all that 
some readers ever want to see, while other readers may want to skip it entirely.

\noindent {\bf In \ref{HomotopyTypeTheory}} we review relevant aspects of 
\emph{homotopy type theory}, the theory of \emph{$\infty$-categories} and
\emph{$\infty$-toposes}, in terms of which all of the following is formulated.
This serves to introduce context and notation and to provide a list of 
technical lemmas which we need in the following, some of which are not, 
or not as explicitly, stated in existing literature. 

\noindent
{\bf In \ref{GeneralAbstractTheory}} we introduce
\emph{cohesive homotopy type theory}, a general abstract theory 
of differential geometry, differential cohomology and Chern-Weil theory 
in terms of universal constructions
in $\infty$-topos theory. This is in the spirit of 
Lawvere's proposals \cite{Lawvere} for axiomatic characterizations of those 
\emph{gros toposes} that serve as contexts for abstract \emph{geometry} 
in general and \emph{differential geometry} in particular: \emph{cohesive toposes}. 
We claim that the 
decisive role of these axioms is realized when generalizing from topos theory
to $\infty$-topos theory and we discuss a fairly long list of 
geometric structures that is induced by the axioms in this case. Notably
we show that every $\infty$-topos satisfying the immediate analog of 
Lawvere's axioms -- every \emph{cohesive $\infty$-topos}-- 
comes with a good intrinsic notion of differential cohomology
and Chern-Weil theory. 

Then we add a further simple set of axioms
to obtain a theory of what we call \emph{differential cohesion},
a refinement of cohesion that axiomatizes the explicit (``synthetic'')
presence of infinitesimal objects. This is closely related to Lawvere's
\emph{other} proposal for axiomatizing toposes for differential geometry, 
called \emph{synthetic differential geometry} \cite{LawvereSynth}, 
but here formulated entirely in terms of higher \emph{closure modalities} as
for cohesion itself. We find that these axioms also capture 
the modern synthetic-differential theory of \emph{D-geometry} \cite{LurieCrystal}.
In particular a differential cohesive $\infty$-topos has an intrinsic notion
of (formally) \emph{{\'e}tale maps}, which makes it an axiomatic geometry
in the sense of \cite{LurieSpaces} and equips it with intrinsic \emph{manifold}
theory.

\noindent
{\bf In \ref{Implementation}} we discuss models of the axioms, 
hence $\infty$-toposes of $\infty$-groupoids which are equipped with a geometric structure
(topology, smooth structure, supergeometric structure, etc.) in a way that 
all the abstract differential geometry theory developed in the previous chapter can be 
realized.
The main model of interest for our applications is the cohesive 
$\infty$-topos $\mathrm{Smooth} \infty \mathrm{Grpd}$
as well as its infinitesimal
thickening $\mathrm{SynthDiff}\infty \mathrm{Grpd}$,
which we construct.
Then we go step-by-step through the list of general abstract structures
in cohesive $\infty$-toposes and unwind what these amount to in 
this model. We demonstrate
that these subsume traditional definitions and constructions
and generalize them 
to higher differential geometry and differential cohomology.

\noindent
{\bf In \ref{Applications}} we discuss applications of 
the general theory 
in the context of smooth $\infty$-groupoids and
their synthetic-differential and super-geometric refinements
to aspects of higher gauge prequantum field theory.
We present a fairly long list of higher $\mathrm{Spin}$-
and $\mathrm{Spin}^c$-structures,
of classes of local action functionals on higher moduli stacks of fields
of higher Chern-Simons type and functionals of higher Wess-Zumino-Witten type, 
that are all naturally induced by higher Chern-Weil theory.
We exhibit a higher analog of geometric prequantization that applies to 
these systems and show that it captures a wealth of structures,
such as notably the local boundary and higher codimension defect structure.
Apart from the new constructions and results, 
this shows that large parts of local prequantum gauge field theory 
are canonically and fundamentally induced by abstract cohesion.

\noindent
{\bf In \ref{MotivicQuantizationApplications}} we close with an outlook
on how the quantization of the local prequantum gauge field theory 
to genuine local quantum field theory
proceeds via higher linear algebra in cohesive $\infty$-tpoposes, namely
via duality of cohesive stable homotopy types.

\newpage

\tableofcontents

\newpage

\section{Introduction}
\label{Introduction}

In \ref{Motivation}
we motivate the formalization of physics within higher differential geometry. 
Then section 
\ref{TheGeometryOfPhysics}
is a lecture-note style introduction to the formulation of physics in terms
of cohesive higher differential geometry. This serves as a warm-up for the general
abstract discussion in the main sections to follow and is meant to illustrate the
connection of cohesive higher geometry to traditional discussion notably
of classical mechanics \ref{BasicClassicalMechanicsByPrequantizedLagrangianCorrespondences} 
and of classical field theory \ref{DeDonderWeylTheoryViaHigherCorrespondences}.

\subsection{Motivation}
\label{Motivation}

In the year 1900, at the {International Congress of Mathematics} in Paris, 
David Hilbert stated his famous list of 23 central open questions of mathematics
\cite{Hi1900}. Among them, the sixth problem  (see \cite{Corry04} for a review) 
has maybe received the least attention
from mathematicians, but is arguably the one that Hilbert himself regarded as the 
most valuable:
``From all the problems in the list, the sixth is the only one that continually engaged 
[Hilbert's] efforts over a very long period, at least between 1894 and 1932.''
\cite{Corry06}.
Hilbert stated the problem as follows\footnote{I am grateful to Hisham Sati and to David Corfield for discussion of this point.} 
(the boldface emphasis is ours):

\vspace{4pt}

\noindent{\bf Hilbert's mathematical problem 6.}
{\it To treat by means of {\bf axioms}, those {\bf physical sciences} in 
	which mathematics plays an important part}.

\vspace{4pt}

Hilbert went on to specify how such axiomatization should proceed:

\vspace{4pt}

\noindent {\it   try first by a {\bf small number of axioms} 
    to include as large a class as possible of physical phenomena, 
    and then by adjoining new axioms to arrive gradually at the more special theories. }

\vspace{4pt}

Finally there one more remark:

\vspace{4pt}

\noindent {\it take account not only of those theories coming near to reality, but also, 
	as in geometry, all {\bf logically possible theories} }. 

\vspace{4pt}

Since then, various aspects of physics have been given a mathematical formulation.
The following table, necessarily incomplete, gives a broad idea of central
concepts in theoretical physics and the mathematics
that captures them.

\vspace{.5cm}

\begin{tabular}{|l|ll|l|}
    \hline
	&& & 
	\\
    & {\bf physics} & {\bf maths} & 
	\\
	&& & 
    \\
	\hline\hline
    & {\it prequantum physics} & {\it differential geometry} & \ref{TheGeometryOfPhysics}
    \\
	\hline\hline
    18xx-19xx & {mechanics} & {symplectic geometry} & \ref{BasicClassicalMechanicsByPrequantizedLagrangianCorrespondences}
	\\
	1910s & {gravity} & 
	{Riemannian geometry} & \ref{FieldsOfGravtySpecialAndGeneralizedGeometry}
	\\
	1950s & {gauge theory} & 
	{Chern-Weil theory} & \ref{InfinityChernWeilHomomorphismIntroduction}
	\\
	2000s & {higher gauge theory} & {differential cohomology} & \ref{GeometryOfPhysicsPrincipalConnections}
	\\
	\hline
	& & & 
	\\
	& & & 
	\\
	\hline \hline
    & {\it quantum physics} & {\it noncommutative algebra} & \ref{MotivicQuantizationApplications}
    \\
	\hline\hline
	1920s & {quantum mechanics} & {operator algebra} & 
	\\
	1960s & {local observables}  & {co-sheaf theory} & 
    \\
    1990s-2000s & {local field theory}
    & {$(\infty,n)$-category theory} & 
	\\
	\hline
\end{tabular}

\vspace{.5cm}

\noindent These are traditional solutions to aspects of Hilbert's sixth  problem. 
Two points are noteworthy: on the one hand the items in the list involve 
crown jewels of mathematics; on the other hand their appearance is somewhat 
unconnected and remains piecemeal.

\newpage

Towards the end of the 20th century, William Lawvere, the founder
of categorical logic and of categorical algebra, aimed for a more
encompassing answer that rests the axiomatization of physics on a decent
unified foundation. He suggested to 
\begin{enumerate}
  \item rest the foundations of mathematics itself in topos theory
     \cite{Lawvere65};
  \item build the foundations of physics \emph{synthetically} inside topos theory by
  \begin{enumerate}
    \item 
  imposing properties on a topos which ensure that the
  objects  have the
  structure of \emph{differential geometric spaces}
  \cite{Lawvere98};
  \item formalizing classical mechanics on this basis by 
        universal constructions
        \\
  (``Categorical dynamics'' \cite{Lawvere67},  
   ``Toposes of laws of motion'' \cite{LawvereSynth}).
\end{enumerate}
\end{enumerate}
\noindent While this is a grandiose plan, we have to note that it falls short in two respects:
\begin{enumerate}
  \item Modern mathematics naturally wants foundations not in 
  topos theory, but in \emph{higher topos theory}
  (\cite{Lurie, HoTT}).
  \item Modern physics needs to refine classical mechanics to
  \emph{quantum mechanics} and \emph{quantum field theory}
  at small length scales/high energy scales (e.g. \cite{Feynman85, SatiSchreiber}).
\end{enumerate}
This book is about
refining Lawvere's synthetic approach on Hilbert's sixth
problem from classical physics formalized in synthetic differential geometry
axiomatized in topos theory 
to
high energy physics formalized in higher differential geometriy 
axiomatized in higher topos theory. 
Moreover, following Hilbert's problem description to consider
``all logically possible theories'', we consider also string theoretic 
physics \cite{DeligneMorgan, PolchinskiBook}.

The central claim for which we accumulate evidence is this:
\begin{claim}
  In cohesive $\infty$-topos theory-foundations 
  fundamental physics 
  is synthetically axiomatized 
  \begin{enumerate}
   \item naturally -- the axioms are simple, elegant and meaningful;
   \item faithfully -- the axioms capture the deep nontrivial phenomena.
  \end{enumerate}
\end{claim}

\medskip

We now give more detailed motivation for this development (see also \cite{sQFT}). 
\begin{itemize}
\item In \ref{MotivationByGaugeTheory} we observe that what in physics is 
called the \emph{gauge principle} is already what in mathematics is the 
principle of \emph{homotopy theory}.

\item While the gauge principle is widely hailed as the central principle of
modern physics, many discussions of mathematical physics in fact restrict attention
to \emph{perturbation theory} where only the infinitesimal
approximation, the tangents to global physical effects are retained.
From this perspective the true global effects of physics appear as
if a departure from the normal and are accordingly called \emph{anomalies}. 
While this is a useful perspective for many computations and often a good 
approximations to a complicated reality, a fundamental formulation
of fundamental physics must instead take all these global effects naturally into
account from the start. For practical matters this is the central reason 
for the application of higher differential geometry and differential cohomology
applied to physics: it is the theory of global effects and of anomaly cancellation
in field theory. This we come to in  \ref{MotivationFromAnomalyCancellation}.

\item When one explores fundamental physics it is noteworthy that 
among all the many possible kinds of action functionals and hence of theories
of physics, some appears more ``naturally'' than others. We observe that 
in higher differential geometry the natural action functionals are induced
from the theory of characteristic classes, primary, secondary, tertiary, etc.
and their differential refinement, hence from higher Chern-Weil theory.
This we consider in \ref{MotivationFromActionFunctionals}.

\item Finally, for the inclined reader,  we offer a more philosophical motivation, 
in \ref{MotivationFromHigherToposTheory}.
\end{itemize}

\subsubsection{The gauge principle and geometric homotopy theory}
\index{gauge theory!overview}
\label{MotivationByGaugeTheory}

Modern physics rests on a few basic principles, 
among them the \emph{gauge principle} (e.g. \cite{Gross92})
and the \emph{principle of locality}. We indicate here how these two principles imply that
spaces of physical fields are \emph{higher moduli stacks}
given by \emph{cohesive $\infty$-groupoids}.

\medskip

We start with the general statement 
and then look at its incarnations following
the history of physics.

\paragraph*{General discussion}
\label{GaugePrincipleGeneralDiscussion}

The gauge principle says that configurations of physical fields may be equivalent without
being equal, there may be \emph{gauge transformations} which turn
one field configuration into a gauge equivalent one. In
mathematical terms this means that fields in physics 
do not quite form a set, but that \emph{fields form a groupoid}
or equivalently that \emph{fields form a homotopy 1-type}. 
In addition, by the variational principle of extremal action, the groupoid
of field configurations is equipped with a differentiable (smooth) structure. In mathematical
terms this says that \emph{fields form a Lie groupoid} or more accurately that
given a field theory there is a \emph{moduli stack of fields} or
a \emph{geometric homotopy 1-type} of fields.

In much of the literature this moduli stack is considered (only) in its
first order infinitesimal approximation, its associated Lie algebroid.
This, or rather its algebra of functions, is famous in the physics literature
as the \emph{BRST complex} of a gauge field theory \cite{HenneauxTeitelboim}.
The cotangents to the gauge transformations, hence to the morphisms in the 
Lie groupoid of fields, are known as the \emph{ghost fields}.

The gauge principle applies also to gauge transformations themselves: there
are field theories where it is wrong to say that two gauge transformation between
two given field configurations are equal or not, and where instead one needs
to ask whether they are related by a \emph{gauge-of-gauge equivalence}. 
The same applies ad infinitum, to ever higher order gauge transformations.
Mathematically this means that field configurations may not form just a groupoid,
but what is called an \emph{$\infty$-groupoid} or \emph{homotopy type}.
Equipped with the smooth structure on the space of fields which is necessary for
speaking about infinitesimal variations of fields, this says that there is 
a \emph{moduli $\infty$-stack} of fields or a \emph{geometric homotopy type}
of fields. 

In much of the literature, again, this higher stack of fields
is considered (only) in its first order infinitesimal approximation,
its associated $L_\infty$-algebroid. This, or rather its algebra of functions,
is famous in the physics literature as the \emph{BV-BRST complex} (see e.g. \cite{Paugam}).
Here the co-tangents to the $k$-th order gauge-of-gauge transformations
(the $k$-morphisms in the $\infty$-groupoid of fields) are known as the 
$k$th order ``ghost-of-ghost fields''. 

In conclusion, the \emph{gauge principle} in physics means
that the collection of physical fields in a field theory is 
what mathematically is called
a \emph{higher moduli stack} or \emph{geometric homotopy type}
or what we will refer to here as a \emph{cohesive $\infty$-groupoid}.

\medskip

Evident as this is, it is not much amplified in traditional physics textbooks.
This has two causes. One is that many discussions of field theory rest on 
the approximation of perturbation theory, 
where many subtleties go away, and where discussion of global effects is
often postponed to the extent
that they are either forgotten or left to the esoteric-seeming literature on 
``anomalies''. Another cause is that often the nature of the gauge principle
is actively misunderstood: often one sees texts claiming that gauge invariance is
just a ``redundancy'' in the description of a physics, insinuating that one might 
just as well pass to the set of gauge equivalence classes.
And this is not true: passing to gauge equivalence classes leads to violation of the
other principle of modern physics, the principle of locality.
For reconstructing non-trivial global gauge field configurations
(often known as ``instantons'' in the physics literature) from local data,
it is crucial to retain all the information about the gauge equivalences,
for it is the way in which these serve to glue local gauge field data
to global data that determines the global field content. This ``gluing''
is what in mathematics is often referred to as ``descent'' and it is the 
hallmark of higher stack theory/higher topos theory. A higher stack is a local assignment of 
$\infty$-groupoids/homotopy types which satisfies descent, hence which 
globally glues. 

\medskip

Therefore the gauge principle and the principle of locality 
combines means that spaces of fields in physics form higher geometric stacks.

\paragraph*{Instances of gauge theory}
\label{GaugePrincipleInstances}

Around 1850 Maxwell realized that the field strength of the electromagnetic 
field is modeled  by what today we call a closed differential 2-form on spacetime. 
In the 1930s Dirac observed that in the presence of 
electrically charged quantum particles such as electrons, 
this 2-form is more precisely the \emph{curvature} 2-form of a 
\emph{$U(1)$-principal bundle with connection}.  

In modern terms this, in turn, means equivalently that the
electromagnetic field is modeled by a degree 2-cocycle in 
(ordinary) \emph{differential cohomology}.
This is a differential refinement of 
the degree-2 integral cohomology that classifies the 
underlying $U(1)$-principal bundles themselves via what 
mathematically is their \emph{Chern class} and what physically 
is the topological
\emph{magnetic charge}. A coboundary in 
degree-2 differential cohomology
is, mathematically, 
a smooth isomorphism of bundles with connection, hence,
physically, is a \emph{gauge transformation} between field configurations.
Therefore classes in differential cohomology characterize
the \emph{gauge-invariant} information encoded in gauge field
configurations, such as the electromagnetic field.

Meanwhile, in 1915, Einstein had identified also the field strength of the field of gravity as the 
$\mathfrak{so}(d,1)$-valued curvature 2-form of the canonical $O(d,1)$-principal bundle with 
connection on a $d+1$-dimensional spacetime Lorentzian manifold. 
This is a cocycle in differential \emph{nonabelian} cohomology: in Chern-Weil theory.

In the 1950s Yang-Mills-theory identified the field strength of all the gauge fields 
in the standard model of particle physics as the $\mathfrak{u}(n)$-valued curvature 2-forms 
of $U(n)$-principal bundles with connection. This is again a cocycle in 
differential nonabelian cohomology.

\begin{center}
\fbox{
  \begin{minipage}{12cm}
 {\bf Entities of ordinary gauge theory}
 \\
 Lie algebra $\mathfrak{g}$ with gauge Lie group $G$ ---
 connection with values in $\mathfrak{g}$ on $G$-principal bundle over a smooth manifold $X$
   \end{minipage}
}
\end{center}

It is noteworthy that already in this mathematical formulation of experimentally well-confirmed fundamental physics the seed of higher differential cohomology is hidden: 
Dirac had not only identified the electromagnetic field as a line bundle with connection, but he also correctly identified (rephrased in modern language) its underlying cohomological Chern class with the (physically hypothetical but formally inevitable) magnetic charge located in spacetime. But in order to make sense of this, he had to resort to removing the support of the magnetic 
charge density from the spacetime manifold, because Maxwell's equations imply that at the support 
of any magnetic charge the 2-form representing the field strength of the electromagnetic field is in fact 
not closed and hence in particular not the curvature 2-form of an ordinary connection on
an ordinary bundle.

In \cite{Freed} Dirac's old argument was improved by refining the model for the 
electromagentic field one more step: Freed observes that the charge current 3-form is itself 
to be regarded as a curvature, but for a connection on a circle 2-bundle with connection -- 
also called a bundle gerbe --, which is a cocycle in degree-3 ordinary differential cohomology. 
Accordingly, the electromagnetic field is fundamentally not quite a line bundle, 
but a \emph{twisted bundle} with connection, with the twist being the magnetic charge 3-cocycle. 
Freed shows that this perspective is inevitable for understanding the quantum anomaly of the 
action functional for electromagnetism is the presence of magnetic charge.

In summary, the experimentally verified models, to date, of fundamental physics are based on 
the notion of (twisted) $U(n)$-principal bundles with connection for the Yang-Mills field 
and $O(d,1)$-principal bundles with connection for the description of gravity, 
hence on nonabelian differential cohomology in degree 2 (possibly with a degree-3 twist).

In attempts to better understand the structure of these two theories and their interrelation, 
theoretical physicists were led to consider variations and generalizations of them that are 
known as \emph{supergravity} and \emph{string theory} \cite{DeligneMorgan}. 
In these theories the notion of gauge field 
turns out to generalize: instead of just Lie algebras, Lie groups and 1-form connections with values in these, 
one finds structures called \emph{Lie 2-algebras}, \emph{Lie 2-groups} and the gauge fields themselves 
are given by differential 2-forms values in these.

\begin{center}
\fbox{
  \begin{minipage}{12cm}
 {\bf Entities of 2-gauge theory}
 \\
 Lie 2-algebra $\mathfrak{g}$ with gauge Lie 2-group $G$ ---
 connection with values in $\mathfrak{g}$ on a $G$-principal 2-bundle/gerbe 
 over an orbifold $X$
   \end{minipage}
}
\end{center}

Notably the fundamental string is charged under a field called the \emph{Kalb-Ramond field} or 
\emph{$B$-field} which is modeled by a $\mathbf{B}U(1)$-principal 2-bundle with connection, 
where $\mathbf{B}U(1)$ is the Lie 2-group delooping of the circle group: 
the circle Lie 2-group. Its Lie 2-algebra $\mathbf{B}\mathfrak{u}(1)$ is given by the differential crossed module 
$[\mathfrak{u}(1) \to 0]$ which has $\mathfrak{u}(1)$ shifted up by one in homological degree.

So far all these differential cocycles were known and understood mostly as concrete constructs, 
without making their abstract home in differential cohomology explicit. It is the next gauge field that 
made Freed and Hopkins propose \cite{FreedHopkins} that the theory of differential cohomology is 
generally the formalism that models gauge fields in physics:

The superstring is charged also under what is called the \emph{RR-field}, a gauge field 
modeled by cocycles in differential K-theory. In even degrees we may think of this as a 
differential cocycle whose curvature form has coefficients in the 
$L_\infty$-algebra $\oplus_{n \in \mathbb{N}} \mathbf{B}^{2n} \mathfrak{u}(1)$. 
Here $\mathbf{B}^{2n} \mathfrak{u}(1)$ is the abelian $2n$-Lie algebra 
whose underlying complex is concentrated in degree $2n$ on $\mathbb{R}$.
So fully generally, one finds \emph{$\infty$-Lie algebras}, 
\emph{$\infty$-Lie groups} and 
gauge fields modeled by connections with values in these.

\begin{center}
\fbox{
  \begin{minipage}{12cm}
 {\bf Entities of general gauge theory}
 \\
 $\infty$-Lie algebra $\mathfrak{g}$ with gauge $\infty$-Lie group $G$ ---
 connection with values in $\mathfrak{g}$ on a $G$-principal $\infty$-bundle 
 over a smooth $\infty$-groupoid $X$
   \end{minipage}
}
\end{center}

Apart from generalizing the notion of gauge Lie groups to Lie 2-groups and further, 
structural considerations in fundamental physics also led theoretical physicists to consider 
models for spacetime that are more general than the notion of a smooth manifold. 
In string theory spacetime is allowed to be more generally an orbifold or a 
generalization thereof, such as an orientifold. The natural mathematical model 
for these generalized spaces are Lie groupoids or, essentially equivalently,
\emph{differentiable stacks}.

It is noteworthy that the notions of generalized gauge groups and the generalized spacetime 
models encountered this way have a natural common context: all of these are examples of
\emph{smooth $\infty$-groupoids}.
There is a natural mathematical concept that serves to describe contexts of such generalized spaces: 
a \emph{cohesive $\infty$-topos}. The notion of 
\emph{differential cohomology in a cohesive $\infty$-topos} provides 
a unifying perspective on the mathematical structure encoding the generalized gauge fields and generalized spacetime models encountered in modern theoretical physics in such a general context.

\subsubsection{Global effects and anomaly cancellation}
\label{MotivationFromAnomalyCancellation}
\index{anomaly cancellation!in introduction}

One may wonder to which extent the higher gauge fields,
that above in \ref{MotivationByGaugeTheory} we said
motivate the theory of higher differential cohomology,
can themselves be motivated within physics. It turns out that 
an important class of examples is 
required already by consistency of the quantum mechanics of 
higher dimensional fermionic (``spinning'')
quantum objects. 

We indicate now how the full description of this
\emph{quantum anomaly cancellation} forces one
to go beyond classical Chern-Weil theory to a 
more comprehensive theory of higher differential cohomology.

\medskip

Consider a smooth manifold $X$. Its tangent bundle $T X$
is a real vector bundle of rank $n = \mathrm{dim} X$.
By the classical theorem which identifies isomorphism classes of
rank-$n$ real vector bundles with homotopy classes of \emph{continuous}
maps to 
the classifying space $B \mathrm{O}(n)$, for $O(n)$ the orthogonal group, 
$$
  \mathrm{VectBund}(X)/_\sim \simeq [X, B O]
  \,,
$$
we have that $T X$ is classified by a continuous map
which we shall denote by the same symbol
$$
  T X : X \to B \mathrm{O}(n)
  \,.
$$ 
Notice that this map takes place after passing
from smooth spaces to just topological spaces. A central theme
of our discussion later on are first \emph{smooth} and then
\emph{differential} refinements of such maps.

A standard question to inquire about $X$ is whether it is orientable.
If so, a \emph{choice} of orienation is, in terms of this classifying map,
given by a lift through the canonical map $B \mathrm{SO}(n) \to B \mathrm{O}(n)$
from the classifying space of the \emph{special} orthogonal group.
Further, we may ask if $X$ admits a 
\emph{$\mathrm{Spin}$-structure}. If so, a choice of Spin-structure corresponds
to a further lift through the canonical map 
$B \mathrm{Spin}(n) \to B \mathrm{O}(n)$ from the classifying space
of the $\mathrm{Spin}$-group, which is the universal simply connected
cover of the special orthogonal group. (Details on these basic notions are
reviewed at the beginning of \ref{Applications} below.)

These lifts of structure groups are just the first 
steps through a whole tower of higher group extensions, called the
\emph{Whitehead tower} of $B \mathrm{O}(n)$, as shown in the following
picture. Here $\mathrm{String}$ is a \emph{topological group} which is the
universal 3-connected cover of $\mathrm{Spin}$, and then 
$\mathrm{Fivebrane}$ is the universal 7-connected cover of $\mathrm{String}$.
$$
  \xymatrix{    
    && B \mathrm{Fivebrane}
	\ar[d]
	&&
	\mbox{fivebrane structure}
    \\
    && B \mathrm{String}
	\ar[r]^{\frac{1}{6}p_2}
	\ar[d]
	&
	K(\mathbb{Z},8)
	&
	\mbox{string structure}
    \\
    && B \mathrm{Spin}
	\ar[r]^{\frac{1}{2}p_1}
	\ar[d]
	&
	K(\mathbb{Z},4)
	&
	\mbox{spin structure}
    \\
    &
	&
	B \mathrm{SO}
	\ar[r]^{w_2}
	\ar[d]
	&
	K(\mathbb{Z}_2,2)
	&
	\mbox{orientation structure}
	\\
	\Sigma 
	\ar[r]^{\phi}
	&
    X 
	\ar[r]^{T X}
	\ar@/^.3pc/@{-->}[ur]
	\ar@/^.4pc/@{-->}[uur]
	\ar@/^.6pc/@{-->}[uuur]
	\ar@/^.8pc/@{-->}[uuuur]
    &
	B \mathrm{O}
	\ar[r]^{w_1}
	&
	K(\mathbb{Z}_2,1)
	&
	\mbox{Riemannian structure}
  }
  \,.
$$
Here all subdiagrams of the form
$$
  \xymatrix{
    B \hat G
	\ar[d]
	\\
	B G
	\ar[r]^{c}
	&
	K(A,n)
  }
$$
are homotopy fiber sequences. This means that $B \hat G$ is the homotopy fiber
of the characteristic map $c$ and $\hat G$ itself is the homotopy fiber
of the looping $\Omega c$ of $c$. By the universal property of the
homotopy pullback, this implies the obstruction theory for the existence of
these lifts. The first two of these are classical.
For instance the orientation structure exists if the
\emph{first Stiefel-Whitney class} $[w_1(T X)] \in H^1(X, \mathbb{Z}_2)$ is trivial.
Then a Spin-structure exists if moreover the 
\emph{second Stiefel-Whitney class} $[w_2(T X)] \in H^2(X, \mathbb{Z}_2)$
is trivial. 

Analogously, a \emph{string structure} exists on $X$ if moreover 
the \emph{first fractional Pontryagin class} 
$[\frac{1}{2}p_1(T X)] \in H^4(X, \mathbb{Z})$ is trivial, and 
if so, a \emph{fivebrane structure} exists if moreover the
\emph{second fractional Pontryagin class} 
$[\frac{1}{6}p_2(T X)] \in H^8(X, \mathbb{Z})$ is trivial.

The names of these structures indicate their role in 
quantum physics. Let $\Sigma$ be a $d+1$-dimensional manifold
and assume now that also $X$ is smooth. Then a smooth map
$\phi : \Sigma \to X$ may be thought of as modelling the 
trajectory of a $d$-dimensional object propagating 
through $X$. For instance for $d = 0$ this would be the
trajectory of a point particle, for $d = 1$ it would be 
the worldsheet of a \emph{string}, and for $d = 5$ the
6-dimensional worldvolume of a \emph{5-brane}. 
The intrinsic ``spin''
of point particles and their higher dimensional analogs is
described by a spinor bundle $S \to \Sigma$ equipped 
for each $\phi : \Sigma \to X$ with a 
Dirac operator  $D_{\phi^* T X}$ that is twisted by 
the pullback of the tangent bundle of $X$ along $\phi$.
The fermionic part of the \emph{path integral}
that gives the quantum dynamics of this setup computes
the analog of the determinant of this Dirac operator,
which is an element in a complex line called the
\emph{Pfaffian line} of $D_{\phi^* T X}$.
As $\phi$ varies, these Pfaffian lines arrange into
a line bundle on the mapping space
$$
  \xymatrix{
    \left\{
	   \mathrm{Pfaff}(D_{\phi^* T X})
	\right\}
	\ar[d]
	\\
	\left\{
	  \phi : \Sigma \to X
	\right\}
	\ar@{=}[r]
	&
	\mathrm{SmthMaps}(\Sigma,X)
	\ar[r]^>>>>>{\mathrm{tg}_\Sigma (c)}
	& 
	K(\mathbb{Z}, 2)
  }
  \,.
$$
Since the result of the fermionic part of the path integral
is therefore a section of this line bundle, the resulting
effective action functional can be a well defined function
only if this line bundle is trivializable, hence if its
Chern class vanishes. Therefore the Chern class of the Pfaffian
line bundle over the bosonic configuration space is called the
\emph{global quantum anomaly} of the system. It is an obstruction 
to the existence of quantum dynamics of $d$-dimensional objects
with spin on $X$. 

Now, it turns out that this Chern class is 
the \emph{transgression} $\mathrm{tg}_\Sigma (c)$ 
of the corresponding class $c$ appearing in the 
picture of the Whitehead tower above. Therefore the vanishing of these
classes implies the vanishing of the quantum anomaly.

For instance a choice of a \emph{spin structure} on $X$ 
cancels the global quantum anomaly of the quantum spinning  particle.
Then a choice of \emph{string structure} cancels the global
quantum anomaly of the quantum spinning string, and a 
choice of \emph{fivebrane structure} cancels the global quantum
anomaly of the quantum spinning 5-brane.

However, the Pfaffian line bundle turns out to be canonically
equipped with more refined differential structure: it carries
a \emph{connection}. Moreover, in order to 
obtain  a consistent quantum theory
it needs to be trivialized as a bundle with connection. 

For the Pfaffian line bundle with connection still to be the 
transgression of the corresponding obstruction class on $X$,
evidently the entire story so far needs to be refined from 
cohomology to a differentially refined notion of cohomology.

Classical Chern-Weil theory achieves this, in parts, for the
first few steps through the Whitehead tower
(see \cite{GHV} for a classical textbook reference and
\cite{HopkinsSinger} for the refinement to differential cohomology
that we need here). 
\index{Chern-Weil theory}
For instance, since maps $X \to B \mathrm{Spin}$ classify
$\mathrm{Spin}$-principal bundles on $X$, and since 
$\mathrm{Spin}$ is a Lie group, it is clear that the 
corresponding differential refinement is given by 
$\mathrm{Spin}$-principal connections. Write
$H^1(X, \mathrm{Spin})_{\mathrm{conn}}$ for the equivalence
classes of these structures on $X$. 

For every $n \in \mathbb{N}$ there is a notion of differential refinement
of $H^n(X, \mathbb{Z})$ to the \emph{differential cohomology group}
$H^n(X, \mathbb{Z})_{\mathrm{conn}}$. These groups fit into 
square diagrams as indicated on the right of the following 
diagram.
$$
 \xymatrix@C=4pt{
    H^1_{\mathrm{conn}}(X, \mathrm{Spin})
	\ar[rrrr]^{[\frac{1}{2}\hat {\mathbf{p}_1}]}
	&&&&
	H^4_{\mathrm{diff}}(X, \mathbb{Z})	
	\ar[dl]|{\mathrm{curvature}}
	\ar[dr]|{\mathrm{top}.\; \mathrm{class}}
	\\
	\,&\,&\,& \Omega^4_{\mathrm{cl}}(X) 
	\ar[dr] && 
	H^4(X, \mathbb{Z})
	\ar[dl]
	\\
	&&&& 
	H^4_{\mathrm{dR}}(X) \simeq H^4(X, \mathbb{R})
  }
  \,.
$$
As shown there, an element in $H^{n}_{\mathrm{diff}}(X, \mathbb{Z})$
involves an underlying ordinary integral class, but also 
a differential $n$-form on $X$ such that both structures
represent the same class in real cohomology (using the de Rham isomorphism
between real cohomology and de Rham cohomology). The differential
form here is to be thought of as a \emph{higher curvature form}
on a higher line bundle corresponding to the given integral
cohomology class.

Finally, the refined form of classical Chern-Weil theory provides
differential refinements for instance of the first fractional 
Pontryagin class $[\frac{1}{2}p_1] \in H^4(X, \mathbb{Z})$
to a differential class $[\frac{1}{2}\hat {\mathbf{p}}_1]$
as shown in the above diagram. This is the differential
refinement that under transgression produces the differential
refinement of our Pfaffian line bundles.

But this classical theory has two problems.
\begin{enumerate}
  \item Beyond the $\mathrm{Spin}$-group, the topological groups
  $\mathrm{String}$, $\mathrm{Fivebrane}$ etc. do not admit the
  structure of finite-dimensional Lie groups anymore, hence 
  ordinary Chern-Weil theory fails to apply.
  \item
   Even in the situation where it does apply, 
   ordinary Chern-Weil theory only works on 
   cohomology classes, not on cocycles. Therefore 
   the differential refinements cannot
   see the homotopy fiber sequences anymore, that crucially
   characterized the obstruction problem of lifting through the
   Whitehead tower.
\end{enumerate}

The source of the first problem may be thought to be 
the evident fact that the category
$\mathrm{Top}$ of topological spaces does not encode smooth structure. 
But the problem goes deeper, even. 
In homotopy theory, $\mathrm{Top}$ is not even about
topological structure. Rather, it is about homotopies and
\emph{discrete} geometric structure.

One way to make this precise is to say that there is a
\emph{Quillen equivalence} between the model category structures
on topological spaces and on simplicial sets.
$$
  \xymatrix{
   \mathrm{Top}
   \ar@{<-}@<+3pt>[r]^{\vert - \vert}
   \ar@<-3pt>[r]_{\mathrm{Sing}}
   &
   \mathrm{sSet}
 }
  \;\;\;\;
  \mathrm{Ho}(\mathrm{Top})
  \simeq
  \mathrm{Ho}(\mathrm{sSet})
  \,.
$$
Here the \emph{singular simplicial complex functor} 
$\mathrm{Sing}$ sends a topological space to the simplicial set
whose $k$-cells are maps from the topological $k$-simplex into $X$.

In more abstract modern language we may restate this as saying that there
is an equivalence
$$
  \xymatrix{
    \mathrm{Top}
    \ar[r]^{\Pi}_{\simeq}
    &
    \infty \mathrm{Grpd}
  }
$$
between the homotopy theory of topological spaces and that of
\emph{$\infty$-groupoids}, exhibited by forming the 
\emph{fundamental $\infty$-groupoid} of $X$.

To break this down into a more basic statement, let 
$\mathrm{Top}_{\leq 1}$ be  the subcategory of
homotopy 1-types, hence of these topological spaces
for which only the 0th and the first homotopy groups may be
nontrivial. Then the above equivalence resticts to an 
equivalence
$$
  \xymatrix{
    \mathrm{Top}_{\leq 1}
    \ar[r]^\Pi_{\simeq}
	&
	\mathrm{Grpd}
  }
$$
with ordinary groupoids. Restricting this even further to 
(pointed) connected 1-types, hence spaces for which only the
first homotopy group may be non-trivial, we obtain an equivalence
$$
  \xymatrix{
    \mathrm{Top}_{1, \mathrm{pt}}
    \ar[r]^{\pi_1}_\simeq
	&
	\mathrm{Grp}
  }
$$
with the category of groups. Under this equivalence a 
connected 1-type topological space is simply identified with
its first fundamental group.

Manifestly, the groups on the right here are just bare groups
with no geometric structure; or rather with \emph{discrete} geometric
structure. Therefore, since the morphism $\Pi$ is an equivalence, also
$\mathrm{Top}_1$ is about \emph{discrete} groups, $\mathrm{Top}_{\leq 1}$
is about \emph{discrete} groupoids and $\mathrm{Top}$ is about
\emph{discrete $\infty$-groupoids}.

\medskip

There is a natural solution to this problem. This solution and
the differential cohomology theory that it supports is the topic
of this book.

The solution is to equip discrete $\infty$-groupoids $A$ with 
\emph{smooth structure} by equipping them with
information about what the \emph{smooth families} of $k$-morphisms in it are.
In other words, to assign to each smooth parameter space $U$ an 
$\infty$-groupoid of smoothly $U$-parameterized families of cells in 
$A$.

If we write $\mathbf{A}$ for $A$ equipped with smooth structure, 
this means that we have an assignment
$$
  \mathbf{A} : U \mapsto \mathbf{A}(U) =: 
  \mathrm{Maps}(U,A)_{\mathrm{smooth}}
  \in
  \infty \mathrm{Grpd}
$$
such that $\mathbf{A}(*) = A$.

Notice that here the notion of smooth
maps into $A$ is not defined before we declare $\mathbf{A}$,
rather it is defined \emph{by} declaring $\mathbf{A}$. A more detailed
discussion of this idea is below in \ref{IntroToposes}.

We can then define the homotopy theory of \emph{smooth}
$\infty$-groupoids by writing
$$
  \mathrm{Smooth}\infty \mathrm{Grpd}
  :=
  L_W
  \mathrm{Funct}(\mathrm{SmoothMfd}^{\mathrm{op}}, 
  \mathrm{sSet})
  \,.
$$
Here on the right we have the category of contravariant functors
on the category of smooth manifolds, such as the $\mathbf{A}$ from above.
In order for this to inform this simple construction about the 
local nature of smoothness, we need to formally invert some 
of the morphisms between such functors, which is indicated 
by the symbol $L_W$ on the left. The set of morphisms $W$ that 
are to be inverted are those natural transformation that are
\emph{stalkwise} weak homotopy equivalences of simplicial sets.

We find that there is a canonical notion of 
\emph{geometric realization} on smooth $\infty$-groupoids
$$
  \vert - \vert
  :
  \mathrm{Smooth}\infty\mathrm{Grpd}
  \stackrel{\Pi}{\to}
  \infty \mathrm{Grpd}
  \stackrel{|-|}{\to}
  \mathrm{Top}
  \,,
$$
where $\Pi$ is the derived left adjoint to 
the embedding 
$$
  \mathrm{Disc} : \infty \mathrm{Grpd}
    \hookrightarrow 
	\mathrm{Smooth}\infty \mathrm{Grpd}
$$
of bare $\infty$-groupoids as discrete smooth
$\infty$-groupoids. We may therefore ask for
\emph{smooth refinements} of given topological spaces
$X$, by asking for smooth $\infty$-groupoids $\mathbf{X}$
such that $\vert \mathbf{X} \vert \simeq X$.

A simple example is obtained from any 
Lie algebra $\mathfrak{g}$. Consider the functor
$\exp(\mathfrak{g}) : \mathrm{SmoothMfd}^{\mathrm{op}} \to \mathrm{sSet}$
given by the assignment
$$
  \exp(\mathfrak{g}) : U \mapsto ([k] \mapsto 
  \Omega^1_{\mathrm{flat}, \mathrm{vert}}{U \times \Delta^k}, \mathfrak{g})
  \,,
$$
where on the right we have the set of differential forms on the
parameter space times the smooth $k$-simplex which are flat
and vertical with respect to the projection $U \times \Delta^k \to U$.

We find that the 1-truncation of this smooth $\infty$-groupoid
is the Lie groupoid
$$
  \tau_1 \exp(\mathfrak{g}) = \mathbf{B}G
$$
that has a single object and whose morphisms form the simply
connected Lie group $G$ that integrates $\mathfrak{g}$.
We may think of this Lie groupoid also as the 
\emph{moduli stack} of smooth $G$-principal bundles.
In particular, this is a smooth refinement of the classifying space for
$G$-principal bundles in that
$$
  \vert \mathbf{B}G \vert \simeq B G 
  \,.
$$
So far this is essentially what classical Chern-Weil theory can already see.
But smooth $\infty$-groupoids now go much further. 

In the next step there is a \emph{Lie 2-algebra}
$\mathfrak{g} = \mathfrak{string}$ such that its exponentiation
$$
  \tau_2 \exp(\mathfrak{string}) = \mathbf{B}\mathrm{String}
$$
is a smooth 2-groupoid, which we may think of as the
\emph{moduli 2-stack of $\mathrm{String}$-principal }  
which is a smooth refinement of the 
$\mathrm{String}$-classifying space 
$$
  \vert \mathbf{B} \mathrm{String} \vert 
    \simeq 
  B \mathrm{String} 
  \,.
$$
Next there is a Lie 6-algebra $\mathfrak{fivebrane}$
such that
$$
  \tau_6 \exp(\mathfrak{fivebrane}) = \mathbf{B}\mathrm{Fivebrane}
$$
with 
$$
  \vert \mathbf{B} \mathrm{Fivebrane} \vert \simeq B \mathrm{Fivebrane} 
  \,.
$$

Moreover, the characteristic maps that we have seen now refine first to 
smooth maps on these moduli stacks, for instance
$$
  \frac{1}{2}\mathbf{p}_1 : \mathbf{B} \mathrm{Spin}
  \to \mathbf{B}^3 U(1)
  \,,
$$
and then further to \emph{differential} refinement of these maps
$$
  \frac{1}{2} \hat{\mathbf{p}}_1 : 
  \mathbf{B} \mathrm{Spin}_{\mathrm{conn}}
  \to 
   \mathbf{B}^3 U(1)_{\mathrm{conn}}
  \,,
$$
where now on the left we have the moduli stack of smooth 
$\mathrm{Spin}$-connections, and on the right the moduli 3-stack
of \emph{circle $n$-bundles with connection}.

A detailed discussion of these constructions is below in 
\ref{FractionalClasses}.

In addition to capturing smooth and differential refinements, these constructions 
have the property that they work not just at the level of cohomology classes,
but at the level of the full cocycle $\infty$-groupoids. 
For instance for $X$ a smooth manifold, postcomposition with $\frac{1}{2}\hat {\mathbf{p}}$
may be regarded not only as inducing a function 
$$
  H^1_{\mathrm{conn}}(X,\mathrm{Spin})
  \to 
  H^4_{\mathrm{conn}}(X)
$$
on cohomology sets, but a morphism
$$
  \frac{1}{2}\hat {\mathbf{p}}(X)
   : 
  \mathbf{H}^1(X,\mathrm{Spin})
  \to 
  \mathbf{H}^3(X, \mathbf{B}^3 U(1)_{\mathrm{conn}})
$$
from the groupoid of smooth principal $\mathrm{Spin}$-bundles with connection
to the 3-groupoid of smooth circle 3-bundles with connection. Here the 
boldface $\mathbf{H} =\mathrm{Smooth}\infty \mathrm{Grpd}$ denotes the ambient 
$\infty$-topos of smooth $\infty$-groupoids and $\mathbf{H}(-,-)$ its hom-functor.

By this refinement to cocycle $\infty$-groupoids we have access to the homotopy fibers 
of the morphism $\frac{1}{2}{\hat {\mathbf{p}}}_1$. Before differential refinement 
the homotopy fiber
$$
  \xymatrix{
    \mathbf{H}(X, \mathbf{B}\mathrm{String})
    \ar[r]
     &
    \mathbf{H}(X, \mathbf{B}\mathrm{Spin}) \ar[r]^{\frac{1}{2}\mathbf{p}_1} & \mathbf{H}(X,B^3 U(1))      
  }
  \,,
$$
is the 2-groupoid 
of smooth $\mathrm{String}$-principal 2-bundles on $X$: smooth \emph{string structures} on $X$.
As we pass to the differential refinement, we obtain \emph{differential string structures} on $X$
$$
  \xymatrix{
    \mathbf{H}(X, \mathbf{B}\mathrm{String}_{\mathrm{conn}})
    \ar[r]
     &
    \mathbf{H}(X, \mathbf{B}\mathrm{Spin}_{\mathrm{conn}}) \ar[r]^{\frac{1}{2}{\hat {\mathbf{p}}}_1} & \mathbf{H}(X,B^3 U(1)_{\mathrm{conn}})      
  }
  \,.
$$
A cocycle in the 2-groupoid $\mathbf{H}(X, \mathbf{B}\mathrm{String}_{\mathrm{conn}})$is naturally identified with 
a tuple consisting of
\begin{itemize}
\item a smooth $\mathrm{Spin}$-principal bundle $P \to X$ with connection $\nabla$;

\item the Chern-Simons 2-gerbe with connection $CS(\nabla)$ induced by this;

\item a choice of trivialization of this Chern-Simons 2-gerbe and its connection.
\end{itemize}
We may think of this as a refinement of secondary characteristic classes: 
the first Pontryagin curvature characteristic form $\langle F_\nabla \wedge F_\nabla \rangle$ 
itself is constrained to vanish, and so the Chern-Simons form 3-connection itself constitutes cohomological data.

More generally, we have access not only to the homotopy fiber over the 0-cocycle, but may pick
one cocycle in each
cohomology class to a total morphism
$H_{\mathrm{diff}}^4(X) \to \mathbf{H}(X, \mathbf{B}^3 U(1)_{\mathrm{conn}})$
and consider the collection of all homotopy fibers over all connected components as the 
homotopy pullback
$$
  \xymatrix{
    \frac{1}{2}\hat {\mathbf{p}}_1 \mathrm{Struc}_{\mathrm{tw}}(X) \ar[r] \ar[d]
    &
    H_{\mathrm{diff}}^4(X) \ar[d]
    \\
    \mathbf{H}(X, \mathbf{B}\mathrm{Spin}_{\mathrm{conn}})
    \ar[r]^{\frac{1}{2}\hat {\mathbf{p}}_1}
    &
    \mathbf{H}(X, \mathbf{B}^3 U(1)_{\mathrm{conn}})    
  }
  \,.
$$
This yields the 2-groupoid of \emph{twisted differential string structure}. These objects,
and their higher analogs given by twisted differential fivebrane structures, appear in 
background field structure of the heterotic string and its magnetic dual, as discussed
in \cite{SSSIII}.

These are the kind of structures that $\infty$-Chern-Weil theory studies.

\subsubsection{Hierarchies of natural action functionals}
\label{MotivationFromActionFunctionals}

We present here a motivation for our constructions, starting from the 
observation that classical Chern-Weil theory induces action functionals
of Chern-Simons type, and observing that this phenomenon ought to have 
certain natural generalizations.

\medskip

First a brief word on the general context of quantum physics.

In recent years the notion of \emph{topological quantum field theory} (TQFT)
from physics 
has been fully formalized and made accessible to strong mathematical 
tools and classifications.
In its refined variant of \emph{fully local} or \emph{extended}
$n$-dimensional TQFT, the fundamental concept is that of a higher
category, denoted $\mathrm{Bord}_n$, whose $(k \leq n)$-cells are 
$k$-dimensional smooth manifolds with boundary and corners, and whose composition
operation is gluing along these boundaries. The disjoint union of manifolds
equips this with a symmetric monoidal structure. Then for another 
symmetric monoidal $n$-category $n\mathrm{Vect}_{\mathrm{fd}}$, whose $k$-cells one thinks
of as higher order linear maps between $n$-categorical analogs of finite dimensional
(or ``fully dualizable'') vector spaces,
an $n$-dimensional extended TQFT is formalized as an $n$-functor
$$
  Z : \mathrm{Bord}_n \to n \mathrm{Vect}
$$
that respects this monoidal structure. 

Here the higher order linear map $Z(\Sigma_{n-1})$ that is assigned to a \emph{closed} 
$(n-1)$-dimensional manifold $\Sigma_{n-1}$ can typically canonically be identified
with a vector space, and be interpreted as the \emph{space of states} of the 
physical system described by $Z$, for field configurations over a space of 
shape $\Sigma_{n-1}$. 
Then for $\Sigma_n$ a cobordism between two such closed $(n-1)$-manifolds, 
$Z(\Sigma_n)$ identifies with a linear map from the space of states over the
incoming to that over the outgoing boundary, and is interpreted as the 
(``time''-)\emph{propagation} of states.

This idea is by now classical. A survey can for instance be found in \cite{Kapustin10}.

But beyond constituting a formalization of 
some concept motivated from physics, it is remarkable that this 
construction is itself entirely rooted in a universal construction in 
higher category theory, and would
have eventually been discovered as such even in the absence of any motivation
from physics. The notion of extended TQFT \emph{derives} from higher category theory.

Namely, according to the celebrated 
result of \cite{LurieTFT}, earlier hypothesized in \cite{BaezDolan},
$\mathrm{Bord}_n$ is a \emph{free construction} -- essentially the 
\emph{free symmetric monoidal $n$-category} generated by just the point.
This means that symmetric monoidal maps
$Z : \mathrm{Bord}_n \to n\mathrm{Vect}_{\mathrm{fd}}$
are equivalently encoded by $n$-functors from the point $Z(*) : {*} \to n\mathrm{Vect}_{\mathrm{fd}}$,
which in turn are, of course, canonically identified simply with $n$-vector spaces,
the \emph{$n$-vector space of states} assigned by $Z$ to the point.
This adjunction is both, an intrinsic characterization of $\mathrm{Bord}_n$, as well
as a full \emph{classification} of extended TQFTs: these are entirely determined by 
their higher space of states. All the assignments on higher dimensional $\Sigma$ are
obtained by forming higher order \emph{traces} on this single higher space of states
over the point.

Here we will not further dwell on extended TQFT as such, but instead use
this state of affairs to motivate an investigation of a \emph{source of examples}
of \emph{natural} TQFTs.
Because the TQFTs that actually appear in fundamental physics, even  
when including the families of theories found in the study of theory space
away from the loci of experimentally observed theories, are far from being
random TQFTs allowed by the above classification. 

First of all, the TQFTs that do appear are
typically theories that arise by a process of \emph{quantization}
from a local \emph{action functional} on a space of field configurations (recalled below).
Secondly, even among all TQFTs arising by quantization from local action functionals
they are special, in that they have a natural formulation in differential geometry,
something that we will make precise below. The typical action functional appearing in practice
is not random, but follows some natural pattern.

One may therefore ask which principle it is that selects from a universal
construction in higher category theory -- that of free symmetric monoidal structure 
-- a certain subclass of ``natural'' geometric examples. We will provide evidence here that 
this is another universal construction, but now in \emph{higher topos theory}:
\emph{cohesion}.

Below in \ref{structures} (specifically in \ref{StrucChern-SimonsTheory} and 
\ref{StrucWZWFunctional}) we show that cohesion in an $\infty$-topos induces,
first, a notion of \emph{differential characteristic maps}, via a generalized
\emph{Chern-Weil theory}, and, second, from each such
the corresponding spaces -- in fact \emph{moduli $\infty$-stacks} -- 
of higher gauge field configurations, and, third,  
canonically equips these with action functionals, via a generalized higher
\emph{Chern-Simons theory}. Moreover, it induces from any such 
a corresponding action functional of one dimension lower, via
a generalized higher \emph{Wess-Zumino-Witten theory}. And finally 
the process of (geometric) quantization of these functionals on moduli stacks
is itself naturally induced in a cohesive context.

\paragraph*{Geometric quantization}
\index{symplectic higher geometry!geometric prequantization!in introduction}
\label{GeometricQuantizationInIntroduction}

For completeness, we briefly recall the basic ideas of \emph{quantization}
in its formalization known as \emph{geometric quantization} 
(which we disccuss in abstract cohesion
below in \ref{StrucGeometricPrequantization} and in the traditional formulation
in differential geometry in \ref{SmoothStrucGeometricPrequantization}). 

\medskip

The input datum is, for a given manifold of the form
$\Sigma = \Sigma_{n-1} \times [0,1]$ 
a smooth space $\mathrm{Conf}(\Sigma_n)$ of 
\emph{field configurations} on $\Sigma$, equipped with a suitably smooth 
map, called the ``action functional'' of the theory,
$$
  S : \mathrm{Conf}(\Sigma_{n}) \to \mathbb{R}
$$
taking values in the real numbers. 

From this input one first obtains the \emph{covariant phase space} of the system,
given as the variational \emph{critical locus} of $S$, schematically
the subspace
$$
  P = \{ \phi \in \mathrm{Conf}(\Sigma) \;|\; (d S)_\phi = 0\}
$$
of field configurations on which the \emph{variational} derivative $d S$ of
$S$ vanishes. These field configurations are said to satisfy the 
\emph{Euler-Lagrange equations of motion} of the dynamics encoded by $S$.

If $S$ is a \emph{local} action functional, in that it depends on the fields
$\phi$ via an integral over $\Sigma$ whose integrand only depends
on finitely many derivatives of $\phi$, then this space canonically carries
a \emph{pre-symplectic form}, a closed 2-form $\omega \in \Omega^2_{\mathrm{cl}}(P)$.

A \emph{symmetry} of the system is a vector field on $P$ which is in the kernel of
$\omega$. The quotient of $P$ by the flows of these symmetries is called the
\emph{reduced phase space}. This quotient is typically very ill-behaved if regarded
in ordinary geometry, but is a natural nice space in higher geometry
(modeled by \emph{BV-BRST formalism}). The pre-symplectic form $\omega$ descends
to a symplectic form $\omega_{\mathrm{red}}$ on the reduced phase space. 

A \emph{geometric prequantization} of the symplectic smooth space
$(P_{\mathrm{red}}, \omega_{\mathrm{red}})$ is now,
if it exists, a choice of line bundle $E \to P_{\mathrm{red}}$ 
with connection $\nabla$, such that
$\omega = F_\nabla$ is the corresponding curvature 2-form. 
This becomes a \emph{geometric quantization} proper
when furthermore equipped with a choice of foliation of $P_{\mathrm{red}}$ 
by Lagrangian submanifolds
(submanifolds of maximal dimension on which $\omega_{\mathrm{red}}$ vanishes).
This foliation is a choice of decomposition of phase space into 
``canonical coordinates and momenta'' of the physical system.

Finally, the quantum space of states, $Z(\Sigma_{n-1})$, that is defined by this 
construction is the vector space of those sections of $E$ that are covariantly
constant along the leaves of the foliation.

The notion of fully local/extended TQFTs suggests that there ought to be 
an analogous fully local/extended version of geometric quantization, which produces
not just the datum $Z(\Sigma_{n-1})$, but $Z(\Sigma_k)$ for all $0 \leq k \leq n$.
By the above classification result it follows that the value for $k = 0$
alone will suffice to define the entire quantum theory.  This should involve not
just line bundles with connection, but higher analogs of these, 
called \emph{circle $(n-k)$-bundles with connection} or 
\emph{bundle $(n-k-1)$-gerbes with connection}. 

We discuss such a 
\emph{higher geometric prequantization} axiomatically 
in \ref{StrucGeometricPrequantization}, 
and discuss examples in \ref{SmoothStrucGeometricPrequantization}
and \ref{PrequantizationApplications}.

\paragraph*{Classical Chern-Weil theory and its shortcomings}
\label{ClassicalChernWeilAndItsShortcomings}

Even in the space of all topological local action functionals, those that 
typically appear in fundamental physics are special. 
The archetypical example of a TQFT is 3-dimensional Chern-Simons theory
(see \cite{FreedCS} for a detailed review).
Its action functional happens to arise from a natural construction in 
classical \emph{Chern-Weil theory}. We now briefly summarize this 
process, which already produces a large family of natural
topological action functionals on gauge equivalence classes of gauge fields. 
We then point out 
deficiencies of this classical theory, which are removed by 
lifting it to higher geometry.

\medskip

A classical problem in topology is the classification of vector bundles
over some topological space $X$. These are continuous maps $E \to X$
such that there is a vector space $V$, and an open cover $\{U_i \hookrightarrow X\}$, 
and such that over each patch we have fiberwise linear identifications
$E|_{U_i} \simeq U_i \times V$.
Examples include
\begin{itemize}
  \item the tangent bundle $T X$ of a smooth manifold $X$;
  \item the canonical $\mathbb{C}$-line bundle over the 2-sphere, 
    $S^3 \times_{S^1} \mathbb{C} \to S^2$ which is
    associated to the Hopf fibration.
\end{itemize}

A classical tool for studying isomorphism classes of vector bundles is
to assign to them simpler \emph{characteristic classes} in the ordinary
integral cohomology of the base space. For vector bundles over the complex
numbers these are the \emph{Chern classes}, which are maps
$$
  [c_1] : \mathrm{VectBund}_{\mathbb{C}}(X)/_\sim \to H^2(X, \mathbb{Z})
$$
$$
  [c_2] : \mathrm{VectBund}_{\mathbb{C}}(X)/_\sim \to H^4(X, \mathbb{Z})
$$
etc. natural in $X$. If two bundles have differing characteristic classes, they
must be non-isomorphic. For instance for $\mathbb{C}$-line bundles the first
Chern-class $[c_1]$ is an isomorphism, hence provides a complete invariant characterization.

In the context of \emph{differential geometry}, where $X$ and $E$ are taken to be 
smooth manifolds and the local identifications are taken to be smooth maps, 
one wishes to obtain \emph{differential} characteristic classes. To that end, 
one can use the canonical inclusion $\mathbb{Z} \hookrightarrow \mathbb{R}$
of coefficients to obtain the map $H^{n+1}(X,\mathbb{Z}) \to H^{n+1}(X, \mathbb{R})$
from integral to real cohomology, and send any integral characteristic class $[c]$ to 
its  real image $[c]_{\mathbb{R}}$. 
Due to the de Rham theorem, which identifies the real cohomology of a smooth manifold
with the cohomology of its complex of differential forms,
$$
  H^{n+1}(X,\mathbb{R}) \simeq H^{n+1}_{\mathrm{dR}}(X)
  \,,
$$
this means that for $[c]_{\mathbb{R}}$ one has representatives given by closed
differential $(n+1)$-forms $\omega \in \Omega^{n+1}_{\mathrm{cl}}(X)$,
$$
  [c]_{\mathbb{R}} \sim [\omega]
  \,.
$$
But since the passage to real cohomology may lose topological information 
(all torsion group elements map to zero), one wishes to keep the information
both of the topological characteristic class $[c]$ as well as of its 
``differential refinement'' $\omega$. This is accomplished by the
notion of \emph{differential cohomology} $H^{n+1}_{\mathrm{diff}}(X)$
(see \cite{HopkinsSinger} for a review). These are
families of cohomology groups equipped with compatible projections both
to integral classes as well as to differential forms
$$
  \xymatrix{
    & H^{n+1}_{\mathrm{diff}}(X)
	\ar[dl]
	\ar[dr]
	\\
	H^{n+1}(X, \mathbb{Z}) 
	\ar[dr]
	  && 
	\Omega^{n+1}_{\mathrm{cl}}(X)
	\ar[dl]
	\\
	& H^{n+1}(X, \mathbb{R})
	\simeq
	H^{n+1}_{\mathrm{dR}}(X)
  }
  \;\;\;\;\;
  \xymatrix{
    & [\hat c]
	\ar@{|->}[dl]
	\ar@{|->}[dr]
	\\
	[c]
	\ar@{|->}[dr]
	&& 
	 \omega
	 \ar@{|->}[dl]
	\\
	& [c]_{\mathbb{R}} \sim [\omega]
	\\
  }
  \,.
$$
Moreover, these differential cohomology groups come equipped with a
notion of \emph{volume holonomy}. For $\Sigma_{n}$ an $n$-dimensional compact manifold,
there is a canonical morphism
$$
  \int_\Sigma \, : H^{n+1}_{\mathrm{diff}}(\Sigma) \to U(1)
$$
to the circle group. 

For instance for $n = 1$, we have that $H^2(X, \mathbb{Z})$ classifies circle bundles
/ complex line bundles over $X$, $H^2_{\mathrm{diff}}(X)$ classifies such bundles
\emph{with connection} $\nabla$, and the map $\int_\Sigma : H^2_{\mathrm{diff}}(\Sigma) \to U(1)$
is the \emph{line holonomy} obtained from the \emph{parallel transport} of $\nabla$
over the 1-dimensional manifold $\Sigma$.

With such differential refinements of characteristic classes in hand, it is desirable to
have them classify differential refinements of vector bundles. These are known as
\emph{vector bundles with connection}. We say a differential refinement of a 
characteristic class $[c]$
is a map $[\hat c]$ fitting into a diagram
$$
  \raisebox{20pt}{
  \xymatrix{
    \mathrm{VectBund}_{\mathrm{conn}}(X)/_\sim 
	\ar[r]^<<<<<<{[\hat c]}
	\ar[d]
	&
	H^{n+1}_{\mathrm{diff}}(X)
	\ar[d]
	\\
    \mathrm{VectBund}(X)/_\sim 
	\ar[r]^{[c]}
	&
	H^{n+1}(X, \mathbb{Z})	
  }
  }\,,
$$
where the vertical maps forget the differential refinement.
Such a $[\hat c]$ contains information even when $[c] = 0$. 
Therefore one also calls $[\hat c]$ a \emph{secondary characteristic class}.

All of this has a direct interpretation in terms of quantum gauge field theory.
\begin{itemize}
  \item the elements in $\mathrm{VectBund}_{\mathrm{conn}}(X)/_\sim$
  are gauge equivalence classes of \emph{gauge fields} on $X$
  (for instance the electromagnetic field, or nuclear force fields);
  \item the differential class $[\hat c]$ defines a canonical 
  \emph{action functional} $S_{[c]}$ on such fields, by composition with the 
  volume holonomy
  $$
    \exp(i S_c(-))
	:
	\mathrm{Conf}(\Sigma)/_{\sim}
	:=
	\mathrm{VectBund}_{\mathrm{conn}}(\Sigma)/_\sim
	 \stackrel{[\hat c]}{\to}
	H^{n+1}_{\mathrm{diff}}(\Sigma)
	 \stackrel{\int_\Sigma}{\to}
	U(1)
	\,.
  $$
\end{itemize}
The action functionals that arise this way are of \emph{Chern-Simons type}. 
If we write $A \in \Omega^1(\Sigma, \mathfrak{u}(n))$ for a differential
form representing locally the connection on a vector bundle, then we
have
\begin{itemize}
  \item $\int_\Sigma {c_1} \;:\; 	A \mapsto \exp(i \int_\Sigma \mathrm{tr}(A))$;
  \item $\int_\Sigma c_2 \; : \; A \mapsto \exp(
    i \int_\Sigma \mathrm{tr}(A \wedge d_{\mathrm{dR} A} + \frac{2}{3}\mathrm{tr}(A \wedge A \wedge A))
  )$
  \item 
   etc.
\end{itemize}
Here the second expression, coming from the second Chern-class, 
is the standard action functional for 3-dimensional Chern-Simons theory.
The first, coming from the first Chern-class, is a 1-dimensional Chern-Simons type theory.
Next in the series is an action functional for a 5-dimensional Chern-Simons theory. 
Later we will see that by generalizing 
here from vector bundles to \emph{higher bundles} of various kinds, a host of 
known action functionals for quantum field theories arises this way.

Despite this nice story, this traditional Chern-Weil theory has several shortcomings.
\begin{enumerate}
  \item 
    It is \emph{not local}, related to the fact that it deals with cohomology classes
	$[c]$ instead of the cocycles $c$ themselves. This means that there is no good 
	obstruction theory and no information about the locality of the resulting
	QFTs.
  \item 
    It does not apply to \emph{higher topological structures}, 
	hence to \emph{higher gauge fields} that take values in higher covers of 
	Lie groups which are not themselves compact Lie groups anymore. 
  \item 
    It is \emph{restricted to ordinary differential geometry} and does not
	apply to variants such as supergeometry, infinitesimal geometry or
	derived geometry, all of which appear in examples of QFTs of interest.
\end{enumerate}

\paragraph*{Formulation in cohesive homotopy type theory}
\label{CohesiveHomotopyInMotivation}

We discuss now these problems in slightly more detail, together with their 
solution in \emph{cohesive homotopy type theory}.

\medskip

The problem with the locality is that every vector bundle is, by definition, 
\emph{locally equivalent} to a trivial bundle. Also, locally on contractible
patches $U \hookrightarrow X$ every integral cocycle becomes cohomologous
to the trivial cocycle. Therefore the restriction of 
a characteristic class to local patches retains no information at all
$$
  \xymatrix{
    \mathrm{VectBund}(X)/_\sim 
	\ar[r]^{[c]}
	\ar[d]^{(-)|_{U}}
	&
	H^{n+1}(X, \mathbb{Z})
	\ar[d]^{(-)|_{U}}
	\\
	{*}
	\ar[r]^{\mathrm{Id}}
	&
	{*}
  }
  \,.
$$
Here we may think of the singleton ${*}$ as the class of the trivial bundle over $U$. 
But even though on $U$ every bundle is equivalent to the trivial bundle, this
has non-trivial gauge automorphisms
$$
  {*} \stackrel{g}{\to} {*}
  \;\;\;\;
  g \in C^\infty(U, G := \mathrm{GL}(V))
  \,.
$$
These are not seen by traditional Chern-Weil theory, as they are not visible
after passing to equivalence classes and to cohomology.

But by collecting this information over each 
$U$, it organizes into a \emph{presheaf of gauge groupoids}. We shall write
$$
  \mathbf{B} G : U \mapsto 
  \left\{
    \xymatrix{
	 {*}
	 \ar[rr]^{g \in C^\infty(U,G)}
	 &&
	 {*}
	}
  \right\}
  \;\;\;
  \in \mathrm{Funct}(\mathrm{SmthMfd}^{\mathrm{op}}, \mathrm{Grpd})
  \,.
$$
In order to retain all this information, we may pass to the 2-category
$$
  \mathbf{H} := L_W \; \mathrm{Func}(\mathrm{SmthMfd}^{\mathrm{op}}, \mathrm{Grpd})
$$
of such groupoid-valued functors, where we formally invert all those morphisms
(natural transformations) in the class $W$ of \emph{stalkwise} equivalences of groupoids.
This is called the \emph{2-topos of stacks} on smooth manifolds. 

For example we have
\begin{itemize}
  \item $\mathbf{H}(U, \mathbf{B}G) \simeq 
      \left\{
    \xymatrix{
	 {*}
	 \ar[rr]^{g \in C^\infty(U,G)}
	 &&
	 {*}
	}
  \right\}
  $
  \item 
  $\pi_0 \mathbf{H}(X, \mathbf{B}G) \simeq \mathrm{VectBund}(X)/_\sim$
\end{itemize}
and hence the object $\mathbf{B}G \in \mathbf{H}$ constitutes a genuine smooth
refinement of the classifying space for rank $n$-vector bundles, which sees not
just their equivalence classes, but also their local smooth transformations.

The next problem of traditional Chern-Weil theory is that it cannot see
beyond groupoids even in cohomology. Namely, under the standard nerve operation, 
groupoids embed into \emph{simplicial sets} 
(described in more detail in \ref{ModelForPrincipalInfinityBundles} below)
$$
  N : \mathrm{Grpd} \hookrightarrow \mathrm{sSet}
  \,.
$$
But simplicial sets model \emph{homotopy theory}. 
\begin{itemize}
  \item 
    There is a notion of homotopy groups $\pi_k$ of simplicial sets;
	\item 
	and there is a notion of \emph{weak homotopy equivalences}, morphisms
	$f : X \to Y$ which induce isomorphisms on all homotopy groups. 
\end{itemize}
Under the above embedding, groupoids yield only (and precisely) those simplicial sets,
up to equivalence, for which only $\pi_0$ and $\pi_1$ are nontrivial. One says that these
are \emph{homotopy 1-types}. A general simplicial set presents what is called a 
\emph{homotopy type} and may contain much more information.

Therefore we are led to refine the above construction and consider the
simplicial category
$$
  \mathbf{H} := L_W \; \mathrm{Func}(\mathrm{SmthMfd}^{\mathrm{op}}, \mathrm{sSet})
$$
of functors that send smooth manifolds to simplicial sets, where now we formally
invert those morphisms that are stalkwise weak homotopy equivalences of simplicial sets.

This is called the \emph{$\infty$-topos of $\infty$-stacks} on smooth manifolds. 

For instance, there are objects $\mathbf{B}^n U(1)$ in this context which are smooth refinements
of higher integral cohomology, in that 
$$
  \pi_0 \mathbf{H}(X, \mathbf{B}^n U(1)) \simeq H^{n+1}(X, \mathbb{Z})
  \,.
$$

Finally, in this construction it is straightforward to change the geometry by changing the
category of geometric test spaces. For instance we many replace smooth manifolds here by
supermanifolds or by formal (synthetic) smooth manifolds. In all these cases
$\mathbf{H}$ describes \emph{homotopy types with differential geometric structure}.
One of our main statements below is the following theorem. 

These $\mathbf{H}$ all satisfy a simple set of axioms for ``cohesive homotopy types'',
which were proposed for 0-types by Lawvere. In the fully homotopical context these 
axioms canonically induce in $\mathbf{H}$
\begin{itemize}
  \item 
    differential cohomology;
  \item 
    higher Chern-Weil theory;
  \item 
     higher Chern-Simons functionals;
  \item 
    higher geometric prequantization.
\end{itemize} 
This is such that it reproduces the traditional notions where they apply, and
otherwise generalizes them beyond the realm of classical applicability.

\paragraph*{Extended higher Chern-Simons theory}

It has become a familiar fact, known from examples as those indicated above, 
that there should be an $n$-dimensional
topological quantum field theory $Z_{\mathbf{c}}$ associated to the following data:
\begin{enumerate}
  \item a \emph{gauge group} $G$: a Lie group such as $U(n)$; or more generally 
  a higher smooth group, such as the smooth \emph{circle $n$-group} $\mathbf{B}^{n-1}U(1)$
  or the \emph{String 2-group} or the smooth \emph{Fivebrane 6-group}
  \cite{SSSIII, FSS};
  \item a universal characteristic class $[c]\in H^{n+1}(BG,\mathbb{Z})$
  and/or its image $\omega$ in real/de Rham cohomology,
\end{enumerate}  
where $Z_{\mathbf{c}}$ is a 
$G$-gauge theory defined naturally over all closed oriented
$n$-dimensional smooth manifolds $\Sigma_n$, and such that whenever $\Sigma_n$
happens to be the boundary of some manifold $\Sigma_{n+1}$ the action fuctional 
on a field configuration $\phi$ is given by the integral of the pullback form
$\hat \phi^* \omega$ (made precise below)
over $\Sigma_{n+1}$, for some extension $\hat \phi$ of $\phi$.
These are \emph{Chern-Simons type} gauge theories. 
See 
\cite{Za} for a gentle introduction to the general idea of Chern-Simons
theories. 

\medskip
Notably for 
$G$ a connected and simply connected simple Lie group, for
$c \in H^4(B G, \mathbb{Z})\simeq \mathbb{Z}$ any integer -- the ``level'' --
and hence for $\omega = \langle -,-\rangle$ the Killing form on the Lie algebra $\mathfrak{g}$,
this quantum field theory is the original and standard Chern-Simons theory
introduced in \cite{WittenCS}. See \cite{FreedCS} for a comprehensive review. 
Familiar as this theory is, there is an interesting aspect of it that has not yet found attention, 
and which is an example of our constructions here. 

\medskip
To motivate this, it is helpful to 
look at the 3d Chern-Simons action functional as follows:
if we write $H(\Sigma_3, \mathbf{B}G_{\mathrm{conn}})$ for the set of gauge equivalence classes of $G$-principal connections $\nabla$ on $\Sigma_3$, then the (exponentiated) action functional of 3d Chern-Simons theory over $\Sigma_3$ is a function of sets
$$
  \exp(i S(-)) : H(\Sigma_3, \mathbf{B}G_{\mathrm{conn}}) \to U(1)
  \,.
$$
Of course this function acts by picking a representative of the gauge 
equivalence class, given by a smooth 1-form $A \in \Omega^1(\Sigma_3, \mathfrak{g})$ 
and sending that to the element $\exp(2 \pi i k \int_{\Sigma_3} \mathrm{CS}(A)) \in U(1)$, 
where $\mathrm{CS}(A) \in \Omega^3(\Sigma_3)$ is the Chern-Simons 3-form of $A$ \cite{ChernSimons},
that gives the whole theory its name. 
That this is well defined is the fact that for every gauge 
transformation $g : A \stackrel{}{\to} A^g$, for $g \in C^\infty(\Sigma_3, G)$, both $A$ 
as well as its gauge transform $A^g$, are sent to the same element of $U(1)$. 
A natural formal way to express this is to consider the \emph{groupoid}
$\mathbf{H}(\Sigma_3, \mathbf{B}G_{\mathrm{conn}})$ whose objects are gauge fields $A$ and whose 
morphisms are gauge transformations $g$ as above. Then the fact that the Chern-Simons action is
defined on individual gauge field configurations while being invariant under gauge transformations
is equivalent the statement that it is a \emph{functor}, hence a morphism of groupoids, 
$$
  \exp(i S(-)) : \mathbf{H}(\Sigma_3, \mathbf{B}G_{\mathrm{conn}}) \to U(1)
  \,,
$$
where the set underlying $U(1)$ is regarded as a groupoid with only identity morphisms.
Hence the fact that $\exp(i S(-))$ has to send every morphism on the left to a
morphism on the right is the gauge invariance of the action.

\medskip
Furthermore, the action functional has the property of being \emph{smooth}. It takes any \emph{smooth family} of gauge fields, over some parameter space $U$, to a corresponding smooth family of elements of $U(1)$ and such that these assignmens are compatible with precomposition of smooth functions $U_1 \to U_2$ between parameter spaces. 
The formal language that expresses this concept is that of 
\emph{stacks on the site of smooth manifolds} (discussed in detail in 
\ref{SmoothInfgrpds} below): 
to say that for every $U$ there is a groupoid, as above, of smooth $U$-families of gauge fields and smooth $U$-families of gauge transformations between them, in a consistent way, is to say that there is a \emph{smooth moduli stack}, denoted $[\Sigma_3,\mathbf{B}G_{\mathrm{conn}}]$, of gauge fields on $\Sigma_3$. Finally, the fact that the Chern-Simons action functional is not only gauge invariant but also smooth is the fact that it refines to a morphism of smooth stacks
$$
  \exp(i \mathbf{S}(-)) : [\Sigma_3, \mathbf{B}G_{\mathrm{conn}}] \to U(1)
  \,,
$$
where now $U(1)$ is regarded as a smooth stack by declaring that a smooth family of elements is a smooth function with values in $U(1)$. 

\medskip
It is useful to think of a smooth stack simply as being a \emph{smooth groupoid}.
Lie groups and Lie groupoids are examples (and are called ``differentiable stacks''
when regarded as special cases of smooth stacks)
but there are important smooth groupoids which are not Lie groupoids in that they have not a
smooth \emph{manifold} but a more general smooth space of objects and of morphisms.
Just as Lie groups have an infinitesimal approximation given by Lie algebras, so 
smooth stacks/smooth groupoids have an infinitesimal approximation given by
\emph{Lie algebroids}.
The smooth moduli stack 
$[\Sigma_3, \mathbf{B}G_{\mathrm{conn}}]$ of gauge field configuration on $\Sigma_3$
is best known in the physics literature in the guise of its 
underlying Lie algebroid: this is the formal dual of the
(off-shell) \emph{BRST complex} of the $G$-gauge theory on $\Sigma_3$: in degree 0
this consists of the functions on the space of gauge fields on $\Sigma_3$, and in 
degree 1 it consists of functions on infinitesimal gauge transformations between these:
the ``ghost fields''.

\medskip
The smooth structure on the action functional is of course crucial in 
field theory: in particular 
it allows one to define the \emph{differential} $d \exp(i \mathbf{S}(-))$ of the action functional and hence its critical locus, characterized by   
the Euler-Lagrange equations of motion. This is the \emph{phase space}
of the theory, which is a substack
$$
  [\Sigma_2, \flat \mathbf{B}G] \hookrightarrow [\Sigma_2, \mathbf{B}G_{\mathrm{conn}}]
$$
equipped with a pre-symplectic 2-form. To formalize 
this, write $\Omega^2_{\mathrm{cl}}(-)$ for the 
smooth stack of closed 2-forms (without gauge transformations), hence the rule that sends
a parameter manifold $U$ to the set $\Omega^2_{\mathrm{cl}}(U)$ of smooth closed 2-forms on $U$.
This may be regarded as the \emph{smooth moduli 0-stack} of closed 2-forms in that for
every smooth manifold $X$ the set of morphisms $X \to \Omega^2_{\mathrm{cl}}(-)$
is in natural bijection to the set $\Omega^2_{\mathrm{cl}}(X)$ of closed 2-forms on $X$.
This is an instance of the \emph{Yoneda lemma}.
Similarly, a smooth 2-form on the moduli stack of field configurations
is a morphism of smooth stacks of the form
$$
  [\Sigma_2, \mathbf{B}G_{\mathrm{conn}}] \to \Omega^2_{\mathrm{cl}}(-)
  \,.
$$
Explicitly, for Chern-Simons theory this morphism sends for each smooth parameter space $U$ a
given smooth $U$-family of gauge fields 
$A  \in \Omega^1(\Sigma_2 \times U, \mathfrak{g})$
to the 2-form
$$
  \int_{\Sigma_2} \langle d_U A \wedge d_U A\rangle \in \Omega^2_{\mathrm{cl}}(U)
  \,.
$$
Notice that if we restrict to \emph{genuine} families $A$ which are functions of $U$
but vanish on vectors tangent to $U$ (technically these are elements in the
\emph{concretification} of the moduli stack) then this 2-form is the 
\emph{fiber integral}
of the Poincar{\'e} 2-form $\langle F_A \wedge F_A\rangle$ along the projection
$\Sigma_2 \times U \to U$, where $F_A := d A + \tfrac{1}{2}[A \wedge A]$ is the 
curvature 2-form of $A$. This is the first sign of a general pattern, 
which we highlight in a moment.

\medskip
There is more fundamental smooth moduli stack equipped with a closed 2-form:
the moduli stack $\mathbf{B}U(1)_{\mathrm{conn}}$ of $U(1)$-gauge fields,
hence of smooth circle bundles with connection. This is the rule
that sends a smooth parameter manifold $U$ to the groupoid 
$\mathbf{H}(U, \mathbf{B}U(1)_{\mathrm{conn}})$ of $U(1)$-gauge fields $\nabla$ on $U$,
which we have already seen above. Since the  curvature 2-form 
$F_\nabla \in \Omega^2_{\mathrm{cl}}(U)$ of a $U(1)$-principal connection
 is gauge invariant, the assignment
$\nabla \mapsto F_\nabla$ gives rise to a morphism of smooth stacks of the form
$$
  F_{(-)} : \mathbf{B}U(1)_{\mathrm{conn}} \to \Omega^2_{\mathrm{cl}}(-)
  \;.
$$
In terms of this morphism the fact that every $U(1)$-gauge field $\nabla$
on some space $X$
has an underlying field strength 2-form $\omega$ is expressed by the
existence of a commuting diagram of smooth stacks of the form
$$
  \raisebox{20pt}{
  \xymatrix{
    & \mathbf{B}U(1)_{\mathrm{conn}}
	\ar[d]^{F_{(-)}} & \mbox{gauge field / differential cocycle}
    \\
    X
	\ar[r]^\omega 
	\ar[ur]^{\nabla}
	&
	\Omega^2_{\mathrm{cl}}(-) & \mbox{ field strength / curvature}\;.
  }
  }
$$
Conversely, if we regard the bottom morphism $\omega$ as given, and regard this 
closed 2-form as a (pre)symplectic form, then a \emph{choice of lift} $\nabla$
in this diagram is a choice of refinement of the 2-form by a circle bundle with connection,
hence the choice of a \emph{prequantum circle bundle} in the language of 
geometric quantization (see for instance section II in \cite{Brylinski} for a review
of geometric quantization).

\medskip
Applied to the case of Chern-Simons theory this means that a smooth (off-shell)
prequantization of the theory is a choice of dashed morphism in a diagram
of smooth stacks of the form
$$
  \raisebox{20pt}{
  \xymatrix{
    && \mathbf{B}U(1)_{\mathrm{conn}}
	\ar[d]^{F_{(-)}}
    \\
    [\Sigma_2, \mathbf{B}G_{\mathrm{conn}}]
	\ar[rr]_-{\int_{\Sigma_2}\langle F_{(-)},F_{(-)}\rangle}
	\ar@{-->}[urr]
	&&
	\Omega^2_{\mathrm{cl}}(-)~~.
  }
  }
$$
Similar statements apply to on-shell geometric (pre)quantization of Chern-Simons theory,
which has been so successfully applied in the original article
\cite{WittenCS}.
In summary, this means that in the context of smooth stacks 
the Chern-Simons action functional and its prequantization are as in the following table:

\medskip
\medskip
\hspace{1cm}
\begin{tabular}{|c|c|c|}
  \hline
  {\bf dimension} & & {\bf moduli stack description }
  \\
  \hline
  $k = 3$ & action functional (0-bundle) 
  & $ \exp(i \mathbf{S} (-)) : [\Sigma_3, \mathbf{B}G_{\mathrm{conn}}] 
   \to \;\;\;U(1)_{\;\;\;\;\;\;\;}$
  \\
  \hline
  $k = 2$ & prequantum circle 1-bundle 
  & \hspace{2.0cm}$
  [\Sigma_2, \mathbf{B}G_{\mathrm{conn}}]  
  \to
  \mathbf{B} U(1)_{\mathrm{conn}}
  $
  \\
  \hline
\end{tabular}

\medskip
\medskip
There is a precise sense, discussed in section \ref{SmoothStrucDifferentialCohomology} below,
in which a $U(1)$-valued function is a \emph{circle $k$-bundle with connection}
for $k = 0$. If we furthermore regard an ordinary $U(1)$-principal bundle
as a \emph{circle 1-bundle} then this table says that in dimension $k$
Chern-Simons theory appears as a \emph{circle $(3-k)$-bundle with connection}
-- at least for $k = 3$ and $k = 2$.

\medskip
Formulated this way, it should remind one of what is called
\emph{extended} or \emph{multi-tiered} topological quantum field theory
(formalized and classified in \cite{LurieTFT}) 
which is the full formalization of \emph{locality}
in the Schr{\"o}dinger picture of quantum field theory. This says that \emph{after quantization}, an 
$n$-dimensional topological field theory should be a rule that 
to a closed manifold of dimension $k$ assigns an $(n-k)$-categorical
analog of a vector space of quantum states. 
Since ordinary geometric quantization of Chern-Simons theory
assigns to a closed $\Sigma_2$ the vector space of \emph{polarized sections}
(holomorphic sections) of the line bundle associated to the above circle 1-bundle, 
this suggests that there should be an \emph{extended} or \emph{multi-tiered}
refinement of geometric (pre)quantization of Chern-Simons theory,
which to a closed oriented manifold of dimension $0 \leq k \leq n$ assigns
a \emph{prequantum circle $(n-k)$-bundle} 
(bundle $(n-k-1)$-gerbe) on the moduli stack of 
field configurations over $\Sigma_k$, modulated by a morphism 
$[\Sigma_k, \mathbf{B}G_{\mathrm{conn}}] \to \mathbf{B}^{(n-k)} U(1)_{\mathrm{conn}}$
to a moduli $(n-k)$-stack of circle $(n-k)$-bundles with connection.

\medskip
In particular for $k = 0$ and $\Sigma_0$ connected, hence $\Sigma_0 = *$ the point,
we have that the moduli stack of fields on $\Sigma_0$ is the 
\emph{universal} moduli stack itself,
$[*, \mathbf{B}G_{\mathrm{conn}}] \simeq \mathbf{B}G_{\mathrm{conn}}$,
and so a  \emph{fully extended prequantization} of 
3-dimensional $G$-Chern-Simons theory would have to involve a 
\emph{universal characteristic} morphism
$$
  \mathbf{c}_{\mathrm{conn}} : \mathbf{B}G_{\mathrm{conn}}
  \to 
  \mathbf{B}^3 U(1)_{\mathrm{conn}}
$$
of smooth moduli stacks, hence a smooth circle 3-bundle with connection
on the universal moduli stack of $G$-gauge fields. This indeed naturally
exists: an explicit construction is given in \cite{FSS}.
This morphism of smooth higher stacks is a differential refinement of a smooth refinement
of the level itself: forgetting the connections and only remembering the 
underlying (higher) gauge bundles, we still have a morphism of smooth higher stacks
$$
  \mathbf{c}: \mathbf{B}G \to \mathbf{B}^3 U(1)
  \,.
$$
This expression should remind one of the continuous map of topological 
spaces
$$
  c : B G \to B^3 U(1) \simeq K(\mathbb{Z},4)
$$
from the classifying space $B G$ to the Eilenberg-MacLane space $K(\mathbb{Z},4)$, which 
represents the level as a class in integral cohomology $H^4(B G, \mathbb{Z}) \simeq \mathbb{Z}$.
Indeed, there is a canonical \emph{derived functor} or \emph{$\infty$-functor}
$$
  {\vert-\vert} : \mathbf{H} \to \mathrm{Top}
$$
from smooth higher stacks to topological spaces (one of the defining properties
of a cohesive $\infty$-topos), 
derived left adjoint to the operation of forming \emph{locally constant higher stacks},
and under this map we have
$$
  \vert \mathbf{c}\vert \simeq c
  \,.
$$
In this sense $\mathbf{c}$ is a \emph{smooth refinement} of $[c] \in H^4(B G, \mathbb{Z})$
and then $\mathbf{c}_{\mathrm{conn}}$ is a further \emph{differential refinement}
of $\mathbf{c}$.

\medskip
However, more is true. Not only is there an extension of the prequantization of 
3d $G$-Chern-Simons theory to the point, but this also induces 
the extended prequantization in every other dimension by \emph{tracing}:
for $0 \leq k \leq n$ and $\Sigma_k$ a closed and oriented smooth manifold,
there is a canonical morphism of smooth higher stacks of the form
$$
  \exp(2 \pi i \int_{\Sigma_k}(-)) : [\Sigma_k, \mathbf{B}^n U(1)_{\mathrm{conn}}]
  \to 
  \mathbf{B}^{n-k}U(1)_{\mathrm{conn}}\;,
$$
which refines the fiber integration of differential forms, that we have
seen above, from curvature $(n+1)$-forms to their entire prequantum 
circle $n$-bundles (we discuss this below in section \ref{FibInt}).
Since, furthermore, the formation of mapping stacks $[\Sigma_k,-]$ is functorial, this means
that from a morphism $\mathbf{c}_{\mathrm{conn}}$ as above we get for every
$\Sigma_k$ a composite morphism as such:
$$
  \exp(2 \pi i \int_{\Sigma_k} [\Sigma_k, \mathbf{c}_{\mathrm{conn}}])
  :
  \xymatrix{
    [\Sigma_k, \mathbf{B}G_{\mathrm{conn}}]
	\ar[rr]^-{[\Sigma_k, \mathbf{c}_{\mathrm{conn}}]}
	&&
	[\Sigma_k, \mathbf{B}^n U(1)_{\mathrm{conn}}]
	\ar[rrr]^-{\exp(2 \pi i\int_{\Sigma_k}(-))}
	&&&
	\mathbf{B}^{n-k}U(1)_{\mathrm{conn}}
  }
  \,.
$$ 
For 3d $G$-Chern-Simons theory and $k = n = 3$ this composite 
\emph{is} the action functional of the theory (down on the set
$H(\Sigma_3, \mathbf{B}G_{\mathrm{conn}})$ this is effectively
the perspective on ordinary Chern-Simons theory amplified in \cite{CJMSW}).
Therefore, for general $k$ we may speak of this as the 
\emph{extended action functional}, with values not in $U(1)$
but in $\mathbf{B}^{n-k}U(1)_{\mathrm{conn}}$.

\medskip
This way we find that the above table, containing the
Chern-Simons action functional together with its prequantum circle 1-bundle,
extends to the following table that reaches all the way from dimension 3
down to dimension 0.\\

\vspace{3mm}
\hspace{-1.2cm}
{\small
\begin{tabular}{|c|c|c|c|}
  \hline
  {\bf dim.} & & {\bf prequantum $(3-k)$-bundle} & 
  \\
  \hline
  $k = 0$ & \begin{tabular}{c}differential fractional \\ first Pontrjagin\end{tabular} & 
  $\mathbf{c}_{\mathrm{conn}} : \mathbf{B}G_{\mathrm{conn}} 
  \to \mathbf{B}^3 U(1)_{\mathrm{conn}}$ &
  \cite{FSS}
  \\
  \hline
  $k = 1$ & \begin{tabular}{c} WZW \\ background B-field \end{tabular} &
  $ \xymatrix{ [S^1, \mathbf{B}G_{\mathrm{conn}}]
   \ar[rr]^-{[S^1,\mathbf{c}_{\mathrm{conn}}]} &&
    [S^1, \mathbf{B}^3 U(1)_{\mathrm{conn}}] 
	 \ar[rr]^-{\exp(2 \pi i \int_{S^1}(-))} && \mathbf{B}^2 U(1)_{\mathrm{conn}}
	 } $
	 &
  \\
  \hline
  $k = 2$ & \begin{tabular}{c} off-shell CS \\ prequantum bundle \end{tabular} & 
  $ \xymatrix{ [\Sigma_2, \mathbf{B}G_{\mathrm{conn}}]
   \ar[rr]^-{[\Sigma_2,\mathbf{c}_{\mathrm{conn}}]} &&
    [\Sigma_2, \mathbf{B}^3 U(1)_{\mathrm{conn}}] 
	 \ar[rr]^-{\exp(2 \pi i \int_{\Sigma_2}(-))} && \mathbf{B}U(1)_{\mathrm{conn}}
	 }
  $
   & 
  \\
  \hline
  $k = 3$ & \begin{tabular}{c} 3d CS \\ action functional \end{tabular} & 
  $ \xymatrix{[\Sigma_3, \mathbf{B}G_{\mathrm{conn}}]
   \ar[rr]^-{[\Sigma_3,\mathbf{c}_{\mathrm{conn}}]}
     &&
    [\Sigma_3, \mathbf{B}^3 U(1)_{\mathrm{conn}}] 
	 \ar[rr]^-{\exp(2 \pi i \int_{\Sigma_3}(-))} && U(1)
	 }
  $  &
  \cite{FSS}
  \\
  \hline
\end{tabular}  
}

\vspace{5mm}
\noindent For each entry of this table one may compute the \emph{total space} object of the
corresponding prequantum $k$-bundle. This is now in general itself
a higher moduli stack. In full codimension $k = 0$ one finds  that 
this is the moduli 2-stack of $\mathrm{String}(G)$-2-connections described in 
\cite{SSSIII, FiorenzaSatiSchreiberI}. This we discuss in section \ref{InfinCSOrdinaryCS} below.

\medskip

It is clear now that this is just the first example
of a general class of theories which we may call
\emph{higher extended prequantum Chern-Simons theories} or just
\emph{$\infty$-Chern-Simons theories}, for short.
These are defined by a choice of
\begin{enumerate}
  \item a smooth higher group $G$;
  \item a smooth universal characteristic map $\mathbf{c} : \mathbf{B}G \to \mathbf{B}^n U(1)$;
  \item a differential refinement $\mathbf{c}_{\mathrm{conn}} : \mathbf{B}G_{\mathrm{conn}} \to
    \mathbf{B}^n U(1)_{\mathrm{conn}}$.
\end{enumerate}
An example of a 7-dimensional such theory on $\mathrm{String}$-2-form gauge fields
is discussed in \cite{FSScfield}, given by a differential refinement
of the second fractional Pontrjagin class to a morphism of smooth moduli 7-stacks
$$
  \tfrac{1}{6}(\mathbf{p}_2)_{\mathrm{conn}}
  :
  \mathbf{B}\mathrm{String}_{\mathrm{conn}}
  \to
  \mathbf{B}^7 U(1)_{\mathrm{conn}}
  \,.
$$
We expect that these $\infty$-Chern-Simons theories are part of a general procedure of 
\emph{extended geometric quantization} (\emph{multi-tiered} geometric quantization)
which proceeds in two steps, as indicated in the following table.

\vspace{5mm}
\hspace{-1cm}
{\small
\begin{tabular}{|c|c|c|}
  \hline
  {\bf classical system}
  &
  {\bf geometric prequantization}
  &
  {\bf quantization}
  \\
  \hline
  \begin{tabular}{c}
    char. class $c$ of deg. $(n+1)$
	\\
	with de Rham image $\omega$:
	\\
	invariant polynomial/
	\\
	$n$-plectic form 
  \end{tabular}
  &
  \begin{tabular}{c}
    prequantum circle $n$-bundle
	\\
	on moduli $\infty$-stack of fields
	\\
	$\mathbf{c}_{\mathrm{conn}} : \mathbf{B}G_{\mathrm{conn}} \to \mathbf{B}^n U(1)_{\mathrm{conn}}$
  \end{tabular}
  &
  \begin{tabular}{c}
    extended quantum field theory 
    \\
    $Z_{\mathbf{c}}: \Sigma_k \mapsto
	 \left\{
	   \begin{array}{l}
	    \mbox{polarized sections of }
	   \\
	    \mbox{prequantum $(n-k)$-bundle}
		\\
	   \exp(2 \pi i \int_{\Sigma_k}[\Sigma_k, \mathbf{c}_{\mathrm{conn}}])
	   \end{array}
	 \right\}
	$
  \end{tabular}
  \\
  \hline
\end{tabular}
}

\vspace{5mm}
\noindent Here we are concerned with the first step, the 
discussion of $n$-dimensional Chern-Simons gauge theories
(higher gauge theories)
in their incarnation as prequantum circle $n$-bundles on their
universal moduli $\infty$-stack of fields. 
A dedicated discussion of 
higher geometric prequantization, including the discussion of 
higher Heisenberg groups, higher quantomorphism groups, 
higher symplectomorphisms and higher
Hamiltonian vector fields,
and their action on higher prequantum spaces of states by higher
Heisenberg operators, is given below. As shown there, plenty of interesting
physical information turns out to be captured by extended
prequantum $n$-bundles. For instance, if one regards the
B-field in type II superstring backgrounds as a prequantum 2-bundle, then 
its extended prequantization knows all about
twisted Chan-Paton bundles, the Freed-Witten anomaly 
cancellation condition for type II superstrings on D-branes and the associated anomaly line bundle
on the string configuration space.

\vspace{3mm}
Generally, all higher Chern-Simons theories that arise from extended action functionals this way 
enjoy a collection of  very good formal properties. Effectively, they may be understood 
as constituting examples of a fairly extensive generalization of 
the \emph{refined} Chern-Weil homomorphism with coefficients in \emph{secondary characteristic cocycles}. Moreover, we have shown previously  that the class of theories arising this way is large and contains not only several  familiar theories, some of which are not traditionally recognized to be of this good form, but also contains various new QFTs that turn out to be of interest  within known contexts, e.g. \cite{FiorenzaSatiSchreiberI,FiorenzaSatiSchreiberI}.  
Here we further enlarge the pool of such examples. 

\medskip

Notably, here we are concerned with examples arising from \emph{cup product}
characteristic classes, hence of $\infty$-Chern-Simons theories which are decomposable
or non-primitive secondary characteristic cocyles, obtained by cup-ing more
elementary characteristic cocycles. 
The most familiar example of these is again ordinary 3-dimensional Chern-Simons theory,
but now for the non-simply connected gauge group $U(1)$. In this case 
a gauge field configuration in $\mathbf{H}(\Sigma_3, \mathbf{B}U(1)_{\mathrm{conn}})$
is not necessarily given by a globally defined 1-form $A \in \Omega^1(\Sigma_3)$,
instead it may have a non-vanishing ``instanton number'', the Chern-class
of the underlying circle bundle.
Only if that happens to vanish is the value of the action functional again given by the 
simple expression $\exp(2 \pi i k \int_{\Sigma_3} A \wedge d_{\mathrm{dR}} A)$ as before.
But in view of the above we are naturally led to ask: which circle 3-bundle
(bundle 2-gerbe)
with connection
over $\Sigma_3$, depending naturally on the $U(1)$-gauge field,
has $A \wedge d_{\mathrm{dR}} A$ as its connection 3-form 
in this special case, so that the correct action functional
in generality is again the \emph{volume holonomy} of this 3-bundle
(see section \ref{VolumeHolonomy} below)? 
The answer is that it is the \emph{differential cup square}
of the gauge field with itself. As a fully extended action functional this is 
a natural morphism of higher moduli stacks of the form
$$
  (-)^{\cup_{\mathrm{conn}}^2} 
    : 
  \mathbf{B}U(1)_{\mathrm{conn}}
  \to
  \mathbf{B}^3 U(1)_{\mathrm{conn}}
  \,.
$$
This morphism of higher stacks is characterized by the
fact that under forgetting the differential refinement and then taking geometric realization as
before, it is exhibited as a differential refinement of the ordinary 
cup square on Eilenberg-MacLane spaces 
$$
  (-)^{\cup^2} : K(\mathbb{Z},2) \to K(\mathbb{Z}, 4)
$$
and hence on ordinary integral cohomology.
By the above general procedure, we obtain a well-defined action functional
for $3d$ $U(1)$-Chern-Simons theory by the expression
$$
  \exp(2 \pi i \int_{\Sigma_3} [\Sigma_3, (-)^{\cup_{\mathrm{conn}}^2}])
  :
  [\Sigma_3, \mathbf{B}U(1)_{\mathrm{conn}}]
  \to
  U(1)
$$
and this is indeed the action functional of the familiar $3d$ $U(1)$-Chern-Simons theory,
also on non-trivial instanton sectors, see section \ref{3dU1CS} below.

\medskip
In terms of this general construction, there is nothing particular to the
low degrees here, and we have generally a differential cup square
/ extended action functional for a $(4k+3)$-dimensional Chern-Simons theory
$$
 (-)^{\cup_{\mathrm{conn}}^2} : 
 \mathbf{B}^{2k+1}U(1)_{\mathrm{conn}}
 \to
 \mathbf{B}^{4k+3}U(1)_{\mathrm{conn}}
$$
for all $k \in \mathbb{N}$, which induces an ordinary action functional
$$
  \exp(2 \pi i \int_{\Sigma_3} [\Sigma_{4k+3}, (-)^{\cup_{\mathrm{conn}}^2}])
  :
  [\Sigma_{4k+3}, \mathbf{B}^{4k+3}U(1)_{\mathrm{conn}}]
  \to
  U(1)
$$
on the moduli $(2k+1)$-stack of $U(1)$-$(2k+1)$-form gauge fields, 
given by the fiber integration on differential cocycles over the 
differential cup product of the fields.
This is discussed in section \ref{4k+3} below.

\vspace{3mm}
Forgetting the smooth structure on $[\Sigma_{4k+3}, \mathbf{B}^{2k+1}U(1)_{\mathrm{conn}}]$ 
and passing to gauge equivalence classes of fields yields the cohomology group
$H^{2k+2}_{\mathrm{conn}}(\Sigma_{4k+3})$. This is what is known as 
\emph{ordinary differential cohomology} and is equivalent to the group of
\emph{Cheeger-Simons differential characters}, a review with further pointers is 
in \cite{HopkinsSinger}.
That gauge equivalence classes of 
higher degree $U(1)$-gauge fields are to be regarded as
differential characters and that the $(4k+3)$-dimensional 
$U(1)$-Chern-Simons action functional on these is given by the 
fiber integration of the cup product is discussed in 
detail in \cite{FP}, also mentioned notably in \cite{Witten96,Witten98}
and expanded on in \cite{Freed}. Effectively this observation led to the general
development of differential cohomology in \cite{HopkinsSinger}. 
Or rather, the main 
theorem there concerns a shifted version of the functional of
$(4k+3)$-dimensional $U(1)$-Chern-Simons theory which allows one to further divide it
by 2. We have discussed the refinement of this to smooth moduli stacks of fields in 
\cite{FiorenzaSatiSchreiberI}. 
These developments were largely motivated from the 
relation of $(4k+3)$-dimensional $U(1)$-Chern-Simons theories 
as the holographic duals to theories of self-dual
forms in dimension $(4k+2)$ (see \cite{BelovMoore} for survey and references):
a choice of conformal structure on a $\Sigma_{4k+2}$ naturally induces
a polarization of the prequantum 1-bundle of the $(4k+3)$-dimensional theory,
and for every choice the resulting space of quantum states is naturally
identified with the corresponding conformal blocks (correlators) of the
$(4k+2)$-dimensional theory. 

\medskip
Therefore we have that regarding the differential cup square on smooth higher moduli 
stacks as an extended action functional yields the following table of familiar notions
under extended geometric prequantization.

\vspace{5mm}
\hspace{-2cm}
{\small
\begin{tabular}{|c|c|c|}
  \hline
  {\bf dim.} & & {\bf prequantum $(4k+3-d)$-bundle} 
  \\
  \hline
  $d = 0$ & \begin{tabular}{c}differential cup square\end{tabular} & 
  $(-)^{\cup_{\mathrm{conn}}^2} : \mathbf{B}^{2k+1} U(1)_{\mathrm{conn}} 
  \to \mathbf{B}^{4k+3} U(1)_{\mathrm{conn}}$ 
  \\
  \hline
   $\vdots$ & $\vdots$ & $\vdots$
   \\
   \hline
   $d = 4k+2 $ & \begin{tabular}{c} ``pre-conformal blocks'' of \\ self-dual $2k$-form field \end{tabular}
   & 
   $ 
    \xymatrix{[\Sigma_{4k+2}, \mathbf{B}^{2k+1}U(1)_{\mathrm{conn}}]
    \ar[rr]^-{[\Sigma_{4k+2},(-)^{\cup_{\mathrm{conn}}^2}]}
      &&
      [\Sigma_{4k+2}, \mathbf{B}^{2k+1} U(1)_{\mathrm{conn}}]  
 	  \ar[rr]^-{\exp(2 \pi i \int_{\Sigma_{4k+2}}(-))} && \mathbf{B}U(1)_{\mathrm{conn}}
	  }
    $   
   \\
  \hline
  $d = 4k+3$ & \begin{tabular}{c} CS \\ action functional \end{tabular} & 
  $ 
   \xymatrix{[\Sigma_{4k+3}, \mathbf{B}^{2k+1}U(1)_{\mathrm{conn}}]
   \ar[rr]^-{[\Sigma_{4k+3},(-)^{\cup_{\mathrm{conn}}^2}]}
     &&
     [\Sigma_{4k+3}, \mathbf{B}^{2k+1} U(1)_{\mathrm{conn}}] 
	 \ar[rr]^-{\exp(2 \pi i \int_{\Sigma_{4k+3}}(-))} && U(1)
	 }
  $ 
  \\
  \hline
\end{tabular}  
}

\vspace{5mm}
This fully extended prequantization of $(4k+3)$-dimensional $U(1)$-Chern-Simons theory
allows for instance to ask for and
compute the total space of the prequantum circle $(4k+3)$-bundle. 
This is now itself
a higher smooth moduli stack. For $k = 0$, hence in $3d$-Chern-Simons theory 
it turns out to be the moduli 
2-stack of \emph{differential T-duality structures}.
This we discuss in section \ref{3dU1CS} below. 

\medskip

More generally, as the name suggests, the \emph{differential cup square} is a specialization
of a general \emph{differential cup product}. As a morphism of bare homotopy types
this is the familiar cup product of Eilenberg-MacLane spaces
$$
  (-)\cup (-) : K(\mathbb{Z},p+1) \times K(\mathbb{Z}, q+1) \to K(\mathbb{Z}, p+q+2)
$$
for all $p,q \in \mathbb{N}$.
Its smooth and then its further differential refinement is a morphism of smooth higher stacks of the
form
$$
  (-)\cup_{\mathrm{conn}} (-)
  :
  \mathbf{B}^p U(1)_{\mathrm{conn}}
  \times
  \mathbf{B}^1 U(1)_{\mathrm{conn}}
  \to
  \mathbf{B}^{p+q+1}U(1)_{\mathrm{conn}}
  \,.
$$

\medskip
By the above discussion this now defines a higher extended gauge theory
in dimension $p+q+1$ of \emph{two different} species of higher 
$U(1)$-gauge fields.
One example of this is the \emph{higher electric-magnetic coupling anomaly} in higher
(Euclidean) $U(1)$-Yang-Mills theory, as explained in section 2 of \cite{Freed}. 
In this example one considers on an oriented smooth manifold $X$ 
(here assumed to be closed, for simplicity) 
an \emph{electric current}
$(p+1)$-form $J_{\mathrm{el}} \in \Omega^{p+1}_{\mathrm{cl}}(X)$
and a \emph{magnetic current} $(q+1)$-form 
$J_{\mathrm{mag}} \in \Omega^{q+1}_{\mathrm{cl}}(X)$, such that
$p + q = \mathrm{dim}(X)$ is the dimension of $X$.
A \emph{prequantization}
of these current forms in our sense of higher geometric quantization 
is a lift to differential cocycles 
$$
  \raisebox{20pt}{
  \xymatrix{
    && \mathbf{B}^p U(1)_{\mathrm{conn}}
	\ar[d]^{F_{(-)}}
    \\
    X
	\ar@{-->}[urr]^{\widehat J_{\mathrm{el}}}
	\ar[rr]^{J_{\mathrm{el}}} 
	&&
	\Omega^{p+1}_{\mathrm{cl}}(-)~,
  }
  }
  \;\;\;~~~~~~~~~
    \raisebox{20pt}{
  \xymatrix{
    && \mathbf{B}^q U(1)_{\mathrm{conn}}
	\ar[d]^{F_{(-)}}
    \\
    X
	\ar@{-->}[urr]^{\widehat J_{\mathrm{mag}}}
	\ar[rr]^{J_{\mathrm{mag}}} 
	&&
	\Omega^{q+1}_{\mathrm{cl}}(-)
  }
  }
$$
and here this amounts to electric and magnetic \emph{charge quantization}, respectively:
the electric charge is the universal integral cohomology class of 
the circle $p$-bundle underlying the electric charge cocycle: its
\emph{higher Dixmier-Doudy class}
$[\widehat J_{\mathrm{el}}] \in H^{p+1}_{\mathrm{cpt}}(X, \mathbb{Z})$
(see section \ref{VolumeHolonomy} below); and similarly
for the magnetic charge.
Accordingly, the higher mapping stack 
$[X, \mathbf{B}^p U(1)_{\mathrm{comm}}\times \mathbf{B}^q U(1)_{\mathrm{conn}}]$
is the smooth higher moduli stack of charge-quantized electric and magnetic currents on 
$X$. Recall that this assigns to a smooth test manifold $U$ the higher groupoid
whose objects are $U$-families of pairs of charge-quantized electric and magnetic 
currents, namely such currents on $X \times U$.
As \cite{Freed} explains in terms of such families of fields, the 
$U(1)$-principal bundle with connection that in the present formulation is 
the one modulated by the morphism
$$
 \nabla_{\mathrm{an}}
 :=
  \exp(2 \pi i \int_X [X, (-)\cup_{\mathrm{conn}}(-)])
  :
  [X, \mathbf{B}^p U(1)_{\mathrm{comm}}\times \mathbf{B}^q U(1)_{\mathrm{conn}}]
  \to
  \mathbf{B}U(1)_{\mathrm{conn}}
$$
is the \emph{anomaly line bundle} of $(p-1)$-form electromagnetism on $X$, 
in the presence of electric and magnetic currents subject to charge quantization.
In the language of $\infty$-Chern-Simons theory as above, this is
equivalently the off-shell prequantum 1-bundle of the higher cup
product Chern-Simons theories on pairs of $U(1)$-gauge $p$-form and $q$-form fields.

\medskip
Regarded as an anomaly bundle, one calls its curvature the
\emph{local anomaly} and its \emph{holonomy} the
``global anomaly''. In our contex the holonomy of $\nabla_{\mathrm{an}}$ is
(discussed again in section \ref{VolumeHolonomy} below) the morphism
$$
 \mathrm{hol}(\nabla_{\mathrm{an}})
 =
  \exp(2 \pi i \int_{S^1} [S^1,  \nabla_{\mathrm{an}}])
  :
  [S^1, [X, \mathbf{B}^p U(1)_{\mathrm{comm}}\times \mathbf{B}^q U(1)_{\mathrm{conn}}
  \to 
  U(1)
$$
from the loop space of the moduli stack of fields to $U(1)$.
By the characteristic universal propery of higher mapping stacks, 
together with the ``Fubini-theorem''-property of fiber integration, this is equivalently
the morphism
$$
  \exp(2 \pi i \int_{X \times S^1} [X \times S^1, (-)\cup_{\mathrm{conn}}(-)])
  :
  [X\times S^1, \mathbf{B}^p U(1)_{\mathrm{comm}}\times \mathbf{B}^q U(1)_{\mathrm{conn}}]
  \to 
  U(1)
  \,.
$$
But from the point of view of $\infty$-Chern-Simons theory  this is the \emph{action functional}
of the higher cup product Chern-Simons field theory induced by $\cup_{\mathrm{conn}}$. 
The situation is now summarized in the following table.

\vspace{3mm}
{\small
\hspace{-2cm}
\begin{tabular}{|c|c|c|}
  \hline
  {\bf dim.} & & {\bf prequantum $(\mathrm{dim}(X)+1-k)$-bundle} 
  \\
  \hline
  $k = 0$ & \begin{tabular}{c}differential cup product\end{tabular} & 
  $(-)^{\cup_{\mathrm{conn}}^2} : 
  \mathbf{B}^{p} U(1)_{\mathrm{conn}} 
  \mathbf{B}^{q} U(1)_{\mathrm{conn}}
  \to \mathbf{B}^{d+2} U(1)_{\mathrm{conn}}$ 
  \\
  \hline
   $\vdots$ & $\vdots$ & $\vdots$
    \\
   \hline
   $k = \mathrm{dim}(X) $ & \begin{tabular}{c} higher E/M-charge \\ anomaly line bundle \end{tabular}
   & 
   $ 
    \xymatrix{
	  \exp(2 \pi i \int_{X} [X,(-)\cup_{\mathrm{conn}} (-)])
	  :
	  [X, \mathbf{B}^p U(1)_{\mathrm{conn}} \times \mathbf{B}^q U(1)_{\mathrm{conn}}]
	  \ar[r]
	  &
	  \mathbf{B}U(1)_{\mathrm{conn}}
	}
    $   
   \\
  \hline
  $k = \mathrm{dim}(X)+1$ & \begin{tabular}{c} global anomaly \end{tabular} & 
   $ 
    \xymatrix{
	  \exp(2 \pi i \int_{X\times S^1} [X\times S^1,(-)\cup_{\mathrm{conn}} (-)])
	  :
	  [X \times S^1, \mathbf{B}^p U(1)_{\mathrm{conn}} \times \mathbf{B}^q U(1)_{\mathrm{conn}}]
	  \to
	  &
	  \hspace{-1.2cm}
	  U(1)
	}
    $   
  \\
  \hline
\end{tabular}  
}

\vspace{3mm}
These higher electric-magnetic anomaly Chern-Simons theories
are of particular interest when the higher electric/magnetic currents
are themselves induced by other gauge fields. Namely if we have any two 
$\infty$-Chern-Simons theories given by extended action functionals
$\mathbf{c}^1_{\mathrm{conn}} : \mathbf{B}G^1_{\mathrm{conn}} \to \mathbf{B}^{p}U(1)_{\mathrm{conn}}$
and 
$\mathbf{c}^2_{\mathrm{conn}} : \mathbf{B}G^2_{\mathrm{conn}} \to \mathbf{B}^{q}U(1)_{\mathrm{conn}}$,
respectively, then composition of these with the differential cup product yields an
extended action functional of the form
$$
  \mathbf{c}_{\mathrm{conn}}^1
  \cup_{\mathrm{conn}}
  \mathbf{c}_{\mathrm{conn}}^2
  :
  \xymatrix{
    \mathbf{B}(G^1 \times G^2)_{\mathrm{conn}}
	\ar[rr]^-{(\mathbf{c}^1_{\mathrm{conn}}, \mathbf{c}^2_{\mathrm{conn}})}
	&&
	\mathbf{B}^p U(1)_{\mathrm{conn}}
	\times
	\mathbf{B}^1 U(1)_{\mathrm{conn}}
	\ar[rr]^-{(-)\cup_{\mathrm{conn} }(-)}
	&&
    \mathbf{B}^{p+q+1} U(1)_{\mathrm{conn}}
  }
  \,,
$$
which describes extended topological field theories in dimension $p+q+1$ on
two species of (possibly non-abelian, possibly higher) gauge fields, 
or equivalently describes the higher electric/magnetic anomaly for
higher electric fields induced by $\mathbf{c}^1$ and higher magnetic
fields induced by $\mathbf{c}^2$.

\medskip
For instance for heterotic string backgrounds $\mathbf{c}^2_{\mathrm{conn}}$
is the differential refinement of the first fractional Pontrjagin class
$\tfrac{1}{2}p_1 \in H^4(B \mathrm{Spin}, \mathbb{Z})$
\cite{SSSIII, FSS} of the form
$$
  \mathbf{c}^2_{\mathrm{conn}}
  =
  \widehat J^{\mathrm{NS5}}_{\mathrm{mag}}
  =
  \tfrac{1}{2}(\mathbf{p}_1)_{\mathrm{conn}}
  :
  \mathbf{B}\mathrm{Spin}_{\mathrm{conn}}
  \to
  \mathbf{B}^3 U(1)_{\mathrm{conn}}
  \,,
$$
formalizing the \emph{magnetic NS5-brane charge} needed to cancel the 
fermionic anomaly of the heterotic string by way of the Green-Schwarz mechanism.
It is curious to observe, going back to the very first 
example of this introduction, that this $\widehat J_{\mathrm{mag}}^{\mathrm{NS5}}$
is at the same time the extended action functional for 3d $\mathrm{Spin}$-Chern-Simons theory.

\medskip

Still more generally, we may differentially cup in this way more than two 
factors. Examples for such \emph{higher order cup product theories}
appear in 11-dimensional supergravity. 
Notably plain classical 11d supergravity contains an 11-dimensional cubic
Chern-Simons term whose extended action functional in our sense is
$$
  (-)^{\cup_{\mathrm{conn}}^3}
  :
  \mathbf{B}^3 U(1)_{\mathrm{conn}}
  \to
  \mathbf{B}^{11} U(1)_{\mathrm{conn}}
  \,.
$$
Here for $X$ the 11-dimensional spacetime, a field in $[X, \mathbf{B}^3 U(1)]$
is a first approximation to a model for the \emph{supergravity $C$-field}.
If the differential cocycle happens to be given by a globally defined 3-form
$C$, then the induced action functional
$\exp(2 \pi i \int_{X} [X, (-)^{\cup_{\mathrm{conn}}^3}])$ sends this 
to element in $U(1)$ given by the familiar expression
$$
  \exp(2 \pi i \int_{X} [X, (-)^{\cup_{\mathrm{conn}}^3}])
    : 
	C
	\mapsto
  \exp(2 \pi i \int_{X} C \wedge d_{\mathrm{dR}} C \wedge d_{\mathrm{dR}} C)
   \,.
$$
More precisely this model receives quantum corrections from an 11-dimensional
Green-Schwarz mechanism. In \cite{FiorenzaSatiSchreiberI,FiorenzaSatiSchreiberI} 
we have discussed in detail relevant corrections to the above extended
cubic cup-product action functional on the moduli stack of flux-quantized
$C$-field configurations.

\paragraph*{Boundaries and long fiber sequences of characteristic classes}
\label{MotivationFromLongFiberSequences}

It is a traditionally familiar fact that short exact sequences of (discrete) groups give rise
to long sequences in cohomology with coefficients in these groups. In fact, before
passing to cohomology, these long exact sequences are refined by corresponding long
fiber sequences of the homotopy types obtained by the higher delooping of these groups:
of the higher classifying spaces of these groups. 

An example for which these long fiber sequences are 
of interest in the context of quantum field theory is 
the universal first fractional Pontryagin class 
$\tfrac{1}{2}p_1$ on the classifying space of Spin-principal bundles. 
The following digram displays the first steps in the long fiber sequence
that it induces, together with an actual Spin-principal bundle $P \to X$
classified by a map $X \to B \mathrm{Spin}$. All squares are homotopy pullback
squares of bare homotopy types.
$$
  \xymatrix{
    B U(1) \ar[rr] \ar[dd] && \mathrm{String} \ar[rr] \ar[dd]|{B U(1)\atop \mathrm{bundle}} && 
	 \hat P \ar[rr] 
	 \ar[dd]^{B U(1) \atop \mathrm{-bundle}}
	 \ar@/_1pc/[dddd]_>>>>>>>>{\mathrm{String} \atop \mathrm{bundle}}
	&& {*} \ar[dd]
    \\
	\\
    {*} \ar[rr] && \mathrm{Spin} \ar[rr] \ar[dd] 
	\ar@/_1pc/[rrrr]_>>>>>{\mathrm{canonical}\atop \mathrm{3-class}}
	&& P \ar[rr] \ar[dd]^>>>>>>{\mathrm{Spin}\atop \mathrm{bundle}} 
	&& B^2 U(1) \ar[rr] \ar[dd] 
	  && {*} \ar[dd] &&
    \\
	\\
    && {*} \ar[rr]^x && X \ar@/_1pc/[rrrr]_>>>>>>{\mathrm{classifies} \atop {\mathrm{Spin}\;\mathrm{bundle}}} \ar@{-->}[rr] 
	\ar@/_1pc/[rrrrdd]_<<<<<<<<<{\mathrm{Pontryagin}\atop \mathrm{class}}
	&& B \mathrm{String} \ar[rr] \ar[dd] && 
	  B \mathrm{Spin} \ar[dd]^{\tfrac{1}{2}p_1} &&
	\\
	\\
	&& && && {*} \ar[rr] && B^3 U(1)
  }
  \,.
$$
The topological group $\mathrm{String}$ which appears here as the loop space object
of the homotopy fiber of $\tfrac{1}{2}p_1$ is the \emph{String group}. We discuss this 
in detail below in \ref{FractionalClasses}. It is a $B U(1)$-extension of the 
$\mathrm{Spin}$-group.

If $X$ happens to be equipped with the structure of a smooth manifold, 
then it is natural to also equip the $\mathrm{Spin}$-principal bundle
$P \to X$ with the structure of a smooth bundle, and hence to lift the classifying
map $X \to B \mathrm{Spin}$ to a morphism $X \to \mathbf{B} \mathrm{Spin}$
into the \emph{smooth moduli stack} of smooth $\mathrm{Spin}$-principal bundles
(the morphism that 
not just classifies but ``modulates'' $P \to X$ as a smooth structure).
An evident question then is: can the rest of the diagram be similarly
lifted to a smooth context?

This indeed turns out to be the case, if we work in the context of \emph{higher} smooth
stacks. For instance there is a smooth moduli 3-stack $\mathbf{B}^2 U(1)$ such that a morphism
$\mathrm{Spin} \to \mathbf{B}^2 U(1)$ not just classifies a $B U(1)$-bundle over 
$\mathrm{Spin}$, but ``modulates'' a smooth \emph{circle 2-bundle} or 
\emph{$U(1)$-bundle gerbe} over $\mathrm{Spin}$. One then gets the following diagram
$$
  \xymatrix{
    \mathbf{B} U(1) \ar[rr] \ar[dd] && \mathrm{String} \ar[rr] \ar[dd]|{\mathrm{WZW}\atop \mathrm{2-bundle}} && 
	 \hat P \ar[rr] 
	 \ar[dd]^{\mathbf{B} U(1) \atop \mathrm{2-bundle}}
	 \ar@/_1pc/[dddd]_>>>>>>>>{\mathrm{String} \atop \mathrm{2-bundle}}
	&& {*} \ar[dd]
    \\
	\\
    {*} \ar[rr] && \mathrm{Spin} \ar[rr] \ar[dd] 
	\ar@/_1pc/[rrrr]_>>>>>{\mathrm{modulates} \atop {\mathrm{WZW} \atop \mathrm{2-bundle}}}
	&& P \ar[rr] \ar[dd]^>>>>>>{\mathrm{Spin}\atop \mathrm{bundle}} 
	&& \mathbf{B}^2 U(1) \ar[rr] \ar[dd] 
	  && {*} \ar[dd] &&
    \\
	\\
    && {*} \ar[rr]^x && X \ar@/_1pc/[rrrr]_>>>>>>{\mathrm{modulates} \atop {\mathrm{Spin}\;\mathrm{bundle}}} \ar@{-->}[rr] 
	\ar@/_1pc/[rrrrdd]_<<<<<<<<<{\mathrm{modulates} \atop {\mathrm{Chern-Simons} \atop \mathrm{3-bundle}}}
	&& \mathbf{B} \mathrm{String} \ar[rr] \ar[dd] && 
	  \mathbf{B} \mathrm{Spin} \ar[dd]^{\tfrac{1}{2}\mathbf{p}_1} &&
	\\
	\\
	&& && && {*} \ar[rr] && \mathbf{B}^3 U(1)
  }
  \,,
$$
where now all squares are homotopy pullbacks of smooth higher stacks.

With this smooth geometirc structure in hand, one can then go further and ask for
\emph{differential} refinements: the smooth $\mathrm{Spin}$-principal bundle
$P \to X$ might be equipped with a principal connection $\nabla$, and if so, this
will be ``modulated'' by a morphism $X \to \mathbf{B}\mathrm{Spin}_{\mathrm{conn}}$
into the smooth moduli stack of $\mathrm{Spin}$-connections. 

One of our central theorems below in \ref{FractionalClasses} is that the 
universal first fractional 
Pontryagin class can be lifted to this situation to a 
\emph{differential smooth} universal morphism of higher moduli stacks, which we write
$\tfrac{1}{2}\hat {\mathbf{p}}_1$. Inserting this into the above diagram and then forming
homotopy pullbacks as before yields further differential refinements. It turns out
that these now induce the Lagrangians of 3-dimensional $\mathrm{Spin}$ Chern-Simons theory
and of the WZW theory on $\mathrm{Spin}$.
$$
  \xymatrix{
    \mathbf{B} U(1) \ar[rr] \ar[dd] && \mathrm{String} \ar[rr] \ar[dd]|{\mathrm{WZW}\atop \mathrm{2-bundle}} && 
	 \hat P 
	 \ar[dd]^{\mathbf{B} U(1) \atop \mathrm{2-bundle}}
	 \ar@/_1pc/[dddd]_>>>>>>>>{\mathrm{String} \atop \mathrm{2-bundle}}
	&& 
    \\
	\\
    {*} \ar[rr] && \mathrm{Spin} \ar[rr] \ar[dd] 
	\ar@/_1pc/[rrrr]_>>>>>{{\mathrm{WZW} \atop \mathrm{Lagrangian}}}
	&& P \ar[rr] \ar[dd]^>>>>>>{\mathrm{Spin}\atop \mathrm{bundle}} 
	&& \mathbf{B}^2 U(1)_{\mathrm{conn}}  \ar[dd] 
	  &&  &&
    \\
	\\
    && {*} \ar[rr]^x && X \ar@/_1pc/[rrrr]_>>>>>>{{\mathrm{Spin} \atop \mathrm{connection}}} \ar@{-->}[rr] 
	\ar@/_1pc/[rrrrdd]_<<<<<<<<<{{\mathrm{Chern-Simons} \atop \mathrm{Lagrangian}}}
	&& \mathbf{B} \mathrm{String}_{\mathrm{conn}} \ar[rr] \ar[dd] && 
	  \mathbf{B} \mathrm{Spin}_{\mathrm{conn}} \ar[dd]^{\tfrac{1}{2}\hat {\mathbf{p}}_1} &&
	\\
	\\
	&& && && {*} \ar[rr] && \mathbf{B}^3 U(1)_{\mathrm{conn}}
  }
  \,.
$$

One way to understand our developments here is as a means to formalize and then analyze
this setup and its variants and generalizations.

\subsubsection{Philosophical motivation}
\label{MotivationFromHigherToposTheory}

Finally we offer a motivation for the development of physics in 
cohesive higher topos theory for readers who can appreciate 
philosophical considerations formalized in higher category theory. 
Other readers should kindly ignore this section.

In \cite{Como} Lawvere refers to cohesive toposes as 
\emph{Categories of Being} and refers to the phenomenon exhibited by the
adjunctions that define them as \emph{Becoming},  thereby following
the terminology of \cite{Hegel} and in effect proposing
a formal interpretation of Hegel's ontology in topos theory.
The following might be regarded as further expanding on this line of thought.

\medskip

The history of theoretical fundamental physics is the story of a search
for the suitable mathematical notions and structural concepts that naturally model 
the physical phenomena in question.
Examples include, roughly in historical order,
\begin{enumerate}
\item  the identification of symplectic geometry as the underlying structure of
classical Hamiltonian mechanics;
\item the identification of (semi-)Riemannian differential geometry as the
underlying structure of gravity;
\item the identification of group and representation theory as the underlying 
structure of the zoo of fundamental particles; 
\item the identification of Chern-Weil theory and differential cohomology
as the underlying structure of gauge theories.
\end{enumerate}
All these examples exhibit the identification of the precise mathematical language
that naturally captures the physics under investigation. Modern theoretical
insight in theoretical fundamental physics is literally \emph{unthinkable} 
without these formulations.

Therefore it is natural to ask whether one can go further. 
Not only have we seen above in \ref{MotivationFromAnomalyCancellation}
that some of these formulations leave open questions that we
would want them to answer. But one is also led to wonder if this list of
mathematical theories cannot be subsumed into a single more fundamental
system altogether. 

\noindent In a philosophical vein we should ask

\medskip
\emph{Where does physics take place, conceptually?}
\medskip

\noindent Such philosophical-sounding questions can be given 
useful formalizations in terms of category theory. In this context
``place'' translates to \emph{topos}, ``taking place''
translates to \emph{internalization} and whatever
it is that takes places is characterized by
a collection of \emph{universal constructions}
(categorical limits and colimits, categorical adjunctions).

So we translate

$$
  \begin{array}{lll}    
    \mbox{Physics} & \mbox{takes} & \mbox{place.}
	\\
	\\
	  \mbox{Certain universal constructions}
    &
	\mbox{internalize}
	&
	\mbox{in a suitable topos}.
  \end{array}
$$
(For the following explanation of what precisely this means the reader
only needs to know the concept of \emph{adjoint functors}.)

\noindent The remaining question is

\medskip
\emph{What characterizes a suitable topos and what are the universal constructions capturing physics.}
\medskip

At the bottom of it there are two aspects to physics, \emph{kinematics}
and \emph{dynamics}. Roughly, kinematics is about the nature of 
\emph{geometric spaces} appearing in physics, dynamics is about
\emph{trajectories} -- paths -- in these spaces. We will argue that 
\begin{itemize}
 \item the notion of a topos of geometric spaces is usefully given by
 what goes by the technical term
\emph{local topos};
 \item the notion of a topos of spaces with trajectories is usefully
 given by what goes by the technical term \emph{$\infty$-connected topos}.
\end{itemize}
A topos that is both local and $\infty$-connected we call \emph{cohesive}.

$$
  \begin{array}{c}
     \mbox{physics}
	 \\
	 \underbrace{
	 \overbrace{
	   \begin{array}{ccc}
	    \mbox{kinematic} && \mbox{dynamics}
		\\
		\\
		\mbox{local topos} && \mbox{$\infty$-connected topos}
	   \end{array}
	 }}
	 \\
	 \mbox{cohesive topos}
  \end{array}
  \,.
$$

\paragraph*{Kinematics -- local toposes.}

With a notion of \emph{bare} spaces given, a notion of geometric spaces
comes with a forgetful functor $\mathrm{GeometricSpaces} \to \mathrm{BareSpaces}$
that forgets geometric structure. The claim is that 
two extra conditions on this functor guarantee
that indeed the structure it forgets is some \emph{geometric structure}.
\begin{itemize}
  \item 
    There is a category $C$ of \emph{local models} such that every
	geometric space is obtained by \emph{gluing} of local models.
	The operation of gluing following a blueprint 
	is left adjoint to the inclusion of geometric spaces into 
	blueprints for geometric spaces.
	
  \item
    Every bare space can canonically be equipped with 
	the two universal cases of geometric structure, 
	\emph{discrete} and \emph{indiscrete} geometric structure.
	(For instance a set can be equipped with discrete topology or
	discrete smooth structure.)
	
	Equipping with these structure is left and right adjoint, respectively,
	to forgetting geometric structure.
\end{itemize}

$$
  \xymatrix@C=35pt{
    \mathrm{BlueprintsOfGeometricSpaces}
	\ar@<+7pt>[rr]^<<<<<<<<<<<<<{\mbox{glue local models}}
	\ar@{<-^{)}}@<-7pt>[rr]
	&&
	\mathrm{GeometricSpaces}
	\ar@{<-^{)}}@<+18pt>[rrrr]^{\mbox{form discrete geometric structure}}
	\ar@<+6pt>[rrrr]|{\mbox{forget geometric structure}}
	\ar@{<-^{)}}@<-6pt>[rrrr]_<<<<<<<<<<<<<<<<<<<<<<<<<<<<<{\mbox{form indiscrete geometric structure}}
	&&&&
	\mathrm{BareSpaces}
  }
  \,.
$$
If we take a bare space to be a set of points, then this 
translates into the following formal statement. 
$$
  \xymatrix{
    \mathrm{Func}(C^{\mathrm{op}}, \mathrm{Set})
	\ar@<+4pt>[rr]^<<<<<<<<<<<{\mathrm{sheafification}}
	\ar@{<-^{)}}@<-4pt>[rr]
	&&
	\mathrm{Sh}(C)
	\ar@{<-^{)}}@<+12pt>[rr]^{\mathrm{Disc}}
	\ar@<+4pt>[rr]|{\Gamma}
	\ar@{<-^{)}}@<-4pt>[rr]_{\mathrm{coDisc}}
	&&
	\mathrm{Set}
  }
  \,.
$$
The category of geometric spaces embeds into the category
of contravariant functors on test spaces, and this embedding has 
a left adjoint. It is a basic fact of topos theory that such
\emph{reflective embeddings} are precisely categories of 
\emph{sheaves} on $C$ with respect to some Grothendieck topology on $C$
(which is defined by the reflective embedding). Therefore the
first demand above says that $\mathrm{GeometricSpaces}$ is to be 
what is called a \emph{sheaf topos}. 

Another basic fact of topos theory says that this already implies
the first part of the second demand, and uniquely so. There is 
unique pair of adjoint functors $(\mathrm{Disc} \dashv \Gamma)$
as indicated. The demand of the further right adjoint embedding
$\mathrm{coDisc}$ is what makes the sheaf topos a
\emph{local topos}.

These and the following axioms are very simple. 
Nevertheless, by the power of category theory, it turns out that they
have rich implications. But we will we show that for them
to have implications \emph{just rich enough} to indeed formalize
the kind of structures mentioned at the beginning, we want to
pass to $\infty$-toposes instead. Then the above becomes

$$
  \xymatrix{
    \infty \mathrm{Func}(C^{\mathrm{op}}, \infty \mathrm{Grpd})
	\ar@<+4pt>[rr]^<<<<<<<<<{\infty\mathrm{-stackification}}
	\ar@{<-^{)}}@<-4pt>[rr]
	&&
	\mathrm{Sh}_\infty(C)
	\ar@{<-^{)}}@<+12pt>[rr]^{\mathrm{Disc}}
	\ar@<+4pt>[rr]|{\Gamma}
	\ar@{<-^{)}}@<-4pt>[rr]_{\mathrm{coDisc}}
	&&
	\infty \mathrm{Grpd}
  }
  \,.
$$

\paragraph*{Dynamics -- $\infty$-connected toposes}

With a notion of \emph{discrete $\infty$-groupoids} inside
geometric $\infty$-groupoids given, we can ask for 
discrete $\infty$-bundles over any $X$ to be characterized
by the \emph{parallel transport} that takes their fibers
into each other, as they move along paths in $X$. 
By the basic idea of \emph{Galois theory} 
(see \ref{StrucGaloisTheory}), this 
completely characterizes a notion of trajectory.

Formally this means that we require a further left adjoint
$(\Pi \dashv \mathrm{Disc})$. 
$$
  \begin{array}{ccc}
    \mathrm{Geometric}\infty \mathrm{Grpd}(X, \mathrm{Disc} K)
	&\simeq&
	\infty\mathrm{Grpd}(\Pi(X), K)
	\\
	\\
	\begin{array}{l}
	  \mbox{bundles of}
	  \\
	  \mbox{discrete $\infty$-groupoids}
	  \\
	  \mbox{on $X$}
	\end{array}
	&
	&
	\begin{array}{l}
	  \mbox{parallel transport}
	  \\
	  \mbox{of discrete $\infty$-groupoids}
	  \\
	  \mbox{along trajectories}
	  \\
      \mbox{in 	$X$}
	\end{array}	
  \end{array}
  \,.
$$

This means that for any $X$ we can think of $\Pi(X)$
as the $\infty$-groupoid of paths in $X$, of paths-between-paths in $X$,
and so on.

In order for this to yield a consistent notion of paths in
the geometric context, we want to demand that there are no non-trivial
paths in the point (the terminal object), hence that
$$
  \Pi(*) \simeq *
  \,.
$$

An ordinary topos for which $\Pi$ exists and satisfies this property
is called \emph{locally connected and connected}. Hence an
$\infty$-topos for which $\Pi$ exists and satisfies this extra condition we
call \emph{$\infty$-connected}. This terminology is good, but a bit subtle,
since it refers to the meta-topology of the \emph{collection of all geometric spaces}
rather than to any that of any topological space itself. The reader is 
advised to regard it just as a technical term for the time being.

\paragraph*{Physics -- cohesive toposes}

An $\infty$-topos that is both local as well as $\infty$-connected
we call \emph{cohesive}. The idea is that the extra adjoints on it 
encode the information of how sets of cells in an $\infty$-groupoid 
are geometrically held together, for instance in that there are
smooth paths between them. In the models of cohesive $\infty$-toposes that
we will construct the local models are \emph{open balls} with geometric structure
and each such open ball can be thought of as a ``cohesive blob of points''.

The axioms on a cohesive topos are simple and fully formal.
They involve essentially just the notion of adjoint functors.

We can ask now for universal constructions such that internalized in 
any cohesive $\infty$-topos they usefully model differental geometry,
differential cohomology, action functionals for physical systems, etc.
Below in \ref{structures} we give a comprehensive discussion of 
an extensive list of such structures. Here we highlight one 
them. Differential forms.

One consequence of the axioms of cohesion is that 
every \emph{connected} object in a cohesive $\infty$-topos $\mathbf{H}$ has
an essentially unique point (whereas in general it may fail to have
a point). We have an equivalence
$$
  \xymatrix{
    \infty\mathrm{Grp}(\mathbf{H})
	 \ar@{<-}@<+3pt>[r]^{\Omega}
	 \ar@<-3pt>[r]_{\mathbf{B}}
	&
	\mathbf{H}_{*, \geq 1}
  }
$$
between group objects $G$ in $\mathbf{H}$ and (uniquely pointed) connected
objects in $\mathbf{H}$.

Define now
$$
  (\mathbf{\Pi} \dashv \mathbf{\flat}) :=
  (\mathrm{Disc} \Pi \dashv \mathrm{Disc} \Gamma)
  \,.
$$
The $(\mathrm{Disc} \dashv \Gamma)$-counit gives a morphism
$$
  \mathbf{\flat} \mathbf{B}G \to \mathbf{B}G
  \,.
$$
We write $\mathbf{\flat}_{\mathrm{dR}} \mathbf{B}G$ for the 
$\infty$-pullback
$$
  \xymatrix{
    \mathbf{\flat}_{\mathrm{dR}} \mathbf{B}G
	\ar[r]
	\ar[d]
	&
	\mathbf{\flat}\mathbf{B}G
	\ar[d]
	\\
	{*}
	\ar[r]
	&
	\mathbf{B}G
  }
  \,.
$$
We show in \ref{SmoothStrucdeRham} that with this construction internalized
in smooth $\infty$-groupoids, the object $\mathbf{\flat}_{\mathrm{dR}}\mathbf{B}G$
is the coefficient object for flat $\mathfrak{g}$-valued differential forms,
where $\mathfrak{g}$ is the $\infty$-Lie algebra of $G$.

Moreover, there is a canonical such form on $G$ itself. This is 
obtained by forming the pasting diagram of $\infty$-pullbacks
$$
  \xymatrix{
    A \ar[r] \ar[d]^{\theta} & {*}\ar[d]	
    \\
    \mathbf{\flat}_{\mathrm{dR}} \mathbf{B}G
	\ar[r]
	\ar[d]
	&
	\mathbf{\flat}\mathbf{B}G
	\ar[d]
	\\
	{*}
	\ar[r]
	&
	\mathbf{B}G
  }
  \,.
$$
We show below in \ref{SmoothStrucCurvature} that this theta
is canonical (Maurer-Cartan) $\mathfrak{g}$-valued form on $G$.
Then in \ref{SmoothStrucDifferentialCohomology} we show that 
for $G$ a shifted abelian group, this form is the 
\emph{universal curvature characteristic}. Flat parallel $G$-valued
transport that is \emph{twisted} by this form encodes
non-flat $\infty$-connections. Gauge fields and higher gauge fields are
examples.

In \ref{SmoothStrucChernSimons} we show that, just as canonically,
action functionals for these higher gauge fields exist in 
$\mathbf{H}$. 

\medskip 

\noindent All this just from a system of adjoint $\infty$-functors.

\newpage

  \subsection{The geometry of physics}
  \label{TheGeometryOfPhysics}

The following is an introduction to the higher differential geometric
structures in the formulation of modern fundamental physics,
in particular in pre-quantized classical mechanics, 
\ref{BasicClassicalMechanicsByPrequantizedLagrangianCorrespondences},
higher pre-quantized local classical field theory,
\ref{DeDonderWeylTheoryViaHigherCorrespondences}
and specifically for twisted and higher gauge fields,
\ref{PrequantumInHigherCodimension}.

To some extent this
is classical material, roughly along the lines of a textbook
such as \cite{Frankel}, but we present it from a modern perspective that 
serves to motivate and prepare for 
the more general abstract developments in section \ref{GeneralAbstractTheory}.
More details on applications are in section \ref{Applications}.

This section has an online counterpart in \cite{SchreiberGeometryOfPhysics} with 
more material and further pointers.

\medskip

\noindent {\bf Geometry}
\begin{itemize}
  \item \ref{GeometryOfPhysicCoordinateSystems} -- Coordinate systems 
  \item \ref{GeometryOfPhysicSmoothSpaces} -- Smooth 0-types
  \item \ref{TheGeometryOfPhysicsDifferentialForms} -- Differential forms
  \item \ref{IntroGeneralAbstractTheory} -- Smooth homotopy types
  \item \ref{SmoothPrincipalnBundles} -- Principal bundles
  \item \ref{GeometryOfPhysicsPrincipalConnections} -- Principal connections
  \item \ref{CharacteristicClassesInLowDegree} -- Characteristic classes
  \item \ref{LInfinityAlgebraicStructures} -- Lie algebras
  \item \ref{InfinityChernWeilHomomorphismIntroduction} -- Chern-Weil homomorphism
\end{itemize}

\noindent {\bf Physics}
\begin{itemize}
  \item \ref{BasicClassicalMechanicsByPrequantizedLagrangianCorrespondences}
    -- Hamilton-Jacobi-Lagrange mechanics via Prequantized Hamiltonian correspondences
  \item \ref{DeDonderWeylTheoryViaHigherCorrespondences}
    -- Hamilton-de Donder-Weyl field theory via Higher correspondences
  \item \ref{PrequantumInHigherCodimension} -- Higher pre-quantum gauge fields
  \item \ref{VariationalCalculusCriticalLoci} -- Variational calculus on higher moduli stacks of fields
  \item \ref{PrequantumGeometry} -- Higher geometric pre-quantum theory
  \item \ref{GeometryOfPhysicsExamplesOfHigherPrequantumFieldTheories} 
    -- Examples of higher prequantum field theories
\end{itemize}

\newpage

\subsubsection{Coordinate systems}
\label{GeometryOfPhysicCoordinateSystems}

Every kind of geometry is modeled on a collection of archetypical basic spaces and 
geometric homomorphisms between them. In differential geometry the archetypical spaces 
are the abstract standard Cartesian coordinate systems, denoted $\mathbb{R}^n$, in every dimension 
$n \in \mathbb{N}$, and the geometric homomorphism between them are smooth functions 
$\mathbb{R}^{n_1} \to \mathbb{R}^{n_2}$, hence smooth (and possibly degenerate) coordinate transformations.

Here we discuss the central aspects of the nature of such abstract coordinate systems in themselves. At this point these are not yet coordinate systems on some other space. That is instead the topic of the next section Smooth spaces.

\paragraph{The continuum real (world-)line}

The fundamental premise of differential geometry as a model of geometry in physics is the following.

{\bf Premise.} {\it The abstract worldline of any particle is modeled 
by the continuum real line $\mathbb{R}$.}

This comes down to the following sequence of premises.

\begin{enumerate}
\item
There is a linear ordering of the points on a worldline: in particular if we 
pick points at some intervals on the worldline we may label these in an order-preserving 
way by integers
$$
  \mathbb{Z}
  \,.
$$
\item
These intervals may each be subdivided into $n$ smaller intervals, for each natural number $n$. 
Hence we may label points on the worldline in an order-preserving way by the rational numbers
$$
  \mathbb{Q}
  \,.
$$
\item
This labeling is dense: every point on the worldline is the supremum of an 
inhabited bounded subset of such labels. This means that a worldline is the 
\emph{real line}, the continuum of real numbers
$$
 \mathbb{R}
 \,.
$$
\end{enumerate}

The adjective``real'' in ``real number'' is a historical shadow of the old idea 
that real numbers are related to observed reality, hence to physics in this way. 
The experimental success of this assumption shows that it is valid at least to 
very good approximation.

Speculations are common that in a fully exact theory of quantum gravity, 
currently unavailable, this assumption needs to be refined. For instance in p-adic physics 
one explores the hypothesis that the relevant completion of the rational numbers as above is 
not the reals, but p-adic numbers $\mathbb{Q}_p$ for some prime number $p \in \mathbb{N}$. 
Or for example in the study of QFT on non-commutative spacetime one explore the idea 
that at small scales the smooth continuum is to be replaced by an object in 
noncommutative geometry. Combining these two ideas leads to the notion of 
non-commutative analytic space as a potential model for space in physics. And so forth.

For the time being all this remains speculation and differential geometry based on 
the continuum real line remains the context of all fundamental model building in 
physics related to observed phenomenology. Often it is argued that these speculations 
are necessitated by the very nature of quantum theory applied to gravity. But, at least 
so far, such statements are not actually supported by the standard theory of quantization: 
we discuss below in Geometric quantization how not just classical physics but also quantum theory, 
in the best modern version available, is entirely rooted in differential geometry based on the 
continuum real line.

This is the motivation for studying models of physics in geometry modeled on the continuum real line. 
On the other hand, in all of what follows our discussion is set up such as to be maximally independent 
of this specific choice (this is what \emph{topos theory} accomplishes for us). If we do desire to 
consider another choice of archetypical spaces for the geometry of physics we can simply ``change the site'', 
as discussed below and many of the constructions, propositions and theorems in the following will 
continue to hold. This is notably what we do below in Supergeometric coordinate systems when we 
generalize the present discussion to a flavor of differential geometry that also formalizes the 
notion of fermion particles: ``differential supergeometry''.

\paragraph{Cartesian spaces and smooth functions}

\begin{definition}
A function of sets $f : \mathbb{R} \to \mathbb{R}$ is called 
a \emph{smooth function} if, coinductively:
\begin{enumerate}
\item the derivative $\frac{d f}{d x} : \mathbb{R} \to \mathbb{R}$ exists;

\item and is itself a smooth function.
\end{enumerate}
\label{SmoothFunctions}
\end{definition}
\begin{definition}
\label{CartesianSpaceAndHomomorphism}
For $n \in \mathbb{N}$, the \emph{Cartesian space} $\mathbb{R}^n$ is the set
$$
  \mathbb{R}^n = \{ (x^1 , \cdots, x^{n}) | x^i \in \mathbb{R} \}
$$
of $n$-tuples of real numbers. For $1 \leq k \leq n$ write
$$
  i^k : \mathbb{R} \to \mathbb{R}^n
$$
for the function such that $i^k(x) = (0, \cdots, 0,x,0,\cdots,0)$ is the tuple whose $k$th entry is $x$ and all whose other entries are $0 \in \mathbb{R}$; and write
$$
  \mathbb{p}^k : \mathbb{R}^n \to \mathbb{R}
$$
for the function such that $p^k(x^1, \cdots, x^n) = x^k$.

A \emph{homomorphism} of Cartesian spaces is a smooth function
$$
  f : \mathbb{R}^{n_1} \to \mathbb{R}^{n_2}
  \,,
$$
hence a function $f : \mathbb{R}^{n_1} \to \mathbb{R}^{n_2}$ such that all partial derivatives exist and are continuous.
\end{definition}

\begin{example}
Regarding $\mathbb{R}^n$ as an $\mathbb{R}$-vector space, 
every linear function $\mathbb{R}^{n_1} \to \mathbb{R}^{n_2}$
is in particular a smooth function.
\end{example}

\begin{remark}
But a homomorphism of Cartesian spaces in def. \ref{CartesianSpaceAndHomomorphism} is 
\emph{not} required to be a linear map. We do \emph{not} regard the Cartesian spaces here 
as vector spaces. 
\end{remark}

\begin{definition}
A smooth function $f : \mathbb{R}^{n_1} \to \mathbb{R}^{n_2}$ is called a 
\emph{diffeomorphism} if there exists another smooth function 
$\mathbb{R}^{n_2} \to \mathbb{R}^{n_1}$ such that the underlying functions of sets are 
inverse to each other
$$
  f \circ g = \mathrm{id}
$$
and

$$
  g \circ f = \mathrm{id}
  \,.
$$
\end{definition}

\begin{proposition}
There exists a diffeomorphism $\mathbb{R}^{n_1} \to \mathbb{R}^{n_2}$ precisely if $n_1 = n_2$.
\end{proposition}

\begin{definition}
We will also say equivalently that
\begin{enumerate}
\item a Cartesian space $\mathbb{R}^n$ is an \emph{abstract coordinate system};

\item a smooth function $\mathbb{R}^{n_1} \to \mathbb{R}^{n_2}$ is an \emph{abstract coordinate transformation};

\item the function $p^k : \mathbb{R}^{n} \to \mathbb{R}$ is the $k$th \emph{coordinate} 
of the coordinate system $\mathbb{R}^n$. We will also write this function as 
$x^k : \mathbb{R}^{n} \to \mathbb{R}$.

\item for $f : \mathbb{R}^{n_1} \to \mathbb{R}^{n_2}$ a smooth function, and $1 \leq k \leq n_2$ we write
\begin{enumerate}
   \item $f^k := p^k\circ f$

   \item $(f^1, \cdots, f^n) := f$.
\end{enumerate}
\end{enumerate}
\end{definition}
\begin{remark}
It follows with this notation that
$$
  \mathrm{id}_{\mathbb{R}^n} = (x^1, \cdots, x^n) : \mathbb{R}^n \to \mathbb{R}^n
  \,.
$$
Hence an abstract coordinate transformation 
$$ 
  f : \mathbb{R}^{n_1} \to \mathbb{R}^{n_2}
$$
may equivalently be written as the tuple
$$
  \left(
    f^1 \left( x^1, \cdots, x^{n_1} \right)
    ,
    \cdots,
    f^{n_2}\left( x^1, \cdots, x^{n_1} \right)
  \right)
  \,.
$$
\end{remark}

\begin{proposition}
\label{CartSpCategory}
Abstract coordinate systems  
form a \emph{category} 
-- to be denoted $\mathrm{CartSp}$ -- whose
\begin{itemize}
\item objects are the abstract coordinate systems $\mathbb{R}^{n}$ 
(the class of objects is the set $\mathbb{N}$ of natural numbers $n$);

\item morphisms $f : \mathbb{R}^{n_1} \to \mathbb{R}^{n_2}$ are the 
abstract coordinate transformations = smooth functions.
\end{itemize}
Composition of morphisms is given by composition of functions.

We have that
\begin{enumerate}
\item The identity morphisms are precisely the identity functions.

\item The isomorphisms are precisely the diffeomorphisms.
\end{enumerate}
\end{proposition}

\begin{definition}
Write CartSp${}^{op}$ for the opposite category of CartSp.

This is the category with the same objects as $CartSp$, 
but where a morphism $\mathbb{R}^{n_1} \to \mathbb{R}^{n_2}$ in $CartSp^{op}$ 
is given by a morphism $\mathbb{R}^{n_1} \leftarrow \mathbb{R}^{n_2}$ in $CartSp$.
\end{definition}

We will be discussing below the idea of exploring smooth spaces by laying out abstract coordinate systems in them in all possible ways. The reader should begin to think of the sets that appear in the following definition as the \emph{set of ways} of laying out a given abstract coordinate systems in a given space. 

\begin{definition}
A functor  $X : CartSp^{op} \to Set$ (a ``presheaf'') is
\begin{enumerate}
\item for each abstract coordinate system $U$ a set $X(U)$

\item  for each coordinate transformation 
$f : \mathbb{R}^{n_1} \to \mathbb{R}^{n_2}$ a function $X(f) : X(\mathbb{R}^{n_1}) \to X(\mathbb{R}^{n_2})$
\end{enumerate}
such that 
\begin{enumerate}
\item identity is respected $X(id_{\mathbb{R}^n}) = id_{X(\mathbb{R}^n)}$;

\item composition is respected $X(f_2)\circ X(f_1) = X(f_2 \circ f_1)$
\end{enumerate}
\end{definition}

\paragraph{The fundamental theorems about smooth functions}

The special properties smooth functions that make them play an 
important role different from other classes of functions are the following.

\begin{enumerate}
\item existence of bump functions and partitions of unity
\item the Hadamard lemma and Borel's theorem
\end{enumerate}

Or maybe better put: what makes smooth functions special is that the first of these properties holds, while the second is still retained.

\subsubsection{Smooth 0-types}
\label{GeometryOfPhysicSmoothSpaces}

We now discuss concretely the definition of smooth sets/smooth spaces and of  homomorphisms between them,  together with basic examples and properties.

\paragraph{Plots of smooth spaces and their gluing}
\label{PlotsOfSmoothSpacesAndTheirGluing}

The general kind of ``smooth space'' that we want to consider is something that can be 
\emph{probed} by laying out coordinate systems  
inside it, and that can be obtained by \emph{gluing} all the possible coordinate systems in it together. 

At this point we want to impose no further conditions on a ``space'' than this. In particular we do not assume that we know beforehand a set of points underlying $X$. Instead, we define smooth spaces $X$ entirely 
\emph{operationally} as something about which we can ask 
``Which ways are there to lay out $\mathbb{R}^n$ inside $X$?'' and such that there is a 
self-consistent answer to this question. The following definitions make precise what we mean by this.

For brevity we will refer ``a way to lay out a coordinate system in $X$'' as a 
\emph{plot} of $X$. The first set of consistency conditions on plots of a space is that they respect 
\emph{coordinate transformations}. This is what the following definition formalizes.

\begin{definition}
\label{SmoothPreSpace}
A {\it smooth pre-space $X$} is 
\begin{enumerate}
\item a collection of sets: 
for each Cartesian space $\mathbb{R}^n$ (hence for each natural number $n$) a set 
   $$
     X(\mathbb{R}^n) \in \mathrm{Set}
   $$  
   -- to be thought of as the \emph{set of ways of laying out $\mathbb{R}^n$ inside $X$};
\item  for each abstract coordinate transformation, hence for each smooth function 
$f : \mathbb{R}^{n_1} \to \mathbb{R}^{n_2}$ a function between the corresponding sets
   $$
     X(f) : X(\mathbb{R}^{n_2}) \to X(\mathbb{R}^{n_1})
   $$

   -- to be thought of as the function that sends a \emph{plot} 
   of $X$ by $\mathbb{R}^{n_2}$ to the correspondingly transformed plot by $\mathbb{R}^{n_1}$ 
   induced by laying out $\mathbb{R}^{n_1}$ inside $\mathbb{R}^{n_2}$.
\end{enumerate}
such that this is compatible with coordinate transformations:
\begin{enumerate}
\item the identity coordinate transformation does not change the plots:
   $$
     X(id_{\mathbb{R}^n}) = id_{X(\mathbb{R}^n)}
     \,,
   $$
\item changing plots along two consecutive coordinate transformations $f_1 \colon \mathbb{R}^{n_1} \to \mathbb{R}^{n_2}$ and $f_2 \colon \mathbb{R}^{n_2} \to \mathbb{R}^{n_3}$ is the same as changing them along the 
composite coordinate transformation $f_2 \circ f_1$:
   $$
     X(f_1) \circ X(f_2)  = X(f_2 \circ f_1)
     \,.
   $$
 \end{enumerate}
\end{definition}
But there is one more consistency condition for a collection of plots to really be probes of some space: it must be true that if we glue small coordinate systems to larger ones, then the plots by the larger ones are the same as the plots by the collection of smaller ones that agree where they overlap. 
We first formalize this idea of ``plots that agree where their coordinate systems overlap''.

\begin{definition}
\label{MatchingFamiliesOfPlots}
Let $X$ be a smooth pre-space, def. \ref{SmoothPreSpace}.
For $\{U_i \to \mathbb{R}^n\}_{i \in I}$ 
a differentially good open cover, def. \ref{DifferentiallyGoodOpenCover}, let 
$$
  \mathrm{GluedPlots}(\{U_i \to \mathbb{R}^n\}, X) \in \mathrm{Set}
$$
be the set of $I$-tuples of $U_i$-plots of $X$ which coincide on all double intersections 
$$  
  \xymatrix{
    & U_i \cap U_j
	\ar[dl]_{\iota_i}
	\ar[dr]^{\iota_j}
	\\
	U_i \ar[dr] && U_J \ar[dl]
	\\
	& \mathbb{R}^n
  }
$$ 
(also called the \emph{matching families} of $X$ over the given cover):
$$
  \mathrm{GluedPlots}(\{U_i \to \mathbb{R}^n\}, X)
  \;:=\;
  \left\{
    \;
    \left(p_i \in X(U_i)\right)_{i \in I}
    \;|\;\;
    \forall_{i,j \in I} \;:\; X(\iota_i)(p_i) = X(\iota_j)(p_j)
    \;
  \right\}
  \,.
$$
\end{definition}

\begin{remark}
In def. \ref{MatchingFamiliesOfPlots} the equation
$$ 
  X(\iota_i)(p_i) = X(\iota_j)(p_j)
$$
says in words:

``The plot $p_i$ of $X$ by the coordinate system $U_i$ inside the bigger coordinate system $\mathbb{R}^n$ coincides with the plot $p_j$ of $X$ by the other coordinate system $U_j$ inside $X$ when both are restricted to the intersection $U_i \cap U_j$ of $U_i$ with $U_j$ inside $\mathbb{R}^n$.''
\end{remark}

\begin{remark}
\label{NaiveDescentMorphism}
For each differentially good open cover $\{U_i \to X\}_{i \in I}$ and each smooth pre-space $X$, 
def. \ref{SmoothPreSpace}, there is a canonical function
$$
  X(\mathbb{R}^n) \to \mathrm{GluedPlots}(\{U_i \to \mathbb{R}^n\}, X)
$$
from the set of $\mathbb{R}^n$-plots of $X$ to the set of tuples of glued plots, which sends a plot $p \in X(\mathbb{R}^n)$ to its restriction to all the $\phi_i \colon U_i \hookrightarrow \mathbb{R}^n$:
$$
  p \mapsto (X(\phi_i)(p))_{i \in I}
  \,.
$$
\end{remark}
If $X$ is supposed to be consistently probable by coordinate systems, then it must be true that the set of ways of laying out a coordinate system $\mathbb{R}^n$ inside it coincides with the set of ways of laying out tuples of glued coordinate systems inside it, for each good cover $\{U_i \to \mathbb{R}^n\}$ as above. Therefore:
\begin{definition}
\label{SmoothSpace}
A smooth pre-space $X$, def. \ref{SmoothPreSpace} is a {\it smooth space} if for all differentially good open covers $\{U_i \to \mathbb{R}^n\}$, def. \ref{DifferentiallyGoodOpenCover}, 
the canonical function of remark \ref{NaiveDescentMorphism} from plots to glued plots is a bijection
$$
  X(\mathbb{R}^n) \stackrel{\simeq}{\to} \mathrm{GluedPlots}(\{U_i \to \mathbb{R}^n\}, X)
  \,.
$$
\end{definition}

\begin{remark}
\label{OnTheNotionOfSmoothSpaces}
We may think of a smooth space as being a kind of space whose 
\emph{local models} (in the general sense discussed at \emph{geometry}) are Cartesian spaces: 

while definition \ref{SmoothSpace} explicitly says that a smooth space is something that is 
\emph{consistently probeable} by such local models; by a general abstract fact 
that is sometimes called the \emph{co-Yoneda lemma}, it follows in fact that smooth spaces are precisely the objects that are obtained by 
\emph{gluing coordinate systems} together.

For instance we will see that two open 2-balls $\mathbb{R}^2 \simeq D^2$ along a common rim yields the smooth space version of the sphere $S^2$, a basic example of a smooth manifold. But before we examine such explicit constructions, we discuss here for the moment more general properties of smooth spaces. 
\end{remark}
\begin{example}
\label{CartesianSpaceAsSmoothSpace}
For $n \in \mathbb{R}^n$, there is a smooth space, def. \ref{SmoothSpace}, whose set of plots over the abstract coordinate systems $\mathbb{R}^k$ is the set
$$
  \mathrm{CartSp}(\mathbb{R}^k, \mathbb{R}^n) \in \mathrm{Set}
$$ 
of smooth functions from $\mathbb{R}^k$ to $\mathbb{R}^n$.

Clearly this is the rule for plots that characterize $\mathbb{R}^n$ itself as a smooth space, and so we will just denote this smooth space by the same symbols ``$\mathbb{R}^n$'':
$$
  \mathbb{R}^n \colon \mathbb{R}^k \mapsto \mathrm{CartSp}(\mathbb{R}^k, \mathbb{R}^n)
  \,.
$$
In particular the real line $\mathbb{R}$ is this way itself a smooth space.
\end{example}
In a moment we find a formal justification for this slight abuse of notation.

Another basic class of examples of smooth spaces are the discrete smooth spaces:

\begin{definition}
\label{DiscreteSmoothSpace}
For $S \in \mathrm{Set}$  a set, write
$$
  \mathrm{Disc} S \in \mathrm{Smooth}0\mathrm{Type}
$$
for the smooth space whose set of $U$-plots for every $U \in \mathrm{CartSp}$ is always $S$.
$$
  \mathrm{Disc} S \colon U \mapsto S
$$
and which sends every coordinate transformation $f \colon \mathbb{R}^{n_1} \to \mathbb{R}^{n_2}$ to the identity function on $S$.

A smooth space of this form we call a {\it discrete smooth space}.
\end{definition}

More examples of smooth spaces can be built notably by intersecting images of two smooth spaces inside a bigger one. In order to say this we first need a formalization of homomorphism of smooth spaces. This we turn to now.

\paragraph{Homomorphisms of smooth spaces}
\label{HomomorphismsOfSmoothSpaces}

We discuss ``functions'' or ``maps'' between smooth spaces, def. \ref{SmoothSpace}, which preserve the smooth space structure in a suitable sense. As with any notion of function that preserves structure, we refer to them as \emph{homomorphisms}.

The idea of the following definition is to say that whatever a homomorphism $f : X \to Y$ between two smooth spaces is, it has to take the plots of $X$ by $\mathbb{R}^n$ to a corresponding plot of $Y$, such that this respects coordinate transformations.

\begin{definition}
\label{HomomorphismOfSmoothSpaces}
Let $X$ and $Y$ be two smooth spaces, def. \ref{SmoothSpace}. Then a homomorphism $f \colon X \to Y$ is
\begin{itemize}
\item
 for each abstract coordinate system $\mathbb{R}^n$ (hence for each $n \in \mathbb{N}$) a function
  $f_{\mathbb{R}^n} : X(\mathbb{R}^n) \to Y(\mathbb{R}^n)$
  that sends $\mathbb{R}^n$-plots of $X$ to $\mathbb{R}^n$-plots of $Y$ 
\end{itemize}
such that 
\begin{itemize}
\item for each smooth function $\phi : \mathbb{R}^{n_1} \to \mathbb{R}^{n_2}$ we have
  $$
    Y(\phi) \circ f_{\mathbb{R}^{n_1}} = f_{\mathbb{R}^{n_2}} \circ X(\phi)
	\,,
  $$   
  hence a commuting diagram
  $$
    \xymatrix{
      X(\mathbb{R}^{n_1}) \ar[r]^{f_{\mathbb{R}^{n_1}}} \ar[d]^{{X(\phi)}} 
	  & Y(\mathbb{R}^{n_1}) \ar[d]^{{Y(\phi)}}
      \\
      X(\mathbb{R}^{n_2}) \ar[r]^{f_{\mathbb{R}^{n_2}}} & Y(\mathbb{R}^{n_1})
    }
    \,.
  $$
\end{itemize}
For $f_1 : X \to Y$ and $f_2 : X \to Y$ two homomorphisms of smooth spaces, their composition $f_2 \circ f_1 \colon X \to Y$ is defined to be the homomorphism whose component over $\mathbb{R}^n$ is the composite of functions of the components of $f_1$ and $f_2$:
$$
  (f_2\circ f_1)_{\mathbb{R}^n} := {f_2}_{\mathbb{R}^n} \circ {f_1}_{\mathbb{R}^n}
  \,.
$$
\end{definition}
\begin{definition}
Write $\mathrm{Smooth}0\mathrm{Type}$ for the category whose objects are 
smooth spaces, def. \ref{SmoothSpace}, and whose morphisms are homomorphisms of smooth spaces, 
def. \ref{HomomorphismOfSmoothSpaces}.
\end{definition}

At this point it may seem that we have now \emph{two different} notions for how to lay out a coordinate system in a smooth space $X$: on the hand, $X$ comes by definition with a rule for what the set $X(\mathbb{R}^n)$ of its $\mathbb{R}^n$-plots is. On the other hand, we can now regard the abstract coordinate system $\mathbb{R}^n$ itself as a smooth space, by example \ref{CartesianSpaceAsSmoothSpace}, and then say that an  $\mathbb{R}^n$-plot of $X$ should be a homomorphism of smooth spaces of the form $\mathbb{R}^n \to X$.

The following proposition says that these two superficially different notions actually naturally coincide. 

\begin{proposition}
\label{YonedaForSmoothSpaces}
Let $X$ be any smooth space, def. \ref{SmoothSpace}, and regard the abstract coordinate system $\mathbb{R}^n$ as a smooth space, by example \ref{CartesianSpaceAsSmoothSpace}. There is a natural bijection
$$
  X(\mathbb{R}^n) \simeq Hom_{\mathrm{Smooth}0\mathrm{Type}}(\mathbb{R}^n, X)
$$
between the \emph{postulated} $\mathbb{R}^n$-plots of $X$ and the 
\emph{actual} $\mathbb{R}^n$-plots given by homomorphism of smooth spaces $\mathbb{R}^n \to X$.
\end{proposition}
\proof
This is a special case of the \emph{Yoneda lemma}.
The reader unfamiliar with this should write out the simple proof explicitly: 
use the defining commuting diagrams in def. \ref{HomomorphismOfSmoothSpaces} to deduce that a homomorphism $f : \mathbb{R}^n \to X$ is uniquely fixed by the image of the identity element in  $\mathbb{R}^n(\mathbb{R}^n) := CartSp(\mathbb{R}^n, \mathbb{R}^n)$ under the component function $f_{\mathbb{R}^n} : \mathbb{R}^n(\mathbb{R}^n) \to X(\mathbb{R}^n)$.
\endofproof

\begin{example}
\label{SmoothFunctionOnSmoothSpace}
Let $\mathbb{R} \in \mathrm{Smooth}0\mathrm{Type}$ denote the real line, 
regarded as a smooth space by def. \ref{CartesianSpaceAsSmoothSpace}. 
Then for $X \in \mathrm{Smooth}0\mathrm{Type}$ any smooth space, a homomorphism of smooth spaces
$$
  f \colon X \to \mathbb{R}
$$
is a \emph{smooth function on $X$} Prop. \ref{YonedaForSmoothSpaces} says here that when $X$ happens to be an abstract coordinate system regarded as a smooth space by def. \ref{CartesianSpaceAsSmoothSpace}, then this general notion of smooth functions between smooth spaces reproduces the basic notion of def, 
\ref{CartesianSpaceAndHomomorphism}.
\end{example}

\begin{definition}
\label{PointsOfASmoothSpace}
The 0-dimensional abstract coordinate system $\mathbb{R}^0$ we also call the 
\emph{point} and regarded as a smooth space we will often write it as
$$
  * \in \mathrm{Smooth}0\mathrm{Type}
  \,.
$$
For any $X \in \mathrm{Smooth}0\mathrm{Type}$, we say that a homomorphism
$$
  x \colon * \to X
$$ 
is a \emph{point of $X$}.
\end{definition}

\begin{remark}
By prop. \ref{YonedaForSmoothSpaces} the points of a smooth space $X$ are naturally identified with its 0-dimensional plots, hence with the ``ways of laying out a 0-dimensional coordinate system'' in $X$:
$$
  Hom(*, X) \simeq X(\mathbb{R}^0)
  \,.
$$
\end{remark}

\paragraph{Products and fiber products of smooth spaces}
\label{ProductsAndFiberProductsOfSmoothSpaces}

\begin{definition}
\label{ProductOfSmoothSpaces}
Let $X, Y \in Smooth0Type$ by two smooth spaces. 
Their {\it product} is the smooth space 
$X \times Y \in \mathrm{Smooth}0\mathrm{Type}$ whose plots are pairs of plots of $X$ and $Y$:
$$
  X\times Y (\mathbb{R}^n) := X(\mathbb{R}^n) \times Y(\mathbb{R}^n)
  \;\;
  \in \mathrm{Set}
  \,.
$$
The {\it projection on the first factor} is the homomorphism
$$
   p_1 \colon X \times Y \to X
$$
which sends $\mathbb{R}^n$-plots of $X \times Y$ to those of $X$ by forming the projection of the cartesian product of sets:
$$
  {p_1}_{\mathbb{R}^n} \colon X(\mathbb{R}^n) \times Y(\mathbb{R}^n) \stackrel{p_1}{\to} X(\mathbb{R}^n)
  \,.
$$
Analogously for the \emph{projection to the second factor}
$$
  p_2 \colon X \times Y \to Y
  \,.
$$
\end{definition}
\begin{proposition}
\label{ProductOfSmoothSpaceWithThePoint}
Let $* = \mathbb{R}^0$ be the point, regarded as a smooth space, def. \ref{PointsOfASmoothSpace}. 
Then for $X \in \mathrm{Smooth}0\mathrm{Type}$ any smooth space the canonical projection homomorphism
$$
  X \times * \to X
$$
is an isomorphism.
\end{proposition}
\begin{definition}
Let $f \colon X \to Z$ and $g \colon Y \to Z$ be two homomorphisms of smooth spaces, 
def. \ref{HomomorphismOfSmoothSpaces}. There is then a new smooth space to be denoted 
$$
  X \times_Z Y \in \mathrm{Smooth}0\mathrm{Type}
$$ 
(with $f$ and $g$ understood), 
called the \emph{fiber product} of $X$ and $Y$ along $f$ and $g$, and defined as follows:

the set of $\mathbb{R}^n$-plots of $X \times_Z Y$ is the set of pairs of plots of $X$ and $Y$ which become the same plot of $Z$ under $f$ and $g$, respectively:
$$
  (X \times_Z Y)(\mathbb{R}^n)
  =
  \left\{
    (p_X \in X(\mathbb{R}^n), p_Y \in Y(\mathbb{R}^n))
    \; |\;
    f_{\mathbb{R}^n}(p_X) = g_{\mathbb{R}^n}(p_Y)
  \right\}
  \,.
$$
\end{definition}

\paragraph{Smooth mapping spaces and smooth moduli spaces}
\label{SmoothMappingSpaces}

\begin{definition}
\label{SmoothFunctionSpace}
Let $\Sigma, X \in Smooth0Type$ be two smooth spaces, def. \ref{SmoothSpace}. 
Then the \emph{smooth mapping space} 
$$
  [\Sigma,X] \in \mathrm{Smooth}0\mathrm{Type}
$$ 
is the smooth space defined by saying that its set of $\mathbb{R}^n$-plots is

$$
  [\Sigma, X](\mathbb{R}^n)
   :=
  \mathrm{Hom}(\Sigma \times \mathbb{R}^n, X)
  \,.
$$
\end{definition}

Here in $\Sigma \times \mathbb{R}^n$ we first regard the abstract coordinate system $\mathbb{R}^n$ as a smooth space by example \ref{CartesianSpaceAsSmoothSpace} and then we form the product smooth space by 
def. \ref{ProductOfSmoothSpaces}.
\begin{remark}
\label{NatureOfPlotsOfMappingSpace}
This means in words that a $\mathbb{R}^n$-plot of the mapping space $[\Sigma,X]$ is a smooth $\mathbb{R}^n$-parameterized collection of homomorphisms $\Sigma \to X$.
\end{remark}
\begin{proposition}
\label{UniversalPropertyOfMappingSpace}
There is a natural bijection
$$
  \mathrm{Hom}(K, [\Sigma, X])
  \simeq
  \mathrm{Hom}(K \times \Sigma, X)  
$$
for every smooth space $K$.
\end{proposition}
\proof
With a bit of work this is straightforward to check explicitly by unwinding the definitions. It follows however from general abstract results once we realize that $[-,-]$ is of course the 
\emph{internal hom} of smooth spaces. 
\endofproof
\begin{remark}
\label{MappingSpaceAsModuliSpace}
This says in words that a smooth function from any $K$ into the mapping space $[\Sigma,X]$ is equivalently a smooth function from $K \times \Sigma$ to $X$. The latter we may regard as 
a \emph{$K$-parameterized smooth collections} of smooth functions $\Sigma \to X$. Therefore in view of the previous remark \ref{NatureOfPlotsOfMappingSpace} this says that smooth mapping spaces have a universal property not just over abstract coordinate systems, but over all smooth spaces.

We will therefore also say that $[\Sigma,X]$ is the \emph{smooth moduli space} of smooth functions from $\Sigma \to X$, because it is such that smooth maps $K \to [\Sigma,X]$ into it \emph{modulate}, as we move around on $K$, a family of smooth functions $\Sigma\to X$, depending on $K$.
\end{remark}
\begin{proposition}
\label{UnderlyingSetOfSmoothMappingSpace}
The set of points, def. \ref{PointsOfASmoothSpace}, of a smooth mapping space $[\Sigma,X]$ is the bare set of homomorphism $\Sigma \to X$: there is a natural isomorphism
$$
  \mathrm{Hom}(*, [\Sigma, X]) \simeq \mathrm{Hom}(\Sigma, X)
  \,.
$$
\end{proposition}
\proof
Combine prop. \ref{UniversalPropertyOfMappingSpace} with prop. \ref{ProductOfSmoothSpaceWithThePoint}.
\endofproof
\begin{example}
\label{SmoothPathSpace}
Given a smooth space $X \in Smooth0Type$, its smooth 
\emph{path space} is the smooth mapping space
$$
  \mathbf{P}X := [\mathbb{R}^1, X]
  \,.
$$ 
By prop. \ref{UnderlyingSetOfSmoothMappingSpace} the points of $P X$ are indeed precisely the smooth trajectories $\mathbb{R}^1 \to X$. But $P X$ also knows how to 
\emph{smoothly vary} such smooth trajectories. 
\end{example}
This is central for variational calculus which determines equations of motion in physics.
\begin{remark}
In physics, if $X$ is a model for spacetime, then $P X$ may notably be interpreted as the smooth space of worldlines \emph{in} $X$, hence the smooth space of paths or \emph{trajectories} of a particle in $X$. 
\end{remark}
\begin{example}
\label{SmoothLoopSpace}
If in the above example \ref{SmoothPathSpace} the path is
constraind to be a loop in $X$, one obtains the \emph{smooth loop space}
$$
  \mathbf{L}X := [S^1, X]
  \,.
$$
\end{example}

\paragraph{The smooth moduli space of smooth functions}
\label{SmoothModuliSpaceOfSmoothFunctions}

In example \ref{SmoothFunctionOnSmoothSpace} we saw that a smooth function on a general smooth space $X$ is a homomorphism of smooth spaces, def. \ref{HomomorphismOfSmoothSpaces}
$$
  f \colon X \to \mathbb{R}
  \,.
$$
The collection of these forms the hom-set 
$\mathrm{Hom}_{\mathrm{Smooth}0\mathrm{Type}}(X, \mathbb{R})$. 
But by the discussion in \ref{SmoothMappingSpaces} such hom-sets are naturally 
refined to smooth spaces themselves.

\begin{definition}
\label{SmoothSpaceOfSmoothFunctions}
For $X \in Smooth0Type$ a smooth space, we say that the 
\emph{moduli space of smooth functions} on $X$ is the smooth mapping space 
(def. \ref{SmoothFunctionSpace}), from $X$ into the standard real line $\mathbb{R}$
$$
  [X, \mathbb{R}] \in \mathrm{Smooth}0\mathrm{Type}
  \,.
$$
We will also denote this by 
$$
  \mathbf{C}^\infty(X) := [X, \mathbb{R}]
  \,,
$$ 
since in the special case that $X$ is a Cartesian space this is the smooth refinement 
of the set $C^\infty(X)$ of smooth functions, def. \ref{SmoothFunctions}, on $X$.
\end{definition}
\begin{remark}
We call this a \emph{moduli space} because by prop. \ref{UniversalPropertyOfMappingSpace} 
above and in the sense of remark \ref{MappingSpaceAsModuliSpace} it is such that smooth functions into 
it \emph{modulate} smooth functions $X \to \mathbb{R}$.

By prop. \ref{UnderlyingSetOfSmoothMappingSpace} a point $* \to [X,\mathbb{R}^1]$ of the moduli space is equivalently a smooth function $X \to \mathbb{R}^1$.
\end{remark}

\paragraph{Outlook}
\label{SmoothSpacesOutlook}

Later we define/see the following:
\begin{itemize}
\item A \emph{smooth manifold}  is a smooth space that is \emph{locally equivalent} to a coordinate system;

\item A \emph{diffeological space} is a smooth space such that every coordinate labels a point in the space. 
In other words, a diffeological space is a smooth space that has an underlying set $X_s \in Set$ of points such that the set of $\mathbb{R}^n$-plots is a subset of the set of all functions:
  $$
    X(\mathbb{R}^n) \hookrightarrow \mathrm{Functions}(\mathbb{R}^n, S_s)
    \,.
  $$
\end{itemize}
We discuss below a long sequence of faithful inclusions 

$\{$coordinate systems $\}$
 $\hookrightarrow$
$\{$smooth manifolds$\}$ 
 $\hookrightarrow$
$\{$diffeological spaces$\}$ 
 $\hookrightarrow$
$\{$smooth spaces$\}$ 
 $\hookrightarrow$
$\{$smooth groupoids$\}$ 
 $\hookrightarrow \cdots $

\subsubsection{Differential forms}
\label{TheGeometryOfPhysicsDifferentialForms}
\index{differential form!on smooth space}

A fundamental concept in differential geometry is that of 
\emph{differential forms}. We here introduce this in the spirit 
of the topos of smooth spaces.

\paragraph{Differential forms on abstract coordinate systems}

We introduce the basic concept of a \emph{smooth differential form} 
on a Cartesian space $\mathbb{R}^n$. Below in 
\ref{SmoothUniversalModuliSpaceOfDifferentialForms} 
we use this to define differential forms on any smooth space.

\begin{definition}
For $n \in \mathbb{N}$ a 
\emph{smooth differential 1-form} $\omega$ 
on the Cartesian space $\mathbb{R}^n$ is an $n$-tuple 
$$
  \left(\omega_i \in \mathrm{CartSp}\left(\mathbb{R}^n,\mathbb{R}\right)\right)_{i = 1}^n
$$
of smooth functions, which we think of equivalently as the coefficients of a formal linear combination
$$
  \omega = \sum_{i = 1}^n f_i \mathbf{d}x^i
$$
on a set $\{\mathbf{d}x^1, \mathbf{d}x^2, \cdots, \mathbf{d}x^n\}$ of cardinality $n$.

Write 
$$
  \Omega^1(\mathbb{R}^k) \simeq \mathrm{CartSp}(\mathbb{R}^k, \mathbb{R})^{\times k}\in 
  \mathrm{Set}
$$
for the set of smooth differential 1-forms on $\mathbb{R}^k$.
 \label{Differential1FormsOnCartesianSpaces}
\end{definition}

\begin{remark}
We think of $\mathbf{d} x^i$ as a measure for infinitesimal displacements 
along the $x^i$-coordinate of a Cartesian space. This idea is made precise 
by the notion of \emph{parallel transport}.
\end{remark}

If we have a measure of infintesimal displacement on some $\mathbb{R}^n$ 
and a smooth function $f \colon \mathbb{R}^{\tilde n} \to \mathbb{R}^n$, 
then this induces a measure for infinitesimal displacement on $\mathbb{R}^{\tilde n}$ 
by sending whatever happens there first with  $f$ to $\mathbb{R}^n$ and 
then applying the given measure there. This is captured by the following definition.

\begin{definition}
For $\phi \colon \mathbb{R}^{\tilde k} \to \mathbb{R}^k$ a smooth function, 
the \emph{pullback of differential 1-forms} along $\phi$ is the function
$$
  \phi^\ast \colon \Omega^1(\mathbb{R}^{k}) \to \Omega^1(\mathbb{R}^{\tilde k})
$$
between sets of differential 1-forms, def. \ref{Differential1FormsOnCartesianSpaces}, which is defined on basis-elements by
$$
  \phi^\ast \mathbf{d} x^i := \sum_{j = 1}^{\tilde k} \frac{\partial \phi^i}{\partial \tilde x^j} \mathbf{d}\tilde x^j
$$
and then extended linearly by
$$
  \begin{aligned}
    \phi^* \omega & = \phi^* \left( \sum_{i} \omega_i \mathbf{d}x^i \right)
    \\
    & :=
     \sum_{i = 1}^k \left(\phi^* \omega\right)_i \sum_{j = 1}^{\tilde k} \frac{\partial \phi^i }{\partial \tilde x^j}  \mathbf{d} \tilde x^j 
    \\
    & = 
     \sum_{i = 1}^k  \sum_{j = 1}^{\tilde k} (\omega_i \circ \phi) \cdot \frac{\partial \phi^i }{\partial \tilde x^j}  \mathbf{d} \tilde x^j 
  \end{aligned}
  \,.
$$
\label{PullbackOfDifferential1FormsOnCartesianSpaces}
\end{definition}

\begin{remark}
The term ``pullback'' in \emph{pullback of differential forms} is not really related, 
certainly not historically, to the term \emph{pullback} in category theory. 
One can relate the pullback of differential forms to categorical pullbacks, but this is not really essential here. The most immediate property that both concepts share is that they take a morphism going in one direction to a map between structures over domain and codomain of that morphism which goes in the other direction, and in this sense one is ``pulling back structure along a morphism'' in both cases.
\end{remark}
Even if in the above definition we speak only about the set $\Omega^1(\mathbb{R}^k)$ 
of differential 1-forms, this set naturally carries further structure.

\begin{definition}
The set $\Omega^1(\mathbb{R}^k)$ is naturally an abelian group with addition given by componentwise addition

 $$
    \begin{aligned}
       \omega + \lambda & =
         \sum_{i = 1}^k \omega_i \mathbf{d}x^i + \sum_{j = 1}^k \lambda_j \mathbf{d}x^j
       \\
       & = \sum_{i = 1}^k(\omega_i + \lambda_i) \mathbf{d}x^j
    \end{aligned}
    \,,
$$

Moreover, the abelian group $\Omega^1(\mathbb{R}^k)$ is naturally equipped with the structure of a module over the ring $C^\infty(\mathbb{R}^k,\mathbb{R}) = \mathrm{CartSp}(\mathbb{R}^k, \mathbb{R})$ of smooth functions, where the action $C^\infty(\mathbb{R}^k,\mathbb{R}) \times\Omega^1(\mathbb{R}^k) \to \Omega^1(\mathbb{R}^k)$ is given by componentwise multiplication

   $$
     f \cdot \omega = \sum_{i = 1}^k( f \cdot \omega_i) \mathbf{d}x^i
     \,.
   $$
   
 \label{ModuleStructureOn1FormsOnRk}
\end{definition}

\begin{remark}
More abstractly, this just says that $\Omega^1(\mathbb{R}^k)$ is the free module over $C^\infty(\mathbb{R}^k)$ on the set $\{\mathbf{d}x^i\}_{i = 1}^k$.
\end{remark}
The following definition captures the idea that if $\mathbf{d} x^i$ is a measure for displacement along the $x^i$-coordinate, and $\mathbf{d}x^j$ a measure for displacement along the $x^j$ coordinate, then there should be a way te get a measure, to be called $\mathbf{d}x^i \wedge \mathbf{d} x^j$, for infinitesimal \emph{surfaces} (squares) in the $x^i$-$x^j$-plane. And this should keep track of the orientation of these squares, whith 

$$
  \mathbf{d}x^j \wedge \mathbf{d}x^i = - \mathbf{d}x^i \wedge \mathbf{d} x^j
$$

being the same infinitesimal measure with orientation reversed.

\begin{definition}
For $k,n \in \mathbb{N}$, the 
\emph{smooth differential forms} on $\mathbb{R}^k$ is the exterior algebra
$$
  \Omega^\bullet(\mathbb{R}^k) 
   :=
  \wedge^\bullet_{C^\infty(\mathbb{R}^k)} \Omega^1(\mathbb{R}^k)
$$
over the ring $C^\infty(\mathbb{R}^k)$ of smooth functions of the module $\Omega^1(\mathbb{R}^k)$ of smooth 1-forms, prop. \ref{ModuleStructureOn1FormsOnRk}.

We write $\Omega^n(\mathbb{R}^k)$ for the sub-module of degree $n$ and call its elements the 
\emph{smooth differential n-forms}.
 \label{DifferentialnForms}
\end{definition}

\begin{remark}
Explicitly this means that a differential n-form $\omega \in \Omega^n(\mathbb{R}^k)$ on $\mathbb{R}^k$ is a formal linear combination over $C^\infty(\mathbb{R}^k)$ of basis elements of the form $\mathbf{d} x^{i_1} \wedge \cdots \wedge \mathbf{d}x^{i_n}$ for $i_1 < i_2 < \cdots < i_n$:

$$
  \omega = \sum_{1 \leq i_1 < i_2 < \cdots < i_n < k} \omega_{i_1, \cdots, i_n} \mathbf{d}x^{i_1} \wedge \cdots \wedge \mathbf{d}x^{i_n}
  \,.
$$
\end{remark}
\begin{remark}
The pullback of differential 1-forms of def. \ref{Differential1FormsOnCartesianSpaces} extends as an $C^\infty(\mathbb{R}^k)$-algebra homomorphism to $\Omega^n(-)$, given for a smooth function $f \colon \mathbb{R}^{\tilde k} \to \mathbb{R}^k$ on basis elements by
$$
  f^* \left(  \mathbf{d}x^{i_1} \wedge \cdots \wedge \mathbf{d}x^{i_n} \right)
  = 
  \left(f^* \mathbf{d}x^{i_1} \wedge \cdots \wedge f^* \mathbf{d}x^{i_n} \right)
  \,. 
$$ 
\label{PullbackOfDifferentialForms}
\end{remark}

\paragraph{Differential forms on smooth spaces}

Above we have defined differential $n$-form on abstract coordinate systems. Here we extend this definition to one of differential $n$-forms on arbitrary smooth spaces. 
We start by observing that the space of \emph{all}3 differential $n$-forms on cordinate systems themselves naturally is a smooth space.

\begin{proposition}
The assignment of differential $n$-forms
$$
  \Omega^n(-) \colon \mathbb{R}^k \mapsto \Omega^n(\mathbb{R}^k)
$$
of def. \ref{DifferentialnForms} together with the pullback of 
differential forms-functions of def. \ref{PullbackOfDifferentialForms}
$$
  \xymatrix{
    \mathbb{R}^{k_1} \ar@{|->}[r] & \Omega^1(\mathbb{R}^{k_1}) \ar[d]^{f^\ast}
    \\
    \mathbb{R}^{k_2} \ar[u]^f \ar@{|->}[r] & \Omega^1(\mathbb{R}^{k_2})
  }
$$
defines a smooth space in the sense of def. \ref{SmoothSpace}:
$$
  \Omega^n(-) \in Smooth0Type
  \,.
$$
\end{proposition}

\begin{definition}
We call this 
$$
  \Omega^n \colon \mathrm{Smooth}0\mathrm{Type}
$$
the \emph{universal smooth moduli space} of differential $n$-forms. 
 \label{SmoothModuliSpaceOfnForms}
\end{definition}
The reason for this terminology is that homomorphisms of smooth spaces into $\Omega^1$ 
\emph{modulate} differential $n$-forms on their domain, 
by prop. \ref{YonedaForSmoothSpaces} (and hence by the Yoneda lemma):

\begin{example}
For the Cartesian space $\mathbb{R}^k$ regarded as a smooth 
space by example \ref{CartesianSpaceAsSmoothSpace}, there is a natural bijection
$$
  \Omega^n(\mathbb{R}^k) \simeq Hom(\mathbb{R}^k, \Omega^1)
$$
between the set of smooth $n$-forms on $\mathbb{R}^n$ according to def. \ref{Differential1FormsOnCartesianSpaces} and the set of homomorphism of smooth spaces, $\mathbb{R}^k \to \Omega^1$, according to def. \ref{HomomorphismOfSmoothSpaces}.
\end{example}

In view of this we have the following elegant definition of smooth $n$-forms 
on an arbitrary smooth space.

\begin{definition}
For $X \in Smooth0Type$ a smooth space, def. \ref{SmoothSpace}, 
a \emph{differential $n$-form} on $X$ is a homomorphism of smooth spaces of the form
$$
  \omega \colon X \to \Omega^n(-)
  \,.
$$
Accordingly we write
$$
  \Omega^n(X) := Smooth0Type(X,\Omega^n)
$$
for the set of smooth $n$-forms on $X$.
 \label{DifferentialnFormOnSmoothSpace}
\end{definition}

We may unwind this definition to a very explicit description of differential forms on smooth spaces. This we do in a moment in remark \ref{DifferentialFormOnSmoothSpaceAsSystemOfDiffFormsOnCoordinates}.

Notice the following
\begin{proposition}
 Differential 0-forms are equivalently smooth $\mathbb{R}$-valued functions:
$$\Omega^0 \simeq \mathbb{R}\,.$$
\end{proposition}

\begin{definition}
For $f \colon X \to Y$ a homomorphism of smooth spaces, def. \ref{HomomorphismOfSmoothSpaces}, the \emph{pullback of differential forms} along $f$ is the function
$$
  f^\ast \colon \Omega^n(Y) \to \Omega^n(X)
$$
given by the hom-functor into the smooth space $\Omega^n$ of def. \ref{SmoothModuliSpaceOfnForms}:
$$
 f^\ast := Hom(-, \Omega^n)
 \,.
$$
This means that it sends an $n$-form $\omega \in \Omega^n(Y)$ which is modulated by a homomorphism $Y \to \Omega^n$ to the $n$-form $f^* \omega \in \Omega^n(X)$ which is modulated by the composition|composite $X \stackrel{f}{\to} Y \to \Omega^n$.
\label{PullbackOfDifferentialFormsOnSmoothSpaces}
\end{definition}
By the Yoneda lemma we find:
\begin{proposition}
For $X = \mathbb{R}^{\tilde k}$ and $Y = \mathbb{R}^{k}$ definition \ref{PullbackOfDifferentialFormsOnSmoothSpaces} reproduces def. \ref{PullbackOfDifferentialForms}.
\end{proposition}

\begin{remark}
Using def. \ref{PullbackOfDifferentialFormsOnSmoothSpaces}
for unwinding def. \ref{DifferentialnFormOnSmoothSpace} yields the following explicit description:

a differential $n$-form $\omega \in \Omega^n(X)$ on a smooth space $X$ is

\begin{enumerate}
\item for each way $\phi \colon \mathbb{R}^k \to X$ of laying out a coordinate system $\mathbb{R}^k$ in $X$ a differential $n$-form

   $$
     \phi^* \omega \in \Omega^n(\mathbb{R}^k)
   $$

   on the abstract coordinate system, as given by def. \ref{DifferentialnForms};

\item for each abstract coordinate transformation $f \colon \mathbb{R}^{k_2} \to \mathbb{R}^{k_1}$ a corresponding compatibility condition between local differential forms $\phi_1 \colon \mathbb{R}^{k_1} \to X$ and $\phi_2 \colon \mathbb{R}^{k_2} \to X$ of the form

   $$
     f^* \phi_1^* \omega = \phi_2^* \omega
     \,.
   $$
\end{enumerate}

Hence a differential form on a smooth space is simply a collection of differential forms on all its coordinate systems such that these glue along all possible coordinate transformations.
\label{DifferentialFormOnSmoothSpaceAsSystemOfDiffFormsOnCoordinates}
\end{remark}

The following adds further explanation to the role of 
$\Omega^n \in \mathrm{Smooth}0\mathrm{Type}$ as a \emph{moduli space}. Notice that since $\Omega^n$ is itself a smooth space, we may speak about differential $n$-forms on $\Omega^n$ itsefl.

\begin{definition}
The \emph{universal differential $n$-forms} is the differential $n$-form
$$
  \omega^n_{univ} \in \Omega^n(\Omega^n)
$$
which is modulated by the identity homomorphism $id \colon \Omega^n \to \Omega^n$.
\label{UniversalDifferentialnForm}
\end{definition}

With this definition we have:
\begin{proposition}
For $X \in \mathrm{Smooth}0\mathrm{Type}$ any smooth space, 
every differential $n$-form on $X$, $\omega \in \Omega^n(X)$ is the pullback of differential forms, def. \ref{PullbackOfDifferentialFormsOnSmoothSpaces}, of the universal differential $n$-form, def. \ref{UniversalDifferentialnForm}, along a homomorphism $f$ from $X$ into the moduli space $\Omega^n$ of differential $n$-forms:
$$
  \omega = f^* \omega^n_{\mathrm{univ}}
  \,.
$$
\end{proposition}
\begin{remark}
This statement is of course in a way a big tautology. Nevertheless it is a very useful tautology to make explicit. The whole concept of differential forms on smooth spaces here may be thought of as simply a variation of the theme of the Yoneda lemma.
\end{remark}

\paragraph{Concrete smooth spaces}

The smooth universal moduli space of differential forms $\Omega^n(-)$
from def. \ref{SmoothModuliSpaceOfnForms}
is noteworthy in that it has a property not shared by 
many smooth spaces that one might think of more naively: 
while evidently being ``large'' (the space of all differential forms!) 
it has ``very few points'' and``very few $k$-dimensional subspaces'' for low $k$.  In fact 

\begin{proposition}
For $k < n$ the smooth space $\Omega^n$ admits only a unique probe by $\mathbb{R}^k$:
$$
  \mathrm{Hom}(\mathbb{R}^k, \mathbf{\Omega}^n)
  \simeq
  \Omega^n(\mathbb{R}^k)
  = 
  \{0\}
  \,.
$$
\end{proposition}

So while $\mathbf{\Omega}^n$ is a large smooth space, 
it is ``not supported on probes'' in low dimensions in as much as one might expect, from more naive notions of smooth spaces.

We now formalize this. The formal notion of an smooth space which 
is \emph{supported on its probes} is that of a \emph{concrete object}. 
There is a univeral map that sends any smooth space to its 
\emph{concretification}. The universal moduli spaces of differential forms turn out 
to be \emph{non-concrete} in that their concetrification is the point.

\begin{definition}
Let $\mathbf{H}$ be a local topos. Write $\sharp \colon \mathbf{H} \to \mathbf{H}$ 
for the corresponding sharp modality, def. \ref{SharpModalityOfLocalTopos}. Then. 

\begin{enumerate}
\item An object $X \in \mathbf{H}$ is called a \emph{concrete object} if
   $$
     \mathrm{DeCoh}_X \colon X \to \sharp X
   $$
   is a monomorphism.

\item For $X \in \mathbf{H}$ any object, 
its  \emph{concretification} $\mathrm{Conc}(X) \in \mathbf{H}$ is 
the image factorization of $\mathrm{DeCoh}_X$, 
hence the factorization into an epimorphism followed by a monomorphism
   $$
     \mathrm{DeCoh}_X : X \to \mathrm{Conc}(X) \hookrightarrow \sharp X
     \,.
   $$
 \end{enumerate}
\label{ConcreteObjectsAndConcretification}
\end{definition}
\begin{remark}
Hence the concretification $\mathrm{Conc}(X)$ of an object $X$ is itself a concrete object and it is universal property|universal with this property. 
\end{remark}

\begin{proposition}
Let $C$ be a site of definition for the local topos
$\mathbf{H}$, with terminal object $*$. Then for 
$X \colon C^{op} \to Set$ a sheaf, $DeCoh_X$
is given over $U \in C$ by

$$
  X(U)
  \stackrel{\simeq}{\longrightarrow}
  \mathbf{H}(U, X)
  \stackrel{\Gamma_{U,X}}{\to}
  Set(\Gamma(U),\Gamma(X))
  \,.
$$
\label{DecohesOverASiteWithTerminalObject}
\end{proposition}

\begin{proposition}
For $n \geq 1$ we have 
$$
  \mathrm{Conc}(\Omega^n) \simeq *
  \,.
$$
\end{proposition}
In this sense the smooth moduli space of differential $n$-forms is 
\emph{maximally non-concrete}.

\paragraph{Smooth moduli spaces of differential forms on a smooth space}

We discuss the smooth space of differential forms \emph{on a fixed smooth space} $X$. 

\begin{remark}
For $X$ a smooth space, the smooth mapping space 
$[X, \Omega^n] \in \mathrm{Smooth}0\mathrm{Type}$ is the smooth space whose $\mathbb{R}^k$-plots are differential $n$-forms on the product $X \times \mathbb{R}^k$
$$
  [X, \Omega^n] \colon \mathbb{R}^k \mapsto \Omega^n(X \times \mathbb{R}^k)
  \,.
$$
This is not \emph{quite} what one usually wants to regard as an $\mathbb{R}^k$-parameterized of differential forms on $X$. That is instead usually meant to be a differential form $\omega$ on $X \times \mathbb{R}^k$ which has ``no leg along $\mathbb{R}^k$''. Another way to say this is that the family of forms on $X$ that is represented by some $\omega$ on $X \times \mathbb{R}^k$ is that which over a point $v \colon * \to \mathbb{RR}^k$ has the value $(id_X,v)^* \omega$. Under this pullback of differential forms any components of $\omega$ with ``legs along $\mathbb{R}^k$'' are identified with the 0 differential form 
\end{remark}

This is captured by the following definition.
\begin{definition}
For $X \in \mathrm{Smooth}0\mathrm{Type}$ and $n \in \mathbb{N}$, 
the \emph{smooth space of differential $n$-forms} $\mathbf{\Omega}^n(X)$ on $X$ is the concretification, def. \ref{ConcreteObjectsAndConcretification}, of the smooth mapping space $[X, \Omega^n]$, def. \ref{SmoothFunctionSpace}, into the smooth moduli space of differential $n$-forms, 
def. \ref{SmoothModuliSpaceOfnForms}:
$$
  \mathbf{\Omega}^n(X) := \mathrm{Conc}([X, \Omega^n])
  \,.
$$
\label{SmoothSpaceOfFormsOnSmoothSpace}
\label{SmoothUniversalModuliSpaceOfDifferentialForms}
\end{definition}

\begin{proposition}
The $\mathbb{R}^k$-plots of $\mathbf{\Omega}^n(\mathbb{R}^k)$
are indeed smooth differential $n$-forms on $X \times \mathbb{R}^k$ which are such that their evaluation on vector fields tangent to $\mathbb{R}^k$ vanish.
\end{proposition}
\proof
By def. \ref{ConcreteObjectsAndConcretification} and 
prop. \ref{DecohesOverASiteWithTerminalObject} the set of plots of $\mathbf{\Omega}^n(X)$ over $\mathbb{R}^k$ is the image of the function
$$
  \Omega^n(X \times \mathbb{R}^k)
  \simeq
   \mathrm{Hom}_{\mathrm{Smooth}0\mathrm{Type}}(\mathbb{R}^k, [X,\Omega^n])
   \stackrel{\Gamma_{ \mathbb{R}^k, [X,\Omega^n] }}{\to}
   \mathrm{Hom}_{\mathrm{Set}}(\Gamma(\mathbb{R}^k), \Gamma [X, \Omega^n])
   \simeq
   \mathrm{Hom}_{\mathrm{Set}}(\mathbb{R}^k_s, \Omega^n(X))
  \,,
$$
where on the right $\mathbb{R}^k_s$ denotes, just for emphasis, the underlying set of $\mathbb{R}^k_s$. This function manifestly sends a smooth differential form $\omega \in \Omega^n(X \times \mathbb{R}^k)$ to the function from points $v$ of $\mathbb{R}^k$ to differential forms on $X$ given by

$$
  \omega \mapsto \left(v \mapsto  (id_X, v)^* \omega \right)
  \,.
$$

Under this function all components of differential forms with a "leg along" $\mathbb{R}^k$ are sent to the 0-form. Hence the image of this function is the collection of smooth forms on $X \times \mathbb{R}^k$ with ``no leg along $\mathbb{R}^k$''.
\endofproof

\begin{remark}
For $n = 0$ we have (for any $X\in \mathrm{Smooth}0\mathrm{Type}$)
$$
  \begin{aligned}
    \mathbf{\Omega}^0(X) 
      & := Conc [X, \Omega^1]
    \\
    & \simeq Conc [X, \mathbb{R}]
    \\
    & \simeq [X, \mathbb{R}]
  \end{aligned}
  \,,
$$
by prop. \ref{SpaceOf0FormsIsRealLine}.
\label{SpaceOf0FormsIsRealLine}
\end{remark}

\subsubsection{Smooth homotopy types}
\label{IntroGeneralAbstractTheory}

Here we give an introduction to and a survey of the  
general theory of cohesive differential geometry
that is developed 
in detail below in \ref{GeneralAbstractTheory} below.
\medskip

The framework of all our constructions is 
\emph{topos theory} \cite{Johnstone} or rather, more generally, 
\emph{$\infty$-topos theory} \cite{Lurie}. 
In \ref{IntroToposes} and \ref{IntroInfinToposes} 
below we recall and survey basic notions with 
an eye towards our central example of an $\infty$-topos:
that of smooth $\infty$-groupoids.
In these sections the reader is assumed to be familiar with basic notions of category theory
(such as adjoint functors) and basic notions of 
homotopy theory (such as weak homotopy equivalences). 
A brief introduction to relevant basic concepts
(such as Kan complexes and homotopy pullbacks) 
is given in section \ref{IntroGeneralAbstractTheory},
which can be read independently of the discussion here.

Then in \ref{IntroCohomology} and \ref{IntroHomotopy} we describe, 
similarly in a leisurely manner, 
the intrinsic notions of cohomology and geometric homotopy in an $\infty$-topos. 
Several aspects of the discussion are fairly well-known, we put them in 
the general perspective of (cohesive) $\infty$-topos theory and then go beyond.

Finally in \ref{IntroDiffCohomology} we indicate how the combination of the intrinsic 
cohomology and geometric homotopy in a locally $\infty$-connected $\infty$-topos 
yields an intrinsic notion of differential cohomology in an $\infty$-topos.

\begin{itemize}
  \item \ref{IntroToposes} -- Toposes;
  \item \ref{IntroInfinToposes} -- $\infty$-Toposes;
  \item \ref{IntroCohomology} -- Cohomology;
  \item \ref{IntroHomotopy} -- Homotopy;
  \item \ref{IntroDiffCohomology} -- Differential cohomology.
\end{itemize}

Each of these topics surveyed here are discussed in technical detail below in 
\ref{GeneralAbstractTheory}.

\paragraph{Toposes}
\label{IntroToposes}
\index{topos!overview}

There are several different perspectives on the notion of \emph{topos}. 
One is that a topos is a category that looks like a category of spaces that 
sit by local homeomorphisms 
over a given base space: all spaces that are locally modeled on a given base space.

The archetypical class of examples are sheaf toposes 
over a topological space $X$ denoted $\mathrm{Sh}(X)$. 
These are equivalently categories of {\'e}tale spaces over $X$: 
topological spaces $Y$ that are equipped with a local homeomorphisms $Y \to X$.
When $X = *$ is the point, this is just the category $\mathrm{Set}$ of all sets: 
spaces that are modeled on the point. This is the archetypical topos itself.

What makes the notion of toposes powerful is the following fact: even though the 
general topos contains objects that are considerably different from and possibly 
considerably richer than plain sets and even richer than {\'e}tale spaces over 
a topological space, the general abstract category theoretic properties of every topos 
are essentially the same as those of $\mathrm{Set}$. 
For instance in every topos all small limits and colimits exist and it is cartesian closed 
(even locally). This means that a large number of constructions in $\mathrm{Set}$ have immediate 
analogs internal to every topos, and the analogs of the statements about these constructions that 
are true in $\mathrm{Set}$ are true in every topos.

This may be thought of as saying that toposes are \emph{very nice categories of 
spaces} in that whatever construction on spaces one thinks of
-- for instance formation of quotients or of intersections or of mapping spaces -- 
the resulting space with 
the expected general abstract properties will exist in the topos. 
In this sense toposes are \emph{convenient categories for geometry} -- 
as in: \emph{convenient category of topological spaces}, but even more convenient than that.

On the other hand, we can de-emphasize the role of the objects of the topos and instead treat the 
topos itself as a "generalized space" (and in particular, a categorified space). We then consider 
the sheaf topos $\mathrm{Sh}(X)$ as a representative of $X$ itself, while toposes not of this form 
are ``honestly generalized'' spaces. This point of view is supported by the fact that the 
assignment $X \mapsto \mathrm{Sh}(X)$ is a full embedding of (sufficiently nice) topological 
spaces into toposes, and that many topological properties of a space $X$ can be detected at the 
level of $\mathrm{Sh}(X)$. 

Here we are mainly concerned with toposes that are far from being akin
to sheaves over a topological space, and instead behave like abstract 
\emph{fat points with geometric structure}. This implies that the objects of these toposes
are in turn generalized spaces modeled locally on this geometric structure. Such toposes
are called \emph{gros toposes} or \emph{big toposes}. There is a formalization of the properties
of a topos that make it behave like a big topos of generalized spaces inside of which there is
geometry: this is the notion of \emph{cohesive toposes}.

\subparagraph{Sheaves}

More concretely, the idea of sheaf toposes formalizes the idea that
any notion of space is typically modeled on a given collection of simple test spaces. 
For instance differential geometry 
is the geometry that is modeled Cartesian spaces $\mathbb{R}^n$, 
or rather on the category $C = \mathrm{CartSp}$ of Cartesian spaces and smooth functions between them.

A presheaf on such $C$ is a functor $X : C^{\mathrm{op}} \to \mathrm{Set}$ from the opposite category of 
$C$ to the category of sets. 
We think of this as a rule that assigns to each test space $U \in C$ the set 
$X(U) :=: \mathrm{Maps}(U,X)$ of structure-preserving maps from the test space $U$ into the 
would-be space $X$ - the \emph{probes} of $X$ by the test space $U$. 
This assignment defines the generalized space $X$ modeled on $C$. 
Every category of presheaves over a small category is an example of a topos. 
But these presheaf toposes, while encoding the \emph{geometry} of generalized spaces 
by means of probes by test spaces in $C$ fail to 
correctly encode the \emph{topology} of these spaces. This is captured by
restricting to \emph{sheaves} among all presheaves.

Each test space $V \in C$ itself specifies presheaf, by forming the 
hom-sets $\mathrm{Maps}(U,V):= \mathrm{Hom}_C(U,V)$ in $C$. 
This is called the \emph{Yoneda embedding} of test spaces into the collection of all 
generalized spaces modeled on them. Presheaves of this form are the \emph{representable presheaves}. 
A bit more general than these are the \emph{locally representable presheaves}: 
for instance on $C = \mathrm{CartSp}$ this are the smooth manifolds $X \in \mathrm{SmoothMfd}$, 
whose presheaf-rule is $\mathrm{Maps}(U,X):=\mathrm{Hom}_{\mathrm{SmoothMfd}}(U,X)$. 
By definition, a manifold is locally isomorphic to a Cartesian space, 
hence is locally representable as a presheaf on $\mathrm{CartSp}$.

These examples of presheaves on $C$ are special in that they are in fact \emph{sheaves}: 
the value of $X$ on a test space $U$ is entirely determined by the restrictions to each $U_i$ 
in a \emph{cover} $\{U_i \to U\}_{i \in I}$ of the test space $U$ by other test spaces $U_i$. 
We think of the subcategory of sheaves $\mathrm{Sh}(C) \hookrightarrow \mathrm{PSh}(C)$ 
as consisting of those special presheaves that are those rules of probe-assignments which 
respect a certain notion of ways in which test spaces $U,V \in C$ 
may glue together to a bigger test space.

One may axiomatize this by declaring that the collections of all covers under consideration 
forms what is called a \emph{Grothendieck topology} on $C$ that makes $C$ a \emph{site}. 
But of more intrinsic relevance is the equivalent fact that categories of sheaves 
are precisely the subtoposes of presheaves toposes
$$
  \xymatrix{
    \mathrm{Sh}(C) 
	  \ar@{<-}@<+4pt>[r]^L
	  \ar@{^{(}->}@<-4pt>[r]
      &
     \mathrm{PSh}(C) 
	 \ar@{=}[r]
	 &
	 [C^{\mathrm{op}}, \mathrm{Set}]
  }
  \,,
$$
meaning that the embedding $\mathrm{Sh}(X) \hookrightarrow \mathrm{PSh}(X)$ has a 
left adjoint functor $L$ that preserves finite limits. This may be taken to be the 
\emph{definition} of Grothendieck toposes. The left adjoint is called the 
\emph{sheafification functor}. It is determined by and determines a Grothendieck topology on $C$. 

For the choice $C = \mathrm{CartSp}$ such is naturally given by the good open cover coverage, 
which says that a bunch of maps $\{U_i \to U\}$ in $C$ exhibit the test object $U$ as being 
glued together from the test objects $\{U_i\}$ if these form a good open cover of $U$. 
With this notion of coverage every smooth manifold is a sheaf on $\mathrm{CartSp}$.

But there are important genenralized spaces modeled on $\mathrm{CartSp}$ 
that are not smooth manifolds: topological spaces for which one can consistently define which 
maps from Cartesian spaces into them count as smooth in a way that makes this assignment a sheaf on 
$\mathrm{CartSp}$, but which are not necessarily locally isomorphic to a Cartsian space: 
these are called \emph{diffeological spaces}. A central example of a space that is naturally 
a diffeological space but not a finite dimensional manifold is a mapping space $[\Sigma,X]$ 
of smooth functions between smooth manifolds $\Sigma$ and $X$: 
since the idea is that for $U$ any Cartesian space the smooth $U$-parameterized families 
of points in $[\Sigma,X]$ are smooth $U$-parameterized families of smooth maps $\Sigma \to X$, 
we can take the plot-assigning rule to be
$$
  [\Sigma, X] : U \mapsto \mathrm{Hom}_{\mathrm{SmoothMfd}}(\Sigma \times U , X)
  \,.
$$
It is useful to relate all these phenomena in the topos $\mathrm{Sh}(C)$ to their 
image in the archetypical topos $\mathrm{Set}$. This is simply the category of sets, 
which however we should think of here as the category $\mathrm{Set} \simeq \mathrm{Sh}(*)$ 
of sheaves on the category $*$ which contains only a single object and no nontrivial morphism: 
objects in here are generalized spaces \emph{modeled on the point}. 
All we know about them is how to map the point into them, 
and as such they are just the sets of all possible such maps from the point.

Every category of sheaves $\mathrm{Sh}(C)$ comes canonically with an essentially unique 
topos morphism to the topos of sets, given by a pair of adjoint functors
$$
  \xymatrix{
    \mathrm{Sh}(C)
    \ar@<-3pt>[r]_<<<<{\Gamma}
    \ar@{<-}@<+3pt>[r]^<<<<{\mathrm{Disc}}
    &
    \mathrm{Sh}(*)
    \simeq \mathrm{Set}
   }
  \,.
$$
Here $\Gamma$ is called the \emph{global sections functor}. If $C$ has a terminal 
object $*$, then it is given by evaluation on that object: the functor $\Gamma$ sends a 
plot-assigning rule $X : C^{\mathrm{op}} \to \mathrm{Set}$ to the set of plots by the point $\Gamma(X) = X(*)$. 
For instance in $C = \mathrm{CartSp}$ the terminal object exists and is the ordinary point $* = \mathbb{R}^0$. 
If $X \in \mathrm{Sh}(C)$ is a smooth manifold or diffeological space as above, then $\Gamma(X) \in \mathrm{Set}$ 
is simply its underlying set of points. So the functor $\Gamma$ can be thought of as forgetting 
the \emph{cohesive structure} that is given by the fact that our generalized spaces are modeled on $C$. 
It remembers only the underlying point-set.

Conversely, its left adjoint functor $\mathrm{Disc}$ takes a set $S$ to the 
sheafification $\mathrm{Disc}(S) = L \mathrm{Const}(S)$ of the constant presheaf $\mathrm{Const} : U \mapsto S$,
which asserts that the set of its plots by any test space is always the same set $S$. 
This is the plot-rule for the  \emph{discrete space} modeled on $C$ given by the set $S$: 
a plot has to be a constant map of the test space $U$ to one of the elements $s \in S$. 
For the case $C = \mathrm{CartSp}$ this interpretation is literally true in the familiar sense: 
the generalized smooth space $\mathrm{Disc}(S)$ is the discrete smooth manifold 
or discrete diffeological space with point set S.

\subparagraph{Concrete and non-concrete sheaves}

The examples for generalized spaces $X$ modeled on $C$ that we considered so far all had 
the property that the collection of plots $U \to X$ into them was a subset of the set of maps 
of sets from $U$ to their underlying set $\Gamma(X)$ of points. These are called 
\emph{concrete sheaves}. Not every sheaf is concrete. The concrete sheaves form a subcategory 
inside the full topos which is itself almost, but not quite a topos: it is the 
\emph{quasitopos} of concrete objects.
$$
  \xymatrix{
    \mathrm{Conc}(C) 
	 \ar@{<-}@<+4pt>[r]
	 \ar@{^{(}->}@<-4pt>[r]
	 &
	\mathrm{Sh}(C)
  }
  \,.
$$
Non-concrete sheaves over $C$ may be exotic as compared to smooth manifolds, but they 
are still usefully regarded as generalized spaces modeled on $C$. For instance for 
$n \in \mathbb{N}$ there is the sheaf $\kappa(n,\mathbb{R})$ given by saying that plots 
by  $U \in \mathrm{CartSp}$ are identified with closed differential $n$-forms on $U$:
$$
  \kappa(n,\mathbb{R}) : U \mapsto \Omega^n_{\mathrm{\mathrm{cl}}}(U)
  \,.
$$
This sheaf describes a very non-classical space, which for $n \geq 1$ has only a single point, 
$\Gamma(\kappa(n,\mathbb{R})) = *$ , only a single curve, a single surface, etc., up to a single 
$(n-1)$-dimensional probe, but then it has a large number of $n$-dimensional probes. 
Despite the fact that this sheaf is very far in nature from the test spaces that it is modeled on, 
it plays a crucial and very natural role: it is in a sense a model for an Eilenberg-MacLane space 
$K(n,\mathbb{R})$. We shall see in \ref{SmoothStrucLieAlgebras} that  
these sheaves are part of an incarnation of the $\infty$-Lie-algebra $b^n \mathbb{R}$ 
and the sense in which it models an Eilenberg-MacLane space is that of Sullivan models 
in rational homotopy theory. In any case, we want to allow ourselves to regard
non-concrete objects such as $\kappa(n, \mathbb{R})$ on the same footing as 
diffeological spaces and smooth manifolds.

\paragraph{$\infty$-Toposes}
\label{IntroInfinToposes}
\index{topos!overview}

While therefore a general object in the sheaf topos $\mathrm{Sh}(C)$ may exhibit a 
considerable generalization of the objects $U \in C$ that it is modeled on, for many 
natural applications this is still not quite general enough: if for instance $X$ is a 
\emph{smooth orbifold} (see for instance \cite{MoerdijkPronk}), 
then there is not just a set, but a \emph{groupoid} of ways of probing it by a Cartesian test 
space $U$: if a probe $\gamma : U \to X$ is connected by an orbifold transformation to another 
probe $\gamma' : U \to X$, then this constitutes a morphism in the groupoid $X(U)$ of probes of $X$ 
by $U$.

Even more generally, there may be a genuine \emph{$\infty$-groupoid} 
of probes of the generalized space $X$ 
by the test space $U$: a set of probes with morphisms between different probes, 2-morphisms between 
these 1-morphisms, and so on. 

Such structures are described in \emph{$\infty$-category theory}: where a category has a 
set of morphisms between any two objects, an $\infty$-category has an $\infty$-groupoid 
of morphisms, whose compositions are defined up to higher coherent homotopy. 
The theory of $\infty$-categories is effectively the combination of category theory and 
homotopy theory. The main fact about it, emphasized originally by Andr{\'e} Joyal and then 
further developed in \cite{Lurie}, is that it behaves formally entirely analogously to 
category theory: there are notions of $\infty$-functors,
$\infty$-limits, adjoint $\infty$-functors etc., that satisfy all the familiar relations 
from category theory.

\subparagraph{$\infty$-Groupoids }
\label{InfinityGroupoidsInIntro}

We first look at bare $\infty$-groupoids and then discuss how to equip these with smooth structure.

An $\infty$-groupoid is first of all supposed to be a structure that has \emph{$k$-morphisms} for all 
$k \in \mathbb{N}$, which for $k \geq 1$ go between $(k-1)$-morphisms. A useful tool for organizing 
such collections of morphisms is the notion of a \emph{simplicial set}. This is a functor on the opposite 
category of the simplex category $\Delta$, whose objects are the abstract cellular $k$-simplices, 
denoted $[k]$ or $\Delta[k]$ for all $k \in \mathbb{N}$, and whose morphisms $\Delta[k_1] \to \Delta[k_2]$ 
are all ways of mapping these into each other. So we think of such a simplicial set given by a functor
$$
  K : \Delta^{\mathrm{op}} \to \mathrm{Set}
$$
as specifying
\begin{itemize}
  \item
    a set $[0] \mapsto K_0$  of \emph{objects};
  \item
    a set $[1] \mapsto K_1$  of \emph{morphisms};
  \item
    a set $[2] \mapsto K_2$  of \emph{2-morphisms};
  \item
    a set $[3] \mapsto K_3$  of \emph{3-morphisms};
\end{itemize}
and generally
\begin{itemize}
  \item
    a set $[k] \mapsto K_k$ of \emph{$k$-morphisms}.
\end{itemize}
as well as specifying
\begin{itemize}
\item functions $([n] \hookrightarrow [n+1]) \mapsto (K_{n+1} \to K_n)$
  that send $n+1$-morphisms to their boundary $n$-morphisms;

\item functionss $([n+1] \to [n]) \mapsto (K_{n} \to K_{n+1})$
  that send $n$-morphisms to identity $(n+1)$-morphisms
  on them.
\end{itemize}
The fact that $K$ is supposed to be a functor enforces that these assignments of sets 
and functions satisfy conditions that make consistent our interpretation of them 
as sets of $k$-morphisms and source and target maps between these. 
These are called the \emph{simplicial identities}.
But apart from this source-target matching, a generic simplicial set does not 
yet encode a notion of \emph{composition} of these morphisms. 

For instance for $\Lambda^1[2]$ the simplicial set consisting of two attached 1-cells 
$$
  \Lambda^1[2] = 
  \left\{
      \raisebox{20pt}{
      \xymatrix{
         & 1 \ar[dr]
         \\
        0 \ar[ur]&& 2
      }
      }
   \right\}
$$
and for $(f,g) : \Lambda^1[2] \to K$ an image of this situation in $K$, 
hence a pair $x_0 \stackrel{f}{\to} x_1 \stackrel{g}{\to} x_2$ of two 
\emph{composable} 1-morphisms in $K$, we want to demand that there exists a third 
1-morphisms in $K$ that may be thought of as the \emph{composition} $x_0 \stackrel{h}{\to} x_2$ 
of $f$ and $g$. But since we are working in higher category theory, we want to identify this 
composite only up to a 2-morphism equivalence
$$
      \raisebox{20pt}{
      \xymatrix{
         & x_1 \ar[dr]^g
         \\
        x_0 \ar[ur]^{f} \ar[rr]_h^{\ }="t" && x_2
        \ar@{=>}^\simeq "t"+(0,5); "t" 
      }
      }
     \,.
$$
From the picture it is clear that this is equivalent to demanding that for 
$\Lambda^1[2] \hookrightarrow \Delta[2]$ the obvious inclusion of the two abstract 
composable 1-morphisms into the 2-simplex we have a diagram of morphisms of simplicial sets
$$
  \xymatrix{
    \Lambda^1[2] \ar[r]^{(f,g)} \ar[d] & K
    \\
    \Delta[2] \ar[ur]_{\exists h}
  }
  \,.
$$
A simplicial set where for all such $(f,g)$ a corresponding such $h$ exists may be thought of as a collection of 
higher morphisms that is equipped with a notion of composition of adjacent 1-morphisms. 

For the purpose of describing groupoidal composition, we now want that this composition operation has all inverses. For that purpose, notice that for 
$$
  \Lambda^2[2] = 
  \left\{
      \raisebox{20pt}{
      \xymatrix{
         & 1 \ar[dr]
         \\
        0 \ar[ur]&& 2
      }
      }
   \right\}
$$
the simplicial set consisting of two 1-morphisms that touch at their end, hence for 
$$
  (g,h) : \Lambda^2[2] \to K
$$
two such 1-morphisms in $K$, then if $g$ had an inverse $g^{-1}$ we could use the above composition operation to compose that with $h$ and thereby find a morphism $f$ connecting the sources of $h$ and $g$. This being the case is evidently equivalent to the existence of diagrams of morphisms of simplicial sets of the form
$$
  \xymatrix{
    \Lambda^2[2] \ar[r]^{(g,h)} \ar[d] & K
    \\
    \Delta[2] \ar[ur]_{\exists f}
  }
  \,.
$$
Demanding that all such diagrams exist is therefore demanding that we have on 1-morphisms 
a composition operation with inverses in $K$. 

In order for this to qualify as an $\infty$-groupoid, this composition operation 
needs to satisfy an associativity law up to 2-morphisms, which means that we can find 
the relevant tetrahedra in $K$. These in turn need to be connected by 
\emph{pentagonators} and ever so on.  It is a nontrivial but true and powerful fact, 
that all these coherence conditions are captured by generalizing the above conditions to 
all dimensions in the evident way:

Let $\Lambda^i[n] \hookrightarrow \Delta[n]$ be the simplicial set -- called the $i$th $n$-horn -- that consists of all cells of the $n$-simplex $\Delta[n]$ except the interior $n$-morphism and the $i$th $(n-1)$-morphism.

Then a simplicial set is called a 
\emph{Kan complex},
\index{Kan complex}
\index{category theory!Kan complex}
\index{groupoid!Kan complex}
if for all images $f : \Lambda^i[n] \to K$ of such horns in $K$, the missing two cells can be found in $K$ -- 
in that we can always find a \emph{horn filler} $\sigma$ in the diagram
$$
  \xymatrix{
    \Lambda^i[n] \ar[r]^{f} \ar[d] & K
    \\
    \Delta[n] \ar[ur]_{\exists \sigma}
  }
  \,.
$$
The basic example is the \emph{nerve} $N(C) \in \mathrm{sSet}$ of an ordinary groupoid $C$, 
which is the simplicial set with $N(C)_k$ being the set of sequences of $k$ composable morphisms in $C$. The nerve operation is a full and faithful functor from 1-groupoids into Kan complexes and hence may be thought of as embedding 1-groupoids in the context of general $\infty$-groupoids.

$$
   \raisebox{24pt}{
   \xymatrix{
   & \mbox{\small Groupoids}
\ar@{_{(}->}[dl]
\ar@{^{(}->}[dr]^N
\\
\mbox{\small Categories} 
\ar@{^{(}->}[dr]^N
&& \mathbf{KanComplexes} 
\ar@{^{(}->}[dl]& 
\\
& \mbox{\small QuasiCategories} \ar@{^{(}->}[d] \ar@{}[r]|{\simeq}& \mbox{\small $\infty$-Categories }
&& 
\\
& \mbox{\small SimplicialSets}
 }
 }
$$

But we need a bit more than just bare $\infty$-groupoids. In generalization to Lie groupoids, 
we need \emph{smooth $\infty$-groupoids}. A useful way to encode that an $\infty$-groupoid has extra structure modeled on geometric test objects that themselves form a category $C$ is to remember the rule which for each test space $U$ in $C$ produces the $\infty$-groupoid of $U$-parameterized families of $k$-morphisms in $K$.  For instance for a smooth $\infty$-groupoid we could test with each Cartesian space $U = \mathbb{R}^n$ and find the $\infty$-groupoids $K(U)$ of smooth $n$-parameter families of $k$-morphisms in $K$.

This data of $U$-families arranges itself into a presheaf with values in Kan complexes
$$
  K : C^{\mathrm{op}} \to \mathrm{KanCplx} \hookrightarrow \mathrm{sSet}
  \,,
$$
hence with values in simplicial sets. This is equivalently a simplicial presheaf of sets. The functor category $[C^{\mathrm{op}}, \mathrm{sSet}]$ on the opposite category of the category of test objects $C$ serves as a model for the 
$\infty$-category of $\infty$-groupoids with $C$-structure.

While there are no higher morphisms in this functor 1-category that could for instance witness that two $\infty$-groupoids are not isomorphic, but still equivalent, it turns out that all one needs in order to reconstruct 
\emph{all} these higher morphisms (up to equivalence!) is just the information of which morphisms of simplicial presheaves would become invertible if we were keeping track of higher morphism. These would-be invertible morphisms are called \emph{weak equivalences} and denoted $K_1 \stackrel{\simeq}{\to} K_2$. 

For common choices of $C$ there is a well-understood way to define the weak 
equivalences $W \subset \mathrm{Mor} [C^{\mathrm{op}}, \mathrm{sSet}]$, and equipped with this information 
the category of simplicial presheaves becomes a \emph{category with weak equivalences}. 
There is a well-developed but somewhat intricate theory of how exactly this 1-categorical data 
models the full higher category of structured groupoids that we are after, but for our purposes here 
we essentially only need to work inside the category of \emph{fibrant} objects of a model structure on 
presheaves, which in practice amounts to the fact that we use the following three basic constructions:

\begin{enumerate}
  \item{\bf $\infty$-anafunctor}
  A morphisms $X \to Y$ between $\infty$-groupoids with $C$-structure is not just a 
  morphism $X\to Y$ in $[C^{\mathrm{op}}, \mathrm{sSet}]$, but is a span of such ordinary morphisms
   $$
     \xymatrix{
       \hat X \ar[r] \ar[d]^\simeq & Y
       \\
       X
     }
     \,,
   $$
   where the left leg is a weak equivalence. 
   This is sometimes called an \emph{$\infty$-anafunctor} from $X$ to $Y$.
 
\item {\bf homotopy pullback} -- For $A \to B \stackrel{p}{\leftarrow} C$ a diagram, the 
  $\infty$-pullback of it is the ordinary pullback in $[C^{\mathrm{op}}, \mathrm{sSet}]$ of a replacement diagram $A \to B \stackrel{\hat p}{\leftarrow} \hat C$, where $\hat p$ is a \emph{good replacement}  of $p$ in the sense of the following factorization lemma. 

\end{enumerate}
\begin{proposition}[factorization lemma]
For $p : C \to B$ a morphism in $[C^{\mathrm{op}}, \mathrm{sSet}]$, 
  a \emph{good replacement} $\hat p : \hat C \to B$ is given by the composite 
  vertical morphism in the ordinary pullback diagram
  $$
    \xymatrix{
      \hat C \ar[r] \ar[d] & C \ar[d]^p
      \\
      B^{\Delta[1]} \ar[r] \ar[d] & B
      \\
      B
    }
    \,,
  $$
  where $B^{\Delta[1]}$ is the path object of $B$: the presheaf that is over each $U \in C$ the
  simplicial path space $B(U)^{\Delta[1]}$.
\end{proposition}

\subparagraph{$\infty$-Sheaves / $\infty$-Stacks }

In particular, there is a notion of $\infty$-presheaves on a category (or $\infty$-category) $C$: 
$\infty$-functors
$$
  X : C^{\mathrm{op}} \to \infty \mathrm{Grpd}
$$
to the $\infty$-category $\infty\mathrm{Grpd}$ of $\infty$-groupoids -- there is 
an $\infty$-Yoneda embedding, and so on. Accordingly, $\infty$-topos theory proceeds 
in its basic notions along the same lines as we sketched above for topos theory:

an $\infty$-topos of $\infty$-sheaves is defined to be a 
reflective sub-$\infty$-category
$$
  \xymatrix{
    \mathrm{Sh}_{(\infty,1)}(C) 
       \ar@<+4pt>@{<-}[r]^{L}
       \ar@{^{(}->}@<-4pt>[r]
    &
    \mathrm{PSh}_{(\infty,1)}(C)
  }
$$
of an $\infty$-category of $\infty$-presheaves.
As before, such is essentially determined by and determines a Grothendieck topology or coverage on $C$. 
Since a $2$-sheaf with values in groupoids is usually called a \emph{stack}, 
an $\infty$-sheaf is often also called an \emph{$\infty$-stack}. 

In the spirit of the above discussion, the objects of the $\infty$-topos of $\infty$-sheaves 
on $C= \mathrm{CartSp}$ we shall think of as \emph{smooth $\infty$-groupoids}. 
This is our main running example. We shall write 
$\mathrm{Smooth}\infty\mathrm{Grpd} := \mathrm{Sh}_\infty(\mathrm{CartSp})$ for the 
$\infty$-topos of smooth $\infty$-groupoids.

Let
\begin{itemize}
  \item $C := \mathrm{SmthMfd}$ be the category of all smooth manifolds
  (or some other site, here assumed to have enough points);
\item $\mathrm{gSh}(C)$ be the category of groupoid-valued sheaves over $C$,
 \\
 for instance $X = \{\xymatrix{X \ar@<+2pt>[r]\ar@<-2pt>[r] & X}\}, \mathbf{B}G = \{\xymatrix{G \ar@<+2pt>[r]\ar@<-2pt>[r] & {*}}\} \in \mathrm{gSh}(C)$;
\item $\mathrm{Ho}_{\mathrm{gSh}(C)}$ the \emph{homotopy category} obtained by universally turning 
 the \emph{stalkwise groupoid-equivalences} into isomorphisms.
\end{itemize}
\noindent {\bf Fact:} \hspace{2cm}$H^1(X,G) \simeq \mathrm{Ho}_{\mathrm{gSh}(C)}(X,\mathbf{B}G)$.
Let
\begin{itemize}
 \item $\mathrm{sSet}(C)_{\mathrm{lfib}} \hookrightarrow \mathrm{Sh}(C, \mathrm{sSet})$ 
   be the stalkwise Kan simplicial sheaves;
 \item $L_W \mathrm{sSh}(C)_{\mathrm{lfib}}$ the \emph{simplicial localization}
   obtained by universally turning 
   \emph{stalkwise homotopy equivalences} into \emph{homotopy equivalences}.
\end{itemize}

\noindent {\bf Definition/Theorem.} This is the $\infty$-category theory analog of the sheaf topos
over $C$, the \emph{$\infty$-stack $\infty$-topos}:
 \hspace{.9cm}
  $\mathbf{H}:= \mathrm{Sh}_\infty(C) \simeq L_W \mathrm{sSh}(C)_{\mathrm{lfib}}$.

\noindent {\bf Example.} 
$\mathrm{Smooth}\infty \mathrm{Grpd} := \mathrm{Sh}_\infty(\mathrm{SmthMfd})$
is the $\infty$-topos of \emph{smooth $\infty$-groupoids}.

\noindent{\bf Proposition.} Every object in $\mathrm{Smooth}\infty \mathrm{Grpd}$
is presented by a simplicial manifold, but not necessarily by a \emph{locally Kan}
simplicial manifold (see below).

But a crucial point of developing our theory in the language of 
$\infty$-toposes is that all constructions work in great generality. By simply passing to 
another site $C$, all constructions apply to the theory of generalized spaces modeled on the 
test objects in $C$. Indeed, to really capture all aspects of $\infty$-Lie theory, we should 
and will adjoin to our running example $C = \mathrm{CartSp}$ that of the slightly larger 
site $C = \mathrm{CartSp}_{\mathrm{synthdiff}}$ of infinitesimally thickened Cartesian spaces. 
Ordinary sheaves 
on this site are the generalized spaces considered in \emph{synthetic differential geometry}: 
these are smooth spaces such as smooth loci that may have infinitesimal extension. For 
instance the first order jet $D \subset \mathbb{R}$ of the origin in the real line exists as an 
infinitesimal space in $\mathrm{Sh}(\mathrm{CartSp}_{\mathrm{synthdiff}})$. Accordingly, $\infty$-groupoids 
modeled on $\mathrm{CartSp}_{\mathrm{synthdiff}}$ are smooth $\infty$-groupoids that may have $k$-morphisms of 
infinitesimal extension. We will see that a smooth $\infty$-groupoid all whose morphisms has infinitesimal 
extension is a Lie algebra or Lie algebroid or generally an $\infty$-Lie algebroid.

While $\infty$-category theory provides a good abstract definition and theory of $\infty$-groupoids 
modeled on test objects in a category $C$ in terms of the $\infty$-category of $\infty$-sheaves 
on $C$, for concrete manipulations it is often useful to have a presentation of the $\infty$-categories 
in question in terms of generators and relations in ordinary category theory. Such a 
generators-and-relations-presentation is provided by the notion of a \emph{model category} structure. 
Specifically, the $\infty$-toposes of $\infty$-presheaves that we are concerned with are 
presented in this way by a model structure on simplicial presheaves, 
i.e. on the functor category $[C^{\mathrm{op}}, \mathrm{sSet}]$ from $C$ to the 
category $\mathrm{sSet}$ of simplicial sets.
In terms of this model, the corresponding $\infty$-category of $\infty$-sheaves is given by 
another model structure on $[C^{\mathrm{op}}, \mathrm{sSet}]$, called the 
\emph{left Bousfield localization} at the set of covers in $C$.

These models for $\infty$-stack $\infty$-toposes have been proposed, known and studied since the 
1970s and are therefore quite well understood. The full description and proof of their abstract 
role in $\infty$-category theory was established in \cite{Lurie}.

As before for toposes, there is an archetypical $\infty$-topos, which is 
$\infty \mathrm{Grpd} = \mathrm{Sh}_{(\infty,1)}(*)$ itself: the collection of 
generalized $\infty$-groupoids that are modeled on the point. All we know about these 
generalized spaces is how to map a point into them and what the homotopies and higher homotopies of 
such maps are, but no further extra structure. So these are bare $\infty$-groupoids without extra 
structure. Also as before, every $\infty$-topos comes with an essentially unique 
geometric morphism to this archetypical $\infty$-topos given by a pair of adjoint $\infty$-functors
$$
  \xymatrix{
    \mathrm{Sh}_{(\infty,1)}(C) 
       \ar@<+3pt>@{<-}[r]^{\mathrm{Disc}}
       \ar@<-3pt>[r]_{\Gamma}
    &
    \infty \mathrm{Grpd}
  }
  \,.
$$
Again, if $C$ happens to have a terminal object $*$, then $\Gamma$ is the operation that 
evaluates an $\infty$-sheaf on the point: it produces the bare $\infty$-groupoid underlying an 
$\infty$-groupoid modeled on $C$. For instance for $C = \mathrm{CartSp}$ a smooth $\infty$-groupoid 
$X \in \mathrm{Sh}_{(\infty,1)}(C)$ is sent by $\Gamma$ to to the underlying 
$\infty$-groupoid that forgets the smooth structure on $X$.

Moreover, still in direct analogy to the 1-categorical case above, the left adjoint $\mathrm{Disc}$ 
is the $\infty$-functor that sends a bare $\infty$-groupoid $S$ to the $\infty$-stackification 
$\mathrm{Disc}S = L \mathrm{Const}S$ of the constant $\infty$-presheaf on $S$. 
This models the discretely structured $\infty$-groupoid on $S$. For instance for 
$C=\mathrm{CartSp}$ we have that $\mathrm{Disc}S$ is a smooth $\infty$-groupoid with discrete 
smooth structure: all smooth families of points in it are actually constant.

\subparagraph{Structured $\infty$-Groups}
\label{GroupsInIntroduction}

It is clear that we may speak of \emph{group objects} in any topos, 
(or generally in any category with finite products): objects $G$ equipped with 
a  multiplication $G \times G \to G$ and a neutral element $* \to G$ such that 
the multiplication is unital, associative and has inverses for each element.
In a sheaf topos, such a $G$ is equivalently a \emph{sheaf of groups}.
For instance every Lie group canonically becomes a group object in 
$\mathrm{Sh}(\mathrm{CartSp})$.

As we pass to an $\infty$-topos the situation is essentially the same,
only that the associativity condition is replaced by 
\emph{associativity up to coherent homotopy} (also called: up to 
\emph{strong homotopy}), and similarly for the unitalness and the
existence of inverses. One way to formalize this is to say
that 
  a group object in an $\infty$-topos $\mathbf{H}$ is an
  $A_\infty$-algebra object $G$ such that its 0-truncation $\tau_0 G$
  is a group object in the underlying 1-topos.
(This is discussed in \cite{LurieAlgebra}.) 

For instance in the $\infty$-topos over $\mathrm{CartSp}$ a Lie 
group still naturally is a group object, but also a \emph{Lie 2-group}
or \emph{differentiable group stack} is. Moreover, every sheaf of
\emph{simplicial groups} presents a group object in the $\infty$-topos,
and we will see that all group objects here have a presentation by sheaves
of simplicial groups.

A group object in $\infty \mathrm{Grpd} \simeq \mathrm{Top}$ we will for emphasis
call an \emph{$\infty$-group}. In this vein a group object in an $\infty$-topos
over a non-trivial site is a \emph{structured $\infty$-group} (for instance
a topological $\infty$-group or a smooth $\infty$-group).

A classical source of $\infty$-groups are \emph{loop spaces}, where
the group multiplication is given by concatenation of based loops in a given
space, the homotopy-coherent associativity is given by reparameterizations of
concatenations of loops, and inverses are given by reversing the parameterization
of a loop. 
A classical result of Milnor says, in this language, that every 
$\infty$-group arises as a loop space this way. This statement
generalizes from discrete $\infty$-groups (group objects in 
$\infty \mathrm{Grpd} \simeq \mathrm{Top}$) to structured $\infty$-groups.

\noindent{\bf Theorem.} (Milnor--Lurie) 
There is an equivalence
$$
  \xymatrix{
    \left\{
	  \mbox{
	  groups in $\mathbf{H}$
	  }
	\right\}
		\ar@{<-}@<+5pt>[rrr]^<<<<<<<<<<<<<{\mbox{\small  looping}\; \Omega}
	\ar@<-5pt>[rrr]_<<<<<<<<<<<<<{\mbox{\small delooping}\; \mathbf{B}}^\simeq
    &&&
	\left\{
	  \mbox{\begin{tabular}{c} pointed connected \\ objects in $\mathbf{H}$ \end{tabular}}
	\right\}
  }
$$
This equivalence is a most convenient tool. In the following we will almost exclusively 
handle $\infty$-groups $G$ in terms of their pointed connected delooping
objects $\mathbf{B}G$. 
We discuss this in more detail below in \ref{StrucInftyGroups}.
This is all the more useful as the objects $\mathbf{B}G$ happen to be the
\emph{moduli $\infty$-stacks} of \emph{$G$-principal $\infty$-bundles}.
We come to this in \ref{HigherPrincipalBundlesInIntroduction}.

\paragraph{Cohomology}
\label{IntroCohomology}
\index{cohomology!overview}

Where the archetypical topos is the category $\mathrm{Set}$, the
archetypical $\infty$-topos is the $\infty$-category $\infty \mathrm{Grpd}$
of $\infty$-groupoids. 
This, in turn, is equivalent by a classical result 
(see \ref{DiscreteInfGroupoids}) to $\mathrm{Top}$, 
the category of topological spaces, regarded as an $\infty$-category by taking the 
2-morphisms to be homotopies between continuous maps, 3-morphisms to be homotopies of 
homotopy, and so forth:
$$
  \infty \mathrm{Grpd} \simeq \mathrm{Top}
  \,.
$$

In $\mathrm{Top}$ it is familiar -- 
from the notion of \emph{classifying spaces} and from the 
\emph{Brown representability theorem} 
(the reader in need of a review of such matter might try \cite{MayConcise})  
-- 
that the cohomology of a topological space $X$ may be identified as the set of 
homotopy classes of continuous maps from $X$ to some coefficient space $A$
$$
  H(X,A) := \pi_0 \mathrm{Top}(X,A)
  \,.
$$
For instance for $A = K(n,\mathbb{Z}) \simeq B^{n} \mathbb{Z}$ 
the topological space called the $n$th \emph{Eilenberg-MacLane space} of the additive group of
integers, we have that
$$
  H(X,A) := \pi_0 \mathrm{Top}(X, B^n \mathbb{Z}) \simeq H^n(X,\mathbb{Z})
$$
is the ordinary integral (singular) cohomology of $X$. 
Also \emph{nonabelian cohomology} 
is famously exhibited this way: for $G$ a (possibly nonabelian) topological group and $A = B G$ its 
classifying space 
(we discuss this construction and its generalization 
in detail in \ref{ETopStrucGeometricRealization}) 
we have that
$$
  H(X,A) := \pi_0 \mathrm{Top}(X,B G) \simeq H^1(X,G)
$$
is the degree-1 nonabelian cohomology of $X$ with coeffients in $G$, which classifies 
\emph{$G$-principal bundles} over $X$ (more on that in a moment).

Since this only involves forming $\infty$-categorical hom-spaces and since this is 
an entirely categorical operation, it makes sense to \emph{define} for $X$, $A$ any two objects in an 
arbitrary $\infty$-topos $\mathbf{H}$ the intrinsic cohomology of $X$ with coefficients in $A$ to be
$$
  H(X,A) := \pi_0 \mathbf{H}(X,A)
  \,,
$$
where $\mathbf{H}(X,A)$ denotes the $\infty$-groupoid of morphism from $X$ to $A$ in $\mathbf{H}$.
This general identification of cohomology with hom-spaces in $\infty$-toposes is
central to our developments here. 
We indicate now two classes of justification for this definition.
\begin{enumerate}
  \item 
    Essentially every notion of cohomology already considered in the literature is an example
	of this definition.
	Moreover, those that are not are often improved on by fixing them to become an example.
  \item
    The use of a good notion of $G$-cohomology on $X$ should be that it \emph{classifies}
	``$G$-structures over $X$'' and exhibits the \emph{obstruction theory}
	for extensions or lifts of such structures.
	We find that it is precisely the context 
	of an ambient $\infty$-topos (precisely: the $\infty$-Giraud axioms
	that characterize an $\infty$-topos)
	that makes such a classification and obstruction theory work.
\end{enumerate}

\medskip

We discuss now a list of examples of $\infty$-toposes
$\mathbf{H}$
together with notions of cohomology whose cocycles are given by morphisms 
$c \in \mathbf{H}(X,A)$
between a domain object $X$ and coefficient object $A$ in this $\infty$-topos.
Some of these examples are evident and classical, modulo our emphasis on the
$\infty$-topos theoretic perspective, others are original. 
Even those cases that are classical receive 
new information from the $\infty$-topos theoetic perspective.

Details are below in the relevant parts of
\ref{Implementation} and \ref{Applications}.

\medskip

In view of the unification that we discuss, some of the traditional
names for notions of cohomology are a bit suboptimal. For instance
the term \emph{generalized cohomology} for theories satisfying the
Eilenberg-Steenrod axioms does not well reflect that it
is a generalization of ordinary cohomology of topological spaces (only)
which is, in a quite precise sense, \emph{orthogonal} to the 
generalizations given by passage to sheaf cohomology or 
to nonabelian cohomology, all of which are subsumed by cohomology
in an $\infty$-topos. In order to usefully distinguish the 
crucial aspects here we will use the following terminology
\begin{itemize}
  \item We speak of \emph{structured cohomology} to indicate
   that a given notion is realized in an $\infty$-topos other than the
   archetypical $\infty \mathrm{Grpd} \simeq \mathrm{Top}$
   (which representes ``discrete structre'' in the precise sense 
   discussed in \ref{DiscreteInfGroupoids}). Hence traditional
   sheaf cohomology is ``structured'' in this sense, while
   ordinary cohomology and Eilenberg-Steenrod cohomology is
   ``unstructured''.
  \item We speak of \emph{nonabelian cohomology} when 
   coefficient objects are not \emph{required} to be abelian (groups)
   or stable (spectra), but may generally be deloopings $A := \mathbf{B}G$ of arbitrary
   (structred) $\infty$-groups $G$.

   More properly this might be called
   \emph{not-necessarily abelian cohomology}, but following
   common practice (as in ``noncommutative geometry'') we stick with the
   slightly imprecise but shorter term. One point that we will dwell on 
   (see the discussion of examples in \ref{TwistedStructures}) is that
   the traditional notion of \emph{twisted cohomology} 
   (already twisted abelian cohomology)
   is naturally
   a special case of nonabelian cohomology.
\end{itemize}
Notice that the ``generalized'' in ``generalized cohomology'' of
Eilenberg-Steenrod type refers to allowing coefficient objects which are 
abelian $\infty$-groups more general than Eilenberg-MacLane objects.
Hence this is in particular subsumed in \emph{nonabelian cohomology}.

In this terminology, the notion of cohomology in $\infty$-toposes
that we are concerned with here is 
\emph{structured nonabelian/twisted generalized cohomology}.

Finally, not only is it natural to allow the coefficient objects $A$ to be 
general objects in a general $\infty$-topos, but also there is
no reason to restrict the nature of the domain objects $X$.
For instance traditional sheaf cohomology always takes $X$, in our language, to be the
\emph{terminal object} $X = *$ of the ambient $\infty$-topos. 
This is also called the \emph{(-2)-truncated object} 
(see \ref{TruncatedObjects} below) of the $\infty$-topos, being the 
unique member of the lowest class
in a hierarchy of \emph{$n$-truncated objects} for $(-2) \leq n \leq \infty$.
As we increase $n$ here, we find that the domain object is generalized to
\begin{itemize}
  \item $n = -1$: subspaces of $X$;
  \item $n = 0$: {\'e}tale spaces over $X$;
  \item $n = 1$: orbifolds / orbispaces / groupoids over $X$;
  \item $n \geq 2$: higher orbifolds / orbispaces / groupoids  
\end{itemize}
One finds then that cohomology of an $n$-truncated object for $n \geq 1$ reproduces
the traditional notion of \emph{equivariant cohomology}. In particular
this subsumes \emph{group cohomology}: ordinary group cohomology in the
unstructured case (in $\mathbf{H} = \infty \mathrm{Grpd}$) and generally
structured group cohomology such as \emph{Lie group cohomology}.

Therefore, strictly speaking, we are here concerned with 
\emph{equivariant structured nonabelian/twisted generalized cohomology}.
All this is neatly encapsulated by just the fundamental notion
of hom-spaces in $\infty$-toposes.

{\bf Cochain cohomology}
 \index{Dold-Kan correspondence!and cochain cohomology}

The origin and maybe the most elementary notion of cohomology is
that appearing in \emph{homological algebra}:
given a \emph{cochain complex} of abelian groups
$$
  V^\bullet
  =
  \left[
  \xymatrix{
    \cdots \ar@{<-}[r]^{d^2} & V^2 \ar@{<-}[r]^{d^1} & V^1 \ar@{<-}[r]^{d^0} & V^0
  }
  \right]
  \,,
$$
its cohomology group in degree $n$ is defined to be the quotient group
$$
  H^n(V) := \mathrm{ker}(d^{n})/\mathrm{im}(d^{n-1})
  \,.
$$
To see how this is a special case of cohomology in an $\infty$-topos,
consider a fixed abelian group $A$ and
suppose that this cochain complex is the $A$-dual of a \emph{chain} complex
$$
  V_\bullet
  =
  \left[
  \xymatrix{
    \cdots \ar[r] & V_2 \ar[r]^{\partial_2} & V_1 \ar[r]^{\partial_1} & V_0
  }
  \right]
  \,,
$$
in that $V^\bullet = \mathrm{Hom}_{\mathrm{Ab}}(V_\bullet, A)$.
For instance if $A = \mathbb{Z}$ and $V_n$ is the free abelian group
on the set of $n$-simplices in some topological space, then $V^n$ is the 
group of \emph{singular $n$-cochains} on $X$.

Write then $A[n]$ (or $A[-n]$, if preferred) for the chain complex concentrated in 
degree $n$ on $A$. In terms of this
\begin{enumerate}
  \item morphisms of chain complexes $c : V_\bullet \to A[n]$ are in natural bijection
  with \emph{closed} elements in $V^n$, hence with $\mathrm{ker}(d^n)$;
  \item chain homotopies $\eta : c_1 \to c_2$ between two such chain morphisms
  are in natural bijection with elements in $\mathrm{im}(d^{n-1})$.
\end{enumerate} 
This way the cohomology group $H^n(V^\bullet)$ is naturally identified with the
\emph{homotopy classes} of maps $V_\bullet \to A[n]$.

Consider then again an  example as that of singular cochains as above, where 
$V_\bullet$ is degreewise a free abelian group in a simplicial set $X$.
Then this cohomology is the group of connected components of a hom-space in an $\infty$-topos.
To see this, one observes that the category of chain complexes $\mathrm{Ch}_\bullet$
is but a convenient presentation for the category of $\infty$-groupoids that are
equipped with 
\emph{strict abelian group structure} in their incarnation as Kan complexes: 
simplicial abelian groups. This
equivalence $\mathrm{Ch}_\bullet \simeq \mathrm{sAb} $ is known as the
\emph{Dold-Kan correspondence}, to be discussed in more detail in 
\ref{SheafAndNonabelianDoldKan}. We write $\Xi(V_\bullet)$ for the 
Kan complex corresponding to a chain complex under this equivalence. 
Moreover, for chain complexes of the form $A[n]$ we write
$$
  \mathbf{B}^n A := \Xi(A[n])
  \,.
$$
With this notation, the $\infty$-groupoid of chain maps $V_\bullet \to A[n]$ is 
equivalently that of $\infty$-functors 
$X \to \mathbf{B}^n A$ and hence the cochain cohomology of $V^\bullet$ is 
$$
  H^n(V^\bullet) \simeq \pi_0\mathbf{H}(X, \mathbf{B}^n A)
  \,.
$$

{\bf Lie group cohomology}

There are some definitions in the literature of cohomology theories that are not special 
cases of this general concept, but in these cases it seems that the failure is with the 
traditional definition, not with the above notion. We will be interested in particular in the 
group cohomology of Lie groups. Originally this was defined using a naive direct generalization 
of the formula for bare group cohomology as
$$
  H^n_{\mathrm{naive}}(G,A) = \{\mbox{smooth maps $G^{\times n} \to A$}\}/\sim
  \,.
$$
But this definition was eventually found to be too coarse: there are structures that ought to be 
cocycles on Lie groups but do not show up in this definition. Graeme Segal therefore proposed a refined 
definition that was later rediscovered by Jean-Luc Brylinski, called 
\emph{differentiable Lie group cohomology} $H^n_{\mathrm{diffbl}}(G,A)$. 
This refines the naive Lie group cohomology in that there is a natural morphism 
$H^n_{\mathrm{naive}}(G,A) \to H^n_{\mathrm{diffbl}}(G,A)$.

But in the $\infty$-topos of smooth $\infty$-groupoids  
$\mathbf{H} = \mathrm{Sh}_{\infty}(\mathrm{CartSp})$ 
we have the natural intrinsic definition of Lie group cohomology as
$$
  H^n_{\mathrm{Smooth}}(G,A) := \pi_0 \mathbf{H}(\mathbf{B}G, \mathbf{B}^n A)
$$
and one finds that this naturally includes the Segal/Brylinski definition
$$
  H^n_{\mathrm{naive}}(G,A)
  \to 
  H^n_{\mathrm{diffrbl}}(G,A)
  \to 
  H^n_{\mathrm{Smooth}}(G,A)
  :=
  \pi_0 \mathbf{H}(\mathbf{B}G, \mathbf{B}^n A)
  \,.
$$
and at least for $A$ a discrete group, or the group of real numbers or a quotient of these such as $U(1) = \mathbb{R}/\mathbb{Z}$, the notions coincide
$$
  H^n_{\mathrm{diffrbl}}(G,A)
  \simeq 
  H^n_{\mathrm{Smooth}}(G,A)
  \,.
$$
Details on this discussion about refined Lie group
cohomology are below in \ref{SmoothStrucLieGroupCohomology}.

For instance one of the crucial aspects 
of the notion of cohomology is that a cohomology class on $X$ \emph{classifies} 
certain structures over $X$.

It is a classical fact that if $G$ is a (discrete) group and 
$BG$ its delooping in Top, then the structure classified by a cocycle 
$g : X \to BG$ is the $G$-principal bundle over $X$ obtained as the 1-categorical 
pullback $P \to X$
$$
  \xymatrix{
     P \ar[r] \ar[d] & E G \ar[d]
     \\
     X \ar[r]^g & B G
  }
$$
of the universal $G$-principal bundle $E G \to B G$. 
But one finds that this pullback construction is just a 1-categorical \emph{model} for 
what intrinsically is something simpler: this is just the 
\emph{homotopy pullback} in $\mathrm{Top}$ of the point
$$
  \xymatrix{
     P \ar[r]_>{\ }="s" \ar[d] & {*} \ar[d]
     \\
     X \ar[r]_g^<{\ }="t" & B G
     \ar@{=>}^\simeq "s"; "t"
  }	
$$
This form of the construction of the $G$-principal bundle classified by a cocycle 
makes sense in any $\infty$-topos $\mathbf{H}$:

We say that for $G \in \mathbf{H}$ a group object in $\mathbf{H}$ and 
$\mathbf{B}G$ its delooping and for $g : X \to \mathbf{B}G$ a cocycle
(any morphism in $\mathbf{H}$) that the 
\emph{$G$-principal $\infty$-bundle} classified by $g$ is the $\infty$-pullback/homotopy pullback
$$
  \xymatrix{
     P \ar[r]_>{\ }="s" \ar[d] & {*} \ar[d]
     \\
     X \ar[r]_g^<{\ }="t" & B G
     \ar@{=>}^\simeq "s"; "t"
  }	
$$
in $\mathbf{H}$. (Beware that usually we will notationally suppress the homotopy filling this square diagram.)

Let $G$ be a Lie group and $X$ a smooth manifold, both regarded naturally as objects in the 
$\infty$-topos of smooth $\infty$-groupoids. Let $g : X \to \mathbf{B}G$ be a morphism in 
$\mathbf{H}$. One finds that in terms of the presentation of $\mathrm{Smooth}\infty\mathrm{Grpd}$ 
by the model structure on simplicial presheaves this is a {\v C}ech 1-cocycle on $X$ with values in 
$G$. The corresponding $\infty$-pullback $P$ is (up to equivalence or course) 
the smooth $G$-principal bundle classified in the usual sense by this cocycle.

The analogous proposition holds for $G$ a Lie 2-group and $P$ a $G$-principal 2-bundle.

Generally, we can give a natural definition of $G$-principal $\infty$-bundle in any 
$\infty$-topos $\mathbf{H}$ over any $\infty$-group object $G \in \mathbf{H}$. 
One finds that it is the Giraud axioms that characterize $\infty$-toposes that ensure that these 
are equivalently classified as the $\infty$-bullbacks of morphisms $g : X \to \mathbf{B}G$. 
Therefore the intrinsic cohomology
$$
  H(X,G):=  \pi_0 \mathbf{H}(X,BG)
$$
in $\mathbf{H}$ classifies $G$-principal $\infty$-bundles over $X$. Notice that $X$ here may 
itself be any object in $\mathbf{H}$.

\paragraph{Homotopy}
\label{IntroHomotopy}
\index{homotopy!overview}

Every $\infty$-sheaf $\infty$-topos $\mathbf{H}$ 
canonically comes equipped with a geometric morphism given by pair of adjoint $\infty$-functors
$$
  (L \mathrm{Const} \dashv \Gamma)
    :
  \xymatrix{
    \mathbf{H}
      \ar@<+3pt>@{<-}[r]^{L \mathrm{Const}}
      \ar@<-3pt>[r]_{\Gamma}
       &
    \infty \mathrm{Grpd}
  }
$$
relating it to the archeytpical $\infty$-topos of $\infty$-groupoids. Here $\Gamma$ produces 
the global sections of an $\infty$-sheaf and $L \mathrm{Const}$ produces the constant $\infty$-sheaf 
on a given $\infty$-groupoid.

In the cases that we are interested in here $\mathbf{H}$ is a big topos of $\infty$-groupoids 
equipped with cohesive structure, notably equipped with smooth structure in our motivating example. 
In this case $\Gamma$ has the interpretation of sending a cohesive $\infty$-groupoid 
$X \in \mathbf{H}$ to its underlying $\infty$-groupoid, after forgetting the cohesive 
structure, and $L \mathrm{Const}$ has the interpretation of forming $\infty$-groupoids 
equipped with discrete cohesive structure. We shall write $\mathrm{Disc} := L \mathrm{Const}$ 
to indicate this.

But in these cases of cohesive $\infty$-toposes there are actually more adjoints to 
these two functors, and this will be essentially the general abstract definition of cohesiveness. 
In particular there is a further left adjoint
$$
  \Pi : \mathbf{H} \to \infty \mathrm{Grpd}
$$
to $\mathrm{Disc}$: the \emph{fundamental $\infty$-groupoid functor on a locally $\infty$-connected $\infty$-topos}. Following the standard terminology of \emph{locally connected toposes} in ordinary topos theory we 
shall say that $\mathbf{H}$ with such a property is a \emph{locally $\infty$-connected $\infty$-topos}. 
This terminology reflects the fact that if $X$ is a locally contractible topological space then 
$\mathbf{H} = \mathrm{Sh}_{\infty}(X)$ is a locally contractible $\infty$-topos. 
A classical result of Artin-Mazur implies, that in this case the value of $\Pi$ on 
$X \in \mathrm{Sh}_{\infty}(X)$ is, up to equivalence, the \emph{fundamental $\infty$-groupoid of $X$}:
$$
  \Pi : (X \in \mathrm{Sh}_{\infty}(X)) \mapsto (\mathrm{Sing} X \in \infty \mathrm{Grpd})
  \,,
$$
which is the $\infty$-groupoid whose
\begin{itemize}
  \item 
    objects are the points of $X$;
  \item
    morphisms are the (continuous) paths in $X$;
  \item
    2-morphisms are the continuous homotopies between such paths;
  \item
    $k$-morphisms are the higher order homotopies between $(k-1)$-dimensional paths.
\end{itemize}
This is the object that encodes all the homotopy groups of $X$ in a canonical fashion, 
without choice of fixed base point.

Also the big $\infty$-topos $\mathrm{Smooth}\infty\mathrm{Grpd} = \mathrm{Sh}_{\infty}(\mathrm{CartSp})$
turns out to be locally  $\infty$-connected
$$
  (\Pi \dashv \mathrm{Disc} \dashv \Gamma) : 
  \xymatrix{
    \mathrm{Smooth}\infty\mathrm{Grpd}
    \ar@<+12pt>[rr]^{\Pi}
    \ar@<+4pt>@{<-}[rr]|{\mathrm{Disc}}
    \ar@<-4pt>[rr]_{\Gamma}
    &&
    \infty \mathrm{Grpd}
  }  
$$
as a reflection of the fact that every Cartesian space $\mathbb{R}^n \in \mathrm{CartSp}$ 
is contractible as a topological space. We find that for $X$ any paracompact smooth manifold, 
regarded as an object of $\mathrm{Smooth}\infty\mathrm{Grpd}$, again 
$\Pi(X) \in \mathrm{Smooth}\infty\mathrm{Grpd}$ 
is the corresponding fundamental $\infty$-groupoid. More in detail, under the 
\emph{homotopy-hypothesis}-equivalence 
$(|-|\dashv \mathrm{Sing}) : \xymatrix{\mathrm{Top} 
\ar@<-4pt>[r]_{\mathrm{Sing}}^{\simeq} \ar@<+4pt>@{<-}[r]^{|-|} & \infty \mathrm{Grpd}} $
we have that the composite
$$
  |\Pi(-)| : \mathbf{H} \stackrel{\Pi}{\to} \infty \mathrm{Grpd} \stackrel{|-|}{\to}
    \mathrm{Top}
$$
sends a smooth manifold $X$ to its homotopy type: 
the underlying topological space of $X$, up to weak homotopy equivalence.

Analogously, for a general object $X \in \mathbf{H}$ we may think of $|\Pi(X)|$ as the 
generalized geometric realization in $\mathrm{Top}$. For instance we find that if 
$X \in \mathrm{Smooth}\infty\mathrm{Grpd}$ is presented by a simplicial paracompact manifold, 
then $|\Pi(X)|$ is the ordinary geometric realization of the underlying simplicial 
topological space of 
$X$.
This means in particular that for $X \in \mathrm{Smooth}\infty\mathrm{Grpd}$ a Lie groupoid, $\Pi(X)$ 
computes its \emph{homotopy groups of a Lie groupoid} as traditionally defined.

The ordinary homotopy groups of $\Pi(X)$ or equivalently of $|\Pi(X)|$ we call the 
\emph{geometric homotopy groups} of $X \in \mathbf{H}$, because these are based on a 
notion of homotopy induced by an intrisic notion of 
geometric paths in objects in $X$. 
This is to be contrasted with the \emph{categorical homotopy groups} of $X$. 
These are the homotopy groups of the underlying $\infty$-groupoid $\Gamma(X)$ of $X$. 
For instance for $X$ a smooth manifold we have that
$$
  \pi_n(\Gamma(X)) \simeq 
   \left\{
      \begin{array}{cl}
         X \in \mathrm{Set} & | n = 0
         \\
         0 & | n > 0
      \end{array}
    \right.
$$
but
$$
  \pi_n(\Pi(X)) \simeq \pi_n(X \in \mathrm{Top})
  \,.
$$
This allows us to give a precise sense to what it means to have a \emph{cohesive refinement} 
(continuous refinement, smooth refinement, etc.) of an object in $\mathrm{Top}$. 
Notably we are interested in smooth refinements of classifying spaces $B G \in \mathrm{Top}$ 
for topological groups $G$ by deloopings $\mathbf{B}G \in \mathrm{Smooth}\infty\mathrm{Grpd}$ 
of $\infty$-Lie groups $G$ and we may interpret this as saying that
$$
  \Pi(\mathbf{B}G) \simeq 	B G
$$
in $\mathrm{Top} \simeq \mathrm{Smooth}\infty\mathrm{Grpd}$.

%
%

\subsubsection{Principal bundles}
\label{SmoothPrincipalnBundles}
\index{principal $\infty$-bundle!in low dimensions}

The following is an exposition of the notion of 
\emph{principal bundles} in higher but low degree.

We assume here that the reader has a working knowledge of 
groupoids and at least a rough idea of 2-groupoids. 
For introductions see for instance \cite{BHS} \cite{Porter}

Below in \ref{ModelForPrincipalInfinityBundles} a discussion of the
formalization of $\infty$-groupoids in terms of Kan complexes is given and is used to  present a systematic way to understand these constructions 
in all degrees.

\medskip

\paragraph{Principal 1-bundles} 
\label{Principal1Bundles}

Let $G$ be a Lie group and $X$ a smooth manifold (all our smooth manifolds are assumed to be finite dimensional 
and paracompact). We give a discussion of smooth $G$-principal bundles on $X$ in a manner that 
paves the way to a straightforward generalization to a description of principal $\infty$-bundles.
From $X$ and $G$ are naturally induced certain Lie groupoids.

From the group $G$ we canonically obtain a groupoid that we write $B G$ and call the 
\emph{delooping groupoid} of $G$. Formally this groupoid is
$$
  B G = (\xymatrix{
     G \ar@<+3pt>[r] 
       \ar@<-3pt>[r] & {*}
  })
$$
with composition induced from the product in $G$. 
A useful depiction of this groupoid is
$$
  B G 
  = 
  \left\{
  \raisebox{20pt}{
  \xymatrix{
     & {*} \ar[dr]^{g_2}
     \\
    {*} \ar[rr]_{g_2 \cdot g_1}^{\ }="t" 
       \ar[ur]^{g_1}
     && {*}
    \ar@{=} "t"+(0,5); "t"
  }
  }
  \right\}
  \,,
$$
where the $g_i \in G$ are elements in the group, and the bottom morphism is labeled by 
forming the product in the group. 
(The order of the factors here is a convention whose choice, once and for all,
does not matter up to equivalence.)

But we get a bit more, even. Since $G$ is a Lie group, there is smooth structure on 
$B G$ that makes it a Lie groupoid, an internal groupoid in the category 
$\mathrm{SmoothMfd}$ of smooth manifolds: its collection of objects (trivially) and of 
morphisms each form a smooth manifold, and all structure maps 
(source, target, identity, composition) are smooth functions. We shall write
$$
  \mathbf{B}G \in \mathrm{LieGrpd}
$$
for $B G$ regarded as equipped with this smooth structure. Here and in the following the 
boldface is to indicate that we have an object equipped with a bit more structure 
-- here: smooth structure -- than present on the object denoted by the same symbols, 
but without the boldface. Eventually we will make this precise by having the boldface symbols denote 
objects in the $\infty$-topos $\mathrm{Smooth}\infty\mathrm{Grpd}$ which are taken by a suitable functor 
to objects in $\infty \mathrm{Grpd}$ denoted by the corresponding non-boldface symbols.

Also the smooth manifold $X$ may be regarded as a Lie groupoid 
-– a groupoid with only identity morphisms. Its depiction is simply
$$
  X = \{\xymatrix{x \ar[r]^{\mathrm{Id}} & x}\}
$$
for all $x \in X$
But there are other groupoids associated with X: let $\{U_i \to X\}_{i \in I}$ 
be an open cover of $X$. To this is canonically associated the {\v C}ech-groupoid 
$C(\{U_i\})$. Formally we may write this groupoid as
$$
  C(\{U_i\}) = 
  \left\{
    \xymatrix{
      \coprod_{i,j} U_i \cap U_j \ar@<+3pt>[r] \ar@<-3pt>[r] & 
       \coprod_i U_i
    }
  \right\}
  \,.
$$
A useful depiction of this groupoid is
$$
  C(\{U_i\})
  = 
  \left\{
  \raisebox{20pt}{
  \xymatrix{
     & (x,j) \ar[dr]^{}
     \\
    (x,i) \ar[rr]_{}^{\ }="t" 
       \ar[ur]^{}
     && (x,k)
    \ar@{=} "t"+(0,5); "t"
  }
  }
  \right\}
  \,,
$$
This indicates that the objects of this groupoid are pairs $(x,i)$ consisting of a point $x \in X$ and 
a patch $U_i \subset X$ that contains $x$, and a morphism is a triple $(x,i,j)$ consisting of a point 
and two patches, that both contain the point, in that $x \in U_i \cap U_j$. 
The triangle in the above depiction symbolizes the evident way in which these morphisms compose. 
All this inherits a smooth structure from the fact that the $U_i$ are smooth manifolds and 
the inclusions $U_i \hookrightarrow X$ are smooth functions. Hence also 
$C(\{U_i\})$ becomes a Lie groupoid.

There is a canonical projection functor
$$
  C(\{U_i\})\to X : (x,i) \mapsto x
  \,.
$$
This functor is an internal functor in $\mathrm{SmoothMfd}$ and moreover it is evidently essentially 
surjective and full and faithful. However, while 
essential surjectivity and full-and-faithfulness implies that the underlying bare functor
has a homotopy-inverse, that homotopy-inverse never 
has itself smooth component maps, unless $X$ itself is a Cartesian space and the chosen cover is trivial.

We do however want to think of $C(\{U_i\})$ as being equivalent to $X$ even as a Lie groupoid. 
One says that a smooth functor whose underlying bare functor is an equivalence of groupoids is
a \emph{weak equivalence} of Lie groupoids, which we write as 
$C(\{U_i\}) \stackrel{\simeq}{\to} X$.
Moreover, we shall think of $C(\{U_i\})$ as a \emph{good} equivalent replacement of $X$ if it comes 
from a cover that is in fact a \emph{good open cover} in that all its non-empty finite intersections 
$U_{i_0, \cdots, i_n} := U_{i_0} \cap \cdots \cap U_{i_n}$ are diffeomorphic to the Cartesian 
space $\mathbb{R}^{\mathrm{dim}X}$.

We shall discuss later in which precise sense this condition makes $C(\{U_i\})$ 
\emph{good} in the sense 
that smooth functors out of $C(\{U_i\})$ model the correct notion of morphism out of $X$ 
in the context  of smooth groupoids (namely it will mean that $C(\{U_i\})$ is cofibrant in a 
suitable model category 
structure on the category of Lie groupoids). The formalization of this statement is what $\infty$-topos 
theory is all about, to which we will come. For the moment we shall be content with accepting this as an ad hoc statement.

Observe that a functor
$$
  g : C(\{U_i\}) \to \mathbf{B}G
$$
is given in components precisely by a collection of smooth functions
$$
  \{g_{i j} : U_{i j} \to G\}_{i,j \in I}
$$
such that on each $U_{i} \cap U_j \cap U_k$ the equality $g_{j k} g_{i j} = g_{i k}$ of functions
holds.

It is well known that such collections of functions characterize $G$-principal bundles on $X$. 
While this is a classical fact, we shall now describe a way to derive it that is true to the 
Lie-groupoid-context and that will make clear how smooth principal $\infty$-bundles work.

First observe that in total we have discussed so far spans of smooth functors of the form
$$
  \raisebox{20pt}{
  \xymatrix{
    C(\{U_i\}) \ar[r]^g \ar[d]^{\simeq} & \mathbf{B}G
    \\
    X
  }
  }
  \,.
$$
Such spans of functors, whose left leg is a weak equivalence, are sometimes known,
essentially equivalently, as 
\emph{Morita morphisms}, as \emph{generalized morphisms} of Lie groupoids, 
as \emph{Hilsum-Skandalis morphisms},
or as \emph{groupoid bibundles} or as 
\emph{anafunctors}. 
\index{anafunctor}
\index{anafunctor!anafunctor}
We are to think of these as concrete models for more intrinsically 
defined direct morphisms $X \to \mathbf{B}G$ in the $\infty$-topos of smooth $\infty$-groupoids.

Now consider yet another Lie groupoid canonically associated with $G$: 
we shall write $\mathbf{E}G$ for the groupoid 
-- the \emph{smooth universal $G$-bundle}
\index{principal $\infty$-bundle!universal principal $\infty$-bundle!smooth, in introduction} -- whose formal description is
$$
  \mathbf{E}G = \left(
    \xymatrix{
       G \times G \ar@<+3pt>[r]^{(-)\cdot(-)} \ar@<-3pt>[r]_{p_1} & G
    }
  \right)
$$
with the evident composition operation. The depiction of this groupoid is
$$
  \left\{
  \raisebox{20pt}{
  \xymatrix{
     & g_2 \ar[dr]^{g_3 g_2^{-1}}
     \\
    g_1 \ar[rr]_{g_3 g_1^{-1}}^{\ }="t" 
       \ar[ur]^{g_2 g_1^{-1}}
     && g_3
    \ar@{=} "t"+(0,5); "t"
  }
  }
  \right\}
  \,,
$$
This again inherits an evident smooth 
structure from the smooth structure of $G$ and hence becomes a Lie groupoid.

There is an evident forgetful functor
$$
  \mathbf{E}G\to \mathbf{B}G
$$
which sends
$$
  (g_1 \to g_2) \mapsto (\bullet \stackrel{g_2 g_1^{-1}}{\to} \bullet)
  \,.
$$
Consider then the pullback diagram
$$
  \xymatrix{
    \tilde P \ar[r]\ar[d] & \mathbf{E}G \ar[d]
    \\
    C(\{U_i\}) \ar[r]^g \ar[d]^{\simeq} & \mathbf{B}G
    \\
    X
  }
$$
in the category $\mathrm{Grpd}(\mathrm{SmoothMfd})$. The object $\tilde P$ is the Lie groupoid
whose depiction is
$$
  \tilde P =
  \left\{
    \xymatrix{
       (x,i,g_1) \ar[r] &(x,j,g_2 = g_{i j}(x) g_1 )
    }
  \right\}
  \,;
$$
where there is a unique morphism as indicated, whenever the group labels match as indicated. 
Due to this uniqueness, this Lie groupoid is weakly equivalent to one that comes just from a 
manifold $P$ (it is 0-truncated)
$$
  \tilde P \stackrel{\simeq}{\to} P
  \,.
$$
This $P$ is traditionally written as
$$
  P = 
    \left(
      \coprod_i U_i \times G
    \right)/\sim
    \,,
$$
where the equivalence relation is precisely that exhibited by the morphisms in $\tilde P$. 
This is the traditional way to construct a $G$-principal bundle from cocycle functions 
$\{g_{i j}\}$. We may think of $\tilde P$ as \emph{being} $P$. 
It is a particular representative of $P$ in the $\infty$-topos of Lie groupoids.

While it is easy to see in components that the $P$ obtained this way does indeed have a 
principal $G$-action on it, for later generalizations it is crucial that we can also 
recover this in a general abstract way. For notice that there is a canonical action
$$
  (\mathbf{E}G) \times G \to \mathbf{E}G
  \,,
$$
given by the group action on the space of objects.
Then consider the pasting diagram of pullbacks
$$
  \xymatrix{
    \tilde P \times G \ar[r] \ar[d] & \mathbf{E}G \times G \ar[d]
    \\
    \tilde P \ar[r] \ar[d] & \mathbf{E}G \ar[d]
    \\
    C(U) \ar[r]^g \ar[d]^{\simeq }& \mathbf{B}G
    \\
    X 
  }
  \,.
$$
Here the morphism $\tilde P \times G \to \tilde P$ exhibits the principal $G$-action of $G$ on $\tilde P$.

In summary we find the following
\begin{observation}
  For $\{U_i\to X\}$ a good open cover,
  there is an equivalence of categories
  $$
    \mathrm{SmoothFunc}(C(\{U_i\}), \mathbf{B}G) \simeq G \mathrm{Bund}(X)
  $$
  between the functor category of smooth functors and smooth natural transformations, and 
  the groupoid of smooth $G$-principal bundles on $X$.
\end{observation}
It is no coincidence that this statement looks akin to the maybe more familiar statement which 
says that equivalence classes of $G$-principal bundles are classified by homotopy-classes of 
morphisms of topological spaces
$$
  \pi_0 \mathrm{Top}(X, B G)
  \simeq
  \pi_0 G \mathrm{Bund}(X)
  \,,
$$
where $B G \in \mathrm{Top}$ is the topological classifying space of $G$. 
What we are seeing here is a first indication of how cohomology of bare $\infty$-groupoids 
is lifted inside a richer $\infty$-topos 
to cohomology of $\infty$-groupoids with extra structure.

In fact, all of the statements that we considered so far becomes conceptually simpler in the $\infty$-topos. 
We had already remarked that the anafunctor span 
$X \stackrel{\simeq}{\leftarrow} C(\{U_i\}) \stackrel{g}{\to} \mathbf{B}G$ 
is really a model for what is simply a direct morphism $X \to \mathbf{B}G$ in the $\infty$-topos. 
But more is true: that pullback of $\mathbf{E}G$ which we considered is just a model for the 
homotopy pullback of just the \emph{point}

$$
  \begin{array}{cccc}
    \xymatrix{
       \vdots & \vdots
       \\
       \tilde P \times G \ar[r] \ar[d] & \mathbf{E}G \times G \ar[d]
       \\
       \tilde P \ar[r] \ar[d] & \mathbf{E}G \ar[d]
       \\
       C(U) \ar[r]^g \ar[d]^\simeq & \mathbf{B}G
       \\
       X
    } 
    &&&
    \xymatrix{
      \vdots & \vdots
      \\
      P \times G \ar[r]_>{\ }="s1" \ar[d] & G \ar[d]
      \\
      P \ar[r]^<{\ }="t1"_>{\ }="s2" \ar[d] & {*} \ar[d]
      \\
      X \ar[r]_g^<{\ }="t2" & \mathbf{B}G
      \\
     \ar@{=>}^\simeq "s1"; "t1"
     \ar@{=>}^\simeq "s2"; "t2"
    }
    \\ 
    \\
    \mbox{in the model category}
      &&&
    \mbox{in the $\infty$-topos}
  \end{array}
  \,.
$$
The traditional statement which identifies the classifying topological space $B G$
as the quotient of the contractible $E G$ by the free $G$-action
$$
  B G \simeq E G / G
$$
becomes afte the refinement to smooth groupoids the statement that 
$\mathbf{B}G$ is the \emph{homotopy quotient} of $G$ acting on the point:
$$
  \mathbf{B}G \simeq {*}/\!/G
  \,.
$$
Generally: 
\begin{definition}
\index{action groupoid!in introduction}
\label{ActionGroupoidInIntroduction}
For $V$ a smooth manifold equipped with a smooth action by 
$G$ (not necessarily free), the \emph{action groupoid} $V/\!/G$ is 
the Lie groupoid whose space of objects is $V$, and whose morphisms
are group elements that connect two points (which may coincide) in $V$.
$$
  V/\!/G
  =
  \left\{
    \xymatrix{
	  v_1 \ar[r]^g & v_2 
	}
	| v_2 = g(v_1)
  \right\}
  \,.
$$
\end{definition}
Such an action groupoid is canonically equipped with a morphism
to $\mathbf{B}G \simeq */\!/G$ obtained by sending all objects to the 
single object and acting as the identity on morphisms.
Below in \ref{StrucRepresentations} we discuss that 
the sequence
$$
  V \to V/\!/ G \to \mathbf{B}G
$$
entirely encodes the action of $G$ on $V$.
Also we will see in \ref{Twisted0BundlesSectionsOfVectorBundles} that the morphism
$V/\!/G \to \mathbf{B}G$ is the smooth refinement of the 
$V$-bundle which is \emph{associated to the universal $G$-bundle}
via the given action. If $V$ is a vector space acted on linearly, then 
this is an associated vector bundle. Its pullbacks along anafunctors
$X \to \mathbf{B}G$ yield all $V$-vector bundles on $X$.

\medskip

\paragraph{Principal 2-bundles and twisted 1-bundles} 
\label{Principal2Bundles}
The discussion above of $G$-principal bundles was all based on the Lie groupoids $\mathbf{B}G$ and 
$\mathbf{E}G$ that are canonically induced by a Lie group $G$. We now discuss the case 
where $G$ is generalized to a \emph{Lie 2-group}. The above discussion will go through essentially 
verbatim, only that we pick up 2-morphisms everywhere. This is the first step towards higher 
Chern-Weil theory. The resulting generalization of the notion of principal bundle is that of 
\emph{principal 2-bundle}. For historical reasons these are known in the literature often as \emph{gerbes} or as \emph{bundle gerbes}\index{bundle gerbe}, even though strictly speaking there are some conceptual differences.

Write $U(1)  = \mathbb{R}/\mathbb{Z}$ for the circle group. We have already seen above the 
groupoid $\mathbf{B}U(1)$ obtained from this. But since $U(1)$ is an abelian group this 
groupoid has the special property that it still has itself the structure of a group object. 
This makes it what is called a \emph{2-group}. Accordingly, we may form its delooping once 
more to arrive at a Lie 2-groupoid $\mathbf{B}^2 U(1)$. Its depiction is
$$
  \mathbf{B}^2 U(1)
  =
  \left\{
  \raisebox{20pt}{
  \xymatrix{
     & {*} \ar[dr]^{\mathrm{Id}}
     \\
    {*} \ar[rr]_{\mathrm{Id}}^{\ }="t" 
       \ar[ur]^{\mathrm{Id}}
     && {*}
    \ar@{=>}^g "t"+(0,5); "t"
  }
  }
  \right\}
$$
for $g \in U(1)$. Both horizontal composition as well as vertical composition of the 2-morphisms is 
given by the product in $U(1)$.

Let again $X$ be a smooth manifold with good open cover $\{U_i \to X\}$. 
The corresponding {\v C}ech groupoid we may also think of as a Lie 2-groupoid,
$$
  C(U)
  = 
  \left(
    \xymatrix{
      \coprod_{i,j,k} U_i \cap U_j \cap U_k
      \ar@<+6pt>[r]
      \ar@<+0pt>[r]
      \ar@<-6pt>[r]
      &
      \coprod_{i,j} U_i \cap U_j
      \ar@<+3pt>[r]
      \ar@<-3pt>[r]
      &
      \coprod_i U_i
    }  
  \right)
  \,.
$$
What we see here are the first stages of the full \emph{{\v C}ech nerve} of the cover. 
Eventually we will be looking at this object in its entirety, since for all degrees this is always 
a \emph{good} replacement of the manifold $X$, as long as $\{U_i \to X\}$ is a good open cover. 
So we look now at 2-anafunctors given by spans
\index{2-anafunctor}
\index{anafunctor!2-anafunctor}
$$
  \xymatrix{
    C(\{U_i\}) \ar[r]^{g} \ar[d]^{\simeq} & \mathbf{B}^2 U(1)
    \\
    X
  }
$$
of internal 2-functors. These will model direct morphisms $X \to \mathbf{B}^2 U(1)$ in the 
$\infty$-topos. It is straightforward to read off the following
\begin{observation}
 \label{CocyclesForCircle2Bundles}
A smooth 2-functor 
$g : C(\{U_i\}) \to \mathbf{B}^2 U(1)$ is given by the data of a 2-cocycle in the 
{\v C}ech cohomology of $X$ with coefficients in $U(1)$. 
\end{observation}
Because on 2-morphisms it specifies an assignment
$$
  g : 
    \left\{
   \raisebox{20pt}{
  \xymatrix{
     & (x,j) \ar[dr]^{}
     \\
    (x,i) \ar[rr]_{}^{\ }="t" 
       \ar[ur]^{}
     && (x,k)
    \ar@{=>} "t"+(0,8); "t"
  }
  }
  \right\}
  \;\;\;\;
   \mapsto
  \;\;\;\;
  \left\{
  \raisebox{20pt}{
  \xymatrix{
     & {*} \ar[dr]^{\mathrm{Id}}
     \\
    {*} \ar[rr]_{\mathrm{Id}}^{\ }="t" 
       \ar[ur]^{\mathrm{Id}}
     && {*}
    \ar@{=>}|<<<<{g_{i j k}(x)} "t"+(0,8); "t"
  }
  }
  \right\}
$$
that is given by a collection of smooth functions
$$
  (g_{i j k} : U_i \cap U_j \cap U_k \to U(1))
  \,.
$$
On 3-morphisms it gives a constraint on these functions, since there are only identity 3-morphisms in 
$\mathbf{B}^2 U(1)$:
$$
  \left(
  \left(
  \raisebox{20pt}{
  \xymatrix{
     (x,j) \ar[r] & (x,k) \ar[d]
     \\
    (x,i) \ar[r] \ar[u] \ar[ur] & (x,l)
  }
  }
  \right)
  \Rightarrow
  \left(
  \raisebox{20pt}{
  \xymatrix{
     (x,j) \ar[r] \ar[dr] & (x,k) \ar[d]
     \\
    (x,i) \ar[r] \ar[u]  & (x,l)
  }
  }
  \right)
  \right)
  \mapsto
  \;\;
  \left(
  \left(
  \raisebox{35pt}{
  \xymatrix{
     {*} \ar[rr] && {*} \ar[dd]
     \\
     \\
    {*} \ar[rr] \ar[uu] \ar[uurr]^{\ }="t"_{\ }="s" && {*}
    \ar@{=>}^<<{g_{i j k}(x)} "t"+(-4,4); "t"
    \ar@{=>}_<<<<<<{g_{i k l}(x)} "s"; "s"+(4,-4)
  }
  }
  \right)
   =
  \left(
  \raisebox{35pt}{
  \xymatrix{
     {*} \ar[rr] \ar[ddrr]^{\ }="t"_{\ }="s" && {*} \ar[dd]
     \\
     \\
    {*} \ar[rr] \ar[uu]  && {*}
    \ar@{=>}_{g_{j k l}(x)\hspace{-10pt}} "t"+(4,4); "t"
    \ar@{=>}^{\hspace{-15pt}g_{i j l}(x)} "s"; "s"+(-4,-4)
  }
  }
  \right)
  \right)
  \,.
$$
This relation
$$
  g_{i j k} \cdot g_{i k l} = g_{i j l} \cdot g_{j k l}
$$
defines degree-2 cocycles in \emph{{\v C}ech cohomology} with coefficients in $U(1)$.

In order to find the circle principal 2-bundle classified by such a cocycle by a pullback operation as before, 
we need to construct the 2-functor 
$\mathbf{E} \mathbf{B} U(1) \to \mathbf{B}^2 U(1)$ that exhibits the universal principal 2-bundle over $U(1)$.
The right choice for $\mathbf{E B} U(1)$ -- 
which we justify systematically in \ref{ModelForPrincipalInfinityBundles} -- is  indicated by
$$
  \mathbf{E B} U(1)
    = 
  \left\{
     \raisebox{20pt}{
       \xymatrix{
           & {*} \ar[dr]^{c_2}
           \\
          {*} \ar[ur]^{c_1}
           \ar[rr]_{c_3 = g c_2 c_1}^{\ }="t"
          &&
          {*}
          \ar@{=>}^g "t"+(0,5); "t" 
       }
     }
  \right\}
$$
for $c_1, c_2, c_3, g \in U(1)$, where all possible composition operations are given by forming the
product of these labels in $U(1)$. The projection $\mathbf{E B}U(1) \to \mathbf{B}^2 U(1)$ is
the obvious one that simply forgets the labels $c_i$ of the 1-morphisms and just remembers the labels 
$g$ of the 2-morphisms.
\begin{definition}
 \label{Circle2BundleTotalSpace}
\index{principal $\infty$-bundle!circle 2-bundle}
With $g : C(\{U_i\}) \to \mathbf{B}^2 U(1)$ a {\v C}ech cocycle as above, the \emph{$U(1)$-principal 2-bundle} or \emph{circle 2-bundle}
that it defines is the pullback
	$$
  \xymatrix{
    \tilde P \ar[r] \ar[d] & \mathbf{E}\mathbf{B}U(1) \ar[d]
    \\
    C(\{U_i\}) \ar[r]^{g} \ar[d]_{\simeq} & \mathbf{B}^2 U(1)
    \\
    X
  }
  \,.
$$
\end{definition}
Unwinding what this means, we see that  $\tilde P$ is the 2-groupoid whose objects are that of 
$C(\{U_i\})$, whose morphisms are finite sequences of morphisms in $C(\{U_i\})$, each equipped with 
a label $c \in U(1)$, and whose 2-morphisms are generated from those that look like
$$
   \raisebox{20pt}{
  \xymatrix{
     & (x,j) \ar[dr]^{c_2}
     \\
    (x,i) \ar[rr]_{c_3}^{\ }="t" 
       \ar[ur]^{c_1}
     && (x,k)
    \ar@{=>}^{g_{i j k}(x)} "t"+(0,5); "t"
  }
  }
$$
subject to the condition that
$$
  c_1 \cdot c_2 = c_3 \cdot g_{i j k}(x)
$$
in $U(1)$. As before for principal 1-bundles $P$, where we saw that the analogous 
pullback 1-groupoid $\tilde P$ was equivalent to the 0-groupoid $P$, 
here we see that this 2-groupoid is equivalent to the 1-groupoid 
$$
  P =
  \left(
     \xymatrix{
        C(U)_1 \times U(1) 
         \ar@<+3pt>[r]
         \ar@<-3pt>[r]
         &
         C(U)
     }
  \right)
$$
with composition law	
$$
  ((x,i) \stackrel{c_1}{\to} (x,j) \stackrel{c_2}{\to} (x,k))
  =
  ((x,i) \stackrel{(c_1 \cdot c_2 \cdot g_{i j k }(x))}{\to}
  (x,k))
  \,.
$$
This is a groupoid central extension
$$
  \mathbf{B}U(1) \to P \to C(\{U_i\}) \simeq X
  \,.
$$
Centrally extended groupoids of this kind are known in the literature as 
\emph{bundle gerbes} 
(over the surjective submersion $Y = \coprod_i U_i \to X$ ). They may 
equivalently be thought of as given by a line bundle
$$
  \xymatrix{
    L \ar[d]
    \\
    (C(U)_1 = \coprod_{i, j} U_i \cap U_j)
    \ar@<+3pt>[r]
    \ar@<+3pt>[r]
    &
    (C(U)_0 = \coprod_i U_i)
    \ar[d]
    \\
    & X
  }
$$
over the space $C(U)_1$ of morphisms, and a line bundle morphism
$$
  \mu_g : \pi_1^* L \otimes \pi_2^* L \to \pi_1^* L
$$
that satisfies an evident associativity law, equivalent to the cocycle codition on $g$.
In summary we find that:
\begin{observation} 
 \label{BundleGerbeIsCircle2Bundle}
\index{bundle gerbe!as principal circle 2-bundle}
\index{principal $\infty$-bundle!bundle gerbe}
Bundle gerbes are presentations of Lie groupoids that are total spaces of $\mathbf{B}U(1)$-principal 2-bundles, def. \ref{Circle2BundleTotalSpace}.
\end{observation}
Notice that, even though there is a close relation, the notion
of \emph{bundle gerbe} is different from the original notion of 
\emph{$U(1)$-gerbe}. This point we discuss in more detail 
below in \ref{UnderlyingGerbeOf2Bundle} and  more abstractly in 
\ref{ETopStrucGerbes}.

\medskip

This discussion of \emph{circle 2-bundles} has a generalization to 2-bundles that are principal over more general \emph{2-groups}.
\begin{definition}
  \label{Strict2GroupInIntroduction}
  \label{crossed module}
  \index{group!2-group!strict}
  \index{group!2-group!crossed module}
  \begin{enumerate}
  \item A smooth \emph{crossed module} of Lie groups is a pair of homomorphisms 
  $\partial : G_1 \to G_0$ and $\rho : G_0 \to \mathrm{Aut}(G_1)$ of 
  Lie groups, such that for all $g \in G_0$ and $h,h_1, h_2 \in G_1$
  we have $\rho(\partial h_1)(h_2) = h_1 h_2 h_1^{-1}$ and
  $\partial \rho(g)(h) = g \partial(h) g^{-1}$.

  \item For $(G_1 \to G_0)$ a smooth crossed module, the corresponding \emph{strict Lie 2-group} is 
  the smooth groupoid 
  $\xymatrix{
    G_0 \times G_1 \ar@<+4pt>[r]_{}\ar@<-4pt>[r]^{} & G_0}$, whose source map is given by projection on 
    $G_0$, whose target map is given by applying $\partial$ to the second factor and then multiplying
    with the first in $G_0$, and whose composition is given by multiplying in $G_1$.
 
   This groupoid has a strict monoidal structure with strict inverses given by equipping 
   $G_0 \times G_1$ with the semidirect product group structure $G_0 \ltimes G_1$ induced
   by the action $\rho$ of $G_0$ on $G_1$.
 
  \item 
   The corresponding one-object strict smooth 2-groupoid we write $\mathbf{B}(G_1 \to G_0)$.
   As a simplicial object (under the Duskin nerve of 2-categories)
  this is of the form
  $$
    \mathbf{B}(G_1 \to G_0) =
    \mathrm{cosk}_3
    \left(
     \xymatrix{
       G_0^{\times 3} \times G_1^{\times 3}\ar@<-6pt>[r]\ar[r]\ar@<+6pt>[r] & G_0^{\times 2} \times G_1 \ar@<-4pt>[r]\ar@<+4pt>[r] & G_0 \ar[r] & {*}
       }
    \right)
    \,.
  $$
  \end{enumerate}
\end{definition}
The infinitesimal analog of a crossed module of groups is a
\emph{differential crossed module}.
\begin{definition}
 \label{DifferentialCrossedModule}
 A \emph{differential crossed module} is a chain complex of vector
 space of length 2 $V_1 \to V_0$ equipped with the structure of
 a dg-Lie algebra.
\end{definition}
\begin{example}
For $G_1 \to G_0$ a smooth crossed module of Lie groups, 
differentiation of all structure maps yields a corresponding
differential crossed module $\mathfrak{g}_1 \to \mathfrak{g}_0$.
\end{example}
\begin{observation}
For $G := [G_1 \stackrel{\delta}{\to} G_0]$ a crossed module, the 2-groupoid 
delooping a 2-group coming from a crossed module is of the form
$$
  \mathbf{B}G
  = 
  \left\{
  \raisebox{20pt}{
  \xymatrix{
     & {*} \ar[dr]^{g_2}
     \\
    {*} \ar[rr]_{\delta(k )g_2 \cdot g_1}^{\ }="t" 
       \ar[ur]^{g_1}
     && {*}
    \ar@{=>}^k "t"+(0,5); "t"
  }
  }
    \;\;
    |
    \;\;
    g_1, g_2 \in G_0, k \in G_1 
  \right\}
  \,,
$$
where the 3-morphisms -- the composition identities -- are
$$
  \left(
  \raisebox{35pt}{
  \xymatrix{
     {*} \ar[rr]^{g_2} && {*} \ar[dd]^{g_3}
     \\
     \\
    {*} \ar[rr]_{} 
	 \ar[uu]^{g_1} 
	  \ar[uurr]^{\ }="t"_{\ }="s" && {*}
    \ar@{=>}^<<{h_1} "t"+(-4,4); "t"
    \ar@{=>}_<<<{h_2} "s"; "s"+(4,-4)
  }
  }
  \right)
  \xymatrix{
    \ar[rrr]^{h_2 \cdot \rho(g_3)(h_1) = h_4 \cdot h_3} &&&
  }
  \left(
  \raisebox{35pt}{
  \xymatrix{
     {*} \ar[rr]^{g_2} \ar[ddrr]^{\ }="t"_{\ }="s" && {*} \ar[dd]^{g_3}
     \\
     \\
    {*} \ar[rr] \ar[uu]^{g_1}  && {*}
    \ar@{=>}^{h_3} "t"+(4,4); "t"
    \ar@{=>}_>>>>>>{h_4} "s"; "s"+(-4,-4)
  }
  }
  \right)
$$
\end{observation}
\begin{remark}
  \label{StrictificationOf2GroupsInIntroduction}
  All ingredients here are functorial, so that the above statements
  hold for presheaves over sites, hence in particular for
  cohesive  2-groups such as smooth 2-groups. 
  Below in corollarly \ref{StrictificationOf2GroupObjects}
  it is shown that every cohesive 2-group has a presentation 
  by a crossed module this way. 
\end{remark}
Notice that there are different equivalent conventions possible
for how to present $\mathbf{B}G$ in terms of the correspondiung crossed
module, given by the choices of order in the group products. Here we are following convention ``LB'' in \cite{RobertsSchreiber}.
\begin{example}[shift of abelian Lie group]
  \label{CrossedModuleOfShiftedAbelianGroup}
  For $K$ an abelian Lie group then $\mathbf{B}K$ is the delooping 2-group coming from the crossed module $[K \to 1]$ and $\mathbf{B}\mathbf{B}K$ is the 2-group coming from the complex $[K \to 1 \to 1]$.
\end{example}
\begin{example}[automorphism 2-group]
    \label{Automorphism2GroupCrossedModule}
    \index{group!2-group!automorphism 2-group}
    \index{2-group!automorphism 2-group}
    For $H$ any Lie group with automorphism Lie group $\mathrm{Aut}(H)$, the morphism
    $H \stackrel{\mathrm{Ad}}{\to} \mathrm{Aut}(H)$ that sends group elements to inner
    automorphisms, together with $\rho = \mathrm{id}$, is a crossed module.
    We write $\mathrm{AUT}(H) := (H \to \mathrm{Aut}(H))$ and speak of the 
    \emph{automorphism 2-group} of $H$.
\end{example}
\begin{example}
    \label{CrossedModuleFromNormalSubgroup}
    \index{group!2-group!from normal subgroup}
    The inclusion of any normal subgroup $N \hookrightarrow G$ with 
	conjugation action of $G$ on $N$ is a crossed module,
	with the canonical induced conjugation action of $G$ on $N$.
\end{example}
\begin{example}[string 2-group]
    \label{String2GroupInIntroduction}
    \index{group!2-group!string 2-group}
    For $G$ a compact, simple and simply connected Lie group,
	write $P G$ for the smooth group of based paths in $G$ and
	$\hat \Omega G$ for the universal central extension of the 
	smooth group of based loops. Then the evident morphism
   $({\hat \Omega} G \to P G)$ equipped with a lift of the 
   adjoint action of paths on loops is a crossed module
   \cite{BCSS}. The corresponding strict 2-group 
   is (a presentation of what is) called the \emph{string 2-group} extension of $G$. The string 2-group we discuss in detail in \ref{SmoothBString}.
\end{example}
It follows immediately that 
\begin{observation}
 \label{CocycleForPrincipal2Bundles}
 \index{principal $\infty$-bundle!principal 2-bundle}
For $G = (G_1 \to G_0)$ 
a 2-group coming from a crossed module, a cocycle
$$
  X \stackrel{\simeq}{\leftarrow} C(U_i) \stackrel{g}{\to}
  \mathbf{B}G
$$
is given by data 
$$
  \{h_{i j} \in C^\infty(U_{i j}, G_0), g_{i j k } \in C^\infty(U_{i j k}, G_1)\}
$$
such that on each $U_{i j k}$ we have
$$
  h_{i k} = \delta(h_{i j k}) h_{j k} h_{i j}
$$
and on each $U_{i j k l}$ we have
$$
  g_{i k l} \cdot \rho(h_{jk})(g_{i j k})
  =
  g_{i j k} \cdot g_{j k l}
  \,.
$$
\end{observation}
Because under the above correspondence between crossed modules
and 2-groups, this is the data that encodes assignments
$$
  g : 
    \left\{
   \raisebox{20pt}{
  \xymatrix{
     & (x,j) \ar[dr]^{}
     \\
    (x,i) \ar[rr]_{}^{\ }="t" 
       \ar[ur]^{}
     && (x,k)
    \ar@{=>} "t"+(0,8); "t"
  }
  }
  \right\}
  \;\;\;\;
   \mapsto
  \;\;\;\;
  \left\{
  \raisebox{20pt}{
  \xymatrix{
     & {*} \ar[dr]^{h_{j k}(x)}
     \\
    {*} \ar[rr]_{h_{i k}(x)}^{\ }="t" 
       \ar[ur]^{h_{i j}(x)}
     && {*}
    \ar@{=>}|<<<<{g_{i j k}(x)} "t"+(0,8); "t"
  }
  }
  \right\}
$$
that satisfy
$$
  \left(
  \raisebox{35pt}{
  \xymatrix{
     {*} \ar[rr]^{h_{j k}} && {*} \ar[dd]^{h_{k l}}
     \\
     \\
    {*} \ar[rr]_{} 
	 \ar[uu]^{h_{i j}} 
	  \ar[uurr]^{\ }="t"_{\ }="s" && {*}
    \ar@{=>}^<<{g_{i j k}} "t"+(-4,4); "t"
    \ar@{=>}_<<<{g_{i k l}} "s"; "s"+(4,-4)
  }
  }
  \right)
  \xymatrix{
    \ar[rr]^{} &&
  }
  \left(
  \raisebox{35pt}{
  \xymatrix{
     {*} \ar[rr]^{h_{j k}} \ar[ddrr]^{\ }="t"_{\ }="s" && {*} \ar[dd]^{h_{k l}}
     \\
     \\
    {*} \ar[rr] \ar[uu]^{h_{i j}}  && {*}
    \ar@{=>}^{g_{j k l}} "t"+(4,4); "t"
    \ar@{=>}_>>>>>>{g_{i j l}} "s"; "s"+(-4,-4)
  }
  }
  \right)
$$
For the case of the crossed module $(U(1) \to 1)$ this recovers
the cocycles for circle 2-bundles from observation \ref{CocyclesForCircle2Bundles}. 

Apart from the notion of \emph{bundle gerbe},
there is also 
the original notion of \emph{gerbe}. The terminology is
somewhat unfortunate, since neither of these concepts is, in general,
a special case of the other. But they are of course closely related.
We consider here the simple cocycle-characterization of gerbes and
the relation of these to cocycles for 2-bundles.
\begin{definition}[$G$-gerbe]
\label{GGerbeAsCocycle}
\index{gerbe!cocycle}
Let $G$ be a smooth group. Then a cocycle for a smooth \emph{$G$-gerbe}
over a manifold $X$ is a cocycle for a $\mathrm{AUT}(G)$-principal
2-bundle, where $\mathrm{AUT}(G)$ is the automorphism 2-group
from example \ref{Automorphism2GroupCrossedModule}.
\end{definition}
\begin{observation}
  \label{UnderlyingGerbeOf2Bundle}
  For every 2-group coming from a 
  crossed module $(G_1 \stackrel{\delta}{\to} G_0, \rho)$  there is a canonical
  morphism of 2-groups
  $$
    (G_1 \to G_0) \to \mathrm{AUT}(G_1)
  $$
  given by the commuting diagram of groups
  $$
    \xymatrix{
	   G_1 \ar[r]^\delta \ar[d]^{\mathrm{id}} & G_0
	   \ar[d]^{\rho}
	   \\
	   G_1 \ar[r]^{\mathrm{Ad}} & \mathrm{Aut}(G_0)
	}
	\,.
  $$
  Accordingly, every $(G_1 \to G_0)$-principal 2-bundle has an 
  underlying $G_1$-gerbe, def. \ref{GGerbeAsCocycle}. 
  But in general the passage to this underlying $G_1$-gerbe
  discards information.
\end{observation}
\begin{example}
  For $G$ a simply connected and compact 
  simple Lie group, let $\mathrm{String} \simeq (\hat \Omega G \to P G)$
  be the corresopnding String 2-group from 
  example \ref{String2GroupInIntroduction}.
  Then by observation \ref{UnderlyingGerbeOf2Bundle}
  every $\mathrm{String}$-principal 2-bundle has an underlying
  $\hat \Omega G$-gerbe. But there is more information in the 
  $\mathrm{String}$-2-bundle than in this gerbe underlying it.
\end{example}

\begin{example}
  \label{2GroupResolutionOfR}
  \index{principal $\infty$-bundle!2-bundle!$(\mathbb{Z}\to \mathbb{R})$-2-bundles}
  \index{principal $\infty$-bundle!2-bundle!$\mathbf{B}\mathbb{Z}$-2-bundles}
  Let $G = (\mathbb{Z} \hookrightarrow \mathbb{R})$ be the crossed module
  that includes the additive group of integers into the additive group of real numbers, with trivial action.
  The canonical projection morphism
  $$
    \mathbf{B}(\mathbb{Z} \to \mathbb{R}) \stackrel{\simeq}{\to}
	\mathbf{B}U(1)
  $$
  is a weak equivalence, by the fact that locally every smooth $U(1)$-valued function
  is the quotient of a smooth $\mathbb{R}$-valued function by a (constant) $\mathbb{Z}$-valued function. This means in particular that up to equivalence, 
  $(\mathbb{Z} \to \mathbb{R})$-2-bundles are the same as ordinary
  circle 1-bundles. But it means a bit more than that:
  
  On a manifold $X$ also $\mathbf{B}\mathbb{Z}$-principal 2-bundles have the same classification as $U(1)$-bundles. But the \emph{morphisms} of $\mathbf{B}\mathbb{Z}$-principal 2-bundles are essentially different from those of $U(1)$-bundles. This means that the 2-groupoid $\mathbf{B}\mathbb{Z} \mathrm{Bund}(X)$ is not, in general equivalent to $U(1)\mathrm{Bund}(X)$. But we do have an equivalence of 2-groupoids
  $$
    (\mathbb{Z} \to U(1)) \mathrm{Bund}(X)
	\simeq
	U(1)\mathrm{Bund}(X)
    \,.
  $$
\end{example}
\begin{example}
  \index{twisted cohomology!twisted bundle}
  \label{OrdinaryTwistedBundleExample}
  \label{BundleGerbeModulesAsSections}
  Let $\hat G \to G$ be a central extension of Lie groups by an abelian 
group $A$. This induces the crossed module $(A \to \hat G)$. 
There is a canonical 2-anafunctor
$$
  \xymatrix{
     \mathbf{B}(A \to \hat G)
	 \ar[r]^c
	 \ar[d]^{\simeq}
	 &
	 \mathbf{B}(A \to 1) = \mathbf{B}^2 A
	 \\
	 \mathbf{B}G
  }
$$  
from $\mathbf{B}G$ to $\mathbf{B}^2 A$. This can be seen to be the \emph{characteristic class}
that classifies the extension 
(see \ref{CharacteristicClassesInLowDegree} below): $\mathbf{B}\hat G \to \mathbf{B}G$ is the $A$-principal 2-bundle classified by this cocycle.

Accordingly, the collection of all $(A \to \hat G)$-principal 2-bundles is, up to equivalence, the same as that of plain $G$-1-bundles. But they exhibit the natural projection to $\mathbf{B}A$-2-bundles. Fixing that projection gives \emph{twisted $G$-1-bundles}.

more in detail: the above 2-anafunctor indiuces a 2-anafunctor on cocycle 2-groupoid
$$
  \xymatrix{
    (A \to \hat G)\mathrm{Bund}(X)
	\ar[r]^{c}
	\ar[d]^\simeq
	&
	\mathbf{B}A \mathrm{Bund}(X)
    \\
    G \mathrm{Bund}(X)
  }
  \,.
$$
If we fix a $\mathbf{B}A$-2-bundle $g$ we can consider the fiber of the characteristic class $c$ over $g$, hence the pullback $G \mathrm{Bund}_{[g]}(X)$ in 
$$
  \xymatrix{
    G \mathrm{Bund}_{[g]}(X) \ar[d] \ar[r] & {*} \ar[d]^{g}
    \\
    (A \to \hat G)\mathrm{Bund}(X)
	\ar[r]^{c}
	\ar[d]^\simeq
	&
	\mathbf{B}A \mathrm{Bund}(X)
    \\
    G \mathrm{Bund}(X)
  }
  \,.
$$
This is the groupoid of \emph{$[g]$-twisted $G$-bundles}. The principal 2-bundle classfied by $g$ is also called the \emph{lifting gerbe} of the $G$-principal bundles underlying the $[g]$-twisted $\hat G$-bundle: because this is the obstruction to lifting the former to a genuine $\hat G$-principal bundle.

If $g$ is given by a {\v C}ech cocycle 
$\{g_{i j k} \in C^\infty(U_{i j k}, A)\}$ then $[g]$-twisted $G$-bundles are given by data $\{h_{i j} \in C^\infty(U_{i j}, G)\}$ which does not quite satisfy the usual cocycle condition, but instead a modification by $g$:
$$
  h_{i k} = \delta(g_{i j k}) h_{j k} h_{i j}
  \,.
$$

For instance 
for the extension $U(1) \to U(n) \to PU(n)$ the corresponding twisted bundles are those that model \emph{twisted K-theory} with $n$-torsion twists (\ref{SmoothStrucTwistedCohomology}).
\end{example}

\medskip

\paragraph{Principal 3-bundles and twisted 2-bundles}
\label{Principal3BundlesAndTwisted2Bundles}
As one passes beyond (smooth) 2-groups and their 2-principal bundles, 
one needs more sophisticated tools for presenting them.
While the crossed modules from
def. \ref{Strict2GroupInIntroduction}
have convenient higher analogs -- called \emph{crossed complexes} --
the higher analog of 
remark \ref{StrictificationOf2GroupsInIntroduction} does not
hold for these: not every (smooth) 3-group is presented by them, much
less every $n$-group for $n > 3$.
Therefore below in \ref{ModelForPrincipalInfinityBundles}
we switch to a different tool for the general situation: simplicial 
groups.

However, it so happens that a wide range of relevant examples
of (smooth) 3-groups and generally of smooth $n$-groups
does have a presentation by a crossed complex after all,
as do the examples which we shall discuss now. 
\begin{definition}
  \label{CrossedComplex}
  \index{crossed complex}
  \index{groupoid!strict!crossed complex}
  A \emph{crossed complex of groupoids}  is a diagram
  $$
    C_\bullet
    =
    \left(
      \raisebox{20pt}{
      \xymatrix{
        \cdots
        \ar[r]^{\delta}
        &
        C_3
        \ar[r]^\delta
        \ar[d]
        &
        C_2
        \ar[r]^\delta
        \ar[d]
        &
        C_1
        \ar@<+3pt>[r]^{\delta_t}
        \ar@<-3pt>[r]_{\delta_s}
        \ar[d]^{\delta_s}
        &
        C_0
        \ar[d]^=
        \\
        \cdots
        \ar[r]_=
        &
        C_0
        \ar[r]_=
        &
        C_0
        \ar[r]_=
        &
        C_0
        \ar[r]_=
        &
        C_0
      }
      }
    \right),
  $$
  where
$\xymatrix{
  C_1
  \ar@<+3pt>[r]^{\delta_t}
  \ar@<-3pt>[r]_{\delta_s}
  &
  C_0
}$
is equipped with the structure of a 1-groupoid, and where 
$\xymatrix{
   C_k
   \ar[r]
   &
   C_{0}
}
$,
for all $k \geq 2$, are bundles of groups,
abelian for $k \geq 2$;
and equipped with an action $\rho$ of the groupoid $C_1$, such that
\begin{enumerate}
\item
 the maps $\delta_k$, $k \geq 2$ are morphisms of groupoids over
 $C_0$ compatible with the action by $C_1$;
\item
  $\delta_{k-1} \circ \delta_k = 0$; k $\geq 3$;
\item 
  $\mathrm{im}(\delta_2) \subset C_1$ acts by conjugation on
  $C_2$ and trivially on $C_k$, $k \geq 3$.
\end{enumerate}
\end{definition}
Surveys of standard material on crossed complexes 
of groupoids are in \cite{BHS}\cite{Porter}. 
We discuss sheaves of crossed complexes, hence \emph{cohesive crossed complexes} 
in more detail below in \ref{SheafAndNonabelianDoldKan}. As mentioned there, 
the key aspect of crossed complexes is that they provide an equivalent encoding 
of precisely those $\infty$-groupoids that are called \emph{strict}.
\begin{definition}
\label{StrictnGroup}
\index{group!strict}
 A \emph{crossed complex of groups} is a crossed complex $C_\bullet$ of groupoids
 with $C_0 = *$.
  If the complex of groups is constant on the trivial group beyond $C_n$, we say this is a \emph{strict $n$-group}.

  Explicitly, a \emph{crossed complex of groups} is 
  a complex of groups of the form
  $$
    \xymatrix{
      \cdots 
	  \ar[r]^{\delta_2}
	  &
	  G_2 
	  \ar[r]^{\delta_1}
	  &
	  G_1 
	  \ar[r]^{\delta_0}
	  &
	  G_0
	}
  $$
  with $G_{k \geq 2}$ abelian (but $G_1$ and $G_0$ not necessarily abelian), 
  together with an action $\rho_k$ of $G_0$ on $G_{k}$ for all $k \in \mathbb{N}$,
  such that 
  \begin{enumerate}
    \item 
	  $\rho_0 $ is the adjoint action of $G_0$ on itself;
	\item 
	  $\rho_1\circ \delta_0$ is the adjoint action of $G_1$ on itself;
	\item 
	  $\rho_k \circ \delta_0$ is the trivial action of $G_1$ on $G_k$
	  for $k > 1$;
	\item 
	  all $\delta_k$ respect the actions.
  \end{enumerate}
  A morphism of crossed complexes of groups
  is a sequence of morphisms of component groups, respecting all this structure.  
\end{definition} 
For $n = 2$ this reproduces the
notion of \emph{crossed module} and \emph{strict 2-group}, def. \ref{Strict2GroupInIntroduction}.
If furthermore $G_1$ and $G_0$ here are abelian and the action of $G_0$
is trivial, then this is an ordinary \emph{complex of abelian groups}
as considered in homological algebra. Indeed, all of homological algebra may be thought of 
as the study of this presentation of abelian $\infty$-groups. (More on this in \ref{SheafAndNonabelianDoldKan} below.)

We consider now examples of strict 3-groups and of the corresponding
principal 3-bundles.
\index{group!3-group!strict}
\begin{example}
  For $A$ an abelian group, the delooping of the 
  3-group given by the 
  complex $(A \to 1 \to 1)$ is the one-object 3-groupoid
  that looks like
  $$
    \mathbf{B}^3 A 
	\, = \,
  \left\{
  \raisebox{35pt}{
  \xymatrix{
     {*} \ar[rr]^{\mathrm{id}} && {*} \ar[dd]^{\mathrm{id}}
     \\
     \\
    {*} \ar[rr]^{\ }="t2" 
	 \ar[uu]^{\mathrm{id}} 
	  \ar[uurr]^{\ }="t"_>>>>>>>>>>>{\ }="s" && {*}
    \ar@{=>}^{\mathrm{id}} "t"+(-4,4); "t"
    \ar@{=>}_{\mathrm{id}} "s"; "t2"
  }
  }
  \xymatrix{
    \ar[rr]^{a \in A} &&
  }
  \raisebox{35pt}{
  \xymatrix{
     {*} \ar[rr]^{\mathrm{id}} 
	 \ar[ddrr]^{\ }="t"_<<<<<<<<<<<{\ }="s" 
	 && {*} \ar[dd]^{\mathrm{id}}
     \\
     \\
    {*} \ar[rr]^{\ }="t2" \ar[uu]^{\mathrm{id}}  && {*}
    \ar@{=>}^{\mathrm{id}} "t"+(4,4); "t"
    \ar@{=>}_>>>>>>{\mathrm{id}} "s"; "t2"
  }
  }
  \right\}
$$
\end{example}
Therefore an $\infty$-anafunctor 
$ X \stackrel{\simeq}{\leftarrow} C(\{U_i\}) 
  \stackrel{g}{\to}
  \mathbf{B}^3 U(1)$ 
   sends 3-simplices in the {\v C}ech groupoid
\begin{small}
$$
    \left\lbrace
      \raisebox{26pt}{
      \xymatrix@R=16pt@C=16pt{
        (x,j)
        \ar[rr]_<{\ }="s1"
        &&
        (x,k)
        \ar[dd]
        \\
        \\
        (x,i)
        \ar[uu]
        \ar[uurr]^{\ }="t1"_>>>>>>>>{\ }="s2"
        \ar[rr]^{\ }="t2"
        &&
        (x,l)
        \ar@{=>} "s1"; "t1"
        \ar@{=>} "s2"; "t2"
      }
      }
      \xymatrix{
        \ar[r] &
      }
      \raisebox{26pt}{
      \xymatrix@R=16pt@C=16pt{
        (x,j)
        \ar[rr]_>{\ }="s1"
        \ar[ddrr]^{\ }="t1"_<<<<<<<<{\ }="s2"
        &&
        (x,k)
        \ar[dd]
        \\
        \\
        (x,i)
        \ar[uu]
        \ar[rr]^{\ }="t2"
        &&
        (x,l)
        \ar@{=>} "s1"; "t1"
        \ar@{=>} "s2"; "t2"
      }
      }
    \right\rbrace
$$
\end{small}
to 3-morphisms in $\mathbf{B}^3 U(1)$ labeled by group elements
$g_{i j k l}(x) \in U(1)$ 
$$
    \left\lbrace
      \raisebox{26pt}{
      \xymatrix@R=16pt@C=16pt{
        \bullet
        \ar[rr]_<{\ }="s1"
        &&
        \bullet
        \ar[dd]
        \\
        \\
        \bullet
        \ar[uu]
        \ar[uurr]^{\ }="t1"_{\ }="s2"
        \ar[rr]^{\ }="t2"
        &&
        \bullet
        \ar@{=>} "s1"; "t1"
        \ar@{=>} "s2"; "t2"
      }
      }
      \xymatrix{
        \ar[r]^{g_{i j k l }(x)} &
      }
      \raisebox{26pt}{
      \xymatrix@R=16pt@C=16pt{
        \bullet
        \ar[rr]_>{\ }="s1"
        \ar[ddrr]^{\ }="t1"_{\ }="s2"
        &&
        \bullet
        \ar[dd]
        \\
        \\
        \bullet
        \ar[uu]
        \ar[rr]^{\ }="t2"
        &&
        \bullet
        \ar@{=>} "s1"; "t1"
        \ar@{=>} "s2"; "t2"
      }
      }
    \right\rbrace
$$
(where all 1-morphisms and 2-morphisms in $\mathbf{B}^3 U(1)$ are necessarily identities).

\vfill

The 3-functoriality of this assignment is given by the 
following identity on all {\v C}ech 4-simplices (x,(h,i,j,k,l)):
\begin{small}
$$
      \raisebox{-96pt}{
      \xy
 (45,0)*{
   \xy
   \xymatrix@C=9pt{
     && \bullet
     \ar[drr]_>{\ }="s3"
     \ar[ddr]^>>>>>{\ }="t3"
     \\
     \bullet
     \ar[urr]_<{\ }="s1"
     &&&& 3
     \ar[dl]
     \\
     &\bullet
     \ar[ul]
     \ar[rr]^{\ }="t2"
     \ar[uur]^<<<<<{\ }="t1"_>>>>{\ }="s2"
     &&
     \bullet
     \ar@{=>} "s1"; "t1"
     \ar@{=>} "s2"; "t2"
     \ar@{=>} "s3"; "t3"
   }
   \endxy
 }="b";
 (0,30)*{
   \xy
   \xymatrix@C=9pt{
     && \bullet
     \ar[drr]_>{\ }="s3"
     \\
     \bullet
     \ar[urr]_<{\ }="s1"
     &&&& \bullet
     \ar[dl]
     \\
     &\bullet
     \ar[ul]
     \ar[rr]^>>{\ }="t3"
     \ar[uur]^<<<<<{\ }="t1"_>>>{\ }="s2"
     \ar[urrr]^{\ }="t2"_>>>>>>>>>{\ }="s3"
     &&
     \bullet
     \ar@{=>} "s1"; "t1"
     \ar@{=>} "s2"; "t2"
     \ar@{=>} "s3"; "t3"
   }
   \endxy
 }="a";
 (90,30)*{
   \xy
   \xymatrix@C=9pt{
     && \bullet
     \ar[drr]_>>>{\ }="s3"
     \ar[ddr]^>>>>>>{\ }="t3"_<<{\ }="s2"
     \\
     \bullet
     \ar[urr]
     \ar[drrr]^{\ }="t2"_<<<<<<<{\ }="s1"
     &&&&
     \bullet
     \ar[dl]
     \\
     &\bullet
     \ar[ul]
     \ar[rr]^<<{\ }="t1"
     &&
     \bullet
     \ar@{=>} "s1"; "t1"
     \ar@{=>} "s2"; "t2"
     \ar@{=>} "s3"; "t3"
   }
   \endxy
 }="c";
 (16,77)*{
   \xy
   \xymatrix@C=9pt{
     && \bullet
     \ar[drr]_>{\ }="s3"
     \\
     \bullet
     \ar[urr]_>{\ }="s3"
     \ar[rrrr]_<<<<{\ }="s2"^{ }="t3"
     &&&&
     \bullet
     \ar[dl]
     \\
     &\bullet
     \ar[ul]
     \ar[rr]^{\ }="t1"
     \ar[urrr]^<<<{\ }="t2"_{\ }="s1"
     &&
     \bullet
     \ar@{=>} "s1"; "t1"
     \ar@{=>} "s2"; "t2"
     \ar@{=>} "s3"; "t3"
   }
   \endxy
 }="d";
 (74,77)*{
   \xy
   \xymatrix@C=9pt{
     && \bullet
     \ar[drr]_>{\ }="s3"
     \\
     \bullet
     \ar[urr]_>{\ }="s3"
     \ar[rrrr]_>>>>>>{\ }="s2"^{ }="t3"
     \ar[drrr]_{\ }="s1"^>>>>{\ }="t2"
     &&&& \bullet
     \ar[dl]
     \\
     &\bullet
     \ar[ul]
     \ar[rr]^{\ }="t1"
     &&
     \bullet
     \ar@{=>} "s1"; "t1"
     \ar@{=>} "s2"; "t2"
     \ar@{=>} "s3"; "t3"
   }
   \endxy
 }="e";
 \ar_{g_{h j k l}(x)} "a"; "b"
 \ar_{g_{h i j l}(x)} "b"; "c"
 \ar^{g_{h i j k}} "a"; "d"
 \ar^{g_{h i k l}} "d"; "e"_{\ }="t"
 \ar^{g_{i j k l}} "e"; "c"
 \ar@{=>}^{=} "b"; "t";
\endxy
  }
$$\end{small}
This means that the cocycle data $\{g_{i j k l}(x)\}$ has to satisfy the equations
$$
  g_{h i j k}(x) g_{h i k l}(x) g_{i j k l}(x)
  =
  g_{h j k l}(x) g_{h i j l}(x)
$$
for all $(h,i,j,k,l)$ and all $x \in U_{h i j k l}$. Since $U(1)$ is abelian
this can equivalently be rearranged to
$$
  g_{h i j k}(x) g_{h i j l}(x)^{-1} g_{h i k l}(x) g_{h j k l}(x)^{-1} g_{i j k l}(x)
  = 1
  \,.
$$
This is the usual form in which a {\v C}ech 3-cocycles with coefficients in 
$U(1)$ are written. 
\begin{definition}
\label{Circle3Bundle}
\index{circle $n$-bundle with connection!circle 3-bundle}
Given a cocycle as above, the total space object 
$\tilde P$ given by the pullback
$$
  \xymatrix{
    \tilde P \ar[r] \ar[d] & \mathbf{E}\mathbf{B}^2 U(1) \ar[d]
    \\
    C(U) \ar[r]^g \ar[d]^{\simeq} & \mathbf{B}^3 U(1)
    \\
    X 
  }
$$
is the corresponding \emph{circle principal 3-bundle}.
\end{definition}
In direct analogy to the argument that leads to 
observation \ref{BundleGerbeIsCircle2Bundle} we find:
\begin{observation}
\label{Bundle2GerbeIsCircle3Bundle}
\index{bundle gerbe!bundle 2-gerbe!as principal 3-bundle}
\index{bundle 2-gerbe!as principal 3-bundle}
\index{principal $\infty$-bundle!bundle 2-gerbe}
 The structures known as \emph{bundle 2-gerbes}
 \cite{Stevenson} are presentations of the 2-groupoids
 that are total spaces of circle principal 2-bundles, as above.
\end{observation}
Again, notice that, despite a close relation, this is
different from the original notion of \emph{2-gerbe}.
More discussion of this point is below in \ref{ETopStrucGerbes}.

The next example is still abelian, but captures basics of 
the central mechanism of twistings of principal 2-bundles by 
principal 3-bundles.
\begin{example}
\label{CocyclePrincipal3Bundle}
Consider a morphism $\delta : N \to A$ of abelian groups and the corresponding
shifted crossed complex $(N \to A \to 1)$. The corresponding 
delooped 3-group looks like
$$
 \mathbf{B}(N \to A \to 1)
   \;\; = \;\;
    \left\lbrace
      \raisebox{26pt}{
      \xymatrix@R=16pt@C=16pt{
        \bullet
        \ar[rr]_<{\ }="s1"
        &&
        \bullet
        \ar[dd]
        \\
        \\
        \bullet
        \ar[uu]
        \ar[uurr]^{\ }="t1"_{\ }="s2"
        \ar[rr]^{\ }="t2"
        &&
        \bullet
        \ar@{=>}^{a_1} "s1"; "t1"
        \ar@{=>}^{a_2} "s2"; "t2"
      }
      }
      \xymatrix{
        \ar[r]^{\delta(n) = a_4 a_3 a_2^{-1} a_1^{-1}} &
      }
      \raisebox{26pt}{
      \xymatrix@R=16pt@C=16pt{
        \bullet
        \ar[rr]_>{\ }="s1"
        \ar[ddrr]^{\ }="t1"_{\ }="s2"
        &&
        \bullet
        \ar[dd]
        \\
        \\
        \bullet
        \ar[uu]
        \ar[rr]^{\ }="t2"
        &&
        \bullet
        \ar@{=>}_{a_3} "s1"; "t1"
        \ar@{=>}_{a_4} "s2"; "t2"
      }
      }
    \right\rbrace
	\,.
$$
A cocycle for a $(N \to A \to 1)$-principal 3-bundle is 
given by data
$$
  \{
    a_{i j k} \in C^\infty(U_{i j k}, A),
	\,
	n_{i j k l} \in C^\infty(U_{i j k l}, N)
  \}
$$
such that 
\begin{enumerate}
\item $
  a_{j k l} a_{i j k}^{-1} a_{i j k} a_{i k l }^{-1}
  = 
  \delta( n_{i j k l} )
$
\item 
$
  n_{h i j k}(x) n_{h i k l}(x) n_{i j k l}(x)
  =
  n_{h j k l}(x) n_{h i j l}(x)
  \,.
$
\end{enumerate}
The first equation on the left is the cocycle for a 2-bundle as in 
observation \ref{CocyclesForCircle2Bundles}.
But the extra term $n_{i j k l}$ on the right ``twists''
the cocycle. This twist itself satisfies a higher order cocycle condition.
\end{example}
Notice that there is a canonical projection
$$
  \mathbf{B}(N \to A \to 1)
  \to 
  \mathbf{B}(N \to 1 \to 1)
  =
  \mathbf{B}^3 N
  \,.
$$
Therefore we can consider the higher analog of the notion  
of twisted bundles in example \ref{OrdinaryTwistedBundleExample}:
\begin{definition}
\label{Twisted2Bundle}
\index{twisted cohomology!twisted 2-bundle!abelian example}
Let $N \to A$ be an inclusion and consider
a fixed $\mathbf{B}^2 N$-principal 3-bundle with cocycle
$g$, let $\mathbf{B}(A/N) \mathrm{Bund}_{[g]}(X)$ be the pullback
in 
$$
  \xymatrix{
    \mathbf{B}(A/N) \mathrm{Bund}_{[g]}(X)
	\ar[r]
	\ar[d]
	&
	{*}
	\ar[d]^{g}
    \\
    \mathbf{B}(N \to A) \mathrm{Bund}(X)
	\ar[r]
	\ar[d]^{\simeq}
	&
	\mathbf{B}^2 N \mathrm{Bund}(X)
	\\
	\mathbf{B}(A/N) \mathrm{Bund}(X)
  }
  \,.
$$
We say an object in this 2-groupoid is a 
\emph{$[g]$-twisted $\mathbf{B}(A/N)$-principal 2-bundle}.
\end{definition}
Below in example \ref{Dixmier-Douady class} we discuss this and
its relation to characteristic classes of 2-bundles in more detail.

We now turn to the most general 3-group that is presented by a 
crossed complex.
\begin{observation}
  \index{group!3-group!strict!delooping}
  For $(L \stackrel{\delta}{\to} H \stackrel{\delta}{\to} G)$
  an arbitrary strict 3-group, def. \ref{StrictnGroup}, the delooping
  3-groupoid looks like
  $$
    \mathbf{B}(L \to H \to G)
	\,\,=
	\,\,
  \left\{
  \raisebox{35pt}{
  \xymatrix{
     {*} \ar[rr]^{g_2} && {*} \ar[dd]^{g_3}
     \\
     \\
    {*} \ar[rr]_{} 
	 \ar[uu]^{g_1} 
	  \ar[uurr]|{\delta(h_1)g_2 g_1}^{\ }="t"_{\ }="s" && {*}
    \ar@{=>}^<<{h_1} "t"+(-4,4); "t"
    \ar@{=>}_<<<{h_2} "s"; "s"+(4,-4)
  }
  }
  \xymatrix{
    \ar[r]^{\lambda \in L} &
  }
  \raisebox{35pt}{
  \xymatrix{
     {*} \ar[rr]^{g_2} \ar[ddrr]|{\delta(h_3)g_2 g_3}^{\ }="t"_{\ }="s" 
	   && {*} \ar[dd]^{g_3}
     \\
     \\
    {*} \ar[rr] \ar[uu]^{g_1}  && {*}
    \ar@{=>}_{h_3} "t"+(4,4); "t"
    \ar@{=>}^>>>>>>{h_4} "s"; "s"+(-4,-4)
  }
  }
  \;\;\vert\;\;
    \begin{array}{l}
    h_4 h_3
	\\
	=
	\\
    \delta(\lambda)  \cdot h_2 \cdot \rho(g_3)(h_1)	
	\end{array}
  \right\}\,,
  $$
  \vfill
  with the 4-cells -- the composition identities -- being
  \begin{small}
$$
      \raisebox{-96pt}{
      \xy
 (45,0)*{
   \xy
   \xymatrix@C=9pt{
     && \bullet
     \ar[drr]^{g_{23}}_>{\ }="s3"
     \ar[ddr]^>>>>>{\ }="t3"
     \\
     \bullet
     \ar[urr]^{g_{12}}_<{\ }="s1"
     &&&& 3
     \ar[dl]^{g_{34}}
     \\
     &\bullet
     \ar[ul]^{g_{01}}
     \ar[rr]_{}^{\ }="t2"
     \ar[uur]^<<<<<{\ }="t1"_>>>>{\ }="s2"
     &&
     \bullet
     \ar@{=>} "s1"; "t1"
     \ar@{=>} "s2"; "t2"
     \ar@{=>} "s3"; "t3"
   }
   \endxy
 }="b";
 (0,30)*{
   \xy
   \xymatrix@C=9pt{
     && \bullet
     \ar[drr]^{g_{23}}_>{\ }="s3"
     \\
     \bullet
     \ar[urr]^{g_{12}}_<{\ }="s1"
     &&&& \bullet
     \ar[dl]^{g_{34}}
     \\
     &\bullet
     \ar[ul]^{g_{01}}
     \ar[rr]^>>{\ }="t3"
     \ar[uur]^<<<<<{\ }="t1"_>>>{\ }="s2"
     \ar[urrr]^{\ }="t2"_>>>>>>>>>{\ }="s3"
     &&
     \bullet
     \ar@{=>} "s1"; "t1"
     \ar@{=>} "s2"; "t2"
     \ar@{=>} "s3"; "t3"
   }
   \endxy
 }="a";
 (90,30)*{
   \xy
   \xymatrix@C=9pt{
     && \bullet
     \ar[drr]^{g_{23}}_>>>{\ }="s3"
     \ar[ddr]^>>>>>>{\ }="t3"_<<{\ }="s2"
     \\
     \bullet
     \ar[urr]^{g_{12}}
     \ar[drrr]^{\ }="t2"_<<<<<<<{\ }="s1"
     &&&&
     \bullet
     \ar[dl]^{g_{34}}
     \\
     &\bullet
     \ar[ul]^{g_{01}}
     \ar[rr]^<<{\ }="t1"
     &&
     \bullet
     \ar@{=>} "s1"; "t1"
     \ar@{=>} "s2"; "t2"
     \ar@{=>} "s3"; "t3"
   }
   \endxy
 }="c";
 (16,77)*{
   \xy
   \xymatrix@C=9pt{
     && \bullet
     \ar[drr]^{g_{23}}_>{\ }="s3"
     \\
     \bullet
     \ar[urr]^{g_{12}}_>{\ }="s3"
     \ar[rrrr]_<<<<{\ }="s2"^{ }="t3"
     &&&&
     \bullet
     \ar[dl]^{g_{34}}
     \\
     &\bullet
     \ar[ul]^{g_{01}}
     \ar[rr]^{\ }="t1"
     \ar[urrr]^<<<{\ }="t2"_{\ }="s1"
     &&
     \bullet
     \ar@{=>} "s1"; "t1"
     \ar@{=>} "s2"; "t2"
     \ar@{=>} "s3"; "t3"
   }
   \endxy
 }="d";
 (74,77)*{
   \xy
   \xymatrix@C=9pt{
     && \bullet
     \ar[drr]^{g_{23}}_>{\ }="s3"
     \\
     \bullet
     \ar[urr]^{g_{12}}_>{\ }="s3"
     \ar[rrrr]_>>>>>>{\ }="s2"^{ }="t3"
     \ar[drrr]_{\ }="s1"^>>>>{\ }="t2"
     &&&& \bullet
     \ar[dl]^{g_{34}}
     \\
     &\bullet
     \ar[ul]^{g_{01}}
     \ar[rr]^{\ }="t1"
     &&
     \bullet
     \ar@{=>} "s1"; "t1"
     \ar@{=>} "s2"; "t2"
     \ar@{=>} "s3"; "t3"
   }
   \endxy
 }="e";
 \ar_{h_{0 2 3 4}} "a"; "b"
 \ar_{\rho(g_{23})(\lambda_{0 1 2 4})} "b"; "c"
 \ar^{\rho(g_{34}) (\lambda_{0 1 2 3})} "a"; "d"
 \ar^{\lambda_{0 1 3 4}} "d"; "e"_{\ }="t"
 \ar^{\lambda_{1 2 3 4}} "e"; "c"
 \ar@{=>}^{=} "b"; "t";
\endxy
  }
$$
\end{small}
\end{observation}
If follows that a cocycle
$$
  X \stackrel{\simeq}{\leftarrow}
  C(U_i)
  \stackrel{(\lambda, h,g)}{\to}
  \mathbf{B}(L \to H \to G)
$$
for a $(L \to H \to G)$-principal 3-bundle is a collection of functions
$$
  \{g_{i j} \in C^\infty(U_{i j}, G), \;
    h_{i j k} \in C^\infty(U_{i j k}, H), \;
	\lambda_{i j k l} \in C^\infty(U_{i j k l}, L)\}
$$
satisfying the cocycle conditions
$$
  g_{i k} = \delta(h_{i j k}) g_{j k} g_{i j} \;\;\;\;
  \mbox{on $U_{i j k}$}
$$
$$
    h_{i j l} h_{j k l}
	=
    \delta(\lambda_{i j k l})  \cdot h_{i k l} \cdot \rho(g_3)(h_{i j k})	  
	\;\;\;\;
	\mbox{on $U_{i j k l}$}
$$
$$
  \lambda_{i j k l} \lambda_{h i k l} \rho(g_{kl})(\lambda_{h i j k})
  =
  \rho(g_{j k})\lambda_{h i j l} \lambda_{h j k l}
  \;\;\;\;
  \mbox{on $U_{h i j k l}$}
  \,.
$$
\begin{definition}
\index{principal $\infty$-bundle!principal 3-bundle!for strict 3-group}
 Given such a cocycle, the pullback 3-groupoid $P$ we call the
 corresponding \emph{principal $(L \to H \to G)$-3-bundle}
 $$
   \xymatrix{
     P \ar[r]\ar[d] & \mathbf{E}\mathbf{B}(L \to H \to G) \ar[d]
	 \\
	 C(U_i) \ar[d]^\simeq\ar[r]^<<<<<{(\lambda, h, g)} & \mathbf{B}(L \to H \to G)
	 \\
	 X
   }
 $$
\end{definition}
We can now give the next higher analog of the notion of
twisted bundles, def. \ref{OrdinaryTwistedBundleExample}.
\begin{definition}
\index{twisted cohomology!twisted 2-bundle}
  Given a $3$-anafunctor
  $$
    \xymatrix{
	  \mathbf{B}(L \to H \to G)
	  \ar[r]
	  \ar[d]^{\simeq}
	  &
	  \mathbf{B}(L \to 1 \to 1)
	  \ar@{=}[r]
	  &
	  \mathbf{B}^3 L
	  \\
	  \mathbf{B}(H/L \to G)
	}
  $$
  then for $g$ the cocycle for an $\mathbf{B}^2 L$-principal 3-bundle
  we say that the pullback $(H \to G) \mathrm{Bund}_g(X)$
  in 
  $$
    \xymatrix{
	   (H \to G) \mathrm{Bund}_g(X)
	   \ar[r]
	   \ar[d]
	   &
	   {*}
	   \ar[d]^g
	   \\
	   (L \to H \to G)\mathrm{Bund}(X)
	   \ar[r]
	   &
	   \mathbf{B}^3 L \mathrm{Bund}(X)
	}
  $$
  is the 3-groupoid of \emph{$g$-twisted $(H \to G)$-principal 2-bundles}
  on $X$.
\end{definition}
\begin{example}
  \label{ResolutionForStringLift}
  \label{TwistedString2Bundle}
  \index{twisted cohomology!twisted 2-bundle!twisted string-2-bundle}
  Let $G$ be a compact and simply connected simple Lie group.
  By example \ref{String2GroupInIntroduction} we have associated with
  this the \emph{string 2-group} crossed module
  $\hat \Omega G \to P G$, where
  $$
    U(1) \to \hat \Omega G \to \Omega G
  $$
  is the Kac-Moody central extension of level 1 of the based loop 
  group of $G$. Accordingly, there is an evident crossed complex
  $$
    U(1) \to \hat \Omega G \to P G
	\,.
  $$
  The evident projection
  $$
    \mathbf{B}(U(1) \to \hat \Omega G \to P G)
	\stackrel{\simeq}{\to}
	\mathbf{B}G
  $$
  is a weak equivalence. This means that 
  $(U(1) \to \hat \Omega G \to P G)$-principal 3-bundles
  are equivalent to $G$-1-bundles. 
  For fixed projection $g$ to a $\mathbf{B}^2 U(1)$-3-bundle
  a $(U(1) \to \hat \Omega G \to P G)$-principal 3-bundles
  may hence be thought of as a $g$-twisted string-principal 2-bundle.

  One finds that these serve as
  a resolution of $G$-1-bundles in attempts to lift to
  string-2-bundles (discussed below in \ref{FractionalClasses}).
  \end{example}

\paragraph{A model for principal $\infty$-bundles}
\label{ModelForPrincipalInfinityBundles}
\index{principal $\infty$-bundle!simplicial presentation in introduction}

We have seen above that the theory of ordinary smooth principal bundles is naturally situated 
within the context of Lie groupoids, and then that the theory of smooth principal 2-bundles 
is naturally situated within the theory of Lie 2-groupoids. This is clearly the beginning of 
a pattern in higher category theory where in the next step we see smooth 3-groupoids and so on. 
Finally the general theory of principal $\infty$-bundles deals with smooth $\infty$-groupoids.
A comprehensive discussion of such smooth $\infty$-groupoids is given in section
\ref{SmoothInfgrpds}. In this introduction here we will just briefly describe principal $\infty$-bundles in this model. 

\medskip

Recall the discussion of $\infty$-groupoids from \ref{InfinityGroupoidsInIntro},
in terms of Kan simplicial sets.
Consider an object $\mathbf{B}G \in [C^{\mathrm{op}}, \mathrm{sSet}]$ which is 
an $\infty$-groupoid with a single object, so that we may think of it as the delooping of an $\infty$-group 
$G$. 
Let $*$ be the point and $* \to \mathbf{B}G$ the unique inclusion map. 
The \emph{good replacement} of this inclusion morphism is the 
\emph{universal $G$-principal $\infty$-bundle} 
\index{principal $\infty$-bundle!universal principal $\infty$-bundle!in introduction}
\index{universal principal $\infty$-bundle!in introduction}
$\mathbf{E}G \to \mathbf{B}G$ given  by the pullback diagram
$$
  \xymatrix{
    \mathbf{E}G \ar[r] \ar[d] & {*} \ar[d]
    \\
    (\mathbf{B}G)^{\Delta[1]} \ar[r] \ar[d] & \mathbf{B}G
    \\
    \mathbf{B}G
  }
  \,.
$$
An $\infty$-anafunctor $X \stackrel{\simeq}{\leftarrow} \hat X \to \mathbf{B}G$ we call a \emph{cocycle} 
\index{anafunctor!as cocycles}
on $X$ with coefficients in $G$, 
and the $\infty$-pullback $P$ of the point along 
this cocycle, which by the above discussion is the ordinary limit
$$
  \xymatrix{
    P \ar[r] \ar[dd]& \mathbf{E}G \ar[r] \ar[d] & {*} \ar[d]
    \\
    & \mathbf{B}G^{\Delta[1]} \ar[r] \ar[d] & \mathbf{B}G
    \\
    \hat X \ar[d]^{\simeq}\ar[r]^g & \mathbf{B}G
    \\
    X
  }
$$
we call the principal $\infty$-bundle $P \to X$ \emph{classified} by the cocycle.
\begin{example}
\index{principal $\infty$-bundle!universal principal 2-bundle}
  A detailed description of the 3-groupoid fibration that constitutes
  the universal principal 2-bundle $\mathbf{E}G$ for $G$ any 
  strict 2-group in given in \cite{RobertsSchreiber}.
\end{example}

It is now evident that our discussion of ordinary smooth principal bundles 
above is the special case of this for $\mathbf{B}G$ the nerve of the one-object 
groupoid associated with the ordinary Lie group $G$.
So we find the complete generalization of the situation that we already indicated there, which is summarized in the following diagram:
$$
  \begin{array}{cccc}
    \xymatrix{
       \vdots & \vdots
       \\
       \tilde P \times G \ar[r] \ar[d] & \mathbf{E}G \times G \ar[d]
       \\
       \tilde P \ar[r] \ar[d] & \mathbf{E}G \ar[d]
       \\
       C(U) \ar[r]^g \ar[d]^\simeq & \mathbf{B}G
       \\
       X
    } 
    &&&
    \xymatrix{
      \vdots & \vdots
      \\
      P \times G \ar[r]_>{\ }="s1" \ar[d] & G \ar[d]
      \\
      P \ar[r]^<{\ }="t1"_>{\ }="s2" \ar[d] & {*} \ar[d]
      \\
      X \ar[r]_g^<{\ }="t2" & \mathbf{B}G
      \\
      \ar@{=>}_\simeq "s1"; "t1" 
      \ar@{=>}_\simeq "s2"; "t2" 
    }
    \\ 
    \\
    \mbox{in the model category}
      &&&
    \mbox{in the $\infty$-topos}
  \end{array}
$$

\paragraph{Higher fiber bundles}
\label{HigherPrincipalBundlesInIntroduction}

We indicate here the natural notion of \emph{principal bundle} in an $\infty$-topos
and how it relates to the intrinsic notion of cohomology discussed above. 

\subparagraph{Ordinary principal bundles}
\label{OrdinaryPrincipalBundles}

For $G$ a group, a \emph{$G$-principal bundle} over some space $X$ is, roughly, a
space $P \to X$ over $X$, which is equipped with a $G$-action over $X$ that is fiberwise
free and transitive (``principal''), hence which after a choice of basepoint
in a fiber looks there like the canonical action of $G$ on itself. A central
reason why the notion of $G$-principal bundles is relevant is that it consistutes a 
``geometric incarnation'' of the degree-1 (nonabelian) cohomology $H^1(X,G)$
of $X$ with coefficients in $G$ (with $G$ regarded as the sheaf of $G$-valued functions on $G$):
$G$-principal bundles are \emph{classified} by $H^1(X,G)$. We will see that this
classical statement is a special case of a natural and much more general fact, where
\emph{principal $\infty$-bundles} incarnate cocycles in the intrinsic cohomology of any
$\infty$-topos.
Before coming to that, here we briefly review aspects of the classical theory
to set the scene.

\medskip

Let $G$ be a topological group and let $X$ be a topological space.
\begin{definition}
  A \emph{topological $G$-principal bundle} over $X$ is a continuous map $p : P \to X$
  equipped with a continuous fiberwise $G$-action $\rho : P \times G \to G$
  $$
    \xymatrix@R=11pt{
	  P \times G
	  \ar@<-3pt>[d]_{p_1}
	  \ar@<+3pt>[d]^{\rho}
	  \\
	  P 
	  \ar[d]^p
	  \\
	  X
	}
  $$
  which is \emph{locally trivial}: there exists a cover $\phi : U \to X$
  and an isomorphism of topological $G$-spaces
  $$
    P|_U \simeq U \times G
  $$
  between the restriction (pullback) of $P$ to $U$ and the trivial bundle $U \times G \to U$
  equipped with the canonical $G$-action given by multiplication in $G$.
\end{definition}
\begin{observation}
  Let $P \to X$ be a topological $G$-principal bundle. An immediate consequence of the definition is
  \begin{enumerate}
  \item The base space $X$ is isomorphic to 
  the quotient of $P$ by the $G$-action,
  and, moreover, under this identitfication $P \to X$ is the quotient 
  projection $P \to P/G$.
  \item
    The \emph{principality condition} is satisfied: the \emph{shear map}
	$$
	  (p_1, \rho) : P \times G \to P \times_X P
	$$
	is an isomorphism.
  \end{enumerate}
\end{observation}
\begin{remark}
 \label{PalaisPrincipalBundle}
Sometimes the quotient property of principal
bundles has been taken to be the defining property. For instance 
\cite{CartanI, CartanII} calls every quotient map 
$P \to P/G$ of a free topological group action
a ``$G$-principal bundle'', \emph{without} requiring it to be 
locally trivial. This is a strictly weaker definition: there are many examples of 
such quotient maps which are not locally trivial. To distinguish the 
notions, \cite{Palais} refers to the weaker definition as that of a
\emph{Cartan principal bundle}. Also for instance the standard textbook
\cite{Husemoeller} takes the definition via quotient maps as fundamental
and explicitly adds the adjective ``locally trivial'' when necessary.

For our purposes the following two points are relevant.
\begin{enumerate}
  \item Local triviality is crucial for the classification of topological $G$-principal 
    bundles by nonabelian sheaf cohomology to work, and so from this perspective
	a \emph{Cartan principal bundle} may be pathological.
	
  \item On the other hand, we see below that this problem is an artefact of 
    considering $G$-principal bundles in the ill-suited context of the
    1-category of topological spaces or manifolds. 
	We find below that after embedding into an 
    $\infty$-topos (for instance that of Euclidean topological $\infty$-groupoids,
	discussed in \ref{ContinuousInfGroupoids}) both definitions in fact coincide.	
	
	The reason is that the Yoneda embedding
	into the higher categorical context of an $\infty$-topos
	``corrects the quotients'': those quotients of $G$-actions that are 
    not locally trivial get replaced,
	while the ``good quotients'' are being preserved by the embedding.
	This statement we make precise in \ref{Principal infinity-bundles presentations} below.
	See also the discussion in \ref{PrincipalBundlesIntroductionAndSurvey} below.
\end{enumerate}
\end{remark}

It is a classical fact that 
for $X$ a manifold and $G$ a topological or Lie group,
regarded as a sheaf of groups $C(-,G)$ on $X$, there is an equivalence
of the following kind\\
\begin{tabular}{|ccc|}
  \hline
  {\bf algebraic data on $X$} && {\bf geometric data on $X$}
  \\
  \hline
  \hline
  $
      \left\{
	  \mbox{
      \begin{tabular}{l}
        degree-1 nonabelian
    	  \\
	    sheaf cohomology
	  \end{tabular}
	  }
    \right\}
  $
  &$\simeq$& 
  $
  	\left\{
	  \mbox{
	    \begin{tabular}{c}
  	      isomorphism classes of \\
	      G-principal bundles over $X$
		\end{tabular}
	  }
	\right\}
  $
  \\
  $H^1(X,G)$
  &&
  $G \mathrm{Bund}(X)$
  \\
  \hline
  \\
  $
   \left\{
    \raisebox{60pt}{
    \xymatrix@C=2pt{
	  & (x,j) \ar[dr]\ar@{|->}[dd]|{\phantom{\mathbf{X} atop x}}
	  &&& {*} \ar[dr]^{g_{jk}(x)}
	  \\
	  (x,i) \ar@{|->}[dr]\ar[rr] \ar[ur]_{\ }="s" && (x,k)\ar@{|->}[dl] &
	  {*} \ar[rr]_{g_{ik}(x)} \ar[ur]^>>>>{g_{i j}(x)}_{\ }="t" && {*} &	  
	  \\
	  & x
	  \ar@{|->}^g "s"; "s"+(30,0)
	  \\
	  & X \ar[rrr]^g_{\mbox{cocycle}} &&& \mathbf{B}G
	}
	}
	\right\}_{/\sim}
  $
  &$\simeq$&
  $
  \left\{
  \raisebox{66pt}{
  \xymatrix@C=-7pt@R=2pt{
    P \times G \ar[rr] \ar@<-3pt>[ddd]_{p_1} \ar@<+3pt>[ddd]^\rho 
	&& E G \times G
	\ar@<-3pt>[ddd]_{p_1} \ar@<+3pt>[ddd]^\rho
    \\
	\\ && & \mbox{\small $G$-actions}
	\\
    P \ar[rr] \ar[ddd] && E G \ar[ddd] & \mbox{\small total spaces}
	\\
	\\ & \mbox{pullback} && 
	\\
	X \ar[rr]_{\vert g\vert } && B G &  \mbox{\small quotient spaces}
	\\
	\mbox{\small \begin{tabular}{c} $G$-principal \\ bundle \end{tabular}}
	\ar@{}[rr]_{\mbox{\small classifying}  \atop \mbox{\small map}}
	&& \mbox{\small \begin{tabular}{c} universal \\ bundle \end{tabular}}
  }}
  \right\}_{/\sim}
 $
 \\
 \hline
\end{tabular}

We give a detailed exposition of the construction indicated in this diagram 
below in \ref{Principal1Bundles}.

\subparagraph{Principal $\infty$-bundles}

Let now $\mathbf{H}$ be an $\infty$-topos, \ref{IntroInfinToposes}, and $G$
a group object in $\mathbf{H}$, \ref{GroupsInIntroduction}. 
Up to the technical issue of formulating homotopy coherence, 
the formulation in $\mathbf{H}$ of the definition of $G$-principal bundles, 
\ref{OrdinaryPrincipalBundles}, in its version as
\emph{Cartan $G$-principal bundle}, remark \ref{PalaisPrincipalBundle}, 
is immediate:
\noindent{\bf Definition.} A {\bf $G$-principal bundle} over $X \in \mathbf{H}$ is 
\begin{itemize}
  \item a morphism $P \to X$; with an $\infty$-action $\rho : P \times G \to P$;
  \item such that $P \to X$ is the $\infty$-quotient map $P \to P/\!/G$.
\end{itemize}
In \ref{PrincipalInfBundle} below we discuss a precise formulation of this definition
and the details of the following central statement about the relation
between $G$-principal $\infty$-bundles and the intrinsic cohomology
of $\mathbf{H}$ with coefficients in the delooping object $\mathbf{B}G$.

\noindent{\bf Theorem.} There is equivalence of $\infty$-groupoids
$
  \xymatrix{
    G \mathrm{Bund}(X) 
	 \ar@<-5pt>[rr]_{\lim\limits_\to}^\simeq
	 \ar@{<-}@<+5pt>[rr]^{\mathrm{hofib}}
	 &&
	\mathbf{H}(X, \mathbf{B}G)
  }
$, where
\vspace{-.4cm}
\begin{enumerate}
  \item $\mathrm{hofib}$ sends a cocycle $X \to \mathbf{B}G$ to its homotopy fiber;
	\item $\lim\limits_{\longrightarrow}$ sends an $\infty$-bundle to the map on 
	$\infty$-quotients 
	 $X \simeq P/\!/G \to */\!/G \simeq \mathbf{B}G$.
\end{enumerate}
In particular, $G$-principal $\infty$-bundles are classified by the intrinsic cohomology
of $\mathbf{H}$
$$
  G \mathrm{Bund}(X)/_{\sim} \simeq H^1(X,G) := \pi_0 \mathbf{H}(X, \mathbf{B}G)
  \,.
$$
\begin{tabular}{l|l}
 \begin{tabular}{l}
 {Idea of Proof.}
  Repeatedly apply two of the \\
  \emph{Giraud-Rezk-Lurie axioms}, def. \ref{GiraudRezkLurieAxioms}, \\
  that characterize $\infty$-toposes:\\
  $\,$\\
  \hspace{.2cm}  1. every $\infty$-quotient is effective;
	\\
   \hspace{.2cm} 2. $\infty$-colimits are preserved \\
   \hspace{.7cm} by $\infty$-pullbacks.
\endofproof
\end{tabular}
 &
 $$
  \raisebox{50pt}{
  \xymatrix@C=-7pt@R=2pt{
    \vdots && \vdots
    \\
    P \times G \times G 
	  \ar[rr] \ar@<-4pt>[ddd] \ar@<+0pt>[ddd] \ar@<+4pt>[ddd] 
	  &&
	G \times G
      \ar@<-4pt>[ddd] \ar@<+0pt>[ddd] \ar@<+4pt>[ddd]	
    \\
	\\
    \\
    P \times G \ar[rr] \ar@<-3pt>[ddd]_{p_1} \ar@<+3pt>[ddd]^\rho 
	&& G
	\ar@<-3pt>[ddd]_{} \ar@<+3pt>[ddd]
    \\
	\\ && & \mbox{\small $G$-$\infty$-actions}
	\\
    P \ar[rr] \ar[ddd] && {*} \ar[ddd] & \mbox{\small total objects}
	\\
	\\ & \mbox{\small $\infty$-pullback} && 
	\\
	X \ar[rr]_{g } &&  \mathbf{B}G &  \mbox{\small quotient objects}
	\\
	\mbox{\small \begin{tabular}{c} $G$-principal \\ $\infty$-bundle \end{tabular}}
	\ar@{}[rr]_{\mbox{\small cocycle}}
	&& \mbox{\small \begin{tabular}{c} universal \\ $\infty$-bundle \end{tabular}}
  }
  }
$$
\end{tabular}
  
This gives a general abstract theory of principal $\infty$-bundles in every $\infty$-topos.
We also have the following explicit presentation.
\noindent {\bf Definition}
  For $G \in \mathrm{Grp}(\mathrm{sSh}(C))$, and $X \in \mathrm{sSh}(C)_{\mathrm{lfib}}$, 
  a \emph{weakly $G$-principal simplicial bundle} is a $G$-action $\rho$ over $X$ such that
  the \emph{principality morphism}
  $(\rho,p_1) : P \times G \to P \times_X P$ is a stalkwise weak equivalence.  

Below in \ref{Principal infinity-bundles presentations} we discuss that this
construction gives a presentation of the $\infty$-groupoid of $G$-principal bundles
as the nerve of the ordinary category of weakly $G$-principal simplicial bundles.
$$
  \mathrm{Nerve}\left\{
    \mbox{
	  \begin{tabular}{c}
	    weakly $G$-principal
		\\
		simplicial bundles
		\\
		over $X$
	  \end{tabular}
	}
  \right\}
  \simeq
  G \mathrm{Bund}(X)
  \,.
$$
For the special case that $X$ is the terminal object over the site 
$C$ and when restricted from cocycle 
$\infty$-groupoids to sets of cohomology classes, this reproduces the statement of
\cite{JardineLuo}. For our applications in \ref{Applications},
in particular for applications in twisted cohomology, \ref{StrucTwistedCohomology},
it is important to have the general statement, where the base space of 
a principal $\infty$-bundle may be an arbitrary $\infty$-stack, and where we
remember the $\infty$-groupoids of gauge transformations between them, instead
of passing to their sets of equivalence classes.

The special case where the site $C$ is trivial, $C = *$, leads to 
the notion of principal $\infty$-bundles in $\infty \mathrm{Grp}$.
These are presented by certain bundles of simplicial sets. This we discuss
below in \ref{DiscStrucPrincipalInfinityBundles}.

\subparagraph{Associated and twisted $\infty$-bundles}

The notion of $G$-principal bundle is a very special case of the following natural
more general notion.
For any $F$, 
an \emph{$F$-fiber bundle} over some $X$ is a space $E \to X$ over $X$ such that there
is a cover $U \to X$ over which it becomes equivalent as a bundle to the trivial $F$-bundle
$U \times F \to U$. 

Principal bundles themselves form but a small subclass of all possible
fiber bundles over some space $X$. 
Even among $G$-fiber bundles the $G$-principal bundles
are special, due to the constraint that the local trivialization has
to respect the $G$-action on the fibers.
However, every $F$-fiber bundle is \emph{associated} to a $G$-principal bundle.

Given a representation $\rho : F \times G \to F$, the
\emph{$\rho$-associated} $F$-fiber bundle is the quotient $P \times_G F$
of the product $P \times F$ by the diagonal $G$-action. Conversely, using that 
the automorphism group $\mathrm{Aut}(F)$ of $F$ canonically acts on $F$, it 
is immediate that every $F$-fiber bundle is associated to an $\mathrm{Aut}(F)$-principal
bundle (a statement which, of course, crucially uses the local triviality clause). 

All of these constructions and statements have their straightforward generalizations
to higher bundles, hence to \emph{associated $\infty$-bundles}. 
Moreover, just as the theory of principal bundles \emph{improves} in the 
context of $\infty$-toposes, as discussed above, so does the theory of
associated bundles.

For notice that by the above classification theorem of 
$G$-principal $\infty$-bundles, every $G$-$\infty$-action
$\rho : V \times G \to G$ has a \emph{classifying map}, which we will denote
by the same symbol:
$$
  \raisebox{20pt}{
  \xymatrix{
    V \ar[r] & V/\!/G 
	\ar[d]^\rho
	\\
	& \mathbf{B}G
  }
  }
  \,.
$$  
One may  observe now that this map $V/\!/G \to \mathbf{B}G$ is the 
\emph{universal $\rho$-associated $V$-$\infty$-bundle}: for every $F$-fiber $\infty$-bundle
$E \to X$ there is a morphism $X \to \mathbf{B}G$ such that $E \to X$ is the 
$\infty$-pullback of this map to $X$.
$$
  \raisebox{20pt}{
  \xymatrix{
    E \ar[d]\ar[r] & V/\!/G \ar[d]^\rho
	\\
	X \ar[r]^g & \mathbf{B}G
  }
  }
  \,.
$$
One implication of this is, by the universal property of the $\infty$-pullback, that
\emph{sections} $\sigma$ of the associated bundle
$$
  \xymatrix{
     E 
	 \ar[d]
	 \\
	 X
	 \ar@/^1pc/[u]^{\sigma}
  }
$$
are equivalently lifts of its classifying map through the universal $\rho$-associated bundle
$$
    \Gamma_X(P \times_G V)
	:=
	\left\{
	  \raisebox{20pt}{
	  \xymatrix{
	    & V /\! / G
		  \ar[d]^\rho
	    \\
	    X \ar[r]^g 
		\ar@{-->}[ur]^{\sigma}
		& \mathbf{B}G
	  }
	  }
	\right\}
	\,.
$$
One observes that by local triviality and by the fact that $V$ is, by the above, the homotopy
fiber of $V/\!/G \to \mathbf{B}G$, it follows that locally over a cover $U \to X$
such a section is identified with a $V$-valued map $U \to V$. Conversely, globally
a section of a $\rho$-associated bundle may be regarded as a \emph{twisted} $V$-valued function.

While this is an elementary and familiar statement for ordinary associated bundles,
this is where the theory of associated $\infty$-bundles becomes considerably
richer than that of ordinary $\infty$-bundles: because here $V$ itself may be a higher
stack, notably it may be a moduli $\infty$-stack $V = \mathbf{B}A$ for $A$-principal
$\infty$-bundles. If so, maps $U \to V$  classify $A$-principal $\infty$-bundles locally
over  the cover $U$ of $X$, and so conversely the section $\sigma$ itself may
globally be regarded as exhibiting a \emph{twisted $A$-principal $\infty$-bundle} over $X$.

We can refine this statement by furthermore observing that the space of all sections
as above is itself the hom-space in an $\infty$-topos, namely in the slice $\infty$-topos
$\mathbf{H}_{/\mathbf{B}G}$. This means that such sections are themselves cocycles in 
a structured nonabelian cohomology theory:
$$
    \Gamma_X(P \times_G V)
	:=
	\mathbf{G}_{/\mathbf{B}G}(g,\rho)
	\,.
$$
This we may call the $g$-\emph{twisted cohomology} of $X$  relative to $\rho$.
We discuss below in \ref{TwistedStructures} how traditional notions of twisted cohomology are special
cases of this general notion, as are many further examples.

Now $\rho$, regarded as an object of the slice $\mathbf{H}_{/\mathbf{B}G}$ is not 
in general a connected object. This means that it is not in general the moduli object
for some principal $\infty$-bundles over the slice. But instead, we find that
we can naturally identify geometric incarnations of such cocycles in the form of
\emph{twisted $\infty$-bundles}.

\noindent {\bf Theorem.} The $g$-twisted cohomology $\mathbf{H}_{/\mathbf{B}G}(g,\rho)$  
classifies 
$P$-\emph{twisted $\infty$-bundles}: twisted $G$-equivariant $\Omega V$-$\infty$-bundles
on $P$:
$$
  \xymatrix{
    Q 
	\ar[d] \ar[r] & {*} \ar[d]
	&&
	\mbox{$P$-twisted $\Omega V$-principal $\infty$-bundle}
    \\
    P \ar[d] \ar[r] & V \ar[r]  \ar[d] & {*} \ar[d] & \mbox{$G$-principal $\infty$-bundle}
    \\
    X \ar[r]^\sigma \ar@/_2pc/[rr]_g & V/\!/G \ar[r]^\rho & \mathbf{B}G & \mbox{section of $\rho$-associated $V$-$\infty$-bundle}
  }
$$

$$
  \left\{
     \mbox{
	    \begin{tabular}{c}
		   sections of \\ $\rho$-associated $V$-$\infty$-bundle
		\end{tabular}
	 }
  \right\}
  \simeq
  \left\{
     \mbox{
	    \begin{tabular}{c}
		   $g$-twisted $\Omega V$-cohomology \\ 
		   relative $\rho$
		\end{tabular}
	 }
  \right\}
  \simeq  
  \left\{
     \mbox{
	    \begin{tabular}{l}
		   $\Omega V$-$\infty$-bundles 
		   \\
		   twisted by $P$
		\end{tabular}
	 }
  \right\}
  $$
  
A survey of classes of examples of twisted $\infty$-bundles
classified by twisted cohomology is below in \ref{TwistedDiffStructures}.
Among them, in particular the classical notion of nonabelian \emph{gerbe} \cite{Giraud}, 
and \emph{2-gerbe} \cite{Breen} is a special case.
  
Namely one see that a (nonabelian/Giraud-)\emph{gerbe} on $X$ is nothing but 
a connected and 1-truncated object
in $\mathbf{H}_{/X}$. Similarly, a 
(nonabelian/Breen) \emph{2-gerbe} over $X$ is just a connected and 2-truncated
object in $\mathbf{H}_{/X}$. Accordingly we may call a general connecte
object in $\mathbf{H}_{/X}$ an \emph{nonabelian $\infty$-gerbe} over $X$.
We say that it is a \emph{$G$-$\infty$-gerbe} if it is 
an $\mathrm{Aut}(\mathbf{B}G)$-associated
$\infty$-bundle. We say its \emph{band} is the underlying $\mathrm{Out}(G)$-principal
$\infty$-bundle. For 1-gerbes and 2-gerbes this reproduces the classical notions.

In terms of this, the above says that $G$-$\infty$-gerbes \emph{bound by a band} 
are classified by 
$(\mathbf{B} \mathrm{Aut}(\mathbf{B}G) \to \mathbf{B}\mathrm{Out}(G))$-twisted cohomology.
This is the generalization of Giraud's original theorem.
We discuss all this in detail below in \ref{StrucInftyGerbes}.

%
%

%
%

%
%

%
%

%
%

\subsubsection{Principal connections}
\label{GeometryOfPhysicsPrincipalConnections}

\paragraph{Parallel $n$-transport for low $n$}
\label{ConnectionsOnPrincipalsnBundles}
\index{parallel transport!in low dimensions}

With a decent handle on principal $\infty$-bundles as described above, we now turn to 
the description of \emph{connections on $\infty$-bundles}. It will turn out that the 
above cocycle-description of $G$-principal $\infty$-bundles in terms of  $\infty$-anafunctors 
$X \stackrel{\simeq}{\leftarrow} \hat X \stackrel{g}{\to} \mathbf{B}G$ has, under mild conditions, a natural generalization where $\mathbf{B}G$ is replaced by a (non-concrete) simplicial presheaf
$\mathbf{B}G_{\mathrm{conn}}$, which we may think of as the $\infty$-groupoid of $\infty$-Lie algebra valued forms. 
This comes with a canonical map $\mathbf{B}G_{\mathrm{conn}} \to \mathbf{B}G$ and 
an $\infty$-connection $\nabla$ on the $\infty$-bundle classified by $g$ is a lift $\nabla$ of $g$ in the 
diagram
$$
  \xymatrix{
    & \mathbf{B}G_{\mathrm{conn}}
         \ar[d]
    \\
    \hat X \ar[r]^g  \ar[d]^\simeq \ar[ur]^{\nabla} & \mathbf{B}G
    \\
    X
  }
  \,.
$$
In the language of $\infty$-stacks we may think of $\mathbf{B}G$ as the $\infty$-stack (on $\mathrm{CartSp}$) 
or $\infty$-prestack (on $\mathrm{SmoothMfd}$) $G \mathrm{TrivBund}(-)$ of \emph{trivial} $G$-principal bundles, 
and of $\mathbf{B}G_{\mathrm{conn}}$ correspondingly as the object $G \mathrm{TrivBund}_{\nabla}(-)$ 
of  trivial $G$-principal bundles with (non-trivial) connection. In this sense the statement that 
$\infty$-connections are cocycles with coefficients in some $\mathbf{B}G_{\mathrm{conn}}$ is a tautology. 
The real questions are:
\begin{enumerate}
  \item 
    What is $\mathbf{B}G_{\mathrm{conn}}$ in concrete formulas?
  \item
    Why are these formulas what they are? What is the general abstract concept of an 
     $\infty$-connection? What are its defining abstract properties?
\end{enumerate}
A comprehensive answer to the second question is provided by the general abstract concepts
discussed in section \ref{GeneralAbstractTheory}. Here in this introduction we will not go into the full 
abstract theory, but using classical tools we get pretty close. What we describe is a 
generalization of the concept of \emph{parallel transport} to 
\emph{higher parallel transport}. As we shall see, this is naturally expressed in terms of 
$\infty$-anafunctors out of path $n$-groupoids. This reflects how the full abstract theory arises 
in the context of an $\infty$-connected $\infty$-topos that comes canonically with a notion of
fundamental $\infty$-groupoid. 

Below we begin the discussion of $\infty$-connections by reviewing the classical theory of 
connections on a bundle in a way that will make its generalization to higher connections relatively 
straightforward.
In an analogous way we can then describe certain classes of connections on a 2-bundle -- 
subsuming the notion of connection on a bundle gerbe.
With that in hand we then revisit the discussion of connections on ordinary bundles. By associating to each bundle with connection its corresponding 
\emph{curvature 2-bundle with connection}\index{curvature!curvature 2-bundle} we obtain a more refined 
description of connections on bundles, one that is naturally adapted to the construction of 
curvature characteristic forms in the Chern-Weil homomorphism.
This turns out to be the kind of formulation of connections on an $\infty$-bundle 
that drops out of the general abstract theory. In classical terms, its full formulation 
involves the description of circle $n$-bundles with connection in terms of Deligne cohomology 
and the description of the $\infty$-groupoid of $\infty$-Lie algebra valued forms 
in terms of dg-algebra homomorphisms. 
The combination of these two aspects yields naturally an explicit model for the Chern-Weil homomorphism and its generalization to higher bundles.

Taken together, these constructions allow us to express a good deal of the general $\infty$-Chern-Weil theory with classical tools. As an example, we describe how the classical {\v C}ech-Deligne cocycle construction of the refined Chern-Weil homomorphism drops out from these constructions.

\subparagraph{Connections on a principal bundle}
\index{connection!ordinary principal connection}

There are different equivalent definitions of the classical notion of a connection. One that is useful for our purposes is that a connection $\nabla$ on a $G$-principal bundle $P \to X$ is a rule $\mathrm{tra}_\nabla$ for 
\emph{parallel transport} along paths: a rule that assigns to each path $\gamma : [0,1] \to X$ a morphism $tra_\nabla(\gamma) : P_x \to P_y$ between the fibers of the bundle above the endpoints of these paths, in a compatible way:
$$
  \xymatrix{  
    P_x
    \ar[r]^{\mathrm{tra}_\nabla(\gamma)}
    &
    P_y
    \ar[r]^{\mathrm{tra}_\nabla(\gamma')}
    &
    P_z && P \ar[d]
    \\
    x \ar[r]^\gamma & y \ar[r]^{\gamma'} & z && X
  }
$$
In order to formalize this, we introduce a (diffeological) Lie groupoid to be called the 
\emph{path groupoid}\index{paths!path groupoid} of $X$. (Constructions and results in this section are from 
\cite{SWI}.
\begin{definition}
  For $X$ a smooth manifold let $[I,X]$ be the set of smooth functions 
  $I = [0,1] \to X$. For $U$ a Cartesian space, we say that 
\emph{a $U$-parameterized smooth family of points in $[I,X]$} 
 is a smooth map $U \times I \to X$. (This makes $[I,X]$ a diffeological space).

Say a path $\gamma \in [I,X]$ has \emph{sitting instants} if it is constant in a 
neighbourhood of the boundary $\partial I$. Let $[I,P]_{\mathrm{\mathrm{si}}} \subset [I,P]$ 
be the subset of paths with sitting instants. 

Let $[I,X]_{\mathrm{\mathrm{si}}} \to [I,X]_{\mathrm{si}}^{\mathrm{th}}$ be the projection to the set of 
equivalence classes where two paths are regarded as equivalent if they are cobounded by a 
smooth thin homotopy.

Say a $U$-parameterized smooth family of points in $[I,X]_{\mathrm{\mathrm{si}}}^{\mathrm{th}}$ is one that 
comes from a $U$-family of representatives in $[I,X]_{\mathrm{\mathrm{si}}}$ under this projection. 
(This makes also $[I,X]_{\mathrm{\mathrm{si}}}^{\mathrm{th}}$ a diffeological space.)
\end{definition}

The passage to the subset and quotient $[I,X]_{\mathrm{\mathrm{si}}}^{\mathrm{th}}$ of the set of 
all smooth paths in the above definition is essentially the minimal adjustment to enforce 
that the concatenation of smooth paths at their endpoints defines the composition operation in a groupoid.

\begin{definition}
The \emph{path groupoid} $\mathbf{P}_1(X)$ is the groupoid
$$
  \mathbf{P}_1(X) = ([I,X]_{\mathrm{si}}^{th} \stackrel{\to}{\to} X)
$$
with source and target maps given by endpoint evaluation and composition given by concatenation of 
classes $[\gamma]$ of paths along any orientation preserving \emph{diffeomorphism} 
$[0,1] \to [0,2] \simeq [0,1] \coprod_{1,0} [0,1]$ of any of their representatives
$$
  [\gamma_2] \circ [\gamma_1] : [0,1]
  \stackrel{\simeq}{\to} [0,1] \coprod_{1,0} [0,1]
  \stackrel{(\gamma_2 , \gamma_1)}{\to}
  X
  \,.
$$
This becomes an internal groupoid in diffeological spaces with the above $U$-families of smooth paths. 
We regard it as a groupoid-valued presheaf, an object in $[CartSp^{op}, \mathrm{Grpd}]$:
$$
  \mathbf{P}_1(X) : U \mapsto (\mathrm{SmoothMfd}(U \times I, X)_{\mathrm{\mathrm{si}}}^{\mathrm{th}} 
  \stackrel{\to}{\to} \mathrm{SmoothMfd}(U,X) )
  \,.
$$
\end{definition}
Observe now that for $G$ a Lie group and $\mathbf{B}G$ its delooping Lie groupoid discussed above, 
a smooth functor $\mathrm{tra} : \mathbf{P}_1(X) \to \mathbf{B}G$ sends each 
(thin-homotopy class of a) path to an element of the group $G$ 
$$
  \mathrm{tra} : (x \stackrel{[\gamma]}{\to} y)
  \mapsto
  (
   \bullet
   \stackrel{\mathrm{tra}(\gamma) \in G}{\to}
   \bullet
  )
$$
such that composite paths map to products of group elements :
$$
  \mathrm{tra}
  :
  \left\{
   \raisebox{20pt}{
  \xymatrix{
     & y \ar[dr]^{[\gamma']}
     \\
    x \ar[rr]_{[\gamma' \circ \gamma]}^{\ }="t" 
       \ar[ur]^{[\gamma]}
     && z
    \ar@{=} "t"+(0,5); "t"
  }
  }
  \right\}
  \;\;\;\;
   \mapsto
  \;\;\;\;
  \left\{
  \raisebox{20pt}{
  \xymatrix{
     & {*} \ar[dr]^{\mathrm{tra}(\gamma')}
     \\
    {*} \ar[rr]_{\mathrm{tra}(\gamma')\mathrm{tra}(\gamma)}^{\ }="t" 
       \ar[ur]^{\mathrm{tra}(\gamma)}
     && {*}
    \ar@{=} "t"+(0,5); "t"
  }
  }
  \right\}
  \,.
$$
and such that $U$-families of smooth paths induce smooth maps $U \to G$ of elements. 

There is a classical construction that yields such an assignment: the \emph{parallel transport} of a 
\emph{Lie-algebra valued 1-form}.
\begin{definition}
Suppose $A \in \Omega^1(X, \mathfrak{g})$ is a degree-1 differential form on $X$ with values in 
the Lie algebra $\mathfrak{g}$ of $G$. Then its parallel transport is the smooth functor
$$
   \mathrm{tra}_A : \mathbf{P}_1(X) \to \mathbf{B}G
$$
given by
$$
  [\gamma] \mapsto P \exp(\int_{[0,1]} \gamma^* A) \; \in G
  \,,
$$
where the group element on the right is defined to be the value at 1 of the unique solution 
$f :  [0,1] \to G$ of the differential equation
$$
  d_{\mathrm{dR}} f + \gamma^*A \wedge f = 0
$$
for the boundary condition $f(0) = e$.
\end{definition}
\begin{proposition}
 \label{GroupoidOfLieAlgebraValuedForms}
 \index{$L_\infty$-algebroid!valued differential forms!Lie algebra valued}
 \index{differential form!with values in Lie algebra}
This construction $A \mapsto \mathrm{tra}_A$ induces an equivalence of categories
$$
  [\mathrm{CartSp}^{\mathrm{op}},\mathrm{Grpd}](\mathbf{P}_1(X), \mathbf{B}G)
  \simeq
  \mathbf{B}G_{\mathrm{conn}}(X)
  \,,
$$
where on the left we have the hom-groupoid of groupoid-valued presheaves, 
and  where on the right we have the 
\emph{groupoid of Lie-algebra valued 1-forms}, whose 
\begin{itemize}
\item objects are 1-forms $A \in \Omega^1(X,\mathfrak{g})$, 
\item morphisms $g : A_1 \to A_2$ are labeled by smooth functions 
$g \in C^\infty(X,G)$ such that $A_2 = g^{-1} A g + g^{-1}d g$.
\end{itemize}
\end{proposition}
This equivalence is natural in $X$, so that we obtain another smooth groupoid.
\begin{definition}
Define $\mathbf{B}G_{\mathrm{conn}} : \mathrm{CartSp}^{\mathrm {op}} \to \mathrm{Grpd}$ to be the (generalized) 
Lie groupoid
$$
  \mathbf{B}G_{\mathrm{conn}} : U \mapsto [\mathrm{CartSp}^{\mathrm{op}}, \mathrm{Grpd}](\mathbf{P}_1(-), \mathbf{B}G)
$$
whose $U$-parameterized smooth families of groupoids form the groupoid of Lie-algebra valued 1-forms on $U$.
\end{definition}
This equivalence in particular subsumes the classical facts that parallel transport $\gamma \mapsto P \exp(\int_{[0,1]} \gamma^* A)$ 
\begin{itemize}
\item is invariant under orientation preserving reparameterizations of paths;
\item sends reversed paths to inverses of group elements.
\end{itemize}
\begin{observation}
There is an evident natural smooth functor $X \to \mathbf{P}_1(X)$ that includes 
points in $X$ as constant paths. This induces a natural morphism 
$\mathbf{B}G_{\mathrm{conn}} \to \mathbf{B}G$ that forgets the 1-forms.
\end{observation}
\begin{definition}
Let $P \to X$ be a $G$-principal bundle that corresponds to a 
cocycle $g : C(U) \to \mathbf{B}G$ under the construction discussed above.  
Then a \emph{connection} $\nabla$ on $P$ is  a lift $\nabla$ of the cocycle through 
$\mathbf{B}G_{\mathrm{conn}} \to \mathbf{B}G$.
$$
  \xymatrix{
     & \mathbf{B}G_{\mathrm{conn}}  \ar[d]
    \\
    C(U) \ar[r]^{g} \ar[ur]^{\nabla} & \mathbf{B}G
  }
$$
\end{definition}
\begin{observation}
 \label{BGConn}
This is equivalent to the traditional definitions.
\end{observation}
A morphism $\nabla : C(U) \to \mathbf{B}G_{\mathrm{conn}}$ is
\begin{itemize}
\item on each $U_i$ a 1-form $A_i \in \Omega^1(U_i, \mathfrak{g})$;
\item on each $U_i \cap U_j$ a function $g_{i j} \in C^\infty(U_i \cap U_j , G)$;
\end{itemize}
such that
\begin{itemize}
\item on each $U_i \cap U_j$ we have $A_j = g_{i j}^{-1}( A + d_{\mathrm{dR}} )g_{i j}$;
\item on each $U_i \cap U_j \cap U_k$ we have $g_{i j} \cdot g_{j k} = g_{i k}$.
\end{itemize}
\begin{definition}
Let $[I,X]_{\mathrm{\mathrm{si}}}^{\mathrm{th}} \to [I,X]^h$ the projection onto the full quotient by smooth homotopy classes of paths.
Write $\mathbf{\Pi}_1(X) = ([I,X]^h \stackrel{\to}{\to} X)$ for the smooth groupoid defined as $\mathbf{P}_1(X)$, but where instead of thin homotopies, all homotopies are divided out.
\end{definition}
\begin{proposition}
The above restricts to a natural equivalence
$$
  [\mathrm{CartSp}^{\mathrm{op}}, \mathrm{Grpd}](\mathbf{\Pi}_1(X), \mathbf{B}G)
  \simeq
  \mathbf{\flat}\mathbf{B}G
  \,,
$$
where on the left we have the hom-groupoid of groupoid-valued presheaves, 
and on the right we have the full sub-groupoid $\mathbf{\flat}\mathbf{B}G \subset \mathbf{B}G_{\mathrm{conn}}$ 
on those $\mathfrak{g}$-valued differential forms whose 
curvature 2-form\index{curvature!curvature 2-form (ordinary)} $F_A = d_{\mathrm{dR}} A + [A \wedge A]$ 
vanishes.

A connection $\nabla$ is \emph{flat} precisely if it factors through the inclusion $\flat \mathbf{B}G \to \mathbf{B}G_{\mathrm{conn}}$.
\end{proposition}
For the purposes of Chern-Weil theory we want a good way to extract the curvature 2-form in a general abstract way from a cocycle $\nabla : X \stackrel{\simeq}{\leftarrow }C(U) \to \mathbf{B}G_{\mathrm{conn}}$. In order to do that, we first need to discuss connections on 2-bundles.

\subparagraph{Connections on a principal 2-bundle}

There is an evident higher dimensional generalization of the definition of connections on 1-bundles in terms of functors out of the path groupoid discussed above. This we discuss now. We will see that, however, the obvious generalization captures not quite all 2-connections. But we will also see a way to recode 1-connections in terms of flat 2-connections. And that recoding then is the right general abstract perspective on connections, which generalizes to principal $\infty$-bundles and in fact which in the full theory follows from first principles.

(Constructions and results in this section are from \cite{SWII}, \cite{SWIII}.)

\begin{definition}
The path \emph{path 2-groupoid}\index{paths!path 2-groupoid} 
$\mathbf{P}_2(X)$ is the smooth strict 2-groupoid analogous to 
$\mathbf{P}_1(X)$, but with nontrivial 2-morphisms given by thin homotopy-classes of disks $\Delta^2_{Diff} \to X$ with sitting instants.

In analogy to the projection $\mathbf{P}_1(X) \to \mathbf{\Pi}_1(X)$ there is a 
projection to $\mathbf{P}_2(X) \to \mathbf{\Pi}_2(X)$ to the 2-groupoid obtained by dividing 
out full homotopy of disks, relative boundary.
\end{definition}
We want to consider 2-functors out of the path 2-groupoid into 
connected 2-groupoids of the form $\mathbf{B}G$, for $G$ a \emph{2-group},
def. \ref{Strict2GroupInIntroduction}.
A smooth 2-functor  $\mathbf{\Pi}_2(X) \to \mathbf{B}G$ now assigns information also to surfaces
$$
  \mathrm{tra}
  :
  \left\{
   \raisebox{20pt}{
  \xymatrix{
     & y \ar[dr]^{[\gamma']}
     \\
    x \ar[rr]_{[\gamma' \circ \gamma]}^{\ }="t" 
       \ar[ur]^{[\gamma]}
     && z
    \ar@{=>}^{[\Sigma]} "t"+(0,5); "t"
  }
  }
  \right\}
  \;\;\;\;
   \mapsto
  \;\;\;\;
  \left\{
  \raisebox{20pt}{
  \xymatrix{
     & {*} \ar[dr]^{\mathrm{tra}(\gamma')}
     \\
    {*} \ar[rr]_{}^{\ }="t" 
       \ar[ur]^{\mathrm{tra}(\gamma)}
     && {*}
    \ar@{=>}|{\mathrm{tra}(\Sigma)} "t"+(0,5); "t"
  }
  }
  \right\}
$$
and thus encodes \emph{higher parallel transport}.

\begin{proposition}
  \label{2GroupoidOfLie2AlgebraValuedForms}
There is a natural equivalence of 2-groupoids
$$
  [\mathrm{CartSp}^{\mathrm{op}}, \mathrm{2Grpd}](\mathbf{\Pi}_2(X), \mathbf{B}G)
  \simeq
  \mathbf{\flat} \mathbf{B}G
$$
where on the right we have the 2-groupoid of Lie 2-algebra valued forms] whose
\begin{itemize}
\item objects are pairs $A \in \Omega^1(X,\mathfrak{g}_1)$, 
  $B \in \Omega^2(X,\mathfrak{g}_2)$ such that
  the 2-form curvature \index{curvature!curvature 2-form (general)}
  $$
    F_2(A,B) := d_{\mathrm{dR}} A + [A \wedge A] + \delta_* B
  $$
  and the 3-form curvature\index{curvature!curvature 3-form}
  $$
    F_3(A,B) := d_{\mathrm{dR}} B + [A \wedge B] 
  $$
  vanish.
\item
  morphisms $(\lambda,a) : (A,B) \to (A',B')$ 
  are pairs $a \in \Omega^1(X,\mathfrak{g}_2)$, 
  $\lambda \in C^\infty(X,G_1)$ such that
  $A' = \lambda A \lambda^{-1} + \lambda d \lambda^{-1} + \delta_* a$
  and $B' = \lambda(B) + d_{\mathrm{dR}} a + [A\wedge a]$
\item
 The description of 2-morphisms we leave to the reader (see \cite{SWII}).
\end{itemize}
\end{proposition}
As before, this is natural in $X$, so that we that we get a presheaf of 2-groupoids
$$
  \mathbf{\flat}\mathbf{B}G : U \mapsto
  [\mathrm{CartSp}^{\mathrm{op}}, 2\mathrm{Grpd}](\mathbf{\Pi}_2(U), \mathbf{B}G)
  \,.
$$
\begin{proposition}
If in the above definition we use $\mathbf{P}_2(X)$ instead of $\mathbf{\Pi}_2(X)$, we obtain the same 2-groupoid, except that the 3-form curvature $F_3(A,B)$ is not required to vanish.
\end{proposition}
\begin{definition}
Let $P \to X$ be a $G$-principal 2-bundle classified by a cocycle 
$C(U) \to \mathbf{B}G$. Then a structure of a \emph{flat connection on a 2-bundle} 
$\nabla $ on it is a lift
$$
  \xymatrix{
     & \mathbf{\flat}\mathbf{B}G \ar[d]
     \\
    C(U) \ar[r]^g \ar[ur]^{\nabla_{\mathrm{flat}}} & \mathbf{B}G
  }
  \,.
$$
For $G = \mathbf{B}A$, a \emph{connection on a 2-bundle} (not necessarily flat) is a lift
$$
  \xymatrix{
     & [\mathbf{P}_2(-), \mathbf{B}^2 A] \ar[d]
     \\
    C(U) \ar[r]^g \ar[ur]^{\nabla_{\mathrm{flat}}} & \mathbf{B}G
  }
  \,.
$$
\end{definition}
We do not state the last definition for general Lie 2-groups $G$. The reason is that for general $G$ 2-anafunctors out of $\mathbf{P}_2(X)$ do not produce the fully general notion of 2-connections that we are after,  but yield a special case in between flatness and non-flatness: the case where precisely the 2-form curvature-components vanish, while the 3-form curvature part is unrestricted. This case is important in itself and discussed in detail below. 
Only for $G$ of the form $\mathbf{B}A$ does the 2-form curvature necessarily vanish anyway, so that in this case the  definition by morphisms out of $\mathbf{P}_2(X)$ happens to already coincide with the proper general one. This serves in the following theorem as an illustration for the toolset that we are exposing, but for the purposes of introducing the full notion of $\infty$-Chern-Weil theory we will rather focus on flat 2-connections, and then show 
below how using these one does arrive at a functorial definition of 1-connections that does generalize to the fully general definition of $\infty$-connections.
\begin{proposition}
  \label{BundleGerbesByCocycles}
Let $\{U_i \to X\}$ be a good open cover, a cocycle $C(U) \to [\mathbf{P}_2(-), \mathbf{B}^2 A]$ 
is a cocycle in {\v C}ech-Deligne cohomology in degree 3. 

Moreover, we have a natural equivalence of bicategories
$$
  [\mathrm{CartSp}^{\mathrm{op}}, 2\mathrm{Grpd}](C(U), [\mathbf{P}_2(-), \mathbf{B}^2 U(1)])
  \simeq
  U(1) \mathrm{Gerb}_\nabla(X)
  \,,
$$
where on the right we have the bicategory of $U(1)$-bundle gerbes with connection \cite{gajer}.

In particular the equivalence classes of cocycles form the degree-3 ordinary differential cohomology of $X$:
$$
  H^3_{\mathrm{diff}}(X, \mathbb{Z}) \simeq \pi_0( [C(U), [\mathbf{P}_2(-), \mathbf{B}^2 U(1))
  \,.
$$
\end{proposition}
A cocycle as above naturally corresponds to a 2-anafunctor
$$
  \xymatrix{
    Q \ar[r] \ar[d]^{\simeq} & \mathbf{B}^2 U(1)
    \\
    \mathbf{P}_2(X)
  }
$$
The value of this on 2-morphisms in $\mathbf{P}_2(X)$ is the higher parallel transport 
of the connection on the 2-bundle. 
This appears for instance in the action functional of the sigma model that describes strings charged under a Kalb-Ramond field.

The following example of a flat nonabelian 2-bundle is very degenerate as far as 2-bundles go, but does contain in it the seed of a full understanding of connections on 1-bundles.
\begin{definition}
For $G$ a Lie group, its inner automorphism 2-group $\mathrm{INN}(G)$ is as a groupoid the universal $G$-bundle 
$\mathbf{E}G$, but regarded as a 2-group with the group structure coming from the crossed module $[G \stackrel{Id}{\to} G]$.
\end{definition}
The depiction of the delooping 2-groupoid $\mathbf{B}\mathrm{INN}(G)$ is
$$
  \mathbf{B}\mathrm{INN}(G)
  = 
  \left\{
  \raisebox{20pt}{
  \xymatrix{
     & {*} \ar[dr]^{g_2}
     \\
    {*} \ar[rr]_{k  g_2 g_1}^{\ }="t" 
       \ar[ur]^{g_1}
     && {*}
    \ar@{=>}^k "t"+(0,5); "t"
  }
  }
    \;\;
    |
    \;\;
    g_1, g_2 \in G, k \in G 
  \right\}
  \,.
$$
This is the Lie 2-group whose Lie 2-algebra $\mathrm{inn}(\mathfrak{g})$ is the one whose 
Chevalley-Eilenberg algebra is the Weil algebra of $\mathfrak{g}$.
\begin{example}
By the above theorem we have that there is a bijection of sets
$$
  \{\mathbf{\Pi}_2(X) \to \mathbf{B} \mathrm{INN}(G)\}
  \simeq
  \Omega^1(X, \mathfrak{g})
$$
of flat $\mathrm{INN}(G)$-valued 2-connections and Lie-algebra valued 1-forms. 
Under the identifications of this theorem this identification works as follows:
\begin{itemize}
\item the 1-form component of the 2-connection is $A$;

\item the vanishing of the 2-form component of the 2-curvature $F_2(A,B) = F_A + B$  
identifies the 2-form component of the 2-connection with the curvature 2-form, $B = - F_A$;

\item the vanishing of the 3-form component of the 3-curvature $F_3(A,B) = d B + [A \wedge B] = d_A + [A \wedge F_A]$ is the Bianchi identity satisfied by any curvature 2-form.
\end{itemize}
\end{example}
This means that 2-connections with values in $\mathrm{INN}(G)$ actually model 1-connections \emph{and} keep track of their curvatures. Using this we see in the next section a general abstract definition of connections on 1-bundles that naturally supports the Chern-Weil homomorphism.

\subparagraph{Curvature characteristics of 1-bundles}

We now describe connections on 1-bundles in terms of their \emph{flat curvature 2-bundles} .

Throughout this section $G$ is a Lie group, $\mathbf{B}G$ its delooping 2-groupoid 
and $\mathrm{INN}(G)$ its inner automorphism 2-group and $\mathbf{B}\mathrm{INN}(G)$ the corresponding delooping Lie 2-groupoid. 
\begin{definition}
Define the smooth groupoid $\mathbf{B}G_{\mathrm{diff}} \in [\mathrm{CartSp}^{\mathrm{op}}, \mathrm{Grpd}]$ 
as the pullback
$$
  \mathbf{B}G_{\mathrm{diff}} = \mathbf{B}G \times_{\mathbf{B}\mathrm{INN}(G)} \mathbf{\flat} \mathbf{B}\mathrm{INN}(G)
  \,.
$$
This is the groupoid-valued presheaf which assigns to $U \in \mathrm{CartSp}$ the groupoid 
whose objects are commuting diagrams

$$
  \xymatrix{
    U \ar[r] \ar[d] & \mathbf{B}G \ar[d]
    \\
    \mathbf{\Pi}_2(U) \ar[r] & \mathbf{B}\mathrm{INN}(G)
  }
  \,,
$$
where the vertical morphisms are the canonical inclusions discussed above, and whose morphisms are compatible pairs of natural transformations 
$$
  \xymatrix{
    U 
     \ar@/^.6pc/[r]_{\ }="st"
     \ar@/_1pc/[r]^{\ }="tt" 
     \ar[d] & \mathbf{B}G \ar[d]
    \\
    \mathbf{\Pi}_2(U) 
        \ar@/^.6pc/[r]_{\ }="sb"
       \ar@/_1pc/[r]^{\ }="tb" 
     & \mathbf{B}\mathrm{INN}(G)
    \ar@{=>} "st"; "tt"
    \ar@{=>} "sb"; "tb"
  }
  \,,
$$
of the horizontal morphisms.
\end{definition}
By the above theorems, we have over any $U \in \mathrm{CartSp}$  that
\begin{itemize}
\item 
  an object in $\mathbf{B}G_{\mathrm{diff}}(U)$ is a 1-form $A \in \Omega^1(U,\mathfrak{g})$;
\item 
  amorphism $A_1 \stackrel{(g,a)}{\to} A_2$ is labeled by a function 
  $g \in C^\infty(U,G)$ and a 1-form $a \in \Omega^1(U,\mathfrak{g})$ such that
  $$
    A_2 = g^{-1}A_1 g + g^{-1}d g  + a
    \,.
  $$  
  Notice that this can always be uniquely solved for $a$, so that the genuine information in this morphism is just the data given by $g$. 
\item
ther are \emph{no} nontrivial 2-morphisms, even though $\mathbf{B}\mathrm{INN}(G)$ is a 2-groupoid: 
since $\mathbf{B}G$ is just a 1-groupoid this is enforced by the commutativity of the above diagram.
\end{itemize}
From this it is clear that
\begin{proposition}
The projection $\mathbf{B}G_{\mathrm{diff}} \stackrel{\simeq}{\to} \mathbf{B}G$ is a weak equivalence.
\end{proposition}
So $\mathbf{B}G_{\mathrm{diff}}$ is a resolution of $\mathbf{B}G$. We will see that it is the resoluton 
that supports 2-anafunctors out of $\mathbf{B}G$ which represent curvature characteristic classes.
\begin{definition}
 \label{PseudoConnection}
For $X \stackrel{\simeq}{\leftarrow}C(U) \to \mathbf{B}U(1)$ a cocycle for a $U(1)$-principal 
bundle $P \to X$, we call a lift $\nabla_{\mathrm{ps}}$ in
$$
  \xymatrix{
      & \mathbf{B}G_{\mathrm{diff}} \ar[d]
     \\
    C(U) \ar[r]^g \ar[ur]^{\nabla_{\mathrm{ps}}} & \mathbf{B}G
  }
$$
a \emph{pseudo-connection} on $P$.
\end{definition}
Pseudo-connections in themselves are not very interesting. But notice that every ordinary connection is in particular a pseudo-connection and we have an inclusion  morphism of smooth groupoids
$$
  \mathbf{B}G_{\mathrm{conn}} \hookrightarrow \mathbf{B}G_{\mathrm{diff}}
  \,.
$$
This inclusion plays a central role in the theory. The point is that while 
$\mathbf{B}G_{\mathrm{diff}}$ is such a boring extension of $\mathbf{B}G$ that it is actually equivalent to $\mathbf{B}G$, there is no inclusion of $\mathbf{B}G_{\mathrm{conn}}$ into $\mathbf{B}G$, but there is into $\mathbf{B}G_{\mathrm{diff}}$. This is the kind of situation that resolutions are needed for.

It is useful to look at some details for the case that $G$ is an abelian group such as the circle group $U(1)$.
In this abelian case the 2-groupoids $\mathbf{B}U(1)$, $\mathbf{B}^2 U(1)$, $\mathbf{B}INN(U(1))$, etc., that so far we noticed are given by crossed complexes are actually given by ordinary chain complexes: we write 
$$
  \Xi : \mathrm{Ch}_\bullet^+ \to s\mathrm{Ab} \to \mathrm{KanCplx}
$$
for the Dold-Kan correspondence map that identifies chain complexes with simplicial abelian group and then 
considers their underlying Kan complexes. Using this map we have the following identifications of our 2-groupoid valued presheaves with complexes of group-valued sheaves
$$
  \mathbf{B}U(1) = \Xi[C^\infty(-,U(1)) \to 0]
$$
$$
  \mathbf{B}^2 U(1) = \Xi[C^\infty(-,U(1))  \to 0 \to 0]
$$
$$
  \mathbf{B} \mathrm{INN} U(1) = \Xi[C^\infty(-,U(1)) \stackrel{\mathrm{Id}}{\to} C^\infty(-,U(1)) \to 0]
 \,.
$$
\begin{observation}
For $G = A$ an abelian group, in particular the circle group, there is a canonical morphism 
$\mathbf{B} \mathrm{INN}(U(1)) \to \mathbf{B}\mathbf{B}U(1)$.
\end{observation}
On the level of chain complexes this is the evident chain map
$$
  \xymatrix{
    [C^\infty(-,U(1)) \ar[r]^{Id} \ar[d]& C^\infty(-,U(1)) 
    \ar[r] \ar[d] & 0\ar[d]
    \\
    [C^\infty(-,U(1)) \ar[r] & 0 \ar[r] & 0]    
  }
  \,.
$$
On the level of 2-groupoids this is the map that forgets the labels on the 1-morphisms
$$
  \left\{
  \raisebox{20pt}{
  \xymatrix{
     & {*} \ar[dr]^{g_2}
     \\
    {*} \ar[rr]_{k g_2 g_1}^{\ }="t" 
       \ar[ur]^{g_1}
     && {*}
    \ar@{=>}^k "t"+(0,5); "t"
  }
  }
  \right\}
  \;\;\;\;
   \mapsto
  \;\;\;\;
  \left\{
  \raisebox{20pt}{
  \xymatrix{
     & {*} \ar[dr]^{\mathrm{Id}}
     \\
    {*} \ar[rr]_{\mathrm{Id}}^{\ }="t" 
       \ar[ur]^{\mathrm{Id}}
     && {*}
    \ar@{=>}^k "t"+(0,5); "t"
  }
  }
  \right\}
$$
In terms of this map $\mathrm{INN}(U(1))$ serves to interpolate between the single and the double 
delooping of $U(1)$. In fact the sequence of 2-functors
$$
  \mathbf{B}U(1) \to \mathbf{B}\mathrm{INN}(U(1)) \to 
  \mathbf{B}^2 U(1)
$$
is a model for the universal $\mathbf{B}U(1)$-principal 2-bundle
$$
  \mathbf{B}U(1) \to  \mathbf{E} \mathbf{B}U(1)
  \to 
  \mathbf{B}^2 U(1)
  \,.
$$
This happens to be an exact sequence of 2-groupoids. Abstractly, what really matters is rather that it is a fiber sequence, meaning that it is exact in the correct sense inside the $\infty$-category $\mathrm{Smooth}\infty\mathrm{Grpd}$. 
For our purposes it is however relevant that this particular model is exact also in the ordinary sense in that we have an ordinary pullback diagram
$$
  \xymatrix{
    \mathbf{B}U(1) \ar[r] \ar[d]& {*} \ar[d]
    \\
    \mathbf{B}\mathrm{INN}(U(1)) \ar[r] & \mathbf{B}^2 U(1)
  }
  \,,
$$
exhibitng $\mathbf{B}U(1)$ as the kernel of $\mathbf{B}\mathrm{INN}(U(1)) \to \mathbf{B}^2 U(1)$.

We shall be interested in the pasting composite of this diagram with the one defining 
$\mathbf{B}G_{\mathrm{diff}}$ over a domain $U$:
$$
  \xymatrix{
    U \ar[r] \ar[d] & \mathbf{B}U(1) \ar[r] \ar[d] & {*} \ar[d]
    \\
    \mathbf{\Pi}_2(U) \ar[r] & \mathbf{B}\mathrm{INN}(U(1)) \ar[r] & \mathbf{B}^2 U(1)
  }
  \,,
$$
The total outer diagram appearing this way is a component of the following (generalized) Lie 2-groupoid.
\begin{definition}
Set
$$
  \mathbf{\flat}_{\mathrm{\mathrm{dR}}} \mathbf{B}^2U(1) 
    := 
  * \times_{\mathbf{B}^2 U(1)} \mathbf{\flat} \mathbf{B}^2 U(1)
  \,.
$$
\end{definition}
Over any $U \in \mathrm{CartSp}$ this is the 2-groupoid whose objects are sets of diagrams
$$
  \xymatrix{
    U \ar[r] \ar[d] & {*} \ar[d]
    \\
    \mathbf{\Pi}_2(U) \ar[r] & \mathbf{B}^2 U(1)
  }
  \,.
$$
This are equivalently just morphisms $\mathbf{\Pi}_2(U) \to \mathbf{B}^2 U(1)$, which by the above theorems we may identify with closed 2-forms $B \in \Omega^2_{\mathrm{\mathrm{cl}}}(U)$.

The morphisms $B_1 \to B_2$ in $\mathbf{\flat}_{\mathrm{dR}} \mathbf{B}^2 U(1)$ over $U$ are compatible pseudonatural transformations of the horizontal morphisms 
$$
  \xymatrix{
    U 
     \ar@/^.6pc/[r]_{\ }="st"
     \ar@/_1pc/[r]^{\ }="tt" 
     \ar[d] & {*} \ar[d]
    \\
    \mathbf{\Pi}_2(U) 
        \ar@/^.6pc/[r]_{\ }="sb"
       \ar@/_1pc/[r]^{\ }="tb" 
     & \mathbf{B}\mathrm{INN}(G)
    \ar@{=>} "st"; "tt"
    \ar@{=>} "sb"; "tb"
  }
  \,,
$$
which means that they are pseudonatural transformations of the bottom morphism whose components over the points of $U$ vanish. These identify with 1-forms $\lambda \in \Omega^1(U)$ such that  $B_2 = B_1 + d_{\mathrm{dR}} \lambda$.
Finally the 2-morphisms would be modifications of these, but the commutativity of the above diagram constrais these to be trivial.

In summary this shows that
\begin{proposition}
Under the Dold-Kan correspondence $\mathbf{\flat}_{\mathrm{\mathrm{dR}}} \mathbf{B}^2 U(1)$ is the sheaf of truncated de Rham complexes
$$
  \mathbf{\flat}_{\mathrm{\mathrm{dR}}} \mathbf{B}^2 U(1)
  =
  \Xi[\Omega^1(-) \stackrel{d_{\mathrm{\mathrm{dR}}}}{\to} \Omega^2_{\mathrm{cl}}(-)]
  \,.
$$
\end{proposition}
\begin{corollary}
Equivalence classes of 2-anafunctors
$$
  X \to \mathbf{\flat}_{\mathrm{dR}} \mathbf{B}^2 U(1)
$$
are canonically in bijection with the degree 2 de Rham cohomology of $X$.
\end{corollary}
Notice that --  while every globally defined closed 2-form $B \in \Omega^2_{\mathrm{\mathrm{cl}}}(X)$ 
defines such a 2-anafunctor -- not every such 2-anafunctor comes from a globally defined closed 2-form. 
Some of them assign closed 2-forms $B_i$ to patches $U_1$, that differ by differentials 
$B_j - B_i = d_{\mathrm{\mathrm{dR}}} \lambda_{i j}$ of 1-forms $\lambda_{i j}$ on double overlaps, 
which themselves satisfy on triple intersections the cocycle condition 
$\lambda_{i j} + \lambda_{j k} = \lambda_{i k}$. But (using a partition of unity) these non-globally 
defined forms are always equivalent to globally defined ones.

This simple technical point turns out to play a role in the abstract definition of connections on $\infty$-bundles: generally, for all $n \in \mathbb{N}$ the cocycles given by globally defined forms in 
$\mathbf{\flat}_{\mathrm{\mathrm{dR}}} \mathbf{B}^n U(1)$ constitute curvature characteristic forms of 
\emph{genuine} connections. The non-globally defined forms \emph{also} constitute curvature invariants, but of pseudo-connections. The way the abstract theory finds the genuine connections inside all pseudo-connections is by the fact that we may find for each cocycle in $\mathbf{\flat}_{\mathrm{\mathrm{dR}}} \mathbf{B}^n U(1)$ 
an equivalent one that does comes from a globally defined form.
\begin{observation}
There is a canonical 2-anafunctor 
${\hat {\mathbf{c}}}_1^{\mathrm{\mathrm{dR}}} : \mathbf{B}U(1) \to \mathbf{\flat}_{\mathrm{\mathrm{dR}}}\mathbf{B}^2 U(1)$

$$
  \xymatrix{
    \mathbf{B}U(1)_{\mathrm{diff}} 
      \ar[r]
      \ar[d]^{\simeq} 
     & 
     \mathbf{\flat}_{\mathrm{\mathrm{dR}}} \mathbf{B}^2 U(1)
    \\
    \mathbf{B}U(1)
  }
  \,,
$$
where the top morphism is given by forming the -composite with the 
universal $\mathbf{B} U(1)$-principal 2-bundle, as described above.
\end{observation}
For emphasis, notice that this span is governed by a presheaf of diagrams that over $U \in CartSp$ is of the form
$$
  \xymatrix{
    U \ar[r] \ar[d] & \mathbf{B}U(1) \ar[d] && \mbox{transition function}
    \\
    \mathbf{\Pi}_2(U) \ar[r] \ar[d] & \mathbf{B}\mathrm{INN}(U) \ar[d] && \mbox{connection}
    \\
    \mathbf{\Pi}_2(U) \ar[r] & \mathbf{B}^2 U(1) && \mbox{curvature}
  }
  \,.
$$
The top morphisms are the components of the presheaf $\mathbf{B}U(1)$. The top squares are those of $\mathbf{B}U(1)_{\mathrm{diff}}$. Forming the bottom square is forming the bottom morphism, 
which necessarily satifies the constraint that makes it a components of $\mathbf{\flat}\mathbf{B}^2 U(1)$. 

The interpretation of the stages is as indicated in the diagram:
\begin{enumerate}
\item
  the top morphism is the transition function of the underlying bundle;
\item
  the middle morphism is a choice of (pseudo-)connection on that bundle;
\item
  the bottom morphism picks up the curvature of this connection.
\end{enumerate}
We will see that full $\infty$-Chern-Weil theory is governed by a slight refinement of presheaves of essentially this kind of diagram. We will also see that the three stage process here is really an incarnation of the computation of a connecting homomorphism, reflecting the fact that behind the scenes the notion of 
\emph{curvature} is exhibited as the obstruction cocycle to lifts from bare bundles to flat bundles.
\begin{observation}
For $X \stackrel{\simeq}{\leftarrow} C(U) \stackrel{g}{\to} \mathbf{B}U(1)$ the cocycle for a $U(1)$-principal bundle as described above, the composition of 2-anafunctors of $g$ with $\hat {\mathbf{c}}_1^{\mathrm{dR}}$ yields a cocycle for a 2-form $\hat {\mathbf{c}}_1^{\mathrm{dR}}(g)$
$$
  \xymatrix{
    & \mathbf{B}U(1)_{\mathrm{conn}} \ar[d]
    \\
    C(V) \ar[ur]^{\nabla} \ar[r] \ar[d]^\simeq & 
      \mathbf{B} U(1)_{\mathrm{diff}} \ar[r] \ar[d]^\simeq & 
      \mathbf{\flat}_{\mathrm{\mathrm{dR}}} \mathbf{B}^2 U(1)
    \\
    C(U) \ar[r]^g \ar[d]^{\simeq} & \mathbf{B}U(1)
    \\
    X 
  }
  \,.
$$
If we take $\{U_i \to X\}$ to be a good open cover, then we may assume $V = U$. 
We know we can always find a pseudo-connection $C(V) \to \mathbf{B}U(1)_{\mathrm{diff}}$ 
that is actually a genuine connection on a bundle in that it factors through the inclusion 
$\mathbf{B}U(1)_{\mathrm{conn}} \to \mathbf{B}U(1)_{\mathrm{diff}}$ as indicated.

The corresponding total map $c_1^{\mathrm{dR}}(g)$ represented by 
$\hat {\mathbf{c}}_1^{\mathrm{\mathrm{dR}}}(\nabla)$ is the cocycle for the curvature 2-form of this connection. This represents the first Chern class of the bundle in de Rham cohomology.
\end{observation}

For $X,A$ smooth 2-groupoids, write $\mathbf{H}(X,A)$ for the 2-groupoid of 2-anafunctors between them. 
\begin{corollary}
Let $H_{\mathrm{\mathrm{dR}}}^2(X) \to \mathbf{H}(X,\mathbf{\flat}_{\mathrm{\mathrm{dR}}} \mathbf{B}^2 U(1))$ 
be a choice of one closed 2-form representative for each degree-2 de Rham cohomology-class of $X$. Then the pullback groupoid $\mathbf{H}_{\mathrm{diff}}(X,\mathbf{B}U(1))$ in 
$$
  \xymatrix{
    \mathbf{H}_{\mathrm{conn}}(X,\mathbf{B}U(1)) \ar[r] \ar[d] & H_{\mathrm{\mathrm{dR}}}^2(X) \ar[d]
    \\
    \mathbf{H}(X, \mathbf{B}U(1)_{\mathrm{diff}}) \ar[r] \ar[d]^\simeq &
    \mathbf{H}(X, \mathbf{\flat}_{\mathrm{\mathrm{dR}}} \mathbf{B}^2 U(1))
    \\
    \mathbf{H}(X,\mathbf{B}U(1)) \simeq U(1) \mathrm{Bund}(X)
  }
$$
is equivalent to disjoint union of groupoids of $U(1)$-bundles with connection whose curvatures are the 
chosen 2-form representatives.
\end{corollary}

\medskip

\subparagraph{Circle $n$-bundles with connection}
\index{connection!circle $n$-bundle with connection!in introduction}
\index{circle $n$-bundle with connection!in introduction}

For $A$ an abelian group there is a straightforward generalization of the above constructions to $(G = \mathbf{B}^{n-1}A)$-principal $n$-bundles with connection for all $n \in \mathbb{N}$. We spell out the ingredients of the construction in a way analogous to the above discussion. A first-principles derivation of the objects we consider here below in \ref{SmoothStrucDifferentialCohomology}.

This is  content that appeared partly in \cite{SSSIII}, \cite{FSS}. 
We restrict attention to the circle $n$-group $G = \mathbf{B}^{n-1}U(1)$. 

\medskip

There is a familiar traditional presentation of ordinary differential cohomology in terms 
of Cech-Deligne cohomology. We briefly recall how this works and then 
indicate how this presentation can be derived along the above lines as a presentation 
of circle $n$-bundles with connection.

\begin{definition}
 \label{DeligneComplex}
For $n \in \mathbb{N}$ the \emph{Deligne-Beilinson complex} is the chain complex of sheaves 
(on $\mathrm{CartSp}$ for our purposes here) of abelian groups given as follows
$$
 \mathbb{Z}(n+1)^\infty_D = 
  \left[
    \raisebox{10pt}{
    \xymatrix@R=3pt{
      C^\infty(-,\mathbb{R}/\mathbb{Z})
       \ar[r]^{d_{\mathrm{\mathrm{dR}}}}
      &
      \Omega^1(-)
       \ar[r]^{d_{\mathrm{\mathrm{dR}}}}
      &
      \cdots
        \ar[r]^{d_{\mathrm{\mathrm{dR}}}}
      &
      \Omega^{n-1}(-)
        \ar[r]^{d_{\mathrm{\mathrm{dR}}}}
      &
      \Omega^n(-)      
      \\
       n & n-1 & \cdots & 1 & 0
    }}
  \right]
  \,.
$$
\end{definition}
This definition goes back to \cite{Deligne} \cite{Beilinson}. 
The complex is similar to the $n$-fold shifted de Rham complex, 
up to two important differences.
\begin{itemize}
\item In degree $n$ we have the sheaf of $U(1)$-valued functions, 
not of $\mathbb{R}$-valued functions (= 0-forms). The action of the de Rham differential 
on this is often written $d \mathrm{log} : C^\infty(-, U(1)) \to \Omega^1(-)$. But if we think 
of $U(1) \simeq \mathbb{R}/\mathbb{Z}$ then it is just the ordinary de Rham differential applied 
to any representative in $C^\infty(-, \mathbb{R})$ of an element in $C^\infty(-, \mathbb{R}/\mathbb{Z})$.

\item In degree 0 we do not have closed differential $n$-forms 
(as one would have for the de Rham complex shifted into non-negative degree), but all 
$n$-forms.
\end{itemize}
As before, we may use of the Dold-Kan correspondence 
$\Xi : \mathrm{Ch}_\bullet^{+} \stackrel{\simeq}{\to} \mathrm{sAb} \stackrel{U}{\to} \mathrm{sSet}$ to identify sheaves of chain complexes with simplicial sheaves. We write
$$
  \mathbf{B}^n U(1)_{\mathrm{conn}} := \Xi \, \mathbb{Z}(n+1)^\infty_D
$$
for the simplicial presheaf corresponding to the Deligne complex.

Then for $\{U_i \to X\}$ a good open cover, 
the Deligne cohomology of $X$ in degree $(n+1)$ is 
$$
  H_{\mathrm{diff}}^{n+1}(X)
  =
  \pi_0 [\mathrm{CartSp}^{\mathrm{op}}, \mathrm{sSet}](
     C(\{U_i\}), \mathbf{B}^n U(1)_{\mathrm{conn}}
  )
  \,.
$$
Further using the Dold-Kan correspondence, this is equivalently the cohomology of the 
{\v C}ech-Deligne double complex. A cocycle in degre $(n+1)$ then is a tuple
$$
  (g_{i_0, \cdots, i_n}, \cdots, A_{i j k}, B_{i j}, C_{i})
$$
with 
\begin{itemize}
\item $C_i \in \Omega^n(U_i)$;

\item $B_{i j} \in \Omega^{n-1}(U_i \cap U_j)$;

\item $A_{i j k } \in \Omega^{n-2}(U_i \cap U_j \cap U_k)$

\item and so on...

\item $g_{i_0, \cdots, i_n} \in C^\infty(U_{i_0} \cap \cdots \cap U_{i_n} , U(1))$
\end{itemize}
satisfying the cocycle condition
$$
  (d_{\mathrm{\mathrm{dR}}} + (-1)^{deg}\delta)
  (g_{i_0, \cdots, i_n}, \cdots, A_{i j k}, B_{i j}, C_{i})
  = 0
  \,,
$$
where $\delta = \sum_{i} (-1)^i p_i^*$ is the alternating sum of the pullback of forms along the 
face maps of the {\v C}ech nerve.

This is a sequence of conditions of the form
\begin{itemize}
\item $C_i - C_j = d B_{i j}$;

\item $B_{i j} - B_{i k} + B_{j k} = d A_{i j k}$;

\item and so on

\item $(\delta g)_{i_0, \cdots, i_{n+1}} = 0$.
\end{itemize}
For low $n$ we have seen these conditions in the dicussion of line bundles and of 
line 2-bundles (bundle gerbes) 
with connection above. Generally, for any $n \in \mathbb{N}$, this is 
{\v C}ech-cocycle data for a \emph{circle $n$-bundle}
\index{principal $\infty$-bundle!circle $n$-bundle with connection} with connection, where
\begin{itemize}
\item $C_i$ are the local connection $n$-forms;

\item $g_{i_0, \cdots, i_n}$ is the transition function of the circle $n$-bundle.
\end{itemize}
We now indicate how the Deligne complex may be derived from differential refinement of cocycles
for circle $n$-bundles along the lines of the above discussions.
To that end, write
$$
  \mathbf{B}^n U(1)_{\mathrm{ch}} := \Xi \, U(1)[n]
  \,,
$$
for the simplicial presheaf given under the Dold-Kan correspondence by the chain complex
$$
  U(1)[n] = 
  \left(
     C^\infty(-,U(1)) \to 0 \to \cdots \to 0
  \right)
$$
with the sheaf represented by $U(1)$ in degree $n$.
\begin{proposition}
For $\{U_i \to X\}$ an open cover of a smooth manifold $X$ and $C(\{U_i\})$ its {\v C}ech nerve,  $\infty$-anafunctors
$$
  \xymatrix{
    C(\{U_i\}) \ar[r]^g \ar[d]^{\simeq} & \mathbf{B}^n U(1)
    \\
    X
  }
$$
are in natural bijection with tuples of smooth functions
$$
  g_{i_0 \cdots i_n} : U_{i_0} \cap \cdots \cap U_{i_n} \to \mathbb{R}/\mathbb{Z}
$$
satisfying 
$$
  (\partial g)_{i_0 \cdots i_{n+1}}
  :=
  \sum_{k = 0}^{n} g_{i_0 \cdots i_{k-1} i_k \cdot i_n}
  = 0
  \,,
$$
that is, with cocycles in degree-$n$ {\v C}ech cohomology on $U$ with values in 
$U(1)$.

Natural transformations
$$
  \xymatrix{
    C(\{U_i\})\cdot \Delta^1 
     \ar[rr]^{(g \stackrel{\lambda}{\to} g')}
     \ar[d]^\simeq
     && 
	 \mathbf{B}^n U(1)
    \\
    X \cdot \Delta^1
  }
$$
are in natural bijection with tuples of smooth functions
$$
  \lambda_{i_0 \cdots i_{n-1}} : U_{i_0} \cap \cdots \cap U_{i_{n-1}} \to \mathbb{R}/\mathbb{Z}
$$
such that
$$
  g'_{i_0 \cdots i_n}
  - 
  g_{i_0  \cdots i_n} 
  = 
  (\delta \lambda)_{i_0 \cdots i_n}
  \,,
$$
that is, with {\v C}ech coboundaries.
\end{proposition}
The $\infty$-bundle $P \to X$ classified by such a cocycle
according to \ref{ModelForPrincipalInfinityBundles} we call a 
\emph{circle n-bundle}. For $n = 1$ this reproduces the 
ordinary $U(1)$-principal bundles that we considered 
before in \ref{Principal1Bundles}, 
for $n =2 $ the bundle gerbes considered in \ref{Principal2Bundles} 
and for $n=3$ the bundle 2-gerbes discussed in \ref{Principal3BundlesAndTwisted2Bundles}.

To equip these circle $n$-bundles with connections, 
we consider the differential refinements of $\mathbf{B}^n U(1)_{\mathrm{ch}}$ 
to be denoted $\mathbf{B}^n U(1)_{\mathrm{diff}}$, 
$\mathbf{B}^n U(1)_{\mathrm{conn}}$ and $\mathbf{\flat}_{\mathrm{\mathrm{dR}}} \mathbf{B}^{n+1}U(1)$.
\begin{definition}
Write
$$
  \mathbf{\flat}_{\mathrm{\mathrm{dR}}}\mathbf{B}^{n+1}U(1)_{\mathrm{chn}} := 
  \Xi\left(
    \Omega^1(-) \stackrel{d_{\mathrm{dR}}}{\to}
    \Omega^2(-) \stackrel{d_{\mathrm{dR}}}{\to}
    \cdots
    \stackrel{d_{\mathrm{dR}}}{\to} \Omega^n_{\mathrm{cl}}(-)
  \right)
$$
-- the \emph{truncated de Rham complex} -- and
$$
  \mathbf{B}^n U(1)_{\mathrm{diff}} = 
  \left\{
    \raisebox{20pt}{
   \xymatrix@C=6pt{
      (-) \ar[r] \ar[d] & \mathbf{B}^n U(1) \ar[d]
     \\
     \mathbf{\Pi}(-) \ar[r] & \mathbf{B}^n \mathrm{INN}(U(1))
   }
   }
  \right\}
  =
  \Xi
  \left(
    \raisebox{20pt}{
    \xymatrix@C=6pt@R=3pt{
      C^\infty(-, \mathbb{R}/\mathbb{Z})
        \ar[r] 
        &
        \Omega^1(-)
        \ar[r]^{d_{\mathrm{\mathrm{dR}}}}
        &
        \cdots
        \ar[r]
        &
        \Omega^n(-)
        \\
        \oplus 
        \\
        \Omega^1(-)
        \ar[uur]_{\mathrm{Id}}
         \ar[r]_{d_{\mathrm{\mathrm{dR}}}}
         &
         \cdots
         \ar[r]^{d_{\mathrm{\mathrm{dR}}}}
         &
         \Omega^n(-)
         \ar[uur]_{\mathrm{Id}}
    }
    }
  \right)
$$
and
$$
  \mathbf{B}^n U(1)_{\mathrm{conn}}
  = 
  \Xi\left(
    C^\infty(-, \mathbb{R}/\mathbb{Z})
    \stackrel{d_{\mathrm{dR}}}{\to} 
    \Omega^1(-) \stackrel{d_{\mathrm{dR}}}{\to}
    \Omega^2(-) \stackrel{d_{\mathrm{dR}}}{\to}
    \cdots
    \stackrel{d_{\mathrm{dR}}}{\to} \Omega^n(-)
  \right)
$$
-- the \emph{Deligne complex}, def. \ref{DeligneComplex}.
\end{definition}
\begin{observation}
We have a pullback diagram
$$
  \xymatrix{
    \mathbf{B}^n U(1)_{\mathrm{conn}}
    \ar[r] \ar[d] &
    \Omega^{n+1}_{\mathrm{\mathrm{cl}}}(-)
     \ar[d]
    \\
    \mathbf{B}^n U(1)_{\mathrm{diff}}
    \ar[r]^{\mathrm{curv}} \ar[d]^\simeq &
    \mathbf{\flat}_{\mathrm{\mathrm{dR}}}\mathbf{B}^{n-1}U(1)
    \\
    \mathbf{B}^n U(1)
  }
$$
in $[\mathrm{CartSp}^{op}, \mathrm{sSet}]$. This models an $\infty$-pullback
$$
  \xymatrix{
    \mathbf{B}^n U(1)_{\mathrm{conn}}
    \ar[r] \ar[d] &
    \Omega^{n+1}_{\mathrm{\mathrm{cl}}}(-)
     \ar[d]
    \\
    \mathbf{B}^n U(1)
    \ar[r] &
    \mathbf{\flat}_{\mathrm{\mathrm{dR}}}\mathbf{B}^{n-1}U(1)
  }
$$
in the $\infty$-topos $\mathrm{Smooth}\infty\mathrm{Grpd}$, and hence for each smooth manifold $X$
(in particular) a homotopy pullback
$$
  \raisebox{20pt}{
  \xymatrix{
    \mathbf{H}(X, \mathbf{B}^n U(1)_{\mathrm{conn}})
    \ar[r] \ar[d] &
    \Omega^{n+1}_{\mathrm{\mathrm{cl}}}(X)
     \ar[d]
    \\
    \mathbf{H}(X,\mathbf{B}^n U(1))
    \ar[r] &
    \mathbf{H}(X,\mathbf{\flat}_{\mathrm{\mathrm{dR}}}\mathbf{B}^{n-1}U(1))
  }
  }
  \,.
$$
We write 
$$
  H_{\mathrm{diff}}^n(X)
  :=
  \mathbf{H}(X, \mathbf{B}^n U(1)_{\mathrm{conn}})
$$
for the group of cohomology classes on $X$ with coefficients in 
$\mathbf{B}^n U(1)_{\mathrm{conn}}$. On these cohomology classes the
above homotopy pullback diagram reduces to the commutative diagram
$$
  \xymatrix{
    & H^{n+1}_{\mathrm{diff}}(X)
	\ar[dl]
	\ar[dr]
	\\
	H^{n+1}(X, \mathbb{Z})
    \ar[dr]	
	 && 
	\Omega^{n+1}_{\mathrm{cl}}(X)
	\ar[dl]
	\\
	& H^{n+1}(X, \mathbb{R}) \simeq H^{n+1}_{\mathrm{dR}}(X)
  }
$$
that had appeared above in \ref{MotivationFromActionFunctionals}.
But notice that the homotopy pullback of the cocycle
$n$-groupoids contains more information than this projection to 
cohomology classes. 
\end{observation}

Objects in $\mathbf{H}(X,\mathbf{B}^n U(1)_{\mathrm{conn}})$ are modeled by
$\infty$-anafunctors 
$X \stackrel{\simeq}{\leftarrow}C(\{U_i\}) \to \mathbf{B}^n U(1)_{\mathrm{conn}}$,
and these are in natural bijection with tuples
$$
  \left(
     C_{i}, B_{i_0 i_1}, A_{i_0 i_1, i_2}, 
     \cdots
     Z_{i_0 \cdots i_{n-1}},
     g_{i_0 \cdots i_{n}}
  \right)
  \,,
$$
where $C_i \in \Omega^n(U_i)$, $B_{i_0 i_1} \in \Omega^{n-1}(U_{i_0} \cap U_{i_1})$, etc., 
such that
$$
  C_{i_0} - C_{i_1} = d B_{i_0 i_1}
$$
and
$$
  B_{i_0 i_1} - B_{i_0 i_2} + B_{i_1 i_2} = d A_{i_0 i_1 i_2}
  \,,
$$
etc. This is a cocycle in {\v C}ech-Deligne cohomology. We may think of this as encoding a 
circle $n$-bundle with connection. The forms $(C_i)$ are the local \emph{connection $n$-forms}.

The definition of $\infty$-connections on $G$-principal $\infty$-bundles for 
nonabelian $G$ may be reduced to this definition, by \emph{approximating} every $G$-cocylce 
$X \stackrel{\simeq}{\leftarrow} C(\{U_i\}) \to \mathbf{B}G$ by abelian cocycles 
in all possible ways,
by postcomposing with all possible \emph{characteristic classes} 
$\mathbf{B}G \stackrel{\simeq}{\leftarrow} \widehat {\mathbf{B}G}\to \mathbf{B}^n U(1)$ 
to extract a circle $n$-bundle from it. This is what we turn to below in
\ref{CharacteristicClassesInLowDegree}.

\subparagraph{Holonomy and canonical action functionals}
\label{HolonomyInIntroduction}
\index{connection!holonomy!in introduction}

We had started out with motivating differential refinements of 
bundles and higher bundles by the notion of higher parallel transport. 
Here we discuss aspects
of this for the circle $n$-bundles

Let $\Sigma$ be a compact smooth manifold of dimension $n$.
For every smooth function $\Sigma \to X$ there is a corresponding 
pullback operation
$$
  H^{n+1}_{\mathrm{diff}}(X) \to H^{n+1}_{\mathrm{diff}}(\Sigma)
$$
that sends circle $n$-connections on $X$ to circle $n$-connections on $\Sigma$.
But due to its dimension, the curvature $(n+1)$-form of any circle $n$-connection
on $\Sigma$ is necessarily trivial. From the definition of homotopy pullback
one can show that this implies that every circle $n$-connection on $\Sigma$
is equivalent to one which is given by a Cech-Deligne cocycle that involves
a globally defined connection $n$-form $\omega$. The integral of this form over
$\Sigma$ produces a real number. One finds that this is well-defined
up to integral shifts. This gives an \emph{$n$-volume holonomy} map
$$
  \int_\Sigma \;:\;
  \mathbf{H}(\Sigma, \mathbf{B}^n U(1)_{\mathrm{conn}})  
  \to U(1)
  \,.
$$
For instance for $n = 1$ this is the map that sense an ordinary connection
on an ordinary circle bundle over $\Sigma$ to its ordinary parallel transport 
along $\Sigma$, its line holonomy.

For $G$ any smooth (higher) group, any morphism
$$
  \hat {\mathbf{c}} : 
  \mathbf{B}G_{\mathrm{conn}} \to \mathbf{B}^n U(1)_{\mathrm{conn}}
$$
from the moduli stack of $G$-connections to that of circle $n$-connections
therefore induces a canonical functional
$$
  \exp(i S_{\mathbf{c}}(-))
  :
  \xymatrix{
    \mathbf{H}(\Sigma, \mathbf{B}G_{\mathrm{conn}})
	\ar[rr]^{\mathbf{H}(\Sigma, \hat {\mathbf{c}}})
	&&
	\mathbf{H}(\Sigma, \mathbf{B}^n U(1)_{\mathrm{conn}})
	\ar[rr]^{\int_\Sigma}
	&&
	U(1)
  }
$$
from the $\infty$-groupoid of $G$-connections on $\Sigma$ to $U(1)$.

\paragraph{Differential cohomology}
\label{IntroDiffCohomology}
\index{differential cohomology!overview}

We now indicate how the combination of the \emph{intrinsic cohomology} and the 
\emph{geometric homotopy} in a locally $\infty$-connected $\infty$-topos yields a 
good notion of \emph{differential cohomology in an $\infty$-topos}.

Using the defining adjoint $\infty$-functors $(\Pi \dashv \mathrm{Disc} \dashv \Gamma)$ 
we may reflect the fundamental $\infty$-groupoid $\Pi : \mathbf{H} \to \infty\mathrm{Grpd}$ 
from $\mathrm{Top}$ back into $\mathbf{H}$ by considering the composite endo-edjunction
$$
  (\mathbf{\Pi} \dashv \mathbf{\flat}) 
   :=
   (\mathrm{Disc} \circ \Pi \dashv \mathrm{Disc} \circ \Gamma)
   : 
  \xymatrix{
     \mathbf{H}
       \ar@<-3pt>[r] \ar@<+3pt>@{<-}[r]
       &
     \mathbf{H}
  }
  \,.
$$
The $(\Pi \dashv \mathrm{Disc})$-unit $X \to \mathbf{\Pi}(X)$
may be thought of as the inclusion of $X$ into its fundamental $\infty$-groupoid 
as the collection of constant paths in $X$.

As always, the boldface $\mathbf{\Pi}$ is to indicate that we are dealing with a cohesive 
refinement of the topological structure $\Pi$. The symbol ``$\mathbf{\flat}$'' (``flat'') 
is to be suggestive of the meaning of this construction:

For $X \in \mathbf{H}$ any cohesive object, we may think of $\Pi(X)$ as its cohesive fundamental 
$\infty$-groupoid. A morphism
$$
  \nabla : \mathbf{\Pi}(X) \to \mathbf{B}G
$$
(hence a $G$-valued cocycle on $\mathbf{\Pi}(X)$) may be interpreted as assigning:
\begin{itemize}
  \item
    to each point $x \in X$ the fiber of the corresponding $G$-principal $\infty$-bundle 
    classified by the composite $g : X \to \mathbf{\Pi}(X) \stackrel{\nabla}{\to} \mathbf{B}G$;
  \item
    to each path in $X$ an equivalence between the fibers over its endpoints;
  \item
    to each homotopy of paths in $X$ an equivalence between these equivalences;
  \item
    and so on.
\end{itemize}
This in turn we may think as being the \emph{flat higher parallel transport} 
of an $\infty$-connection on the bundle classified by $g : X \to \mathbf{\Pi}(X) \stackrel{\nabla}{\to}
\mathbf{B}G$.

The adjunction equivalence allows us to identify $\mathbf{\flat}\mathbf{B}G$ as the coefficient object 
for this flat differential $G$-valued cohomology on $X$:
$$
  H_{\mathrm{flat}}(X,G) := \pi_0 \mathbf{H}(X, \mathbf{\flat}\mathbf{B}G)
    \simeq
  \pi_0 \mathbf{H}(\mathbf{\Pi}(X), \mathbf{B}G)
   \,.
$$
In $\mathbf{H} = \mathrm{Smooth}\infty\mathrm{Grpd}$ and with $G \in \mathbf{H}$ an ordinary Lie group 
and $X \in \mathbf{H}$ an ordinary smooth manifold, we have that $H_{\mathrm{flat}}(X,G)$ 
is the set of equivalence classes of ordinary $G$-principal bundles on $X$ with flat connections.

The $(\mathrm{Disc} \dashv \Gamma)$-counit $\mathbf{\flat}\mathbf{B}G \to \mathbf{B}G$
provides the forgetful morphism
$$
  H_{\mathrm{flat}}(X,G) \to H(X,G)
$$
form $G$-principal $\infty$-bundles with flat connection to their underlying principal $\infty$-bundles. 
Not every $G$-principal $\infty$-bundle admits a flat connection. The failure of this to be true 
-– the obstruction to the existence of flat lifts -– is measured by the homotopy fiber of the counit, 
which we shall denote $\mathbf{\flat}_{\mathrm{\mathrm{dR}}} \mathbf{B}G$, defined by the fact that 
we have a fiber sequence
$$
  \mathbf{\flat}_{\mathrm{\mathrm{dR}}} \mathbf{B}G \to \mathbf{\flat}\mathbf{B}G \to \mathbf{B}G
    \,.
$$
As the notation suggests, it turns out that $\mathbf{\flat}_{\mathrm{\mathrm{dR}}}\mathbf{B}G$ 
may be thought of as the coefficient object for nonabelian generalized de Rham cohomology. 
For instance for $G$ an odinary Lie group regarded as an object in $\mathbf{H} = \mathrm{Smooth}\infty\mathrm{Grpd}$, 
we have that $\mathbf{\flat}_{\mathrm{\mathrm{dR}}} \mathbf{B}G$ is presented by the sheaf 
$\Omega^1_{\mathrm{flat}}(-,\mathfrak{g})$ of Lie algebra valued differential forms with vanishing 
curvature 2-form. And for the circle Lie $n$-group $\mathbf{B}^{n-1}U(1)$ we find that 
$\mathbf{\flat}_{\mathrm{\mathrm{dR}}} \mathbf{B}^n U(1)$ is presented by the complex of sheaves 
whose abelian sheaf cohomology is de Rham cohomology in degree $n$.
(More precisely, this is true for $n \geq 2$. For $n = 1$ we get just the sheaf of closed 1-forms.
This is due to the obstruction-theoretic nature of $\mathbf{\flat}_{\mathrm{\mathrm{dR}}}$: 
as we shall see, in degree 1
it computes 1-form curvatures of groupoid principal bundles, 
and these are not quotiented by exact 1-forms.)
Moreover, in this case our fiber sequence extends not just to the left but also to the right
$$
  \mathbf{\flat}_{\mathrm{\mathrm{dR}}}\mathbf{B}^n U(1)
    \to 
  \mathbf{\flat}\mathbf{B}^n U(1)
    \to
  \mathbf{B}^n U(1)
    \stackrel{\mathrm{curv}}{\to}
  \mathbf{\flat}_{\mathrm{\mathrm{dR}}}\mathbf{B}^{n+1} U(1)  
   \,.
$$
The induced morphism 
$$
  \mathrm{curv}_X : \mathbf{H}(X, \mathbf{B}^n U(1)) \to \mathbf{H}(X,\mathbf{\flat}_{\mathrm{\mathrm{dR}}}\mathbf{B}^{n+1} U(1))
$$
we may think of as equipping an $\mathbf{B}^{n-1}U(1)$-principal $n$-bundle 
(equivalently an $(n-1)$-bundle gerbe) with a connection, and then sending it to the 
higher curvature class of this connection. The homotopy fibers
$$
  \mathbf{H}_{\mathrm{diff}}(X, \mathbf{B}^n U(1))
    \to 
  \mathbf{H}(X, \mathbf{B}^n U(1)) \stackrel{\mathrm{curv}}{\to} \mathbf{H}(X,\mathbf{\flat}_{\mathrm{\mathrm{dR}}}\mathbf{B}^{n+1} U(1))
$$
of this map therefore have the interpretation of being the cocycle $\infty$-groupoids of 
circle $n$-bundles with connection. This is the realization in 
$\mathrm{Smooth}\infty\mathrm{Grpd}$ of our general definition of 
ordinary differential cohomology  in an $\infty$-topos.

All these definitions make sense in full generality for any locally $\infty$-connected $\infty$-topos. 
We used nothing but the existence of the triple of adjoint $\infty$-functors 
$(\Pi \dashv \mathrm{Disc} \dashv \Gamma) : \mathbf{H} \to \infty \mathrm{Grpd}$. 
We shall show for the special case that $\mathbf{H} = \mathrm{Smooth}\infty\mathrm{Grpd}$ 
and $X$ an ordinary smooth manifold, that this general abstract definition reproduces 
ordinary differential cohomology over smooth manifolds as traditionally considered.

The advantage of the general abstract reformulation is that it generalizes the ordinary notion 
naturally to base objects that may be arbitrary smooth $\infty$-groupoids. 
This gives in particular the $\infty$-Chern-Weil homomorphism in an almost tautological form:

for $G \in \mathbf{H}$ any $\infty$-group object and $\mathbf{B}G \in \mathbf{H}$ its delooping, 
we may think of a morphism
$$
  \mathbf{c} : \mathbf{B}G \to \mathbf{B}^n U(1)
$$
as a representative of a characteristic class on $G$, in that this induces a morphism
$$
  [\mathbf{c}(-)] : H(X, G) \to H^n(X, U(1))
$$
from $G$-principal $\infty$-bundles to degree-$n$ cohomology-classes. Since the 
classification of $G$-principal $\infty$-bundles by cocycles is entirely general, 
we may equivalently think of this as the $\mathbf{B}^{n-1}U(1)$-principal $\infty$-bundle 
$P \to \mathbf{B}G$ given as the homotopy fiber of $\mathbf{c}$. A famous 
example is the Chern-Simons circle 3-bundle (bundle 2-gerbe) for $G$ a simply connected Lie group.

By postcomposing further with the canonical morphism 
$\mathrm{curv} : \mathbf{B}^n U(1) \to \mathbf{\flat}_{\mathrm{\mathrm{dR}}} \mathbf{B}^{n+1}U(1)$ 
this gives in total a \emph{differential characteristic class}
$$
  {\mathbf{c}}_{\mathrm{\mathrm{dR}}} : \mathbf{B}G \stackrel{\mathbf{c}}{\to} \mathbf{B}^n U(1)
    \stackrel{\mathrm{curv}}{\to}
   \mathbf{\flat}_{\mathrm{\mathrm{dR}}}\mathbf{B}^{n+1}U(1)
$$
that sends a $G$-principal $\infty$-bundle to a class in de Rham cohomology
$$
  [{\mathbf{c}}_{\mathrm{\mathrm{dR}}}] : H(X,G) \to H_{\mathrm{\mathrm{dR}}}^{n+1}(X)
  \,.
$$
This is the generalization of the plain Chern-Weil homomorphism.associated with 
the characteristic class $c$. 
In cases accessible by traditional theory, it is well known that this may be refined to what 
are called the assignment of \emph{secondary characteristic classes} to $G$-principal bundles 
with connection, taking values in ordinary differential cohomology
$$
  [\hat {\mathbf{c}}] : H_{\mathrm{conn}}(X,G) \to H_{\mathrm{diff}}^{n+1}(X)
  \,.
$$
We will discuss that in the general formulation this corresponds to finding 
objects $\mathbf{B}G_{\mathrm{conn}}$ that lift all curvature characteristic classes 
to their corresponding circle $n$-bundles with connection, in that it fits into the diagram
$$
  \xymatrix{
    \mathbf{H}(-,\mathbf{B}G_{\mathrm{conn}})
      \ar[r]
      \ar[d]
      &
     \prod_i \mathbf{H}_{\mathrm{diff}}(-, \mathbf{B}^{n_i} U(1))
      \ar[r]
      \ar[d]
      &
     \prod_i H_{\mathrm{dR}}^{n_i +1}(-)
     \ar[d]
     \\
     \mathbf{H}(-, \mathbf{B}G)
      \ar[r]
      &
      \prod_i
       \mathbf{H}(-, \mathbf{B}^{n_i}U(1))
      \ar[r]^{\mathrm{curv}}
      &
      \prod_i \mathbf{H}(-, \mathbf{\flat}_{\mathrm{\mathrm{dR}}}\mathbf{B}^{n_i+1}U(1))
  }
$$
The cocycles in $\mathbf{H}_{\mathrm{conn}}(X, \mathbf{B}G) := \mathbf{H}(X, \mathbf{B}G_{\mathrm{conn}})$ 
we may identify with $\infty$-connections on the underlying principal $\infty$-bundles. 
Specifically for $G$ an ordinary Lie group this captures the ordinary notion of 
connection on a bundle, for $G$ Lie 2-group it captures the notion of connection on a 2-bundle/gerbe.

\paragraph{Higher geometric prequantization}

\noindent {\bf Observation.} There is a canonical $\infty$-action 
$\gamma$ of
$\mathrm{Aut}_{\mathbf{H}_{/\mathbf{B}G}}(g)$ on the space of 
$\infty$-sections $\Gamma_X(P \times_G V)$.

\noindent{\bf Claim.} Since $\mathrm{Sh}_\infty(\mathrm{SmthMfd})$ is cohesive,
there is a notion of \emph{differential refinement} of the above discussion, yielding
\emph{connections} on $\infty$-bundles.

\noindent{\bf Example.} Let $\mathbb{C} \to \mathbb{C}/\!/U(1) \to \mathbf{B}U(1)$ be the 
canonical complex-linear circle action. Then \\
\vspace{-.7cm}
\begin{itemize}
 \item $g_{\mathrm{conn}} : X \to \mathbf{B}U(1)_{\mathrm{conn}}$
   classifies a circle bundle with connection, a \emph{prequantum line bundle} of its curvature 2-form;
 \item
   \vspace{-.3cm}
   $\Gamma_X(P \times_{U(1)} \mathbb{C})$ is the corresponding space of smooth sections;
 \item
   \vspace{-.3cm}
   $\gamma$ is the $\exp(\mbox{\small Poisson bracket})$-group action of preqantum operators,
    containing the Heisenberg group action.
\end{itemize}

\noindent{\bf Example.} Let $\mathbf{B} U \to \mathbf{B} \mathrm{PU} \to \mathbf{B}^2 U(1)$
be the canonical 2-circle action. Then \\
\vspace{-.7cm}
\begin{itemize}
 \item $g_{\mathrm{conn}} : X \to \mathbf{B}^2U(1)_{\mathrm{conn}}$
   classifies a circle 2-bundle with connection, a \emph{prequantum line 2-bundle} 
   of its curvature 3-form;
 \item
   \vspace{-.3cm}
   $\Gamma_X(P \times_{\mathbf{B}U(1)} \mathbf{B}U)$ is the corresponding groupoid of smooth sections = twisted bundles;
 \item
   \vspace{-.3cm}
   $\gamma$ is the $\exp(\mbox{\small 2-plectic bracket})$-2-group action of 2-plectic geometry,
   containing the \emph{Heisenberg 2-group} action.
\end{itemize}

\subsubsection{Characteristic classes}
\label{CharacteristicClassesInLowDegree}
\index{principal $\infty$-bundle!characteristic classes!in introduction}
\index{characteristic class!in introduction}

We discuss explicit presentations of \emph{characteristic classes}
of principal $n$-bundles for low values of $n$ and for low
degree of the characteristic class.

\begin{itemize}
  \item General concept
  \item Examples
\begin{itemize}
  \item example \ref{DeterminantLineBundle} -- First Chern class of unitary 1-bundles
  \item example \ref{Dixmier-Douady class} -- Dixmier-Douady class of circle 2-bundles (of bundle gerbes)
  \item example \ref{ClassOfCentralExtension} -- Obstruction class of central extension
  \item example \ref{FirstStiefelWhitneyClass} -- First Stiefel-Whitney class
    of an $\mathrm{O}$-principal bundle
  \item example \ref{SecondStiefelWhineyClass} -- Second Stiefel-Whitney class
    of an $\mathrm{SO}$-principal bundle
  \item example \ref{BocksteinHomomorphism} -- Bockstein homomorphism
  \item example \ref{ThirdIntegralStiefelWhitneyClass} -- Third integral Stiefel-Whitney class
  \item example \ref{PontryaginClassInIntro} -- First Pontryagin class of 
   Spin-1-bundles and twisted string-2-bundles
\end{itemize}
\end{itemize}

\medskip

In the context of higher (smooth) groupoids the notion of characteristic class is conceptually very simple: for $G$ some $n$-group and $\mathbf{B}G$ the corresponding one-object $n$-groupoid, a characteristic class of degree $k \in \mathbb{N}$ with coefficients in some abelian (Lie-)group $A$ is presented simply by a morphism
$$
  c : \mathbf{B}G \to \mathbf{B}^n A
$$
of cohesive $\infty$-groupoids. For instance if $A = \mathbb{Z}$ such a morphism represents a \emph{universal integral characteristic class} on $\mathbf{B}G$. Then for
$$
  g : X \to \mathbf{B}G
$$
any morphism of (smooth) $\infty$-groupoids that classifies a given $G$-principal $n$-bundle $P \to X$, as discussed above in \ref{SmoothPrincipalnBundles}, the corresponding characteristic class of $P$ (equivalently of $g$) is the class of the composite
$$
  c(P) : \xymatrix{
    X \ar[r]^g & \mathbf{B}G \ar[r]^{c} & \mathbf{B}^K A
  }
  \,,
$$
in the cohomology group $H^k(X,A)$ of the ambient $\infty$-topos.

In other words, in the abstract language of cohesive $\infty$-toposes the notion of characteristic classes of cohesive principal $\infty$-bundles is verbatim that of principal fibrations in ordinary homotopy theory. The crucial difference, though, is in the implementation of this abstract formalism.

Namely, as we have discussed previously, all the abstract morphisms $f : A \to B$ of cohesive $\infty$-groupoids here are presented  by \emph{$\infty$-anafunctors}, hence by spans of genuine morphisms of Kan-complex valued presheaves, whose left leg is a weak equivalence that exhibits a resolution of the source object. 

This means that the characteristic map itself is presented by a span
$$
  \xymatrix{
    \widehat {\mathbf{B}G}
	\ar[r]^c
	\ar[d]^{\simeq}
	&
	\mathbf{B}^k A
	\\
	\mathbf{B}G
  }
  \,,
$$
as is of course the cocycle for the principal $n$-bundle
$$
  \xymatrix{
     C(U_i)
	 \ar[r]^g
	 \ar[d]^{\simeq}
	 &
	 \mathbf{B}G
	 \\
	 X
  }
$$
and the characteristic class $[c(P)]$ of the corresponding principal $n$-bundle is presented by a (any) span composite
$$
  \xymatrix{
     C(T_i) \ar[r]^{\hat g}\ar[d]^{\simeq} & 
	 \widehat{\mathbf{B}G}
	 \ar[r]^c
	 \ar[d]^\simeq
	 &
	 \mathbf{B}^k A
     \\
     C(U_i)
	 \ar[r]^g
	 \ar[d]^{\simeq}
	 &
	 \mathbf{B}G
	 \\
	 X
  }
  \,,
$$
where $C(T_i)$ is, if necessary, a refinement of the cover $C(U_i)$ over which the $\mathbf{B}G$-cocycle $g$ lifts to a $\widehat {\mathbf{B}G}$-cocycle as indicated.

Notice the similarity of this situation to that of the discussion of
twisted bundles in example \ref{OrdinaryTwistedBundleExample}. This is not a coincidence: every characteristic class induces a corresponding notion of \emph{twisted $n$-bundles} and, conversely, every notion of twisted $n$-bundles can be understood as arising from the failure of a certain characteristic class to vanish. 

We discuss now a list of examples.
\begin{example}[first Chern class]
 \label{DeterminantLineBundle}
 \index{characteristic class!Chern class!first}
Let $N \in \mathbb{N}$. Consider the unitary group $U(n)$.
By its definition as a matrix Lie group, this comes canonically equipped with the determinant function
$$
  \mathrm{det} : U(n) \to U(1)
$$
and by the standard properties of the determinant, this is in fact a group homomorphism. Therefore this has a delooping to a morphism of Lie groupoids
$$
  \mathbf{B}\mathrm{det} : \mathbf{B}U(n) \to \mathbf{B}U(1)
  \,.
$$
Under geometric realization this maps to a morphism
$$
  |\mathbf{B} \mathrm{det}| : B U(n) \to B U(1) \simeq K(\mathbb{Z},2)
$$
of topological spaces. This is a characteristic class on the 
classifying space $B U(n)$: the ordinary \emph{first Chern class}.
Hence the morphism $\mathbf{B}\mathrm{det}$ on Lie groupoids
is a \emph{smooth refinement} of the ordinary first Chern class.

This smooth refinement acts on smooth $U(n)$-principal bundles as follows. Postcomposition of a {\v C}ech cocycle 
$$
  \xymatrix{
    P : & C(\{U_i\}) \ar[r]^{(g_{i j})} \ar[d]^{\simeq} & \mathbf{B} U(n)
    \\
    & X
  }
$$
for a $U(n)$-principal bundle on a smooth manifold $X$
with this characteristic class yields the cocycle 
$$
  \xymatrix{
    \mathrm{det} P : & C(\{U_i\})
	\ar[d]^\simeq	
	\ar[r]^{(g_{i j})} 
	\ar[r]^{\simeq} & \mathbf{B} U(n)
    \ar[r]^{\mathbf{B}\mathrm{det}} & \mathbf{B}U(1)
    \\
    & X
  }
$$
for a circle bundle (or its associated line bundle)
with transition functions $(\mathrm{det} (g_{i j}))$:
the \emph{determinant line bundle} of $P$. 

We may easily pass to the \emph{differential refinement} of the first
Chern class along similar lines. By prop. \ref{GroupoidOfLieAlgebraValuedForms}
the differential refinement $\mathbf{B} U(n)_{\mathrm{conn}} \to \mathbf{B} U(n)$
of the moduli stack of $U(n)$-principal bundles is given by the groupoid-valued
presheaf which over a test manifold $U$ assigns 
$$
  \mathbf{B} U(n)_{\mathrm{conn}}
   : 
  U \mapsto 
  \left\{
    A \stackrel{g}{\to}  A^g
	\;
	|\;
	A \in \Omega^1(U, \mathfrak{u}(n)); \,
	g \in C^\infty(U, U(n))
  \right\}
  \,.
$$
One checks that $\mathbf{B}\mathrm{det}$ uniquely extends to a 
morphism of groupoid-valued presheaves $\mathbf{B}\mathrm{det}_{\mathrm{conn}}$
$$
  \xymatrix{
    \mathbf{B}U(n)_{\mathrm{conn}}
	\ar[r]^{\mathbf{B}\mathrm{det}_{\mathrm{conn}}}
	\ar[d]
	&
	\mathbf{B}U(1)_{\mathrm{conn}}
	\ar[d]
	\\
    \mathbf{B}U(n)
	\ar[r]^{\mathbf{B}\mathrm{det}}
	&
	\mathbf{B}U(1)	
  }
$$
by sending $A \mapsto \mathrm{tr}(A)$. Here the trace operation on the matrix Lie 
algebra $\mathfrak{u}(n)$ is a unary \emph{invariant polynomial}
$\langle - \rangle : \mathfrak{u}(n) \to \mathfrak{u}(1) \simeq \mathbb{R}$.

Therefore, over a 1-dimensional compact manifold $\Sigma$ (a disjoint union of circles)
the canonical action functional, \ref{HolonomyInIntroduction}, induced by the first Chern class is
$$
  \exp(i S_{\mathbf{c}_1}) : 
  \xymatrix{
    \mathbf{H}(\Sigma, \mathbf{B}U(n)_{\mathrm{conn}})
	\ar[rrr]^{\mathbf{H}(\Sigma, \mathbf{B}\mathrm{det}_{\mathrm{conn}})}
	&&&
	\mathbf{H}(\Sigma, \mathbf{B}U(1)_{\mathrm{conn}})
	\ar[rr]^{\int_\Sigma}
	&&
	U(1)
  }
$$
sending
$$  
  A \mapsto \exp(i \int_\Sigma \mathrm{tr}(A))
  \,.
$$
This is the action functional of 1-dimensional $U(n)$-Chern-Simons theory,
discussed below in \ref{1dCSTheories}.

\medskip

It is a basic fact that the cohomology class of
line bundles can be identified within the second 
\emph{integral cohomology} of $X$. For our purposes here it
is instructive to rederive this fact in terms of anafunctors, 
\emph{lifting gerbes} and twisted bundles.

To that end, consider from example \ref{2GroupResolutionOfR}
the equivalence of the 2-group $(\mathbb{Z} \hookrightarrow \mathbb{R})$
with the ordinary circle group, which supports the 2-anafunctor
$$
  \xymatrix{
    \mathbf{B}(\mathbb{Z} \to \mathbb{R})
	\ar[r]^{c_1}
	\ar[d]^{\simeq}
	&
	\mathbf{B}(\mathbb{Z} \to 1)
	\ar@{=}[r]
	&
	\mathbf{B}^2 \mathbb{Z}
	\\
	\mathbf{B}U(1)
  }
  \,.
$$
We see now that this presents an integral characteristic class 
in degree 2 on $\mathbf{B}U(1)$. Given a cocycle 
$\{h_{i j} \in C^\infty(U_{i j}, U(1))\}$
for any circle bundle, the postcomposition with this 2-anafunctor 
amounts to the following:
\begin{enumerate}
 \item
   refine the cover, if necessary, to a \emph{good} open cover
   (where all non-empty $U_{i_0, \cdots, i_k}$ are contractible)
   -- we shall still write $\{U_i\}$ now for this good cover;
 \item
   choose on each $U_{i j}$ a (any) lift of the circle-valued 
   functor $h_{i j} : U_{i j} \to U(1)$ through the 
   quotient map $\mathbb{R} \to U(1)$ to a function
   $\hat h_{i j} : U_{i j} \to \mathbb{R}$ -- this is always
   possible over the contractible $U_{i j}$;
 \item
   compute the failures of the lifts thus chosen to constitute
   the cocycle for an $\mathbb{R}$-principal bundle: these are the elements
   $$
     \lambda_{i j k} := \hat h_{i k} \hat h_{i j}^{-1} \hat h_{j k}^{-1}
	  \in C^\infty(U_{i j k}, \mathbb{Z})
	  \,,
   $$
   which are indeed $\mathbb{Z}$-valued (hence constant) smooth functions
   due to the fact that the original $\{h_{i j}\}$ satisfied its cocycle law;
  \item
    notice that by observation \ref{CocycleForPrincipal2Bundles} 
	this yields the construction of the cocycle
	for a $(\mathbb{Z} \to \mathbb{R})$-principal 2-bundle
	$$
	  \{ \hat h_{i j} \in C^\infty(U_{i j}, \mathbb{R}),
          \lambda_{i j k} \in C^\infty(U_{i j k}, \mathbb{Z})
	 \}
	 \,,
	$$
	which by example \ref{OrdinaryTwistedBundleExample} we may also read as the
	cocycle for a twisted $\mathbb{R}$-1-bundle, with respect to the
	central extension $\mathbb{Z} \to \mathbb{R} \to U(1)$;
  \item
    finally project out the cocycle for the ``lifting $\mathbb{Z}$-gerbe''
	encoded by this, which is the $\mathbf{B}\mathbb{Z}$-principal 2-bundle given by the $\mathbf{B}\mathbb{Z}$ cocycle
	$$
	  \{ 
          \lambda_{i j k} \in C^\infty(U_{i j k}, \mathbb{Z})
	 \}
	 \,,
	$$	
\end{enumerate} 
This last cocycle is manifestly in degree-2 integral {\v C}ech cohomology, and
hence indeed represents a class in $H^2(X, \mathbb{Z})$. This is the
first Chern class of the circle bundle given by $\{h_{i j}\}$.
If here $h_{i j} = \mathrm{det} g_{i j}$ is the determinant circle bundle
of some unitary bundle, the this is also the first Chern class of that 
unitary bundle.
\end{example}
\begin{example}[Dixmier-Douady class]
 \label{Dixmier-Douady class}
 \index{characteristic class!Dixmier-Douady class}
 The discussion in example \ref{DeterminantLineBundle} of 
 the first Chern class of a circle 1-bundle has an immediate
 generalization to an analogous canonical class of 
 circle 2-bundles, def. \ref{Circle2BundleTotalSpace},
 hence, by observation \ref{BundleGerbeIsCircle2Bundle}, to bundle gerbes.
 As before, while this amounts to a standard and basic fact, 
 for our purposes it shall be instructive to spell this
 out in terms of $\infty$-anafunctors and twisted principal 2-bundles.
 
 To that end, notice that by delooping the equivalence
 $\mathbf{B}(\mathbb{Z} \to \mathbb{R}) \stackrel{\simeq}{\to} 
  \mathbf{B}U(1)$  yields
  $$
  \mathbf{B}^2(\mathbb{Z} \to \mathbb{R}) \stackrel{\simeq}{\to} 
  \mathbf{B}^2 U(1)
    \,.
  $$
  This says that $\mathbf{B}U(1)$-principal 2-bundles/bundle gerbes
  are equivalent to $\mathbf{B}(\mathbb{Z}\to \mathbb{R})$-principal 
  3-bundles, def. \ref{Circle3Bundle}.
  
  As before, this supports a canonical integral characteristic class,
  now in degree 3, presented by the $\infty$-anafunctor
  $$
    \xymatrix{
	  \mathbf{B}^2(\mathbb{Z} \to \mathbb{R})
	  \ar[r]
	  \ar[d]^{\simeq}
	  &
	  \mathbf{B}^2(\mathbb{Z} \to 1)
	  \ar@{=}[r]
	  &
	  \mathbf{B}(\mathbb{Z} \to 1 \to 1)
	  \\
	  \mathbf{B}^2 U(1)
	}
	\,.
  $$
  The corresponding class in $H^3(\mathbf{B}U(1), \mathbb{Z})$
  is the (smooth lift of) the \emph{universal Dixmier-Douady class}.
  
  Explicitly, for $\{g_{i j k} \in C^\infty(U_{i j k}, U(1))\}$
  the {\v C}ech cocycle for a circle-2-bundle, 
  def. \ref{Circle2BundleTotalSpace},
  this class is computed as the composite of spans
  $$
    \xymatrix{
	  C(U_i) \ar[d]^\simeq \ar[r]^{(\hat g, \lambda)}
	  &
	  \mathbf{B}^2(\mathbb{Z} \to \mathbb{R})
	  \ar[r]
	  \ar[d]^{\simeq}
	  &
	  \mathbf{B}^3 \mathbb{Z}
	  \\
	  C(U_i)
	  \ar[r]^{g}
	  \ar[d]^\simeq
	  &
	  \mathbf{B}^2 U(1)
	  \\
	  X
	}
	\,,
  $$
  where we assume for simplicity of notation that the cover $\{U_i \to X\}$
  already has be chosen (possibly after refining another cover)
  such that all patches and their non-empty intersections are 
  contractible.
   
  Here the lifted cocycle data $\{ \hat g_{i j k} : U_{i j k} \to U(1)$\} is through the quotient map $\mathbb{R} \to U(1)$
  to real valued functions. These lifts will, in general, not satisfy 
  the condition of a cocycle for a $\mathbf{B} \mathbb{R}$-principal 2-bundle.
  The failure is uniquely picked up by the functions
  $$
    \lambda_{i j k l} := \hat g_{j k l} g_{i j k}^{-1} g_{i j l} g_{i k l}^{-1}
	\in 
	C^\infty(U_{i j k l}, \mathbb{Z})
	\,.
  $$
  By example \ref{CocyclePrincipal3Bundle}
  this data constitutes the cocycle for a 
  $(\mathbb{Z} \to \mathbb{R} \to 1)$-principal 3-bundle
  or, by def. \ref{Twisted2Bundle} that of a
  \emph{twisted $\mathbf{B}R$-principal 2-bundle}.
  
  The above composite of spans projects out the integral cocycle
  $$
    \lambda_{i j k l} 
	\in 
	C^\infty(U_{i j k l}, \mathbb{Z})
	\,,
  $$
  which manifestly gives a class in $H^3(X, \mathbb{Z})$.
  This is the Dixmier-Douady class of the original circle 3-bundle,
  the higher analog of the Chern-class of a circle bundle.
\end{example}
\begin{example}[obstruction class of central extension]
  \label{ClassOfCentralExtension}
  \index{characteristic class!obstruction class of central extension}
  For $A \to \hat G \to G$ a central extension of Lie groups, 
  there is a long sequence of (deloopings of) Lie 2-groups
  $$
    \mathbf{B}A \to \mathbf{B}\hat G \to \mathbf{B}G
	\stackrel{\mathbf{c}}{\to} \mathbf{B}^2 A
	\,,
  $$
  where the characteristic class $\mathbf{c}$ is presented by the
  $\infty$-anafunctor
  $$
    \xymatrix{
	  \mathbf{B}(A \to \hat G)
	  \ar[r]
	  \ar[d]^{\simeq}
	  &
	  \mathbf{B}(A \to 1)
	  \ar@{=}[r]
	  &
	  \mathbf{B}^2 A
	  \\
	  \mathbf{B}G
	}
 $$
 with $(A \to \hat G)$ the crossed module from example 
 \ref{CrossedModuleFromNormalSubgroup}.
\end{example}
The proof of this is discussed below in 
prop. \ref{GroupCentralExtensionCohomology}.
\begin{example}[first Stiefel-Whitney class]
  \label{FirstStiefelWhitneyClass}
  \index{characteristic class!Stiefel-Whitney class!first}
  The morphism of groups
  $$
    \mathrm{O}(n) \to \mathbb{Z}_2
  $$
  which sends every element in the connected component of the
  unit element of $\mathrm{O}(n)$ to the unit element of
  $\mathbb{Z}_2$ and every other element to the non-trivial
  element of $\mathbb{Z}_2$ induces a morphism of 
  delooping Lie groupoids
  $$
    \mathbf{w}_1 : \mathbf{B}\mathrm{O}(n) 
	\to 
	\mathbf{B}\mathbb{Z}_2
	\,.
  $$
  This represents the universal smooth \emph{first Stiefel-Whitney class}.
\end{example}
The relation of $\mathbf{w}_1$ to orientation structure is 
discussed below in \ref{orientation structure}.
\begin{example}[second Stiefel-Whitney class]
  \label{SecondStiefelWhineyClass}
  \index{characteristic class!Stiefel-Whitney class!second}
  The exact sequence that characterizes the $\mathrm{Spin}$-group
  is
  $$
    \mathbb{Z}_2 \to \mathrm{Spin} \to \mathrm{SO}
  $$
  induces, by example \ref{ClassOfCentralExtension}, 
  a long fiber sequence
  $$
    \mathbf{B}\mathbb{Z}_2 
	  \to 
	\mathbf{B}\mathrm{Spin} 
	   \to 
	\mathbf{B}\mathrm{SO}
	  \stackrel{\mathbf{w}_2}{\to}
	\mathbf{B}^2 \mathbb{Z}_2
	\,.
  $$
  Here the morphism $\mathbf{w}_2$ is presented by the
  $\infty$-anafunctor
  $$
    \xymatrix{
	   \mathbf{B}(\mathbb{Z}_2 \to \mathrm{Spin})
	   \ar[r]
	   \ar[d]^{\simeq}
	   &
	   \mathbf{B}(\mathbb{Z}_2 \to 1)
	   \ar@{=}[r]
	   &
	   \mathbf{B}^2 \mathbb{Z}_2
	   \\
	   \mathbf{B}\mathrm{SO}
	}
	\,.
  $$
  This is a smooth incarnation of the
  \emph{universal second Stiefel-Whitney class}.
  The $\mathbf{B}\mathbb{Z}_2$-principal 2-bundle associated by
  $\mathbf{w}_2$ to any $\mathrm{SO}(n)$-principal bundles
  is dicussed in \cite{MurraySinger} in terms of the corresponding
  bundle gerbe, via. observation \ref{BundleGerbeIsCircle2Bundle}.
\end{example}
\begin{example}[Bockstein homomorphism]
  \label{BocksteinHomomorphism}
  \index{characteristic class!Bockstein homomorphism}
  The exact sequence
  $$
    \mathbb{Z} \stackrel{\cdot 2}{\to}
	\mathbb{Z} \to \mathbb{Z}_2
  $$
  induces, by example \ref{ClassOfCentralExtension},
  for each $n \in \mathbb{N}$
  a characteristic class
  $$
    \beta_2 : \mathbf{B}^n \mathbb{Z}_2
	\to \mathbf{B}^{n+1} \mathbb{Z}
	\,.
  $$
  This is the \emph{Bockstein homomorphism}.
  \end{example}
\begin{example}[third integral Stiefel-Whitney class]
  \index{characteristic class!Stiefel-Whitney class!third integral}
  \label{ThirdIntegralStiefelWhitneyClass}
  The composite of the second Stiefel-Whitney class
  from example \ref{SecondStiefelWhineyClass}
  with the Bockstein homomorphism from 
  example \ref{BocksteinHomomorphism} is the
  \emph{third integral Stiefel-Whitney class}
  $$
    W_3 : 
	\mathbf{B} \mathrm{SO}
	  \stackrel{\mathbf{w}_2}{\to}
	\mathbf{B}^2 \mathbb{Z}_2
      \stackrel{\beta_2}{\to}
    \mathbf{B}^3 \mathbb{Z}	
    \,.	
  $$
   This has a refined factorization through the
  universal Dixmier-Douady class from example \ref{Dixmier-Douady class}:
  $$
    \mathbf{W}_3 : \mathbf{B} \mathrm{SO} \to \mathbf{B}^2 U(1)
	\,.
  $$
  This is discussed in lemma \ref{SmoothLiftOfW3} below.
\end{example}

\begin{example}[first Pontryagin class]
  \label{PontryaginClassInIntro}
  \index{characteristic class!Pontryagin class!first}
  Let $G$ be a compact and simply connected simple Lie group.
  Then the resolution from example \ref{ResolutionForStringLift}
  naturally supports a characteristic class presented by
  the 3-anafunctor
  $$
    \xymatrix{
	  \mathbf{B}(U(1) \to \hat \Omega G \to P G)
	  \ar[r]
	  \ar[d]^{\simeq}
	  &
	  \mathbf{B}(U(1) \to 1 \to 1)
	  \ar@{=}[r]
	  &
	  \mathbf{B}^3 U(1)
	  \\
	  \mathbf{B}G
	}
	\,.
  $$  
  For $G = \mathrm{Spin}$ the spin group, this presents
  one half of the universal \emph{first Pontryagin class}.
  This we dicuss in detail in \ref{FractionalClasses}.
  
  Composition with this class sends $G$-principal bundles
  to circle 2-bundles, \ref{Circle2BundleTotalSpace}, 
  hence by \ref{Bundle2GerbeIsCircle3Bundle} to bundle 2-gerbes.
  Our discussion in \ref{FractionalClasses} shows that these
  are the \emph{Chern-Simons 2-gerbes}.

  The canonical action functional, \ref{HolonomyInIntroduction}, 
  induced by $\frac{1}{2}\mathbf{p}_1$
  over a compact 3-dimensional $\Sigma$
  $$
    \exp(i S_{\frac{1}{2}\mathbf{p}_1})
	: 
	\xymatrix{
  	  \mathbf{H}(\Sigma, \mathbf{B}\mathrm{Spin}_{\mathrm{conn}})
	  \ar[rrr]^{\mathbf{H}(\Sigma, \frac{1}{2}\hat{\mathbf{p}}_1)}
	  &&&
	  \mathbf{H}(\Sigma, \mathbf{B}^3 U(1)_{\mathrm{conn}})
	  \ar[rr]^{\int_\Sigma}
	  &&
	  U(1)
	}
  $$  
  is the action functional of ordinary 3-dimensional Chern-Simons theory,
  refined to the moduli stack of field configurations. This we discuss in 
  \ref{InfinCSOrdinaryCS}.
\end{example}

\newpage

\subsubsection{Lie algebras}
\label{LInfinityAlgebraicStructures}

A Lie algebra is, in a precise sense, the infinitesimal approximation to 
a Lie group. This statement generalizes to \emph{smooth $n$-groups}
(the strict case of which we had seen in definition \ref{CrossedComplex});
their infinitesimal approximation are \emph{Lie $n$-algebras}
which for arbitrary $n$ are known as \emph{$L_\infty$-algebras}.
The statement also generalizes to \emph{Lie groupoids} (discussed in 
\ref{SmoothPrincipalnBundles}); their infinitesimal approximation
are \emph{Lie algeboids}. Both these are special cases of a joint
generalization; where smooth $n$-groupoids have 
\emph{$L_\infty$-algebroids} as their infinitesimal approximation.

The following is an exposition of basic $L_\infty$-algebraic structures,
their relation to smooth $n$-groupoids and the notion of connection
data with coefficients in $L_\infty$-algebras.

\medskip
The following discussion proceeds by these topics: 
\begin{itemize}
  \item $L_\infty$-algebroids;
  \item Lie integration;
  \item Characteristic cocycles from Lie integration;
  \item $L_\infty$-algebra valued connections;
  \item Curvature characteristics and Chern-Simons forms;
  \item $\infty$-Connections from Lie integration;
\end{itemize}

\medskip

\paragraph{$L_\infty$-algebroids}

There is a precise sense in which one may think of a Lie algebra $\mathfrak{g}$ as the infinitesimal 
sub-object of the delooping groupoid $\mathbf{B}G$ of the corresponding Lie group $G$. Without here going into the details, which are discussed in detail below in
\ref{StrucSynthLie},
we want to build certain smooth $\infty$-groupoids from the knowledge of their infinitesimal subobjects: 
these subobjects are \emph{$L_\infty$-algebroids} and specifically \emph{$L_\infty$-algebras}. 

For $\mathfrak{g}$ an $\mathbb{N}$-graded vector space,
write $\mathfrak{g}[1]$ for the same underlying vector space
with all degrees shifted up by one. (Often this is denoted
$\mathfrak{g}[-1]$ instead). Then 
$$
  \wedge^\bullet \mathfrak{g} = \mathrm{Sym}^\bullet 
   (\mathfrak{g}[1])
$$
is the \emph{Grassmann algebra} on $\mathfrak{g}$; the 
free graded-commutative algebra on $\mathfrak{g}[1]$.
\begin{definition}
  \label{LInfinityAlgebra}
  \index{$L_\infty$-algebra}
  \index{Lie algebroid!$L_\infty$-algebra} 
  An \emph{$L_\infty$-algebra} structure on an $\mathbb{N}$-graded
  vector space $\mathfrak{g}$ is a family of multilinear maps
  $$
    [-,\cdots, -]_k : \mathrm{Sym}^k (\mathfrak{g}[1]) \to \mathfrak{g}[1]
  $$
  of degree -1, for all $k \in \mathbb{N}$, 
  such that the 
  \emph{higher Jacobi identities} 
  $$
    \sum_{k+l = n+1}
	\sum_{\sigma \in \mathrm{UnSh}(l,k-1)}
	(-1)^\sigma
	t_{a_1}, \cdots, t_{a_l}], t_{a_{l+1}}, \cdots, t_{a_{k+l-1}}]
	= 0
  $$
  are satisfied
  for all $n \in \mathbb{N}$ and all $\{t_{a_i} \in \mathfrak{g}\}$.
\end{definition}
See \cite{SSSI} for a review and for references.
\begin{example}
   \label{LieAlgebrasAsLInfinityAlgebra}
   \index{$L_\infty$-algebra!Lie algebra}
   If $\mathfrak{g}$ is concentrated in degree 0, then 
   an $L_\infty$-algebra structure on $\mathfrak{g}$ is the
   same as an ordinary Lie algebra structure. The only non-trivial
   bracket is $[-,-]_2 : \mathfrak{g} \otimes \mathfrak{g} \to \mathfrak{g}$
   and the higher Jacobi identities reduce to the ordinary Jacobi identity.
 \end{example}
We will see many other examples of $L_\infty$-algebras. For identifying
these, it turns out to be useful to have the following dual formulation
of $L_\infty$-algebras.
\begin{proposition}
  \label{LInfinityAlgebraFromDGAlgebra}
  \index{$L_\infty$-algebra!characterization by dual dg-algebra}
  Let $\mathfrak{g}$ be a $\mathbb{N}$-graded vector space
  that is degreewise finite dimensional. Write $\mathfrak{g}^*$ for the degreewise dual, also $\mathbb{N}$-graded.
  
  Then dg-algebra structures on the Grassmann algebra 
  $\wedge^\bullet \mathfrak{g}^* = \mathrm{Sym}^\bullet \mathfrak{g}[1]^*$
  are in canonical bijection with $L_\infty$-algebra
  structures on $\mathfrak{g}$, def. \ref{LInfinityAlgebra}.
\end{proposition}
Here the sum is over all $(l,k-1)$\emph{unshuffles}, which means all permutations $\sigma \in \Sigma_{k+l-1}$ that preserves the order within the first $l$ and within the last $k-1$ arguments, respectively, and 
$(-1)^{\mathrm{sgn}}$
is the Koszul-sign of the permutation: the sign picked up by
``unshuffling'' $t^{a_1} \wedge \cdots, \wedge t^{a_{k+l-1}}$ according to $\sigma$.\\
\proof
  Let $\{t_a\}$ be a basis of $\mathfrak{g}[1]$. Write $\{t^a\}$ for the dual basis of $\mathfrak{g}[1]^*$, where $t^a$ is 
  taken to be in the same degree as $t_a$.

  A derivation $d : \wedge^\bullet \mathfrak{g}^* \to \wedge^\bullet \mathfrak{g}^*$ of the Grassmann algebra is fixed by its value on generators, where it determines and is determined by a sequence of brackets graded-symmetric multilinear maps $\{[-,\cdots,-]_k\}_{k=1}^\infty$ by
  $$
    d : t^a \mapsto 
	-\sum_{k = 1}^\infty \frac{1}{k!}
	 [t_{a_1}, \cdots, t_{a_k}]^a
	 \,
	 t^{a_1}\wedge \cdots \wedge t^{a_k}
	 \,,
  $$
  where a sum over repeated indices is understood.
  This derivation is of degree +1 precisely if all the 
  $k$-ary maps are of degree -1.
  It is straightforward to check that the condition $d \circ d = 0$
  is equivalent to the higher Jacobi identities.
\endofproof
\begin{definition}
  \label{CEAlgebraInIntro}
  \index{Chevalley-Eilenberg algebra}
  \index{$L_\infty$-algebra!Chevalley-Eilenberg algebra}
  The dg-algebra corresponding to an $L_\infty$-algebra
  $\mathfrak{g}$ by prop. \ref{LInfinityAlgebraFromDGAlgebra}
  we call the \emph{Chevalley-Eilenberg algebra}
  $\mathrm{CE}(\mathfrak{g})$ of $\mathfrak{g}$.
\end{definition}
\begin{example}
  For $\mathfrak{g}$ an ordinary Lie algebra, as in example
  \ref{LieAlgebrasAsLInfinityAlgebra}, the notion of
  Chevalley-Eilenberg algebra from def. \ref{CEAlgebraInIntro}
  coincides with the traditional notion.
\end{example}
\begin{examples}
 \label{ExamplesForLInfinityAlgebras}
\begin{itemize}
\item
  A \emph{strict} $L_\infty$-algebra algebra is a 
  dg-Lie algebra $(\mathfrak{g}, \partial, [-,-])$ 
   with $(\mathfrak{g}^*, \partial^*)$ a cochain complex in non-negative degree. 
    With $\mathfrak{g}^*$ denoting the degreewise dual, the corresponding CE-algebra is 
   $CE(\mathfrak{g}) = (\wedge^\bullet \mathfrak{g}^*, d_{CE} = [-,-]^* + \partial^*$. 
\item We had already seen above the infinitesimal approximation of a Lie 2-group: 
  this is a Lie 2-algebra\index{$L_\infty$-algebra!Lie 2-algebra}. 
  If the Lie 2-group is a smooth strict 2-group it is encoded equivalently 
by a crossed module of ordinary Lie groups, and the corresponding Lie 2-algebra is given by a differential crossed module of ordinary Lie algebras.
\item For $n \in \mathbb{N}$, $n \geq 1$, the Lie $n$-algebra $b^{n-1}\mathbb{R}$ 
is the infinitesimal approximation to $\mathbf{B}^n U(\mathbb{R})$ and $\mathbf{B}^n \mathbb{R}$. 
Its CE-algebra is the dg-algebra on a single generators in degree $n$, with vanishing differential.
\item For any $\infty$-Lie algebra $\mathfrak{g}$ there is an $L_\infty$-algebra 
$\mathrm{inn}(\mathfrak{g})$\index{$L_\infty$-algebra!inner derivation $L_\infty$-algebra} 
defined by the fact that its CE-algebra is the Weil algebra of $\mathfrak{g}$:
  $$
    \mathrm{CE}(\mathrm{inn}(\mathfrak{g})) = 
    \mathrm{W}(\mathfrak{g}) = 
    (\wedge^\bullet (\mathfrak{g}^* \oplus \mathfrak{g}^*[1]), 
    d_{\mathrm{W}}|_{\mathfrak{g}^*} = d_{CE} + \sigma )
    \,,
  $$
  where $\sigma : \mathfrak{g}^* \to \mathfrak{g}^*[1]$ is the grading shift isomorphism, extended as a derivation.
 \end{itemize}
\end{examples}
\begin{example}
  \label{automorphismLInfinityAlg}
  \index{$L_\infty$-algebra!automorphism $L_\infty$-algebra}
  For $\mathfrak{g}$ an  $L_\infty$-algebra, its
  \emph{automorphism $L_\infty$-algebra} 
  $\mathfrak{der}(\mathfrak{g})$ is the dg-Lie algebra
  whose elements in degree $k$ are the derivations
  $$
    \iota : \mathrm{CE}(\mathfrak{g}) \to \mathrm{CE}(\mathfrak{g})
  $$
  of degree $-k$, whose differential is given by the graded commutator
  $[d_{\mathrm{CE}(\mathfrak{g})},-]$
  and whose Lie bracket is the commutator bracket of derivations.
\end{example}
In the context of rational homotopy theory, this is discussed on p. 312
of \cite{Sullivan}.

One advantage of describing an $L_\infty$-algebra in terms of 
its dual Chevalley-Eilenberg algebra is that in this form the 
correct notion of morphism is manifest.
\begin{definition}
 A morphism of $L_\infty$-algebras $\mathfrak{g} \to \mathfrak{h}$
 is a morphism of dg-algebras $\mathrm{CE}(\mathfrak{g}) \leftarrow
 \mathrm{CE}(\mathfrak{h})$.
 
 The category $L_\infty \mathrm{Alg}$ of $L_\infty$-algebras is therefore the 
 full subcatgeory of the opposite category of dg-algebras
 on those whose underlying graded algebra is free:
 $$
   L_\infty \mathrm{Alg} \stackrel{\mathrm{CE}(-)}{\to}
   \mathrm{dgAlg}_\mathbb{\mathbb{R}}^{\mathrm{op}}
   \,.
 $$
\end{definition}
Replacing in this characterization the 
ground field $\mathbb{R}$ by an algebra of smooth functions on a manifold $\mathfrak{a}_0$, we obtain the notion of an 
\emph{$L_\infty$-algebroid} $\mathfrak{g}$ over $\mathfrak{a}_0$. Morphisms $\mathfrak{a} \to \mathfrak{b}$ 
of such $\infty$-Lie algebroids are dually precisely morphisms of dg-algebras 
$\mathrm{CE}(\mathfrak{a}) \leftarrow \mathrm{CE}(\mathfrak{b})$.
\begin{definition}
\index{$L_\infty$-algebroid}
The category of \emph{$L_\infty$-algebroids}
is the opposite category of the full subcategory of 
$\mathrm{dgAlg}$ 
$$
  \infty\mathrm{LieAlgbd} \subset \mathrm{dgAlg}^{op}
$$
on graded-commutative cochain dg-algebras in non-negative degree whose underlying graded algebra is an 
exterior algebra over its degree-0 algebra, and this degree-0 algebra
is the algebra of smooth functions on a smooth manifold.
\end{definition}
\begin{remark}
  More precisely the above definition is that of 
  \emph{affine} $C^\infty$-$L_\infty$-algebroids. There are
  various ways to refine this to something more encompassing,
  but for the purposes of this introductory discussion the above
  is convenient and sufficient. A more comprehensive discussion
  is in \ref{StrucSynthLie} below.
\end{remark}
\begin{example}
\begin{itemize}
\item  The \emph{tangent Lie algebroid}
\index{$L_\infty$-algebroid!tangent Lie algebroid} $T X$ 
  of a smooth manifold $X$ is the infinitesimal approximation to its fundamental $\infty$-groupoid. Its CE-algebra is the de Rham complex 

  $\mathrm{CE}(T X) = \Omega^\bullet(X)$.
\end{itemize}
\end{example}

\medskip

\paragraph{Lie integration}

We discusss \emph{Lie integration}: a construction that sends an $L_\infty$-algebroid to a 
smooth $\infty$-groupoid of which it is the infinitesimal approximation.

The construction we want to describe may be understood as a generalization
of the following proposition. This is classical, even if maybe not reflected in the standard textbook literature to the extent it deserves to be.
\begin{definition}
For $\mathfrak{g}$ a (finite-dimensional) Lie algebra, 
let $\exp(\mathfrak{g}) \in [\mathrm{CartSp}^{\mathrm{op}}, \mathrm{sSet}]$ 
be the simplicial presheaf given by the assignment
$$
  \exp(\mathfrak{g})  : U
  \mapsto \mathrm{Hom}_{\mathrm{dgAlg}}(\mathrm{CE}(\mathfrak{g}), \Omega^\bullet(U \times \Delta^\bullet)_{\mathrm{\mathrm{vert}}})
  \,,
$$
in degree $k$ of dg-algebra homomorphisms from the Chevalley-Eilenberg algebra of 
$\mathfrak{g}$ to the dg-algebra of vertical differential forms with respect to the trivial bundle 
$U \times \Delta^k \to U$.
\end{definition}
Shortly we will be considering variations of such assignments that are best thought about when writing out the hom-sets on the right here as sets of arrows; as in
$$
  \exp(\mathfrak{g})
   :
   (U,[k])
    \mapsto
    \left\{ 
        \Omega^\bullet_{\mathrm{\mathrm{vert}}}(U \times \Delta^k) 
        \stackrel{A_{\mathrm{\mathrm{vert}}}}{\leftarrow}
        \mathrm{CE}(\mathfrak{g})
    \right\}
   ) 
  \,.
$$
For $\mathfrak{g}$ an ordinary Lie algebra it is an ancient and simple
but important  observation that dg-algebra morphisms  
$\Omega^\bullet(\Delta^k) \leftarrow \mathrm{CE}(\mathfrak{g})$ are in natural bijection with 
Lie-algebra valued 1-forms that are \emph{flat} in that their curvature 2-forms vanish: 
the 1-form itself determines precisely a morphism of the underlying graded algebras, and the respect for the differentials is exactly the flatness condition. It is this elementary but similarly important 
observation that historically led Eli Cartan to Cartan calculus and the algebraic formulation of Chern-Weil theory.

One finds that it makes good sense to generally, for $\mathfrak{g}$ any $\infty$-Lie algebra or even 
$\infty$-Lie algebroid, think of $\mathrm{Hom}_{dgAlg}(\mathrm{CE}(\mathfrak{g}), \Omega^\bullet(\Delta^k))$ 
as the set of $\infty$-Lie algebroid valued differential forms whose curvature forms 
(generally a whole tower of them) vanishes.

\begin{proposition}
Let $G$ be the simply-connected Lie group integrating $\mathfrak{g}$ according to Lie's three theorems 
and $\mathbf{B}G \in [\mathrm{CartSp}^{\mathrm{op}}, \mathrm{Grpd}]$ its delooping Lie groupoid 
regarded as a groupoid-valued presheaf on $\mathrm{CartSp}$. Write $\tau_1(-)$ for the 
truncation operation that 
quotients out 2-morphisms in a simplicial presheaf to obtain a presheaf of groupoids. 

We have an isomorphism
$$
  \mathbf{B}G  
  =
  \tau_1
  \exp(\mathfrak{g})
  \,.
$$
\end{proposition}
To see this, observe that the presheaf $\exp(\mathfrak{g})$ has as 
1-morphisms $U$-parameterized families of $\mathfrak{g}$-valued 1-forms $A_{\mathrm{\mathrm{vert}}}$ on the interval, 
and as 2-morphisms $U$-parameterized families of \emph{flat} 1-forms on the disk, interpolating between 
these. By identifying these 1-forms with the pullback of the Maurer-Cartan form on $G$, 
we may equivalently think of the 1-morphisms as based smooth paths in $G$ and 2-morphisms smooth 
homotopies relative endpoints between them. Since $G$ is simply-connected this means that after dividing out 2-morphisms only the endpoints of these paths remain, which identify with the points in $G$. 

The following proposition establishes the Lie integration of the shifted
1-dimensional abelian $L_\infty$-algebras $b^{n-1} \mathbb{R}$.

\begin{proposition}
\label{IntegrationOfBnR}
For $n \in \mathbb{N}$, $n \geq 1$. Write 
$$
  \mathbf{B}^n \mathbb{R}_{\mathrm{ch}} :=
   \Xi \mathbb{R}[n]
$$
for the simplicial presheaf on $\mathrm{CartSp}$ that is the image of the sheaf of 
chain complexes represented by $\mathbb{R}$ in degree $n$ and 0 in other degrees, 
under the Dold-Kan correspondence $\Xi : \mathrm{Ch}_\bullet^+ \to \mathrm{sAb} \to \mathrm{sSet}$.

Then there is a canonical morphism
$$
  \int_{\Delta^\bullet} : 
   \exp(b^{n-1}\mathbb{R})
   \stackrel{\simeq}{\to}
   \mathbf{B}^n \mathbb{R}_{ch}
$$
given by fiber integration of differential forms along $U \times \Delta^n \to U$ and 
this is an equivalence (a global equivalence in the model structure on simplicial presheaves).
\end{proposition}
The proof of this statement is discussed in \ref{SmoothStrucLieAlgebras}.

This statement will make an appearance repeatedly in the following discussion. 
Whenever we translate a construction given in terms $\exp(-)$ 
into a more convenient chain complex representation.

\medskip

\paragraph{Characteristic cocycles from Lie integration}

We now describe characteristic classes and curvature characteristic forms on $G$-bundles in terms of 
these simplicial presheaves.
For that purpose it is useful for a moment to ignore the truncation issue -- to come back to it later -- 
and consider these simplicial presheaves untruncated. 

To see characteristic classes in this picture, write $\mathrm{CE}(b^{n-1} \mathbb{R})$ 
for the commutative real dg-algebra on a single generator in degree $n$ with vanishing differential. 
As our notation suggests, this we may think as the Chevalley-Eilenberg algebra of a 
\emph{higher Lie algebra}  -- the $\infty$-Lie algebra $b^{n-1} \mathbb{R}$ -- 
which is an Eilenberg-MacLane object in the homotopy theory of $\infty$-Lie algebras, 
representing $\infty$-Lie algebra cohomology in degree $n$ with coefficients in $\mathbb{R}$.

Restating this in elementary terms, this just says that dg-algebra homomorphisms
$$
  \mathrm{CE}(\mathfrak{g}) \leftarrow \mathrm{CE}(b^{n-1}\mathbb{R}) : \mu
$$
are in natural bijection with elements $\mu \in \mathrm{CE}(\mathfrak{g})$ of degree $n$, that are closed, $d_{\mathrm{CE}(\mathfrak{g})} \mu = 0$. This is the classical description of a cocycle in 
the Lie algebra cohomology of $\mathfrak{g}$.
\begin{definition}
Every such $\infty$-Lie algebra cocycle $\mu$ induces a morphism of simplicial presheaves
$$
  \exp(\mu) : \exp(\mathfrak{g}) \to \exp(b^n \mathbb{R})
$$  
given by postcomposition
$$
  \Omega^\bullet_{\mathrm{\mathrm{vert}}}(U \times \Delta^l)
  \stackrel{A_{\mathrm{\mathrm{vert}}}}{\leftarrow}
  \mathrm{CE}(\mathfrak{g})
  \stackrel{\mu}{\leftarrow}
  \mathrm{CE}(b^n \mathbb{R})
  \,.
$$
\end{definition}
\begin{example}
Assume $\mathfrak{g}$ to be a semisimple Lie algebra, let $\langle -,-\rangle$ be the Killing 
form and $\mu = \langle -,[-,-]\rangle$ the corresponding 3-cocycle in Lie algebra cohomology. 
We may assume without restriction that this cocycle is normalized such that its left-invariant 
continuation to a 3-form on $G$ has integral periods. Observe that since $\pi_2(G)$ is trivial 
we have that the 3-coskeleton 
(see around def. \ref{coskeleton} for details on coskeleta) 
of $\exp(\mathfrak{g})$ is equivalent to $\mathbf{B}G$. 
By the inegrality of $\mu$, the operation of $\exp(\mu)$ on $\exp(\mathfrak{g})$ followed by 
integration over simplices descends to an $\infty$-anafunctor from $\mathbf{B}G$ to 
$\mathbf{B}^3 U(1)$, as indicated on the right of this diagram in 
$[\mathrm{CartSp}^{\mathrm{op}}, \mathrm{sSet}]$
$$
  \xymatrix{
     & 
     \exp(\mathfrak{g}) \ar[r]^{\exp(\mu)} \ar[d]
     &
     \exp(b^{n-1}\mathbb{R}) \ar[d]^{\int_{\Delta^\bullet}}
     \\
     C(V) \ar[r]^{\hat g} \ar[d]^\simeq
      & \mathbf{cosk}_3 \exp(\mathfrak{g})
     \ar[r]^{\int_{\Delta^\bullet}\mathbf{cosk}_3 \exp(\mu)}
     \ar[d]^\simeq
     &
     \mathbf{B}^3 \mathbb{R}/\mathbb{Z}
     \\
     C(U) \ar[d]^\simeq \ar[r]^{g}& \mathbf{B}G
     \\
     X
  }
  \,.
$$
Precomposing this -- as indicated on the left of the diagram -- with another $\infty$-anafunctor 
$X \stackrel{\simeq}{\leftarrow}C(U)\stackrel{g}{\to} \mathbf{B}G$ for a $G$-principal bundle, 
hence a collection of transition functions $\{g_{i j} : U_i \cap U_j \to G\}$ amounts to choosing (possibly on a refinement $V$ of the cover $U$ of $X$)
\begin{itemize}
\item 
  on each $V_i \cap V_j$ a lift $\hat g_{i j}$ of $g_{i j}$ to a familly of smooth based paths in $G$  -- $\hat g_{i j} : (V_i \cap V_j) \times \Delta^1 \to G$ -- with endpoints $g_{i j}$;

\item on each $V_i \cap V_j \cap V_k$ a smooth family  $\hat g_{i j k} : (V_i \cap V_j \cap V_k) \times \Delta^2 \to G$ of disks interpolating between these paths;

\item on each $V_i \cap V_j \cap V_k \cap V_l$ a a smooth family  $\hat g_{i j k l} : (V_i \cap V_j \cap V_k \cap V_l) \times \Delta^3 \to G$ of 3-balls interpolating between these disks.
\end{itemize}
On this data the morphism $\int_{\Delta^\bullet} \exp(\mu)$ acts by sending each 3-cell to the number
$$
  \hat g_{i j k l} \mapsto \int_{\Delta^3} \hat g_{i j k l}^* \mu
  \;\;  \mathrm{mod}\; \mathbb{Z}
  \,,
$$
where $\mu$ is regarded in this formula as a closed 3-form on $G$.
\end{example}
We say this is \emph{Lie integration of Lie algebra cocycles}.
\begin{proposition}
For $G = \mathrm{Spin}$, the {\v C}ech cohomology cocycle obtained this way is the first fractional 
\emph{Pontryagin class} of the $G$-bundle classified by $G$.
\end{proposition}
We shall show this below, as part of our $L_\infty$-algebraic reconstruction of the above 
motivating example. In order to do so, we now add differential refinement to this Lie integration 
of characteristic classes.

\medskip
\paragraph{$L_\infty$-algebra valued connections}

In \ref{SmoothPrincipalnBundles} we described ordinary connections on bundles as well as connections 
on 2-bundles in terms of parallel transport over paths and surfaces, and showed how such 
is equivalently given by cocycles with coefficients in Lie-algebra valued differential forms 
and Lie 2-algebra valued differential forms, respectively. 

Notably we saw for the case of ordinary $U(1)$-principal bundles, that the connection and curvature 
data on these is encoded in presheaves of diagrams that over a given test space $U \in \mathrm{CartSp}$  
look like
$$  
  \xymatrix{
    U \ar[r] \ar[d] & \mathbf{B}U(1) \ar[d] && \mbox{transition function}
    \\
    \mathbf{\Pi}(U) \ar[r] \ar[d] & \mathbf{B}\mathrm{INN}(U) \ar[d] && \mbox{connection}
    \\
    \mathbf{\Pi}(U) \ar[r] & \mathbf{B}^2 U(1) && \mbox{curvature}
  }
$$
together with a constraint on the bottom morphism.

It is in the form of such a kind of diagram that the general notion of connections on $\infty$-bundles 
may be modeled. In the full theory in \ref{GeneralAbstractTheory} this follows from first principles, 
but for our present introductory purpose we shall be content with taking this simple situation of 
$U(1)$-bundles together with the notion of Lie integration as sufficient motivation for the 
constructions considered now. 

So we pass now to what is to some extent the reverse construction of the one considered before: 
we define a notion of $L_\infty$-algebra valued differential forms and show how by a variant of 
Lie integration these integrate to coefficient objects for connections on $\infty$-bundles.

\medskip
\paragraph{Curvature characteristics and Chern-Simons forms}

For $G$ a Lie group, we have described above connections on $G$-principal bundles in 
terms of cocycles with coefficients in the Lie-groupoid of Lie-algebra valued forms 
$\mathbf{B}G_{\mathrm{conn}}$
$$
  \xymatrix{
  & \mathbf{B}G_{conn} \ar@{^{(}->}[d] && \mbox{connection}
  \\
  & \mathbf{B}G_{\mathrm{diff}} \ar[d]^{\simeq} && \mbox{pseudo-connection}
  \\
  C(U) 
    \ar[ur]|{\nabla_{\mathrm{ps}}}
    \ar[uur]^{\nabla}
    \ar[r]{g}
    \ar[d]^\simeq
     & \mathbf{B}G && \mbox{transition function}
   \\
   X
  }
  \,.
$$
In this context we had \emph{derived} Lie-algebra valued forms from the parallel transport description $\mathbf{B}G_{\mathrm{conn}} = [\mathbf{P}_1(-), \mathbf{B}G]$. We now turn this around and use 
Lie integration to construct parallel transport from Lie-algebra valued forms. 
The construction is such that it generalizes verbatim to $\infty$-Lie algebra valued forms.
For that purpose notice that another classical dg-algebra associated with $\mathfrak{g}$ is its 
\emph{Weil algebra} $\mathrm{W}(\mathfrak{g})$. 
\begin{proposition}
The Weil algebra $\mathrm{W}(\mathfrak{g})$ 
is the free dg-algebra on the graded vector space $\mathfrak{g}^*$, meaning that there is
a natural bijection
$$
  \mathrm{Hom}_{\mathrm{dgAlg}}(W(\mathfrak{g}), A) \simeq \mathrm{Hom}_{\mathrm{Vect}_{\mathbb{Z}}}(\mathfrak{g}^*, A)
  \,,
$$
which is singled out among the isomorphism class of dg-algebras with this property by the
fact that the projection of graded vector spaces $\mathfrak{g}^* \oplus \mathfrak{g}^*[1] \to \mathfrak{g}^*$
extends to a dg-algebra homomorphism 
$$
  \mathrm{CE}(\mathfrak{g}) \leftarrow W(\mathfrak{g}) : i^*
  \,.
$$
\end{proposition}
(Notice that general the dg-algebras that we are dealing with are \emph{semi-free} dg-algebras 
in that only their underlying graded algebra is free, but not the differential). 

The most obvious realization of the free dg-algebra on $\mathfrak{g}^*$ is $\wedge^\bullet (\mathfrak{g}^* \oplus \mathfrak{g}^*[1])$ equipped with the differential that is precisely the degree shift isomorphism $\sigma : \mathfrak{g}^* \to \mathfrak{g}^*[1]$ extended as a derivation. This is not the Weil algebra on the nose, 
but is of course isomorphic to it. The differential of the Weil algebra on $\wedge^\bullet (\mathfrak{g}^* \oplus \mathfrak{g}^*[1])$ is given on the unshifted generators by the sum of the 
$\mathrm{CE}$-differential with the shift isomorphism
$$
  d_{W(\mathfrak{g})}|_{\mathfrak{g}^*} = d_{\mathrm{CE}(\mathfrak{g})} + \sigma
  \,.
$$
This uniquely fixes the differential on the shifted generators -- 
a phenomenon known (at least after mapping this to differential forms, as we discuss below) 
as the \emph{Bianchi identity}\index{Bianchi identity}\index{curvature!Bianchi identity}.

Using this, we can express also the presheaf 
$\mathbf{B}G_{\mathrm{diff}}$ from above in diagrammatic fashion
\begin{observation}
For $G$ a simply connected Lie group, 
the presheaf $\mathbf{B}G_{\mathrm{diff}} \in [\mathrm{CartSp}^{\mathrm{op}}, \mathrm{Grpd}]$ is isomorphic to
$$
  \mathbf{B}G_{\mathrm{diff}}  =
  \tau_1
  \left(
  \exp(\mathfrak{g})_{\mathrm{diff}}
   :
   (U,[k]) \mapsto 
    \left\{ 
      \raisebox{20pt}{
      \xymatrix{
        \Omega^\bullet_{\mathrm{\mathrm{vert}}}(U \times \Delta^k) 
        \ar@{<-}[r]{A_{\mathrm{vert}}}&
        \mathrm{CE}(\mathfrak{g})
        \\
        \Omega^\bullet(U \times \Delta^k)
        \ar@{<-}[r]{A} \ar[r] \ar[u] &
        \mathrm{W}(\mathfrak{g}) \ar[u]
      }}
    \right\} 
    \right)
  \,
$$
where on the right we have the 1-truncation of the simplicial presheaf of diagrams as indicated, where the vertical morphisms are the canonical ones.
\end{observation}
Here over a given $U$ the bottom morphism in such a diagram is an arbitrary $\mathfrak{g}$-valued 1-form $A$ on 
$U \times \Delta^k$. This we can decompose as $A = A_U + A_{\mathrm{\mathrm{vert}}}$, where $A_U$ vanishes on 
tangents to $\Delta^k$ and $A_{\mathrm{vert}}$ on tangents to $U$. The commutativity of the diagram asserts that 
$A_{\mathrm{\mathrm{vert}}}$ has to be such that the  curvature 2-form $F_{A_{\mathrm{\mathrm{vert}}}}$ vanishes when 
both its arguments are tangent to $\Delta^k$.

On the other hand, there is in the above no further constraint on $A_U$. Accordingly, 
as we pass to the 1-truncation of $\exp(\mathfrak{g})_{\mathrm{diff}}$ we find that morphisms 
are of the form $(A_U)_1 \stackrel{g}{\to} (A_U)_2$ with $(A_U)^i$ arbitrary. This is the definition of $\mathbf{B}G_{\mathrm{diff}}$.

We see below that it is not a coincidence that this is reminiscent to the first condition on an Ehresmann connection on a $G$-principal bundle, which asserts that restricted to the fibers a connection 1-form on the total space 
of the bundle has to be flat. Indeed, the 
simplicial presheaf $\mathbf{B}G_{\mathrm{\mathrm{diff}}}$ may be thought of as 
the $\infty$-sheaf of pseudo-connections on \emph{trivial} $\infty$-bundles. Imposing on 
this also the second Ehresmann condition will force the pseudo-connection to be a genuine connection.

We now want to lift the above construction $\exp(\mu)$ of characteristic classes by 
Lie integration of Lie algebra cocycles $\mu$ from plain bundles classified by 
$\mathbf{B}G$ to bundles with (pseudo-)connection classified by 
$\mathbf{B}G_{\mathrm{diff}}$. 
By what we just said we therefore need to extend $\exp(\mu)$ from a map on just 
$\exp(\mathfrak{g})$ to a map on $\exp(\mathfrak{g})_{\mathrm{diff}}$.
This is evidently achieved by completing a square in $\mathrm{dgAlg}$ of the form
$$
  \xymatrix{
    \mathrm{CE}(\mathfrak{g}) \ar@{<-}[r]{\mu}& 
      \mathrm{CE}(b^{n-1} \mathbb{R})
    \\
    \mathrm{W}(\mathfrak{g}) \ar[u]\ar@{<-}[r]^{cs}& \mathrm{W}(b^{n-1} \mathbb{R})
      \ar[u]
  }
$$
and defining $\exp(\mu)_{\mathrm{diff}} : \exp(\mathfrak{g})_{\mathrm{diff}} 
\to \exp(b^{n-1}\mathbb{R})_{\mathrm{diff}}$ to be the operation of forming pasting composites with this.

Here $\mathrm{W}(b^{n-1}\mathbb{R})$ is the Weil algebra of the 
Lie $n$-algebra $b^{n-1} \mathbb{R}$. This is the dg-algebra on two generators 
$c$ and $k$, respectively, in degree $n$ and $(n+1)$ with the differential given by 
$d_{\mathrm{W}(b^{n-1} \mathbb{R})} : c \mapsto k$.
The commutativity of this diagram says that the bottom morphism takes the degree-$n$ generator $c$ 
to an element $\mathrm{cs} \in \mathrm{W}(\mathfrak{g})$ whose restriction to the unshifted generators 
is the given cocycle $\mu$. 

As we shall see below, any such choice $\mathrm{cs}$ will extend the characteristic cocycle 
obtained from $\exp(\mu)$ to a characteristic differential cocycle, exhibiting the 
$\infty$-Chern-Weil homomorphism. But only for special nice choices of $\mathrm{cs}$ will this 
take genuine $\infty$-connections to genuine $\infty$-connections -- instead of to pseudo-connections. 
As we discuss in the full $\infty$-Chern-Weil theory, this makes no difference in cohomology. 
But in practice it is useful to fine-tune the construction such as to produce nice models of the 
$\infty$-Chern-Weil homomorphism given by genuine $\infty$-connections. 
This is achieved by imposing the following additional constraint on the choice of extension 
$\mathrm{cs}$ of $\mu$: 
\begin{definition}
\label{transgression}
For $\mu \in \mathrm{CE}(\mathfrak{g})$ a cocycle and $\mathrm{cs} \in \mathrm{W}(\mathfrak{g})$ 
a lift of $\mu$ through $\mathrm{W}(\mathfrak{g}) \leftarrow \mathrm{CE}(\mathfrak{g})$, 
we say that $d_{\mathrm{W}(\mathfrak{g})}$ is an invariant polynomial 
\emph{in transgression} with $\mu$ if $d_{\mathrm{W}(\mathfrak{g})}$ 
sits entirely in the shifted generators, 
in that $d_{\mathrm{W}(\mathfrak{g})} \in \wedge^\bullet \mathfrak{g}^*[1] \hookrightarrow W(\mathfrak{g})$.
\end{definition}
\begin{definition}
Write $\mathrm{inv}(\mathfrak{g}) \subset \mathrm{W}(\mathfrak{g})$ 
(or $\mathrm{W}(\mathfrak{g})_{\mathrm{basic}}$) for the sub-dg-algebra on invariant polynomials.
\end{definition}
\begin{observation}
We have $\mathrm{W}(b^{n-1}\mathbb{R}) \simeq \mathrm{CE}(b^n \mathbb{R})$.
\end{observation}
Using this, we can now encode the two conditions on the extension 
$\mathrm{cs}$ of the cocycle $\mu$ as the commutativity of this double square diagram
$$
  \xymatrix{
    \mathrm{CE}(\mathfrak{g}) \ar@{<-}[r]^{\mu}& \mathrm{CE}(b^{n-1} \mathbb{R})
    && \mbox{cocycle}
    \\
    \mathrm{W}(\mathfrak{g}) \ar@{<-}[r]^{\mathrm{cs}} \ar[u] 
     & 
     \mathrm{W}(b^{n-1} \mathbb{R})
     \ar[u]
    && \mbox{Chern-Simons element}
    \\
    \mathrm{inv}(\mathfrak{g})
   \ar[r]^{\langle -\rangle} \ar[u]&
    \mathrm{inv}(b^{n-1} \mathbb{R}) 
    \ar[u]
    && 
    \mbox{invariant polynomial}
  }
  \,.
$$
\begin{definition}
In such a diagram, we call $\mathrm{cs}$ the 
\emph{Chern-Simons element}
\index{Chern-Simons element}
\index{Chern-Simons functionals!Chern-Simons element}
 that exhibits the transgression between $\mu$ and $\langle - \rangle$. 
\end{definition}
We shall see below that under the $\infty$-Chern-Weil homomorphism, 
Chern-Simons elements give rise to the familiar Chern-Simons forms -- 
as well as their generalizations -- as local connection data of secondary characteristic classes 
realized as circle $n$n-bundles with connection. 
\begin{observation}
What this diagram encodes is the construction of the connecting homomorphism for the long exact sequence 
in cohomology that is induced from the short exact sequence
$$
  \mathrm{ker}(i^*) \to \mathrm{W}(\mathfrak{g}) \to \mathrm{CE}(\mathfrak{g})
$$
subject to the extra constraint of basic elements.
$$
  \xymatrix{
     & \langle - \rangle \ar@{<-|}[r]& \langle - \rangle
     \\
     \mu \ar@{<-|}[r]& \mathrm{cs} \ar@{|->}[u]^{d_{W}}
     \\
     \mathrm{CE}(\mathfrak{g}) \ar@{<-}[r]^{i^*} & \mathrm{W}(\mathfrak{g})
     \ar@{<-}[r] &
     \mathrm{inv}(\mathfrak{g})
  }
  \,.
$$
\end{observation}
To appreciate the construction so far, recall the following classical fact
\begin{fact}
For $G$ a compact Lie group, the rationalization $B G \otimes k$ 
of the classifying space $B G$ is the rational space whose Sullivan model 
is given by the algebra $\mathrm{inv}(\mathfrak{g})$ of invariant polynomials on the Lie algebra 
$\mathfrak{g}$.
\end{fact} 
So we have obtained the following picture:
$$
  \xymatrix@C=4pt{
    \mbox{delooped $\infty$-group}
    &&
    \mathbf{B}G 
      \ar[d]
     & 
     \mathfrak{g} 
      \ar[d] 
     & 
    \mathrm{CE}(\mathfrak{g})  && \mbox{Chevalley-Eilenberg algebra}
    \\
    \mbox{\begin{tabular}{c}delooped groupal \\universal $\infty$-bundle\end{tabular}}
    &&
    \mathbf{B E}G 
      \ar[d] 
     & 
     \mathrm{inn}(\mathfrak{g}) 
        \ar[d]
     &
     \mathrm{W}(\mathfrak{g}) = \mathrm{CE}(\mathrm{inn}(\mathfrak{g})) \ar[u] 
      && \mbox{Weil algebra}
    \\
    \mbox{\begin{tabular}{c}rationalized \\ classifying space\end{tabular}}
    &&
    \prod_i \mathbf{B}^{n_i} \mathbb{R} 
    &
    \prod_i b^{n_i-1} \mathbb{R}
    & 
    \mathrm{inv}(\mathfrak{g}) 
     \ar[u]
    &&
    \mbox{\begin{tabular}{c}algebra of \\ invariant polynomials\end{tabular}} 
    \\
    &&
    \ar@{<-}[r]^{\mbox{Lie integration}}& 
  }
$$
\begin{example}
For $\mathfrak{g}$ a semisimple Lie algebra, $\langle -,-\rangle$ 
the Killing form invariant polynomial, there is a Chern-Simons element 
$\mathrm{cs} \in \mathrm{W}(\mathfrak{g})$ witnessing the transgression to the 
cocycle $\mu = - \frac{1}{6} \langle -,[-,-] \rangle$. 
Under a $\mathfrak{g}$-valued form $\Omega^\bullet(X) \leftarrow W(\mathfrak{g}) : A$ 
this maps to the ordinary degree 3 Chern-Simons form
  $$ 
    \mathrm{cs}(A) = \langle A \wedge d A\rangle + \frac{1}{3} \langle A \wedge [A \wedge A]\rangle
    \,.
  $$
\end{example}

\medskip
\paragraph{$\infty$-Connections from Lie integration}
 \label{Infinity-Connections}
 \index{connection!from Lie integration}

For $\mathfrak{g}$ an $L_\infty$-algebroid 
we have seen above the object
$\exp(\mathfrak{g})_{\mathrm{diff}}$ that represents pseudo-connections
on $\exp(\mathfrak{g})$-principal $\infty$-bundles and serves to support the $\infty$-Chern-Weil homomorphism. 
We now discuss the genuine $\infty$-connections among these pseudo-connections. 
A derivation from first principles of the following construction is
given below in \ref{SmoothStrucInfChernWeil}.

The construction is due to \cite{SSSIII} and \cite{FSS}.

\medskip

\begin{definition}
  \label{gValuedFormsAndCurvature}
  \label{definition.a-valued-differential-form}
  \index{$L_\infty$-algebroid!valued forms}
  \index{differential form!with values in $L_\infty$-algebroid}
Let $X$ be a smooth manifold and $\mathfrak{g}$ an $L_\infty$-algebra algebra or more generally 
an $L_\infty$-algebroid.

An \emph{$L_\infty$-algebroid valued differential form} 
  on $X$ is a morphism of dg-algebras
$$
  \Omega^\bullet(X)
  \leftarrow
  \mathrm{W}(\mathfrak{g})
  : 
  A
$$
from the Weil algebra of $\mathfrak{g}$ to the de Rham complex of $X
$. Dually this is a morphism of $L_\infty$-algebroids
$$
  A : T X \to \mathrm{inn}(\mathfrak{g})
$$
from the inner automorphism $\infty$-Lie algebra.

Its \emph{curvature} is the composite of morphisms of graded vector spaces
$$
  \Omega^\bullet(X) \stackrel{A}{\leftarrow}
  \mathrm{W}(\mathfrak{g})
  \stackrel{F_{(-)}}{\leftarrow}
  \mathfrak{g}^*[1]
  : 
  F_{A}
  \,.
$$
Precisely if the curvatures vanish does the morphism factor through the Chevalley-Eilenberg algebra 
$$
  (F_A = 0) 
  \;\;\Leftrightarrow
  \;\;
  \left(
  \raisebox{20pt}{
  \xymatrix{
     & \mathrm{CE}(\mathfrak{g}) \ar[dl]_{\exists A_{\mathrm{flat}}}
     \\
     \Omega^\bullet(X) \ar@{<-}[r]^{A}& W(\mathfrak{g}) \ar[u]     
  }}
  \right)
$$
in which case we call $A$ \emph{flat}.
\end{definition}
\begin{remark}
  \label{ImagesOfdAnddWUnderForms}
   For $\{x^a\}$ a coordinate chart of 
   an $L_\infty$-algebroid $\mathfrak{a}$
  and
  $$
    A^a := A(x^a) \in \Omega^{\mathrm{deg}(x^a)}(X)
  $$
  the differential form assigned to the generator $x^a$ by 
  the $\mathfrak{a}$-valued form $A$, we have the 
  curvature components
  $$
    F_A^a = A(\mathbf{d}x^a) \in \Omega^{\mathrm{deg}(x^a)+1}(X)
    \,.
  $$
  Since $d_{\mathrm{W}}=d_{\mathrm{CE}}+\mathbf{d}$, this can be equivalently written as
  $$
    F_A^a = A(d_{\mathrm{W}}x^a-d_{\mathrm{CE}}x^a)
    \,,
  $$
  so the \emph{curvature} of $A$ precisely measures the ``lack of flatness'' of $A$.
  Also notice that, since $A$ is required to be 
  a dg-algebra homomorphism, we have
  $$
    A(d_{\mathrm{W}(\mathfrak{a})} x^a) = d_{\mathrm{dR}} A^a
    \,,
  $$
   so that
  $$
    A(d_{\mathrm{CE}(\mathfrak{a})} x^a) = d_{\mathrm{dR}} A^a- F_A^a
    \,.
  $$ 
 \end{remark}

Assume now $A$ is a degree 1 $\mathfrak{a}$-valued differential form on the smooth manifold $X$, and that $\mathrm{cs}$ is a Chern-Simons element transgressing an invariant polynomial $\langle-\rangle$ of $\mathfrak{a}$ to some cocycle $\mu$, by def. \ref{transgression}. 
We can then consider the image $A(\mathrm{cs})$ of the Chern-Simons element $\mathrm{cs}$ in $\Omega^\bullet(X)$. Equivalently, we can look at $\mathrm{cs}$ as a map from degree 1 $\mathfrak{a}$-valued differential forms on $X$ to ordinary (real valued) differential forms on $X$.
\begin{definition}
\label{definition.chern-simons-form} 
In the notations above, we write 
$$
  \xymatrix{
    \Omega^\bullet(X)
     \ar@{<-}[r]^{A} &
    \mathrm{W}(\mathfrak{a})
     \ar@{<-}[r]^{\mathrm{cs}} &
    \mathrm{W}(b^{n+1}\mathbb{R})
  }
  :
  \mathrm{cs}(A)
$$ 
for the differential form associated by the Chern-Simons element $\mathrm{cs}$ to the degree 1 $\mathfrak{a}$-valued differential form 
$A$, 
and call this the \emph{Chern-Simons differential form} associated with $A$. 

Similarly, for $\langle -\rangle$ an invariant polynomial on 
$\mathfrak{a}$, we write $\langle F_A \rangle$
for the evaluation
$$
  \xymatrix{
    \Omega^\bullet_{\mathrm{closed}}(X)
     \ar@{<-}[r]^{A} &
    \mathrm{W}(\mathfrak{a})
     \ar@{<-}[r]^{\langle -\rangle} &
    \mathrm{inv}(b^{n+1}\mathbb{R})
  }
  :
  \langle F_A \rangle
  \,.
$$ 
We call this the \emph{curvature characteristic forms}
of $A$.
\end{definition}
\begin{definition}
 \label{InfinityConnectionSimplicialObject}
\index{$L_\infty$-algebra!$L_\infty$-algebra valued forms}
\index{connection!from Lie integration}
For $U$ a smooth manifold, the 
\emph{$\infty$-groupoid of $\mathfrak{g}$-valued forms}
is the Kan complex 
$$
  \exp(\mathfrak{g})_{\mathrm{conn}}(U)
  :
  [k]
  \mapsto
  \left\{
     \Omega^\bullet(U \times \Delta^k)
      \stackrel{A}{\leftarrow}
     \mathrm{W}(\mathfrak{g})
     \;\;
     |
     \;\;
       \forall v \in \Gamma(T \Delta^k) : \iota_v F_A = 0
  \right\}
$$
whose $k$-morphisms are $\mathfrak{g}$-valued forms $A$ on $U \times \Delta^k$ 
with sitting instants, and with the property that their curvature vanishes on vertical vectors.

The canonical morphism
$$
  \exp(\mathfrak{g})_{\mathrm{conn}} \to \exp(\mathfrak{g})
$$
to the untruncated Lie integration of $\mathfrak{g}$ is given by restriction of $A$ 
to vertical differential forms (see below).
\end{definition}
Here we are thinking of $U \times \Delta^k \to U$ as a trivial bundle.

The \emph{first} Ehresmann condition can be identified with the conditions on lifts 
$\nabla$ in $\infty$-anafunctors
$$
  \xymatrix{
    & \exp(\mathfrak{g})_{\mathrm{conn}} \ar[d]
    \\
    C(U) \ar[d]^\simeq \ar[ur]^\nabla \ar[r]^{g} & \exp(\mathfrak{g})
    \\
    X
  }
$$
that define connections on $\infty$-bundles.

\medskip

\subparagraph{Curvature characteristics}
\index{connection!curvature characteristic}

\begin{proposition}
For $A \in \exp(\mathfrak{g})_{\mathrm{conn}}(U,[k])$ 
a $\mathfrak{g}$-valued form on $U \times \Delta^k$ and 
for $\langle - \rangle \in \mathrm{W}(\mathfrak{g})$ 
any invariant polynomial, the corresponding curvature characteristic 
form $\langle F_A \rangle \in \Omega^\bullet(U \times \Delta^k)$ descends down to $U$.
\end{proposition}
To see this, it is sufficient to show that for all $v \in \Gamma(T \Delta^k)$ we have
\begin{enumerate}
\item $\iota_v \langle F_A \rangle = 0$;
\item $\mathcal{L}_v \langle F_A \rangle = 0$.
\end{enumerate}
The first condition is evidently satisfied if already $\iota_v F_A = 0$. The second condition follows 
with Cartan calculus and using that $d_{\mathrm{dR}} \langle F_A\rangle = 0$:
$$
  \mathcal{L}_v \langle F_A \rangle = 
  d \iota_v \langle F_A \rangle
  + 
  \iota_v d \langle F_A \rangle
  = 0
  \,.
$$

Notice that for a general $\infty$-Lie algebra $\mathfrak{g}$ the curvature forms $F_A$ 
themselves are not generally closed (rather they satisfy the more Bianchi identity), 
hence requiring them to have no component along the simplex does not imply that they descend. 
This is different for abelian $\infty$-Lie algebras: for them the curvature forms 
themselves are already closed, and hence are themselves already curvature characteristics that do descent.

It is useful to organize the $\mathfrak{g}$-valued form $A$, together with its restriction 
$A_{\mathrm{\mathrm{vert}}}$ to vertical differential forms and with its curvature characteristic forms in the commuting diagram
$$
    \xymatrix{
      \Omega^\bullet(U \times \Delta^k)_{\mathrm{\mathrm{vert}}}
      \ar@{<-}[r]^{A_{\mathrm{\mathrm{vert}}}}&
      \mathrm{CE}(\mathfrak{g})
      &&
      \mbox{gauge transformation}
      \\
      \Omega^\bullet(U \times \Delta^k)
      \ar@{<-}[r]^{A}
      \ar[u]
      &
      \mathrm{W}(\mathfrak{g})
      \ar[u]
      &&
      \mbox{$\mathfrak{g}$-valued form}
      \\
      \Omega^\bullet(U)
      \ar@{<-}[r]^{\langle F_A\rangle}
      \ar[u]
      &
      \mathrm{inv}(\mathfrak{g}) \ar[u]
      &&
      \mbox{curvature characteristic forms}
    }
$$
in $\mathrm{dgAlg}$.
The commutativity of this diagram is implied by $\iota_v F_A = 0$.
\begin{definition}
Write $\exp(\mathfrak{g})_{\mathrm{CW}}(U)$ for 
the $\infty$-groupoid of $\mathfrak{g}$-valued forms fitting into such diagrams.
$$
  \exp(\mathfrak{g})_{CW}(U)
  :
  [k]
  \mapsto
  \left\{
    \raisebox{40pt}{
    \xymatrix{
      \Omega^\bullet(U \times \Delta^k)_{\mathrm{\mathrm{vert}}}
      \ar@{<-}[r]^{A_{\mathrm{\mathrm{vert}}}}&
      \mathrm{CE}(\mathfrak{g})
      \\
      \Omega^\bullet(U \times \Delta^k)
      \ar@{<-}[r]^{A}
      \ar[u]
      &
      \mathrm{W}(\mathfrak{g})
      \ar[u]
      \\
      \Omega^\bullet(U)
      \ar@{<-}[r]^{\langle F_A\rangle}
      \ar[u]
      &
      \mathrm{inv}(\mathfrak{g}) \ar[u]
    }
    }
  \right\}
  \,.
$$
\end{definition}
We call this the coefficient for $\mathfrak{g}$-valued \emph{$\infty$-connections}
\index{connection!simplicial presentation}

\subparagraph{1-Morphisms: integration of infinitesimal gauge transformations}
\index{connection!gauge transformation}

The 1-morphisms in $\exp(\mathfrak{g})(U)$ may be thought of as 
\emph{gauge transformations} between $\mathfrak{g}$-valued forms. We unwind what these look like concretely.
\begin{definition}
 \label{InfinitesimalGaugeTransfrmation}
 \index{gauge theory!infinitesimal gauge transformation}
Given a 1-morphism in $\exp(\mathfrak{g})(X)$, represented by $\mathfrak{g}$-valued forms
$$
  \Omega^\bullet(U \times \Delta^1) 
  \leftarrow
  \mathrm{W}(\mathfrak{g})
  : 
  A
$$
consider the unique decomposition
$$
  A = A_U + ( A_{\mathrm{\mathrm{vert}}} := \lambda \wedge d t)  \; \; 
  \,,
$$
with $A_U$ the horizonal differential form component and $t : \Delta^1 = [0,1] \to \mathbb{R}$ the canonical coordinate.

We call $\lambda$ the \emph{gauge parameter}. 
This is a function on $\Delta^1$ with values in 0-forms on $U$ for $\mathfrak{g}$ an ordinary Lie algebra, plus 1-forms on $U$ for $\mathfrak{g}$ a Lie 2-algebra, plus 2-forms for a Lie 3-algebra, and so forth.
\end{definition}
We describe now how this encodes a gauge transformation
$$
  A_0(s=0) \stackrel{\lambda}{\to} A_U(s = 1)
  \,.
$$
\begin{observation}
By the nature of the Weil algebra we have
$$
  \frac{d}{d s} A_U
  =
  d_U \lambda + [\lambda \wedge A] + [\lambda \wedge A \wedge A] + \cdots
  + 
  \iota_s F_A
  \,,
$$
where the sum is over all higher brackets of the $\infty$-Lie algebra $\mathfrak{g}$.
\end{observation}
In the Cartan calculus for the case that $\mathfrak{g}$ an ordinary one 
writes the corresponding  \emph{second Ehremsnn condition} $\iota_{\partial_s} F_A = 0$ 
equivalently
$$
  \mathcal{L}_{\partial_s} A = \mathrm{ad}_\lambda A
  \,.
$$
\begin{definition}
Define the \emph{covariant derivative of the gauge parameter} to be
$$
  \nabla \lambda := d \lambda + [A \wedge \lambda] + [A \wedge A \wedge \lambda] + \cdots
  \,.
$$
\end{definition}
\begin{remark}
  \label{HorizontalityAndGaugeTransformation}
In this notation we have
\begin{itemize}
\item
  the general identity 
  \[
    \frac{d}{d s} A_U = \nabla \lambda + (F_A)_s
    \label{ShiftedGaugeTrafo}
  \]

\item 
  the \emph{horizontality} constraint or 
  \emph{second Ehresmann condition} $\iota_{\partial_s} F_A = 0$, the differential equation
  \[
    \frac{d}{d s} A_U = \nabla \lambda
    \label{GaugeTrafo}
    \,.
  \]
\end{itemize}
This is known as the equation for \emph{infinitesimal gauge transformations} of an $\infty$-Lie algebra 
valued form. 
\end{remark}
\begin{observation}
By Lie integration we have that $A_{\mathrm{\mathrm{vert}}}$ -- and hence $\lambda$ -- 
defines an element $\exp(\lambda)$ in the $\infty$-Lie group that integrates $\mathfrak{g}$. 

The unique solution $A_U(s = 1)$ of the above differential equation at $s = 1$ 
for the initial values $A_U(s = 0)$ we may think of as the result of acting on $A_U(0)$ 
with the gauge transformation $\exp(\lambda)$. 
\end{observation}

\newpage

\paragraph{Examples of $\infty$-connections}
\index{connection!examples}

We discuss some examples of $\infty$-groupoids of $\infty$-connections
obtained by Lie integration, as discussed in \ref{Infinity-Connections} above.
\begin{itemize}
  \item
    \ref{OrdinaryConnectionsFromLieIntegration} --
   	Connections on ordinary principal bundles
  \item 
    \ref{String2ConnectionsFromLieIntegration} -- $\mathfrak{string}$-2-connections
\end{itemize}

\subparagraph{Connections on ordinary principal bundles}
\label{OrdinaryConnectionsFromLieIntegration}
\index{connection!ordinary principal connection}

Let $\mathfrak{g}$ be an ordinary Lie algebra and write $G$ 
for the simply connected Lie group integrating it. 
Write $\mathbf{B}G_{\mathrm{conn}}$ the groupoid of Lie algebra-valued forms 
from prop. \ref{GroupoidOfLieAlgebraValuedForms}.
\begin{proposition}
 \label{OrdinaryBGconn}
The 1-truncation of the object $\exp(\mathfrak{g})_{\mathrm{conn}}$
from def. \ref{InfinityConnectionSimplicialObject} is
equivalent to the coefficient object for $G$-principal connections 
from prop. \ref{GroupoidOfLieAlgebraValuedForms}.
We have an equivalence
$$
  \tau_1 \exp(\mathfrak{g})_{\mathrm{conn}} = \mathbf{B}G_{\mathrm{conn}}
$$
\end{proposition}
\proof
To see this, first note that the sheaves of objects on both sides are manifestly isomorphic, both are the sheaf of $\Omega^1(-,\mathfrak{g})$. For morphisms, observe that for a form $\Omega^\bullet(U \times \Delta^1) \leftarrow \mathrm{W}(\mathfrak{g}) : A$ which we may decompose into a horizontal and a 
verical piece as $A = A_U + \lambda \wedge d t$ the condition $\iota_{\partial_t} F_A = 0$ is equivalent to the differential equation
$$
  \frac{\partial}{\partial t} A
  =
  d_U \lambda + [\lambda, A]
  \,.
$$
For any initial value $A(0)$ this has the unique solution
$$
  A(t) = g(t)^{-1} (A + d_{U}) g(t)
  \,,
$$
where $g : [0,1] \to G$ is the parallel transport of $\lambda$:
$$
  \begin{aligned}
    &
    \frac{\partial}{\partial t}
    \left(
       g_(t)^{-1} (A + d_{U}) g(t)
     \right)
     \\
     = &
     g(t)^{-1} (A + d_{U}) \lambda g(t)
     -
     g(t)^{-1} \lambda (A + d_{U}) g(t)    
   \end{aligned}
$$
(where for ease of notation we write actions as if $G$ were a matrix Lie group).

In particular this implies that the endpoints of the path of $\mathfrak{g}$-valued 1-forms are related by the usual cocycle condition in $\mathbf{B}G_{conn}$
$$
  A(1) = g(1)^{-1} (A + d_U) g(1)
  \,.
$$
In the same fashion one sees that given 2-cell in $\exp(\mathfrak{g})(U)$ and any 1-form on $U$ at one vertex, there is a unique lift to a 2-cell in $\exp(\mathfrak{g})_{conn}$, obtained by parallel transporting the form around. The claim then follows from the previous statement of Lie integration that $\tau_1 \exp(\mathfrak{g}) = \mathbf{B}G$.
\endofproof

\subparagraph{$\mathfrak{string}$-2-connections}
\label{String2ConnectionsFromLieIntegration}
\index{connection!string 2-connection}

We discuss the $\mathfrak{string}$ Lie 2-algebra and local differential
form data for $\mathfrak{string}$-2-connections.
A detailed discussion of the corresponding 
$\mathrm{String}$-principal 2-bundles is below in 
\ref{String2Group}, more discussion of the 2-connections and their
twisted generalization is in \ref{HigherSpinStructure}.

\medskip

Let $\mathfrak{g}$ be a semisimple Lie algebra.
Write $\langle -,-\rangle : \mathfrak{g}^{\otimes 2} \to \mathbb{R}$
for its Killing form and 
$$
  \mu = \langle -, [-,-]\rangle : \mathfrak{g}^{\otimes 3} \to \mathbb{R}
$$
for the canonical 3-cocycle.

We discuss two very different looking, but nevertheless equivalent
Lie 2-algebras.
\begin{definition}[skeletal version of $\mathfrak{string}$]
  \label{SkeletalStringLie2Algebra}
  Write $\mathfrak{g}_\mu$ for the Lie 2-algebra
  whose underlying graded vector space is
  $$
    \mathfrak{g}_\mu = \mathfrak{g} \oplus \mathbb{R}[-1]
	\,,
  $$
  and whose nonvanishing brackets are defined as follows.
  \begin{itemize}
    \item 
	  The binary bracket is that of $\mathfrak{g}$ when both 
	  arguments are from $\mathfrak{g}$ and 0 otherwise.
	\item
	  The trinary bracket is the 3-cocycle
	  $$
	    [-,-,-]_{\mathfrak{g}_\mu}
		:=
		\langle -, [-,-]\rangle
		: 
		\mathfrak{g}^{\otimes 3} \to \mathbb{R}
		\,.
	  $$
  \end{itemize}
\end{definition}
\begin{definition}[strict version of $\mathfrak{string}$]
  \label{StrictStringLie2Algebra}
  Write $(\hat \Omega \mathfrak{g} \to P_* \mathfrak{g})$
  for the Lie 2-algebra 
  coming from the differential crossed module,
  def. \ref{DifferentialCrossedModule},
  whose underlying vector space is
  $$
    (\hat \Omega \mathfrak{g} \to P \mathfrak{g})
	= 
	P_* \mathfrak{g} \oplus (\Omega \mathfrak{g} \oplus \mathbb{R})[-1]
	\,,
  $$
  where $P_* \mathfrak{g}$ is the vector space of smooth 
  maps $\gamma : [0,1] \to \mathfrak{g}$ such that $\gamma(0) = 0$,
  and where $\Omega \mathfrak{g}$ is the subspace for which also
  $\gamma(1) = 0$, and whose non-vanishing brackets are defined as follows
  \begin{itemize}
    \item 
	  $[-]_1 = \partial := \Omega \mathfrak{g} \oplus \mathbb{R} \to 
	 \Omega \mathfrak{g} \hookrightarrow P_* \mathfrak{g}$;
	\item
	  $[-,-] : P_* \mathfrak{g} \otimes P_* \mathfrak{g} \to 
	   P_* \mathfrak{g}$
	   is given by the pointwise Lie bracket on $\mathfrak{g}$ as
	   $$
	     [\gamma_1, \gamma_2] = (\sigma \mapsto [\gamma_1(\sigma), \gamma_2(\sigma)])
		 \,;
	   $$
	\item 
	  $[-,-] : P_* \mathfrak{g} \otimes (\Omega \mathfrak{g} \oplus \mathbb{R}) \to 
	   \Omega \mathfrak{g} \oplus \mathbb{R}$
	   is given by pairs
	   \(
	     \label{CocycleInStrictStringLie2Algebra}
	     [\gamma, (\ell, c)]
		 :=
		 \left(
		   [\gamma,\ell],
		   \;
		   2 \int_0^1 \langle \gamma(\sigma), \frac{d \ell}{d \sigma}(\sigma)\rangle
		   d \sigma
		 \right)
		 \,,
	   \)
		 where the first term is again pointwise the Lie bracket in 
		 $\mathfrak{g}$.
  \end{itemize}
\end{definition}
\begin{proposition}
   \label{EquivalenceOfTheTwoStringLie2Algebras}
  The linear map
  $$
    P_* \mathfrak{g} \oplus (\Omega \mathfrak{g} \oplus \mathbb{R})[-1]
	\to 
	\mathfrak{g} \oplus \mathbb{R}[-1]
	\,,
  $$
  which in degree 0 is evaluation at the endpoint
  $$
    \gamma \mapsto \gamma(1)
  $$
  and which in degree 1 is projection onto the $\mathbb{R}$-summand,
  induces a weak equivalence of $L_\infty$algebras
  $$
    \mathfrak{string}
	\simeq
    (\hat \Omega \mathfrak{g} \to P_* \mathfrak{g})
	\simeq
	\mathfrak{g}_\mu
  $$
\end{proposition}
\proof
  This is theorem 30 in \cite{BCSS}.
\endofproof
\begin{definition}
  We write $\mathfrak{string}$ for the \emph{string Lie 2-algebra}
  if we do not mean to specify a specific presentation such as
  $\mathrm{so}_\mu$ or $(\hat \Omega \mathfrak{so} \to P_* \mathfrak{so})$. 
  
  In more technical language we would say that 
  $\mathfrak{string}$ is defined to be the homotopy fiber of
  the morphism of $L_\infty$-algebras
  $\mu_3 : \mathfrak{so} \to b^2 \mathbb{R}$,
  well defined up to weak equivalence.
  \index{string Lie 2-algebra}
\end{definition}
\begin{remark}
  \label{TwoPointsOfViewOnStringConnectionData}
  Proposition \ref{EquivalenceOfTheTwoStringLie2Algebras}
  says that the two Lie 2-algebras 
  $(\hat \Omega \mathfrak{g} \to P_* \mathfrak{g})$ and
  $\mathfrak{g}_{\mu}$, which look quite different, are actually 
  equivalent. Therefore also the local data for a 
  $\mathrm{String}$-2 connection can take two very different
  looking but nevertheless equivalent forms.

  Let $U$ be a smooth manifold. The data of
  $(\hat \Omega \mathfrak{g} \to P_* \mathfrak{g})$-valued
  forms on $X$ is a triple
  \begin{enumerate}
    \item $A \in \Omega^1(U, P \mathfrak{g})$;
	\item $B \in \Omega^2(U, \Omega \mathfrak{g})$;
	\item $\hat B \in \Omega^2(U, \mathbb{R})$
	\,.
  \end{enumerate}
  consisting of a 1-form with values in the path Lie algebra of 
  $\mathfrak{g}$, a 2-form with values in the loop Lie algebra
  of $\mathfrak{g}$, and an ordinary real-valued 2-form that 
  contains the central part of $\hat \Omega \mathfrak{g}
  = \Omega \mathfrak{g} \oplus \mathbb{R}$.
  The curvature data of this is
  \begin{enumerate}
    \item $F = d A + \frac{1}{2}[A \wedge A] + B \in \Omega^2(U, P\mathfrak{g})$;
	\item $H = d(B + \hat B) + [A \wedge (B + \hat B)]
	  \in \Omega^3(U, \Omega \mathfrak{g} \oplus \mathbb{R})$,
	  \,,
  \end{enumerate}
  where in the last term we have the bracket from 
  (\ref{CocycleInStrictStringLie2Algebra}).
  Notice that if we choose a basis $\{t_a\}$ of $\mathfrak{g}$
  such that we have structure constant $[t_b, t_c] = f^a{}_{b c} t_a$,
  then for instance the first equation is
  $$
    F^a(\sigma) = d A^a(\sigma) +  \frac{1}{2}f^a{}_{b c} A^b(\sigma) 
	\wedge A^c(\sigma) 
    + B^a(\sigma)
	\,.
  $$

  On the other hand, the data of 
  forms  on $U$ is a tuple
  \begin{enumerate}
    \item $A \in \Omega^1(U, \mathfrak{g})$;
	  \item $\hat B \in \Omega^2(U, \mathbb{R})$,
  \end{enumerate}
  consisting of a $\mathfrak{g}$-valued form  and a
  real-valued 2-form. The curvature data of this is
  \begin{enumerate}
    \item $F = d A + [A \wedge A] \in \Omega^2(\mathfrak{g})$;
	\item $H = d \hat B +  \langle A \wedge [A \wedge A]\rangle \in 
	  \Omega^3(U)$.
  \end{enumerate}

  While these two sets of data look very different,   
  proposition \ref{EquivalenceOfTheTwoStringLie2Algebras}
  implies that under their respective higher gauge transformations
  they are in fact equivalent. 
  
  Notice that in the first case the 2-form is valued in a 
  nonabelian Lie algebra, whereas in the second case the 2-form is
  abelian, but, to compensate this, a trilinear term appears in the
  formula for the curvatures. By the discussion in 
  section \ref{Infinity-Connections} this means that 
  a $\mathfrak{g}_\mu$-2-connection
  looks simpler on a single patch than an 
  $(\hat \Omega \mathfrak{g} \to P_* \mathfrak{g})$-2-connection,
  it has relatively more complicated behavious on double intersections.  
  
  Moreover, notice that in the second case we see that one part of 
  Chern-Simons term for $A$ occurs, namely 
  $\langle A \wedge [A \wedge A]\rangle$ . The rest of the
  Chern-Simons term appears in this local formula after passing to 
  yet another equivalent version of $\mathfrak{string}$, one which is
  well-adapted to the discussion of twisted $\mathrm{String}$ 2-connections.
  This we discuss in the next section.
\end{remark}
The equivalence of the skeletal and the strict presentation for
$\mathfrak{string}$ corresponds under Lie integration to 
two different but equivalent models of the smooth 
$\mathrm{String}$-2-group.
\begin{proposition}
  The degeewise Lie integration of 
  $\hat \Omega \mathfrak{so} \to P_* \mathfrak{so}$
  yields the strict Lie 2-group
  $(\hat \Omega \mathrm{Spin} \to P_* \mathrm{Spin})$,
  where $\hat \Omega \mathrm{Spin}$ is the level-1 Kac-Moody
  central extension of the smooth loop group of $\mathrm{Spin}$.
\end{proposition}
\proof
  The nontrivial part to check is that the action of 
  $P_* \mathfrak{so}$ on $\hat \Omega \mathfrak{so}$
  lifts to a compatible action of $P_* \mathrm{Spin}$
  on $\hat \Omega \mathrm{Spin}$. This is shown in 
  \cite{BCSS}.
\endofproof
  Below in \ref{String2Group} we show that 
  there is 
  an equivalence of smooth $n$-stacks
  $$
    \mathbf{B}(\hat \Omega \mathrm{Spin} \to P_* \mathrm{Spin})
	\simeq
	\tau_2 \exp(\mathfrak{g}_\mu)
	\,.
  $$

\subsubsection{The Chern-Weil homomorphism }
\label{InfinityChernWeilHomomorphismIntroduction}
\index{Chern-Weil homomorphism!overview}

We now come to the discussion the Chern-Weil homomorphism and its generalization to the $\infty$-Chern-Weil homomorphism.

We have seen in \ref{SmoothPrincipalnBundles} $G$-principal $\infty$-bundles for general smooth 
$\infty$-groups $G$ and in particular for abelian groups $G$. Naturally, the abelian case is easier 
and more powerful statements are known about this case. A general strategy for studying nonabelian 
$\infty$-bundles therefore is to \emph{approximate} them by abelian bundles. This is achieved 
by considering characteristic classes. Roughly, a characteristic class is a map that 
functorially sends $G$-principal $\infty$-bundles to $\mathbf{B}^n K$-principal $\infty$-bundles, 
for some $n$ and some abelian group $K$. In some cases such an assignment may be 
obtained by integration of infinitesimal data. If so, then the assignment refines to one 
of $\infty$-bundles with connection. For $G$ an ordinary Lie group this is then what is called 
the \emph{Chern-Weil homomorphism}. For general $G$ we call it  the \emph{$\infty$-Chern-Weil homomorphism}.

The material of this section is due to \cite{SSSI} and \cite{FSS}.

\medskip
\paragraph{Motivating examples}
\label{ChernWeilMotivatingExamples}

A simple motivating example for characteristic classes and the Chern-Weil homomorphism
is the construction of determinant line bundles 
from example \ref{DeterminantLineBundle}.
This construction directly extends to the case where the bundles carry 
connections. We give an exposition of this
\emph{differential refinement} of the \emph{universal first Chern class},
example \ref{DeterminantLineBundle}. A more formal discussion of this
situation is below in \ref{Twistedc1Structures}.

\medskip

For $N \in \mathbb{N}$ We may canonically identify the Lie algebra 
$\mathfrak{u}(N)$ with the
matrix Lie algebra of skew-hermitian matrices on which we have the
trace operation
$$
  \mathrm{tr} : \mathfrak{u}(N) \to \mathfrak{u}(1) = i \mathbb{R}
  \,.
$$
This is the differential version of the determinant in that when regarding the Lie algebra 
as the infinitesimal neighbourhood of the neutral element in $U(N)$ the determinant becomes 
the trace under the exponential map
$$
  \mathrm{det} (1 + \epsilon A) = 1 + \epsilon \mathrm{tr}(A)
$$
for $\epsilon^2 = 0$.
It follows that for $\mathrm{tra}_\nabla : \mathbf{P}_1(U_i) \to \mathbf{B}U(N)$ 
the parallel transport of a connection on $P$ locally given by a 1-forms 
$A \in \Omega^1(U_i, \mathfrak{u}(N))$ by
$$
  \mathrm{tra}_\nabla(\gamma) = \mathcal{P} \exp \int_{[0,1]} \gamma^* A
$$
the determinant parallel transport
$$
  \mathrm{det} (\mathrm{tra}_\nabla= : \mathbf{P}_1(U_i) \stackrel{\mathrm{tra}_\nabla}{\to}
   \mathbf{B} U(N) \stackrel{\mathrm{det}}{\to} \mathbf{B}U(1)
$$
is locally given by the formula
$$
  \mathrm{det} (\mathrm{tra}_\nabla(\gamma)) = \mathcal{P} \exp \int_{[0,1]} \gamma^* \mathrm{tr} A
  \,,
$$
which means that the local connection forms on the determinant line bundle are obtained from those of the unitary bundle by tracing.
$$
  (\mathrm{det},\mathrm{tr}) : \{(g_{i j}), (A_i)\} \mapsto
   \{(\mathrm{det}  g_{i j}), (\mathrm{tr} A_i)\}
  \,.
$$
This construction extends to a functor
$$
  (\hat {\mathbf{c}}_1) := (\mathrm{det}, \mathrm{tr}) :  
   U(N) \mathrm{Bund}_{\mathrm{conn}}(X) \to U(1) \mathrm{Bund}_{\mathrm{conn}}(X)
$$
natural in $X$, that sends $U(n)$-principal bundles with connection to 
circle bundles with connection, hence to cocycles in degree-2 ordinary differential cohomology.

This assignment remembers of a unitary bundle one inegral class and its differential refinement:
\begin{itemize}
\item the integral class of the determinant bundle is the 
  first Chern class the $U(N)$-principal bundle
  $$
    [{\hat {\mathbf{c}}}_1(P)] = c_1(P)
    \,;
  $$

\item the curvature 2-form of its connection is a representative
  in de Rham cohomology of this class
  $$
    [F_{\nabla_{{\hat {\mathbf{c}}}_1(P)}}] = c_1(P)_{\mathrm{\mathrm{dR}}}
    \,.
  $$
\end{itemize}
$$
  \xymatrix{
    & H_{\mathrm{diff}}^2(X)
    \ar[dl] \ar[dr]
    \\
    H^2(X,\mathbb{Z}) && \Omega^2_{\mathrm{\mathrm{cl}}}(X)
  }
  \;\;\;\;\;
  \xymatrix{
    & {\hat {\mathbf{c}}}_1(P)
    \ar@{|->}[dl] \ar@{|->}[dr]
    \\
    c_1(P) && \mathrm{tr}(F_\nabla)
  }  
$$
Equivalently this assignment is given by postcomposition of cocycles with a morphism 
of smooth $\infty$-groupoids
$$
  \hat {\mathbf{c}}_1 : \mathbf{B}U(N)_{\mathrm{conn}} \to \mathbf{B}U(1)_{\mathrm{conn}}
  \,.
$$
We say that $\hat {\mathbf{c}}_1$ is a \emph{differential characteristic class}, 
the differential refinement of the first Chern class.

In \cite{brylinski-mclaughlin}
an algorithm is given for contructing
differential characteristic classes on {\v C}ech cocycles in this fashion for 
more general Lie algebra cocycles.
For instance these authors give the following construction for the diffrential refinement of the 
first Pontryagin class \cite{bm:pont}.

Let $N \in \mathbb{N}$, write $\mathrm{Spin}(N)$ for the Spin group
and consider the canonical Lie algebra cohomology 3-cocycle
$$
  \mu = \langle -,[-,-]\rangle : \mathfrak{so}(N) \to \mathbf{b}^2 \mathbb{R}
$$
on semisimple Lie algebras, where $\langle -,- \rangle$ is the Killing form invariant polynomial.
Let $(P \to X, \nabla)$ be a $\mathrm{Spin}(N)$-principal bundle with connection. 
Let $A \in \Omega^1(P, \mathfrak{so}(N))$ be the Ehresmann connection 1-form on the total 
space of the bundle.

Then construct a {\v C}ech cocycle for Deligne cohomology in degree 4 as follows:
\begin{enumerate}
\item pick an open cover $\{U_i \to X\}$ such that there is a choice of 
   local sections $\sigma_i : U_i \to P$. Write 
   $$
     (g_{i j}, A_i) := (\sigma_i^{-1} \sigma_j, \sigma_i^* A)
   $$
   for the induced {\v C}ech cocycle.

\item Choose a lift of this cocycle to an assignment 
  \begin{itemize}
   \item of based paths in $\mathrm{Spin}(N)$ to double intersections 
      $$
       \hat g_{i j} : U_{i j}\times \Delta^1 \to Spin(N)
       \,,
     $$
     with
     $\hat g_{i j}(0) = e$ and $\hat g_{i j}(1) = g_{i j}$;
   \item of based 2-simplices between these paths to triple intersections     
     $$
       \hat g_{i j k} : U_{i j k}\times \Delta^2 \to \mathrm{Spin}(N)
       \,;
     $$  
     restricting to these paths in the obvious way;

   \item similarly of based 3-simplices between these paths to quadruple intersections     
     $$
       \hat g_{i j k l} : U_{i j k l}\times \Delta^3 \to \mathrm{Spin}(N)
       \,.
     $$
  \end{itemize}
  Such lifts always exists, because the Spin group is connected 
  (because already $\mathrm{SO}(N)$ is), simply connected 
  (because $\mathrm{Spin}(N)$ is the universal 
  cover of $\mathrm{SO}(N)$) and also has $\pi_2(\mathrm{Spin}(N)) = 0$ 
  (because this is the case for every 
  compact Lie group).
\item
 Define from this a Deligne-cochain by setting
   $$
     \frac{1}{2}{\hat {\mathbf{p}}}_1(P)
     :=
     (g_{i j k l}, A_{i j k}, B_{i j}, C_{i})
     :=
     \left(
      \begin{array}{l}
       \int_{\Delta^3} (\sigma_i \cdot\hat g_{i j k l})^* \mu(A) mod \mathbb{Z},
        \\ 
       \int_{\Delta^2} (\sigma_i\cdot \hat g_{i j k})^* \mathrm{cs}(A),
        \\
       \int_{\Delta^1} (\sigma_i \cdot \hat g_{i j})^* \mathrm{cs}(A),
        \\
       \sigma_i^* \mu(A)
      \end{array}
     \right)
     \,,
   $$
   where $\mathrm{cs}(A) = \langle A \wedge F_A\rangle + c \langle A \wedge [A \wedge A]\rangle  $ 
   is the Chern-Simons form of the connection form $A$ with respect to the cocyle 
   $\mu(A) = \langle A \wedge [A \wedge A]\rangle$.
\end{enumerate}
They then prove:
\begin{enumerate}
\item This is indeed a Deligne cohomology cocycle;

\item it represents the differential refinement of the first 
   fractional Pontryagin class of $P$.
\end{enumerate}
$$
  \xymatrix{
    & H_{\mathrm{diff}}^4(X)
    \ar[dl] \ar[dr]
    \\
    H^4(X,\mathbb{Z}) && \Omega^4_{\mathrm{\mathrm{cl}}}(X)
  }
  \;\;\;\;
  \xymatrix{
    & \frac{1}{2}{\hat {\mathbf{p}}}_1(P)
    \ar@{|->}[dl] \ar@{|->}[dr]
    \\
    \frac{1}{2}p_1(P) && d \mathrm{cs}(A)
  }  
  \,.
$$
In the form in which we have (re)stated this result here the second statement amounts, 
in view of the first statement, to the observation that the curvature 4-form
\index{curvature!curvature 4-form} 
of the Deligne cocycle is proportional to
$$
  d \mathrm{cs}(A) \propto \langle F_A \wedge F_A \rangle \in \Omega^4_{\mathrm{\mathrm{cl}}}(X)
$$
which represents the first Pontryagin class in de Rham cohomology. 
Therefore the key observation is that we have a Deligne cocycle at all. 
This can be checked directly, if somewhat tediously, by hand. 

But then the question remains: 
where does this successful \emph{Ansatz} come from? And is it \emph{natural}? 
For instance: does this construction extend to a morphism of smooth $\infty$-groupoids
$$
  \frac{1}{2}{\hat {\mathbf{p}}}_1 : \mathbf{B} \mathrm{Spin}(N)_{\mathrm{conn}}
   \to \mathbf{B}^3 U(1)_{\mathrm{conn}}
$$
from $\mathrm{Spin}$-principal bundles with connection to circle 3-bundles with connection?

In the following we give a natural presentation of the 
$\infty$-Chern-Weil homomorphism by means of Lie integration of 
$L_\infty$-algebraic data to simplicial presheaves. Among other things, this construction 
yields an understanding of why this construction is what it is and does what it does. 

The construction proceeds in the following broad steps
\begin{enumerate}
\item The infinitesimal analog of a characteristic class
   $\mathbf{c} : \mathbf{B}G \to \mathbf{B}^n U(1)$ is an $L_\infty$-algebra cocycle
   $$
     \mu : \mathfrak{g} \to b^{n-1} \mathbb{R}
     \,.
   $$

\item There is a formal procedure of universal Lie integration which sends this to a morphism of 
smooth $\infty$-groupoids
   $$
     \exp(\mu) : \exp(\mathfrak{g}) \to \exp(b^{n-1} \mathbb{R}) \simeq \mathbf{B}^n \mathbb{R}
   $$
   presented by a morphism of simplicial presheaves on $\mathrm{CartSp}$.

\item By finding a Chern-Simons element $\mathrm{cs}$ that witnesses the transgression of $\mu$ 
to an invariant polynomial on $\mathfrak{g}$ this construction has a differential refinement to a morphism
   $$
     \exp(\mu,\mathrm{cs}) : \exp(\mathfrak{g})_{\mathrm{conn}} \to \mathbf{B}^n \mathbb{R}_{\mathrm{conn}}     
   $$
   that sends $L_\infty$-algebra valued connections to line $n$-bundles with connection.

\item The $n$-truncation $\mathbf{cosk}_{n+1} \exp(\mathfrak{g})$ of the object on the left produces the smooth $\infty$-groups on interest -- $\mathbf{cosk}_{n+1} \exp(\mathfrak{g}) \simeq \mathbf{B}G$ -- and the corresponding truncation of $\exp((\mu,cs))$ carves out the lattice $\Gamma$ of periods in $G$ of the cocycle $\mu$ inside $\mathbb{R}$. The result is the differential characteristic class
   $$
     \exp(\mu,\mathrm{cs}) : \mathbf{B}G_{\mathrm{conn}} \to \mathbf{B}^n \mathbb{R}/\Gamma_{\mathrm{conn}}
     \,.
   $$
   Typically we have $\Gamma \simeq \mathbb{Z}$ such that this then reads
   $$
     \exp(\mu,\mathrm{cs}) : \mathbf{B}G_{\mathrm{conn}} \to \mathbf{B}^n U(1)_{\mathrm{conn}}
     \,.
   $$
\end{enumerate}   
   
\medskip
   
\paragraph{The $\infty$-Chern-Weil homomorphism}

In the full $\infty$-Chern-Weil theory the $\infty$-Chern-Weil homomorphism is conceptually very simple: 
for every $n$ there is canonically a morphism of smooth $\infty$-groupoids $\mathbf{B}^n U(1) \to \mathbf{\flat}_{\mathrm{\mathrm{dR}}}\mathbf{B}^{n+1}U(1)$ where the object on the right classifies 
ordinary de Rham cohomology in degree $n+1$. For $G$ any $\infty$-group and any 
characteristic class $\mathbf{c} : \mathbf{B}G \to \mathbf{B}^{n+1}U(1)$, 
the $\infty$-Chern-Weil homomorphism is the operation that takes a $G$-principal $\infty$-bundle 
$X \to \mathbf{B}G$ to the composite 
$X \to \mathbf{B}G \to \mathbf{B}^n U(1) \to \mathbf{\flat}_{\mathrm{\mathrm{dR}}} \mathbf{B}^{n+1}U(1)$.

All the constructions that we consider here in this introduction serve to \emph{model} this 
abstract operation. The $\infty$-connections that we considered yield resolutions of $\mathbf{B}^n U(1)$ and $\mathbf{B}G$ in terms of which the abstract morphisms are modeled as $\infty$-anafunctors.

\medskip

\subparagraph{$\infty$-Chern-Simons functionals}

If we express $G$ by Lie integration of an $\infty$-Lie algebra $\mathfrak{g}$, then the basic $\infty$-Chern-Weil homomorphism is modeled by composing an $\infty$-connection $(A_{\mathrm{\mathrm{vert}}}, A, \langle F_A\rangle)$ with the transgression of an invariant polynomial $(\mu, \mathrm{cs}, \langle - \rangle)$ as follows
$$
  \left(
    \raisebox{50pt}{
    \xymatrix@C=8pt{
      \Omega^\bullet(U \times \Delta^k)_{\mathrm{\mathrm{vert}}}
      \ar@{<-}[r]^{A_{\mathrm{\mathrm{vert}}}}&
      \mathrm{CE}(\mathfrak{g})
      &
      \mbox{ {\v C}ech cocycle}
      \\
      \Omega^\bullet(U \times \Delta^k)
      \ar@{<-}[r]^{A}
      \ar[u]
      &
      \mathrm{W}(\mathfrak{g})
      \ar[u]
      &
      \mbox{connection}
      \\
      \Omega^\bullet(U)
      \ar@{<-}[r]^{\langle F_A\rangle}
      \ar[u]
      &
      \mathrm{inv}(\mathfrak{g}) \ar[u]
      &
      \mbox{\begin{tabular}{c}curvature \\characteristic forms\end{tabular}}
    }
    }
  \right)
  \circ
  \left(
  \raisebox{50pt}{
  \xymatrix@C=8pt{
    \mathrm{CE}(\mathfrak{g}) \ar@{<-}[r]^{\mu}& \mathrm{CE}(b^{n-1} \mathbb{R})
    & \mbox{cocycle}
    \\
    \mathrm{W}(\mathfrak{g}) \ar@{<-}[r]^{\mathrm{cs}} \ar[u] 
     & 
     \mathrm{W}(b^{n-1} \mathbb{R})
     \ar[u]
    & \mbox{\begin{tabular}{c}Chern-Simons \\element \end{tabular}}
    \\
    \mathrm{inv}(\mathfrak{g})
   \ar@{<-}[r]^{\langle -\rangle} \ar[u]&
    \mathrm{inv}(b^{n-1} \mathbb{R}) 
    \ar[u]
    & 
    \mbox{\begin{tabular}{c}invariant \\polynomial\end{tabular}}
  }
  }
  \right)
$$
$$
  = 
  \;
  \;
  \;
  \left(
    \raisebox{50pt}{
    \xymatrix{
      \Omega^\bullet(U \times \Delta^k)_{\mathrm{\mathrm{vert}}}
      \ar@{<-}[r]^{A_{\mathrm{\mathrm{vert}}}}
      &
      \mathrm{CE}(\mathfrak{g})
      \ar@{<-}[r]^{\mu}
      &
      \mathrm{CE}(b^{n-1} \mathbb{R})
      &
      : \mu(A_{\mathrm{\mathrm{vert}}})
      &
      \mbox{characteristic class}
      \\
      \Omega^\bullet(U \times \Delta^k)
      \ar@{<-}[r]^{A}
      \ar[u]
      &
      \mathrm{W}(\mathfrak{g})
      \ar[u]
      \ar@{<-}[r]^{\mathrm{cs}}
      &
      \mathrm{W}(b^n-1\mathbb{R}) \ar[u]
      &
      : \mathrm{cs}_\mu(A) 
      &
      \mbox{Chern-Simons form}
      \\
      \Omega^\bullet(U)
      \ar@{<-}[r]^{\langle F_A\rangle}
      \ar[u]
      &
      \mathrm{inv}(\mathfrak{g}) \ar[u]
      \ar@{<-}[r]^{\langle - \rangle}
      &
      \mathrm{inv}(b^{n-1}\mathbb{R}) \ar[u]
      &
      : \langle F_A \rangle_\mu
      &
      \mbox{\begin{tabular}{c}curvature \\characteristic forms\end{tabular}}
    }
    }
  \right)
  \,.
$$
This clearly yields a morphism of simplicial presheaves
$$
  \exp(\mu)_{\mathrm{conn}} : \exp(\mathfrak{g})_{\mathrm{conn}} \to \exp(b^{n-1}\mathbb{R})_{\mathrm{conn}} 
$$
and, upon restriction to the top two horizontal layers, a morphism
$$
  \exp(\mu)_{\mathrm{diff}} : \exp(\mathfrak{g})_{\mathrm{diff}} \to 
   \exp(b^{n-1}\mathbb{R})_{\mathrm{diff}} 
  \,.
$$
Projection onto the third horizontal component gives the map to the curvature classes
$$
  \exp(b^{n-1}\mathbb{R})_{\mathrm{diff}} \to \mathbf{\flat}_{\mathrm{\mathrm{dR}}}
   \exp(b^{n} \mathbb{R})_{\mathrm{simp}}
  \,,
$$
In total, this constitutes an $\infty$-anafunctor
$$
  \xymatrix{
    \exp(\mathfrak{g})_{\mathrm{diff}}
    \ar[r]^{\exp(\mu)_{\mathrm{diff}}}
    \ar[d]^\simeq
    &
    \exp(b^{n-1}\mathbb{R})_{\mathrm{diff}}
    \ar[r]&
    \mathbf{\flat}_{\mathrm{\mathrm{dR}}}b^n \mathbb{R}
    \\
    \exp(\mathfrak{g})
  }
$$
Postcomposition with this is the simple $\infty$-Chern-Weil homomorphism: it sends a cocycle
$$
  \xymatrix{
    C(U) \ar[r] \ar[d]^\simeq & \exp(\mathfrak{g})
    \\
    X
  }
$$
for an $\exp(\mathfrak{g})$-principal bundle to the curvature form represented by
$$
  \xymatrix{
    C(V)
    \ar[r]^{(g,\nabla)}
    \ar[d]^\simeq
     & 
      \exp(\mathfrak{g})_{\mathrm{diff}}
      \ar[d]^\simeq     
      \ar[r]^{\exp(\mu)_{\mathrm{diff}}}
     &
      \exp(b^{n-1}\mathbb{R})_{\mathrm{diff}}
    \ar[r]&
    \mathbf{\flat}_{\mathrm{\mathrm{dR}}}b^n \mathbb{R}
    \\
    C(U) \ar[r]^{g}
    \ar[d]^\simeq
    & \exp(\mathfrak{g})
    \\
    X
  }
  \,.
$$
\begin{proposition}
For $\mathfrak{g}$ an ordinary Lie algebra with simply connected Lie group $G$,
the image under $\tau_1(-)$ of this diagram constitutes the ordinary Chern-Weil homomorphism in that:

for $g$ the cocycle for a $G$-principal bundle, any ordinary connection on a bundle constitutes 
a lift $(g,\nabla)$ to the tip of the anafunctor and the morphism represented by that is 
the {\v C}ech-hypercohomology cocycle on $X$ with values in the truncated de Rham complex given by 
the globally defined curvature characteristic form $\langle F_\nabla \wedge \cdots \wedge F_\nabla\rangle$.
\end{proposition}
But evidently we have more information available here. The ordinary Chern-Weil homomorphism refines from a map that assigns curvature characteristic forms, to a map that assigns secondary characteristic classes 
in the sense that it assigns circle $n$-bundles with connection whose curvature 
is this curvature characteristic form.
The local connection forms of these circle bundles are given by the middle horizontal morphisms. These are the Chern-Simons forms 
$$
  \Omega^\bullet(U)
  \stackrel{A}{\leftarrow}
  \mathrm{W}(\mathfrak{g})
  \stackrel{cs}{\leftarrow}
  \mathrm{W}(b^{n-1} \mathbb{R})
  :
  \mathrm{cs}(A)
 \,.
$$

\medskip
\subparagraph{Secondary characteristic classes}

So far we discussed the untruncated coefficient object $\exp(\mathfrak{g})_{\mathrm{conn}}$
of $\mathfrak{g}$-valued $\infty$-connections.
The real object of interest is the $k$-truncated version 
$\tau_k \exp(\mathfrak{g})_{\mathrm{conn}}$ where $k \in \mathbb{N}$ is such 
that $\tau_k \exp(\mathfrak{g}) \simeq \mathbf{B}G$ is the delooping of the $\infty$-Lie group 
in question. 

Under such a truncation, the integrated $\infty$-Lie algebra cocycle $exp(\mu) : exp(\mathfrak{g}) \to exp(b^{n-1}\mathbb{R})$ will no longer be a simplicial map. Instead, the periods of $\mu$ will cut out a lattice $\Gamma$ in $\mathbb{R}$, and $\exp(\mu)$ does descent to the quotient of $\mathbb{R}$ by that lattice
$$
  \exp(\mu) : \tau_k \exp(\mathfrak{g}) \to \mathbf{B}^n \mathbb{R}/\Gamma
  \,.
$$
We now say this again in more detail.

Suppose $\mathfrak{g}$ is such that 
the $(n+1)$-coskeleton 
$\mathbf{cosk}_{n+1} \exp(\mathfrak{g}) \stackrel{\simeq}{\to} \mathbf{B}G$ for the desired $G$. Then the periods of $\mu$ over $(n+1)$-balls 
cut out a lattice $\Gamma \subset \mathbb{R}$ and thus we get an $\infty$-anafunctor
$$
  \xymatrix{
    \mathbf{cosk}_{n+1} \exp(\mathfrak{g})_{\mathrm{diff}}
    \ar[r]
    \ar[d]^\simeq
     &
    \mathbf{B}^{n}\mathbb{R}/\Gamma_{\mathrm{diff}}
    \ar[r]
    &
    \mathbf{\flat}_{\mathrm{\mathrm{dR}}} \mathbf{B}^{n+1} \mathbb{R}/\Gamma
    \\
    \mathbf{B}G
  }
$$
This is \emph{curvature characteristic class}. 
We may always restrict to genuine $\infty$-connections and refine
$$
  \xymatrix{
    \mathbf{cosk}_{n+1} \exp(\mathfrak{g})_{\mathrm{conn}}
    \ar[r] 
    \ar@{^{(}->}[d]
    &
    \mathbf{B}^{n}\mathbb{R}/\Gamma_{\mathrm{conn}}
    \ar@{^{(}->}[d]
    \\
    \mathbf{cosk}_{n+1} \exp(\mathfrak{g})_{\mathrm{diff}}
    \ar[r]
    \ar[d]^\simeq
     &
    \mathbf{B}^{n}\mathbb{R}/\Gamma_{\mathrm{diff}}
    \ar[r]
    &
    \mathbf{\flat}_{\mathrm{\mathrm{dR}}} \mathbf{B}^{n+1} \mathbb{R}/\Gamma
    \\
    \mathbf{B}G
  }
$$

which models the refined $\infty$-Chern-Weil homomorphism with values in ordinary differential cohomology

$$
  H_{\mathrm{conn}}(X,G)
  \to 
  \mathbf{H}_{conn}^{n+1}(X, \mathbb{R}/\Gamma)
  \,.
$$

\begin{example}
Applying this to the discussion of the Chern-Simons circle 3-bundle above, we find a differential refinement
$$
  \xymatrix{
    & 
     \exp(\mathfrak{g})_{\mathrm{diff}} \ar[r]{\exp(\mu)_{\mathrm{diff}}}
     \ar[d]
     & 
     \exp(b^{n-1}\mathbb{R})_{\mathrm{diff}}
     \ar[d]^{\int_{\Delta^\bullet}}
    \\
    C(V) \ar[r]^{(\hat g,\hat \nabla)} 
      \ar[d]^\simeq
     & 
     \mathbf{cosk}_3\exp(\mathfrak{g})_{\mathrm{diff}} 
    \ar[r] \ar[d] &
    \mathbf{B}^3 U(1)_{\mathrm{diff}}
    \\
    C(U) \ar[r]^{(g,\nabla)} \ar[d]^\simeq & \mathbf{B}G_{\mathrm{diff}}
    \\
    X
  }
  \,.
$$
Chasing components through this composite one finds that this descibes the cocycle in Deligne cohomology given by
$$
  (CS(\sigma_i^* \nabla) , \; \int_{\Delta^1} CS(\hat g_{i j}^* \nabla),
  \int_{\Delta^2} CS(\hat g_{i j k}^* \nabla), 
  \int_{\Delta^3} \hat g_{i j k l}^* \mu)
  \,.
$$
This is the cocycle for the circle $n$-bundle with connection. 
\end{example}
This is precisely the form of the {\v C}ech-Deligne cocycle for the first Pontryagin class
given in \cite{brylinski-mclaughlin}, only that here it comes out 
automatically normalized such
as to represent the fractional generator $\frac{1}{2}\mathbf{p}_1$.

By feeding in more general transgressive $\infty$-Lie algebra cocycles through this
machine, we obtain cocycles for more general differential characteristic classes. For
instance the next one is the second fractional Pontryagin class of $\mathrm{String}$-2-bundles
with connection \cite{FSS}.
Moreover, these constructions naturally yield the full cocycle
$\infty$-groupoids, not just their cohomology sets. This 
allows us to form the homotopy fibers of the $\infty$-Chern-Weil
homomorphism and thus define \emph{differential string structures} etc.
and \emph{twisted} differential string structures etc.
\cite{SSSIII}.

\subsubsection{Hamilton-Jacobi-Lagrange mechanics via prequantized Lagrangian correspondences}
\label{BasicClassicalMechanicsByPrequantizedLagrangianCorrespondences}
\label{ClassicalLocalFieldTheory}
\index{classical mechanics}

We show here how classical mechanics 
-- Hamiltonian mechanics, Lagranian mechanics, Hamilton-Jacobi theory, 
see e.g. \cite{Arnold} --
naturally arises from and is accurately captured by 
``pre-quantized Lagrangian correspondences''. 
Since field theory is a refinement of classical mechanics, 
this serves also as a blueprint
for the discussion of De Donder-Weyl-style classical field theory
by higher correspondences
below in \ref{DeDonderWeylTheoryViaHigherCorrespondences}, and more 
generally for the discussion of local prequantum field theory
in \cite{hgp, Nuiten, lpqft}.

The reader unfamiliar with classical mechanics may take the following to be 
a brief introduction to and indeed a systematic derivation of 
the central concepts of classical mechanics from 
the notion of correspondences in slice toposes. 
Conversely, the reader familiar with classical mechanics 
may take the translation of classical mechanics into correspondences
in slice toposes as the motivating example for the formalization of 
prequantum field theory in \cite{lpqft}. 
The translation is summarized
as a diagramatic dictionary below in 
\ref{TheClassicalActionFunctionalPrequantizesLagrangianCorrespondences}.

\medskip

The following sections all follow, in their titles, the pattern
\begin{center}
  \emph{Physical concept} and \emph{mathmatical formalization}
\end{center}
and each first recalls a naive physical concept, then motivates
its mathematical formalization, then discusses this formalization and how it reflects back
on the understanding of the physics.

\medskip

\begin{itemize}
  \item \ref{PhaseSpacesAndSymplecticManifolds} -- Phase spaces and symplectic manifolds;
  \item \ref{CoordinateSystemsAndTheToposOfSmoothSpaces} -- Coordinate systems and the topos of smooth spaces;
  \item \ref{GOPCanonicalTransformationsAndSymplectomorphisms} -- Coordinate transformations and symplectomorphisms;
  \item \ref{GOPTrajectoriesAndLagrangianCorrespondences} -- Trajectories and Lagrangian correspondences;
  \item \ref{ClassicalObservablesAndThePoissonBracket} -- Observables, symmetries, and the Poisson bracket Lie algebra;
  \item \ref{HamiltonianTimeEvolutionTrajectoriesAndHamiltonianCorrespondences} -- Hamiltonian (time evolution) correspondence and Hamiltonian correspondence;
  \item \ref{NoetherSymmetriesAndEquivariantStructures} -- Noether symmetries and equivariant structure;
  \item \ref{GaugeTheorySmoothGroupoidsAndHigherToposes} -- Gauge symmetry, smooth groupoids and higher toposes;
  \item \ref{TheKineticActionPreQuantizationAndDifferentialCohomology} -- The kinetic action, prequantiuation and differential cohomology;
  \item \ref{TheClassicalActionLegendreTransformAndHamuiltonianFlows} -- The classical action, the Legendre transform and Hamiltonian flows;
  \item \ref{TheClassicalActionFunctionalPrequantizesLagrangianCorrespondences} -- The classical action functional pre-quantizes Lagrangian correspondences;
  \item \ref{QuantizationTheHeisenbergGroupAndSliceAutomorphismGroups} -- Quantization, the Heisenberg group and slice automorphism groups;
  \item \ref{IntegrableSystems} -- Integrable systems, moment maps, and maps into the Poisson bracket;
  \item \ref{ClassicalAnomaliesAndProjectiveSymplecticReduction} -- Classical anomalies and projective symplectic reduction;
\end{itemize}

\medskip

\noindent {\bf Historical comment.} 
Much of the discussion here is induced by just the notion of
\emph{pre-quantized Lagrangian correspondences}.
The notion of plain Lagrangian correspondences 
(not pre-quantized) has been observed
already in the early 1970s to usefully capture central aspects of 
Fourier transformation theory \cite{Hoermander71} 
and of classical mechanics \cite{Weinstein71}, notably to unify
the notion of Lagrangian subspaces of phase spaces with 
that of ``canonical transformations'', hence symplectomorphisms, between them.
This observation has since been particularly advertized by Weinstein
(e.g \cite{Weinstein83}), 
who proposed that some kind of \emph{symplectic category}
of symplectic manifolds with Lagrangian correspondences 
between them should be a good domain for a 
formalization of \emph{quantization} along the lines of geometric quantization.
Several authors have since discussed aspects of this idea.
A recent review in the context of field theory is in \cite{CattaneoMnevReshetikhin}.

But geometric quantization proper proceeds not from plain symplectic manifolds
but from a lift of their symplectic form to a cocycle in differential cohomology, called
a \emph{pre-quantization} of the symplectic manifold. 
Therefore it is to be expected that some notion of
pre-quantized Lagrangian correspondences, which put into correspondence
these prequantum bundles and not just their underlying symplectic manifolds,
is a more natural domain for geometric quantization, hence a more accurate
formalization of pre-quantum geometry.

There is an evident such notion of prequantization of Lagrangian correspondences, 
and this is what we introduce and discuss in the
following. While evident, it seems that it
has previously found little attention in the literature, certainly not 
attention comparable to the fame enjoyed by Lagrangian correspondences. 
But it should. As we show now, classical mechanics globally done right 
is effectively identified with the study of prequantized Lagrangian correspondences.

\paragraph{Phase spaces and symplectic manifolds}
\label{PhaseSpacesAndSymplecticManifolds}

Given a physical system, one says that its \emph{phase space} is the
space of its possible (``classical'') histories
or trajectories. 
Newton's second law of mechnanics says that trajectories
of physical systems are (typically) determined by differential equations of 
\emph{second} order, and therefore
these spaces of trajectories are (typically) equivalent to initial 
value data of 0th and of 1st derivatives. In physics this data 
(or rather its linear dual) is referred
to as the \emph{canonical coordinates} and the \emph{canonical momenta},
respectively, traditionally denoted by the symbols ``$q$'' and ``$p$''. 
Being coordinates, these are actually far from being canonical in the
mathematical sense; all that has invariant meaning is, locally, the surface element
$\mathbf{d}p \wedge \mathbf{d}q$ spanned by a change of coordinates and momenta.

Made precise, this says that a physical phase space is 
a sufficiently smooth manifold $X$ which is equipped with a closed
and non-degenerate
differential 2-form $\omega  \in \Omega^2_{\mathrm{cl}}(X)$, 
hence that phase spaces are \emph{symplectic manifolds} $(X,\omega)$.
\begin{example}
  The simplest nontrivial example is the phase space 
  $\mathbb{R}^2 \simeq T^\ast \mathbb{R}$ of a
  single particle propagating on the real line. The standard coordinates
  on the plane are traditionally written $q,p \;:\; \mathbb{R}^2 \longrightarrow \mathbb{R}$
  and the symplectic form is the canonical volume form 
  $\mathbf{d}q \wedge \mathbf{d}p$.
  \label{CanonicalR2PhaseSpace}
\end{example}
This is a special case of the following general and fundamental definition of
\emph{covariant phase spaces}
(whose history is long and convoluted, 
two references being \cite{Zuckerman, CrnkovicWitten}).
\begin{example}[covariant phase space]
  Let $F$ be a smooth manifold -- to be called the 
  \emph{field fiber} -- and write $[\Sigma_1, F]$ for the
  manifold of
  smooth maps from the closed interval $\Sigma_1 := [0,1]\hookrightarrow \mathbb{R}$ into $F$
  (an infinite-dimensional Fr{\'e}chet manifold).
  We think of $F$ as a space of \emph{spatial field configurations}
  and of $[\Sigma_1,F]$ as the space of \emph{trajectories} or 
  \emph{histories} of spatial field configurations. 
  Specifically, we may think of $[\Sigma_1,F]$ as the space of trajectories
  of a particle propagating in a space(-time) $F$.
  
  A smooth function
  $$
    L \;:\; [\Sigma_1,F] \longrightarrow \Omega^1(\Sigma_1)
  $$
  to the space of differential 1-forms on $\Sigma_1$ is called
  a \emph{local Lagrangian} of fields in $F$ if for all $t \in \Sigma_1$ the 
  assignment $\gamma \mapsto L_\gamma(t)$ is a smooth function of 
  $\gamma(t), \dot \gamma(t), \ddot \gamma(t), \cdots$
  (hence of the value of a curve $\gamma : \Sigma_1 \to F$ at $t$ and of the values
  of all its derivatives at $t$). One traditionally writes
  $$
    L 
	  \;:\; 
	\gamma \mapsto L(\gamma, \dot \gamma, \ddot \gamma, \cdots) 
	\wedge\mathbf{d}t
  $$
  to indicate this. 
  In cases of interest typically only first derivatives appear
  $$
    L \;:\; \gamma \mapsto L(\gamma, \dot \gamma) \wedge\mathbf{d}t
  $$
  and we concentrate on this case now for notational simplicity.
  Given such a local Lagrangian, the induced \emph{local action functional} 
  $S : [\Sigma_1, F] \to \mathbb{R}$ is
  the smooth function on trajectory space which is given by integrating
  the local Lagrangian over the interval:
  $$
    S = \int_{\Sigma_1} L \;:\; [\Sigma_1,F] \stackrel{L}{\longrightarrow} \Omega^1(\Sigma_1)
	  \stackrel{\int_I}{\longrightarrow}
	  \mathbb{R}
	  \,.
  $$
  The \emph{variational derivative} of the local Lagrangian is the smooth differential 2-form
  $$
    \mathbf{\delta} L   \in  \Omega^{1,1}([\Sigma_1,F]  \times \Sigma_1)
  $$
  on the product of trajectory space and parameter space, which is
  given by the expression
  $$
    \begin{aligned}
    \mathbf{\delta} L_\gamma
	& = 
	\frac{\partial L}{\partial \gamma} \wedge \mathbf{d}t \wedge \mathbf{\delta \gamma}
	+ 
	\frac{\partial L}{\partial \dot \gamma} \wedge \mathbf{d}t \wedge 
	\frac{d}{dt} \wedge \mathbf{\delta}\gamma
	\\
	& =
	\underbrace{
	\left(
	   \frac{\partial L}{\partial \gamma}
	   -
	   \frac{\partial}{\partial t}
	   \frac{\partial L}{\partial \dot \gamma}
	\right)
	}_{ =: \mathrm{EL}_\gamma  }
	\mathbf{d}t \wedge 
	\mathbf{\delta} \gamma
	+
	\frac{d}{d t}
	\underbrace{
	\left(
	  \frac{\partial L}{\partial \dot \gamma} \wedge \mathbf{\delta} \gamma 
	\right)
	}_{=: \theta_\gamma}
	\mathbf{d} t
	\end{aligned}
	\,.
  $$
  One says that $\mathrm{EL}_\gamma = 0$ (for all $t \in I$) is the
  \emph{Euler-Lagrange equation of motion} induced by the local Lagrangian
  $L$,  and that the 0-locus
  $$
    X := \left\{\gamma \in [\Sigma_1,F] \;|\; \mathrm{EL}_\gamma = 0 \right\}
	\hookrightarrow
	[\Sigma_1,F]
  $$
  (also called the ``shell'')
  equipped with the 2-form
  $$
    \omega := \mathbf{\delta} \theta  
  $$  
  is the \emph{unreduced covariant phase space} $(X,\omega)$ induced by $L$.
  \label{PhaseSpaceFromLocalActionFunctionalOnTheLine}
\end{example}
\begin{example}
  Consider the case that $F = \mathbb{R}$ and 
  that the Lagrangian is of the form
  $$
    \begin{aligned}
    L &:=
    L_{\mathrm{kin}}
    -
    L_{\mathrm{pot}}
   	\\
	& :=
	\left(
	  \tfrac{1}{2}\dot \gamma^2 
	  -
	  V(\gamma)
	\right)
	\wedge
	\mathbf{d}t
	\end{aligned}
	\,,
  $$
  hence is a quadratic form on the first derivatives of the 
  trajectory -- called the \emph{kinetic energy density} -- 
  plus any smooth function $V$ of the trajectory position itself
  -- called (minus) the \emph{potential energy density}.
  Then the corresponding phase space
  is equivalent to $\mathbb{R}^2 \simeq T^\ast \mathbb{R}$
  with the canonical coordinates identified with the initial value data
  $$
    q := \gamma(0)\;\,,\;\; p = \dot \gamma
  $$
  and with 
  $$
    \theta = p \wedge \mathbf{d}q
  $$
  and hence 
  $$
    \omega = \mathbf{d}q \wedge \mathbf{d}p	
	\,.
  $$
  This is the phase space of example \ref{CanonicalR2PhaseSpace}.
  Notice that the symplectic form here is a reflection entirely only of the
  kinetic action, independent of the potential action. This
  we come back to below in \ref{TheKineticActionPreQuantizationAndDifferentialCohomology}.
  \label{StandardLagrangianForMechanics}
\end{example}
\begin{remark}
  The differential 2-form $\omega$ on an unreduced covariant phase space
  in example \ref{PhaseSpaceFromLocalActionFunctionalOnTheLine}
  is closed, even exact, but in general far from non-degenerate, hence far
  from being symplectic. We may say that $(X,\omega)$ is a \emph{pre-symplectic manifold}.
  This is because this differential
  form measures the reaction of the Lagrangian/action functional to 
  variations of the fields, but the action functional may be \emph{invariant}
  under some variation of the fields; one says that it has
  (\emph{gauge}-)\emph{symmetries}. To obtain a genuine symplectic form 
  one needs to quotient out the flow of these symmetries from
  unreduced covariant phase space to obtain the \emph{reduced}
  covariant phase space.
  This we turn to below in \ref{NoetherSymmetriesAndEquivariantStructures}.
\end{remark}  
\begin{remark}
  In the description of the mechanics of just particles,
  the Lagrangian $L$ above has no further more fundamental description,
  it is just what it is.
  But in applications to $n$-dimensional \emph{field} theory 
  the differential 1-forms $L$ and $\theta$ 
  in example \ref{PhaseSpaceFromLocalActionFunctionalOnTheLine}
  arise themselves from integration of differential $n$-forms 
  over space (Cauchy surfaces), hence from \emph{transgression} of 
  higher-degree data in higher codimension. 
  This we describe in example \ref{TransgressionOfKineticLocalLagrangianToCodimension1} below.
  Since transgression in 
  general loses some information, one should really 
  work locally instead of integrating over Cauchy surfaces, 
  hence work with the 
  de-transgressed data and develop classical field theory for that.
  This we turn to below in \ref{DeDonderWeylTheoryViaHigherCorrespondences}
  for classical field theory
  and then more generally for local prequantum field theory 
  in \cite{lpqft}.
  \label{LagrangiansFromTransgression}  
\end{remark}

\paragraph{Coordinate systems and the topos of smooth spaces}
\label{CoordinateSystemsAndTheToposOfSmoothSpaces}
 
When dealing with spaces $X$ that are equipped with extra structure, such as 
a closed differential 2-form
$\omega \in \Omega^2_{\mathrm{cl}}(X)$, then it is useful to have 
a \emph{universal moduli space} for these structures, and this will be central
for our developments here. 
So we need a ``smooth space'' $\mathbf{\Omega}^2_{\mathrm{cl}}$ of sorts,
characterized by the property that  there is a natural bijection between smooth closed differential 
2-forms $\omega \in \Omega^2_{\mathrm{cl}}(X)$ and smooth maps
$
  \xymatrix{
    X \ar[r] & \mathbf{\Omega}^2_{\mathrm{cl}}
  }
$.
Of course such a universal moduli spaces of closed 2-forms does not exist in the 
category of smooth manifolds. But it does exist canonically if we 
slightly generalize the notion of ``smooth space'' suitably
(the following is discussed in more detail below in \ref{GeometryOfPhysicSmoothSpaces}). 
\begin{definition}
  A \emph{smooth space} or \emph{smooth 0-type} $X$ is 
  \begin{enumerate}
  \item an assignment to each $n \in \mathbb{N}$ of a 
  set, to be written $X(\mathbb{R}^n)$ and to be called the 
  \emph{set of smooth maps from $\mathbb{R}^n$ into $X$},
  \item an assignment to each ordinary smooth function 
  $f : \mathbb{R}^{n_1} \to \mathbb{R}^{n_2}$
  between Cartesian spaces
     of 
	a function of sets $X(f) : X(\mathbb{R}^{n_2}) \to X(\mathbb{R}^{n_1})$,
	to be called the \emph{pullback of smooth functions into $X$ along $f$};
  \end{enumerate}
  such that
  \begin{enumerate}
    \item this assignment respects composition of smooth functions;
	\item this assignment respect the covering of Cartesian spaces
	by open disks: for every good open cover 
	$\{\mathbb{R}^n \simeq U_i \hookrightarrow \mathbb{R}^n\}_i$,
	the set $X(\mathbb{R}^n)$ of smooth functions out of $\mathbb{R}^n$
	into $X$ is in natural bijection with the set
	$\left\{ (\phi_i)_i \in \prod_i X(U_i) \;|\;  \forall_{i,j}\; \phi_i|_{U_{i} \cap U_j}  
	 = \phi_j|_{U_{i} \cap U_j} \right\}$
	 of tuples of smooth functions out of the patches of the cover which agree on
	 all intersections of two patches.
  \end{enumerate}
  \label{SmoothSpaceInComponents}
\end{definition}
\begin{remark}
  One may think of definition \ref{SmoothSpaceInComponents} as a formalization of the
  common idea in physics that we understand spaces by charting them with coordinate
  systems. A Cartesian space $\mathbb{R}^n$ is nothing but 
  the standard $n$-dimensional coordinate system and one may think of the set 
  $X(\mathbb{R}^n)$ above as the set of all possible ways 
  (including all degenerate ways) of laying out this
  coordinate system in the would-be space $X$. Moreover, a function
  $f : \mathbb{R}^{n_1} \longrightarrow \mathbb{R}^{n_2}$ is nothing but
  a \emph{coordinate transformation} (possibly degenerate), and hence the
  corresponding functions $X(f) : X(\mathbb{R}^{n_2}) \longrightarrow X(\mathbb{R}^{n_1})$
  describe how the probes of $X$ by coordinate systems change under coordinate transformations.
  Definition \ref{SmoothSpaceInComponents} takes the idea that any space in physics
  should be probe-able by coordinate systems in this way to the extreme, in that it
  \emph{defines} a smooth spaces as a collection of probes by coordinate systems
  equipped with information about all possible coordinate transformations.
\end{remark}
The notion of smooth spaces is maybe more familiar with one little axiom added:
\begin{definition}
  A smooth space $X$ is called \emph{concrete} if there exists a set 
  $X_{\mathrm{disc}} \in \mathrm{Set}$ such that for each $n \in \mathbb{N}$ the 
  set $X(\mathbb{R}^n)$ of smooth functions from $\mathbb{R}^n$
  to $X$ is a subset of the set of \emph{all} functions 
  from the underlying set of $\mathbb{R}^n$ to the set $X_{\mathrm{disc}} \in \mathrm{Set}$.
  \label{ConcreteSmoothSpace}
\end{definition}
This definition of concrete smooth spaces 
goes back to \cite{Chen} in various slight variants, see \cite{Stacey}
for a comparative discussion. 
A comprehensive textbook account
of differential geometry formulated with this definition of smooth
spaces (called ``diffeological spaces'' there) is in \cite{IglesiasZemmour}.

While the formulation of def. \ref{SmoothSpaceInComponents} is designed to make transparent its
geometric meaning, of course equivalently but more abstractly this says the following:
\begin{definition}
  Write $\mathrm{CartSp}$ for the category of Cartesian spaces with 
  smooth functions between them, and consider it 
equipped with the coverage (Grothendieck pre-topology) 
  of good open covers. 
  A \emph{smooth space} or \emph{smooth 0-type} is a sheaf on this site.
  The \emph{topos of smooth 0-types} is the sheaf category
  $$
    \mathrm{Smooth}0\mathrm{Type} := \mathrm{PSh}(\mathrm{CartSp})[\{\mbox{covering maps}\}^{-1}]
	\,.
  $$
\end{definition}
In the following we will abbreviate the notation to
$$
  \mathbf{H} := \mathrm{Smooth}0\mathrm{Type}
  \,.
$$
For the discussion of pre-symplectic manifolds, we need the following two 
examples.
\begin{example}
  Every smooth manifold $X \in \mathrm{SmoothManifold}$ becomes
  a smooth 0-type by the assignment 
  $$
    X : n \mapsto C^\infty(\mathbb{R}^n, X)
	\,.
  $$
  (This defines in fact a concrete smooth space, def. \ref{ConcreteSmoothSpace}, 
  the underlying set $X_{\mathrm{disc}}$ being just the underlying set of 
  points of the given manifold.)
  This construction extends to a full and faithful embedding
  of smooth manifolds into smooth 0-types
  $$
    \xymatrix{
	  \mathrm{SmoothManifold}
	  \ar@{^{(}->}[r]
	  &
	  \mathbf{H}
	}
	\,.
  $$
\end{example}
The other main example is in a sense at an opposite extreme in the space of 
all examples. It is given by smooth moduli space of \emph{differential forms},
see the discussion in \ref{TheGeometryOfPhysicsDifferentialForms}.
\begin{example}
  For $p \in \mathbb{N}$, write $\mathbf{\Omega}^p_{\mathrm{cl}}$
  for the smooth space given by the assignment
  $$
    \mathbf{\Omega}^p_{\mathrm{cl}} : n \mapsto \Omega^p_{\mathrm{cl}}(\mathbb{R}^n)
  $$
  and by the evident pullback maps of differential forms.
  These smooth spaces $\mathbf{\Omega}^n_{\mathrm{cl}}$ are \emph{not}
  concrete, def. \ref{ConcreteSmoothSpace}. In fact they are maximally
  non-concrete in that there is only a single smooth map
  $\ast \to \mathbf{\Omega}^n_{\mathrm{cl}}$ from the point into them. 
  Hence the underlying point set of the smooth space
  $\mathbf{\Omega}^n_{\mathrm{cl}}$ looks like a singleton, and yet 
  these smooth spaces are far from being the trivial smooth space: they
  admit  many smooth maps $X \longrightarrow \mathbf{\Omega}^n_{\mathrm{cl}}$
  from smooth manifolds of dimension at least $n$, 
  as the following prop. \ref{PresymplecticFormsAsMapsIntoASmoothSpace} shows.
\end{example}
This solves the moduli problem for closed smooth differential forms:
\begin{proposition}
  For $p \in \mathbb{N}$
  and $X \in \mathrm{SmoothManifold} \hookrightarrow \mathrm{Smooth}0\mathrm{Type}$, 
  there is a natural bijection
  $$
    \mathbf{H}(X,\mathbf{\Omega}^p_{\mathrm{cl}})
	\simeq
	\Omega^p_{\mathrm{cl}}(X)
	\,.
  $$
  \label{PresymplecticFormsAsMapsIntoASmoothSpace}
\end{proposition}
So a pre-symplectic manifold $(X,\omega)$ is equivalently a map of smooth spaces
of the form
$$
  \omega : \xymatrix{
    X \ar[r] & \mathbf{\Omega}^2_{\mathrm{cl}}
  }
  \,.
$$

\paragraph{Canonical transformations and Symplectomorphisms}
\label{GOPCanonicalTransformationsAndSymplectomorphisms}

An equivalence between two phase spaces, hence a re-expression
of the ``canonical'' coordinates and momenta, is called a 
\emph{canonical transformation} in physics. Mathematically this is
a \emph{symplectomorphism}:
\begin{definition}
 Given two (pre-)symplectic manifolds $(X_1, \omega_1)$ and $(X_2, \omega_2)$
 a \emph{symplectomorphism} 
 $$
   f : (X_1, \omega_1) \longrightarrow (X_2, \omega_2)
 $$
 is a diffeomorphism $f : X_1 \longrightarrow X_2$ 
 of the underlying smooth spaces, which respects the
 differential forms in that 
 $$
   f^\ast \omega_2 = \omega_1
   \,.
 $$
\end{definition}
The formulation above in \ref{CoordinateSystemsAndTheToposOfSmoothSpaces} 
of pre-symplectic manifolds
as maps into a moduli space of closed 2-forms
yields the following equivalent re-formulation of symplectomorphisms, 
which is very simple in itself, 
but contains in it the seed of an important phenomenon:
\begin{proposition}
  Given two symplectic manifolds $(X_1, \omega_1)$ and $(X_2, \omega_2)$,
  a symplectomorphism $\phi : (X_1,\omega_1) \to (X_2, \omega_2)$
  is equivalently a commuting diagram of smooth spaces of the 
  following form:
  $$
    \raisebox{20pt}{
    \xymatrix{
      X_1 
  	  \ar[dr]_{\omega_1} \ar[rr]^-\phi
	  &&
	  X_2
	  \ar[dl]^{\omega_2}
	  \\
	  & \mathbf{\Omega}^2_{\mathrm{cl}}
    }
	}
	\,.
  $$
\end{proposition}
Situations like this are naturally interpreted in the \emph{slice topos}:
\begin{definition}
  For $A \in \mathbf{H}$ any smooth space,
  the \emph{slice topos}  $\mathbf{H}_{/A}$ is the category whose
  objects are objects $X \in \mathbf{H}$ equipped with maps
  $X \to A$, and whose morphisms are commuting diagrams in $\mathbf{H}$
  of the form
  $$
    \raisebox{20pt}{
    \xymatrix{
	  X \ar[rr] \ar[dr] &&  Y \ar[dl]
	  \\
	  & A
	}
	}
	\,.
  $$
  \label{SliceTopos}
\end{definition}
Hence if we write $\mathrm{SymplManifold}$ for the category of smooth pre-symplectic manifolds
  and symplectomorphisms betwen them, then we have the following.
\begin{proposition}
   The construction of
  prop. \ref{PresymplecticFormsAsMapsIntoASmoothSpace} constitutes a
  full embedding
  $$
    \xymatrix{
      \mathrm{SymplManifold} 
	  \ar@{^{(}->}[r] 
	  &
	  \mathbf{H}_{/\mathbf{\Omega}^2_{\mathrm{cl}}}
	}
  $$
  of pre-symplectic manifolds with symplectomorphisms between them into 
  the slice topos of smooth spaces over the smooth moduli space of
  closed differential 2-forms.
  \label{EmbeddingOfSymplecticManifoldsIntoSliceOverOmega2}
\end{proposition}

\paragraph{Trajectories and Lagrangian correspondences}
\label{GOPTrajectoriesAndLagrangianCorrespondences}

A symplectomorphism clearly puts two symplectic manifolds
``in relation'' to each other. It turns out to be useful to say this 
formally. Recall:
\begin{definition}
For $X,Y \in $ Set two sets, a relation $R$ between elements of $X$ and elements of $Y$ is a subset of the Cartesian product set 
$$
  R \hookrightarrow X \times Y
  \,.
$$
More generally, for $X, Y \in \mathbf{H}$ two objects of a topos (such as the topos of smooth spaces), 
then a relation $R$ between them is a subobject of their Cartesian product
$$
  R \hookrightarrow X \times Y
  \,.
$$
\end{definition}

In particular any function induces the relation ``$y$ is the image of $x$'':
\begin{example}
  \label{RelationInducedByFunction}
For $f \;\colon\; X \longrightarrow Y$ a function, its 
\emph{induced relation} is the relation which is exhibited by \emph{graph} of $f$
$$
  \mathrm{graph}(f) 
    :=
  \left\{
    \left(x,y\right) \in X \times Y \;|\; f(x) = y
  \right\}
$$
canonically regarded as a subobject 
$$
  \mathrm{graph}(f) \hookrightarrow X \times Y
  \,.
$$
\end{example}

Hence in the context of classical mechanics, in particular any symplectomorphism 
$f \;\colon\; (X_1, \omega_1) \longrightarrow (X_2, \omega_2)$ induces the relation 
$$
  \mathrm{graph}(f) \hookrightarrow X_1 \times X_2
  \,.
$$
Since we are going to think of $f$ as a kind of``physical process'', it is useful to think of the smooth space $graph(f)$ here as the \emph{space of trajectories} of that process. To make this clearer, notice that we may equivalently rewrite every relation $R \hookrightarrow X \times Y$ as a diagram of the following form:
$$
  \raisebox{20pt}{
  \xymatrix{
    & R
	\ar[dl]
	\ar[dr]
	\\
	X && Y
  }}
  \;\;
  = 
  \;\;
  \raisebox{30pt}{
  \xymatrix{
    & R
	\ar[d]
	\\
	& X \times Y
	\ar[dl]_{p_X}
	\ar[dr]^{p_Y}
	\\
	X && Y
  }}
$$
reflecting the fact that every element $(x \sim y) \in R$ defines an element 
$x = p_X(x \sim y) \in X$ and an element $y = p_Y(x \sim y) \in Y$. 

Then if we think of the space 
$R = \mathrm{graph}(f)$ of example \ref{RelationInducedByFunction}
as being a space of trajectories starting in $X_1$ and ending in $X_2$, 
then we may read the relation as 
``there is a trajectory from an incoming configuration $x_1$ to an outgoing configuration $x_2$'':
$$
  \xymatrix{
    & \mathrm{graph}(f)
	\ar[dl]_{\mathrm{incoming}}
	\ar[dr]^{\mathrm{outgoing}}
    \\
    X_1 && X_2
  }
  \,.
$$
Notice here that the defining property of a relation as a subset/subobject translates into the property of classical physics that there is \emph{at most one trajectory} from some incoming configuration $x_1$ to some outgoing trajectory $x_2$ (for a fixed and small enough parameter time interval at least, we will formulate this precisely in the next section when we genuinely consider Hamiltonian correspondences).

In a more general context one could consider there to be several such trajectories, and even a whole smooth space of such trajectories between given incoming and outgoing configurations. 
Each such trajectory would ``relate'' $x_1$ to $x_2$, but each in a possible different way. 
We can also say that each trajectory makes $x_1$ \emph{correspond} to $x_2$ in a different way, and that is the mathematical term usually used:

\begin{definition}
For $X, Y \in \mathbf{H}$ two spaces, 
a \emph{correspondence} between them is a diagram in $\mathbf{H}$ of the form

$$
  \xymatrix{
     &Z
	 \ar[dl]
	 \ar[dr]
    \\
    X && Y
  }
$$
with no further restrictions. 
Here $Z$ is also called the \emph{correspondence space}. 
\end{definition}

Observe that the graph of a function  $f \colon X \to Y$ is, while defined differently, in fact equivalent to just the space $X$, the equivalence being induced by the map $x \mapsto (x,f(x))$
$$
  X \stackrel{\simeq}{\longrightarrow} \mathrm{graph}(f)
  \,.
$$
In fact the relation/correspondence which expresses 
``$y$ is the image of $f$ under $x$'' may just as well be exhibited by the diagram
$$
  \xymatrix{
    &X
    \ar[dl]_{\mathrm{id}}
	\ar[dr]^f
    \\
    X && Y
  }
  \,.
$$
It is clear that this correspondence with correspondence space $X$ should be regarded as being equivalent to the one with correspondence space 
$\mathrm{graph}(f)$. We may formalize this as follows
\begin{definition}
Given two correspondences $\xymatrix{X & Z_1 \ar[l] \ar[r] & Y}$
and $\xymatrix{X & Z_2 \ar[l] \ar[r] & Y}$ between the same objects
in $\mathbf{H}$,
then an equivalence between them is an equivalence 
$\xymatrix{Z_1 \ar[r]^-\simeq & Z_2}$ in $\mathbf{H}$
which fits into a commuting diagram of the form
$$
  \xymatrix{
     &
	 Z_1
	 \ar[dl]
	 \ar[dr]
	 \ar[dd]^\simeq
    \\
    X && Y
    \\
    & Z_2
	\ar[ur]
	\ar[ul]
  }$$
\end{definition}
\begin{example}
Given an function $f \colon X \longrightarrow Y$
we have the commuting diagram
$$
  \xymatrix{
     &X
	 \ar[dl]_{\mathrm{id}}
	 \ar[dr]^f
	 \ar[dd]^\simeq
    \\
    X && Y
    \\
    & \mathrm{graph}(f)
	\ar[ur]_{i_Y}
	\ar[ul]^{i_X}
  }
$$
exhibiting an equivalence of the correspondence at
the top with that at the bottom.
\label{EquivalenceOfCorrespndencesInducedByFunction}
\end{example}
Correspondences between $X$ any $Y$ with such equivalences between them 
form a \emph{groupoid}.  Hence we write
$$
  \mathrm{Corr}\left(\mathbf{H}\right)(X,Y) \in \mathrm{Grpd}
  \,.
$$
Moreover, if we think of correspondences as modelling spaces of trajectories, then it is clear that their should be a notion of composition:
$$
  \left(
  \raisebox{20pt}{
  \xymatrix{    
    & Y_1 \ar[dl] \ar[dr] && Y_2 \ar[dl] \ar[dr]
    \\
    X_1 &&  X_2 &&  X_3 
  }}
  \right)
  \;\;\;\;
  \mapsto 
  \;\;\;\;
  \left(
  \raisebox{20pt}{
  \xymatrix{    
    & Y_1 \circ_{X_2} Y_2 
	\ar[dl]
	\ar[dr]
    \\
    X_1 &&  X_3 
  }}
  \right)
  \,.
$$
Heuristically, the composite space of trajectories $Y_1 \circ_{X_2} Y_2$ should consist precisely of those pairs of trajectories $( f, g ) \in Y_1  \times Y_2$ such that the endpoint of $f$ is the starting point of $g$. 
The space with this property is precisely the 
\emph{fiber product} of $Y_1$ with $Y_2$ over $X_2$, 
denoted $Y_1 \underset{X_2}{\times} Y_2$ (also called the \emph{pullback} 
of $Y_2 \longrightarrow X_2$ along $Y_1 \longrightarrow X_2$:
$$
  \left(
   \raisebox{20pt}{
  \xymatrix{    
    & Y_1 \circ_{X_2} Y_2 
	\ar[dl]
	\ar[dr]
    \\
    X_1 && X_3 
  }}
  \right)
  \;\;\;
  =
  \;\;\;
  \left(
  \raisebox{42pt}{
  \xymatrix{  
     && Z_1 \underset{Y}{\times} Z_2
	 \ar[dl]
	 \ar[dr]
     \\
      & Z_1 \ar[dl] \ar[dr] &&  Z_2 \ar[dl] \ar[dr]
      \\
    X_1 &&  X_2 &&  X_3
  }}
  \right)
  \,.
$$ 
Hence given a topos $\mathbf{H}$, correspondences between its objects form a category which composition the fiber product operation, where however the collection of morphisms between any two objects is not just a set, but is a groupoid (the groupoid of correspondences between two given objects and equivalences between them).

One says that correspondences form a \emph{$(2,1)$-category}
$$
  \mathrm{Corr}(\mathbf{H}) \in (2,1)\mathrm{Cat}
  \,.
$$

One reason for formalizing this notion of correspondences so much in the present context that it is useful now to apply it not just to the ambient topos $\mathbf{H}$ of smooth spaces, but also to its slice topos $\mathbf{H}_{/\mathbf{\Omega}_{cl}^2}$ over the universal moduli space of closed differential 2-forms.

To see how this is useful in the present context, notice 
the following
\begin{proposition}
  Let $\phi : (X_1, \omega_1) \to (X_2,\omega_2)$ be a 
  symplectomorphism.
  Write 
  $$
    (i_1,i_2) : \mathrm{graph}(\phi) \hookrightarrow X_1 \times X_2
  $$
  for the graph of the underlying diffeomorphsm.
  This fits into a commuting diagram in $\mathbf{H}$ of the form
$$
  \xymatrix{
    & \mathrm{graph}(\phi)
	  \ar[dl]_{i_1}
	  \ar[dr]^{i_2}_{\ }="s"
    \\
    X_1 \ar[dr]_{\omega_1}^{\ }="t" && X_2 \ar[dl]^{\omega_2}
	\\
	& \Omega^2_{\mathrm{cl}}
	\ar@{=} "s"; "t"
  }
  \,.
$$
Conversely, a smooth function $\phi : X_1 \to X_2$ is a symplectomorphism
precisely if its graph makes the above diagram commute.
  \label{SymplectomorphismsAsDiagramsCommutingOverOmega2}
\end{proposition}
Traditionally this is formalized as follows.
\begin{definition}
Given a symplectic manifold $(X,\omega)$, a submanifold $L \hookrightarrow X$ is called a 
\emph{Lagrangian submanifold} if $\omega|_{L} = 0$ and if $L$ has dimension $dim(L) = dim(X)/2$.
\end{definition}
\begin{definition}
  For $(X_1,\omega_1)$ and $(X_2,\omega_2)$ two symplectic manifolds,
  a correspondence $\xymatrix{X_1 & Y \ar[l]_{p_1} \ar[r]^{p_2} & X_2}$ of the underlying manifolds
  is a \emph{Lagrangian correspondence} if the map $Y \to X_1 \times X_2$
  exhibits a Lagrangian submanifold of 
  the symplectic manifold given by $(X_1 \times X_2, p_2^\ast \omega_2 - p_1^\ast \omega_1)$.
  
  Given two Lagrangian correspondence which intersect transversally over one
  adjacent leg, then their \emph{composition} is the correspondence given by the intersection.
  \label{LagrangianCorrespondence}
\end{definition}
But comparison with def. \ref{SliceTopos} shows that Lagrangian 
correspondences are in fact plain correspondences, just not in 
smooth spaces, but in the
slice $\mathbf{H}_{/\mathbf{\Omega}^2_{\mathrm{cl}}}$ of all smooth spaces
over the universal
smooth moduli space of closed differential 2-forms:
\begin{proposition}
  Under the identification of prop. \ref{EmbeddingOfSymplecticManifoldsIntoSliceOverOmega2}
  the construction of the diagrams in prop. \ref{SymplectomorphismsAsDiagramsCommutingOverOmega2}
  constitutes an injection of Lagrangian correspondence between $(X_1,\omega_1)$
  and $(X_2, \omega_2)$ into the Hom-space
  $\mathrm{Corr}\left(\mathbf{H}_{/\mathbf{\Omega}^2_{\mathrm{cl}}}\right)
    \left(
	   \left(X_1,\omega_1\right),
	   \left(X_2,\omega_2\right)
	\right)
  $.
  Moreover, composition of Lagrangian correspondence, when defined, coincides
  under this identification
  with the composition of the respective correspondences. 
  \label{LagrangianCorrespondencesAsCorrespondencesOverOmegs2}
\end{proposition}
\begin{remark}
  The composition of correspondences in the slice topos is always defined. 
  It may just happen the composite is given by a correspondence space
  which is a smooth space but not a smooth manifold. Or better, one
  may replace in the entire discussion the topos of smooth spaces
  with a topos of ``derived'' smooth spaces, modeled not on Cartesian 
  spaces but on Cartesian dg-manifolds. This will then automatically 
  make composition of Lagrangian correspondences take care of 
  ``transversal perturbations''. Here we will not further dwell on this
  possibility. In fact, the formulation of Lagrangian correspondences
  and later of prequantum field theory by correspondences in toposes
  implies a great freedom in the choice of type of geometry in which 
  set up everything. 
  (The bare minimum condition on the topos $\mathbf{H}$ which we need to require
  is that it be \emph{differentially cohesive}, \ref{InfinitesimalCohesion}).
\end{remark}
It is also useful to make the following phenomenon explicit, which is the
first incarnation of a recurring theme in the following discussions. 
\begin{proposition}
  The category $\mathrm{Corr}(\mathbf{H}_{/\mathbf{\Omega}^2_{\mathrm{cl}}})$
  is naturally a symmetric monoidal category, where the tensor product is given by
  $$
    (X_1,\omega_1) \otimes (X_2, \omega_2)
	=
	(X_1 \times X_2, \omega_1 + \omega_2)
	\,.
  $$
  The tensor unit is $(\ast, 0)$.
  With respect to this tensor product, every object is dualizable, with dual
  object given by
  $$
    (X,\omega)^v = (X, - \omega)
	\,.
  $$
  \label{CorrespondencesOverOmegs2IsMonoidalCategory}
\end{proposition}
\begin{remark}
  Duality induces natural equivalences of the form
  $$
    \mathrm{Corr}\left(\mathbf{H}_{/\mathbf{\Omega}^2_{\mathrm{cl}}}\right)\left(
	  \left(X_1,\omega_1\right),
	  \left(X_2,\omega_2\right),	  
	\right)
	\stackrel{\simeq}{\longrightarrow}
    \mathrm{Corr}\left(\mathbf{H}_{/\mathbf{\Omega}^2_{\mathrm{cl}}}\right)\left(
	  \left(\ast,0\right),
	  \left(X_1 \times X_2,\omega_2 - \omega_1\right),	  
	\right)
	\,.
  $$
  Under this equivalence an  isotropic (Lagrangian) correspondences 
  which in $\mathbf{H}$ is given by a diagram 
  as in prop. \ref{SymplectomorphismsAsDiagramsCommutingOverOmega2}
  maps to the diagram of the form
  $$
    \raisebox{20pt}{
    \xymatrix{
	  & \mathrm{graph}(\phi)
	  \ar[dl]
	  \ar[dr]^{(i_1, i_2)}_{\ }="s"
	  \\
	  \ast 
	  \ar[dr]_0^{\ }="t"
	  && X_1 \times X_2
	  \ar[dl]^{\omega_2 - \omega_1}
	  \\
	  & \mathbf{\Omega}^2_{\mathrm{cl}}
	  \ar@{=} "s"; "t"
	}}
	\,.
  $$
  This makes the condition that the pullback of the difference $\omega_2 - \omega_1$
  vanishes on the correspondence space more manifest. It is also the blueprint 
  of a phenomenon that is important in the generalization to field theory
  in the sections to follow, where trajectories map to boundary conditions,
  and vice versa. 
\end{remark}

\paragraph{Observables, symmetries and the Poisson bracket Lie algebra}
\label{ClassicalObservablesAndThePoissonBracket}
\label{Poisson bracket!ordinary}

Given a phase space $(X,\omega)$ of some physical system, then a function
$O : X \longrightarrow \mathbb{R}$ is an assignment of a value to every possible
state (phase of motion) of that system. For instance it might assign to 
every phase of motion its position 
(measured in some units with respect to some reference frame), or its momentum, or its 
energy. The premise of classical physics is that all of these quantitites
may in principle be observed in experiment, 
and therefore functions on phase space are traditionally called
\emph{classical observables}. Often this is abbreviated to
just \emph{observables} if the context is understood 
(the notion of observable in quantum mechanics and quantum field theory is 
more subtle, for a formalization 
of quantum observables in terms of correspondences in cohesive homotopy types 
see \cite{Nuiten}).

While this is the immediate physics heuristics about what functions on phase space are
are, it turns out that a central characteristic of
mechanics and of field theory is an intimiate relation between 
the observables of a mechanical system
and its \emph{infinitesimal symmetry transformations}:
an infinitesimal symmetry transformation of a phase space characterizes that
observable of the system which is invariant under the symmetry transformation. 
Mathematically this relation is captured by a the structure of a 
Lie algebra on the vector space of all observables after relating them
them to their \emph{Hamiltonian vector fields}.
\begin{definition}
  Given a symplectic manifold $(X,\omega)$ and a function $H : X \to \mathbb{R}$,
  its \emph{Hamiltonian vector field} is the unique $v \in \Gamma(T X)$ which 
  satisfies \emph{Hamilton's equation of motion}
  $$
    \mathbf{d}H = \iota_v \omega
	\,.
  $$
  \label{HamiltonEquations}
\end{definition}
\begin{example}
  For $(X,\omega) = (\mathbb{R}^2, \mathbf{d}q \wedge \mathbf{d}p)$
  the 2-dimensional phase space form example \ref{CanonicalR2PhaseSpace},
  and for $t \mapsto (q(t), p(t)) \in X$ a curve, it is a Hamiltonian
  flow line if its tangent vectors 
  $(\dot q(t), \dot p(t)) \in T_{(q(t),p(t))} \mathbb{R}^2 \simeq \mathbb{R}^2$
  satisfy Hamilton's equations in the 
  classical form:
  $$
    \dot q = \frac{\partial H}{\partial p}
	\;;\;\;\;\;
	\dot p = -\frac{\partial H}{\partial q}
	\,.
  $$
\end{example}
\begin{proposition}
  Given a symplectic manifold $(X,\omega)$, every Hamiltonian vector field
  $v$ is an \emph{infinitesimal symmetry} of $(X,\omega)$ --
  an \emph{infinitesimal symplectomorphism}
  -- in that 
  the Lie derivative of the symplectic form along $v$ vanishes
  $$
    \mathcal{L}_v \omega = 0
	\,.
  $$
  \label{HamiltonianVectorFieldsAreInfinitesimalSymplectomorphisms}
\end{proposition}
\proof
  Using Cartan's formula for the Lie derivative
  $$
    \mathcal{L}_v = \mathbf{d} \circ \iota_v + \iota_v \circ \mathbf{d}
  $$ 
  and the defining condition that the symplectic form is closed and  
  that there is a function $H$ with $\mathbf{d} H = \iota_v \omega$,
  one finds that 
  the Lie derivative of $\omega$ along $v$ is 
  given by 
  $$
    \mathcal{L}_v \omega = \mathbf{d}\iota_v \omega + \iota_v \mathbf{d}\omega  
     = 
	 \mathbf{d}^2 H = 0
   \,.
  $$
\endofproof
Since infinitesimal symmetries should form a Lie algebra, this motivates the following
definition.
\begin{definition}[Poisson bracket for symplectic manifolds]
  Let $(X,\omega)$ be a symplectic manifold.
  Given two functions $f,g \in C^\infty(X)$ with Hamiltonian
  vector fields $v$ and $w$, def. \ref{HamiltonEquations}, respectively, their \emph{Poisson bracket}
  is the function obtained by evaluating the symplectic form on these two vector fields
  $$
    \{f,g\} := \iota_w\iota_v\omega
	\,.
  $$
  This operation
  $$
    \{-,-\} : C^\infty(X) \otimes C^\infty(X) \longrightarrow C^\infty(X)
  $$
  is skew symmetric and satisfies the Jacobi identity. Therefore
  $$
    \mathfrak{pois}(X,\omega) := 
	\left(
	  C^\infty(X),\,
	  \{-,-\}
	\right)
  $$
  is a Lie algebra (infinite dimensional in general), called the 
  \emph{Poisson bracket Lie algebra of classical observables} of the symplectic manifold $X$.
  \label{PoissonBracket}
\end{definition}
\begin{remark}
Below in \ref{QuantizationTheHeisenbergGroupAndSliceAutomorphismGroups}
we indicate a general abstract characerization of the Poisson bracket Lie algebra
(which is discussed in moreo detail below in \ref{QuantomorphismAndHeisenbergGroup}):
it is the Lie algebra of ``the automorphism group of any prequantization
of $(X,\omega)$ in the higher slice topos over the moduli stack of circle-principal
connections'' \cite{hgp}. To state this we first need the notion of 
\emph{pre-quantization}
which we come to below in \ref{TheKineticActionPreQuantizationAndDifferentialCohomology}.
In the notation introduced there we will
discuss in \ref{QuantizationTheHeisenbergGroupAndSliceAutomorphismGroups} that the
Poisson bracket is given as
$$
  \mathfrak{pois}(X,\omega)
  = 
  \mathrm{Lie}
  \left(
  \mathbf{Aut}_{/\mathbf{B}U(1)_{\mathrm{conn}}}\left(\nabla\right)
  \right)
  = 
  \left\{
    \raisebox{20pt}{
    \xymatrix{
	  X \ar[rr]^\simeq_{\ }="s" \ar[dr]_{\nabla}^{\ }="t"  &&  X \ar[dl]^\nabla
	  \\
	  & \mathbf{B}U(1)_{\mathrm{conn}}
	  \ar@{=>} "s"; "t"
	}
	}
  \right\}
  \,,
$$
  where $\nabla$ denotes a pre-quantization of $(X,\omega)$.
\end{remark}
This general abstract construction makes sense also for pre-symplectic 
manifolds and shows that the following slight generalization of the above traditonal
definition is good and useful.
\begin{definition}[Poisson bracket for pre-symplectic manifolds]
 \index{Poisson bracket!for pre-symplectic manifolds}
  For $(X,\omega)$ a pre-symplectic manifold, denote by 
  $\mathfrak{pois}(X,\omega)$ the Lie algebra whose underlying 
  vector space is the space of pairs
  of Hamiltonians $H$ with a \emph{choice} of Hamiltonian vector field $v$
  $$
    \left\{
	  (v, H) \in \Gamma(T X) \otimes C^\infty(X)
	  \;|\;
	  \iota_v \omega = \mathbf{d}H
	\right\}
	\,,
  $$
  and whose Lie bracket is given by
  $$
    \left[
	  \left(v_1, H_1\right), \left(v_2, H_2\right)
	\right]
	=
	\left(
	  \left[v_1, v_2\right],
	  \;
	  \iota_{v_1 \wedge v_2} \omega
	\right)
	\,.
  $$
  \label{PoissonBracketForPresymplectic}
\end{definition}
\begin{remark}
  On a smooth manifold $X$ there is a bijection between smooth vector fields
  and derivations of the algebra $C^\infty(X)$ of smooth functions, 
  given by identifying a vector field $v$ with the operation $v(-)$
  of differentiating functions along $v$.
  Under this identification the Hamiltonian vector field $v$ corresponding 
  to a Hamiltonian $H$ is identified with the derivation given by
  forming the Poisson bracket with $H$:
  $$
    v(-) = \{H,-\} \;:\; C^\infty(X) \longrightarrow C^\infty(X)
	\,.
  $$
  \label{HamiltonianVectorFieldAsDerivation}
\end{remark}

In applications in physics, given a phase space $(X,\omega)$ typically
one smooth function $H : X \longrightarrow \mathbb{R}$,
interpreted as the energy observable, is singled out
and called \emph{the} Hamiltonian. Its corresponding Hamiltonian
vector field is then interpreted as giving the infinitesimal 
time evolution of the system, and this is where Hamilton's equations
in def. \ref{HamiltonEquations} originate. 
\begin{definition}
  Given a phase space with Hamiltonian $((X,\omega), H)$, then
  any other classical $O \in C^\infty(X)$,
  it is called an \emph{infinitesimal symmetry} of $((X, \omega), H)$ if 
  the Hamiltonian vector field $v_O$ of $O$ preserves not just
the symplectic form 
(as it automatically does by prop. \ref{HamiltonianVectorFieldsAreInfinitesimalSymplectomorphisms} )
but also the given Hamiltonian, in that  $\iota_{v_O}\mathbf{d}H = 0$.
 \label{InfinitesimalSymmetryOfPhaseSpaceWithTimeEvolution}
\end{definition}
\begin{proposition}[symplectic Noether theorem]
  If a Hamiltonian vector field $v_O$ is an infinitesimal symmetry of 
  a phase space $(X,\omega)$ with time evolution $H$ according 
  to def. \ref{InfinitesimalSymmetryOfPhaseSpaceWithTimeEvolution}, then 
  the corresponding Hamiltonian function $O\in C^\infty(X)$ is 
  a conserved quantity along the time evolution, in that
  $$
    \iota_{v_H} \mathbf{d}O = 0
	\,.
  $$
  Conversely, if a function $O \in C^\infty(X)$ is preserved by the
  time evolution of a Hamiltonian $H$ in this way, then its 
  Hamiltonian vector field $v_O$ is an infinitesimal symmetry
  of $((X,\omega), H)$.
  \label{SymplecticNoetherTheorem}
\end{proposition}
\proof
  This is immediate from the definition \ref{HamiltonEquations}:
  $$
    \begin{aligned}
	  \iota_{v_H} \mathbf{d}O & = \iota_{v_H} \iota_{v_O} \omega
	  \\
	  & = - \iota_{v_O} \iota_{v_H}  \omega
	  \\
	  & = \iota_{v_O} \mathbf{d}H
	\end{aligned}
	\,.
  $$
\endofproof
\begin{remark}
  The utter simplicity of the proof of prop. \ref{SymplecticNoetherTheorem}
  is to be taken as a sign of the power of the symplectic formalism
  in the formalization of physics, not as a sign that the statement
  itself is shallow. On the contrary, under a Legendre
  transform and passage from ``Hamiltonian mechanics'' to 
  ``Lagrangian mechanics'' that we come to 
   below in \ref{TheClassicalActionFunctionalPrequantizesLagrangianCorrespondences},
   the identification of symmetries with preserved observables in 
   prop. \ref{TheClassicalActionFunctionalPrequantizesLagrangianCorrespondences}
   becomes the seminal \emph{first Noether theorem}. See for instance
   \cite{Butterfield} for a review of the Lagrangian Noether theorem
   and its symplectic version in the context of classical mechanics.
   Below in \ref{LocalObservablesAndHigherPoissonBracketHomotopyLieAlgebras} 
   we observe that the same holds true also
   in the full context of classical field theory, if only one
   refines Hamiltonian mechanics to its localization by 
   Hamilton-de Donder-Weyl field theory. The full
   \emph{$n$-plectic Noether theorem} (for all field theory dimensions $n$)
   is prop. \ref{nPlecticNoetherTheorem} below. 
\end{remark}
In the next section we pass from infinitesimal Hamiltonian flows
to their finite version, the Hamiltonian symplectomorphism.

\paragraph{Hamiltonian (time evolution) trajectories and Hamiltonian correspondences}
\label{HamiltonianTimeEvolutionTrajectoriesAndHamiltonianCorrespondences}

We have seen so far transformations of phase space given by ``canonical transformations'',
hence symplectomorphisms. Of central importance in physics are of course those
transformations that are part of a smooth evolution group, notably for time evolution.
These are the ``canonical transformations'' coming from a generating function,
hence the symplectomorphisms which come from a Hamiltonian function
(the energy function, for time evolution),
the \emph{Hamiltonian symplectomorphisms}.
Below in \ref{TheClassicalActionLegendreTransformAndHamuiltonianFlows}
we see that this notion is implied by prequantizing Lagrangian correspondences,
but here it is good to recall the traditional definition.

\begin{definition}
  The flow of a Hamiltonian vector field is called the corresponding
  \emph{Hamiltonian flow}.
\end{definition}
Notice that by prop. \ref{HamiltonianVectorFieldsAreInfinitesimalSymplectomorphisms} we have
\begin{proposition}
  Every Hamiltonian flow is a symplectomorphism.
  \label{HamiltonianFlowIsSymplectomorphism}
\end{proposition}
Those symplectomorphisms arising this way are called the \emph{Hamiltonian symplectomorphisms}.
Notice that the Hamiltonian symplectomorphism depends on the Hamiltonian only
up to addition of a locally constant function.

Using the Poisson bracket $\{-,-\}$ induced by the symplectic form $\omega$,
identifying the derivation $\{H,-\} : C^\infty(X) \longrightarrow C^\infty(X)$ 
with the corresponding Hamiltonian vector field $v$ 
by remark \ref{HamiltonianVectorFieldAsDerivation} 
and the exponent notation 
$\exp(t \{H,-\})$ with the Hamiltonian flow for parameter 
``time'' $t \in \mathbb{R}$, we may write these Hamiltonian symplectomorphisms as 
$$
  \exp( t \{H,-\}) \;\colon\; (X,\omega) \longrightarrow (X,\omega)
  \,.
$$ 
It then makes sense to say that 
\begin{definition}
A Lagrangian correspondence, def. \ref{LagrangianCorrespondence}, 
which is  induced from a Hamiltonian symplectomorphism is a 
\emph{Hamiltonian correspondences}
$$
  \left(
  \raisebox{20pt}{
  \xymatrix{
    & \mathrm{graph}\left(\exp\left(t \left\{H,-\right\}\right)\right)
	\ar[dl]_{i_1}
	\ar[dr]^{i_2}
	\\
	X
	&&
	X
  }}
  \right)
  \;\;\;
  \simeq
  \;\;\;
  \left(
  \raisebox{20pt}{
  \xymatrix{
    & X
	\ar[dl]_{=}
	\ar[dr]^{\exp\left(t \left\{H,-\right\}\right)}
	\\
	X
	&&
	X
  }}
  \right)
  \,.
$$
\end{definition}
\begin{remark}
The smooth correspondence space of a
Hamiltonian correspondence is naturally identified with the space of
\emph{classical trajectories}
$$
  \mathrm{Fields}_{\mathrm{traj}}^{\mathrm{class}}(t) 
    := 
  \mathrm{graph}\left( \exp(t) \{H,-\}\right)
$$
in that 
\begin{enumerate}
\item every point in the space corresponds uniquely to a trajectory of parameter time length $t$ characterized as satisfying the equations of motion as given by Hamilton's equations for $H$;
\item the two projection maps to $X$ send a trajectory to its initial and to its final configuration, respectively.
\end{enumerate}
\end{remark}
group structure is 
\begin{remark}
By construction, Hamiltonian flows form a 1-parameter Lie group. 
By prop. \ref{LagrangianCorrespondencesAsCorrespondencesOverOmegs2} this 
group structure is preserved by the composition of the induced
Hamiltonian correspondences. 
\end{remark}
It is useful to highlight this formally as follows.
\begin{definition}
Write $\mathrm{Bord}^{\mathrm{Riem}}_1$ 
for the category of 1-dimensional cobordisms equipped with Riemannian structure
(hence with a real, non-negative length which is additive under composition), 
regarded as a symmetric monoidal category under disjoint union of cobordisms.
\end{definition}
Then:
\begin{proposition}
The Hamiltonian correspondences induced by a Hamiltonian function 
$H \;:\; X \longrightarrow \mathbb{R}$
are equivalently encoded in a smooth monoidal functor of the form
$$
  \exp((-)\{H,-\}) 
    \;\colon\; 
  \mathrm{Bord}^{\mathrm{Riem}}_1 \longrightarrow \mathrm{Corr}_1(\mathbf{H}_{/\Omega^2})
  \,,
$$
where on the right we use the monoidal structure on 
correspondence of prop. \ref{CorrespondencesOverOmegs2IsMonoidalCategory}.
\label{EvolutionFunctorInducedByHamiltonian}
\end{proposition}
Below the general discussion of prequantum field theory, such monoidal
functors from cobordisms to correspondences of spaces of field configurations
serve as the fundamental means of axiomatization. Whenever one is 
faced with such a functor, it is of particular interest to consider 
its value on \emph{closed} cobordisms. Here in the 1-dimensional case
this is the circle, and the value of such a functor on the circle
would be called its (pre-quantum) \emph{partition function}.
\begin{proposition}
  Given a phase space symplectic manifold $(X,\omega)$ and a 
  Hamiltonian $H : X \longrightarrow \mathbb{R}$, them 
  the prequantum evolution functor of prop. \ref{EvolutionFunctorInducedByHamiltonian}
  sends the circle of circumference  $t$, 
  regarded as a cobordism from the empty 0-manifold to itself
  $$
    \raisebox{20pt}{
    \xymatrix{
	  & S^1
	  \\
	  \emptyset
	  \ar@{^{(}->}[ur]
	  &
	  &
	  \emptyset
	  \ar@{_{(}->}[ul]
	}}
  $$
  and equipped with the constant Riemannian metric of 1-volume $t$,
  to the correspondence
  $$
    \xymatrix{
	   & \left\{ x\in X | \exp(t \{H,-\})(x) = x  \right\}
	   \ar[dl]
	   \ar[dr]
	   \\
	  \ast && \ast
	}
  $$
  which is the smooth space of $H$-Hamiltonian trajectories
  of (time) length $t$ that are closed, hence that come back to their
  initial value, regarded canonically as a correspondence form the 
  point to itself.
\end{proposition}
\proof
  We can decompose the circle of length $t$ as the compositon of
  \begin{enumerate}
    \item The coevaluation map on the point, regarded as a dualizable object
	$\mathrm{Bord}_1^{\mathrm{Riem}}$;
	\item the interval of length $t$;
	\item the evaluation map on the point.
  \end{enumerate}
  The monoidal functor accordingly takes this to the composition
  of correspondences of 
  \begin{enumerate}
    \item the coevaluation map on $X$, regarded as a dualizable object in 
	$\mathrm{Corr}(\mathbf{H})$;
	\item the Hamiltonian correspondence induced by $\exp(t \{H,-\})$;
	\item the evaluation map on $X$.
  \end{enumerate}
  As a diagram in $\mathbf{H}$, this is the following:
  $$
    \raisebox{20pt}{
    \xymatrix{
	   & X \ar[dl] \ar[dr]|{\Delta} && 
	    \mathrm{graph}(\exp(t\{H,-\})) \times X 
		\ar[dl]
		\ar[dr]
		&&
		X
		\ar[dl]|{\Delta}
		\ar[dr]
	   \\
	  \ast && X\times X && X \times X && \ast
	}
	}
	\,.
  $$
  By the definition of composition in $\mathrm{Corr}(\mathbf{H})$, the 
  resulting composite correspondence space is the joint fiber product
  in $\mathbf{H}$
  over these maps. This is essentially verbatim the diagrammatic definition
  of the space of closed trajectories of parameter length $t$.
\endofproof

\paragraph{Noether symmetries and equivariant structure}
\label{NoetherSymmetriesAndEquivariantStructures}

So far we have considered smooth spaces equipped with differential forms, and
correspondences between these. To find genuine classical mechanics and in particular
find the notion of prequantization, we need to bring the notion of 
\emph{gauge symmetry} into the picture. 
We introduce here symmetries in classical field theory
following Noether's seminal analysis
and then point out the crucial notion of \emph{equivariance} of symplectic potentials
necessary to give this global meaning. Below in \ref{GaugeTheorySmoothGroupoidsAndHigherToposes}
we see how building the \emph{reduced phase space} by taking the symmetries
into account makes the first little bit of ``higher differential geometry'' appear
in classical field theory.
 \begin{definition}
   Given a local Lagrangian as in example \ref{PhaseSpaceFromLocalActionFunctionalOnTheLine}
   A \emph{symmetry} of $L$ is a vector field $v \in \Gamma(T P X)$ such that 
   $\iota_v \mathbf{\delta}L = 0$. 
   It is called a \emph{Hamiltonian symmetry} if restricted to phase space
   $v$ is a Hamiltonian vector field, in that 
   the contraction $\iota_v \omega$ is exact. 
 \end{definition}
 By definition of $\theta$ and $\mathrm{EL}$ 
 in example \ref{PhaseSpaceFromLocalActionFunctionalOnTheLine}, it follows that for $v$
 a symmetry, the 0-form 
 $$
   J_v := \iota_v \theta
 $$
 is closed with respect to the time differential
 $$
   \mathbf{d}_t J_v = 0
   \,.
 $$
 \begin{definition}
   The function $J_v$ induced by a symmetry $v$ is called the 
   \emph{conserved Noether charge}
   of $v$.
 \end{definition}
 \begin{example}
   For $Y = \mathbb{R}$ and $L = \tfrac{1}{2}\dot \gamma^2 \mathbf{d}t$
   the vector field $v$ tangent to the flow $\gamma \mapsto \gamma((-) + a)$
   is a symmetry. This is such that $\iota_v \delta \gamma = \dot \gamma$. 
   Hence the conserved quantity is $E := J_v = \dot \gamma^2$, the energy 
   of the system. It is also a Hamiltonian symmetry.
 \end{example}
 Let then $G$ be the group of Hamiltonian symmetries acting on 
 $(\{\mathrm{EL} = 0\}, \omega = \delta \theta)$.
 Write $\mathfrak{g} = \mathrm{Lie}(G)$ for the Lie algebra of the Lie group.
 Given $v \in \mathfrak{g} = \mathrm{Lie}(G)$
 identify it with the corresponding Hamiltonian vector field. Then it 
 follows that the Lie derivative of $\theta$ is exact, hence that for every 
 $v$ one can find an $h$ such that 
 $$
   \mathcal{L}_v \theta = \mathbf{d}h
   \,.
 $$
 The choice of $h$ here is a choice of identification that relates 
 the phase space potential $\theta$ to itself under a different
 but equivalent perspective of what the phase space points are. 
 Such choices of ``gauge equivalences'' are necessary in order to 
 give the (pre-)symplectic form on the unreduced phase space an 
 physical meaning in view of the symmetries of the system. Moreover,
 what is really necessary for this is a coherent choice of such 
 gauge equivalences also for the ``global'' or ``large'' gauge transformations
 that may not be reached by exponentiating Lie algebra elements of the 
 symmetry group $G$.  Such a coherent choice of gauge equivalences
 on $\theta$ reflecting the symmetry of the physical system is 
 mathematically called a \emph{$G$-equivariant structure}.
 \begin{definition}
   Given a smooth space $X$ equipped with the action $\rho : X \times G \longrightarrow X$ 
   of a smooth group, and given a differential 1-form $\theta \in \Omega^1(X)$, 
   and finally given a discrete subgroup $\Gamma \hookrightarrow \mathbb{R}$, then
   a \emph{$G$-equivariant structure} on $\theta$ regarded as a $(\mathbb{R}/\Gamma)$-principal
   connection is
   \begin{itemize}
     \item for each $g \in G$ an equivalence 
	 $$  
	   \eta_g
	   :
	   \xymatrix{
	     \theta \ar[r]^-\simeq & \rho(g)^\ast \theta
	   }
	 $$
	 between $\theta$ and the pullback of $\theta$
	 along the action of $g$, hence a smooth function $\eta_g \in C^\infty(X, \mathbb{R}/\Gamma)$
	 with
	 $$
	   \rho(g)^\ast \theta - \theta = \mathbf{d} \eta_g
	 $$
   \end{itemize}
   such that 
   \begin{enumerate}
     \item the assignment $g \mapsto \eta_g$ is smooth;
	 \item for all pairs $(g_1,g_2) \in G \times G$ there is an equality
	 $$
	   \eta_{g_2} \eta_{g_1} = \eta_{g_2 g_1}
	   \,.
	 $$
   \end{enumerate}
   \label{EquivariantStructureOnConnection1Form}
 \end{definition}
 \begin{remark}
   Notice that the condition $\rho(g)^\ast \theta - \theta = \mathbf{d} \eta_g$
   depends on $\eta_g$ only modulo elements in the discrete group $\Gamma \hookrightarrow \mathbb{R}$,
   while the second condition $\eta_{g_2} \eta_{g_1} = \eta_{g_2 g_1}$ crucially
   depends on the actual representatives in $C^\infty(X,\mathbb{R}/\Gamma)$. 
   For $\Gamma$ the trivial group there is no difference, but in general it is 
   unlikely that in this case the second condition may be satisfied. The
   second condition can in general only be satisfied modulo some subgroup of 
   $\mathbb{R}$. Essentially the only such which yields a regular quotient
   is $\mathbb{Z} \hookrightarrow \mathbb{R}$ (or any non-zero rescaling of this), in which case
   $$
     \mathbb{R}/\mathbb{Z} \simeq U(1)
   $$
   is the circle group. This is the origin of the central role of 
   \emph{circle principal bundles} in field theory (``prequantum bundles''), 
   to which we come below
   in \ref{TheKineticActionPreQuantizationAndDifferentialCohomology}.
 \end{remark}
The point of $G$-equivariant structure is that it makes 
the (pre-)symplectic potential $\theta$ ``descend'' to the 
quotient of $X$ by $G$ (the ``correct quotient'', in fact), 
which is the \emph{reduced phase space}. 
To say precisely what this means, we now introduce the 
concept of smooth groupoids in \ref{GaugeTheorySmoothGroupoidsAndHigherToposes}.
\begin{remark}
 This equivariance on local Lagrangian is one of the motivations 
 for refining the discussion here to \emph{local prequantum field theory}
 in \cite{lpqft}: By remark \ref{LagrangiansFromTransgression} 
 for a genuine $n$-dimensional field theory, the
Lagrangian 1-form $L$ above is the transgression of an $n$-form Lagrangian 
on a moduli space of fields. In local prequantum field theory we impose an
equivariant structure already on this de-transgressed $n$-form Lagrangian
such that under transgression it then induces equivariant structures in 
codimension 1, and hence consistent phase spaces, in fact consistent prequantized
phase spaces.
\end{remark}

\paragraph{Gauge theory, smooth groupoids and higher toposes}
\label{GaugeTheorySmoothGroupoidsAndHigherToposes}

As we mentioned in \ref{MotivationByGaugeTheory} \emph{gauge principle} is 
a deep principle of modern physics, which says that in general two configurations
of a physical system may be nominally different and still be identified by a
\emph{gauge equivalence} between them. In homotopy type theory precisely this
principle is what is captured by \emph{intensional identity types}
(see remark \ref{IdentityTypes}).
One class of example of such gauge equivalences in physics
are the Noether symmetries induced by local Lagrangians which we considered above 
in \ref{NoetherSymmetriesAndEquivariantStructures}.
Gauge equivalences can be  composed (and associatively so) and
can be inverted. All physical statements respect this
gauge equivalence, but it is wrong to identify gauge equivalent field configurations
and pass to their sets of equivalence classes, as some properties depend
on non-trivial auto-gauge transformations.

In mathematical terms what this says is precisely that field configurations and
gauge transformations between them form what is called a \emph{groupoid}
or \emph{homotopy 1-type}. 
\begin{definition}
  A \emph{groupoid} $\mathcal{G}_\bullet$ is a set $\mathcal{G}_0$ 
  -- to be called its set of of \emph{objects}
  or \emph{configurations} --
  and a set $\mathcal{G}_1 = \left\{ \left(x_1 \stackrel{f}{\longrightarrow} x_2\right) | x_1,x_2 \in \mathcal{G}_0 \right\}$ 
  -- to be called the set of \emph{morphisms} or \emph{gauge transformations} -- 
  between these objects, together with 
  a partial composition operation of morphisms over common objects
  $$
    f_2 \circ f_1
	 :
	\xymatrix{
	  x_1 \ar[r]^{f_1} & x_2 \ar[r]^{f_2} & x_3
	}
  $$
  which is associative, and for which every object has a unit 
  (the identity morphism $\mathrm{id}_x : x \to x$) and such that every 
  morphism has an inverse.
\end{definition}
The two extreme examples are:
\begin{example}
  For $X$ any set, it becomes a groupoid by considering for each object
  an identity morphism and no other morphisms.
\end{example}
\begin{example}
  For $G$ a group, there is a groupoid which we denote $\mathbf{B}G$ defined to 
  have a single object $\ast$, one morphism from that object to itself for each 
  element of the group
  $$
    (\mathbf{B}G)_1 = \left\{ \ast \stackrel{g}{\longrightarrow} \ast \;|\; g \in G \right\}
  $$
  and where composition is given by the product operation in $G$.
  \label{DeloopingGroupoid}
\end{example}
The combination of these two examples which is of central interest here is the following.
\begin{example}
  For $X$ a set and $G$ a group with an action $\rho : X \times G \longrightarrow X$
  on $X$, the corresponding \emph{action groupoid} or \emph{homotopy quotient}, 
  denoted $X/\!/G$, is the groupoid whose
  objects are the elements of $X$, and whose morphisms are of the form
  $$
    \xymatrix{
      x_1 \ar[r]^-g & (x_2 = \rho(g)(x_1)) 
	}
  $$
  with composition given by the composition in $G$.
  \label{ActionGroupoid}
\end{example}
\begin{remark}
  The homotopy quotient is a refinement of the actual quotient $X/G$ in which 
  those elements of $X$ which are related by the $G$-action are actually 
  \emph{identified}. In contrast to that, the homotopy quotient makes 
  element which are related by the action of the ``gauge'' group $G$ be
  \emph{equivalent without being equal}. Moreover it remember \emph{how}
  two elements are equivalent, hence which ``gauge transformation'' relates them.
  This is most striking in example \ref{DeloopingGroupoid},
  which is in fact the special case of the homotopy quotient construction
  for the case that $G$ acts on a single element:
  $$
    \mathbf{B}G \simeq \ast /\!/G
	\,.
  $$
\end{remark}
Therefore given an unreduced phase space $X$ as in \ref{PhaseSpacesAndSymplecticManifolds} 
and equipped with an action of a gauge symmetry group 
as in \ref{NoetherSymmetriesAndEquivariantStructures}, then the corresponding
\emph{reduced phase space} should be the homotopy quotient $X/\!/G$, hence the
space of fields with gauge equivalences between them.
But crucially for physics, this is not just a discrete set of points 
with a discrete set of morphisms between them, as in the
above definition, but in addition to the information about field configurations
and gauge equivalences between them carries a \emph{smooth structure}.

We therefore need a definition of \emph{smooth groupoids}, hence of homotopy types
which carry \emph{differential geometric} structure. Luckily, the definition 
in \ref{CoordinateSystemsAndTheToposOfSmoothSpaces} of smooth spaces immediately
generalizes to an analogous definition of smooth groupoids.

First we need the following obvious notion.
\begin{definition}
  Given two groupoids $\mathcal{G}_\bullet$ and $\mathcal{K}_\bullet$,
  a homomorphism $F_\bullet : \mathcal{G}_\bullet \longrightarrow \mathcal{K}_\bullet$ 
  between them (called a \emph{functor}) is a function
  $F_1 : \mathcal{G}_1 \longrightarrow \mathcal{K}_1$ between the sets of morphisms
  such that identity-morphisms are sent to identity morphisms and such that
  composition is respected.
\end{definition}
Groupoids themselves are subject to a notion of gauge equivalence:
\begin{definition}
 A functor $F_\bullet$ is called an \emph{equivalence of groupoids} if 
 its image hits every equivalence class of objects in $\mathcal{K}_\bullet$
 and if for all $x_1, x_2 \in \mathcal{G}_0$ the map $F_1$ restricts to a
 bijection between the morphisms from $x_1$ to $x_2$ in $\mathcal{G}_\bullet$
 and the morphisms between $F_0(x_1)$ and $F_0(x_2)$ in $\mathcal{K}_\bullet$.
 \label{EquivalenceOfGroupoids}
\end{definition}
With that notion we can express coordinate transformations between smooth groupoids
and arrive at the following generalization of def. \ref{SmoothSpaceInComponents}.
\begin{definition}
  A \emph{smooth groupoid} or \emph{smooth homotopy 1-type} $X_\bullet$ is 
 \begin{enumerate}
  \item an assignment to each $n \in \mathbb{N}$ of a 
  groupoid, to be written $X_\bullet(\mathbb{R}^n)$ and to be called the 
  \emph{groupoid of smooth maps from $\mathbb{R}^n$ into $X$ and gauge transformations between these},
  \item an assignment to each ordinary smooth function 
  $f : \mathbb{R}^{n_1} \to \mathbb{R}^{n_2}$
  between Cartesian spaces
     of 
	a functor of groupoids $X(f) : X_\bullet(\mathbb{R}^{n_2}) \to X_\bullet(\mathbb{R}^{n_1})$,
	to be called the \emph{pullback of smooth functions into $X$ along $f$};
  \end{enumerate}
  such that both the components $X_0$ and $X_1$ form a smooth space according to 
  def \ref{SmoothSpaceInComponents}.
  \label{SmoothGroupoidInComponents}
\end{definition}
With this definition in hand we can now form the reduced phase space in a way
that reflects both its smooth structure as well as its gauge-theoretic structure:
\begin{example}
  Given a smooth space $X$ and a smooth group $G$ with a smooth action
  $\rho : X \times G \longrightarrow X$, then the \emph{smooth homotopy quotient}
  of this action is the smooth groupoid, def. \ref{SmoothGroupoidInComponents}. 
  which on each coordinate chart is the homotopy quotient, def. \ref{ActionGroupoid},
  of the coordinates of $G$ acting on the coordinates of $X$, hence the assignment
  $$
    X/\!/G 
	  : 
	\mathbb{R}^n 
	  \mapsto 
	\left( X\left(\mathbb{R}^n\right)\right)/\!/\left(G\left(\mathbb{R}^n\right)\right)
	\,.
  $$
  \label{SmoothHomotopyQuotient}
\end{example}
\begin{remark}
  In most of the physics literature only the infinitesimal approximation to the 
  smooth homotopy quotient $X/\!/G$ is considered, that however is famous: it is the
  \emph{BRST complex} of gauge theory \cite{HenneauxTeitelboim}.
  More in detail, to any Lie group $G$ is associated a Lie algebra $\mathfrak{g}$,
  which is its ``infinitesiamal approximation'' in that it consists of the first 
  order neightbourhood of the neutral element in $G$, equipped with the 
  first linearized group structure, incarnated as the Lie bracket. 
  In direct analogy to this, a smooth grouppoid such as $X/\!/G$ has an infinitesimal
  approximation given by a \emph{Lie algebroid}, a vector bundle on $X$ whose
  fibers form the first order neighbourhood of the smooth space of morphisms at the
  identity morphisms. Moreover, Lie algebroids can equivalently be encoded dually
  by the algebras of functions on these first order neighbourhoods. 
  These are differential graded-commutative algebras and the dgc-algebra
  associated this way to the smooth groupoid $X/\!/G$ is what in the 
  physics literature is known as the BRST complex. 
\end{remark}
To correctly capture the interplay between the differential geometric structure 
and the homotopy theoretic structure in this definition we have to in addition declare the 
following
\begin{definition}
  A homomorphism $f_\bullet: X_\bullet \longrightarrow Y_\bullet$ of smooth groupoids
  is called a \emph{local equivalence} if it is a \emph{stalkwise} equivalence of 
groupoids, hence if for each Cartesian space $\mathbb{R}^n$
  and for each point $x \in \mathbb{R}^n$, there is an open neighbourhood
  $\mathbb{R}^n \simeq U_x \hookrightarrow \mathbb{R}^n$ such that 
  $F_\bullet$ restricted to this open neighbourhood is an equivalence of groupoids
  according to def. \ref{EquivalenceOfGroupoids}.
\end{definition}
\begin{definition}
  The \emph{$(2,1)$-topos of smooth groupoids} is the 
  homotopy theory obtained from the category $\mathrm{Sh}(\mathrm{CartSp}, \mathrm{Grpd})$
  of smooth groupoids by universally turning the local equivalences into actual equivalences,
  def. \ref{HomotopyToposByBousfieldLocalization}.
  \label{21ToposOfSmoothGroupoids}
\end{definition}
This refines the construciton of the topos of smooth spaces form before, and hence we 
find it convenient to use the same symbol for it:
$$
  \mathbf{H} := \mathrm{Sh}(\mathrm{CartSp}, \mathrm{Grpd})[\{\mbox{local equivalences}\}^{-1}]
  \,.
$$

\paragraph{The kinetic action, pre-quantization and differential cohomology}
\label{TheKineticActionPreQuantizationAndDifferentialCohomology}

The refinement of gauge transformations of differential 1-forms to 
coherent $U(1)$-valued functions which we have seen in the construction of the 
reduced phase space above in \ref{NoetherSymmetriesAndEquivariantStructures}
also appears in physics from another angle, which is not explicitly gauge theoretic,
but related to the global definition of the exponentiated action functional.

Given a pre-symplectic form $\omega \in \Omega^2_{\mathrm{cl}}(X) $, 
by the Poincar{\'e} lemma there is a good cover $\{U_i \hookrightarrow X\}_i$
and smooth 1-forms $\theta_i \in \Omega^1(U_i)$ such that 
$\mathbf{d}\theta_i = \omega_{|U_i}$. Physically such a 1-form is 
(up to a factor of 2) a choice
of \emph{kinetic energy density} called a \emph{kinetic Lagrangian} 
$L_{\mathrm{kin}}$:
$$
  \theta_i = 2 L_{\mathrm{kin}, i}
  \,.
$$
\begin{example}
Consider the phase space $(\mathbb{R}^2, \; \omega = \mathbf{d} q \wedge \mathbf{d} p)$ 
of example \ref{CanonicalR2PhaseSpace}. Since $\mathbb{R}^2$ is a contractible topological space we consider the trivial covering ($\mathbb{R}^2$ covering itself) since this is already a good covering in this case. Then all the $\{g_{i j}\}$ are trivial and the data of a prequantization consists simply of a choice of 1-form $\theta \in \Omega^1(\mathbb{R}^2)$ such that 
$$
  \mathbf{d}\theta = \mathbf{d}q \wedge \mathbf{d}p
  \,.
$$
A standard such choice is 
$$
  \theta = - p \wedge \mathbf{d}q
  \,.
$$
Then given a trajectory $\gamma \colon [0,1] \longrightarrow X$ which satisfies Hamilton's equation for a standard kinetic energy term, then $(p \mathbf{d}q)(\dot\gamma)$ is this kinetic energy of the particle which traces out this trajectory.
\label{StandardPrequantizationOfStandardR2PhaseSpace}
\end{example}

Given a path $\gamma : [0,1] \to X$ in phase space, its 
\emph{kinetic action} $S_{\mathrm{kin}}$ 
is supposed to be the integral of $\mathcal{L}_{\mathrm{kin}}$
along this trajectory. In order to make sense of this in generality with the
above locally defined kinetic Lagrangians $\{\theta_i\}_i$,
there are to be transition functions $g_{i j} \in C^\infty(U_i \cap U_j, \mathbb{R})$
such that
$$
  \theta_j|_{U_j} - \theta_i|_{U_i} = \mathbf{d}g_{i j}
  \,.
$$
If on triple intersections these functions satisfy
$$
  g_{ij} + g_{j k} = g_{i k} \;\;\; \mbox{on $U_i \cap U_j \cap U_K$}
$$
then there is a well defined action functional
$$
  S_{\mathrm{kin}}(\gamma) \in \mathbb{R}
$$
obtained by dividing $\gamma$ into small pieces that each map to a single
patch $U_i$, integrating $\theta_i$ along this piece, and adding the 
contribution of $g_{i j}$ at the point where one switches from using
$\theta_i$ to using $\theta_j$.

However, requiring this condition on triple overlaps as an equation between
$\mathbb{R}$-valued functions makes the local patch structure trivial: if this
holds then one can find a single $\theta \in \Omega^1(X)$ and
functions $h_i \in C^\infty(U_i, \mathbb{R})$ such that 
superficially pleasant effect that the action is 
$\theta_i = \theta|_{U_i} + \mathbf{d}h_i$. This has the
simply the integral against this globally defined 1-form, 
$S_{\mathrm{kin}} = \int_{[0,1]} \gamma^\ast L_{\mathrm{kin}}$, but it also
means that the pre-symplectic form $\omega$ is exact, which is 
not the case in many important examples.

On the other hand, what really matters in physics is not the action functional
$S_{\mathrm{kin}} \in \mathbb{R}$ itself, 
but the \emph{exponentiated} action 
$$
  \exp\left( \tfrac{i}{\hbar} S \right) \in \mathbb{R}/(2\pi \hbar)\mathbb{Z}
  \,.
$$
For this to be well defined, one only needs that the equation
$g_{i j} + g_{j k} = g_{i k}$ holds modulo addtion of an integral
multiple of $h = 2\pi \hbar$, which is \emph{Planck's constant}, def. \ref{PlanckConstant}. 
If this is the case, then one says that the data 
$(\{\theta_i\}, \{g_{i j}\})$ defines 
equivalently
\begin{itemize}
  \item a $U(1)$-principal connection;
  \item a degree-2 cocycle in ordinary differential cohomology 
\end{itemize}
on $X$, with \emph{curvature} the given symplectic 2-form $\omega$.

Such data is called a \emph{pre-quantization} of the symplectic manifold 
$(X,\omega)$. Since it is the exponentiated action functional
$\exp(\frac{i}{\hbar} S)$ that enters the quantization of the 
given mechanical system (for instance as the integrand of a path integral),
the prequantization of a symplectic manifold is indeed precisely
the data necessary before quantization.

Therefore, in the spirit of the above discussion of pre-symplectic structures,
we would like to refine the smooth moduli space of closed 
differential 2-forms to a moduli space of prequantized differential 
2-forms. 

Again this does naturally exist if only we allow for a good notion of
``space''. An additional phenomenon to be taken care of now is that
while pre-symplectic forms are either equal or not, their
pre-quantizations can be different and yet be \emph{equivalent}:

because there is still a remaining freedom to change this data without
changing the exponentiated action along a \emph{closed} path:
we say that a choice of functions 
$h_i \in C^\infty(U_i, \mathbb{R}/(2\pi\hbar)\mathbb{Z})$
defines an equivalence between 
$(\{\theta_i\}, \{g_{i j}\})$ and $(\{\tilde \theta_i\}, \{\tilde g_{i j}\})$
if $\tilde \theta_i - \theta_i = \mathbf{d}h_i$
and $\tilde g_{i j} - g_{i j} = h_j - h_i$.

This means that the space of prequantizations of $(X,\omega)$
is similar to an \emph{orbifold}: it has points which are connected by 
gauge equivalences: there is a \emph{groupoid} of pre-quantum structures
on a manifold $X$. Otherwise this space of prequantizations is similar
to the spaces $\mathbf{\Omega}^2_{\mathrm{cl}}$ of differential forms,
in that for each smooth manifold there is a collection of smooth such data
and it may consistently be pullback back along smooth functions of smooth
manifolds.

As before for the pre-symplectic differential forms
in \ref{CoordinateSystemsAndTheToposOfSmoothSpaces} 
it will be useful to find a moduli space for such prequantum structures.
This certainly cannot exist as a smooth manifold, but due to the 
gauge transformations between prequantizations it can also not exist as
a more general smooth space. However, it does exist as a \emph{smooth groupoid}, 
def. \ref{21ToposOfSmoothGroupoids}.
\begin{definition}
  For $X = \mathbb{R}^n$ a Cartesian space, wrrite $\Omega^1(X)$
  for the set of smooth differential 1-forms on $X$ and write 
  $C^\infty(X,U(1))$ for the set of smooth circle-group valued function on $X$. 
  There is an action
  $$
    \rho
	:
    C^\infty(X,U(1)) \times \Omega^1(\mathbb{R}^n )
	\longrightarrow
	\Omega^1(X, U(1))
  $$
  of functions on 1-forms $A$ by gauge transformation $g$, given by the formula
  $$
    \rho(g)(A) := A + \mathbf{d} \mathrm{log}g
	\,.
  $$
  Hence if $g = \exp(i \kappa)$ is given by the exponential of a smooth 
  real valued function (which is always the case on $\mathbb{R}^n$) then 
  this is
  $$
    \rho(g)(A) := A + \mathbf{d} \kappa
	\,.
  $$
\end{definition}
\begin{definition}
Write
$$
  \mathbf{B}U(1)_{\mathrm{conn}}
  \in \mathbf{H}
  \,,
$$
for the smooth groupoid, def. \ref{SmoothGroupoidInComponents}, which for
Cartesian space $\mathbb{R}^n$ has as groupoid of coordinate charts the
homotopy quotient, def. \ref{ActionGroupoid}, of the smooth functions 
on the coordinate chart acting on the 
smooth 1-forms on the coordinate chart.
$$
  \mathbf{B}U(1)_{\mathrm{conn}}
    \;:\;
  \mathbb{R}^n
    \;\mapsto\;
  \Omega^1(\mathbb{R}) /\!/ \C^\infty(\mathbb{R}^n, U(1))
  \,.
$$
Equivalently this is the smooth homotopy quotient, def. \ref{SmoothHomotopyQuotient},
of the smooth group $U(1) \in \mathbf{H}$ acting on the universal smooth moduli space
$\mathbf{\Omega}^1$ of smooth differential 1-forms:
$$
  \mathbf{B}U(1)_{\mathrm{conn}}
  \simeq
  \mathbf{\Omega}^1/\!/U(1)
  \,.
$$
We call this the \emph{universal moduli stack of prequantizations}
or \emph{universal moduli stack of $U(1)$-principal connections}.
\end{definition}
\begin{remark}
This smooth groupoid $\mathbf{B}U(1)_{\mathrm{conn}} \simeq \mathbf{\Omega}^1/\!/U(1)$
is equivalently characterized by the following properties.
\begin{enumerate}
\item 
 For $X$ any smooth manifold, smooth functions
$$
  \xymatrix{
    X \ar[r] 
	 & 
	\mathbf{B}U(1)_{\mathrm{conn}}
  }
$$
are equivalent to prequantum structures $(\{\theta_i\}, \{g_{i j}\})$ on $X$, 
\item
a homotopy 
$$
  \xymatrix{
    X 
	  \ar@/^1pc/[r]_{\ }="s"
	  \ar@/_1pc/[r]^{\ }="t" 
	 & 
	\mathbf{B}U(1)_{\mathrm{conn}}
	\ar@{=>} "s"; "t"
  }
$$
between two such maps is equivalently a gauge transformation $(\{h_i\})$ 
between these prequantizations.
\end{enumerate}
\end{remark}
\begin{proposition}
There is then in $\mathbf{H}$ a morphism
$$
  F
  :
  \xymatrix{
    \mathbf{B}U(1)_{\mathrm{conn}}
	\ar[r]
	&
	\mathbf{\Omega}^2_{\mathrm{cl}}
  }
$$
from this universal moduli stack of prequantizations back to the 
universal smooth moduli space of closed differential 2-form. This is the 
\emph{universal curvature} map in that for $\nabla : X \longrightarrow \mathbf{B}U(1)_{\mathrm{conn}}$
a prequantization datum $(\{\theta_i\}, \{g_{i j}\})$, the composite
$$
  F_{(-)} 
    : 
  \xymatrix{
    X \ar[r]^-\nabla 
	  & 
	 \mathbf{B}U(1)_{\mathrm{conn}} \ar[r]^-{F_{(-)}} &
   \mathbf{\Omega}^2_{\mathrm{cl}}	
  }
$$
is the closed differential 2-form on $X$ characterized by $\omega|_{U_i} = \mathbf{d}\theta_i$,
for every patch $U_i$. Again, this property characterizes the map $F_{(-)}$ and may 
be taken as its definition.
\end{proposition}
Using this language of the $(2,1)$-topos $\mathbf{H}$ of smooth groupoids,
we may then formally capture the above discussion of prequantization as follows:
\begin{definition}
Given a symplectic manifold $(X,\omega)$, 
regarded by prop. \ref{EmbeddingOfSymplecticManifoldsIntoSliceOverOmega2}
as an object $(X \stackrel{\omega}{\longrightarrow} \mathbf{\Omega}^2_{\mathrm{c}})
\in \mathbf{H}_{/\mathbf{\Omega}^2_{\mathrm{cl}}}$, then 
a \emph{prequantization} of $(X, \omega)$ is a lift $\nabla$ in the diagram
$$
  \xymatrix{
    X \ar[dr]_-{\omega}
	\ar@{-->}[r]^-\nabla
	& \mathbf{B}U(1)_{\mathrm{conn}}
	\ar[d]^{F_{(-)}}
	\\
	& \Omega^2_{\mathrm{cl}}
  }
$$
in $\mathbf{H}$, hence is a lift of $(X,\omega)$ through the 
\emph{base change} functor
(see prop. \ref{LocalCartesianClosureViaBaseChange} for this 
terminology) or \emph{dependent sum} functor
(see def. \ref{DependentTypeTheoryLanguage})
$$
  \underset{F_{(-)}}{\sum}
    \;:\;
  \mathbf{H}_{/\mathbf{B}U(1)_{\mathrm{conn}}}
    \longrightarrow
  \mathbf{H}_{/\mathbf{\Omega}^2_{\mathrm{cl}}}
$$
that goes from the slice over the universal moduli stack of prequantizations to the 
slice over the universal smooth moduli space of closed differential 2-forms.
 \label{PrequantizationAsLiftThroughBaseChangeAlongUniversalCurvature}
\end{definition}
Moreover, in this language of geometric homotopy theory we then also find a 
conceptual re-statement of the descent of the (pre-)symplectic potential
to the reduced phase space, from \ref{NoetherSymmetriesAndEquivariantStructures}:
\begin{proposition}
  Given a covariant phase space $X$ with (pre-)symplectic potential $\theta$
  and gauge group action $\rho : G \times X \longrightarrow X$, 
  a $G$-equivariant structure on $\theta$, 
  def. \ref{EquivariantStructureOnConnection1Form}, is equivalently 
  an extension $\nabla_{\mathrm{red}}$ of $\theta$ along the map to 
  the smooth homotopy quotient $X /\!/G$ as a $(\mathbb{R}/\Gamma)$-principal 
  connection, hence a diagram in $\mathbf{H}$ of the form
  $$
    \raisebox{20pt}{
    \xymatrix{
	  X \ar[d]\ar[r]^-{\theta} & \mathbf{B}U(1)_{\mathrm{conn}}
	  \\
	  X/\!/G \ar@{-->}[ur]_{\nabla_{\mathrm{red}}}
	}
	}
	\,.
  $$
\end{proposition}

\paragraph{The classical action, the Legendre transform and Hamiltonian flows}
\label{TheClassicalActionLegendreTransformAndHamuiltonianFlows}

The reason to consider Hamiltonian symplectomorphisms, 
prop. \ref{HamiltonianFlowIsSymplectomorphism} instead of 
general symplectomorphisms, is really because these give homomorphisms 
not just between plain symplectic manifolds, 
but between their prequantizations, def. \ref{PrequantizationAsLiftThroughBaseChangeAlongUniversalCurvature}. 
To these we turn now.

Consider a morphism
$$
  \raisebox{20pt}{
  \xymatrix{
    X \ar[rr]^{\phi}_{\ }="s" \ar[dr]_{\nabla}^{\ }="t" 
	&& X \ar[dl]^{\nabla}
    \\
    & \mathbf{B}U(1)_{\mathrm{conn}}
	\ar@{=>} "s"; "t"
  }
  }
  \,,
$$
hence a morphism in the slice topos $\mathbf{H}_{/\mathbf{B}U(1)_{conn}}$. This has been discussed in detail in \cite{hgp}.

One finds that infinitesimally such morphisms are given by a Hamiltonian 
and its Legendre transform. 

\begin{proposition}
Consider the phase space $(\mathbb{R}^2, \; \omega = \mathbf{d} q \wedge \mathbf{d} p)$ 
of example \ref{CanonicalR2PhaseSpace} equipped with its canonical prequantization by $\theta = p \mathbf{d}q$ from example \ref{StandardPrequantizationOfStandardR2PhaseSpace}. 
Then for $H \colon \mathbb{R}^2 \longrightarrow \mathbb{R}$ a Hamiltonian, and for $t \in \mathbb{R}$ a parameter ("time"), a lift of the Hamiltonian symplectomorphism 
$\exp(t \{H,-\})$ from $\mathbf{H}$ to the slice topos $\mathbf{H}_{/\mathbf{B}U(1)_{conn}}$ 
is given by
$$
  \raisebox{20pt}{
  \xymatrix{       
    X \ar[rr]^{\exp(t \{H,-\})}_{\ }="s" \ar[dr]_{\theta}^{\ }="t" 
	&& X \ar[dl]^{\theta}
    \\
    & \mathbf{B}U(1)_{\mathrm{conn}}
	\ar@{=>}^{\exp( i S_t  )} "s"; "t"
  }
  }
  \,,
$$
where 
\begin{itemize}
\item $S_t \;\colon\; \mathbb{R}^2 \longrightarrow \mathbb{R}$ is the action functional of the classical trajectories induced by $H$,

\item which is the integral $S_t = \int_{0}^t L \, d t$ of the Lagrangian $L \,d t$ induced by $H$,

\item which is the Legendre transform
  $$
    L := p \frac{\partial H}{\partial p} - H \;\colon\; \mathbb{R}^2 \longrightarrow \mathbb{R}
    \,. 
  $$
\end{itemize}
In particular, this induces a functor
$$
  \exp(i S)
  \;\colon\;
  \mathrm{Bord}_1^{\mathrm{Riem}} 
    \longrightarrow 
  \mathbf{H}_{/\mathbf{B}U(1)_{\mathrm{conn}}}
  \,.
$$
Conversely, a symplectomorphism, being a morphism in $\mathbf{H}_{/\mathbf{\Omega}^2_{cl}}$ is a Hamiltonian symplectomorphism precisely if it admits such a lift 
to $\mathbf{H}_{/\mathbf{B}U(1)_{conn}}$.
\label{HamiltonianTransformationIsPrequantizedByTheExponentiatedAction}
\end{proposition}
This is a special case of the discussion in \cite{hgp}.
\proof
The canonical prequantization of $(\mathbb{R}^2, \mathbf{d} q \wedge \mathbf{d} p)$ is the globally defined connection on a bundle|connection 1-form
$$
  \theta := p \, \mathbf{d} q
  \,.
$$
We have to check that on $graph(\exp(t\{H,-\}))$ we have the equation
$$
  p_2 \wedge \mathbf{d} q_2 = p_1 \wedge \mathbf{d} q_1 + \mathbf{d} S 
  \,.
$$
Or rather, given the setup, it is more natural to change notation to
$$
  p_t \wedge \mathbf{d} q_t = p \wedge \mathbf{d} q + \mathbf{d} S
  \,.
$$
Notice here that by the nature of $\mathrm{graph}(\exp(t\{H,-\}))$ we can identify
$$
  \mathrm{graph}(\exp(t\{H,-\}))
  \simeq
  \mathbb{R}^2
$$
and under this identification
$$
  q_t = \exp(t \{H,-\}) q
$$
and
$$
  p_t = \exp(t \{H,-\}) p
  \,.
$$
It is sufficient to check the claim infinitesimal object|infinitesimally. So let $t = \epsilon$ be an infinitesimal, hence such that $\epsilon^2 = 0$. Then the above is Hamilton's equations and reads equivalently
$$
  q_\epsilon = q + \frac{\partial H}{\partial p} \epsilon
$$
and
$$
  p_\epsilon = p - \frac{\partial H}{\partial q} \epsilon
  \,.
$$
Using this we compute
$$
  \begin{aligned}
    \theta_\epsilon - \theta 
     & = 
    p_\epsilon \wedge \mathbf{d} q \epsilon - p \wedge \mathbf{d} q
     \\
      & =
    \left(p - \frac{\partial H}{\partial q} \epsilon \right)
	\wedge
    \mathbf{d}
    \left(
      q + \frac{\partial H}{\partial p} \epsilon
    \right)
    - p \wedge \mathbf{d}q
    \\
    & =
    \epsilon
    \left(
      p \wedge \mathbf{d}\frac{\partial H}{\partial p}
      - 
      \frac{\partial H}{\partial q} \wedge \mathbf{d}q
    \right)
    \\
    & = 
    \epsilon
    \left(
      \mathbf{d}\left( p \frac{\partial H}{\partial p}\right)
      -
      \frac{\partial H}{\partial p} \wedge \mathbf{d} p
      - 
      \frac{\partial H}{\partial q} \wedge \mathbf{d}q
    \right)
    \\
    & =
    \epsilon \mathbf{d}
    \left(
      p \frac{\partial H}{\partial p}
      -
      H
    \right)
  \end{aligned}
  \,.
$$
\endofproof
\begin{remark}
  When one speaks of symplectomorphisms as ``canonical transformations''
  (see e.g. \cite{Arnold}, p. 206), then 
  the function $S$ in prop. \ref{HamiltonianTransformationIsPrequantizedByTheExponentiatedAction}
  is also known as the ``generating function of the canonical transformation'',
  see \cite{Arnold}, chapter 48.
\end{remark}
\begin{remark}
Proposition \ref{HamiltonianTransformationIsPrequantizedByTheExponentiatedAction} 
says that the slice topos $\mathbf{H}_{/\mathbf{B}U(1)_{conn}}$
unifies classical mechanics in its two incarnations as
Hamiltonian mechanics and as Lagrangian mechanics. A morphism 
here is a diagram in $\mathbf{H}$ of the form
$$
  \xymatrix{
    X \ar[rr] \ar[dr] && Y \ar[dl]
    \\
    & \mathbf{B}U(1)_{\mathrm{conn}}
  }
$$
and which may be regarded as having two components: the top horizontal 1-morphism
as well as the homotopy/2-morphism filling the slice. 
Given a smooth flow of these, the horizontal morphism is the flow
of a Hamiltonian vector field for some Hamiltonian function $H$, 
and the 2-morphism is a $U(1)$-gauge transformation given (locally) by 
a $U(1)$-valued function which is the exponentiated action functional
that is the integral of the Lagrangian $L$ which is the Legendre transform
of $H$.

So in a sense the prequantization lift through the base change/dependent sum
along the universal curvature map
$$
  \underset{F_{(-)}}{\sum}
   \;\colon\;
  \mathbf{H}_{/\mathbf{B}U(1)_{conn}}
  \longrightarrow
  \mathbf{H}_{/\mathbf{\Omega}^2_{cl}}
$$
is the Legendre transform which connects Hamiltonian mechanics with Lagrangian mechanics.
\end{remark}

\paragraph{The classical action functional pre-quantizes Lagrangian correspondences}
\label{TheClassicalActionFunctionalPrequantizesLagrangianCorrespondences}

We may sum up these observations as follows.
\begin{definition}
Given a Lagrangian correspondence
$$
  \xymatrix{
    & \mathrm{graph}(\phi)
	  \ar[dl]_{i_1}
	  \ar[dr]^{i_2}_{\ }="s"
    \\
    X_1 \ar[dr]_{\omega_1}^{\ }="t"
	&& X_2 \ar[dl]^{\omega_2}
	\\
	& \Omega^2_{\mathrm{cl}}
	\ar@{=} "s"; "t"
  }
$$
as in prop. \ref{SymplectomorphismsAsDiagramsCommutingOverOmega2}, 
a \emph{prequantization} of it is a lift of this diagram in $\mathbf{H}$
to a diagram of the form
$$
  \xymatrix{
    & \mathrm{graph}(\phi)
	  \ar[dl]_{i_1}
	  \ar[dr]^{i_2}_{\ }="s"
    \\
    X_1 \ar[ddr]_{\omega_1}
	\ar[dr]|{\nabla_1}^{\ }="t" 
	&& X_2 \ar[ddl]^{\omega_2}
     \ar[dl]|{\nabla_2}	
	\\
	& \mathbf{B}U(1)_{\mathrm{conn}}
	 \ar[d]|{F_{(-)}}
	\\
	& \Omega^2_{\mathrm{cl}}
	\ar@{=>} "s"; "t"
  }
$$
\end{definition}
\begin{remark}
This means that a prequantization of a Lagrangian correspondence is a prequantization of the 
source and target symplectic manifolds by prequantum circle bundles as in 
def. \ref{PrequantizationAsLiftThroughBaseChangeAlongUniversalCurvature},
together with a choice of (gauge) equivalence between thes respective
pullback of these two bundles to the correspondence space.
More abstractly, such a prequantization is a lift through the
base change/dependent sum map along the universal curvature morphism
$$
   \mathrm{Corr}\left(\underset{F_{(-)}}{\sum}\right)
   :
   \mathrm{Corr}\left(\mathbf{H}_{/\mathbf{B}U(1)_{\mathrm{conn}}}\right)
     \longrightarrow
   \mathrm{Corr}\left(\mathbf{H}_{/\mathbf{\Omega}^2_{\mathrm{cl}}}\right)
   \,.
$$
\end{remark}
From prop. \ref{HamiltonianTransformationIsPrequantizedByTheExponentiatedAction}
and under the equivalence of example \ref{EquivalenceOfCorrespndencesInducedByFunction}
it follows that smooth 1-parameter groups of prequantized 
Lagrangian correspondences are equivalently Hamiltonian flows, 
and that the prequantizaton of the underlying Hamiltonian correspondences
is given by the classical action funtional.

In summary, the description of classical mechanics
here identifies prequantized Lagrangian correspondences 
schematically as follows:
$$
  \xymatrix{
    & \mathrm{graph}\left(\exp\left(t\{H,-\}\right)\right)
	\ar[dl]
	\ar[dr]_{\ }="s"
	&
	&&
    & \mbox{\begin{tabular}{c}space of \\ trajectories\end{tabular}}
	\ar[dl]_{\mbox{\small \begin{tabular}{c}initial \\ values\end{tabular}}}
	\ar[dr]^{\mbox{\small \begin{tabular}{c}Hamiltonian \\ evolution\end{tabular}}}_{\ }="s2"
	\\
	X \ar[dr]_{\nabla_{\mathrm{in}}}^{\ }="t"
	  && 
	X \ar[dl]^{\nabla_{\mathrm{out}}}
	&&
	\mbox{\begin{tabular}{c}incoming \\ configurations\end{tabular}}
	\ar[dr]_{\mbox{\small \begin{tabular}{c}prequantum \\ bundle \end{tabular}}}^{\ }="t2"
	&&
	\mbox{\begin{tabular}{c}outgoing \\ configurations\end{tabular}}
	\ar[dl]^{\mbox{\small \begin{tabular}{c}prequantum \\ bundle \end{tabular}}}
	\\
	& \mathbf{B}U(1)_{\mathrm{conn}}
	&
	&&
    & \mbox{\begin{tabular}{c}2-group \\ of phases\end{tabular}}
	\ar@{=>}|{\exp\left(\frac{i}{\hbar} S_t\right) = \exp\left(\frac{i}{\hbar}\int^t_0 L dt\right)} "s"; "t"
	\ar@{=>}|{\mbox{\small \begin{tabular}{c} action \\ functional \end{tabular}}} "s2"; "t2"
  }
$$

This picture of classical mechanics as the theory of 
correspondences in higher slices topos is what allows a seamless
generalization to a local discussion of prequantum field theory in \cite{lpqft}.

\paragraph{Quantization, the Heisenberg group, and slice automorphism groups}
\label{QuantizationTheHeisenbergGroupAndSliceAutomorphismGroups}
\label{TraditionalPrequantumGeometryViaSlicing}

While we do not discussion genuine quantization here
(in a way adapted to the perspective here this is discussed in \cite{Nuiten})
it is worthwhile to notice that the perspective of classical mechanics by correspondences 
in slice toposes seamlessly leads over to
\emph{quantization} by recognizing that the slice automorphism groups
of the prequantized phase spaces
are nothing but the ``quantomorphisms groups'' containing the
famous Heisenberg groups of quantum operators. This has been 
developed for higher prequantum field theory in \cite{hgp}, see
\ref{QuantomorphismAndHeisenbergGroup} below. Here we give
an exposition, which re-amplifies some of the structures already found above.

\medskip

Quantization of course was and is motivated by experiment, hence by observation of the observable universe: it just so happens that quantum mechanics and quantum field theory correctly account for experimental observations where classical mechanics and classical field theory gives no answer or incorrect answers (see for instance \cite{Dirac}). A historically important example is the phenomenon called the "ultraviolet catastrophe", a paradox predicted by classical statistical mechanics which is \emph{not} observed in nature, and which is corrected by quantum mechanics.

But one may also ask, independently of experimental input, if there are good formal mathematical reasons and motivations to pass from classical mechanics to quantum mechanics. Could one have been led to quantum mechanics by just pondering the mathematical formalism of classical mechanics? (Hence more precisely: is there a natural ``Synthetic quantum field theory'' \cite{sQFT}).

The following spells out an argument to this effect.  

So to briefly recall, a system of classical mechanics/prequantum field theory|prequantum mechanics is a phase space, formalized as a symplectic manifold $(X, \omega)$. A symplectic manifold is in particular a Poisson manifold, which means that the algebra of functions on phase space $X$, hence the algebra of 
\emph{classical observables}, is canonically equipped with a compatible Lie bracket: the 
\emph{Poisson bracket}. This Lie bracket is what controls dynamics in classical mechanics. For instance if $H \in C^\infty(X)$ is the function on phase space which is interpreted as assigning to each configuration of the system its energy -- the Hamiltonian function -- then the Poisson bracket with $H$ yields the infinitesimal object|infinitesimal time evolution of the system: the differential equation famous as Hamilton's equations.

Something to take notice of here is the \emph{infinitesimal} nature of the Poisson bracket. Generally, whenever one has a Lie algebra $\mathfrak{g}$, then it is to be regarded as the infinitesimal object|infinitesimal approximation to a globally defined object, the corresponding Lie group (or generally smooth group) $G$. One also says that $G$ is a \emph{Lie integration} of $\mathfrak{g}$ and that $\mathfrak{g}$ is the Lie differentiation of $G$.

Therefore a natural question to ask is: \emph{Since the observables in classical mechanics form a Lie algebra under Poisson bracket, what then is the corresponding Lie group?}

The answer to this is of course "well known" in the literature, in the sense that there are relevant monographs which state the answer. But, maybe surprisingly, the answer to this question is not (at time of this writing) a widely advertized fact that has found its way into the basic educational textbooks. The answer is that this Lie group which integrates the Poisson bracket is the "quantomorphism group", an object that seamlessly leads to the quantum mechanics of the system.

Before we spell this out in more detail, we need a brief technical aside: of course Lie integration is not quite unique. There may be different global Lie group objects with the same Lie algebra. 

The simplest example of this is already one of central importance for the issue of quantization, namely, the Lie integration of the abelian line Lie algebra $\mathbb{R}$. This has essentially two different Lie groups associated with it: the simply connected topological space|simply connected translation group, which is just $\mathbb{R}$ itself again, equipped with its canonical additive abelian group structure, and the discrete space|discrete quotient of this by the group of integers, which is the circle group

$$
  U(1) = \mathbb{R}/\mathbb{Z}
  \,.
$$

Notice that it is the discrete and hence "quantized" nature of the integers that makes the real line become a circle here. This is not entirely a coincidence of terminology, but can be traced back to the heart of what is "quantized" about quantum mechanics.

Namely, one finds that the Poisson bracket Lie algebra $\mathfrak{poiss}(X,\omega)$ of the classical observables on phase space is (for $X$ a connected topological space|connected manifold) a Lie algebra extension of the Lie algebra $\mathfrak{ham}(X)$ of Hamiltonian vector fields on $X$ by the line Lie algebra:

$$
  \mathbb{R} \longrightarrow \mathfrak{poiss}(X,\omega) \longrightarrow \mathfrak{ham}(X)
  \,.
$$

This means that under Lie integration the Poisson bracket turns into an central extension of the group of Hamiltonian symplectomorphisms of $(X,\omega)$. And either it is the fairly trivial non-compact extension by $\mathbb{R}$, or it is the interesting central extension by the circle group $U(1)$. For this non-trivial Lie integration to exist, $(X,\omega)$ needs to satisfy a quantization condition which says that it admits a prequantum line bundle. If so, then this $U(1)$-central extension of the group $Ham(X,\omega)$ of Hamiltonian symplectomorphisms exists and is called... the ``{quantomorphism group}'' 
$\mathrm{QuantMorph}(X,\omega)$:
$$
  U(1) \longrightarrow \mathrm{QuantMorph}(X,\omega) \longrightarrow \mathrm{HamSympl}(X,\omega)
  \,.
$$
More precisely, this group is just the slice automorphism group:
\begin{proposition}
  Let $(X,\omega)$ be a symplectic manifold with prequantization 
  $\nabla : X \longrightarrow \mathbf{B}U(1)_{\mathrm{conn}}$, according to 
  def. \ref{PrequantizationAsLiftThroughBaseChangeAlongUniversalCurvature}, 
  then the smooth automorphism group of $\nabla$ regarded as an object
  in the higher slice topos $\mathbf{H}_{/\mathbf{B}U(1)_{\mathrm{conn}}}$
  is the quantomorphism group $\mathrm{QuantMorph}(X,\omega)$
  $$
    \begin{aligned}
    \mathrm{QuantMorph}(X,\omega)
	& \simeq
	\mathbf{Aut}_{\mathbf{H}_{/\mathbf{B}U(1)_{\mathrm{conn}}}}(\nabla)
	\\
	& \simeq
		\mathbf{Aut}_{\mathrm{Corr}\left(\mathbf{H}_{/\mathbf{B}U(1)_{\mathrm{conn}}}\right)}(\nabla)
	\\
	&
	\simeq
	\left\{
	  \raisebox{20pt}{
	  \xymatrix{
	    X \ar[rr]^-\phi_\simeq \ar[dr]_{\nabla}^{\ }="t" && X \ar[dl]_{\ }="s"
		\\
		& \mathbf{B}U(1)_{\mathrm{conn}}
		\ar@{=>}^\simeq "s"; "t"
	  }}
	\right\}
	\end{aligned}
  $$
  in that
  \begin{enumerate}
    \item The Lie algebra of $\mathrm{QuantMorph}(X,\omega)$ is the Poisson bracket Lie 
	algebra of $(X,\omega)$;
	\item This group constitutes a $U(1)$-central extension of the group of Hamiltonian
	symplectomorphisms.
  \end{enumerate}
  \label{QuantomorphismGroup}
\end{proposition}
While important, for some reason this group is not very well known, which is striking because it contains a small subgroup which is famous in quantum mechanics: the \emph{Heisenberg group}.

More precisely, whenever $(X,\omega)$ itself has a Hamiltonian action|compatible group structure, notably if $(X,\omega)$ is just a symplectic vector space (regarded as a group under addition of vectors), then we may ask for the subgroup of the quantomorphism group which covers the (left) action of phase space $(X,\omega)$ on itself. This is the corresponding Heisenberg group $\mathrm{Heis}(X,\omega)$, which in turn is a $U(1)$-central extension of the group $X$ itself:

$$
  U(1) \longrightarrow \mathrm{Heis}(X,\omega) \longrightarrow X
  \,.
$$
\begin{proposition}
  If $(X,\omega)$ is a symplectic manifold that at the same time is a group which 
  acts on itself by Hamiltonian diffeomorphisms, then the Heisenberg group of 
  $(X,\omega)$ is the pullback $\mathrm{Heis}(X,\omega)$ of smooth groups
  in the following diagram in $\mathbf{H}$
  $$
    \raisebox{20pt}{
    \xymatrix{
	  \mathrm{Heis}(X,\omega)
	  \ar[r]
	  \ar[d]
	  &
	  \mathrm{QuantMorph}(X,\omega)
	  \ar[d]
	  \\
	  X \ar[r] & \mathrm{HamSympl}(X,\omega)
	}
	}
	\,.
  $$
  \label{HeisenbergGroup}
\end{proposition}
\begin{remark}
  In other words this exhibits $\mathrm{QuantMorph}(X,\omega)$ as
  a universal $U(1)$-central extension characteristic of quantum mechanics
  from which various other $U(1)$-extension in QM are obtained by pullback/restriction.
  In particular all \emph{classical anomalies} arise this way, discussed below in 
  \ref{ClassicalAnomaliesAndProjectiveSymplecticReduction}.
\end{remark}

At this point it is worth pausing for a second to note how the hallmark of quantum mechanics has appeared as if out of nowhere simply by applying Lie integration to the Lie algebra|Lie algebraic structures in classical mechanics:

if we think of Lie integration|Lie integrating $\mathbb{R}$ to the interesting circle group $U(1)$ instead of to the uninteresting translation group $\mathbb{R}$, then the name of its canonical basis element $1 \in \mathbb{R}$ is canonically "$i$", the imaginary unit. Therefore one often writes the above central extension instead as follows:

$$
  i \mathbb{R} \longrightarrow \mathfrak{poiss}(X,\omega) \longrightarrow \mathfrak{ham}(X,\omega)
$$

in order to amplify this. But now consider the simple special case where $(X,\omega) = (\mathbb{R}^2, d p \wedge d q)$ is the 2-dimensional symplectic vector space which is for instance the phase space of the particle propagating on the line. Then a canonical set of generators for the corresponding Poisson bracket Lie algebra consists of the linear functions $p$ and $q$ of classical mechanics textbook fame, together with the \emph{constant} function. Under the above Lie theoretic identification, this constant function is the canonical basis element of $i \mathbb{R}$, hence purely Lie theoretically it is to be called "$i$".

With this notation then the Poisson bracket, written in the form that makes its Lie integration manifest, indeed reads

$$
  [q,p] = i
  \,.
$$

Since the choice of basis element of $i \mathbb{R}$ is arbitrary, we may rescale here the $i$ by any non-vanishing real number without changing this statement. If we write "$\hbar$" for this element, then the Poisson bracket instead reads

$$
  [q,p] = i \hbar
  \,.
$$

This is of course the hallmark equation for quantum physics, if we interpret $\hbar$ here indeed 
as Planck's constant, def. \ref{PlanckConstant}. We see it arises here merely by considering the non-trivial (the interesting, the non-simply connected) Lie integration of the Poisson bracket. 

This is only the beginning of the story of quantization, naturally understood and indeed "derived" from applying Lie theory to classical mechanics. From here the story continues. It is called the story of 
\emph{geometric quantization}. We close this motivation section here by some brief outlook.

The quantomorphism group which is the non-trivial Lie integration of the Poisson bracket is naturally constructed as follows: given the symplectic form $\omega$, it is natural to ask if it is the curvature 2-form of a $U(1)$-principal connection $\nabla$ on complex line bundle $L$ over $X$ (this is directly analogous to Dirac charge quantization when instead of a symplectic form on phase space we consider the  the field strength 2-form of electromagnetism on spacetime). If so, such a connection $(L, \nabla)$ is called a \emph{prequantum line bundle} of the phase space $(X,\omega)$. The quantomorphism group is simply the automorphism group of the prequantum line bundle, covering diffeomorphisms of the phase space (the Hamiltonian symplectomorphisms mentioned above).

As such, the quantomorphism group naturally acts on the space of sections of $L$. Such a section is like a wavefunction, except that it depends on all of phase space, instead of just on the 
``canonical coordinates''. 
For purely abstract mathematical reasons (which we won't discuss here, but see at 
\emph{motivic quantization} for more) it is indeed natural to choose a "polarization" of phase space into canonical coordinates and canonical momenta and consider only those sections of the prequantum line bundle which depend only on the former. 
These are the actual \emph{wavefunctions} of quantum mechanics, hence the \emph{quantum states}. And the subgroup of the quantomorphism group which preserves these polarized sections is the group of exponentiated quantum observables. For instance in the simple case mentioned before where $(X,\omega)$ is the 2-dimensional symplectic vector space, this is the Heisenberg group with its famous action by multiplication and differentiation operators on the space of complex-valued functions on the real line.

\paragraph{Integrable systems, moment maps and maps into the Poisson bracket}
\label{IntegrableSystems}

\begin{remark}
  Given a phase space (pre-)symplectic manifold $(X,\omega)$,
  and given $n \in \mathbb{N}$, then Lie algebra homomorphisms
  $$
    \mathbb{R}^n \longrightarrow \mathfrak{pois}(X,\omega)
  $$
  from the abelian Lie algebra on $n$ generators
  into the Poisson bracket Lie algebra, def. \ref{PoissonBracketForPresymplectic}
  are equivalently choices of $n$-tuples of Hamiltonians $\{H_i\}_{i = 1}^n$
  (and corresponding Hamiltonian vector fields $v_i$) that pairwise commute with 
  each other under the Poisson bracket, $\forall_{i,j} \{H_i,H_j\} = 0$.
  If the set $\{H_i\}_i$ is maximal with this property and one of the 
  $H_i$ is regarded the time evolution Hamiltonian of a physical system,
  then one calls this system \emph{integrable}.
  
  By the discussion in \ref{QuantizationTheHeisenbergGroupAndSliceAutomorphismGroups}, 
  the Lie integration of the Lie algebra homomorphism 
  $\mathbb{R}^n \longrightarrow \mathfrak{pois}(X,\omega)$ is a morphism of 
  smooth groupoids 
  $$
    \mathbf{B}(\mathbb{R}^n)
	\longrightarrow
	\mathbf{B}\mathbf{Aut}_{/\mathbf{B}U(1)_{\mathrm{conn}}}(\nabla)
	\hookrightarrow
	\mathbf{H}_{/\mathbf{B}U(1)_{\mathrm{conn}}}
  $$
  from the smooth delooping groupoid (def. \ref{DeloopingGroupoid})
  of $\mathbb{R}^n$, now regarded as the translation group of $n$-dimensional
  Euclidean space, to the automorphism group of any pre-quantization of 
  the phase space (its quantomoprhism group).
  \label{CommutingHamiltonians}
\end{remark}
\begin{remark}
  Below in \ref{DonderWeylFlowAndLieIntegrationOfHomotopyLieAlgebra}
  we re-encounter this situation, but in a more refined context. There
  we find that $n$-dimensional classical field theory is encoded
  by a homomorphism of the form
  $$
    \mathbb{R}^n \longrightarrow \mathfrak{pois}(X,\omega)
	\,,
  $$
  where however now $\omega$ is a closed differential form of degree $(n+1)$ 
  and where $\mathfrak{pois}(X,\omega)$ is a homotopy-theoretic refinement
  of the Poisson bracket Lie algebra (a \emph{Lie $n$-algebra} or $(n-1)$-type in
  homotopy Lie algebras). In that context such a homomorphism does not encode
  a set of strictly Poisson-commuting Hamiltonians, but a of 
  Hamiltonian flows in the $n$ spacetime directions of the field theory  
  which commute under an $n$-ary higher bracket only \emph{up to} a
  specified homotopy. That specified homotopy is the de Donder-Weyl-Hamiltonian
  of classical field theory.
\end{remark}

\begin{remark}
  For $\mathfrak{g}$ any Lie algebra and $(X,\omega)$ a (pre-)symplectic
  manifold,
  a Lie algebra homomorphism
  $$
    \mathfrak{g} \longrightarrow \mathfrak{pois}(X,\omega)
  $$
  is called a \emph{moment map}. Equivalently this is an actin of 
  $\mathfrak{g}$ by Hamiltonian vector fields \emph{with} chosen
  Hamiltonians.
  The Lie integration of this is a homomorphism of smooth groups
  $$
    G \longrightarrow \mathbf{Aut}_{/\mathbf{B}U(1)_{\mathrm{conn}}}\simeq
	\mathrm{QuantMorph}(X,\omega)
  $$
  from a Lie group integrating $\mathfrak{g}$ to the quantomorphism group.
  This is called a \emph{Hamiltonian $G$-action}.
  \label{HamiltonianAction}
\end{remark}

\paragraph{Classical anomalies and projective symplectic reduction}
\label{ClassicalAnomaliesAndProjectiveSymplecticReduction}
\index{anomaly cancellation!classical anomaly}

Above in \ref{NoetherSymmetriesAndEquivariantStructures} we saw that
for a gauge symmetry to act consistently on a phase space, it needs
to act by \emph{Hamiltonian diffeomorphisms}, because this
is the data necessary to put a gauge-equivariant structure on the 
symplectic potential (hence on the pre-quantization of the phase space).

Under mild conditions every single infinitesimal gauge transformation comes
from a Hamiltonian. But these Hamiltonians 
may not combine to a genuine
Hamiltonian action, remark \ref{HamiltonianAction}, 
but may be specified only up to addition 
of a locally constant function, and it may happen that these locally constant
``gauges'' may not be chose globally for the whole gauge group such as to 
make the whole gauge group act by Hamiltonians. This is the lifting problem
of pre-quantization discussed above in \ref{TheKineticActionPreQuantizationAndDifferentialCohomology}.

But if the failure of the local Hamiltonians to combine to a global Hamiltonian is 
sufficiently coherent in that it is given by a \emph{group 2-cocycle}, then 
one can at least find a Hamiltonian action by a central extension of the gauge 
group. This phenomenon is known as a \emph{classical anomaly} in field theory:
\begin{definition}
  Let $(X,\omega)$ be a phase space symplectic manifold and let 
  $\rho : G \times X \longrightarrow X$ be a smooth action of a Lie group
  $G$ on the underlying smooth manifold by Hamiltonian symplectomorphisms, hence 
  a group homomorphism
  $$
    \xymatrix{
	  G \ar[r]
	  &
	  \mathrm{HamSympl}(X,\omega)
	}
	\,.
  $$
  Then we say this system has a \emph{classical anomaly} if this morphism
  lifts to the quantomorphism group, prop. \ref{QuantomorphismGroup}, only
  up to a central extension $\widehat G \longrightarrow G$, hence if it
  fits into the following diagram of smooth group, without the dashed diagonal 
  morphism existing:
  $$
    \raisebox{20pt}{
    \xymatrix{
	  \widehat {G}
	  \ar[d]
	  \ar[r] & 
	  \mathrm{QuantMorph}(X,\omega)
	  \ar[d]
	  \\
	  G \ar[r]^-\rho \ar@{-->}[ur] & \mathrm{HamSympl}(X,\omega)
	}
	}
	\,.
  $$
\end{definition}
This is the Lie-integrated version of the Lie-algebraic definition in 
appendix 5 of \cite{Arnold}. For a list of examples of classical 
anomalies in field theories see \cite{Toppan}.
\begin{remark}
Comparison with prop. \ref{HeisenbergGroup} above shows that 
for $(X,\omega)$ a symplectic group acting on itself by Hamiltonian 
symplectomorphism, then its Heisenberg group is the ``universal classical anomaly''.
\end{remark}

\subsubsection{Hamilton-De Donder-Weyl field theory via Higher correspondences}
\label{DeDonderWeylTheoryViaHigherCorrespondences}
\label{De Donder-Weyl field theory via higher correspondences}
\index{classical field theory}

We now turn attention from just classical \emph{mechanics} 
(hence of dynamics along a single parameter, such as the Hamiltonian 
time parameter in \ref{HamiltonianTimeEvolutionTrajectoriesAndHamiltonianCorrespondences} above)  
to, more generally, classical \emph{field theory}, which is dynamics
parameterized by higher dimensional manifolds (``spacetimes'' or ``worldvolumes''). 
Or rather, we turn attention to the \emph{local} description of classical field theory.
See also section \ref{LocalPrequantumFieldTheories} below.

Namely, the situation of example \ref{PhaseSpaceFromLocalActionFunctionalOnTheLine}
above, where a trajectory of a physical system is given by a 1-dimensional curve
$[0,1] \longrightarrow Y$ in a space $Y$ of fields can -- and traditionally is --
also be applied to field theory, if only we allow $Y$ to be a smooth space more general
than a finite-dimensional manifold. 
Specifically, for a field theory on a parameter manifold $\Sigma_n$ of some dimension $n$
(to be thought of as spacetime or as the ``worldvolume of a brane''), and for
$\mathbf{Fields}$ a smooth moduli space of of fields, a \emph{local} field configuration is 
a map 
$$
  \phi : \Sigma_n \longrightarrow \mathbf{Fields}
  \,.
$$ 
If however
$\Sigma_d \simeq \Sigma_{d-1} \times \Sigma_1$ is a cylinder with 
$\Sigma_1 = [0,1]$ over a base manifold $\Sigma_{d-1}$
(a \emph{Cauchy surface} if we think of $\Sigma$ as spacetime), then such a map
is equivalently a map out of the interval into the mapping space of $\Sigma_{d-1}$
into $\mathbf{Fields}$:
$$
  \phi_{\Sigma_{d-1}} : \Sigma_1 \longrightarrow [\Sigma_{d-1}, \mathbf{Fields}]
  \,.
$$
This brings the field theory into the form of example \ref{PhaseSpaceFromLocalActionFunctionalOnTheLine},
but at the cost of making it ``spatially non-local'': for instance 
the energy of the system, as discussed 
in \ref{HamiltonianTimeEvolutionTrajectoriesAndHamiltonianCorrespondences}, would 
at each point of $\Sigma_1$ be the energy contained in the fields over all of 
$\Sigma_{d-1}$,
while the information that this energy arises from integrating contributions
localized along $\Sigma_{d-1}$ is lost.

In more mathematical terms this means that by \emph{transgression to codimension 1}
classical field theory takes the form of classical mechanics
as discussed above in \ref{HamiltonianTimeEvolutionTrajectoriesAndHamiltonianCorrespondences}.
To ``localize'' the field theory again (make it ``extended'' or ``multi-tiered'')
we have to undo this process and ``de-transgress'' classical mechanics to full codimension.

At the level of Hamilton's differential equations, def. \ref{HamiltonEquations},
such a localization is ``well known'', but much less famous than Hamilton's equations:
it is the multivariable variational calculus 
of Carath{\'e}odory, de Donder, and Weyl, 
as reviewed for instance in section 2 of \cite{Helein02}.
Below in \ref{DonderWeylFlowAndLieIntegrationOfHomotopyLieAlgebra} we show that 
the de Donder-Weyl equation secretly describes the Lie integration
of a higher Poisson bracket Lie algebra in direct but higher analogy to how 
in \ref{QuantizationTheHeisenbergGroupAndSliceAutomorphismGroups} we 
saw that the ordinary Hamilton equations exhibit the Lie integration of the 
ordinary Poisson bracket Lie algebra. 

From this one finds that an $n$-dimensional \emph{local} classical field theory 
is described not by a symplectic 2-form as a system of classical mechanics
is, but by a differential $(n+1)$-form which transgresses to the 2-form
after passing to mapping spaces. This point of view has been explored
under the name of ``covariant mechanics'' or ``multisymplectic geometry'' 
(see \cite{ForgerRomero} for a review)
and ``$n$-plectic geometry'', see \ref{SmoothStrucGeometricPrequantization} below. 
Here we show, based on the results in \cite{hgp}, how both of these
approaches are unified and ``pre-quantized'' to a global description of
local classical field theory by systems of higher correspondences in 
higher slices toposes, in higher generalization to the picture
which we 
found in \ref{TheClassicalActionFunctionalPrequantizesLagrangianCorrespondences} 
for classical mechanics.

\medskip

\begin{itemize}
  \item \ref{LocalFieldTheoryLagrangiansAndnPlecticSmoothSpaces} -- Local field theory Lagrangians and $n$-plectic smooth spaces
  \item \ref{LocalObservablesAndHigherPoissonBracketHomotopyLieAlgebras} -- Local observables, conserved currents and higher Poisson brackets
  \item \ref{DonderWeylFlowAndLieIntegrationOfHomotopyLieAlgebra} -- Field equations of motion and Higher Poisson-Maurer-Cartan elements
  \item \ref{SourceTermsOffShell} -- Source terms, off-shell Poisson bracket and Poisson holography
\end{itemize}

\paragraph{Local field theory Lagrangians and $n$-plectic smooth spaces}
\label{LocalFieldTheoryLagrangiansAndnPlecticSmoothSpaces}

Traditionally, a classical field over a \emph{spacetime} $\Sigma$ 
is encoded by a fiber bundle $E \to X$, the \emph{field bundle}.
The fields on $X$ are the sections of $E$. 
\begin{example}
  Let $d \in \mathbb{N}$ and let 
  $\Sigma = \mathbb{R}^{d-1,1}$
  be the $d$-dimensional real vector space, regarded as a pseudo-Riemannian
  manifold with the Minkowski metric $\eta$ (\emph{Minkowski spacetime}).
  Let moreover $F$ be a finite dimensional 
  real vector space -- the \emph{field fiber} -- eqipped with a positive definite
  bilinear form $k$. 
  Consider the bundle $\Sigma \times F \to \Sigma$, to be called the
  \emph{field bundle}, and write  
  $$
    \left(X \to \Sigma\right) := \left(j_\infty^1 \left(\Sigma \times F\right) \to \Sigma\right)
  $$
  for its first jet bundle.
  
  If we denote the canonical coordinates of $\Sigma$ by 
  $\sigma^i : \Sigma \to \mathbb{R}$ for $i \in \{0, \cdots, n-1\}$, 
  and choose a dual basis
  $$
    \phi^a : F \to \mathbb{R}
  $$ 
  of $F$ (hence with $a \in \{1, \cdots, \mathrm{dim}(V)\}$)
  then $X$ is the vector space with canonical dual basis elements labeled by
  $$
    \{\sigma^i\}, \{\phi^a\}, \{\phi^a_{,i}\}
  $$
  and equipped with bilinear form $(\eta \oplus k \oplus (\eta\otimes k))$.
  While all of these are coordinates on $X$, traditionally one says that
  \begin{enumerate}
   \item 
     the functions
	 $$
	   \sigma^i : X \longrightarrow \mathbb{R}
	 $$
	 are the \emph{spacetime coordinates};
    \item the functions
  $$
    \phi^a : X \longrightarrow \mathbb{R}
  $$
  are are the \emph{canonical coordinates} of the $F$-field
  \item  the functions  
  $$
    p_a^i := \eta^{i j}k_{a b}\phi^b_{,j} \;\colon X \longrightarrow \mathbb{R}
  $$
  are the \emph{canonical momenta} of the \emph{free} $F$-field.
  \end{enumerate}
  \label{FreeFieldBundle}
\end{example}
\begin{definition}
  Given a field bundle $X  = j_\infty^1(\Sigma \times F)\to \Sigma$ 
  as in example \ref{FreeFieldBundle},
  the \emph{free field theory local kinetic Lagrangian} is the 
  horizontal differential $n$-form
  $$
    L_{\mathrm{kin}}^{\mathrm{loc}} \in \Omega^{n,0}(X)
  $$
  given by
  $$
    \begin{aligned}
    L^{\mathrm{loc}}_{\mathrm{kin}} 
	  & :=
	  \langle \nabla \phi, \nabla \phi\rangle
	  \wedge 
	  \mathrm{vol}_\Sigma
	  \\
	  &:=
	 \left(
	   \tfrac{1}{2} k_{ab}\eta^{i j}\phi^a_{,i} \phi^{b}_{,j}
	 \right)
	 \wedge
	 \mathbf{d}\sigma^1 \wedge \cdots \wedge \mathbf{d}\sigma^d
	 \end{aligned}
  $$
  (where a sum over repeated indices is understood).  
  Here we regard the volume form of $\Sigma$ canonically as a 
  horizontal differential form on the first jet bundle
  $$
    \mathrm{vol}_\Sigma := \mathbf{d}\sigma^1 \wedge \cdots \wedge \mathbf{d}\sigma^a
	\;\;\in\;\; \Omega^{d,0}_\Sigma(X)
	\,.
  $$
  \label{FreeFieldLocalLagrangian}
\end{definition}
The localized analog of example \ref{PhaseSpaceFromLocalActionFunctionalOnTheLine} is now the following.
\begin{definition}
  Given a free field bundle as in example \ref{FreeFieldBundle}
  and given a horizontal $n$-form
  $$
    L^{\mathrm{loc}} \in \Omega^{n,0}(X)
  $$  
  on its first jet bundle, regarded as a local Lagrangian
  as in def. \ref{FreeFieldLocalLagrangian}, then
  the associated \emph{Lagrangian current} is the $n$-form
  $$
    \theta^{\mathrm{loc}} \in \Omega^{n-1,1}(X)
  $$
  given by the formula
  $$
    \theta^{\mathrm{loc}} 
	  := 
	 \iota_{\partial_i} \left(\frac{\partial}{\partial \phi^a_{,i}}L^{\mathrm{loc}}\right) \wedge
	\mathbf{d}\phi^a
  $$
  (where again a sum over repeated indices is understood).
  We say that the corresponding 
  \emph{pre-symplectic current} or \emph{pre-$n$-plectic form} \cite{LocalObservables}
  is
  $$
    \omega^{\mathrm{loc}} := \mathbf{d} \theta^{\mathrm{loc}}
	\,.
  $$
  \label{LagrangianCurrent}
  \label{nPlectic}
\end{definition}
\begin{remark}
  The formula in def. \ref{LagrangianCurrent} is effectively that for the 
  pre-symplectic current as it arises in 
  the discussion of \emph{covariant phase spaces} in \cite{Zuckerman, CrnkovicWitten}.
  In the coordinates of example \ref{FreeFieldBundle}  the Lagrangian current reads
  $$
    \theta^{\mathrm{loc}}
	 =
	 p_a^i
	 \wedge
	 \mathbf{d}\phi^a
	 \wedge
	 \iota_{\partial_i}\mathrm{vol}_\Sigma
  $$
  and hence the pre-symplectic current reads
  $$
    \omega^{\mathrm{loc}}
	=
	 \mathbf{d}p_a^i
	 \wedge
	 \mathbf{d}\phi^a
	 \wedge
	 \iota_{\partial_i}\mathrm{vol}_\Sigma
  $$
  In this form this is manifestly the $(n-1,1)$-component of the canonical
  ``multisymplectic form'' that is considered in multisymplectic geometry,
  see for instance section 2 of \cite{Helein02}. 
  
  This direct relation
  between the covariant phase space formulation and the multisymplectic
  description of local classical field theory seems not to have been
  highlighted much in the literature. 
  It essentially appears in section 3.2 of \cite{ForgerRomero} and
  in section 2.1 of \cite{RomanRoy}.
\end{remark}
\begin{example}
  Consider the simple case $d = 1$ hence $\Sigma = \mathbb{R}$, and 
  $F = \mathbb{R}$, both equipped with the canonical bilinear form on $\mathbb{R}$
  (given by multiplication). Jet prolongation followed by evaluation
  yields the smooth function
  $$
    \mathrm{ev_\infty}
	:
    [\Sigma,F] \times \Sigma 
	  \stackrel{(j_\infty, \mathrm{id})}{\longrightarrow} 
	\Gamma_\Sigma(X) \times \Sigma 
	\stackrel{\mathrm{ev}}{\longrightarrow} X 
	\,.
  $$
  Then the pullback of the local free field Lagrangian 
  of def. \ref{FreeFieldLocalLagrangian} along this map is the
  kinetic Lagrangian of example \ref{StandardLagrangianForMechanics}:
  $$
    L_{\mathrm{kin}} = \mathrm{ev}_\infty^\ast L^{\mathrm{loc}}_{\mathrm{kin}}
	\,.
  $$
  The pullback of the corresponding Lagrangian current according to def. \ref{LagrangianCurrent}
  is the pre-symplectic potential $\theta$ 
  in example \ref{PhaseSpaceFromLocalActionFunctionalOnTheLine}
  $$
      \theta 
	  =
	  \mathrm{ev}_\infty^\ast \theta^{\mathrm{loc}}
	  \,.
  $$
\end{example}
\begin{definition}
  For $d \in \mathbb{N}$, write 
  $\Sigma = \Sigma_1 \times \Sigma_{d-1}$ for the
  decomposition of Minkowski spacetime into a time axis $\Sigma_1$
  and a spatial slice $\Sigma_{d-1}$,
  hence with $\Sigma_1 = \mathbb{R}$
  the real line. Restrict attention to sections of the field bundle
  which are periodic in all spatial directions, hence pass to the $(d-1)$-torus
  $\Sigma_{d-1} := \mathbb{R}^d/\mathbb{Z}^d$ (in order to have a compact
  spatial slice).
  Then given a free field local Lagrangian as in def. 
  \ref{FreeFieldLocalLagrangian}, say that its \emph{transgression to codimension 1}
  is the pullback of the local Lagrangian $n$-form along
  $$
    \mathrm{ev}_\infty
	 :
    [\Sigma_1, [\Sigma_{d-1},F \times \Sigma_1 \times \Sigma_{d-1}
	\stackrel{\simeq}{\longrightarrow}
    [\Sigma,F] \times \Sigma 
	  \stackrel{(j_\infty, \mathrm{id})}{\longrightarrow} 
	\Gamma_\Sigma(X) \times \Sigma 
	\stackrel{\mathrm{ev}}{\longrightarrow} X 
  $$
  followed by fiber integration $\int_{\Sigma_{d-1}}$ over space $\Sigma_{d-1}$,
  to be denoted
  $$
    L_{\mathrm{kin}} := 
	\int_{\Sigma_{d-1}} \mathrm{ev}_\infty^\ast L^{\mathrm{loc}}_{\mathrm{kin}}
	\,.
  $$  
  Similarly the transgression to codimension 1 of the Lagrangian current, 
  def. \ref{LagrangianCurrent} is
  $$
    \theta 
	  := 
	\int_{\Sigma_{d-1}} \mathrm{ev}_\infty^\ast \theta^{\mathrm{loc}}
	\,.
  $$
  \label{TransgressionOfLagrangianCurrent}
\end{definition}
\begin{remark}
  This is the standard way in which the kinetic Lagrangians in 
  example \ref{PhaseSpaceFromLocalActionFunctionalOnTheLine} arise by transgression of local data.
  \label{TransgressionOfKineticLocalLagrangianToCodimension1}
\end{remark}
It is useful to combine this data as follows. 
\begin{definition}
  Given a first jet bundle $j_\infty^1(\Sigma \times X)$ as in example
  \ref{FreeFieldBundle}, we write
  \begin{enumerate}
    \item $j_\infty^1(\Sigma \times X)^\ast \to  \Sigma \times X$
      for its fiberwise linear densitized dual, as a bundle over the field bundle,
      to be called the \emph{dual first jet bundle};
    \item $j_\infty^1(\Sigma \times X)^\vee \to X \Sigma \times X$
    for the fiberwise affine densitized dual, to be called the 
    \emph{affine dual first jet bundle}.
  \end{enumerate}
  \label{dualFirstJetBundle}
\end{definition}
\begin{remark}
With respect to the canonical coordinates in example \ref{FreeFieldBundle}, the
canonical coordinates of the dual first jet bundle are 
$\{\sigma^i, \phi^a, p^i_a\}$ (spacetime coordinates, fields and canonical field momenta) 
and the canonical coordinates of the
affine dual first jet bundle are $\{\sigma^i, \phi^a, p^i_a, e\}$ with one more
coordinate $e$.
\end{remark}
\begin{definition}
  \begin{enumerate} 
  \item 
    The 
    \emph{canonical pre-$n$-plectic form} on the affine dual first jet bundle, 
    def. \ref{dualFirstJetBundle}, is
  $$
    \omega_e 
      := 
    \mathbf{d}\phi^a \wedge \mathbf{d}p_a^i \wedge \iota_{\partial_{\sigma^i}} \mathrm{vol}_\Sigma
    +
    \mathbf{d}e \wedge \mathrm{vol}_\Sigma
    \;\;\;
    \in \Omega^{n+1}(j^1(\Sigma \times X)^\vee)
    \,.
  $$
    \item 
    Given a function $H \in C^\infty(j^1(\Sigma \times X)^\ast)$ on the linear dual first jet bundle,
    def. \ref{dualFirstJetBundle},
    then
    the corresponding \emph{DWH pre-$n$-plectic form} is
  $$
    \omega_H 
      := 
    \mathbf{d}\phi^a \wedge \mathbf{d}p_a^i \wedge \iota_{\partial_{\sigma^i}} \mathrm{vol}_\Sigma
    +
    \mathbf{d}H \wedge \mathrm{vol}_\Sigma
    \;\;\;
    \in \Omega^{n+1}(j^1(\Sigma \times X)^\ast)
    \,.
  $$
  \end{enumerate}
  \label{HamiltonianNPlecticForm}
\end{definition}
\begin{definition}[local Legendre transform]
  Given a local Lagrangian as in def. \ref{FreeFieldLocalLagrangian}, 
  hende a horzontal $n$-form $L^{\mathrm{loc}} \in \Omega^{(n,0)}(J^1(E))$
  on the jets of the field bundle, its \emph{local Legendre transform}
  is the function
  $$
    \mathbb{F}L^{\mathrm{loc}}
    :
    J^1(X) \longrightarrow (J^1(X))^\vee
  $$
  from jets to the affine dual jet bundle, def. \ref{dualFirstJetBundle}
  which is the first order Taylor series of $L^{\mathrm{loc}}$.
  \label{localLegendreTransform}
\end{definition}
This definition was suggested in section 2.5 of \cite{ForgerRomero}.
It conceptualizes the traditional notion of local Legendre transform:
\begin{example}
  In the local coordinates of example \ref{FreeFieldBundle}, 
  the Legendre transform of a local Lagrangian $L^{\mathrm{loc}}$,
  def. \ref{localLegendreTransform} has affine dual jet bundle 
  coordinates given by
  $$
    p_a^i = \frac{\partial L^{\mathrm{loc}}}{\partial \phi^a_{,i}}
  $$
  and
  $$
    e = L^{\mathrm{loc}} - \frac{\partial L^{\mathrm{loc}}}{\partial \phi^a_{, i}} \phi^a_{,i}
    \,.
  $$
  The latter expression is what is traditionally taken to be the local Legendre transform
  of $L^{\mathrm{loc}}$.
\end{example}
The following observation relates the canonical pre-$n$-plectic form
$\omega_e$ on the affine dual jet bundle to the central ingredients of the
covariant phase space formalism.
\begin{proposition}
  Given a local Lagrangian $L^{\mathrm{loc}} \in \Omega^{(n,0)}(J^1(E))$,
  then the pullback of the canonical pre-$n$-plectic form $\omega_e$,
  def. \ref{HamiltonianNPlecticForm}, along the local Legendre transform
  $\mathbb{F}L^{\mathrm{loc}}$ of def. \ref{localLegendreTransform}
  is the sum of the Euler-Lagrange equation term 
  $\mathrm{EL}_{L^{\mathrm{loc}}} \in \Omega^{(n,1)}(J^1(X))$
  and of the canonical pre-$n$-plectic current 
  $\mathbf{d}_v \theta_{L^{\mathrm{loc}}} \in \Omega^{(n-1,2)}(J^1(X))$
  of def. \ref{LagrangianCurrent}:
  $$
    \begin{aligned}
      \omega_{L^{\mathrm{loc}}}
      & :=
      (\mathbb{F}L^{\mathrm{loc}})^\ast \omega_e
      \\
      & = \mathrm{EL}_{L^{\mathrm{loc}}} + \mathbf{d}_v \theta_{\mathrm{L}^{\mathrm{loc}}}
    \end{aligned}
    \,.
  $$
  \label{ELFormAndPrePlecticCurrentByPullback}
\end{proposition}
This follows with equation (54) and theorem 1 of \cite{ForgerRomero}.\footnote{
  This statement and its formulation in terms of notions in the variational 
  bicomplex as given here has kindly been amplified to us by 
  Igor Khavkine.
} 
In \ref{DonderWeylFlowAndLieIntegrationOfHomotopyLieAlgebra} below 
we see how using this the equations of motion 
of the field theory are naturally expressed.

\medskip

In conclusion, we find that where phase spaces in classical mechanics
are given by smooth spaces equipped with a closed 2-form, phase spaces
in ``de-transgressed'' or ``covariant'' or 
``localized'' classical field theory of dimension $n$
are given by smooth spaces equipped
with a closed $(n+1)$-form. To give this a name we say \cite{hgp}:
\begin{definition}
  For $n \in \mathbb{N}$, a \emph{pre-$n$-plectic smooth space}
  is a smooth space $X$ and a smooth closed $(n+1)$-form
  $$
    \omega : X \longrightarrow \mathbf{\Omega}^{n+1}_{\mathrm{cl}}
	\,,
  $$
  hence an object of the slice topos 
  $$
    (X,\omega) \in \mathbf{H}_{/\mathbf{\Omega}^{n+1}_{\mathrm{cl}}}
	\,.
  $$
  \label{nPlecticManifold}
\end{definition}

\paragraph{Local observables, conserved currents and higher Poisson brackets}
\label{LocalObservablesAndHigherPoissonBracketHomotopyLieAlgebras}
\index{Poisson bracket!higher}

Above in  \ref{ClassicalObservablesAndThePoissonBracket} we discussed how
functions on a phase space are interpreted as observables 
of states of the mechanical system, for instance the energy of the system.
Now in \ref{LocalFieldTheoryLagrangiansAndnPlecticSmoothSpaces} above we 
saw that that notably the energy of an $n$-dimensional field theory at some moment in time
(over some spatial hyperslice of spacetime) is really the integral over
$(d-1)$-dimensional space $\Sigma_{d-1}$ of an \emph{energy density} $(d-1)$-form
$H^{\mathrm{loc}}$, hence by def. \ref{TransgressionOfLagrangianCurrent} the
transgression of an $(n-1)$-form on the localized $n$-plectic phase space:
$$
  H = \int_{\Sigma_{n-1}} \mathrm{ev}_\infty^\ast H^{\mathrm{loc}}
  \,.
$$
Therefore in analogy with the notion of 
observables on a symplectic manifold, 
given an $n$-plectic manifold, def. \ref{nPlecticManifold},
its degree-$(n-1)$ differential forms may be called the \emph{local observables}
of the system. To motivate from physics how exactly to formalize
such local observables (which we do below in def. \ref{PoissonLienAlgebra}),
we first survey how such local observables appear in the physics literature:
\begin{example}[currents in physics as local observables]
  In the situation of example \ref{FreeFieldBundle},
  consider a vector field $j \in \Gamma(T \Sigma_d)$ on the $d$-dimensional Minkowski spacetime
  $\Sigma_d = \mathbb{R}^{d-1,1}$. In physics this represents 
  a quantity which -- for an inertial observer
  characterized by the coordinates chosen in example \ref{FreeFieldBundle} -- 
  has local density $j^0$ at each point in space and time, of a quantity 
  that flows through space as given by the vector $(j^1, \cdots, j^{d-1})$. 
  
  For instance in the description of electric sources
  distributed in spacetime, the component $j^0$
  would be an \emph{electric charge density} and the vector $(j^1, \cdots, j^{d-1})$
  would be the \emph{electric current density}. To emphasize that therefore
  $j$ combines the information of a spatial current with the density of the 
  substance that flows, traditional physics textbooks call $j$ a 
  ``$d$-current'' -- usually a ``4-current'' when identifying $d$ 
  with the number of macroscopic spacetime dimensions of the observable universe.
  But once the spacetime context is understood, one just speaks of $j$ as 
  a \emph{current}.
  
  The currents of interest in physics are those which satisfy a 
  \emph{conservation law}, a law which states that the change in coordinate time $\sigma^0$ 
  of the density $j^0$ is equal to the negative of the divergence of the
  spatial current, hence that the spacetime  divergence of $j$ vanishes:
  $$
    \mathrm{div}(j) = \frac{\partial j^0}{\partial \sigma^0} + \sum_{i = 1}^{n-1}
	\frac{\partial j^i}{\partial \sigma_i} = 0
	\,.
  $$
  If this is the case, one calls the current $j$ a \emph{conserved current}.
  (Beware that the ``conserved'' is so important in applications that it is often taken to be 
  implicit and notationally suppressed.)
  
  In order to formulate the notion of divergence of a vector field intrinsically
  (as opposed with respect to a chosen coordinate system as above), one needs a
  specified volume form $\mathrm{vol}_\Sigma \in \Omega^{d}(\Sigma_d)$ of spacetime. 
  With that given, the divergence $\mathrm{div}(j) \in C^\infty(\Sigma_d)$ 
  of the vector field is
  defined by the equation
  $$
    \mathrm{div}(j) \wedge \mathrm{vol}_{\Sigma_d}
	:=
	\mathcal{L}_j \mathrm{vol}_{\Sigma_d}
	=
	\mathbf{d} \left(\iota_j \mathrm{vol}_\Sigma\right)
	\,.
  $$
  In particular, a current $j$ is a conserved current precisely if the 
  degree-$(n-1)$ differential form
  $$
    J := \iota_j \mathrm{vol}_{\Sigma_d}
  $$
  is a closed differential form
  $$
    \left(
	   \mbox{$j \in \Gamma(T\Sigma_d)$ is a conserved current}
	\right)
	\;\;\Leftrightarrow\;\;
	\left(
	   \mathbf{d}J = 0 
	\right)
	\,.
  $$
  Due to this and related relations, one finds eventually that 
  the degree-$(d-1)$ differential form $J$ itself is the more fundamental
  mathematical reflection of the physical current. 
  But by the above introduction, this is in turn the same 
  as saying that a current is a local observable.
  Accordingly, we will often use the terms ``current'' and ``local observable''
  interchangeably. 
  
  If currents are local observables, then by the above discussion their integral
  over a spatial hyperslice of spacetime is to be the corresponding global observable.
  In the special case of the electromagnetic current $J_{\mathrm{el}}$,
  the laws of electromagnetism in the form of \emph{Maxwell's equation}
  $$
    J_{\mathrm{el}} = \mathbf{d}\star F_{\mathrm{em}}
  $$
  say that this integral
  -- assuming now that $J_{\mathrm{el}}$ is spatially compactly supported --
  is the integral of the Hodge dual electromagnetic field strength $F_{\mathrm{em}}$
  over the boundary of a 3-ball $D^3 \hookrightarrow \Sigma_{d-1}$ 
  enclosing the support of the electromagnetic current. 
  This is the \emph{total electric charge} $Q_{\mathrm{el}}$ in space:
  $$
       Q_{\mathrm{el}} 
	   = \int_{S^2} \ast F_{\mathrm{em}} 
	   = \int_{D^3} J_{\mathrm{el}}
       = \int_{\Sigma_{d-1}} J_{\mathrm{el}}	   
	 \,.
  $$
  Based on this example, in physics one generally speaks of 
  the integral of a spacetime current over space as a \emph{charge}.
  So charges are the global observables of the local observables, which
  are currents.  
  
  Notice that for a \emph{conserved} current the corresponding charge is
  also conserved in that it does not change with time or in fact under
  any isotopy of $\Sigma_{d-1}$ inside $\Sigma_d$, due to Stokes' theorem:
  $$
    \begin{aligned}
      \mathbf{d}_{\Sigma_{1}} Q 
	  & = \mathbf{d}_{\Sigma_1} \int_{\Sigma_{d-1}} J
      \\
      & =
      \int_{\Sigma_{d-1}} \mathbf{d}_{\Sigma_d} J
      \\
      & = 0	  
	\end{aligned}
	\,.
  $$
  
  Therefore currents in physics are necessarily subject of 
  \emph{higher gauge equivalences}:
  if $J$ is a conserved current $(d-1)$-form, then for any $(d-2)$-form $\alpha$
  the sum $J + \mathbf{d}\alpha$ is also a conserved current, which, by Stokes' theorem,
  has the same total charge as $J$ in any $(d-1)$-ball in space, 
  and has the same flux as $J$ through the boundary of that $(d-1)$-ball.
  This means that the conserved currents $J$ and $J + \mathbf{d}\alpha$
  are physically equivalent, while nominally different, hence that 
  $\alpha$ exhibits a \emph{gauge equivalence transformation} between currents
  $$
    \alpha : J \stackrel{\simeq}{\longrightarrow} (J' = J + \mathbf{d}\alpha)
	\,.
  $$
  The analogous consideration holds for $\alpha$ itself: 
  for any $(d-3)$-form $\beta$ also $\alpha + \mathbf{d}\beta$
  exhibits a gauge transformation between the currents $J$ and $J'$ above.
  One says this is a \emph{gauge of gauge}-transformation or a 
  \emph{higher gauge transformation} of second order. This phenomenon
  continues up to the 0-forms (the smooth functions), which therefore are
  $(d-1)$-fold higher gauge transformations between consderved currents
  on a $d$-dimensional spacetime.
    
  Finally notice that in a typical application to physics, a current form $J$ is 
  naturally defined
  also ``off shell'', hence for all field configurations (say of the electromagnetic field),
  but its conservation law only holds ``on shell'', hence when these 
  field configurations satisfy their equations of motion 
  (to which we come below in \ref{DonderWeylFlowAndLieIntegrationOfHomotopyLieAlgebra}).
  Since the $n$-plectic localized phase spaces in the discussion in 
  \ref{LocalFieldTheoryLagrangiansAndnPlecticSmoothSpaces} above a priori contain
  all field configurations, we are not to expect that a local observable 
  $(d-1)$-form $J$ is a
  conserved current only if its differential strictly vanishes, but already
  if its differential vanishes at least on those $d$-tuples of vector fields
  $v_1 \vee \cdots \vee v_d$
  which are tangent to jets of those sections of the field bundle that 
  satisfy their equations of motion:
  $$
    \left(
	  \mbox{$J$ is conserved current}
	\right)
	\Leftrightarrow
	\left(
	  \left(
	  \mbox{$v_1 \vee \cdots \vee v_d$ satisfies field equations of motion}
	  \right)
	  \Rightarrow
	  \iota_{v_1 \vee \cdots \vee v_n} \mathbf{d}J = 0
	\right)
	\,.
  $$  
    This we formalize below by the 
  ``$n$-plectic Noether theorem'', prop. \ref{nPlecticNoetherTheorem}.
  There we will see how such conserved current $(d-1)$-forms arise from vector fields
  $v$ that consitute infinitesimal symmetries of a Hamiltonian function,
  by the evident higher degree generalizatin of Hamilton's equations, namely
  $\mathbf{d}J = \iota_v \omega$.
  \label{CurrentsInPhysics}
\end{example}
In summary, example \ref{CurrentsInPhysics} motivates the 
following definition (first proposed in \cite{Rogers} and
then interpreted in homotopy topos theory in \cite{hgp,LocalObservables}) 
of the localized/higher analog of the
Poisson bracket Lie algebra of observables, defs. 
\ref{PoissonBracket}, \ref{PoissonBracketForPresymplectic}, 
as we pass from global observable on (pre-)symplectic manifolds 
to local observables on (pre-)$n$-plectic manifolds.
The general abstract characterization of the following definition,
(which appeared first in \cite{Rogers:2010nw})
we give below in \ref{HigherGaugeTheorySmoothInfinityGroupoids},
see the dicussion in \ref{SmoothStrucGeometricPrequantization}.
\begin{definition}[higher Poisson bracket of local observables]
  Given a pre-$n$-plectic manifold $(X,\omega)$,
  its vector space of \emph{local Hamiltonian observables} is 
  $$
    \Omega^{n-1}_{\omega}(X)
	:=
	\left\{
	  \left(
	    v, J
	  \right)
	  \in \Gamma(T X)\oplus \Omega^{n-1}(X)
	  \;|\;
	  \iota_v \omega = - \mathbf{d}J
	\right\}
	\,.
  $$
  We say that the de Rham complex ending in these Hamiltonian observables
  is the \emph{complex of local observables} of $(X,\omega)$, denoted
  $$
    \Omega^\bullet_\omega(X)
	:=
    \left(
	  C^\infty(X)
	  \stackrel{\mathbf{d}}{\longrightarrow}
	  \Omega^1(X)
	  \stackrel{\mathbf{d}}{\longrightarrow}
	  \cdots
	  \stackrel{\mathbf{d}}{\longrightarrow}
	  \Omega^{n-2}(X)
	  \stackrel{(0,\mathbf{d})}{\longrightarrow}
	  \Omega^{n-1}_\omega(X)
	\right)
	\,.
  $$
  The \emph{binary higher Poisson bracket} on local Hamiltonian observables
  is the linear map
  $$
    \{-,-\} \;:\; \Omega^{n-1}_\omega(X) \otimes \Omega^{n-1}_\omega(X) \longrightarrow 
	 \Omega^{n-1}_\omega(X)
  $$
  given by the fomula
  $$
    \left[
	  \left(v_1, J_1\right),
	  \left(v_2, J_1\right)	  
	\right]
	:=
	\left[
	  \left(
	    \left[v_1, v_2\right],
		\,
		\iota_{v_1 \vee v_2}\omega
	  \right)
	\right]
	\,;
  $$
  and for $k \geq 3$ 
  the \emph{$k$-ary higher Poisson bracket} is the 
  linear map
  $$
    \{-,\cdots, -\}
	\;:\;
    \left( \Omega^{n-1}_\omega(X)\right)^{\otimes^k}
	\longrightarrow
	\Omega^{n+1-k}(X)
  $$
  given by the formula
  $$
    \left[
	  \left(v_1, J_1\right),\,
	  \cdots,
	  \left(v_k, J_k\right)  
	\right]
	:=
	(-1)^{\lfloor \frac{k-1}{2}\rfloor}
	\iota_{v_1 \vee \cdots \vee v_k}
	\omega
	\,.
  $$
  The chain complex of local observables equipped with these linear maps
  for all $k$ we call 
  the \emph{higher Poisson bracket homotopy Lie algebra} of
  $(X,\omega)$, denoted
  $$
    \mathfrak{pois}(X,\omega)
	:=
	\left(
	  \Omega^\bullet_\omega(X),
	  \left\{-,-\right\}, \left\{-,-,-\right\}, \cdots
	\right)
	\,.
  $$ 
  \label{PoissonLienAlgebra}
 \label{ham-infty}  
\end{definition}
\begin{remark}
  What we call a \emph{homotopy Lie algebra} in def. \ref{PoissonLienAlgebra}
  is what originally was called a \emph{strong homotopy Lie algebra}
  and what these days is mostly called an \emph{$L_\infty$-algebra} or,
  since the above chain complex is concentrated in the lowest $n$ degrees,
  a \emph{Lie $n$-algebra}. These are the structures that are to 
  group-like smooth homotopy types
  as Lie algebras are to smooth groups. The reader can find all further details
  which we need not dwell on here as well as
  pointers to the standard literature in \cite{LocalObservables}.
\end{remark}
\begin{remark}
 For $n = 2$ definition \ref{PoissonLienAlgebra} indeed reproduces
 the definition of the ordinary Poisson bracket Lie algebra, 
 def. \ref{PoissonBracketForPresymplectic}.
\end{remark}

\paragraph{Field equations of motion and Higher Poisson-Maurer-Cartan elements}
\label{DonderWeylFlowAndLieIntegrationOfHomotopyLieAlgebra}

Where in classical mechanics the equations of motion that determine the
physically realized trajectories are Hamilton's equations, def. \ref{HamiltonEquations},
in field theory the equations of motion are typically \emph{wave equations}
on spacetime. But as we localize from \mbox{(pre-)}symplectic phase spaces
to \mbox{(pre-)}$n$-plectic phase spaces as in \ref{LocalFieldTheoryLagrangiansAndnPlecticSmoothSpaces}
above, Hamilton's equations also receive a localization to the 
\emph{Hamilton-de Donder-Weyl} equation. This indeed coincides with the
field-theoretic equations of motion. We briefly review the classical
idea of de Donder-Weyl formalism and then show how it naturally follows
from a higher geometric version of Hamilton's equations in $n$-plectic geometry.
\begin{definition}
  Let $(X,\omega)$ be a pre-$n$-plectic smooth manifold, and let 
  $H \in C^\infty(X)$ be a smooth function, 
  to be called the \emph{de Donder-Weyl Hamiltonian}. Then 
  for $v_i \in \Gamma(T X)$ with $i \in \{1, \cdots, n\}$ an $n$-tuple 
  of vector fields, the \emph{Hamilton-de Donder-Weyl} equation is
  $$
    (\iota_{v_n} \cdots \iota_{v_1}) \omega = \mathbf{d}H
	\,.
  $$
  Generally, for $J \in \Omega^{n-k}(X)$ a smooth differential form
  for $1 \leq k \leq n$, and for $\{v_i\}$ a $k$-tuple of vector fields,
  the \emph{extended Hamilton-deDonder-Weyl equation} is
  $$
    \iota_{v_k} \cdots \iota_{v_1}\omega = \mathbf{d}J
	\,.
  $$
  \label{DWHequation}
\end{definition}
We now first show how this describes equations of motion of field theories.
Then we discuss how this de Donder-Weyl-Hamilton equation is naturally found
in higher differential geometry.
For simplicity of exposition we stick with the simple local situation of 
example \ref{FreeFieldBundle}. The ambitious reader can readily generalize 
all of the following discussion to non-trivial and non-linear field bundles.
\begin{definition}
  Let $\Sigma \times X \to \Sigma$ be a field bundle as in example \ref{FreeFieldBundle}.
  For $\Phi := (\phi^i, p_i^a) : \Sigma \to j^1(\Sigma \times X)^\ast$ a section of the 
  linear dual jet bundle write
  $$
      v^\Phi_i 
	    = 
		 \frac{\partial}{\partial \sigma^i}
		 + 
	  \frac{\partial \phi^a}{\partial \sigma^i} \frac{\partial}{\partial \phi^a}
	   + 
	  \frac{\partial p_a^j}{\partial \sigma^i} \frac{\partial}{\partial p_a^j}
  $$
  for its canonical basis of tangent vector fields.
  Similarly for $\Phi := (\phi^i, p_i^a, e) : \Sigma \to j^1(\Sigma \times X)^\vee$ a section of the 
  affine dual jet bundle write
  $$
      v^\Phi_i 
	    = 
		 \frac{\partial}{\partial \sigma^i}
		 + 
	  \frac{\partial \phi^a}{\partial \sigma^i} \frac{\partial}{\partial \phi^a}
	   + 
	  \frac{\partial p_a^j}{\partial \sigma^i} \frac{\partial}{\partial p_a^j}
	  +
	  \frac{\partial e}{\partial \sigma^i} \frac{\partial}{\partial e}
  $$
  for its canonical basis of tangent vector fields.
  \label{TangentsToASection}
\end{definition}
\begin{proposition}
  For $(\Sigma \times X) \to \Sigma$ a field bundle
  as in example \ref{FreeFieldBundle},  
  let $H \in C^\infty(j^1(\Sigma \times X)^\ast)$ be a function
  on the linear dual (and hence on the affine dual) first jet bundle.
  Then for a section $\Phi$ of the linear dual field bundle 
  the homogeneous (``relativistic'') de Donder-Weyl-Hamilton equation,
  def. \ref{DWHequation}, of the 
  Hamiltonian pre-$n$-plectic form, def. \ref{HamiltonianNPlecticForm},
  $$
    \left(
      \iota^\Phi_{n}
      \cdots
      \iota^\Phi_1
    \right)
    \omega_H
     = 0
  $$
  has a unique lift, up to an addive constant, to a solution of the 
  DWH equation on the affine dual field bundle of the form
  $$
    \left(
      \iota^\Phi_{n}
      \cdots
      \iota^\Phi_1
    \right)
    \omega_e
     = \mathbf{d}(H + e)
     \,.
  $$
  Moreover, both these equations are equivalent to the following system of differential equations
  $$
    \partial_{i} \phi^a = \frac{\partial H}{\partial p_a^i}
	\;\;\;\,;
	\;\;\;\;\;
	\partial_i p^i_a = - \frac{\partial H}{\partial \phi^a}
	\,.
  $$  
  \label{DWHinComponents}
\end{proposition}
The last system of differential equations is the form in which the de Donder-Weyl-Hamilton
equation is traditionally displayed, see for instance theorem 2 \cite{RomanRoy}.
The inhomogeneous version on the affine dual first jet bundle above has been highlighted in 
\cite{Helein02}, around equation (4) there.
\begin{example}
  For a field bundle as in example \ref{FreeFieldBundle}, the
  standard form of an energy density function for a field theory on $\Sigma$ is 
  $$
    H \mathrm{vol}_\Sigma = L_{\mathrm{kin}} + V(\{\phi^a\}) \mathrm{vol}_\Sigma
    \,,
  $$
  where the first summand  is the kinetic energy density from example \ref{FreeFieldLocalLagrangian}
  and where the second is
  any potential term as in example \ref{StandardLagrangianForMechanics}.
  More explicitly this means that
  $$
    H = \langle \nabla \phi, \nabla \phi\rangle + V(\{\phi^a\})
	  =  
	  k^{ab} \eta_{i j} p_a^i p_{b}^j + V(\{\phi^a\})
	\,.
  $$
  For this case the first component of the Hamilton-de Donder-Weyl equation 
  in the form of prop. \ref{DWHinComponents} is 
  the equation
  $$
    \partial_i \phi^a = k^{ab}\eta_{ij} p_b^j
    \,.
  $$
  This identifies the canonical momentum with the actual momentum. More formally,
  this first equation \emph{enforces the jet prolongation}
  in that it forces the section of the dual first jet bundle to the field bundle 
  to be the actual dual jet of an actual section of the field bundle.
    
  Using this, the second component of the DWH equation in the form of
  prop. \ref{DWHinComponents} is equivalently the \emph{wave equation}
  $$
    \eta^{i j}\partial_i \partial_{j} \phi^a = -\frac{\partial V}{\partial \phi^a}
  $$
  with inhomogeneity given by the gradient of the potential. 
  These equations are the hallmark of classical field theory.
\end{example}
In full generality we can express the Euler-Lagrange 
equations of motion of a local Lagrangian 
in Hamilton-de Donder-Weyl form by prop. \ref{ELFormAndPrePlecticCurrentByPullback}.

In order for the Hamilton-de Donder-Weyl equation to qualify as a good
``localization'' or ``de-transgression'' of non-covariant classical field theory
as in example \ref{PhaseSpaceFromLocalActionFunctionalOnTheLine} it should be
true that it reduces to this under transgression. This is indeed the case\footnote{
Again thanks go to Igor Khavkine for discussion of this point.} 
\begin{proposition}
  With $\omega_{L^{\mathrm{loc}}}$ as in prop. \ref{ELFormAndPrePlecticCurrentByPullback},
  we have that for any Cauchy surface $\Sigma_{n-1}$ that
  transgression of $\omega_{L^{\mathrm{loc}}}$ yields the 
  covariant phase space pre-symplectic form of 
  example \ref{PhaseSpaceFromLocalActionFunctionalOnTheLine}.
\end{proposition}

Using the $n$-plectic formulation of the Hamilton-de Donder-Weyl equation,
we naturally obtain now the following $n$-plectic formulation of the refinement
of the ``symplectic Noether theorem'', def. \ref{SymplecticNoetherTheorem}, form 
mechanics to field theory:
\begin{proposition}[$n$-plectic Noether theorem]
  Let $(X,\omega)$ be a pre-$n$-plectic manifold 
  equipped with a function $H \in C^\infty(X)$, to be regarded
  as a de Donder-Weyl Hamiltonian. If a vector field 
  $v \in \Gamma(T X)$ is a symmetry of $H$ in that 
  $$
    \iota_v \mathbf{d}H = 0
	\,,
  $$
  then along any $n$-vector field $ v_1 \vee \cdots \vee v_n$ 
  which solves the 
  Hamilton-de Donder-Weyl equation, def. \ref{DWHequation}, 
  the corresponding current 
  $\mathbf{J}_{v} := \iota_v \omega$
  is conserved, in that
  $$
    \iota_{(v_1, \cdots, v_n)}\mathbf{d}J_v = 0
	\,.
  $$
  Conversely, if a current is conserved on solutions to the 
  Hamilton-de Donder-Weyl equations of motion this way, then
  it generates a symmetry of the de Donder-Weyl Hamiltonian.
  \label{nPlecticNoetherTheorem}
\end{proposition}
\proof
  By the various definitions and assumptions we have
  $$
    \begin{aligned}
      \iota_{v_1 \vee \cdots \vee v_n} \mathbf{d}J_{n+1}
	  & =
      \iota_{v_1 \vee \cdots \vee v_n} \iota_{v} \omega
	  \\
	  & = 
      (-)^n \iota_{v} \iota_{v_1 \vee \cdots \vee v_n}  \omega
	  \\
	  & =
	  \iota_{v} \mathbf{d}H
	  \\
	  & = 0
	\end{aligned}
	\,.
  $$
\endofproof

This shows how the multisymplectic/$n$-plectic analog of the 
symplectic formulation of Hamilton's equations, def. \ref{HamiltonEquations},
serves to encode the equations of motion, the symmetries 
and the conserved currents of classical field theory. But in 
\ref{TheClassicalActionLegendreTransformAndHamuiltonianFlows} and 
\ref{QuantizationTheHeisenbergGroupAndSliceAutomorphismGroups} above we had seen
that the symplectic formulation of Hamilton's equations in turn is equivalently
just an infinitesimal characterization of the automorphisms of a 
pre-quantized phase space $X \stackrel{\nabla}{\longrightarrow} \mathbf{B}U(1)_{\mathrm{conn}}$
in the higher slice topos $\mathbf{H}_{/\mathbf{B}U(1)_{\mathrm{conn}}}$.
This suggests that 
$n$-dimensional Hamilton-de Donder-Weyl flows should characterize 
$n$-fold homotopies in the higher automorphism group of a higher prequantization,
regarded as an object in a higher slice topos to be denoted
$\mathbf{H}_{/\mathbf{B}^n U(1)_{\mathrm{conn}}}$.
This we come to below in \ref{HigherGaugeTheorySmoothInfinityGroupoids}.

Here we now first consider the infinitesimal aspect this statement.
To see what this will look like, observe that the statement for $n = 1$
is that the Lie algebra of slice automorphisms of $\nabla$
is the Poisson bracket Lie algebra $\mathfrak{pois}(X,\omega)$ whose elements, by 
def. \ref{PoissonBracketForPresymplectic}, are precisely the pairs $(v,H)$
that satisfy Hamilton's equation $\iota_v \omega = H$. To say this 
more invariantly: Hamilton's equations on $(X,\omega)$ precisely 
characterize the Lie algebra homomorphisms
of the form
$$
  \mathbb{R} \longrightarrow \mathfrak{pois}(X,\omega)
  \,,
$$
where on the left we have the abelian Lie algebra on a single generator.
This suggests that for a  (pre-)$n$-plectic manifold, we consider 
homotopy Lie algebra homomorphism of the form
$$
  \mathbb{R}^n \longrightarrow \mathfrak{pois}(X,\omega)
  \,,
$$
where now on the left we have the abelian Lie algebra on $n$ generators,
regarded canonically as a homotopy Lie algebra.
In comparison with prop. \ref{EvolutionFunctorInducedByHamiltonian},
this may be thought of as characterizing the infinitesimal approximation
to an evolution $n$-functor from Riemannian $n$-dimensional cobordisms into the 
(delooping of) the higher Lie integration of $\mathfrak{pois}(X,\omega)$
(recall remark \ref{CommutingHamiltonians} above).

Such homomorphisms of homotopy Lie algebras are computed as follows.
\begin{definition}
   Given a pre-$n$-plectic smooth space $(X,\omega)$,
  write
  $$
    \mathfrak{pois}(X,\omega)^{(\Box^n)} 
	:=
	(\wedge^\bullet\mathbb{R}^n) \otimes \mathfrak{pois}(X,\omega)
  $$
  for the homotopy Lie algebra obtained from the Poisson bracket 
  Lie $n$-algebra of def. \ref{PoissonLienAlgebra} by tensoring with
  the Grassmann algebra on $n$ generators, hence the graded-symmetric
  algebra on $n$ generators in degree 1.
  \label{BoxnIntoPois}
\end{definition}
\begin{remark}
A basic fact of homotopy Lie algebra theory implies that 
homomorphisms of the form $\mathbb{R}^n \longrightarrow \mathfrak{pois}(X,\omega)$
are equivalent to elements $\mathcal{J} \in \mathfrak{pois}(X,\omega)^{\Delta^n}$
of degree 1,
which satisfy the \emph{homotopy Maurer-Cartan equation}
$$
  d \mathcal{J} 
   + 
   \tfrac{1}{2}\{\mathcal{J},\mathcal{J}\} 
   + 
   \tfrac{1}{6}\{\mathcal{J},\mathcal{J}, \mathcal{J}\}
  + 
  \cdots
  \; = 0
$$
\end{remark}
\begin{example}
  Write $\{\mathbf{d}\sigma^i\}_{i = 1}^n$
  for the generators of $\wedge^\bullet\mathbb{R}^n$.
  Then a general element of degree 1 in $\mathfrak{pois}(X,\omega)^{(\Box^n)}$ is
  of the form
  $$
    \mathcal{J}
	  = 
	\mathbf{d}\sigma^i \otimes (v_i, J_i)
	+ 
	\mathbf{d}\sigma^i \wedge \mathbf{d}\sigma^j
	\otimes
	J_{i j}
	+ 
	\mathbf{d}\sigma^i \wedge \mathbf{d}\sigma^j \wedge \mathbf{d}\sigma^k
	\otimes
	J_{i j k}
	+ 
	\cdots
	+ 
	(\mathbf{d}\sigma^1 \wedge \cdots \wedge \mathbf{d}\sigma^n)
	\otimes
	H
	\,,
  $$
  where 
  \begin{enumerate}
    \item $v_i \in \Gamma(T X)$ is a vector field and $J_i \in \Omega^{n}(X)$
	is a differential $n$-foms such that $\iota_{v_i}\omega = \mathbf{d}J_i$
	\item $J_{i_1 \cdots i_k} \in \Omega^{n+1-k}(X)$;
	\item $H \in C^\infty(X)$.
  \end{enumerate}
  \label{ComponentsOfElementsInBoxnPois}
\end{example}
From this we deduce the following.
\begin{proposition}
     Given a pre-$n$-plectic smooth space $(X,\omega)$,
 the extended Hamilton-de Donder-Weyl equations,
  def. \ref{DWHequation}, characterize, under the identification of 
  example \ref{ComponentsOfElementsInBoxnPois}, 
  the homomorphims of homotopy Lie algebras from $\mathbb{R}^n$
  into the higher Poisson bracket Lie $n$-algebra of def. \ref{PoissonLienAlgebra}:
  $$
    \left(
	  \mathcal{J} : \mathbb{R}^n \longrightarrow \mathfrak{pois}(X,\omega)
	\right)
	\;\;
	 \Leftrightarrow
	\;\;
	\left\{
	  \begin{array}{ll}
	    \iota_{v_n} \cdots \iota_{v_1}\omega = \mathbf{d}H
		\\
		\iota_{v_{i k}}\cdots \iota_{v_{i_2}}\iota_{v_{i_1}}\omega
		=
		\mathbf{d}J_{i_1 i_2 \cdots i_k} & \forall_{k} \forall_{i_1, \cdots, i_k}
	  \end{array}
	\right.
  $$
  \label{DWHisMC}
\end{proposition}
\begin{remark}
  The Lie integration of the Lie $n$-algebra $\mathfrak{pois}(X,\omega)$
  is the smooth $n$-groupoid whose $n$-cells are Maurer-Cartan elements in 
  $$
    \Omega^\bullet_{\mathrm{si}}(\Delta^n) \otimes \mathfrak{pois}(X,\omega)
	\,,
  $$
  see \cite{FSS} for details.
  The construction in def. \ref{BoxnIntoPois} is a locally constant approximation
  to that. In general there are further $\sigma$-dependent terms. 
\end{remark}

Due to \cite{hgp, LocalObservables} we have that the Lie integration
of $\mathfrak{pois}(X,\omega)$ is the automorphism $n$-group 
$\mathbf{Aut}_{/\mathbf{B}^n U(1)_{\mathrm{conn}}}(\nabla)$ of 
any pre-quantization $\nabla$ of $(X,\omega)$, see \ref{HigherDifferentialGeometryInIntroduction}. 
This means that the above maps
$$
  \mathbb{R}^n \longrightarrow \mathfrak{pois}(X,\omega)
$$
are infinitesimal approximations to something lie $n$-functors of the form
$$
  \mbox{
   ``
   $
  \mathrm{Bord}_n^{\mathrm{Riem}}
  \longrightarrow
  \mathbf{H}_{/\mathbf{B}^n U(1)_{\mathrm{conn}}}
  $''}
$$
in higher dimensional analogy of prop. \ref{EvolutionFunctorInducedByHamiltonian}.
This we come to below.

\paragraph{Source terms, off-shell Poisson bracket and Poisson holography}
\label{SourceTermsOffShell}
\index{holography!1d mechanics/2d Poisson-CS}

We connect now the discussion of mechanics 
in \ref{BasicClassicalMechanicsByPrequantizedLagrangianCorrespondences} 
to that of higher Chern-Simons field
theory in by showing that the space of 
all trajectories of a mechanical system naturally carries 
a Poisson brakcet structure which is foliated by symplectic
leafs that are labled by source terms.\footnote{This phenomenon was kindly pointed out to 
by Igor Khavkine.} 
The corresponding leaf space 
is naturally refined to the symplectic groupoid that is the 
moduli stack of fields of the non-perturbative
2s Poisson-Chern-Simons theory. This yields a precise implementation of the 
``holographic principle'' where the 2d Poisson-Chern-Simons theory in the 
bulk carries on its boundary a 1d field theory (mechanical system) such 
that fields in the bulk correspond to sources on the boundary.

\medskip

Let $(X,\omega)$ be a symplectic manifold. We write 
$$
  \{-,-\} \;\colon\; C^\infty(X)\otimes C^\infty(X) \longrightarrow C^\infty(X)
$$
for the Poisson bracket induced by the symplectic form $\omega$, 
hence by the Poisson bivector $\pi := \omega^{-1}$.

For notational simplicity we will restrict attention to the special case that 
$$
  X = \mathbb{R}^2 \simeq T^\ast \mathbb{R}
$$ 
with canonical coordinates 
$$
  q,p \;\colon\; \mathbb{R}^2 \longrightarrow \mathbb{R}
$$ 
and symplectic form 
$$
  \omega = \mathbf{d}q \wedge \mathbf{d}p
  \,.
$$ 
The general case of the following discussion is a straightforward generalization of this, which is just notationally more inconvenient. 

Write $I := [0,1]$ for the standard interval regarded as a smooth manifold manifold with boundary|with boundary. The mapping space 
$$
  P X := [I, X]
$$
canonically exists as a smooth space, but since $I$ is compact topological space|compact this structure canonically refines to that of a Fréchet manifold. This implies that there is a good notion of tangent space $T P X$. The task now is to construct a certain Poisson bivector as a section $\pi \in \Gamma^{\wedge 2}(T P X)$.

Among the smooth functions on $P X$ are the evaluation maps 

$$
  ev \;\colon\; P X \times I = [I,X] \times I \stackrel{}{\longrightarrow} X
$$

whose components we denote, as usual, for $t \in I$ by

$$
  q(t) := q \circ ev_t \;\colon\; P X \longrightarrow \mathbb{R}
$$

and

$$
  p(t) := p \circ ev_t \;\colon\; P X \longrightarrow \mathbb{R}
  \,.
$$

Generally for $f \colon X \to \mathbb{R}$ any smooth function, we write $f(t) := f \circ ev_t \in C^\infty(P X)$. This defines an embedding

$$
  C^\infty(X) \times I \hookrightarrow C^\infty(P X)
  \,.
$$
Similarly we have 
$$
  \dot q(t) \;\colon\; P X \longrightarrow \mathbb{R}
$$
and
$$
  \dot q(t) \;\colon\; P X \longrightarrow \mathbb{R}
$$
obtained by differentiation of $t \mapsto q(t)$ and $t \mapsto p(t)$.

Let now 
$$
  H \;\colon\; X \times I \longrightarrow \mathbb{R}
$$ 
be a smooth function, to be regarded as a time-dependent Hamiltonian. This induces a time-dependent function on trajectory space, which we denote by the same symbol
$$
  H
  \;\colon\;
  P X \times I
   \stackrel{(ev,id)}{\longrightarrow}
  X \times X
   \stackrel{H}{\longrightarrow}
  \mathbb{R}
  \,.
$$
Hence for $t \in I$ we write
$$
  H(t) 
  \;\colon\;
  P X \times \{t\}
  \stackrel{(ev, id)}{\longrightarrow}
  X \times \{t\}
  \stackrel{H}{\longrightarrow}
  \mathbb{R}
$$
for the function that assigns to a trajectors $(q(-),p(-)) \colon I \longrightarrow X$ its energy at (time) parameter value $t$.
Define then the Euler-Lagrange equation|Euler-Lagrange density induced by $H$ to be the functions
$$
  \mathrm{EL}(t)
  \;\colon\;
  P X
  \longrightarrow 
  \mathbb{R}^2
$$
with components
$$
  \mathrm{EL}(t)
  = 
  \left(
    \begin{array}{l}
      \dot q(t) - \frac{\partial H}{\partial p}(t)
      \\
      \dot p(t) + \frac{\partial H}{\partial p}(t)
    \end{array}
  \right)
  \,.
$$
The trajectories $\gamma \colon I \to X$ on which $EL(t)$ vanishes for all $t \in I$ are equivalently those 
\begin{itemize}
\item for which the tangent vector $\dot \gamma \in T_{\gamma}X$ is a Hamiltonian vector field|Hamiltonian vector for $H$;
\item which satisfy Hamilton's equations equations of motion|of motion for $H$.
\end{itemize}
Since the differential equations 
$\mathrm{EL} = 0$ have a unique solution for given initial data $(q(0), p(0))$, the evaluation map
$$
  \left\{
    \gamma \in P X | \forall_{t \in I}\; EL_\gamma(t) = 0 
  \right\}
  \stackrel{\gamma \mapsto \gamma(0)}{\longrightarrow}
  X
$$
is an equivalence (an isomorphism of smooth spaces).

Write 
$$
  \mathrm{Poly}(P X) \hookrightarrow C^\infty(P X)
$$
for the subalgebra of smooth functions on path space which are
polynomials of integrals over $I$, of the smooth functions in the image of $C^\infty(X) \times I \hookrightarrow C^\infty(P X)$ and all their derivatives along $I$.

Define a bilinear function
$$
  \{-,-\}
  \;\colon\;
  \mathrm{Poly}(P X) \otimes \mathrm{Poly}(P X)
  \longrightarrow \mathrm{Poly}(P X)
$$
as the unique function which is a derivation in both arguments and moreover is a solution to the differential equations
$$
  \frac{\partial}{\partial t_2}
  \left\{f(t_1), q(t_2)\right\}
  =
  \left\{
    f(t_1), \frac{\partial H}{\partial p}(t_2)
  \right\}
$$
$$
  \frac{\partial}{\partial t_2}
  \left\{f(t_1), p(t_2)\right\}
  =
  -
  \left\{
    f(t_1), \frac{\partial H}{\partial q}(t_2)
  \right\}
$$
subject to the initial conditions
$$
  \{f(t), q(t)\} = \{f,q\}
$$
$$
  \{f(t), p(t)\} = \{f,p\}
$$
for all $t \in I$, where on the right we have the original Poisson bracket on $X$.

This bracket directly inherits skew-symmetry and the Jacobi identity from the Poisson bracket of $(X, \omega)$, hence equips the vector space $Poly(P X)$ with the structure of a Lie bracket. Since it is by construction also a derivation of  $Poly(P X)$ as an associative algebra, we have that 
$$
  \left(
    \mathrm{Poly}\left(P X\right),
    \;
    \left\{
      -,-
    \right\}
  \right)
  \;\;\;
  \in P_1 Alg
$$
is a Poisson algebra. This is the ``off-shell Poisson algebra'' 
on the space of trajectories in $(X,\omega)$.

Observe that by construction of the off-shell Poisson bracket, 
specifically by the differential equations defining it, the
Euler-Lagrange equation|Euler-Lagrange function $\mathrm{EL}$
generate a Poisson reduction|Poisson ideal. 

For instance 
$$
 \left(
  \begin{array}{rcl}
  \frac{\partial}{\partial t_2}
  \left\{f(t_1), q(t_2)\right\}
  &=&
  \left\{
    f(t_1), \frac{\partial H}{\partial p}(t_2)
  \right\}
  \\
  \frac{\partial}{\partial t_2}
  \left\{f(t_1), p(t_2)\right\}
  &=&
  -
  \left\{
    f(t_1), \frac{\partial H}{\partial q}(t_2)
  \right\}
  \end{array}
  \right)
  \;\;\;
  \Leftrightarrow
  \;\;\;
  \left(
    \left\{
      f(t_1), \;
      EL(t)
    \right\}
    =  
    0
  \right)
  \,.
$$
Moreover, since $\{EL(t) = 0\}$ are  equations of motion the Poisson reduction defined by this Poisson idea is the subspace of those trajectories which are solutions of Hamilton's equations, hence the "on-shell trajectories".

As remarked above, the initial value map canonically identifies this on-shell trajectory space with the original phase space manifold $X$. Moreover, by the very construction of the off-shell Poisson bracket as being the original Poisson bracket at equal times, hence in particular at time $t = 0$, it follows that restricted to the zero locus $EL = 0$ the off-shell Poisson bracket becomes symplectic manifold|symplectic.

All this clearly remains true with the function $EL$ replaced by the function $EL - J$, for  $J \in C^\infty(I)$ any function of the (time) parameter (since $\{J,-\} = 0$). Any such choice of $J$ hence defines a symplectic subspace 
$$
  \left\{
    \gamma \in P X
    \;|\;
    \forall_{t \in I}\; EL_\gamma(t) = J
  \right\}
$$

of the off-shell Poisson structure on trajectory space. Hence $\left(O X, \left\{-,-\right\}\right)$ has a foliation by symplectic leaves with the leaf space being the smooth space $C^\infty(I)$ of smooth functions on the interval. 

Notice that changing $\mathrm{EL} \mapsto \mathrm{EL} - J$ corresponds changing the time-dependent Hamiltonian $H$ as
$$
  H \mapsto H - J q
  \,.
$$
Such a term linear in the canonical coordinates (the field (physics)|fields) is a 
\emph{source term}. (The action functionals with such source terms added serve as integrands of generating functions for correlators in statistical mechanics and in quantum mechanics.)

Hence in conclusion we find the following statement:

The trajectory space (history space) of a mechanical system carries a natural Poisson manifold|Poisson structure whose symplectic leaves are the subspaces of those trajectories which satisfy the equations of motion with a fixed source term and hence whose symplectic leaf space is the space of possible sources. 
 
Notice what becomes of this statement as we consider the 2d Chern-Simons theory induced by the off-shell Poisson bracket (the non-perturbative field theory|non-pertrbative Poisson sigma-model) whose moduli stack of field (physics)|fields is the symplectic groupoid $SG\left(P X, \left\{-,-\right\}\right)$ induced by the Poisson structure. 

By the discussion at ...  the Poisson space $\left(P X, \left\{-,-\right\}\right)$ defines a  boundary field theory (in the sense of local prequantum field theory) for this 2d Chern-Simons theory, exhibited by a boundary correspondence of the form
$$
  \xymatrix{
    &P X
	\ar[dl]
	\ar[dr]_{\ }="s"
	\\
    \ast \ar[dr]^{\ }="t" && \mathrm{SG}\left(P X, \left\{-,-\right\}\right)
	\ar[dl]
    \\
    & \mathbf{B}^2 U(1)
	\ar[d]
    \\
    & \mathrm{KU} \mathrm{Mod}
	\ar@{=>} "s"; "t"
  }
  \,.
$$
Notice that the symplectic groupoid is a version of the symplectic leaf|symplectic leaf space of the given Poisson manifold (its 0-truncation is exactly the leaf space). Hence in the case of the off-shell Poisson bracket, the symplectic groupoid is the space of \emph{sources} of a mechanical system. At the same time it is the moduli space of field (physics)|fields of the 2d Chern-Simons theory of which the mechanical system is the boundary field theory.
Hence the field (physics)|fields of the bulk field theory are identified with the sources of the boundary field theory. Hence conceptually the above boundary correspondence diagram is of the following form
$$
  \xymatrix{
    &\mbox{Sources}
	\ar[dl]
	\ar[dr]_{\ }="s"
	\\
    \ast \ar[dr]^{\ }="t" 
	&& 
	\mbox{Fields}
	\ar[dl]
    \\
    & \mbox{Phases}
	\ar@{=>} "s"; "t"
  }
  \,.
$$

\subsubsection{Higher pre-quantum gauge fields}
\label{PrequantumInHigherCodimension}
\label{Fields}

We give an introduction and survey to some aspects of the formulation of
higher prequantum field theory in a cohesive $\infty$-topos. 

\medskip

One of the pleasant consequences of formulating the geometry of (quantum) field theory in terms of 
higher stacks, hence in terms of higher topos theory, is that a wealth of constructions find 
a natural and unified formulation, which subsumes varied traditional constructions and generalizes
them to higher geometry. In this last part here we give an outlook of the scope
of field theoretic phenomena that the theory naturally captures or exhibits in the first place.

In the following we write $\mathbf{H}$ for the collection of higher stacks under consideration.
The reader may want to think of the special case that was discussed in the previous sections,
where $\mathbf{H} = \mathrm{Smooth}\infty\mathrm{Grpd}$ is the collection of 
\emph{smooth $\infty$-groupoids}, hence of higher stacks
on the site of smooth manifolds, or, equivalently, its dense subsite of Cartesian spaces.
But one advantage of speaking in terms of higher topos theory is that essentially every
construction considered in the following makes sense much more generally if 
only $\mathbf{H}$ is any higher topos that satisfies a small set of axioms called 
(differential) \emph{cohesion}. This allows one to transport all considerations across
various kinds of geometries. Notably we can speak of higher \emph{supergeometry},
hence of fermionic quantum fields,
simply by refining the site of definition to be that of supermanifolds: 
also the higher topos $\mathbf{H} = \mathrm{SmoothSuper}\infty\mathrm{Grpd}$ is
differentially cohesive.

Therefore we speak in the following in generality of \emph{cohesive maps} when we refer to maps
with geometric structure, be it topological, smooth, analytic, supergeometric or otherwise.
Throughout, this geometric structure is \emph{higher geometric} which we will sometimes
highlight by adding the ``$\infty$-''-prefix as in \emph{cohesive $\infty$-group}, but
which we will often notationally suppress for brevity. Similarly, \emph{all} of the diagrams
appearing in the following are filled with homotopies, but only sometimes we explicitly
display them (as double arrows) for emphasis or in order to label them.

The special case of \emph{geometrically discrete} cohesion is exhibited by the $\infty$-topos
$\infty \mathrm{Grpd}$ of bare $\infty$-groupoids or \emph{homotopy types}. This is 
the context of traditional homotopy theory, presented by topological spaces regarded
up to weak homotopy equivalences (``whe''s):
$\infty \mathrm{Grpd} \simeq L_{\mathrm{whe}} \mathrm{Top}$. One of the axioms satisfied
by a cohesive $\infty$-topos $\mathbf{H}$ is that the inclusion 
$\mathrm{Disc} : \infty \mathrm{Grpd} \hookrightarrow \mathbf{H}$ of bare $\infty$-groupoids
as cohesive $\infty$-groupoids equipped with discrete cohesive structure has 
not only a right adjoint $\infty$-functor
$\Gamma : \mathbf{H} \to \infty \mathrm{Grpd}$ -- the functor that forgets the cohesive structure and
remembers only the underlying bare $\infty$-groupoid --  but also a 
 left adjoint
${\vert -\vert} : \mathbf{H} \to \infty \mathrm{Grpd}$. This is the \emph{geometric realization}
of cohesive $\infty$-groupoids.

\medskip

We discuss first the general notion of (quantum) fields, then that of Lagrangians
and action functionals on spaces of fields and the corresponding \emph{phase spaces},
and finally we discuss the geometric prequantum theory of such data.

\begin{itemize}
  \item \ref{Fields} -- Fields
  \item \ref{PhaseSpaces} -- Phase spaces
  \item \ref{PrequantumGeometry} -- Prequantum geometry
\end{itemize}

The following discussion is based on and in part reviews previous work such as 
\cite{SSSIII, FiorenzaSatiSchreiberIV}. 
Lecture notes that provide an exposition of this material with an emphasis 
on fields as twisted (differential) cocycles are in \cite{TwistedStructuresLecture}.

\medskip

We discuss now how a plethora of species of (quantum) fields are naturally 
and precisely expressed by constructions in the higher topos $\mathbf{H}$.
In fact, it is the \emph{universal moduli stacks} $\mathbf{Fields}$ 
of a given species of fields which are naturally expressed: 
those objects such that maps $\phi : X \to \mathbf{Fields}$ 
into them are equivalently quantum fields of the given species on $X$. 
This has three noteworthy effects on the formulation of the corresponding field theory.

First of all it means that every quantum field theory thus expressed is formally 
analogous to a $\sigma$-model -- the ``target space'' is a higher moduli stack -- which 
brings about a unified treatment of varied types of QFTs. 

Second it means that a differential cocycle on $\mathbf{Fields}$ of degree $(n+1)$ -- 
itself modulated by a map 
$$
  \mathbf{L} : \mathbf{Fields} \to \mathbf{B}^n U(1)_{\mathrm{conn}}
$$
to the moduli stack $n$-form connections -- serves as an \emph{extended} Lagrangian
of a field theory, in the sense that it
expresses a QFT fully locally by Lagrangian data in arbitrary codimenion: for every 
closed oriented worldvolume $\Sigma_k$ of dimension $k \leq n$ there is 
a \emph{transgressed} Lagrangian
$$
  \exp(2 \pi i \int_{\Sigma_k}[\Sigma_k, \mathbf{L}])
  :
  \xymatrix{
    \mathbf{Fields}(\Sigma_k)
	\ar[rr]^{[\Sigma_k, \mathbf{L}]}
	&&
	[\Sigma_k, \mathbf{B}^n \mathbb{C}_{\mathrm{conn}}]
	\ar[rr]^{\exp(2 \pi i \int_{\Sigma_k} (-))}
	&&
	\mathbf{B}^{n-k} \mathbb{C}^\times_{\mathrm{conn}}
  }
$$
which itself is a differential $(n-k)$-form connection on the space of fields on $\Sigma_k$.
In particular, when $n = k$ then $\mathbf{B}^0 U(1)_{\mathrm{conn}} \simeq U(1)$
and the transgressed Lagragian in codimension $0$ is the (exponentiated) \emph{action functional} 
of the theory, $\exp(i S(-)) : \mathbf{Fields}(\Sigma_n) \to U(1)$. 
On the other hand, 
the $(n-k)$-connections in higher codimension are higher (off-shell) \emph{prequantum bundles} of the theory.
This we discuss further below in \ref{PrequantumGeometry}.

Third, it means that the representation of fields by their higher moduli stacks in a higher topos
identifies the notion of quantum field entirely with that of \emph{cocycle} in 
general \emph{cohomology}. This we turn to now in \ref{CocyclesGeneralizedParameterizedTwisted}.

\paragraph{Cocycles: generalized, parameterized, twisted}
\label{CocyclesGeneralizedParameterizedTwisted}

We discuss general aspects of cocycles and cohomology in an $\infty$-topos, as a 
general blueprint for all of the discussion to follow. The reader eager to see explicit
structure genuinely related to (quantum) physics may want, on first reading, to skip ahead to 
\ref{FieldsOfGravtySpecialAndGeneralizedGeometry} and come back here only as need be.

In higher topos theory the
notion of \emph{cocycle} $c$ on some space $X$ with coefficients in some object $A$ 
and with some \emph{cohomology} class $[c]$
is identified simply with that of a map (a morphism) $c : X \to A$ with equivalence class
$$
  [c] \in H(X,A) := \pi_0 \mathbf{H}(X,A)
  \,.
$$
This is traditionally familiar for the case of discrete geometric structure
hence bare homotopy theory $\mathbf{H} = \infty \mathrm{Grpd}$, where for any
Eilenberg-Steenrod-\emph{generalized cohomology theory} the object $E$ is the corresponding spectrum,
as given by the Brown representability theorem. That over non-trivial sites the
same simple formulation subsumes all of \emph{sheaf cohomology} (``parameterized cohomology'') 
is known since \cite{Brown}, but it appears in the literature mostly in a bit of disguise
in terms of some explicit model of a \emph{derived global section functor}, computed by means of 
suitable projective/injective resolutions.)

If here $A = \mathbf{Fields}$ is interpreted as the moduli stack of certain \emph{fields},
then such a cocycle \emph{is} a field configuration on $X$. This is familiar for the 
case that we think of $A = X$ as the target space of a \emph{$\sigma$-model}. But for instance
for $G \in \mathrm{Grp}(\mathbf{H})$ a (higher) group and $A := \mathbf{B}G_{\mathrm{conn}}$
a differential refinement of the universel moduli stack of $G$-principal $\infty$-bundles,
a map $c : X \to \mathbf{B}G_{\mathrm{conn}}$ is on the one hand a cocycle in 
(nonabelian) differential $G$-cohomology on $X$, and on the other hand equivalently
a $G$-\emph{gauge field}  on $X$. In particular this means that in higher topos theory
gauge field theories are unified with $\sigma$-models: an (untwisted) gauge field is a 
$\sigma$-modelfield whose target space is a universal differential moduli stack
$\mathbf{B}G_{\mathrm{conn}}$. 

Indeed, the kinds of fields which are identified as $\sigma$-model fields in higher topos
theory, hence with cocycles in some geometric cohomology theory, 
is considerable richer, still. The reason for this is that with $B \in \mathbf{H}$
any object, the \emph{slice} $\mathbf{H}_{/B}$ is itself again a higher topos. This slice topos
is the collection of morphisms of $\mathbf{H}$ into $B$, where a map between two such morphisms
$f_{1,2} : X_{1,2} \to B$ is 
\begin{enumerate}
\item a map $\phi : X_1 \to X_2$ in $\mathbf{H}$ 
\item  a homotopy
$\eta : \xymatrix{f_1\ar[r]^-\simeq & f_2\circ \phi}$,
\end{enumerate} 
hence a diagram in $\mathbf{H}$ of the form
$
  \raisebox{20pt}{
  \xymatrix{
    X_1 \ar[rr]^\phi_{\ }="s" \ar[dr]_{f_1}^{\ }="t" && X_2 \ar[dl]^{f_2}
	\\
	& B
	\ar@{=>}^\eta "s"; "t"
  }}
  \,.
$
We are particularly interested in the case that $B = \mathbf{B}G$ is a moduli stack of $G$-principal 
$\infty$-bundles (or a differential refinement thereof). 
The fact that $\mathbf{H}$ is \emph{cohesive} implies in particular that 
every morphism $g : X \to \mathbf{B}G$ has a unique global homotopy
fiber $P \to X$. This is the \emph{$G$-principal bundle} over $X$ modulated by $g$, sitting in a
long homotopy fiber sequence of the form
$$
  \xymatrix{
    G \ar[r] & P \ar[d]
	\\
	& X \ar[r]^g & \mathbf{B}G
  }
  \,.
$$
In particular this means that there is an action of $G$ on $P$ 
(precisely: a \emph{homotopy coherent} or $A_\infty$-action) and that 
$$
  P \to P /\!/ G \simeq X 
$$
is the quotient map of this action. Moreover, conversely every action of $G$ on any
object $V \in \mathbf{H}$ arises this way and is modulated by a morphism
$\xymatrix{V/\!/G \ar[r]^{\rho} & \mathbf{B}G}$, sitting in a homotopy fiber sequence of the form
$$
  \raisebox{20pt}{
  \xymatrix{
    V \ar[r] & V/\!/G \ar[d]^\rho
	\\
	& \mathbf{B}G
  }
  }
  \,.
$$
(This and the following facts about $G$-principal $\infty$-bundles in $\infty$-toposes 
and the representation theory
and twisted cohomology of cohesive $\infty$-groups is due to \cite{NSSa}, an account in the 
present context is in section 3.6 here.)
This fiber sequence exhibits $V/\!/G \to \mathbf{B}G$ as the universal $V$-fiber bundle which is
$\rho$-associated to the universal $G$-principal bundle over $\mathbf{B}G$.
For instance the fiber sequence $G \to * \to \mathbf{B}G$ which defines the delooping 
of $G$ corresponds to the action of $G$ on itself by right (or left) multiplication;
the fiber sequence $\xymatrix{V \ar[r] &  V \times \mathbf{B}G \ar[r]^{p_2} &\mathbf{B}G}$
corresponds to the trivial action on any $V$, and the fiber sequence
$\xymatrix{G \ar[r] & \mathcal{L}\mathbf{B}G \ar[r] & \mathbf{B}G}$ of the free loop space
object of $\mathbf{B}G$ corresponds to the adjoint action of $G$ on itself.

Another case of special interest is that where $V \simeq \mathbf{B}A$ 
and $V/\!// G \simeq \mathbf{B}\hat G$ are themselves deloopings of $\infty$-groups.
In this case the above fiber sequence reads
$$
  \xymatrix{\mathbf{B}A \ar[r] & \mathbf{B}\hat G \ar[r] & \mathbf{B}G}
$$
and exhibits an \emph{extension} $\hat G$ of $G$ by $A$. The implied action of $G$
on $\mathbf{B}A$ via $\mathbf{Aut}(\mathbf{B}G) \simeq \mathbf{Aut}_{\mathrm{Grp}}(G)/\!/\mathrm{ad}$
is the datum known from traditional \emph{Schreier theory} of general (nonabelian) group extensions.  
Now the previous discussion implies
that if $A$ is equipped with sufficient abelian structure 
in that also $\mathbf{B}A$ is equipped with $\infty$-group structure
(a ``braided $\infty$-group'')
and such that $\mathbf{B}\hat G \to \mathbf{B}G$ is the quotient projection of a $\mathbf{B}A$-action, 
then the extension is classified by an \emph{$\infty$-group cocycle} 
$\mathbf{c} : \xymatrix{ \mathbf{B}G \ar[r] & \mathbf{B}^2 A}$ in $\infty$-group cohomology
$[\mathbf{c}] \in H^2_{\mathrm{grp}}(G, A)$. Notice that this is \emph{cohesive} group cohomology
in that it does respect and reflect the geometric structure on $G$ and $A$. Notaby in
smooth cohesion and for $G$ a Lie group and $A = \mathbf{B}^n K$ the $n$-fold delooping of 
an abelian Lie group, this reproduces not the naive Lie group cohomology but the refined
Segal-Brylinski Lie group cohomology (this is shown in section 4.4.6.2 here).
This implies that for $G$ a compact Lie group and $A = \mathbf{B}^n U(1)$ we have an equivalence
$$
  H_{\mathrm{Grp}}^n(G,U(1))
  \simeq
  H^{n+1}(B G, \mathbb{Z})
$$
between the refined cohesive group cohomology with coefficients in the circle group and the 
ordinary integral cohomology of the clasifying space $B G \simeq {\vert \mathbf{B}G\vert}$
in one degree higher. In other words this means that every \emph{universal characteristic class}
$c : \xymatrix{B G \ar[r] &  K(\mathbb{Z}, n+1)}$ is cohesively refined essentially uniquely to 
(the instanton sector of) a 
higher gauge field: a cohesive circle $n$-bundle
(bundle $(n-1)$-gerbe) on the universal moduli stack $\mathbf{B}G$. The ``universality'' of 
this higher gauge field is reflected in the fact that this is really the 
(twisting structure underlying) an \emph{extended action function for higher Chern-Simons theory}
controld by the given universal class. This we come back to below in \ref{GaugeFieldHigherTwistedNonAbelian}.

From this higher bundle theory, higher group theory and higher representation theory, 
we obtain a finer interpretation
of maps in the slice $\mathbf{H}_{/\mathbf{B}G}$. First of all one finds that 
$$
  \mathbf{H}_{/\mathbf{B}G}
  \simeq
  G\mathbf{Act}
$$
is indeed the $\infty$-category of $G$-actions and $G$-action homomorphisms.
In particular the base change functors $(\mathbf{G}\phi)_*$ and $(\mathbf{B}\phi)_!$ 
along maps $\mathbf{B}\phi : \mathbf{B}G \to \mathbf{B}G'$ corresponds to 
the (co)induction functors from $G$-representations to $G'$-representations along a
group homomorphism $\phi$. Since all this is homotopy-theoretic (``derived'')
the space of maps in the slice from the trivial representation to 
any given representation $(V, \rho)$ (hence the \emph{derived invariants} of $(V,\rho)$) 
is the cocycle $\infty$-groupoid of the 
\emph{group cohomology} of $G$ \emph{with coefficients} in $V$:
$$
  H_{\mathrm{Grp}}(G,V)
  \simeq
  \pi_0
  \mathbf{H}_{/\mathbf{B}G}(\mathrm{id}_{\mathbf{B}G}, \rho)  
  \,.
$$
We are interested in the generalizations of this to the case where the univeral 
$G$-principal $\infty$-bundle modulated by $\mathrm{id}_{\mathbf{B}G}$ is 
replaced by any $G$-principal bundle modulated by a map $g_X : X \to \mathbf{B}G$.
To see what general cocycles in $\mathbf{H}_{/\mathbf{B}G}(g_X, \rho)$ are like, notice that
every $G$-principal $\infty$-bundle over a given $X$ locally trivializes over a cover
$\xymatrix{U \ar@{->>}[r] & X}$ (an \emph{effective epimorphism} in $\mathbf{H}$)
in that the modulating map becomes null-homotopic on $U$: $g_X|_U \simeq \mathrm{pt}_{\mathbf{B}G}$.
But by the universal property of homotopy fibers this means that a cocycle 
$\sigma : g_X \to \rho$ in $\mathbf{H}_{/\mathbf{B}G}$ is \emph{locally} a cocycle
$\sigma|_{U} : U \to V$ in $\mathbf{H}$ with coefficients in the given $G$-module $V$, 
as shown on the left of the following diagram:
$$
  \raisebox{20pt}{
  \xymatrix{
    U \ar@{->>}[d] \ar[rr]^{\sigma|_U} && V \ar[d]
	\\
	X \ar[dr]_{g_X} \ar[rr]^\sigma   && V /\!/G \ar[dl]^\rho
	\\
	& \mathbf{B}G
  }
  }
  \;\;\;\;
  \simeq
  \;\;\;\;
  \raisebox{20pt}{
  \xymatrix{
    & V \ar[r] &  P \times_G V \ar[r] \ar[d] & V/\!/G \ar[d]^{\rho}
    \\
    U \ar@{->>}[r] \ar[ur]^{\sigma|_U} & X \ar[r]^{\mathrm{id}} \ar[ur]^{\sigma} & X \ar[r]^{g_X}
	&
	\mathbf{B}G
  }
  }
  \,.
$$
This means that $\sigma$ is a cocycle with \emph{local coefficients} in $V$, which 
however globally vary as controlled by $g_X$: it is \emph{twisted} by $g_X$.
On the right hand of the above diagram the same situation is displayed in an equivalent alternative perspective:
since $\rho : V/\!/G \to \mathbf{B}G$ is also the univeral $\rho$-associated $V$-fiber bundle,
it follows that the $V$-fiber bundle $P \times_G V \to X$ associated to $P \to X$ is its
pullback along $g_X$ and then using again the universal property of the homotopy pullback
it follows that $\sigma$ is equivalently a \emph{section} of this associated bundle.
This is the traditional perspective of $g_X$-\emph{twisted $V$-cohomology} as familiar 
notably from twisted $K$-theory, as well as from modern formulations of ordinary cohomology
with local coefficients.

The perspective of twisted cohomology as cohomology in slice $\infty$-topos $\mathbf{H}_{/\mathbf{B}G}$
makes it manifest that what acts on twisted cocycle spaces are \emph{twist homomorphisms},
hence maps
$(\phi, \eta) :  g_Y \to g_X$ in $\mathbf{H}_{/\mathbf{B}G}$. In particular for $g_X$
and given twist its automorphism $\infty$-group $\mathrm{Aut}_{/\mathbf{B}G}(g_X)$ acts
on the twisted cohomology $\mathbf{H}_{/\mathbf{B}G}(g_X, \rho)$ by precomposition
in the slice.

In conclusion we find that cocycles and fields in the slice 
slice $\infty$-topos $\mathbf{H}_{/\mathbf{B}G}$ of a cohesive $\infty$-topos over the delooping
of an $\infty$-group are structures with components as summarized in the following diagram:

$$
  \xymatrix{
    && U \ar@{->>}[d] \ar[rr] && V \ar[d]
    \\
    Y \ar[rr] \ar[drr] && X\ar[d] \ar[rr] && V/\!/G \ar[dll]
	\\
	&& \mathbf{B}G
  }
$$

\fbox{
$$
  \xymatrix{
    && {\mbox{local} \atop \mbox{cover}}
	\ar@{->>}[d]
	\ar[rr]^{\mbox{local} \atop \mbox{cocycle}}
	&&
	  {\mbox{local}
	  \atop 
	  \mbox{coefficients}}
	  \ar[d]
    \\
    Y \ar[rr]_>>{\ }="s" \ar[drr]_{\mbox{twist}}^{\ }="t"  
	&& 
	X  \ar[d]|{\mbox{twist}} \ar[rr]|-{\mbox{twisted} \atop \mbox{cocycle} } 
	&& 
	\mbox{ \begin{tabular}{c} local coefficient \\ bundle \end{tabular} }
	\ar[dll]^{\mbox{twist-action} \atop \mbox{on local coefficients}}
	\\
	&& \mbox{\begin{tabular}{c} twist \\ coefficients \end{tabular}} && 
	\ar@{=>}|{\mbox{twist transf.}} "s"; "t"
  }
$$
}

\medskip

\noindent In the following we list a wide variety of classes of examples of this unified
general abstract picture.

\paragraph{Fields of gravity: special and generalized geometry}
\label{FieldsOfGravtySpecialAndGeneralizedGeometry}

As special cases of the above general discussion, we now discuss moduli $\infty$-stacks
of \emph{fields of gravity} and their generalizations as found in higher dimensional
(super)gravity. 

\medskip

For $X \in \mathrm{Mfd}_n \hookrightarrow \mathbf{H}$ a manifold of dimension $n$,
we may naturally regard it as an object in the slice $\mathbf{H}_{/\mathbf{B}\mathrm{GL}(n)}$
by way of the canonical map $\tau_X : X \to \mathbf{B}\mathrm{GL}(n)$ that modulates its 
frame bundle, the principal $\mathrm{GL}(n)$-bundle to which the tangent bundle $T X$ is associated. 
A map $(\phi,\eta) : \tau_X \to \tau_Y$ 
in $\mathbf{H}_{/\mathbf{B}\mathrm{GL}(n)}$ between two 
manifolds $X,Y$ embedded in this way is equivalently a \emph{local diffeomorphism} 
$\phi: X \to Y$ equipped with an explicit choice $\eta : \phi^* \tau_Y \simeq \tau_X$
of identification of the pullback tangent bundle with that of $X$.

The slice topos $\mathbf{H}_{/\mathbf{B}\mathrm{GL}(n)}$ allows us to express physical
fields which may not be restricted along arbitrary morphisms of manifolds
(or morphisms of whatever kind of test geometries $\mathbf{H}$ is modeled on), 
but only along
local diffeomorphism, such as \emph{metric}/\emph{vielbein} fields or symplectic structures.

For let $\mathbf{OrthStruc}_n : \mathbf{B}O(n) \to \mathbf{B}\mathrm{GL}(n)$ be the 
morphism of moduli stacks induced from the canonical inclusion of the orthogonal 
group into the general linear group, regarded as an object of the slice, 
$\mathbf{OrthStruc}_n \in \mathbf{H}_{/\mathbf{B}\mathrm{GL}(n)}$. Then 
a cocycle/field 
$$
  (o_X, e) :  \tau_X \to \mathbf{OrthStruc}_n
$$ 
is equivalently
\begin{enumerate}
  \item an \emph{orthogonal structure} $o_X$ on $X$ (a choice of \emph{Lorentz frame bundle});
  \item a \emph{vielbein} field $e : \xymatrix{\mathbf{OrthStruc}_n \circ o_X \ar[r] & \tau_X}$ 
  which equips the frame bundle with that orthogonal structure.
\end{enumerate}
Together this is equivalently a \emph{Riemannian metric} field on $X$, hence a field of Euclidean gravity,
and $\mathbf{OrthStruc}_n \in \mathbf{H}_{/\mathbf{B}\mathrm{GL}_n}$ is the universal moduli stack of 
Riemannian metrics in dimension $n$. Notice that this defines a notion of Riemannian metric
for any object in $\mathbf{H}$ as soon as it is equipped with a $\mathrm{GL}(n)$-principal bundle.
We obtain actual pseudo-Riemannian metrics by considering instead the delooped inclusion of
$O(n-1,1)$ into $\mathrm{GL}(n)$ and obtain dS-geometry, AdS-geometry etc. by further varying the
signature. 

This notion of $\mathbf{OrthStruc}_n$-structure in smooth stacks is of course
closely related to the notion of orthogonal structure as considered in traditional homotopy theory.
But there is a crucial difference, which we highlight now. First notice that there is a canonical $\infty$-functor 
$$
  {\vert -\vert} 
  \;:\; 
  \mathbf{H} \to \infty \mathrm{Grpd} \simeq L_{\mathrm{whe}} \mathrm{Top}
$$
which sends every cohesive $\infty$-groupoid/$\infty$-stack to its \emph{geometric realization}.
Under certain conditions on the cohesive $\infty$-group $G$, in particular for Lie groups as considered here,
this takes the moduli stack $\mathbf{B}G$ to the traditional \emph{classifying space} $B G$.
So under this map a choice of vielbein turns into a homotopy lift as shown on the right of
$$
  \raisebox{20pt}{
  \xymatrix{
    & \mathbf{B}O(n) \ar[d]
    \\
    X \ar[r]^{\tau_X} \ar[ur]^{o_X} & \mathbf{B} \mathrm{GL}(n)
  }
  }
  \;\;\;\;
  \xymatrix{\ar@{|->}[rr]^{{\vert-\vert}} && }
  \;\;\;\;
  \raisebox{20pt}{
  \xymatrix{
    & BO(n) \ar[d]^{\simeq}
    \\
    X \ar[r]^{\vert\tau_X\vert} \ar[ur]^{\vert o_X\vert} & B \mathrm{GL}(n)
  }
  }  
  \,.
$$
But since $O(n) \to \mathrm{GL}(n)$ is the inclusion of a maximal compact subgroup, it is
a homotopy equivalence of the underlying topological spaces. Hence under ${\vert - \vert}$
a choice of $\mathbf{OrthStruc}_n$-structure is no choice at all, up to equivalence, there is
no information encoded in this choice. This is of course the familiar statement 
that 
every vector bundle \emph{admits} an orthogonal structure. But only in the context of cohesive 
stacks is the \emph{choice} of this orthogonal structure actually equivalent to 
geometric data, to a choice of Riemannian metric.

Also notice that the 
homotopy fiber of $\mathbf{OrthStruc}_n$ is the cohesive coset $\mathrm{GL}(n)/\mathrm{O}(n)$
(the coset equipped with its smooth manifold structure) in that we have a fiber sequence
$$
  \raisebox{20pt}{
  \xymatrix{
    \mathrm{GL}(n)/O(n)
	\ar[r]
	&
	\mathbf{B}\mathrm{O}(n)
	\ar[rr]^-{\mathbf{OrthStruc}_n}
	&&
	\mathbf{B}\mathrm{GL}(n)
  }
  }
  \,.
$$
in $\mathbf{H}$,
and by the discussion in \ref{CocyclesGeneralizedParameterizedTwisted} above a metric field 
$(o_X, e) : \xymatrix{\tau_X \ar[r] &  \mathbf{OrthStruc}_n}$ is
equivalently a \emph{$\tau_X$-twisted $\mathrm{GL}(n)/O(n)$-cocycle}. This
reproduces the traditional statement that the space of choices of vielbein fields is locally
the space of maps into the coset $\mathrm{GL}(n)/O(n)$ and fails to be globally so
to the extent that the tangent bundle is non-trivial.

Moreover, by the general discussion in \ref{CocyclesGeneralizedParameterizedTwisted} we
find that a twist transformation that may act on orthogonal structures is a morphism
$\tau_Y \to \tau_X$ in the slice $\mathbf{H}_{/\mathbf{B}\mathrm{GL}(n)}$. This
is equivalently a cohesive map $\phi : Y \to X$ in $\mathbf{H}$ equipped with an 
equivalence $\eta : \xymatrix{\phi^* \tau_X \ar[r]^\simeq & \tau_X}$ from the pullback
of the tangent bundle on $X$ to that on $Y$. But such an isomorphism precisely witnesses
$\phi$ as a \emph{local diffeomorphism}. Hence it is the local diffeomorphisms that act as 
twist morphisms on tangent bundles regarded as twists for $\mathrm{GL}(n)/O(n)$-structures.
This statement of course reproduces the traditional fact that metrics pull back along
local diffeomorphisms (but not along more general cohesive maps). Abstractly it is 
reflected in the fact that the moduli stack $\mathbf{OrthStruc}_n$ for metrics in $n$ dimensions
is an object not of the base $\infty$-topos $\mathbf{H}$, but of the slice 
$\mathbf{H}_{/\mathbf{B}\mathrm{GL}(n)}$.

In conclusion, the following diagram summarizes the components of the formulation of metric fields 
as cocycles in the slice over $\mathbf{B}{\mathrm{GL}(n)}$,
displayed as a special case of the general diagram for twisted cocycles that is discussed in 
\ref{CocyclesGeneralizedParameterizedTwisted}.

\vspace{5pt}

\fbox{
$$
  \xymatrix{
    && \mbox{\begin{tabular}{c} local \\ cover \end{tabular}}
	\ar[rr] 
	\ar[d]
	&&
	\mbox{coset}
	\ar[d]
    \\
    Y \ar[rr]^{\mbox{local} \atop \mbox{diffeomorphism}}_>>{\ }="s" 
	\ar[drr]_{\mbox{tangent} \atop \mbox{bundle}}^{\ }="t"  
	&& 
	X  \ar[d]|<<<<<{\mbox{tangent} \atop \mbox{bundle}}^>>>{\ }="t2"
	\ar[rr]^-{\mbox{orthogonal} \atop \mbox{structure} }_{\ }="s2" 
	&& 
	\mbox{ \begin{tabular}{c} \mbox{delooped} \\  \mbox{orthogonal} \\ \mbox{group} \end{tabular} }
	\ar[dll]
	\\
	&& \mbox{\begin{tabular}{c} \mbox{delooped}\\\mbox{general linear} \\\mbox{group} \end{tabular}} && 
	\ar@{=>} "s"; "t"
	\ar@{=>}|{\mbox{vielbein}} "s2"; "t2"
  }
$$
}

\medskip

This discussion of metric structure and vielbein fields of gravity 
is but a special case of \emph{generalized vielbein fields} obtained from 
\emph{reduction of structure groups}. If 
$\mathbf{\mathbf{c}} : K \to G$ is any morphism of groups in $\mathbf{H}$
(typically taken to be a subgroup inclusion if one is speaking of structure group \emph{reduction},
but not necessarily so in general, as for instance the example of the \emph{generalized tangent bundle}, 
discussed in a moment, shows), 
and if 
$\tau_X : X \to \mathbf{B}G$ is the map modulating a given $G$-structure on $X$, then 
a map $(\phi, \eta) : \tau_X \to \mathbf{c}$ in $\mathbf{H}_{/\mathbf{B}G}$
is a generalized vielbein field on $X$ which exhibits the reduction of the structure group
from $G$ to $H$ along $\mathbf{c}$. These $\mathbf{c}$-\emph{geometries} are compatible with pullback
along along twist transformations $\eta : \tau_Y \to \tau_X$, namely along 
maps $\phi : Y \to X$ in $\mathbf{H}$ which are \emph{generalized local diffeomorphisms} in that they are 
equipped with an equivalence $\eta : \xymatrix{\phi^* \mathbf{c} \ar[r]^\simeq & \tau_X}$.

\medskip

Of relevance in the T-duality covariant formulation of 
type II supergravity (``doubled field theory'') is the reduction along the inclusion of the maximal
compact subgroup into the orthogonal group $O(n,n)$ (where $n = 10$ for full type II supergravity),
whose delooping in $\mathbf{H}$ we write
$$
  \mathbf{typeII} : 
  \xymatrix{
    \mathbf{B}\left(O\left(n\right) \!\times\! O\left(n\right)\right) 
	  \ar[r] 
	  &
	\mathbf{B}O\left(n,n\right)
  }
  \,.
$$ 
A spacetime $X$ that is to carry a $\mathbf{typeII}$-field accordingly must carry an $O(n,n)$-structure
in the first place in that it must be equipped with a lift of its tangent bundle
$\tau_X \in \mathbf{H}_{/\mathbf{B}\mathrm{GL}(n)}$ in the slice over $\mathbf{B}\mathrm{GL}(n)$, 
as discussed above, to
an object $\tau_X^{\mathrm{gen}}$ in the slice 
$\mathbf{H}_{/\mathbf{B}O(n,n)}$. Since there is no suitable homomorphism from $O(n,n)$ to 
$\mathrm{GL}(n)$, this lift needs to be through a subgroup of $O(n,n)$ that does map to 
$\mathrm{GL}(n)$. The maximal such group is called the \emph{geometric subgroup} 
$\xymatrix{G_{\mathrm{geom}}(n) \ar@{^{(}->}[r]^\iota & \mathrm{GL}(n)}$. We write
$$
  \raisebox{20pt}{
  \xymatrix{
    \mathbf{B}G_{\mathrm{geom}}(n)
	\ar[r]^{\mathbf{B}\iota}
	\ar[d]^{\mathbf{genTan}_n}
	&
	\mathbf{B}O(n,n)
	\\
	\mathbf{B}\mathrm{GL}(n)
  }
  }
$$
in $\mathbf{H}$.
Then for $X \in \mathrm{Mfd} \hookrightarrow \mathbf{H}$ a spacetime, a map 
$(\tau_X^{\mathrm{gen}}, \eta) :  \xymatrix{\tau_X \ar[r] & \mathbf{genTan}_n}$ 
in $\mathbf{H}_{/\mathbf{B}\mathrm{GL}(n)}$,
hence a diagram
$$
  \raisebox{20pt}{
  \xymatrix{
    X \ar[dr]_{\tau_X}^{\ }="t"
	\ar@{-->}[rr]^{\tau_X^{\mathrm{gen}}}_{\ }="s"
    && \mathbf{B}G_{\mathrm{geom}}(n) \ar[dl]^{\mathbf{genTan}_n}
    \\
    & \mathbf{B}\mathrm{GL}(n)
	\ar@{=>}^{\eta} "s"; "t"
  }
  }
$$
in $\mathbf{H}$,
is called a choice of \emph{generalized tangent bundle} for $X$. Given such,
a map 
$$
  (o_X^{\mathrm{gen}}, e^{\mathrm{gen}}) : \mathbf{B}\iota \circ \tau_X^{\mathrm{gen}} \to \mathbf{typeII}
$$
in the slice $\mathbf{H}_{/\mathbf{B}O(n,n)}$ is equivalent to what is called a
\emph{generalized vielbein field} for \emph{type II geometry} on $X$. This is a model
for the generalized fields of gravity in the T-duality-covariant formulation of type II
supergravity backgounds. 
(See for instance section 2 of \cite{GMPW} for a review and see section 4 here
for discussion in the present context.)
So $\mathbf{typeII} \in \mathbf{H}_{/\mathbf{B}O(n,n)}$ is the 
moduli stack for T-duality covariant \emph{type II gravity} fields. 
 
Similarly, if $X$ is a manifold of even dimension $2n$ equipped with a generalized tangent bundle,
then a map 
$\xymatrix{
  \tau_X^{\mathrm{gen}}
  \ar[r]
  &
  \mathbf{genComplStruc}
}$ 
in the slice with coefficients in the canonical morphism
$$
  \mathbf{genComplStruc} 
   : 
   \xymatrix{
     \mathbf{B} U(n,n) 
	   \ar[r] & 
	 \mathbf{B}O(2n,2n)
   }
$$
in a \emph{generalized complex structures} on $\tau_X$. Such $\mathbf{genComplStruc}$-fields
appear in compactifications of supergravity on \emph{generalized Calabi-Yau manifolds},
such that a global $N=1$ supersymmetry is preserved.

Notice that the homotopy fiber sequence of the local coefficient bundle $\mathbf{typeII}$ is
$$
  \xymatrix{
    O(n)\backslash O(n,n)/O(n)
    \ar[r]
	&
	\mathbf{B}O(n) \times O(n)
	\ar[rr]^-{\mathbf{typeII}}
	&&
	\mathbf{B}O(n,n)
  }
$$
in $\mathbf{H}$. The coset fiber on the left is the familiar local moduli spaces of generalized geometries known 
from the literature on T-duality and generalized geometry.

Notice also that the theory automatically determines what replaces the notion of
\emph{local diffeomorphism} in these generalized type II geometries: the generalized tangent bundles
$\tau_X^{\mathrm{gen}}$ now are the twists, and and a twist transformation 
$(\phi, \eta) : \tau^{\mathrm{gen}}_Y \to \tau^{\mathrm{gen}}_X$ in 
$\mathbf{H}_{/\mathbf{B}G_{\mathrm{geom}}(n)}$ is therefore
a cohesive map $\phi : Y \to X$ equipped with an equivalence 
$\eta : \xymatrix{\phi^* \tau^{\mathrm{gen}}_X \ar[r]^\simeq & \tau^{\mathrm{gen}}_Y}$ in $\mathbf{H}$
between the pullback of the generalized tangent bundle of $Y$ and that of $Y$.

\medskip

One can consider this setup for moduli objects being arbitrary group homomorphisms 
$\mathrm{genGeom} :  \mathbf{B}K \to \mathbf{B}G$ regarded as objects in the slice
$\mathbf{H}_{/\mathbf{B}G}$. For instance the 
delooped inclusion 
$$
  \mathbf{SuGraCompt}_n : 
  \xymatrix{
    \mathbf{B}K_n \ar[r]
	& \mathbf{B}E_{n(n)}
  }
$$ 
of the maxiomal compact subgroup of the 
the exceptional Lie groups produces the
moduli object for $U$-duality covariant fields of supergravity compactified on an $n$-dimensional
fiber. A map $\xymatrix{ \tau_{X}^{\mathrm{gen}} \ar[r] & \mathbf{SuGraCompt}_n}$
is a generalized vielbein field in 
\emph{exceptional generalized geometry} \cite{Hull}. 
Another type of exceptional geometry, that we will come back
to below in \ref{PrequantumGeometry}, is that induced by the delooping 
$$
  \mathbf{G}_2\mathbf{Struc}
  :
  \xymatrix{
    \mathbf{B}G_2 
	\ar[r]
	&
	\mathbf{B}\mathrm{GL}(7)
  }
$$
of the defining inclusion of the exceptional Lie group $G_2$ as the subgroup of those
linear transformations of $\mathbb{R}^7$ which preserves the ``associative 3-form''
$\langle -, (-)\times (-)\rangle$. 
For $X$ a manifold of dimension 7,
a field $\phi : \tau_X \to \mathbf{G}_2 \mathbf{Struc}$ is a 
\emph{$G_2$-structure} on $X$.

\medskip

So far all the groups in the examples have been  ordinary cohesive (Lie) groups, 
hence \emph{0-truncated} cohesive $\infty$-group objects in $\mathbf{H}$. More
generally we have ``reduction'' of structure groups for general $\infty$-groups
exhibited by ``higher vielbein fields'' which are maps into moduli objects in a 
slice $\infty$-topos.

One degree higher, the first example comes from central extensions
$$
  \xymatrix{A \ar[r] &  \hat G \ar[r] &  G}
$$ of ordinary groups. 
These induce long fiber sequences
$$
  \xymatrix{
    A \ar[r] & \hat G \ar[r] & G \ar[r]^-{\Omega \mathbf{c}} & \mathbf{B}A \ar[r] & \mathbf{B}\hat G 
	\ar[r] & \mathbf{B}G \ar[r]^{\mathbf{c}} & \mathbf{B}^2 A
  }
$$
in $\mathbf{H}$. Here $\mathbf{c}$ is the (cohesive) group 2-cocycle that classifies the extension,
exhibited as a $\mathbf{B}A$-2-bundle $\mathbf{B}\hat G \to \mathbf{B}G$.
Generally an object $(X, \phi_X) \in \mathbf{H}_{/\mathbf{B}}$ is an object $X \in \mathbf{H}$
equipped with a $\mathbf{B}A$-2-bundle (an $A$-bundle gerbe) modulated by a map
$\phi_X : X \to \mathbf{B}^2 A$. A field $(\sigma, \eta) : \phi_X \to \mathbf{c}$ in 
$\mathbf{H}_{/\mathbf{B}^2 A}$ is a choice $\sigma$ of a $G$-principal bundle
on $X$ together with an equivalence 
$\eta : \xymatrix{ \sigma^* \mathbf{c} \ar[r]^\simeq & \phi_X }$.

\medskip

Of particular relevance for physics is of course the example of this which is
given by the $\mathrm{Spin}$-extension of the special orthogonal grouop
$$
  \xymatrix{
   \mathbf{B}\mathbb{Z}_2 
    \ar[r]
 	& 
	\mathbf{B}\mathrm{Spin}
	\ar[rr]^{\mathbf{SpinStruc}}
	&&
	\mathbf{B}\mathrm{SO}
	\ar[rr]^{\mathbf{w}_2}
	&&
	\mathbf{B}^2 \mathbb{Z}_2
  }
  \,,
$$
which is classified by the universal second Stiefel-Whitney class
$\mathbf{w}_2$.
(From now on we notationally suppress, for convenience, the dimension $n$ when displaying these groups.)
For $o_X : X \to \mathbf{B}\mathrm{SO}$ an orientation structure on a 
manifold $X$, a map 
$$
  \xymatrix{o_X \ar[r] & \mathbf{SpinStruc} }
$$ 
in  $\mathbf{H}_{/\mathbf{B}\mathrm{SO}}$ is equivalently a choice of 
$\mathrm{Spin}$-structure on $o_X$. Alternatively, if 
$\phi : \xymatrix{X \ar[r] & \mathbf{B}^2 \mathbb{Z}_2}$ is the map modulating a 
given $\mathbb{Z}_2$-2-bundle ($\mathbb{Z}_2$-bundle gerbe) over $X$, then 
a map $\xymatrix{ \phi_X \ar[r] & \mathbf{w}_2 }$ covering $o_X$ is
a \emph{$\phi$-twisted spin structure} on $o_X$. 
An important special case of this is where $\phi = \mathbf{c_1}(E) \,\mathrm{mod}\, 2$
is the mod-2 reduction of the Chern class of a given $U(1)$-principal bundle/complex line bundle
on $X$: a $\mathbf{c}_1(E)$-twisted spin structure is equivalently a $\mathrm{Spin}^c$-structure
on $X$ whose underlying $U(1)$-principal bundle is $E$. More generally, $E$ itself is 
taken to be part of the field content and so we consider the universal Chern-class
$$
  \mathbf{c_1} 
    : 
  \xymatrix{
    \mathbf{B}U(1)
	\ar[r]
	&
	\mathbf{B}^2 \mathbb{Z}
  }
$$
of the universal $U(1)$-principal bundle. There is a diagram
$$
  \xymatrix{
    \mathbf{B}\mathrm{Spin}^c \ar[r] 
	\ar[d]_{\mathbf{Spin}^c\mathbf{Struc}}
	& \mathbf{B} U(1)
	\ar[d]^{\mathbf{c}_1 \mathrm{mod} 2}
	\\
	\mathbf{B} \mathrm{SO}
	\ar[r]^{\mathbf{w}_2}
	&
	\mathbf{B}^2 \mathbb{Z}_2
  }
$$
in $\mathbf{H}$ which exhibits the moduli stack of $\mathrm{Spin}^c$-principal bundles as 
the homotopy fiber product of $\mathbf{c}_1$ with $\mathbf{w}_2$. With this, maps
$$
  \xymatrix{
    o_X \ar[r] & \mathbf{Spin}^c\mathbf{Struc}
  }
$$
in $\mathbf{H}_{\mathbf{B}\mathrm{SO}}$
are equivalently $\mathrm{Spin}^c$-structures on $X$ (for arbitrary underlying $U(1)$-principal bundle).
Notice that the formalism of twist transformations again tells us what the right 
kind of transformations is along which $\mathrm{Spin}$-structures and $\mathrm{Spin}^c$-structures
may be pulled back: these are maps $\xymatrix{ o_Y \ar[r] & o_X}$ in $\mathbf{H}_{/\mathbf{B}\mathrm{SO}}$
and hence \emph{orientation-preserving} local diffeomorphisms.

\medskip

All of this is just a low-degree step in a whole tower of \emph{higher $\mathrm{Spin}$-structures}
and \emph{higher $\mathrm{Spin}^c$-structure} that appear as fields in 
the effective higher supergravity theories underlying superstring theory.
This tower is the  \emph{Whitehead tower} of $\mathbf{B}O$. Its smooth lift
through ${\vert-\vert}$ to a tower of higher moduli stacks
has been constructed in \cite{FSS}
(an interpreted in physics as discussed now in \cite{SSSIII}, 
reviewed in the broader context of cohesive $\infty$-toposes in section 4 here):
$$
  \xymatrix{
    \ar@{..}[d]
    \\
    \mathbf{B}\mathrm{Fivebrane} 
    \ar[d]_{\mathbf{FivebraneStruc}}
	&& 
     \\
    \mathbf{B}\mathrm{String} \ar[rr]^{\tfrac{1}{6}\mathbf{p}_2} 
	\ar[d]_{\mathbf{StringStruc}} 
	&& \mathbf{B}^7 U(1)
    \\
    \mathbf{B} \mathrm{Spin} \ar[rr]^{\tfrac{1}{2}\mathbf{p}_1} 
	\ar[d]_{\mathbf{SpinStruc}} && \mathbf{B}^3 U(1)
    \\
    \mathbf{B}\mathrm{SO} \ar[d]_{\mathbf{OrientStruc}}  \ar[rr]^{\mathbf{w}_2} && \mathbf{B}^2 \mathbb{Z}_2 
    \\
    \mathbf{B}\mathrm{O} \ar[rr]^{\mathbf{w}_1} 
	\ar[d]_{\mathbf{OrthStruc}}
	&& \mathbf{B}\mathbb{Z}_2
	\\
    \mathbf{B}\mathrm{GL}
  }
$$
All of these structures can be further twisted. For instance we have the higher analog of
$\mathrm{Spin}^c$ given by the delooping 2-group of the homotopy fiver product
$$
  \raisebox{20pt}{
  \xymatrix{
    \mathbf{B}\mathrm{String}^{c_2}
	\ar[r]
	\ar[d]_{\mathbf{String}^{c_2}\mathbf{Struc}}
	&
	\mathbf{B} (\mathrm{E}_8 \times E_8)
	\ar[d]^{\mathbf{c}_2}
	\\
	\mathbf{B}\mathrm{Spin}
	\ar[r]^{\tfrac{1}{2}\mathbf{p}_1}
	&
	\mathbf{B}^3 U(1)
  }
  }
$$
of $\tfrac{1}{2}\mathbf{p}_1$ with the smooth universal 
second Chern class $\mathbf{c}_2 : \xymatrix{\mathbf{B}(E_8 \times E_8) \ar[r] & \mathbf{B}^3 U(1)}$.
On manifolds $X$ equipped with a $\mathrm{Spin}$-structure $s_X : X \to \mathbf{B}\mathrm{Spin}$, 
a field
$$
  \xymatrix{
    s_X
	\ar[r]
	&
    \mathbf{String}^{c_2}\mathbf{Struc}
  }
$$
in $\mathbf{H}_{/\mathbf{B}\mathrm{Spin}}$ is a choice of $\mathrm{String}^{c_2}$-structure,
equivalently a choice of $(E_8 \times E_8)$-principal bundle and an equivalence between its
Chern-Simons circle 3-bundle and the Chern-Simons circle 3-bundle of the $\mathrm{Spin}$-structure.
This is the quantum-anomaly-free instanton sector of a gauge field in the effective 
heterotic supergravity underlying the heterotic string \cite{SSSIII}. 
Below in \ref{GaugeFieldHigherTwistedNonAbelian} we discuss how the differential refinement of
$\mathrm{String}^{c_2}$-structures capture the dynamical field of gravity and the gauge field
in heterotic supergravity.

In summary, the specialization of the diagram of \ref{CocyclesGeneralizedParameterizedTwisted}
to the anomaly-free instanton-sector of heterotic supergravity looks as follows.

\fbox{
$$
  \xymatrix{
    && \mbox{\begin{tabular}{c} local \\ cover \end{tabular}}
	\ar[rr] 
	\ar[d]
	&&
	\mbox{
	\begin{tabular}{c}
	  delooped
	  \\
	  String
	  \\
	  2-group
	\end{tabular}
	}
	\ar[d]
    \\
    Y \ar[rr]^{\mbox{local} \atop \mbox{diffeomorphism}}_>>{\ }="s" 
	\ar[drr]_{\mbox{$c_2($gauge field$)$}}^{\ }="t"  
	&& 
	X  \ar[d]|<<<<<{}^>>>{\ }="t2"
	\ar[rr]^-{\mbox{spin} \atop \mbox{structure} }_{\ }="s2" 
	&& 
	\mbox{ \begin{tabular}{c} \mbox{delooped} \\  \mbox{spin} \\ \mbox{group} \end{tabular} }
	\ar[dll]
	\\
	&& \mbox{\begin{tabular}{c} \mbox{delooped}\\\mbox{general linear-} \\ \mbox{times circle-3-} \\ \mbox{group} \end{tabular}} && 
	\ar@{=>} "s"; "t"
	\ar@{=>}|{\mbox{vielbein}\\ \atop \mbox{$\mathrm{String}^{c}$ struc.}} "s2"; "t2"
  }
$$
}

\medskip

There are further variants of all these examples and other further cases of 
gravity-like fields
in physics given by maps in slice toposes. But for the present discussion we leave it at this
and now turn to the other fundamental kind of fields in physics besides gravity: gauge fields.

\paragraph{Gauge fields: higher, twisted, non-abelian}
\label{GaugeFieldHigherTwistedNonAbelian}

The other major kind of (quantum) fields besides the (generalized) fields of gravity that 
we discussed above are of course \emph{gauge fields}. A seminal result of Dirac's old
argument about electric/magnetic \emph{charge quantization} is that a configuration of the plain
\emph{electromagnetic field} is mathematically a \emph{connection} on a $U(1)$-principal bundle. 
Similarly the Yang-Mills field of quantum chromodynamics is mathematically a connection on a
$G$-principal bundle, where $G$ is the corresponding gauge group. The connection itself is 
locally the \emph{gauge potential} traditonally denoted $A$, while the class of the underlying
global bundle is the \emph{magnetic background charge} for the case of electromagnetism
and is the \emph{instanton sector} for the case of $G = \mathrm{SU}(n)$.

Analogously, it has long been known that the background $B$-field to which the string couples
is mathematically a connection on a $U(1)$-principal \emph{2-bundle} 
(often presented as $U(1)$-bundle gerbe), hence a bundle that is principal under the
higher group (2-group) $\mathbf{B}U(1)$. Together with the case of ordinary $U(1)$-principal 
bundles these are the first two (or three) degrees of what are known as cocycles in 
\emph{ordinary differential cohomology}, a refinement of cocyles modulated in 
the coefficient stack $\mathbf{B}^n U(1)$
by \emph{curvature twists} controled by smooth differential form data. 
A general formalization of this based on the underlying topological classifying spaces
$K(\mathbb{Z}, n+1) \simeq {\vert \mathbf{B}^n U(1)\vert }$, 
or in fact any infinite loop space $\vert \mathbf{B}\mathbb{G}\vert$
representing a generalized cohomology theory, has
been given in \cite{HopkinsSinger}.  Here we refine this construction to the cohesive
higher topos case and obtain higher cohesive moduli stacks 
$\mathbf{B}\mathbb{G}_{\mathrm{conn}}$ such that maps 
$X \to \mathbf{B}\mathbb{G}_{\mathrm{conn}}$ with coefficients in these are 
differential $\mathbb{G}$-cocycles and hence equivalently (higher) \emph{gauge fields}
on $X$ for the (higher, cohesive) gauge group $\mathbb{G}$. 

\medskip

An $\infty$-group $\mathbb{G} \in \mathrm{Grp}(\mathbf{H})$ is \emph{abelian}
or $E_\infty$ if it is equipped with an $n$-fold delooping 
$\mathbf{B}^n \mathbb{G} \in \mathbf{H}$
for all $n \in \mathbb{N}$. If it is equipped at least with a second delooping
$\mathbf{B}^2 \mathbb{G}$, then we say it is a \emph{braided $\infty$-group}.
Equivalently this means that the single delooping object $\mathbf{B}\mathbb{G}$
is itself equipped with the structure of an $\infty$-group.
For example the full subcategory of any braided monoidal $\infty$-category on the 
objects that are invertible under the tensor product is a braided $\infty$-group,
hence the name.

For a braided $\infty$-group $\mathbb{G}$ in a cohesive $\infty$-topos, 
the axioms of cohesion induce a canonical map
$$
  \mathrm{curv}_{\mathbb{G}}
  :
  \xymatrix{
    \mathbf{B}\mathbb{G}
	\ar[r]
	&
	\flat_{\mathrm{dR}}\mathbf{B}^2 \mathbb{G}
  }
$$
to the \emph{de Rham coefficient objects} of the group $\mathbf{B}\mathbb{G}$.
On the one hand this may be interpreted as the \emph{Maurer-Cartan form} on th
cohesive group $\mathbf{B}\mathbb{G}$. Equivalently, one finds that this is
the \emph{universal curvature characteristic} of $\mathbb{G}$-principal $\infty$-bundles:
the map can be seen to proceed by equipping a $\mathbb{G}$-principal $\infty$-bundle
with a \emph{pseudo-connection} and then sending that to the coresponding
curvature in the de Rham hypercohomology with coefficients in the $\infty$-Lie algebra
of $\mathbb{G}$. 

In order to pick among those (higher) pseudo-connections with curvature in hypercohomology
those that are genuine (higher) connections characterized by having globally well defined
curvature differential form data, let $\Omega_{\mathrm{cl}}(-,\mathbb{G}) \in \mathbf{H}$ be a
0-truncated object equipped with a map 
$\xymatrix{\Omega_{\mathrm{cl}}(-,\mathbb{G}) \ar[r] & \flat_{\mathrm{dR}}\mathbf{B}^2 \mathbb{G} }$ 
which has the following property:  for every manifold $\Sigma$ the induced map
$$
  \xymatrix{
    [\Sigma, \Omega_{\mathrm{cl}}(-,\mathbb{G})]
    \ar[r]
	&
	[\Sigma, \flat_{\mathrm{dR}} \mathbf{B}^2 \mathbb{G}]
  }
$$
is 1-epimorphism (an effective epimorphism, hence an epimorphism in the sheaf topos under 0-truncation).
This expresses the fact that $\Omega_{\mathrm{cl}}(-, \mathbb{G})$ is a sheaf of 
flat $\mathrm{Lie}(\mathbb{G})$-valued differential forms, in that every de Rham cohomology
class over a manifold is represented by such a form.

(More generally one considers  a suitable  filtration 
$
\xymatrix{\Omega_{\mathrm{cl}}^{\bullet}(-,\mathbb{G}) \ar[r] & \flat_{\mathrm{dR}}\mathbf{B}^2 \mathbb{G} }
$, hence a kind of \emph{universal mixed Hodge structure on $\mathbb{G}$-cohomology}).

Then the moduli object $\mathbf{B}\mathbb{G}_{\mathrm{conn}}$
for \emph{differential $\mathbb{G}$-cocycles} is the homotopy pullback in 
$$
  \xymatrix{
    \mathbf{B}\mathbb{G}_{\mathrm{conn}}
	\ar[rr]
	\ar[d]
	&&
	\Omega^n_{\mathrm{cl}}(-)
	\ar[d]
	\\
	\mathbf{B}\mathbb{G}
	\ar[rr]^{\mathrm{curv}_{\mathbb{G}}}
	&&
	\flat_{\mathrm{dR}}\mathbf{B}^2 \mathbb{G}
  }
  \,.
$$

For example if $\mathbb{G} \simeq \mathbf{B}^{n-1}U(1)$ in smooth $\infty$-groupoids, then 
the object $\mathbf{B}^n U(1)_{\mathrm{conn}}$ defined this way is the $n$-stack which is 
presenented under the Dold-Kan correspondence by the \emph{Deligne-complex} of sheaves.
It modulates ordinary differential cohomology. 

A configuration of the electromagnetic field on a space $X$ is a map $X \to \mathbf{B}U(1)_{\mathrm{conn}}$.
A configuration of the $B$-field background gauge field
of the bosonic string is a map $X \to \mathbf{B}^2 U(1)_{\mathrm{conn}}$.
(For the superstring the situation is a bit more refined, discussed below.)
A configuration of the $C$-field background gauge field of $M$-theory involves
(among other data) a map $X \to \mathbf{B}^3 U(1)_{\mathrm{conn}}$.




{\bf Differential T-duality and $B_n$-geometry}

Above we have seen that the \emph{extended} Lagrangian 
$\mathbf{L} : \mathbf{B}G_{\mathrm{conn}} \to \mathbf{B}^3 U(1)_{\mathrm{conn}}$
for $G= \mathrm{Spin}, \mathrm{SU}$-Chern-Simons 3d gauge field theory also serves as 
the twist that defines the moduli stack $\mathbf{B}\mathrm{String}^{c_2}_{\mathrm{conn}}$
of Green-Scharz anomaly-free heterotic background gauge field configurations.
In view of this it is natural to ask: does the extended Lagrangian of 
$U(1)$-Chern-Simons theory similarly play a role as part of the background
gauge field structure for superstrings? Indeed this turns out to be the case: 
the extended $U(1)$-Chern-Simons Lagrangian encodes the twist that defines
\emph{differential T-duality structures} and \emph{$B_n$-geometry}.

To see this, we observe by direct inspection that what in \cite{KahleValentino} is  
called a \emph{differential $T$-duality structure} on a pair 
of circle-bundles $S^1 \to X_1, X_2 \to Y$ over some base $Y$
and equipped with connections 
$\nabla_1$ and $\nabla_2$, is a trivialization of the corresponding 
cup-product circle 3-bundle, hence of the 
extended Chern-Simons Lagrangian of two-species $U(1)$-Chern-Simons theory
pulled back along the map that modulates the two circle bundles.

We now say this again in more detail. 
Let $T^1$ be a circle and $\tilde T^1 := \mathrm{Hom}(T^1, U(1))$
the dual circle, with the canonical pairing denoted  
$\langle -,-\rangle : T^1 \times \tilde T^1 \to U(1)$. Then 
the first spacetime $X_1 \to Y$ is modulated by a map 
$\mathbf{c}_1 : \xymatrix{Y \ar[r] &  \mathbf{B}T^1_{\mathrm{conn}} }$, 
and its T-dual $\tilde c_1 : X_2 \to Y$ by a map 
$\tilde {\mathbf{c}_1} : Y \to \mathbf{B}\tilde T^1_{\mathrm{conn}}$. 

Now the pairing and  the cup product together
form a universal characteristic map of moduli stacks
$$
  \langle -\cup - \rangle
  :
  \xymatrix{
    \mathbf{B}(T^1\times \tilde T^1)
	\ar[r]
	&
	\mathbf{B}^3 U(1)
  }
  \,.
$$
By the above discussion, this has a differential refinement
$$
  \langle -\cup - \rangle
  :
  \xymatrix{
    \mathbf{B}(T^1\times \tilde T^1)_{\mathrm{conn}}
	\ar[r]
	&
	\mathbf{B}^3 U(1)_{\mathrm{conn}}
  }
$$
which is the extended Lagrangian of $U(1)$-Chern-Simons theory in 3d. 
If instead we regard the same map as a 3-cocycle, it modulates a
higher group extension $\mathrm{String}(T^1\times \tilde T^1) \to T^1 \times \tilde T^1$,
sitting in a long fiber sequence of higher moduli stacks of the form

\hspace{-2cm}
$$
  \xymatrix{
    \cdots
	\ar[r]
	&
	\mathbf{B}U(1)
	\ar[r]
	&
	\mathrm{String}(T^1 \times \tilde T^1)
	\ar[r]
	&
	(T^1 \times \tilde T^1)
    \ar[r]
    &
    \mathbf{B}^2 U(1)
	\ar[r]
    &
    \mathbf{B}\mathrm{String}(T^1 \times \tilde T^1)
	\ar[r]
	&
	\mathbf{B}(T^1 \times \tilde T^1 )
	\ar[r]^{}
	&
	\mathbf{B}^3 U(1)
  }
  \,.
$$

One sees from this that a
\emph{differential T-duality structure} on $(X_1, X_2)$ 
as considered in def. 2.1 of \cite{KahleValentino}
is equivalently -- when refined to the context of smooth higher geometry -- 
a lift of $(\mathbf{c}_1, \tilde {\mathbf{c}}_1)$
through the left vertical projection in the homotopy pullback square
$$
 \raisebox{20pt}{
  \xymatrix{
    \mathbf{B}\mathrm{String}(T^1 \times \tilde T^1)_{\mathrm{conn}}
	\ar[d]
	\ar[rr]
	&&
	\Omega_{\mathrm{cl}}^{4 \leq \bullet \leq 3}
	\ar[d]
	\\
	\mathbf{B}(T^1 \times \tilde T^1)_{\mathrm{conn}}
	\ar[rr]^{\langle -\cup -\rangle}
	&&
	\mathbf{B}^3 U(1)_{\mathrm{conn}}
  }}
  \,,
$$
hence is a map in the slice over $\mathbf{B}^3 U(1)_{\mathrm{conn}}$,
hence is a \emph{differential $\mathrm{String}(T^1 \times \tilde T^1)$-structure}
on the given data. Along the lines of the discussion in \cite{FSS}
one finds, as for the twisted differential String-structures discussed above, 
that such a lift locally corresponds to a choice of 3-form $H$ satisfying
$$
  d_{\mathrm{dR}} H = \langle F_{A_1} \wedge F_{A_2} \rangle
  \,,
$$
where $A_1, A_2$ are the local connection forms of the two circle bundles.
This is the local structure that has been referred to as 
\emph{$B_n$-geometry}, see the corresponding discussion and references given in 
\cite{FiorenzaSatiSchreiberIV}.

Observe that
by the universal property of homotopy fibers, the 
underlying trivialization of the cup product circle 3-bundle corresponds to a
choice of factorization of $(\mathbf{c}_1, \tilde {\mathbf{c}}_1)$
as shown on the bottom of the following diagram
$$
  \raisebox{20pt}{
  \xymatrix{
    X_1 \times_{Y} X_2
	\ar[r]^{\kappa}
	\ar[d]
	&
	\mathbf{B}^2 U(1)
	\ar[r]
	\ar[d]
	&
	{*}
	\ar[d]
    \\
    Y
	\ar[r]
	&
	\mathbf{B}\mathrm{String}(T^1 \times \tilde T^1)
	\ar[r]
	&
	\mathbf{B}(T^1 \times \tilde T^1)
  }
  }
  \,.
$$
Forming the consecutive homotopy pullback of the point inclusion
as given by these two squares, the 
map $X_1 \times_Y X_2 \to \mathbf{B}^2 U(1)$ induced by the universal property
of the homotopy pullback
modulates a circle 2-bundle ($U(1)$-bundle gerbe) on the correspondence space.
This is the bundle gerbe on the correspondence space considered in 2.2, 2.3 of
\cite{KahleValentino}.
Notice that this is just a special case of the general phenomenon of 
twisted higher bundles, as laid out in \cite{NSSa}.

\paragraph{Gauge invariance, equivariance and general covariance}
\label{GaugeInvarianceEquivarianceAndGeneralCovariance}

The notion of \emph{gauge transformation} and \emph{gauge invariance} is
built right into higher geometry. Any object $X \in \mathbf{H}$ in general
contains not just (local) points, but also gauge equivalences between these,
gauge-of-gauge equivalences between those, and so on. A map 
$\exp(i S(-)) : \mathbf{Fields} \to U(1)$ is automatically a \emph{gauge invariant function}
with respect to whatever gauge transformations the species of fields encoded by the
moduli object $\mathbf{Fields}$ encodes.

Specifically, if an $\infty$-group $G$ acts on some $Y$, then a 
\emph{$G$-equivariance} structure on a map $Y \to A$ is an extension 
$$
  \raisebox{20pt}{
  \xymatrix{
    Y \ar[rr] \ar[d] && A
	\\
	Y/\!/G \ar@{-->}[urr]
  }
  }
$$
along the canonical quotient projection.

If $A$ here is a 0-truncated object such that $U(1)$, then the existence of such an extension
is just a property. But if $A$ has itself gauge equivalences, say if $A = \mathbf{B}^n U(1)_{\mathrm{conn}}$
for positive $n$-then a choice of such an extension is genuine extra structure. For $n = 1$ 
this is the familiar structure on an \emph{equivariant bundle}. For higher $n$ it is a
suitable higher order generalization of this notion. 

Equivariance is preserved by transgression. 
If $\mathbf{L} : \mathbf{Fields} \to \mathbf{B}^n U(1)_{\mathrm{conn}}$ is an extended Lagrangian,
hence equivalently a equivariant $n$-connection on the space of fields, then for $\Sigma_k$
any object the mapping space $[\Sigma_k, \mathbf{Fields}]$ contains the gauge equivalences of
the given field species on $\Sigma$ and accordingly the transgressed Lagrangian
$$
  \exp(2\pi i \int_{\Sigma_k}[\Sigma_k, \mathbf{L}]) 
  \; : \; [\Sigma_k, \mathbf{Fields}] 
   \to 
   \mathbf{B}^{n-k}U(1)_{\mathrm{conn}}
$$
is gauge invariant (precisely: carries gauge-equivariant structure).

A particular kind of gauge equivalence/equivariance is the \emph{diffeomorphism equivariance}
of a \emph{generally covariant field theory}. In such a field theory two fields 
$\phi_1, \phi_2 : \Sigma \to \mathbf{Fields}$ are to be regarded as gauge equivalent if there is a 
diffeomorphism, hence an automorphism $\alpha : \xymatrix{\Sigma \ar[r]^\simeq & \Sigma}$ in 
$\mathbf{H}$, such that $\alpha^* \phi_2 \simeq \phi_1$. 

Formally this means that for generally covariant field theries the field space $[\Sigma, \mathbf{Fields}]$
over a given worldvolume $\Sigma$ is to be formed in the slice 
$\mathbf{H}_{/\mathbf{B}\mathbf{Aut}}(\Sigma) \simeq \mathbf{Aut}(\Sigma)\mathrm{Act}$, 
with $\Sigma$ understood as equipped with the defining $\mathbf{Aut}(\Sigma)$-action and with
$\mathbf{Fields}$ equipped with the trivial $\mathbf{Aut}(\Sigma)$-action, we write
$$
  [\Sigma, \mathbf{Fields}]_{/\mathbf{B}\mathbf{Aut}(\Sigma)}
  \in
  \mathbf{H}_{/\mathbf{B}\mathbf{Aut}(\Sigma)}
$$
for emphasis. To see this one observes that generally for $(V_1, \rho_1), (V_2, \rho_2) \in \mathbf{G}\mathbf{Act}$
two objects equipped with $G$-action, their mapping space $[V_1,V_2]_{/\mathbf{B}G}$ formed in the slice
is the absolute mapping space $[V_1, V_2]$ formed in $\mathbf{H}$ and equipped with the 
\emph{conjugation action} of $G$, under which an element $g \in G$ acts on an element 
$f : V_1 \to V_2$ by sending it to $\rho_2(g)^{-1} \circ f \circ \rho_1(g)$.

Hence the mapping space $[\Sigma, \mathbf{Fields}]_{/\mathbf{B}\mathbf{Aut}(\Sigma)}$
formed in the slice corresponds in $\mathbf{H}$ to the fiber sequence
$$
  \xymatrix{
    \Sigma \ar[r] & \mathbf{Aut}(\Sigma) \backslash\!\backslash[\Sigma, \mathbf{Fields}]
	\ar[d]
     \\
	 & 
	 \mathbf{B}\mathbf{Aut}(\Sigma)
	}
$$
and a \emph{generally covariant field theory} for the given species of fields
is one whose configuration spaces are 
$\mathbf{Aut}(\Sigma) \backslash\!\backslash[\Sigma, \mathbf{Fields}]$, 
the action groupoids of the $\infty$-groupoid of field configurations on $\Sigma$
by the diffeomorphism action on $\Sigma$.

Ordinary 3d Chern-Simons theory is strictly speaking to be regared as a generally covarnat 
field theory, but this is often not made explicit, due to a special property of 3d Chern-Simons theory:
if two \emph{on-shell} field configurations are related by a diffeomorphism (connected to the identity), then they
are already gauge equivalent also by a gauge transformation in $[\Sigma, \mathbf{B}G_{\mathrm{conn}}]$.
This holds in fact also for all higher Chern-Simons theories that come
from \emph{binary} invariant polynomials, but it does not hold fully generally. Even when this is
the case, supporessing the general covariance is a dubious move, since while
the gauge equivalence classes may coincide, 
$\tau_{0} [\Sigma, \mathbf{Fields}]_{\mathrm{onshell}} \simeq \tau_{0} 
\mathbf{Aut}(\Sigma) \backslash\!\backslash[\Sigma, \mathbf{Fields}]_{\mathrm{onshell}}$, 
the two full homotopy types still need not be equivalent and hence the corresponding
quantum field theories  may not be equivalent.

\subsubsection{Variational calculus on higher moduli stacks of fields}
\label{VariationalCalculusCriticalLoci}
\label{PhaseSpaces}
\index{structures in a cohesive $\infty$-topos!phase space!in introduction}

Traditionally, the \emph{phase space} of a physical system which is given by an 
action functional $\exp(i S) : \xymatrix{ \mathbf{Fields}(\Sigma) \ar[r] & U(1) }$
is the \emph{variational critical locus} of $\exp(i S)$: the subspace of 
field configurations $\mathbf{Fields}(\Sigma)$ on some manifold $\Sigma$ with boundary,
such that the variation $\mathbf{d} S$ of the action (with fields on $\partial \Sigma$ held fixed)
vanishes when restricted to this subspace. One also calls this the space of solutions of the
\emph{Euler-Lagrange equations} of the system. Often one considers the special case
where $\Sigma = \Sigma_{\mathrm{in}} \times [0,1]$ is the cylinder over a 
closed manifold $\Sigma_{\mathrm{in}}$ and under suitable conditions on $S$, 
solutions to the Euler-Lagrange equations are fixed by their value and first derivative
on $\Sigma_{\mathrm{in}}$, in which case the phase space may be identified with the cotangent
bundle $T^* \mathbf{Field}(\Sigma_{\mathrm{in}})$. This simple special case is
sometimes regarded as the definition of the notion of phase space, and in order to 
distinguish the general notion from this special case one calls the space of 
solutions of the Euler-Lagrange equations also the \emph{covariant phase space}. 
For $S$ a local action functional this space is canonically equipped with a pre-symplectic form.
Quotienting out the (gauge) symmetries makes this a genuine symplectic form on what is called the
\emph{reduced phase space}. But as with all quotients, this quotient makes good sense in general
only when formed in a suitable homotopy-theoretic sense, hence in higher geometry. 
The physics literature knows a formalism for dealing with this as the \emph{BV-BRST formalism}.

In the following we discuss these issues in differential cohesive higher geometry,
for the prequantum theory of $n$-dimensional field theories defined by extended 
Lagrangians $\mathbf{L} : \xymatrix{\mathbf{Fields} \ar[r] & \mathbf{B}^n U(1)_{\mathrm{conn}}}$,
which induce action functionals as above after transgression to the mapping space out of
some $\Sigma$.

Let $\mathbb{G} \in \mathrm{Grp}(\mathbf{H})$ be a cohesive $\infty$-group. 
By the discussion in \ref{GaugeFieldHigherTwistedNonAbelian} above there is
a canonical de Rham cocycle on $\mathbb{G}$, the \emph{$\infty$-Maurer-Cartan form} 
$$
  \theta : \xymatrix{\mathbb{G} \ar[r] & \flat_{\mathrm{dR}}\mathbf{B}\mathbb{G} }
  \,.
$$
A little reflection shows that in the context of higher stacks, this form is also the 
\emph{universal differential} for $\mathbb{G}$-valued functions, in that for
$$
  S : \xymatrix{X \ar[r] & \mathbb{G}}
$$
any map, the composite
$$
  S^{-1}\mathbf{d}S 
   : 
  \xymatrix{X \ar[r]^{S} & \mathbb{G} \ar[r]^-{\theta} & \flat_{\mathrm{dR}} \mathbf{B}\mathbb{G}}
$$
which corresponds to a $\mathrm{Lie}(\mathbb{G})$-valued differential cocycle on $X$, is
the normalized differential of $S$. 

If here $X = \mathbf{Fields}(\Sigma)$ is a space of fields on some manifold $\Sigma$ with bounday
$\partial \Sigma \hookrightarrow \Sigma$, then the \emph{variational differential}
is the restriction of this differential to variations which keep the field configurations
over the boundary $\partial \Sigma$ fixed.
This restriction is given by precomposition with the top horizontal morphism in the 
following homotopy pullback diagram
$$
  \xymatrix{
    \mathbf{Fields}(\Sigma)_{\partial \Sigma}
	\ar[rr]^{\iota}
	\ar[d]
	&&
    \mathbf{Fields}(\Sigma)	\ar[d]
	\\
    \flat \mathbf{Fields}(\partial \Sigma)
	\ar[rr]
	&&
	\mathbf{Fields}(\partial \Sigma)
  }
  \,.
$$
Here $\flat : \mathbf{H} \to \mathbf{H}$ is the \emph{flat modality} given by the cohesion of
$\mathbf{H}$. In summary, the \emph{variational differential} of a map
$S : \xymatrix{\mathbf{Fields} \ar[r] & \mathbb{G} }$ is the composite
$$
  S^{-1} \mathbf{d}_{\mathrm{var}} S
  : 
  \xymatrix{
    \mathbf{Fields}(\Sigma)_{\partial \Sigma}
	\ar[r]^-{\iota}
	&
	\mathbf{Fields}(\Sigma)
	\ar[r]^-{S}
	&
	\mathbb{G}
	\ar[r]^-{\theta}
	&
	\flat_{\mathrm{dR}}\mathbf{B}\mathbb{G}
  }
  \,.
$$

Now the \emph{phase space} or \emph{variational critical locus} or
\emph{solution space of the Euler-Lagrange equations} of $S$ is supposed to be the 
subobject of $\mathbf{Fields}(\Sigma)_{\partial \Sigma}$ ``on which this differential vanishes''.
But one needs to be careful with how to interpret this. For instance the differential vanishes
when the whole expression above is restricted to any point $* \to \mathbf{Fields}(\Sigma)_{\partial \Sigma}$,
simply because any de Rham data on the point is trivial: there is only an essentially unique map
$\xymatrix{{*} \ar[r] & \flat_{\mathrm{dR}}\mathbf{B}\mathbb{G}}$: the 0-form. 
Therefore, what should
really be meant by a point where the differential vanishes is a point such that the differential vanishes
\emph{on every infinitesimal neighbourhood} of it. 

In other words, when testing whether $S^{-1}\mathbf{d}_{\mathrm{var}} S$ vanishes when restricted
to a subspace 
$\phi : \xymatrix{U \ar@{^{(}->}[r] & \mathbf{Fields}(\Sigma)_{\partial \Sigma}}$, we need to ensure 
that $U$ is \emph{infinitesimally spread out} or \emph{infinitesimally open}
in $\mathbf{Fields}(\Sigma)$. Such a \emph{spread-out map} $\phi$ is
commonly known by the French term as an \emph{{\'e}tale map}; and an \emph{infinitesimally spread out}
map is known as a \emph{formally {\'e}tale map} (with ``formal'' as in ``formal power series'' rings,
which are the rings of functions on the infinitesimal neighbourhood of the origin in a linear space).  

The differential cohesion of the ambient $\infty$-topos canonically induces a notion of such
formally {\'e}tale maps: the \emph{infinitesimal path modality} 
$\mathbf{\Pi}_{\mathrm{inf}} : \xymatrix{ \mathbf{H} \ar[r] & \mathbf{H}}$ sends an object
$X$ to what is sometimes called its \emph{de Rham space} $\mathbf{\Pi}_{\mathrm{inf}}(X)$, in which
infinitesimally close points are made equivalent. There is a natural inclusion 
$\xymatrix{X \ar[r] & \mathbf{\Pi}_{\mathrm{inf}}(X)}$ which may alternatively be thought of
as the inclusion of the constant paths in $X$ into the infinitesimal paths in $X$, or as the
quotient map that quotients out the infinitesimal neighbourhood relation. 

Now, a map $f : \xymatrix{X \ar[r] & Y}$ is \emph{formally {\'e}tale} if the naturality square
of this inclusion
$$  
  \raisebox{20pt}{
  \xymatrix{
    X \ar[r] \ar[d]^f & \mathbf{\Pi}_{\mathrm{inf}}(X) \ar[d]^{\mathbf{\Pi}_{\mathrm{inf}}(f)}
	\\
    Y \ar[r] & \mathbf{\Pi}_{\mathrm{inf}}(Y)	
  }
  }
$$
is a homotopy pullback square. This is a generalization to cohesive $\infty$-groupoids 
of the traditional fact that a map $f$ between smooth manifolds is a \emph{local diffeomorphism}
precisely if the square of tangent bundle projections
$$
  \raisebox{20pt}{
  \xymatrix{
     T X \ar[r] \ar[d]^{d f} & X \ar[d]^f
	 \\
	 T Y \ar[r] & Y
  }
  }
$$
is a pullback diagram of smooth manifolds. 
(To see how the general condition above relates to this one, 
let $D \hookrightarrow \mathbb{R}$ be the first order infinitesimal neighbourhood
of the origin in the real line and observe that $X^D \simeq T X$, $f^D \simeq d f$, but that
$(\mathbf{\Pi}_{\mathrm{inf}}(X))^D \simeq \mathbf{\Pi}_{\mathrm{inf}}(X)$.)

For any object $X \in \mathbf{H}$ we then have the $\infty$-category
$$
  \mathrm{Sh}_{\mathbf{H}}(X)
  :=
  \xymatrix{
    (\mathbf{H}_{/X})
	\ar@{^{(}->}[r]-^{\mathrm{fet}}
	\ar@<-4pt>@{<-}[r]_-{\mathrm{Et}}
	&
	\mathbf{H}_{/X}
  }
$$
of formally {\'e}tale maps into $X$. As the notation on the left indicates, this 
may be thought of as the \emph{petit $\infty$-topos of $\infty$-sheaves on $X$},
in generalization of the classical fact of topos theory which identifies
sheaves on a topological space with {\'e}tale topological spaces over it. 

The inclusion of formally {\'e}tale maps into the entire slice $\infty$-topos
$\mathbf{H}_{/X}$ (the \emph{gros $\infty$-topos} of $X$) has a right adjoint reflector
$\mathrm{Et}$, as indicated above. This induces for any object $A \in \mathbf{H}$ the
$\infty$-sheaf on $X$ of \emph{$A$-valued functions on $X$}:
$$
  \mathcal{O}_X(A)
  :=
  \mathrm{Et}(X \times A \stackrel{p_1}{\to} X)
  \in 
  \mathrm{Sh}_{\mathbf{H}}(X)
  \,.
$$
In particular, for $\mathbb{G}$ as above we have the $\infty$-sheaf
$$
  \mathcal{O}_X(\flat_{\mathrm{dR}} \mathbf{B}\mathbb{G})
  \in 
  \mathrm{Sh}_{\mathbf{H}}(X)
$$
of flat $\mathrm{Lie}(\mathbb{G})$-valued differential forms on $X$. The
0-section $0 : * \to \flat_{\mathrm{dR}}\mathbf{B}\mathbb{G}$ induces a 
0-section 
$$
  0 : X \to \mathcal{O}_X(\flat_{\mathrm{dR}} \mathbf{B}\mathbb{G})
$$
in $\mathrm{Sh}_{\mathbf{H}}(X)$, and more generally any map 
$\omega : \xymatrix{X \ar[r] & \flat_{\mathrm{dR}} \mathbf{B}\mathbb{G}}$ induces a section
$$
  \omega : X \to \mathcal{O}_{X}(\flat_{\mathrm{dR}}\mathbf{B}\mathbb{G}  )
$$
in $\mathrm{Sh}_{\mathbf{H}}(X)$. 

But now since in $\mathrm{Sh}_{\mathbf{H}}(X)$ every subspace 
$\xymatrix{U \ar@{^{(}->}[r]  & X}$ is guaranteed to be formally {\'e}tale,
this is the right context to solve the Euler-Lagrange equations of an 
action functional: we say that the \emph{critical locus} of 
$S : \xymatrix{\mathbf{Fields}(\Sigma)_{\partial \Sigma} \ar[r] & \mathbb{G} }$
is the homotopy fiber 
$$
  \underset{\phi \in \mathbf{Fields}(\Sigma)_{\partial \Sigma}}{\sum}
(S^{-1}(\phi) \mathbf{d}_{\mathrm{var}}S(\phi) \simeq 0)
  \in 
  \mathrm{Sh}_{\mathbf{H}}(\mathbf{Fields}(\Sigma)_{\partial \Sigma})
  \,,
$$
sitting in the 
homotopy pullback square
$$
  \raisebox{20pt}{
  \xymatrix{
    \underset{\phi \in \mathbf{Fields}(\Sigma)_{\partial \Sigma}}{\sum}
(S^{-1}(\phi) \mathbf{d}_{\mathrm{var}}S(\phi) \simeq 0)
      \ar[r]
	  \ar[d]
	  &
	  \mathbf{Fields}(\Sigma)_{\partial \Sigma}
	  \ar[d]^0
	  \\
	  \mathbf{Fields}(\Sigma)_{\partial \Sigma}
	  \ar[r]^{S^{-1}\mathbf{d}_{\mathrm{var}}S }
	  &
	  \mathcal{O}_{\mathbf{Fields}(\Sigma)_{\partial \Sigma}}(\flat_{\mathrm{dR}}\mathbf{B}\mathbb{G} )
  }
  }
$$
in $\mathrm{Sh}_{\mathbf{H}}(\mathbf{Fields}(\Sigma)_{\partial \Sigma})$. 

This critical locus is known in traditional literature for the special case
that $\mathbb{G} = \mathbb{R}$ is the additive Lie group of real numbers
and in its \emph{infinitesimal} approximation: the \emph{$\infty$-Lie algebroid}
of the critical locus is known, dually as the \emph{on-shell BRST complex}
of the system (whereas the $\infty$-Lie algebroid of $\mathbf{Fields}(\Sigma)$
itself is the \emph{off-shell BRST complex}). Moreover, if the ambient 
$\infty$-topos $\mathbf{H}$ is not 1-localic, and specifically if it has a site
of definition given by formal duals of simplicial (smooth) algebras, then 
the critical locus as above is also called the \emph{derived crititcal locus}
for emphasis, and its $\infty$-Lie algrboid is dually known as the 
\emph{BV-BRST complex} of the system. 
(For discussion of how the traditional formulation of BV-BRST complexes
models homotopy pullbacks of the above form see \cite{SchreiberBVSeminar}.)

But with the general notion of critical loci in cohesive $\infty$-toposes, we 
obtain examples beyond those discussed in the literature whenever $\mathbb{G}$
is a \emph{higher} group.

Notably when $\mathbb{G} := \mathbf{B}^{n}U(1)$ is the circle $(n+1)$-group,
then the universal differential 
$$
  \theta : \xymatrix{\mathbf{B}^n U(1) \ar[r] & \flat_{\mathrm{dR}}\mathbf{B}^{n+1}U(1)  }
$$  
is equivalently, by the discussion in 
\ref{GaugeFieldHigherTwistedNonAbelian}, the \emph{universal curvature characteristic}
for smooth circle $n$-bundles, and so there are accordingly higher order 
interpretations of phase spaces in \emph{extended} prequantization. 

For example, let $\mathbf{c} : \xymatrix{\mathbf{B}G \ar[r] & \mathbf{B}^n U(1)}$
be a universal characteristic map on the moduli stack of a cohesive $\infty$-group $G$. 
Then 
$$
  S := p_1 \circ [\mathbf{\Pi}(S^1), \mathbf{c}]
  : 
  \xymatrix{
    G /\!/_{\mathrm{ad}}G
	\ar[r]
	&
	\mathbf{B}^{n-1}U(1) /\!/_{\mathrm{ad}} \mathbf{B}^{n-1}U(1)
	\ar[r]
	&
	\mathbf{B}^{n-1}U(1)
  }
$$
is the WZW-$(n-1)$-bundle (equiopped with its ad-equivariant structure) of the corresponding
$n$-dimensional Chern-Simons theory. Regarding this as a $\mathbb{G}$-valued function, we find
that its variational differential
$$
  S^{-1}\mathbf{d}_{\mathrm{var}} S
  : 
  \xymatrix{
    G /\!/_{\mathrm{ad}}G
	\ar[r]^S
	&
	\mathbf{B}^{n-1}U(1)
	\ar[r]^{\theta}
	&
	\flat_{\mathrm{dR}}\mathbf{B}^n U(1)
  }
$$
is the curvature, in de Rham hypercohomology, of the WZW-$(n-1)$-bundle.

\subsubsection{Higher geometric prequantum theory}
\label{PrequantumGeometry}

We had indicated in section \ref{PrequantumLineBundlesOnModuliStacks} how a single
extended Lagrangian, given by a map of universal higher moduli stacks
$\mathbf{L}: {\mathbf{B}G_{\mathrm{conn}} \to \mathbf{B}^n U(1)_{\mathrm{conn}}}$,
induces, by transgression, circle $(n-k)$-bundles with connection
$$
  \mathrm{hol}_{\Sigma_k} \mathbf{Maps}(\Sigma_k, \mathbf{L})
  \;:\;
  \mathbf{Maps}(\Sigma_k, \mathbf{B}G_{\mathrm{conn}}) \longrightarrow \mathbf{B}^{n-k}U(1)_{\mathrm{conn}}
$$
on moduli stacks of field configurations over each closed $k$-manifold $\Sigma_k$.
In codimension 1, hence for $k = n-1$, this reproduces the ordinary \emph{prequantum circle bundle}
of the $n$-dimensional Chern-Simons type theory, as discussed in section 
\ref{TheSymplecticStructureOnModuliSpaceOfConnections}. 
The space of sections of the associated line bundle is the space of 
\emph{prequantum states} of the theory. This becomes the space of genuine quantum states
after choosing a \emph{polarization} 
(i.e., a decomposition of the moduli space of fields into \emph{canonical coordinates}
and \emph{canonical momenta}) and restricting to polarized sections (i.e., those depending only on the 
canonical coordinates).
But moreover, for each $\Sigma_k$ we may regard $\mathrm{hol}_{\Sigma_k} \mathbf{Maps}(\Sigma_k, \mathbf{L})$ 
as a \emph{higher prequantum bundle} of the theory in higher codimension 
and hence consider its prequantum geometry
in higher codimension. 

We discuss now some generalities of such a higher geometric prequantum theory and then 
show how this perspective sheds a useful light on the gauge coupling of the open string,
as part of the transgression of prequantum 2-states of Chern-Simons theory in codimension 2
to prequantum states in codimension 1.

\medskip 

We indicate now the basic concepts of higher extended prequantum theory and how they
reproduce traditional prequantum theory.

Consider a (pre)-$n$-plectic form, given by 
a map
$$
  \omega : X \longrightarrow \Omega^{n+1}(-;\mathbb{R})_{\mathrm{cl}}
$$
in $\mathbf{H}$. A \emph{$n$-plectomorphism} of $(X,\omega)$ is an 
auto-equivalence of $\omega$ regarded as an object in the slice $\mathbf{H}_{/\Omega^{n+1}_{\mathrm{cl}}}$,
hence a diagram of the form
$$
  \raisebox{20pt}{
  \xymatrix{
    X \ar[dr]_{\omega} \ar[rr]^{\simeq} && X \ar[dl]^{\omega}
	\\
	& \Omega^{n+1}(-;\mathbb{R})_{\mathrm{cl}}
  }
  }\,.
$$ 
A \emph{prequantization} of $(X, \omega)$ is a choice of prequantum line bundle, hence a choice of 
lift $\nabla$ in 
$$
  \raisebox{20pt}{
  \xymatrix{
    & \mathbf{B}^n U(1)_{\mathrm{conn}}
	  \ar[d]^{F_{(- )}}
    \\
    X \ar[r]_{\omega} \ar[ur]^{\nabla} & \Omega^{n+1}_{\mathrm{cl}}
  }
  }\,,
$$
modulating a circle $n$-bundle with connection on $X$. We write
$\mathbf{c}(\nabla) : X \xrightarrow{\nabla}\mathbf{B}^n U(1)_{\mathrm{conn}} \to\mathbf{B}^n U(1)$
for the underlying $(\mathbf{B}^{n-1}U(1))$-principal $n$-bundle.
An autoequivalence  
$$
  \hat O : \nabla \stackrel{\simeq}{\longrightarrow} \nabla
$$
of the prequantum $n$-bundle
regarded as an object in the slice $\mathbf{H}_{/\mathbf{B}^n U(1)_{\mathrm{conn}}}$, 
hence a diagram in $\mathbf{H}$ of the form
$$
  \xymatrix{
    X \ar[rr]^\simeq_{\ }="s" \ar[dr]_{\nabla}^{\ }="t" && X \ar[dl]^{\nabla}
	\\
	& \mathbf{B}^n U(1)_{\mathrm{conn}}
	\ar@{=>}^O "s"; "t"
  }
$$
is an (exponentiated) \emph{prequantum operator} or \emph{quantomorphism}  or \emph{regular contact transformation}
of the prequantum geometry $(X, \nabla)$, forming an $\infty$-group in $\mathbf{H}$. 
The $L_\infty$-algebra of this 
\emph{quantomorphism $\infty$-group} 
is the higher \emph{Poisson bracket} Lie algebra of the system. If $X$ is equipped with
abelian group structure then the quantomorphisms covering these translations form the 
\emph{Heisenberg $\infty$-group}. The homotopy labeled $O$ above diagram is the \emph{Hamiltonian}
of the prequantum operator. The image of the quantomorphisms in the symplectomorphisms
(given by composition the above diagram with the curvature morphism 
$F_{(-)} : \mathbf{B}^n U(1)_{\mathrm{conn}} \to \Omega^{n+1}_{\mathrm{cl}}$) is the 
group of \emph{Hamiltonian $n$-plectomorphisms}.  A lift of an $\infty$-group action
$G \to \mathbf{Aut}(X)$ on $X$ from automorphisms of $X$ (diffeomorphism) to quantomorphisms
is a \emph{Hamiltonian action}, infinitesimally (and dually) a \emph{momentum map}.

To define higher prequantum states we fix a representation $(V, \rho)$ of the circle $n$-group $\mathbf{B}^{n-1}U(1)$.
By the general results in \cite{NSSa} this is equivalent to fixing a homotopy fiber sequence of the form
$$
  \raisebox{20pt}{
  \xymatrix{
    \underline{V} \ar[r] & \underline{V}/\!/\mathbf{B}^{n-1}U(1)
	\ar[d]^{\mathbf{\rho}}
	\\
	& \mathbf{B}^n U(1)
  }
  }
$$
in $\mathbf{H}$. The vertical morphism here is the \emph{universal $\rho$-associated $V$-fiber $\infty$-bundle}
and characterizes $\rho$ itself. Given such, a section of the $V$-fiber bundle which is 
$\rho$-associated to $\mathbf{c}(\nabla)$ is equivalently a map
$$
  \Psi : \mathbf{c}(\nabla) \longrightarrow \mathbf{\rho}
$$
in the slice $\mathbf{H}_{/\mathbf{B}^n U(1)}$. This is a higher \emph{prequantum state} of the
prequantum geometry $(X, \nabla)$. Since every prequantum operator $\hat O$ as above
in particular is an auto-equivalence of the underlying prequantum bundle 
$\hat O : \mathbf{c}(\nabla) \stackrel{\simeq}{\longrightarrow} \mathbf{c}(\nabla)$ it
canonically acts on prequantum states given by maps as above simply by precomposition
$$
  \Psi \mapsto \hat O \circ \Psi
  \,.
$$
Notice also that from the perspective of section \ref{FieldsInSlices} all this has
an equivalent interpretation in terms of twisted cohomology: a preqantum state is a cocycle
in twisted $V$-cohomology, with the twist being the prequantum bundle. And a 
prequantum operator/quantomorphism is equivalently a twist automorphism 
(or ``generalized local diffeomorphism'').

For instance if $n = 1$ then $\omega$ is an ordinary (pre)symplectic form 
and $\nabla$ is the connection on a circle bundle. In this case the above notions
of prequantum operators, quantomorphism group, Heisenberg group and Poisson bracket Lie algebra
reproduce exactly all the traditional notions if $X$ is a smooth manifold,
and generalize them to the case that $X$ is for instance an orbifold or even itself a higher moduli stack, 
as we have seen. The canonical representation of the circle group $U(1)$ on the complex numbers
yields a homotopy fiber sequence
$$
  \raisebox{20pt}{
  \xymatrix{
    \underline{\mathbb{C}}
	\ar[r]
	&
	\underline{\mathbb{C}}/\!/\underline{U}(1) \ar[d]^{\mathbf{\rho}}
	\\
	&
	\mathbf{B}U(1)
  }
  }
  \,,
$$
where $\underline{\mathbb{C}}/\!/\underline{U}(1)$ is the stack corresponding to the ordinary action groupoid of the action of $U(1)$ on $\mathbb{C}$, 
and where the vertical map is the canonical functor forgetting the data of the local $\mathbb{C}$-valued functions. 
This is the \emph{universal complex line bundle} associated to the universal $U(1)$-principal bundle.
One readily checks that a prequantum state $\Psi : \mathbf{c}(\nabla) \to \mathbf{\rho}$,
hence a diagram of the form
$$
  \raisebox{20pt}{
  \xymatrix{
    X \ar[rr]^\sigma \ar[dr]_{\mathbf{c}(\nabla)} && \underline{\mathbb{C}}/\!/\underline{U}(1) \ar[dl]^{\mathbf{\rho}}
	\\
	& \mathbf{B}U(1)
  }
  }
$$
in $\mathbf{H}$ is indeed equivalently a section of the complex line bundle canonically associated to 
$\mathbf{c}(\nabla)$ and that under this equivalence the pasting composite
$$
  \xymatrix{
    X \ar[r]^\simeq_>{\ }="s" \ar[dr]_{\mathbf{c}(\nabla)}^{\ }="t" & X \ar[r] \ar[d]|{\mathbf{c}(\nabla)} & \underline{\mathbb{C}}/\!/\underline{U}(1)
	\ar[dl]^{\mathbf{\rho}}
	\\
	& \mathbf{B}U(1)
	\ar@{=>}_O "s"; "t"
  }
$$
is the result of the traditional formula for the action of the prequantum operator $\hat O$ on $\Psi$.

Instead of forgetting the connection on the prequantum bundle in the above composite, 
one can equivalently equip the prequantum state with a differential refinement, namely with its
\emph{covariant derivative} and then exhibit the prequantum operator action directly. 
Explicitly, let 
$\mathbb{C}/\!/U(1)_{\mathrm{conn}}$ denote the quotient stack
  $(\underline{\mathbb{C}}\times \Omega^1(-,\mathbb{R}) )/\!/\underline{U}(1)
  \,,$
with $U(1)$ acting diagonally. This sits in a 
homotopy fiber sequence
$$
  \xymatrix{
   \underline{\mathbb{C}} \ar[r] & 
      \underline{\mathbb{C}}/\!/\underline{U}(1)_{\mathrm{conn}} \ar[d]^{\rho_{\mathrm{conn}}}
	\\
	& \mathbf{B}U(1)_{\mathrm{conn}}
  }
$$
which may be thought of as the differential refinement of the above fiber sequence
$\underline{\mathbb{C}} \to \underline{\mathbb{C}}/\!/\underline{U}(1) \to \mathbf{B}U(1)$. 
(Compare this to section \ref{WithWilsonLoops}, where we had similarly 
seen the differential refinement of the 
fiber sequence $\underline{G}/\underline{T}_\lambda \to \mathbf{B}T_\lambda \to \mathbf{B}G$, 
which analogously characterizes the canonical action of $G$ on the coset space $G/T_{\lambda}$.)
Prequantum states are now equivalently maps
$$
  \widehat{\mathbf{\Psi}}
  :
  \nabla \longrightarrow \mathbf{\rho}_{\mathrm{conn}}
$$
in $\mathbf{H}_{/\mathbf{B}U(1)_{\mathrm{conn}}}$.
This formulation realizes a section of an associated line bundle equivalently as a connection on 
what is sometimes called a groupoid bundle. As such, $\widehat{\mathbf{\Psi}}$
has not just a 2-form curvature (which is that of the prequantum bundle) but also a 1-form
curvature: this is the covariant derivative $\nabla \sigma$ of the section.

Such a relation between sections of higher associated bundles and higher covariant derivatives
holds more generally. 
In the next degree for $n = 2$ one finds that the quantomorphism 2-group is the Lie 2-group
which integrates the \emph{Poisson bracket Lie 2-algebra} of the underlying 2-plectic geometry
as introduced in \cite{Rogers}.
In the next section we look at an example for $n = 2$ in more detail and show how it interplays
with the above example under transgression.

The above higher prequantum theory becomes a genuine quantum theory
after a suitable higher analog of 
a choice of \emph{polarization}.
In particular, for $\mathbf{L} : X \to \mathbf{B}^n U(1)_{\mathrm{conn}}$ an extended Lagrangian
of an $n$-dimensional quantum field theory as discussed in all our examples here, and for $\Sigma_k$ any closed
manifold, the polarized prequantum states of the transgressed prequantum bundle
$\mathrm{hol}_{\Sigma_k}\mathbf{Maps}(\Sigma_k, \mathbf{L})$ 
should form the 
$(n-k)$-vector spaces of higher quantum states in codimension $k$.
These states 
 would be assigned to $\Sigma_k$ by the 
\emph{extended quantum field theory}, in the sense of \cite{LurieTFT}, 
obtained from the extended Lagrangian $\mathbf{L}$ by extended geometric quantization.
There is an equivalent reformulation of this last step for $n = 1$ given simply by the 
push-forward of the prequantum line bundle in K-theory 
(see section 6.8 of \cite{GinzburgGuilleminKarshon}) 
and so one would expect that accordingly
the last step of higher geometric quantization involves similarly a push-forward
of the associated $V$-fiber $\infty$-bundles above
in some higher generalized cohomology theory. But this remains to be investigated.

\subsubsection{Examples of higher prequantum field theories}
\label{GeometryOfPhysicsExamplesOfHigherPrequantumFieldTheories}
\label{PrequantumLineBundlesOnModuliStacks}

We consider now some examples of the higher geometric prequantum field theory
discussed above.

\medskip

\begin{itemize}
  \item \ref{GOPExtended3dCS} -- Extended 3d Chern-Simons theory 
  \item \ref{OpenStringGaugeCoupling} -- The anomaly-free gauge coupling of the open string
\end{itemize}

\paragraph{Extended 3d Chern-Simons theory}
\label{GOPExtended3dCS}

For $G$ a simply connected compact simple Lie group, the above construction
of the refined Chern-Weil homomorphism yields a differential characteristic map of 
moduli stacks 
$$
  \hat {\mathbf{c}} : \xymatrix{\mathbf{B}G_{\mathrm{conn}} & \mathbf{B}^3 U(1)_{\mathrm{conn}}}
$$
which is the smooth and differential refinement of the universal characteristic class
$[c] \in H^4(B G, \mathbb{Z})$.

We discuss now how this serves as the \emph{extended} Lagrangian for 3d Chern-Simons theory
in that its \emph{transgression} to mapping stacks out of $k$-dimensional manifolds yields
all the ``geometric prequantum'' data of Chern-Simons theory in the corresponding dimension, 
in the sense of geometric quantization. For the purpose of this exposition 
we use terms such as ``prequantum $n$-bundle'' freely without formal definition. We expect the 
reader can naturally see at least vaguely the higher prequantum picture alluded to here. 
A more formal survey of these notions is in section \ref{PrequantumInHigherCodimension}.

The following paragraphs draw from \cite{FiorenzaSatiSchreiberCS}.

\medskip

If $X$ is  a compact oriented manifold without boundary, then there is a fiber integration in differential cohomology lifting fiber integration in integral cohomology \cite{HopkinsSinger}:
\[
\xymatrix{
\hat{H}^{n+\dim X}(X\times Y;\mathbb{Z})\ar[d]\ar[r]^{\phantom{mmm}\int_X}& \hat{H}^{n}(Y;\mathbb{Z})\ar[d]\\
{H}^{n+\dim X}(X\times Y;\mathbb{Z})\ar[r]^{\phantom{mmm}\int_X}& {H}^{n}(Y;\mathbb{Z})\;.
}
\]
In \cite{GomiTerashima} Gomi and Terashima describe an explicit lift of this at the level of \v{C}ech-Deligne cocycles. 
Such a lift has a natural interpretation 
as a morphism
\[
\mathrm{hol}_X: \mathbf{Maps}(X,\mathbf{B}^{n+\dim X}U(1)_{\mathrm{conn}})\to \mathbf{B}^{n}U(1)_{\mathrm{conn}}
\]
from the $(n+\dim X)$-stack of moduli of $U(1)$-$(n+\dim X)$-bundles 
with connection over $X$ to the $n$-stack of  $U(1)$-$n$-bundles 
with connection, \ref{SmoothStrucDifferentialCohomology}.
Therefore, if $\Sigma_k$ is a compact 
oriented manifold of dimension $k$ with $0\leq k\leq 3$, we have a composition
\[
\mathbf{Maps}(\Sigma_k,\mathbf{B}G_{\mathrm{conn}})\xrightarrow{\mathbf{Maps}(\Sigma_k,\hat{\mathbf{c}})}\mathbf{Maps}(\Sigma_k,\mathbf{B}^3U(1)_{\mathrm{conn}})\xrightarrow{\mathrm{hol}_{\Sigma_k}} \mathbf{B}^{3-k}U(1)_{\mathrm{conn}}\;.
\]
This is the canonical $U(1)$-$(3-k)$-bundle with connection over the moduli space of principal $G$-bundles with connection over $\Sigma_k$ induced by $\hat {\mathbf{c}}$: the
\emph{transgression} of $\hat {\mathbf{c}}$ to the mapping space. 
Composing on the right with the curvature morphism
we get the underlying canonical closed $(4-k)$-form 
\[
\mathbf{Maps}(\Sigma_k,\mathbf{B}G_{\mathrm{conn}})\to \Omega^{4-k}(-;\mathbb{R})_{\mathrm{cl}}
\] 
on this moduli space. 
In other words, the moduli stack of principal $G$-bundles with connection over $\Sigma_k$ carries a canonical 
\emph{pre-$(3-k)$-plectic structure} 
(the higher order generalization of a symplectic structure, \cite{Rogers}) 
and, moreover, this is equipped with a canonical geometric prequantization:
the above $U(1)$-$(3-k)$-bundle with connection. 

Let us now investigate in more  detail the cases $k=0,1,2,3$.

\subparagraph{$k=0$: the universal Chern-Simons 3-connection $\hat{\mathbf{c}}$}

The connected 0-manifold $\Sigma_0$ is the point and, by definition of $\mathbf{Maps}$,
 one has a canonical identification
\[
\mathbf{Maps}(*,\mathbf{S})\cong \mathbf{S}
\]
for any (higher) stack $\bf S$. Hence the morphism 
\[
\mathbf{Maps}(*,\mathbf{B}G_{\mathrm{conn}})\xrightarrow{\mathbf{Maps}(*,\hat{\mathbf{c}})}\mathbf{Maps}(*,\mathbf{B}^3U(1)_{\mathrm{conn}})
\]
is nothing but the 
universal differential characteristic map $\hat{\mathbf{c}}:\mathbf{B}G_{\mathrm{conn}}\to \mathbf{B}^3U(1)_{\mathrm{conn}}$ that refines the universal characteristic class $c$. 
This map modulates a circle 3-bundle with connection (bundle 2-gerbe) on the universal moduli stack
of $G$-principal connections. For $\nabla : X \longrightarrow \mathbf{B}G_{\mathrm{conn}}$
any given $G$-principal connection on some $X$, the pullback 
$$
  \hat {\mathbf{c}}(\nabla) : 
  \xymatrix{
    X \ar[r]^-\nabla & \mathbf{B}G_{\mathrm{conn}} \ar[r]^-{\hat {\mathbf{c}}} & \mathbf{B}^3 U(1)_{\mathrm{conn}}
  }
$$
is a 3-bundle (bundle 2-gerbe) on $X$ which is sometimes in the literature called the 
\emph{Chern-Simons 2-gerbe} of the given connection $\nabla$. Accordingly, $\hat {\mathbf{c}}$ modulates
the \emph{universal} Chern-Simons bundle 2-gerbe with universal 3-connection.
From the point of view of higher geometric quantization, this is the \emph{prequantum 3-bundle}
of extended prequantum Chern-Simons theory.

This means that the prequantum $U(1)$-$(3-k)$-bundles associated with $k$-dimensional manifolds are all determined by  the prequantum $U(1)$-3-bundle associated with the point, in agreement with the formulation of fully extended topological field theories \cite{FHLT}.
We will denote by the symbol $\omega^{(4)}_{\mathbf{B}G_{\mathrm{conn}}}$ the pre-3-plectic 4-form induced on $\mathbf{B}G_{\mathrm{conn}}$ by the curvature morphism.

\subparagraph{$k=1$: the Wess-Zumino-Witten gerbe}
\label{TheWZWGerbe}

We now come to the transgression of the extended Chern-Simons Lagrangian to the 
closed connected 1-manifold, the circle $\Sigma_1 = S^1$. 
Notice that, on the one hand, we can think  of the mapping stack 
$\mathbf{Maps}(\Sigma_1, \mathbf{B}G_{\mathrm{conn}}) \simeq  \mathbf{Maps}(S^1 , \mathbf{B}G_{\mathrm{conn}})$
 as a kind of moduli stack of $G$-connections on the circle
-- up to the subtlety of differential concretification discussed in \ref{DifferentialModuli}.
On the other hand, we can think of that mapping stack 
 as the \emph{free loop space} of the universal moduli stack $\mathbf{B}G_{\mathrm{conn}}$.

The subtlety here is related to the differential refinement, 
so it is instructive to first discard the differential refinement and consider just the 
smooth characteristic map $\mathbf{c} : {\mathbf{B}G \to \mathbf{B}^3 U(1)}$
which underlies the extended Chern-Simons Lagrangian and which  modulates the universal circle 3-bundle on $\mathbf{B}G$ (without connection). Now, for every pointed stack 
$* \to \mathbf{S}$ we have the corresponding (categorical) \emph{loop space}
$
  \Omega \mathbf{S} := * \times_{\mathbf{S}} *
$,
which is the homotopy pullback of the point inclusion along itself. Applied to 
the moduli stack $\mathbf{B}G$ this recovers the Lie group $G$, identified with the sheaf (i.e, the $0$-stack) of smooth functions with target $G$:
$
  \Omega \mathbf{B}G \simeq \underline{G}
  $.
This kind of looping/delooping equivalence is familiar from the homotopy theory of classifying spaces;
but notice that since we are working with smooth (higher) stacks, the loop space
 $\Omega \mathbf{B}G$ also knows the smooth structure of the group $G$, i.e. it knows $G$ as a Lie group. Similarly,
we have
$$
  \Omega \mathbf{B}^3 U(1)\simeq \mathbf{B}^2 U(1)
$$
and so forth in higher degrees.
Since the looping operation is functorial, we may also apply it to the characteristic map $\mathbf{c}$
itself to obtain a map
$$
  \Omega \mathbf{c} :    \underline{G} \to \mathbf{B}^2 U(1)
$$
which modulates a $\mathbf{B}U(1)$-principal 2-bundle on the Lie group $G$.
This is also known as the \emph{WZW-bundle gerbe}; see for instance \cite{SchweigertWaldorf}. 
The reason, as discussed there and as we will see in a moment, is that
this is the 2-bundle that underlies the 2-connection with surface holonomy over a worldsheet 
given by the Wess-Zumino-Witten action functional.  
However, notice first that there is more structure implied here: 
by the discussion in \ref{CotensoringOfSmoothHigherStacksOverHomotopyTypesOfSmoothManifolds},
for any pointed stack $\mathbf{S}$ there is a natural equivalence
$
  \Omega {\mathbf{S}} \simeq \mathbf{Maps}_*(\Pi(S^1), {\mathbf{S}})
  $,
 between the loop space object $ \Omega {\mathbf{S}}$ and 
 the moduli stack of \emph{pointed maps} from the categorical circle  
 $\mathbf{\Pi}(S^1)\simeq \mathbf{B}\mathbb{Z}$ to $\mathbf{S}$. 
On the other hand, if we do not fix the base point then we obtain the \emph{free loop space object}
$
  \mathcal{L} {\mathbf{S}} \simeq \mathbf{Maps}(\mathbf{\Pi}(S^1),  {\mathbf{S}})
 $. Since a map $\mathbf{\Pi}(\Sigma)\to \mathbf{B}G$ is equivalently 
a  map $\Sigma \to \flat \mathbf{B}G$, i.e., a flat $G$-principal connection on $\Sigma$,
the free loop space $\mathcal{L}\mathbf{B}G$ is equivalently the moduli stack of flat $G$-principal connections on $S^1$.
We will come back to this perspective in section \ref{DifferentialModuli} below.
The homotopies that do not fix the base point act by conjugation on loops and hence
we have, for any smooth  (higher) group, that 
$$
  \mathcal{L}\mathbf{B}G \simeq \underline{G}/\!/_{\mathrm{Ad}}\underline{G}
$$
is the (homotopy) quotient of the adjoint action of $G$ on itself; see \cite{NSSa} for details
on homotopy actions of smooth  higher groups. For $G$ a Lie group this is the familiar adjoint
action quotient stack. But the expression holds fully generally. Notably, we also have
$$
  \mathcal{L}\mathbf{B}^3 U(1) \simeq \mathbf{B}^2 U(1)/\!/_{\mathrm{Ad}}\mathbf{B}^2 U(1)
$$
and so forth in higher degrees.
However, in this case, since the smooth 3-group $\mathbf{B}^2 U(1)$ is abelian (it is a groupal $E_\infty$-algebra)
the adjoint action splits off in a direct factor and we have a projection
$$
  \xymatrix{
    \mathcal{L}\mathbf{B}^3 U(1)
\simeq
	\mathbf{B}^2 U(1) \times ({*}/\!/\mathbf{B}^2 U(1))
	\ar[r]^-{p_1}
	&
	\mathbf{B}^2 U(1)
  }\;.
$$
In summary, this means that the map $\Omega \mathbf{c}$ modulating the WZW 2-bundle over $G$ descends
to the adjoint quotient to the map
$$
  p_1 \circ \mathcal{L}\mathbf{c} 
  :
    \underline{G}/\!/_{\mathrm{Ad}}\underline{G}
	\to
	\mathbf{B}^2 U(1)
  \,,
$$
and this means that the WZW 2-bundle is canonically equipped with the structure of an 
\emph{$\mathrm{ad}_G$-equivariant} bundle gerbe, a crucial feature of the WZW bundle gerbe.

We emphasize that the derivation here is fully general and holds for any smooth  (higher) group $G$
and any smooth characteristic map $\mathbf{c} : {\mathbf{B}G \to \mathbf{B}^n U(1)}$.
Each such pair induces a WZW-type $(n-1)$-bundle on the smooth  (higher) group $G$ modulated by
$\Omega \mathbf{c}$ and equipped with $G$-equivariant structure exhibited by
$p_1 \circ \mathcal{L}\mathbf{c}$. We discuss such higher examples of higher Chern-Simons-type theories
with their higher WZW-type functionals further below in section \ref{CupProductCS}.

We now turn to the differential refinement of this situation. In analogy to the 
above construction, but taking care of the connection data in 
the extended Lagrangian $\hat {\mathbf{c}}$, we find a
homotopy commutative diagram in $\mathbf{H}$ of the form
\[
\hspace{-2mm}
\xymatrix{
&\mathbf{Maps}(S^1;\mathbf{B}G_{\mathrm{conn}})\ar[d]_{\mathrm{hol}}\ar[rr]^{\hspace{-3mm}\mathbf{Maps}(S^1,\hat{\mathbf{c}})}
&& \mathbf{Maps}(S^1;\mathbf{B}^3U(1)_{\mathrm{conn}})\ar[d]^{\mathrm{hol}}&\\
\underline{G}\ar[r]&\underline{G}/\!/_{\mathrm{Ad}}\underline{G}\ar[rr]^{\hspace{-6mm}\mathbf{wzw}\phantom{mmm}}&&\mathbf{B}^2U(1)_{\mathrm{conn}}/\!/_{\mathrm{Ad}}\mathbf{B}^2U(1)_{\mathrm{conn}}\ar[r]&\mathbf{B}^2U(1)_{\mathrm{conn}}\;,
}
\]
where the vertical maps are obtained by forming holonomies of (higher) connections along
the circle.
The lower horizontal row is the differential refinement of 
$\Omega \mathbf{c}$: it modulates the Wess-Zumino-Witten $U(1)$-bundle gerbe with connection
\[
\mathbf{wzw}:\underline{G}\to \mathbf{B}^2U(1)_{\mathrm{conn}}\;.
\]
That $\mathbf{wzw}$ is indeed the correct differential refinement can be seen, 
for instance, by interpreting the construction by Carey-Johnson-Murray-Stevenson-Wang in \cite{CJMSW} in terms of the above diagram. 
That is, choosing a basepoint $x_0$ in $S^1$ one obtains a canonical lift of the leftmost vertical arrow:
\[
\xymatrix{
&\mathbf{Maps}(S^1;\mathbf{B}G_{\mathrm{conn}})\ar[d]^{\mathrm{hol}}\\
\underline{G}\ar[r]\ar[ru]^-{(P_{x_0},\nabla_{x_0})}&\underline{G}/\!/_{\mathrm{Ad}}\underline{G}\;,
}
\]
where $(P_{x_0}\nabla_{x_0})$ is the principal $G$-bundle with connection on the product $G\times S^1$ characterized by the property that the holonomy of $\nabla_{x_0}$ along $\{g\}\times S^1$ 
with starting point $(g,x_0)$ is  the element $g$ of $G$. Correspondingly, we have a homotopy commutative diagram
\[
\hspace{-2mm}
\xymatrix{
&\mathbf{Maps}(S^1;\mathbf{B}G_{\mathrm{conn}})\ar[d]_{\mathrm{hol}}\ar[rr]^{\hspace{-2mm}\mathbf{Maps}(S^1,\hat{\mathbf{c}})}
&& \mathbf{Maps}(S^1;\mathbf{B}^3U(1)_{\mathrm{conn}})\ar[d]^{\mathrm{hol}}\ar[dr]^{~~~\mathrm{hol}_{S^1}}&\\
\underline{G}\ar[ru]^-{(P_{x_0},\nabla_{x_0})}\ar[r]&\underline{G}/\!/_{\mathrm{Ad}}\underline{G}
\ar[rr]^-{\mathbf{wzw}}&&\mathbf{B}^2U(1)_{\mathrm{conn}}/\!/_{\mathrm{Ad}}\mathbf{B}^2U(1)_{\mathrm{conn}}\ar[r]&\mathbf{B}^2U(1)_{\mathrm{conn}}\;.
}
\]
Then Proposition 3.4 from \cite{CJMSW} identifies the upper path (and hence also the lower path) from $\underline{G}$ to $\mathbf{B}^2U(1)_{\mathrm{conn}}$ with the Wess-Zumino-Witten bundle gerbe. 
\par
Passing to equivalence classes of global sections, we see that $\mathbf{wzw}$ induces, for any smooth manifold $X$,
 a natural map $C^\infty(X;G)\to \hat{H}^2(X;\mathbb{Z})$.
In particular, if $X=\Sigma_2$ is a compact Riemann surface, we can further integrate over $X$ to get
\[
wzw:C^\infty(\Sigma_2;G)\to  \hat{H}^2(X;\mathbb{Z})\xrightarrow{\int_{\Sigma_2}} U(1)\;.
\]
This is the \emph{topological term} in the Wess-Zumino-Witten model; 
see \cite{Ga, FreedWitten, CareyJohnsonMurray}.
Notice how the fact that $\mathbf{wzw}$ factors through $\underline{G}/\!/_{\mathrm{Ad}}\underline{G}$ gives the conjugation invariance of the Wess-Zumino-Witten bundle gerbe, and hence of the topological term in the Wess-Zumino-Witten model.

\subparagraph{$k=2$: Symplectic structure on the moduli of flat connections}
\label{TheSymplecticStructureOnModuliSpaceOfConnections}

For $\Sigma_2$ a compact Riemann surface, the transgression of the extended Lagrangian
$\hat {\mathbf{c}}$ yields a map
\[
\mathbf{Maps}(\Sigma_2;\mathbf{B}G_{\mathrm{conn}})\xrightarrow{\mathbf{Maps}(\Sigma_2,\hat{\mathbf{c}})}
 \mathbf{Maps}(\Sigma_2;\mathbf{B}^3U(1)_{\mathrm{conn}})\xrightarrow{\mathrm{hol}_{\Sigma_2}} \mathbf{B}U(1)_{\mathrm{conn}}
 \,,
\]
modulating a circle-bundle with connection on the moduli space of gauge fields on $\Sigma_2$.
The underlying curvature of this connection is the map obtained by composing this with
\[
  \xymatrix{
    \mathbf{B}U(1)_{\mathrm{conn}}
    \ar[r]^{F_{(-)}}
	&
    \Omega^2(-;\mathbb{R})_{\mathrm{cl}}
  }\;,
\]
which gives the canonical pre-symplectic 2-form
\[
 \omega 
 :
 \xymatrix{
   \mathbf{Maps}(\Sigma_2;\mathbf{B}G_{\mathrm{conn}})
   \ar[r]
   &
   \Omega^2(-;\mathbb{R})_{\mathrm{cl}}
  }
\]
on the moduli stack of principal $G$-bundles with connection on $\Sigma_2$. 
Equivalently, this is the transgression of the invariant polynomial 
$\langle -\rangle : \xymatrix{\mathbf{B}G_{\mathrm{conn}} \ar[r] &  \Omega^4_{\mathrm{cl}}}$
to the mapping stack out of $\Sigma_2$. The restriction of this 2-form to the moduli stack $\mathbf{Maps}(\Sigma_2;\flat\mathbf{B}G_{\mathrm{conn}})$ of flat principal $G$-bundles on $\Sigma_2$ induces a canonical symplectic structure  on the moduli space
\[
\mathrm{Hom}(\pi_1(\Sigma_2),G)/_{\mathrm{Ad}}G
\]
of flat $G$-bundles on $\Sigma_2$. 
Such a symplectic structure was identified as the phase space structure of Chern-Simons theory in \cite{Witten98}.

To see more explicitly what this form $\omega$ is, consider any test manifold $U \in \mathrm{CartSp}$. 
Over  this the map of stacks $\omega$ is a function which sends a $G$-principal connection 
$A \in \Omega^1(U \times \Sigma_2)$ (using that every $G$-principal bundle over $U \times \Sigma_2$
is trivializable) to the 2-form 
$$
  \int_{\Sigma_2} \langle F_{A} \wedge F_{A}\rangle
  \in
  \Omega^2(U)
  \,.
$$
Now if $A$ represents a field in the phase space, hence an element in the concretification of 
the mapping stack, then it has no ``leg''
\footnote{That is, when written in local coordinates $(u, \sigma)$ on $U \times \Sigma_2$,
then  
$A=A_i(u, \sigma) du^i + A_j (u, \sigma) d\sigma^j$ reduces to the second summand.}
 along $U$, and so it is a 1-form on $\Sigma_2$ that depends
smoothly on the parameter $U$: it is a $U$-parameterized \emph{variation} of such a 1-form.
Accordingly, its curvature 2-form splits as
$$
  F_{A} = F_A^{\Sigma_2} + d_U A
  \,,
$$
where $F_A^{\Sigma_2} := d_{\Sigma_2} A + \tfrac{1}{2}[A \wedge A]$ 
is the $U$-parameterized collection of curvature forms on $\Sigma_2$. 
The other term is the \emph{variational differential} of the $U$-collection of forms.
Since the fiber integration map $\int_{\Sigma_2} : \Omega^4(U \times \Sigma_2) \to \Omega^2(U)$ 
picks out the component of $\langle F_A \wedge F_A\rangle$ with two legs along $\Sigma_2$
and two along $U$, integrating over the former we have that
$$
  \omega|_U
  = 
  \int_{\Sigma_2} \langle F_A \wedge F_A\rangle
  = 
  \int_{\Sigma_2} \langle d_U A \wedge d_U A \rangle
  \in
  \Omega^2_{\mathrm{cl}}(U)
  \,.
$$
In particular if we consider, without loss of generality,
$(U = \mathbb{R}^2)$-parameterized variations and expand 
$$
  d_U A = (\delta_1 A) du^1 + (\delta_2 A) du^2 \in \Omega^2(\Sigma_2 \times U)
  \,,
$$
then 
$$
  \omega|_U = \int_{\Sigma_2} \langle \delta_1 A, \delta_2 A \rangle
  \,.
$$
In this form the symplectic structure appears, for instance, 
in prop. 3.17 of part I of \cite{FreedCS}
(in \cite{Witten} this corresponds to (3.2)).

In summary, this means that the circle bundle with connection obtained by transgression of the 
extended Lagrangian $\hat {\mathbf{c}}$ is a \emph{geometric prequantization} of the phase space of
3d Chern-Simons theory. Observe that traditionally prequantization involves an
arbitrary \emph{choice}: the choice of prequantum bundle with connection whose 
curvature is the given symplectic form. 
Here we see that in \emph{extended} prequantization this 
choice is eliminated, or at least reduced: while there may be many differential cocycles
lifting a given curvature form, only few of them arise by transgression from a 
higher differential cocycles in top codimension. In other words, the restrictive choice
of the single geometric prequantization of the invariant polynomial 
$\langle -,-\rangle : \mathbf{B}G_{\mathrm{conn}} \to \Omega^4_{\mathrm{cl}}$
by $\hat{\mathbf{c}} : \mathbf{B}G_{\mathrm{conn}}\to \mathbf{B}^3 U(1)_{\mathrm{conn}}$
down in top codimension induces canonical choices of prequantization 
over all $\Sigma_k$ in all lower codimensions $(n-k)$.

\subparagraph{$k=3$: the Chern-Simons action functional}

Finally, for $\Sigma_3$ a compact oriented 3-manifold without boundary,
transgression of the extended Lagrangian $\hat {\mathbf{c}}$ produces the morphism
\[
\mathbf{Maps}(\Sigma_3;\mathbf{B}G_{\mathrm{conn}})\xrightarrow{\mathbf{Maps}(\Sigma_3,\hat{\mathbf{c}})}
 \mathbf{Maps}(\Sigma_3;\mathbf{B}^3U(1)_{\mathrm{conn}})\xrightarrow{\mathrm{hol}_{\Sigma_3}} 
 \underline{U}(1)\;.
\]
Since the morphisms in $\mathbf{Maps}(\Sigma_3; \mathbf{B}G_{\mathrm{conn}})$ are 
\emph{gauge transformations} between field configurations, while $\underline{U}(1)$ has
no non-trivial morphisms, this map necessarily gives a \emph{gauge invariant} $U(1)$-valued function
on field configurations. 
Indeed, evaluating over the point and passing to isomorphism classes
(and hence to gauge equivalence classes), this induces the \emph{Chern-Simons action functional}
\[
S_{\hat{\mathbf{c}}}:\{\text{$G$-bundles with connection on~} \Sigma_3\}/\text{iso}\to U(1)\;.
\]
It follows from the description of $\hat{\mathbf{c}}$  that 
if the principal $G$-bundle $P\to \Sigma_3$ is trivializable then
\[
S_{\hat{\mathbf{c}}}(P,\nabla)=\exp 2\pi i\int_{\Sigma_3} \mathrm{CS}_3(A)\;,
\]
where $A\in \Omega^1(\Sigma_3,\mathfrak{g})$ is the $\mathfrak{g}$-valued 1-form 
on $\Sigma_3$ representing 
the connection $\nabla$ in a chosen trivialization of $P$. This is actually 
always the case, but notice two things: first, in the stacky description one does not need to know a priori that every principal $G$-bundle on a 3-manifold is trivializable; second, the independence of $S_{\hat{\mathbf{c}}}(P,\nabla)$ on the trivialization chosen is automatic from the fact that $S_{\hat{\mathbf{c}}}$ is a morphism of stacks read at the level of equivalence classes.\par

Furthermore, if $(P,\nabla)$ can be extended to a principal $G$-bundle with connection $(\tilde{P},\tilde{\nabla})$ over a compact 4-manifold $\Sigma_4$ bounding $\Sigma_3$, one has
\[
S_{\hat{\mathbf{c}}}(P,\nabla)=\exp 2\pi i\int_{\Sigma_4}\tilde{\varphi}^*\omega^{(4)}_{\mathbf{B}G_{\mathrm{conn}}}=\exp 2\pi i\int_{\Sigma_4}\langle F_{\tilde{\nabla}},F_{\tilde{\nabla}}\rangle\;,
\]
where $\tilde{\varphi}:\Sigma_4\to \mathbf{B}G_{\mathrm{conn}}$ is the morphism corresponding to the extended bundle $(\tilde{P},\tilde{\nabla})$. Notice that the right hand side is independent of the extension chosen. Again, this is always the case, so one can actually take the above equation as a definition of the Chern-Simons action functional, see, e.g., 
\cite{FreedCS}. However, notice how in the stacky approach we do not need a priori to know that the oriented cobordism ring is trivial in dimension 3. Even more remarkably, the stacky point of view tells us that there would be a natural and well-defined 3d Chern-Simons action functional even if the oriented cobordism ring were nontrivial in dimension 3 or that not every $G$-principal bundle on a 3-manifold were trivializable.
An instance of checking a nontrivial 
higher cobordism group vanishes can be found in \cite{KS2}, allowing for the 
application of the construction of Hopkins-Singer \cite{HopkinsSinger}.

\subparagraph{The Chern-Simons action functional with Wilson loops}
\label{WithWilsonLoops}
\index{Chern-Simons-functionals!Wilson loops}
\index{orbit method}

To conclude our exposition of the examples of 1d and 3d Chern-Simons theory in higher geometry,
we now briefly discuss how both unify into the theory of 3d Chern-Simons gauge fields
with Wilson line defects. Namely,
 for every embedded knot
$$\iota : S^1 \hookrightarrow \Sigma_3$$ in the closed 3d worldvolume and every complex linear representation
$R : G \to \mathrm{Aut}(V)$ one can consider the  \emph{Wilson loop observable} $W_{\iota,R}$ mapping a gauge field 
$A : {\Sigma \to \mathbf{B}G_{\mathrm{conn}}}$, to the corresponding
``Wilson loop holonomy''
$$
  W_{\iota,R}:A \mapsto \mathrm{tr}_{R}( \mathrm{hol}(\iota^*A)) \in \mathbb{C}
  \,.
$$
This is the trace, in the given representation, of the parallel transport defined by the connection $A$ around the loop $\iota$
(for any choice of base point).
It is an old observation\footnote{This can be traced back to \cite{BalachandranBorchardtStern};
a nice modern review can be found in section 4 of 
\cite{Beasley}.} 
that this Wilson loop $W(C,A,R)$ is itself the \emph{partition function}
of a 1-dimensional topological $\sigma$-model quantum field theory that describes the topological
sector of a particle charged under the nonabelian background gauge field $A$.
In section 3.3 of \cite{Witten} it was therefore emphasized that  Chern-Simons theory
with Wilson loops should really be thought of as given by a single Lagrangian
which is the sum of the 3d Chern-Simons Lagrangian for the gauge field as above, 
plus that for this topologically charged particle.

We now briefly indicate how this picture is naturally captured by
higher geometry and refined to a single \emph{extended} Lagrangian for coupled 1d and 3d Chern-Simons theory,
given by maps on higher moduli stacks. In doing this, we will also see how
the ingredients of Kirillov's orbit method  and the Borel-Weil-Bott theorem
find a natural rephrasing in the context of smooth differential moduli stacks.
The key observation is that for 
$
  \langle \lambda , -\rangle 
$
an integral weight for our simple, connected, simply connected and compact  Lie group $G$, the
contraction of $\mathfrak{g}$-valued differential forms with $\lambda$ extends
to a morphism of smooth moduli stacks of the form
$$
  \langle \mathbf{\lambda} , - \rangle
  \;:\, 
    \Omega^1(-,\mathfrak{g})//\underline{T}_\lambda
	\to
	\mathbf{B}U(1)_{\mathrm{conn}}
  \,,
$$
where  $T_\lambda \hookrightarrow G$ is the maximal torus of $G$ which is the stabilizer subgroup of
$\langle \lambda, -\rangle$ under the coadjoint action of $G$ on $\mathfrak{g}^*$. 
Indeed, 
this is just the classical statement that exponentiation of $\langle \lambda,-\rangle$ 
induces an isomorphism between the integral weight lattice $\Gamma_{\mathrm{wt}}(\lambda)$ realtive to the maximal torus $T_\lambda$ and the $\mathbb{Z}$-module $\mathrm{Hom}_{\mathrm{Grp}}(T_\lambda,U(1))$ and that under this isomorphism a gauge transformation
of a $\mathfrak{g}$-valued 1-form $A$ turns into that of the $\mathfrak{u}(1)$-valued 1-form
$\langle \lambda, A\rangle$.

This is 
the extended Lagrangian of a 1-dimensional Chern-Simons theory. In fact it is just a slight
variant of the trace-theory discussed there: if we realize $\mathfrak{g}$ as a matrix Lie algebra
and write $\langle \alpha, \beta\rangle = \mathrm{tr}(\alpha \cdot \beta)$ as the matrix trace,
then the above Chern-Simons 1-form is 
given 
by the ``$\lambda$-shifted trace''
$$
  \mathrm{CS}_\lambda(A) := \mathrm{tr}(\lambda \cdot A) \in \Omega^1(-;\mathbb{R})
  \,.
$$
Then, clearly, while the ``plain'' trace is invariant under the adjoint action of all of $G$, 
the $\lambda$-shifted trace is invariant only under the subgroup $T_\lambda$ of $G$ that fixes $\lambda$. 

Notice that the domain of $\langle \mathbf{\lambda}, -\rangle$ 
naturally sits 
inside $\mathbf{B}G_{\mathrm{conn}}$
by the canonical map
$$
    \Omega^1(-,\mathfrak{g})/\!/ \underline{T}_\lambda
	\to
    \Omega^1(-,\mathfrak{g})/\!/ \underline{G}
	\simeq
	\mathbf{B}G_{\mathrm{conn}}
  \,.
$$
One sees that the homotopy fiber of this map to be the \emph{coadjoint orbit} 
$\mathcal{O}_\lambda \hookrightarrow \mathfrak{g}^*$ of $\langle \lambda, -\rangle$,
equipped with the map of stacks
$$
  \mathbf{\theta}
   :
    \mathcal{O}_{\lambda} \simeq \underline{G}/\!/\underline{T}_\lambda 
	\to
	\Omega^1(-,\mathfrak{g})/\!/\underline{T}_\lambda
$$
which over a test manifold $U$ sends $g \in C^\infty(U,G)$ to 
the pullback $g^* \theta_G$ of the Maurer-Cartan form.
Composing this with the above extended Lagrangian $\langle \mathbf{\lambda}, -\rangle$
yields a map
$$
  \langle \mathbf{\lambda}, \mathbf{\theta}\rangle
  : 
  \xymatrix{
    \mathcal{O}_\lambda
	\ar[r]^-{\mathbf{\theta}}
	&
	\Omega^1(-,\mathfrak{g})/\!/\underline{T}_\lambda
	\ar[r]^-{\mathbf{\langle \lambda, -\rangle}}
	&
	\mathbf{B}U(1)_{\mathrm{conn}}
  }
$$
which modulates a canonical $U(1)$-principal bundle with connection on the coadjoint orbit.
One finds that this is the canonical prequantum bundle used in the orbit method 
 \cite{Kirillov}.
 In particular its curvature is 
 the canonical symplectic form on the coadjoint orbit.
%

So far this shows how the ingredients of the orbit method are incarnated in smooth moduli 
stacks. This now immediately induces Chern-Simons theory with Wilson loops by
considering the map $ \Omega^1(-,\mathfrak{g})/\!/ \underline{T}_\lambda\to	\mathbf{B}G_{\mathrm{conn}}$
itself as the target\footnote{This means that here we are secretely moving from the topos of (higher) stacks on smooth manifolds to its \emph{arrow topos}, 
see section \ref{FieldsInSlices} below.} for a field theory defined on knot inclusions $\iota: S^1 \hookrightarrow \Sigma_3$.
%
%
This means that a field configuration
is a diagram of smooth stacks of the form
\[
 \raisebox{20pt}{
  \xymatrix{
     S^1 \ar[rr]^-{(\iota^*A)^g}_>{\ }="s" \ar[d]_\iota^>{\ }="t" 
	 && \Omega^1(-,\mathfrak{g})/\!/\underline{T}_\lambda
	 \ar[d]
	 \\
	 \Sigma_3
	 \ar[rr]_-{A}
	 &&
	~ \mathbf{B}G_{\mathrm{conn}}\;,
	 \ar@{=>}^g "s"; "t"
  }
  }
  \label{FieldsForCSWithWilsonLoop}
\]
i.e., that a field configuration consists of 
\begin{itemize}
  \item a gauge field $A$ in the ``bulk'' $\Sigma_3$;
  \item a $G$-valued function $g$ on the embedded knot
\end{itemize}
such that 
 the
restriction of the ambient gauge field $A$ to the knot is equivalent, via the gauge transformation $g$, to a $\mathfrak{g}$-valued connection on $S^1$ whose local $\mathfrak{g}$-valued 1-forms are related each other by local gauge transformations taking values in the torus $T_\lambda$. 
Moreover, a gauge transformation between two such field configurations $(A,g)$ and $(A',g')$ is a pair $(t_{\Sigma_3},t_{S^1})$ consisting of a 
$G$-gauge transformation $t_{\Sigma_3}$ on $\Sigma_3$ and a $T_\lambda$-gauge transformation
$t_{S^1}$ on $S^1$, intertwining the 
gauge transformations $g$ and $g'$. In particular if the bulk gauge field on $\Sigma_3$ is held
fixed, i.e., if $A=A'$, then $t_{S^1}$ satisfies the equation $g' = g\, t_{S^1}$. This means that 
the Wilson-line components of gauge-equivalence classes
of field configurations are naturally identified with smooth functions $S^1 \to G/T_{\lambda}$, i.e., with smooth functions on the Wilson loop with values in the coadjoint orbit.
This is essentially a rephrasing of the above statement that $G/{T_\lambda}$ is the homotopy
fiber of the inclusion of the moduli stack of Wilson line field configurations into the  moduli stack of bulk field configurations.

We may postcompose the two horizontal maps in this square with our two extended
Lagrangians, that for 1d and that for 3d Chern-Simons theory, to get the diagram
$$
 \raisebox{20pt}{
  \xymatrix{
     S^1 \ar[rr]^-{(\iota^*A)^g}_>{\ }="s" \ar[d]_\iota^>{\ }="t" 
	 && \Omega^1(-,\mathfrak{g})/\!/T
	 \ar[d]
	 \ar[r]^-{\langle \mathbf{\lambda}, -\rangle}
	 &
	 \mathbf{B}U(1)_{\mathrm{conn}}
	 \\
	 \Sigma_3
	 \ar[rr]^-{A}
	 &&
	 \mathbf{B}G_{\mathrm{conn}}
	 \ar[r]^-{\hat {\mathbf{c}}}
	 &
	 \mathbf{B}^3 U(1)_{\mathrm{conn}}\;.
	 \ar@{=>}_g "s"; "t"
  }
  }
$$
Therefore, writing $\mathbf{Fields}_{\mathrm{CS}+ \mathrm{W}}\left(S^1 \stackrel{\iota}{\hookrightarrow} \Sigma_3\right)$ for the moduli stack of field configurations for Chern-Simons theory with Wilson lines, we find
two action functionals as the composite top and left morphisms in the diagram
$$
 \raisebox{40pt}{
  \xymatrix{
    \mathbf{Fields}_{\mathrm{CS}+ \mathrm{W}}\left(S^1 \stackrel{\iota}{\hookrightarrow} \Sigma_3\right)
	\ar[r]
	\ar[d]
	&
	\mathbf{Maps}(\Sigma_3, \mathbf{B}G_{\mathrm{conn}})
	\ar[d]
	\ar[rrr]^-{\mathrm{hol}_{\Sigma_3}\mathbf{Maps}(\Sigma_3, \hat {\mathbf{c}})}
	&&&
	\underline{U}(1)
	\\
	\mathbf{Maps}(S^1, \Omega^1(-,\mathfrak{g})/\!/T_\lambda)
	\ar[r]
	\ar[dd]|{\mathrm{hol}_{S^1} \mathbf{Maps}(S^1, \langle \mathbf{\lambda},-\rangle)}
	&
	\mathbf{Maps}(S^1, \mathbf{B}G_{\mathrm{con}})
	\\
	\\
	\underline{U}(1)
  }
  }
$$
in $\mathbf{H}$, where the top left square is the homotopy pullback that characterizes
maps in $\mathbf{H}^{(\Delta^1)}$ in terms of maps in $\mathbf{H}$.
The product of these is the
action functional
$$
\xymatrix{
\mathbf{Fields}_{\mathrm{CS}+ \mathrm{W}}\left(S^1 \stackrel{\iota}{\hookrightarrow} \Sigma_3\right)\ar[r]& \mathbf{Maps}(\Sigma_3, \mathbf{B}^3 U(1)_{\mathrm{conn}})\times  \mathbf{Maps}(S^1, \mathbf{B} U(1)_{\mathrm{conn}})
\ar[d]\\
&\underline{U}(1)\times \underline{U}(1)\ar[r]^-{\cdot}
&\underline{U}(1)\;.
&
}
$$
where the rightmost arrow is the multiplication in $U(1)$.
Evaluated on a field configuration with components $(A,g)$ as just discussed, this 
is 
$$
  \exp\left(
    2 \pi i 
	\left(
	   \int_{\Sigma_3} \mathrm{CS}_3(A)
	   + 
       \int_{S^1} \langle \lambda, (\iota^*A)^g\rangle 
	\right)
  \right)
  \,.
$$
This is indeed the action functional for Chern-Simons theory with Wilson loop $\iota$
in the representation $R$ correspponding to the integral weight $\langle \lambda, -\rangle$ by
the Borel-Weil-Bott theorem, 
as reviewed for instance 
in Section 4 of \cite{Beasley}.

\medskip

Apart from being an elegant and concise repackaging of this well-known action functional
and the quantization conditions that go into it, the above reformulation in terms of stacks immediately 
leads to prequantum line bundles in Chern-Simons theory with Wilson loops. Namely, by considering the codimension 1 case, one finds the
the symplectic structure and
the canonical prequantization for the moduli stack of field configurations on surfaces with specified singularities at
specified punctures \cite{Witten}.
Moreover, this is just the first example in a general mechanism of (extended) action functionals
with defect and/or boundary insertions. Another example of the same mechanism is the 
gauge coupling action functional of the open string. This we discuss in section \ref{OpenStringGaugeCoupling}
below.

\paragraph{The anomaly-free gauge coupling of the open string}
\label{OpenStringGaugeCoupling}
\index{anomaly cancellation!open string on D-brane}
\index{anomaly cancellation!Freed-Witten-Kapustin anomaly}
\index{brane!D-brane!Freed-Witten-Kapustin anomaly}

As another example of the general phenomena of 
higher prequantum field theory, we close by briefly indicating how the
higher prequantum states of 3d Chern-Simons theory in codimension 2 reproduce the 
\emph{twisted Chan-Paton gauge bundles} of open string backgrounds, and how their
transgression to codimension 1 reproduces the cancellation of the 
Freed-Witten-Kapustin anomaly of the open string.
This section draws from \cite{FiorenzaSatiSchreiberCS}.

By the above, the Wess-Zumino-Witten gerbe 
$\mathbf{wzw} : {G \to \mathbf{B}^2 U(1)_{\mathrm{conn}}}$ as discussed in section
\ref{TheWZWGerbe} may be regarded as the \emph{prequantum 2-bundle} of Chern-Simons theory
in codimension 2 over the circle. Equivalently,  if we consider the WZW $\sigma$-model for the 
string on $G$ and take the limiting TQFT case obtained by 
sending the kinetic term to 0 while keeping
only the gauge coupling term in the action, then it is the extended Lagrangian of the string
$\sigma$-model: its transgression to the mapping space
out of a \emph{closed}  worldvolume $\Sigma_2$ of the string is the topological piece of the exponentiated
WZW $\sigma$-model action. For $\Sigma_2$ with boundary the situation is more interesting,
and this we discuss now.

The canonical representation of the 2-group $B U(1)$ is on the complex K-theory spectrum,
whose smooth (stacky) refinement is given by 
$\mathbf{B}U := \underset{\longrightarrow}{\lim}_n \mathbf{B}U(n)$ in $\mathbf{H}$. 
On any
component for fixed $n$ the action of the smooth 2-group $\mathbf{B}U(1)$ 
is exhibited by the long homotopy fiber sequence
$$
  \xymatrix{
    U(1) \longrightarrow U(n) \to \mathrm{PU}(n) \longrightarrow \mathbf{B}U(1)
	\longrightarrow \mathbf{B}U(n) \longrightarrow \mathbf{B}\mathrm{PU}(n)
    \ar[r]^-{\mathbf{dd}_n} & \mathbf{B}^2 U(1)	
  }
$$
in $\mathbf{H}$, in that $\mathbf{dd}_n$ is the universal $(\mathbf{B}U(n))$-fiber 2-bundle which is
associated by this action to the universal $(\mathbf{B}U(1))$-2-bundle.\footnote{
 The notion of $(\mathbf{B}U(n))$-fiber 2-bundle is equivalently that of 
 nonabelian $U(n)$-\emph{gerbes} in the original sense of Giraud, see \cite{NSSa}.
 Notice that for $n = 1$ this is more general than then notion of $U(1)$-bundle gerbe:
 a $G$-gerbe has structure 2-group $\mathbf{Aut}(\mathbf{B}G)$, but a $U(1)$-bundle gerbe
 has structure 2-group only in the left inlcusion of the fiber sequence
 $\mathbf{B}U(1) \hookrightarrow \mathbf{Aut}(\mathbf{B}U(1)) \to \mathbb{Z}_2$.
}
Using the general higher representation theory 
in $\mathbf{H}$ as developed in \cite{NSSa},  a local section
of the $(\mathbf{B}U(n))$-fiber prequantum 2-bundle which is $\mathbf{dd}_n$-associated to 
the prequantum 2-bundle $\mathbf{wzw}$, hence a local prequantum 2-state, is, equivalently, a map
$$
  \mathbf{\Psi} : 
    \mathbf{wzw}|_Q
	\longrightarrow
	\mathbf{dd}_n
$$
in the slice $\mathbf{H}_{/\mathbf{B}^2 U(1)}$, where $\iota_Q : Q \hookrightarrow G$ is some subspace.
Equivalently (compare with the general discussion in section \ref{FieldsInSlices}), this is a map 
$$
  (\mathbf{\Psi}, \mathbf{wzw}) 
  :
  \iota_Q 
  \longrightarrow
  \mathbf{dd}_n
$$
in $\mathbf{H}^{(\Delta^1)}$, hence a diagram in $\mathbf{H}$ of the form
$$
  \raisebox{20pt}{
  \xymatrix{
    Q \ar[rr]^-{\mathbf{\Psi}}_>{\ }="s" \ar@{^{(}->}[d]_{\iota_Q}^>{\ }="t" 
    && \mathbf{B}\mathrm{PU}(n) \ar[d]^{\mathbf{dd}_n}
	\\
	G \ar[rr]_-{\mathbf{wzw}} 
	&& \mathbf{B}^2 U(1)\;.
	\ar@{=>} "s"; "t"
  }
  }
$$
One finds 
that this equivalently modulates a unitary bundle on $Q$ which is \emph{twisted}
by the restriction of $\mathbf{wzw}$ to $Q$ as in twisted K-theory
(such a twisted bundle is also called a \emph{gerbe module} if $\mathbf{wzw}$ is thought of in terms
of bundle gerbes \cite{CBMMS}). So  
$$
  \mathbf{dd}_n \;\in \mathbf{H}_{/\mathbf{B}^2 U(1)}
$$ 
is the moduli stack for twisted rank-$n$ unitary bundles.
As with the other moduli stacks before,  one finds a differential refinement of this moduli stack,
which we write
$$
  (\mathbf{dd}_n)_{\mathrm{conn}}
  \;:\;
  (\mathbf{B}U(n)/\!/\mathbf{B}U(1))_{\mathrm{conn}}
  \to
  \mathbf{B}^2 U(1)_{\mathrm{conn}}
  \,,
$$ 
and which modulates twisted unitary bundles with twisted connections
(bundle gerbe modules with connection). Hence 
a differentially refined state is a map
$\widehat {\mathbf{\Psi}}
  \;:\;
    \mathbf{wzw}|_Q
	\to
	(\mathbf{dd}_n)_{\mathrm{conn}}
$
in $\mathbf{H}_{/\mathbf{B}^2 U(1)_{\mathrm{conn}}}$; and this is precisely a 
twisted gauge field on a D-brane $Q$ on which open strings in $G$ may end. Hence these are the
\emph{prequantum 2-states} of Chern-Simons theory in codimension 2. 
Precursors of this perspective of Chan-Paton bundles over D-branes 
as extended prequantum 2-states can be found in \cite{Schreiber07, RogersQuantization}.

Notice that by the above discussion, together the discussion in 
section \ref{FieldsInSlices}, an equivalence
$$
  \hat O 
  : \xymatrix{\mathbf{wzw} \ar[r]^\simeq & \mathbf{wzw}} 
$$
in $\mathbf{H}_{/\mathbf{B}^2 U(1)_{\mathrm{conn}}}$
has two different, but equivalent, important interpretations:
\begin{enumerate}
 \item it is an element of the \emph{quantomorphism 2-group} 
 (i.e. the possibly non-linear generalization of the Heisenberg 2-group) of 2-prequantum operators;
 \item it is a twist automorphism analogous to the generalized diffeomorphisms for the fields in 
 gravity.
\end{enumerate}
Moreover, such a transformation is locally a structure well familiar from the literature on D-branes:
it is locally (on some cover) given by a transformation of the B-field 
of the form $B \mapsto B + d_{\mathrm{dR} } a$ for a local 1-form $a$ 
(this is the \emph{Hamiltonian 1-form} in the interpretation of this transformation in
higher prequantum geometry)
and its prequantum operator action on prequantum 2-states, hence on Chan-Paton gauge fields
$\widehat{\mathbf{\Psi}} : \xymatrix{\mathbf{wzw} \ar[r] & (\mathbf{dd}_n)_{\mathrm{conn}}}$ (by precomposition) 
is given by shifting the connection on a twisted Chan-Paton bundle (locally) by this
local 1-form $a$. This local gauge transformation data
$$
    B  \mapsto B + d a\;, \qquad \qquad
	A  \mapsto A + a\;,
$$
is familiar from string theory and D-brane gauge theory 
(see e.g. \cite{PolchinskiBook}). The 2-prequantum operator action
$\Psi \mapsto \hat O \Psi$ which we see here is the fully globalized refinement of this
transformation.

\medskip

The map $\widehat{\mathbf{\Psi}} : (\iota_Q, \mathbf{wzw}) \to (\mathbf{dd}_n)_{\mathrm{conn}}$ above is 
the gauge-coupling part of the extended Lagrangian of the 
\emph{open} string on $G$ in the presence of a D-brane $Q \hookrightarrow G$. We indicate what this
means and how it works. Note that for all of the following the target space $G$ and background gauge field
$\mathbf{wzw}$ could be replaced by any
target space with any circle 2-bundle with connection on it.


The object $\iota_Q$ in $\mathbf{H}^{(\Delta^1)}$
is the target space for the open string. The worldvolume of that string is a smooth compact manifold
$\Sigma$ with boundary inclusion $\iota_{\partial \Sigma} : \partial \Sigma \to \Sigma$, also
regarded as an object in $\mathbf{H}^{(\Delta^1)}$. A field configuration of the 
string $\sigma$-model is then a map
$$
  \phi : \iota_\Sigma \to \iota_Q
$$
in $\mathbf{H}^{(\Delta^1)}$, hence a diagram
$$
  \xymatrix{
    \partial \Sigma \ar[rr] \ar@{^{(}->}[d]_{\iota_{\partial \Sigma}} 
    && Q \ar@{^{(}->}[d]^{\iota_Q}
	\\
	\Sigma \ar[rr]^\phi 
	&& G
  }
$$
in $\mathbf{H}$, hence a smooth function $\phi : \Sigma \to G$ subject to the constraint that the boundary
of $\Sigma$ lands on the D-brane $Q$. Postcomposition with the background gauge field 
$\widehat {\mathbf{\Psi}}$ 
yields the diagram
$$
  \xymatrix{
    \partial \Sigma \ar[rr] \ar@{^{(}->}[d]_{\iota_{\partial \Sigma}} 
	&& Q \ar@{^{(}->}[d]^{\iota_Q} \ar[r]^-{\widehat {\mathbf{\Psi}}} 
	& (\mathbf{B}U(n)/\!/U(1))_{\mathrm{conn}}
	\\
	\Sigma \ar[rr]^\phi && G \ar[r]_-{\mathbf{wzw}} & \mathbf{B}^2 U(1)_{\mathrm{conn}}\;.	
  }
$$
Comparison with the situation of Chern-Simons theory with Wilson lines in section \ref{WithWilsonLoops}
shows that the total action functional for the open string should be the product of the
fiber integration of the top composite morphism with that of the bottom composite morphisms.
 Hence that functional is the product
of the surface parallel transport of the $\mathbf{wzw}$ $B$-field over $\Sigma$ with the 
line holonomy of the twisted Chan-Paton bundle over $\partial \Sigma$. 

This is indeed again true, but for more subtle reasons this time,
since  the fiber integrations
here are \emph{twisted} (we discuss this in detail below in \ref{FiberIntegrationOfOrdinaryDifferentialCocycles}): 
since $\Sigma$ has a boundary, 
parallel transport over $\Sigma$ does not yield a function on the mapping space out of $\Sigma$, but
rather a section of the line bundle on the mapping space out of $\partial \Sigma$, pulled back to this
larger mapping space.

Furthermore, the connection on a twisted unitary bundle  does not quite 
have a well-defined traced holonomy in $\mathbb{C}$, but rather a well defined traced
holonomy up to a coherent twist. More precisely, the transgression of the 
WZW 2-connection to maps
out of the circle as in section \ref{PrequantumLineBundlesOnModuliStacks} 
fits into a diagram of moduli stacks in $\mathbf{H}$ of the form
$$
  \raisebox{20pt}{
  \xymatrix{
    \mathbf{Maps}(S^1, (\mathbf{B}U(n)/\!/\mathbf{B}U(1))_{\mathrm{conn}})
	\ar[dd]|{\mathbf{Maps}(S^1, (\mathbf{dd}_n)_{\mathrm{conn}})}
	\ar[rr]^-{\mathrm{tr}\,\mathrm{hol}_{S^1}}
	&&
	\underline{\mathbb{C}}/\!/\underline{U}(1)_{\mathrm{conn}}
	\ar[dd]
	\\
	\\
	\mathbf{Maps}(S^1 , \mathbf{B}^2 U(1)_{\mathrm{conn}})
	\ar[rr]^-{\mathrm{hol}_{S^1}}
	&&
	\mathbf{B}U(1)_{\mathrm{conn}}\;.
  }
  }
$$
This is a transgression-compatibility of the form that we have already seen in 
section \ref{TheWZWGerbe}. 

In summary, we obtain the transgression of the extended Lagrangian of the 
open string in the background of B-field and Chan-Paton bundles 
as the following pasting diagram of moduli stacks in $\mathbf{H}$
(all squares are filled with homotopy 2-cells, which are notationally suppressed for readability)

$$
\hspace{-1cm}
  \xymatrix{
    \mathbf{Fields}_{\mathrm{OpenString}}(\iota_{\partial \Sigma})
	\ar[rr]
	\ar[dd]
	&&
	\mathbf{Maps}(\Sigma,G)
	\ar[rr]^{\exp(2 \pi i \int_\Sigma [\Sigma, \mathbf{wzw}] )}
	\ar[dd]|{~~~\mathbf{Maps}(\iota_{\partial \Sigma}, G)}
	&&
	\underline{\mathbb{C}}/\!/\underline{U}(1)_{\mathrm{conn}}
	\ar[dddd]
	\\
	\\
	\mathbf{Maps}(S^1, Q)
	\ar[rr]^-{\mathbf{Maps}(S^1, \iota_Q)} 
	\ar[d]|-{~~\mathbf{Maps}(S^1, \widehat{\mathbf{\Psi}})}
	&&
	\mathbf{Maps}(S^1, G)
	\ar[dr]|{\mathbf{Maps}(S^1, \mathbf{wzw})}
	\\
	\mathbf{Maps}(S^1, (\mathbf{B}U(n)/\!/\mathbf{B}U(1))_{\mathrm{conn}})
	\ar[rrr]^{\mathbf{Maps}(S^1, (\mathbf{dd}_n)_{\mathrm{conn}})}
	\ar[d]|{\mathrm{tr}\,\mathrm{hol}_{S^1}}
	&&&
	\mathbf{Maps}(S^1, \mathbf{B}^2 U(1)_{\mathrm{conn}})
	\ar[dr]|{\mathrm{hol}_{S^1}}
	\\
	\underline{\mathbb{C}}/\!/\underline{U}(1)_{\mathrm{conn}}
	\ar[rrrr]
	&&
	&&
	\mathbf{B}U(1)_{\mathrm{conn}}
  }
$$
Here
\begin{itemize}
  \item 
    the top left square is the homotopy pullback square that computes the mapping stack
	$\mathbf{Maps}(\iota_{\partial \Sigma}, \iota_Q)$ in $\mathbf{H}^{(\Delta^1)}$, which here
	is simply the smooth space of string configurations $\Sigma \to G$ which are such that the string boundary
	lands on the D-brane $Q$;
  \item 
    the top right square is the twisted fiber integration of the $\mathbf{wzw}$ background 2-bundle
	with connection: this exhibits the parallel transport of the 2-form connection over the 
	worldvolume $\Sigma$ with boundary $S^1$ as a section of the pullback of the 
	transgression line bundle
	on loop space to the space of maps out of $\Sigma$;
  \item
    the bottom square is the above compatibility between the twisted traced holonomy of twisted unitary bundles
	and the trangression of their twisting 2-bundles.
\end{itemize}
The total diagram obtained this way exhibits a difference between two section of a single complex
line bundle on $\mathbf{Fields}_{\mathrm{OpenString}}(\iota_{\partial \Sigma})$
(at least one of them non-vanishing), hence a
map
$$
  \exp\left(2 \pi i \int_{\Sigma} [\Sigma, \mathbf{wzw}]  \right)
  \cdot
  \mathrm{tr}\,\mathrm{hol}_{S^1}([S^1, \widehat {\mathbf{\Psi}}])
  \;:\;
  \mathbf{Fields}_{\mathrm{OpenString}}(\iota_{\partial \Sigma})
  \longrightarrow
 \underline{ \mathbb{C}}
  \,.
$$
This is the well-defined action functional of the open string with endpoints on the D-brane $Q \hookrightarrow G$,
charged under the background $\mathbf{wzw}$ B-field and under the twisted Chan-Paton gauge bundle
$\widehat {\Psi}$. 

Unwinding the definitions, one finds that this phenomenon is precisely the 
twisted-bundle-part, due to Kapustin \cite{Kapustin}, 
of the Freed-Witten anomaly cancellation for open strings on D-branes, 
hence is the Freed-Witten-Kapustin anomaly cancellation
mechanism either for the open bosonic string or else for the open type II superstring on 
$\mathrm{Spin}^c$-branes. Notice how in the traditional discussion the existence of
twisted bundles on the D-brane is identified just as \emph{some} construction
that happens to cancel the B-field anomaly. Here, in the perspective of extended quantization, we 
see that this choice follows uniquely from the general theory of extended prequantization, once
we recognize that $\mathbf{dd}_n$ above is (the universal associated 2-bundle induced by) 
the canonical representation of the circle 2-group $\mathbf{B}U(1)$, just as in one codimension up
$\mathbb{C}$ is the canonical representation of the circle 1-group $U(1)$.

\newpage

\section{Homotopy type theory}
\label{HomotopyTypeTheory}

We discuss here aspects of \emph{homotopy type theory}, the theory
of locally cartesian closed \emph{$\infty$-categories} and of \emph{$\infty$-toposes}, 
that we need in the following.
Much of this is a review of material available in the 
literature, we just add some facts that we will need and
for which we did not find a citation. 
The reader at least roughly familiar with this theory
can skip ahead to our main contribution, the
discussion of \emph{cohesive $\infty$-toposes} in 
\ref{GeneralAbstractTheory}. We will refer back to these sections
here as needed.

\subsection{$\infty$-Categories}
\label{InfinityCategories}

The natural joint generalization of the notion of \emph{category}
and of \emph{homotopy type} is that of \emph{$\infty$-category}:
a collection of objects, such that between any ordered pair of them
there is a homotopy type of morphisms. 
We briefly survey key definitions and properties in the theory
of $\infty$-categories.

\medskip

\begin{itemize}
  \item \ref{DependentHomotopyTypeTheory} -- Dependent homotopy types and Locally cartesian closed $\infty$-categories;
  \item \ref{PresentationOfInfinityCategoriesBySimplicialSets} -- Presentation by simplicial sets;
  \item \ref{InfinityCategoriesBySimplicialCategories} -- Presentation by simplicially enriched categories.
\end{itemize}

\subsubsection{Dependent homotopy types and Locally cartesian closed $\infty$-categories}
\label{DependentHomotopyTypeTheory}

For the most basic notions of category theory see the
first pages of \cite{Moerdijk} or A.1 in \cite{Lurie}. 
\begin{definition}
  A category $\mathcal{C}$ is called \emph{cartesian closed} if it
  has Cartesian products $X \times Y$ of all objects $X,Y \in \mathcal{C}$
  and if there is for each $X \in \mathcal{C}$ a mapping space functor
  $[X,-] : \mathcal{C} \longrightarrow \mathcal{C}$,
  characterized by the fact that there is a bijection of hom-sets
  $$
    \mathcal{C}(X \times A, Y)
	\simeq
	\mathcal{C}(A, [X,Z])
  $$
  natural in the objects $A,X,Y \in \mathcal{C}$.
  A category $\mathcal{C}$ is called \emph{locally cartesian closed}
  if for each object $X \in \mathcal{C}$ the slice category
  $\mathcal{C}_{/X}$ is a cartesian closed category. 
  \label{LocallyCartesianClosedCategory}
\end{definition}
The main example of locally cartesian closed categories of interest here are
toposes, to which we come below in def. \ref{ToposByLocalizationAtCoverage}.
It is useful to equivalently re-express local cartesian closure in terms of 
\emph{base change}:
\begin{proposition}
  If $\mathcal{C}$ is a locally cartesian closed category, def. \ref{LocallyCartesianClosedCategory},
  then for $f : X \longrightarrow Y$ any morphism in $\mathcal{C}$ there
  exists an adjoint triple of functors  between the slice categories over $X$ and $Y$
  (called \emph{base change} functors)
   $$
     \xymatrix{
	   \mathcal{C}_{/\Gamma_1}
	   \ar@<+8pt>@{->}[rr]|-{f_!}
	   \ar@{<-}[rr]|-{f^\ast}
	   \ar@<-8pt>@{->}[rr]|-{f_*}
	   &&
	   \mathcal{C}_{/\Gamma_2}
	 }
	 \,,
   $$
   where $f^\ast$ is given by pullback along $f$, $f_!$ is its left adjoint and
   $f_*$ its right adjoint. Conversely, if a category
   $\mathcal{C}$ has pullbacks and has for every morphism $f$ a left and right adjoint
   $f_!$ and $f_*$ to the pullback functor $f^\ast$, then it is locally cartesian closed.
   \label{LocalCartesianClosureViaBaseChange}
\end{proposition}
It turns out that base change may usefully be captured syntactically such as
to constitute a flavor of formal logic called \emph{constructive set theory}
or \emph{type theory} \cite{MartinLoef}:
\begin{definition}
Given a locally cartesian closed category $\mathcal{C}$, 
one says equivalently that 
\begin{itemize}
  \item its internal logic is a \emph{dependent type theory};
 \item it provides \emph{categorical semantics} for dependent type theory
\end{itemize}
as follows:
\begin{itemize}
 \item the objects of $\mathcal{C}$ are called the \emph{types};
 \item the objects in a slice $\mathcal{C}_{/\Gamma}$ are called the types
   \emph{in context} $\Gamma$ or \emph{dependent on} $\Gamma$, denoted
   $$
     \Gamma \;\vdash\; X : \mathrm{Type}
   $$
 \item 
   a morphism $\ast \to X$ (from the terminal object into any
   object $X$) in a slice $\mathcal{C}_{\Gamma}$ is called 
   \emph{a term of type $X$} in context $\Gamma$, and denoted
   $$
     \Gamma \;\vdash\; x : X 
   $$
   or more explicitly
   $$
     a : \Gamma \;\vdash\; x(a) : X(a);
   $$   
 \item
   given a morphism $f : \Gamma_1 \longrightarrow \Gamma_2$ in $\mathcal{C}$
   with its induced base change adjoint triple of functors between slice categories
   from prop. \ref{LocalCartesianClosureViaBaseChange}
   $$
     \xymatrix{
	   \mathcal{C}_{/\Gamma_1}
	   \ar@<+8pt>@{->}[rr]|-{f_!}
	   \ar@{<-}[rr]|-{f^\ast}
	   \ar@<-8pt>@{->}[rr]|-{f_*}
	   &&
	   \mathcal{C}_{/\Gamma_2}
	 }
   $$
   then
   \begin{itemize}
     \item given a morphism $(\ast \to X)$ in $\mathcal{C}_{/\Gamma_2}$, 
	 hence a term $\Gamma_2 \;\vdash\; x : X$, then its pullback by $f^\ast$
	 is denoted by \emph{substitution} of variables
	 $$
	   a : \Gamma_1 \;\vdash\; x(f(a)) : X(f(a))
	   \,,
	 $$
     \item given an object $X \in \mathcal{C}_{\Gamma_1}$
	 its image $f_!(X) \in \mathcal{C}_{/\Gamma_2}$ is called the
	 \emph{dependent sum} of $X$ along $f$ and is denoted as
	 $$
	   \Gamma_2 \;\vdash\; \underset{f}{\sum} X \;:\; \mathrm{Type}
	   \,,
	 $$
	 \item
     given an object $X \in \mathcal{C}_{\Gamma_1}$
	 its image $f_*(X) \in \mathcal{C}_{/\Gamma_2}$ is called the
	 \emph{dependent product} of $X$ along $f$ and is denoted as
	 $$
	   \Gamma_2 \;\vdash\; \underset{f}{\prod} X \;:\; \mathrm{Type}
	   \,,
	 $$
   \end{itemize}
   \item the universal property of the adjoints $(f_! \dashv f^\ast \dashv f_\ast)$
   translates to evident rules for introducing and for transforming terms of these
   dependent sum/product types, called \emph{term introduction} and \emph{term elimination}
   rules.
\end{itemize}
\label{DependentTypeTheoryLanguage}
\end{definition}
We consider bundles and base change in more detail below in 
\ref{Bundles}.

When this syntactic translation is properly formalized, it yields an equivalent description
of locally cartesian closed categories:
\begin{proposition}[\cite{Seely, ClairambaultDybjer}]
  There is an equivalence of $2$-categories between locally cartesian closed
  categories and dependent type theories.
\end{proposition}
\begin{remark}
  Given any object $X \in \mathcal{C}_{/\Gamma}$, its
  diagonal $X \longrightarrow X \times X$ regarded as an object of
  $\mathcal{C}_{/(\Gamma \times X\times X)}$ serves as the \emph{identity type}
  of $X$, denoted
  $$
    \Gamma, (x_1,x_2) : X \times X \;\vdash\; (x_1 = x_2) : \mathrm{Type}
	\,.
  $$
  Namely given two terms $x_1, x_2 : X$, then a term
  $\Gamma \;\vdash\; p : (x_1 = x_2)$ is as a morphism in $\mathcal{C}$ an element
  on the diagonal of $X \times X$ and in the type theory is a \emph{proof of equality}
  of $x_1$ and $x_2$. If there is such a proof of equality then it is unique,
  since the diagonal is always a monomorphism.
  
  But consider now the case that $\mathcal{C}$ in addition carries the structure of 
  a \emph{model category} (see A.2 in \cite{Lurie} for a review).
  Then there is for each $X$ a path space object $X^I \longrightarrow X \times X$. 
  Using this as the categorical semantics of identity types, instead of the
  plain diagonal $X \longrightarrow X \times X$, means to 
  make identity behave instead like \emph{higher gauge equivalence} in physics: 
  there are then possibly many equivalences between two terms of a given type,
  and many equivalences between equivalences, and so on.
  If $\mathcal{C}$ is moreover right proper as a model category and such that
  its cofibrations are precisely its monomorphisms, 
  then there exists a variant of the dependent type theory of remark \ref{DependentTypeTheoryLanguage}
  reflecting these homotopy-theoretic identity types. This is called dependent type theory
  \emph{with intensional identity types} or, more recently, 
  \emph{homotopy type theory} \cite{HoTT}.
  At the same time, such a model category is a presentation for the 
  homotopy-theoretic analogy of a locally cartesian closed category:
  a \emph{locally cartesian closed $(\infty,1)$-category} (see A.3 of \cite{Lurie}).
  \label{IdentityTypes}
\end{remark}
The following was maybe first explicitly suggested by \cite{Joyal}.
A proof of the technical details involved appeared in \cite{CisinskiShulman}.
\begin{proposition}
  Up to equivalence, the internal type theory of 
  a locally Cartesian closed $(\infty,1)$-category is homotopy type theory
  (without necessarily univalence) and conversely 
  homotopy type theory (without necessarily univalence) has
  categorical semantics in locally cartesian closed $(\infty,1)$-categories.
\end{proposition}

We now turn to description such $\infty$-categories ``externally'' in terms of
simplicial sets and categories enriched over simplicial sets. We briefly come
back to the ``inernal'' perspective of homotopy type theory below in \ref{InternalCohesion}.

\subsubsection{Presentation by simplicial sets}
\label{PresentationOfInfinityCategoriesBySimplicialSets}

\begin{definition}
  An \emph{$\infty$-category} 
  is a simplicial set $C$ such that all
  horns $\Lambda^i[n] \to C$ that are \emph{inner}, in that
  $0 < i < n$, have an extension to a simplex $\Delta[n] \to C$.
  
  A vertex $c \in C_0$ is an \emph{object}, an edge $f \in C_1$ is
  a \emph{morphism} in $C$.
  
  An \emph{$\infty$-functor} $f : C \to D$ between $\infty$-categories
  $C$ and $D$ is a morphism of the underlying simplicial sets.
  \label{InfinityCategory}
\end{definition}
This definition is due \cite{Joyal}. 
\begin{remark}
For $C$ an $\infty$-category, we think of $C_0$ as its collection of
\emph{objects}, and of $C_1$ as its collection of \emph{morphisms}
and generally of $C_k$ as the collection of 
\emph{$k$-morphisms}. 
The inner horn filling property can be seen to encode 
the existence of composites of $k$-morphisms, well defined up
to coherent $(k+1)$-morphisms.
It also implies that for $k > 1$ these $k$-morphisms are
invertible, up to higher morphisms. 
To emphasize this fact one also says that $C$ is an
\emph{$(\infty,1)$-category}. 
(More generally an \emph{$(\infty,n)$-category}
would have $k$ morphisms for all $k$ such that for $k > n$ these are 
equivalences.)
\end{remark}

The power of the notion of $\infty$-categories is that 
it supports the higher analogs of all the crucial
facts of ordinary category
theory. This is a useful meta-theorem to keep in mind, 
originally emphasized by Andr{\'e} Joyal and Charles Rezk.
\begin{fact} 
In general
\begin{itemize}
\item $\infty$-Category theory parallels category theory;
\item $\infty$-Topos theory parallels topos theory.
\end{itemize}
\end{fact}
More precisely, essentially all the standard constructions 
and theorems have their $\infty$-analogs 
if only we replace \emph{isomorphism} between objects and equalities between morphisms  
consistently by \emph{equivalence}s and coherent 
higher equivalences in an $\infty$-category. 
\begin{proposition}
  For $C$ and $D$ two $\infty$-categories, the internal hom of 
  simplicial sets
  $\mathrm{sSet}(C,D) \in \mathrm{sSet}$ is an $\infty$-category.
\end{proposition}
\begin{definition}
  We write $\mathrm{Func}(C,D)$ for this $\infty$-category
  and speak of the \emph{$\infty$-category of $\infty$-functors}
  between $C$ and $D$.
\end{definition}
\begin{remark}
  The objects of $\mathrm{Func}(C,D)$ are indeed the
  $\infty$-functors from def. \ref{InfinityCategory}.
  The morphisms may be called 
  \emph{$\infty$-natural transformations}.
\end{remark}
\begin{definition}
  The \emph{opposite} $C^{\mathrm{op}}$ of an $\infty$-category
  $C$ is the $\infty$-category corresponding to the 
  opposite of the corresponding $\mathrm{sSet}$-category.
\end{definition}
\begin{definition}
  Let $\mathrm{KanCplx} \subset \mathrm{sSet}$ be the 
  full subcategory of $\mathrm{sSet}$ on the Kan complexes,
  regarded naturally as an $\mathrm{sSet}$-enriched category,
  in fact a Kan-complex enriched category.
  Below in $\ref{InfinityCategoriesBySimplicialCategories}$ we recall
  the \emph{homotopy coherent nerve} construction $N_h$ that sends a 
  Kan-complex enriched category to an $\infty$-category.
  
  We say that
  $$
    \infty \mathrm{Grpd} := N_h \mathrm{KanCplx}
  $$
  is the \emph{$\infty$-category of $\infty$-groupoids}.
\end{definition}
\begin{definition}
  For $C$ an $\infty$-category, we write
  $$
    \mathrm{PSh}_\infty(C) :=
    \mathrm{Func}(C^{\mathrm{op}}, \infty \mathrm{Grpd})
  $$
  and speak of the \emph{$\infty$-category of $\infty$-presheaves}
  on $C$.
  \index{category theory!$\infty$-presheaves}
  \label{InfinityPresheaves}
\end{definition}
The following is the $\infty$-category theory analog of the
Yoneda lemma.
\begin{proposition}
  \index{category theory!$\infty$-Yoneda lemma}
  For $C$ an $\infty$-category, $U \in C$ any object,
  $j(U) \simeq C(-,U) : C^{\mathrm{op}} \to \infty \mathrm{Grpd}$
  an $\infty$-presheaf represented by $U$ we have for every
  $\infty$-presheaf $F \in \mathrm{PSh}_\infty(C)$ a 
  natural equivalence of $\infty$-groupoids
  $$
    \mathrm{PSh}_\infty(C)(j(U), F)
    \simeq
    F(U)
    \,.
  $$
  \label{InfinityYonedaLemma}
\end{proposition}
From this derives 
a notion of $\infty$-limits and of adjoint $\infty$-functors 
and they satisfy the expected properties. This
we discuss below in \ref{LimitsAndColimits}.

\subsubsection{Presentation by simplicially enriched categories}
\label{InfinityCategoriesBySimplicialCategories}

A convenient way of handling $\infty$-categories is 
via $\mathrm{sSet}$-enriched categories:
categories which for each ordered pair of objects has
not just a set of morphisms, but a simplicial set of morphisms 
(see \cite{Kelly} for enriched category theory in general and
section A of \cite{Lurie} for $\mathrm{sSet}$-enriched category
theory in the context of $\infty$-category theory in particular):
\begin{proposition}
  \label{HomotopyCoherentNerve}
  There exists an adjunction between simplicially enriched 
  categories and simplicial sets
  $$
    \xymatrix{
      (\vert - \vert \dashv N_h) 
       : 
      \mathrm{sSet}\mathrm{Cat}
      \ar@<-4pt>[r]_<<<<<{N_h}  
      \ar@{<-}@<+4pt>[r]^<<<<<{\vert-\vert}  
      &
      \mathrm{sSet}
    }
  $$
  such that
  \begin{itemize}
    \item 
  if $S \in \mathrm{sSet}\mathrm{Cat}$ is such that for
  all objects $X, Y \in S$ the simplicial set $S(X,Y)$ is a Kan
  complex, then $N_h(S)$ is an $\infty$-category;
   \item
     the unit of the adjunction is an 
     equivalence of $\infty$-categories
     (see def. \ref{HomotopyCategory} below).
  \end{itemize}
\end{proposition}
This is for instance prop. 1.1.5.10 in \cite{Lurie}.
\begin{remark}
  In particular, for $C$ an ordinary category, regarded as
  an $\mathrm{sSet}$-category with simplicially constant
  hom-objects, $N_h C$ is an $\infty$-category.
  A functor $C \to D$ is precisely an $\infty$-functor
  $N_h C \to N_h D$. In this and similar cases
  we shall often notationally suppress the $N_h$-operation.
  This is justified by the following statements.
\end{remark}
\begin{definition}
  \label{HomotopyCategory}
  For $C$ an $\infty$-category, its \emph{homotopy category}
  $\mathrm{Ho}(C)$ (or $\mathrm{Ho}_C$) is the ordinary category obtained from
  $\vert C \vert$ by taking connected components of all 
  simplicial hom-sets:
  $$
    \mathrm{Ho}_C(X,Y) = \pi_0 (\vert C \vert (X,Y))
    \,.
  $$

  A morphism $f \in C_1$ is called an \emph{equivalence}
  if its image in $\mathrm{Ho}(C)$ is an isomorphism.
  Two objects in $C$ connected by an equivalence are called
  \emph{equivalent objects}.
\end{definition}
\begin{definition}  
  An $\infty$-functor $F : C \to D$ is called an
  \emph{equivalence of $\infty$-categories} if 
  \begin{enumerate}
    \item
      It is \emph{essentially sujective} in that the induced
      functor $\mathrm{Ho}(f) : \mathrm{Ho}(C) \to \mathrm{Ho}(D)$
      is essentially surjective;
    \item 
      and it is \emph{full and faithful} in that for all objects 
      $X,Y$ the induced morphism $ f_{X,Y} : \vert C \vert (X,Y) \to 
      \vert D \vert (X,Y)$ is a weak homotopy equivalence of 
      simplicial sets.
  \end{enumerate}
\end{definition}
For $C$ an $\infty$-category and $X$, $Y$ two of its objects, 
we write
$$
  C(X,Y) := \vert C\vert(X,Y)
$$
and call this Kan complex the \emph{hom-$\infty$-groupoid} of
$C$ from $X$ to $Y$.

The following assertion guarantees that  
$\mathrm{sSet}$-categories
are indeed a faithful presentation of 
$\infty$-categories.
\begin{proposition}
  For every $\infty$-category $C$ the 
  unit of the $(\vert-\vert \dashv N_h)$-adjunction from
  prop. \ref{HomotopyCoherentNerve} is an
  equivalence of $\infty$-categories
  $$
    C \stackrel{\simeq}{\to} N_h \vert C\vert
    \,.
  $$
\end{proposition}
This is for instance theorem 1.1.5.13 
together with remark 1.1.5.17 in \cite{Lurie}.
\begin{definition}
  An $\infty$-groupoid is an $\infty$-category in which 
  all morphisms are equivalences.
\end{definition}
\begin{proposition}
  $\infty$-groupoids in this sense are precisely
  Kan complexes.
\end{proposition}
This is due to \cite{JoyalGroupoids}. 
See also prop. 1.2.5.1 in \cite{Lurie}.

A convenient way of constructing 
$\infty$-categories in terms of $\mathrm{sSet}$-categories
is via categories with weak equivalences. 
\begin{definition}
  A \emph{category with weak equivalences} $(C,W)$ is 
  a category $C$ equipped with a subcategory $W \subset C$
  which contains all objects of $C$ and
  such that $W$ satsifies the \emph{2-out-of-3 property}:
  for every commuting triangle
  $$
    \xymatrix{
       & y \ar[dr]
       \\
       x \ar[ur] \ar[rr] && z
    }
  $$
  in $C$ with two of the three morphisms in $W$, also 
  the third one is in $W$.
\end{definition}
\begin{definition}  
  \label{SimplicialLocalization}
  The \emph{simplicial localization} of 
  a category with weak equivalences $(C,W)$ is the 
  $\mathrm{sSet}$-category 
  $$
    L_W C \in \mathrm{sSet}\mathrm{Cat}
  $$
  (or $L C$ for short, when $W$ is understood) 
  given as follows: the objects are those of $C$; and 
  for $X,Y \in C$ two objects, the simplicial hom-set
  $L C(X,Y)$ is the inductive limit over $n \in \mathbb{N}$ of
  the nerves of the following categories:
  \begin{itemize}
    \item 
      objects are equivalence classes of 
      zig-zags of length $n$ of morphisms
      $$
        \xymatrix{
          X \ar@{<-}[r]^\simeq & K_1 \ar[r] & K_2
          \ar@{<-}[r]^\simeq
          &
          \cdots
          \ar[r] & Y
        }
      $$
      in $C$, such that the left-pointing morphisms are in $W$;
     \item
       morphisms are equivalence classes of transformations of such 
       zig-zags
       $$
         \xymatrix{
            & K_1 \ar[dd]^\simeq 
           \ar[dl]_\simeq \ar[r] 
             & K_2 \ar[dd]^\simeq \ar@{<-}[r]^\simeq & \cdots & \ar[dr]
            \\
            X &&&&& Y
            \\
            & K'_1 \ar[ul]_\simeq \ar[r] & K'_2 \ar@{<-}[r]^\simeq 
            & \cdots
            &\ar[ur]            
         }
         \,,
       $$
       such that the vertical morphisms are in $W$;
       \item
         subject to the equivalence relation that identifies 
         two such (transformations of) zig-zags if one is obtained
         from the other by discarding identity morphisms and 
         then composing
         consecutive morphisms.
  \end{itemize}
\end{definition}
This simplicial ``hammock localization'' is due to 
\cite{DwyerKanComputations}.
\begin{proposition}
  Let $(C,W)$ be a category with weak equivalences and
  $L C$ be its simplicial localization. Then its homotopy category
  in the sense of def. \ref{HomotopyCategory}
  is equivalent to the ordinary homotopy category 
  $\mathrm{Ho}(C,W)$ 
  (the category obtained from $C$ by 
   universally inverting the morphisms in $W$):
  $$
    \mathrm{Ho}L_W C \simeq \mathrm{Ho}(C,W)
    \,.
  $$
\end{proposition}

A convenient way of controlling simplicial localizations
is via $\mathrm{sSet}_{\mathrm{Quillen}}$-enriched model category structures
(see section A.2 of \cite{Lurie} for a good discussion of 
all related issues).
\begin{definition}
  \label{Quillen}
  A \emph{model category} is a category with weak equivalences
  $(C,W)$ that has all limits and colimits and is equipped with
  two further classes of morphisms, 
  $\mathrm{Fib}, \mathrm{Cof} \subset \mathrm{Mor}(C)$ 
  -- the \emph{fibrations} and \emph{cofibrations} -- such that
  $(\mathrm{Cof}, \mathrm{Fib}\cap W)$ and
  $(\mathrm{Cof} \cap W, \mathrm{Fib})$ are two weak
  factorization systems on $C$. Here the elements in $\mathrm{Fib}\cap W$
  are called \emph{acyclic fibrations} and those in $\mathrm{Cof} \cap W$
  are called $\emph{acyclic cofibrations}$.
    An object $X \in C$ is called \emph{cofibrant} if the canonical 
  morphism $\emptyset \to X$ is a cofibration. 
  It is called \emph{fibrant} if the canonical morphism 
  $X \to *$ is a fibration.

  A \emph{Quillen adjunction} between two model categories is
  a pair of adjoint functors between the underlying categories, 
  such that the right adjoint preserves
  cofibrations and acyclic cofibrations, which equivalently
  means that the left adjoint preserves cofibrations and acyclic 
  cofibrations.
  \label{model category}
\end{definition}
\begin{remark}
  The axioms on model categories directly imply that
  every object is weakly equivalent to a fibrant object,
  and to a cofibrant objects and in fact to a fibrant
  and cofibrant objects.
\end{remark}  
\begin{example}
  The category of simplicial sets carries a model category structure, 
  here denoted $\mathrm{sSet}_{\mathrm{Quillen}}$, whose
  weak equivalences are the weak homotopy equivalences, 
  cofibrations are the monomorphisms, and fibrations and the Kan fibrations.
\end{example}
\begin{definition}
  \label{LeftQuillenBifunctor}
  Let $A,B,C$ be model categories. Then a functor
  $$
     F : A \times B \to C
  $$
  is a \emph{left Quillen bifunctor} if 
  \begin{enumerate}
    \item it preserves colimits separately in each argument;
	\item for $i : a \to a'$ and $j : b \to b'$ two cofibrations in $A$ and in $B$, 
	respectively, the canonical induced morphism
	$$
	  F(a', b) \coprod_{F(a,b)} F(a,b') \to F(a',b')
	$$
	is a cofibration in $C$ and is in addition a weak equivalence if $i$ or $j$ is.
  \end{enumerate}
\end{definition}
\begin{remark}
  \label{QuillenBifunctorWithOneArgumentFixedAndCofibrant} 
  In particular, for $F : A \times B \to C$ a left Quillen 
  bifunctor,  if $a \in A$ is cofibrant then 
  $$
    F(a,-) : B \to C
  $$
  is an ordinary left Quillen functor if $F$ is a left Quillen bifunctor, as is
  $$
    F(-,b) : A \to C
  $$
  for $b$ cofibrant.
\end{remark}
\begin{definition}
  A \emph{monoidal model category} is a category equipped both with the
  structure of a model category and with the structure of a monoidal category, such 
  that the tensor product functor of the monoidal structure is a left Quillen
  bifunctor, def. \ref{LeftQuillenBifunctor}, with respect to the 
  model category structure.
\end{definition}
\begin{example}
  The model category $\mathrm{sSet}_{\mathrm{Quillen}}$ is 
  a monoidal model category with respect to its Cartesian monoidal structure.
\end{example}
\begin{definition}
  For $\mathcal{V}$ a monoidal model category,
  an $\mathcal{V}$-\emph{enriched model category}
  is a model category equipped with 
  the structure of
  an $\mathcal{V}$-enriched category which is also
  $\mathcal{V}$-tensored and -cotensored, such that the 
  $\mathcal{V}$-tensoring functor
  is a left Quillen bifunctor, def. \ref{LeftQuillenBifunctor}.
\end{definition}
\begin{remark}
  An $\mathrm{sSet}_{\mathrm{Quillen}}$-enriched model category
  is often called a \emph{simplicial model category}. Notice that, while
  entirely standard, this use of terminology is imprecise:
  first, not every simplicial object in categories is a $\mathrm{sSet}$-enriched
  category, and second, there are other and inequivalent model category structure on 
  $\mathrm{sSet}$ that make it a monoidal model category with respect to its
  Cartesian monoidal structure. 
  \label{SimplicialModelCategory}
\end{remark}
\begin{definition}
  For $C$ an ($\mathrm{sSet}_{\mathrm{Quillen}}$-enriched) model category write
  $$
    C^\circ \in \mathrm{sSet}\mathrm{Cat}
  $$
  for the full $\mathrm{sSet}$-subcategory on the fibrant and 
  cofibrant objects.
\end{definition}
\begin{proposition}
  \label{FullSubcatOnFibrantAndCofibrantObjects}
  Let $C$ be an $\mathrm{sSet}_{\mathrm{Quillen}}$-enriched model 
  category. Then there is an
  equivalence of $\infty$-categories
  $$
    C^\circ \simeq L C
    \,.
  $$
\end{proposition}
This is corollary 4.7 with prop. 4.8 in 
\cite{DwyerKanFunctionComplexes}. 
\begin{proposition}
  \label{DerivedHomSpaceInSimplicialModelCategory}
  The hom-$\infty$-groupoids $(N_h C^\circ)(X,Y)$ 
  are already correctly given by the hom-objects in 
  $C$ from a cofibrant to a fibrant representative of 
  the weak equivalence class of $X$ and $Y$, respectively.
\end{proposition}
In this way $\mathrm{sSet}_{\mathrm{Quillen}}$-enriched
model category structures constitute particularly convenient
extra structure on a category with weak equivalences for
constructing the corresponding $\infty$-category. 

\medskip

In terms of the presentation of $\infty$-categories
by simplicial categories, \ref{InfinityCategoriesBySimplicialCategories},
adjoint $\infty$-functors 
are presented by \emph{simplicial Quillen adjunctions}, def. \ref{Quillen}, between 
simplicial model categories: the restriction of a simplicial Quillen adjunction to 
fibrant-cofibrant objects is the $\mathrm{sSet}$-enriched functor that presents the 
$\infty$-derived functor under the model of $\infty$-categories by 
simplicially enriched categories.
\begin{proposition}  
\label{InfAdjBySimpAdj}
\index{category theory!$\infty$-adjunctions from derived Quillen adjunctions}
Let $C$ and $D$ be simplicial model categories and let
$$
  (L \dashv R) : 
  \xymatrix{
     C 
	  \ar@{<-}@<+3pt>[r]^{L}
	  \ar@<-3pt>[r]_{R}
	  &
	 D
  }
$$
be an $\mathrm{sSet}$-enriched adjunction whose underlying ordinary adjunction is 
a Quillen adjunction. Let $C^\circ$ and $D^\circ$ be the $\infty$-categories presented 
by $C$ and $D$ (the Kan complex-enriched full $\mathrm{sSet}$-subcategories on fibrant-cofibrant objects). Then the Quillen adjunction lifts to a pair of adjoint $\infty$-functors
$$
  (\mathbb{L}L \dashv \mathbb{R}R) : 
  \xymatrix{
     C^\circ
	  \ar@{<-}@<+3pt>[r]^{}
	  \ar@<-3pt>[r]_{}
	  &
	 D^\circ
  }
$$  
On the decategorified level of the homotopy categories these are the total left and right derived functors, respectively, of $L$ and $R$.
\end{proposition}
This is \cite{Lurie}, prop 5.2.4.6.

The following proposition states conditions under which a simplicial Quillen adjunction 
may be detected already from knowing of the right adjoint only that it preserves 
fibrant objects (instead of all fibrations).
\begin{proposition} \label{AdjRecognition}
If $C$ and $D$ are simplicial model categories and $D$ is a left proper model category, 
then for an $\mathrm{sSet}$-enriched adjunction
$$
  (L \dashv R) : 
  \xymatrix{
     C 
	  \ar@{<-}@<+3pt>[r]^{}
	  \ar@<-3pt>[r]_{}
	  &
	 D
  }
$$
to be a Quillen adjunction it is already sufficient that $L$ preserves 
cofibrations and $R$ preserves fibrant objects.
\end{proposition}
This appears as \cite{Lurie}, cor. A.3.7.2.

We will use this for finding simplicial Quillen adjunctions into left Bousfield localizations of left proper model categories: the left Bousfield localization preserves the left properness, and the fibrant objects in the Bousfield localized structure have a good characterization: they are the fibrant objects in the original model structure that are also local objects with respect to the set of morphisms at which one localizes.
Therefore for $D$ the left Bousfield localization of a simplicial left proper model category $E$ at a class $S$ of morphisms, for checking the Quillen adjunction property of $(L \dashv R)$ 
it is sufficient to check that $L$ preserves cofibrations, and that $R$ takes fibrant objects 
$c$ of $C$ to such fibrant objects of $E$ that have the property 
that for all $f \in S$ 
the derived hom-space map $\mathbb{R}\mathrm{Hom}(f,R(c))$ is a weak equivalence.

\subsection{$\infty$-Toposes}
\label{InfinityToposes}
\index{topos theory}

The natural context for discussing the geometry of spaces that are locally
modeled on test spaces in some category $C$ 
(and equipped with a notion of coverings) is the category called the 
\emph{sheaf topos}
$\mathrm{Sh}(C)$  over $C$ \cite{Johnstone}. 
Analogously, the natural context for discussing the \emph{higher}
geometry of such spaces is the $\infty$-category called the
\emph{$\infty$-sheaf topos} $\mathbf{H} = \mathrm{Sh}_\infty(C)$.

The theory of $\infty$-toposes has been given a
general abstract formulation in \cite{Lurie},
using the $\infty$-category theory introduced by \cite{Joyal} 
and building on \cite{Rezk} and \cite{ToenVezzosiTopos}.
One of the central results proven there is that 
the old homotopy theory of simplicial presheaves,
originating around \cite{Brown} and developed
notably in \cite{Jardine} and \cite{Dugger},
is indeed a \emph{presentation} of $\infty$-topos theory.

\medskip

\begin{itemize}
  \item \ref{InfinityToposesGeneralAbstract} -- Abstract $\infty$-category theoretic characterization
  \item \ref{SyntaxOfHomotopyTypeTheory} -- Homotopy type theory with type universes
  \item \ref{InfinityToposPresentation} -- Presentation by simplicial (pre-)sheaves
  \item \ref{PresentationBySimplicialObjectsInTheSite} -- Presentation by simplicial objects in the site
  \item \ref{SheavesDescent} -- $\infty$-Sheaves and descent
  \item \ref{SheafAndNonabelianDoldKan} -- $\infty$-Sheaves with values in chain complexes
\end{itemize}

\subsubsection{Abstract $\infty$-category theoretic characterization}
\label{InfinityToposesGeneralAbstract}

Following \cite{Lurie}, for us ``$\infty$-topos'' means this:
\begin{definition} 
  \label{SheafInfinityTopos}
  \index{topos theory!$\infty$-topos}
An $\infty$-topos is an accessible $\infty$-geometric embedding
$$
  \xymatrix{
    \mathbf{H} \ar@<+4pt>@{<-}[r]^-{L} \ar@<-4pt>@{^{(}->}[r] & 
   \mathrm{Func}(C^{\mathrm{op}}, \infty \mathrm{Grpd})
  }
$$
into an $\infty$-category of $\infty$-presheaves, def. \ref{InfinityPresheaves} 
over some small $\infty$-category $C$, hence a full and faithful embedding functor
which preserves filtered $\infty$-colimits, and has a left adjoint
$\infty$-functor which preserves finite $\infty$-limits.

We say this is an \emph{$\infty$-category of $\infty$-sheaves} 
(as opposed to a hypercompletion of such) if $\mathbf{H}$ is the reflective
localization at the covering sieves of a Grothendieck topology on the 
homotopy category of $C$ (a \emph{topological localization}), 
and then write $\mathbf{H}= \mathrm{Sh}_{\infty}(C)$ 
with the site structure on $C$ understood.

More intrinsically, $\infty$-toposes are characterized as follows
(we review the ingredients of the following statement in 
\ref{LimitsAndColimits} and \ref{StrucInftyGroupoids} below).
\begin{definition}[Giraud-Rezk-Lurie axioms]
  \label{GiraudRezkLurieAxioms}
  An \emph{$\infty$-topos} is a presentable $\infty$-category $\mathbf{H}$ that satisfies the
  following properties.
  \begin{enumerate}
    \item {\bf Coproducts are disjoint.}  For every two objects 
    $A, B \in \mathbf{H}$, the intersection of $A$ and 
    $B$ in their coproduct is the initial object: in other words the diagram
	$$
	  \xymatrix{
	    \emptyset \ar[r] \ar[d] & B \ar[d]
		\\
		A \ar[r] & A \coprod B
	  }
	$$
	is a pullback.
	
	\item {\bf Colimits are preserved by pullback.} 
	  For all morphisms $f\colon X\to B$ in $\mathbf{H}$ and all 
	  small diagrams $A\colon I\to \mathbf{H}_{/B}$, there is an 
	  equivalence 
	  $$
	    \varinjlim_i f^*A_i \simeq f^*(\varinjlim_i A_i)
	  $$	  
	  between the pullback of the colimit and the colimit over the pullbacks of its
	  components.
	\item 
	  {\bf Quotient maps are effective epimorphisms.} Every simplicial object
	  $A_\bullet : \Delta^{\mathrm{op}} \to \mathbf{H}$ that satisfies the
	  groupoidal Segal property (Definition~\ref{GroupoidObject}) is the {\v C}ech nerve of its quotient projection:
	  $$
	    A_n \simeq 
		A_0 \times_{\varinjlim_n A_n} A_0 \times_{\varinjlim_n A_n} \cdots \times_{\varinjlim_n A_n} A_0
		\;\;\;
		\mbox{($n$ factors)}
		\,.
	  $$
  \end{enumerate}
\end{definition}
The equivalence of these two 
definitions is theorem 6.1.0.6 in \cite{Lurie}. 

An ordinary topos is famously characterized by the existence of a classifier object
for monomorphisms, the \emph{subobject classifier}. 
With hindsight, this statement already carries in it the seed 
of the close relation between topos theory and bundle theory, for we may think of 
a monomorphism $E \hookrightarrow X$ as being a \emph{bundle of $(-1)$-truncated fibers}
over $X$. The following axiomatizes the existence of arbitrary universal bundles
\begin{definition}
  An \emph{$\infty$-topos} $\mathbf{H}$ is a presentable $\infty$-category with the following
  properties.
  \begin{enumerate}
	\item {\bf Colimits are preserved by pullback.} 
	\item {\bf There are universal $\kappa$-small bundles.}
	  For every sufficiently large regular cardinal $\kappa$, there exists 
	  a morphism $\widehat {\mathrm{Obj}}_\kappa \to \mathrm{Obj}_\kappa$ in $\mathbf{H}$
      which \emph{represents the core of the $\kappa$-small codomain fibration}	  
	  in that for every object $X$, there is an equivalence
	  $$
	    \mathrm{name}
		: 
	    \mathrm{Core}(\mathbf{H}_{/_{\kappa}X})
		\stackrel{\simeq}{\longrightarrow}
		\mathbf{H}(X, \mathrm{Obj}_\kappa)
	  $$
	  between the $\infty$-groupoid of bundles (morphisms) $E \to X$ which are 
	  relatively $\kappa$-small over $X$
	  and the $\infty$-groupoid of morphisms from $X$ into $\mathrm{Obj}_\kappa$, 
      such that there are $\infty$-pullback squares
      $$
	    \raisebox{20pt}{
	    \xymatrix{
		   E \ar[rr] \ar[d] && \widehat{\mathrm{Obj}_\kappa}
		   \ar[d]
		   \\
		   X \ar[rr]^-{\mathrm{name}(E)} && \mathrm{Obj}_\kappa
		}
		}
		\,.
      $$	  
  \end{enumerate}
  \label{RezkCharacterization}
\end{definition}
These two characterizations of $\infty$-toposes, Definition \ref{GiraudRezkLurieAxioms}
and Definition \ref{RezkCharacterization} are equivalent;
this is due to Rezk and Lurie, appearing as Theorem 6.1.6.8 in \cite{Lurie}.
We find that the second of these axioms 
gives the equivalence between $V$-fiber bundles and 
$\mathbf{Aut}(V)$-principal $\infty$-bundles
in Proposition \ref{VBundleIsAssociated}.

\medskip

For $\mathbf{H}$ an $\infty$-topos we write $\mathbf{H}(X,Y)$ for its 
hom-$\infty$-groupoid between objects $X$ and $Y$ and write 
$H(X,Y) = \pi_0 \mathbf{H}(X,Y)$ for the 
hom-set in the homotopy category.
\end{definition}
The theory of cohesive $\infty$-toposes 
revolves around situations where the 
following fact has a refinement:
\begin{proposition} 
 \index{topos theory!terminal geometric morphism}
 \label{Terminalgeometricmorphism}
 \index{topos theory!global sections}
 \index{topos theory!constant $\infty$-stacks}
For every $\infty$-topos $\mathbf{H}$ there is an essentially 
unique geometric morphism
to the $\infty$-topos $\infty \mathrm{Grpd}$.
$$
  (\Delta \dashv \Gamma) : 
  \xymatrix{
    \mathbf{H} \ar@<+3pt>@{<-}[r]^-{\Delta} \ar@<-3pt>[r]_-\Gamma & \infty \mathrm{Grpd}
  }     
$$
\end{proposition}
This is prop 6.3.41 in \cite{Lurie}.
\begin{proposition}
 \label{GlobalSectionMorphismExplicitly}
Here $\Gamma$ forms global sections, in that $\Gamma(-) \simeq \mathbf{H}(*,-)$, 
and $\Delta$ forms constant $\infty$-sheaves -- $\Delta (-) \simeq L \mathrm{Const}(-)$.
\end{proposition}
\proof
  By prop. \ref{Terminalgeometricmorphism} it is sufficient to exhibit an $\infty$-adjunction 
$(L \mathrm{Const}(-) \dashv \mathbf{H}(*,-))$ such that the left adjoint preserves
finite $\infty$-limits.  The latter follows since $\mathrm{Const} : \infty \mathrm{Grpd} \to 
\mathrm{PSh}_\infty(C)$ preserves all limits (for $C$ some $\infty$-site of definition
for  $\mathbf{H}$) and $L : \mathrm{PSh}(C) \to \mathbf{H}$ by definition preserves
finite $\infty$-limits.
To show the $\infty$-adjunction we use prop. \ref{InfinityGroupoidIsColimitOverItself}, which says  
that every $\infty$-groupoid is the $\infty$-colimit
over itself of the $\infty$-functor constant on the point: 
$S \simeq {\lim\limits_{\longrightarrow}}_S *$.
From this we obtain the natural hom-equivalence 
$$
  \begin{aligned}
     \mathbf{H}(L \mathrm{Const} S, X)
       & \simeq
       \mathrm{PSh}_C(\mathrm{Const} S, X)
       \\
       & \simeq \mathrm{PSh}(\mathrm{Const} {\lim\limits_{\longrightarrow}}_S *, X)
       \\
       & \simeq {\lim\limits_{\longleftarrow}}_S \mathrm{Psh}(\mathrm{Const}*, X)
       \\
       & \simeq {\lim\limits_{\longleftarrow}}_S \mathbf{H}(L \mathrm{Const}*, X)
       \\
       & \simeq {\lim\limits_{\longleftarrow}}_S \mathbf{H}(*, X)
       \\
        & \simeq {\lim\limits_{\longleftarrow}}_S \infty \mathrm{Grpd}(*, \mathbf{H}(*, X))
       \\
         & \simeq \infty \mathrm{Grpd}({\lim\limits_{\longrightarrow}}_S *, \mathbf{H}(*, X))
        \\
        & \simeq \infty \mathrm{Grpd}(S, \mathbf{H}(*,X))
        \,.
  \end{aligned}
  \,.
$$
Here and in the following ``$*$'' always denotes the terminal object in the corresponding
$\infty$-category. We used that $L \mathrm{Const}$ preserves the terminal object
(the empty $\infty$-limit.)
\endofproof

\subsubsection{Syntax of homotopy type theory with type universes}
\label{SyntaxOfHomotopyTypeTheory}
\index{topos theory!syntax}

Above in \ref{DependentHomotopyTypeTheory} we indicated how 
locally cartesian closed $\infty$-categories have an internal homotopy type theory.
In locally cartesian closed
$\infty$-categories which are $\infty$-toposes, the ``object of small objects''
of def. \ref{GiraudRezkLurieAxioms} above is internally the 
\emph{type of types} denoted $\mathrm{Type}$ \cite{HoTT}.  

In this context the type theoretic 
judgement ``$x : X \vdash E(x) : \mathrm{Type}$'' is interpreted in the $\infty$-topos as
the \emph{name} morphism $X \stackrel{\mathrm{name}(E)}{\longrightarrow} \mathrm{Obj}_\kappa$
of a morphism $E \to X$ in the $\infty$-topos, according to 
def. \ref{RezkCharacterization}. If here we declare to abbreviate
$(\vdash E) := \mathrm{name}(E)$ then this means we have the following
disctionary between the symbols used to talk about 
objects of slices in $\infty$-toposes and equivalently dependent
types in homotopy type theory.

{\bf morphisms to sequents}:

\begin{tabular}{|c|rccl|rccl|}
  \hline
  notation in$\backslash$ for& \multicolumn{4}{c}{objects/types} & \multicolumn{4}{c}{elements/terms} 
  \\
  \hline
  \hline
  $\infty$-topos theory & $X$ & $\stackrel{\vdash E}{\longrightarrow}$ && $\mathrm{Obj}_\kappa$
    &
	$X$ & $\stackrel{t}{\longrightarrow}_X$ && $E$
  \\
   homotopy type theory & $x : X$ & $\vdash E(x)$ &$:$& $\mathrm{Type}$ & $x : X$ & $\vdash t(x)$ &$:$& $E(x)$  
   \\
   \hline
\end{tabular}

\subsubsection{Presentation by simplicial (pre-)sheaves}
\label{InfinityToposPresentation}
\label{ModelCategoriesOfSimplicialPresheaves}
\label{ToposByLocalizationAtCoverage}
\index{topos theory!presentations}

For computations it is useful to employ a generators-and-relations 
presentation of presentable $\infty$-categories in general and 
of $\infty$-toposes in particular, given by ordinary 
$\mathrm{sSet}$-enriched categories 
equipped with the structure of combinatorial simplicial model categories. 
These may be obtained by left Bousfield localization of a model structure on 
simplicial presheaves (as reviewed in appendix 2 and 3 of \cite{Lurie}).

We discuss these presentations and then discuss various constructions 
in terms of these presentations that will be useful over and over again
in the following. Much of this material is standard and our discussion 
serves to briefly collect the relevant pieces. 
But we also highlight a few points that are not usually discussed
explicitly in the literature, but which we will need later on.

\begin{definition}
  \label{ModelPresentationsOfInfinityToposes}
  Let $C$ be a small category.
  \begin{itemize}
   \item 
    Write $[C^{\mathrm{op}}, \mathrm{sSet}]$
    for the category of functors $C^{\mathrm{op}} \to \mathrm{sSet}$
    to the category of simplicial sets. This is naturally
    equivalent to the category 
    $[\Delta^{\mathrm{op}},[C^{\mathrm{op}}, \mathrm{Set}$
    of simplicial objects in the category of presheaves on $C$.
    Therefore one speaks of the \emph{category of simplicial presheaves}
    over $C$.
   \item
     For $\{U_i \to U\}$ a covering family in the site $C$, write 
     $$
       C(\{U_i\}) \in [C^{\mathrm{op}}, \mathrm{sSet}]
       :=
       \int^{[k] \in \Delta} \Delta[k] \cdot \coprod_{i_0, \cdots, i_k}
         j(U_{i_0}) \times_{j(U)} \cdots \times_{j(U)} j(U_{i_k})
     $$
     for the corresponding \emph{{\v C}ech nerve} simplicial presheaf.
	 This is in degree $k$ the disjoint union of the $(k+1)$-fold intersections
	 of patches of the cover.
     It is canonically equipped with a morphism $C(\{U_i\}) \to j(U)$.
     (Here $j : C \to [C^{\mathrm{op}}, \mathrm{Set}]$ 
     is the Yoneda embedding.)
   \item 
     The category $[C^{\mathrm{op}}, \mathrm{sSet}]$ is naturally 
     an $\mathrm{sSet}$-enriched category.
     For any two objects $X,A \in [C^{\mathrm{op}}, \mathrm{sSet}]$ 
     write $\mathrm{Maps}(X,A) \in \mathrm{sSet}$ for the 
     simplicial hom-set.
   \item
     Write $[C^{\mathrm{op}}, \mathrm{sSet}]_{\mathrm{proj}}$
     for the category of simplicial presheaves equipped with the 
     following choices of classes of morphisms (which are
     natural transformations between $\mathrm{sSet}$-valued functors):
     \begin{itemize}
       \item the \emph{fibrations} are those morphisms whose 
       component over
       each object $U \in C$ is a Kan fibration of simplicial sets;
       \item the \emph{weak equivalences} are those morphisms
       whose component over each object is a weak equivalence in the
       Quillen model structure on simplicial sets;
       \item the \emph{cofibrations} are the morphisms having the
       right lifting property against th morphisms that are both
       fibrations as well as weak equivalences.
     \end{itemize}
     This makes $[C^{\mathrm{op}}, \mathrm{sSet}]_{\mathrm{proj}}$ into a
     combinatorial simplicial model category.
     \item
     Write $[C^{\mathrm{op}}, \mathrm{sSet}]_{\mathrm{proj}, \mathrm{loc}}$
     for model category structure on simplicial presheaves
     which is the left Bousfield localization of 
     $[C^{\mathrm{op}}, \mathrm{sSet}]_{\mathrm{proj}}$ at the 
     set of morphisms of the form $C(\{U_i\}) \to U$
     for all covering families $\{U_i \to U\}$ of $C$.

     This is called the \emph{projective} local model structure
     on simplicial presheaves \cite{Dugger}.
  \end{itemize}
\end{definition}
\begin{definition}
  \label{HomotopySheaves}
  The operation of forming objectwise simplicial homotopy
  groups extends to functors
  $$
    \pi^{\mathrm{PSh}}_0 : [C^{\mathrm{op}}, \mathrm{sSet}]
      \to 
      [C^{\mathrm{op}}, \mathrm{Set}]
  $$
  and for $n > 1$
  $$
    \pi^{\mathrm{PSh}}_n : [C^{\mathrm{op}}, \mathrm{sSet}]_*
      \to 
      [C^{\mathrm{op}}, \mathrm{Set}]
      \,.
  $$
  These presheaves of homotopy groups may be sheafified. We write
  $$
    \pi_0 : [C^{\mathrm{op}}, \mathrm{sSet}]
      \stackrel{\pi_0^{\mathrm{PSh}}}{\to} 
      [C^{\mathrm{op}}, \mathrm{Set}]
      \to 
      \mathrm{Sh}(C)
  $$
  and for $n > 1$
  $$
    \pi_n : [C^{\mathrm{op}}, \mathrm{sSet}]_*
      \stackrel{\pi_n^{\mathrm{PSh}}}{\to}
      [C^{\mathrm{op}}, \mathrm{Set}]
      \to
      \mathrm{Sh}(C)
      \,.
  $$
\end{definition}
\begin{proposition}
  For $X \in [C^{\mathrm{op}}, \mathrm{sSet}]_{\mathrm{proj}, \mathrm{loc}}$
  fibrant, the homotopy sheaves $\pi_n(X)$ from 
  def. \ref{HomotopySheaves} coincide with the abstractly
  defined homotopy groups of $X \in \mathrm{Sh}_\infty(C)$
  from \cite{Lurie}.
\end{proposition}
\proof
  One may observe that the $\mathrm{sSet}_{\mathrm{Quillen}}$-powering
  of $[C^{\mathrm{op}}, \mathrm{sSet}]_{\mathrm{proj}, \mathrm{loc}}$
  does model the abstract $\infty \mathrm{Grpd}$-powering of
  $\mathrm{Sh}_\infty(C)$.
\endofproof
\begin{definition}
  A site $C$ has \emph{enough points} if 
  a morphism $(A \stackrel{f}{\to} B)\in \mathrm{Sh}(C)$ in its sheaf topos
  is an isomorphism precisely if for every \emph{topos point}, hence for
  every geometric morphism
  $$
    (x^* \dashv x_*) : 
    \xymatrix{
      \mathrm{Set}
      \ar@<+4pt>@{<-}[r]^{x^*}
      \ar@<-4pt>[r]_{x_*}
      &
      \mathrm{Sh}(C)
    }
  $$
  from the topos of sets
  we have that $x^*(f) : x^* A \to x^* B$ is an isomorphism.
  \label{site with enough points}
\end{definition}
  Notice here that, by definition of geometric morphism, the 
  functor $i^*$ is left adjoint to $i_*$ -- hence
  preserves all colimits --
  and in addition preserves all \emph{finite} limits. 
\begin{example}
  The following sites have enough points.
  \begin{itemize}
    \item
       The categories $\mathrm{Mfd}$ ($\mathrm{SmoothMfd}$) of 
      (smooth) finite-dimensional, paracompact manifolds
       and smooth functions between them;
     \item
       the category $\mathrm{CartSp}$ of Cartesian spaces 
       $\mathbb{R}^n$ for $n \in \mathbb{N}$ and continuous (smooth) functions
       between them.
  \end{itemize}
\end{example}
This is discussed in detail below in \ref{ETopStrucStalks}.
We restrict from now on attention to this case.
\begin{assumption}
  \label{CHasEnoughPoints}
  The site $C$ has enough points.
\end{assumption}
\begin{theorem}
  \label{CharacterizationOfLocalWeakEquivalence}
  For $C$ a site with enough points, the weak equivalences in 
  $[C^{\mathrm{op}}, \mathrm{sSet}]_{\mathrm{proj}, \mathrm{loc}}$
  are precisely the stalkwise weak equivalences in 
  $\mathrm{sSet}_{\mathrm{Quillen}}$
\end{theorem}
\proof
  By theorem 17 in \cite{JardineBoolean} and using our assumption
  \ref{CHasEnoughPoints} the statement is true 
  for the local injective model structure. 
  The weak equivalences
  there coincide with those of the local projective model structure.
\endofproof
\begin{definition}
  \label{LocalFibrations}
  We say that a morphism $f : A \to B$ in $[C^{\mathrm{op}}, \mathrm{sSet}]$
  is a \emph{local fibration} or a \emph{local weak equivalence}
  precisely if for all topos points $x$ the morphism 
  $x^* f : x^* A \to x^* B$ is a fibration of weak equivalence, respectively.  
\end{definition}
\noindent{\bf Warning.} While by 
theorem \ref{CharacterizationOfLocalWeakEquivalence}
the local weak equivalences are indeed the weak equivalences
in $[C^{\mathrm{op}}, \mathrm{sSet}]_{\mathrm{proj}, \mathrm{loc}}$,
it is not true that the fibrations in this model structure are
the local fibrations of def. \ref{LocalFibrations}.
\begin{proposition}
 Pullbacks in $[C^{\mathrm{op}}, \mathrm{sSet}]$ along local 
 fibrations preserve local weak equivalences.
\end{proposition}
\proof
  Let 
  $$
    \xymatrix{
       A \ar[r] \ar[d] & C \ar[d] & \ar[l] B\ar[d]
       \\
       A' \ar[r] & C' & \ar[l] B'       
    }
  $$
  be a diagram where the vertical morphisms are local weak equivalences.
  Since the inverse image $x^*$ of a topos point $x$ preserves
  finite limits and in particular pullbacks, we have 
  $$
    x^* ( A \times_C B \stackrel{f}{\to} A' \times_{C'} B')
    =
    (
     x^* A \times_{x^* C} x^* B
    \stackrel{x^* f}{\to}
     x^* A' \times_{x^* C'} x^* B'
    )
    \,.
  $$
  On the right the pullbacks are now by assumption pullbacks 
  of simplicial sets along Kan fibrations. Since $\mathrm{sSet}_{\mathrm{Quillen}}$
  is right proper, these are homotopy pullbacks and therefore preserve
  weak equivalences. So $x^* f$ is a weak equivalence for all
  $x$ and thus $f$ is a local weak equivalence.
\endofproof
The following characterization of $\infty$-toposes
is one of the central statements of \cite{Lurie}. 
For the purposes of our discussion here
the reader can take this to be the \emph{definition} 
of $\infty$-toposes.
\begin{theorem}
  \label{PresentationOfInfinityToposBySimplicialPresheaves}
  \index{topos theory!presentation by simplicial presheaves}
  \index{topos!presentation by simplicial presheaves}  
  For $C$ a site with enough points, the $\infty$-topos
  over $C$ is the simplicial localization, 
  def. \ref{SimplicialLocalization},
  $$
    \mathrm{Sh}_\infty(C) 
     \simeq
     L ([C^{\mathrm{op}}, \mathrm{sSet}]_{\mathrm{proj}, \mathrm{loc}}     
  $$
  of the category of simplicial presheaves on $C$ at the local
  weak equivalences.  
\end{theorem}
In view of prop. \ref{ModelStructureModelsSimplicialLocalization} 
this is prop. 6.5.2.14 in \cite{Lurie}. 

\subsubsection{Presentation by simplicial objects in the site}
\label{PresentationBySimplicialObjectsInTheSite}

We will  
have use of the following different presentation of
$\mathrm{Sh}_\infty(C)$.

\begin{definition}
  Let $C$ be a small site with enough points. Write
  $\bar C \subset [C^{\mathrm{op}}, \mathrm{sSet}]$
  for the free coproduct completion.
  
  Let $(\bar C^{\Delta^{\mathrm{op}}}, W)$
  be the category of simplicial objects in $\bar C$ equipped
  with the stalkwise weak equivalences inherited from the
  canonical embedding
  $$
    i 
    :
    \bar C^{\Delta^{\mathrm{op}}}
    \hookrightarrow
    [C^{\mathrm{op}}, \mathrm{sSet}]
    \,.
  $$
\end{definition}
\begin{proposition}
  \label{ModelStructureModelsSimplicialLocalization}
  The induced $\infty$-functor
  $$
    N_h L \bar C^{\Delta^{\mathrm{op}}}
    \to 
    N_h L [C^{\mathrm{op}}, \mathrm{sSet}]_{\mathrm{proj}, \mathrm{loc}}
  $$
  is an equivalence of $\infty$-categories.
\end{proposition}
This is due to \cite{NSSb}. We prove this after 
noticing the following fact.
\begin{proposition}
  \label{DegreewiseRepresentability}
  Let $C$ be a category and $\bar C$ its free coproduct
  completion.
  
  Every simplicial presheaf over $C$ is equivalent in 
  $[C^{\mathrm{op}}, \mathrm{sSet}]_{\mathrm{proj}}$ to 
  a simplicial object in $\bar C$ (after the degreewise Yoneda
  embedding $j^{\Delta^{\mathrm{op}}} : \bar C^{\Delta^{\mathrm{op}}}
\to [C^{\mathrm{op}}, \mathrm{sSet}] $).

  If moreover $C$ has pullbacks and sequential colimits, then the simplicial object 
  in $\bar C$  can be taken to be globally Kan, hence fibrant in 
  $[C^{\mathrm{op}}, \mathrm{sSet}]_{\mathrm{proj}}$.
\end{proposition}
\proof
  The first statement is prop. 2.8 in \cite{Dugger}, which
  says that for every $X \in [C^{\mathrm{op}}, \mathrm{sSet}]$
  the canonical morphism from the simplicial presheaf
  $$
    (Q X) : [k] \mapsto \coprod_{U_0 \to \cdots \to U_k \to X_k} j(U_0)
	\,,
  $$  
  where the coproduct runs over all sequences of morphisms between representables
  $U_i$ as indicated and using the evident face and degeneracy maps, 
  is a global weak equivalence
  $$
    Q X \stackrel{\simeq}{\to} X
	\,.
  $$
  The second statement follows by postcomposing with Kan's
  fibrant replacement functor (see for instance section 3 in 
  \cite{Jardine})
  $$
    \mathrm{Ex}^\infty : \mathrm{sSet} \to \mathrm{KanCplx} \hookrightarrow
	\mathrm{sSet}
	\,.
  $$
  This functor forms new simplices by subdivision, which only involves
  forming iterated pullbacks over the spaces of the original simplices.
\endofproof
\begin{example}
  Let $C$ be a category of \emph{connected} topological
  spaces with given extra structure and properties 
  (for instance smooth manifolds). Then 
  $\bar C$ is the category of all such spaces
 (with arbitrary many connected components). 
 
  Then the statement is that every $\infty$-stack over
  $C$ has a presentation by a simplicial object in $\bar C$.
  This is true with respect to any Grothendieck topology 
  on $C$, since
  the weak equivalences in the global projective model
  structure that prop. \ref{DegreewiseRepresentability} refers
  to remain weak equivalences in any left Bousfield localization.
  
  If moreover $C$ has all pullbacks (for instance for connected
  topological spaces, but not for smooth manifolds) then
  every $\infty$-stack over $C$ even has a presentation by a
  globally Kan simplicial object in $\bar C$.
\end{example}
\proofoftheorem{ModelStructureModelsSimplicialLocalization}
  Let 
   $Q : [C^{\mathrm{op}}, \mathrm{sSet}] 
    \to     
  \bar C^{\Delta^{\mathrm{op}}}$
  be Dugger's replacement functor from the proof
  of prop. \ref{DegreewiseRepresentability}.  
  In \cite{Dugger} it is shown that for all $X$ the simplicial 
  presheaf  $Q X$
  is cofibrant in $[C^{\mathrm{op}}, \mathrm{sSet}]_{\mathrm{proj}}$
  and that the natural morphism $Q X \to X$ is a weak equivalence.
  Since left Bousfield localization does not affect the cofibrations
  and only enlarges the weak equivalences, the same is still true in 
  $[C^{\mathrm{op}}, \mathrm{sSet}]_{\mathrm{proj}, \mathrm{loc}}$.
    
  Therefore we have a natural transformation
  $$
    i \circ Q \to \mathrm{Id} 
    : 
   [C^{\mathrm{op}}, \mathrm{sSet}]
   \to
   [C^{\mathrm{op}}, \mathrm{sSet}]
  $$
  whose components are weak equivalences. From this the claim
  follows by prop. 3.5 in \cite{DwyerKanComputations}.
\endofproof
\begin{remark}
  \index{topos theory!presentation by simplicial $C$-manifolds}
  \index{topos!presentation by simplicial $C$-manifolds}
  If the site $C$ is moreover equipped with the structure
  of a \emph{geometry} as in \cite{LurieSpaces} then there is 
  canonically the notion of a \emph{$C$-manifold}:
  a sheaf on $C$ that is \emph{locally} isomorphic to a
  representable in $C$. Write
  $$
    \bar C \hookrightarrow C \mathrm{Mfd}
    \hookrightarrow [C^{\mathrm{op}}, \mathrm{Set}]
  $$
  for the full subcategory of presheaves on the $C$-manifolds.
 
  Then the above argument applies verbatim also to 
  the category $C\mathrm{Mfd}^{\Delta^{\mathrm{op}}}$ of
  simplicial $C$-manifolds. Therefore we find that
  the $\infty$-topos over $C$ is presented by the
  simplicial localization of simplicial $C$-manifolds at
  the stalkwise weak equivalences:
  $$
    \mathrm{Sh}_\infty(C) \simeq N_h L C \mathrm{Mfd}^{\Delta^{\mathrm{op}}}
    \,.
  $$
\end{remark}
\begin{example}
  \label{SmoothInfinityGroupoidByManifolds}
  \index{smooth $\infty$-groupoid!presentation by simplicial smooth manifolds}
  Let $C = \mathrm{CartSp}_{\mathrm{smooth}}$ be the 
  full subcategory of the category $\mathrm{SmthMfd}$ of
  smooth manifolds on the Cartesian spaces, $\mathbb{R}^n$,
for $n \in \mathbb{R}$. Then $\bar C \subset \mathrm{SmthMfd}$
is the full subcatgory on manifolds that  are disjoint unions of
Cartesian spaces and $C \mathrm{Mfd} \simeq \mathrm{SmthMfd}$.
Therefore we have an equivalence of $\infty$-categories
$$
  \mathrm{Sh}_\infty(\mathrm{SmthMfd})
    \simeq
  \mathrm{Sh}_\infty(\mathrm{CartSp})
   \simeq
  L \; \mathrm{SmthMfd}^{\Delta^{\mathrm{op}}}
  \,.
$$
\end{example}

\subsubsection{$\infty$-Sheaves and descent}
\label{SheavesDescent}
\index{topos!descent}
\index{sheaf}
\index{stack}

We discuss some details of the notion of \emph{$\infty$-sheaves}
from the point of view of the presentations discussed above in 
\ref{InfinityToposPresentation}.

\medskip

By def. \ref{SheafInfinityTopos} we have, abstractly, 
that an $\infty$-sheaf over some site $C$ is an $\infty$-presheaf that is in the essential image of a given reflective inclusion $\mathrm{Sh}_\infty(C) \hookrightarrow \mathrm{PSh}_\infty(C)$. By 
prop. \ref{PresentationOfInfinityToposBySimplicialPresheaves}
this reflective embedding is presented by the Quillen adjunction that
exhibits the left Bousfield localization of the 
model category of simplicial presheaves at the {\v C}ech covers
$$
  \xymatrix{
    ([C^{\mathrm{op}}, \mathrm{sSet}]_{\mathrm{proj}, \mathrm{loc}})^\circ
	\ar[d]^\simeq
	\ar@{<-}@<+3pt>^{\mathbb{L}\mathrm{Id}}[rr]
	\ar@<-3pt>_{\mathbb{R}\mathrm{Id}}[rr]
	&&
    ([C^{\mathrm{op}}, \mathrm{sSet}]_{\mathrm{proj}})^\circ
	\ar[d]^\simeq
	\\
	\mathrm{Sh}_\infty(C)
	\ar@{<-}@<+3pt>^{L}[rr]
	\ar@{^{(}->}@<-3pt>[rr]
	&&
	\mathrm{PSh}_\infty(X)
  }
  \,.
$$
Since the Quillen adjunction that exhibits left Bousfield localization
is given by identity-1-functors, as indicated, the computation of
$\infty$-sheafification ($\infty$-stackification) $L$ by deriving the left Quillen functor is all in the cofibrant replacement in 
$[C^{\mathrm{op}}, \mathrm{sSet}]_{\mathrm{proj}}$ followed by 
fibrant replacement in $[C^{\mathrm{op}}, \mathrm{sSet}]_{\mathrm{proj}, \mathrm{loc}}$. Since the collection of cofibrations is preserved by left Bousfield
localization, this simply amounts to cofibrant-fibrant replacement in
$[C^{\mathrm{op}}, \mathrm{sSet}]_{\mathrm{proj}, \mathrm{loc}}$. 
Since, finally, the derived hom space $\mathrm{Sh}_\infty(U, A)$
is computed in $[C^{\mathrm{op}}, \mathrm{sSet}]_{\mathrm{proj}, \mathrm{loc}}$ already on a fibrant resolution of $A$ out of a 
cofibrant resolution of $U$, and since every representable is 
necessarily cofibrant, one may effectively identify the $\infty$-sheaf
condition in $\mathrm{PSh}_\infty(C)$ with the fibrancy condition in 
$[C^{\mathrm{op}}, \mathrm{sSet}]_{\mathrm{proj}, \mathrm{loc}}$.

We discuss aspects of this fibrancy condition.
\begin{definition}
  \label{GoodCovers}
  For $C$ a site, we say a covering family $\{U_i \to U\}$
  is a \emph{good cover} if the corresponding
  {\v C}ech nerve 
  $$
    C(U_i) :=
	\int^{[k] \in \Delta}
	 \coprod_{i_0, \cdots, i_k} 
	 j(U_{i_0}) \times_{j(U)} \cdots \times_{j(U)} j(U_k)
	 \;\;
	 \in 
	 [C^{\mathrm{op}}, \mathrm{sSet}]_{\mathrm{proj}}
  $$	
  (where $j : C \to [C^{\mathrm{op}}, \mathrm{sSet}]$ is the Yoneda
  embedding) is degreewise a coproduct of representables, hence
  if all non-empty finite intersections 
   of the $U_i$ are again
  representable:
    $$
     j(U_{i_0, \cdots, i_k}) 
     =
	 U_{i_0} \times_{U} \cdots \times_U U_{i_k}
	 \,.
  $$
\end{definition}
\begin{proposition}
  \label{CechNervesOfGoodCoversAreProjectivelyCofibrant}
  The {\v C}ech nerve $C(U_i)$ of a good cover is cofibrant in
  $[C^{\mathrm{op}}, \mathrm{sSet}]_{\mathrm{proj}}$
  as well as in 
  $[C^{\mathrm{op}}, \mathrm{sSet}]_{\mathrm{proj}, \mathrm{loc}}$.
\end{proposition}
\proof
In the terminology of  \cite{dugger-hollander-isaksen}
the good-ness condition on a cover makes its {\v C}ech nerve
a \emph{split hypercover}. By the result of \cite{Dugger}
this is cofirant in $[C^{\mathrm{op}}, \mathrm{sSet}]_{\mathrm{proj}}$.
Since left Bousfield localization preserves cofibrations, it is
also cofibrant in $[C^{\mathrm{op}}, \mathrm{sSet}]_{\mathrm{proj}, \mathrm{loc}}$.
\endofproof
\begin{definition}
  \label{DescentObject}
  \index{descent!descent object}
 For $A$ a simplicial presheaf with values in Kan complexes and
 $\{U_i \to U\}$ a good cover in the site $C$, we say that
 $$
   \mathrm{Desc}(\{U_i\}, A)
   :=
   [C^{\mathrm{op}}, \mathrm{sSet}](C(U_i), A)
   \,,
 $$
 where on the right we have the $\mathrm{sSet}$-enriched hom
 of simplicial presheaves, is  the \emph{descent object}
 of $A$ over $\{U_i \to U\}$.
\end{definition}
\begin{remark}
  By assumption $A$ is fibrant and $C(U_i)$ is cofibrant (by prop. \ref{CechNervesOfGoodCoversAreProjectivelyCofibrant}) in $[C^{\mathrm{op}}, \mathrm{sSet}]_{\mathrm{proj}}$. Since this is a simplicial model category,
it follows that $\mathrm{Desc}(\{U_i\}, A)$ is a Kan complex,
an $\infty$-groupoid. We may also speak of the 
\emph{descent $\infty$-groupoid}. Below we show that its
objects have the interpretation of \emph{gluing data} or
\emph{descent data} for $A$. See \cite{dugger-hollander-isaksen}
for more details.
\end{remark}
\begin{proposition}
  \index{descent!descent condition}
  \label{DescentConditionForSimplicialPresheaves}
  For $C$ a site whose topology is generated from good covers,
  a simplicial presheaf $A$ is fibrant in 
  $[C^{\mathrm{op}}, \mathrm{sSet}]_{\mathrm{proj}, \mathrm{loc}}$
  precisely if it takes values in Kan complexes and if
  for each generating good cover $\{U_i \to U\}$ the canonical morphism
  $$
    A(U) \to \mathrm{Desc}(\{U_i\}, A)
  $$
  is a weak equivalence of Kan complexes.
\end{proposition}
\proof
   By standard results recalled in A.3.7 of \cite{Lurie}
   the fibrant objects in the local model structure are precisely 
   those which are fibrant in the global model structure
   and which are \emph{local} with respect to the morphisms
   at which one localizes: such that the derived hom out of these
   morphisms into the given object produces a weak equivalence.
   
   By prop. \ref{CechNervesOfGoodCoversAreProjectivelyCofibrant} 
   we have that $C(U_i)$
   is cofibrant for $\{U_i \to U\}$ a good cover. Therefore
   the derived hom is computed already by the enriched hom as in the above
   statement.
\endofproof
\begin{remark}
  \index{sheaf!ordinary sheaf}
  The above condition manifestly generalizes the \emph{sheaf} condition
  on an ordinary sheaf \cite{Johnstone}. One finds that 
  $$
   (\pi_0^{\mathrm{PSh}}(C(U_i)) \to \pi_0^{\mathrm{PSh}}(U))
   =
   (S(U_i) \hookrightarrow U)
  $$
  is the (subfunctor corresponding to the) \emph{sieve}
  associated with the cover $\{U_i \to U\}$. Therefore when $A$ is
  itself just a presheaf of sets (of simplicially constant simplicial sets)
  the above condition reduces to the statement that
  $$
    A(U) \to [C^{\mathrm{op}}, \mathrm{Set}](S(U_i), A)
  $$
  is an isomorphism. This is the standard sheaf condition.
\end{remark}
We discuss the descent object, def. \ref{DescentObject}, in more
detail.
\begin{definition}
  Write
  $$
    \mathrm{coDesc}(\{U_i\}, A) \in \mathrm{sSet}^\Delta
  $$
  for the cosimlicial simplicial set that in degree $k$ 
  is given by the value of $A$ on the $k$-fold intersections:
  $$
    \mathrm{coDesc}(\{U_i\}, A)_k
	=
	\prod_{i_0, \cdots, i_k} A(U_{i_0, \cdots , i_k})
	\,.
  $$
\end{definition}
\begin{proposition}
  \index{descent!descent object!from totalization}
  The descent object from def. \ref{DescentObject}
  is the \emph{totalization} of the codescent object:
  $$
    \begin{aligned}
      \mathrm{Desc}(\{U_i\}, A)  
	  & =
	  \mathrm{tot}(\mathrm{coDesc}(\{U_i\}), A)
	  \\
	  & :=
	  \int_{[k] \in \Delta} \mathrm{sSet}(\Delta[k], \mathrm{coDesc}(\{U_i\},A)_k)
	\end{aligned}
  $$
\end{proposition}
Here and in the following equality signs denote isomorphism 
(such as to distinguish from just weak equivalences of simplicial sets).\\
\proof
  Using $\mathrm{sSet}$-enriched category calculus 
  for the $\mathrm{sSet}$-enriched and $\mathrm{sSet}$-tensored
  category of simplicial presheaves
  (for instance \cite{Kelly} around (3.67)) we compute as follow
  $$
  \begin{aligned}
    \mathrm{Desc}(\{U_i\}, A)
	& :=
    [C^{\mathrm{op}}, \mathrm{sSet}](C(U_i), A)
	\\
	& =
	[C^{\mathrm{op}}, \mathrm{sSet}](\int^{[k] \in \Delta} \Delta[k] \cdot C(U_i)_k, A)
	\\
	& = 
	\int_{[k] \in \Delta} [C^{\mathrm{op}}, \mathrm{sSet}](\Delta[k] \cdot C(U_i), A)
    \\
    & = \int_{[k \in \Delta]} \mathrm{sSet}( \Delta[k], 
	[C^{\mathrm{op}}, \mathrm{sSet}](C(U_i)_k), A)
    \\
    &=
    \int_{[k \in \Delta]} \mathrm{sSet}( \Delta[k], 
	A(C(U_i)_k))
	\\
	& = \mathrm{tot}(A(C(U_i)_\bullet))
	\\
	& = \mathrm{tot} \left(\mathrm{coDesc}(\{C(U_i)\}, A)\right)
	\,.
  \end{aligned}
  $$
  Here we used in the first step that every simplicial set $Y$
  (hence every simplicial presheaf) is the realization of itself, in 
  that
  $$
    Y = \int^{[k] \in \Delta} \Delta[k] \cdot Y_k
	\,,
  $$
  which is effectively a variant of the Yoneda-lemma.
\endofproof
\begin{remark}
\index{descent!cocycles}
This provides a fairly explicit description of the 
objects in $\mathrm{Desc}(\{U_i\}, A)$ by what is called
\emph{nonabelian {\v C}ech hypercohomology}. 

Notice that
an element $c$ of the end $\int_{[k] \in \Delta} \mathrm{sSet}(\Delta[k], \mathrm{coDesc}(\{U_i\}, A))$ is by definition of \emph{end}s 
a collection of morphisms 
$$
  \{c_k : \Delta[k] \to \prod_{i_0, \cdots, i_k}A_k(U_{i_0, \cdots, i_k})\}
$$ 
that makes commuting all parallel diagrams in the following:
$$
  \xymatrix{
     && 
    \\
    \Delta[2]
	 \ar[rr]^{c_2}
	 \ar@{..}[u]
	 &&
	 \prod_{i_0, i_1, i_2}
	 A(U_{i_0, i_1, i_2})
	 \ar@{..}[u]
     \\
     \Delta[1]
	 \ar[rr]^{c_1}
	 \ar@<+9pt>[u]
	 \ar@{<-}@<+6pt>[u]
	 \ar@<+3pt>[u]
	 \ar@{<-}@<+0pt>[u]
	 \ar@<-3pt>[u]
	 \ar@{<-}@<-6pt>[u]
	 \ar@<-9pt>[u]
	 &&
	 \prod_{i_0 i_1}
	 A(U_{i_0, i_1})
	 \ar@<+9pt>[u]
	 \ar@{<-}@<+6pt>[u]
	 \ar@<+3pt>[u]
	 \ar@{<-}@<+0pt>[u]
	 \ar@<-3pt>[u]
	 \ar@{<-}@<-6pt>[u]
	 \ar@<-9pt>[u]
     \\
     \Delta[0] 
	 \ar@<+3pt>[u]
	 \ar@{<-}@<+0pt>[u]
	 \ar@<-3pt>[u]
	 \ar[rr]^{c_0} 
	 && 
	 \prod_{i_0} A(U_{i_0})
	 \ar@<+3pt>[u]
	 \ar@{<-}@<+0pt>[u]
	 \ar@<-3pt>[u]
  }
  \,.
$$
\end{remark}
This says in words that $c$ is
\begin{enumerate}
  \item a collection of objects ${a_i \in A(U_{i})}$ on each 
  patch;
  \item 
    a collection of morphisms $\{g_{i j} \in A_1(U_{i j})\}$
	over each double intersection, such that these go between the
	restrictions of the objects $a_i$ and $a_j$, respectively
	$$
	  \xymatrix{
	     a_i|_{U_{i j}}
		 \ar[rr]^{g_{i j}}
		 &&
		 a_j|_{U_{i j}}
	  }
	$$
\item
  a collection of 2-morphisms $\{h_{i j k} \in A_2(U_{i j k})\}$
  over triple intersections, which go between the corresponding 
  1-morphisms:
  $$
    \xymatrix{
	   & a_j|_{U_{i j k}}
	   \ar[dr]^{g_{jk}|_{U_{i j k}}}
	   \\
	   a_i|_{U_{i j k}}
	   \ar[rr]^{\ }="t"_{g_{i k}|_{U_{i j k}}}
	   \ar[ur]^{g_{i j}|_{U_{i j k}}}
	   &&
	   a_k|_{U i j k}
	   \ar@{=>}|{h_{i j k}} "t"+(0,10); "t"
	}
	\,,
  $$
 \item 
  a collection of 3-morphisms $\{\lambda_{i j k l} \in A_3(U_{i j k l})\}$
  of the form
  $$
    \raisebox{35pt}{
  \xymatrix{
     a_j|_{U_{i j k l}} \ar[rr]^{g_{j k}|_{U_{i j k l}}} 
	 && a_k|_{U_{i j k l}} 
	 \ar[dd]^{g_{k l}|_{U_{i j k l}}}
     \\
     \\
    a_i|_{U_{i j k l}} \ar[rr]_{} 
	 \ar[uu]^{g_{i j}|_{U_{i j k l}}} 
	  \ar[uurr]|{}^{\ }="t"_{\ }="s" 
	   && 
	  a_l|_{U_{i j k l}}
    \ar@{=>}^<<{\hspace{-.4cm}h_{i j k}|_{U_{i j k l}}} "t"+(-4,4); "t"
    \ar@{=>}_<<<{\hspace{+1.5cm}h_{i k l}|_{U_{i j k l}}} "s"; "s"+(4,-4)
  }
  }
  \xymatrix{
    \ar[r]^{\lambda_{i j k l}} &
  }
  \raisebox{35pt}{
  \xymatrix{
     a_j |_{U_{i j k l}} \ar[rr]^{g_{j k}|_{U_{i j k l}}} 
	 \ar[ddrr]|{}^{\ }="t"_{\ }="s" 
	   && a_j|_{U_{i j k l}} \ar[dd]^{g_{kl}|_{U_{i j k l}}}
     \\
     \\
    a_i|_{U_{i j k l}} \ar[rr] \ar[uu]^{g_{i j}|_{U_{i j k l}}}  
	&& 
	a_l |_{U_{i j k l}}
    \ar@{=>}^{h_{j k l}|_{U_{i j k l}}} "t"+(4,4); "t"
    \ar@{=>}_>>>>>>{\hspace{-.4cm}h_{i j l}|_{U_{i j k l}}} "s"; "s"+(-4,-4)
  }
  }
  \,;
  $$
  \item
   and so on.
\end{enumerate}
This recovers the cocycle diagrams that we have discussed
more informally in \ref{SmoothPrincipalnBundles} and generalizes them 
to arbitrary coefficient objects $A$.

\subsubsection{$\infty$-Sheaves with values in chain complexes}
 \label{SheafAndNonabelianDoldKan}
 \index{Dold-Kan correspondence}
 \index{Dold-Kan correspondence!nonabelian}

Many simplicial presheaves appearing in practice are (equivalent to) 
objects in 
sub-$\infty$-categories of $\mathrm{Sh}_{\infty}(C)$ of $\infty$-sheaves with values in
abelian or at least in ``strict'' $\infty$-groupoids. 
These subcategories typically offer convenient and desireable 
contexts for formulating and 
proving statements about special cases of general simplicial presheaves.  

One well-known such notion is given by the \emph{Dold-Kan correspondence} 
(discussed for instance in \cite{GoerssJardine}). 
This identifies chain complexes of abelian groups with strict and 
strictly symmetric monoidal $\infty$-groupoids.
\begin{proposition} 
 \label{EmbeddingOfChainComplexes}
  Let $\mathrm{Ch}^+_{\mathrm{proj}}$ be the standard projective 
  model structure on chain 
  complexes of abelian groups in non-negative degree and let $\mathrm{sAb}_{\mathrm{proj}}$
  be the standard projective model structure on simplicial abelian groups. 
  Let $C$ be any small category.
  There is a composite Quillen adjunction
  $$
    ((N_\bullet F)_* \dashv \Xi)
    :
    \xymatrix{
      [C^{\mathrm{op}}, Ch^+_{\mathrm{proj}}]_{\mathrm{proj}}
      \ar@{<-}@<+4pt>[r]^{(N_\bullet)_*}
      \ar@<-4pt>[r]_{\Gamma_*}^{\simeq}
      &
      [C^{\mathrm{op}}, \mathrm{sAb}_{\mathrm{proj}}]_{\mathrm{proj}}
      \ar@{<-}@<+4pt>[r]^{F_*}
      \ar@<-4pt>[r]_{U_*}
      &
      [C^{\mathrm{op}}, \mathrm{sSet}_{\mathrm{Quillen}}]_{\mathrm{proj}}
    }   
    \,,
  $$
  where the first is given by postcomposition 
  with the Dold-Puppe-Kan correspondence
  and the second by postcomposition with the degreewise 
  free-forgetful adjunction 
  for abelian groups over sets.
\end{proposition}
We also write $\mathrm{DK} := \Xi$ for this Dold-Kan map.
Dropping the condition on symmetric monoidalness we obtain a more general such inclusion, a kind of non-abelian Dold-Kan correspondence: the identification of \emph{crossed complexes}, 
def. \ref{CrossedComplex}, with 
strict $\infty$-groupoids (see \cite{BHS}\cite{Porter} for details).
\begin{definition}
\index{groupoid!strict}
\label{StrictInfinityGroupoids}
\label{StrictInfinityGroupoid}
  A \emph{globular set} $X$ is a collection of sets $\{X_n\}_{n \in \mathbb{N}}$
  equipped with functions $\{s_n,t_n : X_{n+1} \to X_n\}_{n \in \mathbb{N}}$
  such that $\forall_{n \in \mathbb{N}} (s_{n} \circ s_{n+1} = s_{n} \circ t_{n+1})$
  and $\forall_{n\in \mathbb{N}}( t_n \circ s_{n+1} = t_n \circ t_{n+1} )$.
  (These relations ensure that for every pair $k_1 < k_2 \in \mathbb{N}$
  there are uniquely defined functions $s,t : \xymatrix{X_{k_2}} \to X_{k_1}$.)
  A \emph{strict $\infty$-groupoid} is a globular set $X_\bullet$ equipped for
  each $k \geq 1$ with the structure of a groupoid on 
  $\xymatrix{X_k \ar@<+3pt>[r]^{s} \ar@<-3pt>[r]_{t} & X_0}$ such that 
  for all $k_1 < k_2 \in \mathbb{N}$ this induces the structure of 
  a strict 2-groupoid on 
  $$
    \xymatrix{
	  X_{k_2} \ar@<+3pt>[r]^{s} \ar@<-3pt>[r]_{t}
	  &
	  X_{k_1} \ar@<+3pt>[r]^{s} \ar@<-3pt>[r]_{t}
	  &
	  X_0
	}
	\,.
  $$
\end{definition}
\begin{remark}
We have a sequence of (non-full) inclusions
$$
  \xymatrix{
     \mathrm{ChainComplex} \ar[r] \ar[d]^\simeq 
     & 
	 \mathrm{CrossedComplex} 
	 \ar[r] \ar[d]^\simeq & 
	 \mathrm{KanComplex}
       \ar[d]^\simeq
     \\
     \mathrm{StrAb Str}\infty \mathrm{Grpd} 
     \ar[r] &
     \mathrm{Str} \infty \mathrm{Grpd}
     \ar[r] &
     \infty \mathrm{Grpd} 
  }
$$
of strict $\infty$-groupoids into all $\infty$-groupoids, where in the top row we list the explicit
presentation and in the bottom row the abstract notions.
\end{remark}
We state a useful theorem for the computation of descent for presheaves,
prop. \ref{DescentConditionForSimplicialPresheaves}, 
with values in strict $\infty$-groupoids. 

Suppose that $\mathcal{A} : C^{\mathrm{op}} \to \mathrm{Str}\infty \mathrm{Grpd}$ is a presheaf 
with values in strict $\infty$-groupoids. In the context of strict $\infty$-groupoids the standard $n$-simplex is given by the $n$th \emph{oriental} $O(n)$ \cite{Street}. 
This allows us to perform a construction that looks like a descent object in $\mathrm{Str}\infty \mathrm{Grpd}$:
\begin{definition}[Street 04]
 \label{StreetDescent}
 \index{descent!descent object!for strict $\infty$-groupoids}
The descent object for $\mathcal{A} \in [C^{\mathrm{op}}, \mathrm{Str} \infty \mathrm{Grpd}]$ 
relative to $Y \in [C^{\mathrm{op}}, \mathrm{sSet}]$ is
$$
  \mathrm{Desc}_{\mathrm{Street}}(Y,\mathcal{A}) := \int_{[n] \in \Delta} \mathrm{Str}\infty \mathrm{Cat}(O(n), \mathcal{A}(Y_n))
  \;\in \mathrm{Str} \infty \mathrm{Grpd}
  \,,
$$
where the end is taken in $\mathrm{Str} \infty \mathrm{Grpd}$. 
\end{definition}
This object had been suggested by Ross Street to be the right descent object for strict $\infty$-category-valued presheaves in \cite{Street}.

Canonically induced by the orientals is the $\omega$-nerve
$$
  N : \mathrm{Str}\omega\mathrm{Cat} \to \mathrm{sSet}
$$
Applying this to the descent object of prop. \ref{StreetDescent}
yields the simplicial set $N \mathrm{Desc}(Y,\mathcal{A})$. On the other hand, 
applying the $\omega$-nerve componentwise to $\mathcal{A}$ yields a simplicial presheaf 
$N \mathcal{A}$ to which the ordinary simplicial descent from 
def. \ref{DescentObject}
applies. The following theorem asserts that under certain conditions 
the $\infty$-groupoids presented by both these simplicial sets are equivalent.
\begin{proposition}[Verity 09]
 \label{VerityTheorem}
 \index{descent!for strict $\infty$-groupoid valued presheaves}
If $\mathcal{A} : C^{\mathrm{op}}, \mathrm{Str} \infty \mathrm{Grpd}$ 
and $Y : C^{\mathrm{op}} \to \mathrm{sSet}$ are such that $N \mathcal{A}(Y_\bullet) : \Delta \to sSet$ 
is fibrant in the Reedy model structure $[\Delta, \mathrm{sSet}_{\mathrm{Quillen}}]_{\mathrm{Reedy}}$, then 
$$ 
  N \mathrm{Desc}_{\mathrm{Street}}(Y,\mathcal{A}) \stackrel{\simeq}{\to} \mathrm{Desc}(Y, N \mathcal{A})
$$
is a weak homotopy equivalence of Kan complexes.
\end{proposition}
This is proven in \cite{Verity}. In our applications the assumptions of this theorem
are usually satisfied:
\begin{corollary}
If $Y \in [C^{\mathrm{op}}, \mathrm{sSet}]$ is such that $Y_\bullet : \Delta \to [C^{\mathrm{op}}, \mathrm{Set}] \hookrightarrow [C^{\mathrm{op}}, \mathrm{sSet}]$ is cofibrant in $[\Delta, [C^{\mathrm{op}}, \mathrm{sSet}]_{\mathrm{proj}}]_{\mathrm{Reedy}}$ 
then for $\mathcal{A} : C^{\mathrm{op}} \to \mathrm{Str} \infty \mathrm{Grpd}$ we have
a weak equivalence
$$ 
  N \mathrm{Desc}(Y,\mathcal{A}) \stackrel{\simeq}{\to} \mathrm{Desc}(Y, N \mathcal{A})
  \,.
$$
\end{corollary}
\proof
If $Y_\bullet$ is Reedy cofibrant, then by definition the canonical morphisms
$$
  \lim_{\to}( ([n] \stackrel{+}{\to} [k]) \mapsto Y_k ) \to Y_n
$$
are cofibrations in $[C^{\mathrm{op}}, \mathrm{sSet}]_{\mathrm{proj}}$. 
Since the latter is an $\mathrm{sSet}_{\mathrm{Quillen}}$-enriched model category 
and $N \mathcal{A}$ is fibrant in $[C^{\mathrm{op}}, \mathrm{sSet}]_{\mathrm{proj}}$, 
it follows that the hom-functor 
$[C^{\mathrm{op}}, \mathrm{sSet}](-, N \mathcal{A})$ sends cofibrations to fibrations, so that
$$
  N\mathcal{A}(Y_n)
  \to
  \lim_{\leftarrow}( [n]\stackrel{+}{\to} [k] \mapsto N\mathcal{A}(Y_k))
$$ 
is a Kan fibration. But this says that $N \mathcal{A}(Y_\bullet)$ is Reedy fibrant, so 
that the assumption  of prop. \ref{VerityTheorem} is met.
\endofproof

\subsection{Universal constructions}
\label{LimitsAndColimits}

We discuss some basic abstract properties and some presentations of universal constructions in 
$\infty$-category theory that we will refer to frequently.

\subsubsection{General abstract}

\paragraph{$\infty$-Colimits in $\infty\mathrm{Grpd}$}

The following proposition says that every $\infty$-groupoid
is the $\infty$-colimit over itself, regarded as a diagram, of the
$\infty$-functor constant on the point in $\infty \mathrm{Grpd}$.

\begin{proposition}
  For $S \in \infty \mathrm{Grpd}$, the $\infty$-colimit of the
  $\infty$-functor $S \to \infty \mathrm{Grpd}$ constant on the terminal object is
equivalent to $S$:
  $$
    \underset{\longrightarrow_{S}}{\lim} {*} \simeq S
	\,.
  $$  
  \label{InfinityGroupoidIsColimitOverItself}
\end{proposition}
This is essentially corollary 4.4.4.9  in \cite{Lurie}.

\paragraph{$\infty$-Pullbacks}

We will have have ample application for the following 
immediate $\infty$-category theoretic generalization of a basic 1-categorical fact.
\begin{proposition} [pasting law for $\infty$-pullbacks] 
  \label{PastingLawForPullbacks}
  \index{pullback!pasting law}
  \index{category theory!pasting law for $\infty$-pullbacks}
  Let 
  $$
    \xymatrix{
       a \ar[r] \ar[d] & b \ar[r]\ar[d] & c \ar[d]
       \\
       d \ar[r] & e \ar[r] & f
    }
  $$
  be a diagram in an $\infty$-category and suppose that the right square is an 
  $\infty$-pullback. Then the left square is an $\infty$-pullback precisely if the 
  outer rectangle is.
\end{proposition}
This appears as \cite{Lurie}, lemma 4.4.2.1. 
Notice that here and in the following we do
not explicitly display the 2-morphisms/homotopies that do fill these diagrams in 
the given $\infty$-category.

\paragraph{Effective epimorphisms}
\label{effective epimorphisms}

We briefly record the definition and main properties of effective epimorphisms
in an $\infty$-topos from \cite{Lurie}, section 6.2.3. 

\medskip

\begin{definition}
  \label{EffectiveEpimorphism}
  \index{effective epimorphism}
  A morphism $Y \to X$ in an $\infty$-topos is an 
  \emph{effective epimorphism} if it exhibits the $\infty$-colimit
  over the simplicial diagram that is its {\v C}ech nerve:
  $$
    Y \simeq \lim\limits_{\longrightarrow_n}\, Y^{\times^n_X}
	\,.
  $$
\end{definition}
See for instance below cor. 6.2.3.5 in \cite{Lurie}.
\begin{remark}
  \label{atlas}
  \index{atlas}
  In view of the discussion of groupoid objects below in 
  \ref{StrucInftyGroupoids} (see remark \ref{GroupoidObjectsAreGroupoidsWithAnAtlas} there) 
  we also speak of an effective epimorphism
  $\xymatrix{U \ar@{->>}[r] & X}$ as being an \emph{atlas},
  or, more explicitly, as \emph{exhibiting $U$ as an atlas of $X$}.
\end{remark}
\begin{proposition}
  Effective epimorphisms are preserved by $\infty$-pullback.
  \label{1EpimorphismsArePreservedByPullback}
\end{proposition}
This is prop. 6.2.3.15 in  \cite{Lurie}.
\begin{proposition}
   \label{EffectiveEpiDetectedOnoTruncation}
    \item A morphism $p : X \to Y$ is an effective epimorphism
	precisely if its 0-truncation $\tau_0 p : \tau_0 X \to \tau_0 Y$, 
	def. \ref{truncated object}, is an effective epimorphism, hence
 	an epimorphism, in the 
	1-topos of 0-truncated objects.
\end{proposition}
This is prop. 7.2.1.14 in \cite{Lurie}.
\begin{example}
  A morphism in $\infty\mathrm{Grpd}$ is effective epi precisely
  if it induces an epimorphism $\pi_0(X) \to \pi_0(Y)$
  of sets of connected components.
\end{example}

\subsubsection{Presentations}

We discuss presentations of various classes of $\infty$-limits
and $\infty$-colimits in an $\infty$-category by \emph{homotopy limits}
and \emph{homotopy colimits} in categories with weak equivalences
presenting them.

\paragraph{$\infty$-Pullbacks}
\label{InfinityPullbackAndHomotopyPullback}

We discuss here tools for computing $\infty$-pullbacks in an $\infty$-category
$\mathbf{H}$ in terms of homotopy pullbacks in a homotopical 1-category
presenting it.

\medskip

\begin{proposition}
  \label{ConstructionOfHomotopyLimits}
  Let $A \to C \leftarrow B$ be a cospan diagram in a model category, 
  def. \ref{model category}.
  Sufficient conditions for the ordinary pullback $A \times_C B$ to be 
  a homotopy pullback are
  \begin{itemize}
    \item 
	   one of the two morphisms is a fibration and all three objects are fibrant;
	\item 
	   one of the two morphisms is a fibration and the model structure is right proper.
  \end{itemize}
\end{proposition}
This appears for instance as prop. A.2.4.4 in \cite{Lurie}.

It remains to have good algorithms for identifying fibrations and 
for resolving morphisms by fibrations. 
A standard recipe for constructing fibration resolutions is
\begin{proposition}[factorization lemma]
  \label{FactorizationLemma}
  Let $B \to C$ be a morphism between fibrant objects in a model category
  and let $\xymatrix{C \ar[r]^\simeq & C^I \ar@{->>}[r] & C \times C}$ be a path object
  for $B$. Then the composite vertical morphism in 
  $$
    \xymatrix{
	  C^I \times_C B \ar[d] \ar[r] \ar@{->>}@/_1pc/[dd] & B \ar[d]
	  \\
	  C^I \ar[d]\ar[r] & C
	  \\
	  C
	}
  $$
  is a fibrantion replacement of $B \to C$. 
\end{proposition}
This appears for instance on p. 4 of \cite{Brown}.
\begin{corollary}
  For $A \to C \leftarrow B$ a diagram of fibrant objects in a model category, 
  its homotopy pullback is presented by the ordinary limit $A \times_C^h B$
  in
  $$
    \raisebox{35pt}{
    \xymatrix{
	  A \times_C^h B \ar[r] \ar[dd] 
	   & 
	   C^I \times_C B \ar[r] \ar[d]
	  & B
	  \ar[d]
	  \\
	  & C^I \ar[r] \ar[d] & C
	  \\
	  A \ar[r] & C
	}
	}
	\,,
  $$
  which is, up to isomorphism, the same as the ordinary pullback in 
  $$
    \raisebox{20pt}{
    \xymatrix{
	  A \times_C^h B \ar[r] \ar[d] & C^I \ar[d]
	  \\
	  A \times B \ar[r] & C \times C
	}
	}
	\,.
  $$
\end{corollary}
\begin{remark}
  For the special case of ``abelian'' objects 
  another useful way of constructing fibrations is 
  via  the \emph{Dold-Kan correspondence}, wich we discuss
  in \ref{SheafAndNonabelianDoldKan}. As described there,
  a morphism between simplicial presheaves that 
  arise from presheaves of chain complexes is a fibration
  (in the projective model structure on simplicial presheaves)
  if it arises from a degreewise surjection of chain complexes.
\end{remark}

\paragraph{Finite $\infty$-limits of $\infty$-sheaves}
\label{FiniteInfinityLimitsOfInfinitySheaves}

We discuss presentations for finite $\infty$-limits specifically 
in $\infty$-toposes.

\medskip

\begin{proposition}
  \label{PullbacksAlongLocalFibrationsAreHomotopyPullbacks}
  Let $C$ be a site with enough points, def. \ref{site with enough points}. Write
  $\mathbf{H} \simeq (\mathrm{Sh}(C)^{\Delta^{\mathrm{op}}}, W)$
  for the hypercomplete $\infty$-topos over $C$, where $W$ is the class of 
  local weak equivalences, theorem \ref{CharacterizationOfLocalWeakEquivalence}.

  Then pullbacks in $\mathrm{Sh}(C)^{\Delta^{\mathrm{op}}}$ along 
  local fibrations, def. \ref{LocalFibrations}, 
  are homotopy pullbacks, hence present $\infty$-pullbacks
  in $\mathbf{H}$.  
\end{proposition}
\proof
  Let 
  $\xymatrix{
    A \ar@{->>}[r]^{\mathrm{loc}} & C \ar@{<-}[r] & B
  }$
  be a cospan with the left leg a local fibration. 
  By the existence of the projective local model structure
  $[C^{\mathrm{op}}, \mathrm{sSet}]_{\mathrm{proj}, \mathrm{loc}}$
  there exists a morphism of diagrams
  $$
    \raisebox{20pt}{
    \xymatrix{
    A \ar@{->>}[r]^{\mathrm{loc}} \ar[d]^\simeq & C \ar[d]^{\simeq} \ar@{<-}[r] & B \ar[d]^{\simeq}
	\\
    A' \ar@{->>}[r]^{} & C' \ar@{<-}[r] & B'
  }}
  \,,
  $$
  where the bottom cospan is a fibrant diagram with respect to the projective local 
  model structure, hence a cospan of
  genuine fibrations between fibrant objects, so that the ordinary
  pullback $A' \times_{C'} B'$ is a presentation of the homotopy pullback
  of the original diagram.
  Here the vertical morphisms
  are weak equivalences, and by theorem \ref{CharacterizationOfLocalWeakEquivalence}
  this means that they are stalkwise weak equivalences of simplicial sets.
  Moreover, by the nature of left Bousfield localization, the genuine
  fibrations are in particular global projective fibrations, hence in particular
  are stalkwise fibrations.
  
  Now for $p : \mathrm{Set} \to \mathrm{Sh}(C)$ any topos point, 
  the stalk functor $p^*$ preserves
  finite limits and hence preserves (the sheafification of) the above pullbacks. So by the asumption 
  that $A \to C$ is a local fibration, the simplicial set
  $p^*(A \times_C B)$ is a pullback of simplicial sets along a Kan fibration,
  hence, by the right properness of $\mathrm{sSet}_{\mathrm{Quillen}}$, 
  and using prop. \ref{ConstructionOfHomotopyLimits}, is
  a homotopy pullback there. Moreover, the induced morphism
  $p^*(A \times_C B) \to p^* (A' \times_{C'} B')$ is therefore
  a morphism of homotopy pullbacks along a weak equivalence of diagrams.
  This means that it is itself a weak equivalence. Since this is true 
  for all topos points, it follows that $A \times_C B \to A' \times_{C'} B'$
  is a stalkwise weak equivalence, hence a weak equivalence, hence that $A \times_C B$
  is itself already a model for the homotopy pullback.
\endofproof

The following proposition establishes the model category 
analog of the statement that 
by left exactness of $\infty$-sheafification,
finite $\infty$-limits of $\infty$-sheafified $\infty$-presheaves may be computed as the 
$\infty$-sheafification of the finite $\infty$-limit of the $\infty$-presheaves.
\begin{proposition} 
  \label{FiniteHomotopyLimitsInPresheaves}
  \index{pullback!presentation by homotopy limit}
Let $C$ be a site and
$F : D \to [C^{\mathrm{op}}, \mathrm{sSet}]$ be a finite diagram.

Write $\mathbb{R}_{\mathrm{glob}}\lim\limits_{\leftarrow} F \in [C^{op}, \mathrm{sSet}]$ 
for (any representative of) the homotopy limit over $F$ computed in the global model 
structure $[C^{\mathrm{op}}, \mathrm{sSet}]_{\mathrm{proj}}$, 
well defined up to isomorphism in the homotopy category. 

Then $\mathbb{R}_{\mathrm{glob}}\lim\limits_{\leftarrow} F \in [C^{\mathrm{op}},\mathrm{sSet}]$ 
presents also the homotopy limit of $F$ in the local model structure 
$[C^{\mathrm{op}}, \mathrm{sSet}]_{\mathrm{proj},\mathrm{loc}}$.
\end{proposition}
\proof
By \cite{Lurie}, theorem 4.2.4.1, we have that the homotopy limit
$\mathbb{R}\lim\limits_{\leftarrow}$ computes the corresponding 
$\infty$-limit.
Since $\infty$-sheafification $L$ is by definition a left exact
$\infty$-functor it preserves these finite $\infty$-limits:
$$
  \xymatrix{
     ([D, [C^{\mathrm{op}}, \mathrm{sSet}]_{\mathrm{proj}, \mathrm{loc}}]_{\mathrm{inj}})^\circ
     \ar[d]^{\mathbb{R}\lim\limits_{\longleftarrow}}
     \ar@{<-}[r]^{L_*}
     &
     ([D, [C^{\mathrm{op}}, \mathrm{sSet}]_{\mathrm{proj}}]_{\mathrm{inj}})^\circ
     \ar[d]^{\mathbb{R}\lim\limits_{\longleftarrow}}
     \\
     ([C^{\mathrm{op}}, \mathrm{sSet}]_{\mathrm{proj},\mathrm{loc}})^\circ
     \ar@{<-}[r]^{L \simeq \mathbb{L} \mathrm{Id}}
     &
     ([C^{op}, \mathrm{sSet}]_{\mathrm{proj}})^\circ
  }
 \,.
$$
Here $L \simeq \mathbb{L} \mathrm{Id}$ is the left derived functor of the identity for the 
left Bousfield localization. Therefore for $F$ a finit diagram in simplicial presheaves, its
homotopy limit in the local model structure $\mathbb{R}\lim_\leftarrow L_* F$ 
is equivalently computed by $\mathbb{L}\mathrm{Id} \mathbb{R}\lim_\to F$, with 
$\mathbb{R}\lim_\leftarrow F$ the homotopy limit in the global model structure. 
\endofproof
Together with \ref{InfinityPullbackAndHomotopyPullback}, this provides an efficient
algorithm for computing presentations of $\infty$-pullbacks in a model
structure on simplicial presheaves.
\begin{remark}
  \label{ComputingHomotopyPullbacks}
  Taken together, prop. \ref{FiniteHomotopyLimitsInPresheaves}, 
  prop. \ref{ConstructionOfHomotopyLimits} and definition 
  \ref{ModelPresentationsOfInfinityToposes} imply
  that we may compute
  $\infty$-pullbacks in an $\infty$-topos by the following algorithm:
  \begin{enumerate}
    \item Present the
      $\infty$-topos by a local \emph{projective} model structure on simplicial presheaves;
	\item 
     find a presentation of the morphisms to be pulled back 
     such that one of them is over each object of the site a Kan fibration
	 of simplicial sets;
	\item 
     then form the ordinary pullback of simplicial presheaves,
      which in turn is over each object the ordinary pullback of simplicial sets.
  \end{enumerate}
  The resulting object presents the $\infty$-pullback of $\infty$-sheaves. 
\end{remark}

\paragraph{$\infty$-Colimits}
\label{InfinityColimits}

We collect some standard facts and tools concerning the computation
of homotopy colimits.

\medskip

\begin{proposition}
  Let $C$ be a combinatorial model category and let $J$ be a 
  small category. Then the colimit over $J$-diagrams in $C$ 
  is a left Quillen functor
  for the projective model structure on functors on $J$:
  $$
    \lim\limits_{\longrightarrow} : [J,C]_{\mathrm{proj}} \to C
	\,.
  $$
\end{proposition}
\proof
  For $C$ combinatorial, the projective model structure exists
  by \cite{Lurie} prop. A.2.8.2.
  The right adjoint to the colimit
  $$
    \mathrm{const} : C \to [J,C]_{\mathrm{proj}}
  $$
  is manifestly right Quillen for the projective model structure.
\endofproof
\begin{example}
  \label{CotowerCofibrancy}
  Write 
  $$
    (\mathbb{N}, \leq) := 
	\{
	  \xymatrix{
	    0 \ar[r] & 1 \ar[r] & 2 \ar[r] & \cdots
	  }
	\}
  $$
  for the \emph{cotower category}. A cotower
  $
    X_0 \to X_1 \to A_2 \to \cdots
  $
  in a model category $C$ is projectively cofibrant precisely if
  \begin{enumerate}
    \item every morphism $X_i \to X_{i+1}$ is a cofibration in $C$;
	\item the first object $X_0$, and hence all objects $X_i$, are cofibrant in $C$.
  \end{enumerate}
  Therefore a sequential $\infty$-colimit over a cotower is presented by the
  ordinary colimit of a presentation of this cotower where all morphisms are
  cofibrations and all objects are cofibrant.
\end{example}
 This is a simple example, but since we will need 
 details of this at various places, we spell out the proof for the record.\\
\proof
  Given a cotower $X_\bullet$ with properties as stated, we need to check that for
  $p_\bullet : A_\bullet \to B_\bullet$ a morphism of cotowers such that for all
  $n \in \mathbb{N}$ the morphism $ p_n : A_n \to B_n$ is an acyclic fibration
  in $C$, and for $f_\bullet : X_\bullet \to B_\bullet$ any morphism, there is a
  lift $\hat f_\bullet$ in
  $$
    \raisebox{20pt}{
    \xymatrix{
	  & A_\bullet
	  \ar@{->>}[d]^{p_\bullet}
	  \\
	  X_\bullet \ar[r]^{f_\bullet} \ar@{-->}[ur]^{\hat f_\bullet} & B_\bullet
	}
	}
	\,.
  $$
  This lift we can construct by induction on $n$. For $n = 0$ we only need a lift in
  $$
    \raisebox{20pt}{
    \xymatrix{
	  & A_0
	  \ar@{->>}[d]^{p_0}
	  \\
	  X_0 \ar[r]^{f_0} \ar@{-->}[ur]^{\hat f_0} & B_0
	}
	}
	\,,
  $$
  which exists by assumption that $X_0$ is cofibrant. 
  Assume then that a lift has been for $f_{\leq n}$.
  Then the next lift $\hat f_{n+1}$ needs to make the diagram
  $$
    \xymatrix{
	  && A_n \ar[dr]
	  \ar[d]
	  \\
	  X_n
	  \ar@{^{(}->}[dr] \ar[urr]^{\hat f_n}
	  \ar[rr]
	  &&
	  B_n
	  &
	  A_{n+1} \ar@{->>}[d]
	  \\
	  & X_{n+1} \ar[rr]_{f_{n+1}} \ar@{-->}[urr]|{\hat f_{n+1}} && B_{n+1}
	}
  $$
  commute. Such a lift exists now by assumption that $X_n \to X_{n+1}$ is a cofibration.
  
  Conversely, assume that $X_\bullet$ is projectively cofibrant.
  Then first of all it has the left lifting property against all
cotower morphisms of the form 
  $$
    \xymatrix{
	  A_0 \ar[r] \ar@{->>}[d]^\simeq & {*} \ar[d] \ar[r] & {*} \ar[r] \ar[d] & \cdots
      \\
	  B_0 \ar[r] & {*} \ar[r] & {*} \ar[r] & \cdots      	  
	}
	\,.
  $$  
  Such a lift is equivalent to a lift of $X_0$ against 
  $\xymatrix{A_0 \ar@{->>}[r]^\simeq & B_0}$ and hence $X_0$ is cofibrant in $C$.
  To see that every morphism $X_n \to X_{n+1}$ is a cofibration, notice that for
  every lifting problem in $C$ of the form
  $$
    \xymatrix{
	  X_n \ar[d] \ar[r] & A \ar@{->>}[d]^\simeq
	  \\
	  X_{n+1} \ar[r] & B
	}
  $$
  the cotower lifting problem of the form
  $$
    \xymatrix{
	  & X_0 \ar[r] \ar@{=}[d] & \cdots \ar[r] & X_n \ar@{=}[d] \ar[r] 
	    & A \ar[d] \ar[r] & {*} \ar@{=}[d] \ar[r] & {*} \ar[r] \ar@{=}[d] & \cdots
	  \\
	  & X_0 \ar[r] & \cdots \ar[r] & X_n \ar[r] & B \ar[r] & {*} \ar[r] & {*} \ar[r] & \cdots
	  \\
	  X_0 \ar@{=}[ur]
	  \ar[r]
	  &
	  \cdots
	  \ar[r]
	  &
	  X_n
	  \ar@{=}[ur]
	  \ar[r]
	  &
	  X_{n+1}
	  \ar[ur]
	  \ar[r]
	  &
	  \cdots
	}
  $$
  is equivalent.
\endofproof
For less trivial diagram categories it quickly becomes hard to obtain
projective cofibrant resolutions.
In these cases it is often it is useful to compute the (homotopy) colimit 
instead as a special 
case of a (homotopy) coend.\begin{proposition}
  \label{CoendOverQuillenBifunctIsQuillenBifunct}
  Let $F : A \times B \to C$  be a Quillen bifunctor, def. \ref{LeftQuillenBifunctor}, 
  and let $J$ be a Reedy category, then
  the coend over $F$ (see \cite{Kelly})
  $$
    \int^S F(-,-) : [J, A]_{\mathrm{Reedy}} \times [J^{\mathrm{op}}, B]_{\mathrm{Reedy}}
	\to C
  $$
  is a Quillen bifunctor from the product of the Reedy model categories 
  on functors with values in $A$ and $B$,
  respectively, to $C$.
  
  Similarly, if $A$ and $B$ are combinatorial model categories and $J$ is any small category,
  then the coend
  $$
    \int^S F(-,-) : [J, A]_{\mathrm{proj}} \times [J^{\mathrm{op}}, B]_{\mathrm{inj}}
	\to C
  $$
  is a Quillen bifunctor.
\end{proposition}
This appears in \cite{Lurie} as prop. A.2.9.26 and remark A.2.9.27.

\begin{proposition}
  \label{HomotopyColimitByDerivedCoend}
  If $\mathcal{V}$ is a closed monoidal model category, $C$ is a $\mathcal{V}$-enriched
  model category, and $J$ is a small category which is Reedy, then the homotopy colimit
  of $J$-shaped diagrams in $C$
  is presented by the left derived functor of
  $$
    \int^J (-)\cdot Q_{\mathrm{Reedy}}(I) : [J,C]_{\mathrm{Reedy}} \to C
	\,,
  $$
  where $Q_{\mathrm{Reedy}}(I)$ is a cofibrant replacement of the 
  functor constant in the tensor unit in 
  $[J^{\mathrm{op}}, \mathcal{V}]_{\mathrm{Reedy}}$, and where
  $$
    (-) \cdot (-)  : C \times \mathcal{V} \to C
  $$  
  is the given $\mathcal{V}$-tensoring of $C$.
  Similarly, if $J$ is not necessarily Reedy, but $\mathcal{V}$ and $C$
  are combinatorial, then the homotopy colimit is also 
  given by the left derived functor of 
  $$
    \int^J (-)\cdot Q_{\mathrm{proj}}(I) : [J,C]_{\mathrm{inj}} \to C
	\,,
  $$
  where now $Q_{\mathrm{proj}}(I)$ is a cofibrant resolution of the tensor unit
  in $[J^{\mathrm{op}}, \mathcal{V}]_{\mathrm{proj}}$.
\end{proposition}
This is nicely discussed in \cite{Gambino}.\\
\proof
  By definition of enriched category, the $\mathcal{V}$-tensoring operation is a left
  Quillen bifunctor. With this the statement follows from 
  prop. \ref{CoendOverQuillenBifunctIsQuillenBifunct}.
\endofproof
Various classical facts of model category theory are special cases of 
these formulas.

\paragraph{$\infty$-Colimits over simplicial diagrams}
\label{InfinityColimitOverSimplicialDiagram}

We discuss here a standard presentation of 
\emph{homotopy colimits over simplical diagrams} given by the 
\emph{diagonal simplicial set} or the \emph{total simplicial set} 
associated with a bisimplicial set.

\medskip

\begin{proposition}
  \label{TheSimplexAndTheFatSimplex}
  Write $[\Delta, \mathrm{sSet}]$ for the category of cosimplicial simplicial sets.
  For $\mathrm{sSet}$ equipped with its cartesian monoidal structure, the
  tensor unit is the terminal object $*$.
  \begin{itemize}
    \item The \emph{simplex functor}
	  $$
	    \Delta : [n] \mapsto \Delta[n] := \Delta(-,[n])
	  $$
	  is a cofibrant resolution of $*$ in $[\Delta, \mathrm{sSet}_{\mathrm{Quillen}}]_{\mathrm{Reedy}}$;
    \item 
     the \emph{fat simplex functor}
     $$
	   \mathbf{\Delta} : [n] \mapsto N(\Delta/[n])
     $$	 
	 is a cofibrant resolution of $*$ in $[\Delta, \mathrm{sSet}_{\mathrm{Quillen}}]_{\mathrm{proj}}$.
  \end{itemize}
\end{proposition}
\begin{proposition}
  \label{BousfieldKanFormula}
  Let $C$ be a simplicial model category and $F : \Delta^{\mathrm{op}} \to C$
  a simplicial diagram
  \begin{enumerate}
    \item 
	  If every monomorphism in $C$ is a cofibration, then the homotopy colimit over
	  $F$ is given by the realization
	  $$
	    \mathbb{L}\lim\limits_{\to} F \simeq \int^{[n] \in \Delta} F([n]) \cdot \Delta[n]
		\,.
	  $$
	\item
	  If $F$ takes values in cofibrant objects, then the homotopy colimit 
	  over $F$ is given by the fat realization
	  $$
	    \mathbb{L}\lim\limits_{\to} F \simeq \int^{[n] \in \Delta} F([n]) \cdot \mathbf{\Delta}[n]
		\,.
	  $$
	\item If $F$ is Reedy cofibrant, then the canonical morphism
	$$
	  \int^{[n] \in \Delta} F([n]) \cdot \mathbf{\Delta}[n]
	  \to 
	  \int^{[n] \in \Delta} F([n]) \cdot \Delta[n]
	$$
	(the \emph{Bousfield-Kan map})
	is a weak equivalence.
  \end{enumerate}
\end{proposition}
\proof
  If every monomorphism is a cofibration, then $F$ is 
  necessarily cofibrant in $[\Delta^{\mathrm{op}}, C]_{\mathrm{Reedy}}$. 
  The first statement then follows from prop. \ref{HomotopyColimitByDerivedCoend}
  and the first item in prop. \ref{TheSimplexAndTheFatSimplex}.
  On the other hand, if $F$ takes values in cofibrant objects, then it is
  cofibrant in $[\Delta^{\mathrm{op}}, C]_{\mathrm{inj}}$,
  and so the second statement follows from 
  prop. \ref{HomotopyColimitByDerivedCoend} and the second item in 
  prop. \ref{TheSimplexAndTheFatSimplex}.
  
  Notice that projective cofibrancy implies
  Reedy cofibrancy, so that $\mathbf{\Delta}$ is also Reedy cofibrant.
  Therefore the morphism in the last item of the proposition is,
  by remark \ref{QuillenBifunctorWithOneArgumentFixedAndCofibrant},
  the image under a left Quillen functor of a weak equivalence between
  cofibrant objects and therefore itself a weak equivalence.
\endofproof
An important example of this general situation is the following.
\begin{proposition}
  Every simplicial set, and more generally every simplicial presheaf
  is the homotopy colimit over its simplicial diagram of cells.
  Precisely, let $C$ be a small site, and let 
  $[C^{\mathrm{op}}, \mathrm{sSet}_{\mathrm{Quillen}}]_{\mathrm{inj}, \mathrm{loc}}$
  be the corresponding local injective model structure on simplicial presheaves.
  Then for any $X \in [C^{\mathrm{op}}, \mathrm{sSet}]$, with 
  $$
    X_\bullet : \Delta^{\mathrm{op}} \to [C^{\mathrm{op}}, \mathrm{Set}]
	 \hookrightarrow
	 [C^{\mathrm{op}}, \mathrm{sSet}_{\mathrm{Quillen}}]
  $$
  its simplicial diagram of components, we have
  $$
    X \simeq \mathbb{L}\lim\limits_{\longrightarrow} X_\bullet
	\,.
  $$
  \label{SimplicialSetIfHocolimOverItsCompnentDiagram}
\end{proposition}
\proof
  By prop. \ref{BousfieldKanFormula} the homotopy colimit is given by the coend
  $$
    \mathbb{L}\lim\limits_{\longrightarrow} X_\bullet
	\simeq
	\int^{[n] \in \Delta} X_n \times \Delta[n]
	\,.
  $$
  By basic properties of the coend, this is isomorphic to $X$.
\endofproof

\begin{proposition}
  \label{SimplicialHocolimGivenByDiagonal}
  The homotopy colimit of a simplicial diagram in $\mathrm{sSet}_{\mathrm{Quillen}}$,
  or more generally of a simplicial diagram of simplicial presheaves, is given by
  the diagonal of the corresponding bisimplicial set / bisimplicial presheaf.
  
  More precisely, for
  $$
    F : \Delta^{\mathrm{op}} \to [C^{\mathrm{op}}, \mathrm{sSet}_{\mathrm{Quillen}}]_{\mathrm{inj}, \mathrm{log}}
  $$
  a simplicial diagram, its homotopy colimit is given by
  $$
    \mathbb{L} \lim\limits_{\longrightarrow} F_\bullet 
	  \simeq 
	  d F : ([n] \mapsto (F_n)_n)
	\,.
  $$
\end{proposition}
\proof
  By prop. \ref{BousfieldKanFormula}
  the homotopy colimit is given by the coend
  $$
    \mathbb{L}\lim\limits_{\longrightarrow} F_\bullet \simeq 
	\int^{[n] \in \Delta} F_n \cdot \Delta[n]
	\,.
  $$
  By a standard fact (e.g. exercise 1.6 in \cite{GoerssJardine}), 
  this coend is in fact isomorphic to the diagonal.
\endofproof

\begin{definition}
  \label{TotalSimplicialSetAndTotalDecalage}
  Write $\Delta_a$ for the \emph{augmented simplex category},
  which is the simplex category with an initial object adjoined, denoted
  $[-1]$.
  
  This is a symmetric monoidal category with tensor product
  being the \emph{ordinal sum} operation
  $$
    [k], [l] \mapsto [k+l +1]
	\,.
  $$
  Write
  $$
    \sigma : \Delta \times \Delta \to \Delta
  $$
  for the restriction of this tensor product along the canonical 
  inclusion $\Delta \hookrightarrow \Delta_a$. Write
  $$
    \sigma^* : \mathrm{sSet} \to [\Delta^{\mathrm{op}}, \mathrm{sSet}]
  $$
  for the operation of precomposition with this functor. By right
  Kan extension this induces an adjoint pair of functors
  $$
    (\mathrm{Dec} \dashv T)
	:
    \xymatrix{
	  [\Delta^{\mathrm{op}}, \mathrm{sSet}]
	  \ar@<+4pt>@{<-}[r]^>>>>{\sigma^*}
	  \ar@<-2pt>@{->}[r]_>>>>{\sigma_*}
	  &
	  \mathrm{sSet}
	}
	\,.
  $$
  \begin{itemize}
    \item $\mathrm{Dec} := \sigma^*$ is called the \emph{total d{\'e}calage} functor;
	\item $T := \sigma_*$ is called the \emph{total simplicial set} functor.
  \end{itemize}
\end{definition}
The total simplicial set functor was introduced in \cite{ArtinMazur2}.
Details are in \cite{Stevenson2}.
\begin{remark}
  By definition, for  $X \in [\Delta^{\mathrm{op}}, \mathrm{sSet}]$,
  its total d{\'e}calage is the bisimplicial set given by  
  $$
    (\mathrm{Dec} X)_{k,l}  = X_{k+l+1}
	\,.
  $$
\end{remark}
\begin{remark}
  \label{Total simplicial object is built from finite limits}
  For $X \in [\Delta^{\mathrm{op}}, \mathrm{sSet}]$, the simplicial set
  $T X$ is in each degree given by an equalizer of maps between finite products
  of components of $X$. Hence forming $T$ is compatible with sheafification
  and other processes that preserve finite limits.
\end{remark}
See \cite{Stevenson2}, equation (2).
\begin{proposition}
  \label{TotalSimpSetEquivalentToDiagonal}
  For every $X \in [\Delta^{\mathrm{op}}, \mathrm{sSet}]$
  \begin{itemize}
    \item 
	  the canonical morphism 
	  $$ 
	    d X \to T X
	  $$
	  from the diagonal to the total simplicial set
	  is a weak equivalence in $\mathrm{sSet}_{\mathrm{Quillen}}$;
	\item
	  the adjunction unit
	  $$
	    X \to T \mathrm{Dec} X
	  $$
	  is a weak equivalence in $\mathrm{sSet}_{\mathrm{Quillen}}$.
  \end{itemize}
  For every $X \in \mathrm{sSet}$
  \begin{itemize}
    \item 
	   there is a natural isomorphism $T \mathrm{const} X \simeq X$.
  \end{itemize}
\end{proposition}
This is due to \cite{CegarraRemedios}\cite{Stevenson2}.
\begin{corollary}
  \label{SimplicialHomotopyColimitByCodiagonal}
  For 
  $$
    F : \Delta^{\mathrm{op}} \to [C^{\mathrm{op}}, \mathrm{sSet}_{\mathrm{Quillen}}]_{\mathrm{inj}, \mathrm{loc}}
  $$ a simplicial object in simplicial presheaves, its homotopy colimit is given by
  applying objectwise over each $U \in C$ the total simplicial set functor
  $$
    \mathbb{L} \lim\limits_{\longrightarrow} F \simeq (U \mapsto T F(U))
	\,.
  $$
\end{corollary}
\proof
  By prop. \ref{TotalSimpSetEquivalentToDiagonal} this follows from prop. 
  \ref{SimplicialHocolimGivenByDiagonal}.
\endofproof
\begin{remark}
  The use of the total simplicial set instead of the diagonal simplicial set in the 
  presentation of simplicial homotopy colimits is useful and 
  reduces to various traditional notions in particular
  in the context of group objects and action groupoid objects. This we discuss below in 
  \ref{InfinityGroupPresentations} and \ref{Universal princial bundles}.
\end{remark}

\paragraph{Effective epimorphisms, atlases and d{\'e}calage}

We  discuss  apsects of the presentation of
effective epimorphisms, def. \ref{EffectiveEpimorphism}, with respect to presentations of the 
ambient $\infty$-topos by categories of simplicial presheaves,
\ref{InfinityToposPresentation}.

\medskip

\begin{observation}
  \label{CanonicalAtlasOfSimplicialPresheaf}
  If the $\infty$-topos $\mathbf{H}$ is presented by a category of
  simplicial presheaves, \ref{InfinityToposPresentation},
  then for $X$ a simplicial presheaf the canonical morphism
  of simplicial presheaves
  $\mathrm{const} X_0 \to X$ that includes the presheaf of 
  0-cells as a simplicially constant simplicial presheaf
  presents an effective epimorphism in $\mathbf{H}$.
\end{observation}
\proof
    By prop. \ref{EffectiveEpiDetectedOnoTruncation}.
\endofproof
\begin{remark}
  In practice the presentation of an $\infty$-stack by a simplicial presheaf is
  often taken to be understood, and then observation \ref{CanonicalAtlasOfSimplicialPresheaf}
  induces also a canonical atlas.
\end{remark}

We now discuss a fibration resolution of the canonical atlas.
Let $\sigma : \Delta \times \Delta \to \Delta$ the functor
from def. \ref{TotalSimplicialSetAndTotalDecalage}, defining
\emph{total d{\'e}calage}.
\begin{definition}
  \label{Decalage}
  Write
  $$
    \mathrm{Dec}_0 : \mathrm{sSet} \to \mathrm{sSet}
  $$
  for the functor given by precomposition with $\sigma(-,[0]) : \Delta \to \Delta$,
  and 
  $$
    \mathrm{Dec}^0 : \mathrm{sSet} \to \mathrm{sSet}
  $$
  for the functor given by precomposition with $\sigma([0],-) : \Delta \to \Delta$.
  This is called the plain \emph{d{\'e}calage functor} or \emph{shifting functor}.
\end{definition}
This functor was introduced in \cite{Illusie2}. A discussion in the present 
context is in section 2.2 of \cite{Stevenson2}.
\begin{proposition}
  \label{MorphismsOutOfPlainDecalage}
  The d{\'e}calage of $X$ is isomorphic to the simplicial set
  $$
    \mathrm{Dec}_0 X = \mathrm{Hom}( \Delta^\bullet \star \Delta[0], X)
	\,,
  $$
  where $(-)\star (-) : \mathrm{sSet} \times \mathrm{sSet} \to \mathrm{sSet}$
  is the join of simplicial sets. 
  The canonical inclusions $\Delta[n], \Delta[0] \to \Delta[n] \star \Delta[0]$
  induce two canonical morphisms
  $$
    \xymatrix{
	   \mathrm{Dec}_0 X \ar[d]^\simeq \ar@{->>}[r] & X
	   \\
	   \mathrm{const} X_0
	}
	\,,
  $$
  where 
  \begin{itemize}
    \item the horizontal morphism is given in degree $n$ by $d_{n+1} : X_{n+1} \to X_n$;
	\item the horizontal morphism is a Kan fibration;
	\item the vertical morphism is a weak homotopy equivalence;
	\item a weak homotopy inverse is given by the morphism that is degreewise given by
	  the degeneracy morphisms in $X$.
  \end{itemize}
\end{proposition}
\proof
  The relation to the join of simplicial sets is nicely discussed around page 
  7 of \cite{RobertsStevenson}. The weak homotopy equivalence is classical,
  see for instance \cite{Stevenson2}.
  
  To see that $\mathrm{Dec}_0 X \to X$ is 
  a Kan fibration,
  notice that for all $n \in \mathbb{N}$ we have
  $(\mathrm{Dec}_0 X)_n = \mathrm{Hom}(\Delta[c] \star \Delta[0], X)$,
  where $(-)\star (-) : \mathrm{sSet} \times \mathrm{sSet} \to \mathrm{sSet}$
  is the join of simplicial sets. Therefore the lifting problem
  $$
    \xymatrix{
	  \Lambda^i[n] \ar[r] \ar[d] & \mathrm{Dec}_0 X \ar[d]
	  \\
	  \Delta[n] \ar[r] & X
	}
  $$
  is equivalently the lifting problem
  $$
   \raisebox{20pt}{
    \xymatrix{
	  (\Lambda^i[n] \star \Delta[n] ) \coprod_{\Lambda^i[n]} \Delta[n]
	  \ar[r]
	  \ar[d]
	  &
	  X
	  \ar[d]
	  \\
	  \Delta[n] \star \Delta[0]
	  \ar[r]
	  &
	  {*}
	}
	}\,.
  $$
  Here the left moprhism is a anodyne morphism, in fact is an $(n+1)$-horn 
  inclusion. Hence a lift exists if $X$ is a Kan complex. 
  (Alternatively, 
  notice that $\mathrm{Dec}_0 X$ is the disjoint union of
  slices $X_{/x}$ for $x \in X_0$. By cor. 2.1.2.2 in \cite{Lurie} 
  the projection $X_{/x} \to X$ is a left fibration if $X$ is Kan fibrant, and
  by prop. 2.1.3.3 there this implies that it is a Kan fibration).
 \endofproof
\begin{corollary}
  \label{DecalageIsFibrationResolution}
 For $X$ in $[C^{\mathrm{op}}, \mathrm{sSet}]_{\mathrm{proj}}$ 
 fibrant, a fibration resolution of the 
 canonical effective epimorphism $\mathrm{const} X_0 \to X$
 from observation \ref{CanonicalAtlasOfSimplicialPresheaf}
 is given by the d{\'e}calage morphism
 $\mathrm{Dec}_0 X \to X$, def. \ref{Decalage}.
\end{corollary}
\proof
  It only remains to observe that we have a commuting diagram
  $$
    \raisebox{20pt}{
    \xymatrix{
	  \mathrm{const}X_0 \ar[r]^s \ar[d] & \mathrm{Dec}_0 X \ar[d]
      \\
      X \ar[r]^= & X	  
	}
	}
	\,,
  $$
  where the top morphism, given degreewise by the degeneracy maps in $X$,
  is a weak homotopy equivalence by classical results. 
\endofproof

\newpage

\section{Cohesive and differential homotopy type theory}
\label{GeneralAbstractTheory}

We discuss here the general abstract theory of 
\emph{cohesive $\infty$-toposes} and of \emph{differential cohesive $\infty$-toposes}
and of the homotopical, cohomological, geometrical and differential structures internal to them.

Below in \ref{Implementation} we construct models of these axioms.

\subsection{Introduction and survey}
\index{cohesive $\infty$-topos!contents}

A topos or $\infty$-topos may be viewed both as a category or, respectively, $\infty$-category
\emph{of} generalized spaces -- then also called a ``\emph{gros topos}'' --
or as a generalized space itself -- then also called a ``\emph{petit topos}''.
The duality relation between these two perspectives is 
given by prop. \ref{ToposEquivalentToItsEtaleToposes}, which says
that every $\infty$-topos regarded as a generalized space is equivalent to 
the $\infty$-category of generalized {\'e}tale spaces \emph{over} it,
while, conversely, every collection of generalized spaces encoded by an
$\infty$-topos may be understood as being those generalized spaces 
equipped with local equivalences to a fixed generalized model space.

From this description it is intuitively clear that 
the ``smaller'' an $\infty$-topos is when regarded as a generalized space,
the ``larger'' is the collection of generalized spaces locally modeled on it,
and vice versa. 
If by ``size'' we mean ``dimension'', there are two notions of 
\emph{dimension of an $\infty$-topos} $\mathbf{H}$ that coincide with 
the ordinary notion of dimension of a manifold $X$ when $\mathbf{H} = \mathrm{Sh}_\infty(X)$,
but which may be different in general. These are
\begin{itemize}
  \item homotopy dimension (see def. \ref{HomotopyDimension} below);
  \item cohomology dimension (\cite{Lurie}, section 7.2.2).
\end{itemize}
If by ``size'' we mean ``nontriviality of homotopy groups'', hence
nontriviality of \emph{shape} of a space, there is the notion of 
\begin{itemize}
  \item shape of an $\infty$-topos (\cite{Lurie}, section 7.1.6);
\end{itemize}  
which coincides with 
the topological shape of $X$ in the case that $\mathbf{H} = \mathrm{Sh}_\infty(X)$, as above.
Finally, if by ``small size'' we just mean \emph{finite dimensional}, then
the property of $\infty$-toposes reflecting that is
\begin{itemize}
  \item hypercompleteness (\cite{Lurie}, section 6.5.2).
\end{itemize} 

For the description of higher geometry and higher differential geometry,
we are interested in $\infty$-toposes that are ``maximally \emph{gros}''
and``minimally \emph{petit}'': regarded as generalized spaces they 
should look like \emph{fat points} or \emph{contractible blobs} 
being the abstract blob of \emph{geometry} that every object in 
them is supposed to be
locally modeled on, but that otherwise do not make these objects be
parameterized over a nontrivial space. 

The following notions of \emph{local $\infty$-topos}, 
\emph{$\infty$-connected $\infty$-topos}, \emph{cohesive $\infty$-topos},
and \emph{differential cohesive $\infty$-topos} describe extra properties of the global section
geometric morphism of an $\infty$-topos that imply that some or all 
of the measures of ``size'' of the $\infty$-topos vanish, hence that make
the $\infty$-topos be far from being a non-trivial generalized space
itself, and instead be genuinely a collection \emph{of} generalized spaces
modeled on some notion of local geometry.

All these properties are equivalently encoded in terms of
\emph{idempotent $\infty$-(co)monads} on the $\infty$-topos $\mathbf{H}$
$$
  \Box, \Diamond : \mathbf{H} \to \mathbf{H}
  \,.
$$ 
Internally, on the homotopy type theory language of $\mathbf{H}$, 
these are (higher) \emph{closure operators} or \emph{modalities} on the type system
(more on this is below in \ref{InternalCohesion}).
Externally, these structures correspond to adjunctions
$$
  (L \dashv R)
  :
  \xymatrix{
    \mathbf{H}
	\ar@{<-}@<+3pt>[r]^L
	\ar@<-3pt>[r]_R
	&
	\mathbf{B}
  }
$$
such that $L$ or $R$ is a fully faithful $\infty$-functor, by 
$\Box  \simeq L \circ R$ and $\Diamond \simeq R \circ L$, or the other way around.
\begin{proposition}
  Let $(L \dashv R) : \xymatrix{ \mathcal{C} \ar@<-3pt>[r]_{R} \ar@<+3pt>@{<-}[r]^{L} & \mathcal{D} }$
  be a pair of adjoint $\infty$-functors. Then
  \begin{enumerate}
    \item The left adjoint $\infty$-functor $L$ is fully faithful precisely if 
	 the adjunction unit is an equivalence 
	 $\xymatrix{\mathrm{id}_{\mathcal{D}} \ar[r]^\simeq & R \circ L }$.
    \item The right adjoint $\infty$-functor $R$ is fully faithful precisely if 
	 the adjunction counit is an equivalence 
	 $\xymatrix{ L \circ R \ar[r]^\simeq & \mathrm{id}_{\mathcal{C}}}$.
  \end{enumerate}
  \label{FullyFaithfulAdjunctions}
\end{proposition}
\proof
This is \cite{Lurie}, p. 308 or follows directly from it.
\endofproof

For encoding ``gros'' geometry in the above sense, here the comonadic
$\Box$ is itself to be part of an 
adjunction with the monadic $\Diamond$, as 
$\Box \dashv \Diamond$ or $\Diamond \dashv \Box$. 
Such a situation corresponds externally to adjoint triples of $\infty$-functors
$$
  (f_! \dashv f^* \dashv f_*)
  :
  \xymatrix{
    \mathbf{H}
	\ar@{->}@<+9pt>[r]^{f_!}
	\ar@{<-}@<+3pt>[r]|{f^*}
	\ar@<-3pt>[r]_{f_*}
	&
	\mathbf{B}
  }
  \;\;\;\;\;\;\;\;
  \mbox{or}
  \;\;\;\;\;\;\;\;
  (f^* \dashv f_* \dashv f^!)
  :
  \xymatrix{
    \mathbf{H}
	\ar@{<-}@<+3pt>[r]^{f^*}
	\ar@{->}@<-3pt>[r]|{f_*}
	\ar@{<-}@<-9pt>[r]_{f^!}
	&
	\mathbf{B}
  }$$
such that the middle functor or the two outer functors are fully faithful:
$$
  (\Diamond \dashv \Box) \simeq (f^* f_! \dashv f^* f_*)
  \;\;\;\;
  \mbox{or}
  \;\;\;\;
  (\Box \dashv \Diamond) \simeq ( f^* f_* \dashv f^! f_*)
  \,.
$$

All that matters for the nature of the induced modalities is in which 
direction these functors go and which of them are fully faithful. Moreover, 
both direction and fully faithfulness are necessarily alternating through the 
adjoint triple, so what really matters is only which functor we regard as the
direct image, the number of adjoints it has to the left and to the right, 
and whether it is itself fully faithful or its adjoints are. 
To bring that basic information out more clearly it may be helpful
to introduce the following condensed notation.

Let 
$ \xymatrix@C=20pt{
    \ar@{.}[rrr]
     &
    \ar@<+6pt>@{-}[r]
    \ar@<+6.4pt>@{-}[r]
    \ar@<+6.8pt>@{-}[r]
    \ar@<+7.2pt>@{-}[r]
    \ar@<+7.6pt>@{-}[r]
    \ar@<+0pt>@{-}[r]
    \ar@<+.4pt>@{-}[r]
    \ar@<+.8pt>@{-}[r]
    \ar@<+1.2pt>@{-}[r]
    \ar@<+1.6pt>@{-}[r]
	&&
  }$
stand for an adjoint pair where the direct image $f_*$ points from $\mathbf{H}$ to $\mathbf{B}$,
(this is the bar on the dotted baseline) and such that it has a single left adjoint $f^*$
(the second bar on top). Accordingly, if there is a further left adjoint $f_!$ then we
draw a further bar on top
$ \xymatrix@C=20pt{
    \ar@{.}[rrr]
     &
    \ar@<+12pt>@{-}[r]
    \ar@<+12.4pt>@{-}[r]
    \ar@<+12.8pt>@{-}[r]
    \ar@<+13.2pt>@{-}[r]
    \ar@<+13.6pt>@{-}[r]
    \ar@<+6pt>@{-}[r]
    \ar@<+6.4pt>@{-}[r]
    \ar@<+6.8pt>@{-}[r]
    \ar@<+7.2pt>@{-}[r]
    \ar@<+7.6pt>@{-}[r]
    \ar@<+0pt>@{-}[r]
    \ar@<+.4pt>@{-}[r]
    \ar@<+.8pt>@{-}[r]
    \ar@<+1.2pt>@{-}[r]
    \ar@<+1.6pt>@{-}[r]
	&&
  }$.  If there is a further right adjoint $f^!$ then we draw a further bar on the bottom
  $ \xymatrix@C=20pt{
    \ar@{.}[rrr]
     &
    \ar@<+6pt>@{-}[r]
    \ar@<+6.4pt>@{-}[r]
    \ar@<+6.8pt>@{-}[r]
    \ar@<+7.2pt>@{-}[r]
    \ar@<+7.6pt>@{-}[r]
    \ar@<+0pt>@{-}[r]
    \ar@<+.4pt>@{-}[r]
    \ar@<+.8pt>@{-}[r]
    \ar@<+1.2pt>@{-}[r]
    \ar@<+1.6pt>@{-}[r]
     \ar@<-6pt>@{-}[r]
    \ar@<-6.4pt>@{-}[r]
    \ar@<-6.8pt>@{-}[r]
    \ar@<-5.2pt>@{-}[r]
    \ar@<-5.6pt>@{-}[r]
	&&
  }$. And so forth: bars on top are left adjoint to bars below them, and the direction 
  is left-to-right for the bar on the base line and for every second bar next to it, while
  it is right-to-left for every other bar. Finally, we mark the fully faithful functors
  by breaking the corresponding bar. For instance the notation
  $ \xymatrix@C=20pt{
    \ar@{.}[rrr]
     &
    \ar@<+6pt>@{-}[r]|{\ }
    \ar@<+6.4pt>@{-}[r]|{\ }
    \ar@<+6.8pt>@{-}[r]|{\ }
    \ar@<+7.2pt>@{-}[r]|{\ }
    \ar@<+7.6pt>@{-}[r]|{\ }
    \ar@<+0pt>@{-}[r]
    \ar@<+.4pt>@{-}[r]
    \ar@<+.8pt>@{-}[r]
    \ar@<+1.2pt>@{-}[r]
    \ar@<+1.6pt>@{-}[r]
	&&
  }$
  means that the inverse image is fully faithful, hence is shorthand for an adjunction
  of the form
  $
    \xymatrix{
	  \mathbf{H}
	  \ar@{<-^{)}}@<+3pt>^{f^*}[r]
	  \ar@<-3pt>_{f_*}[r]
	  &
	  \mathbf{B}
	}
  $, and so forth.

\newpage
  
The following table lists, in the above notation, the possibilities for adjoint higher
modalities together with the name of the corresponding attribute 
of $\mathbf{H}$ as an $\infty$-topos over the base $\mathbf{B}$.

\medskip
\medskip
\medskip

\noindent {\bf Locality} $(\flat \dashv \sharp)$ (section \ref{LocalToposes}).

$$
  \xymatrix@C=20pt{
    &&
	\ar@{}[r]|{\mbox{locally} \atop \mbox{local}}
	&&&&
	\ar@{}[r]|{\mbox{local}}
	&&&&
	\ar@{}[r]|{{\mbox{locally} \atop \mbox{local}} \atop \mbox{embedded}}
	&&&&
	\ar@{}[r]|{\mbox{discrete}}
	&&&
    \\
	\\
    \ar@{.}[rrrrrrrrrrrrrrrrr]
     &&
    \ar@<+6pt>@{-}[r]
    \ar@<+6.4pt>@{-}[r]
    \ar@<+6.8pt>@{-}[r]
    \ar@<+7.2pt>@{-}[r]
    \ar@<+7.6pt>@{-}[r]
    \ar@<+0pt>@{-}[r]
    \ar@<+0.4pt>@{-}[r]
    \ar@<+0.8pt>@{-}[r]
    \ar@<+1.2pt>@{-}[r]
    \ar@<+1.6pt>@{-}[r]
    \ar@<-6pt>@{-}[r]
    \ar@<-6.4pt>@{-}[r]
    \ar@<-6.8pt>@{-}[r]
    \ar@<-5.2pt>@{-}[r]
    \ar@<-5.6pt>@{-}[r]
	&&&&
    \ar@<+6pt>@{-}[r]|{\ }
    \ar@<+6.4pt>@{-}[r]|{\ }
    \ar@<+6.8pt>@{-}[r]|{\ }
    \ar@<+7.2pt>@{-}[r]|{\ }
    \ar@<+7.6pt>@{-}[r]|{\ }
    \ar@<+0pt>@{-}[r]
    \ar@<+0.4pt>@{-}[r]
    \ar@<+0.8pt>@{-}[r]
    \ar@<+1.2pt>@{-}[r]
    \ar@<+1.6pt>@{-}[r]
    \ar@<-6pt>@{-}[r]|{\ }
    \ar@<-6.4pt>@{-}[r]|{\ }
    \ar@<-6.8pt>@{-}[r]|{\ }
    \ar@<-5.2pt>@{-}[r]|{\ }
    \ar@<-5.6pt>@{-}[r]|{\ }
	 &&
	 &&
    \ar@<+6pt>@{-}[r]
    \ar@<+6.4pt>@{-}[r]
    \ar@<+6.8pt>@{-}[r]
    \ar@<+7.2pt>@{-}[r]
    \ar@<+7.6pt>@{-}[r]
    \ar@<+0pt>@{-}[r]|{\ }
    \ar@<+0.4pt>@{-}[r]|{\ }
    \ar@<+0.8pt>@{-}[r]|{\ }
    \ar@<+1.2pt>@{-}[r]|{\ }
    \ar@<+1.6pt>@{-}[r]|{\ }
    \ar@<-6pt>@{-}[r]
    \ar@<-6.4pt>@{-}[r]
    \ar@<-6.8pt>@{-}[r]
    \ar@<-5.2pt>@{-}[r]
    \ar@<-5.6pt>@{-}[r]
	&&&&
    \ar@<+6pt>@{-}[r]|{\ }
    \ar@<+6.4pt>@{-}[r]|{\ }
    \ar@<+6.8pt>@{-}[r]|{\ }
    \ar@<+7.2pt>@{-}[r]|{\ }
    \ar@<+7.6pt>@{-}[r]|{\ }
    \ar@<+0pt>@{-}[r]|{\ }
    \ar@<+0.4pt>@{-}[r]|{\ }
    \ar@<+0.8pt>@{-}[r]|{\ }
    \ar@<+1.2pt>@{-}[r]|{\ }
    \ar@<+1.6pt>@{-}[r]|{\ }
    \ar@<-6pt>@{-}[r]|{\ }
    \ar@<-6.4pt>@{-}[r]|{\ }
    \ar@<-6.8pt>@{-}[r]|{\ }
    \ar@<-5.2pt>@{-}[r]|{\ }
    \ar@<-5.6pt>@{-}[r]|{\ }
	&&&&&&&&&&&&&
  }
$$

\medskip
\medskip

\noindent {\bf $\infty$-Connectedness} $(\mathbf{\Pi} \dashv \flat)$ (section \ref{InfinityConnectedToposes}).

$$
  \xymatrix@C=20pt{
    &&
	\ar@{}[r]|{\mbox{locally} \atop \mbox{$\infty$-connected}}
	&&&&
	\ar@{}[r]|{\mbox{$\infty$-connected}}
	&&&&
	\ar@{}[r]|{\mbox{essentially}\atop \mbox{embedded}}
	&&&&
	\ar@{}[r]|{\mbox{discrete}}
	&&&
    \\
	\\
    \ar@{.}[rrrrrrrrrrrrrrrrr]
     &&
    \ar@<+12pt>@{-}[r]
    \ar@<+12.4pt>@{-}[r]
    \ar@<+12.8pt>@{-}[r]
    \ar@<+13.2pt>@{-}[r]
    \ar@<+13.6pt>@{-}[r]
    \ar@<+6pt>@{-}[r]
    \ar@<+6.4pt>@{-}[r]
    \ar@<+6.8pt>@{-}[r]
    \ar@<+7.2pt>@{-}[r]
    \ar@<+7.6pt>@{-}[r]
    \ar@<+0pt>@{-}[r]
    \ar@<+.4pt>@{-}[r]
    \ar@<+.8pt>@{-}[r]
    \ar@<+1.2pt>@{-}[r]
    \ar@<+1.6pt>@{-}[r]
	 &&
	 &&
    \ar@<+12pt>@{-}[r]
    \ar@<+12.4pt>@{-}[r]
    \ar@<+12.8pt>@{-}[r]
    \ar@<+13.2pt>@{-}[r]
    \ar@<+13.6pt>@{-}[r]
    \ar@<+6pt>@{-}[r]|{\ }
    \ar@<+6.4pt>@{-}[r]|{\ }
    \ar@<+6.8pt>@{-}[r]|{\ }
    \ar@<+7.2pt>@{-}[r]|{\ }
    \ar@<+7.6pt>@{-}[r]|{\ }
    \ar@<+0pt>@{-}[r]
    \ar@<+.4pt>@{-}[r]
    \ar@<+.8pt>@{-}[r]
    \ar@<+1.2pt>@{-}[r]
    \ar@<+1.6pt>@{-}[r]
	 &&
	 &&
    \ar@<+12pt>@{-}[r]|{\ }
    \ar@<+12.4pt>@{-}[r]|{\ }
    \ar@<+12.8pt>@{-}[r]|{\ }
    \ar@<+13.2pt>@{-}[r]|{\ }
    \ar@<+13.6pt>@{-}[r]|{\ }
    \ar@<+6pt>@{-}[r]
    \ar@<+6.4pt>@{-}[r]
    \ar@<+6.8pt>@{-}[r]
    \ar@<+7.2pt>@{-}[r]
    \ar@<+7.6pt>@{-}[r]
    \ar@<+0pt>@{-}[r]|{\ }
    \ar@<+.4pt>@{-}[r]|{\ }
    \ar@<+.8pt>@{-}[r]|{\ }
    \ar@<+1.2pt>@{-}[r]|{\ }
    \ar@<+1.6pt>@{-}[r]|{\ }
	&&&&
    \ar@<+12pt>@{-}[r]|{\ }
    \ar@<+12.4pt>@{-}[r]|{\ }
    \ar@<+12.8pt>@{-}[r]|{\ }
    \ar@<+13.2pt>@{-}[r]|{\ }
    \ar@<+13.6pt>@{-}[r]|{\ }
    \ar@<+6pt>@{-}[r]|{\ }
    \ar@<+6.4pt>@{-}[r]|{\ }
    \ar@<+6.8pt>@{-}[r]|{\ }
    \ar@<+7.2pt>@{-}[r]|{\ }
    \ar@<+7.6pt>@{-}[r]|{\ }
    \ar@<+0pt>@{-}[r]|{\ }
    \ar@<+.4pt>@{-}[r]|{\ }
    \ar@<+.8pt>@{-}[r]|{\ }
    \ar@<+1.2pt>@{-}[r]|{\ }
    \ar@<+1.6pt>@{-}[r]|{\ }
	&&&&&&&&&&&&&
  }
$$

\medskip
\medskip

\noindent {\bf Cohesion} $(\mathbf{\Pi} \dashv \flat \dashv \sharp)$ (section \ref{CohesiveToposes}).

$$
  \xymatrix@C=20pt{
    &&
	\ar@{}[r]|{ \mbox{cohesive} }
	&&&&
	\ar@{}[r]|{\mbox{infinitesimally} \atop \mbox{embedded}}
	&&&&
	&&&
    \\
	\\
    \ar@{.}[rrrrrrrrrrrrrrr]
     &&
    \ar@<+12pt>@{-}[r]
    \ar@<+12.4pt>@{-}[r]
    \ar@<+12.8pt>@{-}[r]
    \ar@<+13.2pt>@{-}[r]
    \ar@<+13.6pt>@{-}[r]
    \ar@<+6pt>@{-}[r]|{\ }
    \ar@<+6.4pt>@{-}[r]|{\ }
    \ar@<+6.8pt>@{-}[r]|{\ }
    \ar@<+7.2pt>@{-}[r]|{\ }
    \ar@<+7.6pt>@{-}[r]|{\ }
    \ar@<+0pt>@{-}[r]
    \ar@<+0.4pt>@{-}[r]
    \ar@<+0.8pt>@{-}[r]
    \ar@<+1.2pt>@{-}[r]
    \ar@<+1.6pt>@{-}[r]
    \ar@<-6pt>@{-}[r]|{\ }
    \ar@<-6.4pt>@{-}[r]|{\ }
    \ar@<-6.8pt>@{-}[r]|{\ }
    \ar@<-5.2pt>@{-}[r]|{\ }
    \ar@<-5.6pt>@{-}[r]|{\ }
	&&&&
    \ar@<+12pt>@{-}[r]|{\ }
    \ar@<+12.4pt>@{-}[r]|{\ }
    \ar@<+12.8pt>@{-}[r]|{\ }
    \ar@<+13.2pt>@{-}[r]|{\ }
    \ar@<+13.6pt>@{-}[r]|{\ }
    \ar@<+6pt>@{-}[r]
    \ar@<+6.4pt>@{-}[r]
    \ar@<+6.8pt>@{-}[r]
    \ar@<+7.2pt>@{-}[r]
    \ar@<+7.6pt>@{-}[r]
    \ar@<+0pt>@{-}[r]|{\ }
    \ar@<+0.4pt>@{-}[r]|{\ }
    \ar@<+0.8pt>@{-}[r]|{\ }
    \ar@<+1.2pt>@{-}[r]|{\ }
    \ar@<+1.6pt>@{-}[r]|{\ }
    \ar@<-6pt>@{-}[r]
    \ar@<-6.4pt>@{-}[r]
    \ar@<-6.8pt>@{-}[r]
    \ar@<-5.2pt>@{-}[r]
    \ar@<-5.6pt>@{-}[r]
	 &&
	 &&
	&&&&
	&&&&&&&&&&&&&
  }
$$

\medskip
\medskip

\noindent {\bf Differential cohesion} $(\mathbf{Red}\dashv \mathbf{\Pi}_{\mathrm{inf}} \dashv \mathbf{\flat}_{\mathrm{inf}})$ (section \ref{InfinitesimalCohesion}).

$$
  \xymatrix@C=3pt{
    &&&&
	\ar@{}[rrrrrrr]|{ \mbox{infinitesimally} \atop \mbox{cohesive}   }
	&&&&&&&&&&&&
	\ar@{}[rrrrrrr]|{ \mbox{differentially} \atop \mbox{cohesive}   }	
	&&&&&&&&&&&&
    \\
	\\
    \ar@{.}[rrrrrrrrrrrrrrrrrrrrrrrrrrrrrrrrrrrrrrrrrrrrrr]
     &&&&
    \ar@<+12pt>@{-}[rrr]|{\ }
    \ar@<+12.4pt>@{-}[rrr]|{\ }
    \ar@<+12.8pt>@{-}[rrr]|{\ }
    \ar@<+13.2pt>@{-}[rrr]|{\ }
    \ar@<+13.6pt>@{-}[rrr]|{\ }
    \ar@<+6pt>@{-}[rrr]
    \ar@<+6.4pt>@{-}[rrr]
    \ar@<+6.8pt>@{-}[rrr]
    \ar@<+7.2pt>@{-}[rrr]
    \ar@<+7.6pt>@{-}[rrr]
    \ar@<+0pt>@{-}[rrr]|{\ }
    \ar@<+0.4pt>@{-}[rrr]|{\ }
    \ar@<+0.8pt>@{-}[rrr]|{\ }
    \ar@<+1.2pt>@{-}[rrr]|{\ }
    \ar@<+1.6pt>@{-}[rrr]|{\ }
    \ar@<-6pt>@{-}[rrr]
    \ar@<-6.4pt>@{-}[rrr]
    \ar@<-6.8pt>@{-}[rrr]
    \ar@<-5.2pt>@{-}[rrr]
    \ar@<-5.6pt>@{-}[rrr]
	&&&&
    \ar@<+12pt>@{-}[rrr]
    \ar@<+12.4pt>@{-}[rrr]
    \ar@<+12.8pt>@{-}[rrr]
    \ar@<+13.2pt>@{-}[rrr]
    \ar@<+13.6pt>@{-}[rrr]
    \ar@<+6pt>@{-}[rrr]|{\ }
    \ar@<+6.4pt>@{-}[rrr]|{\ }
    \ar@<+6.8pt>@{-}[rrr]|{\ }
    \ar@<+7.2pt>@{-}[rrr]|{\ }
    \ar@<+7.6pt>@{-}[rrr]|{\ }
    \ar@<+0pt>@{-}[rrr]
    \ar@<+0.4pt>@{-}[rrr]
    \ar@<+0.8pt>@{-}[rrr]
    \ar@<+1.2pt>@{-}[rrr]
    \ar@<+1.6pt>@{-}[rrr]
    \ar@<-6pt>@{-}[rrr]|{\ }
    \ar@<-6.4pt>@{-}[rrr]|{\ }
    \ar@<-6.8pt>@{-}[rrr]|{\ }
    \ar@<-5.2pt>@{-}[rrr]|{\ }
    \ar@<-5.6pt>@{-}[rrr]|{\ }
     &&&&&&&&
    \ar@<+18pt>@{-}[rrr]|{\ }
    \ar@<+18.4pt>@{-}[rrr]|{\ }
    \ar@<+18.8pt>@{-}[rrr]|{\ }
    \ar@<+19.2pt>@{-}[rrr]|{\ }
    \ar@<+19.6pt>@{-}[rrr]|{\ }
    \ar@<+12pt>@{-}[rrr]
    \ar@<+12.4pt>@{-}[rrr]
    \ar@<+12.8pt>@{-}[rrr]
    \ar@<+13.2pt>@{-}[rrr]
    \ar@<+13.6pt>@{-}[rrr]
    \ar@<+6pt>@{-}[rrr]|{\ }
    \ar@<+6.4pt>@{-}[rrr]|{\ }
    \ar@<+6.8pt>@{-}[rrr]|{\ }
    \ar@<+7.2pt>@{-}[rrr]|{\ }
    \ar@<+7.6pt>@{-}[rrr]|{\ }
    \ar@<-0pt>@{-}[rrr]
    \ar@<0.4pt>@{-}[rrr]
    \ar@<0.8pt>@{-}[rrr]
    \ar@<1.2pt>@{-}[rrr]
    \ar@<1.6pt>@{-}[rrr]
	&&&&
    \ar@<+12pt>@{-}[rrr]
    \ar@<+12.4pt>@{-}[rrr]
    \ar@<+12.8pt>@{-}[rrr]
    \ar@<+13.2pt>@{-}[rrr]
    \ar@<+13.6pt>@{-}[rrr]
    \ar@<+6pt>@{-}[rrr]|{\ }
    \ar@<+6.4pt>@{-}[rrr]|{\ }
    \ar@<+6.8pt>@{-}[rrr]|{\ }
    \ar@<+7.2pt>@{-}[rrr]|{\ }
    \ar@<+7.6pt>@{-}[rrr]|{\ }
    \ar@<+0pt>@{-}[rrr]
    \ar@<+0.4pt>@{-}[rrr]
    \ar@<+0.8pt>@{-}[rrr]
    \ar@<+1.2pt>@{-}[rrr]
    \ar@<+1.6pt>@{-}[rrr]
    \ar@<-6pt>@{-}[rrr]|{\ }
    \ar@<-6.4pt>@{-}[rrr]|{\ }
    \ar@<-6.8pt>@{-}[rrr]|{\ }
    \ar@<-5.2pt>@{-}[rrr]|{\ }
    \ar@<-5.6pt>@{-}[rrr]|{\ }	 
    \ar@<-5.6pt>@{-}[rrr]|{\ }	 
	&&
	 &&
	&&&&
	&&&&&&&&&&&&&&&&&&&&&&&&&&&&&&&&&&&&&&&
  }
$$

\medskip
\medskip

We discuss a list of structures that may be formulated internal to such $\mathbf{H}$:

\begin{itemize}
  \item \ref{Structures in a topos} -- Structures in an $\infty$-topos;
  \item \ref{Structures in a local topos} -- Structures in a local $\infty$-topos;
  \item \ref{Structures in a locally infinity-connected topos} -- 
     Structures in an $\infty$-connected $\infty$-topos;
  \item \ref{structures} -- Structures in a cohesive $\infty$-topos;
  \item \ref{StructuresInInfinitesimalCohesiveNeighbourhood} -- Structures in a differential $\infty$-topos.
\end{itemize}

\subsection{Local $\infty$-toposes}
\label{LocalToposes}

The following definition is the direct generalization of the notion 
of \emph{local topos} \cite{JohnstoneMoerdijk}.
\begin{definition}
  \label{LocalInfinityTopos}
  An $\infty$-topos $\mathbf{H}$ is called \emph{locally local}
  if the global section geometric morphism has a right adjoint.
  $$
    \xymatrix{
	  \mathbf{H}
	    \ar@{<-}@<+5pt>[r]^-{\mathrm{Disc}}
	    \ar[r]|-{\Gamma}
	    \ar@{<-}@<-5pt>[r]_-{\mathrm{coDisc}}
		&
	  \infty \mathrm{Grpd}
	}
	\,.
  $$
  It is called \emph{local} if that right adjoint is in addition fully faithful.
\end{definition}
\begin{proposition} 
   \label{PointlikePropertyOfLocalInfinityTopos}
   A local $\infty$-topos 
  \begin{enumerate}
    \item has homotopy dimension 0 (see def. \ref{HomotopyDimension} below);
    \item has cohomological dimension 0 (\cite{Lurie}, section 7.2.2);
	\item is hypercomplete.
  \end{enumerate}
\end{proposition}
\proof
  The first statement is cor. \ref{HomotopyDimensionOfCohesiveInfinityTopos} below.
  The second is a consequence of the first
  by \cite{Lurie}, cor. 7.2.2.30. The third follows from the 
  second by \cite{Lurie}, cor. 7.2.1.12.
\endofproof

\subsection{Locally $\infty$-connected $\infty$-toposes}
\label{InfinityConnectedToposes}

We discuss $\infty$-toposes satisfying a higher geometric connectedness condition.

\subsubsection{General abstract}

The following definition is the direct generalization standard notion of a 
\emph{locally/globally connected topos} \cite{Johnstone}: 
a topos whose terminal geometric
morphism has an extra left adjoint that computes geometric connected components, hence
a geometric notion of $\pi_0$. We will see in \ref{StructuresInLocallyInfinityConnectedTopos}, 
that as we pass to 
$\infty$-toposes, the extra left adjoint provides a good definition of all geometric 
homotopy groups. 
\begin{definition} 
 \label{DefConnectedTopos}
 \index{topos!locally $\infty$-connected}
 \index{topos!$\infty$-connected}
An $\infty$-topos $\mathbf{H}$ we call 
\emph{locally $\infty$-connected} if the (essentially unique) 
global section $\infty$-geometric morphism from prop. \ref{Terminalgeometricmorphism} is an 
\emph{essential $\infty$-geometric morphism} in that it has a further left adjoint $\Pi$:
$$
  (\Pi \dashv \Delta \dashv \Gamma) : 
  \xymatrix{
    \mathbf{H}
      \ar@<+12pt>[r]^-{\Pi}
      \ar@<+4pt>@{<-}[r]|-{\Delta}
      \ar@<-4pt>[r]_-{\Gamma}
      &
    \infty \mathrm{Grpd}
  }
  \,.
$$
If in addition $\Delta$ is fully faithful, then we say that 
$\mathbf{H}$ is in addition an \emph{$\infty$-connected} 
or \emph{globally $\infty$-connected} $\infty$-topos. 
\end{definition}
\begin{remark}
  \label{ConstantShape}
  Meanwhile, a locally $\infty$-connected $\infty$-topos as above 
  has been called an $\infty$-topos
  \emph{of constant shape} in \cite{LurieAlgebra}, section A.1.
  Some of the following statements now overlap with the discussion there.
\end{remark}
\begin{proposition}
  For $\mathbf{H}$ a locally/globally $\infty$-connected $\infty$-topos, the underlying
  1-topos $\tau_{\leq 0}\mathbf{H}$ of 0-truncated objects (def. \ref{truncated object}) is a 
  \emph{locally/globally connected topos} (as in \cite{Johnstone} C1.5, C3.3).
  \label{ForLocallyInfinityConnectedInfinityToposUnderlying1ToposIsLocallyConnected}
\end{proposition}
\proof
  By prop. \ref{GlobalSectionMorphismExplicitly} and by the very definition 
  of truncated objects $\Gamma$ takes 0-truncated objects in $\mathbf{H}$ to 0-truncated
  objects in $\infty\mathrm{Grpd}$, hence the restriction $\Gamma|_{\tau_{\leq}}$ 
  factors through the inclusion
  $\mathrm{Set} \simeq \tau_{\leq 0} \infty \mathrm{Grpd} \hookrightarrow \infty \mathrm{Grpd}$.
 
  Similarly the restriction $\Delta|_{\leq 0}$ factors through the
  inclusion $\tau_{\leq 0}\mathbf{H} \hookrightarrow \mathbf{H}$:  by definition
  this is the case if for all $S \in \mathrm{Set}$ and all $X \in \mathbf{H}$ the
  hom-$\infty$-groupoid $\mathbf{H}(X, \Delta S) \in \infty \mathrm{Grpd}$ is equivalently a set.
  But by the defining right-adjointness of $\Delta$ this is equivalently 
  $$
    \mathbf{H}(X,\Delta S) \simeq \infty \mathrm{Grpd}(\Pi(X),S) \simeq \mathrm{Set}(\tau_{\leq 0}\Pi(X), S)
	\in \mathrm{Set} \hookrightarrow \infty \mathrm{Grpd}
	\,,
  $$
  which is a set by assumption that $S$ is 0-truncated.
  
  By uniqueness of adjoints and the fact that $\tau_{\leq 0} : \infty \mathrm{Grpd} \to \mathrm{Set}$ 
  is left adjoint to the inclusion, this means that 
  $\Delta|_{\leq 0} : \xymatrix{ \mathrm{Set} \ar@{^{(}->}[r] & \infty \mathrm{Grpd} \ar[r]^\Delta & \mathbf{H}}$
  has a left adjoint 
  $$
    \Pi_0 := \tau_{\leq} \circ \Pi
	\,.
  $$ 
  Finally  $\tau_{\leq 0}$ preserves finite products by \cite{Lurie}, lemma 6.5.1.2.
  and if $\Pi$ preserves the terminal object then so does $\Pi_0$.   
\endofproof
\begin{proposition} 
\label{DeltaIsFF}
  A locally $\infty$-connected topos 
  $(\Pi \dashv \Delta \dashv \Gamma) : \mathbf{H} \to \infty \mathrm{Grpd}$ 
  is globally $\infty$-connected precisely if 
  the  following equivalent conditions hold.
  \begin{enumerate}
    \item The inverse image $\Delta$ is a fully faithful $\infty$-functor.
	\item The extra left adjoint $\Pi$ preserves the terminal object.
  \end{enumerate}
\end{proposition}
\proof
  This follows verbatim the proof for the familiar statement about connected toposes, since
  all the required properties have $\infty$-analogs:
  we have that
  \begin{itemize}
    \item 
    $\Delta$ is fully faithful precisely if the $(\Pi \dashv \Delta)$-adjunction 
	unit is an equivalence, by prop. \ref{FullyFaithfulAdjunctions}.
    \item
      every $\infty$-groupoid $S$ is the $\infty$-colimit over itself of the $\infty$-functor
      constant on the point, by prop. \ref{InfinityGroupoidIsColimitOverItself}:
      $$
        S\simeq \underset{\longrightarrow_S}{\lim} *
        \,.
      $$
  \end{itemize}
  Therefore if $\Delta$ is fully faithful, then 
  $$
    \begin{aligned}
	  \Pi({*}) & \simeq \Pi \Delta({*})
	  \\
	  & \simeq {*}
	\end{aligned}
  $$
  and hence $\Pi$ preserves the terminal object. 
  Conversely, if $\Pi$ preserves the terminal object then for any
    $S \in \infty \mathrm{Grpd}$ we have that
  $$
    \begin{aligned}
      \Pi \Delta S & \simeq \Pi \Delta {\lim\limits_{\to}}_S *
        \\
          & \simeq {\lim\limits_{\longrightarrow}}_S \Pi \Delta *
        \\
          & \simeq {\lim\limits_{\longrightarrow}}_S *
        \\
          & \simeq S
    \end{aligned}
    \,.
  $$
  and hence $\Delta$ is fully faithful.
\endofproof
\begin{proposition}
  \label{PointlikePropertyOfLocallyInfinityConnectedTopos}
  A locally $\infty$-connected $\infty$-topos 
  \begin{enumerate}
    \item has the shape of $\Pi(*)$;
	\item hence has the shape of the point if it is globally $\infty$-connected.
  \end{enumerate}
\end{proposition}
\proof
  By inspection of the definitions.
\endofproof

\subsubsection{Presentations}

We discuss presentations of locally and globally $\infty$-connected $\infty$-toposes,
def. \ref{DefConnectedTopos}, by categories of simplicial presheaves over a suitable site of
definition.

\medskip

\begin{definition} 
 \label{ConnectedSite}
We call a site (a small category equipped with a coverage) 
 \emph{locally and globally $\infty$-connected} 
if
\begin{enumerate}
\item it has a terminal object $*$;
\item for every generating covering family $\{U_i \to U\}$ in $C$ 
\begin{enumerate}
  \item $\{U_i \to U\}$ is a \emph{good covering}, def. \ref{GoodCovers}: 
  the {\v C}ech nerve $C(\{U_i\}) \in [C^{\mathrm{op}}, \mathrm{sSet}]$ is degreewise 
       a coproduct of representables;

  \item 
     the colimit $\lim\limits_{\longrightarrow} : [C^{\mathrm{op}}, \mathrm{sSet}] \to \mathrm{sSet}$ 
     of $C(\{U_i\})$ is weakly contractible

     $$
       \lim\limits_{\longrightarrow} C(\{U_i\}) \stackrel{\simeq}{\to} *
       \,.
     $$
\end{enumerate}  
\end{enumerate}
\end{definition}
\begin{proposition} 
  \label{ToposOnInfConnectedSiteIsInfConnected}
  For $C$ a locally and globally $\infty$-connected site, the $\infty$-topos $Sh_{\infty}(C)$
  is locally and globally $\infty$-connected.
\end{proposition}
We prove this after noting two lemmas.
\begin{lemma}
 \label{LemmaCUIsCofib}
For $\{U_i \to U\}$ a covering family in the $\infty$-connected site $C$, 
the {\v C}ech nerve
$C(\{U_i\}) \in [C^{\mathrm{op}}, \mathrm{sSet}]$ is a cofibrant resolution of $U$ both in the 
global projective model structure $[C^{\mathrm{op}}, \mathrm{sSet}]_{\mathrm{proj}}$ as well as in the 
local model structure $[C^{\mathrm{op}}, \mathrm{sSet}]_{\mathrm{proj},\mathrm{loc}}$.
\end{lemma}
\proof
By assumption on $C$ we have that $C(\{U_i\})$ is a split hypercover
\cite{dugger-hollander-isaksen}.
This implies that $C(\{U_i\})$ is cofibrant in the global model structure. 
By general properties of left Bousfield localization we have that the cofibrations 
in the local model structure as the same as in the global one. 
Finally that $C(\{U_i\}) \to U$ is a weak equivalence in the local model structure holds effectively by definition (since we are localizing at these morphisms).
\endofproof
\begin{proposition}
  On a locally and globally $\infty$-connected site $C$, the global section
  $\infty$-geometric morphsm 
  $(\Delta \dashv \Gamma) : \mathrm{Sh}_{\infty}(C) \to \infty \mathrm{Grpd}$
  is presented under prop. \ref{InfAdjBySimpAdj} by the simplical Quillen adjunction
  $$
    (\mathrm{Const} \dashv \Gamma)
     :
  \xymatrix{
    [C^{\mathrm{op}}, \mathrm{sSet}]_{\mathrm{proj}, \mathrm{loc}}
      \ar@<+3pt>@{<-}[r]^<<<<{\mathrm{Const}}
      \ar@<-3pt>[r]_<<<<{\Gamma}
      &
      \mathrm{sSet}_{\mathrm{Quillen}}
   }
      \,,
  $$
  where $\Gamma$ is the functor that evaluates on the terminal object, $\Gamma(X) = X(*)$,
  and where $\mathrm{Const}$ is the functor that assigns constant presheaves 
  $\mathrm{Const} S : U \mapsto S$.
\end{proposition}
\proof
  That we have a 1-categorical adjunction $(\mathrm{Const} \dashv \Gamma)$ follows
  by noticing that since $C$ has a terminal object we have that $\Gamma = \lim\limits_{\longleftarrow}$
  is given by the limit operation.
  
  To see that we have a Quillen adjunction first notice that we have a Quillen adjunction
  on the global model structure
  $$
    (\mathrm{Const} \dashv \Gamma)
     :
     \xymatrix{
    [C^{\mathrm{op}}, \mathrm{sSet}]_{\mathrm{proj}}
      \ar@<+3pt>@{<-}[r]^{\mathrm{Const}}
      \ar@<-3pt>[r]_{\Gamma}
      &
      \mathrm{sSet}_{\mathrm{Quillen}}
    }
    \,,
  $$
  since $\Gamma$ manifestly preserves fibrations and 
  acyclic fibrations there. Because 
  $[C^{\mathrm{op}}, \mathrm{sSet}]_{\mathrm{proj,\mathrm{loc}}}$ 
  is left proper and has the same cofibrations as the global model structure, 
  it follows with prop. \ref{AdjRecognition}
  that for this to descend to a Quillen adjunction on the local model structure it is 
  sufficient that $\Gamma$ preserves locally fibrant objects. But every fibrant object in the 
  local structure is in particular fibrant in the global structure, 
  hence in particular fibrant over the terminal object of $C$. 
  
  The left derived functor $\mathbb{L}\mathrm{Const}$ of 
  $\mathrm{Const}: \mathrm{sSet}_{\mathrm{Quillen}} \to [C^{\mathrm{op}}, \mathrm{sSet}]$ 
  preserves $\infty$-limits (because $\infty$-limits in an $\infty$-category of $\infty$-presheaves 
  are computed objectwise), and moreover $\infty$-stackification, being the left derived functor of 
  $\mathrm{Id} 
    : 
     [C^{\mathrm{op}}, \mathrm{sSet}]_{\mathrm{proj}}
      \to 
     [C^{\mathrm{op}}, \mathrm{sSet}]_{\mathrm{proj}}
   $,
  is a left exact $\infty$-functor, therefore the left derived functor 
  of
  $\mathrm{Const}: \mathrm{sSet}_{\mathrm{Quillen}} \to [C^{\mathrm{op}}, \mathrm{sSet}]_{\mathrm{proj}, \mathrm{loc}}$
  preserves finite $\infty$-limits.
  
  This means that our Quillen adjunction does model an  $\infty$-geometric morphism 
  $\mathrm{Sh}_{\infty}(C) \to \infty \mathrm{Grpd}$.
  By prop. \ref{Terminalgeometricmorphism}
  this is indeed a representative of the terminal geometric morphism as claimed. 
\endofproof
\proofoftheorem{ToposOnInfConnectedSiteIsInfConnected}
By general abstract facts the $\mathrm{sSet}$-functor 
$\mathrm{Const} : \mathrm{sSet} \to [C^{\mathrm{op}}, \mathrm{sSet}]$ 
given on $S \in \mathrm{sSet}$ by $\mathrm{Const}(S) : U \mapsto S$ 
for all $U \in C$ has an  $\mathrm{sSet}$-left adjoint 
$$
  \Pi : X \mapsto \int^U X(U) = \lim\limits_{\longrightarrow} X
$$ 
naturally in $X$ and $S$, given by the colimit operation. Notice that since 
$\mathrm{sSet}$ is itself a category of presheaves (on the simplex category), 
these colimits are degreewise colimits in $\mathrm{Set}$. 
Also notice that the colimit over a representable functor is the point 
(by a simple Yoneda lemma-style argument).

Regarded as a functor  
$\mathrm{sSet}_{\mathrm{Quillen}} \to [C^{\mathrm{op}}, \mathrm{sSet}]_{\mathrm{proj}}$ 
the functor $\mathrm{Const}$ manifestly preserves fibrations and acyclic fibrations and hence
$$
  (\Pi \dashv \mathrm{Const}) 
   : 
   \xymatrix{
     [C^{\mathrm{op}}, \mathrm{sSet}]_{\mathrm{proj}} 
      \ar@<+3pt>@{->}[r]^-{\lim\limits_{\longrightarrow}}
      \ar@<-3pt>@{<-}[r]_-{\mathrm{Const}}
      &
    \mathrm{sSet}_{\mathrm{Quillen}} 
   }
$$
is a Quillen adjunction, in particular 
   $\Pi : [C^{\mathrm{op}},\mathrm{sSet]}_{\mathrm{proj}} 
  \to \mathrm{sSet}_{\mathrm{Quillen}}$ preserves cofibrations. 
  Since by general properties of left 
   Bousfield localization  the cofibrations of 
  $[C^{\mathrm{op}},\mathrm{sSet}]_{\mathrm{proj},\mathrm{loc}}$ 
  are the same, also 
   $\Pi : [C^{\mathrm{op}}, \mathrm{sSet}]_{\mathrm{proj},\mathrm{loc}} 
   \to 
    \mathrm{sSet}_{\mathrm{Quillen}}$ 
   preserves cofibrations. 

Since $\mathrm{sSet}_{\mathrm{Quillen}}$ is a left proper model category 
 it follows with prop. \ref{AdjRecognition} that for 
$$
  (\Pi \dashv \mathrm{Const}) 
   : 
   \xymatrix{
     [C^{\mathrm{op}}, \mathrm{sSet}]_{\mathrm{proj}, \mathrm{loc}} 
      \ar@<+3pt>@{->}[r]^-{\lim\limits_{\longrightarrow}}
      \ar@<-3pt>@{<-}[r]_-{\mathrm{Const}}
      &
    \mathrm{sSet}_{\mathrm{Quillen}} 
    }
$$
to be a Quillen adjunction, it suffices now that $\mathrm{Const}$ 
preserves fibrant objects. This means that constant simplicial presheaves satisfy 
descent along covering families in the $\infty$-cohesive site $C$: for every covering 
family $\{U_i \to U\}$ in $C$ and every simplicial set $S$ it must be true that
$$
  [C^{\mathrm{op}}, \mathrm{sSet}](U, \mathrm{Const} S) 
   \to 
   [C^{\mathrm{op}},  \mathrm{sSet}](C(\{U_i\}), \mathrm{Const} S)
$$
is a homotopy equivalence of Kan complexes. (Here we use that $U$, being a 
representable, is cofibrant, that $C(\{U_i\})$ is cofibrant by the lemma 
\ref{LemmaCUIsCofib} and 
that $\mathrm{Const} S$ is fibrant in the projective structure by the assumption that 
$S$ is fibrant. So the simplicial hom-complexes in the above equaltion really are the 
correct derived hom-spaces.)

But that this is the case follows by the condition on the $\infty$-connected 
site $C$ by which $\lim\limits_{\longrightarrow} C(\{U_i\}) \simeq *$: using this we have that
$$
  [C^{\mathrm{op}}, \mathrm{sSet}](C(\{U_i\}), \mathrm{Const} S) = 
   \mathrm{sSet}(\lim\limits_{\longrightarrow} C(\{U_i\}), S) 
   \simeq 
  \mathrm{sSet}(*, S) = S
  \,.
$$

So we have established that $(\lim\limits_{\longrightarrow} \dashv \mathrm{Const})$ is also a Quillen adjunction
on the local model structure.

It is clear that the left derived functor of $\lim\limits_{\longrightarrow}$ preserves the terminal object: 
since that is representable by assumption on $C$, it is cofibrant in 
$[C^{\mathrm{op}}, \mathrm{sSet}]_{\mathrm{proj},\mathrm{loc}}$, hence 
$\mathbb{L} \lim\limits_{\longrightarrow} {*} \simeq \lim\limits_{\longrightarrow} {*} = {*}$.
\endofproof

\subsection{Cohesive $\infty$-toposes}
\label{CohesiveToposes}

We now combine the notions of local $\infty$-toposes and $\infty$-connected $\infty$-toposes
to that of cohesive $\infty$-toposes

\subsubsection{General abstract}
\label{CohesionGeneralAbstract}

We give the definition and basic properties of cohesive $\infty$-toposes
first externally, in \ref{DefinitionCohesiveToposes} in terms of 
properties of the global section geometric morphism, and then
internally, in the language of the internal type theory of an $\infty$-topos,
in \ref{InternalCohesion}.

\paragraph{External formulation}
\label{DefinitionCohesiveToposes}

The following definition is the direct generalization of 
the main axioms in the definition of \emph{topos of cohesion} from \cite{Lawvere}.
\begin{definition} 
  \label{CohesiveInfinToposDefinition}
   \index{topos!cohesive}
   \index{cohesive $\infty$-topos}
  A \emph{cohesive $\infty$-topos} $\mathbf{H}$ is  
  \begin{enumerate}
    \item
       a locally and globally $\infty$-connected topos $\mathbf{H}$, def \ref{DefConnectedTopos},
    \item
      which in addition is a  \emph{local $\infty$-topos}, def. \ref{LocalInfinityTopos};
	 \item 
	   and such that the extra left adjoint preserves not just the terminal object,
	   but all finite products.
  \end{enumerate}
\end{definition}
\begin{remark}
The two conditions say in summary that an $\infty$-topos is 
cohesive precisely if it admits quadruple of 
adjoint $\infty$-functors
$$
    (\Pi \dashv \Delta \dashv \Gamma \dashv \nabla)
    :
  \xymatrix{
    \mathbf{H}
    \ar@{->}@<+12pt>[rr]|<\times^-{\Pi}
    \ar@{<-^{)}}@<+4pt>[rr]|-{\Delta}
    \ar@{->}@<-4pt>[rr]|-{\Gamma}
    \ar@{<-^{)}}@<-12pt>[rr]_-{\nabla}
    &&
    \infty \mathrm{Grpd}
   }
$$
such that $\Pi$ preserves finite products. 
\end{remark}
We may think of these axioms as encoding properties 
that characterize those $\infty$-toposes of $\infty$-groupoids 
that are equipped with extra \emph{cohesive structure}. 
In order to reflect this geometric interpretation notationally
we will from now on write
$$
    (\Pi \dashv \mathrm{Disc} \dashv \Gamma \dashv \mathrm{coDisc})
    :
  \xymatrix{
    \mathbf{H}
    \ar@{->}@<+12pt>[rr]|<\times^{\Pi}
    \ar@{<-^{)}}@<+4pt>[rr]|{\mathrm{Disc}}
    \ar@{->}@<-4pt>[rr]|{\Gamma}
    \ar@{<-^{)}}@<-12pt>[rr]_{\mathrm{coDisc}}
    &&
    \infty \mathrm{Grpd}
   }
$$
for the defining $\infty$-connected and $\infty$-local geometric morphism and say
for $S \in \infty \mathrm{Grpd}$ that
\begin{itemize}
  \item $\mathrm{Disc} S \in \mathbf{H}$ is 
  a \emph{discrete object} of $\mathbf{H}$ or a 
  \emph{discrete cohesive $\infty$-groupoid}
  obtained by equipping $S$ with \emph{discrete cohesive structure};
  \item $\mathrm{coDisc} S \in \mathbf{H}$ is 
  a \emph{codiscrete object} of $\mathbf{H}$ or a 
  \emph{codiscrete cohesive $\infty$-groupoid}, obtained 
  by equipping $S$ with \emph{indiscrete cohesive structure};
\end{itemize}
and for $X \in \mathbf{H}$ that
\begin{itemize}
  \item $\Gamma (X) \in \infty \mathrm{Grpd}$ is the \emph{underlying $\infty$-groupoid} of $X$;
  \item $\Pi(X)$ is the \emph{fundamental $\infty$-groupoid}\index{paths!fundamental $\infty$-groupoid} or 
     \emph{geometric path $\infty$-groupoid} of $X$.
\end{itemize}
A simple but instructive toy example illustrating these interpretations is given by the
\emph{Sierpinski $\infty$-topos}, discussed below in example \ref{SierpinskiTopos}.
A detailed discussion of these geometric interpretations in various models is in 
\ref{Implementation}.
For emphasis we summarie the following list of
properties of a cohesive $\infty$-topos $\mathbf{H}$ that show that 
regarded as a generalized space itself, $\mathbf{H}$ looks like one 
fat point, to be thought of as the archetypical cohesive blob.
\begin{proposition} 
   \label{PointlikeProperty}
   \label{CohesiveIsHypercomplete}
   \index{cohesive $\infty$-topos!pointlike properties}
  A  cohesive $\infty$-topos 
  \begin{enumerate}
    \item has homotopy dimension 0;
    \item has cohomological dimension 0;
    \item has the shape of the point;
	\item is hypercomplete.
  \end{enumerate}
\end{proposition}
\proof
  By prop. \ref{PointlikePropertyOfLocalInfinityTopos} 
  and prop. \ref{PointlikePropertyOfLocallyInfinityConnectedTopos}.
\endofproof

\medskip

Every adjoint quadruple of functors induces an adjoint triple of 
endofunctors:
\begin{definition}
  \label{TheAdjointTripleOfEndofunctors}
  On any cohesive $\infty$-topos $\mathbf{H}$ define
  the adjoint triple of functors
  $$
    (
	  \mathbf{\Pi}
	  \dashv
	  \flat
	  \dashv 
	  \sharp
	)
	:
	\xymatrix{
	  \mathbf{H}
	  \ar@<+8pt>[rr]|<\times^{\Pi}
	  \ar@{<-^{)}}@<+0pt>[rr]|{\mathrm{Disc}}
	  \ar@<-8pt>[rr]_{\Gamma}
	  &&
	  \infty \mathrm{Grpd}
	  \ar@{^{(}->}@<+8pt>[rr]^{\mathrm{Disc}}
	  \ar@{<-}@<+0pt>[rr]|{\Gamma}
	  \ar@{^{(}->}@<-8pt>[rr]_{\mathrm{coDisc}}
	  &&
	  \mathbf{H}
	}
	\,.
  $$
\end{definition}
\begin{remark}
The geometric interpretation of these three functors is discussed below
in \ref{StrucPaths}, \ref{StrucFlatDifferential} and \ref{StrucConcrete},
respectively: 
\begin{itemize}
  \item $\mathbf{\Pi}$ is the \emph{shape modality}, the \emph{geometric path} or \emph{geometric homotopy} functor
  or \emph{fundamental $\infty$-groupoid} functor;
  \item $\flat$ is the \emph{flat modality}, for $A \in \mathbf{H}$ we may pronounce $\flat A$ as 
  ``flat $A$'', it is the coefficient object for \emph{flat cohomology} with
coefficients in $A$;
  \item $\sharp$ is the \emph{sharp modality}, for $A \in \mathbf{H}$ we may pronounce $\sharp A$ 
as ``sharp $A$'', it is the classifying object 
 for ``sharply varying'' $A$-principal $\infty$-bundles, those that need not
 be geometric (not continuous).
\end{itemize}
\label{SharpModalityOfLocalTopos}
\end{remark}

\medskip

Every adjoint triple $(\Pi \dashv \mathrm{Disc} \dashv \Gamma)$ induces
a canonical transformation:
\begin{definition}
  For $\mathbf{H}$ a cohesive $\infty$-topos with modalities
  $(\mathbf{\Pi} \dashv \flat \dashv \sharp)$, we say that the composite
  transformation
  $$
    \left(
      \flat X \longrightarrow \mathbf{\Pi} X
    \right)
    :=
    \left(
      \flat X \longrightarrow X \longrightarrow \mathbf{\Pi}X
    \right)
  $$
  of the $(\mathrm{Disc} \dashv \Gamma)$-counit followed by the
  $(\Pi \dashv \mathrm{Disc})$-unit,
  natural in $X \in \mathbf{H}$,
  is the \emph{pieces-to-points transform}.
  \label{pointstopiecestransform}
\end{definition}
Given the geometric interpretation of $\mathbf{\Pi}$ and $\flat$, this 
map may be thought of as sending each point of a cohesive space $X$ to the
\emph{cohesive piece} that it sits in. This is a central conceptual insight in 
\cite{Lawvere}.
\begin{proposition}
  There is a natural equivalence of natural transformations
  $$
    \left(
      \flat A \longrightarrow \mathbf{\Pi}X
    \right)
    \simeq
    \mathrm{Disc}
    \left(
      \Gamma X \longrightarrow \Pi X
    \right)
    \,,
  $$
  where
  $$
    \left(
      \Gamma X \longrightarrow \Pi X
    \right)
    :=
    \left(
      \Gamma X
      \longrightarrow
      \Gamma \mathrm{Disc} \Pi X
      \stackrel{\simeq}{\longrightarrow}
      \Pi X
    \right)
  $$
  and where on the right we have the composite of the image under $\Gamma$
  of the $(\Pi \dashv \mathrm{Disc})$-unit followed by the 
  $(\mathrm{Disc} \dashv \Gamma)$-counit applied to $\Pi X$.
  In particular the points-to-pieces transform $\flat \to \mathbf{\Pi}$ 
  is an equivalence on $X \in \mathbf{H}$ precisely if
  $\Gamma \to \Pi$ is.
  \label{adjunctofpointstopieces}
\end{proposition}
\proof
  By the formula for $\infty$-adjuncts and the fully faithfulness of 
  $\mathrm{Disc}$.
\endofproof
\begin{definition}
  \label{PiecesHavePoints}
  Given an object $X \in \mathbf{H}$ of a cohesive $\infty$-topos
  over $\infty\mathrm{Grpd}$, 
   we say that
    \begin{enumerate}
	  \item 
     \emph{pieces have points} in $X$ if 
     the points-to-pieces transform
      is an effective epimorphism, def. \ref{EffectiveEpimorphism},
      $\xymatrix{\flat X \ar@{->>}[r] & \mathbf{\Pi}X}$;
	 \item 
	   $X$ has \emph{one point per piece} the points-to-pieces transform
	   is an equivalence, $\xymatrix{\flat X \ar[r]^\simeq & \mathbf{\Pi}X}$.
	\end{enumerate}
\end{definition} 
\begin{example}
For the class of cohesive $\infty$-toposes constructed below in 
\ref{CohesionPresentations} from $\infty$-cohesive sites, 
it is true for all their objects that 
\emph{pieces have points}. A class of (relative) cohesive $\infty$-toposes
for which this is not the case is discussed in \ref{DiagramToposes}.
\end{example}
\begin{example}
The property \emph{one point per piece} is 
a characteristic feature of \emph{infinitesimally thickened points}
(often called ``formal points'') and indeed we find examples
of this below in 
in Goodwillie-tangent cohesion, \ref{DiagramsOfCohesiveGroupoids},
in synthetic differential cohesion, \ref{SynthDiffInfGrpd},
and in supergeometric cohesion, \ref{SuperInfinityGroupoids}.
\end{example}
\begin{remark}
  The pieces-to-points transformation appears as part of the 
  canonical \emph{stable differential cohomology diagram} 
  (prop. \ref{StableDifferentialCohomologyDiagram} below) which exists for every
  object in Goodwillie-tangent cohesive $\infty$-toposes
  (def. \ref{TangentInfinityTopos} below).
\end{remark}
Therefore it is useful to introduce the following terminology.\footnote{
I am grateful to Mike Shulman for discussion of this notion.
In def. 1 of \cite{Lawvere} essentially this notion is referred to as a ``quality type''.}
\begin{definition}
  A cohesive $\infty$-topos $\mathbf{H}$ for which the 
  points-to-pieces transform, def. \ref{pointstopiecestransform}, is an equivalence
  $$
    \flat \stackrel{\simeq}{\longrightarrow} \mathbf{\Pi}
  $$
  we call an \emph{infinitesimal cohesive $\infty$-topos}.
\end{definition}
Infinitesimal cohesive $\infty$-toposes are typically simple in themselves,
but in examples they are relevant as alternative base $\infty$-toposes
over which richer $\infty$-toposes are cohesive. We will appeal repeatedly
to the following elementary fact.
\begin{proposition}
  If $\mathcal{C}$ is a small $\infty$-category with a zero-object 
  (an object which is both initial as well as terminal), then 
  the $\infty$-presheaf $\infty$-category $\mathrm{PSh}_\infty(\mathcal{C})$,
  def. \ref{InfinityPresheaves},
  is infinitesimally cohesive, def. \ref{InfinitesimalCohesion}.
  \label{InfinitesimalCohesiveSite}
\end{proposition}
\proof
  The constant $\infty$-presheaf $\infty$-functor
  $\mathrm{Disc} : \infty \mathrm{Grpd} \to \mathrm{PSh}_\infty(\mathrm{Grpd})$
  has a left adjoint $\Pi$ and a right adjoint $\Gamma$, given by forming $\infty$-limits
  and $\infty$-colomits of $\infty$-functors on $\mathcal{C}$, respectively. 
  Due to the assumption of a zero object $\ast$, both of these are given by
  evaluation on that zero object. This first of all implies that 
  $\Gamma \mathrm{Disc} \simeq \mathrm{id}$, hence that $\Gamma$ is full and 
  faithful, and that $\Pi$ preserves all $\infty$-limits, hence finite $\infty$-products,
  so that $\mathrm{PSh}_\infty(C)$ is indeed cohesive. 
  Second it implies that the unit $\mathrm{id} \longrightarrow \mathrm{Disc}\Pi$
  is given on generators in $\mathcal{C} \hookrightarrow \mathrm{PSh}_\infty(\mathcal{C})$
  by sending each of them to the zero object, and hence that 
  $\Gamma \to \Gamma \mathrm{Disc}\Pi$ is an equivalence. 
  By prop. \ref{adjunctofpointstopieces} this implies the claim.
\endofproof
\begin{remark}
Below in \ref{DefinitionOfInfinitesimalCohesion} we re-encounter infinitesimal
cohesion in the more general context of \emph{differential cohesion}.
\end{remark}

\paragraph{Internal formulation}
\label{InternalCohesion}

The above discussion in \ref{DefinitionCohesiveToposes} looks at an $\infty$-topos 
``from the outside'', namely as an object of the $\infty$-category
of all $\infty$-toposes, and characterizes its cohesion in terms of additional
properties of functors defined \emph{on} it. But in 
\ref{DependentHomotopyTypeTheory} we saw that an $\infty$-topos
also comes with its \emph{internal homotopy type theory} \cite{HoTT},
which describes it ``from inside''. Mike Shulman has shown how one may
formulate the axioms of cohesion in this internal homotopy type theory,
to obtain \emph{cohesive homotopy type theory}. 
An exposition of this is in \cite{ScSh}, where pointers to the full details are given.

The crucial insight of Mike Shulman \cite{Shulman11}
is that to implement cohesion fully formally
in homotopy type theory one is to regard the sharp modality $\sharp$,
remark \ref{SharpModalityOfLocalTopos}, as the fundamental axiom that serves
to exhibit the external base $\infty$-topos as an internal sub-system of homotopy types.
Then the flat modality and the shape modality are axiomatized based on the
existence of the sharp modality.

While traditional topos theory (hence: 1-topos theory) had had an emphasis on the internal logic
provided by toposes from the very beginning \cite{Lawvere65},
the formulation of constructions in higher topos theory in general and of 
cohesive higher topos theory in particular in terms of the internal language of
homotopy type theory has only just begun to be explored. But it is clear
that it can provide considerable advantages. For instance the whole theory of
relative Postnikov-Whitehead towers in $\infty$-toposes 
(see \ref{StrucEpi} below), which in 
\cite{Lurie} takes a fairly lenghty list of lemmas to establish,
follows elegantly with a few simple proofs from homotopy type theory, 
see chapter 7 of \cite{HoTT} (some of this goes back to \cite{Rijke}). 
Combined with the richness of the formal consequences
of the axioms of cohesion, for instance in the derivation of the
long fiber sequences in stable differential cohomology
in \ref{BundlesOfCohesiveSpectra} below, this opens interesting
perspectives.

In the following we briefly sketch how one begins going about re-formulating
the axioms of cohesion in terms of structure internal to the ambient $\infty$-topos.
For more details we refer the reader to \cite{ScSh} and the pointers given there.

\medskip

\begin{theorem}
  \label{SemiInternalFormulationOfCohesion}
  Let $\mathbf{H}$ be an $\infty$-topos.
  The inclusion of a 
  full sub-$\infty$-category 
  $$
    \mathrm{Disc} : \mathbf{B}_{\mathrm{disc}} \hookrightarrow \mathbf{H}
  $$ 
  -- to be called the \emph{discrete objects} --
  and of a full sub-$\infty$-category
  $$
    \mathrm{coDisc} : \mathbf{B}_{\mathrm{cod}} \hookrightarrow \mathbf{H}
  $$ 
  -- to be called the \emph{codiscrete objects} --
  satisfies $\mathbf{B}_{\mathrm{disc}} \simeq \mathbf{B}_{\mathrm{cod}}$ 
  and extends to an adjoint quadruple of the form
  $$
    \xymatrix{
	  \mathbf{H}
	   \ar@{->}@<+12pt>[rr]|<\times|{\Pi}
	   \ar@{<-^{)}}@<+4pt>[rr]|{\mathrm{Disc}}
	   \ar@{->}@<-4pt>[rr]|{\Gamma}
	   \ar@{<-^{)}}@<-12pt>[rr]|{\mathrm{coDisc}}
	   &&
	  \mathbf{B}
	}
  $$
  as in def. \ref{CohesiveInfinToposDefinition}
  precisely if for every object $X \in \mathbf{H}$
  \begin{enumerate}
    \item 
	  there exists, with notation from def. \ref{TheAdjointTripleOfEndofunctors},
	  \begin{enumerate}
	    \item a morphism 
		 $X \to \mathbf{\Pi} X$ to a discrete object;
	    \item a morphism 
		 $\flat X \to X$ from a discrete object;
	    \item a morphism 
		 $X \to \sharp X$ to codiscrete object;
	  \end{enumerate}
	\item 
	  such that for all discrete $Y$ and codiscrete $\tilde Y$ the induced morphisms
	  \begin{enumerate}
	    \item 
		  $\mathbf{H}(\mathbf{\Pi}X, Y) \to \mathbf{H}(X,Y)$;
		\item 
		  $\mathbf{H}(Y, \flat X) \to \mathbf{H}(Y,X)$;
		\item
		  $\mathbf{H}(\sharp X, \tilde Y) \to \mathbf{H}(X, \tilde Y)$;
        \item
          $\sharp (\flat X \to X)$;
        \item
          $\flat(X \to \sharp X)$
	  \end{enumerate}
	  are equivalences.
  \end{enumerate}
  Finally, $\Pi$ preserves the terminal object if 
  the morphism ${*} \to \mathbf{\Pi}*$ is an equivalence.
\end{theorem}
\proof
  Prop. 5.2.7.8 in \cite{Lurie} asserts that
  a full sub-$\infty$-category $\mathbf{B}\hookrightarrow \mathbf{H}$
  is reflectively embedded precisely if for every object $X \in \mathbf{H}$
  there is a morphism 
  $$
    \mathrm{loc}_X : X \to \mathbf{L} X
  $$ to an object
  $\mathbf{L}X \in \mathbf{H} \hookrightarrow \mathbf{H}$
  such that for all $Y \in \mathbf{B} \hookrightarrow \mathbf{H}$
  the morphism
  $$
    \mathbf{H}(\mathrm{loc}_X, Y) : \mathbf{H}(\mathbf{L}X, Y)
	\to 
	\mathbf{H}(X,Y)
  $$
  is an equivalence. In this case $\mathbf{L}$ is the 
composite of the embedding and its left adjoint.  
  Accordingly, a dual statement
  holds for coreflective embeddings. This gives the structure
  and the first three properties of the above assertion. We
  identify therefore
  $$
    (\mathbf{\Pi} \dashv \flat \dashv \sharp)
	:=
	(\mathrm{Disc}\,\Pi \dashv \mathrm{Disc}\,\Gamma \dashv \mathrm{coDisc}\,\Gamma)
	\,.
  $$
  
  It remains to show that the last two properties say precisely that
  the sub-$\infty$-categories of discrete and codiscrete objects
  are equivalent and that under this equivalence their coreflective
  and reflective embedding, respectively, fits into a single adjoint triple.
  It is clear that if this is the case then the last two properties hold.
  We show the converse.
  
  First notice that the two embeddings always combine into an
  adjunction of the form
  $$
    \xymatrix{
	  \mathbf{B}_{\mathrm{disc}}
	   \ar@{^{(}->}@<+4pt>^{\mathrm{Disc}}[rr]
	   \ar@{<-}@<-4pt>_{\Gamma}[rr]	   
	  &&
	  \mathbf{H}
	   \ar@{->}@<+4pt>^{\tilde \Gamma}[rr]	   
	   \ar@{<-^{)}}@<-4pt>_{\mathrm{coDisc}}[rr]
	  &&
	  \mathbf{B}_{\mathrm{cod}}
	}
	\,.
  $$
  The equivalence $\sharp(\flat X \to X)$ applied to $X := \mathrm{coDisc} A$
  gives that $\mathrm{coDisc}$ applied to the counit of this 
  composite adjunction is an equivalence
  $$
    \mathrm{coDisc}\, \tilde \Gamma\, \mathrm{Disc}\, \Gamma \mathrm{coDisc} A
	\stackrel{\simeq}{\longrightarrow}
	\mathrm{coDisc}\, \tilde \Gamma\, \mathrm{coDisc} A
	\stackrel{\simeq}{\longrightarrow}
	\mathrm{coDisc} A
  $$
  and since $\mathrm{coDisc}$ is full and faithful, so is the
  composite counit itself. Dually, the equivalence 
  $\flat (X \to \sharp X)$ implies that the unit of this composite
  adjunction is an equivalence. Hence the adjunction itself
  is an equivalence, and so 
  $\mathbf{B}_{\mathrm{disc}} \simeq \mathbf{B}_{\mathrm{cod}}$.
  Using this we obtain a composite equivalence
  $$
    \mathrm{Disc}\,\tilde \Gamma X 
	\stackrel{\simeq}{\to}
	\mathrm{Disc}\,\Gamma \mathrm{coDisc}\, \tilde \Gamma X
	\stackrel{\simeq}{\to}
	\mathrm{Disc}\,\Gamma X
	\,,
  $$
  where the left morphism is the image under $\mathrm{Disc}$
  of the ave composite adjunction on the codiscrete object
  $\tilde \Gamma X$, and where the second is a natural inverse
  of $\flat(X \to \sharp X)$. Since $\mathrm{Disc}$ is full and
  faithful, this implies that
  $$
    \Gamma \simeq \tilde \Gamma
	\,.
  $$
\endofproof
This formulation of cohesion is not entirely internal yet, since
it still refers to the external hom $\infty$-groupoids $\mathbf{H}$.
But cohesion also implies that the external $\infty$-groupoids
can be re-internalized.
\begin{proposition}
  The statement of theorem \ref{SemiInternalFormulationOfCohesion}
  remains true with items 2. a) -  2. b) replaced by
  \begin{enumerate}
    \item[2. (a')] $\sharp [\mathbf{\Pi}X, Y] \to \sharp[X, Y]$;
	  \item[2. (b')] $\sharp [Y, \flat X] \to \sharp [Y,X]$;
	  \item[2. (c')] $[\sharp X, \tilde Y] \to [X, \tilde Y]$;
  \end{enumerate}
  where $[-,-]$ denotes the internal hom in $\mathbf{H}$.
\end{proposition}
\proof
By prop. \ref{codiscreteObjectsStableUnderExponentiation}
we have for codiscrete $\tilde Y$ equivalences $[X, \tilde Y] \simeq
\mathrm{coDisc} \mathbf{H}(X, \tilde Y)$. Since $\mathrm{coDisc}$
is full and faithful, the morphism 
$\mathbf{H}(\sharp X, \tilde Y) \to 
\mathbf{H}(X, \tilde Y)$ is an equivalence precisely if
$[\sharp X, \tilde Y] \to [X, \tilde Y]$ is.

Generally, we have $\Gamma [X,Y] \simeq \mathbf{H}(X,Y)$.
With the full and faithfulness of $\mathrm{coDisc}$ this 
similarly gives the remaining statements.
\endofproof

\subsubsection{Presentation}
\label{CohesionPresentations}

We discuss presentations of cohesive $\infty$-toposes, in the 
sense of presentation of $\infty$-toposes as discussed in 
\ref{InfinityToposPresentation}. 
In \ref{CohesiveSites} we consider sites such that the
$\infty$-topos of $\infty$-sheaves over them is cohesive. 
In \ref{CohesiveFibrancy} we analyze fibrancy and descent over these sites.
These considerations serve as the basis for the construction of models of 
cohesion below in \ref{Implementation}.

\paragraph{Presentation over $\infty$-cohesive sites}
\label{CohesiveSites}

We discuss a class of sites with the property that the 
$\infty$-toposes of $\infty$-sheaves over them (\ref{InfinityToposPresentation})
are cohesive, def. \ref{CohesiveInfinToposDefinition}.

\medskip

\begin{definition} 
  \label{CohesiveSite}
  \index{cohesive site}
  \index{cohesive $\infty$-topos!site of definition}
  An \emph{$\infty$-cohesive site} is a site such that
  \begin{enumerate}
    \item 
      it has finite products;
    \item
      every object $U \in C$ has at least one point: $C(*,U) \neq \emptyset$;
    \item
      for every covering family $\{U_i \to U\}$ its
     {\v C}ech nerve $C(\{U_i\}) \in [C^{\mathrm{op}}, \mathrm{sSet}]$ 
     is degreewise a coproduct of representables 
    \item
     the canonical morphisms $C(\{U_i\}) \to U$ are taken to weak equivalences 
     by both limit and colimit $[C^{\mathrm{op}}, \mathrm{sSet}] \to \mathrm{sSet}$:
     $$
     \begin{aligned}
       &\lim\limits_{\longrightarrow} C(\{U_i\}) \stackrel{\simeq}{\to} \lim\limits_{\longrightarrow} U_i
       \\
       &\lim\limits_{\longleftarrow} C(\{U_i\}) \stackrel{\simeq}{\to} \lim\limits_{\longleftarrow} U_i
     \end{aligned}
      \,.
     $$
  \end{enumerate}
\end{definition}
Notice that for the representable $U$ we have $\lim_\to U \simeq {*}$ and that since $C$
is assumed to have finite products and hence in particular a terminal object
$\lim_\leftarrow U = C(*,U)$.
\begin{proposition} 
  \index{cohesive $\infty$-topos!presentation over $\infty$-cohesive site}
  The $\infty$-sheaf $\infty$-topos over an $\infty$-cohesive site is 
  a cohesive $\infty$-topos in which for all objects \emph{pieces have points},
  def. \ref{PiecesHavePoints}.
  \label{InfSheavesOverCohesiveSiteAreCohesive}
\end{proposition}
\proof
  Since an $\infty$-cohesive site is in particular a locally and globally
  $\infty$-connected site (def. \ref{ConnectedSite}) it follows with
  theorem \ref{ToposOnInfConnectedSiteIsInfConnected} that $\Pi$ exists
  and preserves the terminal object. Moreover, by the discussion there $\Pi$
  acts by sending a fibrant-cofibrant simplicial presheaf $F : C^{\mathrm{op}} \to \mathrm{sSet}$ 
  to its colimit. Since
  $C$ is assumed to have finite products, $C^{\mathrm{op}}$ has finite coproducts, hence
  is a sifted category. Therefore taking colimits of functors on $C^{\mathrm{op}}$ commutes
  with taking products of these functors. Since the $\infty$-product of $\infty$-presheaves
  is modeled by the ordinary product on fibrant simplicial presheaves, it follows that 
  over an $\infty$-cohesive site
  $\Pi$ indeed exhibits a strongly $\infty$-connected $\infty$-topos.
  
  Using the notation and results of the proof of theorem \ref{ToposOnInfConnectedSiteIsInfConnected}, 
  we show that the further right adjoint $\Delta$ exists by
  exhibiting a suitable right Quillen adjoint to 
  $\Gamma : [C^{\mathrm{op}}, \mathrm{sSet}] \to \mathrm{sSet}$, which is given by
  evaluation on the terminal object.
  Its $\mathrm{sSet}$-enriched right adjoint is given by 
  $$
    \nabla S : U \mapsto \mathrm{sSet}(\Gamma(U), S)
  $$
  as confirmed by the following end/coend computation:
 $$
  \begin{aligned}
    [C^{op}, \mathrm{sSet}](X, \nabla(S))
    & =
    \int_{U \in C} \mathrm{sSet}(X(U), \mathrm{sSet}(\Gamma(U), S)
    \\
    & = \int_{U \in C} \mathrm{sSet}(X(U) \times \Gamma(U), S)
    \\
    & = \mathrm{sSet}( \int^{U \in C} X(U) \times \Gamma(U), \;\; S )
    \\
    & = \mathrm{sSet}( \int^{U \in C } X(U) \times \mathrm{Hom}_C(*, U), \;\; S)
    \\
    & = \mathrm{sSet}(X(*), S)
    \\
    & = \mathrm{sSet}(\Gamma(X), S)
  \end{aligned}
  \,,
$$
We have that
$$
  (\Gamma \dashv \nabla)
    : 
  [C^{\mathrm{op}}, \mathrm{sSet}]_{\mathrm{proj}} \stackrel{\overset{\Gamma}{\to}}{\underset{\nabla}{\leftarrow}}
  \mathrm{sSet}_{\mathrm{Quillen}}
$$
is a Quillen adjunction, since $\nabla$ manifestly preserves fibrations and acyclic fibrations. 
Since $[C^{\mathrm{op}}, \mathrm{sSet}]_{\mathrm{proj},\mathrm{loc}}$ is a 
left proper model category, to see that this descends to a Quillen adjunction on the 
local model structure it is sufficient by prop. \ref{AdjRecognition} 
to check that $\nabla : \mathrm{sSet}_{\mathrm{Quillen}} \to  [C^{\mathrm{op}}, \mathrm{sSet}]_{\mathrm{proj},\mathrm{loc}}$ preserves fibrant objects, 
in that for $S$ a Kan complex we have that $\nabla S$ satisfies descent along 
{\v C}ech nerves of covering families. 

This is implied by the second defining condition on the $\infty$-local site $C$, that 
$\lim\limits_{\longleftarrow} C(\{U_i\}) = \mathrm{Hom}_C(*,C(\{U_i)\}) \simeq 
  \mathrm{Hom}_C(*,U) = \lim\limits_{\longleftarrow} U$
is a weak equivalence. 
Using this we have for fibrant $S \in \mathrm{sSet}_{\mathrm{Quillen}}$
 the descent weak equivalence
$$
  \begin{aligned}
   {[C^{\mathrm{op}}, \mathrm{sSet}]}(U, \nabla S)
   &=
  \mathrm{sSet}(\mathrm{Hom}_C(*,U), S)  
  \\
   &\simeq
  \mathrm{sSet}(\mathrm{Hom}_C(*,C(U)), S)
  \\
   &=
  [C^{\mathrm{op}}, \mathrm{sSet}](C(U), \nabla S)
  \end{aligned}
	\,,
$$
where we use in the middle step that 
$\mathrm{sSet}_{\mathrm{Quillen}}$ is a simplicial model category so that 
homming the weak equivalence between cofibrant objects into the fibrant object $S$ 
indeed yields a weak equivalence.

It remains to show that \emph{pieces have points}, def. \ref{PiecesHavePoints}, 
in $\mathrm{Sh}_{\infty}(C)$.
For the first statement we use the cofibrant replacement theorem from \cite{Dugger} for 
$[C^{\mathrm{op}}, \mathrm{sSet}]_{\mathrm{proj, \mathrm{loc}}}$ 
which says that for $X$ any simplicial presheaf, a 
functorial projective cofibrant replacement is
given by the object
$$
  Q X :=
  \left(
  \xymatrix{
     \cdots
     \ar@<+4pt>[r]
     \ar@<+0pt>[r]
     \ar@<-4pt>[r]
     &
     \coprod_{U_0 \to U_1 \to X_1} U_0
     \ar@<+2pt>[r]
     \ar@<-2pt>[r]
     &
     \coprod_{U_0 \to X_0} U_0
  }
  \right)
  \,,
$$
where the coproducts are over the set of morphisms of presheaves from representables $U_i$
as indicated. By the above discussion, the presentations of $\Gamma$ and $\Pi$
by left Quillen functors $\lim\limits_{\leftarrow}$ and $\lim\limits_{\longrightarrow}$ 
takes this to the morphism $\lim\limits_{\leftarrow} QX \to \lim\limits_{\to} Q X$ 
induced in components by
$$
  \xymatrix{
     \cdots
     \ar@<+4pt>[r]
     \ar@<+0pt>[r]
     \ar@<-4pt>[r]
     &
     \coprod_{U_0 \to U_1 \to X_1} C(*.U_0)
     \ar@<+2pt>[r]
     \ar@<-2pt>[r]
     \ar[d]
     &
     \coprod_{U_0 \to X_0} C(*,U_0) \ar[d]
     \\
     \cdots
     \ar@<+4pt>[r]
     \ar@<+0pt>[r]
     \ar@<-4pt>[r]
     &
     \coprod_{U_0 \to U_1 \to X_1} {*}
     \ar@<+2pt>[r]
     \ar@<-2pt>[r]
     &
     \coprod_{U_0 \to X_0} {*}
  }
  \,.
$$
By assumption on $C$ we have that all sets $C(*,U_0)$ are non-empty, so that this is
componentwise an epimorphism and hence induces in particular an epimorphism on connected components.

Finally, for $S$ a Kan complex we have by the above that $\mathrm{Disc} S$ is the presheaf constant
on $S$. Its homotopy sheaves are the presheaves constant on the homotopy groups of $S$.
The inclusion of these into the homotopy sheaves of $\mathrm{coDisc}S$ is over each 
$U \in C$ the diagonal injection
$$
  \pi_n(S,x) \hookrightarrow \pi_n(S,x)^{C(*,U)}
  \,.
$$
Therefore also \emph{discrete objects are concrete} in the $\infty$-topos over the $\infty$-cohesive site 
$C$.
\endofproof
Below in \ref{Implementation} we discuss in detail the following examples.
\begin{examples}
  The following sites are $\infty$-cohesive.
  \begin{itemize}
    \item The site $\mathrm{CartSp}_{\mathrm{top}}$ of Cartesian spaces, continuous maps between them and good open covers (prop. \ref{CartSpTopIsCohesive}).
	\item The site $\mathrm{CartSp}_{\mathrm{smooth}}$ of Cartesian spaces, smooth maps between them and good open covers (prop. \ref{CartSpSmoothIsCohesive}),
	\item The site $\mathrm{CartSp}_{\mathrm{SynthDiff}}$ of Cartesian spaces with infinitesimal thickening, smooth maps between the and good open covers that are the identity on the thickening (prop. \ref{SynthDiffIsCohesive}).
	\item The site $\mathrm{CartSp}_{\mathrm{super}}$ of super-Cartesian spaces, morphisms of supermanifolds between them and good open covers.
  \end{itemize}
\end{examples}

We record a fact that is expected to hold quite generally for
$\infty$-toposes, but for which we currently have a proof
only over $\infty$-connected sites.
\begin{theorem}[parameterized $\infty$-Grothendieck construction]
  \label{ParameterizedGrothendieckConstruction}
  Let $\mathbf{H}$ be an $\infty$-topos with an $\infty$-connected
  site of definition, def. \ref{ConnectedSite},
  and let $A \in \infty \mathrm{Grpd}$ be any
  $\infty$-groupoid. 
  Then there is an equivalence of $\infty$-categories
  $$
    \mathbf{H}/_{\mathrm{Disc} A} \simeq \mathbf{H}^A
  $$
  between the slice $\infty$-topos 
  of $\mathbf{H}$ over the discrete cohesive $\infty$-groupoid on $A$
  and the $\infty$-category of $\infty$-functors $A \to \mathbf{H}$.
\end{theorem}
\proof
  For the case that the site of definition is terminal, hence that
  $\mathbf{H} \simeq \infty \mathrm{Grpd}$ this statement is the
  \emph{$\infty$-Grothendieck construction} from section 2 of 
  \cite{Lurie}. There the equivalence of $\infty$-categories
  $$
    \infty \mathrm{Grpd}_{/ A} \simeq \infty \mathrm{Grpd}^A
  $$
  which takes a fibration to an $\infty$-functor that assigns its
  fibers
  is presented by a Quillen equivalence of model categories
  $$
    \xymatrix{
      \mathrm{sSet}^+/A 
	  \ar@<+3pt>[r]
	  \ar@{<-}@<-3pt>[r]
	  &
	  [w(A)^{\mathrm{op}}, \mathrm{sSet}]_{\mathrm{proj}}
	}
  $$
  between a model structure on \emph{marked simplicial sets} $\mathrm{sSet}^+$
  over a Kan complex $A$ and the global projective model structure on
  enriched presheaves on the simplically enriched category $w(A)$
  corresponding to $A$ by the discussion in section 1.1.5 of \cite{Lurie}. 
  
  Now for $C$ an $\infty$-connected site and
  $\mathbf{H} \simeq 
  ([C^{\mathrm{op}}, \mathrm{sSet}]_{\mathrm{proj}, \mathrm{loc}})^\circ$ 
  we have by the proof of prop. \ref{ToposOnInfConnectedSiteIsInfConnected} 
  that
  with $A$ a Kan complex, the constant simplicial presheaf
  $\mathrm{const} A : C^{\mathrm{op}} \to \mathrm{sSet}$ is a 
  fibrant presentation in $[C^{\mathrm{op}}, \mathrm{sSet}]_{\mathrm{proj}, \mathrm{loc}}$ of $\mathrm{Disc}A$. Therefore the $\infty$-categorical slice 
  $\mathbf{H}_{/\mathrm{Disc} A}$ is presented by the induced model structure
  on the 1-categorical slice category
  $$
    \mathbf{H}_{/\mathrm{Disc} A}
	\simeq
	\left(
	   ([C^{\mathrm{op}}, \mathrm{sSet}]_{/{\mathrm{const} A}})_{\mathrm{proj}, \mathrm{loc}/\mathrm{const}A}
	 \right)^\circ
	 \,.
  $$
  We have an evident equivalence of 1-categories
  $$
    [C^{\mathrm{op}}, \mathrm{sSet}]_{/\mathrm{const} A}
	\simeq
    [C^{\mathrm{op}}, \mathrm{sSet}_{/A}]
  $$
  under which the above slice model structure is seen to become 
  the model structure on presheaves with values in the slice model structure
  $(\mathrm{sSet}_{/A})_{\mathrm{Quillen}/A}$, hence
  $$
    \mathbf{H}_{/\mathrm{Disc} A}
	\simeq
	\left(
	   [C^{\mathrm{op}}, (\mathrm{sSet}_{/A})_{\mathrm{Quillen}/A}]_{\mathrm{proj}, \mathrm{loc}}
	\right)^\circ
	\,.
  $$
  Since $A$ is fibrant in the Quillen model structure, the slice model structure
  here presents the $\infty$-categorical slice of $\infty$-groupoids
  $$
     \infty \mathrm{Grpd}_{/A} 
	 \simeq
 	\left(
	  (\mathrm{sSet}_{/A})_{\mathrm{Quillen}/A}
	\right)^\circ
	\,.
   $$
   By the above presentation of the $\infty$-Grothendieck construction 
   by marked simplicial sets, this is equivalently
   $$
	\cdots \simeq
	\left(
	  \mathrm{sSet}^+/A
	\right)^\circ 
	\simeq
	\left(
	  [w(A)^{\mathrm{op}}, \mathrm{sSet}]_{\mathrm{proj}}
	\right)^\circ 
	\,.
  $$
  Since all model categories appearing here are combinatorial, it
  follows with prop. 4.2.4.4 in \cite{Lurie} that we have an equivalence
  of $\infty$-categories
  $$
    \mathbf{H}_{/\mathrm{Disc}A}
	\simeq
    \left(
      [
	    C^{\mathrm{op}}, 
		[w(A)^{\mathrm{op}}, \mathrm{sSet}]_{\mathrm{proj}}
	  ]_{\mathrm{proj}, \mathrm{loc}}  
	\right)^\circ
  $$
  and hence
  $$
    \cdots
	\simeq
    \left(
	  [w(A)^{\mathrm{op}},
	    [C^{\mathrm{op}}, \mathrm{sSet}]_{\mathrm{proj}, \mathrm{loc}}
	  ]_{\mathrm{proj}}
	\right)^\circ	
	\simeq
	\mathbf{H}^A
	\,.
  $$
\endofproof

\paragraph{Fibrancy over $\infty$-cohesive sites}
\label{CohesiveFibrancy}

The condition on an object 
$X \in [C^{\mathrm{op}}, \mathrm{sSet}]_{\mathrm{proj}}$
to be fibrant models the fact that $X$ is an 
\emph{$\infty$-presheaf of $\infty$-groupoids}. 
The condition that $X$ is also fibrant as an object in
$[C^{\mathrm{op}}, \mathrm{sSet}]_{\mathrm{proj}, \mathrm{loc}}$
models the higher analog of the sheaf condition: it makes $X$ an
\emph{$\infty$-sheaf}.
For generic sites $C$ fibrancy in the local model structure is a
property rather hard to check or establish concretely. But 
often a given site can be replaced by another site on which 
the condition is easier to control, without changing the
corresponding $\infty$-topos, up to equivalence. Here we discuss
for cohesive sites, def. \ref{CohesiveSite} explicit conditions for
a simplicial presheaf over them to be fibrant.

\medskip

In order to discuss descent over $C$ it is convenient to 
introduce the following notation for ``cohomology over the site $C$''.
For the moment this is just an auxiliary technical notion. Later
we will see how it relates to an intrinsically defined notion of 
cohomology.
\begin{definition}
 For $C$ an  $\infty$-cohesive site,
 $A \in [C^{\mathrm{op}}, \mathrm{Set}]_{\mathrm{proj}}$ 
 fibrant, and $\{U_i \to U\}$ a good cover in $U$, we write 
 $$
   H^n_C(\{U_i\},A) := \pi_0 \mathrm{Maps}(C(\{U_i\}), A)
   \,.
 $$
 Moreover, if $A$ is equipped with (abelian) group structure
 we write
 $$
   H^n_C(\{U_i\},A) := \pi_0 \mathrm{Maps}(C(\{U_i\}), \Wbar^n A)
   \,.
 $$
\end{definition}
\begin{definition}
  \label{CAcyclic}
  An object $A \in [C^{\mathrm{op}}, \mathrm{sSet}]$
  is called \emph{$C$-acyclic} if 
  \begin{enumerate}
    \item it is fibrant in $[C^{\mathrm{op}}, \mathrm{sSet}]_{\mathrm{proj}}$;
    \item for all $n \in \mathbb{N}$
  the homotopy group presheaves $\pi_n^{\mathrm{PSh}}$
  from def. \ref{HomotopySheaves}
  are already sheaves $\pi_n(A) \in \mathrm{Sh}(C)$;
    \item 
      for $n=1$ and $k = 1$ as well as $n \geq 2$ and $k \geq 1$ we have
    $
      H^{k}_C(\{U_i\},\pi_n(A)) \simeq *
    $
    for all good covers $\{U_i \to U\}$.
  \end{enumerate}
\end{definition}
\begin{remark}
  This definition can be formulated and the following statements about it
  are true over any site whatsoever. 
  However, on generic sites $C$ the $C$-acyclic objects are not 
  very interesting. On $\infty$-cohesive sites on the other hand they
  are of central importance.   
\end{remark}
\begin{observation}
  \label{LoopingOfHCAcyclicObjects}
  If $A$ is $C$-acyclic then for every point $x : * \to A$ 
  also $\Omega_x A$ is $C$-acyclic (for any model of the loop space 
  object in $[C^{\mathrm{op}}, \mathrm{sSet}]_{\mathrm{proj}}$).
\end{observation}
\proof
   The standard statement in $\mathrm{sSet}_{\mathrm{Quillen}}$
   $$
     \pi_n \Omega X \simeq \pi_{n+1} X
   $$
   directly prolongs 
   to $[C^{\mathrm{op}}, \mathrm{sSet}]_{\mathrm{proj}}$.
\endofproof
\begin{theorem}
  \label{CAcyclicityAndLocalFibrancy}
  \label{SufficientConditionsForFibrancyOverCohesives}
  Let $C$ be an $\infty$-cohesive site.
  Sufficient conditions for an object 
  $A \in [C^{\mathrm{op}}, \mathrm{sSet}]$ 
  to be fibrant in the local 
  model structure 
  $[C^{\mathrm{op}}, \mathrm{sSet}]_{\mathrm{proj}, \mathrm{loc}}$
  are
  \begin{itemize}
    \item 
      $A$ is 0-truncated and $C$-acyclic;
    \item 
      $A$ is connected and $C$-acyclic;
    \item
      $A$ is a group object and $C$-acyclic.
  \end{itemize}
\end{theorem}
Here and in the following ``truncated'' and ``connected'' are as simplicial presheaves
(not after sheafification of homotopy presheaves).

We demonstrate this statement in several stages.
\begin{proposition}
  \label{DescentOf0TruncatedObjects}
  A 0-truncated object is fibrant in
  $[C^{\mathrm{op}}, \mathrm{sSet}]_{\mathrm{proj}, \mathrm{loc}}$
  precisely if it is fibrant in 
  $[C^{\mathrm{op}}, \mathrm{sSet}]_{\mathrm{proj}}$
  and weakly equivalent to a sheaf: to an object in the image of the 
  canonical inclusion
  $$
   \mathrm{Sh}_C \hookrightarrow
  [C^{\mathrm{op}}, \mathrm{Set}] \hookrightarrow
  [C^{\mathrm{op}}, \mathrm{sSet}]
    \,.
  $$
\end{proposition}
\proof
  From general facts of left Bousfield localization we have 
  that the fibrant objects in the local model structure
  are necessarily fibrant also in the global structure.
  
  Since moreover $A \to \pi_0(A)$ is a weak equivalence in the 
  global model structure by assumption, we have for every
  covering $\{U_i \to U\}$ in $C$
  a sequence of weak equivalences
  $$
    \mathrm{Maps}(C(\{U_i\}), A)
     \stackrel{\simeq}{\to}
    \mathrm{Maps}( C(\{U_i\}), \pi_0(A))    
     \stackrel{\simeq}{\to}
    \mathrm{Maps}( \pi_0 C(\{U_i\}), \pi_0(A))    
     \stackrel{\simeq}{\to}
    \mathrm{Sh}_C(S(\{U_i\}), \pi_0(A))    
    \,,
  $$
  where $S(\{U_i\}) \hookrightarrow U$ is the sieve corresponding to the
  cover. Therefore the descent condition
  $$
    \mathrm{Maps}(U, A)\stackrel{\simeq}{\to}
    \mathrm{Maps}(C(\{U_i\}), A)
  $$
  is precisely the sheaf condition for $\pi_0(A)$.
\endofproof
\begin{proposition}
  \label{DescentOnPi0AndLooping}
  A connected fibrant object $A \in [C^{\mathrm{op}}, \mathrm{sSet}]_{\mathrm{proj}}$ 
  is fibrant in 
  $[C^{\mathrm{op}}, \mathrm{sSet}]_{\mathrm{proj}, \mathrm{loc}}$ 
  if for all objects $U \in C$
  \begin{enumerate}
    \item 
      $H_C(U, A) \simeq *$;
    \item
      $\Omega A$ is fibrant in 
        $[C^{\mathrm{op}}, \mathrm{sSet}]_{\mathrm{proj}, \mathrm{loc}}$\,,
  \end{enumerate}
  where $\Omega A$ is any fibrant object in 
  $[C^{\mathrm{op}}, \mathrm{sSet}]_{\mathrm{proj}}$ representing the
  looping of $A$.
\end{proposition}
\proof
  For $\{U_i \to U\}$ a covering we need to show that the 
  canonical morphism
  $$
    \mathrm{Maps}(U, A)
    \to
    \mathrm{Maps}(C(\{U_i\}), A)
  $$
  is a weak homotopy equivalence. This is equivalent to 
  the two morphisms
  \begin{enumerate}
    \item 
      $\pi_0 \mathrm{Maps}(U, A)
      \to
      \pi_0\mathrm{Maps}(C(\{U_i\}), A)$
    \item
       $\Omega \mathrm{Maps}(U, A)
       \to
      \Omega \mathrm{Maps}(C(\{U_i\}), A)$
  \end{enumerate}
  being weak equivalences.
  Since $A$ is connected the first of these says that there is 
  a weak equivalence
  $* \stackrel{\simeq}{\to} H_C(U,A)$.
  The second condition is equivalent to 
       $\mathrm{Maps}(U, \Omega A)
       \to
      \mathrm{Maps}(C(\{U_i\}), \Omega A)$,
      being a weak equivalence, hence to the descent of $\Omega A$.
\endofproof
\begin{proposition}
  \label{DescentFor1Truncated}
  An object $A$ which is connected, 1-truncated and $C$-acyclic is
  fibrant in $[C^{\mathrm{op}}, \mathrm{sSet}]_{\mathrm{proj}, \mathrm{loc}}$.
\end{proposition}
\proof
  Observe that for a connected and 1-truncated objects
  we have a weak equivalence $A \simeq \Wbar \pi_1(A)$
  in $[C^{\mathrm{op}}, \mathrm{sSet}]_{\mathrm{proj}}$.
  The first condition of prop. \ref{DescentOnPi0AndLooping}
  is then implied by $C$-connectedness.
  The second condition there is that $\pi_1(A)$ satisfies descent.
  By $C$-acyclicity this is a sheaf and it is 0-truncated
  by assumption, therefore it  
  satisfies descent by prop \ref{DescentOf0TruncatedObjects}.
\endofproof
\begin{proposition}
  \label{DescentOfConnectedHCacyclicObjects}
  Every connected and $C$-acyclic object 
  $A \in [C^{\mathrm{op}}, \mathrm{sSet}]_{\mathrm{proj}}$ 
 is fibrant in 
  $[C^{\mathrm{op}}, \mathrm{sSet}]_{\mathrm{proj}, \mathrm{loc}}$.
\end{proposition}
\proof
  We first show the statement for truncated $A$
  and afterwards for the general case.

  The $k$-truncated case in turn we consider by 
  induction over $k$.
  If $A$ is 1-truncated the proposition holds by
  prop. \ref{DescentFor1Truncated}.
  Assuming then that the statement has been shown for 
  $k$-truncated $A$, we need to show it for $(k+1)$-truncated $A$.

  This we do by decomposing $A$ into its canonical Postnikov tower
  def. \ref{PostnikovTower}:
  For $n \in \mathbb{N}$ let
  $$
    A(n) := A/_{\sim_n}
  $$
  be the quotient simplicial presheaf where two cells 
  $$
    \alpha,\beta : \Delta^n \times U \to A
  $$
  are identified, $\alpha \sim_n \beta$, 
  precisely if they agree on their $n$-skeleton:
  $$
    \mathrm{sk}_n \alpha = \mathrm{sk}_n \beta : 
    \mathrm{sk}_n \Delta \hookrightarrow \Delta^n
    \to 
    A(U)
    \,.
  $$
  It is a standard fact 
  (shown in \cite{GoerssJardine}, theorem VI 3.5 for simplicial 
  sets, which generalizes immediately to the global model structure
  $[C^{\mathrm{op}}, \mathrm{sSet}]_{\mathrm{proj}}$ ) that 
  for all $n > 1$ we have sequences
  $$
    K(n) \to A(n) \to A(n-1)
    \,,
  $$
  where
  $A(n-1)$ is $(n-1)$-truncated with homotopy groups in degree
  $\leq n-1$ those of $A$, and 
  where the right morphism is a Kan fibration and the left 
  morphism is its kernel, such that 
  $$
    A = \lim\limits_{\longleftarrow_n} A(n)
    \,.
  $$
  Moreover, there are canonical weak homotopy equivalences
  $$
    K(n) \to \Xi ((\pi_{n-1} A)[n])
  $$
  to the Eilenberg-MacLane object on the $n$th homotopy group in degree $n$.

  Since $A(n-1)$ is $(n-1)$-truncated and connected, the 
  induction assumption implies that it is fibrant in the 
  local model structure.
  
  Moreover we see that $K(n)$ is fibrant in 
  $[C^{\mathrm{op}}, \mathrm{sSet}]_{\mathrm{proj}, \mathrm{loc}}$:
  the first condition of \ref{DescentOnPi0AndLooping}
  holds by the assumption that $A$ is $C$-connected. The
  second condition is implied again by the induction hypothesis, 
  since $\Omega K(n)$ is $(n-1)$-truncated, connected and 
  still $C$-acyclic, by observation \ref{LoopingOfHCAcyclicObjects}.

  Therefore in the diagram (where $\mathrm{Maps}(-,-)$ denotes the 
  simplicial hom complex)
 $$
   \xymatrix{
     \mathrm{Maps}(U,K(n))
     \ar[r]
     \ar[d]^\simeq
     &
     \mathrm{Maps}(U,A(n))
     \ar[r]
     \ar[d]
     &
     \mathrm{Maps}(U,A(n-1))
     \ar[d]^\simeq
     \\
     \mathrm{Maps}(C(\{U_i\}),K(n))
     \ar[r]
     &
     \mathrm{Maps}(C(\{U_i\}),A(n))
     \ar[r]
     &
     \mathrm{Maps}(C(\{U_i\}),A(n-1))     
   }
 $$
 for $\{U_i \to U\}$ any good cover in $C$
 the top and bottom rows are fiber sequences 
 (notice that all simplicial sets in the top row are connected because $A$ is connected) 
 and the left and
 right vertical morphisms are weak equivalences in 
 $[C^{\mathrm{op}}, \mathrm{sSet}]_{\mathrm{proj}}$
 (the right one since $A(n-1)$ is fibrant in the local model structure
 by induction hypothesis, as remarked before,
 and the left one by $C$-acyclicity of $A$). It follows
 that also the middle morphism is a weak equivalence.
 This shows that $A(n)$ is fibrant in 
 $[C^{\mathrm{op}}, \mathrm{sSet}]_{\mathrm{proj}, \mathrm{loc}}$.
 By completing the induction the same then follows for the 
 object $A$ itself.
 
 This establishes the claim for truncated $A$. To demonstrate
 the claim for general $A$ notice that the limit over a sequence
 of fibrations between fibrant objects 
 is a homotopy limit (by example \ref{CotowerCofibrancy}). 
 Therefore we have
 $$
   \raisebox{20pt}{
   \xymatrix{
      \mathrm{Maps}(U, A)
      \ar[d]
      & \simeq &
      \lim\limits_{\longleftarrow_n} \mathrm{Maps}(U, A(n))
      \ar[d]^{\simeq}
      \\
      \mathrm{Maps}(C(\{U_i\}), A)
      & \simeq &
      \lim\limits_{\longleftarrow_n}\mathrm{Maps}(C(\{U_i\}), A(n))
   }
   }
   \,,
 $$
 where the right vertical morphism is a morphism between homotopy
 limits in $[C^{\mathrm{op}}, \mathrm{sSet}]_{\mathrm{proj}}$
 induced by a weak equivalence of diagrams, hence is itself a weak equivalence.
 Therefore $A$ is fibrant in 
 $[C^{\mathrm{op}}, \mathrm{sSet}]_{\mathrm{proj}, \mathrm{loc}}$.
\endofproof
\begin{lemma}
  \label{CanonicalFiberSequenceForSimplicialGroup}
  For $G \in [C^{\mathrm{op}}, \mathrm{sSet}]$ 
  a group object, the canonical sequence
  $$
    G_0 \to G \to G/G_0
  $$
  is a homotopy fiber sequence in 
  $[C^{\mathrm{op}}, \mathrm{sSet}]_{\mathrm{proj}}$.
\end{lemma}
\proof
  Since homotopy pullbacks of presheaves are computed objectwise,
  it is sufficient to show this for $C = *$, hence in
  $\mathrm{sSet}_{\mathrm{Quillen}}$.
  One checks that generally, for $X$ a Kan complex and $G$ a simplicial group
  acting on $X$, the quotient morphism $X \to X/G$ is a Kan fibration.
  Therefore the homotopy fiber of $G \to G/G_0$ is presented by the
  ordinary fiber in $\mathrm{sSet}$. 
  Since the action of $G_0$ on $G$ is free, this is indeed $G_0 \to G$.   
\endofproof
\begin{proposition}
  Every $C$-acyclic group object 
  $G \in [C^{\mathrm{op}}, \mathrm{sSet}]_{\mathrm{proj}}$ 
  for which $G_0$ is a sheaf
  is fibrant in 
  $[C^{\mathrm{op}}, \mathrm{sSet}]_{\mathrm{proj}, \mathrm{loc}}$. 
\end{proposition}
\proof
  By lemma \ref{CanonicalFiberSequenceForSimplicialGroup}
  we have a fibration sequence
  $$  
     G_0 \to G \to G/G_0
     \,.
  $$
  Since $G_0$ is assumed to be a sheaf it is fibrant in the local
  model structure by 
  prop. \ref{DescentOf0TruncatedObjects}.
  Since $G/G_0$ is evidently connected and $C$-acyclic 
  it is fibrant in the local model structure by 
  prop. \ref{DescentOfConnectedHCacyclicObjects}.
  As before in the proof there this implies that also
  $G$ is fibrant in the local model structure.
\endofproof

We discuss some examples.
\begin{proposition}
  \label{FibrancyOfWCrossedModule}
  \index{group!2-group!crossed module!fibrancy}
  Let $(\delta : G_1 \to G_0)$ be a crossed module, 
  def. \ref{Strict2GroupInIntroduction},
  of sheaves over an $\infty$-cohesive site $C$. Then the simplicial delooping
  $\bar W (G_1 \to G_0)$ is fibrant in 
  $[C^{\mathrm{op}}, \mathrm{sSet}]_{\mathrm{proj}, \mathrm{loc}}$
  if the image factorization of $G_0 \times G_1 \stackrel{}{\to} G_0 \times G_0$ has sections over each $U \in C$ and if the presheaf 
  $\mathrm{ker} \delta$ is a sheaf.
\end{proposition}
\proof
  The existence of the lift ensures that the homotopy presheaf
  $\pi^{\mathrm{PSh}}_1 \bar W G$ is a sheaf.
  Notice that $\pi^{\mathrm{PSh}}_2 \bar W G = \mathrm{ker}(\delta)$.
  Since moreover $\bar W G$ is manifestly connected, the claim
  follows with theorem \ref{SufficientConditionsForFibrancyOverCohesives}.
\endofproof

\subsection{Differential cohesive $\infty$-toposes}
\label{InfinitesimalCohesion}
\index{infinitesimal cohesion}
\index{cohesive $\infty$-topos!infinitesimal cohesion}

We discuss extra structure on a cohesive $\infty$-topos
that encodes a refinement of the corresponding notion of cohesion to 
a notion of what may be called \emph{infinitesimal cohesion} or 
\emph{differential cohesion}. With respect to such 
it makes sense to ask if an object in the topos has \emph{infinitesimal extension}.

A basic class of examples of objects with infinitesimal extension are
\emph{infinitesimal intervals} $\mathbb{D}$ that arise,
in the presence of infinitesimal cohesion, from 
\emph{line objects} $\mathbb{A}$ as
the subobjects $\mathbb{D} \hookrightarrow \mathbb{A}$ of elements that
square to 0 (in the internal logic of the topos)
$$
  \mathbb{D} = \{x \in \mathbb{A} | x \cdot x = 0\}
  \,.
$$
These objects co-represent tangent spaces, in that for any other object
$X$ the internal hom object $T X := [\mathbb{D}, X]$ plays the role of the
\emph{tangent bundle} of $X$.

A well-known proposal for an axiomatic characterization of infinitesimal objects in a
1-topos goes by the name \emph{synthetic differential geometry} 
\cite{LawvereSynth}, where infinitesimal extension is characterized
by algebraic properties of dual function algebras, as above. From the point of view and in the
presence of cohesion in an $\infty$-topos, however, there is a more immediate geometric 
characterization: an object $\mathbb{D}$ in a cohesive $\infty$-topos $\mathbf{H}$ behaves
like a possibly infinitesimally thickened point if 
\begin{enumerate}
  \item it is geometrically contractible, $\Pi (\mathbb{D}) \simeq * $;
  \item it has a single global point, $\Gamma (\mathbb{D}) \simeq *$.
\end{enumerate}
This axiomatization we discuss in the following. We show that it formalizes a modern
refinement of infinitesimal calculus called \emph{$\mathcal{D}$-geometry}
\cite{BeilinsonDrinfeld} \cite{LurieCrystal}.

More precisely, we consider geometric inclusions
$\mathbf{H} \hookrightarrow \mathbf{H}_{\mathrm{th}}$ of cohesive 
$\infty$-toposes that exhibit the objects of $\mathbf{H}_{\mathrm{th}}$
as infinitesimal cohesive neighbourhoods of objects in 
$\mathbf{H}$. Equivalently, if the cohesive $\infty$-topos $\mathbf{H}$ 
is itself regarded as a fat point by prop. \ref{PointlikeProperty}, then
$\mathbf{H}_{\mathrm{th}}$ is an infinitesimal thickening of that fat point
itself. Below in \ref{InfStrucFormalInfinityGroupoid} we furthermore consider
the $\infty$-cofiber $\mathbf{H}_{\mathrm{inf}}$ of this inclusion
$$
  \raisebox{20pt}{
  \xymatrix{
    \mathbf{H} \ar@{^{(}->}[r] \ar[d] & \mathbf{H}_{\mathrm{th}} \ar[d]
	\\
	\infty \mathrm{Grpd} \ar@{^{(}->}[r] & \mathbf{H}_{\mathrm{inf}}
  }
  }
  \,.
$$
This cofiber is interpreted accordingly as the respective infinitesimal thickening
of the absolute point. We observe in \ref{LInfinityAlgebrasInSynthDiff} that 
the sub-$\infty$-category of globally trivial objects of $\mathbf{H}_{\mathrm{inf}}$
is equivalent to that of $L_\infty$-algebras, by the theory of 
``formal moduli problems'' of \cite{LurieFormal}. Moreover, the
reflection along
$$
  \mathrm{Grp}(\mathbf{H}_{\mathrm{th}})
  \simeq
  (\mathbf{H}_{\mathrm{th}})^{*/}_{\geq 1}
  \longrightarrow
  (\mathbf{H}_{\mathrm{inf}})^{*/}_{\geq 1}  
$$ 
is Lie differentiation, sending a cohesive $\infty$-group to the $L_\infty$-algebra
that approximates it infinitesimally.

\medskip

Below in \ref{StructuresInInfinitesimalCohesiveNeighbourhood}
we discuss a list of structures that are canonically present in infinitesimal
cohesive neighbourhoods.

Further below in \ref{SynthDiffInfGrpd} we discuss a model for these axioms
by \emph{synthetic differential $\infty$-groupoids} which is an
$\infty$-categorical generalization of a topos that is a model for 
\emph{synthetic differential geometry}.
In this model the above $\infty$-cofiber sequence of cohesive $\infty$-toposes
reads
$$
  \xymatrix{
    \mathrm{Smooth}\infty \mathrm{Grpd}
    \ar@^{^{(}->}[r]
    &
    \mathrm{SynthDiff}\infty \mathrm{Grpd}
    \ar@^{->>}[r]
    &
    \mathrm{Inf}\infty\mathrm{Grpd}
  }
  \,,
$$
where the on the right we have ``infinitesimal $\infty$-groupoids'' 
(essentially the ``formal moduli problems'' of \cite{LurieFormal}), 
which are infinitesimally cohesive. 
This is prop. \ref{PushoutCharacterizationOfInfinitesimalGroupoids} below.

A similar model, differing by the existence of a grading on the 
infinitesimals, is that of supergeometric $\infty$-groupoids,
discussed below in  in \ref{SuperInfinityGroupoids}.
There the $\infty$-cofiber sequence of cohesive $\infty$-toposes
reads
$$
  \xymatrix{
    \mathrm{Smooth}\infty \mathrm{Grpd}
    \ar@^{^{(}->}[r]
    &
    \mathrm{SmoothSuper}\infty \mathrm{Grpd}
    \ar@^{->>}[r]
    &
    \mathrm{Super}\infty\mathrm{Grpd}
  }
  \,,
$$
where on the right we have bare but ``super'' $\infty$-groupoids, an
infinitesimally cohesive $\infty$-topos whose internal algebra is superalgebra.
This is prop. \ref{PushoutForSuperCohesion} below.

\subsubsection{General abstract}
\label{DefinitionOfInfinitesimalCohesion}
\index{infinitesimal cohesion!definition}
\index{cohesive $\infty$-topos!infinitesimal cohesion!definition}

\begin{definition} 
 \label{InfinitesimalCohesiveNeighbourhood}
\index{infinitesimal cohesive neighbourhood}
Given a cohesive $\infty$-topos $\mathbf{H}$ we say that 
an \emph{infinitesimal cohesive neighbourhood} of $\mathbf{H}$
is a geometric embedding $i : \mathbf{H} \hookrightarrow \mathbf{H}_{\mathrm{th}}$ 
into another cohesive $\infty$-topos $\mathbf{H}_{\mathrm{th}}$,
such that there is an extra left adjoint $i_!$ (necessarily also full 
and faithful) and an extra right
adjoint $i^!$
$$
  (i_! \dashv i^* \dashv i_* \dashv i^!) : 
  \xymatrix{
    \mathbf{H}
     \ar@{^{(}->}@<+12pt>[r]^{i_!}
     \ar@{<-}@<+4pt>[r]|{i^{*}}
     \ar@{^{(}->}@<-4pt>[r]|{i_{*}}
     \ar@{<-}@<-12pt>[r]_{i^!}
     &
    \mathbf{H}_{\mathrm{th}}
   }
$$
and such that $i_!$ preserves finite products. 

When we think of this as exhibiting extra structure on $\mathbf{H}_{\mathrm{th}}$,
we call $\mathbf{H}_{\mathrm{th}}$ equipped with this embedding a 
\emph{differental cohesive $\infty$-topos} or \emph{differential $\infty$-topos} for short.
\end{definition}
\begin{remark}
This definition captures the characterization of infinitesimal objects as having a single
global point surrounded by an infinitesimal neighbourhood: as we discuss in  detail 
below in \ref{InfStrucDeRhamSpace}, the $\infty$-functor $i^*$ may be thought of as contracting 
away any infinitesimal extension of an object. Thus $X$ being an infinitesimal object
amounts to  $i^* X \simeq *$, and the $\infty$-adjunction $(i_! \dashv i^*)$ then implies
that $X$ has only a single global point, since 
$$
  \begin{aligned}
    \mathbf{H}_{\mathrm{th}}(*, X) 
      & \simeq \mathbf{H}_{\mathrm{th}}(i_! *, X) 
      \\
      & \simeq \mathbf{H}(*, i^* X)
      \\
      & \simeq \mathbf{H}(*, *)
      \\
      & \simeq *
 \end{aligned}
  \,.
$$
\end{remark}
\begin{observation} \label{InfinitesimalNeighbourhoodIsOverInfGroupoid}
The inclusion into the infinitesimal neighbourhood is necessarily
a morphism of $\infty$-toposes over $\infty\mathrm{Grpd}$.
$$
  \xymatrix{
     \mathbf{H} \ar[rr]^{(i^* \dashv i_*)} 
       \ar[dr]_{\Gamma_{\mathbf{H}}}
     && \mathbf{H}_{\mathrm{th}}
     \ar[dl]^{\Gamma_{\mathbf{H}_{\mathrm{th}}}}
     \\
     & \infty \mathrm{Grpd} 
  }
$$
as is the induced $\infty$-geometric morphism
$(i_* \dashv i^!) : \mathbf{H}_{\mathrm{th}} \to \mathbf{H}$:
$$
  \xymatrix{
     \mathbf{H}_{\mathrm{th}} \ar[rr]^{(i_* \dashv i^!)}
      \ar[dr]_{\Gamma_{\mathbf{H}_{\mathrm{th}}}}
      && \mathbf{H}     
      \ar[dl]^{\Gamma_{\mathbf{H}}}
     \\
     & \infty \mathrm{Grpd}
  }
  \,.
$$
\end{observation}
\proof
By essential uniqueness of the terminal global section geometric morphism, 
prop. \ref{Terminalgeometricmorphism}.
In both cases the direct image functor has as left adjoint
that preserves the terminal object. Therefore we compute in the first case
$$
  \begin{aligned}
    \Gamma_{\mathbf{H}_{\mathrm{th}}}( i_* X )
    & 
    \simeq
    \mathbf{H}_{\mathrm{th}}(*, i_* X)
    \\
    & \simeq \mathbf{H}(i^* *, X)
    \\
    & 
    \simeq \mathbf{H}(*, X)
    \\
    & \simeq \Gamma_{\mathbf{H}}(X)
  \end{aligned}
$$
and analogously in the second.
\endofproof
\begin{definition}
  \label{PiInf}
For $(i_! \dashv i^* \dashv i_* \dashv i^!) : \mathbf{H} \to \mathbf{H}_{\mathrm{th}}$ an
infinitesimal neighbourhood of a cohesive $\infty$-topos, we write
$$
  (\Pi_{\mathrm{inf}} \dashv \mathrm{Disc}_{\mathrm{inf}} \dashv \Gamma_{\mathrm{inf}})
  := 
  (i^* \dashv i_* \dashv i^!)
  \,,
$$
so that the locally connected terminal geometric morphism of $\mathbf{H}_{\mathrm{th}}$ factors
as
$$
  (\Pi_{\mathbf{H}_{\mathrm{th}}} \dashv \mathrm{Disc}_{\mathbf{H}_{\mathrm{th}}} 
     \dashv \mathbf{\flat}_{\mathbf{H}_{\mathrm{th}}})
   :
  \xymatrix{
    \mathbf{H}_{\mathrm{th}}
    \ar@{<-^{)}}@<+20pt>[rr]|-{\mathrm{Red}}
    \ar@{->}@<+12pt>[rr]|-{\Pi_{\mathrm{inf}}}
    \ar@{<-^{)}}@<+4pt>[rr]|-{\mathrm{Disc}_{\mathrm{inf}}}
    \ar@{->}@<-4pt>[rr]|-{\Gamma_{\mathrm{inf}}}
    &&
    \mathbf{H}
    \ar@{->}@<+12pt>[rr]|-{\Pi_{\mathbf{H}}}
    \ar@{<-^{)}}@<+4pt>[rr]|-{\mathrm{Disc}_{\mathbf{H}}}
    \ar@{->}@<-4pt>[rr]|-{\Gamma_{\mathbf{H}}}
    \ar@{<-^{)}}@<-12pt>[rr]|-{\mathrm{coDisc}}
    &&
    \infty \mathrm{Grpd}
  }
  \,.
$$
\end{definition}	
The interrelation between overlapping adjoint triples here is discussed in more detail below
in \ref{InifnitesimalPathInfinityGroupoid}.

As a simple class of examples we record:
\begin{proposition}
  If $\mathbf{H}$ is an infinitesimally cohesive $\infty$-topos over
  $\infty\mathrm{Grpd}$
  def. \ref{InfinitesimalCohesiveSite}, then it is also enjoys
  differential cohesion relative to $\infty \mathrm{Grpd}$. 
\end{proposition} 
\proof
  By the properties of infinitesimal cohesion the composite
  $$
    \xymatrix{
      \infty \mathrm{Grpd}
      \ar@<+16pt>@{^{(}->}[rr]
      \ar@<+8pt>@{<-}[rr]
      \ar@{^{(}->}[rr]
      \ar@<-8pt>@{<-}[rr]
      &&
      \mathbf{H}
      \ar@<+16pt>@{->}[rr]
      \ar@<+8pt>@{<-^{)}}[rr]
      \ar@{<-^{)}}[rr]
      \ar@<-8pt>@{->}[rr]
      &&
      \infty \mathrm{Grpd}
    }
  $$
  is the identity adjoint quadruple, which is the one that exhibits the
  discrete cohesion of $\infty \mathrm{Grpd}$ over itself.
\endofproof

\subsubsection{Presentations}
\label{InfinitesimalNeighourhoodsOfCohesiveSite}
\index{infinitesimal cohesion!infinitesimal neighbourhood site}
\index{cohesive $\infty$-topos!infinitesimal cohesion!site of definition}

We establish a presentation of differential $\infty$-toposes, def. 
\ref{InfinitesimalCohesiveNeighbourhood},
in terms of categories of simplicial presheaves over suitable
neighbourhoods of $\infty$-cohesive sites.

\medskip

\begin{definition} 
  \label{InfinitesimalNeighBourhoodSite}
  \index{infinitesimal cohesive site}
  Let $C$ be an $\infty$-cohesive site, def. \ref{CohesiveSite}. We say a site $C_{\mathrm{th}}$ 
\begin{itemize}
\item equipped with a co-reflective embedding
  $$
    (i \dashv p) : 
    \xymatrix{
      C
      \ar@<+4pt>@{^{(}->}[r]^i 
      \ar@<-4pt>@{<-}[r]_p
      &
      C_{\mathrm{th}}
    }
  $$
\item such that 
\begin{enumerate}
\item $i$ preserves finite products;
\item $i$ preserves pullbacks along morphisms in covering families;

\item both $i$ and $p$ send covering families to covering families;
\item
    for all $\mathbf{U} \in C_{\mathrm{th}}$ and 
    for all covering families $\{U_i \to p(\mathbf{U})\}$ 
    in $C$ there is
    a lift through $p$ to a covering family
    $\{\mathbf{U}_i \to \mathbf{U}\}$ in $C_{\mathrm{th}}$
\end{enumerate}
\end{itemize}
is an \emph{infinitesimal neighbourhood site} of $C$.
\end{definition} \label{InfinitesimalNeighbourhoodFromInfinitesimalSite}
\begin{proposition} 
 \label{InfinitesimalNeighbourhoodFromSites}
Let $C$ be an $\infty$-cohesive site and let       
$(i \dashv p) : C \stackrel{\overset{i}{\hookrightarrow}}{\underset{p}{\leftarrow}} C_{\mathrm{th}}$ 
be an infinitesimal neighbourhood site. 

Then the $\infty$-category of $\infty$-sheaves on $C_{\mathrm{th}}$ is a cohesive 
$\infty$-topos and the restriction $i^*$ along $i$ exhibits it as an infinitesimal neighbourhood 
of the cohesive $\infty$-topos over $C$.
$$
 ( i_! \dashv i^* \dashv i_* \dashv i^! )
  :
  \mathrm{Sh}_{\infty}(C)
    \to
  \mathrm{Sh}_{\infty}(C_{\mathrm{th}})
  \,.
$$
Moreover, $i_!$ restricts on representables to the $\infty$-Yoneda embedding factoring through $i$:
$$
  \xymatrix{
    C \ar@{^{(}->}[r] \ar[d]^{i} & \mathrm{Sh}_{\infty}(C) \ar[d]^{i_!}
    \\
    C_{\mathrm{th}} \ar@{^{(}->}[r]& \mathrm{Sh}_{\infty}(C_{\mathrm{th}})
  }
  \,.
$$  
\end{proposition}
\proof
We demonstrate this in the model category presentation of $\mathrm{Sh}_{\infty}(C_{\mathrm{th}})$
as in the proof of prop. \ref{InfSheavesOverCohesiveSiteAreCohesive}.

Consider the right Kan extension 
$\mathrm{Ran}_i : [C^{\mathrm{op}}, \mathrm{sSet}] 
   \to [C_{\mathrm{th}}^{\mathrm{op}},\mathrm{sSet}]$ of simplicial presheaves 
along the functor $i$. On an object $\mathbf{K} \in C_{\mathrm{th}}$ 
it is given by 
$$
  \begin{aligned}
    \mathrm{Ran}_{i} F : \mathbf{K} 
    & \mapsto 
    \int_{U \in C} \mathrm{sSet}( C_{\mathrm{th}}(i(U), \mathbf{K})  , F(U))
    \\
    & \simeq 
    \int_{U \in C} \mathrm{sSet}( C(U, p(\mathbf{K}))  , F(U))
    \\
    & \simeq 
    F(p(\mathbf{K})
  \end{aligned}
  \,,
$$
where in the last step we use the Yoneda reduction-form of the Yoneda lemma. 

This shows that the right adjoint to $(-)\circ i$ is itself given by precomposition 
with a functor, and hence has itself a further right adjoint, which gives us a total of four adjoint functors
$$
  \xymatrix{
  [C^{\mathrm{op}}, \mathrm{sSet}]  
    \ar@{->}@<+12pt>[rr]^{\mathrm{Lan}_i}
    \ar@{<-}@<+4pt>[rr]|{(-)\circ i}
    \ar@{->}@<-4pt>[rr]|{(-)\circ p}
    \ar@{<-}@<-12pt>[rr]_{\mathrm{Ran}_p}    
    &&
  [C_{\mathrm{th}}^{\mathrm{op}}, \mathrm{sSet}]
  }
  \,.
$$
From this are induced the corresponding simplicial Quillen adjunctions 
on the global projective and injective model structure on simplicial presheaves
$$
  (\mathrm{Lan}_i \dashv (-) \circ i) : 
  \xymatrix{
    [C^{\mathrm{op}}, \mathrm{sSet}]_{\mathrm{proj}}
     \ar@<+4pt>@{->}[r]^{\mathrm{Lan}_i}
     \ar@<-4pt>@{<-}[r]_{(-) \circ i}
     &
    [C_{\mathrm{th}}^{\mathrm{op}}, \mathrm{sSet}]_{\mathrm{proj}}
  }
  \,;
$$
$$
  ((-)\circ i \dashv (-) \circ p) : 
  \xymatrix{
    [C^{\mathrm{op}}, \mathrm{sSet}]_{\mathrm{proj}}
     \ar@<+4pt>@{<-}[r]^{(-)\circ i}
     \ar@<-4pt>@{->}[r]_{(-) \circ p}
     &
   [C_{\mathrm{th}}^{\mathrm{op}}, \mathrm{sSet}]_{\mathrm{proj}}
  }
  \,;
$$
$$
  ((-) \circ p \dashv \mathrm{Ran}_p) : 
  \xymatrix{
   [C^{\mathrm{op}}, \mathrm{sSet}]_{\mathrm{inj}}
     \ar@<+4pt>@{->}[r]^{(-) \circ p}
     \ar@<-4pt>@{<-}[r]_{\mathrm{Ran}_p}
     &
   [C_{\mathrm{th}}^{\mathrm{op}}, \mathrm{sSet}]_{\mathrm{inj}}
  }
  \,.
$$
By prop. \ref{AdjRecognition}, for these Quillen adjunctions to descend to the {\v C}ech-local model 
structure on simplicial presheaves it suffices that the right adjoints preserve locally fibrant objects. 

We first check that $(-) \circ i$ sends locally fibrant objects to locally fibrant objects.  
To that end, let $\{U_i \to U\}$ be a covering family in $C$. Write 
$\int^{[k] \in \Delta} \Delta[k] \cdot \coprod_{i_0, \cdots, i_k} (j(U_{i_0}) \times_{j(U)} j(U_{i_1}) \times_{j(U)} \cdots \times_{j(U)} j(U_k))$ for its {\v C}ech nerve, where $j$ denotes the Yoneda embedding. 
Recall by the definition of the $\infty$-cohesive site $C$ that all the fiber products of 
representable presheaves here are again themselves representable, 
hence $\cdots = \int^{[k] \in \Delta} \Delta[k] \cdot \coprod_{i_0, \cdots, i_k} (j(U_{i_0} \times_U U_{i_1} \times_U \cdots \times_U U_k))$. Using that the left adjoint $\mathrm{Lan}_i$ preserves the coend and 
tensoring, that it restricts on representables to $i$ and by the assumption that $i$ preserves 
pullbacks along covers we have that
$$
  \begin{aligned}
    \mathrm{Lan}_i 
    C(\{U_i \to U\})
    & 
    \simeq 
     \int^{[k] \in \Delta} \Delta[k] \cdot \coprod_{i_0, \cdots, i_k} \mathrm{Lan}_i (j(U_{i_0} \times_U U_{i_1} \times_U \cdots \times_U U_k))    
     \\
     & \simeq
     \int^{[k] \in \Delta} \Delta[k] \cdot \coprod_{i_0, \cdots, i_k} 
  j (i (U_{i_0} \times_U U_{i_1} \times_U \cdots \times_U U_k))    
    \\
    & \simeq
     \int^{[k] \in \Delta} \Delta[k] \cdot \coprod_{i_0, \cdots, i_k} 
  j (i(U_{i_0}) \times_{i(U)} i(U_{i_1}) \times_{i(U)} \cdots \times_{i(U)} i(U_k))      
  \end{aligned}
   \,.
$$ 
By the assumption that $i$ preserves covers, this is the {\v C}ech nerve of a covering family in 
$C_{\mathrm{th}}$. Therefore for $F \in [C_{\mathrm{th}}^{\mathrm{op}}, \mathrm{sSet}]_{\mathrm{proj},\mathrm{loc}}$ fibrant we have for all coverings $\{U_i \to U\}$ in $C$ that the descent morphism
$$
  i^* F(U) = F(i(U))
   \stackrel{\simeq}{\to}
  [C_{\mathrm{th}}^{\mathrm{op}}, \mathrm{sSet}](C(\{i(U_i)\}), F)
  =
  [C^{\mathrm{op}}, \mathrm{sSet}](C(\{U_i\}), i^* F)
$$
is a weak equivalence.

To see that $(-) \circ p$ preserves locally fibrant objects, we apply the analogous reasoning after 
observing that its left adjoint $(-)\circ i$ preserves all limits and colimits of simplicial presheaves 
(as these are computed objectwise) and by observing that for 
$\{\mathbf{U}_I \stackrel{p_i}{\to} \mathbf{U}\}$ a covering family in $C_{\mathrm{th}}$ 
we have that its image under 
$(-) \circ i$ is its image under $p$, by the Yoneda lemma:
$$
  \begin{aligned}
     {[C^{\mathrm{op}}, \mathrm{sSet}]}(K, ((-)\circ i)(\mathbf{U}))
     & \simeq
     C_{\mathrm{th}}(i(K), \mathbf{U})
     \\
     & \simeq
      C(K, p(\mathbf{U}))
  \end{aligned}
$$
and using that $p$ preserves covers by assumption.

Therefore $(-) \circ i$ is a left and right local Quillen functor with left local Quillen 
adjoint $\mathrm{Lan}_i$ and right local Quillen adjoint $(-)\circ p$.

Finally to see by the above reasoning that also $\mathrm{Ran}_p$ preserves locally fibant objects
notice that for every covering family $\{U_i \to U\}$ in $C$ 
and every morphism $\mathbf{K} \to p^* U$ in 
$C_{\mathrm{th}}$ we may find a covering $\{\mathbf{K}_j \to \mathbf{K}\}$ 
such that we have commuting diagrams as on the left of
$$
  \raisebox{20pt}{
  \xymatrix{
    \mathbf{K}_j \ar[r] \ar[d]& p^* U_{i(j)} \ar[d]
    \\
    \mathbf{K} \ar[r] & p^* U
  }
  }
  \;\;\;\;
  \leftrightarrow
  \raisebox{20pt}{
  \xymatrix{
    p(\mathbf{K}_j) \ar@{=}[r] \ar[d] &
    i^*(\mathbf{K}_j) \ar[r] \ar[d]&  U_{i(j)} \ar[d]
    \\
    p(\mathbf{K}) \ar@{=}[r] & i^*(\mathbf{K}) \ar[r] &  U
  }
  }
  \,,
$$
because by the $(i^* \dashv p^*)$ adjunction established above these correspond to the
diagrams as indicated on the right, which exist by definition of coverage and the
fact that, by definition, in $C_{\mathrm{th}}$ covers lift through $p$.

This implies that $\{p^* U_i \to p^* U\}$ is a \emph{generalized cover} in the terminology 
of \cite{dugger-hollander-isaksen}, which by the discussion there implies that the corresponding 
{\v C}ech nerve projection $C(\{p^* U_i\}) \to p^* U$ is a weak equivalence in 
$[C_{\mathrm{th}}^{\mathrm{op}}, \mathrm{sSet}]_{\mathrm{proj}, \mathrm{loc}}$.

This establishes the quadruple of adjoint $\infty$-functors as claimed. 

To see that $\mathrm{Lan}_i$ preserves products, use that, by the local 
formula for the left Kan extension, it is sufficient that for each 
$K \in C_{\mathrm{th}}$ the functor 
$$
  X \mapsto \lim\limits_{\to}(p^{\mathrm{op}}/K \to C^{\mathrm{op}} \stackrel{X}{\to} \mathrm{sSet})
$$
preserves finite products. By a standard fact this is the case precisely if 
the slice category $p^{\mathrm{op}}/K$ is sifted. A sufficient condition for
this is that it has coproducts. This is equivalent to $K/p$ having products,
and this is finally true due to the assumption that $p$ preserves products. 

It remains to see that $i_!$ is a full and faithful $\infty$-functor.
For that notice the general fact that left 
Kan extension along a full and faithful functor $i$ satisfies 
$\mathrm{Lan}_i \circ i \simeq \mathrm{id}$. 
It only remains to observe that since $(-)\circ i$ is not only right but also left Quillen 
by the above, we have that $i^* \circ \mathrm{Lan}_i$ applied to a cofibrant object is already the 
derived functor of the composite.
\endofproof

\newpage

\subsection{Structures in an $\infty$-topos}
\label{Structures in a topos}

We discuss here a list of fundamental homotopical and cohomological 
structures that exist in every $\infty$-topos but are particularly
expressive in a \emph{local} $\infty$-topos, def. \ref{LocalInfinityTopos},
or rather: over a base $\infty$-topos that is local. As we discuss below
in \ref{ConnectedObjects}, every local $\infty$-topos has
the \emph{homotopy dimension} of the point and hence
\emph{all gerbes are delooped groups}. This means that group objects
in a local $\infty$-topos, discussed in \ref{StrucInftyGroups} below,
behave as \emph{absolute structured groups} rather than as 
$\infty$-sheaves of groups that vary over a fixed nontrivial space.
This is the first central property of the \emph{gros} toposes $\mathbf{H}$ that 
we are interested in here. For every object $X \in \mathbf{H}$ the 
slice $\infty$-topos $\mathbf{H}_{/X} \to \mathbf{H}$ is an $\infty$-topos
relative to its local base $\mathbf{H}$, but is itself in general not local.
Group objects in the slice are groups parameterized over $X$ and 
pointed connected objects in the slice are the \emph{$\infty$-gerbes} over $X$.
This we discuss below in \ref{StrucInftyGerbes}.

Structures entirely specific to local $\infty$-toposes we discuss
below in \ref{Structures in a local topos}.
Additional structures that are present if we assume that $\mathbf{H}$
is locally $\infty$-connected are discussed below 
in \ref{StructuresInLocallyInfinityConnectedTopos},
and those in an actual cohesive $\infty$-topos below in \ref{structures}.

\medskip

\begin{itemize}
  \item \ref{Bundles} -- Bundles
  \item \ref{TruncatedObjects} -- Truncated objects and Postnikov towers
  \item \ref{StrucEpi} -- Epi-/mono-morphisms, images and relative Postnikov systems
  \item \ref{CompactObjects} -- Compact objects
  \item \ref{Homotopy} -- Homotopy
  \item \ref{ConnectedObjects} -- Connected objects
  \item \ref{StrucInftyGroupoids} -- Groupoids
  \item \ref{StrucInftyGroups} -- Groups 
  \item \ref{StrucCohomology} -- Cohomology 
  \item \ref{PrincipalInfBundle} -- Principal bundles
  \item \ref{AssociatedBundles} -- Associated fiber bundles
  \item \ref{StrucTwistedCohomology} -- Sections and twisted cohomology
  \item \ref{StrucRepresentations} -- Representations and group cohomology
  \item \ref{ExtensionsOfCohesiveInfinityGroups} -- Extensions and twisted bundles
  \item \ref{StrucInftyGerbes} -- Gerbes
  \item \ref{StrucRelativeCohomology} -- Relative cohomology
\end{itemize}

\subsubsection{Bundles}
\label{Bundles}
\index{structures in a cohesive $\infty$-topos!$\infty$-bundles}

We discuss the general notion of \emph{bundles} or
\emph{objects in a slice} in an $\infty$-topos. In the following sections
this general notion is specialized to \emph{principal bundles},
\ref{PrincipalInfBundle}, and \emph{associated fiber bundles}, \ref{AssociatedBundles}.

\paragraph{General abstract}

For $X \in \mathbf{H}$ an object, a \emph{bundle} over $X$ is, in full generality,
nothing but a morphism
$$
  \xymatrix{
    T
	\ar[d]^p
	\\
	X
  }
$$
in $\mathbf{H}$ with codomain $X$, and a \emph{homomorphism of bundles} over $X$ is a
diagram of the form
$$
  \raisebox{20pt}{
  \xymatrix{
    T_1 \ar[rr]_{\ }="s" \ar[dr]^{\ }="t" && T_2 \ar[dl]
	\\
	& X
	\ar@{=>} "s"; "t"
  }
  }
$$
in $\mathbf{H}$.
The full $\infty$-category of bundles over $X$ in $\mathbf{H}$
is also called the \emph{slice} of $\mathbf{H}$ over $X$:
\begin{definition}
  For $\mathbf{H}$ an $\infty$-category and for $X \in \mathbf{H}$ an object,
  the \emph{slice $\infty$-category} $\mathbf{H}_{/X}$ is the $\infty$-pullback
  $$
    \mathbf{H}_{/X} := \mathbf{H}^{\Delta[1]} \underset{\mathbf{H}}{\times} \{X\}
  $$
  in the diagram of $\infty$-categories
  $$
    \xymatrix{
	  \mathbf{H}_{/X} 
	    \ar[r] \ar@/^1.5pc/[rr]^{\sum_X}
		\ar[d]
	    &
	  \mathbf{H}^{\Delta[1]} \ar[r]^{\mathrm{dom}} 
	  \ar[d]^{\mathrm{cod}}
	  & \mathbf{H}
	  \\
	  {*} \ar[r]^{\vdash X} &  \mathbf{H}
	}
	\,.
  $$
  \label{Slice}
\end{definition}
\begin{proposition}
  For $\mathbf{H}$ an $\infty$-topos and $X \in \mathbf{H}$, also the slice
  $\mathbf{H}_{/X}$, def. \ref{Slice}, is an $\infty$-topos. 
  Moreover, the forgetful $\infty$-functor $\sum_X$ in def. \ref{Slice} 
  is the extra left adjoint in an essential geometric morphism of $\infty$-toposes
  $$
  \left(
    \sum_X \dashv X^*  \dashv \prod_X
  \right)
   :
  \xymatrix{
	 \mathbf{H}_{/X}
	 \ar@<+12pt>[rr]|{\sum_X}
	 \ar@{<-}@<+2pt>[rr]|{X \times(-)}
	 \ar@<-8pt>[rr]|{\prod_X}
	 &&
	 \mathbf{H}
  }
$$
called the \emph{{\'e}tale geometric morphism} of $\mathbf{H}_{/X}$.

Here $\prod_X$ is also called the \emph{dependent product} over $X$ and $\sum_X$ is also
called the \emph{dependent sum} over $X$, see \ref{DependentHomotopyTypeTheory} above.

Finally , $X \times (-)$ is a cartesian closed $\infty$-functor, which 
equivalently means that it satisfies \emph{Frobenius reciprocity}: 
for $U \in \mathbf{H}$ and $E \in \mathbf{H}_{/X}$  there is a natural equivalence
$$
  \xymatrix{
   \underset{X}{\sum} \left( E \times_X \left(X \times U\right)\right)
    \ar[r]^-\simeq
	&
   \left(\underset{X}{\sum}E\right) \times U
  }
$$
exhibited by the canonical morphism.
 \label{BaseChange}
\end{proposition}
This is prop. 6.3.5.1 in \cite{Lurie}.
\begin{example}
  The terminal object of the slice $\mathbf{H}_{/X}$ is 
  given by the identity
  morphism on $X$ in $\mathbf{H}$.
  \label{TerminalObjectInSlice}
\end{example}
\begin{remark}
  The interpretation of these base change functors is as follows: 
  an object in the slice $\mathbf{H}_{/X}$ corresponds to a morphism
  into $X$ in $\mathbf{H}$. The functor $\sum_X$ picks out the domain 
  of these morphisms: it forms the ``sum (union) of all the fibers''. Therefore 
  an object $E \in \mathbf{H}_{/X}$ in the slice corresponds to a morphism
  of the form
  $$
    \xymatrix{
	  \sum_A E
	  \ar[d]
	  \\
	  X
	}
  $$
  in $\mathbf{H}$. More generally, a morphism $f : E_1 \to E_2$ in the slice corresponds to a 
  diagram of the form
  $$
    \raisebox{20pt}{
    \xymatrix{
	  \sum_A E_1 \ar[rr]^{\sum_A f}_{\ }="s" \ar[dr]^{\ }="t" && \sum_A E_2 \ar[dl]
	  \\
	  & X
	  \ar@{=>}^f "s"; "t"
	}
	}
  $$
  in $\mathbf{H}$.
  
  On the other hand, the right adjoint $\prod_A$ forms internal spaces of \emph{sections} of these morphisms.
  With $E \in \mathbf{H}_{/X}$ as above we have 
  $$
    \prod_X E \;\simeq\; [X,\sum_X E] \underset{[X,X]}{\times} \{\mathrm{id}\}
	\,,
  $$
  which says that $\prod_X E$ is the homotopy fiber of the projection
  $[X,\sum_X E ] \to [X,X]$ from the internal hom space of maps from the base $X$ to the
  domain $\sum_A E$, picking those morphisms in there which go to the identity on $X$, up to homotopy,
  when postcomposed with $E$, regarded as a morphism in $\mathbf{H}$.
\end{remark}
This kind of relation also holds externally:
\begin{proposition}
  For $E_1, E_2 \in \mathbf{H}_{/X}$ two objects in a slice $\infty$-topos over $X \in \mathbf{H}$, 
  the hom $\infty$-groupoid $\mathbf{H}_{/X}(E_1, E_2)$ between them is characterized as the
  homotopy fiber product
  $$
    \mathbf{H}_{/X}(E_1, E_2) 
	  \simeq
	\mathbf{H}\left(\sum_X E_1, \sum_X E_2\right) \underset{\mathbf{H}(\sum_X E_1, X)}{\times} \{E_1\}
  $$
  of hom-$\infty$-groupoids in $\mathbf{H}$, sitting in the $\infty$-pullback diagram
  $$  
    \raisebox{20pt}{
    \xymatrix{
	  \mathbf{H}_{/X}(E_1, E_2)
	  \ar[rr]
	  \ar[d]
	  &&
	  {*}
	  \ar[d]^{\vdash E_1}
	  \\
	  \mathbf{H}(\sum_X E_1, \sum_X E_2)
      \ar[rr]^{E_2 \circ (-)}
      &&	
      \mathbf{H}(\sum_X E_1, X)	  
	}
	}
	\,.
  $$
  \label{SliceHomAsHomotopyFiber}
\end{proposition}
This appears as prop. 5.5.5.12 in \cite{Lurie}.

Therefore the slice $\infty$-topos $\mathbf{H}_{/X}$ may be regarded not only
as living over the canonical base $\infty$-topos $\infty\mathrm{Grpd}$, but
also as living over $\mathbf{H}$. As such its $\mathbf{H}$-valued hom
is the dependent product of its interal hom.
\begin{definition}
  For $X \in \mathbf{H}$ and $E_1, E_2 \in \mathbf{H}_{/X}$ we write
  $$
    [E_1,E_2]_{\mathbf{H}} := \prod_X [E_1, E_2]
  $$
  and speak of the \emph{$\mathbf{H}$-valued hom} between $E_1$ and $E_2$ in the slice.
  \label{HValuedMappingSpace}
\end{definition}
\begin{remark}
  A global element of $\prod_X [E_1, E_2]$ corresponds again to a diagram of the form
  $$
    \xymatrix{
	  \sum_A E_1 \ar[rr]_{\ }="s" \ar[dr]^{\ }="t" && \sum_A E_2 \ar[dl]
	  \\
	  & X
	  \ar@{=>} "s"; "t"
	}
  $$
  in $\mathbf{H}$. 
  The morphism of prop. \ref{CanonicalMorphismFromSliceInternalHomToBaseInternalHom}
  sends such a global element to the top horizontal morphism $\sum_A E_1 \to \sum_A E_2$, 
  regarded as a global element of $[\sum_A E_1, \sum_A E_2]$.
  \label{GlobalElementsInHValuedSliceInternalHom}
\end{remark}
\begin{proposition}
  The $\infty$-groupoid of global points of $[E_1, E_2]_{\mathbf{H}}$ is the 
  slice hom $\mathbf{H}_{/X}(E_1, E_2)$:
  $$
    \mathbf{H}_{/X}(E_1, E_2)
	\simeq
    \Gamma \left(\left[E_1, E_2\right]_{\mathbf{H}}\right)
	\simeq
	 \mathbf{H}\left( {*}, \left[E_1, E_2\right]_{\mathbf{H}} \right)
	 \,.
  $$
  \label{SliceHomIsGlobalSectionsOfHValuedHom}
\end{proposition}
\proof
  We compute
  $$
    \begin{aligned}
	  \mathbf{H}\left(*, \left[E_1, E_2\right]_{\mathbf{H}}\right)
	  & 
	  \simeq
	  \mathbf{H}_{/X}\left( \left(*\times X\right), \left[E_1, E_2\right]\right)
	  \\
	  & \simeq \mathbf{H}_{/X}\left( X \times_X E_1, E_2 \right)
	  \\
	  & \simeq \mathbf{H}_{/X}\left(E_1, E_2\right)
	\end{aligned}
	\,.
  $$
  Here the first equivalence is that of the defining 
  $\left( (-)\times_X E_1 \dashv \underset{X}{\prod} \right)$-adjunction of the 
  dependent product, def. \ref{BaseChange}, the second is that of the
  $\left( \left(-\right) \times_X E_1 \dashv \left[E_1, -\right] \right)$-adjunction and the last
  one finally uses that $X$ is the terminal object in $\mathbf{H}_{/X}$.
\endofproof
We may compare the internal hom in the slice with that in the base by the 
following comparison morphism.
\begin{proposition}
  For $X \in \mathbf{H}$ and $E_1, E_2 \in \mathbf{H}_{/X}$, there is a natural morphism
  $$
    p_{X}
    :
    \prod_X \left[E_1, E_2\right]
	\to 
	\left[
	   \sum_X E_1, \sum_X E_2
	\right]
	\,.
  $$
  \label{CanonicalMorphismFromSliceInternalHomToBaseInternalHom}
\end{proposition}
\proof
  Let $U \in \mathbf{H}$ be any object. Consider then the 
  morphism of $\infty$-groupoids given by the composite
  $$
    \begin{aligned}
	  \mathbf{H}\left(U, \prod_X \left[ E_1, E_2\right]\right)
	  & \simeq
	  \mathbf{H}_{/X}\left( X^* U, \left[ E_1, E_2\right] \right)
	  \\
	  & \simeq \mathbf{H}_{/X}\left(X^* U \times E_1, E_2 \right)
	  \\
	  & \stackrel{}{\to} \mathbf{H}\left( \sum_X\left(f^* U \times E_1\right), \sum_X E_2\right)
	  \\
	  & \simeq \mathbf{H}\left(U \times \sum_X E_1, \sum_X E_2\right)
	  \\
	  & \simeq \mathbf{H}\left(U , \left[ \sum_X E_1, \sum_X E_2\right]\right)
	\end{aligned}
	\,.
  $$
  Here the first and last equivalences are the adjunction properties, the morphism in the
  middle is the relevant component of the $\infty$-functor $\sum_X : \mathbf{H}_{/X} \to \mathbf{H}$
  and the step after that uses the \emph{Frobenius reciprocity} property of the dependent sum
  (reflecting that $X*$ is a cartesian closed morphism). Since this morphism of $\infty$-groupoids
  is natural in $U$, the $\infty$-Yoneda lemma asserts that it is given by homming $U$
  into a morphism $\prod_X[E_1, E_2] \to [\sum_X E_1, \sum_X E_2]$ in $\mathbf{H}$. 
\endofproof
\begin{proposition}
  For $E_1, E_2 \in \mathbf{H}_{/X}$,
  there is an $\infty$-pullback diagram in $\mathbf{H}$ of the form
  $$
    \raisebox{20pt}{
    \xymatrix{
	  \left[E_1, E_2\right]_{\mathbf{H}}
	  \ar[r]
	  \ar[d]^{p_X}
	  &
	  {*}
	  \ar[d]^{\vdash E_1}
	  \\
	  \left[\underset{X}{\sum}E_1, \underset{X}{\sum}E_2\right]
	  \ar[r]^-{E_2 \circ (-)}
	  &
	  \left[\underset{X}{\sum}E_1, X\right]
	}
	}
	\,,
  $$
  where the left vertical projections is the morphism of 
  prop. \ref{CanonicalMorphismFromSliceInternalHomToBaseInternalHom}.
  \label{HValuedHomAsHomotopyPullback}
\end{proposition}
\proof
  We may check this on a set $U \in \mathbf{H}$ of generators of $\mathbf{H}$
  (for instance the objects in a small $\infty$-site of definition).
  Since $\mathbf{H}(U,-)$ preserves $\infty$-limits (and detects them as $U$ ranges
  over the set of generators), applying it to the above diagram 
  (and using the definition $[E_1,E_2]_{\mathbf{H}} := \underset{X}{\prod}[E_1,E_2]$) 
  yields the diagram
    $$
    \raisebox{20pt}{
    \xymatrix{
	  \mathbf{H}_{/X}(U \times X, [E_1, E_2])
	  \ar[rr]
	  \ar[dd]
	  &&
	  {*}
	  \ar[dd]^{\vdash (U \times X) \times_X E_1}
	  \\
	  \\
	  \mathbf{H}\left(U \times \underset{X}{\sum}E_1, \underset{X}{\sum}E_1\right)
	  \ar[rr]^-{\mathbf{H}(U \times \underset{X}{\sum}, E_1 )}
	  &&
	  \mathbf{H}\left(U \times \underset{X}{\sum}E_1, X\right)
	}
	}
	\,.
  $$
  Here in the top left we can apply the 
  $( (-) \times_X E_1  \dashv  [E_1, -] )$-adjunction equivalence
  $$
    \mathbf{H}_{/X}(U \times X, [E_1, E_2])
	\simeq
	\mathbf{H}_{/X}((U \times X) \times_X E_1, E_2)
	\,,
  $$
  and moreover by Frobenius reciprocity
  $$
    \underset{X}{\sum}\left( (U \times X) \times_X E_1 \right)
	\simeq
	U \times \underset{X}{\sum}E_1
	\,.
  $$
  Therefore the above diagrams are $\infty$-pullbacks by prop. \ref{SliceHomAsHomotopyFiber}. 
\endofproof

Accordingly there is a $\mathrm{Grp}(\mathbf{H})$-valued automorphism group construction:
\begin{definition}
  For $X \in \mathbf{H}$ and $E \in \mathbf{H}_{/X}$ we say that
  the \emph{$\mathbf{H}$-valued automorphism group} of $E$ is the
  dependent product, def \ref{BaseChange},
  $$
    \mathbf{Aut}_{\mathbf{H}}(E)
	:=
	\prod_X 
	\mathbf{Aut}(E)
  $$
  of the automorphism group of $E$ in $\mathbf{H}_{/X}$, def. \ref{AutomorphismGroup}.
  \label{HValuedAutomorphismGroup}
\end{definition}
\begin{proposition}
  For $E \in \mathbf{H}_{/X}$ the object $\mathbf{Aut}_{\mathbf{H}}(E) \in \mathbf{H}$
  of def. \ref{HValuedAutomorphismGroup} sits in an $\infty$-pullback diagram of the form
  $$
    \xymatrix{
	  \mathbf{Aut}_{\mathbf{H}}(E)
	  \ar[rr]
	  \ar[d]
	  &&
	  {*}
	  \ar[d]^{\vdash E}
	  \\
	  \mathbf{Aut}(X)
	  \ar[rr]^{E \circ (-)}
	  &&
	  \left[\underset{X}{\sum}E, X\right]
	}
	\,.
  $$
  \label{SliceAutomorphismGroupInHInAHomotopyPullback}
\end{proposition}
\proof
By prop. \ref{HValuedHomAsHomotopyPullback}.
\endofproof

More generally we have the following.
\begin{proposition}
  For $\mathbf{H}$ an $\infty$-topos and for $f : X \to Y$ a morphism
  in $\mathbf{H}$, the functor
  $$
    \sum_f := f \circ (-) : \mathbf{H}_{/X} \to \mathbf{H}_{/Y}
  $$
  between the slices over the domain and codomain given by postcomposition with $f$
  is the extra left adjoint in an essential geometric morphism
  $$
    (\sum_f \dashv f^* \dashv \prod_f) 
	  : 
    \xymatrix{
	   \mathbf{H}_{/X}
	   \ar@<+12pt>[rr]|{\sum_f}
	   \ar@{<-}@<+2pt>[rr]|{f^* }
	   \ar@<-8pt>[rr]|{\prod_f}
	   &&
	   \mathbf{H}
    }
	\,,
  $$
  called the \emph{base change} geometric morphism. Here $f^*$ is given by forming the 
  $\infty$-pullback in $\mathbf{H}$ along $f$. As before $\sum_f$ is called the
  \emph{dependent sum} along $f$ and $\prod_{f}$ the \emph{dependent product} along $f$.
  \label{BaseChangeGeomMorphism}
  \label{EtaleGeometricMorphism}
  \index{base change}
  \index{topos theory!base change}
\end{proposition}
This is prop. 6.3.5.1, remark 6.3.5.10 of \cite{Lurie}.
\begin{proposition}
  \label{ToposEquivalentToItsEtaleToposes}
  For $\mathbf{H}$ an $\infty$-topos, 
  the $\infty$-functor 
  $$
    \mathbf{H}_{/(-)} : \mathbf{H} \to \infty\mathrm{Topos}^{\mathrm{et}}/_{\mathbf{H}}
  $$
  given by prop. \ref{EtaleGeometricMorphism}, constitutes 
  an equivalence of $\infty$-categories between $\mathbf{H}$ and the full sub-$\infty$-category
  of the slice of $\infty$-toposes and geometric morphisms over $\mathbf{H}$
  on the {\'e}tale geometric morphisms.
\end{proposition}
This is \cite{Lurie}, remark 6.3.5.10.

The internal hom in the slice is closely related to the dependent product:
\begin{proposition}
  For $\mathbf{H}$ an $\infty$-topos and $X \in \mathbf{H}$ an object,
  let $E_1, E_2 \in \mathbf{H}_{/X}$ be two object in the slice, 
  corresponding to morphisms $E_i : \underset{X}{\sum} E_i  \to X$ in $\mathbf{H}$. Then
  there is a natural equivalence
  $$
    [E_1, E_2]
	\simeq
	\underset{f_1}{\prod} f_1^* E_2
	\,.
  $$
  \label{PullProdIsSliceHom}
\end{proposition}
\proof
  The product in the slice $\mathbf{H}_{/X}$ is given by the fiber product in $\mathbf{H}$
  over $X$. Hence for $E\in \mathbf{H}_{/X}$ the product functor is
  $$
    (-)\times E \simeq \underset{f}{\sum} f^* 
	\,.
  $$
  Since the internal hom is right adjoint to this functor, the statement follows
  by the defining adjoint tripe $(\prod_f \dashv f^* \dashv \sum_f)$.
\endofproof
\begin{proposition}
  For $\mathbf{H}$ an $\infty$-topos, $X \in \mathbf{H}$ an object and 
  $E \in \mathbf{H}_{/X}$ a slice, the $\infty$-fiber of the morphism
  $p_X$ from def. \ref{CanonicalMorphismFromSliceInternalHomToBaseInternalHom}
  over the identity 
  $\xymatrix{{*} \ar[r]^-{\vdash \mathrm{id}_{\sum_X E}} & [\sum_X E, \sum_X E]}$ 
  is $\Omega_E [\sum_X E, X]$: there is a fiber sequence of the form
  $$
	\xymatrix{
	  \Omega_E [\sum_X E, X]
	  \ar@{^{(}->}[r]
	  &
	  \prod_X[E,E]
	  \ar@{->}[r]^-{p_X}
	  &
	  \left[
	    \sum_X E, \sum_X E
	  \right]
	}
	\,.
  $$
  \label{FiberOfMapFromSliceMappingToBaseMapping}
\end{proposition}
\proof
  This follows directly with prop. \ref{HValuedHomAsHomotopyPullback}
  and the pasting law, prop. \ref{PastingLawForPullbacks}.
  
  More explicitly, 
  by the proof of prop. \ref{CanonicalMorphismFromSliceInternalHomToBaseInternalHom}
  the morphism $p_X$  is 
  for any $U \in \mathbf{H}$ characterized, up to equivalence, as being the
  forgetful morphism
  $$
    \mathbf{H}(U, p) 
	  : 
	\xymatrix{
	  \mathbf{H}_{/X}( U \times E, E)
	  \ar[r]
	  &
	  \mathbf{H}(U \times X, X)
    }	
  $$
  that sends a morphism in the slice over $X$
  to the morphism obtained by forgetting the maps to $X$.
  Since $\mathbf{H}(U,-)$ preserves $\infty$-limits, 
  it is sufficient to show that the homotopy fiber of this morphism (in $\infty\mathrm{Grpd}$) is 
  $\mathbf{H}(U, \Omega_{E}[\sum_X E, X])$, 
  naturally for each $U$. To that end, notice that $\mathbf{H}(U,p_X)$
  is the middle vertical morphism in the following 
  diagram, where the right square is the $\infty$-pullback
  diagram that exhibits the hom space in the slice by prop. \ref{SliceHomAsHomotopyFiber}:
  $$
    \raisebox{20pt}{
	\xymatrix{
	  \mathbf{H}(U, \Omega_E [\sum_X E, X])
	  \ar[r]
	  \ar[d]
	  &
	  \mathbf{H}_{/X}( U \times E, E)
	  \ar[d]^{\mathbf{H}(U,p_X)}
	  \ar[rr]
	  &&
	  {*}
	  \ar[d]^{\vdash U \times E}
	  \\
	  {*}
	  \ar[r]
	  &
	  \mathbf{H}(U \times \sum_X E, \sum_X E)
       \ar[rr]^{E \circ (-)}
	   &&
	   \mathbf{H}(U \times \sum_X E, X)
	  }	
	  }
	  \,.
  $$
  With the left square now denoting the $\infty$-pullback in question, we obtain the
  fiber in the top left by the pasting law for $\infty$-pullbacks, which says that 
  also the total rectangle here is an $\infty$-pullback. But this total pullback rectangle 
  is by example \ref{LoopSpaceObject} the one that characterizes the loop space object
  and hence identifies the top left item in the above diagram as claimed.
 \endofproof

\paragraph{Presentations}
\label{SlicingOfModelCategories}

We discuss presentations of slice $\infty$-categories, def. \ref{Slice},
by simplicial model categories, remark \ref{SimplicialModelCategory}.

\begin{proposition}
  \label{SliceModelStructure}
  For $C$ a model category and $X \in C$ an object, the slice category (overcategory) $C_{/X}$
  as well as the co-slice category (undercategory) $C^{X/}$ inherit model category structures
  whose fibrations, cofibrations and weak equivalences are precisely those
  of $C$ under the canonical forgetful functors $C_{/X} \to C$
  and $C^{X/} \to C$, respectively.
\end{proposition}
\begin{proposition}
  \label{PropertiesInheritedBySliceModelStructures}
  If the model category $C$ is 
  \begin{itemize}
    \item cofibrantly generated;
	\item or proper;
	\item or cellular
  \end{itemize}
  then so are the (co)-slice model structures of prop. \ref{SliceModelStructure},
  for every object $X \in C$.
\end{proposition}
This is shown in \cite{PH}. 
\begin{proposition}
  If the model category $C$ is combinatorial, then so is the slice model structure
  $C_{/X}$, for every object $X \in C$.
\end{proposition}
\proof
  With prop. \ref{PropertiesInheritedBySliceModelStructures}
  this follows form the fact that
  the slice of a locally presentable category is again locally presentable, 
  (e.g. remark 3 in \cite{CRV}).
\endofproof
\begin{proposition}
  If $C$ is a simplicial model category, then so is its slice $C_{/X}$,
  for every object $X \in C$.
\end{proposition}
\begin{proposition}
  \label{SliceModelPresentsInfinitySlicing}
  Let $C$ be a simplicial model category and write $\mathcal{C}$ for the 
  $\infty$-category that it presents. 
  If $X$ is fibrant in $C$, 
  then the slice model structure $C_{/X}$ is a presentation of the
  $\infty$-categorical slicing $\mathcal{C}_{/X}$. 
  If $X$ is cofibrant in $C$, then the co-slice model structure
  $C^{X/}$ is a presentation of the $\infty$-categorical co-slicing
  $\mathcal{C}^{X/}$.
\end{proposition}
\proof
  We discuss the first case. The other one is dual.
  We need to check that the derived hom-spaces are the correct
  $\infty$-categorical hom-spaces.
  Let $A \stackrel{a}{\to}X$ and $B \stackrel{b}{\to}X$ be two 
  objects of $\mathcal{C}_{/X}$. By prop. \ref{SliceHomAsHomotopyFiber}
  the hom $\mathcal{C}_{/X}(a,b)$ is the $\infty$-pullback
  $$
    \mathcal{C}_{/X}(a,b) \simeq \mathcal{C}(A,B) \times_{\mathcal{C}(A,X)} \{a\}
  $$
  in $\infty\mathrm{Grpd}$.
  Now write $a$ for a cofibrant representative of this object in $C_{/X}$
and $b$ for a fibrant representative. 
The $\mathrm{sSet}$-hom object
in $C_{/X}$ is the ordinary pullback
$$
  C_{/X}(a,b) \simeq C(A,B) \times_{C(A,X)} \{a\}
$$
in $\mathrm{sSet}$. One finds that $a$ being cofibrant in 
$C_{/X}$ means that $A$ is cofibrant in 
$C$ and $b$ being fibrant in $C_{/X}$ means that it is a fibration in $C$.  
Since by assumption $X$ is fibrant in $C$, it follows that also $B$ is
fibrant in $C$.
By the fact that $\mathrm{sSet}_{\mathrm{Quillen}}$ is 
itself a simplicial model category, it follows 
with prop. \ref{DerivedHomSpaceInSimplicialModelCategory}
that the simplicial hom-objects
appearing in the above pullback are the correct hom-spaces, and that the 
pullback is along a fibration.
Together this means by prop. \ref{ConstructionOfHomotopyLimits} that 
the ordinary pullback is indeed a model for the above 
$\infty$-pullback.
\endofproof

\subsubsection{Truncated objects and Postnikov towers}
 \label{TruncatedObjects}
  \label{PostnikovDecomposition}
 \index{structures in a cohesive $\infty$-topos!truncated objects}
  \index{structures in a cohesive $\infty$-topos!Postnikov towers}
 \index{Postnikov decomposition}

We discuss general notions and presentations of truncated objects and
Postnikov towers in an $\infty$-topos. 

\paragraph{General abstract}

\begin{definition}
  \label{truncated object}
  For $n \in \mathbb{N}$
  an $\infty$-groupoid $X \in \infty \mathrm{Grpd}$ 
  is called \emph{$n$-truncated} or a \emph{homotopy $n$-type} 
  if all its homotopy groups 
  in degree $> n$ are trivial. It is called \emph{$(-1)$-truncated}
  if it is either empty or contractible. It is called \emph{$(-2)$-truncated}
  if it is non-empty and contractible.
  
  For $\mathbf{H}$ an $\infty$-topos, and object $A \in \mathbf{H}$
  is called \emph{$n$-truncated} for $-2 \leq n \leq \infty$ if
  for all $X \in \mathbf{H}$ the hom $\infty$-groupoid 
  $\mathbf{H}(X, A)$ is $n$-truncated.

    An $\infty$-functor between $\infty$-groupoids is 
  called \emph{$k$-truncated} for $-2 \leq k \leq \infty$ if all
  its homotopy fibers are $k$-truncated.
   A morphism $f : A \to B$ in an $\infty$-topos $\mathbf{H}$
   is $k$-truncated if for all objects $X \in \mathbf{H}$ the
   induced $\infty$-functor $\mathbf{H}(X,f) : \mathbf{H}(X,A) \to \mathbf{H}(X,B)$
   is $k$-truncated.  
\end{definition}
This appears as \cite{Rezk} 7.1  and \cite{Lurie} def. 5.5.6.8.
\begin{remark}
  \begin{itemize}  
    \item A morphism is $(-2)$-truncated precisely if it is an equivalence.
	\item A morphism between $\infty$-groupoids that is $(-1)$-truncated 
	is a \emph{full and faithful $\infty$-functor}. A general morphism
	that is $(-1)$-truncated is an \emph{$\infty$-monomorphism}.
  \end{itemize}
\end{remark}
\begin{proposition}
  \label{truncation}
  For all $(-2) \leq n \leq \infty$ the full sub-$\infty$-category
  $\mathbf{H}_{\leq n}$
  of $\mathbf{H}$ on the $n$-truncated objects is reflective in $\mathbf{H}$
  in that the inclusion functor has a left adjoint $\infty$-functor $\tau_n$
  $$
    \xymatrix{
	  \mathbf{H}_{\leq n}
	  \ar@{<-}@<+4pt>[r]^{\tau_n}
	  \ar@{^{(}->}@<-4pt>[r]
	  &
	  \mathbf{H}
	}
	\,.
  $$
  Moreover, $\tau_n$ preserves finite products 
\end{proposition}
This is \cite{Lurie} prop. 5.5.6.18, lemma 6.5.1.2.
\begin{definition}
  For an object $X \in \mathbf{H}$ in an $\infty$-topos, 
  we say that the canonical sequence
  $$
    \xymatrix{
	  & & 
	  X
	  \ar[dl]^{\ }="s"_{\ }="t1" 
	  \ar[dr]_{\ }="t"
	  \ar@{}[dll]^{\ }="s1"
	  \ar[drr]
	  \
	  \\
	  \cdots \ar[r] & \tau_{n}X  \ar[r] & \cdots \ar[r] & \tau_0 X \ar[r] & \tau_{-1}X
	 \ar@{..} "s"; "t"
	}
  $$
  induced from the reflectors of prop. \ref{truncation} is the
  \emph{Postnikov tower} of $X$.
  
  We say that the Postnikov tower \emph{converges} if the above diagram
  exhibits $X$ as the $\infty$-limit over its Postnikov tower
  $$
    X \simeq \lim\limits_{\longleftarrow_n} \tau_n X
	\,.
  $$
  \label{PostnikovTower}
\end{definition}
This is def. 5.5.6.23 in \cite{Lurie}.

\begin{remark}
  Postnikov towers are a special cases of towers of higher \emph{images}.
  This we discuss further below in \ref{StrucEpi}. 
\end{remark}

\paragraph{Presentations}

\begin{proposition}
  \label{PresentationOfTruncationSubcategory}
  Let $C$ be a small site of definition of an $\infty$-topos $\mathbf{H}$, 
  so that 
  $$
    \mathbf{H} \simeq L_W [C^{\mathrm{op}}, \mathrm{sSet}]_{\mathrm{proj},\mathrm{loc}}
  $$
  according to theorem \ref{PresentationOfInfinityToposBySimplicialPresheaves}.
  Let $[C^{\mathrm{op}}, \mathrm{sSet}]_{\mathrm{proj,\mathrm{loc}, \leq n}}$
  be the left Bousfield localization of the local projective model structure
  on simplicial presheaves at the set of morphisms
  $$
    \{ \partial \Delta[k+1] \hookrightarrow U \to \Delta[k+1] \cdot U \;|\;
	 U \in C; k > n\}
	 \,.
  $$
  
  This is a presentation of the sub-$\infty$-category of $n$-truncated objects
  $$
    \mathbf{H}_{\leq n} \simeq ([C^{\mathrm{op}}, \mathrm{sSet}]_{\mathrm{proj},\mathrm{loc}, \leq n})^\circ
  $$
  and the canonical Quillen adjunction
  $$
    \xymatrix{
     [C^{\mathrm{op}}, \mathrm{sSet}]_{\mathrm{proj},\mathrm{loc}}
	 \ar@{<-}@<+3pt>[r]^-{\mathrm{id}}
	 \ar@<-3pt>[r]_-{\mathrm{id}}
	 &
     [C^{\mathrm{op}}, \mathrm{sSet}]_{\mathrm{proj}, \mathrm{loc}, \leq n}
	}	
  $$
  presents the reflection, $\tau_n \simeq \mathbb{L}\mathrm{id}$.
\end{proposition}
This appears in the proof of \cite{Rezk}, prop. 7.5.

We now discuss an explicit presentation for $n$-truncation and Postnikov
decompositions, def. \ref{PostnikovTower}, 
in terms of the projective model structure on simplicial presheaves.
First recall the following classical notions, reviewed for instance in 
\cite{GoerssJardine}.
\begin{definition}
  \index{coskeleton}
  \label{coskeleton}
  Let $\iota_{n+1} :  \Delta_{\leq n+1} \hookrightarrow \Delta$ 
  be the full subcatgeory of the simplex
  category on the objects $[k]$ for $k \leq n+1$.
  Write $\mathrm{sSet}_{\leq n+1} := \mathrm{Func}(\Delta^{\mathrm{op}}_{\leq n+1}, \mathrm{Set})$
  for the category of $(n+1)$-stage simplicial sets.
  
  Finally, write 
  $$
    \mathbf{cosk}_{n+1} 
	  : 
	\xymatrix{
      \mathrm{sSet}
 	   \ar[r]^-{\iota_{n+1}^*}
	  &
	  \mathrm{sSet}_{\leq n+1}
	   \ar@{^{(}->}[r]^-{\mathrm{cosk}_{n+1}}
	  &
	  \mathrm{sSet}
	}
  $$
  for the composite of the pullback along $\iota_{n+1}$ with its 
  \emph{right adjoint} $\mathrm{cosk}_{n+1}$.
  
  For $X \in \mathrm{sSet}$ we say that $\mathbf{cosk}_{n+1} X$ is it
  \emph{$(n+1)$-coskeleton}.

   All of these constructions prolong to simplicial presheaves.
\end{definition}
\begin{theorem}
  For $X \in \mathrm{sSet}$ a Kan complex, the tower of $\mathbf{cosk}$-units
  $$
    \cdots \to \mathbf{cosk}_3 X \to \mathbf{cosk}_2 X \to \mathbf{cosk}_1 X
  $$
  presents the Postnikov decomposition of $X$ in $\infty \mathrm{Grpd}$.
\end{theorem}
This is a classical result due to \cite{dwyer-kan}.
\begin{proposition}
  \label{TruncationByCoskeleton}
  For $C$ the site of definition of a hypercomplete $\infty$-topos,
  let $X \in [C^{\mathrm{op}}, \mathrm{sSet}]_{\mathrm{proj}, \mathrm{loc}}$
  be a fibrant simplicial presheaf. Then the tower of $\mathbf{cosk}$-units
  $$
    \cdots \to \mathbf{cosk}_3 X \to \mathbf{cosk}_2 X \to \mathbf{cosk}_1 X
  $$
  presents the Postnikov decomposition of $X$ in $\mathrm{Sh}_\infty(X)$.
\end{proposition}  
\proof  
  It is sufficient to show that $X \to \mathbf{cosk}_{n+1} X$
  presents the $n$-truncation $X \to \tau_n X$ in $\mathrm{Sh}_\infty(X)$.
  For this, in turn, it is sufficient to observe that this morphism
  exhibits a fibrant resolution in 
  $[C^{\mathrm{op}}, \mathrm{sSet}]_{\mathrm{proj}, \mathrm{loc}, \leq n}$.
  By standard facts about left Bousfield localizations, $\mathbf{cosk}_{n+1} X$
  is indeed fibrant in that model structure, since it is fibrant in the original
  structure by assumption and is local with respect to higher
  sphere inclusions by the nature of the coskeleton construction. 
  
  So it
  remains to see that the morphism $X \to \mathbf{cosk}_{n+1} X$ is
  a weak equivalence in the localized model structure. 
  We notice that by assumption of hypercompleteness,
  the homotopy category is also computed by the derived hom in 
  the truncation-localization of the 
  Jardine model structure \cite{Jardine}. By the nature of $\mathbf{cosk}$,
  the morphism induces an isomorphism on all homotopy sheaves in degree
  $\leq n$ (since the homotopy presheaves of $X$ and $\mathbf{cosk}_{n+1} X$ 
  in these degrees are manifestly equal and $X \to \mathbf{cosk}_{n+1}$ is the identity
  on cells in these degrees). Since by prop. \ref{PresentationOfTruncationSubcategory}
  also the localized Jardine structure presents the full sub-$\infty$-category
  on $n$-truncated objects, the morphisms which are isos on homotpy groups
  in degree $\leq n$ are already equivalences here.
\endofproof

\subsubsection{Epi-/mono-morphisms, images and relative Postnikov systems}
\label{StrucEpi}

In an $\infty$-topos there is an infinite tower of notions of epimorphisms and monomorphisms:
the $(n-2)$-connected and $(n-2)$-truncated morphisms for all $n \in \mathbb{N}$
\cite{Rezk, Lurie}. Accordingly, factorization through these induces a notion of 
$n$-images of morphisms in an $\infty$-topos, for each $n \in \mathbb{N}$.
The case when $n = -1$ is in some sense the most direct generalization of the 1-categorical notion.

\paragraph{General abstract}

\begin{definition}
  For $f : X \to Y$ a morphism in an $\infty$-topos $\mathbf{H}$ and for
  $n \in \mathbb{N}$, the \emph{$(n-2)$-connected/$(n-2)$-truncated factorization}
  of $f$ is the $(n-2)$-truncation of $f$, def. \ref{truncated object}, 
  as an object in the slice $\mathbf{H}_{/Y}$,
  def. \ref{Slice}:
  $$
    \xymatrix{
	   X \ar[rr] \ar[dr]_f 
	   && 
	   \underset{Y}{\sum} \tau_{n-2} f \ar[dl]|{\tau_{n-2} f}
	   \\
	   & 
	   Y
	}
	\,.
  $$  
  We write
  $$
    \mathrm{im}_n(f) := \underset{Y}{\sum} \tau_{n-2} f
  $$
  and call this the \emph{$n$-image} of $f$. We also say that
  $$
    \mathrm{im}_\infty(f) := X
  $$
  is the \emph{1-image} of $f$.
  \label{nImage}
  \label{ImagesInIntroduction}
\end{definition}
\begin{definition}
  A morphism $f : X \to Y$ is called
  \begin{itemize}  
    \item an \emph{$n$-epimorphism} if its $n$-image injection $\mathrm{im}_n(f) \to Y$
	is an equivalence;
	\item an \emph{$n$-monomorphism} if its $n$-image projection $X \to \mathrm{im}_n(f)$ 
	is an equivalence.
  \end{itemize}
\end{definition}
\begin{proposition}
  For all $n$, the classes $( \mathrm{Epi}_n(\mathbf{H}), \mathrm{Mono}_n(\mathbf{H}) )$
  constitute an orthogonal factorization system.
  \label{EpiMonoFactorizationSystem}
\end{proposition}
This is Proposition 8.5 in \cite{Rezk} and Example 5.2.8.16 in \cite{Lurie}.

Moreover:
\begin{proposition}
  The factorization systems of prop. \ref{EpiMonoFactorizationSystem} are
  \emph{stable}: for all $n$, the class of $n$-monomorphisms is preserved by $\infty$-pullback.
  \label{EpiMonoFactorizationIsStable}
\end{proposition}
This is \cite{Lurie}, prop. 6.1.5.16(6). 
\begin{remark}
By  prop. \ref{1EpimorphismsArePreservedByPullback} also 1-epimorphisms are preserved by
$\infty$-pullback (as are 0-epimorphisms = equivalences), but the class of 
$n$-epimorphisms for $n > 1$ is in general not preserved by $\infty$-pullback.
\end{remark}
\begin{proposition}
  A morphism $f : X \to Y$ is an $n$-monomorphism, precisely if its diagonal
  $X \to X \underset{Y}{\times} X$ is an $(n-1)$-monomorphism.
  \label{CharacterizationOfMonoByDiagonal}
\end{proposition}
This is \cite{Lurie}, lemma 5.5.6.15.

Of particular interest are 1-epimorphisms/1-monomorphisms. 
\begin{definition}
  For $f : X \to Y$ a morphism in $\mathbf{H}$, we write its
  1-epi/1-mono factorization given by Proposition \ref{EpiMonoFactorizationSystem}
  as
  $$
    f : 
    \xymatrix{
	  X \ar@{->>}[r] & \mathrm{im}_1(f)\  \ar@{^{(}->}[r] & Y 
	}
  $$
  and we call $\xymatrix{\mathrm{im}_1(f)\ \ar@{^{(}->}[r] & Y}$ the \emph{1-image}  
  (or just \emph{image}, for short)
  of $f$.
  \label{image}
\end{definition}
Equivalently the 1-image is the $(-1)$-truncation of $f : X \to Y$ regarded as
an object in the slice $\infty$-topos.
\begin{definition}
Let $\mathbf{H}$ be an $\infty$-topos.  
For $X \to Y$ any morphism in $\mathbf{H}$, there is a
  simplicial object $\check{C}(X \to Y)$ in $\mathbf{H}$ (the {\em \v{C}ech 
  nerve} of $f\colon X\to Y$) which in degree $n$ is the 
  $(n+1)$-fold $\infty$-fiber product of $X$ over $Y$ with itself
  $$
    \check{C}(X \to Y) : [n] \mapsto X^{\times^{n+1}_Y}
  $$
  A morphism $f : X \to Y$ in $\mathbf{H}$ is an \emph{effective epimorphism}
  if it is the colimiting cocone under its own {\v C}ech nerve:
  $$
    f : X \to \varinjlim \check{C}(X\to Y)
	\,.
  $$  
  Write $\mathrm{Epi}(\mathbf{H}) \subset \mathbf{H}^I$ for the collection of 
  effective epimorphisms.
  \label{EffectiveEpi}
\end{definition}
\begin{proposition}
  A morphism $f : X \to Y$ in the $\infty$-topos $\mathbf{H}$
  is an effective epimorphism precisely if its 0-truncation
  $\tau_0 f : \tau_0 X \to \tau_0 Y$ is an epimorphism (necessarily effective)
  in the 1-topos $\tau_{\leq 0} \mathbf{H}$.
  \label{EffectiveEpiIsEpiOn0Truncation}
\end{proposition}
This is Proposition 7.2.1.14 in \cite{Lurie}.
\begin{proposition}
  The effective epimorphisms of def.  \ref{EffectiveEpi}
  are equivalently the 1-epimorphisms of def. \ref{nImage}.
  In partiicular, for $f : X \to Y$ any morphism, its 1-image, def. \ref{image},
  is given by the $\infty$-colimit over its {\v C}ech nerve, def. \ref{EffectiveEpi}:
  $$
    \mathrm{im}_1(f)
	\simeq
	\underset{\longrightarrow}{\lim}_n \left( X^{\times_Y^{n+1}}\right)
	\,.
  $$
  \label{1ImageByInfinityColimitOverNerve}
\end{proposition}
\proof
  Let $f : \xymatrix{ X \ar@{->>}[r] & \mathrm{im}_1(f) \ar[r] & Y }$ be
  the essentially unique 1-image factorization. Then by prop. \ref{CharacterizationOfMonoByDiagonal}
  the diagram exhibiting the $\infty$-fiber product of this morphism with itself
  decomposes into a pasting diagram of $\infty$-pullbacks of the form
  $$
    \raisebox{30pt}{
    \xymatrix{
	  X \underset{Y}{\times}X 
	  \ar@{}[r]|\simeq
	  & X \underset{\mathrm{im}_1(f)}{\times} X
	  \ar[r]
	  \ar[d]
	  &
	  X
	  \ar[r]^\simeq
	  \ar@{->>}[d]
	  &
	  X
	  \ar@{->>}[d]
	  \ar@/^1.3pc/[dd]^f
	  \\
	  & X \ar@{->>}[r] 
	  \ar[d]^\simeq 
	  & 
	  \mathrm{im}_1(f) 
	  \ar[r]^\simeq 
	  \ar[d]^\simeq 
	  & 
	  \mathrm{im}_1(f) 
	  \ar@{^{(}->}[d]
	  \\
	  & X \ar[r] \ar@/_1pc/[rr]_f & \mathrm{im}_1(f) \ar@{^{(}->}[r] & Y
	}
	}
	\,.
  $$
  By the pasting law, prop. \ref{PastingLawForPullbacks} this identifis the
  $\infty$-fiber product of $f$ with itself over $Y$ with its product over 
  $\mathrm{im}_1(f)$, as indicated, and hence the {\v C}ech nerve of
  $f$ is equivalently that of its image projection $\xymatrix{X \ar@{->>}[r] & \mathrm{im}_1(f)}$.
  Finally be the Giraud-Rezk-Lurie axiom, def. \ref{GiraudRezkLurieAxioms}, 
  satisfied by the ambient $\infty$-topos, the $\infty$-colimit over the {\v C}ech nerve of
  $\xymatrix{X \ar@{->>}[r] & \mathrm{im}_1(f)}$ is that morphism itself.
\endofproof

The following is a simple consequence which we will need.
\begin{proposition}
For 
$$
  \iota
  :
  \xymatrix{
    A \ar@{^{(}->}[r] & B
  }
$$
a 1-monomorphism in $\mathbf{H}$ and for $X \in \mathbf{H}$
any object, the image of $\phi$ under the internal hom 
$[X,-] : \mathbf{H} \to \mathbf{H}$ is again a 1-monomorphism.
$$
  [X,\iota]
  :
  \xymatrix{
    [X, A]
	\ar@{^{(}->}[r]
	&
	[X,B]
  }
$$
\end{proposition}
\proof
 By prop. \ref{CharacterizationOfMonoByDiagonal} 
 a morphism is a 1-monomorphism precisely if the $\infty$-fiber product
 with itself reproduces its domain. Since $[X,-]$ preserves $\infty$-limits,
 this implies the claim.
\endofproof
\begin{proposition}
  For $\iota : \xymatrix{X\ar@{^{(}->}[r]& {*}}$
  a 1-monomorphism (exhibiting $X$ as a \emph{subterminal object}), 
  and for $E_1, E_2 \in \mathbf{H}_{/X}$ two objects in the slice, the canonical map 
  $$
    p_X
	:
	\underset{X}{\prod} \left[E_1, E_2\right] 
	\to 
	\left[
	  \underset{X}{\sum} E_1,
	  \underset{X}{\sum} E_2
	\right]
  $$
  of prop. \ref{CanonicalMorphismFromSliceInternalHomToBaseInternalHom}
  is an equivalence. 
  \label{InternalSliceHomOverSubterminal}
\end{proposition}
\proof
  By the proof of 
  prop. \ref{CanonicalMorphismFromSliceInternalHomToBaseInternalHom} 
  it suffices to show that the analogous statement
  holds for the external hom, hence that 
  we have that the canonical map 
  $$
    \xymatrix{
      \mathbf{H}_{/X}(E_1, E_2)
	  \ar[r]
	  &
	 \mathbf{H}(\underset{X}{\sum}E_1, \underset{X}{\sum}E_1)
	}
  $$
  of prop. \ref{SliceHomAsHomotopyFiber} is an equivalence. 
  That morphism sits in the $\infty$-pullback on the left of the diagram
  $$
    \xymatrix{
	  \mathbf{H}_{/X}(E_1, E_2)
	  \ar[r]
	  \ar[d]
	  &
	  {*}
	  \ar[d]^{\vdash E_1}
	  \ar[dr]
	  \\
	  \mathbf{H}(\underset{X}{\sum}E_1, \underset{X}{\sum}E_1)
	  \ar[r]^{E_2\circ (-)}
	  &
	  \mathbf{H}(\underset{X}{\sum}E_1, X)
	  \ar@{^{(}->}[r]_-{\mathbf{H}(X,\iota)}
	  &
	  {*}
	}
  $$
  in $\infty \mathrm{Grpd}$. Here $\mathbf{H}(\underset{X}{\sum}E_1, X)$
  is subterminal and inhabited, hence is terminal. Therefore the right vertical
  morphism is an equivalence and hence so is the left vertical morphism.
\endofproof
  By taking $\mathbf{H}$ in prop. \ref{InternalSliceHomOverSubterminal} itself
  to be a slice of another $\infty$-topos, the statement implies the following
  seemingly more general statement:
\begin{proposition}
  for $f : \xymatrix{X \ar@{^{(}->}[r] & Y}$ a 1-monomorphism in an $\infty$-topos
  $\mathbf{H}$ and for 
  $E_1, E_2 \in \mathbf{H}_{/X}$ two objects in the slice over $X$, 
  the canonical morphism
  $$
    \underset{Y}{\prod} p_f
	:
	\left[E_1, E_2\right]_{\mathbf{H}}
	\to
	\left[\underset{f}{\sum}E_1, \underset{f}{\sum}E_2\right]_{\mathbf{H}}
  $$  
  between the $\mathbf{H}$-valued slice homs of def. \ref{HValuedMappingSpace} 
  is an equivalence.
  \label{SliceHomComparedAlongAMonomorphism}
\end{proposition}
The following is another simple fact that we will need.
\begin{proposition}
  For $f : X \to Y$ any morphism in $\mathbf{H}$ its homotopy fiber
  over any global point of $Y$ in the image of $f$ is equivalent to the
  homotopy fiber over the corresponding point in $\mathrm{im}_1(f)$.
  \label{FiberOverPointIsFiberIn1Image}
\end{proposition}
\proof
  By the pasting law, prop. \ref{PastingLawForPullbacks}
  the homotopy fiber sits in a pasting diagram of $\infty$-pullbacks.
  $$
    \raisebox{30pt}{
    \xymatrix{
	  \mathrm{fib}_y(f) \ar[r] \ar[d] & X \ar@{->>}[d]
	  \\
	  {*} \ar[d] \ar[r]^y & \mathrm{im}_1(f) \ar@{^{(}->}[d]
	  \\
	  {*} \ar[r]^y & \mathrm{Y}
	}}
	\,.
  $$
  That $y$ is in the image of $f$ precisely says that we have the bottom square
  and the fact that that the bottom right morphism is a 1-monomorphism says that 
  the bottom square is an $\infty$-pullback. 
  This identifies the middle row of the digram as
  indicated.
  (For instance one can check this 
  by applying $\mathbf{H}(U,-)$ to the diagram where $U$ ranges over a set
  of generators and then using that the only suobobjects in $\infty\mathrm{Grpd}$
  of $* \simeq \mathbf{H}(U,*)$ are $\emptyset$ and $*$ itself).
  \endofproof

Now we turn to discusson of the \emph{towers} of $n$-image factorizations as $n$ ranges, 
which are the \emph{relative Postnikov towers} in an $\infty$-topos.
\begin{remark}
  The $n$-images for all $n$ form a tower 
  $$
    \xymatrix{
	  X \ar[r]^-= 
	  \ar@/_1.2pc/[rrrrrr]_{f}
	  & \mathrm{im}_\infty(f) \ar[r] & \cdots \ar[r] & \mathrm{im}_2(f) \ar[r] & \mathrm{im}_1(f) \ar[r] & \mathrm{im}_0(f) \ar[r]^-{\simeq} & Y
	}
	\,,
  $$
  also called the \emph{relative Postnikov tower} of $f$. For $Y \simeq *$ the terminal object
  this is the (absolute) \emph{Postnikov tower} of the object $X$. 
  For $X \simeq {*}$ the terminal object, this is the \emph{Whitehead tower} of $Y$.
  Conversely, the
  relative Postnikov tower of $f$ in $\mathbf{H}$ is equivalently the absolute Postnikov
  tower of $f$ regarded as an object of the slice $\mathbf{H}_{/Y}$.
  \label{RelativePostnikovTower}
\end{remark}
\begin{remark}
 For $f : X \to *$ a terminal morphism, the $n$-image coincides with the $(n-2)$-truncation of $X$:
 $$
   \tau_{n-2}X \simeq \mathrm{im}_n(X \to *)
   \,.
 $$
\end{remark}
\begin{proposition}
  Let $f : X \to Y$ be a morphism in an $\infty$-topos $\mathbf{H}$ and let $x : * \to X$
  be a base point. Then for all $n \in \mathbb{N}$, forming $n$-images commutes with 
  forming loop space objects up to a shift in image-degree, in that there is a natural equivalence
  $$
    \Omega \left(\mathrm{im}_n(f)\right) \simeq \mathrm{im}_{n-1}(\Omega f)
	\,.
  $$
  \label{ImagesCommuteWithLooping}
\end{proposition}
\proof
  The corresponding statement in homotopy type theory is shown in \cite{Rijke}. The above
  statement is the categorical semantics of that. 
\endofproof

\paragraph{Presentations}
  
We discuss presentations of $n$-images in $\infty$-toposes by constructions on simplicial presheaves.
  
In $\mathbf{H} = \infty \mathrm{Grpd}$, the general notion of relative Postnikov
towers, remark \ref{RelativePostnikovTower}, reproduces the traditional one.
  
\begin{definition}
  For $X, Y\in \mathrm{sSet}$ two simplicial sets, let  $f : X \to Y$ be a Kan fibration.
  For $n \in \mathbb{N}$ define an equivalence relation $\sim_n$ on $X_\bullet$
  by declaring that two $k$-simplices $\sigma_1, \sigma_2 : \Delta^k \to X$
  of $X$ are equivalent if
  \begin{enumerate}
    \item they have the same $n$-skeleton 
	$\xymatrix{\mathrm{sk}_n \Delta^k \ar[r] & \Delta^k \ar[r]^{\sigma_1,\sigma_2} & X}$
	\item and $f(\sigma_1) = f(\sigma_2)$.
  \end{enumerate}
  Write then
  $$
    \mathrm{im}_{n+1}(f) := X/\sim_n
  $$
  for the quotient simplicial set. This comes equipped with canonical morphisms
  of simplicial sets
  $$
    \xymatrix{
	  X \ar[r]\ar@/_1.2pc/[rr] & \mathrm{im}_{n+1}(f) \ar[r] & Y
	}
	\,.
  $$
  \label{PostnikovRelativeInsSet}
\end{definition}
This appears for instance as def. VI 2.9 in \cite{GoerssJardine}. 
\begin{proposition}
  Under the equivalence $\infty \mathrm{Grpd} \simeq L_{\mathrm{whe}}\mathrm{sSet}$,
  the construction of def. \ref{PostnikovRelativeInsSet}
  is a presentation of the relative Postnikov tower, remark \ref{RelativePostnikovTower}, in 
 $\mathbf{H} = \infty \mathrm{Grpd}$. 
\end{proposition}  
This is essentially the statement of theorem VI 2.11 in \cite{GoerssJardine}.

For maps between low truncated objects, we have the following simple identification of 
their $n$-images.
\begin{proposition}
  \label{nTruncated1Functors}
  A 1-functor between 1-groupoids is $n$-truncated as 
  a morphism of $\infty$-groupoids precisely if
  \begin{itemize}
    \item for $n = -2$ it is an equivalence of categories;
	\item for $n = -1$ it is a full and faithful functor;
	\item for $n = 0$ it is a faithful functor.
  \end{itemize}
\end{proposition}
\proof
  We consider the case $n = 0$.
  A functor $f : X \to Y$ between groupoids being faithful is 
  equivalent to the induced morphisms on first homotopy groups
  being monomorphisms.
  Therefore for $F \to X \to Y$ the homotopy fiber over any point
  of $Y$, the long exact sequence of homotopy groups yields
  $$
    \cdots \to \pi_1(F) \to \pi_1(X) 
	  \stackrel{f_*}{\hookrightarrow} \pi_1(Y) \to \cdots
  $$
  and hence realizes $\pi_1(F)$ as the kernel of an injective map.
  Therefore $\pi(F) \simeq *$ and hence $F$ is 0-truncated
  for every basepoint. This is the defining condition for $f$ being 0-truncated.
\endofproof
\begin{proposition}
  \label{0TruncatedStackMorphisms}
  Let $C$ be a site and let $f : X \to Y$ be a morphism of
  presheaves of groupoids on $C$ which, under the nerve, are 
  fibrant objects in 
  $[C^{\mathrm{op}}, \mathrm{sSet}]_{\mathrm{proj}, \mathrm{loc}}$.
  If  $f$ is objectwise a) an equivalence, b) full and faithful
  or c) faithful,
  then the morphism presented by $f$ in $\mathbf{H} := \mathrm{Sh}_{\infty}(X)$
  is a) -2-truncated, b) (-1)-truncated, c) 0-truncated, respectively.
\end{proposition}
\proof
  We need to compute for every $A \in \mathbf{H}$
  the homotopy fibers of $\mathbf{H}(A,f)$. Since by assumption $X$ and $Y$
are fibrant presentations, we may pick any cofibrant presentation of
$A$ and obtain this morphism as $[C^{\mathrm{op}}, \mathrm{sSet}](A,f)$.  
This is the nerve of a functor of groupoids which is a) an equivalence,
b) full and faithful or c) faithful, respectively. 
The statement then follows with observation \ref{nTruncated1Functors}.
\endofproof
\begin{proposition}
  \label{EsoFullPresheafFunctorsAre0Connected}
  Let $f : X \to Y$ be a morphism between presheaves of groupoids
  that are fibrant as objects of $[C^{\mathrm{op}}, \mathrm{sSet}]_{\mathrm{proj}}$,
  and such that $f$ is objectwise an essentially surjective and full functor.
  
  Then $f$ presents a 0-connected morphism in $\mathrm{Sh}_\infty(C)$.
\end{proposition}
\proof
  One checks that functors between 1-groupoids are 0-connected 
  as morphisms in $\infty \mathrm{Grpd}$ precisely if they are
  essentially surjective and faithful.
  
  The direction (eso+full) $\Rightarrow$ 0-connected of this 
  argument goes through objectwise.
\endofproof

More generally, we obtain a similarly simple and concrete presentation of 
$n$-image factorization of morphisms in the case that they are presented by 
homomorphisms of \emph{strict} $\infty$-groupoids, def. \ref{StrictInfinityGroupoid}.

\begin{proposition}
  Let $f : X \to Y$ be a morphism in $\infty \mathrm{Grpd}$ which is in the essential image
  of the inclusion
  $$
    \mathrm{Str}\infty\mathrm{Grpd} \hookrightarrow \mathrm{KanCplx} \to L_{\mathrm{whe}} \mathrm{sSet}
	\simeq \infty \mathrm{Grpd}
  $$
  of a morphism strict $\infty$-groupoids, given by an underlying morphism of globular sets 
  $f_\bullet : X_\bullet, Y_\bullet$. Then for $n \in \mathbb{N}$ the $n$-image
  factorization def. \ref{nImage} of $f$ is presented under this inclusion by the
  strict $\infty$-groupoid $\mathrm{im}_n(f)$ whose underlying globular set is
  $$
    \left( \mathrm{im}_n(f)\right)_k
	:=
	\left\{
	  \begin{array}{cc}
	    X_k & \forall k < n-1
		\\
		\mathrm{im}(X_{n-1}) \subset Y_{n-1} & \forall k = n-1
		\\
		Y_k & \forall k \geq n
	  \end{array}
	\right.
  $$
  equipped with the evident composition operations induced from those on $X_\bullet$ and $Y_\bullet$,
  and with the evident morphisms
  $$
    \xymatrix{
	   X_\bullet 
	   \ar[r]
	   &
	   \mathrm{im}_n(f)_\bullet
	   \ar[r]
	   &
	   Y_\bullet
	}
	\,,
  $$
 the left one being the identity in degree $k < n-1$, the quotent projection in degree $n-1$ and $f$ in degree 
  $k \geq n$, and the right one being $f$ in degree $k < n-1$, the image inclusion in degree $n-1$ and the identity in degree $k \geq n$.
  \label{PostnikovFactorizationOfMorphismOfStrictInfinityGroupoids}
\end{proposition}
\proof
  The homotopy groups of a strict globular $\infty$-groupoid in any degree $k$ are simply given 
  by the groups of $k$-automorphisms of the identity $(k-1)$-morphism on a given baspoint 
  modulo $(k+1)$-morphisms (hence the homology of the corresponding crossed complex, 
  def. \ref{CrossedComplex} 
  in that degree). Therefore it is clear from the construction of $im_n(f)$ above that 
  $X \to im_n(f)$ is surjective on $\pi_0$ and an isomorphism on $\pi_{k < n-1}$, and that $im_n(f)$ is a monomorphism on $\pi_{n-1}$ and an isomorphism on $\pi_{k \geq n}$.
\endofproof
\begin{remark}
  For the case $Y = *$ the content of 
  prop. \ref{PostnikovFactorizationOfMorphismOfStrictInfinityGroupoids}  is
  discussed in \cite{BFGM}.
\end{remark}

\subsubsection{Compact objects}
 \label{CompactObjects}
 \index{structures in a cohesive $\infty$-topos!compact objects}
  
Traditionally there are two notions referred to as \emph{compactness}
of a space, which are closely related but subtly different.
\begin{enumerate}
  \item 
On the one hand a space is called compact if regarded as an object of 
a certain \emph{site} each of its covering families has a finite
subfamily that is still covering. 
 \item
On the other hand, an object in a category with colimits is called compact if 
the hom-functor out of that object commutes with all filtered colimits.
Or more generally in the $\infty$-category context: if the hom-$\infty$-functor
out of the objects commutes with all filtered $\infty$-colimits 
(section 5.3 of \cite{Lurie}).
\end{enumerate}
For instance in the site of topological spaces or 
of smooth manifolds, equipped with the usual open-cover coverage, the first
definition reproduces the traditional definition of \emph{compact topological space} 
and of \emph{compact smooth manifold}, respectively. But the notion of compact object in the 
category of topological spaces in the sense of the second definition is not quite
equivalent. For instance the two-element set equipped with the indiscrete topology
is compact in the first sense, but not in the second.

The cause of this mismatch, as we will discuss in detail below, becomes clearer
once we generalize beyond 1-category theory to $\infty$-topos theory: in that context it is
familiar that locality of morphisms out of an object $X$ into an $n$-truncated object
$A$ (an $n$-stack) is no longer controled by just the notion of \emph{covers} of $X$,
but by the notion of \emph{hypercover of height $n$}, which reduces to the 
ordinary notion of cover for $n = 0$. Accordingly it is clear that 
the ordinary condition on a compact topological space to admit finite refinement
of any cover is just the first step in a tower of conditions: we may say an object 
is \emph{compact of height $n$} if every hypercover of height $n$ over the object
is refined by a ``finite hypercover'' in a suitable sense.

Indeed, the condition on a \emph{compact object} in a 1-category to distribute over
filtered colimits turns out to be a compactness condition of \emph{height 1}, 
which conceptually explains why it is stronger than the existence of finite refinements 
of covers. This state of affairs in the first two height levels has been known,
in different terms, in topos theory, where one distinguishes between a topos being
\emph{compact} and being \emph{strongly compact} \cite{MoerdijkVermeulen}:

\begin{definition}
  A 1-topos 
  $(\Delta \dashv \Gamma) :  
   \xymatrix{\mathcal{X} \ar@{<-}@<+3pt>[r]\ar@<-3pt>[r] & \mathrm{Set}}$ is called
  \begin{enumerate}
    \item 
	  a \emph{compact topos} if the global section functor 
	  $\Gamma$ preserves filtered colimits of 
	   subterminal objects (= (-1)-truncated objects);
	\item
	  a \emph{strongly compact topos} if $\Gamma$ preserves all filtered colimits
	   (hence of all 0-truncated objects).
  \end{enumerate}
\end{definition}
Clearly these are the first two stages in a tower of notions which continues as follows.
\begin{definition}
 For $(-1) \leq n \leq \infty$,  an $\infty$-topos 
   $(\Delta \dashv \Gamma) :  
   \xymatrix{\mathcal{X} \ar@{<-}@<+3pt>[r]\ar@<-3pt>[r] & \infty \mathrm{Grpd}}$ is 
 called \emph{compact of height $n$} if $\Gamma$ preserves filtered
 $\infty$-colimits of $n$-truncated objects.
\end{definition}
Since therefore the traditional terminology concerning ``compactness'' is not
quite consistent across fields, with the category-theoretic ``compact object''
corresponding, as shown below, to the topos theoretic ``strongly compact'',
we introduce for definiteness the following terminology.
\begin{definition}
  For $C$ a subcanonical site, call an object 
  $X \in C \hookrightarrow \mathrm{Sh}(C) \hookrightarrow \mathrm{Sh}_\infty(C)$ 
  \emph{representably compact} if every covering family $\{U_\alpha \to X\}_{i \in I}$
  has a finite subfamily $\{U_j \to X\}_{j \in J \subset I}$ which is still covering.
  \label{representably compact}
\end{definition}

The relation to the traditional notion of compact spaces and compact objects is given by the following
\begin{proposition}
  Let $\mathbf{H}$ be a 1-topos and $X \in \mathbf{H}$ an object. Then
  \begin{enumerate}
    \item if $X$ is representably compact, def. \ref{representably compact}, with respect to the 
  	  canonical topology, then the slice topos
	$\mathbf{H}_{/X}$, def. \ref{Slice} is a compact topos;
	\item the slice topos $\mathbf{H}_{/X}$ is strongly compact precisely if
	 $X$ is a compact object.
  \end{enumerate}
\end{proposition}
\proof
  Use that the global section functor $\Gamma$ on the slice topos is given by 
  $$
    \Gamma([E \to X]) = \mathbf{H}(X,E) \times_{\mathbf{H}(X,X)} \{\mathrm{id}_X\}
  $$
  and that colimits in the slice are computed as colimits in $\mathbf{H}$:
  $$
    \lim\limits_{\longrightarrow_i} [E_i \to X] \simeq 
	 [(\lim\limits_{\longrightarrow_i} E_i) \to X]
	 \,.
  $$
  
  For the first statement, observe that the subterminal objects of $\mathbf{H}_{/X}$ are the
  monomorphisms in $\mathbf{H}$. Therefore $\Gamma$ sends all subterminals to the 
  empty set except the terminal object itself, which is sent to the singleton set.
  Accordingly, if $U_\bullet : I \to \mathbf{H}_{/X}$
  is a filtered colimit of subterminals then 
  \begin{itemize}
    \item either the $\{U_\alpha\}$ do not cover, hence in particular none of the $U_\alpha$ is $X$ 
	 itself, and hence both $\Gamma( \lim\limits_{\longrightarrow_i} U_\alpha)$ as well as
	 $\lim\limits_{\longrightarrow_i} \Gamma(  U_\alpha)$ are the empty set;
	\item
	 or the $\{U_\alpha\}_{i \in I}$ do cover. Then by assumption on $X$ there is a finite subcover
	 $J \subset I$,
	 and then by assumption 
	 that  $U_\bullet$ is filtered
	 the cover contains the finite union $\lim\limits_{\longrightarrow \atop {i \in J}} U_\alpha = X$
	 and hence both $\Gamma( \lim\limits_{\longrightarrow_i} U_\alpha)$ as well as
	 $\lim\limits_{\longrightarrow_i} \Gamma(  U_\alpha)$ are the singleton set.
  \end{itemize}
  
  For the second statement, assume first that $X$ is a compact object. Then 
   using that colimits in a topos  are preserved by pullbacks, it follows for all
   filtered diagrams $[E_\bullet \to X]$ in $\mathbf{H}_{/X}$ that
   $$
     \begin{aligned}
	   \Gamma(\lim\limits_{\longrightarrow_i} [E_i \to X])
	   &
	   \simeq
	   \mathbf{H}(X, \lim\limits_{\longrightarrow_i}E_i) \times_{\mathbf{H}(X,X)} \{\mathrm{id}\}
	   \\
	   & \simeq
	   (\lim\limits_{\longrightarrow_i}\mathbf{H}(X, E_i)) \times_{\mathbf{H}(X,X)} \{\mathrm{id}\}
	   \\
	   & \simeq
	   \lim\limits_{\longrightarrow_i}(\mathbf{H}(X, E_i) \times_{\mathbf{H}(X,X)} \{\mathrm{id}\})
	   \\
	   & \simeq
	   \lim\limits_{\longrightarrow_i} \Gamma[E_i \to X] 
	 \end{aligned}
	 \,,
   $$
   and hence $\mathbf{H}_{/X}$ is strongly compact. 
   
   Conversely, assume that $\mathbf{H}_{/X}$ is strongly compact. 
   Observe that for every object $F \in \mathbf{H}$ we have a natural isomorphism
   $\mathbf{H}(X, F) \simeq \Gamma( [X \times F \to X])$.
   Using this, we obtain for every filtered diagram $F_\bullet$ in $\mathbf{H}$
   that
   $$
     \begin{aligned}
	   \mathbf{H}(X, \lim\limits_{\longrightarrow_i} F_i)
	   &
	   \simeq
	   \Gamma( [X \times (\lim\limits_{\longrightarrow_i} F_i) \to X])
	   \\
	   & \simeq
	   \Gamma( \lim\limits_{\longrightarrow_i}[X \times F_i \to X])
	   \\
	   & \simeq 
	   \lim\limits_{\longrightarrow_i} \Gamma( [X \times F_i \to X])
	   \\
	   & \simeq
	   \lim\limits_{\longrightarrow_i} \mathbf{H}(X,F_i)
	 \end{aligned}
   $$
   and hence $X$ is a compact object.
\endofproof

Notice that a diagram of subterminal objects necessarily consists 
only of monomorphisms.
We show now that a representably compact object 
generally distributes over such \emph{monofiltered colimits}.
\begin{definition}
  Call a filtered diagram $ A : I \to D$ in a category $D$ \emph{mono-filtered}
  if for all morphisms $i_1 \to i_2$ in the diagram category $I$ the morphism
  $A(i_1 \to i_2)$ is a monomorphism in $D$.  
\end{definition}
\begin{lemma}
  \label{MonofilteredColimitsSendSheavesToSeparated}
  For $C$ a site and $A : I \to \mathrm{Sh}(C) \hookrightarrow \mathrm{PSh}(C)$
  a monofiltered diagram of sheaves, its colimit 
  $\lim\limits_{\longrightarrow_i} A_i \in \mathrm{PSh}(C)$
  is a separated presheaf.
\end{lemma}
\proof
  For $\{U_\alpha \to X\}$ any covering family in $C$ with 
  $S(\{U_\alpha\}) \in \mathrm{PSh}(C)$ the corresponding sieve, 
  we need to show that
$$
  \lim\limits_{\longrightarrow_i} A_i(X) 
    \to 
  \mathrm{PSh}_C(S(\{U_\alpha\}), \underset{\longrightarrow_i}{\lim} A_i)
$$
is a monomorphism. An element on the left is represented by a pair $(i \in I, a \in A_i(X))$. 
Given any other such element, we may assume by filteredness that they are both represented over the same index $i$. So let $(i,a)$ and $(i,a')$ be two such elements. 
Under the above function, $(i,a)$ is mapped to the collection 
$\{ i, a|_{U_\alpha} \}_\alpha$ and $(i,a')$ to $\{ i, a'|_{U_\alpha} \}_\alpha$. If $a$ is different from $a'$, then these families differ at stage $i$, hence
at least one pair $a|_{U_\alpha}$, $a'|_{U_\alpha}$ is different 
at stage $i$. Then by mono-filteredness, this pair differs also at all 
later stages, hence the corresponding families 
$\{U_\alpha \to \lim\limits_{\longrightarrow_i} A_i\}_\alpha$ differ.
\endofproof
\begin{proposition}
  For $X \in \mathrm{C} \hookrightarrow \mathrm{Sh}(C)$ a representably
  compact object, def. \ref{representably compact},
  $\mathrm{Hom}_{\mathrm{Sh}(C)}(X,-)$ commutes with all mono-filtered
  colimits.
\end{proposition}
\proof
  Let $A : I \to \mathrm{Sh}(C) \hookrightarrow \mathrm{PSh}(C)$ 
  be a mono-filtered diagram of sheaves,  regarded as a diagram of presheaves.
  Write $\lim\limits_{\longrightarrow_i} A_i$ for its colimit. So with
  $L : \mathrm{PSh}(C) \to \mathrm{Sh}(C)$ denoting sheafification,
  $L \lim\limits_{\longrightarrow_i} A_i$ is the colimit of sheaves in question.
  By the Yoneda lemma and since colimits of presheaves are computed
  objectwise, it is sufficient to show that for $X$ a 
  representably compact object, 
  the value of the sheafified colimit is the colimit of the values of the 
  sheaves on $X$
  $$
    (L \underset{\longrightarrow _i}{\lim} A_i)(X) 
	  \simeq
	(\underset{\longrightarrow _i}{\lim} A_i)(X)
	  =
	\underset{\longrightarrow _i}{\lim} A_i(X)
	\,.
  $$
  To see this, we evaluate the sheafification by the plus construction. 
  By lemma \ref{MonofilteredColimitsSendSheavesToSeparated}, the presheaf 
  $\lim\limits_{\longrightarrow_i} A_i$ is already separated, so we obtain its 
  sheafification by applying the plus-construction just \emph{once}.

  We observe now that \emph{over a representably compact object} $X$ the single plus-construction
  acts as the identity on the presheaf $\lim\limits_{\longrightarrow_i} A_i$.
  Namely the single plus-construction over $X$ takes the colimit of the value of
  the presheaf on sieves 
  $$
    S(\{U_\alpha\}) := 
	\underset{\longrightarrow}{\lim}( 
	  \xymatrix{
	    \coprod_{\alpha, \beta} U_{\alpha,\beta} 
   		 \ar@<+2pt>[r]
   		 \ar@<-2pt>[r]
         &		 
		\coprod_\alpha U_\alpha
	  }
	) 
  $$ 
  over the opposite of the category of covers 
  $\{U_\alpha \to X\}$ of $X$. By the very definition of compactness, 
  the inclusion of (the opposite category of) the category of finite covers of $X$ into that of all covers is a final functor. Therefore we may 
  compute the plus-construction over $X$ by the colimit over just the collection of finite covers. 
  On a finite cover we have
  $$
    \begin{aligned}
	  \mathrm{PSh}(S(\{U_\alpha\}), \lim\limits_{\longrightarrow_i} A_i)
	  & :=
      \mathrm{PSh}( \lim\limits_{\longrightarrow}( 
	    \xymatrix{
	     \coprod_{\alpha,\beta} U_{\alpha \beta} 
		 \ar@<+2pt>[r]
		 \ar@<-2pt>[r]
		 &
		 \coprod_\alpha  U_\alpha
		 }
		 ), 
 	\lim\limits_{\longrightarrow_i} A_i)
	\\
	   & \simeq
     \lim\limits_{\longleftarrow}
     (	 
	   \xymatrix{
 	    \prod_{\alpha} \lim\limits_{\longrightarrow_i} A_i(U_{\alpha })
         \ar@<+2pt>[r]
		 \ar@<-2pt>[r]
         &
 	    \prod_{\alpha, \beta } \lim\limits_{\longrightarrow_i} A_i(U_{\alpha, \beta})
	   }
	  )
	  \\
	  & \simeq
     \lim\limits_{\longrightarrow_i}\lim\limits_{\longleftarrow}     (	 
	  \xymatrix{
 	    \prod_{\alpha } A_i(U_{\alpha })
         \ar@<+2pt>[r]
		 \ar@<-2pt>[r]
         &
        \prod_{\alpha, \beta } A_i(U_{\alpha, \beta})
	   }
	  )	  
	  \\ & \simeq
	  \lim\limits_{\longrightarrow_i} A_i(X)
	\end{aligned}  
	\,,
  $$
  where in the second but last step we used that filtered colimits commute with 
  finite limits, and in the last step we used that each $A_i$ is a sheaf.

So in conclusion, for $X$ a representably compact object and $A : I \to \mathrm{Sh}(C)$ 
a monofiltered diagram, we have found that
$$
  \begin{aligned}
    \mathrm{Hom}_{\mathrm{Sh}(C)}(X, L \lim\limits_{\longrightarrow_i} A_i)
    & \simeq
    (\lim\limits_{\longrightarrow_i} A_i)^+ (X)
    \\
    & \simeq \lim\limits_{\longrightarrow_i} A_i(X)
    \\
    & \simeq \lim\limits_{\longrightarrow_i} \mathrm{Hom}_{\mathrm{Sh}(C)}(X, A_i)
  \end{aligned}
$$
\endofproof
The discussion so far suggests that there should be conditions 
for ``representantably higher compactness'' on objects in a site that 
imply that the Yoneda-embedding of these objects into the $\infty$-topos
over the site distribute over larger classes of filtered $\infty$-colimits.
\begin{definition}
  For $C$ a site, say that an object $X \in C$
  is \emph{representably paracompact} if each bounded hypercover over 
  $X$ can be refined by the {\v C}ech nerve of an ordinary cover.
  \label{representaby paracompact}
\end{definition}
The motivating example is
\begin{proposition}
  Over a paracompact topological space, every bounded hypercover 
  is refined by the {\v C}ech nerve of an ordinary open cover.
  \label{HypercoversOverParacompactaAreRefinedByCechNerves}
\end{proposition}
\proof
  Let $Y \to X$ be a bounded hypercover. 
  By lemma 7.2.3.5 in \cite{Lurie} we may find for each $k \in \mathbb{N}$
  a refinement of the cover given by $Y_0$ such that the 
  non-trivial $(k+1)$-fold
  intersections of this cover factor through $Y_{k+1}$.
  Let then $n \in \mathbb{N}$ be a bound for the height of $Y$
  and form the intersection of the covers obtained by this lemma for
  $0 \leq k \leq n$. Then the resulting {\v C}ech nerve projection
  factors through $Y \to X$.
\endofproof
\begin{proposition}
Let $X \in C \hookrightarrow \mathrm{Sh}_\infty(C) =: \mathbf{H}$ be an
object which is 
\begin{enumerate}
  \item representably paracompact, def. \ref{representaby paracompact};
  \item representably compact, def. \ref{representably compact}
\end{enumerate} 
then it distributes over sequential $\infty$-colimits $A_\bullet : I \to \mathrm{Sh}_\infty(C)$
of $n$-truncated objects for every $n \in \mathbb{N}$.
\label{RepresentablyCompactAndParacompactImpliesDistributivityOverFilteredColimits}
\end{proposition}
\proof
  Let $A_\bullet : I \to [C^{\mathrm{op}}, \mathrm{sSet}]$ be a presentation
  of a given sequential diagram in $\mathrm{Sh}_\infty(\mathrm{Mfd})$, such that it is  fibrant and
  cofibrant in $[I, [C^{\mathrm{op}}, \mathrm{sSet}]_{\mathrm{proj},\mathrm{loc}}]_{\mathrm{proj}}$.
  Note for later use that this implies in particular that
  \begin{itemize}
    \item The ordinary colimit $\lim\limits_{\longrightarrow_i} A_i \in [C^{\mathrm{op}}, \mathrm{sSet}]$
	is a homotopy colimit.
	\item Every $A_i$ is fibrant in $[C^{\mathrm{op}}, \mathrm{sSet}]_{\mathrm{proj}, \mathrm{loc}}$
	  and hence also in $[C^{\mathrm{op}}, \mathrm{sSet}]_{\mathrm{proj}}$.
	\item Every morphism $A_i \to A_j$ is (by example \ref{CotowerCofibrancy})
	 a cofibration in $[C^{\mathrm{op}}, \mathrm{sSet}]_{\mathrm{proj}, \mathrm{loc}}$,
	 hence in $[C^{\mathrm{op}}, \mathrm{sSet}]_{\mathrm{proj}}$, hence
	 in particular in $[C^{\mathrm{op}}, \mathrm{sSet}]_{\mathrm{inj}}$, hence
	 is over each $U\in C$ a monomorphism.
  \end{itemize}
  
  Observe that $\lim\limits_{\longrightarrow_i} A_i$ is still fibrant in $[C^{\mathrm{op}}, \mathrm{sSet}]_{\mathrm{proj}}$:
  since the colimit is taken in presheaves, it is computed objectwise, and since it is
  filtered, we may find the lift against horn inclusions (which are inclusions of 
  degreewise finite simplicial sets) at some stage in the colimit, where it exists by
  assumption that $A_\bullet$ is projectively fibrant, so that each $A_i$ is projectively
  fibrant in the local and hence in particular in the global model structure.
  
  Since $X$, being representable, is cofibrant in 
  $[C^{\mathrm{op}}, \mathrm{sSet}]_{\mathrm{proj},\mathrm{loc}}$, it 
  also follows 
  by this reasoning that the diagram
  $$
    \mathbf{H}(X, A_\bullet) : I \to \infty \mathrm{Grpd}
  $$ 
  is presented by 
  $$
    A_\bullet(X) : I \to \mathrm{sSet}
	\,.
  $$
  
  Since the functors
  $$
    \xymatrix{
    [I, [C^{\mathrm{op}}, \mathrm{sSet}]_{\mathrm{proj}, \mathrm{loc}}]_{\mathrm{proj}}
	  \ar[r]^{\mathrm{id}}
	  &
    [I,[C^{\mathrm{op}}, \mathrm{sSet}]_{\mathrm{proj}}]_{\mathrm{proj}}
	  \ar[r]^{\mathrm{id}}
	  &
	[I,[C^{\mathrm{op}}, \mathrm{sSet}]_{\mathrm{inj}}]_{\mathrm{proj}}
	  \ar[r]^{\mathrm{id}}
	  &
	[I,\mathrm{sSet}_{\mathrm{Quillen}}]_{\mathrm{proj}}
	}
  $$
  all preserve cofibrant objects, it follows that $A_\bullet(X)$ is cofibrant
  in $[I,\mathrm{sSet}_{\mathrm{Quillen}}]_{\mathrm{proj}}$. Therefore also
its ordinary colimit presents the corresponding $\infty$-colimit.

This means that the equivalence which we have to establish can be 
written in the form
$$
  \mathbb{R}\mathrm{Hom}(X, \lim\limits_{\longrightarrow_i} A_i)
  \simeq
  \lim\limits_{\longrightarrow_i} A_i(X)
  \,.
$$  
If here $\lim\limits_{\longrightarrow_i} A_i$ were fibrant in 
$[C^{\mathrm{op}}, \mathrm{sSet}]_{\mathrm{proj}, \mathrm{loc}}$, then the derived
hom on the left would be given by the simplicial mapping space and the equivalence would
hold trivially. So the remaining issue is now to deal with the fibrant replacement:
the $\infty$-sheafification of $\lim\limits_{\longrightarrow_i} A_i$.

We want to appeal to theorem 7.6 c) in \cite{dugger-hollander-isaksen} to 
compute the derived hom into this $\infty$-stackification by 
a colimit over hypercovers of the ordinary simplicial 
homs out of these hypercovers into $\lim\limits_{\longrightarrow_i} A_i$ itself.
To do so, we now argue that by the assumptions on $X$, we may in fact 
replace the hypercovers here with finite {\v C}ech covers.

So consider the colimit
$$
  \lim\limits_{\{U_\alpha \to X\}_{\mathrm{finite}}}
  [C^{\mathrm{op}}, \mathrm{sSet}](\check{C}(\{U_\alpha\}), \lim\limits_{\longrightarrow_i} A_i)
$$
over all finite covers of $X$. Since by representable compactness of $X$ these are cofinal in 
all covers of $X$, this is isomorphic to the colimit over all {\v C}ech covers
$$
  \cdots 
   = 
  \lim\limits_{\{U_\alpha \to X\}}
  [C^{\mathrm{op}}, \mathrm{sSet}](\check{C}(\{U_\alpha\}), \lim\limits_{\longrightarrow_i} A_i)
  \,.
$$
Next, by representable paracomopactness of $X$, the {\v C}ech covers in turn are cofinal in all
bounded hypercovers $Y \to X$, so that, furthermore, this is isomorphic to the colimit
over all bounded hypercovers
$$
  \cdots 
   = 
  \lim\limits_{Y \to X}
  [C^{\mathrm{op}}, \mathrm{sSet}](Y, \lim\limits_{\longrightarrow_i} A_i)
  \,.
$$
Finally, by the assumption that the $A_i$ are $n$-truncated, the colimit here
may equivalently be taken over all hypercovers. 

We now claim that the canonical morphism
$$
  \lim\limits_{\{U_\alpha \to X\}_{\mathrm{finite}}}
  [C^{\mathrm{op}}, \mathrm{sSet}](\check{C}(\{U_\alpha\}), \lim\limits_{\longrightarrow_i} A_i)
  \to 
  \mathbb{R}\mathrm{Hom}(X, \lim\limits_{\longrightarrow_i} A_i)
$$
is a weak equivalence.
Since the category of covers is filtered, we may first compute
homotopy groups and then take the colimit. With the above
isomorphisms, the statement is then
given by theorem 7.6 c) in \cite{dugger-hollander-isaksen}.

Now to conclude: since maps out of the finite Cech nerves pass through the filtered colimit, 
we have
$$
  \begin{aligned}
    \mathbb{R}\mathrm{Hom}(X, \lim\limits_{\longrightarrow_i} A_i)
	&
	\simeq
	\lim\limits_{\{U_\alpha \to X\}_{\mathrm{finite}}} 
	 [C^{\mathrm{op}}, \mathrm{sSet}](\check{C}(\{U_\alpha\}), \lim\limits_{\longrightarrow_i} A_i)
	 \\
	 &\simeq
	\lim\limits_{\{U_\alpha \to X\}_{\mathrm{finite}}} 
	\lim\limits_{\longrightarrow_i}
	 [C^{\mathrm{op}}, \mathrm{sSet}](\check{C}(\{U_\alpha\}),  A_i)
     \\
	 &\simeq
	 	\lim\limits_{\longrightarrow_i}
	\lim\limits_{\{U_\alpha \to X\}_{\mathrm{finite}}} 
	 [C^{\mathrm{op}}, \mathrm{sSet}](\check{C}(\{U_\alpha\}),  A_i)
	 \\
	 & \simeq
	 	\lim\limits_{\longrightarrow_i}
		A_i(X) 
  \end{aligned}
  \,.
$$
Here in the last step we used that each single $A_i$ is fibrant in 
$[C^{\mathrm{op}}, \mathrm{sSet}]_{\mathrm{proj}, \mathrm{loc}}$, so that 
for each $i \in I$
$$
  [C^{\mathrm{op}}, \mathrm{sSet}](X, A_i)
    \to
  [C^{\mathrm{op}}, \mathrm{sSet}](\check{C}(\{U_\alpha\}), A_i)
$$
is a weak equivalence. Moreover, the diagram
$[C^{\mathrm{op}}, \mathrm{sSet}](\check{C}(\{U_\alpha\}),  A_\bullet)$ in 
$\mathrm{sSet}$ is still projectively cofibrant, by example \ref{CotowerCofibrancy},
since all morphisms are cofibrations in $\mathrm{sSet}_{\mathrm{Quillen}}$,
and so the colimit in the second but last line
is still a homotopy colimit and thus preserves these weak equivalences.
\endofproof

\subsubsection{Homotopy}
 \label{Homotopy}
 \index{structures in a cohesive $\infty$-topos!homotopy}

\paragraph{General abstract}

\begin{definition}
  \label{HomotopyGroup}
  Let $\mathbf{H}$ an $\infty$-topos and $X \in \mathbf{H}$
  an object. For $n \in \mathbb{N}$ write
  $$
    (X^{(* \to \partial\Delta[n+1])} : X^{\Delta[n]} \to X) \in \mathbf{H}_{/X}
  $$
  for the cotensoring of $X$ by the point inclusion into the 
  simplicial $n$-sphere, regarded as an object in the slice of $\mathbf{H}$
  over $X$, def. \ref{Slice}. The \emph{$n$th homotopy group} of $X$ is the image
  of this under 0-truncation, prop. \ref{truncation}
  $$
    \pi_n(X) := \tau_0( X^{* \to \partial \Delta[n+1]} ) \in \tau_0 (\mathbf{H}_{/X})
	\,.
  $$
\end{definition}
This appears as def. 6.5.1.1 in \cite{Lurie}.
\begin{remark}
  Since truncation preserves finite products by prop. \ref{truncation}
  we have that $\pi_n(X)$ is indeed a group object in the 1-topos
  $\tau_0(\mathbf{})$ for $n \geq 1$ and
  is an abelian group object for $n \geq 2$.
\end{remark}
\begin{remark}
  For $\mathbf{H} = \infty \mathrm{Grpd} \simeq \mathrm{Top}$ 
  and $x : * \to X \in \infty \mathrm{Grpd}$
  a pointed object, we have for all $n \in \mathbb{N}$ that 
  $$
    \pi_n(X,x) := x^* \pi_n(X) \in \tau_0 \infty \mathrm{Grpd}_{/*} \simeq \mathrm{Set}
  $$
  is the $n$th homotopy group of $X$ at $x$ as traditionally defined. 
\end{remark}
In \cite{Lurie} this is remark 6.5.1.6.

\paragraph{Presentations}

(...)

\subsubsection{Connected objects}
 \label{ConnectedObjects}
 \index{structures in a cohesive $\infty$-topos!connected objects}

We discuss objects in an $\infty$-topos which are connected
or higher connected in that their first non-trivial homotopy group,
\ref{Homotopy}, is in some positive degree. 

In a local $\infty$-topos
and hence in particular in a cohesive $\infty$-topos, these
are precisely the \emph{deloopings} of \emph{group objects}, discussed below in 
\ref{StrucInftyGroups}. In a more general $\infty$-topos,
such as a slice of a cohesive $\infty$-topos, these are the 
(nonabelian/Giraud-)\emph{gerbes},
discussed below in \ref{StrucInftyGerbes}.

\paragraph{General abstract}
\label{Connected objects general}

\begin{definition}
  \label{ConnectedObject}
  Let $n \in \mathbb{Z}$, with $-1 \leq n$. 
  An object $X \in \mathbf{H}$ is called \emph{$n$-connected} if 
  \begin{enumerate}
    \item the terminal morphism $X \to *$ is an effective epimorphism, def. \ref{EffectiveEpimorphism};
	\item all categorical homotopy groups $\pi_k(X)$, def. \ref{HomotopyGroup}, 
	for $k \leq n$ are trivial.
  \end{enumerate}
  One also says
  \begin{itemize}
    \item \emph{inhabited} or \emph{well-supported} for (-1)-connected;
    \item \emph{connected} for 0-connected;
	\item \emph{simply connected} for 1-connected;
	\item \emph{$(n+1)$-connective} for $n$-connected.
  \end{itemize}
  A morphism $f : X \to Y$ in $\mathbf{H}$ is called $n$-connected if
  it is $n$-connected regarded as an object of $\mathbf{H}_{/Y}$. 
\end{definition}
This is def. 6.5.1.10 in \cite{Lurie}.
\begin{example}
  An object $X \in \infty \mathrm{Grpd} \simeq \mathrm{Top}$ is 
  $n$-connected precisely if it is $n$-connected in the traditional
  sense of higher connectedness of topological spaces.
  (A morphism in $\infty\mathrm{Grpd}$ is effective epi
  precisely if it induces an epimorphism on sets of connected
  components.)
\end{example}
\begin{example}
  For $C$ an $\infty$-site, a connected object in $\mathrm{Sh}_\infty(C)$
  may also be called an (``nonabelian'' or ``Giraud''-) 
  \emph{$\infty$-gerbe} over $C$. This we discuss below in
  \ref{StrucInftyGerbes}.
\end{example}
\begin{definition}
  \label{HomotopyDimension}
  An $\infty$-topos $\mathbf{H}$ has \emph{homotopy dimension} $n \in \mathbb{N}$
  if $n$ is the smallest number such that 
  every $(n-1)$-connected object $X \in \mathbf{H}$ admits a morphism
  $* \to X$ from the terminal object
\end{definition}
\begin{remark}
  A morphism $* \to X$ is a \emph{section} of the terminal
  geometric morphism. So in an $\infty$-topos of homotopy dimension $n$
  every $(n-1)$-connected object $X$ has a section. For such $X$ 
  the terminal geometric morphism is 
  therefore in fact a \emph{split epimorphism}.
\end{remark}
\begin{example}
  The trivial $\infty$-topos $\mathbf{H} = *$ is, up to equivalence, the unique 
  $\infty$-topos of homotopy dimension 0.
\end{example}
This is example 7.2.1.2 in \cite{Lurie}.
\begin{proposition}
  \label{HomotopyDimensionByGlobalSections}
   An $\infty$-topos $\mathbf{H}$ has homotopy dimension $\leq n$ precisely if the global 
   section geometric morphism $\Gamma : \mathbf{H} \to \infty \mathrm{Grpd}$,
   def. \ref{Terminalgeometricmorphism}, sends $(n-1)$-connected morphisms
   to $(-1)$-connected morphisms (effective epimorphisms).
\end{proposition}
\proof
  This is essentially lemma 7.2.1.7 in \cite{Lurie}. 
  The proof there shows a bit more, even.
\endofproof
\begin{proposition}
  \label{HomotopyDimensionOfLocalInfinityTopos}
  A local $\infty$-topos, def. \ref{LocalInfinityTopos}, has homotopy dimension 0.
\end{proposition}
\proof
  By prop. \ref{HomotopyDimensionByGlobalSections} it is sufficient
  to show that effective epimorphisms are sent to effective epimorphisms.
  Since for a local $\infty$-topos the global section functor
  is a left adjoint, it preserves not only the $\infty$-limits
  involved in the characterization of effective epimorphisms, def. \ref{EffectiveEpimorphism}, 
  but also 
  the $\infty$-colimits.
\endofproof
\begin{remark}
  In particular an $\infty$-presheaf $\infty$-topos over an $\infty$-site
  with a terminal object is local. For this special case the
  statement of prop. \ref{HomotopyDimensionOfLocalInfinityTopos} 
  is example. 7.2.1.2 in \cite{Lurie}, the argument above being effectively the 
  same as the one given there.
\end{remark}
\begin{corollary}
  \label{HomotopyDimensionOfCohesiveInfinityTopos}
  A cohesive $\infty$-topos, def. \ref{CohesiveInfinToposDefinition}, has homotopy dimension 0.
\end{corollary}
\proof
  By definition, a cohesive $\infty$-topos is in particular a local 
  $\infty$-topos.
\endofproof

In an ordinary topos every morphism has a unique factorization
into an epimorphism followed by a monomorphism, the 
\emph{image factorization}.
$$
  \xymatrix{
    X \ar[rr]^f
    \ar[dr]_{\mathrm{epi}}	
	&& A
	\\
	&
	\mathrm{im}(f)
	\ar[ur]_{\mathrm{mono}}
  }
  \,.
$$
In an $\infty$-topos this notion generalizes to a tower
of factorizations.

\begin{proposition}  
  \label{HomotopyImageFactorization}
  In an $\infty$-topos $\mathbf{H}$ for any $-2 \leq k \leq \infty$,
  every morphism $f : X \to Y$ admits a factorization
$$
  \xymatrix{
    X \ar[rr]^f
    \ar[dr]_{}	
	&& A
	\\
	&
	\mathrm{im}_{k+1}(f)
	\ar[ur]_{}
  }
$$
into a $k$-connected morphism, def. \ref{ConnectedObject} 
followed by a $k$-truncated morphism, def. \ref{truncated object},
and the space of choices of such factorizations is contractible.
\end{proposition}
This is \cite{Lurie}, example 5.2.8.18.
\begin{remark}
  For $k = -1$ this is the immediate generalization 
  of the (epi,mono) factorization system in ordinary toposes. 
  In particular, the 0-image factorization of a
  morphism between 0-truncated objects is the ordinary image 
  factorization.
  
  For $k = 1$ this is the generalization of the 
  (essentially surhective and full, faithful) factorization 
  system for functors between groupoids.
\end{remark}

\paragraph{Presentations}
\label{Connected objects presentations}

We discuss presentations of connected and \emph{pointed}
connected objects in an $\infty$-topos 
by presheaves of pointed or reduced simplicial sets.

\medskip

\begin{observation}
  Under the presentation $\infty \mathrm{Grpd} \simeq (\mathrm{sSet}_{\mathrm{Quillen}})^\circ$,
  a Kan complex $X \in \mathrm{sSet}$ presents an $n$-connected $\infty$-groupoid
  precisely if
  \begin{enumerate}  
    \item $X$ is inhabited (not empty);
	\item all simplicial homotopy groups of $X$ in degree $k \leq n$ are trivial.
  \end{enumerate}
\end{observation}
\begin{definition}
  For $n \in \mathbb{N}$ a simplicial set $X \in \mathrm{sSet}$
  is \emph{$n$-reduced} if it has a single $k$-simplex for all $k \leq n$, 
  hence if its $n$-skeleton is the point
  $$
    \mathrm{sk}_n X = *
	\,.
  $$
  For \emph{0-reduced} we also just say \emph{reduced}. 
  Write
  $$
    \mathrm{sSet}_n \hookrightarrow \mathrm{sSet}
  $$
  for the full subcategory of $n$-reduced simplicial sets.
\end{definition}
\begin{proposition}
  \label{PropertiesOfTheReducedsSetInclusion}
  The $n$-reduced simplicial sets form a reflective subcategory
  $$
    \xymatrix{
	  \mathrm{sSet}_n
	  \ar@{<-}@<+4pt>[r]^{\mathrm{red}_n}
	  \ar@{^{(}->}@<-4pt>[r]
	  &
	  \mathrm{sSet}
	}
  $$
  of that of simplicial sets, 
  where the reflector $\mathrm{red}_n$ identifies all the $n$-vertices
  of a given simplicial set, in other words $\mathrm{red}_n(X) = X/\mathrm{sk}_n X$ 
  for $X$ a simplicial set.
  
  The inclusion $\mathrm{sSet}_n \hookrightarrow \mathrm{sSet}$ 
  uniquely factors through the forgetful functor $\mathrm{sSet}^{*/} \to \mathrm{sSet}$
  from pointed simplicial sets, and that factorization is co-reflective
  $$
    \xymatrix{
	  \mathrm{sSet}_n
	  \ar@{^{(}->}@<+4pt>[r]^{}
	  \ar@{<-}@<-4pt>[r]_{E_{n+1}}
	  &
	  \mathrm{sSet}^{*/}
	}
	\,.
  $$
  Here the coreflector $E_{n+1}$ sends a pointed simplicial 
  set $* \stackrel{x}{\to} X$ to the sub-object $E_{n+1}(X,x)$ 
  -- the \emph{$(n+1)$-Eilenberg subcomplex} (e.g. def. 8.3 in \cite{May}) --
  of cells whose $n$-faces coincide with the base point, hence to the fiber
  \[
   \xymatrix{ 
     E_{n+1}(X,x) \ar[r] \ar[d] & X \ar[d] 
	  \\ 
     \{\ast\} \ar[r] & \mathrm{cosk}_n X } 
  \]
  of the projection to the $n$-coskeleton.
  
  For $(* \to X) \in \mathrm{sSet}^{*/}$ such that $X \in \mathrm{sSet}$ is 
  Kan fibrant and $n$-connected,
  the counit $E_{n+1}(X,*) \to X$ is a homotopy equivalence.
\end{proposition}
The last statement appears for instance as part of theorem 8.4  in \cite{May}.
\begin{proposition}
  \label{PresentationOfPointedConnectedObjectsByPresheavesOfReducedsSets}
  Let $C$ be a site with a terminal object and let 
  $\mathbf{H} := \mathrm{Sh}_\infty(C)$.
  Then under the presentation
  $\mathbf{H} \simeq ([C^{\mathrm{op}}, \mathrm{sSet}]_{\mathrm{proj}, \mathrm{loc}})^\circ$ 
  every pointed $n$-connected object in $\mathbf{H}$ is presented by a
  presheaf of $n$-reduced simplicial sets, under the canonical inclusion
  $[C^{\mathrm{op}}, \mathrm{sSet}_n]
	\hookrightarrow
    [C^{\mathrm{op}}, \mathrm{sSet}]$.
\end{proposition}
\proof
  Let $X \in [C^{\mathrm{op}}, \mathrm{sSet}]$ be a simplicial presheaf presenting
  the given object. 
  Then its objectwise Kan fibrant replacement
  $\mathrm{Ex}^\infty X$ is still a presentation, fibrant in the 
  global projective model structure. 
  Since the terminal object in $\mathbf{H}$ is presented by the 
  terminal simplicial presheaf and since by assumption on $C$
  this is representable
  and hence cofibrant in the projective model structure, the point inclusion is
  presented by a morphism of simplicial presheaves $* \to \mathrm{Ex}^\infty X$,
  hence by a presheaf of pointed simplicial sets 
  $(* \to \mathrm{Ex}^\infty X) \in [C^{\mathrm{op}}, \mathrm{sSet}^{*/}]$. 
  So with observation
  \ref{PropertiesOfTheReducedsSetInclusion}
  we obtain the presheaf of $n$-reduced simplicial sets
  $$
    E_{n+1}(\mathrm{Ex}^\infty X, *) 
	 \in 
	[C^{\mathrm{op}}, \mathrm{sSet}_n] 
	  \hookrightarrow
	[C^{\mathrm{op}}, \mathrm{sSet}]
  $$
  and the inclusion $E_{n+1}(\mathrm{Ex}^\infty X,*) \to \mathrm{Ex}^\infty X$
  is a global weak equivalence, hence a local weak equivalence, hence
  exhibits $E_{n+1}(\mathrm{Ex}^\infty X,*)$ as another presentation of the
  object in question.
\endofproof
\begin{proposition}
  \label{ModelStructureOnReducedSimplicialSets}
  The category $\mathrm{sSet}_{0}$ of reduced simplicial sets 
  carries a left proper combinatorial model category structure
  whose weak equivalences and cofibrations are those in $\mathrm{sSet}_{\mathrm{Quillen}}$
  under the inclusion $\mathrm{sSet}_{0} \hookrightarrow \mathrm{sSet}$.
\end{proposition}
\proof
  The existence of the model structure itself is prop. V.6.2 in \cite{GoerssJardine}.
  That this is left proper combinatorial follows for instance from 
  prop. A.2.6.13 in \cite{Lurie}, taking the set $C_0$ there to be
  $$
    C_0 := \{ \mathrm{red}(\Lambda^k[n] \to \Delta[n])\}_{n \in \mathbb{N}, 0 \leq k \leq n}
	\,,
  $$
  the image under 
  of the horn inclusions (the generating cofibrations in 
  $\mathrm{sSet}_{\mathrm{Quillen}}$) 
  under the left adjoint, from observation \ref{PropertiesOfTheReducedsSetInclusion},
  to the inclusion functor.
\endofproof
\begin{lemma}
  \label{HomotopyPropertyOfInclusionOfReducedsSets}
  Under the inclusion $\mathrm{sSet}_0 \to \mathrm{sSet}$
  a fibration with respect to the model structure from 
  prop. \ref{ModelStructureOnReducedSimplicialSets}
  maps to a fibration in $\mathrm{sSet}_{\mathrm{Quillen}}$ precisely if
  it has the right lifting property against the morphism
  $(* \to S^1) := \mathrm{red}(\Delta[0] \to \Delta[1])$.
  
  In particular it maps fibrant objects to fibrant objects.
\end{lemma}
The first statement appears as lemma 6.6. in \cite{GoerssJardine}. 
The second (an immediate consequence) as corollary 6.8.
\begin{proposition}
  The adjunction
  $$
    \xymatrix{
	  \mathrm{sSet}_0
	  \ar@{^{(}->}@<+4pt>[r]^{i}
	  \ar@{<-}@<-4pt>[r]_{E_1}
	  &
	  \mathrm{sSet}_{\mathrm{Quillen}}^{*/}
	}
  $$
  from observation \ref{Connected objects presentations}
  is a Quillen adjunction between the model structure form 
  prop. \ref{ModelStructureOnReducedSimplicialSets}
  and the co-slice model structure, prop. \ref{SliceModelStructure},
  of $\mathrm{sSet}_{\mathrm{Quillen}}$ under the point. 
  This presents the 
  full inclusion
  $$
    \infty \mathrm{Grpd}^{*/}_{\geq 1} \hookrightarrow
	\infty \mathrm{Grpd}^{*/}
  $$
  of connected pointed $\infty$-groupoids into all pointed $\infty$-groupoids.
\end{proposition}
\proof
  It is clear that the inclusion preserves cofibrations 
  and acyclic cofibrations, in fact all weak equivalences.
  Since the point is necessarily cofibrant in $\mathrm{sSet}_{\mathrm{Quillen}}$, 
  the model structure on the 
  right is by prop. \ref{SliceModelPresentsInfinitySlicing} 
  indeed a presentation of $\infty \mathrm{Grpd}^{*/}$.
  
  We claim now that the derived $\infty$-adjunction of this 
  Quillen adjunction presents a homotopy full and faithful inclusion whose
  essential image consists of the connected pointed objects.
  For homotopy full- and faithfulness it is 
  sufficient to show that for the derived functors there is a natural weak equivalence
  $$
    \mathrm{id} \simeq \mathbb{R}E_1\circ \mathbb{L}i
	\,.
  $$
  This is the case, because by prop. \ref{HomotopyPropertyOfInclusionOfReducedsSets} 
  the composite derived functors are computed by
  the composite ordinary functors precomposed with a fibrant replacement 
  functor $P$, so that
  we have a natural morphism
  $$
    X \stackrel{\simeq}{\to} P X = E_1 \circ i (P X) \simeq 
	(\mathbb{R}E_1)\circ (\mathbb{L}i) (X)
    \,.	
  $$
  Hence $\mathbb{L} i$ is homotopy full-and faithful and 
  by prop. \ref{PresentationOfPointedConnectedObjectsByPresheavesOfReducedsSets}
  its essential image consists of the connected pointed objects. 
\endofproof

\subsubsection{Groupoids}
\label{StrucInftyGroupoids}

In any $\infty$-topos $\mathbf{H}$ we may consider groupoids \emph{internal}
to $\mathbf{H}$, in the sense of internal category theory 
(as exposed for instance in the introduction of \cite{LurieGoodwillie}). 

Such a \emph{groupoid object} 
$\mathcal{G}$ in $\mathbf{H}$ is an $\mathbf{H}$-object $\mathcal{G}_0$ ``of $\mathcal{G}$-objects''
together with an $\mathbf{H}$-object $\mathcal{G}_1$ ``of $\mathcal{G}$-morphisms''
equipped with source and target assigning morphisms $s,t : \mathcal{G}_1 \to \mathcal{G}_0$,
an identity-assigning morphism $i : \mathcal{G}_0 \to \mathcal{G}_1$ and a composition
morphism $\mathcal{G}_1 \times_{\mathcal{G}_0} \mathcal{G}_1 \to \mathcal{G}_1$
that all satisfy the axioms of a groupoid (unitalness, associativity, existence of
inverses) up to coherent homotopy in $\mathbf{H}$. One way to formalize what it
means for these axioms to hold up to coherent homotopy is the following.

One notes that
ordinary groupoids, i.e. groupoid objects internal to $\mathrm{Set}$, are 
characterized by the fact that their nerves are simplicial objects 
$\mathcal{G}_\bullet : \Delta^{\mathrm{op}} \to \mathrm{Set}$ 
in $\mathrm{Set}$ such that all groupoidal Segal maps 
(see def. \ref{GroupoidObject} below) are isomorphisms. This turns out
to be a characterization that makes sense generally internal to higher categories: 
a groupoid object in $\mathbf{H}$ is an $\infty$-functor 
$\mathcal{G} : \Delta^{\mathrm{op}} \to \mathbf{H}$ such that all groupoidal 
Segal morphisms are equivalences in $\mathbf{H}$.
This defines an $\infty$-category $\mathrm{Grpd}(\mathbf{H})$
of groupoid objects in $\mathbf{H}$.

Here a subtlety arises that is the source of a lot of interesting structure
in higher topos theory: by the discussion in \ref{InfinityToposes}
the very objects of $\mathbf{H}$ are already to be regarded
as ``structured $\infty$-groupoids'' themselves. Indeed, there is a 
full embedding $\mathrm{const} : \mathbf{H} \hookrightarrow \mathrm{Grpd}(\mathbf{H})$
that forms constant simplicial objects and thus regards every object $X \in \mathbf{H}$
as a groupoid object which, even though it has a trivial object of morphisms, already
has a structured $\infty$-groupoid of objects. This embedding is in fact 
reflective, with the reflector given by forming the $\infty$-colimit
over a simplicial diagram
$$
  \xymatrix{
    \mathbf{H}
      \ar@{<-}@<+4pt>[rr]^{\lim\limits_{\longrightarrow}}
	  \ar@{^{(}->}@<-4pt>[rr]_{\mathrm{const}}
	  &&
	  \mathrm{Grpd}(\mathbf{H})
  }
  \,.
$$
For $\mathcal{G}$ a groupoid object in $\mathbf{H}$, the object 
$\lim\limits_{\longrightarrow} \mathcal{G}_\bullet$ in $\mathbf{H}$ 
may be thought of as the 
$\infty$-groupoid obtained from ``gluing together the object of objects of
$\mathcal{G}$ along the object of morphisms of $\mathcal{G}$''. 
This idea that groupoid objects in an $\infty$-topos are 
like structured $\infty$-groupoids together with gluing information 
is formalized by the theorem that groupoid objects in 
$\mathbf{H}$ are equivalent to the \emph{effective epimorphisms} 
$\xymatrix{Y \ar@{->>}[r] & X}$ in $\mathbf{H}$, the intrinsic notion of
\emph{cover} (of $X$ by $Y$) in $\mathbf{H}$. The effective epimorphism / cover
corresponding to a groupoid object $\mathcal{G}$ is the colimiting cocone
$\xymatrix{\mathcal{G}_0 \ar@{->>}[r] & \lim\limits_{\longrightarrow} \mathcal{G}_\bullet}$.
This state of affairs is a fundamental property of $\infty$-toposes,
and as such part of the \emph{$\infty$-Giraud axioms} def. 
\ref{GiraudRezkLurieAxioms}.

The following statement refines
the third $\infty$-Giraud axiom, Definition \ref{GiraudRezkLurieAxioms}.
\begin{theorem}
   There is a natural equivalence of $\infty$-categories
  $$
    \mathrm{Grpd}(\mathbf{H})
     \simeq
    (\mathbf{H}^{\Delta[1]})_{\mathrm{eff}}
    \,,
  $$ 
  where $(\mathbf{H}^{\Delta[1]})_{\mathrm{eff}}$ 
  is the full sub-$\infty$-category of the 
  arrow category $\mathbf{H}^{\Delta[1]}$ 
  of $\mathbf{H}$ on the effective epimorphisms, Definition \ref{EffectiveEpi}.
  \label{NaturalThirdGiraud}
  \label{1EpisAreEquivalentToGroupoidObjects}
\end{theorem}
This appears below Corollary 6.2.3.5 in \cite{Lurie}.

\paragraph{General abstract}

We briefly recall the notion of \emph{groupoid objects} in an $\infty$-topos
from \cite{Lurie} with a note on how this notion axiomatizes that of 
$\infty$-groupoids with geometric structure and \emph{equipped with an atlas} 
(a choice of \emph{object of objects}) in \ref{GroupoidsAtlases}. 
Then we discuss the notion of the $\infty$-group of \emph{bisections} associated
to such a choice of atlas in \ref{GroupoidsBisections} and how
these arrange to \emph{Lie-Rinehart pairs} 
describing $\infty$-groupoids with atlases. 
Finally, by the 1-image factorization
every morphism in an $\infty$-topos induces an atlas on its 1-image $\infty$-groupoid.
This universal construction we identify as a generalization of the traditional
notion of Atiyah groupoids, which we discuss in \ref{AtiyahGroupoids}.

\begin{itemize}
  \item \ref{GroupoidsAtlases} -- Atlases;
  \item \ref{GroupoidsBisections} -- Group of bisections;
  \item \ref{AtiyahGroupoids} -- Atiyah groupoids.
\end{itemize}

\subparagraph{Atlases}
\label{GroupoidsAtlases}

On the one hand, \emph{every} object in an $\infty$-topos $\mathbf{H}$ may be thought of as being
an $\infty$-groupoid equipped with certain structure, notably with geometric
or cohesive structure. On the other hand, traditional notions of geometric groupoids,
such as \emph{Lie groupoids} (discussed in detail in \ref{SmoothStrucGroupoids} below), 
typically involve (often implicitly) more data: the additional choice of an \emph{atlas},
def. \ref{atlas}. 
An extreme example is the \emph{pair groupoid} on some space $X$, which 
we discuss as example \ref{PairGroupoid} below. As just an 
object of $\mathbf{H}$ every pair groupoid is trivial: it is equivalent to the point;
but what traditonal literature really means (often implicitly) by the pair groupoid
is the groupoid-with-atlas $X \to *$ with $X$ regarded as an atlas of the point.

Abstractly, an atlas on an $\infty$-groupoid in $\mathbf{H}$ is just a 1-epimorphism
in $\mathbf{H}$. Here we discuss this notion of \emph{$\infty$-groupoids with atlas}.
This gives us occasion to put one of the Giraud-Rezk-Lurie axioms, 
def. \ref{GiraudRezkLurieAxioms}, into a higher geometric context and to establish some 
perspetive on $\infty$-groupoids which is crucial in the succeeding discussion.

\medskip

\begin{definition}
  \label{GroupoidObject}
  A \emph{groupoid object} in an $\infty$-topos $\mathbf{H}$ is 
  a simplicial object
  $$
    \mathcal{G} : \Delta^{\mathrm{op}} \to \mathbf{H}
  $$
  such that all its groupoidal Segal maps are equivalences: for every
  $n \in \mathbb{N}$
  and every partition $[k] \cup [k'] \to [n]$ into two subsets with
  exactly one joint element $\{*\} = [k] \cap [k'] $, the canonical diagram
  $$
    \xymatrix{
      \mathcal{G}[n] \ar[r] \ar[d] & \mathcal{G}[k] \ar[d]
      \\
      \mathcal{G}[k'] \ar[r] & \mathcal{G}[{*}]
    }
  $$
  is an $\infty$-pullback diagram.

  Write
  $$
    \mathrm{Grpd}(\mathbf{H}) \subset \mathrm{Func}(\Delta^{\mathrm{op}}, \mathbf{H})
  $$
  for the full subcategory of the $\infty$-category of simplicial
  objects in $\mathbf{H}$ on the groupoid objects. 
\end{definition}
This is def. 6.1.2.7 of \cite{Lurie}, using prop. 6.1.2.6.
\begin{example}
  For $Y \to X$ any morphism in $\mathbf{H}$, there is a
  groupoid object $\check{C}(Y \to X)$ which in degree $n$ is the 
  $(n+1)$-fold $\infty$-fiber product of $Y$ over $X$ with itself
  $$
    \check{C}(Y \to X) : [n] \mapsto Y^{\times^{n+1}_X}
  $$
\end{example}
This appears in \cite{Lurie} as prop. 6.1.2.11.
The following statement strengthens the third $\infty$-Giraud axiom
of def. \ref{GiraudRezkLurieAxioms}.
\begin{theorem}
  \index{topos!Giraud-Lurie axioms}
  \label{InfinityGiraudAxioms}
  \label{GroupoidObjectsAreEffective}
  \label{ColimitsAreUniversal}
  In an $\infty$-topos $\mathbf{H}$ we have
  \begin{enumerate}
    \item
  Every groupoid object in $\mathbf{H}$ is
  \emph{effective}: the canonical morphism 
  $\mathcal{G}_0 \to \lim\limits_{\longrightarrow} \mathcal{G}_\bullet$
  is an effective epimorphism, and $\mathcal{G}$ is 
  equivalent to the {\v C}ech nerve
  of this effective epimorphism.
  
  Moreover, this extends to a natural equivalence of $\infty$-categories
  $$
    \mathrm{Grpd}(\mathbf{H})
     \simeq
    (\mathbf{H}^{\Delta[1]})_{\mathrm{eff}}
    \,,
  $$ 
  where on the right we have the full sub-$\infty$-category of the 
  arrow category of $\mathbf{H}$ on the effective epimorphisms.
 
  \item
    The $\infty$-pullback along any morphism preserves
    $\infty$-colimits
    $$
      \xymatrix{
         \lim\limits_{\to i} f^* P_i 
              \ar@{}[r]|\simeq
           & f^* \lim\limits_{\to_i} P_i
         \ar[d]  
         \ar[r]
         & \lim\limits_{\to_i} P_i
         \ar[d]
         \\
         & Y \ar[r]^f &  X
      }
    $$
  \end{enumerate}
\end{theorem}
This are two of the \emph{Giraud-Rezk-Lurie axioms}, def. \ref{GiraudRezkLurieAxioms},
that characterize $\infty$-toposes. (The equivalence of 
$\infty$-categories in the first point follows with 
the remark below corollary 6.2.3.5 of \cite{Lurie}.)

\begin{remark}
  If geometric structure is understood (as in a cohesive $\infty$-topos),
  there is a slight ambiguity in the word \emph{groupoid} as usually
  used:
  in one sense every object of an $\infty$-topos itself is already
  a \emph{parameterized $\infty$-groupoid} 
  (an $\infty$-sheaf of $\infty$-groupoids, def. \ref{SheafInfinityTopos}). 
  However, for instance the
  literature on
  \emph{Lie groupoid} theory often (and often implicitly) takes a choice of \emph{object of objects}
  as part of the data of a Lie groupoid. For instance the notion of 
  \emph{group of bisection} of a Lie groupoid $X$ or of its associated \emph{Lie algebroid}
  both require that the inclusion of a manifold of objects is specified, a morphism
  $X_0 \to X$. This choice is genuine extra structure on $X$, as it is not in general 
  preserved by equivalences on $X$. The main technical requirement on this choice is that
  it indeed captures ``all objects'' of the groupoid, up to equivalence. 
  One often says that the inclusion has to be an \emph{atlas} of $X$. In the general
  abstract terms of $\infty$-topos theory this means simply that $X_0 \to X$ is a
  \emph{1-epimorphism}, remark \ref{atlas}.
  
  In view of this we interpret theorem \ref{InfinityGiraudAxioms}: if
  we follow remark \ref{atlas} and call a 1-epimorphism in an $\infty$-topos 
  an \emph{atlas} of its codomain parameterized $\infty$-groupoid, then the 
  \emph{groupoid objects} of def. \ref{GroupoidObject} are really the
  ``parameterized $\infty$-groupoids equipped with a choice of atlas''.
  (In traditional geometric groupoid theory the atlas (the domain object) is usually 
  required to be 0-truncated, and this is often the choice of interest, 
  also in applications of higher geometry, but in general every 1-epimorphism
  qualifies as an \emph{atlas} in this sense.)
  \label{GroupoidObjectsAreGroupoidsWithAnAtlas}
\end{remark}
With this understood, the following definitions axiomatize and generalize 
standard constructions in traditional geometric groupoid theory. That
they indeed reduce to these traditional notions is shown below in \ref{SmoothStrucGroupoids}.

\begin{example}
  For $G \in \mathrm{Grp}(\mathbf{H})$ an $\infty$-group, \ref{StrucInftyGroups},
  its delooping $\mathbf{B}G$ is essentially uniquely pointed, and this point 
  inclusion $\xymatrix{{*} \ar@{->>}[r] & \mathbf{B}G}$ 
  is a 1-epimorphism (for instance by prop. \ref{EffectiveEpiIsEpiOn0Truncation}).
  Hence this is the canonical incarnation of the delooping of $G$ as an
  $\infty$-groupoid with atlas. In terms of this we may read 
  theorem \ref{DeloopingTheorem} as saying that
  \emph{$\infty$-groups are equivalent to their delooping $\infty$-groupoids with canonical atlases}.
\end{example}
\begin{example}
  By def. \ref{ConnectedObject} an object $X \in \mathbf{H}$ is called \emph{inhabited} 
  if the canonical morphism to the terminal object is a 1-epimorphism. 
  Therefore for $X$ inhabited the map $\xymatrix{X \ar@{->>}[r] & {*}}$
  may be regarded as an $\infty$-groupoid with atlas. 
  To see what this means consider its {\v C}ech nerve, which is 
  of course of the form
  $$
    \left(
	  \xymatrix{
	    \cdots
	    \ar@<+5pt>[r]
	    \ar@<+0pt>[r]
	    \ar@<-5pt>[r]
	    &
	    X \times X
	    \ar@<+3pt>[r]^{p_1}
	    \ar@<-3pt>[r]_{p_2}
	    &
	    X
	  }
	\right)
	\in
	\mathbf{H}^{\Delta^{\mathrm{op}}}
	\,.
  $$
  This is a groupoid object whose objects are the points of $X$, whose
  morphisms are ordered pairs of points in $X$, and where composition is
  given in the evident way. This is what in the literature is known as the \emph{pair groupoid}
  of $X$. 
  $$
    \mathrm{Pair}(X)
	:=
	\left(\xymatrix{X \ar@{->>}[r] & {*}}\right)
	\in 
	(\mathbf{H}^{\Delta^1})_{\mathrm{eff}}
	\simeq
	\mathrm{Grpd}(\mathbf{H})
	\,.
  $$
  \label{PairGroupoid}
\end{example}
Almost trivial as it may seem, the pair groupoid plays an important role for
instance in the theory of Atiyah groupoids, discussed below in \ref{AtiyahGroupoids}.

As these examples show, often it is more convenient to work with the atlas
than with the groupoid object that it equivalently corresponds to. The 
following propositions shows how to compute $\infty$-limits in this perspective.
\begin{proposition}
  An $\infty$-limit of a diagram in in $(\mathbf{H}^{\Delta^1})_{\mathrm{eff}}$ is given by
  the (-1)-truncation projection of the $\infty$-limit of the underlying diagram
  in $\mathbf{H}^{\Delta^1}$. Hence if $A : J \to (\mathbf{H}^{\Delta^1})_{\mathrm{eff}}$
  is a diagram with underlying diagrams $X := \partial_1\circ A $ and $Y := \partial_2 \circ A$
  in $\mathbf{H}$, then
  $$
   \underset{\longleftarrow}{\lim}_j A_j
   \simeq
   \left(
   \underset{\longleftarrow}{\lim}_j X_j
   \to 
   \mathrm{im}_1\left(
     \xymatrix{
	   \underset{\longleftarrow}{\lim}_j X_j
	   \ar[r]
	   &
	   \underset{\longleftarrow}{\lim}_j Y_j	   
	 }
   \right)
   \right)
   \,.
  $$
  \label{LimitsInEffectiveEpimorphisms}
\end{proposition}
\proof
 One checks the defining universal property
 by the orthogonal factorization system of prop. \ref{EpiMonoFactorizationSystem}.
\endofproof

\subparagraph{Group of Bisections}
\label{GroupoidsBisections}
\index{group!of bisections}

We discuss here the description of $\infty$-groupoids $X \in \mathbf{H}$ equipped with \emph{atlases}
$\xymatrix{X_0 \ar@{->>}[r] & X}$ in terms of their $\infty$-groups $\mathbf{Aut}_X(X_0)$ of autoequivalences of
$X_0$ over $X$. In the case that $\mathbf{H}$ is the $\infty$-topos of smooth cohesion
described below in \ref{SmoothInfgrpds} and for the example that $X$ is presented by a
traditional \emph{Lie groupoid} this is the group which is traditionally known as the \emph{group of bisections}
of $X$, this we discuss in \ref{SmoothStrucGroupOfBisections} below. 
Since this is a good descriptive term also in the general case, we here generally
speak of $\mathbf{Aut}_X(X_0)$ as the \emph{$\infty$-group of bisections}.

Due to their special construction, groups of bisections have special properties. In the 
traditional literature these are best known after Lie differentiation: again for $X$ a Lie
groupoid, the pair $(C^\infty(X_0), \mathrm{Lie}(\mathbf{Aut}_X(X_0)))$ consisting of the
associative algebra of smooth functions on $X_0$ and the Lie algebra of the group of 
bisections is known as the \emph{Lie-Rinehart algebra pair} associated with the groupoid. 
It enjoys the special property that each of the two algebras is equipped with an action
of the other algebra in a compatible way. This is an equivalent way of encoding the 
\emph{Lie algebroid} associated with the Lie groupoid $X$.

\medskip

\begin{definition}
  For $X_\bullet \in \mathbf{H}^{\Delta^{\mathrm{op}}}$ a groupoid object in an 
  $\infty$-topos, def. \ref{GroupoidObject}, with 
  $\phi_X : \xymatrix{X_0 \ar@{->>}[r] & X}$ the corresponding 1-epimorphism
  by theorem \ref{InfinityGiraudAxioms} (the \emph{atlas} 
  by remark \ref{GroupoidObjectsAreGroupoidsWithAnAtlas}), we say that the
  \emph{group of bisections} $\mathbf{BisSect}(\phi_X) \in \mathrm{Grp}(\mathbf{H})$
  of $X_\bullet$
  (also written $\mathbf{BiSect}_X(X_0)$ if the morphism $p_X$ is understood) is the 
  relative automorphism group,  def. \ref{HValuedAutomorphismGroup}, of $X_0$ over $X$:
  $$
    \mathbf{BiSect}_X(X_0)
	:=
	\mathbf{Aut}_{\mathbf{H}}(p_X)
	:=
	\underset{X}{\prod} \mathbf{Aut}(p_X)
	\,.
  $$
  \label{GroupOfBisections}
  \index{structures in a cohesive $\infty$-topos!groupoids!group of bisections}
\end{definition}
\begin{remark}
  We discuss how this general abstract notion reduces to that of the 
  group of bisections of a Lie groupoid as traditionally defined below in 
  prop. \ref{GroupOfBisectionsOfLieGroupoidFromGeneralAbstract}.
\end{remark}
\begin{definition}
  The \emph{atlas automorphisms} $\mathbf{AtlasAut}_X(X_0)$
  of the atlas $\phi_X : \xymatrix{X_0 \ar@{->>}[r] & X}$
  is the 1-image of the morphism $p_{X}$ 
  of def. \ref{CanonicalMorphismFromSliceInternalHomToBaseInternalHom},
  hence the factorization of $p_{X}$ as
  $$
	\xymatrix{
	  \mathbf{BiSect}_X(X_0)
	  \ar@{->>}[r]^p
	  &
	  \mathbf{AtlasAut}_X(X_0)
	  \ar@{^{(}->}[r]
	  &
	  \mathbf{Aut}(X)
	}
	\,.
  $$
  \label{MapFromGroupOfBisectionsToAutomorphismsOfX0}
\end{definition}
\begin{proposition}
  For $X_\bullet \in \mathbf{H}^{\Delta^{\mathrm{op}}}$ a groupoid object in an 
  $\infty$-topos, def. \ref{GroupoidObject}, with 
  $\phi_X : \xymatrix{X_0 \ar@{->>}[r] & X}$ the corresponding 1-epimorphism
  by theorem \ref{InfinityGiraudAxioms},  we have a fiber sequence
  $$
	\xymatrix{
	  \Omega_{\phi_X} [X_0, X]
	  \ar@{^{(}->}[r]
	  &
	  \mathbf{BiSect}_X(X_0)
	  \ar@{->>}[r]^-p
	  &
	  \mathbf{AtlasAut}_X(X_0)
	}
  $$
  in $\mathrm{Grp}(\mathbf{H})$
  which exhibits $\mathbf{BiSect}_X(X_0)$ as an $\infty$-group extension
  of $\mathbf{AtlasAut}_X(X_0)$ by the automorphism $\infty$-group of the 
  atlas $X_0$ inside $X$.
  \label{TheAtiyahSequenceOfGroups}
\end{proposition}
\proof
  Since $\mathbf{AtlasAut}_X(X_0)$ is by definition the 1-image of the morphism 
  $p : \mathbf{BiSect}_X(X_0) \to \mathbf{Aut}(X)$ the statement is equivalent to
  the diagram
  $$
	\xymatrix{
	  \Omega_\nabla [X, X]
	  \ar@{^{(}->}[r]
	  &
	  \mathbf{BisSect}_X(X_0)
	  \ar@{->>}[r]^-p
	  &
	  \mathbf{Aut}(X)
	}
  $$
  being a fiber sequence, since, by the pasting law, with the bottom square in the following
  diagram being an $\infty$-pullback, the top square is precisely so if the outer rectangle is.  
  $$
    \xymatrix{
	  \Omega_\nabla [X_0, X]
	  \ar[r] \ar[d] & 
	  \mathbf{BiSect}_X(X_0) \ar@{->>}[d]^{p}
	  \\
	  {*} \ar[d] \ar[r] & \mathbf{AtlasAut}_X(X_0) \ar@{^{(}->}[d]
	  \\
	  {*} \ar[r] & \mathbf{Aut}(X)
	}
  $$
  That the outer rectangle is an $\infty$-pullback is the statement of 
  prop. \ref{FiberOfMapFromSliceMappingToBaseMapping}.
 \endofproof
\begin{remark}
  The sequence of prop. \ref{TheAtiyahSequenceOfGroups} 
  is actually the sequence of bisection groups induced by a fiber sequence of
  $\infty$-groupoids with atlases: the generalized \emph{Atiyah sequence}.
  This we discuss below in \ref{AtiyahGroupoids}.   
\end{remark}

\begin{example}
 For $X \in \mathbf{H}$ inhabited, the group of bisections of the 
 pair groupoid $\mathrm{Pair}(X)$, example \ref{PairGroupoid}, is 
 canonically equivalent to $\mathbf{Aut}(X)$:
 $$
   \mathbf{BiSect}(\mathrm{Pair}(X))
   \simeq
   \mathbf{Aut}(X)
   \,.
 $$
 \label{GroupOfBisectionsOfPairGroupoid}
\end{example}
\begin{example}
 For $X \in \mathbf{H} \stackrel{\mathrm{const}}{\hookrightarrow} \mathrm{Grpd}(\mathbf{H})$
 the constant groupoid object on $X$, its group of bisections is the trivial group
 $$
   \mathbf{BiSect}(\mathrm{const} X) \simeq {*}
   \,.
 $$
 \label{GroupOfBisectionsOfConstantGroupoid}
\end{example}
\proof
  By example \ref{TerminalObjectInSlice} 
  the identity morphism on $X$ is the terminal object in the slice
  $\infty$-topos $\mathbf{H}_{/X}$.
\endofproof

\subparagraph{Atiyah groupoids}
\label{AtiyahGroupoids}
\label{HigherAtiyahGroupoidsInIntroduction}
\index{Atiyah groupoid}

By the 1-image factorization, def. \ref{nImage}, every morphism in an $\infty$-topos
induces an atlas for an $\infty$-groupoid, in the sense discussed above 
in \ref{GroupoidsAtlases}.
If the codomain is a pointed connected object, hence of the form $\mathbf{B}G$
for some $\infty$-group $G$, then we may equivalently think of 
this $\infty$-groupoid with atlas as associated
to the corresponding $G$-principal $\infty$-bundle over the domain, 
discussed below in \ref{PrincipalInfBundle}. One finds that this
construction generalizes the traditional notion of the Lie groupoid which 
Lie integrates the \emph{Atiyah Lie algebroid} of a smooth principal bundle
(this traditional example we discuss in \ref{SmoothStrucAtiyahGroupoids} below).
Therefore we generally speak of \emph{Atiyah $\infty$-groupoids}.

A special case this construction relevant for codomains that are 
moduli $\infty$-stacks specifically for \emph{differential} cocycles
are \emph{Courant groupoids} which we discuss below in \ref{CourantGroupoids}.

{\bf Note.} This section partly refers to definitions and results in the theory
of principal $\infty$-bundles which we discuss only below in \ref{PrincipalInfBundle}.
We nevertheless group the discussion of Atiyah groupoids here since one of the key
aspects of their general definition in $\infty$-toposes is that they apply
much more generally than just to principal $\infty$-bundles.

\medskip

A fundamental construction in the traditional theory of $G$-principal bundles $P \to X$ 
is that of the corresponding \emph{Atiyah Lie algebroid} and that of the Lie groupoid
which integrates it, which we will call the \emph{Atiyah groupoid} $\mathrm{At}(P)$.
In words this is the Lie groupoid whose manifold of objects is $X$, and whose
morphisms between two points are the $G$-equivariant maps between the fibers of $P$
over these points. Observing that a $G$-equivariant map between two $G$-torsors over the
point is fixed by its image on any one point, this groupoid is usually written as
on the left of
$$
  \begin{array}{ccc}
    \mathrm{At}(P) &\to& \mathrm{Pair}(X)
	\\
	= && =
	\\
	 \left(
	   \raisebox{20pt}{
       \xymatrix{
        (P \times P)/_{\mathrm{diag}} G 
	    \ar@<+4pt>[d]
	    \ar@{<-}@<+0pt>[d]
	    \ar@<-4pt>[d]
	    \\
	    X
       }}
    \right)
	&&
	 \left(
	   \raisebox{20pt}{
       \xymatrix{
        X \times X 
	    \ar@<+4pt>[d]
	    \ar@{<-}@<+0pt>[d]
	    \ar@<-4pt>[d]
	    \\
	    X
       }
	   }
    \right)
  \end{array}
  \,.
$$
There is a conceptual simplification to this construction when expressed in 
terms of the smooth moduli stack $\mathbf{B}G$ of $G$-principal bundles
(in the smooth model for cohesion, discussed below in \ref{SmoothInfgrpds}): if 
$\nabla^0 : X \to \mathbf{B}G$ is the map which modulates $P \to X$, then 
\begin{proposition}
The space
of morphisms of $\mathrm{At}(P)$ is naturally identified with the \emph{homotopy fiber product}
of $\nabla^0$ with itself:
$$
  (P \times P)/_{\mathrm{diag}}G
  \simeq
  X \underset{\mathbf{B}G}{\times} X
  \,.
$$
Moreover, the canonical \emph{atlas} of the Atiyah groupoid, given by the
canonical inclusion $p_{\mathrm{At}(P)} : \xymatrix{X \ar@{->>}[r] & \mathrm{At}(P)}$, is equivalently the 
homotopy-colimiting cocone under the full {\v C}ech nerve of the classifying map $\nabla^0$:
$$
  \xymatrix{
    \ar@{..}[r]
    &
    X \underset{\mathbf{B}G}{\times} X \underset{\mathbf{B}G}{\times}
	\ar@<+12pt>[r]
	\ar@{<-}@<+8pt>[r]
	\ar@<+4pt>[r]
	\ar@{<-}@<-0pt>[r]
	\ar@<-4pt>[r]
	\ar@{<-}@<-8pt>[r]
	\ar@<-12pt>[r]
    &
    X \underset{\mathbf{B}G}{\times}X
	\ar@<+4pt>[r]
	\ar@{<-}@<-0pt>[r]
	\ar@<-4pt>[r]
	&
	X
	\ar@{->>}[rr]^-{p_{\mathrm{At}(P)}}
	&&
	\left(\underset{\longrightarrow}{\lim}_n X^{\times^{n+1}_{\mathbf{B}G}}\right)
	\simeq
	\mathrm{At}(P)
  }
  \,.
$$
\label{TraditionalAtiyahGroupoidIsHocolimOverNerveOfModulatingMap}
\end{proposition}
This is by direct verification, the details of this example are discussed below in 
\ref{SmoothStrucAtiyahGroupoids}.
In terms of groups of bisections the above proposition 
\ref{TraditionalAtiyahGroupoidIsHocolimOverNerveOfModulatingMap} becomes:
\begin{proposition}
  The Atiyah groupoid $\mathrm{At}(P)$ of a smooth $G$-principal bundle $P\to X$ 
  is the Lie groupoid which is universal
  with the property that its group of bisections is naturally equivalent to the
  group of automorphisms of the modulating map $\nabla^0$ of $P\to X$ in the slice:
  $$
    \begin{array}{ccc}
	  \mathbf{BiSect}(\mathrm{At}(P))
	  &\simeq&
	  \mathbf{Aut}_{\mathbf{H}}(\nabla^0)
	  \\
	  = && =
	  \\
      \left\{
        \raisebox{20pt}{
        \xymatrix{
	      X \ar[dr]_{p_{\mathrm{At}(P)}}^{\ }="t" \ar[rr]^{\phi}|\simeq_{\ }="s" 
		   && X \ar[dl]^{p_{\mathrm{At}(P)}}
	     \\
	     & \mathrm{At}(P)
	     \ar@{=>} "s"; "t"
	    }
	    }
      \right\}
	  &&
      \left\{
        \raisebox{20pt}{
        \xymatrix{
	      X \ar[dr]_{\nabla^0}^{\ }="t" \ar[rr]^{\phi}|\simeq_{\ }="s" 
		   && X \ar[dl]^{\nabla^0}
	     \\
	     & \mathbf{B}G
	     \ar@{=>} "s"; "t"
	    }
	    }
      \right\}
	\end{array}
	\,.
  $$
  \label{AtiyahBisectionsAreAutomorphismOfModulatingMapInSlice}
\end{proposition}
In terms of 1-image factorizations we may naturally understand 
proposition \ref{TraditionalAtiyahGroupoidIsHocolimOverNerveOfModulatingMap}
as saying that (the atlas of) the Atiyah groupoid provides the essentially
unique factorization
$$
  \nabla^0 : \xymatrix{
    X \ar@{->>}[rr]^{p_{\mathrm{At}(P)}}
	&&
	\mathrm{At}(P)
	\ar@{^{(}->}[rr]
	&&
	\mathbf{B}G
  }
$$
of the modulating map $\nabla^0$ of $P \to X$ by a 1-epimorphism of stacks followed by a
1-monomorphism, namely the \emph{first relative Postnikov stage} of $\nabla^0$, 
in the context of smooth stacks. As for traditional relative Postnikov theory
in traditional homotopy theory, this characterizes $\mathrm{At}(P)$ uniquely as
receiving an epimorphism on smooth connected components from $X$ (the atlas $p_{\mathrm{At}(P)}$),
while at the same time having a \emph{fully faithful embedding} into $\mathbf{B}G$.
This being fully faithful directly implies that the components of any natural transformation
from $\nabla^0$ to itself necessarily factor through this fully faithful inclusion:
$$
  \begin{array}{ccc}
    \left\{
      \raisebox{30pt}{
      \xymatrix{
        X \ar[ddr]_{\nabla^0}^{\ }="t" \ar[rr]^\phi|\simeq_{\ }="s"  && X \ar[ddl]^{\nabla^0}
	    \\
	    \\
	    & \mathbf{B}G
	    \ar@{=>} "s"; "t"
     }
    }
  \right\}
   &\simeq&
   \left\{
     \raisebox{33pt}{
	   \xymatrix{
	     X \ar[ddr]_{\nabla^0} \ar@{->>}[dr]|{p}^{\ }="t" \ar[rr]^\phi|\simeq_{\ }="s" 
		 && X
		 \ar[ddl]^{\nabla^0} \ar@{->>}[dl]|{p}
		 \\
		 & \mathrm{At}(P)
		  \ar@{^{(}->}[d]
		 \\
		 & \mathbf{B}G
		 \ar@{=>} "s"; "t"
	   }
	 }
   \right\}
  \end{array}
  \,.
$$
This relation translates to a proof of prop. \ref{AtiyahBisectionsAreAutomorphismOfModulatingMapInSlice}.

\medskip

This discussion of Atiyah groupoids of traditional $G$-principal bundles generalizes
directly now to bundles in an $\infty$-topos.

\begin{definition}
  Let $\phi : X \to \mathbf{F}$ a morphism in $\mathbf{H}$.
  We say that its 1-image projection, def. \ref{nImage},
  $$
    \xymatrix{
	  X \ar@{->>}[r] & \mathrm{im}_1(\phi)
	}
	\,,
  $$
  regarded as an $\infty$-groupoid $\mathrm{im}_1(\phi)$
  with atlas $X$ by remark \ref{atlas}, 
  is the \emph{Atiyah groupoid} $At(\phi) \in \mathrm{Epi}_1(\mathbf{H})$ of $\phi$.
  \label{AtiyahGroupoid}
\end{definition}
Here for the direct generalization of the traditional notion of Atiyah groupoids
we set $\mathbf{F} = \mathbf{B}G$ the delooping of some $\infty$-group. But
the definition and many of its uses does not depend on this restriction. 
An exception os the following fact, which generalizes a standard theorem
about Atiyah groupoids known from textbooks on differential geometry.
\begin{proposition}
  For $G \in \mathrm{Grp}(\mathbf{H})$ an $\infty$-group, every $G$-principal $\infty$-bundle
  $P \to X$ in $\mathbf{H}$, def. \ref{principalbundle}, 
  over an inhabited object $X$, def. \ref{ConnectedObject},
  is equivalently the source-fiber of a transitive 
  higher groupoid $\mathcal{G} \in \mathrm{Grpd}(\mathbf{H})$ with 
  vertex $\infty$-group $G$. Here in particular we can set $\mathcal{G} = \mathrm{At}(P)$.
\end{proposition}
\proof
  For $P \to X$ a $G$-principal $\infty$-bundle, write $g : X \to \mathbf{B}G$
  for the map that modulates it by theorem \ref{PrincipalInfinityBundleClassification}.
  Then the outer rectangle of
  $$
   \raisebox{20pt}{
    \xymatrix{
	  P \ar@{->>}[r] \ar[d] & {*} \ar[r]^\simeq \ar@{->>}[d]^{x} & {*} \ar@{->>}[d]
	  \\
	  X
	  \ar@{->>}[r]
	  \ar@/_1pc/[rr]_g
	  &
	  \mathrm{At}(P)
	  \ar@{^{(}->}[r]
	  &
	  \mathbf{B}G
	}
	}
  $$
  is an $\infty$-pullback by that theorem \ref{PrincipalInfinityBundleClassification}.
  Also the right sub-square is an $\infty$-pullback (for any global point $x \in X$) 
  because by $\infty$-pullback stability
  of 1-epimorphisms (prop. \ref{1EpimorphismsArePreservedByPullback}) 
  and 1-monomorphisms (prop. \ref{EpiMonoFactorizationIsStable}), 
  the top right morphism is a 1-monomorphism from
  an inhabited object to the terminal object, hence is not just a 1-mono but also a 1-epi
  and hence is an equivalence.
  Now by the pasting law for $\infty$-pullbacks, prop. \ref{PastingLawForPullbacks}, 
  also the left sub-square is an $\infty$-pullback
  and this exhibits $P$ as the source fiber of $\mathrm{At}(P)$ over $x \in X$.
\endofproof

\begin{proposition}
  For $\phi : X \to \mathbf{F}$ a morphism, there is a canonical equivalence
  $$
    \mathbf{BiSect}(\mathrm{At}(\phi))
	\simeq
	\mathbf{Aut}_{\mathbf{H}}(\phi)
  $$
  between the $\infty$-group of bisections, def. \ref{GroupOfBisections}, 
  of the higher Atiyah groupoid of $\phi$, def. \ref{AtiyahGroupoid},
  and the $\mathbf{H}$-valued automorphism $\infty$-group of $\phi$
  
  Moreover, the $\infty$-group of bisections of the higher Atiyah $\infty$-groupoid
  sits in a homotopy fiber sequence of $\infty$-groups of the form
$$
  \xymatrix{
    \Omega_{\phi} [X, \mathbf{F}]
	\ar[r]
	&
	\mathbf{BiSect}(\mathbf{At}(\phi))
	\ar[r]
	\ar@{}[d]|\simeq
	&
	\mathbf{Aut}(X)
	\\
	& \mathbf{Aut}_{\mathbf{H}}(\phi)
  }
  \,,
$$
where on the right we have the canonical forgetful map.
  \label{SliceAutomorphismsInvariantUnderImageFactorization}
\end{proposition}
\proof
  This is the restriction of the statement of 
  prop. \ref{SliceHomComparedAlongAMonomorphism} to those endomorphisms that 
  are equivalences.
\endofproof

\begin{definition}[Atiyah sequence]
  For $\phi : X \to \mathbf{B}G$ a cocycle, 
  write
  $$
    \xymatrix{
	   \mathrm{At}(\phi)
	   \ar[r]^p
	   &
	   \mathrm{Pair}(X)
	}
  $$
  for the morphism of groupoid objects to the \emph{pair groupoid} of $X$, 
  example \ref{PairGroupoid}, given by the canonical map of atlases
  $$
    \raisebox{20pt}{
    \xymatrix{
	  X \ar[r]^{\mathrm{id}} \ar[d] & X \ar[d]
	  \\
	  \mathrm{im}_1(\phi)
	  \ar[r]
	  &
	  {*}
	}
	}
	\,.
  $$
  We say that the $\infty$-fiber sequence of this morphism over $X$ 
  $$
    \xymatrix{
	  \mathrm{ad}(\phi)
	  \ar[r]
	  &
	  \mathrm{At}(\phi)
	  \ar[r]
	  &
	  \mathrm{Pair}(X)
	}
	\,,
  $$
  is the \emph{Atiyah sequence} of $\phi$, hence the sequence
  given by the $\infty$-pullback diagram
  $$
    \raisebox{20pt}{
    \xymatrix{
	  \mathrm{ad}(\phi) \ar[r] \ar[d] & X \ar[d]
	  \\
	  \mathrm{At}(\phi) \ar[r]^p & \mathrm{Pair}(X)
	}
	}
	\,.
  $$
\end{definition}
\begin{proposition}
  Given $\phi : X \to \mathbf{B}G$, the induced sequence of groups of bisections, 
  def. \ref{GroupOfBisections}, is the sequence of prop. \ref{TheAtiyahSequenceOfGroups}.
\end{proposition}
\proof
  By prop. \ref{GroupOfBisectionsOfPairGroupoid} and prop. \ref{SliceAutomorphismsInvariantUnderImageFactorization}
  the morphism of groupoid objects $\mathrm{At}(\phi) \to \mathrm{Pair}(X)$ induces
  the morphism of groups of bisections 
  $\mathbf{Aut}(\phi) \to \mathbf{Aut}(X)$. Therefore it remains to show that 
  $\mathrm{ad}(\phi) \to \mathrm{At}(\phi)$ is as claimed.

  By prop. \ref{LimitsInEffectiveEpimorphisms} we obtain $\mathrm{ad}(\phi)$
  as the 1-image factorization of the limit in $\mathbf{H}^{\Delta^1}$ over
  $$
    \raisebox{20pt}{
    \xymatrix{
	  X \ar[r]^{\mathrm{id}} \ar@{->>}[d] & X \ar@{->>}[d] \ar@{<-}[r] 
	  & X   \ar@{->>}[d]^{\mathrm{id}}
	  \\
	  \mathrm{im}_1(\phi)
	  \ar[r]
	  &
	  {*}
	  \ar@{<-}[r] & X
	}
	}\,,
  $$
  hence the 1-image factorization of the diagonal
  $\xymatrix{ X \ar@{->>}[r] & X \times \mathrm{im}_1(\phi)}$.
  Moreover by prop. \ref{SliceAutomorphismsInvariantUnderImageFactorization}
  the group of bisections of this image factorization is equivalently that of the 
  morphism itself.  Now a bisection of the diagonal, hence a diagram
  $$
    \raisebox{20pt}{
    \xymatrix{
	  X \ar[rr]^\simeq_{\ }="s" \ar[dr]^{\ }="t" && X \ar[dl]
	  \\
	  & X \times \mathrm{im}_1(\phi)
	  \ar@{=>} "s"; "t"
	}}
  $$
  is equivalently a pair of a diagrams of the form
  $$
    \raisebox{20pt}{
    \xymatrix{
	  X \ar[rr]^f|\simeq_{\ }="s" \ar[dr]_{\mathrm{id}}^{\ }="t" && X \ar[dl]^{\mathrm{id}}
	  \\
	  & X 
	  \ar@{=>} "s"; "t"
	}
	}
	\;\;\;\;
	\,,
	\;\;\;\;
	\raisebox{20pt}{
    \xymatrix{
	  X \ar[rr]^f|\simeq_{\ }="s" \ar[dr]_{}^{\ }="t" && X \ar[dl]^{}
	  \\
	  & \mathrm{im}_1(\phi)
	  \ar@{=>} "s"; "t"
	}
	}
  $$
  that share the top horizontal morphism, as indicated. 
  By example \ref{GroupOfBisectionsOfConstantGroupoid} the $\infty$-groupoid of diagrams 
  as on the left is contractible, hence up to essentially unique equivalence we have $f = \mathrm{id}$.
  This reduces the diagram on the right to an automorphism of $\phi$, as claimed. 
\endofproof

The Atiyah groupoid acts on sections of the corresponding bundle and its
associated bundles:
\begin{definition}
  For $G \in \mathrm{Grp}(\mathbf{H})$ an $\infty$-group, for
  $P \to X$ a $G$-principal $\infty$-bundle modulated by 
  a map $g : X \to \mathbf{B}G$, and for $\rho : V/\!/G \to \mathbf{B}G$
  an action of $G$ on some $V \in \mathbf{H}$, write
  $$
    (P \times_G V)/\!/\mathrm{At}(P) \to \mathrm{At}(P)
  $$
  for the $\infty$-pullback of $\rho$ along the defining 1-monomorphism from the 
  Atiyah groupoid of $P$. Then by the pasting law, prop. \ref{PastingLawForPullbacks},
  and by the characterization of the universal $\rho$-associated bundle, 
  prop. \ref{UniversalAssociatedBundle}, we have an $\infty$-pullback square
  as on the left of the following diagram:
  $$
    \raisebox{20pt}{
    \xymatrix{  
	  P \times_G V 
	    \ar[r] \ar[d]
	  & (P \times_G V)/\!/\mathrm{At}(P) 
	    \ar[r] \ar[d] 
	  & V/\!/G
	  \ar[d]^\rho
	  \\
	  X 
	     \ar@{->>}[r] \ar@/_1.5pc/[rr]_g 
	   & \mathrm{At}(P) 
	     \ar@{^{(}->}[r] 
		& \mathbf{B}G
	}
	}\,.
	$$
	This exhibits
	$(P \times_G V)/\!/\mathrm{At}(P)$ as a groupoid action
	of $\mathrm{At}(P)$ on the associated $V$-fiber bundle $P \times_G V \to X$.
	This we call the \emph{canonical Atiyah-groupoid action on sections}.
	\label{ActionOfAtiyahGroupoidOnSections}
\end{definition}






\subsubsection{Groups}
\label{StrucInftyGroups}
\index{structures in a cohesive $\infty$-topos!cohesive $\infty$-groups}

Every $\infty$-topos $\mathbf{H}$ 
comes with a notion of \emph{$\infty$-group objects} that generalizes the
ordinary notion of group objects in a topos as well as that of
grouplike $A_\infty$-spaces in $\mathrm{Top} \simeq \infty \mathrm{Grpd}$
\cite{StasheffSpaces}.
Operations of \emph{looping} and \emph{delooping} identify $\infty$-group
objects with pointed connected objects. If moreover $\mathbf{H}$ is cohesive
then it follows that every connected object is canonically pointed, and hence
every connected object uniquely corresponds to an $\infty$-group object.

This section to a large extent collects and reviews general facts about 
$\infty$-group objects in $\infty$-toposes from \cite{Lurie} and \cite{LurieAlgebra}.
We add some observations that we need later on.

\paragraph{General abstract}
\label{Infinity groups -- general abstract}

\begin{definition}
  Write 
  \begin{itemize}
    \item $\mathbf{H}^{*/}$ for the
  $\infty$-category of pointed objects in $\mathbf{H}$;
    \item  $\mathbf{H}_{\geq 1}$ for 
    for the full sub-$\infty$-category of $\mathbf{H}$ on the 
    connected objects;
    \item
      $\mathbf{H}^{*/}_{\geq 1}$ for the full sub-$\infty$-category
      of the pointed and connected objects.
   \end{itemize}
\end{definition}
\begin{definition}
  \label{loop space object}
  \label{LoopSpaceObject}
  \index{fiber sequence!loop space object} 
  \index{loop space object}
  Write 
  $$
    \Omega : \mathbf{H}^{*/} \to \mathbf{H}
  $$
  for the $\infty$-functor that sends a pointed object $* \to X$
  to its \emph{loop space object}: the $\infty$-pullback
  $$
   \raisebox{20pt}{
   \xymatrix{
      \Omega X \ar[r]\ar[d] & {*} \ar[d]
      \\
      {*} \ar[r] & X
   }
   }
   \,.
  $$
\end{definition}
\begin{definition}
  An \emph{$\infty$-group} in $\mathbf{H}$
  is an $A_\infty$-algebra $G$ in $\mathbf{H}$ such that
  $\pi_0(G)$ is a group object. 
  Write
  $\mathrm{Grp}(\mathbf{H})$ for the $\infty$-category
  of $\infty$-groups in $\mathbf{H}$.
\end{definition}
This is def. 5.1.3.2 in \cite{LurieAlgebra}, together with
remark 5.1.3.3.
\begin{theorem}
  \label{DeloopingTheorem}
  \label{delooping}
  Every loop space object canonically has the structure of an 
  $\infty$-group, and this construction extends to an 
  $\infty$-functor
  $$
    \Omega : \mathbf{H}^{*/} \to \mathrm{Grp}(\mathbf{H})
    \,.
  $$
  This constitutes an equivalence of $\infty$-categories
  $$
    (\Omega \dashv \mathbf{B})
    :
    \xymatrix{
      \mathrm{Grp}(\mathbf{H})
       \ar@{<-}@<+5pt>[r]^<<<<{\Omega}
       \ar@<-5pt>[r]_<<<<{\mathbf{B}}^\simeq
      &
      \mathbf{H}^{*/}_{\geq 1}
    }  
  $$  
  of $\infty$-groups with connected pointed objects in $\mathbf{H}$.
\end{theorem}
This is lemma 7.2.2.1 in \cite{Lurie}. 
(See also theorem 5.1.3.6 of \cite{LurieAlgebra} 
where this is the equivalence denoted $\phi_0$ in the proof.)
\begin{definition}
We call the inverse 
$\mathbf{B} : \mathrm{Grp}(\mathbf{H}) \to \mathbf{H}^{*/}_{\geq 1}$
the \emph{delooping} functor of $\mathbf{H}$. By convenient abuse
of notation we write $\mathbf{B}$ also for the composite
$\mathbf{B} : \infty \mathrm{Grpd}(\mathbf{H}) \to \mathbf{H}^{*/}_{\geq 1} 
\to \mathbf{H}$ with the functor that forgets the basepoint and the
connectedness.
\end{definition}
\begin{remark}
  While by prop. \ref{PointlikeProperty}
  every connected object in a cohesive $\infty$-topos has a unique point,
  nevertheless the homotopy type of the full hom-$\infty$-groupoid 
  $\mathbf{H}^{*/}(\mathbf{B}G, \mathbf{B}H)$ 
  of pointed objects in general differs
  from the hom $\infty$-groupoid $\mathbf{H}(\mathbf{B}G, \mathbf{B}H)$
  of the underlying unpointed objects. 
  
  For instance let $\mathbf{H} := \infty \mathrm{Grpd}$ and let $G$ be 
  an ordinary group, regarded as a group object in $\infty \mathrm{Grpd}$.
  Then $\mathbf{H}^{*/}(\mathbf{B}G, \mathbf{B}G) \simeq \mathrm{Aut}(G)$
  is the ordinary automorphism group of $G$, but
  $\mathbf{H}(\mathbf{B}G, \mathbf{B}G) = \mathrm{AUT}(G)$ is the 
  automorphism 2-group, example \ref{Automorphism2GroupCrossedModule}.
\end{remark}
The more deloopings an $\infty$-group admits, the ``more abelian'' it is:
\begin{definition}
  A \emph{braided} $\infty$-group in $\mathbf{H}$ is 
  an $\infty$-group $G \in \mathrm{Grp}(\mathbf{H})$
  equipped with the following equivalent additional structures:
  \begin{enumerate}
	  \item a lift of the groupal $A_\infty \simeq E_1$-algebra structure to an $E_2$-algebra structure;
	  \item the structure of an $\infty$-group on the delooping $\mathbf{B}G$;
	  \item a choice of double delooping $\mathbf{B}^2 G$.
  \end{enumerate}
  \label{BraidedInfinityGroup}
  \label{BraidedGroups}
\end{definition}
\begin{definition}
  An \emph{abelian} $\infty$-group in $\mathbf{H}$ is 
  an $\infty$-group $G \in \mathrm{Grp}(\mathbf{H})$
  equipped with the following equivalent additional structures:
  \begin{enumerate}
	  \item a lift of the groupal $A_\infty \simeq E_1$-algebra structure to an $E_\infty$-algebra structure;
	  \item coinductively: a choice of abelian $\infty$-group structure on its delooping $\mathbf{B}G$.
  \end{enumerate}
  \label{AbelianInfinityGroup}
\end{definition}
\begin{proposition}
  \label{InfinityGroupObjectsAsGroupoidObjects}
  $\infty$-groups $G$ in $\mathbf{H}$ are equivalently 
  those groupoid objects, def. \ref{GroupoidObject}, $\mathcal{G}$ in $\mathbf{H}$
  for which $\mathcal{G}_0 \simeq *$. 
\end{proposition}
This is the statement of the compound equivalence
$\phi_3\phi_2\phi_1$ in the proof of theorem 5.1.3.6 in 
\cite{LurieAlgebra}.
\begin{remark}
  \label{PointIntoBGIsEffectiveEpimorphism}
  This means that for $G$ an $\infty$-group object 
  the {\v C}ech nerve extension of its delooping fiber
  sequence $G \to * \to \mathbf{B}G$ is the simplicial 
  object
  $$
    \xymatrix{
       \cdots
       \ar@<+6pt>[r] \ar@<+2pt>[r] \ar@<-2pt>[r] \ar@<-6pt>[r]
       &
       G \times G
       \ar@<+4pt>[r] 
       \ar[r] 
        \ar@<-4pt>[r] 
         & 
         G 
         \ar@<+2pt>[r] \ar@<-2pt>[r] 
       & 
       {*}
       \ar@{->>}[r]
       &
       \mathbf{B}G
    }
  $$
  that exhibits $G$ as a groupoid object over $*$.
  In particular it means that for $G$ an $\infty$-group, the
  essentially unique morphism $* \to \mathbf{B}G$
  is an effective epimorphism.
  \end{remark}

\begin{definition}
  For $f : Y \to Z$ any morphism in $\mathbf{H}$
  and $z : * \to Z$ a point, the \emph{$\infty$-fiber} or
  \emph{homotopy fiber} of $f$ over this point is the $\infty$-pullback 
  $ X := {*} \times_Z Y$
  $$
    \xymatrix{
        X \ar[r] \ar[d]& {*} \ar[d]
        \\
        Y \ar[r]^f & Z
    }
    \,.
  $$
\end{definition}
\begin{observation}
  Suppose that also $Y$ is pointed and $f$ is a morphism of pointed objects.
  Then the $\infty$-fiber of an $\infty$-fiber is the loop object of the base.
\end{observation}
This means that we have a diagram
  $$
    \xymatrix{
        \Omega_z Z  \ar[d] \ar[r] & X \ar[r] \ar[d]& {*} \ar[d]
        \\
        {*} \ar[r] & Y \ar[r]^f & Z
    }
    \,.
  $$
where the outer rectangle is an $\infty$-pullback if the left square is an 
$\infty$-pullback. This follows from the pasting law prop. \ref{PastingLawForPullbacks}.

\paragraph{Presentations}
\label{InfinityGroupPresentations}

We discuss presentations of the notion of $\infty$-groups, 
\ref{Infinity groups -- general abstract}, by simplicial groups in 
a category with weak equivalences.

\medskip

\begin{definition}
  \label{BarWAsCompositeWithTotal}
  One writes $\Wbar$ for the composite functor from simplicial
  groups to simplicial sets given by
  $$
    \Wbar
	:
    \xymatrix{
       [\Delta^{\mathrm{op}}, \mathrm{Grpd}]
	   \ar[r]^{[\Delta^{\mathrm{op}}, \mathbf{B}]}
	   &
 	  [\Delta^{\mathrm{op}}, \mathrm{Grpd}]
	  \ar[r]^<<<<<{[\Delta^{\mathrm{op}}, N]}
	  &
	  [\Delta^{\mathrm{op}}, \mathrm{sSet}]
	  \ar[r]^<<<<<{T}
	  &
	 \mathrm{sSet}
	}
	\,,
  $$
  where
  \begin{itemize}
     \item  $[\Delta^{\mathrm{op}},\mathbf{B}] : [\Delta^{\mathrm{op}}, \mathrm{Grp}] \to 
  [\Delta^{\mathrm{op}}, \mathrm{Grpd}]$  is the functor from simplicial groups
  to simplicial groupoids that sends degreewise a group to the corresponding one-object 
  groupoid;
    \item 
	  $	 T : [\Delta^{\mathrm{op}}, \mathrm{sSet}] \to \mathrm{sSet}$
	  is the total simplicial set functor, def. \ref{TotalSimplicialSetAndTotalDecalage}.
  \end{itemize}
\end{definition}
This simplicial delooping $\Wbar$ was originally introduced in components in 
\cite{EilenbergMacLane}, now a classical construction. 
The above formulation is due to \cite{Duskin}, see lemma 15 in \cite{Stevenson2}.
\begin{remark}
  This functor takes values in \emph{reduced} simplicial sets
  $\mathrm{sSet}_{\geq 1} \hookrightarrow \mathrm{sSet}$, those with precisely
  one vertex.
\end{remark}
\begin{remark}
  For $G$ a simplicial group, the simplicial set $\Wbar G$ is,
  by corollary \ref{SimplicialHomotopyColimitByCodiagonal}, the
  homotopy colimit over a simplicial diagram in simplicial sets. 
  Below in \ref{Principal infinity-bundles presentations} we see that 
  this simplicial diagram is that presenting the groupoid object 
  $*/\!/G$ which is the action groupoid of $G$ acting trivially on the point.
\end{remark}
\begin{proposition}
  \label{ModelStructureOnSimplicialGroups}
  The category $\mathrm{sGrpd}$ of simplicial groups carries a cofibrantly
  generated model structure for which the fibrations and the weak equivalences
  are those of $\mathrm{sSet}_{\mathrm{Quillen}}$ under the forgetful functor
  $\mathrm{sGrpd} \to \mathrm{sSet}$.
\end{proposition}
\proof
  This is theorem 2.3 in \cite{GoerssJardine}. Since model structure is
  therefore transferred along the forgetful functor, it inherits 
  generating (acyclic) cofibrations from those of $\mathrm{sSet}_{\mathrm{Quillen}}$.
\endofproof
\begin{theorem}
  \label{SimplicialLoopingQuillenEquivalence}
  The functor $\Wbar$ is the right adjoint of a Quillen equivalence
$$
  (L \dashv \Wbar) 
    :
   \xymatrix{ 
    \mathrm{sGrp} 
      \ar@<-3pt>[r]_{\Wbar}
      \ar@{<-}@<+3pt>[r]^{L}
      &
    \mathrm{sSet}_{\geq 1}
   }
  \,,
$$
with respect to the model structures of prop. \ref{ModelStructureOnSimplicialGroups}
and prop. \ref{ModelStructureOnReducedSimplicialSets}.
In particular 
\begin{itemize}
  \item the adjunction unit is a weak equivalence 
$$
  Y \stackrel{\simeq}{\to} \Wbar L Y  
$$
for every $Y \in \mathrm{sSet}_0 \hookrightarrow \mathrm{sSet}_{\mathrm{Quillen}}$ 
\item  $\Wbar L Y$ is always a Kan complex.
\end{itemize}
\end{theorem}
This is discussed for instance in chapter V of \cite{GoerssJardine}.
A new proof is given in \cite{Stevenson2}.
\begin{definition}
  \label{WGToWbarG}
  For $G$ a simplicial group, write
  $$
    W G \to \Wbar G
  $$
  for the d{\'e}calage, def. \ref{Decalage}, on $\Wbar G$.
\end{definition}
This characterization by d{\'e}calage of the object going by the classical name $W G$ 
is made fairly explicit on p. 85 of \cite{Duskin}. The fully explicit statement is in
\cite{RobertsStevenson}.
\begin{proposition}
  \label{WGToWbarGIsFibrationResolution}
  The morphism $W G \to \Wbar G$ is a Kan fibration resolution of the point inclusion
  ${*} \to \Wbar G$. 
\end{proposition}
\proof
  This follows directly from the characterization of $W G \to \Wbar G$ by d{\'e}calage.
\endofproof
Pieces of this statement appear in \cite{May}: lemma 18.2 there gives the
fibration property, prop. 21.5 the contractibility of $W G$. 
\begin{corollary}
  \label{GToWGToWbarGPresentsLoopingFiberSequence}
  For $G$ a simplicial group, the sequence of simplicial sets
  $$
    \xymatrix{
      G \ar[r] & W G \ar@{->>}[r] & \Wbar G
	}
  $$
  is a presentation in $\mathrm{sSet}_{\mathrm{Quillen}}$ 
  by a pullback of a Kan fibration of the looping fiber sequence, 
  theorem. \ref{DeloopingTheorem},
  $$
    G \to * \to \mathbf{B}G
  $$
  in $\infty \mathrm{Grpd}$.
\end{corollary}
\proof
  One finds that $G$ is the 1-categorical fiber of $W G \to \Wbar G$.
  The statement then follows using prop. \ref{WGToWbarGIsFibrationResolution} in 
  prop. \ref{ConstructionOfHomotopyLimits}.
\endofproof
The explicit statement that the sequence $G \to W G \to \Wbar G$ is a model for
the looping fiber sequence appears on p. 239 of \cite{Porter}.
The universality of $W G \to \Wbar G$ for $G$-principal simplicial bundles is the 
topic of section 21 in \cite{May}, where however it is not made explicit that the
``twisted cartesian products'' considered there are precisely the models for the 
pullbacks as above. This is made explicit for instance on page 148 of \cite{Porter}.
\begin{corollary}
  \label{SimplicialLoopingModelsInfinityLooping}
  The Quillen equivalence $(L \dashv \Wbar)$ from theorem 
  \ref{SimplicialLoopingQuillenEquivalence} is a presentation of the looping/delooping
  equivalence, theorem \ref{DeloopingTheorem}.
\end{corollary}

We now lift all these statements from simplicial sets to simplicial presheaves.
\begin{proposition} 
 \label{InftyGroupsBySimplicialGroups}
If the cohesive $\infty$-topos $\mathbf{H}$ has site of definition $C$ 
with a terminal object, then 
\begin{itemize}
\item every $\infty$-group object has a presentation by a presheaf of simplicial groups 
  $$
    G \in [C^{\mathrm{op}}, \mathrm{sGrp}] \stackrel{U}{\to} [C^{\mathrm{op}}, \mathrm{sSet}]
  $$ 
  which is fibrant in $[C^{\mathrm{op}}, \mathrm{sSet}]_{\mathrm{proj}}$;
\item
   the corresponding delooping object is presented by the presheaf 
  $$
    {\Wbar} G \in [C^{\mathrm{op}}, \mathrm{sSet}_0] \hookrightarrow [C^{\mathrm{op}}, \mathrm{sSet}]
  $$
  which is given over each $U \in C$ by ${\Wbar} (G(U))$ .
\end{itemize}
\end{proposition}
\proof
By theorem \ref{DeloopingTheorem} every $\infty$-group is the loop
space object of a pointed connected object. By 
prop. \ref{PresentationOfPointedConnectedObjectsByPresheavesOfReducedsSets}
every such is presented by a presheaf $X$ of reduced simplicial sets.
By the 
simplicial looping/delooping Quillen equivalence, 
theorem \ref{SimplicialLoopingQuillenEquivalence},
the presheaf 
$$
  \Wbar L X \in [C^{\mathrm{op}}, \mathrm{sSet}]_{\mathrm{proj}}
$$
is weakly equivalent to the simplicial presheaf $X$.
From this the statement follows with corollary \ref{GToWGToWbarGPresentsLoopingFiberSequence},
combined with prop. \ref{FiniteHomotopyLimitsInPresheaves}, which together say that
the presheaf $L X$ of simplicial groups presents the given $\infty$-group.
\endofproof
\begin{remark}
  We may read this as saying that every $\infty$-group may be 
  \emph{strictified}.
\end{remark}
\begin{example}
  \label{StrictificationOf2GroupObjects}
  Every 2-group in $\mathbf{H}$ (1-truncated group object) 
  has a presentation by a crossed module,
  def. \ref{crossed module}, in simplicial presheaves.
\end{example}

\subsubsection{Cohomology}
\label{StrucCohomology}
\index{structures in a cohesive $\infty$-topos!cohomology}

There is an intrinsic notion of \emph{cohomology} in every
$\infty$-topos. It is the joint generalization of the definition 
of cohomology in $\mathrm{Top}$ in terms of maps into 
classifying spaces and of \emph{sheaf cohomology} over 
any site of definition of the $\infty$-topos. 

For the case of abelian coefficients, as discussed in \ref{SheafAndNonabelianDoldKan},
this perspective of (sheaf) cohomology as the cohomology intrinsic
to an $\infty$-topos is essentially made explicit already in \cite{Brown}.
In more modern language analogous discussion is in section 7.2.2 of 
\cite{Lurie}. 

Here we review central concepts and discuss further aspects 
that will be needed later on.

\paragraph{General abstract}

\begin{definition}
  \label{cohomology}
 \index{cohomology}
For $X,A \in \mathbf{H}$ two objects, we say that
$$
  H(X,A) := \pi_0 \mathbf{H}(X,A)
$$
is the \emph{cohomology set}\index{cohomology!general abstract} of $X$ with coefficients in $A$. 
If $A = G$ is an  $\infty$-group we write
$$
  H^1(X,G) := \pi_0 \mathbf{H}(X, \mathbf{B}G)
$$
for cohomology with coefficients in its delooping. 
Generally, if  $K \in \mathbf{H}$ has a $p$-fold delooping for some $p \in \mathbb{N}$, we write
$$
  H^p(X,K) := \pi_0 \mathbf{H}(X, \mathbf{B}^p K)
  \,.
$$
\end{definition}
In the context of cohomology on $X$ wth coefficients in $A$ we we say that
\begin{itemize}
\item
 the hom-space $\mathbf{H}(X,A)$ is the \emph{cocycle $\infty$-groupoid}\index{cohomology!cocycle};
\item
  a morphism $g : X \to A$ is a \emph{cocycle};
\item 
  a 2-morphism : $g \Rightarrow h$ is a \emph{coboundary} between cocycles.
\item 
  a morphism $c : A \to B$ represents the \emph{characteristic class}\index{characteristic class!general abstract}
  $$
    [c] : H(-,A) \to H(-,B)
    \,.
  $$
\end{itemize}
\begin{remark}
  Traditionally attention is often concentrated on the case that $K \in \tau_{0}\mathrm{Grp}(\mathbf{H})$
  is an abelian 0-truncated group object and $A := \mathbf{B}^p K$ is the Eilenber-MacLane object
  with $K$ in degree $p$. The corresponding cohomology $H^p(-,K) \simeq \pi_0 \mathbf{H}(-, \mathbf{B}^p K)$
  is sometimes called \emph{ordinary cohomology} with coefficients in $K$, to distinguish it from the
  generalizations obtained by allowing more general $K$, which traditionally go by the 
  term \emph{hypercohomology} (if $K$ is not necessarily concentrated in a single degree but is
  still an abelian $\infty$-group, def. \ref{AbelianInfinityGroup}) and more generally
  \emph{nobabelian cohomology} (if $A$ is allowed to be any homotopy type).
  \label{OrdinaryCohomology}
\end{remark}
Below in \ref{PrincipalInfBundle} we discuss the notion of an \emph{$\infty$-group $G$ acting on a space $X$}
and the corresponding (homotopy) quotient $X/\!/G$. Then we say
\begin{definition}
 The cohomology of $X/\!/G$ is the $G$-\emph{equivariant cohomology} of $X$
 with respect to the given action.
\index{equivariant cohomology!in introduction}
\index{cohomology!equivariant cohomology}.
 \label{EquivariantCohomology}
\end{definition}
\begin{remark}
 \label{CohomologyOverX}
 \index{cohomology!in petit $\infty$-topos}
There is also a notion of cohomology in the \emph{petit} $\infty$-topos of $X \in \mathbf{H}$, the slice of $\mathbf{H}$ over $X$
\index{topos!petit}
$$
  \mathcal{X} := \mathbf{H}_{/X}
  \,.
$$ 
This is canonically equipped with the 
{\'e}tale geometric morphism, prop. \ref{EtaleGeometricMorphism}
$$
  (X_! \dashv X^* \dashv X_*)
  :
  \xymatrix{
     \mathbf{H}/X
	 \ar@<+12pt>[r]^<<<<<{X_!}
	 \ar@{<-}@<+4pt>[r]|<<<<<{X^*}
	 \ar@<-4pt>[r]_<<<<<{X_*}
	 &
	 \mathbf{H}
  }  
  \,,
$$
where $X_!$ simply forgets the morphism to $X$ and where $X^* = X \times (-)$ forms the product with $X$. Accordingly $X^* (*_{\mathbf{H}}) \simeq *_{\mathcal{X}} =: X$ and $X_! (*_{\mathcal{X}}) = X \in \mathbf{H}$. Therefore cohomology over $X$ with coefficients of the form $X^* A$ is equivalently the cohomology in $\mathbf{H}$ of $X$ with coefficients in $A$:
$$
  \mathcal{X}(X, X^* A) \simeq \mathbf{H}(X,A)
  \,.
$$ 
For a general coeffcient object $A \in \mathcal{X}$ the $A$-cohomology over $X$ in $\mathcal{X}$ is a \emph{twisted} cohomology of $X$ in $\mathbf{H}$, discussed below in \ref{StrucTwistedCohomology}.
\end{remark}
Typically one thinks of a morphism $A \to B$ in $\mathbf{H}$ as presenting a \emph{characteristic class} of $A$ if
$B$ is ``simpler'' than $A$, notably if $B$ is an Eilenberg-MacLane object $B = \mathbf{B}^n K$ for
$K$ a 0-truncated abelian group in $\mathbf{H}$. In this case the characteristic class may
be regarded as being in the degree-$n$ $K$-cohomology of $A$
$$
  [c] \in H^n(A,K)
  \,.
$$
\begin{definition}
 \label{fiber sequence}
 \label{LongFiberSequence}
 \index{fiber sequence}
For every morphism $c : \mathbf{B}G \to \mathbf{B}H \in \mathbf{H}$ define the 
\emph{long fiber sequence to the left}
$$
   \cdots 
   \to 
   \Omega G
  \to 
  \Omega H 
   \to 
   F
    \to 
   G 
     \to  
   H 
     \to 
    \mathbf{B} F 
     \to 
    \mathbf{B}G
      \stackrel{c}{\to} 
   \mathbf{B}H
$$
to be given by the consecutive pasting diagrams of $\infty$-pullbacks
$$
  \xymatrix{
      & \ar@{..}[d] & \ar@{..}[d]
      \\
      \ar@{..}[r] & F \ar[d]\ar[r]& G \ar[r] \ar[d] & {*} \ar[d]
      \\
      & {*} \ar[r] & H \ar[r] \ar[d] & \mathbf{B}F \ar[r] \ar[d] & {*} \ar[d]
      \\
      & & {*} \ar[r] & \mathbf{B}G \ar[r]^c & \mathbf{B}H
  }
  \,.
$$
\end{definition}
\begin{proposition}
  \label{FormOfLongFiberSequences}
  This is well-defined, in that the objects in the fiber sequence
  are indeed as indicated.
\end{proposition}
\proof
  Repeatedly apply the pasting law \ref{PastingLawForPullbacks}
  and definition \ref{loop space object}.
\endofproof
\begin{proposition}
\begin{enumerate}
\item The long fiber sequence to the left of $c : \mathbf{B}G \to \mathbf{B}H$ becomes constant on the point after $n$ iterations if $H$ is $n$-truncated. 
\item For every object $X \in \mathbf{H}$ we have a long exact sequence of pointed cohomology sets
  $$  
    \cdots \to H^0(X,G) \to H^0(X,H) \to H^1(X,F) \to H^1(X,G) \to H^1(X,H)
   \,.
  $$
\end{enumerate}
\end{proposition}
\proof
The first statement follows from the observation that a 
loop space object $\Omega_x A$ is a fiber of the free loop space object $\mathcal{L} A$
and that this may equivalently be computed by the 
$\infty$-powering $A^{S^1}$, where $S^1 \in \mathrm{Top} \simeq \infty \mathrm{Grpd}$ is the circle. 

The second statement follows by observing that the 
$\infty$-hom-functor $\mathbf{H}(X,-)$ preserves all 
$\infty$-limits, so that we have $\infty$-pullbacks
$$ 
  \xymatrix{
    \mathbf{H}(X,F) \ar[r] \ar[d] & {*} \ar[d]
    \\
    \mathbf{H}(X,G) \ar[r] & \mathbf{H}(X,H)
  }
$$
etc. in $\infty \mathrm{Grpd}$ at each stage of the fiber sequence. 
The statement then follows with the familiar long exact sequence for homotopy groups 
in $\mathrm{Top} \simeq \infty \mathrm{Grpd}$.
\endofproof
\begin{remark}
To every cocycle $g : X \to \mathbf{B}G$ is canonically associated its homotopy fiber 
$P \to X$, the $\infty$-pullback
$$
  \raisebox{20pt}{
  \xymatrix{
    P \ar[r] \ar[d]& {*} \ar[d]
    \\
    X \ar[r] ^g & \mathbf{B}G
    \,.
  }
  }
  \,.
$$
We discuss below in \ref{PrincipalInfBundle} that such $P$ 
canonically has the structure of a 
\emph{$G$-principal $\infty$-bundle} and that 
$\mathbf{B}G$ is the fine moduli space 
-- the moduli $\infty$-stack -- for 
$G$-principal $\infty$-bundles.
\end{remark}

\begin{proposition}[Mayer-Vietoris fiber sequence]
 \label{MayerVietorisFiberSequence}
 \index{Mayer-Vietoris sequence}
  Let $\mathbf{H}$ be an $\infty$-topos with a 
  1-site of definition
  (for instance an $\infty$-cohesive site as in def. \ref{CohesiveSite})
  and let $B$ be an $\infty$-group object in $\mathbf{H}$. 
  Then for any two morphisms $f : X \to B$ and $g : Y \to B$
  the $\infty$-pullback $X \times_B Y$ is equivalently the
  $\infty$-pullback
  $$
    \raisebox{20pt}{
    \xymatrix{
	   X \times_B Y \ar[r] \ar[d] & {*} \ar[d]
	   \\
	   X \times Y
	   \ar[r]^<<<<<<{f \cdot g^{-1}} & B
	}
	}
	\,,
  $$
  where the bottom morphism is the composite
  $$
    f \cdot g^{-1}
	: 
	\xymatrix{
	  X \times Y 
	  \ar[r]^{(f,g)}
	  &
	  B \times B
	  \ar[rr]^{(\mathrm{id}, (-)^{-1})}
	  \ar@/_1pc/[rrr]_{-}
	  &&
	  B \times B
	  \ar[r]^{\cdot}
	  &
	  B
	}
  $$
  of the pair $(f,g)$ with the morphism that inverts the second factor and the
morphism that exhibits the group product on $B$.  
  
We have then a fiber sequence that starts out as
$$
  \xymatrix{
    \cdots 
	\ar[r]
    &
	\Omega B 
    \ar[r]
	&
	X \times_B Y 
    \ar[r]
	&
	X \times Y \ar[r]^<<<<<<<{f \cdot g^{-1}}
	&
	B
  }
  \,.
$$
\end{proposition}
\proof
  By prop \ref{InftyGroupsBySimplicialGroups} there is a presheaf of simplicial
  groups presenting $B$ over the site $C$, which we shall denote by the same symbol,
  $B \in [C^{\mathrm{op}}, \mathrm{sGrp}] \to [C^{\mathrm{op}}, \mathrm{sSet}]$.
  In terms of this the morphism $- : B \times B \to B$ is, objectwise over
  $U \in C$, given by the simplicial morphism $-_U : B(U) \times B(U) \to B(U)$
  that sends $k$-cells $(a,b) : \Delta[k] \to B(U) \times B(U)$ to 
  $a \cdot b^{-1}$, using the degreewise group structure.
  
  We observe first that this morphism is objectwise a Kan fibration and hence 
  a fibration in $[S^{\mathrm{op}}, \mathrm{sSet}]_{\mathrm{proj}}$.
  To see this, let
  $$
    \xymatrix{
	  \Lambda[k]_i \ar[d]^j \ar[r]^<<<<<{(ha, hb)} & B(U) \times B(U) \ar[d]^{-}
	  \\
	  \Delta[k] \ar[r]^{\sigma} & B(U)
	}
  $$
  be a lifting problem. Since $B(U)$, being the simplicial set
  underlying a simplicial group, is a Kan complex, there is a filler 
  $b : \Delta[k] \to B(U)$ of the horn $hb$. Define then a $k$-cell
  $$
    a := \sigma \cdot b
	\,.
  $$
  This is a filler of $ha$, since the face maps are group homomorphisms:
  $$
    \begin{aligned}
	   \delta_l a & = \delta_l(\sigma \cdot b)
	   \\
	    &= \delta_l(\sigma) \cdot \delta_l(b)
		\\
		& = \delta_l(\sigma) \cdot (hb)_l
		\\
		&= (ha)_l
	\end{aligned}
	\,.
  $$
  So we have a filler
  $$
    \xymatrix{
	  \Lambda[k]_i \ar[d]^j \ar[r]^<<<<<<<{(ha, hb)} & B(U) \times B(U) \ar[d]^{-}
	  \\
	  \Delta[k] \ar[r]^{\sigma} 
	  \ar[ur]^{(a,b)}
	  & B(U)
	}
	\,.
  $$
  Observe then that there is a pullback diagram of simplicial presheaves
  $$
    \raisebox{20pt}{
    \xymatrix{
	  B \ar[r] \ar[d]^{\Delta_B} & {*}\ar[d]^{e}
	  \\
	  B \times B
	  \ar[r]^{-}
	  &
	  B
	}
	}
	\,,
  $$
  where the left morphism is the diagonal on $B$ and where the right morphism
  picks the neutral element in $B$. Since, by the above, the bottom morphism
  is a  fibration, this presents a homotopy pullback. 
  
  Next, by the \emph{factorization lemma}, lemma \ref{FactorizationLemma},
  and using prop. \ref{FiniteHomotopyLimitsInPresheaves}, the
  homotopy pullback of $f$ along $g$ is presented by the ordinary
  pullback of simplicial presheaves
  $$
    \raisebox{20pt}{
    \xymatrix{
	  Q \ar[r] \ar[d]& B^{\Delta[0]} \ar[d]
	  \\
	  X \times Y
	  \ar[r]^{(f,g)}
	  &
	  B \times B
	}
	}
	\,,
  $$
  where the right morphism is endpoint evaluation out of the canonical path object
  of $B$, which is a fibration replacement of the diagonal $\Delta_B$.
  Therefore this presents an $\infty$-pullback
  $$
    \raisebox{20pt}{
    \xymatrix{
	  Q \ar[r] \ar[d]& B \ar[d]^{\Delta_B}
	  \\
	  X \times Y
	  \ar[r]^{(f,g)}
	  &
	  B \times B
	}
	}
	\,.
  $$
  Now by the pasting law, prop. \ref{PastingLawForPullbacks}, 
  $Q$ is also an $\infty$-pullback for the total outer diagram in
  $$
    \raisebox{20pt}{
    \xymatrix{
	  Q \ar[r] \ar[d]& B \ar[d]^{\Delta_B} \ar[r] & {*} \ar[d]^e
	  \\
	  X \times Y
	  \ar[r]^{(f,g)}
	  \ar@/_1pc/[rr]_{f \cdot g^{-1}}
	  &
	  B \times B
	  \ar[r]^{-} 
	  & 
	  B
	}
	}
	\,.
  $$
\endofproof

\paragraph{Presentations}
\label{GroupoidsOfMorphisms}

We discuss explicit presentations of cocycles, cohomology classes
and fiber sequences in an $\infty$-topos.

\medskip

\subparagraph{Cocycle $\infty$-groupoids and cohomology classes}

We discuss a useful presentation of cocycle $\infty$-groupoids and of
cohomology classes by a construction that exists when the 
ambient $\infty$-topos is presented by a category with weak
equivalences that is equipped with the structure of a 
\emph{category of fibrant objects} \cite{Brown}.

\begin{definition}[Brown]
  \label{CategoryOfFibrantObjects}
A \emph{category of fibrant objects} is a category 
equipped with two distinguished classes 
of morphisms, called \emph{fibrations} and \emph{weak equivalences}, such that
\begin{enumerate}
\item the category has a terminal object $*$ and finite products;

\item fibrations and weak equivalences form 
subcategories that contain all isomorphisms;
weak equivalences moreover satisfy the 2-out-of-3 property;

\item for any object $B$ the map $B \to *$ is a fibration;

\item the classes of fibrations and of \emph{acyclic fibrations}
(the fibration that are also weak equivalences)
are stable under pullback. That means: given a 
diagram $A \xrightarrow g C \xleftarrow f B$ 
where $f$ is a (acyclic) fibration then the pullback 
$A \times_C B$ exists and the morphism 
$A \times_C B  \to A$ is again a (acyclic) fibration. 

\item For every object $B$ there is a path object $B^I$, i.e.\ 
a factorization of the diagonal $\Delta\colon B\to B\times B$ into  
\begin{equation*}
 \xymatrix{
   B \ar[r]^\simeq & B^I \ar@{->>}[r]  & B \times B
 }
\end{equation*}
such that left map is weak equivalence and the right 
map a fibration. We assume here moreover for 
simplicity that this $B^I$ can be chosen functorial in $B$.
\end{enumerate}
\end{definition}
Given a category of fibrant objects, we will
denote the class of weak equivalence by $W$ and the class of fibrations by $F$. 
\begin{examples}
  \label{BasicExamplesOfCatsOfFibObjects}
We have the following well known examples of categories of fibrant objects.  
\begin{itemize}
\item For any model category (with functorial factorization) the full subcategory of fibrant objects is a category of fibrant objects. 

\item The category of stalkwise Kan simplicial 
presheaves on any site with enough points. 
In this case the fibrations are the stalkwise fibrations 
and the weak equivalences are the stalkwise weak equivalences.
\end{itemize}
\end{examples}
\begin{remark}
 \label{StalkwiseFibrationsAreNotModelStructureFibrations}
Notice that (over a non-trivial site) the second example above is \emph{not} 
a special case of the first:
while there are model structures on categories of simplicial presheaves whose
weak equivalences are the stalkwise weak equivalences, their fibrations 
(even between fibrant objects) are much more 
restricted than just being stalkwise fibrations.  
\end{remark}

\begin{theorem}
  \label{SimplicialLocalizationOfCatOfFibrantObjects}
  Let the $\infty$-category $\mathbf{H}$ be presented by 
  a category with weak equivalences $(\mathcal{C}, W)$ that carries
  a compatible structure of a category of fibrant objects, def. \ref{CategoryOfFibrantObjects}.
  
  Then for $X, A$ and two objects in $\mathcal{C}$, presenting two objects
  in $\mathbf{H}$, the $\infty$-groupoid $\mathbf{H}(X,A)$ is presented
  in $\mathrm{sSet}_{\mathrm{Quillen}}$ by the nerve of the category
  whose
  \begin{itemize}
    \item objects are spans (cocycles / $\infty$-anafunctors) 
	  $$
	    \xymatrix{
		  X \ar@{<<-}[r]^\simeq & \hat X \ar[r]^g & A 
		}
	  $$
	  in $\mathcal{C}$;
	\item
	  morphisms $f : (\hat X, g) \to (\hat X', g')$ are given by morphisms $f : \hat X \to \hat X'$
	  in $\mathcal{C}$ such that the diagram
	  $$
	    \xymatrix@R=7pt{
		  & \hat X \ar@{->>}[dl]_\simeq \ar[dd]^f \ar[dr]^{g}
		  \\
		  X && A
		  \\
		  & \hat X' \ar@{->>}[ul]^\simeq \ar[ur]_{g'}
		}
	  $$
	  commutes.
  \end{itemize}
\end{theorem}
This appears for instance as prop. 3.23 in \cite{Cisinski}.
\begin{example}
  By the discussion in \ref{InfinityToposPresentation}, 
  if $\mathbf{H}$ has a 1-site of definition $C$ with enough 1-topos
  points, then it is presented by the category $\mathrm{Sh}(C)^{\Delta^{\mathrm{op}}}$
  of simplicial sheaves on $C$ with weak equivalences the stalkwise weak equivalences
  of simplicial sets, and equivalently by its full subcategory of stalkwise
  Kan fibrant simplicial sheaves. With the local fibrations, def. \ref{LocalFibrations}
  as fibrations, this is a category of fibrant objects. So in this case the
  cocycle $\infty$-groupoid $\mathbf{H}(X,A)$ is presented by the 
  Kan fibrant replacement of the category whose objects 
  are spans 
  	  $$
	    \xymatrix{
		  X \ar@{<<-}[r]^\simeq & \hat X \ar[r]^g & A 
		}
	  $$
  for $\hat X \to X$ a stalkwise acyclic Kan fibration, and whose morphisms
  are as above.
\end{example}

\subparagraph{Fiber sequences}
\label{PresentationOfFiberSequences}

We discuss explicit presentations of certain fiber sequences,
def. \ref{fiber sequence}, in an $\infty$-topos.

\medskip

\begin{proposition}
  \label{LongFiberSequenceFromCentralExtensionOfGroups}
  Let $A \to \hat G \to G$ be a central extension of (ordinary) groups. 
  Then there is a long fiber sequence in $\infty \mathrm{Grpd}$ of the form
  $$
    \xymatrix{
	  A \ar[r] & \hat G \ar[r] & G 
	  \ar[r]^{\Omega \mathbf{c}} & \mathbf{B}A
	  \ar[r] & \mathbf{B} \hat G \ar[r] & \mathbf{B} G
	  \ar[r]^{\mathbf{c}} & \mathbf{B}^2 A
	}
	\,,
  $$
  where the connecting homomorphism is presented by the 
  correspondence of crossed modules, def. \ref{Strict2GroupInIntroduction},
  given by
  $$
    \xymatrix{
	  (1 \to G)
	  \ar@{<-}[r]^\simeq
	  &
	  (A \to \hat G)
	  \ar[r] &
	  (A \to 1)
	}
	\,.
  $$  
  Here in the middle appears the crossed module defined by the central extension,
  def. \ref{CrossedModuleFromNormalSubgroup}.
\end{proposition}

\subsubsection{Principal bundles}
\label{section.PrincipalInfinityBundle}
\label{PrincipalInfBundle}
\index{structures in a cohesive $\infty$-topos!principal $\infty$-bundles}

For $G$ an $\infty$-group object in a cohesive $\infty$-topos
$\mathbf{H}$ and
$\mathbf{B}G$ its delooping in $\mathbf{H}$,
as discussed in \ref{StrucInftyGroups}, the cohomology 
over an object $X$ with coefficients
in $\mathbf{B}G$, as in \ref{StrucCohomology}, classifies 
maps $P \to X$ that are equipped with a $G$-action that is
\emph{principal}. We discuss here these 
\emph{$G$-principal $\infty$-bundles}. 

\paragraph{Introduction and survey}
\label{PrincipalBundlesIntroductionAndSurvey}

We give an exposition of some central ideas and phenomena of
higher principal bundles, discussed in detail below. 

This section draws from \cite{NSSa}.

\medskip

Let $G$ be a topological group, or Lie group or 
some similar such object. The traditional 
definition of \emph{$G$-principal bundle} is the following:  
there is a map  
$$
  P \to X := P/G
$$ 
which is the quotient projection
induced by a \emph{free} action 
$$
  \rho : P \times G \to P
$$ 
of $G$ on a space (or manifold, depending on context) $P$,
such that there is a cover $U \to X$ over which the quotient projection is isomorphic
to the trivial one $U \times G \to U$.
 
In higher geometry, if $G$ is a topological or smooth 
$\infty$-group, the quotient projection must be 
replaced by the $\infty$-quotient (homotopy quotient) 
projection 
\[
P\to X := P/\!/ G
\]
for the action of $G$ on a topological or smooth $\infty$-groupoid 
(or $\infty$-stack) $P$.  It is a remarkable fact that this 
single condition on the map $P\to X$ 
already implies that $G$ acts freely on $P$ and that $P\to X$ 
is locally trivial, when the latter notions are understood in the 
context of higher geometry.  We will therefore define 
a $G$-principal $\infty$-bundle to be such a map $P\to X$.


As motivation for this, notice that if a Lie group $G$ acts properly, 
but not freely, then the quotient $P \to X := P/G$ differs from the homotopy quotient.
Specifically, if precisely the subgroup $G_{\mathrm{stab}} \hookrightarrow G$ acts trivially, then  
the homotopy quotient is 
instead the \emph{quotient stack} $X /\!/ G_{\mathrm{stab}}$ 
(sometimes written $[X/\!/G_{\mathrm{stab}}]$, 
which is an orbifold if $G_{\mathrm{stab}}$ is finite). The ordinary 
quotient coincides with the homotopy quotient if and only if the stabilizer subgroup
$G_{\mathrm{stab}}$ is trivial, and hence if and only if the action of $G$ is free.

Conversely this means that in the context of higher geometry a non-free action
may also be principal: with respect not to a base space, but with respect to a base groupoid/stack.
In the example just discussed, we have that the projection $P \to X/\!/ G_{\mathrm{stab}}$
exhibits $P$ as a $G$-principal bundle over the action groupoid 
$P /\!/ G \simeq X/\!/ G_{\mathrm{stab}}$. For instance if $P = V$ is a
vector space equipped with a $G$-representation, then $V \to V/\!/ G$ is a
$G$-principal bundle over a groupoid/stack.
In other words, the traditional requirement of freeness in a principal action is not so much
a characterization of principality as such, as rather a condition that ensures that the
base of a principal action is a 0-truncated object in higher geometry.

Beyond this specific class of 0-truncated examples, this means that we have the following 
noteworthy general statement: in higher geometry \emph{every} $\infty$-action 
is principal with respect to
\emph{some} base, namely with respect to its $\infty$-quotient. In this sense
the notion of principal bundles is (even) more fundamental to higher geometry than
it is to ordinary geometry. 
Also, several constructions in ordinary geometry that are
traditionally thought of as conceptually different from the notion of principality
turn out to be special cases of principality in higher geometry. For instance
a central extension of groups $A\to \hat G \to G$ turns out to be 
equivalently a higher principal bundle, namely a $\mathbf{B}A$-principal 2-bundle
of moduli stacks $\mathbf{B}\hat G \to \mathbf{B}G$. Following this
through, one finds that the topics 
of principal $\infty$-bundles, of $\infty$-group extensions (\ref{ExtensionsOfCohesiveInfinityGroups}), 
of $\infty$-representations (\ref{StrucRepresentations}), 
and of $\infty$-group cohomology
are all different aspects of just one single concept in higher geometry.

More is true: in the context of an $\infty$-topos 
every $\infty$-quotient projection of an $\infty$-group action 
is locally trivial, with respect to
the canonical intrinsic notion of cover, hence of locality. Therefore
also the condition of local triviality in the classical definition of principality
becomes automatic. This is a direct consequence of the third
$\infty$-Giraud axiom, Definition~\ref{GiraudRezkLurieAxioms}
that ``all $\infty$-quotients are effective''.
This means that the projection map $P \to P /\!/ G$ is always a cover
(an \emph{effective epimorphism}) and so, since every $G$-principal $\infty$-bundle
trivializes over itself, it exhibits a local trivialization of itself; 
even without explicitly requiring it to be locally trivial.

As before, this means that the local triviality clause appearing in the
traditional definition of principal bundles is not so much a characteristic of
principality as such, as rather a condition that ensures that a given quotient 
taken in a category of geometric spaces coincides with the ``correct'' quotient
obtained when regarding the situation in the ambient $\infty$-topos. 

Another direct consequence of the $\infty$-Giraud axioms
is the equivalence of the definition of principal bundles as quotient maps, which 
we discussed so far, with the other main definition of principality: the condition 
that the ``shear map'' $ (\mathrm{id}, \rho) :  P \times G \to P \times_X P$ is an equivalence. 
It is immediate to verify in traditional 1-categorical contexts that this is 
equivalent to the action being properly free and exhibiting $X$ as its quotient
(we discuss this in detail in \cite{NSSc}). 
Simple as this is, one may observe, in view of the above discussion, 
that the shear map being an equivalence is much more fundamental even: notice
that $P \times G$ is the first stage of the \emph{action groupoid object}
$P/\!/G$, and that $P \times_X P$ is the first stage of the \emph{{\v C}ech nerve groupoid object}
$\check{C}(P \to X)$ of the corresponding quotient map. Accordingly, the shear map equivalence
is the first stage in the equivalence of groupoid objects in the $\infty$-topos
$$
  P /\!/G \simeq \check{C}(P \to X)
  \,.
$$
This equivalence is just the explicit statement of the fact mentioned before: the groupoid object
$P/\!/G$ is effective -- as is any groupoid object in an $\infty$-topos -- and, equivalently,
its principal $\infty$-bundle map $P \to X$ is an effective epimorphism.

Fairly directly from this fact, finally, springs the classification theorem of 
principal $\infty$-bundles. For we have a canonical morphism of groupoid objects
$P /\!/G \to * / \!/G$ induced by the terminal map $P \to *$. By the $\infty$-Giraud
theorem the $\infty$-colimit over this sequence of morphisms of groupoid objects
is a $G$-cocycle on $X$ (Definition~\ref{cohomology}) canonically induced by $P$:
$$
  \varinjlim  \left(\check{C}(P \to X)_\bullet \simeq (P /\!/G)_\bullet \to (* /\!/G)_\bullet \right) 
    = 
  (X \to \mathbf{B}G) 
  \;\;\;
  \in \mathbf{H}(X, \mathbf{B}G)
  \,.
$$
Conversely, from any such $G$-cocycle one finds that one obtains a $G$-principal 
$\infty$-bundle simply by forming its $\infty$-fiber: the $\infty$-pullback of
the point inclusion ${*} \to \mathbf{B}G$. We show in \cite{NSSb} that in presentations
of the $\infty$-topos theory by 1-categorical tools, the computation of this homotopy
fiber is \emph{presented} by the ordinary pullback of a big resolution of the point, 
which turns out to be nothing but the universal $G$-principal bundle. 
This appearance of the universal $\infty$-bundle as just 
a resolution of the point inclusion may be understood in light of the above discussion 
as follows.   
The classical characterization of the 
universal $G$-principal bundle $\mathbf{E}G$ is as a space that is homotopy equivalent
to the point and equipped with a \emph{free} $G$-action. But by the above, freeness of the
action is an artefact of 0-truncation and not a characteristic of principality in higher
geometry. Accordingly, in higher geometry the universal $G$-principal $\infty$-bundle
for any $\infty$-group $G$ may be taken to \emph{be} the point, equipped with the 
trivial (maximally non-free) $G$-action. As such, it is a bundle not over the 
classifying \emph{space} $B G$ of $G$, but over the full moduli $\infty$-stack $\mathbf{B}G$.

This way we have natural assignments of $G$-principal $\infty$-bundles to 
cocycles in $G$-nonabelian cohomology, and vice versa. We find (see 
Theorem~\ref{PrincipalInfinityBundleClassification} below) that
precisely the second $\infty$-Giraud axiom of Definition~\ref{GiraudRezkLurieAxioms}, 
namely the fact that in an $\infty$-topos $\infty$-colimits are preserved by
$\infty$-pullback, 
implies that these 
constructions constitute an equivalence of $\infty$-groupoids, hence 
that $G$-principal $\infty$-bundles are classified by $G$-cohomology.

The following table summarizes the relation between
$\infty$-bundle theory and the $\infty$-Giraud axioms as indicated above, and as 
proven in the following section.

\medskip
\begin{center}
\begin{tabular}{c|c}
 {\bf $\infty$-Giraud axioms} & {\bf principal $\infty$-bundle theory}
 \\
 \hline
 \hline
 quotients are effective & 
   \begin{tabular}{c} \\ every $\infty$-quotient $P \to X := P/\!/ G$ \\is principal \\ \ \end{tabular}
 \\
 \hline
  colimits are preserved by pullback & 
  \begin{tabular}{c}\\ $G$-principal $\infty$-bundles \\ are classified by $\mathbf{H}(X,\mathbf{B}G)$\\
   \end{tabular}
\end{tabular}
\end{center}

\paragraph{Definition and classification}
\label{Principal infinity-bundles general abstract}

We discuss the general definition and the central classification theorem of
principal $\infty$-bundles. 

This section draws from \cite{NSSa}.

\medskip

\begin{definition} 
  \label{ActionInPrincipal}
  \label{RepresentationOfInfinityGroup}
  For $G \in \mathrm{Grp}(\mathbf{H})$ a group object,
  we say a \emph{$G$-action} on an object $P \in \mathbf{H}$
  is a groupoid object $P/\!/G$ (Definition~\ref{GroupoidObject}) of the form
  $$
    \xymatrix{
       \cdots
       \ar@<+6pt>[r] \ar@<+2pt>[r] \ar@<-2pt>[r] \ar@<-6pt>[r]
       &
       P \times G \times G
       \ar@<+4pt>[r] 
       \ar[r] 
        \ar@<-4pt>[r] 
         & 
         P \times G 
         \ar@<+2pt>[r]^<<<<<{\rho := d_0 } \ar@<-2pt>[r]_{d_1} 
       & P
    }
  $$
  such that $d_1 : P \times G \to P$ is the projection, and such that 
  the degreewise projections 
  $P \times G^n \to G^n $ constitute a morphism of groupoid
  objects
  $$
    \xymatrix{
       \cdots
       \ar@<+6pt>[r] \ar@<+2pt>[r] \ar@<-2pt>[r] \ar@<-6pt>[r]
       &
       P \times G \times G
       \ar[d]
       \ar@<+4pt>[r] 
       \ar[r] 
        \ar@<-4pt>[r] 
         & 
         P \times G 
         \ar[d]
         \ar@<+2pt>[r] \ar@<-2pt>[r] 
       & P
       \ar[d]
       \\
       \cdots
       \ar@<+6pt>[r] \ar@<+2pt>[r] \ar@<-2pt>[r] \ar@<-6pt>[r]
       &
       G \times G
       \ar@<+4pt>[r] 
       \ar[r] 
        \ar@<-4pt>[r] 
         & 
         G 
         \ar@<+2pt>[r] \ar@<-2pt>[r] 
       & {*}       
    }
  $$
where the lower simplicial object exhibits $G$ as a groupoid 
object over $\ast$. 
  
  With convenient abuse of notation we also write
  $$ 
    P/\!/G := \varinjlim (P \times G^{\times^\bullet})\;\; 
    \in \mathbf{H}
  $$
  for the corresponding $\infty$-colimit object, the \emph{$\infty$-quotient} of this 
  action.   
  
  Write
  $$
    G \mathrm{Action}(\mathbf{H}) \hookrightarrow \mathrm{Grpd}(\mathbf{H})_{/({*}/\!/G)}
  $$
  for the full sub-$\infty$-category of groupoid objects over $*/\!/G$
  on those that are $G$-actions. 
\end{definition}
\begin{remark}
  \label{ActionMapEncodedInGBundle}
  The remaining face map $d_0$
  $$
    \rho := d_0 : P \times G \to P
  $$
  is the action itself.
\end{remark}
\begin{remark}
  Using this notation in Proposition~\ref{InfinityGroupObjectsAsGroupoidObjects}
  we have
  $$
    \mathbf{B}G \simeq */\!/G
    \,.
  $$
\end{remark}
We list examples of $\infty$-actions below as Example \ref{ExamplesOfActions}.
This is most conveniently done after establishing the theory of 
principal $\infty$-actions, to which we now turn.

\begin{definition}
  \label{principalbundle}
  Let $G \in \infty \mathrm{Grp}(\mathbf{H})$ be an 
  $\infty$-group and let $X$ be an object of $\mathbf{H}$.   
  A {\em $G$-principal $\infty$-bundle} over $X$ 
  (or \emph{$G$-torsor over $X$}) 
  is 
  \begin{enumerate}
    \item a morphism $P \to X$ in $\mathbf{H}$;
	\item together with a $G$-action on $P$;
  \end{enumerate}  
  such that $P \to X$ is the colimiting cocone exhibiting the quotient map
  $X \simeq P/\!/G$ (Definition \ref{ActionInPrincipal}).
    
  A \emph{morphism} of $G$-principal $\infty$-bundles over $X$ is a morphism of $G$-actions 
  that fixes $X$; the $\infty$-category of $G$-principal $\infty$-bundles over $X$
  is the homotopy fiber of $\infty$-categories
  $$
    G \mathrm{Bund}(X) := G \mathrm{Action}(\mathbf{H}) \times_{\mathbf{H}} \{X\}
  $$
  over $X$ of the quotient map
  $$
    \xymatrix{
      G \mathrm{Action}(\mathbf{H}) 
	  \ar@{^{(}->}[r] & \mathrm{Grpd}(\mathbf{H})_{/(*/\!/G)} 
	  \ar[r] &
	  \mathrm{Grpd}(\mathbf{H}) 
	  \ar[r]^-{\varinjlim}
	  &
	  \mathbf{H}
	}
	\,.
  $$
\end{definition}
\begin{remark}
  \label{GBundlesAreEffectiveEpimorphisms}
  By the third $\infty$-Giraud axiom (Definition~\ref{GiraudRezkLurieAxioms})
  this means in particular that a 
  $G$-principal $\infty$-bundle $P \to X$ is an 
  effective epimorphism in $\mathbf{H}$.
\end{remark}
\begin{remark}
  Even though $G \mathrm{Bund}(X)$ is by definition a priori an $\infty$-category,
  Proposition \ref{MorphismsOfInfinityBundlesAreEquivalences} below says
  that in fact it happens to be $\infty$-groupoid: all
  its morphisms are invertible.
\end{remark}
\begin{proposition}
  \label{PrincipalityCondition}
  A $G$-principal $\infty$-bundle $P \to X$ satisfies the
  \emph{principality condition}: the canonical morphism
  $$
    (\rho, p_1)
	:
    \xymatrix{
	  P \times G 
	  \ar[r]^{\simeq}
	  &
	  P \times_X P
	}
  $$
  is an equivalence, where $\rho$ is the $G$-action.
\end{proposition}
\proof
  By the third $\infty$-Giraud axiom (Definition~\ref{GiraudRezkLurieAxioms}) the groupoid object
  $P/\!/G$ is effective, which means that it is equivalent
  to the   {\v C}ech nerve of $P \to X$. In first degree this implies
  a canonical equivalence $P \times G \to P \times_X P$. Since
  the two face maps $d_0, d_1 : P \times_X P \to P$ in the
  {\v C}ech nerve are simply the projections out of the fiber product, 
  it follows that the two components of this canonical equivalence
  are the two face maps $d_0, d_1 : P \times G \to P$ of $P/\!/G$.
  By definition, these are the projection onto the first factor
  and the action itself.   
\endofproof
\begin{proposition}
  \index{principal $\infty$-bundle!construction from cocycle}
  \label{BundleStructureOnInfinityFiber}
  \label{PrincipalBundleAsHomotopyFiber}
  For $g : X \to \mathbf{B}G$ any morphism, its homotopy fiber
  $P \to X$ canonically carries the structure of a 
  $G$-principal $\infty$-bundle over $X$.
\end{proposition}
\proof
  That $P \to X$ is the fiber of $g : X \to \mathbf{B}G$
  means that we have an $\infty$-pullback diagram
  $$
    \raisebox{20pt}{
    \xymatrix{
      P \ar[d]\ar[r] & {*}\ar[d]
      \\
      X \ar[r]^g & \mathbf{B}G.
    }
	}
  $$
  By the pasting law for $\infty$-pullbacks, Proposition~\ref{PastingLawForPullbacks},
  this induces a compound diagram
  $$
    \xymatrix{
       \cdots
       \ar@<+6pt>[r] \ar@<+2pt>[r] \ar@<-2pt>[r] \ar@<-6pt>[r]
       &
       P \times G \times G
       \ar[d]
       \ar@<+4pt>[r] 
       \ar[r] 
        \ar@<-4pt>[r] 
         & 
         P \times G 
         \ar[d]
         \ar@<+2pt>[r] \ar@<-2pt>[r] 
       & 
       P
       \ar[d]
       \ar@{->>}[r]
       &
       X
       \ar[d]^g
       \\
       \cdots
       \ar@<+6pt>[r] \ar@<+2pt>[r] \ar@<-2pt>[r] \ar@<-6pt>[r]
       &
       G \times G
       \ar@<+4pt>[r] 
       \ar[r] 
        \ar@<-4pt>[r] 
         & 
         G 
         \ar@<+2pt>[r] \ar@<-2pt>[r] 
       & 
       {*}       
       \ar@{->>}[r]
       &
       \mathbf{B}G
    }
  $$
  where each square and each composite rectangle is an 
  $\infty$-pullback. 
  This exhibits the $G$-action on $P$. 
  Since $* \to \mathbf{B}G$
  is an effective epimorphism, so is its $\infty$-pullback
  $P \to X$. Since, by the $\infty$-Giraud theorem, $\infty$-colimits are preserved
  by $\infty$-pullbacks we have that $P \to X$ exhibits the 
  $\infty$-colimit $X \simeq P/\!/G$.  
\endofproof
\begin{lemma}
  For $P \to X$ a $G$-principal $\infty$-bundle
  obtained as in Proposition~\ref{BundleStructureOnInfinityFiber}, and for
  $x : * \to X$ any point of $X$, we have 
  a canonical equivalence
  $$
    \xymatrix{
	    x^* P \ar[r]^{\simeq} & G
	}
  $$
  between the fiber $x^*P$ and the $\infty$-group object $G$.
\end{lemma}
\proof
  This follows from the pasting law for $\infty$-pullbacks, which 
  gives the diagram
  $$
    \xymatrix{
      G \ar[d] \ar[r] & P \ar[d]\ar[r] & {*} \ar[d]
      \\
      {*} \ar[r]^x & X \ar[r]^g & \mathbf{B}G
    }
  $$
  in which both squares as well as the total rectangle are
  $\infty$-pullbacks.
\endofproof
\begin{definition}
  \label{TrivialGBundle}
  The \emph{trivial} $G$-principal $\infty$-bundle $(P \to X) \simeq (X \times G \to X)$
  is, up to equivalence, the one obtained via Proposition~\ref{BundleStructureOnInfinityFiber}
  from the morphism $X \to * \to \mathbf{B}G$. 
\end{definition}
\begin{observation}
  \label{PullbackOfInfinityTorsors}
  For $P \to X$ a $G$-principal $\infty$-bundle and $Y \to X$ any morphism, the
  $\infty$-pullback $Y \times_X P$ naturally inherits the structure of 
  a $G$-principal $\infty$-bundle.
\end{observation}
\proof
  This uses the same kind of argument as in Proposition~\ref{BundleStructureOnInfinityFiber}
  (which is the special case of the pullback of what we will see is the
  universal $G$-principal $\infty$-bundle $*\to \mathbf{B}G$ below in 
  Proposition~\ref{LocalTrivialityImpliesCocycle}).
\endofproof
\begin{definition}
  \label{GInfinityBundleAndMorphism}
  \label{GPrincipalInfinityBundle}
  A $G$-principal $\infty$-bundle $P \to X$ is called
  \emph{locally trivial}
  if there exists an effective epimorphism $\xymatrix{U \ar@{->>}[r] & X}$
  and an equivalence of $G$-principal $\infty$-bundles
  $$
    U \times_X P \simeq U \times G
  $$
  from the pullback of $P$ (Observation~\ref{PullbackOfInfinityTorsors})
  to the trivial $G$-principal $\infty$-bundle over $U$ (Definition~\ref{TrivialGBundle}).
\end{definition}
\begin{proposition}
  \label{EveryGBundleIsLocallyTrivial}
  Every $G$-principal $\infty$-bundle is locally trivial.
\end{proposition}
\proof
  For $P \to X$ a $G$-principal $\infty$-bundle, it is, by 
  Remark~\ref{GBundlesAreEffectiveEpimorphisms}, itself an effective
  epimorphism. The pullback of the $G$-bundle to its 
  own total space along this morphism is trivial,
   by the principality condition
  (Proposition~\ref{PrincipalityCondition}). Hence setting $U := P$ 
  proves the claim.
\endofproof
\begin{remark}
  This means that every $G$-principal $\infty$-bundle is in particular a
  $G$-fiber $\infty$-bundle (in the evident sense of Definition~\ref{FiberBundle} below).
  But not every $G$-fiber bundle is $G$-principal, since the local trivialization
  of a fiber bundle need not respect the $G$-action. 
\end{remark}
\begin{proposition}
  \label{LocalTrivialityImpliesCocycle}
  For every $G$-principal $\infty$-bundle $P \to X$ the square
  $$
    \xymatrix{
       & P \ar[d] \ar[r] & {*} \ar[d]
       \\
       X \ar@{}[r]|<<<\simeq
       & \varinjlim_n (P \times G^{\times_n})
       \ar[r]
       &
       \varinjlim_n G^{\times_n}
       \ar@{}[r]|\simeq
       &
       \mathbf{B}G
    }
  $$
  is an $\infty$-pullback diagram.
\end{proposition}
\proof
  Let $U \to X$ be an effective epimorphism
  such that $P \to X$ pulled back to $U$ becomes the trivial $G$-principal
  $\infty$-bundle. By Proposition~\ref{EveryGBundleIsLocallyTrivial} this exists.
  By definition of morphism of $G$-actions and 
  by functoriality of the $\infty$-colimit, this induces 
  a morphism in ${\mathbf{H}^{\Delta[1]}}_{/(* \to \mathbf{B}G)}$ 
  corresponding to the diagram
  $$
    \raisebox{20pt}{
    \xymatrix{
	  U \times G \ar@{->>}[r] \ar@{->>}[d] & P \ar[r] \ar@{->>}[d] & {*} \ar@{->>}[d]^{\mathrm{pt}}
	  \\
	  U \ar@{->>}[r] & X \ar[r] & \mathbf{B}G
	}
	}
	\;\;
	\simeq
	\;\;
    \raisebox{20pt}{
    \xymatrix{
	  U \times G \ar@{->>}[rr] \ar@{->>}[d] & & {*} \ar@{->>}[d]^{\mathrm{pt}}
	  \\
	  U \ar[r] & {*} \ar[r]^{\mathrm{pt}} & \mathbf{B}G
	}
	}
  $$
  in $\mathbf{H}$.  
  By assumption, in this diagram the outer rectangles and the square on the very left 
  are $\infty$-pullbacks. We need to show that
  the right square on the left is also an $\infty$-pullback.
  
  Since $U \to X$ is an effective epimorphism by assumption, and since these are
  stable under $\infty$-pullback, $U \times G \to P$ is also an effective epimorphism,
  as indicated. This means that 
  $$
    P \simeq {\varinjlim_n}\, (U \times G)^{\times^{n+1}_P}
	\,.
  $$
  We claim that for all $n \in \mathbb{N}$ the fiber products in the colimit on the right
  are naturally equivalent to $(U^{\times^{n+1}_X}) \times G$. For $n = 0$ this is
  clearly true. Assume then by induction that 
  it holds for some $n \in \mathbb{N}$. Then
  with the pasting law (Proposition~\ref{PastingLawForPullbacks}) we find  an 
  $\infty$-pullback diagram of the form
  $$
    \raisebox{30pt}{
    \xymatrix{
	  (U^{\times^{n+1}_X}) \times G 	  
	  \ar@{}[r]|\simeq
	  & 
	  (U \times G)^{\times^{n+1}_P}
	  \ar[r] \ar[d] 
	  &  
	  (U \times G)^{\times^n_P}
      \ar[d]
	  \ar@{}[r]|{\simeq}
	  &
	  (U^{\times^{n}_X}) \times G 
	  \\
	  & U \times G \ar[r] \ar[d] & P \ar[d]
	  \\
	  & U \ar[r] & X.
	}
	}
  $$
  This completes the induction.
  With this the above expression for $P$ becomes
  $$
    \begin{aligned}
      P & \simeq {\varinjlim_n}\,  (U^{\times^{n+1}_X}) \times G
	   \\
	    & \simeq {\varinjlim_n}  \,\mathrm{pt}^* \, (U^{\times^{n+1}_X}) 
	   \\
	    & \simeq \mathrm{pt}^* \, {\varinjlim_n}\,    (U^{\times^{n+1}_X}) 	  
	  \\
	    & \simeq \mathrm{pt}^* \, X,
	\end{aligned}
  $$
  where we have used that by the second $\infty$-Giraud axiom (Definition~\ref{GiraudRezkLurieAxioms})
  we may take the $\infty$-pullback out of the $\infty$-colimit and 
  where in the last step we 
  used again the assumption that $U \to X$ is an effective epimorphism.
\endofproof
\begin{example}
  The fiber sequence 
  $$
    \xymatrix{
	  G \ar[r] & {*} \ar[d]
	  \\
	  & \mathbf{B}G
	}
  $$
  which exhibits the delooping $\mathbf{B}G$ of $G$ according to 
  Theorem \ref{DeloopingTheorem} is a $G$-principal $\infty$-bundle
  over $\mathbf{B}G$, with \emph{trivial} $G$-action on its total space
  $*$. Proposition \ref{LocalTrivialityImpliesCocycle} says that this is 
  the \emph{universal $G$-principal $\infty$-bundle} in that every
  other one arises as an $\infty$-pullback of this one. 
  In particular, $\mathbf{B}G$ is a classifying object for $G$-principal
  $\infty$-bundles. 
  
  Below in Theorem \ref{ClassificationOfTwistedGEquivariantBundles} 
  this relation is strengthened:
  every \emph{automorphism} of a $G$-principal $\infty$-bundle, and in fact
  its full automorphism $\infty$-group arises from pullback of the above
  universal $G$-principal $\infty$-bundle: $\mathbf{B}G$ is the fine
  \emph{moduli $\infty$-stack} of $G$-principal $\infty$-bundles.
  
  The traditional definition of universal $G$-principal bundles in terms of  
  contractible objects equipped with a free $G$-action has no intrinsic
  meaning in higher topos theory. Instead this appears in 
  \emph{presentations} of the general theory in model categories
  (or categories of fibrant objects)
  as \emph{fibrant representatives}
  $\mathbf{E}G \to \mathbf{B}G$ of the above point inclusion.
  This we discuss in \cite{NSSb}.  
  \label{UniversalPrincipal}
\end{example}
The main classification Theorem \ref{PrincipalInfinityBundleClassification} below
implies in particular that every morphism in $G\mathrm{Bund}(X)$ is an equivalence.
For emphasis we note how this also follows directly:
\begin{lemma}
  \label{EquivalencesAreDetectedOverEffectiveEpimorphisms}
  Let $\mathbf{H}$ be an $\infty$-topos and let $X$ be an 
  object of $\mathbf{H}$.  A morphism 
  $f\colon A\to B$ in $\mathbf{H}_{/X}$ 
  is an equivalence if and only if $p^*f$ is an equivalence in 
  $\mathbf{H}_{/Y}$ for any effective epimorphism $p\colon Y\to X$ in 
  $\mathbf{H}$.

\end{lemma}
\proof
  It is clear, by functoriality, that $p^* f$ is a weak equivalence if $f$ is. 
  Conversely, assume that $p^* f$ is a weak equivalence. 
  Since effective epimorphisms as well as 
  equivalences are preserved by pullback 
  we get a simplicial diagram of the form
  $$
    \raisebox{20pt}{
    \xymatrix{
       \cdots
       \ar@<+4pt>[r]
       \ar@<+0pt>[r]
       \ar@<-4pt>[r]
       &
       p^* A \times_A p^* A
       \ar@<+2pt>[r]
       \ar@<-2pt>[r]
       \ar[d]^\simeq
       &
       p^* A
       \ar[d]^\simeq
       \ar@{->>}[r]
       & 
       A
       \ar[d]^f
       \\
       \cdots
       \ar@<+4pt>[r]
       \ar@<+0pt>[r]
       \ar@<-4pt>[r]
       &
       p^* B \times_B p^* B
       \ar@<+2pt>[r]
       \ar@<-2pt>[r]
       &
       p^* B
       \ar@{->>}[r]
       & 
       B
    }
	}
  $$
  where the rightmost horizontal morphisms are effective epimorphisms, as indicated.
  By definition of effective epimorphisms this exhibits
  $f$ as an $\infty$-colimit over equivalences, hence as
  an equivalence.
\endofproof
\begin{proposition}
  \label{MorphismsOfInfinityBundlesAreEquivalences}
  Every morphism between $G$-actions over $X$ that are 
  $G$-principal $\infty$-bundles over $X$ is an equivalence.
\end{proposition}
\proof
  Since a morphism of $G$-principal bundles
  $P_1 \to P_2$ is a morphism of {\v C}ech nerves that fixes 
  their $\infty$-colimit $X$, up to equivalence, 
  and since $* \to \mathbf{B}G$ is an effective 
  epimorphism,
  we are, by Proposition~\ref{LocalTrivialityImpliesCocycle}, in the situation of  
  Lemma~\ref{EquivalencesAreDetectedOverEffectiveEpimorphisms}.
\endofproof
\begin{theorem}
  \label{PrincipalInfinityBundleClassification}
  For all $X, \mathbf{B}G \in \mathbf{H}$ there is a natural
  equivalence of $\infty$-groupoids
  $$
    G \mathrm{Bund}(X)
    \simeq
    \mathbf{H}(X, \mathbf{B}G)
  $$ 
  which on vertices is the construction of Definition~\ref{BundleStructureOnInfinityFiber}:
  a bundle $P \to X$ is mapped to a morphism
  $X \to \mathbf{B}G$ such that $P \to X \to \mathbf{B}G$ is a fiber
  sequence.
\end{theorem}
We therefore say
\begin{itemize}
  \item $\mathbf{B}G$ is the \emph{classifying object} 
  or \emph{moduli $\infty$-stack} for 
  $G$-principal $\infty$-bundles;
  \item a morphism $c : X \to \mathbf{B}G$ is a \emph{cocycle}
  for the corresponding $G$-principal $\infty$-bundle and its class 
  $[c] \in \mathrm{H}^1(X,G)$ is its 
 \emph{characteristic class}.
\end{itemize}
\proof
  By Definitions~\ref{ActionInPrincipal} 
  and~\ref{principalbundle} and using 
  the refined statement of the third $\infty$-Giraud axiom  
  (Theorem~\ref{NaturalThirdGiraud}), the
  $\infty$-groupoid of $G$-principal $\infty$-bundles over $X$ is 
  equivalent to the fiber over $X$
  of the
  sub-$\infty$-category 
  of the slice
  of the arrow $\infty$-topos 
  on those squares
  $$
    \xymatrix{
	  P \ar[r] \ar@{->>}[d] & {*} \ar@{->>}[d]
	  \\
	  X \ar[r] & \mathbf{B}G
	}
  $$
  that exhibit $P \to X$ as a $G$-principal $\infty$-bundle.  By 
  Proposition~\ref{BundleStructureOnInfinityFiber} and
  Proposition~\ref{LocalTrivialityImpliesCocycle} these are 
  the $\infty$-pullback squares 
      $
    \mathrm{Cart}({\mathbf{H}^{\Delta[1]}}_{/{(* \to \mathbf{B}G)}})
    \hookrightarrow {\mathbf{H}^{\Delta[1]}}_{/{(* \to \mathbf{B}G)}}
  $, hence
  $$
    G \mathrm{Bund}(X) \simeq 
	 \mathrm{Cart}({\mathbf{H}^{\Delta[1]}}_{/{(* \to \mathbf{B}G)}}) \times_{\mathbf{H}} \{X\}
	 \,.
  $$
  By the universality of the $\infty$-pullback
  the morphisms between these are fully determined by their value on $X$,
  so that the above is equivalent to 
  $$
  \mathbf{H}_{/\mathbf{B}G} \times_{\mathbf{H}} \{X\}
	\,.
  $$
  (For instance in terms of model categories: choose a model structure for
  $\mathbf{H}$ in which all objects are cofibrant, choose a fibrant representative
  for $\mathbf{B}G$ and a fibration resolution $\mathbf{E}G \to \mathbf{B}G$
  of the universal $G$-bundle. Then the slice model structure of the arrow model structure
  over this presents the slice in question and the statement follows from the analogous
  1-categorical statement.)
  This finally is equivalent to 
  $$
	\mathbf{H}(X, \mathbf{B}G)
	\,.
  $$
  (For instance in terms of quasi-categories: the projection 
  $\mathbf{H}_{/\mathbf{B}G} \to \mathbf{H}$ is a fibration by 
  Proposition 2.1.2.1 and 4.2.1.6 in \cite{Lurie}, hence the homotopy fiber
  $\mathbf{H}_{/\mathbf{B}G} \times_{\mathbf{X}} \{X\}$ 
  is the ordinary fiber of quasi-categories. This is manifestly 
  the
  $\mathrm{Hom}^R_{\mathbf{H}}(X, \mathbf{B}G)$ from Proposition 1.2.2.3 of \cite{Lurie}.
  Finally, by Proposition 2.2.4.1 there, this is equivalent to $\mathbf{H}(X,\mathbf{B}G)$.)
\endofproof
\begin{corollary}
  Equivalence classes of $G$-principal $\infty$-bundles over $X$ are
  in natural bijection with the degree-1 $G$-cohomology of $X$:
  $$
    G \mathrm{Bund}(X)_{/\sim} \simeq H^1(X, G)
	\,.
  $$
\end{corollary}
\proof
  By Definition \ref{cohomology} this is the restriction of 
  the equivalence $G \mathrm{Bund}(X) \simeq \mathbf{H}(X, \mathbf{B}G)$ to
  connected components.
\endofproof

\paragraph{Universal principal $\infty$-bundles and the Borel construction}
\label{Universal princial bundles}
\index{Borel construction}
\index{universal principal $\infty$-bundle}

By prop. \ref{InftyGroupsBySimplicialGroups} every $\infty$-group
in an $\infty$-topos over an $\infty$-cohesive site is
presented by a (pre-)sheaf of simplicial groups, hence by a strict 
group object $G$ in a 1-category of simplicial (pre-)sheaves. We have seen in 
\ref{InfinityGroupPresentations} that for such a presentation
the delooping $\mathbf{B}G$ is presented by $\bar W G$. By the above discussion in 
\ref{Principal infinity-bundles general abstract} the theory of
$G$-principal $\infty$-bundles is essentially that of homotopy fibers
of morphisms into $\mathbf{B}G$, hence into $\bar W G$. 
By prop. \ref{ConstructionOfHomotopyLimits} such homotopy fibers are
computed as ordinary pullbacks of fibration resolutions of the point inclusion
into $\bar W G$. Here we discuss these fibration resolutions. They turn out 
to be the classical \emph{universal simplicial principal bundles} $W G \to \bar W G$.

This section draws from \cite{NSSb}.

\medskip

By prop. \ref{InftyGroupsBySimplicialGroups} every $\infty$-group
in an $\infty$-topos over an $\infty$-cohesive site is
presented by a (pre-)sheaf of simplicial groups, hence by a strict 
group object $G$ in a 1-category of simplicial (pre-)sheaves. We have seen in 
\ref{InfinityGroupPresentations} that for such a presentation
the delooping $\mathbf{B}G$ is presented by $\bar W G$. By the above discussion in 
\ref{Principal infinity-bundles general abstract} the theory of
$G$-principal $\infty$-bundles is essentially that of homotopy fibers
of morphisms into $\mathbf{B}G$, hence into $\bar W G$. 
By prop. \ref{ConstructionOfHomotopyLimits} such homotopy fibers are
computed as ordinary pullbacks of fibration resolutions of the point inclusion
into $\bar W G$. Here we discuss these fibration resolutions. They turn out 
to be the classical \emph{universal simplicial principal bundles} $W G \to \bar W G$.

\medskip

Let $C$ be some site. We consider group objects in the category of simplicial presheaves
$[C^{\mathrm{op}}, \mathrm{sSet}]$. Since sheafification preserves finite limits,
all of the following statements hold verbatim also in the category 
$\mathrm{Sh}(C)^{\Delta^{\mathrm{op}}}$
of simplicial sheaves over $C$.

\begin{definition}
 \label{BisimplicialActionGroupoid}
For $G$ be a group object in $[C^{\mathrm{op}}, \mathrm{sSet}]$ 
and for $\rho : P \times G \to P$
a $G$-action, its \emph{action groupoid object} is the simplicial object
$$
  P/\!/G \in [\Delta^{\mathrm{op}},[C^{\mathrm{op}}, \mathrm{sSet}
$$
whose value in degree $n$ is
\[
  (P/\!/G)_n :=  P\times G^{\times^n} \;\;\in [C^{\mathrm{op}}, \mathrm{sSet}]
  \,,
\]
whose face maps are given by 
\[
d_i(p,g_1,\ldots, g_n) = \begin{cases} 
(pg_1,g_2,\ldots, g_n) & \text{if}\ i=0, \\ 
(p,g_1,\ldots, g_ig_{i+1},\ldots, g_n) & \text{if}\ 1\leq i\leq n-1, \\ 
(p,g_1,\ldots, g_{n-1}) & \text{if}\ i=n, 
\end{cases} 
\]
and whose degeneracy maps are given by 
\[
s_i(p,g_1,\ldots, g_n) = (p,g_1,\ldots, g_{i-1},1,g_i,\ldots, g_n) 
\,.
\]
\end{definition}
\begin{definition}
\label{WeakQuotient}
For $\rho : P \times G \to P$ an action, write 
$$
  P/_h G := T(P/\!/G)\in [C^{\mathrm{op}}, \mathrm{sSet}]
$$
for the corresponding total simplicial object, def. \ref{TotalSimplicialSetAndTotalDecalage}.
\end{definition}
\begin{remark}
  According to corollary \ref{SimplicialHomotopyColimitByCodiagonal}
  the object $P/_h G$ presents the homotopy colimit over the simplicial object
  $P/\!/G$. We say that $P/_h G$ is the \emph{homotopy quotient} of $P$ by the
  action of $G$.
\end{remark}
\begin{example}
  \label{ActionOfSimplicialGroupOnPointAndOnItself}
  The unique trivial action of a group object $G$ on the terminal object $*$
  gives rise to a canonical action groupoid $*/\!/G$. 
  According to def. \ref{BarWAsCompositeWithTotal} we have
  $$
    * /_h G = \Wbar G
	\,.
  $$
  The multiplication morphism $\cdot : G \times G \to G$ regarded as an 
  action of $G$ on itself gives rise to a canonical action groupoid $G/\!/G$.
  The terminal morphism $G \to *$ induces a morphism of simplicial objects
  $$
    G/\!/G \to * /\!/G
	\,.
  $$
  Defined this way $G/\!/G$ carries a \emph{left} $G$-action relative to this
morphism. To stay with our convention that actions on bundles are from the right,
  we consider in the following instead the right action of $G$ on itself given by
  $$
    \xymatrix{
      G \times G \ar[r]^{\sigma} & 
	   G \times G \ar[rr]^{(\mathrm{(-)^{-1}, \mathrm{id}})}
	   &&
	   G \times G
	   \ar[r]^{\cdot}
	   &
	   G
	}
	\,,
  $$
  where $\sigma$ exchanges the two cartesian factors
  $$
    (h,g) \mapsto g^{-1} h
	\,.
  $$
  With respect to this action, the action groupoid object $G/\!/G$ 
  is canonically equipped with the right $G$-action
  by multiplication from the right. Whenever in the following we write
  $$
    G/\!/G \to */\!/G
  $$
  we are referring to this latter definition.
\end{example}
\begin{definition}
   \label{UniversalSimplicialPrincipalBundle}
   Given a group object in $[C^{\mathrm{op}}, \mathrm{sSet}]$,
   write
   $$
     (W G \to \bar W G) := (G/_h G \to */_h G) \;\; \in [C^{\mathrm{op}}, \mathrm{sSet}]
   $$
   for the morphism induced on homotopy quotients, def. \ref{WeakQuotient}, 
   by the morphism of canonical
   action groupoid objects of example \ref{ActionOfSimplicialGroupOnPointAndOnItself}.
   
   We will call this the \emph{universal weakly $G$-principal bundle}.
\end{definition}
   This term will be justified by prop. \ref{PropertiesOfUniversalGPrincipalBundle},
   remark \ref{TheUniversalBundleIsTheUniversalBundle}
   and theorem \ref{Classification theorem for weakly G-principal bundles} below.
   We now discuss some basic properties of this morphism.
\begin{definition}
  For $\rho : P \times G \to P$ a $G$-action in $[C^{\mathrm{op}}, \mathrm{sSet}]$, 
  we write 
  $$
    P \times_G W G := (P \times W G)/G \in [C^{\mathrm{op}}, \mathrm{sSet}]
  $$
  for the quotient by the diagonal $G$-action with respect to the
  given right $G$ action on $P$ and the canonical right $G$-action
  on $W G$ from prop. \ref{PropertiesOfUniversalGPrincipalBundle}. 
  We call this quotient the \emph{Borel construction} of the $G$-action 
  on $P$.
\end{definition}
\begin{proposition}
  \label{TotalSimplicialObjectByBorelConstruction}
For $P \times G \to P$ an action in $[C^{\mathrm{op}}, \mathrm{sSet}]$, there is an isomorphism
$$
  P/_h G
  \simeq
  P\times_G WG
  \,,
$$
between the homotopy quotient, def. \ref{WeakQuotient},
and the Borel construction.
In particular, for all $n \in \mathbb{N}$ there are ismorphisms
\[
  (P/_h G)_n \simeq P_n\times G_{n-1} \times \cdots \times G_0 
  \,.
\]
\end{proposition}
\begin{proof}
  This follows by a straightforward computation.
\end{proof}
\begin{lemma}
  \label{PropertiesOfSimplicialQuotientsBySimplicialGroups}
  Let $P$ be a Kan complex, $G$ a simplicial group and $\rho : P \times G \to P$
  an action. The following holds.
  \begin{enumerate}
    \item 
	  The qotient map $P \to P/G$ is a Kan fibration.
	\item 
	  If the action is free, then $P/G$ is a Kan complex. 
  \end{enumerate}
\end{lemma}
The second statement is for instance lemma V3.7 in \cite{GoerssJardine}.
\begin{lemma}
 \label{KanSimplicialHomotopyQuotientIsKan}
 For $P$ a Kan complex and $P \times G \to P$ an action by a group object, 
 the homotopy quotient $P /_h G$, def. \ref{WeakQuotient}, is itself a Kan complex.
\end{lemma}
\proof
  By prop. \ref{TotalSimplicialObjectByBorelConstruction} the homotopy
  quotient is isomorphic to the Borel construction. Since
  $G$ acts freely on $WG$ it acts freely on $P \times WG$.
  The statement then follows with lemma \ref{PropertiesOfSimplicialQuotientsBySimplicialGroups}.
\endofproof
\begin{proposition}
  \label{PropertiesOfUniversalGPrincipalBundle}
  For $G$ a group object in $[C^{\mathrm{op}}, \mathrm{sSet}]$, the 
  morphism $W G \to \Wbar G$ from def. \ref{UniversalSimplicialPrincipalBundle}
  has the following properties.
  \begin{enumerate}
    \item It is isomorphic to the traditional morphism denoted by these symbols, e.g. 
	  \cite{May}.
	 \item It is isomorphic to the d{\'e}calage morphism $\mathrm{Dec}_0 \Wbar G \to \Wbar G$,
	   def. \ref{Decalage}.
	 \item It is canonically equipped with a right $G$-action over $\Wbar G$
	   that makes it a weakly $G$-princial bundle (in fact the shear map is an isomorphism).
	 \item It is an objectwise Kan fibration replacement of the point inclusion $* \to \bar W G$.
  \end{enumerate}
\end{proposition}
This is lemma 10 in \cite{RobertsStevenson}.
\begin{remark}
  \label{TheUniversalBundleIsTheUniversalBundle}
  \label{UniversalPrincipalInfinityBundle}
  Let $\hat X \to \bar W G$ be a morphism in $[C^{\mathrm{op}}, \mathrm{sSet}]$,
  presenting, by prop. \ref{InftyGroupsBySimplicialGroups}, a morphism 
  $X \to \mathbf{B}G$
  in the $\infty$-topos $\mathbf{H} = \mathrm{Sh}_\infty(C)$.
  By prop. \ref{LocalTrivialityImpliesCocycle} every $G$-principal
  $\infty$-bundle over $X$ arises as the homotopy fiber of such a morphism.  By using 
  prop. \ref{PropertiesOfUniversalGPrincipalBundle} in prop. \ref{ConstructionOfHomotopyLimits}
  it follows that the principal $\infty$-bundle classified by $\hat X \to \bar W G$ is 
  presented by the
  ordinary pullback of $W G \to \bar W G$. This is the defining property of the
  universal principal bundle. 
\end{remark}
In \ref{Principal infinity-bundles presentations} below we show how this
observation leads to a complete presentation of the theory of principal $\infty$-bundles
by simplical weakly principal bundles.

\paragraph{Presentation in locally fibrant simplicial sheaves}
\label{Principal infinity-bundles presentations}

We discuss a presentation of the general notion of principal $\infty$-bundles,
\ref{Principal infinity-bundles general abstract} by weakly principal bundles
in a 1-category of simplicial sheaves.

\medskip

Let $\mathbf{H}$ be a hypercomplete $\infty$-topos
(for instance a cohesive $\infty$-topos), such that 
it admits a 1-site $C$ with enough points. 
\begin{observation}
By prop. \ref{CharacterizationOfLocalWeakEquivalence}
a category with weak equivalences that presents $\mathbf{H}$
under simplicial localization, def. \ref{SimplicialLocalization}, 
is the category of simplicial 1-sheaves 
on $C$, $\mathrm{sSh}(C)$, with the weak
equivalences $W \subset \mathrm{Mor}(\mathrm{sSh}(C))$ 
being the stalkwise weak equivalences:
$$
  \mathbf{H} \simeq L_W \mathrm{sSh}(C)
  \,.
$$
Also the full subcategory
$$
  \mathrm{sSh}(C)_{\mathrm{lfib}} \hookrightarrow \mathrm{sSh}(C)
$$
on the locally fibrant objects is a presentation.
\label{SimplicialSheavesWithStalkwiseWeakEquivalencesModel1LocalicHypercompleteInfinityTopos}
\end{observation}
\begin{corollary}
  Regard $\mathrm{sSh}(C)_{\mathrm{lfib}}$ as a category
  of fibrant objects, def. \ref{CategoryOfFibrantObjects}, 
  with weak equivalences and fibrations 
  the stalkwise weak equivalences and firations in $\mathrm{sSet}_{\mathrm{Quillen}}$,
  respectively, as in example \ref{BasicExamplesOfCatsOfFibObjects}.
  
  Then for any two objects $X, A \in \mathbf{H}$ there are simplicial sheaves, 
  to be denoted by the same symbols, such that the hom $\infty$-groupoid
  in $\mathbf{H}$ from $X$ to $A$ is presented in $\mathrm{sSet}_{\mathrm{Quillen}}$ 
  by the Kan complex of 
  cocycles \ref{GroupoidsOfMorphisms}.
\end{corollary}
\proof
  By theorem \ref{SimplicialLocalizationOfCatOfFibrantObjects}.
\endofproof
We now discuss for the general theory of principal $\infty$-bundles in 
$\mathbf{H}$ from \ref{Principal infinity-bundles general abstract}
a corresponding realization in the presentation for $\mathbf{H}$
given by $(\mathrm{sSh}(C), W)$.

By prop. \ref{InftyGroupsBySimplicialGroups} every $\infty$-group
in $\mathbf{H}$ is presented by an ordinary group in $\mathrm{sSh}(C)$.
It is too much to ask
that also every $G$-principal $\infty$-bundle is presented by a
principal bundle in $\mathrm{sSh}(C)$.
But something close is true: every principal $\infty$-bundle
is presented by a \emph{weakly principal} bundle in 
$\mathrm{sSh}(C)$. 

\medskip

\begin{definition}
  \label{WeaklyGPrincipalBundle}
Let $X \in \mathrm{sSh}(C)$ be any object, and let 
$G \in \mathrm{sSh}(C)$ 
be equipped with the structure of a group object. 
A {\em weakly $G$-principal bundle} is
\begin{itemize}
\item an object $P \in \mathrm{sSh}(C)$ (the {\em total space});

\item a local fibration $\pi\colon P\to X$ (the {\em bundle projection});

\item a right action 
  $$ 
    \raisebox{10pt}{
    \xymatrix{
      P \times G \ar[dr] \ar[rr]^{\rho} && P \ar[dl]
	  \\
	  & X
	}
	}
  $$ 
  of $G$ on $P$ over $X$
\end{itemize} 
such that 
\begin{itemize}
\item the action of $G$ is {\em weakly principal} in that the \emph{shear map}
\[
 (p_1, \rho) :  P \times G \to P \times_X P \qquad (p,g) \mapsto (p,p g)
\]
is a local weak equivalence.
\end{itemize}
\end{definition}

\begin{remark}
We do not ask the $G$-action to be degreewise free as in \cite{JardineLuo}, 
where a similar notion is considered. However we show in 
Corollary \ref{befreiung} below that each weakly $G$-principal bundle 
is equivalent to one with free $G$-action.
\end{remark}

\begin{definition}
A morphism of weakly $G$-principal bundles $(\pi,\rho) \to (\pi',\rho')$ over $X$ 
is a morphism $f : P \to P'$ in $\mathrm{sSh}(C)$ 
that is $G$-equivariant and 
commutes with the bundle projections, hence such that it makes this diagram 
commute:
$$
  \raisebox{20pt}{
  \xymatrix{
    P \times G
	\ar[rr]^{(f, \mathrm{id})}
	\ar[d]^{\rho}
	&&
	P' \times G
	\ar[d]^{\rho'}
	\\
    P \ar[dr]_{\pi} \ar[rr]^{f} && P' \ar[dl]^{\pi'}
	\\
	& X
  }
  }
  \,.
$$
Write 
$$
  \mathrm{w}G\mathrm{Bund}(X)
  \in 
  \mathrm{sSet}_{\mathrm{Quillen}}
$$
for the nerve of the category of weakly $G$-principal bundles and morphisms
as above. The $\infty$-groupoid that this presents under 
$\infty \mathrm{Grpd} \simeq (\mathrm{sSet}_{\mathrm{Quillen}})^\circ$
we call the \emph{$\infty$-groupoid of weakly $G$-principal bundles over $X$}.
\end{definition}

\begin{lemma}
 \label{multishear}
Let $\pi : P \to X$ be a weakly $G$-principal bundle.  Then the following statements are true: 
\begin{enumerate}
\item For any point $p : * \to P$ the action of $G$ induces a weak equivalence
\begin{equation*}
G \longrightarrow P_x  
\end{equation*}
where $x = \pi p$ and where $P_x$ is the fiber of $P\to X$ over $x$.
\item
For all $n \in \mathbb{N}$, the multi-shear maps 
\begin{equation*}
P \times G^n  \to P^{\times^{n+1}_X} \qquad (p,g_1,...,g_n) \mapsto (p,p g_1,...,p g_n)
\end{equation*}
are weak equivalences.
\end{enumerate}
\end{lemma}
\begin{proof}
We consider the first statement.  Regard the weak equivalence 
$P \times G \xrightarrow{\sim} P \times_X P$ 
as a morphism over $P$ where in both cases the map to $P$ is given by projection onto the first factor. 
By basic properties of categories of fibrant objects, both of these morphisms are
fibrations.
Therefore, by prop. \ref{PullbacksAlongLocalFibrationsAreHomotopyPullbacks}
the pullback of the shear map along $p$ is still a weak equivalence. 
But this pullback is just the map 
$G\to P_x$, which proves the claim.     

For the second statement, we use induction on $n$.  Suppose that 
$P\times G^n\to P^{\times^{n+1}_X}$ is a weak equivalence.  By
prop. \ref{PullbacksAlongLocalFibrationsAreHomotopyPullbacks}, 
the pullback 
$P^{\times^n_X}\times_X (P\times G)\to P^{\times^{n+2}_X}$ 
of the shear map itself along $P^{\times^n_X} \to X$ is again a 
weak equivalence, as is 
the product $P\times G^n\times G\to P^{\times^{n+1}_X}\times G$
of the $n$-fold shear map with $G$.
The composite of these two weak equivalences is the multi-shear map
$ P \times G^{n+1} \to P^{\times^{n+2}_X}$, which is hence a also weak equivalence.
\end{proof}

\begin{proposition}
Let $P \to X$ be a weakly $G$-principal bundle and let $f: Y \to X$ be an arbitrary morphism. 
Then the pullback $f^*P \to Y$ exists and is also 
canonically a weakly $G$-principal bundle. This 
operation extends to define a pullback morphism
$$
  f^* : \mathrm{w}G\mathrm{Bund}(X) \to \mathrm{w}G\mathrm{Bund}(Y)
  \,.
$$
\end{proposition}
\begin{proof}
By basic properties of a category of fibrant objects:

The pullback $f^*P$ exists  
and the morphism $f^*P\to Y$ is again a local fibration. 
Thus it only 
remains to show that $f^*P$ is weakly principal, i.e.\  that the 
morphism $f^*P \times G \to f^*P \times_Y f^*P$ is a weak 
equivalence. This follows from prop. \ref{PullbacksAlongLocalFibrationsAreHomotopyPullbacks}. 
\end{proof}

\begin{remark}
The functor $f^*$ associated to the 
map $f\colon Y\to X$ above is the restriction of a functor 
$f^*\colon \mathrm{sSh}(C)/X \to \mathrm{sSh}(C)/Y $ mapping from simplicial sheaves over 
$X$ to simplicial sheaves over $Y$.  This functor $f^*$ has a left 
adjoint 
$f_!\colon \mathrm{sSh}(C)/Y \to \mathrm{Sh}^{\Delta^{\mathrm{op}}}/X $
given by composition 
along $f$, in other words 
\[
f_!(E\to Y) = E\to Y\xrightarrow{f} X.
\]  
Note that the functor $f_!$ does not usually restrict to a functor 
$f_!\colon \mathrm{w}G\mathrm{Bund}(Y) \to \mathrm{w}G\mathrm{Bund}(X)$. 
But when it does, we say that 
principal $\infty$-bundles {\em satisfy descent along $f$}.  In this situation, 
if $P$ is a weakly $G$-principal bundle on $Y$, then $P$ is weakly equivalent 
to the pulled-back principal $\infty$-bundle $f^*f_!P$ on $Y$, in other words 
$P$ `descends' to $f_!P$.  
\end{remark}
The next result 
says that weakly $G$-principal bundles satisfy descent along 
local acyclic fibrations (hypercovers).  
 
\begin{proposition}
  \label{PushforwardOfGBundlesAlongHypercovers}
Let $p: Y \to X$ be a local acyclic fibration in $\mathrm{sSh}(C)$.  
Then the functor $p_!$ defined above restricts to 
a functor $p_!\colon \mathrm{w}G\mathrm{Bund}(Y) \to \mathrm{w}G\mathrm{Bund}(X)$, 
left adjoint to $p^*\colon \mathrm{w}G\mathrm{Bund}(X) \to 
\mathrm{w}G\mathrm{Bund}(Y)$, hence to a homotopy equivalence in $\mathrm{sSet}_{\mathrm{Quillen}}$.
\end{proposition}
\begin{proof}
Given a weakly $G$-principal bundle $P \to Y$, 
the first thing we have to check is that the map $P \times G \to P \times_X P$ 
is a weak equivalence. This map can be factored as 
$P \times G \to P\times_Y P \to P \times_X P$. 
Hence it suffices to show that the map $P \times_Y P \to P \times_X P$ 
is a weak equivalence. 
But this follows by prop. \ref{PullbacksAlongLocalFibrationsAreHomotopyPullbacks}, 
since both pullbacks are along local fibrations and $Y \to X$ is a local weak equivalence
by assumption. 

This establishes the existence of the functor $p_!$. 
It is easy to see that it is left adjoint to $p^*$. This implies that it induces a homotopy equivalence in $\mathrm{sSet}_{\mathrm{Quillen}}$. 
\end{proof} 
\begin{corollary}
For $f: Y \to X$ a local weak equivalence,
the induced functor $f^*: \mathrm{w}G\mathrm{Bund}(X) \to \mathrm{w}G\mathrm{Bund}(Y)$ 
is a homotopy equivalence.
\end{corollary}
\begin{proof}
By lemma \ref{FactorizationLemma} we can factor the weak equivalence $f$ 
into a composite of a local acyclic fibration and a left inverse to a local acyclic fibration. 
Therefore, by prop. \ref{PushforwardOfGBundlesAlongHypercovers}, $f^*$
may be factored as the composite of two homotopy equivalences, hence is itself a homotopy
equivalence.
\end{proof}

\medskip

We discuss now how weakly $G$-principal bundles arise from the universal
$G$-principal bundle, def. \ref{UniversalSimplicialPrincipalBundle} by pullback, and how
this establishes their equivalence with $G$-ccoycles.

\begin{proposition}
  For $G$ a group object in $\mathrm{sSh}(C)$,
  the map $W G \to \Wbar G$ from def. \ref{UniversalSimplicialPrincipalBundle}
  equipped with the $G$-action of prop. \ref{PropertiesOfUniversalGPrincipalBundle}
  is a weakly $G$-principal bundle.
\end{proposition}
Indeed, it is a strictly $G$-principal bundle. This is a classical fact,
for instance around lemma V4.1 in \cite{GoerssJardine}.In terms of
the total simplicial set functor it is  
observed in section 4 of  \cite{RobertsStevenson}. \newline
\begin{proof}
  By inspection one finds that 
  $$
    \xymatrix{
	  (G /\!/G) \times G \ar[d]\ar[r]  & G /\!/ G \ar[d]
	  \\
	  G /\!/G \ar[r] & {*}/\!/G
	}
  $$
  is a pullback diagram in $[\Delta^{\mathrm{op}}, \mathrm{sSh}(C)]$. 
  Since the total simplicial object functor $T$ of def. \ref{TotalSimplicialSetAndTotalDecalage} 
  is right
  adjoint it preserves this pullback. This shows the principality of the shear map.
\end{proof}
\begin{definition}
  \label{CechNerve}
  For $Y \to X$ a morphism in $\mathrm{sSh}(C)$, write
  $$
    \check{C}(Y)
	\in [\Delta^{\mathrm{op}}, \mathrm{sSh}(C)]
  $$
  for its \emph{{\v C}ech nerve}, 
  given in degree $n$ by the $n$-fold fiber product of
  $Y$ over $X$
  $$
    \check{C}(Y)_n := Y^{\times_X^{n+1}}
	\,.
  $$
\end{definition}
\begin{observation}
  \label{CanonicalMorphismOutOfChechNerveHocolim}
  The canonical morphism of simplicial objects $\check{C}(Y) \to X$, with $X$
  regarded as a constant simplicial object induces under 
  totalization, def. \ref{TotalSimplicialSetAndTotalDecalage}, 
  and by prop. \ref{TotalSimpSetEquivalentToDiagonal} a 
  canonical morphism
  $$
    T \check{C}(Y) \to X\;\; \in \mathrm{sSh}(C)
	\,.
  $$
\end{observation}
\begin{lemma}
  \label{CechNerveProjectionIsWeakEquivalence}
  For $p : Y \to X$ a local acyclic fibration, the morphism 
  $T \check{C}(Y) \to X$ from observation \ref{CanonicalMorphismOutOfChechNerveHocolim}
  is a local weak equivalence.
\end{lemma}
\begin{proof}
  By pullback stability of local acylic fibrations, for each $n \in \mathbb{N}$
  the morphism $Y^{\times^n_X} \to X$ is a local weak equivalence. By
  remark. \ref{Total simplicial object is built from finite limits}
  and prop. \ref{TotalSimpSetEquivalentToDiagonal} this degreewise local weak 
  equivalence is preserved by the functor $T$.
\end{proof}
The main statement now is the following.
\begin{theorem}
  \label{HomotopyQuotientWPrincBundleLoAcyclicFibration}
   For $P \to X$ a weakly $G$-principal bundle in $\mathrm{sSh}(C)$, 
   the canonical morphism
  $$
    P \!/_h G  \longrightarrow X
  $$
  is a local acyclic fibration.
\end{theorem}
\begin{proof}
To see that the morphism is a local weak equivalence,
factor $P/\!/G \to X$ in $[\Delta^{\mathrm{op}}, \mathrm{sSh}(C)]$ 
via the multi-shear maps from lemma \ref{multishear} through the
{\v C}ech nerve, def. \ref{CechNerve}, as  
\[
  P/\!/G \to \check{C}(P) \to X
  \,.
\]
Applying to this the total simplicial object functor $T$, 
def. \ref{TotalSimplicialSetAndTotalDecalage}, 
yields a factorization
$$
  P \!/_h G  \to  T \check{C}(P)  \to X
  \,.
$$
The left morphism is a weak equivalence because, by lemma \ref{multishear}, 
the multi-shear maps are weak equivalences and by 
corollary \ref{SimplicialHomotopyColimitByCodiagonal} 
$T$ preserves sends degreewise weak equivalences to weak equivalences. 
The right map is a weak equivalence by lemma \ref{CechNerveProjectionIsWeakEquivalence}.

We now prove that $P/_h G\to X$ is a local fibration.  
We need to show that for each topos point $p$ of $\mathrm{Sh}(C)$
the morphism  of stalks $p(P/_h G)\to p(X)$ is a Kan fibration of simplicial sets.
By prop. \ref{TotalSimplicialObjectByBorelConstruction} this means equivalently
that the morphism
$$
  p( P \times_G W G ) \to p(X) 
$$
is a Kan fibration.
By definition of topos point, $p$ commutes with all the finite products 
and colimits involved here. Therefore equivalently we need to show that
$$
  p(P) \times_{p(G)} W p(G)  \to p(X) 
$$
is a Kan fibration for all topos points $p$.

Observe that this morphism factors the projection 
$p(P) \times W (p(G))\to p(X)$ as 
$$
  p(P)\times W(p(G))\to p(P) \times_{p(G)} W (p(G))\to p(X) 
$$
in $\mathrm{sSet}$. Here the first morphism is a Kan fibration by
lemma \ref{PropertiesOfSimplicialQuotientsBySimplicialGroups}, which 
in particular is also surjective on vertices. Also the total composite morphism is a Kan 
fibration, since $W (p(G))$ is Kan fibrant.  
From this the desired result follows with the next lemma \ref{RightCancelSurjectiveFibFromFib}.  
\end{proof}
\begin{lemma} 
  \label{RightCancelSurjectiveFibFromFib}
Suppose that $X\xrightarrow{p} Y \xrightarrow{q} Z$ is a 
diagram of simplicial sets such that $p$ is a Kan fibration surjective 
on vertices and $qp$ is a Kan 
fibration.  Then $q$ is also a Kan fibration.  
\end{lemma} 

\begin{proof} 
Consider a lifting problem of the form 
\[
\xymatrix{ 
\Lambda^k[n] \ar[r] \ar[d] & Y \ar[d]^-q \\ 
\Delta[n] \ar[r] & Z.  } 
\]
Choose a 0-simplex of $X$ which projects to the 0-simplex 
of $Y$ corresponding to the image of the vertex 0 under the 
map $\Lambda^k[n]\to Y$.  Since 
$\Delta[0]\to \Lambda^k[n]$ is an acyclic cofibration, we may 
choose a map $\Lambda^k[n]\to 
X$ such that the diagram 
\[
\xymatrix{ 
\Delta[0] \ar[d] \ar[r] & X \ar[d]^-p \\ 
\Lambda^k[n] \ar[r] \ar[ur] & Y }
\]
commutes.  This map gives rise to a commutative diagram 
\[
\xymatrix{ 
\Lambda^k[n] \ar[r] \ar[d] & X \ar[d]^-{qp} \\ 
\Delta[n] \ar[r] & Z  } 
\]
and any diagonal filler in this diagram gives a solution of the 
original lifting problem.  
\end{proof}

\medskip

We now discuss the equivalence between weakly $G$-principal bundles and $G$-cocycles.
For $X, A \in \mathrm{sSh}(C)$, write $\mathrm{Cocycle}(X,A)$
for the category of cocycles from $X$ to $A$, according to \ref{GroupoidsOfMorphisms}.
\begin{definition}
 \label{FunctorsBetweenBundlesAndCocycles}
   Let $X,G \in \mathrm{sSh}(C)$ with $G$ equipped with the
   structure of a group object (hence necessarily locally fibrant) and also
   with $X$ being locally fibrant.

   Define a functor
   $$
     \mathrm{Extr} : \mathrm{w}G\mathrm{Bund}(X) \to \mathrm{Cocycle}(X, \Wbar G)
   $$
   (``extracting'' a cocycle) 
   on objects by sending a weakly $G$-principal bundle $P \to X$ to the cocycle  
$$
  \xymatrix{
    X \ar@{<<-}[r]^\sim & P/_h G \ar[r] & \Wbar G
  }
  \,,
$$
where the left morphism is the local acyclic fibration from 
theorem \ref{HomotopyQuotientWPrincBundleLoAcyclicFibration}, 
and where the right morphism
is the image under the total simplicial object functor, 
def. \ref{TotalSimplicialSetAndTotalDecalage}, of the 
canonical morphism $P/\!/G \to */\!/G$ of simplicial objects.

Define also a functor 
\[
  \mathrm{Rec} : \mathrm{Cocycle}(X,\Wbar G) \to \mathrm{w}G\mathrm{Bund}(X) 
\]
(``reconstruction'' of the bundle)
which on objects takes a cocycle $X\xleftarrow{\pi} Y \xrightarrow{g} \Wbar G$ to the 
weakly $G$-principal bundle
$$
  g^* W G \to Y \stackrel{\pi}{\to} X
  \,,
$$
which is the pullback of the universal $G$-principal bundle, 
def. \ref{UniversalSimplicialPrincipalBundle},
along $g$,
and which on morphisms takes a coboundary to the morphism between pullbacks induced from
the corresponding morphism of pullback diagrams.
\end{definition}
\begin{observation}  
  \label{TheUniversalCocycle}
  The functor $\mathrm{Extr}$ sends the universal $G$-principal bundle
  $W G \to \Wbar G$ to the cocycle
  $$
    \Wbar G \simeq * \times_G W G \stackrel{\simeq}{\leftarrow}
	W G \times_G W G \stackrel{\simeq}{\to} W G \times_G * \simeq \Wbar G
	\,.
  $$
  Write 
  $$
    q : \mathrm{Cocycle}(X,  \Wbar G) \to \mathrm{Cocycle}(X,  \Wbar G)
  $$
  for the functor given by postcomposition with this universal cocycle.
  This has an evident left and right adjoint $\bar q$. Therefore 
  under the simplicial nerve these
  functors induce homotopy equivalences in $\mathrm{sSet}_{\mathrm{Quillen}}$.
\end{observation}
\begin{theorem}
 \label{Classification theorem for weakly G-principal bundles}
  The functors $\mathrm{Extr}$ and $\mathrm{Rec}$ from def. \ref{FunctorsBetweenBundlesAndCocycles}
  induce weak equivalences
 $$
   N \mathrm{w}G\mathrm{Bund}(X) \simeq N \mathrm{Cocycle}(X, \Wbar G)
   \;\;
   \in \mathrm{sSet}_{\mathrm{Quillen}}
 $$
 between the simplicial nerves of the category of weakly $G$-principal bundles
 and of cocycles, respectively.
\end{theorem}
\begin{proof}
We construct natural transformations
$$
  \mathrm{Extr} \circ \mathrm{Rec} \Rightarrow q
$$
and
$$
  \mathrm{Rec} \circ \mathrm{Extr} \Rightarrow \mathrm{id}
  \,,
$$
where $q$ is the homotopy equivalence from observation \ref{TheUniversalCocycle}.

For
$$
  X \xleftarrow{\pi} Y \xrightarrow{f} \Wbar G. 
$$
a cocycle, its image under $\mathrm{Extr} \circ \mathrm{Rec} $ is  
\[
  X \leftarrow (f^* WG) /_h G \to \Wbar G
  \,.
\] 
The morphism $(f^*WG) /_hG $ factors through $Y$ by construction, 
so that the left triangle in the diagram
\[
\xymatrix@R=0.7pc {
  & & (f^*WG) /_hG \ar[dd]\ar@{->>}[lld]_\sim\ar[rrd] & & 
  \\
  X & & & & \Wbar G 
  \\
  & & Y \ar@{->>}[llu]^\sim\ar[rru]^{q(f)} & & 
}
\]
commutes. The top right morphism is by definition the image under the 
total simplicial set functor, def. \ref{TotalSimplicialSetAndTotalDecalage}, of 
$(f^* WG) /\!/ G \to * /\!/ G$. This factors the top horizontal morphism in
$$
  \raisebox{20pt}{
  \xymatrix{
    (f^* W G) /\!/ G \ar[d] \ar[r] & (W G)/\!/G \ar[d] \ar[r] & {*} /\!/G 
	\\
	Y \ar[r]^f & \Wbar G 
  }
  }
  \,.
$$
Applying the total simplicial object functor to this diagram gives the above commuting
triangle on the right. Clearly this construction is natural and hence provides 
a natural transformation $\mathrm{Extr} \;\mathrm{Rec} \Rightarrow q$.

For the other natural trasformation, 
let now $P \to X$ be a weakly $G$-principal bundle. This induces the following
commutative diagram of simplicial objects (with $P$ and $X$ regarded as constant
simplicial objects)
$$
  \raisebox{20pt}{
  \xymatrix{
    P \ar@{<-}[r] \ar[d] & P \times_X (P /\!/G) \ar[d] &
    (P \times G) /\!/ G \ar[l]^<<<<\sim_<<<<{\phi} \ar[r]  \ar[d] &	G /\!/ G \ar[d]	  
    \\
    X \ar@{<-}[r] & P/\!/G  \ar@{=}[r] & P/\!/G \ar[r] & {*} /\!/ G
  }
  }
  \,,
$$
where the left and the right square are pullbacks, and where the 
top horizontal morphism $\phi$ is 
the degreewise local weak equivalence which is degreewise 
induced by the shear map, composed with exchange of the two factors. 

Explicitly, in degree 0 the morphism $\phi$ is given on generalized elements by
$$
  \xymatrix{
     (p', g) & \ar@{|->}[l]_{\phi_0} (p' g , p')
  }
$$
and in degree 1 by
$$
  \raisebox{20pt}{
  \xymatrix{
    (p'g, (p',h))
	\ar@{|->}[d]^{d_0}
     &
    ((p', g),h)
	\ar@{|->}[d]^{d_0}
	\ar@{|->}[l]_{\phi_1}
	\\
	(p'g, p' h)
	&
	((p' h,h^{-1} g)
	\ar@{|->}[l]_{\phi_0}
  }
  }
  \,,
$$
etc. Here the top horizontal morphisms also respect the right $G$-actions $\rho$ induced from
the weakly $G$-principal bundle structure on $P \to X$ and on $G/\!/G \to */\!/G$.
For instance the respect of the right $G$-action of $\phi$ in degree 0 is on
elements verified by
$$
  \raisebox{20pt}{
  \xymatrix{
    ((p'g, p'), k)
	\ar@{|->}[d]^{\rho}
     &
    ((p', g), k)
	\ar@{|->}[d]^{\rho}
	\ar@{|->}[l]_{\phi_0}
	\\
	(p'g k, p' )
	&
	((p' , g k )
	\ar@{|->}[l]_{\phi_0}
  }
  }
  \,.	
$$

The image of the above diagram under the total simplicial object functor, which 
preserves all the pullbacks and weak equivalences involved, is
$$
  \raisebox{20pt}{
  \xymatrix{
    P \ar@{<<-}[r]^<<<<<<<\sim \ar@{->>}[d] & P \times_X P /_h G \ar@{->>}[d] &
    (P \times G) /_h G \ar[l]_\sim \ar[r]  \ar@{->>}[d] &	W G \ar@{->>}[d]	  
    \\
    X \ar@{<<-}[r]^\sim & P/_hG  \ar@{=}[r] & P/_hG \ar[r] & \Wbar G
  }
  }
  \,.
$$
Here the total bottom span is the cocycle $\mathrm{Extr}(P)$, and so 
the object $(P \times G)/_h G$ over $X$ is $\mathrm{Rec}(\mathrm{Extr}(P))$. 
Therefore this exhibits a natural morphism $\mathrm{Rec} \,\mathrm{Extr} P \to P$.
\end{proof}
\begin{remark}
  \label{ClassificationTheoremRelatedToCocyclesSpaces}
  By theorem \ref{SimplicialLocalizationOfCatOfFibrantObjects}
  the simplicial set $N \mathrm{Cocycle}(X, \Wbar G)$ is a presentation of the 
  intrinsic cocycle $\infty$-groupoid $\mathbf{H}(X, \mathbf{B}G)$
  of the hypercomplete $\infty$-topos
 $\mathbf{H} = \mathrm{Sh}_\infty^{\mathrm{hc}}(C)$.
 Therefore the equivalence of theorem \ref{Classification theorem for weakly G-principal bundles}
 is a presentation of that of
 theorem \ref{PrincipalInfinityBundleClassification},
 $$
   G \mathrm{Bund}_\infty(X) \simeq \mathbf{H}(X, \mathbf{B}G)
 $$
 between the $\infty$-groupoid of $G$-principal $\infty$-bundles in $\mathbf{H}$
  and 
 the intrinsic cocycle $\infty$-groupoid of $\mathbf{H}$.
\end{remark}
\begin{corollary}
 \label{befreiung}
For each weakly $G$-principal bundle $P \to X$ there is a weakly $G$-principal
bundle $P^{f}$ with a levelwise free $G$-action and a weak equivalence 
$P^{f} \xrightarrow\sim P$ of weakly $G$-principal bundles over $X$. 
In fact, the assignment $P \mapsto P^f$ is an homotopy inverse to the full inclusion
of weakly $G$-principal bundles with free action into all weakly $G$-principal bundles.
\end{corollary}
\begin{proof}
Note that the universal bundle $WG \to \Wbar G$ carries a free $G$-action, in the sense that the levelwise action of $G_n$ on $(W G)_n$ is free. 
This means that the functor $\mathrm{Rec}$ 
from the proof of theorem \ref{Classification theorem for weakly G-principal bundles} 
indeed takes values in weakly $G$-principal budles with free action.
 Hence we can set 
 $$
   P^f 
     := 
    \mathrm{Rec}(\mathrm{Extr}(P)) = (P \times G) /_h G
	\,.
 $$
 By the discussion there we have a natural morphism $P^f \to P$ and one checks
 that this exhibits the homomotopy inverse.
\end{proof}

\subsubsection{Associated fiber bundles}
\label{AssociatedBundles}
\index{structures in a cohesive $\infty$-topos!$\infty$-group actions}
\index{structures in a cohesive $\infty$-topos!associated bundles}

We discuss the notion of representations/actions/modules of 
$\infty$-groups in an $\infty$-topos and the 
structures directly induced by this: the corresponding
twisted cohomology is cohomology with coefficients in 
\emph{modules} (the generalization of group cohomology with
coefficients in a module) and the corresponding notion of 
\emph{associated $\infty$-bundles}.

\paragraph{General abstract}

This section draws from \cite{NSSa}.

\medskip

Let $\mathbf{H}$ be an $\infty$-topos, $G \in \mathrm{Grp}(\mathbf{H})$
an $\infty$-group.
Fix an action $\rho : V \times G \to V$ (Definition \ref{ActionInPrincipal}) on an object $V\in \mathbf{H}$.
We discuss the induced notion of \emph{$\rho$-associated $V$-fiber $\infty$-bundles}.
We show that there is a \emph{universal} $\rho$-associated $V$-fiber bundle over 
$\mathbf{B}G$ and observe that under Theorem \ref{PrincipalInfinityBundleClassification}
this is effectively identified with the action itself. Accordingly, we also further discuss
$\infty$-actions as such.

\medskip

\begin{definition}
  For $V,X \in \mathbf{H}$ any two objects, 
a \emph{$V$-fiber $\infty$-bundle} over $X$ is a morphism $E \to X$, 
such that there is an effective epimorphism
$\xymatrix{U \ar@{->>}[r] & X}$ and an $\infty$-pullback of the form
$$
  \raisebox{20pt}{
  \xymatrix{
    U \times V \ar[r] \ar[d] & E \ar[d]
	\\
	U \ar@{->>}[r] & X\, .
  }
  }
$$  
 \label{FiberBundle}
\end{definition}
We say that $E \to X$ locally trivializes with respect to $U$.
As usual, we often say \emph{$V$-bundle} for short.

\begin{definition}
  For $P \to X$ a $G$-principal $\infty$-bundle, we write
  $$ 
    P \times_G V := (P\times V)/\!/G 
  $$ 
  for the $\infty$-quotient of the diagonal $\infty$-action of $G$ on $P \times V$.
  Equipped with the canonical morphism
  $P \times_G V \to X$ we call this the $\infty$-bundle \emph{ $\rho$-associated} to $P$.
  \label{AssociatedBundle}
\end{definition}
\begin{remark}
  The diagonal $G$-action on $P \times V$ is the product in 
  $G \mathrm{Action}(\mathbf{H})$ of the given actions on $P$ and on $V$.
  Since $G\mathrm{Action}(\mathbf{H})$ is a full sub-$\infty$-category of a slice
  category of a functor category, the product is given by a degreewise
  pullback in $\mathbf{H}$:
  $$
    \raisebox{20pt}{
    \xymatrix{
	  P \times V \times G^{\times_n} 
	  \ar[r]
	  \ar[d]
	  &
	  V \times G^{\times_n}
	  \ar[d]
	  \\
	  P \times G^{\times_n}
	  \ar[r]
	  &
	  G^{\times_n}\,.
	}
	}
  $$
  and so
  $$
    P \times_G V \simeq \varinjlim_n (P \times V \times G^{\times_n})
	\,.
  $$
  The canonical bundle morphism of the corresponding $\rho$-associated
  $\infty$-bundle  is the realization of the left morphism of this diagram:
  $$
    \raisebox{20pt}{
    \xymatrix{
	  P \times_G V
	  \ar@{}[r]|<<<<{:=}
	  \ar[d]
	  &
	  \varinjlim_n (P \times V \times G^{\times_n})
	  \ar[d]
	  \\
	  X \ar@{}[r]|<<<<<<<<<{\simeq} & 
	  \varinjlim_n (P \times G^{\times_n})\,.
	}
	}
  $$
  \label{ProductActionByPullback}
\end{remark}
\begin{example}
By Theorem \ref{PrincipalInfinityBundleClassification} every $\infty$-group action
$\rho : V \times G \to V$ has a classifying morphism $\mathbf{c}$ defined on its homotopy
quotient, which fits into a fiber sequence of the form
$$
  \raisebox{20pt}{
  \xymatrix{
     V \ar[r] & V/\!/G \ar[d]^{\mathbf{c}} 
     \\	 
	 & \mathbf{B}G\,.
  }
  }
$$

  Regarded as an 
  $\infty$-bundle, this is
  $\rho$-associated to the universal $G$-principal $\infty$-bundle
  $\xymatrix{{*} \ar[r] & \mathbf{B}G}$ from Example \ref{UniversalPrincipal}:
  $$
    V/\!/G \simeq {*} \times_G V
	\,.
  $$
  \label{ActionGroupoidIsRhoAssociated}
\end{example}
\begin{lemma}
  The realization functor $\varinjlim : \mathrm{Grpd}(\mathbf{H}) \to \mathbf{H}$
  preserves the $\infty$-pullback of Remark \ref{ProductActionByPullback}:
  $$
    P \times_G V \simeq \varinjlim_n (P \times V \times G^{\times_n})
	\simeq
	(\varinjlim_n P \times G^{\times_n}) \times_{(\varinjlim_n G^{\times_n})} (\varinjlim_n V \times G^{\times_n})
	\,.
  $$
  \label{RealizationPreservesProductOfRepresentations}
\end{lemma}
\proof
  Generally, let $X \to Y \leftarrow Z \in \mathrm{Grpd}(\mathbf{H})$ be a
  diagram of groupoid objects, such that in the induced diagram
  $$
    \xymatrix{
	  X_0 \ar[r] \ar@{->>}[d] & Y_0 \ar@{<-}[r] \ar@{->>}[d] & Z_0 \ar@{->>}[d]
	  \\
	  \varinjlim_n X_n \ar[r] & \varinjlim_n Y_n \ar@{<-}[r] & \varinjlim_n Z_n
	}
  $$
  the left square is an $\infty$-pullback. By the third 
  $\infty$-Giraud axiom (Definition~\ref{GiraudRezkLurieAxioms}) the vertical 
  morphisms are effective epi, as indicated. 
  By assumption we have a pasting of $\infty$-pullbacks as shown on the
  left of the following diagram, and by
  the pasting law (Proposition \ref{PastingLawForPullbacks}) this is equivalent to
  the pasting shown on the right:
  $$
    \raisebox{38pt}{
    \xymatrix{
	  X_0 \times_{Y_0} Z_0 \ar[r] \ar[d] & Z_0 \ar[d]
	  \\
	  X_0 \ar[r] \ar[d] & Y_0 \ar[d]
	  \\
	  \varinjlim_n X_n \ar[r] & \varinjlim_n Y_n
	}
	}
	\;\;\;
	\simeq
	\;\;\;
    \raisebox{38pt}{
    \xymatrix{
	  X_0 \times_{Y_0} Z_0 \ar[r] \ar@{->>}[d] & Z_0 \ar@{->>}[d]
	  \\
	  (\varinjlim_n X_n) \times_{(\varinjlim_n Y_n)} (\varinjlim_n Z_n) \ar[r] \ar[d] & 
	  \varinjlim_n Z_n \ar[d]
	  \\
	  \varinjlim_n X_n \ar[r] & \varinjlim_n Y_n.
	}
	}
  $$
Since effective epimorphisms are stable under $\infty$-pullback, this identifies 
the canonical morphism 
$$
  X_0 \times_{Y_0} Z_0
  \to 
  (\varinjlim_n X_n) \times_{(\varinjlim_n Y_n)} (\varinjlim_n Z_n)
$$
as an effective epimorphism, as indicated. 

Since $\infty$-limits commute over each other, the {\v C}ech nerve of this morphism 
is the groupoid object $[n] \mapsto X_n \times_{Y_n} Z_n$.
Therefore the third $\infty$-Giraud axiom now says that $\varinjlim$ preserves the
$\infty$-pullback of groupoid objects:
$$
  \varinjlim (X \times_Y Z) 
   \simeq  
  \varinjlim_n (X_n \times_{Y_n} Z_n )
   \simeq
  (\varinjlim_n X_n) \times_{(\varinjlim_n Y_n)} (\varinjlim_n Z_n)
  \,.
$$

Consider this now in the special case that $X \to Y \leftarrow Z$ is 
$(P \times G^{\times_\bullet}) \to G^{\times_\bullet} \leftarrow (V \times G^{\times_\bullet})$.
Theorem \ref{PrincipalInfinityBundleClassification} implies that the initial assumption above is 
met, in that $P \simeq (P/\!/G) \times_{*/\!/G} {*} \simeq X \times_{\mathbf{B}G} {*}$, 
and so the claim follows.
\endofproof
\begin{proposition}
  For $g_X : X \to \mathbf{B}G$ a morphism and $P \to X$
  the corresponding $G$-principal $\infty$-bundle according to Theorem 
  \ref{PrincipalInfinityBundleClassification},  
  there is a natural equivalence
  $$
    g_X^*(V/\!/G) \simeq P \times_G V
  $$
  over $X$, between the pullback of the
  $\rho$-associated $\infty$-bundle
  $\xymatrix{V/\!/G \ar[r]^{\mathbf{c}} & \mathbf{B}G}$
  of Example \ref{ActionGroupoidIsRhoAssociated}
  and the $\infty$-bundle $\rho$-associated to $P$ by Definition \ref{AssociatedBundle}.
  \label{UniversalAssociatedBundle}
  \label{AssociatedBundleByRho}
  \label{TwistingCocycleAsAssociatedBundle}
\end{proposition}
\proof
 By Remark \ref{ProductActionByPullback} the product action is given by the 
 pullback 
 $$
   \xymatrix{
     P \times V \times G^{\times_\bullet}
	 \ar[r]
	 \ar[d]
	 &
	 V \times G^{\times_\bullet}
	 \ar[d]
	 \\
	 P \times G^{\times_\bullet} \ar[r] & G^{\times_\bullet}
   }
 $$
 in $\mathbf{H}^{\Delta^{\mathrm{op}}}$. 
 By Lemma $\ref{RealizationPreservesProductOfRepresentations}$ the realization functor
 preserves this $\infty$-pullback. By 
 Remark \ref{ProductActionByPullback} it sends the left morphism to the 
 associated bundle, and by Theorem \ref{PrincipalInfinityBundleClassification}
 it sends the bottom morphism to $g_X$. Therefore it produces an $\infty$-pullback
 diagram of the form
 $$
   \raisebox{20pt}{
   \xymatrix{
     V \times_G P \ar[r] \ar[d] & V/\!/G \ar[d]^{\mathbf{c}}
	 \\
	 X \ar[r]^{g_X} & \mathbf{B}G\,.
   }
   }
 $$ 
\endofproof
\begin{remark}
  This says that $\xymatrix{V/\!/G \ar[r]^{\mathbf{c}} & \mathbf{B}G}$ is both, 
  the $V$-fiber $\infty$-bundle 
  $\rho$-associated to the universal $G$-principal $\infty$-bundle, Observation
  \ref{ActionGroupoidIsRhoAssociated},  
  as well as the universal $\infty$-bundle for $\rho$-associated $\infty$-bundles.
  \label{RhoAssociatedToUniversalIsUniversalVBundle}
\end{remark}
\begin{proposition}
  Every $\rho$-associated $\infty$-bundle is a $V$-fiber $\infty$-bundle, 
  Definition \ref{FiberBundle}.
  \label{AssociatedIsFiberBundle}
\end{proposition}
\proof
  Let $P \times_G V \to X$ be a $\rho$-associated $\infty$-bundle.
  By the previous Proposition \ref{UniversalAssociatedBundle} it is 
  the pullback $g_X^* (V/\!/G)$ of the universal $\rho$-associated bundle.
  By Proposition \ref{EveryGBundleIsLocallyTrivial} there exists an 
  effective epimorphism $\xymatrix{U \ar@{->>}[r] & X}$ over which
  $P$ trivializes, hence such that $g_X|_U$ factors through the point, up
  to equivalence. In summary and by the pasting law, Proposition \ref{PastingLawForPullbacks},
  this gives a pasting of $\infty$-pullbacks of the form
  $$
    \raisebox{20pt}{
    \xymatrix@R=8pt{
	  U \times V
	  \ar[dd]
	  \ar[r]
	  &
	  P \times_G V
	  \ar[r]
	  \ar[dd]
	  &
	  V/\!/G
	  \ar[dd]
	  \\
	  \\
	  U \ar@{->>}[r] 
	  \ar[dr]
	  & 
	  X
	  \ar[r]^{g_X}
	  &
	  \mathbf{B}G
	  \\
	  & {*}
	  \ar[ur]
	}
	}
  $$
  which exhibits $P \times_G V \to X$ as a $V$-fiber bundle by a local trivialization
  over $U$.
\endofproof

So far this shows that every $\rho$-associated $\infty$-bundle is a
$V$-fiber bundle. We want to show that, conversely, every $V$-fiber bundle
is associated to a principal $\infty$-bundle.
\begin{definition}
  Let $V \in \mathbf{H}$ be a $\kappa$-compact object, for some regular cardinal $\kappa$. 
  By the characterization of Definition \ref{RezkCharacterization}, there exists
  an $\infty$-pullback square in $\mathbf{H}$ of the form
  $$
    \raisebox{20pt}{
    \xymatrix{
	  V \ar[r] \ar[d] & \widehat {\mathrm{Obj}}_\kappa \ar[d]
	  \\
	  {*} \ar[r]^<<<<<<{\vdash V} & \mathrm{Obj}_\kappa
	}
	}
  $$
  Write
  $$
    \mathbf{B}\mathbf{Aut}(V) := \mathrm{im}(\vdash V)
  $$
  for the 1-image, Definition \ref{image}, 
  of the classifying morphism $\vdash V$ of $V$. 
  By definition this comes with an effective epimorphism 
  $$
    \xymatrix{
	  {*} \ar@{->>}[r] & \mathbf{B}\mathbf{Aut}(V)
	  \ar@{^{(}->}[r] & \mathrm{Obj}_\kappa
	}
	\,,
  $$
  and hence, by Proposition \ref{InfinityGroupObjectsAsGroupoidObjects},
  it is the delooping of an $\infty$-group 
  $$
    \mathbf{Aut}(V) \in \mathrm{Grp}(\mathbf{H})
  $$
  as indicated. We call this the \emph{internal automorphism $\infty$-group} of $V$.
  
  By the pasting law, Proposition \ref{PastingLawForPullbacks}, 
  the image factorization gives a pasting
  of $\infty$-pullback diagrams of the form
  $$
    \xymatrix{
	  V \ar[r] \ar[d] & V/\!/\mathbf{Aut}(V) \ar[r] \ar[d]^{\mathbf{c}_V} & 
	  \widehat {\mathrm{Obj}}_\kappa \ar[d]
	  \\
	  {*} \ar@{->>}[r]^<<<<<<{\vdash V} & \mathbf{B}\mathbf{Aut}(V) \ar@{^{(}->}[r] & 
	  \mathrm{Obj}_\kappa
	}
  $$
  By Theorem~\ref{PrincipalInfinityBundleClassification} this defines a canonical 
  $\infty$-action 
  $$
    \rho_{\mathbf{Aut}(V)} : V \times \mathbf{Aut}(V) \to V
  $$
  of $\mathbf{Aut}(V)$ on $V$ with homotopy quotient $V/\!/\mathbf{Aut}(V)$
  as indicated. 
  \label{InternalAutomorphismGroup}
  \label{AutomorphismInfinityGroup}
  \label{AutomorphismGroup}
\end{definition}
\begin{proposition}
  Every $V$-fiber $\infty$-bundle is $\rho_{\mathbf{Aut}(V)}$-associated to an 
  $\mathbf{Aut}(V)$-principal $\infty$-bundle.
  \label{VBundleIsAssociated}
\end{proposition}
\proof
  Let $E \to V$ be a $V$-fiber $\infty$-bundle. 
  By Definition \ref{FiberBundle} there exists an effective epimorphism
  $\xymatrix{U \ar@{->>}[r] & X}$  along which the bundle trivializes locally.
  It follows 
  by the second Axiom in Definition \ref{RezkCharacterization}
  that on $U$ the morphism $\xymatrix{X \ar[r]^<<<<<{\vdash E} & \mathrm{Obj}_\kappa}$
  which classifies $E \to X$ factors through the point
  $$
    \raisebox{20pt}{
    \xymatrix@R=8pt{
	  U \times V \ar[r]\ar[dd] & E \ar [r] \ar[dd] & \widehat{\mathrm{Obj}}_\kappa \ar[dd]
	  \\
	  \\
	  U \ar@{->>}[r] \ar[dr] & X \ar[r]^<<<<<{\vdash E} & \mathrm{Obj}_\kappa.
	  \\
	  & {*} \ar[ur]_<<<<<{\vdash V}
	}
	}
  $$
  Since the point inclusion, in turn, factors through its 1-image 
  $\mathbf{B}\mathbf{Aut}(V)$, Definition \ref{InternalAutomorphismGroup},
  this yields the outer commuting diagram of the following form
  $$
    \raisebox{20pt}{
    \xymatrix{
	  U \ar[r] \ar@{->>}[d] & {*} \ar[r] &  \mathbf{B}\mathbf{Aut}(V) \ar@{^{(}->}[d]
	  \\
	  X \ar[rr]_{\vdash E} 
	   \ar@{-->}[urr]^{g}
	  && \mathrm{Obj}_\kappa
	}}
  $$
  By the epi/mono factorization system of Proposition \ref{EpiMonoFactorizationSystem}
  there is a diagonal lift $g$ as indicated. Using again the 
  pasting law and by Definition \ref{InternalAutomorphismGroup}
  this factorization induces a pasting of $\infty$-pullbacks of the form
  $$
    \raisebox{20pt}{
    \xymatrix{
	  E \ar[r] \ar[d] & V/\!/\mathbf{Aut}(V) \ar[r] \ar[d]^{\mathbf{c}_V} & 
	  \widehat {\mathrm{Obj}}_\kappa \ar[d]
	  \\
	  X \ar[r]^<<<<<g & \mathbf{B}\mathbf{Aut}(V) \ar@{^{(}->}[r] & \mathrm{Obj}_\kappa
	}
	}
  $$
  Finally, by Proposition \ref{UniversalAssociatedBundle}, this 
  exhibits $E \to X$ as being $\rho_{\mathbf{Aut}(V)}$-associated to the 
  $\mathbf{Aut}(V)$-principal $\infty$-bundle with class $[g] \in H^1(X,G)$.
\endofproof
\begin{theorem}
  $V$-fiber $\infty$-bundles over $X \in \mathbf{H}$ are classified by 
  $H^1(X, \mathbf{Aut}(V))$.
  \label{VBundleClassification}
\end{theorem}
  Under this classification, the $V$-fiber $\infty$-bundle corresponding to 
  $[g] \in H^1(X, \mathbf{Aut}(V))$ is identified,
  up to equivalence, with the $\rho_{\mathbf{Aut}(V)}$-associated $\infty$-bundle 
  (Definition~\ref{AssociatedBundle}) to the $\mathbf{Aut}(V)$-principal $\infty$-bundle
  corresponding to $[g]$ by Theorem~\ref{PrincipalInfinityBundleClassification}.
 \\
\proof
  By Proposition~\ref{VBundleIsAssociated} every morphism 
  $\xymatrix{
    X \ar[r]^<<<<<{\vdash E} & \mathrm{Obj}_\kappa
  }$
  that classifies a small $\infty$-bundle $E \to X$ 
  which happens to be a $V$-fiber $\infty$-bundle factors via some $g$
  through 
  the moduli for $\mathbf{Aut}(V)$-principal $\infty$-bundles
  $$ 
    \xymatrix{
	  X
	  \ar[r]^<<<<<{g}
	  \ar@/_1pc/[rr]_{\vdash E}
	  &
      \mathbf{B}\mathbf{Aut}(V) 
	  \ar@{^{(}->}[r]
	  & \mathrm{Obj}_\kappa
     }
	 \,.
   $$ 
   Therefore it only remains to show that also every homotopy 
   $(\vdash E_1) \Rightarrow (\vdash E_2)$ factors 
   through a homotopy $g_1 \Rightarrow g_2$. 
   This follows by applying the epi/mono lifting property of 
   Proposition \ref{EpiMonoFactorizationSystem} to the diagram
   $$
     \xymatrix{
	  X \coprod X \ar[r]^<<<<<{(g_1, g_2)} \ar@{->>}[d] & \mathbf{B}\mathbf{Aut}(V) 
	  \ar@{^{(}->}[d]
	  \\
	  X \ar[r] \ar@{-->}[ur] & \mathrm{Obj}_\kappa
	 }
   $$
   The outer diagram exhibits the original homotopy. The left morphism is 
   an effective epi (for instance immediately by Proposition \ref{EffectiveEpiIsEpiOn0Truncation}), 
   the right morphism is a monomorphism by construction. Therefore the dashed
   lift exists as indicated and so the top left triangular diagram exhibits
   the desired factorizing homotopy.
\endofproof
\begin{remark}
  In the special case that $\mathbf{H} = \mathrm{Grpd}_{\infty}$, the classification 
  Theorem \ref{VBundleClassification}
  is classical \cite{Stasheff,May}, traditionally
  stated in (what in modern terminology is) the 
  presentation of $\mathrm{Grpd}_{\infty}$ by simplicial sets
  or by topological spaces. Recent discussions include \cite{BlomgrenChacholski}.
  For $\mathbf{H}$ a general 1-localic $\infty$-topos (meaning: with a 1-site of definition), 
  the statement of Theorem \ref{VBundleClassification} appears in \cite{Wendt}, 
  formulated there in terms of the presentation of $\mathbf{H}$ by simplicial presheaves.
  (We discuss the relation of these presentations to the above general abstract result 
  in \cite{NSSb}.)
  Finally, one finds that the classification of \emph{$G$-gerbes} \cite{Giraud} and 
  \emph{$G$-2-gerbes} in \cite{Breen} is the special case of the general statement,
  for $V = \mathbf{B}G$ and $G$ a 1-truncated $\infty$-group. 
  This we discuss below in Section~\ref{StrucInftyGerbes}.
  \label{ReferencesOnClassificationOfVBundles}
\end{remark}

We close this section with a list of some fundamental classes 
of examples of $\infty$-actions, or equivalently, 
by Remark \ref{RhoAssociatedToUniversalIsUniversalVBundle}, 
of universal associated $\infty$-bundles. 
For doing so we use again that, 
by Theorem \ref{PrincipalInfinityBundleClassification}, to give an 
$\infty$-action of $G$ on $V$ is equivalent to giving a fiber
sequence of the form $V \to V/\!/G \to \mathbf{B}G$.

\begin{example}
    \label{ExamplesOfActions}
  The following is a list of examples for $\infty$-actions of 
  $\infty$-groups $G \in \mathrm{Grp}(\mathbf{H})$ on objects in $\mathbf{H}$. 

  We display the universal associated $\infty$-bundles, remark \ref{RhoAssociatedToUniversalIsUniversalVBundle}, over the moduli
  $\mathbf{B}G$ of $G$-principal $\infty$-bundles, 
  that characterize these $\infty$-actions
  according to theorem \ref{PrincipalInfinityBundleClassification}, 
  as discussed in \ref{AssociatedBundles}.
  
  So an $\infty$-action of some $\infty$-group $G$ on an object $V$ is displayed as 
  $$
    \raisebox{20pt}{
    \xymatrix{
	  V \ar[r] & V/\!/ G \ar[d]
	  \\
	  & \mathbf{B}G
	}
	}
	\;\;\;\;\;\;\leftrightarrow\;\;\;\;\;\;
	\raisebox{20pt}{
    \xymatrix{
	  \mbox{Representation space} \ar[r] & \mbox{\begin{tabular}{l}Quotient space/\\ total space of 
	  \\universal associated $V$-bundle\end{tabular}} \ar[d]
	  \\
	  & \mbox{Moduli of $G$-principal bundles}
	}
	}
  $$
  The examples are listed roughly ordered by generality. The first are classes
  of examples that exist in every $\infty$-topos. The more axioms on the ambient
  $\infty$-topos are needed, the further down the list the example appears.
  \begin{enumerate}
    \item 
	  For every $V \in \mathbf{H}$, the fiber sequence
	  $$
	   \raisebox{20pt}{
	   \xymatrix{
	      V 
		  \ar[rr]^-{(\mathrm{id}_V, \mathrm{pt}_{\mathbf{B}G})}
		  &&
		  V \times \mathbf{B}G
		  \ar[d]^{p_2}
		  \\
		  &&
		  \mathbf{B}G
		}
		}
	  $$
	  is the \emph{trivial $\infty$-action} of $G$ on $V$.
    \item
	 For every $G \in \mathrm{Grp}(\mathbf{H})$, the fiber sequence
	 $$
	   \raisebox{20pt}{
	   \xymatrix{
	     G \ar[r]
		 &
		 {*} \ar[d] 
		 \\
		 & \mathbf{B}G
	   }
	   }
	 $$
	 which defines $\mathbf{B}G$ by Theorem \ref{DeloopingTheorem}
	 induces the \emph{right action of $G$ on itself}
	 $$
	   * \simeq G/\!/G
	   \,.
	 $$
	 At the same time
	 this sequence, but now regarded as a bundle over $\mathbf{B}G$, 
	 is the universal $G$-principal $\infty$-bundle, Remark \ref{UniversalPrincipal}.
   \item
     For every object $X \in \mathbf{H}$ write 
	 $$
	   \mathbf{L}X := X \times_{X \times X} X
	 $$ 
	 for its \emph{free loop space} object, the $\infty$-fiber product of the 
	 diagonal on $X$ along itself
	 $$
	   \raisebox{20pt}{
	   \xymatrix{
	     \mathbf{L}X \ar[r] \ar[d]_{\mathrm{ev}_{*}} & X \ar[d]
		 \\
		 X \ar[r] & X \times X
	   }
	   }.
	 $$
     For every $G \in \mathrm{Grp}(\mathbf{H})$ there is a fiber sequence
	 $$
	   \raisebox{20pt}{
	   \xymatrix{
	     G 
		 \ar[r]
		 &
		 \mathbf{L}\mathbf{B}G
		 \ar[d]^{\mathrm{ev}_{*}}
		 \\
		 &
		 \mathbf{B}G
	   }
	   }
	   \,.
	 $$
	 This exhibits the \emph{adjoint action of $G$ on itself}
	 $$
	   \mathbf{L}\mathbf{B}G \simeq G/\!/_{\mathrm{ad}} G
	   \,.
	 $$
	\item
	  For every $V \in \mathbf{H}$ there is the canonical 
	  \emph{$\infty$-action by automorphisms} of the \emph{automorphism $\infty$-group}
	  $\mathbf{Aut}(V)$,  def. \ref{InternalAutomorphismGroup}, on $V$, exhibited by the fiber 
      sequence on the left of the pasting diagram of $\infty$-pullbacks
	  $$
	    \raisebox{20pt}{
	    \xymatrix{
		  V
		  \ar[r]
		  \ar[d]
		  &
		  V/\!/\mathbf{Aut}(V)
		  \ar[d]
		  \ar[r]
		  &
		  \widehat{\mathrm{Obj}}
		  \ar[d]
		  \\
		  {*}
		  \ar@{->>}[r]
		  \ar@/_1.2pc/[rr]_{\vdash V}
		  &
		  \mathbf{B}\mathbf{Aut}(V)
		  \ar@{^{(}->}[r]
		  &
		  \mathrm{Obj}
		}
		}
		\,,
	  $$
	\item
	  For $\rho_1, \rho_2 \in \mathbf{H}_{/\mathbf{B}G}$ two 
	  $G$-$\infty$-actions on objects $V_1, V_2 \in \mathbf{H}$, respectively, 
	  their internal hom $[\rho_1, \rho_2] \in \mathbf{H}_{/\mathbf{B}G}$ in the slice over $\mathbf{B}G$ is a $G$-$\infty$-action on the internal hom $[V_1, V_2] \in \mathbf{H}$:
	  $$
	    \xymatrix{
		  [V_1, V_2]
		  \ar[r]
		  &
		  [V_1, V_2]/\!/ G \ar@{}[r]|\simeq 
		  \ar[d]
		  & \underset{\mathbf{B}G}{\sum}[\rho_1, \rho_2]
		  \\
		  & \mathbf{B}G
		}
		\,,
	  $$
	  hence $[V_1, V_2]/\!/G \simeq \sum_{\mathbf{B}G}[\rho_1, \rho_2]$
	  (this follows by the fact that the inverse image of base change along 
	  $\mathrm{pt}_{\mathbf{B}G} : * \to \mathbf{B}G$ is a cartesian closed $\infty$-functor
	  and hence preserves internal homs\footnote{U.S. thanks Mike Shulman for discussion of this point.})
	  This is the \emph{conjugation $\infty$-action} of $G$ on morphisms $V_1 \to V_2$ by
	  pre- and postcomposition with the action of $G$ on $V_1$ and $V_2$, respectively. 
	  \item
	   The \emph{precomposition action} of the automorphism $\infty$-group $\mathbf{Aut}(V)$
	   on a mapping space $[V,A]$ is given by
	   $$
	     \raisebox{20pt}{
	     \xymatrix{
		   [V,A] \ar[r] & \underset{\mathbf{B}G}{\sum}\left[ \rho_{\mathrm{aut}} , \mathbf{B}G^* A\right]
		   \ar[d]
		   \\
		   & \mathbf{B}\mathbf{Aut}(V)
		 }
		 }
		 \,.
	   $$
	   \item 
	   Let now $\mathbf{H}$ be a differential cohesive $\infty$-topos, \ref{GeneralAbstractTheory}.
	   Let $\mathbb{G} \in \mathrm{Grp}(\mathbf{H})$ be a braided $\infty$-group, def. \ref{BraidedInfinityGroup},
	   and write $\Omega^2_{\mathrm{cl}}(-,\mathbb{G}) \to \mathbf{B} \mathbb{G}_{\mathrm{conn}}$
	   for the corresponding moduli of $\mathbb{G}$-differential cocycles, 
	   \ref{GeneralAbstractDifferentialCohomologyWithDifferentialFormCurvature}. 
	   
	   Let furthermore
	   $$
	     \xymatrix{
		   & \mathbf{B}\mathbb{G}_{\mathrm{conn}}
		   \ar[d]^{F_{(-)}}
		   \\
		   X \ar[r]_\omega \ar[ur]^\nabla & \mathbf{B}\mathbb{G}
		 }
	   $$
	   a pre-symplectic structure $\omega$ with prequantization $\nabla$, \ref{StrucGeometricPrequantization}
	   and let $\rho$ be an action of $\mathbb{G}$ on some $V$. Then the 
	   (higher Heisenberg group-)
	   \emph{$\infty$-action of higher prequantum operators on the space $\mathbf{\Gamma}_X(E)$ of higher prequantum states} is 
	   
	   $$
	     \xymatrix{
		   \mathbf{\Gamma}_X(E)
		   \ar[r]
		   &
		   \underset{\mathbf{B}\mathbb{G}}{\prod}
		   \left(
		     \left[
			   \underset{U}{\sum}\nabla, \rho
			 \right]
			 /\!/\underset{U}{\prod}\mathbf{Aut}(\nabla)
			 \right)
			 \ar[d]
			 \\
			 & \mathbf{Aut}_\mathbf{H}(\nabla)
		 }
		 \,,
	   $$
	   where $E$ is the $\rho$-associated $V$-bundle to $\underset{U}{\sum}\nabla$.
	   \item
	    Let specifically $\mathbf{H} = \mathrm{Smooth}\infty\mathrm{Grpd}$, \ref{SmoothInfgrpds}.
		
		There we have 
		\begin{enumerate}
		\item 
		the $\infty$-action of the moduli of circle principal bundles
		(the circle 2-group) $\mathbf{B}U(1)$ on the moduli of unitary bundles, 
		\ref{Twisted1BundlesTwistedKTheory},
		$$
		  \raisebox{20pt}{
		  \xymatrix{
		    \mathbf{B}U(n) \ar[r] & \mathbf{B}\mathrm{PU}(n)
			\ar[d]^{\mathbf{dd}_n}
			\\
			& \mathbf{B}^2 U(1)
		  }
		  }
		$$
		\item 
		the $\infty$-action of the moduli of circle principal 2-bundles (the circle 3-group)
		$\mathbf{B}^2 U(1)$ on the moduli for $\mathrm{String}$-principal 2-bundles, 
		\ref{String2Group},
		$$
		  \raisebox{20pt}{
		  \xymatrix{
		    \mathbf{B}\mathrm{String} \ar[r] & \mathbf{B}\mathrm{Spin}
			\ar[d]^{\tfrac{1}{2}\mathbf{p}_1}
			\\
			& \mathbf{B}^3 U(1)
		  }
		  }
		$$
		\item 
		the $\infty$-action of the moduli of circle principal 6-bundles (the circle 7-group)
		$\mathbf{B}^6 U(1)$ on the moduli for $\mathrm{Fivebrane}$-principal 6-bundles,
		\ref{FivebraneSixGroup},
		$$
		  \raisebox{20pt}{
		  \xymatrix{
		    \mathbf{B}\mathrm{Fivebrane} \ar[r] & \mathbf{B}\mathrm{String}
			\ar[d]^{\tfrac{1}{6}\mathbf{p}_2}
			\\
			& \mathbf{B}^7 U(1)
		  }
		  }
		$$
		\end{enumerate}
  \end{enumerate}
\end{example}
For more examples along these lines see \ref{TwistedDiffStructures}.

\paragraph{Presentation in locally fibrant simplicial sheaves}

We discuss associated $\infty$-bundles 
in an $\infty$-topos $\mathbf{H} = \mathrm{Sh}_\infty(C)$ 
in terms of the presentation of $\mathbf{H}$ by locally fibrant
simplicial sheaves, corresponding to the respective presentation
of principal $\infty$-bundles from \ref{Principal infinity-bundles presentations}.

This section draws from \cite{NSSb}.

\medskip

Let $C$ be a site with terminal object. 

By prop. \ref{InftyGroupsBySimplicialGroups}
every $\infty$-group over $C$ has a presentation by a 
sheaf of simplicial groups $G \in \mathrm{Grp}(\mathrm{sSh}(C)_{\mathrm{lfib}})$. 
Moreover, by theorem \ref{Classification theorem for weakly G-principal bundles}
every $\infty$-action of $G$ on an object $V$, def. \ref{RepresentationOfInfinityGroup},
is exhibited by a weakly principal simplicial bundle 
$$
  \raisebox{20pt}{
  \xymatrix{
    V \ar[r] & V /_h G
	  \ar[d]^{\rho}
	\\
	& \Wbar G
  }
  }
  \,.
$$
By example \ref{UniversalAssociatedBundle} this is a presentation for the
\emph{universal $\rho$-associated $V$-bundle}. 

We now spell out what this means in the presentation.
\begin{lemma}
  \label{CanonicalCocycleIsFibration}
  The morphism $V /_h G \to \Wbar G$
  is a local fibration.
\end{lemma}
\proof
  By the same argument as in
  the proof of theorem \ref{HomotopyQuotientWPrincBundleLoAcyclicFibration}.
\endofproof
\begin{proposition}
  Let $P \to X$ in $\mathrm{sSh}(C)_{\mathrm{lfib}}$ be a weakly $G$-principal bundle
  with classifying cocycle 
  $X \stackrel{\simeq}{\leftarrow} \hat X \stackrel{g}{\to} \Wbar G$.
  Then the corresponding $\rho$-associated $\infty$-bundle, def. \ref{AssociatedBundleByRho},
  is presented by the ordinary $V$-associated bundle $P \times_G V$
  formed in $\mathrm{sSh}(C)_{\mathrm{lfib}}$.
\end{proposition}
\proof
  By def. \ref{AssociatedBundleByRho} the associated $\infty$-bundle is
  the $\infty$-pullback of $V /\!/G \to \mathbf{B}G$ along 
  $g$. Using lemma \ref{CanonicalCocycleIsFibration} 
  in prop. \ref{PullbacksAlongLocalFibrationsAreHomotopyPullbacks} we find
  that this is presented already by the ordinary pullback of
  $V /_h G \to \Wbar G$ along $g$. By prop. \ref{TotalSimplicialObjectByBorelConstruction}
  this in turn is isomorphic to the pullback of
  $V \times_G W G \to \Wbar G$. Since $\mathrm{sSh}(C)$ is a 1-topos, 
  pullbacks preserve quotients, and so this pullback finally is
  $$
    g^* (W G \times_G V) \simeq (g^* W G) \times_G V \simeq P \times_G W G
	\,.
  $$  
\endofproof

\subsubsection{Sections and twisted cohomology}
\label{StrucTwistedCohomology}
\index{structures in a cohesive $\infty$-topos!twisted cohomology}

We discuss here how the general notion of cohomology in an 
$\infty$-topos considered above in \ref{StrucCohomology}, already subsumes the notion of
\emph{twisted cohomology} and we discuss the corresponding
geometric structure classified by twisted cohomology: 
\emph{twisted $\infty$-bundles}.

Where ordinary cohomology is given by a derived hom-$\infty$-groupoid, 
twisted cohomology is given by the $\infty$-groupoid of 
\emph{sections of 
a local coefficient bundle} in an $\infty$-topos.
This is a geometric and unstable variant of the picture   
of twisted cohomology developed in \cite{AndoBlumbergGepner} \cite{MaySigurdsson}.
It is fairly immediate that given a \emph{universal coefficient bundle},
the induced twisted cohomology is equivalently the ordinary
cohomology in the corresponding slice $\infty$-topos. This
identification provides a clean formulation of the contravariance
of twisted cocycles.
Finally, we observe that twisted cohomology in an $\infty$-topos 
equivalently classifies extensions of structure groups
of principal $\infty$-bundles. 

\medskip

This section draws from \cite{NSSa} and \cite{NSSb}.

\paragraph{General abstract}
\label{TwistedCohomologyGeneral}

\begin{definition}
  Let $p : E \to X$ be any morphism in $\mathbf{H}$, to be regarded as an 
  $\infty$-bundle over $X$. A \emph{section} of $E$ is a diagram
  $$
    \raisebox{20pt}{
    \xymatrix@!C=40pt@!R=30pt{
	   & E \ar[d]^p
	   \\
	  X \ar[r]_{\mathrm{id}}^{\ }="t" 
	   \ar[ur]^{\sigma}_{\ }="s"
	  & X
	  \ar@{=>}^{\simeq} "s"; "t"
	}
	}
  $$
  (where for emphasis we display the presence of the homotopy filling the diagram).
  The \emph{$\infty$-groupoid of sections} of $E \stackrel{p}{\to} X$ is the homotopy fiber
  $$
    \Gamma_X(E) := \mathbf{H}(X,E) \times_{\mathbf{H}(X,X)} \{\mathrm{id}_X\}
  $$
  of the space of all morphisms $X \to E$ on those that cover the identity on $X$.
  \label{Sections}
\end{definition}
We record two elementary but important observations about spaces of sections.
\begin{observation}
  There is a canonical identification
  $$
    \Gamma_X(E) \simeq \mathbf{H}_{/X}(\mathrm{id}_X, p)
  $$
  of the space of sections of $E \to X$ with the hom-$\infty$-groupoid in the 
  slice $\infty$-topos $\mathbf{H}_{/X}$ between the identity on $X$ and the bundle map $p$.
  \label{SectionBySliceMaps}
  \label{TwistedCohomologyBySections}
\end{observation}
\proof
  By prop. \ref{SliceHomAsHomotopyFiber}.
\endofproof
\begin{lemma}
  Let 
  $$
    \xymatrix{
	  E_1 \ar[r] \ar[d]^{p_1} & E_2 \ar[d]^{p_2}
	  \\
	  B_1 \ar[r]^{f} & B_2
	}
  $$ 
  be an $\infty$-pullback diagram in $\mathbf{H}$
  and let $\xymatrix{X \ar[r]^{g_X} & B_1}$ be any morphism. Then 
  post-composition with $f$ induces
  a natural equivalence of hom-$\infty$-groupoids
  $$
    \mathbf{H}_{/B_1}(g_X, p_1)
	\simeq
	\mathbf{H}_{/B_2}(f\circ g_X, p_2)
	\,.
  $$
  \label{SliceMapsIntoPullbacks}
  \label{PullbackCharacterizationOfTwistedCohomology}
\end{lemma}
\proof
  By Proposition \ref{SliceHomAsHomotopyFiber}, the left hand side is given by the homotopy pullback
  $$
    \raisebox{20pt}{
    \xymatrix{
	  \mathbf{H}_{/B_1}(g_X, p_1) \ar[r] \ar[d] & \mathbf{H}(X, E_1) 
	    \ar[d]^{\mathbf{H}(X,p_1)}
	  \\
	  \{g_X\} \ar[r] & \mathbf{H}(X,B_1)\,.
	}
	}
  $$
  Since the hom-$\infty$-functor
  $\mathbf{H}(X,-) : \mathbf{H} \to \mathrm{Grpd}_{\infty}$ 
  preserves the $\infty$-pullback $E_1 \simeq f^* E_2$, this
  extends to a pasting of $\infty$-pullbacks, which by the pasting law 
  (Proposition \ref{PastingLawForPullbacks}) is
  $$
    \raisebox{20pt}{
    \xymatrix{
	  \mathbf{H}_{/B_1}(g_X, p_1) \ar[r] \ar[d] 
	     & 
	  \mathbf{H}(X, E_1) \ar[d]^{\mathbf{H}(X,p_1)} \ar[r]
	  &
	  \mathbf{H}(X, E_2)
	  \ar[d]^{\mathbf{H}(X, p_2)}
	  \\
	  \{g_X\} \ar[r] & \mathbf{H}(X,B_1)
	  \ar[r]_-{\mathbf{H}(X, f)} &
      \mathbf{H}(X, B_2)	  
	}
	}
	\;\;\;
	\simeq
	\;\;\;
	\raisebox{20pt}{
	\xymatrix{
	  \mathbf{H}_{/B_2}(f \circ g_X, p_2)
	  \ar[r]
	  \ar[d]
	  &
	  \mathbf{H}(X, E_2)
	  \ar[d]^{\mathbf{H}(X, p_2)}
	  \\
	  \{f\circ g_X\}
	  \ar[r]
	  &
	  \mathbf{H}(X, B_2).
	}
	}
  $$
\endofproof
Fix now an $\infty$-group $G \in \mathrm{Grp}(\mathbf{H})$ and an 
$\infty$-action $\rho : V \times G \to V$. Write 
$$
  \xymatrix{
    V \ar[r] & V/\!/G \ar[d]^{\mathbf{c}}
	\\
	& \mathbf{B}G
  }
$$
for the corresponding \emph{universal $\rho$-associated $\infty$-bundle} 
as discussed in Section~\ref{StrucRepresentations}.
\begin{proposition}
  For $g_X : X \to \mathbf{B}G$ a cocycle and $P \to X$ the corresponding
$G$-principal $\infty$-bundle according to Theorem \ref{PrincipalInfinityBundleClassification},
there is a natural equivalence
$$
  \Gamma_X(P \times_G V) \simeq \mathbf{H}_{/\mathbf{B}G}(g_X, \mathbf{c})
$$
between the space of sections of the corresponding $\rho$-associated $V$-bundle 
(Definition~\ref{AssociatedBundle}) and the hom-$\infty$-groupoid of the slice
$\infty$-topos of $\mathbf{H}$ over $\mathbf{B}G$, between $g_X$ and $\mathbf{c}$. 
Schematically:
\[
\left\{ 
	\begin{xy} 
		(10,10)*+{E}="1";
		(-10,-10)*+{X}="2";
		(10,-10)*+{X}="3"; 
		(5,0)*+{}="4";
		(3,-5)*+{}="5"; 
		{\ar^-{p} "1";"3"};
		{\ar^-{\sigma} "2";"1"}; 
		{\ar_-{\mathrm{id}} "2";"3"};	
		{\ar@{=>}^-{\simeq} "4";"5"};
	\end{xy} 
\right\} 
 \;\;
    \simeq
  \;\;
\left\{
\begin{xy} 
		(10,10)*+{V/\!/G}="1";
		(-10,-10)*+{X}="2";
		(10,-10)*+{\mathbf{B}G}="3"; 
		(5,0)*+{}="4";
		(3,-5)*+{}="5"; 
		{\ar^-{\mathbf{c}} "1";"3"};
		{\ar^-{\sigma} "2";"1"}; 
		{\ar_-{\mathrm{g_X}} "2";"3"};	
		{\ar@{=>}^-{\simeq} "4";"5"};
	\end{xy} 
\right\} 
\]
\label{SectionsAndSliceHoms}
\end{proposition}
\proof
  By Observation \ref{SectionBySliceMaps} and Lemma \ref{SliceMapsIntoPullbacks}.
\endofproof
\begin{observation}
  If in the above the cocycle $g_X$ is trivializable, in the sense that it factors through the
  point $* \to \mathbf{B}G$ (equivalently if its class $[g_X] \in H^1(X,G)$ is
  trivial) then there is an equivalence
  $$
    \mathbf{H}_{/\mathbf{B}G}(g_X, \mathbf{c})
	\simeq
	\mathbf{H}(X,V)
	\,.
  $$
  \label{TwistedCohomologyIsLocallyVCohomology}
\end{observation}
\proof
  In this case the homotopy pullback on the right in the proof of
  Proposition \ref{SectionsAndSliceHoms} is
  $$
	\raisebox{20pt}{
	\xymatrix{
	  \mathbf{H}_{/\mathbf{B}G}(g_X, \mathbf{c})
	  \ar@{}[r]|{\simeq}
	  &
	  \mathbf{H}(X, V)
	  \ar[r]
	  \ar[d]
	  &
	  \mathbf{H}(X, V/\!/G)
	  \ar[d]^{\mathbf{H}(X, \mathbf{c})}
	  \\
	  \{g_X\}
	  \ar@{}[r]|\simeq
	  &
	  \mathbf{H}(X, {*})
	  \ar[r]
	  &
	  \mathbf{H}(X, \mathbf{B}G)
	}
	}
  $$
  using that $V \to V/\!/G \stackrel{\mathbf{c}}{\to} \mathbf{B}G$ is a fiber
  sequence by definition, and that $\mathbf{H}(X,-)$ preserves this fiber sequence.
\endofproof
\begin{remark}
  Since by Proposition~\ref{EveryGBundleIsLocallyTrivial} 
  every cocycle $g_X$ trivializes locally over some cover 
  $\xymatrix{U \ar@{->>}[r] & X}$
  and equivalently, by Proposition \ref{AssociatedIsFiberBundle}, every
  $\infty$-bundle $P \times_G V$ trivializes locally, 
  Observation \ref{TwistedCohomologyIsLocallyVCohomology} says that
  elements $ \sigma \in \Gamma_X(P \times_G V) \simeq \mathbf{H}_{/\mathbf{B}G}(g_X, \mathbf{c})$
  \emph{locally} are morphisms $\sigma|_U : U \to V$ with values in $V$.
   They fail to be so \emph{globally} to the extent that $[g_X] \in H^1(X, G)$ is non-trivial,
   hence to the extent that $P \times_G V \to X$ is non-trivial.
\end{remark}
This motivates the following definition.
\begin{definition}
  We say that 
  the $\infty$-groupoid
  $\Gamma_X(P \times_G V) \simeq \mathbf{H}_{/\mathbf{B}G}(g_X, \mathbf{c})$
  from Proposition \ref{SectionsAndSliceHoms}
  is the $\infty$-groupoid of \emph{$[g_X]$-twisted cocycles} with values in $V$, with respect to the
  \emph{local coefficient $\infty$-bundle} $V/\!/G \stackrel{\mathbf{c}}{\to} \mathbf{B}G$.
  
  Accordingly, its set of connected components we call the 
  \emph{$[g_X]$-twisted $V$-cohomology} with respect to the local coefficient bundle $\mathbf{c}$
  and write:
  $$
    H^{[g_X]}(X, V) := \pi_0 \mathbf{H}_{/\mathbf{B}G}(g_X, \mathbf{c})
	\,.
  $$
  \label{TwistedCohomologyInOvertopos}
\end{definition}
\begin{remark}
  The perspective that twisted cohomology is the theory of sections of 
  associated bundles whose fibers are classifying spaces is maybe most famous
  for the case of twisted K-theory, where it was described in this form in 
  \cite{Rosenberg}. But already the old theory of 
  \emph{ordinary cohomology with local coefficients} is of this form, 
  as is made manifest in \cite{BFG} (we discuss this in detail in \cite{NSSc}).
  
  A proposal for a comprehensive theory in terms of bundles of topological spaces is
  in \cite{MaySigurdsson} and a systematic formulation 
  in $\infty$-category theory and
  for the case of multiplicative generalized cohomology theories is in 
  \cite{AndoBlumbergGepner}. The formulation above refines this, unstably, 
  to geometric cohomology
  theories/(nonabelian) sheaf hypercohomology, 
  hence from bundles of classifying spaces to 
  $\infty$-bundles of moduli $\infty$-stacks. 
  
  A wealth of examples and applications of such geometric nonabelian twisted cohomology
  of relevance in quantum field theory and in string theory is discussed in 
  \ref{TwistedDifferentialStructures}. 
  \label{ReferencesOnTwisted}
\end{remark}
\begin{remark}
  Of special interest is the case where 
  $V$ is pointed connected, hence (by Theorem \ref{DeloopingTheorem}) 
  of the form $V = \mathbf{B}A$ for some $\infty$-group $A$, and so 
  (by Definition \ref{cohomology}) the coefficient for degree-1 $A$-cohomology, 
  and hence itself (by Theorem \ref{PrincipalInfinityBundleClassification}) the moduli $\infty$-stack for
  $A$-principal $\infty$-bundles. 
  In this case $H^{[g_X]}(X, \mathbf{B}A)$ is
  \emph{degree-1 twisted $A$-cohomology}. Generally, if $V = \mathbf{B}^n A$
  it is \emph{degree-$n$ twisted $A$-cohomology}. In analogy with Definition 
  \ref{cohomology} this is sometimes written
  $$
    H^{n+[g_X]}(X,A) := H^{[g_X]}(X,\mathbf{B}^n A)
	\,.
  $$
  
  Moreover, in this case $V/\!/G$ is itself pointed connected, hence of the form
  $\mathbf{B}\hat G$ for some $\infty$-group $\hat G$, and so the universal local 
  coefficient bundle
  $$
    \xymatrix{
	  \mathbf{B}A \ar[r] & \mathbf{B}\hat G \ar[d]^{\mathbf{c}}
	  \\
	  & \mathbf{B}G
	}
  $$
  exhibits $\hat G$ as an \emph{extension of $\infty$-groups} of $G$ by $A$.
  This case we discuss below in Section \ref{ExtensionsOfCohesiveInfinityGroups}.
  \label{PointedConnectedLocalCoefficients}
\end{remark}
In this notation the local coefficient bundle $\mathbf{c}$ is left implicit.
This convenient abuse of notation is justifed to some extent by the fact that there
is a \emph{universal local coefficient bundle}:
\begin{example}
  The classifying morphism of the $\mathbf{Aut}(V)$-action on some 
  $V \in \mathbf{H}$ from Definition \ref{InternalAutomorphismGroup} according to 
  Theorem \ref{PrincipalInfinityBundleClassification} yields a local coefficient 
  $\infty$-bundle of the form
  $$
    \raisebox{20pt}{
    \xymatrix{
	  V \ar[r] & V/\!/\mathbf{Aut}(V)\ar[d]
	  \\
	  & \mathbf{B}\mathbf{Aut}(V)
	}
	}
  $$
  which we may call the \emph{universal local $V$-coefficient bundle}.
  In the case that $V$ is pointed connected and hence of the form
  $V = \mathbf{B}G$
  $$
    \raisebox{20pt}{
    \xymatrix{
	  \mathbf{B}G \ar[r] & (\mathbf{B}G)/\!/\mathbf{Aut}(\mathbf{B}G)\ar[d]
	  \\
	  & \mathbf{B}\mathbf{Aut}(\mathbf{B}G)
	}
	}
  $$
  the universal twists of the corresponding twisted $G$-cohomology are 
  the \emph{$G$-$\infty$-gerbes}. These we discuss below in section
  \ref{StrucInftyGerbes}.
  \label{UniversalTwistByAutomorphisms}
\end{example}

We now internalize the formulation of spaces of sections, to obtain objects of
sections in the ambient $\infty$-topos.
\begin{definition}
  For $p : E \to X$ a $\rho$-associated $V$-fiber bundle, its
  object of sections is the dependent product, def. \ref{BaseChange}:
  $$
    \mathbf{\Gamma}_X(E)
	\simeq
	\prod_X p
	\,.
  $$
  \label{SectionByDependentProduct}
\end{definition}
\begin{proposition}
  For $p : E \to X$ a $\rho$-associated $V$-fiber bundle, its
object of sections is equivalently given by
$$
  \mathbf{\Gamma}_X(E) \simeq \underset{\mathbf{B}G}{\prod} [g, \rho]
  \,,
$$  
where $g : X \to \mathbf{B}G$ is the modulus of the $G$-principal bundle to which $E$
is associated.
\label{ObjectOfSectionsAsMappingSpace}
\end{proposition}
\proof
By functoriality we have
$$
  \begin{aligned}
    \underset{X}{\prod} g^* \rho
    & \simeq
    \underset{\mathbf{B}G}{\prod} \underset{g}{\prod} g^* \rho
	\\
	& \simeq \underset{\mathbf{B}G}{\prod} [g,\rho]
  \end{aligned}
  \,,
$$
where the second step is prop. \ref{PullProdIsSliceHom}.
\endofproof

\paragraph{Presentations}

\begin{remark}
  \label{ExtrinsicDefinitionOfTwistsInTwistedCohomology}
When the $\infty$-topos $\mathbf{H}$ is presented by a model structure on simplicial presheaves as in \ref{InfinityToposPresentation}
and presentations for $X$ and $C$ have been chosen, 
then the cocycle $\infty$-groupoid $\mathbf{H}(X,C)$ is presented by an explicit 
simplicial set $\mathbf{H}(X,C)_{\mathrm{simp}} \in \mathrm{sSet}$. 
Once these choices are made, 
there is therefore the inclusion of simplicial presheaves
$$
  \mathrm{const} (\mathbf{H}(X,C)_{\mathrm{simp}})_0 \to \mathbf{H}(X,C)_{\mathrm{simp}}
  \,,
$$
where on the left we have the simplicially constant object on the vertices 
of $\mathbf{H}(X,C)_{\mathrm{simp}}$. This morphism, in turn, presents a morphism in 
$\infty \mathrm{Grpd}$ that in general contains a multitude of copies of the components of any 
$H(X,C) \to \mathbf{H}(X,C)$,
a multitude of representatives of twists for each cohomology class of twists. 
Since the twisted cohomology does not depend, 
up to equivalence, on the choice of representative of the twist, 
the corresponding $\infty$-pullback yields 
in general a larger coproduct of $\infty$-groupoids as the corresponding twisted cohomology. 
This however just contains copies of the homotopy types already present in 
$\mathbf{H}_{\mathrm{tw}}(X,A)$ as defined above and therefore constitutes
no additional information. 

However, the choice of effective epimorphism $H(X,C) \to \mathbf{H}(X,C)$, 
while unique up to equivalence, can usually not be made functorially
in $X$. Therefore twisted cohomology can have a 
\emph{representing object} only if one does consider multiple twist
representatives in a suitable way. An example of this situation appears
in the discussion of differential cohomology below in 
\ref{StrucDifferentialCohomology}.
\end{remark}

\subsubsection{Representations and group cohomology}
\label{StrucRepresentations}
\label{GroupCohomology}
\index{cohomology!group cohomology}

We further discuss the notion of representations/actions/modules of 
$\infty$-groups in an $\infty$-topos and the related notions of
quotients, invariants and group cohomology

\paragraph{General abstract.}

Let $G \in \mathrm{Grp}(\mathbf{H})$ be a group object. By the 
discussion in \ref{AssociatedBundles} we may identify the slice
$\infty$-topos over its delooping with the $\infty$-category of $G$-actions:
\begin{proposition}
We have an equivalence of $\infty$-categories
$$
  G \mathrm{Act} \simeq \mathbf{H}_{/\mathbf{B}G}
  \,,
$$
under which an action of $G$ on some $V \in \mathbf{H}$ is identified
with a morphism $V/\!/G \to \mathbf{B}G$, regarded as an object in 
$\mathbf{H}_{/\mathbf{B}G}$, whose $\infty$-fiber is $V$:
$$
  \xymatrix{
    V \ar[r] & V/\!/G \ar[r] & \mathbf{B}G
  }
  \,.
$$ 
\label{CategoryOfActionsEquivalentToSlice}
\end{proposition}
It is useful to identify the structure seen here more formally: write
$$
  \xymatrix{
    \mathbf{H}_{/\mathbf{B}G}
	\ar@<-5pt>[rr]|-{\prod_{\mathbf{B}G}}
	\ar@<+5pt>@{<-}[rr]|-{(\mathbf{B}G) \times (-)}
	\ar@<+15pt>[rr]|-{\sum_{\mathbf{B}G}}
	&&
	\mathbf{H}
  }
$$
for the induced {\'e}tale geometric morphism, prop. \ref{BaseChangeGeomMorphism}.
We introduce some basic terminology on $G$-actions and analyze some properties.

\begin{definition}
  For $\rho \in \mathbf{H}_{/\mathbf{B}G}$ a $G$-action on some $V \in \mathbf{H}$, we say that
  \begin{enumerate}
    \item its dependent sum $\sum_{\mathbf{B}G} \rho \in \mathbf{H}$
	is the \emph{quotient object} of the action;
	\item its dependent product $\prod_{\mathbf{B}G} \rho \in \mathbf{H}$ is the 
	\emph{object of invariants} of the action.
  \end{enumerate}
  Moreover, for $V \in \mathbf{H}$ any object, we say that 
  $(\mathbf{B}G)^* V \in \mathbf{H}_{/\mathbf{B}G}$ is the
  \emph{trivial action} of $G$ on $V$.
  \label{DependentSumAndProductOfActionAsQuotientAndInvariants}.
\end{definition}
\begin{proposition}
  \begin{enumerate}
    \item The quotient object in the sense of def. \ref{DependentSumAndProductOfActionAsQuotientAndInvariants} coincides with the quotient in the sense of def. \ref{AssociatedBundle}:
	  $$\sum_{\mathbf{B}G} \rho \simeq V /\!/G \,.$$
	\item 
	  The object of invariants coincides with the object of sections of 
	  the universal $V$-associated bundle, def. \ref{SectionsAndSliceHoms}:
	   $$\prod_{\mathbf{B}G} \rho \simeq \mathbf{\Gamma}_{\mathbf{B}G}(V/\!/G)\,.$$	  
  \end{enumerate}
\end{proposition}
\begin{definition}
  \label{ConjugationAction}
  For $\rho_1, \rho_2 \in \mathbf{H}_{/\mathbf{B}G}$
  two $G$-actions on objects $V_1, V_2 \in \mathbf{H}$, respectively, 
  write $[\rho_1,\rho_2] \in \mathbf{H}_{/\mathbf{B}G}$
  for their internal hom in the slice. This we call the
  \emph{conjugation action} of $G$ on morphisms $V_1 \to V_2$. 
  We say its direct image under the above {\'e}tale geometric morphism
  is the object of \emph{action homomorphisms} and write
  $$
    \mathbf{Hom}_G(\rho_1, \rho_2)
	:=
	\prod_{\mathbf{B}G} [\rho_1, \rho_2]
	\in \mathbf{H}
	\,.
  $$
\end{definition}
\begin{remark}
  \label{ActionHomomorphismIfInvariantOfConjugationAction}
  In words this says that a $G$-action homomorphism is a morphism $V_1 \to V_2$
  which is an invariant (up to homotopy) of the conjugation action of $G$.
\end{remark}
\begin{proposition}
  The conjugation action $[\rho_1, \rho_2]$,
  def. \ref{ConjugationAction}, is a $G$-action on the internal hom object
  $[V_1,V_2] \in \mathbf{H}$.  
\end{proposition}
\proof
  By def. \ref{AssociatedBundle} we need to show that the internal hom $[\rho_1, \rho_2]$ in the slice 
  sits in a fiber sequence in $\mathbf{H}$ of the form
  $$
    \xymatrix{
	   [V_1,V_2] \ar[r] & \sum_{\mathbf{B}G}[\rho_1, \rho_2] \ar[d]
	    \\
		& \mathbf{B}G
	}
	\,.
  $$
  Observe that forming the homotopy fiber is applying the inverse image of base change
  along the point inclusion $\mathrm{pt}_{\mathbf{B}G} : * \to \mathbf{B}G$ and that base change inverse images
  are cartesian closed functors\footnote{Thanks to Mike Shulman for discussion of this point.}, 
  hence preserve fibers. Using this we compute
  $$
    \begin{aligned}
	   (\mathrm{pt}_{\mathbf{B}G})^* [\rho_1, \rho_2]
	   &
	   \simeq [ (\mathrm{pt}_{\mathbf{B}G})^* (V_1, \rho_1), (\mathrm{pt}_{\mathbf{B}G})^* (V_2, \rho_2) ]
	   \\
	    & \simeq [ V_1, V_2 ]
	\end{aligned}
	\,.
  $$
\endofproof
\begin{definition}
  For $G_1, G_2 \in \mathrm{Grp}(\mathbf{H})$ two groups and
  $f : G_1 \to G_2$ a group homomorphism, hence 
  $\mathbf{B}f : \mathbf{B}G_1 \to \mathbf{B}G_2$ a morphism in $\mathbf{H}$
  we say that
  \begin{enumerate}
    \item 
	  the base change
	  $$(\mathbf{B}f)^* : \xymatrix{\mathrm{Act}(G_2) \simeq \mathbf{H}_{/\mathbf{B}G_2} 
  	   \ar[r] & \mathbf{H}_{/\mathbf{B}G_1} \simeq \mathrm{Act}(G_1) } $$
      is the \emph{pullback representation} functor (or \emph{restricted representation} functor
	  if $f$ is a monomorphism);
   \item 
      the dependent sum
      $$\underset{\mathbf{B}f}{\sum} : \xymatrix{\mathrm{Act}(G_1) \simeq \mathbf{H}_{/\mathbf{B}G_1} 
  	   \ar[r] & \mathbf{H}_{/\mathbf{B}G_2} \simeq \mathrm{Act}(G_2) } $$
	   is the \emph{induced representation} functor.
	\item the dependent product
      $$\underset{\mathbf{B}f}{\prod} : \xymatrix{\mathrm{Act}(G_1) \simeq \mathbf{H}_{/\mathbf{B}G_1} 
  	   \ar[r] & \mathbf{H}_{/\mathbf{B}G_2} \simeq \mathrm{Act}(G_2) } $$
	  is the \emph{coinduced representation} functor.
  \end{enumerate}
  \label{InducedRepresentation}
\end{definition}
\begin{remark}
  For the case of permutation representations of discrete groups, this identification
  of dependent sum/dependent product along contexts of pointed connected discrete groupoids
  has been mentioned on p. 14 of \cite{LawvereAdjointness}.
\end{remark}
\begin{example}
  For $X \in \mathbf{H}$ any object, 
  the automorphism group $\mathbf{Aut}(X)$ of def. \ref{AutomorphismGroup}
  has a canonical action $\rho_{\mathrm{aut}(X)}$ on $X$, given by
  the pasting of $\infty$-pullback diagrams
  $$
    \raisebox{20pt}{
    \xymatrix{
	  X \ar[r] \ar[d] & V/\!/\mathbf{Aut}(X) \ar[r] \ar[d]|{\rho_{\mathrm{aut}(X)}} & \widehat{\mathrm{Obj}} \ar[d]
	  \\
	  {*} \ar@{->>}[r] \ar@/_1pc/[rr]_{\vdash X} 
	  & \mathbf{B}\mathbf{Aut}(X) \ar@{^{(}->}[r] & \mathrm{Obj}
	}}
	\,,
  $$
  where the morphism on the right is the universal small object bundle.
\end{example}
\begin{example}
  For $X, Y \in \mathbf{H}$ two objects, the automorphism group $\mathbf{Aut}(X)$
  of $X$, def. \ref{AutomorphismGroup} has a canonical action $\rho_{\mathrm{prec}}$ 
  by \emph{precomposition} on the internal hom $[X,Y] \in \mathbf{H}$, given itself
  by the internal hom
  $$
    \rho_{\mathrm{prec}} := \left[ \rho_{\mathrm{aut}(X)}, \mathbf{B}\mathbf{Aut}(X)^* Y \right]
  $$
  in $\mathrm{Act}(\mathbf{Aut}(X))$, hence by the congugation action on morphisms
  from $X$ to $Y$ with $Y$ regarded as equipped with the trivial $\mathbf{Aut}(X)$-action;
  we have a fiber sequence
  $$
    \raisebox{20pt}{
    \xymatrix{
	  [X,Y] \ar[r] & [X,Y]/\!/\mathbf{Aut}(X) 
	  \ar[d]^{\rho_{\mathrm{prec}}}
	  \\
	  & \mathbf{B}\mathbf{Aut}(X)
	}
	}
  $$
  in $\mathbf{H}$.
  \label{AutXActionOnXY}
\end{example}

\begin{definition}
  \label{GroupCohomologyWithCoefficients}
  For $*$ the point equipped with the (necessarily) trivial $G$-action, 
  and for $(V,\rho) \in \mathbf{H}_{/\mathbf{B}G}$ we say that 
  $$
    \mathbf{Hom}_G(*,V) \in \mathbf{H}
  $$
  is the \emph{cocycle $\infty$-groupoid} of $G$-group cohomology with
  coefficients in $V$. We say that
  $$
    H_{\mathrm{Grp}}(G,V)
	:=
	\pi_0 \mathbf{Hom}_G(*, V)
  $$  
  is the \emph{group cohomology} of $G$ with coefficients in $V$.
\end{definition}
\begin{remark}
  By remark \ref{ActionHomomorphismIfInvariantOfConjugationAction} 
  and since the action on $*$ is trivial, this says in words
  that group cohomology with coefficients in $V$ is the collection of equivalence classes
  of invariants of $V$.
\end{remark}

\paragraph{Presentations.}

\begin{remark}
  In the case that $V \in \mathbf{H}$ is presented by a chain complex under the Dold-Kan
  correspondence, def. \ref{EmbeddingOfChainComplexes} and that $G \in \mathrm{Grp}(\mathbf{H})$
  is a 0-truncated group, 
  def. \ref{GroupCohomologyWithCoefficients} of group cohomology of $G$ with coefficients in $V$
  manifestly reduces to the traditional definition of group cohomology in homological algebra,
  given by the derived functor of the invariants functor of $G$-modules.
\end{remark}

\subsubsection{Extensions and twisted bundles}
\label{ExtensionsOfCohesiveInfinityGroups}
\label{GroupExtension}
\index{structures in a cohesive $\infty$-topos!extensions of cohesive $\infty$-groups}

We discuss the notion of \emph{extensions} of $\infty$-groups
(see Section~\ref{StrucInftyGroups}), generalizing the traditional notion of 
group extensions. 
This is in fact a special case of the notion of 
principal $\infty$-bundle, Definition~\ref{principalbundle}, for base space objects that 
are themselves deloopings of $\infty$-groups. 
For every extension of $\infty$-groups, there is the 
corresponding notion of 
\emph{lifts of structure $\infty$-groups} of principal $\infty$-bundles. 
These are classified equivalently by trivializations of an \emph{obstruction class}
and by the twisted cohomology with coefficients in the extension itself, regarded
as a local coefficient $\infty$-bundle.

Moreover, we show
that principal $\infty$-bundles with an extended structure $\infty$-group
are equivalent to principal $\infty$-bundles with unextended structure $\infty$-group
but carrying a principal $\infty$-bundle for the \emph{extending} $\infty$-group on their
total space, which on fibers restricts to the given $\infty$-group extension.
We formalize these \emph{twisted (principal) $\infty$-bundles}
and observe that they are classified by twisted cohomology, 
Definition~\ref{TwistedCohomologyInOvertopos}.

\medskip  

\begin{definition}
  \label{ExtensionOfInfinityGroups}
We say a sequence of $\infty$-groups,
\[
  A \to \hat G \to G
\]
in $\mathrm{Grp}(\mathbf{H})$
\emph{exhibits $\hat G$ as an extension of $G$ by $A$} if the delooping
$$
  \mathbf{B}A \to \mathbf{B}\hat G \to \mathbf{B}G
$$
is a fiber sequence in $\mathbf{H}$, def. \ref{fiber sequence}.
\end{definition}
\begin{remark}
  By continuing the fiber sequence to the left 
  $$
    A \to \hat G \to G \to \mathbf{B}A \to \mathbf{B}\hat G \to \mathbf{B}G
  $$
  this implies by theorem  \ref{PrincipalInfinityBundleClassification} that $\hat G \to G$ is an $A$-principal
  bundle and that 
  $$
    G \simeq \hat G \to A
  $$
  is the quotient of the $A$-action.
  \label{CentralExtensionsAsPrincipalBundles}
\end{remark}
\begin{definition}
  For $A$ a braided $\infty$-group, def. \ref{BraidedInfinityGroup}, a
  \emph{central extension} $\hat G$ of $G$ by $A$ is an extension $A \to \hat G \to G$,
  such that the defining delooping extends one step further to the right:
$$
  \xymatrix{
    \mathbf{B}A \ar[r] &  \mathbf{B}\hat G \ar[r]^{\mathbf{p}} &  \mathbf{B}G
	\ar[r]^{\mathbf{c}} & \mathbf{B}^2 A
	\,.
  }
$$
We write 
$$
  \mathrm{Ext}(G,A) := \mathbf{H}(\mathbf{B}G, \mathbf{B}^2 A)
  \simeq (\mathbf{B}A)\mathrm{Bund}(\mathbf{B}G)
$$
for the \emph{$\infty$-groupoid of extensions} of $G$ by $A$.
  \label{CentralExtensionOfInfinityGroups}
\end{definition}
\begin{definition}
Given an $\infty$-group extension  
$\xymatrix{A \ar[r] & \hat G \ar[r]^{\Omega \mathbf{c}} & G}$ 
and given a $G$-principal $\infty$-bundle $P \to X$ in $\mathbf{H}$, we say that a 
\emph{lift} $\hat P$ of $P$ to a $\hat G$-principal $\infty$-bundle is a lift 
$\hat g_X$ of its classifying 
cocycle $g_X : X \to \mathbf{B}G$, under the equivalence of Theorem
\ref{PrincipalInfinityBundleClassification}, through the extension:
$$
  \raisebox{20pt}{
  \xymatrix{
      & \mathbf{B} {\hat G}
      \ar[d]^{\mathbf{p}}
    \\
    X \ar@{-->}[ur]^{\hat g_X} \ar[r]_-{g_X} & \mathbf{B}G.
  }
  }
$$
Accordingly, the \emph{$\infty$-groupoid of lifts} of $P$ with respect to 
$\mathbf{p}$ is 
$$
  \mathrm{Lift}(P,\mathbf{p}) := \mathbf{H}_{/\mathbf{B}G}(g_X, \mathbf{p})
  \,.
$$
\label{PrincipalInfinityBundleExtension}
\end{definition}
\begin{observation}
  By the universal property of the $\infty$-pullback, a lift exists precisely if the 
  cohomology class
  $$
    [\mathbf{c}(g_X)] := [\mathbf{c}\circ g_X] \in H^2(X, A)
  $$
  is trivial. 
  \label{ObstructionClassObstructs}
\end{observation}

This is implied by Theorem \ref{ExtensionsAndTwistedCohomology}, to which we turn after
introducing the following terminology.
\begin{definition}
  In the above situation, 
  we call $[\mathbf{c}(g_X)]$ the \emph{obstruction class} to the extension;
  and we call $[\mathbf{c}] \in H^2(\mathbf{B}G, A)$ the
  \emph{universal obstruction class} of extensions through $\mathbf{p}$.
  
  We say that a \emph{trivialization} of the obstruction cocycle
  $\mathbf{c}(g_X)$ is a morphism $\mathbf{c}(g_X) \to *_X$ in $\mathbf{H}(X, \mathbf{B}^2 A)$,
  where ${*}_X : X \to * \to \mathbf{B}^2 A$ is the trivial cocycle. Accordingly, the
  \emph{$\infty$-groupoid of trivializations of the obstruction} is
  $$
    \mathrm{Triv}(\mathbf{c}(g_X)) := \mathbf{H}_{/\mathbf{B}^2 A}(\mathbf{c}\circ g_X, *_X)
	\,.
  $$
  \label{Obstructions}
\end{definition}
We give now three different characterizations of spaces of extensions of $\infty$-bundles.
The first two, by spaces of twisted cocycles and by spaces of trivializations 
of the obstruction class, are immediate consequences of the previous discussion:
\begin{theorem}
  Let $P \to X$ be a $G$-principal $\infty$-bundle corresponding by
  Theorem \ref{PrincipalInfinityBundleClassification} to a cocycle $g_X : X \to \mathbf{B}G$.
  \begin{enumerate}
    \item 
	  There is a natural equivalence 
	 $$
	    \mathrm{Lift}(P, \mathbf{p}) \simeq \mathrm{Triv}(\mathbf{c}(g_X))
	 $$
	 between the $\infty$-groupoid of lifts of $P$ through $\mathbf{p}$, 
     Definition \ref{PrincipalInfinityBundleExtension},	 
	 and the $\infty$-groupoid of trivializations 
	 of the obstruction class, Definition \ref{Obstructions}.
	 \item 
	 There is a natural equivalence
	 $\mathrm{Lift}(P, \mathbf{p}) \simeq \mathbf{H}_{/\mathbf{B}G}(g_X, \mathbf{p})$
	 between the $\infty$-groupoid of lifts and the $\infty$-groupoid of 
	 $g_X$-twisted cocycles relative to $\mathbf{p}$, Definition \ref{TwistedCohomologyInOvertopos},
	 hence a classification
	 $$
	   \pi_0 \mathrm{Lift}(P, \mathbf{P}) \simeq H^{1+[g_X]}(X, A)
	 $$
	 of equivalence classs of lifts by the $[g_X]$-twisted $A$-cohomology of $X$
	 relative to the local coefficient bundle
    $$
      \raisebox{20pt}{
      \xymatrix{
         \mathbf{B}A \ar[r] & \mathbf{B}\hat G \ar[d]^{\mathbf{p}}  
    	 \\
	     & \mathbf{B}G\,.
       }
      }
    $$  
\end{enumerate}
 \label{ExtensionsAndTwistedCohomology}
\end{theorem}
\proof
  The first statement is the special case of Lemma \ref{SliceMapsIntoPullbacks}
  where the $\infty$-pullback $E_1 \simeq f^* E_2$ in the notation there is identified
  with $\mathbf{B}\hat G \simeq \mathbf{c}^* {*}$.
  The second is evident after unwinding the definitions.
\endofproof
\begin{remark}
  For the special case that $A$ is 0-truncated, we may, by the discussion in 
  \cite{NikolausWaldorf, NSSc}, identify 
  $\mathbf{B}A$-principal $\infty$-bundles with $A$-\emph{bundle gerbes},
  \cite{Mur}.
  Under this identification the $\infty$-bundle classified by the 
  obstruction class $[\mathbf{c}(g_X)]$ above is what is called the
  \emph{lifting bundle gerbe} of the lifting problem, see for instance 
  \cite{CBMMS} for a review. 
  In this case 
  the first item of Theorem \ref{ExtensionsAndTwistedCohomology} reduces to
  Theorem 2.1 in \cite{Waldorf} and Theorem A (5.2.3) in \cite{NikolausWaldorf2}. 
  The reduction of this statement to 
  connected components, 
  hence the special case of Observation \ref{ObstructionClassObstructs},
  was shown in \cite{BreenBitorseurs}.
  \label{ReferencesOnLiftings}
\end{remark}
While, therefore, the discussion of extensions of $\infty$-groups and of 
lifts of structure $\infty$-groups
is just a special case of the discussion in the previous sections, this special case
admits geometric representatives of cocycles in the corresponding twisted cohomology by
twisted principal $\infty$-bundles. This we turn to now. 
\begin{definition}
  \label{TwistedBundle}
  Given an extension of $\infty$-groups $A \to \hat G \xrightarrow{\Omega \mathbf{c}} G$
  and given a $G$-principal $\infty$-bundle $P \to X$, with class $[g_X] \in H^1(X,G)$, 
  a \emph{$[g_X]$-twisted $A$-principal $\infty$-bundle} on $X$ is an $A$-principal
  $\infty$-bundle $\hat P \to P$ such that the cocycle $q : P \to\mathbf{B}A$
  corresponding to it under Theorem \ref{PrincipalInfinityBundleClassification}
  is a morphism of $G$-$\infty$-actions.
  
  The \emph{$\infty$-groupoid of $[g_X]$-twisted $A$-principal $\infty$-bundles on $X$} is 
  $$
    A\mathrm{Bund}^{[g_X]}(X)
	:=
	G \mathrm{Action}( P, \mathbf{B}A )
	\subset
	\mathbf{H}(P, \mathbf{B}A)
	\,.
  $$
\end{definition}
\begin{observation}
  \label{BundleOnTotalSpaceFromExtensionOfBundles}
  Given an $\infty$-group extension $A \to \hat G \stackrel{\Omega \mathbf{c}}{\to} G$,
  an extension of a $G$-principal $\infty$-bundle $P \to X$
  to a $\hat G$-principal $\infty$-bundle, Definition \ref{PrincipalInfinityBundleExtension}, induces
  an $A$-principal $\infty$-bundle $\hat P \to P$ 
  fitting into a pasting diagram of $\infty$-pullbacks of the form
$$
  \raisebox{30pt}{
  \xymatrix{
    \hat G \ar[r] \ar[d]^{\Omega \mathbf{c}} & \hat P \ar[r] \ar[d] &  {*} \ar[d]
    \\
    G \ar[r] \ar[d] & P \ar[r]^{q} \ar[d]& \mathbf{B}A \ar[r] \ar[d] & {*} \ar[d]
    \\
    {*} \ar[r]^{x}&
    X \ar@/_1pc/[rr]_g \ar[r]^{\hat g} & \mathbf{B}\hat G
    \ar[r]^{\mathbf{c}}& \mathbf{B}G.
  }
  }
$$
  In particular, it has the following properties: 
  \begin{enumerate}
    \item $\hat P \to P$ is a $[g_X]$-twisted $A$-principal bundle, Definition \ref{TwistedBundle};
	\item for all points $x : * \to X$ the restriction of $\hat P \to P$ to the 
	fiber $P_x$ is equivalent to the $\infty$-group extension $\hat G \to G$.
  \end{enumerate}  
  \label{ExtendedBundlesByBundlesOfExtensions}
\end{observation}
\proof
This follows from repeated application of the pasting law for $\infty$-pullbacks, 
Proposition~\ref{PastingLawForPullbacks}.

The bottom composite $g : X \to \mathbf{B}G$ is a cocycle for the given $G$-principal $\infty$-bundle 
$P \to X$ and it factors through $\hat g : X \to \mathbf{B}\hat G$ by assumption of the existence of 
the extension $\hat P \to P$. 

Since also the bottom right square is an $\infty$-pullback by the given $\infty$-group extension, 
the pasting law asserts that the square over $\hat g$ is also an  $\infty$-pullback, and then that so 
is the square over $q$. This exhibits $\hat P$ as an $A$-principal $\infty$-bundle over $P$ classified
by the cocycle $q$ on $P$. By Proposition~\ref{ClassificationOfTwistedGEquivariantBundles}
this $\hat P \to P$ is twisted $G$-equivariant.

Now choose any point $x : {*} \to X$ of the base space 
as on the left of the diagram. Pulling this back upwards 
through the diagram and using the pasting law and the 
definition of loop space objects $G \simeq \Omega 
\mathbf{B}G \simeq * \times_{\mathbf{B}G} *$ the diagram 
completes by $\infty$-pullback squares on the left as indicated, which proves the claim.
\endofproof

\begin{theorem}
  The construction of Observation \ref{BundleOnTotalSpaceFromExtensionOfBundles}
  extends to an equivalence of $\infty$-groupoids
  $$
    A\mathrm{Bund}^{[g_X]}(X)
	\simeq
	\mathbf{H}_{/\mathbf{B}G}(g_X, \mathbf{c})
  $$
  between that of $[g_X]$-twisted $A$-principal bundles on $X$, 
  Definition \ref{TwistedBundle}, and the cocycle $\infty$-groupoid of
  degree-1 $[g_X]$-twisted $A$-cohomology, Definition \ref{TwistedCohomologyInOvertopos}.
  
  In particular the classification of $[g_X]$-twisted $A$-principal bundles is
  $$
    A\mathrm{Bund}^{[g_X]}(X)_{/\sim}
	\simeq
	H^{1+[g_X]}(X, A)
	\,.
  $$
  \label{ClassificationOfTwistedGEquivariantBundles}
\end{theorem}
\proof
  For $G = *$ the trivial group, the statement reduces to  
  Theorem~\ref{PrincipalInfinityBundleClassification}. The general proof
  works along the same lines as the proof of that theorem.
  The key step is the generalization of the proof of Proposition~\ref{LocalTrivialityImpliesCocycle}.
  This proceeds verbatim as there, only with $\mathrm{pt} : * \to \mathbf{B}G$ generalized to
  $i : \mathbf{B}A \to \mathbf{B}\hat G$. 
  The morphism of $G$-actions
  $P \to \mathbf{B}A$ and a choice of effective epimorphism $U \to X$ over
  which $P \to X$ trivializes gives rise to a morphism in 
  $\mathbf{H}^{\Delta[1]}_{/(* \to \mathbf{B}G)}$ which involves the diagram
    $$
    \raisebox{20pt}{
    \xymatrix{
	  U \times G \ar@{->>}[r] \ar[d] & P \ar[r] \ar[d] & \mathbf{B}A \ar[d]^{i}
	  \\
	  U \ar@{->>}[r] & X \ar[r] & \mathbf{B}\hat G
	}
	}
	\;\;
	\simeq
	\;\;
    \raisebox{20pt}{
    \xymatrix{
	  U \times G \ar@{->>}[rr] \ar[d] & & \mathbf{B}A \ar[d]^{i}
	  \\
	  U \ar[r] & {*} \ar[r]^{\mathrm{pt}} & \mathbf{B}\hat G
	}
	}
  $$
  in $\mathbf{H}$. (We are using that for the 0-connected object $\mathbf{B}\hat G$
  every morphism $* \to \mathbf{B}G$ factors through $\mathbf{B}\hat G \to \mathbf{B}G$.)
  Here the total rectangle and the left square on the left are $\infty$-pullbacks,
  and we need to show that the right square on the left is then also an 
  $\infty$-pullback. Notice that by the pasting law the rectangle on the
  right is indeed equivalent to the pasting of $\infty$-pullbacks
  $$    
    \raisebox{20pt}{
    \xymatrix{
	  U \times G \ar@{->}[r] \ar[d] & G \ar[r] \ar[d] &  \mathbf{B}A \ar[d]^{i}
	  \\
	  U \ar[r] & {*} \ar[r]^{\mathrm{pt}} & \mathbf{B}\hat G
	}
	}
  $$
  so that the relation
  $$
    U^{\times^{n+1}_X} \times G
	\simeq
	i^* (U^{\times^{n+1}_X})
  $$
  holds. With this the proof finishes as in the proof of 
  Proposition~\ref{LocalTrivialityImpliesCocycle}, with $\mathrm{pt}^*$
  generalized to $i^*$. 
\endofproof

\begin{remark}
  Aspects of special cases of this theorem can be identified in the literature.
  For the special case of ordinary extensions of ordinary Lie groups,
  the equivalence of the corresponding extensions of a principal bundle with 
  certain equivariant structures on its total space is essentially the content
  of \cite{Mackenzie, Androulidakis}. 
  In particular the  twisted unitary bundles or \emph{gerbe modules}
  of twisted K-theory \cite{CBMMS}
  are equivalent to such structures.

  For the case of $\mathbf{B}U(1)$-extensions of Lie groups, such as the
  $\mathrm{String}$-2-group, the equivalence of the corresponding 
  $\mathrm{String}$-principal 2-bundles, 
  by the above theorem, to certain bundle gerbes on the total spaces of 
  principal bundles underlies constructions such as in \cite{Redden}.
  Similarly the bundle gerbes on double covers considered in 
  \cite{SSW} are $\mathbf{B}U(1)$-principal 2-bundles on $\mathbb{Z}_2$-principal
  bundles arising by the above theorem from the extension 
  $\mathbf{B}U(1) \to \mathbf{Aut}(\mathbf{B}U(1)) \to \mathbb{Z}_2$,
  a special case of the extensions that we consider in the next Section
  \ref{StrucInftyGerbes}.
  
  These and more examples we discuss in detail below.
\end{remark}

\subsubsection{Gerbes}
\label{StrucInftyGerbes}
\index{structures in a cohesive $\infty$-topos!$\infty$-gerbes}

We discuss the general notion of (nonabelian) \emph{gerbes} and
higher gerbes in an $\infty$-topos. 

This section draws from \cite{NSSa}.

\medskip

Remark \ref{PointedConnectedLocalCoefficients} above indicates that of special relevance are those
$V$-fiber $\infty$-bundles $E \to X$ in an $\infty$-topos $\mathbf{H}$ 
whose typical fiber $V$ is pointed connected,
and hence is the moduli $\infty$-stack $V = \mathbf{B}G$ of $G$-principal 
$\infty$-bundles for some $\infty$-group $G$. 
Due to their local triviality, 
when regarded as objects in the slice $\infty$-topos $\mathbf{H}_{/X}$,
these $\mathbf{B}G$-fiber $\infty$-bundles are themselves \emph{connected objects}.
Generally, for $\mathcal{X}$ an $\infty$-topos regarded as an $\infty$-topos
of $\infty$-stacks over a given space $X$, it makes sense to consider
its connected objects as $\infty$-bundles over $X$. Here we discuss these
\emph{$\infty$-gerbes}.

\medskip

In the following discussion it is useful to consider two $\infty$-toposes:
\begin{enumerate}
  \item an ``ambient'' $\infty$-topos $\mathbf{H}$ as before, to be thought of 
    as an $\infty$-topos ``of all geometric homotopy types'' for a given notion of geometry,  
	in which $\infty$-bundles are given by \emph{morphisms} and the terminal 
	object plays the role of the geometric point $*$;
  \item an $\infty$-topos $\mathcal{X}$, to be thought of as the topos-theoretic
   incarnation of a single geometric homotopy type (space) $X$, hence as  
   an $\infty$-topos of
  ``geometric homotopy types {\'e}tale over $X$'', in which an $\infty$-bundle over $X$
  is given by an \emph{object} and the terminal object plays the role of the base space $X$. 
  
  In practice, $\mathcal{X}$ is the slice
  $\mathbf{H}_{/X}$ of the previous  ambient $\infty$-topos over $X \in \mathbf{H}$, or
  the smaller $\infty$-topos $\mathcal{X} = \mathrm{Sh}_\infty(X)$ of (internal)
  $\infty$-stacks over $X$. 
\end{enumerate}
In topos-theory literature the role of $\mathbf{H}$ above is sometimes referred to as that of a 
\emph{gros} topos and then the role of $\mathcal{X}$ is referred to as that of a \emph{petit} topos. 
The reader should beware that much of the classical literature on gerbes is written
from the point of view of only the \emph{petit} topos $\mathcal{X}$.

\medskip

The original definition of a \emph{gerbe} 
on $X$ \cite{Giraud} is: a stack $E$ (i.e.\  a 
1-truncated $\infty$-stack) over $X$ that is 
1. \emph{locally non-empty} and 2. \emph{locally connected}. 
In the more intrinsic language of higher topos theory, these 
two conditions simply say that $E$ is 
a \emph{connected object} (Definition 6.5.1.10 in \cite{Lurie}): 
1. the terminal morphism $E \to *$ is an 
effective epimorphism and 2. the 0th homotopy sheaf is trivial, 
$\pi_0(E) \simeq *$. This reformulation is 
made explicit in the literature for instance in 
Section 5 of \cite{JardineLuo} and in Section 7.2.2 of \cite{Lurie}. Therefore:
\begin{definition}
  For $\mathcal{X}$ an $\infty$-topos, a \emph{gerbe} in $\mathcal{X}$ is an object $E \in \mathcal{X}$ which is
  \begin{enumerate}
    \item connected;
	\item 1-truncated.
  \end{enumerate}
  For $X \in \mathbf{H}$ an object, a \emph{gerbe $E$ over $X$} is a gerbe in the slice
  $\mathbf{H}_{/X}$. This is an object $E \in \mathbf{H}$ together with an effective epimorphism $E \to X$ such that $\pi_i(E) = X$ for all $i \neq 1$. 
  \label{Gerbe}
  \label{1Gerbe}
\end{definition}
\begin{remark}
  Notice that conceptually this is different from the
  notion of \emph{bundle gerbe} introduced in \cite{Mur} 
  (see \cite{NikolausWaldorf} for a review). 
  We discuss in \cite{NSSc} that 
  bundle gerbes are presentations of \emph{principal} $\infty$-bundles 
  (Definition \ref{principalbundle}).
  But gerbes -- at least the \emph{$G$-gerbes}
  considered in a moment in Definition \ref{G-Gerbe} -- 
  are $V$-fiber $\infty$-bundles (Definition \ref{FiberBundle})
  hence \emph{associated} to principal $\infty$-bundles 
  (Proposition \ref{VBundleIsAssociated}) with the special 
  property of having pointed connected fibers. 
  By Theorem \ref{VBundleClassification}  
  $V$-fiber $\infty$-bundles may be identified with their underlying $\mathbf{Aut}(V)$-principal 
  $\infty$-bundles and so one may identify $G$-gerbes with nonabelian 
  $\mathrm{Aut}(\mathbf{B}G)$-bundle gerbes
  (see also around Proposition~\ref{ClassificationOfGGerbes} below), 
  but considered generally, neither of these
  two notions is a special case of the other. Therefore the terminology is 
  slightly unfortunate, but it is standard.
\end{remark}
Definition \ref{Gerbe} has various obvious generalizations. The following is considered in \cite{Lurie}.
\begin{definition}
  \index{gerbe!EM $n$-gerbe}
 For $n \in \mathbb{N}$, an \emph{EM $n$-gerbe} is an object $E \in \mathcal{X}$ which is
 \begin{enumerate}
   \item $(n-1)$-connected;
   \item $n$-truncated.
 \end{enumerate}
\end{definition}
\begin{remark}
This is almost the definition of an 
\emph{Eilenberg-Mac Lane object} in $\mathcal{X}$, 
only that the condition requiring a global section 
$* \to E$ (hence $X \to E$) is missing. Indeed, the 
Eilenberg-Mac Lane objects of degree $n$ in 
$\mathcal{X}$ are precisely the EM $n$-gerbes of 
\emph{trivial class}, according to Proposition~\ref{ClassificationOfGGerbes} below.
\end{remark}
There is also an earlier established definition of 
\emph{2-gerbes} in the literature \cite{Breen}, which is 
more general than EM 2-gerbes. Stated in the above fashion it reads as follows.
\begin{definition}[Breen \cite{Breen}]
  \index{gerbe!2-gerbe}
  \index{2-gerbe}
  A \emph{2-gerbe} in $\mathcal{X}$ is an object $E \in \mathcal{X}$ which is
  \begin{enumerate}
    \item connected;
	\item 2-truncated.
  \end{enumerate}
\end{definition}
This definition has an evident generalization to arbitrary degree, which we adopt here.
\begin{definition}
  \label{nGerbe}
  An \emph{$n$-gerbe} in $\mathcal{X}$ is an object $E \in \mathcal{X}$ which is 
   \begin{enumerate}
      \item connected;
	  \item $n$-truncated.
   \end{enumerate}
In particular an \emph{$\infty$-gerbe} is a connected object.
\end{definition}
The real interest is in those $\infty$-gerbes which have a prescribed
typical fiber:
\begin{remark}
By the above, $\infty$-gerbes (and hence EM $n$-gerbes 
and 2-gerbes and hence gerbes) are much like deloopings 
of $\infty$-groups (Theorem \ref{DeloopingTheorem}) 
only that there is no requirement that there 
exists a global section. An $\infty$-gerbe for which there exists a  
global section $X \to E$ is called \emph{trivializable}. By 
Theorem~\ref{DeloopingTheorem} trivializable $\infty$-gerbes are equivalent to $\infty$-group objects in $\mathcal{X}$ 
(and the $\infty$-groupoids of all of these are equivalent 
when transformations are required to preserve the canonical global section).
\end{remark}
But \emph{locally} every $\infty$-gerbe $E$ is of this form. For let
$$
  (x^* \dashv x_*) :
  \xymatrix{
     \mathrm{Grpd}_{\infty}
	    \ar@{<-}@<+3pt>[r]^-{x^*}
	    \ar@<-3pt>[r]_-{x_*}
		&
     \mathcal{X}	 
  }
$$
be a topos point. Then the stalk $x^* E \in \mathrm{Grpd}_{\infty}$ 
of the $\infty$-gerbe is connected: because inverse images 
preserve the finite $\infty$-limits involved in the definition of homotopy sheaves, and preserve the terminal object. Therefore
$$
  \pi_0\,  x^* E \simeq x^* \pi_0 E \simeq x^* * \simeq *
  \,.
$$
Hence for every point $x$ we have a stalk $\infty$-group $G_x$ and an equivalence 
$$
  x^* E \simeq B G_x 
  \,.
$$
Therefore one is interested in the following notion.
\begin{definition}
  For $G \in \mathrm{Grp}(\mathcal{X})$ an $\infty$-group 
  object, a \emph{$G$-$\infty$-gerbe} is an $\infty$-gerbe $E$ such that there exists 
  \begin{enumerate}
    \item an effective epimorphism $\xymatrix{U \ar@{->>}[r] & X}$;
	\item an equivalence $E|_U \simeq \mathbf{B} G|_U$.
  \end{enumerate}
  Equivalently: a $G$-$\infty$-gerbe is a $\mathbf{B}G$-fiber $\infty$-bundle,
  according to Definition \ref{FiberBundle}.
  \label{G-Gerbe}
  \label{GGerbe}
\end{definition}
In words this says that a $G$-$\infty$-gerbe is one that locally looks like the 
moduli $\infty$-stack of $G$-principal $\infty$-bundles.
\begin{example}
  For $X$ a topological space and $\mathcal{X} = 
  \mathrm{Sh}_\infty(X)$ the $\infty$-topos of $\infty$-sheaves over it, these notions reduce to the following.
  \begin{itemize}
    \item a 0-group object $G \in \tau_{0}\mathrm{Grp}(\mathcal{X}) 
    \subset \mathrm{Grp}(\mathcal{X})$ is a sheaf of groups on $X$ 
    (here $\tau_0\mathrm{Grp}(\mathcal{X})$ denotes the 0-truncation of 
    $\mathrm{Grp}(\mathcal{X})$;
	\item for $\{U_i \to X\}$ any open cover, the canonical 
	morphism $\coprod_i U_i \to X$ is an effective epimorphism to the terminal object;
	\item $(\mathbf{B}G)|_{U_i}$ is the stack of 
	   $G|_{U_i}$-principal bundles ($G|_{U_i}$-torsors).
  \end{itemize}
\end{example}
It is clear that one way to construct a $G$-$\infty$-gerbe 
should be to start with an $\mathbf{Aut}(\mathbf{B}G)$-principal 
$\infty$-bundle, Remark \ref{UniversalTwistByAutomorphisms},
and then canonically \emph{associate} a fiber $\infty$-bundle to it.
\begin{example}
  \label{automorphism2GroupAbstractly}
  For $G \in \tau_{0}\mathrm{Grp}(\mathrm{Grpd}_{\infty})$ an 
  ordinary group, $\mathbf{Aut}(\mathbf{B}G)$ is usually called the 
  \emph{automorphism 2-group} of $G$. Its underlying groupoid is equivalent to
  \[
  \mathbf{Aut}(G) \times G\rightrightarrows 
  \mathbf{Aut}(G), 
  \]
  the action groupoid for the action of $G$ on $\mathbf{Aut}(G)$ 
  via the homomorphism $\mathrm{Ad}\colon G\to \mathbf{Aut}(G)$.  
\end{example}
\begin{corollary}
  Let $\mathcal{X}$ be a  1-localic $\infty$-topos 
  (i.e.\ one that has  a 1-site of definition). 
  Then for $G \in \mathrm{Grp}(\mathcal{X})$ any 
  $\infty$-group object, $G$-$\infty$-gerbes are classified by $\mathbf{Aut}(\mathbf{B}G)$-cohomology:
  $$
    \pi_0 G \mathrm{Gerbe}
	\simeq
	\pi_0 \mathcal{X}(X,\mathbf{B}\mathbf{Aut}(\mathbf{B}G))
	=:
	H^1_{\mathcal{X}}(X,\mathbf{Aut}(\mathbf{B}G))
	\,.
  $$
  \label{ClassificationOfGGerbes}
\end{corollary}
\proof
  This is the special case of Theorem \ref{VBundleClassification}
  for $V = \mathbf{B}G$.
\endofproof
For the case that $G$ is 0-truncated (an ordinary group object) 
this is the content of Theorem 23 in \cite{JardineLuo}.
\begin{example}
  For $G \in \mathrm{Grp}(\mathcal{X}) \subset 
  \tau_{\leq 0}\mathrm{Grp}(\mathcal{X})$ an ordinary 1-group object, 
  this reproduces the classical result of \cite{Giraud}, 
  which originally motivated the whole subject: 
  by Example~\ref{automorphism2GroupAbstractly} in this case $\mathbf{Aut}(\mathbf{B}G)$ is the traditional automorphism 2-group and 
  $H^1_{\mathcal{X}}(X, \mathbf{Aut}(\mathbf{B}G))$
  is Giraud's nonabelian $G$-cohomology that classifies $G$-gerbes
  (for arbitrary \emph{band}, see Definition \ref{BandOfInfinityGerbe} below).
  
  For $G \in \tau_{\leq 1}\mathrm{Grp}(\mathcal{X}) \subset 
  \mathrm{Grp}(\mathcal{X})$ a 2-group, we recover the classification of 2-gerbes
  as in \cite{Breen,BreenNotes}.
\end{example}
\begin{remark}
  In Section 7.2.2 of \cite{Lurie} the special case that here
  we called \emph{EM-$n$-gerbes} is considered. Beware 
  that there are further differences: for instance the notion 
  of morphisms between $n$-gerbes as defined in \cite{Lurie} is more 
  restrictive than the notion considered here. For instance with our definition 
  (and hence also that in \cite{Breen}) each group automorphism of 
  an abelian group object $A$ induces an automorphism of the 
  trivial $A$-2-gerbe $\mathbf{B}^2 A$. But, except for the identity, 
  this is not admitted in \cite{Lurie} (manifestly so by the diagram 
  above Lemma 7.2.2.24 there). Accordingly, the classification 
  result in \cite{Lurie} is different: it involves  the cohomology group  
  $H^{n+1}_{\mathcal{X}}(X, A)$. Notice that there is a canonical morphism
  $$
    H^{n+1}_{\mathcal{X}}(X, A) \to H^1_{\mathcal{X}}(X, \mathbf{Aut}(\mathbf{B}^n A))
  $$  
  induced from the morphism $\mathbf{B}^{n+1}A \to \mathbf{Aut}(\mathbf{B}^n A)$.
\end{remark}

We now discuss how the $\infty$-group extensions, Definition \ref{ExtensionOfInfinityGroups},
given by the Postnikov stages of $\mathbf{Aut}(\mathbf{B}G)$ induces the notion of
\emph{band} of a gerbe, and how the corresponding twisted cohomology,
according to Remark \ref{ExtensionsAndTwistedCohomology}, reproduces 
the original definition of nonabelian cohomology in \cite{Giraud} and generalizes
it to higher degree.
\begin{definition}
 \label{outerAutomorphismInfinityGroup}
 Fix $k \in \mathbb{N}$. For $G \in \infty\mathrm{Grp}(\mathcal{X})$ a 
 $k$-truncated $\infty$-group object (a $(k+1)$-group), write
 $$
   \mathbf{Out}(G) := \tau_{k}\mathbf{Aut}(\mathbf{B}G)
 $$
 for the $k$-truncation of $\mathbf{Aut}(\mathbf{B}G)$. (Notice 
 that this is still an $\infty$-group, since by Lemma 6.5.1.2 in 
 \cite{Lurie} $\tau_n$ preserves all $\infty$-colimits and additionally 
 all products.) We call this the \emph{outer automorphism $n$-group} of $G$.
 
 In other words, we write
 $$
   \mathbf{c} : \mathbf{B}\mathbf{Aut}(\mathbf{B}G) \to \mathbf{B}\mathbf{Out}(G)
 $$
 for the top Postnikov stage of $\mathbf{B}\mathbf{Aut}(\mathbf{B}G)$.
\end{definition}
\begin{example}
  Let $G \in \tau_0\mathrm{Grp}(\mathrm{Grpd}_{\infty})$ be 
  a 0-truncated group object, an 
  ordinary group,.
  Then by Example \ref{automorphism2GroupAbstractly}, 
  $\mathbf{Out}(G) = \mathbf{Out}(G)$ is the coimage of 
  $\mathrm{Ad} : G \to \mathrm{Aut}(G)$, which is the traditional group of outer automorphisms of $G$.
\end{example}
\begin{definition}
  Write $\mathbf{B}^2 \mathbf{Z}(G)$ for the $\infty$-fiber of 
  the morphism $\mathbf{c}$ from Definition \ref{outerAutomorphismInfinityGroup}, 
  fitting into a fiber sequence
  $$
    \xymatrix{
      \mathbf{B}^2 \mathbf{Z}(G) \ar[r] & 
	    \mathbf{B}\mathbf{Aut}(\mathbf{B}G) \ar[d]^{\mathbf{c}}
	  \\
	  & \mathbf{B}\mathbf{Out}(G)
	}
	\,.
  $$
  We call $\mathbf{Z}(G)$ the \emph{center} of the $\infty$-group $G$.
  \label{Center}
\end{definition}
\begin{example}
  For $G$ an ordinary group, so that $\mathbf{Aut}(\mathbf{B}G)$ 
  is the automorphism 2-group from Example \ref{automorphism2GroupAbstractly},
  $\mathbf{Z}(G)$ is the center of $G$ in the traditional sense.
\end{example}
By theorem \ref{ClassificationOfGGerbes} there is an  induced morphism
$$
  \mathrm{Band} 
    : 
  \pi_0 G \mathrm{Gerbe} \to H^1(X, \mathbf{Out}(G))
   \,.
$$
\begin{definition}
  \label{BandOfInfinityGerbe}
  For $E \in G \mathrm{Gerbe}$ we call $\mathrm{Band}(E)$ the \emph{band} of $E$.

  By using Definition \ref{Center} in Definition \ref{TwistedCohomologyInOvertopos}, 
  given a band $[\phi_X] \in H^1(X, \mathbf{Out}(G))$, 
  we may regard it as a twist for twisted $\mathbf{Z}(G)$-cohomology,
  classifying $G$-gerbes with this band:
  $$
    \pi_0 G \mathrm{Gerbe}^{[\phi_X]}(X) \simeq H^{2+[\phi_X]}(X, \mathbf{Z}(G))
	\,.
  $$
\end{definition}
\begin{remark}
  The original definition of \emph{gerbe with band} in \cite{Giraud} is slightly
  more general than that of \emph{$G$-gerbe} (with band) in \cite{Breen}: in the former
  the local sheaf of groups whose delooping is locally equivalent to the gerbe need not
  descend to the base. These more general Giraud gerbes are 1-gerbes in the sense of
  Definition \ref{nGerbe}, but only the slightly more restrictive $G$-gerbes of Breen
  have the good property of being connected fiber $\infty$-bundles. From our
  perspective this is the decisive property of gerbes, and the notion of band is 
  relevant only in this case.
\end{remark}
\begin{example}
  For $G$ a 0-group this reduces to the notion of band as introduced in  \cite{Giraud},
  for the case of $G$-gerbes as in \cite{Breen}.
\end{example}

\subsubsection{Relative cohomology}
\label{StrucRelativeCohomology}
\index{structures in a cohesive $\infty$-topos!relative cohomology}

We discuss the notion of \emph{relative cohomology} internal to 
any $\infty$-topos $\mathbf{H}$.

\medskip

\begin{definition}
  \label{RelativeCohomology}
  Let $i : Y \to X$ and $f : B \to A$ be two morphisms in $\mathbf{H}$.
  We say that the $\infty$-groupoid of \emph{relative cocycles} on $i$
  with coefficients in $f$ is the hom $\infty$-groupoid
  $\mathbf{H}^I(i,f)$, where $\mathbf{H}^I := \mathrm{Funct}(\Delta[1], \mathbf{H})$.  
  The corresponding set of equivalence classes / homotopy classes
  we call the \emph{relative cohomology}
  $$
    H_Y^B(X,A) := \pi_0 \mathbf{H}^I(i,f)
	\,.
  $$
  
  When $A$ is understood to be a pointed object, 
  $B = *$ is the terminal object and $f : B \simeq * \to A$
  is the point inclusion, we speak for short of the 
  \emph{cohomology of $X$ with coefficients in $A$ relative to $Y$}
  and write
  $$
    H_Y(X,A) := H_Y^*(X,A)
	\,.
  $$
\end{definition}
\begin{proposition}
  The $\infty$-groupoid of relative cocycles fits into an
  $\infty$-pullback diagram of the form
  $$
    \raisebox{20pt}{
    \xymatrix{
	  \mathbf{H}^I(i,f) \ar[r] \ar[d] & \mathbf{H}(X,A) \ar[d]^{i^*}
	  \\
	  \mathbf{H}(Y,B)
	  \ar[r]^{f_*}
	  &
	  \mathbf{H}(Y,A)
	}
	}
	\,.
  $$
\end{proposition}
\proof
  Let $C$ be an $\infty$-site of definition of $\mathbf{H}$ and
  $$
    \mathbf{H} \simeq ([C^{\mathrm{op}}, \mathrm{sSet}]_{\mathrm{proj}, \mathrm{loc}})^\circ
  $$
  be a presentatin by simplicial presheaves as in \ref{InfinityToposPresentation}.
  Then $\mathbf{H}^I$ is presented by the, say, Reedy model structure on 
  simplicial functors from $\Delta[1]$ to simplicial presheaves
  $$
    \mathbf{H}^I \simeq 
	(
	  [\Delta[1], \, [C^{\mathrm{op}}, \mathrm{sSet}]_{\mathrm{proj}, \mathrm{loc}}]_{\mathrm{Reedy}}
	)^\circ
	\,.
  $$
  We may find for $i : Y \to X$ in $\mathbf{H}$ a presentation by a cofibration between 
  cofibrant objects in $[C^{\mathrm{op}}, \mathrm{sSet}]_{\mathrm{proj}, \mathrm{loc}}$,
  and similarly for $f : B \to A$ a presentation by a fibration between fibrant objects.
  Let these same symbols now denote these presentations. Then $i$ is also cofibrant
  in the above presentation for $\mathbf{H}^I$ and similarly $f$ is fibrant there. 
  
  This implies that the $\infty$-categorical hom space in question is given by the 
  hom-simplicial set
  $$
    \mathbf{H}^I(i,f) \simeq 
	 [\Delta[1], \, [C^{\mathrm{op}}, \mathrm{sSet}(i,f)
	 \,.
  $$
  This in turn is computed as the 1-categorical pullback of simplicial sets
  $$
    \xymatrix{
	   [\Delta[1], \, [C^{\mathrm{op}}, \mathrm{sSet}(i,f)
	   \ar[r] \ar[d] & 
	   [C^{\mathrm{op}}, \mathrm{sSet}](X,A)
	   \ar[d]^{i^*}
	   \\
	   [C^{\mathrm{op}}, \mathrm{sSet}](Y,A)
	   \ar[r]^{f_*}
	   &
	   [C^{\mathrm{op}}, \mathrm{sSet}](Y,A)
	}
	\,.
  $$
  Since $[C^{\mathrm{op}}, \mathrm{sSet}]$ is a simplicial model category,
  and by assumption on our presentations for $i $ and $f$, here the bottom and the
  right morphism are Kan fibrations. Therefore by prop. \ref{ConstructionOfHomotopyLimits}
  this presents a homotopy pullback diagram, which proves the claim.
\endofproof
\begin{remark}
  This says in words that a cocycle relative to $Y \to X$ with coefficients in
  $B \to A$ is an $A$-cocycle on $X$ whose pullback to $Y$ is 
  equipped with a coboundary to a $B$-cocycle. In particular, in the case that
  $B \simeq *$ it is an $A$-cocycle on $X$ equipped with a trivialization of its
  pullback to $Y$.
\end{remark}
In the case that $B$ is not trivial, this definition of relative cohomology is 
a generalization of the twisted cohomology discussed in \ref{StrucTwistedCohomology}.
\begin{observation}
  Let $\mathbf{c} : X \to A$ be a fixed $A$-cocycle on $X$. Then the
  fiber of the $\infty$-groupoid of $(i,f)$-relative cocycles over 
  $\mathbf{c}$ is equivalently the $\infty$-groupoid of $[i^* \mathbf{c}]$-twisted 
  cohomology on $Y$, according to def. \ref{TwistedCohomologyInOvertopos}.
\end{observation}
\proof
  By the pasting law, prop. \ref{PastingLawForPullbacks}
  the relative cocycles over $\mathbf{c}$ sitting in the top $\infty$-pullback
  square of
  $$
    \xymatrix{
	  \mathbf{H}^I(i,f)|_{\mathbf{c}}
	  \ar[r]
	  \ar[d]
	  &
	  {*}
	  \ar[d]^{\mathbf{c}}
	  \\
	  \mathbf{H}^I(i,f) 
	  \ar[r]
	  \ar[d]
	  &
	  \mathbf{H}(X,A)
	  \ar[d]^{i^*}
	  \\
	  \mathbf{H}(Y,B)
	  \ar[r]^{f_*}
	  &
	  \mathbf{H}(Y,A)
	}
  $$
  also form the $\infty$-pullback of the total rectangle, which is the 
  $\infty$-groupoid of $[i^* \mathbf{c}]$-twisted cocycles on $Y$.
\endofproof
\begin{remark}
  In the special case that the coefficients $B$ and $A$ have a presentation 
  by sheaves of chain complexes in the 
  image of the Dold-Kan correspondence, prop. \ref{EmbeddingOfChainComplexes}, 
  the morphism $i^* : \mathbf{H}(X, A) \to \mathbf{H}(Y,A)$ has a presentation
  by a morphism of cochain complexes and 
  the above $\infty$-pullback may be computed in terms of the dual
  mapping cone on this morphism. Specicially in the case that $B \simeq *$
  the homotopy pullback is presented by that dual mapping cone itself, 
  and hence the relative cohomology is the cochain cohomology of the
  mapping cone on $i^*$. In this form relative cohomology is traditionally
  defined in the literature.
\end{remark}

\subsection{Structures in a local $\infty$-topos}
\label{Structures in a local topos}

We discuss structures present in a \emph{local $\infty$-topos},
def. \ref{LocalInfinityTopos}.

\begin{itemize}
  \item \ref{StrucCodiscrete} -- Codiscrete objects;
  \item \ref{StrucConcrete} -- Concrete objects.
\end{itemize}

\subsubsection{Codiscrete objects}
\label{StrucCodiscrete}
 \index{structures in a cohesive $\infty$-topos!codiscrete objects}

\begin{observation}
  The cartesian internal hom 
  $[-,-]: \mathbf{H}^{\mathrm{op}} \times \mathbf{H} \to \mathbf{H}$
  is related to the external hom
  $\mathbf{H}(-,-): \mathbf{H}^{\mathrm{op}} \times \mathbf{H} \to \infty \mathrm{Grpd}$ by
  $$
    \mathbf{H}(-,-) \simeq \Gamma [-,-]
	\,..
  $$
\end{observation}
\proof
  The $\infty$-Yoneda lemma implies, by the same argument as for
  1-categorical sheaf toposes, that the internal hom
  is the $\infty$-stack given on any test object $U$ by
  $$
    [X,A](U) \simeq \mathbf{H}(U, [X,A]) \simeq \mathbf{H}(X \times U, A).
  $$
  By prop. \ref{GlobalSectionMorphismExplicitly} the global section
  functor $\Gamma$ is given by evaluation on the point, so that
  $$
    \Gamma([X,A])
	\simeq
	\mathbf{H}(*, [X,A])
	\simeq
	\mathbf{H}(X \times *, A)
	\simeq
	\mathbf{H}(X,A)
	\,.
  $$
\endofproof
\begin{proposition}
  \label{codiscreteObjectsStableUnderExponentiation}
  The codiscrete objects in a local $\infty$-topos, 
  hence in a cohesive $\infty$-topos,
  $\mathbf{H}$ are stable under internal exponentiation: 
  for all $X \in \mathbf{H}$ and $A \in \infty \mathrm{Grpd}$ we have
  $$
    [X, \mathrm{coDisc}A] \in \mathbf{H}
  $$ 
  is codiscrete. Specifically, the
  internal hom into a codiscrete object
  is the codiscretificartion of the external hom
  $$
    [X, \mathrm{coDisc}A] 
	  \simeq 
	\mathrm{coDisc} \mathbf{H}(X, \mathrm{coDisc} A)
	\,.
  $$
\end{proposition}
\proof
  The internal hom is the $\infty$-stack given by the assignment
  $$
    [X, \mathrm{coDisc} A] : U \mapsto \mathbf{H}(X \times U, \mathrm{coDisc}A)
	\,.
  $$
  By the $(\Gamma \dashv \mathrm{Disc})$-adjunction the right hand is
  $$
    \simeq \infty \mathrm{Grpd}(\Gamma(X \times U), A)
	\,.
  $$
  Since $\Gamma$ is also a right adjoint it preserves the product, so that
  $$
    \cdots \simeq \infty \mathrm{Grpd}(\Gamma(X) \times \Gamma(U), A)
	\,.
  $$
  Using the cartesian closure of $\infty \mathrm{Grpd}$ this is
  $$
    \cdots \simeq \infty \mathrm{Grpd}(\Gamma(U), [\Gamma(X), A])
	\,.
  $$
  Using again the $(\Gamma \dashv \mathrm{coDisc})$-adjunction this 
  is
  $$
    \cdots \simeq \mathbf{H}(U, \mathrm{coDisc}[\Gamma(X), A]).
  $$
  Since all of these equivalence are 
  natural, with the $\infty$-Yoneda lemma it finally follows that
  $$
    [X, \mathrm{coDisc}A ] \simeq \mathrm{coDisc} \infty\mathrm{Grpd}(\Gamma(X), A)
	\simeq \mathrm{coDisc} \mathbf{H}(X, \mathrm{coDisc} A)
	\,.
  $$
\endofproof

\subsubsection{Concrete objects}
\label{StrucConcrete}
 \index{structures in a cohesive $\infty$-topos!concrete objects}
 \index{concrete cohesive $\infty$-groupoid}
 \index{structures in a cohesive $\infty$-topos!concrete objects}

The cohesive structure on an object in a cohesive $\infty$-topos need not be
supported by points. We discuss a general abstract characterization of  
objects that do have an interpretation as bare $n$-groupoids equipped with cohesive structure.
Further refinements of these constructions are discussed further below
in \ref{StrucDifferentialModuli} for objects that serve as moduli of differetial cocycles.

The content of this section is partly taken from \cite{CarchediSchreiber}.

\paragraph{General abstract}

\begin{proposition} 
  \label{NablaIsEmbedding}
  On a cohesive $\infty$-topos $\mathbf{H}$ both $\mathrm{Disc}$ and $\mathrm{coDisc}$ 
  are full and faithful
  $\infty$-functors and $\mathrm{coDisc}$ exhibits $\infty \mathrm{Grpd}$ 
  as a sub-$\infty$-topos of $\mathbf{H}$ by an  $\infty$-geometric embedding
  $$
    \xymatrix{
      \infty \mathrm{Grpd}
        \ar@{<-}@<+4pt>[r]^<<<<{\Gamma}
        \ar@{^{(}->}@<-4pt>[r]_<<<<{\mathrm{coDisc}}
        &
       \mathbf{H}
    }
    \,.
  $$
\end{proposition}
\proof
    The full and faithfulness of $\mathrm{Disc}$ was shown in prop. \ref{DeltaIsFF}
    and that for $\mathrm{coDisc}$ follows from the same kind of argument. Since $\Gamma$
    is also a right adjoint it preserves in particular finite $\infty$-limits, so that
    $(\Gamma \dashv \mathrm{coDisc})$ is indeed an $\infty$-geometric morphism.
\endofproof
Recall that we write 
$$
  \sharp := \mathrm{coDisc}\circ \Gamma
$$
\begin{corollary} 
  \label{SubtoposOfcodiscreteObjects}
  The $\infty$-topos $\infty \mathrm{Grpd}$ is equivalent to the full
  sub-$\infty$-category of $\mathbf{H}$ on those objects $X \in \mathbf{H}$ for which the 
  canonical morphism
  $X \to \sharp X$
  is an equivalence.
\end{corollary}
\proof
  This follows by general facts about reflective sub-$\infty$-categories
  (\cite{Lurie}, section 5.5.4).  
\endofproof
\begin{proposition}
  Let $\mathbf{H}$ be the $\infty$-topos over an $\infty$-cohesive site $C$.
  For a 0-truncated object $X$ in $\mathbf{H}$ the morphism
  $$
    X \to \sharp X
  $$
  is a monomorphism precisely if $X$ is a \emph{concrete sheaf} in the 
  traditional sense of \cite{Dubuc}.
  \label{ConcreteSheavesAreConcreteSheaves}
\end{proposition}
\proof
  Monomorphisms of sheaves are detected objectwise. So by the Yoneda lemma and 
  using the $(\Gamma \dashv \mathrm{coDisc})$-adjunction we have that 
  $X \to \mathrm{coDisc}\, \Gamma\, X$ is a monomorphism precisely if for all
  $U \in C$ the morphism
  $$
    X(U) \simeq \mathbf{H}(U,X) \to \mathbf{H}(U, \mathrm{coDisc}\, \Gamma\, X)
    \simeq \mathbf{H}(\Gamma(U), \Gamma(X))
  $$
  is a monomorphism. This is the traditional definition.
\endofproof
\begin{definition}
 For $X \in \mathbf{H}$, write
 $$
   \xymatrix{
     X =: \sharp_\infty X
	 \ar[r]
	 &
	 \cdots
	 \ar[r]
	 &
	 \sharp_2 X
	 \ar[r]
	 &
	 \sharp_1 X
	 \ar[r]
	 &
	 \sharp_0 X := \sharp X
   }
 $$
 for the tower of $n$-image factorizations, def. \ref{nImage}, of $X \to \sharp X$,
 hence with
 $$
   \sharp_n X := \mathrm{im}_n(X\to \sharp X)
 $$
 for all $n \in \mathbb{N}$.
 \label{ImagesOfXToSharpX}
\end{definition}
\begin{definition} 
  \label{ConcreteObjects}
  \index{concrete objects}
  \index{concrete objects!concretification}
  \label{kConcretification}
   For $n \in \mathbb{N}$ and $X \in \mathbf{H}$ an $n$-truncated object, 
we say that $X \to \sharp_{n+1}X$ is its \emph{$n$-concretification}.   
 If this is an equivalence we say that $X$ is \emph{$n$-concrete}.
 \end{definition}

\paragraph{Presentations}

We discuss presentations of $n$-concrete objects for low $n$.

\begin{proposition}
  \label{1ConcreteOverInfinityCohesiveSite}
  Let $C$ be an $\infty$-cohesive site, \ref{CohesiveSites}, 
  and let $A \in \mathrm{Sh}_\infty(C)$ be a 1-truncated object 
  that has a presentation
  by a groupoid-valued presheaf on $C$ which is fibrant
  as a simplicial presheaf. Then it is 1-concrete if
  in degree 1 this is a concrete sheaf. Moreover, its 1-concretification,
  def. \ref{kConcretification}, has a presentation by a 
  presheaf of groupoids which in degree 1 is a concrete sheaf.
\end{proposition}
\proof
  Any functor $f: X \to Y$ between groupoids has a 
  factorization $X \to \mathrm{im}_1 f \to Y$, where
  the groupoid $\mathrm{im}_1 f$ has the same objects
  as $X$ and has as morphisms equivalence classes $[\xi]$ of 
  morphisms $\xi$ in $X$ under the relation 
  $[\xi_1] = [\xi_2]$ precisely if $f(\xi_1) = f(\xi)_2$.
  The evident functor $\mathrm{im}_1 f \to Y$ is manifestly
  faithful and this factorization is natural. Therefore if
  now $f$ is a morphism of presheaves of groupoids,
  it, too, has a factorization wich is objectwise of this form.
  
  By the discussion in \ref{CohesiveSites}, over an 
  $\infty$-cohesive site the units $\eta_X : X \to \sharp X$
  of the $(\Gamma \dashv \mathrm{coDisc})$-$\infty$-adjunction
  are presented for fibrant simplicial presheaf representatives $X$ 
  by morphisms of simplicial presheaves that 
  object- and degreewise send the value set of a presheaf to the 
  set of concrete values. By the previous paragraph and 
  prop. \ref{0TruncatedStackMorphisms} it follows that the
  1-image factorization $X \to \mathrm{im}_1 \eta_X \to \sharp X$
  is in the second morphism objectwise a faithful functor. 
  This means that the hom-presheaf $(\mathrm{im}_1 \eta_X)_1$
  is a concrete sheaf on $C$.
\endofproof

\subsection{Structures in a locally $\infty$-connected $\infty$-topos}
\label{StructuresInLocallyInfinityConnectedTopos}
\label{Structures in a locally infinity-connected topos}

We discuss here homotopical, cohomological and geometrical structures 
that are canonically present in a locally $\infty$-connected $\infty$-topos
$\mathbf{H}$,
\ref{DefConnectedTopos}. The existence of the extra left adjoint
$\Pi$ for a locally $\infty$-connected $\infty$-topos encodes
an intrinsic notion of \emph{geometric paths} in the objects of $\mathbf{H}$.

If $\mathbf{H}$ is in addition \emph{cohesive}, then these
$\Pi$-geometric structures combine with the cohomological structures
of a local $\infty$-topos, discussed in \ref{Structures in a local topos} to produce
differential geometry and differential cohomological structures.
This we discuss below in \ref{structures}.

\medskip

\begin{itemize}
  \item \ref{StrucGeometricHomotopy} -- Geometric homotopy  / {\'E}tale homotopy
  \item \ref{Concordance} -- Concordance
  \item \ref{StrucGeometricPostnikov} -- Paths and geometric Postnikov towers
  \item \ref{StrucGeometricWhitehead} --  Universal coverings and geometric Whitehead towers
  \item \ref{StrucFlatDifferential} -- Flat connections and local systems
  \item \ref{StrucGaloisTheory} -- Galois theory
\end{itemize}

\subsubsection{Geometric homotopy / {\'E}tale homotopy}
\label{StrucGeometricHomotopy}
\index{structures in a cohesive $\infty$-topos!geometric homotopy}

We discuss internal realizations of the notions of 
\emph{geometric realization},
and \emph{geometric homotopy} in any cohesive
$\infty$-topos $\mathbf{H}$.

\medskip

\begin{definition}
For $\mathbf{H}$ a locally $\infty$-connected $\infty$-topos and 
$X \in \mathbf{H}$ an object, we call $\Pi(X) \in \infty \mathrm{Grpd}$  the 
\emph{fundamental $\infty$-groupoid}\index{fundamental $\infty$-groupoid!general abstract} of $X$. 

The ordinary homotopy groups of $\Pi(X)$ we call the 
\emph{geometric homotopy groups}\index{homotopy!geometric} of $X$
$$
  \pi_\bullet^{\mathrm{geom}}(X \in \mathbf{H}) := \pi_\bullet(\Pi (X \in \infty \mathrm{Grpd}))
  \,.
$$
\end{definition}
\begin{definition}
\label{GeometricRealization}
For $\vert - \vert : \infty \mathrm{Grpd} \stackrel{\simeq}{\to}\mathrm{Top}$  
the canonical equivalence of $\infty$-toposes, we write
$$
  \vert X \vert := \vert \Pi X \vert \in \mathrm{Top}
$$
and call this the \emph{geometric realization}\index{geometric realization!general abstract} of $X$.
\end{definition}
\begin{remark}
In presentations of $\mathbf{H}$ by simplicial presheaves, as in 
prop. \ref{InfSheavesOverCohesiveSiteAreCohesive}, 
aspects of this abstract notion are more or less implicit in the literature. 
See for instance around remark 2.22 of \cite{SimpsonTeleman}. The key 
insight is already in \cite{ArtinMazur}, if somewhat implicitly. 
This we discuss in detail in \ref{ETopStrucHomotopy}.
\end{remark}
  
In some applications we need the following characterization of geometric homotopies in
a cohesive $\infty$-topos.
\begin{definition} \label{GeometricHomotopy}
We say a \emph{geometric homotopy} between two morphisms $f,g : X \to Y$
in $\mathbf{H}$ is a diagram
$$
  \xymatrix{
    X \ar[d]_{(\mathrm{Id},i)} \ar[dr]^f 
    \\
    X \times I \ar[r]^\eta & Y
    \\
    X \ar[u]^{(\mathrm{Id},o)} \ar[ur]_{g}
  }
$$
such that $I$ is geometrically connected, $\pi_0^{geom}(I) = *$.
\end{definition}
\begin{proposition}  \label{PiPreservesGeometricHomotopy}
If two morphism $f,g : X\to Y$ in a cohesive $\infty$-topos $\mathbf{H}$ 
are geometrically homotopic  then their images $\Pi(f), \Pi(g)$ are 
equivalent in $\infty Grpd$.
\end{proposition}
\proof
By the condition that $\Pi$ preserves products
in a strongly $\infty$-connected $\infty$-topos we have that the image of the
geometric homotopy in $\infty \mathrm{Grpd}$ is a diagram of the form
$$
  \xymatrix{
    \Pi(X) \ar[d]_{(\mathrm{Id},\Pi(i))} \ar[dr]^{\Pi(f)} 
    \\
    \Pi(X) \times \Pi(I) \ar[r]^{\Pi(\eta)} & \Pi(Y)
    \\
    \Pi(X) \ar[u]^{(\mathrm{Id},\Pi(o))} \ar[ur]_{\Pi(g)}
  }
$$
Since $\Pi(I)$ is connected by assumption, there is a diagram
$$
  \xymatrix{
     & {*} \ar[d]^{\Pi(i)}
     \\
    {*} \ar[ur] \ar[dr]\ar[r] & \Pi(I)
    \\ 
    & {*} \ar[u]_{\Pi(o)}
  }
$$
in $\infty \mathrm{Grpd}$ (filled with homotopies,  which we do not display, as usual, 
that connect the three points in $\Pi(I)$).
Taking the product of this diagram with $\Pi(X)$ and pasting the result
to the above image $\Pi(\eta)$ of the geometric homotopy constructs
the equivalence $\Pi(f) \Rightarrow \Pi(g)$ in $\infty \mathrm{Grpd}$.
\endofproof

We consider a refinement of these kinds of considerations below in \ref{A1HomotopyGeneralAbstract}.

\begin{proposition} 
  \label{ConnectedOverToposes}
For $\mathbf{H}$ a locally $\infty$-connected $\infty$-topos, also all its objects 
$X \in \mathbf{H}$ are locally $\infty$-connected, in the sense that their 
over-$\infty$-toposes $\mathbf{H}/X$ are locally $\infty$-connected
$(\Pi_X \dashv \Delta_X \dashv \Gamma_X) : \mathbf{H}/X \to \infty \mathrm{Grpd}$.

The two notions of fundamental $\infty$-groupoids of any object $X$ induced this way do agree,
in that  there is a natural equivalence
$$
  \Pi_X(X \in \mathbf{H}/X) \simeq \Pi(X \in \mathbf{H})
  \,.
$$
\end{proposition}
\proof
By the general properties of over-$\infty$-toposes (\cite{Lurie}, prop 6.3.5.1)
we have a a composite essential $\infty$-geometric morphism
$$
  (\Pi_X \dashv \Delta_X \dashv \Gamma_X) : 
  \xymatrix{
    \mathbf{H}/X
      \ar@<+12pt>[r]^<<<<<{X_!}
      \ar@<+4pt>@{<-}[r]|<<<<<{X^*}
      \ar@<-4pt>[r]_<<<<<{X_*}
      &
    \mathbf{H}
      \ar@<+12pt>[r]^<<<<<{\Pi}
      \ar@<+4pt>@{<-}[r]|<<<<<{\Delta}
      \ar@<-4pt>[r]_<<<<<{\Gamma}
      &
    \infty \mathrm{Grpd}
  }
$$
and $X_!$ is given by sending $(Y \to X) \in \mathbf{H}/X$ to $Y \in \mathbf{H}$.
\endofproof

\subsubsection{Concordance}
\label{Concordance}
\index{concordance}
\index{structures in a cohesive $\infty$-topos!concordance}

We formulate the notion of \emph{concordance} (of bundles or cocycles)
abstractly internal to a cohesive $\infty$-topos.

\medskip

\begin{definition}
  For $\mathbf{H}$ a cohesive $\infty$-topos and $X,A \in \mathbf{H}$
  two objects, we say that the $\infty$-groupoid of \emph{concordances}
  from $X$ to $A$ is 
  $$
    \mathrm{Concord}(X,A) := \Pi [X, A]
	\,,
  $$
  where $[-,-] : \mathbf{H}^{\mathrm{op}} \times \mathbf{H} \to \mathbf{H}$
  is the internal hom.
\end{definition}
\begin{observation}
  For $X, A, B \in \mathbf{H}$
  three objects, there is a canonical composition $\infty$-functor
  of concordances between them
  $$
    \mathrm{Concord}(X,A) \times \mathrm{Concord}(A,B)
	\to
	\mathrm{Concord}(X,B)
	\,.
  $$
  Using that, by the axioms of cohesion, $\Pi$ preserves products, this 
  is the image under $\Pi$ of the composition on internal homs
  $$
    [X,A] \times [A,B] \to [X,B]
	\,.
  $$
\end{observation}

\subsubsection{Paths and geometric Postnikov towers}
\label{StrucGeometricPostnikov}
\label{StrucPaths}
\index{structures in a cohesive $\infty$-topos!paths}
\index{structures in a cohesive $\infty$-topos!geometric Postnikov tower}

The fundamental $\infty$-groupoid $\Pi X$ of objects $X$ in 
$\mathbf{H}$ may be reflected back into 
$\mathbf{H}$, where it gives a notion of \emph{geometric homotopy path $n$-groupoids} and a  geometric notion of Postnikov towers of objects in $\mathbf{H}$.

\medskip

Recall from def. \ref{TheAdjointTripleOfEndofunctors} the 
pair of adjoint endofunctors
$$
  (\mathbf{\Pi} \dashv \flat)
  :
  \mathbf{H}
  \to
  \mathbf{H}
$$
on any locally connected $\infty$-topos $\mathbf{H}$.

We say for any $X, A \in  \mathbf{H}$
\begin{itemize}
\item $\mathbf{\Pi}(X)$ is the \emph{path $\infty$-groupoid}
\index{paths!path $\infty$-groupoid} of $X$ -- 
the reflection of the fundamental $\infty$-groupoid from \ref{StrucGeometricHomotopy} 
back into the cohesive context of $\mathbf{H}$;

\item $\mathbf{\flat} A$ (``flat $A$'') is the coefficient object for 
  \emph{flat differential $A$-cohomology}
  or for \emph{$A$-local systems} (discussed below in \ref{StrucFlatDifferential}).
\end{itemize}
Write
$$
  (\tau_n \dashv i_n)
  :
  \mathbf{H}_{\leq n}
  \stackrel{\overset{\tau_{n}}{\leftarrow}}{\underset{i}{\hookrightarrow}}
  \mathbf{H}
$$
for the reflective sub-$\infty$-category of $n$-truncated objects 
(\cite{Lurie}, section 5.5.6)
and 
$$
  \mathbf{\tau}_n : 
    \mathbf{H} \stackrel{\tau_n}{\to} \mathbf{H}_{\leq n}
  \hookrightarrow
  \mathbf{H}
$$
for the localization funtor. We say
$$
  \mathbf{\Pi}_n : \mathbf{H} \stackrel{\mathbf{\Pi}_n}{\to}
   \mathbf{H} \stackrel{\mathbf{\tau}_n}{\to}
  \mathbf{H}
$$
is the \emph{homotopy path $n$-groupoid} functor. 
The (truncated) components of the $(\Pi \dashv \mathrm{Disc})$-unit
$$
  X \to \mathbf{\Pi}_n(X)
$$
we call the \emph{constant path inclusion}. Dually we have canonical morphisms
$$
  \mathbf{\flat}A \to A
$$
natural in $A \in \mathbf{H}$.

\begin{definition}
For $X \in \mathbf{H}$ we say that the \emph{geometric Postnikov tower}
of $X$ is the categorical Postnikov tower (\cite{Lurie} def. 5.5.6.23) 
of $\mathbf{\Pi}(X) \in \mathbf{H}$ :
$$
  \mathbf{\Pi}(X) \to \cdots
   \to 
   \mathbf{\Pi}_2(X) \to \mathbf{\Pi}_1(X) \to \mathbf{\Pi}_0(X)
  \,.
$$
\end{definition}
The main purpose of geometric Postnikov towers for us is the notion of
\emph{geometric Whitehead towers} that they induce, discussed in the next section.
	
\subsubsection{Universal coverings and geometric Whitehead towers}
\label{StrucGeometricWhitehead}
\index{structures in a cohesive $\infty$-topos!universal coverings}
\index{structures in a cohesive $\infty$-topos!geometric Whitehead tower}

We discuss an intrinsic notion of Whitehead towers in a locally $\infty$-connected 
$\infty$-topos $\mathbf{H}$.

\medskip

\begin{definition}
 \label{GeometricWhiteheadTower}
For $X \in \mathbf{H}$ a pointed object, the 
\emph{geometric Whitehead tower} of $X$ is the sequence of objects 
$$
  X^{\mathbf{(\infty)}} \to \cdots \to X^{\mathbf{(2)}} \to X^{\mathbf{(1)}} \to X^{\mathbf{(0)}} \simeq X
$$
in $\mathbf{H}$, where for each $n \in \mathbb{N}$ the object $X^{(n+1)}$ is the homotopy fiber of 
the canonical morphism $X \to \mathbf{\Pi}_{n+1} X$ to the path $(n+1)$-groupoid of $X$
(\ref{StrucGeometricPostnikov}). 
We call $X^{\mathbf{(n+1)}}$ the $(n+1)$-fold 
\emph{universal covering space} of $X$.
We write $X^{\mathbf{(\infty)}}$ for the homotopy fiber of the untruncated constant path inclusion.
$$
  X^{\mathbf{(\infty)}} \to X \to \mathbf{\Pi}(X)
  \,.
$$

Here the morphisms $X^{\mathbf{(n)}} \to X^{\mathbf{n-1}}$ are those induced from this 
pasting diagram of $\infty$-pullbacks
$$
  \xymatrix{
    X^{\mathbf{(n)}} \ar[r] \ar[d] & {*} \ar[d]
    \\
    X^{\mathbf{(n-1)}} \ar[d]\ar[r] & \mathbf{B}^n \mathbf{\pi}_n(X) \ar[d]\ar[r] & {*} \ar[d]
    \\
    X \ar[r] & \mathbf{\Pi}_n(X) \ar[r]^{\mathbf{\tau}_{\leq (n-1)}} & \mathbf{\Pi}_{(n-1)}(X)
  }
  \,,
$$
where the object $\mathbf{B}^n \mathbf{\pi}_n(X)$ is defined as the homotopy fiber of the bottom right morphism.
\end{definition}
\begin{proposition}
Every object $X$ in a cohesive $\infty$-topos 
$\mathbf{H}$ is covered by objects of the form $X^{\mathbf{(\infty)}}$ 
for different choices of base points in $X$, in the sense that every $X$ is the 
$\infty$-colimit over a diagram whose vertices are of this form.
\end{proposition}
\proof
Consider the diagram
$$
  \xymatrix{
    {\lim\limits_{\longrightarrow}}_{s \in \Pi(X)} (i^{*} {*}_s)
      \ar[r] 
      \ar[d]^\simeq
    & 
    {\lim\limits_{\longrightarrow}}_{s \in \Pi(X)} {*}_s
    \ar[d]^\simeq
    \\
    X \ar[r]^i & \mathbf{\Pi}(X)
  }
  \,.
$$
The bottom morphism is the constant path inclusion, the $(\Pi \dashv \mathrm{Disc})$-unit. 
The right morphism is the equivalence that is the image under $\mathrm{Disc}$ of the 
decomposition ${\lim\limits_{\longrightarrow}}_S {*} \stackrel{\simeq}{\to} S$ of every $\infty$-groupoid 
as the $\infty$-colimit over itself of the $\infty$-functor constant on the point.
The left morphism is the $\infty$-pullback along $i$ of this equivalence, hence itself an equivalence. 
By universality of $\infty$-colimits in the $\infty$-topos $\mathbf{H}$, the top left object is 
the $\infty$-colimit over the single homotopy fibers $i^* *_s$  of the form $X^{\mathbf{(\infty)}}$ as indicated. 
\endofproof
We would like claim that moreover each of the patches $i^* {*}$ of the object $X$ 
in a cohesive $\infty$-topos is geometrically contractible,
thus exhibiting a generic cover of any object by contractibles. 
The following states something slightly weaker.
\begin{proposition}
The inclusion $\Pi(i^* {*}) \to \Pi(X)$ of the fundamental $\infty$-groupoid $\Pi(i^* *)$ 
of each of these patches into $\Pi(X)$ is homotopic to the point.
\end{proposition}
\proof
We apply $\Pi(-)$ to the above diagram over a single vertex $s$ and attach the 
$(\Pi \dashv \mathrm{Disc})$-counit to get
$$
  \xymatrix{
    \Pi(i^* *)
      \ar[rr]
      \ar[d]
      && 
      {*}
      \ar[d]
    \\
    \Pi(X) \ar[r]^{\Pi(i)}& 
   \Pi \, \mathrm{Disc} \, \Pi(X)
     \ar[r] & \Pi(X)
  }
  \,.
$$
Then the bottom morphism is an equivalence by the $(\Pi \dashv \mathrm{Disc})$-zig-zag-identity.
\endofproof
This implies that in a cohesive $\infty$-topos every principal

\subsubsection{Flat connections and local systems}
\label{StrucFlatDifferential}
\index{structures in a cohesive $\infty$-topos!flat $\infty$-connections}
\index{structures in a cohesive $\infty$-topos!local system}
\index{parallel transport!flat $\infty$-transport}

We describe for a locally $\infty$-connected 
$\infty$-topos $\mathbf{H}$ a canonical intrinsic notion 
of \emph{flat connections on $\infty$-bundles}, \emph{flat higher parallel transport} 
and \emph{$\infty$-local systems}.

\medskip

Let $\mathbf{\Pi}: \mathbf{H} \to \mathbf{H}$ be the path $\infty$-groupoid functor from 
def. \ref{TheAdjointTripleOfEndofunctors}, discussed in \ref{StrucPaths}.
\begin{definition}
  \label{FlatDifferentialCohomology}
For $X, A \in \mathbf{H}$ we write
$$
  \mathbf{H}_{\mathrm{flat}}(X,A) := \mathbf{H}(\mathbf{\Pi}X, A)
$$
and call $H_{\mathrm{flat}}(X,A) := \pi_0 \mathbf{H}_{\mathrm{flat}}(X,A)$ the 
\emph{flat (nonabelian) differential cohomology} of $X$ with coefficients in $A$.
We say a morphism $\nabla : \mathbf{\Pi}(X) \to A$ is a 
\emph{flat $\infty$-connnection} on the principal $\infty$-bundle corresponding to 
$X \to \mathbf{\Pi}(X) \stackrel{\nabla}{\to} A$, or an \emph{$A$-local system} on $X$.

The induced morphism
$$
  \mathbf{H}_{\mathrm{flat}}(X,A) \to \mathbf{H}(X,A)
$$
we say is the forgetful functor that \emph{forgets flat connections}.
\end{definition}
The object $\mathbf{\Pi}(X)$ has the interpretation of the path $\infty$-groupoid of $X$: 
it is a cohesive $\infty$-groupoid whose $k$-morphisms may be thought of as generated from 
the $k$-morphisms in $X$ and $k$-dimensional cohesive paths in $X$.
Accordingly a mophism $\mathbf{\Pi}(X) \to A$ may be thought of as assigning
\begin{itemize}
\item 
  to each point of $X$ a fiber in $A$;
\item 
  to each path in $X$ an equivalence between these fibers;
\item 
  to each disk in $X$ a 2-equivalalence between these equivaleces associated 
to its boundary
\item and so on.
\end{itemize}
This we think of as encoding a flat \emph{higher parallel transport} on $X$, 
coming from some flat $\infty$-connection and \emph{defining} this flat $\infty$-connection.
\begin{observation}
By the $(\mathbf{\Pi} \dashv \mathbf{\flat})$-adjunction we have a natural equivalence
$$
  \mathbf{H}_{\mathrm{flat}}(X,A) \simeq \mathbf{H}(X,\mathbf{\flat}A)
  \,.
$$
A cocycle $g : X \to A$ for a principal $\infty$-bundle on $X$ is in the image of 
$$
  \mathbf{H}_{\mathrm{flat}}(X,A) \to \mathbf{H}(X,A)
$$
precisely if there is a lift $\nabla$ in the diagram
$$
  \xymatrix{
     & \mathbf{\flat}A \ar[d]
     \\
     X \ar[ur]^\nabla \ar[r]^g & A
  }
  \,.
$$
\end{observation}
We call $\mathbf{\flat}A$ the \emph{coefficient object for flat $A$-connections}.

\begin{proposition}
For $G := \mathrm{Disc}(G_0) \in \mathbf{H}$ 
discrete $\infty$-group (\ref{StrucInftyGroups}) the canonical morphism 
$\mathbf{H}_{\mathrm{flat}}(X,\mathbf{B}G) \to \mathbf{H}(X,\mathbf{B}G)$ 
is an equivalence.
\end{proposition}
\proof
  This follows by definition 
  \ref{TheAdjointTripleOfEndofunctors} $\mathbf{\flat} = \mathrm{Disc} \, \Gamma$ 
  and using that $\mathrm{Disc}$
  is full and faithful.
\endofproof
This says that for discrete structure $\infty$-groups $G$ there is an essentially 
unique flat $\infty$-connection on any $G$-principal $\infty$-bundle. Moreover, 
the further equivalence
$$
  \mathbf{H}(\mathbf{\Pi}(X), \mathbf{B}G)
  \simeq
  \mathbf{H}_{\mathrm{flat}}(X, \mathbf{B}G)
  \simeq
  \mathbf{H}(X, \mathbf{B}G)
$$
may be read as saying that the $G$-principal $\infty$-bundle for discrete $G$
is entirely characterized by the flat higher parallel transport
of this unique $\infty$-connection.

Below in \ref{StrucGaloisTheory}
we discuss in more detail 
the total spaces classified by $\infty$-local systems.

\subsubsection{Galois theory}
\label{StrucGaloisTheory}
\index{structures in a cohesive $\infty$-topos!Galois theory}

We discuss a canonical internal realization of 
\emph{locally constant $\infty$-stacks} and their
classification by
\emph{Galois theory} inside any cohesive $\infty$-topos.

\emph{Classical Galois theory} is the classification of certain extensions of a field $K$.
Viewing the formal dual $\mathrm{Spec}(K)$ as a space, this generalizes
to \emph{Galois theory of schemes}, which classifies $\kappa$-compact 
{\'e}tale morphisms $E \to X$
over a connected scheme $X$ by functors 
$$
  {\Pi}_1(X) \simeq \mathbf{B} \pi_1(X) \to \mathrm{Set}_\kappa
$$
from the classifying groupoid of the fundamental group of $X$ (defined thereby) to the category
of $\kappa$-small sets. See for instance \cite{Lenstra} for an account. 

From the point of view of topos theory over the {\'e}tale site, $\kappa$-compact
{\'e}tale morphisms are equivalently sheaves 
(namely the sheaves of local sections of the {\'e}tale morphism) 
that are locally constant on $\kappa$-small sets. The notion 
of locally constant sheaves of course exists over any
site and in any topos whatsoever, and hence \emph{topos theoretic Galois theory}
more generally classifies locally constant sheaves. A general abstract 
category theoretic discussion of such generalized Galois theory 
is given by Janelidze, whose construction
in the form of \cite{CJKP} we generalize below to locally connected $\infty$-toposes.

A generalization of Galois theory from topos theory to $\infty$-topos theory
as a classification of \emph{locally constant $\infty$-stacks} was envisioned by
Grothendieck and, for the special case over topological spaces, first formalized in 
\cite{Toen00},
where it is shown that the homotopy type of a connected locally contractible topological 
space $X$ is the automorphism $\infty$-group of the fiber functor on locally constant
$\infty$-stacks over $X$. Similar discussion appeared later in \cite{PoleselloWaschkies}
and \cite{Shulman}.

We show below that this central statement of \emph{higher Galois theory} holds generally 
in every $\infty$-connected $\infty$-topos.

\medskip

For $\kappa$ an uncountable regular cardinal, write 
$$
  \mathrm{Core}\, \infty \mathrm{Grpd}_\kappa \in \infty \mathrm{Grpd}
$$ 
for the $\infty$-groupoid of $\kappa$-small $\infty$-groupoids,
def. \ref{CoreInfinityGroupoidKappa}.
\begin{definition}
  \label{LocallyConstantInfinityStacks}
For $X \in \mathbf{H}$ write
$$
  \mathrm{LConst}(X) := \mathbf{H}(X, \mathrm{Disc} (\mathrm{Core}\,\infty \mathrm{Grpd}_\kappa))
$$
for the cocycle $\infty$-groupoid on $X$ with coefficients in the discretely cohesive
$\infty$-groupoid on the $\infty$-groupoid of $\kappa$-small $\infty$-groupoids.
We call this the $\infty$-groupoid of \emph{locally constant $\infty$-stacks}
\index{locally constant $\infty$-stack} on $X$.
\end{definition}
\begin{observation}
 \label{LocallyConstantStacksAsPrincipalBundles}
Since $\mathrm{Disc}$ is left adjoint and right adjoint, it commutes with 
coproducts and with delooping, def. \ref{delooping}, so that by remark 
\ref{CoreInfinityGroupoidAsCorpoductOfAutomorphisms} we have
$$
  \mathrm{Disc}( \mathrm{Core}\, \infty \mathrm{Grpd}_\kappa )
    \simeq 
  \coprod_i \mathbf{B}\, \mathrm{Disc}( \mathrm{Aut}(F_i))
  \,.
$$
Therefore, by the discussion in \ref{section.PrincipalInfinityBundle}, 
a locally constant $\infty$-stack 
$P \in \mathrm{LConst}(X)$ may be identified on each 
geometric connected component of $X$ with 
the total space of a  $\mathrm{Disc}\, \mathrm{Aut}(F_i)$-principal 
$\infty$-bundle $P \to X$. 

Moreover, by the discussion in \ref{StrucRepresentations}, to 
each such $\mathrm{Aut}(F_i)$-principal $\infty$-bundle is canonically
associated a $\mathrm{Disc}(F_i)$-fiber $\infty$-bundle $E \to X$.
This is the $\infty$-pullback
$$
  \raisebox{20pt}{
  \xymatrix{
    E \ar[r] \ar[d] & \mathrm{Disc}(F_i) /\!/ \mathrm{Disc}(\mathrm{Aut}(F_i))
	\ar[d]
	\\
	X \ar[r] & \mathbf{B} \mathrm{Disc}(\mathrm{Aut}(F_i))
  }
  }
  \,.
$$

Since by corollary \ref{CharacterizationOfDiscreteInfinityBundles} every 
discrete $\infty$-bundle with $\kappa$-small fibers over connected $X$ 
arises this way, essentially uniquely,
we may canonically identify the morphism $E \to X$
with an object $E \in \mathbf{H}_{/X}$ in the little topos over $X$,
which interprets as the $\infty$-topos of $\infty$-stacks over $X$,
as discussed at the beginning of \ref{StrucInftyGerbes}.
This way the objects of $\mathrm{LConst}(X)$ are indeed identified 
with $\infty$-stacks over $X$.
\end{observation}

The following proposition says that the central statement of 
Galois theory\index{Galois theory} 
holds for the notion of locally constant $\infty$-stacks 
in a cohesive $\infty$-topos.
\begin{proposition}
For $\mathbf{H}$ locally and globally $\infty$-connected, we have
\begin{enumerate}
\item 
  a natural equivalence 
  $$
    \mathrm{LConst}(X) \simeq \infty \mathrm{Grpd}(\Pi(X), \infty \mathrm{Grpd}_\kappa)
  $$
  of locally constant $\infty$-stacks on $X$ with 
   $\infty$-permutation representations of the fundamental   
   $\infty$-groupoid of $X$ (\emph{local systems} on $X$);

\item for every point $x : * \to X$ a natural equivalence of the endomorphisms of 
   the fiber functor 
   $$
     x^* : \mathrm{LConst}(X) \to \infty \mathrm{Grpd}_\kappa
   $$ 
   and the loop space of $\Pi(X)$ at $x$
  $$
    \mathrm{End}( x^* ) \simeq \Omega_x \Pi(X)
    \,.
  $$
\end{enumerate}
\end{proposition}
\proof
The first statement is essentially the $(\Pi \dashv \mathrm{Disc})$-adjunction :
$$
  \begin{aligned}
     \mathrm{LConst}(X) & := 
	   \mathbf{H}(X, \mathrm{Disc}( \mathrm{Core}\, \infty \mathrm{Grpd}_\kappa))
     \\
     &  \simeq \infty \mathrm{Grpd}(\Pi(X), \mathrm{Core}\, \infty \mathrm{Grpd}_\kappa)
     \\
     &  \simeq \infty \mathrm{Grpd}(\Pi(X), \infty \mathrm{Grpd}_\kappa)     
  \end{aligned}
  \,.
$$
Using this and that $\Pi$ preserves the terminal object, so that the adjunct of
$({*} \to X \to \mathrm{Disc}\, \mathrm{Core}\, \infty \mathrm{Grpd}_\kappa)$ is
$({*} \to \Pi(X) \to \infty \mathrm{Grpd}_\kappa)$,
the second statement follows with an iterated application of the $\infty$-Yoneda lemma: 

The fiber functor $x^* : \mathrm{Func}_\infty(\Pi(X), \infty \mathrm{Grpd}) \to \infty \mathrm{Grpd}$ 
evaluates an $\infty$-presheaf on $\Pi(X)^{\mathrm{op}}$ at $x \in \Pi(X)$. 
By the $\infty$-Yoneda lemma this is the same as homming out of $j(x)$, 
where $j : \Pi(X)^{\mathrm{op}} \to \mathrm{Func}(\Pi(X), \infty \mathrm{Grpd})$ is the $\infty$-Yoneda embedding:
$$
  x^* \simeq \mathrm{Hom}_{\mathrm{PSh}(\Pi(X)^{op})}(j(x), -)
  \,.
$$
This means that $x^*$ itself is a representable object in 
$\mathrm{PSh}_\infty(\mathrm{PSh}_\infty(\Pi(X)^{\mathrm{op}})^{\mathrm{\mathrm{op}}})$. 
If we denote by $\tilde j : \mathrm{PSh}_\infty(\Pi(X)^{\mathrm{op}})^{\mathrm{\mathrm{op}}} \to \mathrm{PSh}_\infty(\mathrm{PSh}_\infty(\Pi(X)^{\mathrm{op}})^{\mathrm{op}})$ 
the corresponding Yoneda embedding, then 
$$
  x^* \simeq {\tilde j} (j (x))
  \,.
$$
With this, we compute the endomorphisms of $x^*$ by applying the $\infty$-Yoneda lemma two more times:
$$
  \begin{aligned}
    \mathrm{End}(x^*)
     & \simeq
    \mathrm{End}_{\mathrm{PSh}(\mathrm{PSh}(\Pi(X)^{\mathrm{op}})^{\mathrm{op}})} ({\tilde j}(j (x)))
     \\
     & \simeq
     \mathrm{End}(\mathrm{PSh}(\Pi(X))^{\mathrm{op}}) (j(x))
     \\
     & \simeq
     \mathrm{End}_{\Pi(X)^{\mathrm{op}}}(x,x)
    \\
     & \simeq
     \mathrm{Aut}_x \Pi(X)
    \\
     & =: \Omega_x \Pi(X)
  \end{aligned}
   \,.
$$
\endofproof

Next we discuss how this intrinsic Galois theory in a cohesive 
$\infty$-topos is in line with the \emph{categorical Galois theory}
of Janelidze, as treated in \cite{CJKP}.
This revolves around factorization systems associated with 
the path functor $\mathbf{\Pi}$ from \ref{StrucGeometricPostnikov}.
\begin{theorem}
  \label{PiPreservesPullbacksOverDiscretes}
  If $\mathbf{H}$ has an 
  $\infty$-cohesive site of definition, def. \ref{CohesiveSite}, 
  the functor
  $\Pi : \mathbf{H} \to \infty \mathrm{Grpd}$ preserves
  $\infty$-pullbacks over discrete objects.
\end{theorem}
\noindent This was pointed out by Mike Shulman.\\
\proof
  By prop. 5.2.5.1 in \cite{Lurie}
  the $(\Pi \dashv \mathrm{Disc})$-adjunction 
  passes for each $A \in \infty \mathrm{Grpd}$ to the slice
  as 
  $$
    (\Pi_{/\mathrm{Disc} A}
	\dashv \mathrm{Disc}_{/\mathrm{Disc} A})
	:
	\mathbf{H}_{/\mathrm{Disc}A} \to \infty \mathrm{Grpd}_{/A}
	\,.
  $$
  Under the parameterized
  $\infty$-Grothendieck construction,  
  prop. \ref{ParameterizedGrothendieckConstruction},
  we have that $\Pi_{/\mathrm{Disc} A}$ becomes
  $$
    \Pi^A : \mathbf{H}^A \to \infty \mathrm{Grpd}^A
	\,.
  $$
  Since $\infty$-limits of functor $\infty$-categories
  are computed objectwise, and since $\Pi$ preserves finite products
  by the axioms of cohesion, $\Pi^A$ preserves finite products
  and hence so does $\Pi_{/\mathrm{Disc} A}$. Since a binary product in 
  $\mathbf{H}_{/\mathrm{Disc}A}$ is an $\infty$-pullback over
  $\mathrm{Disc}A $ in $\mathbf{H}$, this completes the proof.
\endofproof
\begin{remark}
We find below that over some $\infty$-cohesive
sites of interest $\Pi$ preserves further $\infty$-pullbacks. See
prop. \ref{GeometricRealizationOfHomotopyFibers}.
\end{remark}
\begin{definition}
  \label{PiClosure}
  For $f : X \to Y$ a morphism in $\mathbf{H}$, write
  $$
     c_{\mathbf{\Pi}}f  := Y \times_{\mathbf{\Pi}Y} \mathbf{\Pi}(X) \to Y
  $$ 
  for the $\infty$-pullback in 
  $$
    \raisebox{20pt}{
    \xymatrix{
	  c_{\mathbf{\Pi}}f \ar[r] \ar[d] & \mathbf{\Pi} X \ar[d]^{\mathbf{\Pi}f}
	  \\
	  Y \ar[r] & \mathbf{\Pi} Y
	}
	}
	\,,
  $$
  where the bottom morphism is the $(\Pi \dashv \mathrm{Disc})$-unit.
  We say that $c_{\mathbf{\Pi}} f$ is the \emph{$\mathbf{\Pi}$-closure} of $f$,
  and that $f$ is \emph{$\mathbf{\Pi}$-closed} if $X \simeq c_{\mathbf{\Pi}} f$.
\end{definition}	
\begin{remark}
  In the discussion of \emph{differential} cohesion below in \ref{InfinitesimalCohesion} we see that 
  the \emph{infinitesimal} analog of $\mathbf{\Pi}$-closesness is 
  \emph{formal {\'e}taleness}, see def. \ref{FormallEtaleMorphismInHth} below.
  There is a close conceptual relation: as we now discuss 
  (prop. \ref{LocallyConstantInfinityStacksAreThePiClosedMorphisms} below) morphisms 
  $X \stackrel{f}{\to} Y$ that 
  are $\mathbf{\Pi}$-closed may be identified with the total space  projections of 
  \emph{locally constant $\infty$-stacks over $Y$}. Accordingly in a context
  of differential cohesion, $\mathbf{\Pi}_{\mathrm{inf}}$-closed such morphisms
  may be interpreted as projections out of total spaces of general $\infty$-stacks
  over $Y$.
\end{remark}
\begin{definition}
  \label{PiEquivalence}
  Call a morphism $f : X \to Y$ in $\mathbf{H}$ a \emph{$\Pi$-equivalence}
  if $\Pi(f)$ is an equivalence in $\infty \mathrm{Grpd}$.
\end{definition}
\begin{remark}
  Since $\mathrm{Disc} : \infty \mathrm{Grpd} \to \mathbf{H}$ is
  full and faithful, we may equivalently speak of 
  \emph{$\mathbf{\Pi}$-equivalences}.
\end{remark}
\begin{proposition}
  \label{GaloisFactorization}
  If $\mathbf{H}$ has an $\infty$-connected site of definition,
  then every morphism $f : X \to Y$ in $\mathbf{H}$ factors as
  $$
    \xymatrix{
	  X \ar[rr]^f \ar[dr]_{f'} && Y
	  \\
	  & c_{\mathbf{\Pi} f} \ar[ur]_{}
	}
	\,,
  $$
  where $f'$ is a $\mathbf{\Pi}$-equivalence.
\end{proposition}
\proof
  The naturality of the adjunction unit together with the universality of 
  the $\infty$-pullback that defines $c_{\mathbf{\Pi}} f$ gives the factorization
  $$
    \xymatrix{
	  X
	  \ar@{-->}[r]^<<<<<{f'}
	  \ar[dr]_f
	  & Y \times_{\mathbf{\Pi} Y} \mathbf{\Pi} X
	  \ar[r] 
	  \ar[d]^{} & \mathbf{\Pi} X \ar[d]^{\mathbf{\Pi} f}
	  \\
	  &Y \ar[r] & \mathbf{\Pi} Y
	}
	\,.
  $$
  By theorem \ref{PiPreservesPullbacksOverDiscretes} 
  the functor $\Pi$ preserves the above $\infty$-pullback. 
  Since $\Pi(X \to \mathbf{\Pi}X)$ is an equivalence, it follows that 
  $\mathbf{\Pi}X$ is also a pullback of the $\mathbf{\Pi}$-image of the diagram,
  and hence $\mathbf{\Pi}(f')$ is an equivalence.
\endofproof
\begin{proposition}
  For $\mathbf{H}$ with an $\infty$-cohesive site of definition,
  the pair of classes of morphisms
  $$
    (\mbox{$\mathbf{\Pi}$-equivalences}, \; \mbox{$\mathbf{\Pi}$-closed morphisms})
	\subset
	\mathrm{Mor}(\mathbf{H}) \times \mathrm{Mor}(\mathbf{H})
  $$
  constitutes an orthogonal factorization system (5.2.8 in \cite{Lurie}).
  \label{PiEquivalencePiClosedFactorizationSystem}
\end{proposition}
\proof
  The factorization is given by prop. \ref{GaloisFactorization}. It remains
  to check orthogonality. 
  
  Let therefore
  $$
    \xymatrix{
	  A \ar[r] \ar[d] & X \ar[d] 
	  \\
	  B \ar[r] & Y
	}
  $$  
  be any commuting diagram in $\mathbf{H}$, where the left morphism is
  a $\mathbf{\Pi}$-equivalence and the right morphism is $\mathbf{\Pi}$-closed.
  Then, by assumption, there exists a pullback diagram on the right in
  $$
    \raisebox{20pt}{
    \xymatrix{
	  A \ar[r] \ar[d] & X \ar[d] \ar[r] & \mathbf{\Pi} X \ar[d]
	  \\
	  B \ar[r] & Y \ar[r] & \mathbf{\Pi} Y
	}
	}
	\,.
  $$  
  By the naturality of the $(\Pi \dashv \mathrm{Disc})$-unit,
  the outer rectangle above is equivalent to the outer rectangle of 
  $$
    \raisebox{20pt}{
    \xymatrix{
	  A \ar[r] \ar[d] & \mathbf{\Pi} A \ar[d]^{\simeq} \ar[r] & \mathbf{\Pi} X \ar[d]
	  \\
	  B \ar[r] & \mathbf{\Pi}B \ar[r] & \mathbf{\Pi} Y
	}
	}
	\,,
  $$  
  where now, again by assumption, the middle vertical morphism is an equivalence.
  Therefore there exists an essentially unique lift in the right square of this 
  diagram. This induces a lift in the outer rectangle. By the universality
  of the adjunction unit, such lifts factor essentially uniquely through 
  $\mathbf{\Pi} B$ and hence this lift, too,
  is essentially unique. Finally by the universal property of the pullback
  $X \simeq c_{\mathbf{\Pi}} f$, this
  gives the required essentially unique lift on the left of 
  $$
    \raisebox{20pt}{
    \xymatrix{
	  A \ar[r] \ar[d] & X \ar[d] \ar[r] & \mathbf{\Pi} X \ar[d]
	  \\
	  B \ar@{-->}[ur]\ar[r] & Y \ar[r] & \mathbf{\Pi} Y
	}
	}
	\,.
  $$  
\endofproof
We now identifiy the $\mathbf{\Pi}$-closed morphisms with covering spaces,
hence with total spaces of locally constant $\infty$-stacks.
\begin{observation}
  For $f : X \to Y$ a $\mathbf{\Pi}$-closed morphism, its fibers $X_y$
  over global points $y : * \to Y$ are discrete objects.
\end{observation}
\proof
  By assumption and using the pasting law, prop. \ref{PastingLawForPullbacks},
  it follows that the fibers of $f$
  are the fibers of $\mathbf{\Pi}f$. Since the terminal object
  is discrete and since $\mathrm{Disc}$
  preserves $\infty$-pullbacks, these are the images under $\mathrm{Disc}$
  of fibers of $\Pi f$, and hence are discrete.
\endofproof
Conversely we have:
\begin{example}
  Let $X \in \mathbf{H}$ be any object, and let 
  $A \in \infty\mathrm{Grpd} $ be any discrete $\infty$-groupoid. 
  Then the projection morphism $p : X \times \mathrm{Disc}(A) \to X$ out of the product
  is $\mathbf{\Pi}$-closed.
\end{example}
\proof
  Since $\mathbf{\Pi}$ preserves products, by the axioms of cohesion,
  and $\mathrm{Disc}$ preserves products as a right adjoint and is 
  moreover full and faithful,
  we have that $\mathbf{\Pi}(p)$ is the projection
  $$
    \mathbf{\Pi}(p) : \mathbf{\Pi}(X) \times \mathrm{Disc}(A) \to \mathbf{\Pi}(X)
	\,.
  $$
  Since $\infty$-limits commute with $\infty$-limits, it follows that 
  $$ 
    \xymatrix{
	  X \times \mathrm{Disc}(A) 
	  \ar[r]
	  \ar[d]
	  &
	  \mathbf{\Pi}(X) \times \mathrm{Disc}(A)
	  \ar[d]
	  \\
	  X \ar[r] & \mathbf{\Pi}(X)
	}
  $$
  is an $\infty$-pullback.
\endofproof
\begin{remark}
Morphisms of the form $X \times \mathrm{Disc}(A) \to X$
fit into pasting diagrams of $\infty$-pullbacks of the form
$$
  \raisebox{20pt}{
  \xymatrix{
    X \times \mathrm{Disc}(A)
	\ar[r]
    \ar[d]	
	& 
	\mathrm{Disc}(A)
	\ar[r]
	\ar[d]
	&
	\mathrm{Disc}(A/\!/ \mathrm{Aut}(A))
	\ar[d]
	\\
	X 
	\ar[r]
	&
	{*}
	\ar[r]
	&
	\mathbf{B} \mathrm{Disc}(\mathrm{Aut}(A))
  }
  }
  \,,
$$
where the square on the right is the universal discrete $A$-bundle,
by the discussion in \ref{StrucRepresentations}. According to
def. \ref{LocallyConstantInfinityStacks} the composite morphism
on the bottom classifies the 
\emph{trivial} locally constant $\infty$-stack with fiber $A$ over $X$,
hence the \emph{constant} $\infty$-stack with fiber $A$ over $X$.
Therefore the above $\infty$-pullback exhibits $X \times \mathrm{Disc}(A) \to X$
as the total space incarnation of that constant $\infty$-stack on $X$.
\end{remark}
The following proposition generalizes this statement to all 
locally constant $\infty$-stacks over $X$.
\begin{proposition}
  \label{LocallyConstantInfinityStacksAreThePiClosedMorphisms}
  Let $\mathbf{H}$ have an $\infty$-cohesive site of definition, \ref{CohesiveSites}.
  Then for any $X \in \mathbf{H}$ the locally constant $\infty$-stacks 
  $E \in \mathrm{LConst}(X)$, regarded as $\infty$-bundle morphisms
  $p : E \to X$ by
  observation \ref{LocallyConstantStacksAsPrincipalBundles},
  are precisely the $\mathbf{\Pi}$-closed morphisms into $X$.
\end{proposition}
\proof
  We may without restriction of generality assume that 
  $X$ has a single geometric connected component. 
  Then $E \to X$ is given by an $\infty$-pullback of the form
  $$
    \xymatrix{
	  E \ar[r] \ar[d]^p & \mathrm{Disc}(F_i /\!/ \mathrm{Aut}(F_i)) \ar[d]
	  \\
	  X \ar[r]^<<<<<g & \mathbf{B}\mathrm{Disc} \mathrm{Aut}(F_i)
	}
	\,.
  $$
  By theorem \ref{PiPreservesPullbacksOverDiscretes} the functor $\Pi$ preserves this 
  $\infty$-pullback, so that also
  $$
    \xymatrix{
	  \mathbf{\Pi}E \ar[r] \ar[d] & \mathrm{Disc}(F_i /\!/\mathrm{Aut}(F_i)) \ar[d]
	  \\
	  \mathbf{\Pi}X \ar[r]^<<<<<{\mathbf{\Pi} g} & \mathbf{B}\mathrm{Disc}\, \mathrm{Aut}(F_i)
	}
  $$
  is an $\infty$-pullback, where we used that,
  by the axioms of cohesion, $\mathbf{\Pi}$ sends discrete objects to themselves.
  
  By def. \ref{PiClosure} the factorization in question is given by forming 
  the $\infty$-pullback on the left of   
  $$
    \raisebox{20pt}{
    \xymatrix{
	  X \times_{\mathbf{\Pi}X} \mathbf{\Pi} E \ar[r] \ar[d] & \mathbf{\Pi}E \ar[r] \ar[d] & 
	  \mathrm{Disc}(F_i /\!/ \mathrm{Aut}(F_i)) \ar[d]
	  \\
	  X \ar[r] & \mathbf{\Pi}X \ar[r]^<<<<<{\mathbf{\Pi} g} & 
	  \mathbf{B} \mathrm{Disc} \mathrm{Aut}(F_i)
	}
	}
	\,.
  $$
  By the universal property of the $(\Pi \dashv \mathrm{Disc})$-reflection,
  the bottom composite is again equivalent to $g$, hence by the pasting law,
  prop. \ref{PastingLawForPullbacks}, it follows that the pullback on the left is 
  equivalent to $E \to X$.
  
  Conversely, if the $\infty$-pullback diagram on the left is given, it follows
  with prop. \ref{ObjectClassifierInInfinityGroupoid} and using,
  by definition of cohesion, that $\mathrm{Disc}$
  is full and faithful, that an $\infty$-pullback square as on the right exists. 
  Again by the pasting law, this implies that the morphism on the left is 
  the total space projection of a locally constant $\infty$-stack over $X$.
\endofproof
\begin{remark}
  In the ``1-categorical Galois theory'' of \cite{CJKP} only the trivial 
  discrete $\infty$-bundles arise as pullbacks this way, and much of the theory
deals with getting around this restriction.  In our language, this is 
because in the context of 1-categorical cohesion, as in \cite{Lawvere},
the $\infty$-functor $\mathbf{\Pi}$ reduces to the 1-functor 
$\mathbf{\Pi}_0 \simeq \tau_0 \circ \mathbf{\Pi}$, discussed in 
\ref{StrucGeometricPostnikov},  
on a locally connected and connected 1-topos, which assigns only the set of 
connected components, instead of the full path $\infty$-groupoid. 
 
Clearly, the pullback over an object of the form $\mathbf{\Pi}_0 K$ is indeed a
locally constant $\infty$-stack that is trivial as a discretely fibered $\infty$-bundle.
But this restriction is lifted by passing from cohesive 1-toposes 
to cohesive $\infty$-toposes.
\end{remark}
We now characterize locally constant $\infty$-stacks over $X$ as precisely the
``relatively discrete'' objects over $X$.
To that end, recall, by prop. \ref{ConnectedOverToposes}, that for $\mathbf{H}$ a locally
$\infty$-connected $\infty$-topos also all the slice $\infty$-toposes
$\mathcal{X} := \mathbf{H}_{/X}$ for all objects $X \in \mathbf{H}$ are locally $\infty$-connected.
\begin{definition}
  \label{SlicedTerminalGeometricMorphism}
  For $X \in \mathbf{H}$ an object in a cohesive $\infty$-topos $\mathbf{H}$
  and 
  $$
    \xymatrix{
	  \mathbf{H}_{/X}
	  \ar@<+9pt>@{->}[rr]^{p_!}
	  \ar@<+3pt>@{<-}[rr]|{p^*}
	  \ar@<-3pt>@{->}[rr]_{p_*}
	  &&
	  \infty \mathrm{Grpd}
	}
  $$
  the corresponding locally $\infty$-connected terminal geometric morphism,
  write
  $$
    \xymatrix{
	  \mathbf{H}_{/X}
	  \ar@<+4pt>@{->}[rr]^{p_!/X}
	  \ar@<-4pt>@{<-}[rr]|{p^*/X}
	  &&
	  \infty \mathrm{Grpd}_{/\Pi(X)}
	}
  $$
  for the induced $\infty$-adjunction on the slices,
  by prop. 5.2.5.1 in \cite{Lurie}, where the left adjoint 
  $p_!/X$ sends $(E \to X)$ to $(\Pi(E) \to \Pi(X))$.
\end{definition}
\begin{proposition}
  \label{LocallyConstantIsIndeedAbstractlyLocallyConstant}
  Let the cohesive $\infty$-topos $H$ have an $\infty$-cohesive site of
  definition, def. \ref{CohesiveSite} and let $X \in \mathbf{H}$ be any object.
  
  The full sub-$\infty$-category of $\mathbf{H}_{/X}$
  on the $\mathbf{\Pi}$-closed morphisms into $X$, def. \ref{PiClosure},
  hence on the locally constant $\infty$-stacks over $X$, 
  prop. \ref{LocallyConstantInfinityStacksAreThePiClosedMorphisms}, is
  equivalent to the image of the moprhism
  $p^* /X : \infty \mathrm{Grpd}_{/\Pi(X)} \to \mathbf{H}_{/X}$.
\end{proposition}
\proof
  By prop 5.2.5.1 in \cite{Lurie}, the $\infty$-functor $p^* /X$ is the composite
  $$
    p^*/X 
	:
	\xymatrix{
	   \infty \mathrm{Grpd}_{/\Pi(X)}
	   \ar[rr]^{\mathrm{Disc}}
	   &&
	   \mathbf{H}_{/\mathbf{\Pi}}
	   \ar[rr]^{X \times_{\mathbf{\Pi}(X)}(-)}
	   &&
	   \mathbf{H}_{/X}
	}
	\,.
  $$
  This sends a morphism $Q \to \Pi(X)$ to the pullback on the left of the
  pullback square
  $$
   \raisebox{20pt}{
    \xymatrix{
	  E \ar[r] \ar[d] & \mathrm{Disc}(Q) \ar[d]
	  \\
	  X \ar[r] & \mathbf{\Pi}(X)
	}
	}
	\,.
  $$
  Since $\Pi$ preserves this $\infty$-pullback, by theorem 
  \ref{PiPreservesPullbacksOverDiscretes}, and sends $X \to \mathbf{\Pi}(X)$
  to an equivalence, it follows that $\Pi(E \to X)$
  is equivalent to $Q \to \Pi(X)$ and hence the above pullback diagram looks like
  $$
    \raisebox{20pt}{
    \xymatrix{
	  E \ar[r] \ar[d] & \mathbf{\Pi}(E) \ar[d]
	  \\
	  X \ar[r] & \mathbf{\Pi}(X)
	}
	}
	\,.
  $$
  The naturality of the $(\Pi \dashv \mathrm{Disc})$-unit and the 
  universality of the pullback imply that the top horizontal morphism
  here is indeed the $E$-component of the $(\Pi \dashv \mathrm{Disc})$ unit. 
  
  This shows that, up to equivalence, precisely the $\mathbf{\Pi}$-closed 
  morphism $E \to X$ arise this way.
\endofproof
\begin{remark}
  A definition of locally constant objects in general $\infty$-toposes
  is given in section A.1 of \cite{LurieAlgebra}. The above 
  prop. \ref{LocallyConstantIsIndeedAbstractlyLocallyConstant}
  together with theorem A.1.15 in \cite{Lurie} shows that restricted to
  the slices $\mathbf{H}_{/X}$ it coincides with the definition discussed here.
\end{remark}

\subsection{Structures in a cohesive $\infty$-topos}
\label{structures}
\index{structures in a cohesive $\infty$-topos}
\index{cohesive $\infty$-topos!structures in a cohesive $\infty$-topos}

We discuss differential geometric and differential cohomological 
structures that exist in any \emph{cohesive $\infty$-topos}, 
def. \ref{CohesiveInfinToposDefinition}. These are obtained from 
the $\Pi$-geometric structures of a locally $\infty$-connected $\infty$-topos, 
discussed in \ref{Structures in a locally infinity-connected topos} by interpreting them
in the \emph{gros} cohomological context of a local $\infty$-topos,
dscussed in \ref{Structures in a local topos}.

\medskip

\begin{itemize}
  \item \ref{A1HomotopyGeneralAbstract} -- $\mathbb{A}^1$-Homotopy / The Continuum
  \item \ref{Strucmanifolds} -- Manifolds
  \item \ref{StrucDeRham} -- de Rham cohomology
  \item \ref{StrucLieAlgebras} -- Exponentiated Lie algebras
  \item \ref{StructCurvatureCharacteristic} -- Maurer-Cartan forms and curvature characteristic forms
  \item \ref{StrucDifferentialCohomology} -- Differential cohomology
  \item \ref{StrucChern-WeilHomomorphism} -- Chern-Weil homomorphism
  \item \ref{TwistedDifferentialStructures} -- Twisted differential structures
  \item \ref{StrucFiberIntegration} -- Higher holonomy
  \item \ref{StrucTransgression} -- Transgression
  \item \ref{StrucChern-SimonsTheory} -- Chern-Simons functionals
  \item \ref{StrucWZWFunctional} -- Wess-Zumino-Witten functionals
  \item \ref{StrucGeometricPrequantization} -- Prequantum geometry
  \item \ref{LocalPrequantumFieldTheories} -- Local prequantum field theory
\end{itemize}

\subsubsection{$\mathbb{A}^1$-Homotopy / The Continuum}
\label{A1HomotopyGeneralAbstract}
\index{structures in a cohesive $\infty$-topos!$\mathbb{A}^1$-homotopy}

We formalize in a cohesive $\infty$-topos $\mathbf{H}$ the notion of \emph{the continuum} in 
the sense in which the standard real line $\mathbb{R}$ is traditionally called
\emph{the continuum}. Abstractly this is an object $\mathbb{A}^1 \in \mathbf{H}$ which, when 
regarded as a \emph{line object}, \emph{induces} the geometric homotopy in $\mathbf{H}$
as discussed in \ref{StrucGeometricHomotopy}. Explicitly this means that 
$\mathbf{\Pi} : \xymatrix{\mathbf{H} \ar[r]^\Pi & \infty \mathrm{Grpd} \ar@{^{(}->}[r] & \mathbf{H}} $ exhibits the  \emph{localization} of $\mathbf{H}$ which inverts all those morphisms that are products
of an object with the terminal morphism $\mathbb{A}^1 \to *$.
Since by cohesion $\mathbf{\Pi}(*) \simeq *$, this means in particular that 
such an $\mathbb{A}^1$ is a geometrically contractible object in that
$\mathbf{\Pi}(\mathbb{A}^1) \simeq *$. Together this are the characterizing property 
of the archetypical ``continuum'' $\mathbb{R}$. Below in \ref{Strucmanifolds}
we discuss how a continuum line object induces a notion of \emph{manifold} objects
in $\mathbf{H}$.

\begin{remark}
 The  $\infty$-topos $\mathbf{H}$,
 being in particular a presentable $\infty$-category, admits a choice
of a small set $\{c_i \in \mathbf{H}\}_i$ of generating objects, and 
every small set of morphisms in $\mathbf{H}$
induces a full reflective sub-$\infty$-category of objects that are 
\emph{local} with respect to these morphisms.
\end{remark}
This is \cite{Lurie}, section 5.
\begin{definition}
  \label{ILocalization}
  For $\mathbf{H}$ a cohesive $\infty$-topos, we say an object $I \in \mathbf{H}$
  is an \emph{continuum line object exhibiting the cohesion} of  $\mathbf{H}$ 
  if the reflective inclusion
  of the discrete objects
  $$
    (\Pi \dashv \mathrm{Disc}) : 
	\xymatrix{
	  \infty \mathrm{Grpd}
	  \ar@{<-}@<+3pt>[r]^<<<<<{\Pi}
	  \ar@{^{(}->}@<-3pt>[r]_<<<<<{\mathrm{Disc}}
	  &
	  \mathbf{H}
	}
  $$
  is induced by the localization at the set of morphisms 
  $$
     S := \{c_i \times (I \to *)\}_i, 
	 \,,
  $$ 
  for $\{c_i\}_i$ some small set of generators of $\mathbf{H}$.
\end{definition}
\begin{remark}
  In this situation, for $X \in \mathbf{H}$ we may think of $\Pi(X)$
  also as the \emph{$I$-localization} of $X$. 
\end{remark}
A class of examples of this situation is the following.
\begin{proposition}
  \label{IntervalOverLawvereCohesiveSite}
  Let $C$ be an $\infty$-cohesive site, def. \ref{CohesiveSite},
  which moreover is the syntactic category of a Lawvere algebraic theory
  (see chapter 3, volume 2 of \cite{Borceux}), in that 
  it has finite products and 
  there is an object 
  $$
     \mathbb{A}^1 \in C
  $$ 
  such that every other object is
  isomorphic to an $n$-fold cartesian product $\mathbb{A}^n = (\mathbb{A}^1)^n$.
  
  Then $\mathbb{A}^1 \in C \hookrightarrow \mathrm{Sh}_{\infty}(C)$ is
  a geometric interval exhibiting the cohesion of the $\infty$-topos over $C$.
\end{proposition}
\proof
  A set of generating objects of $\mathbf{H} = \mathrm{Sh}_\infty(C)$
  is given by the set of isomorphism classes of objects of $C$, hence,
  by assumption,  by $\{\mathbb{A}^n\}_{n \in \mathbb{N}}$.
  The set of localizing morphisms is therefore
  $$
    S := \{ \mathbb{A}^{n+1} \to \mathbb{A}^n \;|\; n \in \mathbb{N}\}
	\,.
  $$
  By prop. \ref{InfSheavesOverCohesiveSiteAreCohesive},
  $\mathbf{H}$
  is presented by the model category $[C^{\mathrm{op}}, \mathrm{sSet}]_{\mathrm{proj}, \mathrm{loc}}$.
  By the proof of \cite{Lurie} cor. A.3.7.10  the localization of $\mathbf{H}$
  as $S$ is presented by the left Bousfield localization of this model category at $S$,
  given by a Quillen adjunction to be denoted
  $$
    (L_{\mathbb{A}^1} \dashv R_{\mathbb{A}^1})
	 :
    \xymatrix{
    [C^{\mathrm{op}}, \mathrm{sSet}]_{\mathrm{proj}, \mathrm{loc}, \mathbb{A}1}
	  \ar@{<-}@<+4pt>[rr]^{\mathrm{id}}
	  \ar@<-4pt>[rr]_{\mathrm{id}}
	  &&
	[C^{\mathrm{op}}, \mathrm{sSet}]_{\mathrm{proj}, \mathrm{loc}}
	}
	\,.
  $$
  Observe that we also have a Quillen
  adjunction
  $$
    (\mathrm{const} \dashv (-)_{*})
	:
    \xymatrix{
     [C^{\mathrm{op}}, \mathrm{sSet}]_{\mathrm{proj}, \mathrm{loc}, \mathbb{A}^1}
	  \ar@{<-}@<+4pt>[rr]^<<<<<<<{\mathrm{const}}
	  \ar@<-4pt>[rr]_<<<<<<<{(-)_{*}}
	 &&
	 \mathrm{sSet}_{\mathrm{Quillen}}
	}
	\,,
  $$
  where the right adjoint evaluates at the terminal object $\mathbb{A}^0$, and
  where the left adjoint produces constant simplicial presheaves.
  This is because the two functors are clearly a Quillen adjunction before localization
  (on $[C^{\mathrm{op}}, \mathrm{sSet}]_{\mathrm{proj}}$) and so by 
  \cite{Lurie} cor. A.3.7.2 it is sufficient to observe that on the local 
  structure the right adjoint still preserves fibrant objects, which it does
  because the fibrant objects in the localization are in particular fibrant 
  in the unlocalized structure.
  
  Moreover, we claim that $(\mathrm{const} \dashv (-)_{*})$ is in fact a 
  Quillen equivalence, by observing that the derived adjunction unit and
  counit are equivalences.
  For the derived adjunction unit, notice that by the proof of 
  prop. \ref{InfSheavesOverCohesiveSiteAreCohesive} a
  constant simplicial presheaf is fibrant in 
  $[C^{\mathrm{op}}, \mathrm{sSet}]_{\mathrm{proj}, \mathrm{loc}}$, and so 
  it is clearly fibrant in $[C^{\mathrm{op}}, \mathrm{sSet}]_{\mathrm{proj}, \mathrm{loc}, \mathbb{A}^1}$. Therefore the plain adjunction unit, which is the identity, 
  is already the derived adjunction unit.
  For the derived counit, let 
  $X \in [C^{\mathrm{op}}, \mathrm{sSet}]_{\mathrm{proj}, \mathrm{loc}, \mathbb{A}1}$
  be fibrant.
  Then also the adjunction counit
  $$
    \eta : \mathrm{const} (X(\mathbb{A}^0)) \to  X 
  $$
  is already the derived counit (since $X(\mathbb{A}^1) \in \mathrm{sSet}_{\mathrm{Quillen}}$
  is necessarily cofibrant).
  At every $\mathbb{A}^n \in C$ it is isomorphic to 
  the sequence of morphisms
  $$
    \eta(\mathbb{A}^n) ;  X(\mathbb{A}^0) \to X(\mathbb{A}^1) \to \cdots \to X(\mathbb{A}^n)
	\,,
  $$
  each of which is a weak equivalence by the $\mathbb{A}^1$-locality of $X$.
  
  Now observe that we have an equivalence of $\infty$-functors
  $$
    \mathrm{Disc} \simeq \mathbb{R}R_{\mathbb{A}^1} \circ \mathbb{L}\mathrm{const}
	: 
	\infty \mathrm{Grpd} \to \mathbf{H}
	\,.
  $$
  Because for $A \in \mathrm{sSet}$ fibrant,
  $\mathbb{L}\mathrm{const}(A) \simeq A$ is still fibrant, by the proof of
  prop. \ref{InfSheavesOverCohesiveSiteAreCohesive}, and so 
  $(\mathbb{R}R_{\mathbb{A}^1}) ((\mathbb{L}\mathrm{const})(A)) \simeq \mathrm{const} A$
  is  presented simply by the constant simplicial presheaf on $A$, which
  indeed is a presentation for $\mathrm{Disc} A$, again by the proof of
  prop. \ref{InfSheavesOverCohesiveSiteAreCohesive}.

  Finally, since by the above $\mathbb{L} \mathrm{const}$ is in fact an equivalence, 
by essential uniqueness of $\infty$-adjoints it follows now that $\mathbb{L} L_{\mathbb{A}^1}$
is left adjoint to the $\infty$-functor $\mathrm{Disc}$, and this proves the claim.
\endofproof

\begin{remark}
  Below in \ref{ETopStrucIHomotopy} we show that in the models of Euclidean-topological cohesion and
  of smooth cohesion the standard real line is indeed the continuum line object in the
  above abstract sense.
\end{remark}

\subsubsection{Manifolds (unseparated)}
\label{Strucmanifolds}
\index{structures in a cohesive $\infty$-topos!manifolds}

We discuss a general abstract realization of the notion of \emph{unseparated manifolds}
internal to a cohesive $\infty$-topos. In order to formalize separated manifolds 
(Hausdorff manifolds) we need the extra axioms of differential cohesion. This is
discussed below in \ref{DifferentialStrucmanifolds}.

\medskip

\begin{remark}
The theory of principal $\infty$-bundles in \ref{PrincipalInfBundle} 
extensively used two of the three Giraud-Rezk-Lurie axioms characterizing 
$\infty$-toposes, def. \ref{GiraudRezkLurieAxioms}
(universal coproducts and effective groupoid objects). Here we now use the third one,
that \emph{coproducts are disjoint}. 
\end{remark}
\begin{proposition}
  If $A \in \mathbf{H}$ is 0-truncated, def. \ref{truncated object}
  is geometrically connected in that $\Pi(A) \in \infty\mathrm{Grpd}$
  is connected, then morphisms $A \to X \coprod Y$ into a coproduct 
  of 0-truncated objects in 
  $\mathbf{H}$ factor through one of the two inclusions $X \hookrightarrow X \coprod Y$
  or $Y \hookrightarrow X \coprod Y$.
  \label{MapsFromConnectedIntoCoproductFactorThroughInclusion}
\end{proposition}
\proof
  The 1-topos $\tau_{\leq 0} \mathbf{H}$ of 0-truncated objects of a locally
  $\infty$-connected $\infty$-topos is a locally connected 1-topos by 
  prop. \ref{ForLocallyInfinityConnectedInfinityToposUnderlying1ToposIsLocallyConnected}.
  Under this identification, $A \in \tau_0 \mathbf{H}$
  as above is a connected object, and hence is in particular not a 
  coproduct of two non-initial objects. Since moreover coproducts in 
  $\mathbf{H}$ and in $\tau_{\leq 0}\mathbf{H}$ are disjoint and since 
  truncation (being a left adjoint) preserves them, the statement 
  reduces to a standard fact in topos theory (for instance \cite{Johnstone}, p. 34).
\endofproof

Let now $\mathbb{A}^1 \in \mathbf{H}$ be a continuum line 
object that \emph{exhibits the cohesion of $\mathbf{H}$} in the sense of
def. \ref{ILocalization}. For $n \in \mathbb{N}$, write 
$$
  \mathbb{A}^n := \underbrace{ \mathbb{A}^1 \times \cdots \times \mathbb{A}^1 }_{\mbox{$n$ factors}}
  \,.
$$
\begin{proposition}
  For all $n \in \mathbb{N}$ the objects $\mathbb{A}^n \in \mathbf{H}$
  are geometrically connected.
   \label{AnIsGeometricallyConnected}
\end{proposition}
\proof
  By cohesion, $\Pi : \mathbf{H} \to \infty \mathrm{Grpd}$ preserves finite products
and so the statement reduces to the fact that the product of two connected $\infty$-groupoids
is itself a connected $\infty$-groupoid.  
\endofproof

\begin{definition}
  Given an object $\mathbb{A}^1 \in \mathbf{H}$ exhibiting the cohesion of 
  the cohesive topos  $\mathbf{H}$,  
  an object $X \in \mathbf{H}$ is  an \emph{unseparated $\mathbb{A}$-manifold}
  of \emph{dimension} $n \in \mathbb{N}$ if there exists 
  a small set of monomorphisms of the form
  $$
    \{\mathbb{A}^n \stackrel{\phi_j}{\hookrightarrow} X\}_{j}
  $$
  such that for the corresponding 
  $$
    \phi : 
    \xymatrix{
	  \coprod_{j} \mathbb{A}^n \ar@{->>}[rr]^{(\phi_j)_j} && X
	}
  $$
  we have
  \begin{enumerate}
    \item $\phi$ is an effective epimorphism, def. \ref{EffectiveEpimorphism};
    \item the nerve simplicial object $C_\bullet(\phi)$ of $\phi$ is degreewise a 
  coproduct of copies of $\mathbb{A}^n$.
   \end{enumerate}
  \label{IntrinsicManifold}
\end{definition}
\begin{remark}
  Since monomorphisms are stable under pullback and since by the Giraud-Rezk-Lurie 
  axioms coproducts are preserved under pullback, it follows that the simplicial 
  object in def. \ref{IntrinsicManifold} is 
      such that all components $\mathbb{A}^n \to \mathbb{A}^n$ of all face maps 
	 (given by prop. \ref{MapsFromConnectedIntoCoproductFactorThroughInclusion} 
	 and prop. \ref{AnIsGeometricallyConnected})
     are monomorphisms.
\end{remark}
\begin{remark}
  Below in \ref{ETopStrucmanifolds} and \ref{SmoothStrucmanifolds} is discussed that in the standard model of 
  Euclidean-topological and of smooth cohesion
  this abstract definition reproduces the traditional definition of topological and of 
  smooth manifolds, respectively. 
\end{remark}

\subsubsection{de Rham cohomology}
\label{StrucDeRham}
\index{structures in a cohesive $\infty$-topos!de Rham cohomology}
\index{cohomology!de Rham (general abstract)}

We discuss how 
in every locally $\infty$-connected $\infty$-topos $\mathbf{H}$ there is an intrinsic notion of 
\emph{nonabelian de Rham cohomology}.

\medskip
 
We have already seen the notions of \emph{Principal bundles}, 
\ref{section.PrincipalInfinityBundle}, and of flat $\infty$-connections 
on principal $\infty$-bundles, 
\ref{StrucFlatDifferential}, in any locally $\infty$-connected $\infty$-topos.
In traditional differential geometry, flat connection on the \emph{trivial} 
principal bundle
may be canonically identified with flat differential 1-forms on the base space. 
In the following we 
take this idea to be the \emph{definition} of flat $\infty$-group/$\infty$-Lie algebra
valued forms: flat $\infty$-connections on trivial principal $\infty$-bundles.

\medskip

\begin{definition} 
Let $\mathbf{H}$ be a locally $\infty$-connected $\infty$-topos.
For $X \in \mathbf{H}$ an object, write 
$\mathbf{\Pi}_{\mathrm{\mathrm{dR}}}X := {*} \coprod_X \mathbf{\Pi}X$
for the $\infty$-pushout
$$
  \xymatrix{
    X \ar[r] \ar[d]& {*} \ar[d]
    \\
    \mathbf{\Pi}(X) \ar[r]& \mathbf{\Pi}_{\mathrm{dR}}X
  }
  \,.
$$
We call this the \emph{cohesive de Rham homotopy type} of $X$ 
(see remark \ref{DeRhamHomotopyType} below).

For $\mathrm{pt}_A : * \to A$ any pointed object in $\mathbf{H}$, write 
$\mathbf{\flat}_{\mathrm{\mathrm{dR}}} A := {*} \prod_A \mathbf{\flat}A$ 
for the $\infty$-pullback
$$
  \xymatrix{
    \mathbf{\flat}_{\mathrm{dR}} A \ar[r] \ar[d]& \mathbf{\flat} A  \ar[d]
    \\
    {*} \ar[r] & A
  }
  \,.
$$
We call this the \emph{de Rham coefficient object} of $\mathrm{pt}_A : * \to A$.
\label{deRhamCoefficientObject}
\end{definition}

\begin{proposition}
This construction yields a pair of adjoint $\infty$-functors 
$$
  (\mathbf{\Pi}_{\mathrm{\mathrm{dR}}} \dashv \mathbf{\flat}_{\mathrm{\mathrm{dR}}} )
  : 
  \xymatrix{
    {*}/\mathbf{H}
      \ar@{<-}@<+4pt>[r]^{\mathbf{\Pi}_{\mathrm{\mathrm{dR}}}}
      \ar@<-4pt>[r]_{\mathbf{\flat}_{\mathrm{\mathrm{dR}}}}
      &
    \mathbf{H}
  }
  \,.
$$
\end{proposition}
\proof
We check the defining natural hom-equivalence
$$
  {*}/\mathbf{H}(\mathbf{\Pi}_{\mathrm{\mathrm{dR}}}X,A)
  \simeq
  \mathbf{H}(X, \mathbf{\flat}_{\mathrm{\mathrm{dR}}}A)
  \,.
$$
The hom-space in the under-$\infty$-category $*/\mathbf{H}$ is computed 
by prop. \ref{SliceHomAsHomotopyFiber} as the $\infty$-pullback
$$
  \xymatrix{
     {*}/\mathbf{H}(\mathbf{\Pi}_{\mathrm{\mathrm{dR}}}X, A)
      \ar[r] \ar[d] &
     \mathbf{H}(\mathbf{\Pi}_{\mathrm{\mathrm{dR}}}X, A) \ar[d]
     \\
     {*} \ar[r]^{\mathrm{pt}_A}& \mathbf{H}(*,A)
  }
  \,.
$$
By the fact that the hom-functor 
$\mathbf{H}(-,-) : \mathbf{H}^{\mathrm{op}} \times \mathbf{H} \to \infty \mathrm{Grpd} $ 
preserves $\infty$-limits in both arguments we have a natural equivalence
$$
  \begin{aligned}
     \mathbf{H}(\mathbf{\Pi}_{\mathrm{\mathrm{dR}}} X, A)
     & :=
     \mathbf{H}( {*} \coprod_{X} \mathbf{\Pi}(X), A  )     
     \\
     & \simeq 
     \mathbf{H}({*},A) \prod_{\mathbf{H}(X,A)} 
      \mathbf{H}(\mathbf{\Pi}(X),A)
  \end{aligned}
  \,.
$$
We paste this pullback to the above pullback diagram to obtain
$$
  \xymatrix{
     {*}/\mathbf{H}(\mathbf{\Pi}_{\mathrm{\mathrm{dR}}}X, A)
      \ar[r] \ar[d]&
     \mathbf{H}(\mathbf{\Pi}_{\mathrm{\mathrm{dR}}}X, A) \ar[r]\ar[d]&
      \mathbf{H}(\mathbf{\Pi}(X),A) \ar[d]
     \\
     {*} \ar[r]^{\mathrm{pt}_A} & \mathbf{H}(*,A)
     \ar[r] & 
      \mathbf{H}(X,A)
  }
  \,.
$$
By the pasting law for $\infty$-pullbacks, prop. \ref{PastingLawForPullbacks}, 
the outer diagram is still a pullback. We may evidently rewrite the bottom composite as in
$$
  \xymatrix{
     {*}/\mathbf{H}(\mathbf{\Pi}_{\mathrm{\mathrm{dR}}}X, A)
      \ar[rr] \ar[d]&
     &
      \mathbf{H}(\mathbf{\Pi}(X),A) \ar[d]
     \\
     {*} \ar[r]^{\simeq} & \mathbf{H}(X,{*})
     \ar[r]^{(\mathrm{pt}_A)_*} & 
      \mathbf{H}(X,A)
  }
  \,.
$$
This exhibits the hom-space as the pullback
$$
  \begin{aligned}
    {*}/\mathbf{H}(\mathbf{\Pi}_{\mathrm{\mathrm{dR}}}(X),A)
    \simeq
     \mathbf{H}(X,{*}) \prod_{\mathbf{H}(X,A)} \mathbf{H}(X,\mathbf{\flat} A)
   \end{aligned}
  \,,
$$
where we used the $(\mathbf{\Pi} \dashv \mathbf{\flat})$-adjunction. Now using again that $\mathbf{H}(X,-)$ preserves pullbacks, this is
$$
  \cdots \simeq \mathbf{H}(X, {*} \prod_A \mathbf{\flat}A )
  \simeq \mathbf{H}(X , \mathbf{\flat}_{\mathrm{\mathrm{dR}}}A)
  \,.
$$
\endofproof
\begin{observation} 
 \label{TripleOfDeRhamAdjunctions}
If $\mathbf{H}$ is also local, then there is a further right adjoint 
$\mathbf{\Gamma}_{\mathrm{\mathrm{dR}}}$
$$
  (\mathbf{\Pi}_{\mathrm{\mathrm{dR}}} \dashv \mathbf{\flat}_{\mathrm{\mathrm{dR}}} 
  \dashv \mathbf{\Gamma}_{\mathrm{\mathrm{dR}}})
  :
  \xymatrix{
    \mathbf{H}
      \ar@{->}@<+12pt>[r]|{\mathbf{\mathbf{\Pi}}_{\mathrm{\mathrm{dR}}}}
      \ar@{<-}@<+4pt>[r]|{\mathbf{\mathbf{\flat}}_{\mathrm{\mathrm{dR}}}}
      \ar@<-4pt>[r]_{\mathbf{\Gamma}_{\mathrm{\mathrm{dR}}}}
    &
    {*}/\mathbf{H}
  }
$$
given by
$$
  \mathbf{\Gamma}_{\mathrm{dR}} X 
    :=
    {*} \coprod_{X} \mathbf{\Gamma}(X)
  \,.
$$
\end{observation}
\begin{definition}
  \label{IntrinsicDeRhamCohomology}
For $X, A \in \mathbf{H}$ we write
$$
  \mathbf{H}_{\mathrm{\mathrm{dR}}}(X,A)
  :=
  \mathbf{H}(\mathbf{\Pi}_{\mathrm{\mathrm{dR}}}X, A)
  \simeq
  \mathbf{H}(X, \mathbf{\flat}_{\mathrm{dR}} A)
  \,.
$$
A cocycle $\omega : X \to \mathbf{\flat}_{\mathrm{\mathrm{dR}}}A$ 
we call a \emph{flat $A$-valued differential form} on $X$.

We say that $H_{\mathrm{\mathrm{dR}}}(X,A) {:=} \pi_0 \mathbf{H}_{\mathrm{\mathrm{dR}}}(X,A)$
is the \emph{de Rham cohomology} of $X$ with coefficients in $A$.
\end{definition}
\begin{observation}
A cocycle in de Rham cohomology
$$
  \omega : \mathbf{\Pi}_{\mathrm{\mathrm{dR}}}X \to A
$$
is precisely a flat $\infty$-connection on a 
\emph{trivializable} $A$-principal $\infty$-bundle. 
More precisely, $\mathbf{H}_{\mathrm{\mathrm{dR}}}(X,A)$ is the homotopy fiber of the 
forgetful functor from $\infty$-bundles with flat $\infty$-connection to $\infty$-bundles: 
we have an $\infty$-pullback diagram
$$
  \xymatrix{
    \mathbf{H}_{\mathrm{\mathrm{dR}}}(X,A) \ar[r] \ar[d] & {*} \ar[d]
    \\
    \mathbf{H}_{\mathrm{flat}}(X,A) \ar[r] & \mathbf{H}(X,A)
  }  
  \,.
$$
\end{observation}
\proof
This follows by the fact that the hom-functor $\mathbf{H}(X,-)$ 
preserves the defining $\infty$-pullback for $\mathbf{\flat}_{\mathrm{\mathrm{dR}}} A$.
\endofproof
Just for emphasis, notice the dual description of this situation: by the universal property of the $\infty$-colimit that defines $\mathbf{\Pi}_{\mathrm{\mathrm{dR}}} X$ we have that 
$\omega$ corresponds to a diagram
$$
  \xymatrix{
    X \ar[r] \ar[d] & {*} \ar[d]
    \\
    \mathbf{\Pi}(X) \ar[r]^\omega & A
  }
  \,.
$$
The bottom horizontal morphism is a flat connection on the 
$\infty$-bundle which in turn is given by the composite cocycle 
$X \to \mathbf{\Pi}(X) \stackrel{\omega}{\to} A$. The diagram says that this is equivalent to the trivial bundle given by the trivial cocycle $X \to * \to A$.
\begin{proposition} \label{deRhamWithDiscCoeffsIsTrivial}
The de Rham cohomology with coefficients in discrete objects is trivial: 
for all $S \in \infty \mathrm{Grpd}$ we have
$$
  \mathbf{\flat}_{\mathrm{\mathrm{dR}}} \mathrm{Disc} S \simeq *
  \,.
$$
\end{proposition}
\proof
Using that in a $\infty$-connected $\infty$-topos the functor 
$\mathrm{Disc}$ is a full and faithful $\infty$-functor so that unit 
$\mathrm{Id} \to \Gamma \mathrm{Disc} $ is an equivalence and using that by the 
zig-zag identity the counit component
$\mathbf{\flat} \mathrm{Disc} S := \mathrm{Disc} \Gamma \mathrm{Disc} S \to \mathrm{Disc} S$ is also an equivalence, 
we have
$$
  \begin{aligned}
     \mathbf{\flat}_{\mathrm{\mathrm{dR}}} \mathrm{Disc} S 
     & {:=}
     * \prod_{\mathrm{Disc} S} \mathbf{\flat} \mathrm{Disc} S
     \\
     & \simeq {*} \prod_{\mathrm{Disc} S} \mathrm{Disc} S
     \\
     & \simeq {*}
  \end{aligned}
  \,,
$$
since the pullback of an equivalence is an equivalence.
\endofproof
\begin{proposition}
For every $X$ in a cohesive $\infty$-topos $\mathbf{H}$, the object 
$\mathbf{\Pi}_{\mathrm{dR}} X $ is
globally connected in that $\pi_0 \mathbf{H}(*, \mathbf{\Pi}_{\mathrm{\mathrm{dR}}}X) = *$.

If $X$ has at least one point ($\pi_0(\Gamma X) \neq \emptyset $) 
and is geometrically connected ($\pi_0 (\Pi X) = {*}$) then 
$\mathbf{\Pi}_{\mathrm{\mathrm{dR}}}(X)$ is also locally connected: 
$\tau_0 \mathbf{\Pi}_{\mathrm{\mathrm{dR}}} \simeq {*} \in \mathbf{H}$.
\end{proposition}
\proof
Since $\Gamma$ preserves $\infty$-colimits in a cohesive $\infty$-topos we have
$$
  \begin{aligned}
    \mathbf{H}(*, \mathbf{\Pi}_{\mathrm{\mathrm{dR}}}X)
    & \simeq
    \Gamma \mathbf{\Pi}_{\mathrm{\mathrm{dR}}} X
    \\  
    & \simeq
    * \coprod_{\Gamma X} \mathbf{\Gamma \mathbf{\Pi}X}  
    \\
    & \simeq
    * \coprod_{\Gamma X} \mathbf{\Pi X}  
  \end{aligned}
  \,,
$$
where in the last step we used that $\mathrm{Disc}$ is 
full and faithful, so that there is an equivalence 
$\Gamma \mathbf{\Pi}X := \Gamma \mathrm{Disc} \Pi X \simeq \Pi X$.

To analyse this $\infty$-pushout we present it by a homotopy
pushout in $\mathrm{sSet}_{\mathrm{Quillen}}$. Denoting by $\Gamma X$ and $\Pi X$
any represetatives in $\mathrm{sSet}_{\mathrm{Quillen}}$ of the objects of the same name in 
$\infty \mathrm{Grpd}$, this may be computed by the ordinary pushout 
of simplicial sets
$$
  \xymatrix{
    \Gamma X \ar[r] \ar[d] & (\Gamma X) \times \Delta[1] \coprod_{\Gamma X} {*}
      \ar[d]
    \\
    \Pi X \ar[r] & Q
  }
  \,,
$$
where on the right we have inserted the cone on $\Gamma X$ in order to turn the
top morphism into a cofibration. From this ordinary pushout it is clear that the
connected components of $Q$ are obtained from those of $\Pi X$ by identifying all those
in the image of a connected component of $\Gamma X$. So if the left morphism is surjective on
$\pi_0$ then $\pi_0(Q) = *$. This is precisely the condition that \emph{pieces have points}
in $\mathbf{H}$.

For the local analysis we consider the same setup objectwise in the injective model structure
$[C^{\mathrm{op}}, \mathrm{sSet}]_{\mathrm{inj},\mathrm{loc}}$. For any $U \in C$
we then have the pushout $Q_U$ in
$$
  \xymatrix{
    X(U) \ar[r] \ar[d] & (X(U)) \times \Delta[1] \coprod_{X(U)} {*}
      \ar[d]
    \\
    \mathrm{sSet}(\Gamma(U), \Pi X) \ar[r] & Q_U
  }
  \,,
$$
as a model for the value of the simplicial presheaf presenting $\mathbf{\Pi}_{\mathrm{\mathrm{dR}}}(X)$.
If $X$ is geometrically connected then $\pi_0 \mathrm{sSet}(\Gamma(U), \Pi(X)) = *$
and hence for the left morphism to be surjective on $\pi_0$ it suffices that the top left
object is not empty. Since the simplicial set $X(U)$ contains at least the vertices 
$U \to * \to X$ of which there is by assumption at least one, this is the case.
\endofproof
\begin{remark}
In summary we see that in any cohesive $\infty$-topos 
the objects $\mathbf{\Pi}_{\mathrm{dR}}(X)$ of def. \ref{deRhamCoefficientObject} have
the essential abstract properties of pointed \emph{geometric de Rham homotopy types} 
(\cite{toen}, section 3.5.1).
In section \ref{Implementation} we will see that, indeed, 
the intrinsic de Rham cohomology of the cohesive $\infty$ -topos $\mathbf{H} = \mathrm{Smooth}\infty\mathrm{Grpd}$
$$H_{\mathrm{\mathrm{dR}}}(X, A) {:=} \pi_0 \mathbf{H}(\mathbf{\Pi}_{\mathrm{\mathrm{dR}}} X, A)$$ 
reproduces ordinary de Rham cohomology in degree $d > 1$.

In degree 0 the intrinsic de Rham cohomology is necessarily trivial, while in degree 1 we find 
that it reproduces closed 1-forms, not divided out by exact forms. This difference to ordinary de Rham cohomology in the lowest two degrees may be understood in terms of the obstruction-theoretic meaning of de Rham cohomology by which we essentially characterized it above: we have that the intrinsic $H_{\mathrm{\mathrm{dR}}}^n(X,K)$ is the home for the obstructions to flatness of 
$\mathbf{B}^{n-2}K$-principal $\infty$-bundles. For $n = 1$ this are groupoid-principal 
bundles over the \emph{groupoid} with $K$ as its space of objects. But the 1-form curvatures of groupoid bundles are not to be regarded modulo exact forms.  
 \label{DeRhamHomotopyType}
\end{remark}

We turn now to identifying certain de Rham cocycles that are adapted to intrinsic
manifolds, as discussed in \ref{Strucmanifolds}. In general a cocycle 
$\omega : X \to \flat_{\mathrm{dR}}\mathbf{B}A$ is to be thought of as
what traditionally is called a cocycle in de Rham \emph{hypercohomology}.
The following definition models the idea of picking in de Rham hypercohomology
over a manifold those cocycles that are given by globally defined differential forms.

Fix a line object $\mathbb{A}^1 \in \mathbf{H}$ which \emph{exhibits the cohesion}
of $\mathbf{H}$ in the sense of def. \ref{ILocalization}.

\begin{definition}
  For $A \in \mathrm{Grp}(\mathbf{H})$ an $\infty$-group, 
  a choice of \emph{$A$-valued differential forms} is a morphism
  $$
    \Omega_{\mathrm{cl}}(-,A)
	 \to 
	 \flat_{\mathrm{dR}}\mathbf{B}A
  $$
  in $\mathbf{H}$, which is an \emph{atlas over manifolds} of $\flat_{\mathrm{dR}}\mathbf{B}A$,
  in that:
  \begin{enumerate}
    \item $\Omega_{\mathrm{cl}}(-,A)$ is 0-truncated;
	\item for each intrinsic $\mathbb{A}^1$-manifold $\Sigma$, def. \ref{IntrinsicManifold},
	 the morphism $[\Sigma, \Omega^n_{\mathrm{cl}}(-,A)] \to [\Sigma,\flat_{\mathrm{dR}} \mathbf{B}^n A]$
	 is an effective epimorphism, def. \ref{EffectiveEpimorphism}.
  \end{enumerate}
  \label{DifferentialFormObject}
\end{definition}
\begin{remark}
  We discuss below in \ref{SmoothClosedFormsAreDifferentialFormObject} 
  how in the standard model of smooth cohesion
  this notion reproduces the traditional notion of smooth differential forms. 
\end{remark}

\subsubsection{Exponentiated $\infty$-Lie algebras}
\label{StrucLieAlgebras}
\index{structures in a cohesive $\infty$-topos!exponentiated $\infty$-Lie algebras}
We consider an intrinsic notion of \emph{exponentiated} $\infty$-Lie algebras in every
cohesive $\infty$-topos. In order to have a general abstract notion of the $\infty$-Lie 
algebras themselves we need the further axiomatics of \emph{infinitesimal cohesion},
discussed below in \ref{InfinitesimalCohesion} and \ref{InfStrucFormalInfinityGroupoid}.
\begin{definition}
For a connected object $\mathbf{B}\exp(\mathfrak{g}) $ in $\mathbf{H}$ that is 
\emph{geometrically contractible}
$$
  \Pi (\mathbf{B}\exp(\mathfrak{g})) \simeq *
$$
we call its loop space object (see \ref{StrucInftyGroups}) 
$\exp(\mathfrak{g}) := \Omega_* \mathbf{B}\exp(\mathfrak{g})$ a 
\emph{Lie integrated $\infty$-Lie algebra} in $\mathbf{H}$. 
\end{definition}
\begin{definition}
Set
$$
  \exp\mathrm{Lie}
  := 
  \mathbf{\Pi}_{\mathrm{\mathrm{dR}}} \circ \mathbf{\flat}_{\mathrm{\mathrm{dR}}} 
  : {*}/\mathbf{H} \to {*}/\mathbf{H}
  \,.
$$
\end{definition}
\begin{observation} \label{LieAsALeftAdjoint}
If $\mathbf{H}$ is cohesive, then $\exp\mathrm{Lie}$ is a left adjoint.
\end{observation}
\proof
  By the construction in def. \ref{TheAdjointTripleOfEndofunctors}.
\endofproof
\begin{example}
For all $X \in \mathbf{H}$ the object $\mathbf{\Pi}_{\mathrm{\mathrm{dR}}}(X)$ is geometrically contractible.
\end{example}
\proof
Since on the locally $\infty$-connected and $\infty$-connected $\mathbf{H}$ 
the functor $\Pi$ preserves $\infty$-colimits and the terminal object, we have
$$
  \begin{aligned}
    \Pi \mathbf{\Pi}_{\mathrm{\mathrm{dR}}} X
    & {:=}
     \Pi (*) \coprod_{\Pi X} \Pi \mathbf{\Pi} X
    \\
    & \simeq
      {*} \coprod_{\Pi X} \Pi \mathrm{Disc} \Pi X   
    \\
    & \simeq 
      {*} \coprod_{\Pi X} \Pi X
    & \simeq
      {*}
  \end{aligned}
  \,,
$$
where we used that on the $\infty$-connected $\mathbf{H}$ the functor 
$\mathrm{Disc}$ is full and faithful.
\endofproof
\begin{corollary}
We have for every $(* \to A) \in {*}/\mathbf{H}$ that $\exp \mathrm{Lie} A$ is geometrically contractible.
\end{corollary}
We shall write $\mathbf{B}\exp(\mathfrak{g})$ for $\exp \mathrm{Lie} \mathbf{B}G$, when the context is clear.
\begin{proposition} \label{LieValuesofDeRham}
Every de Rham cocycle (\ref{StrucDeRham}) $\omega : \mathbf{\Pi}_{\mathrm{dR}} X \to \mathbf{B}G$ 
factors through the Lie integrated $\infty$-Lie algebra of $G$
$$
  \xymatrix{
     & \mathbf{B}\exp(\mathfrak{g})
       \ar[d]
     \\
     \mathbf{\Pi}_{\mathrm{\mathrm{dR}}}X 
     \ar[r]^{\omega} \ar[ur]&
     \mathbf{B}G
  }
  \,.
$$
\end{proposition}
\proof
By the universality of the $(\mathbf{\Pi}_{\mathrm{\mathrm{dR}}} \dashv \mathbf{\flat}_{\mathrm{\mathrm{dR}}})$-counit we have 
that $\omega$ factors through the counit 
$\epsilon : \exp \mathrm{Lie} \mathbf{B}G \to \mathbf{B}G$
$$
  \xymatrix{
    & \mathbf{\Pi}_{\mathrm{\mathrm{dR}}}X
    \ar[dr]^{\omega}
    \ar[dl]_{\mathbf{\Pi}_{\mathrm{\mathrm{dR}}} \tilde \omega}
    \\
    \mathbf{\Pi}_{\mathrm{\mathrm{dR}}} \mathbf{\flat}_{\mathrm{\mathrm{dR}}} \mathbf{B}G
    \ar[rr]^\epsilon    
    &&
    \mathbf{B}G
  }
  \,,
$$
where $\tilde \omega : X \to \mathbf{\flat}_{\mathrm{\mathrm{dR}}} \mathbf{B}G$ is the adjunct of $\omega$.
\endofproof
Therefore instead of speaking of a $G$-valued de Rham cocycle, 
it is less  redundant to speak of an $\exp(\mathfrak{g})$-valued de Rham cocycle. 
In particular we have the following.
\begin{corollary}
Every morphism $\mathbf{B}\exp(\mathfrak{h}) := 
 \exp \mathrm{Lie} \mathbf{B}H \to \mathbf{B}G$ from a 
Lie integrated $\infty$-Lie algebra to an $\infty$-group factors through the 
Lie integrated $\infty$-Lie algebra of that 
$\infty$-group
$$
  \xymatrix{
    \mathbf{B}\exp(\mathfrak{h}) \ar[r]\ar[dr]& \mathbf{B}\exp(\mathfrak{g}) \ar[d]
    \\
    & \mathbf{B}G
  }
  \,.
$$
\end{corollary}

\subsubsection{Maurer-Cartan forms and curvature characteristic forms}
\label{StructCurvatureCharacteristic}
\index{structures in a cohesive $\infty$-topos!Maurer-Cartan forms}
\index{structures in a cohesive $\infty$-topos!curvature characteristic forms}

In the intrinsic de Rham cohomology of 
the cohesive $\infty$-topos $\mathbf{H}$ there exist canonical cocycles that 
we may identify with \emph{Maurer-Cartan forms} and with 
universal \emph{curvature characteristic forms}.

\begin{definition} 
  \label{UniversalFormOnInftyGroup}
  \index{Maurer-Cartan form!general abstract}
For $G \in \mathrm{Group}(\mathbf{H})$ an $\infty$-group in the cohesive $\infty$-topos
$\mathbf{H}$, write
$$
  \theta : G \to \mathbf{\flat}_{\mathrm{\mathrm{dR}}} \mathbf{B}G
$$
for the $G$-valued de Rham cocycle on $G$ induced by this 
pasting of $\infty$-pullbacks
$$
  \xymatrix{
     G \ar[r] \ar[d]_{\theta}  & {*} \ar[d]
     \\
     \mathbf{\flat}_{\mathrm{\mathrm{dR}}} \mathbf{B}G 
      \ar[r]
      \ar[d] 
       & 
     \mathbf{\flat}\mathbf{B}G
      \ar[d]
     \\
     {*} \ar[r] & \mathbf{B}G 
  }
$$
using prop. \ref{LieValuesofDeRham}.

We call $\theta$ the \emph{Maurer-Cartan form} on $G$.
\end{definition}
\begin{remark}
For any object $X$, postcomposition the Maurer-Cartan form sends  $G$-valued functions on 
$X$ to $\mathfrak{g}$-valued forms on $X$
$$
  [\theta_*] : H^0(X,G) \to H^1_{\mathrm{\mathrm{dR}}}(X,G)
  \,.
$$
\end{remark}
\begin{remark} 
For $G = \mathbf{B}^n A$ an Eilenberg-MacLane object, we also write 
$$
  \mathrm{curv} : \mathbf{B}^n A \to \mathbf{\flat}_{\mathrm{\mathrm{dR}}} \mathbf{B}^{n+1} A
$$
for its intrinsic Maurer-Cartan form and call this the intrinsic 
\emph{universal curvature characteristic form} on $\mathbf{B}^n A$.

These curvature characteristic forms serve to define differential cohomology in the 
next section.
\end{remark}

\subsubsection{Differential cohomology}
\label{StrucDifferentialCohomology}
\label{DifferentiaCoefficients}
\index{structures in a cohesive $\infty$-topos!differential cohomology}

We discuss an intrinsic realization of \emph{differential cohomology}
(see for instance \cite{BunkeDifferentialCohomology})
with coefficients in braided $\infty$-groups in any cohesive $\infty$-topos.

We first give a general discussion in \ref{DifferentialCohomologyGeneral} and then 
consider a special class of cases in \ref{GeneralAbstractDifferentialCohomologyWithDifferentialFormCurvature}.
Finally we discuss issues of constructing differential moduli objects in 
\ref{StrucDifferentialModuli}.

In the case that the homotopy type is not just braided, hence twice deloopable, 
but is in fact stable (a spectrum object), then there is a strengthening
of the theory of differential cohomology to differential stable cohomology,
which enjoys very good properties. This we come to below 
in the discussion of the models of Goodwillie-tangent cohesion 
\ref{BundlesOfCohesiveSpectra}.

Notice that for many of the applications in \ref{Applications}
it is crucial to have available also generally the non-stable 
differential cohomology discussed here. This is necessary specifically
for the discussion of Wess-Zumino-Witten-type prequantum field theory
in \ref{WZWApplications}.

\paragraph{General }
 \label{DifferentialCohomologyGeneral}
 
\begin{definition}
  For $\mathbb{G}$ a braided $\infty$-group, def. \ref{BraidedInfinityGroup}, 
  write 
  $$
    \mathrm{curv}_{\mathbb{G}} := \theta_{\mathbf{B}{\mathbb{G}}} 
	 : 
	 \mathbf{B}{\mathbb{G}} \to \flat_{\mathrm{dR}}\mathbf{B}^2 {\mathbb{G}}
  $$
  for the Maurer-Cartan form, def. \ref{UniversalFormOnInftyGroup}, on 
  its delooping  $\infty$-group $\mathbf{B}{\mathbb{G}}$.
  We call this the \emph{universal curvature characteristic} of ${\mathbb{G}}$.
  
  We say that the cohomology in the slice 
  $\infty$-topos $\mathbf{H}_{/\flat_{\mathrm{dR}} \mathbf{B}^2 {\mathbb{G}}}$
  with coefficients in $\mathrm{curv}_{\mathbb{G}}$ is the \emph{differential cohomology} with coefficients
  in $\mathbf{B}{\mathbb{G}}$.
  \label{UniversalCurvatureCharacteristic}
  \index{characteristic class!curvature}
  \index{curvature!characteristic form (general abstract)}
\end{definition}
\begin{remark}
  A domain object $(X,F) \in \mathbf{H}_{/\flat_{\mathrm{dR}}\mathbf{B}^2 {\mathbb{G}}}$
  is an object $X \in \mathbf{H}$ equipped with a de Rham cocycle 
  $F : X \to \flat_{\mathrm{dR}}\mathbf{B}^2 {\mathbb{G}}$, 
  to be thought of as a prescribed \emph{curvature differential form}.

  A differential cocycle 
  $\nabla \in \mathbf{H}_{/\flat_{\mathrm{dR}}\mathbf{B}^2 {\mathbb{G}}}((X,F), \mathrm{curv}_G)$
  on such a pair is
  a diagram of the form
  $$
    \xymatrix{
	  X \ar[rr]^g_{\ }="s" \ar[dr]_F^{\ }="t" && \mathbf{B}G \ar[dl]^{\mathrm{curv}_{\mathbb{G}}}
	  \\
	  & \flat_{\mathrm{dR}\mathbf{B}^2 {\mathbb{G}}}
	  \ar@{=>}^\nabla "s"; "t"
	}
  $$
  in $\mathbf{H}$. This is 
  \begin{enumerate}
    \item a cocycle $g : X \to \mathbf{B}{\mathbb{G}}$ 
	 in $\mathbf{H}$ for a ${\mathbb{G}}$-principal $\infty$-bundle over $X$;
	\item 
    a choice of equivalence
  $$
    \xymatrix{ g^* \mathrm{curv}_{\mathbb{G}} \ar[r]^-\nabla_-\simeq & F }
  $$
  between the pullback of the universal ${\mathbb{G}}$-curvature characteristic, 
  def. \ref{UniversalCurvatureCharacteristic}
  and the prescribed curvature differential form. 
  \end{enumerate}
  This choice of equivalence is
  to be interpreted as a \emph{connection} on the ${\mathbb{G}}$-principal bundle modulated by $g$.
  \label{GeneralDifferentialCocyclesExpressedInH}
\end{remark}

Often one is of interested in demanding that the curvature 
$F : X \to \flat_{\mathrm{dR}}\mathbf{B}^2 {\mathbb{G}}$ 
in the above factors through a prescribed morphism $C \to \flat_{\mathrm{dR}}\mathbf{B}^2 {\mathbb{G}}$,
notably through an inclusion of differential forms as in def. \ref{DifferentialFormObject}. This means
that one is interested in cocycles as in remark \ref{GeneralDifferentialCocyclesExpressedInH} 
above which factor as diagrams
$$
  \xymatrix{
    X \ar[rr]^g_{\ }="s" \ar[d]_F^{\ }="t" && \mathbf{B}G \ar[dl]^{\mathrm{curv}_{\mathbb{G}}}
	\\
	C \ar[r] & \flat_{\mathrm{dR}}\mathbf{B}^2 {\mathbb{G}}
	\ar@{=>}^\nabla "s"; "t"
  }
  \,.
$$
This in turn 
means equivalently that the cocycle is given by a morphism 
$\nabla : X \to \mathbf{B}{\mathbb{G}}_{\mathrm{conn}}$
into the $\infty$-pullback 
$\mathbf{B}{\mathbb{G}}_{\mathrm{conn}} 
 \simeq C 
 \times_{\flat_{\mathrm{dR}}\mathbf{B}^2 {\mathbb{G}}} \mathbf{B}G$.
This object we may then regard as a \emph{moduli stack for differential cohomology} with 
coefficients in $A$ and curvatures in $C$.

This we now discuss in \ref{GeneralAbstractDifferentialCohomologyWithDifferentialFormCurvature} below. 

\paragraph{Global curvature forms}
 \label{GeneralAbstractDifferentialCohomologyWithDifferentialFormCurvature}
 
We consider the subcase of the general notion of differential cohomology
as in \ref{DifferentialCohomologyGeneral} above, 
where now the curvatures are required to be globally defined differential forms
according to def. \ref{DifferentialFormObject}.
The resulting definition essentially reproduces that of
differential cohomology in terms of homotopy pullbacks  as discussed in 
\cite{HopkinsSinger}, but is formulated entirely
internal to a cohesive $\infty$-topos. Therefore it refines
the construction of \cite{HopkinsSinger} in two ways\footnote{
 After we had proposed this refinement, in \cite{Hopkins} it says that this is
 the context to which the article \cite{HopkinsSinger} was intended to be
 refined.
}:
\begin{enumerate}
 \item The
coefficient object may be a cohesive $\infty$-groupoid, 
where in \cite{HopkinsSinger} it is just a 
topological space, hence, as explained below in 
\ref{DiscreteInfGroupoids}, a \emph{discrete} $\infty$-groupoid. 
\item 
 The domain object may also be a cohesive $\infty$-groupoid, where in
\cite{HopkinsSinger} it is restricted to be a manifold. In particular it 
can be an orbifold, or itself a moduli stack.
\end{enumerate}
We give below an intrinsic characterization of domain objects that
are manifolds in the sense of def. \ref{IntrinsicManifold}. 
On more general objects our definition
subsumes also a notion of \emph{equivariant} differential cohomology. 

\medskip

\begin{definition}
For $\mathbb{G}$ a braided $\infty$-group in $\mathbf{H}$, def. \ref{BraidedInfinityGroup}, 
the \emph{moduli of closed 2-forms} with values in $\mathbb{G}$ is a morphism
$$
  \xymatrix{
     \Omega^2_{\mathrm{cl}}(-,\mathbb{G}) \ar[r] &  \flat_{\mathrm{dR}}\mathbf{B}^2 \mathbb{G}
  }
$$
characterized as follows:
\begin{enumerate}
\item $\Omega^2_{\mathrm{cl}}(-,\mathbb{G}) \in \mathbf{H}$ is 0-truncated;
\item for every $\mathbb{A}^1$-manifold $\Sigma \in \mathbf{H}$, def. \ref{BraidedInfinityGroup},
we have that
$$
  [\Sigma, \iota] 
    :  
  \xymatrix{
     [\Sigma, \Omega^2_{\mathrm{cl}}(-,\mathbb{G})] 
	 \ar[r]
	 &
	 [\Sigma, \flat_{\mathrm{dR}}\mathbf{B}^2 \mathbb{G}]
  }
$$
is an epimorphism
\item
$\iota$ is universal with the above two properties.
\end{enumerate}
A morphism $\omega X : \Omega^2_{\mathrm{cl}}(-,\mathbb{G})$
we call a \emph{closed $\mathrm{Lie}(\mathbb{G})$-valued differential 2-form} on $X$,
or a \emph{pre-symplectic structure} on $X$, with values in $\mathrm{Lie}(\mathbb{G})$.
 \label{Closed2FormModuli}
\end{definition}
\begin{definition}
  For $\mathbb{G}$ a braided $\infty$-group, we write
  $$
    \mathbf{B}\mathbb{G}_{\mathrm{conn}} := \mathbf{B}\mathbb{G} \underset{\flat_{\mathrm{dR}}\mathbf{B}^2 \mathbb{G}}{\times} \Omega^2(-,\mathbb{G}) 
  $$
  for the $\infty$-fiber product in 
  $$
    \raisebox{20pt}{
    \xymatrix{
	  \mathbf{B}\mathbb{G}_{\mathrm{conn}}
	  \ar[d]^{U}
	  \ar[r]^{F_{(-)}}
	  &
	  \Omega^2(-,\mathbb{G})
	  \ar[d]
	  \\
	  \mathbf{B}\mathbb{G}
	  \ar[r]^{\mathrm{curv}_{\mathbb{G}}}
	  &
	  \flat_{\mathrm{dR}}\mathbf{B}^2 \mathbb{G}
	}
	}\,.
  $$
  We say that 
  \begin{enumerate}
    \item $\mathbf{B}\mathbb{G}_{\mathrm{conn}}$ is the \emph{moduli object} for 
	\emph{differential cocycles with coefficients in $\mathbb{G}$} or equivalently
    for \emph{$\mathbb{G}$-principal connections};
	\item For $\nabla : X \to \mathbf{B}\mathbb{G}_{\mathrm{conn}}$ we say that
	\begin{enumerate}
	\item $F_{\nabla} : X \to \Omega^2(-,\mathbb{G})$ is the \emph{curvature form} of $\nabla$
	\item that $U(\nabla) : X \to \mathbf{B}\mathbb{G}$ is (the morphism modulation) the
	\emph{underlying} $\mathbb{G}$-principal bundle of $\nabla$.
	\end{enumerate}
  \end{enumerate}
\end{definition}

\begin{proposition}
  For $\mathbb{G} \in \mathrm{Grp}(\mathbf{H})$ a braided $\infty$-group,
  the loop space object, def. \ref{loop space object}, of  
  $\mathbf{B}\mathbb{G}_{\mathrm{conn}}$ 
  is equivalent to the flat coefficient object $\flat G$
  $$
    \Omega \mathbf{B}\mathbb{G}_{\mathrm{conn}}
	\simeq
	\flat \mathbb{G}
	\,.
  $$
  \label{LoopSpaceObjectOfBGconn}
\end{proposition}
\proof
Using that $\Omega_{\mathrm{cl}}(-,\mathbb{G})$ is 0-truncated by
definition, using that $\flat$ is right adjoint and hence commutes with 
$\infty$-pullbacks and repeatedly using the pasting law, prop. \ref{PastingLawForPullbacks},
we find a pasting diagram of $\infty$-pullbacks of the form
$$
  \xymatrix{
    \flat \mathbb{G} \ar[r] \ar[d] & \mathbb{G} \ar[r] \ar[d] & {*} \ar[d]
    \\
    {*} \ar[r] \ar[d] &  \flat_{\mathrm{dR}}\mathbf{B}\mathbb{G}\ar[r] \ar[d]
	& \flat \mathbf{B}\mathbb{G} \ar[r] \ar[d]  & {*} \ar[d]
	\\
	{*} \ar[r] &  \ar[r] \ar[d] 
	& \mathbf{B}\mathbb{G}_{\mathrm{conn}} \ar[r] \ar[d] & \Omega_{\mathrm{cl}}(-,\mathbb{G}) \ar[d]
	\\
	& {*} \ar[r] & \mathbf{B}\mathbb{G} \ar[r] & \flat_{\mathrm{dR}}\mathbf{B}^2 \mathbb{G}
  }
  \,.
$$
\endofproof

\paragraph{Ordinary differential cohomology}
 \label{GeneralAbstractOrdinaryDifferentialCohomology}

We now spell out the constructions of \ref{GeneralAbstractDifferentialCohomologyWithDifferentialFormCurvature} in more detail for the
special case that $\mathbb{G}$ is an Eilenberg-MacLane object, 
hence for the case there is a
0-truncated abelian group object 
$A \in \mathrm{Grp}(\tau_{\leq 0} \mathbf{H}) \hookrightarrow \mathbf{H}$
and $n \in \mathbb{N}$ such that 
$$
  \mathbf{B}\mathbb{G} \simeq \mathbf{B}^n A
  \,.
$$ 
This is the case of \emph{ordinary differential cohomology} that refines 
what the \emph{ordinary cohomology} with coefficients in $A$, according to 
remark \ref{OrdinaryCohomology}. The explicit realization of this 
construction in smooth cohesion is discussed below in \ref{SmoothStrucDifferentialCohomology}.

\medskip

By the discussion in \ref{StrucInftyGroups}
we have for all 
$n \in \mathbf{N}$ the corresponding 
Eilenberg-MacLane object $\mathbf{B}^n A$. By the discussion in 
\ref{section.PrincipalInfinityBundle} this classifies $\mathbf{B}^{n-1}A$-principal
$\infty$-bundles in that for any $X \in \mathbf{H}$ we have an equivalence
of $n$-groupoids
$$
  \mathbf{B}^{n-1}A \mathrm{Bund}(X)
  \simeq
  \mathbf{H}(X, \mathbf{B}^n A)
$$
whose objects are $\mathbf{B}^{n-1}A$-principal $\infty$-bundles on $X$, whose
morphisms are gauge transformations between these, and so on.
The following definition refines this by equipping these $\infty$-bundles
with the structure of a \emph{connection}.

Let $\mathbb{A}^1 \in \mathbf{H}$ be a line object \emph{exhibiting the cohesion}
of $\mathbf{H}$ in the sense of def. \ref{ILocalization}. Let then furthermore for each 
$n \in \mathbb{N}$
$$
  \Omega^n_{\mathrm{cl}}(-,A) \to \flat_{\mathrm{dR}}\mathbf{B}^n A
$$
be a choice of differential form objects, according to def. \ref{DifferentialFormObject}.

 \begin{definition}
 \label{OrdinaryDiffCohomology}
For $X \in \mathbf{H}$ any object and $n \geq 1$ write
$$
  \mathbf{H}_{\mathrm{diff}}(X,\mathbf{B}^n A) 
   := 
    \mathbf{H}(X,\mathbf{B}^n A) \prod_{\mathbf{H}_{\mathrm{\mathrm{dR}}}(X,\mathbf{B}^n A)}
    H_{\mathrm{\mathrm{dR}}}^{n+1}(X,A)
$$ 
for the cocycle $\infty$-groupoid of \emph{twisted cohomology}, \ref{StrucTwistedCohomology}, 
of $X$ with coefficients in 
$A$ relative to the canonical curvature characteristic morphism
$\mathrm{curv} : \mathbf{B}^n A \to \mathbf{\flat}_{\mathrm{dR}}\mathbf{B}^{n+1}A$ (\ref{StructCurvatureCharacteristic}). 
By prop. \ref{PullbackCharacterizationOfTwistedCohomology} this is the $\infty$-pullback
$$
  \raisebox{20pt}{
  \xymatrix{
    \mathbf{H}_{\mathrm{diff}}(X,\mathbf{B}^n A) 
     \ar[r]^{[F]}
     \ar[d]^c
      & 
     H_{\mathrm{\mathrm{dR}}}^{n+1}(X,A)
     \ar[d]
    \\
    \mathbf{H}(X,\mathbf{B}^n A) 
        \ar[r]^{\mathrm{curv}_*}& 
     \mathbf{H}_{\mathrm{\mathrm{dR}}}(X,\mathbf{B}^{n+1} A)
  }
  }
  \,,
$$
where the right vertical morphism $\pi_0 \mathbf{H}_{\mathrm{\mathrm{dR}}}(X,\mathbf{B}^{n+1}A)
 \to \mathbf{H}_{\mathrm{\mathrm{dR}}}(X,\mathbf{B}^{n+1}A)$ is 
 the unique, up to equivalence, effective epimorphism
 out of a 0-truncated object: a choice of cocycle 
representative in each cohomology class, equivalently 
a choice of point in every connected component.

We call
$$
  H_{\mathrm{diff}}^n(X,A) {:=} \pi_0 \mathbf{H}_{\mathrm{diff}}(X, \mathbf{B}^{n} A)
$$
the degree-$n$ \emph{differential cohomology}\index{differential cohomology!general abstract}
\index{twisted cohomology!differential cohomology} of $X$ with coefficient in $A$.

For $\nabla \in \mathbf{H}_{\mathrm{diff}}(X,\mathbf{B}^n A)$ a cocycle, we call
\begin{itemize}
\item $[c(\nabla)] \in H^n(X,A)$ the \emph{characteristic class} of the 
  \emph{underlying $\mathbf{B}^{n-1} A$-principal $\infty$-bundle};
\item
  $[F](\nabla) \in H_{\mathrm{\mathrm{dR}}}^{n+1}(X,A)$ the \emph{curvature} class of $c$
  (this is the \emph{twist}).
\end{itemize}
We also say that $\nabla$ is an \emph{$\infty$-connection} on the principal 
$\infty$-bundle $\eta(\nabla)$.
\end{definition}
\begin{observation} \label{DiffCohIsWellDefined}
The differential cohomology 
$H_{\mathrm{diff}}^n(X,A)$ does not depend 
on the choice of morphism 
$H_{\mathrm{\mathrm{dR}}}^{n+1}(X,A) \to \mathbf{H}_{\mathrm{\mathrm{dR}}}(X, \mathbf{B}^{n+1}A)$ 
(as long as it is an isomorphism on $\pi_0$, as required).
In fact, for different choices the corresponding cocycle $\infty$-groupoids $\mathbf{H}_{\mathrm{diff}}(X,\mathbf{B}^n A)$ are equivalent.
\end{observation}
\proof
The set  
$$
  H_{\mathrm{\mathrm{dR}}}^{n+1}(X,A) = \coprod_{H_{\mathrm{\mathrm{dR}}}^{n+1}(X,A)} {*} 
$$ 
is, as a 0-truncated $\infty$-groupoid, an $\infty$-coproduct of the terminal object 
$\infty \mathrm{Grpd}$. By universal colimits in this $\infty$-topos we have that 
$\infty$-colimits are preserved by $\infty$-pullbacks, 
so that $\mathbf{H}_{\mathrm{diff}}(X, \mathbf{B}^n A)$ is the coproduct 
$$
  \mathbf{H}_{\mathrm{diff}}(X,\mathbf{B}^n A) 
   \simeq
  \coprod_{H_{\mathrm{\mathrm{dR}}}^{n+1}(X,A)} 
   \left(
   \mathbf{H}(X,\mathbf{B}^n A) 
      \underset{\mathbf{H}_{\mathrm{\mathrm{dR}}}\times (X,\mathbf{B}^{n+1}A)} 
    {*}
  \right)
$$
of the homotopy fibers of $\mathrm{curv}_*$ over each of the chosen points 
${*} \to \mathbf{H}_{\mathrm{\mathrm{dR}}}(X,\mathbf{B}^{n+1}A)$. 
These homotopy fibers only  depend, up to equivalence, 
on the connected component over which they are taken.
\endofproof
\begin{proposition} 
  \label{DiffCohomologyRestrictedToVanishingCurvature}
When restricted to vanishing curvature, differential cohomology coincides with 
flat differential cohomology, \ref{StrucFlatDifferential},
$$
  H_{\mathrm{diff}}^n (X,A)|_{[F] = 0} \simeq H_{\mathrm{flat}}(X,\mathbf{B}^n A)
  \,.
$$
Moreover this is true at the level of cocycle $\infty$-groupoids
$$
 \left(
  \mathbf{H}_{\mathrm{diff}}(X, \mathbf{B}^n A) 
  \underset{H_{\mathrm{\mathrm{dR}}}^{n+1}(X,A)}{\times} \{[F] = 0\}
  \right)
  \simeq
  \mathbf{H}_{\mathrm{flat}}(X,\mathbf{B}^n A)
  \,,
$$
hence there is a canonical embedding by a full and faithful morphism
$$
  \xymatrix{
    \mathbf{H}_{\mathrm{flat}}(X, \mathbf{B}^n A)
    \ar@{^{(}->}[r]
    &	
    \mathbf{H}_{\mathrm{diff}}(X, \mathbf{B}^n A)
  }
$$
\end{proposition}
\proof
By the pasting law for $\infty$-pullbacks, prop. \ref{PastingLawForPullbacks}, 
the claim is equivalently that 
we have a pasting of $\infty$-pullback diagrams
$$
  \xymatrix{
    \mathbf{H}_{\mathrm{flat}}(X, \mathbf{B}^n A) \ar[r] \ar[d] & {*} \ar[d]^{[F] = 0}
    \\
    \mathbf{H}_{\mathrm{diff}}(X,\mathbf{B}^n A) \ar[r]^-{[F]} \ar[d]^\eta & 
     H_{\mathrm{\mathrm{dR}}}^{n+1}(X,A) \ar[d]
    \\
    \mathbf{H}(X,\mathbf{B}^n A) 
        \ar[r]^-{\mathrm{curv}_{*}}
     & 
     \mathbf{H}_{\mathrm{\mathrm{dR}}}(X,\mathbf{B}^{n+1} A)
  }
  \,.
$$
By definition of flat cohomology, def. \ref{FlatDifferentialCohomology} 
and of intrinsic de Rham cohomology, def. \ref{IntrinsicDeRhamCohomology}, in $\mathbf{H}$,
the outer rectangle is 
$$
  \xymatrix{
    \mathbf{H}(X,\mathbf{\flat}\mathbf{B}^n A) \ar[r] \ar[d] & {*} \ar[d]
    \\
    \mathbf{H}(X, \mathbf{B}^n A) 
      \ar[r]^-{\mathrm{curv}_*}
      &
    \mathbf{H}(X, \mathbf{\flat}_{\mathrm{\mathrm{dR}}}\mathbf{B}^{n+1} A)
  }
  \,.
$$
Since the hom-functor $\mathbf{H}(X,-)$ preserves $\infty$-limits this is a pullback if
$$
  \xymatrix{
    \mathbf{\flat} \mathbf{B}^n A \ar[r] \ar[d] &  {*} \ar[d]
    \\
    \mathbf{B}^n A 
      \ar[r]^-{\mathrm{curv}}& 
    \mathbf{\flat}_{\mathrm{\mathrm{dR}}} \mathbf{B}^{n+1} A
  }
$$
is. Indeed, this is one step in the fiber sequence 
$$
   \cdots
   \to 
   \mathbf{\flat} \mathbf{B}^n A
   \to
   \mathbf{B}^n A
   \stackrel{\mathrm{curv}}{\to}
   \mathbf{\flat}_{\mathrm{\mathrm{dR}}} \mathbf{B}^{n+1}A
  \to
  \mathbf{\flat} \mathbf{B}^{n+1} A
   \to
  \mathbf{B}^{n+1} A
$$
that defines $\mathrm{curv}$ (using that $\mathbf{\flat}$ preserves limits and hence looping and delooping).

Finally, $\xymatrix{{*} \ar[r]^<<<{[F] = 0} & H^{n-1}_{\mathrm{dR}}(X,A)}$
is, trivially, a monomorphism of sets, hence a full and faithfull
morphism of $\infty$-groupoids, and since these are stable under
$\infty$-pullback, it follows that the canonical inclusion of
flat $\infty$-connections into all $\infty$-connections is 
also full and faithful.
\endofproof
The following establishes the characteristic short exact sequences that characterizes intrinsic differential cohomology as an extension of curvature forms by flat $\infty$-bundles and of bare $\infty$-bundles by connection forms.
\begin{proposition}  
 \label{CurvatureExactSequence}
Let $\mathrm{im} F \subset H_{\mathrm{dR}}^{n+1}(X, A)$ be the image of the curvatures. 
Then the differential cohomology group $H_{\mathrm{diff}}^n(X,A)$ fits into a short exact sequence
$$
  0 \to H^n_{\mathrm{flat}}(X, A) \to H^n_{\mathrm{diff}}(X,A) \to \mathrm{im} F \to 0
$$
\end{proposition}
\proof
Form the long exact sequence in homotopy groups of the fiber sequence 
$$
  \mathbf{H}_{\mathrm{flat}}(X, \mathbf{B}^n A)
    \to 
  \mathbf{H}_{\mathrm{diff}}(X, \mathbf{B}^n A)
   \stackrel{[F]}{\to}
  H_{\mathrm{dR}}^{n+1}(X,A)
$$
of prop. \ref{DiffCohomologyRestrictedToVanishingCurvature} and use that 
$H_{\mathrm{dR}}^{n+1}(X,A)$ is, as a set -- a homotopy 0-type -- to get the short exact sequence
on the bottom of this diagram
$$
  \xymatrix{
    \pi_1(H_{\mathrm{dR}}(X,A))  
     \ar[r] \ar@{=}[d]& 
    \pi_0(\mathbf{H}_{\mathrm{flat}}(X, \mathbf{B}^n A))
     \ar[r] \ar@{=}[d]& 
    \pi_0(\mathbf{H}_{\mathrm{diff}}(X, \mathbf{B}^n A))
     \ar[r]^{[F]} \ar@{=}[d] &
    \pi_0(H_{\mathrm{dR}}^{n+1}(X,A))
    \ar[d]
    \\
    0 
      \ar[r] & 
    H_{\mathrm{flat}}^n(X, A) 
      \ar[r] &  
    H_{\mathrm{diff}}^n(X,A)
      \ar[r] &
    \mathrm{im} [F]
  }
  \,.
$$
\endofproof
\begin{proposition}
  \label{ShortExactSequenceForIntrinsicOrdinaryDiffCohomology}
The differential cohomology group 
$H_{\mathrm{diff}}^n(X,A)$ fits into a short exact sequence of abelian groups
$$
  0 
   \to 
  H_{\mathrm{\mathrm{dR}}}^n(X,A)/H^{n-1}(X,A) 
   \to 
  H_{\mathrm{diff}}^n(X,A) 
   \stackrel{c}{\to}
  H^n(X,A)
   \to 
  0
  \,.
$$
\end{proposition}
\proof
We claim that for all $n \geq 1$ we have a fiber sequence
$$
  \mathbf{H}(X, \mathbf{B}^{n-1}A)
  \to 
  \mathbf{H}_{\mathrm{\mathrm{dR}}}(X, \mathbf{B}^n A)
  \to 
  \mathbf{H}_{\mathrm{diff}}(X, \mathbf{B}^n A)
  \to
  \mathbf{H}(X, \mathbf{B}^n A)
$$
in $\infty \mathrm{Grpd}$. 
This implies the short exact sequence using that 
by construction the last morphism is surjective on connected components 
(because in the defining $\infty$-pullback for $\mathbf{H}_{\mathrm{diff}}$ the right 
vertical morphism is by assumption surjective on connected components).

To see that we do have the fiber sequence as claimed, consider the 
pasting composite of $\infty$-pullbacks
$$
  \xymatrix{
    \mathbf{H}_{\mathrm{\mathrm{dR}}}(X,\mathbf{B}^{n-1} A) 
    \ar[r]  \ar[d]& 
      \mathbf{H}_{\mathrm{diff}}(X,\mathbf{B}^n A)  
    \ar[r] \ar[d] & 
     H_{\mathrm{\mathrm{dR}}}(X, \mathbf{B}^{n+1} A) \ar[d]
    \\
    {*} 
      \ar[r]& 
    \mathbf{H}(X, \mathbf{B}^n A) 
      \ar[r]^{\mathrm{curv}}
      &
    \mathbf{H}_{\mathrm{\mathrm{dR}}}(X, \mathbf{B}^{n+1} A)
  }
  \,.
$$
The square on the right is a pullback by def. \ref{OrdinaryDiffCohomology}.
Since also the square on the left is assumed to be an $\infty$-pullback
it follows by the pasting law for $\infty$-pullbacks, prop. \ref{PastingLawForPullbacks}, 
that the top left object is the $\infty$-pullback of the total rectangle diagram. 
That total diagram is 
$$
  \raisebox{20pt}{
  \xymatrix{
    \Omega \mathbf{H}(X, \mathbf{\flat}_{\mathrm{\mathrm{dR}}} \mathbf{B}^{n+1}A) 
    \ar[r]
      \ar[d]
     & 
     H(X,\mathbf{\flat}_{\mathrm{\mathrm{dR}}} \mathbf{B}^{n+1} A)
     \ar[d]
    \\
    {*} \ar[r] & \mathbf{H}(X,\mathbf{\flat}_{\mathrm{\mathrm{dR}}} \mathbf{B}^{n+1} A)
  }}
  \,,
$$
because, as before, this $\infty$-pullback is the coproduct of the 
homotopy fibers, and they are empty over the connected components not
in the image of the bottom morphism and are the loop space object
over the single connected component that is in the image. 

Finally using that
$$
  \Omega \mathbf{H}(X,\mathbf{\flat}_{\mathrm{\mathrm{dR}}} \mathbf{B}^{n+1}A) 
    \simeq 
   \mathbf{H}(X,\Omega \mathbf{\flat}_{\mathrm{\mathrm{dR}}} \mathbf{B}^{n+1}A)
$$ 
and
$$ 
  \Omega \mathbf{\flat}_{\mathrm{\mathrm{dR}}} \mathbf{B}^{n+1}A 
    \simeq 
   \mathbf{\flat}_{\mathrm{\mathrm{dR}}} \Omega \mathbf{B}^{n+1}A
$$
since both $\mathbf{H}(X,-)$ as well as $\mathbf{\flat}_{\mathrm{\mathrm{dR}}}$
preserve $\infty$-limits and hence formation of loop space objects,
the claim follows.
\endofproof

\medskip

Often it is desireable to restrict attention to differential cohomology over
domains on which the twisting cocycles can be chosen functorially. This
we consider now.

\begin{definition}
  \label{BnAconn}
    For any $n \in \mathbb{N}$ write $\mathbf{B}^n A_{\mathrm{conn}}$
	for the $\infty$-pullback
	$$
	  \xymatrix{
	    \mathbf{B}^n A_{\mathrm{conn}}
		\ar[r]
		\ar[d]
		&
		\Omega^{n+1}_{\mathrm{cl}}(-,A)
		\ar[d]
		\\
		\mathbf{B}^n A
		\ar[r]^{\mathrm{curv}}
		&
		\mathbf{\flat}_{\mathrm{dR}}\mathbf{B}^{n+1} A
	  }
	$$
	in $\mathbf{H}$.
	
	For $X$ an $A$-dR-projective object we write
	$$
	  H_{\mathrm{conn}}^n(X, A)
	  :=
	  \pi_0 \mathbf{H}(X, \mathbf{B}^n A_{\mathrm{conn}})
	$$
	for the cohomology group on $X$ with coefficients in 
	$\mathbf{B}^n A_{\mathrm{conn}}$.
\end{definition}
The objects $\mathbf{B}^n A_{\mathrm{conn}}$ represent differential cohomology
in the following sense.
\begin{observation}
  \label{IncludingAbelianDiffIntoAbelianConn}
  For every $A$-dR-projective object $X$
  there is a full and faithful morphism
  $$
    \mathbf{H}_{\mathrm{diff}}(X, \mathbf{B}^n A)
	\hookrightarrow
	\mathbf{H}(X, \mathbf{B}^n A_{\mathrm{conn}})
	\,,
  $$
  hence in particular an inclusion
  $$
    H_{\mathrm{diff}}^n(X,A)
	\hookrightarrow
	H_{\mathrm{conn}}^n(X,A)
	\,.
  $$
\end{observation}
\proof
  Since $\Omega^{n+1}_{\mathrm{cl}}(X,A) \to H^{n+1}_{\mathrm{dR}}(X,A)$
  is a surjection by definition,
  there exists a factorization
  $$
    H^{n+1}_{\mathrm{dR}}(X,A)
	\hookrightarrow
	\Omega^{n+1}_{\mathrm{cl}}(X,A)
	\to
	\mathbf{H}(X, \mathbf{\flat}_{\mathrm{dR}} \mathbf{B}^{n+1} A)
  $$
  of the canonical effective epimorphism (well defined up to homotopy),
  where the first morphism is an injection of sets, hence a monomorphism
  of $\infty$-groupoids. Since these are stable under $\infty$-pullback,
  it follows that also the top left morphism in the pasting diagram
  of $\infty$-pullbacks
  $$
    \xymatrix{
	  \mathbf{H}_{\mathrm{diff}}(X,\mathbf{B}^n A)
	  \ar[r]
	  \ar@{^{(}->}[d]
	  &
	  H^{n+1}_{\mathrm{dR}}(X,A)
	  \ar@{^{(}->}[d]
	  \\
	  \mathbf{H}(X, \mathbf{B}^n A_{\mathrm{conn}})
	  \ar[r]
	  \ar[d]
	  &
	  \Omega^{n+1}_{\mathrm{cl}}(X,A)
	  \ar[d]
	  \\
	  \mathbf{H}(X,\mathbf{B}^n A)
	  \ar[r]^{\mathrm{curv}}
	  &
	  \mathbf{H}(X, \mathbf{\flat}_{\mathrm{dR}} \mathbf{B}^{n+1}A)
	}
  $$
  is a monomorphism. 
  
  Notice that here the bottom square is indeed an $\infty$-pullback,
  by def. \ref{BnAconn}  combined with the fact that the hom-functor 
  $\mathbf{H}(X,-) : \mathbf{H} \to \infty\mathrm{Grpd}$ preserves 
  $\infty$-pullbacks, and that with the top square defined to be an
  $\infty$-pullback the total outer rectangle is an $\infty$-pullback
  by prop. \ref{PastingLawForPullbacks}. This identifies the 
  top left object as $\mathbf{H}_{\mathrm{diff}}(X, \mathbf{B}^n A)$
  by def. \ref{OrdinaryDiffCohomology}.
\endofproof
The reason that prop. \ref{IncludingAbelianDiffIntoAbelianConn}
gives in inclusion is that $H_{\mathrm{conn}}^n(X,A)$ 
contains connections for all possible curvature forms, while
$H_{\mathrm{diff}}^n(X, A)$ contains only connections for 
one curvature representative in each de Rham cohomology class.
This is made precise by the following refinement of the
exact sequences from prop. \ref{CurvatureExactSequence} and 
prop. \ref{ShortExactSequenceForIntrinsicOrdinaryDiffCohomology}.
\begin{definition}
  Write 
  $$
    \Omega^n_{\mathrm{cl}, \mathrm{int}}(-,A)
  	  \hookrightarrow
    \Omega^n_{\mathrm{cl}}(-,A)
  $$
  for the image factorization of the canonical
  morphism $\mathbf{B}^n A _{\mathrm{conn}} \to \Omega^n_{\mathrm{cl}}(-,A)$
  from def. \ref{BnAconn}.
\end{definition}
\begin{proposition}
  For $X$ an $A$-dR-projective object we have a short exact sequence
  of groups
  $$
    \xymatrix{
	  H_{\mathrm{flat}}^{n}(X,A)
	  \ar[r]
	  &
      H_{\mathrm{conn}}^n(X,A)
	  \ar[r]^{\mathrm{curv}}
	  &
	  \Omega^{n+1}_{\mathrm{cl}, \mathrm{int}}(X,A)
	}
	\,.
  $$
\end{proposition}
\proof
  As in the proof of prop. \ref{DiffCohomologyRestrictedToVanishingCurvature} 
  we have a pasting diagram of $\infty$-pullbacks
  $$
    \xymatrix{
	   {*}
	   \ar[r]
	   \ar[d]
	   &
	   \mathbf{H}(X, \mathbf{\flat}\mathbf{B}^n A)
	   \ar[r]
	   \ar[d]
	   & {*} \ar[d]^0
	   \\
	   {*}
	   \ar[r]
	   &
	   \mathbf{H}(X, \mathbf{B}^n A_{\mathrm{conn}})
	   \ar[r]
	   \ar[d]
	   &
	   \Omega^{n+1}_{\mathrm{cl}, \mathrm{int}}(X,A)
	   \ar@{^{(}->}[r]
	   &
	   \Omega^{n+1}_{\mathrm{cl}}(X,A)
	   \ar[d]
	   \\
	   &
	   \mathbf{H}(X,\mathbf{B}^n A)
	   \ar[rr]^{\mathrm{curv}}
	   &&
	   \mathbf{H}(X, \mathbf{\flat}_{\mathrm{dR}}\mathbf{B}^{n+1}A)
	}
	\,.
  $$
  After passing to connected components, this implies the claim.
\endofproof

Details on how traditional ordinary differential cohomology 
is recovered by implementing the above in 
the context of smooth cohesion are discussed in
\ref{SmoothStrucDifferentialCohomology}.

\paragraph{Differential moduli}
\label{StrucDifferentialModuli}
\label{StrucGeneralDifferentialModuli}
\label{DifferentialModuli}
\label{DifferentialModuliObject}
\label{TheCorrectU1DifferentialModuli}

We discuss issues related to the formulation of \emph{moduli objects} 
in a cohesive $\infty$-topos for 
differential cocycles as discussed above, over a fixed base object. 

To motivate this consider the following. 
Given a coefficient object $\mathbf{B}\mathbb{G}_{\mathrm{conn}} \in \mathbf{H}$
for differential cohomology as discussed above, and given any object $X \in \mathbf{H}$,
the mapping space object $[X, \mathbf{B}\mathbb{G}_{\mathrm{conn}}] \in \mathbf{H}$
is a kind of moduli object for $\mathbb{G}$-differential cocycles on $X$, in that its
global points are precisely such cocycles. However, for any $U \in \mathbf{H}$
a $U$-plot of $[X, \mathbf{B}\mathbb{G}_{conn}]$ may be more general than 
just a cohesively parameterized $U$-collection of such cocycles on $X$, because it is 
actually a differential cocycle on $U \times X$ and hence may contain nontrivial
differential/connection data along $U$, not just along $X$.

In some applications this behaviour of $[X, \mathbf{B}\mathbb{G}_{\mathrm{conn}}]$ is 
exactly what is needed. This is notably the case for the construction of 
extended Chern-Simons action functionals in all codimensions, discussed below in \ref{StrucChern-SimonsTheory}.
But in other applications, such as the construction of the extended phase spaces of
Chern-Simons functionals, one rather needs to have a 
object of genuine \emph{differential moduli}, which is such that its $U$-plots 
are genuine $U$-parameterized collections of differential cocycles (and their gauge transformations)
just on $X$. This issue is discussed in more detail with illustrative examples in the model of smooth cohesion
below in \ref{SmoothStructDifferentialModuli}.

Here we discuss how such differential moduli objects are obtained general abstractly
in a cohesive $\infty$-topos
from a \emph{degreewise concretification} of the mapping space objects 
$[X, \mathbf{B}\mathbb{G}_{\mathrm{conn}}]$
in the sense of \ref{StrucConcrete}.

\begin{definition}
  Let $\mathbb{G} \in \mathrm{Grp}(\mathbf{H})$ be a braided $\infty$-group, 
  def. \ref{BraidedInfinityGroup}, which is exactly $n-1$-truncated, def. \ref{truncated object}. 
  Then for $k \leq n+1 \in \mathbb{N}$
  write $\mathbf{B}\mathbb{G}_{\mathrm{conn}^k}$ for the $\infty$-pullback in 
  $$
    \xymatrix{
	  \mathbf{B}\mathbb{G}_{\mathrm{conn}^k}
	  \ar[r]
	  \ar[d]
	  &
	  \Omega^{n+1 \leq \bullet \leq k}(-, \mathbb{G})
	  \ar[d]
	  \\
	  \mathbf{B}\mathbb{G}
	  \ar[r]^-{\mathrm{curv}_{\mathbb{G}}}
	  &
	  \flat_{\mathrm{dR}}\mathbf{B}^2 \mathbb{G}
	}
  $$
  \label{BGconnk}
\end{definition}
\begin{remark}
  For $A$ a 0-truncated abelian group and $\mathbb{G} \simeq \mathbf{B}A$, 
  the objects $\mathbf{B}^2 A_{\mathrm{conn}^1}$ of def. \ref{BGconnk} modulates what in the 
literature is oftn known as a \emph{bundle gerbe with connective data but without curving}.
In this context then the structures modulated by 
$\mathbf{B}^2_{\mathrm{conn}^2} \simeq \mathbf{B}^2 A_{\mathrm{conn}}$
would be called \emph{bundles gerbes with connective data and with curving}.
We discuss this in more detail in \ref{SmoothStrucDifferentialCohomology} below.
\end{remark}
\begin{remark}
  The objects $\mathbf{B}\mathbb{G}_{\mathrm{conn}^k}$ of def. \ref{BGconnk} play two key roles:
  \begin{enumerate}  
    \item They appear as an ingredient in the construction of differential moduli stacks
	in def. \ref{DifferentialModuliByIteratedPullback} below. Here their role is mainly
	a technical one: the object of interest is really $\mathbf{B}\mathbb{G}_{\mathrm{conn}}$
	and the $\mathbf{B}\mathbb{G}_{\mathrm{conn}^k}$ just serve to refine its structure.
	\item They appear as variant differential coefficients in their own right in 
	 various contexts, for instance in the context of higher Atiyah Lie algebroids
	 and Courant Lie algebroids in \ref{CourantGroupoids} below. 
  \end{enumerate}
\end{remark}
\begin{remark}
  By the universal property of the $\infty$-pullback, the canonical tower of morphisms
  $$
    \xymatrix{
	  \Omega_{\mathrm{cl}}^{n+1}
	  \ar[r]
	  &
	  \Omega_{\mathrm{cl}}^{n+1 \leq \bullet \leq n}
	  \ar[r]
	  &
	  \cdots
	  \ar[r]
	  &
	  \Omega_{\mathrm{cl}}^{n+1 \leq \bullet \leq 1}
	  \ar[r]^-\simeq
	  &
      \flat_{\mathrm{dR}}\mathbf{B}^{2}\mathbb{G}
	}
  $$
  induces a tower of morphisms
  $$
    \xymatrix{
	  \mathbf{B}\mathbb{G}_{\mathrm{conn}}
	  \ar[r]^\simeq
	  &
	  \mathbf{B}\mathbb{G}_{\mathrm{conn}^n}
	  \ar[r]
	  &
	  \mathbf{B}\mathbb{G}_{\mathrm{conn}^{n-1}}
	  \ar[r]
	  &
	  \cdots
	  \ar[r]
	  &
	  \mathbf{B}\mathbb{G}_{\mathrm{conn}^0}
	  \ar[r]^\simeq
	  &
	  \mathbf{B}\mathbb{G}
	}
	\,.
  $$
  \label{CanonicalProjectionsOfRestrictedConnections}
\end{remark}
\begin{definition}
  For $X \in \mathbf{H}$ and $n \in \mathbb{N}$, $n \geq 1$, 
  $\mathbb{G} \in \mathrm{Grp}(\mathbf{H})$ a braided $\infty$-group which is 
  precisely $(n-1)$-truncated, then
  the
  \emph{moduli of $\mathbb{G}$-principal connections} on $X$ is 
  the iterated $\infty$-fiber product
  $$
    \begin{aligned}
    &\mathbb{G}\mathbf{Conn}(X)
	\\
	&
	:=
	\sharp_1 [X, \mathbf{B}\mathbb{G}_{\mathrm{conn}^n}]
	\underset{\sharp_1 [X, \mathbf{B}\mathbb{G}_{\mathrm{conn}^{n-1}}]}{\times}
	\sharp_2 [X, \mathbf{B}\mathbb{G}_{\mathrm{conn}^{n-1}}]
	\underset{\sharp_2 [X, \mathbf{B}\mathbb{G}_{\mathrm{conn}^{n-2}}]}{\times}
	\cdots
	\underset{\sharp_{n} [X, \mathbf{B}\mathbb{G}_{\mathrm{conn}^0}]}{\times}	
	[X, \mathbf{B}\mathbb{G}_{\mathrm{conn}^0}]
	\end{aligned}
	\,,
  $$
  of the morphisms
  $$
    \xymatrix{
      \sharp_k [X,\mathbf{B}\mathbb{G}_{\mathrm{conn}^{n-k+1}}]
	  \ar[r]
	  &
	  \sharp_k [X,\mathbf{B}\mathbb{G}_{\mathrm{conn}^{n-k}}]
	}
  $$ 
  which are the image under $\sharp_k$, def. \ref{ImagesOfXToSharpX},
  of the image under the internal hom $[X, -]$ of the canonical projections of
  remark \ref{CanonicalProjectionsOfRestrictedConnections}, and of the morphisms
  $$
    \xymatrix{
      \sharp_{k+1} [X,\mathbf{B}^n U(1)_{\mathrm{conn}^{n-k}}]
	  \ar[r]
	  &
	  \sharp_k [X,\mathbf{B}^n U(1)_{\mathrm{conn}^{n-k}}]
	}
  $$ 
  of def. \ref{ImagesOfXToSharpX}.
  \label{DifferentialModuliByIteratedPullback}
\end{definition}
 \begin{remark}
   By the universal property of the $\infty$-pullback, the commuting naturality diagrams
   $$
     \raisebox{20pt}{
     \xymatrix{
	   \sharp_{k_2}[X, \mathbf{B}G_{\mathrm{conn}^{n_2}}]
	   \ar[rr]^{}
	   \ar[d]
	   &&
	   \sharp_{k_2}[X, \mathbf{B}G_{\mathrm{conn}^{n_1}}]
	   \ar[d]
	   \\
	   \sharp_{k_1} [X, \mathbf{B}G_{\mathrm{conn}^{n_2}}]
	   \ar[rr]^{}
	   &&
	   \sharp_{k_1} [X, \mathbf{B}G_{\mathrm{conn}^{n_1}}]	   
	 }}
   $$
   induce a canonical projection
   $$
     \mathrm{conc}
	 :
     \xymatrix{
	    [X, \mathbf{B}G_{\mathrm{conn}}]
		\ar[r]
		&
		G\mathbf{Conn}(X)
	}
   $$
   from the mapping space object into the object of differential moduli.
   We call this \emph{differential concretification}.
   \label{ProjectionFromMappingSpaceIntoBGconnToDifferentialModuli}
   \label{DifferentialConcretificationMap}
 \end{remark}
 We need the analogous construction also for the $\mathbf{B}\mathbb{G}_{\mathrm{conn}^k}$
 regarded as coefficient objects themselves. The following straightforwardly generalizes def. 
 \ref{DifferentialModuliByIteratedPullback} from $k = n$ to arbitrary $k \leq n$.
 \begin{definition}
  For $X \in \mathbf{H}$ and $n \in \mathbb{N}$, $n \geq 1$, 
  $0 \leq k \leq n$, 
  $\mathbb{G} \in \mathrm{Grp}(\mathbf{H})$ a braided $\infty$-group which is 
  precisely $(n-1)$-truncated, then
  the
  \emph{moduli of $\mathbb{G}$-principal $k$-connections} on $X$ is 
  the iterated $\infty$-fiber product
  $$
    \begin{aligned}
    &\mathbb{G}\mathbf{Conn}_k(X)
	\\
	&
	:=
	\sharp_{n-k+1} [X, \mathbf{B}\mathbb{G}_{\mathrm{conn}^k}]
	\underset{\sharp_{n-k+1} [X, \mathbf{B}\mathbb{G}_{\mathrm{conn}^{k-1}}]}{\times}
	\sharp_{n-k+2} [X, \mathbf{B}\mathbb{G}_{\mathrm{conn}^{k-1}}]
	\underset{\sharp_{n-k+2} [X, \mathbf{B}\mathbb{G}_{\mathrm{conn}^{k-2}}]}{\times}
	\cdots
	\underset{\sharp_{n} [X, \mathbf{B}\mathbb{G}_{\mathrm{conn}^0}]}{\times}	
	[X, \mathbf{B}\mathbb{G}_{\mathrm{conn}^0}]
	\end{aligned}
	\,.
  $$
  \label{DifferentialModuliForkTruncatedConnectionDataByIteratedPullback}
\end{definition}
\begin{remark}
  The projection maps out of the iterated $\infty$-pullbacks
  induce a canonical sequence of projections  
  $$
    \xymatrix{
      \mathbb{G}\mathbf{Conn}(X)
	  \simeq
	  \mathbb{G}\mathbf{Conn}_{n}(X)
	  \ar[r]
	  &
	  \mathbb{G}\mathbf{Conn}_{n-1}(X)
	  \ar[r]
	  &
	  \cdots
	  \ar[r]
	  &
	  \mathbb{G}\mathbf{Conn}_{1}(X)
	  \ar[r]
	  &
	  \mathbb{G}\mathbf{Conn}_0(X)
	  \simeq
	  \mathbf{B}\mathbb{G}
	 }
	  \,.
  $$
  \label{CanonicalProjectionsOfModuliForTruncatedDifferentiaCocycles}
\end{remark}
 
\paragraph{Flat Differential moduli}
\label{StrucFlatDifferentialModuli}
\label{TheCorrectU1DifferentialModuliFlat}
 
We now turn to defining moduli for \emph{flat} differential cocycles. 

\medskip

\begin{definition}
  For $\mathbb{G}$ a braided $\infty$-group which is precisely
  $(n-1)$-truncated, and for any $X \in \mathbf{H}$, 
  we call the iterated $\infty$-fiber product
  $$
    \begin{aligned}
    &\mathbb{G}\mathbf{FlatConn}(X)
	\\
	&
	:=
	\sharp [X, \flat\mathbf{B}\mathbb{G}]
	\underset{\sharp [X, \Omega(\mathbf{B}\mathbb{G}_{\mathrm{conn}^{n-1}}])}{\times}
	\sharp_1 [X, \Omega(\mathbf{B}\mathbb{G}_{\mathrm{conn}^{n-1}})]
	\underset{\sharp_1 [X, \Omega(\mathbf{B}\mathbb{G}_{\mathrm{conn}^{n-2}})]}{\times}
	\cdots
	\underset{\sharp_{n} [X, \Omega(\mathbf{B}\mathbb{G}_{\mathrm{conn}^0})]}{\times}	
	[X, \mathbb{G}]
	\end{aligned}
  $$
  the \emph{moduli object for flat $\mathbb{G}$-connections on $X$}.
  \label{FlatDifferentialModuli}
\end{definition}
\begin{proposition}
  For $\mathbf{B}\mathbb{G}$ a truncated braided $\infty$-group  
  we have a natural equivalence
  $$
    \mathbb{G}\mathbf{FlatConn}\left(X\right)
	\simeq
	 \Omega_0 \left(\left(\mathbf{B}\mathbb{G}\right)\mathbf{Conn}\left(X\right)\right)
	 \,.
  $$
  Moreover, if $\mathbf{H}$ has a set of generators being concrete objects
  (in particular if is has an $\infty$-cohesive site of definition, def. \ref{CohesiveSite})
  then for $\mathbb{G}$ a 0-truncated $\infty$-group and $X$ geometrically connected
  (meaning that $\tau_0 \Pi(X) \simeq *$), we have
  $$
    \mathbb{G} \simeq \Omega_0 \left( \mathbb{G}\mathbf{Conn}(X)\right)
  $$
  \label{FlatDiffModuliAsLoops}
\end{proposition}
\proof
Since forming loops is an $\infty$-pullback operation, it commutes with the 
iterated $\infty$-fiber product. Moreover, 
by prop. \ref{ImagesCommuteWithLooping} it passes through the $\sharp_k$, while
lowering their degree by one. Finally 
by prop. \ref{LoopSpaceObjectOfBGconn} we have
$$
  \Omega \left(\mathbf{B}^2 G_{\mathrm{conn}}\right)
  \simeq
  \flat \mathbf{B}\mathbb{G}
  \,.
$$
This gives the first claim. For the second, observe that with the same reasoning we
obtain
$$
  \begin{aligned}
    \Omega \left( \mathbb{G}\mathbf{Conn}(X) \right)
	 &
	 \simeq
	 \Omega
	 \left(
	   \sharp_1[X, \mathbf{B}\mathbb{G}_{\mathrm{conn}}]
	   \underset{\sharp_1 [X, \mathbf{B}\mathbb{G}]}{\times}
	   [X, \mathbf{B}\mathbb{G}]
	 \right)
	 \\
	 & \simeq
	   \sharp[X, \flat \mathbb{G}]
	   \underset{\sharp [X, \mathbb{G}]}{\times}
	   [X, \mathbb{G}]	 
  \end{aligned}
  \,.
$$
Hence for any concrete $U \in \mathbf{H}$ we have
$$
  \begin{aligned}
    \mathbf{H}(U, \Omega( \mathbb{G}\mathbf{Conn}(X) ))
	&
	\simeq
	\infty \mathrm{Grpd}(\Gamma(U), \mathbf{H}(X, \flat \mathbb{G}))
	\underset{\infty\mathrm{Grpd}(\Gamma(U), \mathbf{H}(X, \mathbb{G}))}{\times}
	\mathbf{H}(U \times X, \mathbb{G})
	\\
	& \simeq
	\infty \mathrm{Grpd}(\Gamma(U) \times \Pi(X), \Gamma(\mathbb{G}))
	\underset{\infty \mathrm{Grpd}(\Gamma(U), \mathbf{H}(X, \mathbb{G}) ) }{\times}
	\mathbf{H}(U \times X, \mathbb{G})
	\\
	& \simeq
	\mathrm{Set}(\tau_0\Gamma(U), \Gamma(\mathbb{G}))
	\underset{\mathrm{Set}(\tau_0\Gamma(U), \mathbf{H}(X, \mathbb{G}) ) }{\times}
	\mathbf{H}(U \times X, \mathbb{G})
    \\
     & \simeq \mathbf{H}(U,\mathbb{G})	
  \end{aligned}
  \,.
$$
Here we used the defining adjunctions of cohesion and that $\mathbb{G}$ is 0-truncated by
assumption, so that $\mathbf{H}(-,\mathbb{G})$ takes values in sets. 
In the last step we used that $U$ is concrete so that maps out of it are determined by
their value on all global points of $U$. So the second but last row says in words
``those maps out of $U \times X$ which for every point of $U$ are independent of $X$''
and the last equivalence identifies that with the maps out of just $U$.
Since these equivalences are all natural in $U$ the claim follows by the 
assumption that the $U$s range over a set of generators 
(hence with the 
$\infty$-Yoneda lemma, prop. \ref{InfinityYonedaLemma}, if the $U$s range over
the objects of a site of definition).
\endofproof

\subsubsection{Chern-Weil theory}
\label{StrucChern-WeilHomomorphism}
\label{InfinityConnections}
\index{structures in a cohesive $\infty$-topos!Chern-Weil homomorphism}
  \index{Chern-Weil theory!$\infty$-Chern-Simons homomorphism}
  \index{Chern-Weil homomorphism!general abstract}

We discuss an intrinsic realization of the Chern-Weil homomomorphism 
\cite{GHV} in cohesive $\infty$-toposes.

\medskip

\begin{definition}
  \label{CurvatureCharClass}
For $G$ an $\infty$-group and 
$$
  \mathbf{c} : \mathbf{B}G \to \mathbf{B}^n A
$$
a representative of a characteristic class $[\mathbf{c}] \in H^n(\mathbf{B}G, A)$ 
we say that the composite
$$
  \mathbf{c}_{\mathrm{\mathrm{dR}}} 
   : 
   \mathbf{B}G \stackrel{\mathbf{c}}{\to}
    \mathbf{B}^n A 
     \stackrel{\mathrm{curv}}{\to}
    \mathbf{\flat}_{\mathrm{\mathrm{dR}}} \mathbf{B}^{n+1} A
$$
represents the \emph{curvature characteristic class} 
$[\mathbf{c}_{\mathrm{\mathrm{dR}}}] \in H_{\mathrm{\mathrm{dR}}}^{n+1}(\mathbf{B}G, A)$.
The induced map on cohomology
$$
  (\mathbf{c}_{\mathrm{\mathrm{dR}}})_* 
   : 
  H^1(-,G)
  \to
  H^{n+1}_{\mathrm{\mathrm{dR}}}(-,A)
$$
we call the (unrefined) \emph{$\infty$-Chern-Weil homomorphism} induced by $\mathbf{c}$.
\end{definition}
The following construction universally lifts the 
$\infty$-Chern-Weil homomorphism from taking values in 
the de Rham cohomology
to values in the differential cohomology of $\mathbf{H}$.
\begin{definition}
\index{connection!general abstract}
For $X \in \mathbf{H}$ any object, define the $\infty$-groupoid 
$\mathbf{H}_{\mathrm{conn}}(X,\mathbf{B}G)$ as the $\infty$-pullback
$$
  \xymatrix{
    \mathbf{H}_{\mathrm{conn}}(X, \mathbf{B}G) 
     \ar[r]^<<<<{({\hat {\mathbf{c}}}_i)_i}
     \ar[d]^{\eta}
     & 
     \prod\limits_{[\mathbf{c}_i] \in H^{n_i}(\mathbf{B}G,A); i \geq 1} 
     \mathbf{H}_{\mathrm{diff}}(X,\mathbf{B}^{n_i} A)
     \ar[d]
    \\
   \mathbf{H}(X, \mathbf{B}G) 
     \ar[r]^<<<<<<{({ {\mathbf{c}}}_i)_i} &
     \prod\limits_{[\mathbf{c}_i] \in H^{n_i}(\mathbf{B}G,A); i \geq 1} 
   \mathbf{H}(X,\mathbf{B}^{n_i} A)
  }
  \,.
$$
We say 
\begin{itemize}
\item
   a cocycle in $\nabla \in \mathbf{H}_{\mathrm{conn}}(X, \mathbf{B}G)$ 
   is an \emph{$\infty$-connection} 
\item
   on the principal $\infty$-bundle $\eta(\nabla)$;
\item 
  a morphism in $\mathbf{H}_{\mathrm{conn}}(X, \mathbf{B}G)$ 
  is a \emph{gauge transformation} of connections;
\item
  for each $[\mathbf{c}] \in H^n(\mathbf{B}G, A)$ the morphism
  $$
    [\hat {\mathbf{c}}] :  H_{\mathrm{conn}}(X,\mathbf{B}G) \to H_{\mathrm{diff}}^n(X, A)
  $$
  is the (full/refined) \emph{$\infty$-Chern-Weil homomorphism} induced
  by the characteristic class $[\mathbf{c}]$.
\end{itemize}
\end{definition}
\begin{observation}
Under the curvature projection 
$[F] : H_{\mathrm{diff}}^n (X,A) \to H_{\mathrm{dR}}^{n+1}(X,A)$ the refined Chern-Weil homomorphism 
for $\mathbf{c}$ projects to the unrefined Chern-Weil homomorphism.
\end{observation}
\proof
This is due to the existence of the pasting composite
$$
  \xymatrix{
    \mathbf{H}_{\mathrm{conn}}(X, \mathbf{B}G) 
     \ar[r]^<<<<{({\hat {\mathbf{c}}}_i)_i}
     \ar[d]^{\eta}
     & 
     \prod\limits_{[\mathbf{c}_i] \in H^{n_i}(\mathbf{B}G,A); i \geq 1} 
     \mathbf{H}_{\mathrm{diff}}(X,\mathbf{B}^{n_i} A)
    \ar[r]^{[F]}
    \ar[d]
     &
     \prod\limits_{[\mathbf{c}_i] \in H^{n_i}(\mathbf{B}G,A); i \geq 1} 
     H_{\mathrm{\mathrm{dR}}}^{n_i+1}(X,A)
     \ar[d]
   \\
   \mathbf{H}(X, \mathbf{B}G) 
     \ar[r]^<<<<<<{({ {\mathbf{c}}}_i)_i} &
     \prod_{[\mathbf{c}_i] \in H^{n_i}(\mathbf{B}G,A); i \geq 1} 
   \mathbf{H}(X,\mathbf{B}^{n_i} A)
    \ar[r]^{\mathrm{curv}_*}&
     \prod_{[\mathbf{c}_i] \in H^{n_i}(\mathbf{B}G,A); i \geq 1} 
    \mathbf{H}_{\mathrm{dR}}(X, \mathbf{B}^{n_i+1},A)
  }
$$
of the defining $\infty$-pullback for 
$\mathbf{H}_{\mathrm{conn}}(X,\mathbf{B}G)$ 
with the products of the definition $\infty$-pullbacks for the 
$\mathbf{H}_{\mathrm{diff}}(X, \mathbf{B}^{n_i}A)$.
\endofproof

As before for abelian differential cohomology in 
\ref{StrucDifferentialCohomology}, nonabelian differential
cohomology is in general not representable, but becomes 
representable on a suitable collection of domains. 
To reflect this we expand def. \ref{BnAconn} as follows.
\begin{definition}
  \label{BGconn}
  Let $\mathbf{c} : \mathbf{B}G \to \mathbf{B}^n A$ be a characteristic 
  map, and let $\mathbf{B}^n A_{\mathrm{conn}}$ be a differential
  refinement as in def. \ref{BnAconn}. Then we write
  $\mathbf{B}G_{\mathrm{conn}}$ for an object that fits into
  a factorization
  $$
    \xymatrix{
	  \mathbf{\flat}\mathbf{B}G
	  \ar[r]^{\mathbf{\flat}\mathbf{c}}
	  \ar[d]
	  &
	  \mathbf{\flat}\mathbf{B}^n A
	  \ar[d]
	  \\
	  \mathbf{B}G_{\mathrm{conn}}
	  \ar[r]^{\hat {\mathbf{c}}}
	  \ar[d]
	  &
	  \mathbf{B}^n A_{\mathrm{conn}}
	  \ar[d]
	  \\
	  \mathbf{B}G
	  \ar[r]^{\mathbf{c}}
	  &
	  \mathbf{B}^n A
	}
  $$
  of the naturality diagram of the $(\mathrm{Disc} \dashv \Gamma)$-counit.
\end{definition}
\begin{warning}
  The object $\mathbf{B}G_{\mathrm{conn}}$ here depends,
  in general, on the choices involved. But for the moment we find it
  convenient not to indicate this in the notation but have it be implied by
  the corresponding context. 
\end{warning}

\subsubsection{Twisted differential structures}
\label{TwistedDifferentialStructures}
\index{structures in a cohesive $\infty$-topos!twisted differential structures}

We discuss the differential refinement of \emph{twisted cohomology},
def. \ref{StrucTwistedCohomology}. Following \cite{SSSIII} we speak
of \emph{twisted differential $\mathbf{c}$-structures}.

\begin{definition}
  \label{TwistedCStructures}
  \index{twisted cohomology!twisted $\mathbf{c}$-structures}
  For $\mathbf{c} : \mathbf{B}G \to \mathbf{B}^n A$ a characteristic map
  in a cohesive $\infty$-topos  $\mathbf{H}$, define for any
  $X \in \mathbf{H}$ the
  $\infty$-groupoid $\mathbf{c}\mathrm{Struc}_{\mathrm{tw}}(X)$
  to be the $\infty$-pullback
  $$
    \xymatrix{
	  \mathbf{c}\mathrm{Struc}_{\mathrm{tw}}(X)
	  \ar[r]^{\mathrm{tw}} 
	  \ar[d]
	  & H^n(X, A)
	  \ar[d]
	  \\
	  \mathbf{H}(X, \mathbf{B}G)
	  \ar[r]^{\mathbf{c}}
	  &
	  \mathbf{H}(X, \mathbf{B}^n A)
	}
	\,,
  $$
  where the vertical morphism on the right is the essentially unique
  effective epimorphism that picks on point in every connected component.
\end{definition}
Let now $\mathbf{H}$ be a cohesive $\infty$-topos that canonically
contains the circle group $A = U(1)$, such as 
$\mathrm{Smooth}\infty \mathrm{Grpd}$ and its variants. 
Then by
\ref{SmoothStrucDifferentialCohomology} the intrinsic differential
cohomology with $U(1)$-coefficients reproduces traditional ordinary
differential cohomology and by \ref{SmoothStrucInfChernWeil}
we have models for the $\infty$-connection coefficients
$\mathbf{B}G_{\mathrm{conn}}$.
Using this we consider the differential refinement of 
def. \ref{TwistedCStructures} as follows.
\begin{definition}
  \label{twisteddifferentialcstructures}
  \index{twisted cohomology!twisted differential $\mathbf{c}$-structures}
  For $\mathbf{c} : \mathbf{B}G \to \mathbf{B}^n U(1)$
  a characteristic map as above, and for
  $\hat {\mathbf{c}} : \mathbf{B}G_{\mathrm{conn}}
  \to \mathbf{B}^n U(1)_{\mathrm{conn}}$ a differential refinement,
  we write $\hat {\mathbf{c}}\mathrm{Struc}_{\mathrm{tw}}(X)$ for the 
  corresponding twisted cohomology, def. \ref{TwistedCohomologyInOvertopos},
  $$
    \xymatrix{
      \hat {\mathbf{c}}\mathrm{Struc}_{\mathrm{tw}}(X) \ar[r]^{\mathrm{tw}}\ar[d]^{\chi} & 
       H^n_{\mathrm{diff}}(X, U(1)) \ar[d]
      \\
      \mathbf{H}(X, \mathbf{B}G_{\mathrm{conn}})
        \ar[r]^{\hat {\mathbf{c}}}
      &
      \mathbf{H}(X, \mathbf{B}^n U(1)_{\mathrm{conn}})
    }
	\,,
  $$  
  We say $\hat {\mathbf{c}}\mathrm{Struc}_{\mathrm{tw}}(X)$
  is the $\infty$-groupoid of
  \emph{twisted differential $\mathbf{c}$-structures} on $X$.
\end{definition}

\subsubsection{Higher holonomy}
\label{StrucFiberIntegration}
\index{structures in a cohesive $\infty$-topos!holonomy}
\index{holonomy!general abstract}

The notion of $\infty$-connections in a 
cohesive $\infty$-topos induces a notion of \emph{higher holonomoy}.

\medskip

\begin{definition}
 \label{CohomologicalDimensionOfAnObject}
We say an object $\Sigma \in \mathbf{H}$ has 
\emph{cohomological dimension} $\leq n \in \mathbb{N}$ 
if for all Eilenberg-MacLane objects $\mathbf{B}^{n+1}A$ the corresponding cohomology on $\Sigma$ is trivial
$$
  H(\Sigma, \mathbf{B}^{n+1}A ) \simeq *
  \,.
$$
Let $\mathrm{dim}(\Sigma)$ be the maximum $n$ for which this is true.
\end{definition}
\begin{observation}
If $\Sigma$ has cohomological dimension $\leq n$ then its de Rham cohomology,
def. \ref{IntrinsicDeRhamCohomology},
vanishes in degree $k > n$
$$
  H_{\mathrm{\mathrm{dR}}}^{k > n}(\Sigma, A) \simeq *
  \,.
$$
\end{observation}
\proof
Since $\mathbf{\flat}$ is a right adjoint it preserves delooping and 
hence $\mathbf{\flat} \mathbf{B}^k A \simeq \mathbf{B}^k \mathbf{\flat}A$. 
It follows that 
$$
  \begin{aligned}
     H_{\mathrm{\mathrm{dR}}}^{k}(\Sigma,A)
     & :=
     \pi_0 \mathbf{H}(\Sigma, \mathbf{\flat}_{\mathrm{\mathrm{dR}}} \mathbf{B}^k A)
    \\
    & \simeq
     \pi_0 \mathbf{H}(\Sigma, {*} \prod_{\mathbf{B}^k A} \mathbf{B}^k \mathbf{\flat}A)
    \\
    & \simeq
    \pi_0
     \left( 
     \mathbf{H}(\Sigma,{*}) \prod_{\mathbf{H}(\Sigma, \mathbf{B}^k A)}
      \mathbf{H}(\Sigma, \mathbf{B}^k \mathbf{\flat}A)
     \right)
   \\
   & \simeq
    \pi_0 (*)
  \end{aligned}
  \,.
$$
\endofproof
Let now $A$ be fixed as in \ref{StrucDifferentialCohomology}.
\begin{definition} 
Let $\Sigma \in \mathbf{H}$, $n \in \mathbf{N}$ with
$\mathrm{dim} \Sigma \leq n$. 
We say that the composite
$$
  \int_\Sigma 
  : 
  \xymatrix{
  \mathbf{H}_{\mathrm{flat}}(\Sigma, \mathbf{B}^n A)
   \ar[r]^-{\simeq}
   &
  \infty \mathrm{Gprd}(\Pi(\Sigma), \Pi(\mathbf{B}^n A))
   \ar[rr]^-{\tau_{\leq n-dim(\Sigma)}}
   &&
  \tau_{n-dim(\Sigma)}
    \infty \mathrm{Gprd}(\Pi(\Sigma), \Pi(\mathbf{B}^n A))
  }
$$ 
of the adjunction equivalence followed by truncation as indicated is the 
\emph{flat holonomy} operation on flat $\infty$-connections.

More generally, let
\begin{itemize}
\item $\nabla \in \mathbf{H}_{\mathrm{diff}}(X, \mathbf{B}^n A)$ 
be a differential coycle on some $X \in \mathbf{H}$
\item $\phi : \Sigma \to X$ a morphism.
\end{itemize}
Write
$$
  \phi^* : \mathbf{H}_{\mathrm{diff}}(X, \mathbf{B}^{n+1} A) 
    \to 
  \mathbf{H}_{\mathrm{diff}}(\Sigma, \mathbf{B}^n A)
   \simeq
  \mathbf{H}_{\mathrm{flat}}(\Sigma, \mathbf{B}^n A)
$$
(using prop. \ref{DiffCohomologyRestrictedToVanishingCurvature}) 
for the morphism on $\infty$-pullbacks induced by the morphism of diagrams
$$
  \raisebox{20pt}{
  \xymatrix{  
     \mathbf{H}(X, \mathbf{B}^n A) \ar[r] \ar[d]^{\phi^*}
     &
     \mathbf{H}_{\mathrm{\mathrm{dR}}}(X, \mathbf{B}^{n+1} A)
     \ar@{<-}[r]
     \ar[d]^{\phi^*}
     &
     H_{\mathrm{\mathrm{dR}}}^{n+1}(X, A)
     \ar[d]
     \\
     \mathbf{H}(\Sigma, \mathbf{B}^n A) 
     \ar[r] &
     \mathbf{H}_{\mathrm{\mathrm{dR}}}(X, \mathbf{B}^{n+1} A)
     \ar@{<-}[r]
     &
     {*}     
  }
  }
$$
The \emph{holonomomy} of $\nabla$ over $\sigma$ is the flat holonomy of $\phi^* \nabla$:
$$
  \int_\phi \nabla := \int_{\Sigma} \phi^* \nabla
  \,.
$$ 
\label{IntegrationAndHolonomy}
\end{definition}
This is a special case of the more general notion of 
transgression, \ref{StrucTransgression}.

\subsubsection{Transgression}
\label{StrucTransgression}
\index{structures in a cohesive $\infty$-topos!transgression}

We discuss an intrinsic notion of \emph{transgression} of 
differential cocycles to mapping spaces. This generalizes the 
notion of holonomy from \ref{StrucFiberIntegration} to the case of higher
codimension.

\medskip

Let $A \in \infty \mathrm{Grp}(\mathbf{H})$ be an abelian group object
and $\mathbf{B}^n A_{\mathrm{conn}}$ a differential coefficient object,
as in \ref{StrucDifferentialCohomology}, for $n \in \mathbb{N}$.

Let $\Sigma \in \mathbf{H}$ be of cohomological dimension
$k \leq n$, def. \ref{CohomologicalDimensionOfAnObject}.

\begin{definition}
  For $\hat {\mathbf{c}} : \mathbf{B}G_{\mathrm{conn}} \to \mathbf{B}^n A_{\mathrm{conn}}$
  a differentia characteristic map as in def. \ref{BGconn}, we
  say that the \emph{transgression} of $\hat {\mathbf{c}}$ to 
  $[\Sigma, \mathbf{B}G_{\mathrm{conn}}]$ is the composite
  $$
    \mathrm{tg}_\Sigma \hat {\mathbf{c}}
	:
    \xymatrix{
      [\Sigma, \mathbf{B}G_{\mathrm{conn}}]
   	    \ar[r]^{[\Sigma, \hat {\mathbf{c}}]}
		&
  	  [\Sigma, \mathbf{B}^n A_{\mathrm{conn}}]
	    \ar[r]^{}
	  &
	  \mathrm{conc}_{n-k}\tau_{n-k}
	  [\Sigma, \mathbf{B}^n A_{\mathrm{conn}}]
	}
	\,,
  $$
  where $[-,-] : \mathbf{H} \times \mathbf{H} \to \mathbf{H}$ is the 
  cartesian internal hom, where $\tau_{n-k}$ is $(n-k)$-truncation, 
  prop. \ref{PostnikovTower}, and where $\mathrm{conc}_{n-k}$ is
   $(n-k)$-concretification from def. \ref{kConcretification}.  
\end{definition}

\begin{remark}
  In the models we consider we find inclusions
  $$
	\mathbf{B}^{n-k} A_{\mathrm{conn}}
	\hookrightarrow
    \mathrm{conc}_{n-k}\tau_{n-k}
	  [\Sigma, \mathbf{B}^n A_{\mathrm{conn}}]	
	\,.
  $$
  In these cases truncation takes $A$-principal $n$-connections
  $\hat {\mathbf{c}}$
  on $\mathbf{B} G_{\mathrm{conn}}$ to $A$-principal $(n-k)$-connections
  $\mathrm{tg}_\Sigma \hat {\mathbf{c}}$ on $[\Sigma, \mathbf{B}G_{\mathrm{conn}}]$.

  In particular for $k = n$ in this case the transgression is of the form
  $$
    \mathrm{tg}_\Sigma \hat {\mathbf{c}}
	 : 
    [\Sigma, \mathbf{B}G_{\mathrm{conn}}]
	\to 
	A
	\,.
  $$  
\end{remark}

\subsubsection{Chern-Simons functionals}
\label{StrucChern-SimonsTheory}
\index{structures in a cohesive $\infty$-topos!Chern-Simons functional}
  \index{Chern-Simons functionals!general abstract}
  \index{parallel transport!holonomy (general abstract)}

Combining the refined $\infty$-Chern-Weil homomorphism,
\ref{StrucChern-WeilHomomorphism} with the 
higher holonomy, \ref{StrucFiberIntegration}, of the 
resulting $\infty$-connections produces 
a notion of higher \emph{Chern-Simons functionals} internal to 
any cohesive $\infty$-topos.
For a review of standard Chern-Simons functionals see 
\cite{FreedCS}.

\medskip

\begin{definition}
 \label{CanonicalChernSimonsFunctional}
Let $\Sigma \in \mathbf{H}$ be of cohomological dimension 
$\mathrm{dim}\Sigma = n \in \mathbb{N}$ and let $\mathbf{c} : X \to \mathbf{B}^n A$ 
a representative of a characteristic class $[\mathbf{c}] \in H^n(X, A)$ 
for some object $X$. We say that the composite
$$
  \exp(S_{\mathbf{c}}(-))
  : 
  \mathbf{H}(\Sigma, X)
   \stackrel{\hat {\mathbf{c}}}{\to}
  \mathbf{H}_{\mathrm{diff}}(\Sigma, \mathbf{B}^n A)
   \stackrel{\simeq}{\to}
  \mathbf{H}_{flat}(\Sigma, \mathbf{B}^n A)
   \stackrel{\int_\Sigma}{\to}
  \tau_{\leq 0}
    \infty \mathrm{Grpd}(\Pi(\Sigma), \Pi \mathbf{B}^n A)
$$
is the \emph{$\infty$-Chern-Simons functional} 
induced by $\mathbf{c}$ on $\Sigma$.
\end{definition}
Here  $\hat {\mathbf{c}}$ denotes the refined Chern-Weil homomorphism,
\ref{StrucChern-WeilHomomorphism},
induced by $\mathbf{c}$, and $\int_\Sigma$ is the holonomy
over $\Sigma$, \ref{StrucFiberIntegration}, 
of the resulting $n$-bundle with connection.

\begin{remark}
In the language of $\sigma$-model quantum field theory the ingredients of this definition have the following interpretation
\begin{itemize}
\item $\Sigma$ is the \emph{worldvolume of a fundamental $(\mathrm{dim}\Sigma-1)$-brane}	 ;

\item $X$ is the \emph{target space};

\item $\hat {\mathbf{c}}$ is the \emph{background gauge field} on $X$;

\item the external hom $\mathbf{H}_{\mathrm{conn}}(\Sigma,X)$ is the 
 \emph{space of worldvolume field configurations} $\phi : \Sigma \to X$ or 
   \emph{trajectories} of the brane in $X$;

\item $\exp(S_{\mathbf{c}}(\phi)) =  \int_\Sigma \phi^* \hat {\mathbf{c}}$ 
 is the value of the action functional on the field configuration $\phi$.
\end{itemize}

Traditionally, $\sigma$-models have been considered for $X$ an ordinary 
(Riemannian) manifold, or at most an orbifold, see for instance \cite{DeligneMorgan}.
The observation that it makes sense to allow target objects $X$ to be more generally
a gerbe, \ref{StrucInftyGerbes}, is
explored in \cite{PantevSharpe} \cite{HellermanSharpe}. 
Here we see that once we pass
to fully general (higher) stacks, then also all (higher) gauge theories
are subsumed as $\sigma$-models.

For if there is an $\infty$-group $G$ such that
the target space object $X$ is the moduli $\infty$-stack
of $G$-$\infty$-connections, def. \ref{BGconn},
 $X \simeq \mathbf{B}G_{\mathrm{conn}}$, 
then a ``trajectory'' $\Sigma \to X \simeq \mathbf{B}G_{\mathrm{conn}}$
is in fact a $G$-gauge field on $\Sigma$. Hence in the context of 
$\infty$-stacks, the notions of gauge theories and of $\sigma$-models
unify. 
\end{remark}
More in detail, 
assume that $\mathbf{H}$ has a canonical line object $\mathbb{A}^1$ and a natural numbers object 
$\mathbb{Z}$. Then the action functional $\exp(i S(-))$ may lift to the internal hom with 
respect to the canonical cartesian closed monoidal structure on any $\infty$-topos to a 
morphism of the form
$$
  \exp(i S_{\mathbf{c}}(-)) : [\Sigma,\mathbf{B}G_{\mathrm{conn}}] 
     \to 
  \mathbf{B}^{n-\mathrm{dim} \Sigma}\mathbb{A}^1/\mathbb{Z}
  \,.
$$
We call the internal hom $[\Sigma, \mathbf{B}G_{\mathrm{conn}}]$ the 
\emph{moduli $\infty$-stack} of field configurations on $\Sigma$ of the 
\emph{$\infty$-Chern-Simons theory} defined by $\mathbf{c}$ and $\exp(i S_\mathbf{c}(-))$ 
the action functional in codimension  $(n-\mathrm{dim}\Sigma)$ defined on it. 

A list of examples of Chern-Simons action functionals defined on moduli stacks 
obtained this way is given in \ref{SmoothStrucChernSimons}.

\subsubsection{Wess-Zumino-Witten functionals}
\label{StrucWZWFunctional}
\index{structures in a cohesive $\infty$-topos!WZW functional}
\index{Wess-Zumino-Witten functionals!general abstract}

We discuss a canonical realization of Wess-Zumino-Witten
action functionals and their higher analogs in every cohesive $\infty$-topos. 

For a review of traditional WZW functionals see for instance \cite{Gawedzki}
and see below in \ref{WZWApplications}.

\medskip

In higher (super-)differential geometry every (super-) $L_\infty$-algebra
$\mathfrak{g}$ has \emph{Lie integrations} to higher smooth (super-)groups
$G$; see \cite{FSS} for details. 
For instance, the abelian $L_\infty$-algebra $\mathbb{R}[n]$ integrates
to the \emph{circle n+1-group} $\mathbf{B}^n U(1)$. This is at the same
time the higher \emph{moduli stack} for circle $n$-bundles 
(also called $(n-1)$-bundle gerbes). 

\medskip
Recall then from the Introduction that 
 a perturbative higher WZW model of dimension $n$ is all encoded by a 
morphism of \mbox{(super-)}$L_\infty$-algebras of the form
$$
  \mu \;:\; \xymatrix{ \mathfrak{g} \ar[r] &  \mathbb{R}[n]}
  \,.
$$
Therefore, its non-perturbative refinement is to be an $n$-form connection on 
a circle $n$-bundle over the higher group $G$. The latter is given by a morphism of 
higher smooth \mbox{(super-)}groups the form
$$
  \Omega \mathbf{c}
  :
  \xymatrix{
    G \ar[r] & \mathbf{B}^n U(1)
  }
  \,.
$$
(This is the higher and smooth analog of the canonical morphism $G\to K(\mathbb{Z},3)$ defining the fundamental class $[\omega_G]\in H^3(G;\mathbb{Z})$ for a compact, simple 
and simply connected Lie group $G$, in the traditional WZW model.)
Equivalently, this is a morphism of the corresponding delooping stacks
$$
  \mathbf{c}
  :
  \xymatrix{
    \mathbf{B}G \ar[r] & \mathbf{B}^{n+1} U(1)
  }\;.
$$
It is shown in \cite{FSS} that this always and canonically exists,
it is just the Lie integration $\mathbf{c} = \exp(\mu)$ of the original
$L_\infty$-cocycle.
\footnote{Here and in the following we use $U(1) = \mathbb{R}/\mathbb{Z}$ for brevity,
but in general what appears is $\mathbb{R}/\Gamma$, for $\Gamma\hookrightarrow \mathbb{R}$
the discrete subgroup of \emph{periods} of $\mu$; see \cite{FSS} for details.}

\medskip
Now, as indicated in the Introduction, 
the local Lagrangian for the non-perturbative WZW model is to be an
\emph{$n$-connection} on this $n$-bundle whose curvature $n+1$-form
is $\mu(\theta_{\mathrm{global}})$, the value of the original cocycle
applied to a \emph{globally defined} Maurer-Cartan form on $G$.
Every higher group in cohesive geometry 
does carry a higher Maurer-Cartan form 
(see also \cite{hgp}), given by a canonical map
$\theta_G : G \to \flat_{\mathrm{dR}}\mathbf{B}G$
with values in the (nonabelian) \emph{de Rham hypercohomology} stack
$\flat_{\mathrm{dR}}\mathbf{B}G$. 
Exactly as $[\omega_G]$ for a Lie group 
is represented by the closed left-invariant 3-form $\omega_G=\mu(\theta_G\wedge\theta_G\wedge\theta_G)$, where $\theta_G$ is the Maurer-Cartan form of $G$, the morphism $\mathbf{\Omega}\mathbf{c}$ has a canonical factorization
\[
\xymatrix{
   G
     \ar[drr]_{\mathbf{\Omega}\mathbf{c}}\ar[r]^-{\theta_G}
	 &
	 \flat_{\mathrm{dR}}\mathbf{B}G
	 \ar[r]^-{\flat_{\mathrm{dR}}\mathbf{c}}
	 &
	 \flat_{\mathrm{dR}}
	 \mathbf{B}^{n+1} U(1)\\
&& \mathbf{B}^{n}U(1)\;,
\ar[u]_{\mathrm{curv}}}
\]
where
$\flat_{\mathrm{dR}}\mathbf{B}G$ and $\flat_{\mathrm{dR}}\mathbf{B}^nU(1)$ are the higher smooth stacks of flat $G$-valued and of flat $\mathbf{B}^{n}U(1)$-valued differential forms, respectively, $\theta_G$ is the Maurer-Cartan form, and 
$\mathrm{curv}:\mathbf{B}^{n}U(1)\to \flat_{\mathrm{dR}}\mathbf{B}^{n+1} U(1)$ is the canonical curvature morphism (see \cite{FSS,hgp} for details).

\medskip
There is, however, a fundamental difference between the general case of a higher smooth group and the classical case of a compact Lie group. Namely, the higher Maurer-Cartan form $\theta_G:\mathbf{B}G\to \flat_{\mathrm{dR}}\mathbf{B}G$ will not, in general, be represented by a globally defined flat differential form with coefficients in the $L_\infty$-algebra $\mathfrak{g}$. In other words, we do not have, in general, a factorization
\[
\xymatrix{
& \Omega^1_{\mathrm{flat}}(-;\mathfrak{g})\ar[d]\\
G\ar@{..>}[ru]\ar[r]^{\theta_G}&\flat_{\mathrm{dR}}\mathbf{B}G
}
\] 
as in the case of compact Lie groups. 
Rather, in general $\theta_G$ is a genuine hyper-cocycle: a collection 
of local differential forms on an atlas for $G$, with gauge transformations where their
domain of definition overlaps and higher gauge transformations 
on higher intersections. 
The universal way to force a globally defined curvature form is 
to consider the smooth stack $\tilde{G}$ which is the universal solution to the above factorization problem. That is, we consider the (higher) smooth stack $\tilde{G}$ 
defined as the following homotopy pullback
\[
\xymatrix{
\tilde{G}\ar[rr]^-{\theta_{\mathrm{global}}} \ar[rr]_<{\rfloor}\ar[d]&& \Omega^1_{\mathrm{flat}}(-;\mathfrak{g})\ar[d]\\
G\ar[rr]^-{\theta_G}&&\flat_{\mathrm{dR}}\mathbf{B}G
}
\]
in higher supergeometric smooth stacks.
In conclusion then the
non-perturbative WZW-model induced by the cocycle $\mu$ is to be
an $n$-connection local Lagrangian of the form
$$
  \mathcal{L}_{\mathrm{WZW}}
  :
  \xymatrix{
    \tilde G \ar[r] & \mathbf{B}^n U(1)_{\mathrm{conn}}\;,
  }
$$
satisfying two conditions:
\begin{enumerate}
  \item its curvature $(n+1)$-form is the evaluation of $\mu$ on the globally
  defined Maurer-Cartan form;
  \item the underlying $n$-bundle is the higher group cocycle $\Omega\mathbf{c}$ 
  given by Lie integration of $\mu$.
\end{enumerate}
The following proposition now asserts that this indeed exists canonically and
is essentially uniquely. 
\begin{proposition}
  On $\tilde G$ there is an essentially unique factorization
  of the globally defined invariant form $\mu(\theta_{\mathrm{global}})$
  through an extended WZW action functional $\mathcal{L}_{\mathrm{WZW}}$
  $$
    \xymatrix{
	  \tilde G
	  \ar@{-->}[dr]_{\mathcal{L}_{\mathrm{WZW}}}
	  \ar[r]^-{\theta_{\mathrm{global}}}
	  &
	  \Omega_{\mathrm{flat}}(-,\mathfrak{g})
	  \ar[r]^{\mu}
	  &
	  \Omega^{n+1}_{\mathrm{cl}}
	  \\
	  & \mathbf{B}^n U(1)_{\mathrm{conn}}
	  \ar[ur]_{F_{(-)}}\;,
	}
  $$
  such that the underlying smooth class $G \to \mathbf{B}^n U(1)$
  is the looping of the exponentiated cocycle $\mathbf{c} = \exp(\mu)$.
  \label{ConstructionOfTheFullWZWTerm}
\end{proposition}
\proof
One considers the smooth stacks 
$\flat\mathbf{B}G$ and $\flat\mathbf{B}^{n+1}U(1)$ 
of  $G$-principal bundles and $U(1)$-principal $(n+1)$-bundles with flat connections, respectively, together with the canonical morphisms 
$\flat_{\mathrm{dR}}\mathbf{B}G\to \flat\mathbf{B}G$ and 
$\flat_{\mathrm{dR}}\mathbf{B}^{n+1} U(1)\to \flat\mathbf{B}^{n+1} U(1)$ (again, see \cite{FSS,hgp} for definitions). By naturality of these morphisms one has a homotopy commutative diagram
of the form
\[
\xymatrix{
\flat_{\mathrm{dR}}\mathbf{B}G\ar[r]^-{\flat_{\mathrm{dR}}\mathbf{c}}\ar[d]&\flat_{\mathrm{dR}}\mathbf{B}^{n+1}U(1)\ar[d]\\
\flat\mathbf{B}G\ar[r]^-{\flat\mathbf{c}}&\mathbf{B}^{n+1}U(1)\;.
}
\]
Then, by naturally of the inclusions 
$\Omega^1_{\mathrm{flat}}(-;\mathfrak{g})\to \flat_{\mathrm{dR}}\mathbf{B}G$ and $\Omega^{n+1}_{\mathrm{cl}}=\Omega^1_{\mathrm{flat}}(-;\mathbb{R}[n])\to \flat_{\mathrm{dR}}
\mathbf{B}^{n+1} U(1)$, one has a homotopy commutative diagram
\[
\xymatrix{
\Omega^1_{\mathrm{flat}}(-;\mathfrak{g})\ar[r]^{\mu}\ar[d]&\Omega^{n+1}_{\mathrm{cl}}\ar[d]\\
\flat_{\mathrm{dR}}\mathbf{B}G\ar[r]^-{\flat_{\mathrm{dR}}\mathbf{c}}&\flat_{\mathrm{dR}}
 \mathbf{B}^{n+1} U(1)\;.
}
\]
Finally, since by definition $\flat_{\mathrm{dR}}\mathbf{B}G$ is the homotopy fiber of the forgetful morphism $
\flat\mathbf{B}G\to \mathbf{B}G$, we have a homotopy pullback diagram of the form
\[
\xymatrix{
\hskip-1emG\simeq \mathbf{\Omega}\mathbf{B}G\ar[r]\ar[d]&\flat_{\mathrm{dR}}\mathbf{B}G\ar[d]\\
\ast\ar[r]&\flat\mathbf{B}G\;.
}
\]
Pasting together the above three diagrams 
and the homotopy commutative diagram defining $\tilde{G}$ we obtain the big homotopy commutative diagram
\[
\xymatrix{
	  && \tilde G 
	  \ar[dl]
	  \ar[dr]^{\theta_{\mathrm{global}}}
	  \\
	  & G 
	  \ar[dl]
	  \ar[dr]^-\theta
	  \ar[dl]
	  &&
	  \Omega^1_{\mathrm{flat}}(-,\mathfrak{g})
	  \ar[dl]
	  \ar[dr]^{\mu}
	  \\
	  \ast
	  \ar[dr]
	  &&  \flat_{\mathrm{dR}}\mathbf{B}G
	  \ar[dl]
	  \ar[dr]^{\flat_{\mathrm{dR}}\mathbf{c}}
	  &&
	  \Omega^{n+1}_{\mathrm{cl}}\;,
	  \ar[dl]
	  \\
	  & \flat \mathbf{B}G
	  \ar[dr]_{\flat \mathbf{c}}
	  &&
	  \flat_{\mathrm{dR}}\mathbf{B}^{n+1}U(1)	  \ar[dl]
	  \\
	  && 
	  \flat \mathbf{B}^{n+1}U(1)
	}
\]
and hence the homotopy commutative diagram
\[
\xymatrix{
	  & \tilde G 
	  \ar[dl]
	  \ar[dr]^{\mu(\theta_{\mathrm{global}})}
	  \\
	   \ast 
	  \ar[dr]_{0}
	  &&
	  \Omega^{n+1}_{\mathrm{cl}}
	  \ar[dl]
	  \\
	  &  \flat\mathbf{B}^{n+1} U(1)
	 	}
	\,
\]
as the outermost part of the above big diagram. Then, by the universal property of the homotopy pullback, this factors essentially uniquely as 
  $$
    \raisebox{49pt}{
    \xymatrix{
	  & \tilde G
	    \ar@/_1pc/[ddl]
	    \ar@/^1pc/[ddr]^{\mu(\theta_{\mathrm{global}})}
		\ar@{-->}[d]^{\mathcal{L}_{\mathrm{WZW}}}
	  \\
	  & \mathbf{B}^n U(1)_{\mathrm{conn}}
	  \ar[dl]
	  \ar[dr]^{F_{(-)}}_{\ }="s"
	  \\
	  \ast 
	    \ar[dr]_0^{\ }="t"
	  && \Omega^{n+1}_{\mathrm{cl}}\;,
	  \ar[dl]
	  \\
	  & \flat \mathbf{B}^{n+1}U(1)
	  \ar@{=>} "s"; "t"
	}
	}
	  $$
where we have used the fact that the stack $\mathbf{B}^nU(1)_{\mathrm{conn}}$ of $U(1)$-$n$-bundles with connection is naturally the homotopy fiber of the inclusion $\Omega^{n+1}_{\mathrm{cl}}\to \flat \mathbf{B}^{n+1}U(1)$; see \cite{FSS}.
\endofproof  

\begin{remark}
The above proposition has been stated having in mind a cocycle with integral periods, so that $\mathbb{R}/\mathbb{Z}\cong U(1)$. The generalization to an arbitrary subgroup of periods $\Gamma\hookrightarrow \mathbb{R}$ is immediate. 
\end{remark}  

 \begin{remark}
  The construction of the full higher WZW term 
  $\mathcal{L}_{\mathrm{WZW}}$ in Prop. \ref{ConstructionOfTheFullWZWTerm}
  turns out to canonically exhibit the higher WZW-type theory as the boundary theory
  of a higher Chern-Simons-type theory, in the precise sense of 
 Def. Prop. \ref{BoundaryConditions}.  
  To see this, first recall  that 
  an $(n+1)$-dimensional local Chern-Simons-type prequantum field theory
  for a cocycle $\mathbf{c} : \mathbf{B}G \to \mathbf{B}^{n+1}U(1)$
  as above is a map of smooth higher moduli stacks of the form
  $$
    \mathcal{L}_{\mathrm{CS}} : \mathbf{B}G_{\mathrm{conn}} 
	\to \mathbf{B}^{n+1}U(1)_{\mathrm{conn}}
  $$
  which fits into a homotopy commutative diagram of the form
  $$
    \raisebox{30pt}{
    \xymatrix{
	  \flat \mathbf{B}G \ar[rr]^{\flat \mathbf{c}} 
	  \ar[d] && \flat\mathbf{B}^{n+1}U(1) \ar[d]
	  \\
	  \mathbf{B}G_{\mathrm{conn}}
	  \ar[d]
	  \ar[rr]^-{\mathcal{L}_{\mathrm{CS}}}
	  &&
	  \mathbf{B}^{n+1}U(1)_{\mathrm{conn}}
	  \ar[d]
	  \\
	  \mathbf{B}G
	  \ar[rr]^-{\mathbf{c}}
	  &&
	  \mathbf{B}^{n+1}U(1)\;.
	}
	}
  $$
This  hence is a refinement to differential cohomology that respects both the inclusion
  of flat higher connections as well as the underlying universal principal 
  $n$-bundles. In \cite{FSS} is given a general construction 
  of such $\mathcal{L}_{\mathrm{CS}}$ by a stacky/higher version of Chern-Weil theory,
  which applies whenever the cocycle $\mu$ is in transgression with an 
  invariant polynomial on the $L_\infty$-algebra $\mathfrak{g}$. 
  For instance ordinary 3d Chern-Simons
  theory is induced this way from the transgressive 3-cocycle 
  $\langle -,[-,-]\rangle$ on a semisimple Lie algebra, and the 
  nonabelian 7d Chern-Simons theory on String 2-connections 
  which appears in quantum corrected 11d supergravity is induced
  by the corresponding 7-cocycle \cite{FiorenzaSatiSchreiberI}.
  
  \medskip
  Now by pasting this 
  diagram 
  below 
  the diagram 
$$
  \raisebox{30pt}{
  \xymatrix{
     \tilde G \ar[r]^-{\theta_{\mathrm{global}}}_<{\rfloor} \ar[d] 
	 & \Omega^1_{\mathrm{flat}}(-,\mathfrak{g}) 
	 \ar[d]
	 \ar[rr]^-{\mu}
	 &&
	 \Omega^{n+1}_{\mathrm{cl}}
	 \ar[d]
	 \\
	 G \ar[r]^-{\theta_G}
	 & \flat_{\mathrm{dR}}\mathbf{B}G
	 \ar[rr]^-{\flat_{\mathrm{dR}} \mathbf{c}}
	 &&
	 \flat_{\mathrm{dR}}\mathbf{B}^{n+1} U(1)
  }
  }
$$
appearing
  in the proof of Prop. \ref{ConstructionOfTheFullWZWTerm} 
  we obtain the homotopy commutative diagram
  of smooth higher moduli stacks
  $$
  \raisebox{30pt}{
  \xymatrix{
     \tilde G \ar[r]^-{\theta_{\mathrm{global}}}_<{\rfloor} \ar[d] 
	 & \Omega^1_{\mathrm{flat}}(-,\mathfrak{g}) 
	 \ar[d]
	 \ar[rr]^-{\mu}
	 &&
	 \Omega^{n+1}_{\mathrm{cl}}
	 \ar[d]
	 \\
	 G \ar[r]^-{\theta_G}_<{\rfloor} 
	 \ar[d]
	 & \flat_{\mathrm{dR}}\mathbf{B}G
	 \ar[rr]^-{\flat_{\mathrm{dR}} \mathbf{c}} 
	 \ar[d]
	 &&
	 \flat_{\mathrm{dR}}\mathbf{B}^{n+1} U(1)
	 \ar[d]
	 \\
	 \ast \ar[r] 
	 \ar[d]
	 &
     \flat \mathbf{B}G	 
	 \ar[rr]^-{\flat \mathbf{c}}
	 \ar[d]
	 &&
	 \flat \mathbf{B}^{n+1} U(1)
	 \ar[d]
	 \\
	 \ast 
	 \ar[r]
	 &
	 \mathbf{B}G_{\mathrm{conn}}
	 \ar[rr]^-{\mathcal{L}_{\mathrm{CS}}}
	 &&
	 \mathbf{B}^{n+1}U(1)_{\mathrm{conn}}\;.
  }
  }
$$
\begin{remark}
Inside the above diagram 
one reads the 
correspondence 
$$
  \raisebox{20pt}{
  \xymatrix{
    & \tilde G
	\ar[dl]
	\ar[dr]^{\theta_{\mathrm{global}}}_{\ }="s"
	\\
	\ast 
	\ar[dr]_0^{\ }="t"
	&& 
	\mathbf{B}G_{\mathrm{conn}}\;,
	\ar[dl]^{\mathcal{L}_{\mathrm{CS}}}
	\\
	& \mathbf{B}^{n+1} U(1)_{\mathrm{conn}}
	\ar@{=>}^{\mathcal{L}_{\mathrm{WZW}}} "s"; "t"
  }
  }
$$
which equivalently expresses the higher WZW term as a cocycle in 
degree $n$ differential cohomology twisted by the 
Chern-Simons term 
evaluated on the
globally defined Maurer-Cartan
form. 
  According to definition \ref{BoundaryConditions} this precisely exhibits $\mathcal{L}_{\mathrm{WZW}}$
as a boundary condition for $\mathcal{L}_{\mathrm{CS}}$.
\label{WZWAsBoundaryofCS}
\index{Wess-Zumino-Witten functionals!as boundaries of Chern-Simons field theories}
\end{remark}
 \medskip
  This general mathematical statement seems to be well in 
  line with the relation between higher Chern-Simons terms and
  higher WZW models found in \cite{Witten}.
  Notice that with $\mathcal{L}_{\mathrm{WZW}}$ realized as a boundary theory
  of $\mathcal{L}_{\mathrm{CS}}$
  this way, any further boundary of $\mathcal{L}_{\mathrm{WZW}}$,
  notably as in Def. \ref{BoundaryCondition}, makes that a \emph{corner}
  of $\mathcal{L}_{\mathrm{CS}}$. In fact, in \cite{lpqft} is
  shown that $\mathcal{L}_{\mathrm{CS}}$ itself is already naturally
  a boundary theory for a topological field theory of yet one dimension
  more, namely a universal higher topological Yang-Mills theory.
  Hence we find here a whole cascade of \emph{corner field theories}
  of arbitrary codimension. For instance from the results above we have
  the sequence of higher order corner theories that looks like
  $$
    \xymatrix{
	  \mbox{M2-brane}
	  ~\ar@{^{(}->}[rr]^-{\mbox{\tiny ends on}} && 
	  \mbox{M5-brane}
	  ~\ar@{^{(}->}[rrr]^-{\mbox{\tiny WZW boundary of}} &&& \mbox{7d CS in 11d Sugra}
	  ~\ar@{^{(}->}[rr]^-{\mbox{\tiny boundary of}}
	  && 
	  \mbox{8d tYM}
	}
	\,.
  $$
  Such hierarchies of higher order corner field theories have 
  previously been recognized and amplified in string theory and M-theory 
  \cite{SatiC, F}. More discussion of the above formalization of these
  hierarchies in local (multi-tiered) prequantum field theory is in \cite{lpqft}.
  Closely related considerations have appeared in \cite{Freed432}.
\end{remark}
To further appreciate the abstract construction of the higher
WZW term $\mathcal{L}_{\mathrm{WZW}}$ in Prop. \ref{ConstructionOfTheFullWZWTerm}, 
it is helpful
to notice the following two basic examples, which are in a way
at opposites ends of the space of all examples.
\begin{example}
  For $\mathfrak{g}$ an ordinary (super-)Lie algebra and $G$ an ordinary 
  (super-)Lie group integrating it, we have 
  $\flat_{\mathrm{dR}}\mathbf{B}G \simeq \Omega^1_{\mathrm{flat}}(-,\mathfrak{g})$. 
  This implies that in this case $\tilde G \simeq G$, 
  hence that there is no extra ``differential extension''. 
  Now for $\mu$ a 3-cocycle, the induced $\mathcal{L}_{\mathrm{WZW}}$
  is the traditional WZW term, refined to a Deligne 2-cocycle/bundle gerbe
  with connection as in \cite{Ga, FreedWitten}.
  \label{FullWZWOnLieGroup}
\end{example}
\begin{example}
  For $\mathfrak{g} = \mathbb{R}[n]$ we can take the smooth higher group
  integrating it to be the $(n+1)$-group 
  $G = \mathbf{B}^nU(1)$. 
  In this case the definition of $\tilde G$ is precisely the 
  characterization of the moduli $n$-stack of $U(1)$-$n$-bundles with connections,
  by def. \ref{SmoothDiffCohWithFormTwist}, so that
  $$
    \tilde G \simeq \mathbf{B}^n U(1)_{\mathrm{conn}}
	\,
  $$
  in this case.
  Then for $\mu : \mathfrak{g} \to \mathbb{R}[n]$
  the canonical cocycle (the identity), it follows that 
  $\mathcal{L}_{\mathrm{WZW}}$ is the identity, hence is the 
  canonical $U(1)$-$n$-connection on the moduli $n$-stack of all
  $U(1)$-$n$-connections.
  This describes the extreme case of a higher WZW-type field theory with 
  \emph{no} $\sigma$-model fields and \emph{only} a ``tensor field''
  on its worldvolume, and whose action functional is simply the 
  higher volume holonomy of that higher gauge field.
  \label{FullWZWOnCirclenGroup}
\end{example}
Generic examples of higher WZW theories are twisted products
of the above two basic examples:
\begin{example}
  Consider $K$ a higher (super-)group extension of a Lie (super-)group $G$ of the form
  $$
    \xymatrix{
	  \mathbf{B}^n U(1)
	  \ar[r]
	  &
	  K
	  \ar[r]
	  &
	  G
	}
	\,.
  $$
  For instance $G$ may be a translation super-group $\mathbb{R}^{d;N}$
  and $K$ the Lie integration of one of the extended superspaces
  such as $\mathfrak{m}2\mathfrak{brane}$ considered above 
  (spacetime filled with a brane condensate, Remark \ref{BraneCondensates}).
  This means that $K$ is a \emph{twisted product}  of 
  the (super-)Lie group $G$ and the $(n+1)$-group $\mathbf{B}^n U(1)$, 
  which appear in examples \ref{FullWZWOnLieGroup} and
  \ref{FullWZWOnCirclenGroup} above. Since the construction of
  $\mathcal{L}_{\mathrm{WZW}}$ in the proof of Prop. \ref{ConstructionOfTheFullWZWTerm}
  suitably respects products, it follows that 
  the field content of a higher WZW model on the higher smooth (super-)group $K$ is
  a tuple consisting of
  \begin{enumerate}
    \item a $\sigma$-model field with values in $G$;
	\item an $n$-form higher gauge field,
  \end{enumerate}
  both subject to a twisting condition which gives the
  higher gauge field a twisted Bianchi identity depending on the 
  $\sigma$-model fields.
  
  \medskip
  In particular, for the extended spacetime given by an M2-brane condensate in 
  11-dimensional $(N=1)$-super spacetime, this says that the M5-brane
  higher WZW model according to Section \ref{The5brane}
  has fields given by a multiplet consisting of embedding fields into
  spacetime and a 2-form higher gauge field (``tensor field'') on its
  worldvolume. Notice that the higher gauge transformations of the 2-form
  field are correctly taken into account by this 
  full (in particular non-perturbative) construction of the WZW term
  as a higher prequantum bundle.
\end{example}

\subsubsection{Prequantum geometry}
\label{StrucGeometricPrequantization}
\label{HigherDifferentialGeometryInIntroduction}
 \index{structures in a cohesive $\infty$-topos!geometric prequantum theory}
 \index{symplectic higher geometry!geometric prequantization}
 \index{prequantum geometry!general abstract}

Traditional \emph{prequantum geometry} (see for instance \cite{EE} for a standard account) 
is the differential geometry of smooth manifolds
which are ``twisted'' by circle-principal bundles and circle-principal connections
-- thought of as
``prequantum bundles'' -- or equivalently is the \emph{contact geometry} \cite{Etnyre} of the 
total spaces of these bundles thought of as \emph{regular contact manifolds} \cite{BoothbyWang}. 
Prequantum geometry studies the automorphisms of prequantum bundles covering diffeomorphisms of the base
-- the \emph{prequantum operators} -- and the action of these on the space of sections
of the associated line bundle -- the \emph{prequantum states}. This is an intermediate
step in the genuine \emph{geometric quantization} of the curvature 2-form of these bundles,
which is obtained by dividing the above data in half, important for instance in the \emph{orbit method}. 
But prequantum geometry is of interest already in its own right. 
For instance the above automorphism group naturally provides the Lie integration
of the \emph{Poisson Lie algebra} of the underlying symplectic manifold. Moreover, it is canonically
included into the group of bisections of the Lie integration of the 
Atiyah Lie algebroid of the given circle bundle.

We now formulate \emph{geometric prequantum theory}
internally to any cohesive $\infty$-topos to obtain \emph{higher prequantum geometry}.

This section draws from \cite{hgp}.

\medskip

\begin{itemize}
  \item \ref{PrequantumGeometrySurvey} -- Introduction and Survey
  \item \ref{Prequantization} -- Prequantization;
  \item \ref{Symplectomorphisms} -- Symplectomorphism group;
  \item \ref{ContactTransformations} -- Contactomorphism group;
  \item \ref{QuantomorphismGroup} -- Quantomorphism group and Heisenberg group;
  \item \ref{CourantGroupoids} -- Courant Lie algebroid;
  \item \ref{PrequantumStatesInCohesion} -- Prequantum states;
  \item \ref{PrequantumOperatorsInCohesion} -- Prequantum operators.
\end{itemize}

\paragraph{Introduction and survey}
\label{PrequantumGeometrySurvey}
\label{TraditionalPrequantumGeometryViaSlicing}

Traditional prequantum geometry  
is the differential geometry of smooth manifolds
which are equipped with a {twist}  
in the form of a $U(1)$-principal bundle with a $U(1)$-principal connection.
(See section II of \cite{BrylinskiLoop} for a modern account.)
In the context of {geometric quantization} \cite{GQ}
of symplectic manifolds these 
arise as \emph{prequantizations} (whence the name): lifts of the symplectic 
form from de Rham cocycles to {differential cohomology}. 
Equivalently, prequantum geometry is the \emph{contact geometry} of the 
total spaces of these bundles, equipped with their Ehresmann connection 1-form \cite{BoothbyWang}. 
Prequantum geometry studies the automorphisms of prequantum bundles covering diffeomorphisms of the base
-- the {prequantum operators} or {contactomorphisms} -- and the action of these on the space of sections
of the associated line bundle -- the {prequantum states}. This is an intermediate
step in the genuine {geometric quantization} of symplectic manifolds,
which is obtained by ``dividing the above data in half'' by a choice of polarization. 
While polarizations do play central role in geometric quantum theory, 
for instance in the orbit method in geometric representation theory 
\cite{Kirillov}, to name just one example,
geometic prequantum theory is of interest in its own right. 
For instance the quantomorphism group naturally provides a non-simply connected Lie integration
of the {Poisson bracket} Lie algebra of the underlying symplectic 
manifold and the pullback of this extension along Hamiltonian actions
induces central extensions of infinite-dimensional Lie groups (see for instance \cite{RS,Viz}).
Moreover, the quantomorphism group comes equipped with a canonical
injection into the group of bisections of the groupoid which integrates the  
{Atiyah Lie algebroid} associated with the given principal bundle
(this we discuss below in \ref{HigherCourantQuantomorphimsGroupoids}). 
These are fundamental objects in the study of principal bundles over manifolds.

We observe now that all this has a simple natural reformulation in terms of the maps into 
the {smooth moduli stacks}  that classify 
-- better: {modulate} -- principal bundles and principal connections. 
This reformulation exhibits an abstract characterization of 
prequantum geometry which immediately generalizes 
to higher geometric contexts richer than traditional differential geometry.

\medskip

To start with, if we write $\Omega^2_{\mathrm{cl}}$
for the sheaf of smooth closed differential 2-forms (on the site of all smooth manifolds), then
by the Yoneda lemma a closed (for instance symplectic) 2-form $\omega$ on a smooth manifold $X$ is
equivalently a map of sheaves
$
  \omega : \xymatrix{
    X \ar[r] & \Omega^2_{\mathrm{cl}}
  }
$.
It is useful to think of this as a simple first instance of moduli stacks:
$\Omega^2_{\mathrm{cl}}$ is the {universal moduli stack of smooth closed 2-forms}.

Similarly but more interestingly, there is a smooth moduli stack 
of circle-principal connections, def. \ref{SmoothDiffCohWithFormTwist}. 
This we denote by
$\mathbf{B}U(1)_{\mathrm{conn}}$ in order to indicate that it is a differential 
refinement of the universal moduli stack $\mathbf{B}U(1)$ of just $U(1)$-principal connections,
which in turn is a smooth refinement of the traditional classifiying space 
$B U(1) \simeq K(\mathbb{Z},2)$ of just equivalence classes of such bundles.
Hence $\mathbf{B}U(1)_{\mathrm{conn}}$ is the ``smooth homotopy 1-type'' which is
uniquely characterized by the fact that maps $X \to \mathbf{B}U(1)_{\mathrm{conn}}$
from a smooth manifold $X$ are equivalently circle-principal connections on $X$, and that
homotopies between such maps are equivalently smooth gauge transformations between such
connections. This is a refinement of $\Omega^2_{\mathrm{cl}}$:
the map which sends a circle-principal connection to its curvature 2-form constitutes
a map of universal moduli stacks 
$F_{(-)} : \xymatrix{ \mathbf{B}U(1)_{\mathrm{conn}} \ar[r] & \Omega^2_{\mathrm{cl}}}$, 
hence a universal invariant 2-form on $\mathbf{B}U(1)_{\mathrm{conn}}$.
This {universal curvature form} characterizes traditional prequantization: 
for $\omega \in \Omega^2_{\mathrm{cl}}(X)$ 
a (pre-)symplectic form as above, a \emph{prequantization} of $(X,\omega)$ is equivalently 
a lift $\nabla$ in the diagram
$$
  \raisebox{20pt}{
    \xymatrix{
      X \ar@{-->}[rr]^\nabla \ar[dr]_{\omega} && \mathbf{B}U(1)_{\mathrm{conn}} \ar[dl]^{F_{(-)}}
  	  \\
	  & \Omega^2_{\mathrm{cl}}
    }}
  \,,
$$
where the commutativity of the diagram expresses the traditional prequantization condition
$\omega = F_{\nabla}$.

A triangular diagram as above may naturally be interpreted as exhibiting a map 
\emph{from $\omega$ to $F_{(-)}$ in the slice topos over $\Omega^2_{\mathrm{cl}}$}.
This means that the map $F_{(-)}$ is itself a universal moduli stack
-- the universal moduli stack of prequantizations. As such, $F_{(-)}$ lives not
in the topos over all smooth manifolds, but in its {slice} over $\Omega^2_{\mathrm{cl}}$,
which is the topos of smooth stacks equipped with a map into $\Omega^2_{\mathrm{cl}}$.

Now given a prequantization $\nabla$, then a \emph{quantomorphism} 
or \emph{integrated prequantum operator} is traditionally defined to be 
a pair $(\phi, \eta)$, consisting of a 
diffeomorphism $\phi : \xymatrix{X \ar[r]^\simeq & X}$ together with an equivalence 
of prequantum connections $\eta : \xymatrix{\phi^* \nabla \ar[r]^\simeq & \nabla}$. A moment of reflection
shows that such a pair is equivalently again a triangular diagram, now as on the right of
$$  
  \mathbf{QuantMorph}(\nabla)
  =
  \left\{
    {\phi \in \mathrm{Diff}(X)}\,, \atop {\eta : \phi^* \nabla \stackrel{\simeq}{\to} \nabla}
  \right\}
  \simeq
  \left\{
    \raisebox{20pt}{
    \xymatrix{
	  X \ar[rr]^\phi|\simeq_{\ }="s" \ar[dr]_\nabla^{\ }="t" && X \ar[dl]^\nabla
	  \\
	  & \mathbf{B}U(1)_{\mathrm{conn}}
	  \ar@{=>}^\eta "s"; "t"
	}
	}
  \right\}
  \,.
$$
This also makes the group structure on these pairs manifest 
-- the \emph{quantomorphism group}: it is given by the evident pasting of 
triangular diagrams.
In this form, the quantomorphism group is realized as an example of a very general
construction that directly makes sense also in higher geometry: it is the 
{automorphism group} of a modulating morphism regarded as an object in the {slice topos}
over the corresponding moduli stack -- a {relative automorphism group}. 
Also in this form the central property of the quantomorphism group -- the
fact that over a connected manifold it is a $U(1)$-extension of the group of Hamiltonian 
symplectomorphisms -- is revealed to be just a special case of a very general 
extension phenomenon, expressed by the schematic diagrams below:
\[
  \begin{array}{ccccc}
  U(1) &\to& \mathbf{QuantMorph}(\nabla) &\to& \mathbf{HamSympl}(\nabla)
  \\
  \\
  \left\{
    \raisebox{20pt}{
    \xymatrix{
	  X 
	  \ar@/^1pc/[d]^\nabla_{\ }="s"
	  \ar@/_1pc/[d]_\nabla^{\ }="t"
	  \\
	  \mathbf{B}U(1)_{\mathrm{conn}}
	  \ar@{=>} "s"; "t"
	}
	}
  \right\}
  &\to&
  \left\{
    \raisebox{20pt}{
    \xymatrix{
	  X \ar[dr]_{\nabla}^{\ }="t" \ar[rr]^\simeq_{\ }="s" && X \ar[dl]^{\nabla}
	  \\
	  & \mathbf{B}U(1)_{\mathrm{conn}}
	  \ar@{=>} "s"; "t"
	}
	}
  \right\}
  &\to&
  \left\{
    \raisebox{20pt}{
    \xymatrix@R=1pt{
	  \\
	  \\
	  X  \ar[rr]^\simeq_{\ }="s" && X 
	  \\
	}
	}
  \right\}
  \end{array}
  \,.
\]
Our main theorems in \ref{TheCentralTheorems}
below are a general account of canonical extensions induced
by (higher) automorphism groups in slices over (higher, differential) moduli stacks in this fashion.

This $U(1)$-extension is the hallmark of quantization: under Lie differentiation the
above sequence of (infinite-dimensional) Lie groups turns into the extension of Lie algebras
$$
  \begin{array}{ccccc}
    i \mathbb{R} &\to& \mathfrak{Poisson}(X, \omega) & \to &  \mathcal{X}_{\mathrm{Ham}}(X, \omega)
  \end{array}
$$
that exhibits the Poisson bracket Lie algebra of the symplectic manifold as an 
$i\mathbb{R} \simeq \mathrm{Lie}(U(1))$-extension of the Lie algebra of Hamiltonian vector fields
on $X$ -- the \emph{Kostant-Souriau extension} (e.g section 2.3 of \cite{BrylinskiLoop}). 
If we write $\hbar \in \mathbb{R}$ for the canonical basis element 
(``Planck's constant'') then this expresses the
quantum deformation of ``classical commutators'' in $\mathcal{X}_{\mathrm{Ham}}(X,\omega)$ by 
the central term $i \hbar$.

More widely known than the quantomorphism groups of all prequantum operators
are a class of small subgroups of them,
the \emph{Heisenberg groups} of translational prequantum operators: 
if $(X, \omega)$ is a symplectic
vector space of dimension $2n$, regarded as a symplectic manifold, then the translation
group $\mathbb{R}^{2n}$ canonically acts on it by Hamiltonian symplectomorphisms,
hence by a group homomorphism $\mathbb{R}^{2n} \to \mathbf{HamSympl}(\nabla)$. 
The pullback of the above quantomorphism group extension along this map yields a
$U(1)$-extension of $\mathbb{R}^{2n}$, and this is the traditional Heisenberg group
$H(n,\mathbb{R})$. More generally, for $(X,\omega)$ any (prequantized) symplectic manifold
and $G$ any Lie group, one considers \emph{Hamiltonian $G$-actions}: smooth group
homomorphisms $\phi : G \to \mathbf{HamSympl}(\nabla)$. Pulling back the quantomorphism
group extension now yields a $U(1)$-extension of $G$ and this we may call, more generally,
the Heisenberg group extension induced by the Hamiltonian $G$-action:
$$
  \begin{array}{ccccc}
  U(1) &\to& \mathbf{Heis}_\phi(\nabla) &\to& G
  \end{array}
  \,.
$$

The crucial property of the
quantomorphism group and any of its Heisenberg subgroups, 
at least for the purposes of geometric quantization, 
is that these are canonically equipped with an action on the space
of prequantum states (the space of sections of the complex line bundle which is associated to the
prequantum bundle), this is the action of the \emph{exponentiated prequantum operators}.
Under an \emph{integrated moment map}, 
-- a group homomorphism $G \to \mathbf{QuantMorph}(\nabla)$ covering a Hamiltonian $G$-action --
this induces a representation of $G$ on the space of prequantum states.
After a choice of polarization this is the 
construction that makes geometric quantization a valuable tool in geometric representation theory.

This action of prequantum operators on prequantum states is naturally interpreted in 
terms of slicing, too:
A prequantum operator is traditionally defined to be a function $H \in C^\infty(X)$ with action on 
prequantum states $\psi$ traditionally given by the fomula
$$
  O_H : \psi \mapsto i \nabla_{v_H}\psi + H \cdot \psi
  \,,
$$
where the first term is the covariant derivative of the prequantum connection along the Hamiltonian vector field corresponding to $H$. To see how this formula  together with its Lie integration,
falls out naturally from the perspective of the slice over the moduli stack,
write $\mathbb{C}/\!/U(1)$ for the quotient stack of the canonical 
1-dimensional complex representation of the circle group, 
and observe that this comes equipped with a canonical map 
$\rho : \xymatrix{ \mathbb{C}/\!/U(1) \ar[r] & \ast/\!/U(1) \simeq \mathbf{B}U(1) }$
to the moduli stack of circle-principal bundles. This is the {universal}
complex line bundle over the moduli stack of $U(1)$-principal bundles, 
and it has a differential refinement compatible with that of
its base stack to a map
$\rho_{\mathrm{conn}} : 
\xymatrix{ \mathbb{C}/\!/U(1)_{\mathrm{conn}} \ar[r] & \mathbf{B}U(1)_{\mathrm{conn}} }$.
Now one can work out that maps $\psi : \nabla \to \rho_{\mathrm{conn}}$ in the slice over 
$\mathbf{B}U(1)_{\mathrm{conn}}$ 
are equivalently sections of
the complex line bundle $P \times_{U(1)} \mathbb{C}$ which is $\rho$-associated to the 
$U(1)$-principal prequantum bundle:
$$
  \Gamma_X\left(P \times_{U(1)}\mathbb{C}\right)
  \simeq
  \left\{
   \raisebox{20pt}{
  \xymatrix{
    X \ar[dr]_{\nabla}^{\  }="t" \ar[rr]^\psi_{\ }="s" && \mathbb{C}/\!/U(1) \ar[dl]^{\rho_{\mathrm{conn}}}
	\\
	& \mathbf{B}U(1)_{\mathrm{conn}}
	\ar@{=>} "s"; "t"
   }
   }
  \right\}
  \,.
$$

With this identification, the action of quantomorphisms on prequantum states 
$$
  (O_h, \psi) \mapsto O_h(\psi)
$$
is 
simply the precomposition action in the slice $\mathbf{H}_{/\mathbf{B}U(1)}$, 
hence the action by pasting
of triangular diagrams in $\mathbf{H}$:
$$
  \hspace{-1cm}
  \left(
   \raisebox{20pt}{
  \xymatrix{
    X \ar[dr]_{\nabla}^{\  }="t" \ar[rr]^\phi_{\ }="s" && X \ar[dl]^{\nabla}
	\\
	& \mathbf{B}U(1)_{\mathrm{conn}}
	\ar@{=>}^{O_h} "s"; "t"
   }
   }
   \;\,,\;
   \raisebox{20pt}{
  \xymatrix{
    X \ar[dr]_{\nabla}^{\  }="t" \ar[rr]^\psi_{\ }="s" && \mathbb{C}/\!/U(1)_{\mathrm{conn}} 
	\ar[dl]^{\rho_{\mathrm{conn}}}
	\\
	& \mathbf{B}U(1)_{\mathrm{conn}}
	\ar@{=>} "s"; "t"
   }
   }
  \right)
  \mapsto
    \raisebox{20pt}{
    \xymatrix{
	  X \ar[r]^\phi|\simeq_>>>>{\ }="s1" \ar[dr]_{\nabla}^{\ }="t1" 
	   & 
	   X \ar[d]|{\nabla}^{\ }="t2" \ar[r]^\psi_{\ }="s2" & \mathbb{C}/\!/U(1)_{\mathrm{conn}}
	   \ar[dl]^{\rho_{\mathrm{conn}}}
	  \\
	  & \mathbf{B}U(1)_{\mathrm{conn}}
	  \ar@{=>} "s1"; "t1"
	  \ar@{=>} "s2"; "t2"
	}
	}
$$

\medskip

Once formulated this way as the geometry of stacks in the higher slice topos over
the smooth moduli stack of principal connections, it is clear that there is a natural generalization of
traditional prequantum geometry, hence of regular contact geometry, 
obtained by 
interpreting these diagrams in {higher differential geometry} with smooth moduli stacks 
of principal bundles and principal connections refined 
to 
{higher smooth moduli stacks}. 
Morever, 
by carefully abstracting the minimum number of axioms on the ambient toposes actually needed 
in order to express the relevant
constructions (this we discuss in \ref{HigherDifferentialGeometryInIntroduction}) one
obtains generalizations to various other flavors of higher/derived geometry, such as 
higher/derived supergeometry.

Just as traditional prequantum geometry and contact geometry is of interest in itself, 
this natural refinement to higher geometry is of interest in itself, and is one motivation for
studying higher prequantum geometry. For instance 
in \ref{PrequantizationApplications} we indicate how various higher central extensions 
of interest in {string geometry} can be constructed as higher Heisenberg-group 
extensions in higher prequantum geometry.

But the strongest motivation for studying traditional prequantum geometry is, as the name indicates,
as a means in quantum mechanics and quantum field theory. This we come to below
in \ref{MotivicQuantizationApplications}.

\hspace{-1.4cm}
\begin{tabular}{|r|c|l|}
  \hline
   \begin{tabular}{c}
  {\bf $\infty$-geometric} \\ {\bf quantization}  
   \end{tabular} & 
   \begin{tabular}{c} {\bf cohesive homotopy type theory}   
	 \end{tabular} & 
	 \begin{tabular}{c}
    \begin{tabular}{l}{\bf twisted hyper-} \\ {\bf sheaf cohomology}
	\end{tabular}  
	\end{tabular}
  \\
  \hline
  \begin{tabular}{l} pre-$n$-plectic 
  \\cohesive $\infty$-groupoid 
  \end{tabular} & $\omega : X \to \Omega^{2}_{\mathrm{cl}}(-, \mathbb{G})\;\;$
  \mbox{(e.g. $\mathbb{G} = \mathbf{B}^{n-1}U(1)$ or $= \mathbf{B}^{n-1}\mathbb{C}^\times$)}
  & 
    \begin{tabular}{l}twisting cocycle \\in de Rham cohomology \end{tabular}
  \\
  \hline
  \begin{tabular}{l}
  symplectomorphisms  
  \end{tabular}&
  $
    \mathbf{Aut}_\mathbf{H}(\omega)
	=
	\left\{
	  \raisebox{20pt}{
	  \xymatrix{
	     X \ar[rr]^{\simeq} \ar[dr]_{\omega} && X \ar[dl]^{\omega}
		 \\
		 & \Omega^{2}_{\mathrm{cl}}(-,\mathbb{G})
	  }
	  }
	\right\}
  $
  &
  \begin{tabular}{l}
    twist automorphism
	\\
	$\infty$-group
  \end{tabular}
  \\
  \hline
  prequantum bundle & 
  $
    \raisebox{20pt}{
    \xymatrix{
	   & \mathbf{B} \mathbb{G}_{\mathrm{conn}} \ar[d]^-{F_{(-)}}
	   \\
	  X \ar[r]^-\omega \ar[ur]^-{\nabla} & \Omega^{2}_{\mathrm{cl}}(-, \mathbb{G})
	}
	}
  $
  &
  \begin{tabular}{l}
    twisting cocycle in \\
	differential cohomology
  \end{tabular}
  \\
  \hline
  Planck's constant $\hbar$ & 
  $\tfrac{1}{\hbar} \nabla : X \to \mathbf{B}^n \mathbb{G}_{\mathrm{conn}}$
  & \begin{tabular}{l} divisibility \\ of twist class \end{tabular}
  \\
  \hline
  \begin{tabular}{c}
  quantomorphism group 
  \\
  $\supset$
  \\
  Heisenberg group
  \end{tabular}
  & 
  $
    \mathbf{Aut}_{\mathbf{H}}(\nabla)
	=
	\left\{
	  \raisebox{20pt}{
	  \xymatrix{
	     X \ar[rr]^{\simeq}_{\ }="s" 
		 \ar[dr]_{\nabla}^{\ }="t" && X \ar[dl]^{\nabla}
		 \\
		 & \mathbf{B}^n \mathbb{G}_{\mathrm{conn}}
		 \ar@{=>}^\simeq "s"; "t"
	  }
	  }
	\right\}
  $
  &
  \begin{tabular}{l} twist automorphism \\ $\infty$-group
  \end{tabular}
  \\
  \hline
  Hamiltonian $G$-action
  &
  $\mu : \mathbf{B}G \to \mathbf{Aut}_{\mathbf{H}}(\nabla)$
  &
  \begin{tabular}{c}
    $G$-$\infty$-action
	\\
	on the twisting cocycle
  \end{tabular}
  \\
  \hline
  gauge reduction & $\nabla/\!/G \;:\; X/\!/G \to \mathbf{B} \mathbb{G}_{\mathrm{conn}}$
  &
  \begin{tabular}{c}
    $G$-$\infty$-quotient
	\\
	of the twisting cocycle
  \end{tabular}
  \\
  \hline
  \begin{tabular}{l} 
    Hamiltonian observables
	\\
	with Poisson bracket
  \end{tabular}
  &
  $\mathrm{Lie}(\mathbf{Aut}_{\mathbf{H}}(\nabla))$
  &
  \begin{tabular}{l}
    infinitesimal
	\\
	twist automorphisms
  \end{tabular}
  \\
  \hline
  \begin{tabular}{c}
  Hamiltonian 
    \\
  symplectomorphisms
  \end{tabular}
  &
  $
  \mathrm{image}\left(
    \xymatrix{
	  \mathbf{Aut}_{\mathbf{H}}(\nabla)
	  \ar[r]
	  &
	  \mathbf{Aut}(X)
	}
  \right)
  $
  &
  \begin{tabular}{c} 
    twists in 
	\\ de Rham cohomology
	\\
	that lift to
	\\
	differential cohomology
  \end{tabular}
  \\
  \hline
  \begin{tabular}{c}
    $\mathbb{G}$-representation
  \end{tabular}  
  &
  $\raisebox{20pt}{\xymatrix{ V \ar[r] & V/\!/\mathbb{G} \ar[d]^{\rho} \\ & \mathbf{B} \mathbb{G}}}$
  &
  local coefficient $\infty$-bundle
  \\
  \hline
  prequantum space of states
  &
  $\mathbf{\Gamma}_X(E) 
	=
	\left\{
	  \raisebox{20pt}{
	  \xymatrix{
	    X \ar[rr]^{\sigma}_{\ }="s" \ar[dr]_{\mathbf{c}}^{\ }="t" & & 
		V/\!/ \mathbb{G} \ar[dl]^{\rho}
		\\
		& \mathbf{B}\mathbb{G}
		\ar@{=>}^\simeq "s"; "t"
	  }
	  }
	\right\}
	$
	&
	\begin{tabular}{c}
	  cocycles in 
	  \\
	  $[\mathbf{c}]$-twisted cohomology
	\end{tabular}
	\\
	\hline
	prequantum operator action
	&
	 $\widehat{(-)}
	  :
	  \mathbf{\Gamma}_X(E) \times \mathbf{Aut}_\mathbf{H}
	  \to
	  \mathbf{\Gamma}_X(E)
	 $
	&
	\begin{tabular}{c}
	  $\infty$-action of 
	  \\
	  twist automorphisms
	  \\
	  on twisted cocycles
	\end{tabular}
	\\
	\hline
	\hline
	  transgression
	&
	\begin{tabular}{c}
	composition with:
	\\
	$
	 \raisebox{20pt}{
	 \xymatrix{
	  [S^1, V/\!/\mathbb{G}_{\mathrm{conn}}]
	  \ar[d]^-{\rho^{V}_{\mathrm{conn}}}
	  \ar[rr]^-{\mathrm{tr}\,\mathrm{hol}_{S^1}}
	  &&
	  V/\!/\Omega\mathbb{G}_{\mathrm{conn}}
	  \ar[d]^-{\rho^{V}_{\mathrm{conn}}}
	  \\
	  \mathbf{B}\mathbb{G}_{\mathrm{conn}}
	  \ar[rr]^-{\exp(2 \pi i \int_{S^1}(-))}
	  &&
	  \mathbb{G}_{\mathrm{conn}}
	}}$
	\end{tabular}
	&
	\begin{tabular}{c}
      fiber integration in
	  \\
	  (nonabelian)
	  \\
	  differential cohomology
	\end{tabular}
	\\
	\hline
\end{tabular}

\newpage

\paragraph{Prequantization}
\index{structures in a cohesive $\infty$-topos!prequantum geometry!prequantization}
\label{Prequantization}

Let $X \in \mathbf{H}$ be a cohesive homotopy type. Let $\mathbb{G} \in \mathrm{Grp}(\mathbf{H})$ 
be a braided cohesive group, def. \ref{BraidedInfinityGroup}.
In the present context we say

\begin{definition}
\begin{enumerate}
\item A morphism (def. \ref{Closed2FormModuli})
$$
  \omega : \xymatrix{X \ar[r] & \Omega^2_{\mathrm{cl}}(-,\mathbb{G})} 
$$
is a \emph{pre-symplectic structure} on $X$.
\item Given a pre-symplectic structure, a lift $\nabla$ in 
$$
  \xymatrix{
    & \mathbf{B}\mathbb{G}_{\mathrm{conn}}
    \ar[d]^{F_{(-)}}_{\ }="s"
    \\
    X 
    \ar@{->}[ur]^{\nabla}
    \ar[r]_-{\omega}^{\ }="t"
    &
    \Omega^2_{\mathrm{cl}}(-,\mathbb{G})
    \ar@{=>} "s"; "t"
  }
$$
is a \emph{prequantization} of $(X,\omega)$.
\end{enumerate}
\label{PrequantizationLift}
\end{definition}

\paragraph{Symplectomorphisms}
\index{structures in a cohesive $\infty$-topossymplectomorphisms}
\label{Symplectomorphisms}

Let $X \in \mathbf{H}$ be a cohesive homotopy type. Let $\mathbb{G} \in \mathrm{Grp}(\mathbf{H})$ 
be a braided cohesive group, def. \ref{BraidedInfinityGroup}.
Let
$$
  \omega : \xymatrix{X \ar[r] & \Omega^2_{\mathrm{cl}}(-,\mathbb{G})} 
  \,.
$$
be a pre-symplectic structure, def. \ref{Closed2FormModuli}.

\begin{definition}
  The \emph{symplectomorphism group} $\mathbf{Sympl}(\omega)$ of the pre-symplectic geometry
  $(X,\omega)$ is the $\mathbf{H}$-valued automorphism group, def. \ref{HValuedAutomorphismGroup},
  of $\omega \in \mathbf{H}_{/\Omega^2_{\mathrm{cl}}(-,\mathbb{G})}$:
  $$
    \mathbf{Sympl}(\omega)
	:=
	\mathbf{Aut}_{\mathbf{H}}(\omega)
	:=
	\prod_{\Omega^2_{\mathrm{cl}}(-,\mathbb{G})}
	\mathbf{Aut}(\omega)
	\,.
  $$
\end{definition}
\begin{remark}
  According to remark \ref{GlobalElementsInHValuedSliceInternalHom} a global element of
  $\mathbf{Sympl}(\omega)$ corresponds to a diagram in $\mathbf{H}$ of the form
  $$
    \raisebox{20pt}{
    \xymatrix{
	  X \ar[rr]^\phi_{\simeq}\ar[dr]_{\omega} && X \ar[dl]^{\omega}
	  \\
	  & \Omega^2_{\mathrm{cl}}(-,\mathbb{G})
	}
	}
	\,.
  $$
  This is a diffeomorphism $\phi$ of $X$ which preserves the pre-symplectic structure in that
  $$
    \phi^* \omega = \omega
	\,.
  $$
\end{remark}
\begin{definition}
  Write
  $$
    p_{\Omega^2_{\mathrm{cl}}(-,\mathbb{G})}
	:
	\xymatrix{
	  \mathbf{Sympl}(\omega)
	  \ar[r]
	  &
	  \mathbf{Aut}(X)
	}
  $$
  for the canonical morphism induced by restriction of the morphism of 
  prop. \ref{CanonicalMorphismFromSliceInternalHomToBaseInternalHom}.
  \label{MapFromSymplectomorphismsToDiffeomorphisms}
\end{definition}
\begin{proposition}
  The morphism $p_{\Omega^2_{\mathrm{cl}}(-,\mathbb{G})}$ of 
  def. \ref{MapFromSymplectomorphismsToDiffeomorphisms} is a monomorphism
\end{proposition}
\proof
  By direct generalization of the proof of prop. \ref{FiberOfMapFromSliceMappingToBaseMapping} 
  we find that for each $U \in \mathbf{H}$
  the fibers of $p_{\Omega^2_{\mathrm{cl}}(-,\mathbb{G})}$ are path space objects
  of $[X, \Omega^2_{\mathrm{cl}}(-,\mathbb{G})]$. But since 
  $\Omega^2_{\mathrm{cl}}(-,\mathbb{G})$ is 0-truncated by def. \ref{Closed2FormModuli},
  also $[X, \Omega^2_{\mathrm{cl}}(-,\mathbb{G})]$ is 0-truncated, and so its
  path spaces are either contractible or empty. 
\endofproof

\paragraph{Contactomorphisms}
\index{structures in a cohesive $\infty$-topos!prequantum geometry!contactomorphisms}
\label{ContactTransformations}

\begin{definition}
Given two $\mathbb{G}$-principal connections 
$\nabla_1 : X_1 \to \mathbf{B}\mathbb{G}_{\mathrm{conn}}$ 
and
$\nabla_2 : X_2 \to \mathbf{B}\mathbb{G}_{\mathrm{conn}}$, 
a (strict) \emph{contactomorphism between regular contact spaces} 
from $\nabla_1$ to $\nabla_2$ is a morphism
between them in the slice $\mathbf{H}_{/\mathbf{B}\mathbb{G}_{\mathrm{conn}}}$.
The \emph{$\infty$-groupoid of contactomorphisms} betwen $\nabla_1$ and $\nabla_2$ is
$$
  \mathrm{ContactMorph}(\nabla_1, \nabla_2)
  := 
  \Gamma\left([\nabla_1, \nabla_2]_{\mathbf{H}}\right)
  :=
  \Gamma
  \underset{\mathbf{B}\mathbb{G}_{\mathrm{conn}}}{\prod} [\nabla_1, \nabla_2]
  \,,
$$
\label{Contactomorphisms}
\end{definition}
\begin{remark}
  This means that a single contactomorphism from $\nabla_1$ to $\nabla_2$ is given by a
  diagram
  $$
    \raisebox{20pt}{
    \xymatrix{
	  X \ar[rr]^\simeq_{\ }="s" \ar[dr]_{\nabla_1}^{\ }="t" && X \ar[dl]^{\nabla_2}
	  \\
	  & \mathbf{B}\mathbb{G}_{\mathrm{conn}}
	  \ar@{=>} "s"; "t"
	}
	}
  $$
  in $\mathbf{H}$. 
  However, in order to obtain the correct cohesive structure on the collection
  of all contactomorphisms we need to \emph{concretify} the object 
  $[\nabla_1, \nabla_2]_{\mathbf{H}}$, as in the discussion at \ref{StrucDifferentialModuli}.
\end{remark}

\paragraph{Quantomorphism group and Heisenberg group}
\index{structures in a cohesive $\infty$-topos!prequantum geometry!quantomorphism group}
\label{QuantomorphismGroup}
\index{structures in a cohesive $\infty$-topos!prequantum geometry!Heisenberg group}
\index{Kostant-Souriau-extension!general abstract}
\label{HeisenbergGroup}
 \label{QuantomorphismAndHeisenbergGroup}
 \label{TheQuantomorphismExtension}
 \label{HigherGaugeTheorySmoothInfinityGroupoids}
 \label{QuantomoprhismGroup}
 \label{TheCentralTheorems}
 \label{TheLInfinityExtension}
 \label{HeisenbergLieExtension}

Famously, quantum theory is governed by the appearance of a group of 
quantum observables/operators called a \emph{Heisenberg group}. But in fact 
the Heisenberg group is but the subgroup on \emph{linear} translations
in phase space of the full group of prequantum operators. 
In standard textbooks on geometric quantization the latter is
called the \emph{quantomorphism group}. While standard in geometric
quantization, that term is rather less wide-spread in the physics literature.
Many physics textbooks know the quantomorphism group, if at all, just as 
\emph{the Fr{\'e}chet-Lie group which integrates the Poisson bracket}.

Here we take the opposite perspective: we give a general abstract formalization
of quantomorphism $\infty$-groups in a cohesive $\infty$-topos. We work out concrete differential-geometric
incarnations of this in the context of smooth cohesion in section \ref{SmoothStrucQuantomorphisms} Then, 
further below in section \ref{PoissonAndHeisenbergLieAlgebra}, we 
\emph{define} the \emph{Poisson $\infty$-Lie algebra} to be the $\infty$-Lie algebra
of the quantomorphism $\infty$-group. Our main result below says that this 
reduces in the appropriate special cases not only to the traditional 
Poisson bracket Lie algebra in symplectic geometry, but also to the 
Poisson Lie-$n$-algebras of $n$-plectic geometry.

Let $\mathbf{H}$ be a cohesive $\infty$-topos (such as that of 
smooth $\infty$-groupoids), let 
$\mathbb{G} \in \mathrm{Grp}_2(\mathbf{H})$ be a braided $\infty$-group
in $\mathbf{H}$, let $\mathbf{B}\mathbb{G}_{\mathrm{conn}}$ be the universal 
moduli $\infty$-stack of $\mathbb{G}$-principal connections.
 
\begin{definition}
  For $\nabla : X \to \mathbf{B}\mathbb{G}_{\mathrm{conn}}$ the map
  modulating 
  a $\mathbb{G}$-principal connection, the corresponding
  \emph{higher quantomorphism groupoid} $\mathrm{At}(\nabla)_\bullet \in \mathrm{Grpd}(\mathbf{H})$
  or \emph{higher contactomorphism groupoid} induced by $\nabla$
  is the corresponding higher Atiyah-groupoid,
  hence under the equivalence of prop. \ref{1EpisAreEquivalentToGroupoidObjects}
  is the $\infty$-groupoid with atlas which is the 1-image projection
  $$
    \xymatrix{
	  X \ar@{->>}[r] & \mathrm{At}(\nabla) := \mathrm{im}_1(\nabla)
	}
  $$
  of $\nabla$.
  \label{QuantomorphismGroupoid}
\end{definition}
\begin{remark}
The unconcretified $\infty$-group of bisections of 
the higher quantomorphism groupoid $\mathrm{At}(\nabla)_\bullet$ 
of def. \ref{QuantomorphismGroupoid}
sits in a homotopy fiber sequence of the form
$$
  \xymatrix{
     \mathbf{BiSect}(\mathrm{At}(\nabla)_\bullet)
	 \ar[r]
	 &
	 \mathbf{Aut}(X)
	 \ar[r]^-{\nabla \circ (-)}
	 &
	 [X,\mathbf{B}\mathbb{G}_{\mathrm{conn}}]
  }\,,
$$
with the object on the right taken to be pointed by $\nabla$. 
But now that we are considering a differential cocycle, not just from a bundle cocycle, 
one finds that this $\infty$-group of bisections does have the correct global points,
but does not quite have the geometric structure on these that one would typically
need in applications (such as in the theorems below in \ref{TheCentralTheorems}).
Instead, one wants the \emph{differentially concretified} version of
$\mathbf{BiSect}(\mathrm{At}(\nabla)_\bullet)$, 
along the lines of the above discussion around def. \ref{DifferentialModuliObject}.
\end{remark}
But in view of the above fiber sequence, there is a natural candidate of such differential
concretification:
\begin{definition}
  The \emph{quantomorphism $\infty$-group} of a $\mathbb{G}$-principal connection
  $\nabla$ is the homotopy fiber $\mathbf{QuantMorph}(\nabla) \in \mathrm{Grp}(\mathbf{H})$
  in 
$$
  \raisebox{20pt}{
  \xymatrix{
     \mathbf{QuantMorph}(\nabla)
	 \ar[r]
	 &
	 \mathbf{Aut}(X)
	 \ar[r]^-{\nabla \circ (-)}
	 &
	 \mathbb{G}\mathbf{Conn}(X)
  }
  }\,,
$$
where the right morphism is the composite of $\nabla \circ (-)$ with the 
differential concretification projection 
$\xymatrix{[X, \mathbf{B}\mathbb{G}_{\mathrm{conn}}] \ar[r] & \mathbb{G}\mathbf{Conn}(X)}$
of remark \ref{DifferentialConcretificationMap}.
 \label{HigherQuantomorphismGroupInIntroduction}
\end{definition}
\begin{remark}
The canonical forgetful map 
$u_{\mathbf{B}\mathbb{G}} : \mathbf{B}\mathbb{G}_{\mathrm{conn}} \to \mathbf{B}\mathbb{G}$ induces
a morphism from the higher quantomorphism groupoid to the Atiyah groupoid of the underlying
$\mathbb{G}$-principal bundle
$$
  \xymatrix{
    \mathrm{At}(\nabla)_\bullet \ar[r] &  \mathrm{At}(\nabla^0)_\bullet
  }
$$
which is the identity on objects. This in turn induces a canonical homomorphism
$$
  u_{\mathbf{B}\mathbb{G}} \circ (-)
  :
  \xymatrix{
    \mathbf{QuantMorph}(\nabla) \ar[r] &  \mathbf{BiSect}(\mathrm{At}(P)_\bullet)
  }
$$
from the quantomorphism $\infty$-group, def. \ref{HigherQuantomorphismGroupInIntroduction}, 
into that of bisections of the  Atiyah groupoid, 
prop. \ref{AtiyahBisectionsAreAutomorphismOfModulatingMapInSlice}. 
Thereby 
the quantomorphism $\infty$-group acts on the
space of sections of any associated $V$-fiber $\infty$-bundle to $\nabla^0$. This is the 
\emph{higher prequantum operator} action. It is the global 
version of the canonical action of the higher quantomorphism groupoid itself,
in the sense of groupoid actions 
which is exhibited
by the left square in the following pasting diagram of $\infty$-pullbacks:
  $$
    \xymatrix{
	  P \times_{\mathbb{G}} V \ar[r] \ar[d] 
	  & (P \times_{\mathbb{G}} V)/\!/\mathrm{Qu}(\nabla) \ar[d] 
	  \ar[rr] && V/\!/G
	  \ar[d]^\rho
	  \\
	  X \ar@{->>}[r] 
	   & \mathrm{Qu}(\nabla)
	  \ar@{^{(}->}[r]
	  \ar[dr]
	  &
	  \mathbf{B}\mathbb{G}_{\mathrm{conn}}
	  \ar[r]
	  &
	  \mathbf{B}\mathbb{G}
	  \\
	  && \mathrm{At}(\nabla^0)
	  \ar@{^{(}->}[ur]
	}
  $$ 
\end{remark}  

Given all of the above, we now have the following list of evident generalizations of traditional notions
in prequantum theory.
\begin{definition}
  Let $\nabla : X \to \mathbf{B}\mathbb{G}_{\mathrm{conn}}$ be given,
  regarded as a prequantum $\infty$-bundle. Then 
  \begin{enumerate}
    \item the \emph{Hamiltonian symplectomorphism group} 
	  $\mathbf{HamSympl}(\nabla) \in \mathrm{Grp}(\mathbf{H})$ is the sub-$\infty$-group
	  of the automorphisms of $X$ which is the 1-image, def. \ref{ImagesInIntroduction}, 
	  of the quantomorphisms:
	  $$
	    \xymatrix{
		   \mathbf{QuantMorph}(\nabla)
		   \ar@{->>}[r]
		   &
		   \mathbf{HamSympl}(\nabla)
		   \ar@^{^{(}->}[r]
		   &
		   \mathbf{Aut}(X)
		}
	  $$
	 \item
       for $G \in \mathrm{Grp}(\mathbf{H})$ an $\infty$-group,
       a \emph{Hamiltonian $G$-action} on $X$ is an $\infty$-group homomorphim
	   $$\xymatrix{G \ar[r]^-\phi & \mathbf{HamSympl}(\nabla) \ar@{^{(}->}[r] &\mathbf{Aut}(X)}\,;$$
	 \item
	   an \emph{integrated $G$-momentum map} is an action by quantomorphisms
	   $$
	     \xymatrix{G \ar[r]^-{\hat \phi} & \mathbf{QuantMorph}(\nabla) \ar@{^{(}->}[r] & \mathbf{Aut}(X) }
		 \,;
	   $$
	 \item
	   given a Hamiltonian $G$-action $\phi$, the corresponding 
	   \emph{Heisenberg $\infty$-group} $\mathbf{Heis}_\phi(\nabla)$ is the
	   homotopy fiber product in
	   $$
	     \raisebox{20pt}{
	     \xymatrix{
		    \mathbf{Heis}_\phi(\nabla) \ar[r] \ar[d] & \mathbf{QuantMorph}(\nabla) \ar[d]
			\\
			G \ar[r]^-\phi & \mathbf{HamSympl}(\nabla)
		 }
		 }\,.
	   $$
  \end{enumerate}
  \label{HamiltonianActionMomentumMapAndHeisenbergGroup}
\end{definition}

As in the discussion in \ref{DifferentiaCoefficients}, 
let $\mathbf{H}$ be a cohesive $\infty$-topos 
(such as $\mathrm{Smooth}\infty\mathrm{Grpd}$),
let  $\mathbb{G} \in \mathrm{Grp}_2(\mathbf{H})$
a braided $\infty$-group, def. \ref{BraidedGroups}, let $X \in \mathbf{H}$ any object, 
let $\omega : X \to \Omega^2_{\mathrm{cl}}(-,\mathbb{G})$
be a flat differential form datum
and let $\nabla : X \to \mathbf{B}\mathbb{G}_{\mathrm{conn}}$ a $\mathbb{G}$-prequantization of it.
Then we have the following characterization of the corresponding
quantomorphism $\infty$-group of def. \ref{HigherQuantomorphismGroupInIntroduction}.
\begin{theorem}
There is a long homotopy fiber sequence in $\mathrm{Grp}(\mathbf{H})$ of the form
\begin{itemize}
\item if $\mathbb{G}$ is 0-truncated:
$$
  \xymatrix{
    \mathbb{G}
	\ar[r]
	&
	\mathbf{QuantMorph}(\nabla)
	\ar[r]
	&
	\mathbf{HamSympl}(\nabla)
	\ar[r]^-{\nabla \circ (-)}
	&
	\mathbf{B}\left(
	  \mathbb{G}\mathbf{ConstFunct}(X)
	\right)
  }
$$
\item otherwise:
$$
  \xymatrix{
     (\Omega \mathbb{G})\mathbf{FlatConn}(X)
	 \ar[r]
	 &
	 \mathbf{QuantMorph}(\nabla)
	 \ar[r]
	 &
	 \mathbf{HamSympl}(\nabla)
	 \ar[r]^-{\nabla \circ (-)}
	 &
	 \mathbf{B}
	 \left(
	   \left(\Omega\mathbb{G}\right)
	   \mathbf{FlatConn}(\nabla)
	 \right)
  }
  \,,
$$
\end{itemize}
which hence exhibits the quantomorphism group $\mathbf{QuantMorph}(\nabla) \in \mathrm{Grp}(\mathbf{H})$
as an $\infty$-group extension, \ref{ExtensionsOfCohesiveInfinityGroups} 
of the $\infty$-group of Hamiltonian symplectomorphisms,
def. \ref{HamiltonianActionMomentumMapAndHeisenbergGroup}, 
by  the differential moduli of flat $\Omega \mathbb{G}$-principal connections on $X$, 
def. \ref{FlatDifferentialModuli}, classified by an $\infty$-group cocycle
which is given by postcomposition with $\nabla$ itself.
 \label{TheLongHomotopyFiberSequenceOfTheQuantomorphimsGroup}
\end{theorem}
\proof
  This is an immediate variant, under the differential concretification,
  def. \ref{DifferentialModuliByIteratedPullback},
  of the
  higher Atiyah sequence 
  Consider the natural 1-image factorization of the
  horizontal maps in the defining $\infty$-pullback 
  of def. \ref{HigherQuantomorphismGroupInIntroduction}:
  $$
    \raisebox{20pt}{
    \xymatrix{
      \mathbf{QuanMorph}(\nabla)
	  \ar@{->>}[r]
	  \ar[d]
	  &
	  \mathbf{HamSympl}(\nabla)
	  \ar@{^{(}->}[r]
	  \ar[d]^{\nabla \circ (-)}
	  &
	  \mathbf{Aut}(X)
	  \ar[d]^{\nabla \circ (-)}
	  \\
	  {*}
	  \ar@{->>}[r]
	  \ar@/_1pc/[rr]_{\vdash \nabla}
	  &
	  \mathbf{B}\left( \Omega_{\nabla} \left( \mathbb{G}\mathbf{Conn}(X)\right) \right)
	  \ar@{^{(}->}[r]
	  &
	  \mathbb{G}\mathbf{Conn}(X)
	}
   }
   \,.
  $$
  By homotopy pullback stability of both 1-epimorphisms and 1-monomorphisms and by 
  essential uniqueness of 1-image factorizations this is a pasting diagram of homotopy pullback
  squares. The claim then follows 
  .
\endofproof
The analogous statement also holds for Heisenberg $\infty$-groups:
\begin{corollary}
If $\phi : G \to \mathbf{HamSympl}(\nabla) \hookrightarrow \mathbf{Aut}(X)$
is any Hamiltonian $G$-action, def. \ref{HamiltonianActionMomentumMapAndHeisenbergGroup},
then the corresponding Heisenberg $\infty$-group sits in 
the $\infty$-fiber sequence
$$
  \xymatrix{
     (\Omega \mathbb{G})\mathbf{FlatConn}(X)
	 \ar[r]
	 &
	 \mathbf{Heis}_\phi(\nabla)
	 \ar[r]
	 &
	 G
	 \ar[r]^-{\nabla \circ (-)}
	 &
	 \mathbf{B}
	 \left(
	   \left(\Omega\mathbb{G}\right)
	   \mathbf{FlatConn}(\nabla)
	 \right)
  }
  \,,
$$
\end{corollary}
\proof
  By the pasting law for homotopy pullbacks.
\endofproof
\begin{example} 
  For $\mathbb{G} = U(1) \in \mathrm{Grp}(\mathrm{Smooth}\infty\mathrm{Grpd})$
  the smooth circle group
  and for $X \in \mathrm{SmoothMfd} \hookrightarrow \mathrm{Smooth}\infty\mathrm{Grpd}$
  a connected smooth manifold, theorem \ref{TheLongHomotopyFiberSequenceOfTheQuantomorphimsGroup}
  reproduces the traditional quantomorphism group as a $U(1)$-extension of the 
  traditional group of Hamiltonian symplectomorphisms, as discussed for instance in 
  \cite{RS, Viz}.
\end{example}
In order to put the higher generalizations of the quantomorphism extensions
into this context, we notice the following basic fact.
\begin{proposition}
  For $\mathbb{G} = \mathbf{B}U(1) \in \mathrm{Grp}(\mathrm{Smooth}\infty\mathrm{Grpd})$
  the smooth circle 2-group
  consider $X \in \mathrm{SmoothMfd} \hookrightarrow \mathrm{Smooth}\infty\mathrm{Grpd}$
  a connected and simply connected smooth manifold. Then from 
  prop. \ref{TheCorrectU1DifferentialModuli} and example \ref{TheCorrectU1DifferentialModuliFlat}
  one obtains an equivalence of smooth group stacks
  $$
    U(1)\mathbf{FlatConn}(X) \simeq  \mathbf{B}U(1)
	\,.
  $$
  Generally, for $n \geq 1$ 
  and for $\mathbb{G} = \mathbf{B}^n U(1) \in \mathrm{Grp}(\mathrm{Smooth}\infty\mathrm{Grpd})$
  the smooth circle $(n+1)$-group, there is for $X$ 
  an $n$-connected smooth manifold an equivalence of smooth $\infty$-groups
  $$
    (\mathbf{B}^{n-1}U(1))\mathbf{FlatConn}(X) \simeq \mathbf{B}^n U(1)
	\,.
  $$
  \label{FlatConnectionsOnSimplyConnectedManifolds}
\end{proposition}
\proof
  We use the description of $U(1)\mathbf{FlatConn}(X)$ given 
  by prop. \ref{TheCorrectU1DifferentialModuli} and example \ref{TheCorrectU1DifferentialModuliFlat}.
  First notice then that on a simply connected manifold there is up to equivalence just a single
   flat connection, hence $U(1)\mathbf{FlatConn}(X)$ is pointed connected.
  Moreover, an auto-gauge transformation from that single flat connection (any one)
to itself is a $U(1)$-valued function which is \emph{constant on X}.
But therefore by prop. \ref{TheCorrectU1DifferentialModuli} the $U$-plots of 
the first homotopy sheaf of $U(1)\mathbf{FlatConn}(X)$ are
smoothly $U$-parameterized collections of constant $U(1)$-valued functions on $X$,
  hence are smoothly $U$-parameterized collections of elements in $U(1)$, hence are
  smooth $U(1)$-valued functions on $U$. These are, by definition,
  equivalently the $U$-plots of automorphisms of the point in $\mathbf{B}U(1)$.
  
  The other cases work analogously.
\endofproof
\begin{remark}
  Therefore in the situation of prop. \ref{FlatConnectionsOnSimplyConnectedManifolds} 
  the quantomorphism $\infty$-group is a smooth 2-group
  extension by the circle 2-group $\mathbf{B}U(1)$. The archetypical example
  of $\mathbf{B}U(1)$-extensions is the smooth \emph{String 2-group},
  def. \ref{SmoothBString}.
  Indeed, this occurs as the {Heisenberg 2-group extension}
  of the {WZW sigma-model} regarded as a local 2-dimensional quantum field theory. 
  This we turn to in \ref{ExtendedWZW} below.
\end{remark}

\paragraph{Courant groupoids}
 \label{CourantGroupoids}
 \label{HigherCourant}
 \label{HigherCourantQuantomorphimsGroupoids}
 
 Given a $\mathbb{G}$-principal $\infty$-connection
$$
  \xymatrix{
    & \mathbf{B}\mathbb{G}_{\mathrm{conn}}
	\ar[d]^{u_{\mathbf{B}\mathbb{G}}}
    \\
    X \ar[ur]^{\nabla} \ar[r]^{\nabla^0} & \mathbf{B}\mathbb{G}
  }
$$
we have considered in \ref{HigherAtiyahGroupoidsInIntroduction} the
corresponding higher Atiyah groupoid $\mathrm{At}(\nabla^0)_{\bullet}$
and in \ref{QuantomorphismAndHeisenbergGroup} the higher quantomorphism 
groupoid $\mathrm{At}(\nabla)$ equipped with a canonical map
$
  \xymatrix{
    \mathrm{At}(\nabla)_\bullet
	\ar[r]
	&
	\mathrm{At}(\nabla^0)_\bullet
  }
  \,.
$
But in view of the towers of differential coefficients discussed in 
\ref{DifferentiaCoefficients} this has a natural generalization 
to towers of higher groupoids interpolating between the 
higher Atiyah groupoid and the higher quantomorphism groupoid.

In particular, let $\mathbb{G} \in \mathrm{Grp}_3(\mathbf{H})$ a sylleptic
$\infty$-group, def. \ref{BraidedGroups}, with compatibly chosen factorization of differential form coefficients
and induced factorization of differential coefficients
$$
  \xymatrix{
    \mathbf{B}^2 \mathbb{G}_{\mathrm{conn}}
	\ar[r]
	&
	\mathbf{B}(\mathbf{B}\mathbb{G}_{\mathrm{conn}})
	\ar[r]
	&
	\mathbf{B}^2\mathbb{G}
  }
  \,.
$$
Then in direct analogy with def. \ref{QuantomorphismGroupoid} we set:
\begin{definition}
  For $\nabla^{n-1} : X \to \mathbf{B}(\mathbf{B}\mathbb{G}_{\mathrm{conn}})$
  a $\mathbb{G}$-principal connection without top-degree connection data
  as in def. \ref{PrincipalConnectionWithoutTopDegreeForms}, we say 
  that the corresponding \emph{higher Courant groupoid} is the 
  corresponding higher Atiyah groupoid 
  $\mathrm{At}(\nabla^{n-1})_\bullet \in \mathrm{Grpd}(\mathbf{H})$,
  hence the groupoid object which by prop. \ref{1EpisAreEquivalentToGroupoidObjects}
  is equivalent to the $\infty$-groupoid with atlas given by the 1-image
  factorization of $\nabla^{n-1}$
  $$
    \xymatrix{
	  X \ar@{->>}[r]
	  &
	  \mathrm{At}(\nabla^{n-1})
	  :=
	  \mathrm{im}_1(\nabla^{n-1})
	}
	\,.
  $$
  \label{HigherCourantGroupoid}
\end{definition}
\begin{example}
  If $\mathbf{H} = \mathrm{Smooth}\infty\mathrm{Grpd}$
  is the $\infty$-topos of smooth $\infty$-groupoids 
  and $\mathbb{G} = \mathbf{B}U(1) \in \mathrm{Grp}_\infty(\mathbf{H})$ is the
  smooth circle 2-group
  and if finally $X \in \mathrm{SmoothMfd} \hookrightarrow \mathrm{Smooth}\infty\mathrm{Grpd}$
  is a smooth manifold, 
  then by def. \ref{BundleGerbeWithConnectionButWithoutCurving} a
  map $\nabla^{1} : X \to \mathbf{B}(\mathbf{B}U(1)_{\mathrm{conn}})$
  is equivalently a ``$U(1)$-bundle gerbe with connective structure but without curving''
  on $X$. 
  
  In this case the higher Courant groupoid according to def. \ref{HigherCourantGroupoid}
  is a smooth 2-groupoid and its $\infty$-group of 
  bisections $\mathbf{BiSect}(\mathrm{At}(\nabla^{1})_\bullet)$
    is a smooth 2-group. The points of this
  2-group are
  equivalently pairs $(\phi, \eta)$ consisting of a diffeomorphism 
  $\phi : \xymatrix{X \ar[r]^\simeq & X}$ and an equivalence of bundle 
  gerbes with connective structure but without curving 
  of the form $\eta : \xymatrix{\phi^* \nabla^{n-1} \ar[r]^-\simeq & \nabla^{n-1}}$.
  A homotopy of bisections between two such pairs 
  $(\phi_1, \eta_1) \to (\phi_2, \eta_2)$ exists if $\phi_1 = \phi_2$
  and is then given by a higher gauge equivalence $\kappa$
  $\kappa : \xymatrix{ \eta_1 \ar[r]^\simeq & \eta_2 }$. 
  Moreover, with prop. \ref{TheCorrectU1DifferentialModuli} the smooth
  structure on the differentially concretified 2-group of such bisections is the expected
  one, where $U$-plots are smooth $U$-parameterized collections of diffeomorphisms
  and of bundle gerbe gauge transformations.
  
  Precisely these smooth 2-groups have been studied in \cite{BC}. There it was shown that 
  the Lie 2-algebras that correspond to them under Lie differentiation 
  are the Lie 2-algebras of sections of 
  the \emph{Courant Lie 2-algebroid} which is traditionally associated with
  a bundle gerbe with connective structure. (See the citations in \cite{BC} 
  for literature on Courant Lie 2-algebroids.) Therefore the
  abstractly defined smooth higher Courant groupoid $\mathrm{At}(\nabla^{n-1})$
  according to def. \ref{HigherCourantGroupoid} indeed is a Lie integration
  of the traditional Courant Lie 2-algebroid assigned to $\nabla^{n-1}$, hence is the
  \emph{smooth Courant 2-groupoid}.
  \label{TheTraditionalCourant2Groupoid}
\end{example}
\begin{example}
  More generally, in the situation of example \ref{TheTraditionalCourant2Groupoid}
  consider now for some $n \geq 1$ the smooth circle $n$-group
  $\mathbb{G} = \mathbf{B}^{n-1}U(1)$.
  Then a map
  $$
    \nabla^{n-1} : 
	\xymatrix{X \ar[r] & \mathbf{B}(\mathbf{B}^{n-1}U(1)_{\mathrm{conn}})}
  $$
  is equivalently a Deligne cocycle on $X$ in degree $(n+1)$ without $n$-form data. 
  
  To see what the corresponding smooth higher Courant groupoid
  $\mathrm{At}(\nabla^{n-1})$ is like, consider first the local case in which 
  $\nabla^{n-1}$ is trivial. In this case a bisection of $\mathrm{At}(\nabla^{n-1})$
  is readily seen to be a pair consisting of a diffeomorphism 
  $\phi \in \mathrm{Diff}(X)$ together with an $(n-1)$-form $H \in \Omega^{n-1}(X)$,
  satisfying no further compatibility condition. This means that there
  is an $L_\infty$-algebra representing the Lie differentiation of 
  the higher Courant groupoid $\mathrm{At}(\nabla^{n-1})_\bullet$ 
  which in lowest degree is the space of sections
  of a bundle on $X$ which is locally the sum $T X \oplus \wedge^{n-1} T^* X$
  of the tangent bundle with the $(n-1)$-form bundle.
  This is precisely what the sections of higher Courant Lie $n$-algebroids
  are supposed to be like, see for instance \cite{Zambon}. 
\end{example}

Finally, if we are given a tower of differential refinements of $\mathbb{G}$-principal
bundles as discussed in \ref{DifferentiaCoefficients}
$$
  \xymatrix{
	&& \mathbf{B}\mathbb{G}_{\mathrm{conn}} \ar[d]
	\\
	&& \mathbf{B}\mathbb{G}_{\mathrm{conn}^{n-1}}
	   \ar[d]
	\\
	&& \vdots \ar[d]
	\\
    X \ar[rr]|{\nabla^0}^{\ }="t"
	 \ar[uurr]|{\nabla^{n-1}}_{\ }="s"
	 \ar[uuurr]^{\nabla}
	&& \mathbf{B}\mathbb{G}
	\ar@{..} "s"; "t"
  }
$$
then there is correspondingly a tower of higher gauge groupoids:
$$
  \xymatrix{
    \mbox{\begin{tabular}{c}higher \\ Quantomorphism \\ groupoid \end{tabular}}
	&
	\mbox{\begin{tabular}{c} higher \\ Courant \\ groupoid \end{tabular}}
	& \cdots & 
	\mbox{\begin{tabular}{c} intermediate \\ differential \\ higher \\ Atiyah \\ groupoid \end{tabular}}
	& \cdots
	&
	\mbox{\begin{tabular}{c} higher \\ Atiyah \\ groupoid \end{tabular}}
    \\
    \mathrm{At}(\nabla)_\bullet
	\ar[r]
	&
	\mathrm{At}(\nabla^{n-1})_\bullet
	\ar[r]
	&
	\cdots
	\ar[r]
	&
	\mathrm{At}(\nabla^k)
	\ar[r]
	&
	\cdots
	\ar[r]
	&
	\mathrm{At}(\nabla^0)
  }
  \,.
$$
The further intermediate stages appearing here seem not to correspond to 
anything that has already been given a name in traditional literature. 
We might call them \emph{intermediate higher differential gauge groupoids}. 
These structures are an integral part of higher prequantum geometry.

\paragraph{Poisson and Heisenberg Lie algebra}
\index{structures in a cohesive $\infty$-topos!Poisson $\infty$-Lie algebra}
\index{structures in a cohesive $\infty$-topos!Heisenberg $\infty$-Lie algebra}
\label{PoissonAndHeisenbergLieAlgebra}

We consider now the $\infty$-Lie algebras of these $\infty$-groups in prequantum geometry.

\begin{definition}
\label{HamiltonianVectorFieldsOnGrpd}
\index{symplectic higher geometry!Poisson $\infty$-Lie algebra}
\index{symplectic higher geometry!Hamiltonian vector fields}
\begin{itemize}
\item The  $\infty$-Lie algebra 
$$
  \mathfrak{poisson}(X, \hat \omega)
  :=
  \mathrm{Lie}(\mathbf{QuantMorph}(\nabla))
$$
of the quantomorphism group
we call the \emph{Poisson $\infty$-Lie algebra} of the prequantum geometry $(X,\nabla)$.

\item The $\infty$-Lie algebra of the Hamiltonian symplectomorphisms
$$
  \mathcal{X}_{\mathrm{Ham}}(X, \hat \omega)
  :=
  \mathrm{Lie}(\mathbf{HamSympl}(\nabla))
$$ 
we call the $\infty$-Lie algebra of 
 \emph{Hamiltonian vector fields} of the prequantum geometry.
\end{itemize}
\end{definition}
\begin{remark}
 \label{HeisenbergInfinityGroup}
 \index{symplectic higher geometry!Heisenberg $\infty$-Lie algebra}
 \index{symplectic higher geometry!Heisenberg $\infty$-group}
  If $X$ has a linear structure (the structure of a vector space) 
  and $\omega$ is constant on $X$, then 
  we can consider the sub $\infty$-Lie algebra of $\mathfrak{poisson}(X, \hat \omega)$
  on the constant and linear elements. We discuss realizations of this below 
  in \ref{SmoothStrucGeometricPrequantizationOrdinary}. 
  This sub $\infty$-Lie algebra we call the \emph{Heisenberg $\infty$-Lie algebra}
  $$
    \mathfrak{heis}(\nabla) \hookrightarrow \mathfrak{poisson}(\nabla)
	\,.
  $$
  The corresponding sub-$\infty$-group we call the \emph{Heisenberg $\infty$-group}
  $$
    \mathbf{Heis}(\nabla) \hookrightarrow \mathbf{QuantMorph}(\nabla)
	\,.
  $$
\end{remark}

\paragraph{Prequantum states}
\index{structures in a cohesive $\infty$-topos!prequantum states}
\label{PrequantumStatesInCohesion}

Given a prequantum geometry 
$$\xymatrix{X \ar[r]^-\nabla & \mathbf{B}\mathbb{G}_{\mathrm{conn}} \ar[r]^-{F_{(-)}} & \Omega^2_{\mathrm{cl}}(-,\mathbb{G})}$$ 
as above, choose now finally a representation, def. \ref{RepresentationOfInfinityGroup},
of $\mathbb{G}$, hence a fiber sequence in $\mathbf{H}$ of the form
$$
  \xymatrix{
    V \ar[r] & V/\!/\mathbb{G} \ar[r]^\rho & \mathbf{B}\mathbb{G}
  }
  \,.
$$

For $U_{\mathbf{B}\mathbb{G}} : \xymatrix{\mathbf{B}\mathbb{G}_{\mathrm{conn}} \ar[r] & \mathbf{B}\mathbb{G}}$
the forgetful morphism, we obtain from the prequantum connection
$\nabla \in \mathbf{H}_{/\mathbf{B}\mathbb{G}_{\mathrm{conn}}}$ 
the underlying modulus  
$$
  \underset{U_{\mathbf{B}\mathbb{G}}}{\sum} {\nabla}
  \in
  \mathbf{H}_{/\mathbf{B}\mathbb{G}}
$$
of the prequantum bundle proper.

\begin{definition}
  The $\rho$-associated $V$-fiber bundle 
  $$
    E := 
	 \left(
	    \underset{U_{\mathbf{B}\mathbb{G}}}{\sum}{\nabla} 
	 \right)^*
	 \rho 
	 \in \mathbf{H}_{/X}
  $$
  to $\underset{U_{\mathbf{B}\mathbb{G}}}{\sum}{\nabla}$,
  def. \ref{AssociatedBundleByRho}, we call the \emph{prequantum $V$-bundle}
  (or just \emph{prequantum line bundle} if $V$ is equipped compatibly with a ring structure).
\end{definition}
\begin{remark}
  If we write $P \to X$ for the 
  total space projection of the prequantum bundle, sitting in the
  $\infty$-pullback diagram 
  $$
    \raisebox{20pt}{
    \xymatrix{
	  P \ar[rr] \ar[d] && {*} \ar[d]
	  \\
	  X \ar[rr]|-{\underset{U_{\mathbf{B}\mathbb{G}}}{\sum}{\nabla}} && \mathbf{B}\mathbb{G}
	}
	}
	\,,
  $$
  then by prop. \ref{AssociatedBundleByRho} 
  the total space projection of the prequantum line bundle is the left morphism
  in the $\infty$-pullback diagram
  $$
    \raisebox{20pt}{
    \xymatrix{
	  P \times_{\mathbb{G}} V \ar[rr] \ar[d] && V/\!/\mathbb{G} \ar[d]^{\rho}
	  \\
	  X \ar[rr]|-{\underset{U_{\mathbf{B}\mathbb{G}}}{\sum}{\nabla}} && \mathbf{B}\mathbb{G}
	}
	}
	\,.
  $$
\end{remark}
\begin{definition}
  The space of sections, def. \ref{TwistedCohomologyBySections}, of the prequantum line bundle
  $$
    \mathbf{\Gamma}_X(E) 
	\in \mathbf{H}
  $$
  we call the \emph{prequantum space of states}.
  \label{PrequantumStates}
\end{definition}
\begin{remark}
  By prop. \ref{ObjectOfSectionsAsMappingSpace} the prequantum space of 
  states is equivalently expressed as
  $$
    \mathbf{\Gamma}_X(E)
	\simeq
	\underset{\mathbf{B}\mathbb{G}}{\prod}
	\left[
	  \underset{U}{\sum} \nabla, \rho
	\right]
	\,.
  $$
  \label{PrequantumStatesAsMappingSpace}
\end{remark}

\paragraph{Prequantum operators}
\index{structures in a cohesive $\infty$-topos!prequantum operator}
\label{PrequantumOperatorsInCohesion}

\begin{definition}
  The \emph{prequantum operator action} of the quantomorphism group $\mathbf{QuantMorph}(\nabla)$, def. \ref{QuantomoprhismGroup},
  on the space of prequantum states $\mathbf{\Gamma}_X(E)$, def. \ref{PrequantumStates},
  is the action, def. \ref{RepresentationOfInfinityGroup},  
  $$
    \xymatrix{
	  \mathbf{\Gamma}_X(E) \ar[r] & \mathbf{\Gamma}_X(E) /\!/ \mathbf{QuantMorph}(\nabla)
	  \ar[d]^{\rho_{\mathrm{prequant}}}
	  \\
	  & \mathbf{B}\mathbf{QuantMorph}(\nabla)
	}
  $$
  given by the canonical precomposition action, example \ref{AutXActionOnXY}, 
  of $\mathbf{Aut}_{\mathbf{H}}(\underset{U}{\sum}\nabla)$ on 
  $\mathbf{\Gamma}_X(E) \simeq \underset{\mathbf{B}\mathbb{G}}{\prod} \left[\underset{U}{\sum}\nabla, \rho\right]_{\mathbf{H}}$ (remark \ref{PrequantumStatesAsMappingSpace})
  restricted to a $\mathbf{QuantMorph}(\nabla) := \mathbf{Aut}_{\mathbf{H}}(\nabla)$-action, 
  def. \ref{InducedRepresentation}, along the canonical morphism
  $p_U : \mathbf{Aut}_{\mathbf{H}}(\nabla) \to \mathbf{Aut}_{\mathbf{H}}(\underset{U}{\sum}\nabla)$.
  \label{PrequantumOperatorAction}
  \label{PrequantumOperators}
\end{definition}
\begin{remark}
  The prequantum operator action of def. \ref{PrequantumOperatorAction}
  is exhibited by the following pasting diagram of $\infty$-pullback squares.
  $$
    \xymatrix{
	 \mathbf{\Gamma}_X(E) \simeq
	  \underset{\mathbf{B}\mathbb{G}}{\prod}[\underset{U}{\sum}\nabla, \rho]
	  \ar[r]
	  \ar[dd]
	  &
	  \underset{\mathbf{B}\mathbb{G}}{\prod} 
	  \left(
	    \left[\underset{U}{\sum}\nabla, \rho\right]
	    /\!/\underset{U}{\prod}\mathbf{Aut}(\nabla)
	  \right)
	  \ar[rr]
	  \ar[d]
	  && \underset{\mathbf{B}\mathbb{G}}{\prod} 
	   \left(
	     \left[\underset{U}{\sum}\nabla, \rho\right]
	     /\!/ \mathbf{Aut}(\underset{U}{\sum} \nabla)
        \right)
		\ar[d]
	  \\
	  & \mathbf{B} \underset{\mathbf{B}\mathbb{G}}{\prod} 
	  \left(\underset{U}{\prod} \mathbf{Aut}\left(\nabla\right)\right)
	  \ar[rr]|{\mathbf{B}\underset{\mathbf{B}\mathbb{G}}{\prod} (p_{U})}
	  \ar@{=}[d]
	  &&
	  \mathbf{B}
	  \underset{\mathbf{B}\mathbb{G}}{\prod}
	  \left(
	    \mathbf{Aut}\left( \underset{U}{\sum} \nabla\right)
	  \right)
	  \ar@{=}[d]
	  \\
	  {*}
	  \ar[r]
	  \ar@{=}[d]
	  &
	  \mathbf{B}\left(\mathbf{Aut}_{\mathbf{H}}(\nabla)\right)
	  \ar[rr]
	  \ar@{=}[d]
	  &&
	  \mathbf{B}\left(\mathbf{Aut}_{\mathbf{H}}(\underset{U}{\sum}\nabla)\right)
	  \\
	  {*} \ar[r]& \mathbf{B}\left(\mathbf{QuantMorph}(\nabla)\right)
	}
	\,.
  $$
  This uses that the dependent product is right adjoint and hence preserves $\infty$-pullbacks
  (as well as group structure).
\end{remark}
\begin{remark}
A prequantum state is given by a diagram
$$
  \xymatrix{
    X \ar[rr]^\psi_{\ }="s" \ar[dr]_{\underset{U}{\sum}{\nabla}}^{\ }="t" && V/\!/\mathbb{G} \ar[dl]^{\rho}
	\\
	& \mathbf{B}\mathbb{G}
	\ar@{=>} "s"; "t"
  }
$$
and a prequantum operator by a diagram
$$
  \xymatrix{
    X \ar[rr]^\phi_{\ }="s" \ar[dr]_{\nabla}^{\ }="t" && X \ar[dl]^{\rho}
	\\
	& \mathbf{B}\mathbb{G}_{\mathrm{conn}}
	\ar@{=>}^O "s"; "t"
  }
  \,.
$$
Then the result of the action is the new prequantum state $O(\psi)$ given by the 
pasting diagram
$$
  \xymatrix{
    X \ar[r]^\phi_{\ }="s1" \ar[dr]|\nabla^{\ }="t1" 
	\ar[ddr]_{\underset{U}{\sum}\nabla}
	& X \ar[d]|{\nabla} 
	 \ar[r]^{\psi}_{\ }="s2" & V/\!/\mathbb{G} \ar[ddl]^{\rho}
	\\
	& \mathbf{B}\mathbb{G}_{\mathrm{conn}} \ar[d]|U^{\ }="t2" 
	\\
	& \mathbf{B}\mathbb{G}
  }
$$
(where all the 2-cells are notationally suppressed, for readability).
\end{remark}

\subsubsection{Local prequantum field theory}
\label{LocalPrequantumFieldTheories}
\index{prequantum field theory}

We discuss now a formalization \emph{local prequantum field theory}
(see \ref{BasicClassicalMechanicsByPrequantizedLagrangianCorrespondences}
and \ref{DeDonderWeylTheoryViaHigherCorrespondences} in the introduction) 
in cohesive $\infty$-toposes.

The contents of this section draw from discussion with Domenico Fiorenza.

\medskip

\begin{itemize}
  \item \ref{StructureslpqftIntroduction} -- Introduction;
  \item \ref{BulkFieldTheory} -- Bulk field theory;
  \item \ref{BulkFieldTheory2} -- Local action functionals for the bulk field theory;
  \item \ref{BoundaryFieldTheory} -- Boundary field theory
  \item \ref{CornerFieldTheory} -- Corner field theory
  \item \ref{DefectTheory} -- Defect field theory
\end{itemize}

\paragraph{Introduction}
\label{StructureslpqftIntroduction}

The quantum field theories (QFTs) of interest, both in nature as well as theoretically,
are typically not generic examples of the axioms of quantum field theory 
(see  \cite{QFTIntroduction} for a survey of modern formalizations of QFT)
but rather are special in two respects:
\begin{enumerate}
 \item they arise 
from geometric data -- the Lagrangian and action functional -- 
via some process of quantization, 
and notably from \emph{higher geometric data}
such as Lagrangian densities, pre-symplectic currents and higher gauge fields,
subject to gauge equivalences, 
and higher order gauge of gauge transformations;
\item 
  they are \emph{local} in that the spaces of configurations (states)
  which they assign to a piece of worldvolume/spacetime are determined 
  from gluing the data assigned to pieces of any decomposition of the
  worldvolume/spacetime.
\end{enumerate}

While quantized field theories (topological QFTs as well as non-topological boundary 
quantum field theories) are axiomatically characterized  
by the cobordism theorem \cite{LurieTFT}
(see \cite{Bergner} for a brief survey), here we are after understanding
the axiomatization the local higher geometric \emph{pre-quantum} data of
those quantum field theories that arise from quantization. This 
also proceeds by the cobordism theorem, but with the ``linear''
$n$-categorical coefficients appropriate for quantum field theories replaced
by non-linear geometric $n$-categorical coefficients. Since the natural context for
higher geometry are higher toposes \cite{Lurie} and
specifically cohesive higher toposes  and since, as we will discuss,
local action functionals are naturally objects in slices of such higher toposes
over differential coefficient objects, the $n$-categorical coefficients
that we consider are higher correspondences in such higher slice toposes.

For the case that the ambient higher topos encodes \emph{discrete geometry}
(suitable for the discussion of finite gauge theories such as Dijkgraaf-Witten theory)
the definition of local prequantum field theory that we consider is that indicated in
section 3 of \cite{FHLT}. 

One goal here is to show that by allowing the ambient
higher topos to be more general and in particular by choosing differentially cohesive
higher toposes, the genuine differential geometric data familiar from
general field theories is naturally captured and usefully analyzed.

\subparagraph{Action functionals and correspondences}
\label{ActionFunctionalsAndCorrespondences}

Traditionally in physics one considers 
(smooth) spaces of trajectories of physical fields 
(``spaces of histories''), which we will denote by 
$\mathbf{Fields}_{\mathrm{traj}}$, and considers smooth functions on these
spaces valued in the circle group, called the \emph{exponentiated action functionals}
or the \emph{phases}
$$
  \exp\left(
    \tfrac{i}{\hbar}
    S
  \right)
  \;:\;
  \mathbf{Fields}_{\mathrm{traj}}
  \longrightarrow
  U(1)
  \,,
$$
where $2 \pi \hbar \in \mathbb{R}$ denotes the choice of isomorphism
$$
  U(1) \simeq \mathbb{R}/2\pi\hbar \mathbb{Z}
  \,,
$$
of the circle group with the quotient of the additive group of real numbers by 
a copy of the integers, which physically is ``Planck's constant'', 
see def. \ref{PlanckConstant} below.
By the \emph{principle of extremal action} the critical locus of such functionals
encodes those trajectories which are realized in macroscopic physics (classical physics), 
while  integrals over trajectory space (``path integrals'') of such functionals
are to produce the integral kernels that enocde the microscopic dynamics (quantum mechanics).

For example for $X$ a smooth manifold to be thought of as spacetime, 
and for $\nabla$ a circle-principal connection on $X$, to be thought of 
as an electromagnetic field, then the Lorentz force inter-action between
a charged particle that travels around loops $S^1 \longrightarrow X$ in 
spacetime and the background electromagnetic field is encoded by the
holonomy functional
$$
  \exp\left(
    \tfrac{i}{\hbar}S_{\mathrm{Lor}}^\nabla
  \right)
  :=
  \mathrm{hol}_\nabla
  :
  [S^1, X]
  \longrightarrow
  U(1)
  \,,
$$
where $[S^1, X]$ denotes the loop space of $X$ regarded as a smooth space
(for instance as a Fr{\'e}chet manifold or as a diffeological space)
and where $\mathrm{hol}_\nabla$ is the function that sends a curve in $X$
to its holonomy under $\nabla$.

More generally, action functionals are in fact not $U(1)$-valued functions,
but are sections of $U(1)$-principal bundles. 
To say this more formally, we introduce the notation $\mathbf{B}U(1)$ for the 
universal moduli stack of smooth $U(1)$-principal bundles. This is characterized as being the
object such that for $X$ any smooth manifold then homomorphisms $X \longrightarrow \mathbf{B}U(1)$
are equivalent to smooth $U(1)$-principal bundles on $X$ and homotopies between such 
are equivalently to smooth isomorphisms/gauge transformations between those.
For an introduction into the language of smooth (moduli) stacks that we are using here
see \cite{FiorenzaSatiSchreiberCS}.

As the notation suggests, the characteristic feature of $\mathbf{B}U(1)$ is that 
it is the \emph{delooping} of the group $U(1)$, and the boldface $\mathbf{B}$ is to 
indicate that we consider this with everything equipped with its smooth geometric structure.
This means that $U(1)$ as a smooth Lie group is the homotopy fiber product of the point with
itself inside $\mathbf{B}U(1)$. By the universal property of the homotopy fiber construction
this in turn means that an exponentiated action functional as above is equivalently 
a homotopy from the pullback of the trivial $U(1)$-principal bundle to itself, as follows:
$$
  \raisebox{20pt}{
  \xymatrix{
   & \mathbf{Fields}_{\mathrm{traj}}
   \ar[dr]_{\ }="s"
   \ar[dl]
   \\
   \ast
   \ar[dr]^{\ }="t"
    && \ast
    \ar[dl]
   \\
   & \mathbf{B}U(1)
   \ar@{=>} "s"; "t"
  }}
  \;\;
  \simeq
  \;\;
  \raisebox{20pt}{
  \xymatrix{
   & \mathbf{Fields}_{\mathrm{traj}}
   \ar[dr]_{\ }="s"
   \ar[dl]
   \ar@{-->}[d]|{\exp(\tfrac{i}{\hbar} S)}
   \\
   \ast
   \ar[dr]^{\ }="t"
    & U(1) \ar[l] \ar[r]_{\ }="s2" & \ast \ar[dl]
   \\
   & \mathbf{B}U(1)
   \ar@{=>} "s2"; "t"
  }}
  \,.
$$
Indeed, in the above example of the electromagnetic interaction, if instead
of closed particle trajectories of the shape of a circle we consider trajectories
of the shape of the interval $I := [0,1]$, regarded as a smooth manifold with
boundary $\partial I = \ast \coprod \ast$, then the inter-action functional is not
given by the holonomy but more generally by the \emph{parallel transport} of the connection 
$\nabla$ along paths,
which is not a function but is a section of the oriented pullback of the background bundle
along the path endpoint evaluation map, in that it is a homotopy diagram like this:
$$
  \raisebox{20pt}{
  \xymatrix{
    & [I,X]
    \ar[dl]_{(-)|_0}
    \ar[dr]^{(-)|_{1}}_{\ }="s"
    \\
    X \ar[dr]_{\chi(\nabla)}^{\ }="t" && X \ar[dl]^{\chi(\nabla)}
    \\
    & \mathbf{B}U(1)
    \ar@{=>}^{\mathrm{tra}_\nabla} "s"; "t"
  }
  }
  \,.
$$
Here $\chi(\nabla)$ is the class (or rather the \emph{modulus}/\emph{cocycle}) 
of the $U(1)$-principal bundle underlying the 
$U(1)$-principal connection $\nabla$. We discuss this and its higher dimensional
generalization below in \ref{FiberIntegrationOfOrdinaryDifferentialCocycles}.

This diagram is a \emph{correspondence} from the background field $\chi(\nabla)$ to itself,
regarded as an object in the \emph{slice topos} over $\mathbf{B}U(1)$. 
Since $U(1)$ is the ``group of phases'' in traditional formulations of physics, 
$\mathbf{B}U(1)$ here plays the role of a \emph{higher group of phases}. 
Below we see that such correspondences in slices over higher groups of phases 
serve to encode local pre-quantum field theory quite generally.

\medskip

Another archetypical example for such correspondences 
-- which is \emph{almost} familiar from traditional
literature -- are pre-quantizations of \emph{Lagrangian correspondences} in 
symplectic geometry \cite{Weinstein71, Weinstein83}. 
In this context, consider $(X,\omega)$  a symplectic manifold, 
to be thought of as the phase space of some physical system. In the spirit of the
above discussion we stick to representing all extra structure on spaces in terms
of maps into moduli stacks of these structures, and hence we think of the
symplectic differential 2-form $\omega \in \Omega^2(X)$ here a morphism
$$
  \omega : X \longrightarrow \mathbf{\Omega}^2_{cl}
$$
to the smooth moduli space of closed differential 2-forms (technically this is 
simply the sheaf of closed 2-forms on the site of all smooth manifolds).
In terms of such maps we have for instance that a diffeomorphism 
$\phi : X \longrightarrow X$ is a \emph{symplectomorphism} precisely if
it makes the following diagram of smooth spaces commute:
$$
  \raisebox{20pt}{
  \xymatrix{
    X \ar[rr]^\phi \ar[dr]_\omega && X \ar[dl]^{\omega}
    \\
    & \mathbf{\Omega}^2
  }
  }
  \,.
$$
Equivalently, if we write $(\mathrm{id},\phi) :  \mathrm{graph}(\phi) \hookrightarrow X \times X$
for the graph of the function $\phi$, then $\phi$ is a symplectomorphism precisely
if it induces a correspondence from $\omega$ to itself
regarded as an object in the slice topos over $\mathbf{\Omega}^2$ as follows:
$$
  \raisebox{20pt}{
  \xymatrix{
    & \mathrm{graph}(\phi)
    \ar[dl]_{p_1}
    \ar[dr]^{p_2}
    \\
    X \ar[dr]_{\omega} && X \ar[dl]^{\omega}
    \\
    & \mathbf{\Omega}^2
  }
  }
  \,.
$$
While such \emph{Lagrangian correspondence} have long been studied
and have been proposed as a foundation for geometric quantization
\cite{Weinstein83}, it is well known that a symplectic manifold is 
too crude a model for a physical phase space, and that more accurately
a physical phase space is a ``pre-quantization'' of a symplectic manifold, 
namely a choice of $U(1)$-principal connection $\nabla$ whose curvature
2-form coincides with the symplectic form $F_\nabla = \omega$. 
(See the introduction of \cite{hgp} for a review of geometric prequantization
and for further pointers to the literature.)

In order to see the effect of this refinement on the above discussion,
observe that sending a $U(1)$-principal connection $\nabla$ to its
curvature 2-form $F_\nabla$ is a natural operation, compatible with 
gauge equivalences, 
and hence is given by a universal morphism of stacks
$$
  F_{(-)} \;:\; \mathbf{B}U(1)_{\mathrm{conn}} \longrightarrow \mathbf{\Omega}^2_{\mathrm{cl}}
$$
from the universal moduli stack $\mathbf{B}U(1)_{\mathrm{conn}}$ of $U(1)$-principal
connections to the univseral smooth space of closed differential 2-forms.
In terms of this a prequantization of a symplectic manifold $(X,\omega)$
is a lift $\nabla$ in the diagram
$$
  \raisebox{20pt}{
  \xymatrix{
    X
    \ar[dr]_\omega
    \ar@{-->}[r]^-\nabla
    &
    \mathbf{B}U(1)_{\mathrm{conn}}
    \ar[d]^{F_{(-)}}
    \\
    & \mathbf{\Omega}^2
  }
  }
  \,.
$$
In view of this it is clear what a \emph{pre-qauntized Lagrangian correspondence}
should be: this is a lift of the above Lagrangian correspondence through the
universal curvature homomorphism to a correspondence in the slice over 
$\mathbf{B}U(1)_{\mathrm{conn}}$
of the form
$$
  \xymatrix{
    & \mathrm{graph}(\phi)
    \ar[dl]
    \ar[dr]_{\ }="s"
    \\
    X \ar[dr]|{\nabla}^{\ }="t" 
    \ar@/_1pc/[ddr]_{\omega}
    && 
    X \ar[dl]|\nabla
    \ar@/^1pc/[ddl]^{\omega}
    \\
    & \mathbf{B}U(1)_{\mathrm{conn}}
    \ar[d]^{F_{(-)}}
    \\
    & \mathbf{\Omega}^2
    \ar@{=>} "s"; "t"
  }
  \,.
$$
While this is an obvious refinement of the traditional notion of Lagrangian 
correspondence, it does not seem to have found due attention in the existing
literature. Its relevance may be seen from the following observation \cite{hgp}
which we discuss in more detail below in \ref{ClassicalLocalFieldTheory}:
Smooth 1-parameter flows of prequantized Lagrangian correspondences
as above are given precisely by choices of smooth functions 
$H \in C^\infty(X)$, where such a function induces the flow that 
sends $t \in \mathbb{R}$ to the correspondence
$$
  t 
  \;\;\;
   \mapsto
  \;\;\;\;\;\;
  \raisebox{30pt}{
  \xymatrix{
    & \mathrm{graph}(\exp(t\{H,-\}))
    \ar[dl]
    \ar[dr]_{\ }="s"
    \\
    X \ar[dr]_{\nabla}^{\ }="t" && X \ar[dl]^\nabla
    \\
    & \mathbf{B}U(1)_{\mathrm{conn}}
    \ar@{=>}^{\exp(i S_t)} "s"; "t"
  }}
  \,,
$$
where $\exp(t\{H,-\})$ is the \emph{Hamiltonian flow} induced by $H$
and $S_t = \int_0^t L d t$ is its \emph{Hamilton-Jacobi action}, namely
the integral of the \emph{Lagrangian} $L$ which is the Legendre transform
of $H$. Hence the notion of flows of Lagrangian correspondences unifies
a fair bit of traditional classical mechanics \cite{Arnold}.
We survey in \ref{ClassicalLocalFieldTheory} how when this is lifted to 
Lagrangian correspondences between prequantum $n$-bundles
for $n \in \mathbb{N}$ 
as in \cite{hgp}, then $n$-dimensional flows in $n$-fold correspondences
encode the equations of motion of local Lagrangians on jet spaces
in deDonder-Weyl-Hamiltonian (``multisymplectic'') formulation.

In summary, the description of classical mechanics
here identifies prequantized Lagrangian correspondences 
schematically as follows:
$$
  \xymatrix{
    & \mathrm{graph}\left(\exp\left(t\{H,-\}\right)\right)
	\ar[dl]
	\ar[dr]_{\ }="s"
	&
	&&
    & \mbox{\begin{tabular}{c}space of \\ trajectories\end{tabular}}
	\ar[dl]_{\mbox{\small \begin{tabular}{c}initial \\ values\end{tabular}}}
	\ar[dr]^{\mbox{\small \begin{tabular}{c}Hamiltonian \\ evolution\end{tabular}}}_{\ }="s2"
	\\
	X \ar[dr]_{\nabla_{\mathrm{in}}}^{\ }="t"
	  && 
	X \ar[dl]^{\nabla_{\mathrm{out}}}
	&&
	\mbox{\begin{tabular}{c}phase space\\ of incoming fields\end{tabular}}
	\ar[dr]_{\mbox{\small \begin{tabular}{c}prequantum \\ bundle \end{tabular}}}^{\ }="t2"
	&&
	\mbox{\begin{tabular}{c}phase space of \\ outgoing fields\end{tabular}}
	\ar[dl]^{\mbox{\small \begin{tabular}{c}prequantum \\ bundle \end{tabular}}}
	\\
	& \mathbf{B}U(1)_{\mathrm{conn}}
	&
	&&
    & \mbox{\begin{tabular}{c}higher group \\ of phases\end{tabular}}
	\ar@{=>}|{\exp\left(\frac{i}{\hbar} S_t\right) = \exp\left(\frac{i}{\hbar}\int^t_0 L dt\right)} "s"; "t"
	\ar@{=>}|{\mbox{\small \begin{tabular}{c} action \\ functional \end{tabular}}} "s2"; "t2"
  }
$$
This state of affairs turns out to be essentially a blueprint for the formulation of local prequantum
field theory that we obtain below in \ref{LocalPrequantumFieldTheories}, via maps from cobordisms to $n$-fold correspondences
in higher slices toposes.

\subparagraph{Local Lagrangians and higher differential cocycles}

To see the need for passing from traditional symplectic geometric
and prequantum bundles to prequantum $n$-bundles, first observe the traditional
formulation of higher dimensional field theory along the above lines. 
Let $\mathbf{Fields}$ be a moduli space/moduli stack of fields of some field
theory -- for instance $\mathbf{Fields} = \mathbf{B}G_{\mathrm{conn}}$ the 
universal moduli stack of $G$-principal connections of some Lie group $G$,
for the case of $G$-gauge theory.  Then over a closed manifold 
$\Sigma_{n-1}$ of dimension $(n-1)$, to be thought of as a spatial slice of spacetime,
the space of field configurations on $\Sigma_{n-1}$ is the mapping stack
$\mathbf{Field}(\Sigma_{n-1}) = [\Sigma_{n-1}, \mathbf{Fields}]$
(or some slight variant of this, such as its ``differential concretification'' \cite{hgp},
see the examples below in \ref{ExamplesAndApplications} for more).
Now the evolution of fields on $\Sigma_{n-1}$ in time is a trajectory 
given by a map
$$
  I \longrightarrow [\Sigma_{n-1}, \mathbf{Fields}]
$$
which by the internal hom-adjunction is equivalently a field configuration
$$
  \phi \in [\Sigma_{n-1}\times I, \mathbf{Fields}]
$$
on the cylinder over $\Sigma_{n-1}$. Hence the $n$-dimensional field theory 
\emph{transgressed} to maps out of $\Sigma_{n-1}$ looks like a mechanical system
with space of fields being the mapping space $[\Sigma_{n-1}, \mathbf{Fields}]$.

For instance if the field theory is the $(n = p+1)$-dimensional 
worldvolume theory of a $p$-brane which 
is charged under a $(p+1)$-form connection $\nabla$, then the action functional
over such cylinders is of the same general form as that for electrically charged
particles above
$$
  \raisebox{20pt}{
  \xymatrix{
    & \left[I,\left[\Sigma_{p}, \mathbf{Fields}\right]\right]
    \ar[dl]_{(-)|_0}
    \ar[dr]^{(-)|_{1}}_{\ }="s"
    \\
    [\Sigma_{p},\mathbf{Fields}] 
    \ar[dr]_{\chi\left(\exp(\tfrac{i}{\hbar}\int_{\Sigma_{n-1}}\nabla\right)}^{\ }="t" 
    && 
    [\Sigma_{p}, \mathbf{Fields}] 
    \ar[dl]^{\chi\left(\exp(\tfrac{i}{\hbar}\int_{\Sigma_{n-1}}\nabla\right)}
    \\
    & \mathbf{B}U(1)
    \ar@{=>}^{\mathrm{tra}_\nabla} "s"; "t"
  }
  }
  \,,
$$
where now $\exp\left(\tfrac{i}{\hbar}\int_{\Sigma_p}\nabla\right)$ is an ordinary 1-form connection
on the mapping space $[\Sigma_p, \mathbf{Fields}]$, obtained by \emph{transgression}
of the given $(p+1)$-form connection $\nabla$ on the moduli space of fields itself.


While in this fashion all $n$-dimensional field theories may be thought of in terms of
mechanics (1-dimensional field theory) on the space of fields over $(n-1)$-dimensional
spatial slices, restricting to this perspective alone loses the manifest \emph{locality} of the
theory: the data for codimension-1 manifolds $\Sigma_{n-1}$ is not necessarily represented as 
obtained by gluing data on smaller patches.
In physics terminology, essentially this problem is known as the
problem of the \emph{non-covariance of canonical quantization}, referring to the 
explicit and non-natural choice of $(n-1)$-dimensional spatial slices 
$\Sigma_{n-1}$ of spacetime.

Imposing locality then amounts to requiring that all the data of the $n$-dimensional theory can be 
reconstructed by the data for codimension-$n$ manifolds, hence
for collections of just points. To continue the pattern of phases $U(1)$
and higher phases $\mathbf{B}U(1)_{\mathrm{conn}}$ that we have seen emerging 
in codimension-0 and 1, 
one sees that the natural codimension-$k$ datum for a $n$-dimensional prequantum theory is that of a 
morphism of stacks of the form
\[
[\Sigma_{n-k},\mathbf{Fields}] \longrightarrow \mathbf{B}^{n-k}U(1)_{\mathrm{conn}}\;,
\]
where on the right we have the $(n-k)$-stack of 
$(n-k)$-form connections on higher $(n-k)$-circle bundles
(bundle $(n-k-1)$-gerbes with connection). An introduction to this perspective is in 
\cite{FiorenzaSatiSchreiberCS}.

Going down to codimension $n$ 
and observing that if $\ast$ denotes the 1-point manifold then $[*,\mathbf{Fields}]\cong\mathbf{Fields}$, we see that imposing locality on a prequantum theory means that the whole theory, in any codimension, is determined by a single datum: 
a morphism of higher stacks of the form
\[
  \mathbf{L} \;:\; \mathbf{Fields}\longrightarrow \mathbf{B}^{n}U(1)_{\mathrm{conn}}
  \,.
\]
Notice that such an $n$-connection on the moduli stack of fields is 
locally given by a differential $n$-form. Moreover, this being an $n$-form
on a stack means that for each test manifold $\Sigma$ this is an $n$-form (locally)
on $\Sigma$, depending on the field configurations on $\Sigma$. Such a form is
familiar in, and central to, traditional prequantum field theory. It is the 
\emph{Lagrangian} of the theory; whence the choice of symbol ``$\mathbf{L}$''.

\medskip
Indeed, once such an $\mathbf{L}$ is given, all the codimension-$k$ prequantum $(n-k)$-$U(1)$-bundles with connections on the moduli stacks $[\Sigma_{n-k},\mathbf{Fields}]$ 
are naturally obtained by transgression
of $n$-bundles (fiber integration/push-forward on cocycles in differential cohomology):
\[
  \exp\left( 
    \tfrac{i}{\hbar} \underset{\Sigma_{n-k}}{\int} \mathbf{L}	
   \right)
   :
  [\Sigma_{n-k},\mathbf{Fields}]
    \xrightarrow{[\Sigma_{n-k},\mathbf{L}]} 
  [\Sigma_{n-k},\mathbf{B}^{n}U(1)_{\mathrm{conn}}]
   \xrightarrow{\exp\left( 
     \tfrac{i}{\hbar} \underset{\Sigma_{n-k}}{\int} (-)
    \right)}
  \mathbf{B}^{n-k}U(1)_{\mathrm{conn}}\;.
\]
The rightmost map here is fiber integration in Deligne cohomology, seen as morphism of smooth stacks, 
this we describe below in 
\ref{FiberIntegrationOfOrdinaryDifferentialCocycles}.  In particular, for $k=0$ one recovers the action functional as
$$
  \exp\left(\tfrac{i}{\hbar} S_{\Sigma_n}\right)
    =
  \exp\left( \tfrac{i}{\hbar} \int_{\Sigma_n} \mathbf{L} \right)
  \;:\;
 [\Sigma_n, \mathbf{Fields}]\longrightarrow \mathbf{B}^0 U(1)_{\mathrm{conn}} \simeq U(1)\;.
$$

\medskip
The universal curvature morphisms 
\[
  \mathrm{curv}
   \;:\;
    \mathbf{B}^{n-k} U(1)_{\mathrm{conn}}
    \longrightarrow 
    \mathbf{\Omega}^{n-k+1}_{\mathrm{cl}}
\]
endow the moduli spaces of field configurations with canonical closed degree $n-k+1$ differential forms.
In the traditional case, if $\mathbf{Fields}$ here is the jet bundle to a field bundle, then 
this is the \emph{pre-symplectic current density} known from the
``covariant phase space'' formulation of classical field theory \cite{Zuckerman, CrnkovicWitten}.
The pre-quantum theory of such ``multisymplectic'' or ``$n$-plectic'' structure
has been described systematically in \cite{hgp}.
For $k=1$ this is the traditional pre-symplectic structure on $[\Sigma_{n-1},\mathbf{Fields}]$, 
so the ``local prequantization'' can be seen as a  \emph{de-transgression} of this
pre-symplectic structure to a pre-$n$-plectic structure on the stack of fields.

In this fashion we consider here differential $n$-cocycles 
$\mathbf{L} : \mathbf{Fields} \longrightarrow \mathbf{B}^n U(1)_{\mathrm{conn}}$
on higher moduli stacks as pre-quantized local Lagrangians for $n$-dimensional
field theories. More precisely, these define ``bulk'' field theories
on $n$-dimensional worldvolumes/spacetimes without physical boundaries
or other singularities (``defects'').

\subparagraph{Boundary field theory and twisted relative cohomology}
\index{prequantum field theory!boundary field theory}
\index{boundary field theory}

We observe in  \ref{BoundaryFieldTheory} below that by the full cobordism theorem
in the presence of boundaries and singularities, a codimension-1 boundary condition for
a local prequantum field theory $\mathbf{L} : \mathbf{Fields} \to \mathbf{B}^n U(1)_{\mathrm{conn}}$ 
as above is equivalently the data 
of a correspondence of the form
$$
  \raisebox{20pt}{
  \xymatrix{
    & \mathbf{Fields}_{\mathrm{bdr}}
    \ar[dl]
    \ar[dr]_{\ }="s"
    \\
    \ast \ar[dr]_0^{\ }="t" 
    && 
    \mathbf{Fields} 
      \ar[dl]^{\mathbf{L}} 
    \\
    & \mathbf{B}^n U(1)_{\mathrm{conn}}
    \ar@{=>} "s"; "t"
  }
  }
  \,,
$$
hence a choice of boundary fields $\mathbf{Fields}_{\mathrm{bdr}}$, a choice of map
from boundary fields into bulk fields, and a choice of trivialization of the 
pre-quantized bulk field Lagrangian after restriction to the boundary fields.

The prototypical example of this is the relation 
between 3d \emph{Chern-Simons theory} and 4d ``univresal topological Yang-Mills theory'',
which we discuss below in \ref{HigherCSAsLocalPrequantumFT}.
That 3d Chern-Simons theory is a theory which ultimately 
deals with boundaries of 4-manifolds is something coming from the very origin 
of the theory \cite{ChernSimons}.  In the language of smooth moduli stacks \cite{hgp} 
this is completely formalized and summarized in the following (homotopy) commutative diagram

  $$
    \raisebox{20pt}{
    \xymatrix{
	  & \mathbf{B}G_{\mathrm{conn}}
	  \ar@/_.3pc/[ddl]
	  \ar@/^.3pc/[ddr]^{\langle F_{(-)}\wedge F_{(-)}\rangle}
	  \ar@{-->}[d]^{\mathbf{cs}}
	  \\
	  & \mathbf{B}^3 U(1)_{\mathrm{conn}}
	  \ar[dl]
	  \ar[dr]|-{F_{(-)}}
	  \\
	  \ast \ar[dr]_0 &
	  \mbox{\tiny (pb)}
	  & \Omega^{4}_{\mathrm{cl}}~~\;, \ar[dl]^{ \mathbf{L}_{\mathrm{tYM}} }
	  \\
	  & \flat \mathbf{B}^{4}U(1)
	}
     }	
  $$
  where $\mathbf{B}G_{\mathrm{conn}}$ is the stack of principal $G$-bundles with connection for a compact simple and simply connected Lie group $G$, 
  $$
    \langle F_{(-)}\wedge  F_{(-)}\rangle
	: 
	\mathbf{B}G_{\mathrm{conn}} \to \Omega^{4}_{\mathrm{cl}}
  $$
   is the 
  \emph{Chern-Weil} 4-form representing the fundamental degree four characteristic class of $G$, and 
  \[
  \mathbf{cs}:\mathbf{B}G_{\mathrm{conn}}\to \mathbf{B}^3U(1)_{\mathrm{conn}}
  \]
  is the Chern-Simons action functional lifted to a morphism of stacks from $\mathbf{B}G_{\mathrm{conn}}$ 
  to the 3-stack of $U(1)$-3-bundles with connection (see \cite{FiorenzaSatiSchreiberCS} for details). 
  In the lower part of the diagram,
  \[
    \mathbf{L}_{\mathrm{tYM}}
     \;:\; \Omega^{4}_{\mathrm{cl}} \longrightarrow \flat \mathbf{B}^{4}U(1)
  \]
  is the canonical embedding of closed 4-forms into the stack of flat $U(1)$-4-bundles with connection. 
  Here we are denoting it by the symbol 
  $\mathbf{L}_{\mathrm{tYM}})$ since we are physically interpreting it as the 
  the Lagrangian of
  \emph{topological 4d Yang-Mills theory}. The lower part of the diagram is what exhibits 3d Chern-Simons as a boundary theory for 4d topological Yang-Mills. More precisely, since the lower part of the diagram is a homotopy pullback, it exhibits $\mathbf{B}^3U(1)_{\mathrm{conn}}$ as the \emph{universal} boundary condition for 4d topological Yang-Mills. we will come back to this in detail in Section \ref{tYM}.

\medskip
Finally, in a fully extended field theory, going from the bulk to the boundary is only the 
first step: one can go in higher codimension to boundaries of boundaries (or corners) or 
consider high codimension submanifolds of the bulk. For instance, in 4d topological Yang-Mills, 
this is the way Wess-Zumino-Witten theory and and Wilson loop actions appears as a 
codimension-2 corner theory and as codimension-3 defects, respectively. We will recover these 
as examples of more general corner and defect theories in 
Section \ref{HigherWZWLocalPrequantumFieldTheory}. 

A list of examples of twisted boundary fields is discussed in detail below in \ref{TwistedStructures}.

\paragraph{The local bulk fields}
\label{BulkFieldTheory}

%
\begin{definition}
  For $n \in \mathbb{N}$, we write 
  $
    \mathrm{Bord}_n^\otimes 
  $
  for the symmetric monoidal $(\infty,n)$-category of $n$-dimensional framed cobordism.
   For $\mathcal{C}^\otimes$ any symmetric monoidal $(\infty,n)$-category, 
  a \emph{local topological field theory} in dimension $n$
  with coefficients in $\mathcal{C}$
  is a symmetric monoidal $(\infty,n)$-functor 
  $$
    Z : \mathrm{Bord}_n^\otimes \to \mathcal{C}^\otimes
	\,.
  $$
\end{definition}
By the cobordism hypothesis, $\mathrm{Bord}_n^\otimes$ is the free symmetric monoidal $(\infty,n)$-category with full duals generated by a single object $*$. This means that an $n$-dimensional local topological field theory is completely determined by the object $Z(*)$ in $\mathcal{C}^\otimes$ and that this object is necessarily a fully dualizable object.

\begin{remark}
  Note the slight notational difference to \cite{LurieTFT}:
  there the undecorated symbols ``$\mathrm{Bord}_n$'' denote 
  \emph{framed} cobordisms.
\end{remark}

%
%

The  following definition is sketched in section 3.2 of \cite{LurieTFT}
(there written ``$\mathrm{Fam}_n$'' instead of ``$\mathrm{Corr}_n$''.) 
\begin{definition}\label{def.spanH}
  Write
  $$
    \mathrm{Corr}_1 := \left\{ \xymatrix{i \ar@{<-}[r] & c \ar@{->}[r] & o }\right\}
  $$
  for the {\it category free on a single correspondence}, i.e. consisting of three objects 
  and two non-identity morphisms from one to the other two. 
  For $n \in \mathbb{N}$ write
  $$
    \mathrm{Corr}_n := (\mathrm{Corr}_1)^{\times^n}
  $$
  for the $n$-fold cartesian product of this category with itself.
Finally, $\mathrm{Corr}_n(\mathbf{H})$ 
  is the $\infty$-groupoid of functors from $ \mathrm{Corr}_n$ to $\mathbf{H}$. 
 \end{definition} 
  \begin{remark}
  Under composition of correspondences by fiber product of maps to a common face,
  this naturally carries the structure of an $n$-fold category object in 
  $\mathbf{\infty\textbf{-}Grpd}$, hence
  of an $(\infty,n)$-category. Moreover, 
from the 
  cartesian product in $\mathbf{H}$ the $(\infty,n)$-category
  $\mathrm{Corr}_n(\mathbf{H})$ inherits a natural structure of  
  symmetric monoidal $(\infty,n)$-category, which we will denote $
    \mathrm{Corr}_n(\mathbf{H})^\otimes$
\label{remark SpanH}
\end{remark}
\begin{example}
By definition, $\mathrm{Corr}_n$ is the terminal category, so a 0-morphism (i.e., an object) in $\mathrm{Corr}_n(\mathbf{H})$ is just an object in $\mathbf{H}$. 
  A 1-morphism in $\mathrm{Corr}_n(\mathbf{H})$ is a diagram in $\mathbf{H}$
  of the form
  $$
    \xymatrix{
	  A_{i}
	  &
	  A_c
	  \ar[l]
	  \ar[r]
	  &
	  A_o
	}
	\,.
  $$
  In the application to prequantum field theory such a diagram is typically 
  interpreted as follows: $A_i$ is a moduli stack of fields on an {\it i}ncoming
  piece of worldvolume and $A_o$ that of field on an {\it o}utgoing piece.
  The object $A_c$ is that of fields on a piece of worldvolume {\it c}onnecting
  these two pieces, putting them in {\it c}orrespondence, hence $A_c$ is the 
  collection of \emph{trajectories} of field configurations from the incoming to the outgoing piece. The left map sends such a trajectory to its initial configuration, the right
  one to its final configuration. A 2-morphism in $\mathrm{Corr}_n(\mathbf{H})$ is a diagram in $\mathbf{H}$
  of the form
   $$
  \raisebox{20pt}{
    \xymatrix{
      A_{ii} & A_{ic} \ar[l] \ar[r] & A_{io}
	  \\
	  A_{ci} \ar[u] \ar[d] & A_{cc} \ar[u] \ar[d] \ar[l] \ar[r] & A_{co}
	  \ar[u] \ar[d]
	  \\
	  A_{oi} & A_{oc} \ar[l] \ar[r] & A_{oo}\;,
    }
  }
  $$
  and so on.
  Composition of morphisms is via homotopy fiber products in $\mathbf{H}$. For instance, the composition of the two 1-morphisms
  \[
    \xymatrix{
	  X
	  &
	  Y
	  \ar[l]
	  \ar[r]
	  &
	  Z
	}
	\,\quad\text{and}\quad
	 \xymatrix{
	  Z
	  &
	  S
	  \ar[l]
	  \ar[r]
	  &
	  T
	}
  \]
  is the 1-morphism
  \[
  \xymatrix{
	  X
	  &
	  Y\times_Z S
	  \ar[l]
	  \ar[r]
	  &
	  T
	}
	\,. 
  \]
In the above interpretation of these correspondences in prequantum field theory, this
  operation corresponds to gluing or concatenating trajectories of field configurations
  whenever they match over their outgoing/ingoing pieces of worldvolume, respectively.  The compositions of higher morphisms are defined analogously.
%
\end{example}

\medskip
The following proposition appears essentially as
remark 3.2.3 in \cite{LurieTFT}. We spell of some details of the proof.
\begin{proposition}\label{prop.everything-dualizable}
  For all $n \in \mathbb{N}$,
  every object $ X \in \mathrm{Corr}_n(\mathbf{H})^\otimes$ is fully dualizable and 
  is in fact its
  own full dual. The $k$-dimensional trace of the identity on 
  $X$ in $\mathrm{Corr}_n(\mathbf{H})^\otimes$ is its
  free $k$-sphere space object:
  $$
    \mathrm{dim}_k(X) \simeq [\Pi(S^k), X]
	\,,
  $$
  seen as a $k$-fold correspondence from the terminal object to itself.  \label{EveryObjectinSpanIsDualizable}
\end{proposition}
\proof
  Let $X \in \mathbf{H} \hookrightarrow \mathrm{Corr}_n(\mathbf{H})$
  be any object.
  
  The first step is to exhibit $X$ as the ordinary dual of itself.
  For this,
  take the co-evaluation and evaluation morphisms
  $\epsilon : \mathbb{I} \to X \times X$ and $\eta : X \times X \to \mathbb{I}$
  to be given by the ``\textsf{C}'' and by the ``\reflectbox{\textsf{C}}", i.e.
  in $\mathbf{H}$ by the 
  correspondences
  \[
   \xymatrix{
	  \ast
	  &
	  X
	  \ar[l]
	  \ar[r]^-{\Delta_X}
	  &
	  X\times X
	}
  \qquad
  \text{and}
  \qquad
  \xymatrix{
	  X\times X
	  &
	  X
	  \ar[l]_-{\Delta_X}
	  \ar[r]	  &
	  \ast
	}\;,
   \]
%
  where $\Delta_X$ denotes the diagonal map for $X$. 
  Notice that this 
  diagonal map is equivalent to the evaluation at the two endpoints of the 
  interval (1-disk) $\Pi(D^1)$ in the mapping space $[\Pi(D^1), X]$, 
  so that 
  $\epsilon$ is equivalent to 
  $$
    \xymatrix{\ast & [\Pi(D^1), X] \ar[l] \ar[rr]^{({\rm ev}_0 , {\rm ev}_1)} &&X\times X}\;,
  $$
  and similarly for $\eta$. 
  
  For $\epsilon$ and $\eta$ to exhibit a self-duality, the zig-zig-identities
  $$
    \xymatrix{
	  X 
	    \ar[r]^-{X \times \epsilon}_>{\ }="s"
		\ar@/_1pc/[rr]_-{\mathrm{id}_X}^{\ }="t"
		&
	  X \times X \times X
	  \ar[r]^-{\eta \times X}
	  &
	  X
	  %
	}
	 \quad {\rm and}
	\quad
	\xymatrix{
	  X \ar[r]^-{\epsilon \times X}
	  \ar@/_1pc/[rr]_{\mathrm{id}_X}
	  &
	  X \times X \times X
	  \ar[r]^-{\eta \times X}
	  &
	  X
	}
  $$
  have to hold as diagram in $\mathrm{Corr}_n(\mathbf{H})$. 
  Indeed, as a composite of correspondences this is given in $\mathbf{H}$ by
  $$
    \xymatrix{
	  && X
	  \ar@/_1.7pc/[ddll]_{\mathrm{id}_X}
	  \ar@/^1.7pc/[ddrr]^{\mathrm{id}_X}
	  \ar[dl]|{\Delta_X}
	  \ar[dr]|{\Delta_X}
	  \\
	  & X \times X \ar[dl]|{p_1} \ar[dr]|{(\mathrm{id}_X, \Delta_X)} 
	  &
	  \mbox{\tiny (pb)}
	  & 
	  X \times X
	  \ar[dl]|{(\Delta_X, \mathrm{id}_X)}
	  \ar[dr]|{p_2}
	  \\
	  X
	  &&
	  X \times X \times X
	  &&
	  X\;, 
	}
  $$
  and similarly for the other composite. As a consequence, the trace of the identity of $X$ 
  $$
    \mathrm{tr}(\mathrm{id}_X) := 
	\xymatrix{
	  \mathbb{I}
	  \ar[r]^-\epsilon 
	  & 
	  X \times X
	  \ar[r]^-\eta
	  &
	  \mathbb{I}
	}
  $$
  is given by the correspondence
  $$
    \raisebox{20pt}{
    \xymatrix{
	  && \mathcal{L}X=[\Pi(S^1), X]
	  \ar[dl]
	  \ar[dr]
	  \\
	  & X=[\Pi(D^1), X] \ar[dl]
	  \ar[dr]|{\Delta_X} 
	  &\mbox{\tiny (pb)}&
	  X=[\Pi(D^1), X] \ar[dl]|{\Delta_X} \ar[dr]
	  \\
	  \ast && X \times X && \ast\;,
	}
	}
	\,.
  $$
  Hence
  \[
  \mathrm{dim}_1(X) \simeq \mathcal{L}X\simeq [\Pi(S^1), X]\;,
  \]
  which amounts to the pictorial identity
  $
  \reflectbox{\textsf{C}}\circ \textsf{C} \cong  \textsf{O}$.

  Next, to exhibit the self-duality $(\epsilon,\eta)$ on $X$ 
  as a full duality, we need
  to produce full adjoints $\epsilon^\ast$ and $\eta^\ast$
  of $\epsilon$ and of $\eta$, respectively 
  with units
  $$
  \xymatrix{
    \mathbb{I} 
      \ar[r]^-\epsilon 
      \ar@/^3pc/[rr]^{\ }_{}="s" 
    &
  X \times X
  \ar[r]^-{\epsilon^\ast}
  &
    \mathbb{I}
     \ar@{<=} "s"; "s"+(0,-10);
  }
  \;\;\,,\;\;
  \xymatrix{
    \mathbb{I} 
      \ar[r]^-{\eta^\ast} 
      \ar@/^3pc/[rr]^{\ }_{}="s" 
    &
  X \times X
  \ar[r]^-{\eta}
  &
    \mathbb{I}
     \ar@{<=} "s"; "s"+(0,-10);
  }
  $$
  and similar co-units.
  Here we may choose $\eta^\ast := \epsilon$ and $\epsilon^\ast := \eta$
  and we take their unit and its dual
  $$
    \xymatrix{
      \mathbb{I}
      \\
      X \times X
      \ar[u]^{\eta}
      \\
      \mathbb{I}
      \ar[u]^\epsilon
      \ar@/^2.4pc/[uu]^{\mathrm{id}}|{\ }="s"
      \ar@{<=>} "s"; "s"+(5,0)
    }
  $$
  to be the given by the 2-fold correspondences in $\mathbf{H}$ 
  which exhibit the ``cap'' and the ``cup'':
   $$
  \raisebox{40pt}
  {
    \xymatrix{
     \ast & \ast \ar[l] \ar[r] & \ast
	  \\
    \ast \ar[u] \ar[d] & X \ar[u] \ar[d] \ar[l] \ar[r] & [\Pi(S^1), X]
    \ar[u] \ar[d]
	  \\
    \ast & \ast \ar[l] \ar[r] & \ast
    }
  }
  \qquad\text{and}\qquad
   \raisebox{40pt}{
    \xymatrix{
      \ast & \ast \ar[l] \ar[r] & \ast
	  \\
	  [\Pi(S^1), X]\ar[u] \ar[d] & X \ar[u] \ar[d] \ar[l] \ar[r] & \ast
	  \ar[u] \ar[d]
	  \\
	  \ast & \ast \ar[l] \ar[r] & \ast 
    }
  }
  $$
  and take the co-unit and its dual
  $$
    \xymatrix{
      X \times X
      \ar[r]^-\eta
      \ar@/_2pc/[rr]_{\mathrm{id}}^{\ }="t"
      & \mathbb{I}
        \ar[r]^-{\epsilon}
      &
      X\times X
      \ar@{<=>} "t"+(0,5); "t"
    }
  $$
  to be given by the ``saddle'' correspondence\footnote{The picture of the saddle here has
  been stolen from Aaron Lauda's website. Thanks to Domenico Fiorenza and to Hisham Sati for 
  lending a hand with the typesetting here.},
  $$
    \raisebox{78pt}{
    \xymatrix{
	  X \times X&& X \times X
	  \ar[ll]_{\Delta_X \circ p_1} \ar[rr]^{\Delta_X \circ p_2}&& X \times X
	  \\
	  X \times X\ar[u]^{\mathrm{id}_{X \times X}}\ar[d]_{\mathrm{id}_{X \times X}}
	  &&
	  X \ar[ll]|{\Delta_X }
	  \ar[rr]|{\Delta_X }
	  \ar[u]|{\Delta_X}
	  \ar[d]|{\Delta_X}
	  &&
	  X \times X\ar[u]_{\mathrm{id}_{X \times X}}\ar[d]^{\mathrm{id}_{X \times X}}
	  \\
	 X \times X && X \times X
	  \ar[ll]_{\mathrm{id}_{X \times X}} \ar[rr]^{\mathrm{id}_{X \times X}}&& X \times X
	}
	}\;
	\;\;\;\;\;\;\;\;\;\;\;\;
	\raisebox{68pt}{
  \xy 
(20,2)*{}="RU"; (16,-3)*{}="RD"; 
(-16,2)*{}="LU"; (-20,-3)*{}="LD"; 
"RU";"RD" **\crv{(4,2) & (4,-1)}; 
"LD";"LU" **\crv{(-4,-2) & (-4,1)}; 
(7.5,0)*{}="x1"; (-7.5,0)*{}="x2"; 
"x1"; "x2" **\crv{(7,-10) & (-7,-10)}; 
(16,-20)*{}="RDD"; (-20,-20)*{}="LDD"; 
(20,-15)*{}="RUD"; (-16,-15)*{}="LUD"; 
(-16,-2.5)*{}="A"; (16,-15)*{}="B"; 
"RD"; "RDD" **\dir{-}; 
"LD"; "LDD" **\dir{-}; 
"A"; "LUD" **\dir{.}; 
"RDD"; "LDD" **\dir{-}; 
"RU"; "RUD" **\dir{-}; 
"LU"; "A" **\dir{-}; 
"B"; "RUD" **\dir{-}; 
"B"; "LUD" **\dir{.}; 
(0,-8)*{\bullet}; 
\endxy 
  }
  \,.
  $$
 Notice that here the top row of the diagram arises from the fiber product
 composition of correspondences given by
 $$
 \xymatrix{
 && X \times X 
 \ar[dl]_{p_1}
 \ar[dr]^{p_2}
 &&
 \\
 X \times X 
 &
 X 
 \ar[r]
 \ar[l]_-\Delta
 &
 \ast
 &
 X
 \ar[l] 
 \ar[r]^-\Delta 
 &
 X \times X\;.
 }
 $$
 The zig-zag identity for these
$$
\raisebox{0pt}{
\xymatrix{ 
  \mathbb{I}
    \ar[r]^-\epsilon
    \ar@/^3pc/[rr]^{\quad}_{}="0" 
& X \times X
\ar[r]^-{\eta} 
\ar@/_3pc/[rr]^{\quad}_{}="2" 
\ar@{=>}"0"; {}  
& \mathbb{I}
\ar[r]^-\epsilon
\ar@{=>}{}; "2"  
& X \times X
}}
\qquad
 \simeq
 \qquad
\raisebox{0pt}{
\xymatrix{
\mathbb{I}
\ar@/^2pc/[rr]_{\quad}^{\epsilon}="1" 
\ar@/_2pc/[rr]_{\epsilon}="2" 
&& X \times X
\ar@{}"1";"2"|(.2){\,}="7" 
\ar@{}"1";"2"|(.8){\,}="8" 
\ar@{=>}"7" ;"8"^{\mathrm{id}} }   
}
$$   
 in indeed satisfied, as exhibited by the equivalence of following diagram
 in $\mathbf{H}$, formed from pasting the above diagrams, 
  \[
  \xymatrix{
 \ast
	  &
	  X
	  \ar[l]
	  \ar[rr]|{\Delta_X}
	  && X \times X && X \times X
	  \ar[ll]_{\mathrm{id}_{X \times X}} \ar[rr]^{\mathrm{id}_{X \times X}}&& X \times X
	  \\
	\ast\ar[u]\ar[d]
	  &
	  X\ar[u]|{\mathrm{id}_X}\ar[d]|{\mathrm{id}_X}
	  \ar[l]
	  \ar[rr]|{\Delta_X}
	  && X \times X\ar[u]|{\mathrm{id}_{X \times X}}\ar[d]|{\mathrm{id}_{X \times X}}
	  &&
	  X \ar[ll]|{\Delta_X }
	  \ar[rr]|{\Delta_X }
	  \ar[u]|{\Delta_X}
	  \ar[d]|{\Delta_X}
	  &&
	  X \times X\ar[u]_{\mathrm{id}_{X \times X}}\ar[d]^{\mathrm{id}_{X \times X}}
	  \\
	  \ast
	  &
	  X
	  \ar[l]
	  \ar[rr]|{\Delta_X}
	  && X \times X&& X \times X
	  \ar[ll]|{\Delta_X \circ p_1} \ar[rr]|{\Delta_X \circ p_2}&& X \times X\\
	   \ast\ar@{=}[u]
	  &
	  X\ar@{=}[u]
	  \ar[l]
	  && [\Pi(S^1),X]\times X\ar[rr] \ar[ll]\ar@{}[u]|{\mbox{\tiny (pb)}}&& X \times X\ar@{=}[u]
	  \ar[r]|-{p_2}&X\ar[r]|-{\Delta_X}& X \times X\ar@{=}[u]
	  \\
	   \ast\ar[u]\ar[d]
	  &
	  && X\times X\ar[u]\ar[d]|{p_2}\ar[rrr]|{p_2} \ar[lll]
	 &&&X\ar[u] \ar[d]|{\mathrm{id}_X}\ar[r]|-{\Delta_X}& X \times X\ar[u]_{\mathrm{id}_{X \times X}}\ar[d]^{\mathrm{id}_{X \times X}}
	 \\
 \ast
	  &
	  && X\ar[rrr]|-{\mathrm{id}_X} \ar[lll]
	 &&&X \ar[r]|-{\Delta_X}& X \times X	 
	 }
  \]
%
  with the ``vertical identity'' 2-correspondence
  \[
  \raisebox{78pt}{
    \xymatrix{
      \ast & X \ar[l] \ar[rr]|{\Delta_X} && X\times X
	  \\
	 \ast \ar[u] \ar[d] & X \ar[u]|{\mathrm{id}_X} \ar[d]|{\mathrm{id}_X} \ar[l] \ar[rr]|{\Delta_X} && X\times X
	  \ar[u]_{\mathrm{id}_{X\times X}} \ar[d]^{\mathrm{id}_{X\times X}}
	  \\
	  \ast & X \ar[l] \ar[rr]|{\Delta_X} && X\times X
    }
    }\;,
  \]
   by the universal property of the homotopy pullback enjoyed by  
   $[\Pi(S^1),X]$. Checking of the other zig-zag identities is completely analogous.
%
  
  \medskip
  In this fashion we proceed by induction. The $k$-fold 
  units and their adjoints are given in $\mathbf{H}$
  by $k$-fold correspondences 
  of correspondences with tips given by
       $$
   {
    \xymatrix{
	 \ast  & X
	  \ar[l] 
	  \ar[r]&
	  [\Pi(S^k), X]
	}}
	\;\qquad\text{and}\qquad
	{
    \xymatrix{
	  [\Pi(S^k), X]& X
	  \ar[l] 
	  \ar[r] & \ast 
	}}
	\,.  $$
  By proposition \ref{nPluOneSphereSpaceFromnSphereSpace}
  the $k$-fold trace on the identity then is indeed
  $$
    \xymatrix{
	  && [\Pi(S^{k+1}), X]	  \ar[dl]
	  \ar[dr]
	  \\
	  & X \ar[dl] 
	    \ar[dr] 
		&& 
		X
		\ar[dl]
		\ar[dr]
	  \\
	  \ast && [\Pi(S^k), X] && \ast \;.
	}
  $$
\endofproof

By the classification of local topological field theories \cite{LurieTFT} we, therefore, 
have the following
\begin{proposition}
  Fully extended topological field theory with coefficients in $ \mathrm{Corr}_n(\mathbf{H})$
  are equivalent to objects $\mathbf{Fields} \in \mathbf{H}$
  $$
    Z_{\mathbf{Fields}} 
     : 
    \mathrm{Bord}_n^\otimes \longrightarrow \mathrm{Corr}_n(\mathbf{H})^\otimes
	\,,
  $$ 
 via $Z_{\mathbf{Fields}}(*)\cong \mathbf{Fields}$.
 \label{prop.Fields-to-Z}
\end{proposition}
Therefore we will mostly just write this as
$$
  \mathbf{Fields}
  :
  \mathrm{Bord}_n^\otimes \longrightarrow \mathrm{Corr}_n(\mathbf{H})^\otimes  
  \,,
$$
for short.

\begin{remark}
  By handle decomposition of smooth manifolds 
  it follows that
  the symmetric monoidal functor $\mathbf{Fields}$ sends a closed
  manifold $\Sigma_k$ of dimension $k$ to the mapping stack 
  $[\Pi(\Sigma_k), \mathbf{Fields}]$,
 seen as a  $k$-fold correspondence of correspondences between the terminal object and itself.  
 Generally, a cobordism $\Sigma$ with incoming boundary $\Sigma_{\mathrm{in}}$
 and outgoing boundary $\Sigma_{\mathrm{out}}$ is sent to the correspondence
$$
  \left(\Sigma_{\rm in} \hookrightarrow  \Sigma \hookleftarrow \Sigma_{\rm out}
\right)
\quad 
\longmapsto 
\qquad
\left(
\raisebox{20pt}{
\xymatrix{
&
\left[ \Pi (\Sigma), {\bf Fields}  \right]
\ar[dl]_{(-)|_{\mathrm{in}}} 
\ar[dr]^{(-)|_{\mathrm{out}}}
&
\\
\left[ \Pi (\Sigma_{\rm in}), {\bf Fields}\right] 
&&
\left[ \Pi (\Sigma_{\rm out}), {\bf Fields} \right]
}
}
\right)
$$ 
in $\mathbf{H}$.
  \label{AssignmentToCobordismInducedByObjectInSpans}
\end{remark}

\paragraph{Local action functionals for the bulk field theory}
\label{BulkFieldTheory2}
  
  In addition to field configurations, prequantum field theory encodes the
  local \emph{action functionals} or \emph{Lagrangians} on these. This involves
  equipping all the objects described above with maps to a given space ``of phases'',
  a suitable higher version of the group $U(1)$ in which traditional action 
  functionals 
  take values. For instance, in the introduction we considered Lagrangians of the form
  $\mathbf{L}:\mathbf{Fields}\to \mathbf{B}^nU(1)_{\mathrm{conn}}$, in which the space of phases was the $n$-stack of $U(1)$ $n$-bundles with connection. More generally, 
  we will choose the space of phases to be a commutative group object $\mathbf{Phases}$ in $\mathbf{H}$. Clearly, since we are working in a higher categorical setting, ``commutative''
   here means ``commutative up to coherent homotopies'', and the same consideration applies to the group structure of the space of phases. 
   That is $\mathbf{Phases}$ is an $E_\infty$-group object in $\mathbf{H}$.
  
  \begin{remark}
  The fact that here we consider $\mathbf{Phases}$ to be group object in $\mathbf{H}$ instead of in 
  a more general
  stack of symmetric monodical $(\infty,n)$-categories is related to the fact that here
  we are considering pre-quantum field theory as opposed to 
  quantum field theory. For the latter one chooses a representation
  $\mathbf{Phases} \to \mathcal{C}$ of the space of phases on a genuine $(\infty,n)$-category and postcomposes
  the Lagrangian with this, see \cite{Nuiten}.
\end{remark}
  
  The general mechanism to describe local action functionals is based on the following simple observation. 
  
  \begin{remark}\label{rem.symmetric-monoidal}
  The commutative group structure on $\mathbf{Phases}$ endows the slice topos $\mathbf{H}_{/\mathbf{Phases}}$ with a natural tensor product lifting the cartesian product of $\mathbf{H}$ by 
  $$
    \left[
	  \raisebox{20pt}{
	  \xymatrix{
	    X
		\ar[d]^{F_1}
		\\
		B
	  }}
	\right]
	\otimes
    \left[
	  \raisebox{20pt}{
	  \xymatrix{
	    Y
		\ar[d]^{F_2}
		\\
		B
	  }}
	\right]
	:=
    \left[
	  \raisebox{20pt}{
	  \xymatrix{
	    X \times Y
		\ar[d]^{\pi_X^*F_1  + \pi_Y^*F_2}
		\\
		\mathbf{Phases}
	  }}
	\right]
	:=
    \left[
		\raisebox{41pt}{
	  \xymatrix{
	    X \times Y
		\ar[d]|{(\pi_X^*F_1 , \pi_Y^*F_2)}
		\\
		\mathbf{Phases} \times \mathbf{Phases}
		\ar[d]^{+}
		\\
		\mathbf{Phases}
	  }}
	\right]	
	\,,
  $$
  where on the right we use the group structure on $\mathbf{Phases}$. 
  Here $\pi_X$ and $\pi_Y$ are the corresponding projections. 
  The tensor unit is the unit inclusion:
$$
  \mathbb{I} = 
  \left[
    \raisebox{23pt}{
    \xymatrix{
	  \ast \ar[d]^-0
	  \\
	  \mathbf{Phases}
	}}
  \right]
  \,.
  $$
\end{remark}
We can therefore lift Definition \ref{def.spanH} and Remark \ref{remark SpanH}
from fields to fields equipped with action functionals as follows.
\begin{definition}
 The  symmetric monoidal $(\infty,n)$-category $\mathrm{Corr}_n(\mathbf{H}_{/\mathbf{Phases}})^\otimes$ is the $(\infty,n)$-category structure on the 
 $\infty$-groupoid of functors from $ \mathrm{Corr}_n$ to $\mathbf{H}_{/\mathbf{Phases}}$ with compositions of correspondences by fiber product of maps to a common face; 
 and with symmetric monoidal product induced by 
  the symmetric monoidal category structure on $\mathbf{H}_{/\mathbf{Phases}}$ described in Remark \ref{rem.symmetric-monoidal}.
\end{definition}
\begin{remark}
  When $\mathbf{H} = \infty \mathrm{Grpd}$ 
  (geometrically discrete $\infty$-groupoids), 
  and for $\mathbf{Phases} \in \infty \mathrm{Grpd}$ any 
  $\infty$-groupoid equipped with symmetric group structure,
  then we may regard this equivalently as 
  a symmetric monoidal $(\infty,n)$-category whose
  underlying $(\infty,n)$-category happens to be an 
  $(\infty,0)$-category.
  Therefore $\mathrm{Corr}_n(\infty \mathrm{Grpd}_{/\mathbf{Phases}})$
  in this case
  is an example of the class of $(\infty,n)$-categories considered
  in \cite{LurieTFT} around prop. 3.2.8 there.
  Notice that when $\mathbf{H}$ is not the canonical $\infty$-topos
  $\infty \mathrm{Grpd}$ of geometrically discrete $\infty$-groupoids, then
  the analogous generalization would allow $\mathcal{C}$ to be not just 
  a bare $(\infty,n)$-category, but an $(\infty,n)$-category 
  internal to $\mathbf{H}$, hence a stack of $(\infty,n)$-categories
  on an $\infty$-site of definition for $\mathbf{H}$.
\end{remark}

Notice that the forgetful morphism $\mathbf{H}_{/\mathbf{Phases}} \to \mathbf{H}$, which forgets the map to the space of 
phases,  induces a natural forgetful monoidal contravariant functor 
\[
  \mathrm{Corr}_n(\mathbf{H}_{/\mathbf{Phases}})^\otimes
  \longrightarrow 
  \mathrm{Corr}_n(\mathbf{H})^\otimes
  \,.
\]
Thanks to the commutative group structure on the space of phases, we have the following generalization of Proposition \ref{prop.everything-dualizable}.
\begin{proposition}
Every object $\mathbf{L}: X\xrightarrow{}\mathbf{Phases}$ 
in $\mathrm{Corr}_n(\mathbf{H}_{/\mathbf{Phases}})^\otimes$
  is fully dualizable, the full dual being $-\mathbf{L}$.
  \label{EveryMapToAGroupObjectIsFullyDualizableInSpansInSlice}
\end{proposition}
\proof
  Observe that we may take the co-evaluation map $\mathbb{I} \to \mathbf{L}\otimes (-\mathbf{L})$ 
  and evaluation map $\mathbf{L}\otimes (-\mathbf{L})\to \mathbb{I}$
  to be given by
  $$
	\raisebox{40pt}
	{
    \xymatrix{
	  & X
	  \ar[dr]^-{\Delta_X}
	  \ar[dl]
	  \\
	  \ast
	  \ar[dr]_-0
	  &&
	  X \times X
	  \ar[dl]^-{p_1^*\mathbf{L} - p_2^*\mathbf{L} }
	  \\
	  &\mathbf{Phases}
	}
	}
	\;\;\;\;\mbox{and}\;\;\;\;
    \raisebox{40pt}
    {
    \xymatrix{
	  & X
	  \ar[dl]_-{\Delta_X}
	  \ar[dr]
	  \\
	  X \times X
	  \ar[dr]_-{\hspace{-4mm}p_1^*\mathbf{L} - p_2^*\mathbf{L}}
	  &&
	  \ast
	  \ar[dl]^-0
	  \\
	  & \mathbf{Phases}
	}}
	\,,
  $$
  respectively. Here $p_1$ and $p_2$ denote projection to the first and second factors, respectively, and 
   the squares are filled by the canonical equivalence $p_1\circ\Delta_X\cong p_2\circ\Delta_X$. 
  From here on the argument proceeds just as in the proof of
 Proposition  \ref{prop.everything-dualizable}.
\endofproof
Therefore  we have the following analogue of Proposition \ref{prop.Fields-to-Z}.
\begin{proposition}
  A morphism $\mathbf{L}:\mathbf{Fields}\longrightarrow \mathbf{Phases}$ in $\mathbf{H}$ 
  equivalently determines a fully extended topological field theory with 
  coefficients in $ \mathrm{Corr}_n(\mathbf{H}_{/\mathbf{Phases}})$,
  $$
    \exp\left(
      \tfrac{i}{\hbar}S_{\mathbf{L}}
    \right)
    \;:\; 
    \mathrm{Bord}_n^\otimes 
      \longrightarrow
    \mathrm{Corr}_n(\mathbf{H}_{/\mathbf{Phases}})^\otimes
	\,,
  $$ 
 characterized by the condition 
 $\exp\left(\tfrac{i}{\hbar} S_{\mathbf{L}}\right)(\ast) \simeq \mathbf{L}$.
 \label{LocalActionFunctionalsAsMonoidalBordismReps}
\end{proposition}
\begin{definition}
  In view of the fully extended TQFT it defines, 
  we call a morphism $\mathbf{L}:\mathbf{Fields}\to \mathbf{Phases}$ 
  a \emph{local action functional} or \emph{local Lagrangian} 
  for the TQFT with $\mathbf{Fields}$ as stack of fields.
  Notice how the notion of local action and local Lagrangian unify here:
  the local Lagrangian is value of the local (extended) action functional
  on the point.
\end{definition}
\begin{remark}
  Since fully extended topological field theories 
  are completely determined by their value on the point, a local 
  action functional on a prescribed moduli stack of fields 
  $\mathbf{Fields}$ is equivalent to the 
  datum of a symmetric monoidal lift
  $$
    \xymatrix{
	  & \mathrm{Corr}_n(\mathbf{H}_{/\mathbf{Phases}})
	  \ar[d]
	  \\
	  \mathrm{Bord}_n
	  \ar[r]_-{\mathbf{Fields}}
	  \ar[ur]^-{\hspace{-4mm} \exp\left(\tfrac{i}{\hbar} S_{\mathbf{L}}\right)}
	  &
	  \mathrm{Corr}_n(\mathbf{H})\;,
	}
	\,.
  $$
\label{LocalPrequantumFieldTheory}
\end{remark}
This  is the the
perspective in section 3 of \cite{FHLT} from 
field theories with geometrically discrete
to those with cohesively geometric moduli stacks of fields.
\begin{example}
Given a local action functional $\exp(\tfrac{i}{\hbar}S_{\mathbf{L}})$
as in prop. \ref{LocalActionFunctionalsAsMonoidalBordismReps},
then to the circle $S^1$, regarded as a 1-morphism in $\mathrm{Bord}_n$,
is assigned the following correspondence in $\mathbf{H}_{/\mathbf{Phases}}$
$$
    \xymatrix{
	  && [\Pi(S^1), \mathbf{Fields}]
	  \ar[dl] \ar[dr]_{\ }="s1"
	  \\
	  & \mathbf{Fields} \ar[dl]
  	   \ar[dr]|{\Delta_{\mathbf{Fields}\phantom{\int}}}^{\ }="t1"  && 
	   \mathbf{Fields}
	   \ar[dl]|{\Delta_{\mathbf{Fields}\phantom{\int}}} \ar[dr]
	  \\
	  \ast \ar[drr]_0 && \mathbf{Fields} \times \mathbf{Fields} 
	  \ar[d]|{p_1^*\mathbf{L}-p_2^*\mathbf{L}}
	  &&
	  \ast\;,
	  \ar[dll]^0
	  \\
	  && \mathbf{Phases}
	  \ar@{} "s1"; "t1"|{\mbox{\tiny (pb)}}
	}
  $$
  where the top two morphisms are restrictions to the left and right 
  semicircles (hemispheres) of $S^1$ which are both homotopic to the point. 
  By the universal property of the pullback, this induces a morphism
  \[
    [\Pi(S^1), \mathbf{Fields}] \longrightarrow \Omega\mathbf{Phases},
  \]
  into the loop space object of the stack $\mathbf{Phases}$ of higher phases.
  Notice that since $\mathbf{Phases}$ is an abelian group object in $\mathbf{H}$
  then so is $\Omega\mathbf{Phases}$.
  
  Unwinding this in components 
  shows that the displayed homotopy in the middle
  exhibits the circle by two semi-circles  that start and end at the same point.
  The whiskering with the vertical map evaluates the action functional 
  on the first semi-circle and minus the action functional on the second,
  hence evaluates the action functional itself on one full copy of the circle.
  So this is the transgression of the Lagrangian to an action functional
  on the loop space.
  \label{InducinglocalPrequantumFTfromPoint}
\end{example}

\paragraph{Boundary field theory}
\label{BoundaryFieldTheory}
\index{boundary field theory}
%
We now turn to the discussion of boundary data for a local prequantum field theory. 
  
  Notice that the cobordism theorem in the version of theorem 2.4.6 in 
  \cite{LurieTFT} essentially says that $\mathrm{Bord}_n^\otimes$ is the
  symmetric monoidal $(\infty,n)$-category with fully dualizable objects
  which is freely generated from a single object:
  $$  
    \mathrm{Bord}_n \simeq \mathrm{FreeSMD}(\{\ast\})
    \,.
  $$
  Under this equivalence that single object is indeed identified with the 
  manifold $\mathbb{R}^0$, which in the above discussion is what locally 
  supports a \emph{bulk field theory}. But theorem 4.3.11 in \cite{LurieTFT}
  provides a considerable generalization of this situation. This theorem
  essentially says that for any collection of $(\infty,n)$-categorical 
  generating cells,
  there is a notion of smooth manifolds \emph{with singularities}
  such that the $(\infty,n)$-category ${\mathrm{Bord}_n^{\mathrm{sing}}}^\otimes$
  of $n$-dimensional cobordisms of manifolds with such singularities
  is the symmetric monoidal $(\infty,n)$-category with fully dualizable objects
  which is free on the given collection of cells.
  
  We consider this now for a singularity that corresponds to a 1-morphism
  of the form
  $$
    \emptyset \longrightarrow \ast
    \,,
  $$
  hence a morphism from the tensor unit to a generating object.
  Regarded as a cobordism, this is going to be interpreted as a cobordism
  that is much like the edge $[0,1] : \ast \longrightarrow \ast$,
  only that to the left it is not possible to sew further edges to this.
  Hence under the cobordism theorem for manifolds with singularities,
  the above 1-cell is interpreted as a cobordism of the form  
 \[
 \xymatrix{{\vert}\!\!\! \ar@{-}[r] & \!\!\!\ast}\,,
  \] 
  hence by a 1-dimensional cobordism that has a constrained boundary on the left.
  
\begin{definition}
 Write
 $$
   {\mathrm{Bord}_n^{\partial}}^\otimes
   :=
   \mathrm{FreeSMD}(\{\emptyset \to \ast\})
 $$
 for the symmetric monoidal $\infty$-category of cobordisms of manifolds
 with codimension-1 boundaries, correspoding to the 1-cell datum $\{\emptyset \to \ast\}$
 under theorem 4.3.11 in \cite{LurieTFT}.
\end{definition}    
Notice that by free-ness and by construction, there is a canonical inclusion
$$
  {\mathrm{Bord}_n}^\otimes 
     \longrightarrow
  {\mathrm{Bord}_n^\partial}^\otimes 
$$
\begin{definition}
Let $\mathbf{Fields} : \mathrm{Bord}_n^\otimes \to \mathrm{Corr}_n(\mathbf{H})$ 
be a choice of bulk fields according to prop. \ref{prop.Fields-to-Z}, 
then a choice of \emph{boundary fields} for these bulk fields
is a choice of extension $\mathbf{Fields}^{\partial}$:
\[
\xymatrix{
  {\mathrm{Bord}_n}^\otimes\ar[d]\ar[rr]^-{Z_{\mathbf{Fields}}} 
  &&
  \mathrm{Corr}_n(\mathbf{H})^\otimes 
  \\
  {\mathrm{Bord}_n^{\partial}}^\otimes 
  \ar[rru]_-{Z_{\mathbf{Fields}_{\partial}}}
}\,.
\]
\end{definition}
The following immediate consequence is worth recording.
\begin{proposition}
A choice of boundary fields for $\mathbf{Fields}$ is equivalently a 
choice of moduli stack $\mathbf{Fields}_\partial \in \mathbf{H}$
together with a choice of of morphism
\[
\mathbf{Fields}_{\partial}\to \mathbf{Fields}
\]
in $\mathbf{H}$.
\label{BoundaryDatum1}
\end{proposition}  
\proof
 Since $\mathrm{Bord}_n^{\partial}$ is free symmetric monoidal with duals
  on a single morphism out of the unit object, 
  a symmetric monoidal functor 
  ${\mathrm{Bord}_n^\partial}^\otimes \to \mathrm{Corr}_n(\mathbf{H})$ 
  is equivalent to the datum of a 1-morphism in $\mathrm{Corr}_n(\mathbf{H})$ out of $\ast$. 
  Requiring this to be an extension of the bulk fields
  amounts to asking that this 1-morphism in $\mathrm{Corr}_n(\mathbf{H})$ has target $\mathbf{Fields}$, 
  and so it is a correspondence in $\mathbf{H}$ of the form
 \[
    \xymatrix{
	 \ast  & \mathbf{Fields}_{\partial}
	  \ar[l] 
	  \ar[r]&
	  \mathbf{Fields}
	}\,.
	\]
Since $\ast$ is the terminal object in $\mathbf{H}$, this is equivalent to the datum of the morphism $\mathbf{Fields}_{\partial}\to \mathbf{Fields}$.
\endofproof  
\begin{remark}
Therefore we will write $(\mathbf{Fields}_\partial \to \mathbf{Fields})$
for $Z_{\mathbf{Fields}_\partial}$.
Notice that hence the \emph{$\infty$-category of boundary fields} for given bulk $\mathbf{Fields}$
is the slice $\infty$-topos $\mathbf{H}_{/\mathbf{Fields}}$. 
\end{remark}
The boundary field theory version of remark \ref{AssignmentToCobordismInducedByObjectInSpans} 
about the bulk field theory is now the following (this was pointed out by Domenico Fiorenza).
\begin{proposition}\label{prop.boundary-configurations}
  A boundary field assignment
  $$
    (\mathbf{Fields}_{\partial}\to \mathbf{Fields})
	\;:\;
    (\mathrm{Bord}_n^{\partial})^\otimes \to \mathrm{Corr}_n(\mathbf{H})^{\otimes}
  $$
  sends cobordisms $({\partial}\Sigma \hookrightarrow \Sigma) \in \mathrm{Bord}_n^\partial$
  with marked boundary $\partial \Sigma$ to
  $$
    (\mathbf{Fields}_{\partial}\to \mathbf{Fields})
	\;:\;
	({\partial}\Sigma \hookrightarrow \Sigma)
	\;\mapsto\;
	[\Pi({ \partial} \Sigma), \mathbf{Fields}_{{\partial}}]
	\underset{[\Pi({\partial}\Sigma), \mathbf{Fields}]}{\times}
	[\Pi(\Sigma), \mathbf{Fields}]\,,
  $$
  hence
  to the stack of diagrams in $\mathbf{H}$ of the form
  $$
   \xymatrix{
    \Pi(\partial \Sigma) \ar[rr]^{\phi_\partial} 
    ~\ar@{^{(}->}[d]
    && \mathbf{Fields}_{\partial}
    \ar[d]
    \\
    \Pi(\Sigma) 
    \ar[rr]^\phi && \mathbf{Fields}\;. 
  }\,.
  $$  
  \label{RelativeBoundaryFieldConfigurations}
\end{proposition}
\proof
  Every cobordism $\Sigma$ with marked boundary component
  ${\partial} \Sigma$ decomposes as the gluing of the 
cylinder   
$(\xymatrix{{\vert}\!\!\! \ar@{-}[r] & \!\!\!\ast})
	\times
	{\partial}\Sigma$
	with $\Sigma$ regarded as a manifold with unmarked boundary.
 Since $\xymatrix{{\vert}\!\!\! \ar@{-}[r] & \!\!\!\ast}$ is mapped to the corespondence
 \[
 \xymatrix{
	  \ast & \mathbf{Fields}_{\partial} \ar[l] \ar[r] & \mathbf{Fields}
	  }
 \]
 in  $\mathbf{H}$,
 we find that $(\xymatrix{{\vert}\!\!\! \ar@{-}[r] & \!\!\!\ast})
	\times
	{\partial}\Sigma$ is mapped to
  $$
	\xymatrix{
	  \ast 
	  & 
	  [\Pi({\partial}\Sigma), \mathbf{Fields}_{\partial}] \ar[r] \ar[l] 
	  &  [\Pi(\partial\Sigma), \mathbf{Fields}] 
	}
	\,.
  $$
On the other hand, on the ``piece'' given by $\Sigma$ with unmarked boundary $\partial\Sigma$ the 
field theory reduces to the one associated with the stack $\mathbf{Fields}$, and we know from  Remark \ref{AssignmentToCobordismInducedByObjectInSpans} that $\partial\Sigma\hookrightarrow \Sigma$ is 
mapped by $\mathbf{Fields}$ to
  $$
	\xymatrix{
	 [\Pi(\partial\Sigma),\mathbf{Fields}] 
	  & 
	 [\Pi(\Sigma),\mathbf{Fields}] \ar[r] \ar[l] 
	  & \ast 
	}	
	\,.
  $$
 The composite of these two contributions is
 \[
\xymatrix{
	\ast 
	  & 
	 [\Pi({\partial}\Sigma), \mathbf{Fields}_{\partial}]
	 \underset{[\Pi({\partial}\Sigma), \mathbf{Fields}]}{\times}
	 [\Pi(\Sigma),\mathbf{Fields}] \ar[r] \ar[l] 
	  & \ast 
	}\;,
 \]
 as claimed.
\endofproof
\begin{remark}[twisted relative cohomology]
In words this says that for the boundary field theory
$\mathbf{Fields}_{\partial}\to \mathbf{Fields}$, 
a field configurations on a manifold $\Sigma$ with constrained boundary 
${{\partial}}\Sigma$
is a bulk field configuration on $\Sigma$ together with a boundary field configuration 
on ${{\partial}}\Sigma$ and an equivalence of the boundary field configuration 
with the restriction of the bulk field configuration to the boundary. 
These data are equivalently those of a \emph{twisted cocycle} with local
  coefficient bundle $\mathbf{Fields}_{\partial} \to \mathbf{Fields}$,
  \emph{relative} to the boundary inclusion.  In particular, 
  when $\mathbf{Fields}_{\partial} \simeq \ast$ then these are equivalently cocycles in \emph{relative cohomology}
  with coefficients in $\mathbf{Fields}$.
\end{remark}
We now add local action functionals with boundary conditions to the 
boundary fields.
\begin{definition}
Let $\exp\left(\tfrac{i}{\hbar}S\right):\mathbf{Fields}\to \mathbf{Phases}$ 
be a local action functional for a 
bulk prequantum field theory according to prop. \ref{LocalActionFunctionalsAsMonoidalBordismReps},
then a \emph{boundary condition} (or \emph{boundary extension}) for $\mathbf{L}$ 
is an extension
\[
\xymatrix{
  {\mathrm{Bord}_n}^\otimes
  \ar[d]\ar[rr]^-{\exp\left(\tfrac{i}{\hbar}S_{\mathbf{L}}\right)} 
   &&
   \mathrm{Corr}_n(\mathbf{H}_{/\mathbf{Phases}})^\otimes 
   \\
  {\mathrm{Bord}_n^{\partial}}^\otimes 
   \ar[rru]_-{\exp\left(\tfrac{i}{\hbar} S_{\mathbf{L}}^\partial\right)}
}\,,
\]
\end{definition}
\begin{proposition}
A boundary condition for a local Lagrangian $\mathbf{L}$ 
with respecto to boundary fields $\mathbf{Fields}_\partial \to \mathbf{Fields}$ 
is equivalently a choice of homotopy in 
  $$
    (\xymatrix{{\vert}\!\!\! \ar@{-}[r] & \!\!\!\ast})
	\;\;\;\;
	\mapsto
	\;\;\;\;
    \raisebox{33pt}{
    \xymatrix{
	  & \mathbf{Fields}_{\partial}
	  \ar[dl]
	  \ar[dr]_{\ }="s"
	  \\
	  \ast \ar[dr]_0^{\ }="t" && \mathbf{Fields} \ar[dl]^{\exp\left(\tfrac{i}{\hbar}S_{\mathbf{L}}\right)}
	  \\
	  & \mathbf{Phases}
	  \ar@{=>} "s"; "t"
	}}
	\,.
  $$
  \label{BoundaryDatum}
  in $\mathbf{H}$,
  which in turn is equivalently a choice of morphism
\[
\mathbf{Fields}_{\partial}\to \mathrm{fib}(\mathbf{L})
\]
in $\mathbf{H}$, where $\mathrm{fib}(\mathbf{L})$ 
is the homotopy fiber of $\mathbf{L}:\mathbf{Fields}\to \mathbf{Phases}$ 
on the zero element of the commutative group stack of phases.  
\label{BoundaryDatum}
\end{proposition}  
\proof
 Since $\mathrm{Bord}_n^{\partial}$ is free symmetric monoidal with duals
  on a single morphism out of the unit object, 
  a symmetric monoidal functor $\exp\left(\tfrac{i}{\hbar}S_{\mathbf{L}}^\partial\right)$ 
  is equivalent to the datum of a 1-morphism 
  in $\mathrm{Corr}_n(\mathbf{H}_{/\mathbf{Phases}})$ out of $\ast\xrightarrow{0}\mathbf{Phases}$. 
\endofproof  
Therefore we set:
\begin{definition}
  The \emph{$\infty$-category of boundary conditions} 
  for $\mathbf{L}:\mathbf{Fields}\to\mathbf{Phases}$
  is the slice $\infty$-topos $\mathbf{H}_{/\mathrm{fib}(\mathbf{L})}$. 
  For $n \in \mathbb{N}$ and 
  $\exp\left(\tfrac{i}{\hbar}S_{\mathbf{L}}\right) \in \mathbf{H}_{/\mathbf{Phases}}$, 
  we call
  $$
    \mathrm{Bdr}\left(
       \exp\left( \tfrac{i}{\hbar}S_{\mathbf{L}} \right)
    \right) 
     := 
	 \mathrm{Corr}_1\left(
	   \mathbf{H}_{/\mathbf{Phases}}
	 \right)
	 \left(
	   0, \exp\left( \tfrac{i}{\hbar} S_{\mathbf{L}} \right)
	 \right)
  $$
  the \emph{$\infty$-category of boundary conditions} of the local action functional
  $\exp\left(\tfrac{i}{\hbar}S_{\mathbf{L}}\right)$.
  \label{CategoryOfBoundaryConditions}
\end{definition}
\begin{definition}
  For $\exp\left(\tfrac{i}{\hbar}S_{\mathbf{L}}\right)$
  a local bulk prequantum field theory, by prop. \ref{LocalActionFunctionalsAsMonoidalBordismReps},
  we say that its \emph{universal boundary condition}
  is that which is given via remark \ref{BoundaryDatum} 
  by the square exhibiting the homotopy fiber of $S$
  in $\mathbf{H}$
  $$
    \raisebox{20pt}{
    \xymatrix{
	  & \mathrm{fib}\left(\exp\left(\tfrac{i}{\hbar}S\right)\right)
	  \ar[dl]
	  \ar[dr]_{\ }="s"
	  \\
	  \ast 
	   \ar[dr]_0^{\ }="t" 
	   && 
	   \mathbf{Fields} 
	   \ar[dl]^{\exp\left(\tfrac{i}{\hbar}S\right)}
	  \\
	  & \flat \mathbf{B}^n U(1)
	  \ar@{=>} "s"; "t"
	}}
	\,.
  $$
  \label{TerminalBoundaryCondition}
\end{definition}
The following immediate consequence is relevant.
\begin{proposition}
  The universal boundary condition is the terminal object in the 
  $\infty$-category
  $\mathrm{Bdr}\left(\exp\left(\tfrac{i}{\hbar}S_{\mathbf{L}}\right)\right)$ 
  of boundary conditions, def. \ref{CategoryOfBoundaryConditions}.
  A general boundary condition with moduli stack 
  $\mathbf{Fields}_{\partial}$ is equivalently a morphism
  $\mathbf{Fields}_{\partial} \to \mathrm{fib}(\exp(i S))$:
  there is a natural equivalence
  $$
    \mathrm{Bdr}\left(\exp\left(\tfrac{i}{\hbar}S\right)\right)
	\simeq
	\mathbf{H}_{/\mathrm{fib}\left(\exp\left(\tfrac{i}{\hbar}S\right)\right)}
  $$
  between the $\infty$-category of boundary conditions for $\exp(i S)$
  and the slice $\infty$-topos of $\mathbf{H}$ over $\mathrm{fib}(\exp(i S))$.
  \label{TerminalBoundaryConditionIsTerminal}
  \label{BoundaryConditonsAsConeCategory}
\end{proposition}
\proof
 The $\infty$-category $\mathrm{Bdr}\left(\exp\left(\tfrac{i}{\hbar}S_{\mathbf{L}}\right)\right)$ 
 is equivalently the $\infty$-category
  of cones over the diagram ${}_0\vee_{\exp(i S)} : \mathrm{cosp} \to \mathbf{H}$
  from the free cospan category which exhibits the diagram 
  $$
    \left\{
	  \raisebox{20pt}{
	  \xymatrix{
	    \ast \ar[dr]_-0 && \mathbf{Fields} \ar[dl]^-{\exp(i S)}
		\\
		& \flat \mathbf{B}^n U(1)
	  }}
	\right\}
	\,.
  $$
  In the notation of section 1.2.9 \cite{Lurie} this means
  $$
    \mathrm{Bdr}\left(
         \exp\left( \tfrac{i}{\hbar}S_{\mathbf{L}} \right)
     \right) 
     \simeq 
      \mathbf{H}_{/ ( {}_0 \vee_{ \exp\left( \tfrac{i}{\hbar}S_{\mathbf{L}} \right) } ) }
    \,.
  $$
  Let then
  $
    \xymatrix{
	  \ast~ \ar@{^{(}->}[r] & \Box 
       \ar@{<-^{)}}[r]
       &	   
	  ~ \mathrm{cosp}
	 }
  $
  be the inclusion of the point as the initial object of the 
  box-shaped diagram $\infty$-category
  $$
    \Box = 
	\left\{
	  \raisebox{20pt}{
	  \xymatrix{
	     0 \ar[r]_>{\ }="s" \ar[d]^>{\ }="t" & 01 \ar[d]
		 \\
		 10 \ar[r] & 11
		 \ar@{=>} "s"; "t"
	  }
	  }
	\right\}
	\,,
  $$
  and the inclusion of the underlying cospan, respectively. Let then 
  $
    \widehat{ {}_0 \vee_{\exp(i S)} } : \Box \to \mathbf{H}
  $
  be the homotopy pullback diagram that exhibits 
  the homotopy fiber $\mathrm{fib}(\exp(i S))$
  and write
  $
    {}_0 \vee_{\exp(i S)} : \mathrm{cosp} \to \mathbf{H}
  $
  for its restriction to the underlying cospan, as in 
  remark \ref{BoundaryConditonsAsConeCategory}.  This induces 
  a diagram of $\infty$-functors
  $$
    \raisebox{20pt}{
    \xymatrix{
	  \mathbf{H}_{/\mathrm{fib}(\exp(i S))}
	  &
	  \mathbf{H}_{/\widehat{{}_0 \vee_{\exp(i S)}}}
	  \ar[r]^-\simeq
	  \ar[l]_-{\simeq}
	  &
	  \mathbf{H}_{/{}_0\vee_{\exp(i S)}}
	  \simeq \mathrm{Bdr}(\exp(i S))\;.
	}}
  $$
  The equivalence on the far right is that of remark \ref{BoundaryConditonsAsConeCategory}.
  The functor in the middle is an equivalence by finality of the $\infty$-limiting
  cones, as for instance in the proof of  
  prop. 1.2.13.8 in \cite{Lurie}. 
  And finally -- since the inclusion of an initial object is an op-final $\infty$-functor
  by prop. 4.1.3.1 in \cite{Lurie} --
  also the left functor, being the restriction of slices along an op-final functor,
  is an equivalence, by prop. 4.1.1.8 in \cite{Lurie}.
\endofproof

\paragraph{Corner field theory}
\label{CornerFieldTheory}

We now consider singularities of codimension 2 at which two boundaries
of codimension 1 meet, a \emph{corner} singularity.

\begin{definition}
 Write
 $$
   {\mathrm{Bord}_n^{\partial_1 \partial_2}}^\ast
   :=
   \mathrm{FreeSMD}
   \left(
     (\xymatrix{{\vert}\!\!\! \ar@{-}[r] & \!\!\!\ast})
	 \times
	 \left(
	 \raisebox{20pt}{
	 \xymatrix{
	   { -}
	   \ar@{-}[d]
	   \\
	   {\ast}
	 }
	 }
	 \right)
	 :
	 \raisebox{20pt}{
	   \xymatrix{
	     \emptyset \ar[r]^{\mathrm{id}} \ar[d]_{\mathrm{id}} & \emptyset \ar[d]
         \\
         \emptyset \ar[r] & \ast \;. 	 
	   }
	   }
   \right)
 $$
 for the symmetric monoidal $(\infty,n)$-category with fully dualizable objects
 which is free on a 2-cell as show on the right, considered as the
 $(\infty,n)$-category of cobordisms with two types of marked codimension-1
 boundaries and one kind of corner between these, by theorem 4.3.11 in 
 \cite{LurieTFT}.
	 \label{BordWithBoundaryAndCorners}
\end{definition}
As an immediate consequence, we have:
\begin{proposition} 
A symmetric monoidal $(\infty,n)$-functor
$$
  Z_{\mathbf{Fields}_{\partial_1\partial_2}}
  \;:\;
  (\mathrm{Bord}^{\partial_1\partial_2}_n)^\otimes 
  \longrightarrow 
  \mathrm{Corr}_n(\mathbf{H})^\otimes
$$
is equivalently the datum of
\begin{enumerate}
  \item a moduli stack $\mathbf{Fields} \in \mathbf{H}$
   of \emph{bulk fields};
  \item two moduli stacks 
  $\mathbf{Fields}_{\partial_1}, \mathbf{Fields}_{\partial_1}$
  of \emph{boundary fields};
  \item 
   a moduli stack $\mathbf{Fields}_{\partial_1\partial_2}$ 
  of \emph{corner fields} or 
  \emph{defect fields};
  \item
  a homotopy diagram
  $$
    \xymatrix{
	  \mathbf{Fields}_{\partial_1 \partial_2}
	  \ar[r]_>{\ }="s"
	  \ar[d]
	  &
	  \mathbf{Fields}_{\partial_1}
	  \ar[d]
	  \\
	  \mathbf{Fields}_{\partial_2}
	  \ar[r]^<{\ }="t"
	  &
	  \mathbf{Fields}
	  \ar@{=>}^\simeq "s"; "t"
	}
	\,
  $$
  in $\mathbf{H}$.
\end{enumerate}
A lift of that to correspondences in the slice 
$$
  \xymatrix{
    (\mathrm{Bord}^{\partial_1\partial_2}_n)^\otimes 
    \ar[drr]_{Z_{\mathbf{Fields}_{\partial_1\partial_2}}}
    \ar[rr]^{  \exp\left(\tfrac{i}{\hbar} S_{\mathbf{L}}\right)  }
    &&
    \mathrm{Corr}_n(\mathbf{H}_{/\mathbf{Phases}})^\otimes
    \ar[d]
    \\
    && \mathrm{Corr}_n(\mathbf{H})^\otimes
  }
$$
is a choice of extension of the above homotopy commutative diagram in $\mathbf{H}$ as
$$
  \xymatrix{
    &&\ast \ar@{..>}[dddd]
    & \ast \ar[rr] \ar[l] 
    \ar@{..>}[ddddl]
    && \ast \ar@/^1pc/[ddddlll]^0_{\ }="t3"
    \\
    &\ast \ar[dl] 
   \ar[ur]
  \ar@{..>}[dddr]& 
  \mathbf{Fields}_{\partial_1 \partial_2} 
  \ar[ur]
  \ar[l]
  \ar[rr]_>{\ }="s" 
  \ar[dl]
  && 
  \mathbf{Fields}_{\partial_2} 
  \ar[ur]
  \ar[dl]^{\ }="s3"
  \\
  \ast 
  \ar[ddrr]_0^-{\ }="t2"  
  & \mathbf{Fields}_{\partial_1} 
  \ar[l]
  \ar[rr]^<{\ }="t"_-{\ }="s2" 
  && \mathbf{Fields}~~~ \ar[ddl]|{\mathbf{L}} & \\
  && ~~~~~\Rrightarrow &&&\\
 && \mathbf{Phases} &&
  \ar@{=>}^\simeq "s"; "t"
  \ar@{=>}^\simeq "s2"; "t2"
  \ar@{=>}^\simeq "s3"; "t3"
}
$$
\label{prop mesh corner}
\end{proposition}
\begin{remark}
  This means that for two boundary conditions which are given by
  relative boundary trivializations of their local action functionals
  as in the previous section, a corner defect condition for them 
  is a further homotopy between the pullback of these two trivializations
  to the moduli stack of corner field configurations.
\end{remark}

\paragraph{Defect field theory}
\label{DefectTheory}
\index{prequantum field theory!defect field theory}
\index{defect field theory}

Finally, let us sketch a few lines on general pre-quantum defect field theory
(see for instance \cite{DKR} for general considerations about extended defect 
field theory). 
These correspond to adding another piece to the
 picture of framed cobordism, namely that of a punctured $k$-disk, seen as a morphism from the vacuum 
 to the $(k-1)$-sphere. In more formal terms, since a $k$-disk is homotopically trivial, this amounts to the following.
\begin{definition}
  Given a bulk field $\mathbf{Fields}$ in $\mathbf{H}$,
  a codimension-$k$ defect datum is a $k$-fold correspondence of the form
  $$ 
    \xymatrix{
    \mathbf{Fields}_{\mathrm{ins}}
	  \ar@{<-}[r]
	  &
	  \mathbf{Fields}_{\mathrm{def}}
	  \ar[r]
	 	  & 
	 	  [\Pi(S^{k-1}),\mathbf{Fields}]
	 }
	\,.
  $$
  \label{CodimensionkDefect}
\end{definition}
Examples of such defects and further comments on how to think of them appear as 
Example \ref{WZWDefect} and Example \ref{TerminalWilsonDefect}
below.

\newpage

\subsection{Structures in a differentially cohesive $\infty$-topos}
\label{StructuresInInfinitesimalCohesiveNeighbourhood}
\index{infinitesimal cohesion!structures}
\index{cohesive $\infty$-topos!infinitesimal cohesion!structures}

We discuss a list of differential geometric notions that can be formulated
in the presence of the axioms for infinitesimal cohesion, \ref{structures}.
These structures parallel the structures in a general cohesive $\infty$-topos, \ref{structures}.

\begin{itemize}
  \item \ref{InfStrucDeRhamSpace} -- Infinitesimal path $\infty$-groupoid and de Rham spaces;
  \item \ref{InfStrucFLat} -- Crystalline cohomology, flat infinitesimal $\infty$-connections and local systems;
  \item \ref{JetBundles} -- Jet $\infty$-bundles;
  \item \ref{FormallySmooth} -- Infinitesimal Galois theory / Formally {\'e}tale morphisms;
  \item \ref{InfStrucEtaleGroupoid} -- Formally {\'etale} groupoids;
  \item \ref{DifferentialStrucmanifolds} -- Manifolds (separated);
  \item \ref{CriticalLoci} -- Critical loci, variational calculus and BV-BRST complexes;
  \item \ref{InfStrucFormalInfinityGroupoid} -- Formal cohesive $\infty$-groupoids.
\end{itemize}

\medskip

\subsubsection{Infinitesimal path $\infty$-groupoid and de Rham spaces}
\label{InifnitesimalPathInfinityGroupoid}
\label{InfStrucDeRhamSpace}
\index{structures in a cohesive $\infty$-topos!infinitesimal path $\infty$-groupoid}
\index{structures in a cohesive $\infty$-topos!de Rham space}

We discuss the infinitesimal analog of the \emph{path $\infty$-groupoid},
\ref{StrucGeometricPostnikov}, which exists in a context of infinitesimal cohesion,
def. \ref{DefinitionOfInfinitesimalCohesion}.

\medskip

Let $(i_! \dashv i^* \dashv i_* \dashv i^1) : \mathbf{H} \to \mathbf{H}_{\mathrm{th}}$ be an
infinitesimal neighbourhood of a cohesive $\infty$-topos.
\begin{definition} 
 \label{InfinitesimalPathsAndReduction}
 \index{infinitesimal path $\infty$-groupoid}
 \index{paths!infinitesimal path $\infty$-groupoid}
 \index{de Rham space / de Rham $\infty$-groupoid}
 Write
$$
 (\mathbf{Red} \dashv \mathbf{\Pi}_{\mathrm{inf}} \dashv \mathbf{\flat}_{\mathrm{inf}})
 : 
 (i_! i^* \dashv i_* i^* \dashv i_* i^! ) 
  :
 \mathbf{H}_{\mathrm{th}} 
  \to 
 \mathbf{H}_{\mathrm{th}}
$$
for the adjoint triple induced by the adjoint quadruple that defines the differential cohesion.
For $X\in \mathbf{H}_{\mathrm{th}}$ we say that
\begin{itemize}
\item $\mathbf{\Pi}_{\mathrm{inf}}(X)$ is the \emph{infinitesimal path $\infty$-groupoid} 
  of $X$;

  The $(i^* \dashv i_*)$-unit 
  $$
    X \to \mathbf{\Pi}_{\mathrm{inf}}(X)
  $$
  we call the \emph{constant infinitesimal path inclusion}.

\item $\mathbf{Red}(X)$ is the \emph{reduced cohesive $\infty$-groupoid} underlying
  $X$.

  The $(i_* \dashv i^*)$-counit 
  $$
    \mathbf{Red} X \to X
  $$
  we call the \emph{inclusion of the reduced part} of $X$.
 \end{itemize}
\end{definition}
\begin{remark}
This is an abstraction of the setup considered in \cite{SimpsonTeleman}.
In traditional contexts as considered there, the object $\mathbf{\Pi}_{\mathrm{inf}}(X)$ 
is called the \emph{de Rham space} of $X$ or the \emph{de Rham stack} of $X$.
Here we may tend to avoid this terminology, since by 
\ref{StrucDeRham} we have a good notion of intrinsic de Rham cohomology in every 
cohesive $\infty$-topos already without equipping it with infinitesimal cohesion, 
which, over some $X \in \mathbf{H}$ 
is co-represented by the object $\mathbf{\Pi}_{\mathrm{dR}}(X)$, 
the cohesive de Rham homotopy type of remark \ref{DeRhamHomotopyType}. 
On the other hand, $\mathbf{\Pi}_{\mathrm{inf}}$ co-represents instead
what is called \emph{crystalline cohomology}, \ref{CrystallineCohomology} below.
\label{deRhamSpaceAndInfinitesimalPaths}
\end{remark}
\begin{proposition} 
  \label{InclusionOfConstantIntoInfinitesimalIntoAllPaths}
In the notation of def. \ref{PiInf}, there is a canonical natural transformation
$$
  \mathbf{\Pi}_{\mathrm{inf}}(X) \to \mathbf{\Pi}(X)
$$
that factors the finite path inclusion through the infinitesimal path inclusion
$$
  \raisebox{20pt}{
  \xymatrix{
    & \mathbf{\Pi}_{\mathrm{inf}}(X) \ar[d]     
    \\
    X \ar[ur] \ar[r] &  \mathbf{\Pi}(X)
  }
  }
  \,.
$$
Dually there is a canonical natural transformation
$$
  \flat A \to \flat A
$$
that factors the $\flat$-counits
$$
 \raisebox{20pt}{
  \xymatrix{
    \flat A
	\ar[d] \ar[dr]
	\\
	\flat_{\mathrm{inf}} A
	\ar[r]
	& A
  }
  }
  \,.
$$
\end{proposition} 
\proof
By def. \ref{PiInf} this is just the formula for the unit of the composite adjunction
$$
  (\mathbf{\Pi}_{\mathbf{H}_{\mathrm{th}}} \dashv \mathbf{\flat}_{\mathbf{H}_{\mathrm{th}}})
  :
  \xymatrix{
    \mathbf{H}_{\mathrm{th}}
    \ar@<+4pt>[r]^{\Pi_{\mathrm{inf}}}
    \ar@{<-^{)}}@<-4pt>[r]_{\mathrm{Disc}_{\mathrm{inf}}}
    &
    \mathbf{H}
    \ar@<+4pt>[r]^<<<<{\Pi}
    \ar@{<-^{)}}@<-4pt>[r]_<<<<{\mathrm{Disc}}
    &
    \infty \mathrm{Grpd}
  }  
  \,,
$$
more explicitly given by
$$
  \raisebox{20pt}{
  \xymatrix{
     & & 	  \mathrm{Disc}_{\mathrm{inf}}\circ \Pi_{\mathrm{inf}} (X) \ar[d]
    \\
    X \ar[rr] 
	   \ar[urr]
	 & & 
	  \mathrm{Disc}_{\mathrm{inf}}\circ \mathrm{Disc}_{\mathrm{H}} 
	  \circ \Pi_{\mathbf{H}} \circ \Pi_{\mathrm{inf}} (X)
  }
  }\,.
$$
The case for $\flat$ is formally dual. 
\endofproof

\subsubsection{Crystalline cohomology, flat infinitesimal connections and local systems}
 \label{InfStrucFLat}
\label{CrystallineCohomology}

\begin{definition}
  For $X \in \mathbf{H}_{\mathrm{th}}$ an object, we call the 
  cohomology, def. \ref{cohomology} of $\mathbf{\Pi}_{\mathrm{inf}}(X)$
  the \emph{crystalline cohomology} of $X$.
\end{definition}

We discuss now the infinitesimal analog of intrinsic flat cohomology, \ref{StrucFlatDifferential}.

\begin{definition}
  \label{InfinitesimalFlatCohomology}
  For $X \in \mathbf{H}_{\mathrm{th}}$ an object, we call the 
  cohomology, def. \ref{cohomology} of $\mathbf{\Pi}_{\mathrm{inf}}(X)$
  the \emph{crystalline cohomology} of $X$.
More specificaly, for $A \in \mathbf{H}_{\mathrm{th}}$ we say that
$$
  H_{\mathrm{infflat}}(X,A) := \pi_0 \mathbf{H}(\mathbf{\Pi}_{\mathrm{inf}}(X), A)
   \simeq
   \pi_0 \mathbf{H}(X, \mathbf{\flat}_{\mathrm{inf}}A)
$$
is the \emph{infinitesimal flat cohomology} of $X$ with coefficient in $A$.
\end{definition}
\begin{remark}
  \label{SyntheticDeRham}
  That traditional crystalline cohomology is the cohomology of the 
  ``de Rham stack'', see remark \ref{deRhamSpaceAndInfinitesimalPaths} above
  with coefficients in a suitable stack
  is discussed in \cite{LurieCrystal}, above theorem 0.4.
The relation to de Rham cohomology in traditional contexts is discussed for
instance in \cite{SimpsonTeleman}.
\end{remark}
\begin{remark}
By observation \ref{InclusionOfConstantIntoInfinitesimalIntoAllPaths}
we have canonical natural morphisms
$$
  \mathbf{H}_{\mathrm{flat}}(X,A) 
    \to  
  \mathbf{H}_{\mathrm{infflat}}(X,A)
    \to
  \mathbf{H}(X,A)
$$
The objects on the left are principal $\infty$-bundles equipped with flat $\infty$-connection. 
The first morphism forgets their higher parallel transport along finite volumes and just 
remembers the parallel transport along infinitesimal volumes. 
The last morphism finally forgets also this connection information.
\end{remark}
\begin{definition}
  \label{deRhamTheorem}
   \index{structures in a cohesive $\infty$-topos!de Rham theorem}
For $A \in \mathbf{H}_{\mathrm{th}}$ a 0-truncated abelian $\infty$-group object we say that the 
\emph{de Rham theorem} for $A$-coefficients holds in $\mathbf{H}_{\mathrm{th}}$ 
if for all $X \in \mathbf{H}_{\mathrm{th}}$ the infinitesimal path inclusion
of observation \ref{InclusionOfConstantIntoInfinitesimalIntoAllPaths}
$$
  \mathbf{\Pi}_{\mathrm{inf}}(X) \to \mathbf{\Pi}(X)
$$
is an equivalence in $A$-cohomology, hence if for all $n \in \mathbb{N}$ we have that 
$$
  \pi_0 \mathbf{H}_{\mathrm{th}}(\mathbf{\Pi}(X), \mathbf{B}^n A)
  \to 
  \pi_0 \mathbf{H}_{\mathrm{th}}(\mathbf{\Pi}_{\mathrm{inf}}(X), \mathbf{B}^n A)
$$
is an isomorphism.
\end{definition}
If we follow the notation of remark \ref{SyntheticDeRham} and moreover write 
$\vert X \vert = \vert \Pi X \vert$ for the intrinsic geometric realization,
def. \ref{GeometricRealization}, then this becomes
$$
  H^{\bullet}_{\mathrm{dR}, \mathrm{th}}(X,A) \simeq H^\bullet(|X|, A_{\mathrm{disc}})
  \,,
$$
where on the right we have ordinary cohomology in $\mathrm{Top}$ 
(for instance realized as singular cohomology) with coefficients in the 
discrete group $A_{\mathrm{disc}} := \Gamma A$ underlying the cohesive group $A$.

In certain contexts of infinitesimal neighbourhoods of cohesive $\infty$-toposes the de Rham theorem 
in this form has been considered in \cite{SimpsonTeleman}. We discuss a realization below in
\ref{StrucSynthCohomology}.

\subsubsection{Jet bundles}
\label{JetBundles}

In the presence of infinitesimal cohesion there is a canonical higher
analog notion of \emph{jet bundles}: the generalization of tangent bundles
to higher order infinitesimals (higher order tangents).

\medskip

\begin{definition}
 \label{JetBundle}
 \index{jet $\infty$-bundle}
 \index{structures in a cohesive $\infty$-topos!jet $\infty$-bundles}
For any object $X \in \mathbf{H}$ write
$$
  \mathrm{Jet} 
    :
  \xymatrix{
    \mathbf{H}_{/X} 
	  \ar@<-3pt>[rr]_{i_*}
	  \ar@{<-}@<+3pt>[rr]^{i^*}
	  &&
    \mathbf{H}_{/\mathbf{\Pi}_{\mathrm{inf}}(X)}
  }
$$
for the base change geometric morphism, prop. \ref{BaseChangeGeomMorphism}, 
induced by the constant infinitesimal path inclusion $i : X \to \mathbf{\Pi}_{\mathrm{inf}}(X)$, 
def. \ref{InfinitesimalPathsAndReduction}.

For $(E \to X) \in \mathbf{H}_{/X}$ we call $\mathrm{Jet}(E) \to \mathbf{\Pi}_{\mathrm{inf}}(X)$ as 
well as its pullback $i^* \mathrm{Jet}(E) \to X$ (if the context is clear) the \emph{jet $\infty$-bundle} 
of $E \to X$.
\end{definition}
\begin{remark}
In the context over an algebraic site the construction of 
def. \ref{JetBundle} reduces to the construction in 
section 2.3.2 of \cite{BeilinsonDrinfeld}, see \cite{Paugam} for a review.
\end{remark}

\subsubsection{Infinitesimal Galois theory / Formally {\'e}tale morphisms }
\label{FormallySmooth}
\index{structures in a cohesive $\infty$-topos!formally smooth/{\'e}tale/unramified morphisms}

In every context of infinitesimal cohesion, there are canonical induced notions
of morphisms being \emph{formally {\'e}tale}, meaning that at least on infinitesimal
neighbourhoods of every point they behave like the analog of what in topology is 
a \emph{local homeomorphism}/\emph{{\'e}tale map}. Close cousins of this are the
notions of \emph{formally smooth} and of \emph{formally unramified} morphisms.

We first discuss formal {\'e}taleness in $\mathbf{H}$. 
Below in def. \ref{FormallEtaleMorphismInHth}
we discuss the notion more generally in $\mathbf{H}_{\mathrm{th}}$.
\begin{definition}
 \label{FormalSmoothness}
We say an object $X \in \mathbf{H}_{\mathrm{th}}$ is 
\emph{formally smooth} if the constant infinitesimal path inclusion, 
def. \ref{InfinitesimalPathsAndReduction},
$$
  X \to \mathbf{\Pi}_{\mathrm{inf}}(X)
$$
is an effective epimorphism, def. \ref{EffectiveEpimorphism}.
\end{definition}
\begin{remark}
In this form this is the direct $\infty$-categorical analog of 
the characterization of formal smoothness in \cite{SimpsonTeleman}.
The following equivalent reformulation corresponds in turn to the discussion
in section 4.1 of \cite{RosenbergKontsevich}.
\end{remark}
\begin{definition}
  Write
  $$
    \phi : i_! \to i_*
  $$
  for the canonical natural transformation given as the composite
  $$
    \xymatrix{
      i_!
      \ar[r]^-{\eta i_!}
	  &
      \mathbf{\Pi}_{\mathrm{inf}} i_! 
      \ar@{-}[r]^{:=} 
	  &
	  i_* i^* i_!
      \ar[r]^-{\simeq}
	  &
      i_*
    }	  
  \,.
$$
 \label{TheCanonicalNaturalTransformationOfInfinitesimalCohesion}
\end{definition}
Since the last composite on the right here is an equivalence due to $i_!$ being 
fully faithful we have:
\begin{proposition}
  \label{FormalSmoothnessByCanonicalMorphism}
An object $X \in \mathbf{H} \stackrel{i_!}{\hookrightarrow} \mathbf{H}_{\mathrm{th}}$ 
is formally smooth according to 
def. \ref{FormalSmoothness} precisely if the canonical morphism
$$
  \phi : i_! X \to i_{*} X
$$
is an effective epimorphism.
\end{proposition}
\begin{remark} 
In this form this characterization of formal smoothness is the evident generalization 
of the condition given in section 4.1 of \cite{RosenbergKontsevich}. 
(Notice that the notation there is related to the one used here by $u^* = i_!$, $u_* = i^*$ and $u^! = i_*$.)
\end{remark}
Therefore with \cite{RosenbergKontsevich} we have the following more general definitions.
\begin{definition}
  \label{FormalRelativeSmoothnessByCanonicalMorphism}
For $f : X \to Y$ a morphism in $\mathbf{H}$, we say that 
\begin{enumerate}
\item $f$ is a \emph{formally smooth morphism}\index{formally smooth morphism}
 if the canonical morphism
   $$
     i_! X 
      \to 
     i_! Y \prod_{i_* Y} i_* Y
   $$
   is an effective epimorphism;

\item $f$ is a \emph{formally {\'e}tale morphism}\index{formally {\'e}tale morphism} 
  if this morphism is an equivalence, equivalently 
 if the naturality square
   $$
      \xymatrix{
        i_! X   \ar[r]^{i_! f} \ar[d]^{\phi_X} & i_! Y \ar[d]^{\phi_Y}
        \\
        i_{*} X \ar[r]^{i_* f} & i_{*} Y
      }
   $$
   is an  $\infty$-pullback square.

\item $f$ is a \emph{formally unramified morphism} if this is a (-1)-truncated morphism. 
  More generally,  $f$ is an \emph{order-$k$ formally unramified morphisms} 
  for $(-2) \leq k \leq \infty$ if this is a $k$-truncated morphism (\cite{Lurie}, 5.5.6).
\end{enumerate}
\end{definition}
\begin{remark}
An order-$(-2)$ formally unramified morphism is equivalently a formally {\'e}tale morphism.
Only for 0-truncated $X$ does formal smoothness together with formal unramifiedness imply 
formal {\'e}taleness.
\end{remark}
\begin{remark}
  \label{EtalenessByPullbackInLiterature}
  The idea of characterizing {\'e}tale morphisms 
  with respect to a notion of \emph{infinitesimal extension} 
  as those making certain naturality 
  squares into pullback squares goes back to lectures by Andr{\'e} Joyal in the 1970s,
  as is recalled in the introduction of \cite{DubucEtale}. 
  Notice that in sections 
  3 and 4 there the analog of our functor $i_!$ is assumed to be the inverse image of
  a geometric morphism, whereas here we only require it to be a left adjoint and to 
  preserve finite products, as opposed to all finite limits. Indeed, it will fail
  to preserve general pullbacks in most models for infinitesimal cohesion of interest, 
  such as the one discussed
  below in \ref{SynthDiffInfGrpd}. 
  In \cite{JoyalMoerdijk} a different kind of axiomatization, by way of closure properties.
  This we discuss further below, see remark \ref{AsOpenMaps}.
  
  The characterization of formal {\'e}taleness by cartesian naturality squares
  induced specifically by adjoint triples of functors, 
  as in our def. \ref{FormalSmoothness}, appears around 
  prop. 5.3.1.1 of \cite{RosenbergKontsevich}.
\end{remark}
But in view of prop. \ref{FormalSmoothness}, which applies to objects in 
$\mathbf{H}_{\mathrm{th}}$ not necessarily in the image of the inclusion 
$i_!$, and in view of def. \ref{TheCanonicalNaturalTransformationOfInfinitesimalCohesion} 
it is natural to generalize further:
\begin{definition}
  A morphism $f : X \to Y$ in $\mathbf{H}_{\mathrm{th}}$ is
  a \emph{formally {\'e}tale morphism} if the naturality diagram
  $$
    \xymatrix{
	  X \ar[r] \ar[d]^f & \mathbf{\Pi}_{\mathrm{inf}}(X) \ar[d]^{\mathbf{\Pi}_{\mathrm{inf}}(f)}
	  \\
	  Y \ar[r] & \mathbf{\Pi}_{\mathrm{inf}}(Y)
	}
  $$
  of the infinitesimal path inclusion, def. \ref{InfinitesimalPathsAndReduction}, is an $\infty$-pullback.
  \label{FormallEtaleMorphismInHth}
\end{definition}
\begin{remark}
  Def. \ref{FormallEtaleMorphismInHth} is compatible with 
  def. \ref{FormalRelativeSmoothnessByCanonicalMorphism} in that a morphism $f \in \mathbf{H}$
  is formally {\'e}tale in the sense of the former precisely if $i_! f \in \mathbf{H}_{\mathrm{th}}$ is 
  formally {\'e}tale in the sense of the latter.
\end{remark}
\begin{remark}
  This condition is the immediate infinitesimal analog of the notion of 
  \emph{$\mathbf{\Pi}$-closure} in def. \ref{PiClosure}: we may say equivalently that
  a morphism $f \in \mathbf{H}_{\mathrm{th}}$ is formally {\'e}tale precisely
  if it is \emph{$\mathbf{\Pi}_{\mathrm{inf}}$-closed}. 
  Moreover, by the discussion in \ref{StrucGaloisTheory} the $\mathbf{\Pi}$-closed
  morphisms into some $X$ are interpreted as the total space projections of 
  \emph{locally constant $\infty$-stacks} over $X$ by general abstract 
  Galois theory. 
  Accordingly here 
  we may think of $\mathbf{\Pi}_{\mathrm{inf}}$-closed morphisms into $X$ as
  total space projections of more general $\infty$-stacks over $X$
  by what we may call general abstract \emph{infinitesimal Galois theory}. 
  This perspective we develop below in \ref{StructureSheaves}.
  \index{structures in a cohesive $\infty$-topos!Galois theory!infinitesimal}
\end{remark}
In particular, we have the following immediate infinitesimal analogs of 
properties of $\mathbf{\Pi}$-closure.
\begin{definition}
  Call a morphism $f : X \to Y$ in $\mathbf{H}_{\mathrm{th}}$ a 
  \emph{$\mathbf{\Pi}_{\mathrm{inf}}$-equivalence} if $\mathbf{\Pi}_{\mathrm{inf}}(f)$
  is an equivalence.
\end{definition}
\begin{proposition}
  For $i : \mathbf{H} \to \mathbf{H}_{\mathrm{th}}$ a differentially cohesive $\infty$-topos,
  the pair of classes of morphisms
  $$
    \left(
	  \mbox{$\mathbf{\Pi}_{\mathrm{inf}}$-equivalences},
	  \;
	  \mbox{formally {\'e}tale morphisms}
	\right)
	\subset \mathrm{Mor}(\mathbf{H}_{\mathrm{th}}) \times \mathrm{Mor}(\mathbf{H}_{\mathrm{th}})
  $$
  constitutes an orthogonal factorization system.
  \label{PiEquivPiClosedFactorization}
\end{proposition}
\proof
  Since $\mathbf{\Pi}_{\mathrm{inf}}$ has the left adjoint $\mathbf{Red}$
  it preserves all $\infty$-pullbacks and hence in particular those over
  objects of the form $\mathbf{\Pi}_{\mathrm{inf}}(X)$. Therefore 
  factorization follows
  as in the proof of prop. \ref{GaloisFactorization}. Accordingly, 
  orthogonality follows as in the proof of prop. \ref{PiEquivalencePiClosedFactorizationSystem}. 
\endofproof
This and the fact that $\mathbf{\Pi}_{\mathrm{inf}}$ preserves $\infty$-limits
implies a wealth of stability properties of formally {\'e}tale maps.
\begin{corollary}
  Formally {\'e}tale morphisms in $\mathbf{H}_{\mathrm{th}}$, def. \ref{FormallEtaleMorphismInHth},
  satisfy the following stability properties
\begin{enumerate}
\item 
    Every equivalence is formally {\'e}tale.
\item
    The composite of two formally {\'e}tale morphisms is itself formally {\'e}tale.
\item If
   $$
     \xymatrix{
       & Y \ar[dr]^g
       \\
       X \ar[ur]^f \ar[rr]^h  && Z
     }
   $$
   is a diagram such that $g$ and $h$ are formally {\'e}tale, then also $f$ is formally {\'e}tale.
\item Any retract of a formally {\'e}tale morphisms is itself formally {\'e}tale.
\item The $\infty$-pullback of a formally {\'e}tale morphisms is formally {\'e}tale.
\end{enumerate}
\label{StabilityOfEtaleMorphismsInHth}
\end{corollary}
But since the embedding functor $i_!$ does not preserve $\infty$-limits in general,
closure under pullback in $\mathbf{H}$ requires a condition on the codomain:
\begin{proposition} 
 \label{PropertiesOfFormallyEtaleMorphisms}
The collection of formally {\'e}tale morphisms in $\mathbf{H}$, def. 
\ref{FormalRelativeSmoothnessByCanonicalMorphism}, is closed under the following operations.
\begin{enumerate}
\item 
    Every equivalence is formally {\'e}tale.
\item
    The composite of two formally {\'e}tale morphisms is itself formally {\'e}tale.

\item If
   $$
     \xymatrix{
       & Y \ar[dr]^g
       \\
       X \ar[ur]^f \ar[rr]^h  && Z
     }
   $$
   is a diagram such that $g$ and $h$ are formally {\'e}tale, then also $f$ is formally {\'e}tale.

\item Any retract of a formally {\'e}tale morphisms is itself formally {\'e}tale.

\item The $\infty$-pullback of a formally {\'e}tale morphisms is formally {\'e}tale if the pullback 
 is preserved by $i_!$.
\end{enumerate}
\end{proposition}
\begin{remark}
The statements about closure under composition and pullback appears as
prop. 5.4, prop. 5.6 in \cite{RosenbergKontsevich}. The extra assumption that 
$i_!$ preserves the pullback is implicit in their setup.
\end{remark}
\proof
The first statement follows trivially because $\infty$-pullbacks are well defined up
to equivalence.
The second two statements follow by the pasting law for $\infty$-pullbacks, 
prop. \ref{PastingLawForPullbacks}: 
let $f : X \to Y$ and $g : Y \to Z$ be two 
morphisms and consider the pasting diagram 
$$
  \xymatrix{
    i_! X \ar[r]^{i_! f } \ar[d] & i_! Y \ar[r]^{i_! g} \ar[d] &  Z \ar[d]
    \\
    i_* X \ar[r]^{i_* f } & i_* Y \ar[r]^{i_* g} & i_* Z    
  }
  \,.
$$
If $f$ and $g$ are formally {\'e}tale then both small squares are pullback squares. 
Then the pasting law says that so is the outer rectangle and hence $g \circ f$ is formally {\'e}tale. 
Similarly, if $g$ and $g \circ f$ are formally {\'e}tale then the right square and the total reactangle 
are pullbacks, so the pasting law says that also the left square is a pullback and so also $f$ 
is formally {\'e}tale.

For the fourth claim, let $\mathrm{Id} \simeq (g \to f \to g)$ be a retract in he arrow $\infty$-category
$\mathbf{H}^I$. By applying the natural transformation $\phi : i_! \to i_*$ this becomes a
retract
$$
  \mathrm{Id} \simeq ((i_! g \to i_*g) \to (i_! f \to i_*f) \to (i_! g \to i_*g))
$$
in the category of squares $\mathbf{H}^{\Box}$. By assumption the middle square is an
$\infty$-pullback square and we need to show that the also the outer square is. This follows
generally: a retract of an $\infty$-limiting cone is itself $\infty$-limiting. To 
see this, we invoke the presentation of $\infty$-limits by \emph{derivators}
(thanks to Mike Shulman for this argument): we have
\begin{enumerate}
  \item $\infty$-limits in $\mathbf{H}$ are computed by homotopy limits in an
    presentation by a model category $K := [C^{\mathrm{op}}, \mathrm{sSet}]_{\mathrm{loc}}$
    \cite{Lurie};
  \item for $j : J \to J^{\triangleleft}$ the inclusion of a diagram into its cone
   (the join with an initial element), the homotopy limit over $C$ is given by forming
   the right Kan extension $j_* : \mathrm{Ho}(K^J(W^{J})^{-1}) \to \mathrm{Ho}(K^{J^{\triangleleft}}(W^{J^{\triangleleft}})^{-1})$,
  \item
    a $J^{\triangleleft}$-diagram $F$ is a homotopy limiting cone precisely if the unit
    $$
      F \to j_* j^* F
    $$
    us an isomorphism.
\end{enumerate}
Therefore we have a retract in $[\Delta[1], [\Box, K]]$
$$
  \xymatrix{
    (i_! g \to i_! g) \ar[r]\ar[d] & (i_! f \to i_! f) \ar[r]\ar[d] & (i_! g \to i_! g) \ar[d]
    \\
    j^* j_*(i_! g \to i_! g) \ar[r] & j^* j_*(i_! f \to i_! f) \ar[r] & 
   j^* j_*(i_! g \to i_! g)     
  }
  \,,
$$
where the middle morphism is an isomorphism. Hence so is the outer morphism and therefore also
$g$ is formally {\'e}tale.

For the last claim, consider an $\infty$-pullback diagram
$$
  \xymatrix{
    A \times_Y X \ar[d]^p \ar[r] & X \ar[d]^f
    \\
    A \ar[r]& Y
  }
$$
where $f$ is formally {\'e}tale.
Applying the natural transformation $\phi : i_! \to i_*$ to this yields a square of squares. 
Two sides of this are the pasting composite
$$
  \xymatrix{
    i_! A \times_Y X \ar[r] \ar[d]^{i_! p} & i_! X \ar[r]^{\phi_X} \ar[d]^{i_! f} &  
      i_{*} X \ar[d]^{i_* f}
    \\
    i_! A \ar[r] & i_! Y \ar[r]^{\phi_Y}& i_* Y
  }
$$
and the other two sides are the pasting composite
$$
  \xymatrix{
     i_! A \times_Y X \ar[r]^{\phi_{A \times_Y X}} \ar[d]^{i_! p} 
      & i_* A \times_Y A
      \ar[r] \ar[d]^{i_* p} & i_* X \ar[d]^{i_* f}
     \\
     i_! A \ar[r]^{\phi_A} & i_* A \ar[r] & i_* Y
  }
  \,.
$$
Counting left to right and top to bottom, we have that
\begin{itemize}
\item the first square is a pullback by assumption that $i_!$ preserves the given pullback;

\item the second square is a pullback, since $f$ is formally {\'e}tale.

\item the total top rectangle is therefore a pullback, by the pasting law;

\item the fourth square is a pullback since $i_*$ is right adjoint and so also preserves pullbacks;

\item also the total bottom rectangle is a pullback, since it is equal to the top  total rectangle;

\item therefore finally the third square is a pullback, by the other clause of the pasting law. 
  Hence  $p$ is formally {\'e}tale.
\end{itemize}
\endofproof

We consider now types of $\infty$-pullbacks that are preserved by $i_!$.
\begin{proposition}
  If $\xymatrix{
    U \ar@{->>}[r] & X
  }$
  is an effective epimorphism in $\mathbf{H}$ that it is addition
  formally {\'e}tale, def. \ref{FormalRelativeSmoothnessByCanonicalMorphism}, 
  then also its image $i_!U \to i_! X$ in $\mathbf{H}_{\mathrm{th}}$
  is an effective epimorphism.
\end{proposition}
\proof
  Because $i_*$ is left and right adjoint it preserves all small
  $\infty$-limits and $\infty$-colimits and therefore preserves
  effective epimorphisms. Since these are stable under $\infty$-pullback, 
  it follows by definition of formal {\'e}taleness that with 
  $i_* U \to i_* X$ also $i_! U \to i_! X$ is an effective epimorphism.
\endofproof
\begin{proposition}
  \label{iPreservesPullbacksOverDiscreteObjects}
  If in a differentially cohesive $\infty$-topos 
  $i : \mathbf{H} \hookrightarrow \mathbf{H}_{\mathrm{th}}$
  both $\mathbf{H}$ as well as $\mathbf{H}_{\mathrm{th}}$
  have an $\infty$-cohesive site of definition, 
  then the functor $i_!$ preserves pullbacks over discrete objects.
\end{proposition}
\proof
  Since it preserves finite products by assumption, the claim follows
  as in the proof of theorem \ref{PiPreservesPullbacksOverDiscretes}.
\endofproof
\begin{proposition}
  If in an infinitesimal cohesive neighbourhood 
  $i : \mathbf{H} \hookrightarrow \mathbf{H}_{\mathrm{th}}$
  both $\mathbf{H}$ as well as $\mathbf{H}_{\mathrm{th}}$
  have an $\infty$-cohesive site of definition, 
  then the morphism $E \to X$ in $\mathbf{H}$ out of the total space 
  of a locally constant $\infty$-stack
  over $X$, \ref{StrucGaloisTheory}, is formally {\'e}tale.
\end{proposition}
\proof
  First observe that every discrete morphism $\mathrm{Disc}(A \stackrel{f}{\to} B)$
  is formally {\'e}tale: since every discrete $\infty$-groupoid is
  an $\infty$-colimit over the $\infty$-functor constant on the point,
  $\phi_* : i_! * \to i_* *$ is an equivalence, and $i_! \to i_*$ preserves
  $\infty$-colimits, so we have that $\phi_{\mathrm{Dic}(A)}$ and $\phi_{\mathrm{Disc}(B)}$
  are equivalences. Therefore the relevant diagram is an $\infty$-pullback.
  
  Next, by definition, $E \to X$ is a pullback of a discrete morphism.
  By prop. \ref{iPreservesPullbacksOverDiscreteObjects} this pullback
  is preserved by $i_!$ and so by prop. \ref{PropertiesOfFormallyEtaleMorphisms}
  also $E \to X$ is locally {\'e}tale.
\endofproof
\begin{remark}
  \label{AsOpenMaps}
The properties listed in prop. \ref{StabilityOfEtaleMorphismsInHth} 
imply in particular that {\'e}tale morphisms in $\mathbf{H}_{\mathrm{th}}$
are ``admissible maps'' modelling a notion of \emph{local homeomorphism} 
in a \emph{geometry for structured $\infty$-toposes} according to def. 1.2.1 of 
\cite{LurieSpaces}. In the terminology used there this means that 
$\mathbf{H}_{\mathrm{th}}$ equipped with its canonical topology and with this 
notion of admissible maps is a \emph{geometry}, see remark \ref{HthWithEtaleMapsIsGeometry} below.

Another proposal for an axiomatization of \emph{open maps} and
{\'e}tale maps has been proposed in \cite{JoyalMoerdijk}, and the above list of properties
covers most, but not necessarily all of these axioms.
\end{remark}

\medskip

In order to interpret the notion of formal smoothness, we close by further discussion of infinitesimal reduction.
\begin{observation} 
 \label{RedIsIdempotent}
The operation $\mathbf{Red}$ is an idempotent projection of
$\mathbf{H}_{\mathrm{th}}$ onto the image of $\mathbf{H}$ under $i_!$:
$$
  \mathbf{Red} \, \mathbf{Red} \simeq \mathbf{Red}
  \,.
$$
Accordingly also
$$
  \mathbf{\Pi}_{\mathrm{inf}} \mathbf{\Pi}_{\mathrm{inf}} \simeq \mathbf{\Pi}_{\mathrm{inf}}
$$
and
$$
  \mathbf{\flat}_{\mathrm{inf}} \mathbf{\flat}_{\mathrm{inf}} \simeq 
   \mathbf{\flat}_{\mathrm{inf}}
  \,.
$$
\end{observation}
\proof
By definition of infinitesimal neighbourhood we have that
$i_!$ is a full and faithful $\infty$-functor. It follows that 
$i^* i_! \simeq \mathrm{id}$ and hence
$$
  \begin{aligned}
    \mathbf{Red} \mathbf{Red}
    & \simeq
    i_! i^* i_! i^* 
    \\
    & \simeq i_! i^*
    \\
    & \simeq \mathbf{Red}
  \end{aligned}
  \,.
$$
\endofproof
\begin{observation}
  For every $X \in \mathbf{H}_{\mathrm{th}}$, we have that 
$\mathbf{\Pi}_{\mathrm{inf}}(X)$ is formally smooth according to 
def. \ref{FormalSmoothness}.
\end{observation}
\proof
  By prop. \ref{RedIsIdempotent} we have that 
$$
  \mathbf{\Pi}_{\mathrm{inf}}(X) \to \mathbf{\Pi}_{\mathrm{inf}} \mathbf{\Pi}_{\mathrm{inf}}(X)
$$
is an equivalence. As such it is in particular an 
effective epimorphism.
\endofproof

\subsubsection{Formally {\'e}tale groupoids}
 \label{InfStrucEtaleGroupoid}

We discuss an intrinsic realization of 
the notion of \emph{formally {\'e}tale groupoids} internal to a
differential $\infty$-topos. In typical models, for instance 
that discussed below in \ref{SynthDiffInfGrpd}, formal {\'e}taleness
automatically implies global {\'e}taleness, and so the following
formulation captures the notion of \emph{{\'e}tale groupoid} objects
in a differential $\infty$-topos. 
For a classical texts on {\'e}tale 1-groupoids see \cite{MoerdijkMrcun}.

\medskip

Recall from \ref{StrucInftyGroupoids} that groupoid objects
$\mathcal{G}$ in an $\infty$-topos $\mathbf{H}$ are equivalent
to effective epimorphisms $\xymatrix{U \ar@{->>}[r]^p & X}$ in $\mathbf{H}$, which we
think of as being an \emph{atlas} for $X \in \mathbf{H}$.

\begin{definition}
  \label{FormallyEtaleGroupoid}
  For $\mathbf{H} \stackrel{i}{\hookrightarrow} \mathbf{H}_{\mathrm{th}}$ 
  a differential $\infty$-topos, def. \ref{InfinitesimalCohesiveNeighbourhood},
  we say that a groupoid object is \emph{formally {\'e}tale} if
  the corresponding atlas $\xymatrix{U \ar@{->>}[r]^p & X}$ is a
  formally {\'e}tale morphism, def. \ref{FormalRelativeSmoothnessByCanonicalMorphism}.
\end{definition}
\begin{remark}
  When $\mathbf{H}$ is presented by a category of simplicial (pre)sheaves,
  \ref{InfinityToposPresentation},
  then for any simplicial presheaf $X$ there is, 
  by remark \ref{CanonicalAtlasOfSimplicialPresheaf},
  a canonical atlas, given by the inclusion $\mathrm{const} X_0 \to X$.
  If the presentation of $X$ and the induced canonical 
  atlas is understood explicitly, we often speak just of $X$ itself being
  a formally {\'e}tale groupoid or a \emph{formally {\'e}tale $\infty$-stack}.
\end{remark}

\begin{observation}
  If $\xymatrix{U \ar@{->>}[r]^p & X}$ is a formally {\'e}tale groupoid,
  then both $\xymatrix{i_* U \ar@{->>}[r]^{i_* p} & i_* X}$
  and 
  $\xymatrix{i_! U \ar@{->>}[r]^{i_! p} & i_! X}$
  are effective epimorphisms in $\mathbf{H}_{\mathrm{th}}$.
\end{observation}
\proof
  Since $i_*$ is both left and right $\infty$-adjoint, it preserves
  all the $\infty$-limits and $\infty$-colimits that define
  effective epimorphisms. Then since these are stable under 
  $\infty$-pullback, and since $p : U \to X$ being formally {\'e}tale
  by definition means that $i_! p$ is an $\infty$-pullback
  of $i_*$, it follows that also $i_! p$ is an effective epimorphism.
\endofproof

\subsubsection{Manifolds (separated)}
\label{DifferentialStrucmanifolds}
\index{structures in a cohesive $\infty$-topos!manifolds!separated}

We discuss a formalization of the notion of \emph{separated manifold} 
(Hausdorff manifold) in a context of differential cohesion.

\medskip

Let $\mathbb{A}^1 \in \mathbf{H}$ be a line object exhibiting the cohesion of
$\mathbf{H}$ according to def. \ref{ILocalization}.

\begin{definition}
  An (unseparated) manifold $X \in \mathbf{H}_{\mathrm{th}}$, def. \ref{IntrinsicManifold},
  is \emph{separated} if it admits a defining cover $\phi : \coprod_j \mathbb{A}^n \to X$
  such that the induced {\v C}ech nerve is a formally {\'e}tale grouoid
  over $\coprod_j \mathbb{A}^n$, def. \ref{FormallyEtaleGroupoid}.
  \label{IntrinsicSeparatedManifold}
\end{definition}
\begin{remark}
  In the standard synthetic differential model for differential cohesion, 
  $\mathbb{A}^n \simeq \mathbb{R}^n$ is the standard Cartesian space
  (by prop. \ref{ReallLineExhibitsEuclideanTopologicalCohesion}) and
  formal {\'e}taleness makes the components of the face maps be 
  local diffeomorphisms (prop. \ref{SubmersionImmersioOfManifoldsByFormallySmoothUnramified} below).
  These are in particular open maps, which ensures that the corresponding 
  space $X$ is a smooth Hausdorff manifold in the traditional sense.
  This is prop. \ref{SyntheticDifferentialIntrinsicManifoldsAreOrdinaryHausdorffManifolds} below.
\end{remark}

\subsubsection{Structure sheaves}
\label{StructureSheaves}
\label{CotangentBundles}

For $X \in \mathbf{H}_{\mathrm{th}}$ an object in a differential cohesive $\infty$-topos,
we formulate
\begin{itemize}
  \item the $\infty$-topos $\mathrm{Sh}_{\mathbf{H}}(\mathcal{X})$
  of \emph{$\infty$-sheaves over $X$}, or rather of \emph{formally {\'e}tale maps into $X$};
  \item the \emph{structure sheaf} $\mathcal{O}_X$ of $X$.
\end{itemize}
The resulting pair $(\mathrm{Sh}_{\mathbf{H}}, \mathcal{O}_X)$ 
is essentially a $\mathbf{H}_{\mathrm{th}}$-\emph{structured $\infty$-topos}
in the sense of \cite{LurieSpaces}.

\medskip

One way to  motivate the following construction, is to notice that for 
$G \in Grp(\mathbf{H}_{\mathrm{th}})$ a differential cohesive $\infty$-group with 
de Rham coefficient object 
$\flat_{dR}\mathbf{B}G$ and for $X \in \mathbf{H}_{\mathrm{th}}$, def. \ref{deRhamCoefficientObject} 
any differential homotopy type, the product projection 
$$
  X \times \flat_{\mathrm{dR}} \mathbf{B}G \to X
$$
regarded as an object of the slice $\infty$-topos $(\mathbf{H}_{\mathrm{th}})_{/X}$ 
\emph{almost} qualifies as a ``bundle of flat $\mathfrak{g}$-valued differential forms'' 
over $X$: for $U \to X$ a cover (a 1-epimorphism) regarded in $(\mathbf{H}_{\mathrm{th}})_{/X}$, 
a $U$-plot of this product projection is a $U$-plot of $X$ together with a 
flat $\mathfrak{g}$-valued de Rham cocycle on $X$.

This is indeed what the sections of a corresponding bundle of differential forms over $X$ 
are supposed to look like -- but only \emph{if} $U \to X$ is sufficiently ``spread out'' 
over $X$, hence sufficiently {\'e}tale. Because, on the extreme, 
if $X$ is the point (the terminal object), then there should be no non-trivial 
section of differential forms relative to $U$ over $X$, 
but the above product projection instead reproduces all the sections 
of $\flat_{\mathrm{dR}} \mathbf{B}G$.

In order to obtain the correct cotangent-like bundle from the product 
with the de Rham coefficient object, it needs to be \emph{restricted} to plots out 
of sufficiently {\'e}tale maps into $X$. In order to correctly test differential 
form data, ``suitable'' here should be ``formally'', namely infinitesimally. 
Hence the restriction should be along the full inclusion
$$
  (\mathbf{H}_{\mathrm{th}})_{/X}^{\mathrm{fet}} \hookrightarrow (\mathbf{H}_{\mathrm{th}})_{/X}
$$
of the formally {\'e}tale maps  into $X$. 
Since on formally {\'e}tale covers the sections should be those given 
by $\flat_{\mathrm{dR}}\mathbf{B}G$, one finds that the corresponding 
\emph{sheaf of flat forms} $\mathcal{O}_X(\flat_{\mathrm{dR}}\mathbf{B}G)$ 
must be the \emph{coreflection} of the given projection along this map.

\medskip

\begin{definition}
  For $X \in \mathbf{H}_{\mathrm{th}}$ an object, write 
  $$
    \xymatrix{
      (\mathbf{H}_{\mathrm{th}})_{/X}^{\mathrm{fet}}
	  \ar@{^{(}->}[r]
	  &
	  (\mathbf{H}_{\mathrm{th}})_{/X}
	}
  $$
  for the full sub-$\infty$-category of the slice over $X$, def. \ref{Slice},
  on the formally {\'e}tale morphisms into $X$, def. \ref{FormallEtaleMorphismInHth}.
  \label{InclusionOfEtaleMorphismsInSlice}
\end{definition}
\begin{proposition}
  The inclusion of def. \ref{InclusionOfEtaleMorphismsInSlice} is
  both reflective as well as coreflective: we have a left and a right adjoint
  $$
    \xymatrix{
      (\mathbf{H}_{\mathrm{th}})_{/X}^{\mathrm{fet}}
	  \ar@<+6pt>@{<-}[r]
	  \ar@{^{(}->}[r]
	  \ar@<-6pt>@{<-}[r]_{\mathrm{Et}}
	  &
	  (\mathbf{H}_{\mathrm{th}})_{/X}
	}
	\,.
  $$
  \label{CoReflectivityOfEtaleInclusion}
\end{proposition}
\proof
  The reflection is given by the factorization of prop. \ref{PiEquivPiClosedFactorization}.
  This exhibits $(\mathbf{H}_{\mathrm{th}})_{/X}^{\mathrm{fet}}$ as a 
  presentable $\infty$-category and hence, by the adjoint $\infty$-functor theorem,
  the coreflection exists precisely if the inclusion preserves all small $\infty$-colomits.
  Since the inclusion is full, for this it is sufficient to show that an $\infty$-colimit
  in $(\mathbf{H}_{\mathrm{th}})_{/X}$ of a diagram $A$ that factors through the inclusion,
  $$
    A : I \to 
      (\mathbf{H}_{\mathrm{th}})_{/X}^{\mathrm{fet}}
	  \hookrightarrow
	  (\mathbf{H}_{\mathrm{th}})_{/X}
	  \,,
  $$
  is again in the inclusion. 
  Since moreover $\infty$-colimits in a slice are preserved and detected by 
  the dependent sum, prop. \ref{BaseChange}, we are, by def. \ref{FormallEtaleMorphismInHth},
  reduced to showing that for the above diagram the square
  $$
    \raisebox{20pt}{
    \xymatrix{
	  \underset{\longrightarrow}{\lim}_{i \in I} A_i \ar[r] \ar[d]
	  &  \mathbf{\Pi}_{\mathrm{inf}}\underset{\longrightarrow}{\lim}_{i \in I} A_i \ar[d]
	  \\
	  X \ar[r] & \mathbf{\Pi}_{\mathrm{inf}}(X)
	}
	}
  $$
  is an $\infty$-pullback square in $\mathbf{H}_{\mathrm{th}}$. 
  Since $\mathbf{\Pi}_{\mathrm{inf}}$ is a left adjoint
  by def. \ref{InfinitesimalPathsAndReduction}, this square is equivalent to 
  $$
    \raisebox{20pt}{
    \xymatrix{
	  \underset{\longrightarrow}{\lim}_{i \in I} A_i \ar[r] \ar[d]
	  &  \underset{\longrightarrow}{\lim}_{i \in I} \mathbf{\Pi}_{\mathrm{inf}} A_i \ar[d]
	  \\
	  X \ar[r] & \mathbf{\Pi}_{\mathrm{inf}}(X)
	}
	}
	\,.
  $$
  Now that this square is an $\infty$-pullback follows since $\infty$-colimits
  are preserved by $\infty$-pullback in the $\infty$-topos $\mathbf{H}_{\mathrm{th}}$,
  def. \ref{GiraudRezkLurieAxioms}, and the fact that every component square
  $$
    \raisebox{20pt}{
    \xymatrix{
	  A_i \ar[r] \ar[d]
	  &   \mathbf{\Pi}_{\mathrm{inf}} A_i \ar[d]
	  \\
	  X \ar[r] & \mathbf{\Pi}_{\mathrm{inf}}(X)
	}
	}
  $$
  is an $\infty$-pullback by the assumption that the diagram factored through the
  inclusion of the {\'e}tale morphisms into the slice.
\endofproof
\begin{proposition}
  For $X \in \mathbf{H}_{\mathrm{th}}$, the $\infty$-category
  $(\mathbf{H}_{\mathrm{th}})_{/X}^{\mathrm{fet}}$ of def. \ref{InclusionOfEtaleMorphismsInSlice}
  is an $\infty$-topos, and the defining inclusion into the slice $(\mathbf{H}_{\mathrm{th}})_{/X}$
  is a geometric embedding. 
\end{proposition}
\proof
  By prop. \ref{CoReflectivityOfEtaleInclusion} the $\infty$-category
  $\mathrm{Sh}_{\mathbf{H}}(X)$ is the sub-slice induced by a reflective factorization
  system. This is a stable factorization system (in that the left class of 
  $\mathbf{\Pi}_{\mathrm{inf}}$-equivalences is stable under $\infty$-pullback)
  and reflective factorization systems are stable precisely if the corresponding 
  reflector preserves finite $\infty$-limits. Hence the embedding is a geometric
  embedding of a sub-$\infty$-topos.
\endofproof
\begin{definition}
  \index{structures in a cohesive $\infty$-topos!structure sheaf}
  \index{structures in a cohesive $\infty$-topos!petit $\infty$-topos of an object}
  For $\mathbf{H}_{\mathrm{th}}$ a differential cohesive $\infty$-topos and
  $X \in \mathbf{H}_{\mathrm{th}}$, we call the $\infty$-topos 
  $$
    \mathrm{Sh}_{\mathbf{H}}(X) := (\mathbf{H}_{\mathrm{th}})_{/X}^{\mathrm{fet}}
  $$
  the \emph{petit $\infty$-topos} of $X \in \mathbf{H}_{\mathrm{th}}$. An object
  of $\mathrm{Sh}_{\mathbf{H}}(X)$ we also call an \emph{$\infty$-sheaf over $X$}.
  The composite functor
  $$
    \mathcal{O}_X 
	  : 
	\xymatrix{
	  \mathbf{H}_{\mathrm{th}}
	  \ar[rr]^-{(-) \times X}
	  &&
	  (\mathbf{H}_{\mathrm{th}})_{/X}
	  \ar[rr]^-{\mathrm{Et}}
	  &&
	  (\mathbf{H}_{\mathrm{th}})_{/X}
	  =:
	  \mathrm{Sh}_{\mathbf{H}}(X)
	}
	\,,
  $$
  with $\mathrm{Et}$ the right adjoint of prop. \ref{CoReflectivityOfEtaleInclusion},
  we call the \emph{structure $\infty$-sheaf} of $X$. For $A \in \mathbf{H}_{\mathrm{th}}$
  we say that 
  $$
    \mathcal{O}_X(A) \in \mathrm{Sh}_{\mathbf{H}}(X)
  $$
  is the \emph{$\infty$-sheaf of $A$-valued functions on $X$}.
  \label{PetitToposAndStructureSheaf}
\end{definition}
\begin{proposition}
  The functor $\mathcal{O}_X$ is right adjoint to the forgetful functor
  $$
    \xymatrix{
      \mathrm{Sh}_{\mathbf{H}}(X)
	  :=
	  (\mathbf{H}_{\mathrm{th}})_{/X}^{\mathrm{fet}}
	  \ar@{^{(}->}[r]
	  &
	  (\mathbf{H}_{\mathrm{th}})_{/X}
      \ar[r]^-{\sum_X}
	  &
	  \mathbf{H}_{\mathrm{th}}
	 }
	 \,.
  $$
  In particular it preserves all small $\infty$-limits.
\end{proposition}
\proof
  By essential uniqueness of $\infty$-adjoints, it is sufficient to observe that the
  the component maps are pairwise adjoint. For the first this is prop. \ref{BaseChange},
  for the second it is prop. \ref{CoReflectivityOfEtaleInclusion}.
\endofproof
\begin{remark}
  The triple $(\mathbf{H}_{\mathrm{th}}, \mathrm{can},  \mathrm{fet})$ of the differential cohesive 
  $\infty$-topos equipped with 
  \begin{enumerate}
    \item its \emph{canonical topology} (a collection $\{U_i \to X\}_i$ of morphisms 
	in $\mathbf{H}_{\mathrm{th}}$ is covering precisely if $\coprod_i U_i \to X$ is a 
	1-epimorphism, def. \ref{EffectiveEpimorphism});
	\item its class of formally {\'e}tale morphisms, def. \ref{FormallEtaleMorphismInHth}.
  \end{enumerate}
  is a (large) \emph{geometry} in the sense of \cite{LurieSpaces}. 
  For $X \in \mathbf{H}_{\mathrm{th}}$, the pair $(\mathrm{Sh}_{\mathbf{H}}(X), \mathcal{O}_X)$
  of def. \ref{PetitToposAndStructureSheaf} is a \emph{structured $\infty$-topos}
  with respect to this geometry in the sense of \cite{LurieSpaces}. 
  In fact, it is essentially the structured $\infty$-topos associated to $X$ in the 
  geometry $\mathbf{H}_{\mathrm{th}}$ by def. 2.2.9 there.
  \label{HthWithEtaleMapsIsGeometry}
\end{remark}

We close this section by making explicit the 
special case of $\infty$-sheaves of \emph{flat de Rham coefficients} over $X$.
\begin{definition}
  For $G \in \mathrm{Grp}(\mathbf{H}_{\mathrm{th}})$ a differential cohesive $\infty$-group
  and for $X \in \mathbf{H}_{\mathrm{th}}$ any object, we say that the 
  \emph{$\infty$-sheaf of flat $\exp(\mathfrak{g})$-valued differential forms} over $X$ is
  $$
    \mathcal{O}_X(\flat_{\mathrm{dR}}\mathbf{B}G)
	\in
	(\mathbf{H}_{\mathrm{th}})_{/X}^{\mathrm{fet}}
	\hookrightarrow
	(\mathbf{H})_{\mathrm{th}})_{/X}
	\,,
  $$
  where $\mathcal{O}_X$ is given by def. \ref{PetitToposAndStructureSheaf} and where
  $\flat_{\mathrm{dR}}\mathbf{B}G$ is given by def. \ref{deRhamCoefficientObject}.
  \label{CotangentBundleByEt}
\end{definition}
\begin{definition}
  The canonical point $0 : {*} \to\flat_{\mathrm{dR}}\mathbf{B}G$ induces
  a section 
  $$
    (\mathrm{id}_X, 0) :  X \to X \times \flat_{\mathrm{dR}}\mathbf{B}G
  $$
  of the projection map. The image of this section under the coreflection
  of prop. \ref{CoReflectivityOfEtaleInclusion}
  $$
    \xymatrix{
	  & \mathcal{O}_X(\flat_{\mathrm{dR}}\mathbf{B}G)
	  \ar[d]
	  \\
	  X \ar[r]^= \ar[ur]^{0 := \mathrm{Et}(\mathrm{id},0)} & X
	}
  $$
  we call the \emph{0-section} of the $\infty$-sheaf of flat differential forms.
  \label{TheZeroSection}
\end{definition}

\subsubsection{Critical loci, variational calculus and BV-BRST complexes}
\label{CriticalLoci}
\index{structures in a cohesive $\infty$-topos!phase space}
\index{structures in a cohesive $\infty$-topos!critical locus}

We give a general abstract formulation of the notion of \emph{critical locus}
of a function, the local of its domain where its first derivative vanishes.
Applied to functions that are regarded as \emph{action functionals}
and with a constraint that the differential is trivial on certain boundaries,
this critical locus is known as the space of solutions of the
\emph{Euler-Lagrange equations} of the action functional. 
If the ambient cohesive $\infty$-topos is $\infty$-localic, then this
critical locus is what is called a \emph{derived critical locus},
whose complex of functions is known as the \emph{BV-BRST complex}
of the action functional

\medskip

Let $G \in \mathrm{Grp}(\mathbf{H}_{\mathrm{th}})$ be a differential cohesive $\infty$-group.
Write
$$
  \theta_G : G \to \flat_{\mathrm{dR}}\mathbf{B}G
$$
for its canonical differential form, def. \ref{UniversalFormOnInftyGroup}.
\begin{definition}
 For $X \in \mathbf{H}_{\mathrm{th}}$ any differential cohesive homotopy type and
 for 
 $$
   S : \xymatrix{ X \ar[r] & G}
 $$
 any morphism, write
 $$
   \mathbf{d} S := S^* \theta : \xymatrix{X \ar[r]^-{S} & G \ar[r]^-{\theta_G} & \flat_{\mathrm{dR}}\mathbf{B}G }
 $$
 for its composite with the canonical differential form on $G$, def. \ref{UniversalFormOnInftyGroup}.
 We call this the \emph{de Rham derivative} of $S$.
 
 By def. \ref{CotangentBundleByEt} this corresponds to a section
 $$
   \xymatrix{
     & \mathcal{O}_X(\flat_{\mathrm{dR}}\mathbf{B}G)
	 \ar[d]
     \\
     X \ar[r]^= \ar[ur]^{\mathbf{d}S} & X
   }
 $$
 of the $\infty$-sheaf of flat $G$-valued forms over $X$, which we denote by the same symbols.
 \label{deRhamDifferential}
\end{definition}
\begin{definition}
  The \emph{critical locus} of $S : X \to G$ is the object
  $$
    \underset{x : X}{\sum} (\mathbf{d}S(x) \simeq 0) \in \mathbf{H}_{\mathrm{th}}
  $$ 
  in the $\infty$-pullback
  $$
    \xymatrix{
	   \underset{x : X}{\sum} (\mathbf{d}S(x) \simeq 0)
	   \ar[r]
	   \ar[d]
	   &
	   X
	   \ar[d]^0
	   \\
	   X \ar[r]^-{\mathbf{d}S} & \mathcal{O}_X(\flat_{\mathrm{dR}}\mathbf{B}G)
	}
  $$
  in $\mathrm{Sh}_{\mathbf{H}_{\mathrm{th}}}(X)$, where the
  horizontal section is the de Rham differential of $S$ from def. \ref{deRhamDifferential},
  and where the right vertical morphism is the 0-section of 
  def. \ref{TheZeroSection}.
  \label{CriticalLocus}
\end{definition}

If $X$ here is itself a space of functions, then for \emph{variational calculus}
one wants to constrain the differential of $S$ to vary the data in $X$ only
away from the boundary. This is what the following construction achieves.

\begin{definition}
  Let $\Sigma \in \mathbf{H}_{\mathrm{th}}$ be a manifold, def. \ref{IntrinsicSeparatedManifold},
  with boundary $\partial \Sigma \hookrightarrow \Sigma$. Let $A \in \mathbf{H}_{\mathrm{th}}$
  be any object. Then the \emph{variational domain}
  $$
    [\Sigma, A]_{\partial \Sigma} \in \mathbf{H}_{\mathrm{th}}
  $$
  is the $\infty$-pullback in
  $$
   \raisebox{20pt}{
    \xymatrix{
	  [\Sigma,A]_{\partial \Sigma}
	  \ar[r] \ar[d]
	  & \flat [\partial \Sigma, A]
	  \ar[d]
	  \\
	  [\Sigma,A]
	  \ar[r]
	  &
	  \flat [\Sigma,A]
	}
	}\,.
  $$
  For
  $$
    S : [\Sigma, A]_{\partial \Sigma} \to G
  $$
  a map, we say that its critical locus, def. \ref{CriticalLocus}
  $$
    \underset{\phi : \Sigma \to A}{\sum}
	(\mathbf{d}S(\phi) \simeq 0)
  $$
  is the space of solutions to the \emph{Euler-Lagrange equations} of $S$.
\end{definition}

\subsubsection{Formal groupoids}
  \label{InfStrucFormalInfinityGroupoid}
   \index{structures in a cohesive $\infty$-topos!formal $\infty$-groupoids}

The infinitesimal analog of an exponentiated $\infty$-Lie algebra, \ref{StrucLieAlgebras},
is a formal cohesive $\infty$-group.
\begin{definition} 
 \label{InfinitesimalObject}
An object $X \in \mathbf{H}_{\mathrm{th}}$ is a 
\emph{formal cohesive $\infty$-groupoid}\index{Lie algebroid!formal $\infty$-groupoid}
\index{formal $\infty$-groupoid} if 
$\mathbf{\Pi}_{\mathrm{inf}} X \simeq *$.

An $\infty$-group object $\mathfrak{g} \in \mathbf{H}_{\mathrm{th}}$ 
that is infinitesimal we call a \emph{formal $\infty$-group}. 

For $X \in \mathbf{H}$ any object, we say 
$\mathfrak{a} \in \mathbf{H}_{\mathrm{th}}$ is an 
\emph{formal cohesive $\infty$-groupoid over $X$} if 
$\mathbf{\Pi}_{\mathrm{inf}}(\mathfrak{a}) \simeq \mathbf{\Pi}_{\mathrm{inf}}(X)$; 
equivalently: if there is a morphism
$$
  \mathfrak{a} \to \mathbf{\Pi}_{\mathrm{inf}}(X)
$$
equivalent to the infinitesimal path inclusion, def. \ref{InfinitesimalPathsAndReduction}, 
for $\mathfrak{a}$.
\end{definition}
\begin{proposition}
An infinitesimal cohesive $\infty$-groupoid, def. \ref{InfinitesimalObject} -- 
$X \in \mathbf{H}_{\mathrm{th}}$ with $\mathbf{\Pi}_{\mathrm{inf}}(X) \simeq *$ --
is both geometrically contractible and has as underlying discrete $\infty$-groupoid the point:
\begin{itemize}
\item $\Pi X \simeq *$
\item $\Gamma X \simeq {*}$.
\end{itemize}
\end{proposition}
\proof
The first statement is implied by the fact  
both $i_!$ as well as $i_*$ are full and faithful, by definition of
infinitesimal neighbourhood. This means that
if $\mathbf{\Pi}_{\mathrm{inf}}(X) \simeq *$ then already $i^* X = \Pi_{\mathrm{inf}}(X) \simeq *$.
Since $\Pi_{\mathbf{H}_{\mathrm{th}}} \simeq \Pi_{\mathbf{H}} \Pi_{\mathrm{inf}}$ and 
$\Pi_{\mathbf{H}}$ preserves the terminal object by cohesiveness, this implies the first claim.
 
The second statement follows by 
$$
  \begin{aligned}
    \Gamma X & \simeq \mathbf{H}_{\mathrm{th}}(*,X)
     \\
     & \simeq \mathbf{H}_{\mathrm{th}}(\mathbf{Red}*, X)
     \\
     & \simeq \mathbf{H}_{\mathrm{th}}(*, \mathbf{\Pi}_{\mathrm{inf}}(X))
     \\
     & \simeq \mathbf{H}_{\mathrm{th}}(*,*)
     \\
     & \simeq *
  \end{aligned}
  \,.
$$
\endofproof
\begin{observation}
For all $X \in \mathbf{H}$, we have that $X$ and 
$\mathbf{\Pi}_{\mathrm{inf}}(X)$ are formal cohesive $\infty$-groupoids over $X$, $X$ by the 
constant infinitesimal path inclusion and $\mathbf{\Pi}_{\mathrm{inf}}(X)$ by the identity.
\end{observation}
\proof
For $X$ this is tautological, for $\mathbf{\Pi}(X)$ it follows from prop. 
\ref{RedIsIdempotent} and the $(i^* \dashv i_*)$-zig-zag-identity.
\endofproof
\begin{proposition}
The delooping $\mathbf{B}\mathfrak{g}$ of a formal $\infty$-group $\mathfrak{g}$,
def. \ref{InfinitesimalObject}, is a formal $\infty$-groupoid over the point.
\end{proposition}
\proof
Since both $i^*$ and $i_*$ are right adjoint, $\mathbf{\Pi}_{\mathrm{inf}}$ 
commutes with delooping. Therefore 
$$
  \begin{aligned}
    \mathbf{\Pi}_{\mathrm{inf}} \mathbf{B}\mathfrak{g}
    & \simeq
    \mathbf{B} \mathbf{\Pi}_{\mathrm{inf}} \mathfrak{g}
    \\
    & \simeq \mathbf{B} *
    \\
    & \simeq *
    \\
    & \simeq \mathbf{\Pi}_{\mathrm{inf}} *
  \end{aligned}
  \,.
$$
\endofproof

\newpage

\section{Models} 
\label{Implementation}
\label{Models}
\index{cohesive $\infty$-topos!models}

In this section we construct specific cohesive $\infty$-toposes, \ref{CohesiveToposes},
and differential cohesive $\infty$-toposes, \ref{InfinitesimalCohesion},
and discuss the realization of 
the general abstract structures of \ref{structures} in these models.

We start with a generic class of models
\begin{itemize}
 \item \ref{DiagramsOfCohesiveGroupoids} -- parameterized cohesive homotopy types;
\end{itemize}
which construct a new cohesive $\infty$-topos $T \mathbf{H}$ from a given one $\mathbf{H}$,
the \emph{Goodwillie-tangent $\infty$-topos} of $\mathbf{H}$. Where
a generic $\mathbf{H}$ is a cohesive version of homotopy theory and
\emph{non-abelian} cohomology, its tangent $\infty$-topos $T \mathbf{H}$
extends $\mathbf{H}$ by its stabilization given by stable cohesive homotopy types 
(cohesive spectrum objects) and hence also accommodates the cohesive
\emph{stable homotopy theory} and stable (meaning: generalized Eilenberg-Steenrod-type)
cohesive cohomology. This construction can be considered in particular for all
of the specific models to follow.

Next we discuss the following specific kinds of geometric cohesion:
\begin{itemize}
 \item \ref{DiscreteInfGroupoids} -- discrete cohesion;
 \item \ref{ContinuousInfGroupoids} -- Euclidean-topological cohesion;
 \item \ref{SmoothInfgrpds} -- smooth cohesion;
 \item \ref{SynthDiffInfGrpd} -- synthetic differential cohesion;
 \item \ref{SuperInfinityGroupoids} -- super- and supergeometric cohesion.
\end{itemize}

These six cohesive $\infty$-toposes fit into a diagram of geometric
morphisms of the following form:
$$
  \raisebox{20pt}{
  \xymatrix{
    & \mbox{\bf cohesion} & \mbox{\bf differential cohesion} 
	& \mbox{\begin{tabular}{c}{\bf base} \\{\bf$\infty$-topos}\end{tabular}}
    \\
    \mbox{\bf supergeometry} & \mathrm{SmoothSuper}\infty\mathrm{Grpd}
	\ar@{^{(}->}[r]
	\ar[d]
	&
	\mathrm{SynthDiffSuper}\infty\mathrm{Grpd}
	\ar[r] \ar[d]
	&
	\mathrm{Super}\infty\mathrm{Grpd}
	\ar[d]
	\\
    \mbox{\bf differential geometry} & \mathrm{Smooth}\infty\mathrm{Grpd}
	\ar@{^{(}->}[r]
	&
	\mathrm{SynthDiff}\infty\mathrm{Grpd}
	\ar[r]
	&
	\infty \mathrm{Grpd}
  }
  }
  \,.
$$

In the bottom right we have plain $\infty$-groupoids, modelling \emph{discrete} cohesion, 
\ref{DiscreteInfGroupoids}. The bottom left is the cohesive $\infty$-topos of 
\emph{smooth $\infty$-groupoids},
\ref{SmoothInfgrpds} and the middle entry on the bottom is the cohesive $\infty$-topos
\emph{synthetic differential cohesion}, \ref{SynthDiffInfGrpd}. The total bottom 
row exhibits the latter as a model for \emph{differential cohesion} in the sense of 
\ref{InfinitesimalCohesion}. This we regard as the standard model for 
\emph{higher differential geometry}. The top row shows the supergeometric refinement
of this situation. See below in \ref{SuperInfinityGroupoids} for more discussion of the 
top row of this diagram.

\newpage

\subsection{Parameterized cohesive homotopy theory}
\label{DiagramsOfCohesiveGroupoids}
\index{!parameterized cohesive homotopy theory}

We discuss here, given any cohesive $\infty$-topos $\mathbf{H}$, new 
$\infty$-toposes of objects parameterized over those of $\mathbf{H}$, which are
cohesive over $\mathbf{H}$.

\begin{itemize}
 \item \ref{DiagramToposes} -- Bundles of cohesive homotopy types
 \item \ref{BundlesOfCohesiveSpectra} -- Bundles of cohesive stable homotopy types
\end{itemize}

The first of these is just the arrow category $\mathbf{H}^{\Delta[1]}$ of $\mathbf{H}$. 
While simple in itself, this is conceptually noteworthy as the $\infty$-topos 
whose intrinsic cohomology is \emph{twisted nonabelian cohomology} in $\mathbf{H}$
according to the discussion in \ref{StrucTwistedCohomology}, 
and because it serves an illustrative purpose: it is a simple but non-trivial model
of cohesion that illuminates the central notions, such as cohesive homotopy types, 
by elementary combinatorial reasoning. 

The second of these is the ``fiberwise stabilization'' of the first, the tangent $\infty$-topos
$T \mathbf{H}$ of \emph{parameterized spectrum objects} in $\mathbf{H}$.
This is the class of cohesive $\infty$-toposes whose 
intrinsic intrinsic differential cohomology 
accommodates the \emph{stable} (hence: generalized Eilenberg-Steenrod-type)
differential cohomology in $\mathbf{H}$ in the sense of \cite{HopkinsSinger}
and generally is the \emph{twisted differential stable cohomology}
developed in \cite{BunkeNikolausVoelkl}.

There is in fact a whole tower of cohesive $\infty$-toposes interpolating between these two examples
$$
  \xymatrix{
    \mathbf{H}^{\Delta[1]}
    \ar[r]
    &
    \cdots
    \ar[r]
    &
    J^n \mathbf{H}
    \ar[r]
    &
    \cdots
    \ar[r]
    &
    T \mathbf{H}
    \ar[d]
    \\
    &&&& \mathbf{H}
  }
  \,,
$$
where $J^n \mathbf{H} \simeq \mathrm{Exc}^n(\infty \mathrm{Grpd}^{\ast/}, \mathbf{H})$ is the 
$\infty$-category of $n$-excisive $\infty$-endofunctors. 
(This goes back to the observation in section 35 of 
\cite{JoyalLogoi}, it follows with theorem 1.8 in \cite{Goodwillie03} and more explicitly with 
theorem 7.1.1.10, remark 7.1.1.11 in \cite{LurieAlgebra}.
\footnote{Thanks to Charles Rezk for discussion of this point.}) In terms of 
intrinsic cohomology this chain interpolates stagewise between general non-abelian
twisted differential cohomology in $\mathbf{H}$ on the left and twisted stable (generalized Eilenberg-Steenrod-type)
differential cohomology in $\mathbf{H}$ on the right. Since the higher Chern-Weil theory
discussed here may be regarded as approximating the former by the latter, one may think
of the intermediate stages here as the home of a tower of intermediate higher Chern-Weil
theory. But for the moment we do not explore this further.

\subsubsection{Bundles of cohesive homotopy types}
\label{DiagramToposes}

We discuss a class of examples of cohesive $\infty$-toposes that 
are obtained from a given cohesive $\infty$-topos $\mathbf{H}$ by passing to the 
$\infty$-topos $\mathbf{H}^D$ of interval-shaped diagrams in it. The
cohesive interpretation of an object in $\mathbf{H}^D$
is as a bundle of $\mathbf{H}$-cohesive $\infty$-groupoids
all whose fibers are regarded as being geometrically contractible.

\medskip

\begin{proposition}
  Let $\mathbf{H}$ be a cohesive $\infty$-topos. Let 
  $D$ be a small category with initial object $\bot$ and terminal
  object $\top$.
  
  There is an adjoint triple of $\infty$-functors
  $$
    \xymatrix{
	   D 
	   \ar@<+6pt>[r]^{\bot}
	   \ar@{<-}[r]|{p}
	   \ar@<-6pt>[r]_{\top}
	   &
	   {*}
	}
  $$
  obtained from the inclusion of the terminal and the initial object. 
  
  The $\infty$-functor $\infty$-category $\mathbf{H}^D$ 
  ($D$-shaped diagrams in $\mathbf{H}$) is a cohesive $\infty$-topos,
  exhibited by the composite adjoint quadruple
  $$
   (\Pi \dashv \mathrm{Disc} \dashv \Gamma \dashv \mathrm{coDisc})
    \; : \;
    \xymatrix{
	   \mathbf{H}^D
	   \ar@{->}@<+16pt>[rr]^{\top^*}
	   \ar@{<-}@<+8pt>[rr]|{p^*}
	   \ar@{->}@<-0pt>[rr]|{\bot^*}
	   \ar@{<-^{)}}@<-8pt>[rr]_{\bot_*}
	   &&
	   \mathbf{H}
	   \ar@{->}@<+16pt>[rr]^{\Pi_{\mathbf{H}}}
	   \ar@{<-}@<+8pt>[rr]|{\mathrm{Disc}_{\mathbf{H}}}
	   \ar@{->}@<-0pt>[rr]|{\Gamma_{\mathbf{H}}}
	   \ar@{<-^{)}}@<-8pt>[rr]_{\mathrm{coDisc}_{\mathbf{H}}}
	   &&
	   \infty \mathrm{Grpd}
	}
	\,.
  $$
\end{proposition}
\proof
  Each of the first three functors induces an adjoint triple
  $(p_! \dashv p^* \dashv p_*)$, etc., where $p^*$ is given by 
  precomposition, $p_!$ by left $\infty$-Kan extension and $p_*$ by
  right  $\infty$-Kan extension (use for instance \cite{Lurie}, A.2.8).
  In particular therefore $\top^*$ preserves finite products
  (together with all small $\infty$-limits).
  The adjointness $(\bot \dashv p \dashv \top)$ implies that
  $p_! \simeq \top^*$ and $\bot_! \simeq p^*$. This yields
  the adjoint quadruple as indicated.
  Finally it is clear that $\top^* p^* \simeq \mathrm{id}$,
  which means that $p^*$ is full and faithful, and by adjointness
  so is $\bot_*$.
\endofproof
The following simple example not only illustrates the above
proposition, but also serves as a useful toy example for the 
notion of cohesion itself. 
\begin{example}
  \label{SierpinskiTopos}
  \index{Sierpinski $\infty$-Topos}
  \index{cohesive $\infty$-topos!Sierpinski $\infty$-topos}

  For $\mathbf{H}$ any cohesive $\infty$-topos, also its 
  arrow category $\mathbf{H}^{\Delta[1]}$ is cohesive.
  
  In particular, for $\mathbf{H} = \infty \mathrm{Grpd}$
  (see \ref{DiscreteInfGroupoids} for a discussion of 
  $\infty\mathrm{Grpd}$ as a cohesive $\infty$-topos), the arrow
  $\infty$-category $\infty \mathrm{Grpd}^{\Delta[1]}$ is
  cohesive. This is equivalently the $\infty$-category
  of $\infty$-presheaves on the interval $\Delta[1]$, which in turn is 
  equivalent to the $\infty$-category of $\infty$-sheaves
  on the topological spaces called the \emph{Sierpinski space}
  $$
     \mathrm{Sierp}
	 =
	 \left(
	   \{0,1\}, \mathrm{Opens} = (\emptyset \hookrightarrow \{1\} \hookrightarrow \{0,1\})
	 \right)
  $$
  (see for instance \cite{Johnstone}, B.3.2.11):
  $$
    \infty \mathrm{Grpd}^{\Delta[1]}
	 \simeq
	\mathrm{PSh}_{\infty}(\Delta[1])
	 \simeq
	\mathrm{Sh}_\infty(\mathrm{Sierp})
	\,.
  $$
  We call this the \emph{Sierpinski $\infty$-topos}.
  
  Notice that the Sierpinski space, as a topological space,
  \begin{enumerate}
    \item is contractible;
	\item is locally contractible;
	\item has a focal point (a point whose only open neighbourhood is the entire space).
  \end{enumerate}
  The Sierpinski $\infty$-topos is 0-localic, being the image of the 
  Sierpinski space under the embedding of topological spaces into 
  $\infty$-toposes. Accordingly the cohesion of $\mathrm{Sh}_{\infty}(\mathrm{Sierp})$
  may be traced back to these three properties, which imply, in this order,
  that $\mathrm{Sh}_\infty(\mathrm{Sierp})$ is, as an $\infty$-topos,
  \begin{enumerate}
    \item $\infty$-connected;
	\item locally $\infty$-connected;
	\item local.
  \end{enumerate}
  So the Sierpinski space is the ``abstract cohesive blob'' on which the
  cohesion of $\mathrm{Sh}_\infty(\mathrm{Sierp})$ is modeled: it is 
  the abstract ``point with an open neighbourhood''.
  
  While the cohesion encoded by the Sierpinski $\infty$-topos is very simple, 
  it may be instructive to make the geometric interpretation fully explicit
  (the reader may want to compare the following with the more detailed discussions
  of the meaning of the functor $\Pi$ on a cohesive $\infty$-topos below 
  in \ref{StrucGeometricHomotopy}):
  
  an object of $\mathrm{Sh}_\infty(\mathrm{Sierp})$ is a morphism $[P \to X]$
  in $\infty \mathrm{Grpd}$. The functor $\Pi$ sends this to its domain
  $$
    \Pi([P \to X]) \simeq X
	 \,.
  $$
  In particular
  $$
    \Pi([P \to *]) \simeq *
	\,.
  $$
  Therefore $\Pi$ sees $[P \to *]$ as being cohesively/geometrically contractible
  and sees a bundle $[P \to X]$ as having cohesively/geometrically 
  contractible fibers. At the same time, for $X \in \infty \mathrm{Grpd}$, 
  we have
  $$
    \mathrm{Disc}(X) \simeq [X \stackrel{id}{\to} X]
	\,,
  $$
  which says that the base of such a bundle is regarded by the
  cohesion of the Sierpinski $\infty$-topos as being 
  discrete. Accordingly, we may interpret $[P \to X]$ as describing a 
  discrete $\infty$-groupoid $X$ to which are attached cohesively contractible
  blobs, being the fibers of the morphism $P \to X$.
  
  Even though they are geometrically contractible, these fibers have inner structure:
  this is seen by $\Gamma$, which takes the underlying $\infty$-groupoid to be the
  total space of the bundle
  $$
    \Gamma([P \to X]) \simeq P
	\,.
  $$
  Finally a codiscrete object is one of the form
  $$
    \mathrm{coDisc}(Q) \simeq [Q \to *]
	\,,
  $$
  which is entirely cohesively contractible, for any inner structure.
\end{example}
\begin{observation}
  Let $\mathbf{H}$ be a cohesive $\infty$-topos and 
  regard the Sierpinski $\infty$-topos $\mathbf{H}^{I}$, def. \ref{SierpinskiTopos},
  as a cohesive $\infty$-topos over $\mathbf{H}$. Then 
  \begin{enumerate}
    \item the full sub-$\infty$-category of $\mathbf{H}^I$ on those
	 objects for which \emph{pieces have points}, def. \ref{PiecesHavePoints},
	 is canonically identified with the $\infty$-category
	 of effective epimorphisms in $\mathbf{H}$, hence with the 
	 $\infty$-category of groupoid objects in $\mathbf{H}$, def. \ref{GroupoidObject};
	\item 
	  the full sub-$\infty$-category of $\mathbf{H}^I$ on those
	 objects which have \emph{one point per piece}, def. \ref{PiecesHavePoints},
	 is canonically identified with $\mathbf{H}$ itself.
  \end{enumerate}
\end{observation}

\subsubsection{Bundles of cohesive stable homotopy types}
\label{BundlesOfCohesiveSpectra}
\index{structures in a cohesive $\infty$-topos!twisted cohomology!stable (Eilenberg-Steenrod-type)}  

We discuss here how given a cohesive $\infty$-topos $\mathbf{H}$,
there is its \emph{tangent $\infty$-topos} $T \mathbf{H}$
which is itself cohesive over $T \infty \mathrm{Grpd}$
and which is an extension of $\mathbf{H}$ by the stabilization
$\mathrm{Stab}(\mathbf{H})$ of $\mathbf{H}$, hence by the 
$\infty$-category of spectrum objects in $\mathbf{H}$ \cite{LurieAlgebra}.
We observe that this is the class of $\infty$-toposes whose intrinsic
cohomology is \emph{twisted stable cohomology}
and that the \emph{stable} homotopy types inside $T \mathbf{H}$
\emph{all} canoninically sit in the system of homotopy fiber
sequences characteristic of (stable) differential cohomology
(an observation due to \cite{BunkeNikolausVoelkl}).

\medskip

The following goes back to theorem 1.8 in \cite{Goodwillie03},
see  section 7.1.1 and section 8.3 in \cite{LurieAlgebra}. 
We present it in the fashion of section 35 of \cite{JoyalLogoi}.
\begin{definition}
  Let $\mathrm{seq}$ be the diagram $\infty$-category of the form
  $$
    \mathrm{seq}
    :=
    \left\{
    \raisebox{75pt}{
    \xymatrix{
      & \vdots \ar[d]
      \\
      \cdots \ar[r] & x_{n-1} \ar[r]_>{\ }="s1" \ar[d]^>{\ }="t1" & \ast \ar[d]  \ar[dr]^{\mathrm{id}}
      \\
      & \ast \ar[dr]_{\mathrm{id}} \ar[r] & x_{n} \ar[r]_>{\ }="s" \ar[d]^>{\ }="t" 
      &  \ast \ar[d]
      \\
      & & \ast \ar[r] & x_{n+1} \ar[r] \ar[d] & \cdots
      \\
      & & & \vdots
      \ar@{=>} "s"; "t"
      \ar@{=>} "s1"; "t1"
    }}
    \right\}
    \,,
  $$
  where $n$ ranges over $\mathbb{Z}$.
  For $\mathcal{C}$ an $\infty$-category, 
  write
  $$
    \mathcal{C}^{\mathrm{seq}} := \mathrm{Func}(\mathrm{seq}, \mathcal{C})
  $$
  for the $\infty$-category of $\infty$-functors from $\mathrm{seq}$.
  \label{PretangentCategory}
\end{definition}
\begin{remark}
  For $\mathcal{C}$ an $\infty$-category with finite $\infty$-limits,
  an $\infty$-functor $E_\bullet : \mathrm{seq} \longrightarrow \mathcal{C}$
  is equivalently
  \begin{enumerate}
    \item a choice of object $B \in \mathcal{C}$ (the image of the zero-object of $\mathrm{seq}$);
    \item a collection $\{E_n \in \mathcal{C}_{/B}\}_{n \in \mathbb{Z}}$ of objects 
      in the slice of $\mathcal{C}$ over $B$, def. \ref{Slice} 
      (the images of the $x_n \in \mathrm{seq}$);
    \item 
      for each $n \in \mathbb{Z}$ a choice of homotopy from the zero-map 
      $0_n : E_n \longrightarrow E_{n+1}$ to itself, which by 
      the universal property of the $\infty$-fiber product is equivalently a 
      map
      $$
        E_n \longrightarrow \Omega_B E_{n+1}
      $$
      into the loop space object, def. \ref{loop space object}, 
      of $E_{n+1} \in \mathcal{C}_{/C}$. 
  \end{enumerate}
  One might call such a collection of data a \emph{spectrum object} 
  over $B$, but better to call it a \emph{pre-spectrum object} over $B$.
  \label{DataInPrespectrumObjects}
\end{remark}
\begin{definition} 
  For $\mathcal{C}$ an $\infty$-category with finite $\infty$-limits,
  an object $E_\bullet \in \mathcal{C}^{\mathrm{seq}}$, def. \ref{PretangentCategory},
  over $B \in \mathcal{C}$
  for which the morphisms 
  of remark \ref{DataInPrespectrumObjects} are equivalences
  $$
    E_n \stackrel{\simeq}{\longrightarrow} \Omega_B E_{n+1}
    \;\;,
    \;\;\;\;
    n \in \mathbb{Z}
  $$
  we call an \emph{$\Omega$-specturm object} 
  over $B$ or just \emph{spectrum object} over $B$. We write
  $$
    T \mathcal{C} \hookrightarrow \mathcal{C}^{\mathrm{seq}}
  $$  
  for the full sub-$\infty$-category of $\mathcal{C}^{\mathrm{seq}}$,
  def. \ref{PretangentCategory}, on the $\Omega$-spectrum objects and
  call this the \emph{Goodwillie-tangent $\infty$-category} of 
  $\mathcal{C}$, or just \emph{tangent $\infty$-category}, for short.
  \label{TangentInfinityTopos}
\end{definition}
The following observation is originally due to Georg Biedermann, 
see section 35 of \cite{JoyalLogoi}.
\begin{proposition}
  For $\mathbf{H}$ an $\infty$-topos,
  the inclusion $T \mathbf{H} \hookrightarrow \mathbf{H}^{\mathrm{seq}}$
  is left exact reflective, hence it has a left adjoint $\infty$-functor
  (``spectrification'')
  which preserves finite $\infty$-limits
  $$
    \xymatrix{
      T \mathbf{H}
      \ar@<+8pt>@{<-}[r]^{\mathrm{lex}}
      \ar@{^{(}->}[r]
      &
      \mathbf{H}^{\mathrm{seq}}
    }
    \,.
  $$
  \label{SpectrificationIsLexLeftAdjoint}
\end{proposition}
\proof
  By a small object argument in the presentable $\infty$-category $\mathbf{H}$,
  one finds that the left adjoint exists and 
  is given by a sufficiently long transfinite composite
  of looping maps $\mathrm{id} \longrightarrow \Omega$. This transfinite
  composition is an example of a filtered $\infty$-colimit and in 
  an $\infty$-topos these preserve finite $\infty$-limits,
  for instance by example 7.1.1.8 in \cite{LurieAlgebra}.
\endofproof
It therefore follows that
\begin{proposition}
  For $\mathbf{H}$ an $\infty$-topos also 
  its tangent $\infty$-category $T \mathbf{H}$,
  def. \ref{TangentInfinityTopos},
  is an $\infty$-topos, to be called its \emph{tangent $\infty$-topos}.
  \label{TangentInfinityTopos}
\end{proposition}
\proof
  By prop. \ref{SpectrificationIsLexLeftAdjoint} $T \mathbf{H}$
  is a left exact reflective sub-$\infty$-category of an $\infty$-topos,
  and so by the very definition 
  \ref{SheafInfinityTopos} is itself an $\infty$-topos.
\endofproof
\begin{proposition}
  If $\mathbf{H}$ is an $\infty$-topos which is cohesive, def. \ref{CohesiveInfinToposDefinition},
  then its tangent $\infty$-topos
  $T \mathbf{H}$, prop. \ref{TangentInfinityTopos}, 
  is cohesive over $T \infty \mathrm{Grpd}$
  and infinitesimally cohesive def. \ref{InfinitesimalCohesion}, 
  over $\mathbf{H}$.
  Moreover, the cohesive structure maps fit into a diagram of the form
$$
  \xymatrix{
    &&\mathrm{Stab}(\mathbf{H})
    \ar@<+18pt>@{->}[rr]|{\Pi^{\mathrm{sp}}}
    \ar@<+8pt>@{<-^{)}}[rr]|{\mathrm{Disc}^{\mathrm{sp}}}
    \ar@<-2pt>[rr]|{\Gamma^{\mathrm{sp}}}
    \ar@<-12pt>@{<-^{)}}[rr]|{\mathrm{coDisc}^{\mathrm{sp}}}
    \ar@{^{(}->}[dd]
    &&
    \mathrm{Spectra}
    \ar@{^{(}->}[dd]
    \\
    \\
    \mathbf{H} 
    \ar@<+8pt>@{->}[rr]|-{d}
    \ar@<-2pt>@{<-}[rr]|-{\mathrm{tot}}
    &&
    T\mathbf{H}
    \ar@<+18pt>@{->}[rr]|{T\Pi}
    \ar@<+8pt>@{<-^{)}}[rr]|{T\mathrm{Disc}}
    \ar@<-2pt>[rr]|{T\Gamma}
    \ar@<-12pt>@{<-^{)}}[rr]|{T\mathrm{coDisc}}
    \ar@<-10pt>[dd]|{\mathrm{base}}
    \ar@<+0pt>@{<-^{)}}[dd]
    &&
    T\infty \mathrm{Grpd}
    \ar[dd]|{\mathrm{base}}
    \ar@<+10pt>@{<-^{)}}[dd]
    \\
    \\
    &&
    \mathbf{H}
    \ar@<+18pt>@{->}[rr]|{\Pi}
    \ar@<+8pt>@{<-^{)}}[rr]|{\mathrm{Disc}}
    \ar@<-2pt>[rr]|{\Gamma}
    \ar@<-12pt>@{<-^{)}}[rr]|{\mathrm{coDisc}}
    &&
    \infty \mathrm{Grpd}
  }
  \,,
  $$
  where
  \begin{itemize}
    \item $\mathrm{Stab}(\mathbf{H})$ is the stabilization of $\mathbf{H}$, 
    the stable $\infty$-category of spectrum object in $\mathbf{H}$ \cite{LurieAlgebra};
    \item $\mathrm{Spectra} = \mathrm{Stab}(\infty \mathrm{Grpd})$ is the stable
    $\infty$-category of spectra;
    \item $\mathrm{base}$ is the $\infty$-functor that sends a bundle of spectra to its base
    homotopy type, exhibiting the infinitesimal cohesion of $T \mathbf{H}$ over $\mathbf{H}$;
    \item its left and also right adjoint is the $\infty$-functor that assigns the 0-bundle of spectra
    to a given base homotopy type;
    \item $\mathrm{tot}$ is the $\infty$-functor which sends a bundle $E_\bullet$ of spectra 
    in a slice of $\mathbf{H}$
    to $\Omega^\infty E_\bullet = E_0$:
    \item $d$ is its left adjoint
  \end{itemize}
\end{proposition}
\proof
  To see that $T \mathbf{H}$ is cohesive over $\mathbf{H}$ observe that the
  prolongation of the right adjoints $(\mathrm{Disc} \dashv \Gamma \dashv \mathrm{coDisc})$
  to presheaves over $\mathrm{seq}$, as in the proof of prop. \ref{TangentInfinityTopos},
  immediately descent to $T \mathbf{H}$, since they preserve $\infty$-limits and hence
  the loop space objects involved in the definition of spectrum objects. 
  The prolongation of $\Pi$ may fail to preserve these but by the lex reflection
  of spetrum objects inside pre-spectrum objects it follows that the composition of 
  the prolongation of $\Pi$ with spectrification is left adjoint to the prolongation of 
  $\mathrm{Disc}$ and does preserve finite $\infty$-limits and hence finite $\infty$-products.
  This establishes the cohesion $(T \Pi \dashv T \mathrm{Disc} \dashv T \Gamma \dashv T \mathrm{coDisc})$.
  
  That $T \mathbf{H}$ is infinitesimal cohesive over $\mathbf{H}$ follows from the fact 
  that spectrum objects contain a zero-object.
  
  Finally the left adjoint $d$ to $\mathrm{tot}$ is due to section 7.3 of \cite{LurieAlgebra}.
\endofproof
\begin{remark}
  In section 7.3 of \cite{LurieAlgebra} the left adjoint $d : \mathcal{C} \to \mathbf{T}\mathcal{C}$
  of the total space $\infty$-functor is identified as the \emph{co-tangent complex $\infty$-functor}
  if the objects of the $\infty$-category $\mathcal{C}$ are interpreted as \emph{algebras} of some kind.
  But in our case the objects of $\mathbf{H}$ are instead to be interpreted as
  \emph{spaces} of some kind, while it would be
  the objects of the opposite category $\mathcal{C} = \mathbf{H}^{\mathrm{op}}$ that 
  behave like generalized algebras. Therefore 
  in the above $d$ should instead be thought of as a
  \emph{tangent complex} $\infty$-functor.
\end{remark}
To capture the fact that tangent cohesion involves stable homotopy theory, it 
is useful to introduce the following terminology (following Joyal)
\begin{definition}
  Given an $\infty$-topos $\mathcal{E}$, then an object 
  $X \in \mathcal{E}$ is called a \emph{stable homotopy type} or just
  \emph{stable} if the canonical morphism
  $$
    X \longrightarrow \Omega \Sigma X
  $$
  into the loop space objects, def. \ref{loop space object}, of 
  its suspension object $\Sigma X := \ast \underset{X}{\coprod} \ast$
  is an equivalence.
  \label{StableHomotopyType}
\end{definition}
\begin{example}
  In $\infty \mathrm{Grpd} \simeq \mathrm{Top}[\{\mbox{weak hom. equiv.}^{-1}\}]$
  the only stable homotopy type is the point.
\end{example}
\begin{example}
  In a tangent $\infty$-topos $T \mathbf{H}$ all the objects in the inclusion
  $\mathrm{Stab} \hookrightarrow T \mathbf{H}$ are stable homotopy types.
  \label{StableHomotopyTypesInTangentInfinityTopos}
\end{example}

\medskip

We now discuss the various general abstract structures induced by cohesion, \ref{structures},
realized in Goodwillie-tangent cohesion.

\begin{itemize}
  \item \ref{TangentCohesionCohomology} -- Cohomology
  \item \ref{TangentCohesionDifferentialCohomology} -- Differential cohomology
\end{itemize}

\paragraph{Cohomology}
\label{TangentCohesionCohomology}
\index{cohomology!stable}

We discuss the notion of intrinsic cohomology, \ref{StrucCohomology}, realized
in parameterized stable cohesive homotopy types.

\medskip

The following proposition says that the intrinsic cohomology
of tangent $\infty$-toposes, as discussed generally in \ref{StrucCohomology},
is \emph{twisted stable cohomology}, the stable version of the twisted
cohomology discussed in \ref{StrucTwistedCohomology}.
\begin{proposition}
  For $T \mathbf{H}$ a tangent $\infty$-topos, prop. \ref{TangentInfinityTopos},
  and for
  \begin{itemize}
    \item $X \in \mathbf{H} \hookrightarrow T \mathbf{H}$ a homotopy type
    \item $E \in \mathrm{Stab}(\mathbf{H}) \hookrightarrow T \mathbf{H}$ a stable homotopy
    type
  \end{itemize}
  then the internal hom 
  $$
   [X,E]_{T \mathbf{H}} \in T \mathbf{H}
  $$
  is equivalent to the mapping spectrum
  $$
    [X,E]_{T \mathbf{H}}
    \simeq
    [\Sigma^\infty X, E]_{\mathrm{Stab}(\mathbf{H})}
    \in 
    \mathrm{Stab}(\mathbf{H}) \hookrightarrow T \mathbf{H}
    \,.
  $$
  \label{MappingSpectrumOutOfUnstableAsInternalHomInTangentTopos}
\end{proposition}
\proof
  With $T \mathbf{H} \hookrightarrow \mathbf{H}^{\mathrm{seq}}$
  as in the proof of prop. \ref{TangentInfinityTopos}, $X \in \mathbf{T}H$ is the constant
  $\mathrm{seq}$-diagram on $X \in \mathbf{H}$, while $E$ is a diagram with base point the
  terminal object. From this the statement follows from the general formula
  for internal homs of diagram $\infty$-categories and the $\infty$-Yoneda lemma.
\endofproof
More generally, one is interested in \emph{local coefficients} of spectra, 
def. \ref{TwistedCohomologyInOvertopos}, as follows.
\begin{example}
 Let $E \in \mathrm{Stab}(\mathbf{H})$ be equipped with the structure of an
  $E_\infty$-ring \cite{LurieAlgebra}, and write 
  $\mathrm{GL}_1(E) \in \mathrm{Stab}(\mathbf{H})$ for
  the abelian $\infty$-group (connective spectrum) of units of $E$
  \cite{ABGHR}.
  Then there is the universal associated bundle of 
  stable homotopy types, as discussed in \ref{AssociatedBundles},
  $$
    \left[
      \raisebox{20pt}{
      \xymatrix{
        E /\!/\mathrm{GL}_1(E)
        \ar[d]
        \\
        \mathbf{B}\mathrm{GL}_1(E)
      }
      }
    \right]
    \in
    \mathrm{Stab}(\mathbf{H}_{/\mathbf{B}\mathrm{GL}_1(E)})
    \hookrightarrow
    T\mathbf{H}
    \,.
  $$
  This is the \emph{universal $E$-line $\infty$-bundle} \cite{AndoBlumbergGepner}.
  \label{UniversalELineBundle}
\end{example}
\begin{proposition}
 For $X \in \mathbf{H} \hookrightarrow T \mathbf{H}$ a cohesive homotopy type
 and for $E/\!/\mathrm{GL}_1(E) \in T\mathbf{H}$ a universal $E$-line $\infty$-bundle
 as in prop. \ref{UniversalELineBundle}, then the internal mapping space
 $$
   [X,E/\!/\mathrm{GL}_1(E)]_{T\mathbf{H}} \in T \mathbf{H}
 $$ 
 is the bundle of spectra in $\mathbf{H}$ whose base homotopy type 
 is the space $[X, \mathbf{B}\mathrm{GL}_1(E)]_{\mathbf{H}}$ of twists of $E$-cohomology
 on $X$ (as discussed in \ref{StrucTwistedCohomology}) and whose total space
 is the collection of all twisted $E$-cohomology spectra $E^\bullet(X)$ of $X$
 where the fiber over a twist $\chi \in [X,\mathbf{B}\mathrm{GL}_1(E)]$ is
 $E^\chi(X)$:
 $$
   [X,E/\!/\mathrm{GL}_1(E)]_{T\mathbf{H}}
   \simeq
   \left[
     \raisebox{20pt}{
     \xymatrix{
       E^\bullet(X)
       \ar[d]
       \\
       [X,\mathbf{B}\mathrm{GL}_1(E)]
     }
     }
   \right]
 $$
\end{proposition}
\proof
  This follows with a variation of the argument in the proof of 
  prop. \ref{MappingSpectrumOutOfUnstableAsInternalHomInTangentTopos}. 
  An elegant formal homotopy type-theoretic proof has been written out by 
  Mike Shulman in \cite{nLabTangentCohesion}.
\endofproof

\paragraph{DifferentialCohomology}
\label{TangentCohesionDifferentialCohomology}
  
We discuss the realization of the general abstract notion of
differential cohomology, def. \ref{StrucDifferentialCohomology}, 
realized in tangent cohesive $\infty$-toposes.

\medskip

The following is the central formal observation of \cite{BunkeNikolausVoelkl}, 
there considered in 
$\mathrm{Stab}(\mathbf{H})$ for $\mathbf{H} = \mathrm{Smooth}\infty\mathrm{Grpd}$
as in \ref{SmoothInfgrpds} below.
\begin{proposition}
  For $\mathbf{H}$ a cohesive $\infty$-topos,
  stable homotopy type (def. \ref{StableHomotopyTypesInTangentInfinityTopos})
  $$
    E \in \mathrm{Stab}(\mathbf{H}) \hookrightarrow T \mathbf{H}
  $$
  in $T \mathbf{H}$
  sits in a diagram of the form
  $$
    \xymatrix{
      & \Pi_{\mathrm{dR}}\Omega E 
       \ar[dr] 
       \ar[rr]
      && 
      \flat_{\mathrm{dR}}\Sigma E
       \ar[dr]
      \\
      \flat\Pi_{\mathrm{dR}}\Omega E 
       \ar[ur]
       \ar[dr]
      && E 
       \ar[ur]|{\theta_E}
       \ar[dr]      
      && \Pi \flat_{\mathrm{dR}}\Sigma E
      \\
      & 
      \flat E 
        \ar[ur]
        \ar[rr]
        && 
      \Pi E 
      \ar[ur]|{\Pi\theta_E}
    }
    \,,
  $$
  where
  \begin{itemize}
    \item $\Pi$ and $\flat$ are the cohesion modalities of $T \mathbf{H}$ 
    and $\Pi_{\mathrm{dR}}$ and $\flat_{\mathrm{dR}}$
    are the de Rham modalities of $T \mathbf{H}$ as defined in \ref{StrucDeRham};
    \item the diagonals are the homotopy fiber sequences of the 
    Maurer-Cartan form on $E$, \ref{StructCurvatureCharacteristic},
    (using that $E$ is a stable homotopy type by
    example \ref{StableHomotopyTypesInTangentInfinityTopos});
    \item the two squares are $\infty$-pullback squares;
    \item the bottom morphism is the points-to-pieces transform,
    def. \ref{pointstopiecestransform}.
  \end{itemize}
  \label{StableDifferentialCohomologyDiagram}
\end{proposition}
\proof
  The diagram exists as a homotopy-commutative diagram
  by the naturality of the $\Pi$-unit and the $\flat$-counit.
  To see that the right square, the $\Pi$-naturality square of the
  Maurer-Cartan form of $E$, is an $\infty$-pullback, observe that it 
  extends to a diagram of the form
  $$
    \xymatrix{
      \flat E \ar[d]^{\simeq}\ar[r] & E \ar[d]\ar[r]^-{\theta_E} & \flat_{\mathrm{dR}} \Sigma E
      \ar[d]
      \\
      \Pi(\flat E) \ar[r] & \Pi E \ar[r]^-{\Pi\theta_E} & \Pi(\flat_{\mathrm{dR}}\Sigma E )
    }
    \,,
  $$
  where, by stability of $\mathrm{Stab}(\mathbf{H})$ and using that $\Pi$ 
  preserves $\infty$-colimits, both rows are homotopy fiber sequences 
  def. \ref{fiber sequence}. But by cohesion  the morphism
  $\flat E \longrightarrow \Pi \flat E$ is an equivalence, and hence
  by the homotopy-fiber characterization of homotopy pullbacks 
  exhibits the naturality square on the right as a homotopy pullback.
  The argument for the other square is dual this reasoning.
\endofproof
\begin{remark}
  By the discussion of higher Galois theory in \ref{StrucGaloisTheory},
  we find that the right diagram in prop. \ref{StableDifferentialCohomologyDiagram}
  says equivalently that the Maurer-Cartan form, \ref{StructCurvatureCharacteristic},
  exhibits every stable cohesive homotopy type 
  as a locally constant $\infty$-stack over its de Rham coefficient homotopy type.
\end{remark}
\begin{remark}
  Diagrams as in prop. \ref{StableDifferentialCohomologyDiagram} have been known to 
  be characteristic of 
  differential cohomology theories, see for instance 
  prop. 4.57 in \cite{BunkeDifferentialCohomology}, where this is referred to 
  as ``the differential cohomology diagram''. 
  Prop. \ref{StableDifferentialCohomologyDiagram} shows that this diagram is 
  naturally and generally induced for every stable cohesive homotopy type,
  just by the axioms of cohesion and stability.
  
  The existence of this diagram for every stable homotopy type 
  makes the concepts of ``cohesive'' 
  (e.g. ``smooth'', \ref{SmoothInfgrpds}) and 
  ``differential'' merge into a single concept for stable homotopy types:
  it says that every cohesive stable homotopy type $E$ is the differential coefficients
  of some differential cohomology theory whose underlying 
  Eilenberg-Steenrod type cohomology theory is represented by the spectrum $\Pi(E)$
  and whose de Rham coefficients are $\flat_{\mathrm{dR}}\Sigma E$.
\end{remark}

\newpage

\subsection{Geometrically discrete $\infty$-groupoids}
\label{DiscreteInfGroupoids}
\index{cohesive $\infty$-topos!models!discrete cohesion}

For completeness, and because it serves to put some concepts into a useful perspective,
we record aspects of the case of \emph{discrete} cohesion, hence of
plain $\infty$-groupoids explicitly regarded as \emph{geometrically discrete}
$\infty$-groupoids.

\medskip

\begin{observation}
The terminal $\infty$-sheaf $\infty$-topos $\infty\mathrm{Grpd}$ 
is trivially a cohesive $\infty$-topos, where each of the defining four 
$\infty$-functors $(\Pi \dashv \mathrm{Disc} \dashv \Gamma \dashv \mathrm{coDisc}) 
  : \infty Grpd \to \infty Grpd$ is an equivalence of $\infty$-categories. 
\end{observation}
\begin{definition}
In the context of cohesive $\infty$-toposes we say that $\infty \mathrm{Grpd}$ 
defines \emph{discrete cohesion} and refer to its objects as 
\emph{discrete $\infty$-groupoids}.

More generally, given any other cohesive $\infty$-topos 
$$
  (\Pi \dashv \mathrm{Disc} \dashv \Gamma \dashv \mathrm{coDisc}) : \mathbf{H}  
  \to   \infty \mathrm{Grpd}
$$
the inverse image $\mathrm{Disc}$ of the global section functor is a full and faithful $\infty$-functor 
and hence embeds $\infty \mathrm{Grpd}$ as a full sub-$\infty$-category of $\mathbf{H}$. 
We say $X \in \mathbf{H}$ is a \emph{discrete $\infty$-groupoid} if it is in the image of 
$\mathrm{Disc}$.
\end{definition}
This generalizes the traditional use of the terms \emph{discrete space} and \emph{discrete group}: 
\begin{itemize}
\item a \emph{discrete space} is equivalently a 0-truncated discrete $\infty$-groupoid;
\item a \emph{discrete group} is equivalently a 0-truncated group object in discrete $\infty$-groupoids.
\end{itemize}

We now discuss some of the general abstract structures in cohesive $\infty$-toposes,
\ref{structures}, in the context of discrete cohesion.

\begin{itemize}
  \item \ref{DiscStrucHomotopy} -- Geometric homotopy
  \item \ref{DiscStrucCohesiveGroups} -- Groups
  \item \ref{DiscStrucCohomology} -- Cohomology
  \item \ref{DiscStrucPrincipalInfinityBundles} -- Principal bundles
  \item \ref{DiscStrucTwistedCohomology} -- Twisted cohomology
  \item \ref{DiscStrucRepresentatios} -- Representations and associated bundles
\end{itemize}

\subsubsection{Geometric homotopy} 
 \label{TopAsCohesiveTopos}
 \label{DiscStrucHomotopy}
 \index{structures in a cohesive $\infty$-topos!geometric homotopy!discrete}

We discuss geometric homotopy and path $\infty$-groupoids, \ref{StrucGeometricHomotopy},
in the context of discrete cohesion, \ref{DiscreteInfGroupoids}. 
Using $\mathrm{sSet}_{\mathrm{Quillen }}$
as a presentation for $\infty \mathrm{Grpd}$ this is entirely trivial, but for the
equivalent presentation by $\mathrm{Top}_{\mathrm{Quillen}}$ it becomes effectively
a discussion of the classical Quillen equivalence 
$\mathrm{Top}_{\mathrm{Quillen}} \simeq \mathrm{sSet}_{\mathrm{Quillen}}$ from the point of view of cohesive $\infty$-toposes. It may be useful to make this
explicit.

\medskip

By the homotopy hypothesis-theorem\index{homotopy hypothesis theorem} the $\infty$-toposes $\mathrm{Top}$ and 
$\infty \mathrm{Grpd}$ are equivalent, hence indistinguishable by general abstract constructions 
in $\infty$-topos theory. However, in practice it can be useful to distinguish them as two 
different presentations for an equivalence class of $\infty$-toposes.
For that purposes consider the following
\begin{definition}
Define the quasi-categories
$$
  \mathrm{Top} := N(\mathrm{Top}_{\mathrm{Quillen}})^\circ 
$$
and
$$
  \infty \mathrm{Grpd} := N(\mathrm{sSet}_{\mathrm{Quillen}})^\circ 
  \,.
$$
\end{definition}
Here on the right we have the standard model structure on topological spaces, 
$\mathrm{Top}_{\mathrm{Quillen}}$, and the standard model structure on simplicial sets, 
$\mathrm{sSet}_{\mathrm{Quillen}}$,  and $N((-)^\circ)$ denotes the homotopy coherent nerve of the 
simplicial category given by the full $\mathrm{sSet}$-subcategory of these simplicial model categories 
on fibrant-cofibrant objects.

For 
$$
  ({|-| \dashv \mathrm{Sing}}) : 
    \xymatrix{
       \mathrm{Top}_{\mathrm{Quillen}} 
         \ar@{<-}@<+4pt>[r]^{|-|}
         \ar@<-4pt>[r]_{\mathrm{Sing}}
         &
      \mathrm{sSet}_{\mathrm{Quillen}}
     }
$$
the standard Quillen equivalence given by the singular simplicial complex-functor and geometric realization, 
write
$$
  (\mathbb{L} {|-|} \dashv \mathbb{R}\mathrm{Sing}) : 
    \xymatrix{
       \mathrm{Top} 
         \ar@{<-}@<+4pt>[r]^{\mathbb{L}|-|}
         \ar@<-4pt>[r]_{\mathbb{R}\mathrm{Sing}}
         &
      \infty \mathrm{Grpd}
     }
$$
for the corresponding derived $\infty$-functors (the image under the homotopy coherent nerve of the restriction 
of ${|-|}$ and $\mathrm{Sing}$ to fibrant-cofibrant objects followed by functorial fibrant-cofibrant replacement) 
that constitute a pair of adjoint $\infty$-functors modeled as morphisms of quasi-categories.

Since this is an equivalence of $\infty$-categories either functor serves as the left adjoint and right 
$\infty$-adjoint and so we have
\begin{observation}
$\mathrm{Top}$ is exhibited as a cohesive $\infty$-topos by
$$
  (\Pi \dashv \mathrm{Disc} \dashv \Gamma \dashv \mathrm{coDisc})
  :
   \xymatrix@C=30pt{
    \mathrm{Top}
      \ar@<+12pt>@{->}[rr]^<<<<<<<<<<{\mathbb{L} \mathrm{Sing}}
      \ar@<+4pt>@{<-}[rr]|<<<<<<<<<<<{\mathbb{R}|-|}
      \ar@<-4pt>@{->}[rr]|<<<<<<<<<<<{\mathrm{L}\mathrm{Sing}}
      \ar@<-12pt>@{<-}[rr]_<<<<<<<<<<<{\mathbb{R}|-|}
      &&
    \infty \mathrm{Grpd}
   }
$$
In particular a presentation of the intrinsic fundamental $\infty$-groupoid  is given by the familiar 
singular simplicial complex\index{fundamental $\infty$-groupoid!singular simplicial complex} construction
$$
  \Pi(X) \simeq \mathbb{R} \mathrm{Sing} X
  \,.
$$
\end{observation}
Notice that the topology that enters the explicit construction of the objects in 
$\mathrm{Top}$ here does \emph{not} show up as cohesive structure. A topological space here 
is a model for a \emph{discrete} $\infty$-groupoid, the topology only serves to allow the 
construction of $\mathrm{Sing} X$.
For discussion of $\infty$-groupoids equipped with genuine \emph{topological cohesion} 
see \ref{ContinuousInfGroupoids}.

\subsubsection{Groups} 
\label{DiscStrucCohesiveGroups}

Discrete $\infty$-groups may be presented by simplicial groups.
See \ref{InfinityGroupPresentations}.

\subsubsection{Cohomology} 
 \label{DiscStrucCohomology}
\index{structures in a cohesive $\infty$-topos!cohomology!discrete}

We discuss the general  notion of cohomology
in cohesive $\infty$-toposes, \ref{StrucCohomology}, in the context of 
discrete cohesion.

Cohomology in $\mathrm{Top}$ is the ordinary notion of
(nonabelian) cohomology. The equivalence to $\infty \mathrm{Grpd}$
makes manifest in which way this is equivalently the
\emph{cohomology of groups} for connected, homotopy 1-types, the 
\emph{cohomology of groupoids} for general 1-types
and generally, of course, the cohomology of $\infty$-groups.

\paragraph{Group cohomology}
\label{DiscStrucGroupCohomology}
\index{cohomology!group cohomology!of discrete groups}

\begin{proposition}
  \label{GroupCohomologyAsIntrinsicCohomology}
  For $G$ a (discrete) group, $A$ a (discrete) abelian group,
  the group cohomology of $G$ with coefficients in the trivial $G$-module $A$
  is
  $$
    H^n_{\mathrm{grp}}(G, A)
	\simeq
	\pi_0 \mathrm{Disc}\infty\mathrm{Grpd}(\mathbf{B}G, \mathbf{B}^n A)
	\,.
  $$
\end{proposition}
The case of group cohomology with coefficients in a non-trivial 
module is a special case of \emph{twisted cohomology}
in $\mathrm{Disc}\infty\mathrm{Grpd}$. This is discussed below in 
\ref{Group cohomology with coefficients in nontrivial modules}.

\subsubsection{Principal bundles} 
 \label{DiscStrucPrincipalInfinityBundles}
\index{simplicial principal bundle!discrete}
\index{principal $\infty$-bundle!discrete}
\index{structures in a cohesive $\infty$-topos!principal $\infty$-bundles!discrete}

We discuss the general notion of principal $\infty$-bundles 
in cohesive $\infty$-toposes, \ref{section.PrincipalInfinityBundle}, 
in the context of discrete cohesion.

There is a traditional theory of \emph{strictly} principal Kan simplicial bundles,
i.e. simplicial bundles with $G$ action for which the shear map is an \emph{isomorphism}
instead of more generally a weak equivalence.
A classical reference for this is \cite{May}. A standard modern reference is
section V of \cite{GoerssJardine}. We now compare this classical theory 
of strictly principal simplicial bundles to the theory of weakly principal
simplicial bundles from \ref{Principal infinity-bundles presentations}.

\medskip

\begin{definition}
  Let $G$ be a simplicial group and $X$ a Kan simplicial set. A
  \emph{strictly $G$-principal bundle} over $X$ is a morphism of simplicial 
  sets $P \to X$ equipped with a $G$-action on $P$ over $X$ such that
  \begin{enumerate}
    \item the $G$ action is degreewise free;
	\item the canonical morphism $P/G \to X$ out of the ordinary (1-categorical)
	  quotient is an isomorphism of simplicial sets.
  \end{enumerate}
  A morphism of stricly $G$-principal bundles over $X$ is a map $P \to P'$
  respecting both the $G$-action as well as the projection to $X$.
  
  Write $\mathrm{s}G\mathrm{Bund}(X)$ for the category of strictly $G$-principal bundles.
\end{definition}
In \cite{GoerssJardine} this is definition V3.1, V3.2.
\begin{lemma}
  \label{MorphismsOfStrictlyPrincipalSimplicialBundlesAreIsos}
  Every morphism in $\mathrm{s}G\mathrm{Bund}(X)$ is an isomorphism.
\end{lemma}
In \cite{GoerssJardine} this is remark V3.3.
\begin{observation}
  \label{InclusionOfStrictlyPrincipalSimplicialBundles}
  Every strictly $G$-principal bundle is evidently also a weakly $G$-principal
  bundle, def. \ref{WeaklyGPrincipalBundle}. In fact the strictly principal $G$-bundles
  are precisely those weakly $G$-principal bundles for which the shear map is an isomorphism.
  This identification induces a full inclusion of categories
  $$
    \mathrm{s}G\mathrm{Bund}(X) \hookrightarrow \mathrm{w}G\mathrm{Bund}(X)
	\,.
  $$
\end{observation}
\begin{lemma}
  \label{MorphismsOfWeaklyPrincipalSimplicialBundlesAreWeakEquivalences}
  Every morphism of weakly principal Kan simplicial bundles is a weak
  equivalence on the underlying Kan complexes.
\end{lemma}
\begin{proposition}
  \label{ModelStructureOnGSimplicialSets}
  For $G$ a simplicial group, the category $\mathrm{sSet}_G$ of $G$-actions on
  simplicial sets and $G$-equivariant morphisms carries the structure of a 
  simplicial model category where the fibrations and weak equivalences are
  those of the underlying simplicial sets.
\end{proposition}
This is theorem V2.3 in \cite{GoerssJardine}.
\begin{corollary}
  \label{SliceModelStructureOnGSimplicialSetsOverX}
  For $G$ a simplicial group and $X$ a Kan complex, the slice category
  $\mathrm{sSet}_G/X$ carries a simplicial model structure where the fibrations
  and weak equivalences are those of the underlying simplicial sets after
  forgetting the map to $X$.
\end{corollary}
\begin{lemma}
  \label{WeaklyPrincipalSimplicialBundlesSitHomotopyFFInModelStructureOnGSimplicialSets}
  Let $G$ be a simplicial group and $P \to X$ a weakly $G$-principal simplicial bundle.
  Then the loop space $\Omega_{(P \to X)} \mathrm{Ex}^\infty N (\mathrm{w}G\mathrm{Bund}(X))$   
  has the same homotopy type as the derived hom space
  $\mathbb{R}\mathrm{Hom}_{\mathrm{sSet}_G/X}(P,P)$.
\end{lemma}
\proof
  By theorem V2.3 of \cite{GoerssJardine} and 
  lemma \ref{MorphismsOfWeaklyPrincipalSimplicialBundlesAreWeakEquivalences} 
  the free resolution $P^f$ of $P$
  from corollary \ref{befreiung} is a cofibrant-fibrant resolution of $P$ in 
  the slice model structure of corollary \ref{SliceModelStructureOnGSimplicialSetsOverX}.
  Therefore the derived hom space is presented by the simplicial set
  of morphisms $\mathrm{Hom}_{\mathrm{sSet}_G/X}(P^f \cdot \Delta^\bullet, P^f)$
  and all these morphisms are equivalences. Therefore by prop. 2.3 in 
  \cite{DwyerKanClassification} this simplicial set is equivalent to the 
  loop space of the nerve of the subcategory of $\mathrm{sSet}_G/X$ on the 
  weak equivalences connected to $P^f$. 
  By lemma \ref{MorphismsOfWeaklyPrincipalSimplicialBundlesAreWeakEquivalences}
  this subcategory is 
  equivalent (isomorphic even) to the connected component of $\mathrm{w}G\mathrm{Bund}(X)$
  on $P$.
\endofproof
\begin{proposition}
  \label{StrictlyPrincipalBundlesSurjectOnPi0}
  Under the simplicial nerve, the inclusion of 
  observation \ref{InclusionOfStrictlyPrincipalSimplicialBundles} yields a morphism
  $$
    N \mathrm{s}G\mathrm{Bund}(X) \to N \mathrm{w}G\mathrm{Bund}(X)
	\;\;
	\in \mathrm{sSet}_{\mathrm{Quillen}}
  $$
  which is 
  \begin{itemize}
    \item for all $G$ and $X$ an isomorphism on connected components;
	\item not in general a weak equivalence.
  \end{itemize}
\end{proposition}
\proof
  Let $P \to X$ be a weakly $G$-principal bundle. 
  To see that it is connected in $\mathrm{w}G\mathrm{Bund}(X)$ to some
  strictly $G$-principal bundle, first observe that by corollary
  \ref{befreiung} it is 
  connected via a morphism $P^f \to P$ to the bundle 
  $$
    P^f := \mathrm{Rec}(X \leftarrow P/_h G \stackrel{f}{\to} \Wbar G)
	\,,
  $$
  which has free $G$-action, but does not necessarily satisfy strict principality.
  Since, by theorem \ref{HomotopyQuotientWPrincBundleLoAcyclicFibration}, 
  the morphism $P/_h G \to X$ is an acyclic fibration of simplicial sets
  it has a section $\sigma : X \to P /_h G$ (every simplicial set
  is cofibrant in $\mathrm{sSet}_{\mathrm{Quillen}}$). The bundle 
  $$
    P^s := 
	  \mathrm{Rec}(X \stackrel{\mathrm{id}}{\leftarrow} X \stackrel{f\circ \sigma}{\to} \Wbar G)
  $$
  is strictly $G$-principal, and with the morphism
  $$
    (P^s \to P^f)
	:=
    \mathrm{Rec}
	\left(
	   \raisebox{44pt}{
	   \xymatrix{
	     & P /_h G
		 \ar@{->>}[dl]_\sim \ar[dr]^f
	     \\ 
	     X && \Wbar G
		 \\
		 & X \ar@{->>}[ul]^{\mathrm{id}} \ar[uu]_\sigma \ar[ur]_{f \circ \sigma}
	   }
	   }
	\right)
  $$
  we obtain (non-naturally, due to the choice of section) in total a morphism
  $P^s \to P^f \to P$
  of weakly $G$-principal bundles from a strictly $G$-principal replacement $P^s$ to $P$.
  
  To see that the full embedding of strictly $G$-principal bundles is also injective on connected
  components, notice that by lemma 
  \ref{WeaklyPrincipalSimplicialBundlesSitHomotopyFFInModelStructureOnGSimplicialSets}
  if a weakly $G$-principal bundle $P$ with degreewise free $G$-action is connected
  by a zig-zag of morphisms to some other weakly $G$-principal bundle $P$, then there is 
  already a direct morphism $P \to P'$. Since all strictly $G$-principal bundles
  have free action by definition, this shows that two of them that are
  connected in $\mathrm{w}G\mathrm{Bund}(X)$ are already connected in 
  $\mathrm{s}G\mathrm{Bund}(X)$.
  
  To see that in general $N\mathrm{s}G\mathrm{Bund}(X)$ 
  nevertheless does not have the 
  correct homotopy type,  it is sufficient to notice that 
  the category $\mathrm{s}G\mathrm{Bund}(X)$ is always a groupoid,
  by lemma \ref{MorphismsOfStrictlyPrincipalSimplicialBundlesAreIsos}. Therefore 
  $N\mathrm{s}G\mathrm{Bund}(X)$ it is always a homotopy 1-type. 
  But by theorem \ref{Classification theorem for weakly G-principal bundles}
  the object $N\mathrm{w}G\mathrm{Bund}(X)$ is not an $n$-type if $G$ is not an
  $(n-1)$-type. 
\endofproof
\begin{corollary}
 \label{StrictlyPrincipalSimplicialBundlesModelCohomologyClasses}
 For all Kan complexes $X$ and simplicial groups $G$ there is an isomorphism
 $$
	\pi_0 N \mathrm{s}G\mathrm{Bund} \simeq  H^1(X,G) := \pi_0 \infty\mathrm{Grpd}(X, \mathbf{B}G)
 $$
 between the isomorphism classes of strictly $G$-principal bundles over $X$ and
 the first nonabelian cohomology of $X$ with coefficients in $G$.
 
  But this isomorphism on cohomology does not in general lift to an equivalence
  on cocycle spaces.
\end{corollary}
\proof
  By prop. \ref{StrictlyPrincipalBundlesSurjectOnPi0} and remark
  \ref{ClassificationTheoremRelatedToCocyclesSpaces}.
\endofproof
\begin{remark}
The first statement of corollary \ref{StrictlyPrincipalSimplicialBundlesModelCohomologyClasses} 
is the classical classification result for strictly principal simplicial bundles,
for instance theorem V3.9 in \cite{GoerssJardine}. 
\end{remark}

\subsubsection{Twisted cohomology} 
 \label{DiscStrucTwistedCohomology}
\index{structures in a cohesive $\infty$-topos!twisted cohomology!discrete}
\label{Group cohomology with coefficients in nontrivial modules}
\index{twisted cohomology!group cohomology!with coefficients}

We discuss the notion of twisted cohomology, 
\ref{StrucTwistedCohomology}, in the context of discrete cohesion.

Specifically, we discuss here $\infty$-group cohomology 
for discrete $\infty$-groups with coefficients in 
a module according to \ref{StrucRepresentations}.

\medskip

For $G$ a (discrete) group and $A$ a (discrete) group
equipped with a $G$-action, write $\mathbf{B}^n A/\!/G$
for the $n$-groupoid which is given by the crossed complex,
def. \ref{CrossedComplex} of groups
$$
 \mathbf{B}^n A/\!/G := [A \to 1 \to \cdots \to 1 \to G]
$$
coming from the given $G$-action on $A$.
There is a canonical morphism 
$$
 \mathbf{B}^n A/\!/G \to \mathbf{B}G
 \,.
$$
\begin{proposition}
  We have a fiber sequence
  $$
    \mathbf{B}^n A \to \mathbf{B}^n A/\!/G \to \mathbf{B} G
  $$
  in $\mathrm{Disc}\infty\mathrm{Grpd}$.
\end{proposition}
In view of remark \ref{TwistingCocycleAsAssociatedBundle} this 
fiber sequence exhibits a $\mathbf{B}^n A$-fiber bundle which is 
associated to the universal $G$-principal $\infty$-bundle, 
\ref{DiscStrucPrincipalInfinityBundles}.

In generalization of prop. \ref{GroupCohomologyAsIntrinsicCohomology} 
we have
\begin{proposition}
  The group cohomology of $G$ with coefficients in the 
  module $A$ is naturally identified with the 
  $\mathrm{id}$-twisted cohomology of $\mathbf{B}G$, relative to
  $\mathbf{B}^n A/\!/G$,
  $$
    H^n_{\mathrm{grp}}(G, A)
	\simeq
	\pi_0 \mathrm{Disc}\infty\mathrm{Grpd}_{[\mathrm{id}]}(\mathbf{B}G, \mathbf{B}^n A/\!/G)
	\,.
  $$
\end{proposition}
\begin{remark}
  Equivalently this says that group cohomology with coefficients
  in nontrivial modules $A$ describes the sections of the 
  bundle $\mathbf{B}^n A/\!/G$.
\end{remark}

\newpage

\subsubsection{Representations and associated bundles}
\label{DiscStrucRepresentatios}

We discuss canonical representations of automorphism $\infty$-groups
in $\mathrm{Disc}\infty \mathrm{Grpd}$, following \ref{StrucRepresentations}.

\medskip

For all of the following, fix a regular uncountable cardinal $\kappa$.
\begin{definition}
  \label{CoreInfinityGroupoidKappa}
  Write $\mathrm{Core}\,\infty \mathrm{Grpd}_\kappa$ for the core
  (the maximal $\infty$-groupoid inside) the full sub-$\infty$-category
  of $\infty \mathrm{Grpd}$ on the $\kappa$-small $\infty$-groupoids,
  \cite{Lurie} def. 5.4.1.3. We regard this canonically as an object
  $$
    \mathrm{Core}\,\infty \mathrm{Grpd}_\kappa \in \infty \mathrm{Grpd}
	\,.
  $$
\end{definition}
\begin{remark}
 \label{CoreInfinityGroupoidAsCorpoductOfAutomorphisms}
We have 
$$
    \mathrm{Core}\,\infty \mathrm{Grpd}_\kappa 
    \simeq
  \coprod_i \mathbf{B} \mathrm{Aut}(F_i)
  \,,
$$
where the coproduct ranges over all $\kappa$-small homotopy types $[F_i]$ 
and where $\mathrm{Aut}(F_i)$ is the automorphism $\infty$-group 
of any representative $F_i$ of $[F_i]$.
\end{remark}
\begin{lemma}
  \label{CharacterizationSmallObjectsInInfinityGroupoids}
  For $X $ a $\kappa$-small $\infty$-groupoid,
  and $f : Y \to X$ a morphism in $\infty \mathrm{Grpd}$,
  the following are equivalent
  \begin{enumerate}
	\item for all objects $x \in X$ 
	   the homotopy fiber $Y_{x} := Y \times_X \{x\}$ of $f$ is $\kappa$-small;
    \item $Y$ is $\kappa$-small.
  \end{enumerate}
\end{lemma}
\proof
  The implication $1. \Rightarrow 2.$ is stated for $\infty$-categories,
  and assuming that $f$ is presented by a
  Cartesian fibration of simplicial sets, as  prop. 5.4.1.4 in \cite{Lurie}.
  But by prop. 2.4.2.4 there, every Cartesian fibration between Kan complexes
  is a right fibration; and by prop. 2.1.3.3 there over a Kan complex every
  right fibration is a Kan fibration. Finally, by the Quillen model structure
  every morphism of $\infty$-groupoids is presented by a Kan fibration. Therefore
  the condition that $f$ be presented by a Cartesian morphism is automatic
  in our case.
  
  For the converse, assume that all homotopy fibers are $\kappa$-small.
  We may write $X$ as the $\infty$-colimit of the functor constant on the point,
  over itself (\cite{Lurie}, corollary 4.4.4.9 )
  $$
    X \simeq {\lim\limits_{\longrightarrow}}_{x \in X} \{x\}
	\,.
  $$
   Since $\infty\mathrm{Grpd}$ is an $\infty$-topos, its $\infty$-colimits
   are preserved by $\infty$-pullback. Therefore we have an $\infty$-pullback
   diagram
   $$
     \xymatrix{
	   {\lim\limits_{\longrightarrow}}_{x \in X} Y_x \ar[r]^<<<<<{\simeq} \ar[d]^f & Y \ar[d]^f
	   \\
	   {\lim\limits_{\longrightarrow}}_{x \in X} \{x\} \ar[r]^<<<<<{\simeq} &  X
	 }
	 \,.
   $$
   that exhibits $Y$ as the $\infty$-colimit over $X$ of the homotopy fibers of $f$.
   By corollary 5.4.1.5 in \cite{Lurie}, the $\kappa$-small $\infty$-groupoids are
   precisely the $\kappa$-compact objects of $\infty \mathrm{Grpd}$.
   By corollary 5.3.4.15 there, $\kappa$-compact objects are closed under
   $\kappa$-small $\infty$-colimits. Therefore the above $\infty$-colimit
   exhibits $Y$ as a $\kappa$-small $\infty$-groupoid.
\endofproof
\begin{definition}
  Write $\widehat{\mathrm{Core}\infty\mathrm{Grpd}_\kappa} \to 
  \mathrm{Core}\,\infty\mathrm{Grpd}_\kappa$ for the 
  $\infty$-pullback
  $$
    \xymatrix{
	  \widehat{\mathrm{Core}\infty\mathrm{Grpd}_\kappa} 
	  \ar[r]
	  \ar[d]
	  &
	  Z|_{\infty \mathrm{Grpd}}
	  \ar[d]
	  \\
	  \mathrm{Core}\,\infty\mathrm{Grpd}
	  \ar[r]
	  &
	  \infty \mathrm{Grpd}
	}
  $$
  of the universal right fibration $Z|_{\infty \mathrm{Grpd}} \to \infty \mathrm{Grpd}$, 
  as in \cite{Lurie} above prop. 3.3.2.5.,
  along the canonical map that embeds $\kappa$-small $\infty$-groupoids
  into all $\infty$-groupoids.
\end{definition}
\begin{proposition}
  \label{ObjectClassifierInInfinityGroupoid}
  \index{object classifier!in $\infty\mathrm{Grpd}$}
  The morphism $\widehat{\mathrm{Core}\infty\mathrm{Grpd}_\kappa} \to 
  \mathrm{Core}\,\infty\mathrm{Grpd}_\kappa$ is
  the $\kappa$-compact object-classifier,
  section 6.1.6 of \cite{Lurie},  in $\infty \mathrm{Grpd}$.
\end{proposition}
\proof
  By prop. 3.3.2.5 in \cite{Lurie} the universal right fibration
  classifies right fibrations; and for $[X] : * \to \infty \mathrm{Grpd}$
  the name of an $\infty$-groupoid $X$, the homotopy fiber 
  $$
    Z \times_{\infty \mathrm{Grpd}} \{[X]\} \simeq X
  $$
  is equivalent to $X$.
  As in the proof of lemma \ref{CharacterizationSmallObjectsInInfinityGroupoids},
  every morphism between $\infty$-groupoids is represented by a Cartesian fibration.
  Since moreover every morphism out of an $\infty$-groupoid into $\infty\mathrm{Grpd}$
  factors essentially unqiquely through $\mathrm{Core}\,\infty \mathrm{Grpd}$ it follows
  that $\widehat{\mathrm{Core}\infty\mathrm{Grpd}_\kappa} \to 
  \mathrm{Core}\,\infty\mathrm{Grpd}_\kappa$ classifies morphisms of 
  $\infty$-groupoids with $\kappa$-small homotopy fibers. 
  By lemma \ref{CharacterizationSmallObjectsInInfinityGroupoids} 
  and using again that $\kappa$-compact objects in $\infty\mathrm{Grpd}$
  are $\kappa$-small $\infty$-groupoids,  these
  are precisely the relatively $\kappa$-compact morphisms from 
  def. 6.1.6.4 of \cite{Lurie}.
\endofproof
\begin{remark}
  By remark \ref{CoreInfinityGroupoidAsCorpoductOfAutomorphisms}
  we have that $\widehat{\mathrm{Core}\infty\mathrm{Grpd}_\kappa} \to 
  \mathrm{Core}\,\infty\mathrm{Grpd}_\kappa$ decomposes as a 
  coproduct of morphisms $\coprod_{[F_i]} \rho_i$ 
  indexed by the $\kappa$-small homotopy
  types. According to prop. \ref{ObjectClassifierInInfinityGroupoid}
  the (essentially unique) homotopy fiber of $\rho_i$ is 
  equivalent to the $\kappa$-small $\infty$-groupoid $F_i$ itself.
  Therefore by def. \ref{RepresentationOfInfinityGroup} 
  we may write 
  $$
    \rho_i : F_i /\!/ \mathrm{Aut}(F_i) \to \mathbf{B}\mathrm{Aut}(F_i)
  $$
  and identify this with the canonical representation of $\mathrm{Aut}(F_i)$
  on $F_i$, exhibited, by example \ref{UniversalAssociatedBundle}, 
  as the universal $F_i$-fiber bundle which is $\rho_i$-associated 
  to the universal $\mathrm{Aut}(F_i)$-principal bundle.
\end{remark}
In terms of this perspective we have the following classical result.
\begin{corollary}
  \label{CharacterizationOfDiscreteInfinityBundles}
  For $X$ a connected $\infty$-groupoid, every morphism $P \to X$
  in $\infty \mathrm{Grpd}$ with $\kappa$-small small homotopy fibers
  $F$ (over one and hence, up to equivalence, over each object $x \in X$) arises 
  as the $F$-fiber bundle $\rho$-associated to an 
  $\mathrm{Aut}(F)$-principal $\infty$-bundle, \ref{section.PrincipalInfinityBundle},
  given by an $\infty$-pullback of the form
  $$
    \
    \xymatrix{
	  P \ar[r] \ar[d] & F /\!/ \mathrm{Aut}(F) \ar[d]
	  \\
	  X \ar[r] & \mathbf{B}\mathrm{Aut}(F)
	}
	\,.
  $$
\end{corollary}

More discussion of discrete principal and discrete associated $\infty$-bundles
is in \ref{StrucGaloisTheory} and \ref{DiscStrucPrincipalInfinityBundles}.

\subsection{Euclidean-topological $\infty$-groupoids}
\label{ContinuousInfGroupoids}
\index{cohesive $\infty$-topos!models!Euclidean-topological cohesion}

We discuss here \emph{Euclidean-topological cohesion}, 
modeled on Euclidean topological spaces and continuous maps
between them. This subsumes the homotopy theory of 
simplicial topological spaces.

\medskip

\begin{definition} 
  \label{CartSpTop}
Let $\mathrm{CartSp}_{\mathrm{top}}$ be the site whose underlying category has as objects the 
Cartesian spaces $\mathbb{R}^n$, $n \in \mathbb{N}$ equipped with the standard Euclidean topology 
and as morphisms the continuous maps between them; and whose coverage is given by 
good open covers.
\end{definition}
\begin{proposition} 
  \label{CartSpTopIsCohesive}
  The site $\mathrm{CartSp}_{\mathrm{top}}$ is an $\infty$-cohesive site 
  (def \ref{CohesiveSite}).
\end{proposition}
\proof
  Clearly $\mathrm{CartSp}_{\mathrm{loc}}$ has finite products, given by
  $\mathbb{R}^k \times \mathbb{R}^l \simeq \mathbb{R}^{k+l}$, and clearly every object has 
  a point $* = \mathbb{R}^0 \to \mathbb{R}^n$. In fact $\mathrm{CartSp}_{\mathrm{top}}(*, \mathbb{R}^n)$
  is the underlying set of the Cartesian space $\mathbb{R}^n$.
  
  Let $\{U_i \to U\}$ be a good open covering family in  $\mathrm{CartSp}_{\mathrm{top}}$. 
  By the very definition of \emph{good cover} it follows
  that the {\v C}ech nerve $C(\coprod_i U_i \to U) \in [\mathrm{CartSp}^{\mathrm{op}}, \mathrm{sSet}]$ 
  is degreewise a coproduct of representables.

  The condition $\lim\limits_{\longrightarrow} C(\coprod_i U_i) \stackrel{\simeq}{\to} \lim\limits_{\longrightarrow} U  = *$
  follows from the nerve theorem \cite{borsuk}, which asserts that 
$\lim\limits_{\longrightarrow} C(\coprod_i U_i \to U) \simeq \mathrm{Sing} U$, 
  and using that, as a topological space, every Cartesian space is contractible.

 The condition 
$\lim\limits_{\longleftarrow} C(\coprod_i U_i) \stackrel{\simeq}{\to} \lim\limits_{\longleftarrow} U  = 
  \mathrm{CartSp}_{\mathrm{loc}}(*,U)$ is immediate. Explicitly, for 
$(x_{i_0} \in U_{i_0}, \cdots, x_{i_n} \in U_{i_n})$ a sequence of points in the covering
patches of $U$ such that
any two consecutive ones agree in $U$, then they all agree in $U$. So the morphism 
of simplicial sets in question
has the right lifting property against all boundary inclusions
$\partial \Delta[n] \to \Delta[n]$ and is therefore is a weak equivalence.
\endofproof
\begin{definition} 
 \label{ETopInftyGrpd}
Define
$$
  \mathrm{ETop} \infty \mathrm{Grpd} := \mathrm{Sh}_{\infty}(\mathrm{CartSp}_{\mathrm{top}})
$$
to be the $\infty$-category of $\infty$-sheaves on $\mathrm{CartSp}_{\mathrm{top}}$.
\end{definition}
\begin{proposition}
  The $\infty$-category $\mathrm{ETop}\infty \mathrm{Grpd}$ is a cohesive $\infty$-topos.
\end{proposition}
\proof
  This follows with prop. \ref{CartSpTopIsCohesive} by prop. 
  \ref{InfSheavesOverCohesiveSiteAreCohesive}.
\endofproof
\begin{definition}
  We say that $\mathrm{ETop}\infty \mathrm{Grpd}$ defines \emph{Euclidean-topological cohesion}.
  An object in $\mathrm{ETop}\infty \mathrm{Grpd}$ we call a \emph{Euclidean-topological $\infty$-groupoid}.
\end{definition}
\begin{definition} \label{TopologicalManifolds}
  Write $\mathrm{TopMfd}$ for the category whose objects are 
  topological manifolds that are
  \begin{itemize}
    \item finite-dimensional;
    \item paracompact;
    \item with an arbitrary set of connected components (hence not assumed to be second-countable);
  \end{itemize} 
  and 
  whose morphisms are continuous functions between these. Regard this as a (large) site with the
  standard open-cover coverage.
\end{definition}
\begin{proposition} 
  \label{ToposOverTopMfd}
  The $\infty$-topos $\mathrm{ETop}\infty Grpd$ is equivalently that of 
  hypercomplete $\infty$-sheaves (\cite{Lurie}, section 6.5)
  on $\mathrm{TopMfd}$
  $$
    \mathrm{ETop}\infty \mathrm{Grpd} \simeq \hat {\mathrm{Sh}}_{\infty}(\mathrm{TopMfd})
    \,.
  $$
\end{proposition}
\proof
   Since every topological manifold admits an cover by open balls homeomorphic to 
   a Cartesian space, we have that $\mathrm{CartSp}_{\mathrm{top}}$ is a dense sub-site
   of $\mathrm{TopMfd}$. By theorem C.2.2.3 in \cite{Johnstone} it follows
   that the sheaf toposes agree
   $$
     \mathrm{Sh}(\mathrm{CartSp}_{\mathrm{top}}) \simeq \mathrm{Sh}(\mathrm{TopMfd})
     \,.
   $$
   From this it follows directly that the Joyal model structures on simplicial sheaves over both sites
   (see \cite{Jardine}) are Quillen equivalent. By \cite{Lurie}, prop 6.5.2.14, these present the
   hypercompletions 
   $$
     \hat {\mathrm{Sh}}_\infty(\mathrm{CartSp}_{\mathrm{top}}) \simeq \hat{\mathrm{Sh}}_{\infty}(\mathrm{TopMfd})
     \,.
   $$
  of the corresponding $\infty$-sheaf $\infty$-toposes. But by corollary \ref{CohesiveIsHypercomplete}
  we have that $\infty$-sheaves on $\mathrm{CartSp}_{\mathrm{top}}$ are already hypercomplete, so that
   $$
     {\mathrm{Sh}_\infty}(\mathrm{CartSp}_{\mathrm{top}}) \simeq \hat{\mathrm{Sh}}_{\infty}(\mathrm{TopMfd})
     \,.
   $$  
\endofproof
\begin{definition} \label{TopCGH}
  Let $\mathrm{Top}_{\mathrm{cgH}}$ be the 1-category of compactly generated and Hausdorff
  topological spaces and continuous functions between them.
\end{definition}
\begin{proposition}
  The category $\mathrm{Top}_{\mathrm{cgH}}$ is cartesian closed.
\end{proposition}
See \cite{Steenrod}. We write 
$[-,-] : 
  \mathrm{Top}_{\mathrm{cgH}}^{\mathrm{op}} \times \mathrm{Top}_{\mathrm{cgH}} \to \mathrm{Top}_{\mathrm{cgH}}$
for the corresponding internal hom-functor.
\begin{definition}
  There is an evident functor
  $$
    j : \mathrm{Top}_{\mathrm{cgH}} \to \mathrm{ETop}\infty \mathrm{Grpd}
  $$
  that sends each topological space $X$ to the 0-truncated $\infty$-sheaf (ordinary sheaf)
  represented by it 
  $$
    j(X) : (U \in \mathrm{CartSp}_{\mathrm{top}}) \mapsto \mathrm{Hom}_{\mathrm{Top}_{}\mathrm{cgH}}(U,X) 
      \in \mathrm{Set} \hookrightarrow \infty \mathrm{Grpd}
     \,.
  $$
\end{definition}
\begin{corollary}
  The functor $j$ exhibits $\mathrm{TopMfd}$ as a full sub-$\infty$-category of 
  $\mathrm{ETop}\infty \mathrm{Grpd}$
  $$
    j : \mathrm{TopMfd} \hookrightarrow \mathrm{ETop}\infty \mathrm{Grpd}
  $$
\end{corollary}
\proof
  By prop. \ref{ToposOverTopMfd} this is a special case of the $\infty$-Yoneda lemma.
\endofproof
\begin{remark}
 \label{MoreFibrantObjectsOverCartSpTop}
While, according to prop. \ref{ToposOverTopMfd}, the model categories 
$[\mathrm{CartSp}_{\mathrm{top}}^{\mathrm{op}}, \mathrm{sSet}]_{\mathrm{proj}, \mathrm{loc}}$
and\\
$[\mathrm{TopMfd}^{\mathrm{op}}, \mathrm{sSet}]_{\mathrm{proj}, \mathrm{loc}}$ 
are both presentations of $\mathrm{ETop}\infty \mathrm{Grpd}$, they lend themselves
to different computations: in the former there are more fibrant objects, fewer cofibrant objects than in the latter, and vice versa. 
\end{remark}
In \ref{CohesiveFibrancy} we gave a general discussion concerning this point,
here we amplify specific detail for the present case.
\begin{proposition} \label{SeparatedAndFibrantOnCartSp}
Let $X \in [\mathrm{TopMfd}^{\mathrm{op}}, \mathrm{sSet}]$ 
be an object that is \emph{globally fibrant}, \emph{separated} and \emph{locally trivial}, meaning that
\begin{enumerate}
\item $X(U)$ is a non-empty Kan complex for all $U \in \mathrm{TopMfd}$;
\item for every covering $\{U_i \to U\}$ in $\mathrm{TopMfd}$ the descent morphism
   $X(U) \to [\mathrm{TopMfd}^{\mathrm{op}}, \mathrm{sSet}](C(\{U_i\}), X)$ 
   is a full and faithful $\infty$-functor;
\item for contractible $U$ we have $\pi_0[\mathrm{TopMfd}^{\mathrm{op}}, \mathrm{sSet}](C(\{U_i\}), X) \simeq *$.
\end{enumerate}
Then the restriction of $X$ along $\mathrm{CartSp}_{\mathrm{top}} \hookrightarrow \mathrm{TopMfd}$ 
is a fibrant object in the local model structure $[\mathrm{CartSp}_{\mathrm{top}}^{\mathrm{op}}, \mathrm{sSet}]_{\mathrm{proj},\mathrm{loc}}$.
\end{proposition}
\proof
The fibrant objects in the local model structure are precisely those that are Kan complexes 
over every object and for which the descent morphism is an equivalence for all covers.
The first condition is given by the first assumption. The second and third assumptions 
imply the second condition over contractible manifolds, such as the Cartesian spaces.
\endofproof
\begin{example}
Let $G$ be a topological group, regarded as the presheaf over $\mathrm{TopMfd}$ 
that it represents. Write ${\bar W G}$ 
for the simplicial presheaf on $\mathrm{TopMfd}$ given by the nerve of the topological groupoid 
$(G \stackrel{\to}{\to} *)$. (We discuss this in more detail in \ref{ETopStrucGroups} below.)

The fibrant resolution of ${\bar W} G$ in $[\mathrm{TopMfd}^{\mathrm{op}}, \mathrm{sSet}]_{\mathrm{proj},\mathrm{loc}}$ is (the rectification of) its stackification: the stack $G \mathrm{Bund}$ of topological $G$-principal bundles. 
But the canonical morphism
$$
  \bar W G \to G \mathrm{Bund}
$$
is a full and faithful functor (over each object $U \in \mathrm{TopMfd}$): 
it includes the single object of $\bar W G$ as the trivial $G$-principal bundle. 
The automorphisms of the single object in $\bar W G$ over $U$ are $G$-valued continuous functions on $U$, 
which are precisely the automorphisms of the trivial $G$-bundle. 
Therefore this inclusion is full and faithful, the presheaf $\bar W G$ is a separated prestack.

Moreover, it is locally trivial: every {\v C}ech cocycle for a $G$-bundle over a Cartesian space is 
equivalent to the trivial one. Equivalently, also $\pi_0 G \mathrm{Bund}(\mathbb{R}^n) \simeq *$.
Therefore $\bar W G$, when restricted $\mathrm{CartSp}_{\mathrm{top}}$, 
does become a fibrant object in $[\mathrm{CartSp}_{\mathrm{top}}^{\mathrm{op}}, \mathrm{sSet}]_{\mathrm{proj},\mathrm{loc}}$.

On the other hand, let $X \in \mathrm{TopMfd}$ be any non-contractible manifold. 
Since in the projective model structure on simplicial presheaves every representable 
is cofibrant, this is a cofibrant object in $[\mathrm{Mfd}^{\mathrm{op}}, \mathrm{sSet}]_{\mathrm{proj},\mathrm{loc}}.$However, it 
fails to be cofibrant in $[\mathrm{CartSp}_{\mathrm{top}}^{\mathrm{op}}, \mathrm{sSet}]_{\mathrm{proj},\mathrm{loc}}$. Instead, there a cofibrant replacement is given by the {\v C}ech nerve $C(\{U_i\})$ of any 
good open cover $\{U_i \to X\}$.

This yields two different ways for computing the first nonabelian cohomology 
$$
  H^1_{\mathrm{ETop}}(X,G) := \pi_0 \mathrm{ETop}\infty \mathrm{Grpd} (X, \mathbf{B}G)
$$
in $\mathrm{ETop}\infty \mathrm{Grpd}$ on $X$ with coefficients in $G$:
\index{cohomology!{\v C}ech cocycles versus $\infty$-stackification}
\begin{enumerate}
\item $\cdots \simeq \pi_0 [\mathrm{Mfd}^{\mathrm{op}}, \mathrm{sSet}](\mathrm{X}, G \mathrm{Bund}) 
  \simeq \pi_0 G \mathrm{Bund}(X)$;
\item
 $\cdots \simeq \pi_0 [\mathrm{CartSp}_{\mathrm{top}}^{\mathrm{op}}, \mathrm{sSet}](C(\{U_i\}), \bar W G) 
  \simeq H^1(X,G)$.
\end{enumerate}
In the first case we need to construct the fibrant replacement $G \mathrm{Bund}$. 
This amounts to constructing $G$-principal bundles over \emph{all} paracompact manifolds and then 
evaluate on the given one, $X$, by the 2-Yoneda lemma.
In the second case however we cofibrantly replace $X$ by a good open cover, 
and then find the {\v C}ech cocycles with coefficients in $G$ on that. 

For ordinary $G$-bundles the difference between the two computations may be irrelevant in practice, 
because ordinary $G$-principal bundles are very well understood. However, for more general 
coefficient objects, for instance general topological simplicial groups $G$, the first approach 
requires to find the full $\infty$-sheafification to the $\infty$-sheaf of all principal $\infty$-bundles, 
while the second approach requires only to compute specific coycles over one specific base object. 
In practice the latter is often all that one needs.
\end{example}

We discuss a few standard techniques for constructing \emph{cofibrant} resolutions
in $[\mathrm{CartSp}_{\mathrm{top}}^{\mathrm{op}}, \mathrm{sSet}]_{\mathrm{proj}, \mathrm{loc}}$.

\begin{proposition}
  Let
  $$
    X \in \mathrm{TopMfd}
	\hookrightarrow 
	 [\mathrm{CartSp}_{\mathrm{top}}^{\mathrm{op}}, \mathrm{sSet}]_{\mathrm{proj}, \mathrm{loc}}
  $$
  be a topological manifold and let $\{U_i \to X\}$ be a good open cover.
  Then the {\v C}ech nerve
  $$
    C(\{U_i\}) 
	   := 
	\int^{[n] \in \Delta} \Delta[n] \cdot \coprod_{i_0, \cdots, i_n} 
	j(U_{i_0}) \cap \cdots \cap j(U_{i_n})
  $$
  (where $j : \mathrm{TopMfd} \hookrightarrow [\mathrm{CartSp}^{\mathrm{op}}, \mathrm{sSet}]$
  is the Yoneda embedding)
  equipped with the canonical projection $C(\{U_i\}) \to X$ is
  a cofibrant resolution of $X$.
  \label{ResolutionOfTopManifoldsByGoddOpenCovers}
\end{proposition}
\proof  
  The morphism is clearly a stalkwise weak equivalence. Therefore it is
  a weak equivalence in the local model structure by theore, 
  \ref{CharacterizationOfLocalWeakEquivalence}.
  
  Moreover, by the very definition of \emph{good} open cover
  the non-empty finite intersections of the $U_i$ are themselves
  represented by objects in $\mathrm{CartSp}^{\mathrm{op}}$. 
  Therefore the {\v C}ech nerve is degreewise a coproduct of 
  representables. Also, its degeneracies split off as a direct summand
  in each degree. By \cite{Dugger} this means that it is cofibrant in the
  global projective model structure. But the cofibrations do not change
  under left Bousfield localization to the local model structure, therefore
  it is cofibrant also there.
\endofproof
\begin{proposition}
  $$
    X_\bullet \in \mathrm{TopMfd}^{\Delta^{\mathrm{op}}}
	\hookrightarrow 
	 [\mathrm{CartSp}_{\mathrm{top}}^{\mathrm{op}}, \mathrm{sSet}]_{\mathrm{proj}, \mathrm{loc}}
  $$
  be a simplicial manifold, such that there is a choice $\mathcal{U}$ 
  of good open covers
  $\{U_{n,i} \to X_n\}_i$ in each degree which are simplicially compatible in that
  they arrange into a morphism of bisimplicial presheaves
  $$
    C(\mathcal{U})_{\bullet, \bullet} \to X_\bullet
	\,.
  $$
  Then 
  $$
    \int^{[n] \in \Delta} \mathbf{\Delta}[n] \cdot C(\mathcal{U})_{n,\bullet}
	\;
	\to
	\;
	X_\bullet
	\,,
  $$
  where $\mathbf{\Delta} : \Delta^{\mathrm{op}} \to \mathrm{sSet}$
  is given by $\mathbf{\Delta}[n] := N(\Delta/[n])$,
  is a cofibrant resolution in 
  $[\mathrm{CartSp}_{\mathrm{top}}^{\mathrm{op}}]_{\mathrm{proj}, \mathrm{loc}}$.
  \label{ResolutionOfSimplicialManifolds}
\end{proposition}
\proof
  First consider 
  $$
    \int^{[n] \in \Delta} \Delta[n] \cdot C(\mathcal{U})_{n,\bullet}
	\;
	\to
	\;
	X_\bullet
  $$
  with the ordinary simplex in the integrand. 
  Over ach object $U \in \mathrm{CartSp}_{\mathrm{top}}$ 
  the coend appearing here is
  isomorphic to the diagonal of the given bisimplicial set.
  Since the diagonal sends degreewise weak equivalences to 
  weak equivalences, prop. \ref{ResolutionOfTopManifoldsByGoddOpenCovers}
  implies that this is a weak equivalence in the local model structure.
  
  Let $\mathbf{\Delta} \to \Delta$ be the canonical projection. 
  We claim that the induced morphism
  $$
    \int^{[n] \in \Delta} \mathbf{\Delta}[n] \cdot C(\mathcal{U})_{n,\bullet}
	  \to
    \int^{[n] \in \Delta} \Delta[n] \cdot C(\mathcal{U})_{n,\bullet}
  $$
  is a global projective weak equivalence, and hence in particular also 
  a local projective weak equivalence. 
  This follows from the fact that 
  $$
    \int^{\Delta} (-) \cdot (-) :
	[\Delta, \mathrm{sSet}_{\mathrm{Quillen}}]_{\mathrm{Reedy}}
	\times
	[\Delta^{\mathrm{op}}, [\mathrm{CartSp}^{\mathrm{op}_{\mathrm{op}}}, \mathrm{sSet}]_{\mathrm{inj}}]_{\mathrm{Reedy}}
	\to
	[\mathrm{CartSp}^{\mathrm{op}_{\mathrm{op}}}, \mathrm{sSet}]_{\mathrm{inj}}]_{\mathrm{Reedy}}
  $$
  is a left Quillen bifunctor prop. \ref{CoendOverQuillenBifunctIsQuillenBifunct}. 
  Since every object in $[\Delta^{\mathrm{op}}, [\mathrm{CartSp}^{\mathrm{op}_{\mathrm{op}}}, \mathrm{sSet}]_{\mathrm{inj}}]_{\mathrm{Reedy}}$ is cofibrant,
  and since $\mathbf{\Delta} \to \Delta$ is a Reedy equivalence between Reedy cofibrant
  objects, 
  the coend over the tensoring preserves this weak equivalence and produces
  a global injective weak equivalence which is also a global projective weak equivalence.
  
  This shows that the morphism is question is a weak equivalence. To see that
  it is a cofibrant resolution use that $\mathbf{\Delta}$ is also 
  cofibrant in $[\Delta, \mathrm{sSet}]_{\mathrm{proj}}$ and that
  also
  $$
    \int^{\Delta} (-) \cdot (-) :
	[\Delta, \mathrm{sSet}_{\mathrm{Quillen}}]_{\mathrm{proj}}
	\times
	[\Delta^{\mathrm{op}}, [\mathrm{CartSp}^{\mathrm{op}_{\mathrm{op}}}, \mathrm{sSet}]_{\mathrm{proj}}]_{\mathrm{inj}}
	\to
	[\mathrm{CartSp}^{\mathrm{op}_{\mathrm{op}}}, \mathrm{sSet}]_{\mathrm{proj}}]
  $$  
  is a left Quillen bifunctor, prop. \ref{CoendOverQuillenBifunctIsQuillenBifunct}. 
  By prop. \ref{ResolutionOfTopManifoldsByGoddOpenCovers}
  we have a cofibration
  $\emptyset \hookrightarrow C(\mathcal{U})_{\bullet, \bullet}$ in
  $[\Delta^{\mathrm{op}}, [\mathrm{CartSp}^{\mathrm{op}_{\mathrm{op}}}, \mathrm{sSet}]_{\mathrm{proj}}]_{\mathrm{inj}}$, which is therefore preserved by 
  $\int^{\Delta} \mathbf{\Delta} \cdot (-)$. Again using that global projective
  cofibrations are also local projective cofibrations, the claim follows.
\endofproof

\newpage

We now discuss some of the general abstract structures in any cohesive $\infty$-topos,
\ref{structures}, realized in $\mathrm{ETop}\infty \mathrm{Grpd}$.
\begin{itemize}
  \item \ref{ETopStrucStalks} -- Stalks
  \item \ref{ETopStrucGroups} -- Groups
  \item \ref{ETopStrucHomotopy} -- Geometric homotopy
  \item \ref{ETopStrucIHomotopy} -- $\mathbb{R}^1$-homotopy / The standard continuum
  \item \ref{ETopStrucmanifolds} -- Manifolds
  \item \ref{ETopStrucPathAndGeometricPostnikov} -- Paths and geometric Postnikov towers
  \item \ref{ETopStrucCohomology} -- Cohomology
  \item \ref{ETopStrucPrincipalInfinityBundles} -- Principal $\infty$-bundles  
  \item \ref{ETopStrucWhitehead} -- Universal coverings and geometric Whitehead towers
\end{itemize}

\subsubsection{Stalks} 
\label{ETopStrucStalks}

We discuss the points of $\mathrm{ETop}\infty\mathrm{Grpd}$.

\medskip

\begin{proposition}
 For every $n \in \mathbb{N}$ there is a topos point
 $$
   p(n) : 
    \xymatrix{
     \mathrm{Set}
     \ar@{<-}@<+4pt>[rr]^{p(n)^*}
     \ar@<-4pt>[rr]_{p(n)_*}
	 &&
	 \mathrm{Sh}(\mathrm{Mfd})
	}
 $$
 as well as a corresponding $\infty$-topos point
 $$
   p(n) : 
    \xymatrix{
     \infty\mathrm{Grpd}
     \ar@{<-}@<+4pt>[rr]^{p(n)^*}
     \ar@<-4pt>[rr]_{p(n)_*}
	 &&
	 \mathrm{ETop}\infty\mathrm{Grpd}
	}
	\,,
 $$
 where the inverse image $p(n)^*$ forms the stalk at the origin of $\mathbb{R}^n$:
 $$
   p(n)^* : X \mapsto \lim\limits_{\longrightarrow \atop {k \in \mathbb{N}}}
   X(D^n(1/k))
   \,.
 $$
 Here for $r \in \mathbb{R}_{\geq 0}$ we denote by  $D^n(r) \hookrightarrow \mathbb{R}^n$
 the inclusion of the standard open $n$-disk of radius $r$.
 In particular
 $$
   p(0) \simeq (\Gamma \dashv \mathrm{coDisc})
   \,.
 $$
 The collection of topos points $\{p(n)\}_{n \in \mathbb{N}}$ exhibits
 the topos $\mathrm{Sh}(\mathrm{Mfd})$
 and the $\infty$-topos $\mathrm{ETop}\infty\mathrm{Grpd}$ 
 (hence the sites $\mathrm{CartSp}$ and $\mathrm{Mfd}$) as having 
 \emph{enough points}, def. \ref{site with enough points}.
 
 These points form a tower of retractions
 $$
   \xymatrix{
     p(0)
	 \ar@{<-}@<+4pt>[r]
	 \ar@{^{(}->}@<-4pt>[r]
	 \ar[drr]
	 &
	 p(1)
	 \ar@{<-}@<+4pt>[r]
	 \ar@{^{(}->}@<-4pt>[r]
	 \ar[dr]
	 &
	 \cdots
	 \ar@{<-}@<+4pt>[r]
	 \ar@{^{(}->}@<-4pt>[r]
	 &
	 p(n)
	 \ar@{<-}@<+4pt>[r]
	 \ar@{^{(}->}@<-4pt>[r]
	 \ar[dl]
	 &
	 \cdots
	 \\
	 & & p(\infty)
   }
   \,.
 $$
 The inductive limit $p(\infty) := \lim\limits_{\longrightarrow \atop n} p(n)$
 over the tower of inclusions is the topos point whose inverse image is given by
 $$
   p(\infty)^* X = \lim\limits_{\longrightarrow \atop n}\lim\limits_{\longrightarrow\atop k}
   X(D^n(1/k))
   \,.
 $$
 This point alone forms a set of enough points: a morphism $f : X \to Y$
 is an equivalence precisely if $p(\infty)^* f$ is.
\end{proposition}
\proof
  For convenience, we discuss this in terms of the 1-topos. The
  discussion for the $\infty$-topos is verbatim the same.
  
  First it is clear that for all $n \in \mathbb{N}$ 
  the functor $p(n)^*$ is indeed the inverse image of 
  a geometric morphism: being given by a filtered colimit, it commutes
  with all colimits and with finite limits.

  To see that these points are enough to detect isomorphisms of sheaves,
  notice the following construction.
  For $A \in \mathrm{Sh}(\mathrm{Mfd})$ and $X \in \mathrm{Mfd}$,
  we obtain a sheaf $\tilde A \in \mathrm{Sh}(\mathrm{Mfd}/_{\mathrm{op}}X)$ 
  on the slice site of open embeddings into $X$ by restriction of $A$. 
  The topos
  $\mathrm{Sh}(\mathrm{Mfd}/_{\mathrm{op}}X)$ 
  clearly has enough points, given by the 
  ordinary stalks at the ordinary points $x \in X$, formed as
  $$
    p_x(n)^* \tilde A = \lim\limits_{\longrightarrow_{k}} \tilde A(D^n_x(1/k))
	\,,
  $$
  where $D^n_x(r) \hookrightarrow \mathbb{R}^n \stackrel{\phi}{\hookrightarrow} X$ is a 
  disk of radius $r$ around $x$ in any coordinate patch $\phi$ containing $X$.
  (Because if a morphism of sheaves on $\mathrm{Mfd}/_{\mathrm{op}} X$ is an isomorphism
  on an open disk around every point of $X$, then it is an isomorphism on the covering 
  given by the union of all these disks, hence is an isomorphism of sheaves).
  Notice that by defintion of $\tilde A$ the above stalk is 
  in fact independent of the point $x$
  and coincides with $p(n)^*$ applied to the original $A$:
  $$
    \cdots \simeq \lim\limits_{\longrightarrow_{k}} A(D^n(1/k))
	=: p(n)^* A
	\,.
  $$
  So if for a morphism $f : A \to B$ in $\mathrm{Sh}(\mathrm{Mfd})$ 
  all the $p(n)^* f$ are isomorphisms, then for every $X \in \mathrm{Mfd}$
  the induced morphism $\tilde f : \tilde A \to \tilde B$ is an isomorphism,
  hence is an isomorphism $\tilde f(X) = f(X)$ on global sections. 
  Since this is true for all $X$, it
  follows that $f$ is already an isomorphism.
  This shows that $\{p(n)\}_{n \in \mathbb{N}}$ is a set of 
  enough points of $\mathrm{Sh}(\mathrm{Mfd})$.
  
  To see that these points sit in a sequence of retractions as stated,
  choose a tower of inclusions
  $$
    \mathbb{R}^0 \hookrightarrow \mathbb{R}^1 \hookrightarrow \mathbb{R}^2 \hookrightarrow \cdots
    \;\;\; \in \mathrm{Mfd}\,,	
  $$
  where each morphism is isomorphic to $\mathbb{R}^n \times \mathbb{R}^0 \stackrel{(\mathrm{id}, 0)}{\to} \mathbb{R}^n \times \mathbb{R}^1$.
  
  This induces for each $n \in \mathbb{N}$ and $r \in \mathbb{R}$
  an inclusion of disks $D^n(r) \to D^{n+1}(r)$, which regards 
  $D^n(r)$ as an equatorial plane of $D^{n+1}(r)$, and it 
  induces a projection $D^{n+1}(r)$, which together exhibit a
  retraction
  $$
    \xymatrix{
	  D^n 
	  \ar[r]
	  \ar@/_1pc/[rr]_{\mathrm{id}}
	  &
	  D^{n+1}
	  \ar[r]
	  &
	  D^n
	}
	\,.
  $$
  All this is
  natural with respect to the inclusions $D^n(\tfrac{1}{k+1}) \to D^n(\tfrac{1}{k})$.
  Therefore we have induced morphisms
  $$
    \xymatrix{
      \lim\limits_{\longrightarrow_k} X(D^{n}(1/k))
  	  \ar[r]
	  \ar@/_1.4pc/[rr]_{\mathrm{id}}
	  &
      \lim\limits_{\longrightarrow_k} X(D^{n+1}(1/k))
	  \ar[r]
	  &
      \lim\limits_{\longrightarrow_k} X(D^{n}(1/k))
	}
	\,.
  $$
  Since these are natural in $X$, they consistute natural transformations
  $$
    \xymatrix{
	  p(n)^*
	  \ar[r]
	  \ar@/_1pc/[rr]_{\mathrm{id}}
	  &
	  p(n+1)^*
	  \ar[r]
	  &
	  p(n)^*
	}
  $$
  of inverse images, hence morphisms
  $$
    \xymatrix{
	  p(n)
	  \ar[r]
	  \ar@/_1pc/[rr]_{\mathrm{id}}
	  &
	  p(n+1)
	  \ar[r]
	  &
	  p(n)
	}
  $$
  of geometric morphisms.
  
  Finally, since equivalences are stable under retract, it follows that 
  $p(n)^* f$ is an equivalence if $p(n+1)^*$ is. 
  Similarly, for every  $n \in \mathbb{N}$ we have a retract
  $$
    \xymatrix{
      p(n) \ar[r] \ar@/_1.2pc/[rr]_{\mathrm{id}} & p(\infty) \ar[r] & p(n)
	}
  $$
  seen by noticing that each $p(n)$ naturally forms a co-cone under the
  above tower of inclusions.
  So an isomorphism under $p(\infty)^*$ implies one under all
the $p(n)$.
\endofproof

\subsubsection{Groups} 
\label{ETopStrucGroups}
\index{structures in a cohesive $\infty$-topos!cohesive $\infty$-groups!Euclidean-topological}

We discuss cohesive $\infty$-group objects, def \ref{StrucInftyGroups}, realized in 
$\mathrm{ETop}\infty \mathrm{Grpd}$: \emph{Euclidean-topological $\infty$-groups}.

\medskip

Recall that by prop. \ref{InftyGroupsBySimplicialGroups} every $\infty$-group object in
$\mathrm{ETop}\infty\mathrm{Grpd}$ has a presentation by a presheaf of simplicial groups.
Among the presentations for concrete $\infty$-groups in $\mathrm{ETop}\infty\mathrm{Grpd}$
are therefore \emph{simplicial topological groups}. 

Write $\mathrm{sTop}_{\mathrm{cgH}}$ for the category of simplicial objects in 
$\mathrm{Top}_{\mathrm{cgH}}$, def. \ref{TopCGH}.
For $X,Y \in \mathrm{sTop}_{\mathrm{cgH}}$, write
$$
  \mathrm{sTop}_{\mathrm{cgH}}(X,Y) := \int_{[k] \in \Delta} [X_k, Y_k] \;\; \in \mathrm{Top}_{\mathrm{cgH}}
$$
	for the hom-object, where in the integrand of the end 
$[-,-]$ is the internal hom of $\mathrm{Top}_{\mathrm{cgH}}$.

\begin{definition} \label{GloballyKanSimplicialTopologicalSpace}
We say a morphism $f :  X \to Y$ of simplicial topological spaces is a 
\emph{global Kan fibration} if for all $n \in \mathbb{N}$ 
and $0 \leq k \leq n$ the canonical morphism
$$
  X_n \to Y_n \;\times_{\mathrm{sTop}_{\mathrm{cgH}}(\Lambda[n]_i, Y)}\; \mathrm{sTop}_{\mathrm{cgH}}(\Lambda[n]_i, X)
$$
in $\mathrm{Top}_{\mathrm{cgH}}$ has a section, where 
$\Lambda[n]_i \in \mathrm{sSet} \hookrightarrow \mathrm{sTop}_{\mathrm{cgH}}$ 
is the $i$th $n$-horn regarded as a discrete simplicial topological space.

We say a simplicial topological space $X_\bullet$ is a \emph{(global) Kan simplicial space} 
if the unique morphism $X_\bullet \to *$ is a global Kan fibration, hence if for all 
$n \in \mathbb{N}$ and all $0 \leq i \leq n$ the canonical continuous function
$$
  X_n \to \mathrm{sTop}_{\mathrm{cgH}}(\Lambda[n]_i, X)
$$
into the topological space of $i$th $n$-horns admits a section.
\end{definition}
This global notion of topological Kan fibration is considered for instance 
in \cite{BrownSzczarba}, def. 2.1, def. 6.1. In fact there a stronger condition is imposed: 
a Kan complex in $\mathrm{Set}$ automatically has the lifting property not only against all full horn 
inclusions but also against sub-horns; and in \cite{BrownSzczarba} all these fillers are required to 
be given by global sections. This ensures that with $X$ globally Kan also the internal hom 
$[Y,X] \in \mathrm{sTop}_{\mathrm{cgH}}$ is globally Kan, for any simplicial topological space $Y$. 
This is more than we need and want to impose here. For our purposes it is sufficient to observe that 
if $f$ is globally Kan in the sense of \cite{BrownSzczarba}, def. 6.1, then it is so also in the 
above sense.

For $G$ a simplicial group, 
there is a standard presentation of its universal 
simplicial bundle by a morphism of Kan complexes traditionally denoted 
$W G \to \bar W G$. This
construction has an immediate analog for simplicial topological groups. A review is in \cite{RobertsStevenson}. 

\begin{proposition} 
 \label{SimplicialTopologicalUniversalBundle}
 \index{principal $\infty$-bundle!universal principal $\infty$-bundle!topological}
Let $G$ be a simplicial topological group. Then 
\begin{enumerate}
\item $G$ is a globally Kan simplicial topological space;
\item $\bar W G$ is a globally Kan simplicial topological space;
\item $W G \to \bar W G$ is a global Kan fibration.
\end{enumerate}
\end{proposition}
\proof
The first and last statement appears as \cite{BrownSzczarba}, theorem 3.8 and lemma 6.7, respectively, 
the second is noted in \cite{RobertsStevenson}.
\endofproof
Let for the following $\mathrm{Top}_s \subset \mathrm{Top}_{\mathrm{cgH}}$ 
be any small full subcategory. 
Under the degreewise Yoneda embedding 
$\mathrm{sTop}_s \hookrightarrow [\mathrm{Top}_s^{\mathrm{op}}, \mathrm{sSet}]$ 
simplicial topological spaces embed into the category of simplicial presheaves on $\mathrm{Top}_s$.
We equip this with the projective model structure on simplicial presheaves 
$[\mathrm{Top}_s^{\mathrm{op}}, \mathrm{sSet}]_{\mathrm{proj}}$.
\begin{proposition} 
 \label{GlobalKanFibImpliesProjectiveFib}
Under this embedding a global Kan fibration, def. \ref{GloballyKanSimplicialTopologicalSpace}, 
$f : X \to Y$ in $\mathrm{sTop}_s$ maps to a fibration in $[\mathrm{Top}_s^{\mathrm{op}}, \mathrm{sSet}]_{\mathrm{proj}}$.
\end{proposition}
\proof
By definition, a morphism $f : X \to Y$ in 
$[\mathrm{Top}_s^{\mathrm{op}}, \mathrm{sSet}]_{\mathrm{proj}}$ is a fibration if for all 
$U \in \mathrm{Top}_s$ and all $n \in \mathbb{N}$ and $0 \leq i \leq n$ diagrams of the form
$$
  \xymatrix{
    \Lambda[n]_i \cdot U \ar[r] \ar[d] & X \ar[d]^f
    \\
    \Delta[n] \cdot U \ar[r] & Y
  }
$$
have a lift. This is equivalent to saying that the function
$$
  \mathrm{Hom}(\Delta[n]\cdot U, X)
   \to 
  \mathrm{Hom}(\Delta[n]\cdot U,Y) 
    \times_{\mathrm{Hom}(\Lambda[n]_i \cdot U, Y)}
  \mathrm{Hom}(\Lambda[n]_i \cdot U, X)
$$
is surjective. Notice that we have
$$
  \begin{aligned}
    \mathrm{Hom}_{[\mathrm{Top}_s^{\mathrm{op}}, \mathrm{sSet}]}(\Delta[n]\cdot U, X)
    & = 
    \mathrm{Hom}_{\mathrm{sTop}_s}(\Delta[n]\cdot U, X)
    \\
     & = \int_{[k] \in \Delta} \mathrm{Hom}_{\mathrm{Top}_s}( \Delta[n]_k \times U, X_k)
   \\
    & = \int_{[k] \in \Delta} \mathrm{Hom}_{\mathrm{Top}_s}(U, [\Delta[n]_k, X_k])
   \\
    & = \mathrm{Hom}_{\mathrm{Top}}(U, \int_{[k] \in \Delta} [\Delta[n]_k, X_k])
   \\
    & = \mathrm{Hom}_{\mathrm{Top}_s}(U, \mathrm{sTop}(\Delta[n], X))
   \\
    & = \mathrm{Hom}_{\mathrm{Top}_s}(U, X_n)
  \end{aligned}
$$
and analogously for the other factors in the above morphism. 
Therefore the lifting problem equivalently says that the function
$$
  \mathrm{Hom}_{\mathrm{Top}}(U, \; X_n \to Y_n \times_{\mathrm{sTop}_s(\Lambda[n]_i, Y)} \mathrm{sTop}_s(\Lambda[n]_i,X) \;)
$$
is surjective. But by the assumption that $f : X \to Y$ is a global Kan fibration of simplicial 
topological spaces, def. \ref{GloballyKanSimplicialTopologicalSpace}, we have a 
section $\sigma : Y_n \times_{\mathrm{sTop}_s(\Lambda[n]_i), Y} \mathrm{sTop}_s(\Lambda[n]_i,X) \to X_n$. 
Therefore $\mathrm{Hom}_{\mathrm{Top}_s}(U, \sigma)$ is a section of our function.
\endofproof
In section \ref{ETopStrucHomotopy} we use this in the discussion of geometric realization of simplicial topological groups.

In summary, we find that $W G \to \bar W G$ is a presentation of the 
universal $G$-principal $\infty$-bundle, \ref{ModelForPrincipalInfinityBundles}.
\index{universal principal $\infty$-bundle!for topological simplicial groups}).
\begin{proposition}
  \label{barWGIsPresentationOfBG}
  Let $G \in \mathrm{ETop}\infty \mathrm{Grpd}$ be a group object presented 
  in $[\mathrm{CartSp}_{\mathrm{top}}^{\mathrm{op}}, \mathrm{sSet}]_{\mathrm{proj}, \mathrm{loc}}$ 
  by
  a simplicial topological group (to be denoted by the same symbol) which is degreewise
  a topological manifold. 
  Then
  its delooping $\mathbf{B}G$, def. \ref{delooping}, is presented by $\bar W G$.
\end{proposition}
\proof
  By prop. \ref{SimplicialTopologicalUniversalBundle} and 
  prop. \ref{GlobalKanFibImpliesProjectiveFib} the morphism $W G \to \bar W G$
  is a fibration presentation of $* \to \mathbf{B}G$ in 
  $[\mathrm{CartSp}_{\mathrm{top}}^{\mathrm{op}}, \mathrm{sSet}]_{\mathrm{proj}}$.
  Since $\bar W G$ is evidently connected, and 
  since we have an ordinary pullback diagram
  $$
    \raisebox{20pt}{
    \xymatrix{
	  G \ar[r] \ar[d] & W G \ar[d]
	  \\
	  {*}
	  \ar[r]
	  &
	  \bar W G
	}
	}
	\,,
  $$
  it follows with the discussion in \ref{InfinityPullbackAndHomotopyPullback}
  that this presents in $\mathrm{ETop}\infty\mathrm{Grpd}$ the $\infty$-pullback
  $$
    \xymatrix{
	   G \ar[r] \ar[d] & {*} \ar[d]
	   \\
	   {*}\ar[r] & \mathbf{B}G
	}
  $$
  that defines the delooping $\mathbf{B}G$.
\endofproof

\subsubsection{Representations}
\label{ETopStrucInfinityGroupRepresentations}
\index{structures in a cohesive $\infty$-topos!$\infty$-group representations!topological}

We discuss the intrinsic notion of $\infty$-group representations, \ref{StrucRepresentations},
realized in the context $\mathrm{ETop}\infty\mathrm{Grpd}$.

\medskip

We make precise the role of \emph{topological action groupoids}, 
introduced informally in \ref{Principal1Bundles}. 
\begin{proposition}
  \label{ActionGroupoidForTopologicalGroups}
  Let $X$ be a toplogical manifold, and $G$ a topological group. Then the 
  category of continuous $G$-actions on $X$ in the traditional sense
  is equivalent to the category of $G$-actions on $X$ in 
  the cohesive $\infty$-topos $\mathrm{ETop}\infty\mathrm{Grpd}$, according to 
  def. \ref{RepresentationOfInfinityGroup}.
\end{proposition}
\proof
  For $\rho : X \times G \to X$ a given $G$-action, define the \emph{action groupoid}
  $$
    X/\!/G := 
	(
	  \xymatrix{
	     X \times G
		 \ar@<+4pt>[r]^{\rho}
		 \ar@<-4pt>[r]_{p_1}
		 &
		 X
	  }
	)
  $$
  with the evident composition operation. This comes with the evident 
  morphism of topological groupoids
  $$
    X/\!/G \to */\!/G \simeq \mathbf{B}G
	\,,
  $$
  with $\mathbf{B}G$ as in prop. \ref{DeloopedLieGroup}. It is immediate
  that regarding this as a morphism in $[\mathrm{CartSp}_{\mathrm{top}}^{\mathrm{op}}, \mathrm{sSet}]_{\mathrm{proj}}$ in the canonical way, this is a fibration.
  Therefore, by \ref{FiniteHomotopyLimitsInPresheaves}, 
  the homotopy fiber of this morphism in $\mathrm{Smooth}\infty\mathrm{Grpds}$ 
  is given by the ordinary fiber of this morphism in simplicial presheaves. This is 
  manifestly $X$. 
  
  Accordingly this construction constitutes an embedding of the traditional $G$ actions
  on $X$ into the category $\mathrm{Rep}_G(X)$ from def. \ref{RepresentationOfInfinityGroup}.
  By turning this argument around, one finds that this embedding is essentially surjective.  
\endofproof

\begin{remark}
  \label{TopologicalActionGroupoidSimplicially}
  Let $X, \in \in \mathrm{TopMfd}$, $G$ a topological group, and let 
  $\rho : X \times G \to X$ be a continuous action. Write
  $X /\!/G \in \mathrm{ETop}\infty\mathrm{Grpd}$ for the 
  corresponding action groupoid. As a simplicial topological space
  the action groupoid is
  $$
    X/\!/G
	= 
	\left(
	  \xymatrix{
  	    \ar@{..}[r]
	    &
	    X \times G \times G
	    \ar@<+6pt>[rr]^{(\rho, \mathrm{id})}
	    \ar@<-0pt>[rr]|{(\mathrm{id}, \cdot)}
	    \ar@<-6pt>[rr]_{(p_1, p_2)}
	    &&
	    X \times G
	    \ar@<+3pt>[rr]^{\rho}
	    \ar@<-3pt>[rr]_{p_1}
		&&
	    X
	   }
	  \right)
  $$
\end{remark}

\subsubsection{Geometric homotopy } 
\label{ETopStrucHomotopy}
\index{structures in a cohesive $\infty$-topos!geometric homotopy!Euclidean topological}

We discuss the intrinsic geometric homotopy, \ref{StrucGeometricHomotopy}, in 
$\mathrm{ETop}\infty \mathrm{Grpd}$.

\paragraph{Geometric realization of topological $\infty$-groupoids} 
\label{ETopStrucGeometricRealization}
\index{structures in a cohesive $\infty$-topos!geometric homotopy!geometric realization}

\index{geometric realization!of topological $\infty$-groupoids}

We start by recalling some facts about geometric realization of simplicial topological spaces.

\begin{definition} 
  \label{def rrw}
  \index{geometric realization}
  For $X_\bullet \in \mathrm{sTop}_{\mathrm{cgH}}$ a simplicial topological space, write
  \begin{itemize}
    \item
      $\vert X_\bullet\vert := \int^{[k] \in \Delta} \Delta^k_{\mathrm{Top}} \times X_k$
      for its \emph{geometric realization};
    \item
      $\Vert X_\bullet\Vert := \int^{[k] \in \Delta_+} \Delta^k_{\mathrm{Top}} \times X_k$
      for its \emph{fat geometric realization},
  \end{itemize}
  where in the second case the coend is over the subcategory $\Delta_+ \hookrightarrow \Delta$ 
spanned by the face maps.
\end{definition}
See \cite{RobertsStevenson} for a review.
\begin{proposition}
  \label{RealizationPreservesPullbacks}
  Ordinary geometric realization $\vert-\vert : \mathrm{sTop}_{\mathrm{cgH}} \to \mathrm{Top}_{\mathrm{cgH}}$ 
   preserves pullbacks.
  Fat geometric realization preserves pullbacks when regarded as a functor
  $\Vert-\Vert : \mathrm{sTop}_{\mathrm{cgH}} \to \mathrm{Top}_{\mathrm{cgH}}/\Vert * \Vert$.
\end{proposition}
\begin{definition} 
 \label{GoodAndWellPointed}
  We say
  \begin{itemize}
    \item a simplicial topological space $X \in \mathrm{sTop}_{\mathrm{cgH}}$, def. \ref{TopCGH},
   is \emph{good} if all degeneracy maps 
  $s_i : X_n \to X_{n+1}$ are closed Hurewicz cofibrations;
    \item
      a simplicial topological group $G$ is \emph{well pointed} if all units $i_n : * \to G_n$
      are closed Hurewicz cofibrations.
   \end{itemize}
\end{definition}
The notion of good simplicial topological spaces goes back to \cite{Segal73}. 
For a review see \cite{RobertsStevenson}.
\begin{proposition} 
  \label{RealizationOfGoodSimplicialSpacesIsHomotopyColimit}
  For $X \in \mathrm{sTop}_{s}$ a good simplicial topological space, its
  ordinary geometric realization is equivalent to its homotopy colimit, when regarded
  as a simplicial diagram:
  $$
    \xymatrix{
       \mathrm{sTop}_s 
    \ar@{^{(}->}[r]
	& [\mathrm{Top}_s^{\mathrm{op}}, \mathrm{sSet}]_{\mathrm{proj}} 
  \ar[rr]^{\mathrm{hocolim}}
     && 
	 \mathrm{Top}_{\mathrm{Quillen}}
	 }
    \,.
  $$
\end{proposition}
\proof
  Write $\Vert-\Vert$ for the fat geometric realization. By standard facts about 
  geometric realization of simplicial topological spaces \cite{Segal} we have the following 
  zig-zag of weak homotopy equivalences 
  $$
    \xymatrix{
       \Vert X_\bullet \Vert
         \ar[d]^\simeq 
         & \Vert \;\vert\mathrm{Sing}(X_\bullet)\vert  \;\Vert \ar[l]_\simeq \ar[d]^\simeq
       \ar[d]^{\simeq} 
       \\
       \vert X_\bullet \vert 
       & 
       \vert \;\vert \mathrm{Sing}(X_\bullet)\vert  \;\vert
       \ar@{=}[r]_{\mathrm{iso}} 
       & 
       \vert \mathrm{diag} \mathrm{Sing}(X_\bullet)_\bullet \vert
       \ar[r]^\simeq
       &
       \vert \mathrm{hocolim}_n \mathrm{Sing} X_n\vert
    }
    \,.
  $$
  By the Bousfield-Kan map, the object on the far right is manifestly a model for the homotopy
  colimit $\mathrm{hocolim}_n X_n$. 
\endofproof
\begin{proposition} 
  \label{ResolutionOfManifoldsByGoodCovers}
  For $X \in \mathrm{TopMfd}$ and $\{U_i \to X\}$ a good open cover, the {\v C}ech nerve
  $C(\{U_i\}) := \int^{[k] \in \Delta} \Delta[k] \cdot \coprod_{i_0,\cdots, i_n} U_{i_0} \times_X \cdots \times 
  U_{i_n}$ is cofibrant in $[\mathrm{CartSp}_{\mathrm{top}}^{\mathrm{op}}, \mathrm{sSet}]_{\mathrm{proj}, \mathrm{loc}}$
  and the canonical projection $C(\{U_i\}) \to X$ is a weak equivalence.
\end{proposition}
\proof
  Since the open cover is good, the {\v C}ech nerve is degreewise a coproduct of representables,
  hence is a \emph{split hypercover} in the sense of \cite{dugger-hollander-isaksen}, def. 4.13.
  Moreover $\coprod_i U_i \to X$ is directly seen to be a \emph{generalized cover} in the sense used there
  (below prop. 3.3) By corollary A.3 there, $C(\{U_i\}) \to X$ is a weak equivalence.
\endofproof
\begin{proposition} 
 \label{FundGroupoidOfParacompact}
Let $X$ be a paracompact topological space that admits a good open cover by open balls
(for instance a topological manifold). Write $i(X) \in \mathrm{ETop}\infty \mathrm{Grpd}$ for
its incarnation as a 0-truncatd Euclidean-topological $\infty$-groupoid.
Then $\Pi(X) := \Pi(i(X)) \in \infty \mathrm{Grpd}$ is equivalent to the standard fundamental 
$\infty$-groupoid of $X$, presented by the singular simplicial complex 
$\mathrm{Sing}X : [k] \mapsto \mathrm{Hom}_{\mathrm{Top}_{\mathrm{cgH}}}(\Delta^k, X)$
$$
  \Pi(X) \simeq \mathrm{Sing} X
  \,.
$$
Equivalently, under geometric realization $\mathbb{L}|-| : \infty \mathrm{Grpd} \to \mathrm{Top}$ 
we have that there is a weak homotopy equivalence
$$
  X \simeq |\Pi(X)|
  \,.
$$ 
\end{proposition}
\proof
By the proof of prop. \ref{InfSheavesOverCohesiveSiteAreCohesive}
we have an equivalence $\Pi(-) \simeq \mathbb{L} \lim\limits_{\longrightarrow}$ 
to the derived functor of the $\mathrm{sSet}$-colimit functor 
$\lim\limits_{\longrightarrow} : [\mathrm{CartSp}^{\mathrm{op}}, \mathrm{sSet}]_{\mathrm{proj},\mathrm{loc}} 
\to \mathrm{sSet}_{\mathrm{Quillen}}$. 

To compute this derived functor, let $\{U_i \to X\}$ be a good open cover by open balls, hence
homeomorphically by Cartesian spaces. By goodness of the cover the {\v C}ech nerve 
$C(\coprod_i U_i \to X) \in [\mathrm{CartSp}^{\mathrm{op}}, \mathrm{sSet}]$ 
is degreewise a coproduct of representables, hence a split hypercover.
By \cite{dugger-hollander-isaksen} we have that in this case the canonical morphism
$$
  C(\coprod_i U_i \to X) \to X
$$
is a cofibrant resolution of $X$ in $[\mathrm{CartSp}^{\mathrm{op}}, \mathrm{sSet}]_{\mathrm{proj},\mathrm{loc}}$. Accordingly we have
$$
  \Pi(X) \simeq (\mathbb{L} \lim\limits_{\longrightarrow}) (X) \simeq \lim\limits_{\longrightarrow} C(\coprod_i U_i \to X)
  \,.
$$
Using the equivalence of categories $[\mathrm{CartSp}^{\mathrm{op}}, \mathrm{sSet}] 
\simeq [\Delta^{op}, [\mathrm{CartSp}^{\mathrm{op}}, \mathrm{Set}]$ and that colimits in 
presheaf categories are computed objectwise, and finally using that the colimit of a representable 
functor is the point (an incarnation of the Yoneda lemma) we have that $\Pi(X)$ is presented 
by the Kan complex that is obtained by contracting in the {\v C}ech nerve $C(\coprod_i U_i)$ 
each open subset to a point.

The classical nerve theorem \cite{borsuk} asserts that this implies the claim.
\endofproof
Regarding $\mathrm{Top}$ itself as a cohesive $\infty$-topos by 
\ref{TopAsCohesiveTopos}, the above proposition may be stated as saying 
that for $X$ a paracompact topological space with a good covering, we have
$$
  \Pi_{\mathrm{ETop}\infty \mathrm{Grpd}}(X) \simeq \Pi_{\mathrm{Top}}(X)
  \,.
$$
\begin{proposition} 
  \label{FundGroupoidOfSimplicialParacompact}
Let $X_\bullet$ be a 
good simplicial topological space that is degreewise paracompact
and degreewise admits a good open cover, regarded naturally as an object 
$X_\bullet \in \mathrm{sTop}_{\mathrm{cgH}} \to \mathrm{ETop} \infty \mathrm{Grpd}$. 

We have that the intrinsic $\Pi(X_\bullet) \in \infty \mathrm{Grpd}$ coincides under geometric realization 
$\mathbb{L}|-| : \infty \mathrm{Grpd} \stackrel{\simeq}{\to} \mathrm{Top}$ 
with the ordinary geometric realization of simplicial topological spaces 
$|X_\bullet|_{\mathrm{Top}^{\Delta^{\mathrm{op}}}}$ from def. \ref{RealizationPreservesPullbacks}:
$$
  |\Pi(X_\bullet)| \simeq |X_\bullet|
  \,.
$$
\end{proposition}
\proof
Write $Q$ for Dugger's cofibrant replacement functor, prop. \ref{DegreewiseRepresentability},
on $[\mathrm{CartSp}^{\mathrm{op}}, \mathrm{sSet}]_{\mathrm{proj},\mathrm{loc}}$. 
On a simplicially constant simplicial presheaf $X$ it is given by 
$$
 Q X := \int^{[n] \in \Delta} \Delta[n] \cdot 
    \left(
      \coprod_{U_0 \to \cdots \to U_n \to X} U_0
    \right)
   \,,
$$
where the coproduct in the integrand of the coend is over all sequences of morphisms from 
representables $U_i$ to $X$ as indicated. On a general simplicial presheaf 
$X_\bullet$ it is given by
$$
  Q X_\bullet := \int^{[k] \in \Delta} \Delta[k] \cdot Q X_k
  \,,
$$ 
which is the simplicial presheaf that over any $\mathbb{R}^n \in \mathrm{CartSp}$ takes as value the diagonal of the bisimplicial set whose $(n,r)$-entry is
$\coprod_{U_0 \to \cdots \to U_n \to X_k} \mathrm{CartSp}_{\mathrm{top}}(\mathbb{R}^n,U_0)$.
Since coends are special colimits, the colimit functor itself commutes with them and we find
$$
  \begin{aligned}
     \Pi(X_\bullet) 
     & \simeq 
      (\mathbb{L} \lim\limits_{\longrightarrow}) X_\bullet
     \\
     & \simeq \lim\limits_{\longrightarrow} Q X_\bullet
     \\
      & \simeq \int^{[n] \in \Delta} \Delta[k] \cdot \lim\limits_{\longrightarrow} (Q X_k)
    \,.
  \end{aligned}
$$
By general facts about the Reedy model structure on bisimplicial sets,
this coend is a homotopy colimit over the simplicial diagram 
$\lim\limits_{\longrightarrow} Q X_\bullet : \Delta \to \mathrm{sSet}_{\mathrm{Quillen}}$
$$
  \cdots \simeq \mathrm{hocolim}_\Delta \lim\limits_{\longrightarrow} Q X_\bullet
  \,.
$$
By prop. \ref{FundGroupoidOfParacompact} we have for each $k \in \mathbb{N}$ weak equivalences 
$\lim\limits_{\longrightarrow} Q X_k \simeq (\mathbb{L} \lim\limits_{\longrightarrow}) X_k \simeq \mathrm{Sing} X_k$, 
so that
$$
  \begin{aligned}
    \cdots &\simeq \mathrm{hocolim}_\Delta \mathrm{Sing} X_\bullet
    \\
     & \simeq \int^{[k] \in \Delta} \Delta[k] \cdot \mathrm{Sing} X_k
    \\
     & \simeq \mathrm{diag}\, \mathrm{Sing}(X_\bullet)_\bullet
  \end{aligned}
  \,.
$$
By prop. \ref{RealizationOfGoodSimplicialSpacesIsHomotopyColimit}  
this is the homotopy colimit of the simplicial topological space $X_\bullet$, 
given by its geometric realization if $X_\bullet$ is proper. 
\endofproof

\paragraph{Examples}

We discuss some examples related to the geometric realization of
topological $\infty$-groupoids.

\medskip

\begin{proposition}
  \label{HomotopyEquivalenceOfTopologicalGroupsDeloopsToPiEquivalence}
  Let $K$ and $G$ be topological groups whose underlying topological space is 
  a manifold. Consider a morphism of topological groups $f : K \to G$ that 
  is a homotopy equivalence of the underlying topological manifolds.
  Then 
  $$
    \Pi \mathbf{B}f
	:
    \xymatrix{
      \Pi(\mathbf{B}K) \ar[r] & \Pi(\mathbf{B}G)
	}
  $$
  is a weak equivalence.
\end{proposition}
\proof
  By prop. \ref{barWGIsPresentationOfBG} the delooping 
  $\mathbf{B}G$ is presented in 
  $[\mathrm{CartSp}_{\mathrm{top}^{\mathrm{op}}}, \mathrm{sSet}]_{\mathrm{proj}, \mathrm{loc}}$
  by $(\mathbf{B}G_{\mathrm{ch}}) : n \mapsto G^{\times n}$.
  Therefore $\Pi(K^{\times n}) \to \Pi(G^{\times n})$ is an equivalence in 
  $\infty \mathrm{Grpd}$. By the discussion in 
  \ref{StrucInftyGroups} we have that the delooping $\mathbf{B}K$
  is the $\infty$-colimit
  $$
    \mathbf{B}K \simeq {\lim\limits_{\to}}_{n} K^{\times n}
  $$
  and similarly for $\mathbf{B}G$. The morphism of moduli stacks is 
  the $\infty$-colimit  of the component inclusions
  $$
    \mathbf{c}
	\simeq
	{\lim\limits_{\to}}_n( K^{\times n} \to G^{\times n})
	\,.
  $$
  Since $\Pi$ is left adjoint, it commutes with these colimits,
  so that $\Pi(\mathbf{c})$ is exhibited as an $\infty$-colimit
  over equivalences, hence as an equivalence.
\endofproof

\begin{proposition}
  \label{BorelConstruction}
  \index{Borel construction}
  \index{structures in a cohesive $\infty$-topos!cohomology!equivariant cohomology!Borel construction}
  Let $X$ be a topological manifold, equipped with a continuous action
  $\rho : X \times G \to X$ of a group in $\mathrm{TopMfd}$.
  Then the geometric realization of the corresponding action groupoid,
  def. \ref{ActionGroupoidForTopologicalGroups},
  is the Borel space 
  $$
    \Pi(X/\!/G)
	\simeq
	 {\vert X/\!/G\vert}
	 =
     X \times_G E G 
	 \,.
  $$
\end{proposition}
\proof
  By remark \ref{TopologicalActionGroupoidSimplicially}
  the action groupoid as an object in 
  $\mathrm{TopMfd}^{\Delta^{\mathrm{op}}}\hookrightarrow [\mathrm{CartSp}_{\mathrm{Top}}, \mathrm{sSet}]$ is 
  $$
    X/\!/G
	= 
	\left(
	  \xymatrix{
	    \ar@{..}[r]
	    &
	    X \times G \times G
	    \ar@<+6pt>[rr]^{(\rho, \mathrm{id})}
	    \ar@<-0pt>[rr]|{(\mathrm{id}, \cdot)}
	    \ar@<-6pt>[rr]_{(p_1, p_2)}
	    &&
	    X \times G
	    \ar@<+3pt>[r]^{\rho}
	    \ar@<-3pt>[r]_{p_1}
		&
	    X
	  }
	\right)
	\,.
  $$
  Accordingly 
  $$
    \mathbf{E}G := G/\!/G
	= 
	\left(
	  \xymatrix{
	    \ar@{..}[r]
	    &
	    G \times G \times G
	    \ar@<+6pt>[rr]^-{(\cdot, \mathrm{id})}
	    \ar@<-0pt>[rr]|-{(\mathrm{id}, \cdot)}
	    \ar@<-6pt>[rr]_-{(p_1, p_2)}
	    &&
	    G \times G
	    \ar@<+3pt>[r]^{\cdot}
	    \ar@<-3pt>[r]_{p_1}
		&
	    X
	 }
	\right)
	\,.
  $$
  Therefore we have an isomorphism
  $$
    X/\!/G = X \times_G \mathbf{E}G
	\,.
  $$
  By prop. \ref{RealizationPreservesPullbacks} geometric realization
  preserves the product involved here, and, being given by a coend, it
  preserves the quotient involved, so that we have isomorphisms
  $$
    \vert X/\!/G\vert
	=
	\vert X \times_G \mathbf{E}G\vert
	=
	X \times_G E G
	\,.
  $$
\endofproof
Below in \ref{TopologicalEquivariantCohomology} we discuss how the cohomology
of the Borel space is related to the equivariant cohomology of $X$.

\subsubsection{$\mathbb{R}^1$-homotopy / The standard continuum} 
\label{ETopStrucIHomotopy}
\index{structures in a cohesive $\infty$-topos!$\mathbb{A}^1$-homotopy!$\mathbb{R}^1$-homotopy}

We discuss that the standard continuum real line 
$\mathbb{R} \in \mathrm{SmthMfd} \hookrightarrow \mathrm{ETop}\infty\mathrm{Grpd}$
regarded in Euclidean-topological cohesion 
is indeed a contiuum $\mathbb{A}^1$-line object in the general abstract sense of 
\ref{StrucGeometricHomotopy}.

\medskip

\begin{proposition}
  \label{ReallLineExhibitsEuclideanTopologicalCohesion}
  The real line 
  $\mathbb{R}^1 \in \mathrm{TopMfd} \hookrightarrow \mathrm{ETop}\infty\mathrm{Grpd}$
  is a geometric interval, def. \ref{ILocalization}, 
  exhibiting the cohesion of $\mathrm{ETop}\infty\mathrm{Grpd}$.
\end{proposition}
\proof
  Since $\mathrm{CartSp}_{\mathrm{top}}$ is a site of definition
  for $\mathrm{ETop}\infty\mathrm{Grpd}$ and is both $\infty$-cohesive
  (prop. \ref{CartSpTopIsCohesive})
  and the syntactic category of a Lawvere algebraic theory, with 
  $$
    \mathbb{A}^1 = \mathbb{R}^1
	\,,
  $$
  the claim follows with prop. \ref{IntervalOverLawvereCohesiveSite}.
\endofproof
\begin{remark}
 The statement of prop. \ref{ReallLineExhibitsEuclideanTopologicalCohesion} 
 is the central claim
 of the notes \cite{DuggerSheavesAndHomotopy}, where it essentially appears
 stated as theorem 3.4.3. 
\end{remark}

\subsubsection{Manifolds} 
\label{ETopStrucmanifolds}
\index{structures in a cohesive $\infty$-topos!manifolds!topological manifolds}

We discuss the realization of the general abstract notion of manifolds in a
cohesive $\infty$-topos in \ref{Strucmanifolds} realized in Euclidean-topological cohesion.

\medskip

With $\mathbb{A} := \mathbb{R} \in \mathrm{TopMfd} \hookrightarrow \mathrm{ETop}\infty\mathrm{Grpd}$
the standard line object exhibiting the cohesion of $\mathrm{ETop}\infty\mathrm{Grpd}$
according to prop. \ref{ReallLineExhibitsEuclideanTopologicalCohesion}, def. \ref{IntrinsicManifold}
is equivalent to the traditional definition of topological manifolds.

\subsubsection{Paths and geometric Postnikov towers} 
 \label{ETopStrucPathAndGeometricPostnikov} 
 \index{structures in a cohesive $\infty$-topos!paths!Euclidean topological}
 \index{path!path $\infty$-groupoid!Euclidean-topological}

We discuss the general abstract notion of path $\infty$-groupoid,
\ref{StrucGeometricPostnikov}, realized in $\mathrm{ETop}\infty\mathrm{Grpd}$.

\medskip

\begin{proposition}
  Let $X$ be a paracompact topological space, canonically regarded as an 
  object of $\mathrm{ETop}\infty \mathrm{Grpd}$, then 
  the path $\infty$-groupoid $\mathbf{\Pi}(X)$ is presented by the simplicial
  presheaf $\mathrm{Disc}\, \mathrm{Sing} X \in [\mathrm{CartSp}^{\mathrm{op}}, \mathrm{sSet}]$ which is constant on the singular simplicial complex of $X$:
  $$
    \mathrm{Disc}\,\mathrm{Sing}X :	(U,[k]) \mapsto 
	\mathrm{Sing} X
	\,.
  $$
\end{proposition}
\proof
  By definition we have $\mathbf{\Pi}(X) = \mathrm{Disc}\, \Pi(X)$. 
  By prop. \ref{FundGroupoidOfParacompact} $\Pi(X) \in \infty \mathrm{Grpd}$
  is presented by $\mathrm{Sing} X$. By prop. \ref{InfSheavesOverCohesiveSiteAreCohesive}
  the $\infty$-functor $\mathrm{Disc}$ is presented by the left
  derived functor of the constant presheaf functor. Since 
  every object in $\mathrm{sSet}_{\mathrm{Quillen}}$ is cofibrant 
  this is  just the plain constant presheaf functor.
\endofproof 
A more natural presentation of the idea of a topological 
path $\infty$-groupoid may be one that remembers the topology
on the space of $k$-dimensional paths:
\begin{definition}
  For $X$ a paracompact topological space, write 
  $\mathbf{Sing}X \in [\mathrm{CartSp}^{\mathrm{op}}, \mathrm{sSet}]$
  for the simplicial presheaf given by
  $$
    \mathbf{Sing} X : (U,[k]) \mapsto 
	 \mathrm{Hom}_{\mathrm{Top}}(U \times \Delta^k, X)
	 \,.
  $$
\end{definition} 
\begin{proposition}
  \label{TopologicalPathsPresentPaths}
  Also $\mathrm{Sing} X$ is a presentation of $\mathbf{\Pi}X$.
\end{proposition}
\proof
  For each $U \in \mathrm{CartSp}$ the canonical inclusion 
  of simplicial sets
  $$
    \mathrm{Sing} X \to \mathbf{Sing}(X)(U)
  $$
  is a weak homotopy equivalence, because $U$ is continuously contractible.
  Therefore the canonical inclusion of simplicial presheaves
  $$
    \mathrm{Disc}\, \mathrm{Sing} X \to \mathbf{Sing} X
  $$
  is a weak equivalence in
  $[\mathrm{CartSp}^{\mathrm{op}}, \mathrm{sSet}]_{\mathrm{proj}, \mathrm{loc}}$.
\endofproof
\begin{remark}
  Typically one is interested in mapping out of $\mathbf{\Pi}(X)$.
  While $\mathrm{Disc}\,\mathrm{Sing}X$ is always cofibrant in 
  $[\mathrm{CartSp}^{\mathrm{op}}, \mathrm{sSet}]_{\mathrm{proj}}$,
  the relevant resolutions of $\mathbf{Sing}(X)$ may be harder to
  determine.
\end{remark}

\subsubsection{Cohomology} 
 \label{ETopStrucCohomology}
 \index{principal $\infty$-bundle!topological}
 \index{structures in a cohesive $\infty$-topos!cohomology!Euclidean-topological}

We dicuss aspects of the intrinsic cohomology (\ref{StrucCohomology})
in $\mathrm{E Top} \infty \mathrm{Grpd}$.

\medskip

\paragraph{{\v C}ech cohomology}
\label{CechCohomology}
\index{cohomology!{\v C}ech cohomology}

We expand on the way that the intrinsic cohomology in $\mathrm{ETop}\infty\mathrm{Grpd}$
is expressed in terms of traditional {\v C}ech cohomology
over manifolds, further specializing the general discussion of \ref{SheavesDescent}. 

\medskip

\begin{proposition}
   \label{CohomologyByCovers}
  For $X \in \mathrm{TopMfd}$ and $A \in [\mathrm{CartSp}^{\mathrm{op}}, \mathrm{sSet}]_{\mathrm{proj}, \mathrm{loc}}$
  a fibrant representative of an object in $\mathrm{ETop}\infty \mathrm{Grpd}$, the intrinsic
  cocycle $\infty$-groupoid $\mathrm{ETop}\infty \mathrm{Grpd}$ is given by the 
  {\v C}ech cohomology cocycles on $X$ with coefficients in $A$.
\end{proposition}
\proof
  Let $\{U_i \to X\}$ be a good open cover. By prop. \ref{ResolutionOfManifoldsByGoodCovers} 
  its {\v C}ech nerve 
  $C(\{U_i\}) \stackrel{\simeq}{\to} X$ is a cofibrant replacement for $X$
  (it is a split hypercover \cite{Dugger} and hence cofibrant because the cover is good, 
  and it is a weak equivalence because it is a \emph{generalized cover} in the sense of
  \cite{dugger-hollander-isaksen}). 
  Since $[\mathrm{CartSp}^{\mathrm{op}}, \mathrm{sSet}]_{\mathrm{proj}, \mathrm{loc}}$
  is a simplicial model category, it follows that the cocycle $\infty$-groupoid in question is
  given by the Kan complex $[\mathrm{CartSp}^{\mathrm{op}}, \mathrm{sSet}](C(\{U_i\}), A)$.
  One checks that its vertices are {\v C}ech cocycles as claimed, its edges are 
  {\v C}ech homotopies, and so on.
\endofproof

\paragraph{Nonabelian cohomology with constant coefficients}

\begin{definition}
Let $A \in \infty \mathrm{Grpd}$ be any discrete $\infty$-groupoid. 
Write $|A| \in \mathrm{Top}_{\mathrm{cgH}}$  for its geometric realization. 
For $X$ any topological space, the nonabelian cohomology of $X$ with coefficients in 
$A$ is the set of homotopy classes of maps $X \to |A|$
$$ 
  H_{\mathrm{Top}}(X,A) := \pi_0 \mathrm{Top}(X,|A|)
  \,.
$$
We say $\mathrm{Top}(X,|A|)$ itself is the cocycle $\infty$-groupoid for 
$A$-valued nonabelian cohomology on $X$.

Similarly, for $X, \mathbf{A} \in \mathrm{ETop} \infty \mathrm{Grpd}$ 
two Euclidean-topological $\infty$-groupoids, write
$$
  H_{\mathrm{ETop}}(X,\mathbf{A}) := \pi_0 \mathrm{ETop}\infty \mathrm{Grpd}(X,\mathbf{A})
$$
for the intrinsic cohomology of $\mathrm{ETop} \infty \mathrm{Grpd}$ on $X$ with coefficients in $\mathbf{A}$.
\end{definition}
\begin{proposition}
  \label{TopologicalCohomologyWithConstantCoefficients}
Let $A \in \infty \mathrm{Grpd}$ , write 
$\mathrm{Disc} A \in \mathrm{ETop} \infty \mathrm{Grpd}$ 
for the corresponding discrete topological $\infty$-groupoid. 
Let $X$ be a paracompact topological space admitting a good open cover, regarded as 
0-truncated Euclidean-topological $\infty$-groupoid. 

We have an isomorphism of cohomology sets
$$ 
  H_{\mathrm{Top}}(X,A) \simeq H_{\mathrm{ETop}}(X,\mathrm{Disc} A)
$$
and in fact an equivalence of cocycle $\infty$-groupoids
$$
  \mathrm{Top}(X,|A|) \simeq \mathrm{ETop}\infty \mathrm{Grpd}(X, \mathrm{Disc} A)
  \,.
$$
\end{proposition}
\proof
By the $(\Pi \dashv \mathrm{Disc})$-adjunction of the locally $\infty$-connected 
$\infty$-topos $\mathrm{ETop} \infty \mathrm{Grpd}$ we have
$$
  \mathrm{ETop}\infty \mathrm{Grpd}(X, \mathrm{Disc} A) \simeq 
   \infty \mathrm{Grpd}(\Pi(X), A)
  \xymatrix{ \ar[r]_{\simeq}^{|-|} & }
  \mathrm{Top}(|\Pi X|, |A|)
  \,.
$$
From this the claim follows by prop. \ref{FundGroupoidOfParacompact}.
\endofproof

\paragraph{Equivariant cohomology}
\label{TopologicalEquivariantCohomology}
\index{structures in a cohesive $\infty$-topos!cohomology!equivariant cohomology!topological}

\begin{proposition}
  Given an action $\rho : X \times G \to X$
  of a topological group $G$ on a topological manifold $X$,
  as in prop. \ref{BorelConstruction}, $n \in \mathbb{N}$ and 
  $K$ a discrte group, abelian if $n \geq 2$, then the $G$-equivariant cohomology, 
  def. \ref{EquivariantCohomology},
  of $X$ with coefficients in $K$ is the cohomology of the Borel space, 
  prop. \ref{BorelConstruction},  with 
  values in $K$
  $$
    H_G^n(X,K) \simeq H^n(X \times_G E G, K)
	\,.
  $$
\end{proposition}
\proof
  The equivariant cohomology is the cohomology of the action groupoid
  $$
    H^n_G(X, K)
	\simeq
	\pi_0 \mathrm{ETop}\infty\mathrm{Grpd}(X/\!/G, \mathbf{B}^n K)
	\,.
  $$
  Since $K$ is assumed discrete, this is equivalently, as in 
  prop. \ref{TopologicalCohomologyWithConstantCoefficients},
  $$
    \cdots \simeq \pi_0 \infty \mathrm{Grpd}(\Pi(X /\! / G), \mathbf{B}^n K)
	\,
  $$
  By prop. \ref{BorelConstruction} this is
  $$
    \cdots \simeq \pi_0 \mathrm{Top}(X \times_G E G, B^n K)
	\simeq
	H^n(X \times_G E G, K)
	\,.
  $$
\endofproof

\subsubsection{Principal bundles} 
 \label{ETopStrucPrincipalInfinityBundles}
 \index{principal $\infty$-bundle!topological}
 \index{structures in a cohesive $\infty$-topos!principal $\infty$-bundles!Euclidean-topological}
\index{simplicial principal bundle!topological}

We  discuss principal $\infty$-bundles, \ref{section.PrincipalInfinityBundle}, 
with topological structure and presented by 
topological simplicial principal bundles.

\medskip

\begin{proposition} 
 \label{BarWGIsGoodIfGIsWellSectioned}
If $G$ is a well-pointed simplicial topological group, def. \ref{GoodAndWellPointed}, then both 
$W G$ and $\bar W G$ are good simplicial topological spaces.
\end{proposition}
\proof
For $\bar W G$  this is \cite{RobertsStevenson} prop. 19. 
For $W G$ this follows with their lemma 10, lemma 11, which says that 
$W G = \mathrm{Dec}_0 \bar W G$ and the observations in the proof of 
prop. 16 that $\mathrm{Dec}_0 X$ is good if $X$ is.
\endofproof
\begin{proposition} \label{RealizationSimplicialTopologicalUniversalBundle}
For $G$ a well-pointed simplicial topological group, the geometric realization of the universal 
simplicial principal bundle $W G \to \bar W G$
$$
  {\vert W G \vert} \to {\vert \bar W G \vert}
$$
is a fibration resolution in $\mathrm{Top}_{\mathrm{Quillen}}$ of the point inclusion 
$* \to B{|G|}$ into the classifying space of the geometric realization of $G$.
\end{proposition}
This is \cite{RobertsStevenson}, prop. 14.
\begin{proposition} \label{SimplicialTopolgicalBundleIsGood} 
Let $X_\bullet$ be a good simplicial topological space and 
$G$ a well-pointed simplicial topological group. Then for every morphism
$$
  \tau :  X \to \bar W G
$$
the corresponding topological simplicial principal bundle $P$ over $X$ is itself a good 
simplicial topological space.
\end{proposition}
\proof
The bundle is the pullback $P = X \times_{\bar W G} W G$ in $\mathrm{sTop}_{\mathrm{cgH}}$
$$
  \xymatrix{
     P \ar[r]\ar[d] & \bar W G \ar[d]
     \\
     X \ar[r]^\tau & \bar W G
  }
  \,.
$$
By assumption on $X$ and $G$ and using 
prop. \ref{BarWGIsGoodIfGIsWellSectioned} we have that 
$X$, $\bar W G$ and $W G$ are all good simplicial spaces.
This means that the degeneracy maps of $P_\bullet$ are induced degreewise by morphisms 
between pullbacks in $\mathrm{Top}_{\mathrm{cgH}}$ that are degreewise closed cofibrations, 
where one of the morphisms in each pullback is a fibration. 
This implies that also these degeneracy maps of $P_\bullet$ are closed cofibrations.
\endofproof
\begin{proposition} \label{HocolimOfGoodSimplicialTopMorphismsPreservesHomotopyFibers}
The homotopy colimit operation
$$
  \mathrm{sTop}_s \hookrightarrow [\mathrm{Top}_s^{\mathrm{op}}, \mathrm{sSet}]_{\mathrm{proj}} 
  \stackrel{\mathrm{hocolim}}{\to} \mathrm{Top}_{\mathrm{Quillen}}
$$
preserves homotopy fibers of morphisms $\tau \colon  X \to \bar W G$ with $X$ good 
and $G$ well-pointed
(def. \ref{GoodAndWellPointed}) and globally Kan 
(def. \ref{GloballyKanSimplicialTopologicalSpace}).
\end{proposition}
\proof
By prop. \ref{SimplicialTopologicalUniversalBundle} 
and prop. \ref{GlobalKanFibImpliesProjectiveFib} we have that $W G \to \bar W G$ is a 
fibration resolution of the point inclusion $* \to \bar W G$ in 
$[\mathrm{Top}^{\mathrm{op}}, \mathrm{sSet}]_{\mathrm{proj}}$.
By general properties of homotopy limits this means that the homotopy fiber of 
a morphism $\tau \colon  X \to \bar W G$ is computed as the ordinary pullback $P$ in 
$$
  \xymatrix{
    P \ar[r] \ar[d] & W G \ar[d]
    \\
    X \ar[r]^\tau & \bar W G
  }
$$
(since all objects $X$,  $\bar W G$ and $W G$ are fibrant and at least one of the two morphisms 
in the pullback diagram is a fibration) and hence
$$
  \mathrm{hofib}(\tau) \simeq P
  \,.
$$
By prop. \ref{SimplicialTopologicalUniversalBundle} 
and prop. \ref{SimplicialTopolgicalBundleIsGood} 
it follows that all objects here 
are good simplicial topological spaces. 
Therefore by prop. \ref{RealizationOfGoodSimplicialSpacesIsHomotopyColimit} we have
$$
  \mathrm{hocolim} P_\bullet \simeq {|P_\bullet|}
$$
in $\mathrm{Ho}(\mathrm{Top}_{\mathrm{Quillen}})$. 
By prop. \ref{RealizationPreservesPullbacks} we have that 
$$
  \cdots = {|X_\bullet|} \times_{|\bar W G|} {|W G|}
  \,.
$$
But prop. \ref{RealizationSimplicialTopologicalUniversalBundle} 
says that this is again the presentation of a homotopy pullback/homotopy fiber by an ordinary pullback
$$
  \xymatrix{
    {|P|} \ar[r] \ar[d] &  {|W G|} \ar[d]
    \\
    {|X|} \ar[r]^\tau & {|\bar W G|}
  }
  \,,
$$
because $|W G| \to |\bar W G|$ is again a fibration resolution of the point inclusion. Therefore
$$
  \mathrm{hocolim} P_\bullet \simeq \mathrm{hofib}( {|\tau|} )
  \,.
$$
Finally by prop. \ref{RealizationOfGoodSimplicialSpacesIsHomotopyColimit} 
and using the assumption that $X$ and $\bar W G$ are both good, this is
$$
  \cdots \simeq \mathrm{hofib} (\mathrm{hocolim} \tau)
  \,.
$$
In total we have shown
$$
  \mathrm{hocolim} (\mathrm{hofib} (\tau)) \simeq \mathrm{hofib} (\mathrm{hocolim} (\tau))
  \,.
$$
\endofproof

We now generalize the model of \emph{discrete} principal $\infty$-bundles
by simplicial principal bundles over simplicial groups, from \ref{DiscStrucCohomology}, to Euclidean-topological cohesion.

Recall from theorem \ref{PiPreservesPullbacksOverDiscretes}
that over any $\infty$-cohesive site 
$\Pi$ preserves homotopy pullbacks over discrete objects. The
following proposition says that on $\mathrm{ETop}\infty\mathrm{Grpd}$
it preserves also a large class of $\infty$-pullbacks over
non-discrete objects.
\begin{theorem} 
  \label{GeometricRealizationOfHomotopyFibers}
  Let $G$ be a well-pointed simplicial group object in $\mathrm{TopMfd}$. 
  Then the $\infty$-functor $\Pi : \mathrm{ETop}\infty\mathrm{Grpd} \to \infty \mathrm{Grpd}$
  preserves homotopy fibers of all morphisms of the form $X \to \mathbf{B}G$
  that are presented in $[\mathrm{CartSp}_{\mathrm{top}}^{\mathrm{op}}, \mathrm{sSet}]_{\mathrm{proj}}$
  by morphism of the form $X \to \bar W G$ with $X$ fibrant
  $$
    \Pi( \mathrm{hofib}(X \to \bar W G) )
    \simeq
    \mathrm{hofib}(\Pi(X \to \bar W G))
    \,.
  $$
\end{theorem}
\proof 
  By prop. \ref{FiniteHomotopyLimitsInPresheaves} we may 
  discuss the homotopy fiber in the global model structure on simplicial presheaves.
  Write $Q X \stackrel{\simeq}{\to} X$ for the global cofibrant resolution
  given by $Q X : [n] \mapsto \coprod\limits_{\{U_{i_0} \to \cdots \to U_{i_n} \to X_n\}} U_{i_0}$,
  where the $U_{i_k}$ range over $\mathrm{CartSp}_{\mathrm{top}}$ \cite{Dugger}.
  This has degeneracies splitting off as direct summands, and hence is a good 
  simplicial topological space that is degreewise in $\mathrm{TopMfd}$. 
  Consider then the pasting of two pullback diagrams of 
  simplicial presheaves
  $$
    \xymatrix{
      P' \ar@{->>}[d] \ar[r]^\simeq & P \ar[r] \ar@{->>}[d] & W G \ar@{->>}[d]
      \\
      Q X \ar[r]^\simeq& X \ar[r] & \bar W G
    }
    \,.
  $$
  Here the top left morphism is a global weak equivalence because 
  $[\mathrm{CartSp}_{\mathrm{top}}^{\mathrm{op}}, \mathrm{sSet}]_{\mathrm{proj}}$ is right proper.
  Since the square on the right is a pullback of fibrant objects with one morphism being a 
  fibration, $P$ is a presentation of the homotopy fiber of $X \to \bar W G$. Hence so is
  $P'$, which is moreover the pullback of a diagram of good simplicial spaces.
  By prop. \ref{FundGroupoidOfSimplicialParacompact} we have that on the outer
  diagram $\Pi$ is presented by geometric realization of simplicial topological spaces $|-|$.
  By prop. \ref{RealizationSimplicialTopologicalUniversalBundle} we have 
  a pullback in $\mathrm{Top}_{\mathrm{Quillen}}$
  $$
    \xymatrix{
      |P| \ar[r] \ar[d] & |W G| \ar@{->>}[d]
      \\
      |Q X| \ar[r] & |\bar W G|
    }
  $$
  which exhibits $|P|$ as the homotopy fiber of $|Q X| \to |\bar W G|$. But this is
  a model for $|\Pi(X \to \bar W  G)|$.
\endofproof

\subsubsection{Gerbes} 
\label{ETopStrucGerbes}
\index{structures in a cohesive $\infty$-topos!$\infty$-gerbes!Euclidean-topological}
\index{gerbe!over a topological space}

We discuss $\infty$-gerbes, \ref{StrucInftyGerbes}, in the 
context of Euclidean-topological cohesion, with respect to the cohesive $\infty$-topos $\mathbf{H} := \mathrm{ETop}\infty \mathrm{Grpd}$ 
from def. \ref{ETopInftyGrpd}.

For $X \in \mathrm{TopMfd}$ write 
$$
  \mathcal{X} := \mathbf{H}/X 
$$
for the slice of $\mathbf{H}$ over $X$, as in remark \ref{CohomologyOverX}. This is equivalently the $\infty$-category of $\infty$-sheaves on $X$ itself
$$
  \mathcal{X} \simeq \mathrm{Sh}_\infty(X)
  \,.
$$ 
By remark \ref{CohomologyOverX} this comes with the canonical {\'e}tale essential geometric morphism
$$
  (X_! \dashv X^* \dashv X_*)
  :
  \xymatrix{
     \mathbf{H}/X
	 \ar@<+12pt>[r]^<<<<<{X_!}
	 \ar@{<-}@<+4pt>[r]|<<<<<{X^*}
	 \ar@<-4pt>[r]_<<<<<{X_*}
	 &
	 \mathbf{H}
  }  
  \,.
$$
Any topological group $G$ is naturally an object 
$G \in \mathrm{Grp}(\mathbf{H})\subset \infty \mathrm{Grp}(\mathbf{H})$
and hence as an object 
$$
  X^* G \in \mathrm{Grp}(\mathcal{X})
  \,.
$$
Under the identification $\mathcal{X} \simeq \mathrm{Sh}_\infty(X)$ this is 
the sheaf of grpups which assigns sets of continuous functions from open 
subsets of $X$ to $G$:
$$
  X^* G : (U \subset X) \mapsto C(U, G) 
  \,.
$$
Since the inverse image $X^*$ commutes with looping and delooping,
we have
$$
  X^* \mathbf{B}G \simeq \mathbf{B} X^* G
  \,.
$$
On the left $\mathbf{B}G$ is the abstract stack of topological $G$-principal bundles, regarded over $X$, on the right is the stack over $X$ of $X^* G$-torsors.

More generally, an arbitrary group object $G \in \mathrm{Grp}(\mathcal{X})$ is (up to equivalence) any sheaf of groups on $X$, and $\mathbf{B}G \in \mathcal{X}$ is the corresponding stack of $G$-torsors over $X$. (A detailed discussion of these is for instance in \cite{BreenNotes}. )
\begin{definition}
  \label{CircleNGerbe}
  \label{CircleNGroup}
  \index{gerbe!circle $n$-gerbe}
  Let $G = U(1) := \mathbb{R}/\mathbb{Z}$ and $n \in \mathbb{N}$, 
  $n \geq 1$. Write 
  $\mathbf{B}^{n-1}U(1) \in \infty \mathrm{Grp}(\mathbf{H})$ for the topological
  \emph{circle $n$-group}.

  A $\mathbf{B}^{n-1}U(1)$-$n$-gerbe we call a \emph{circle $n$-gerbe}.
\end{definition}

\begin{proposition}
  \index{gerbe!circle $n$-gerbe}
  The automorphism $\infty$-groups, def. \ref{AutomorphismInfinityGroup},
  of the circle $n$-groups, def. \ref{CircleNGroup}, are given by the following crossed complexes (def. \ref{StrictnGroup})
  $$
    \mathrm{AUT}(U(1)) \simeq [U(1) \stackrel{0}{\to} \mathbb{Z}_2]
	\,,
  $$
  $$
    \mathrm{AUT}(\mathbf{B}U(1)) \simeq [U(1) \stackrel{0}{\to} U(1) \stackrel{0}{\to} \mathbb{Z}_2]
	\,.
  $$
  Here $\mathbb{Z}_2$ acts on the $U(1)$ by the canonical action via
  $\mathbb{Z}_2 \simeq \mathrm{Aut}_{\mathrm{Grp}}(U(1))$.
  
  The outer automorphism $\infty$-groups, def. \ref{outerAutomorphismInfinityGroup} are
  $$
    \mathrm{Out}(U(1)) \simeq \mathbb{Z}_2\,;
  $$
  $$
    \mathrm{Out}(\mathbf{B}U(1)) \simeq [U(1) \stackrel{0}{\to} \mathbb{Z}_2]
	\,.
  $$
  Hence both $\infty$-groups are, of course, their own center.
  \end{proposition}
With prop. \ref{ClassificationOfGGerbes} it follows that 
$$
  \pi_0 U(1) \mathrm{Gerbe}(X) \simeq H^1(X,[U(1) \stackrel{0}{\to} \mathbb{Z}_2) 
$$
$$
  \pi_0 \mathbf{B}U(1) \mathrm{Gerbe}(X) \simeq H^1(X,[U(1) \stackrel{0}{\to} U(1) \stackrel{0}{\to} \mathbb{Z}_2) 
  \,.
$$
Notice that this classification is different (is richer) than that of $U(1)$ \emph{bundle gerbes} and $U(1)$ \emph{bundle 2-gerbes}. These are really models for $\mathbf{B}U(1)$-principal 2-bundles and $\mathbf{B}^2 U(1)$-principal 3-bundles on $X$, and hence instead have the classification of prop. \ref{PrincipalInfinityBundleClassification}:
$$
  \pi_0 \mathbf{B}U(1) \mathrm{Bund}(X) \simeq H^1(X, [U(1) \to 1])
  \simeq H^2(X, U(1))
  \,,
$$
$$
  \pi_0 \mathbf{B}^2U(1) \mathrm{Bund}(X) \simeq H^1(X, [U(1) \to 1 \to 1])
  \simeq H^3(X, U(1))
  \,.
$$
Alternatively, this is the classification of the $U(1)$-1-gerbes
and $\mathbf{B}U(1)$-2-gerbes with trivial \emph{band}, def. \ref{BandOfInfinityGerbe}, in $H^1(X, \mathrm{Out}(U(1)))$ and 
$H^1(X, \mathrm{Out}(\mathbf{B}U(1)))$.
$$
  \pi_0 U(1)\mathrm{Gerbe}_{* \in H^1(X,\mathrm{Out}(U(1)))}(X) 
  \simeq H^2(X, U(1))
  \,,
$$
$$
  \pi_0 \mathbf{B} U(1) \mathrm{Gerbe}_{* \in H^1(X,\mathrm{Out}(U(1)))}(X)   \simeq H^3(X, U(1))
  \,.
$$

\subsubsection{Universal coverings and geometric Whitehead towers} 
\label{ETopStrucWhitehead}
\index{structures in a cohesive $\infty$-topos!geometric Whitehead tower!Euclidean-topological}

We discuss geometric Whitehead towers (\ref{StrucGeometricWhitehead}) in 
$\mathrm{ETop}\infty \mathrm{Grpd}$.

\begin{proposition}
Let $X$ be a pointed paracompact topological space that admits a good open cover.  
Then its ordinary Whitehead tower 
$X^{\mathbf{(\infty)}} \to \cdots X^{(2)} \to X^{(1)} \to X^{(0)} = X$ in 
$\mathrm{Top}$ coincides with the image under the intrinsic fundamental $\infty$-groupoid functor 
$|\Pi(-)|$ of its geometric Whitehead tower 
$* \to \cdots X^{\mathbf{(2)}} \to X^{\mathbf{(1)}} \to X^{\mathbf{(0)}} = X$ in 
$\mathrm{ETop} \infty \mathrm{Grpd}$:
$$
  \begin{aligned} 
     |\Pi(-)| & :   (X^{\mathbf{(\infty)}} \to \cdots X^{\mathbf{(2)}} \to X^{\mathbf{(1)}} \to X^{\mathbf{(0)}} = X) 
      \in \mathrm{ETop}\infty \mathrm{Grpd}
    \\
     & \mapsto 
      (* \to \cdots X^{(2)} \to X^{(1)} \to X^{(0)} = X) 
       \in \mathrm{Top}
  \end{aligned}
  \,.
$$
\end{proposition}
\proof
The geometric Whitehead tower is characterized for each $n$ by the fiber sequence 
$$
  X^{\mathbf{(n)}} \to X^{\mathbf{(n-1)}} \to 
  \mathbf{B}^n \mathbf{\pi}_n(X)
  \to 
  \mathbf{\Pi}_n(X) \to \mathbf{\Pi}_{(n-1)}(X)
  \,.
$$ 
By the above prop. \ref{FundGroupoidOfParacompact} we have 
that $\mathbf{\Pi}_n(X) \simeq \mathrm{Disc} (\mathrm{Sing} X)$. Since 
$\mathrm{Disc}$ is right adjoint and hence preserves homotopy fibers this 
implies that 
$\mathbf{B} \mathbf{\pi}_n(X) \simeq \mathbf{B}^n \mathrm{Disc} \pi_n(X)$, 
where $\pi_n(X)$ is the ordinary $n$th homotopy group of the pointed topological space $X$. 

Then by prop. \ref{GeometricRealizationOfHomotopyFibers} we have that 
under $|\Pi(-)|$ the space $X^{\mathbf{(n)}}$ maps to the homotopy fiber of 
$|\Pi(X^{\mathbf{(n-1)}})| \to B^n |\mathrm{Disc} \pi_n(X)| = B^n \pi_n(X)$.

By induction over $n$ this implies the claim.
\endofproof

\newpage

\subsection{Smooth $\infty$-groupoids}
\label{SmoothInfgrpds}
\index{cohesive $\infty$-topos!models!smooth cohesion}

We discuss \emph{smooth} cohesion.

\medskip

\begin{definition}
  Write $\mathrm{SmoothMfd}$ for the category whose objects are smooth manifolds that are
  \begin{itemize}
    \item finite-dimensional;
    \item paracompact;
    \item with arbitrary set of connected components;
  \end{itemize}
  and whose morphisms are smooth functions between these.
\end{definition}
Notice the evident forgetful functor
$$
  i : \mathrm{SmoothMfd} \to \mathrm{TopMfd}
$$
to the category of topological manifolds, from def. \ref{TopologicalManifolds}.
\begin{definition} 
  For $X \in \mathrm{SmoothMfd}$, say an open cover $\{U_i \to X\}$ is a
  \emph{differentiably good open cover} if each non-empty finite intersection of the
  $U_i$ is \emph{diffeomorphic} to a Cartesian space $\mathbb{R}^n$.
  \label{DifferentiablyGoodOpenCover}
  \label{DifferentiallyGoodOpenCover}
\end{definition}
\begin{proposition} 
   \label{diffGoodOpenCovExists}
  Every paracompact smooth manifold admits a differentiably good open cover.
\end{proposition}
\proof
  This is a folk theorem. A detailed proof is in the appendix of \cite{FSS}.
\endofproof
Notice that the statement here is a bit stronger than the familiar statement about topologically
good open covers, where the intersections are only required to be homeomorphic to a ball.
\begin{definition} 
  \label{CartSpSmooth}
Regard $\mathrm{SmoothMfd}$ as a large site equipped with the coverage of differentiably good open covers.
Write $\mathrm{CartSp}_{\mathrm{smooth}} \hookrightarrow \mathrm{SmoothMfd}$ for the full sub-site
on Cartesian spaces.
\end{definition}
\begin{observation}
  Differentiably good open covers do indeed define a coverage and the Grothendieck topology generated from 
  it is the standard open cover topology.
\end{observation}
\proof
  For $X$ a paracompact smooth manifold, $\{U_i \to X\}$ an open cover and $f : Y \to X$ any
  smooth function from a paracompact manifold $Y$, the inverse images $\{f^{-1}(U_i) \to Y\}$
  form an open cover of $Y$. Since $\coprod_i f^{-1}(U_1)$ is itself a paracompact smooth manifold,
  there is a differentiably good open cover $\{K_j \to \coprod_i U_i\}$, hence a differentiably
  good open cover $\{K_j \to Y\}$ such that for all $j$ there is an $i(j)$ such that we 
  have a commuting square
  $$
    \xymatrix{
      K_j \ar[r]\ar[d] & U_{i(j)} \ar[d]
      \\
      Y \ar[r]^f & X
    }
    \,.
  $$
\endofproof
\begin{proposition} 
  \label{CartSpSmoothIsCohesive}
  $\mathrm{CartSp}_{\mathrm{smooth}}$ is an $\infty$-cohesive site.
\end{proposition}
\proof
  By the same kind of argument as in prop. \ref{CartSpTopIsCohesive}.
\endofproof
\begin{definition}
  The $\infty$-topos of \emph{smooth $\infty$-groupoids} is the $\infty$-sheaf $\infty$-topos
  on $\mathrm{CartSp}_{\mathrm{smooth}}$:
  $$
    \mathrm{Smooth}\infty \mathrm{Grpd} := \mathrm{Sh}_{\infty}(\mathrm{CartSp}_{\mathrm{smooth}})
    \,.
  $$
\end{definition}
Since $\mathrm{CartSp}_{\mathrm{smooth}}$ is similar to the site 
$\mathrm{CartSp}_{\mathrm{top}}$ from def. \ref{CartSpTop}, various properties of
$\mathrm{Smooth}\infty\mathrm{Grpd}$ are immediate analogs of the corresponding
properties of $\mathrm{ETop}\infty\mathrm{Grpd}$ from def. \ref{ETopInftyGrpd}.
\begin{proposition}
  $\mathrm{Smooth} \infty \mathrm{Grpd}$ is a cohesive $\infty$-topos.
\end{proposition}
\proof
 With prop. \ref{CartSpSmoothIsCohesive} this follows by prop. \ref{InfSheavesOverCohesiveSiteAreCohesive}.
\endofproof
\begin{proposition} 
  \label{HypercompInfSheavesOnSmoothMfd}
  $\mathrm{Smooth} \infty \mathrm{Grpd}$ is equivalent to the hypercompletion
  of the $\infty$-sheaf $\infty$-topos over $\mathrm{SmoothMfd}$:
  $$
    \mathrm{Smooth} \infty \mathrm{Grpd} \simeq
    {\hat {\mathrm{Sh}}}_{\infty}(\mathrm{SmoothMfd})
    \,.
  $$
\end{proposition}
\proof
  Observe that $\mathrm{CartSp}_{\mathrm{smooth}}$ is a small dense sub-site of
  $\mathrm{SmoothMfd}$. With this the claim follows as 
  in prop. \ref{ToposOverTopMfd}.
\endofproof
\begin{corollary} 
  \label{SmoothManifoldsInSmoothInfGrpds}
 The canonical embedding of smooth manifolds as 0-truncated objects of 
$\mathrm{Smooth}\infty \mathrm{Grpd}$ extends to a full and faithful $\infty$-functor
$$
  \mathrm{SmoothMfd} \hookrightarrow \mathrm{Smooth}\infty\mathrm{Grpd}.
$$
\end{corollary}
\proof
  With prop. \ref{HypercompInfSheavesOnSmoothMfd} this follows from the 
  $\infty$-Yoneda lemma. 
\endofproof
\begin{remark}
  \label{SmoothInfinityGroupoidsAndSimplicialManifolds}
  By example \ref{SmoothInfinityGroupoidByManifolds}
  there is an equivalence of $\infty$-categories
  $$
    \mathrm{Smooth}\infty \mathrm{Grpd}
    \simeq L_W \mathrm{SmthMfd}^{\Delta^{\mathrm{op}}}
    \,,
  $$
  where on the right we have the simplicial localization
  of the category of simplicial smooth manifolds 
  (with arbitrary set of connected components) at the
  stalkwise weak equivalences.
  
  This says that every smooth $\infty$-groupoid has
  a presentation by a simplicial smooth
  manifold (not in general a locally Kan simplicial manifold, though) 
  and that this identification is even homotopy-full and faithful.
\end{remark}
Consider the canonical forgetful functor
$$
  i : \mathrm{CartSp}_{\mathrm{smooth}} \to \mathrm{CartSp}_{\mathrm{top}}
$$
to the site of definition for the cohesive $\infty$-topos $\mathrm{ETop}\infty \mathrm{Grpd}$
of Euclidean-topological $\infty$-groupoids, def. \ref{ETopInftyGrpd}.
\begin{proposition} \label{RelativeTopologicalCohesion}
The functor $i$ extends to an essential geometric morphism
$$
  (i_! \dashv i^* \dashv i_*)
  : 
  \xymatrix{
  \mathrm{Smooth}\infty \mathrm{Grpd}
    \ar@{->}@<+8pt>[rr]^{i_!}
    \ar@{<-}@<+4pt>[rr]|{i^*}
    \ar@{->}@<-4pt>[rr]_{i_*}
	&&
  \mathrm{ETop}\infty \mathrm{Grpd}
  }
$$
such that the $\infty$-Yoneda embedding is factored through the induced 
inclusion $\mathrm{SmoothMfd} \stackrel{i}{\hookrightarrow} \mathrm{Mfd}$ as
$$
  \xymatrix{
    \mathrm{SmoothMfd} \ar@{^{(}->}[r] \ar[d]^i
     & \mathrm{Smooth}\infty \mathrm{Grpd} \ar[d]^{i_!}
    \\
    \mathrm{Mfd} \ar@{^{(}->}[r]& \mathrm{ETop}\infty \mathrm{Grpd}
  }
  \,
$$
\end{proposition}
\proof
Using the observation that $i$ preserves coverings and pullbacks along morphism in covering families, 
the proof follows the steps of the proof of prop. \ref{InfinitesimalNeighbourhoodIsOverInfGroupoid}.
\endofproof
\begin{corollary} \label{UnderlyingETopologicalInftyGroupoids}
The essential global section $\infty$-geometric morphism of 
$\mathrm{Smooth} \infty \mathrm{Grpd}$ factors through that of $\mathrm{ETop}\infty\mathrm{Grpd}$
$$
  (\Pi_{\mathrm{Smooth}} \dashv \mathrm{Disc}_{\mathrm{Smooth}} \dashv \Gamma_{\mathrm{Smooth}})
   : 
   \xymatrix{
      \mathrm{Smooth} \infty \mathrm{Grpd}
       \ar@{->}@<+9pt>[r]^<<<<<{i_!}
       \ar@{<-}@<+3pt>[r]|<<<<<{i^*}
       \ar@{->}@<-3pt>[r]_<<<<<{i_*}
      &
      \mathrm{ETop}\infty \mathrm{Grpd}
       \ar@{->}@<+9pt>[rr]^<<<<<<<<<{\Pi_{\mathrm{ETop}}}
       \ar@{<-}@<+3pt>[rr]|<<<<<<<<<{\mathrm{Disc}_{\mathrm{ETop}}}
      \ar@{->}@<-3pt>[rr]_<<<<<<<<<{\Gamma_{\mathrm{ETop}}}
     &&
      \infty \mathrm{Grpd} 
   }  
$$
\end{corollary}
\proof
This follows from the essential uniqueness of the global section $\infty$-geometric morphism,
prop \ref{Terminalgeometricmorphism},  and of adjoint $\infty$-functors. 
\endofproof
The functor $i_!$ here is the forgetful functor that \emph{forgets smooth structure} and only 
\emph{remembers Euclidean topology-structure}.

\newpage

We now discuss the various general abstract structures in a cohesive $\infty$-topos,
\ref{structures}, realized in $\mathrm{Smooth}\infty\mathrm{Grpd}$.
\begin{itemize}
  \item \ref{SmoothStrucConcreteObjects} -- Concrete objects
  \item \ref{SmoothStrucCohesiveInfiniGroups} -- Groups
  \item \ref{SmoothStrucGroupoids} -- Groupoids
  \item \ref{SmoothStrucHomotopy} -- Geometric homotopy
  \item \ref{SmoothStrucPathsAndGeometricPostnikov} -- Paths and geometric Postnikov towers
  \item \ref{SmoothStrucCohomology} -- Cohomology
  \item \ref{SmoothStrucPrincipalInfinityBundles} -- Principal $\infty$-bundles
  \item \ref{SmoothStrucTwistedCohomology} -- Twisted cohomology
  \item \ref{SmoothStrucInfinityGroupRepresentations} -- $\infty$-Group representations
  \item \ref{SmoothStrucAssociatedBundles} -- Associated bundles
  \item \ref{SmoothStrucmanifolds} -- Manifolds
  \item \ref{SmoothStrucFlatDifferential} -- Flat $\infty$-connections and local systems
  \item \ref{SmoothStrucdeRham} -- de Rham cohomology
  \item \ref{SmoothStrucLieAlgebras} -- Exponentiated $\infty$-Lie algebras
  \item \ref{SmoothStrucCurvature} -- Maurer-Cartan forms and curvature characteristic forms
  \item \ref{SmoothStrucDifferentialCohomology} -- Differential cohomology
  \item \ref{SmoothStrucInfChernWeil} -- $\infty$-Chern-Weil homomorphism
  \item \ref{SmoothStrucHolonomy} -- Higher holonomy
  \item \ref{SmoothStrucChernSimons} -- $\infty$-Chern-Simons functionals
  \item \ref{SmoothStrucGeometricPrequantization} -- Prequantum geometry
\end{itemize}

\subsubsection{Concrete objects}
\label{SmoothStrucConcreteObjects}
\index{structures in a cohesive $\infty$-topos!concrete objects!smooth}

We discuss the general notion of \emph{concrete objects} in a cohesive $\infty$-topos, 
\ref{StrucConcrete}, realized in $\mathrm{Smooth}\infty \mathrm{Grpd}$.

The following definition generalizes the notion of smooth manifold
and has been used as a convenient context for differential geometry.
It goes back to \cite{Souriau} and, in a slight variant, 
to \cite{Chen}. The formulation of differential geometry in this context is carefully exposed in
\cite{IglesiasZemmour}. The sheaf-theoretic formulation of the definition that we state is amplified in 
\cite{BaezHoffnung}. 
\begin{definition} 
  \label{DiffeologicalSpace}
  \index{diffeological space}
  A sheaf $X$ on $\mathrm{CartSp}_{\mathrm{smooth}}$ is a \emph{diffeological space}
  if it is a \emph{concrete sheaf} in the sense of \cite{Dubuc}: 
  if for every $U \in \mathrm{CartSp}_{\mathrm{smooth}}$
  the canonical function
  $$
    X(U) \simeq \mathrm{Sh}(U,X) \stackrel{\Gamma}{\to} \mathrm{Set}(\Gamma(U), \Gamma(X))
  $$
  is an injection.
\end{definition}
The following observations are due to \cite{CarchediSchreiber}.
\begin{proposition}
 \label{Concrete0TruncatedSmoothInfinityGroupodsAreDiffeologicalSpaces}
 \index{concrete cohesive $\infty$-groupoid!diffeological space}
Write $\mathrm{Conc}(\mathrm{Smooth} \infty \mathrm{Grpd})_{\leq 0}$ 
for the full subcategory on the 0-truncated concrete objects, according to def. 
\ref{ConcreteObjects}. This is equivalent to the full subcategory of 
$\mathrm{Sh}(\mathrm{CartSp}_{\mathrm{smooth}})$ on the diffeological spaces:
$$
  \mathrm{DiffeolSpace} \simeq \mathrm{Conc}(\mathrm{Smooth}\infty \mathrm{Grpd})_{\leq 0}
  \,.
$$
\end{proposition}
\proof
Let 
$X \in \mathrm{Sh}(\mathrm{CartSp}_{\mathrm{smooth}}) 
\hookrightarrow \mathrm{Smooth} \infty \mathrm{Grpd}$ be a sheaf. 
The condition for it to be a concrete object according to def. \ref{ConcreteObjects} is that the 
$(\Gamma \dashv \mathrm{coDisc})$-unit
$$
  X \to \mathrm{coDisc} \Gamma X
$$
is a monomorphism. Since monomorphisms of sheaves are detected objectwise 
this is equivalent to the statement that for all $U \in \mathrm{CartSp}_{\mathrm{smooth}}$ 
the morphism
$$
  X(U) \simeq \mathrm{Smooth}\infty \mathrm{Grpd}(U, X) \to 
    \mathrm{Smooth} \infty \mathrm{Grpd}(U, \mathrm{coDisc} \Gamma X)
  \simeq
  \infty \mathrm{Grpd}(\Gamma U, \Gamma X)
$$
is a monomorphism of sets, where in the first step we used the $\infty$-Yoneda lemma and in 
the last one the $(\Gamma \dashv \mathrm{coDisc})$-adjunction. 
This is manifestly the defining condition for concrete sheaves that define diffeological spaces.
\endofproof
\begin{corollary}
The canonical embedding $\mathrm{SmoothMfd} \hookrightarrow \mathrm{Smooth} \infty \mathrm{Grpd}$
from prop. \ref{SmoothManifoldsInSmoothInfGrpds} factors through 
diffeological spaces: we have a sequence of full and faithful $\infty$-functors
$$
  \mathrm{SmoothMfd} \hookrightarrow \mathrm{DiffeolSpace} \hookrightarrow \mathrm{Smooth} \infty \mathrm{Grpd}
  \,.
$$
\end{corollary}
\begin{definition}
  \index{diffeological groupoid}
  Write $\mathrm{DiffeolGrpd} \hookrightarrow \mathrm{SmoothGrpd}$ for the full
  sub-$\infty$-category on those smooth $\infty$-groupoids that are represented by a 
  groupoid object internal to diffeological spaces.
\end{definition}
\begin{proposition}
 \index{concrete cohesive $\infty$-groupoid!diffeological groupoid}
  There is a canonical equivalence
  $$
    \mathrm{DiffeolGrpd} \simeq \mathrm{Conc}(\mathrm{Smooth}\infty\mathrm{Grpd})_{\leq 1}
  $$
  identifying diffeological groupoids with the concrete 1-truncated smooth $\infty$-groupoids.
\end{proposition}
\proof
  By definition, an object $X \in \mathrm{Smooth}\infty\mathrm{Grpd}$ is concrete precisely
  if there exists a 0-concrete object $U$, and an effective epimorphism $U \to X$ such that 
  $U \times_X U$ is itself 0-concrete. By prop. \ref{Concrete0TruncatedSmoothInfinityGroupodsAreDiffeologicalSpaces}
  both $U$ and $U \times_X U$ are equivalent to diffeological spaces.
  Therefore the groupoid object $(\xymatrix{ U \times_X U \ar@<-3pt>[r]\ar@<+3pt>[r] & U })$ 
  internal to $\mathrm{Smooth}\infty\mathrm{Grpd}$
  comes from a groupoid object internal to diffeological spaces. By Giraud's axioms for $\infty$-toposes,
  $X$ is equivalent to (the $\infty$-colimit over) this groupoid object:
  $$
    X \simeq \lim_\to (\xymatrix{ U \times_X U \ar@<-3pt>[r]\ar@<+3pt>[r] & U })
    \,.
  $$
\endofproof

\subsubsection{Groups} 
 \label{SmoothStrucCohesiveInfiniGroups}
 \index{structures in a cohesive $\infty$-topos!cohesive $\infty$-groups!smooth}

We discuss some cohesive $\infty$-group objects, according to \ref{StrucInftyGroups},
in $\mathrm{Smooth}\infty \mathrm{Grpd}$.

\medskip

Let $G \in \mathrm{SmoothMfd}$ be a Lie group. Under the embedding
$\mathrm{SmoothMfd} \hookrightarrow \mathrm{Smooth} \infty \mathrm{Grpd}$ 
this is canonically identifed as a  0-truncated $\infty$-group object in 
$\mathrm{Smooth} \infty \mathrm{Grpd}$.
Write $\mathbf{B}G \in \mathrm{Smooth} \infty \mathrm{Grpd}$ for the corresponding 
delooping object.
\begin{proposition} 
  \label{DeloopedLieGroup}
A fibrant presentation of the delooping object $\mathbf{B}G$
in the projective local model structure on simplicial presheaves
$[\mathrm{CartSp}^{\mathrm{op}}_{\mathrm{smooth}}, \mathrm{sSet}]_{\mathrm{proj}, \mathrm{loc}}$ 
is given by the simplicial presheaf that is the nerve
of the one-object Lie groupoid
$$
  \mathbf{B}G_{\mathrm{ch}} := (G \stackrel{\to}{\to} * ) 
$$
regarded as a simplicial manifold and canonically embedded into 
simplicial presheaves:
$$
  \mathbf{B}G_{\mathrm{ch}} : U \mapsto N(C^\infty(U,G) \stackrel{\to}{\to} *)
  \,.
$$
\end{proposition}
\proof
This is essentially a special case of prop. \ref{SeparatedAndFibrantOnCartSp}.
The presheaf is clearly objectwise a Kan complex, being objectwise
the nerve of a groupoid. It satisfies descent along good open covers
$\{U_i \to \mathbb{R}^n\}$ of Cartesian spaces, because the descent 
$\infty$-groupoid $[\mathrm{CartSp}_{\mathrm{smooth}}^{\mathrm{op}}, \mathrm{sSet}](C(\{U_i\}), \mathbf{B}G)$ 
is $\cdots \simeq G \mathrm{Bund}(\mathbb{R}^n) \simeq G \mathrm{TrivBund}(\mathbb{R}^n)$: 
an object is a {\v C}ech 1-cocycle with coefficients in $G$, 
a morphism a {\v C}ech coboundary. This yields the groupoid of $G$-principal bundles 
over $U$, which for the Cartesian space $U$ is however equivalent to the groupoid of 
trivial $G$-bundles over $U$.

To show that $\mathbf{B}G$ is indeed the delooping object of $G$
it is sufficient by prop. \ref{FiniteHomotopyLimitsInPresheaves}
to compute the $\infty$-pullback 
$G \simeq * \times_{\mathbf{B}G} * \in \mathrm{Smooth}\infty \mathrm{Grpd}$ 
in the global model structure $[\mathrm{CartSp}^{\mathrm{op}}, \mathrm{sSet}]_{\mathrm{proj}}$.
This is accomplished by the ordinary pullback of the 
fibrant replacement diagram 
$$
  \xymatrix{
    G \ar[r] \ar[d]
       & N(G\times G \stackrel{\overset{p_1 \cdot p_2}{\to}}{\underset{p_1}{\to}} G)
      \ar[d]^{p_2}
    \\
    {*} \ar[r] & N(G \stackrel{\to}{\to} {*})
  }
  \,.
$$
\endofproof
\begin{proposition}
   For $G$ a Lie group, $\mathbf{B}G$ is 
   a 1-concrete object in $\mathbf{H}$.
\end{proposition}
\proof
  Since $\mathbf{B}G_{\mathrm{ch}}$ is fibrant
  in $[\mathrm{CartSp}^{\mathrm{op}}, \mathrm{sSet}]_{\mathrm{proj}, \mathrm{loc}}$
  and since $G$ presents a concrete sheaf, this follows with 
  prop. \ref{1ConcreteOverInfinityCohesiveSite}.
\endofproof
\begin{definition} 
 \label{Circlengroup}
Write equivalently
$$
  U(1) = S^1 = \mathbb{R}/\mathbb{Z}
$$ 
for the \emph{circle Lie group}, regarded as a 0-truncated $\infty$-group object in 
$\mathrm{Smooth}\infty \mathrm{Grpd}$ under the embedding prop. \ref{SmoothManifoldsInSmoothInfGrpds}.

For $n \in \mathbb{N}$ the $n$-fold delooping 
$\mathbf{B}^n U(1) \in \mathrm{Smooth} \infty \mathrm{Grpd}$ we call the circle 
\emph{Lie $(n+1)$-group}\index{group!circle Lie $n$-group}.
\end{definition}
Write 
$$
  U(1)[n] := [\cdots \to 0 \to C^\infty(-,U(1)) \to 0 \to \cdots \to 0]
  \in
  [\mathrm{CartSp}_{\mathrm{smooth}}^{\mathrm{op}}, \mathrm{Ch}_{\bullet \geq 0}]
$$
for the chain complex of sheaves concentrated in degree $n$ on $U(1)$.
Recall the right Quillen functor $\Xi : [\mathrm{CartSp}_{\mathrm{smooth}}^{op}, 
  \mathrm{Ch}^+]_{\mathrm{proj}} \to [\mathrm{CartSp}_{\mathrm{smooth}}^{\mathrm{op}}, \mathrm{sSet}]_{\mathrm{proj}}$ 
from prop. \ref{EmbeddingOfChainComplexes}.
\begin{proposition} \label{ChainCircleLieGroupModelIsFibrant}
The simplicial presheaf $\Xi(U(1)[n])$ is a fibrant representative in 
$[\mathrm{CartSp}_{\mathrm{smooth}}^{\mathrm{op}},\mathrm{sSet}]_{\mathrm{proj},\mathrm{loc}}$
of the circle Lie $(n+1)$-group $\mathbf{B}^n U(1)$.
\end{proposition}
\proof
First notice that since $U(1)[n]$ is fibrant in $[\mathrm{CartSp}_{\mathrm{smooth}}^{\mathrm{op}}, \mathrm{Ch}_\bullet]_{\mathrm{proj}}$ 
we have that $\Xi U(1)[n]$ is fibrant in the global model structure $[\mathrm{CartSp}^{\mathrm{op}}, \mathrm{sSet}]_{\mathrm{proj}}$. 
By prop. \ref{FiniteHomotopyLimitsInPresheaves} 
we may compute the $\infty$-pullback that defines the loop space object 
in $\mathrm{Smooth}\infty \mathrm{Grpd}$ in terms of a homotopy pullback in this global model structure.

To that end, consider the global fibration resolution of the point inclusion 
$* \to \Xi(U(1)[n])$ given under $\Xi$ by the morphism of chain complexes
$$
  \xymatrix{
    [C^\infty(-,U(1)) \ar[r]^{\mathrm{Id}} \ar[d]^{\mathrm{Id}} 
      & C^\infty(-,U(1)) \ar[r] \ar[d] 
      & 0 \ar[r] \ar[d]
      & \cdots \ar[r] 
      & 0] \ar[d] 
    \\
    [C^\infty(-,U(1)) \ar[r] 
       &  \ar[r] 0  
       &  \ar[r] 0  
       & \cdots \ar[r] 
       & 0]
  }
  \,.
$$ 
The underlying morphism of chain complexes is clearly degreewise surjective, 
hence a projective fibration, hence its image under $\Xi$ is a projective fibration. 
Therefore the homotopy pullback in question is given by the ordinary pullback
$$
  \xymatrix{
    \Xi[0 \to C^\infty(-,U(1)) \to 0 \to \cdots \to 0]
    \ar[r]
    \ar[d]
    &
    \Xi [C^\infty(-,U(1)) \stackrel{\mathrm{Id}}{\to} C^\infty(-,U(1)) \to 0 \to \cdots \to 0]
    \ar[d]
    \\
    \Xi [0 \to 0  \to 0 \to \cdots \to 0]
    \ar[r] &
    \Xi [C^\infty(-,U(1)) \to 0  \to 0 \to \cdots \to 0]
  }
  \,,
$$ 
computed in $[\mathrm{CartSp}^{\mathrm{op}}, \mathrm{Ch}^+]$ and then using that 
$\Xi$ is the right adjoint and hence preserves pullbacks. 
This shows that the loop object $\Omega \Xi(U(1)[n])$ is indeed presented by 
$\Xi (U(1)[n-1])$.

Now we discuss the fibrancy of $U(1)[n]$ in the local model structure. 
We need to check that for all differentiably good open covers $\{U_i \to U\}$ of 
a Cartesian space $U$ we have that the mophism
$$
  C^\infty(U,U(1))[n] \to [\mathrm{CartSp}^{\mathrm{op}}, \mathrm{sSet}](C(\{U_i\}), \Xi (U(1)[n]))
$$
is an equivalence of Kan complexes, where $C(\{U_i\})$ is the {\v C}ech nerve of the cover. 
Observe that the Kan complex on the right is that whose vertices are cocycles 
in degree-$n$ {\v C}ech cohomology (see \cite{FSS} for more on this) 
with coefficients in $U(1)$ and whose morphisms are coboundaries between these.

We proceed by induction on $n$. For $n = 0$ the condition is just that 
$C^\infty(-,U(1))$ is a sheaf, which clearly it is.
For general $n$ we use that since $C(\{U_i\})$ is cofibrant, the above is the 
derived hom-space functor which commutes with homotopy pullbacks and hence 
with forming loop space objects, so that
$$
  \pi_1 [\mathrm{CartSp}_{\mathrm{smooth}}^{\mathrm{op}}, \mathrm{sSet}](C(\{U_i\}), \Xi (U(1)[n]))
  \simeq
  \pi_0 [\mathrm{CartSp}_{\mathrm{smooth}}^{\mathrm{op}}, \mathrm{sSet}](C(\{U_i\}), \Xi (U(1)[n-1]))
$$
by the above result on delooping. So we find that for all $0 \leq k \leq n$ 
that $\pi_k [\mathrm{CartSp}^{\mathrm{op}}, \mathrm{sSet}](C(\{U_i\}), \Xi(U(1)[n]))$ 
is the {\v C}ech cohomology of $U$ with coefficients in $U(1)$ in degree $n-k$.
By standard facts about {\v C}ech cohomology 
(using the short exact sequence of abelian groups $\mathbb{Z} \to U(1)\to \mathbb{R}$ and the 
fact that the cohomology with coefficients in $\mathbb{R}$ vanishes in positive degree, 
for instance by a partition of unity argument) 
we have that this is given by the integral cohomology groups
$$
  \pi_0 [\mathrm{CartSp}^{\mathrm{op}}, \mathrm{sSet}](C(\{U_i\}), \Xi (U(1)[n]))
  \simeq
  H^{n+1}(U, \mathbb{Z})
$$ 
for $n \geq 1$. For the contractible Cartesian space all these cohomology groups vanish.

So we find that $\Xi(U(1)[n])(U)$ and 
$[\mathrm{CartSp}_{\mathrm{smooth}}^{\mathrm{op}}, \mathrm{sSet}](C(\{U_i\}), \Xi U(1)[n])$ 
both have homotopy groups concentrated in degree $n$ on $U(1)$. 
The above looping argument together with the fact that $U(1)$ is a sheaf also shows 
that the morphism in question is an isomorphism on this degree-$n$ homotopy group, 
hence is indeed a weak homotopy equivalence.
\endofproof
Notice that in the equivalent presentation of $\mathrm{Smooth}\infty \mathrm{Grpd}$ 
by simplicial presheaves on the large site $\mathrm{SmoothMfd}$ the objects $\Xi (U(1)[n])$ are far 
from being locally fibrant. Instead, their locally fibrant replacements are given by 
the $n$-stacks of circle $n$-bundles.

\subsubsection{Groupoids}
\label{SmoothStrucGroupoids}

We discuss aspects of the general abstract theory of \emph{groupoid objects},
\ref{StrucInftyGroupoids}, realized in the context of smooth cohesion.

\paragraph{Group of bisections}
\label{SmoothStrucGroupOfBisections}

We discuss the general notion of groups of bisections of \ref{GroupoidsBisections},
realized in smooth cohesion.

\medskip

  Let 
  $$
    X = \xymatrix{ X_1 \ar@<-3pt>[r] \ar@<+3pt>[r] & X_0 }
	\in
	\mathrm{Grpd}(\mathrm{SmthMfd})
	\hookrightarrow
	\mathrm{Smooth}\infty\mathrm{Grpd}
  $$
  be a Lie groupoid, regarded canonically as smooth $\infty$-groupoid
  and equipped with the atlas given by the canonical inclusion
  $$
    i_X : \xymatrix{X_0 \ar@{->>}[r] & X}
  $$ 
  of the manifold of objcts. 
\begin{proposition}
The group of
  bisections $\mathbf{BiSect}_X(X_0) \in \mathrm{Grp}(\mathrm{Smooth}\infty\mathrm{Grpd})$
  of this groupoid object, according to def, \ref{GroupOfBisections},
  is equivalent to the traditional diffeological group of bisections of Lie groupoid theory
  and the canonical morphism of def. \ref{MapFromGroupOfBisectionsToAutomorphismsOfX0}.
  \label{GroupOfBisectionsOfLieGroupoidFromGeneralAbstract}
\end{proposition}
\proof
  First observe that the hom-groupoid $\mathbf{Smooth}\infty \mathrm{Grpd}_{X}(X_0,X_0)$
  is equivalently given by that of $\mathrm{Grpd}(\mathrm{SmoothMfd})_{/X}(X_0,X_0)$.
  This follows for instance from prop. \ref{SliceHomAsHomotopyFiber}, according to which
  we have a homotopy pullback diagram
  $$
    \raisebox{20pt}{
    \xymatrix{
	  \mathbf{H}_{/X}(U \times X_0,X_0)
	  \ar[r]
	  \ar[d]
	  &
	  \mathbf{H}(U \times X_0,X_0)
	  \ar[d]^{\mathbf{H}(U \times X_0, i_X)}
	  \\
	  {*}
	  \ar[r]^{\vdash i_X}
	  &
	  \mathbf{H}(U \times X_0,X)
	}
	}
  $$
  for each $U \in \mathrm{CartSp} \hookrightarrow \mathrm{Smooth}\infty\mathrm{Grpd}$.
  Here the top right morphism set is equivalent to $\mathrm{SmoothMfd}(U \times X_0, X_0)$.
  The bottom right morphism set is a priori given by morphisms out of the Cech nerve of 
  a good open over of $U \times X_0$. But since the right and bottom morphism both hit
  elements in there which come from direct maps out of $U \times X_0$, also the 
  gauge transformations between them are given by globally defined smooth functions
  $U \times X_0 \to X_1$.
 
  With this now it remains to observe that a diagram
  $$
    \xymatrix{
	  U \times X_0 \ar[rr]^{\phi}_{\ }="s" \ar[dr]_{i_X}^{\ }="t" && X_0 \ar[dl]^{i_X}
	  \\
	  & X_0
	  \ar@{=>} "s"; "t"
	}
  $$
  of smooth groupoids is equivalently 
  \begin{enumerate}
    \item a smoothly $U$-paramaterized collection of smooth function $\phi_u : X_0 \to X_0$;
	\item for each such a smooth choice of morphisms $x \to \phi(x)$ in $X_1$for all
	$x \in x_0$.
  \end{enumerate}
  This is precisely the traditional description of the group of bisections of $X$.
\endofproof

\paragraph{Atiyah groupoids}
\label{SmoothStrucAtiyahGroupoids}

We discuss the general notion of Atiyah groupoids, \ref{AtiyahGroupoids}, realized in smooth cohesion.

\medskip

Let $G \in \mathrm{Grp}(\mathrm{Top}) \hookrightarrow \mathrm{Grp}(\mathrm{Smooth}\infty\mathrm{Grpd})$
be a Lie group, and write $\mathbf{B}G \in \mathrm{ETop}\infty \mathrm{Grpd}$
for its internal delooping, as in \ref{SmoothStrucCohesiveInfiniGroups} above.
Let $X \in \mathrm{SmthMfd} \hookrightarrow \mathrm{Smooth}\infty\mathrm{Grpd}$
be a smooth manifold. Let $P \to X$ be any $G$-principal bundle over $X$ and write
$g : X \to \mathbf{B}G$ for the, essentially unique, morphism that modulates it
(discussed in more detail in \ref{SmoothStrucPrincipalInfinityBundles} below).

The following definition is traditional
\begin{definition}
  The \emph{Atiyah Lie groupoid} of the $G$-principal bundle $P \to X$ is the 
  Lie groupoid
  $$
    \mathrm{At}(P)
	:=
	\left(
	   \xymatrix{
	     P \times_{G} P 
		 \ar@<+3pt>[r]
		 \ar@<-3pt>[r]
		 &
		 X
	   }
	\right)
	\,,
  $$
  with composition defined by the evident composition of pairs of representatives.
  $[s_2,s_3] \circ [s_1, s_2] := [s_1,s_3]$.
  \label{TraditionalAtiyahGroupoid}
\end{definition}
\begin{remark}
  Here $P \times_{U(1)} P = (P \times P)/U(1)$ is the quotient of the cartesian product
  of the total space of the bundle with itself by the diagonal action of $G$ on 
  both factors. So if $(x_1,x_2) \in X \times X$ is fixed then 
  the morphisms in $\mathrm{At}(P)_{x_1,x_2}$ with this source and target form the 
  space $(P_{x_1} \times P_{x_2})/G$. But this is canonically isomorphic to the space
  of $G$-torsor homomorphisms (over the point) $P_{x_1} \to P_{x_2}$:
  $$
    \mathrm{At}(P)_{x_1,x_2} = G \mathrm{Tor}(P_{x_1}, P_{x_2})
	\,.
  $$  
  \label{NatureOfTraditonalAtiyahGroupoid}
\end{remark}
We now discuss that this traditional construction is indeed a special case of the general
discussion in \ref{AtiyahGroupoids}.
\begin{proposition}
  For $P \to X$ a smooth $G$-principal bundle with modulating map
  $g : X \to \mathbf{B}G$ as above, its Atiyah groupoid 
  in $\mathrm{Smooth}\infty \mathrm{Grpd}$ in the sense of 
  def. \ref{AtiyahGroupoid} is canonically represented by the 
  traditional Atiyah groupoid construction of def. \ref{TraditionalAtiyahGroupoid},
  under the canonical embedding $\mathrm{LieGrpd} \to \mathrm{Smooth}\infty \mathrm{Grpd}$.
\end{proposition}
\proof
  By prop. \ref{1ImageByInfinityColimitOverNerve}
  we have that $\mathrm{im}_1(g)$ is given by the $\infty$-colimit over 
  its {\v C}ech nerve. Since $X \in \mathrm{Smooth}\infty\mathrm{Grpd}$ is 0-truncated 
  and $\mathbf{B}G \in \mathrm{Smooth}\infty \mathrm{Grpd}$ is 1-truncated, this
  {\v C}ech nerve is given by a 2-coskeletal simplicial smooth manifold:
  $$
    \mathrm{im}_1(g)
	\simeq
	\underset{\longrightarrow}{\lim}
	\left(
	  \xymatrix{
	    \cdots
		\ar@<+5pt>[r]
		\ar@<-0pt>[r]
		\ar@<-5pt>[r]
		&
	    X \underset{\mathbf{B}G}{\times} X
		\ar@<+3pt>[r]
		\ar@<-3pt>[r]
		&
		X
	  }
	\right)
	\,.
  $$
  Therefore by prop. \ref{SimplicialSetIfHocolimOverItsCompnentDiagram}
  this simplicial diagram, regarded under the embedding
  $\mathrm{SmthMfd}^{\Delta^{\mathrm{op}}} \to \mathrm{Smooth}\infty \mathrm{Grpd}$,
  is equivalently the 1-image of $g$. 
  It is then sufficient to observe that
  $$
    X \underset{\mathbf{B}G}{\times} X
	\simeq
	P \times_G P
	\,.
  $$
  To see this, observe that (since 
  the $\infty$-hom functor $\mathbf{H}(U,-)$ preserves homotopy limits)
  for every $U \in \mathrm{CartSp}$ the $U$-plots of the object on the left
  are equivalently pairs of smooth functions 
  $r,l : U \to X$ equipped with a morphism of $G$-principal bundles
  $l^* P \to r^* P$. By remark \ref{NatureOfTraditonalAtiyahGroupoid} this
  are equivalently the $U$-plots of $P \times_G P$. 
\endofproof

\subsubsection{Geometric homotopy} 
 \label{SmoothStrucHomotopy}
 \index{geometric realization!of  smooth $\infty$-groupoids}
 \index{structures in a cohesive $\infty$-topos!geometric homotopy!smooth}

We discuss the intrinsic fundamental $\infty$-groupoid construction, 
\ref{StrucGeometricHomotopy}, and the induced notion of geometric realization,
realized in $\mathrm{Smooth}\infty\mathrm{Grpd}$.

\paragraph{Geometric realization of simplicial smooth spaces}

\begin{proposition} \label{UnderlyingSimplicialTopologicalSpace}
If $X \in \mathrm{Smooth}\infty \mathrm{Grpd}$ is presented by 
$X_\bullet \in \mathrm{SmoothMfd}^{\Delta^{\mathrm{op}}} \hookrightarrow [\mathrm{CartSp}_{\mathrm{smooth}}^{\mathrm{op}}, \mathrm{sSet}]$, 
then its image $i_!(X) \in $ $\mathrm{ETop}\infty \mathrm{Grpd}$ 
under the relative topological cohesion morphism, prop. \ref{RelativeTopologicalCohesion}, 
is presented by the underlying simplicial topological space 
$X_\bullet \in \mathrm{TopMfd}^{\Delta^{\mathrm{op}}} 
\hookrightarrow [\mathrm{CartSp}_{\mathrm{top}}^{\mathrm{op}}, \mathrm{sSet}]$.
\end{proposition}
\proof
Let first $X \in \mathrm{SmoothMfd} \hookrightarrow \mathrm{SmoothMfd}^{\Delta^{\mathrm{op}}}$ 
be simplicially constant. Then there is a differentiably good open cover, 
\ref{diffGoodOpenCovExists}, 
$\{U_i \to X\}$ such that the {\v C}ech nerve projection
$$
  \left(
  \int^{[k] \in \Delta}
    \Delta[k] \cdot \coprod_{i_0, \cdots, i_k}
     U_{i_0} \times_X \cdots \times_X U_{i_k}
  \right)
   \stackrel{\simeq}{\to}
  X
$$
is a cofibrant resolution in $[\mathrm{CartSp}_{\mathrm{smooth}}^{\mathrm{op}}, \mathrm{sSet}]_{\mathrm{proj},\mathrm{loc}}$ which is degreewise a coproduct of representables. 
That means that the left derived functor $\mathbb{L} \mathrm{Lan}_i$ on $X$ is computed by the 
application of $\mathrm{Lan}_i$ on this coend, which by the fact that this is defined to be the 
left Kan extension along $i$ is given degreewise by $i$, and since $i$ preserves pullbacks along covers, 
this is
$$
  \begin{aligned}
    (\mathbb{L} \mathrm{Lan}_i) X 
    & \simeq
    \mathrm{Lan}_i 
    \left(
      \int^{[k] \in \Delta}
      \Delta[k] \cdot \coprod_{i_0, \cdots, i_k}
       U_{i_0} \times_X \cdots \times_X U_{i_k}
    \right)
    \\
    & = 
    \int^{[k] \in \Delta}
    \Delta[k] \cdot \coprod_{i_0, \cdots, i_k}
         \mathrm{Lan}_i 
           (U_{i_0} \times_X \cdots \times_X U_{i_k})
   \\
    & \simeq
  \int^{[k] \in \Delta}
    \Delta[k] \cdot \coprod_{i_0, \cdots, i_k}
         i 
           (U_{i_0} \times_X \cdots \times_X U_{i_k})
   \\
    & \simeq
  \int^{[k] \in \Delta}
    \Delta[k] \cdot \coprod_{i_0, \cdots, i_k}
           (i(U_{i_0}) \times_{i(X)} \cdots \times_{i(X)} i(U_{i_k}))
   \\
   & \simeq i(X)
  \end{aligned}
  \,,
$$
The last step follows from observing that we have manifestly the {\v C}ech nerve as before, 
but now of the underlying topological spaces of the $\{U_i\}$ and of $X$. 

The claim then follows for general simplicial spaces by observing that 
$X_\bullet = \int^{[k] \in \Delta} \Delta[k] \cdot X_k \in [\mathrm{CartSp}_{\mathrm{smooth}}^{\mathrm{op}}, \mathrm{sSet}]_{\mathrm{proj},\mathrm{loc}}$ presents the $\infty$-colimit over 
$X_\bullet : \Delta^{\mathrm{op}} \to \mathrm{SmoothMfd} \hookrightarrow \mathrm{Smooth} \infty \mathrm{Grpd}$ 
and the left adjoint $\infty$-functor $i_!$ preserves these. 
\endofproof
\begin{corollary} 
  \label{SmoothMfdDeltaRealized}
If $X \in \mathrm{Smooth}\infty \mathrm{Grpd}$ is presented by 
$X_\bullet \in \mathrm{SmoothMfd}^{\Delta^{\mathrm{op}}} \hookrightarrow [\mathrm{CartSp}_{\mathrm{smooth}}^{\mathrm{op}}, \mathrm{sSet}]$,
then the image of $X$ under the fundamental $\infty$-groupoid functor, \ref{StrucGeometricHomotopy},
$$
  \xymatrix{
    \mathrm{Smooth} \infty \mathrm{Grpd} 
     \ar[r]^<<<<<{\Pi}
     &
  \infty \mathrm{Grpd}
   \ar[r]^{|-|}_\simeq
   &
  \mathrm{Top}
  }
$$
is weakly homotopy equivalent to the geometric realization of 
(a Reedy cofibrant replacement of) the underlying simplicial topological space
$$
  |\Pi(X)| \simeq |Q X_\bullet|
  \,.
$$
In particular if $X$ is an ordinary smooth manifold then 
$$
  \Pi(X) \simeq \mathrm{Sing} X
$$
is equivalent to the standard fundamental $\infty$-groupoid of $X$.
\end{corollary}
\proof
By prop. \ref{UnderlyingETopologicalInftyGroupoids} the functor $\Pi$ factors as 
$\Pi X \simeq \Pi_{\mathrm{ETop}} i_! X$. 
By prop. \ref{UnderlyingSimplicialTopologicalSpace} 
this is $\Pi_{\mathrm{Etop}}$ applied to the underlying simplicial topological space. 
The claim then follows with prop. \ref{FundGroupoidOfSimplicialParacompact}.
\endofproof
\begin{corollary} \label{SmoothPiPreservesSomeHomotopyFibers}
  The $\infty$-functor $\Pi : \mathrm{Smooth}\infty\mathrm{Grpd} \to \infty \mathrm{Grpd}$
  preserves homotopy fibers of morphisms that are presented in 
  $[\mathrm{CartSp}_{\mathrm{smooth}}^{\mathrm{op}}, \mathrm{sSet}]_{\mathrm{proj}}$ by morphisms
  of the form $X \to \bar W G$ with $X$ fibrant and $G$ a simplicial group in $\mathrm{SmoothMfd}$.
\end{corollary}
\proof
  By prop. \ref{UnderlyingETopologicalInftyGroupoids} the functor factors as
  $\Pi_{\mathrm{Smooth}} \simeq \Pi_{\mathrm{ETop}} \circ i_!$.
  By prop. \ref{UnderlyingSimplicialTopologicalSpace} $i_!$ assigns the underlying 
  topological spaces. If we can show that this preserves the homotopy fibers in question, then 
  the claim follows with prop. \ref{GeometricRealizationOfHomotopyFibers}.
  We find this as in the proof of the latter proposition, by considering the pasting 
  diagram of pullbacks of simplicial presheaves
  $$
    \xymatrix{
      P' \ar@{->>}[d] \ar[r]^\simeq & P \ar[r] \ar@{->>}[d] & W G \ar@{->>}[d]
      \\
      Q X \ar[r]^\simeq& X \ar[r] & \bar W G
    }
    \,.
  $$
  Since the component maps of the right vertical morphisms are surjective, the degreewise pullbacks
  in $\mathrm{SmoothMfd}$ that define $P'$ are all along transversal maps, and thus the underlying
  objects in $\mathrm{TopMfd}$ are the pullbacks of the underlying topological manifolds.
  Therefore the degreewise  forgetful functor $\mathrm{SmoothMfd} \to \mathrm{TopMfd}$ presents
  $i_!$ on the outer diagram and sends this homotopy pullback to a homotopy pullback.
\endofproof

\paragraph{Co-Tensoring of smooth $\infty$-Stacks over homotopy types of manifolds}
\label{CotensoringOfSmoothHigherStacksOverHomotopyTypesOfSmoothManifolds}

 \begin{example}
 There is a natural equivalence $[\Pi(S^1),X]\cong \mathcal{L}
 X$ between the moduli stack of maps from the homotopy type of the 
 circle $S^1$ to $X$ and the \emph{free loop space object} of $X$. 
 Namely, the free loop space object $\mathcal{L}X$ is defined as the 
 homotopy pullback of its diagonal map along itself
  $$
    \mathcal{L}X := X \underset{X \times X}{\times} X
	\,,
  $$
 i.e., as the object defined by the homotopy pullback diagram   $$
    \raisebox{20pt}{
    \xymatrix{
	  \mathcal{L}X \ar[r] \ar[d] & X \ar[d]^{\Delta_{X}}
	  \\
	  X \ar[r]_-{\Delta_X} & X \times X\;.
	}}
  $$
  \label{FreeLoopSpaceObject}
  One then notices that $S^1$ is obtained by gluing two segments (which are contractible) along 
  their endpoints, which amount to saying that at the level of homotopy types we have an equivalence
  $$
    \Pi(S^1) 
	\simeq
	\ast \underset{\ast \coprod \ast}{\coprod} \ast\;,
  $$
  and uses the fact that $[-,X]$ 
  preserves homotopy limits. Here the top $\coprod$ denotes pushout, while the bottom 
  one denotes disjoint union (itself viewed as an instance of  a pushout). One can similarly 
  see that the $\infty$-groupoid corresponding to the 2-sphere $S^2$ can be viewed as 
   $$
    \Pi(S^2) 
	\simeq
	\ast 
	\underset{\ast  \underset{\ast \coprod \ast}{\coprod} \ast}{\coprod} 
	\ast\;.
  $$
  One can iterate in an obvious way to get the higher cases. 
 \end{example}
The above example immediately generalizes from circles to arbitrary $n$-spheres.
\begin{definition}
  For $X$ an object in $\mathbf{H}$ and for $n \in \mathbb{N}$, 
  the \emph{free $n$-sphere space object} of $X$ is 
  $$
   [\Pi(S^n), X]
	\,.
  $$
  \label{FreenSphereBundle}
\end{definition}
An $(n+1)$-sphere is obtained by gluing two $(n+1)$-disks along their common boundary, which is an $n$-sphere. Since the disks are contractible, from a homotopy type point of view, this amount to the natural equivalence
 $$
    \Pi(S^{n+1})
	 \simeq
	 \ast \underset{\Pi(S^n)}{\coprod} \ast
	 \,.
  $$
  Applying the internal homs to $X$ to this equivalence and recalling that $[-,X]$ preserves homotopy limits, one obtains the following result.
\begin{proposition}
  For all $n \in \mathbb{N}$  we have a natural homotopy pullback square 
  $$
    \raisebox{20pt}{
    \xymatrix{  
	   [\Pi(S^{n+1}), X]
	  \ar[d] \ar[r] & X \ar[d]
	  \\
	  X \ar[r] &  [\Pi(S^n), X]\;.
	}
	}
  $$
  \label{nPluOneSphereSpaceFromnSphereSpace}
\end{proposition}
\proof
  We may use that for all $n$ we have
  $$
    \Pi(S^{n+1})
	 \simeq
	 \ast \underset{\Pi(S^n)}{\coprod} \ast
	 \,.
  $$
  From this the statement follows by using that 
  $[-,X] : \mathbf{H}^{\mathrm{op}} \to \mathbf{H}$ preserves homotopy limits.
  The above statement is standard, one can see it for instance by 
  presenting the situation in the standard model structure on simplicial 
  sets, there replacing one of the maps from the $n$-sphere to the point by the
  cofibration given by the inclusion of the $n$-phere as the boundary of 
  the $(n+1)$-disk, and finally computing the ordinary (1-categorical)
  cofiber of that.
\endofproof

\subsubsection{Paths and geometric Postnikov towers} 
 \label{SmoothStrucPathsAndGeometricPostnikov} 
 \index{structures in a cohesive $\infty$-topos!paths!smooth}
 \index{path!path $\infty$-groupoid!smooth}

We discuss the general abstract notion of path $\infty$-groupoid,
\ref{StrucGeometricPostnikov}, realized in $Smooth\infty\mathrm{Grpd}$.

The presentation of $\mathbf{\Pi}(X)$ in 
$\mathrm{ETop}\infty \mathrm{Grpd}$, \ref{ETopStrucPathAndGeometricPostnikov}
has a direct refinement to smooth cohesion:
\begin{definition}
  For $X \in \mathrm{SmthMfd}$ write
  $\mathbf{Sing} X \in [\mathrm{CartSp}^{\mathrm{op}}, \mathrm{sSet}]$
  for the simplicial presheaf given by
  $$
    \mathbf{Sing} X : (U,[k]) \mapsto 
	\mathrm{Hom}_{\mathrm{SmthMfd}}(U \times \Delta^k, X)
	\,.
  $$
\end{definition}
\begin{proposition}
  The simplicial presheaf $\mathbf{Sing}X$ is a presentation of 
  $\mathbf{\Pi}(X) \in \mathrm{Smooth}\infty \mathrm{Grpd}$.
\end{proposition}
\proof
  This reduces to the argument of prop. \ref{TopologicalPathsPresentPaths}
  after using the Steenrod approximation theorem \cite{Wockel} to 
  refine continuous paths to smooth paths
\endofproof

\subsubsection{Cohomology} 
\label{SmoothStrucCohomology}
\index{structures in a cohesive $\infty$-topos!cohomology!smooth}

We discuss the intrinsic cohomology, \ref{StrucCohomology}, in $\mathrm{Smooth}\infty\mathrm{Grpd}$.

\medskip

\begin{itemize}
  \item \ref{SmoothStrucCohomologyWithConstantCoeffs}
    -- Cohomology with constant coefficients;
  \item \ref{SmoothStrucLieGroupCohomology}
    -- Refined Lie group cohomology.
\end{itemize}

\paragraph{Cohomology with constant coefficients} 
\label{SmoothStrucCohomologyWithConstantCoeffs}
\index{structures in a cohesive $\infty$-topos!cohomology!with constant coefficients}

\begin{proposition} 
 \label{SmoothCohomologyWithConstantCoefficients}
Let $A \in \infty \mathrm{Grpd}$, write $\mathrm{Disc} A \in \mathrm{Smooth} \infty \mathrm{Grpd}$ 
for the corresponding discrete smooth $\infty$-groupoid. Let 
$X \in \mathrm{SmoothMfd} \stackrel{i}{\hookrightarrow} \mathrm{Smooth} \infty \mathrm{Grpd}$ 
be a paracompact topological space regarded as a 0-truncated 
Euclidean-topological $\infty$-groupoid. 

We have an isomorphism of cohomology sets
$$ 
  H_{\mathrm{Top}}(X,A) \simeq H_{\mathrm{Smooth}}(X,\mathrm{Disc} A)
$$
and in fact an equivalence of cocycle $\infty$-groupoids
$$
  \mathrm{Top}(X,|A|) \simeq \mathrm{Smooth}\infty \mathrm{Grpd}(X, \mathrm{Disc} A)
  \,.
$$
More generally, for $X_\bullet \in \mathrm{SmoothMfd}^{\Delta^{op}}$ 
presenting an object $X \in \mathrm{Smooth} \infty \mathrm{Grpd}$ we have
$$
  H_{\mathrm{Smooth}}(X_\bullet, \mathrm{Disc} A)
  \simeq
  H_{\mathrm{Top}}(|X|, |A|)
  \,.
$$
\end{proposition}
\proof
This follows from the $(\Pi \dashv \mathrm{Disc})$-adjunction 
and prop. \ref{SmoothMfdDeltaRealized}. 
\endofproof

\paragraph{Refined Lie group cohomology} 
\label{SmoothStrucLieGroupCohomology}
\index{structures in a cohesive $\infty$-topos!cohomology!of Lie groups}
\label{SegalBrylinski}

The cohomology of a Lie group $G$ with coefficients in 
a Lie group $A$ was historically originally 
defined in terms of cocycles given by smooth functions $G^{\times n} \to A$,
by naive analogy with the situation discussed in \ref{DiscStrucGroupCohomology}.
In the language of simplicial presheaves on $\mathrm{CartSp}$ these are
morphisms of simplicial 
presheaves of the form $\mathbf{B}G_{\mathrm{ch}} \to \mathbf{B}^n A$,
with the notation as in \ref{SmoothStrucCohesiveInfiniGroups}.
This is clearly not a good definition, in general, since 
while $\mathbf{B}^n A$ will be fibrant in 
$[\mathrm{CartSp}^{\mathrm{op}}, \mathrm{sSet}]_{\mathrm{proj}, \mathrm{loc}}$,
the object
$\mathbf{B}G_{\mathrm{ch}}$ in general fails to be cofibrant, hence the
above naive definition in general misses cocycles.

A refined definition of Lie group cohomology was proposed in 
\cite{Segal} and later independently in \cite{Brylinski}. The following
theorem asserts that the definitions given there do coincide with the
intrinsic cohomology of the stack $\mathbf{B}G$ in the 
cohesive $\infty$-topos $\mathrm{Smooth}\infty\mathrm{Grpd}$.

\begin{theorem} 
  \label{LieGroupCohomologyInSmoothInfinGroupoid}
  For $G \in \mathrm{SmoothMfd} \hookrightarrow \mathrm{Smooth}\infty \mathrm{Grpd}$ 
  a Lie group and $A$ either 
  \begin{enumerate}
    \item a discrete abelian group 
    \item the additive Lie group of real numbers $\mathbb{R}$
  \end{enumerate}
the intrinsic cohomology of $G$ in $\mathrm{Smooth}\infty \mathrm{Grpd}$ coincides with the refined 
  Lie group cohomology of Segal \cite{Segal}\cite{Brylinski}
$$
  H^n_{\mathrm{Smooth}\infty \mathrm{Grpd}}(\mathbf{B}G, A) \simeq H^n_{\mathrm{Segal}}(G,A)
  \,.
$$
In particular we have in general
$$
  H^n_{\mathrm{Smooth}\infty \mathrm{Grpd}}(\mathbf{B}G, \mathbb{Z})
  \simeq
  H^{n}_{\mathrm{Top}}(B G, \mathbb{Z})
$$
and for $G$ compact and $n \geq 1$ also
$$
  H^n_{\mathrm{Smooth}\infty \mathrm{Grpd}}(\mathbf{B}G, U(1))
  \simeq
  H^{n+1}_{\mathrm{Top}}(B G, \mathbb{Z})
  \,.
$$
\end{theorem}
\proof
  The statement about constant coefficients is a special case of prop. 
  \ref{SmoothCohomologyWithConstantCoefficients}. The statement about real coefficients
  is a special case of a more general statement in the context of 
  synthetic differential $\infty$-groupoids that will be proven as 
  prop. \ref{RealCohomologyOfCompactLieGroup}.
  The last statement finally follows from this using that
  $H^{n}_{\mathrm{Segal}}(G, \mathbb{R}) \simeq 0$ 
  for positive $n$ and $G$ compact and using the fiber sequence, 
  def. \ref{LongFiberSequence}, induced by the short sequence
  $\mathbb{Z} \to \mathbb{R} \to \mathbb{R}/\mathbb{Z} \simeq U(1)$.
\endofproof

\subsubsection{Principal bundles} 
\label{SmoothStrucPrincipalInfinityBundles}
\index{principal $\infty$-bundle!smooth}
\index{structures in a cohesive $\infty$-topos!principal $\infty$-bundles!smooth}

We discuss principal $\infty$-bundles, \ref{section.PrincipalInfinityBundle}, realized in smooth $\infty$-groupoids.

\medskip

The following proposition asserts that the notion of smooth principal $\infty$-bundle
reproduces traditional notions of smooth bundles and smooth higher bundles.
\begin{proposition}
  \label{ReproducingOrdinaryPrincipalBundles}
  \index{principal $\infty$-bundle!ordinary smooth principal bundles}
  \index{principal bundle!as smooth principal $\infty$-bundle}
  For $G$ a Lie group and $X \in \mathrm{SmoothMfd}$, we have that
  $$
    \mathrm{Smooth}\infty \mathrm{Grpd}(X, \mathbf{B}G)
      \simeq
    G \mathrm{Bund}(X)
  $$ 
  is equivalent to the groupoid of smooth principal $G$-bundles and smooth morphisms between
  these, as traditionally defined, where the equivalence is  established by 
  sending a morphism $g : X \to \mathbf{B}G$ in $\mathrm{Smooth}\infty \mathrm{Grpd}$ 
  to  the corresponding principal $\infty$-bundle $P \to X$ according to prop. \ref{PrincipalBundleAsHomotopyFiber}.

  For $n \in \mathbb{N}$ and $G = \mathbf{B}^{n-1}U(1)$ the circle Lie $n$-group,
  def. \ref{Circlengroup},
  and $X \in \mathrm{SmoothMfd}$, we have that
  $$
    \mathrm{Smooth}\infty \mathrm{Grpd}(X, \mathbf{B}^n U(1))
      \simeq
    U(1) (n-1)\mathrm{BundGerb}(X)
  $$ 
  is equivalent to the $n$-groupoid of smooth $U(1)$-bundle $(n-1)$gerbes.
\end{proposition}
\proof
  Presenting $\mathrm{Smooth}\infty \mathrm{Grpd}$ by the local projective model structure
  $[\mathrm{CartSp}^{\mathrm{op}}, \mathrm{sSet}]_{\mathrm{proj,loc}}$ on simplicial presheaves
  over the site of Cartesian spaces, we have that $\mathbf{B}G$ is fibrant, by prop. \ref{DeloopedLieGroup},
  and that a cofibrant replacement for $X$ is given by the {\v C}ech nerve $C(\{U_i\})$ of any 
  differentiably good open cover $\{U_i \to X\}$. The cocycle $\infty$-groupoid in question is
  then presented by the simplicial set $[\mathrm{CartSp}^{\mathrm{op}}, \mathrm{sSet}](C(\{U_i\}), \mathbf{B}G)$
  and this is readily seen to be the groupoid of {\v C}ech cocycles with coefficients
  in $\mathbf{B}G$ relative to the chosen cover. 
  
  This establishes that the two groupoids are equivalent. That the equivalence is indeed 
  established by forming homotopy fibers of morphisms has been discussed in 
  \ref{SmoothPrincipalnBundles} (observing that by the discussion in 
  \ref{ModelForPrincipalInfinityBundles} the 
  ordinary pullback of the morphism $\mathbf{E}G \to \mathbf{B}G$ serves as a presentation for
  the homotopy pullback of ${*} \to \mathbf{B}G$).
\endofproof
This establishes the situation for smooth nonabelian cohomology in degree 1 and
smooth abelian cohomology in arbitrary degree. We turn now to a 
discussion of smooth nonabelian cohomology ``in degree 2'', 
the case where $G$ is a \emph{Lie 2-group}:
\emph{$G$-principal 2-bundles}. 

When $G = \mathrm{AUT}(H)$ the \emph{automorphism 2-group} of a Lie group $H$ (see below)
these structures have the same classification as smooth $H$-1-gerbes,
def. \ref{GGerbe}.
To start with, note the general abstract notion of smooth 2-groups:
\begin{definition}
   \index{2-group}
   \index{group!2-group}
   A \emph{smooth 2-group} is a 1-truncated group object in 
   $\mathbf{H} = \mathrm{Sh}_{\infty}(\mathrm{CartSp})$. These are equivalently given by
   their (canonically pointed) delooping 2-groupoids $\mathbf{B}G \in \mathbf{H}$, which are
   precisely, up to equivalence, the connected 2-truncated objects of $\mathbf{H}$.
   
   For $X \in \mathbf{H}$ any object, $G 2\mathrm{Bund}_{\mathrm{smooth}}(X) :=
   \mathbf{H}(X, \mathbf{B}G)$ is the 2-groupoid of smooth $G$-principal 2-bundles on $G$.
\end{definition}
We consider the presentation of smooth 2-groups by Lie 
crossed modules, def. \ref{Strict2GroupInIntroduction}, according to
prop. \ref{StrictificationOf2GroupObjects}.
Write $[G_1 \stackrel{\delta}{\to} G_0]$
for the \emph{2-group} which is the groupoid 
  $$
    \xymatrix{
	   G_0 \times G_1 
	   \ar@<-3pt>[rr]_{p_1}
	   \ar@<+3pt>[rr]^{p_1(-) \cdot \delta(p_2(-))}
	   &&
	   G_0
	 }
  $$  
  equipped with a strict group structure given by the semidirect
  product group structure on $G_0 \times G_1$ that is induced from the action $\rho$. The commutativity of the above two diagrams is precisely the 
  condition for this to be consistent.
Recall the examples of crossed modules, 
starting with example \ref{CrossedModuleOfShiftedAbelianGroup}.

We discuss sufficient conditions for the delooping
of a crossed module of
presheaves to be fibrant in the projective model structure. Recall also the conditions from prop. \ref{FibrancyOfWCrossedModule}.
\begin{proposition}
  Suppose that the smooth crossed module $(G_1 \to G_0)$ 
  is such that the quotient $\pi_0 G = G_0/G_1$ is a smooth manifold and the
  projection $G_0 \to G_0/G_1$ is a submersion.
  
  Then $\mathbf{B}(G_1 \to G_0)$ is fibrant in 
  $[\mathrm{CartSp}^{\mathrm{op}}, \mathrm{sSet}]_{\mathrm{proj}, \mathrm{loc}}$.
\end{proposition}
\proof
  We need to show that for $\{U_i \to \mathbb{R}^n\}$ a good open cover, the canonical 
  descent morphism
  $$
    B(C^\infty(\mathbb{R}^n, G_1) \to C^\infty(\mathbb{R}^n, G_0))
    \to 
    [\mathrm{CartSp}^{\mathrm{op}}, \mathrm{sSet}](C(\{U_i\}), \mathbf{B}(G_1 \to G_0))
  $$
  is a weak homotopy equivalence. 
  The main point to show is that, since the Kan complex on the left is connected by 
  construction, also the Kan complx on the right is. 

  To that end, notice that the category $\mathrm{CartSp}$ equipped with the open cover topology is a 
  \emph{Verdier site} in the sense of section 8 of \cite{dugger-hollander-isaksen}.
  By the discussion there it follows that every hypercover over $\mathbb{R}^n$ can be refined
  by a split hypercover, and these are cofibrant resolutions of $\mathbb{R}^n$ in both the 
  global and the local model structure
  $[\mathrm{CartSp}^{\mathrm{op}}, \mathrm{sSet}]_{\mathrm{proj}, \mathrm{loc}}$. 
  Since also $C(\{U_i\}) \to \mathbb{R}^n$ is a cofibrant resolution and  since
  $\mathbf{B}G$ is clearly fibrant in the \emph{global} structure, it follows from the
  existence of the global model structure that morphisms out of $C(\{U_i\})$ into 
  $\mathbf{B}(G_1 \to G_0)$ capture all cocycles over any hypercover over $\mathbb{R}^n$,
  hence  that 
  $$
    \pi_0 [\mathrm{CartSp}^{\mathrm{op}}, \mathrm{sSet}](C(\{U_i\}), \mathbf{B}(G_1 \to G_0))
    \simeq
    H^1_{\mathrm{smooth}}(\mathbb{R}^n, (G_1 \to G_0))
  $$
  is the standard {\v C}ech cohomology of $\mathbb{R}^n$, defined as a colimit over refinements of
  covers of equivalence classes of {\v C}ech cocycles.
  
  Now by prop. 4.1 of \cite{NikolausWaldorf} (which is the smooth refinement of the 
  statement of \cite{baezstevenson} in the continuous context)
  we have that under our assumptions on $(G_1 \to G_0)$ 
  there is a
  topological classifying space for this smooth {\v C}ech cohomology set. Since $\mathbb{R}^n$
  is topologically contractible, it follows that this is the
  singleton set and hence the above descent morphism is indeed an isomorphism on $\pi_0$.
  
  Next we can argue that it is also an isomorphism on $\pi_1$, by reducing to the analogous
  local trivialization statement for ordinary principal bundles: a loop in 
  $[\mathrm{CartSp}^{\mathrm{op}}, \mathrm{sSet}](C(\{U_i\}), \mathbf{B}(G_1 \to G_0))$
  on the trivial cocycle
  is readily seen to be a $G_0 /\!/ (G_0 \ltimes G_1)$-principal groupoid bundle, over the
  action groupoid as indicated. The underlying $G_0 \ltimes G_1$-principal bundle
  has a trivialization on the contractible $\mathbb{R}^n$ (by classical results or, in fact,
  as a special case of the previous argument), and so equivalence classes of such loops are 
  given gy $G_0$-valued smooth functions on $\mathbb{R}^n$. The descent morphism
  exhibits an isomorphism on these classes. 
  
  Finally the equivalence classes of spheres on both sides are directly seen to be 
  smooth $\mathrm{ker}(G_1 \to G_0)$-valued
  functions on both sides, identified by the descent morphism.
\endofproof
\begin{corollary}
  \index{principal $\infty$-bundle!gerbe}
  \index{gerbe!relation to principal $\infty$-bundle}
  For $X \in \mathrm{SmoothMfd} \subset \mathbf{H}$ a paracompact smooth manifold, and
  $(G_1 \to G_0)$ as above, we have for any good open cover $\{U_i \to X\}$ that
  the 2-groupoid of smooth $(G_1 \to G_0)$-principal 2-bundles is
  $$
    (G_1 \to G_0)\mathrm{Bund}(X) := \mathbf{H}(X, \mathbf{B}(G_1))
    \simeq 
    [\mathrm{CartSp}^{\mathrm{op}}, \mathrm{sSet}](C(\{U_i\}), \mathbf{B}(G_1 \to G_0))
  $$
  and its set of connected components is naturally isomorphic to the nonabelian {\v C}ech
  cohomology
  $$
    \pi_0 \mathbf{H}(X, \mathbf{B}(G_1 \to G_0)) \simeq H^1_{\mathrm{smooth}}(X, (G_1 \to G_0))
    \,.
  $$
  
  In particular, for $G = \mathrm{AUT}(H)$, $\mathbf{B}G \in \mathbf{H}$ is the 
  moduli 2-stack for smooth $H$-gerbes, def. \ref{1Gerbe}.
\end{corollary}

\begin{proposition}
  \label{GroupCentralExtensionCohomology}
  For $A \to \hat G \to G$ a central extension of Lie groups
  such that $\hat G \to G$ is a locally trivial $A$-bundle,
  we have a long fiber sequence in $\mathrm{Smooth}\infty \mathrm{Grpd}$
  of the form
  $$
    A 
	  \to 
    \hat G
      \to 
    G
     \to 
    \mathbf{B}A
     \to
    \mathbf{B} \hat G
	  \to 
    \mathbf{B}G 
	   \stackrel{\mathbf{c}}{\to}
	\mathbf{B}^2 A
	\,,
  $$
  where the morphism $\mathbf{c}$ is presented by the
  span of simplicial presheaves 
  $$
    \xymatrix{
	  \mathbf{B}(A \to \hat G)_c
	  \ar[r]
	  \ar[d]^{\simeq}
	  &
	  \mathbf{B}(A \to 1)_c
	  \ar@{=}[r]
	  &
	  \mathbf{B}^2 A_c
	  \\
	  \mathbf{B}G_{\mathrm{ch}}
	}
  $$
  coming from crossed complexes, def. \ref{CrossedComplex},
  as indicated.
\end{proposition}
\proof
  We need to show that
  $$
    \xymatrix{
	   \mathbf{B}\hat G_{\mathrm{ch}}
	   \ar[d]
	   \ar[r]
	   &
	   {*}
	   \ar[d]
	   \\
	   \mathbf{B}G_{\mathrm{ch}}
	   \ar[r]^{\mathbf{c}}
	   &
	   \mathbf{B}^2 A
	}
  $$
  is an $\infty$-pullback. To that end, we notice 
  that we have an equivalence
  $$
    \mathbf{B}(A \to \hat G)_c 
	   \stackrel{\simeq}{\to}
	\mathbf{B}G_{\mathrm{ch}}
  $$
  and
  that the morphism
  of simplicial presheaves
  $\mathbf{B}(A \stackrel{\mathrm{id}}{\to} A)_c \to \mathbf{B}^2 A_c$ is a fibration
  replacement of $* \to \mathbf{B}^2 A_c$, both in 
  $[\mathrm{CartSp}^{\mathrm{op}}, \mathrm{sSet}]_{\mathrm{proj}}$.  
  
  By prop. \ref{FiniteHomotopyLimitsInPresheaves} it is therefore
  sufficient to observe the ordinary pullback diagram
  $$
    \xymatrix{
	  \mathbf{B}(1 \to A)_c
	  \ar[d]
	  \ar[r]
	  &
	  \mathbf{B}(A \stackrel{\mathrm{id}}{\to} A)_c
	  \ar[d]
	  \\
	  \mathbf{B}(A \to \hat G)
	  \ar[r]
	  &
	  \mathbf{B}(A \to 1)_c
	}
	\,.
  $$
\endofproof

\subsubsection{Twisted cohomology and twisted bundles}
 \label{SmoothStrucTwistedCohomology}
 \index{structures in a cohesive $\infty$-topos!twisted cohomology!smooth}

We give an extensive discussion of twisted cohomology, \ref{StrucTwistedCohomology},
and the corresponding twisted principal $\infty$-bundles, realized in
$\mathrm{Smooth}\infty\mathrm{Grpd}$, below in \ref{TwistedStructures}. 
Most of the discussion there which does not involve differential refinement
also  goes through verbatim in $\mathrm{ETop}\infty \mathrm{Grpd}$, \ref{ContinuousInfGroupoids}.

Notably in \ref{Twisted0BundlesSectionsOfVectorBundles} we discuss as 
a simple consistency check that the general theory of twisted $\infty$-bundles
as sections of associated $\infty$-bundles reproduces the ordinary notion of
smooth sections of a vector bundle. Then in 
\ref{Twisted1BundlesTwistedKTheory} we discuss that twisted vector bundles
and hence twisted K-cocycles do arise as 2-sections of certain 
canonically associated 2-bundles to circle 2-bundles. This serves to 
show how the case of twisted cohomology that traditionally is at the focus
the attention is reproduced. After that we discuss in \ref{TwistedStructures}
a wealth of further examples.

\subsubsection{$\infty$-Group representations}
\label{SmoothStrucInfinityGroupRepresentations}
\index{structures in a cohesive $\infty$-topos!$\infty$-group representations!smooth}

We discuss the intrinsic notion of $\infty$-group representations, \ref{StrucRepresentations},
realized in the context $\mathrm{Smooth}\infty\mathrm{Grpd}$.

\medskip

We make precise the role of \emph{action Lie groupoids}, 
introduced informally in \ref{Principal1Bundles}. 
\begin{proposition}
  \label{ActionGroupoidForLieGroups}
  Let $X$ be a smooth manifold, and $G$ a Lie group. Then the 
  category of smooth $G$-actions on $X$ in the traditional sense
  is equivalent to the category of $G$-actions on $X$ in 
  the cohesive $\infty$-topos $\mathrm{Smooth}\infty\mathrm{Grpd}$, according to 
  def. \ref{RepresentationOfInfinityGroup}.
\end{proposition}
\proof
  For $\rho : X \times G \to X$ a given $G$-action, define the \emph{action Lie groupoid}
  $$
    X/\!/G := 
	(
	  \xymatrix{
	     X \times G
		 \ar@<+4pt>[r]^{\rho}
		 \ar@<-4pt>[r]_{p_1}
		 &
		 X
	  }
	)
  $$
  with the evident composition operation. This comes with the evident 
  morphism of Lie groupoids
  $$
    X /\!/ G \to * /\!/ G \simeq \mathbf{B}G
	\,,
  $$
  with $\mathbf{B}G$ as in prop. \ref{DeloopedLieGroup}. It is immediate
  that regarding this as a morphism in $[\mathrm{CartSp}^{\mathrm{op}}, \mathrm{sSet}]_{\mathrm{proj}}$ in the canonical way, this is a fibration.
  Therefore, by \ref{FiniteHomotopyLimitsInPresheaves}, 
  the homotopy fiber of this morphism in $\mathrm{Smooth}\infty\mathrm{Grpds}$ 
  is given by the ordinary fiber of this morphism in simplicial presheaves. This is 
  manifestly $X$. 
  
  Accordingly this construction constitutes an embedding of the traditional $G$ actions
  on $X$ into the category $\mathrm{Rep}_G(X)$ from def. \ref{RepresentationOfInfinityGroup}.
  By turning this argument around, one finds that this embedding is essentially surjective.  
\endofproof

\subsubsection{Associated bundles}
\label{SmoothStrucAssociatedBundles}
\index{structures in a cohesive $\infty$-topos!associated bundles!2-vector bundle}
\index{structures in a cohesive $\infty$-topos!associated bundles!2-line bundle}

We discuss aspects of the general notion of \emph{associated $\infty$-bundles}, 
\ref{AssociatedBundles}, realized in the context of smooth cohesion.

\medskip

We have been discussing the $n$-stacks $\mathbf{B}^n U(1)$ of 
\emph{circle $n$-bundles} in \ref{SmoothStrucDifferentialCohomology}, but without any substantial
change in the theory we could also use the $n$-stacks $\mathbf{B}^n \mathbb{C}^\times$
which are the $n$-fold delooping in $\mathbf{H}$ of the cohesive mutliplicative group of non-zero 
complex numbers. Under geometric realization 
${\vert -\vert} : \xymatrix{\mathbf{H} \ar[r] & \infty \mathrm{Grpd}}$
the canonical map $\mathbf{B}^n U(1) \to \mathbf{B}^n \mathbb{C}^\times$ becomes an equivalence. 
Nevertheless, some constructions are more naturally expressed in terms of $U(1)$-principal
$n$-bundles, while other are more naturally expressed in terms of $\mathbb{C}^\times$-principal 
$n$-bundles (bundle $(n-1)$-gerbes).  
Notably the latter is naturally identified with the 2-stack 
$2 \mathbf{Line}_{\mathbb{C}}$ of \emph{complex line 2-bundles}.

To interpret this, we say that for $R$ a ring (or more generally an $E_\infty$-ring), 
a \emph{2-vector space} over $R$ is, if it admits a 2\emph{2-basis}, a
category $A \mathrm{Mod}$ of modules over an $R$-algebra $A$ (the algebra $A$ is the 
given 2-basis), and that a \emph{2-linear map} between 2-vector space is a functor 
$A \mathrm{Mod} \to B \mathrm{Mod}$ which is induced by tensoring with a $B$-$A$-bimodule.
This identifies a 2-category $2\mathrm{Vect}_R$ of algebras, bimodules and 
bimodule homomorphisms which we call the 2-category of 2-vector spaces over $R$ 
(appendix A of \cite{Schreiber08}, section 4.4. of \cite{SWIII}, section 7 of \cite{FHLT}). 
This 2-category is naturally braided monoidal. 
Write then
$$
  \xymatrix{
    2 \mathrm{Line}_R
    \ar@{^{(}->}[r]
    &
    2 \mathrm{Vect}_R	
  }
$$
for the full sub-2-category on those objects which are invertible under this tensor product:
the \emph{2-lines} over $R$. This is necessarily a 2-groupoid, the \emph{Picard 2-groupoid}
over $R$, and with the inherited monoidal structure it is a 3-group, 
the \emph{Picard 3-group} of $R$. Its homotopy groups have
a familiar algebraic interpretation:
\begin{itemize}
  \item $\pi_0(2 \mathrm{Line}_R)$ is the \emph{Brauer group} of $R$;
  \item $\pi_1(2 \mathrm{Line}_R)$ is the ordinary \emph{Picard group} of $R$ (of ordinary $R$-lines);
  \item $\pi_2(2\mathrm{Line}_R) \simeq R^\times $ is the \emph{group of units}.
\end{itemize}
If we take the base ring $R$ to be the ring of suitable $k$-valued functions on some space $X$, then 
$2 \mathrm{Vect}_R$ is the 2-category of $k$-2-vector spaces over that vary over $X$, hence of 
complex \emph{2-vector bundles}. This construction is natural in $R$, hence in $X$, and it restricts to
2-lines and hence to \emph{2-line bundles} over $k$. Hence there is a 2-stack
$2\mathbf{Line}_{k} \in \mathbf{H}$ of 2-line bundles over $k$.
If $k$ here is algebraically closed, such 
as $k = \mathbb{C}$, then there is, up to equivalence, only a single 2-line, and only a single
invertible bimodule, and hence we find that
$2 \mathbf{Line}_k \simeq \mathbf{B}^2 k^\times$
In particular we have an equivalence
$$
  2\mathbf{Line}_{\mathbb{C}} \simeq \mathbf{B}^2 \mathbb{C}^\times
  \,.
$$

\medskip

Therefore the 2-stack $2 \mathbf{Line}_{\mathbb{C}}$ is of interest in particular in situations where
this equivalence no longer holds. This is notably so in the context of supergeometric cohesion;
this is discussed below in \ref{SupergeomStrucAssociatedBundles}.

\subsubsection{Manifolds} 
\label{SmoothStrucmanifolds}
\index{structures in a cohesive $\infty$-topos!manifolds!smooth manifolds}

We discuss the realization of the general abstract notion of manifolds in a
cohesive $\infty$-topos in \ref{Strucmanifolds} realized in smooth cohesion.

\medskip

With $\mathbb{A} := \mathbb{R} \in \mathrm{SmthMfd} \hookrightarrow \mathrm{Smooth}\infty\mathrm{Grpd}$
the standard line object exhibiting the cohesion of $\mathrm{Smooth}\infty\mathrm{Grpd}$
according to prop. \ref{ReallLineExhibitsEuclideanTopologicalCohesion}, def. \ref{IntrinsicManifold}
is equivalent to the traditional definition of smooth manifolds.

\subsubsection{Flat connections and local systems} 
 \label{SmoothStrucFlatDifferential}
 \index{structures in a cohesive $\infty$-topos!flat $\infty$-connections!smooth}

We discuss the intrinsic notion of flat $\infty$-connections, \ref{StrucFlatDifferential}, 
realized in
$\mathrm{Smooth} \infty \mathrm{Grpd}$.

\begin{proposition}
  \label{FlatSmoothCohomology}
  Let $X, A\in \mathrm{Smooth} \infty \mathrm{Grpd}$ be any two objects
  and write $|X| \in \mathrm{Top}$ for the intrinsic geometric realization,
  def. \ref{GeometricRealization}. We have that the flat cohomolog in 
  $\mathrm{Smooth} \infty \mathrm{Grpd}$ of $X$ with coefficients 
  in $A$ is equivalent to the ordinary cohomology in $\mathrm{Top}$ of $|X|$
  with coefficients in underlying discrete object of $A$:
  $$
    H_{\mathrm{Smooth},\mathrm{flat}}(X,A)
    \simeq
    H(|X|, |\Gamma A|)
    \,.
  $$
\end{proposition}
\proof
  By definition we have
  $$
    H_{\mathrm{flat}}(X,A) \simeq H(\mathbf{\Pi}X, A) \simeq H(\mathrm{Disc} \Pi X, A)
    \,.
  $$
  Using the $(\mathrm{Disc}) \dashv \Gamma$-adjunction this is
  $$
    \cdots \pi_0 \infty \mathrm{Grpd}(\Pi X, \Gamma A)
    \,.
  $$
  Finally applying the equivalence $\vert \cdot \vert : \infty \mathrm{Grpd} \to \mathrm{Top}$
  this is
  $$
    \cdots \simeq H(\vert \Pi X\vert, \vert \Gamma A \vert)
    \,.
  $$
  The claim hence follows 
  as in prop. \ref{SmoothCohomologyWithConstantCoefficients}.
\endofproof

Let $G$ be a Lie group regarded as a 0-truncated $\infty$-group
in $\mathrm{Smooth} \infty \mathrm{Grpd}$. Write $\mathfrak{g}$ for its Lie algebra.
Write $\mathbf{B}G \in \mathrm{Smooth} \infty \mathrm{Grpd}$ for its delooping.
Recall the fibrant presentation 
$\mathbf{B}G_{\mathrm{ch}} \in [\mathrm{CartSp}_{\mathrm{smooth}}^{\mathrm{op}}, \mathrm{sSet}]_{\mathrm{proj},\mathrm{loc}}$ 
from prop. \ref{DeloopedLieGroup}.
\begin{proposition} 
  \label{LieGroupFibrantFlatInclusion}
The object 
$\mathbf{\flat}\mathbf{B}G \in \mathrm{Smooth} \infty \mathrm{Grpd}$ has a fibrant 
presentation 
$\mathbf{\flat} \mathbf{B}G_{\mathrm{ch}} \in [\mathrm{CartSp}^{\mathrm{op}}, \mathrm{sSet}]_{\mathrm{proj},\mathrm{loc}}$ 
given by the groupoid of Lie-algebra valued forms
$$
  \mathbf{\flat}\mathbf{B}G _{\mathrm{ch}}
  =
  N
  \left(
    \xymatrix{
      C^\infty(-,G)\times \Omega^1_{\mathrm{flat}}(-,\mathfrak{g}) 
     \ar@<+4pt>[rr]^<<<<<<<<<{\mathrm{Ad}_{p_1}(p_2)+ p_1^{-1} d p_1}
     \ar@<-4pt>[rr]_{p2}
     &&
    \Omega^1_{\mathrm{flat}}(-,\mathfrak{g})
     }
  \right)
$$
and this is such that the canonical morphism 
$\mathbf{\flat} \mathbf{B}G \to \mathbf{B}G$ is presented by the canonical 
morphism of simplicial presheaves 
$\mathbf{\flat}\mathbf{B}G_{\mathrm{ch}} \to \mathbf{B}G_{\mathrm{ch}}$ which is a fibration
in $[\mathrm{CartSp}_{\mathrm{smooth}}^{\mathrm{op}}, \mathrm{sSet}]_{\mathrm{proj}}$.
\end{proposition}
\begin{remark}
This means that a $U$-parameterized family of objects of 
$\mathbf{\flat}\mathbf{B}G_{\mathrm{ch}}$ is given by a 
Lie-algebra valued 1-form 
$A \in \Omega^1(U)\otimes \mathfrak{g}$ whose curvature 2-form 
$F_A = d_{\mathrm{dR}} A + [A ,\wedge A]  = 0$ vanishes,  
and a $U$-parameterized family of morphisms $g : A \to A'$ is given by a 
smooth function $g \in C^\infty(U,G)$ such that $A' = \mathrm{Ad}_g A + g^{-1} d g$, 
where $\mathrm{Ad}_g A = g^{-1} A g$ is the adjoint action of $G$ on its 
Lie algebra, and where $g^{-1} d g := g^* \theta$ is the pullback of the 
Maurer-Cartan form on $G$ along $g$.
\end{remark}
\proof
By the proof of prop. \ref{InfSheavesOverCohesiveSiteAreCohesive} we have that 
$\mathbf{\flat} \mathbf{B}G$ is presented by the simplicial presheaf
that is constant on the nerve of the one-object groupoid
$$
  \xymatrix{
    G_{\mathrm{disc}} 
	\ar@<+3pt>[r]
	\ar@<-3pt>[r]
	&
	{*}
  }
  \,,
$$
for the discrete group underlying the Lie group $G$.
The canonical morphism of that into $\mathbf{B}G_{\mathrm{ch}}$ is however 
not a fibration. We claim that the canonical inclusion
$N(\xymatrix{G_{\mathrm{disc}} \ar@<+3pt>[r] \ar@<-3pt>[r] & {*} }) \to \mathbf{\flat} \mathbf{B}G_{c}$
factors the inclusion into $\mathbf{B}G_{\mathrm{ch}}$ by a weak equivalence followed
by a global fibration.

To see the weak equivalence, notice that it is objectwise an equivalence of groupoids: 
it is essentially surjective since every flat $\mathfrak{g}$-valued 1-form on the 
contractible $\mathbb{R}^n$ is of the form $g d g^{-1}$ for some function 
$g : \mathbb{R}^n \to G$ (let $g(x) = P \exp(\int_{0}^x) A$ be the parallel transport of $A$ 
along any path from the origin to $x$). Since the gauge transformation automorphism 
of the trivial $\mathfrak{g}$-valued 1-form are precisely given by the constant $G$-valued functions, 
this is also objectwise a full and faithful functor. 
Similarly one sees that the map $\mathbf{\flat}\mathbf{B}G_{\mathrm{ch}} \to \mathbf{B}G$ is a fibration.

Finally we need to show that $\mathbf{\flat}\mathbf{B}G_{\mathrm{ch}}$ is fibrant in 
$[\mathrm{CartSp}_{\mathrm{smooth}}^{\mathrm{op}}, \mathrm{sSet}]_{\mathrm{proj},\mathrm{loc}}$. This is implied by theorem \ref{SufficientConditionsForFibrancyOverCohesives}.
More explicitly, this can be seen by observing that this sheaf is the coefficient object that in 
{\v C}ech cohomology computes $G$-principal bundles with flat connection and then reasoning as above: 
every $G$-principal bundle with flat connection 
on a Cartesian space is equivalent to a trivial $G$-principal bundle whose connection is 
given by a globally defined $\mathfrak{g}$-valued 1-form. Morphisms between these are precisely 
$G$-valued functions that act on the 1-forms by gauge transformations as in the 
groupoid of Lie-algebra valued forms.
\endofproof

Let now $\mathbf{B}^n U(1)$ be the circle $(n+1)$-Lie group, def. \ref{Circlengroup}.
Recall the notation and  model category presentations as discussed there.
\begin{proposition} 
 \label{FlatCohomologyWithCoeffsInCircle}
For $n \geq 1$ a fibration presentation in 
$[\mathrm{CartSp}^{\mathrm{op}}, \mathrm{sSet}]_{\mathrm{proj}}$ 
of the  canonical morphism $\mathbf{\flat} \mathbf{B}^n U(1) \to \mathbf{B}^n U(1)$
in $\mathrm{Smooth} \infty \mathrm{Grpd}$ 
is given by the image under $\Xi : [\mathrm{CartSp}^{\mathrm{op}}, \mathrm{Ch}^+] 
  \to [\mathrm{CartSp}^{\mathrm{op}}, \mathrm{sSet}]$ of the morphism of chain complexes
$$
  \xymatrix{
    C^\infty(-,U(1)) \ar[r]^{d_{\mathrm{\mathrm{dR}}}} 
      \ar[d]
      & 
      \Omega^1(-) \ar[r]^{d_{\mathrm{\mathrm{dR}}}} 
      \ar[d]
      & 
      \cdots \ar[r]^{d_{\mathrm{\mathrm{dR}}}} 
      & \Omega^n_{\mathrm{cl}}(-)
      \ar[d]
    \\
    C^\infty(-, U(1)) \ar[r]
    & 
    0 \ar[r]
    &
    \cdots \ar[r]
    &
    0    
  }
  \,,
$$
where at the top we have the \emph{flat} Deligne complex. 
\end{proposition}
\proof
It is clear that the morphism of chain complexes is an objectwise 
surjection and hence maps to a projective fibration under $\Xi$.
It remains to observe that the flat Deligne complex is a presentation
of $\mathbf{\flat} \mathbf{B}^n U(1)$:

By the proof of prop. \ref{InfSheavesOverCohesiveSiteAreCohesive} we have that
$\mathbf{\flat} = \mathrm{Disc}  \circ \Gamma$ is presented in the model category on fibrant objects by first 
evaluating on the point  and then extending back to a constant simplicial presheaf. 
Since $\Xi U(1)[n]$ is indeed globally fibrant, a fibrant 
presentation of $\mathbf{\flat} \mathbf{B}^n U(1)$ is given by the 
\emph{constant} presheaf $U(1)_{\mathrm{const}} [n] : U \mapsto \Xi(U(1)[n])$.

The inclusion $U(1)_{\mathrm{const}}[n] \to U(1)[n]$ is not yet a fibration. 
But by a basic fact of abelian sheaf cohomology -- 
using the Poincar{\'e} lemma -- we have a global weak equivalence 
$U(1)_{\mathrm{const}}[n] \stackrel{\simeq}{\to} 
[C^\infty(-,U(1)) \stackrel{d_{\mathrm{\mathrm{dR}}}}{\to} \cdots \stackrel{d_{\mathrm{\mathrm{dR}}}}{\to} \Omega^n_{\mathrm{cl}}(-)]$ that factors this inclusion by the above fibration.
This completes the proof.

\medskip
For emphasis, we repeat this argument in more detail. 
The factorization of $U(1)_{\mathrm{const}}[n] \to U(1)[n]$
into a weak equivalence followed by a fibration
that we are looking at is over each object $\mathbb{R}^q \in \mathrm{CartSp}$
in the site given by the morphisms of chain complexes
whose components are show on the following diagram.
$$
  \xymatrix{
    U(1) \ar[r] \ar@{^{(}->}[d] & 0 \ar[d] \ar[r] & 0 \ar[d] \ar[r] & \cdots \ar[r] & 0 \ar[d]
	\\
	C^\infty(\mathbb{R}^q, U(1))
	\ar[r]^<<<<{d_{\mathrm{dR}} \mathrm{log}}
	\ar[d]^{\mathrm{id}}
	& 
	\Omega^1(\mathbb{R}^q)
	\ar[r]^{d_{\mathrm{dR}}}
	\ar[d]
	&
	\Omega^2(\mathbb{R}^q)
	\ar[r]^{d_{\mathrm{dR}}}
	\ar[d]
	&
	\cdots
	\ar[r]^{d_{\mathrm{dR}}}
	&
	\Omega^n_{\mathrm{cl}}(\mathbb{R}^q)
	\ar[d]
	\\
	C^\infty(\mathbb{R}^q, U(1))
	\ar[r]
	&
	0
	\ar[r]
	&
	0
	\ar[r]
	&
	\cdots
	\ar[r]
	&
	0
  }
  \,.
$$
It is clear that this commutes. It is also clear that the lower vertical morphisms
are all surjections, so the lower row exhibits a fibration of chain complexes.
In order for the top row to exhibit a weak equivalence 
of chain complexes -- a quasi-isomorphism -- we need it to induce an isomorphism
on all chain homology groups. 

The chain homology of the top complex is evidently concentrated in degree $n$,
where it is $U(1)$, as a discrete group. 

The chain homology of the middle complex in degree $n$ is the kernel of 
the differential $d_{\mathrm{dR}} \mathrm{log} : C^\infty(\mathbb{R}^q, U(1)) 
\to \Omega^1(\mathbb{R}^q)$. This kernel manifestly consists of the constant $U(1)$-valued functions.
Since $\mathbb{R}^q$ is connected, these are naturally identified with the 
group $U(1)$ itself. This identification is indeed what the top left vertical
morphism exhibits.

The chain homology of the middle complex in degree $0 \leq k < n$ is the de Rham cohomology
$H_{\mathrm{dR}}^{n-k}(\mathbb{R}^q)$. But this vanishes, since $\mathbb{R}^q$ is smoothly
contractible (the Poincar{\'e} lemma). 

Therefore the homology groups of the top
and of the middle chain complex coincide. And by this discussion, the top vertical
morphisms induce isomorphisms on these homology groups.
\endofproof

We discuss presentations of $\flat \mathbf{B}G$ for $G$ more generally the Lie integration
of an $L_\infty$-algebra $\mathfrak{g}$ further below in \ref{FLatCoefficientsForExponentiatedLInfinityAlgebras}.

\subsubsection{de Rham cohomology}
 \label{SmoothStrucdeRham}
 \index{structures in a cohesive $\infty$-topos!de Rham cohomology!smooth}

We discuss intrinsic notion of de Rham cohomology in a cohesive $\infty$-topos, 
\ref{StrucDeRham}, realized in the context $\mathrm{Smooth} \infty \mathrm{Grpd}$.
Here it reproduces the traditional notion of de Rham cohomology 
with abelian and nonabelian group coefficients, as well as its
equivariant and simplicial refinements.

\medskip

Let $G$ be a Lie group. Write $\mathfrak{g}$ for its 
Lie algebra.
\begin{proposition} 
  \label{LieGroupDeRhamCoefficients}
The object $\mathbf{\flat}_{\mathrm{\mathrm{dR}}}\mathbf{B}G \in \mathrm{Smooth} \infty \mathrm{Grpd}$
has a fibrant presentation in $[\mathrm{CartSp}_{\mathrm{smooth}}^{\mathrm{op}}, \mathrm{sSet}]_{\mathrm{proj},\mathrm{loc}}$ 
by the sheaf 
$\mathbf{\flat}\mathbf{B}G_{\mathrm{ch}} := \Omega^1_{\mathrm{flat}}(-, \mathfrak{g})$
of  flat Lie algebra-valued forms
$$
  \mathbf{\flat}\mathbf{B}G_{\mathrm{ch}}
   :
  U \mapsto \Omega^1_{\mathrm{flat}}(U,\mathfrak{g})
  \,.
$$
\end{proposition}
\proof
By prop. \ref{LieGroupFibrantFlatInclusion} we have a
fibration $\mathbf{\flat}\mathbf{B}G_{\mathrm{ch}} \to \mathbf{B}G_{\mathrm{ch}}$
in $[\mathrm{CartSp}_{\mathrm{smooth}}^{\mathrm{op}}, \mathrm{sSet}]_{\mathrm{proj}}$ 
given by the morphism of sheaves of groupoids
$$
  \raisebox{20pt}{
  \xymatrix{
    C^\infty(-,G) \ar[r]^{(-)^* \theta} \ar[d]^{\mathrm{id}} 
	 & 
	 \Omega^1_{\mathrm{flat}}(-,\mathfrak{g})
	 \ar[d]
	\\
	C^\infty(-,G) \ar[r] & 0
  }
  }\,,
$$
which models the canonical 
inclusion $\mathbf{\flat}\mathbf{B}G \to \mathbf{B}G$. Therefore
by prop. \ref{ConstructionOfHomotopyLimits} we obtain a presentation for the defining $\infty$-pullback
$$
  \mathbf{\flat}_{\mathrm{\mathrm{dR}}}\mathbf{B}G  := * \times_{\mathbf{B}G} \mathbf{\flat} \mathbf{B}G
$$
in $\mathrm{Smooth} \infty \mathrm{Grpd}$ by the ordinary pullback
$$
  \mathbf{\flat}_{\mathrm{\mathrm{dR}}}\mathbf{B}G_{\mathrm{ch}} 
  \simeq * \times_{\mathbf{B}G_{\mathrm{ch}}} \mathbf{\flat} \mathbf{B}G_{\mathrm{ch}}
$$
in $[\mathrm{CartSp}^{\mathrm{op}}, \mathrm{sSet}]_{\mathrm{proj}}$.
This is manifestly equal to $\Omega^1_{\mathrm{flat}}(-,\mathfrak{g})$.
This is fibrant in 
$[\mathrm{CartSp}^{\mathrm{op}}, \mathrm{sSet}]_{\mathrm{proj},\mathrm{loc}}$ because it is a sheaf.
\endofproof
\begin{remark}
  Another equivalent way to compute the homotopy fiber in 
  prop. \ref{LieGroupDeRhamCoefficients} is to produce the
  fibration resolution specifically by the factorization lemma, prop. \ref{FactorizationLemma}. 
  This yields for the de Rham coefficients of the Lie group $G$ the presentation
  $$
    \flat_{\mathrm{dR}}\mathbf{B}G
	\simeq
	G/(G_{\mathrm{disc}})
	\,,
  $$
  where on the right we have the quotient (of sheaves, hence in $\mathrm{Smooth}\infty\mathrm{Grpd}$)
  of the Lie group $G$ (the sheaf $C^\infty(-,G)$) by the underlying \emph{geometrically discrete}
  group (the sheaf constant on the underyling set of $G$). In other words, over a 
  $U \in \mathrm{CartSp}$ the value of $G/(G_{\mathrm{disc}})$ is the set of equivalence
  classes of smooth functions $g : U \to G$, where two are regarded as equivalent if they
  differ by multiplication with a \emph{constant} such function.
  
  By the general theory this sheaf must be equivalent, hence isomorphic, to the one of 
  prop. \ref{LieGroupDeRhamCoefficients}. Indeed, $G_{\mathrm{disc}}$ is the kernel of the
  map $(-)^* \theta : \xymatrix{C^\infty(-,G) \ar[r] & \Omega^1_{\mathrm{flat}}}(-,\mathfrak{g})$
  which sends $g : U \to G$ to the pullback of the Maurer-Cartan form along $g$, often writtenm
  $g^{-1} d_{\mathrm{dR}} g^{}$. Moreover this map is surjective, since for 
  $A \in \Omega^1_{\mathrm{flat}}(U,\mathfrak{g})$ any flat $\mathfrak{g}$-valued form
  the function $P \exp(\int_{x_0}^{(-)} A) : U \to G$ that sends a point $x \in U$
  to the parallel transport of $A$ along any path from any fixed basepoint $x_0 \in U$
  is a preimage. Hence we have the image factorization
  $$
    (-)^* \theta :
	\xymatrix{
	  G
	  \ar@{->>}[r]
	  &
	  G/(G_{\mathrm{disc}})
	  \ar[r]^-{\simeq}
	  &
	  \Omega^1_{\mathrm{flat}}(-,\mathfrak{g})
	}
	\,.
  $$  
  In words this says that a flat differential Lie-algebra valued form on a Cartesian space
  $\mathbb{R}^k$ is equivalently a smooth function from that space to $G$ 
  ``without remembering the origin of this function''. What is noteworthy about this is 
  that this second, equivalent, description, no longer refers to \emph{differentials}.
  \label{DeRhamCoefficientsOf0TruncatedGroupAsQuotientOfGroupByDiscreteGroup}
\end{remark}
Indeed, this second description of the de Rham coefficient object of a group object
is valid for any site, in particular for instance for the Euclidean-topological cohesion 
of \ref{ContinuousInfGroupoids}.

\medskip
  
For $n \in \mathbb{N}$, let now $\mathbf{B}^n U(1)$ be the circle Lie $(n+1)$-group of def. \ref{Circlengroup}. 
Recall the notation and model category presentations from the discussion there.
\begin{proposition} 
  \label{OrdinaryDeRham}
  \index{cohomology!de Rham (ordinary)}
A fibrant representative in $[\mathrm{CartSp}^{\mathrm{op}}, \mathrm{sSet}]_{\mathrm{proj},\mathrm{loc}}$ of 
the de Rham coefficent object $\mathbf{\flat}_{\mathrm{\mathrm{dR}}} \mathbf{B}^n U(1)$ from 
def. \ref{deRhamCoefficientObject} is given by the truncated ordinary de Rham complex of smooth 
differential forms
$$
  \mathbf{\flat}_{\mathrm{\mathrm{dR}}}\mathbf{B}^n U(1)_{\mathrm{chn}}
    :=
  \Xi[\Omega^1(-) \stackrel{d_{\mathrm{\mathrm{dR}}}}{\to} \Omega^2(-) \stackrel{d_{\mathrm{\mathrm{dR}}}}{\to}\cdots \to \Omega^{n-1}(-) \stackrel{d_{\mathrm{\mathrm{dR}}}}{\to}\Omega^n_{\mathrm{\mathrm{cl}}}(-)]
  \,.
$$
\end{proposition}
\proof
By definition and using prop. \ref{FiniteHomotopyLimitsInPresheaves}
the object $\mathbf{\flat}_{\mathrm{dR}}\mathbf{B}^n U(1)$
is given by the homotopy pullback in $[\mathrm{CartSp}^{\mathrm{op}}, Ch_{\bullet \geq 0}]_{\mathrm{proj}}$
of the inclusion $U(1)_{\mathrm{const}}[n] \to U(1)[n]$ along the point inclusion
$* \to U(1)[n]$. We may compute this as the ordinary pullback
after passing to a resolution of this inclusion by a fibration.
By prop. \ref{FlatCohomologyWithCoeffsInCircle} such a 
fibration replacement is given by the map from the flat Deligne
complex. Using this we find the ordinary pullback diagram
$$
  \xymatrix{
    \Xi[0 \to \Omega^1(-) \to \cdots \to \Omega^n_{\mathrm{\mathrm{cl}}}(-)]
    \ar[r] \ar[d] &
    \Xi[C^\infty(-,U(1)) \to \Omega^1(-) \to \cdots \to \Omega^n_{\mathrm{\mathrm{cl}}}(-)]
    \ar[d]
    \\
    \Xi[0 \to 0 \to \cdots \to 0]
    \ar[r] &
    \Xi[C^\infty(-,U(1)) \to 0 \to \cdots \to 0]
  }
  \,.
$$
\endofproof
\begin{proposition} 
  \label{OrdinaryFromIntrinsicDeRham}
Let $X$ be a smooth manifold regarded under the embedding
$\mathrm{SmoothMfd} \hookrightarrow \mathrm{Smooth} \infty \mathrm{Grpd}$. 
Write $H^n_{\mathrm{\mathrm{dR}}}(X)$ for the ordinary de Rham cohomology of $X$.

For $n \in \mathbb{N}$ we have isomorphisms
$$
  \pi_0 \mathrm{Smooth} \infty \mathrm{Grpd}(X, \mathbf{\flat}_{\mathrm{\mathrm{dR}}} \mathbf{B}^n U(1))
  \simeq
  \left\{
    \begin{array}{ll}
      H^n_{\mathrm{\mathrm{dR}}}(X) &| n \geq 2
      \\
      \Omega^1_{\mathrm{\mathrm{cl}}}(X) &| n = 1
      \\
      0 & | n = 0
     \end{array}
   \right.
$$
\end{proposition}
\proof
Let $\{U_i \to X\}$ be a differentiably good open cover. 
The {\v C}ech nerve $C(\{U_i\}) \to X$ is a cofibrant resolution of $X$ in 
$[\mathrm{CartSp}^{\mathrm{op}}, \mathrm{sSet}]_{\mathrm{proj},\mathrm{loc}}$. 
Therefore we have for all $n \in \mathbb{N}$

$$
  \mathrm{Smooth}\infty \mathrm{Grpd}(X,\mathbf{\flat}_{\mathrm{\mathrm{dR}}} \mathbf{B}^n U(1))
  \simeq
  [\mathrm{CartSp}^{\mathrm{op}}, \mathrm{sSet}](C(\{U_i\}), \Xi[\Omega^1(-) \stackrel{d_{\mathrm{\mathrm{dR}}}}{\to} \cdots \to \Omega^n_{\mathrm{cl}}(-)])
  \,.
$$
The right hand is the $\infty$-groupoid of cocylces in the {\v C}ech hypercohomology of 
the truncated complex of sheaves of differential forms. A cocycle is given by a collection
$$
  (C_i, B_{i j}, A_{i j k}, \cdots  , Z_{i_1, \cdots, i_n})
$$
of differential forms, with $C_i \in \Omega^n_{\mathrm{\mathrm{cl}}}(U_i)$, 
$B_{i j} \in \Omega^{n-1}(U_i \cap U_j)$, etc. , 
such that this collection is annihilated by the total differential 
$D = d_{\mathrm{dR}} \pm \delta$, where $d_{\mathrm{\mathrm{dR}}}$ is the de Rham differential and 
$\delta$ the alternating sum of the pullbacks along the face maps of the {\v C}ech nerve.

It is a standard result of abelian sheaf cohomology that such cocycles represent classes in de Rham cohomology
of $n \geq 2$. For $n = 1$ and $n= 0$ our truncated de Rham complex degenerates to 
$\mathbf{\flat}_{\mathrm{\mathrm{dR}}}\mathbf{B}U(1)_{\mathrm{chn}} = \Xi[\Omega^1_{\mathrm{\mathrm{cl}}}(-)]$
and
$\mathbf{\flat}_{\mathrm{dR}}U(1)_{\mathrm{chn}} = \Xi[0]$, respectively, which obviously has the cohomology as 
claimed above.
\endofproof
\begin{remark}
Recall from the discussion in \ref{StrucDeRham} that
the failure of the intrinsic de Rham cohomology of $\mathrm{Smooth}\infty$ to coincide with traditional
de Rham cohomology in degree 0 and 1 is due to the fact that the intrinsic de Rham cohomology 
in degree $n$ is the home for curvature classes of circle $(n-1)$-bundles. For $n=1$ these curvatures
are not to be taken modulo exact forms. And for $n = 0$ they vanish.
\end{remark}

\begin{definition}
  For $n \in \mathbb{N}$, write 
  $\Omega^n_{\mathrm{cl}} \in \mathrm{Sh}(\mathrm{CartSp}) \hookrightarrow \mathrm{Smooth}\infty\mathrm{Grpd}$
  for the ordinary sheaf of smooth closed differential $n$-forms. By prop. \ref{OrdinaryDeRham}
  this has a canonical morphism
  $$
    \Omega^n_{\mathrm{cl}} \to \flat_{\mathrm{dR}}\mathbf{B}^n U(1)
  $$
  into the de Rham coefficient object for $\mathbf{B}^{n-1}U(1)$, given in the presentation of 
  the latter as a simplicial presheaf according to prop. \ref{OrdinaryDeRham} by the 
  inclusion of the simplicial presheaf that is simplicially constant on the degree-0 component.
  \label{SheavesOfDifferentialFormsAsDifferentialFormObjects}
\end{definition}
\begin{proposition}
  The morphisms of def. \ref{SheavesOfDifferentialFormsAsDifferentialFormObjects} 
  are differential form objects in the sense of  def. \ref{DifferentialFormObject}
  with respect to the standard line object $\mathbb{R}$.
  \label{SmoothClosedFormsAreDifferentialFormObject}
\end{proposition}
\proof
  By the discussion in \ref{SmoothStrucmanifolds} the $\mathbb{R}^1$-manifolds are 
  precisely the objects in the inclusion 
  $\mathrm{SmthMfd} \hookrightarrow \mathrm{Sh}_\infty(\mathrm{SmthMfd}) \simeq \mathrm{Smooth}\infty\mathrm{Grpd}$. This means by def. \ref{DifferentialFormObject} that we need to check that for each 
  smooth manifold $\Sigma$ the morphism
  $$
    [\Sigma, \Omega^n_{\mathrm{cl}}] \to [\Sigma, \flat_{\mathrm{dR}}\mathbf{B}^n U(1)]
  $$  
  is an effective epimorphism. By prop. \ref{EffectiveEpiDetectedOnoTruncation} this 
  is equivalent to the 0-truncation of the moprhism being an epimorphism in the 
  sheaf topos $\mathrm{Sh}(\mathrm{CartSp})$. By the characterization of internal
  homs in turn, for this it is sufficient that for each $U \in \mathrm{CartSp}$
  the function
  $\Omega^n_{\mathrm{cl}}(\Sigma \times U) \to \pi_0 \mathbf{H}(\Sigma \times U, \flat_{\mathrm{dR}}\mathbf{B}^n U(1))$
  is a surjection. This is the case by prop. \ref{OrdinaryFromIntrinsicDeRham}.
\endofproof

We discuss the equivariant version of 
smooth de Rham cohomology.
\begin{proposition}
 \label{EquivariantDeRhamCohomology}
 Let $X$ be a smooth manifold equipped with a smooth action by a Lie group $G$. 
 Write $X/\!/G$ for the corresponding action Lie groupoid, prop. \ref{ActionGroupoid}.
 Then for $n \geq 2$ we have an isomorphism
 $$
   \pi_0 \mathrm{Smooth}\infty\mathrm{Grpd}(X/\!/G, \flat_{\mathrm{dR}} \mathbf{B}^n \mathbb{R})
   \simeq
   H_{\mathrm{dR}, G}^n(X)
   \,,
 $$
 where on the right we have ordinary $G$-equivariant de Rham cohomology of $X$.
\end{proposition}

\subsubsection{Exponentiated $\infty$-Lie algebras}
 \label{SmoothStrucLieAlgebras}
 \index{structures in a cohesive $\infty$-topos!exponentiated $\infty$-Lie algebras!smooth}

We discuss the intrinsic notion of exponentiated $\infty$-Lie algebras, 
\ref{StrucLieAlgebras}, realized in $\mathrm{Smooth}\infty \mathrm{Grpd}$.

Recall the characterization of $L_\infty$-algebras, 
def. \ref{LInfinityAlgebra}, 
by dual dg-algebras,
prop. \ref{LInfinityAlgebraFromDGAlgebra} -- 
their \emph{Chevalley-Eilenberg algebras}--, and the characterization
of the category $L_\infty \mathrm{Alg}$ as the full subcategory
$$
  L_\infty \stackrel{\mathrm{CE}}{\hookrightarrow} \mathrm{dgAlg}^{\mathrm{op}}
  \,.
$$

We describe now a presentation of the exponentiation of an $L_\infty$ algebra to a smooth $\infty$-group. 
The following somewhat technical definition serves to control the 
smooth structure on  these exponentiated objects.
\begin{definition}
For $k \in \mathbb{N}$ regard the $k$-simplex $\Delta^k$ as a smooth manifold with corners in the 
standard way. We think of this embedded into the Cartesian space $\mathbb{R}^k$ in the standard way 
with maximal rotation symmetry about the center of the simplex, 
and equip $\Delta^k$ with the metric space structure induced this way.

A smooth differential form $\omega$ on $\Delta^k$ we say has \emph{sitting instants} along the boundary 
if, for every $(r < k)$-face $F$ of $\Delta^k$ there is an open neighbourhood $U_F$ of $F$ 
in $\Delta^k$ such that $\omega$ restricted to $U$ is constant in the directions perpendicular to the 
$r$-face on its value restricted to that face.

More generally, for any $U \in \mathrm{CartSp}$ a smooth differential form $\omega$ on $U \times\Delta^k$ 
is said to have sitting instants if there is $0 < \epsilon \in \mathbb{R}$ such that for all points 
$u : * \to U$ the pullback along  $(u, \mathrm{Id}) : \Delta^k \to U \times \Delta^k$ is a form with 
sitting instants on $\epsilon$-neighbourhoods of faces.

Smooth forms with sitting instants form a sub-dg-algebra of all smooth forms. 
We write $\Omega^\bullet_{\mathrm{\mathrm{si}}}(U \times \Delta^k)$ for this sub-dg-algebra.

We write $\Omega_{\mathrm{\mathrm{si}},\mathrm{\mathrm{vert}}}^\bullet(U \times \Delta^k)$ for the further sub-dg-algebra 
of vertical differential forms with respect to the projection $p : U \times \Delta^k \to U$, 
hence the coequalizer
$$
  \xymatrix{
   \Omega^{\bullet\geq 1}(U)
    \ar@<+3pt>[r]^{p^*}
    \ar@<-3pt>[r]_{0}
    &
    \Omega^\bullet_{\mathrm{\mathrm{si}}}(U \times \Delta^k)
    \ar[r]
    &
    \Omega^\bullet_{\mathrm{\mathrm{si}}, \mathrm{\mathrm{vert}}}(U \times \Delta^k)  
    }
  \,.
$$
\end{definition}
\begin{definition} 
  \label{ExponentiatedLInftyAlgbra}
  \index{Lie integration}
  \index{$L_\infty$-algebra!exponentiation/Lie integration}
For $\mathfrak{g} \in L_\infty$
write $\exp(\mathfrak{g}) \in [\mathrm{CartSp}_{\mathrm{smooth}}^{\mathrm{op}}, \mathrm{sSet}]$ 
for the simplicial presheaf defined over $U \in \mathrm{CartSp}$  and 
$n \in \mathbb{N}$ by
$$
  \exp(\mathfrak{g}) 
   :
  (U, [n]) \mapsto
  Hom_{\mathrm{dgAlg}}(\Omega_{\mathrm{\mathrm{si}},\mathrm{\mathrm{vert}}}^\bullet(U \times \Delta^n), \mathrm{CE}(\mathfrak{g}))
$$
with the evident structure maps given by pullback of differential forms.
\end{definition}
This definition of the $\infty$-groupoid associated to an $L_\infty$-algebra
realized in the smooth context appears in \cite{FSS} and in similar form in \cite{Roytenberg} 
as the evident generalization 
of the definition in Banach spaces in \cite{Henriques} and for discrete $\infty$-groupoids
in \cite{getzler}, which in turn goes back to \cite{hinich}.
\begin{proposition}
The objects $\exp(\mathfrak{g}) \in [\mathrm{CartSp}_{\mathrm{smooth}}^{\mathrm{op}}, \mathrm{sSet}]$ are 
\begin{enumerate}
\item connected;
\item Kan complexes over each $U \in \mathrm{CartSp}$.
\end{enumerate}
\end{proposition}
\proof
That $\exp(\mathfrak{g})_0 = {*}$ follows from degree-counting: 
$\Omega^\bullet_{\mathrm{\mathrm{si}},\mathrm{\mathrm{vert}}}(U \times \Delta^0) = C^\infty(U)$ 
is entirely in degree 0 and  $\mathrm{CE}(\mathfrak{g})$ is in degree 0 the ground field $\mathbb{R}$.

To see that $\exp(\mathfrak{g})$ has all horn-fillers over each $U \in \mathrm{CartSp}$ 
observe that the standard continuous horn retracts $f : \Delta^k \to \Lambda^k_i$ 
are smooth away from the preimages of the $(r < k)$-faces of $\Lambda[k]^i$.

For $\omega \in \Omega^\bullet_{\mathrm{\mathrm{si}},\mathrm{\mathrm{vert}}}(U \times \Lambda[k]^i)$ 
a differential form with sitting instants on $\epsilon$-neighbourhoods, 
let therefore $K \subset \partial \Delta^k$ be the set of points of distance $\leq \epsilon$ 
from any subface. Then we have a smooth function 
$$
  f : \Delta^k \setminus K \to \Lambda^k_i \setminus K
  \,.
$$
The pullback $f^* \omega \in \Omega^\bullet(\Delta^k \setminus K)$ may be extended constantly back to a form with sitting instants on all of $\Delta^k$. The resulting assignment
$$
  (\mathrm{CE}(\mathfrak{g}) \stackrel{A}{\to} \Omega^\bullet_{\mathrm{\mathrm{si}},\mathrm{\mathrm{vert}}}(U \times \Lambda^k_i))
  \mapsto
  (\mathrm{CE}(\mathfrak{g}) \stackrel{A}{\to} \Omega^\bullet_{\mathrm{\mathrm{si}},\mathrm{\mathrm{vert}}}(U \times \Lambda^k_i) \stackrel{f^*}{\to} \Omega^\bullet_{\mathrm{\mathrm{si}},\mathrm{\mathrm{vert}}}(U \times \Delta^n))
$$
provides fillers for all horns over all $U \in \mathrm{CartSp}$.
\endofproof
\begin{definition}
We say that the loop space object $\Omega \exp(\mathfrak{g})$ is the 
\emph{smooth $\infty$-group} exponentiating $\mathfrak{g}$.
\end{definition}
\begin{proposition}
The objects $\exp(\mathfrak{g}) \in \mathrm{Smooth} \infty \mathrm{Grpd}$ are
geometrically contractible:
$$
  \Pi \exp(\mathfrak{g}) \simeq *
	\,.
$$ 
\end{proposition}
\proof
Observe that every simplicial presheaf $X$ is the homotopy colimit over its component 
presheaves $X_n \in [\mathrm{CartSp}_{\mathrm{smooth}}^{\mathrm{op}}, \mathrm{Set}] 
\hookrightarrow [\mathrm{CartSp}_{\mathrm{smooth}}^{\mathrm{op}}, \mathrm{sSet}]$
$$
  X \simeq \mathbb{L} {\lim\limits_{\to}}_n X_n
  \,.
$$
(Use for instance the injective model structure for which $X_\bullet$ is 
cofibrant in the Reedy model structure $[\Delta^{op},[\mathrm{CartSp}_{\mathrm{smooth}}^{\mathrm{op}}, \mathrm{sSet}]_{\mathrm{inj},\mathrm{loc}}]_{\mathrm{Reedy}}$ ).
Therefore it is sufficient to show that in each degree $n$ the 
0-truncated object $\exp(\mathfrak{g})_{n}$ is geometrically contractible. 

To exhibit a geometric contraction, def. \ref{GeometricHomotopy}, 
choose for each $n \in \mathbb{N}$, a smooth retraction
$$
  \eta_n : \Delta^n \times [0,1] \to \Delta^n  
$$
of the $n$-simplex: a smooth map such that $\eta_n(-,1) = \mathrm{Id}$ and $\eta_n(-,0)$ factors through the point.
We claim that this induces a diagram of presheaves
$$
  \raisebox{30pt}{
  \xymatrix{
     \exp(\mathfrak{g})_n
     \ar[d]_{(\mathrm{id},1)} \ar[dr]^{\mathrm{id}}
     \\
     \exp(\mathfrak{g})_n \times [0,1] \ar[r]^{\eta_n^*}& \exp(\mathfrak{g})_n
     \\
     \exp(\mathfrak{g})_n \ar[r] \ar[u]_{(\mathrm{id},0)} & {*} \ar[u]
  }
  }
  \,,
$$
where over $U \in \mathrm{CartSp}$ the middle morphism is given by
$$
  \eta_n^*  
     : 
     (\alpha, f)
      \mapsto 
    (f,\eta_n)^* \alpha
  \,,
$$
where 
\begin{itemize}
\item $\alpha : \mathrm{CE}(\mathfrak{g}) \to \Omega^\bullet_{\mathrm{\mathrm{si}}, \mathrm{\mathrm{vert}}}(U \times \Delta^n)$ 
  is an element of the set $\exp(\mathfrak{g})_n(U)$, 
\item $f$ is an element of $[0,1](U)$;
\item $(f,\eta_n)$ is the composite morphism
  $$
    U \times \Delta^n  
       \stackrel{(\mathrm{id},f)\times \mathrm{id}}{\to}
    U \times [0,1] \times \Delta^n
      \stackrel{(\mathrm{id},\eta_n)}{\to}
    U \times \Delta^n
    \;
  $$
\item $(f,\eta)^* \alpha$ is the postcomposition of $\alpha$ with
  the image of $(f,\eta_n)$ under $\Omega^\bullet_{\mathrm{\mathrm{vert}}}(-)$.
\end{itemize}
Here the last item is well defined given the coequalizer definition of 
$\Omega^\bullet_{\mathrm{\mathrm{vert}}}$ 
because $(f,\eta_n)$ is a morphism of bundles over $U$
$$
  \xymatrix{
    U \times \Delta^n  
       \ar[r]^<<<<<{(\mathrm{id},f) \times \mathrm{id}} \ar[d] &
    U \times [0,1] \times \Delta^n
      \ar[r]^<<<<<{\mathrm{id} \times \eta_n} \ar[d]&
    U \times \Delta^n \ar[d]
    \\
    U \ar[r]^{\mathrm{id}} & U \ar[r]^{\mathrm{id}} & U
  }
  \,.
$$
Similarly, for $h : K \to U$ any morphism in $\mathrm{CartSp}_{\mathrm{smooth}}$ 
the naturality condition for a morphism of presheaves follows from the fact that the 
composites of bundle morphisms
$$
  \xymatrix{
    K \times \Delta^n
     \ar[r]^{h \times \mathrm{id}} \ar[d] &
    U \times \Delta^n  
       \ar[r]^<<<<<{(\mathrm{id},f)\times \mathrm{id}} \ar[d] &
    U \times [0,1] \times \Delta^n
      \ar[r]^<<<<<{(\mathrm{id},\eta_n)} \ar[d] 
     &
    U \times \Delta^n \ar[d]
    \\
    K \ar[r]^h & U \ar[r]^{\mathrm{id}} & U \ar[r]^{\mathrm{id}} & U
  }
$$
and
$$
  \xymatrix{
    K \times \Delta^n  
       \ar[r]^{((\mathrm{id},f \circ h) \times \mathrm{id} } \ar[d] &
    K \times [0,1] \times \Delta^n
      \ar[r]^{\mathrm{id} \times \eta_n} \ar[d] &
    K \times \Delta^n
      \ar[r]^{h \times \mathrm{id}}  \ar[d] &
    U \times \Delta^n \ar[d]
    \\
    K \ar[r]^{\mathrm{id}} & K \ar[r]^{\mathrm{id}}& K
     \ar[r]^{h}& 
   U
  }
$$
coincide.

Moreover, notice that the lower morphism in our diagram of presheaves indeed factors through 
the point as indicated, because for an $L_\infty$-algebra $\mathfrak{g}$ we have that 
the Chevalley-Eilenberg algebra $\mathrm{CE}(\mathfrak{g})$ is in degree 0 the ground 
field algebra algebra $\mathbb{R}$, so that there is a unique morphism 
$\mathrm{CE}(\mathfrak{g}) \to \Omega^\bullet_{\mathrm{\mathrm{vert}}}(U \times \Delta^0) \simeq 
C^\infty(U)$ in $\mathrm{dgAlg}$.

Finally, since $[0,1]$ is a contractible paracompact manifold, we have that 
$\Pi([0,1]) \simeq *$ 
by prop. \ref{FundGroupoidOfParacompact}. Therefore the above diagram of 
presheaves presents a geometric homotopy in $\mathrm{Smooth} \infty \mathrm{Grpd}$ 
from the identity map to a map that factors through the point.
It follows by prop \ref{PiPreservesGeometricHomotopy} that 
$\Pi(\exp(\mathfrak{g})_n) \simeq *$ for all $n \in \mathbb{N}$.
And since $\Pi$ preserves the homotopy colimit 
$\exp(\mathfrak{g}) \simeq \mathbb{L} {\lim\limits_{\longrightarrow}}_n \exp(\mathfrak{g})_n$ 
we have that $\Pi(\exp(\mathfrak{g})) \simeq *$, too.
\endofproof
We may think of $\exp(\mathfrak{g})$ as the smooth 
geometrically \emph{$\infty$-simply connected Lie integration}
of $\mathfrak{g}$. Notice however that $\exp(\mathfrak{g}) \in \mathrm{Smooth} \infty \mathrm{Grpd}$
in general has nontrivial and interesting homotopy sheaves.
The above statement says that its 
\emph{geometric homotopy groups} vanish .

\medskip

\paragraph{Examples}

Let $\mathfrak{g} \in L_\infty$ be an ordinary (finite dimensional) Lie algebra. 
Standard Lie theory provides a simply connected Lie group $G$ integrating $\mathfrak{g}$.
Write $\mathbf{B}G \in \mathrm{Smooth}\infty \mathrm{Grpd}$ for its delooping. 
According to prop. \ref{DeloopedLieGroup} this is presented by the 
simplicial presheaf $\mathbf{B}G_{\mathrm{ch}} \in [\mathrm{CartSp}_{\mathrm{smooth}}^{\mathrm{op}}, \mathrm{sSet}]$.
\begin{proposition} \label{IntegrationToSimplyConnectedLieGroup}
The operation of parallel transport $P \exp(\int -) : \Omega^1([0,1], \mathfrak{g}) \to G$ 
yields a weak equivalence (in $[\mathrm{CartSp}_{\mathrm{smooth}}^{\mathrm{op}}, \mathrm{sSet}]_{\mathrm{proj}}$)
$$
  P \exp(\int - )
  :
  \mathbf{cosk}_3 \exp(\mathfrak{g}) 
  \simeq 
   \mathbf{cosk}_2 \exp(\mathfrak{g}) \simeq \mathbf{B}G_{\mathrm{ch}}
  \,.
$$
\end{proposition}
\proof
  Notice that a flat smooth $\mathfrak{g}$-valued 1-form on a contractible space $X$ is after a choice of
  basepoint canonically identified with a smooth function $X \to G$. The claim then follows from the
  observation that by the fact that $G$ is simply connected any two paths with coinciding endpoints 
  have a continuous homotopy between them, and that for smooth paths this may be chose to be smooth,
  by the Steenrod approximation theorem \cite{Wockel}.
\endofproof
Let now $n \in \mathbb{N}$, $n \geq 1$.
\begin{definition} 
 \label{LineLieNAlgebra}
Write 
$$
  b^{n-1} \mathbb{R} \in L_\infty
$$ 
for the $L_\infty$-algebra whose Chevalley-Eilenberg algebra is given by a single generator 
in degree $n$ and vanishing differential. 
We call this the \emph{line Lie $n$-algebra}\index{$L_\infty$-algebra!line Lie $n$-algebra}.
\end{definition}
\begin{observation}
The discrete $\infty$-groupoid underlying $\exp(b^{n-1} \mathbb{R})$ is given by the 
Kan complex that in degree $k$ has the set of closed differential $n$-forms with sitting instants 
on the $k$-simplex
$$
  \Gamma(\exp(b^{n-1} \mathbb{R}))
  : 
  [k] \mapsto \Omega^n_{\mathrm{\mathrm{si}},\mathrm{\mathrm{cl}}}(\Delta^k)
$$
\end{observation}
\begin{definition}
 We write equivalently
$$
  \mathbf{B}^n \mathbb{R}_{\mathrm{smp}} := \exp(b^{n-1}\mathbb{R})
  \in
  [\mathrm{CartSp}_{\mathrm{smooth}}^{\mathrm{op}}, \mathrm{sSet}]
  \,.
$$
\end{definition}
\begin{proposition} \label{LieIntegrationToLineNGroup}
We have that $\mathbf{B}^n \mathbb{R}_{\mathrm{smp}}$ is indeed a presentation
of the smooth line $n$-group $\mathbf{B}^{n} \mathbb{R}$, from \ref{Circlengroup}.

Concretely, with 
$\mathbf{B}^n \mathbb{R}_{\mathrm{chn}} \in [\mathrm{CartSp}_{\mathrm{smooth}}^{\mathrm{op}}, \mathrm{sSet}]$ 
the standard presentation given under the Dold-Kan correspondence by the chain complex of sheaves concentrated in degree $n$ on $C^\infty(-, \mathbb{R})$ the equivalence is induced by the fiber 
integration of differential $n$-forms over the $n$-simplex:
$$
  \int_{\Delta^\bullet} 
   : 
  \mathbf{B}^n \mathbb{R}_{\mathrm{smp}}
   \stackrel{\simeq}{\to}
  \mathbf{B}^{n} \mathbb{R}_{\mathrm{smp}}
  \,.
$$
\end{proposition}
\proof
  First we observe that the map
$$
  \int_{\Delta^\bullet}
  :
  (\omega \in \Omega^n_{\mathrm{\mathrm{si}},\mathrm{\mathrm{vert}},\mathrm{\mathrm{cl}}}(U \times \Delta^k))
  \mapsto
  \int_{\Delta^k} \omega \in C^\infty(U, \mathbb{R})
$$ 
is indeed a morphism of simplicial presheaves $\exp(b^{n-1} \mathbb{R}) \to \mathbf{B}^{n}\mathbb{R}_{\mathrm{chn}}$
on. Since it goes between presheaves of abelian simplicial groups, by the Dold-Kan correspondence it is 
sufficient to check that we have a morphism of chain complexes of presheaves on the corresponding 
normalized chain complexes.

The only nontrivial degree to check is degree $n$. 
Let $\lambda \in \Omega_{\mathrm{\mathrm{si}},\mathrm{\mathrm{vert}},\mathrm{\mathrm{cl}}}^n(\Delta^{n+1})$. 
The differential of the normalized chains complex sends this to the signed sum of its restrictions 
to the $n$-faces of the $(n+1)$-simplex. Followed by the integral over $\Delta^n$ this is the 
piecewise integral of $\lambda$ over the boundary of the $n$-simplex. 
Since $\lambda$ has sitting instants, there is $0 < \epsilon \in \mathbb{R}$ such that there are 
no contributions to this integral in an $\epsilon$-neighbourhood of the $(n-1)$-faces. 
Accordingly the integral is equivalently that over the smooth surface inscribed into the $(n+1)$-simplex.
Since $\lambda$ is a closed form on the $n$-simplex, this surface integral vanishes, by the Stokes theorem. Hence $\int_{\Delta^\bullet}$ is indeed a chain map.

It remains to show that $\int_{\Delta^\bullet} : \mathbf{cosk}_{n+1} \exp(b^{n-1} \mathbb{R}) \to \mathbf{B}^{n}\mathbb{R}_{\mathrm{chn}}$ is an isomorphism on simplicial homotopy groups over 
each $U \in \mathrm{CartSp}$. This amounts to the statement that 
\begin{itemize}
\item a smooth family of closed $n < k $-forms 
with sitting instants on the boundary of $\Delta^{k+1}$ may be extended to a smooth family of 
closed forms with sitting instants on $\Delta^{k+1}$ 
\item a smooth family of closed $n $-forms 
with sitting instants on the boundary of $\Delta^{n+1}$ may be extended to a smooth family of 
closed forms with sitting instants on $\Delta^{n+1}$ precisely if their smooth family of
integrals over $\partial Delta^{n+1}$ vanishes.
\end{itemize} 
To demonstrate this, we want to work with forms on the $(k+1)$-ball instead of the $(k+1)$-simplex. 
To achieve this, choose again $0 < \epsilon \in \mathbb{R}$ and construct the diffeomorphic image 
of $S^k \times [1-\epsilon,1]$ inside the $(k+1)$-simplex as indicated by the above construction: 
outside an $\epsilon$-neighbourhood of the corners the image is a rectangular $\epsilon$-thickening of 
the faces of the simplex. Inside the $\epsilon$-neighbourhoods of the corners it bends smoothly. 
By the Steenrod-approximation theorem \cite{Wockel} the diffeomorphism from this $\epsilon$-thickening 
of the smoothed boundary of the simplex to $S^k \times [0,1]$ extends to a smooth function from the 
$(k+1)$-simplex to the $(k+1)$-ball. 
By choosing $\epsilon$ smaller than each of the sitting instants of the given $n$-form on 
$\partial \Delta^k$, we have that this $n$-form vanishes on the $\epsilon$-neighbourhoods of the corners 
and is hence entirely determined by its restriction to the smoothed simplex, identified with the $(k+1)$-ball.

It is now sufficient to show: a smooth family of smooth $n$-forms 
$\omega \in \Omega^n_{\mathrm{\mathrm{vert}},\mathrm{\mathrm{cl}}}(U \times S^k)$ extends to a smooth family of closed 
$n$-forms $\hat \omega \in \Omega^n_{\mathrm{\mathrm{vert}},\mathrm{\mathrm{cl}}}(U \times B^{n+1})$ 
that is radially constant in a neighbourhood of the boundary  
for all $n < k$ and for $n = k$ precisely if its smooth family of integrals
 $\int_{S^n} \omega = 0 \in C^\infty(U, \mathbb{R})$ vanishes.

Notice that over the point this is a direct consequence of the de Rham theorem: 
all $k < n$ forms are exact on $S^k$ and $n$-forms are exact precisely if their integral vanishes. 
In that case there is an $(n-1)$-form $A$ with $\omega = d A$. Choosing any smoothing function 
$f : [0,1] \to [0,1]$ (smooth, surjective non,decreasing and constant in a neighbourhood of the boundary) 
we obtain a $n$-form $f \wedge A$ on $(0,1] \times S^n$, vertically constant in a neighbourhood of the ends 
of the interval, equal to $A$ at the top and vanishing at the bottom. Pushed forward along the 
canonical $(0,1] \times S^n \to D^{n+1}$ this defines a form on the $(n+1)$-ball, that we 
denote by the same symbol $f \wedge A$. Then the form $\hat \omega := d (f \wedge A)$ solves the problem.

To complete the proof we have to show that this argument does extend to smooth families of forms 
in that we can find suitable smooth families of the form $A$ in the above discussion. This 
may be accomplished for instance by invoking Hodge theory: If we equip $S^k$ with a Riemannian metric
then the refined form of the Hodge theorem says that we have an equality
$$
  \mathrm{id} - \pi_{\mathcal{H}} = [d, d^* G]
  \,,
$$
of operators on differential forms, where $\pi_{\mathcal{H}}$ is the orthogonal projection on 
harmonic forms and $G$ is the Green operator of the Hodge-Laplace operator. For $\omega$ an
exact form its harmonic projection vanishes so that this gives a homotopy
$$
  \omega = d (d^* G \omega)
  \,.
$$
This operation $\omega \mapsto d^* G \omega$ depends smoothly on $\omega$.
\endofproof

\paragraph{Flat coefficient objects for exponentiated $L\infty$-algebras.}
\label{FLatCoefficientsForExponentiatedLInfinityAlgebras}

We consider now the flat coefficient object, \ref{StrucFlatDifferential}, 
$\mathbf{\flat} \exp(\mathfrak{g})$ of exponentiated $L_\infty$ algebras
$\exp(\mathfrak{g})$, \ref{SmoothStrucLieAlgebras}.\index{$L_\infty$-algebra!flat cofficients}

\medskip

\begin{definition} 
\label{FlatCoefficientsForLieNGroupSimplicial}
Write $\mathbf{\flat}\exp(\mathfrak{g})_{\mathrm{smp}}$ 
or equivalentl $\exp(\mathfrak{g})_{\mathrm{flat}}$ for the simplicial presheaf given by
$$
  \mathbf{\flat}\exp(\mathfrak{g})_{\mathrm{smp}}
  :
  (U,[n]) \mapsto
  \mathrm{Hom}_{\mathrm{dgAlg}}(\mathrm{CE}(\mathfrak{g}), \Omega_{\mathrm{\mathrm{si}}}^\bullet(U \times \Delta^n))
  \,.
$$
\end{definition}
\begin{proposition} \label{FactorizatonForExponentiatedFlatInclusion}
The canonical morphism 
$\mathbf{\flat} \mathbf{B}^n \mathbb{R} \to \mathbf{B}^n \mathbb{R}$
in $\mathrm{Smooth} \infty \mathrm{Grpd}$ is presented in $[\mathrm{CartSp}_{\mathrm{smooth}}^{\mathrm{op}}, \mathrm{sSet}]$ by the composite
$$
  \xymatrix{
    \mathrm{const}\, \Gamma \, \exp(b^{n-1} \mathbb{R})
  \ar[r]^{\simeq}
  &
  \mathbf{\flat} \exp(b^{n-1} \mathbb{R})_{\mathrm{smp}}
  \ar@{->>}[r]
  &
  \exp(b^{n-1} \mathbb{R})
  }
  \,,
$$
where the first morphism is a weak equivalence and the second a 
fibration in $[\mathrm{CartSp}_{\mathrm{smooth}}^{\mathrm{op}}, \mathrm{sSet}]_{\mathrm{proj}}$.
\end{proposition}
We discuss the two morphisms in the composite separately in two
lemmas.
\begin{lemma}
\label{PresentationOfFlatCoefficientsByexp}
The canonical inclusion
$$
  \mathrm{const} \Gamma(\exp(\mathfrak{g}))
   \to
  \mathbf{\flat}\exp(\mathfrak{g})_{\mathrm{smp}}
$$
is a weak equivalence in $[\mathrm{CartSp}^{\mathrm{op}}, \mathrm{sSet}]_{\mathrm{proj}}$.
\end{lemma}
\proof
The morphism in question is on each object $U \in \mathrm{CartSp}$ the morphism of simplicial sets
$$
  \mathrm{Hom}_{\mathrm{dgAlg}}(\mathrm{CE}(\mathfrak{g}), \Omega_{\mathrm{\mathrm{si}}}^\bullet(\Delta^k))
  \to
  \mathrm{Hom}_{\mathrm{dgAlg}}(\mathrm{CE}(\mathfrak{g}), \Omega_{\mathrm{\mathrm{si}}}^\bullet(U \times \Delta^k))
  \,,
$$
which is given by pullback of differential forms along the projection $U \times \Delta^k \to \Delta^k$.

To show that for fixed $U$ this is a weak equivalence in the standard model structure on simplicial sets 
we produce objectwise a left inverse 
$$
  F_U
  :
  \mathrm{Hom}_{\mathrm{dgAlg}}(\mathrm{CE}(\mathfrak{g}), \Omega_{\mathrm{\mathrm{si}}}^\bullet(U \times \Delta^\bullet))
  \to
  \mathrm{Hom}_{\mathrm{dgAlg}}(\mathrm{CE}(\mathfrak{g}), \Omega_{\mathrm{\mathrm{si}}}^\bullet(\Delta^\bullet))
$$
and show that this is an acyclic fibration of simplicial sets. The statement then follows by the
2-out-of-3-property of weak equivalences.

We take $F_U$ to be given by evaluation at $0: * \to U$, i.e. by postcomposition with the morphisms
$$
  \Omega^\bullet(U \times \Delta^k )
  \stackrel{Id \times 0^*}{\to}
  \Omega^\bullet(* \times \Delta^k )
  =
  \Omega^\bullet(\Delta^k)
  \,.
$$
(This is, of course, not natural in $U$ and hence does not extend to a morphism of simplicial presheaves. But for our argument here it need not.)
The morphism $F_U$ is an acyclic Kan fibration precisely if all diagrams of the form
$$  
  \xymatrix{
    \partial \Delta[n] \ar[r] \ar[d] &
    \mathrm{Hom}(\mathrm{CE}(\mathfrak{g}), \Omega_{\mathrm{\mathrm{si}}}^\bullet(U \times \Delta^\bullet ))
    \ar[d]^{F_U}
    \\
    \Delta[n]
    \ar[r] &
    \mathrm{Hom}(\mathrm{CE}(\mathfrak{g}), \Omega_{\mathrm{\mathrm{si}}}^\bullet(\Delta^\bullet))
  }
$$
have a lift. 
Using the Yoneda lemma over the simplex category and since the differential forms on the simplices 
have sitting instants, we may, as above, equivalently reformulate this in terms of spheres as follows:
for every morphism $\mathrm{CE}(\mathfrak{g}) \to \Omega^\bullet_{\mathrm{\mathrm{si}}}(D^n)$ 
and morphism $\mathrm{CE}(\mathfrak{g}) \to \Omega^\bullet_{\mathrm{\mathrm{si}}}(U \times S^{n-1})$ 
such that the diagram
$$
  \xymatrix{
    \mathrm{CE}(\mathfrak{g}) \ar[r] \ar[d] 
      & \Omega^\bullet(U \times S^{n-1}) 
     \ar[d]
    \\
    \Omega_{\mathrm{\mathrm{si}}}^\bullet(D^n) \ar[r] & \Omega^\bullet(S^{n-1})
  }
$$
commutes, this may be factored as
$$
  \xymatrix{
   \mathrm{CE}(\mathfrak{g})
   \ar[dr]
   \\
   &\Omega_{\mathrm{\mathrm{si}}}^\bullet(U \times D^n)
      \ar[r]
      \ar[d] 
     & \Omega^\bullet(U \times S^{n-1})
     \ar[d]
    \\
    &\Omega^\bullet(D^n) \ar[r] & \Omega^\bullet(S^{n-1})
  }
  \,.
$$
(Here the subscript ``${}_{\mathrm{\mathrm{si}}}$'' denotes differential forms on the disk 
that are radially constant in a neighbourhood of the boundary.)

This factorization we now construct. 
Let first $f : [0,1] \to [0,1]$ be any smoothing function, 
i.e. a smooth function which is surjective, non-decreasing, and constant in a neighbourhood 
of the boundary. Define a smooth map
$
  U \times [0,1] \to U 
$
by 
$
  (u,\sigma) \mapsto u \cdot f(1-\sigma)
$,
where we use the multiplicative structure on the Cartesian space $U$. 
This function is the identity at $\sigma = 0$ and is the constant map to 
the origin at $\sigma = 1$. It exhibits a smooth contraction of $U$.

Pullback of differential forms along this map produces a morphism
$$
  \Omega^\bullet(U \times S^{n-1}) 
  \to 
  \Omega^\bullet(U \times S^{n-1} \times [0,1])
$$
which is such that a form $\omega$ is sent to a form which in a 
neighbourhood $(1-\epsilon,1]$ of $1 \in [0,1]$ is constant along $(1-\epsilon,1] \times U$ on the value $(0 , Id_{S^{n-1}})^* \omega$. 

Let now $0 < \epsilon \in \mathbb{R}$ some value such that the given forms $CE(\mathfrak{g}) \to \Omega^\bullet_{\mathrm{si}}(D^k)$ are constant a distance $d \leq \epsilon$ from the boundary of the disk. 
Let $q : [0,\epsilon/2] \to [0,1]$ be given by multiplication by $1/(\epsilon/2)$ and $h : D_{1-\epsilon/2}^k \to D_1^n$ the injection of the $n$-disk of radius $1-\epsilon/2$ into the unit $n$-disk.

We can then glue to the morphism
$$
  \mathrm{CE}(\mathfrak{g}) \to
  \Omega^\bullet(U \times S^{n-1})
  \to
  \Omega^\bullet(U \times [0,1] \times S^{n-1})
  \stackrel{id \times q^* \times id}{\to_\simeq}
  \Omega^\bullet(U \times [0,\epsilon/2] \times S^{n-1})   
$$
to the morphism
$$
  \mathrm{CE}(\mathfrak{g}) \to \Omega^\bullet(D^n)
  \to
  \Omega^\bullet(U \times \{1\} \times D^n)
  \stackrel{h^*}{\to_\simeq}
  \Omega^\bullet(U \times \{1\} \times D^n_{1-\epsilon/2})
$$
by smoothly identifying the union $[0,\epsilon/2] \times S^{n-1} \coprod_{S^{n-1}} D^n_{1-\epsilon/2}$ 
with $D^n$ (we glue a disk into an annulus to obtain a new disk) to obtain in total a morphism
$$
  \mathrm{CE}(\mathfrak{g}) \to \Omega^\bullet(U \times D^n)
$$
with the desired properties: at $u = 0$ the homotopy that we constructed is constant and the 
above construction hence restricts the forms to radius $\leq 1-\epsilon/2$ and then 
extends back to radius $\leq 1$ by the constant value that they had before. 
Away from 0 the homotopy in the rmaining $\epsilon/2$ bit smoothly interpolates to the boundary value.
\endofproof
\begin{lemma}
The canonical morphism
$$
  \mathbf{\flat}\exp(\mathfrak{g})_{\mathrm{smp}}
  \to 
  \exp(\mathfrak{g})
$$
is a fibration in $[\mathrm{CartSp}_{\mathrm{smooth}}^{\mathrm{op}}, \mathrm{sSet}]_{\mathrm{proj}}$.
\end{lemma}
\proof
Over each $U \in \mathrm{CartSp}$
the morphism is induced from the morphism of dg-algebras
$$
  \Omega^\bullet(U) \to C^\infty(U)
$$
that discards all differential forms of non-vanishing degree. 

It is sufficient to show that for 
$$
  \mathrm{CE}(\mathfrak{g}) 
    \to 
   \Omega_{\mathrm{\mathrm{si}},\mathrm{\mathrm{vert}}}^\bullet(
     U \times (D^n \times [0,1])
   )
$$
a morphism and 
$$
  \mathrm{CE}(\mathfrak{g}) \to \Omega^\bullet_{\mathrm{\mathrm{si}}}(U \times D^n )
$$
a lift of its restriction to $\sigma = 0 \in [0,1]$ we have an extension to a 
lift 
$$
  \mathrm{CE}(\mathfrak{g}) \to \Omega_{\mathrm{\mathrm{si}},\mathrm{\mathrm{vert}}}^\bullet(U \times (D^n \times [0,1]))
  \,.
$$
From these lifts all the required lifts are obtained by precomposition with some evident smooth retractions.

The lifts in question are obtained from solving differential equations with boundary conditions, 
and exist due to the existence of solutions of first order systems of partial differential equations 
and the identity $d_{\mathrm{\mathrm{dR}}}^2  = 0$.
\endofproof
We have discussed now two different presentations for the flat coefficient 
object $\mathbf{\flat}\mathbf{B}^n \mathbb{R}$:
\begin{enumerate}
\item $\mathbf{\flat} \mathbf{B}^n \mathbb{R}_{\mathrm{chn}}$ -- prop. \ref{FlatCohomologyWithCoeffsInCircle};
\item $\mathbf{\flat} \mathbf{B}^n \mathbb{R}_{\mathrm{smp}}$ -- prop. \ref{FactorizatonForExponentiatedFlatInclusion};
\end{enumerate}
There is an evident degreewise map
$$
  (-1)^{\bullet+1}
  \int_{\Delta^\bullet} :
  \mathbf{\flat} \mathbf{B}^n \mathbb{R}_{\mathrm{simp}}
  \to  
  \mathbf{\flat} \mathbf{B}^n \mathbb{R}_{\mathrm{chn}}
$$
that sends a closed $n$-form $\omega \in \Omega^n_{\mathrm{cl}}(U \times \Delta^k)$ to $(-1)^{k+1}$ 
times its fiber integration $\int_{\Delta^k} \omega$.
\begin{proposition}
This map yields a morphism of simplicial presheaves
$$
  \int : 
   \mathbf{\flat} \mathbf{B}^n \mathbb{R}_{\mathrm{smp}}
     \to
   \mathbf{\flat} \mathbf{B}^n \mathbb{R}_{\mathrm{chn}}
$$
which is a weak equivalence in $[\mathrm{CartSp}^{\mathrm{op}}, \mathrm{sSet}]_{\mathrm{proj}}$.
\end{proposition}
\proof
First we check that we have a morphism of simplicial sets over each 
$U \in \mathrm{CartSp}$. Since both objects are abelian simplicial groups we may, 
by the Dold-Kan correspondence, check the statement for sheaves of normalized chain complexes.

Notice that the chain complex differential on the forms 
$\omega \in \Omega^n_{\mathrm{cl}}(U \times \Delta^k)$ on simplices
sends a form to the alternating sum of its restriction to the faces of the
simplex. Postcomposed with the integration map this is the operation 
$\omega \mapsto \int_{\partial \Delta^k} \omega$ of integration over the boundary.

Conversely, first integrating over the simplex and then applying the de Rham differential on $U$ yields
$$
  \begin{aligned}
     \omega \mapsto (-1)^{k+1} d_U \int_{\Delta^k} \omega
      &= 
      - \int_{\Delta^k} d_U \omega
      \\
      & = \int_{\Delta^k} d_{\Delta^k} \omega
      \\
      & =  \int_{\partial \Delta^k} \omega
  \end{aligned}
  \,,
$$
where we first used that $\omega$ is closed, 
so that $d_{\mathrm{\mathrm{dR}}} \omega = (d_U + d_{\Delta^k}) \omega = 0$, 
and then used Stokes' theorem.
Therefore we have indeed objectwise a chain map.

By the discussion of the two objects we already know that both present the homotopy 
type of $\mathbf{\flat} \mathbf{B}^n \mathbb{R}$. Therefore it suffices to show that 
the integration map is over each $U \in \mathrm{CartSp}$ an isomorphism on the simplicial 
homotopy group in degree $n$.

Clearly the morphism
$$
  \int_{\Delta^n} : \Omega^\bullet_{\mathrm{si},\mathrm{cl}}(U  \times \Delta^n)
  \to 
  C^\infty(U, \mathbb{R})
$$
is surjective on degree $n$ homotopy groups: for $f : U \to * \to \mathbb{R}$  constant, a preimage is $f \cdot \mathrm{vol}_{\Delta^n}$, the normalized volume form of the $n$-simplex times $f$.
Moreover, these preimages clearly span the whole homotopy group 
$\pi_n (\mathbf{\flat} \mathbf{B}^n \mathbb{R}) \simeq \mathbb{R}_{\mathrm{disc}}$ 
(they are in fact the images of the weak equivalence 
$\mathrm{const} \Gamma \exp(b^{n-1}\mathbb{R}) \to \mathbf{\flat} \mathbf{B}^n \mathbb{R}_{\mathrm{smp}}$ ) 
and the integration map is injective on them. Therefore it is an isomorphism on the homotopy groups in degree $n$.
\endofproof

\medskip

\paragraph{de Rham coefficients}

We now consider the de Rham coefficient object 
$\mathbf{\flat}_{\mathrm{\mathrm{dR}}} \exp(\mathfrak{g})$, \ref{StrucDeRham}, of 
exponentiated $L_\infty$ algebras $\exp(\mathfrak{g})$, def \ref{ExponentiatedLInftyAlgbra}.
\index{$L_\infty$-algebra!de Rham coefficients}
\begin{proposition} \label{DifferentialCoefficientsOfLieInt}
For $\mathfrak{g} \in L_\infty$ a representive in 
$[\mathrm{CartSp}^{\mathrm{op}}, \mathrm{sSet}]_{\mathrm{proj}}$ 
of the de Rham coefficient object
$\mathbf{\flat}_{\mathrm{\mathrm{dR}}} \exp(\mathfrak{g})$ is given by the presheaf
$$
  \mathbf{\flat}_{\mathrm{\mathrm{dR}}} \mathbf{B}^n \mathbb{R}_{\mathrm{smp}}
  :
  (U,[n])
  \mapsto
   \mathrm{Hom}_{\mathrm{dgAlg}}(
     \mathrm{CE}(\mathfrak{g}),
     \Omega_{\mathrm{\mathrm{si}}}^{\bullet\geq 1,\bullet}(U \times \Delta^n)
  )
  \,,
$$
where the notation on the right denotes the dg-algebra of differential forms on 
$U \times\Delta^n$ that 
(apart from having sitting instants on the faces of $\Delta^n$) are along $U$ 
of non-vanishing degree.
\end{proposition}
\proof
By the prop. \ref{FactorizatonForExponentiatedFlatInclusion}
we may present the defining $\infty$-pullback 
$\mathbf{\flat}_{\mathrm{\mathrm{dR}}} \mathbf{B}^n \mathbb{R} := 
* \times_{\mathbf{B}^n \mathbb{R}} \mathbf{\flat} \mathbf{B}^n \mathbb{R}$
in $\mathrm{Smooth} \infty \mathrm{Grpd}$ by the ordinary pullback
$$
  \xymatrix{
    \mathbf{\flat}_{\mathrm{\mathrm{dR}}}\mathbf{B}^n \mathbb{R}_{\mathrm{smp}} 
      \ar[r] \ar[d] &
      \mathbf{\flat}\mathbf{B}^n \mathbb{R}_{\mathrm{smp}} \ar[d]
    \\
    {*} \ar[r] & \mathbf{B}^n \mathbb{R}   
  }
$$
in $[\mathrm{CartSp}_{\mathrm{smooth}}^{\mathrm{op}}, \mathrm{sSet}]$.
\endofproof
We have discussed now two different presentations for the de Rham coefficient 
object $\mathbf{\flat}\mathbf{B}^n \mathbb{R}$:
\begin{enumerate}
\item $\mathbf{\flat}_{\mathrm{\mathrm{dR}}} \mathbf{B}^n \mathbb{R}_{\mathrm{chn}}$ -- prop. \ref{OrdinaryDeRham};
\item $\mathbf{\flat}_{\mathrm{\mathrm{dR}}} \mathbf{B}^n \mathbb{R}_{\mathrm{smp}}$ -- 
   prop \ref{DifferentialCoefficientsOfLieInt};
\end{enumerate}
There is an evident degreewise map
$$
  (-1)^{\bullet+1}
  \int_{\Delta^\bullet} :
  \mathbf{\flat}_{\mathrm{\mathrm{dR}}} \mathbf{B}^n \mathbb{R}_{\mathrm{smp}}
  \to  
  \mathbf{\flat}_{\mathrm{\mathrm{dR}}} \mathbf{B}^n \mathbb{R}_{\mathrm{chn}}
$$
that sends a closed $n$-form $\omega \in \Omega^n_{\mathrm{\mathrm{cl}}}(U \times \Delta^k)$ to $(-1)^{k+1}$ times  its fiber integration $\int_{\Delta^k} \omega$.
\begin{proposition} \label{FiberIntegrationAsWeakEquivalenceForDeRhamCoefficientPresentations}
This map yields a morphism of simplicial presheaves
$$
  \int : 
   \mathbf{\flat}_{\mathrm{\mathrm{dR}}}\mathbf{B}^n \mathbb{R}_{\mathrm{smp}}
     \to
   \mathbf{\flat}_{\mathrm{\mathrm{dR}}} \mathbf{B}^n \mathbb{R}_{\mathrm{chn}}
$$
which is a weak equivalence in $[\mathrm{CartSp}^{\mathrm{op}}, \mathrm{sSet}]_{\mathrm{proj}}$.
\end{proposition}
\proof
This morphism is the morphism on pullbacks induced from the weak equivalence of diagrams
$$
  \xymatrix{
    {*} \ar[r] \ar[d]^= 
     & \exp(b^{n-1}\mathbb{R}) \ar@{<-}[r] \ar[d]^{\int}_\simeq
     & \mathbf{\flat}\mathbf{B}^n \mathbb{R}_{\mathrm{smp}} 
     \ar[d]^{\int}_\simeq
    \\
    {*} \ar[r] & \mathbf{B}^n \mathbb{R}_{\mathrm{chn}} 
      \ar@{<-}[r]
     & \mathbf{\flat}\mathbf{B}^n \mathbb{R}_{\mathrm{chn}}
  }
  \,.
$$
Since both of these pullbacks are homotopy pullbacks by the above discussion, the induced morphism between the pullbacks is also a weak equivalence.
\endofproof

\subsubsection{Maurer-Cartan forms and curvature characteristic forms}
\label{SmoothStrucCurvature}
 \index{structures in a cohesive $\infty$-topos!curvature characteristic forms!smooth}
 \index{structures in a cohesive $\infty$-topos!Maurer-Cartan forms!smooth}

We discuss the universal curvature forms, \ref{StructCurvatureCharacteristic}, 
in $\mathrm{Smooth} \infty \mathrm{Grpd}$.

Specifically, we discuss the canonical Maurer-Cartan form
on the following special cases of (presentations of) smooth $\infty$-groups.
\begin{itemize}
\item
  \ref{CanonicalFormOnAnOrdinaryLieGroup} -- ordinary Lie groups:
\item
  \ref{CanonicalFormOnCirclenGroup} -- circle $n$-groups $\mathbf{B}^{n-1}U(1)$;
\item
  \ref{CanoncalFormOnSimplicialLieGroup}
   --  simplicial Lie groups.
\end{itemize}
Notice that, by the discussion in \ref{SheafAndNonabelianDoldKan},
the case of simplicial Lie groups also subsumes the case
of crossed modules of Lie groups, def. \ref{Strict2GroupInIntroduction}, 
and generally of crossed complexes of Lie groups, 
def. \ref{CrossedComplex}.

\paragraph{Canonical form on an ordinary Lie group}
\label{CanonicalFormOnAnOrdinaryLieGroup}

\begin{proposition} 
 \label{StandardMaurerCartanForm}
 \index{Maurer-Cartan form!ordinary}
Let $G$ be a Lie group with Lie algebra $\mathfrak{g}$.

Under the identification
$$
  \mathrm{Smooth} \infty \mathrm{Grpd}(X, \mathbf{\flat}_{\mathrm{\mathrm{dR}}}\mathbf{B}G)
  \simeq
  \Omega^1_{\mathrm{flat}}(X,\mathfrak{g})
$$
from prop. \ref{LieGroupDeRhamCoefficients}, 
for $X \in \mathrm{SmoothMfd}$, we have that the canonical morphism
$$
  \theta : G \to \mathbf{\flat}_{\mathrm{\mathrm{dR}}} \mathbf{B}G
$$
in $\mathrm{Smooth} \infty \mathrm{Grpd}$ corresponds to the 
ordinary Maurer-Cartan form on $G$.
\end{proposition}
\proof
We compute the defining double $\infty$-pullback
$$
  \xymatrix{
    G \ar[r] \ar[d]_\theta & {*} \ar[d]
    \\
    \mathbf{\flat}_{\mathrm{\mathrm{dR}}} \mathbf{B}G \ar[r] \ar[d] & 
    \mathbf{\flat} \mathbf{B}G
    \ar[d]
    \\
    {*} \ar[r] & \mathbf{B}G
  }
$$
in $\mathrm{Smooth} \infty \mathrm{Grpd}$ as a homotopy pullback
in $[\mathrm{CartSp}_{\mathrm{smooth}}^{\mathrm{op}}, \mathrm{sSet}]_{\mathrm{proj}}$.
In prop. \ref{LieGroupDeRhamCoefficients} we already modeled the lower 
$\infty$-pullback square by the ordinary pullback 
$$
  \xymatrix{
    \mathbf{\flat}_{\mathrm{\mathrm{dR}}}\mathbf{B}G_{\mathrm{ch}}
      \ar[r]
      \ar[d]
    &
    \mathbf{\flat}\mathbf{B}G_{\mathrm{ch}}
    \ar[d]
    \\
    {*} \ar[r] & \mathbf{B}G_{\mathrm{ch}}
  }
  \,.
$$
A standard fibration replacement of the point inclusion
$* \to \mathbf{\flat}\mathbf{B}G $ is given by replacing the point by 
the presheaf that assigns groupoids of the form
$$
  Q : 
  U \mapsto 
  \left\{
    \raisebox{20pt}{
    \xymatrix{
       & A_0 = 0
       \ar[dl]_{g_1} \ar[dr]^{g_2}
       \\
       A_1 \ar[rr]^h && A_2
    }
    }
  \right\}
  \,,
$$
where on the right the commuting triangle is in 
$(\mathbf{\flat}_{\mathrm{\mathrm{dR}}}\mathbf{B}G_{\mathrm{ch}})(U)$ and here regarded as 
a morphism from $(g_1,A_1)$ to $(g_2,A_2)$. 
And the fibration $Q \to \mathbf{\flat}\mathbf{B}G_{\mathrm{ch}}$ is given by projecting 
out the base of these triangles.

The pullback of this along
$\mathbf{\flat}_{\mathrm{\mathrm{dR}}}\mathbf{B}G_{\mathrm{ch}} \to \mathbf{\flat}\mathbf{B}G_{\mathrm{ch}}$ is over each 
$U$  the restriction of the groupoid $Q(U)$ to its set of objects, hence is the sheaf
$$
  U \mapsto 
  \left\{
    \raisebox{20pt}{
    \xymatrix{
       A_0 = 0
       \ar[d]^g
       \\
       g^* \theta
    }
    }
  \right\}
  \simeq 
  C^\infty(U,G)
  = G(U)
  \,,
$$
equipped with the projection
$$
 t_U :  G \to \mathbf{\flat}_{\mathrm{\mathrm{dR}}} \mathbf{B}G_{\mathrm{ch}}
$$
given by
$$
  t_U : (g : U \to G) \mapsto g^* \theta
  \,.
$$
Under the Yoneda lemma (over $\mathrm{SmoothMfd}$) this 
identifies the morphism $t$ with the Maurer-Cartan form
$\theta \in \Omega^1_{\mathrm{flat}}(G,\mathfrak{g})$.
\endofproof

\paragraph{Canonical form on the circle $n$-group}
\label{CanonicalFormOnCirclenGroup}

We consider now the canonical differential form 
on the circle Lie $(n+1)$-group, def. \ref{Circlengroup}.
Below in \ref{SmoothStrucDifferentialCohomology} this serves as
the \emph{universal curvature class} on the universal circle $n$-bundle.

\medskip

\begin{definition}
  For $n \in \mathbb{N}$, write
  $$
  \mathbf{B}^n U(1)_{\mathrm{diff}, \mathrm{chn}}
  :=
  \mathrm{DK}
  \left(
  \raisebox{20pt}{
  \xymatrix@R=1pt{
    & U(1) \ar[r]^{d_{\mathrm{dR}}} 
	& \Omega^1 \ar[r] 
	& \cdots \ar[r] 
	& \Omega^{n-1}  \ar[r]^{d_{\mathrm{dR}}} 
	& \Omega^n 
	\\
    & \oplus & \oplus && \oplus
    \\
    0 \ar[r] \ar[uur] 
	& \Omega^1 \ar[r]_{d_{\mathrm{dR}}} \ar[uur]|{-\mathrm{id}} 
	& \Omega^2 \ar[r] 
	&  \cdots \ar[r]_{d_{\mathrm{dR}}} 
	& 
	\Omega^n 
	\ar[uur]|{(-1)^{n}\mathrm{id}} 
   }
   }
   \right)
   \;\;
   \in [\mathrm{CartSp}^{\mathrm{op}}, \mathrm{sSet}]
   $$
   for the simplicial presheaf which is the image under the Dold-Kan map,
   prop. \ref{EmbeddingOfChainComplexes}, of the 
   chain complex on the right as induicated. (Here we display morphisms between direct sums
   of presheaves of chain complexes by their matrix components, as usual).
   Write moreover
   $$
     \mathrm{curv}_{\mathrm{chn}} 
	 :
	 \mathbf{B}^n U(1)_{\mathrm{diff}, \mathrm{chn}}
	 \to
	 \flat_{\mathrm{dR}}\mathbf{B}^{n+1}U(1)_{\mathrm{chn}}
   $$
   for the morphism of simplicial presheaves which is the image under the Dold-Kan
   map, prop. \ref{EmbeddingOfChainComplexes} 
   of the morphism of sheaves of chain complexes which in components is given by
   $$
    \raisebox{30pt}{
    \xymatrix{
	  \mathbf{B}^n U(1)_{\mathrm{diff},\mathrm{chn}}
	  \ar[dd]^{\mathrm{curv}_{\mathrm{chn}}}
	  \\
	  \\
	  \flat_{\mathrm{dR}}\mathbf{B}^{n+1}U(1)_{\mathrm{chn}}
	}
	}
	\;\;
	:=
	\;\;
	\mathrm{DK}
	\left(
	\raisebox{52pt}{
  \xymatrix@R=1pt{
    & U(1) \ar[r]^{d_{\mathrm{dR}}} 
	& \Omega^1 \ar[r] 
	& \cdots \ar[r] 
	& \Omega^{n-1}  \ar[r]^{d_{\mathrm{dR}}} 
	& \Omega^n 
	\ar[ddddddddddd]|{d_{\mathrm{dR}}}
	\\
    & \oplus & \oplus && \oplus
    \\
    0 \ar[r] \ar[uur] \ar[ddddddddd] 
	& \Omega^1 \ar[r]_{d_{\mathrm{dR}}} \ar[uur]|{-\mathrm{id}} \ar[ddddddddd]|{(-1)^n\mathrm{id}} 
	& \Omega^2 \ar[r] \ar[ddddddddd]|{(-1)^n\mathrm{id}} 
	&  \cdots \ar[r]_{d_{\mathrm{dR}}} 
	& 
	\Omega^n 
	\ar[uur]|{(-1)^{n}\mathrm{id}} \ar[ddddddddd]|{(-1)^n\mathrm{id}}  
	& 
	\\
	\\
	\\
	&&&&&&
    \\
	&&&&&&
	\\
	\\
	\\
	\\
	\\
	0 \ar[r]  
	& \Omega^1 \ar[r]^{d_{\mathrm{dR}}} 
	&
	\Omega^2 \ar[r] 
	&
	\cdots \ar[r] 
	& \Omega^n \ar[r]^{d_{\mathrm{dR}}} 
	& \Omega^{n+1}_{\mathrm{cl}} 
  }
  }
  \right)
$$
\label{BnU1connFromDK}
\label{PrincipalConnectionWithoutTopDegreeForms}
\end{definition}
\begin{proposition}
  The evident projection morphism
  $$
    \xymatrix{
	   \mathbf{B}^n U(1)_{\mathrm{diff},\mathrm{chn}} \ar[r]^\simeq 
	   &
	   \mathbf{B}^n U(1)_{\mathrm{chn}}
	}
  $$
  is a weak equivalence in $[\mathrm{CartSp}, \mathrm{sSet}]_{\mathrm{proj}}$. 
  Moreover, the span
  $$
    \xymatrix{
	  \mathbf{B}^n U(1)_{\mathrm{diff}, \mathrm{chn}}
	  \ar[rr]^-{\mathrm{curv}_{\mathrm{chn}}}
	  \ar[d]^\simeq
	  &&
	  \flat_{\mathrm{dR}}\mathbf{B}^{n+1} U(1)_{\mathrm{chn}}
	  \\
	  \mathbf{B}^n U(1)_{\mathrm{chn}}
	}
  $$
  is a presentation in $[\mathrm{CartSp}^{\mathrm{op}}, \mathrm{sSet}]_{\mathrm{proj}, \mathrm{loc}}$
  of the universal curvature characteristic, def. \ref{UniversalCurvatureCharacteristic},
  $\mathrm{curv} : \mathbf{B}^n U(1) \to \flat_{\mathrm{dR}}\mathbf{B}^{n+1}U(1)$
  in $\mathrm{Smooth}\infty\mathrm{Grpd}$.
  \label{PresentationOfUniversalU1CurvatureCharacteristic}
  \label{CurvatureCharOnBnU1}
\end{proposition}
\proof
By prop. \ref{FiniteHomotopyLimitsInPresheaves} we may present 
the defining $\infty$-pullback
$$
  \xymatrix{
    \mathbf{B}^n U(1) \ar[r] \ar[d]_{\mathrm{curv}} & {*} \ar[d]
    \\
    \mathbf{\flat}_{\mathrm{\mathrm{dR}}} \mathbf{B}^{n+1}U(1) 
      \ar[d] \ar[r] &
    \mathbf{\flat} \mathbf{B}^{n+1} U(1)
      \ar[d]
    \\
    {*} \ar[r] & \mathbf{B}^{n+1} U(1)
  }
$$
in $\mathrm{Smooth}\infty \mathrm{Grpd}$
by a homotopy pullback in $[\mathrm{CartSp}^{\mathrm{op}}, \mathrm{sSet}]_{\mathrm{proj}}$. 
We claim that there is a commuting diagram
$$ 
\hspace{-1cm}
  \xymatrix{
    [0\stackrel{}{\to}
    {C^\infty(-,U(1)) \atop \oplus \Omega^1(-)}
    \stackrel{d_{\mathrm{\mathrm{dR}}} - \mathrm{Id}}{\to}
    {\Omega^1(-) \atop \oplus \Omega^2(-)}
    \stackrel{d_{\mathrm{\mathrm{dR}}} + \mathrm{Id}}{\to}
    \cdots
    \stackrel{d_{\mathrm{\mathrm{dR}}} + \mathrm{Id}}{\to}
    \Omega^n(-)
	]
    \ar[r] 
    \ar[d]^{(p_2, p_2, \cdots, d_{\mathrm{\mathrm{\mathrm{dR}}}})}
    &
    [{\small C^\infty(-,U(1))} 
	  \stackrel{ d_{\mathrm{\mathrm{dR}}} + \mathrm{Id}  }{\to}
    {C^\infty(-,U(1)) \atop \oplus \Omega^1(-)}
    \stackrel{d_{\mathrm{\mathrm{dR}}} - \mathrm{Id}}{\to}
    \cdots
    { \Omega^{n-1}(-) \atop \oplus \Omega^n(-)}
    \stackrel{d_{\mathrm{\mathrm{dR}}} + \mathrm{Id}}{\to}
    \Omega^n(-)
	]
    \ar[d]^{(\mathrm{\mathrm{Id}}, p_2, p_2, \cdots, p_2, d_{\mathrm{\mathrm{\mathrm{dR}}}})}
    \\
    \; 
	[0 \stackrel{}{\to} \Omega^1(-) 
    \stackrel{d_{\mathrm{\mathrm{dR}}}}{\to}
    \Omega^2(-)
    \stackrel{d_{\mathrm{\mathrm{dR}}}}{\to} \cdots \stackrel{d_{\mathrm{\mathrm{dR}}}}{\to} 
    \Omega_{\mathrm{\mathrm{cl}}}^{n+1}(-) ]   
    \ar[r] \ar[d]&
    [C^\infty(-,U(1)) \stackrel{d_{\mathrm{\mathrm{dR}}}}{\to} \Omega^1(-) 
    \stackrel{d_{\mathrm{\mathrm{dR}}}}{\to} 
    \Omega^2(-)
    \stackrel{d_{\mathrm{\mathrm{dR}}}}{\to}
    \cdots \stackrel{d_{\mathrm{\mathrm{dR}}}}{\to} 
    \Omega_{\mathrm{\mathrm{cl}}}^{n+1}(-)
	]
    \ar[d]
    \\
    [0 \to 0 \to 0 \to \cdots \to 0]
    \ar[r] &
    [ C^\infty(-,U(1)) \to 0 \to 0 \to \cdots \to 0]
  }
$$
in $[\mathrm{CartSp}^{\mathrm{op}},\mathrm{Ch}^+]_{\mathrm{proj}}$, where
\begin{itemize}
\item the objects are fibrant models for the corresponding objects in the
  above $\infty$-pullback diagram;
\item the two right vertical morphisms are fibrations;
\item the two squares are pullback squares.
\end{itemize}
This implies that under the right adjoint $\Xi$ we have a homotopy pullback as claimed.
In full detail, the diagram of morphisms of sheaves that exhibits this diagram of
morphisms of complexes of sheaves is
$$
 \hspace{-1cm}
  \xymatrix@R=1pt{
    & U(1) \ar[r]^{d_{\mathrm{dR}}} \ar@{..>}[ddddrrrrrr]
	& \Omega^1 \ar[r] 
	& \cdots \ar[r] \ar@{..>}[ddddrrrrrr]
	& \Omega^{n-1}  \ar[r]^{d_{\mathrm{dR}}} \ar@{..>}[ddddrrrrrr]
	& \Omega^n \ar@{..>}[ddddrrrrrr]
	\ar[ddddddddddd]|{d_{\mathrm{dR}}}
	\\
    & \oplus & \oplus && \oplus
    \\
    0 \ar[r] \ar[uur] \ar[ddddddddd] \ar@{..>}[ddddrrrrrr]
	& \Omega^1 \ar[r]_{d_{\mathrm{dR}}} \ar[uur]|{-\mathrm{id}} \ar[ddddddddd]|{(-1)^n\mathrm{id}} \ar@{..>}[ddddrrrrrr]
	& \Omega^2 \ar[r] \ar[ddddddddd]|{(-1)^n\mathrm{id}} \ar@{..>}[ddddrrrrrr]
	&  \cdots \ar[r]_{d_{\mathrm{dR}}} 
	& 
	\Omega^n 
	\ar[uur]|{(-1)^{n}\mathrm{id}} \ar[ddddddddd]|{(-1)^n\mathrm{id}}  \ar@{..>}[ddddrrrrrr]
	& 
	\\
	\\
    &&&&&&
    & U(1) \ar[r]^{d_{\mathrm{dR}}} & \Omega^1 \ar[r] & \cdots \ar[r] & \Omega^{n-1}  \ar[r]^{d_{\mathrm{dR}}} & \Omega^n
	\ar[ddddddddddd]|{d_{\mathrm{dR}}}
	\\
	&&&&&&
    & \oplus & \oplus && \oplus
    \\
	&&&&&&
    U(1) \ar[r]_{d_{\mathrm{dR}}} \ar[uur]|{\mathrm{id}} \ar[ddddddddd]|{(-1)^n\mathrm{id}}
	& \Omega^1 \ar[r]_{d_{\mathrm{dR}}} \ar[uur]|{-\mathrm{id}} \ar[ddddddddd]|{(-1)^n\mathrm{id}}
	& \Omega^2 \ar[r] \ar[ddddddddd]|{(-1)^n\mathrm{id}}
	&  \cdots \ar[r]_{d_{\mathrm{dR}}} & 
	\Omega^n 
	\ar[uur]|{(-1)^{n}\mathrm{id}} \ar[ddddddddd]|{(-1)^n\mathrm{id}} & 
	\\
	\\
	\\
	\\
	\\
	0 \ar[r]  \ar[ddddddddd] \ar@{..>}[ddddrrrrrr]
	& \Omega^1 \ar[r]^{d_{\mathrm{dR}}} \ar[ddddddddd] \ar@{..>}[ddddrrrrrr]
	&
	\Omega^2 \ar[r] \ar[ddddddddd] \ar@{..>}[ddddrrrrrr]
	&
	\cdots \ar[r] 
	& \Omega^n \ar[r]^{d_{\mathrm{dR}}} \ar[ddddddddd] \ar@{..>}[ddddrrrrrr]
	& \Omega^{n+1}_{\mathrm{cl}} \ar[ddddddddd] \ar@{..>}[ddddrrrrrr]
	\\
	\\
	\\
	\\
	&&&&&&
	U(1) \ar[r]^{d_{\mathrm{dR}}} \ar[ddddddddd]|{\mathrm{id}}
	& \Omega^1 \ar[r]^{d_{\mathrm{dR}}} \ar[ddddddddd]
	&
	\Omega^2 \ar[r] \ar[ddddddddd]
	&
	\cdots \ar[r] 
	& \Omega^n \ar[r]^{d_{\mathrm{dR}}} \ar[ddddddddd]
	& \Omega^{n+1}_{\mathrm{cl}} \ar[ddddddddd]
	\\
	\\
	\\
	\\
	\\
	0 \ar[r] \ar@{..>}[ddddrrrrrr]& 
	0 \ar[r] \ar@{..>}[ddddrrrrrr] & 
	0 \ar[r] \ar@{..>}[ddddrrrrrr] & 
	\cdots \ar[r]  & 
	0 \ar[r] \ar@{..>}[ddddrrrrrr] & 
	0 \ar@{..>}[ddddrrrrrr] 
	\\
	\\
	\\
	\\
	&&&&&&
	U(1) \ar[r]  & 0 \ar[r] \ar[r] 
	&
	0 \ar[r]
	&
	\cdots \ar[r] & 0 \ar[r]  & 0
  }
$$
That the lower square here is a pullback is prop. \ref{OrdinaryDeRham}. 
For the upper square the same type of reasoning applies. 
The main point is to find the chain complex in the top right such that it is a resolution 
of the point and maps by a fibration onto our model for $\mathbf{\flat}\mathbf{B}^n U(1)$. 
This is the mapping cone of the identity on the Deligne complex, as indicated.
The vertical morphism out of it is
manifestly surjective (by the Poincar{\'e} lemma applied to each object $U \in \mathrm{CartSp}$) 
hence this is a fibration.
\endofproof

In prop. \ref{DifferentialCoefficientsOfLieInt} we had discussed an 
alternative equivalent 
presentation of de Rham coefficient objects above. 
We now formulate the curvature characteristic in this alternative form.
\begin{observation}
We may write the simplicial presheaf $\mathbf{\flat}_{\mathrm{\mathrm{dR}}}\mathbf{B}^{n+1} \mathbb{R}_{\mathrm{smp}}$ 
from prop.\ref{DifferentialCoefficientsOfLieInt} equivalently as follows
$$
  \mathbf{\flat}_{\mathrm{\mathrm{dR}}}\mathbf{B}^{n+1} \mathbb{R}_{\mathrm{smp}}
  :
  (U, [k]) 
   \mapsto
  \left\{
    \raisebox{20pt}{
    \xymatrix{
      \Omega^\bullet_{\mathrm{\mathrm{si}},\mathrm{\mathrm{vert}}}(U \times \Delta^k)
       & 0
       \ar[l]
       \\
       \Omega_{\mathrm{\mathrm{si}}}^\bullet(U \times \Delta^k)
        \ar[u]
       &  \ar[l]
          \ar[u]
       \mathrm{CE}(b^{n}\mathbb{R})
    }    
    }
  \right\}
  \,,
$$
where on the right we have the set of commuting diagrams in 
$\mathrm{dgAlg}$ of the given form, with the vertical morphisms being the canonical projections.
\end{observation}
\begin{definition}
Write $\mathrm{W}(b^{n-1}\mathbb{R}) \in \mathrm{dgAlg}$ for the Weil algebra of the 
line Lie $n$-algebra, defined to be free commutative dg-algebra on a single generator in degree $n$, 
hence the graded commutative algebra on a generator in degree $n$ and a generator in degree 
$(n+1)$ equipped with the differential that takes the former to the latter.

We write also $\mathrm{inn}(b^{n-1})$ for the $L_\infty$-algebra corresponding to the Weil
algebra
$$
  \mathrm{CE}(\mathrm{inn}(b^{n-1})) := \mathrm{W}(b^{n-1}\mathbb{R})
$$ 
\end{definition}
\begin{observation} \label{PropertiesOfWeilAlgebraOfLineLienAlgebra}
We have the following properties of $\mathrm{W}(b^{n-1}\mathbb{R})$
\begin{enumerate}
\item There is a canonical natural isomorphism
$$
  \mathrm{Hom}_{\mathrm{dgAlg}}(\mathrm{W}(b^{n-1} \mathbb{R}), \Omega^\bullet(U))
  \simeq
  \Omega^n(U)
$$
between dg-algebra homomorphisms $A : \mathrm{W}(b^{n-1}\mathbb{R}) \to \Omega^\bullet(X)$ 
from the Weil algebra of $b^{n-1}\mathbb{R}$ to the de Rham complex and degree-$n$ differential 
forms, not necessarily closed.

\item There is a canonical dg-algebra homomorphism 
$\mathrm{W}(b^{n-1}\mathbb{R}) \to \mathrm{CE}(b^{n-1}\mathbb{R})$ and the differential 
$n$-form corresponding to $A$ factors through this morphism preciselly if the curvature 
$d_{\mathrm{\mathrm{dR}}} A$ of $A$ vanishes.

\item The image under $\exp(-)$
  $$
    \exp(\mathrm{inn}(b^{n-1})\mathbb{R}) \to \exp(b^{n}\mathbb{R})
  $$
  of the canonical morphism $\mathrm{W}(b^{n-1}\mathbb{R}) \leftarrow \mathrm{CE}(b^n \mathbb{R})$
  is a fibration in $[\mathrm{CartSp}_{\mathrm{smooth}}^{\mathrm{op}}, \mathrm{sSet}]_{\mathrm{proj}}$
  that presents the point inclusion ${*} \to \mathbf{B}^{n+1}\mathbb{R}$ in 
   $\mathrm{Smooth}\infty \mathrm{Grpd}$.
\end{enumerate}
\end{observation}
\begin{definition}
Let $\mathbf{B}^n \mathbb{R}_{\mathrm{diff},\mathrm{smp}} \in [\mathrm{CartSp}_{\mathrm{smooth}}^{\mathrm{op}}, \mathrm{sSet}]$ be the simplicial presheaf defined by
$$
  \mathbf{B}^n \mathbb{R}_{\mathrm{diff},\mathrm{smp}} : (U,[k])
  \mapsto
  \left\{
    \raisebox{20pt}{
    \xymatrix{
      \Omega^\bullet_{\mathrm{\mathrm{si}},\mathrm{\mathrm{vert}}}(U \times \Delta^k)
       &
       \ar[l]_{{A_{\mathrm{vert}}}}
       \mathrm{CE}(b^{n-1}\mathbb{R})
       \\
       \Omega_{\mathrm{si}}^\bullet(U \times \Delta^k) \ar[u]
        &
        \ar[l]_A
       \mathrm{W}(b^{n-1}\mathbb{R}) \ar[u]
    }
    }
  \right\}
  \,,
$$ 
where on the right we have the set of commuting diagrams in $\mathrm{dgAlg}$
as indicated.
\end{definition}
This means that an element of $\mathbf{B}^n \mathbb{R}_{\mathrm{diff},\mathrm{smp}}(U)[k]$ 
is a smooth $n$-form $A$ (with sitting instants) on $U \times \Delta^k$ such that its 
curvature $(n+1)$-form $d A$ vanishes when restricted in all arguments to vector fields 
tangent to $\Delta^k$. We may write this condition as 
$d_{\mathrm{\mathrm{dR}}} A \in \Omega^{\bullet \geq 1, \bullet}_{\mathrm{\mathrm{si}}}(U \times \Delta^k)$.
\begin{observation} \label{LineCurvatureCharacteristicFromExp}
There are canonical morphisms 
$$
  \xymatrix{
    \mathbf{B}^n \mathbb{R}_{\mathrm{diff},\mathrm{smp}}
     \ar[r]^{\mathrm{curv}_{\mathrm{smp}}}
     \ar[d]^{\simeq}
    & 
     \mathbf{\flat}_{\mathrm{\mathrm{dR}}}\mathbf{B}^n \mathbb{R}_{\mathrm{smp}}
    \\
    \mathbf{B}^n \mathbb{R}_{\mathrm{smp}}
  }
$$
in $[\mathrm{CartSp}_{\mathrm{smooth}}^{\mathrm{op}}, \mathrm{sSet}]$, 
where the vertical map is given by remembering only the top horizontal morphism in 
the above square diagram, and the horizontal morphism is given by
forming the pasting composite
$$
  \begin{aligned}
  \mathrm{curv}_{\mathrm{smp}} 
   &: 
  \left\{
    \raisebox{20pt}{
    \xymatrix{
      \Omega^\bullet_{\mathrm{\mathrm{si}},\mathrm{\mathrm{vert}}}(U \times \Delta^k)
       & \ar[l]_{A_{\mathrm{vert}}}
       \mathrm{CE}(b^{n-1}\mathbb{R})
       \\
       \Omega_{\mathrm{si}}^\bullet(U \times \Delta^k) \ar[u]
       &
       \ar[l]_{A}
       \mathrm{W}(b^{n-1}\mathbb{R})
       \ar[u]
    }
    }
  \right\}
  \\
  & \mapsto
  \;\;
  \left\{
    \raisebox{20pt}{
    \xymatrix{
      \Omega^\bullet_{\mathrm{\mathrm{si}},\mathrm{\mathrm{vert}}}(U \times \Delta^k)
       &
       \ar[l]_{A_{\mathrm{vert}}}
        \mathrm{CE}(b^{n-1}\mathbb{R})
       &\ar[l]  0
       \\
       \Omega_{\mathrm{\mathrm{si}}}^\bullet(U \times \Delta^k)
       \ar[u]
        & \ar[l]_A \ar[u]
       \mathrm{W}(b^{n-1}\mathbb{R})
        &
        \ar[l] \ar[u]
       \mathrm{CE}(b^{n}\mathbb{R})
    }
    }
  \right\}  
  \end{aligned}
  \,.
$$
\end{observation}
\begin{proposition} \label{CurvSmp}
This span is a presentation in 
$[\mathrm{CartSp}_{\mathrm{smooth}}^{\mathrm{op}}, \mathrm{sSet}]$ of the 
universal curvature characteristics 
$\mathrm{curv} : \mathbf{B}^n \mathbb{R} \to \mathbf{\flat}_{\mathrm{\mathrm{dR}}} \mathbf{B}^{n+1} \mathbb{R}$,
def. \ref{UniversalCurvatureCharacteristic}, 
in $\mathrm{Smooth} \infty \mathrm{Grpd}$.
\end{proposition}
\proof
We need to produce a fibration resolution of the point inclusion
${*} \to \mathbf{\flat} \mathbf{B}^{n+1} \mathbb{R}_{\mathrm{smp}}$ 
in $[\mathrm{CartSp}_{\mathrm{smooth}}^{\mathrm{op}}, \mathrm{sSet}]_{\mathrm{proj}}$ 
and then show that
the above is the ordinary pullback of this along 
$\mathbf{\flat}_{\mathrm{\mathrm{dR}}} \mathbf{B}^{n+1} \mathbb{R}_{\mathrm{smp}} \to \mathbf{\flat} \mathbf{B}^{n+1} \mathbb{R}_{\mathrm{smp}} $.

We claim that this is achieved by the morphism
$$
  (U,[k]) 
  :   
  \{
    \Omega^\bullet_{\mathrm{\mathrm{si}}}(U \times \Delta^k)
    \leftarrow
    \mathrm{W}(b^{n-1} \mathbb{R})
  \}
  \mapsto 
  \{
    \Omega^\bullet_{\mathrm{\mathrm{si}}}(U \times \Delta^k)
    \leftarrow
    \mathrm{W}(b^{n-1} \mathbb{R})
    \leftarrow
    \mathrm{CE}(b^{n} \mathbb{R})
  \}
  \,.
$$
Here the simplicial presheaf on the left is that which assigns the set of arbitrary 
$n$-forms (with sitting instants but not necessarily closed) on $U \times \Delta^k$ 
and the map is simply given by sending such an $n$-form $A$ to the 
$(n+1)$-form $d_{\mathrm{\mathrm{dR}}} A$.

It is evident that the simplicial presheaf on the left resolves the point: since there 
is no condition on the forms every form on $U \times \Delta^k$ is in the image of 
the map of the normalized chain complex of a form on $U \times \Delta^{k+1}$: such is given 
by any form that is, up to a sign, equal to the given form on one $n$-face and 0 on all the other faces. 
Clearly such forms exist.

Moreover, this morphism is a fibration in 
$[\mathrm{CartSp}_{\mathrm{smooth}}^{\mathrm{op}}, \mathrm{sSet}]_{\mathrm{proj}}$, 
for instanxce because its image under the normalized chains complex functor is a 
degreewise surjection, by the Poincar{\'e} lemma.

Now we observe that we have over each $(U,[k])$ a double pullback diagram in Set
$$
  \begin{array}{ccc}
    \left\{
      \raisebox{20pt}{
      \xymatrix{
        \Omega^\bullet_{\mathrm{\mathrm{si}}, \mathrm{\mathrm{vert}}}(U \times \Delta^k)
        & \ar[l]_{A_{\mathrm{vert}}}
        \mathrm{CE}(b^{n-1}\mathbb{R})
        \\
        \Omega^\bullet_{\mathrm{si}}(U \times \Delta^k) \ar[u]
        &\ar[l]_{A}
        W(b^{n-1}\mathbb{R}) \ar[u]
      }
      }
    \right\}
    &\to&
    \left\{
      \raisebox{20pt}{
      \xymatrix{
        \Omega^\bullet_{\mathrm{\mathrm{si}}, \mathrm{\mathrm{vert}}}(U \times \Delta^k)
        &\ar[l]
        \mathrm{W}(b^{n-1}\mathbb{R})
        \\
        \Omega^\bullet_{\mathrm{si}}(U \times \Delta^k)\ar[u]
        &\ar[l]
        \mathrm{W}(b^{n-1} \mathbb{R})\ar[u]_{\mathrm{id}}
      }
      }
    \right\}
    \\
    \\
    \downarrow && \downarrow
    \\
    \\
    \left\{
      \raisebox{20pt}{
      \xymatrix{
        \Omega^\bullet_{\mathrm{\mathrm{si}}, \mathrm{\mathrm{vert}}}(U \times \Delta^k)
        & \ar[l]
        0
        \\
        \Omega^\bullet_{\mathrm{\mathrm{si}}}(U \times \Delta^k) \ar[u]
        & \ar[l]
        \mathrm{CE}(b^{n} \mathbb{R}) \ar[u]
      }
      }
    \right\}
    &\to&
    \left\{
      \raisebox{20pt}{
      \xymatrix{
        \Omega^\bullet_{\mathrm{\mathrm{si}},\mathrm{\mathrm{vert}}}(U \times \Delta^k)
        &\ar[l]
        \mathrm{CE}(b^{n} \mathbb{R})
        \\
        \Omega^\bullet_{\mathrm{si}}(U \times \Delta^k) \ar[u]
        & \ar[l]
        \mathrm{CE}(b^{n} \mathbb{R})    \ar[u]_{\mathrm{id}}
      }
      }
    \right\}
    \\
    \\
    \downarrow && \downarrow
    \\
    \\
    \left\{
      \raisebox{20pt}{
      \xymatrix{
        \Omega^\bullet_{\mathrm{\mathrm{si}},\mathrm{\mathrm{vert}}}(U \times \Delta^k)
        \ar@{<-}[r] & 0
        \\
        \Omega^\bullet_{\mathrm{\mathrm{si}}}(U \times \Delta^k) \ar[u]
        \ar@{<-}[r] &
        0 \ar[u]
      }
      }
    \right\}    
     &\to&
    \left\{
      \raisebox{20pt}{
      \xymatrix{
        \Omega^\bullet_{\mathrm{\mathrm{si}},\mathrm{\mathrm{vert}}}(U \times \Delta^k)
        &\ar[l]
         \mathrm{CE}(b^{n} \mathbb{R})
        \\
        \Omega^\bullet_{\mathrm{si}}(U \times \Delta^k) \ar[u]
        & \ar[l]
        0 \ar[u]
      }
      }
    \right\}    
  \end{array}
  \,,
$$
hence a corresponding pullback diagram of simplicial presheaves, that we claim 
is a presentation for the defining double $\infty$-pullback for $\mathrm{curv}$.

The bottom square is the one we already discussed for the de Rham coefficients. 
Since the top right vertical morphism is a fibration, also the top square is a homotopy 
pullback and hence exhibits the defining 
$\infty$-pullback for curv.
\endofproof
\begin{corollary}
The degreewise map
$$
  (-1)^{\bullet+1}
  \int_{\Delta^\bullet} :
  \mathbf{B}^n \mathbb{R}_{\mathrm{diff},\mathrm{smp}}
  \to  
  \mathbf{B}^n \mathbb{R}_{\mathrm{diff},\mathrm{chn}}
$$
that sends an $n$-form $A \in \Omega^n(U \times \Delta^k)$ and its 
curvature $d A$ to $(-1)^{k+1}$ times  its fiber integration $(\int_{\Delta^k} A, \int_{\Delta^k} d A)$ is a weak equivalence in 
$[\mathrm{CartSp}_{\mathrm{smooth}}^{\mathrm{op}}, \mathrm{sSet}]_{\mathrm{proj}}$.
\end{corollary}
\proof
Since under homotopy pullbacks a weak equivalence of diagrams is sent to a weak equivalence. 
See the analagous argument in the proof of prop. 
\ref{FiberIntegrationAsWeakEquivalenceForDeRhamCoefficientPresentations}.
\endofproof

\paragraph{Canonical form on a simplicial Lie group}
\label{CanoncalFormOnSimplicialLieGroup}

Above we discussed the canonical differential form on smooth $\infty$-groups $G$ for the special cases where 
$G$ is a Lie group and where $G$ is a circle Lie $n$-group. 
These are both in turn special cases of the situation where $G$ is a 
\emph{simplicial Lie group}. This we discuss now.

\begin{proposition}
For $G$ a simplicial Lie group
the flat de Rham coefficient object 
$\mathbf{\flat}_{\mathrm{dR}}\mathbf{B}G$ 
is presented by 
the simplicial presheaf which in degree $k$ is given by
$\Omega^1_{\mathrm{flat}}(-, \mathfrak{g}_k)$, 
where $\mathfrak{g}_k = \mathrm{Lie}(G_k)$ is the Lie algebra 
of $G_k$.
\end{proposition}
\proof
Let 
$$
  \Omega^1_{\mathrm{flat}}(-,\mathfrak{g}_\bullet)
  /\!/G_\bullet
  =
  \left(
  \Omega^1_{\mathrm{flat}}(-,\mathfrak{g}_\bullet) 
    \times
  C^\infty(-,G_\bullet) 
   \stackrel{\to}{\to} 
  \Omega^1_{\mathrm{flat}}(-,\mathfrak{g}_\bullet)
  \right)
$$
be the presheaf of simplicial groupoids which in degree $k$ is the groupoid of Lie-algebra valued forms with values in $G_k$ from 
theorem. \ref{GroupoidOfLieAlgebraValuedForms}. 
As in the proof of prop. \ref{LieGroupDeRhamCoefficients} 
we have that under the degreewise nerve this is a degreewise fibrant resolution of presheaves of bisimplicial sets
$$
  N \left(  \Omega^1_{\mathrm{flat}}(-,\mathfrak{g}_\bullet)
    /\!/
    G_\bullet
  \right)
   \to
  N */\!/G_\bullet
  =
  N B (G_{\mathrm{disc}})_\bullet
$$
of the standard presentation of the delooping of the discrete group underlying $G$. By basic properties of bisimplicial sets 
\cite{GoerssJardine}
we know that under taking the diagonal 
$$
  \mathrm{diag} : \mathrm{sSet}^\Delta \to \mathrm{sSet}
$$
the object on the right is a presentation for 
$\mathbf{\flat}_{\mathrm{dR}} \mathbf{B}G$, because
(see the discussion of simplicial groups around prop. \ref{InftyGroupsBySimplicialGroups} )
$$
  \mathrm{diag} N B (G_{\mathrm{disc}})_\bullet
   \stackrel{\simeq}{\to}
  \bar W (G_{\mathrm{disc}})
  \simeq
  \mathbf{\flat}\mathbf{B}G
  \,.
$$
Now observe that the morphism
$$
  \mathrm{diag} (N  \Omega^1_{\mathrm{flat}}(-,\mathfrak{g}_\bullet)
    /\!/
    G_\bullet
  )
   \to
  \mathrm{diag} N */\!/ G_{\mathrm{disc}}
$$
is a fibration in the global model structure. 
This is in fact true for every
morphism of the form
$$
  \mathrm{diag} N (S_\bullet/\!/G_\bullet) \to \mathrm{diag} */\!/G_\bullet
$$
for $S_\bullet/\!/G_\bullet \to */\!/G_\bullet$ a simlicial action groupoid projection with $G$ a simplicial group acting on a 
Kan complex $S$: we have that 
$$
  (\mathrm{diag} N (S/\!/G))_k 
   = 
  S_k \times (G_k)^{\times_k}
  \,.
$$
On the second factor the horn filling condition is simply that of the identity map $\mathrm{diag} N B G \to \mathrm{diag} N B G$ which is evidently solvable, whereas on the first factor it amounts to $S \to *$ being a Kan fibration, hence to $S$ being Kan fibrant.

But the simplicial presheaf 
$\Omega^1_{\mathrm{flat}}(-,\mathfrak{g}_\bullet)$ 
is indeed Kan fibrant: for a given $U \in \mathrm{CartSp}$ we may use parallel transport to (non-canonically) identify
$$
  \Omega^1_{\mathrm{flat}}(U, \mathfrak{g}_k)
   \simeq
  \mathrm{SmoothMfd}_*(U, G_k)
  \,,  
$$
where on the right we have smooth functions that send the origin of $U$ to the neutral element. But since $G_\bullet$ is Kan fibrant and has smooth global fillers also $\mathrm{SmoothMfd}_*(U,G_\bullet)$ 
is Kan fibrant.

In summary this means that the defining homotopy pullback 
$$
  \mathbf{\flat}_{\mathrm{dR}} \mathbf{B}G
  :=
  \mathbf{\flat} \mathbf{B}G \times_{\mathbf{B}G} *
$$
is presented by the ordinary pullback of simplicial presheaves
$$
  \mathrm{diag} N \Omega^1_{flat}(-,\mathfrak{g}_\bullet)
   \times
  \mathrm{diag} N B G_\bullet *
  =
  \Omega^1(-, \mathfrak{g}_\bullet)
  \,.
$$
\endofproof
\begin{proposition}
\index{Maurer-Cartan form!on simplicial Lie group}
For $G$ a simplicial Lie group the 
canonical differential form, def. \ref{UniversalFormOnInftyGroup},
$$
  \theta : G \to \mathbf{\flat}_{\mathrm{dR}} \mathbf{B}G
$$
is presented in terms of the above presentation for
$\mathbf{\flat}_{\mathrm{dR}} \mathbf{B}G$ 
by the morphism of simplicial presheaves
$$
  \theta_\bullet 
   : 
  G_\bullet
  \to 
  \Omega^1_{\mathrm{flat}}(-, \mathfrak{g}_\bullet)
$$
which is in degree $k$ 
the presheaf-incarnation of the Maurer-Cartan form of the ordinary Lie group $G_k$ as in prop. \ref{StandardMaurerCartanForm}.
\end{proposition}
\proof
Continuing with the strategy of the previous proof
we find a fibration resolution of 
the point inclusion $* \to \mathbf{\flat} \mathbf{B}G$ 
by applying the construction of the proof of 
prop. \ref{StandardMaurerCartanForm} 
degreewise and then applying $\mathrm{diag} \circ N$.

The defining homotopy pullback
$$
  \xymatrix{
     G \ar[r]\ar[d] & {*} \ar[d]
     \\
     \mathbf{\flat}_{\mathrm{dR}}
    \ar[r] &
    \mathbf{\flat} \mathbf{B}G
  }
$$
for $\theta$ is this way presented by the ordinary pullback
$$
  \xymatrix{
     G_\bullet \ar[r]\ar[d]&
     \mathrm{diag} N 
   ( \Omega^1_{\mathrm{flat}}(-, \mathfrak{g}_\bullet))_{\mathrm{triv}} /\!/ G_\bullet )
   \ar[d]
     \\
     \Omega^1_{\mathrm{flat}}(-, \mathfrak{g}_\bullet)
     \ar[r] &
     \mathrm{diag} N (\Omega^1_{\mathrm{flat}}(-,\mathfrak{g}_\bullet)/\!/G_\bullet)
  }
$$
of simplicial presheaves, where 
$\Omega^1_{\mathrm{flat}}(-,\mathfrak{g}_k)$ is the set of flat $\mathfrak{g}$-valued forms $A$ equipped with a gauge transformation $0 \stackrel{g}{\to} A$. As in the above proof one finds that the right vertical morphism is a fibration, hence indeed a resolution of the point inclusion. The pullback is degreewise that from the case of ordinary Lie groups and thus the result follows.
\endofproof

We can now give a simplicial description of the canonical curvature form $\theta : \mathbf{B}^n U(1) \to \mathbf{\flat}_{dR} \mathbf{B}^{n+1} U(1)$ that 
above in prop. \ref{CurvatureCharOnBnU1} we obtained by a chain complex model:
\begin{example}
The canonical form on the circle Lie $n$-group
$$
  \theta : \mathbf{B}^{n-1}U(1)
  \to 
  \mathbf{\flat}_{\mathrm{dR}} \mathbf{B}^n U(1)
$$
is presented by the simplicial map 
$$
  \Xi( U(1)[n-1] )
  \to 
  \Xi( \Omega^1_{cl}(-)[n-1] )
$$
which is simply the Maurer-Cartan form on $U(1)$ in degree $n$.

The equivalence to the model we obtained before is given by noticing the equivalence in hypercohomology of chain complexes of abelian sheaves
$$
  \Omega^1_{\mathrm{cl}}(-)[n]
   \simeq
  (\Omega^1(-) \stackrel{d_{\mathrm{dR}}}{\to}
   \cdots
   \stackrel{d_{\mathrm{dR}}}{\to}
   \Omega^n_{\mathrm{cl}}(-)
  ) 
$$ 
on $\mathrm{CartSp}$.
\end{example}

\subsubsection{Differential cohomology}
\label{SmoothStrucDifferentialCohomology}
\index{principal $\infty$-bundle!circle $n$-bundle with connection}
\index{connection!circle $n$-bundle with connection}
 \index{structures in a cohesive $\infty$-topos!differential cohomology!smooth}

We discuss the intrinsic differential cohomology, 
defined in \ref{StrucDifferentialCohomology} for any cohesive $\infty$-topos, 
realized in the context $\mathrm{Smooth} \infty \mathrm{Grpd}$, 
with coefficients in the circle Lie $(n+1)$-group
$\mathbf{B}^n U(1)$, def. \ref{Circlengroup}.\index{circle $n$-bundle with connection} 

We show that here the general concept reproduces the Deligne-Beilinson complex, 
\ref{DeligneComplex}, and generalizes it to 
a complex for equivariant differential cohomology for ordinary and 
twisted notions of equivariance.

\begin{itemize}
  \item \ref{CircleNBundlesWithConnection} -- The $n$-groupoid of circle-principal $n$-connections;
  \item \ref{UniversalModuliOfCirclenConnection} -- The universal moduli $n$-stack of circle-principal $n$-connections;
  \item \ref{EquivariantCircleNBundlesWithConnection} -- Equivariant circle $n$-bundles with connection;
\end{itemize}

The disucssion here proceeds in the un-stabilized cohesive $\infty$-topos
$\mathrm{Smooth}\infty \mathrm{Grpd}$. By embedding this into its 
tangent cohesive $\infty$-topos $T \mathrm{Smooth}\infty \mathrm{Grpd}$,
def. \ref{TangentInfinityTopos}, one obtains the characteristic curvature long exact sequences
discussed below from the general abstract discussion of 
prop. \ref{TangentCohesionDifferentialCohomology}

\medskip

\paragraph{The smooth $n$-groupoid of circle-principal $n$-connections}
\label{CircleNBundlesWithConnection}

Here we discuss some basic facts about differential cohomology with 
coefficients in the circle $n$-group, def. \ref{CircleNGroup}, that
are independent of a notion of manifolds and global differential form
objects as in \ref{GeneralAbstractDifferentialCohomologyWithDifferentialFormCurvature}. 
Further below in \ref{UniversalModuliOfCirclenConnection} we do consider
these structures and show that $\mathbf{B}^n U(1)_{\mathrm{conn}}$
is presented by the Deligne complex.

\medskip

Here we discuss first that
intrinsic differential cohomology in $\mathrm{Smooth}\infty \mathrm{Grpd}$ has
the abstract properties of traditional ordinary differential cohomology, 
\cite{HopkinsSinger},
then we establish that both notions indeed coincide in cohomology. The 
intrinsic definition refines this ordinary differential cohomology to 
moduli $\infty$-stacks.

By def. \ref{OrdinaryDiffCohomology} we are to consider the $\infty$-pullback
$$
  \xymatrix{
    \mathbf{H}_{\mathrm{diff}}(X,\mathbf{B}^n U(1)) 
     \ar[r]
     \ar[d] 
     & H_{\mathrm{\mathrm{dR}}}(X,\mathbf{B}^{n+1} U(1))
     \ar[d]
    \\
    \mathbf{H}(X,\mathbf{B}^n U(1))
    \ar[r]^{\mathrm{curv}} &
    \mathbf{H}_{\mathrm{\mathrm{dR}}}(X, \mathbf{B}^{n+1} U(1))
  }
  \,,
$$
where the right vertical morphism picks one point in each connected component.
Moreover, using prop. \ref{OrdinaryDeRham} in def. \ref{BnAconn},
we are entitled to the following bigger object.
\begin{definition}
 \label{SmoothDiffCohWithFormTwist}
 For $n \in \mathbb{N}$ write $\mathbf{B}^n U(1)_{\mathrm{conn}}$
 for the $\infty$-pullback
 $$
   \raisebox{20pt}{
   \xymatrix{
     \mathbf{B}^{n}U(1)_{\mathrm{conn}}
	 \ar[r]
	 \ar[d]
	 &
	 \Omega^{n+1}_{\mathrm{cl}}(-)
	 \ar[d]
	 \\
	 \mathbf{B}^n U(1)
	 \ar[r]^{\mathrm{curv}}
	 &
	 \mathbf{\flat}_{\mathrm{dR}}\mathbf{B}^{n+1}U(1)
   }
   }
 $$
in $\mathrm{Smooth}\infty\mathrm{Grpd}$. 
The cocycle $\infty$-groupoid over some $X \in \mathrm{Smooth}\infty \mathrm{Grpd}$ 
with coefficients in $\mathbf{B}^n U(1)_{\mathrm{conn}}$ is the $\infty$-pullback
$$
  \xymatrix{
    \mathbf{H}(X, \mathbf{B}^n U(1)_{\mathrm{conn}})
	\ar@{}[r]|\simeq
    &
    \mathbf{H}'_{\mathrm{diff}}(X,\mathbf{B}^n U(1)) 
     \ar[r]^{F}
     \ar[d]^{\mathbf{c}} 
     & \Omega_{\mathrm{cl}}^{n+1}(X)
     \ar[d]
    \\
    &\mathbf{H}(X,\mathbf{B}^n U(1))
    \ar[r]^{\mathrm{curv}} &
    \mathbf{H}_{\mathrm{\mathrm{dR}}}(X, \mathbf{B}^{n+1} U(1))
  }
  \,.
$$
\label{DiffModuliStack}
\end{definition}
We call $\mathbf{H}_{\mathrm{diff}}(X,\mathbf{B}^n U(1))$ and its primed version the
cocycle $\infty$-groupoid for \emph{ordinary smooth differential cohomology} in degree $n$
\index{differential cohomology!smooth}.
\begin{proposition}
  \label{TraditionalDiffcohomologyExactSequencesReproduced}
  For $n \geq 1$ and $X \in \mathrm{SmoothMfd}$, 
  the abelian group ${H'}^n_{\mathrm{diff}}(X)$ sits in the following short exact 
  sequences of abelian  groups
  \begin{itemize}
    \item the \emph{curvature exact sequence}\index{fiber sequence!curvature}
    $$
      0 
        \to 
      H^n(X, U(1)_{\mathrm{disc}}) 
        \to
      {H'}^n_{\mathrm{diff}}(X,U(1))
        \stackrel{F}{\to}
      \Omega^{n+1}_{\mathrm{cl}, \mathrm{int}}(X)
        \to
      0
    $$
    \item the \emph{characteristic class exact sequence}
    $$
      0 
        \to 
      \Omega_{\mathrm{cl}}^{n}/\Omega^n_{\mathrm{cl},\mathrm{int}}(X)
        \to 
      {H'}^n_{\mathrm{diff}}(X,U(1))
        \stackrel{\mathbf{c}}{\to} 
      H^{n+1}(X, \mathbb{Z}) 
        \to 
      0 
      \,.
    $$
  \end{itemize}
  Here $\Omega^n_{\mathrm{cl}, \mathrm{int}}$ denotes closed forms with integral periods.
\end{proposition}
\proof
    For the curvature exact sequence we invoke prop. \ref{CurvatureExactSequence},
    which yields (for $H_{\mathrm{diff}}$ as for $H'_{\mathrm{diff}}$)
    $$
      0 
        \to 
      H^n_{\mathrm{flat}}(X, U(1)) 
        \to
      {H'}^n_{\mathrm{diff}}(X,U(1))
        \stackrel{F}{\to}
      \Omega^{n+1}_{\mathrm{cl}, \mathrm{int}}(X)
        \to
      0
      \,.
    $$     
    The claim then follows by using prop. \ref{FlatSmoothCohomology}
    to get $H^n_{\mathrm{flat}}(X, U(1)) \simeq H^n(X, U(1)_{\mathrm{disc}})$.
    
  For the characteristic class exact sequence, we have with 
  \ref{ShortExactSequenceForIntrinsicOrdinaryDiffCohomology}
  for the 
  smaller group $H^n_{\mathrm{diff}}$ (the fiber over the vanishing curvature $(n+1)$-form $F = 0$) 
  the sequence
    $$
      0 
        \to 
      H_{\mathrm{dR}}^n(X)/\Omega^n_{\mathrm{cl},\mathrm{int}}(X)
        \to
      {H'}^n_{\mathrm{diff}}(X,U(1))
        \stackrel{c}{\to}
      H^{n+1}(X, \mathbb{Z})
        \to
      0
      \,
    $$  
    where we used prop. \ref{OrdinaryFromIntrinsicDeRham} to identify the de Rham cohomology
    on the left, and the fact that $X$ is paracompact to 
    identify the integral cohomology on the right.
    Since $\Omega^n_{\mathrm{cl},\mathrm{int}}(X)$ contains the exact forms (with all periods
   being $0 \in \mathbb{Z}$), the leftmost term is equivalently
   $\Omega^n_{\mathrm{cl}}(X)/\!/\Omega^n_{\mathrm{cl},\mathrm{int}}(X)$. As we pass
   from $H_{\mathrm{diff}}$ to the bigger $H'_{\mathrm{diff}}$, we get a copy of a 
   torsor over this group, for each closed form $F$, trivial in de Rham cohomology, to a total of
   $$
     \coprod_{F \in \Omega^{n+1}_{\mathrm{cl}}(X)} \{\omega | d \omega = F\}/\Omega^n_{\mathrm{cl}, \mathrm{int}}
     \simeq
     \Omega^n(X)/\Omega^n_{\mathrm{cl}, \mathrm{int}}(X)
     \,.
   $$
   This yields the curvature exact sequence as claimed.   
\endofproof
If we invoke standard facts about Deligne cohomology, then 
prop. \ref{TraditionalDiffcohomologyExactSequencesReproduced} is also implied
by the following proposition, which asserts that in 
$\mathrm{Smooth}\infty \mathrm{Grpd}$ the groups ${H'}^\bullet_{\mathrm{diff}}$ not only
share the above abstract properties of ordinary differential cohomology, but indeed coincide with it.
\begin{theorem} 
 \label{DeligneCohomologyTheorem}
For $X \in \mathrm{SmoothMfd} \hookrightarrow \mathrm{Smooth} \infty \mathrm{Grpd}$ 
a paracompact smooth manifold we have that the connected components of the
object $\mathbf{H}_{\mathrm{diff}}(X, \mathbf{B}^n U(1))$
are given by
$$
  H^n_{\mathrm{diff}}(X, U(1))
  \simeq
  \left( 
     \;\; H(X,\mathbb{Z}(n+1)_D^\infty) \;\;
  \right)
  \times_{\Omega_{\mathrm{cl}}^{n+1}(X)} H_{\mathrm{\mathrm{dR}}, \mathrm{int}}^{n+1}(X)
  \,.
$$
Here on the right we have the subset of Deligne cocycles that picks for each integral de Rham cohomology class of 
$X$ only one curvature form representative.

For the connected components of $\mathbf{H}'_{\mathrm{diff}}(X, \mathbf{B}^n U(1))$ we get 
the complete ordinary Deligne cohomology of $X$ in degree $n+1$:
$$
   {H'}^n_{\mathrm{diff}}(X,U(1))
  \simeq
     \;\; H(X,\mathbb{Z}(n+1)_D^\infty) \;\;
$$
\end{theorem}
\proof
Choose a differentiably good open cover, def. \ref{DifferentiablyGoodOpenCover}, 
$\{U_i \to X\}$ and let $C(\{U_i\}) \to X$  in $[\mathrm{CartSp}^{\mathrm{op}}, \mathrm{sSet}]_{\mathrm{proj}}$ 
be the corresponding {\v C}ech nerve projection, a cofibrant resolution of $X$.
 
Since the presentation of prop. \ref{CurvatureCharOnBnU1} for the universal curvature class 
$\mathrm{curv}_{\mathrm{chn}} : 
\mathbf{B}^n U(1)_{\mathrm{diff},\mathrm{chn}} \to \mathbf{\flat}_{\mathrm{\mathrm{dR}}} \mathbf{B}^{n+1}U(1)_{\mathrm{chn}}$
is a global fibration and $C(\{U_i\})$ is cofibrant, also 
$$
  [\mathrm{Cartp}^{\mathrm{op}}, \mathrm{sSet}](C(\{U_i\}), \mathbf{B}^n_{\mathrm{diff}}U(1))
  \to
  [\mathrm{Cartp}^{\mathrm{op}}, \mathrm{sSet}](C(\{U_i\}), \mathbf{\flat}_{\mathrm{\mathrm{dR}}}\mathbf{B}^n U(1))
$$
is a Kan fibration by the fact that $[\mathrm{CartSp}^{\mathrm{op}}, \mathrm{sSet}]_{\mathrm{proj}}$ 
is an $\mathrm{sSet}_{\mathrm{Quillen}}$-enriched model category. Therefore the homotopy pullback 
in question is computed as the ordinary pullback of this morphism.

By prop. \ref{OrdinaryDeRham} we can assume that
the morphism $H_{\mathrm{\mathrm{dR}}}^{n+1}(X) \to [\mathrm{CartSp}^{\mathrm{op}}, \mathrm{sSet}](C(\{U_i\}), \mathbf{\flat}_{\mathrm{\mathrm{dR}}}\mathbf{B}^{n+1})$ picks only cocycles represented by 
globally defined closed differential forms $F \in \Omega^{n+1}_{\mathrm{\mathrm{cl}}}(X)$.
We see that the elements in the fiber over such a globally defined $(n+1)$-form $F$ 
are precisely the cocycles with values only in the upper row complex of 
$\mathbf{B}^{n}U(1)_{\mathrm{diff},\mathrm{chn}}$
$$
  C^\infty(-,U(1)) \stackrel{d_{\mathrm{\mathrm{dR}}}}{\to} \Omega^1(-)
  \stackrel{d_{\mathrm{\mathrm{dR}}}}{\to}
  \cdots 
  \stackrel{d_{\mathrm{\mathrm{dR}}}}{\to}
  \Omega^n(-)
  \,,
$$
such that $F$ is the de Rham differential of the last term.
This is the Deligne-Beilinson complex, def. \ref{DeligneComplex},
for Deligne cohomology in degree $(n+1)$.
\endofproof
In terms of def. \ref{BnAconn} we have the object 
$\mathbf{B}^n U(1)_{\mathrm{conn}}$ 
-- the \emph{moduli $n$-stack of circle $n$-bundles with connection} --
which presents $\mathbf{H}'_{\mathrm{diff}}(-,\mathbf{B}^n U(1))$ 
$$
  \mathbf{H}'_{\mathrm{diff}}(-, \mathbf{B}^n U(1))
  \simeq
  \mathbf{H}(-, \mathbf{B}^n U(1)_{\mathrm{conn}})
  \,.
$$

\paragraph{The universal moduli $n$-stack of circle-principal $n$-connections}
\label{UniversalModuliOfCirclenConnection}

\begin{definition}
  For $n \in \mathbb{N}$ and $k \leq n$ write
  $$
    \Omega^{k \leq \bullet \leq n}_{\mathrm{cl}}
	:=
	\mathrm{DK}
	\left(
	  \xymatrix{
	    \Omega^{k} 
		\ar[r]^{d_{\mathrm{dR}}}
		&
		\Omega^{k+1}
		\ar[r]
		&
		\cdots
		\ar[r]^{d_{\mathrm{dR}}}
		&
		\Omega^{n-1}
		\ar[r]^{d_{\mathrm{dR}}}
		&
		\Omega^{n}_{\mathrm{cl}}
	  }
	\right)
	\,.
  $$
  Write 
  $$
    \mathbf{B}^n U(1)_{\mathrm{conn}^k,\mathrm{chn}}
	:=
	\mathrm{DK}
	\left(
	 \xymatrix{
	  U(1)
	  \ar[r]^{d_{\mathrm{dR}}}
	  \Omega^1
	  \ar[r]
	  &
	  \cdots
	  \ar[r]^{d_{\mathrm{dR}}}
	  &
	  \Omega^k
	  \ar[r]
	  &
	  0
	  \ar[r]
	  &
	  \cdots
	  \ar[r]
	  &
	  0
	 }
	\right)
  $$
  for the simplicial presheaf which is the image under the Dold-Kan map of the
  chain complex concentrated in degrees $n$ through $(n-k)$, as indicated.
  Notice that 
  $$
    \mathbf{B}^n U(1)_{\mathrm{conn}^0, \mathrm{chn}} = \mathbf{B}^n U(1)_{\mathrm{chn}}
	\,,
  $$
  and we write
  $$
    \mathbf{B}^n U(1)_{\mathrm{conn},\mathrm{chn}} := \mathbf{B}^n U(1)_{\mathrm{conn}^n,\mathrm{chn}}
	\,.
  $$
  \label{DKPresentationOfU1Coefficients}
\end{definition}

\begin{proposition}
  The object $\mathbf{B}^n U(1)_{\mathrm{conn}}^k \in \mathrm{Smooth}\infty\mathrm{Grpd}$
  is presented in $[\mathrm{CartSp}^{\mathrm{op}}, \mathrm{sSet}]_{\mathrm{proj},\mathrm{loc}}$
  by $\mathbf{B}^n U(1)_{\mathrm{conn}^k,\mathrm{chn}}$.
\end{proposition}
\proof
By prop. \ref{PresentationOfUniversalU1CurvatureCharacteristic} the defining $\infty$-pullback
$$
  \xymatrix{
    \mathbf{B}^n U(1)_{\mathrm{conn}^k}
	\ar[r]^{F_{(-)}}
	\ar[d]
	&
	\Omega^{k \leq \bullet \leq n}_{\mathrm{cl}}
	\ar[d]
	\\
	\mathbf{B}^n U(1)
	\ar[r]^-{\mathrm{curv}}
	&
	\flat_{\mathrm{dR}}\mathbf{B}^{n+1}U(1)
  }
$$
is presented by the homotopy pullback of presheaves of chain complexes
$$
  \xymatrix{
    \mathbf{B}^n U(1)_{\mathrm{diff},\mathrm{chn}} 
	\ar[d]^{\mathrm{curv}_{\mathrm{chn}}}
	\ar@{<-}[r] & 
	\mathbf{B}^n U(1)_{\mathrm{conn}^k, \mathrm{chn}}
	\ar[d]
	\\
	\flat_{\mathrm{dR}}\mathbf{B}^{n+1}U(1)_{\mathrm{chn}}
	\ar@{<-}[r]
	&
	\Omega^{k \leq \bullet \leq n}_{\mathrm{cl}}
  }
$$
(rotated here just for readability in the following)
which in components is given as follows
$$
 \hspace{-1cm}
  \xymatrix@R=1pt{
    U(1) \ar[r]^{d_{\mathrm{dR}}} 
	& \cdots \ar[r]^{d_{\mathrm{dR}}} 
	& \Omega^1 \ar[r] 
	& \cdots \ar[r] \ar@{<..}[ddddrrrrrr]
	& \Omega^{n-1}  \ar[r]^{d_{\mathrm{dR}}} \ar@{<..}[ddddrrrrrr]
	& \Omega^n \ar@{<..}[ddddrrrrrr]
	\ar[ddddddddddd]|{d_{\mathrm{dR}}}
	\\
    \oplus &  & \oplus && \oplus
    \\
    \Omega^1 \ar[r]^{d_{\mathrm{dR}}}  \ar[ddddddddd] \ar@{<..}[ddddrrrrrr]
	 \ar[ddddddddd]|{(-1)^n\mathrm{id}}
	& \cdots \ar[r]_{d_{\mathrm{dR}}}   
	& \Omega^k \ar[r] \ar[ddddddddd]|{(-1)^n\mathrm{id}} \ar@{<..}[ddddrrrrrr]
	&  \cdots \ar[r]_{d_{\mathrm{dR}}} 
	& 
	\Omega^n 
	\ar[uur]|{(-1)^{n}\mathrm{id}} \ar[ddddddddd]|{(-1)^n\mathrm{id}}  \ar@{<..}[ddddrrrrrr]
	& 
	\\
	\\
    &&&&&&
	U(1)
	\ar[r]^{d_{\mathrm{dR}}}
    & \cdots \ar[r]^{d_{\mathrm{dR}}} 
	& \Omega^k \ar[r] 
	& \cdots \ar[r] 
	& \Omega^{n-1}  \ar[r]^{d_{\mathrm{dR}}} & \Omega^n
	\ar[ddddddddddd]|{d_{\mathrm{dR}}}
	\\
	&&&&&&
    \oplus &  & \oplus && \oplus
    \\
	&&&&&&
    0 \ar[r]  \ar[ddddddddd]
	& \cdots \ar[r]  
	& \Omega^{k+1} \ar[r]^{d_{\mathrm{dR}}}  \ar[ddddddddd]|{(-1)^n\mathrm{id}}
	&  \cdots \ar[r]_{d_{\mathrm{dR}}} & 
	\Omega^n 
	\ar[uur]|{(-1)^{n}\mathrm{id}} \ar[ddddddddd]|{(-1)^n\mathrm{id}} & 
	\\
	\\
	\\
	\\
	\\
	\Omega^1 \ar[r]^{d_{\mathrm{dR}}}   \ar@{<..}[ddddrrrrrr]
	& \cdots \ar[r]^{d_{\mathrm{dR}}}  \ar@{<..}[ddddrrrrrr]
	&
	\Omega^{k+1} \ar[r]^{d_{\mathrm{dR}}}  \ar@{<..}[ddddrrrrrr]
	&
	\cdots \ar[r] 
	& \Omega^n \ar[r]^{d_{\mathrm{dR}}}  \ar@{<..}[ddddrrrrrr]
	& \Omega^{n+1}_{\mathrm{cl}}  \ar@{<..}[ddddrrrrrr]
	\\
	\\
	\\
	\\
	&&&&&&
	0 \ar[r] 
	& \cdots \ar[r] 
	&
	\Omega^{k+1}
	\ar[r]^{d_{\mathrm{dR}}}
	&
	\cdots \ar[r]^{d_{\mathrm{dR}}} 
	& \Omega^n \ar[r]^-{d_{\mathrm{dR}}} 
	& \Omega^{n+1}_{\mathrm{cl}} 
  }
  \,.
$$
This shows that $\mathbf{B}^n U(1)_{\mathrm{conn}^k}$ is presented by the 
chain complex appearing on the top right here. The canonical projection morphism
from this pullback to $\mathbf{B}^n U(1)_{\mathrm{conn}^k,\mathrm{chn}}$ is 
clearly a weak equivalence. 
\endofproof
\begin{remark}
  In particular this means that $\mathbf{B}^n U(1)_{\mathrm{conn}}$ is presented by the 
  Deligne complex
  $$
    \mathbf{B}^n U(1)_{\mathrm{conn}}
	\simeq
	\mathrm{DK}
	\left(
	  \xymatrix{
	    U(1)
		\ar[r]^{d_{\mathrm{dR}}}
		&
		\Omega^1
		\ar[r]^{d_{\mathrm{dR}}}
		&
		\cdots
		\ar[r]
		&
		\Omega^{n-1}
		\ar[r]^{d_{\mathrm{dR}}}
		&
		\Omega^{n}
	  }
	\right)
  $$
\end{remark}

The above proof of theorem \ref{DeligneCohomologyTheorem} 
makes a statement not only about cohomology classes, but about the
full moduli $n$-stacks:
\begin{proposition}
  \label{BnU1conn}
  The object $\mathbf{B}^n U(1)_{\mathrm{conn}} \in \mathbf{H}$
  from def. \ref{SmoothDiffCohWithFormTwist}
  is presented by the simplicial presheaf which is the image under
  the 
  Dold-Kan map $\Xi$, def. \ref{EmbeddingOfChainComplexes}, 
  of the Deligne complex in the corresponding degree.
  
  The canonical morphism $\mathbf{B}^n U(1)_{\mathrm{conn}} \to \mathbf{B}^n U(1)$
  is similarly presented via Dold-Kan of the evident morphism of chain 
  complexes of sheaves
  $$
    \raisebox{20pt}{
    \xymatrix{
	   C^\infty(-, U(1))
	   \ar[r]^{d_{\mathrm{dR}} \mathrm{log}}
	   \ar[d]^{\mathrm{id}}
	   &
	   \Omega^1(-)
	   \ar[r]^{d_{\mathrm{dR}}}
	   \ar[d]
	   &
	   \cdots
	   \ar[r]^{d_{\mathrm{dR}}}
	   &
	   \Omega^n(-)
	   \ar[d]
	   \\
	   C^\infty(-, U(1))
	   \ar[r]
	   &
	   0
	   \ar[r]
	   &
	   \cdots
	   \ar[r]
	   &
	   0
	}
	}
	\,.
  $$
\end{proposition}

\begin{proposition}
  The moduli stack $\mathbf{B}U(1)_{\mathrm{conn}}$ 
  of circle bundles (i.e. circle 1-bundles) with connection is 1-concrete, 
  def. \ref{ConcreteObjects}.
\end{proposition}
\proof
 Observing that the presentation by the Deligne complex under the
 Dold-Kan map is fibrant in 
 $[\mathrm{CartSp}^{\mathrm{op}}, \mathrm{sSet}]_{\mathrm{proj}, \mathrm{loc}}$
 and is the concrete sheaf presented by $U(1)$ in degree 1, this follows with
 prop. \ref{1ConcreteOverInfinityCohesiveSite}.
\endofproof

\paragraph{The smooth moduli of connections over a given base}
\label{SmoothStructDifferentialModuli}
\index{connection!circle $n$-bundle with connection!differential moduli}
 \index{structures in a cohesive $\infty$-topos!differential cohomology!differential moduli}

We discuss the \emph{moduli stacks} of higher principal connections, 
over a fixed $X \in \mathrm{Smooth}\infty \mathrm{Grpd}$, following the general abstract
discussion in \ref{StrucDifferentialModuli}.

For $n \in \mathbb{N}$ and with $\mathbf{B}^n U(1)_{\mathrm{con}n} \in \mathrm{Smooth}\infty\mathrm{Grpd}$  
the universal moduli stack for circle $n$-bundles with connection, def. \ref{BnU1connFromDK}, 
and for $X \in \mathrm{Smooth}\infty \mathrm{Grpd}$, one may be tempted to regard the internal hom/mapping space 
$[X, \mathbf{B}^n U(1)_{\mathrm{conn}}]$ as the moduli stack of circle $n$-bundles 
with connection on $X$. However, for $U \in \mathrm{CartSp}$  an abstract coordinate system, 
$U$-plots and their $k$-morphisms in  $[X, \mathbf{B}^n U(1)_{\mathrm{conn}}]$ 
are circle principal $n$-connections and their $k$-fold gauge transformations 
on $U \times X$, and this is not generally what one would want the $U$-plots 
of the moduli stack of such connections on $X$ to be. Rather, that moduli stack should have
\begin{enumerate}
 \item as $U$-plots smoothly $U$-parameterized collections $\{\nabla_u\}$ of $n$-connections on $X$;
\item  as $k$-morphisms smoothly $U$-parameterized collections $\{\phi_u\}$ of gauge 
transformations between them.
\end{enumerate}
The first item is equivalent to: a single $n$-connection on $U \times X$ such that 
its local connection $n$-forms have no legs along $U$. 
This is essentially the situation of moduli of differential forms which we have discussed above (...).

But the second item is different: a gauge transformation of a 
single $n$-connection $\nabla$ on $U \times X$ needs to respect the curvature 
of the connection along $U$, but a family $\{\phi_u\}$ of gauge tranformations 
between the restrictions $\nabla|_u$ of $\nabla$ to points of the coordinate patch $U$ need not.

In order to capture this correctly, the concretification-process that yields the moduli spaces of differential forms is to be refined to a process that concretifies the higher stack 
$[X, \mathbf{B}^n U(1)_{\mathrm{conn}}]$ degreewise in stages.

\begin{definition}
  For $n,k\in \mathbb{N}$ and $k \leq n$ write
  $\mathbf{B}^n U(1)_{\mathrm{conn}^k}$ for the $\infty$-pullback in 
  $$
    \xymatrix{
	  \mathbf{B}^n U(1)_{\mathrm{conn}^k}
	  \ar[d]
	  \ar[r]
	  &
	  \Omega^{n+1 \leq \bullet \leq k}_{\mathrm{cl}}
	  \ar[d]
	  \\
	  \mathbf{B}^n U(1)
	  \ar[r]^-{\mathrm{curv}}
	  &
	  \flat_{\mathrm{dR}}\mathbf{B}^{n+1}U(1)
	}
	\,.
  $$
  By the universal property of the $\infty$-pullback, the canonical tower of morphisms
  $$
    \xymatrix{
	  \Omega_{\mathrm{cl}}^{n+1}
	  \ar[r]
	  &
	  \Omega_{\mathrm{cl}}^{n+1 \leq \bullet \leq n}
	  \ar[r]
	  &
	  \cdots
	  \ar[r]
	  &
	  \Omega_{\mathrm{cl}}^{n+1 \leq \bullet \leq 1}
	  \ar[r]^-\simeq
	  &
      \flat_{\mathrm{dR}}\mathbf{B}^{n+1}U(1)
	}
  $$
  induces a tower of morphisms
  $$
    \xymatrix{
	  \mathbf{B}^n U(1)_{\mathrm{conn}}
	  \ar[r]^\simeq
	  &
	  \mathbf{B}^n U(1)_{\mathrm{conn}^n}
	  \ar[r]
	  &
	  \mathbf{B}^n U(1)_{\mathrm{conn}^{n-1}}
	  \ar[r]
	  &
	  \cdots
	  \ar[r]
	  &
	  \mathbf{B}^n U(1)_{\mathrm{conn}^0}
	  \ar[r]^\simeq
	  &
	  \mathbf{B}^n U(1)
	}
	\,.
  $$
  \label{BnU1k}
  \label{BundleGerbeWithConnectionButWithoutCurving}
\end{definition}
\begin{proposition}
  We have 
  $$
    \mathbf{B}^n U(1)_{\mathrm{conn}^k}
	\simeq
	\mathrm{DK}
	\left(
	  \xymatrix{
	  U(1) \ar[r]^{d_{\mathrm{dR}}}
	  &
	  \Omega^1 
	  \ar[r]^{d_{\mathrm{dR}}}
	  &
	  \cdots
	  \ar[r]^{d_{\mathrm{dR}}}
	  &
	  \Omega^k
	  \ar[r]^{d_{\mathrm{dR}}}
	  &
	  0
	  \ar[r]
	  &
	  \cdots
	  \ar[r]
	  &
	  0
	  }
	\right)
	\,
  $$
  where the chain complex on the right is concentrated in degrees $n$ through $n-k$.
  Under this equivalence the canonical morphism 
  $\mathbf{B}^n U(1)_{\mathrm{conn}^{k+1}}\to \mathbf{B}^n U(1)_{\mathrm{conn}^{k}}$
  is equivalent to the image under $\mathrm{DK}$ to the chain map
  $$
  \raisebox{20pt}{
  \xymatrix{
	  U(1) \ar[r]^{d_{\mathrm{dR}}}
	  \ar[d]^{\mathrm{id}}
	  &
	  \Omega^1 
	  \ar[r]^{d_{\mathrm{dR}}}
	  \ar[d]^{\mathrm{id}}
	  &
	  \cdots
	  \ar[r]^{d_{\mathrm{dR}}}
	  \ar[d]^{\mathrm{id}}
	  &
	  \Omega^{k+1}
	  \ar[r]^{d_{\mathrm{dR}}}
	  \ar[d]^{\mathrm{id}}
	  &
	  \Omega^k
	  \ar[r]
	  \ar[d]
	  &
	  0
	  \ar[r]
	  \ar[d]
	  &
	  \cdots
	  \ar[r]
	  \ar[d]
	  &
	  0
	  \ar[d]
	  \\
	  U(1) \ar[r]^{d_{\mathrm{dR}}}
	  &
	  \Omega^1 
	  \ar[r]^{d_{\mathrm{dR}}}
	  &
	  \cdots
	  \ar[r]^{d_{\mathrm{dR}}}
	  &
	  \Omega^{k+1}
	  \ar[r]^{d_{\mathrm{dR}}}
	  &
	  0
	  \ar[r]
	  &
	  0
	  \ar[r]
	  &
	  \cdots
	  \ar[r]
	  &
	  0
	  }
	  }
	$$
\end{proposition}
\proof
  By the presentation of $\mathrm{curv}$ as in prop. \ref{PresentationOfUniversalU1CurvatureCharacteristic}.
\endofproof

\begin{definition}
  For $X \in \mathbf{H}$ and $n \in \mathbb{N}$, $n \geq 1$, the
  \emph{moduli of circle-principal $n$-connections} on $X$ is 
  the iterated $\infty$-fiber product
  $$
    \begin{aligned}
    &(\mathbf{B}^{n-1}U(1))\mathbf{Conn}(X)
	\\
	&
	:=
	\sharp_1 [X, \mathbf{B}^n U(1)_{\mathrm{conn}^n}]
	\underset{\sharp_1 [X, \mathbf{B}^n U(1)_{\mathrm{conn}^{n-1}}]}{\times}
	\sharp_2 [X, \mathbf{B}^n U(1)_{\mathrm{conn}^{n-1}}]
	\underset{\sharp_2 [X, \mathbf{B}^n U(1)_{\mathrm{conn}^{n-2}}]}{\times}
	\cdots
	\underset{\sharp_{n} [X, \mathbf{B}^n U(1)_{\mathrm{conn}^0}]}{\times}	
	[X, \mathbf{B}^n U(1)_{\mathrm{conn}^0}]
	\end{aligned}
	\,,
  $$
  of the morphisms
  $$
    \xymatrix{
      \sharp_k [X,\mathbf{B}^n U(1)_{\mathrm{conn}^{n-k+1}}]
	  \ar[r]
	  &
	  \sharp_k [X,\mathbf{B}^n U(1)_{\mathrm{conn}^{n-k}}]
	}
  $$ 
  which are the image under $\sharp_k$, def. \ref{ImagesOfXToSharpX},
  of the image under the internal hom $[X, -]$ of the canonical projections of
  prop. \ref{BnU1k}, and of the morphisms
  $$
    \xymatrix{
      \sharp_{k+1} [X,\mathbf{B}^n U(1)_{\mathrm{conn}^{n-k}}]
	  \ar[r]
	  &
	  \sharp_k [X,\mathbf{B}^n U(1)_{\mathrm{conn}^{n-k}}]
	}
  $$ 
  of def. \ref{ImagesOfXToSharpX}.
  \label{DifferentialU1ModuliByIteratedPullback}
\end{definition}

\subparagraph{Moduli of smooth principal 1-connections}
\label{TheModuliOfCircleConnections}

We discuss the general notion of 
moduli of $G$-principal connections, def. \ref{DifferentialModuliByIteratedPullback}
for the special case that $G$ is a 0-truncated group.

For $G = U(1)$ the circle group, the special case of def. \ref{DifferentialU1ModuliByIteratedPullback} 
is the following.
\begin{definition}
 For $X \in \mathrm{Smooth}\infty\mathrm{Grpd}$, 
 the \emph{moduli of circle-principal connections} is given by the $\infty$-pullback
 $$
 \xymatrix{
    U(1)\mathbf{Conn}(X)
	\ar[rr]
	\ar[d]
	&&
	\sharp_2 [X, \mathbf{B}U(1)]
	\ar[d]
	\ar@{}[r]|\simeq & [X, \mathbf{B}U(1)]
	\\
	\sharp_1[X, \mathbf{B}U(1)_{\mathrm{conn}}]
	\ar[rr]^{\sharp_1 [X, U_{\mathbf{B}U(1)}] }
	&&
	\sharp_1 [X, \mathbf{B}U(1)]
  }
  \,,
  $$
  where $U_{\mathbf{B}U(1)} : \mathbf{B}U(1)_{\mathrm{conn}} \to \mathbf{B}U(1)$
  is the canonical forgetful morphism.
  \label{U1ConnXByFiberProductOfSharpn}
 \end{definition}
Of course we have the analogous construction for $G$ any Lie group:
\begin{definition}
 For $X \in \mathrm{Smooth}\infty\mathrm{Grpd}$, 
 the \emph{moduli of circle-principal connections} is given by the $\infty$-pullback
 $$
 \xymatrix{
    G\mathbf{Conn}(X)
	\ar[rr]
	\ar[d]
	&&
	\sharp_2 [X, \mathbf{B}G]
	\ar[d]
	\ar@{}[r]|\simeq & [X, \mathbf{B}G]
	\\
	\sharp_1[X, \mathbf{B}G_{\mathrm{conn}}]
	\ar[rr]^{\sharp_1 [X,U_{\mathbf{B}G}]}
	&&
	\sharp_1 [X, \mathbf{B}G]
  }
  \,,
  $$
  where $U_{\mathbf{B}G} : \mathbf{B}G_{\mathrm{conn}} \to \mathbf{B}G$
  is the canonical forgetful morphism.
  \label{GLieGroupBGConnXByFiberProductOfSharpn}
 \end{definition}

\begin{proposition}
\label{U1ConnXIsIndeedDifferentialModuli}
For $X \in \mathrm{SmthMfd} \hookrightarrow \mathrm{Smooth}\infty\mathrm{Grpd}$, 
the smooth groupoid $U(1)\mathbf{Conn}(X)$ of def. \ref{U1ConnXByFiberProductOfSharpn} 
is indeed the smooth moduli object/moduli stack of circle-principal connections on $X$; 
in that its $U$-plots of  are smoothly $U$-parameterized collections of 
smooth circle-principal connections on $X$ and its morphisms of $U$-plots 
are smoothly $U$-parameterized collections of smooth gauge transformation between these, on $X$.
\end{proposition}
\proof

By the discussion of $n$-image and using arguments as for the concretification of moduli of differential forms above, we have:
\begin{itemize}
\item $\sharp_1 [X, \mathbf{B}U(1)_{\mathrm{conn}}]$ has as $U$-plots 
smoothly $U$-parameterized $U(1)$-principal connections on $X$ that have a lift
to a $U(1)$-principal connection on $U \times X$, and morphisms are discretely 
$\Gamma(U)$-parameterized collections of gauge transformations of these connections on $X$.
	
\item $\sharp_1 [X, \mathbf{B}U(1)]$ looks similarly, just without the connection information;

\item $\sharp_1 [X, U_{\mathbf{B}U(1)_{\mathrm{conn}}}]$ simply forgets the connection data on 
the collections of bundles-with-connection; the point to notice is that oveach each chart $U$ 
it is a fibration(isofibration): given a $\Gamma(U)$-parameterized collection of 
gauge transformations out of a smoothly $U$-parameterized collection of bundles and 
then a smooth choice of smooth connections on these bundles, the $\Gamma(U)$ collection 
of gauge transformations of course also acts on these connections;

\item  $\sharp_2 [X, \mathbf{B} U(1)] \simeq [X, \mathbf{B} U(1)]$ 
(because if two gauge transformations of bundles on $U \times X$ coincide on each point of $U$ as gauge tranformations on $X$, then they were already equal).
\end{itemize}
From the third item it follows that we may compute equivalently simply the pullback in the 1-category of groupoid-valued presheaves on $\mathrm{CartSp}$. This means that a $U$-plot of the pullback is a smoothly $U$-parameterized collection $\{\nabla_u\}$ of $U(1)$-principal connections on $X$ which 
admits a lift to a $U(1)$-principal connection on $U \times X$, 
and that a morphism between such as a $\Gamma(U)$-parameterized collection of 
gauge transformations $\{\phi_u\}$ of connections, such that their underlying collection 
of gauge transformations of bundles is a smoothly $U$-parameterized family. 
But gauge transformations of 1-connections are entirely determined by the underlying 
gauge transformation of the underlying bundle, and so this just means that also the 
morphism of $U$-plots of the pullback are smoothly $U$-paramezerized 
collections of gauge transformations. 

Consider then the functor from $U(1)\mathbf{Conn}(X)_U$ to this pullback
which forgets the lift to a connection on $U \times X$. This is natural in $U$ 
and hence to complete the proof we need to see that for each $U$ 
it is an equivalence of groupoids. 
By the above it is clearly fully faithful, so it remains to see that it is essentially
surjective, hence that every smoothly $U$-parameterized collection of connections on $X$
comes from a single connection on $X \times U$. To this end, consider a smoothly $U$-parameterized collection
$\{\nabla_u\}_{u \in U}$ of $U(1)$-principal connections on $X$. Choosing
a differentiably good open cover $\{U_i \to X\}$ of $X$ the collection of connections
is equivalently given by a collection of cocycle data 
$$
  \{g_{i j}^u \in C^\infty(U_i \cap U_j, U(1)), A_i^u \in \Omega^1(U_i)\}_{u \in U}
$$
with $A_j^u = A_i^u + d_X \mathrm{log} g_{i j}^u$ on $U_i \cap U_j$ for all $i,j$ in the 
index set and all $u \in U$. To see that this is the restriction of a single such
cocycle datum on $\{U_i \times U \to X \times U \}$ we use the standard formula
for the existence of connections on a given bundle represented by a given cocycle, 
but applied just to the $U$-factor. 
So let $\{\rho_i \in C^\infty(U_i \times U)\}$ be a partition of unity on $X \times U$ subordinate to the
chosen cover and define $A_i \in \Omega^1(U_i \times U)$ by
$$
  A_i(u) := A_i^u + \sum_{i_0}  \rho_{i_0} d_U \mathrm{log} g_{i_0 i}(u)
$$
for each $u \in U$.
This is clearly a lift on each patch, and it does constitute a cocycle for a connection 
on $X \times U$ since on each $U \times (U_i \cap U_j)$ we have:
$$
  \begin{aligned}
    A_j(u) - A_i(u) & = 
	\sum_{i_0} \rho_{i_0} \left( A_j^u + d_U \mathrm{log} g_{i_0 j}(u)
	 - A_i^u - d_U \mathrm{log} g_{i_0 i}(u) \right)
	 \\
	 & = 
	 A_j^u - A_i^u + 
	 \sum_{i_0} \rho_{i_0} d_U \mathrm{log}( g_{i i_0}(u) g_{i_0 j}(u) )
	 \\
	 & = d_X \mathrm{log} g_{i j}(u)
	 +
	 d_U \mathrm{log} g_{j i}(u)
	 \\
	 & = d \mathrm{log} g_{i j}(u)
  \end{aligned}
  \,.
$$
\endofproof

\begin{proposition}
  For $G \in \mathrm{Grp}(\mathrm{Smth}\infty\mathrm{Grp})$ 
  a 0-truncated group object and for $X \in \mathrm{Smth}\infty\mathrm{Grpd}$, we have an equivalence
  $$
    \Omega \left( G\mathbf{Conn}(X)\right)
	\simeq
	G
  $$
  in $\mathrm{Smooth}\infty\mathrm{Grpd}$, between the loop space object of the 
  moduli object of $G$ def. \ref{GLieGroupBGConnXByFiberProductOfSharpn}, and $G$ itself.
\end{proposition}
\proof
 For $X$ a smooth manifold and $G$ a Lie group, this is straightforward to check by inspection of the 
 stack $\Omega \left( G\mathbf{Conn}(X)\right)$. Its $U$-plots are the
 smoothly $U$-parameterized collections of gauge transformations of the trivial 
 $G$-principal connection on $X$. Any such is a constant $G$-valued function on $X$,
 hence an element of $G$, and so these form the set $C^\infty(U,G)$ of $U$-plots of $G$.
 
 Generally, the statement follows abstractly from 
 prop. \ref{ImagesCommuteWithLooping}. By that proposition and using 
 that $\Omega$ commutes over $\infty$-fiber products (since both are
 $\infty$-limits) we have
 $$
   \begin{aligned}
     \Omega \left( G\mathbf{Conn}(X)\right)
	 & \simeq
	 \Omega \sharp_1 [X, \mathbf{B}G_{\mathrm{conn}}]
	 \underset{\Omega \sharp_1 [X, \mathbf{B}G]}{\times}
	 \Omega [X, \mathbf{B}G]
	 \\
	 & \simeq
	 \sharp \Omega[X, \mathbf{B}G_{\mathrm{conn}}]
	 \underset{\sharp \Omega [X, \mathbf{B}G]}{\times}
	 \Omega[X, \mathbf{B}G]
	 \\
	 & \simeq 
	 \sharp [X, \Omega\mathbf{B}G_{\mathrm{conn}}]
	 \underset{\sharp [X, \Omega\mathbf{B}G]}{\times}
	 [X, \Omega\mathbf{B}G]
	 \\
	 & \simeq
	 \sharp [X, \flat G]
	 \underset{\sharp [X, G]}{\times}
	 [X, G]
	 \\
	 & \simeq \sharp G \underset{\sharp [X,G]}{\times} [X,G]
   \end{aligned}
   \,.
 $$
 This last $\infty$-fiber product is one of 0-truncated object hence is the
 ordinary fiber products of the corresponding sheaves. The $U$-plots of the 
 left factor are discretely $\Gamma(U)$-parameterized collections of elements of $G$,
 the inclusion of these into $\sharp[X,G]$ is as $\Gamma(U)$-parameterized collections
 of constant $G$-valued functions on $G$, and the right factor picks out among these
 those that are smoothly parameterized over $X \times U$, hence over $U$. 
 This is the statement to be shown.
\endofproof

\subparagraph{Moduli of smooth principal 2-connections}

We discuss the general notion of 
moduli of $G$-principal connections, def. \ref{DifferentialModuliByIteratedPullback}
for the special case that $G$ is a 1-truncated group.

\begin{proposition}
  Given $X \in \mathrm{SmthMfd} \hookrightarrow \mathrm{Smooth}\infty \mathrm{Grpd}$,
  the moduli 2-stack $(\mathbf{B}U(1))\mathbf{Conn}(X)$ of circle 2-bundles with connection
  on $X$, given by the $\infty$-limit in
$$
  \xymatrix{
    (\mathbf{B}U(1))\mathbf{Conn}(X)
	\ar[rr]
	\ar[dd]
    && [X, \mathbf{B}^2 U(1)]
	\ar[d]
    \\ 
	& \sharp_2	 [X, \mathbf{B}^2 U(1)_{\mathrm{conn}^1}]
	\ar[d]
	\ar[r] 
    & \sharp_2  [X, \mathbf{B}^2 U(1)]	
    \\
    \sharp_1 [X, \mathbf{B}^2 U(1)_{\mathrm{conn}}]
	\ar[r]
	&
	\sharp_1 [X, \mathbf{B}^2 U(1)_{\mathrm{conn}^1}]
  }
$$
is equivalent to the 2-stack which assigns to any $U \in \mathrm{CartSp}$ the 2-groupoid 
whose objects, morphisms, and 2-morphisms are smoothly $U$-parameterized collections
of circle-principal connections and their gauge transformations on $X$.
\end{proposition}
\proof
    By a variant of the pasting law,
	we may compute the given $\infty$-limit as the pasting composite of three 
	$\infty$-pullbacks:
$$
  \xymatrix{
    (\mathbf{B}U(1))\mathbf{Conn}(X)
	\ar[r]
	\ar[d]
    &
	\ar[r]
	\ar[d]
	& 
	[X, \mathbf{B}^2 U(1)]
	\ar[d]
    \\
    \ar[r] \ar[d] 
	& \sharp_2	 [X, \mathbf{B}^2 U(1)_{\mathrm{conn}^1}]
	\ar[d]
	\ar[r] 
    & \sharp_2  [X, \mathbf{B}^2 U(1)]	
    \\
    \sharp_1 [X, \mathbf{B}^2 U(1)_{\mathrm{conn}}]
	\ar[r]
	&
	\sharp_1 [X, \mathbf{B}^2 U(1)_{\mathrm{conn}^1}]
  }
  \,.
$$
	
    Since this is a finite $\infty$-limit, we may compute it in $\infty$-presheaves
	over $\mathrm{CartSp}$, hence as a homotopy pullback in 
	$[\mathrm{CartSp}^{\mathrm{op}}, \mathrm{sSet}]_{\mathrm{proj}}$. 
	For $\{U_i \to X\}_{i \in I}$ any choice of 
	differentiable good open cover of $X$, our standard model for the 
	mapping stacks appearing in the diagram are given by the Deligne complex, 
	according to prop. \ref{BnU1conn}. Since this takes values,
	under the Dold-Kan map, in strict $\infty$-groupoids, we find the 
	$\sharp$-images by prop. \ref{PostnikovFactorizationOfMorphismOfStrictInfinityGroupoids}.
	In this standard presentation 
	all simplicial presheaves appearing in the diagram are fibrant and the 
	two horizontal morphisms are fibrations. Therefore we conclude that 
	the $\infty$-limit in question is in fact given by the pasting composite
	of three 1-categorical pullbacks of these presheaves of strict 2-groupoids.
    Using that pullbacks of presheaves of 2-groupoids are computed objectwise and
    degreewise, we find that the pullback presheaf is over $U \in \mathrm{CartSp}$
	given by the following strict 2-groupoid:
	\begin{itemize}
	  \item objects are smoothly $U$-parameterized 
	  collections of Deligne cocycles $\{ B_i^u, A_{i j}^u, g_{i j k}^u \}_{i,j,k \in I,u \in \Gamma(U)}$  on $X$,
	  such that there \emph{exists} a lift to a single cocycle on $X \times U$ 
	  (this is the structure of the objects of $\sharp_1 [X, \mathbf{B}^2 U(1)_{\mathrm{conn}}]$)
	  and
	  \emph{equipped} with a choice of lift of the restricted $\mathbf{B}U(1)_{\mathrm{conn}^1}$-cocycle 
	  $\{ 0 , A_{i j}^u, g_{i j k}^u \}_{u \in U}$
	  to a single restricted cocycle $\{0, A_{i j}, g_{i j k}\}$ on $U \times X$
	  (this is the structure of the objects of $\sharp_2[X, \mathbf{B}^2 U(1)_{\mathrm{conn}^1}]$);
      \item
      morphisms are smoothly $U$-parameterized collections of morphisms of cocycles on $X$
     such that there exists a lift to a morphism of restricted cocycles on $X \times U$;
     \item
      2-morphisms are smoothly $U$-parameterized collections of 2-gauge transformations,
      hence 2-gauge transformations on $X \times U$.
	\end{itemize}
	This is almost verbatim the 2-groupoid claimed in the proposition, except for
	the appearance of the existence and choice of lifts. We need to show that up to 
	equivalence these drop out.
		
	Consider therefore the canonical 2-functor from the 2-groupoid 
	thus described to the one consisting degreewise of genuine smoothly $U$-parameterized
	collections of cocycles and transformations,
	which forgets the lift and the existence of lifts. 
	This 2-functor is clearly natural in $U$, hence is a morphism of 
	simplicial presheaves. It is now sufficient to show that over each $U$ this is an 
	equivalence of 2-groupoids. 
	
	To see that this 2-functor is fully faithful, notice that by the strict abelian group
	structure on all objects we may restrict to considering the homotopy groups
	that are based at the 0-cocycle. But the automorphism groupoid of the trivial
	circle-principal 2-connection is that of flat circle-principal 1-connections. Hence fully
	faithfulness of this 2-functor amouts to the statement of 
	prop. \ref{U1ConnXIsIndeedDifferentialModuli}.
	
	Therefore it remains to check essential surjectivity of the forgetful 2-functor.
	To this end, observe that the underlying circle-principal 2-bundles of 
	a collection of 2-connections smoothly parameterized by a Cartesian 
	(hence topologically contractible) space
	necessarily have the same class at all points $u \in U$ and so every object
	in the pullback 2-groupoid is equivalent to one for which $\{g_{i j k }^u\}$
	is in fact independent of $u$. It is then sufficient to show that any such
	is in the image of the above forgetful 2-functor. 
	
	So consider a smoothly parameterized collection of Deligne cocycles on $\{U_i \to X\}_{i \in I}$ 
	of the form
	$\{ B_i^u, A_{i j}^u, g_{i j k} \}_{u \in U}$. Since now $g_{i j k}$ is constant on $U$, we can 
	obtain a lift of the 1-form part simply by defining for $i,j \in I$ 
	a 1-form $A_{i j} \in \Omega^1(U \times (U_i \cap U_j))$ by
	declaring that at $u \in U$ it is given by
	$$
	  A_{i j}(u) := A_{i j}^u
      \,.	  
	$$
	Next we need to similarly find a lift $\{B_i \in \Omega^2(U \times U_i)\}_{i \in I}$.
	For that, choose now a partition of unity
	$\{\rho_i \in C^\infty(U_i)\}_i$ of $X$, subordinate to the given cover and set	
	$$
	  B_i(u) := B_i^u + \sum_{i_0} \rho_{i_0} d_U A_{i_0 i}(u) 
	  \,.
	$$
	This is clearly patchwise a lift and we check that it satisfies the cocycle 
	condition by computing for each $i,j \in I$, $u \in U$:
	$$
	  \begin{aligned}
	     B_j(u) - B_i(u)
		 &=
		 B_j^u - B_i^u + \sum_{i_0} \rho_{i_0} d_U ( A_{i_0 j} - A_{i_0 i })(u)
		 \\
		 & = 
		 d_X A_{i j}(u)
		 +
		 \sum_{i_0} \rho_{i_0} d_U ( A_{i j} - d_X \mathrm{log} g_{i_0 i j})(u)
		 \\
		 & = d_{U \times X} A_{ i j}
	  \end{aligned}
	  \,,
	$$
	where in the second but last step we used that at each $u$ the $A_{i j}^u$ satisfy their cocycle condition
	and 
	where in the last step we used again that $g_{\cdots}$ is constant on $U$
	on $X$.
	
	So the 2-functor is also essentially surjective and this completes the proof.
\endofproof

\paragraph{Equivariant circle $n$-bundles with connection}
\label{EquivariantCircleNBundlesWithConnection}

We highlight some aspects of the \emph{equivariant}
version, def. \ref{EquivariantCohomology}, of 
smooth differential cohomology.

\medskip

\begin{observation}
  Let $G$ be a Lie group acting on a smooth manifold $X$. 
  Then the Deligne complex, def. \ref{DeligneComplex},
  computes the correct $G$-equivariant differential cohomology
  on $X$ if and only if the $G$-equivariant de Rham cohomology of $X$, 
  prop. \ref{EquivariantDeRhamCohomology},
  coincides with the $G$-\emph{invariant} Rham cohomology of $X$.
\end{observation}
\proof
  By prop. \ref{EquivariantDeRhamCohomology} we have that the $G$-equivariant de Rham 
  cohomology of $X$ is given for $n \geq 1$ by 
  $$
    H_{\mathrm{dR},G}^{n+1}(X)
	\simeq
    \pi_0 \mathbf{H}(X/\!/G, \flat_{\mathrm{dR}}\mathbf{B}^{n+1}\mathbb{R})
	\,.
  $$
  Observe that $\pi_0 \mathbf{H}(X/\!/G, \Omega^n_{\mathrm{cl}}(-))$
  is set of $G$-\emph{invariant} closed differential $n$-forms on $X$
  (which are in particular equivariant, but in general do not exhaust the 
  equivariant cocycles).
  By prop. \ref{DeligneCohomologyTheorem} the Deligne complex
  presents the homotopy pullback of $\Omega^{n+1}_{\mathrm{cl}}(-) \to 
  \flat_{\mathrm{dR}} \mathbf{B}^{n+1} \mathbb{R}$ along the universal
  curvature map on $\mathbf{B}^n U(1)$. If therefore the inclusion
  $\pi_0 \mathbf{H}(X/\!/G, \Omega^{n+1}_{\mathrm{cl}}(-)) \to
  \pi_0 \mathbf{H}(X/\!/G, \flat_{\mathrm{dR}} \mathbf{B}^{n+1}\mathbb{R})$
  of invariant into equivariant de Rham cocycles is not surjective,
  then there are differential cocycles on $X/\!/G$ not presented by the
  Deligne complex.
\endofproof
In other words, if the $G$-invariant de Rham cocycles do not exhaust the
equivariant cocycles, then $X/\!/G$ is not \emph{de Rham-projective}, 
and hence the representable variant, 
def. \ref{BnAconn}, of differential cohomology does not apply.
The correct definition of differential cohomology in this case is the more
general one from def. \ref{OrdinaryDiffCohomology}, which allows the
curvature forms themselves to be in equivariant cohomology.

\subsubsection{$\infty$-Chern-Weil homomorphism}
\label{SmoothStrucInfChernWeil}
\label{HomotopyTheoreticRefinementOfChernWeil}
  \index{Chern-Weil homomorphism!presentation by Lie integration}
  \index{structures in a cohesive $\infty$-topos!Chern-Weil homomorphism!smooth}
  \index{structures in a cohesive $\infty$-topos!holonomy!smooth}

We discuss the general abstract notion of Chern-Weil homomorphism, 
\ref{StrucChern-WeilHomomorphism}, 
realized in $\mathrm{Smooth} \infty \mathrm{Grpd}$.

Recall that for $A \in \mathrm{Smooth} \infty \mathrm{Grpd}$ 
a smooth $\infty$-groupoid regarded as a coefficient object for cohomology, 
for instance the delooping $A = \mathbf{B}G$ of an $\infty$-group $G$ we have 
general abstractly that
\begin{itemize}
\item a characteristic class on $A$ with coefficients in the 
circle Lie $n$-group, \ref{Circlengroup}, is represented by a morphism
  $$
    \mathbf{c} : A \to \mathbf{B}^n U(1)
    \,;
  $$

\item the (unrefined) Chern-Weil homomorphism induced from this is the differential 
characteristic class given by the composite
  $$
    {\mathbf{c}_{\mathrm{\mathrm{dR}}}} : 
     A \stackrel{\mathbf{c}}{\to}
     \mathbf{B}^n U(1) \stackrel{\mathrm{curv}}{\to}
     \mathbf{\flat}_{\mathrm{\mathrm{dR}}} \mathbf{B}^{n+1} \mathbb{R}
  $$
  with the universal curvature characteristic, \ref{StructCurvatureCharacteristic}, 
  on $\mathbf{B}^n U(1)$, or rather: is the morphism on cohomology
  $$
    H^1_{\mathrm{Smooth}}(X,G)
    :=
    \pi_0 \mathrm{Smooth}\infty \mathrm{Grpd}(X, \mathbf{B}G)
      \stackrel{\pi_0({(\mathbf{c}_{\mathrm{\mathrm{dR}}})_*})}{\to}
    \pi_0 \mathrm{Smooth}\infty \mathrm{Grpd}(X, \mathbf{\flat}_{\mathrm{\mathrm{dR}}} \mathbf{B}^{n+1} \mathbb{R})
    \simeq
    H_{\mathrm{\mathrm{dR}}}^{n+1}(X)
  $$
  induced by this.
\end{itemize}
By prop. \ref{LineCurvatureCharacteristicFromExp}
we have a presentation of the 
universal curvature class 
$\mathbf{B}^n \mathbb{R} \to \mathbf{\flat}_{\mathrm{\mathrm{dR}}} \mathbf{B}^{n+1}\mathbb{R}$ 
by a span
$$
  \xymatrix{
    \mathbf{B}^n \mathbb{R}_{\mathrm{diff},\mathrm{smp}}
    \ar[r]^{\mathrm{curv}_{\mathrm{smp}}}
    \ar[d]^{\simeq}
    &
    \mathbf{\flat}_{\mathrm{\mathrm{dR}}}\mathbf{B}^{n+1} \mathbb{R}_{\mathrm{smp}}  
    \\
    \mathbf{B}^n \mathbb{R}_{\mathrm{smp}}
  }
$$
in the model structure on simplicial presheaves 
$[\mathrm{CartSp}_{\mathrm{smooth}}^{\mathrm{op}}, \mathrm{sSet}]_{\mathrm{proj}}$, 
given by maps of smooth families of differential forms.
We now insert this in the above general abstract definition of the $\infty$-Chern-Weil homomorphism to 
deduce a presentation of that in terms of smooth families $L_\infty$-algebra valued differential forms.

The main step is the construction of a well-suited composite of two spans of morphisms of 
simplicial presheaves (of two $\infty$-anafunctors): we consider presentations 
of characteristic classes $\mathbf{c} : \mathbf{B}G \to \mathbf{B}^n U(1)$ in the 
image of the $\exp(-)$ map, def. \ref{ExponentiatedLInftyAlgbra},  
and presented by trunactions and quotients of morphisms of simplicial presheaves of the form
$$
    \exp(\mathfrak{g})
    \stackrel{\exp(\mu)}{\to}
    \exp(b^{n-1}\mathbb{R})
  \,.
$$
Then, using the above, the composite differential characteristic class 
$\mathbf{c}_{\mathrm{\mathrm{dR}}}$ is presented by the zig-zag
$$
  \xymatrix{
    & \mathbf{B}^n \mathbb{R}_{\mathrm{diff},\mathrm{smp}}
    \ar[r]^{\mathrm{curv}_{\mathrm{smp}}} 
    \ar[d]^\simeq
    &
     \mathbf{\flat}_{\mathrm{\mathrm{dR}}}\mathbf{B}^{n+1} \mathbb{R}_{\mathrm{smp}}
    \\
    \exp(\mathfrak{g})
    \ar[r]^{\exp(\mu)}
    &
    \mathbf{B}^n \mathbb{R}_{\mathrm{smp}}
  }
$$
of simplicial presheaves. In order to efficiently compute which morphism in 
$\mathrm{Smooth} \infty \mathrm{Grpd}$ this presents we need to construct, 
preferably naturally in the $L_\infty$-algebra $\mathfrak{g}$, 
a simplicial presheaf $\exp(\mathfrak{g})_{\mathrm{diff}}$ that fills this 
diagram as follows:
$$
  \xymatrix{
    \exp(\mathfrak{g})_{\mathrm{diff}} 
    \ar[r]^{\exp(\mu,\mathrm{cs})}
    \ar[d]^\simeq
    & 
    \mathbf{B}^n \mathbb{R}_{\mathrm{diff},\mathrm{smp}}
    \ar[r]^{\mathrm{curv}_{\mathrm{smp}}}
    \ar[d]^{\simeq}
     &
     \mathbf{\flat}_{\mathrm{\mathrm{dR}}}\mathbf{B}^{n+1} \mathbb{R}_{\mathrm{smp}}
    \\
    \exp(\mathfrak{g})
    \ar[r]^{\exp(\mu)} 
    &
    \mathbf{B}^n \mathbb{R}_{\mathrm{smp}}
  }
  \,.
$$
Given this, $\exp(\mathfrak{g})_{\mathrm{diff},\mathrm{smp}}$ serves as a new resolution 
of $\exp(\mathfrak{g})$ for which the composite differential characteristic class 
is presented by the ordinary composite of morphisms of simplicial presheaves 
$\mathrm{curv}_{\mathrm{smp}}\circ \exp(\mu, cs)$.

This object $\exp(\mathfrak{g})_{\mathrm{diff}}$ we shall see may be interpreted 
as the coefficient for \emph{pseudo}-$\infty$-connections with values in $\mathfrak{g}$. 

There is however still room to adjust this presentation such as to yield in each cohomology class special 
nice cocycle representatives. This we will achieve by finding naturally a 
subobject $\exp(\mathfrak{g})_{\mathrm{conn}} \hookrightarrow \exp(\mathfrak{g})_{\mathrm{diff}}$ 
whose inclusion is an isomorphism on connected components and restricted to 
which the morphism $\mathrm{curv}_{\mathrm{smp}} \circ \exp(\mu,cs)$ yields nice 
representatives in the de Rham hypercohomology encoded by 
$\mathbf{\flat}_{\mathrm{\mathrm{dR}}} \mathbf{B}^{n+1} \mathbb{R}_{\mathrm{smp}}$, 
namely globally defined differential forms. On this object the differential 
characteristic classes we will show factors naturally through the refinements to 
differential cohomology, and hence $\exp(\mathfrak{g})_{\mathrm{conn}}$ is finally 
identified as a presentation for the coefficient object for 
$\infty$-connections with values in $\mathfrak{g}$.

Let $\mathfrak{g} \in L_\infty \stackrel{\mathrm{CE}}{\hookrightarrow} \mathrm{dgAlg}^{\mathrm{op}}$
be an $L_\infty$-algebra, def. \ref{LInfinityAlgebra}.
\begin{definition} 
 \label{LInfinityCocycle}
A $L_\infty$-algebra cocycle on $\mathfrak{g}$ in degree $n$ is a morphism
$$
  \mu : \mathfrak{g} \to b^{n-1} \mathbb{R}
$$
to the line Lie $n$-algebra.
\end{definition}
\begin{observation} \label{LInfinityCocyclesAreCECocycles}
  Dually this is equivalently a morphism of dg-algebras
  $$
    \mathrm{CE}(\mathfrak{g}) \leftarrow \mathrm{CE}(b^{n-1}\mathbb{R}) : \mu
    \,,
  $$
  which we denote by the same letter, by slight abuse of notation. 
  Such a morphism is naturally
  identified with its image of the single generator of $\mathrm{CE}(b^{n-1}\mathbb{R})$,
  which is a closed element
  $$
    \mu \in \mathrm{CE}(\mathfrak{g})
  $$
  in degree $n$, that we also denote by the same letter. 
  Therefore $L_\infty$-algebra cocycles are precisely the ordinary cocycles
  of the corresponding Chevalley-Eilenberg algebras.
\end{observation}
\begin{remark}
  After the injection of smooth $\infty$-groupoids into synthetic differential
  $\infty$-groupoids, discussed below in \ref{SynthDiffInfGrpd}, there is an intrinsic 
  abstract notion of cohomology of $\infty$-Lie algebras. 
  Proposition \ref{IntrinsicRealCohomologyByCECohomology}
  below asserts that the above definition is indeed a presentation of that abstract cohomological
  notion.
\end{remark}
\begin{definition} 
  \label{gmu}
  \index{$L_\infty$-algebra!higher central extension}
  For $\mu : \mathfrak{g} \to b^{n-1}\mathbb{R}$ an $L_\infty$-algebra cocycle with 
  $n \geq 2$,
  write $\mathfrak{g}_\mu$ for the $L_\infty$-algebra whose Chevalley-Eilenberg algebra
  is generated from the generators of $\mathrm{CE}(\mathfrak{g})$ and one single further
  generator $b$ in degree $(n-1)$, with differential defined by
  $$
    d_{\mathrm{CE}(\mathfrak{g}_\mu)}|_{\mathfrak{g}^*} = d_{\mathrm{CE}(\mathfrak{g})}
    \,,
  $$
  and
  $$
    d_{\mathrm{CE}(\mathfrak{g}_\mu)} : b \mapsto \mu
    \,,
  $$
  where on the right we regard $\mu$ as an element of $\mathrm{CE}(\mathfrak{g})$, hence of
  $\mathrm{CE}(\mathfrak{g}_\mu)$, by observation \ref{LInfinityCocyclesAreCECocycles}.
\end{definition}
\begin{remark}
  Below in prop. \ref{gmuIsIndeedHomotopyFiber} we show that, 
  in the context of \emph{synthetic differential cohesion} \ref{SynthDiffInfGrpd}, 
  $\mathfrak{g}_\mu$ is indeed the extension of $\mathfrak{g}$
  classified by $\mu$ in the general sense of \ref{ExtensionsOfCohesiveInfinityGroups}.
\end{remark}
\begin{definition}
 \label{WeilAlgebra}
 For $\mathfrak{g} \in L_\infty \mathrm{Alg}$ an $L_\infty$-algebra, its
 \emph{Weil algebra} $\mathrm{W}(\mathfrak{g}) \in \mathrm{dgAlg}$ is the unique 
representative of the free dg-algebra on the dual cochain complex  
underlying $\mathfrak{g}$ such that the canonical projection 
$\mathfrak{g}_\bullet^*[1] \oplus \mathfrak{g}_\bullet^*[2] \to \mathfrak{g}_\bullet^*[1]$ 
extends to a dg-algebra homomorphism
$$
  \mathrm{CE}(\mathfrak{g}) \leftarrow \mathrm{W}(\mathfrak{g})
  \,.
$$
Since $\mathrm{W}(\mathfrak{g})$ is itself in $L_\infty \mathrm{Alg}^{\mathrm{op}} \hookrightarrow
\mathrm{dgAlg}$ we can idenntify it with the Chevalley-Eilenberg algebra of an $L_\infty$-algebra.
That we write $\mathrm{inn}(\mathfrak{g})$ or $e \mathfrak{g}$:
$$
  \mathrm{W}(\mathfrak{g}) :=: \mathrm{CE}(e \mathfrak{g})
  \,.
$$
In terms of this the above canonical morphism reads
$$
  \mathfrak{g} \to e \mathfrak{g}
  \,.
$$
\end{definition}
\begin{remark}
This notation reflects the fact that $e \mathfrak{g}$ may be regarded as the infinitesimal 
groupal model of the universal $\mathfrak{g}$-principal $\infty$-bundle.
\end{remark}
\begin{proposition}
  For $n \in \mathbb{N}$, $n \geq 2$ we have a pullback in $L_\infty \mathrm{Alg}$
  $$
    \xymatrix{
      b^{n-1}\mathbb{R} \ar[r] \ar[d] & e b^{n-1}\mathbb{R} \ar[d]
      \\
      {*} \ar[r] & b b^{n-1}\mathbb{R}
    }
    \,.
  $$
\end{proposition}
\proof
  Dually this is the pushout diagram of dg-algebras that is free on the short exact sequence of 
  cochain complexes concentrated in degrees $n$ and $n+1$ as follows:
  $$
    \left(
      \raisebox{20pt}{
      \xymatrix{
        0_{n+1}
        \\
        \langle c \rangle_{n}
        \ar[u]^{d_{\mathrm{CE}(\mathfrak{b^{n-1}\mathbb{R}})}}
      }
      }
    \right)
    \leftarrow
    \left(
      \raisebox{20pt}{
      \xymatrix{
        \langle d \rangle_{n+1}
        \\
        \langle c \rangle_{n}
        \ar[u]^{d_{\mathrm{CE}(\mathfrak{e b^{n-1}\mathbb{R}})}}_\simeq
      }
      }
    \right)
    \leftarrow
    \left(
      \raisebox{20pt}{
      \xymatrix{
        \langle d \rangle_{n+1}
        \\
        0_{n}
        \ar[u]^{d_{\mathrm{CE}(\mathfrak{b b^{n-1}\mathbb{R}})}}
      }
      }
    \right)    
    \,.
  $$
\endofproof
\begin{proposition} 
  \label{PullBackCharacterizationOfgmu}
  \index{$L_\infty$-algebra!shifted central extension}
  \index{$L_\infty$-algebra!cocycle!homotopy fiber}
  The $L_\infty$-algebra $\mathfrak{g}_\mu$ from def. \ref{gmu}
  fits into a pullback diagram in $L_\infty \mathrm{Alg}$
  $$
    \xymatrix{
      \mathfrak{g}_\mu \ar[r] \ar[d] & e b^{n-2} \mathbb{R} \ar[d]
      \\
      \mathfrak{g} \ar[r]^\mu & b b^{n-2}\mathbb{R}
    }
    \,.
  $$
\end{proposition}
\begin{proposition} \label{ExponentiatedHomotopyFibers}
 Let $\mu : \mathfrak{g} \to b^n \mathbb{R}$ be a degree-$n$ cocycle on an $L_\infty$-algebra
 and $\mathfrak{g}_\mu$ the $L_\infty$-algebra from def. \ref{gmu}.
 
 We have that $\exp(\mathfrak{g}_\mu) \to \exp(\mathfrak{g})$ presents the
 homotopy fiber of $\exp(\mu) : \exp(\mathfrak{g}) \to \exp(b^{n-1}\mathbb{R})$
 in $[\mathrm{CartSp}^{\mathrm{op}}, \mathrm{sSet}]_{\mathrm{proj,\mathrm{loc}}}$. 

\end{proposition}
 Since $\exp(b^{n-1}\mathbb{R}) \simeq \mathbf{B}^n \mathbb{R}$ by prop. 
 \ref{LieIntegrationToLineNGroup}, this means that $\exp(\mathfrak{g}_\mu)$ is the
 $\mathbf{B}^{n-1}\mathbb{R}$-principal $\infty$-bundle classified by $\exp(\mu)$
 in that we have an $\infty$-pullback
 $$
   \xymatrix{
     \exp(\mathfrak{g}_\mu) \ar[r] \ar[d] & {*} \ar[d]
     \\
     \exp(\mathfrak{g}) \ar[r]^{\exp(\mu)} & \mathbf{B}^n \mathbb{R}
   }
 $$
 in $\mathrm{Smooth}\infty \mathrm{Grpd}$.
\\
\proof
  Since $\exp : L_\infty \mathrm{Alg} \to [\mathrm{CartSp}^{\mathrm{op}}, \mathrm{sSet}]$
  preserves pullbacks (being given componentwise by a hom-functor) it follows from
  \ref{PullBackCharacterizationOfgmu} that we have a pullback diagram
  $$
    \xymatrix{
      \exp(\mathfrak{g}_\mu) \ar[r] \ar[d] & \exp(e b^{n-1}\mathbb{R}) \ar[d]
      \\
      \exp(\mathfrak{g}) \ar[r]^{\exp(\mu)} & \exp(b^{n-1}\mathbb{R})
    }
    \,.
  $$
  The right vertical morphism is a fibration resolution of the point inclusion
  $* \to \exp(b^{n-1}\mathbb{R})$. Hence this is a homotopy pullback in 
  $[\mathrm{CartSp}^{\mathrm{op}}, \mathrm{sSet}]_{\mathrm{proj}}$ and the claim 
  follows with prop. \ref{FiniteHomotopyLimitsInPresheaves}.
\endofproof
We now come to the definition of differential refinements of exponentiated
$L_\infty$-algebras.
\begin{definition}
  \label{expdiff}
For $\mathfrak{g} \in L_\infty$ define the simplicial presheaf 
$\exp(\mathfrak{g})_{\mathrm{diff}} \in [\mathrm{CartSp}_{\mathrm{smooth}}^{\mathrm{op}}, \mathrm{sSet}]$ by
$$
  \exp(\mathfrak{g})_{\mathrm{diff}}
  :
  (U, [k])
  \mapsto
  \left\{
    \raisebox{20pt}{
    \xymatrix{
      \Omega^\bullet_{\mathrm{\mathrm{si}},\mathrm{\mathrm{vert}}}(U \times \Delta^k) 
        &\ar[l] \mathrm{CE}(\mathfrak{g})
      \\
      \Omega^\bullet(U \times \Delta^k) \ar[u]
       & \ar[l]
      \mathrm{W}(\mathfrak{g}) \ar[u]
    }
    }
  \right\}
  \,,
$$
where on the left we have the set of commuting diagrams in 
$\mathrm{dgAlg}$ as indicated, with the vertical morphisms being the canonical projections.
\end{definition}
\begin{proposition}
The canonical projection
$$
  \exp(\mathfrak{g})_{\mathrm{diff}} \to \exp(\mathfrak{g})
$$
is a weak equivalence in $[\mathrm{CartSp}_{\mathrm{smooth}}^{\mathrm{op}}, \mathrm{sSet}]_{\mathrm{proj}}$.

Moreover, for every $L_\infty$-algebra cocycle it fits into a commuting diagram
$$
  \xymatrix{
    \exp(\mathfrak{g})_{\mathrm{diff}} 
      \ar[r]^{\exp(\mu)_{\mathrm{diff}}}
      \ar[d]^\simeq
     &
    \exp(b^{n-1}\mathbb{R})_{\mathrm{diff}} 
      \ar[d]^\simeq
    \ar@{=}[r] & 
    \mathbf{B}^n \mathbb{R}_{\mathrm{diff},\mathrm{smp}}
      \ar[d]^\simeq
    \\
   \exp(\mathfrak{g}) 
      \ar[r]^{\exp(\mu)}
     &
   \exp(b^{n-1}\mathbb{R})
     \ar@{=}[r] 
     &
   \mathbf{B}^n \mathbb{R}_{\mathrm{smp}}
  }
$$
for some morphism $exp(\mu)_{\mathrm{diff}}$.
\end{proposition}
\proof
Use the contractibility of the Weil algebra.
\endofproof
\begin{definition}
Let $G \in \mathrm{Smooth} \infty \mathrm{Grpd}$ be a smooth $n$-group given by 
Lie integration, \ref{SmoothStrucLieAlgebras}, 
of an $L_\infty$ algebra $\mathfrak{g}$, in that the delooping object 
$\mathbf{B}G$ is presented by the $(n+1)$-coskeleton simplicial 
presheaf  $\mathbf{cosk}_{n+1}\exp(\mathfrak{g})$, def. \ref{coskeleton}.

Then for $X \in [\mathrm{CartSp}_{\mathrm{smooth}}, \mathrm{sSet}]_{\mathrm{proj}}$ 
any object and $\hat X$ a cofibrant resolution, we say that 
$$
  [\mathrm{CartSp}_{\mathrm{smooth}}^{\mathrm{op}},\mathrm{sSet}]( 
    \hat X, \mathbf{cosk}_{n+1}\exp(\mathfrak{g})_{\mathrm{diff}})
$$
is the Kan complex of \emph{pseudo-$n$-connections} on $G$-principal $n$-bundles.
\end{definition}
We discuss now subobjects that pick out genuine $\infty$-connections.
\begin{definition}
 \label{InvariantPolynomial}
 \index{invariant polynomial}
 \index{Chern-Simons functionals!invariant polynomial}
An \emph{invariant polynomial}
on an $L_\infty$-algebra $\mathfrak{g}$ is an element 
$\langle - \rangle \in \mathrm{W}(\mathfrak{g})$ in the Weil algebra,
such that
\begin{enumerate}
\item $d_{\mathrm{W}(\mathfrak{g})} \langle -,-\rangle = 0$;
\item $\langle - \rangle \in \wedge^\bullet \mathfrak{g}^*[1] \hookrightarrow W(\mathfrak{g})$;
\end{enumerate}
hence such that it is a closed element built only from shifted generators of $\mathrm{W}(\mathfrak{g})$.
\end{definition}
\begin{proposition}
 \label{EquivalenceOfInvPolyOnOrdinaryLieAlgToTraditionalDef}
For $\mathfrak{g}$ an ordinary Lie algebra, 
this definition of invariant polynomial is equivalent to the traditional one 
(for instance \cite{AzIz}).
\end{proposition}
\proof
Let $\{t^a\}$ be a basis of $\mathfrak{g}^*$ and $\{r^a\}$ 
the corresponding basis of $\mathfrak{g}^*[1]$. Write $\{C^a{}_{b c}\}$ for the structure 
constants of the Lie bracket in this basis. 

Then for $P = P_{(a_1 , \cdots , a_k)} r^{a_1} \wedge \cdots \wedge r^{a_k} \in \wedge^{r} \mathfrak{g}^*[1]$ an element in the shifted generators, the condition that
its image under $d_{\mathrm{W}(\mathfrak{g})}$ is in the shifted copy is equivalent to
$$
  C^{b}_{c (a_1} P_{b, \cdots, a_k)} t^c \wedge r^{a_1} \wedge \cdots \wedge r^{a_k} = 0
  \,,
$$
where the parentheses around indices denotes symmetrization, so that this is equivalent to
$$
  \sum_{i} C^{b}_{c (a_i} P_{a_1 \cdots a_{i-1} b a_{i+1} \cdots, a_k)}
  = 0
$$ 
for all choice of indices. This is the component-version of the defining invariance statement 
$$
  \sum_i P(t_1, \cdots, t_{i-1}, [t_c, t_i], t_{i+1}, \cdots , t_k)
  = 0
$$
for all $t_\bullet \in  \mathfrak{g}$.
\endofproof
\begin{observation}
\label{InvariantPolynomialsOfLineLienAlgebra}
For the line Lie $n$-algeba we have
$$
  \mathrm{inv}(b^{n-1}\mathbb{R}) \simeq \mathrm{CE}(b^n \mathbb{R})
  \,.
$$
This allows us to identify an invariant polynomial $\langle - \rangle$ of degree $n+1$ with a morphism
$$
  \mathrm{inv}(\mathfrak{g}) \stackrel{\langle - \rangle}{\leftarrow}
  \mathrm{inv}(b^{n-1}\mathbb{R})
$$
in $\mathrm{dgAlg}$.
\end{observation}
\begin{remark}
  Write $\iota : \mathfrak{g} \to \mathrm{Der}_\bullet(\mathrm{W}(\mathfrak{g}))$
  for the identification of elements of $\mathfrak{g}$ with inner graded derivations of the 
  Weil-algebra, induced by contraction. For $v \in \mathfrak{g}$ write
  $$
    \mathcal{L}_x := [d_{\mathrm{W}(\mathfrak{g})}, \iota_v] \in \mathrm{der}_\bullet (\mathrm{W}(\mathfrak{g}))
  $$
  for the induced Lie derivative. Then the fist condition on an invariant polynomial $\langle - \rangle$ in
  def. \ref{InvariantPolynomial} is equivalent to
  $$
    \iota_v \langle -\rangle  = 0 \;\;\;\; \forall v \in \mathfrak{g}
  $$
  and the second condition implies that
  $$
    \mathcal{L}_v \langle - \rangle  = 0 \;\;\;\; \forall v \in \mathfrak{g}
    \,.
  $$
  In Cartan calculus \cite{CartanI}\cite{CartanII} elements satisfying these two conditions
  are called \emph{basic elements} or \emph{basic forms}. By prop. 
  \ref{EquivalenceOfInvPolyOnOrdinaryLieAlgToTraditionalDef} on an ordinary Lie algebra 
  the basic forms are precisely the invariant polynomials. But on a general 
  $L_\infty$-algebra there can be non-closed basic forms. Our definition 
  of invariant polynomials hence
  picks the \emph{closed basic forms} on an $L_\infty$-algebra.
\end{remark}
\begin{definition} 
  \label{TransgressionAndCSElements}
  \index{invariant polynomial!transgression via Chern-Simons element}
  \index{Chern-Simons element}
  \index{Chern-Simons functionals!Chern-Simons element}
We say that an invariant polynomial $\langle - \rangle$ on 
$\mathfrak{g}$ is \emph{in transgression} with an $L_\infty$-algebra cocycle 
$\mu : \mathfrak{g} \to b^{n-1} \mathbb{R}$ if 
there is a morphism $\mathrm{cs} : \mathrm{W}(b^{n-1}\mathbb{R}) \to \mathrm{W}(\mathfrak{g})$
such that we have a commuting diagram
$$
  \xymatrix{
    \mathrm{CE}(\mathfrak{g}) \ar@{<-}[r]^{\mu} &
    \mathrm{CE}(b^{n-1}\mathbb{R})
    \\
    \mathrm{W}(\mathfrak{g}) \ar@{<-}[r]^{\mathrm{cs}} \ar[u]&
    \mathrm{W}(b^{n-1}\mathbb{R}) \ar[u]
    \\
    \mathrm{inv}(\mathfrak{g}) \ar@{<-}[r]^{\langle- \rangle} \ar[u] &
    \mathrm{inv}(b^{n-1}\mathbb{R}) \ar@{=}[r] \ar[u]&  \mathrm{CE}(b^n \mathbb{R})
  }
$$
hence such that
\begin{enumerate}
  \item $d_{\mathrm{W}(\mathfrak{g})} \mathrm{cs} = \langle -\rangle$;
  \item $\mathrm{cs}|_{\mathrm{CE}(\mathfrak{g})} = \mu$.
\end{enumerate}
We say that $\mathrm{cs}$ is a \emph{Chern-Simons element}
exhibiting the transgression between $\mu$ and $\langle - \rangle$.

We say that an $L_\infty$-algebra cocycle is \emph{transgressive}
\index{$L_\infty$-algebra!cocycle!transgressive}
if it is in transgression with some invariant polynomial.
\end{definition}
\begin{observation}
 \label{BasicPropertiesOfTransgressionofInvariantPolynomials}
We have
\begin{enumerate}
\item There is a transgressive cocycle for every invariant polynomial.
\item Any two $L_\infty$-algebra cocycles in transgression 
   with the same invariant polynomial are cohomologous.
\item Every decomposable invariant polynomial 
   (the wedge product of 
   two non-vanishing invariant polynomials) transgresses 
   to a cocycle cohomologous to 0.
\end{enumerate}
\end{observation}
\proof
\begin{enumerate}
\item By the fact that the Weil algebra  is free, its
cochain cohomology vanishes and hence the definition
property $d_{\mathrm{W}(\mathfrak{g})} \langle -\rangle = 0$ implies
that there is some element $\mathrm{cs} \in W(\mathfrak{g})$ such that
$d_{\mathrm{W}(\mathfrak{g})} \mathrm{cs} = \langle - \rangle$. Then the image
of $\mathrm{cs}$ along the canonical dg-algebra homomorphism
$\mathrm{W}(\mathfrak{g}) \to \mathrm{CE}(\mathfrak{g})$ is 
$d_{\mathrm{CE}(\mathfrak{g})}$-closed hence is a cocycle on 
$\mathfrak{g}$.
This is by construction in transgression with $\langle - \rangle$.

\item Let $\mathrm{cs}_1$ and $\mathrm{cs}_2$ be Chern-Simons elements for the 
   to given $L_\infty$-algebra cocycles. Then by assumption
   $d_{(\mathfrak{g})} (\mathrm{cs}_1 - \mathrm{cs}_2) = 0 $. 
   By the acyclicity of
   $\mathrm{W}(\mathfrak{g})$ there is then $\lambda \in \mathrm{W}(\mathfrak{g})$
   such that $\mathrm{cs}_1 = \mathrm{cs}_2 + d_{\mathrm{W}(\mathfrak{g})} \lambda$.  
   Since 
   $\mathrm{W}(\mathfrak{g}) \to \mathrm{CE}(\mathfrak{g})$ is a dg-algebra 
   homomorphism this implies that also
   $\mu_1 = \mu_2 + d_{\mathrm{CE}(\mathfrak{g})} \lambda|_{\mathrm{CE}(\mathfrak{g})}$.

\item Given two nontrivial invariant polynomials 
   $\langle - \rangle_1$ and $\langle - \rangle_2$ let
   $\mathrm{cs}_1 \in \mathrm{W}(\mathfrak{g})$ be any element such that
   $d_{\mathrm{W}(\mathfrak{g})}\mathrm{cs}_1 = \langle - \rangle_1$.
   Then  $\mathrm{cs}_{1,2} := \mathrm{cs}_1 \wedge \langle -\rangle_2$ satisfies
   $d_{\mathrm{W}(\mathfrak{g})} \mathrm{cs}_{1,2} = 
     \langle - \rangle_1 \wedge \langle -\rangle_2$. By the
   first observation the restriction of $\mathrm{cs}_{1,2}$ to 
   $\mathrm{CE}(\mathfrak{g})$ is therefore a cocycle in transgression
   with $\langle - \rangle_1 \wedge \langle -\rangle_2$. But
   by the definition of invariant polynomials the restriction of
   $\langle - \rangle_2$ vanishes, and hence so does that 
   of $\mathrm{cs}_{1,2}$.
   The claim the follows with the second point above.
 \end{enumerate}
\endofproof
The following notion captures the equivalence relation induced by 
lifts of cocycles to Chern-Simons
elements on invariant polynomials.
\begin{definition}
  \label{InvariantPolynomialHorizontalEquivalence}
  \index{invariant polynomial!horizontal equivalence}
  We say two invariant polynomials $\langle -\rangle_1, \langle - \rangle_2 \in 
  \mathrm{W}(\mathfrak{g})$ are \emph{horizontally equivalent} if there exists
  $\omega \in \mathrm{ker}(\mathrm{W}(\mathfrak{g}) \to \mathrm{CE}(\mathfrak{g}))$
  such that 
  $$
    \langle - \rangle_1 = \langle - \rangle_2 + d_{\mathrm{W}(\mathfrak{g})} \omega
    \,.
  $$
\end{definition}
\begin{observation}
  \label{DecomposableInvPolysAreHorizontallyTrivial}
  Every decomposable invariant polynomial is horizontally equivalent to 0.
\end{observation}
\proof
  By the argument of prop. \ref{BasicPropertiesOfTransgressionofInvariantPolynomials}, item iii):
  for $\langle - \rangle = \langle -\rangle_1 \wedge \langle -\rangle_2$ let $\mathrm{cs}_1$
  be a Chern-Simons element for $\langle - \rangle_1$. Then $\mathrm{cs}_1 \wedge \langle - \rangle_2$
  exhibits a horizontal equivalence $\langle -\rangle \sim 0$.
\endofproof
\begin{proposition}
  \label{InvPolynomialsOfHigherCentralExtension}
  \index{$L_\infty$-algebra!higher central extension!invariant polynomials}
  For $\mathfrak{g}$ an $L_\infty$-algebra, $\mu : \mathfrak{g} \to b^n \mathbb{R}$
  a cocycle in transgression to an invariant polynomial $\langle \rangle$ on $\mathfrak{g}$ 
  and $\mathfrak{g}_\mu$ the corresponding shifted central extension, \ref{gmu},
  we have that 
  \begin{enumerate}
    \item $\langle -\rangle$ defines an invariant polynomial also on $\mathfrak{g}_\mu$, by
    the defining identification of generators;
    \item but on $\mathfrak{g}_\mu$ the invariant polynomial $\langle - \rangle$
    is horizontally trivial.
  \end{enumerate}
\end{proposition}
\proof
\endofproof
\begin{definition}
  \index{invariant polynomial!dg-algebra of invariant polynomials}
  For $\mathfrak{g}$ an $L_\infty$-algebra we write $\mathrm{inv}(\mathfrak{g})$
  for the free graded algebra on horizontal equivalence classes of
  invariant polynomials. We regard this as a dg-algebra with trivial 
  differential This comes with an inclusion of dg-algebras
  $$
    \mathrm{inv}(\mathfrak{g}) \to W(\mathfrak{g})
  $$
  given by a choice of representative for each class.
\end{definition}
\begin{observation}
  The algebra $\mathrm{inv}(\mathfrak{g})$ is generated from indecomposable
  invariant polynomials.
\end{observation}
\proof
    By observation \ref{DecomposableInvPolysAreHorizontallyTrivial}.
\endofproof
\begin{definition} 
 \label{ChW}
Define the simplicial presheaf 
$\exp(\mathfrak{g})_{\mathrm{ChW}} \in [\mathrm{CartSp}_{\mathrm{smooth}}^{\mathrm{op}}, \mathrm{sSet}]$
by the assignment
$$
  \exp(\mathfrak{g})_{\mathrm{ChW}} : 
   (U , [k])
   \mapsto
   \left\{
     \raisebox{40pt}{
     \xymatrix{
       \Omega^\bullet_{\mathrm{\mathrm{si}},\mathrm{\mathrm{vert}}}(U \times \Delta^k )
       \ar@{<-}[r]^>>>{A_{\mathrm{vert}}}&
       \mathrm{CE}(\mathfrak{g})
       \\
       \Omega^\bullet_{\mathrm{\mathrm{si}}}(U \times \Delta^k )
       \ar@{<-}[r]^{A} \ar[u] &
       \mathrm{W}(\mathfrak{g}) \ar[u]        
       \\
       \Omega^\bullet(U)
       \ar[u]
       \ar@{<-}[r]^{\langle F_A\rangle} &
       \mathrm{inv}(\mathfrak{g})
       \ar[u]
     }
     }
   \right\} 
   \,,
$$
where on the right we have the set of horizontal morphisms in 
$\mathrm{dgAlg}$ making commuting diagrams with the canonical vertical morphisms as indicated.

We call $\langle F_A \rangle$ the 
\emph{curvature characteristic forms}
of $A$.
\end{definition}
Let
$$
  \xymatrix{
    \exp(\mathfrak{g})_{\mathrm{diff}}
      \ar[rr]^<<<<<<<<{(\exp(\mu_i,\mathrm{cs}_i))_i} 
     \ar[d]^\simeq
       &&
    \prod_{i}
     \exp(b^{n_i-1}\mathbb{R})_{\mathrm{diff}}
     \ar[rr]^{((\mathrm{curv}_i)_{\mathrm{smp}})}
       &&
      \prod_i \mathbf{\flat}_{\mathrm{\mathrm{dR}}}\mathbf{B}^{n_i}_{\mathrm{smp}}
   \\
   \exp(\mathfrak{g})
  }
$$
be the presentation, as above, of the product of all
differental refinements of characteristic classes on
$\exp(\mathfrak{g})$ induced from Lie integration of 
transgressive $L_\infty$-algebra cocycles.
\begin{proposition} 
  \label{CharacterizationOfexpChW}
  \index{connection!in $\mathrm{Smooth}\infty\mathrm{Grpd}$}
We have that $\exp(\mathfrak{g})_{\mathrm{ChW}}$ is the 
pullback in $[\mathrm{CartSp}_{\mathrm{smooth}}^{\mathrm{op}}, \mathrm{sSet}]$ 
of the globally defined closed forms along the curvature characteristics induced by all 
transgressive $L_\infty$-algebra cocycles:
$$
  \xymatrix{
    \exp(\mathfrak{g})_{\mathrm{ChW}} \ar[r]^{\exp(\mu,cs)}
     \ar[d]
     &
    \prod_{n_i} \Omega^{n_i + 1}_{\mathrm{\mathrm{cl}}}(-)
    \ar[d]
    \\
    \exp(\mathfrak{g})_{\mathrm{diff,smp}} 
      \ar[r]^{(\mathrm{curv}_i)_i}
    \ar[d]^{\simeq}
     & 
    \prod_i \mathbf{\flat}_{\mathrm{\mathrm{dR}}} \mathbf{B}^{n_i + 1} \mathbb{R}_{\mathrm{smp}}
    \\
    \exp(\mathfrak{g})
  }
  \,.
$$
\end{proposition}
\proof
By prop. \ref{CurvSmp} we have that the bottom
horizontal morphims sends over each $(U,[k])$ 
and for each $i$ an element
$$
  \xymatrix{
    \Omega^\bullet_{\mathrm{\mathrm{si}},\mathrm{\mathrm{vert}}}(U \times \Delta^k) 
      \ar@{<-}[r]^{A_{\mathrm{\mathrm{vert}}}} &
    \mathrm{CE}(\mathfrak{g})
    \\
    \Omega^\bullet_{\mathrm{si}}(U \times \Delta^k) \ar[u]
     \ar@{<-}[r]^{A}
     &
    \mathrm{W}(\mathfrak{g}) \ar[u]
  }
$$
of $\exp(\mathfrak{g})(U)_k$ to the composite
$$
 \left( \;
  \Omega^\bullet_{\mathrm{si}}(U \times \Delta^k)
  \stackrel{A}{\leftarrow}
  \mathrm{W}(\mathfrak{g})
  \stackrel{\mathrm{cs}_i}{\leftarrow}
  \mathrm{W}(b^{n_i-1} \mathbb{R})
  \stackrel{}{\leftarrow}
  \mathrm{inv}(b^{n_i} \mathbb{R}) = \mathrm{CE}(b^{n_i}\mathbb{R}))
  \; \right)
$$
$$
  = \left( \;
      \Omega^\bullet_{\mathrm{\mathrm{si}}}(U \times \Delta^k)
      \stackrel{\langle F_A\rangle_i}{\leftarrow}
      \mathrm{CE}(b^{n_i}\mathbb{R})
    \; \right)
$$
regarded as an element in 
$\mathbf{\flat}_{\mathrm{\mathrm{dR}}} \mathbf{B}^{n_i+1}_{\mathrm{smp}}(U)_k$. The
right vertical morphism 
$\Omega^{n_i + 1}(U) \to \mathbf{\flat}_{\mathrm{\mathrm{dR}}}\mathbf{B}^{n_i+1}\mathbb{R}_{\mathrm{smp}}(U)$ 
from the constant simplicial set of 
closed $(n_i+1)$-forms on $U$ 
picks precisely those of these elements for which
$\langle F_A\rangle$ is a basic form on the $U \times \Delta^k$-bundle in that it is in the image of the pullback
$\Omega^\bullet(U) \to \Omega^\bullet_{\mathrm{\mathrm{si}}}(U \times \Delta^k)$.
\endofproof

This way the abstract differential refinement recovers the 
notion of $\infty$-connections from Lie integration
discussed before in \ref{Infinity-Connections}.

\subsubsection{Higher holonomy and parallel transport -- Fiber integration in differential cohomology }
\label{FiberIntegrationOfOrdinaryDifferentialCocycles}
\index{connection!circle $n$-bundle with connection!fiber integration}
\index{connection!circle $n$-bundle with connection!transgression}
\index{structures in a cohesive $\infty$-topos!differential cohomology!fiber integration}
\index{structures in a cohesive $\infty$-topos!differential cohomology!transgression}
 \label{SmoothStrucHolonomy}
 \index{holonomy!in $\mathrm{Smooth}\infty\mathrm{Grpd}$}
\index{structures in a cohesive $\infty$-topos!holonomy!smooth}
\index{holonomy!with boundaries}

We discuss the general notion of higher holonomy, \ref{SmoothStrucHolonomy},
realized in smooth cohesion.

\medskip

Let $n,k \in \mathbb{N}$ with $k \leq n$.
For
$$
  \Sigma_k \in \mathrm{SmoothMfd} \hookrightarrow \mathrm{Smooth}\infty\mathrm{Grpd}
$$ 
a closed manifold equipped with an orientation, the ordinary fiber integration
of differential forms
$$
  \xymatrix{
    \int_{\Sigma_k} : \Omega^n(\Sigma_k \times U)
	\ar[r]
	&
	\Omega^{n-k}(U)
  }
$$
is natural in $U \in \mathrm{CartSp} \in \mathrm{SmoothMfd}$ and hence
comes from a morphism of smooth spaces
$$
  \xymatrix{
    \int_{\Sigma_k} : [\Sigma_k, \Omega^n]
	\ar[r]
	&
	\Omega^{n-k}
  }
$$
in $\mathrm{Smooth}\infty\mathrm{Grpd}$. Similarly, transgression in ordinary
cohomology constitutes a morphism in 	$\mathrm{Smooth}\infty\mathrm{Grpd}$.
This induces a fiber integration formula also on cocoycles 
in $\mathbf{B}^n U(1)_{\mathrm{conn}}$. 
The following statement expresses this situtation in detail. 
This is theorem 3.1 of \cite{GomiTerashima}, where we observe that 
under the Dold-Kan correspondence it induces the following statement about smooth 
moduli stacks.
\begin{definition}[Planck's constant]
  We label embeddings of abelian groups
  $$
    \tfrac{1}{2 \pi \hbar} : \mathbb{Z} \hookrightarrow \mathbb{R}
  $$
  by 
  $$
    \hbar \in \mathbb{R} - \{0\}
  $$
  such that the embedding sends $1 \in \mathbb{Z}$ to $\tfrac{1}{2 \pi \hbar} \in \mathbb{R}$. 
  \label{PlanckConstant}
  \index{Planck's constant}
\end{definition}
\begin{remark}
  This constant $2 \pi \hbar$ is what in physics is called \emph{Planck's constant}.
  With this constant chosen and under the canonical  identification 
  $\mathbb{R}/\mathbb{Z} \simeq U(1)$
  the corresponding quotient map is
  $$
    \xymatrix{
      \mathbb{R}
      ~\ar@{^{(}->}[r]
      &
      \mathbb{R}
      ~\ar@{->>}[rr]^-{\exp\left(\tfrac{i}{\hbar}(-)\right) }
      &&
      U(1)
      \,.
    }
  $$
\end{remark}  

\begin{proposition}[fiber integration of differential cocycles]
  For $\Sigma_k$ a closed oriented manifold, 
  we have horizontal morphisms making the following
  diagram commute
  $$
    \raisebox{20pt}{
    \xymatrix{
	  [\Sigma_k, \mathbf{B}^n U(1)_{\mathrm{conn}}]
	  \ar[rrr]^-{\exp\left( \tfrac{i}{\hbar} \int_{\Sigma_k} \left(-\right)\right)}
	  \ar[d]
	  &&&
	  \mathbf{B}^{n-k}U(1)_{\mathrm{conn}}
	  \ar[d]
	  \\
	  [\Sigma_k, \Omega^{n+1}_{\mathrm{cl}}]
	  \ar[rrr]^-{\int_{\Sigma_k} (-)}
	  \ar[d]_{[ \Sigma_k, \mathbf{L}_{\mathrm{tYM}}^{n+1}) ]}
	  &&&
	  \Omega^{n+1-k}_{\mathrm{cl}}
	  \ar[d]
	  \\
	  [\Sigma_k, \flat \mathbf{B}^{n+1} U(1)]
	  \ar[rrr]^-{\exp\left( \tfrac{i}{\hbar} \int_{\Sigma_k} \left(-\right)\right)}
	  &&&
	  \flat \mathbf{B}^{n+1-k} U(1)\;.
	}
	}
  $$
  Moreover, for $\Sigma_{k}$ an oriented manifold with boundary 
  $\partial \Sigma_{k}$ of dimension $(k-1)$ we have a diagram
  $$
    \raisebox{20pt}{
    \xymatrix@R=7pt@C=7pt{
      && [\Sigma_k, \mathbf{B}^n U(1)_{\mathrm{conn}}]
  	  \ar[ddll]_{(-)|_{\partial \Sigma}}
	  \ar[ddrr]^{\omega_\Sigma}_{\ }="s"
      \\
      \\
      [\partial \Sigma_k, \mathbf{B}^n U(1)_{\mathrm{conn}}]
	  \ar[ddrr]_{\hspace{-13mm}\exp\left(\tfrac{i}{\hbar} \int_{\partial \Sigma} \left(-\right) \right) }^{\ }="t"
	  &&&&
	  \Omega^{n-k+1}\;,
	  \ar[ddll]
	  \\
	  \\
	  &&
	  \mathbf{B}^{n-k+1}U(1)_{\mathrm{conn}}
	  \ar@{=>}|{~~~\exp\left(\tfrac{i}{\hbar} \int_\Sigma \left(-\right)\right)} "s"; "t"
    }}
  $$
  such that when $\partial \Sigma_k = \emptyset$ the homotopy filling this
  diagram coincides with the above integration map under the identification 
  $$
    \mathbf{B}^{n-k} U(1)_{\mathrm{conn}}
	\simeq
	\ast \underset{\mathbf{B}^{n-k+1}U(1)_{\mathrm{conn}}}{\times} \Omega^{n-k+1}_{\mathrm{cl}}
	\,;
  $$
  hence for the case of no boundary $\partial \Sigma_k = \emptyset$ we have
  $$
    \raisebox{20pt}{
    \xymatrix@R=7pt@C=7pt{
      && [\Sigma_k, \mathbf{B}^n U(1)_{\mathrm{conn}}]
  	  \ar[ddll]
	  \ar[ddrr]^{\omega_\Sigma}_{\ }="s"
      \\
      \\
      \ast
	  \ar[ddrr]^{\ }="t"
	  &&&&
	  \Omega^{n-k+1}
	  \ar[ddll]
	  \\
	  \\
	  &&
	  \mathbf{B}^{n-k+1}U(1)_{\mathrm{conn}}
	  \ar@{=>}|{~~\exp\left( \tfrac{i}{\hbar} \int_\Sigma \left(-\right) \right)} "s"; "t"
    }}
  \;\;\;
  \simeq
  \;\;\;
  \raisebox{33pt}{
  \xymatrix{
    &[\Sigma_k, \mathbf{B}^n U(1)_{\mathrm{conn}}]
	\ar@/^1.5pc/[ddr]^{\omega_\Sigma}
	\ar@/_1.5pc/[ddl]
	\ar@{-->}[d]|{\exp\left( \tfrac{i}{\hbar} \int_{\Sigma_k}\left(-\right)\right)}
	\\
	&\mathbf{B}^{n-k}U(1)_{\mathrm{conn}}
	\ar[dl]
	\ar[dr]^{F_{(-)}}
	\\
	\ast \ar[dr]_0 && \Omega^{n-k+1} \ar[dl]\;.
	\\
	& \mathbf{B}^{n-k+1}U(1)_{\mathrm{cl}}
  }}
  $$
  \label{FiberIntegrationDiagrams}
\end{proposition}
\proof
  For the first statement,  
  we need to produce for each $U \in \mathrm{CartSp} \hookrightarrow \mathrm{SmoothMfd}$
  a map
  $$
    \mathbf{H}(U \times \Sigma_k, \mathbf{B}^n U(1)_{\mathrm{conn}})
	\longrightarrow \mathbf{H}(U , \mathbf{B}^{n-k} U(1)_{\mathrm{conn}})
  $$
  such that this is natural in $U$. By the general discussion of 
  $\mathbf{B}^n U(1)_{\mathrm{conn}}$, after a choice of 
  good open cover $\mathcal{U}$ of $\Sigma_k$ 
  (inducing the good cover $\mathcal{U} \times U$ of $\Sigma_k \times U$) this is given,
  under the Dold-Kan correspondence $\mathrm{DK}(-)$, by a 
  chain map of the form
  $$
    \hspace{-.3cm}
    \xymatrix{
	  C^n(\mathcal{U}\times U, \underline{U}(1)\to \cdots \to \Omega^n)
	  \ar[r]^-D
	  \ar[d]^{\int_\Sigma}
	  &
	  \cdots
	  \ar[r]^-D
	  \ar@{}[d]|{\cdots}
	  &
	  C^{1}(\mathcal{U}\times U, \underline{U}(1)\to \cdots \to \Omega^n)
	  \ar[r]^-D 
	  \ar[d]^{\int_{\Sigma}}
	  & 
	  Z^0(\mathcal{U}\times U, \underline{U}(1)\to \cdots \to \Omega^n)
	  \ar[d]^{\int_{\Sigma}}
	  \\
	  0 \ar[r] & \cdots
	  \ar[r] & 
	  C^{1}(U, \underline{U}(1)\to \cdots \to \Omega^{n-k})
	  \ar[r]^-{D}
	  &
	  Z^{0}(U, \underline{U}(1)\to \cdots \to \Omega^{n-k})\;.	  
	}
  $$
  In \cite{GomiTerashima} a map $\int_{\Sigma}$ as above is defined 
  and theorem 2.1 there asserts that it satisfies the equation
  $$
    \int_\Sigma \;\circ\; D - (-1)^k \;D \;\circ\; \int_\Sigma 
	\;\;=\;\; 
	\int_{\partial \Sigma} \;\circ \; (-)|_{\partial \Sigma}
	\hspace{3cm}
	(\star)
  $$
  in (and this is important) the chain complex 
  $C^\bullet(\mathcal{U}\times U, \underline{U}(1) \to \cdots \to \Omega^{n-k})$.
  For 
  $\partial \Sigma_k = \emptyset$ this asserts that $\int_\Sigma$ is a chain map
  as needed for the above.
  
  Next, for the more general statement in the presence of a boundary,
  we are instead in interpreting formula $(\star)$ as a chain homotopy taking place in
  $C^{\bullet}(\mathcal{U}\times U, \underline{U}(1)\to \cdots \to \Omega^{n-k+1})$
    $$
    \hspace{-.3cm}
    \xymatrix{
	  \cdots
	  \ar[r]^D
	  &
	  C^{1}(\mathcal{U}\times U, \underline{U}(1)\to \cdots \to \Omega^n)
	  \ar[r]^-D 
	  \ar[dl]|{\int_\Sigma}
	  & 
	  Z^0(\mathcal{U}\times U, \underline{U}(1)\to \cdots \to \Omega^n)
	  \ar[dl]|{\int_\Sigma}
	  \\
	  C^{2}(U, \underline{U}(1)\to \cdots \to \Omega^{n-k+1})
	  \ar[r]^-{D}
	  &
	  C^{1}(U, \underline{U}(1)\to \cdots \to \Omega^{n-k+1})	  
	  \ar[r]^-{D}
	  &
	  Z^{0}(U, \underline{U}(1)\to \cdots \to \Omega^{n-k+1})\;.
	}
  $$
  The subtlety to be taken care of now is that the equation in theorem 2.1 of
  \cite{GomiTerashima} holds in the chain complex
  $C^\bullet(\mathcal{U}\times U, \underline{U}(1)\to \cdots \to \Omega^{n-k})$
  instead of in
  $C^\bullet(\mathcal{U}\times U, \underline{U}(1)\to \cdots \to \Omega^{n-k+1})$
  as we need it here.
  But the difference is only that in the latter complex the Deligne differential of an
  $(n-k)$-form on single patches differs from that in the former by the de Rham
  differential $d$ of that differential form, which is by definition absent in the former case. 
  But by degree-counting 
  this difference appears only in the map
  $$
    D
	:
  	C^1(U, \underline{U}(1)\to \cdots \to \Omega^{n-k+1})	  
	\to  
	Z^0(U, \underline{U}(1)\to \cdots \to \Omega^{n-k+1})
	=
	\Omega^{n-k+1}(U)
	\,.
  $$
  Therefore, we may absorb it by modifying the integration chain map in degree 0.
  To that end, notice that for 
  $\mathcal{A} \in Z^0(\mathcal{U}\times U, \underline{U}(1)\to \cdots \to \Omega^{n-k})$
  we have that 
  $$
    (0, \dots, 0, (\int_{\partial \Sigma} \mathcal{A}|_{\partial \Sigma})_i 
	- d (\int_{\Sigma}\mathcal{A})_i)
	\;\;\;\in \;\;
	Z^0(U, \underline{U}(1)\to \cdots \to \Omega^{n-k+1})
  $$
  (hence that the difference is a globally well defined differential form),
  since 
  $$
    \delta (\int_{\partial \Sigma} \mathcal{A}|_{\partial \Sigma})_{i j}
	=
	\pm
	d (\int_{\partial \Sigma} \mathcal{A}|_{\partial \Sigma})_{i j}
	\,,
  $$
  this being the $(i j)$-component of the identity
  $D(\int_{\partial \Sigma} \mathcal{A}|_{\partial \Sigma}) = 0$
  given by the version of $(\star)$ without boundary applied to the boundary, and since 
  also
  $$
    \delta (d (\int_{\Sigma}\mathcal{A}))_{ij}
	= 
	d (\delta (\int_{\Sigma}\mathcal{A})_{ij})
	= 
	d (\int_{\partial \Sigma} \mathcal{A}|_{\partial \Sigma})_{i j}
	\,,
  $$
  this being the image under $d$ of the $(i j)$-component of 
  $(\star)$ applied to the cocycle $\mathcal{A}$, which gives
  $D\int_\Sigma \mathcal{A} = \int_{\partial \Sigma} \mathcal{A}|_{\partial \Sigma}$.
 Therefore, there is a natural chain map  
    $$
    \hspace{-.5cm}
    \xymatrix{
	  \cdots
	  \ar[r]^-D
	  &
	  C^1(\mathcal{U}\times U, \underline{U}(1)\to \cdots \to \Omega^n)
	  \ar[r]^-D 
	  \ar[d]
	  & 
	  Z^0(\mathcal{U}\times U, \underline{U}(1)\to \cdots \to \Omega^n)
	  \ar[d]^{\mathcal{A} \mapsto (\int_{\partial \Sigma} \mathcal{A}|_{\partial \Sigma})_i 
	- d (\int_{\Sigma}\mathcal{A})_i)}
	  \\
	  \cdots
	  \ar[r]
	  &
	  0
	  \ar[r]
	  \ar[d]
	  &
	  \Omega^{n-k+1}(U)
	  \ar[d]^=
	  \\
	  \cdots
	  \ar[r]^-{D}
	  &
	  C^1(U, \underline{U}(1)\to \cdots \to \Omega^{n-k+1})	  
	  \ar[r]^-{D}
	  &
	  Z^0(U, \underline{U}(1)\to \cdots \to \Omega^{n-k+1})\;,
	}
  $$
  which under $\mathrm{DK}(-)$ presents the map denoted
  $$
    \xymatrix{
      [\Sigma_k, \mathbf{B}^n U(1)_{\mathrm{conn}}] 
	  \ar[r]^-{\omega_\Sigma} 
	  &\Omega^{n-k+1} 
	  \ar[r]
	  &
	  \mathbf{B}^{n-k+1}U(1)_{\mathrm{conn}}
	}
  $$
  in the above statement. 
  This is now manifestly so that adding its negative to the right of
  equation $(\star)$ makes this equation define a chain homotopy in 
  $C^\bullet(\mathcal{U}\times U, \underline{U}(1) \to \cdots \to \Omega^{n-k+1})$
  of the form
  $$
    [D,\int_\Sigma] \;:\; \int_{\partial_\Sigma}(-)|_{\partial \Sigma} 
	~~\Rightarrow ~~
	\omega_\Sigma\;.
  $$
\endofproof
\begin{remark}
  These maps express the relative higher \emph{holonomy} and \emph{parallel transport}
  of $n$-form connections, respectively. The second statement says that the
  parallel transport of an $n$-connection over a $k$-dimensional manifold with
  boundary is a section of the $\mathbf{B}^{n-k}U(1)$-principal 
  bundle underlying the transgression of the underlying 
  $\mathbf{B}^{n-1}U(1)$-principal connection to the mapping space out 
  of the boundary $\partial \Sigma_k$.
  The section trivializes that underlying bundle and hence identifies
  a globally defined connection $(n-k+1)$-form. This is the form $\omega_\Sigma$
  in the above diagram.
\end{remark}
\begin{definition}
  For 
  $$
    \mathbf{L} : \mathbf{Fields} \to \mathbf{B}^n U(1)_{\mathrm{conn}}
  $$
  and $\Sigma_k \in \mathrm{SmoothMds} \hookrightarrow \mathrm{Smooth}\infty \mathrm{Grpd}$
  an oriented 
  smooth manifold of dimension $k \leq n$ with boundary
  $\partial \Sigma_k$, we say that the \emph{transgression}
  $\exp\left( \tfrac{i}{\hbar} \int_\Sigma \mathbf{L}_{\mathrm{CS}}\right)$ 
  of $\mathbf{L}$ to the mapping space out of $\Sigma$  
  is the diagram obtained by composing the mapping space construction
  $[\Sigma,-] : \mathbf{H} \to \mathbf{H}$ with the
  fiber integration 
  $\exp\left( \tfrac{i}{\hbar} \int_{\Sigma}\left(-\right)\right)$ 
  of Prop. \ref{FiberIntegrationDiagrams}:
  $$
    \hspace{-1.6cm}
    \raisebox{33pt}{
    \xymatrix{
	    & [\Sigma_k, \mathbf{Fields}]
	    \ar[dr]_{\ }="s"
	    \ar[dl]_{(-)|_{\partial \Sigma}}
	    \\
	    [\partial \Sigma_k, \mathbf{Fields}]
	    \ar[dr]_{\hspace{-16mm}\exp\left( \tfrac{i}{\hbar} \int_{\partial \Sigma_k} \mathbf{L} \right)}^{\ }="t"
	    &&
	    \Omega^{n-k+1}
	    \ar[dl]
	    \\
	    & \mathbf{B}^{n-k+1}U(1)_{\mathrm{conn}}
	    \ar@{=>}|{\exp\left( \tfrac{i}{\hbar} \int_{\Sigma_k} \mathbf{L} \right) } "s"; "t"
	}
	}
	:=
	  \hspace{-.7cm}
    \raisebox{65pt}{
    \xymatrix@R=7pt@C=7pt{
	  && [\Sigma_k, \mathbf{Fields}]
	  \ar[ddll]_{(-)|_{\partial \Sigma_k}}
	  \ar[ddrr]^{[\Sigma_k, \nabla]}
	  \\
	  \\
	  [\partial \Sigma_k, \mathbf{Fields}]
	  \ar[ddrr]_{\hspace{-2mm}[\partial \Sigma_k, \nabla]}
      && && [\Sigma_k, \mathbf{B}^n U(1)_{\mathrm{conn}}]
  	  \ar[ddll]_{(-)|_{\partial \Sigma}}
	  \ar[ddrr]_{\ }="s"
      \\
      \\
      &&[\partial \Sigma_k, \mathbf{B}^n U(1)_{\mathrm{conn}}]
	  \ar[ddrr]_{\hspace{-8mm}\exp\left( \tfrac{i}{\hbar} \int_{\partial \Sigma} \left(-\right)\right) }^{\ }="t"
	  &&&&
	  \Omega^{n-k+1}\;.
	  \ar[ddll]
	  \\
	  \\
	  && &&
	  \mathbf{B}^{n-k+1}U(1)_{\mathrm{conn}}
	  \ar@{=>}^{~~\exp\left( \tfrac{i}{\hbar} \int_\Sigma \left(-\right)\right)} "s"; "t"
    }}
  $$
  \label{Transgression}
\end{definition}
\begin{example}
  If $X \in \mathrm{SmoothMfd} \hookrightarrow \mathrm{Smooth}\infty\mathrm{Grpd}$
  is a smooth manifold and 
  $
    \nabla : X \to \mathbf{B}^n U(1)_{\mathrm{conn}}
  $
  is an $n$-connection on $X$, and for $\Sigma_n$ a closed oriented $n$-dimensional manifold,
  then the transgression
  $$
    \exp\left( \tfrac{i}{\hbar} \int_{\Sigma} \nabla \right)
	\;:\;
	[\Sigma,X] \to U(1)
  $$
  is the \emph{$n$-volume holonomy} function of $\nabla$. For $n = 1$, hence
  $\nabla$ is a $U(1)$-principal connection, and $\Sigma = S^1$, this is the 
  traditional notion of holonomy function of a principal connection along
  closed curves in $X$.
\end{example}

\medskip

We now relate this construction to the abstract characterization of 
higher holonomy of def. \ref{IntegrationAndHolonomy}
\begin{theorem} 
 \label{IntrinsicIntegrationInSmooth}
 \index{Chern-Simons functionals!intrinsic integration}
 \index{parallel transport!holonomy (in $\mathrm{Smooth}\infty \mathrm{Grpd}$)}
If $\Sigma \hookrightarrow \mathrm{SmoothMfd} \hookrightarrow \mathrm{Smooth}\infty \mathrm{Grpd}$ 
is a closed manifold of dimension $ dim \Sigma \leq n$ then the intrinsic integration by
truncation, def. \ref{IntegrationAndHolonomy}, takes values in 
$$
  \tau_{\leq n-\mathrm{dim} \Sigma} \mathbf{H}(\Sigma, \mathbf{B}^n U(1)_{\mathrm{conn}})
  \simeq
  B^{n - \mathrm{dim}\Sigma} U(1) \simeq K(U(1), n- \mathrm{dim}(\Sigma))
  \;\;\;\;\;
  \in 
  \infty \mathrm{Grpd}
  \,.
$$
Moreover, in the case $\mathrm{dim} \Sigma = n$, then the morphism
$$
  \exp(i S_{\mathbf{c}}(-))
   : 
  \mathbf{H}(\Sigma, A_{\mathrm{conn}})
   \to
  U(1)
$$
is obtained from the Lagrangian $\mathbf{L}_{\mathbf{c}}$ by forming the 
volume holonomy of circle $n$-bundles with connection (fiber integration in Deligne cohomology)
$$
  S_{\mathbf{c}}(-) = \int_\Sigma L_{\mathbf{c}}(-)
  \,.
$$
\end{theorem}
\proof
Since $dim \Sigma \leq n$ we have by 
prop. \ref{OrdinaryFromIntrinsicDeRham}
that $H(\Sigma, \mathbf{\flat}_{\mathrm{dR}} \mathbf{B}^{n+1} \mathbb{R}) \simeq H^{n+1}_{\mathrm{dR}}(\Sigma) \simeq *$. 
It then follows by prop. \ref{DiffCohomologyRestrictedToVanishingCurvature} 
that we have an equivalence
$$
  \mathbf{H}_{\mathrm{diff}}(\Sigma, \mathbf{B}^n U(1))
  \simeq
  \mathbf{H}_{\mathrm{flat}}(\Sigma, \mathbf{B}^n U(1))
  =:
  \mathbf{H}(\mathbf{\Pi}(\Sigma), \mathbf{B}^n U(1))
$$
with the flat differential cohomology on $\Sigma$, 
and by the $(\Pi \dashv \mathrm{Disc} \dashv \Gamma)$-adjunction it follows that this is equivalently
$$
  \begin{aligned}
    \cdots 
      & \simeq \infty \mathrm{Grpd}(\Pi(\Sigma), \Gamma \mathbf{B}^n U(1))
      \\
      & \simeq \infty \mathrm{Grpd}(\Pi(\Sigma), B^n U(1)_{\mathrm{disc}})
  \end{aligned}
  \,,
$$
where $B^n U(1)_{\mathrm{disc}}$ is an Eilenberg-MacLane space $\cdots \simeq K(U(1), n)$. 
By prop. \ref{UnderlyingSimplicialTopologicalSpace} we have under 
$|-| : \infty \mathrm{Grpd} \simeq \mathrm{Top}$ a weak homotopy equivalence 
$|\Pi(\Sigma)| \simeq \Sigma$. Therefore the cocycle $\infty$-groupoid is that of 
ordinary cohomology
$$
  \cdots \simeq C^n(\Sigma, U(1))
  \,.
$$
By general abstract reasoning it follows that we have for the homotopy groups an isomorphism
$$
  \pi_i \mathbf{H}_{\mathrm{diff}}(\Sigma, \mathbf{B}^n U(1))
   \stackrel{\simeq}{\to} 
  H^{n-i}(\Sigma, U(1))
  \,.
$$
Now we invoke the universal coefficient theorem. This asserts 
that the morphism
$$
  \int_{(-)}(-)
   :
  H^{n-i}(\Sigma,U(1))
   \stackrel{}{\to}
  \mathrm{Hom}_{\mathrm{Ab}}(H_{n-i}(\Sigma,\mathbb{Z}),U(1))
$$
which sends a cocycle $\omega$ in singular cohomology with coefficients in $U(1)$ to the pairing map
$$
  [c] \mapsto \int_{[c]} \omega
$$
sits inside an exact sequence
$$
  0
  \to 
  \mathrm{Ext}^1(H_{n-i-1}(\Sigma,\mathbb{Z}),U(1))
  \to 
  H^{n-i}(\Sigma,U(1))
   \stackrel{}{\to}
  \mathrm{Hom}_{\mathrm{Ab}}(H_{n-i}(\Sigma,\mathbb{Z}),U(1))
  \to 0
  \,,
$$
But since $U(1)$ is an injective $\mathbb{Z}$-module we have 
$$
  \mathrm{Ext}^1(-,U(1))=0
  \,.
$$  
This means that the integration/pairing map 
$\int_{(-)}(-)$ is an isomorphism
$$
  \int_{(-)}(-)
  :
  H^{n-i}(\Sigma,U(1))
  \simeq 
  \mathrm{Hom}_{\mathrm{Ab}}(H_{n-i}(\Sigma,\mathbb{Z}),U(1))
  \,.
$$
For $i < (n-\mathrm{dim} \Sigma)$, the right hand is zero, so that 
$$
  \pi_i \mathbf{H}_{\mathrm{diff}}(\Sigma, \mathbf{B}^n U(1)) =0 \;\;\;\;
  \mbox{for}\; i < (n-\mathrm{dim} \Sigma)
  \,. 
$$ 
For $i=(n-\mathrm{dim} \Sigma)$, instead, 
$H_{n-i}(\Sigma,\mathbb{Z})\simeq \mathbb{Z}$, since $\Sigma$ is a closed 
$\mathrm{dim} \Sigma$-manifold and so 
$$
  \pi_{(n-\mathrm{dim}\Sigma)} \mathbf{H}_{\mathrm{diff}}(\Sigma, \mathbf{B}^n U(1)) \simeq U(1)
  \,.
$$
\endofproof

More generally, using fiber integration in Deligne hypercohomology as in 
\cite{GomiTerashima}, we get for compact oriented closed smooth manifolds 
$\Sigma$ of dimension $k$ a natural morphism
$$
  \exp(2 \pi i \int_\sigma(-))
  :
  [\Sigma, \mathbf{B}^n U(1)_{\mathrm{conn}}]
  \to
  \mathbf{B}^{n-k} U(1)_{\mathrm{conn}}
  \,.
$$

\subsubsection{Chern-Simons functionals}
 \label{SmoothStrucChernSimons}
 \index{Chern-Simons functionals!in $\mathrm{Smooth}\infty\mathrm{Grpd}$}
 \index{structures in a cohesive $\infty$-topos!Chern-Simons functional!smooth}

We discuss the realization of the intrinsic notion of 
Chern-Simons functionals, \ref{StrucChern-SimonsTheory}, in 
$\mathrm{Smooth}\infty\mathrm{Grpd}$.

\medskip

The proof of theorem \ref{IntrinsicIntegrationInSmooth} 
shows that for $\mathrm{dim} \Sigma = n$ 
and $\exp(i L) : A_{conn} \to \mathbf{B}^n U(1)_{\mathrm{conn}}$
an (Chern-Simons) Lagrangian, we may think of the composite
$$
  \exp(i S) : 
  \mathbf{H}(\Sigma, A_{\mathrm{conn}})
   \stackrel{\exp(i L)}{\to}
  \mathbf{H}(\Sigma, \mathbf{B}^n U(1)_{conn})
    \stackrel{\int_{[\Sigma]}(-)}{\to}
  U(1)
$$
as being indeed given by integrating the Lagrangian over $\Sigma$ in order to obtain the action
$$
  S(-) = \int_\Sigma L(-)
  \,.
$$
We consider precise versions of this statement in \ref{ChernSimonsFunctional}.

\subsubsection{Prequantum geometry}
 \label{SmoothStrucGeometricPrequantization}
 \label{SymplecticPrequantum}
 \label{SmoothStrucQuantomorphisms}
 \index{structures in a cohesive $\infty$-topos!geometric prequantization!smooth}
 \index{geometric prequantization!smooth}
 \index{symplectic higher geometry!prequantum circle $n$-bundles}
 \index{prequantum geometry!smooth}

We discuss the notion of cohesive prequantization, \ref{StrucGeometricPrequantization}, 
realized in the model of smooth cohesion.

\medskip

What is traditionally called \emph{(geometric) prequantization} is the refinement of 
symplectic 2-forms to curvature 2-forms on line bundles with connection. 
Formally: for
$$
  \xymatrix{
    H_{\mathrm{diff}}^2(X) 
	\ar[r]^{\mathrm{curv}}
	&
	\Omega^2_{\mathrm{int}}(X)
	\ar@{^{(}->}[r]
	&
	\Omega^2_{\mathrm{cl}}(X)
  }
$$
the morphism that sends a class in degree-2 differential cohomology
over a smooth manifold $X$ to its curvature 2-form, geometric 
prequantization of some $\omega \in \Omega^2_{\mathrm{cl}}(X)$
is a choice of lift $\hat \omega \in H_{\mathrm{diff}}^2(X)$ through this 
morphism. One says that $\hat \omega$ is (the class of) a 
\emph{prequantum line bundle} or \emph{quantization line bundle} with connection
for $\omega$. See for instance \cite{WeinsteinXu}.

By the curvature exact sequence for differential
cohomology, prop. \ref{TraditionalDiffcohomologyExactSequencesReproduced},
a lift $\hat \omega$ exists precisely if $\omega$ is an 
\emph{integral} differential 2-form. This is called the 
\emph{quantization condition} on $\omega$. 
If it is fulfilled, the group of possible choices of 
lifts is the topological (for instance singular) 
cohomology group $H^1(X, U(1))$.
Notice that the extra non-degeneracy condition that makes a closed
2-form a symplectic form does not appear in \emph{pre}quantization.

The concept of geometric prequantization
has an evident generalization to closed forms of degree $n+1$
for any $n \in \mathbb{N}$. For $\omega \in \Omega^{n+1}_{\mathrm{cl}}(X)$
a closed differential $(n+1)$-form on a manifold $X$, 
a geometric prequantization is a lift of $\omega$ through
the canonical morphism
$$
  \xymatrix{
    H_{\mathrm{diff}}^{n+1}(X) 
	\ar[r]^{\mathrm{curv}}
	&
	\Omega^{n+1}_{\mathrm{int}}(X)
	\ar@{^{(}->}[r]
	&
	\Omega^{n+1}_{\mathrm{cl}}(X)
  }
  \,.
$$
Since the elements of the higher differential cohomology group
$H^{n+1}_{\mathrm{diff}}(X)$ are classes of \emph{circle $n$-bundles
with connection} (equivalently \emph{circle bundle $(n-1)$-gerbes with connection})
on $X$, we may speak of such a lift as a 
\emph{prequantum circle $n$-bundle}. Again, the lift exists precisely
if $\omega$ is integral and the group of possible choices is
$H^n(X,U(1))$. Higher geometric prequantization for $n = 2$
has been considered in \cite{Rogers}.
By the discussion in \ref{SmoothStrucDifferentialCohomology} we
may consider circle $n$-bundles with connection not just over
smooth manifolds, but over any smooth $\infty$-groupoid
(smooth $\infty$-stack) and hence consider, generally, geometric
prequantization of higher forms on higher smooth stacks.

This section draws from \cite{LocalObservables}.

\medskip

\begin{itemize}
  \item \ref{nPlecticManifolds} -- $n$-Plectic manifolds and their Hamiltonian vector fields
  \item \ref{ThePoissonBracketAlgebra} -- The $L_\infty$-algebra of local observables
  \item \ref{TheHeisenbergLInfinityCocycle} -- The Kostant-Souriau $L_\infty$-cocycle
  \item \ref{SmoothStructKostantHeisenbergLieExtension} 
     -- The Kostant-Souriau-Heisenberg $L_\infty$-extension
  \item \ref{SmoothStrucGeometricPrequantizationOrdinary}
    -- Ordinary symplectic geometry and its prequantization;
  \item \ref{SmoothStrucGeometricPrequantizationOf3Forms}
    -- 2-Plectic geometry and its prequantization.
\end{itemize}

\paragraph{$n$-Plectic manifolds and their Hamiltonian vector fields}
 \label{nPlecticManifolds}

In \cite{Baez:2008bu} the following terminology has been introduced.

\begin{definition}
\label{n-plectic_def}
  A \emph{pre-$n$-plectic manifold} $(X,\omega)$ is a smooth manifold $X$
  equipped with a closed $(n+1)$-form $\omega \in \Omega^{n+1}_{\mathrm{cl}}(X)$.
   If the contraction map $\hat{\omega} \maps TX \to \Lambda^{n} T^{\ast}X$
  is injective, then $\omega$ is called \emph{non-degenerate} or \emph{$n$-plectic}
  and $(X,\omega)$ is called an \emph{$n$-plectic manifold}.
\end{definition}
\begin{example}
  For $n = 1$ an $n$-plectic manifold is equivalently an ordinary symplectic manifold.
\end{example}
\begin{example}
Let $G$ be a compact connected simple Lie group. 
Equipped with its canonical left invariant differential 3-form
$\omega := \langle -,[-,-]\rangle$ this is a 2-plectic manifold. 
  \label{CompactSimpleLieGroup2Plectic}
\end{example}

\begin{definition}
Let $(X,\omega)$ be a pre-$n$-plectic manifold. If a vector field $v$ and an $(n-1)$-form $H$ are related by 
\[
\iota_v\omega+dH=0
\] 
then we say that $v$ is a Hamiltonian field for $H$ and that $H$ is a Hamiltonian form for $v$. 
\end{definition}
\begin{definition}
  We denote by 
  $$
   {\mathrm{Ham}}^{n-1}(X)\subseteq \mathfrak{X}(X)\oplus\Omega^{n-1}(X)
  $$ 
  the subspace of pairs $(v,H)$ such that $\iota_v\omega+dH=0$. We call this the
  space of \emph{Hamiltonian pairs}.
\label{VectorFieldsWithHamiltonian}
\label{HamiltonianVectorFields}
The image $\mathfrak{X}_{\mathrm{Ham}}(X)\subseteq \mathfrak{X}(X)$ of the projection ${\mathrm{Ham}}^{n-1}(X)\to \mathfrak{X}(X)$ is called the space of \emph{Hamiltonian vector fields} of $(X,\omega)$.
\end{definition}

\begin{remark}
Given a pre-$n$-plectic manifold $(X,\omega)$
We have a short exact sequence
of vector spaces
\[
0\to \Omega^{n-1}_{\mathrm{cl}}(X)\to {\mathrm{Ham}}^{n-1}(X)\to \mathfrak{X}_{\mathrm{Ham}}(X)\to 0,
\]
i.e., closed $(n-1)$-forms are Hamiltonian, with zero Hamiltonian vector field.
\end{remark}
\begin{remark}\label{remark.hamiltonian}
  It is immediate from the definition that Hamilton vector fields
  preserve the pre-$n$-plectic form $\omega$, i.e.,
  $\mathcal{L}_v\omega=0$. Indeed, since $\omega$ is closed, we have
  $\mathcal{L}_v\omega=d\iota_v\omega=-d^2H_v=0$. Therefore the
  integration of a Hamiltonian vector field gives a diffeomorphism of
  $X$ preserving the pre-$n$-plectic form: a
  \emph{Hamiltonian $n$-plectomorphism}.
\end{remark}
\begin{lemma}\label{lemma.hamiltonian}
The subspace $\mathfrak{X}_{\mathrm{Ham}}(X)$ is a Lie subalgebra of $\mathfrak{X}(X)$. 
\end{lemma}

\begin{remark}
 Hamiltonian vector fields on a pre-$n$-plectic manifold $(X,\omega)$ are by definition 
 those vector fields $v$ such that $\iota_v\omega$ is exact. One may relax this 
 condition and consider \emph{symplectic vector fields} instead, i.e., those vector fields $v$ such that $\iota_v\omega$ is closed. Then the arguments in Remark \ref{remark.hamiltonian} and in Lemma \ref{lemma.hamiltonian} show that symplectic vector fields form a Lie subalgebra $\mathfrak{X}_{\mathrm{symp}}(X)$ of $\mathfrak{X}(X)$ and that $\mathfrak{X}_{\mathrm{symp}}(X)\subseteq \mathfrak{X}_{\mathrm{Ham}}(X)$ is a Lie ideal.
\end{remark}

\begin{definition}
 Let $(X,\omega)$ be a pre-$n$-plectic manifold. A
\emph{prequantization} of $(X,\omega)$ is a lift
\[
\xymatrix{
&\mathbf{B}^{n}U(1)_{\mathrm{conn}}\ar[d]^F\\
X\ar[r]^-{\omega}\ar[ru]^{\nabla}&\Omega^{n+1}(-)_{\mathrm{cl}}.
}
\]
We call the triple $(X,\omega,\nabla)$ a prequantized pre-$n$-plectic manifold.
  \label{PrequantizationOfnPlectic}
\end{definition}

\paragraph{The $L_\infty$-algebra of local observables}
 \label{ThePoissonBracketAlgebra}

We consider now the Lie differentiation of the Quantomorphism 
$\infty$-group of a pre-quantized $n$-plectic smooth manifold.

\medskip

\begin{definition}\label{def.local-observables}
We call the Lie $n$-algebra $L_{\infty}(X,\omega)$ 
of def. \ref{ham-infty}  the \emph{$L_\infty$-algebra of local observables} 
on $(X,\omega)$.
\label{TheLooXOmega}
\end{definition} 
\begin{remark}
  The projection map of def. \ref{HamiltonianVectorFields}
  uniquely extends to a morphism of $L_\infty$-algebras of the form
  $$
    \raisebox{20pt}{
    \xymatrix{
	  L_\infty(X,\omega)
	  \ar[d]^{\pi_L}
	  \\
	  \mathfrak{X}_{\mathrm{Ham}}(X)
	}}
	\,,
  $$
i.e., local observables of $(X,\omega)$ cover Hamiltonian vector fields. Below in \ref{TheHeisenbergLInfinityCocycle} 
we turn to the classification of this map by an $L_\infty$-algebra cocycle.
  \label{MapFromLocalObservablesToHamiltonianVectorFields}
\end{remark}

\begin{example}
 If $n = 1$ then $(X,\omega)$ is a pre-symplectic manifold and the 
 chain complex underlying $L_\infty(X,\omega)$ is
\[
{\mathrm{Ham}}^0(X)=\{v+H\in \mathfrak{X}(X)\oplus C^\infty(X;\mathbb{R})\,|\, \iota_v\omega+dH=0\},
\]
and the Lie bracket is 
\[
[v_1+H_1,v_2+H_2]= [v_1,v_2]+\iota_{v_1\wedge v_2}\omega.
\]
If moreover $\omega$ is non-degenerate so that $(X,\omega)$ is symplectic, 
then the projection $v+H \mapsto H$ is a linear isomorphism 
${\mathrm{Ham}}^0(X) \stackrel{\simeq}{\to} C^\infty(X;\mathbb{R})$.
It is easy to see that under this isomorphism 
$L_\infty(X,\omega)$ is the underlying Lie algebra of the usual
Poisson algebra of functions.  See also Prop.\ 2.3.9 in \cite {BrylinskiLoop}.
  \label{OrdinaryPoissonBracket}
\end{example}

\paragraph{The Kostant-Souriau $L_\infty$-cocycle}
\label{TheHeisenbergLInfinityCocycle}

\begin{definition}
For $X$ a smooth manifold, denote by $\mathbf{B}\mathbf{H}(X,\flat\mathbf{B}^{n-1}\mathbb{R})$ 
the abelian Lie $(n+1)$-algebra given by the chain complex
\[
\Omega^0(X)\xrightarrow{d}\Omega^1(X)\xrightarrow{d}\cdots \xrightarrow{d}\Omega^{n-1}(X)\xrightarrow{d}d\Omega^{n-1}(X),
\]
with $d\Omega^{n-1}(X)$ in degree zero.
\label{ResolvedDeloopedShiftedTruncatedDeRhamComplex}
\end{definition}
\begin{remark}
  The complex of def. \ref{ResolvedDeloopedShiftedTruncatedDeRhamComplex} 
  serves as a resolution of the cocycle complex 
 \[
\Omega^0(X)\xrightarrow{d}\Omega^1(X)\xrightarrow{d}\cdots \xrightarrow{d}\Omega_{\mathrm{cl}}^{n-1}(X)\xrightarrow{} 0
\,,
\]
{for the de Rham cohomology of $X$ up to degree $n-1$}
once delooped (i.e., shifted). 
  \label{CoefficientsOvernConnectedSpace}
\end{remark}
\begin{proposition}
  \label{prop.just-above} 
  Let $(X,\omega)$ be a pre-$n$-plectic manifold. The multilinear maps
\[
  \omega_{[1]} : v\mapsto -\iota_v\omega; 
  \qquad 
  \omega_{[2]} : v_1\wedge v_2\mapsto \iota_{v_1\wedge v_2}\omega;
  \qquad 
     \cdots
     \qquad 
  \omega_{[n+1]} : v_1\wedge v_2\wedge\cdots  v_{n+1}\mapsto -(-1)^{\binom{n+1}{2}}
\iota_{v_1\wedge v_2\wedge \cdots \wedge v_{n+1}}\omega
\]
define an $L_\infty$-morphism 
\[
  \omega_{[\bullet]}
  : 
\mathfrak{X}_{\mathrm{Ham}}(X)\xrightarrow{} \mathbf{B} \mathbf{H}(X,\flat\mathbf{B}^{n-1}\mathbb{R})\,,
\] 
and hence an $L_\infty$-algebra $(n+1)$-cocycle on the Lie algebra of
Hamiltonian vector fields, def. \ref{HamiltonianVectorFields}, with
values in the abelian $(n+1)$-algebra of
def. \ref{ResolvedDeloopedShiftedTruncatedDeRhamComplex}.
\end{proposition}
This is due to \cite{LocalObservables}.

\begin{definition}
The degree $(n+1)$  \emph{higher Kostant-Souriau $L_\infty$-cocycle}
associated to the pre-$n$-plectic manifold $(X,\omega)$ is the $L_{\infty}$-morphism
  $$
    \omega_{[\bullet]} : 
	\xymatrix{
	  \mathfrak{X}_{\mathrm{Ham}}(X)
	  \ar[r]
	  &
	  \mathbf{B}\mathbf{H}(X,\flat\mathbf{B}^{n-1}\mathbb{R})
	}
  $$
  given in Prop.\ \ref{prop.just-above}.
  \label{HigherHeisenbergCocycle}
\end{definition}
If $\rho \colon \mathfrak{g} \to \mathfrak{X}_{\mathrm{Ham}}(X)$ is
an $L_\infty$-morphism encoding an action of an $L_\infty$-algebra
$\mathfrak{g}$ on $(X,\omega)$ by Hamiltonian vector fields, then we call 
the composite $\rho^* \omega_{[\bullet]}$ the corresponding
\emph{Heisenberg $L_\infty$-algebra cocycle}. This terminology is motivated by 
example \ref{CocycleOvernPlecticVectorSpaces} below. 

\begin{example}
   Let $V$ be a vector space equipped with a skew-symmetric multilinear
  form $\omega: \Lambda^{n+1} V \to \R$. Since $V$ is an abelian Lie
  group, we obtain via left-translation of $\omega$ a unique closed invariant form, 
  which we also denote as $\omega$. By identifying $V$ with left-invariant
  vector fields on $V$, the Poincare lemma implies that we have a
  canonical inclusion 
$$
    j_V : V \hookrightarrow \mathfrak{X}_{\mathrm{Ham}}(V)
  $$
  of $V$ regarded as an abelian Lie algebra into the Hamiltonian vector fields on $(V,\omega)$
  regarded as a pre $n$-plectic manifold. 
  Since $V$ is contractible as a topological manifold, we have, 
  by remark \ref{CoefficientsOvernConnectedSpace}, a quasi-isomorphism
  $$
    \xymatrix{
	  \mathbf{B}\mathbf{H}(V;\flat\mathbf{B}^{n-1}\mathbb{R})
	  \ar[r]^-{\simeq}
	  &
	  \mathbb{R}[n]
	}
  $$
  of abelian $L_\infty$-algebras, given by evaluation at $0$.  
  Under this equivalence the restriction of the 
  $L_\infty$-algebra cocycle $\omega_{[\bullet]}$ of def. \ref{HigherHeisenbergCocycle} 
  along $j_V$ is an $L_\infty$-algebra map of the form
  $$
    j_V^* \omega_{[\bullet]}
	:
    \xymatrix{
	  V 
	  \ar[r]
	  &
	  \mathbb{R}[n]
	}
  $$  
  whose single component is the linear map
  $$
    \iota_{(-)} \omega : \wedge^{n+1} V \to \mathbb{R}
	\,.
  $$
  For $n=1$ and $(V,\omega)$ an ordinary symplectic vector space 
the map
  $\iota_{(-)}\omega : V \wedge V \to \mathbb{R}$ is the traditional 
  \emph{Heisenberg cocycle}.
  \label{CocycleOvernPlecticVectorSpaces}
\end{example}

\paragraph{The Kostant-Souriau-Heisenberg $L_\infty$-extension}
\label{SmoothStructKostantHeisenbergLieExtension}
 \label{TheKSExtension}
\index{Kostant-Souriau-extension!$L_\infty$-extension}

We consider here the cohesive quantomorphism and Heinseberg group extensions
from \ref{QuantomorphismGroup} after Lie differentiation as extensions
of $L_\infty$-algebras.

\medskip

\begin{proposition}
If $(X,\omega)$ is a pre-$n$-plectic manifold, then the projection map
$\pi_L \maps L_{\infty}(X,\omega) \to {\mathcal{X}_{\mathrm{ham}}}(X)$  
(remark \ref{MapFromLocalObservablesToHamiltonianVectorFields}) and the 
higher Kostant-Souriau $L_\infty$-cocycle $\omega_{[\bullet]}$
(def. \ref{HigherHeisenbergCocycle}) form a homotopy fiber sequence of
$L_\infty$-algebras, and hence fit into a homotopy pullback
 diagram of the form
\[
\xymatrix{
  L_\infty(X,\omega)
    \ar[d]^{\pi_L}
    \ar[r]
	&
    0\ar[d]
  \\
  \mathfrak{X}_{\mathrm{Ham}}(X)
    \ar[r]^-{\omega_{[\bullet]}}
	& 
  \mathbf{B}\mathbf{H}(X,\flat\mathbf{B}^{n-1}\mathbb{R}).
}
\] 
\label{TheExtensionClassifiedByTheCocycle}
\end{proposition}
This is due to \cite{LocalObservables}

If a Lie algebra $\mathfrak{g}$ acts on an 
$n$-plectic manifold by Hamiltonian vector fields,
then the Kostant-Souriau $L_\infty$-extension of $\mathcal{X}_{\mathrm{Ham}}(X)$, 
discussed above in \ref{TheKSExtension},
restricts to an $L_\infty$-extension of $\mathfrak{g}$. This is a generalization 
{of Kostant's construction \cite{Kostant} of central extensions of Lie algebras
  to the context of $L_\infty$-algebras. Perhaps the most famous of
  these central extensions is the Heisenberg Lie algebra, which is the
  inspiration behind the following terminology:}


\begin{definition}
  Let $(X,\omega)$ be a pre-$n$-plectic manifold and let $\rho :
  \mathfrak{g} \to \mathfrak{X}_{\mathrm{Ham}}(X)$ be a Lie algebra
  homomorphism encoding an action of $\mathfrak{g}$ on $X$ by
  Hamiltonian vector fields. The corresponding \emph{Heisenberg
    $L_\infty$-algebra extension} $\mathfrak{heis}_\rho(\mathfrak{g})$
  of $\mathfrak{g}$ is the extension classified by the composite
  $L_{\infty}$-morphism $\omega_{[\bullet]}\circ \rho$,  i.e.\ the
  homotopy pullback on the left of
  $$
    \raisebox{20pt}{
    \xymatrix{
	  \mathfrak{heis}_\rho(\mathfrak{g})
	  \ar[r]
	  \ar[d]
	  &
	  L_\infty(X,\omega)
	  \ar[r]
	  \ar[d]
	  &
	  0
	  \ar[d]
	  \\
	  \mathfrak{g}
	  \ar[r]^-\rho
	  &
	  \mathfrak{X}_{\mathrm{Ham}}(X)
	  \ar[r]^-{\omega_{[\bullet]}}
	  &
	  \mathbf{B}\mathbf{H}(X,\flat \mathbf{B}^{n-1}\mathbb{R})
	}
	}
	\,.
  $$
\end{definition}
\begin{remark}
  It is natural to call an $L_\infty$-morphism with values in the
  $L_\infty$-algebra of observables of a pre-$n$-plectic manifold
  $(X,\omega)$ an `$L_\infty$ {co-moment map}', which
  generalizes the familiar notion in symplectic geometry. Hence, one
  could say that an action $\rho$ of a Lie algebra $\mathfrak{g}$ on a
  pre-$n$-plectic manifold $(X,\omega)$ via Hamiltonian vector fields
  naturally induces such a co-moment map from the Heisenberg
  $L_\infty$-algebra $\mathfrak{heis}_\rho(\mathfrak{g})$.
 \end{remark}

\paragraph{Ordinary symplectic geometry and its prequantization}
 \label{SmoothStrucGeometricPrequantizationOrdinary}
 \index{structures in a cohesive $\infty$-topos!geometric prequantization!ordinary}
 \index{geometric prequantization!ordinary}
 \index{symplectic higher geometry!prequantum circle bundle}

We discuss how the general abstract notion of higher geometric prequantization
reduces to the traditional notion of geometric prequatization
when interpreted in the smooth context and for $n = 1$. 

The following is essentially a
re-derivation of the discussion in section II.3 and II.4 of 
\cite{BrylinskiLoop} (based on \cite{Kostant}) from the abstract point of view of 
\ref{StrucGeometricPrequantization}.

\medskip

The traditional definition of Hamiltonian vector fields is the 
following.
\begin{definition}
Let $(X, \omega)$ be a smooth symplectic manifold.
A \emph{Hamiltonian vector field} on $X$ is a vector 
field $v \in \Gamma(T X)$ whose contraction with the symplectic form $\omega$ 
yields an exact form, hence such that
$$
  \exists \, h \in C^\infty(X) \;:\; 
  \iota_v \omega = d_{\mathrm{dR}} h
  \,.
$$
Here a choice of function $h$ is called a \emph{Hamiltonian}
for $v$.
\index{symplectic higher geometry!Hamiltonian vector fields!ordinary}
\label{TraditionalHamiltonianVectorFields}
\end{definition}
\begin{proposition}
\label{OrdinaryHamiltonianVectorFields}
\index{symplectic higher geometry!Hamiltonian vector fields!ordinary}
Let $X$ be a smooth manifold which is simply connected, and let
$\omega \in \Omega^2(X)_{\mathrm{int}}$ be an integral symplectic form on $X$.
Then regarding $(X,\omega)$ as a symplectic 0-groupoid in $\mathrm{Smooth}\infty \mathrm{Grpd}$, 
the general definition \ref{HamiltonianVectorFieldsOnGrpd} reproduces the standard notion of Hamiltonian vector fields, def. \ref{TraditionalHamiltonianVectorFields} on the
symplectic manifold $(X, \omega)$.
\end{proposition}
\proof
A Hamiltonian symplectomorphism is an equivalence 
$\phi : X\to X$ that fits into a diagram
$$
  \xymatrix{
     X \ar[rr]^{\phi}_>>>>{\ }="s" \ar[dr]_{\hat \omega}^{\ }="t" 
	 && X \ar[dl]^{\hat \omega}
     \\
     & \mathbf{B} U(1)_{\mathrm{conn}}
	 \ar@{=>}^\alpha "s"; "t"
  }
$$
in $\mathrm{Smooth}\infty\mathrm{Grpd}$.
To compute the Lie algebra of the group of these diffeomorphisms, 
we need to consider smooth 1-parameter 
families of such and differentiate them. 

Assume first that the connection 1-form in $\hat \omega$ is globally defined 
$A \in \Omega^1(X)$ with $d A = \omega$. Then the 
existence of the above diagram is equivalent to the condition
$$
  (\phi(t)^* A - A) = d \alpha(t)
  \,,
$$
where $\alpha(t) \in C^\infty(X)$. Differentiating this at 0 yields the Lie derivative
$$
  \mathcal{L}_v A = d \alpha'
  \,,
$$
where $v$ is the vector field of which $t \mapsto \phi(t)$ is the flow
and where $\alpha' := \frac{d}{dt} \alpha$.
By Cartan calculus this is equivalently
$$
  d_{\mathrm{dR}} \iota_v A + \iota_v d_{dR} A  = d \alpha'
$$
and using that $A$ is the connection on a prequantum circle bundle
for $\omega$
$$
  \iota_v \omega = d (\alpha' - \iota_v A)
  \,.
$$
This says that for $v$ to be Hamiltonian, its contraction with $\omega$ must be exact. This is precisely the definition of Hamiltonian vector fields. The corresponding Hamiltonian function $h$ here is $\alpha'-\iota_v A$.

We now discuss the general case, 
where the prequantum bundle is not necessarily trivial. 
After a choice of cover that is compatible with the flows of vector fields,
the argument proceeds by slight generalization of the previous argument.

We may assume without restriction of generality that $X$ is connected.
Choose then any base point $x_0 \in X$ and let 
$$
 P_* X := [I,X] \times_{X} \{x_0\}
$$
be the based smooth path space of $X$, 
regarded as a diffeological space, def. \ref{DiffeologicalSpace},
where $I \subset \mathbb{R}$ is the standard closed interval.
This comes equipped with the smooth endpoint evaluation map
$$
  p : P_* X \to X
  \,.
$$
Pulled back along this map, every circle bundle has a trivialization, since
$P_*X$ is topologically contractible.
The corresponding {\v C}ech nerve $C(P_* X \to X)$ 
is the simplicial presheaf that starts out as
$$
  \xymatrix{
    \cdots
	\ar@<+4pt>[r]
	\ar@<-0pt>[r]	
	\ar@<-4pt>[r]	
	&
    P_* X \times_X P_*X 
	\ar@<+3pt>[r]^<<<{p_1}
	\ar@<-3pt>[r]_<<<{p_2}
	&
	P_* X 
  }
  \,,
$$
where in first degree we have a certain smooth version of the based loop space of $X$.
Any diffeomorphism $\phi = \exp(v) : X \to X$ lifts to an automorphism
of the {\v C}ech nerve by letting
$$
  P_* \phi : P_*X \to P_*X 
$$
be given by
$$
  P_* \phi(\gamma) : (t \in [0,1]) \mapsto \exp(t v)(\gamma(t))
$$
and similarly for $P_* \phi : P_*X \times_X P_*X \to P_*X \times_X P_* X$.
If $\phi = \exp(t v)$ for $v$ a vector field on $X$, we will 
write $v$ also for the vector fields induced this way on the components
of the {\v C}ech nerve.  

With these preparations, every elements of the group in question is presented by
a diagram of simplicial presheaves of the form
$$
  \xymatrix{
     C(P_{*}X \to X) 
	 \ar[rr]^{P_{*}\phi}_>>>>{\ }="s" 
	 \ar[dr]_{\hat \omega}^{\ }="t" 
	 && C(P_{*}X \to X) 
	  \ar[dl]^{\hat \omega}
     \\
     & \mathbf{B} U(1)_{\mathrm{conn}}
	 \ar@{=>}^\alpha "s"; "t"
  }
  \,.
$$
Here the vertical (diagonal) morphisms now exhibit {\v C}ech-Deligne cocycles 
with transition function 
$$
  g \in C^\infty(P_* X \times_X P_* X)
$$
and connection 1-form 
$$
  A \in \Omega^1(P_*X)
  \,,
$$ 
satisfiying 
$$
  p_2^* A - p_1^* A = d_{\mathrm{dR}}\mathrm{log} g
  \,.
$$

For $\phi(t) = \exp(t v)$ a 1-parameter family of diffeomorphisms,
the homotopy in this diagram is a gauge transformation 
given by a function 
$\alpha(t) \in C^\infty(P_*X, U(1))$ such that 
$$
  p_2^* \alpha(t) \cdot g \cdot p_1^* \alpha(t)^{-1}
  =
  \exp(tv)^* g
$$
and
$$
  \exp(t v)^* A - A = d_{\mathrm{dR}} \mathrm{log} \alpha(t)
  \,.
$$
Differentiating this at $t = 0$ and writing $\alpha' := \alpha'(0)$ as before,
this yields
$$
  p_2^* \alpha' - p_1^* \alpha' = \mathcal{L}_v \mathrm{log} g
$$
and
$$
  \mathcal{L}_v A = d_{\mathrm{dR}} \alpha'
  \,.
$$
The latter formula says that on $P_*X$ $\iota_v \omega$ is exact
$$
  \iota_v p^*\omega = d_{\mathrm{dR}}( \alpha' - \iota_v A  )
  \,.
$$
But in fact the function on the right descends down to $X$, because by the
formulas above we have
$$
  \begin{aligned}
    p_2^* (\alpha' - \iota_v A)
	-
	p_1^* (\alpha' - \iota_v A)
	& =
	\mathcal{L}_v \mathrm{log} g
	-
	\iota_v (p_2^*A - p_1^* A)
	\\
	&= 0
	\,.
  \end{aligned}
  \,.
$$
Write therefore $h \in C^\infty(X)$ for the unique function such that 
$p^* h = \alpha'  -\iota_v A$, then this satisfies
$$
  \iota_v \omega = d h
$$ 
on $X$. 
\endofproof
The traditional definition of the Poisson-bracket Lie algebra associated with a 
symplectic manifold $(X, \omega)$ is the following.
\begin{definition}
  Let $(X, \omega)$ be a smooth symplectic manifold. Then its
  \emph{Poisson-bracket Lie algebra} is the Lie algebra whose underlying 
  vector space is $C^\infty(X)$, the space of smooth function on $X$,
  and whose Lie bracket is given by
  $$
    [h_1,h_2] := \iota_{v_2}\iota_{v_1} \omega
  $$
  for all $h_1, h_2 \in C^\infty(X)$ and for $v_1, v_2$ the corresponding
  Hamiltonian vector fields, def. \ref{TraditionalHamiltonianVectorFields}.
  \label{TraditionalDefinitionPoissonLieAlgebra}
\end{definition}
\begin{proposition}
  The general definition of Poisson $\infty$-Lie algebra, 
  def. \ref{HamiltonianVectorFieldsOnGrpd}, applied
  to the symplectic manifold $(X, \omega)$ regarded as a symplectic smooth 0-groupoid,
  reproduces the traditional definition of the Lie algebra underlying the Poisson algebra
  of $(X, \omega)$.
\end{proposition}
\proof
  The smooth group $\mathbf{Aut}_{\mathbf{H}_{/\mathbf{B}U(1)_{\mathrm{conn}}}}(\hat \omega)$ is manifestly a subgroup of the semidirect product group
  $\mathrm{Diff}(X) \ltimes C^\infty(X)$, where the group structure on the second
  factor is given by addition, and the action of the first factor on the second is
  the canonical one by pullback. Accordingly, its Lie algebra may be identified
  with that of pairs $(v,\alpha)$ in $\Gamma(T X) \times C^\infty(X)$ 
  such that, with the notation as in the proof of prop. \ref{OrdinaryHamiltonianVectorFields},
  $\alpha - \iota_v A$ is a Hamiltonian for $v$; and the Lie bracket is
  given by
  $$
    [\, (v_1,\alpha_1)\,,\,(v_2, \alpha_2) \,]
	=
	([v_1, v_2]\,,\, \mathcal{L}_{v_2} \alpha_1 - \mathcal{L}_{v_1} \alpha_2  )\,.
  $$
  Notice that these pairs are redundant in that $v$ is entirely determined by $\alpha$,
  we just use them to make explicit the embedding into the semidirect product.
   
  It remains to check that with this bracket the map
  $$
     \phi : \alpha \mapsto \alpha - \iota_v A
  $$
  is a Lie algebra isomorphism to the Poisson-bracket Lie algebra,
  def. \ref{TraditionalDefinitionPoissonLieAlgebra}.
  For this first notice the equation
  $$
    \begin{aligned}
	  2  \iota_{v_2} \iota_{v_1} \omega 
	  & =
	  \iota_{v_2} d_{\mathrm{dR}} h_1 - \iota_{v_1} d_{\mathrm{dR}} h_2
	  \\
	  & = 
	  \mathcal{L}_{v_2}(\alpha_1 - \iota_{v_1}A )
	  -
	  \mathcal{L}_{v_1}(\alpha_2 - \iota_{v_2}A )
	  \\
	  & =
	  \mathcal{L}_{v_2} \alpha_1 
	  -
	  \mathcal{L}_{v_1} \alpha_2 
	  +
	  \iota_{v_2}\iota_{v_1} d_{\mathrm{dR}} A - \iota_{[v_1,v_2]}A	  
	  \,,
	\end{aligned}
  $$
  where in the last step we used the identity
  $$
    \iota_{v_2} \iota_{v_1} d_{\mathrm{dR}} A = 
      \mathcal{L}_{v_1} \iota_{v_2}A 
   	   -
      \mathcal{L}_{v_2} \iota_{v_1}A
      -
	  \iota_{[v_1,v_2]}A
	  \,.
  $$
  Subtracting $\iota_{v_2} \iota_{v_1} \omega = \iota_{v_2}\iota_{v_1} d_{\mathrm{dR}} A$
  on both sides yields
  $$
    [h_1,h_2] = \mathcal{L}_{v_2}\alpha_1 - \mathcal{L}_{v_1} \alpha_2 -\iota_{[v_1,v_2]}A
	\,.
  $$
  This is equivalently the equation
  $$
    \begin{aligned}
	  {[\phi(\alpha_1), \phi(\alpha_2)]}
	  & =
	  \mathcal{L}_{v_2} \alpha_1 
	  -
	  \mathcal{L}_{v_1} \alpha_2 
      -
	  \iota_{[v_1,v_2]}A
	  \\
	  & =
       \phi([\alpha_1, \alpha_2])
	\end{aligned}
	\,,
  $$  
  which exhibits $\phi$ as a Lie algebra homomorphism.
\endofproof
We recover the following traditional facts from the general
notions of \ref{StrucGeometricPrequantization}.
\begin{observation}
The \emph{Poisson-bracket group} of the symplectic manifold $(X, \hat \omega)$
according to def. \ref{HamiltonianVectorFieldsOnGrpd} is a 
central extension by $U(1)$ of the group of hamiltonian 
symplectomorphisms: we have a short exact sequence of smooth groups
$$
  U(1) \to \mathrm{Poisson}(X, \hat \omega) \to \mathrm{HamSympl}(X, \hat \omega)
  \,.
$$
On Lie algebras this exhibits the Poisson-bracket Lie algebra as a central extension of the
Lie algebra of Hamiltonian vector fields.
$$
  \mathbb{R}
  \to
  \mathfrak{poisson}(X, \hat \omega)
  \to 
  \mathcal{X}_{\mathrm{ham}}(X, \hat \omega)
  \,.
$$
If $(X, \omega)$ is a \emph{symplectic vector space} in that $X$ is a vector space
and the symplectic differential form $\omega$ is constant with respect to 
(left or right) translation along $X$, then the \emph{Heisenberg Lie algebra}
is the sub Lie algebra 
$$
  \mathfrak{heis}(X, \hat \omega) \hookrightarrow \mathfrak{poisson}(X, \hat \omega)
$$
on the constant and the linear functions, see remark \ref{HeisenbergInfinityGroup}.

Traditional literature knows different conventions about which Lie group to 
pick by default as the one integrating a Heisenberg Lie algebra 
(the unique simply-connected
one or one of its discrete quotients).  
By remark \ref{HeisenbergInfinityGroup} the inclusion
$$
  \mathrm{Heis}(X, \hat \omega) \hookrightarrow \mathrm{Poisson}(X, \hat \omega)
$$
picks the one where the central part is integrated to the circle
group:
$$
  \mathrm{Heis}(X, \hat \omega) \simeq X \times U(1)
  \,.
$$
If in this decomposition we write the canonical generator in 
$$
  \mathfrak{heis}(X, \hat \omega) \simeq X \oplus \mathfrak{u}(1)
$$ 
of the summand
$\mathfrak{u}(1) = \mathrm{Lie}(U(1))$ as ``$\mathrm{i}$'' then the Lie bracket
on $\mathfrak{heis}(X, \hat \omega)$ is given on any two $f,g \in X$ by
$$
  [f,g] = \mathrm{i} \omega(f,g)
  \,.
$$
Specifically for the special case $X = \mathbb{R}^2$ 
with canonical basis vectors denoted $\hat q$ and $\hat p$, 
and with $\omega$ the canonical symplectic form, 
the only nontrivial bracket in $\mathfrak{heis}(X, \hat \omega)$ 
among these generators is 
$$
  [\hat q,\hat p]_{\mathfrak{heis}} = \mathrm{i}
  \,.
$$
The image of this equation under the map $\mathfrak{heis}(X, \hat \omega) \to \mathcal{X}_{\mathrm{Ham}}(X, \hat \omega)$ is
$$
  [q,p]_{\mathcal{X}} = 0
  \,,
$$
where now $q,p$ denote the Hamiltonian vector fields associated with $\hat q$ and $\hat p$, 
respectively. The lift from the latter to the former equation is, historically,
the archetypical hallmark of quantization.
\end{observation}

\begin{proposition}
  For $(X,\omega)$ an ordinary prequantizable symplectic manifold
  and $\nabla : X \to \mathbf{B}^1 U(1)$ any choice of 
  prequantum bundle, def. \ref{PrequantizationLift},
  let $V := \mathbb{C}$ and let $\rho$ be the canonical 
  representation of $U(1)$.
  
  Then def. \ref{PrequantumOperators} 
  reduces to the traditional definition to prequantum operators
  in geometric quantization.
  \index{symplectic higher geometry!prequantum operator!ordinary}
\end{proposition}
\proof
  According to the discussion in \ref{Twisted0BundlesSectionsOfVectorBundles}
  the space of sections $\Gamma_X(E)$ is that of the ordinary sections
  of the ordinary associated line bundle.
  
  Notice that part of the statement there is that the standard presentation
  of $\rho : V/\!/U(1) \to \mathbf{B}U(1)$ by a morphism 
  of simplicial presheaves
  $V/\!/U(1)_{\mathrm{ch}} \to \mathbf{B}U(1)_{\mathrm{ch}}$
  is a fibration.  In particular this means, as used there,
  that the $\infty$-groupoid of 
  sections \emph{up to homotopy} is presented already by the 
  Kan complex (which here is just a set) of strict sections $\sigma$
  $$
    \xymatrix{
	   & V/\!/U(1)_{\mathrm{ch}}
	   \ar[d]^\rho
	   \\
	   C(\{U_i\})
	   \ar[r]^{\mathbf{c}}
	   \ar[ur]^{\sigma}
	   \ar[d]^{\simeq}
	   &
	   \mathbf{B}G_{\mathrm{ch}}
	   \\
	   X
	}
  $$
  and it is these that directly identify with the ordinary sections
  of the line bundle $E \to X$.
  
  Now, a Hamiltonian diffeomorphism 
  in the general sense of def. \ref{PrequantumOperators}
  takes such a section $\sigma$ to the pasting composite
  $$
    \xymatrix{
	  && V/\!/U(1)_{\mathrm{conn}}
	  \ar[dd]^{\rho_{\mathrm{conn}}}
	  \\
	  & X
	  \ar[dr]|\nabla_<{\ }="s"
	  \ar[ur]^\sigma
	  \\
	  X 
	    \ar[rr]_\nabla^{\ }="t"
		\ar[ur]^{\phi}
	  && \mathbf{B}U(1)_{\mathrm{conn}}
	  \ar@{=>}^\alpha "s"; "t"
	}
	\,.
  $$
  By the above, to identify this with a section of the line bundle
  in the ordinary sense, we need to find an equivalent homotopy-section
  whose homotopy is, however, trivial, hence a strict section which is
  equivalent to this as a homotopy section.
  
  Inspection shows that there is a unique such equivalence
  whose underlying natural transformations has components induced by the
  inverse of $\alpha$. 
  Then for $h : X \to \mathbb{C}$ a given function and 
  $t \mapsto (\phi(t),\alpha(t))$ the family of 
  Hamiltonian diffeomorphism associated to it by 
  prop. \ref{OrdinaryHamiltonianVectorFields},
  the proof of that proposition shows that
  the infinitesimal difference between the original section $\sigma$
  and this new section is 
  $$
    i \nabla_{v_h} \sigma + h \cdot \sigma
	\,,
  $$
  where $v_h$ is the ordinary Hamiltonian vector field
  induced by $h$.
  This is the traditional formula for the 
  action of the prequantum operator $\hat h$ on prequantum states.
\endofproof

\paragraph{2-Plectic geometry and its prequantization}
 \label{SmoothStrucGeometricPrequantizationOf3Forms}
 \index{structures in a cohesive $\infty$-topos!geometric prequantization!of 3-forms}

We consider now the general notion of 
higher geometric prequantization, \ref{StrucGeometricPrequantization}, 
specialized to the case of closed 3-forms on smooth manifolds, 
canonically regarded in $\mathrm{Smooth}\infty \mathrm{Grpd}$. 
We show that this reproduces the \emph{2-plectic geometry} 
and its prequantization studied in \cite{Rogers}.

\medskip

The following two definitions are from \cite{Rogers}, def. 3.1, prop. 3.15.
\begin{definition}
  A \emph{2-plectic structure}  on a smooth manifold $X$ is
  a smooth closed differential 3-form $\omega \in \Omega^3_{\mathrm{cl}}(X)$,
  which is non-degenerate in that
  the induced morphism
  $$
    \iota_{(-)}\omega : \Gamma(T X) \to \Omega^2(X)
  $$
  has trivial kernel.
  \index{2-plectic structure}
\end{definition}
\begin{definition}
  Let $(X, \omega)$ be a 2-plectic manifold. Then a 1-form
  $h \in \Omega^1(X)$ is called \emph{Hamiltonian}
  if there exists a vector field $v \in \Gamma(T X)$ such that
  $$
    d_{\mathrm{dR}} h = \iota_v \omega
	\,.
  $$
  If this vector field exists, then it is unique and is called the \emph{Hamiltonian vector field}
  corresponding to $\alpha$. We write $v_h$ to indicate this. We write
  $$
    \Omega^1(X)_{\mathrm{Ham}} \hookrightarrow \Omega^1(X)
  $$
  for the vector space of Hamiltonian 1-forms on $(X, \omega)$.
  
  The \emph{Lie 2-algebra of Hamiltonian vector fields} $L_\infty(X, \omega)$
  is the (infinite-dimensional) $L_\infty$-algebra, 
  def. \ref{LInfinityAlgebra}, whose underlying chain complex is
  $$
    \xymatrix{
      \cdots \ar[r] & 0 \ar[r] & C^\infty(X) \ar[r]^<<<<<{d_{\mathrm{dR}}} & \Omega^1_{\mathrm{Ham}}(X)
	}
	\,,
  $$
  whose non-trivial binary bracket is 
  $$
    [-,-] : (h_1, h_2) \mapsto \iota_{v_{h_2}} \iota_{v_{h_1}} \omega
  $$
  and whose non-trivial trinary bracket is
  $$
    [-,-,-] : (h_1, h_2, h_3) 
	 \mapsto 
	 \iota_{v_{h_1}} \iota_{v_{h_2}} \iota_{v_{h_3}} \omega
	\,.
  $$
\index{symplectic higher geometry!Hamiltonian vector fields!2-plectic}
  \label{Hamiltonian1Forms}
\end{definition}

\begin{proposition}
  Let $(X, \omega)$ be a 2-plectic smooth manifold, canonically regarded in
  $\mathrm{Smooth}\infty\mathrm{Grpd}$. 
  Then for $\hat \omega : X \to \mathbf{B}^2 U(1)_{\mathrm{conn}}$
  any prequantum circle 2-bundle with connection (see \ref{SmoothStrucDifferentialCohomology}) 
  for $\omega$, its Poisson Lie 2-algebra, 
  def. \ref{HamiltonianVectorFieldsOnGrpd}, is equivalent to 
  the Lie 2-algebra $L_\infty(X, \omega)$ from def. \ref{Hamiltonian1Forms}:
  $$
    \mathfrak{poisson}(X, \hat \omega) \simeq L_\infty(X, \omega)
	\,.
  $$
\end{proposition}
\proof
  As in the proof of prop. \ref{OrdinaryHamiltonianVectorFields}, 
  we first consider the case that $\omega$ is exact, so that there
  exists a globally defined 2-form $A \in \Omega^2(X)$ with $d_{\mathrm{dR}} A = \omega$.
  The general case follows from this by working on the 
  path fibration surjective submersion, in straightforward generalization of the
  strategy in the proof of prop. \ref{OrdinaryHamiltonianVectorFields}.

  By def. \ref{HamiltonianVectorFieldsOnGrpd}, 
  an object of the smooth 2-group $\mathrm{Poisson}(X, \hat \omega)$ is 
  a diagram of smooth 2-groupoids
  $$
  \xymatrix{
     X \ar[rr]^{\phi}_>>>>{\ }="s" \ar[dr]_{A}^{\ }="t" 
	 && X \ar[dl]^{A}
     \\
     & \mathbf{B}^2 U(1)_{\mathrm{conn}}
	 \ar@{=>}^\alpha "s"; "t"
  }
  \,,
$$
such that map $\phi$ is a diffeomorphism. Given $\phi$, such diagrams correspond 
to $\alpha \in \Omega^1(X)$  such that
\(
  (\phi^* A - A) = d_{\mathrm{dR}} \alpha
  \,.
  \label{IntegralConditionOnHamiltonian1Forms}
\)
Morphisms in the 2-group may go between
two such objects $(f) : (\phi,\alpha_1) \to (\phi, \alpha_2)$  with the same $\phi$ and are 
given by $f \in C^\infty(X, U(1))$ such that 
$$
  \alpha_2 = \alpha_1 + d_{\mathrm{dR}} \mathrm{log} f
  \,.
$$
Under the 2-group product the objects $(\phi,\alpha)$ 
form a genuine group with multiplication given by
$$
  (\phi_1, \alpha_1) \cdot (\phi_2, \alpha_2)
  =
  (\phi_2 \circ \phi_1,
  \alpha_1 + \phi_1^* \alpha_2 )
  \,.
$$
Similarly the group product on two morphisms $(f_1), (f_2) : (\phi, \alpha_1) \to (\phi,\alpha_2)$
is given by
$$
  (f_1) \cdot (f_2) = f_1 \cdot \phi^* f_2
  \,.
$$
Therefore this is a
 \emph{strict} 2-group, def. \ref{Strict2GroupInIntroduction}, 
  given by the subobject of the crossed module
  $$
    \xymatrix{
      C^\infty(X, U(1))  \ar[rr]^<<<<<<<<<{(0,d_{\mathrm{dR}}\mathrm{log})} 
	  &&  \mathrm{Diff}(X) \ltimes \Omega^1(X)
	}
  $$
  on those pairs of vector fields and 1-forms that satisfy 
  (\ref{IntegralConditionOnHamiltonian1Forms}).
  Here $\mathrm{Diff}(X) \ltimes \Omega^1(X)$ is the semidirect product
  group induced by the pullback action on the additive group of 1-forms, 
  and its action
  on $C^\infty(X, U(1))$ is again by the pullback action of the $\mathrm{Diff}(X)$-factor.

  Therefore the $L_\infty$-algebra $\mathfrak{poisson}(X, \hat \omega)$ 
  may be identified with the subobject of the 
  corresponding strict Lie 2-algebra given by the differential crossed module, 
  def. \ref{DifferentialCrossedModule},
  $$
    \xymatrix{
      C^\infty(X) \ar[r]^<<<<<{d_{\mathrm{dR}}}
	  &
	  \Gamma(TX) \oplus \Omega^1(X)
	}
  $$
  on those pairs $(v,\alpha) \in \Gamma(T X) \times \Omega^1(X)$ for which 
  $$
    \mathcal{L}_v A = d_{\mathrm{dR}} \alpha
	\,,
  $$
  hence, by Cartan's formula, for which
  $$
    h := \alpha - \iota_v A
  $$ 
  is a Hamiltonian 1-form for $v$, def. \ref{Hamiltonian1Forms}. 
  Here $\Gamma(T X)\oplus \Omega^1(X)$ is the 
  semidirect product Lie algebra with bracket
  $$
    [(v_1, \alpha_1), (v_2,\alpha_2)] = ([v_1,v_2], \mathcal{L}_{v_2} \alpha_1 - \mathcal{L}_{v_1} \alpha_2)
  $$
  and 
  its action on $f \in C^\infty(X)$
  is by Lie derivatives of the $\Gamma(T X)$-summand:
  $$
    [(v,\alpha), f] = -\mathcal{L}_v f
	\,.
  $$  
  For emphasis, we write $\Omega^1_{\mathrm{Ham}, p} \subset \Gamma(T X) \oplus \Omega^1(X)$ 
  for the vector space of pairs $(v,\alpha)$ with $\alpha - \iota_v A$ Hamiltonian. 
  The map $\phi : (\alpha, v) \mapsto \alpha - \iota_v A$ consistutes
  a vector space isomorphism
  $$
    \phi : \Omega^1_{\mathrm{Ham}, p} \stackrel{\simeq}{\to} \Omega^1_{\mathrm{Ham}}
  $$  
  and for the moment it is useful to keep this around explicitly.
  So $\mathfrak{poisson}(X, \hat \omega)$ is given by the differential crossed module
  on the top of the diagram
  $$
    \raisebox{20pt}{
    \xymatrix{
      C^\infty(X) \ar[r]^<<<<<{d_{\mathrm{dR}}}
	  \ar[d]^{=}
	  &
	  \Omega^1_{\mathrm{Ham},p}(X)
	  \ar@{^{(}->}[d]
	  \\
      C^\infty(X) \ar[r]^<<<<<{d_{\mathrm{dR}}}
	  &
	  \Gamma(T X) \oplus \Omega^1(X)
	}
	}
	\,,
  $$
  with brackets induced by this inclusion into the crossed module on the bottom.
  
  We need to check that with these brackets the chain map
  $$
    \xymatrix{
	  C^\infty(X) \ar[r]^{\mathrm{id}} 
	  \ar[d]^{d_{\mathrm{dR}}}
	  & C^\infty(X)
	  \ar[d]^{d_{\mathrm{dR}}}
	  \\
	  \Omega^1(X)_{\mathrm{Ham},p}
	  \ar[r]^{\phi}
	  &
	  \Omega^1(X)_{\mathrm{Ham}}
	  \\
	  [-,-]
	  &
	  ([-,-]', J)
	}
  $$
  is a Lie 2-algebra equivalence from the strict brackets
  $[-,-]$ to the brackets $([-,-]',[-,-,-]')$ of 
  def. \ref{Hamiltonian1Forms}.
  
  To that end, first notice the equation
  $$
    \begin{aligned}
	  2  \iota_{v_2} \iota_{v_1} \omega 
	  & =
	  \iota_{v_2} d_{\mathrm{dR}} h_1 - \iota_{v_1} d_{\mathrm{dR}} h_2
	  \\
	  & = 
	  \mathcal{L}_{v_2}(\alpha_1 - \iota_{v_1}A )
	  -
	  \mathcal{L}_{v_1}(\alpha_2 - \iota_{v_2}A )
	  + 
	  d_{\mathrm{dR}}(\iota_{v_1} h_2 - \iota_{v_2}h_1)
	  \\
	  & =
	  \mathcal{L}_{v_2} \alpha_1 
	  -
	  \mathcal{L}_{v_1} \alpha_2 
	  +
	  \iota_{v_2}\iota_{v_1} d_{\mathrm{dR}} A - \iota_{[v_1,v_2]}A	  
	  +
	  d_{\mathrm{dR}}(\iota_{v_1} h_2 - \iota_{v_2}h_1 - \iota_{v_2} \iota_{v_1} A)
	  \,,
	\end{aligned}
  $$
  where in the last step we used the identity
  $$
    \iota_{v_2} \iota_{v_1} d_{\mathrm{dR}} A = 
      \mathcal{L}_{v_1} \iota_{v_2}A 
   	   -
      \mathcal{L}_{v_2} \iota_{v_1}A
      -
	  \iota_{[v_1,v_2]}A
	  +
	  d_{\mathrm{dR}} \iota_{v_2} \iota_{v_1} A
	  \,.
  $$
  Subtracting $\iota_{v_2} \iota_{v_1} \omega = \iota_{v_2}\iota_{v_1} d_{\mathrm{dR}} A$
  on both sides yields
  $$
    \begin{aligned}
	  \iota_{v_2} \iota_{v_1} \omega 
	  & =
	  \mathcal{L}_{v_2} \alpha_1 
	  -
	  \mathcal{L}_{v_1} \alpha_2 
	  - 
	  \iota_{[v_1,v_2]}A	  
	  +
	  d_{\mathrm{dR}}(\iota_{v_1} h_2 - \iota_{v_2}h_1 - \iota_{v_2} \iota_{v_1} A)
	  \,,
	\end{aligned}
  $$  
  Here on the left we have the bracket of $h_1$ with $h_2$ in 
  def. \ref{Hamiltonian1Forms}, which we will write 
  $[h_1, h_2]' := [\phi(v_1, \alpha_1), \phi(v_2, \alpha_2)]'$,
  whereas the first three terms on the right are the image under
  $\phi$ of the bracket from above, to be written $\phi[(v_1, \alpha_1), (v_2, \alpha_2)]$.
  Therefore this equation says that
  \(
    [\phi(v_1, \alpha_1), \phi(v_2, \alpha_2)]'
	=
	\phi([(v_1,\alpha_1), (v_2,\alpha_2)])
	+
	d_{\mathrm{dR}}(\iota_{v_1} \phi(v_2,\alpha_2) - \iota_{v_2}\phi(v_1,\alpha_1) - \iota_{v_2} \iota_{v_1} A)	
	\,.
	\label{EquationInComputingPoissonLie2Algebra}
  ´\)
  In view of the exact term on the far right, this implies that 
  the map
  $$
    \Phi : \Omega^1(X)_{\mathrm{Ham},ü} \otimes \Omega^1(X)_{\mathrm{Ham},p} \to C^\infty(X)
  $$
  given by 
  $$
    \Phi : (h_1 = \alpha_1 - \iota_{v_1}A, h_2 = \alpha_2 - \iota_{v_2}A)  
	 \mapsto 
	 \iota_{v_1} h_2 - \iota_{v_2}h_1 - \iota_{v_2} \iota_{v_1} A
  $$
  should be a chain homotopy between the binary brackets
  $$
    \raisebox{20pt}{
    \xymatrix{
	   (\Omega^1(X)_{\mathrm{Ham},p} \otimes C^\infty(X) )
	   \oplus 
	   (C^\infty(X) \otimes \Omega^1(X)_{\mathrm{Ham},p})
	   \ar[rrr]^<<<<<<<<<<<<{[-,-]' - [-,-]}
	   \ar[d]|{(\mathrm{id}\otimes d_{\mathrm{dR}}) \oplus (d_{\mathrm{dR}} \otimes \mathrm{id})}
	   &&&
	   C^\infty(X)
	   \ar[d]^{d_{\mathrm{dR}}}
	   \\
	   \Omega^1(X)_{\mathrm{Ham},p} \otimes \Omega^1(X)_{\mathrm{Ham},p}
	   \ar[rrr]_{[\phi(-),\phi(-)]' - \phi([-,-])}
	   \ar[urrr]|\Phi
	   &&&
	   \Omega^1_{\mathrm{Ham}}(X)
	}
	}
	\,.
  $$
  Indeed, the bottom right triangle commutes manifestly, 
  by equation (\ref{EquationInComputingPoissonLie2Algebra}). 
  For the top left triangle notice that $[-,-]'$ vanishes here, by definition, 
  and $[-,-]$ is given by
  $$
    [(v,\alpha), f] = -\mathcal{L}_v f
	\,.
  $$
  On the other hand, since the Hamiltonian vector field of $d_{\mathrm{dR} }f$ vanishes,
  we also have
  $$
    \begin{aligned}
      \Phi((v,\alpha), (0,d_{\mathrm{dR}} f))
	  &=
      \iota_v d_{\mathrm{dR}} f
	  \\
	  & = \mathcal{L}_v f
	\end{aligned}
	\,.
  $$

  It remains to check that $\Phi$ respects the Jacobiator, sending the trivial one on
  $\Omega^1(X)_{\mathrm{Ham},p}$ to the nontrivial one of def. \ref{Hamiltonian1Forms}.
  From now on we leave the isomorphism 
  $\phi : \Omega^1(X)_{\mathrm{Ham},p} \stackrel{\simeq}{\to} \Omega^1(X)_{\mathrm{Ham}}$
  implicit, regarding $[-,-]'$ and $[-,-]$ as two different brackets on the same
  vector space.

  Observe that generally, with a chain homotopy of binary brackets
  $\Phi$ given as above, setting
  $$
    J(h_1, h_2, h_3) := \Phi(h_1, [h_2,h_3]) + \mathrm{cyc}
  $$
  for all $h_1, h_2, h_3$ makes the collection of brackets $([-,-]', J)$ 
  (extended by 0 to $C^\infty(X)$) a Lie 2-algebra structure on 
  $C^\infty(X) \to \Omega^1(X)_{\mathrm{Ham}}$ such that $(\phi,\Phi)$ a Lie 2-algebra
  equivalence.
  Notice that we may equivalently write
  $$
    J(h_1, h_2, h_3) = -\Phi(D(h_1 \vee h_2 \vee h_3))
	\,,
  $$
  where $(\vee^\bullet \Omega^1(X)_{\mathrm{Ham}}, D)$ is the differential
  coalgebra incarnation of the Lie algebra $[-,-]$.  

  Indeed, $J$ vanishes on the image of $d_{\mathrm{dR}}$, because
  $$
    \begin{aligned}
      \Phi(d_{\mathrm{dR}} f, [h_2,h_3])  
  	  +
      \Phi(h_2, [h_3, d_{\mathrm{dR}} f]) 
	  + 
	  \Phi(h_3, [d_{\mathrm{dR}} f, h_2])
	  &=
	  -
	  d_{\mathrm{dR}}
	  \left(
	  [f, [h_2, h_3
	  + 
	  [h_2, [h_3,f
	  + 
	  [h_3,[f,h_2
	  \right)
	  \\
	  & = 0
	\end{aligned}
	\,,
  $$
  where we used the chain homotopy property of $\phi$ and the identities of the
  differential crossed module $[-,-]$.
  
  Using this, the coherence law of the Jacobiator, which a priori involves
  $[-,-]'$, is equivalently formulated in terms of $[-,-]$
  (because the two differ by something in the image of $d_{\mathrm{dR}}$), where it 
  then reads
  $$
    J(D(h_1 \vee h_2 \vee h_3 \vee h_4)) = 0
	\,,
  $$
  with $(\vee^\bullet \Omega^1(X)_{\mathrm{Ham}}, D)$ as before. This equation
  follows now due to $D^2 = 0$.
  
   Finally, to see that $J$ as above indeed is a Jacobiator
   for $[-,-]'$ we compute
  $$
    \begin{aligned}
      \,[h_1,[h_2,h_3]']' + \mathrm{cyc}
	  & = 
	  [h_1, [h_2,h_3 + d_{\mathrm{dR}} \Phi(h_2,h_3)]' + \mathrm{cyc}
	  \\
	  & =
	  [h_1,[h_2,h_3 
	  + 
	  [h_1, d_{\mathrm{dR}} \Phi(h_2,h_3)]
	  + d_{\mathrm{dR}}\Phi(h_1, [h_2,h_3] +  d_{\mathrm{dR}} \Phi(h_2,h_3))
	  + 
	  \mathrm{cyc}
	  \\
	  & =
	  d_{\mathrm{dR}}\Phi(h_1, [h_2,h_3]) + \mathrm{cyc}
	\end{aligned}
	\,,
  $$
  where in the last step the first summand disappears due to the
  Jacobi identity satisfied by $[-,-]$, and where we used the 
  chain homotopy propoerty of $\Phi$ to cancel
  two terms.
  
  This way we have produced an equivalence of Lie 2-algebras
  $$
    (\phi, \Phi) :  \mathfrak{poisson}(X, \hat \omega)
	\to
	((C^\infty(X)\to \Omega^1(X)_{\mathrm{Ham}}), [-,-]', J)
	\,,
  $$
  where on the right the binary bracket is that of def. \ref{Hamiltonian1Forms}.
  The last thing to check is that the Jacobiator $J$ is indeed that of
  def. \ref{Hamiltonian1Forms}. But since the differential in the 
  Lie 2-algebra is $d_{\mathrm{dR}}$, any two Jacobiators for the same
  binary bracket must differ by a constant function on $X$. Since at the same
  time the Jacobiators are linear, that constant must be 0, and hence the
  two Jacobiators must coincide.
\endofproof
\newpage

\subsection{Synthetic differential $\infty$-groupoids}
\label{SynthDiffInfGrpd}
\index{cohesive $\infty$-topos!models!synthetic-differential cohesion}

We discuss $\infty$-groupoids equipped with \emph{synthetic differential cohesion},
a version of smooth cohesion in which an explicit notion of 
smooth \emph{infinitesimal} spaces exists.

\medskip

Notice that the category $\mathrm{CartSp}_{\mathrm{smooth}}$, 
def. \ref{CartSpSmooth},  is 
(the syntactic category of) a finitary algebraic theory: a \emph{Lawvere theory} 
(see chapter 3, volume 2 of \cite{Borceux}). 
\begin{definition} \label{SmoothAlgebras}
Write
  $$
    \mathrm{SmoothAlg} := \mathrm{Alg}(\mathrm{CartSp}_{\mathrm{smooth}})
  $$
  for the category of algebras over the algebraic theory $\mathrm{CartSp}_{\mathrm{smooth}}$:
  the category of product-preserving functors $\mathrm{CartSp}_{\mathrm{smooth}} \to \mathrm{Set}$.
\end{definition}
These algebras are traditionally known as \emph{$C^\infty$-rings} or \emph{$C^\infty$-algebras} 
\cite{KainzKrieglMichor}.
\begin{proposition} 
  \label{SmoothManifoldsAsSmoothLoci}
  The map that sends a smooth manifold $X$ to the product-preserving functor
  $$
    C^\infty(X) : \mathbb{R}^k \mapsto \mathrm{SmoothMfd}(X, \mathbb{R}^k)
  $$
  extends to a full and faithful embedding
  $$
    \mathrm{SmoothMfd} \hookrightarrow \mathrm{SmoothAlg}^{\mathrm{op}}
    \,.
  $$
\end{proposition} 
\begin{proposition} 
  \label{InfinitesimalSmoothAlgebras}
  Let $A$ be an ordinary (associative) $\mathbb{R}$-algebra that as an $\mathbb{R}$-vector space
  splits as $\mathbb{R}\oplus V$ with $V$ finite dimensional as an $\mathbb{R}$-vector space and
  nilpotent with respect to the algebra structure: 
  $(v \in V \hookrightarrow A) \Rightarrow (v^2 = 0)$.
  
  There is a unique lift of $A$ through the forgetful functor $\mathrm{SmoothAlg} \to \mathrm{Alg}_{\mathbb{R}}$.
\end{proposition}
\proof
  Use Hadamard's lemma.
\endofproof
\begin{remark}
  In the context of synthetic differential geometry the algebras of prop.
  \ref{InfinitesimalSmoothAlgebras} are usually called \emph{Weil algebras}.
  In other contexts however the underlying rings are known as 
  \emph{Artin rings}, see for instance \cite{LurieFormal}.
\end{remark}
\begin{definition}
  Write
  $$
    \mathrm{InfSmoothLoc} \hookrightarrow \mathrm{SmoothAlg}^{\mathrm{op}}
  $$
  for the full subcategory of the opposite of smooth algebras on those of the form of 
  prop. \ref{InfinitesimalSmoothAlgebras}. We call this the category of 
  \emph{infinitesimal smooth loci} or of \emph{infinitesimally thickened points}.

  Write 
  $$
    \mathrm{CartSp}_{\mathrm{synthdiff}} 
    := \mathrm{CartSp}_{\mathrm{smooth}} \ltimes \mathrm{InfSmoothLoc}
    \hookrightarrow \mathrm{SmoothAlg}^{\mathrm{op}}
  $$
  for the full subcategory of the opposite of smooth algebras on those that are products
  $$
    X \simeq U \times D
  $$
  in $\mathrm{SmoothAlg}^{\mathrm{op}}$ of an object $U$ in the image of 
  $\mathrm{CartSp}_{\mathrm{smooth}} \hookrightarrow \mathrm{SmoothMfd} \hookrightarrow
   \mathrm{SmoothAlg}^{\mathrm{op}}$ and an object $D$ in the image of
   $\mathrm{InfSmoothLoc} \hookrightarrow \mathrm{SmoothAlg}^{\mathrm{op}}$.

  Define a coverage on $\mathrm{CartSp}_{\mathrm{synthdiff}}$ whose covering families
  are precisely those of the form $\{ U_i \times D \stackrel{(f_i, \mathrm{id})}{\to} U \times D\}$
  for $\{U_i \stackrel{f_i}{\to} U\}$ a covering family in $\mathrm{CartSp}_{\mathrm{smooth}}$.
  \label{InfinitesimalSmoothLoci}
\end{definition}
\begin{remark}
This definition appears in \cite{Kock}, following \cite{DubucCahiers}. The sheaf topos 
$\mathrm{Sh}(\mathrm{CartSp}_{\mathrm{synthdiff}})\hookrightarrow \mathrm{SynthDiff}\infty\mathrm{Grpd}$ 
over this site is equivalent to the 
\emph{Cahiers topos} \cite{DubucCahiers} which is a model of some set of axioms 
of \emph{synthetic differential geometry} (see \cite{LawvereSynth} for the abstract idea,
where also the relation to the axiomatics of cohesion is vaguely indicated).
Therefore the following definition may be thought of as describing the \emph{$\infty$-Cahiers topos}
providing a higher geometry version of this model of synthetic differential smooth geometry.
 \label{0TruncatedInSynthDiffIsCahiers}
\end{remark}
\begin{definition}
  The $\infty$-topos of \emph{synthetic differential smooth $\infty$-groupoids} is
  $$
    \mathrm{SynthDiff}\infty \mathrm{Grpd}
    :=
    \mathrm{Sh}_{(\infty,1)}(\mathrm{CartSp}_{\mathrm{synthdiff}})
    \,.
  $$
\end{definition}
\begin{proposition} 
  \label{SynthDiffIsCohesive}
  $\mathrm{SynthDiff}\infty \mathrm{Grpd}$ is a cohesive $\infty$-topos.
\end{proposition}
\proof
  Using that the covering families of $\mathrm{CartSp}_{\mathrm{synthdiff}}$ 
  do by definition not depend on the infinitesimal
  smooth loci $D$ and that these each have a single point, one finds that
  $\mathrm{CartSp}_{\mathrm{synthdiff}}$ is an $\infty$-cohesive site, def. 
  \ref{CohesiveSite}, by reducing to the argument as for 
  $\mathrm{CartSp}_{\mathrm{top}}$, prop. \ref{CartSpTopIsCohesive}.
  The claim then follows with prop. \ref{InfSheavesOverCohesiveSiteAreCohesive}.
\endofproof
\begin{definition}
  Write $\mathrm{FSmoothMfd}$ for the category of \emph{formal smooth manifolds} -- 
  manifolds modeled on 
  $\mathrm{CartSp}_{\mathrm{synthdiff}}$, equipped with the induced site structure.
  \label{FormalSmoothManifolds}
\end{definition}
\begin{proposition} \label{SynthSmoothEquivalentlyOverFormalMfd}
  We have an equivalence of $\infty$-categories
  $$
    \mathrm{SynthDiff} \infty \mathrm{Grpd} \simeq
      \hat{\mathrm{Sh}}_{(\infty,1)}(\mathrm{FSmoothMfd})
  $$
  with the hypercomplete $\infty$-topos over formal smooth manifolds.
\end{proposition}
\proof
  By definition $\mathrm{CartSp}_{\mathrm{synthdiff}}$ is a dense sub-site of $\mathrm{FSmoothMfd}$.
  The statement then follows as in prop. \ref{ToposOverTopMfd}.
\endofproof
Write $i : \mathrm{CartSp}_{\mathrm{smooth}} \hookrightarrow 
\mathrm{CartSp}_{\mathrm{synthdiff}}$ for the canonical embedding.
\begin{proposition} \label{InfinitesimalnessofSynthDiff}
  The functor $i^*$ given by restriction along 
 $i$ exhibits  $\mathrm{SynthDiff}\infty \mathrm{Grpd}$ as an 
  infinitesimal cohesive neighbourhood, def. \ref{InfinitesimalCohesiveNeighbourhood},
  of $\mathrm{Smooth}\infty \mathrm{Grpd}$, 
  in that we have a quadruple of adjoint $\infty$-functors
$$
 ( i_! \dashv i^* \dashv i_* \dashv i^! )
  :
  \mathrm{Smooth} \infty \mathrm{Grpd}
   \to 
    \mathrm{SynthDiff} \infty \mathrm{Grpd}
  \,,
$$
such that $i_!$ is full and faithful and preserves the terminal object.
\end{proposition}
\proof
  We observe that $\mathrm{CartSp}_{\mathrm{smooth}} \hookrightarrow \mathrm{CartSp}_{\mathrm{synthdiff}}$
  is an infinitesimal neighbourhood of sites, according to def. \ref{InfinitesimalNeighBourhoodSite}.
  The claim then follows with prop. \ref{InfinitesimalNeighbourhoodFromSites}.
\endofproof

\newpage

We now discuss the general abstract structures in cohesive $\infty$-toposes, 
\ref{structures} and \ref{InfinitesimalCohesion}, 
realized in $\mathrm{SynthDiff}\infty \mathrm{Grpd}$

\begin{itemize}
 \item \ref{StrucSynthLie} -- $\infty$-Lie algebroids;
 \item \ref{StrucSynthDiffSeparatedmanifolds} -- Manifolds
 \item \ref{StrucSynthCohomology} -- Cohomology;
 \item \ref{StrucSynthPostnikov} -- Paths and geometric Postnikov towers;
 \item \ref{FormallySmoothInSynthDiff} -- Formally smooth/{\'e}tale/unramified morphisms;
 \item \ref{StrucSynthDiffFormallyEtaleGroupoid} -- Formally {\'e}tale groupoids;
 \item \ref{StrucSynthChernWeil} -- Chern-Weil theory.
\end{itemize}

\subsubsection{$\infty$-Lie algebroids}
 \label{StrucSynthLie}
 \index{Lie algebroid!in $\mathrm{SynthDiff}\infty\mathrm{Grpd}$}
  \index{structures in a cohesive $\infty$-topos!exponentiated $\infty$-Lie algebras!synthetic differential}

We discuss explicit presentations for first order 
formal cohesive $\infty$-groupoids, \ref{InfStrucFormalInfinityGroupoid}, 
realized in $\mathrm{SynthDiff}\infty \mathrm{Grpd}$.
We call these \emph{$L_\infty$-algebroids}, 
subsuming the traditional notion of \emph{$L_\infty$-algebras} \cite{lada-markl}.

In the standard presentation of $\mathrm{SynthDiff}\infty \mathrm{Grpd}$ 
by simplicial presheaves over formal smooth manifolds 
these $L_\infty$-algebroids are 
presheaves in the image of the \emph{monoidal Dold-Kan correspondence}
\cite{CasCor} of
semi-free differential graded algebras. 
This construction 
amounts to identifying the traditional description of Lie algebras, Lie algebroids and 
$L_\infty$-algebras by their Chevalley-Eilenberg algebras, def. \ref{LInfinityAlgebra}, 
as a convenient characterization of the corresponding cosimplicial algebras whose formal dual 
simplicial presheaves are manifest presentations of infinitesimal smooth $\infty$-groupoids.

\begin{itemize}
  \item \ref{LInfinityAlgebroidsAndSmoothCommutativeDGAlgebras}
    --  $L_\infty$-Algebroids and smooth commutative dg-algebras;
  \item
    \ref{InfinitesimalSmoothInfinityGroupoids}
	 -- Infinitesimal smooth $\infty$-groupoids;
  \item
    \ref{Lie1AlgebroidsAsInfinitesimalSimplicialPresheaves}
	  -- Lie 1-algebroids as infinitesimal simplicial presheaves
\end{itemize}

\paragraph{$L_\infty$-Algebroids and smooth commutative dg-algebras}
\label{LInfinityAlgebroidsAndSmoothCommutativeDGAlgebras}
\index{$L_\infty$-algebroid}

Recall the characterization of $L_\infty$-algebra structures in terms of dg-algebras from prop. \ref{LInfinityAlgebraFromDGAlgebra}.
\begin{definition} 
  \label{LInfinityAlgebrasSubAlgebroids}
  \label{LInftyGlgebroid}
  \label{LInfinityAlgebroids}
  \index{$L_\infty$-algebroid!definition}
Let
$$
  \mathrm{CE} : L_\infty \mathrm{Algd} \hookrightarrow \mathrm{cdgAlg}_{\mathbb{R}}^{\mathrm{op}}
$$
be the full subcategory on the opposite category of cochain dg-algebras over $\mathbb{R}$ 
on those dg-algebras that are
\begin{itemize}
\item graded-commutative;
\item concentrated in non-negative degree (the differential being of degree +1 );
\item in degree 0 of the form $C^\infty(X)$ for $X \in \mathrm{SmoothMfd}$ ;

\item semifree: their underlying graded algebra is isomorphic to an exterior algebra on 
  an $\mathbb{N}$-graded locally free projective $C^\infty(X)$-module;

\item of finite type;
\end{itemize}
We call this the category of \emph{$L_\infty$-algebroids}
\index{$L_\infty$-algebroid}
\index{Lie algebroid!$L_\infty$-algebroid} over smooth manifolds. 
\end{definition}
More in detail, an object $\mathfrak{a} \in L_\infty \mathrm{Algd}$ may be identified 
(non-canonically) with a pair $(\mathrm{CE}(\mathfrak{a}), X)$, where
\begin{itemize}
\item $X \in \mathrm{SmoothMfd}$ is a smooth manifold -- called the 
  \emph{base space} of the $L_\infty$-algebroid ;

\item $\mathfrak{a}$ is the module of smooth sections of an $\mathbb{N}$-graded vector bundle of 
 degreewise finite rank;

\item $\mathrm{CE}(\mathfrak{a}) = (\wedge^\bullet_{C^\infty(X)} \mathfrak{a}^*, d_{\mathfrak{a}})$ 
  is a semifree dg-algebra on $\mathfrak{a}^*$ -- a Chevalley-Eilenberg algebra -- where
  $$
    \wedge^\bullet_{C^\infty(X)}\mathfrak{a}^* = 
      C^\infty(X) 
       \; \oplus \;
      \mathfrak{a}^*_0
       \; \oplus \;
      \left(
       \left(\mathfrak{a}^*_0 \wedge_{C^\infty(X)} \mathfrak{a}^*_0\right)
       \oplus
      \mathfrak{a}^*_1
      \right)
      \; \oplus \; \cdots
  $$
  with the $k$th summand on the right being in degree $k$.
\end{itemize}
\begin{definition}
An $L_\infty$-algebroid with base space $X = *$ the point is an $L_\infty$-algebra $\mathfrak{g}$, 
def. \ref{LInfinityAlgebra},
or rather is the pointed delooping of an $L_\infty$-algebra. We write $b \mathfrak{g}$ for 
$L_\infty$-algebroids over the point. They form the full subcategory
$$
  L_\infty \mathrm{Alg} \hookrightarrow L_\infty \mathrm{Algd}
  \,.
$$
\end{definition}
The following fact is standard and straightforward to check.
\begin{proposition}
\begin{enumerate}
\item The full subcategory $L_\infty \mathrm{Alg} \hookrightarrow L_\infty \mathrm{Algd}$ 
from def. \ref{LInfinityAlgebrasSubAlgebroids} is equivalent to the traditional definition of the category of 
$L_\infty$-algebras and ``weak morphisms'' / ``sh-maps'' between them.

\item The full subcategory $\mathrm{LieAlgd} \hookrightarrow L_\infty\mathrm{Algd}$ on the 
1-truncated objects is equivalent to the traditional category of Lie algebroids 
(over smooth manifolds).

\item In particular the joint intersection 
$\mathrm{Lie Alg} \hookrightarrow L_\infty \mathrm{Alg}$ on the 1-truncated $L_\infty$-algebras 
is equivalent to the category of ordinary Lie algebras.
\end{enumerate}
\end{proposition}

We now construct an embedding of $L_\infty \mathrm{Algd}$ into $\mathrm{SynthDiff}\infty \mathrm{Grpd}$.
Below in \ref{InfinitesimalSmoothInfinityGroupoids} we show that this embedding exhibitsthe above 
algebraic data as a presentation of synthetic differential $\infty$-groupoids
which are infinitesimal objects in the abstract intrinsic sense 
of \ref{InfinitesimalSmoothInfinityGroupoids}.
 
\begin{remark}
The functor
$$
  \Xi : \mathrm{Ch}^\bullet_+(\mathbb{R}) \to \mathrm{Vect}_{\mathbb{R}}^{\Delta}
$$
of the Dold-Kan correspondence, prop. \ref{EmbeddingOfChainComplexes},
from non-negatively graded cochain complexes of vector spaces to 
cosimplicial vector spaces is a lax monoidal functor and hence induces a functor 
(which we will denote by the same symbol)
$$
  \Xi : \mathrm{dgAlg}_{\mathbb{R}}^+ \to \mathrm{Alg}_{\mathbb{R}}^{\Delta}
$$
from non-negatively graded commutative cochain dg-algebras to cosimplicial commutative algebras (over $\mathbb{R}$).
\label{DoldKanFunctorIsLaxMonoidal}
\end{remark}
\begin{definition} 
  \label{PresentationByMonoidalDoldKan}
  \index{$L_\infty$-algebroid!simplicial presentation}
Write
$$
  \Xi \mathrm{CE} : L_\infty \mathrm{Algd} \to (\mathrm{CAlg}_{\mathbb{R}}^\Delta)^{\mathrm{op}} 
$$
for the restriction of the functor $\Xi$ from remark \ref{DoldKanFunctorIsLaxMonoidal} 
along the defining inclusion 
$\mathrm{CE} : L_\infty \mathrm{Algd} \hookrightarrow \mathrm{dgAlg}^{\mathrm{op}}_{\mathbb{R}}$.
\end{definition}
There are several different ways to present $\Xi \mathrm{CE}$ explicitly in 
components. Below we make use of the following fact,
pointed out in \cite{CasCor} (see the discussion around equations (26) and (49) there).
\begin{proposition}
The functor $\Xi \mathrm{CE}$ from def. \ref{PresentationByMonoidalDoldKan} is given as follows.

For $\mathfrak{a} \in L_\infty \mathrm{Algd}$, the underlying cosimplicial vector space of 
$\Xi \mathrm{CE}(\mathfrak{a})$ is
$$
  \Xi \mathrm{CE}(\mathfrak{a})
   : 
  [n]
    \mapsto
  \bigoplus_{i = 0}^{n}
  \mathrm{CE}(\mathfrak{a})_i \otimes \wedge^i \mathbb{R}^n 
  \,.
$$
The product of the $\mathbb{R}$-algebra structure on this space in degree $n$ is given on 
homogeneous elements 
$(\omega,x), (\lambda,y) \in \mathrm{CE}(\mathfrak{a})_i \otimes \wedge^i \mathbb{R}^n$ 
in the tensor product by
$$
  (\omega , x)\cdot (\lambda ,y) = (\omega \wedge \lambda , x \wedge y)
  \,.
$$
(Notice that $\Xi \mathfrak{a}$ is indeed a \emph{commutative} cosimplicial algebra, since $\omega$ and $x$ in $(\omega,x)$ are by definition in the same degree.)

To define the cosimplicial structure, let 
$\{v_j\}_{j = 1}^n$ be the canonical basis of $\mathbb{R}^n$ and consider 
and set $v_0 := 0$ to obtain a set of vectors $\{v_j\}_{j = 0}^n$.
Then for $\alpha : [k] \to [l]$ a morphism in the simplex category, set
$$
  \alpha : v_j \mapsto v_{\alpha(j)} - v_{\alpha(0)}
$$
and extend this skew-multilinearly to a map 
$\alpha : \wedge^\bullet \mathbb{R}^k \to \wedge^\bullet \mathbb{R}^l$. 
In terms of all this the action of $\alpha$ on homogeneous elements 
$(\omega,x)$ in the cosimplicial algebra is defined by
$$
  \alpha : 
  (\omega, x) 
    \mapsto
  (\omega, \alpha x) + (d_\mathfrak{a} \omega , v_{\alpha(0)}\wedge \alpha(x))
$$
\label{ExplicitVersionOfMonoidalDoldKan}
\end{proposition}
\begin{remark}
  The commutative algebras appearing here 
  may be understood geometrically as being algebras of functions 
  on spaces of infinitesimal based simplices. This we discuss in more detail
  in \ref{Lie1AlgebroidsAsInfinitesimalSimplicialPresheaves} below,
  see prop. \ref{FunctionsOnInfinitesimalSimplicesAreAsInMonoidalDoldKan}
  there.
\end{remark}

We now refine the image of $\Xi$ to cosimplicial \emph{smooth} algebras, def. \ref{SmoothAlgebras}.
Notice that there is a canonical forgetful functor
$$
  U : \mathrm{SmoothAlg} \to \mathrm{CAlg}_{\mathbb{R}}
$$
from the category of smooth algebras 
to the category of commutative associative algebras over the real numbers.
\begin{proposition} 
 \label{SmoothMonoidalDoldKan}
There is a unique factorization of the functor 
$\Xi \mathrm{CE}: L_\infty \mathrm{Algd} \to (\mathrm{CAlg}_{\mathbb{R}}^\Delta)^{\mathrm{op}}$ 
from def. \ref{PresentationByMonoidalDoldKan} through the forgetful functor
$(\mathrm{SmoothAlg}_{\mathbb{R}}^\Delta)^{\mathrm{op}} \to (\mathrm{CAlg}_{\mathbb{R}}^\Delta)^{\mathrm{op}}$ 
such that for any $\mathfrak{a}$ over base space $X$ the degree-0 algebra of smooth functions 
$C^\infty(X)$ lifts to its canonical structure as a smooth algebra
$$
  \xymatrix{
    & (\mathrm{SmoothAlg}^{\Delta})^{\mathrm{op}}
       \ar[d]^U
     \\
    L_\infty \mathrm{Algd} \ar[ur]^{\Xi \mathrm{CE}} \ar[r] 
	& (\mathrm{CAlg}_{\mathbb{R}}^\Delta)^{\mathrm{op}}
  }
  \,.
$$
\end{proposition}
\proof
Observe that for each $n$ the algebra $(\Xi \mathrm{CE}(\mathfrak{a}))_n$ is a finite nilpotent extension 
of $C^\infty(X)$. The claim then follows with the fact that 
$C^\infty : \mathrm{SmoothMfd} \to \mathrm{CAlg}_{\mathbb{R}}^{\mathrm{op}}$ is 
faithful and using Hadamard's lemma for the nilpotent part. 
\endofproof

\begin{proposition}
  The functor $\Xi \mathrm{CE}$ preserves limits
  of $L_\infty$-algebras. It preserves pullbacks of 
  $L_\infty$-algebroids if the two morphisms in degree 0
  are transveral maps of smooth manifolds.
  \label{LInfinityEmbeddingPreservesSomeLimits}
\end{proposition}
\proof
  The functor $\Xi : \mathrm{cdgAlg}^+_{\mathbb{R}} \to \mathrm{CAlg}_\mathbb{R}^\Delta$ 
  evidently preserves colimits. This gives the first statement. 
  The second follows by observing that the functor from smooth manifolds
  to the opposite of smooth algebras preserves transversal pullbacks.
\endofproof

\paragraph{Infinitesimal smooth groupoids}
\label{InfinitesimalSmoothInfinityGroupoids}

We discuss how the $L_\infty$-algebroids from def. \ref{LInftyGlgebroid}
serve to present the intrinsically defined infinitesimal smooth $\infty$-groupoids 
from \ref{InfStrucFormalInfinityGroupoid}.

\medskip

\begin{definition} 
  \label{EmbeddingOfThePresentation}
Write $i : L_\infty \mathrm{Algd} \to \mathrm{SynthDiff}\infty \mathrm{Grpd}$ 
for the composite $\infty$-functor
$$
  \xymatrix{
  L_\infty \mathrm{Algd} 
     \ar[r]^<<<<{\Xi \mathrm{CE}}
	 &
  (\mathrm{SmoothAlg}^{\Delta})^{\mathrm{op}}
    \ar[r]^<<<<{j}
    &	
  [\mathrm{CartSp}_{\mathrm{synthdiff}}^{\mathrm{op}}, \mathrm{sSet}]
   \ar[r]^<<<<{P Q}
   &
  ([\mathrm{CartSp}_{\mathrm{synthdiff}}^{\mathrm{op}}, \mathrm{sSet}]_{\mathrm{loc}})^\circ
  \ar[r]^<<<<\simeq
  &
  \mathrm{SynthDiff}\infty \mathrm{Grpd}
  }
  \,,
$$
where the first morphism is the monoidal Dold-Kan correspondence as in 
prop. \ref{SmoothMonoidalDoldKan}, the second is degreewise the external Yoneda embedding 
$$
  \mathrm{SmoothAlg}^{\mathrm{op}} \to [\mathrm{CartSp}_{\mathrm{synthdiff}}, \mathrm{Set}]
  \,,
$$
and 
$P Q$ is any fibrant-cofibrant resolution functor in the local model structure on simplicial presheaves. 
\end{definition}

We discuss now that $L_\infty\mathrm{Algd}$ is indeed a presentation for objects in
$\mathrm{SynthDiff}\infty\mathrm{Grpd}$
satisfying the abstract axioms from \ref{InfStrucFormalInfinityGroupoid}.
\begin{lemma} 
  \label{CofibrantResolutionOfLinfinityAlgebroid}
For $\mathfrak{a} \in L_\infty \mathrm{Algd}$ and 
$i(\mathfrak{a}) \in [\mathrm{FSmoothMfd}^{\mathrm{op}}, \mathrm{sSet}]_{\mathrm{proj},\mathrm{loc}}$
its image in the presentation for $\mathrm{SynthDiff}\infty\mathrm{Grpd}$,  we have that
$$
  \left(
  \int^{[k]\in \Delta}
    \mathbf{\Delta}[k] \cdot i(\mathfrak{a})_k
  \right)
   \stackrel{\simeq}{\to}
  i(\mathfrak{a})
$$
is a cofibrant resolution, where $\mathbf{\Delta} : \Delta \to sSet$ is the 
\emph{fat simplex}.
\end{lemma}
\proof
The coend over the tensoring 
$$
  \int^{[k] \in \Delta}
  (-)\cdot (-)
  : 
  [\Delta, \mathrm{sSet}_{\mathrm{Quillen}}]_{\mathrm{proj}} 
    \times
  [\mathrm{\Delta}^{\mathrm{op}}, [\mathrm{FSmoothMfd}^{\mathrm{op}}, \mathrm{sSet}]_{\mathrm{proj},\mathrm{loc}} ]_{\mathrm{inj}}
  \to 
  [\mathrm{FSmoothMfd}^{\mathrm{op}}, \mathrm{sSet}]_{\mathrm{proj},\mathrm{loc}}
$$
for the projective and injective global model structure on functors on the simplex category and its opposite is a left Quillen bifunctor, prop. \ref{CoendOverQuillenBifunctIsQuillenBifunct}. 
We have moreover
\begin{enumerate}
\item The fat simplex is cofibrant in $[\Delta, \mathrm{sSet}_{\mathrm{Quillen}}]_{\mathrm{proj}}$,
 prop. \ref{TheSimplexAndTheFatSimplex}. 

\item 
  The object 
  $i(\mathfrak{a})_\bullet \in [\Delta^{\mathrm{op}}, [\mathrm{FSmoothMfd}^{\mathrm{op}}, \mathrm{sSet}]_{\mathrm{proj},\mathrm{loc}} ]_{\mathrm{inj}}$ is cofibrant, 
  because every representable 
$\mathrm{FSmoothMfd} \hookrightarrow [\mathrm{FSmoothMfd}^{\mathrm{op}}, \mathrm{sSet}]_{\mathrm{proj},\mathrm{loc}}$ is cofibrant. 
\end{enumerate}
\endofproof
\begin{proposition}
Let $\mathfrak{g}$ be an $L_\infty$-algebra, regarded as an $L_\infty$-algebroid 
$b \mathfrak{g} \in L_\infty \mathrm{Algd}$ over the point
by the embedding of def. \ref{LInfinityAlgebrasSubAlgebroids}.
Then $i(b \mathfrak{g}) \in \mathrm{SynthDiff}\infty\mathrm{Grpd}$  
is an infinitesimal object, def. \ref{InfinitesimalObject}, 
in that it is geometrically contractible
$$
  \Pi b \mathfrak{g} \simeq *
$$
and has as underlying discrete $\infty$-groupoid the point
$$
  \Gamma b \mathfrak{g} \simeq *
  \,.
$$
\end{proposition}
\proof
We present now $\mathrm{SynthDiff}\infty\mathrm{Grpd}$ 
by $[\mathrm{CartSp}_{\mathrm{synthdiff}}^{\mathrm{op}}, \mathrm{sSet}]_{\mathrm{proj},\mathrm{loc}}$. 
Since $\mathrm{CartSp}_{\mathrm{synthdiff}}$ is an $\infty$-cohesive site by 
prop. \ref{SynthDiffIsCohesive},
we have by the proof of prop. \ref{InfSheavesOverCohesiveSiteAreCohesive} 
that $\Pi$ is presented by the left derived functor $\mathbb{L} \lim\limits\to$ 
of the degreewise colimit and $\Gamma$ is presented by the left derived functor of evaluation on the point.

With lemma \ref{CofibrantResolutionOfLinfinityAlgebroid} we can evaluate 
$$
  \begin{aligned}
     (\mathbb{L} \lim_\to) i(b\mathfrak{g})
     & \simeq
    \lim_\to \int^{[k] \in \Delta} \mathbf{\Delta}[k] \cdot 
     (b \mathfrak{g})_{k} 
    \\
     & \simeq
    \int^{[k] \in \Delta} \mathbf{\Delta}[k] \cdot 
     \lim_\to (b \mathfrak{g})_{k}  
    \\
     & =
    \int^{[k] \in \Delta} \mathbf{\Delta}[k] \cdot *
  \end{aligned}
  \,,
$$ 
because each $(b \mathfrak{g})_n \in \mathrm{InfPoint} \hookrightarrow \mathrm{CartSp}_{\mathrm{smooth}}$ 
is an infinitesimally thickened point, hence representable and hence sent to the point by the colimit functor.

That this is equivalent to the point follows from the fact that $\emptyset \to \mathbf{\Delta}$ is an acylic cofibration in $[\Delta, \mathrm{sSet}_{\mathrm{Quillen}}]_{\mathrm{proj}}$, and that
$$
  \int^{[k] \in \Delta}
  (-)\times (-)
  : 
  [\Delta, \mathrm{sSet}_{\mathrm{Quillen}}]_{\mathrm{proj}}
  \times
  [\Delta^{\mathrm{op}}, \mathrm{sSet}_{\mathrm{Qillen}}]_{\mathrm{inj}}
  \to 
  \mathrm{sSet}_{\mathrm{Quillen}}
$$
is a Quillen bifunctor, using that $* \in [\Delta^{\mathrm{op}}, \mathrm{sSet}_{\mathrm{Quillen}}]_{\mathrm{inj}}$ is cofibrant.

Similarly, we have degreewise that 
$$
  \mathrm{Hom}(*, (b \mathfrak{g})_n) = *
$$
by the fact that an infinitesimally thickened point has a single global point. 
Therefore the claim for $\Gamma$ follows analogously.
\endofproof
\begin{proposition}
Let $\mathfrak{a} \in L_\infty \mathrm{Algd} \hookrightarrow [\mathrm{CartSp}_{\mathrm{synthdiff}}, \mathrm{sSet}]$
be an $L_\infty$-algebroid,
def. \ref{LInftyGlgebroid}, over a 
smooth manifold $X$, regarded as a simplicial presheaf 
and hence as a presentation for an object in $\mathrm{SynthDiff} \infty \mathrm{Grpd}$
according to def. \ref{EmbeddingOfThePresentation}.

We have an equivalence
$$
  \mathbf{\Pi}_{\mathrm{inf}}(\mathfrak{a}) \simeq \mathbf{\Pi}_{\mathrm{inf}}(X)
  \,.
$$
\end{proposition}
\proof
Let first $X = U \in \mathrm{CartSp}_{\mathrm{synthdiff}}$ be a representable. 
Then according to prop. \ref{CofibrantResolutionOfLinfinityAlgebroid} we have that 
$$
  \hat {\mathfrak{a}}
  :=
  \left(
    \int^{k \in \Delta} \mathbf{\Delta}[k] \cdot \mathfrak{a}_k 
  \right)
  \simeq
  \mathfrak{a}
$$ 
is cofibrant in
$[\mathrm{CartSp}_{\mathrm{synthdiff}}^{op}, \mathrm{sSet}]_{\mathrm{proj}}$. 
Therefore, by prop. \ref{InfinitesimalNeighbourhoodFromSites},
we compute the derived functor
$$
  \begin{aligned}
    \mathbf{\Pi}_{\mathrm{inf}}(\mathfrak{a}) & \simeq i_* i^* \mathfrak{a} 
    \\
     & \simeq \mathbb{L} ((-) \circ p) \mathbb{L} ((-) \circ i) \mathfrak{a} 
    \\
    & \simeq ((-) \circ i p ) \hat {\mathfrak{a} }
  \end{aligned}
$$
with the notation as used there.
In view of def. \ref{PresentationByMonoidalDoldKan} we have for all $k \in \mathbb{N}$ 
that  $\mathfrak{a}_k = X \times D$ where $D$ is an infinitesimally thickened point. 
Therefore $((-) \circ i p ) \mathfrak{a}_k = ((-) \circ i p ) X$ for all $k$ and 
hence $((-) \circ i p ) \hat {\mathfrak{a}} \simeq \mathbf{\Pi}_{\mathrm{inf}}(X)$.

For general $X$ choose first a cofibrant resolution by a split hypercover that is degreewise a coproduct of representables (which always exists, by the cofibrant replacement theorem of \cite{Dugger}), 
then pull back the above discussion to these covers.
\endofproof
\begin{corollary}
  \label{LInfinityAlgebrboidsAreFormalInfinityGroupoids}
Every $L_\infty$-algebroid in the sense of 
def. \ref{LInftyGlgebroid} under the embedding of 
def. \ref{EmbeddingOfThePresentation} is indeed 
a formal cohesive $\infty$-groupoid in the sense of def. \ref{InfinitesimalObject}.
\end{corollary}

\paragraph{Lie 1-algebroids as infinitesimal simplicial presheaves}
\label{Lie1AlgebroidsAsInfinitesimalSimplicialPresheaves}

We characterize Lie 1-algebroids $(E \to X, \rho, [-,-])$ as precisely those synthetic differential 
$\infty$-groupoids that under the presentation of def. \ref{EmbeddingOfThePresentation} 
are locally, on any chart $U \to X$ of their base space, 
given by simplicial smooth loci of the form
$$
  \xymatrix{
      \ar@{..}[r]
      &
      U \times \tilde D(k, 2)
      \ar@<-6pt>[r]
      \ar[r]
      \ar@<+6pt>[r]
     &
     U \times \tilde D(k, 1)
      \ar@<-3pt>[r]
      \ar@<+3pt>[r]
     &
     U
  }
$$
where $k = \mathrm{rank}(E)$ is the dimension of the fibers
of the Lie algebroid and where $\tilde D(k,n)$ is the smooth locus of 
\emph{infinitesimal $k$-simplices} based at the origin in 
$\mathbb{R}^n$. 
(These smooth loci have been highlighted in section 1.2 of \cite{KockBook}).

\medskip

The following definition may be either taken as an informal but instructive definition -- in which case the 
next definition \ref{FunctionsOnTwiddleD} is to be taken as the precise one --  or in fact it may be 
already itself be taken as the fully formal and precise definition if one reads it in the internal logic 
of any smooth topos with line object $R$ -- which for the present purpose is the 
\emph{Cahiers topos} \cite{DubucCahiers} $\mathrm{Sh}(\mathrm{CartSp}_{\mathrm{synthdiff}})$ 
with line object $R$, 
remark \ref{0TruncatedInSynthDiffIsCahiers}. 
\begin{definition} 
For $k,n \in \mathbb{N}$, an \emph{infinitesimal $k$-simplex} in $R^n$ based at the origin is 
a collection $(\vec \epsilon_a \in R^n)_{a = 1}^k$ of points in $R^n$, such that each is an 
infinitesimal neighbour of the origin
$$
  \forall a : \;\; \vec \epsilon_a \sim 0
$$ 
and such that all are infinitesimal neighbours of each other
$$
  \forall a,a': \;\; (\vec \epsilon_a - \vec \epsilon_{a'}) \sim 0
  \,.
$$
Write $\tilde D(k,n) \subset R^{k \cdot n}$ for the space of all such infinitesimal $k$-simplices in $R^n$.
\end{definition}
Equivalently:
\begin{definition} \label{FunctionsOnTwiddleD}
For $k,n \in \mathbb{N}$, the smooth algebra
$$
  C^\infty(\tilde D(k,n)) \in \mathrm{SmoothAlg}
$$ 
is the unique lift through the forgetful functor 
$U : \mathrm{SmoothAlg} \to \mathrm{CAlg}_{\mathbb{R}}$ of the commutative $\mathbb{R}$-algebra 
generated from $k \times n$ many generators 
$$
  (\epsilon_a^j)_{1 \leq j \leq n, 1 \leq a \leq k}
$$ 
subject to the relations
$$
  \forall a, j,j' : \;\; \epsilon_a^{j} \epsilon_a^{j'} = 0
$$
and
$$
  \forall a,a',j,j'   : 
   \;\;\; 
  (\epsilon_a^j - \epsilon_{a'}^j) (\epsilon_a^{j'} - \epsilon_{a'}^{j'}) = 0
  \,.
$$
\end{definition}
\begin{remark}
In the above form these relations are the manifest analogs of the conditions 
$\vec \epsilon_a \sim 0$ and $(\vec \epsilon_a - \vec \epsilon_{a'}) \sim 0$.
But by multiplying out the latter set of relations and using the former, we find that jointly they are equivalent to the single set of relations
$$
  \forall a,a',j,j' : \;\;\;
  \epsilon_a^j \epsilon_{a'}^{j'} 
  + \epsilon_{a'}^j \epsilon_{a}^{j'} = 0
  \,,
$$
which of course is equivalent to
$$
  \forall a,a',j,j' : \;\;\;
  \epsilon_a^j \epsilon_{a'}^{j'} 
  + \epsilon_{a}^{j'} \epsilon_{a'}^j  = 0
  \,.
$$
In this expression the roles of the two sets of indices is manifestly symmetric. Hence another equivalent way to state the relations is to say that 
$$
  \forall a,a', j: \;\;\;  \epsilon_a^{j} \epsilon_{a'}^j = 0
$$
and
$$
  \forall a,a',j,j' : 
    \;\;\;
    (\epsilon_a^j - \epsilon_a^{j'})(\epsilon_{a'}^j - \epsilon_{a'}^{j'})
  = 0
$$
\label{EquivalentPresentationsOfFunctionsOnInfinitesimalSimplices}
\end{remark}
This appears around (1.2.1) in \cite{KockBook}.

The following proposition identifies these algebras of functions
on spaces of infinitesimal based simplices with the algebras
that appear in the component expression of the monoidal Dold-Kan correspondence, 
as displayed in prop. \ref{ExplicitVersionOfMonoidalDoldKan}.
\begin{proposition} \label{TwiddleDsAsDOldKan}
For all $k,n \in \mathbb{N}$ we have a natural isomorphism of real commutative and 
hence of smooth algebras
$$
  \phi 
  : 
  \xymatrix{
    C^\infty(\tilde D(k,n))
    \ar[r]^-\simeq
	&
    \oplus_{i = 0}^n (\wedge^i \mathbb{R}^k) \otimes (\wedge^i \mathbb{R}^n)
  }
  \,,
$$
where on the right we have the algebras that appear degreewise in 
def. \ref{PresentationByMonoidalDoldKan}, where the product is given on homogeneous elements by
$$
  (\omega, x) \cdot (\lambda, y) = (\omega \wedge \lambda , x \wedge y)
  \,.
$$
\label{FunctionsOnInfinitesimalSimplicesAreAsInMonoidalDoldKan}
\end{proposition}
\proof
Let $\{t_a\}$ be the canonical basis for $\mathbb{R}^k$ and $\{e^i\}$ the canonical basis 
for $\mathbb{R}^n$. We claim that an isomorphism is given by the assignment
which on generators is
$$
  \phi : \epsilon^i_a \mapsto (t_a , e^i)
  \,.
$$
To see that this defines indeed an algebra homomorphism we need to check that it respects the relations on the generators. By remark \ref{EquivalentPresentationsOfFunctionsOnInfinitesimalSimplices}
for this it is sufficient to observe that for all pairs of pairs of indices
we have
$$
  \begin{aligned}
    \phi(\epsilon_a^i \epsilon_{a'}^{i'})
    & =
   (t_a \wedge t_{a'}, e^i \wedge e^{i'})
    \\
    & = 
    -(t_{a'} \wedge t_{a}, e^i \wedge e^{i'})
    \\
    & = -\phi(\epsilon_{a'}^i \epsilon_{a}^{i'})
  \end{aligned}
  \,.
$$
\endofproof
\begin{remark}
  The proof of prop. \ref{FunctionsOnInfinitesimalSimplicesAreAsInMonoidalDoldKan}
  together with remark \ref{EquivalentPresentationsOfFunctionsOnInfinitesimalSimplices}
  may be interpreted as showing how the skew-linearity which is the hallmark
  of traditional Lie theory arises in the synthetic differential geometry
  of infinitesimal simplices. 
  In the context of the tangent Lie algebroid,
  discussed as example \ref{TangentLieAlgebroidAsSimplicialObjectOfInfinitesimalSimplices} below, 
  this pleasant aspect of Kock's ``combinatorial differential forms''
  had been amplified in \cite{BreenMessingCombinatorialForms}. See also \cite{Stel}.
\end{remark}
\begin{proposition} 
  \label{SimplicialSmoothLocusOfLieAlgebroid}
  \index{Lie algebroid!simplicial presentation}
For $\mathfrak{a} \in L_\infty \mathrm{Alg}$ a 1-truncated object, 
hence an ordinary Lie algebroid of rank $k$ over a base manifold $X$, 
its image under the map $i : L_\infty \mathrm{Alg} \to (\mathrm{SmoothAlg}^\Delta)^{op}$, 
def. \ref{EmbeddingOfThePresentation}, is such that its restriction to any chart 
$U \to X$ is, up to isomorphism, of the form
$$
  i(\mathfrak{a})|_U : [n] \mapsto U \times \tilde D(k,n)
  \,.
$$
\end{proposition}
\proof
Apply prop. \ref{TwiddleDsAsDOldKan} in def. \ref{PresentationByMonoidalDoldKan}, using that by definition $\mathrm{CE}(\mathfrak{a})$ is given by the exterior algebra on locally free $C^\infty(X)$ modules, so that
$$
  \begin{aligned}
    \mathrm{CE}(\mathfrak{a}|_U) 
      & \simeq 
     (\wedge^\bullet_{C^\infty(U)} \Gamma(U\times \mathbb{R}^k)^*, d_{\mathfrak{a}|_U})
    \\
    & \simeq
     (C^\infty(U) \otimes \wedge^\bullet \mathbb{R}^k, d_{\mathfrak{a}|_U})
  \end{aligned}
  \,.
$$
\endofproof

\begin{example}[Lie algebra as infinitesimal simplicial complex]
  For $G$ a Lie group, consider the simplicial manifold
  $$
    \mathbf{B}G_{\mathrm{ch}}
	=
	\left(
	  \xymatrix{
	    \ar@{..}[r]
		&
	    G \times G \ar@<+4pt>[r]\ar@<-0pt>[r] \ar@<-4pt>[r] & G \ar@<+3pt>[r]\ar@<-3pt>[r] &{*}
	  }
	\right)
	\in 
	\mathrm{SmthMfd}^{\Delta^{\mathrm{op}}}
	\hookrightarrow
	[\mathrm{CartSp_{\mathrm{synthdiff}}}, \mathrm{sSet}]
  $$  
  which presents the internal delooping $\mathbf{B}G$ by prop. \ref{barWGIsPresentationOfBG}.
  Consider then the subobject (as simplicial formal manifolds)
  $$
    \xymatrix{
	  \ar@{..}[d]
	  &
	  \ar@{..}[d]	  
	  \\
	  \tilde D(k,2) \ar@{^{(}->}[r]^{i_2}
	  \ar@<+4pt>[d]
	  \ar@<+0pt>[d]
      \ar@<-4pt>[d]	  
	  &
	  G \times G
	  \ar@<+4pt>[d]
	  \ar@<+0pt>[d]
      \ar@<-4pt>[d]	  
	  \\
	  \tilde D(k,1) \ar@{^{(}->}[r]^{i_1} 
	  \ar@<+3pt>[d]
      \ar@<-3pt>[d]	  
	  & G
	  \ar@<+3pt>[d]
      \ar@<-3pt>[d]	  
	  \\
	  {*}
	  \ar[r]^{i_0}
	  &
	  {*}
	  \\
	  (\mathbf{B}\mathfrak{g})_{\mathrm{ch}} \ar@{^{(}->}[r] & (\mathbf{B}G)_{\mathrm{ch}}
	}
	\,,
  $$
  where $k = \mathrm{dim}(G)$, defined as follows:
  \begin{enumerate}
    \item $i_1$ includes the first order infinitesimal neighbourhood of the neutral element
	of $G$, hence synthetically $\{g \in G | g \sim_1 0 \}$.
	\item $i_2$ includes the space of pairs of points in $G$ which are first order
	neighbours of the neutral element and of each other:
	$\{(g_1,g_2) \in G \times G | g_1 \sim_1 e, g_2 \sim_1 e, g_1 \sim g_2\}$.
  \end{enumerate}
  This is implicitly the inclusion that is used in \cite{KockBook}
  in the discussion of Lie algebras in synthetic differential geometry.
  By the above discussion the above identifies 
  $\tilde D(k,1) \simeq \mathfrak{g} = T_e(G)$ as the Lie algebra of $G$
  and $\tilde D(k,2) \simeq \mathfrak{g}\wedge \mathfrak{g}$. 
  Then formula 6.8.2 in \cite{KockBook}together with theorem 6.6.1 there
  show how the group product on the right turns into the Lie bracket on the left.
  
  More in detail, formula 6.8.2 in \cite{KockBook} says that for $g_1, g_2 \sim_1 e$ and $g_1 \sim_1 g_2$
  we have
  $$
    g_1 \cdot g_2 = g_1 + g_2 + \frac{1}{2}\{g_1,g_2\} - \frac{3}{2}e
	\,,
  $$
  where $\{g_1,g_2\} = g_1 g_2 g_1^{-1} g_2^{-1}$ is the group commutator. 
  Theorem 6.6.1 in \cite{KockBook} identifies this 
  on the given elements infinitesimally close to $e$ with the Lie bracket on these elements.
\end{example}
\begin{example}[tangent Lie algebroid as infinitesimal simplicial complex]
  \index{Lie algebroid!synthetic tangent Lie algebroid}
  \index{$L_\infty$-algebra!synthetic $L_\infty$-algebra valued forms}
  For $X$ a smooth manifold and $T X$ its tangent Lie algebroid, its incarnation
  as a simplicial smooth locus via 
  def. \ref{EmbeddingOfThePresentation},  prop. \ref{SimplicialSmoothLocusOfLieAlgebroid}
  is the simplicial complex of \emph{infinitesimal simplices} 
  $\{(x_0, \cdots, x_n) \in X^n | \forall i,j : x_i \sim x_j \}$ in $X$. 
  The normalized cosimplicial function algebra of this complex is called the algebra
  of \emph{combinatorial differential forms} in \cite{KockBook}.
  The corresponding normalized chain dg-algebra is observed there to be 
  isomorphic to the de Rham complex of $X$, which here is a direct consequence
  of the monoidal Dold-Kan correspondence.
  This is made explicit in \cite{Stel}.
  
  Notice that accordingly for  $\mathfrak{g}$ any $L_\infty$-algebra,
  flat $\mathfrak{g}$-valued differential forms are equivalently morphisms of dg-algebras
  $$
    \Omega^\bullet(X) \leftarrow \mathrm{CE}(\mathfrak{g}) : A
  $$
  as well as (``synthetically'') morphisms
  $$
    T X \to \mathfrak{g}
  $$
  of simplicial objects in the Cahiers topos $\mathrm{Sh}(\mathrm{CartSp}_{\mathrm{synthdiff}})$.
  \label{TangentLieAlgebroidAsSimplicialObjectOfInfinitesimalSimplices}
\end{example}

\paragraph{$\infty$-Lie differentiation}
\label{LInfinityAlgebrasInSynthDiff}

We comment on how the operation of \emph{Lie differentiation} is
realized in synthetic differential cohesion, the process that 
sends a pointed connected synthetic differential homotopy type to its infinitesimal
approximation by an higher Lie algebra, an $L_\infty$-algebra.

\medskip

\begin{definition}
  Write 
  $$
    \mathrm{Inf}\infty\mathrm{Grpd} := \mathrm{PSh}_\infty(\mathrm{InfSmoothLoc})
  $$
for the $\infty$-category of $\infty$-presheaves
on the site of infinitesimal smooth loci of 
def. \ref{InfinitesimalSmoothLoci} (formal duals of Weil algebras/Artin algebras).
Write
$$
  \mathrm{Inf}\infty\mathrm{Grpd}_1 
    \hookrightarrow
  \mathrm{Inf}\infty\mathrm{Grpd}   
$$
for the reflective localization at the effective epimorphisms in $\mathrm{InfSmoothLoc}$.
\end{definition}
\begin{proposition}
  We have an $\infty$-pushout diagram of $\infty$-toposes of the form
  $$
    \raisebox{20pt}{
    \xymatrix{
	  \mathrm{Smooth}\infty\mathrm{Grpd}
	  \ar[r]^{i_*}
	  \ar[d]_\Gamma
	  &
	  \mathrm{SynthDiff}\infty\mathrm{Grpd}
	  \ar[d]
	  \\
	  \infty\mathrm{Grpd}
	  \ar[r]
	  &
	  \mathrm{Inf}\infty\mathrm{Grpd}
	}
	}
	\,.
  $$
  \label{PushoutCharacterizationOfInfinitesimalGroupoids}
\end{proposition}
\proof
  By prop. 6.3.2.3 of \cite{Lurie} $\infty$-pushouts of $\infty$-toposes
  are computed as $\infty$-limits of $\infty$-categories with respect to 
  the corresponding inverse image functors. 
  Hence we have to show that there is an $\infty$-pullback diagram
  of $\infty$-categories of the form
    $$
    \raisebox{20pt}{
    \xymatrix{
	  \mathrm{Smooth}\infty\mathrm{Grpd}
	  \ar@{<-}[r]^{i^\ast}
	  \ar@{<-}[d]_{\mathrm{Disc}}
	  &
	  \mathrm{SynthDiff}\infty\mathrm{Grpd}
	  \ar@{<-}[d]
	  \\
	  \infty\mathrm{Grpd}
	  \ar@{<-}[r]
	  &
	  \mathrm{Inf}\infty\mathrm{Grpd}
	}
	}
	\,.
  $$
  Since inverse images preserve
  $\infty$-colimits in the $\infty$-topos, we may compute this kernel on generators,
  hence on objects in the site, under the 
  $\infty$-Yoneda embedding. 
  By prop. \ref{InfSheavesOverCohesiveSiteAreCohesive} and
  prop. \ref{InfinitesimalNeighbourhoodFromSites} this reduces to the 
  diagram
  $$
    \raisebox{20pt}{
    \xymatrix{
	   \mathrm{CartSp}_{\mathrm{smooth}}
	   &
	   \mathrm{CartSp}_{\mathrm{synthdiff}}
	   \ar[l]_p
	   \\
	   {*}
	   \ar[u]^{\ast}
	   &
	   \mathrm{InfSmoothLoc}
	   \ar@{_{(}->}[u]
	   \ar[l]
	}
	}
	\,.
  $$
  This is an evident pullback of categories, exhibiting the infinitesimal smooth loci
  as the objects in the kernel of the map that forgets infinitesimal thickening.
\endofproof
\begin{proposition}
 Write 
 $$
   L\infty \mathrm{Alg}
   \hookrightarrow
   \mathrm{Inf}\infty\mathrm{Grpd}_1
 $$
 for the full sub-$\infty$-category on those objects
 which are sent by $\Gamma$ to the point. This is 
 the $\infty$-category of $L_\infty$-algebras.
\end{proposition}
\proof
  By the central result of \cite{LurieFormal}.
\endofproof
Therefore we say that
\begin{definition}
  The composite $\infty$-functor
  $$
    \mathrm{Lie}
    \;:\;
    \mathrm{Grp}(\mathrm{Smooth}\infty\mathrm{Grpd})
	\simeq
	\mathrm{Smooth}\infty\mathrm{Grpd}^{*/}_{\geq 1}
	\stackrel{i_!^{*/}}{\longrightarrow}
	\mathrm{Smooth}\infty\mathrm{Grpd}^{*/}_{\geq 1}
	\stackrel{j^*}{\longrightarrow}
	L_\infty \mathrm{Alg}
  $$
  is \emph{$\infty$-Lie differentiation}.
\end{definition}

\subsubsection{Manifolds}
\label{StrucSynthDiffSeparatedmanifolds}

We discuss the general abstract notion of \emph{separated manifolds}, \ref{DifferentialStrucmanifolds},
realized in the model of synthetic differential cohesion.

\medskip

Let $\mathbb{A}^1 := \mathbb{R}^1$ be the standard line object of $\mathrm{Smooth}\infty\mathrm{Grpd}$
exhibiting its cohesion, by prop. \ref{ReallLineExhibitsEuclideanTopologicalCohesion}.

\begin{proposition}
  The full subcategory of $\mathrm{Smooth}\infty\mathrm{Grpd}$
  on the separated $\mathbb{R}$-manifolds, def. \ref{IntrinsicSeparatedManifold}
  is equivalently that of smooth Hausdorff paracompact manifolds
  $$
    \mathrm{SmthMfd} \hookrightarrow \mathrm{Smooth}\infty\mathrm{Grpd}
	\,.
  $$
  \label{SyntheticDifferentialIntrinsicManifoldsAreOrdinaryHausdorffManifolds}
\end{proposition}

\subsubsection{Cohomology}
\label{StrucSynthCohomology}
 \index{structures in a cohesive $\infty$-topos!cohomology!synthetic differential}
 \index{structures in a cohesive $\infty$-topos!principal $\infty$-bundles!synthetic differential}

We discuss aspects of the intrinsic cohomology, \ref{StrucCohomology}, 
in $\mathrm{SynthDiff}\infty \mathrm{Grpd}$.

\begin{itemize}
  \item \ref{CohomologyLocalizationInSynthDiff} -- Cohomology localization;
  \item \ref{LieGroupCohomologyInSynthDiff} -- Lie group cohomology
  \item \ref{InfinityLieAlgebroidCohomology} -- $\infty$-Lie algebroid cohomology
  \item \ref{InfinityLieAlgebroidExtension} --
    Infinitesimal principal $\infty$-bundles / extensions of $L_\infty$-algebroids
\end{itemize}

\medskip

\paragraph{Cohomology localization}
\label{CohomologyLocalizationInSynthDiff}

\begin{observation}
The canonical line object of the Lawvere theory ${\mathrm{CartSp}}_{\mathrm{smooth}}$ 
(the free algebra on the singleton) is the real line
$$
  \mathbb{A}^1_{\mathrm{CartSp}_{\mathrm{smooth}}} = \mathbb{R}
  \,.
$$ 
\end{observation}
We shall write $\mathbb{R}$ also for the underlying 
additive group
$$
  \mathbb{G}_a = \mathbb{R}
$$
regarded canonically as an abelian $\infty$-group object in
$\mathrm{SynthDiff}\infty \mathrm{Grpd}$. For $n \in \mathbb{N}$ write
$\mathbf{B}^n \mathbb{R} \in \mathrm{SynthDiff}\infty \mathrm{Grpd}$ 
for its $n$-fold delooping.
For $n \in \mathbb{N}$ and $X \in \mathrm{SynthDiff}\infty \mathrm{Grpd}$ 
write 
$$
  H^n_{\mathrm{shdiff}}(X, \mathbb{R}) := \pi_0 \mathrm{SynthDiff}\infty \mathrm{Grpd}(X,\mathbf{B}^n \mathbb{R})
$$
for the cohomology group of $X$ with coefficients in the canonical line object in degree $n$.
\begin{definition}
Write
$$
  \mathbf{L}_{\mathrm{sdiff}} \hookrightarrow \mathrm{SynthDiff} \infty \mathrm{Grpd}
$$
for the cohomology localization of 
$\mathrm{SynthDiff}\infty \mathrm{Grpd}$ at $\mathbb{R}$-cohomology: 
the full sub-$\infty$-category on the $W$-local objects with respect to the class $W
$ of morphisms that induce isomorphisms in all $\mathbb{R}$-cohomology groups.
\end{definition}
\begin{proposition}
  Let $\mathrm{Ab}^{\Delta}_{\mathrm{proj}}$ be the model structure
  on cosimplicial abelian groups, whose fibrations are the degreewise surjections and 
  whose weak 
  equivalences the quasi-isomorphisms under the normalized cochain functor.
  
  The transferred model structure along the forgetful functor 
  $$
    U : \mathrm{SmoothAlg}^{\Delta} \to \mathrm{Ab}^{\Delta}
  $$
  exists and yields a cofibrantly generated simplicial model category structure
  on cosimplicial smooth algebras (cosimplicial $C^\infty$-rings).
\end{proposition}
See \cite{Stel} for an account.
\begin{proposition} \label{PresentationOfCohomologyLocalization}
\index{cohomology!cohomology localization}
Let 
$j : (\mathrm{SmoothAlg}^{\Delta})^{\mathrm{op}} \to 
[\mathrm{CartSp}_{\mathrm{synthdiff}}, \mathrm{sSet}]$ be the prolonged external Yoneda embedding. 
\begin{enumerate}
\item This constitutes the right adjoint of a simplicial Quillen adjunction
   $$
    (\mathcal{O} \dashv j) 
    :
     \xymatrix{  
       (\mathrm{SmoothAlg}^\Delta)^{\mathrm{op}}
      \ar@{<-}@<+3pt>[r]^<<<<{\mathcal{O}}
      \ar@<-3pt>[r]_<<<<{j}
      &
     [\mathrm{CartSp}_{\mathrm{synthdiff}}, \mathrm{sSet}]_{\mathrm{proj},\mathrm{loc}}
     }
     \,,
   $$
   where the left adjoint $\mathcal{O}(-) = C^\infty(-,\mathbb{R})$
   degreewise forms the algebra of functions obtained by homming presheaves into the line
   object $\mathbb{R}$.

\item Restricted to simplicial formal smooth manifolds of finite truncation along
   $$
      \mathrm{FSmoothMfd}^{\Delta^{\mathrm{op}}}_{\mathrm{fintr}}
      \hookrightarrow
      (\mathrm{SmoothAlg}^\Delta)^{\mathrm{op}}
   $$
   the right derived functor of $j$ is a full and faithful $\infty$-functor 
  that factors through the cohomology localization and thus identifies a full reflective 
  sub-$\infty$-category
   $$
     (\mathrm{FSmoothMfd}^{\Delta^{op}})^\circ_{\mathrm{fintr}}
     \hookrightarrow
     \mathbf{L}_{\mathrm{sdiff}}
     \hookrightarrow
     \mathrm{SynthDiff}\infty \mathrm{Grpd}
     \,.
   $$
\item The intrinsic $\mathbb{R}$-cohomology of any object 
$X \in \mathrm{SynthDiff}\infty \mathrm{Grpd}$ is computed by the ordinary 
cochain cohomology of the Moore cochain complex underlying the cosimplicial abelian group 
of the image of the left derived functor $(\mathbb{L}\mathcal{O})(X)$ 
under the Dold-Kan correspondence:
   $$
      H_{\mathrm{SynthDiff}}^n(X, \mathbb{R}) 
       \simeq 
      H^n_{\mathrm{cochain}}(N^\bullet(\mathbb{L}\mathcal{O})(X))
      \,.
   $$
\end{enumerate}
\end{proposition}
\proof
By prop. \ref{SynthSmoothEquivalentlyOverFormalMfd}
we may equivalently work over the site $\mathrm{FSmoothMfd}$.
The proof there is given in \cite{Stel}, following \cite{toen}. 
\endofproof

\medskip
\paragraph{Lie group cohomology}
\label{LieGroupCohomologyInSynthDiff}

\begin{proposition} \label{RealCohomologyOfCompactLieGroup}
 \index{cohomology!of Lie groups}
Let $G \in \mathrm{SmoothMfd} \hookrightarrow \mathrm{Smooth}\infty \mathrm{Grpd} 
  \hookrightarrow \mathrm{SynthDiff}\infty \mathrm{Grpd} $ be a Lie group. 

Then the intrinsic group cohomology in $\mathrm{Smooth}\infty \mathrm{Grpd}$ and in 
$\mathrm{SynthDiff}\infty \mathrm{Grpd}$ of $G$ with coefficients in
\begin{enumerate}
\item discrete abelian groups $A$;
\item the additive Lie group $A = \mathbb{R}$ 
\end{enumerate}
coincides with  Segal's refined Lie group cohomology \cite{Segal}, \cite{Brylinski}.
$$
  H^n_{\mathrm{Smooth}}(\mathbf{B}G, A)
  \simeq
  H^n_{\mathrm{SynthDiff}}(\mathbf{B}G, A)
  \simeq
  H^n_{\mathrm{Segal}}(G,A)
  \,.
$$
\end{proposition}
\proof
For discrete coefficients this is shown in 
theorem \ref{LieGroupCohomologyInSmoothInfinGroupoid} for
$H_{\mathrm{Smooth}}$, which by the full and faithful embedding then also holds in 
$\mathrm{SynthDiff}\infty\mathrm{Grpd}$.

Here we demonstrate the equivalence for $A = \mathbb{R}$
by obtaining a presentation for 
$H^n_{\mathrm{SynthDiff}}(\mathbf{B}G, \mathbb{R})$ that coincides explicitly with a formula 
for Segal's cohomology observed in \cite{Brylinski}. 

Let therefore 
$\mathbf{B}G_{\mathrm{ch}} 
 \in 
 [\Delta^{\mathrm{op}}, 
   [\mathrm{CartSp}_{\mathrm{synthdiff}}^{\mathrm{op}}, \mathrm{Set}$ 
 be the standard presentation of 
$\mathbf{B}G \in \mathrm{SynthDiff}\infty \mathrm{Grpd}$ by the nerve of the Lie groupoid 
$(G \stackrel{\to}{\to} *)$ as discussed in \ref{SmoothStrucCohesiveInfiniGroups}. We may write this as
$$
  \mathbf{B}G_{\mathrm{ch}} = \int^{[k] \in \Delta} \Delta[k] \cdot G^{\times_k}
  \,.
$$
By prop. \ref{PresentationOfCohomologyLocalization} the intrinsic $\mathbb{R}$-cohomology of 
$\mathbf{B}G$ is computed by the cochain cohomology of the cochain complex 
of the underlying simplicial abelian group of the value
$(\mathbb{L}\mathcal{O})\mathbf{B}G_{\mathrm{ch}}$ of the left derived functor of $\mathcal{O}$.

In order to compute this we shall build and compare various resolutions,
as in prop. \ref{ResolutionOfSimplicialManifolds}, moving back and forth
through the Quillen equivalences
$$
  \xymatrix{
    [\Delta^{\mathrm{op}},D]_{\mathrm{inj}}
      \ar@{<-}@<+3pt>[r]^{\mathrm{id}}
      \ar@<-3pt>[r]_{\mathrm{id}}
    &
    [\Delta^{\mathrm{op}},D]_{\mathrm{Reedy}}
      \ar@{<-}@<+3pt>[r]^{\mathrm{id}}
      \ar@<-3pt>[r]_{\mathrm{id}}
    &
    [\Delta^{\mathrm{op}},D]_{\mathrm{proj}}
  }
$$
between injective, projective and Reedy model structures on functors with values in a
combinatorial model category $D$,
with $D$ either $\mathrm{sSet}_{\mathrm{Quillen}}$ or with $D$ itself the injective or
projective model structure on simplicial presheaves over $\mathrm{CartSp}_{\mathrm{synthdiff}}$.

To begin with, let
$(\xymatrix{ Q \mathbf{B}G_{\mathrm{ch}})_\bullet \ar@{->}[r]^{\simeq} & (G^{\times_\bullet}})
 \in [\Delta^{\mathrm{op}},[\mathrm{CartSp}^{\mathrm{op}}, \mathrm{sSet}]_{\mathrm{proj}}]_{\mathrm{Reedy}}$
be a Reedy-cofibrant resolution of the simplicial presheaf $\mathbf{B}G_{\mathrm{ch}}$
with respect to the projective model structure. This is in particular degreewise a weak equivalence
of simplicial presheaves, hence
$$
  \int^{[k] \in \Delta}
    \Delta[k] \cdot (Q \mathbf{B}G_{\mathrm{ch}})_k
  \stackrel{\simeq}{\to} 
  \int^{[k] \in \Delta}
    \Delta[k] \cdot G^{\times_k}
  = 
  \mathbf{B}G_{c}  
$$
exists and is a weak equivalence in 
$[\mathrm{CartSp}_{\mathrm{synthdiff}}^{\mathrm{op}}, \mathrm{sSet}]_{\mathrm{inj}}$, 
hence in
$[\mathrm{CartSp}_{\mathrm{synthdiff}}^{\mathrm{op}}, \mathrm{sSet}]_{\mathrm{proj}}$, hence in $[\mathrm{CartSp}_{\mathrm{synthdiff}}^{\mathrm{op}}, \mathrm{sSet}]_{\mathrm{proj},\mathrm{loc}}$, because
\begin{enumerate}
\item $\Delta \in [\Delta, \mathrm{sSet}_{\mathrm{Quillen}}]_{\mathrm{Reedy}}$ 
  is cofibrant in the Reedy model structure;

\item every simplicial presheaf $X$ is Reedy cofibrant when regarded as an object 
  $X_\bullet \in [\Delta^{\mathrm{op}}, [\mathrm{CartSp}^{\mathrm{op}}, \mathrm{sSet}]_{\mathrm{inj}}]_{\mathrm{Reedy}}$;

\item the coend over the tensoring
   $$
     \int^\Delta \;:\; [\Delta, \mathrm{sSet}_{\mathrm{Quillen}}]_{\mathrm{Reedy}} \times
      [\Delta^{\mathrm{op}}, [\mathrm{CartSp}_{\mathrm{synthdiff}}^{\mathrm{op}}, \mathrm{sSet}]_{\mathrm{inj}}]_{\mathrm{Reedy}}
     \to
      [\mathrm{CartSp}_{\mathrm{synthdiff}}^{\mathrm{op}}, \mathrm{sSet}]_{\mathrm{inj}}
   $$
   is a left Quillen bifunctor (\cite{Lurie}, prop. A.2.9.26 ), hence 
   in particular a left Quillen functor in one argument
   when the other argument is fixed on a cofibrant object, hence preserves weak equivalences
   between cofibrant objects in that case.
\end{enumerate}
To make this a projective cofibrant resolution we further pull back along the 
Bousfield-Kan fat simplex projection 
$\mathbf{\Delta} \to \Delta$ with $\mathbf{\Delta} := N(\Delta/(-))$ 
to obtain
$$
  \int^{[k] \in \Delta}
    \mathbf{\Delta}[k] \cdot (Q \mathbf{B}G_{\mathrm{ch}})_k
  \stackrel{\simeq}{\to} 
  \int^{[k] \in \Delta}
    \Delta[k] \cdot (Q \mathbf{B}G_{\mathrm{ch}})_k
  \stackrel{\simeq}{\to}
  \mathbf{B}G_{\mathrm{ch}}  
  \,,
$$
which is a weak equivalence again due to the left Quillen bifunctor property 
of $\int^{\Delta}(-)\cdot (-)$, now applied with the second argument fixed, 
and the fact that $\mathbf{\Delta} \to \Delta$ 
is a weak equivalence between cofibrant objects in $[\Delta, \mathrm{sSet}]_{\mathrm{Reedy}}$. 
(This is the \emph{Bousfield-Kan map}).
Finally, that this is indeed cofibrant in $[\mathrm{CartSp}^{\mathrm{op}}, \mathrm{sSet}]_{\mathrm{proj}}$ 
follows from
\begin{enumerate}
\item the fact that the Reedy cofibrant $(Q \mathbf{B}G_{\mathrm{ch}})_\bullet$ is also
cofibrant in
$[\Delta^{\mathrm{op}}, [\mathrm{CartSp}^{\mathrm{op}}, \mathrm{sSet}]_{\mathrm{proj}}]_{\mathrm{inj}}$;
\item the left Quillen bifunctor property of
   $$
     \int^\Delta \;:\; [\Delta, \mathrm{sSet}_{\mathrm{Quillen}}]_{\mathrm{proj}} \times
      [\Delta^{\mathrm{op}}, [\mathrm{CartSp}_{\mathrm{synthdiff}}^{\mathrm{op}}, \mathrm{sSet}]_{\mathrm{proj}}]_{\mathrm{inj}}
     \to
      [\mathrm{CartSp}_{\mathrm{synthdiff}}^{\mathrm{op}}, \mathrm{sSet}]_{\mathrm{proj}}
	  \,;
   $$
\item the fact that the fat simplex is cofibrant in $[\Delta, \mathrm{sSet}]_{\mathrm{proj}}$.
\end{enumerate}
The central point so far is that in order to obtain a projective cofibrant resolution of 
$\mathbf{B}G_{\mathrm{ch}}$ we may form a compatible degreewise projective cofibrant resolution but then need to 
form not just the naive diagonal $\int^{\Delta} \Delta[-] \cdot (-)$ but the fattened diagonal
$\int^{\Delta} \mathbf{\Delta}[-] \cdot (-)$. In the remainder of the proof we observe that for
computing the left derived functor of $\mathcal{O}$, the fattened diagonal is not necessary after all.

For that observe that the functor 
$$
  [\Delta^{\mathrm{op}}, \mathcal{O}] : 
  [\Delta^{\mathrm{op}}, [\mathrm{CartSp}_{\mathrm{synthdiff}}^{\mathrm{op}}, \mathrm{sSet}]_{\mathrm{proj},\mathrm{loc}}]
  \to
  [\Delta^{\mathrm{op}}, (\mathrm{SmoothAlg}^{\Delta})^{\mathrm{op}}]
$$
preserves Reedy cofibrant objects, because the left Quillen functor $\mathcal{O}$ preserves 
colimits and cofibrations and hence the property that the morphisms $L_k X \to X_k$ 
out of latching objects ${\lim\limits_{\longrightarrow}}_{s \stackrel{+}{\to} k} X_s$ are cofibrations.
Therefore we may again apply the Bousfield-Kan map after application of $\mathcal{O}$ to find that 
there is a weak equivalence
$$
  (\mathbb{L}\mathcal{O})(\mathbf{B}G_{\mathrm{ch}})
  \simeq
  \int^{[k] \in \Delta}
    \mathbf{\Delta}[k] \cdot \mathcal{O}((Q\mathbf{B}G_{\mathrm{ch}})_k)
    \simeq
  \int^{[k] \in \Delta}
    \Delta[k] \cdot \mathcal{O}((Q\mathbf{B}G_{\mathrm{ch}})_k)
$$
in $(\mathrm{SmoothAlg}^{\Delta})^{\mathrm{op}}$ to the object where the fat simplex is replaced 
back with the ordinary simplex. Therefore by prop. \ref{PresentationOfCohomologyLocalization} the
$\mathbb{R}$-cohomology that we are after is equivalently computed as the cochain cohomology
of the image under the left adjoint
$$
  (N^\bullet)^{\mathrm{op}} U^{\mathrm{op}} : (\mathrm{SmoothAlg}^\Delta)^{\mathrm{op}} \to (\mathrm{Ch}^\bullet)^{\mathrm{op}}
$$
(where $U : \mathrm{SmoothAlg}^\Delta \to \mathrm{Ab}^\Delta$  is the forgetful functor) of
$$
    \int^{[k] \in \Delta}
   \Delta[k] \cdot   \mathcal{O}(Q\mathbf{B}G_{\mathrm{ch}})_k  
   \in (\mathrm{SmoothAlg}^{\Delta})^{\mathrm{op}}
   \,,
$$
which is
$$
   (N^\bullet)^{\mathrm{op}}
    \int^{[k] \in \Delta}
   \Delta[k] \cdot    U^{\mathrm{op}}\mathcal{O}((Q\mathbf{B}G_{\mathrm{ch}})_k)  
   \in (\mathrm{Ch}^\bullet)^{\mathrm{op}}
   \,,
$$
Notice that
\begin{enumerate}
\item for $S_{\bullet,\bullet}$ a bisimplicial abelian group we have that the coend 
$\int^{[k] \in \Delta} \Delta[k] \cdot S_{\bullet, k} \in (\mathrm{Ab}^{\Delta})^{\mathrm{op}}$ 
is isomorphic to the diagonal simplicial
abelian group and that forming diagonals of bisimplicial abelian groups 
sends degreewise weak equivalences to weak equivalences;
\item the Eilenberg-Zilber theorem asserts that the cochain complex of the diagonal is the 
total complex of the cochain bicomplex:
  $N^\bullet \mathrm{diag} S_{\bullet, \bullet} \simeq \mathrm{tot} C^\bullet(S_{\bullet,\bullet})$;
\item
  the complex $N^\bullet \mathcal{O} (Q\mathbf{B}G_{\mathrm{ch}})_k)$ -- being the correct derived hom-space 
  between $G^{\times_k}$ and $\mathbb{R}$ -- is related by a zig-zag of weak equivalences to 
  $\Gamma(G^{\times_k}, I_{(k)})$,
  where $I_{(k)}$ is an injective resolution of the sheaf of abelian groups $\mathbb{R}$
\end{enumerate}
Therefore finally we have
$$
  H_{\mathrm{SynthDiff}}^n(G, \mathbb{R})
  \simeq
  H^n_{\mathrm{cochain}} \mathrm{Tot} \Gamma(G^{\times_\bullet}, I^{\bullet}_\bullet)
  \,.
$$
On the right this is manifestly $H_{\mathrm{Segal}}^n(G,\mathbb{R})$, 
as observed in \cite{Brylinski}.
\endofproof
\begin{corollary}
For $G$ a compact Lie group we have for $n \geq 1$ that
$$
  H_{\mathrm{SynthDiff}\infty\mathrm{Grpd}}^{n}(G, U(1)) 
  \simeq
  H_{\mathrm{Smooth}\infty\mathrm{Grpd}}^{n}(G, U(1)) 
  \simeq 
  H_{\mathrm{Top}}^{n+1}(B G, \mathbb{Z})
  \,.
$$
\end{corollary}
\proof
  For $G$ compact we have, by \cite{Blanc}, that 
  $H^n_{\mathrm{Segal}}(G,\mathbb{R}) \simeq 0$. The claim then follows with 
  prop. \ref{RealCohomologyOfCompactLieGroup} and 
  theorem \ref{LieGroupCohomologyInSmoothInfinGroupoid}
  applied to the long exact sequence in cohomology induced by the short 
  exact sequence $\mathbb{Z} \to \mathbb{R} \to \mathbb{R}/\mathbb{Z} = U(1)$.
\endofproof

\medskip

\paragraph{$\infty$-Lie algebroid cohomology}
\label{InfinityLieAlgebroidCohomology}

We discuss the intrinsic cohomology, \ref{StrucCohomology}, 
of $\infty$-Lie algebroids, \ref{StrucSynthLie}, 
in $\mathrm{SynthDiff}\infty \mathrm{Grpd}$.
\begin{proposition} 
 \label{IntrinsicRealCohomologyByCECohomology}
\index{cohomology!of $L_\infty$-algebroids}
Let $\mathfrak{a} \in L_\infty \mathrm{Algd}$ be an $L_\infty$-algebroid. 
Then its intrinsic real cohomoloogy in $\mathrm{SynthDiff}\infty \mathrm{Grpd}$
$$
  H^n(\mathfrak{a}, \mathbb{R})
  :=
  \pi_0 \mathrm{SynthDiff}\infty \mathrm{Grpd}(\mathfrak{a}, \mathbf{B}^n \mathbb{R})
$$
coincides with its ordinary $L_\infty$-algebroid cohomology: 
the cochain cohomology of its Chevalley-Eilenberg algebra
$$
  H^n(\mathfrak{a}, \mathbb{R})
   \simeq
  H^n(\mathrm{CE}(\mathfrak{a}))
  \,.
$$
\end{proposition}
\proof
By prop. \ref{PresentationOfCohomologyLocalization} we have that 
$$
  H^n(\mathfrak{a}, \mathbb{R}) \simeq
  H^n N^\bullet(\mathbb{L}\mathcal{O})(i(\mathfrak{a}))
  \,.
$$
By lemma \ref{CofibrantResolutionOfLinfinityAlgebroid} this is
$$
  \cdots \simeq
  H^n N^\bullet
  \left(
    \int^{[k] \in \Delta} \mathbf{\Delta}[k] \cdot \mathcal{O}(i(\mathfrak{a})_k)
  \right)
  \,.
$$
Observe that $\mathcal{O}(\mathfrak{a})_\bullet$ is cofibrant in the Reedy model structure 
$[\Delta^{\mathrm{op}}, (\mathrm{SmoothAlg}^\Delta_{\mathrm{proj}})^{\mathrm{op}}]_{\mathrm{Reedy}}$ 
relative to the opposite of the projective model structure on cosimplicial algebras:  
the map from the latching object in degree $n$ in 
$\mathrm{SmoothAlg}^\Delta)^{\mathrm{op}}$ is dually in 
$\mathrm{SmoothAlg} \hookrightarrow \mathrm{SmoothAlg}^\Delta$ the projection 
$$
  \oplus_{i = 0}^n \mathrm{CE}(\mathfrak{a})_i \otimes \wedge^i \mathbb{R}^n
  \to
  \oplus_{i = 0}^{n-1} \mathrm{CE}(\mathfrak{a})_i \otimes \wedge^i \mathbb{R}^n
$$
hence is a surjection, hence a fibration in 
$\mathrm{SmoothAlg}^\Delta_{\mathrm{proj}}$ and therefore indeed a cofibration in 
$(\mathrm{SmoothAlg}^\Delta_{\mathrm{proj}})^{\mathrm{op}}$.

Therefore using the Quillen bifunctor property of the coend over the tensoring in reverse to 
lemma \ref{CofibrantResolutionOfLinfinityAlgebroid} the above is equivalent to
$$
  \cdots 
  \simeq
  H^n N^\bullet
  \left(
    \int^{[k] \in \Delta} \Delta[k] \cdot \mathcal{O}(i(\mathfrak{a})_k)
  \right)
$$
with the fat simplex replaced again by the ordinary simplex. But in brackets this is now by definition the image under the monoidal Dold-Kan correspondence of the Chevalley-Eilenberg algebra
$$
  \cdots 
  \simeq
  H^n( N^\bullet \Xi \mathrm{CE}(\mathfrak{a}) )
  \,.
$$
By the Dold-Kan correspondence we have hence
$$
  \cdots \simeq
  H^n(\mathrm{CE}(\mathfrak{a}))
  \,.
$$
\endofproof
\begin{remark}
It follows that an intrinsically defined degree-$n$ $\mathbb{R}$-cocycle on $\mathfrak{a}$ is 
indeed presented by a morphism in $L_\infty \mathrm{Algd}$
$$
  \mu : \mathfrak{a} \to b^n \mathbb{R} 
  \,,
$$
as in def. \ref{LInfinityCocycle}.
Notice that if $\mathfrak{a} = b \mathfrak{g}$ is the delooping of an $L_\infty$-
algebra $\mathfrak{g}$ this is equivalently a morphism of $L_\infty$-algebras
$$
  \mu : \mathfrak{g} \to b^{n-1} \mathbb{R} 
  \,.
$$
\end{remark}

\subsubsection{Extensions of $L_\infty$-algebroids}
\label{InfinityLieAlgebroidExtension}

We discuss the general notion of extensions of cohesive $\infty$-groups,
\ref{ExtensionsOfCohesiveInfinityGroups}, for infinitesimal
objects in $\mathrm{SynthDiff}\infty \mathrm{Grpd}$: 
extensions of $L_\infty$-algebras, def. \ref{LInfinityAlgebrasSubAlgebroids}.

\medskip

\begin{proposition}
  Let $\mu : b\mathfrak{g} \to b^{n+1} \mathbb{R}$ be an $(n+1)$-cocycle
  on an $L_\infty$-algebra $\mathfrak{g}$. 
  Then under the embedding of def. \ref{EmbeddingOfThePresentation}
  the $L_\infty$-algebra $\mathfrak{g}_\mu$ of def. \ref{gmu}
  is the extension classified by $\mu$, according to the general 
  definition \ref{ExtensionOfInfinityGroups}.
  \label{gmuIsIndeedHomotopyFiber}
\end{proposition}
\proof
  We need to show that 
  $$
     b \mathfrak{g}_\mu \to \mathfrak{g} \stackrel{\mu}{\to} b^{n+1} \mathbb{R}
  $$  
  is a fiber sequence in $\mathrm{SynthDiff}\infty\mathrm{Grpd}$.
  By prop. \ref{PullBackCharacterizationOfgmu} this sits in 
  a pullback diagram of $L_\infty$-algebras (connected $L_\infty$-algebroids)
  $$
    \xymatrix{
	  b \mathfrak{g}_\mu \ar[r] \ar[d] & e b^n \mathbb{R} \ar[d]
	  \\
	  b\mathfrak{g} \ar[r]^\mu & b^{n+1}\mathbb{R}
	}
	\,.
  $$
  By prop. \ref{LInfinityEmbeddingPreservesSomeLimits} this pullback
  is preserved by the embedding into 
  $[\mathrm{CartSp}_{\mathrm{synthdiff}}^{\mathrm{op}}, \mathrm{sSet}]_{\mathrm{proj}}$.
  Here the right vertical morphism is found to be a fibration replacement of
  the point inclusion $* \to b^{n+1}\mathbb{R}$. By the discussion in 
  \ref{InfinityPullbackAndHomotopyPullback} this identifies $b \mathfrak{g}_\mu$
  as the homotopy fiber of $\mu$.
\endofproof

\subsubsection{Infinitesimal path groupoid and de Rham spaces}
\label{StrucSynthPostnikov}
 \index{structures in a cohesive $\infty$-topos!infinitesimal path $\infty$-groupoid!synthetic differential}

We discuss the intrinsic notion of infinitesimal geometric paths 
in objects in a $\infty$-topos of infinitesimal cohesion, 
\ref{InfStrucDeRhamSpace}, realized in $\mathrm{SynthDiff}\infty \mathrm{Grpd}$.

\medskip

\begin{observation}
For $U \times D \in \mathrm{CartSp}_{\mathrm{smooth}} \ltimes \mathrm{InfinSmoothLoc} = \mathrm{CartSp}_{\mathrm{synthdiff}} \hookrightarrow 
\mathrm{SynthDiff}\infty \mathrm{Grpd}$
we have that
$$
  \mathbf{Red}(U \times D) \simeq U
$$
is the \emph{reduced smooth locus}: the formal dual of the smooth algebra obtained by quotienting 
out all nilpotent elements in the smooth algebra $C^\infty(K \times D) \simeq C^\infty(K) \otimes C^\infty(D)$.
\end{observation}
\proof
By the model category presentation of 
$\mathbf{Red} = \mathbb{L}\mathrm{Lan}_i \circ \mathbb{R}i^*$ of the proof of 
 prop. \ref{InfinitesimalnessofSynthDiff} and using that every representable is cofibrant and fibrant 
in the local projective model structure on simplicial presheaves we have
$$
  \begin{aligned}
    \mathbf{Red}(U \times D)
    & 
    \simeq
     (\mathbb{L}\mathrm{Lan}_i) (\mathbb{R}i^*) (U \times D)
    \\
     &\simeq
     (\mathbb{L}\mathrm{Lan}_i) i^* (U \times D)
     \\
     & \simeq
     (\mathbb{L}\mathrm{Lan}_i) U
     \\
     & \simeq
     \mathrm{Lan}_i U
     \\
    & \simeq U
  \end{aligned}
  \,,
$$
where we are using again that $i$ is a full and faithful functor.
\endofproof
\begin{corollary} \label{PiInfYieldsDeRahmSpace}
For $X \in \mathrm{SmoothAlg}^{\mathrm{op}}  \to \mathrm{SynthDiff} \infty \mathrm{Grpd}$ 
a smooth locus, we have that $\mathbf{\Pi}_{\mathrm{inf}}(X)$ is the corresponding 
\emph{de Rham space}, the object characterized by 
$$
  \mathrm{SynthDiff}\infty\mathrm{Grpd}(U \times D, \mathbf{\Pi}_{\mathrm{inf}}(X)) 
  \simeq \mathrm{SmoothAlg}^{\mathrm{op}}(U, X)
  \,.
$$
\end{corollary}
\proof
By the $(\mathbf{Red} \dashv \mathbf{\Pi}_{\mathrm{inf}})$-adjunction relation we have
$$
  \begin{aligned}
    \mathrm{SynthDiff} \infty \mathrm{Grpd}(U \times D, \mathbf{\Pi}_{\mathrm{inf}}(X))
    & \simeq 
    \mathrm{SynthDiff} \infty \mathrm{Grpd}( \mathbf{Red}(U \times D), X)
    \\
    & \simeq
     \mathrm{SynthDiff} \infty \mathrm{Grpd}( U, X )
  \end{aligned}
  \,.
$$
\endofproof

\subsubsection{Formally smooth/{\'e}tale/unramified morphisms}
\label{FormallySmoothInSynthDiff}
\index{structures in a cohesive $\infty$-topos!formally smooth/{\'e}tale/unramified morphisms!synthetic differential}

We discuss the general notion of formally smooth/{\'e}tale/unramified morphisms,
\ref{FormallySmooth}, realized in 
the differential $\infty$-topos 
$i : \mathrm{Smooth}\infty\mathrm{Grpd} \hookrightarrow\mathrm{SynthDiff}\infty\mathrm{Grpd}$.
given by prop. \ref{InfinitesimalnessofSynthDiff}.

\medskip

\begin{proposition}
  A morphism $f : X \to Y$ in $\mathrm{SynthDiff}\infty\mathrm{Grpd}$
  is formally {\'e}tale in the general sense of def. \ref{FormallEtaleMorphismInHth}
  precisely if for all infinitesimall thickened points 
  $D \in \mathrm{InfSmoothLoc} \hookrightarrow \mathrm{SynthDiff}\infty\mathrm{Grpd}$ the canonical diagrams
  $$
    \raisebox{20pt}{
    \xymatrix{
	  X^D \ar[rr]^-{f^D} \ar[d] && Y^D \ar[d]
	  \\
	  X \ar[rr]^-f && Y
	}
	}
  $$
  (where the vertical morphism are induced by the unique point inclusion
  $* \to D$) are $\infty$-pullbacks under $i^*$. 
\end{proposition}
\proof
  We will write $i : \mathbf{H} \hookrightarrow \mathbf{H}_{\mathrm{th}}$
  as shorthand for 
  $i : \mathrm{Smooth}\infty\mathrm{Grpd} \hookrightarrow \mathrm{SynthDiff}\infty\mathrm{Grpd}$.
  The defining $\infty$-pullback diagram of def. \ref{FormallEtaleMorphismInHth}
  induces and is detected by $\infty$-pullback diagrams for all 
  $U\times D \in \mathrm{CartSp}_{\mathrm{synthdiff}}$ of the form
  $$
    \raisebox{20pt}{
    \xymatrix{
	  \mathbf{H}_{\mathrm{th}}(U\times D, X)\ar[d] \ar[rr] && \mathbf{H}_{\mathrm{th}}(U \times D, Y)
	  \ar[d]
	  \\
	  \mathbf{H}_{\mathrm{th}}(U \times D, i_* i^* X) \ar[rr] 
	   &&
      \mathbf{H}_{\mathrm{th}}(U \times D, i_* i^* Y)	   
	}
	}
	\,.
  $$
  By the $\infty$-Yoneda lemma, the $(i^* \dashv i_*)$-adjunction, the definition of 
  $i$ and the formula
  for the internal hom, this is equivalent to the diagram
  $$
    \raisebox{20pt}{
    \xymatrix{
	  \mathbf{H}(U, i^* X^D)\ar[d] \ar[rr] && \mathbf{H}(U , i^* Y^D)
	  \ar[d]
	  \\
	  \mathbf{H}(U ,  i^* X) \ar[rr] 
	   &&
      \mathbf{H}(U ,  i^* Y)	   
	}
	}
  $$  
  being an $\infty$-pullback for all $U \in \mathrm{CartSp}$. By one more application
  of the $\infty$-Yoneda lemma this is the statement to be proven.
\endofproof
\begin{remark}
  Since $i^*$ is right adjoint and hence preserves $\infty$-pullbacks, 
  it is sufficient for a morphism $f \in \mathrm{SynthDiff}\infty\mathrm{Grpd}$
  to be formally {\'e}tale that 
  $$
    \raisebox{20pt}{
    \xymatrix{
	  X^D \ar[rr]^-{f^D} \ar[d] && Y^D \ar[d]
	  \\
	  X \ar[rr]^-f && Y
	}
	}
  $$
  is an $\infty$-pullback in $\mathrm{SynthDiff}\infty\mathrm{Grpd}$. In this
  form, when restricted to 0-truncated objects,
  formally {\'e}tale morphisms are axiomatized in \cite{KockBookA}, 
  around p. 82, in a topos for synthetic differential geometry, such as the
  Cahier topos $\tau_{\leq 0}\mathrm{SynthDiff}\infty\mathrm{Grpd}\simeq \mathrm{Sh}(\mathrm{CartSp})$
  considered here. 
\end{remark}
We now discuss in more detail the special case of formally {\'e}tale maps between objects
that are presented by simplicial smooth manifolds.
\begin{proposition}
  Let $X \in \mathrm{Smooth}\infty\mathrm{Grpd}$ be presented by a
  simplicial smooth manifold under the canonical inclusion
  $
    X_\bullet \in \mathrm{SmthMfd}^{\Delta^{\mathrm{op}}}
	\hookrightarrow [\mathrm{CartSp}_{\mathrm{smooth}}^{\mathrm{op}}, \mathrm{sSet}]
  $.
  Then $i_! X$ is presented by the same simplicial smooth manifold,
  under the canonical inclusion
  $$
    X_\bullet \in \mathrm{SmthMfd}^{\Delta^{\mathrm{op}}}
	\hookrightarrow [\mathrm{CartSp}_{\mathrm{synthdiff}}^{\mathrm{op}}, \mathrm{sSet}]
	\,.
  $$
  \label{PresentationofInjectionSmthMfdIntoSynthDiff}
\end{proposition}
\begin{proposition}
 \label{SubmersionImmersioOfManifoldsByFormallySmoothUnramified}
\index{formally {\'e}tale morphism!in $\mathrm{Smooth}\infty\mathrm{Grpd}$}
Let $f : X \to Y$ be a morphism in $\mathrm{SmthMfd}$, 
a smooth function between finite dimensional paracompact smooth manifolds,
regarded, by cor. \ref{SmoothManifoldsInSmoothInfGrpds}, as a morphism in 
$\mathrm{Smooth}\infty\mathrm{Grpd}$.
Then 
\begin{itemize}
  \item $f$ is a submersion $\;\Leftrightarrow\;$ $f$ is formally $i$-smooth;
  \item $f$ is a local diffeomorphism $\;\Leftrightarrow\;$ $f$ is formally $i$-{\'e}tale;
  \item $f$ is an immersion $\;\Leftrightarrow\;$ $f$ is formally $i$-unramified;
\end{itemize}
where on the left we have the traditional notions, and on the right those of
def. \ref{FormalRelativeSmoothnessByCanonicalMorphism}. 
\end{proposition}
\proof
  By lemma \ref{PresentationofInjectionSmthMfdIntoSynthDiff}
  the canonical diagram
  $$
    \xymatrix{
	  i_! X \ar[r]^{i_! f} \ar[d] & i_! Y \ar[d]
	  \\
	  i_* X \ar[r]^{i_* f} & i_* Y
	}
  $$
  in $\mathrm{SynthDiff}\infty \mathrm{Grpd}$ is presented in 
  $[\mathrm{CartSp}_{\mathrm{synthdiff}}^{\mathrm{op}}, \mathrm{sSet}]_{\mathrm{proj}, \mathrm{loc}}$
  by the diagram of presheaves
  $$
    U \times D
	\;\;\mapsto \;\;
    \raisebox{20pt}{
    \xymatrix{
	  \mathrm{FSmthMfd}(U \times D, X)  
	   \ar[rrr]^{\mathrm{FSmthMfd}(U \times D, f)} 
	   \ar[d] 
	   &&& 
	  \mathrm{FSmthMfd}(U \times D, Y)
	  \ar[d]
	  \\
	  \mathrm{FSmthMfd}(U, X)
	  \ar[rrr]^{\mathrm{FSmthMfd}(U,f)}
	  &&&
	  \mathrm{FSmthMfd}(U, Y)
	}
	}
	\,,
  $$
  where $\mathrm{FSmthMfd}$ is the category of formal smooth manifolds
  from def. \ref{FormalSmoothManifolds}, $U$ is an ordinary smooth manifold
  and $D$ an infinitesimal smooth loci, def. \ref{InfinitesimalSmoothLoci}.
  
  Consider this first for the case that $D := \mathbb{D} \hookrightarrow \mathbb{R}$
is the first order infinitesimal neighbourhood of the origin in the real line.
 Restricted to this case the above diagram of presheaves is that 
represented on $\mathrm{SmthMfd}$ by the diagram of smooth manifolds
$$
  \raisebox{20pt}{
  \xymatrix{
    T X \ar[r]^{d f} \ar[d] & T Y \ar[d]
	\\
	X \ar[r]^f & Y
  }
  }
  \,,
$$ 
where on the top we have the tangent bundles of $X$ and $Y$ and the differential
of $f$ mapping between them.

Since pullbacks of presheaves are computed objectwise, $f$ being
formally smooth/{\'e}tale/unramified implies that the canonical morphism
$$
  T X \to X \times_Y T Y = f^* T Y
$$
is an epi-/iso-/mono-morphism, respectively. This by definition
means that $f$ is a submersion/local diffeomorphism/immersion, respectively.

Conversely, by standard facts of differential geometry,
$f$ being a submersion means that it is locally
a projection, $f$ being a local isomorphism means that it is in particular 
{\'e}tale, and $f$ being an immersion means that it is locally an embedding.
This implies that also for $D$ any other infinitesimal smooth locus, 
so that $X^D$, $Y^D$ are bundles of possibly higher order formal curves,
the morphism
$$
  X^D \to X \times_Y Y^D
$$
is an epi-/iso-/mono-morphism, respectively.
\endofproof

\subsubsection{Formally {\'e}tale groupoids}
\label{StrucSynthDiffFormallyEtaleGroupoid}

We discuss the general notion of formally {\'e}tale groupoids
in a differential $\infty$-topos, \ref{InfStrucEtaleGroupoid},
realized in 
$\mathrm{Smooth}\infty\mathrm{Grpd} \stackrel{i}{\hookrightarrow} \mathrm{SynthDiff}\infty\mathrm{Grpd}$.

\medskip

\begin{definition}
  \label{EtaleSimplicialManifold}
  Call a simplicial smooth manifold $X \in \mathrm{SmoothMfd}^{\Delta^{\mathrm{op}}}$
  an \emph{{\'e}tale simplicial smooth manifold} if it is fibrant as an 
  object of $[\mathrm{CartSp}^{\mathrm{op}}, \mathrm{sSet}]_{\mathrm{proj}}$
  and if moreover all face and degeneracy morphisms are {\'e}tale morphisms.
\end{definition}
\begin{example}
  The nerve of an {\'e}tale Lie groupoid in the traditional sense is
  an {\'e}tale simplicial smooth manifold.
\end{example}

\begin{proposition}
  Let $X \in \mathrm{SmthMfd}^{\Delta^{\mathrm{op}}}$ be an {\'e}tale simplicial
  manifold, def. \ref{EtaleSimplicialManifold}. Then equipped with its canonical
  atlas, observation \ref{CanonicalAtlasOfSimplicialPresheaf},
  it presents a formally {\'e}tale groupoid object in 
  $\mathrm{Smooth}\infty \mathrm{Grpd} \stackrel{i}{\hookrightarrow}
  \mathrm{SynthDiff}\infty\mathrm{Grpd}$, according to def. 
  \ref{FormallyEtaleGroupoid}.
\end{proposition}
\proof
  We need to check that $i_! X_0$ is the $\infty$-pullback
  $i_* X_0 \times_{i_* X} i_! X$. By 
  prop. \ref{FiniteHomotopyLimitsInPresheaves}, 
  lemma \ref{PresentationofInjectionSmthMfdIntoSynthDiff}
  and prop. \ref{DecalageIsFibrationResolution} it is sufficient to 
  show for the d{\'e}calage replacement 
  $\mathrm{Dec}_0 X \to X$ of the atlas, that
   $i_! \mathrm{Dec}_0 X$ is the ordinary pullback of simplicial presheaves
  $(i_* \mathrm{Dec}_0 X) \times_{i_* X} i_! X$. 
  Since pullbacks of simplicial presheaves are computed degreewise,
  this is the case by prop. \ref{SubmersionImmersioOfManifoldsByFormallySmoothUnramified}
  if for all $n \in \mathbb{N}$ the morphism
  $(\mathrm{Dec}_0 X)_n \to X_n$ is an {\'e}tale morphism of smooth manifolds,
  in the traditional sense. By prop.  \ref{MorphismsOutOfPlainDecalage}
  this morphism is the face map $d_{n+1}$ of $X$. This is
  indeed {\'e}tale by the very assumption that $X$ is an 
  {\'e}tale simplicial smooth manifold.
\endofproof

\subsubsection{Chern-Weil theory}
\label{StrucSynthChernWeil}
\index{structures in a cohesive $\infty$-topos!Chern-Weil homomorphism!synthetic-differential}

We discuss the notion of $\infty$-connections, \ref{SmoothStrucInfChernWeil},
in the context $\mathrm{SynthDiff}\infty\mathrm{Grpd}$.

\paragraph{$\infty$-Cartan connections}

A \emph{Cartan connection} on a smooth manifold is a principal connection
subject to an extra constraint that identifies a component of
the connection at each point with the tangent space of the base
manifold at that point. The archetypical application of this 
notion is to the formulation of the field theory of \emph{gravity},
\ref{FirstOrderFormulationOfGravity}.

We indicate a notion of Cartan $\infty$-connections.

\medskip

The following notion is classical, see for instance
section 5.1 of \cite{Sharpe}.
\begin{definition}
  \label{CartanConnection}
  \index{Cartan connection}
  \index{connection!Cartan connection}
  Let $(H \hookrightarrow G)$ be an inclusion of Lie groups
  with Lie algebras $(\mathfrak{h} \hookrightarrow \mathfrak{g})$.
  A \emph{$(H \to G)$-Cartan connection} on a smooth manifold $X$ is
  \begin{enumerate}
    \item 
	  a $G$-principal bundle $P \to X$ equipped with a 
	  connection $\nabla$;
	\item
      such that
      \begin{enumerate}
	     \item
		   the structure group of $P$ reduces to $H$, hence the classifying
		   morphism factors as $X \to \mathbf{B}H \to \mathbf{B}G$;
		 \item
		   for each point $x \in X$ and any local trivialization of
           $(P, \nabla)$ in some neighbourhood of $X$, the canonical 
		   linear map
		   $$
		     \xymatrix{
		       T_x X 
			   \ar[r]^{\nabla}
			   &
			   \mathfrak{g}
			   \ar[r]
			   &
			   \mathfrak{g}/\mathfrak{h}
			 }
		   $$
		   is an isomorphism,
      \end{enumerate}	  
  \end{enumerate}
\end{definition}
Here $(\mathfrak{h} \to \mathfrak{g})$ are the Lie algebras of the 
given Lie groups and $\mathfrak{g}/\mathfrak{h}$ is the 
quotient of the underlying vector spaces.

\newpage

\subsection{Supergeometric $\infty$-groupoids }
 \label{SuperInfinityGroupoids}
 \index{cohesive $\infty$-topos!models!super cohesion}

We discuss $\infty$-groupoids equipped with \emph{discrete super cohesion}, with 
\emph{smooth super cohesion} and \emph{synthetic differential super cohesion},
where ``super'' is in the sense of \emph{superalgebra} and 
\emph{supergeometry} 
(see for instance \cite{DeligneMorganSupergeometry} 
for a review of traditional superalgebra and supergeometry).

\subsubsection{Survey}

We first introduce \emph{discrete super $\infty$-groupoids} which have super-grading but no smooth structure.
This is the canonical context in which (higher) \emph{superalgebra} takes place: an
$\mathbb{R}$-module internal to super $\infty$-groupoids is externally a chain complex of
\emph{super vector spaces} and an $\mathbb{R}$-algebra internal to super $\infty$-groupoids
is externally a real \emph{superalgebra}. Then we add smooth structure by passing further to
\emph{smooth super $\infty$-groupoids}. This is the canonical context for supergeometry. Notably
the traditional category of smooth supermanifolds faithfully embeds into smooth super $\infty$-groupoids.
Finally we further refine to \emph{synthetic differential super $\infty$-groupoids}
where the smooth structure is refined by explicit commutative infinitesimals in addition to
the super/graded infinitesimals of supergeometry. In summary, this 
yields a super-refinement of three cohesive structures discussed before:

$$
  \begin{tabular}{r|rc}
    \begin{tabular}{c}supergeometric \\ refinement\end{tabular} & \begin{tabular}{c} differential \\ geometry\end{tabular} & \begin{tabular}{c}discussed \\ in section \end{tabular}
	\\
	\hline\hline
    $\mathrm{Super}\infty \mathrm{Grpd}$ & $\mathrm{Disc}\infty \mathrm{Grpd}$ & \ref{DiscreteInfGroupoids}
	\\
    \hline
	$\mathrm{SmoothSuper}\infty\mathrm{Grpd}$ & $\mathrm{Smooth}\infty \mathrm{Grpd}$ & \ref{SmoothInfgrpds}
	\\
	\hline
	$\mathrm{SynthDiffSuper}\infty \mathrm{Grpd}$ & $\mathrm{SynthDiff}\infty\mathrm{Grpd}$ & \ref{SynthDiffInfGrpd}
  \end{tabular}
$$

Accordingly, the canonical site of definition of the most 
inclusive of these cohesive $\infty$-toposes, which is $\mathrm{SynthDiffSuper}\infty\mathrm{Grpd}$, 
contains objects denoted $\mathbb{R}^{p\oplus s|q}$ --
\emph{synthetic differential super Cartesian space} -- that have three gradings:
\begin{itemize}
  \item an ordinary dimension $p$;
  \item an order $s$ of their infinitesimal thickening;
  \item an odd super dimension $q$.
\end{itemize}
In terms of the formally dual function algebras $C^\infty(\mathbb{R}^{p\oplus k|l})$ on these objects, 
$k$ is the number of \emph{commuting} nilpotent generators, while $q$ is the number of 
\emph{graded-commuting} nilpotent generators. In this sense supergeometry may be understood as
a $\mathbb{Z}_2$-graded variant of synthetic differential geometry. This is a perspective that had been 
explored in \cite{YetterSuper} and more recently in \cite{CarchediRoytenberg}.

Of course $\infty$-groupoids $X$ over this synthetic supergeometric site have furthermore their homotopy theoretic degree, their simplicial grading when modeled by simplicial presheaves
$$
  X : (\mathbb{R}^{p\oplus s|q}, \Delta^k) \mapsto X_k(\mathbb{R}^{p\oplus s|q}) \in \mathrm{Set}
  \,.
$$

While for some applications it is useful to regard all these ``kinds of dimension''
as being on equal footing, for other applications it is useful to order them more
hierarchically. 
Specifically the role played by supergeometry in applications is well reflected by
the perspective where smooth/synthetic differential supergeometry is regarded as ordinary 
smooth/synthetic differential geometry but \emph{internal} to the ``bare super context'', 
which is the context parameterized over just the \emph{superpoints} $\mathbb{R}^{0|q}$.
This perspective on supergeometry had been proposed independently in 1984 in 
\cite{Schwarz}, \cite{Molotkov} and \cite{Voronov}. A review is in the appendix of 
\cite{KonechnySchwarz}, whose main part discusses aspects of those synthetic differential
superspaces in this language.

In terms of (higher) topos theory this perspective means that passing from higher differential
geometry to higher supergeometry means to change the \emph{base $\infty$-topos} from 
that of ordinary geometrically discrete $\infty$-groupoids
$\mathrm{Disc}\infty \mathrm{Grpd} \simeq \infty\mathrm{Grpd} \simeq L_{\mathrm{whe}}\mathrm{Top}$ 
to that of ``super $\infty$-groupoids'' 
$\mathrm{Super}\infty\mathrm{Grp} := \mathrm{Sh}_\infty(\{\mathbb{R}^{0|q}\}_q)$
which still have no finite continuous/smooth geometric structure but which do  
have super-grading. 

We find below the
$\infty$-toposes for differential-, synthetic differential- and supergeometry
to arrange in a diagram of geometric morphisms of the form
$$
  \raisebox{20pt}{
  \xymatrix{
    \mathrm{SmoothSuper}\infty\mathrm{Grpd}
	\ar@{^{(}->}[r]
	\ar[d]
	&
	\mathrm{SynthDiffSuper}\infty\mathrm{Grpd}
	\ar[r] \ar[d]
	&
	\mathrm{Super}\infty \mathrm{Grpd}
	\ar[d]
	\\
    \mathrm{Smooth}\infty\mathrm{Grpd}
	\ar@{^{(}->}[r]
	&
	\mathrm{SynthDiff}\infty\mathrm{Grpd}
	\ar[r]
	&
	\infty \mathrm{Grpd}
  }
  }
  \,.
$$
Here the bottom line is the differential cohesion over the base of discrete $\infty$-groupoids
discussed in \ref{SynthDiffInfGrpd}. The top line is the super-refinement 
exhibited by differential cohesion, but now over the base $\mathrm{Super}\infty\mathrm{Grpd}$
of discrete but ``super'' $\infty$-groupoids. This diagram of $\infty$-toposes we present
by a diagram of sites which, with the above notation for synthetic differential super Cartesian spaces, 
looks as follows.
$$
  \raisebox{20pt}{
  \xymatrix{
    \{\mathbb{R}^{p|q}\}_{p,q}
	\ar@{^{(}->}[r]
	\ar[d]
	&
	\{ \mathbb{R}^{p\oplus s| q} \}_{p,s,q}
	\ar[r] \ar[d]
	&
	\{ \mathbb{R}^{0|q} \}_q
	\ar[d]
	\\
    \{\mathbb{R}^p\}_p
	\ar@{^{(}->}[r]
	&
	\{\mathbb{R}^{p\oplus s}\}_{p,s}
	\ar[r]
	&
	\{{\ast}\}
  }
  }
  \,.
$$

\subsubsection{The $\infty$-topos of supergeometric $\infty$-stacks}

\begin{definition}
  \label{SuperPoint}
  Let $\mathrm{GrassmannAlg}_{\mathbb{R}}$ be the category whose objects are finite dimensional
  free $\mathbb{Z}_2$-graded commutative $\mathbb{R}$-algebras (Grassmann algebras).
  Write 
  $$
    \mathrm{SuperPoint} := \mathrm{GrassmannAlg}_{\mathbb{R}}^{\mathrm{op}}
  $$
  for its opposite category. For $q \in \mathbb{N}$ we write $\mathbb{R}^{0|q} \in \mathrm{SuperPoint}$
  for the object corresponding to the free $\mathbb{Z}_2$-graded commutative algebra on 
  $q$ generators and speak of the \emph{superpoint} of order $q$.
  
  We think of $\mathrm{SuperPoint}$ as a site by equipping it with the trivial coverage.
\end{definition}
\begin{definition}
  \label{SuperSet}
  Write
  $$
    \mathrm{SuperSet} 
	  := 
	\mathrm{Sh}(\mathrm{SuperPoint})
	\simeq \mathrm{PSh}(\mathrm{SuperPoint})
  $$
\end{definition}
for the topos of presheaves over $\mathrm{SuperPoint}$.
\begin{definition}  
  Write
  $$
    \mathrm{Super}\infty \mathrm{Grpd}
    :=
    \mathrm{Sh}_{\infty}(\mathrm{SuperPoint})
    \simeq 
    \mathrm{PSh}_{\infty}(\mathrm{SuperPoint})
  $$
  for the $\infty$-topos of $\infty$-sheaves over $\mathrm{SuperPoint}$. We say an 
  object $X \in \mathrm{Super}\infty \mathrm{Grpd}$ is a 
  \emph{super $\infty$-groupoid}.
\end{definition}
\begin{proposition}
  The $\infty$-topos $\mathrm{Super}\infty\mathrm{Grpd}$ is infinitesimal 
  cohesive, def. \ref{InfinitesimalCohesion} over $\infty\mathrm{Grpd}$.
  \label{SuperpointsIsInfinitesimalCohesive}
\end{proposition}
\proof
  The ordinary point in $\mathrm{SuperPoints}$ is both the terminal
  object but also the initial object, since superpoints are
  infinitesimally thickened points in that they only have one global actual point.
  Therefore the statement follows with prop. \ref{InfinitesimalCohesiveSite}.
\endofproof
We regard higher superalgebra and higher supergeometry as being the higher
algebra and geometry \emph{over the base $\infty$-topos} (\cite{Johnstone}, chapter B3) 
$\mathrm{Super}\infty \mathrm{Grpd}$ instead of over the canonical base $\infty$-topos
$\infty \mathrm{Grpd}$. Except for the topos-theoretic rephrasing, this perspective 
has originally been suggested in \cite{Schwarz} and \cite{Molotkov}.
\begin{proposition}
  The $\infty$-topos $\mathrm{Super}\infty \mathrm{Grpd}$
  is cohesive, def. \ref{CohesiveInfinToposDefinition}.
  $$
    \xymatrix{
      \mathrm{Super}\infty \mathrm{Grpd}
         \ar@<+12pt>@{->}[r]^-{\Pi}
         \ar@<+4pt>@{<-^{)}}[r]|-{\mathrm{Disc}}
         \ar@<-4pt>@{->}[r]|-{\mathrm{\Gamma}}
         \ar@<-12pt>@{<-^{)}}[r]_-{\mathrm{coDisc}}
        &
      \infty \mathrm{Grpd}
    }   
    \,.
  $$
\end{proposition}
\proof
  The site $\mathrm{SuperPoint}$ is $\infty$-cohesive, according to 
  def. \ref{CohesiveSite}. Hence the claim follows by prop. \ref{InfSheavesOverCohesiveSiteAreCohesive}.
\endofproof
\begin{proposition}
  \label{SuperGroupoidAsInfinitesimalCohesiveNeighbourhood}
  The inclusion $\mathrm{Disc} : \infty \mathrm{Grpd} \hookrightarrow \mathrm{Super}\infty\mathrm{Grpd}$
  exhibits the collection of super $\infty$-groupoids as forming an infinitesimal cohesive neighbourhood,
  def. \ref{InfinitesimalCohesiveNeighbourhood},
  of the discrete $\infty$-groupoids, \ref{DiscreteInfGroupoids}.
\end{proposition}
\proof
  Observe that the point inclusion $i : \mathrm{Point} := * \hookrightarrow \mathrm{SuperPoint}$
  is both left and right adjoint to the unique projection $p : \mathrm{SuperPoint} \to \mathrm{Point}$.
  Therefore we have even a periodic sequence of adjunctions
  $$
    (\cdots \dashv i^* \dashv p^* \dashv i^* \dashv p^* \dashv \cdots ) :
    \mathrm{Super}\infty \mathrm{Grpd}
    \to 
    \infty \mathrm{Grpd}
    \,,
  $$
  and $p^* \simeq \mathrm{Disc} \simeq \mathrm{coDisc}$ is full and faithful.
\endofproof
\begin{definition}
  \label{RInSuperSet}
  Write $\mathbb{R} \in \mathrm{Super}\infty \mathrm{Grpd}$ for 
  the presheaf $\mathrm{SuperPoint}^{\mathrm{op}} \to \mathrm{Set} \hookrightarrow \infty \mathrm{Grpd}$
  given by
  $$
    \mathbb{R} : \mathbb{R}^{0|q} \mapsto C^\infty(\mathbb{R}^{0|q}) := (\Lambda_q)_{\mathrm{even}}
    \,,
  $$
  which sends the order-$q$ superpoint to the underlying set of the 
  even subalgebra of the Grassmann algebra on $q$
  generators.
\end{definition}
\begin{remark}
  The object $\mathbb{R} \in \mathrm{Super}\infty \mathrm{Grpd}$ is canonically equipped
  with the structure of an internal ring object. Morever, under both $\mathrm{\Pi}$ and
  $\Gamma$ it maps to the ordinary real line $\mathbb{R} \in \mathrm{Set} \hookrightarrow
  \infty \mathrm{Grpd}$ while respecting the ring structures on both sides.
\end{remark}
The following observation is due to \cite{Molotkov}. 
\begin{proposition}
  \label{SuperalgebrasAsInternalAlgebra}
  The theory of ordinary (linear) 
  $\mathbb{R}$-algebra internal to 
  the 1-topos 
  $$
    \mathrm{SuperSet} = \mathrm{Super}0\mathrm{Grpd}\hookrightarrow 
   \mathrm{Super}\infty \mathrm{Grpd}
   $$ 
   is equivalent to the theory of $\mathbb{R}$-superalgebra
   in $\mathrm{Set}$.
\end{proposition}
\begin{definition}
  Write $\mathrm{sCartSp}$ for the full subcategory of that of supermanifolds
  on those that are super Cartesian spaces: $\{\mathbb{R}^{p|q}\}_{p,q \in \mathbb{N}}$.
  Regard this as a site by equipping it with the coverage whose covering families
  are of the form $\{ U_i \times \mathbb{R}^{0|q}  \stackrel{(p_1,\mathrm{id})}{\longrightarrow}  \mathbb{R}^{p|q}\}$
  for $\{U_i \to \mathbb{R}^{p}\}$ a differentiably good open cover, 
  def. \ref{DifferentiablyGoodOpenCover}.
  \label{SuperCartesianSpaces}
\end{definition}
\begin{remark}
  A morphism $\mathbb{R}^{p_1|q_1} \to \mathbb{R}^{p_2|q_2}$ in $\mathrm{CartSp}_{\mathrm{super}}$
  is equivalently a tuple consisting of $p_2$ even elements and $q_2$ odd elements of 
  the superalgebra $C^\infty(\mathbb{R}^{p_1|q_1})$.  
  In particular, under the restricted Yoneda embedding the line of 
  def. \ref{RInSuperSet} is $\mathbb{R} \simeq \mathbb{R}^{1|0}$. 
\end{remark}
\begin{definition}
  Write
  $$
    \mathrm{SmoothSuper}\infty \mathrm{Grpd}
    :=
    \mathrm{Sh}_\infty(\mathrm{sCartSp}_{\mathrm{smooth}})
    \,.
  $$
  An object in this $\infty$-topos we call a \emph{smooth super $\infty$-groupoid}.
\end{definition}
\begin{proposition}
  We have a commuting diagram of cohesive $\infty$-toposes
  which exhibits $\mathrm{Super}\infty\mathrm{Grpd}$ 
  as an $\infty$-pushout
  $$
    \xymatrix{
       \mathrm{SmoothSuper}\infty \mathrm{Grpd}
       \ar@<+12pt>@{->}[rr]|<\times^-{\Pi_{\mathrm{super}}}
       \ar@<+4pt>@{<-^{)}}[rr]|-{\mathrm{Disc}_{\mathrm{super}}}
       \ar@<-4pt>@{->}[rr]|-{\Gamma_{\mathrm{super}}}
       \ar@<-12pt>@{<-^{)}}[rr]_-{\mathrm{coDisc}_{\mathrm{super}}}
       \ar@<-14pt>@{<-_{)}}[dd]_{i_!}
       \ar@<-6pt>@{->}[dd]|{i^\ast}
       \ar@<+2pt>@{<-_{)}}[dd]|{i_*}
       \ar@<+10pt>@{->}[dd]^{i^!}
       &&
       \mathrm{Super}\infty \mathrm{Grpd}
       \ar@<-14pt>@{<-_{)}}[dd]_{i_!}
       \ar@<-6pt>@{->}[dd]|{i^\ast}
       \ar@<+2pt>@{<-_{)}}[dd]|{i_*}
       \ar@<+10pt>@{->}[dd]^{i^!}
       \\
       \\
       \mathrm{Smooth}\infty \mathrm{Grpd}
       \ar@<+9pt>@{->}[rr]|<\times^-{\Pi}
       \ar@<+1pt>@{<-^{)}}[rr]|-{\mathrm{Disc}}
       \ar@<-7pt>@{->}[rr]|-{\Gamma}
       \ar@<-15pt>@{<-^{)}}[rr]_-{\mathrm{coDisc}}
       &&
       \infty \mathrm{Grpd}
    }
    \,.
  $$
  \label{PushoutForSuperCohesion}
\end{proposition}
\proof
  By def. \ref{SuperCartesianSpaces} the arguments of 
  \ref{SmoothInfgrpds} apply verbatim at the stage of each fixed superpoint,
  and this gives the cohesion over $\mathrm{Super}\infty \mathrm{Grpd}$,
  hence the top vertical adjoint quadruple in the above.
  The right vertical morphisms exhibit infinitesimal cohesion by 
  prop. \ref{SuperpointsIsInfinitesimalCohesive}. 
  That the resulting diagram is an $\infty$-pushout 
  follows now with the same argument as in the proof of 
  prop. \ref{PushoutCharacterizationOfInfinitesimalGroupoids}. 
\endofproof
For emphasis we shall refer to the objects of $\mathrm{Super}\infty \mathrm{Grp}$
as \emph{discrete super $\infty$-groupoids}: these refine discrete $\infty$-groupoids, 
\ref{DiscreteInfGroupoids} with super-cohesion and are themselves further refined by
smooth super $\infty$-groupoids with smooth cohesion.

\medskip

We now discuss the various general abstract structures in a cohesive $\infty$-topos,
\ref{structures}, realized in $\mathrm{Super}\infty\mathrm{Grpd}$
and $\mathrm{SmoothSuper}\infty\mathrm{Grpd}$.
\begin{itemize}
  \item \ref{SupergeomStrucAssociatedBundles} -- Associated bundles
  \item \ref{SuperStrucLieAlg} -- Exponentiated $\infty$-Lie algebras
\end{itemize}

\subsubsection{Associated bundles}
\label{SupergeomStrucAssociatedBundles}
\index{structures in a cohesive $\infty$-topos!associated bundles!super 2-line bundle}

We discuss aspects of the general notion of associated fiber $\infty$-bundles,
\ref{AssociatedBundles}, realized in the context of supergeometric cohesion. 

\medskip

In \ref{SmoothStrucAssociatedBundles} above we discussed the 2-stack $2 \mathbf{Line}_{\mathbb{C}}$
of smooth complex line 2-bundles. 
Since the $B$-field that the bosonic string is charged under has moduli in the differential
refinement $\mathbf{B}^2 \mathbb{C}^\times_{\mathrm{conn}}$, we may hence say that it is given by 
2-connections on complex \emph{2-line bundles}. However, a careful analysis
(due \cite{DistlerFreedMoore} and made more explicit in \cite{FreedLectures}) shows that
for the superstring the background $B$-field is more refined. Expressed in the language of 
higher stacks the statement is that it is a 
connection on a complex \emph{super}-2-line bundle. Precisely, in the language of stacks
for supergeometry we are to pass to the higher topos 
$\mathrm{SmoothSuper\infty\mathrm{Grpd}} \simeq \mathrm{Sh}_\infty(\mathrm{SuperMfd})$
on the site of smooth supermanifolds, \ref{SuperInfinityGroupoids}. Internal to that
the term \emph{algebra} now means \emph{superalgebra} and hence the 2-stack
$$
  2\mathbf{sLine}_{\mathbb{C}} \in \mathrm{SmoothSuper}\infty\mathrm{Grpd}
$$
now has global points that are identified with complex Azumaya \emph{super}algebras. Of these
it turns out there is, up to equivalence, not just one, but two: the canonical super 2-line and its 
``superpartner''. Moreover, there are now, up to equivalence, two different invertible 2-linear
maps from each of these super-lines to itself. In summary, the homotopy sheaves of 
the super 2-stack of super line 2-bundles are
\begin{itemize}
  \item $ \pi_0(2\mathbf{sLine}_{\mathbb{C}}) \simeq \mathbb{Z}_2$,
  \item $\pi_1(2\mathbf{sLine}_{\mathbb{C}}) \simeq \mathbb{Z}_2$,
  \item $\pi_2(2\mathbf{sLine}_{\mathbb{C}}) \simeq \mathbb{C}^\times \in \mathrm{Sh}(\mathrm{SuperMfd})$.
\end{itemize}
(where in the last line we emphasize that the \emph{homotopy sheaf} is that represented by
$\mathbb{C}^\times$ as a smooth (super-)manifold). With the discussion in \ref{StrucGeometricHomotopy}
it follows that the geometric realization of this 2-stack has homotopy groups
\begin{itemize}
  \item $ \pi_0(\vert2\mathbf{sLine}_{\mathbb{C}}\vert) \simeq \mathbb{Z}_2$,
  \item $\pi_1(\vert2\mathbf{sLine}_{\mathbb{C}}\vert) \simeq \mathbb{Z}_2$,
  \item $\pi_2(\vert2\mathbf{sLine}_{\mathbb{C}}\vert) \simeq 0$,
  \item $\pi_3(\vert2\mathbf{sLine}_{\mathbb{C}}\vert) \simeq \mathbb{Z}$.
\end{itemize}
These are precisely the correct coefficients for the twists of complex K-theory, witnessing the
fact that the B-field background of the superstring twists the Chan-Paton bundles on the D-branes.

The braided monoidal structure of $2\mathrm{sVect}_{\mathbb{C}}$ induces on 
$2\mathbf{sLine}_{\mathbb{C}}$ the structure of a \emph{braided 3-group}. Therefore the 
above general abstract definition of universal moduli for differential cocycles/higher connections
produces a moduli 3-stack $\mathbf{B}(2\mathbf{sLine}_{\mathbb{C}})_{\mathrm{conn}}$ which is the 
supergeometric refinement of the coefficient object $\mathbf{B}^3 \mathbb{C}^{\times}_{\mathrm{conn}}$
for the extended Lagrangian of bosonic 3-dimensional Chern-Simons theory.  
Therefore for $G$ a super-Lie group a super-Chern-Simons theory that induces the
super-WZW action functional on $G$ is given by an extended Lagragian which is a map
of higher moduli stacks of the form
$$
  \mathbf{L}
    : 
  \xymatrix{
    \mathbf{B}G_{\mathrm{conn}}
    \ar[r]
    &
    \mathbf{B}(2\mathbf{sLine}_{\mathbb{C}})_{\mathrm{conn}} 
  }
  \,.
$$
By the canonical inclusion 
$\mathbf{B}^3 \mathbb{C}^\times_{\mathrm{conn}}
\to \mathbf{B}(2\mathbf{sLine}_{\mathbb{C}})_{\mathrm{conn}}$ every bosonic 
extended Lagrangian of 3-d Chern-Simons type induces such a supergeometric theory
with trivial super-grading part.

\subsubsection{Exponentiated $\infty$-Lie algebras}
 \label{SuperStrucLieAlg}
 \index{structures in a cohesive $\infty$-topos!exponentiated $\infty$-Lie algebras!synthetic differential}

According to prop. \ref{SuperalgebrasAsInternalAlgebra} the following definition
is justified.
\begin{definition}
  \label{SuperLInfinityAlgebra}
  \index{$L_\infty$-algebra!super}
  A \emph{super $L_\infty$-algebra}
  is an $L_\infty$-algebra, def. \ref{LInfinityAlgebra}, 
  internal to 
  the topos $\mathrm{SuperSet}$, def. \ref{SuperSet}, 
  over the ring object $\mathbb{R}$
  from def. \ref{RInSuperSet}.
\end{definition}
\begin{observation}
  The Chevalley-Eilenberg algebra $\mathrm{CE}(\mathfrak{g})$, 
  def. \ref{CEAlgebraInIntro}, of 
  a super $L_\infty$-algebra $\mathfrak{g}$ is externally 
  \begin{itemize}
    \item a graded-commutative algebra over $\mathbb{R}$ on generators
    of bigree in $(\mathbb{N}_+, \mathbb{Z}_2)$ -- the \emph{homotopical degree}
  $\mathrm{deg}_h$
  and the \emph{super degree} $\mathrm{deg}_s$;
  \item such that for 
  any two generators $a,b$ the product satisfies
  $$
    a b = (-1)^{\mathrm{def}_h(a) \mathrm{deg}_h(b) + \mathrm{def}_s(a) \mathrm{deg}_s(b)} \;   b a
    \,;
  $$
  \item
    and equipped with a differential $d_{\mathrm{CE}}$ of bidegree $(1,\mathrm{even})$
    such that $d_{\mathrm{CE}}^2 = 0$.
  \end{itemize}
\end{observation}
\begin{examples}
  \begin{itemize}
    \item Every ordinary $L_\infty$-algebra is canonically a
	 super $L_\infty$-algebra where all element are of even superdegree.
	\item
	  Ordinary super Lie algebras are canonically identified with
	  precisely the super Lie 1-algebras.
	\item
	  For every $n \in \mathbb{N}$ there is the 
	  \emph{super line super Lie $(n+1)$-algebra}
	  $b^n \mathbb{R}^{0|1}$ characterized by the fact that
	  its Chevalley-Eilenberg algebra has trivial differential
	  and a single generator in bidegree $(n,\mathrm{odd})$.
	\item
	  For $\mathfrak{g}$ any super $L_\infty$-algebra and
	  $\mu : \mathfrak{g} \to b^n \mathbb{R}$ a cocycle, its
	  homotopy fiber is the super $L_\infty$-algebra extension
	  of $\mathfrak{g}$, as in def. \ref{gmu}.
	  
	  Below in \ref{SuperPoincareAndExtensions} 
	  we discuss in detail a class of super $L_\infty$-algebras
	  that arise by higher extensions from a super Poincar{\'e} Lie algebra.
  \end{itemize}
\end{examples}
\begin{observation}
  \label{LieIntegrationOfSuper}
  \index{$L_\infty$-algebra!exponentiation/Lie integration!super}
  The Lie integration 
  $$
    \exp(\mathfrak{g}) \in [\mathrm{CartSp}_{\mathrm{smooth}} \times \mathrm{SuperPoint}, \mathrm{sSet}]
      = [\mathrm{SuperPoint}, [\mathrm{CartSp}_{\mathrm{smooth}}, \mathrm{sSet}
  $$ 
  of a super $L_\infty$-algebra $\mathfrak{g}$ according to \ref{SmoothStrucLieAlgebras} 
  is a system of Lie integrated ordinary $L_\infty$-algebras
  $$
    \exp(\mathfrak{g}) : \mathbb{R}^{0|q} \mapsto
      \exp((\mathfrak{g} \otimes_{\mathbb{R}} \Lambda_q)_{\mathrm{even}})
    \,,
  $$
  where $\Lambda_q = C^\infty(\mathbb{R}^{0|q})$ is the Grassmann algebra on $q$ generators.
  
  Over each $U \in \mathrm{CartSp}$ this is the discrete super $\infty$-groupoid given by
  $$
    \exp(\mathfrak{g})_U : \mathbb{R}^{0|q} \mapsto 
    \mathrm{Hom}_{\mathrm{dgsAlg}}(\mathrm{CE}(\mathfrak{g} \otimes \Lambda_q)_{\mathrm{even}},
     \Omega_{\mathrm{vert}}^\bullet(U \times \mathbb{R}^{0|q}\times \Delta^\bullet))
     \,,
  $$
  where on the right we have super differential forms vertical with respect to the projection
  $U \times \mathbb{R}^{0|q} \times \Delta^n \to U \times \mathbb{R}^{0|q}$ of supermanifolds.
\end{observation}
\proof
The first statement holds by the proof of prop. \ref{SuperalgebrasAsInternalAlgebra}.
The second statement is an example of a stadard mechanism in superalgebra:
Using that the category $\mathrm{sVect}$ of finite-dimensional super vector space is a 
compact closed category, we compute
$$
  \begin{aligned}
    \mathrm{Hom}_{\mathrm{dgsAlg}}(\mathrm{CE}(\mathfrak{g}), \Omega^\bullet_{\mathrm{vert}}(U \times \mathbb{R}^{0|q} \times \Delta^n))
    & \simeq
    \mathrm{Hom}_{\mathrm{dgsAlg}}(
     \mathrm{CE}(\mathfrak{g}), 
     C^\infty(\mathbb{R}^{0|q}) \otimes \Omega_{\mathrm{vert}}^\bullet( U \times \Delta^n)
    )
    \\
    & \simeq 
    \mathrm{Hom}_{\mathrm{dgsAlg}}(
      \mathrm{CE}(\mathfrak{g}), 
      \Lambda_q \otimes \Omega_{\mathrm{vert}}^\bullet( U \times \Delta^n)
    )    
    \\
    & \subset
    \mathrm{Hom}_{\mathrm{Ch}^\bullet(\mathrm{sVect})}(\mathfrak{g}^*[1] , \Lambda_q \otimes \Omega_{\mathrm{vert}}^\bullet( U  \times \Delta^n))
    \\
    & \simeq
    \mathrm{Hom}_{\mathrm{Ch}^\bullet(\mathrm{sVect})}(\mathfrak{g}^*[1]\otimes (\Lambda^q)^* ,  \Omega_{\mathrm{vert}}^\bullet( U \times \Delta^n))
    \\
    & \simeq
    \mathrm{Hom}_{\mathrm{Ch}^\bullet(\mathrm{sVect})}((\mathfrak{g} \otimes \Lambda_q)^*[1] ,  \Omega_{\mathrm{vert}}^\bullet( \Delta^n))
    \\
    & \simeq
    \mathrm{Hom}_{\mathrm{Ch}^\bullet(\mathrm{sVect})}((\mathfrak{g} \otimes \Lambda_q)^*[1]_{\mathrm{even}} ,  \Omega_{\mathrm{vert}}^\bullet( U \times \Delta^n))
    \\
    & \supset
    \mathrm{Hom}_{\mathrm{dgsAlg}}( \mathrm{CE}((\mathfrak{g}\otimes_k \Lambda_q)_{\mathrm{even}}),  \Omega_{\mathrm{vert}}^\bullet( U \times \Delta^n))
  \end{aligned}
  \,.
$$
Here in the third step we used that the underlying dg-super-algebra of $\mathrm{CE}(\mathfrak{g})$ 
is free to find the space of morphisms of dg-algebras inside that of super-vector spaces (of generators) as indicated. Since the differential on both sides is $\Lambda_q$-linear, the claim follows.
\endofproof

\newpage

\section{Applications -- Local prequantum higher gauge field theory} 
\label{Applications}
\label{ExamplesAndApplications}

We discuss here applications of the general theory of higher
differential geometry and of the models 
developed above to local prequantum higher gauge field theory and string theory
\cite{SatiSchreiber}.

\medskip

\begin{itemize}
  \item \ref{FractionalClasses} -- Higher $\mathrm{Spin}$-structures
  \item \ref{TwistedStructures} -- Higher prequantum fields
  \item \ref{HigherSymplecticGeometry} -- Higher symplectic geometry
  \item \ref{PrequantizationApplications} -- Higher geometric prequantization
  \item \ref{ChernSimonsFunctional} -- Higher Chern-Simons field theory
  \item \ref{WZWApplications} -- Higher Wess-Zumino-Witten field theory
  \item \ref{LocalBoundaryAndDefectPrequantumFieldTheory} -- Local boundary and defect prequantum field theory
\end{itemize}

\subsection{Higher $\mathrm{Spin}$-structures} 
\label{FractionalClasses}
\index{higher spin structures}

For any $n \in \mathbb{N}$, 
the Lie group $\mathrm{Spin}(n)$ is the universal 
simply connected cover of the
special orthogonal group $\mathrm{SO}(n)$. Since 
$\pi_1 \mathrm{SO}(n) \simeq \mathbb{Z}_2$, it is an extension 
of Lie groups of the form
$$
  \mathbb{Z}_2 \to \mathrm{Spin}(n) \to \mathrm{SO}(n)
  \,.
$$
The lift of an $\mathrm{SO}(n)$-principal bundle through
this extension to a $\mathrm{Spin}(n)$-principal bundle is
a called a choice of \emph{spin structure}.
A classical textbook on the geometry of spin structures is
\cite{LawsonMichelson}.

We discuss how this construction is only one step in  
a whole tower of analogous constructions involving 
smooth $n$-groups for various $n$. These are higher
smooth analogs of the $\mathrm{Spin}$-group and define higher
analogs of smooth spin structures.

The $\mathrm{Spin}$-group carries its name due to the 
central role that it plays in the description of the physics
of quantum \emph{spinning particles}. In \ref{MotivationFromAnomalyCancellation}
we indicated how the higher spin structures to be discussed here
are similarly related to spinning quantum strings and 5-branes.
More in detail, this requires \emph{twisted} higher spin structures,
which we turn to below in \ref{TwistedDifferentialStructures}.

\medskip

\begin{itemize}
  \item \ref{SmoothAndDifferentialWhiteheadtower} -- Overview: the smooth and differential Whitehead tower of $B O$
  \item \ref{orientation structure} -- Orienation structure
  \item \ref{spin structure} -- Spin structure
  \item \ref{String2Group} -- Smooth string structure and the $\mathrm{String}$-2-group
  \item \ref{FivebraneSixGroup} -- Smooth fivebrane structure and the $\mathrm{Fivebrane}$-6-group
  \item \ref{HigherSpinCStructures} -- Higher $\mathrm{Spin}^c$-structures
  \item \ref{SpinCAsHomotopyFiberProduct} -- $\mathrm{Spin}^c$ as a homotopy fiber product in $\mathrm{Smooth}\infty \mathrm{Grpd}$
  \item \ref{SmoothStringC2} -- Smooth $\mathrm{String}^{\mathbf{c}_2}$
\end{itemize}

\subsubsection{Overview: the smooth and differential Whitehead tower of $B O$}
\label{SmoothAndDifferentialWhiteheadtower}

We survey the constructions and results about the smooth 
and differential refinement of the Whitehead tower of $B O$, to be discussed in the following.

\medskip

By definition \ref{GeometricWhiteheadTower} applied in $\infty \mathrm{Grpd} \simeq \mathrm{Top}$,
the first stages of the Whitehead tower of the classifying space $B O$ 
of the orthogonal group, together with the corresponding obstruction
classes is constructed by iterated pasting of homotopy pullbacks as in the following diagram:
$$
  \xymatrix{
    \vdots
    \\
    B \mathrm{Fivebrane}
	\ar[d]
	\ar[r]
	&
	\cdots
	\ar[r]
	&
	{*}
	\ar[d]
    \\
    B \mathrm{String}
	\ar[d]
	\ar@{-}[r]
	&
	\cdots
	\ar[r]^<<<<<{\tfrac{1}{6}p_2}
	&
	B^8 \mathbb{Z}
	\ar[r]
	\ar[r]
	\ar@{-}[d]
	&
	{*}
	\ar[d]
	&&&&&&&
    \\
    B \mathrm{Spin}
	\ar[d]
	\ar@{-}[r]
	&
	\cdots
	\ar[rr]^<<<<<{\tfrac{1}{2}p_1}
	&
	\ar@{-}[d]
	&
	B^4 \mathbb{Z}
	\ar[r]
	\ar@{-}[d]
	&
	{*}
	\ar[d]
	&&&&&&&&&&&
    \\
    B SO
	\ar[d]
	\ar@{-}[r]
	&
	\cdots
	\ar[rrr]^<<<<<{w_2}
	&
	\ar[d]&
	\ar[d]
	&
	B^2 \mathbb{Z}_2
	\ar[r]
	\ar[d]
	&
	{*}
	\ar[d]
	&&&&&&&&
    \\
    B O
	\ar[r]
	\ar[d]
	\ar@/_1pc/[rrrrr]_<<<<<<<<<<<<<{w_1}
	&
	\cdots
	\ar[r]
	&
	\tau_{\leq 8} B O
	\ar[r]
	&
	\tau_{\leq 4} B O
	\ar[r]
	&
	\tau_{\leq 2} B O
	\ar[r]
	&
	\tau_{\leq 1} B O \simeq B \mathbb{Z}_2
	&&
	\\
	B \mathrm{GL}
  }
  \,.
$$
Here the bottom horizontal tower is the Postnikov tower, def. \ref{PostnikovTower}, of $B O$
and all rectangles are homotopy pullbacks.

For $X$ a smooth manifold, there is a canonically given map $X \to B \mathrm{GL}$,
which classifies the tangent bundle $T X$. The lifts of this classifying map
through the above Whitehead tower correspond to structures on $X$ as indicated
in the following diagram:
$$
  \xymatrix{
    &&& B \mathrm{Fivebrane}  \ar[d]
    \\  
    &&& B \mathrm{String} \ar[d]\ar[rr]^{\tfrac{1}{6}p_2} && B^7 U(1) \ar@{}[r]|\simeq & K(\mathbb{Z},8)    
	& 
    \mbox{second fractional Pontryagin class}    
    \\
    &&& B \mathrm{Spin} \ar[d]\ar[rr]^{\tfrac{1}{2}p_1} && B^3 U(1) \ar@{}[r]|\simeq & K(\mathbb{Z},4)
	& 
    \mbox{first fractional Pontryagin class}    
    \\
    &&& B \mathrm{SO} \ar[d]\ar[rr]^{w_2} && B^2 \mathbb{Z}_2 \ar@{}[r]|\simeq & K(\mathbb{Z}_2,2)
	&
	\mbox{second Stiefel-Whitney class}
    \\
    &&& B \mathrm{O} \ar[d]\ar[rr]^{w_1} \ar[d]^{\simeq} && B \mathbb{Z}_2 \ar@{}[r]|\simeq & K(\mathbb{Z}_2,1)
	&
	\mbox{first Stiefel-Whitney class}
    \\
    X 
	  \ar[rrr]|{T X } 
	  \ar@{->}[urrr]|{}
	  \ar@{-->}[uurrr]|{\mathrm{orientation}\;\mathrm{structure}}
	  \ar@/^.4pc/@{-->}[uuurrr]|{\mathrm{spin}\;\mathrm{structure}}
	  \ar@/^1.4pc/@{-->}[uuuurrr]|{\mathrm{string}\;\mathrm{structure}}
	  \ar@/^2.9pc/@{-->}[uuuuurrr]|{\mathrm{fivebrane}\;\mathrm{structure}}
	&&& B \mathrm{GL}
  }
  \,.
$$
Here the horizontal morphisms denote representatives of
universal characteristic classes, such that each sub-diagram of the shape
$$
  \xymatrix{
     B \hat G
	 \ar[d]
	 \\
	 B G \ar[rr]^c && B^n K
  }
$$
is a fiber sequence, def. \ref{LongFiberSequence}. 

The lifting problem presented by each of these steps is exemplified in terms of 
a smooth manifold $X$, which comes with a canonical map $X \to B \mathrm{GL}$ that
classifies the tangent bundle $T X$ of $X$. 

In the first step, since the
$B \mathrm{O} \to B \mathrm{GL}$ is a weak equivalence in $\mathrm{Top} \simeq \infty \mathrm{Grpd}$,
we may always factor $X \to B \mathrm{GL}$, up to homotopy, through $B \mathrm{O}$.
The homotopy class of the 
resulting composite $X \to B \mathrm{O} \stackrel{w_1}{\to} B \mathbb{Z}_2$
is the first Stiefel-Whitney class of the manifold. The fact that $B \mathrm{SO}$
is the homotopy fiber of $w_1$ means, by the universal property of the
homotopy pullback, that the further lift to a map $X \to B \mathrm{SO}$
exists precisely if the first Stiefel-Whitney class vanishes. 
While this is a classical fact, it is useful to make its relation to 
homotopy pullbacks explicit here, since this illuminates the following
steps in this tower as well as all the steps in the smooth and differential
refinements to follow.

Next, if the first Stiefel-Whitney class of $X$ vanishes, then any \emph{choice}
of orientation, hence any choice of lift $X \to B \mathrm{SO}$ induces the
composite map $X \to B \mathrm{SO} \stackrel{w_2}{\to} B^2 \mathbb{Z}_2$,
whose homotopy class is the second Stiefel-Whitney class of $X$ equipped with that
orientation. If that class vanishes, there exists a choice of lift
$X \to B \mathrm{Spin}$, which is a choice of spin structure on $X$.
The resulting composite $X \to B \mathrm{Spin} \stackrel{\tfrac{1}{2}p_1}{\to} B^3 U(1)$
is a representative of the \emph{first fractional Pontryagin class}. 
If this vanishes, there exists a choice of lift $X \to B \mathrm{String}$,
which equips $X$ with a \emph{string structure}. The induced composite
$X\to B \mathrm{String} \stackrel{\tfrac{1}{6}p_2}{\to}B^7 U(1)$
is a representative of the second fractional Pontryagin class of $X$. If that
vanishes, there exists a choice of lift $X \to B \mathrm{Fivebrane}$, which 
is a choice of \emph{fivebrane srructure} on $X$.

In this or slightly different terminology, this is a classical construction in 
homotopy theory. We show in the following that this tower has a
\emph{smooth lift} from topological spaces 
through the geometric realization functor, \ref{SmoothStrucHomotopy},
$$
  \xymatrix{
    \mathrm{Smooth}\infty\mathrm{Grpd}
	\ar[r]^<<<<<\Pi
	& 
	\infty \mathrm{Grpd}
	\ar[r]^{\vert-\vert}_\simeq
	&
	\mathrm{Top}
  }
$$
to smooth $\infty$-groupoids, of the form
$$
  \raisebox{30pt}{
  \xymatrix{
    &&& \mathbf{B} \mathrm{Fivebrane}  \ar[d]
    \\  
    &&& \mathbf{B} \mathrm{String} \ar[d]\ar[rr]^{\tfrac{1}{6}\mathbf{p}_2} && \mathbf{B}^7 U(1) 
    \\
    &&& \mathbf{B} \mathrm{Spin} \ar[d]\ar[rr]^{\tfrac{1}{2}\mathbf{p}_1} && \mathbf{B}^3 U(1) 
    \\
    &&& \mathbf{B} \mathrm{SO} \ar[d]\ar[rr]^{\mathbf{w}_2} && \mathbf{B}^2 \mathbb{Z}_2 
    \\
    &&& \mathbf{B} \mathrm{O} \ar[d]\ar[rr]^{\mathbf{w}_1} \ar[d]^{} && \mathbf{B} \mathbb{Z}_2 
    \\
    X 
	  \ar[rrr]|{T X } 
	  \ar@{-->}[urrr]|{\mathrm{metric}\;\mathrm{structure}}
	  \ar@{-->}[uurrr]|{\mathrm{orientation}\;\mathrm{structure}}
	  \ar@/^.4pc/@{-->}[uuurrr]|{\mathrm{spin}\;\mathrm{structure}}
	  \ar@/^1.4pc/@{-->}[uuuurrr]|{\mathrm{string}\;\mathrm{structure}}
	  \ar@/^2.9pc/@{-->}[uuuuurrr]|{\mathrm{fivebrane}\;\mathrm{structure}}
	&&& \mathbf{B} \mathrm{GL}
  }
  }
$$
Here  $\mathbf{B}^n U(1)$ is the smooth circle $(n+1)$-group, def. \ref{Circlengroup},
the smooth classifying $n$-stack of smooth circle $n$-bundles.
This is such that still all diagrams of the form
$$
  \raisebox{20pt}{
  \xymatrix{
     \mathbf{B} \hat G
	 \ar[d]
	 \\
	 \mathbf{B} G \ar[rr]^{\mathbf{c}} && \mathbf{B}^n K
  }
  }
$$
are fiber sequences, now in the cohesive $\infty$-topos $\mathrm{Smooth}\infty \mathrm{Grpd}$,
exhibiting the smooth moduli $\infty$-stack $\mathbf{B}\hat G$ as the homotopy fiber
of the smooth universal characteristic map $\mathbf{c}$ which is a smooth refinement
of the corresponding ordinary characteristic map $c$.

The corresponding choices of lifts now are more refined than before, as they 
correspond to \emph{smooth structures}. In the first step, the choice of
lift from a morphism $X \to \mathbf{B} \mathrm{GL}$ to a morphism 
$X \to \mathbf{B}\mathrm{SO}$ encodes now genuine information, namely a choice of
\emph{Riemannian metric} on $X$. This is discussed in \ref{OrthogonalRiemannianStructureByTwistedCohomology}
below. 

Further up, a choice of lift $X \to \mathbf{B} \mathrm{Spin}$ is a choice of
smooth $\mathrm{Spin}$-principal bundle on $X$. Next, 
the object denoted $\mathrm{String}$ is a 
smooth 2-group, and a lift $X \to \mathbf{B}\mathrm{String}$ is a choice of 
smooth $\mathrm{String}$-principal 2-bundle on $X$. The object denoted
$\mathrm{Fivebrane}$ is a smooth 6-group and a choice of lift
$X \to \mathbf{B}\mathrm{Fivebrane}$ is a choice of smooth 
$\mathrm{Fivebrane}$-principal 6-bundle.

One consequence of the smooth refinement, which is important for the 
\emph{twisted} such structures discussed below in \ref{TwistedDifferentialStructures},
is that the spaces of choices of lifts are much more refined than those of
the ordinary non-smooth case. 
Another consequence is that it allows us to proceed and next consider a \emph{differential} refinement, def. \ref{BGconn}:

we show that the above smooth Whitehead tower further lifts to a 
\emph{differential Whitehead tower} of the form
$$
  \raisebox{30pt}{
  \xymatrix{
    &&& \mathbf{B} \mathrm{Fivebrane}_{\mathrm{conn}}  \ar[d]
    \\  
    &&& \mathbf{B} \mathrm{String}_{\mathrm{conn}} \ar[d]\ar[rr]^{\tfrac{1}{6}\hat {\mathbf{p}}_2} && \mathbf{B}^7 U(1)_{\mathrm{conn}} 
    \\
    &&& \mathbf{B} \mathrm{Spin}_{\mathrm{conn}} \ar[d]\ar[rr]^{\tfrac{1}{2}\hat {\mathbf{p}}_1} && \mathbf{B}^3 U(1)_{\mathrm{conn}} 
    \\
    &&& \mathbf{B} \mathrm{SO}_{\mathrm{conn}} \ar[d]\ar[rr]^{\mathbf{w}_2} && \mathbf{B}^2 \mathbb{Z}_2 
    \\
    &&& \mathbf{B} \mathrm{O}_{\mathrm{conn}} 
	\ar[d]\ar[rr]^{\mathbf{w}_1} \ar[d]^{} && \mathbf{B} \mathbb{Z}_2 
    \\
    X 
	  \ar[rrr]|{T X } 
	  \ar@{-->}[urrr]|{\mathrm{metric}\;\mathrm{and}\;\mathrm{affine}\;\mathrm{connection}}
	  \ar@{-->}[uurrr]|{}
	  \ar@/^.4pc/@{-->}[uuurrr]|{\mathrm{spin}\;\mathrm{connection}}
	  \ar@/^1.4pc/@{-->}[uuuurrr]|{\mathrm{string}\;2-\mathrm{connection}}
	  \ar@/^3.4pc/@{-->}[uuuuurrr]|{\mathrm{fivebrane}\;6-\mathrm{connection}}
	&&& \mathbf{B} \mathrm{GL}_{\mathrm{conn}}
  }
  }
  \,,
$$
where $\mathbf{B}^n U(1)_{\mathrm{conn}}$ is the moduli $n$-stack of 
circle $n$-bundles with connection, according to \ref{SmoothStrucDifferentialCohomology}.
Still, all diagrams of the form
$$
  \raisebox{20pt}{
  \xymatrix{
     \mathbf{B} \hat G_{\mathrm{conn}}
	 \ar[d]
	 \\
	 \mathbf{B} G_{\mathrm{conn}} \ar[rr]^{\hat {\mathbf{c}}} && \mathbf{B}^n K_{\mathrm{conn}}
  }
  }
$$
are fiber sequences in $\mathrm{Smooth}\infty\mathrm{Grpd}$, exhibiting the smooth moduli
$\infty$-stack $\mathbf{B}\hat G_{\mathrm{conn}}$, def. \ref{BGconn}, 
of higher $\hat G$-connections
as the homotopy fiber of the differential refinement $\hat {\mathbf{c}}$
of the given characteristic map $c$.
Choices of lifts through this tower correspond to choices of smooth higher connections
on smooth higher bundles. 

\subsubsection{Orienation structure}
\label{orientation structure}
\index{orientation structure}

Before going to higher degree beyond the $\mathrm{Spin}$-group,
it is instructive to first consider a \emph{lower} degree. 
The special orthogonal Lie group itself
is a kind of extension of the orthogonal Lie group. 
To see this clearly, consider the
smooth delooping
$\mathbf{B}\mathrm{SO}(n) \in \mathrm{Smooth}\infty \mathrm{Grpd}$
according to \ref{SmoothStrucCohesiveInfiniGroups}.
\begin{proposition}
 \index{characteristic class!Stiefel-Whitney!first and orientation structure}
 The canonical morphism $\mathrm{SO}(n) \hookrightarrow \mathrm{O}(n)$
 induces a long fiber sequence in $\mathrm{Smooth}\infty \mathrm{Grpd}$
 of the form
 $$
   \mathbb{Z}_2 
     \to 
   \mathbf{B}\mathrm{SO}(n)
    \to 
   \mathbf{B}O(n)
   \stackrel{\mathbf{w}_1}{\to}
   \mathbf{B}\mathbb{Z}_2
   \,,
 $$
 where $\mathbf{w}_1$ is the universal smooth first 
 Stiefel-Whitney class from example \ref{FirstStiefelWhitneyClass}.
\end{proposition}
\proof
  It is sufficient to show that the
  homotopy fiber of $\mathbf{w}_1$ is $\mathbf{B}\mathrm{SO}(n)$.
  This implies the rest of the statement by prop. \ref{FormOfLongFiberSequences}.
  
  To see this, notice that by the discussion in 
  \ref{StrucCohomology} we are to compute the
  $\mathbb{Z}_2$-principal bundle over 
  the Lie groupoid $\mathbf{B}\mathrm{SO}(n)$
  that is classified by the above injection. 
  By  observation \ref{UniversalPrincipalInfinityBundle}
  this is accomplished by forming a 
  1-categorical pullback of 
  Lie groupoids 
  $$
    \xymatrix{
	  \mathbb{Z}_2/\!/\mathrm{O}(n)
	  \ar[d]
	  \ar[r]
	  & \mathbb{Z}_2/\!/\mathbb{Z}_2 
	  \ar[d]
	  \\
	  {*}/\!/ \mathrm{O}(n) \ar[r] & {*} /\!/ \mathbb{Z}_2
	}
	\,.
  $$
  One sees that the canonical projection
  $$
    \mathbb{Z}_2/\!/ \mathrm{O}(n) \stackrel{\simeq}{\to}
	*/\!/ \mathrm{SO}(n)
  $$
  is a weak equivalence (it is an essentially surjective and
  full and faithful functor of groupoids).
\endofproof
\begin{definition}
  \label{OrientationStructure}
  \index{orientation structure}
  For $X \in \mathrm{Smooth}\infty \mathrm{Grpd}$
  any object equipped with a morphism 
  $r_X : X \to \mathbf{B}O(n)$, we say a lift $o_X$ of
  $r$ through the above extension
  $$
    \xymatrix{
	  & \mathbf{B} \mathrm{SO}(n) \ar[d]
	  \\
	  X \ar[r]^{r} \ar@{-->}[ur]^{o_X} 
	  & \mathbf{B}O(n)
	}
  $$
  is an \emph{orientation structure} on $(X, r_X)$.
\end{definition}

\subsubsection{Spin structure}
\label{spin structure}

\begin{proposition}
  \index{characteristic class!Stiefel-Whitney!second and spin structure}
  The classical sequence of Lie groups
  $\mathbb{Z}_2 \to \mathrm{Spin} \to \mathrm{SO}$
  induces a long fiber sequence in $\mathrm{Smooth}\infty \mathrm{Grpd}$
  of  the form
  $$
    \mathbb{Z}_2
	\to
    \mathrm{Spin}
    \to
    \mathrm{SO}
    \to
    \mathbf{B}\mathbb{Z}_2 
	\to
	\mathbf{B} \mathrm{Spin}
	\to 
	\mathbf{B}\mathrm{SO}
	  \stackrel{\mathbf{w}_2}{\to}
	\mathbf{B}^2 \mathbb{Z}_2
	\,,
  $$
  where $\mathbf{w}_2$ is the universal smooth second 
  Stiefel-Whitney class from example \ref{SecondStiefelWhineyClass}.
\end{proposition}
\proof
  It is sufficient to show that the
  homotopy fiber of $\mathbf{w}_2$ is $\mathbf{B}\mathrm{Spin}(n)$.
  This implies the rest of the statement by prop. \ref{FormOfLongFiberSequences}.

  To see this notice that the top morphism in the stanard
  anafunctor that presents $\mathbf{w}_2$
  $$
   \xymatrix{
     \mathbf{B}(\mathbb{Z}_2 \to \mathrm{O}(n))_{\mathrm{ch}}
	 \ar[r]
	 \ar[d]^{\simeq}
	 &
	 \mathbf{B}(\mathbb{Z}_2 \to 1)_{\mathrm{ch}}
	 \ar@{=}
	 &
	 \mathbf{B}^2 \mathbb{Z}_2
	 \\
	 \mathbf{B}\mathrm{SO}(n)
   }
  $$
  is a fibration in 
  $[\mathrm{CartSp}^{\mathrm{op}}, \mathrm{sSet}]_{\mathrm{proj}}$.
  By proposition \ref{FiniteHomotopyLimitsInPresheaves}
  this means that the homotopy fiber is given by the
  1-categorical pullback of simplicial presheaves
  $$
    \xymatrix{
	  \mathbf{B}(\mathbb{Z}_2 \to \mathrm{O}(n))_{\mathrm{ch}}
	  \ar[r]
	  \ar[d]
	  &
	  {*}
	  \ar[d]
	  \\
	  \mathbf{B}(\mathbb{Z}_2 \to \mathrm{O}(n))_{\mathrm{ch}}
	  \ar[r]^{\mathbf{w}_2}
	  &
	  \mathbf{B}(\mathbb{Z}_2 \to 1)_{\mathrm{ch}}
	}
	\,.
  $$
  The canonical projection
  $$
    \mathbf{B}(\mathbb{Z}_2 \to \mathrm{O}(n))_{\mathrm{ch}}
	\stackrel{\simeq}{\to}
	\mathbf{B}\mathrm{SO}(n)_{\mathrm{ch}}
  $$
  is seen to be a weak equivalence.
\endofproof
\begin{definition}
  \label{SpinStructure}
  For $X \in \mathrm{Smooth}\infty \mathrm{Grpd}$
  an object equipped with orientation structure 
  $o_X : X \to \mathbf{B}\mathrm{SO}(n)$,
  def. \ref{OrientationStructure}, we say a choice of lift
  $\hat o_X$ in
  $$
    \xymatrix{
	  & \mathbf{B} \mathrm{Spin}
	  \ar[d]
	  \\
	  X \ar[r]^{o_X}
	  \ar@{-->}[ur]^{\hat o_X}
	  &
	  \mathbf{B} \mathrm{SO}(n)
	}
  $$
  equips $(X,o_X)$ with \emph{spin structure}.
\end{definition}

\subsubsection{Smooth string structure and the $\mathrm{String}$-2-group}
\label{String2Group}
\index{string structure}
\index{string 2-group}

The sequence of Lie groupoids
$$
  \cdots
  \to
  \mathbf{B}\mathrm{Spin}(n)
  \to 
  \mathbf{B}\mathrm{SO}(n)
  \to
  \mathbf{B}\mathrm{O}(n)
$$
discussed in \ref{orientation structure} and
\ref{spin structure}
is a smooth refinement of the first two steps of the
\emph{Whitehead tower} of $B O(n)$.
We discuss now the next step. 
This is no longer presented by Lie groupoids, but by
smooth 2-groupoids.

Write $\mathfrak{so}(n)$ for the special orthogonal Lie algebra in dimension $n$.
We shall in the following notationally suppress the dimension and just write
$\mathfrak{so}$. The simply connected Lie group integrating $\mathfrak{so}$ is the $\mathrm{Spin}$-group .

\begin{proposition}
  \label{FirstFracPontryagin}
  Pulled back to $B \mathrm{Spin}$ the 
  universal first Pontryagin class $p_1 : B O \to B^4 \mathbb{Z}$
  is 2 times a generator $\frac{1}{2}p_1$ of $H^4(B \mathrm{Spin}, \mathbb{Z})$
  $$
    \xymatrix{
      B \mathrm{Spin} \ar[r]^{\frac{1}{2}p_1} \ar[d] & B^4 \mathbb{Z} \ar[d]^{\cdot 2}
      \\
      B O \ar[r]^{p_1} & B^4 \mathbb{Z}
    }
    \,.
  $$
  We call $\frac{1}{2}p_1$ the \emph{first fractional Pontryagin class}
   \index{Pontryagin class!first fractional!bare}\index{characteristic class!Pontryagin class!first fractional}.
\end{proposition}
This is due to \cite{Bott}.  See \cite{SSSII} for a review.
\begin{definition} 
  \label{TopologicalStringGroup}
  Write $B \mathrm{String}$ for the homotopy fiber in $\mathrm{Top} \simeq \infty \mathrm{Grpd}$ of the
  first fractional Pontryagin class
  $$
    \xymatrix{
      B \mathrm{String} \ar[r] \ar[d] & {*} \ar[d]
      \\
      B \mathrm{Spin} \ar[r]^{\frac{1}{2}p_1} & B^4 \mathbb{Z}
    }
    \,.
  $$
  Its loop space is the \emph{string group}\index{string group!bare $\infty$-group}
  $$
    \mathrm{String} := O\langle 7\rangle := \Omega B \mathrm{String}
    \,.
  $$
\end{definition}
This is defined up to equivalence as an $\infty$-group object,  
but standard methods give a presentation by a genuine topological group and often the term
\emph{string group} is implicitly reserved for such a topological group model. See also the review
in \cite{Schommer-Pries}. 

We now discuss smooth refinements of $\frac{1}{2}p_1$ and of $\mathrm{String}$ as lifts through
the intrinsic geometric realization, def. \ref{GeometricRealization}, 
$\Pi : \mathrm{Smooth} \infty \mathrm{Grpd} \to \infty \mathrm{Grpd}$ in 
$\mathrm{Smooth} \infty \mathrm{Grpd}$, \ref{SmoothInfgrpds}.
\begin{proposition} 
  \label{IntegrationToSpinGroup}
  We have a weak equivalence
  $$
    \mathbf{cosk}_3(\exp(\mathfrak{so})) \stackrel{\simeq}{\to}
    \mathbf{B}\mathrm{Spin}_c
  $$
  in $[\mathrm{CartSp}_{\mathrm{smooth}}^{\mathrm{op}}, \mathrm{sSet}]_{\mathrm{proj}}$,
  between the Lie integration, \ref{SmoothStrucLieAlgebras}, of $\mathfrak{so}$ and the
  standard presentation, \ref{SmoothStrucCohesiveInfiniGroups}, of $\mathbf{B}\mathrm{Spin}$.
\end{proposition}
\proof
  By prop. \ref{IntegrationToSimplyConnectedLieGroup}.
\endofproof
\begin{corollary} \label{DeloopingOfTheSpinGroup}
  The image of $\mathbf{B}\mathrm{Spin} \in \mathrm{Smooth}\infty \mathrm{Grpd}$
  under the fundamental $\infty$-groupoid/geometric realization functor $\Pi$, 
  \ref{ETopStrucHomotopy}, is the classifying space $B \mathrm{Spin}$ of the topological 
  $\mathrm{Spin}$-group 
  $$
    |\Pi \mathbf{B}\mathrm{Spin}| \simeq B \mathrm{Spin}
    \,.
  $$
\end{corollary}
\proof
  By prop. \ref{FundGroupoidOfSimplicialParacompact} applied to 
  prop. \ref{DeloopedLieGroup}.
\endofproof
\begin{theorem} 
  \label{FirstFractionalDifferentialPontrjagin}
  \index{Pontryagin class!first fractional!smooth}
  \index{Pontryagin class!first fractional!differential}
  The image under Lie integration, \ref{SmoothStrucLieAlgebras}, 
of the canonical Lie algebra 3-cocycle
$$
  \mu = \langle -,[-,-]\rangle : \mathfrak{so} \to b^2 \mathbb{R}
$$
on the semisimple Lie algebra $\mathfrak{so}$ of the $\mathrm{Spin}$ group 
is a morphism in $\mathrm{Smooth}\infty \mathrm{Grpd}$ of the form
$$
  \frac{1}{2} \mathbf{p}_1
  :=
  \exp(\mu) 
  : 
  \mathbf{B}\mathrm{Spin}
    \to
  \mathbf{B}^3 U(1)
$$
whose image under the fundamental $\infty$-groupoid $\infty$-functor/ geometric realization, 
\ref{ETopStrucHomotopy}, 
$\Pi : \mathrm{Smooth} \infty \mathrm{Grpd} \to \infty \mathrm{Grpd}$ 
is the ordinary fractional Pontryagin class
$\frac{1}{2}p_1 : B \mathrm{Spin} \to B^4 \mathbb{Z}$
in $\mathrm{Top}$, and up to equivalence $\exp(\mu)$ is the unique lift
of $\frac{1}{2}p_1$ from $\mathrm{Top}$ to $\mathrm{Smooth}\infty \mathrm{Grpd}$ with codomain
$\mathbf{B}^3 U(1)$. We write $\frac{1}{2}\mathbf{p}_1 := \exp(\mu)$
and call it the \emph{smooth first fractional Pontryagin class}.

Moreover, the corresponding 
refined differential characteristic class, \ref{SmoothStrucInfChernWeil},  
$$
  \frac{1}{2}\hat {\mathbf{p}}_1 : 
  \mathbf{H}_{\mathrm{conn}}(-,\mathbf{B}\mathrm{Spin})
  \to
  \mathbf{H}_{\mathrm{diff}}(-, \mathbf{B}^3 U(1))
  \,,
$$
wich we call the \emph{fractional Pontryagin class}\index{Pontryagin class!first fractional!differential},
is in cohomology the corresponding ordinary refined Chern-Weil homomorphism 
\cite{HopkinsSinger}
$$
  [\frac{1}{2}\hat {\mathbf{p}}_1] : 
  H^1_{\mathrm{Smooth}}(X,\mathrm{Spin}) \to H_{\mathrm{diff}}^4(X)
$$
with values in ordinary differential cohomology that corresponds to the 
Killing form invariant polynomial $\langle - , - \rangle $ on $\mathfrak{so}$.
\end{theorem}
\proof
  This is shown in \cite{FSS}.

  Using corollary. \ref{IntegrationToSpinGroup} and 
  unwinding all the definitions and using the characterization of 
  smooth de Rham coefficient objects, \ref{SmoothStrucdeRham}, and
  smooth differential coefficient objects, \ref{SmoothStrucDifferentialCohomology}, 
  one finds that the postcomposition with $\exp(\mu,\mathrm{cs})_{\mathrm{diff}}$ 
  induces on {\v C}ech cocycles precisely the operation considered in 
  \cite{brylinski-mclaughlin}, and hence the conclusion follows essentially as by the
  reasoning there:
  one reads off the 4-curvature of the circle 3-bundle assigned to a Spin bundle with 
  connection $\nabla$ to be $\propto \langle F_\nabla \wedge F_\nabla \rangle$, with the
  normalization such that this is the image in de Rham cohomology of the generator of
  $H^4(B \mathrm{Spin}) \simeq \mathbb{Z} \simeq \langle \frac{1}{2}p_1\rangle$.

  Finally that $\frac{1}{2}\mathbf{p}_1$ is the unique smooth lift of $\frac{1}{2}p_1$
  follows from theorem \ref{LieGroupCohomologyInSmoothInfinGroupoid}.
\endofproof
By the unique smooth refinement of the first fractional Pontryagin class, 
\ref{FirstFractionalDifferentialPontrjagin}, we obtain a smooth refinement of the
String-group, def. \ref{TopologicalStringGroup}.
\begin{definition} 
  \label{SmoothBString}
    \index{group!smooth string 2-group}
    \index{string group!smooth 2-group}
    \index{2-group!string 2-group}
  Write $\mathbf{B}\mathrm{String}$ for the homotopy fiber in $\mathrm{Smooth}\infty \mathrm{Grpd}$
  of the smooth refinement of the first fractional Pontryagin class from  
  prop. \ref{FirstFractionalDifferentialPontrjagin}:
  $$
    \xymatrix{
      \mathbf{B}\mathrm{String} \ar[r] \ar[d] & {*} \ar[d]
      \\
      \mathbf{B}\mathrm{Spin}
       \ar[r]^{\frac{1}{2}\mathbf{p}_1}
      &
      \mathbf{B}^3 U(1)
    }
    \,.
  $$
  We say its loop space object is the \emph{smooth string 2-group}
  $$
    \mathrm{String}_{\mathrm{smooth}} := \Omega \mathbf{B} \mathrm{String}
    \,.
  $$
\end{definition}
We speak of a smooth \emph{2-group} because $\mathrm{String}_{\mathrm{smooth}}$ is a 
categorical homotopy 1-type in $\mathrm{Smooth}\infty \mathrm{Grpd}$, being an extension
$$
  \mathbf{B}U(1) \to \mathrm{String}_{\mathrm{smooth}} \to \mathrm{Spin}
$$
of the categorical 0-type $\mathrm{Spin}$ by the categorical 1-type $\mathbf{B}U(1)$
in $\mathrm{Smooth}\infty \mathrm{Grp}$. 
\begin{proposition}
  The categorical homotopy groups of the smooth String 2-group, 
$\pi_n(\mathbf{B}\mathrm{String}) \in \mathrm{Sh}(\mathrm{CartSp})$, are
  $$
    \pi_1(\mathbf{B}\mathrm{String}) \simeq \mathrm{Spin}
  $$
  and
  $$
    \pi_2(\mathbf{B}\mathrm{String}) \simeq U(1)
    \,.
  $$
  All other categorical homotopy groups are trivial.
\end{proposition}
\proof
  Notice that by construction the non-trivial categorical homotopy groups of
  $\mathbf{B}\mathrm{Spin}$ and $\mathbf{B}^3 U(1)$ are 
  $\pi_1 \mathbf{B} \mathrm{Spin} = \mathrm{Spin}$ and
  $\pi_3 \mathbf{B}^3 U(1) = U(1)$, respectively.
  Using the long exact sequence of homotopy sheaves
  (use \cite{Lurie} remark 6.5.1.5,with $X = *$ the base point)
  applied to def. \ref{SmoothBString}, we obtain the long exact sequence of 
  pointed objects in $\mathrm{Sh}(\mathrm{CartSp})$
  $$
     \cdots
     \to
    \pi_{n+1}(\mathbf{B}^3 U(1))    
      \to
    \pi_n(\mathbf{B}\mathrm{String})
     \to
    \pi_n(\mathbf{B}\mathrm{Spin})
     \to
    \pi_n(\mathbf{B}^3 U(1))    
     \to 
    \pi_{n-1}(\mathbf{B}\mathrm{String})
    \to
    \cdots
  $$
  this yields for $n = 0$
  $$
    0 \to \pi_1(\mathbf{B}\mathrm{String}) \to \mathrm{Spin} \to 0
  $$
  and for $n = 2$
  $$
    0 \to U(1) \to \pi_2(\mathbf{B}\mathrm{String}) \to 0
  $$
  and for $n \geq 3$ 
  $$
    0 \to \pi_n(\mathbf{B}\mathrm{String}) \to 0
    \,.
  $$
\endofproof
However the \emph{geometric} homotopy type, \ref{StrucGeometricHomotopy},
of $\mathbf{B}\mathrm{String}$ is not bounded, in fact it coincides with that 
of the topological string group:
\begin{proposition} 
  \label{SmoothStringRealizesToTopologicalString}
  Under intrinsic geometric realization, \ref{SmoothStrucHomotopy},
  $|-| : \mathrm{Smooth}\infty \mathrm{Grpd} \stackrel{\Pi}{\to} \infty \mathrm{Grp} \stackrel{|-|}{\to} 
  \mathrm{Top}$  
  the smooth string 2-group maps to the topological string group 
  $$
    |\mathrm{String}_{\mathrm{smooth}}| \simeq \mathrm{String}
    \,.
  $$
\end{proposition}
\proof
  Since $\mathbf{B}^3 U(1)$ has a presentation by a simplicial object in $\mathrm{SmoothMfd}$,
  prop. \ref{SmoothPiPreservesSomeHomotopyFibers} asserts that 
  $$
    |\mathrm{String}_{\mathrm{smooth}}| \simeq \mathrm{hofib} |\frac{1}{2} \mathbf{p}_1|
    \,.
  $$
  The claim then follows with prop. \ref{FirstFractionalDifferentialPontrjagin}
  $$
    \cdots \simeq \mathrm{hofib} \frac{1}{2}p_1
  $$
  and def. \ref{TopologicalStringGroup}
  $$
    \cdots \simeq \mathrm{String}
    \,.
  $$
\endofproof
Notice the following important subtlety:
\begin{proposition}
  There exists an infinite-dimensional Lie group $\mathrm{String}_{1\mathrm{smooth}}$
  whose underlying topological group is a model for the String group in $\mathrm{Top}$,
  def. \ref{TopologicalStringGroup}.
\end{proposition}
This is due to \cite{NSW}, by a refinement of a construction in 
\cite{Stolz}. 
\begin{remark}
  However, $\mathbf{B} \mathrm{String}_{1\mathrm{smooth}}$ 
itself is not a model for def. \ref{SmoothBString}, because it is an internal 1-type
in $\mathrm{Smooth}\infty \mathrm{Grpd}$, hence because 
$\pi_2 \mathbf{B} \mathrm{String}_{\mathrm{smooth}} = 0$. In \cite{NSW} a smooth 2-group
with the correct internal homotopy groups based on $\mathrm{String}_{1\mathrm{smooth}}$
is given, but it is not clear yet whether or not this is a model for def. \ref{SmoothBString}.
\end{remark}

We proceed by discussing concrete presentations of the smooth string 2-group.
\begin{definition} 
  \label{StringLie2Algebra}
  \index{string Lie 2-algebra}
  \index{$L_\infty$-algebra!string Lie 2-algebra}
  Write 
  $$
    \mathfrak{string} := \mathfrak{so}_\mu
  $$ 
  for the $L_\infty$-algebra extension of $\mathfrak{so}$
  induced by $\mu$ according to def \ref{gmu}.
  
  We call this the \emph{string Lie 2-algebra}
\end{definition}
\begin{observation}
  The indecomposable invariant polynomials on $\mathfrak{string}$ are those of $\mathfrak{so}$
  except for the Killing form:
  $$
    \mathrm{inv}(\mathfrak{string}) = \mathrm{inv}(\mathfrak{so})/(\langle-,-\rangle )
    \,.
  $$ 
\end{observation}
\proof
  As a special case of prop. \ref{InvPolynomialsOfHigherCentralExtension}.
\endofproof
\begin{proposition} 
  \label{StringModelByExpgmu}
  The smooth $\infty$-groupoid that is the Lie integration of $\mathfrak{so}_\mu$ 
  is a model for the smooth string 2-group
  $$
    \mathbf{B} \mathrm{String} \simeq \mathbf{cosk}_3 \exp(\mathfrak{so}_\mu)
    \,.
  $$
\end{proposition} 
Notice that this statement is similar to, but different from, the statement about the
untruncated exponentiated $L_\infty$-algebras in prop. \ref{ExponentiatedHomotopyFibers}.
\\
\proof
  By prop. \ref{FirstFractionalDifferentialPontrjagin} 
  an explicit presentation for $\mathbf{B}\mathrm{String}$ is given by the pullback
  $$
    \xymatrix{
      \mathbf{B}\mathrm{String}_c \ar [rr] \ar[d] && \mathbf{E}\mathbf{B}^2 U(1)_{c} \ar[d]
      \\
      \mathbf{cosk}_3 \exp(\mathfrak{so})
      \ar[rr]^{\int_{\Delta^\bullet}\exp(\mu)}
      &&
      \mathbf{B}^3 U(1)_c
    }
  $$
  in $[\mathrm{CartSp}^{\mathrm{op}}, \mathrm{sSet}]$, where $\mathbf{B}^3 U(1)_c$ is
  the simplicial presheaf whose 3-cells form the space $U(1)$, and where 
  $\mathbf{E}B^2 U(1)$ is the simplicial presheaf whose 2-cells form $U(1)$ and whose
  3-cells form the space of arbitrary quadruples of elements in $U(1)$. The right vertical
  morphism forms the oriented sum of these quadruples.

  Since all objects are 
  3-truncated, it is sufficient to consider the pullback of the simplices in degrees 
  0 to 3. In degrees 0 to 1 the morphism $\mathbf{E}\mathbf{B}^2 U(1) \to \mathbf{B}^3 U(1)_c$ 
  is the identity, hence in these degrees $\mathbf{B}\mathrm{String}_c$ coincides
  with $\mathbf{cosk}_3 \exp(\mathfrak{so})$. In degree 2 the pullback is the product of
  $\mathbf{cosk}_3(\mathfrak{so})_2$ with $U(1)$, hence the 2-cells of 
  $\mathbf{B}\mathrm{String}_c$ are pairs $(f,c)$ consisting of a smooth map $f : \Delta^2 \to \mathrm{Spin}$
  (with sitting instants) and an elemement $c \in U(1)$. 
  Finally a 3-cell in $\mathbf{B}\mathrm{String}_c$ is a 
  pair $(\sigma,\{c_i\})$ of a smooth map $\sigma : \Delta^3 \to \mathrm{Spin}$
  and four labels $c_{i} \in U(1)$, subject to the condition that the sum of the labels
  is the integral of the cocycle $\mu$ over $\sigma$:
  $$
    c_{4} c_2 c_1^{-1} c_3^{-1} = \int_{\Delta^3} \sigma^* \mu(\theta) \; \mathrm{mod}\mathbb{Z}
    \,,
  $$
  (with $\theta$ the Maurer-Cartan form on $\mathrm{Spin}$).

  The description of the cells in $\mathbf{cosk}_3 \exp(\mathfrak{g}_\mu)$ is similar:
  a 2-cells is a pair $(f,B)$ consisting of a smooth function $f : \Delta^2 \to \mathrm{Spin}$
  and a smooth 2-form $B \in \Omega^2(\Delta^2)$ (both with sitting instants), and a 3-cell
  is a pair consisting of a smooth function $\sigma : \Delta^3 \to \mathrm{Spin}$   
  and a 2-form $\hat B  \in \Omega^2(\Delta^3)$ such that
  $d \hat B = \sigma^* \mu(\theta)$.
  
  There is an evident morphism
  $$
    p : \int_{\Delta^\bullet} : \mathbf{cosk}_3(\mathfrak{so}_\mu) \to \mathbf{B}\mathrm{String}_c
  $$
  that is the identity on the smooth maps from simplices into the $\mathrm{Spin}$-group and which
  sends the 2-form labels to their integral over the 2-faces
  $$
    p_2 : (f,B) \mapsto (f, (\int_{\Delta^2} B) \mathrm{mod}\mathbb{Z})
    \,.
  $$
  We claim that this is a weak equivalence. 
  The first simplicial homotopy group on both sides is $\mathrm{Spin}$ itself
  (meaning: the presheaf on $\mathrm{CartSp}$ represented by $\mathrm{Spin}$).
  The nontrivial simplicial homotopy group to check
  is the second. Since $\pi_2(\mathrm{Spin}) = 0$ every pair $(f,B)$ on $\partial \Delta^3$
  is homotopic to one where $f$ is constant. It follows from prop. \ref{LieIntegrationToLineNGroup}
  that the homotopy classes of such pairs where also the homotopy involves a constant map
  $\partial \Delta^3 \times \Delta^1 \to \mathrm{Spin}$ are given by $\mathbb{R}$, being the
  integral of the 2-forms. But then moreover there are the non-constant homotopies in 
  $\mathrm{Spin}$ from the constant 2-sphere to itself. 
  Since $\pi_3(\mathrm{Spin}) = \mathbb{Z}$
  and $\mu(\theta)$ is an integral form, this reduces the homotopy classes to 
  $U(1) = \mathbb{R}/\mathbb{Z}$. This are the same as in $\mathbf{B}\mathrm{String}_c$
  and the integration map that sends the 2-forms to elements in $U(1)$ is an isomorphism
  on these homotopy classes.
\endofproof
\begin{remark}
  Propositions \ref{StringModelByExpgmu} and \ref{SmoothStringRealizesToTopologicalString}
  together imply that the geometric realization 
  $|\mathbf{cosk}_3\exp(\mathfrak{so}_\mu)|$ is a model for $B \mathrm{String}$ in $\mathrm{Top}$
  $$
    |\exp(\mathfrak{so}_\mu)| \simeq B \mathrm{String}.
  $$
  With slight differences in the technical realization of
  $\exp(\mathfrak{g}_mu)$ this was originally shown in \cite{Henriques}, theorem 8.4.
  For the following discussion however the above perspective, 
  realizing $\mathbf{cosk}_3 \exp(\mathfrak{so}_\mu)$
  as a presentation of the homotopy fiber of the smooth first fractional Pontryagin class,
  def \ref{SmoothBString}, is crucial.
\end{remark}
We now discuss three equivalent but different models of the smooth String 2-group
by diffeological \emph{strict} 2-groups, hence by crossed modules of  
diffeological groups. See \cite{BCSS} for the general notion of strict 
Fr{\'e}chet-Lie 2-groups
and for discussion of one of the following models.
\begin{definition}
  For $(G_1 \to G_0)$ a crossed module of diffeological groups 
  (groups of concrete sheaves on $\mathrm{CartSp}$) write
  $$
    \Xi(G_1 \to G_0) \in [\mathrm{CartSp}^{\mathrm{op}}, \mathrm{sSet}]
  $$
  for the corresponding presheaf of simplicial groups.
\end{definition}
  There is an evident strictification of $\mathbf{B} \mathrm{String}_c$ from
  the proof of prop \ref{StringModelByExpgmu} given by the following definition.
  For the notion of thin homotopy classes of paths and disks see \cite{SWII}.
\begin{definition} \label{ThinPathModelForString} \label{Stringprime}
  Write
  $$
    \hat \Omega_{\mathrm{th}}\mathrm{Spin} \to P_{\mathrm{th}} \mathrm{Spin}
    \,,
  $$
  for the crossed module where
  \begin{itemize}
    \item
      $P_{\mathrm{th}}\mathrm{Spin}$ is the group whose elements are \emph{thin-homotopy}
      classes of based smooth paths in $G$ and whose product is
      obtained by rigidly translating one path so that its basepoint matches
      the other path's endpoint and then concatenating;
    \item
     $\hat \Omega_{\mathrm{th}}\mathrm{Spin}$ 
     is the group whose elements are equivalence classes of pairs $(d,x)$ consisting of
     \emph{thin homotopy}
     classes of disks $d : D^2 \to G$ in $G$ with sitting instant at a chosen point on the boundary, together with an element $x \in \mathbb{R}/\mathbb{Z}$.
     Two such pairs are taken to be equivalent if the 
     boundary of the disks has the same thin homotopy classes and if the labels
     $x$ and $x'$ differ, in $\mathbb{R}/\mathbb{Z}$, by the integral
     $\int_{D^3} f^* \mu(\theta)$ over any 3-ball $f : D^3 \to G$
     cobounding the two disks.
     The product is given by translating and then \emph{gluing} of disks at their basepoint 
	 (so that their boundary paths are being concatenated, hence multiplied in 
	 $P_{\mathrm{th}}\mathrm{Spin}$) and adding the labels in $\mathbb{R}/\mathbb{Z}$.

	 The map from $\hat \Omega_{\mathrm{th}}\mathrm{Spin}$ to 
	 $P_{\mathrm{th}}\mathrm{Spin}$ is given by sending a disk to its boundary path.
	 
	 The action of $P_{\mathrm{th}}\mathrm{Spin}$ on $\Omega_{\mathrm{th}}\mathrm{Spin}$
	 is given by whiskering a disk by a path and its reverse path.
  \end{itemize}
\end{definition}
\begin{proposition}
  Let 
  $$
    \mathbf{B}\mathrm{String}_c \to 
    \mathbf{B}\Xi(\hat \Omega_{\mathrm{th}}\mathrm{Spin} \to P_{\mathrm{th}} \mathrm{Spin})
  $$
  be the morphism that sends maps to $\mathrm{Spin}$ to their thin-homotopy class.
  This is a weak equivalence in $[\mathrm{CartSp}^{\mathrm{op}}, \mathrm{sSet}]_{\mathrm{proj}}$.
\end{proposition}
We produce now two equivalent crossed modules that are both obtained as central extensions
of path groups. This is joint with Danny Stevenson, based on results in 
\cite{MurrayStevenson}.

The following proposition is standard.
\begin{proposition}
  \label{2grp from cent}
    Let $H \subset G$ be a normal subgroup of some group $G$ and
    lat $\hat H \to H$ be a central extension of groups
    such that the conjugation action of $G$ on $H$ lifts to an
    automorphism action $\alpha : G \to \mathrm{Aut}(\hat H)$
    on the central extension.
    Then
    $
      (\hat H \to G)
    $
    with this $\alpha$ is a crossed module.
\end{proposition}
We construct classes of examples of this type from central extensions of path groups.
\begin{proposition}
  \label{2-grp from central ext}

  Let $G \subset \Gamma$ be a simply connected normal Lie subgroup
of a Lie group $\Gamma$.
Write $PG$ for the based path group of $G$ whose elements
  are smooth maps $[0,1] \to G$ starting at the neutral element and
  whose product is given by the pointwise product in $G$. Consider
  the complex with differential $d \pm \delta$ of simplicial forms
  on $\mathbf{B}G_{\mathrm{ch}}$.
 Let $(F,a,\beta)$ be a triple where

 \noindent i.     $a \in \Omega^1(G\times G)$ such that
      $
        \delta a = 0
      $;

 \noindent ii.       $F$ is a closed integral 2-form on $G$ such that
      $
        \delta F  = da
      $;

 \noindent iii.        $\beta : \Gamma \to \Omega^1(G)$ such that, for
 all $\gamma, \gamma_1, \gamma_2 \in \Gamma$,
       \begin{itemize}
         \item
           $\gamma^* F = F + d\beta_\gamma$;
         \item
           $(\gamma_1)^* \beta_{\gamma_2} - \beta_{\gamma_1 \gamma_2} + \beta_{\gamma_1} =
           0$;
         \item
           $a = \gamma^* a + \delta(\beta_\gamma)$;
         \item
            for all based paths $f : [0,1] \to G$,
           $f^*\beta_\gamma = (f,\gamma^{-1})^* a + (\gamma,f\gamma^{-1})^* a$.
       \end{itemize}
\noindent {\bf 1.}  Then the map
  $
    c :
    PG \times PG
    \to
    U(1)
  $
  given by
  $
    c : (f,g) \mapsto c_{f,g} := \exp\left(
      2 \pi i
        \int\limits_{0,1} (f,g)^* a
    \right)
  $
  is a group 2-cocycle leading to a central extension
  $\widehat PG = PG \ltimes U(1)$ with product
  $
    (\gamma_1, x_1) \cdot (\gamma_2, x_2) =
    (\gamma_1 \cdot \gamma_2, x_1 x_2 c_{\gamma_1,\gamma_2})
    $.

\noindent {\bf 2.}  Since $G$ is simply connected every loop in $G$
bounds a disk $D$.
  There is a normal subgroup $N \subset \widehat PG$ consisting of
  pairs $(\gamma,x)$ with $\gamma(1) = e$ and $x = \exp(2\pi i \int_D F)$
  for any disk $D$ in $G$ such that $\partial D = \gamma$.

\noindent {\bf 3.}  Finally,
  $
    \tilde G := \widehat {PG}/N
  $
  is a central extension of $G$ by $U(1)$ and the conjugation action
  of $\Gamma$ on $G$ lifts to $\tilde G$ by setting
  $
    \alpha(\gamma)(f,x)
    :=
    (\alpha(\gamma)(f), \; x \exp(\in_f \beta_\gamma))
  $
  such that
  $
    \mathrm{Cent}(G,\Gamma,F,a,\beta) :=
    (\tilde G \to \Gamma)
  $
  is a Lie crossed module and hence a strict Lie 2-group of the type in
  prop. \ref{2grp from cent}.
\end{proposition}
\proof
  All statements about the central extension $\hat G$ can be found in
  \cite{MurrayStevenson}. It remains to check that the action
  $\alpha : \Gamma \to \mathrm{Aut}(\tilde G)$ satisfies the required
  axioms of a crossed module, in particular the condition $\alpha(t(h))(h') =
  hh'h^{-1}$.
For this we have to show that
$$
\a(h(1))([f,z]) = [h,1][f,z]\left[h^{-1},\exp(-\int_{(h,h^{-1})}a)\right]\;,
$$
where $h$ denotes a based path in $P\cG$, so that $[h,1]$ represents
an element of $\tilde{\cG}$. By definition of the product in
$\tilde{\cG}$, the right hand side is equal to
$$
\left[hfh^{-1},z\exp\left(\int_{(h,f)}a + \int_{(hf,h^{-1})}a -
\int_{(h,h^{-1})}a\right)\right]\;.
$$
This is not exactly in the form we want, since the left hand side is
equal to
$
\left[h(1)fh(1)^{-1},z\exp(\int_f \b_h)\right]
$.
Therefore, we want to replace $hfh^{-1}$ with the homotopic path
$h(1)fh(1)^{-1}$.  An explicit homotopy between these two paths is
given by
$
H(s,t) = h((1-s)t + s)f(t)h((1-s)t + s)^{-1}
$.
Therefore, we have the equality
\begin{multline*}
\left[hfh^{-1},z\exp\left(\int_{(h,f)}a + \int_{(hf,h^{-1})}a - \int_{(h,h^{-1})}a\right)\right] \\
= \left[h(1)fh(1)^{-1},z\exp\left(\int_{(h,f)}a + \int_{(hf,h^{-1})}a -
\int_{(h,h^{-1})}a + \int H^*F\right)\right]\;.
\end{multline*}
Using the relation $\delta (F) = da$ and the fact that the pullback of
$F$ along the maps $[0,1]\times [0,1]\to G$, $(s,t)\mapsto h((1-s)t
+ s)$ vanish, we see that
$$
\int H^*F = \int_{(f,h(1)^{-1})}a - \int_{(f,h^{-1})}a +
\int_{(h,h^{-1})} a + \int_{(h(1),fh(1)^{-1})}a - \int_{(h,fh^{-1})}
a
\;.
$$
Therefore the sum of integrals
$$
\int_{(h,f)}a + \int_{(hf,h^{-1})}a - \int_{(h,h^{-1})}a + \int H^*F
$$
can be written as
$$
\int_{(h,f)}a + \int_{(hf,h^{-1})}a - \int_{(h,h^{-1})}a + \int_{(f,h(1)^{-1})}a
- \int_{(f,h^{-1})}a + \int_{(h,h^{-1})} a +
\int_{(h(1),fh(1)^{-1})}a - \int_{(h,fh^{-1})} a\;.
$$
Using the condition $\delta(a) = 0$, we see that this simplifies
down to
$
 \int_{(f,h(1)^{-1})}a
   + \int_{(h(1),fh(1)^{-1})}a
$.
Therefore, a sufficient condition to have a crossed module is the
equation
$
f^*\b_h = (f,h(1))^*a + (h(1),fh(1)^{-1})^*a \,.
$
\endofproof
\begin{proposition}
  \label{iso of central ext 2grp}
  Given triples $(F,a,\beta)$ and $(F',a',\beta')$ as above and
  given $b \in \Omega^1(G)$ such that
  \begin{equation}
    \label{first condition on b}
    F' = F + db
    \,,
  \end{equation}
  \begin{equation}
    \label{second condition on b}
    a' = a + \delta(b)
  \end{equation}
  and for all $\gamma \in \Gamma$
  \begin{equation}
    \label{third condition on b}
    \beta_\gamma + \gamma^* b = b + \beta'_\gamma\;,
  \end{equation}
  then there is an isomorphism
  $
    \mathrm{Cent}(G,\Gamma,F,a,\beta)
    \simeq
    \mathrm{Cent}(G,\Gamma,F',a',\beta')
    $\;.
\end{proposition}
In \cite{BCSS} the following special case of this general
construction was considered.
\begin{definition}
  \label{String BCSS}
  Let $G$ be a compact, simple and simply-connected Lie group with
  Lie algebra $\mathfrak{g}$. 
  Let $\langle \cdot, \cdot \rangle $
  be the Killing form invariant polynomial on $\mathfrak{g}$, normalized such that the
  Lie algebra 3-cocycle
  $
    \mu := \langle \cdot, [\cdot,\cdot]\rangle
  $
  extends left invariantly to a 3-form on $G$ which is the image in
  deRham cohomology of one of the two generators of $H^3(G,\mathbb{Z}) =
  \mathbb{Z}$.
  Let $\Omega G$ be the based loop group of $G$ whose elements are smooth
  maps $\gamma : [0,1] \to G$ with $\gamma(0) = \gamma(1) = e$ and
  whose product is by pointwise multiplication of such maps.
  Define $F \in \Omega^2(\Omega
  G)$, $a \in \Omega^1(\Omega G \times \Omega G)$ and $\beta : \Gamma \to \Omega^1(\Omega G)$
  \bea
F(\gamma,X,Y) &:=&
\int^{2\pi}_0 \langle X,Y'\rangle dt
\nonumber
\\
a(\gamma_1,\gamma_2,X_1,X_2) &:=&
\int^{2\pi}_0 \langle X_1,\dot{\c}_2\c_2^{-1}\rangle dt
\nonumber
\\
\b(p)(\gamma,X) &:=&
\int^{2\pi}_0 \langle p^{-1}\dot{p},X\rangle dt \nonumber \eea
This satisfies the axioms of prop. \ref{2-grp from central ext}
and we write
  $$
    \mathrm{String}_{\mathrm{BCSS}}(G) :=
    \Xi\mathrm{Cent}(\Omega G, PG, F, \a, \b)
  $$
for the corresponding diffeological strict 2-group. If $G = \mathrm{Spin}$ we write just
$\mathrm{String}_{\mathrm{BCS}}$ for this.
\end{definition}
There is a variant of this example, using another cocycle on loop groups that
was given in \cite{Mickelsson}.
\begin{definition}
  \label{StringMick}
  With all assumptions as in definition \ref{String BCSS}
   define now
\bea
F(\gamma,X,Y) &:=&
\frac{1}{2}
\int^{2\pi}_0 \langle \c^{-1}\dot{\c},[X,Y]\rangle dt
\nonumber
 \\
a(\gamma_1,\gamma_2,X_1,X_2) &:=&
\frac{1}{2}
\int^{2\pi}_0 \left( \langle
X_1,\dot{\c}_2\c_2^{-1}\rangle - \langle \c_1^{-1}\dot{\c}_1,\c_2
X_2\c_2^{-1}\rangle \right) dt \nonumber
 \\
\b(p)(\gamma,X) &:=&
\frac{1}{2}
\int^{2\pi}_0 \langle \c^{-1}p^{-1}\dot{p}\c +
p^{-1}\dot{p},X\rangle dt \nonumber \eea
This satisfies the axioms of proposition \ref{2-grp from central ext}
and we write
  $$
    \mathrm{String}_{\mathrm{Mick}}(G) :=
    \Xi \mathrm{Cent}(\Omega G, PG, F, \a, \b)
 $$
for the corresponding 2-group. 
If $G= \mathrm{Spin}$ we write just $\mathrm{String}_{\mathrm{Mick}}$ for this.
\end{definition}
\begin{proposition}
  \label{equiv StringBCSS StringMick}
  There is an isomorphism of 2-groups
  $
    \xymatrix{
      \mathrm{String}_{\mathrm{BCSS}}(G)
      \ar^{\simeq}[r]
      &
      \mathrm{String}_{\mathrm{Mick}}(G)
    }$.
\end{proposition}
\proof We show that $ b \in
  \Omega^1(\Omega G)§$
defined by
$
  b(\gamma,X) := \frac{1}{4\pi} \int_0^{2\pi} \langle  \gamma^{-1}\dot \gamma,
  X\rangle dt
$
satisfies the conditions of prop. \ref{iso of central ext
2grp} and hence defines the desired isomorphism.

\begin{itemize}
\item  \underline{Proof of equation \ref{first condition on b}}:
We calculate the exterior derivative
$db$.  To do this we first calculate the derivative $Xb(y)$: if
$\c_t = \c e^{tX}$ then to first order in $t$, $\c_t^{-1}\dot{\c}_t$
is equal to
$
\c^{-1}\dot{\c} + t[\c^{-1}\dot{\c},X] + tX'
$.
Therefore
$$
Xb(Y) =
\frac{1}{2}\int^{2\pi}_0 \left(\langle \c^{-1}\dot{\c},[X,Y]\rangle
+ \langle X',Y\rangle \right) dt\;.
$$
Hence $db$ is equal to
$$
\frac{1}{2} \int^{2\pi}_0 \left( \langle
\c^{-1}\dot{\c},[X,Y]\rangle + \langle X',Y\rangle + \langle
\c^{-1}\dot{c},[X,Y]\rangle - \langle Y',X\rangle - \langle
\c^{-1}\dot{\c},[X,Y]\rangle \right)\;,
$$
which is easily seen to simplify down to
$$
-
\int^{2\pi}_0 \langle X,Y\rangle dt +
\frac{1}{2} \int^{2\pi}_0 \langle \c^{-1}\dot{\c},[X,Y]\rangle dt
\,.
$$

\item  \underline{Proof of equation \ref{second condition on b}}: We get
\begin{multline*}
\frac{1}{2}\int^{2\pi}_0\left\{ \langle \c_2\dot{\c}_2^{-1},X_2\rangle - \langle \c_2^{-1}\c_1^{-1}\dot{\c}_1\c_2,\c_2^{-1}X_1\c_2\rangle - \langle \c_2^{-1}\c_1^{-1}\dot{\c}_1\c_2,X_2\rangle \right. \\
\left. - \langle \c_2^{-1}\dot{\c}_2,\c_2^{-1}X_1\c_2\rangle -
\langle \c_2^{-1}\dot{\c}_2,X_2\rangle + \langle
\c_1^{-1}\dot{\c}_1,X_1\rangle \right\} dt\;,
\end{multline*}
which is equal to
$$
\frac{1}{2} \int^{2\pi}_0\left\{ - \langle
\c_1^{-1}\dot{\c}_1,\c_2X_2\c_2^{-1}\rangle - \langle
\dot{\c}_2\c_2^{-1},X_1\rangle \right\}dt\;,
$$
which in turn equals
$$
\frac{1}{2}\int^{2\pi}_0 \left\{ \langle
X_1,\dot{\c}_2\c_2^{-1}\rangle - \langle
\c_1^{-1}\dot{\c}_1,\c_2X_2\c_2^{-1}\rangle \right\} dt -
\frac{1}{2\pi}\int^{2\pi}_0 \langle X_1,\dot{\c}_2\c_2^{-1}\rangle
dt\;.
$$

\item  \underline{Proof of equation \ref{third condition on b}}:
we get
\bea
p^*b(\c;\c X)  &=& b(p\c p^{-1};p\c p^{-1}(pXp^{-1}))
\nonumber
\\
&=&
\frac{1}{2}\int^{2\pi}_0 \langle p \c
p^{-1}(\dot{p}\c p^{-1} + p\dot{\c}p^{-1} - p\c
p^{-1}\dot{p}p^{-1},pXp^{-1}\rangle dt \nonumber
\\
 &=&
\frac{1}{2}
 \int^{2\pi}_0 \langle p\c^{-1} p^{-1}\dot{p}\c p^{-1} + p\c^{-1}\dot{\c}p^{-1} - \dot{p}p^{-1},pXp^{-1}\rangle dt
 \nonumber
 \\
 &=&
 \frac{1}{2}
 \int^{2\pi}_0 \langle \c^{-1}p^{-1}\dot{p}\c + \c^{-1}\dot{\c} - p^{-1}\dot{p},X\rangle dt
 \nonumber
 \\
 &=& b(\c,\c X) +
 \frac{1}{2}
 \int^{2\pi}_0 \langle \c^{-1}p^{-1}\dot{p}\c - p^{-1}\dot{p},X\rangle dt
 \nonumber
  \\
 &=& b(\c,\c X) +
 \frac{1}{2}
 \int^{2\pi}_0\langle \c^{-1}p^{-1}\dot{p}\c + p^{-1}\dot{p},X\rangle dt - \frac{1}{2\pi}\int^{2\pi}_0 \langle p^{-1}\dot{p},X\rangle dt
 \nonumber
\eea

\end{itemize}
The three conditions in proposition \ref{iso of central ext 2grp}
are satisfied and, therefore, the desired isomorphism is established.
\endofproof
\begin{proposition}
  \label{ana-equivalence String and Stingprimeprime}
  The strict 2-group $\mathrm{String}_{\mathrm{Mick}}$
  from definition \ref{StringMick}
  is equivalent to the
  model $\Xi (\hat \Omega_{\mathrm{th}}\mathrm{Spin} \to P_{\mathrm{th}})\mathrm{Spin}$ 
  from def. \ref{ThinPathModelForString}. 
\end{proposition}
\proof
  We define  a morphism  $F :
  \mathbf{B}\mathrm{String}_{\mathrm{Mick}} \to 
  \mathbf{B}\Xi(\hat \Omega_{\mathrm{th}}\mathrm{Spin} \to P_{\mathrm{th}})\mathrm{Spin}$.
  Its action on 1- and 2-morphisms is obvious:
  it sends parameterized paths $\gamma : [0,1] \to G = \mathrm{Spin}$.
  to their thin-homotopy equivalence class
  $$
    F : \gamma \mapsto [\gamma]
  $$
  and similarly for parameterized disks. On the $\mathbb{R}/\mathbb{Z}$-labels of
  these disks it acts as the identity.

  The subtle part is the compositor measuring the coherent failure of
  this assignment to respect composition:
  Define the components of this compositor for any two parameterized
  based paths $\gamma_1, \gamma_2 : [0,1] \to G$ with pointwise
  product $(\gamma_1 \cdot \gamma_2) : [0,1] \to G$ and images
  $[\gamma_1], [\gamma_2], [\gamma_1 \cdot \gamma_2]$ in thin homotopy classes
  to be represented by a parameterized disk in $G$
  $$
    \xymatrix{
      &
      \ar[dr]^{\gamma_2}
      \\
      \ar[ur]^{\gamma_1}
      \ar[rr]_{\gamma_1 \cdot \gamma_2}^{\ }="t"
      &&
      \ar@{=>}|{d_{\gamma_1,\gamma_2}} "t"+(0,7); "t"
    }
  $$
  equipped with a label $  x_{\gamma_1,\gamma_2}
  \in \mathbb{R}/\mathbb{Z}$ to be determined. Notice that this triangle is a diagram in
  $\Xi(\hat \Omega_{\mathrm{th}}\mathrm{Spin} \to P_{\mathrm{th}})\mathrm{Spin}$, so that composition 
  of 1-morphisms is concatenation
  $\gamma_1 \circ \gamma_2$ of paths. A suitable
  disk in $G$ is obtained via the map
  $$
    \xymatrix{
      D^2 \ar[r]^{a} & [0,1]^2 \ar[rrrr]^{(s_1,s_2) \mapsto \gamma_1(s_1) \cdot \gamma_2(s_2)} &&&& G
    }
    \,,
  $$
  where $a$ is a smooth surjection onto the triangle
  $\{(s_1,s_2) | s_2 \leq s_1\} \subset [0,1]^2$ such that the
  lower semi-circle of $\partial D^2 = S^1$ maps to the hypotenuse
  of this triangle.
  The coherence law for this compositor
  for all triples of parameterized paths $\gamma_1, \gamma_2, \gamma_3 : [0,1] \to G$
  amounts to the following:
  consider the map
  $$
    \xymatrix{
       D^3 \ar[r]^a & [0,1]^3 \ar[rrrrr]^{(s_1,s_2,s_3) \mapsto
          \gamma_1(s_1) \cdot \gamma_2(s_2) \cdot \gamma_3(s_3)} &&&&& G
    }\;,
  $$
  where the map $a$ is a smooth surjection onto the tetrahedron
  $
    \{(s_3 \leq s_2 \leq s_1 )\} \subset [0,1]^3
    $\,.
  Then the coherence condition
  $$
  \raisebox{26pt}{
      \xymatrix@R=30pt@C=30pt{
        \bullet
        \ar[rr]^{\gamma_2}_<{\ }="s1"
        &&
        \bullet
        \ar[dd]^{\gamma_3}
        \\
        \\
        \bullet
        \ar[uu]^{\gamma_1}
        \ar[uurr]|{\gamma_1 \cdot \gamma_2}^{\ }="t1"_{\ }="s2"
        \ar[rr]_{\gamma_1 \cdot \gamma_2 \cdot \gamma_3}^{\ }="t2"
        &&
        \bullet
        \ar@{=>}|{\left\{{{s_3 = 0} \atop { s_2 \leq s_1}}\right\}} "s1"; "t1"
        \ar@{=>}|{\left\{{{s_1 = s2} \atop { s_3 \leq s_1}}\right\}} "s2"; "t2"
      }
      }
      \hspace{7pt}
        =
      \hspace{7pt}
      \raisebox{26pt}{
      \xymatrix@R=30pt@C=30pt{
        \bullet
        \ar[rr]^{\gamma_2}_>{\ }="s1"
        \ar[ddrr]|{\gamma_2 \cdot \gamma_3}^{\ }="t1"_{\ }="s2"
        &&
        \bullet
        \ar[dd]_{\gamma_3}
        \\
        \\
        \bullet
        \ar[uu]^{\gamma_1}
        \ar[rr]_{\gamma_1 \cdot \gamma_2 \cdot \gamma_3}^{\ }="t2"
        &&
        \bullet
        \ar@{=>}|{\left\{{{s_1 = 1} \atop { s_3 \leq s_2 }}\right\}} "s1"; "t1"
        \ar@{=>}|{\left\{{{ s_2 = s_3 } \atop { s_2 \leq  s_1}}\right\}} "s2"; "t2"
      }
      }
  $$
  requires that the integral of the
  canonical 3-form on $G$ pulled back to the 3-ball along these maps
  accounts for the difference in the chosen labels of the disks involved:
  $$
    \int_{D^3} (b \circ a)^* \mu
    =
    \int_{s_3 \leq s_2 \leq s_1} (\gamma_1 \cdot \gamma_2 \cdot
    \gamma_3)^* \mu
    =
    x_{\gamma_1,\gamma_2} + x_{\gamma_1\cdot \gamma_2, \gamma_3}
    -
    x_{\gamma_1,\gamma_2 \cdot \gamma_3}
    -
    x_{\gamma_2,\gamma_3}
    \hspace{10pt}
    \in
    \mathbb{R}/\mathbb{Z}
    \,.
  $$
  (Notice that there is no further twist on the right hand side because whiskering in
  $\mathbf{B}\Xi(\hat \Omega_{\mathrm{th}}G \to P_{\mathrm{th}}G)$ does not affect the labels of the disks.)
  To solve this condition, we need a 2-form to integrate over the
  triangles. This is provided by the degree 2 component
of the simplicial realization
  $(\mu,\nu) \in \Omega^3(G)\times \Omega^2(G \times G)$ of the
  first Pontryagin form as a simplicial form on $\mathbf{B}G_{\mathrm{ch}}$:
  
   for $\gg$ a semisimple Lie algebra, the image of the normalized invariant
  bilinear polynomial $\langle \cdot, \cdot\rangle$ under the
  Chern-Weil map is
  $
    (\mu, \nu) \in \Omega^3(G) \times \Omega^2(G \times G)
  $
  with
  $$
    \mu := \langle \theta \wedge [\theta \wedge \theta]\rangle
  $$
  and
  $$
    \nu := \langle \theta_1 \wedge \bar \theta_2\rangle
    \,,
  $$
  where $\theta$ is the left-invariant canonical $\gg$-valued 1-form on
  $G$ and
  $\bar \theta$ the right-invariant one.

      So, define the label assigned by our compositor to the disks
  considered above by
  $$
    x_{\gamma_1,\gamma_2} :=
     \int_{s_2 \leq s_1} (\gamma_1,\gamma_2)^* \nu
     \,.
  $$
  To show that this assignment satisfies the above condition,
  use the closedness of $(\mu,\nu)$ in the complex of simplicial
  forms on $\mathbf{B}G_{\mathrm{ch}}$: $\delta \mu = d \nu$ and $\delta \nu = 0$.
  From this one obtains
  $$
    (\gamma_1 \cdot \gamma_2 \cdot \gamma_3)^* \mu
    =
    - d (\gamma_1 \cdot \gamma_2,\gamma_3)^*\nu
    =
    - d (\gamma_1 , \gamma_2 \cdot \gamma_3)^*\nu
  $$
  and
  $$
    (\gamma_1,\gamma_2 \cdot \gamma_3)^* \nu
    =
    (\gamma_1 \cdot \gamma_2 , \gamma_3)^* \nu
    +
    (\gamma_1, \gamma_2)^* \nu
    -
    (\gamma_2, \gamma_3)^* \nu
    \,.
  $$
  Now we compute as follows: Stokes' theorem gives
  $$
    \int\limits_{s_3 \leq s_2 \leq s1}
    (\gamma_1 \cdot \gamma_2 \cdot \gamma_3)^* \mu
    =
    \left(
     \int\limits_{s_3 = 0, s_2 \leq s_1}
     +
     \int\limits_{s_1 = s_2, s_3 \leq s_1}
     -
     \int\limits_{s_1 = 1, s_3 \leq s_2}
     -
     \int\limits_{s_2 = s_3, s_2 \leq s_1}
    \right)
    (\gamma_1 , \gamma_2 \cdot \gamma_3)^* \nu
    \,.
  $$
  The first integral is manifestly equal to $x_{\gamma_1,\gamma_2}$.
  The last integral is manifestly equal to $-x_{\gamma_1,\gamma_2 \cdot
  \gamma_3}$.
  For the remaining two integrals we rewrite
  $$
    \cdots =
    x_{\gamma_1,\gamma_2}
    -
    x_{\gamma_1, \gamma_2 \cdot \gamma_3}
    +
    \left(
      \int\limits_{s_1 = s_2, s_3 \leq s_1}
      -
      \int\limits_{s_1 = 1, s_3 \leq s_2}
    \right)
    \left(
      (\gamma_1 \cdot \gamma_2 , \gamma_3)^* \nu
      +
      (\gamma_1,\gamma_2)^* \nu
      -
      (\gamma_2,\gamma_3)^* \nu
    \right)
    \,.
  $$
  The first term in the integrand now manifestly yields
  $x_{\gamma_1\cdot \gamma_2, \gamma_3} - x_{\gamma_2,\gamma_3}$.
  The second integrand vanishes on the integration domain. The third
  integrand, finally, gives the same contribution under both
  integrals and thus drops out due to the relative sign.
  So in total what remains is indeed
  $$
    \cdots =
    x_{\gamma_1,\gamma_2}
    -
    x_{\gamma_1, \gamma_2 \cdot \gamma_3}
    +
    x_{\gamma_1\cdot \gamma_2, \gamma_3} - x_{\gamma_2,\gamma_3}
    \,.
  $$
  This establishes the coherence condition for
  the compositor.

\vspace{3mm}
  Finally we need to show that the compositor is
  compatible with the horizontal composition of 2-morphisms. We
  consider this in two steps, first
  for the horizontal composition of two 2-morphisms both
  starting at the identity 1-morphism in
  $\mathbf{B}\mathrm{String}_{\mathrm{Mick}}(G)$ -- this is the product in
  the loop group $\hat \Omega G$ centrally extended using Mickelsson's cocycle
   --
  then
  for the horizontal composition of an
  identity 2-morphism in $\mathbf{B}\mathrm{String}_{\mathrm{Mick}}(G)$ with a
  2-morphism starting at the identity 1-morphisms  -- this is the action of $PG$ on
  $\hat \Omega G$. These two cases then imply the general case.

  \begin{itemize}
    \item
      Let $(d_1,x_1)$ and $(d_2,x_2)$ represent two 2-morphisms in
      $\mathbf{B}\mathrm{String}_{\mathrm{Mick}}$ starting at the identity
      1-morphisms. So
      $$
        d_i : [0,1] \to \Omega G
      $$
      is a based path in loops in $G$ and $x_i \in U(1)$. We need to
      show that
      $$
        \xymatrix{
          \bullet
          \ar@/^1pc/[rr]^{\mathrm{Id}}_{\ }="s1"
          \ar@/_2pc/[rr]|{\gamma_1}^{\ }="t1"
          \ar@/_4pc/[rrrr]_{\gamma_1 \cdot \gamma_2}^{\ }="t3"
          &&
          \bullet
          \ar@/^1pc/[rr]^{\mathrm{Id}}_{\ }="s2"
          \ar@/_2pc/[rr]|{\gamma_2}^{\ }="t2"
          &&
          \bullet
          \ar@{=>}|{(d_1,x_1)} "s1"; "t1"
          \ar@{=>}|{(d_2,x_2)} "s2"; "t2"
          \ar@{=>}|{(d_{\gamma_1,\gamma_2},x_{\gamma_1,\gamma_2})} "t3"+(0,9); "t3"
        }
        \hspace{10pt}
          =
        \hspace{10pt}
        \xymatrix{
          \bullet
          \ar@/^1pc/[rr]^{\mathrm{Id}}_{\ }="s1"
          \ar@/_4pc/[rrrr]_{\gamma_1 \cdot \gamma_2}^{\ }="t3"
          &&
          \bullet
          \ar@/^1pc/[rr]^{\mathrm{Id}}_{\ }="s2"
          &&
          \bullet
          \ar@{=>}|<<<<<<{(d_1\cdot d_2, x_1 + x_2 + \rho(d_1,d_2))} "t3"+(0,13); "t3"
        }
      $$
      as a pasting diagram equation in 
      $\mathbf{B}\Xi(\hat \Omega_{\mathrm{th}}G \to P_{\mathrm{th}}G)$. Here on the
      left we have gluing of disks in $G$ along their boundaries and addition of their labels,
      while on the right we have the pointwise product
      from definition \ref{StringMick} of labeled disks as representing the
      product of elements $\hat \Omega G$.

     There is an obvious 3-ball interpolating between the disk on
     the left and on the right of the above equation:
     $$
       (\{s_2 \leq s_1\} \subset [0,1]^3) \to G
     $$
     $$
       (s_1,s_2,t) \mapsto (d_1(t,s_1) \cdot d_2(t,s_2))
     $$
      $$
        \xymatrix{
          \bullet
          \ar@/^1pc/[rr]^{\mathrm{Id}}_{\ }="s1"
          \ar@/_2pc/[rr]_{\gamma_1}^{\ }="t1"
          \ar@/_4pc/[rrrr]_{\gamma_1 \cdot \gamma_2}^{\ }="t3"
          &&
          \bullet
          \ar@/^1pc/[rr]^{\mathrm{Id}}_{\ }="s2"
          \ar@/_2pc/[rr]_{\gamma_2}^{\ }="t2"
          &&
          \bullet
          \ar@{=>}|{\{s_2 = 0\}} "s1"; "t1"
          \ar@{=>}|{\{s_1 = 0\}} "s2"; "t2"
          \ar@{=>}|{\{t = 1\}} "t3"+(0,9); "t3"
        }
        \hspace{10pt}
        \,,
        \hspace{10pt}
        \xymatrix{
          \bullet
          \ar@/^1pc/[rr]^{\mathrm{Id}}_{\ }="s1"
          \ar@/_4pc/[rrrr]_{\gamma_1 \cdot \gamma_2}^{\ }="t3"
          &&
          \bullet
          \ar@/^1pc/[rr]^{\mathrm{Id}}_{\ }="s2"
          &&
          \bullet
          \ar@{=>}|{\{s_1 = s_2\}} "t3"+(0,13); "t3"
        }
        \,.
      $$
      The compositor property demands that the integral of the
      canonical 3-form over this ball accounts for the difference
      between $x_{\gamma_1,\gamma_2}$ and $\rho(\gamma_1,\gamma_2)$
      $$
         \rho(d_1,d_2)
         =
        \int\limits_{{s_2 \leq s_1}\atop{ 0 \leq t \leq 1}}
        (d_1\cdot d_2)^* \mu
        +
        \int\limits_{s_2 \leq s_1} (\gamma_1,\gamma_2)^* \nu
        \,.
      $$
      Now use again the relation between $\mu$ and $d \nu$ to
      rewrite this as
      $$
        \cdots =
        \int\limits_{{s_2 \leq s_1}\atop{ 0 \leq t \leq 1}}
        \left(
          (d_1)^* \mu
          +
          (d_2)^* \mu
          -
          d (d_1,d_2)^* \nu
        \right)
        +
        \int\limits_{s_2 \leq s_1} (\gamma_1,\gamma_2)^* \nu
        \,.
      $$
      The first two integrands vanish. The third one leads to
      boundary integrals
      $$
        \cdots =
        -
        \left(
          \int\limits_{s_2 = 0}
          +
          \int\limits_{s_1 = 0}
        \right)
        (d_1,d_2)^* \nu
        -
        \int\limits_{{t=1}\atop{s_2 \leq s_1}}
        (d_1,d_2)^* \nu
        +
        \int\limits_{s_2 \leq s_1} (\gamma_1,\gamma_2)^* \nu
        +
        \int\limits_{{0 \leq t \leq 1}\atop{s_1 = s_2}}
        (d_1,d_2)^* \nu
        \,.
      $$
      The first two integrands vanish on their integration domain.
      The third integral cancels with the fourth one. The remaining
      fifth one is indeed the 2-cocycle on $P \Omega G$ 
      from definition \ref{StringMick}.
     \item
       The second case is entirely analogous: for $\gamma_1$ a path
       and $(d_2,x_2)$ a centrally extended loop we need to show
       that
      $$
        \xymatrix{
          \bullet
          \ar@/^1pc/[rr]^{\gamma_1}_{\ }="s1"
          \ar@/_2pc/[rr]|{\gamma_1}^{\ }="t1"
          \ar@/_4pc/[rrrr]_{\gamma_1 \cdot \gamma_2}^{\ }="t3"
          &&
          \bullet
          \ar@/^1pc/[rr]^{\mathrm{Id}}_{\ }="s2"
          \ar@/_2pc/[rr]|{\gamma_2}^{\ }="t2"
          &&
          \bullet
          \ar@{=>}|{\mathrm{Id}} "s1"; "t1"
          \ar@{=>}|{(d_2,x_2)} "s2"; "t2"
          \ar@{=>}|{(d_{\gamma_1,\gamma_2},x_{\gamma_1,\gamma_2})} "t3"+(0,9); "t3"
        }
        \hspace{10pt}
          =
        \hspace{10pt}
        \xymatrix{
          \bullet
          \ar@/^1pc/[rr]^{\gamma_1}_{\ }="s1"
          \ar@/_4pc/[rrrr]_{\gamma_1 \cdot \gamma_2}^{\ }="t3"
          &&
          \bullet
          \ar@/^1pc/[rr]^{\mathrm{Id}}_{\ }="s2"
          &&
          \bullet
          \ar@{=>}|<<<<<<{(\gamma_1\cdot d_2, x_1 + x_2 + \lambda(\gamma_1,d_2))} "t3"+(0,13); "t3"
        }
        \,.
      $$
     There is an obvious 3-ball interpolating between the disk on
     the left and on the right of the above equation:
     $$
       (\{s_2 \leq s_1\} \subset [0,1]^3) \to G
     $$
     $$
       (s_1,s_2,t) \mapsto (\gamma_1(s_1) \cdot d_2(t,s_2))
     $$
      $$
        \xymatrix{
          \bullet
          \ar@/^1pc/[rr]^{\gamma_1}_{\ }="s1"
          \ar@/_2pc/[rr]_{\gamma_1}^{\ }="t1"
          \ar@/_4pc/[rrrr]_{\gamma_1 \cdot \gamma_2}^{\ }="t3"
          &&
          \bullet
          \ar@/^1pc/[rr]^{\mathrm{Id}}_{\ }="s2"
          \ar@/_2pc/[rr]_{\gamma_2}^{\ }="t2"
          &&
          \bullet
          \ar@{=>}|{\{s_2 = 0\}} "s1"; "t1"
          \ar@{=>}|{\{s_1 = 0\}} "s2"; "t2"
          \ar@{=>}|{\{t = 1\}} "t3"+(0,9); "t3"
        }
        \hspace{10pt}
        \,,
        \hspace{10pt}
        \xymatrix{
          \bullet
          \ar@/^1pc/[rr]^{\gamma_1}_{\ }="s1"
          \ar@/_4pc/[rrrr]_{\gamma_1 \cdot \gamma_2}^{\ }="t3"
          &&
          \bullet
          \ar@/^1pc/[rr]^{\mathrm{Id}}_{\ }="s2"
          &&
          \bullet
          \ar@{=>}|{\{s_1 = s_2\}} "t3"+(0,13); "t3"
        }
        \,.
      $$

      The compositor property demands that the integral of the
      canonical 3-form over this ball accounts for the difference
      between $x_{\gamma_1,\gamma_2}$ and $\lambda(\gamma_1,\gamma_2)$
      $$
         \lambda(\gamma_1,d_2)
         =
        \int\limits_{{s_2 \leq s_1}\atop{ 0 \leq t \leq 1}}
        (d_1\cdot d_2)^* \mu
        +
        \int\limits_{s_2 \leq s_1} (\gamma_1,\gamma_2)^* \nu
        \,.
      $$
      This is essentially the same computation as before, so that
      the result is
      $$
        \cdots =
        \int\limits_{{0\leq t \leq 1}\atop{s_1 = s_2}}
        (\gamma_1, d_2)^* \nu
        \,.
      $$
      This is indeed the quantity from definition \ref{StringMick}.
      \end{itemize}
      \endofproof
Applied to the case $G = \mathrm{Spin}$ in summary 
this shows that all these strict smooth 2-groups are indeed presentations
of the abstractly defined smooth String 2-group from def. \ref{SmoothBString}.
\begin{theorem}
 \label{EquivaleneOfString2GroupModels}
 We have equivalences of smooth 2-groups
$$
  \mathrm{String}
  \simeq
  \Omega \mathbf{cosk}_3\exp(\mathfrak{so}_\mu)
  \simeq
  \mathrm{String}_{\mathrm{BCSS}}
  \simeq
  \mathrm{String}_{\mathrm{Mick}}
  \,.
$$
\end{theorem}
Notice that all the models on the right are degreewise diffeological and in 
fact Fr{\'e}chet, but not degreewise finite dimensional. This means that neither of these
models is a differentiable stack or Lie groupoid in the traditional sense, even though
they are perfectly good models for objects in $\mathrm{Smooth}\infty \mathrm{Grpd}$. 
Some authors found this to be a deficiency. Motivated by this 
it has been shown in \cite{Schommer-Pries} that there exist finite dimensional models
of the smooth String-group. Observe however the following:
\begin{enumerate}
  \item If one allows arbitrary disjoint unions of finite dimensional manifolds, 
  then by prop.  \ref{DegreewiseRepresentability}
    \emph{every} object in $\mathrm{Smooth}\infty \mathrm{Grpd}$ has a presentation
    by a simplicial object that is degreewise of this form, 
    even a presentation which is degreewise a union of 
    just Cartesian spaces.
  \item
    Contrary to what one might expect, it is not the 
    degreewise finite dimensional models
    that seem to lend themselves most directly to differential refinements and differential 
    geometric computations with objects in $\mathrm{Smooth}\infty \mathrm{Grpd}$,
    but the models of the form $\mathbf{cosk}_n \exp(\mathfrak{g})$. 
    See also the discussion in \ref{HigherSpinStructure} below.
\end{enumerate}

\subsubsection{Smooth fivebrane structure and the $\mathrm{Fivebrane}$-6-group}
\label{FivebraneSixGroup}

We now climb up one more step in the smooth Whitehead tower of the orthogonal group,
to find a smooth and differential refinement of the \emph{Fivebrane group} \cite{SSSII}.
\begin{proposition}
  \label{SecondFracPontryagin}
  Pulled back along $B \mathrm{String} \to B O$ the second Pontryagin class
  is 6 times a generator $\frac{1}{6}p_2$ of 
  $H^8(B \mathrm{String}, \mathbb{Z}) \simeq \mathbb{Z}$:
  $$
    \xymatrix{
      B \mathrm{String} \ar[r]^{\frac{1}{6}p_2} \ar[d] & B^8 \mathbb{Z} \ar[d]^{\cdot 6}
      \\
      B \mathrm{Spin} \ar[r]^{p_2} & B^8 \mathbb{Z}
    }
    \,.
  $$
\end{proposition}
This is due to \cite{Bott}.  We call $\frac{1}{6}p_2$ the 
\emph{second fractional Pontryagin class} \index{Pontryagin class!second fractional!bare}
\index{characteristic class!Pontryagin class!second fractional}.
\begin{definition}
  Write $B \mathrm{Fivebrane}$ for the homotopy fiber of the second fractional 
  Pontryagin class in $\mathrm{Top} \simeq \infty \mathrm{Grpd}$
  $$
    \xymatrix{
      B \mathrm{Fivebrane} \ar[r] \ar[d] & {*} \ar[d]
      \\
      B \mathrm{String} \ar[r]^{\frac{1}{6}p_2} & B^8 \mathbb{Z}
    }
    \,.
  $$  
  Write
  $$
    \mathrm{Fivebrane} := \Omega B \mathrm{Fivebrane}
  $$
  for its loop space, the topological \emph{fivebrane $\infty$-group}\index{group!bare fivebrane $\infty$-group}.
\end{definition}
This is the next step in the topological Whitehead tower of $O$ after $\mathrm{String}$, 
often denoted $O\langle 7\rangle$.
For a discussion of its role in the physics of super-Fivebranes that gives it its name 
here in analogy to $\mathrm{String} = O\langle 3\rangle$ 
see \cite{SSSII}. See also \cite{DHH}, around remark 2.8. We now construct
smooth and then differential refinements of this object.
\begin{theorem} 
  \label{SecondFractionalDifferentialPontryagin}
  \index{Pontryagin class!second fractional!smooth}
  \index{Pontryagin class!second fractional!differential}
The image under Lie integration, 
\ref{SmoothStrucLieAlgebras}, 
of the canonical Lie algebra 7-cocycle
$$
  \mu_7 = \langle -,[-,-],[-,-], [-,-]\rangle 
    : 
  \mathfrak{so}_{\mu_3} \to b^6 \mathbb{R}
$$
on the string Lie 2-algebra $\mathfrak{so}_{\mu_3}$, def. \ref{StringLie2Algebra},
is a morphism in $\mathrm{Smooth}\infty \mathrm{Grpd}$ of the form
$$
  \frac{1}{6} \mathbf{p}_2 : 
  \mathbf{B}\mathrm{String}
    \to
  \mathbf{B}^7 U(1)
$$
whose image under the fundamental $\infty$-groupoid $\infty$-functor/ geometric realization, 
\ref{ETopStrucHomotopy}, 
$\Pi : \mathrm{Smooth} \infty \mathrm{Grpd} \to \infty \mathrm{Grpd}$ 
is the ordinary second fractional Pontryagin class
$\frac{1}{6}p_2 : B \mathrm{String} \to B^8 \mathbb{Z}$
in $\mathrm{Top}$. 
We call $\frac{1}{6}{\hat {\mathbf{p}_2}} := \exp(\mu_7)$ the
\emph{second smooth fractional Pontryagin class}

Moreover, the corresponding 
refined differential characteristic cocycle, \ref{SmoothStrucInfChernWeil},  
$$
  \frac{1}{6}\hat {\mathbf{p}}_2 : 
  \mathbf{H}_{\mathrm{conn}}(-,\mathbf{B}\mathrm{Spin})
  \to
  \mathbf{H}_{\mathrm{diff}}(-, \mathbf{B}^7 U(1))
  \,,
$$
induces in cohomology the ordinary refined Chern-Weil homomorphism 
\cite{HopkinsSinger} 
$$
  [\frac{1}{6}\hat {\mathbf{p}}_2] : 
  H^1_{\mathrm{Smooth}}(X,\mathrm{String}) \to H_{\mathrm{diff}}^4(X)
$$
of $\langle -,-,-,-\rangle $ restricted to those
$\mathrm{Spin}$-principal bundles $P$ that have $\mathrm{String}$-lifts
$$
  [P]
   \in
  H^1_{\mathrm{smooth}}(X, \mathrm{String})
  \hookrightarrow
  H^1_{\mathrm{smooth}}(X, \mathrm{Spin})
  \,.
$$
\end{theorem}
\proof
  This is shown in \cite{FSS}.
  The proof is analogous to that of prop. \ref{FirstFractionalDifferentialPontrjagin}.
\endofproof
\begin{definition}
  Write $\mathbf{B}\mathrm{Fivebrane}$ for the homotopy fiber in $\mathrm{Smooth}\infty \mathrm{Grpd}$
  of the smooth refinement of the second fractional Pontryagin class, 
  prop. \ref{SecondFractionalDifferentialPontryagin}:
  $$
    \xymatrix{
      \mathbf{B}\mathrm{Fivebrane} \ar[r] \ar[d] & {*} \ar[d]
      \\
      \mathbf{B}\mathrm{String}
       \ar[r]^{\frac{1}{6}\mathbf{p}_2}
      &
      \mathbf{B}^7 U(1)
    }
    \,.
  $$
  We say its loop space object is the \emph{smooth fivebrane 6-group}\index{group!smooth fivebrane 6-group}
  $$
    \mathrm{Fivebrane}_{\mathrm{smooth}} := \Omega \mathbf{B} \mathrm{Fivebrane}
    \,.
  $$
\end{definition}
This has been considered in \cite{SSSIII}. Similar discussion as for the smooth
String 2-group applies.

\newpage

\subsubsection{Higher $\mathrm{Spin}^c$-structures} 
\label{HigherSpinCStructures}

In \ref{FractionalClasses} we saw that the classical 
extension 
$$
  \mathbb{Z}_2 \to \mathrm{Spin}(n) \to \mathrm{SO}(n)
$$
is only the first step in a tower of \emph{smooth} higher
spin groups.

There is another classical extension of $\mathrm{SO}(n)$, 
not by $\mathbb{Z}_2$ but by the circle group \cite{LawsonMichelson}:
$$
  U(1) \to \mathrm{Spin}^c(n) \to \mathrm{SO}(n)
  \,.
$$
Here we discuss higher smooth analogs of this construction.

This section draws form \cite{FiorenzaSatiSchreiberI}.

\medskip
We find below that $\mathrm{Spin}^c$ is a special case of the
following simple general notion, that turns out to be useful
to identify and equip with a name.
\begin{definition}
  \label{HatGC}
  Let $\mathbf{H}$ be an $\infty$-topos, $G \in \infty \mathrm{Grp}(\mathbf{H})$
  an $\infty$-group object, let $A$ be an abelian group object and let
  $$
    \mathbf{p} : \mathbf{B}G \to \mathbf{B}^{n+1} A
  $$
  be a characteristic map. Write $\hat G \to G$ for the extension
  classified by $\mathbf{p}$, exhibited by a fiber sequence
  $$
    \mathbf{B}^n A \to \hat G \to G
  $$
  in $\mathbf{H}$. Then for $H \in \infty \mathrm{Grp}(\mathbf{H})$,
  any other $\infty$-group with characteristic map of the same form
  $$
    \mathbf{c} : \mathbf{B}H \to \mathbf{B}^{n+1}A
  $$
  we write 
  $$
    \hat {G}^{\mathbf{c}} 
	 := 
	 \Omega\left(\mathbf{B}G {}_{\mathbf{p}}\times_{\mathbf{c}} \mathbf{B}H\right)
	\in \infty \mathrm{Grp}(\mathbf{H})
  $$
  for the loop space object of the $\infty$-pullback
  $$
    \xymatrix{
	  \mathbf{B}\hat G^{\mathbf{c}}
	  \ar[r]
      \ar[d]	  
	  &
	  \mathbf{B} H
	  \ar[d]^{\mathbf{c}}
	  \\
	  \mathbf{B}G
	  \ar[r]^{\mathbf{p}}
	  &
	  \mathbf{B}^{n+1} A
	}
	\,.
  $$
\end{definition}
\begin{remark}
  Since the Eilenberg-MacLane object $\mathbf{B}^{n+1} A$ is tself an 
  $\infty$-group object,
  by the Mayer-Vietoris fiber sequence
  in $\mathbf{H}$, prop. \ref{MayerVietorisFiberSequence}, 
  the object $\mathbf{B}\hat G^{\mathbf{c}}$
  is equivalently the homotopy fiber of the 
  difference $(\mathbf{p}- \mathbf{c})$ of the two characteristic maps
  $$
    \raisebox{20pt}{
    \xymatrix{
	  \mathbf{B}\hat G^{\mathbf{c}}
	  \ar[rr]
	  \ar[d]
	  &&
	  {*}
	  \ar[d]
	  \\
	  \mathbf{B}G \times \mathbf{B}H
	  \ar[rr]^{\mathbf{p}- \mathbf{c}}
	  &&
	  \mathbf{B}^n A
	}
	}
	\,.
  $$
\end{remark}

\subsubsection{$\mathrm{Spin}^c$ as a homotopy fiber product in $\mathrm{Smooth}\infty \mathrm{Grpd}$}
\label{SpinCAsHomotopyFiberProduct}
\index{higher $\mathrm{Spin}^c$-structures}

A classical definition of $\mathrm{Spin}^c$ is the following
(for instance \cite{LawsonMichelson}).
\begin{definition}
 \label{SpinCAsLieGroup}
For each $n \in \mathbb{N}$ the Lie group $\mathrm{Spin}^c(n)$
is the fiber product of Lie groups
$$
  \begin{aligned}
    \mathrm{Spin}^c(n) 
	 &:= \mathrm{Spin}(n) \times_{\mathbb{Z}_2} U(1)
	 \\
	 & = (\mathrm{Spin}(n) \times U(1))/\mathbb{Z}_2
	 \,,
  \end{aligned}
$$
  where the quotient is by the canonical subgroup embeddings.
\end{definition}
We observe now that in the context of $\mathrm{Smooth}\infty\mathrm{Grpd}$
this Lie group has the following intrinsic characterization.
\begin{proposition}
  \label{SpinCAsHomotopyFiberProductOfU1AndSO}
  \index{higher $\mathrm{Spin}^c$-structures!$\mathrm{Spin}^c$ itself as an example}
  In $\mathrm{Smooth}\infty \mathrm{Grpd}$
  we have an $\infty$-pullback diagram of the form
  $$
    \raisebox{20pt}{
    \xymatrix{
	  \mathbf{B} \mathrm{Spin}^c 
	  \ar[r]
	  \ar[d]
	  &
	  \mathbf{B}U(1)
	  \ar[d]^{\mathbf{c}_1 \mathrm{mod} 2}
	  \\
	  \mathbf{B}\mathrm{SO}
	  \ar[r]^{\mathbf{w}_2}
	  &
	  \mathbf{B}^2 \mathbb{Z}_2
	}
	}
	\,,
  $$
  where the right morphism is the smooth universal first Chern class,
  example \ref{DeterminantLineBundle}, composed with the mod-2 reduction
  $\mathbf{B}\mathbb{Z} \to \mathbf{B}\mathbb{Z}_2$,  
  and where $\mathbf{w}_2$ is the smooth universal
  second Stiefel-Whitney class, example \ref{SecondStiefelWhineyClass}.
\end{proposition}
\proof
  By the discussion at these examples, these universal smooth classes
  are represented by spans of simplicial presheaves
  $$
    \xymatrix{
	  \mathbf{B}(\mathbb{Z} \to \mathbb{R})_{\rm ch}
	  \ar[r]^{\mathbf{c}_1}
	  \ar[d]^{\simeq}
	  &
	  \mathbf{B}(\mathbb{Z} \to 1)_{\rm ch}
	  \ar@{=}[r]
	  &
	  \mathbf{B}^2 \mathbb{Z}
	  \\
	  \mathbf{B} U(1)_{\rm ch}
	}
  $$
  and
  $$
    \xymatrix{ 
	  \mathbf{B}(\mathbb{Z}_2 \to \mathrm{Spin})_{\rm ch}
	  \ar[r]
	  \ar[d]^{\simeq}
	  &
	  \mathbf{B}(\mathbb{Z}_2 \to 1)_{\rm ch}
	  \ar@{=}[r]
	  &
	  \mathbf{B}^2 (\mathbb{Z}_2)_{\rm ch}
	  \\
	  \mathbf{B}\mathrm{SO}_{\rm ch}
	}
	\,.
  $$
  Here both horizontal morphism are fibrations in 
  $[\mathrm{CartSp}_{\mathrm{smooth}}^{\mathrm{op}}, \mathrm{sSet}]_{\mathrm{proj}}$. Therefore by prop. \ref{FiniteHomotopyLimitsInPresheaves}
  the $\infty$-pullback in question is given by the ordinary fiber 
  product of these two morphisms. This is
  $$
    \xymatrix{
	  \mathbf{B}(\mathbb{Z} \to \mathrm{Spin} \times \mathbb{R})_{\rm ch}
	  \ar[r]
	  \ar[d]
	  &
	  \mathbf{B}(\mathbb{Z} \to \mathbb{R})_{\rm ch}
	  \ar[d]
	  \\
	  \mathbf{B}(\mathbb{Z} \stackrel{\mathrm{mod} 2}{\to} \mathrm{Spin})_{\rm ch}
	  \ar[d]
	  \ar[r] & 
	  \mathbf{B}(\mathbb{Z}\to 1)_{\rm ch}
	  \ar[d]
	  \\
	  \mathbf{B}(\mathbb{Z}_2 \to \mathrm{Spin})_{\rm ch}
	  \ar[r]
	  &
	  \mathbf{B}(\mathbb{Z}_2 \to 1)_{\rm ch}
	  \
	}
	\,,
  $$
  where the crossed module $(\mathbb{Z} \stackrel{\partial}{\to} \mathrm{Spin} \times \mathbb{R})$
  is given by
  $$
    \partial : n \mapsto (n~ \mathrm{mod} ~2, n)
	\,.
  $$
  Since this is a monomorphism, including (over the neutral element) 
  the fiber of a locally trivial bundle we have an equivalence
  $$
    \mathbf{B}(\mathbb{Z} \to \mathrm{Spin} \times \mathbb{R})
	\stackrel{\simeq}{\to}
    \mathbf{B}(\mathbb{Z}_2 \to \mathrm{Spin} \times U(1))
	\stackrel{\simeq}{\to}
	\mathbf{B} (\mathrm{Spin} \times_{\mathbb{Z}_2} U(1))
  $$
  in $[\mathrm{CartSp}^{\mathrm{op}}, \mathrm{sSet}]_{\mathrm{proj}}$.
  On the right is, by def. \ref{SpinCAsLieGroup}, the delooping of
  $\mathrm{Spin}^c$.
\endofproof
\begin{remark}
  Therefore by def. \ref{HatGC} we have
  $$
    \mathrm{Spin}^c \simeq
	\mathrm{Spin}^{\mathbf{c}_1 \mathrm{mod} 2}
	\,,
  $$
  which is the very motivation for the notation in that
definition.  
\end{remark}
\begin{remark}
  From prop. \ref{SpinCAsHomotopyFiberProductOfU1AndSO}
  we obtain the following characterization of $\mathrm{Spin}^c$-structures
  in $\mathbf{H} = \mathrm{Smooth}\infty\mathrm{Grpd}$
  over a smooth manifold expressed in terms of 
  traditional {\v C}ech cohomology, \ref{CechCohomology}.
  
  For $X \in \mathrm{SmthMfd}$, the fact that $\mathbf{H}(X,-)$
preserves $\infty$-limits implies from prop. \ref{SpinCAsHomotopyFiberProductOfU1AndSO} 
 that we have an $\infty$-pullback of cocycle $\infty$-groupoids
    $$
    \raisebox{20pt}{
    \xymatrix{
	  \mathbf{H}(X,\mathbf\mathbf{B} \mathrm{Spin}^c) 
	  \ar[r]
	  \ar[d]
	  &
	  \mathbf{H}(X,\mathbf{B}U(1))
	  \ar[d]^{\mathbf{c}_1 \mathrm{mod} 2}
	  \\
	  \mathbf{H}(X,\mathbf{B}\mathrm{SO})
	  \ar[r]^{\mathbf{w}_2}
	  &
	  \mathbf{H}(X,\mathbf{B}^2 \mathbb{Z}_2)
	}
	}
	\,.
  $$
  Picking any choice of differentiably good open cover $\{U_i \to X\}$ of $X$
  and using the standard presentation of the coeffcient moduli stacks 
  appearing here
  by sheaves of groupoids as discussed in \ref{SmoothStrucCohesiveInfiniGroups},
  each of the four $\infty$-groupoids appearing here is canonically 
  identified with the groupoid (or 2-groupoid in the bottom right) of 
  {\v C}ech cocycles and {\v C}ech coboundaries with respect to the given cover
  and with coefficients in the given group. Moreover, in this 
  presentation the right vertical morphism of the above diagram is clearly a fibration,
  and so by prop. \ref{ConstructionOfHomotopyLimits} the ordinary pullback
  of these groupoids 
  is already the correct $\infty$-pullback, hence is the
  groupoid $\mathbf{H}(X, \mathbf{B}\mathrm{Spin}^c)$
  of $\mathrm{Spin}^c$-structure on $X$. So we read off from the diagram 
  and the construction in the above proof:
  given a  {\v C}ech 1-cocycle for an $\mathrm{SO}$-structure on $X$ the corresponding
  $\mathrm{Spin}^c$-structure is a lift to a $(\mathbb{Z} \to \mathbb{R})$-valued 
  {\v C}ech cocycle
  of the $\mathbb{Z}_2$-valued {\v C}ech 2-cocycle that represents the
  second Stiefel-Whitney class, as described in \ref{SecondStiefelWhineyClass}, through the 
  evident projection $(\mathbb{Z} \to \mathbb{R}) \to (\mathbb{Z}_2 \to *)$
that by example. \ref{DeterminantLineBundle} presents the universal first Chern class.  
\end{remark}

\subsubsection{Smooth $\mathrm{String}^{\mathbf{c}_2}$ }
\label{SmoothStringC2}
\index{higher $\mathrm{Spin}^c$-structures!$\mathrm{String}^{\mathbf{c}_2}$}

We consider smooth 2-groups of the form $\mathrm{String}^{\mathbf{c}}$,
according to def. \ref{HatGC}, where 
$\mathbf{B} U(1) \to \mathrm{String} \to \mathrm{Spin}$
in $\mathrm{Smooth}\infty \mathrm{Grpd}$
is the smooth String-2-group extension 
of the $\mathrm{Spin}$-group from def. \ref{SmoothBString}.

In \cite{Sati10Twist} the following notion is introduced.
\begin{definition}
  \index{group!$\mathrm{String}^{c}$}
 Let 
 $$
   p_1^c : B \mathrm{Spin}^c \to B \mathrm{Spin} \stackrel{\frac{1}{2}p_1}{\to}
   K(\mathbb{Z},4)
 $$
 in $\mathrm{Top} \simeq \infty \mathrm{Grpd}$,
 where the first map is induced on classifying spaces
 by the defining projection, def. \ref{SpinCAsLieGroup},
 and where the second represents the fractional first Pontryagin class
 from prop. \ref{FirstFracPontryagin}.
 
 Then write $\mathrm{String}^c$ for the topological group, 
 well defined up to weak homotopy equivalence,
 that models the loop space of the homotopy pullback
 $$
   \xymatrix{
     B \mathrm{String}^c
	 \ar[r]
	 \ar[d]
	 &
	 (B U(1)) \times (B U(1))
	 \ar[d]^{c_1 \cup c_1}
	 \\
	 B \mathrm{Spin}^c
	 \ar[r]^{p_1^c}
	 &
	 K(\mathbb{Z},4)
   }
 $$
 in $\mathrm{Top}$.
\end{definition}
This construction, and the role it plays in \cite{Sati10Twist},
is evidently an example of general structure of def. \ref{HatGC},
the notation of which is motivated from this example.
We consider now smooth and differential refinements of such objects.

To that end, recall from theorem. \ref{FirstFractionalDifferentialPontrjagin}
the smooth refinement of the first fractional Pontryagin class
$$
  \frac{1}{2}\mathbf{p}_1 :
  \mathbf{B}\mathrm{Spin}
  \to 
  \mathbf{B}^3 U(1)
$$
and from def. \ref{SmoothBString} the defining fiber sequence
$$
  \xymatrix{
    \mathbf{B}\mathrm{String} 
     \ar[r]
     &	 
    \mathbf{B} \mathrm{Spin}
     \ar[r]^{\frac{1}{2}\mathbf{p}_1}
	 &
    \mathbf{B}^3 U(1)	
  }
  \,.
$$
The proof of theorem \ref{FirstFractionalDifferentialPontrjagin} 
rests only on the fact 
that $\mathrm{Spin}$ is a compact and simply connected simple Lie group.
The same is true for the special unitary group $\mathrm{SU}$ and the exceptional
Lie group $E_8$. 
\begin{proposition}
  \label{HomotopyGroupsOfBE8}
  The first two non-vanishing homotopy groups of 
  $E_8$ are 
  $$
    \pi_3(E_8) \simeq \mathbb{Z}
  $$
  and
  $$
    \pi_{15}(E_8) \simeq \mathbb{Z}
	\,.
  $$
\end{proposition}
This is a classical fact\cite{BottSamelson}.
It follows with the Hurewicz theorem that 
$$
  H^4(B E_8, \mathbb{Z}) \simeq \mathbb{Z}
  \,.
$$
Therefore the generator of this group  is, 
up to sign, a canonical characteristic class, which we write
$$
  [a]\in H^4(B E_8, \mathbb{Z})
$$
corresponding to a characeristic map $a : B E_8 \to K(\mathbb{Z},4)$.
Hence we obtain analogously the following statements.
\begin{corollary}
  \label{SmoothSecondChernClass}
  \index{characteristic class!Chern class!second, smooth}
  The second Chern-class
  $$
    c_2 : B \mathrm{SU} \to K(\mathbb{Z},4)
  $$
  has an essentially unique lift through 
  $\Pi : \mathrm{Smooth}\infty \mathrm{Grpd} \to 
  \infty \mathrm{Grpd} \simeq \mathrm{Top}$
to a morphism of the form
  $$
     \mathbf{c}_2 : \mathbf{B}\mathrm{SU} \to \mathbf{B}^3 U(1)
  $$  
  and a representative is provided by the Lie integration
  $\exp(\mu_3^{\mathfrak{su}})$ of the canonical Lie algebra 3-cocycle 
  $\mu_3^{\mathfrak{su}} : \mathfrak{su} \to b^2 \mathbb{R}$
  $$
    \mathbf{c}_2 \simeq \exp(\mu_3^{\mathfrak{su}})
	\,.
  $$
   
  Similarly the characteristic map 
  $$
    a : B E_8 \to \mathbb{K}(\mathbb{Z},4)
  $$
  has an essentially unique lift through 
  $\Pi : \mathrm{Smooth}\infty \mathrm{Grpd} \to 
  \infty \mathrm{Grpd} \simeq \mathrm{Top}$
  to a morphism of the form
  $$
     \mathbf{a} : \mathbf{B}E_8 \to \mathbf{B}^3 U(1)
  $$  
  and a representative is provided by the Lie integration
  $\exp(\mu_3^{\mathfrak{e}_8})$ of the canonical Lie algebra 3-cocycle 
  $\mu_3^{\mathfrak{e}_8} : \mathfrak{e}_8 \to b^2 \mathbb{R}$
  $$
    \mathbf{a} \simeq \exp(\mu_3^{\mathfrak{e}_8})
	\,.
  $$
\end{corollary}
Therefore we are entitled to the following special case of
def. \ref{HatGC}.
\begin{definition}
  \label{smoothStringC}
  \index{group!$\mathrm{String}^{\mathbf{c}_2}$}
  \index{group!$\mathrm{String}^{\mathbf{a}}$}
  The smooth 2-group
  $$
    \mathrm{String}^{\mathbf{c}_2}
	\in \infty \mathrm{Grp}(\mathrm{Smooth}\infty \mathrm{Grpd})
  $$
  is the loop space object of the $\infty$-pullback  
  $$
    \raisebox{20pt}{
    \xymatrix{
	  \mathbf{B}\mathrm{String}^{\mathbf{c}_2}
	  \ar[r]
	  \ar[d]
	  &
	  \mathbf{B} \mathrm{SU}
	  \ar[d]^{\mathbf{c}_2}
	  \\
	  \mathbf{B} \mathrm{Spin}
	  \ar[r]^{\frac{1}{2}\mathbf{p}_1}
	  &
	  \mathbf{B}^3 U(1)
	}
	}
	\,.
  $$
  Analogously, 
  the smooth 2-group
  $$
    \mathrm{String}^{\mathbf{a}}
	\in \infty \mathrm{Grp}(\mathrm{Smooth}\infty \mathrm{Grpd})
  $$
  is the loop space object of the $\infty$-pullback  
  $$
    \raisebox{20pt}{
    \xymatrix{
	  \mathbf{B}\mathrm{String}^{\mathbf{a}}
	  \ar[r]
	  \ar[d]
	  &
	  \mathbf{B} E_8
	  \ar[d]^{\mathbf{a}}
	  \\
	  \mathbf{B} \mathrm{Spin}
	  \ar[r]^{\frac{1}{2}\mathbf{p}_1}
	  &
	  \mathbf{B}^3 U(1)
	}
	}
	\,.
  $$
\end{definition} 
\begin{remark}
  \label{MayerVietorisFormOfStringa}
  By prop. \ref{MayerVietorisFiberSequence}, $\mathrm{String}^{\mathbf{a}}$
  is equivalently is the homotopy fiber of the difference
  $\frac{1}{2}\mathbf{p}_1 - \mathbf{a}$
  $$
    \raisebox{20pt}{
    \xymatrix{
	  \mathbf{B}\mathrm{String}^{\mathbf{a}}
	  \ar[rr]
	  \ar[d]
	  &&
	  {*}
	  \ar[d]
	  \\
	  \mathbf{B}(
	   \mathrm{Spin}
	  \times
	  E_8
	  )
	  \ar[rr]^{\frac{1}{2}\mathbf{p}_1 - \mathbf{a}}
	  &&
	  \mathbf{B}^3 U(1)
	}
	}
	\,.
  $$
\end{remark}
We consider now a presentation of $\mathrm{String}^{\mathbf{a}}$
by Lie integration, as in \ref{SmoothStrucLieAlgebras}. 
\begin{definition}
  Let 
  $$
    (\mathfrak{so}\otimes \mathfrak{e}_8)_{\mu_3^{\mathfrak{so}}-\mu_3^{\mathfrak{e}_8}}
	\in 
	L_\infty \mathrm{Alg}
  $$
  be the $L_\infty$-algebra extension, according to def. \ref{gmu}, 
  of the tensorproduct Lie algebra
  $\mathfrak{so} \otimes \mathfrak{e}_8$
  by the difference of the canonical 3-cocycles on 
  the two factors.
\end{definition}
\begin{proposition}
  \label{String2aByLieIntegration}
 The Lie integration, def. \ref{ExponentiatedLInftyAlgbra}, of
 the Lie 2-algebra 
 $(\mathfrak{so}\otimes \mathfrak{e}_8)_{\mu_3^{\mathfrak{so}}-\mu_3^{\mathfrak{e}_8}}$
 is a presentation of $\mathrm{String}^{\mathbf{a}}$:
 $$
   \mathrm{String}^{\mathbf{a}}
   \simeq
   \tau_2 \exp\left(\mathfrak{so}\otimes \mathfrak{e}_8)_{\mu_3^{\mathfrak{so}}-\mu_3^{\mathfrak{e}_8}}\right)
 $$
\end{proposition}
\proof
  With remark \ref{MayerVietorisFormOfStringa}
  this is directly analogous to prop. 
  \ref{StringModelByExpgmu}.
\endofproof
\begin{remark}
  \label{FieldContentOfStringaConnection}
 Therefore a 2-connection on a $\mathrm{String}^{\mathbf{a}}$-principal
 2-bundle is locally given by
 \begin{itemize}
   \item an $\mathfrak{so}$-valued 1-form $\omega$;
   \item an $\mathfrak{e}_8$-valued 1-form $A$;
   \item a 2-form $B$;
 \end{itemize}
 such that the 3-form curvature of $B$ is, locally,
 the sum of the de Rham differential of $B$ with the
 difference of the Chern-Simons forms of $\omega$
 and $A$, respectively: 
 $$
   H_3 = d B + \mathrm{cs}(\omega) - \mathrm{cs}(A)
   \,.
 $$
 We discuss the role of such 2-connections in string theory
 below in \ref{HeteroticGreenSchwarz} and \ref{7dCSInSugraOnAdS7}.
\end{remark}

\newpage

\newpage
\subsection{Higher prequantum fields}
\label{TwistedStructures}
\index{cohomology!twisted differential cohomology}
\index{twisted cohomology!twisted differential $\mathbf{c}$-structures}

We discuss various examples of \emph{twisted $\infty$-bundles},
\ref{ExtensionsOfCohesiveInfinityGroups}, and the corresponding
\emph{twisted differential structures}, \ref{TwistedDifferentialStructures},
and interpret these examples as twisted 
prequantum (boundary) fields, \ref{BoundaryFieldTheory},
that may appear in in local prequantum field theory, \ref{LocalPrequantumFieldTheories},
and notably in string theory \cite{TwistedStructuresLecture}.

Most of these appear in various guises in string theory, which
we survey in
\begin{itemize}
\item \ref{TwistedTopologicalStructuresInStringTheory}
 -- Twisted topological $c$-structures in String theory.
\end{itemize}
Below we discuss the following differential refinements and applications.
\begin{itemize}
  \item 
    \ref{TwistedDiffStructures}
	  -- Definition and overview
  \item 
    \ref{ReductionOfTheStructureGroup} -- Reduction of structure groups
  \begin{itemize}
    \item 
      \ref{OrthogonalRiemannianStructureByTwistedCohomology}
	    -- Orthogonal/Riemannian structure
    \item 
      \ref{TypeIIGeometryByTwistedCohomology}
	    -- Type II generalized geometry
    \item 
      \ref{UDualityGeometry}
	    -- U-duality geometry / exceptional generalized geometry
  \end{itemize}
  \item  
    \ref{OrientifoldCircleNBundlesWithConnection}
	 --Orientifolds and higher orientifolds
  \item 
    \ref{TwistedTopologicalStructuresInStringTheory}
	  -- Twisted topological structures in quantum anomaly cancellation
  \item 
    \ref{TwistedDiffStructuresForAnomalyCancellation}
    -- Tisted differential structures in quantum anomaly cancellation 
	\begin{itemize}
      \item
       \ref{Twistedc1Structures}     
    	--
       Twisted differential $\mathbf{c}_1$-structures 
      \item
        \ref{TwistedSpinCStructures}  
	    --
        Twisted differential $\mathrm{spin}^c$-structures 
      \item 
        \ref{HigherSpinStructure}
         --
         Higher differential spin structures: string and fivebrane structures 
   \end{itemize}
   \item 
     \ref{supergravityCField}
	 --
     The supergravity $C$-field       
   \item 
     \ref{DifferentialTDuality}
	 --
     Differential T-duality
\end{itemize}

The discussion in this section draws from \cite{FiorenzaSatiSchreiber},
which in turn draws from the examples discussed in \cite{SSSIII},
\cite{FiorenzaSatiSchreiberI}.

\subsubsection{Introduction and Overview}
\label{TwistedDiffStructures}
\label{FieldsInSlices}
\index{twisted cohomology!twisted fields!overview}

We start with an exposition and overview of the notion of twisted fields in local
prequantum field theory. This section is taken from \cite{FiorenzaSatiSchreiberCS}.
See also the lecture notes \cite{TwistedStructuresLecture}.

\medskip

While twisted higher (gauge) fields embody much of the subtle structure
in string theory backgrounds, actually basic example of them secretly 
appear all over the place in traditional field theory.
For instance the field of gravity in general relativity is a (pseudo-)Riemannian metric on spacetime, 
and there is no such thing as a moduli stack of (pseudo-)Riemannian metrics on the siteö
of smooth manifolds 
This is nothing but the elementary fact that a 
(pseudo-)Riemannian metric cannot be pulled back along an arbitrary smooth morphism between manifolds, 
but only along local diffeomorphisms. Translated into the language of stacks, this tells us that (pseudo-)Riemannian metrics is a stack on the \'etale site of smooth manifolds, but not on the smooth site.
\footnote{See \cite{carchedi} for a comprehensive treatment of the \'etale site of smooth manifolds 
and of the 
higher topos of higher stacks over it.}
 Yet we can still look at (pseudo-)Riemannian metrics on a smooth $n$-dimensional 
 manifold $X$ from the perspective of the topos $\mathbf{H}$ of stacks over the smooth site, 
 and indeed this is
the more comprehensive point of view. Namely, 
working in $\mathbf{H}$ also means to work with all its \emph{slice toposes} 
(or \emph{over-toposes}) $\mathbf{H}/_{\mathbf{S}}$ over the various objects $\mathbf{S}$ in $\mathbf{H}$. For the field of 
gravity this means working in the slice $\mathbf{H}/_{\mathbf{B}GL(n;\mathbb{R})}$ over the stack $\mathbf{B}GL(n;\mathbb{R})$.

Notice that this terminology is just a concise and rigorous way of expressing a familiar fact from Riemannian geometry: endowing a smooth $n$-manifold $X$ with a pseudo-Riemannian metric of signature $(p,n-p)$ is equivalent to performing a
 reduction of the structure group of the tangent bundle of $X$ to $O(p,n-p)$. Indeed, one can look at the tangent bundle (or, more precisely, at the associated frame bundle) as a morphism $\tau_X: X \to \mathbf{B}GL(n;\mathbb{R})$. 
 
 \paragraph{Example: Orthogonal structures.}
 The above reduction is then the datum of a homotopy lift of $\tau_X$
 \[
  \xymatrix{
    & \mathbf{B}O(p,n-p) \ar[d]
    \\
    X \ar[r]_-{\tau_X}^{\ }="t" \ar[ur]^{o_X}_{\ }="s" & \mathbf{B} \mathrm{GL}(n;\mathbb{R})\;,
	\ar@{=>}^{e} "s"; "t"
  }
\]
where the vertical arrow 
$$
  \mathbf{OrthStruc}_n
  :
  \xymatrix{
    \mathbf{B}O(p,n-p)
	\ar[r]
	& 
	\mathbf{B} \mathrm{GL}(n;\mathbb{R})
  }
$$
is induced by the inclusion of groups 
$O(p,n-p)\hookrightarrow GL(n;\mathbb{R})$. 
Such a commutative diagram is precisely a map 
$$
  (o_X, e)
   : 
  \xymatrix{
    \tau_X 
	\ar[r]
	&
	\mathbf{OrthStruc}_n
  }
$$
in the slice $\mathbf{H}/_{\mathbf{B}GL(n;\mathbb{R})}$. The homotopy $e$ appearing in the 
above diagram is
precisely the \emph{vielbein field} (frame field) which exhibits the reduction, hence 
which induces the Riemannian metric. So the moduli stack of Riemannian metrics
in $n$ dimensions is $\mathbf{OrthStruc}_n$, not as an object of the ambient cohesive topos
$\mathbf{H}$, but of the slice $\mathbf{H}_{/\mathbf{B}\mathrm{GL}(n)}$. Indeed, a 
map between manifolds regarded in this slice, namely a map 
$(\phi, \eta) : \tau_Y  \to \tau_X$, is equivalently a smooth map 
$\phi : Y \to X$ in $\mathbf{H}$, but equipped with an equivalence
$\eta : \phi^* \tau_X \to \tau_Y$. This precisely exhibits $\phi$ as a local
diffeomorphism. In this way the slicing formalism automatically knows along which kinds of maps 
metrics may be pulled back.

\paragraph{Example: (Exceptional) generalized geometry.} If we replace in the above
 example the map 
$\mathbf{OrthStruc}_n$ with inclusions of other maximal
compact subgroups, we similarly obtain the moduli stacks for \emph{generalized} geometry
(metric and B-field)
as appearing in type II superstring backgrounds (see, e.g., \cite{Hi}), given by
$$
  \mathbf{typeII} 
    : 
  \xymatrix{ 
     \mathbf{B}(O(n)\times O(n)) \ar[r] & \mathbf{B}O(n,n)
  }
  \;\;\in \mathbf{H}_{/\mathbf{B}O(n,n)}
$$
and of \emph{exceptional generalized geometry} appearing in compactifications of 11-dimensional 
supergravity \cite{Hull}, given by
$$
  \mathbf{ExcSugra}_n : \xymatrix{
    \mathbf{B}K_{n}
	\ar[r]
	&
	\mathbf{B}E_{n(n)}
  }
  \;\;
  \in 
  \mathbf{H}_{/\mathbf{B}E_{n(n)}},
$$
where $E_{n(n)}$ is the maximally non-compact real
form of the Lie group of rank $n$
with $E$-type Dynkin diagram, and $K_n\subseteq E_{n(n)}$ is a maximal compact subgroup.
For instance, a manifold $X$ in type II-geometry is represented by 
$\tau_X^{\mathrm{gen}} : X \to \mathbf{B}O(n,n)$ 
in the slice $\mathbf{H}_{/\mathbf{B}O(n,n)}$,
which is the map modulating what is called the \emph{generalized tangent bundle}, and a field
of generalized type II gravity is a map 
$(o_X^{\mathrm{gen}}, e) : \tau_X^{\mathrm{gen}} \to \mathbf{typeII}$
to the moduli stack in the slice. One checks that the homotopy $e$ is now precisely
what is called the \emph{generalized vielbein field} in type II geometry. 
We read off the kind of maps along which such fields may be pulled back: a map 
$(\phi,\eta) : \tau_Y^{\mathrm{gen}}\to \tau_X^{\mathrm{gen}}$
is a \emph{generalized} local diffeomorphism: a smooth map 
$\phi : Y \to X$
equipped with an equivalence of generalized tangent bundles 
$\eta : \phi^* \tau_X^{\mathrm{gen}} \to \tau_Y^{\mathrm{gen}}$.
A directly analogous discussion applies to the exceptional generalized geometry.

Furthermore,
 various topological structures are generalized fields in this sense, and become
fields in the more traditional sense after differential refinement. 

\paragraph{Example: Spin structures.}
The map
$\mathbf{SpinStruc} :  \mathbf{B}\mathrm{Spin} \to \mathbf{B}\mathrm{GL}$
is, when regarded as an object of $\mathbf{H}_{/\mathbf{B}\mathrm{GL}}$, the moduli 
stack of spin structures. Its differential refinement
$\mathbf{SpinStruc}_{\mathrm{conn}} :  \mathbf{B}\mathrm{Spin}_{\mathrm{conn}} \to 
 \mathbf{B}\mathrm{GL}_{\mathrm{conn}}$ is such that a domain object 
$\tau_X^{\nabla} \in \mathbf{H}_{/\mathbf{GL}_{\mathrm{conn}}}$ is given by an affine connection,
and a map $(\nabla_{\mathrm{Spin}} ,e) : \tau_X^{\nabla} \to \mathbf{SpinStruc}_{\mathrm{conn}}$
is precisely a \emph{Spin connection} and a Lorentz frame/vielbein which identifies $\nabla$
with the corresponding Levi-Civita connection.  

\par 
This example is the first in a whole tower of \emph{higher Spin structure} fields
\cite{SSSI, SSSII, SSSIII}, each of 
which is directly related to a corresponding higher Chern-Simons theory. 
The next higher example in this tower is the following. 

\paragraph{Example: Heterotic fields.}
\index{anomaly cancellation!Green-Schwarz anomaly}
For $n\geq 3$, let $\mathbf{Heterotic}$ be the map
$$
  \mathbf{Heterotic} 
    : 
  \xymatrix{
    \mathbf{B}\mathrm{Spin}(n) \ar[rr]^-{(p, \tfrac{1}{2}\mathbf{p}_1{})}
    &&
    \mathbf{B}\mathrm{GL}(n;\mathbb{R}) \times \mathbf{B}^3 U(1)
  }
$$
regarded as an object in the slice $\mathbf{H}_{/\mathbf{B}\mathrm{GL}(n;\mathbb{R}) \times \mathbf{B}^3 U(1)}$. Here $p$ is the morphism induced by
\[
\mathrm{Spin}(n)\to O(n)\hookrightarrow GL(n;\mathbb{R})
\]
while $\frac{1}{2}\mathbf{p}_1:\mathbf{B}\mathrm{Spin}(n)\to \mathbf{B}^3 U(1)$ is the morphism of 
stacks underlying the first fractional Pontrjagin class, \ref{SmoothStringC2}.
To regard a smooth manifold $X$ as an object in the  slice $\mathbf{H}_{/\mathbf{B}\mathrm{GL}(n;\mathbb{R}) \times \mathbf{B}^3 U(1)}$ means to equip it with a 
$U(1)$-3-bundle  $\mathbf{a}_X : X \to \mathbf{B}^3 U(1)$ in addition to the tangent bundle $\tau_X: X \to \mathbf{B}GL(n;\mathbb{R})$.  
A Green-Schwarz anomaly-free background field configuration in heterotic string theory 
is (the differential refinement of) 
a map $(s_X,\phi) : (\tau_X, \mathbf{a}_X) \to \mathbf{Heterotic}$, i.e., a homotopy commutative diagram
$$
  \xymatrix{
    X 
	\ar[dr]_{\hspace{-4mm}(\tau_X, \mathbf{a}_X)}^{\ }="t"
	\ar[rr]^{s_X}_{\ }="s" 
	 && 
	\mathbf{B}\mathrm{Spin}
	\ar[dl]^{\mathbf{\phantom{mm}Heterotic}}
	\\
	&
	\mathbf{B}\mathrm{GL}(n)
	\times
	\mathbf{B}^3 U(1)\;.
	\ar@{=>}^{\phi} "s"; "t"
  }
$$
The 3-bundle  $\mathbf{a}_X$ serves as a twist: when $\mathbf{a}_X$ is trivial then we are in presence 
of a String structure on $X$; so it is customary to refer to $(s_X,\phi)$ as to an 
$\mathbf{a}_X$-\emph{twisted String structure} on $X$, in the sense of \cite{Wang, SSSIII}. The Green-Schwarz
anomaly cancellation condition is then imposed by requiring that  $\mathbf{a}_X$ 
(or rather its differential refinement) factors as
\[
  \xymatrix{
    X \longrightarrow
	\mathbf{B}SU
	\ar[r]^-{\mathbf{c}_2} 
	&
	\mathbf{B}^3 U(1)
  }\;,
\]
where $\mathbf{c}_2(E)$ is the morphism of stacks underlying the second Chern class. 
Notice that this says that the extended Lagrangians of $\mathrm{Spin}$- and $\mathrm{SU}$-Chern-Simons
theory in 3-dimensions, as discussed above, at the same time serve as the twists that control the
higher background gauge field structure in heterotic supergravity backgrounds.

\paragraph{Example: Dual heterotic fields.} 
\index{Green-Schwarz mechanism!dual}
\index{anomaly cancellation!Green-Schwarz anomaly!dual}
Similarly, the morphism
$$
  \mathbf{DualHeterotic}
  :
  \xymatrix{
    \mathbf{B}\mathrm{String}(n)
    \ar[rr]^{(p, \tfrac{1}{6}\mathbf{p}_2)\phantom{mmm}}
    &&
	\mathbf{B}\mathrm{GL}(n;\mathbb{R}) \times \mathbf{B}^7 U(1)
  }
$$
governs field configurations for the dual heterotic string.
These examples, in their differentially refined version, have been discussed in \cite{SSSIII}.
The last example above is governed by the extended Lagrangian of the 7-dimensional Chern-Simons-type
higher gauge field theory of $\mathrm{String}$-2-connections. This has been discussed in 
\cite{FiorenzaSatiSchreiberI}.

There are many more examples of (quantum) fields modulated by objects in slices of a cohesive higher topos.
To close this brief discussion, notice that the previous example has an evident analog
in one lower degree: a central extension of Lie groups $A \to \hat G \to G$
induces a long fiber sequence 
$$
  \xymatrix{\underline{A} \longrightarrow \hat{\underline{G}} \longrightarrow \underline{G} \longrightarrow \mathbf{B}A \longrightarrow \mathbf{B}\hat G 
  \longrightarrow 
  \mathbf{B} G \ar[r]^-{\mathbf{c}} & \mathbf{B}^2 A}
$$
in $\mathbf{H}$, where $\mathbf{c}$ is the group 2-cocycle that classifies the extension. If 
we regard this as a coefficient object in the slice $\mathbf{H}_{/\mathbf{B}^2 A}$, then 
regarding a manifold $X$ in this slice means to equip it with an $(\mathbf{B}A)$-principal 2-bundle 
(an $A$-bundle gerbe) modulated by a map $\tau_X^{A} : {X \to \mathbf{B}^2A}$; 
and a field
$(\phi, \eta) : { \tau_X^A \to \mathbf{c} }$
is equivalently a $G$-principal bundle $P \to  X$ equipped with an equivalence 
$\eta : \mathbf{c}(E) \simeq \tau_X^A$ with the 2-bundle which obstructs
its lift to a $\hat G$-principal bundle (the ``lifting gerbe''). 
The differential refinement of this setup similarly yields $G$-gauge fields equipped with such 
an equivalence. A concrete example for this is discussed below in section \ref{PrequantumInHigherCodimension}.


This special case of fields in a slice is called a 
\emph{twisted (differential) $\hat G$-structure} in \cite{SSSIII}
In more generality, 
the terminology \emph{twisted (differential) $\mathbf{c}$-structures} 
is used in \cite{SSSIII} 
to denote spaces of fields of the form $\mathbf{H}/_{\mathbf{S}}(\sigma_X,\mathbf{c})$ for some slice topos 
$\mathbf{H}/_{\mathbf{S}}$ and some coefficient object (or ``twisting object'') $\mathbf{c}$; see also  the  exposition in \cite{TwistedStructuresLecture}. In fact in full generality (quantum) fields in slice toposes
are equivalent to cocycles in (generalized and parameterized and possibly non-abelian and differential) 
\emph{twisted cohomology}. The constructions on which the above discussion is 
built is given in some generality in \cite{NSSa}.

\par
In many examples of twisted (differential) structures/fields in slices the twist 
is constrained to have a certain factorization. For instance the twist of 
the (differential) String-structure
in a heterotic background is constrained to be the (differential) second Chern-class of a 
(differential) $E_8 \times E_8$-cocycle; or
for instance  the gauging of the 1d Chern-Simons fields
on a knot in a 3d Chern-Simons theory bulk is constrained to be the restriction of the bulk gauge field,
as discussed in section \ref{WithWilsonLoops}. Another example is the twist of the Chan-Paton bundles
on D-branes, discussed below in section \ref{PrequantumInHigherCodimension}, which is constrained to be the
restriction of the ambient Kalb-Ramond field to the D-brane.
In all these cases the fields may be thought of as being maps in the slice topos
that arise 
from maps in the \emph{arrow topos} $\mathbf{H}^{\Delta^{1}}$.
A moduli stack here is a map of moduli stacks 
$$
  \mathbf{Fields}_{\mathrm{bulk}+\mathrm{def}} 
  : 
  \xymatrix{
    \mathbf{Fields}_{\mathrm{def}} \ar[r] & \mathbf{Fields}_{\mathrm{bulk}} 
  }
$$
in $\mathbf{H}$; and a domain on which such fields may be defined is an object 
$\Sigma_{\mathrm{bulk}} \in \mathbf{H}$ equipped  with a map (often, but not necessarily, an inclusion) 
${\Sigma_{\mathrm{def}} \to \Sigma_{\mathrm{bulk}}} $, and a field configuration is
a square of the form
$$
  \raisebox{20pt}{
  \xymatrix{
    \Sigma_{\mathrm{def}} \ar[rr]^-{\phi_{\mathrm{def}}}_{\ }="s"
	\ar[d]^>{\ }="t"
	&& \mathbf{Fields}_{\mathrm{def}}
	\ar[d]^{\mathbf{Fields}}
	\\
	\Sigma_{\mathrm{bulk}} \ar[rr]_-{\phi_{\mathrm{bulk}}} 
	&& \mathbf{Fields}_{\mathrm{bulk}}
	\ar@{=>}^\simeq "s"; "t"
  }
  }
$$
in $\mathbf{H}$. If we now fix $\phi_{\mathrm{bulk}}$ then $(\phi_{\mathrm{bulk}})|_{\Sigma_{\mathrm{def}}}$
serves as the twist, in the above sense, for $\phi_{\mathrm{def}}$.
If  $\mathbf{Fields}_{\mathrm{def}}$ is trivial (the point/terminal object), 
then such a field is a cocycle in \emph{relative cohomology}: a cocycle $\phi_{\mathrm{bulk}}$ on 
$\Sigma_{\mathrm{bulk}}$ equipped with a trivialization $(\phi_{\mathrm{bulk}})|_{\Sigma_{\mathrm{def}}}$
of its restriction to $\Sigma_{\mathrm{def}}$.

The fields in Chern-Simons theory with Wilson loops displayed in section \ref{FieldsForCSWithWilsonLoop}
clearly constitute an example of this phenomenon. Another example is the field content of type II string theory on 
a 10-dimensional spacetime $X$ with D-brane $Q \hookrightarrow X$,  
for which the above diagram reads
$$
  \raisebox{20pt}{
  \xymatrix{
    Q \ar[rr]_{\ }="s" \ar@{^{(}->}[d]^>{\ }="t" 
    && \mathbf{B}\mathrm{PU}_{\mathrm{conn}}
	\ar[d]^{\mathbf{dd}_{\mathrm{conn}}}
	\\
	X
	\ar[rr]^-B
	&&
	\mathbf{B}^2 U(1)_{\mathrm{conn}}\;,
	\ar@{=>} "s"; "t"
  }}
$$
discussed further below in section \ref{PrequantumInHigherCodimension}. 
In \ref{supergravityCField} we discuss how the supergravity C-field over
an 11-dimensional Ho{\v r}ava-Witten background with 10-dimensional boundary $X \hookrightarrow Y$
is similarly a relative cocyle, with the coefficients controled, once more, by the
extended Chern-Simons Lagrangian 
$$
  \hat {\mathbf{c}} : \xymatrix{ \mathbf{B}(E_8 \times E_8)_{\mathrm{conn}} \ar[r] & \mathbf{B}^3 U(1)_{\mathrm{conn}} }
  \,,
$$
now regarded in $\mathbf{H}^{(\Delta^1)}$.

\medskip

The following table lists some of main (classes of) examples. The left
column displays a given extension of smooth $\infty$-groups, to be regarded
as a bundle of coefficients with typical $\infty$-fiber shown on the far left.
The middle column names the principal $\infty$-bundles, or equivalently the
nonabelian cohomology classes, that are classified by the base of these
extensions. These are to be thought of as twisting cocycles. The right column
names the corresponding twisted $\infty$-bundles, or eqivalently the 
corresponding twisted cohomology classes. 

\medskip 

\begin{tabular}{|rcc|}
  \hline
  \begin{tabular}{c}
    {\bf extension }/ \\ {\bf $\infty$-bundle of coefficients}
  \end{tabular}
  &
  \begin{tabular}{c}
    {\bf twisting $\infty$-bundle} /
	\\
	{\bf twisting cohomology}
  \end{tabular}
  &
  \begin{tabular}{c}
    {\bf twisted $\infty$-bundle} / 
	\\
	{\bf twisted cohomology }
  \end{tabular}
  \\
  \hline\hline  
  $\raisebox{20pt}{\xymatrix{
    V \ar[r] & V/\!/G
	\ar[d]^{\mathbf{\rho}}
	\\
	& \mathbf{B} G
  }}$
  &
  \begin{tabular}{c}
    $\rho$-associated  \\ $V$-$\infty$-bundle
  \end{tabular}  
  &
  \begin{tabular}{c}
    section
  \end{tabular}  
  \\
  \hline
   $\raisebox{20pt}{\xymatrix{
    \mathrm{GL}(d)/O(d) \ar[r] & \mathbf{B} O(d)
	\ar[d]^{\mathbf{}}
	\\
	& \mathbf{B} \mathrm{GL}(d)
  }}$
  &
  \begin{tabular}{c}
    tangent bundle 
  \end{tabular}  
  &
  \begin{tabular}{c}
    orthogonal structure / \\
	Riemannian geometry
  \end{tabular}
  \\
  \hline
  $\raisebox{20pt}{
   \xymatrix@C=8pt{
    O(d)\backslash O(d,d)/O(d) \ar[r] & \mathbf{B} (O(d) \times O(d))
	\ar[d]^{\mathbf{}}
	\\
	& \mathbf{B} O(d,d)
  }}$
  &
  \begin{tabular}{c}
    generalized \\
	tangent bundle 
  \end{tabular}  
  &
  \begin{tabular}{c}
    generalized (type II) \\
	Riemannian geometry
  \end{tabular}  \\
  \hline
  $\raisebox{20pt}{\xymatrix{
    \mathbf{B}U(n)  \ar[r]& \mathbf{B} \mathrm{PU}(n)
	\ar[d]^{\mathbf{dd}}
	\\
	& \mathbf{B}^2 U(1)
  }}$
  &
  \begin{tabular}{c}
    circle 2-bundle / \\
	bundle gerbe
  \end{tabular}
  &
  \begin{tabular}{c}
    twisted vector bundle / \\
	bundle gerbe module
  \end{tabular}  
  \\
  \hline
  $\raisebox{20pt}{\xymatrix{
    \mathbf{B}^2 U(1) \ar[r] & \mathbf{B} \mathrm{Aut}(\mathbf{B}U(1))
	\ar[d]^{\mathbf{}}
	\\
	& \mathbf{B} \mathbb{Z}_2
  }}$
  &
  \begin{tabular}{c}
    double cover
  \end{tabular}  
  &
  \begin{tabular}{c}
    orientifold structure / \\
	Jandl bundle gerbe
  \end{tabular}
  \\
  \hline
  $\raisebox{20pt}{\xymatrix{
    \mathbf{B}^2 \mathrm{ker}(G)  \ar[r] & \mathbf{B} \mathrm{Aut}(\mathbf{B}G)
	\ar[d]^{\mathbf{}}
	\\
	& \mathbf{B} \mathrm{Out}(G)
  }}$
  &
  \begin{tabular}{c}
    band (\emph{lien})
  \end{tabular}  
  &
  \begin{tabular}{c}
    nonabelian (Giraud-Breen)
    \\	
	$G$-$\infty$-gerbe 
  \end{tabular}  
  \\
  \hline
  $\raisebox{20pt}{\xymatrix{
    \mathbf{B}\mathrm{String} \ar[r] & \mathbf{B} \mathrm{Spin}
	\ar[d]^{\tfrac{1}{2}\mathbf{p}_1}
	\\
	& \mathbf{B}^3 U(1)
  }}$
  &
  \begin{tabular}{c}
    circle 3-bundle / \\
	bundle 2-gerbe
  \end{tabular}
  &
  \begin{tabular}{c}
    twisted \\ $\mathrm{String}$ 2-bundle
  \end{tabular}  
  \\
  \hline
  $\raisebox{20pt}{\xymatrix{
    Q \ar[r] & \mathbf{B} (\mathbb{T} \times \mathbb{T}^*)
	\ar[d]^{\langle \mathbf{c}_1 \cup \mathbf{c}_1\rangle}
	\\
	& \mathbf{B}^3 U(1)
  }}$
  &
  \begin{tabular}{c}
    circle 3-bundle / \\
	bundle 2-gerbe
  \end{tabular}
  &
  \begin{tabular}{c}
    twisted \\ T-duality structure
  \end{tabular}  
  \\
  \hline
  $\raisebox{20pt}{\xymatrix{
    \mathbf{B}\mathrm{Fivebrane} \ar[r] & \mathbf{B} \mathrm{String}
	\ar[d]^{\tfrac{1}{6}\mathbf{p}_2}
	\\
	& \mathbf{B}^7 U(1)
  }}$
  &
  \begin{tabular}{c}
    circle 7-bundle 
  \end{tabular}
  &
  \begin{tabular}{c}
    twisted \\ $\mathrm{Fivebrane}$ 6-bundle
  \end{tabular}  
  \\
  \hline
  \hline  
    $\raisebox{20pt}{
   \xymatrix{
    \mathbf{\flat}\mathbf{B}^n U(1) \ar[r] & \mathbf{B}^n U(1)
	\ar[d]^{\mathrm{curv}}
	\\
	& \mathbf{\flat}_{\mathrm{dR}}\mathbf{B}^{n+1}U(1)
  }}$
  &
  \begin{tabular}{c}
    curvature
	\\
  $(n+1)$-form
  \end{tabular}
  &
  \begin{tabular}{c}
    circle $n$-bundle 
	\\
	with connection
  \end{tabular}  
  \\
  \hline
  \end{tabular}

\newpage

\noindent The following table lists smooth twisting $\infty$-bundles $\mathbf{c}$
that become \emph{identities under geometric realization}, def. \ref{def rrw},
(the last one on 15-coskeleta). This means
that the twists are purely geometric, the underlying topological structure being untwisted.

\hspace{-1.4cm}
\begin{tabular}{|c|c|c|}
  \hline
  \phantom{$\bigl(^{a}_{b}$}
  universal twisting $\infty$-bundle & twisted cohomology & relative twisted cohomology
  \\
  \hline
  \hline
  $
    \raisebox{10pt}{
    \xymatrix@R=10pt{
	  \mathbf{B} O(d)
	  \ar[d]
	  \\
	  \mathbf{B}\mathrm{GL}(d)
	}}
  $
  &
  \begin{tabular}{l}
   Riemannian geometry, \\ 
   orthogonal structure
   \end{tabular}
   &
  \\
  \hline
  $
    \raisebox{10pt}{
    \xymatrix@R=10pt{
	  \mathbf{B} O(d) \times O(d)
	  \ar[d]
	  \\
	  \mathbf{B}O(d,d)
	}}
  $
  &
  type II NS-NS generalized geometry
  &
  \\
  \hline
  $
    \raisebox{10pt}{
    \xymatrix@R=10pt{
	  \mathbf{B} H_n
	  \ar[d]
	  \\
	  \mathbf{B} E_{n(n)}
	}}
  $
  &
  \begin{tabular}{c}
    U-duality geometry,  \\ 
	exceptional generalized geometry
  \end{tabular}
  &
  \\
  \hline
  \hline
  $
    \raisebox{10pt}{
    \xymatrix@R=10pt{
	  \mathbf{B} \mathrm{PU}(\mathcal{H})
	  \ar[d]^{\mathbf{dd}}
	  \\
	  \mathbf{B}^2 U(1)
	}}
  $
  &
    twisted $U(n)$-principal bundles
  &
  \begin{tabular}{l}
    Freed-Witten anomaly cancellation 
	\\ on $\mathrm{Spin}^c$-branes:
	\\
	$B$-field \\ with twisted gauge bundles on D-branes
  \end{tabular}
  \\
  \hline
  $
    \raisebox{10pt}{
    \xymatrix@R=10pt{
	  \mathbf{B} E_8
	  \ar[d]^{2\mathbf{a}}
	  \\
	  \mathbf{B}^3 U(1)
	}}
  $
  &
    twisted $\mathrm{String}(E_8)$-principal 2-bundles
  &
  \begin{tabular}{l}
    M5-brane anomaly cancellation:
	\\
	$C$-field \\ with twisted gauge 2-bundles on M5-branes
  \end{tabular}  
  \\
  \hline
\end{tabular}

\medskip
\noindent The following table lists smooth twisted $\infty$-bundles that control
various quantum anomaly cancellations in string theory.

\hspace{-1.4cm}
\begin{tabular}{|c|c|c|}
  \hline
  \phantom{$\bigl(^{a}_{b}$}
  universal twisting $\infty$-bundle & twisted cohomology & relative twisted cohomology
  \\
  \hline
  \hline
    $
    \raisebox{10pt}{
    \xymatrix@R=10pt{
	  \mathbf{B} \mathrm{SO}
	  \ar[d]^{\mathbf{W}_3}
	  \\
	  \mathbf{B}^2 U(1)
	}}
  $  
  &
  \begin{tabular}{c}
    twisted $\mathrm{Spin}^c$-structure
  \end{tabular}
  &
  \\
  \hline
    $
    \raisebox{10pt}{
    \xymatrix@R=10pt{
	  \mathbf{B} \mathrm{PU}(\mathcal{H}) \times \mathrm{SO}
	  \ar[d]^{\mathbf{dd} - \mathbf{W}_3}
	  \\
	  \mathbf{B}^2 U(1)
	}}
  $  
  &
  &
  \begin{tabular}{l}
    general Freed-Witten anomaly cancellation:
	\\
	B-field
	\\
	with twisted gauge bundles on D-branes
  \end{tabular}
  \\
  \hline
  $
    \raisebox{10pt}{
    \xymatrix@R=10pt{
	  \mathbf{B} \mathrm{Spin}
	  \ar[d]^{\tfrac{1}{2}\mathbf{p}_1}
	  \\
	  \mathbf{B}^3 U(1)
	}}
  $  
  &
  \begin{tabular}{l}
    twisted $\mathrm{String}$-2-bundles;
	\\
	heterotic Green-Schwarz \\ anomaly cancellation
  \end{tabular} 
  &  
  \\
  \hline
  $
    \raisebox{10pt}{
    \xymatrix@R=10pt{
	  \mathbf{B} \mathrm{String}
	  \ar[d]^{\tfrac{1}{6}\mathbf{p}_2}
	  \\
	  \mathbf{B}^7 U(1)
	}}
  $  
  &
  \begin{tabular}{l}
    twisted $\mathrm{Fivebrane}$-7-bundles;
	\\
	dual heterotic Green-Schwarz \\ anomaly cancellation
  \end{tabular}  
  &
  \\
  \hline
\end{tabular}

\newpage
\noindent
The following table lists twisting $\infty$-bundles that encode 
geometric structure preserving higher supersymmetry.

\hspace{-1.4cm}
\begin{tabular}{|c|c|c|}
  \hline
  \phantom{$\bigl(^{a}_{b}$}
  universal twisting $\infty$-bundle & twisted cohomology & relative twisted cohomology
  \\
  \hline
  \hline
  $
    \raisebox{10pt}{
    \xymatrix@R=10pt{
	  \mathbf{B} \mathrm{U}(d,d)
	  \ar[d]
	  \\
	  \mathbf{B}\mathrm{O}(2d,2d)
	}}
  $
  &
  generalized complex geometry
  &
  \\
  \hline
  \hline
  $
    \raisebox{10pt}{
    \xymatrix@R=10pt{
	  \mathbf{B} \mathrm{SU}(3) \times \mathrm{SU}(3)
	  \ar[d]
	  \\
	  \mathbf{B}\mathrm{O}(6,6)
	}}
  $
  &
  $d = 6$, $N=2$ type II compactification
  &
  \\
  \hline
  $
    \raisebox{10pt}{
    \xymatrix@R=10pt{
	  \mathbf{B} \mathrm{SU}(7)
	  \ar[d]
	  \\
	  \mathbf{B}E_{7(7)}
	}}
  $
  &
  $d = 7$, $N=1$ 11d sugra compactification
  &
  \\
  \hline
\end{tabular}

\newpage

\subsubsection{Sections of vector bundles -- twisted 0-bundles}
\label{Twisted0BundlesSectionsOfVectorBundles}
\index{twisted cohomology!sections of vector bundles}

We discuss here for illustration purposes twisted $\infty$-bundles
in \emph{lower} degree than traditionally considered, namely 
\emph{twisted 0-bundles}. This degenerate case is in itself simple,
but all the more does it serve to illustrate by familiar example 
the general notions of twisted $\infty$-bundles.

So we consider coefficient $\infty$-bundles such as
$$ 
  \raisebox{20pt}{
  \xymatrix{
    \mathbb{C} \ar[r] & \mathbb{C}/\!/ U(1) \ar[d]^{}
	\\
	& \mathbf{B} U(1)
  }
  }
  \;\;
  \,,
$$
where
\begin{itemize}
  \item $\mathbf{B}U(1)$ is the smooth moduli stack of smooth circle bundles;
  \item $\mathbb{C}$ is the complex plane, regarded as a smooth manifold.
\end{itemize}
By \ref{StrucRepresentations} this corresponds equivalently to a representation
of the Lie group $U(1)$ on $\mathbb{C}$, and this we take to be the canonical 
such representation. Accordingly, the above bundle is indeed the 
\emph{universal complex line bundle} over the base space of the universal
$U(1)$-principal bundle.

It will be meaningful and useful to think of $\mathbb{C}$ itself as a
moduli $\infty$-stack: it is the smooth \emph{moduli 0-stack of complex 0-vector bundles}, 
where, therefore, a complex 0-vector bundle on a smooth space $X$ is simply a smooth function
$\in C^\infty(X, \mathbb{C})$. Accordingly, we should find that such 0-vector bundles
can be twisted by a principal $U(1)$-bundle and indeed, by feeding the above
coefficient $\infty$-bundle through the definition of twisted $\infty$-bundles
in \ref{ExtensionsOfCohesiveInfinityGroups}, one finds, as we discuss
below, that a \emph{twisted 0-bundle} is a smooth section of the 
\emph{associated line bundle}, hence, by local triviality of the line bundle, locally
a complex-valued function, but globally twisted by the twisting circle bundle.

\medskip

\medskip

Let $G$ be a Lie group, $V$ a vector space and $\rho : V \times G \to V$ a 
smooth representation of $G$ on $V$ in the traditional sense.
We discuss how this is an $\infty$-group representation in the sense
of def. \ref{RepresentationOfInfinityGroup}.
\begin{definition}
  \index{action groupoid}
  \label{ActionGroupoid}
  Write 
  $$
    V/\!/G
	:=
	\xymatrix{
	  V \times G
	  \ar@<+3pt>[r]^{p_1}
	  \ar@<-3pt>[r]_{\rho}
	  &
	  V
	}
  $$ 
  for the \emph{action groupoid} of $\rho$, the weak quotient of $V$
  by $G$, regarded as a smooth $\infty$-groupoid
  $V/\!/G \in \mathrm{Smooth}\infty \mathrm{Grpd}$.
\end{definition}
Notice that this is equipped with a canonical morphism
$V/\!/G \to \mathbf{B}G$
and a canonical inclusion $V \to V/\!/ G$.
\begin{proposition}
  \label{GroupRepresentationByFiberSequence}
  We have a fiber sequence
  $$
    V \to V/\!/G \to \mathbf{B}G
  $$
  in $\mathrm{Smooth}\infty \mathrm{Grpd}$.
\end{proposition}
\proof
  One finds that in the canonical presentation by simplicial presheaves
  as in \ref{SmoothStrucCohesiveInfiniGroups},
  the morphism $V/\!/G_{\mathrm{ch}} \to \mathbf{B}G_{\mathrm{ch}}$ 
  is a fibration in 
  $[\mathrm{CartSp}^{\mathrm{op}}, \mathrm{sSet}]_{\mathrm{proj}}$.
  Therefore by prop. \ref{FiniteHomotopyLimitsInPresheaves}
  the homotopy fiber is given by the ordinary fiber of this
  presentation. This ordinary fibe is $V$.
\endofproof
\begin{remark}
  By remark \ref{TwistingCocycleAsAssociatedBundle} we may think of
  the fiber sequence
  $$
    \xymatrix{
	  V \ar[r] \ar[d] & V/\!/G \ar[d]
	  \\
	  {*}
	  \ar[r]
	  &
	  \mathbf{B}G
	}
  $$
  as the vector bundle over the classifying stack $\mathbf{B}G$
  which is $\rho$-associated to the universal $G$-principal bundle.
\end{remark}
More formally, the next proposition shows that the 
$\rho$-associated bundles according to def. \ref{AssociatedBundleByRho}
are the ordinary associated vector bundles.
\begin{proposition}
  \label{AssociatedVectorBundlesByPullbackOfReps}
  Let $X$ be a smooth manifold and $P \to X$ be a smooth $G$-principal 
  bundle. If $g : X \to \mathbf{B}G$ is a cocycle for $P$ 
  as in \ref{SmoothStrucPrincipalInfinityBundles},
  then the $\rho$-associated vector bundle $P \times_G V \to X$
  is equivalent to the homotopy pullback of 
  $V/\!/G \to \mathbf{B}G$ along $G$:
  $$
    \raisebox{20pt}{
    \xymatrix{
	  P \times_G V
	  \ar[r]
	  \ar[d]
	  &
	  V/\!/G
	  \ar[d]
	  \\
	  X \ar[r]^g & \mathbf{B}G
	}
	}
	\,.
  $$
\end{proposition}
\proof
  By the discussion in \ref{SmoothStrucPrincipalInfinityBundles}
  we may present $g$ by a morphism in
  $[\mathrm{CartSp}^{\mathrm{op}}, \mathrm{sSet}]_{\mathrm{proj}, \mathrm{loc}}$
  of the form
  $$
    \xymatrix{
	   C(\{U_i\}) \ar[r]^g \ar[d]^{\simeq} & \mathbf{B}G_{\mathrm{ch}}
	   \\
	   X
	}
	\,,
  $$
  where $C(\{U_i\})$ is the {\v C}ech nerve of a good open cover of $X$.
  Since $V/\!/G_{\mathrm{ch}} \to \mathbf{B}G_{\mathrm{ch}}$ is 
  a fibration in 
  $[\mathrm{CartSp}^{\mathrm{op}}, \mathrm{sSet}]_{\mathrm{proj}}$,
  by prop. \ref{FiniteHomotopyLimitsInPresheaves} its ordinary pullback
  of simplicial presheaves along $g$ presents the homotopy pullback
  in question. By inspection one finds that this is the Lie groupoid
  whose space of objects is $\coprod_i U_i \times V$ and which has
  a unique  morphism
  from $(x \in U_i, \sigma_i(x) \in V)$ to $(x \in U_j, \sigma_j(x))$ if
  $\sigma_j(x) = \rho(g_{i j}(x))(\sigma_i(x))$.
  
  Due to the uniqueness of morphisms, 
  the evident projection from this Lie groupoid to the smooth manifold
  $P \times_G V$ which is the total space of the $V$-bundle $\rho$-accociated to 
  $P$ is a weak equivalence in $[\mathrm{CartSp}^{\mathrm{op}}, \mathrm{sSet}]_{\mathrm{proj}}$, 
  hence in 
  $[\mathrm{CartSp}^{\mathrm{op}}, \mathrm{sSet}]_{\mathrm{proj}, \mathrm{loc}}$.
  So $P \times_G V$ is indeed (one representative of) the homotopy pullback in 
  question.
\endofproof
Since therefore all the information about $\rho$ is encoded in the
bundle $V \hookrightarrow V/\!/G \to \mathbf{B}G$, we may 
identify that bundle with the action. Accordingly we write
$$
  \rho : V/\!/G \to \mathbf{B}G
  \,.
$$
Regarding $\rho$ then as a universal local coefficient bundle, we obtain the
corresponding twisted cohomology, \ref{StrucTwistedCohomology},
and twisted $\infty$-bundles, \ref{ExtensionsOfCohesiveInfinityGroups}.
We show now that the general statement of prop. \ref{TwistedCohomologyBySections}
on twisted cohomology in terms of sections of associated
$\infty$-bundles reduces for twists relative to $\rho$
to the standard notion of spaces of sections.
\begin{proposition}
  Let $P \to X$ be a $G$-principal bundle over a smooth manifold $X$. Then 
  the $\infty$-groupoid of $P$-twisted cocycles relative to $\rho$, equivalently
  the $\infty$-groupoid of $P$-twisted $V$-0-bundles is equivalent to 
  the ordinary set of sections of the vector bundle $E \to X$ which is 
  $\rho$-associated to $P$:
  $$
    \Gamma_X(E) \simeq \mathbf{H}_{/\mathbf{B}G}(
	  g, \rho
	)
	\,.
  $$
  Here $g : X \to \mathbf{B}G$ is the morphism classifying $P$.
\end{proposition}
\proof
  The hom $\infty$-groupoid of the slice $\infty$-topos over $\mathbf{B}G$
  is the $\infty$-pullback
  $$
    \raisebox{20pt}{
    \xymatrix{
	  \mathbf{H}_{/\mathbf{B}G}(g,\rho)
	  \ar[r]
	  \ar[d]
	  &
	  \mathbf{H}(X, V/\!/G)
	  \ar[d]
	  \\
	  {*}
	  \ar[r]^{[g]}
	  &
	  \mathbf{H}(X, \mathbf{B}G)
	}
	}
	\,.
  $$
  Since the {\v C}ech nerve $C(\{U_i\})$ of the good cover
  $\{U_i \to X\}$ is a cofibrant representative of
  $X$ in $[\mathrm{CartSp}^{\mathrm{op}}, \mathrm{sSet}]_{\mathrm{proj}, \mathrm{loc}}$,
  and since $\mathbf{B}G_{\mathrm{ch}}$ and $V/\!/G_{\mathrm{ch}}$ from above are
  fibrant representatives of $\mathbf{B}G$ and $V/\!/G$, respectively, by the
  properties of simplicial model categories the right vertical morphism here is 
  presented by the morphism of Kan complexes.
  $$
    [\mathrm{CartSp}^{\mathrm{op}}, \mathrm{sSet}](C(\{U_i\}), V/\!/G_{\mathrm{ch}})
	\to
    [\mathrm{CartSp}^{\mathrm{op}}, \mathrm{sSet}](C(\{U_i\}), \mathbf{B}G_{\mathrm{ch}})
	\,.
  $$
  Moreover, since this is the simplicial hom out of a cofibrant object into a fibration,
  the properties of simplicial model categories imply that this morphism is indeed a Kan 
  fibration. It follows with prop. \ref{ConstructionOfHomotopyLimits}
  that the ordinary fiber of this morphism over $[g]$ is a Kan complex
  that presents the twisted cocycle $\infty$-groupoid in question.
  
  Since $V/\!/G_{\mathrm{ch}} \to \mathbf{B}G_{\mathrm{ch}}$ is a faithful functor
  of groupoids, this fiber is a set, meaning a constant simplicial set. 
  A $V/\!/G_{\mathrm{ch}}$-valued cocycle is a collection of smooth functions
  $\{\sigma_i : U_i \to V\}_{i}$ and smooth functions
  $\{g_{i j} : U_{i, j} \to G\}_{i,j}$, satisfying the condition that on all $U_{i j}$ we have
  $\sigma_j = \rho(g_{i j})(\sigma_i)$. This is a vertex in the fiber 
  precisely if the second set of functions is that given by the cocycle $g$
  which classifies $P$. In this case this condition is precisely that which 
  identifies the $\{\sigma_i\}_i$ as a section of the associated vector bundle,
  expressed in terms of the local trivialization that corresponds to $g$.
  
  In conclusion, this shows that $\mathbf{H}_{/\mathbf{B}G}(g,\rho)$ is an 
  $\infty$-groupoid equivalent to set of sections of the vector bundle 
  $\rho$-associated to $P$.
\endofproof

\subsubsection{Sections of 2-bundles -- twisted vector bundles and twisted K-classes}
\label{Twisted1BundlesTwistedKTheory}
\index{twisted cohomology!twisted K-theory} 
\index{twisted cohomology!twisted bundle}

We construct now a coefficient $\infty$-bundle of the form
$$
  \raisebox{20pt}{
  \xymatrix{
    \mathbf{B} U \ar[r]  & (\mathbf{B}U)/\!/ \mathbf{B}U(1) \ar[d]^{\mathbf{dd}}
	\\
	& \mathbf{B}^2 U(1)
  }
  }
  \;\;
  \,,
$$
where
\begin{itemize}
  \item $\mathbf{B}^2 U(1)$ is the smooth moduli 2-stack for smooth circle 2-bundles /
  bundle gerbes;
  \item $\mathbf{B}U = \lim\limits_{\longrightarrow_n} \mathbf{B} U(n)$ is the
   inductive $\infty$-limit over the smooth moduli stacks of smooth unitary rank-$n$ vector bundles
   (equivalently: $U(n)$-principal bundles).
\end{itemize}
Equivalently, this is a smooth $\infty$-action of the smooth circle 2-group $\mathbf{B}U(1)$
on the smooth $\infty$-stack $\mathbf{B}U$. 

This may be thought of as the 
canonical 2-representation of the circle 2-group $\mathbf{B}U(1)$, def. \ref{CircleNGroup}, 
being the
higher  analogue to the canonical representation of the circle group $U(1)$
on the complex plane $\mathbb{C}$, discussed above in \ref{Twisted0BundlesSectionsOfVectorBundles}.

We show that the notion of twisted cohomology induced by this local coefficient bundle
according to \ref{StrucTwistedCohomology} is reduced \emph{twisted K-theory}
and that the notion of twisted $\infty$-bundles induced by it according to 
\ref{ExtensionsOfCohesiveInfinityGroups} are ordinary \emph{twisted vector bundles}
also known as \emph{bundle gerbe modules}. 
(See for instance chapter 24 of \cite{MayConcise} 
for basics of K-theory that we need here, and see for instance \cite{CBMMS}
for a discussion of twisted K-theory in terms of twisted bundles.)

This  not only shows how the traditional notion of twisted K-theory is reproduced
from the perspective of cohomology in an $\infty$-topos. It also refines the 
traditional constructions to the smooth context. Notice that there is
a slight clash of terminology, as traditionally the term \emph{smooth K-theory}
is often used synonymously with \emph{differential K-theory}. However, there is
a geometric refinement in between bare (twisted) K-classes and differential (twisted)
K-classes, namely smooth cocycle spaces of smooth (twisted) vector bundles
and \emph{smooth} gauge transformations between them. This is the smooth 
refinement of the situation that we find here, by regarding (twisted) K-theory
as (twisted) cohomology internal to the $\infty$-topos $\mathrm{Smooth}\infty \mathrm{Grpd}$.

\medskip

The construction of the traditional topological classifying space
for reduced $K^0$ proceeds as follows.
For $n \in \mathbb{N}$, let $B U(n)$ be the classifying space of the 
unitary group in complex dimension $n$. The inclusion of groups
$U(n) \to U(n+1)$ induced by the inclusion $\mathbb{C}^n \to \mathbb{C}^{n+1}$
by extension by 0 in the, say, last coordinate gives an inductive system
of topological spaces
$$
  \xymatrix{
    {*} \ar[r] & \cdots \ar[r] B U(n) \ar[r] & B U(n+1) \ar[r] & \cdots 
  }
  \,.
$$
\begin{definition}
  Write
  $$
     B U := \lim\limits_{{\longrightarrow}_n} B U(n)
  $$
  for the homotopy colimit in $\mathrm{Top}_{\mathrm{Quillen}}$.
\end{definition}

Notice that by prop. \ref{DeloopedLieGroup} 
and prop. \ref{FundGroupoidOfSimplicialParacompact} we have, for 
each $n \in \mathbb{N}$, a smooth refinement
of 
$B U(n) \in \mathrm{Top} \simeq \infty \mathrm{Grpd}$ to a smooth moduli
stack $\mathbf{B} U(n) \in \mathrm{Smooth}\infty \mathrm{Grpd}$.
This refines the set $[X, B U(n)]$ of equivalences classes of rank-$n$
unitary vector bundles to the groupoid $\mathbf{H}(X, \mathbf{B}U(n))$
of unitary bundles and smooth gauge transformations between them.

We therefore consider now similarly a smooth refinement to moduli $\infty$-stacks
of the inductive limit $B U$.
\begin{definition}
  Write
  $$
    \mathbf{B} U := \lim\limits_{\longrightarrow_n} \mathbf{B}U(n)
  $$
  for the $\infty$-colimit in $\mathrm{Smooth}\infty\mathrm{Grpd}$
  over the smooth moduli stacks of smooth $U(n)$-principal bundles.
 \end{definition}
\begin{proposition}
  \label{BUIsDeloopingofStableUnitaryGroup}
  The canonical morphism
  $$
    \lim\limits_{\longrightarrow_n} \mathbf{B} U(n)
	\to 
	\mathbf{B}\lim\limits_{\longrightarrow_n} U(n)
  $$
  is an equivalence in $\mathrm{Smooth}\infty\mathrm{Grpd}$.
\end{proposition}
\proof
  Write $\mathbf{B} U(n)_{\mathrm{ch}} := 
  N(\xymatrix{
    U(n)
	\ar@<+2pt>[r]
	\ar@<-2pt>[r]
	&
	{*}
  })
  \in [\mathrm{CartSp}^{\mathrm{op}}, \mathrm{sSet}]
  $
  for the standard presentation of the delooping, prop. \ref{DeloopedLieGroup}. 
  Observe then that the 
  diagram $n \mapsto \mathbf{B} U(n)_{\mathrm{ch}}$ is cofibrant when 
  regarded as an object of 
  $[(\mathbb{N}, \leq), [\mathrm{CartSp}^{\mathrm{op}}, \mathrm{sSet}]_{\mathrm{inj}, \mathrm{loc}}]_{\mathrm{proj}}$,
  because,
  by example \ref{CotowerCofibrancy},
  a cotower is projectively cofibrant if it consists of monomorphisms
  and if the first object, and hence all objects, are cofibrant. Therefore the
  $\infty$-colimit is presented by the ordinary colimit over this diagram.
  Since this is a filtered colimit, it commutes with finite limits of simplicial presheaves:
  $$
    \begin{aligned}
      \lim\limits_{\longrightarrow_n} \mathbf{B} U(n)_{\mathrm{ch}}
	  &=
	  \lim\limits_{\longrightarrow_n}   N(\xymatrix{
      U(n)
	  \ar@<+2pt>[r]
	  \ar@<-2pt>[r]
 	  &
	  {*}
	  })
	  \\
	  & = 
	  N(\xymatrix{
      \lim\limits_{\longrightarrow_n} U(n)
	   \ar@<+2pt>[r]
	   \ar@<-2pt>[r]
 	   &
	   {*}
	  })
	  \\
	  & = (\mathbf{B} \lim\limits_{\longrightarrow_n} U(n))_{\mathrm{ch}}\,.
	\end{aligned}
  $$
\endofproof
\begin{proposition}
  The smooth object $\mathbf{B}U$ is a smooth refinement of the topological 
  space $B U$ in that it reproduces the latter under geometric realization, 
  \ref{ETopStrucGeometricRealization}:
  $$
    \vert \mathbf{B}U \vert \simeq B U
	\,.
  $$
\end{proposition}
\proof
  By prop. \ref{FundGroupoidOfParacompact} for every $n \in \mathbb{N}$ we have
  $$
    \vert \mathbf{B}U(n) \vert \simeq B U(n)
	\,.
  $$
  Moreover, by the discussion at \ref{ETopStrucGeometricRealization},
  up to the equivalence $\mathrm{Top} \simeq \infty \mathrm{Grpd}$
  the geometric realization is given by applying the functor
  $\Pi : \mathrm{Smooth}\infty \mathrm{Grpd} \to \infty\mathrm{Grpd}$.
  That is a left $\infty$-adjoint and hence preserves $\infty$-colimits:
  $$
    \begin{aligned}
	  \vert \mathbf{B}U \vert
	  & \simeq
	  \vert \lim\limits_{\longrightarrow_n} \mathbf{B}U(n) \vert
	  \\
	  & \simeq 
	  \lim\limits_{\longrightarrow_n} \vert \mathbf{B}U(n) \vert
	  \\
	  & \simeq \lim\limits_{\longrightarrow_n}  B U(n)
      \\
      & \simeq B U
      \,.	  
	\end{aligned}
  $$
\endofproof
\begin{corollary}
  \label{CohomologyInbfBUIsReducedK}
  For $X \in \mathrm{SmthMfd} \hookrightarrow \mathrm{Smooth}\infty \mathrm{Grpd}$,
  the intrinsic cohomology of $X$ with coefficients in the smooth stack 
  $\mathbf{B}U$ is the reduced K-theory $\tilde K(X)$:
  $$
    H^1_{\mathrm{smooth}}(X, U)
	:=
	\pi_0 \mathbf{H}(X, \mathbf{B}U)
	\simeq
	\tilde K(X)
	\,.
  $$
\end{corollary}
\proof
  By prop. \ref{CohomologyByCovers} 
  the set $\pi_0 \mathbf{H}(X, \mathbf{B}U)$ is the {\v C}ech cohomology
  of $X$ with coefficients in the stable unitary group $U$. By
  classification theory (as discussed in \cite{RobertsStevenson})
  this is isomorphic to the set of homotopy classes of maps
  $\pi_0 \mathrm{Top}(X, B U)$ into the classifying space $B U$
  for reduced K-theory.
\endofproof
\begin{proposition}
  Let $X$ be a compact smooth manifold. Then 
  $$
    \mathbf{H}(X, \mathbf{B}U) \simeq \lim\limits_{\longrightarrow_n} \mathbf{H}(X, \mathbf{B}U(n))
  $$
  and
  $$
    \mathbf{H}(X, \mathbf{B}PU) \simeq \lim\limits_{\longrightarrow_n} 
	\mathbf{H}(X, \mathbf{B}PU(n))
	\,.
  $$
\end{proposition}
\proof
  That $X$ is a compact manifold means by def. \ref{representably compact}
  that it is a \emph{representably compact object} in the site $\mathrm{SmthMfd}$.
  Since $X$ is in particular paracompact, 
  prop. \ref{HypercoversOverParacompactaAreRefinedByCechNerves} says that it
  is also a \emph{representably paracompact object} in the site, 
  def. \ref{representaby paracompact}. With this the statement is
  given by prop. \ref{RepresentablyCompactAndParacompactImpliesDistributivityOverFilteredColimits}.
\endofproof

We now discuss twisted bundles
induced by the  local coefficient bundles $\mathbf{dd}_n : \mathbf{B} PU(n) \to \mathbf{B}^2 U(1)$
for every $n \in \mathbb{N}$. This is immediately generalized to
general central extensions.

So let $U(1) \to \hat G \to G$ be any $U(1)$-central extension of a Lie group $G$
and let $\mathbf{c} : \mathbf{B}G \to \mathbf{B}^2 U(1)$ the classifying morphism
of moduli 2-stacks,
according to prop. \ref{LongFiberSequenceFromCentralExtensionOfGroups}, sitting
in the fiber sequence
$$
  \xymatrix{
    \mathbf{B}\hat G \ar[r] & \mathbf{B}G  \ar[d]^{\mathbf{c}}
	\\	 
	& \mathbf{B}^2 U(1)
	}
	\,.
$$

\begin{proposition}
  \label{CechCocyclesForTwisted1Bundles}
Let $U(1) \to \hat G \to G$ be a group extension of Lie groups.
Let $X \in \mathrm{SmoothMfd} \hookrightarrow \mathrm{Smooth}\infty \mathrm{Grpd}$
be a smooth manifold with differentiably good open cover $\{U_i \to X\}$.
\begin{enumerate}
\item Relative to this data every twisting cocycle $[\alpha] \in H^2_{\mathrm{Smooth}}(X, U(1))$ 
is a {\v C}ech-cohomology representative given by a collection of functions
   $$
     \{ \alpha_{i j k} : U_i \cap U_j \cap U_k \to U(1) \}
   $$
   satisfying on every quadruple intersection the equation
   $$
     \alpha_{i j k} \alpha_{i k l} = \alpha_{j k l} \alpha_{i j l}
     \,.
   $$
\item  In terms of this cocycle data, the twisted cohomology $H^1_{[\alpha]}(X, \hat G)$ is given by equivalence 
  classes of cocycles consisting of
  \begin{enumerate}
   \item collections of functions
      $$
        \{g_{i j} : U_i \cap U_j \to \hat G \}
      $$
      subject to the condition that on each triple overlap the equation
      $$
        g_{i j} \dot g_{j k} = g_{i k} \cdot \alpha_{i j k}
      $$
      holds, where on the right we are injecting 
      $\alpha_{i j k}$ via $U(1) \to \hat G$ into $\hat G$  
      and then form the product there;

   \item subject to the equivalence relation that identifies two such collections of cocycle data $\{g_{i j}\}$ and $\{g'_{i j}\}$ if there exists functions
      $$
        \{h_i : U_i \to \hat G\}
      $$
      and
      $$
        \{\beta_{i j} : U_i \cap U_j \to \hat U(1)\}
      $$
      such that
      $$
        \beta_{i j} \beta_{j k} = \beta_{i k}
      $$
      and
      $$
        g'_{i j} = h_i^{-1} \cdot g_{i j} \cdot h_j \cdot \beta_{i j}   
        \,.
      $$
  \end{enumerate}
\end{enumerate}
\end{proposition}
\proof
We pass to the standard presentation of $\mathrm{Smooth}\infty \mathrm{Grpd}$ by the projective local 
model structure on simplicial presheaves over the site $\mathrm{CartSp}_{\mathrm{smooth}}$. 
There we compute the defining $\infty$-pullback by a homotopy pullback, according to
remark \ref{ComputingHomotopyPullbacks}.

Write $\mathbf{B}\hat G_{\mathrm{ch}}, \mathbf{B}^2 U(1)_{\mathrm{ch}} 
\in [\mathrm{CartSp}^{\mathrm{op}}, \mathrm{sSet}]$ etc. 
for the standard models of the abstract objects of these names by simplicial presheaves, as 
discussed in \ref{SmoothStrucCohesiveInfiniGroups}. 
Write accordingly $\mathbf{B}(U(1) \to {\hat G})_{\mathrm{ch}}$ 
for the delooping of the crossed module 2-group associated to the central extension $\hat G \to G$.

By prop. \ref{LongFiberSequenceFromCentralExtensionOfGroups},
in terms of this the characteristic class $\mathbf{c}$ is represented by the  $\infty$-anafunctor
$$
  \xymatrix{
    \mathbf{B}(U(1) \to \hat G)_{\mathrm{ch}} \ar[r]^<<<<<{\mathbf{c}}  \ar[d]^{\simeq} 
     & 
     \mathbf{B}(U(1) \to 1)_{\mathrm{ch}}
     =
     \mathbf{B}^2 U(1)_{\mathrm{ch}}
    \\
    \mathbf{B}G_{\mathrm{ch}}
  }
  \,,
$$
where the top horizontal morphism is the evident projection onto the $U(1)$-labels.
Moreover, the {\v C}ech nerve of the good open cover $\{U_i \to X\}$ forms a cofibrant resolution
$$
  \emptyset \hookrightarrow C(\{U_i\}) \stackrel{\simeq}{\to} X
$$
and so $\alpha$ is presented by an $\infty$-anafunctor
$$
  \xymatrix{
     C(\{U_i\}) \ar[r]^{\alpha} \ar[d]^{\simeq} & \mathbf{B}^2 U(1)_c
     \\
     X
  }
  \,.
$$
Using that $[\mathrm{CartSp}^{\mathrm{op}}, \mathrm{sSet}]_{\mathrm{proj}}$ is a simplicial model category 
this means in conclusion that the homotopy pullback in question is given by the ordinary pullback of simplicial sets
$$
  \xymatrix{
     \mathbf{H}^1_{[\alpha]}(X,\hat G) \ar[r] \ar[d]& {*} \ar[d]^{\alpha}
     \\
     [\mathrm{CartSp}^{\mathrm{op}}, \mathrm{sSet}](C(\{U_i\}), \mathbf{B}(U(1) \to \hat G)_c)
     \ar[r]^{\mathbf{c}_*} &
     [\mathrm{CartSp}^{\mathrm{op}}, \mathrm{sSet}](C(\{U_i\}), \mathbf{B}^2 U(1)_c)
  }
  \,.
$$
An object of the resulting simplicial set is then seen to be a simplicial map 
$g : C(\{U_i\}) \to \mathbf{B}(U(1) \to \hat G)_c$ that assigns
$$
  g
  \;\;
  :
  \;\;
  \raisebox{20pt}{
  \xymatrix{
    & (x,j) \ar[dr]
    \\
    (x,i)
    \ar[rr]^{\ }="t"
    \ar[ur]
    &&
    (x,k)
    \ar@{=} "t"+(0,5); "t"
  }}
  \;\;\;\;
  \mapsto
  \;\;\;\;
  \raisebox{20pt}{
  \xymatrix{
    & {*} \ar[dr]^{g_{j k}(x)}
    \\
    {*}
    \ar[rr]_{g_{i k}(x)}^{\ }="t"
    \ar[ur]^{g_{i j}(x)}
    &&
    {*}
    \ar@{=>}|{\alpha_{i j k}(x)} "t"+(0,5); "t"
  }
  }
$$
such that projection out along $\mathbf{B}(U(1) \to \hat G)_c \to \mathbf{B}(U(1) \to 1)_c = \mathbf{B}^2 U(1)_c$ produces $\alpha$.

Similarily for the morphisms. Writing out what these diagrams in $\mathbf{B}(U(1) \to \hat G)_c$ mean in equations, one finds the formulas claimed above.
\endofproof

\subsubsection{Reduction of structure groups}
\label{ReductionOfTheStructureGroup}
\index{twisted cohomology!reduction of structure group}

We discuss the traditional notion of \emph{reduction} of a structure
group in terms of 
twisted differential nonabelian cohomology. This perspective
turns out to embed this standard notion seamlessly into more general
notion of
twisted differential structures, def. \ref{TwistedCStructures}. 
Conversely, this prespective shows that 
the general notion of twisted differential structures may be thought
of as a generalization of the classical notion of reduction of
structure groups from principal bundles to 
principal $\infty$-bundles.

\medskip

Let $G$ be a Lie group and let $K \hookrightarrow G$ be a Lie 
subgroup. Write 
$$
  \mathbf{c} : \mathbf{B}K \to \mathbf{B}G
$$
for the induced morphism of smooth moduli stacks of 
smooth principal bundles,
according to prop. \ref{DeloopedLieGroup}.
\begin{observation}
  \label{HomotopyFiberOfDeloopedLieGroupInclusion}
  The action groupoid $G /\!/K$, def. \ref{ActionGroupoidInIntroduction},
  is 0-truncated, hence the canonical morphism to the 
  smooth manifold quotient
  $$
    G /\!/K \stackrel{\simeq}{\to} G/K
  $$
  is an equivalence in $\mathrm{Smooth}\infty\mathrm{Grpd}$.
  
  We have a fiber sequence of smooth stacks
  $$ 
    G / K \to \mathbf{B}K \to \mathbf{B}G
	\,.
  $$
  This is presented by the evident sequence of simplicial presheaves
  $$
    G /\!/K \to * /\!/ K \to * /\!/ G
	\,.
  $$
\end{observation}
\proof
   The equivalence follows because the action of a subgroup is free.
   The fiber sequence may be computed for instance with the
   factorization lemma, prop. \ref{FactorizationLemma}.
\endofproof
In applications, an important class of examples is the following.
\begin{observation}
  \label{InclusionOfMaximalCompactSubgroup}
 For $G$ a conneced Lie group, let $K \hookrightarrow G$ be the 
 inclusion of its maximal compact subgroup. Then
 $\mathbf{c} : \mathbf{B}K \to \mathbf{B}G$ is a $\Pi$-equivalence,
 def. \ref{PiEquivalence} (hence becomes an equivalence under
 geometric realization, def. \ref{GeometricRealization}).
 Therefore, while the groupoids of $K,G$-principal bundles
 are different and
 $$
   \mathbf{H}(X, \mathbf{B}K) \to \mathbf{H}(X, \mathbf{B}G)
 $$
 is not an equivalence, unless $G$ is itself already compact,
 it does induce an isomorphism on connected components (nonabelian cohomology sets)
 $$
   H^1(X, K) \stackrel{\simeq}{\to} H^1(X, G)
   \,.
 $$
 In the following discussion this difference between the 
 classifying spaces $B G \simeq \Pi (\mathbf{B}G) \simeq \Pi(\mathbf{B}K) \simeq B K$ and their 
 smooth refinements is crucial.
 
 Theorem \ref{GeometricRealizationOfHomotopyFibers} 
 in the present case says that
 $\Pi(G/K) \simeq *$ contractible. This recovers the classical 
 statement that, as a topological space, $G$ is a product of its maximal compact
 subgroup with a contractible space.
\end{observation}
\proof
  It is a classical fact that the maximal compact subgroup 
  inclusion $K \hookrightarrow G$ is a homotopy equivalence
  on the underlying topological spaces. 
  The statement then follows by prop. 
  \ref{HomotopyEquivalenceOfTopologicalGroupsDeloopsToPiEquivalence}.
\endofproof

Given a subgroup inclusion $K \hookrightarrow G$ and a $G$-principal
bundle $P$, a standard question is whether the structure group
of $P$ may be reduced to $K$.  
\begin{definition}
  \label{GroupoidOfKReductions}
  Let $K \hookrightarrow G$ be an inclusion of Lie groups 
  and let $X \in \mathrm{Smooth}\infty\mathrm{Grpd}$
  be any object (for instance a smoot manifold).  Let 
  $g : X \to \mathbf{B}G$
 be a smooth classifying morphism for a $G$-principal bundle $P \to X$. 
 
 A choice of \emph{reduction of the structure group} of $G$ along $K \hookrightarrow G$ 
  (or \emph{$K$-reduction} for short) is a 
 choice of lift $g_{\mathrm{red}}$ and a 
choice of homotopy (gauge transformation) $\eta$ of smooth stacks in the diagram
$$
  \raisebox{20pt}{
  \xymatrix{
    & \mathbf{B}K
	\ar[d]^{\mathbf{c}}_<<<{\ }="s"
    \\
    X
	\ar[r]_g^{\ }="t"
	\ar[ur]^{g_{\mathrm{red}}}
	&
	\mathbf{B}G
	\ar@{=>}^\eta|\simeq "s"; "t"
  }
  }
  \,.
$$
For $(g_{\mathrm{red}}, \eta)$ and $(g'_{\mathrm{red}}, \eta')$ two $K$-reductions
of $P$, an \emph{isomorphism} of $K$-reductions from the first to the second
is a natural transformation of morphisms of smooth stacks
$$  
  \xymatrix{
    X 
	\ar@/^1pc/[rr]^{g_{\mathrm{red}}}_{\ }="s"
	\ar@/_1pc/[rr]_{g'_{\mathrm{red}}}^{\ }="t"
	&&
	\mathbf{B}K
	\ar@{=>}^{\rho} "s"; "t"
  }
  \,,
$$
hence a choice of gauge transformation between the corresponding $K$-principal bundles,
such that 
$$
  \raisebox{20pt}{
  \xymatrix{
    & \mathbf{B}K
	\ar[d]^{\mathbf{c}}_<<<{\ }="s"
    \\
    X
	\ar[r]_g^{\ }="t"
	\ar[ur]|{g'_{\mathrm{red}}}^{\ }="t2"
	\ar@/^2pc/[ur]^{g_{\mathrm{red}}}_{\ }="s2"
	&
	\mathbf{B}G
	\ar@{=>}^{\eta'}|\simeq "s"; "t"
	\ar@{=>}^{\rho}|\simeq "s2"; "t2"
  }
  }
  \;\;\;
  =
  \;\;\;
  \raisebox{20pt}{
  \xymatrix{
    & \mathbf{B}K
	\ar[d]^{\mathbf{c}}_<<<{\ }="s"
    \\
    X
	\ar[r]_g^{\ }="t"
	\ar[ur]^{g_{\mathrm{red}}}
	&
	\mathbf{B}G
	\ar@{=>}^\eta|\simeq "s"; "t"
  }
  }
  \,.
$$
With the obvious notion of composition of such isomorphisms, 
this defines a \emph{groupoid of $K$-reductions} of $P$.
\end{definition}
\begin{remark}
  \label{TheRoleOfEtaInKReduction}
  The crucial information is in the \emph{choice} of the smooth transformation
  $\eta$. Notably in the case that $K \hookrightarrow G$ is the
  inclusion of a maximal compact subgroup 
  as in observation \ref{InclusionOfMaximalCompactSubgroup} the
  underlying reduction problem after geometric 
  realization in the homotopy theory of topological spaces is
  trivial: all bundles involved in the above are equivalent. 
  The important information in $\eta$ is about \emph{how}
  they are chosen to be equivalent, and smoothly so.
  
  Below in \ref{OrthogonalRiemannianStructureByTwistedCohomology}
  we see that in the case that $P = T X $ is the tangent bundle 
  of a manifold, $\eta$ is identified with a choice of \emph{vielbein}
  or \emph{soldering form}.
\end{remark}

Comparison with the discussion in \ref{StrucTwistedCohomology} reveals that
therefore structure group reduction is a topic in 
\emph{twisted nonabelian cohomology}. In particular, we may
apply def. \ref{TwistedCStructures} to form the 
groupoid of all choices of reductions.
\begin{proposition}
  \label{ReductionGroupoidAsTwistedCohomology}
  For $g : X \to \mathbf{B}G$ (the cocycle for) a $G$-principal bundle $P \to X$,
  the \emph{groupoid of $K$-reductions}
  of $P$ according to def. \ref{GroupoidOfKReductions}
  is the groupoid of $[g]$-twisted $\mathbf{c}$-structures, def. \ref{TwistedCStructures}, 
  hence the
  homotopy pullback $\mathbf{c}\mathrm{Struc}_{[g]}(X)$
  in
  $$
    \raisebox{20pt}{
    \xymatrix{
	  \mathbf{c}\mathrm{Struc}_{[g]}(X)
	  \ar[d]
	  \ar[rr]
	  &&
	  {*}
	  \ar[d]^{g}
	  \\
	  \mathbf{H}(X, \mathbf{B} K)
	  \ar[rr]^{\mathbf{H}(X, \mathbf{c})}
	  &&
	  \mathbf{H}(X, \mathbf{B}G)
	}
	}
	\,,
  $$
  where 
  $$
    \mathbf{c} : \mathbf{B}K \to \mathbf{B}G
  $$
  is the induced morphism of smooth moduli stacks.
\end{proposition}
\proof
  Using that $\mathbf{B}K$ and $\mathbf{B}G$
  are 1-truncated objects in $\mathbf{H} := \mathrm{Smooth}\infty\mathrm{Grpd}$,
  by construction, one sees that the groupoid defined in 
  def. \ref{GroupoidOfKReductions} is equivalently the 
  hom-groupoid $\mathbf{H}_{/\mathbf{B}G}(g,\mathbf{c})$ 
  in the slice $\infty$-topos $\mathbf{H}_{/\mathbf{B}G}$.
  Using this, the statement is a special case 
  of prop. \ref{PullbackCharacterizationOfTwistedCohomology}.
\endofproof
\begin{remark}
  By observation \ref{HomotopyFiberOfDeloopedLieGroupInclusion} we may equivalently 
  speak of $\mathbf{c}\mathrm{Struc}_{g}(X)$ as the 
  \emph{groupoid of twisted $G/\!/K$-structures}
  on $X$ (where the latter is given by a corresponding groupoid-principal bundle).
  
  If we think, according to remark \ref{TheRoleOfEtaInKReduction}, 
  of a choice of $K$-reduction as a choice of \emph{vielbein} or
  \emph{soldering form}, then this says that \emph{locally} their moduli 
  space is the cose $G/K$ (while globally there may be a twist).
\end{remark}
The morphism $\mathbf{c}$ as above always has a canonical differential refinement
$$
  \hat {\mathbf{c}} : \mathbf{B}K_{\mathrm{conn}} \to \mathbf{B}G_{\mathrm{conn}}
$$
given by prop. \ref{GroupoidOfLieAlgebraValuedForms}. Accordingly,
we may also apply def. \ref{twisteddifferentialcstructures} to the case of 
structure group reduction.
\begin{definition}
  \label{GroupoidOfDifferentialKReductions}
  For $K \to G$ a Lie subgroup inclusion, 
  and for $\nabla : X \to \mathbf{B}G_{\mathrm{conn}}$
  (a cocycle for ) a $G$-principal bundle with conneciton on $X$,
  we say the \emph{groupoid of $K$-reductions} of $\nabla$
  is the groupoid $\hat {\mathbf{c}}\mathrm{Struc}_{[\nabla]}(X)$
  of \emph{twisted differential $\hat {\mathbf{c}}$-structures},
  given as the homotopy pullback
  $$
    \raisebox{20pt}{
    \xymatrix{
	  \hat {\mathbf{c}}\mathrm{Struc}_{[\nabla]}(X)
	  \ar[d]
	  \ar[rr]
	  &&
	  {*}
	  \ar[d]^{\nabla}
	  \\
	  \mathbf{H}(X, \mathbf{B} K_{\mathrm{conn}})
	  \ar[rr]^{\mathbf{H}(X, \hat {\mathbf{c}})}
	  &&
	  \mathbf{H}(X, \mathbf{B}G_{\mathrm{conn}})
	}
	}
	\,.
  $$
\end{definition}
However, here the differential refinement does not change the 
homotopy type of the twisted cohomology
\begin{proposition}
  \label{DifferentialAndBareReductionsAreEquivalent}
  For $P$ a $G$-principal bundle with connection $\nabla$ the
  groupoid of $K$-reductions of $\nabla$ is equivalent to the 
  groupoid of $K$-reductions of just $P$
  $$
    \hat {\mathbf{c}}\mathrm{Struc}_{[\nabla]}(X)
	\simeq
	\mathbf{c}\mathrm{Struc}_{[P]}(X)
	\,.
  $$
\end{proposition}
\begin{remark}
  This degeneracy of notions does not hold for twisted
  structures controled by higher groups. That it holds in the
  special case of ordinary $K$-reductions is an incarnation
  of a classical fact in differential geometry: as we will
  see in \ref{OrthogonalRiemannianStructureByTwistedCohomology}
  below, for reductions of tangent bundle structure it comes
  down to the fact that for every choice of 
  Riemannian metric and torsion
  there is a unique metric-compatible connection with that 
  torsion. Prop. \ref{DifferentialAndBareReductionsAreEquivalent}
  may be understood as stating this in the fullest
  generality of $G$-principal bundles for $G$ a Lie group.
\end{remark}

\paragraph{Orthogonal/Riemannian structure}
\label{OrthogonalRiemannianStructureByTwistedCohomology}
\index{twisted cohomology!orthogonal structure}
\index{twisted cohomology!Riemannian structure}

For $X$ a smooth manifold, we discuss the traditional notion of \emph{Riemannian} structure
or equivalently of \emph{orthogonal structure} on $X$ as a special case of 
$\mathbf{c}$-twisted cohomology for suitable $\mathbf{c}$. This perspective
on ordinary Riemannian geometry proves to be a useful starting point for
generalizations.

\medskip

Let $X$ be a smooth manifold of dimension $d$. Its tangent bundle $T X$ is 
associated to an essentially canonical $\mathrm{GL}(d)$-principal bundle. We write
$$
  T X : X \to \mathbf{B}\mathrm{GL}(d)
$$
for the corresponding classifying morphism, where $\mathbf{B}\mathrm{GL}(d)$
is the smooth moduli stack of smooth $\mathrm{GL}(d)$-principal bundles. 

Consider the defining inclusion of Lie groups
$$
  \mathrm{O}(d) \hookrightarrow \mathrm{GL}(d)
$$
and the induced morphism of the corresponding moduli stacks
$$
  \mathbf{orth} : \mathbf{B} \mathrm{O}(d) \to \mathbf{B} \mathrm{GL}(d)
  \,.
$$
The general observation \ref{HomotopyFiberOfDeloopedLieGroupInclusion} here reads
\begin{observation}
  \label{HomotopyFiberOfOrth}
  The homotopy fiber of $\mathbf{orth}$ is the quotient manifold 
  $\mathrm{GL}(d)/ \mathrm{O(d)}$. We have a fiber sequence of smooth stacks
  $$
    \xymatrix{
      \mathrm{GL}(d)/ \mathrm{O(d)}
	  \ar[r]
	  &
	  \mathbf{B}\mathrm{O}(d)
	  \ar[r]^{\mathbf{orth}}
	  &
	  \mathbf{B}\mathrm{GL}(d)
	}
	\,.
  $$
\end{observation}
Notice that $\mathrm{O}(d) \hookrightarrow \mathrm{GL}(d)$ is a 
maximal compact subgroup inclusion, so that observation
\ref{InclusionOfMaximalCompactSubgroup} applies.

Definition \ref{ReductionGroupoidAsTwistedCohomology} now becomes
\begin{definition}
  Write $\mathbf{orth}\mathrm{Struc}_{T X}$
  for the groupoid of $T X$-twisted $\mathbf{orth}$-structures on $X$, 
  hence the homotopy pullback in 
  $$
    \raisebox{20pt}{
    \xymatrix{
	   \mathbf{orth}\mathrm{Struc}(X)
	   \ar[rr] 
	   \ar[d]^>{\ }="t"
	   &&
	   {*}
	   \ar[d]^{T X}_<{\ }="s"
	   \\
	   \mathbf{H}(X, \mathbf{B} \mathrm{O}(d))
	   \ar[rr]_{\mathbf{H}(X,\mathbf{orth})}
	   &&
	   \mathbf{H}(X, \mathbf{B} \mathrm{GL}(d))
	   \ar@{=>}_\simeq "s"; "t"
	}
	}
	\,.
  $$\
\end{definition}
\begin{proposition}
  \label{VielbeinAsOrthogonalStructure}
  The groupoid $\mathbf{orth}\mathrm{Struc}_{T X}(X)$
  is naturally identified with the groupoid of choices of vielbein fields
  (soldering forms) on $T X$.
\end{proposition}
\proof
  Let $\{U_i \to X\}$ be any good open cover of $X$ by coordinate
  patches $\mathbb{R}^d \simeq U_i$. Let $C(\{U_i\})$ be the corresponding
  {\v C}ech groupoid. There is then a canonical span 
  of simplicial presheaves
  $$
    \raisebox{20pt}{
    \xymatrix{
      C(\{U_i\}) \ar[r]^{T X_{\mathrm{ch}}} \ar[d]^\simeq  &  \mathbf{B} \mathrm{GL}(d)_{\mathrm{ch}}
	  \\
	  X
	}
	}
	\,.
  $$
  presenting $T X$. Moreover, every morphism $g : X \to \mathbf{B} \mathrm{O}(d)$
  has a presentation by a similar span $g_{\mathrm{ch}}$ with values in 
  $\mathbf{B}\mathrm{O}(d)$. 
  
  An object in $\mathbf{orth}\mathrm{Struc}_{T X}(X)$ is
  \begin{enumerate}
    \item 
	  a cocycle $g_{\mathrm{ch}}$ for an $\mathrm{O}(d)$-principal bundle as above;
	\item
	  over each $U_i$ an element $e|_{U_i} \in C^\infty(U_i , \mathrm{GL}(d))$
  \end{enumerate}
  such that $e$ is compatible, on double overlaps, with the left $\mathrm{O}(d)$-action
  by the transition functions $g_{\mathrm{ch}}$ and the right $\mathrm{GL}(d)$-action
  by the transitiuon functions $T X_{\mathrm{ch}}$. 
  
  A morphism $e \to e'$ in $\mathbf{orth}\mathrm{Struc}_{T X}(X)$ is a 
  gauge transformation $g_{\mathrm{ch}} \to g'_{\mathrm{ch}}$ of 
  $\mathrm{O}(d)$-principal bundles whose left action takes $e$ to $e'$.
  
  From this it is clear that 
  $$
    e = \{e^a{}_{\mu}\}_{a,\mu \in \{1, \cdots, d\}}
  $$  
  is a choice of vielbein. 
\endofproof
There is an evident differential refinement of $\mathbf{orth}$
$$
  \hat {\mathbf{orth}} : \mathbf{B} \mathrm{O}(d)_{\mathrm{conn}}
    \to \mathbf{B} \mathrm{GL}(d)_{\mathrm{conn}}
	\,.
$$
\begin{definition}
  \label{GroupoidOfConnectionsOnTangentBundle}
  Let $\mathrm{Conn}T X \to \mathbf{H}(X, \mathbf{B}\mathrm{GL}(d)_{\mathrm{conn}})$ 
  be the left vertical morphism in the homotopy pullback
  $$
    \raisebox{20pt}{
    \xymatrix{
	  \mathrm{Conn} T X 
	  \ar[r]
	  \ar[d]
	  &
	  {*}
	  \ar[d]^{T X}
	  \\
	  \mathbf{H}(X, \mathbf{B}\mathrm{GL}(d)_{\mathrm{conn}})
	  \ar[r]
	  &
	  \mathbf{H}(X, \mathbf{B}\mathrm{GL}(d))
	}
	}
	\,,
  $$
  where the bottom map is the morphism that forgets the connection.
\end{definition}
This morphism may be thought of as the inclusion of 
connections on the tangent bundle into the groupoid of all
$\mathrm{GL}(d)$-principal connections.
\begin{proposition}
  \label{VielbeinWithSpinConnectionFromTwistedCohomology}
  The homotopy pullback in 
  $$  
    \raisebox{20pt}{
	\xymatrix{
	   \hat {\mathbf{orth}}\mathrm{Struc}_{TX, \mathrm{conn}}(X)
	   \ar[d]
	   \ar[rr]
	   &&
	   \mathrm{Conn} T X
	   \ar[d]
	   \\
	   \mathbf{H}(X, \mathbf{B}\mathrm{O}(d)_{\mathrm{conn}})
	   \ar[rr]^{\mathbf{H}(X, \hat {\mathbf{orth}})}
	   &&
	   \mathbf{H}(X, \mathbf{B}\mathrm{GL}(d)_{\mathrm{conn}})
	}
	}
  $$
  or equivalently that in
  $$  
    \raisebox{20pt}{
	\xymatrix{
	   \hat {\mathbf{orth}}\mathrm{Struc}_{TX, \mathrm{conn}}(X)
	   \ar[d]
	   \ar[rr]
	   &&
	   {*}
	   \ar[d]^{T X}
	   \\
	   \mathbf{H}(X, \mathbf{B}\mathrm{O}(d)_{\mathrm{conn}})
	   \ar[rr]^{}
	   &&
	   \mathbf{H}(X, \mathbf{B}\mathrm{GL}(d))
	}
	}
  $$
  is equivalent to the set of pairs of Riemannian metrics on $X$ 
  and correspondingly metric-compatible connections on $T X$.
\end{proposition}
\proof 
  The two pullbacks are equivalent by def. \ref{GroupoidOfConnectionsOnTangentBundle}
  and the pasting law, prop. \ref{PastingLawForPullbacks}.
  
  Consider the 
  first version. As in the proof of prop. \ref{VielbeinAsOrthogonalStructure}
  an object in the groupoid has an underlying choice of vielbein
  $e$. This now being a morphism of bundles with connection, 
  it related, locally on each $U_i$, the goven connection form
  $\Gamma$ on $T X$ with a connection form $\omega$ on the 
  $\mathrm{O}(d)$-principal bundle, via
  $$
    \omega^a{}_b = 
	  e^a{}_{\alpha} \Gamma^\alpha{}_\beta (e^{-1})^b{}_{\beta}
	 + 
	 e^a{}_{\alpha} d_{\mathrm{dR}} (e^{-1})^b{}_{\beta}
	 \,.
  $$
  But since $\omega$ is by definition an orthogonal connection, by this
  isomorphism $\Gamma$ is a metric-compatible connection. 
\endofproof

\paragraph{Type II NS-NS generalized geometry}
\label{TypeIIGeometryByTwistedCohomology}
\index{twisted cohomology!type II generalized geometry}

The target space geometry for type II superstrings 
in the NS-NS sector is naturally
encoded by a variant of ``generalized complex geometry'' with 
metric structure, discussed for instance in \cite{GMPW}.
We discuss here how this \emph{type II NS-NS generalized geometry}
is a special case of twisted $\mathbf{c}$-structures
as in \ref{ReductionOfTheStructureGroup}.

\medskip

\begin{definition}
 \label{InclusionForTypeIIGeometry}
Consider  the Lie group inclusion
$$
  \mathrm{O}(d) \times \mathrm{O}(d)
  \to 
  \mathrm{O}(d,d)
$$
of those orthogonal transformations, that preserve the positive definite part
or the negative definite part of the bilinear form of signature $(d,d)$, respectively.

If $\mathrm{O}(d,d)$ is presented as the group of $2d \times 2d$-matrices that 
preserve the bilinear form given by the $2d \times 2d$-matrix
$$
  \eta :=
  \left( 
     \begin{array}{cc}
	    0 & \mathrm{id}_d
		\\
		\mathrm{id}_d & 0
	 \end{array}
  \right)
$$
then this inclusion sends a pair $(A_+, A_-)$ of orthogonal $n \times n$-matrices
to the matrix
$$
  (A_+ , A_-) 
    \mapsto 
  \frac{1}{\sqrt{2}}
  \left(
    \begin{array}{cc}
	   A_+ + A_- & A_+ - A_-
	   \\
	   A_+ - A_- & A_+ + A_-
	\end{array}
  \right)
  \,.
$$
\end{definition}
The inclusion of Lie groups induces the corresponding morphism of smooth moduli stacks
of principal bundles
$$
  \mathbf{TypeII}
  :
  \mathbf{B}(\mathrm{O}(d) \times \mathrm{O}(d))
  \to 
  \mathbf{B} \mathrm{O}(d,d)  
  \,.
$$
Observation \ref{HomotopyFiberOfDeloopedLieGroupInclusion} here becomes
\begin{observation}
  There is a fiber sequence of smooth stacks
  $$
    \xymatrix{
      \mathrm{O}(d,d)/(\mathrm{O}(d) \times \mathrm{O}(d))
	   \ar[r] 
	   &
  \mathbf{B}(\mathrm{O}(d) \times \mathrm{O}(d))
    \ar[rr]^{\mathbf{TypeII}}
	 &&
    \mathbf{B} \mathrm{O}(d,d)  	
	}
  \,.
  $$
\end{observation}
\begin{definition}
There is a canonical embedding
$$
  \mathrm{GL}(d) \hookrightarrow \mathrm{O}(d,d)
  \,.
$$
In the above matrix presentation this is given by sending
$$
  a
  \mapsto 
  \left( 
     \begin{array}{cc}
	    a & 0
		\\
		0 & a^{-T}
	 \end{array}
  \right)
  \,,
$$
where in the bottom right corner we have the transpose of the inverse matrix of the invertble
matrix $a$.
\end{definition}
\begin{observation}
  We have a homotopy pullback of smooth stacks
  $$
    \raisebox{20pt}{
    \xymatrix{
	   \mathrm{GL}(d) \backslash \!\backslash \mathrm{O}(d,d)/\!/( \mathrm{O}(d)\times  \mathrm{O}(d))
	   \ar[r]
	   \ar[d]
	   & \mathbf{B} \mathrm{GL}(d)
	   \ar[d]
	   \\
	   \mathbf{B} (\mathrm{O}(d) \times \mathrm{O}(d))
	   \ar[r]
	   &
	   \mathbf{B} \mathrm{O}(d,d)
	}
	}
	\,.
  $$
\end{observation}
\begin{definition}
Under inclusion def. \ref{InclusionForTypeIIGeometry} the tangent bundle of a $d$-dimensional 
manifold $X$ defines an $\mathrm{O}(d,d)$-cocycle
$$
  T X \otimes T^* X
  :
  \xymatrix{
    X \ar[r]^<<<<{T X} 
	&
	 \mathbf{B}\mathrm{GL}(d) \ar[r] 
	& 
	\mathbf{B} \mathrm{O}(d,d) 
  }
  \,.
$$
\end{definition}
The vector bundle canonically associated to this composite cocycles may 
canonically be identified with
the tensor product vector bundle $T X \otimes T^* X$, and so we will
refer to this cocycle by these symbols, as indicated. 

Therefore we may canonically consider the groupoid of 
$T X \otimes T^* X$-twisted $\mathbf{TypeII}$-structures, 
according to def. \ref{ReductionGroupoidAsTwistedCohomology}:
\begin{definition}
  Write $\mathbf{TypeII}\mathrm{Struc}_{T X \otimes T^* X}(X)$
  for the homotopy pullback
  $$
    \raisebox{20pt}{
    \xymatrix{
	  \mathbf{TypeII}\mathrm{Struc}_{T X \otimes T^* X}(X)
	  \ar[rr]
	  \ar[d]
	  &&
	  {*}
	  \ar[d]^{T X \otimes T^* X}
	  \\
	  \mathbf{H}(X, \mathbf{B}(O(d) \times O(d)))
	  \ar[rr]^{\mathbf{H}(X, \mathbf{TypeII})}
	  &&
	  \mathbf{H}(X, \mathbf{B} O(d,d))
	}
	}
	\,.
  $$
\end{definition}
\begin{proposition}
  The groupoid $\mathbf{TypeII}\mathrm{Struc}_{T X \otimes T^* X}(X)$
  is that of ``generalized vielbein fields'' on $X$, as considered for instance 
  around equation (2.24) of \cite{GMPW}
  (there only locally, but the globalization is evident).
  
  In particular, its set of equivalence classes is the set of 
  type-II generalized geometry structures on $X$.
\end{proposition}
\proof
  This is directly analogous to the proof of prop. \ref{VielbeinAsOrthogonalStructure}.
\endofproof
Over a local patch $\mathbb{R}^d \simeq U_i \hookrightarrow X$, the most general such generalized vielbein
(hence the most general $\mathrm{O}(d,d)$-valued function)
may be parameterized as
$$
  E = 
  \frac{1}{2}
  \left(
    \begin{array}{cc}
	  (e_+ + e_-) + (e_+^{-T} - e_-^{-T})B & (e_+^{-T} - e_-^{-T})
	  \\
	  (e_+ - e_-) - (e_+^{-T} + e_-^{-T})B & (e_+^{-T} + e_-^{-T})
	\end{array}
  \right)
  \,,
$$
where $e_+, e_- \in C^\infty(U_i, \mathrm{O}(d))$ are thought of as two ordinary vielbein
fields, and where $B$ is any smooth skew-symmetric $n \times n$-matrix valued function on 
$\mathbb{R}^d \simeq U_i$. 

By an $\mathrm{O}(d) \times \mathrm{O}(d)$-transformation this can always be brought
into a form where $e_+ = e_- =: \tfrac{1}{2}e$ such that
$$
  E = 
  \left(
    \begin{array}{cc}
	  e & 0
	  \\
	  - e^{-T}B  & e^{-T}
	\end{array}
  \right)
  \,.
$$
The corresponding ``generalized metric'' over $U_i$ is
$$
  E^T 
  E
  = 
  \left(
    \begin{array}{cc}
	  e^T & B e^{-1}
	  \\
	  0  & e^{-1}
	\end{array}
  \right)
  \left(
    \begin{array}{cc}
	  e & 0
	  \\
	  - e^{-T}B  & e^{-T}
	\end{array}
  \right)
  = 
  \left(
    \begin{array}{cc}
	  g - B g^{-1} B & B g^{-1}
	  \\
	  - g^{-1} B  & g^{-1}
	\end{array}
  \right)
  \,,
$$
where
$$
  g := e^T e
$$
is the metric (over $\mathbb{R}^q \simeq U_i$ a smooth function with values in symmetric $n \times n$-matrices) 
given by the ordinary vielbein $e$.

\paragraph{U-duality geometry / exceptional generalized geometry }
\label{UDualityGeometry}
\index{twisted cohomology!exceptional generalized geometry}
\index{twisted cohomology!U-duality geometry}
\index{U-duality}

The scalar and bosonic fields of 11-dimensional supergravity
compactified on tori to dimension $d$ \emph{locally}
have moduli spaces identified with the quotients $E_{n(n)}/H_n$
of the split real form $E_{n(n)}$ in the E-series of exceptional 
Lie groups by their maximal compact subgroups $H_n$, where 
$n = 11 - d$. 
The canonical 
action of $E_{n(n)}$ on this coset space -- or of a certain discrete
subgroup $E_{n(n)}(\mathbb{Z}) \hookrightarrow E_{n(n)}$ -- is called
the \emph{U-duality} global symmetry of the supergravity, or of 
its string UV-completion, respectively \cite{HullTownsend}.

In \cite{Hull} it was pointed out that therefore the geometry of the
field content of compactfied supergravity should be encoded by a 
\emph{exceptional generalized geometry} which in direct analogy
to the variant of \emph{generalized complex geometry} that 
controls the NS-NS sector of type II strings, as discussed above in 
\ref{TypeIIGeometryByTwistedCohomology}, is encoded by vielbein fields
that exhibit reduction of a structure group along the inclusion
$H_n \hookrightarrow E_{n(n)}$.

By the general discussion in \ref{ReductionOfTheStructureGroup},
we have that all these geometries are encoded by twisted differential
$\mathbf{c}$-structures, where 
$$
  \mathbf{c} : \mathbf{B} H_n \to \mathbf{B} E_{n(n)}
$$
is the induced morphism of smooth moduli stacks.

\subsubsection{Orientifolds and higher orientifolds}
\label{higher orientifolds}
\label{OrientifoldCircleNBundlesWithConnection}
\index{orientifold}
\index{twisted cohomology!orientifold}

We discuss the notion of circle $n$-bundles with connection over double covering spaces with 
\emph{orientifold}\index{orientifold} structure (see \cite{SSW} and 
\cite{DistlerFreedMoore} for the notion of orientifolds for 2-bundles).

\medskip

\begin{proposition}
  \label{ExtensionOfZTwobyUOne}
The smooth automorphism 2-group of the circle group $U(1)$ is that corresponding to 
the smooth crossed module (as discussed in \ref{SheafAndNonabelianDoldKan})
$$
  \mathrm{AUT}(U(1)) \simeq [U(1) \to \mathbb{Z}_2]
  \,,
$$
where the differential $U(1) \to \mathbb{Z}_2$ is trivial and where the action of 
$\mathbb{Z}_2$ on $U(1)$ is given under the identification of $U(1)$ with the unit circle in the plane 
by reversal of the sign of the angle.

This is an extension of smooth $\infty$-groups, def. \ref{ExtensionOfInfinityGroups}, of 
$\mathbb{Z}_2$ by the circle 2-group $\mathbf{B}U(1)$:
$$
  \mathbf{B}U(1) \to \mathrm{AUT}(U(1)) \to \mathbb{Z}_2
  \,.
$$
\end{proposition}
\proof
The nature of $\mathrm{AUT}(U(1))$ is clear by definition. 
Let $\mathbf{B}U(1) \to \mathrm{AUT}(U(1))$ be the evident inclusion. 
We have to show that its delooping is the homotopy fiber of 
$\mathbf{B}\mathrm{AUT}(U(1)) \to \mathbf{B}\mathbb{Z}_2$.

Passing to the presentation of $\mathrm{Smooth}\infty \mathrm{Grpd}$ by
the model structure on simplicial presheaves 
$[\mathrm{CartSp}_{\mathrm{smooth}}^{\mathrm{op}}, \mathrm{sSet}]_{\mathrm{proj},\mathrm{loc}}$
and using prop. \ref{FiniteHomotopyLimitsInPresheaves}, it is sufficient to show 
that the simplicial presheaf $\mathbf{B}^2 U(1)_c$ from \ref{SmoothStrucCohesiveInfiniGroups} 
is equivalent to the  ordinary pullback of simplicial presheaves 
$\mathbf{B}\mathrm{AUT}(U(1))_c\times_{\mathbf{B}\mathbb{Z}_2} \mathbf{E}\mathbb{Z}_2$ 
of the $\mathbb{Z}_2$-universal principal bundle, as discussed in \ref{SmoothPrincipalnBundles}.

This pullback is the 2-groupoid whose 
\begin{itemize}
\item objects are elements of $\mathbb{Z}_2$;
\item morphisms $\sigma_1 \to \sigma_2$ are labeled by $\sigma \in \mathbb{Z}_2$ such that 
$\sigma_2 = \sigma \sigma_1$;
\item all 2-morphisms are endomorphisms, labeled by $c \in U(1)$;
\item vertical composition of 2-morphisms is given by the group operation in $U(1)$, 
\item horizontal composition of 1-morphisms with 1-morphisms is given by the group operation in $\mathbb{Z}_2$
\item horizontal composition of 1-morphisms with 2-morphisms (\emph{whiskering}) is given by the action of $\mathbb{Z}_2$ on $U(1)$.
\end{itemize}
Over each $U \in \mathrm{CartSp}$ this 2-groupoid has vanishing $\pi_1$, and $\pi_2 = U(1)$. 
The inclusion of $\mathbf{B}^2 U(1)$ into this pullback is 
given by the evident inclusion of elements in $U(1)$ as endomorphisms
of the neutral element in $\mathbb{Z}_2$. This is manifestly an isomorphism on $\pi_2$ 
and trivially an isomorphism on all other homotopy groups. Therefore it is a weak equivalenc.
\endofproof
\begin{observation}
A $U(1)$-gerbe in the full sense Giraud (see \cite{Lurie}, section 7.2.2) 
as opposed to a $U(1)$-bundle gerbe / circle 2-bundle is equivalent to an 
$\mathrm{AUT}(U(1))$-principal 2-bundle, not in general to a circle 2-bundle, which is only a special case.
\end{observation}
More generally we have:
\begin{proposition} 
 \label{ExtensionOfZTwobyHigherUOne}
For every $n \in \mathbb{N}$ the automorphism $(n+1)$-group of $\mathbf{B}^n U(1)$ is given by the crossed complex
(as discussed in \ref{SheafAndNonabelianDoldKan})
$$
  \mathrm{AUT}(\mathbf{B}^n U(1))
  \simeq
  [U(1) \to 0 \to \cdots \to 0 \to \mathbb{Z}_2]
$$
with $U(1)$ in degree $n+1$ and $\mathbb{Z}_2$ acting by automorphisms.
This is an extension of smooth $\infty$-groups
$$
  \xymatrix{
     \mathbf{B}^{n+1} U(1) 
	 \ar[r]
	 &
	 \mathrm{AUT}(\mathbf{B}^n U(1)) 
	 \ar[r]
	 &
	 \mathbb{Z}_2
  }
  \,.
$$
\end{proposition}
With slight abuse of notation we also write
$$
  \mathbf{B}^n U(1)/\!/\mathbb{Z}_2
  :=
  \mathbf{B} \mathrm{AUT}(\mathbf{B}^{n-1}U(1))
  \,.
$$
\begin{definition}
  Write
  $$
    \mathbf{J}_n : \mathbf{B}^{n+1}U(1)/\!/\mathbb{Z}_2 \to \mathbf{B}\mathbb{Z}_2
  $$
  for the corresponding universal characteristic map.
\end{definition}
\begin{definition}
  \index{differential cohomology!orientifold}
  \index{circle $n$-bundle with connection!orientifold}
For $X \in \mathrm{Smooth}\infty \mathrm{Grpd}$, a \emph{double cover} 
$\hat X \to X$ is a $\mathbb{Z}_2$-principal bundle. 

For $n \in \mathbb{N}$, $n \geq 1$, an 
\emph{orientifold circle $n$-bundle (with connection)}
 is an $\mathrm{AUT}(\mathbf{B}^{n-1}U(1))$-principal $\infty$-bundle (with $\infty$-connection) on $X$ that 
extends $\hat X \to X$ (by def. \ref{ExtensionOfInfinityGroups}) with respect to the extension 
of $\mathbb{Z}_2$ by $\mathrm{AUT}(\mathbf{B}^n U(1))$, prop. \ref{ExtensionOfZTwobyHigherUOne}.
\end{definition}
This means that relative to a cocycle $g : X \to \mathbf{B}\mathbb{Z}^2$ for 
a double cover $\hat X$, 
the structure of an orientifold circle $n$-bundle is a factorization of this cocycle as
$$
  g : X \stackrel{\hat g}{\to} \mathbf{B} \mathrm{AUT}(\mathbf{B}^{n-1} U(1)) \to \mathbf{B}\mathbb{Z}^2
$$
where $\hat g$ is the cocycle for the corresponding $\mathrm{AUT}(\mathbf{B}^n U(1))$-principal $\infty$-bundle.
\begin{proposition}
Every orientifold circle $n$-bundle (with connection) on $X$ induces an ordinary 
circle $n$-bundle (with connection) $\hat P \to \hat X$ on the given double cover $\hat X$ 
such that restricted to any fiber of $\hat X$ this is equivalent to 
$\mathrm{AUT}(\mathbf{B}^{n-1}U(1)) \to \mathbb{Z}_2$.
\label{HigherOrientifoldsDecomposed}
\end{proposition}
\proof
  There is a
  pasting diagram of $\infty$-pullbacks of the form
  $$
    \raisebox{30pt}{
    \xymatrix{
	  (U(1) \to \cdots \to \mathbb{Z}_2)^\rho
	  \ar[r]
	  \ar[d]
	  &
	  P
	  \ar[r]
	  \ar[d]
	  &
	  {*}
	  \ar[d]
	  \\
	  \mathbb{Z}_2
	  \ar[r]
	  \ar[d]
	  &
	  \hat X
	  \ar[r]
	  \ar[d]
	  &
	  \mathbf{B}^n U(1)
	  \ar[r]
	  \ar[d]
	  &
	  {*}
	  \ar[d]
	  \\
	  {*}
	  \ar[r]^x
	  &
	  X \ar[r]^<<<<<g 
	  &
	  \mathbf{B}^n U(1)/\!/\mathbb{Z}_2
	  \ar[r]^<<<<{\mathbf{J}_{n-1}}
	  &
	  \mathbf{B}\mathbb{Z}_2
	}
	}
  $$
\endofproof
\begin{proposition}
 \index{twisted cohomology!orientifold!string-}
Orientifold circle 2-bundles over a smooth manifold are equivalent to the 
\emph{Jandl gerbes} introduced in \cite{SSW}. 
\label{JandlGerbes}
\end{proposition}
\proof
  By prop. \ref{CohomologyByCovers} we have that $[U(1) \to \mathbb{Z}_2]$-principal $\infty$-bundles 
  on $X$ are given by {\v C}ech cocycles relative to any good open cover of $X$ with coefficients 
  in the sheaf of 2-groupoids $\mathbf{B}[U(1) \to \mathbb{Z}_2]$. Writing this out in components 
  it is straightforward to check that this coincides with the data of a Jandl gerbe 
  (with connection) over this cover.
\endofproof
\begin{remark}
Orientifold circle $n$-bundles are not $\mathbb{Z}_2$-equivariant circle $n$-bundles: in the latter case the orientation reversal acts by an equivalence between the bundle and its pullback along the orientation reversal, whereas for an orientifold circle $n$-bundle the orientation reversal acts by an equivalence to the 
\emph{dual} of the pulled-back bundle.
\end{remark}
\begin{proposition}
The geometric realization, def. \ref{GeometricRealization},
$$
  \tilde R := |\mathbf{B}[U(1) \to \mathbb{Z}_2]|
$$
of $\mathbf{B}[U(1) \to \mathbb{Z}]$ is the homotopy 3-type with homotopy groups
$$
  \pi_0(\tilde R)  = 0
  \,;
$$
$$
  \pi_1(\tilde R)  = \mathbb{Z}_2
  \,;
$$
$$
  \pi_2(\tilde R)  = 0
  \,;
$$
$$
  \pi_3(R')  = \mathbb{Z}
$$
and nontrivial action of $\pi_1$ on $\pi_3$.
\end{proposition}
\proof
By prop. \ref{UnderlyingSimplicialTopologicalSpace} and the results of 
\ref{ETopStrucCohomology} we have
\begin{enumerate}
\item specifically 
 \begin{enumerate}
   \item $|\mathbf{B} \mathbb{Z}_2| \simeq B \mathbb{Z}_2$; 

   \item $|\mathbf{B}^2 U(1)| \simeq B^2 U(1) \simeq K(\mathbb{Z};3)$;
 \end{enumerate}
   where on the right we have the ordinary classifying spaces going by these names;
\item
generally geometric realization preserves fiber sequences of nice enough objects, 
such as those under consideration, so that we have a fiber sequence
   $$
     K(\mathbb{Z},3) \to \tilde R \to B \mathbb{Z}_2
   $$ 
   in $\mathrm{Top}$.
\end{enumerate}
Since $\pi_3(K(\mathbb{Z}), 3) \simeq \mathbb{Z}$ and $\pi_1(B \mathbb{Z}_2) \simeq \mathbb{Z}_2$ and all other homotopy groups of these two spaces are trivial, the homotopy groups of $\tilde R$ follow by the long exact sequence of homotopy groups associated to our fiber sequence.

Finally, since the action of $\mathbb{Z}_2$ in the crossed module is nontrivial, $\pi_1(\tilde R)$ must act notriviall on $\pi_3(\mathbb{Z})$. It can only act nontrivial in a single way, up to homotopy.
\endofproof
The space 
$$
  R := \mathbb{Z}_2 \times \tilde R
$$
is taken to be the coefficient object for orientifold (differential) cohomology as appearing 
in string theory in \cite{DistlerFreedMoore}. 

The following definition gives the differential refinement of 
$\mathbf{B}\mathrm{AUT}(\mathbf{B}^{n-1}U(1))$. With slight abuse of
notation we will also write
$$
  \mathbf{B}^n U(1)/\!/\mathbb{Z}_2
  :=
  \mathbf{B}\mathrm{AUT}(\mathbf{B}^{n-1}U(1))
  \,.
$$
\begin{definition}
  For $n \geq  2$ write
  $\mathbf{B}^n U(1)_{\mathrm{conn}}/\!/\mathbb{Z}_2$
  for the smooth $n$-stack presented by the presheaf of $n$-groupoids which 
  is given by the presheaf of crossed complexes of groupoids
  $$
    \xymatrix@C=3.3em{
      \Omega^n(-)\times C^\infty(-,U(1))
	  \ar[rr]^{\phantom{mm}(\mathrm{id},d_{\mathrm{dR}} \mathrm{log})}
	  &&
	  \Omega^n(-)\times \Omega^1(-)
	  \ar[r]^{\phantom{mmm}(\mathrm{id},d_{\mathrm{dR}})} &
	  \cdots 
	  \ar[r]^{(\mathrm{id},d_{\mathrm{dR}})\phantom{mmmmm}}
	  &
	  \Omega^n(-) \times \Omega^{n-2}(-)
	  \ar[r]^{\phantom{mmmmm}(\mathrm{id}, d_{\mathrm{dR}})}&
	  }
  $$
 $$
    \xymatrix@C=3.3em{
    \ar[r]^{(\mathrm{id}, d_{\mathrm{dR}})\phantom{mmmmmmmm}}
	  &
	  \Omega^n(-)\times \Omega^{n-1}(-) \times \mathbb{Z}_2
	  \ar@<-3pt>[r]
	  \ar@<+3pt>[r]
	  &
	  \Omega^n(-)
	}
	\,,
  $$
  where 
  \begin{enumerate}
    \item 
    the groupoid on the right has as morphisms $(A,\sigma) :  B \to B'$
	between two $n$-forms $B,B'$ pairs consisting of an $(n-1)$-form $A$
    and an element $\sigma \in \mathbb{Z}_2$, such that
	$(-1)^{\sigma} B' = B + d A$;
	\item
	  the bundles of groups on the left are all trivial as bundles;
	\item 
	  the $\Omega^1(-)\times\mathbb{Z}_2$-action is 
	  by the $\mathbb{Z}_2$-factor only and on forms given by multiplication 
	  by $\pm 1$ and on $U(1)$-valued functions by complex conjugation
	  (regarding $U(1)$ as the unit circle in the complex plane).
  \end{enumerate}
\end{definition}
\begin{remark}
  A detailed discussion of $\mathbf{B}^2 U(1)_{\mathrm{conn}}/\!/\mathbb{Z}_2$
  is in \cite{SWII} and \cite{SWIII}. 
\end{remark}
We now discuss differential cocycles with coefficients in 
$\mathbf{B}^n U(1)_{\mathrm{conn}}/\!/\mathbb{Z}_2$ 
over $\mathbb{Z}_2$-quotient stacks / orbifolds. Let $Y$ be a smooth manifold 
equipped with a smooth $\mathbb{Z}_2$-action $\rho$. Write 
  $Y /\!/ \mathbb{Z}_2$ for the corresponding global orbifold and 
  $\rho : Y/\!/\mathbb{Z}_2 \to \mathbf{B}\mathbb{Z}_2$
  for its classifying morphism, hence for the morphism that fits into 
  a fiber sequence of smooth stacks
  $$
    \xymatrix{
	  Y \ar[r] 
	  & 
	  Y/\!/\mathbb{Z}_2
	  \ar[r]
	  &
	  \mathbf{B}\mathbb{Z}_2
	}
	\,.
  $$
\begin{definition}
   An \emph{$n$-orientifold} structure $\hat G_{\rho}$ on 
   $(Y, \rho)$ is a \emph{$\rho$-twisted 
  $\hat {\mathbf{J}}_n$-structure}
  on $Y/\!/\mathbb{Z}_2$, def. \ref{TwistedCStructures},
  hence a dashed morphism in the diagram
  $$  
    \xymatrix{
	  & \mathbf{B}^{n+1} U(1)_{\mathrm{conn}}/\!/\mathbb{Z}_2
	  \ar[d]^{\hat {\mathbf{J}}_n}
	  \\
	  Y/\!/\mathbb{Z}_2
	  \ar@{-->}[ur]^{\hat G_{\rho}}
	  \ar[r]^{\rho}
	  &
	  \mathbf{B}\mathbb{Z}_2
	}
	\,.
  $$
\end{definition}
\begin{observation}
  By corollary \ref{HigherOrientifoldsDecomposed},
  an $n$-orientifold structure decomposes into 
  an ordinary $(n+1)$-form connection $\hat G$ on a circle $(n+1)$-bundle over
  $Y$, subject to a $\mathbb{Z}_2$-twisted $\mathbb{Z}_2$-equivariance
  condition
  $$
    \raisebox{20pt}{
    \xymatrix{
	  Y
	  \ar[r]^<<<<<{\hat G}
	  \ar[d]
	  &
	  \mathbf{B}^{n+1} U(1)_{\mathrm{conn}}
	  \ar[r]
	  \ar[d]
	  &
	  {*}
	  \ar[d]
	  \\
	  Y/\!/\mathbb{Z}_2
	  \ar[r]^<<<{\hat G_\rho}
	  \ar@/_1pc/[rr]_{\rho}
	  &
	  \mathbf{B}^{n+1} U(1)_{\mathrm{conn}}/\!/\mathbb{Z}_2
	  \ar[r]^<<<<<{\hat {\mathbf{J}}}
	  &
	  \mathbf{B}\mathbb{Z}_2\;\;.
	}
	}
  $$
  For $n = 1$ this reproduces, via observation \ref{JandlGerbes},
  the \emph{Jandl gerbes with connection} from \cite{SSW}, hence
  ordinary string orientifold backgrounds, as discussed there.
  For $n = 2$ this reproduces background structures 
  for membranes as discussed below in \ref{MembraneOrientifolds}.
  \label{DifferentialJStructreDecomposed}
\end{observation}

\subsubsection{Twisted topological structures in quantum anomaly cancellation}
\label{TwistedTopologicalStructuresInStringTheory}
\index{twisted cohomology!twisted topological structures!examples}

We discuss here cohomological conditions arising from
anomaly cancellation in string theory, for various $\sigma$-models. 
In each case we introduce a corresponding notion of  
topological \emph{twisted structures} and interpret the anomaly 
cancellation condition in terms of these.
This prepares the ground for the material in 
the following sections,
where the differential refinement of these twisted structures is considered
and the \emph{differential} anomaly-free 
field configurations are derived from these.

\begin{itemize}
  \item \ref{Type II superstring and twisted Spinc structures}
    -- The type II superstring and twisted $\mathrm{Spin}^c$-structures;
  \item \ref{HeteroticStringAndTwistedStringStructures}
    -- The heterotic/type I superstring and twisted $\mathrm{String}$-structures;
  \item \ref{TwistedString2aStructures}
    -- The M2-brane and twisted $\mathrm{String}^{2a}$-structures;
  \item \ref{NS5 and twisted Fivebrane Structures}
    -- The NS-5-brane and twisted $\mathrm{Fivebrane}$-structures;
  \item \ref{Twisted Fivebrane2a2aStructures}
    -- The M5-brane and twisted $\mathrm{Fivebrane}^{2 a \cup 2 a}$-structures
\end{itemize}

\medskip

The content of this section is taken from \cite{SSSIII}.

\medskip

The physics of all the cases we consider involves a 
manifold $X$ -- 
the \emph{target space} -- or a submanifold $Q \hookrightarrow X$ 
 thereof-- a \emph{$D$-brane} --, 
equipped with 
\begin{itemize}
\item two principal bundles with their canonically associated vector bundles:
\begin{itemize}
\item a $\mathrm{Spin}$-principal bundle underlying the tangent bundle
$T X$ (and we will write $T X$ also to denote that $\mathrm{Spin}$-principal bundle), 
\item and a complex vector bundle $E \to X$
-- the ``gauge bundle'' --  associated to a $SU(n)$-principal bundle or
to an $E_8$-principal bundle with respect to a unitary representation of $E_8$;
\end{itemize}
  \item and an $n$-gerbe / circle $(n+1)$-bundle with class $ H^{n+2}(X,\mathbb{Z})$
    -- the higher background gauge field -- denoted $[H_3]$ or $[G_4]$ 
	or similar in the following.
\end{itemize}
All these structures are equipped with a suitable notion of 
\emph{connetions}, locally given by some differential-form data.
The connection on the $\mathrm{Spin}$-bundle encodes the field of 
gravity, that on the gauge bundle a Yang-Mills field and that 
on the $n$-gerbe a higher analog of the electromagnetic field.

The $\sigma$-model quantum field theory of a super-brane 
propagating in such 
a background (for instance the superstring, or the super 5-brane)
has an effective action functional on its bosonic 
worldvolume fields that takes values, in general, in the fibers of 
the Pfaffian line bundle of a worldvolume Dirac operator,
tensored with a line bundle that remembers the electric and magnetic charges
of the higher gauge field. Only if this 
tensor product \emph{anomaly line bundle} is trivializable is the
effective bosonic action a 
well-defined starting point for
quantization of the $\sigma$-model. Therefore the Chern-class of this
line bundle over the bosonic configuration space is called the 
 \emph{global anomaly} of the system.
Conditions on the background gauge fields that ensure that this class
vanishes are called \emph{global anomaly cancellation conditions}.
These turn out to be conditions on cohomology classes that are characteristic 
of the above background fields. This is what we discuss now.

But moreover, the anomaly line bundle is canonicaly equipped with a 
\emph{connection},
induced from the connections of the background gauge
fields, hence induced from their \emph{differential cohomology} data. 
The curvature 2-form of this connection over
the bosonic configuration space is called the \emph{local anomaly}
of the $\sigma$-model. Conditions on the differential data of the 
background gauge field that canonically induce a trivialization of this
2-fom are called \emph{local anomaly cancellation conditions}. These we
consider below in section \ref{Twisted differential String- and Fivebrane structures}.

\medskip

The phenomenon of anomaly line bundles of $\sigma$-models 
induced from background field differential cohomology is 
classical in the physics literature, if only in broad terms. 
A clear exposition is in \cite{Freed}. Only recently 
the special case of the heterotic string $\sigma$-model for trivial 
background gauge bundle has been made fully precise in 
\cite{Bunke}, using a certain model \cite{Waldorf} 
for the differential string structures
that we discuss in section 
\ref{Twisted differential String- and Fivebrane structures}.

\paragraph{The type II superstring and twisted $\mathrm{Spin}^c$-structures} 
\label{Type II superstring and twisted Spinc structures}

The open type II string propagating on a Spin-manifold $X$ in the presence of 
a background $B$-field with class $[H_3] \in H^3(X,\mathbb{Z})$
and with endpoints fixed on a D-brane
given by an oriented submanifold $Q \hookrightarrow X$,
has a global worldsheet anomaly that vanishes
if \cite{FreedWitten} and only if \cite{ES} the condition
\(
  [W_3(Q)] + [H_3]|_Q = 0 \;\; \in H^3(Q;\Z)
  \;,
  \label{Freed Witten condition}
\)
holds.
Here $[W_3(Q)]$ is the third integral Stiefel-Whitney class of the tangent bundle $TQ$
of the brane and $[H_3]_Q$ denotes the restriction of $[H_3]$ to $Q$.

Notice that $[W_3(Q)]$ is the obstruction to lifting the orientation structure on 
$Q$ to a $\mathrm{Spin}^c$-structure. More precisely, in terms of homotopy theory
this is formulated as follows, \ref{SpinCAsHomotopyFiberProduct}. 
There is a homotopy pullback diagram
\(
  \raisebox{20pt}{
  \xymatrix{
    B \mathrm{Spin}^c \ar[r] \ar[d] & {*} \ar[d]
    \\
    B \mathrm{SO} \ar[r]^{W_3} & B^2 U(1)
  }
  }
  \label{SpincAsHomotopyFiber}
\)
of topological spaces, where $B \mathrm{SO}$ is the classifying space of the
special orthogonal group, where $B^2 U(1) \simeq K(\mathbb{Z},3)$ is 
homotopy equivalent to the Eilenberg-MacLane space that classifies degree-3
integral cohomology, and where the continuous map denoted $W_3$ is a representative
of the universal class $[W_3]$ under this classification. 
This homotopy pullback exhibits the 
classifying space of the group $\mathrm{Spin}^c$ as the homotopy fiber of $W_3$.
The universal property of the homotopy pullback says that the space of continuous maps
$Q \to B \mathrm{Spin}^c$ is the same (is homotopy equivalent to) the space of
maps $o_Q : Q \to B \mathrm{SO}$ that are equipped with a homotopy from the composite
$\xymatrix{
  Q \ar[r]^{o_Q} &   B \mathrm{SO} \ar[r]{W_3} & B^3 U(1)
}$ to the trivial cocycle 
$Q \to {*} \to B^3 U(1)$. In other words, for every choice of homotopy
filling the outer diagram of
$$
  \xymatrix{
    Q 
    \ar@{-->}[dr]
    \ar@/^1pc/[drr]
    \ar@/_1pc/[ddr]_{o_Q}
    \\
    & B \mathrm{Spin}^c \ar[r] \ar[d] & {*} \ar[d]
    \\
    & B \mathrm{SO} \ar[r]^{W_3} & B^2 U(1)
  }
$$
there is a contractible space of 
choices for the dashed arrow such that everything commutes up to
homotopy. Since a choice of map $o_Q : Q \to B \mathrm{SO}$ is
an \emph{orienation structure} on $Q$, and a choice of map $Q \to B \mathrm{Spin}^c$ is
a \emph{$\mathrm{Spin}^c$-structure}, this implies that $[W_3(o_Q)]$ is the obstruction
to the existence of a $\mathrm{Spin}^c$structure on $Q$ (equipped with  $o_Q$.

Moreover, since $Q$ is a manifold, 
the functor $\mathrm{Maps}(Q,-)$ that forms mapping spaces out of $Q$ preserves
homotopy pullbacks. Since $\mathrm{Maps}(Q, B \mathrm{SO})$
is the \emph{space} of orientation structures, we can refine the discussion so far by 
noticing that the \emph{space of $\mathrm{Spin}^c$-structures on $Q$}, 
$\mathrm{Maps}(Q, B \mathrm{Spin}^c)$, is itself the homotopy pullback in the diagram
\(
  \raisebox{20pt}{
  \xymatrix{
    \mathrm{Maps}(Q,B \mathrm{Spin}^c) \ar[rr] \ar[d] && {*} \ar[d]
    \\
    \mathrm{Maps}(Q,B \mathrm{SO}) \ar[rr]^{\mathrm{Maps}(Q,W_3)} && 
     \mathrm{Maps}(Q,B^2 U(1))
  }
  }
  \label{SpaceOfSpincStructures}
  \,.
\)
A variant of this characterization will be crucial for the definition of 
(spaces of) \emph{twisted} such structures below.

These kinds of arguments, even though elementary in homotopy theory, are of importance
for the interpretation of anomaly cancellation conditions that we consider here. Variants
of these arguments (first for other topological structures, then with twists, 
then refined to smooth
and differential structures) will appear over and over again in our discussion

So in the case that the class of the $B$-field vanishes on the D-brane, 
$[H_3]|_Q = 0$, hence that its representative $H_3 : Q \to K(\mathbb{Z},3)$
factors through the point, up to homotopy,  
condition (\ref{Freed Witten condition}) states that the oriented D-brane $Q$ must 
admit a  $\mathrm{Spin}{}^c$-structure, namely a choice of null-homotopy $\eta$ in
\(
  \raisebox{20pt}{
  \xymatrix{
     Q
     \ar[rr]^{o_Q}_>>>{\ }="s"
     \ar[drr]_{{H_3}|_Q \simeq {*}}^{\ }="t"
     &&
     B \mathrm{SO}
     \ar[d]^{W_3}
     \\
     &&
     K(\mathbb{Z},3)
     \ar@{=>}^\eta "s"; "t"
  }
  }
  \,.
\)
(Beware that there are such homotopies filling \emph{all} our diagrams, but only in some
cases, such as here, do we want to make them explicit and given them a name.)
If, generally, $[H_3]_Q$ does not necessarily vanish, 
then condition (\ref{Freed Witten condition})
still is equivalent to the existence of a homotopy $\eta$ in a diagram of the above form:
\(
  \raisebox{20pt}{
  \xymatrix{
     Q
     \ar[rr]^{o_Q}_>>>{\ }="s"
     \ar[drr]_{{H_3}|_Q }^{\ }="t"
     &&
     B \mathrm{SO}
     \ar[d]^{W_3}
     \\
     &&
     K(\mathbb{Z},3)
     \ar@{=>}^\eta "s"; "t"
  }
  }
  \label{HomotopyOfTwistedSpinCStructure}
  \,.
\)
We may think of this as saying that $\eta$ still ``trivializes'' $W_3(o_Q)$, but not
with respect to the canonical trivial cocycle, but with respect to the given reference background 
cocycle ${H_3}|_Q$ of the $B$-field. Accordingly, following \cite{Wang}, 
we may say that such an $\eta$ exhibits not a $\mathrm{Spin}^c$-structure on $Q$, but
an \emph{$[H_3]_Q$-twisted $\mathrm{Spin}^c$-structure}.

For this notion to be useful, we need to say what an equivalence or homotopy
between two twisted $\mathrm{Spin}^c$-structures is, what a homotopy between such
homotopies is, etc., hence what the \emph{space} of twisted 
$\mathrm{Spin}^c$-structures is. 
But by generalization of (\ref{SpaceOfSpincStructures}) 
we naturally have such a space.
\begin{definition}
\label{SpaceOfSPinCStructures}
For $X$ a manifold and $[c] \in H^3(X, \mathbb{Z})$ a degree-3 cohomology
class, we say that the space
$W_3\mathrm{Struc}(Q)_{[c]}$ defined as the homotopy pullback
\(
 \raisebox{30pt}{
  \xymatrix{
    W_3 \mathrm{Struc}(Q)_{[H_3]|_Q}
	\ar[rr]_<{\hpull}
	\ar[d]
	&&
	{*}
	\ar[d]^{c}
	\\
	\mathrm{Maps}(Q, B \mathrm{SO})
	\ar[rr]^{\mathrm{Maps}(Q,W_3)}
	&&
	\mathrm{Maps}(Q, B^2 U(1))
  }
  }
  \label{space of twisted Spinc-structures}
  \,,
\)
is the \emph{space of $[c]$-twisted $\mathrm{Spin}^c$-structures} on $X$,
where the right vertical morphism picks any representative
$c : X \to B^2 U(1) \simeq K(\mathbb{Z},3)$ of $[c]$.
\end{definition}

In terms of this notion, the anomaly cancellation condition
(\ref{Freed Witten condition}) is now read as encoding  
\emph{existence of structure}:
\begin{observation}
On an oriented manifold $Q$, condition (\ref{Freed Witten condition}) 
precisely guarantees the existence of 
\emph{$[H_3]|_Q$-twisted $W_3$-structure}, 
provided by a lift of the orientation structure $o_Q$
on $T Q$ through the left vertical morphism in 
def. \ref{space of twisted Spinc-structures}.
\end{observation}
This makes good sense, because that extra structure is the extra 
structure of the background field  of the 
$\sigma$-model background, subjected to the
condition of anomaly freedom. This we will see in more detail in the
following examples, and then 
in section \ref{Twisted differential String- and Fivebrane structures}.

\paragraph{The heterotic/type I superstring and twisted $\mathrm{String}$-structures} 
\label{HeteroticStringAndTwistedStringStructures}
\index{anomaly cancellation!Green-Schwarz anomaly}

The heterotic/type I string, propagating on a $\mathrm{Spin}$-manifold
$X$ and coupled to a gauge field given by a Hermitean complex vector bundle $E \to X$,
has a global anomaly that vanishes if the 
\emph{Green-Schwarz anomaly cancellation condition} \cite{GS}
\(
  \frac{1}{2}p_1(TX) - {\rm ch}_2(E)=0 \;\;\in H^4(X;\Z)
  \;.
  \label{Green Schwarz anomaly}
\)
holds. Here $\frac{1}{2}p_1(T X)$ is the \emph{first fractional Pontryagin class}
of the $\mathrm{Spin}$-bundle, and  $\mathrm{ch}_2(E)$ is the second
Chern-class of $E$. 

As before, this means that at the level of cocycles a certain homotopy
exists. Here it is this homotopy which is the representative of the 
$B$-field that the string couples to.

In detail, write $\frac{1}{2}p_1 : B \mathrm{Spin} \to B^3 U(1)$ for 
a representative of the universal first fractional Pontryagin class, 
prop. \ref{FirstFracPontryagin}, 
and similarly $\mathrm{ch}_2 : B \mathrm{SU} \to B^3 U(1)$ for a representative of the
universal second Chern class, where now
$B^3 U(1) \simeq K(\mathbb{Z},4)$ is equivalent to the 
Eilenberg-MacLane space that classifies degree-4 integral cohomology. 
Then if $T X : X \to B \mathrm{Spin}$ is a classifying
map of the $\mathrm{Spin}$-bundle and $E : X \to B \mathrm{SU}$ one of the
gauge bundle, the anomaly cancellation condition above says that there
is a homotopy, denoted $H_3$, in the diagram
\(
  \raisebox{20pt}{
  \xymatrix{
    X \ar[r]^{E}_>{\ }="s" \ar[d]_{T X} & B \mathrm{SU} \ar[d]^{\mathrm{ch}_2}
	\\
	B \mathrm{Spin}
	\ar[r]_{\frac{1}{2}p_1}^<{\ }="t"
	&
	B^3 U(1)
  \ar@{=>}^{H_3} "s"; "t"
  } }
  \label{homotopy in twisted string structure}
  \,.
\)
Notice that if both $\frac{1}{2}p_1(T X)$ as well as $\mathrm{ch}_2(E)$
happen to be trivial, such a homotopy is equivalently a map 
$H_3 : X \to \Omega B^3 U(1) \simeq B^2 U(1)$. So in this special case the 
B-field in the background of the heterotic string is a
$U(1)$-gerbe, a circle 2-bundle, as in the previous case of the type II string
in section \ref{Type II superstring and twisted Spinc structures}.
Generally, the homotopy $H_3$ in the above diagram exhibits
the B-field as a \emph{twisted} gerbe, whose twist is the
difference class $[\frac{1}{2}p_1(TX)]- [\mathrm{ch}_2(E)]$.
This is essentially the perspective adopted in \cite{Freed}.

For the general discussion of interest here it is useful to slightly
shift the perspective on the twist. 
Recall that a \emph{String structure}, \ref{String2Group}, on 
the Spin bundle $T X : X \to B \mathrm{Spin}$ is a homotopy 
filling the outer square of
$$
  \raisebox{30pt}{
  \xymatrix{
    X
	\ar@{-->}[dr]
	\ar@/^1pc/[drr]
	\ar@/_1pc/[ddr]_{T X}
    \\
    &
	B \mathrm{String} 
	\ar[r]
	\ar[d]
	&
	{*}
	\ar[d]
	\\
	& B \mathrm{Spin}
	\ar[r]^{\frac{1}{2}p_1}
	&
	B^3 U(1)
  }}
  \,,
$$
or, which is equivalent by the universal property of homotopy pullbacks,
a choice of dashed morphism filling the interior of this square, as indicated.

Therefore, now by analogy with (\ref{HomotopyOfTwistedSpinCStructure}),
we say that a $[\mathrm{ch}_2(E)]$-\emph{twisted string structure}
is a choice of homotopy $H_3$ filling the diagram
(\ref{homotopy in twisted string structure}).

This notion of twisted string structures was originally
suggested in \cite{Wang}. For it to be useful, we need to 
say what homotopies of twisted $\mathrm{String}$-structures
are, homotopies between these, etc. Hence we need to say
what the \emph{space} of twisted $\mathrm{String}$-structures
is. This is what the following definition provides, analogous
to \ref{space of twisted Spinc-structures}.
\begin{definition}
  For $X$ a manifold,  
  and for $[c] \in H^4(X,\mathbb{Z})$ a degree-4 cohomology class,
  we say that the space of \emph{$c$-twisted $\mathrm{String}$-structures}
  on $X$ is the homotopy pullback $\frac{1}{2}p_1 \mathrm{Struc}_{[c]}(X)$ in
  $$
    \raisebox{20pt}{
    \xymatrix{
	  \frac{1}{2}p_1 \mathrm{Struc}_{[c]}(X)
	  \ar[rr]
	  \ar[d]
	  &&
	  {*}
	  \ar[d]^{c}
	  \\
	  \mathrm{Maps}(X, B \mathrm{Spin})
	  \ar[rr]^{\mathrm{Maps}(X,\frac{1}{2}p_1)}
      &&
      \mathrm{Maps}(X, B^3 U(1))	  
	}
	}
	\,,
  $$
  where the right vertical morphism picks a representative $c$ of $[c]$.
  \label{SpaceOfTwistedStringStructures}
\end{definition}
In terms of this then, we find
\begin{observation} 
 The anomaly cancellation condition 
(\ref{Green Schwarz anomaly}) is, for a fixed
gauge bundle $E$, precisely the condition that 
ensures a lift of the given $\mathrm{Spin}$-structure to a
$[\mathrm{ch}_2(E)]$-twisted $\mathrm{String}$-structure on $X$,
through the left vertical morphism of def. \ref{SpaceOfTwistedStringStructures}.
\end{observation}
Of course the full background field content involves more than just 
this topological data, it also consists of local differential form data,
such as a 1-form connection on the bundles $E$ and on $T X$ 
and a connection 2-form on the 2-bundle $H_3$. Below in 
section \ref{Twisted differential String- and Fivebrane structures} we
identify this \emph{differential} anomaly-free field content
with a \emph{differential} twisted $\mathrm{String}$-structure.

\paragraph{The M2-brane and twisted $\mathrm{String}^{2a}$-structures}
\label{TwistedString2aStructures}

The string theory backgrounds discussed above have
lifts to 11-dimensional supergravity/M-theory, where the 
bosonic background field content consists of just the 
$\mathrm{Spin}$-bundle $T X$ as well as the $C$-field, which has underlying
it a 2-gerbe -- or \emph{circle 3-bundle} -- with class $[G_4] \in H^4(X, \mathbb{Z})$.
The M2-brane that couples to these background fields has an anomaly that
vanishes \cite{WittenFluxQuantization} if
\(
  2[G_4] = [\frac{1}{2}p_1(T X)] - 2 [a(E)]
  \;\;
  \in H^4(X, \mathbb{Z})
  \label{WittenQuantizationCondition}
  \,,
\)
where $E \to X$ is an auxiliary $E_8$-principal bundle, whose class is
defined by this condition. 

Since $E_8$ is 15-coskeletal, this condition is equivalent to demanding 
that $[\frac{1}{2}p_1(T X)] \in H^4(X, \mathbb{Z})$ is further divisible by 2. 
In the absence of smooth or differential structure,
one could therefore replace the $E_8$-bundle here by a circle 2-gerbe, hence by
a $B^2 U(1)$-principal bundle, and
replace condition (\ref{WittenQuantizationCondition}) by
$$
  2[G_4] = [\frac{1}{2}p_1(T X)] - 2 [\mathrm{DD}_{2}]
  \,,
$$
where $[\mathrm{DD}_2]$ is the canonical 4-class of this 2-gerbe 
(the ``second Dixmier-Douady class'').
While topologically this condition is equivalent, over an 11-dimensional $X$,
to (\ref{WittenQuantizationCondition}), the spaces of solutions of 
smooth refinements of these
two conditions will differ, because the space of smooth gauge transformations
between $E_8$ bundles is quite different from that of smooth gauge transformations
between circle 2-bundles.
In the Ho{\v r}ava-Witten reduction \cite{HW} of the 11-dimensional theory 
down to the  heterotic string in 10 dimensions, this difference is supposed to be 
relevant, since the heterotic string in 10 dimensions sees the
smooth $E_3$-bundle with connection.

In either case, we can understand the situation as a refinement of that described by
(twisted) $\mathrm{String}$-structures via a higher analogue of the passage from
$\mathrm{Spin}$-structures to $\mathrm{Spin}^{c}$-structures. To that end 
recall prop. \ref{SpinCAsHomotopyFiberProductOfU1AndSO},
which provides an alternative perspective on 
(\ref{SpincAsHomotopyFiber}).

Due to the universal property of the homotopy
pullback, this says, in particular, 
that a lift from an orientation structure to a $\mathrm{Spin}^c$-structure
is a cancelling by a Chern-class of the class obstructing a $\mathrm{Spin}$-structure. 
In this way lifts from orientation structures to  
$\mathrm{Spin}^c$-structures are analogous to the divisibility condition
(\ref{WittenQuantizationCondition}), since in both cases the obstruction to 
a further lift through the Whitehead tower of the orthogonal group is
absorbed by a universal ``unitary'' class.

In order to formalize this we make the following definition.
\begin{definition}
 For $G$ some topological group, and $c : B G \to K(\mathbb{Z},4)$
 a universal 4-class, we say that $\mathrm{String}^c$ is the 
loop group of the homotopy pullback
 $$
   \xymatrix{
     B \mathrm{String}^c \ar[r] \ar[d] & B G \ar[d]^{c}
     \\
     B \mathrm{Spin} \ar[r]^{\frac{1}{2}p_1} & B^3 U(1)
   }
 $$
of $c$ along the first fractional Pontryagin class.
 \label{Stringc}
\end{definition}

For instance for $c = \mathrm{DD}_2$ we have
that a $\mathrm{Spin}$-structure lifts to a $\mathrm{String}^{2\mathrm{DD}_2}$-structure
precisely if $\frac{1}{2}p_1$ is further divisible by 2. 
Similarly, with $a : B E_8 \to B^3 U(1)$ the canonical universal 4-class on 
$E_8$-bundles and $X$ a manifold of dimension $\mathrm{dim}X \leq 14$ we have
that a
$\mathrm{Spin}$-structure on $X$ lifts to a $\mathrm{String}^{2a}$-structure
precisely if $\frac{1}{2}p_1$ is further divisible by 2. 
\(
  \xymatrix{
   & B \mathrm{String}^{2a} \ar[d] \ar[dr]^{\frac{1}{4}p_1} \ar[r] & B E_8 \ar[d]^{2 a}
   \\
   X \ar[r] \ar@{-->}[ur] & B \mathrm{Spin} \ar[r]^{\frac{1}{2}p_1} & B^3 U(1)
  }
  \label{F}
  \,.
\)
Using this we can now reformulate the anomaly cancellation condition 
(\ref{WittenQuantizationCondition})
as follows.
\begin{definition}
  \label{TwistedString2sStructures}
  For $X$ a manifold and for 
  $[c] \in H^4(X,\mathbb{Z})$ a cohomology class, the space
  $(\frac{1}{2}p_1-2a)\mathrm{Struc}_{[c]}(X)$ 
  of \emph{$[c]$-twisted $\mathrm{String}^{2a}$-structures} on $X$
  is the homotopy pullback
  $$
    \raisebox{20pt}{
	\xymatrix{
	  (\frac{1}{2}p_1-2a)\mathrm{Struc}_{[c]}(X)
	  \ar[rr]
	  \ar[d]
	  &&
	  {*}
	  \ar[d]^c
	  \\
	  \mathrm{Maps}(X, B \mathrm{Spin} \times E_8 )
	  \ar[rr]^{\frac{1}{2}p_1 - 2a}
	  &&
	  \mathrm{Maps}(X, B^3 U(1) )
	}
	}
	\,,
  $$
  where the right vertical map picks a cocycle $c$ representing the class $[c]$.
\end{definition}
  In terms of this definition, we have
\begin{observation}
 Condition (\ref{WittenQuantizationCondition}) 
  is precisely the condition guaranteeing a 
  lift of the given $\mathrm{Spin}$- and the given $E_8$-principal
  bundle to a \emph{$[G_4]$-twisted $\mathrm{String}^{2a}$-structure}
  along the left vertical map from def. \ref{TwistedString2sStructures}.
\end{observation}
There is a further variation of this situation, that is of interest.
In the Ho{\v r}ava-Witten reduction of this situation in 11 dimensions
down to the sitation of the heterotic string in 10 dimensions, $X$ has
a boundary, $Q := \partial X \hookrightarrow X$, and there is a boundary
condition on the $C$-field, saying that the restriction of its
4-class to the boundary has to vanish,
$$
  [G_4]|_Q = 0
  \,.
$$
This implies that over $Q$ the anomaly-cancellation conditon 
(\ref{WittenQuantizationCondition}) becomes
$$
  [\frac{1}{2}p_1(TX)]|_Q = 2 [a(E)]|_Q \;\; \in H^4(Q,\mathbb{Z})
  \label{CFieldConditionOverBoundary}
  \,.
$$
Notice that this is the Green-Schwarz anomaly cancellation condition
(\ref{Green Schwarz anomaly}) of the heterotic string, but 
refined by a further cohomological divisibility condition.
The following statement says that this may equivalently be reformulated 
in terms of $\mathrm{String}^{2a}$ structures.
\begin{proposition}
  For $E \to X$ a fixed $E_8$-bundle, 
  we have an equivalence
  $$
    \mathrm{Maps}(X, B \mathrm{String}^{2a})|_{E}
    \simeq
    (\frac{1}{2}p_1)\mathrm{Struc}(X)_{[2a(E)]}
  $$
  between, on the right, the space of $[2a(E)]$-twisted $\mathrm{String}$-structures
from def. \ref{SpaceOfTwistedStringStructures}, and, on the left, the space of 
 $\mathrm{String}^{2a}$-structures with fixed class $2a$, hence the homotopy pullback
 $\mathrm{Maps}(X, B \mathrm{String}^{2a}) \times_{\mathrm{Maps}(X, B E_8)} \{E\}$.
\label{EquivalenceOfString2aAndTwistedStringStructure}
\end{proposition}
\proof
  Consider the diagram
  $$
    \xymatrix{   
	  \mathrm{Maps}(X, \mathrm{String}^{2a})|_E
      \ar[rr]
	  \ar[d]
	  &&
	  {*}
	  \ar[d]^{E}
	  \\
	  \mathrm{Maps}(X, \mathrm{String}^{2a})
	  \ar[rr]
	  \ar[d]
	  &&
	  \mathrm{Maps}(X, B E_8)
	  \ar[d]^{\mathrm{Maps}(X, 2a)}
	  \\
	  \mathrm{Maps}(X, B \mathrm{Spin})
	  \ar[rr]^{\mathrm{Maps}(X, \frac{1}{2}p_1)}
	  &&
	  \mathrm{Maps}(X, B^3 U(1))
	}
  $$
  The top square is a homotopy pullback by definition.
  Since $\mathrm{Maps}(X,-)$ preserves homotopy pullbacks
  (for $X$ a manifold, hence a CW-complex), 
  the bottom square is a homotopy pullback
  by definition \ref{Stringc}. Therefore, by the pasting law,
also the total rectangle is a homotopy pullback. With
 def. \ref{SpaceOfTwistedStringStructures} this implies the claim.    
\endofproof
Therefore the boundary anomaly cancellation condition for the M2-brane
has the following equivalent formulation.
\begin{observation}
  For $X$ a $\mathrm{Spin}$-manifold equipped with a
  complex vector bundle $E \to X$, condition 
  (\ref{CFieldConditionOverBoundary}) precisely guarantees the
  existence of a lift to a $\mathrm{String}^{2a}$-structure
  through the left vertical map in the proof of prop.
  \ref{EquivalenceOfString2aAndTwistedStringStructure}.
\end{observation}

\paragraph{The NS-5-brane and twisted $\mathrm{Fivebrane}$-structures}
\label{NS5 and twisted Fivebrane Structures}

The magnetic dual of the (heterotic) string is the NS-5-brane.
Where the string is electrically charged under the $B_2$-field
with class $[H_3] \in H^3(X,\mathbb{Z})$, the NS-5-brane is electrically
charged under the $B_6$-field with class $[H_7] \in H^7(X, \mathbb{Z})$
\cite{Cham}.
In the presence of a $\mathrm{String}$-structure,
hence when $[\frac{1}{2}p_1(T X)] = 0$, the anomaly
of the 5-brane $\sigma$-model vanishes \cite{SS} \cite{Gates} 
if the background fields satisfy
\( 
   [\frac{1}{6}p_2(T X)]
   =   
   8[{\rm ch}_4(E)]
   \;\;
   \in H^8(X, \mathbb{Z})
  \label{NS5BraneAnomalyCancellationCondition} 
  \,,
\) 
where $\frac{1}{6}p_2(T X)$ is the second fractional Pontryagin class
of the $\mathrm{String}$-bundle $T X$. 

It is clear now that a discussion entirely analogous to that of
section \ref{HeteroticStringAndTwistedStringStructures} applies. 
For the untwisted case the following terminology was introduced in
\cite{SSSII}.
\begin{definition}
  Write $\mathrm{Fivebrane}$ for the loop group of the homotopy
  fiber $B \mathrm{Fivebrane}$ of a representative $\frac{1}{6}p_2$
  of the universal second fractional Pontryagin class
  $$
    \raisebox{20pt}{
    \xymatrix{ 
	  B \mathrm{Fivebrane} \ar[r] \ar[d] & {*} \ar[d]
	  \\
	  B \mathrm{String}
	  \ar[r]^{\frac{1}{6}p_2}
	  &
	  B^7 U(1)
	}
	}
	\,.
  $$  
\end{definition}
In direct analogy with def. \ref{SpaceOfTwistedStringStructures} we therefore have the following
notion.
\begin{definition}
  For $X$ a manifold and $[c] \in H^8(X,\mathbb{Z})$ a class, 
  we say that the 
  \emph{space of $[c]$-twisted $\mathrm{Fivebrane}$-structures}
  on $X$, denoted $(\frac{1}{6}p_2)\mathrm{Struc}_{[c]}(X)$, is the 
  homotopy pullback
  $$
    \raisebox{20pt}{
    \xymatrix{
      (\frac{1}{6}p_2)\mathrm{Struc}_{[c]}(X)
      \ar[rr]
      \ar[d]
      &&
      {*}
      \ar[d]^{c}
      \\
      \mathrm{Maps}(X, B\mathrm{String})
      \ar[rr]^{\mathrm{Maps}(X,\frac{1}{6}p_2)}
      &&
      \mathrm{Maps}(X, B^7 U(1))
    }
	}
    \,,
  $$
\end{definition}
In terms of this we have
\begin{observation}
 For $X$ a manifold with $\mathrm{String}$-structure
 and with a background gauge bundle $E \to X$ fixed, 
 condition (\ref{NS5BraneAnomalyCancellationCondition})
 is precisely the condition for the existence of 
\emph{$[8\,\mathrm{ch}(E)]$-twisted $\mathrm{Fivebrane}$-structure} on $X$.
\end{observation}

\paragraph{The M5-brane and twisted $\mathrm{Fivebrane}^{2 a \cup 2 a}$-structures}
\label{Twisted Fivebrane2a2aStructures}
\index{M-theory!M5-brane}

The magnetic dual of the M2-brane is the M5-brane. Where the M2-brane is electrically charged
under the $C_3$-field with class $[G_4] \in H^4(X, \mathbb{Z})$,
the M5-brane is electrically charged under the dual $C_6$-field with 
class $[G_8] \in H^8(X,\mathbb{Z})$.

If $X$ admits a $\mathrm{String}$-structure, 
then one finds a relation for the background fields analogous to (\ref{WittenQuantizationCondition})
which reads
\(
  8 [G_8] = 4 [a(E)] \cup [a(E)] -  [\frac{1}{6}p_2(T X)]
  \label{M5braneAnomalyCancellationCondition}                                                       
  \,.
\)
The $\mathrm{Fivebrane}$-analog of $\mathrm{Spin}^c$ is then the following.
\begin{definition}
  For $G$ a topological group and $[c] \in H^8(B G, \mathbb{Z})$
  a universal 8-class, we say that $\mathrm{Fivebrane}^c$ is the loop group
  of the homotopy pullback
  $$
    \xymatrix{
	  B \mathrm{Fivebrane}^c \ar[r] \ar[d] & B G \ar[d]^c
	  \\
	  B \mathrm{String}
	  \ar[r]^{\frac{1}{6}p_2}
	  &
	  B^3 U(1)
	}
	\,.
  $$
\end{definition}
In analogy with def. \ref{TwistedString2sStructures}
we have a notion of twisted $\mathrm{Fivebrane}^c$-structures.
\begin{definition}
  \label{TwistedFivebrane2a2aStructures}
  For $X$ a manifold and for 
  $[c] \in H^8(X,\mathbb{Z})$ a cohomology class, the space
  $(\frac{1}{6}p_2- 2a\cup 2a)\mathrm{Struc}_{[c]}(X)$ 
  of \emph{$[c]$-twisted $\mathrm{Fivebrane}^{2a \cup 2a}$-structures} on $X$
  is the homotopy pullback
  $$
    \raisebox{20pt}{
	\xymatrix{
	  (\frac{1}{6}p_2-2a\cup 2a)\mathrm{Struc}_{[c]}(X)
	  \ar[rr]
	  \ar[d]
	  &&
	  {*}
	  \ar[d]^c
	  \\
	  \mathrm{Maps}(X, B \mathrm{String} \times E_8 )
	  \ar[rr]^{\frac{1}{6}p_2 - 2a\cup 2a}
	  &&
	  \mathrm{Maps}(X, B^7 U(1) )
	}
	}
	\,,
  $$
  where the right vertical map picks a cocycle $c$ representing the class $[c]$.
\end{definition}
In terms of these notions we thus see that 
\begin{observation}
Over a manifold $X$ with $\mathrm{String}$-structure
and with a fixed gauge bundle $E$,
condition (\ref{M5braneAnomalyCancellationCondition}) 
is precisely the condition that guarantees 
existence of a lift to 
\emph{$[8G_8]$-twisted $\mathrm{Fivebrane}^{2a \cup 2a}$-structure}
through the left vertical morphism in def. 
\ref{TwistedFivebrane2a2aStructures}.
\end{observation}

\subsubsection{Twisted differential structures in quantum anomaly cancellation}
\label{TwistedDiffStructuresForAnomalyCancellation}
\index{twisted cohomology!twisted differential structures in quantum anomaly cancellation}

We discuss now the differential refinements of the 
twisted topological structures from \ref{TwistedTopologicalStructuresInStringTheory}.

This section draws from \cite{SSSIII}.

\paragraph{Twisted differential $\mathbf{c}_1$-structures}
\label{Twistedc1Structures}
\index{twisted cohomology!twisted differential $\mathbf{c}_1$-structures}
\index{characteristic class!Chern class!first differential}

We discuss the differential refinement 
$\hat {\mathbf{c}}_1$ of the universal first Chern class,
indicated before in \ref{ChernWeilMotivatingExamples}.
The corresponding $\hat {\mathbf{c}}_1$-structures are simply
$\mathrm{SU}(n)$-principal connections, but the derivation of this
fact may be an instructive warmup for the examples to follow.

\medskip

For any $n \in \mathbb{N}$,
let $\mathbf{c}_1 : \mathbf{B}U(n) \to \mathbf{B}U(1)$
in $\mathbf{H} = \mathrm{Smooth}\infty \mathrm{Grpd}$
be the canonical representative of the 
universal smooth first Chern class, described in \ref{DeterminantLineBundle}.
In terms of the standard presentations 
$\mathbf{B}U(n)_{\mathrm{ch}}, \; \mathbf{B}U(1)_{\mathrm{ch}}
\in [\mathrm{CartSp}^{\mathrm{op}}, \mathrm{sSet}]$
of its domain and codomain
from prop. \ref{DeloopedLieGroup} this is given by the 
determinant function, which over any $U \in \mathrm{CartSp}$ sends
$$
  \mathrm{det} : C^\infty(U, U(n)) \to C^\infty(U, U(1))
  \,.
$$
Write $\mathbf{B}U(n)_{\mathrm{conn}}$ for the 
differential refinement from prop. \ref{GroupoidOfLieAlgebraValuedForms}.
Over a test space $U \in \mathrm{CartSp}$ the set of objects is
the set of $\mathfrak{u}(n)$-valued differential forms
$$
  \mathbf{B}U(n)_{\mathrm{conn}}(U)_0
  =
  \Omega^1(U, \mathfrak{u}(n))
$$
and the set of morphisms is that of smooth $U(n)$-valued differential forms,
acting by gauge transformations on the $\mathfrak{u}(n)$-valued 
1-forms
$$
  \mathbf{B}U(n)_{\mathrm{conn}}(U)_1
  =
  \Omega^1(U, \mathfrak{u}(n)) \times C^\infty(U, U(n))
  \,.
$$
\begin{proposition}
  \label{PresentationOfDifferentialFirstChernClass}
  The smooth universal first Chern class has a differential refinement
  $$  
    \hat {\mathbf{c}}_1
	: 
	\mathbf{B} U(n)_{\mathrm{conn}}
	 \to
	\mathbf{B} U(1)_{\mathrm{conn}}
  $$
  given on $\mathfrak{u}(n)$-valued 1-forms by taking the trace
  $$
    \mathrm{tr} : \mathfrak{u}(n) \to \mathfrak{u}(1)
	\,.
  $$
\end{proposition}
The existence of this refinement allows us to consider
differential and twisted differential $\hat {\mathbf{c}}_1$-structures.
\begin{lemma}
  \label{HomotopyFiberOfHatc1}
  There is  an $\infty$-pullback diagram
  $$   
    \xymatrix{
	  \mathbf{B}\mathrm{SU}(n)_{\mathrm{conn}}
	  \ar[d]
	  \ar[r]
	  &
	  {*}
	  \ar[d]
	  \\
	  \mathbf{B}U(n)_{\mathrm{conn}}
	  \ar[r]
	  &
	  \mathbf{B}U(1)_{\mathrm{conn}}
	}
  $$
  in $\mathrm{Smooth}\infty \mathrm{Grpd}$.
\end{lemma}
\proof
  We use the factorization lemma, \ref{FactorizationLemma},
  to resolve the right vertical morphism by a fibration 
  $$
     \mathbf{E}U(1)_{\mathrm{conn}} 
	 \to 
	 \mathbf{B}U(1)_{\mathrm{conn}}
  $$
  in 
  $[\mathrm{CartSp}^{\mathrm{op}}, \mathrm{sSet}]_{\mathrm{proj}}$.
  This gives that an object in $\mathbf{E}U(1)_{\mathrm{conn}}$
  over some test space $U$
  is a morphism of the form 
  $\xymatrix{
    0 \ar[r]^{g} & g^{-1} d_U g
  }$
  for $g \in C^\infty(U,U(1))$, and a morphism in 
  $\mathbf{E}U(1)_{\mathrm{conn}}$
  is given by a commuting diagram
  $$  
    \mathbf{E}U(1)_{\mathrm{conn}}
	=
	\left\{
	  \raisebox{20pt}{
	  \xymatrix{
	    & 0
		\ar[dl]_{g_1} \ar[dr]^{g_2}
		\\
		g_1^{-1} d_U g_1
		\ar[rr]^{h}
		&&
		g_2^{-1} d_U g_2
	  }}
	\right\}
	\,,
  $$
  where on the right we have $h \in C^\infty(U, U(1))$ such that
  $h g_1 = g_2$.
  The morphism to $\mathbf{B}U(1)_{\mathrm{conn}}$ is given by
  the evident projection onto the lower horizontal part of these
  triangles.
  
  Then the ordinary 1-categorical pullback of $\mathbf{E} U(1)_{\mathrm{conn}}$
  along $\hat {\mathbf{c}}_1$ yields the smooth groupoid
  $\hat {\mathbf{c}_1}^* \mathbf{E} U(1)_{\mathrm{conn}}$
  given over any test space $U$ as follows.
  \begin{itemize}  
    \item 
	  objects are pairs consisting of a 
	  $\mathfrak{u}(n)$-valued 1-form 
	  $A \in \Omega^1(U, \mathfrak{u}(n))$ and a smooth function
	  $\rho \in C^\infty(U, U(1))$ such that
	  $$
	    \mathrm{tr} A = \rho^{-1} d \rho
		\,;
	  $$
	\item
	  morphisms $g : (A_1, \rho_1) \to (A_2, \rho_2)$ are labeled by
	  a smooth function $g \in C^\infty(U, U(n))$ such that
	  $A_2 = g^{-1}(A_1 + d_U)g$.
  \end{itemize}
  Therefore there is a canonical functor
  $$
    \mathbf{B} \mathrm{SU}(n)_{\mathrm{conn}}
	\to
	\hat {\mathbf{c}_1}^* \mathbf{E} U(1)_{\mathrm{conn}}
  $$
  induced from the defining inclusion $\mathrm{SU}(n) \to \mathrm{U}(n)$,
  which hits precisely the objects for which $\rho$ is the constant function
  on $1 \in \mathrm{U}(1)$ and which is a bijection to the morphisms
  between these objects, hence is full and faithful. 
  The functor is also essentially surjective, since every 
  1-form of the form $h^{-1} d h$ is gauge equivalent to the 
  identically vanishing 1-form. Therefore it is a weak equivalence
  in $[\mathrm{CartSp}^{\mathrm{op}}, \mathrm{sSet}]_{\mathrm{proj}}$.
  By prop. \ref{FiniteHomotopyLimitsInPresheaves} this proves the
  claim.  
\endofproof
\begin{proposition}
  For $X$ a smooth manifold, we have an $\infty$-pullback
  of smooth groupoids
  $$
    \xymatrix{
	  \mathrm{SU}(n)\mathrm{Bund}_\nabla(X)
	  \ar[r]
	  \ar[d]
	  &
	  {*}
	  \ar[d]
	  \\
	  \mathrm{U}(n)\mathrm{Bund}_\nabla(X)
	  \ar[r]^{\hat {\mathbf{c}}_1}
	  &
	  \mathrm{U}(1)\mathrm{Bund}_\nabla(X)
	}
	\,.
  $$
\end{proposition}
\proof
  This follows from lemma \ref{HomotopyFiberOfHatc1} and the
  facts that for a Lie group $G$ we have
  $\mathbf{H}(X, \mathbf{B}G_{\mathrm{conn}})
  \simeq G \mathrm{Bund}_\nabla(X)$ and 
  that the hom-functor $\mathbf{H}(X,-)$ preserves
  $\infty$-pullbacks.
\endofproof

\paragraph{Twisted differential $\mathrm{spin}^c$-structures}
\label{TwistedSpinCStructures}
\index{twisted cohomology!twisted $\mathrm{spin}^c$-structures}

As opposed to the $\mathrm{Spin}$-group, which is a 
$\mathbb{Z}_2$-extension of the special orthogonal group, 
the $\mathrm{Spin}^c$-group, def. \ref{SpinCAsLieGroup}, 
is a $U(1)$-extension of $\mathrm{SO}$. 
This means
that twisted $\mathrm{Spin}^c$-structures 
have interesting smooth refinements. These we discuss here.

Two standard properties of $\mathrm{Spin}^c$ are
the following (see \cite{LawsonMichelson}).
\begin{observation}
  There is a short exact sequence
  $$
    U(1) \to \mathrm{Spin}^c \to \mathrm{SO}
  $$
  of Lie groups,
  where the first morphism is the canonical inclusion.
\end{observation}
\begin{proposition}
  \label{TopologicalFiberSequenceForSpinC}
  There is a fiber sequence
  $$
    B \mathrm{Spin}^c(n) 
	   \to 
	B \mathrm{SO}(n)
	 \stackrel{W_3}{\to}
	K(\mathbb{Z},3)
  $$
  of classifying spaces in $\mathrm{Top}$, 
  where $W_3$ is a representative of the universal 
  third integral Stiefel-Whitney class.
\end{proposition}
Here $W_3$ is a classical definition, but, as we will
show below, the reader can think of it as being defined
as the geometric realization of the smooth characteristic class
$\mathbf{W}_3$ from example \ref{ThirdIntegralStiefelWhitneyClass}.
Before turning to that, we record the notion of twisted structure
induced by this fact:
\begin{definition}
  For $X$ an oriented manifold of dimension $n$, 
  a \emph{$\mathrm{Spin}^c$-structure} on $X$ is a 
  trivialization
  $$
    \eta : * \stackrel{\simeq}{\to}  W_3(o_X)
	\,,
  $$
  where $o_X : X \to B \mathrm{SO}$ is the given orientation 
  structure.  
\end{definition}
\begin{observation}
  This is equivalently a lift $\hat o_X$ of $o_X$:
  $$
    \xymatrix{ 
	   & B \mathrm{Spin}^c \ar[d]
	   \\
	   X \ar[r]_{o_X}
         \ar@{-->}[ur]^{\hat o_X}	   
	   & B \mathrm{SO}
	}
	\,.
  $$
\end{observation}
\proof
    By prop. \ref{TopologicalFiberSequenceForSpinC}
  and the univsersal property of the homotopy pullback:
  $$
    \xymatrix{ 
	  X 
	  \ar@/^1pc/[drr]
	  \ar@/_1pc/[ddr]_{o_X}
	  \ar@{-->}[dr]^{\hat o _X}
	  \\
	  & B \mathrm{Spin}^c \ar[d]\ar[r] & {*} \ar[d]
	  \\
	  & B \mathrm{SO} \ar[r]^{W_3} & K(\mathbb{Z},3)
	}
	\,.
  $$
\endofproof 
From the general reasoning of twisted cohomology,
def. \ref{TwistedCohomologyInOvertopos}, in the language of 
twisted $\mathbf{c}$-structures,
def. \ref{TwistedCStructures}, we are therefore led to consider
the following.
\begin{definition}
  \index{twisted cohomology!twisted $\mathrm{spin}^c$-structure}
  \label{BareTwistedSpinCStructures}
  The $\infty$-groupoid of \emph{twisted $\mathrm{spin}^c$-structures}
  on $X$ is $W_3 \mathrm{Struc}_{\mathrm{tw}}(X)$.
\end{definition}
\begin{remark}
  It follows from the definition that twisted $\mathrm{spin}^c$-structures
  over an orientation structure $o_X$, def. \ref{OrientationStructure},
  are naturally identified with
  equivalences (homotopies)
  $$
    \eta : c \stackrel{\simeq}{\to} W_3(o_X)
	\,,
  $$
  where $c \in \infty\mathrm{Grpd}(X, B^2 U(1))$ is a given 
  twisting cocycle.
\end{remark}
In this form twisted $\mathrm{spin}^c$-structures 
have been considered in \cite{Douglas} and in \cite{Wang}. 
We  now establish a smooth refinement of this situation.

\begin{observation}
  \label{SmoothLiftOfW3}
  There is an essentially unique lift in $\mathrm{Smooth}\infty\mathrm{Grpd}$
  of $W_3$ 
  through the geometric realization 
  $$
    \vert -\vert
	:
	\mathrm{Smooth}\infty\mathrm{Grpd}
	 \stackrel{\Pi}{\to}
	\infty \mathrm{Grpd}
	 \stackrel{\simeq}{\to}
	\mathrm{Top}
  $$  
  (discussed in \ref{SmoothStrucHomotopy})
  of the form
  $$
    \mathbf{W}_3 : \mathbf{B}\mathrm{SO} \to \mathbf{B}^2 U(1)
	\,,
  $$
  where $\mathbf{B}\mathrm{SO}$ is the delooping of the Lie
  group $\mathrm{SO}$ in $\mathrm{Smooth}\infty \mathrm{Grpd}$
  and $\mathbf{B}^2 U(1)$ that of the smooth circle 2-group,
  as in \ref{SmoothStrucCohesiveInfiniGroups}.
\end{observation}
\proof
  This is a special case of theorem \ref{LieGroupCohomologyInSmoothInfinGroupoid}.
\endofproof
\begin{theorem}
  \label{BSpinCIsHomotopyFiberOfSmoothW3}
  In $\mathrm{Smooth}\infty \mathrm{Grpd}$ we have a 
  fiber sequence of the form
  $$
    \mathbf{B}\mathrm{Spin}^c
	  \to 
	\mathbf{B} \mathrm{SO}
	  \stackrel{\mathbf{W}_3}{\to}
	\mathbf{B}^2 U(1)
	\,,
  $$
  which refines the sequence of prop. \ref{TopologicalFiberSequenceForSpinC}.
\end{theorem}
We consider first a lemma.
\begin{lemma}
  \label{ExplicitSmoothLiftOfW3}
  \index{characteristic class!Stiefel-Whitney class!third integral, smooth}
  A presentation of the essentially unique smooth lift 
  of $W_3$ from observation \ref{SmoothLiftOfW3},
  is given by the morphism of simplicial presheaves
  $$
    \mathbf{W}_3 : 
	\mathbf{B} \mathrm{SO}_{\rm ch} 
	 \stackrel{\mathrm{w}_2}{\to}
	\mathbf{B}^2 \mathbb{Z}_2
	  \stackrel{\boldsymbol{\beta}_2}{\to}
	\mathbf{B}^2 U(1)_{\rm ch}
	\,,
  $$
  where the first morphism is that of example \ref{SecondStiefelWhineyClass}
  and where the second morphism is the one induced from the canonical 
  subgroup embedding.
\end{lemma}
\proof
  The bare Bockstein homomorphism is presented, by example \ref{BocksteinHomomorphism}, by the $\infty$-anafunctor
  $$
    \xymatrix{
	  \mathbf{B}^2 (\mathbb{Z} \stackrel{\cdot 2}{\to} \mathbb{Z})
	  \ar[d]^{\simeq}
	  \ar[r]
	  &
	  \mathbf{B}^2 (\mathbb{Z} \to 1)
	  \ar@{=}[r]
	  &
	  \mathbf{B}^3 \mathbb{Z}
	  \\
	  \mathbf{B}^2 \mathbb{Z}_2
	}
	\,.
  $$
  Accordingly we need to consider the lift of 
  the morphism 
  $$
    \boldsymbol{\beta}_2 : \mathbf{B}^2 \mathbb{Z}_2 \to 
	\mathbf{B}^2 U(1)
  $$ 
  induced form subgroup inclusion to
  to a comparable $\infty$-anafunctor. This is accomplished by
  $$
    \xymatrix{
	  \mathbf{B}^2 (\mathbb{Z} \stackrel{\cdot 2}{\to} \mathbb{Z})
	  \ar[r]^{\hat {\boldsymbol{\beta}_2}}
	  \ar[d]^{\simeq}
	  &
	  \mathbf{B}^2 (\mathbb{Z} \stackrel{\cdot 2}{\to}\mathbb{R})
	  \ar[d]^\simeq 
	  \\
	  \mathbf{B}^2 \mathbb{Z}_2
	  \ar[r]^{
	  \mathbf{\beta_2}}
	  &
	  \mathbf{B}^2 U(1)
	}
	\,.
  $$
  Since $\mathbb{R}$ is contractible, we have indeed under
  geometric realization, \ref{ETopStrucHomotopy}, an equivalence
  $$
    \xymatrix{
	  \vert \mathbf{B}^2 (\mathbb{Z} \stackrel{\cdot 2}{\to} \mathbb{Z})\vert 
	  \ar[r]^{\vert \hat {\boldsymbol{\beta}}_2\vert}
	  \ar[d]^{\simeq}
	  &
	  \vert \mathbf{B}^2(\mathbb{Z} \stackrel{\cdot 2}{\to} \mathbb{R})\vert
	  \ar[d]^{\simeq}
	  \\
	  \vert \mathbf{B}^2(\mathbb{Z} \stackrel{\cdot 2}{\to} \mathbb{Z})\vert
	  \ar[r]
	  \ar[d]^{\simeq}
	  &
	  \vert \mathbf{B}^2 (\mathbb{Z} \to 1)\vert
	  \ar[d]^{\simeq}
	  \\
	  \vert \mathbf{B}^2 \mathbb{Z}_2\vert
	  \ar[r]^{|\beta_2|}
	  &
	  \vert \mathbf{B}^3 \mathbb{Z}\vert
	}
	\,,
  $$
  where $|\beta_2|$ is the geometric realization of $\mathbb{\beta}_2$, according to definition
  \ref{def rrw}. 
  
 \endofproof
\proofoftheorem{BSpinCIsHomotopyFiberOfSmoothW3}
  Consider the pasting diagram in $\mathrm{Smooth}\infty \mathrm{Grpd}$
  $$
    \xymatrix{
	  \mathbf{B}\mathrm{Spin}^c 
	  \ar[r]
	  \ar[d]
	  &
	  \mathbf{B} U(1)
	  \ar[r]
	  \ar[d]|{{\bf c}_1~{\rm mod}~ 2}
	  &
	  {*}
	  \ar[d]
	  \\
	  \mathbf{B}\mathrm{Spin}
	  \ar[r]^{\mathbf{w}_2}
	  &
	  \mathbf{B}^2 \mathbb{Z}_2
	  \ar[r]^{\boldsymbol{\beta}_2}
	  &
	  \mathbf{B}^2 U(1)
	}
	\,.
  $$
  The square on the right is an $\infty$-pullback by 
  prop. \ref{GroupCentralExtensionCohomology}. 
  The square on the left is an $\infty$-pullback by
  proposition \ref{SpinCAsHomotopyFiberProductOfU1AndSO}.
  Therefore by the pasting law \ref{PastingLawForPullbacks}
  the total outer rectanle is an $\infty$-pullback.
  By lemma \ref{ExplicitSmoothLiftOfW3} the composite bottom 
  morphism is indeed the smooth lift $\mathbf{W}_3$
  from observation \ref{SmoothLiftOfW3}.
\endofproof
Therefore we are entitled to the following 
smooth refinement of def. \ref{BareTwistedSpinCStructures}.

\begin{remark}
$\mathbf{B}{\rm Spin}^c$ is the moduli stack of Spin${}^c$-structures,
or, equivalently Spin${}^c$-principal bundles. 
\end{remark}

\begin{definition}
  \index{twisted cohomology!twisted $\mathrm{spin}^c$-structure!smooth}
  For any $X \in \mathrm{Smooth}\infty \mathrm{Grpd}$, 
  the 1-groupoid of smooth \emph{twisted $\mathrm{spin}^c$-structures}
  $\mathbf{W}_3 \mathrm{Struc}_{\mathrm{tw}}(X)$
  is the homotopy pullback
  $$
    \xymatrix{
	  \mathbf{W}_3 \mathrm{Struc}_{\mathrm{tw}}(X)
	  \ar[d]
	  \ar[r] & 
	  H^3(X, \mathbb{Z})
	  \ar[d]
	  \\
	  \mathrm{Smooth}\infty\mathrm{Grpd}(X, \mathbf{B}\mathrm{SO})
	  \ar[r]^{\mathbf{W}_3}
	  &
	  \mathrm{Smooth}\infty \mathrm{Grpd}(X, \mathbf{B}^2 U(1))
	}
	\,.
  $$
\end{definition}
We briefly discuss an application of smooth twisted $\mathrm{spin}^c$-structures 
in physics.
\begin{remark}
  The action functional of the  
  $\sigma$-model of the open type II superstring 
  on a 10-dimensional target $X$
  has in general an anomaly, in that it is not a function, but
  just a section of a possibly non-trivial line bundle over the
  bosonic configuration space. In \cite{FreedWitten} 
  it was shown that in the case that the D-branes $Q \hookrightarrow X$ that
  the open string ends on carry a rank-1 Chan-Paton bundle,
  this anomaly vanishes precisely if this Chan-Paton bundle
  is a twisted line bundle exhibiting an equivalence
  $\mathbf{W}_3(\mathbf{o}_Q) \simeq H|_Q$ between the lifting gerbe
  of the $\mathrm{spin}^c$-structure and the restriction of the
  background Kalb-Ramond 2-bundle to $Q$. By the above discussion
  we see that this is precisely the datum of a smooth 
  twisted $\mathrm{spin}^c$-structure on $Q$, where the Kalb-Ramond
  field serves as the twist.   
  Below in \ref{HeteroticGreenSchwarz} we shall see that the
  quantum anomaly cancellation for the closed \emph{heterotic}
  superstring is analogously given by twisted $\mathrm{string}$-structures,
  which follow the same general pattern of twisted $\mathbf{c}$-structures,
  but in one degree higher.
\end{remark}
But in general this quantum anomaly cancellation involves
twists mediated by a higher rank twisted bundle. This situation we turn 
to now.
\begin{definition}
  \index{twisted cohomology!twisted $\mathrm{spin}^c$-structure!smooth!weakly twisted}
  \index{twisted cohomology!twisted bundle!by twisted $\mathrm{spin}^c$-structure}
  For $X$ equipped with orientation structure $o_X$, 
def. \ref{OrientationStructure},  and
  $c \in \mathbf{H}(X, \mathbf{B}^2 U(1))$ a twisting circle
  2-bundle, we say that the 2-groupoid of 
  \emph{weakly $c$-twisted $\mathrm{spin}^c$-structures}
  on $X$ is $(W_3(o_X)-c)$-twisted cohomology with respect to the
  morphism $\mathbf{c} : \mathbf{B} PU \to \mathbf{B}^2 U(1)$
  discussed in \ref{SmoothStrucTwistedCohomology}.
  \end{definition}
\begin{remark}
  \index{twisted cohomology!twisted bundle!on D-brane}
  By the discussion in \ref{SmoothStrucTwistedCohomology}
  in weakly twisted $\mathrm{spin}^c$-structure the two cocycles
  $W_3(o_X)$ and $c$ are not equivalent, but their difference is
  an  $n$-\emph{torsion} class (for some $n$) in $H^3(X, \mathbb{Z})$ 
  which twists
  a unitary rank-$n$ vector bundle on $X$
\end{remark}
\begin{remark}
  By a refinement of the discussion of \cite{FreedWitten} in 
  \cite{Kapustin} this structure is precisely what removes the
  quantum anomaly from the action functional of the type II superstring
  on oriented D-branes that carry a rank $n$ Chan-Paton bundle.
  A review is in \cite{Laine}.

  Notice that 
  for $i : Q \to X$ a $\mathrm{Spin}^c$-D-brane inclusion into spacetimes
  $X$, the 2-groupoid of $B$-field and brane gauge field bundles is
  the  \emph{relative} ($\mathbf{B}\mathrm{PU} \to \mathbf{B}^2 U(1)$)-cohomology
  on $i$, according to def. \ref{RelativeCohomology}.
 \end{remark}

\paragraph{Twisted differential string structures}
\label{HigherSpinStructure}
\label{Twisted differential String- and Fivebrane structures}

We consider now the obstruction theory for lifts through the smooth and differential refinement,
from \ref{FractionalClasses}, of the Whitehead tower of $O$.

\begin{definition}
  For $X$ a Riemannian manifold, equipping it with
  \begin{enumerate}
    \item orientation
    \item topological spin structure
    \item topological string structure
    \item topological fivebrane structure
  \end{enumerate}
  means equipping it with choices of (homotopy classes of) lifts of the classifying map
  $T X : X\to B O$ of its tangent bundle through the respective steps of the Whitehead tower
  of $B O$
  $$
    \xymatrix{
       & B \mathrm{Fivebrane} \ar[d] && \mbox{fivebrane structure}
      \\
       & B \mathrm{String} \ar[d] && \mbox{string structure}
      \\
       & B \mathrm{Spin} \ar[d] && \mbox{spin structure}
      \\
       & B \mathrm{SO} \ar[d] && \mbox{orientation}
      \\
      X \ar[r]^{T X} 
      \ar@{-->}[ur]
      \ar@{-->}[uur]
      \ar@{-->}[uuur]
      \ar@{-->}[uuur]
      \ar@/^.4pc/@{-->}[uuuur]
      & B \mathrm{O} && \mbox{Riemannian structure}
    }
    \,.
  $$
More in detail:
\begin{enumerate}
  \item 
    The set (homotopy 0-type) of orientations of a Riemannian manifold is the homotopy fiber of
    the first Stiefel-Whitney class
    $$
      (w_1)_* : \mathrm{Top}(X,B \mathrm{O}) \to \mathrm{Top}(X, B \mathbb{Z}_2)
      \,.
    $$
  \item
    The groupoid (homotopy 1-type) of topological spin structures of an oriented manifold is the homotopy fiber of
    the second Stiefel-Whitney class
    $$
      (w_2)_* : \mathrm{Top}(X,B \mathrm{SO}) \to \mathrm{Top}(X, B^2 \mathbb{Z}_2)
      \,.
    $$
  \item
    The 3-groupoid (homotopy 3-type) of topological string structures of a spin manifold is the homotopy fiber of
    the first fractional Pontryagin class
    $$
      (\frac{1}{2}p_1)_* : \mathrm{Top}(X,B \mathrm{Spin}) \to \mathrm{Top}(X, B^4 \mathbb{Z})
      \,,
    $$
  \item
    The 7-groupoid (homotopy 7-type) of topological fivebrane structures of a string manifold is 
    the homotopy fiber of
    the second fractional Pontryagin class
    $$
      (\frac{1}{6}p_2)_* : \mathrm{Top}(X,B \mathrm{String}) \to \mathrm{Top}(X, B^8 \mathbb{Z})
      \,,
    $$  
\end{enumerate}
\end{definition}
See \cite{SSSII} for background and the notion of fivebrane structure.
Using the results of \ref{FractionalClasses} we may 
lift this setup from discrete $\infty$-groupoids to smooth $\infty$-groupoids
and discuss the twisted cohomology, \ref{StrucTwistedCohomology},
relative to the smooth fractional Pontryagin classes 
$\frac{1}{2}\mathbf{p}_1$
and $\frac{1}{6}\mathbf{p}_2$ and their differential refinements 
$\frac{1}{2}{\hat {\mathbf{p}}}_1$
and $\frac{1}{6}{\hat {\mathbf{p}}}_2$
\begin{definition}
  \label{StringStructuresTwistedAndDifferential}
  Let $X \in \mathrm{Smooth}\infty \mathrm{Grpd}$ be any object.
  \begin{enumerate}
    \item
      The 2-groupoid of \emph{smooth string structures} on $X$ is the 
      homotopy fiber of the lift of the first fractional Pontryagin class $\frac{1}{2}\mathbf{p}_1$
      to $\mathrm{Smooth}\infty \mathrm{Grpd}$, prop. \ref{FirstFractionalDifferentialPontrjagin}:
      $$
        \mathbf{String}(X) \to \mathrm{Smooth}\infty\mathrm{Grpd}(X,\mathbf{B}\mathrm{Spin})
          \stackrel{(\frac{1}{2}\mathbf{p}_1)}{\to}
          \mathrm{Smooth}\infty \mathrm{Grpd}(X, \mathbf{B}^3 U(1))
          \,.
      $$
    \item
      The 6-groupoid of \emph{smooth fivebrane stuctures} on $X$ is the 
      homotopy fiber of the lift of the second fractional Pontryagin class $\frac{1}{6}\mathbf{p}_2$
      to $\mathrm{Smooth}\infty \mathrm{Grpd}$, prop. \ref{SecondFractionalDifferentialPontryagin}:
      $$
        \mathbf{Fivebrane}(X) \to \mathrm{Smooth}\infty\mathrm{Grpd}(X,\mathbf{B}\mathrm{String})
          \stackrel{(\frac{1}{6}\mathbf{p}_2)}{\to}
          \mathrm{Smooth}\infty \mathrm{Grpd}(X, \mathbf{B}^7 U(1))
          \,.
      $$
  \end{enumerate}
  More generally, 
  \begin{enumerate}
    \item
      The 2-groupoid of \emph{smooth twisted string sructures}\index{twisted cohomology!twisted string structure} on $X$ is the 
      $\infty$-pullback
      $$
        \xymatrix{
          \mathbf{String}_{\mathrm{tw}}(X) \ar[r]^{\mathrm{tw}} \ar[d] & H_{\mathrm{smooth}}^3(X,U(1)) \ar[d] 
          \\
          \mathrm{Smooth}\infty\mathrm{Grpd}(X,\mathbf{B}\mathrm{Spin})
          \ar[r]^{(\frac{1}{2}\mathbf{p}_1)}[r]
          &
          \mathrm{Smooth}\infty \mathrm{Grpd}(X, \mathbf{B}^3 U(1))
          }
      $$
      in $\infty \mathrm{Grpd}$.
    \item
      The 6-groupoid of \emph{smooth twisted fivebrane stuctures}\index{twisted cohomology!twisted fivebrane structure} on $X$ is the 
      $\infty$-pullback
      $$
        \xymatrix{
          \mathbf{Fivebrane}_{\mathrm{tw}}(X) \ar[r]^{\mathrm{tw}} \ar[d] & H_{\mathrm{smooth}}^7(X,U(1)) \ar[d] 
          \\
          \mathrm{Smooth}\infty\mathrm{Grpd}(X,\mathbf{B}\mathrm{String})
          \ar[r]^{(\frac{1}{6}\hat{\mathbf{p}}_2)}[r]
          &
          \mathrm{Smooth}\infty \mathrm{Grpd}(X, \mathbf{B}^7 U(1))
          }
      $$
    in $\infty \mathrm{Grpd}$.
  \end{enumerate}
  Finally, with $\frac{1}{2}\hat {\mathbf{p}_1}$ and $\frac{1}{4}\hat {\mathbf{p}_2}$
  the differential characteristic classes, \ref{StrucChern-WeilHomomorphism}, 
  we set
  \begin{enumerate}
    \item
      The 2-groupoid of \emph{smooth twisted differential string sructures}
	  \index{twisted cohomology!twisted differential string structure}
	  \index{differential string structure} on $X$ is the 
      $\infty$-pullback
      $$
        \xymatrix{
          \mathbf{String}_{\mathrm{tw}, \mathrm{diff}}(X) 
      \ar[r]^{\mathrm{tw}} \ar[d] & H_{\mathrm{diff}}^4(X) \ar[d] 
          \\
          \mathrm{Smooth}\infty\mathrm{Grpd}(X,\mathbf{B}\mathrm{Spin}_{conn})
          \ar[r]^{(\frac{1}{2}\hat {\mathbf{p}}_1)}[r]
          &
          \mathrm{Smooth}\infty \mathrm{Grpd}(X, \mathbf{B}^3 U(1)_{conn})
          }
      $$
      in $\infty \mathrm{Grpd}$.
    \item
      The 6-groupoid of \emph{smooth twisted differential fivebrane stuctures}
      \index{twisted cohomology!twisted differential fivebrane structure} 
      \index{connection!fivebrane 6-connection}
      on $X$ is the 
      $\infty$-pullback
      $$
        \xymatrix{
          \mathbf{Fivebrane}_{\mathrm{tw}, \mathrm{diff}}(X) \ar[r]^{\mathrm{tw}} \ar[d] 
                 & H_{\mathrm{diff}}^8(X) \ar[d] 
          \\
          \mathrm{Smooth}\infty\mathrm{Grpd}(X,\mathbf{B}\mathrm{String}_{\mathrm{conn}})
          \ar[r]^{(\frac{1}{6}{\hat {\mathbf{p}}}_2)}
          &
          \mathrm{Smooth}\infty \mathrm{Grpd}(X, \mathbf{B}^7 U(1)_{\mathrm{conn}})
          }
      $$
    in $\infty \mathrm{Grpd}$.
  \end{enumerate}  
  The image of a twisted (differential) String/Fivebrane structure under $\mathrm{tw}$ is 
  its \emph{twist}. The restriction to twists whose underlying class vanishes
  we also call \emph{geometric string structures} and \emph{geometric fivebrane structures}.
\end{definition}
\begin{observation}
\begin{enumerate}
\item These $\infty$-pullbacks are, up to equivalence, independent of the choise of the
  right vertical morphism, as long as this hits precisely one cocycle in each cohomology class.
\item The restriction of the $n$-groupoids of twisted structures to vanishing twist reproduces the
untwisted structures.
\end{enumerate}
\end{observation}
The local $L_\infty$-algebra valued form
data of differential twisted string- and
fivebrane structures has been considered in \cite{SSSIII}, as we explain in 
\ref{CocyclesForDifferentialStringStructures}.
Differential string structures for twists with underlying trivial class
(\emph{geometric string structures}) have been considered in 
\cite{Waldorf} modeled on bundle 2-gerbes. 

We have the following immediate consequences of the definition:
\begin{observation}
The spaces of choices of string structures extending a given spin structure $S$ are as follows
\begin{itemize}
\item if $[\frac{1}{2}\mathbf{p}_1(S)] \neq 0$ it is empty: $\mathrm{String}_S(X) \simeq \emptyset$;

\item if $[\frac{1}{2}\mathbf{p}_1(S)] = 0$ it is $\mathrm{String}_S(X) \simeq \mathbf{H}(X, \mathbf{B}^2 U(1))$.
\end{itemize}
In particular the set of equivalence classes of string structures lifting $S$ is the cohomology set 
  $$
    \pi_0 \mathrm{String}_S(X) \simeq H^2_{\mathrm{Smooth}}(X, \mathbf{B}^2 U(1))
    \,.
  $$
If $X$ is a smooth manifold, then this is $\simeq H^3(X, \mathbb{Z})$.
\end{observation}
\proof
Apply the pasting law for $\infty$-pullbacks, prop. \ref{PastingLawForPullbacks} on the diagram
$$
  \xymatrix{  
    \mathrm{String}_S(X) \ar[r] \ar[d] & \mathrm{String}(X) \ar[r] \ar[d] & {*} \ar[d]
    \\
    {*} \ar[r]^{S} & \mathbf{H}(X, \mathbf{B} \mathrm{Spin}(n))
    \ar[r]^{\frac{1}{2}\mathbf{p}_1}&
    \mathbf{H}(X, \mathbf{B}^3 U(1))
  }
  \,.
$$
The outer diagram defines the loop space object of $\mathbf{H}(X, \mathbf{B}^3 U(1))$. 
Since $\mathbf{H}(X,-)$ commutes with forming loop space objects we have
$$
  \mathrm{String}_S(X) \simeq \Omega \mathbf{H}(X, \mathbf{B}^3 U(1))
   \simeq
  \mathbf{H}(X, \mathbf{B}^2 U(1))
  \,.
$$
\endofproof
Sometimes it is useful to express string structures on $X$ in terms of circle 2-bundles/bundle gerbes
on the total space of the given spin bundle $P \to X$ \cite{Redden}:
\begin{proposition}
A smooth string structure on $X$ over a smooth $\mathrm{Spin}$-principal bundle
$P \to X$ induces a circle 2-bundle $\hat P$ on $P$ which restricted to any fiber $P_x \simeq \mathrm{Spin}$
is equivalent to the String 2-group extension $\mathrm{String} \to \mathrm{Spin}$.
\end{proposition}
\proof
 By prop. \ref{ExtendedBundlesByBundlesOfExtensions}.
\endofproof

\subparagraph{$L_\infty$-{\v C}ech cocycles for differential string structures}
\label{CocyclesForDifferentialStringStructures}
 \index{connection!string 2-connection!twisted}

We use the presentation of the $\infty$-topos $\mathrm{Smooth}\infty \mathrm{Grpd}$ 
by the local model structure on simplicial presheaves 
$[\mathrm{CartSp}_{\mathrm{smooth}}^{\mathrm{op}}, \mathrm{sSet}]_{\mathrm{proj},\mathrm{loc}}$ 
to give an explicit construction of twisted differential 
string structures in terms of 
{\v C}ech-cocycles with coefficients in $L_\infty$-algebra valued differential forms.
We will find a twisted version of the $\mathfrak{string}$-2-connections
discussed above in \ref{String2ConnectionsFromLieIntegration}.

\medskip

We need the following fact from \cite{FSS}.
\begin{proposition}
The differential fractional Pontryagin class $\frac{1}{2} \hat {\mathbf{p}}_1$ 
is presented in $[\mathrm{CartSp}_{\mathrm{smooth}}^{\mathrm{op}}, \mathrm{sSet}]_{\mathrm{proj}}$ 
by the top morphism of simplicial presheaves in
$$
  \xymatrix{
    \mathbf{cosk}_3\exp(\mathfrak{so})_{\mathrm{ChW},\mathrm{smp}}
     \ar[r]^{\exp(\mu, \mathrm{cs})}
     \ar[d]
      &
    \mathbf{B}^3 \mathbb{R}/\mathbb{Z}_{\mathrm{ChW},\mathrm{smp}}     
    \ar[d]
    \\
    \mathbf{cosk}_3\exp(\mathfrak{so})_{\mathrm{diff},\mathrm{smp}}
     \ar[r]^{\exp(\mu, \mathrm{cs})} 
     \ar[d]^\simeq
    &
    \mathbf{B}^3 \mathbb{R}/\mathbb{Z}_{\mathrm{smp}} 
    \\
    \mathbf{B}\mathrm{Spin}_{c}
  }
  \,.
$$
\end{proposition}
Here the middle morphism is the direct Lie integration of the $L_\infty$-algebra cocycle, 
\ref{SmoothStrucLieAlgebras}, 
while the top morphisms is its restriction to coefficients for $\infty$-connections,
\ref{SmoothStrucInfChernWeil}.

In order to compute the homotopy fibers of 
$\frac{1}{2}\hat {\mathbf{p}}_1$ we now find a resolution of this morphism 
$\exp(\mu,\mathrm{cs})$ by a fibration in 
$[\mathrm{CartSp}_{\mathrm{smooth}}^{\mathrm{op}}, \mathrm{sSet}]_{\mathrm{proj}}$. By the fact that this is a simplicial model category then also the hom of any cofibrant object into 
this morphism, computing the cocycle $\infty$-groupoids, is a fibration, and therefore, 
by the general natur of homotopy pullbacks, we obtain the homotopy fibers as the ordinary fibers of 
this fibration.

We start by considering such a factorization before differential refinement, 
on the underlying characteristic class $\exp(\mu)$.
To that end, we replace the Lie algebra $\mathfrak{g} = \mathfrak{so}$ by an equivalent but bigger 
Lie 3-algebra (following \cite{SSSIII}). We need the following notation:
\begin{itemize}
\item $\mathfrak{g} = \mathfrak{so}$, 
  the special orthogonal Lie algebra (the Lie algebra of the spin group);
\item  $b^2 \mathbb{R}$, the line Lie 3-algebra, def. \ref{LineLieNAlgebra}, the single generator in degee 3 
of its Chevalley-Eilenberg algebra we denote $c \in CE(b^2 \mathbb{R})$, $d c = 0$.

\item $\langle -,-\rangle \in \mathrm{W}(\mathfrak{g})$ 
  is the Killing form invariant polynomial, regarded as an element of the Weil algebra of 
$\mathfrak{so}$;

\item $\mu := \langle -,[-,-]\rangle \in \mathrm{CE}(\mathfrak{g})$, the degree 3 Lie algebra cocycle, 
identified with a morphism
  $$
    \mathrm{CE}(\mathfrak{g}) \leftarrow \mathrm{CE}(b^2 \mathbb{R}) : \mu
  $$
  of Chevalley-Eilenberg algebras; and normalized such that its continuation to a 3-form on 
  $\mathrm{Spin}$ is the image in  de Rham cohomology of $\mathrm{Spin}$ of a generator of 
   $H^3(\mathrm{Spin},\mathbb{Z}) \simeq \mathbb{Z}$;

\item $\mathrm{cs} \in \mathrm{W}(\mathfrak{g})$ is a Chern-Simons element, 
 def. \ref{TransgressionAndCSElements}, 
interpolating between 
 the two;

\item $\mathfrak{g}_\mu$, the string Lie 2-algebra, 
def. \ref{StringLie2Algebra}.
\end{itemize}
\begin{definition}
Let $(b\mathbb{R} \to \mathfrak{g}_\mu)$ denote the 
$L_\infty$-algebra whose Chevalley-Eilenberg algebra is
$$
  \mathrm{CE}(b\mathbb{R} \to \mathfrak{g}_\mu) = 
  (\wedge^\bullet(  \mathfrak{g}^* \oplus \langle b\rangle \oplus \langle c \rangle ), d)
  \,,
$$
with $b$ a generator in degree 2, and $c$ a generator in degree 3, and with differential 
defined on generators by
$$
  \begin{aligned}
    d|_{\mathfrak{g}^*} & = [-,-]^*
    \\
    d  b & = - \mu + c
    \\
    d c & =  0
  \end{aligned}
  \,.
$$
\end{definition}
\begin{observation} 
 \label{FactorizationOfTheCocycle}
The 3-cocycle $\mathrm{CE}(\mathfrak{g}) \stackrel{\mu}{\leftarrow} \mathrm{CE}(b^2 \mathbb{R})$ 
factors as 
$$
  \xymatrix{
    \mathrm{CE}(\mathfrak{g})
     \ar@{<-}[rr]^{(c \mapsto \mu, b \mapsto 0)}
   &&
    \mathrm{CE}(b\mathbb{R} \to \mathfrak{g})
   \ar@{<-}[r]^{(c \mapsto c)}{\leftarrow}
   &
   \mathrm{CE}(CE(b^2 \mathbb{R})
  }
  :
  \mu
  \,,
$$
where the morphism on the left (which is the identity when restricted to $\mathfrak{g}^*$ 
and acts on the new generators as indicated) is a quasi-isomorphism.
\end{observation}
\proof
To see that we have a quasi-isomorphism, notice that the dg-algebra is somorphic 
to the one with generators $\{t^a, b, c'\}$ and differentials
$$
  \begin{aligned}
    d|_{\mathfrak{g}^*} & = [-,-]^*
    \\
    d  b & = c'
    \\
    d c' & =  0
  \end{aligned}
  \,,
$$
where the isomorphism is given by the identity on the $t^a$s and on $b$ and by
$$
  c \mapsto c' + \mu
  \,.
$$
The primed dg-algebra is the tensor product 
$\mathrm{CE}(\mathfrak{g}) \otimes \mathrm{CE}( inn(b \mathbb{R}))$, 
where the second factor is manifestly cohomologically trivial.
\endofproof
The point of introducing the resolution $(b \mathbb{R} \to \mathfrak{g}_\mu)$ in the above way is that 
it naturally supports the obstruction theory of lifts from $\mathfrak{g}$-connections to 
string Lie 2-algebra 2-connections
\begin{observation}
 \label{LongFiberSequenceOnLieAlgebras}
The defining projection $\mathfrak{g}_\mu \to \mathfrak{g}$
factors through the above quasi-isomorphism 
$(b \mathbb{R} \to \mathfrak{g}_\mu) \to \mathfrak{g}$ by the canonical inclusion
$$
  \mathfrak{g}_\mu \to (b \mathbb{R} \to \mathfrak{g}_\mu)
  \,,
$$
which dually on $CE$-algebras is given by 
$$
  t^a \mapsto t^a
$$
$$
  b \mapsto - b
$$
$$
  c \mapsto 0
  \,.
$$
In total we are looking at a convenient presentation of the long fiber sequence of the 
string Lie 2-algebra extension:
$$
  \xymatrix{   
    & & (b \mathbb{R} \to \mathfrak{g}_\mu) \ar[r] \ar[d]^{\simeq} & b^2 \mathbb{R}
    \\
    b \mathbb{R} \ar[r] & \mathfrak{g}_\mu \ar[ur] \ar[r] & \mathfrak{g}
  }
  \,.
$$
\end{observation}
(The signs appearing here are just unimportant convention made in order for some of the formulas below to come out nice.)
\begin{proposition} 
  \label{BareFibration}
The image under Lie integration of the above factorization is
$$
  \exp(\mu) 
  :
   \mathbf{cosk}_3\exp(\mathfrak{g})
   \to 
   \mathbf{cosk}_3\exp(b \mathbb{R} \to \mathfrak{g}_\mu)
   \to 
   \mathbf{B}^3 \mathbb{R}/\mathbb{Z}_c
$$
where the first morphism is a weak equivalence followed by a fibration in the model structure 
on simplicial presheaves $[\mathrm{CartSp}_{\mathrm{smooth}}^{\mathrm{op}}, \mathrm{sSet}]_{\mathrm{proj}}$.
\end{proposition}
\proof
To see that the left morphism is objectwise a weak homotopy equivalence, notice that 
a $[k]$-cell of $\exp(b \mathbb{R} \to \mathfrak{g}_\mu)$ is identified with a pair consisting of a 
based smooth function $f : \Delta^k \to \mathrm{Spin}$ and a vertical 2-form 
$B \in \Omega^2_{\mathrm{si},\mathrm{vert}}(U \times \Delta^k)$, (both suitably with sitting instants perpendicular to the boundary of the simplex). Since there is no further condition on the 2-form, it can always be extended from the boundary of the 
$k$-simplex to the interior (for instance simply by radially rescaling it smoothly to 0). Accordingly the simplicial homotopy groups of $\exp(b \mathbb{R} \to \mathfrak{g}_\mu)(U)$ are the same as those of $\exp(\mathfrak{g})(U)$. The morphism between them is the identity in $f$ and picks $B = 0$ and is hence clearly an isomorphism on homotopy groups. 

We turn now to discussing that the second morphism is a fibration. The nontrivial degrees of the lifting problem
$$
  \xymatrix{
    \Lambda[k]_i \ar[r] \ar[d]
     & 
     \exp(b\mathbb{R} \to \mathfrak{g}_\mu)(U)
     \ar[d]
    \\
    \Delta[k] \ar[r] & \mathbf{B}^3 \mathbb{R}/\mathbb{Z}_c(U)
  }
$$
are $k = 3$ and $k = 4$.

Notice that a $3$-cell of $\mathbf{B}^3 \mathbb{R}/ \mathbb{Z}_c(U)$ is a smooth function $c : U \to \mathbb{R}/\mathbb{Z}$ and 
that the morphism $\exp(b\mathbb{R} \to \mathfrak{g}_\mu) \to \mathbf{B}^3 \mathbb{R}/\mathbb{Z}_c$ sends the pair $(f,B)$ to the fiber integration $\int_{\Delta^3}(f^* \langle \theta \wedge [\theta \wedge \theta]\rangle + d B)$. 

Given our lifting problem in degree 3,
we have given a function $c : U \to \mathbb{R}/\mathbb{Z}$ and a smooth function (with sitting instants at the subfaces) $U \times \Lambda^3_i \to \mathrm{Spin}$ together with a 2-form $B$ on the horn $U \times \Lambda^3_i$.

By pullback along the standard continuous retract $\Delta^3 \to \Lambda^3_i$ which is non-smooth only where $f$ has sitting instants,  we can always extend $f$ to a smooth function 
$f' : U \times \Delta^3 \to \mathrm{Spin}$ with the property that $\int_{\Delta^3} (f')^* \langle \theta \wedge [\theta \wedge \theta]\rangle = 0$. (Following the general discussion at Lie integration.)

In order to find a horn filler for the 2-form component, consider any smooth 2-form with sitting instants and non-vanishing integeral on $\Delta^2$, regarded as the missing face of the horn. By multiplying it with a suitable smooth function on $U$ we can obtain an extension $\tilde B \in \Omega^3_{\mathrm{si},\mathrm{vert}}(U \times \partial \Delta^3)$ of $B$ to all of $U \times \partial \Delta^3$ with the property that its integral over $\partial \Delta^3$ is the given $c$. By Stokes' theorem it remains to extend $\tilde B$ to the interior of $\Delta^3$ in any way, as long as it is smooth and has sitting instants.

To that end, we can find in a similar fashion a smooth $U$-parameterized family of closed 3-forms $C$ with sitting instants on $\Delta^3$, whose integral over $\Delta^3$ equals $c$. Since by sitting instants this 3-form vanishes in a neighbourhood of the boundary, the standard formula for the Poincare lemma applied to it produces a 2-form $B' \in \Omega^2_{\mathrm{si}, \mathrm{vert}}(U \times \Delta^3)$ with $d B' = C$ that itself is radially constant at the boundary. By construction the difference $\tilde B - B'|_{\partial \Delta^3}$ has vanishing surface integral. 
By the argument in the proof of prop. \ref{LieIntegrationToLineNGroup} 
it follows that the difference extends smoothly and with sitting instants to a closed 2-form $\hat B \in \Omega^2_{\mathrm{si},\mathrm{vert}}(U \times \Delta^3)$. Therefore the sum 
$B' + \hat B \in \Omega^2_{\mathrm{\mathrm{si}},\mathrm{\mathrm{vert}}}(U \times \Delta^3)$
equals $B$ when restricted to $\Lambda^k_i$ and has the property that its integral over $\Delta^3$ equals $c$.
Together with our extension $f'$, this constitutes a pair that solves the lifting problem.

The extension problem in degree 4 amounts to a similar construction: by coskeletalness the condition is that for a given $c : U \to \mathbb{R}/\mathbb{Z}$ and a given vertical 2-form on $U \times \partial \Delta^3$ such that its integral equals $c$, as well as a function $f : U \times \partial \Delta^3 \to Spin$, we can extend the 2-form and the functionalong $U \times \partial \Delta^3 \to U \times \Delta^3$. The latter follows from the fact that 
$\pi_2 \mathrm{Spin} = 0$ which guarantees a continuous filler (with sitting instants), and using the Steenrod-Wockel approximation theorem
\cite{Wockel} to make this smooth. 
We are left with the problem of extending the 2-form, which is the same problem we discussed above after the choice of $\tilde B$.
\endofproof
We now proceed to extend this factorization to the exponentiated differential coefficients, 
\ref{SmoothStrucInfChernWeil}. The direct  idea would be to use the evident factorization of
differential $L_\infty$-cocycles of the form
$$
  \xymatrix{
    \mathrm{CE}(\mathfrak{so}) \ar@{<-}[r] & 
    \mathrm{CE}(b \mathbb{R} \to \mathfrak{string}) \ar@{<-}[r] &
    \mathrm{CE}(b^2 \mathbb{R})
    \\
    \mathrm{W}(\mathfrak{so}) \ar@{<-}[r] \ar[u] & 
    \mathrm{W}(b \mathbb{R} \to \mathfrak{string}) \ar@{<-}[r] \ar[u] &
    \mathrm{W}(b^2 \mathbb{R}) \ar[u]
    \\
    \mathrm{inv}(\mathfrak{so}) \ar@{<-}[r] \ar[u] & 
    \mathrm{inv}(b \mathbb{R} \to \mathfrak{string}) \ar@{<-}[r] \ar[u] &
    \mathrm{inv}(b^2 \mathbb{R})  \ar[u]   
  }
  \,.
$$
For computations we shall find it convenient to consider this after a change of basis.
\begin{observation}
The Weil algebra $\mathrm{W}(b\mathbb{R} \to \mathfrak{g}_\mu)$ of 
$(b^2 \mathbb{R} \to \mathfrak{g})$ is given on the extra shifted generators 
$\{r^a = \sigma t^a, h = \sigma b , g = \sigma c\}$ by
$$
\begin{aligned}
 d t^a & = C^a{}_{b c} t^b \wedge t^c + r^a
 \\
 d r^a  & = - C^a{}_{b c} t^b \wedge r^a
 \\
  d b & = - \mu + c + h
 \\
 d h & = \sigma \mu - g
 \\
  d c & = g
\end{aligned}
$$
(where $\sigma$ is the shift operator extended as a graded derivation).
\end{observation}
\begin{definition}
Define $\tilde {\mathrm{W}}(b\mathbb{R} \to \mathfrak{g}_\mu)$ to be the dg-algebra
with the same underlying graded algebra as $\mathrm{W}(b\mathbb{R} \to \mathfrak{g}_\mu)$ 
but with the differential modified as follows
$$
\begin{aligned}
  d t^a & = C^a{}_{b c} t^b \wedge t^c + r^a
  \\
  d r^a  & = - C^a{}_{b c} t^b \wedge r^a
  \\
  d b & = - \mathrm{cs} + c + h
  \\
  d h & = \langle -,-\rangle - g
  \\
  d c & = g
\end{aligned}
 \,.
$$
Moreover, define $\tilde {\mathrm{inv}}(b \mathbb{R} \to \mathfrak{string})$ to be the 
dg-algebra
$$
  \tilde {\mathrm{inv}}(b \mathbb{R} \to \mathfrak{string})
  :=
  (\mathrm{inv}(\mathfrak{so})\otimes \langle g,h\rangle)/(d h = \langle-,-\rangle - g)
  \,.
$$
\end{definition}
\begin{observation}
 \label{DifferentialfactorizationOfStringLInfinityCocycle}
We have a commutative diagram of dg-algebras
$$
  \xymatrix{
    \mathrm{CE}(\mathfrak{so}) \ar@{<-}[r]_<<<<{\simeq} & 
    \mathrm{CE}(b \mathbb{R} \to \mathfrak{string}) \ar@{<-}[r] &
    \mathrm{CE}(b^2 \mathbb{R})
    \\
    \mathrm{W}(\mathfrak{so}) \ar@{<-}[r]_<<<<{\simeq} \ar[u] & 
    \tilde {\mathrm{W}}(b \mathbb{R} \to \mathfrak{string}) \ar@{<-}[r]\ar[u] &
    \mathrm{W}(b^2 \mathbb{R}) \ar[u]
    \\
    \mathrm{inv}(\mathfrak{so}) \ar@{<-}[r]_<<<<\simeq \ar[u] & 
    \tilde {\mathrm{inv}}(b \mathbb{R} \to \mathfrak{string}) \ar@{<-}[r] \ar[u] &
    \mathrm{inv}(b^2 \mathbb{R})  \ar[u]   
  }
$$
where $\tilde W(b \mathbb{R} \to \mathfrak{string}) \to W(\mathfrak{so})$
acts as
$$
  \begin{array}{l}
    t^a \mapsto t^a
    \\r^a \mapsto r^a
    \\ b \mapsto 0
    \\ c \mapsto \mathrm{cs} 
    \\ h \mapsto 0 
    \\ g \mapsto \langle -,-\rangle
  \end{array}
$$
and we identify $W(b^2 \mathbb{R}) = (\wedge^\bullet \langle c,g\rangle, d c = g)$.
The left horizontal morphisms are quasi-isomorphisms, as indicated.
\end{observation}
\begin{definition}
  We write $\exp(b \mathbb{R} \to \mathfrak{string})_{\tilde {\mathrm{ChW}}}$
  for the simplicial presheaf defined as
  $\exp(b \mathbb{R} \to \mathfrak{string})_{{\mathrm{ChW}}}$, but using 
  $\mathrm{CE}(b \mathbb{R} \to \mathfrak{string}) \leftarrow
    \tilde {\mathrm{W}}(b \mathbb{R} \to \mathfrak{string})
    \leftarrow
    \tilde {\mathrm{inv}}(b \mathbb{R} \to \mathfrak{string})
  $
  instead of the untwiddled version of these algebras.
\end{definition}
\begin{proposition} 
  \label{PresentationByFibration}
Under differential Lie integration 
the above factorization, observation \ref{DifferentialfactorizationOfStringLInfinityCocycle},
maps to a factorization
$$
  \exp(\mu, \mathrm{cs})
   :
  \mathbf{cosk}_3 \exp(\mathfrak{g})_{\mathrm{ChW}} 
   \stackrel{\simeq}{\to} 
  \mathbf{cosk}_3 \exp((b \mathbb{R} \to \mathfrak{g}_\mu))_{\tilde {\mathrm{ChW}}}
   \to
  \mathbf{B}^3 U(1)_{\mathrm{ChW},\mathrm{ch}}
$$
of $\exp(\mu,\mathrm{cs})$ in $[\mathrm{CartSp}^{\mathrm{op}}, \mathrm{sSet}]_{\mathrm{proj}}$, 
where the first morphism is a weak equivalence and the second a fibration.
\end{proposition}
\proof
We discuss that the first morphism is an equivalence. Clearly it is injective on homotopy groups: if a sphere of $A$-data cannot be filled, then also adding the $(B,C)$-data does not yield a filler. So we need to check that it is also surjective on homotopy groups: any two choices of $(B,C)$-data on a sphere are homotopic:
we may interpolate $B$ in any smooth way and then solve the equation $d B = - \mathrm{cs}(A)+ C + H$ 
for the interpolation of $C$. 

We now check that the second morphism is a fibration. It is itself the composite

$$
  \mathbf{cosk}_{3} \exp(b \mathbb{R} \to \mathfrak{g}_\mu)_{\mathrm{ChW}}
  \to 
  \exp(b^2 \mathbb{R})_{\mathrm{ChW}}/\mathbb{Z}
  \stackrel{\int_{\Delta^\bullet}}{\to}
  \mathbf{B}^3 \mathbb{R}/\mathbb{Z}_{\mathrm{ChW},\mathrm{ch}}
  \,.
$$
Here the second morphism is a degreewise surjection of simplicial abelian groups, hence a degreewise surjection under the normalized chain complex functor, hence is itself already a projective fibration. Therefore it is sufficient to show that the first morphism here is a fibration.

In degree $k = 0$ to $k = 3$ the lifting problems
$$
  \xymatrix{
     \Lambda[k]_i 
       \ar[r]
       \ar[d]
       & 
      \exp(b\mathbb{R} \to \mathfrak{g}_{\mu})_{\tilde {\mathrm{ChW}}}(U)
      \ar[d]
      \\
      \Delta[k] 
        \ar[r] &
      \exp(b^2 \mathbb{R})_{\mathrm{ChW}}/\mathbb{Z}(U)
  }
$$
may all be equivalently reformulated as lifting against a cylinder 
$D^k \hookrightarrow D^k \times [0,1]$ by using the sitting instants of all forms.

We have then a 3-form $H \in \Omega^3_{\mathrm{\mathrm{si}}}(U \times D^{k-1}\times [0,1])$ 
and differential form data $(A,B,C)$ on $U \times D^{k-1}$ given. We may always extend 
$A$ along the cylinder direction $[0,1]$ (its vertical part is equivalently a based 
smooth function to $\mathrm{Spin}$ which we may extend constantly). 
$H$  has to be horizontal so is already constantly extended along the cylinder.

We can then use the kind of formula that proves the Poincar{\'e} lemma to extend 
$B$. Let $\Psi : (D^k \times [0,1]) \times [0,1] \to (D^k  \times [0,1])$ be a 
smooth contraction. Then while $d(H - \mathrm{CS}(A) - C)$ may be non-vanishing, 
by horizonatlity of their curvature characteristic forms we still have that 
$\iota_{\partial_t} \Psi_t^* d(H - CS(A) - C)$ vanishes 
(since the contraction vanishes).

Therefore the 2-form
$$
  \tilde B := \int_{[0,1]} \iota_{\partial_t} \Psi_t^*(H - \mathrm{CS}(A)-C)
$$
satisfies $d \tilde B = (H - CS(A) - C)$. It may however not coincide with our 
given $B$ at $t = 0$. But the difference $B - \tilde B	_{t = 0}$ is a closed form on 
the left boundary of the cylinder. We may find some closed 2-form on the other boundary such 
that the integral around the boundary vanishes. Then the argument from the proof of the Lie integration of the line Lie n-algebra applies and we find an extension $\lambda$ to a closed 2-form on the interior. The sum
$$
  \hat B := \tilde B + \lambda
$$
then still satisfies $d \hat B = H - CS(A) - C$ and it coincides with $B$ on the left boundary.

Notice that here $\tilde B$ indeed has sitting instants: since $H$, $\mathrm{CS}(A)$ and $C$ have 
sitting instants they are constant on their value at the boundary in a neighbourhood perpendicular to the boundary, which means for these 3-forms in the degrees $\leq 3$ that they \emph{vanish} in a neighbourhood of the boundary, hence that the above integral is towards the boundary over a vanishing integrand.

In degree 4 the nature of the lifting problem 
$$
  \xymatrix{
  \Lambda[4]_i \ar[r] \ar[d]& \mathbf{cosk}_3\exp(b\mathbb{R} \to \mathfrak{g}_\mu)(U)
  \ar[d]
  \\
  \Delta[4] \ar[r] & \mathbf{B}^3 \mathbb{R}/\mathbb{Z}_{\mathrm{ChW},\mathrm{ch}}
  }
$$
starts out differently, due to the presence of $\mathbf{cosk}_3$, but it then ends up amounting 
to the same kind of argument:

We have four functions $U \to \mathbb{R}/\mathbb{Z}$ which we may realize as the fiber 
integration of a 3-form $H$ on $U \times (\partial \Delta[4] \setminus \delta_i \Delta[3])$,
and we have a lift to $(A,B,C, H)$-data on $U \times (\partial \Delta[4]\setminus \delta_i(\Delta[3]))$ 
(the boundary of the 4-simplex minus one of its 3-simplex faces). 

We observe that we can 
\begin{itemize}
\item always extend $C$ smoothly to the remaining 3-face such that its fiber integration there reproduces the signed difference of the four given functions corresponding to the other faces (choose any smooth 3-form with sitting instants and with non-vanishing integral and rescale smoothly);

\item fill the $A$-data horizonatlly due to the fact that $\pi_2 (\mathrm{Spin}) = 0$. 

\item the $C$-form is already horizontal, hence already filled. 
\end{itemize}
Moreover, by the fact that the 2-form $B$ already is defined on all of $\partial \Delta[4] \setminus \delta_i(\Delta[3])$ its fiber integral over the boundary $\partial \Delta[3]$ coincides with the fiber integral of $H - \mathrm{cs}(A) - C$ over $\partial \Delta[4] \setminus \delta_i (\Delta[3])$). 
But by the fact that we have lifted $C$ and the fact that $\mu(A_{\mathrm{vert}}) = \mathrm{cs}(A)|_{\Delta^3}$ 
is an integral cocycle, it follows that this equals the fiber integral of $C - \mathrm{cs}(A)$ 
over the remaining face.

Use then as above the vertical Poincar{\'e} lemma-formula to find $\tilde B$ on $U \times \Delta^3$ with sitting instants that satisfies the equation $d B = H - \mathrm{cs}(A) - C$ there. Then extend the closed difference $B - \tilde B|_{0}$ to a closed smooth 2-form on $\Delta^3$. As before, the difference
$$
  \hat B := \tilde B  + \lambda
$$
is an extension of $B$ that constitutes a lift.
\endofproof
\begin{corollary}
For any $X \in \mathrm{SmoothMfd} \hookrightarrow \mathrm{Smooth}\infty \mathrm{Grpd}$, 
for any choice of differentiaby good open cover with corresponding cofibrant presentation 
$\hat X = C(\{C_i\})\in [\mathrm{CartSp}_{\mathrm{smooth}}^{\mathrm{op}}, \mathrm{sSet}]_{\mathrm{proj}}$ 
we have that the 2-groupoids of twisted differential string structures 
are presented by the ordinary fibers of the morphism of Kan complexes
$$
  [\mathrm{CartSp}^{\mathrm{op}}, \mathrm{sSet}](\hat X,\exp(\mu,\mathrm{cs}))
$$
$$  
  [\mathrm{CartSp}^{op}, \mathrm{sSet}](\hat X, \mathbf{cosk}_3 \exp(b \mathbb{R} 
   \to \mathfrak{g}_\mu)_{\mathrm{ChW}})
  \to 
  [\mathrm{CartSp}^{\mathrm{op}}, \mathrm{sSet}](\hat X, \mathbf{B}^3 U(1)_{\mathrm{ChW}})
  \,.
$$
over any basepoints in the connected components of the Kan complex on the right, which correspond to the elements  
$[\hat {\mathbf{C}}_3] \in H_{\mathrm{diff}}^4(X)$ in the ordinary differential cohomology  of $X$.
\end{corollary}
\proof
Since $[\mathrm{CartSp}_{\mathrm{smooth}}^{\mathrm{op}}, \mathrm{sSet}]_{\mathrm{proj}}$ 
is a simplicial model category the morphism 
$[\mathrm{CartSp}^{\mathrm{op}}, \mathrm{sSet}](\hat X,\exp(\mu,\mathrm{cs}))$ 
is a fibration because $\exp(\mu,\mathrm{cs})$ is and $\hat X$ is cofibrant.

It follows from the general theory of homotopy pullbacks that the ordinary pullback of simplicial presheaves
$$
  \xymatrix{
    \mathbf{String}_{\mathrm{diff},\mathrm{tw}}(X) \ar[r] \ar[d] & H_{\mathrm{diff}}^4(X) \ar[d]
    \\
    [\mathrm{CartSp}^{\mathrm{op}}, \mathrm{sSet}](\hat X, \mathbf{cosk}_3 \exp(b \mathbb{R} 
     \to \mathfrak{g}_\mu)_{\mathrm{ChW}})
    \ar[r] & 
  [\mathrm{CartSp}^{\mathrm{op}}, \mathrm{sSet}](\hat X, \mathbf{B}^3 U(1)_{\mathrm{ChW}})
  }
$$
is a presentation for the defining $\infty$-pullback for $\mathbf{String}_{\mathrm{diff},\mathrm{tw}}(X)$.
\endofproof
We unwind the explicit expression for a twisted differential string structure under this equivalence.
Any twisting cocycle is in the above presentation given by a 
{\v C}ech-Deligne-cocycle, as discussed at \ref{SmoothStrucDifferentialCohomology}.
$$
  \hat {\mathbf{H}}_3 = ((H_3)_i, \cdots) 
$$
with local connection 3-form $(H_3)_i \in \Omega^3(U_i)$ and globally defined 
curvature 4-form\index{curvature!curvature 4-form} 
$\mathcal{G}_4 \in \Omega^4(X)$.
\begin{observation}
  \label{LocalDataOfTwistedDifferentialStringStruc}
A twisted differential string structure on $X$, twisted by this cocycle, is on patches $U_i$ a morphism
$$
  \Omega^\bullet(U_i)
  \leftarrow
  \tilde {\mathrm{W}}(b\mathbb{R}\to \mathfrak{g}_\mu)
$$
in $\mathrm{dgAlg}$, subject to some horizontality constraints.
The components of this are over each $U_i$ a collection of differential forms of the following structure
$$
  \left(
    \begin{array}{ll}
      F_\omega =& d \omega + \frac{1}{2}[\omega\wedge \omega]
      \\
      H_3 =& \nabla B := d B + CS(\omega) - C_3 
      \\
      \mathcal{G}_4 =& d C_3
      \\
      d F_\omega =& - [\omega \wedge F_\omega]
      \\
      d H_3 =& \mathcal{G}_4 - \langle F_\omega \wedge F_\omega \rangle
      \\
      d \mathcal{G}_4 =& 0
    \end{array}
  \right)_i
  \;\;\;\;
  \stackrel{
    \begin{array}{ll}
       t^a & \mapsto \omega^a
       \\
       r^a & \mapsto F^a_\omega
       \\
       b & \mapsto B
       \\
       c & \mapsto C_3
       \\
       h & \mapsto H_3
       \\
       g & \mapsto \mathcal{G}_4
    \end{array}
  }{\xymatrix{\ar@{<-|}[rrr]&&&}}
  \;\;\;\;
  \left(
    \begin{array}{ll}
       r^a  =& d t^a + \frac{1}{2}C^a{}_{b c} t^b \wedge t^c  
       \\
       h = & d b + \mathrm{cs} - c     
       \\
       g =& d c
       \\
       d r^a  =&  - C^a{}_{b c} t^b \wedge r^a
       \\
       d h =& \langle -,-\rangle - g
       \\
       d g =& 0
    \end{array}
  \right)
  \,.
$$
\end{observation}
Here we are indicating on the right the generators and their relation in $\tilde W(b\mathbb{R} \to \mathfrak{g}_\mu)$ and on the left their images and the images of the relations in $\Omega^\bullet(U_i)$.   This are first the definitions of the curvatures themselves and then the 
Bianchi identities\index{Bianchi identity!for twisted string connections} 
satisfied by these.

By  prop. \ref{CharacterizationOfexpChW} we have that for $\mathfrak{g}$ an 
$L_\infty$-algebra and 
$$
  \mathbf{B}G := \mathbf{cosk}_{n+1} \exp(\mathfrak{g})
$$
the delooping of the smooth Lie $n$-group obtained from it by Lie integration, 
def. \ref{ExponentiatedLInftyAlgbra}
the coefficient for $\infty$-connections on $G$-principal $\infty$-bundles is 
$$  
  \mathbf{B}G_{\mathrm{conn}} := \mathbf{cosk}_{n+1} \exp(\mathfrak{g})_{\mathrm{conn}}
  \,.
$$
\begin{proposition}
 \label{StringConnectionsFromDiffStringStructures}
 \index{string 2-connection}
 \index{connection!string 2-connection}
 The 2-groupoid of entirely untwisted differential string structures, 
def. \ref{StringStructuresTwistedAndDifferential},
on $X$ (the twist being $0 \in H^4_{\mathrm{diff}}(X)$) is equivalent to that of 
principal 2-bundles with 2-connection
 over the string 2-group, 
def. \ref{SmoothBString}, as discussed in \ref{String2ConnectionsFromLieIntegration}:
$$
  \mathrm{String}_{\mathrm{diff}, \mathrm{tw} = 0}(X) \simeq \mathrm{String} 2\mathrm{Bund}_{\nabla}(X)
  \,.
$$
\end{proposition}
\proof
By \ref{CocyclesForDifferentialStringStructures} we compute 
$\mathrm{String}_{\mathrm{diff}, \mathrm{tw} = 0}(X)$ as the ordinary fiber of the morphism of 
simplicial presheaves
$$
  [\mathrm{CartSp}^{\mathrm{op}}, \mathrm{sSet}](
  C(\{U_i\}),  \mathbf{cosk}_3 \exp(b \mathbb{R} \to \mathfrak{g}_\mu))
  \to 
  [\mathrm{CartSp}^{\mathrm{op}}, \mathrm{sSet}](
  C(\{U_i\}), \mathbf{B}^3 U(1)_{\mathrm{diff}})
$$
over the identically vanishing cocycle.

In terms of the component formulas of observation \ref{LocalDataOfTwistedDifferentialStringStruc}, 
this amounts to restricting to those cocyles for which over each $U \times \Delta^k$ 
the equations

$$
  C = 0
$$
$$
  G = 0
$$
hold.  Comparing this to the explicit formulas for 
$\exp(b \mathbb{R} \to \mathfrak{g}_\mu)$ and $\exp(b \mathbb{R} \to \mathfrak{g}_\mu)_{\mathrm{conn}}$ 
in \ref{CocyclesForDifferentialStringStructures} we see that these cocycles are exactly 
those that factor through the canonical inclusion
$$
  \mathfrak{g}_\mu \to (b \mathbb{R} \to \mathfrak{g}_\mu)
$$
from observation \ref{LongFiberSequenceOnLieAlgebras}.
\endofproof

\subparagraph{The Green-Schwarz mechanism in heterotic supergravity}
\label{HeteroticGreenSchwarz}
\index{Green-Schwarz mechanism}
\index{gauge theory!of heterotic supergravity}
\index{anomaly cancellation!Green-Schwarz anomaly}

Local differential form data as in observation \ref{LocalDataOfTwistedDifferentialStringStruc}
is known in theoretical physics in the context of the 
Green-Schwarz mechanism for 10-dimensional supergravity.
We conclude with some comments on the meaning and application of this result
(for background and references on the physics story see for instance \cite{SSSII}).

The standard action functionals of higher dimensional supergravity theories are
generically \emph{anomalous} in that instead of being functions
on the space of field configurations, they are just sections of a line bundle over these spaces.
In order to get a well defined action principle as input for a path-integral quantization
to obtain the corresponding quantum field theories, one needs to prescribe in addition the
data of a \emph{quantum integrand}. This is a choice of trivialization of these line
bundles, together with a choice of flat connection. For this to be possible the 
line bundle has to be trivializable and flat in the first place. Its failure to be tivializable
-- its Chern class -- is called the \emph{global anomaly}, and its failure to be flat -- its
curvature 2-form -- is called its local anomaly.

But moreover, the line bundle in question is the tensor product of two different line bundles
with connection. One is a Pfaffian line bundle induced from the fermionic degrees of freedom
of the theory, the other is a line bundle induced from the higher form fields of the theory
in the presence of higher \emph{electric and magnetic charge}. The Pfaffian line bundle is fixed by 
the requirement of supersymmetry, but there is freedom in choosing the background higher 
electric and magnetic charge. Choosing these appropriately such as to ensure that 
the tensor product of the two anomaly line bundles produces a flat trivializable line bundle
is called an \emph{anomaly cancellation} by a \emph{Green-Schwarz mechanism}.

Concretely, the higher gauge background field of 10-dimensional heterotic supergravity 
is the Kalb-Ramond field, which in the absence of \emph{fivebrane magnetic charge} is
modeled by a circle 2-bundle (bundle gerbe) with connection and curvature 3-form 
$H_3 \in \Omega^3_{\mathrm{cl}}(X)$, satisfying the higher \emph{Maxwell equation}
$$
  d H_3 = 0
  \,.
$$
Notice that we may think of a circle 2-bundle as a homotopy from the trivial circle
3-bundle to itself.

In order to cancel the relevant quantum anomaly it turns out that a magnetic background charge
density is to be added to the system whose differential form representative is the difference
$j_{\mathrm{mag}} := \langle F_{\nabla_{\mathrm{SU}}} \wedge F_{\nabla_{\mathrm{SU}}} \rangle
 - \langle F_{\nabla_{\mathrm{Spin}}} \wedge F_{\nabla_{\mathrm{Spin}}}\rangle$
 between the Pontryagin forms of the Spin-tangent bundle and a given $\mathrm{SU}$-gauge bundle.
This modifies the above Maxwell equation locally, on a patch $U_i \subset X$ to 
$$
  d H_i = \langle F_{A_i} \wedge F_{A_i} \rangle
 - \langle F_{\omega_i} \wedge F_{\omega_i}\rangle
   \,.
$$ 
Comparing with prop. \ref{LocalDataOfTwistedDifferentialStringStruc} and identifying
the curvature of the twist with $\mathcal{G}_4 = \langle F_{A_i} \wedge F_{A_i} \rangle$ 
we see that, while such $H_i$ can no longer be the curvature 3-form of a circle 2-bundle,
it can be the local 3-form component of a \emph{twisted} circle 3-bundle that is part of the 
data of a twisted differential string-structure. The above differential form
equation exhibits a de Rham homotopy between the two Pontryagin forms. This is
the local differential aspect of the very defnition of a twisted differential string-structure:
a homotopy from the Chern-Simons circle 3-bundle of the Spin-tangent bundle to a given
twisting circle 3-bundle.

    For many years the  anomaly cancellation for the heterotic superstring
    was known at the level of precision used in the physics community,
    based on a seminal article by Killingback.
      Recently \cite{Bunke} has given a rigorous proof in the special 
      case that underlying topological class of the
   twisting gauge bundle is trivial.  This proof used the 
   model of  twisted differential string structures with topologically tivial twist 
   given in \cite{Waldorf}.
   This model is explicitly constructed in terms of bundle 2-gerbes and doees not
   exhibit the homotopy pullback property of def. \ref{TwistedDifferentialStructures} explicitly. 
   However, the author shows that his model satisfies the
  abstract properties following from the universal property of the homotopy pullback.

When we take into account also gauge transformations of the gauge bundle, we should replace 
the homotopy pullback defining twisted differential string structurs this by the full
homotopy pullback
$$
  \xymatrix{
    \mathrm{GSBackground}(X) 
      \ar[r] 
      \ar[d]
     & \mathbf{H}_{\mathrm{conn}}(X, \mathbf{B}U)
        \ar[d]^{\hat {\mathbf{c}}_2}
    \\
    \mathbf{H}_{\mathrm{conn}}(X,\mathbf{B}\mathrm{Spin})
     \ar[r]^{\frac{1}{2}\hat {\mathbf{p}}_1}&
    \mathbf{H}_{\mathrm{dR}}(X, \mathbf{B}^3 U(1))
  }
  \,.
$$
The look of this diagram makes manifest how in this situation we are looking at the structures that homotopically cancel the differential classes $\frac{1}{2}\hat {\mathbf{p}}$ and $\hat {\mathbf{c}}_2$ against each other.

Since $\mathbf{H}_{dR}(X, \mathbf{B}^3 U(1))$ is abelian, we may also consider the corresponding 
Mayer-Vietoris sequence by realizing $\mathrm{GSBackground}(X)$ equivalently as the homotopy 
fiber of the difference of differential cocycles $\frac{1}{2}\hat {\mathbf{p}}_1 - \hat {\mathbf{c}}_2$.

$$
  \xymatrix{
    \mathrm{GSBackground}(X) \ar[rr] \ar[d] && {*} \ar[d]
    \\
    \mathbf{H}_{\mathrm{conn}}(X,\mathbf{B}\mathrm{Spin} \times \mathbf{B}U)
     \ar[rr]^{\frac{1}{2}\hat {\mathbf{p}}_1-\hat {\mathbf{c}}_2}&&
    \mathbf{H}_{\mathrm{dR}}(X, \mathbf{B}^4 U(1))
  }
  \,.
$$

\newpage

\subsubsection{Classical supergravity}
 \label{Supergravity}
 \index{gauge theory!supergravity as a gauge theory}
 \index{supergravity}

Action functionals of \emph{supergravity} are extensions to 
super-geometry, \ref{SuperInfinityGroupoids}, of the 
\emph{Einstein-Hilbert action functional} that models the
physics of \emph{gravity}. While these action functionals
are not themselves, generally, of higher Chern-Simons type,
\ref{StrucChern-SimonsTheory}, or of higher Wess-Zumino-Witten type,
\ref{StrucWZWFunctional}, some of them are low-energy
effective actions of \emph{super string field theory} action functionals, that
are of this type, as we discuss below in \ref{CSFTAction}.
Accordingly, supergravity action functionals typically exhibit
rich Chern-Simons-like substructures. 

A traditional introduction to the general topic can be found in
\cite{DeligneMorgan}. A textbook that aims for a more systematic
formalization is \cite{CDF}. Below in \ref{SugraFieldsAsInfinityConnection}
we observe that the discussion of supergravity there is secretly in terms of 
$\infty$-connections, \ref{Infinity-Connections}, 
with values in super $L_\infty$-algebras, \ref{SuperStrucLieAlg}.

\begin{itemize}
  \item 
    \ref{FirstOrderFormulationOfGravity} --
	  First-order/gauge theory formulation of gravity
  \item \ref{SuperPoincareAndExtensions} -- Higher extensions of the super 
    Poincar{\'e} Lie algebra;
  \item \ref{SugraFieldsAsInfinityConnection} --
    Supergravity fields are super $L_\infty$-connections
\end{itemize}

Much of this discussion we re-encounter when we consider super-Minkowski
spacetime as a target space for higher WZW models below in \ref{SuperBranesAndTheirIntersectionLaws}.

\paragraph{First-order/gauge theory formulation of gravity}
\label{FirstOrderFormulationOfGravity}
 \index{gauge theory!gravity as a gauge theory}
 \index{gravity}

The field theory of gravity (``general relativity'')
has a natural \emph{first order formulation} where a field configuration
over a given $(d+1)$-dimensional manifold $X$ is given by a 
$\mathfrak{iso}(d,1)$-valued Cartan connection, 
def. \ref{CartanConnection}.
The following statements briefly review this and
related facts (see for instance also the review in the 
introduction of \cite{Zanelli}).
\begin{definition}
  \label{PoincareGroup}
For $d \in \mathbb{N}$, the \emph{Poincar{\'e} group}
$\mathrm{ISO}(d,1)$ is the group of auto-isometries of the 
Minkowski space $\mathbb{R}^{d,1}$ equipped with its canonical 
pseudo-Riemannian metric $\eta$.

This is naturally a Lie group. Its Lie 
algebra is the \emph{Poincar{\'e} Lie algebra} $\mathfrak{iso}(d,1)$.
\end{definition}
We recall some standard facts about the Poincar{\'e} group.
\begin{observation}
The Poncar{\'e} group is the semidirect product 
$$
  \mathrm{ISO}(d,1)
  \simeq
  \mathrm{O}(d,1) \ltimes \mathbb{R}^{d+1}
$$
of the \emph{Lorentz group} $\mathrm{O}(d,1)$ of \emph{linear}
auto-isometries of $\mathbb{R}^{d,1}$, and the abelian translation group
in $(d+1)$ dimensions, with respect to the defining action of 
$\mathrm{O}(d,1)$ on $\mathbb{R}^{d,1}$. Accordingly there is a 
canonical embedding of Lie groups
$$  
  O(d,1) \hookrightarrow \mathrm{ISO}(d,1)
$$
and the corresponding coset space is Minkowski space
$$
  \mathrm{ISO}(d,1)/\mathrm{O}(d,1) \simeq \mathbb{R}^{d,1}
  \,.,
$$
Analogously the Poincar{\'e} Lie algebra is the semidirect product
$$
  \mathfrak{iso}(d,1)
  \simeq
  \mathfrak{so}(d,1)\ltimes \mathbb{R}^{d,1}
  \,,
$$
Accordingly there is a 
canonical embedding of Lie algebras
$$  
  \mathfrak{so}(d,1) \hookrightarrow \mathfrak{iso}(d,1)
$$
and the corresponding quotient of vector spaces is 
Minkowski space
$$
  \mathfrak{iso}(d,1)/\mathfrak{so}(d,1) \simeq \mathbb{R}^{d,1}
  \,.
$$
\end{observation}
Minkowski space $\mathbb{R}^{d,1}$ is the local model for 
\emph{Lorentzian manifolds}.
\begin{definition}
  A \emph{Lorentzian manifold} is a pseudo-Riemannian manifold
  $(X,g)$ such that each tangent space $(T_x X, g_x)$ 
  for any $x \in X$ is
  isometric to a Minkowski space $(\mathbb{R}^{d,1}, \eta)$.
\end{definition}
\begin{proposition}
Equivalence classes of  
$(O(d,1) \hookrightarrow \mathrm{ISO}(d,1))$-valued
Cartan connections, def. \ref{CartanConnection}, on a smooth manifold
$X$ are in canonical bijection with Lorentzian manifold
structures on $X$.
\end{proposition}
This follows from the following observations.
\begin{observation}
Locally over a patch $U \to X$ a 
$\mathfrak{iso}(d,1)$ connection is given by a
1-form 
$$
  A = (E, \Omega) \in \Omega^1(U, \mathfrak{iso}(d,1))
$$
with a component
$$
  E \in \Omega^1(U, \mathbb{R}^{d+1})
$$
and a component
$$  
  \Omega \in \Omega^1(U, \mathfrak{so}(d,1))
  \,.
$$
If this comes from a 
$(\mathrm{O}(d,1) \to \mathrm{ISO}(d,1))$-\emph{Cartan connection}
then $E$ is non-degenerate in that for all $x \in X$ the induced linear map
$$
  E : T_x X \to \mathbb{R}^{d+1}
$$
is a linear isomorphism. In this case $X$ is equipped 
with the Lorentzian metric
$$
  g := E^* \eta
$$
and $\Omega$ is naturally identified with a 
compatible metric connection on $T X$. 
Then curvature 2-form of the connection
$$
  F_A = (F_\Omega, F_E) \in \Omega^2(U, \mathfrak{iso}(d,1))
$$
has as components the \emph{Riemann curvature}
$$
  F_\Omega = d \Omega + \frac{1}{2}[\Omega \wedge \Omega]
  \in 
  \Omega^2(U, \mathfrak{so}(d,1))
$$
of the metric connection, as well as the \emph{torsion}
$$
  F_E = d E + [\Omega \wedge E]
  \in
  \Omega^2(U, \mathbb{R}^{d,1})
  \,.
$$
Therefore precisely if in addition the torsion vanishes is $\Omega$
uniquely fixed to be the Levi-Civita connection on $(X,g)$.
\end{observation}

Therefore the configuration space of gravity on a smooth manifold
$X$ may be identified with the moduli space of 
$\mathfrak{iso}(d,1)$-valued Cartan connections on $X$.
The field content of \emph{supergravity} is obtained from this
by passing from the to Poincar{\'e} Lie algebra
to one of its \emph{super Lie algeba extensions}, 
a \emph{super Poincar{\'e}} Lie algebra.

There are different such extensions. All involve some
spinor representation of the Lorentz Lie algebra $\mathfrak{so}(d,1)$
as odd-degree elements in the super Lie algebra
The choice of number $N$ of irreps in this representation.
But there are in general more choices, given 
by certain exceptional \emph{polyvector extensions} of 
such super-Poincar{\'e}-Lie algebras which contain 
also new even-graded elements. 

Below we show that these
Lie superalgebra polyvector extensions , in turn, are induced
from canonical \emph{super $L_\infty$-algebra extensions}
given by exceptional super Lie algebra cocycles, and that
the configuration spaces of higher dimensional supergravity 
may be identified with moduli spaces
of $\infty$-connections, \ref{LInfinityAlgebraicStructures},
withvalues in a super $L_\infty$-algebra, def. \ref{SuperLInfinityAlgebra}.
that arise as higher central extensions, def. \ref{gmu},
of a super Poincar{\'e} Lie algebra.

\newpage
\paragraph{$L_\infty$-extensions of the super Poincar{\'e} Lie algebra }
\label{SuperPoincareAndExtensions}

The super-Poincar{\'e} Lie algebra is the local gauge algebra of 
supergravity. It inherits the cohomology of the 
special orthogonal or Lorentz Lie algebra
$\mathfrak{so}(d,1)$, but crucially it exhibits a finite number 
of exceptional $\mathfrak{so}(d,1)$-invariant cocycles on its
super-translation algebra. The super $L_\infty$-algebra extensions
induced by these cocycles control the structure of higher dimensional
supergravity fields as well as of super-$p$-brane $\sigma$-models 
that propagate in a supergravity background.

\begin{itemize}
  \item \ref{TheSuperPoincareLieAlgebra} -- The super Poincar{\'e} Lie algebra;
  \item \ref{SuperM2Brane} -- M2-brane Lie 3-algebra and the M-theory Lie algebra;
  \item \ref{ExceptionalsisoCocyclesAndBraneScan} -- Exceptional cocycles and the brane scan.
\end{itemize}

\paragraph{The super Poincar{\'e} Lie algebra}
\label{TheSuperPoincareLieAlgebra}

\begin{definition}
  \label{SuperPoincareLieAlgebra}
  \index{$L_\infty$-algebra!super Poincar{\'e} Lie algebra}
  For $n \in \mathbb{N}$ and $S$ a spinor representation of $\mathfrak{so}(n,1)$,
  the corresponding \emph{super Poincar{\'e} Lie algebra} 
  $\mathfrak{sIso}(n,1)$ 
  is the super Lie algebra
  whose Chevalley-Eilenberg algebra $\mathrm{CE}(\mathfrak{sIso}(10,1)$ is 
  generated from 
  \begin{enumerate}
    \item generators $\{\omega^{a b}\}$ in degree $(1,\mathrm{even})$ dual to the standard
    basis of $\mathfrak{so}(n,1)$, 
    \item generators  $\{e^a\}$ in degree $(1,\mathrm{even})$ 
    \item
      and generators $\{\psi^\alpha\}$ in degree $(1,\mathrm{odd})$,
  dual to the spinor representation $S$
  \end{enumerate}
  with differential defined by
  $$
    d_{\mathrm{CE}} \omega^{a}{}_b = \omega^a{}_{c} \wedge \omega^{c}{}_d
  $$
  $$
    d_{\mathrm{CE}} e^a = \omega^a{}_b \wedge e^b + \frac{i}{2}\bar \psi \wedge \Gamma^a  \psi
  $$
  $$
    d_{\mathrm{CE}} \psi = \frac{1}{4} \omega^{a b} \Gamma_{a b} \psi
    \,,
  $$
  where $\{\Gamma^a\}$ is the corresponding representation of the Clifford algebra
  $\mathrm{Cl}_{n,1}$ on $S$, and here and in the following $\Gamma^{a_1 \cdots a_k}$ is shorthand
  for the skew-symmetrization of the matrix product $\Gamma^{a_1} \cdot \cdots \cdot \Gamma^{a_k}$
  in the $k$ indices.
\end{definition}

\subparagraph{M2-brane Lie 3-algebra and the M-theory Lie algebra}
\label{SuperM2Brane}
\index{M-theory!M2-brane}
\index{brane!M2-brane}
\index{M-theory!M-theory super Lie algebra}
\index{Wess-Zumino-Witten functionals!super $p$-branes!M2-brane}

We discuss an exceptional extension of the super Poincar{\'e}
Lie algebra in 11-dimensions by a super Lie 3-algebra and further
by super Lie 6-algebra. We show that the corresponding automorphism 
$L_\infty$-algebra contains the polyvector extension called the
\emph{M-theory super Lie algebra}.

\medskip

\begin{proposition}
  For $(n,1) = (10,1)$ and $S$ the canonical spinor representation,
  we have an exceptional super Lie algebra cohomology class in degree 4 
  $$
    [\mu_4] \in  H^{2,2}(\mathfrak{sIso}(10,1))
  $$
  with a representative given by
  $$
    \mu_4 := \frac{1}{2} \bar \psi \wedge \Gamma^{a b} \psi\wedge e_a \wedge e_b
    \,.
  $$
\end{proposition}
This is due to \cite{DF}. 
\begin{definition}
  \label{ElevenDSupergravityLie3Algebra}
  \index{$L_\infty$-algebra!M2-brane Lie 3-algebra}
  \index{supergravity!supergravity Lie 3-algebra} 
  The \emph{M2-brane super Lie 3-algebra}
  $\mathfrak{m}2\mathfrak{brane}_{\mathrm{gs}}$ is the 
  $b \mathbb{R}$-extension of $\mathfrak{sIso}(10,1)$ classified by $\mu_4$, 
  according to prop. \ref{PullBackCharacterizationOfgmu}
  $$
    b^2 \mathbb{R} \to \mathfrak{m}2\mathfrak{brane}_{\mathrm{gs}} \to \mathfrak{siso}(10,1)
    \,.
  $$
\end{definition}
In terms of its Chevalley-Eilenberg algebra this extension was first considered in 
\cite{DF}.
\begin{definition}
  \label{11dPoincarePolyvectorExtension}
  The \emph{polyvector extension} \cite{ACDP} of $\mathfrak{sIso}(10,1)$
  -- called the \emph{M-theory Lie algebra} --
  is the super Lie algebra obtained by adjoining to $\mathfrak{sIso}(10,1)$ generators
  $\{Q_\alpha, Z^{ab}\}$ that transform as spinors with respect to the 
  existing generators, and whose non-vanishing brackets among themselves are
  $$
    [Q_\alpha, Q_\beta] = i(C \Gamma^a)_{\alpha \beta} P_a
	+ (C \Gamma_{a b}) Z^{a b}
  $$
  $$
    [Q_\alpha, 	Z^{ab}] = 2 i (C \Gamma^{[a})_{\alpha \beta} Q^{b]\beta}
	\,.
  $$
\end{definition}
\begin{proposition}
  The automorphism super $L_\infty$-algebra
  $\mathfrak{der}(\mathfrak{m}2\mathfrak{brane}_{\mathrm{gs}})$, 
  def. \ref{automorphismLInfinityAlg},
  contains the polyvector extension of the 11d-super Poincer{\'e} algebra,
  def. \ref{11dPoincarePolyvectorExtension} precisely as its
  graded Lie algebra of exact elements.
\end{proposition}
\proof
  One can see that this is secretly what \cite{Castellani} shows.
\endofproof

\begin{proposition}
  There is a nontrivial degree-7 class 
  $[\mu_7] \in H^{5,2}(\mathfrak{m}2\mathfrak{brane}_{\mathrm{gs}})$
  in the super-$L_\infty$-algebra cohomology of the M2-brane Lie 3-algebra, 
  a cocycle representative of which is
  $$
    \mu_7 :=
    -\frac{1}{2}\bar \psi \wedge \Gamma^{a_1 \cdots a_5} \psi \wedge e_{a_1} \wedge \cdots \wedge e_{a_5}
    - \frac{13}{2} \bar \psi \wedge \Gamma^{a_1 a_2} \psi \wedge e_{a_1} \wedge e_{s_2} \wedge c_3
    \,,
  $$
  where $c_3$ is the extra generator of degree 3 in 
  $\mathrm{CE}(\mathfrak{m}2\mathfrak{brane}_{\mathrm{gs}})$.
\end{proposition}
This is due to \cite{DF}.
\begin{definition}
  \label{ElevenDSupergravityLie6Algebra}
  The \emph{M5-brane Lie 6-algebra}
  \index{supergravity!M5-brane Lie 6-algebra}
  \index{$L_\infty$-algebra!supergravity Lie 6-algebra} 
  $\mathfrak{m}5\mathfrak{brane}_{\mathrm{gs}}$ is the 
  $b^5 \mathbb{R}$-extension of $\mathfrak{m}2\mathfrak{brane}_{\mathrm{gs}}$ 
  classified by $\mu_7$, 
  according to prop. \ref{PullBackCharacterizationOfgmu}
  $$
    b^5 \mathbb{R} \to \mathfrak{m}5\mathfrak{brane}_{\mathrm{gs}} 
	  \to 
	   \mathfrak{m}2\mathfrak{brane}_{\mathrm{gs}}
    \,.
  $$  
\end{definition}

\subparagraph{Exceptional cocycles and the brane scan}
\label{ExceptionalsisoCocyclesAndBraneScan}

The exceptional cocycles discussed above are part of 
a pattern which traditionally goes by the name \emph{brane scan}
\cite{Duff87}.
\begin{proposition}
  \label{10dPoincare3Cocycle}
  For $d, p \in \mathbb{N}$, let $\mathfrak{sIso}(d,1)$ be the 
  super Poincar{\'e} Lie algebra, def. \ref{SuperPoincareLieAlgebra},
  and consider the element
  $$
    \bar \psi \Gamma_{a_0, \cdots, a_{p+1}} \wedge \psi
	\wedge
	e^{a_0}\wedge \cdots \wedge e^{a_{p+1}}
	\in
	\mathrm{CE}(\mathfrak{sIso}(d,1))
  $$
  in degree $p+2$ of the Chevalley-Eilenberg algebra. 
  This is closed, hence is a cocycle, for the  
  combinations of $D := d+1$ and $p \geq 1$ precisely where 
  there are non-empty and non-parenthesis entries
  in the following table.\\
  
  \begin{tabular}{c|ccccccccc}
     & $p = 1$ & $2$ & $3$ & $4$ & $5$
	 \\
	 \hline
	 $D = 11$ & & $\mathfrak{m}2\mathfrak{brane}_{\mathrm{gs}}$ & 
	 \hspace{30pt} & \hspace{30pt} & 
	    ($\mathfrak{m}5\mathfrak{brane}_{\mathrm{gs}}$)
	 \\
	 $10$ & $\mathfrak{string}_{\mathrm{gs}}$ & & & & 
	  $\mathfrak{ns}5\mathfrak{brane}_{\mathrm{gs}}$
	 \\
	 $9$ & & & & ${\ast}$
	 \\
	 $8$  & & & ${\ast}$
	 \\
	 $7$  & & ${\ast}$
	 \\
	 $6$  & ${\ast}$ & & $\ast$
	 \\
	 $5$ &  & $\ast$
	 \\
	 $4$  & $\ast$ & $\ast$
	 \\
	 $3$  & $\ast$
  \end{tabular}
\end{proposition}
The entries in the top two rows are labeled by the name of the extension of
$\mathfrak{sIso}(d,1)$ that the corresponding cocycle classifies.
By prop. \ref{ElevenDSupergravityLie3Algebra} the 7-cocycle that
defines $\mathfrak{m}5\mathfrak{brane}_{\mathrm{gs}}$ does not live on the Lie algebra
$\mathfrak{sIso}(10,1)$, but only on its Lie 3-algebra extension
$\mathfrak{m}2\mathfrak{brane}_{\mathrm{gs}}$. 
This is why in the context of the brane scan it does not appear
in the classical literature, which does not know about higher Lie algebras.

An explicitly Lie-theoretic discussion of these cocycles is in chapter
8 of \cite{AzIz}.
The extension 
$$
  b \mathbb{R} \to \mathfrak{string}_{\mathrm{gs}} \to \mathfrak{sIso}(9,1)
$$ 
and its Lie integration has been considered in \cite{Huerta}.

\paragraph{Supergravity fields are super $L_\infty$-connections}
\label{SugraFieldsAsInfinityConnection}

Among the varied literature in theoretical physics on the topic of
\emph{supergravity}
the book \cite{CDF} and the research program that it summarizes, 
starting with \cite{DF}, 
stands out as an  attempt to identify and make use
of a systematic mathematical structure controlling the general theory.
By careful comparison one can see that the notions considered
in that book may be translated into notions considered here under the following
dictionary
\begin{itemize}
  \item ``FDA'': the Chevalley-Eilenberg algebra $\mathrm{CE}(\mathfrak{g})$ 
     of a super $L_\infty$-algebra $\mathfrak{g}$ (def. \ref{SuperLInfinityAlgebra}), 
         def. \ref{LInftyGlgebroid};
  \item ``soft group manifold'': the Weil algebra $\mathrm{W}(\mathfrak{g})$ 
       of $\mathfrak{g}$, def. \ref{WeilAlgebra}
  \item ``field configuration'':  $\mathfrak{g}$-valued $\infty$-connection, def. \ref{Infinity-Connections}
  \item ``field strength'': curvature of $\mathfrak{g}$-valued $\infty$-connection, def. \ref{gValuedFormsAndCurvature}
  \item ``horizontality condition'': second $\infty$-Ehresmann condition, remark \ref{HorizontalityAndGaugeTransformation}
  \item ``cosmo-cocycle condition'': characterization of $\mathfrak{g}$-Chern-Simons elements, 
    def. \ref{TransgressionAndCSElements}, to first order in the curvatures;
\end{itemize}
All the super $L_\infty$-algebras $\mathfrak{g}$ appearing in \cite{CDF} are 
higher shifted central extensions, in the sense of prop. \ref{PullBackCharacterizationOfgmu},
of the super-Poincar{\'e} Lie algebra.

\subparagraph{The graviton and the gravitino}

\begin{example}
  \label{sisoConnectionContent}
  \index{graviton}
  \index{gravitino}
  For $X$ a supermanifold and 
  $\mathfrak{g} = \mathfrak{sIso}(n,1)$ the super Poincar{\'e} Lie algebra
  from def. \ref{SuperPoincareLieAlgebra},
  $\mathfrak{g}$-valued differential form data
  $$
    A : T X \to \mathfrak{siso}(n,1)
  $$
  consists of
  \begin{enumerate}
    \item
      an $\mathbb{R}^{n+1}$-valued even 1-form $E \in \Omega^1(X,\mathbb{R}^{n+1})$ -- the 
        \emph{vielbein}, identified as the propagating part of the \emph{graviton} field;
    \item
      an $\mathfrak{so}(n,1)$-valued even 1-form $\Omega \in \Omega^1(X, \mathfrak{so}(n,1))$ --
      the \emph{spin connection}, identified as the non-propagating auxiliary part of the graviton field;
    \item
      a spin-representaton -valued odd 1-form $\Psi \in \Omega^1(X,S)$ -- 
     identified as the \emph{gravitino field}.
  \end{enumerate}
\end{example}

\subparagraph{The 11d supergravity $C_3$-field}
\label{SupergravityCFieldInSupergravity}

\begin{example}
  \label{sugra3connectionForm}
  \index{supergravity!dual $C$-field}
  \index{supergravity!$C$-field}  
  For $\mathfrak{g} = \mathfrak{m}2\mathfrak{brane}_{\mathrm{gs}}$
  the Lie 3-algebra from 
  def. \ref{ElevenDSupergravityLie3Algebra},
  a $\mathfrak{g}$-valued form
  $$
    A : T X \to \mathfrak{sugra}_3(10,1)
  $$
  consists in addition to the field content of a $\mathfrak{siso}(10,1)$-connection
  from example \ref{sisoConnectionContent}
  of
  \begin{itemize}
    \item a 3-form $C_3 \in \Omega^3(X)$.
  \end{itemize}
\end{example}
This 3-form field is the local incarnation 
of what is called the \emph{supergravity $C_3$-field}.
The global nature of this field is discussed in 
\ref{supergravityCField}.

\subparagraph{The magnetic dual 11d supergravity $C_6$-field}
\label{DualSupergravityCFieldInSupergravity}

\begin{example}
  For  $\mathfrak{g} = \mathfrak{m}5\mathfrak{brane}_{\mathrm{gs}}$ the
  11d-supergravity Lie 6-algebra,
  def. \ref{ElevenDSupergravityLie6Algebra},
  a $\mathfrak{g}$-valued form
  $$
    A : T X \to \mathfrak{sugra}_6(10,1)
  $$
  consists in addition to the field content of a $\mathfrak{sugra}_3(10,1)$-connection
  given in remark \ref{sugra3connectionForm}
  of
  \begin{itemize}
    \item a 6-form $C_6 \in \Omega^3(X)$ -- the dual \emph{supergravity C-field}.
  \end{itemize}  
\end{example}
The identification of this field content is also due to 
the analysis of \cite{DF}.

\subsubsection{The supergravity $C$-field}
\label{supergravityCField}
\index{supergravity!$C$-field}

 We consider a slight variant of twisted differential 
 $\mathbf{c}$-structures, where instead of having the twist 
 directly in differential cohomology, it is instead first considered
 just in de Rham cohomology but then supplemented by a lift of the
 structure $\infty$-group.

 We observe that when such a twist is by the sum of the 
 first fractional Pontryagin class with the second Chern class, 
 and when the second of these two steps is considered over the boundary
 of the base manifold, then the differental structures obtained this
 way exhibit some properties that a differential cohomological
 description of the \emph{$C_3$-field} in 
 \emph{11-dimensional supergravity}, \ref{SupergravityCFieldInSupergravity}, 
 is expected to have.

  This section draws from \cite{FSScfield} and \cite{FiorenzaSatiSchreiberI}.
  
\medskip

 The supergravity C-field is subject to a certain $\mathbb{Z}_2$-twist
 \cite{Witten96} \cite{WittenFluxQuantization}, due to a quadratic 
 refinement of its action functional, which we review below in 
 \ref{CSWithBackgroundCharge}. A formalization of this twist in
 abelian differential cohomology for fixed background spin structure
 has been given in \cite{HopkinsSinger}, in terms of 
 \emph{differential integral Wu structures}. 
 These we review in \ref{DifferentialIntegralWuStructures}
 and refine them from $\mathbb{Z}_2$-coefficients to circle $n$-bundles.
 Then we present a natural moduli 3-stack of C-field configurations that 
 refines this model to nonabelian differential cohomology,
 generalizing it to dynamical gravitational background fields,
 in \ref{TheModuli3StackOfTheCField}.
 We discuss a natural boundary coupling of these fields to 
 $E_8$-gauge fields in \ref{CFieldRestrictionToTheBoundary}.

\paragraph{Higher abelian Chern-Simons theories with background charge} 
\label{CSWithBackgroundCharge}

The supergravity $C$-field is an example of a general phenomenon of 
higher abelian Chern-Simons QFTs in the presence of 
\emph{background charge}. This phenomenon was originally noticed
in \cite{Witten96} and then made precise in \cite{HopkinsSinger}.
The holographic dual of this phenomenon is that of
self-dual higher gauge theories, which for the supergravity $C$-field
is the nonabelian 2-form theory on the M5-brane \cite{FiorenzaSatiSchreiberI}.
We review the idea in a way that will smoothly lead over to our
refinements to nonabelian higher gauge theory in  
section \ref{supergravityCField}.

\medskip

Fix some natural number $k \in \mathbb{N}$ and an oriented manifold
(compact with boundary) $X$ of dimension $4 k + 3$. 
The gauge equivalence class of a $(2k+1)$-form gauge field $\hat G$ on $X$ is an
element in the differential cohomology group $\hat H^{2k+2}(X)$.
The cup product $\hat G \cup \hat G \in \hat H^{4k+4}(X)$ of this with itself 
has a natural higher holonomy over $X$, denoted
\bea
  \exp(i S (-)) : \hat H^{2k+2}(X) &\to & U(1)
\nonumber\\
  \hat G ~~ &\mapsto & \exp(i \int_X \hat G \cup \hat G)
  \nonumber
  \,.
\eea
This is the exponentiated action functional for bare $(4k+3)$-dimensional
abelian Chern-Simons theory. For $k = 0$ this reduces to ordinary 3-dimensional
abelian Chern-Simons theory. 
Notice that, even in this case, this is a bit more subtle
that Chern-Simons theory for a simply-connected gauge group $G$. In the latter
case all fields can be assumed to be globally defined forms. But in the non-simply-connected
case of $U(1)$, instead the fields are in general cocycles in differential
cohomology. If, however, we restrict attention to fields $C$ in the inclusion
$H^{2k+1}_{\mathrm{dR}}(X) \hookrightarrow \hat H^{2k+2}(X)$, then on these the above
action reduces to the familiar expression
$$
  \exp(i S(C)) = \exp(i \int_X C \wedge d_{\mathrm{dR} } C)
  \,.
$$
Observe now that
the above action functional may be regarded as a \emph{quadratic form}
on the group $\hat H^{2k+2}(X)$. The corresponding bilinear form is the 
(``secondary'', since $X$ is of dimension $4k+3$ instead of $4k+4$)
\emph{intersection pairing}
$$
  \langle -,-\rangle : \hat H^{2k+2}(X) \times \hat H^{2k+2}(X) \to U(1)
$$
$$
  (\hat a_1 , \hat a_2) \mapsto \exp(i \int_X \hat a_1 \cup \hat a_2 )
  \,.
$$
But note that from $\exp(i S(-))$ we do \emph{not} obtain a \emph{quadratic refinement} 
of the pairing. A quadratic refinement is, by definition, a function
$$
  q : \hat H^{2k+2}(X) \to U(1)
$$
(not necessarily homogenous of degree 2 as $\exp(i S(-))$ is), for which
the intersection pairing is obtained via the polarization formula
$$
  \langle \hat a_1, \hat a_2\rangle 
    = 
  q(\hat a_1 + \hat a_2)
  q(\hat a_1)^{-1}
  q(\hat a_2)^{-1}
  q(0)
  \,.
$$
If we took $q := \exp(i S(-))$, then the above formula would yield not 
$\langle -,-\rangle$, but the square $\langle -,-\rangle^2$, given by
the exponentiation of \emph{twice} the integral. 

\vspace{3mm}
The observation in \cite{Witten96} was 
that for the correct holographic physics, we need instead an action functional
which is indeed a genuine quadratic refinement of the intersection pairing.
But since the differential classes in $\hat H^{2k+2}(X)$ refine 
\emph{integral} cohomology, we cannot in general simply divide by 2 and
pass from $\exp( i \int_X \hat G \cup \hat G)$ to 
$\exp( i  \int_X \frac{1}{2} \hat G \cup \hat G)$. The integrand in the
latter expression does not make sense in general in differential cohomology.
If one tried to write it out in the ``obvious'' local formulas one would
find that it is a functional on fields which is not gauge invariant.
The analog of this fact is familiar from nonabelian $G$-Chern-Simons theory
with simply-connected $G$, where also the theory is consistent only at 
interger \emph{levels}. The ``level'' here is nothing but the 
underlying integral class $G \cup G$. 
Therefore the only way to obtain a square root of the quadratic form
$\exp(i S(-))$ is to \emph{shift it}. Here we think of 
the analogy with a
quadratic form 
$$
  q : x \mapsto x^2
$$ 
on the real numbers (a parabola in the plane).
Replacing this by 
$$
  q^{\lambda} : x \mapsto x^2 - \lambda x
$$ 
for some real number
$\lambda$ means keeping the shape of the form, but shifting its minimum from 0 to
$\frac{1}{2}\lambda$. If we think of this as the potential term for a scalar field
$x$ then its ground state is now at $x = \frac{1}{2}\lambda$. We may say that there is
a \emph{background field} or \emph{background charge} that pushes
the field out of its free equilibrium.

\vspace{3mm}
To lift this reasoning to our action quadratic form 
$\exp(i S(-))$ on differential cocycles, we need a 
differential class $\hat \lambda \in H^{2k+2}(X)$ such that 
for every $\hat a \in H^{2k+2}(X)$ the composite class
$$
  \hat a \cup \hat a - \hat a \cup \hat \lambda
  \in 
  H^{4k+4}(X)
$$
is even, hence is divisible by 2. Because then we could define a shifted
action functional
$$
  \exp(i S^\lambda(-)) : \hat a \mapsto \exp\left(
  i \int_X \frac{1}{2}(\hat a \cup \hat a - \hat a \cup \hat \lambda)\right)
  \,,
$$
where now the fraction $\frac{1}{2}$ in the integrand does make sense. 
One directly sees that if this exists, then this shifted action is indeed a
quadratic refinement of the intersection pairing:
$$
  \exp(i S^\lambda(\hat a + \hat b))
  \exp(i S^\lambda(\hat a))^{-1}
  \exp(i S^\lambda(\hat b))^{-1}
  \exp(i S^\lambda(0))
  = 
  \exp(i \int_X \hat a \cup \hat b)
  \,.
$$
The condition on the existence of $\hat \lambda$ here means, equivalently, that the
image of the underlying integral class vanishes under the map
$$
  (-)_{\mathbb{Z}_2} : H^{2k+2}(X, \mathbb{Z}) \to H^{2k+2}(X, \mathbb{Z}_2)
$$
to $\mathbb{Z}_2$-cohomology:
$$
   (a)_{\mathbb{Z}_2} \cup (a)_{\mathbb{Z}_2} 
     - 
   (a)_{\mathbb{Z}_2} \cup (\lambda)_{\mathbb{Z}_2}  
   = 
   0 \in H^{4k+4}(X, \mathbb{Z}_2)
  \,.
$$
Precisely such a class $(\lambda)_{\mathbb{Z}_2}$ does uniquely exist on 
every oriented manifold. It is called the \emph{Wu class} 
$\nu_{2k+2} \in H^{2k+2}(X,\mathbb{Z}_2)$, and may be \emph{defined} by this condition.
Moreover, if $X$ is a $\mathrm{Spin}$-manifold, then every second
Wu class, $\nu_{4k}$, has a pre-image in integral cohomology, 
hence $\lambda$ does exist as required above
$$
  (\lambda)_{\mathbb{Z}_2}
    = 
  \nu_{2k+2}
  \,.
$$
It is given by 
polynomials in the Pontrjagin classes of $X$
(discussed in section E.1 of \cite{HopkinsSinger}). For instance
the degree-4 Wu class (for $k = 1$) is refined by the first fractional Pontrjagin class
$\frac{1}{2}p_1$
$$
  (\tfrac{1}{2}p_1)_{\mathbb{Z}_2} = \nu_4
  \,.
$$
In the present context, this was observed in \cite{Witten96} (see around eq. (3.3) there).

Notice that the equations of motion of the shifted action $\exp(i S^\lambda(\hat a))$
are no longer $\mathrm{curv}(\hat a) = 0$, but are now 
$$
  \mathrm{curv}(\hat a) =  \tfrac{1}{2}\mathrm{curv}(\hat \lambda)
  \,.
$$
We therefore think of
$\exp(i S^\lambda(-))$ as the exponentiated action functional for
\emph{higher dimensional abelian Chern-Simons theory with background charge 
$\frac{1}{2}\lambda$}.

With respect to the shifted action functional it makes sense to introduce the
shifted field
$$
  \hat G  := \hat a  - \tfrac{1}{2}\hat \lambda
  \,.
$$
This is simply a re-parameterization such that the 
Chern-Simons equations of motion again look homogenous, namely $G = 0$.
In terms of this shifted field the action $\exp(i S^\lambda(\hat a))$
from above equivalently reads
$$
  \exp(i S^\lambda(\hat G)) = 
  \exp(
    i \int_X \tfrac{1}{2}(\hat G \cup \hat G - (\tfrac{1}{2}\hat \lambda)^2)
  )
  \,.
$$
For the case $k = 1$, 
this is the form of the action functional for the 7d Chern-Simons dual 
of the 2-form gauge field on the 5-brane
first given as (3.6) in \cite{Witten96}

\medskip

In the language of twisted cohomological structures,
def. \ref{TwistedCStructures}, we may summarize this situation
as follows: 
{\it In order for the action functional of higher abelian Chern-Simons theory 
to be correctly divisible, the images of the fields in $\mathbb{Z}_2$-cohomology
need to form a \emph{twisted Wu-structure}, \cite{II}.Therefore the fields themselves
need to constitute a \emph{twisted $\lambda$}-structure. For $k = 1$ this is
a \emph{twisted String-structure} \cite{SSSIII} and explains the quantization
condition on the $C$-field in 11-dimensional supergravity. }

In \cite{HopkinsSinger} a formalization of the above situation
has been given in terms of a notion there called 
\emph{differential integral Wu structures}. In the following
section we explain how this follows from the notion of 
twisted Wu structures with the twist taken in $\mathbb{Z}_2$-coefficients.
Then we refine this to a formalization to \emph{twisted differential Wu structures}
with the twist taken in smooth circle $n$-bundles.

\paragraph{Differential integral Wu structures}
\label{DifferentialIntegralWuStructures}
\index{twisted cohomology!differential integral Wu structures}

We discuss some general aspects of smooth and differential refinements
of $\mathbb{Z}_2$-valued universal characteristic classes. For the special 
case of \emph{Wu classes} we show how these notions reduce to the 
definition of \emph{differential integral Wu structures} given in 
\cite{HopkinsSinger}. We then construct a refinement of these structures
that lifts the twist from $\mathbb{Z}_2$-valued cocycles to smooth circle $n$-bundles.
This further refinement of integral Wu structures 
is what underlies the model for the supergravity
C-field in section \ref{supergravityCField}.

\medskip

Recall from prop. \ref{SpinCAsHomotopyFiberProductOfU1AndSO} the
characterization of $\mathrm{Spin}^c$ as the 
loop space object of the homotopy pullback
$$
  \raisebox{20pt}{
  \xymatrix{
    \mathbf{B}\mathrm{Spin}^c \ar[r]
    \ar[d]	
	& 
	  \mathbf{B} U(1) \ar[d]^{\mathbf{c}_1\, \mathrm{mod}\, 2}
	\\
	\mathbf{B} \mathrm{SO}
	\ar[r]^{\mathbf{w}_2}
	&
	\mathbf{B}^2 \mathbb{Z}_2
  }}
  \,.
$$
For general $n \in \mathbb{N}$ the analog of the first Chern
class mod 2 appearing here is the higher Dixmier-Douady class
mod 2
$$
  \mathbf{DD}_{\mathrm{mod}\, 2}
  : 
  \xymatrix{
  \mathbf{B}^n U(1)
  \ar[r]^{\mathrm{DD}}
  &
  \mathbf{B}^{n+1} \mathbb{Z}
  \ar[r]^{\mathrm{mod}\, 2}
  &
  \mathbf{B}^{n+1} \mathbb{Z}_2
  }
  \,.
$$
Let now 
$$
  \nu_{n+1} : \mathbf{B} \mathrm{SO}  \to \mathbf{B}^{n+1} \mathbb{Z}_2
$$
be a representative of the universal smooth \emph{Wu class} in degree $n+1$, 
the $(\Pi \dashv \mathrm{Disc})$-adjunct of the topological universal Wu class
using that $\mathbf{B}^{n+1}\mathbb{Z}$ is discrete as a smooth $\infty$-groupoid,
and using that $\Pi(\mathbf{B}\mathrm{SO}) \simeq B \mathrm{SO}$ is the ordinary
classifying space, by prop. \ref{FundGroupoidOfSimplicialParacompact}. 
\begin{definition}
  \label{IntegralUniversalWuClasses}
Let $\mathrm{Spin}^{\nu_{n+1}}$ be the loop space object of the 
homotopy pullback
$$
  \raisebox{20pt}{
  \xymatrix{
    \mathbf{B} \mathrm{Spin}^{\nu_{n+1}}
	\ar[r]
    \ar[d]^{\nnu_{n+1}^{\mathrm{int}}}
	& \mathbf{B}\mathrm{SO}
	\ar[d]^{\nu_{n+1}}
	\\
	\mathbf{B}^n U(1)
	 \ar[r]^{\mathrm{mod}\,2}
	 &
	\mathbf{B}^{n+1} \mathbb{Z}_2
  }
  }
  \,.
$$
We call the left vertical morphism $\nnu_{n+1}$ appearing here
the \emph{universal smooth integral Wu structure} in degree $n+1$.
\end{definition}
A morphism of stacks
$$
  \nnu_{n+1} : X \to \mathbf{B} \mathrm{Spin}^{\nu_{n+1}}
$$
is a choice of orientation structure on $X$ together with a choice of
smooth integral Wu structure lifting the corresponding Wu class
$\nu_{n+1}$.
\begin{example}
  \label{FirstPontrjaginAsIntegralWu}
  The smooth first fractional Pontrjagin class $\tfrac{1}{2}\mathbf{p}_2$, 
  prop. \ref{FirstFracPontryagin}, 
  fits into a diagram
  $$
  \raisebox{20pt}{
  \xymatrix{
    \mathbf{B} \mathrm{Spin}
	\ar@/^1pc/[rrd]
	\ar@/_1pc/[ddr]_{\tfrac{1}{2}\mathbf{p}_1}
	\ar@{-->}[dr]^{u}
    \\
    &\mathbf{B} \mathrm{Spin}^{\nu_{4}}
	\ar[r]
    \ar[d]^{\nnu_{4}^{\mathrm{int}}}
	& \mathbf{B}\mathrm{SO}
	\ar[d]^{\nu_{4}}
	\\
	& \mathbf{B}^3 U(1)
	 \ar[r]^{\mathrm{mod}\,2}
	 &
	\mathbf{B}^{4} \mathbb{Z}_2
  }
  }
  \,.
  $$
  In this sense we may think of $\tfrac{1}{2}\mathbf{p}_1$ as being the 
  integral and, moreover, smooth refinement of the universal degree-4 Wu class on 
  $\mathbf{B}\mathrm{Spin}$.
\end{example}
\proof
  Using the defining property of $\tfrac{1}{2}\mathbf{p}_1$, this 
  follows with the results discussed in appendix E.1 of \cite{HopkinsSinger}.
\endofproof
\begin{proposition}
  \label{RelationToHopkinsSingerDifferentialWu}
  Let $X$ be a smooth manifold equipped with orientation
  $$
    o_X : X \to \mathbf{B} \mathrm{SO}
  $$
  and consider its Wu-class $[\nu_{n+1}(o_X)] \in H^{n+1}(X, \mathbb{Z}_2)$
  $$
    \nu_{n+1}(o_X) : 
    \xymatrix{
	  X \ar[r]^{o_X} 
	  & 
	  \mathbf{B}\mathrm{SO}
	   \ar[r]^{\nu_{n+1}}
	  &
	  \mathbf{B}^{n+1}\mathbb{Z}_2
	}	
	\,.
  $$  
  The $n$-groupoid 
  $\hat {\mathbf{DD}}_{\mathrm{mod}2 }\mathrm{Struc}_{[\nu_{2k}]}(X)$
  of $[\nu_{n+1}]$-twisted differential 
  $\mathbf{DD}_{\mathrm{mod} 2}$-structures, according to def. \ref{TwistedCStructures},
  hence the homotopy pullback
  $$
    \raisebox{20pt}{
    \xymatrix{
	  \hat {\mathbf{DD}}_{\mathrm{mod}2 }\mathrm{Struc}_{[\nu_{n+1}]}(X)
	  \ar[rr]
	  \ar[d]
	  &&
	  {*}
	  \ar[d]^{\nu_{n+1}(o_X)}
	  \\
	  \mathbf{H}(X, \mathbf{B}^3 U(1)_{\mathrm{conn}})
	  \ar[rr]^{\hat {\mathbf{DD}}_{\mathrm{mod}\, 2}}
	  &&
	  \mathbf{H}(X, \mathbf{B}^{n+1} \mathbb{Z}_2)
	}
	}
	\,,
  $$
  categorifies the groupoid $\hat {\mathcal{H}}^{n+1}_{\nu_{n+1}}(X)$
  of \emph{differential integral Wu structures} as in def. 2.12 of \cite{HopkinsSinger}:
  its 1-truncation is equivalent to the groupoid defined there
  $$
    \tau_1 \hat {\mathbf{DD}}_{\mathrm{mod}2 }\mathrm{Struct}_{[\nu_{n+1}]}(X)
	\simeq
	\hat  {\mathcal{H}}^{n+1}_{\nu_{n+1}}(X)
	\,.
  $$
\end{proposition}
\proof
  By prop. \ref{BnU1conn}, the canonical presentation  
  of $\mathbf{DD}_{\mathrm{mod} 2}$
  via the Dold-Kan correspondence is given by an epimorphism of chain complexes of sheaves,
  hence by a fibration in $[\mathrm{CartSp}^{\mathrm{op}}, \mathrm{sSet}]_{\mathrm{proj}}$.
  Precisely,  the composite
  $$
    \hat {\mathbf{DD}}_{\mathrm{mod}\, 2}
	:
	\xymatrix{
	  \mathbf{B}^n U(1)_{\mathrm{conn}}
	  \ar[r]
	  &
	  \mathbf{B}^n U(1)
	  \ar[r]^{\mathrm{DD}}
	  &
	  \mathbf{B}^{n+1} \mathbb{Z}
	  \ar[r]^{\mathrm{mod}\,2}
	  &
	  \mathbf{B}^{n+1} \mathbb{Z}_2
	}
  $$
  is presented by the vertical sequence of morphisms of chain complexes
  $$
    \raisebox{30pt}{
    \xymatrix{
	  \mathbb{Z} \ar@{^{(}->}[r]
	  \ar[d]
	  &
	  C^\infty(-, \mathbb{R})
	  \ar[r]^{d_{\mathrm{dR}} \mathrm{log}}
	  \ar[d]
	  &
	  \Omega^1(-)
	  \ar[r]^{d_{\mathrm{dR}}}
	  \ar[d]
	  &
	  \cdots
	  \ar[r]^{d_{\mathrm{dR}}}
	  &
	  \Omega^n(-)
	  \ar[d]
	  \\
	  \mathbb{Z} \ar@{^{(}->}[r]
	  \ar[d]
	  &
	  C^\infty(-, \mathbb{R})
	  \ar[r]
	  \ar[d]
	  &
	  0
	  \ar[r]
	  \ar[d]
	  &
	  \cdots
	  \ar[r]
	  &
	  0
	  \ar[d]	  
	  \\
	  \mathbb{Z} \ar[r]
	  \ar[d]
	  &
	  0
	  \ar[r]
	  \ar[d]
	  &
	  0
	  \ar[r]
	  \ar[d]
	  &
	  \cdots
	  \ar[r]
	  &
	  0
	  \ar[d]	  
	  \\
	  \mathbb{Z}_2 \ar[r]
	  &
	  0
	  \ar[r]
	  &
	  0
	  \ar[r]
	  &
	  \cdots
	  \ar[r]
	  &
	  0
	}
	}
	\,.
  $$  
  By remark \ref{ComputingHomotopyPullbacks} we may therefore compute the defining 
  homotopy pullback for $\hat {\mathbf{DD}}_{\mathrm{mod}2 }\mathrm{Struct}_{[\nu_{n+1}]}(X)$
  as an ordinary fiber product of the corresponding simplicial sets of 
  cocycles. The claim then follows by inspection.
\endofproof
\begin{remark}
  Explicitly, a cocycle in 
  $
    \tau_1 \hat {\mathbf{DD}}_{\mathrm{mod}2 }\mathrm{Struct}_{[\nu_{n+1}]}(X)$
 is identified with a {\v C}ech cocycle with coefficients in
 the Deligne complex
 $$
  (
   \xymatrix{
	  \mathbb{Z} \ar@{^{(}->}[r]
	  &
	  C^\infty(-, \mathbb{R})
	  \ar[r]^{d_{\mathrm{dR}} \mathrm{log}}
	  &
	  \Omega^1(-)
	  \ar[r]^{d_{\mathrm{dR}}}
	  &
	  \cdots
	  \ar[r]^{d_{\mathrm{dR}}}
	  &
	  \Omega^n(-)
   }
   )
  $$
  such that the underlying $\mathbb{Z}[n+1]$-valued cocycle modulo 2
  equals the given cocycle for $\nu_{n+1}$. A coboundary between two such
  cocycles is a gauge equivalence class of ordinary {\v C}ech-Deligne cocycles
  such that their underlying $\mathbb{Z}$-cocycle vanishes modulo 2. 
  Cocycles of this form are precisely those that arise by multiplication with 
  2 or arbitrary {\v C}ech-Deligne cocycles. 
  
  This is the groupoid structure discussed on p. 14 of \cite{HopkinsSinger},
  there in terms of singular instead of {\v C}ech cohomology.
  \label{UnwindingDifferentialWuStructureClasses}
\end{remark}
We now consider another twisted differential structure, which 
refines these twisting integral Wu structures 
to \emph{smooth} integral Wu structures, def. \ref{IntegralUniversalWuClasses}.
\begin{definition}
  For $n \in \mathbb{N}$, write $\mathbf{B}^n U(1)_{\mathrm{conn}}^{\nu_{n+1}}$
  for the homotopy pullback of smooth moduli $n$-stacks
  $$
    \raisebox{20pt}{
    \xymatrix{
	  \mathbf{B}^n U(1)_{\mathrm{conn}}^{\nu_{n+1}}
	  \ar[rrr]
	  \ar[d]
	  &&&
	  \mathbf{B}^n U(1)_{\mathrm{conn}}
	  \ar[d]
	  \\
	  \mathbf{B} \mathrm{Spin}^{\nu_{n+1}}
		 \times
	   \mathbf{B}^n U(1)
	  \ar[rrr]^{\nnu_{n+1}^{\mathrm{int}} -  2 \mathbf{DD}}
	  &&&
	  \mathbf{B}^n U(1)
	}
	}
	\,,
  $$
  where $\nnu_{n+1}^{\mathrm{int}}$ is the universal smooth integral Wu class from 
  def. \ref{IntegralUniversalWuClasses}, and where
  $2 \mathbf{DD} : \mathbf{B}^n U(1) \to \mathbf{B}^n U(1)$ is the canonical 
  smooth refinement of the operation of multiplication by 2 on integral cohomology.
  
  We call this the moduli $n$-stack of \emph{smooth differential Wu-structures}.
  \label{SmoothDifferentialWuStructures}
\end{definition}
By construction, a morphism $X \to \mathbf{B}^n U(1)_{\mathrm{conn}}^{\nu_{n+1}}$
classifies also all possible orientation structures and 
smooth integral lifts of their Wu structures. In applications one typically
wants to fix an integral Wu structure lifting a given Wu class. 
This is naturally formalized by
the following construction.
\begin{definition}
  For $X$ an oriented manifold, and 
  $$
    \nnu_{n+1} : X \to \mathbf{B}\mathrm{Spin}^{\nu_{n+1}}
  $$
  a given smooth integral Wu structure, def. \ref{IntegralUniversalWuClasses},
  write
  $\mathbf{H}_{\nnu_{n+1}}(X, \mathbf{B}^n U(1)_{\mathrm{conn}}^{\nu_{n+1}})$
  for the $n$-groupoid of cocycles whose underlying 
  smooth integral Wu structure is $\nnu_{n+1}$, hence for the homotopy pullback
  $$
    \xymatrix{
	  \mathbf{H}_{\nnu_{n+1}}(X, \mathbf{B}^n U(1)_{\mathrm{conn}}^{\nu_{n+1}})
	  \ar[r]
	  \ar[d]
	  &
	  \mathbf{H}(X, \mathbf{B}^n U(1)_{\mathrm{conn}}^{\nu_{n+1}})
	  \ar[d]
	  \\
	  \mathbf{H}(X, \mathbf{B}^n U(1))
	  \ar[r]^<<<<<<<{(\nnu_{n+1}, \mathrm{id})}
	  \ar[d]
	  & 
	  \mathbf{H}(X, \mathbf{B}\mathrm{Spin}^{\nu_{n+1}} \times \mathbf{B}^n U(1) )
	  \ar[d]
	  \\
	  {*} \ar[r]^{\nnu_{n+1}}& \mathbf{H}(X, \mathbf{B}\mathrm{Spin}^{\nu_{n+1}} )
	}
	\,.
  $$
  \label{RestrictionOfDifferentialWuStructuresToFixedSmoothWuStructure}
\end{definition}
\begin{proposition}
 \label{RefinementOfDifferentialWuClassesTonBundles}
  Cohomology with coefficients in $\mathbf{B}^n U(1)_{\mathrm{conn}}^{\nu_{n+1}}$
  over a given smooth integral Wu structure coincides with the corresponding
  differential integral Wu structures:
  $$
    \hat H_{\nu_{n+1}}^{n+1}(X) \simeq H_{\nnu_{n+1}}(X, \mathbf{B}^n U(1)_{\mathrm{conn}}^{\nu_{n+1}})
	\,.
  $$
\end{proposition}
\proof
  Let $C(\{U_i\})$ be the {\v C}ech-nerve of a good open cover of $X$.
  By prop. \ref{BnU1conn} the canonical 
  presentation of $\mathbf{B}^n U(1)_{\mathrm{conn}} \to \mathbf{B}^n U(1)$ 
  is a projective fibration. Since $C(\{U_i\})$ 
  is projectively cofibrant and $[\mathrm{CartSp}^{\mathrm{op}}, \mathrm{sSet}]_{\mathrm{proj}}$ 
  is a simplicial model category, the morphism of {\v C}ech cocycle simplicial sets
  $$
    [\mathrm{CartSp}^{\mathrm{op}}, \mathrm{sSet}](C(\{U_i\}), \mathbf{B}^n U(1)_{\mathrm{conn}})
	\to
    [\mathrm{CartSp}^{\mathrm{op}}, \mathrm{sSet}](C(\{U_i\}), \mathbf{B}^n U(1))
  $$
  is a Kan fibration.
  Hence, by remark \ref{ComputingHomotopyPullbacks}, its
  homotopy pullback 
  may be computed as the ordinary pullback of simplicial sets of this map. 
  The claim then follows by inspection.
  
  Explicitly, in this presentation a cocycle in the pullback is
  a pair $(a, \hat G)$ of a cocycle $a$ for a circle $n$-bundle and a 
  Deligne cocycle $\hat G$ with underlying bare cocycle $G$, 
  such that there is an equality of degree-n
  {\v C}ech $U(1)$-cocycles
  $$
    G = \nnu_{n+1} - 2 a
	\,.
  $$
  A gauge transformation between two such cocycles is a pair of {\v C}ech cochains
  $\hat \gamma$, $\alpha$ such that $\gamma = 2 \alpha$ (the cocycle 
  $\nnu_{n+1}$ being held fixed). 
  This means that the gauge transformations acting on 
  a given $\hat G$ solving the above constraint
  are precisely the all Deligne cocychains, but multiplied by 2.
  This is again the explicit description of $\hat H_{\nu_{n+1}}(X)$
  from remark \ref{UnwindingDifferentialWuStructureClasses}.
  \endofproof

\paragraph{Twisted differential $\mathrm{String}(E_8)$-structures}
\label{StringE8Structure}

We discuss smooth and differential refinements of the 
canonical degree-4 universal characteristic class 
$$
  a : B E_8 \to K(\mathbb{Z},4)
$$ 
for $E_8$ the largest of the exceptional semimple Lie algebras. 

\medskip

\begin{proposition}
  There exists a differential refinement of the canonical integral 4-class
  on $B E_8$
  to the smooth moduli stack of $E_8$-connections
  with values in the smooth moduli 3-stack of circle 3-bundles with 
  3-connection
  $$
    \hat {\mathbf{a}}
	  :
    \xymatrix{	  
  	  (\mathbf{B}E_8)_{\mathrm{conn}}
	  \ar[r]
	  &
	  \mathbf{B}^3 U(1)_{\mathrm{conn}}
	}
	\,.
  $$
  \label{aDifferentially}
\end{proposition}
Using the $L_\infty$-algebraic data provided in 
\cite{SSSI}, this was constructed in \cite{FSS}.
\begin{proposition}
  Under geometric realization, prop. \ref{GeometricRealization},
  the smooth class $\mathbf{a}$ becomes an equivalence
  $$
    \vert \mathbf{a}\vert
	: 
	B E_8
	\simeq_{16}
	B^3 U(1)
	\simeq 
	K(\mathbb{Z},4)
  $$
  on 16-coskeleta.
  \label{aIsMaxCompactSubgroup}
\end{proposition}
\proof
  By \cite{BottSamelson} the 15-coskeleton of the topological space
  $E_8$ is a $K(\mathbb{Z}, 4)$.
  By \cite{FSS}, $\mathbf{a}$ is a smooth refinement of
  the generator $[a] \in H^4(B E_8, \mathbb{Z})$. 
  By the Hurewicz theorem this is identified with 
  $\pi_4 (B E_8) \simeq \mathbb{Z}$. Hence in cohomology
  $\mathbf{a}$ induces an isomorphism
  $$
    \xymatrix@C=5pt{
	  \pi_4(B E_8)
      \simeq     
	  [S^4, B E_8]  
	  \simeq
	  H^1(S^4, E_8) 
      \ar[rrr]^{\vert \mathbf{a} \vert}|\simeq	  
	  &&&
	  H^4(S^4, \mathbb{Z}) 
	  \simeq
	  [S^4, K(\mathbb{Z},4)]
	  \simeq
	  \pi_4(S^4)
	  }
	\,.
  $$  
  Therefore $\vert \mathbf{a} \vert$ is a weak homotopy equivalence on 16 coskeleta.
\endofproof

\paragraph{The moduli 3-stack of the $C$-field}
\label{TheModuli3StackOfTheCField}

As we have reviewed above in 
section \ref{CSWithBackgroundCharge}, 
the flux quantization condition for the $C$-field
derived in \cite{WittenFluxQuantization} is the equation
\(
\label{eq.congruent.mod.2}
  [G_4] = \tfrac{1}{2}p_1 \mod 2
  \;\;\;
 \text{in}\quad  H^4(X, \mathbb{Z})
\)
in integral cohomology,  
where $[G_4]$ is the cohomology class 
of the $C$-field itself, and $\tfrac{1}{2}p_1$ is the first fractional Pontrjagin 
class of the Spin manifold $X$. One can equivalently rewrite \eqref{eq.congruent.mod.2} as
\(
\label{eq.integral}
  [G_4] = \tfrac{1}{2}p_1+2a
  \;\;\;
 \text{in}\quad  H^4(X, \mathbb{Z}),
\)
where $a$ is some degree 4 integral cohomology class on $X$. 
By the discussion in section \ref{DifferentialIntegralWuStructures}, 
the correct formalization of this for \emph{fixed} spin structure
is to regard the gauge equivalence class of the 
C-field as a differential integral Wu class 
relative to the integral Wu class $\nu_4^{\mathrm{int}} = \tfrac{1}{2}p_1$,
example \ref{FirstPontrjaginAsIntegralWu}, of that spin structure. 
By prop. \ref{RefinementOfDifferentialWuClassesTonBundles} 
and prop. \ref{FirstFractionalDifferentialPontrjagin}, the natural
refinement of this to a smooth moduli 3-stack of C-field configurations
and arbitrary spin connections is the homotopy pullback of  smooth 3-stacks
$$
  \raisebox{20pt}{
  \xymatrix{
    \mathbf{B}^n U(1)_{\mathrm{conn}}^{\nu_{n+1}}
	\ar[rr]
	\ar[d]
	&& \mathbf{B}^3 U(1)_{\mathrm{conn}}
	\ar[d]
    \\
    \mathbf{B} \mathrm{Spin}_{\mathrm{conn}}
	\times 
	\mathbf{B}^3 U(1)
	\ar[rr]^<<<<<<<<<{\tfrac{1}{2}\hat {\mathbf{p}}_1 + 2 \mathbf{DD}}
	&&
	\mathbf{B}^3 U(1)
  }
  }
  \,.
$$
Here the moduli stack in the bottom left is that of the field of gravity
(spin connections) together with an auxiliary circle 3-bundle / 2-gerbe.
Following the arguments in \cite{FiorenzaSatiSchreiberI} 
(the traditional ones as well as the new ones presented there),
we take this auxiliary circle 3-bundle to be 
the Chern-Simons circle 3-bundle of an $E_8$-principal bundle.
According to prop. \ref{aDifferentially} this is formalized 
on smooth higher moduli stacks by further pulling back along the smooth refinement
$$
  \mathbf{a} : \mathbf{B}E_8 \to \mathbf{B}^3 U(1)
$$
of the canonical universal 4-class $[a] \in H^4(B E_8, \mathbb{Z})$.
Therefore we are led to formalize the
\emph{$E_8$-model for the $C$-field} as follows. 
\begin{definition}
The \emph{smooth moduli 3-stack of spin connections and $C$-field configurations} 
in the $E_8$-model is the homotopy pullback $\mathbf{CField}$ of the moduli
$n$-stack of smooth differential Wu structures $\mathbf{B}^n U(1)_{\mathrm{conn}}^{\nu_4}$,
def. \ref{SmoothDifferentialWuStructures},
to spin connections and $E_8$-instanton configurations, hence the homotopy pullback
\(
  \raisebox{20pt}{
  \xymatrix{
    \mathbf{CField} \ar[rr]^{} \ar[d] 
	 &&  \mathbf{B}^3 U(1)_{\mathrm{conn}}^{\nu_4}
	 \ar[d]
    \\
   \mathbf{B}\mathrm{Spin}_{\mathrm{conn}}
	  \times
	  \mathbf{B}{E_8}
    \ar[rr]^{
	  (u, \mathbf{a})
	}
	&&
   \mathbf{B}\mathrm{Spin}^{\nu_4} \times \mathbf{B}^3 U(1)
	}
	}
  \,,
\)
where $u$ is the canonical morphism from example \ref{FirstPontrjaginAsIntegralWu}.
\label{CField}
\end{definition}
\begin{remark}
By the pasting law, prop. \ref{PastingLawForPullbacks},
$\mathbf{CField}$ is equivalently given as the homotopy pullback
\(
  \raisebox{20pt}{
  \xymatrix{
    \mathbf{CField} \ar[rrr]^{\hat{\mathbf{G}}_4} \ar[d] 
	 &&&  \mathbf{B}^3 U(1)_{\mathrm{conn}}\ar[d]
    \\
   \mathbf{B}\mathrm{Spin}_{\mathrm{conn}}
	  \times
	  \mathbf{B}{E_8}
    \ar[rrr]^{
	  \frac{1}{2}\mathbf{p}_1 + 2 \mathbf{a}
	}
	&&&
   \mathbf{B}^3 U(1)
	}
	}
  \,.
\)
Spelling out this definition, a $C$-field configuration 
$$
  (\nabla_{\mathfrak{so}}, \nabla_{b^2 \mathbb{R}}, P_{E_8})
  :
  X \to \mathbf{CField}
$$
on a smooth manifold $X$  is the datum of 
{\it 
\begin{enumerate}
 \item a principal $\mathrm{Spin}$-bundle with $\mathfrak{so}$-connection $(P_{\mathrm{Spin}},\nabla_{\mathfrak{so}})$ on $X$;
\item a principal $E_8$-bundle $P_{E_8}$  on $X$;
\item a $U(1)$-2-gerbe with connection $(P_{\mathbf{B}^2U(1)},\nabla_{\mathbf{B}^2U(1)})$ on $X$;
\item a choice of equivalence of $U(1)$-2-gerbes between  
  between $P_{\mathbf{B}^2U(1)}$ and the image of 
  $P_{\mathrm{Spin}}\times_XP_{E_8}$ via $\frac{1}{2}\mathbf{p}_1 + 2 \mathbf{a}$.
\end{enumerate}
}
\label{CFieldAs3Bundle}
\end{remark}
It is useful to observe that there is the following further equivalent reformulation of this 
definition.
\begin{proposition}
  The moduli 3-stack $\mathbf{CField}$ from def. \ref{CField} is equivalently
  the homotopy pullback
\(
  \raisebox{20pt}{
  \xymatrix{
    \mathbf{CField} \ar[rrr] \ar[d] 
	 &&& \Omega^4_{\mathrm{cl}}\ar[d]
    \\
   \mathbf{B}\mathrm{Spin}_{\mathrm{conn}}
	  \times
	  \mathbf{B}{E_8}
    \ar[rrr]^{
	  (\frac{1}{2}\mathbf{p}_1 + 2 \mathbf{a})_{\mathrm{dR}}
	}
	&&&
  \mathbf{\flat}_{\mathrm{dR}} \mathbf{B}^4 \mathbb{R}
	  }
	  }
  \,,
\)
where 
the bottom morphism of higher stacks is presented by the
correspondence of simplicial presheaves
\(
  \raisebox{20pt}{
  \xymatrix{
    \mathbf{B}\mathrm{Spin}_{\mathrm{conn}}
	\times
	(\mathbf{B}E_{8})_{\mathrm{diff}}
	\ar[r]
    \ar@{->>}[d]^\wr
      &
    \mathbf{B}\mathrm{Spin}_{\mathrm{diff}}
	\times
	(\mathbf{B}E_{8})_{\mathrm{diff}}
	  \ar[rr]^<<<<<{(\tfrac{1}{2}\mathbf{p}_1 + 2 \mathbf{a})_{\mathrm{diff}}} 
    \ar@{->>}[d]^\wr	
	  && 
	  \mathbf{B}^3 U(1)_{\mathrm{diff}}
    \ar[r]^{\mathrm{curv}}
    \ar@{->>}[d]^\wr
    &
    \mathbf{\flat}_{\mathrm{dR}} \mathbf{B}^{4}\mathbb{R} 
    \\
   \mathbf{B}\mathrm{Spin}_{\mathrm{conn}}\times
	  \mathbf{B}{E_8} 
	  \ar[r] 
	  &
	  \mathbf{B}\mathrm{Spin}\times \mathbf{B}{E_8}
	  \ar[rr]^{\tfrac{1}{2}\mathbf{p}_1 + 2 \mathbf{a}} 
	  && 
	  \mathbf{B}^3 U(1)
  }
  }
  \;.
\)
Moreover, it is equivalently the homtopy pullback
\(
  \raisebox{20pt}{
  \xymatrix{
    \mathbf{CField} \ar[rrr] \ar[d] 
	 &&& \Omega^4_{\mathrm{cl}}\ar[d]
    \\
   \mathbf{B}\mathrm{Spin}_{\mathrm{conn}}
	  \times
	  \mathbf{B}{E_8}
    \ar[rrr]^{
	  (\frac{1}{4}\mathbf{p}_1 + \mathbf{a})_{\mathrm{dR}}
	}
	&&&
  \mathbf{\flat}_{\mathrm{dR}} \mathbf{B}^4 \mathbb{R}
	  }
	  }
  \,,
\)
where now the bottom morphism is the composite of the bottom morphism before,
postcomposed with the morphism
$$
  \tfrac{1}{2} 
    : 
   \mathbf{\flat}_{\mathrm{dR}} \mathbf{B}^4 \mathbb{R}
     \to 
   \mathbf{\flat}_{\mathrm{dR}} \mathbf{B}^4 \mathbb{R}
$$
that is given, via Dold-Kan, by division of differential forms by 2.
\end{proposition}
\proof
 By the pasting law 
 for homotopy pullbacks, prop. \ref{PastingLawForPullbacks}, 
 the first homotopy pullback above may be computed as
 two consecutive homotopy pullbacks
 $$
  \raisebox{20pt}{
  \xymatrix{
    \mathbf{CField} \ar[rr] \ar[d] 
	&&
	\mathbf{B}^n U(1)_{\mathrm{conn}}
	\ar[r]
	\ar[d]
	 & 
	 \Omega^4_{\mathrm{cl}}\ar[d]
    \\
   \mathbf{B}\mathrm{Spin}_{\mathrm{conn}}
	  \times
	  \mathbf{B}{E_8}
    \ar[rr]^{
	  \tfrac{1}{2} {\mathbf{p}}_1 + 2  {\mathbf{a}}
	}
	&&
	\mathbf{B}^3 U(1)
	\ar[r]^{\mathrm{curv}}
	&
    \mathbf{\flat}_{\mathrm{dR}} \mathbf{B}^4 \mathbb{R}
	  }
	  }
  \,,
 $$
 which exhibits on the right the defining pullback of def. \ref{BnU1conn},
 and thus on the left the one from def. \ref{CField}.
 The statement about the second homotopy pullback above 
 follows analogously after noticing that
 \(
  \xymatrix{
     \Omega^4_{\mathrm{cl}}\ar[rrr]^{1/2} \ar[d] 
	 &&& \Omega^4_{\mathrm{cl}}\ar[d]
    \\
     \mathbf{\flat}_{\mathrm{dR}} \mathbf{B}^4 \mathbb{R}    \ar[rrr]^{
	  1/2}
	&&&
  \mathbf{\flat}_{\mathrm{dR}} \mathbf{B}^4 \mathbb{R}
  }
  \,.
\)
is a homotopy pullback.
\endofproof
It is therefore useful to introduce labels as follows.
\begin{definition}
We label the structure morphism of the above composite homotopy pullback as
$$
  \raisebox{20pt}{
  \xymatrix{
    \mathbf{CField} 
	 \ar[rr]^{\hat G_4}_>{\ }="s1" 
	 \ar[d]
	 && 
    \mathbf{B}^3 U(1)_{\mathrm{conn}}
     \ar[r]^{\mathcal{G}_4}_>{\ }="s2" 
	  \ar[d]^{G_4}
	  &
     \Omega^4_{\mathrm{cl}} 
	 \ar[d]
   \\
	 \mathbf{B}\mathrm{Spin}_{\mathrm{conn}}
	 \times
	 \mathbf{B}{E_8}
   \ar[rr]_<<<<<<<<<{\frac{1}{2}\mathbf{p}_2 + 2 \mathbf{a}}^<{\ }="t1"
    &&
   \mathbf{B}^3 U(1)
    \ar[r]_<<<<<<{\mathrm{curv}}^<{\ }="t2"
	&
   \mathbf{\flat}_{\mathrm{dR}} \mathbf{B}^4 U(1)
   \ar@{=>}_{H_3}^\simeq "s1"; "t1"
   \ar@{=>}^\simeq "s2"; "t2"
  }
  }
  \,.
$$
Here $\hat G_4$ sends a C-field configuration to an underlying circle 3-bundle
with connection, whose curvature 4-form is $\mathcal{G}_4$.
\end{definition}
\begin{remark}
 \label{CommentOnTheTwoIncarnationsOfDefOfCField}
  These equivalent reformulations show two things.
  \begin{enumerate}
    \item 
	  The C-field model may be thought of as containing
	  $E_8$-\emph{pseudo-connections}. That is, there is a higher gauge in which
	  a field configuration consists of an $E_8$-connection on an $E_8$-bundle
	   -- even though there is no dynamical $E_8$-gauge field in 11d supergravity --
	  but where gauge transformations are allowed to freely shift these connections.
    \item
	  There is a precise sense in which imposing 
	  the quantization condition (\ref{eq.integral})
	  on integral cohomology is equivalent to imposing the condition
	  $[G_4]/2 = \tfrac{1}{4}p_1 + a$ in de Rham cohomology / real singular cohomology.
  \end{enumerate}
\end{remark}
\begin{observation}
  When restricted to a fixed Spin-connection, gauge equivalence classes 
  of configurations classified by $\mathbf{CField}$ naturally form a torsor
  over the ordinary degree-4 differential cohomology $H^4_{\mathrm{diff}}(X)$.
  \label{RestrictionToFixedSpinConnection}
\end{observation}
\proof
  By the general discussion of differential integral Wu-structures
  in section \ref{DifferentialIntegralWuStructures}.
\endofproof

\paragraph{The homotopy type of the moduli stack}
\label{TheHomotopyTypeOfTheCFieldModuliStack}

We discuss now the homotopy type of the 3-groupoid 
$$
  \mathbf{CField}(X)
  :=
  \mathbf{H}(X, \mathbf{CField})
$$
of C-field configurations over a given spacetime manifold $X$. 
In terms of gauge theory, 
its 0-th homotopy group is the set of \emph{gauge equivalence classes}
of field configurations, its first homotopy group is the 
set of \emph{gauge-of-gauge equivalence classes} of 
auto-gauge transformations of a given configuration, and so on.

\medskip

\begin{definition}
For $X$ a smooth manifold, 
let 
$$
  \raisebox{20pt}{
  \xymatrix{
 	& \mathbf{B}\mathrm{Spin}_{\mathrm{conn}}
    \ar[d]
	\\
    X 
	 \ar[r]^{P_{\mathrm{Spin}}}
	 \ar[ur]^{\nabla_{\mathfrak{so}}}
	 &
	\mathbf{B}\mathrm{Spin}
  }
  }
$$
be a fixed spin structure with fixed spin connection. 
The restriction of $\mathbf{CField}(X)$ to this fixed spin connection is the
homotopy pullback
$$
  \raisebox{20pt}{
  \xymatrix{
    \mathbf{CField}(X)_{P_{\mathrm{Spin}}}
	\ar[rrr]
	\ar[d]
	&&&
	\mathbf{CField}(X)
	\ar[d]
	\\
	\mathbf{H}(X, \mathbf{B}E_8)
	\ar[rrr]^<<<<<<<<<<<<<<<{((P_{\mathrm{Spin}},\nabla_{\mathfrak{so}}), \mathrm{id})}
	&&&
	\mathbf{H}(X, \mathbf{B}\mathrm{Spin}_{\mathrm{conn}} \times\mathbf{B}E_8)
  }
  }
  \,.
$$
\end{definition}
\begin{proposition}
  The gauge equivalence classes of $\mathbf{CField}(X)_{P_{\mathrm{Spin}}}$
  naturally surjects onto the differential integral Wu structures on $X$, relative to 
  $\tfrac{1}{2}p_1(P_{\mathrm{Spin}})\,\mathrm{mod}\,2$,
  (example \ref{FirstPontrjaginAsIntegralWu}):
  $$
    \xymatrix{
      \pi_0 \mathbf{CField}(X)_{P_{\mathrm{Spin}}}
	  \ar@{->>}[r]
	  &
	  \hat H^{n+1}_{\tfrac{1}{2}p_1(P_{\mathrm{Spin}})}(X)
	}
	\,.
  $$
  The gauge-of-gauge equivalence classes of the auto-gauge transformation
  of the trivial C-field configuration naturally surject onto $H^2(X, U(1))$:
  $$
    \xymatrix{
      \pi_1 \mathbf{CField}(X)_{P_{\mathrm{Spin}}}
	  \ar@{->>}[r]
	  &
	  H^2(X, U(1))
	}    
	\,.
  $$
\end{proposition}
\proof
By def. \ref{CField} and the pasting law, prop. \ref{PastingLawForPullbacks},
we have a pasting diagram of homotopy pullbacks of the form
$$
  \raisebox{20pt}{
  \xymatrix@C=10pt{
    \mathbf{CField}(X)_{P_{\mathrm{Spin}}}
	\ar@{->>}[rr]
	\ar[d]
	&&
	\mathbf{H}_{\tfrac{1}{2}\mathbf{p}_1(P_{\mathrm{Spin}})}(X, 
	\mathbf{B}^n U(1)_{\mathrm{conn}}^{\nu_4})
	\ar[rrrr]
	\ar[d]
	&&&&
	\mathbf{H}(X,
  	  \mathbf{B}^n U(1)_{\mathrm{conn}}^{\nu_4}
	)
	\ar[d]
    \\
    \mathbf{H}(X, \mathbf{B}E_8)
	\ar@{->>}[rr]^{\mathbf{H}(X, \mathbf{a})}
	&&
	\mathbf{H}(X, \mathbf{B}^3 U(1))
	\ar[rr]^<<<<<<<<{(\nabla_{\mathfrak{so}}, \mathrm{id})}
	&&
	\mathbf{H}(X, \mathbf{B}\mathrm{Spin}_{\mathrm{conn}} \times \mathbf{B}^3 U(1))
	\ar[rr]^{(u, \mathrm{id})}
	&&
	\mathbf{H}(X, \mathbf{B}\mathrm{Spin}^{\nu_4} \times \mathbf{B}^3 U(1))
  }
  }
  \,,
$$
where in the middle of the top row we identified, by 
def. \ref{RestrictionOfDifferentialWuStructuresToFixedSmoothWuStructure}, the
$n$-groupoid of smooth differential Wu structures lifting the smooth
Wu structure $\tfrac{1}{2}\mathbf{p}_1(P_{\mathrm{Spin}})$.

Due to prop. \ref{RefinementOfDifferentialWuClassesTonBundles}
we are therefore reduced to showing that the top left morphism is 
surjective on $\pi_0$.

But the bottom left morphism is surjective on $\pi_0$, by 
prop. \ref{aIsMaxCompactSubgroup}. Now, the morphisms surjective on
$\pi_0$ are precisely the \emph{effective epimorphisms} in 
$\infty \mathrm{Grpd}$,
and these are stable under pullback. Hence the first claim follows.

For the second, we use that 
$$
  \pi_1 \mathbf{CField}(X)_{P_{\mathrm{Spin}}}
  \simeq
   \pi_0 \Omega \mathbf{CField}(X)_{P_{\mathrm{Spin}}}
$$
and that forming loop space objects (being itself a homotopy pullback)
commutes with homotopy pullbacks and with 
taking cocycles with coefficients in higher stacks, $\mathbf{H}(X,-)$.

Therefore the image of the left square in the above under $\Omega$ is the
homotopy pullback
$$
  \raisebox{20pt}{
  \xymatrix{
    \Omega \mathbf{CField}(X)_{P_{\mathrm{Spin}}}
	\ar@{->>}[rr]
	\ar[d]
	&&
	\mathbf{H}_{\tfrac{1}{2}\mathbf{p}_1(P_{\mathrm{Spin}})}(X, 
	\mathbf{B}^n U(1)_{\mathrm{conn}}^{\nu_4})
	\ar[d]
	\\
	C^\infty(X, E_8)
	\ar@{->>}[rr]^{\mathbf{H}(X, \Omega \mathbf{a})}
	&&
	\mathbf{H}(X, \mathbf{B}^2 U()1)
  }
  }
  \,,
$$
where in the bottom left corner we used
$$
  \begin{aligned}
  \Omega \mathbf{H}(X, \mathbf{B}E_8)
  &\simeq
  \mathbf{H}(X, \Omega \mathbf{B}E_8)
  \\
  &\simeq
  \mathbf{H}(X, E_8)
  \\
  &\simeq
  C^\infty(X, E_8)
  \end{aligned}
  \,,
$$
and similarly for the bottom right corner. 
This identifies the bottom morphism on connected components as the
morphism that sends a smooth function $X \to E_8$ to its homotopy class
under the homotopy equivalence $E_8 \simeq_{15} B^2 U(1) \simeq K(\mathbb{Z},3)$,
which holds over the 11-dimensional $X$.

Therefore the bottom morphism is again surjective on $\pi_0$, and so is
the top morphism. The claim then follows with prop. \ref{RelationToHopkinsSingerDifferentialWu}.
\endofproof

\paragraph{Boundary moduli of the C-field}
\label{CFieldRestrictionToTheBoundary}

We consider now $\partial X$ (a neighbourhood of) the boundary of spacetime $X$,
and discuss a variant of the moduli stack $C \mathrm{Field}$ that encodes
the boundary configurations of the supergravity $C$ field. 

\medskip

Two different kinds of boundary conditions for the $C$-field appear in the
literature.
\begin{itemize}
  \item On an M5-brane boundary, the integral class underlying the $C$-field vanishes.
    (For instance page 24 of \cite{Witten96}).
  \item On the fixed points of a 3-bundle-\emph{orientifold}, def. \ref{OrientifoldCircleNBundlesWithConnection}, 
for the case that $X$ has an $S^1 /\!/\mathbb{Z}_2$-orbifold factor, the
$C$-field vanishes entirely.
\index{differential cohomology!orientifold!Ho{\v r}ava-Witten backgrounds}
(This is considered in \cite{HoravaWitten}. See section 3.1 of \cite{Falkowski} for details.)
\end{itemize}
 We construct higher moduli stacks for both of these conditions in the following.
 In addition to being restricted, the supergravity fields on a boundary
 also pick up additional degrees of freedom
\begin{itemize}
   \item  The $E_8$-principal bundle over the boundary is equipped with a connection.
\end{itemize}

We present now a sequence of natural morphisms of 3-stacks
$$
  \xymatrix{
    C \mathrm{Field}^{\mathrm{bdr}'}
     \ar[r]
	 \ar@/_1pc/[rr]_{\iota'}
    &	 
    C \mathrm{Field}^{\mathrm{bdr}} 
     \ar[r]^\iota
     &	 
    C \mathrm{Field}
  }
$$
into the moduli stack of bulk $C$-fields,
such that $C$-field configurations on $X$ with the above behaviour 
over $\partial X$
correspond to the \emph{relative cohomology}, def. \ref{RelativeCohomology}, 
with coefficients in $\iota$ or $\iota'$,respectively, hence to
commuting diagrams of the form
$$
  \raisebox{20pt}{
  \xymatrix{
    \partial X \ar[r]^<<<<<{\phi_{\mathrm{bdr}}} 
	\ar@{^{(}->}[d]
	  & C \mathrm{Field}^{\mathrm{bdr}} \ar[d]^{\iota}	  
	\\
	X \ar[r]^<<<<<<{\phi} & C \mathrm{Field}
  }
  }
  \,,
$$
and analogously for the primed case.
(This is directly analogous to the characterization of type II supergravity field
configurations in the presence of $D$-branes as discussed in \ref{TwistedSpinCStructures}.)

To this end, recall the general diagram of moduli stacks from def. \ref{BGconn} 
that relates the characteristic map
$\frac{1}{2}\mathbf{p}_1 + 2 \mathbf{a}$ with its differential 
refinement $\frac{1}{2}\hat {\mathbf{p}}_1 + 2 \hat {\mathbf{a}}$:
$$
  \raisebox{20pt}{
  \xymatrix{
    \mathbf{B}(\mathrm{Spin} \times E_8)
	\ar[rr]^{\mathbf{\flat}\frac{1}{2} {\mathbf{p}}_1 + 2  \mathbf{\flat}{\mathbf{a}}}
	\ar[d]
	&&
	\mathbf{\flat}\mathbf{B}^3 U(1)
	\ar[d]
	\\
    \mathbf{B}(\mathrm{Spin} \times E_8)_{\mathrm{conn}}
	\ar[rr]^{\frac{1}{2}\hat {\mathbf{p}}_1 + 2 \hat {\mathbf{a}}}
	\ar[d]
	&&
	\mathbf{B}^3 U(1)_{\mathrm{conn}}
	\ar[d]
    \\
    \mathbf{B}(\mathrm{Spin} \times E_8)
	\ar[rr]^{\frac{1}{2} {\mathbf{p}}_1 + 2  {\mathbf{a}}}
	&&
	\mathbf{B}^3 U(1)
  }
  }
  \,.
$$
The defining $\infty$-pullback diagram for $C \mathrm{Field}$
factors the lower square of this diagram as follows
$$
  \raisebox{30pt}{
  \xymatrix{    
    \mathbf{B}(\mathrm{Spin} \times E_8)_{\mathrm{conn}}
	\ar@{-->}[dr]
	\ar[ddr]
	\ar@/^1pc/[drrrr]^{\frac{1}{2}\hat {\mathbf{p}}_1 + 2 \hat {\mathbf{a}}}
    \\
    & C \mathrm{Field}
	\ar[d]
	\ar[rrr]^{\hat G_4}
	&&&
	\mathbf{B}^3 U(1)_{\mathrm{conn}}
	\ar[d]
    \\
    & \mathbf{B}\mathrm{Spin}_{\mathrm{conn}}
	\times
	\mathbf{B} E_8
	\ar[r]
	&
    \mathbf{B}\mathrm{Spin}
	\times
	\mathbf{B} E_8
	\ar[rr]^{\frac{1}{2} {\mathbf{p}}_1 + 2  {\mathbf{a}}}
	&&
	\mathbf{B}^3 U(1)
  }
  }
  \,.
$$
Here the dashed morphism is the universal morphism induced from the
commutativity of the previous diagram together with the pullback property of 
the 3-stack $C \mathrm{Field}$.
This morphism is the natural map of moduli which induces the relative cohomology
that makes the $E_8$-bundle pick up a connection on the boundary. 

It therefore remains to model the condition that $G_4$ or even $\hat G_4$
vanishes on the boundary.  This condition is realized by further pulling 
back along the sequence 
$$
  \xymatrix{
    {*} 
    \ar[r]^0 
	&
	\Omega^3(-)
	\ar[r]
	& 
	\mathbf{B}^3 U(1)_{\mathrm{conn}}
  }
  \,.
$$
\begin{definition}
  \label{CFieldBdr}
  Write $C \mathrm{Field}^{\mathrm{bdr}}$ 
  and $C \mathrm{Field}^{\mathrm{bdr}'}$, respectively, for the
  moduli 3-stacks which arise as  
  homotopy pullbacks in the top rectangles of
  $$
    \raisebox{20pt}{
    \xymatrix{
	  C \mathrm{Field}^{\mathrm{bdr}'}
	  \ar[rr]
	  \ar@/_5.0pc/[ddd]_{\iota'}
	  \ar[d]
	  &&
	  {*}
	  \ar[d]^0
	  \\
	  C \mathrm{Field}^{\mathrm{bdr}}
	  \ar[rr]
	  \ar@/_3.5pc/[dd]_\iota
	  \ar[d]
	  &&
	  \Omega^3(-)
	  \ar[d]
	  \\
	  \mathbf{B}(\mathrm{Spin} \times E_8)_{\mathrm{conn}}
	  \ar[rr]^{\frac{1}{2}\hat {\mathbf{p}}_1 + 2 \hat {\mathbf{a}}}
	  \ar@{-->}[d]
	  &&
	  \mathbf{B}^3 U(1)_{\mathrm{conn}}
	  \ar@{=}[d]
	  \\
	  C\mathrm{Field}
	  \ar[rr]^{\hat G_4}
	  &&
	  \mathbf{B}^3 U(1)_{\mathrm{conn}}
	}
	}
	\,.
  $$
  For $X$ a smooth manifold with boundary, we say that the 
  3-groupoid of \emph{$C$-field configurations with boundary data} on $X$ is
  the hom $\infty$-groupoid
  $$
    \mathbf{H}^I(
	   \partial X \to X
		   \;,\; 
	   C \mathrm{Field}^{\mathrm{bdr}} \stackrel{\iota}{\to} C \mathrm{Field})
	\,,
  $$
  in the arrow category of the ambient $\infty$-topos 
  $\mathbf{H} = \mathrm{Smooth}\infty \mathrm{Grp}$,
  where on the right we have the composite morphism indicated by the curved arrow above,
  and analogously for the primed case.
\end{definition}
\begin{observation}
  \label{BoundaryCFieldInTermsOfStringE8}
  The moduli 3-stack $C \mathrm{Field}^{\mathrm{brd}}$ is equivalent to
  is the moduli 3-stack of twisted 
  $\mathrm{String}^{2\mathbf{a}}$-2-connections whose underlying twist has trivial
  class. The moduli 3-stack $C \mathrm{Field}^{\mathrm{bdr}'}$
  is equivalent to the moduli 3-stack of untwisted $\mathrm{String}^{2\mathbf{a}}$-2-connections
  $$
    C \mathrm{Field}^{\mathrm{bdr}'} 
	  \simeq
	\mathrm{String}^{2\mathbf{a}}_{\mathrm{conn}}
	\,.
  $$
  This is presented via Lie integration of $L_\infty$-algebras as
  $$
    C \mathrm{Field}^{\mathrm{bdr}'}
	\simeq
    \mathbf{cosk}_3
	  \exp(  (\mathfrak{so} \oplus \mathfrak{e}_8)_{\mu_3^{\mathfrak{so}}+\mu_3^{\mathfrak{e}_8} }  
	  )_{\mathrm{conn}}
	  \,.
  $$
  The presentation of $C \mathrm{Field}^{\mathrm{bdr}}$ by Lie integration is 
  locally given by 
$$
  \left(
    \begin{array}{ll} 
      F_A =& d A + \frac{1}{2}[A \wedge A]
      \\
      H_3 =& \nabla B := d B + \mathrm{CS}(A) - C_3 
      \\
      \mathcal{G}_4  =& d C_3
      \\
      d F_A =& - [A \wedge F_A]
      \\
      d H_3 =&  \langle F_A \wedge F_A\rangle - \mathcal{G}_4
      \\
      d \mathcal{G}_4 =& 0
    \end{array}
  \right)_i
  \;\;\;\;
  \stackrel{
    \begin{array}{ll}
       t^a & \mapsto A^a
       \\
       r^a & \mapsto F^a_A
       \\
       b & \mapsto B
       \\
       c & \mapsto C_3
       \\
       h & \mapsto H_3
       \\
       g & \mapsto \mathcal{G}_4
    \end{array}
  }{\xymatrix{\ar@{<-|}[rrr]&&&}}
  \;\;\;\;
  \left(
    \begin{array}{ll}  
       r^a  =& d t^a + \frac{1}{2}C^a{}_{b c} t^b \wedge t^c + 
       \\
       h = & d b + cs - c     
       \\
       g  =& d c
       \\
       d r^a  =&  - C^a{}_{b c} t^b \wedge r^a
       \\
       d h =& \langle -,-\rangle - g 
       \\
       d g =&  0 
    \end{array}
  \right)
  \,,
$$
where
$$
  \mathfrak{g} = \mathfrak{so} \oplus \mathfrak{e}_8
$$
and hence
$$
  A = \omega + A_{\mathfrak{e}_8}
  \,.
$$
\end{observation}
\proof
  By definition \ref{twisteddifferentialcstructures}
  and prop. \ref{String2aByLieIntegration}.
\endofproof
\begin{remark}
  Notice that with respect to $\mathrm{String}$-connections, there are two
  levels of twists here:
  \begin{enumerate}
    \item The $C$-field 3-form twists the $\mathrm{String}^{2\mathbf{a}}$-2-connections.
	\item For vanishing $C$-field 3-form, a $\mathrm{String}^{2\mathbf{a}}$-2-connection
	is still a twisted $\mathrm{String}$-2-connection, where the twist is now
	by the Chern-Simons 3-bundle with connection of the underlying $E_8$-bundle
	with connection.
  \end{enumerate}
\end{remark}

\paragraph{Ho{\v r}ava-Witten boundaries are membrane orientifolds}
\index{twisted cohomology!orientifold!membrane-}
\label{MembraneOrientifolds}

We now discuss a natural formulation of the origin of the 
Ho{\v r}ava-Witten boundary conditions \cite{HoravaWitten} in terms of 
higher stacks and nonabelian differential cohomology,
specifically, in terms of what we call \emph{membrane orientifolds}.
From this we obtain a corresponding refinement of the 
moduli 3-stack of C-field configurations which now explicitly 
contains the twisted $\mathbb{Z}_2$-equivariance of the
Ho{\v r}ava-Witten background.

\medskip

Recall the notion of higher orientifolds and 
their identification with twisted differential $\mathbf{J}_n$-structures
from \ref{OrientifoldCircleNBundlesWithConnection}.

\begin{observation}
  Let $U/\!/\mathbb{Z}_2 \hookrightarrow Y/\!/\mathbb{Z}_2$
  be a patch on which a given $\hat {\mathbf{J}}_n$-structure 
  has a trivial underlying integral class, such that it is
  equivalent to a globally defined $(n+1)$-form $C_U$ on $U$. Then 
  the components of this 
  this 3-form orthogonal to the $\mathbb{Z}_2$-action are 
  \emph{odd} under the action.
  In particular, if $U \hookrightarrow Y$ sits in the fixed point
  set of the action, then these components vanish. 
  This is the Ho{\v r}ava-Witten boundary condition on the $C$-field
  on an 11-dimensional spacetime $Y = X \times S^1$ equipped with 
  $\mathbb{Z}_2$-action on the circle. See for instance section 3
  of \cite{Falkowski} for an explicit discussion of the $\mathbb{Z}_2$
  action on the C-field in this context.
\end{observation}
We therefore have a natural construction of the moduli 3-stack of 
Ho{\v r}ava-Witten C-field configurations as follows
\begin{definition}
 Let $\mathbf{CField}_J(Y)$ be the homotopy pullback in 
 $$
   \raisebox{20pt}{
  \xymatrix{
    \mathbf{CField}_J(Y)
    \ar[dd]
	\ar[rr]
    &&
    \hat {\mathbf{J}} \mathrm{Struc}_{\rho}(Y/\!/\mathbb{Z}_2)
	\ar[d]
	\\
	&& \mathbf{H}(Y, \mathbf{B}^3 U(1)_{\mathrm{conn}})
	\ar[d]
	\\
	\mathbf{H}(Y, \mathbf{B}\mathrm{Spin}_{\mathrm{conn}} \times \mathbf{B}E_8)
	 \ar[rr]^<<<<<<<<<<{\mathbf{H}(Y, \tfrac{1}{2}\mathbf{p}_1 + 2\mathbf{a})} 
	&& \mathbf{H}(Y, \mathbf{B}^3 U(1))\;\;,
  }
  }
 $$
 where the top right morphism is the map $\hat G_\rho \mapsto \hat G$
 from remark \ref{DifferentialJStructreDecomposed}.
\end{definition}
The objects of $\mathbf{CField}_J(Y)$ are C-field configurations
on $Y$ that not only satisfy the flux quantization condition, but also
the Ho{\v r}ava-Witten twisted equivariance condition
(in fact the proper globalization of that condition from 3-forms to 
full differential cocycles). This is formalized by the following.
\begin{observation}
  There is a canonical morphism $\mathbf{CField}_J(Y) \to \mathbf{CField}(Y)$,
  being the dashed morphism in
 $$
   \raisebox{20pt}{
  \xymatrix{
    \mathbf{CField}_J(Y)
    \ar@{-->}[d]
	\ar[rr]
    &&
    \hat {\mathbf{J}} \mathrm{Struc}_{\rho}(Y/\!/\mathbb{Z}_2)
	\ar[d]
	\\
	\mathbf{CField}(Y) 
	  \ar[rr]
	  \ar[d]
	&& \mathbf{H}(Y, \mathbf{B}^3 U(1)_{\mathrm{conn}})
	\ar[d]
	\\
	\mathbf{H}(Y, \mathbf{B}\mathrm{Spin}_{\mathrm{conn}} \times \mathbf{B}E_8)
	 \ar[rr]^<<<<<<{~~~~~~~\mathbf{H}(Y, \tfrac{1}{2}\mathbf{p}_1 + 2\mathbf{a})} 
	&& \mathbf{H}(Y, \mathbf{B}^3 U(1))\;\;,
  }
  }
  \,
 $$
  which is given by the universal property of the defining homotopy pullback
  of $\mathbf{CField}$, remark \ref{CFieldAs3Bundle}.
\end{observation}
A supergravity field configuration presented by a morphism $Y \to \mathbf{CField}$
into the moduli 3-stack of configurations that satisfy the flux quantization
condition in addition satisfies the Ho{\v r}ava-Witten boundary condition
if, as an element of $\mathbf{CField}(Y) := \mathbf{H}(Y, \mathbf{CField})$
it is in the image of $\mathbf{CField}_J(Y) \to \mathbf{CField}(Y)$.
In fact, there may be several such pre-images. A choice of one is a choice
of membrane orientifold structure.

\subsubsection{Differential T-duality}
\label{DifferentialTDuality}
\index{T-duality} \index{twisted cohomology!T-duality}

In \cite{KahleValentino} (see also the review in section 7.4 of \cite{BunkeSchick10})
a formalization of the differential refinement of topological T-duality  
is given. We discuss here how this is naturally an example of 
the twisted differential $\mathbf{c}$-structures, \ref{TwistedDifferentialStructures}.

(...)

\newpage

\subsection{Higher symplectic geometry}
\label{HigherSymplecticGeometry}
\index{symplectic higher geometry}

The notion of \emph{symplectic manifold} formalizes in physics the concept of a \emph{classical mechanical system}. The notion of \emph{geometric quantization},
\ref{StrucGeometricPrequantization}, 
of a symplectic manifold is one formalization of the general concept in physics of \emph{quantization} of such a system to a \emph{quantum mechanical system}. 

Or rather, the notion of symplectic manifold does not quite capture the most general systems of classical mechanics. One generalization requires passage to \emph{Poisson manifolds}. The original methods of geometric quantization become meaningless on a Poisson manifold that is not symplectic. 
However, a Poisson structure on a manifold $X$ is equivalent to the structure of a Poisson Lie algebroid $\mathfrak{P}$ over $X$. This is noteworthy, because the latter \emph{is} again symplectic, as a Lie algebroid, even if the underlying Poisson manifold is not symplectic: it is a \emph{symplectic Lie 1-algebroid}, prop. \ref{PoissonLieAlgebroid}.

Based on related observations it was suggested, \cite{Weinstein} that a notion of \emph{symplectic groupoid} should naturally replace that of \emph{symplectic manifold} for the purposes of geometric quantization to yield a notion of \emph{geometric quantization of symplectic groupoids}. 
Since a symplectic manifold can be regarded as a symplectic Lie 0-algebroid, prop. \ref{SymplecticPoissonCourant}, and also as a symplectic smooth 0-groupoid this step amounts to a kind of categorification of symplectic geometry. 

More or less implicitly, there has been evidence that this shift in perspective is substantial: the \emph{deformation quantization} of a Poisson manifold famously turns out \cite{Kontsevich} to be constructible in terms of correlators of the 2-dimensional TQFT called the \emph{Poisson $\sigma$-model}, \ref{section.PoissonSigmaModel}, associated with the corresponding Poisson Lie algebroid. The fact that this is 2-dimensional and not 1-dimensional, as the quantum mechanical system that it thus encodes, is a direct reflection of this categorification shift of degree.

On general abstract grounds this already suggests that it makes sense to pass via higher categorification further to symplectic Lie $n$-algebroids, def. \ref{SymplecticLooAlgebroid}, as well as to symplectic 2-groupoids, symplectic 3-groupoids, etc. up to symplectic $\infty$-groupoids, def. \ref{SymplecticInfinityGroupoids}.

Formal hints for such a generalization had been noted in \cite{Severa}
(in particular in its concluding table). More indirect -- but all the more noteworthy -- hints came from quantum field theory, where it was observed that a generalization of symplectic geometry to \emph{multisymplectic geometry} \cite{Helein} of degree $n$ more naturally captures the description of $n$-dimensional QFT (notice that quantum mechanics may be understood as $(0+1)$-dimensional QFT). For, observe that the symplectic form on a symplectic Lie $n$-algebroid is, while always ``binary'', nevertheless a representative of de Rham cohomology in degree $n+2$.

There is a natural formalization of these higher symplectic structures in the context of any cohesive $\infty$-topos. Moreover, by \ref{SymplecticLienAlgebroids} symplectic forms on $L_\infty$-algebroids have a natural interpretation in $\infty$-Lie theory: they are $L_\infty$-invariant polynomials. This means that the $\infty$-Chern-Weil homomorphism applies to them. 

\begin{observation}
  From the perspective of $\infty$-Lie theory, 
  a smooth manifold $\Sigma$ equipped with 
  a symplectic form $\omega$ is equivalently
  a Lie 0-algebroid equipped with a quadratic
  and non-degenerate \emph{$L_\infty$-invariant polynomial}
  (def. \ref{InvariantPolynomial}).
\end{observation}
This observation implies
\begin{enumerate}
  \item
    a direct $\infty$-Lie theoretic analog of symplectic manifolds:
    \emph{symplectic Lie $n$-algebroids} and their Lie integration
	to \emph{symplectic smooth $\infty$-groupoids}
  \item
    the existence of a canonical $\infty$-Chern-Weil homomorphism
    for every symplectic Lie $n$-algebroid.
\end{enumerate}
This is spelled out below in \ref{section.symplectic_dg-geometry}, \ref{SymplecticLienAlgebroids}, \ref{SymplecticGroupoid}.
The $\infty$-group extensions,
def. \ref{ExtensionOfInfinityGroups}, that are induced by the
unrefined $\infty$-Chern-Weil homomorphism, \ref{StrucChern-WeilHomomorphism}, on a symplectic $\infty$-groupoid are their \emph{prequantum circle $(n+1)$-bundles}, the higher analogs of prequantum line bundles in the geometric quantization of symplectic manifolds. This we discuss in \ref{SymplecticPrequantum}. Further below in \ref{InfinCSAKSZ} we show that the \emph{refined}
$\infty$-Chern-Weil homomorphism, \ref{StrucChern-SimonsTheory}, on a symplectic $\infty$-groupoid constitutes the action functional of the 
corresponding \emph{AKSZ $\sigma$-model} (discussed below in \ref{InfinCSAKSZ}).

\medskip
\begin{itemize}
 \item \ref{section.symplectic_dg-geometry} -- Symplectic dg-geometry;
 \item \ref{SymplecticLienAlgebroids}-- Symplectic $L_\infty$-algebroids;
 \item \ref{SymplecticGroupoid} -- Symplectic smooth $\infty$-groupoids;
\end{itemize}

The parts \ref{section.symplectic_dg-geometry} and \ref{SymplecticLienAlgebroids} 
are taken from \cite{frs}.

\subsubsection{Symplectic dg-geometry}
\label{section.symplectic_dg-geometry}

In \ref{SynthDiffInfGrpd} we considered a general 
abstract notion of infinitesimal thickenings in higher
differential geometry and showed how 
from the point of view of $\infty$-Lie theory this leads 
to the notion of $L_\infty$-algebroids,
def. \ref{LInfinityAlgebroids}. As is evident from that definition,
these can also be regarded as objects in 
\emph{dg-geometry} \cite{ToenVezzosi}. We make explicit now
some basic aspects of this identification.

The following definitions formulate a simple notion of
\emph{affine smooth graded manifolds} and 
\emph{affine smooth dg-manifolds}. 
Despite their simplicity
these definitions capture in a precise sense all the relevant structure: namely the
\emph{local} smooth structure. Globalizations of these definitions
can be obtained, if desired, by general abstract constructions.

\begin{definition}
  \label{GradedManifold}
  The category of \emph{affine smooth $\mathbb{N}$-graded manifolds} 
  -- here called \emph{smooth graded manifolds} for short --
  is the full subcategory
  $$
    \mathrm{SmoothGrMfd} \subset \mathrm{GrAlg}_{\mathbb{R}}^{\mathrm{op}}
  $$
  of the opposite category of 
  $\mathbb{N}$-graded-commutative
  $\mathbb{R}$-algebras on those isomorphic to Grassmann algebras
  of the form
  $$
    \wedge^\bullet_{C^\infty(X_0)} \Gamma(V^*)
    \,,
  $$
  where $X_0$ is an ordinary smooth manifold, $V \to X_0$ is an 
  $\mathbb{N}$-graded smooth vector bundle over $X_0$ degreewise of 
  finite rank, and
  $\Gamma(V^*)$ is the graded $C^\infty(X)$-module of
  smooth sections of the dual bundle.
  
  For a smooth graded manifold $X \in \mathrm{SmoothGrMfd}$, we write 
  $C^\infty(X) \in \mathrm{cdgAlg}_{\mathbb{R}}$ for its corresponding
  dg-algebra \emph{of functions}.
\end{definition}
\noindent {\bf Remarks.}
\begin{itemize}
  \item The full subcategory of these objects is equivalent to that
  of all objects isomorphic to one of this form. We may therefore
  use both points of view interchangeably. 
  \item Much of the theory works just as well 
   when $V$ is allowed to be $\mathbb{Z}$-graded. This is the
   case that genuinely corresponds to \emph{derived} (instead of just higher)
   differential geometry. An important class of examples for this case are
   BV-BRST complexes which motivate much of the literature.
   For the purpose of this short note, we shall be content with the 
   $\mathbb{N}$-graded case. 
  \item
   For an $\mathbb{N}$-graded $C^\infty(X_0)$-module $\Gamma(V^*)$ we have
   $$
     \wedge^\bullet_{C^\infty} \Gamma(V^*)
     =
     C^\infty(X_0)
      \oplus
     \Gamma(V_0^*)
       \oplus
     \left(
       \Gamma(V_0^*) \wedge_{C^\infty(X_0)} \Gamma(V_0^*)
       \oplus
       \Gamma(V_1^*)
     \right)
     \oplus
     \cdots
     \,,
   $$
   with the leftmost summand in degree 0, the next one in degree 1,
   and so on.
   \item There is a canonical functor
   $$
     \mathrm{SmoothMfd} \hookrightarrow \mathrm{SmthGrMfd}
   $$
   which identifies an ordinary smooth manifold $X$ with the 
   smooth graded manifold whose function algebra is the 
   ordinary algebra of smooth functions $C^\infty(X_0) := C^\infty(X)$
   regarded as a graded algebra concentrated in degree 0.
   This functor is full and faithful and hence exhibits a full
   subcategory.
\end{itemize}
All the standard notions of differental geometry apply to 
differential graded geometry. 
For instance for $X \in \mathrm{SmoothGrMfd}$, 
there is the graded vector space $\Gamma(T X)$ of vector fields on $X$, 
where a vector field is identified with a
graded \emph{derivation} $v : C^\infty(X) \to C^\infty(X)$.
This is naturally a graded (super) Lie algebra with 
super Lie bracket the graded commutator of derivations.
Notice that for $v \in \Gamma(T X)$ of odd degree we have
$[v,v] = v\circ v + v \circ v = 2 v^2 : C^\infty(X) \to C^\infty(X)$.
\begin{definition}
  \label{DGManifolds}
  The category  of (affine, $\mathbb{N}$-graded) 
  \emph{smooth differential-graded manifolds}
  is the full subcategory 
  $$
     \mathrm{SmoothDgMfd} \subset \mathrm{cdgAlg}_{\mathbb{R}}^{\mathrm{op}}
  $$ 
  of the opposite of differential graded-commutative $\mathbb{R}$-algebras
  on those objects whose underlying graded algebra comes from
  $\mathrm{SmoothGrMfd}$.
\end{definition}
This is equivalently the category whose objects 
are  pairs $(X,v)$ consisting of a smooth graded manifold
$X \in \mathrm{SmoothGrMfd}$
and a grade 1 vector field $v \in \Gamma(T X)$, 
such that $[v,v] = 0$, and whose morphisms $(X_1,v_1) \to (X_2, v_2)$ are 
morphisms $f : X_1 \to X_2$ such that $v_1 \circ f^*  = f^* \circ v_2$.
\begin{remark}
  \label{SmoothDgAlgebras}
  The dg-algebras appearing here are special in that their
  degree-0 algebra is naturally not just
  an $\mathbb{R}$-algebra, but a \emph{smooth algebra}
  (a ``$C^\infty$-ring'', see \cite{Stel} for review and
  discussion). 
\end{remark}
\begin{definition}
  \label{deRhamComplex}
  The \emph{de Rham complex functor}
  $$
    \Omega^\bullet(-) 
      : 
    \mathrm{SmoothGrMfd}
     \to 
    \mathrm{cdgAlg}^{\mathrm{op}}_{\mathbb{R}}
  $$
  sends a dg-manifold $X$ with 
  $C^\infty(X) \simeq \wedge^\bullet_{C^\infty(X_0)}
  \Gamma(V^*)$ to the 
  Grassmann algebra over $C^\infty(X_0)$ on the graded
  $C^\infty(X_0)$-module
  $$
    \Gamma(T^* X) \oplus \Gamma(V^*) \oplus \Gamma(V^*[-1])
    \,,
  $$
  where $\Gamma(T^* X)$ denotes the ordinary smooth 1-form fields
  on $X_0$ and where $V^*[-1]$ is $V^*$ with the grades \emph{increased}
  by one. This is equipped with the differential $\mathbf{d}$
  defined on generators as follows:
  \begin{itemize}
   \item $\mathbf{d}|_{C^\infty(X_0)} = d_{\mathrm{dR}}$ is the 
  ordinary de Rham differential with values in $\Gamma(T^* X)$;
    \item 
      $\mathbf{d}|_{\Gamma(V^*)} \to \Gamma(V^*[-1])$ is the degree-shift
  isomorphism
   \item and $\mathbf{d}$ vanishes on all remaining generators.
   \end{itemize}
\end{definition}
\begin{definition}
\label{TangentLieAlgebroidAsDgManifold}
Observe that $\Omega^\bullet(-) $ evidently factors through the
defining inclusion 
$\mathrm{SmoothDgMfd}$ $\hookrightarrow$ $\mathrm{cdgAlg}_{\mathbb{R}}$.
Write
$$
  \mathfrak{T}(-) : \mathrm{SmoothGrMfd} \to \mathrm{SmoothDgMfd}
$$
for this factorization.
\end{definition}
 \label{GradedTangentBundle}
The dg-space $\mathfrak{T}X$ is often called the 
\emph{shifted tangent bundle} of $X$ and denoted $T[1]X$.
\begin{observation}
  \label{FormsFromGrManifoldMaps}
  For $\Sigma$ an ordinary smooth manifold and 
  for $X$ a graded manifold corresponding to a vector bundle
  $V \to X_0$, there is a natural bijection
  $$
    \mathrm{SmoothGrMfd}(\mathfrak{T}\Sigma, X)
    \simeq
    \Omega^\bullet(\Sigma,V)
    \,
  $$
  where on the right we have the set of $V$-valued smooth differential forms
  on $\Sigma$: tuples consisting of a smooth function 
  $\phi_0 : \Sigma \to X_0$, and for each $n > 1$ 
  an ordinary differential $n$-form 
  $\phi_n \in \Omega^n(\Sigma, \phi_0^* V_{n-1})$ with values in the pullback
  bundle of $V_{n-1}$ along $\phi_0$.
\end{observation}
The standard Cartan calculus of differential geometry
generalizes directly to graded smooth manifolds.
For instance, given a vector field $v \in \Gamma(T X)$
on $X \in \mathrm{SmoothGrMfd}$, 
there is the \emph{contraction derivation}
$$
  \iota_v : \Omega^\bullet(X) \to \Omega^{\bullet}(X)
$$
on the de Rham complex of $X$, and hence the \emph{Lie derivative}
$$
  \mathcal{L}_v := [\iota_{v},\mathbf{d}] : 
  \Omega^\bullet(X) \to \Omega^\bullet(X)
  \,.
$$
\begin{definition}
  For $X \in \mathrm{SmoothGrMfd}$ the \emph{Euler
  vector field} $\epsilon \in \Gamma(T X)$ is defined over 
  any coordinate patch $U \to X$ to be
  given by the formula
  $$
    \epsilon|_U 
      := 
    \sum_a
     \mathrm{deg}(x^a) x^a \frac{\partial}{\partial x^a}
    \,,
  $$
  where $\{x^a\}$ is a basis of generators and 
  $\mathrm{deg}(x^a)$ the degree of a generator.
    The \emph{grade} of a homogeneous element $\alpha$ in 
  $\Omega^\bullet(X)$ is the unique natural number $n \in \mathbb{N}$
  with 
  $$
    \mathcal{L}_\epsilon \alpha = n \alpha
    \,.
  $$
\end{definition}
\noindent{\bf Remarks.}
\begin{itemize}
\item This implies that for $x^i$ an element of grade $n$ on $U$, the
1-form $\mathbf{d}x^i$ is also of grade $n$. This is why we speak
of \emph{grade} (as in ``graded manifold'') instead of \emph{degree} here. 
\item Since coordinate transformations on a graded manifold are
  grading-preserving, the Euler vector field is indeed
  well-defined. Note that the degree-0 coordinates 
  do not appear in the Euler vector field.
\end{itemize}
The existence of $\epsilon$ implies the following 
useful statement (amplified in \cite{RoytenbergCourant}),
which is a trivial variant of what in grade 0
would be the standard Poincar{\'e} lemma.
\begin{observation}
  \label{GradedPoincareLemma}
  On a graded manifold, every closed differential form
  $\omega$ of positive grade $n$ is exact: the form 
  $$
    \lambda := \frac{1}{n} \iota_\epsilon \omega
  $$
  satisfies
  $$
    \mathbf{d}\lambda = \omega
    \,.
  $$
\end{observation}
\begin{definition}
  \label{SymplecticDgManifold}
  A \emph{symplectic dg-manifold} of grade $n \in \mathbb{N}$ is 
  a dg-manifold $(X,v)$ equipped with 
  2-form $\omega \in \Omega^2(X)$ which is
  \begin{itemize}
    \item non-degenerate;
    \item closed;
  \end{itemize}
  as usual for symplectic forms, and
  in addition
  \begin{itemize}
    \item of grade $n$;
    \item $v$-invariant: $\mathcal{L}_v \omega = 0$.
  \end{itemize}
\end{definition}
In a local chart $U$ with coordinates $\{x^a\}$ we may find functions
$\{\omega_{ab} \in C^\infty(U)\}$ such that
$$
  \omega|_U = \frac{1}{2}  \mathbf{d}x^a\, \omega_{ab}\wedge \mathbf{d}x^b
  \,,
$$
where summation of repeated indices is implied. We say that 
$U$ is a \emph{Darboux chart} for $(X,\omega)$ if the $\omega_{ab}$
are constant.
\begin{observation}  
  The function algebra of a symplectic dg-manifold 
  $(X,\omega)$ of grade $n$
  is naturally equipped with a Poisson bracket
  $$
    \{-,-\} : 
    C^\infty(X)\otimes C^\infty(X)
    \to C^\infty(X)
  $$
  which decreases grade by $n$. On a local coordinate patch $\{x^a\}$ this
  is given by
  $$
    \{f,g\} = 
    \frac{f \reflectbox{\small $\partial$}}{x^a \reflectbox{\small $\partial$}}
    \omega^{a b}
    \frac{\partial g}{\partial x^b}
    \,,
  $$
  where $\{\omega^{a b}\}$ is the inverse matrix to $\{\omega_{a b}\}$,
  and where the graded differentiation in the left factor is to 
  be taken from the right, as indicated.
\end{observation}
\begin{definition}
  \label{Hamiltonians}
  For $\pi\in C^\infty(X)$ and $v \in \Gamma(T X)$, we say that
  \emph{$\pi$ is a Hamiltonian for $v$}, or equivalently, that
  \emph{$v$ is the  of $\pi$}
  if
  $$
    \mathbf{d}\pi = \iota_v \omega
    \,.
  $$
  \end{definition}
Note that the convention $(-1)^{n+1} \mathbf{d}\pi = \iota_v \omega$
is also  frequently used for defining Hamiltonians in the 
context of graded geometry.
\begin{remark}
\label{PartialPiByV}
In a local coordinate chart $\{x^a\}$ the defining equation
$\mathbf{d}\pi = \iota_v \omega$
becomes 
$$
    \mathbf{d}x^a \frac{\partial \pi}{\partial x^a}
    =
    \omega_{a b} v^a \wedge \mathbf{d} x^b
    = 
    \omega_{a b} \mathbf{d}x^a \wedge v^b
  \,,
$$
implying that 
$$
  \omega_{ab}v^b = \frac{\partial \pi}{\partial x^a}
  \,.
$$  
\end{remark}

\subsubsection{Symplectic $L_\infty$-algebroids}
\label{SymplecticLienAlgebroids}
\index{symplectic higher geometry!symplectic $L_\infty$-algebroids}
\index{$L_\infty$-algebroid!symplectic} 
\index{symplectic $L_\infty$-algebroid}

Here we discuss $L_\infty$-algebroids, def. \ref{LInfinityAlgebrasSubAlgebroids}, equipped with \emph{symplectic structure}, which we conceive of as:
equipped with de Rham cocycles that are \emph{invariant polynomials},
def. \ref{InvariantPolynomial}.

\begin{definition}
 \label{SymplecticLooAlgebroid}
 \index{symplectic $L_\infty$-algebroid!definition}
A \emph{symplectic Lie $n$-algebroid}
$(\mathfrak{P}, \omega)$ is a 
Lie $n$-algebroid $\mathfrak{P}$ equipped with a quadratic non-degenerate invariant 
polynomial $\omega \in W(\mathfrak{P})$ of degree $n+2$.
\end{definition}
This means that 
\begin{itemize}
\item on each chart $U \to X$ of the base manifold $X$ of 
$\mathfrak{P}$, there is a 
basis $\{x^a\}$ for $\mathrm{CE}(\mathfrak{a}|_U)$ such that
$$
  \omega =  \frac{1}{2}\mathbf{d}x^a \,\omega_{a b}\wedge \mathbf{d}x^b
$$
with $\{\omega_{a b} \in \mathbb{R} \hookrightarrow C^\infty(X)\}$ 
and $\mathrm{deg}(x^a) + \mathrm{deg}(x^b) = n$;
\item the coefficient matrix $\{\omega_{a b}\}$ has an inverse;
\item we have
$$
  d_{\mathrm{W}(\mathfrak{P})} \omega
  = 
  d_{\mathrm{CE}(\mathfrak{P})} \omega 
  +
  \mathbf{d} \omega = 0
  \,.
$$
\end{itemize}
The following observation essentially goes back to \cite{Severa}
and \cite{RoytenbergCourant}.
\begin{proposition}
  \label{SymplDgSpaceAsLAlgd}
  There is a full and faithful embedding of symplectic dg-manifolds
  of grade $n$
  into symplectic Lie $n$-algebroids.
\end{proposition}
\proof
  The dg-manifold itself is identified with an $L_\infty$-algebroid
  by def. \ref{LInfinityAlgebroids}.
  For $\omega \in \Omega^2(X)$ a symplectic form, the conditions
  $\mathbf{d} \omega = 0$ and $\mathcal{L}_v \omega = 0$
  imply $(\mathbf{d}+ \mathcal{L}v)\omega = 0$ 
  and hence that under the identification 
  $\Omega^\bullet(X) \simeq \mathrm{W}(\mathfrak{a})$ 
  this is an invariant polynomial on $\mathfrak{a}$.
  
  It remains to observe that the $L_\infty$-algebroid $\mathfrak{a}$
  is in fact a Lie $n$-algebroid. This is implied by the 
  fact that $\omega$ is of grade $n$ and non-degenerate: 
  the former condition implies that it has no components in elements
  of grade
  $> n$ and the latter then implies that all such elements vanish.
\endofproof
The following characterization may be taken as a definition 
of Poisson Lie algebroids and Courant Lie 2-algebroids.
\begin{proposition}
  \label{SymplecticPoissonCourant}
  \label{PoissonLieAlgebroid}
  Symplectic Lie $n$-algebroids are equivalently:
  \begin{itemize}
    \item for $n = 0$: ordinary symplectic manifolds;
    \item for $n = 1$: Poisson Lie algebroids;
    \item for $n = 2$: Courant Lie 2-algebroids.
  \end{itemize}
\end{proposition}
See \cite{RoytenbergCourant, Severa} for more discussion.
\begin{proposition}
 \label{CanonicalCocycleOfSymplecticLieAlgebroid}
Let $(\mathfrak{P},\omega)$ be a symplectic Lie $n$-algebroid 
for positive $n$ in the image of the embedding of 
proposition \ref{SymplDgSpaceAsLAlgd}.
Then it carries the canonical $L_\infty$-algebroid cocycle 
 $$
   \pi := \frac{1}{n+1} \iota_\epsilon \iota_v \omega
   \in 
   \mathrm{CE}(\mathfrak{P})
 $$ 
 which moreover is the Hamiltonian, 
 according to definition \ref{Hamiltonians}, of $d_{\mathrm{CE}(\mathfrak{P})}$.
\end{proposition} 
 \proof Since $\mathbf{d}\omega=\mathcal{L}_v\omega=0$, we have
 $$
 \begin{aligned} 
\mathbf{d} \iota_\epsilon \iota_v \omega
&= \mathbf{d} \iota_v \iota_\epsilon  \omega\\
&=(\iota_v\mathbf{d}-\mathcal{L}_v)\iota_\epsilon  \omega\\
&=\iota_v\mathcal{L}_\epsilon\omega-[\mathcal{L}_v,\iota_\epsilon]\omega\\
&=n\iota_v\omega-\iota_{[v,\epsilon]}\omega\\
&=(n+1)\iota_v\omega,
\end{aligned}
$$
where Cartan's formula $[\mathcal{L}_v,\iota_\epsilon]=\iota_{[v,\epsilon]}$ and the identity $[v,\epsilon]=-[\epsilon,v]=-v$ have been used. Therefore $\pi:=\frac{1}{n+1}\iota_\epsilon \iota_v \omega$ satisfies the defining equation $\mathbf{d}\pi=\iota_v\omega$
from definition \ref{Hamiltonians}.
\endofproof

\begin{remark}
  \label{remark.local_hamiltonian}
On a local chart 
with coordinates $\{x^a\}$ 
we have
$$
    \pi\bigr\vert_U 
    =
    \frac{1}{n+1}\omega_{ab}\;\deg(x^a) x^a\, \wedge v^b
  \,.
$$
\end{remark}
Our central observation now is the following.
\begin{proposition}
 \label{TheCSElement}
The cocycle $\frac{1}{n} \pi$ 
from prop. \ref{CanonicalCocycleOfSymplecticLieAlgebroid}
is in transgression with 
the invariant polynomial $\omega$.
A Chern-Simons element witnessing the transgression 
according to def. \ref{TransgressionAndCSElements} is
$$
  \mathrm{cs} = \frac{1}{n}\left(\iota_\epsilon \omega + \pi\right)
  \,.
$$
\end{proposition}
\proof
It is clear that $i^* \mathrm{cs} = \frac{1}{n}\pi$. So it remains to 
check that $d_{\mathrm{W}(\mathfrak{P})} \mathrm{cs} = \omega$.
As in the proof of proposition \ref{CanonicalCocycleOfSymplecticLieAlgebroid}, we use $\mathbf{d}\omega=\mathcal{L}_v\omega=0$ and Cartan's identity $ [\mathcal{L}_v, \iota_\epsilon]
  =
  \iota_{[v,\epsilon]}
  =
  - \iota_{v}
$.
By these, the first summand in
$d_{\mathrm{W}(\mathfrak{P})} ( \iota_{\epsilon} \omega + \pi )$ is
$$
  \begin{aligned}
    d_{\mathrm{W}(\mathfrak{P})} \iota_{\epsilon} \omega
     & = 
     (
       \mathbf{d}
       + \mathcal{L}_v
     )
     \iota_\epsilon \omega
     \\
     &=
     [\mathbf{d}
       +\mathcal{L}_v,\iota_\epsilon]\omega\\
       &= n\omega - \iota_v \omega
     \\
     & = n \omega - \mathbf{d}\pi
  \end{aligned}
  \,.
$$
The second summand is simply 
$$
  d_{\mathrm{W}(\mathfrak{P})} \pi =  \mathbf{d}\pi
$$
since $\pi$ is a cocycle. 
\endofproof
\begin{remark}
 \label{remark.local_cs}
In a coordinate patch $\{x^a\}$ the Chern-Simons element is
$$
 \mathrm{cs}\bigr\vert_U
 =
\frac{1}{n}
   \left(
     \omega_{a b} \,\mathrm{deg}(x^a) x^a\, \wedge \mathbf{d}x^b + \pi
   \right)
   \,.
 $$
 In this formula one can substitute $\mathbf{d} = d_{\mathrm{W}}- d_{\mathrm{CE}}$, and this kind of substitution will be 
crucial for the proof our main statement in proposition 
\ref{TheAKSZActionFromCS} below.
Since $d_{\mathrm{CE}} x^i=v^i$ and using
remark \ref{remark.local_hamiltonian} we find
$$
   \sum_a \omega_{a b} \mathrm{deg}(x^a) x^a \wedge d_{\mathrm{CE}} x^b
   = 
  (n+1) \pi
  \,,
$$
and hence
$$
  \mathrm{cs}\bigr\vert_U
   =
  \frac{1}{n}
  \left(
    \mathrm{deg}(x^a) \,
    \omega_{a b} x^a \wedge d_{\mathrm{W}(\mathfrak{P})} x^b
    -
    n \pi
  \right)
  \,.
$$
\end{remark}
In the section \ref{InfinCSAKSZ} we show that this
transgression element $\mathrm{cs}$ \emph{is} the 
AKSZ-Lagrangian.

\subsubsection{Symplectic smooth $\infty$-groupoids}
\label{SymplecticGroupoid}
\index{smooth $\infty$-groupoid!symplectic}
\index{symplectic higher geometry!symplectic smooth $\infty$-groupoid}

We define \emph{symplectic smooth $\infty$-groupoids} in terms of their 
underlying symplectic $L_\infty$-algebroids. 

Recall that
for any $n \in \mathbb{N}$, a \emph{symplectic Lie $n$-algebroid} 
$(\mathfrak{P}, \omega)$ is (def. \ref{SymplecticLooAlgebroid}) 
an $L_\infty$-algebroid $\mathfrak{P}$ that is equipped with a quadratic and non-degenerate $L_\infty$-invariant polynomial.
Under Lie integration, def. \ref{ExponentiatedLInftyAlgbra}, 
$\mathfrak{P}$ integrates to a smooth $n$-groupoid 
$\tau_n \exp(\mathfrak{P}) \in \mathrm{Smooth}\infty\mathrm{Grpd}$. 
Under the $\infty$-Chern-Weil homomorphism, \ref{SmoothStrucInfChernWeil},
the invariant polynomial induces a
differential form on the smooth $\infty$-groupoid, \ref{StrucDeRham}:
$$
  \omega : \tau_n \exp(\mathfrak{P}) \to \flat_{\mathrm{dR}} \mathbf{B}^{n+2} \mathbb{R}
$$
representing a class $[\omega] \in H^{n+2}_{dR}(\tau_n \exp(\mathfrak{P}))$.

\begin{definition}
  \label{SymplecticInfinityGroupoids}
Write 
$$
  \mathrm{SymplSmooth}\infty \mathrm{Grpd}
  \hookrightarrow
  \mathrm{Smooth}\infty \mathrm{Grpd}/(\coprod_{n}\mathbf{\flat}_{\mathrm{dR}}\mathbf{B}^{n+2}\mathbb{R})
$$
for the full sub-$\infty$-category of the over-$\infty$-topos 
of $\mathrm{Smooth}\infty\mathrm{Grpd}$ over the de Rham coefficient objects on those objects in the image of this construction.

We say an object on $\mathrm{SymplSmooth}\infty \mathrm{Grpd}$ is a \emph{symplectic smooth $\infty$-groupoid}.
\end{definition}
\begin{remark}
There are evident variations of this for the ambient $\mathrm{Smooth}\infty\mathrm{Grpd}$ replaced by some variant, such as $\mathrm{SynthDiffInfGrpd}\infty\mathrm{Grpd}$, or $\mathrm{SmoothSuper}\infty\mathrm{Grpd}$, \ref{SuperInfinityGroupoids}).
\end{remark}

We now spell this out for $n = 1$. The following notion was introduced in \cite{Weinstein} in the study of geometric quantization.
\begin{definition}
  A \emph{symplectic groupoid} is a Lie groupoid $\mathcal{G}$
  equipped with a differential 2-form $\omega_1 \in \Omega^2(\mathcal{G}_1)$ which is 
  \begin{enumerate}
    \item a symplectic 2-form on $\mathcal{G}_1$;
	\item closed as a simplicial form: 
	$$
	  \delta \omega_1 = \partial_0^* \omega_1 - \partial_1^* \omega_1 + \partial_2^* \omega_1 = 0
	  \,,
	$$
	where $\partial_i : \mathcal{G}_2 \to \mathcal{G}_1$ are the face maps in the nerve of $\mathcal{G}$.
  \end{enumerate}
\end{definition}
\begin{example}
  \label{SymplecticPathGroupoidFromSymplecticManifold}
  Let $(X, \omega)$ be an ordinary symplectic manifold.
  Then its fundamental groupoid $\Pi_1(X)$ canonically is a 
  symplectic groupoid with $\omega_1 := \partial_1^* \omega - \partial_0^* \omega$.
\end{example}
\begin{proposition}
  Let $\mathfrak{P}$ be the symplectic Lie 1-algebroid (Poisson Lie algebroid),
  def. \ref{SymplecticLooAlgebroid}, induced by the Poisson manifold structure
  corresponding to $(X,\omega)$. Write
  $$
    \mathbf{\omega} : \mathfrak{T}\mathfrak{P} \to \mathfrak{T}b^3 \mathbb{R}
  $$
  for the canonical invariant polynomial. 
  
  Then the corresponding $\infty$-Chern-Weil homomorphism
  according to \ref{SmoothStrucInfChernWeil}
  $$
    \exp(\mathbf{\omega}) : 
	\exp(\mathfrak{P})_{\mathrm{diff}} 
	  \to \mathbf{B}^3_{\mathrm{dR}} \mathbb{R}
  $$
  exhibits the symplectic groupoid from example 
  \ref{SymplecticPathGroupoidFromSymplecticManifold}.
\end{proposition}
\proof
We start with the simple situation where $(X,\omega)$ has a global Darboux coordinate chart $\{x^i\}$. Write $\{\omega_{i j}\}$ for the components of the symplectic form in these coordinates, and $\{\omega^{i j}\}$ for the components of the inverse.

Then the Chevalley-Eilenberg algebra $\mathrm{CE}(\mathfrak{P})$ is generated from $\{x^i\}$ in degree 0 and $\{\partial_i\}$ in degree 1, 
with differential given by
$$
  d_{\mathrm{CE}} x^i = - \omega^{i j} \partial_j
$$
$$
  d_{\mathrm{CE}} \partial_i = \frac{\partial \pi^{j k}}{\partial x^i} \partial_j \wedge \partial_k = 0
  \,.
$$
The differential in the corresponding Weil algebra is hence
$$
  d_{\mathrm{W}} x^i = - \omega^{i j} \partial_j + \mathbf{d}x^i
$$
$$
  d_{\mathrm{W}} \partial_i = \mathbf{d} \partial_i
  \,.
$$
By prop. \ref{SymplecticPoissonCourant}, the symplectic invariant polynomial is
$$
  \mathbf{\omega} = \mathbf{d} x^i \wedge \mathbf{d} \partial_i \in W(\mathfrak{P})
  \,.
$$
Clearly it is useful to introduce a new basis of generators with
$$
  \partial^i := -\omega^{i j} \partial_j
  \,.
$$
In this new basis we have a manifest isomorphism
$$
  \mathrm{CE}(\mathfrak{P}) = \mathrm{CE}(\mathfrak{T}X)
$$
with the Chevalley-Eilenberg algebra of the tangent Lie algebroid of $X$. 

Therefore the Lie integration of $\mathfrak{P}$ is the fundamental groupoid of $X$, which, since we have assumed global Darboux oordinates and hence contractible $X$, is just the pair groupoid:
$$
  \tau_1 \exp(\mathfrak{P}) = \Pi_1(X) = 
   (
     \xymatrix{
	   X \times X 
         \ar@<+3pt>[r]
         \ar@<-3pt>[r]
		 &
	   X
	  }
	)
  \,.
$$
It remains to show that the symplectic form on $\mathfrak{P}$ makes this a symplectic groupoid.

Notice that in the new basis the invariant polynomial reads
$$
  \begin{aligned}
    \mathbf{\omega} 
      &= 
     - \omega_{i j} \mathbf{d}x^i \wedge \mathbf{d} \partial^j 
     \\
      & = \mathbf{d}( \omega_{i j} \partial^i \wedge \mathbf{d}x^j)
   \end{aligned}
   \,.
$$
The corresponding $\infty$-Chern-Weil homomorphism, 
\ref{SmoothStrucInfChernWeil}, that we need to compute is given by the $\infty$-anafunctor
$$
  \xymatrix{
    \exp(\mathfrak{P})_{\mathrm{diff}}
     \ar[r]^{\exp(\mathbf{\omega})} 
     \ar[d]^\simeq
	   &
     \exp(b^3 \mathbb{R})_{\mathrm{dR}}
	 \ar[r]^{\int_{\Delta^\bullet}}
	 &
    \mathbf{\flat}_{dR}\mathbf{B}^3 \mathbb{R} 
    \\
    \exp(\mathfrak{P})
  }
  \,.
$$
Over a test space $U \in \mathrm{CartSp}$ and in degree 1 an element in 
$\exp(\mathfrak{P})_{\mathrm{diff}}$ is a pair
$(X^i, \eta^i)$
$$
  X^i \in C^\infty(U \times \Delta^1)
$$
$$
  \eta^i \in \Omega^1_{\mathrm{vert}}(U \times \Delta^1)
$$
subject to the constraint that along $\Delta^1$ we have
$$
  d_{\Delta^1} X^i + \eta^i_{\Delta^1} = 0
  \,.
$$
The vertical morphism $\exp(\mathfrak{P})_{\mathrm{diff}} \to \exp(\mathfrak{P})$ has in fact a section whose image is given by those pairs for which $\eta^i$ has no leg along $U$. We therefore find the desired form on $\exp(\mathfrak{P})$ by evaluating the top morphism on pairs of this form. 

Such a pair is taken by the top morphism to
$$
  \begin{aligned}
    (X^i, \eta^j) 
     & \mapsto 
   \int_{\Delta^1}
    \omega_{i j} F_{X^i} \wedge F_{\partial^j}
   \\
    & =
  \int_{\Delta^1}
    \omega_{i j} (d_{dR} X^i + \eta^i) \wedge d_{\mathrm{dR}} \eta^j
  \in 
  \Omega^3(U)
  \end{aligned}
  \,.
$$
Using the above constraint and the condition that $\eta^i$ has no leg along $U$, this becomes
$$
  \cdots = 
  \int_{\Delta^1}
  \omega_{i j}
    d_U X^i 
  \wedge
    d_U d_{\Delta^1} X^j 
  \,.
$$
By the Stokes theorem the integration over $\Delta^1$ yields
$$
  \begin{aligned}
   \cdots 
     &=
   \omega_{i j} d_{\mathrm{dR}} X^i \wedge d_{\mathrm{dR}}X^j |_{0}
    -
   \omega_{i j} d_{\mathrm{dR}} X^i \wedge d_{\mathrm{dR}}X^j|_{1}
   \\
   & = 
   \partial_1^* \omega - \partial_0^* \omega
  \end{aligned}
  \,.
$$  
\endofproof

\newpage

\subsection{Higher prequantum geometry}
\label{PrequantizationApplications}
\index{prequantum geometry!Examples}

We discuss here the application 
of cohesive higher prequantum geometry, \ref{StrucGeometricPrequantization},
to the natural action functionals that we consider in 
\ref{ChernSimonsFunctional} and \ref{WZWApplications}.

This section draws from \cite{hgp}.

 Since in higher prequantum theory 
 local Lagrangians are ``fully de-transgressed'' to higher prequantum bundles, 
 conversely every example induces its corresponding transgressions. 
 In the following we always start with a higher extended Chern-Simons-type theory
 and consider then its first transgression.
 As in the discussion in \cite{FiorenzaSatiSchreiberCS} this first transgression 
 is the higher prequantum bundle of the topological sector 
 of a higher extended Wess-Zumino-Witten type theory. In this way our examples appear 
 at least in 
 pairs as shown in the following table: 

 \begin{center}
 \begin{tabular}{|c||c|c|}
   \hline
   & Higher CS-type theory & higher WZW-type theory 
   \\
   \hline \hline
   \ref{ExtendedWZW} & 3d $G$-Chern-Simons theory & 2d WZW-model on $G$
   \\
   \hline 
   \ref{HigherCSFromLieIntegration} & $\infty$-CS theory from $L_\infty$-integration & 
   \\
   \hline
   \ref{DeformationQuantizationInQuantumMechanics} & 2d Poisson Chern-Simons theory & 1d quantum mechanics
   \\
   \hline
   \ref{AnExtended6dSomething} &  7d $\mathrm{String}$-Chern-Simons theory & 6d theory related to M5-brane
   \\
   \hline
 \end{tabular}
 \end{center}
 
\subsubsection{Higher prequantum 2d WZW model and the smooth $\mathrm{string}$ 2-group}
 \label{ExtendedWZW}

In \ref{StrucGeometricPrequantization}
we remarked that an old motivation for what we call higher prequantum geometry here
is the desire to ``de-transgress'' the traditional construction of 
positive energy loop group representations
of simply connected compact Lie groups $G$ by, in our terminology, regarding the
canonical $\mathbf{B}U(1)$-2-bundle on $G$ (the ``WZW gerbe'') as a prequantum 2-bundle.
Here we discuss how prequantum 2-states for the WZW sigma-model provide
at least a partial answer to this question. Then we analyze the quantomorphism
2-group of this model.

\medskip

For $G$ a connected and simply connected compact Lie group such as $G = \mathrm{Spin}(n)$
for $n \geq 3$ or $G = \mathrm{SU}(n)$, the first nontrivial cohomology class of the classifying space $BG$
is in degree 4: $H^4(B G, \mathbb{Z}) \simeq \mathbb{Z}$. For $\mathrm{Spin}(n)$ the
generator here is known as the \emph{first fractional Pontryagin class} $\tfrac{1}{2}p_1$,
while for $\mathrm{SU}(n)$ it is the second Chern class $c_2$. In 
\cite{FSS} was constructed a smooth and differential lift of this class to the
$\infty$-topos $\mathrm{Smooth}\infty\mathrm{Grpd}$, 
namely a diagram of smooth 
higher moduli stacks of the form

\begin{tabular}{c|c|c}
  \xymatrix{
    \mathbf{B}\mathrm{Spin}_{\mathrm{conn}}
	\ar[rr]^{\tfrac{1}{2}\widehat{\mathbf{p}}_1}
	\ar[d]^{u_{\mathbf{B}\mathrm{Spin}}}
	&&
	\mathbf{B}^3 U(1)_{\mathrm{conn}}
	\ar[d]^{u_{\mathbf{B}^2 U(1)}}
    \\
    \mathbf{B}\mathrm{Spin}
	\ar[rr]^{\tfrac{1}{2}\mathbf{p}_1}
	\ar[d]^\int
	&&
	\mathbf{B}^3 U(1)
	\ar[d]^\int
    \\
    B \mathrm{Spin}
	\ar[rr]^{\tfrac{1}{2}p_1}
	&&
	K(\mathbb{Z},4)
  }
  & \;\;\; \xymatrix{ \nabla_{\mathrm{CS}} \\ \nabla^0_{\mathrm{CS}} \\  \int \nabla^0_{\mathrm{CS}}} \;\;\;&
  \xymatrix{
    \mathbf{B}\mathrm{SU}_{\mathrm{conn}}
	\ar[rr]^{\widehat{\mathbf{c}}_2}
	\ar[d]^{u_{\mathbf{B}\mathrm{SU}}}
	&&
	\mathbf{B}^3 U(1)_{\mathrm{conn}}
	\ar[d]^{u_{\mathbf{B}^2 U(1)}}
    \\
    \mathbf{B}\mathrm{SU}
	\ar[rr]^{\mathbf{c}_2}
	\ar[d]^\int
	&&
	\mathbf{B}^3 U(1)
	\ar[d]^\int
    \\
    B \mathrm{SU}
	\ar[rr]^{c_2}
	&&
	K(\mathbb{Z},4)
  }
\end{tabular}

Here  $\int$ is the geometric realization map,
and $u_{(-)}$ is the forgetful map from the higher moduli stacks of 
higher principal connections to that of higher principal bundles of def. \ref{DiffModuliStack}.

In \ref{3dCSTheories} we discuss that this 
3-connection on the smooth moduli stack of $G$-principal connections 
-- which for unspecified $G$ we now denote by $\nabla$ -- is the full 
de-transgression of the (off-shell) prequantum 1-bundle of $G$-Chern-Simons theory, hence is the
localized incarnation of 3d $G$-Chern-Simons theory in higher prequantum theory.
In particular it is a $\mathbf{B}^2 U(1)$-prequantization,
according to def. \ref{DiffModuliStack},
of the Killing form invariant polynomial
$\langle -,-\rangle$ of $G$, which is a differential 4-form 
(hence a \emph{pre-3-plectic form} in the sense of def. \ref{nPlectic}) 
on the moduli stack of fields:
\[
  \raisebox{20pt}{
  \xymatrix{
    && \mathbf{B}^3 U(1)_{\mathrm{conn}} \ar[d]^{F_{(-)}}
    \\
    \mathbf{B}G_{\mathrm{conn}}
	\ar[rr]_-{\langle F_{(-)}, F_{(-)}\rangle}
	\ar[urr]^{\nabla_{\mathrm{CS}}}
	&&
	\Omega^4_{\mathrm{cl}}
  }
  }
  \,.
\]
This 3-connection on the moduli stack of $G$-principal connections does 
not descend to the moduli stack $\mathbf{B}G$ of just $G$-principal bundles; 
it does however descend \cite{WaldorfCS} as a ``3-connection without 
top-degree forms'' as in def. \ref{PrincipalConnectionWithoutTopDegreeForms}:
\[
  \raisebox{20pt}{
  \xymatrix{
    && \mathbf{B}(\mathbf{B}^2 U(1)_{\mathrm{conn}})
	\ar[d]^{\mathbf{B}F_{(-)}}
    \\
    \mathbf{B}G
	\ar[rr]^{\mathbf{}}
	\ar[urr]^{\nabla^2_{\mathrm{CS}}}
	&&
	\mathbf{B}\Omega^3_{\mathrm{cl}}
	}
	}
	\,.
	\label{nabla1CS}
\]
Therefore over the universal moduli stack of Chern-Simons fields $\mathbf{B}G_{\mathrm{conn}}$
we canonically have a higher quantomorphism groupoid $\mathrm{At}(\nabla_{\mathrm{CS}})_\bullet$
as in \ref{QuantomorphismAndHeisenbergGroup}, while over the 
univeral moduli stack of just the ``instanton sectors'' of fields we have just a 
Courant 3-groupoid $\mathrm{At}(\nabla^2_{\mathrm{CS}})_\bullet$ as in \ref{HigherCourant}. 
This kind of phenomenon we re-encounter below in \ref{DeformationQuantizationInQuantumMechanics}.

By the above and
following \cite{CJMSW},
the transgression of $\nabla_{\mathrm{CS}}$ to maps out of the circle $S^1$
is found to be the ``WZW gerbe'', the canonical circle 2-bundle with connection 
$\nabla_{\mathrm{WZW}}$ on  the Lie group $G$ itself. We may obtain this
either as the fiber integration of $\nabla_{\mathrm{CS}}$ restricted 
along the inclusion of $G$ as the constant $\mathfrak{g}$-connection on the circle
\[
  \nabla_{\mathrm{WZW}}
  :
  \xymatrix{
    G
	\ar[r]
	&
	[S^1, \mathbf{B}G_{\mathrm{conn}}]
	\ar[rr]^-{[S^1, \nabla_{\mathrm{CS}}]}
	&&
	[S^1, \mathbf{B}^3 U(1)_{\mathrm{conn}}]
	\ar[rr]^-{\exp(2 \pi i \int_{S^1}(-))}
	&&
	\mathbf{B}^2 U(1)_{\mathrm{conn}}
  }
\]
or equivalently we obtain it as the looping of $\nabla_{\mathrm{CS}}^2$:
\[
  \nabla_{\mathrm{WZW}}
  :
  \xymatrix{
    G
	\simeq
	\Omega \mathbf{B}G
	\ar[rr]^-{\Omega \nabla_{\mathrm{CS}}^2}
	&&
	\Omega \mathbf{B}(\mathbf{B}^2 U(1)_{\mathrm{conn}})
	\simeq
	\mathbf{B}^2 U(1)_{\mathrm{conn}}
  }
  \,.
\]
This $\nabla_{\mathrm{WZW}}$ is the background gauge field of the 2d Wess-Zumino-Witten
sigma-model, the ``Kalb-Ramon B-field'' under which the string propagating on $G$ is charged.
We now regard this as the $\mathbf{B}U(1)$-prequantization
(def. \ref{DiffModuliStack}) 
of the canonical
3-form $\langle -,[-,-]\rangle$ on $G$ (a 2-plectic form):
$$
  \raisebox{20pt}{
  \xymatrix{
    && \mathbf{B}^2 U(1)_{\mathrm{conn}}
	\ar[d]^{F_{(-)}}
    \\
    G
	\ar[urr]^{\nabla_{\mathrm{WZW}}}
	\ar[rr]_-{\langle -,[-,-]\rangle}
	&&
	\Omega^3_{\mathrm{cl}}
  }
  }
  \,.
$$
By example \ref{BundleGerbeModulesAsSections} the prequantum 2-states
of the prequantum 2-bundle $\nabla_{\mathrm{WZW}}$ 
are  twisted unitary bundles with connection
(twisted K-cocycles, after stabilization): the \emph{Chan-Paton gauge fields}.
More explicitly,
with the notation as introduced there, a prequantum 2-state $\Psi$
of the WZW model supported over a D-brane submanifold $Q \hookrightarrow G$
is a map $\Psi : \nabla_{\mathrm{WZW}}|_Q \to \mathbf{dd}_{\mathrm{conn}}$
in the slice over $\mathbf{B}^2 U(1)_{\mathrm{conn}}$, hence a diagram of the form
\[
  \raisebox{20pt}{
  \xymatrix{
    Q
	\ar@{^{(}->}[d]
	\ar[rr]^{\Psi}
	&&
	\coprod_n (\mathbf{B}U(n)/\!/\mathbf{B}U(1))_{\mathrm{conn}}
	\ar[d]
	\\
	G
	\ar@/^1pc/[rr]^-{\nabla_{\mathrm{WZW}}}
	\ar[r]
	\ar[d]
	&
	[S^1, \mathbf{B}G_{\mathrm{conn}}]
	\ar[r]_-{\exp(\int_{S^1}[S^1, \nabla])}
	\ar[d]^{\mathrm{conc.}}
	&
	\mathbf{B}^2 U(1)_{\mathrm{conn}}
	\\
	G/\!/_{\!\mathrm{ad}}G
	\ar[r]^-\simeq
	&
	G \mathbf{Conn}(S)^1
  }
  }
  \,.
\]
Here we have added at the bottom the map to the differential concretification
of the transgressed moduli stack of fields, 
according to example \ref{TheModuliOfCircleConnections}.
As indicated, this exhibits $G$ as fibered over its homotopy quotient
by its adjoint action. The D-brane inclusion $Q \to G$ in the diagram is the homotopy fiber
over a full point of $G/\!/_{\mathrm{ad}}G$ precisely if it is a conjugacy class
of $G$, hence a ``symmetric D-brane'' for the WZW model. In summary this means that this single diagram exhibiting WZW prequantum-2-states
as slice maps encodes all the WZW D-brane data as discussed in the literature 
\cite{Gawedzki04}.
In particular, in \cite{FiorenzaSatiSchreiberCS} we showed that the transgression of these
prequantum 2-states $\Psi$ to prequantum 1-states over the loop group $L G$ 
naturally encodes the anomaly cancellation of the open bosonic string
in the presence of D-branes (the Kapustin-part of the Freed-Witten-Kapustin quantum anomaly cancellation).

\medskip

We may now study the quantomorphism 2-group
of $\nabla_{\mathrm{WZW}}$, def. \ref{HigherQuantomorphismGroupInIntroduction}, 
on these 2-states, hence, in the language of twisted cohomology, the 2-group
of twist automorphism.
First, one sees that by inspection that this is the action that integrates 
and globalizes the D-brane gauge transformations 
which are familiar from the string theory literature, 
where the local connection 1-form $A$ on the
twisted bundle is shifted and the local connection 2-form on the prequantum bundle transforms as
$$
  A \mapsto A + \lambda
  \,,
  \;\;\;\;
  B \mapsto B + d \lambda
  \,.
$$
In order to analyze the quantomorphism 2-group here in more detail,
notice that  since the 2-plectic form $\langle -,[-,-]\rangle \in \Omega^3_{\mathrm{cl}}(G)$ 
is a left invariant form (by definition),  
the left action of $G$ on itself is Hamiltonian, in the sense 
of def. \ref{HamiltonianActionMomentumMapAndHeisenbergGroup}, and so
we have the corresponding Heisenberg 2-group $\mathbf{Heis}(G,\nabla_{\mathrm{WZW}})$
of def. \ref{HamiltonianActionMomentumMapAndHeisenbergGroup} inside
the quantomorphism 2-group.
By theorem \ref{TheLongHomotopyFiberSequenceOfTheQuantomorphimsGroup}
this is a 2-group extension of $G$ of the form
$$
  \xymatrix{
    U(1)\mathbf{FlatConn}(G)
	\ar[r]
	&
	\mathbf{Heis}(\nabla_{\mathrm{WZW}})
	\ar[r]
	&
	G	
  }
  \,.
$$
Since $G$ is connected and simply connected, 
there is by prop. \ref{FlatConnectionsOnSimplyConnectedManifolds} an equivalence
of smooth 2-groups
$
  U(1)\mathbf{FlatConn}(G) \simeq \mathbf{B}U(1)
$
and so the WZW Heisenberg 2-group is in fact a smooth circle 2-group extension
$$
  \xymatrix{
    \mathbf{B}U(1)
	\ar[r]
	&
	\mathbf{Heis}(\nabla_{\mathrm{WZW}})
	\ar[r]
	&
	G	
  }
$$
classified by a cocycle $\mathbf{B}(\nabla_{\mathrm{WZW}} \circ (-)) :\mathbf{B}G \to \mathbf{B}^3 U(1)$.
If $G$ is compact and simply connected, 
then, by the discussion in \ref{SmoothStrucLieGroupCohomology}, 
$\pi_0 \mathbf{H}(\mathbf{B}G, \mathbf{B}^3 U(1)) \simeq H^4(B G, \mathbb{Z}) 
\simeq \mathbb{Z}$. This integer is the \emph{level}, the cocycle corresponding
to the generator $\pm 1$ is $\tfrac{1}{2}\mathbf{p}_1$ for $G = \mathrm{Spin}$
and $\mathbf{c}_2$ for $G = SU$. The corresponding extension is the 
String 2-group extension, def. \ref{SmoothBString}
$$
  \xymatrix{
    \mathbf{B}U(1)
	\ar[r]
	&
	\mathrm{String}_G
	\ar[r]
	&
	G	
  }
  \,.
$$
\index{string 2-group!as automorphisms of WZW term}
Accordingly, under Lie differentiation, one finds, 
that the
Heisenberg Lie 2-algebra extension
of theorem \ref{TheLInfinityExtension}
combined with def. \ref{HeisenbergLieExtension} is the 
\emph{string Lie 2-algebra} extension 
$$
  \xymatrix@R=5pt{
    \mathbf{B}\mathbb{R}
	\ar[r]
	&
	\mathfrak{Heis}_{\langle -,[-,-]\rangle}(\mathfrak{g})
	\ar[r]
	\ar@{}[d]|\simeq
	&
	\mathfrak{g}
	\\
	& \mathfrak{string}_{\mathfrak{g}}
  }
  \,.
$$
More in detail, using the results of \ref{SmoothStructKostantHeisenbergLieExtension}:
\begin{example}
  Let  $G$ be a {(connected)} compact simple Lie group, regarded as a 2-plectic manifold
  with its canonical 3-form $\omega := \langle -,[-,-]\rangle$
  as in example \ref{CompactSimpleLieGroup2Plectic}.  {The
    infinitesimal generators of the action of $G$ on
    itself by right translation are the left invariant vector fields $\gg$, which are 
    Hamiltonian.}
{We have $H^{1}_{\mathrm{dR}}(G) \cong
  H^{1}_{\mathrm{CE}}(\gg,\R)=0$, and therefore a weak equivalence:}
  $$
    \xymatrix{
	  \mathbf{B} \mathbf{H}(G, \flat\mathbf{B} \mathbb{R})
	  \ar[r]^-\simeq
	  &
	  \mathbb{R}[2]
	}
  $$
  {given by the evaluation at the identity element of $G$.}
  The resulting composite cocycle 
  $$
    \langle -,[-,-]\rangle
	:
    \xymatrix{
	  \mathfrak{g}
	  \ar[r]^-\rho
	  &
	  \mathfrak{X}_{\mathrm{Ham}}(X)
	  \ar[r]^-{\omega_{[\bullet]}}
	  &
	  \mathbb{R}[2]
	}
  $$
  is exactly the 3-cocycle which classifies the String Lie-2-algebra,
  namely just $\langle -,[-,-]\rangle$ regarded as a Lie algebra 3-cocycle.
  The String Lie 2-algebra, def. \ref{StringLie2Algebra}, is the homotopy fiber 
  of this cocycle, in that we have a homotopy pullback square of $L_\infty$-algebras
  $$
    \raisebox{20pt}{
    \xymatrix{
	  \mathfrak{string}_{\mathfrak{g}}
	  \ar[r]
	  \ar[d]
	  &
	  0
	  \ar[d]
	  \\
	  \mathfrak{g}
	  \ar[r]^-{\langle -,[-,-]\rangle}
	  &
	  \mathbb{R}[2]
	}}
	\,.
  $$
  Hence the String Lie 2-algebra is the Heisenberg Lie 2-algebra
  of the 2-plectic manifold $(G,\langle -,[-,-]\rangle)$ with its canonical
  $\mathfrak{g}$-action $\rho$:
  $$
    \mathfrak{heis}_\rho(\mathfrak{g}) \simeq \mathfrak{string}_{\mathfrak{g}}
	\,.
  $$
  \label{StringAsHeisenberg}
  The relationship between $\mathfrak{string}_{\mathfrak{g}}$ and $L_{\infty}(G,\omega)$
  was first explored in \cite{RogersString}.
\end{example}
\begin{remark}
  \index{Green-Schwarz mechanism!by parameterized WZW models}
  \index{Wess-Zumino-Witten functionals!parameterized}
  This result means that given a $G$-principal bundle 
  $$
    \raisebox{20pt}{
    \xymatrix{
      G \ar[r] & P \ar[d]
      \\
      & X \ar[r]^-g & \mathbf{B}G
    }}
    \,,
  $$
  then a lift of the modulating map $g$ through the String 2-group extension
  is precisely the structure needed to construct a circle 2-connection
  $\nabla_{\mathrm{glob}}$ on the total space $P$ such that it restricts on
  each fiber to the WZW-2-conenction
  $$
    \raisebox{20pt}{
    \xymatrix{
      G \ar[r] 
      \ar@/^1pc/[rrr]^{\nabla_{\mathrm{WZW}}}
      & P \ar[d] \ar[rr]|-{\nabla_{\mathrm{glob}}} && \mathbf{B}^2 U(1)_{\mathrm{conn}}
      \\
      & X \ar[r] \ar@/^1pc/[rr]^-g & \mathbf{B}\mathrm{String}_G \ar[r] & \mathbf{B}G
    }}
    \,.
  $$
  At the level of the induced action functionals, essentially this was observed
  \cite{DistlerSharpe}.
  If $G = \mathrm{Spin} \times (E_8\times E_8)$ or similar, and if $g$ is modulates the tangent bundle
  of $X$ 
  and a gauge bundle,
  then the obstruction to such a lift is, by \ref{FractionalClasses}, 
  the combination of $\tfrac{1}{2}p_1$
  and $c_2$, by the discussion in \ref{FractionalClasses}. 
  One may interpret the bundle of WZW models on $P$ 
  as the internal degrees of freedom of a heteric string on spacetime $X$
  and recovers a (another) geometric interpretation of the 
  Green-Schwarz anomaly, \ref{FractionalClasses}.
\end{remark}

\subsubsection{Higher prequantum $n$d Chern-Simons-type theories and $L_\infty$-algebra cohomology}
 \label{HigherCSFromLieIntegration}
 
 The construction of the higher prequantum bundle $\nabla_{\mathrm{CS}}$ 
 for Chern-Simons field theory in 
 \ref{ExtendedWZW} above follows a general procedure 
 -- which might be called \emph{differential Lie integration of $L_\infty$-cocycles} - 
 that produces
 a whole class of examples of natural higher prequantum geometries: 
 namely those \emph{extended higher Chern-Simons-type field theories} 
 which are encoded
 by an $L_\infty$-invariant polynomial on an $L_\infty$-algebra, 
 in generalization of how
 ordinary $G$-Chern-Simons theory for a simply connected simple Lie group $G$
 is all encoded by the Killing form invariant polynomial
 (and as opposed to for instance to the cup product higher $U(1)$-Chern-Simons theories.
 Since also the following two examples in  \ref{DeformationQuantizationInQuantumMechanics}
 and \ref{AnExtended6dSomething} are naturally obtained this way, we here briefly
 recall this construction, due to \cite{FSS}, with an eye towards its interpretation in higher prequantum
 geometry.
 
 \medskip
 
 Given an $L_\infty$-algebra $\mathfrak{g} \in L_\infty$, there is a natural notion 
 of sheaves of (flat) $\mathfrak{g}$-valued smooth differential forms
 $$
   \Omega_{\mathrm{flat}}(-,\mathfrak{g}) \hookrightarrow \Omega(-,\mathfrak{g})
   \in \mathrm{Sh}(\mathrm{SmthMfd})
   \,,
 $$
 and this is functorial in $\mathfrak{g}$ (for the correct (``weak'') homomorphisms
 of $L_\infty$-algebras). Therefore there is a functor -- 
 denoted $\exp(-)$ in \cite{FSS} -- which assigns to an $L_\infty$-algebra
 $\mathfrak{g}$ the presheaf of Kan complexes that over a test manifold $U$ 
 has as set of $k$-cells the set of those smoothly $U$-parameterized collections of 
 flat $\mathfrak{g}$-valued differential forms on the $k$-simplex $\Delta^k$
 which are sufficiently well behaved towards the boundary of the simplex 
 (have ``sitting instants'').
 Under the presentation 
 $L_{\mathrm{lhe}}[\mathrm{SmoothMfd}^{\mathrm{op}}] \simeq \mathrm{Smooth}\infty\mathrm{Grpd}$
 of the $\infty$-topos of smooth $\infty$-groupoids
 this yields a Lie integration construction from $L_\infty$-algebras to smooth
 $\infty$-groupoids.
 (So far this is the fairly immediate stacky and smooth refinement of a standard
 construction in rational homotopy theory and deformation theory, see the
 references in \cite{FSS} for a list of predecessors of this construction.)
 
 In higher analogy to ordinary Lie integration, 
 one finds that $\exp(\mathfrak{g})$ is the ``geometrically $\infty$-connected'' Lie integration
 of $\mathfrak{g}$: the geometric realization $\int \exp(\mathfrak{g})$, of 
 $\exp(\mathfrak{g}) \in L_{\mathrm{lhe}}[\mathrm{SmoothMfd}^{\mathrm{op}}, \mathrm{KanCplx}]
 \simeq \mathrm{Smooth}\infty\mathrm{Grpd}$ is always contractible. 
 For instance for $\mathfrak{g} = \mathbb{R}[-n+1] = \mathbf{B}^{n-1}\mathbb{R}$ the abelian $L_\infty$-algebra
 concentrated on $\mathbb{R}$ in the $n$th degree, we have
 $$
   \exp(\mathbb{R}[-n+1]) \simeq \mathbf{B}^n \mathbb{R}
   \in 
   \mathrm{Smooth}\infty\mathrm{Grpd}
 $$
 and it follows that
 $\int \mathbf{B}^n \mathbb{R}\simeq B^n \mathbb{R} \simeq {*}$.
 Geometrically non-$\infty$-connected Lie integrations of $\mathfrak{g}$
 arise notably as truncations of the $\infty$-stack $\exp(\mathfrak{g})$,
 \ref{PostnikovDecomposition}. For
 instance for $\mathfrak{g}_1$ an ordinary Lie algebra, then the 1-truncation of the
 $\infty$-stack $\exp(\mathfrak{g}_1)$ to a stack of 1-groupoids reproduces (the internal delooping of)
 the simply connected Lie group $G$ corresponding to $\mathfrak{g}$ by ordinary Lie theory:
 $$
   \tau_1 \exp(\mathfrak{g}_1) \simeq \mathbf{B}G
   \;\;\;
   \in \mathrm{Smooth}\infty\mathrm{Grpd}
   \,.
 $$
 Similarly for $\mathfrak{string} \in L_\infty \mathrm{Alg}$ the string Lie 2-algebra,
 def. \ref{SkeletalStringLie2Algebra},
 the 2-truncation of its universal Lie integration to a stack of 2-groupoids
 reproduces the moduli stack of String-principal 2-bundles:
 $$
   \tau_2 \exp(\mathfrak{string}) \simeq \mathbf{B}\mathrm{String}
   \;\;\;
   \in \mathrm{Smooth}\infty\mathrm{Grpd}
   \,.
 $$
 Now the simple observation that yields the analgous Lie integration of 
 $L_\infty$-cocycles
 is that a degree-$n$ $L_\infty$-cocycle $\mu$
 on an $L_\infty$-algebra $\mathfrak{g}$ is equivalently a map of $L_\infty$-algebras
 of the form
 $$
   \mu : \mathbf{B}\mathfrak{g} \to \mathbf{B}^n \mathbb{R}
   \,;
 $$
 and since $\exp(-)$ is a functor, every such cocycle immediately integrates
 to a morphism 
 $$
   \exp(\mu) : \exp(\mathfrak{g}) \to \mathbf{B}^n \mathbb{R}
 $$
 in $\mathrm{Smooth}\infty\mathrm{Grpd}$, hence 
 to a universal cocycle on the smooth moduli $\infty$-stack $\exp(\mathfrak{g})$.
 Moreover, this cocycle descends to the $n$-truncation of its domain
 as a $\mathbb{R}/\Gamma$ cocycle on the resulting moduli $n$-stack
 $$
   \exp(\mu) : \tau_n \exp(\mathfrak{g}) \to \mathbf{B}^n (\mathbb{R}/\Gamma)
   \,,	
 $$
 where $\Gamma \hookrightarrow \mathbb{R}$ is the period lattice of the cocycle
 $\mu$. 
 
 For instance for 
 $$
   \langle -,[-,-]\rangle : \mathbf{B}\mathfrak{g}_1 \to \mathbf{B}^3 \mathbb{R}
 $$
 the canonical 3-cocycle on a semisimple Lie algebra (where $\langle -,-\rangle$
 is the Killing form invariant polynomial as before), its period group is
 $\pi_3(G) \simeq \mathbb{Z}$ of the simply connected Lie group $G$ integrating $\mathfrak{g}_1$,
 and hence the Lie integration of the 3-cocycle yields a map of smooth $\infty$-stacks
 of the form
 $$
   \exp(\langle -,[-,-]\rangle) 
     :  
   \xymatrix{
     \mathbf{B}G 
	 \ar[r]^-\simeq
	 &
	 \tau_3\exp(\mathfrak{g}_1)
	 &
	 \mathbf{B}^3 (\mathbb{R}/\mathbb{Z})
	 =
	 \mathbf{B}^3 U(1)
   }
   \,,
 $$
 where we use that for the connected and simply connected Lie group $G$ not only the 
 1-truncation but also still the 3-truncation of 
 $\exp(\mathfrak{g}_1)$ gives the delooping stack: 
 $\tau_3 \exp(\mathfrak{g}_1)\simeq \tau_2 \exp(\mathfrak{g}_1)\simeq \tau_1 \exp(\mathfrak{g}_1)
 \simeq \mathbf{B}G$.
 
 Indeed, this is what yields the refinement of the generator 
 $c : B G \to K(\mathbb{Z},4)$ to smooth cohomology, which we used above in \ref{ExtendedWZW},
 for instance for $\mathfrak{g}_1 = \mathfrak{so}$ the Lie algebra of the Spin group, 
 the Lie integration of its canonical Lie 3-cocycle
 $$
   \exp(\langle -,[-,-]\rangle_{\mathfrak{so}}) \simeq \tfrac{1}{2}\mathbf{p}_1
 $$
 yields the smooth refinement of the first fractional universal Pontryagin class.
 
 This is shown in \cite{FSS} by further refining the $\exp(-)$-construction
 to one that yields not just moduli $\infty$-stacks of $G$-principal $\infty$-bundles, 
 but yields their differential refinements. The key 
 to this construction is the observation that an invariant polynomial 
 $\langle -, \cdots, -\rangle$ on 
 a Lie algebra and more generally on an $L_\infty$-algebra $\mathfrak{g}$ yields
 a \emph{globally} defined (hence invariant) differential form on the
 moduli $\infty$-stack $\mathbf{B}G_{\mathrm{conn}}$:
 $$
   \langle F_{(-)}, \cdots , F_{(-)}\rangle
   : 
   \mathbf{B}G_{\mathrm{conn}} \to \Omega^{n+1}_{\mathrm{cl}}
   \,.
 $$
 In components this is simply given, as the notation is supposed to indicate,
 by sending a $G$-principal connection $\nabla$ first to its $\mathfrak{g}$-valued
 curvature form $F_{\nabla}$ and then evaluating that in the invariant polynomial.
 In fact this property is part of the \emph{definition} of $\mathbf{B}G_{\mathrm{conn}}$
 for the non-braided $\infty$-groups $G$. 
 This we think of as a higher analog of Chern-Weil theory in higher differential
 geometry. We may also usefully think of 
 the invariant polynomial $\langle F_{(-)},\cdots, F_{(-)}\rangle$ as 
 being a pre $n$-plectic form on the moduli stack $\mathbf{B}G_{\mathrm{conn}}$,
 in evident generalization of the terminiology for smooth manifolds 
 in def. \ref{nPlectic}.
 
 Using this, there is a differential refinement $\exp(-)_{\mathrm{conn}}$
 of the $\exp(-)$-construction, which lifts this pre-$n$-plectic form
 to differential cohomology and hence provides its pre-quantization, according to
 def. \ref{DiffModuliStack}:
 $$
   \raisebox{20pt}{
   \xymatrix{
      && \mathbf{B}^n (\mathbb{R}/\Gamma)_{\mathrm{conn}}
	   \ar[d]^{F_{(-)}}
      \\
      \tau_n \exp(\mathfrak{g})_{\mathrm{conn}}
	  \ar[rr]^-{\langle F_{(-)}\wedge \cdots F_{(-)}\rangle}
	  \ar[urr]^-{\exp(\mu)_{\mathrm{conn}}}
	  &&
	  \Omega^{n+1}_{\mathrm{cl}}
   }
   }
   \,.
 $$
 Here the higher stack $\exp(-)_{\mathrm{conn}}$ assigns to a test manifold
 $U$ smoothly $U$-parameterized collections
 of simplicial $L_\infty$-Ehresmann connections: 
 the $k$-cells of $\exp(\mathfrak{g})_{\mathrm{conn}}$ are $\mathfrak{g}$-valued differential 
 forms $A$ on $U \times \Delta^k$ (now not necessarily flat) satisfying an $L_\infty$-analog of the
 two conditions on a traditional Ehresmann connection 1-form: restricted to the fiber
 (hence the simplex) the $L_\infty$-form datum becomes flat, and 
 moreover the curvature
 invariants $\langle F_A \wedge \cdots \wedge F_A\rangle$ 
 obtained by evaluating the $L_\infty$-curvature forms in all 
 $L_\infty$-invariant polynomials descends down the simplex bundle $U \times \Delta^k \to U$.
 
 For example the differential refinement of 
 the prequantum 3-bundle of 3d $G$-Chern-Simons theory
 $\tfrac{1}{2}\mathbf{p}_1 \simeq \tau_3 \exp(\langle -,[-,-]\rangle)$ 
 obtained this way is the universal Chern-Simons 3-connection
 $$
   \exp(\langle -,[-,-]\rangle_{\mathfrak{so}})_{\mathrm{conn}}
   \simeq
   \tfrac{1}{2}\hat{\mathbf{p}}_1
   : 
   \mathbf{B}\mathrm{Spin}_{\mathrm{conn}}
   \to 
   \mathbf{B}^3 U(1)_{\mathrm{conn}}
 $$
 whose transgression to codimension 0 is the standard Chern-Simons action
 functional, as discussed above in \ref{ExtendedWZW}.
 Analgously, the differential Lie integration of the next cocycle, the 
 canonical 7-cocycle, but now regarded as a cocycle on $\mathfrak{string}$,
 yields a prequantum 7-bundle on the moduli stack of $\mathrm{String}$-principal
 2-connections:
 $$
   \exp(\langle -,[-,-], [-,-], [-,-]\rangle_{\mathfrak{so}})_{\mathrm{conn}}
   \simeq
   \tfrac{1}{6}\hat{\mathbf{p}}_2
   \;:\; 
   \mathbf{B}\mathrm{String}_{\mathrm{conn}}
   \to 
   \mathbf{B}^7 U(1)_{\mathrm{conn}}
   \,.
 $$
 This defines a 7-dimensional nonabelian Chern-Simons theory, which 
 we come to below in \ref{AnExtended6dSomething}.
 
 In conclusion this means that $L_\infty$-algebra cohomology is a direct source
 of higher smooth $(\mathbf{B}^{n-1}(\mathbb{R}/\Gamma))$-prequantum geometries 
 on higher differential moduli stacks.
 For $\mu$ any degree-$n$ $L_\infty$-cocycle 
 on an $L_\infty$-algebra $\mathfrak{g}$, differential Lie integration yields
 the higher prequantum bundle
 $$
   \exp(\mu)_{\mathrm{conn}}
   :
   \tau_n \exp(\mathfrak{g})_{\mathrm{conn}} \to \mathbf{B}^n (\mathbb{R}/\Gamma)
   \,.
 $$
 Moreover, these are by construction higher prequantum bundles for higher
 Chern-Simons-type higher gauge theories in that their transgression to 
 codimension 0
 $$
   \exp\left( \int_{\Sigma_n} [\Sigma_n, \exp(\mu)_{\mathrm{conn}}] \right)
   :
   \xymatrix{
     [\Sigma_n, \tau_n \exp(\mathfrak{g})_{\mathrm{conn}}]
	 \ar[r]
	 &
	 \mathbb{R}/\Gamma
   }
 $$
 is an action functional on the stack of $\mathfrak{g}$-gauge fields $A$ on 
 a given closed oriented manifold $\Sigma_n$ which is locally given by the 
 integral of a Chern-Simons $(n-1)$-form $\mathrm{CS}_\mu(A)$
 (with respect to the corresponding $L_\infty$-invariant polynomial)
 and globally given by a higher-gauge consistent globalization of such integrals.
 
 
 All of this discussion generalizes verbatim from $L_\infty$-algebras to \emph{$L_\infty$-algebroids}, too. 
 In \cite{frs} it was observed that therefore all the perturbative
 field theories known as \emph{AKSZ sigma-models} have a Lie integration 
 to what here we call higher prequantum bundles for higher Chern-Simons type field theories:
 these are precisely the cases as above where $\mu$ transgresses to a
 binary invariant polynomial $\langle-,- \rangle$ on the $L_\infty$-algebroid which non-degenerate.
 In the next section \ref{DeformationQuantizationInQuantumMechanics}
 we consider one low-dimensional example in this family and observe that 
 its higher geometric prequantum and quantum theory has 
 secretly been studied in some detail already -- but in 1-geometric disguise.
 
 For higher Chern-Simons action functionals $\exp(\mu)_{\mathrm{conn}}$
 as above, one finds that their variational differential at a field configuration
 $A$ given by globally defined differential form data is proportional to
 $$
   \delta \exp\left( \int_{\Sigma_{n-1}} [\Sigma_n, \exp(\mu)_{\mathrm{conn}}] \right)
   \;\propto\;
   \int_{\Sigma}\langle F_A \wedge \dots F_A \wedge \delta A \rangle
   \,.
 $$
 Therefore the Euler-Lagrange equations of motion of the corresponding $n$-dimensional
 Chern-Simons theory assert that 
 $$
   \langle F_A \wedge \cdots F_A , -\rangle = 0
   \,.
 $$
 (Notice that in general $F_A$ is an inhomogenous differential form, so that 
this equation in general consists of several independent components.) 
In particular, if the invariant polynomial is binary, hence of the form
$\langle -,-\rangle$, and furthermore non-degenerate (this is precisely the case
in which the general $\infty$-Chern-Simons theory reproduces the AKSZ $\sigma$-models), 
then the above equations of motion reduce to
$$
  F_A = 0
$$
and hence assert that the critical/on-shell field configurations are precisely
those $L_\infty$-algbroid valued connections which are flat.

In this case the higher moduli stack $\tau_n \exp(\mathfrak{g})$, which in general
is the moduli stack of instanton/charge-sectors underlying the 
topologically nontrivial $\mathfrak{g}$-connections, acquires also a different
interpretation. By the above discussion, its $(n-1)$-cells are equivalently
flat $\mathfrak{g}$-valued connections on the $(n-1)$-disks and its $n$-cells
implement gauge equivalences between such data. But since the equations of motion
$F_A = 0$ are first order differential equations, flat connections on 
$D^{n-1}$ bijectively correspond to critical field configuration on the cylinder
$D^{n-1} \times [-T,T]$. Therefore the collection of $(n-1)$-cells of 
$\tau_n \exp(\mathfrak{g})$ is the higher/extended \emph{covariant phase space} for
``open genus-0 $(n-1)$-branes'' in the model. Moreover, the $n$-cells between
these $(n-1)$-cells
implement the gauge transformations on such initial value data and hence 
$\tau_n \exp(\mathfrak{g})$ is, in codimension 1, 
the higher/extended \emph{reduced phase space}
of the model in codimension 1. 
For $n = 2$ this perspective was amplified in \cite{CattaneoFelder},
we turn to this special case below in \ref{DeformationQuantizationInQuantumMechanics}).

As an example, from this perspective the construction of the WZW-gerbe 
by looping as discussed above in \ref{HigherCSFromLieIntegration}
is equivalently the construction of the on-shell prequantum 2-bundle in
codimension 2 for ``Dirichlet boundary conditions'' for the open Chern-Simons 
membrane. Namely $\mathbf{B}G$ is now the extended reduced phase space,
and so the extended phase space of membranes stretching between the unique point is 
the homotopy fiber product of the two point inclusions
$\xymatrix{Q_0 \ar[r] & \mathbf{B}G \ar@{<-}[r] & Q_1}$, with $Q_0, Q_1 = \ast$, 
hence is 
$\Omega \mathbf{B}G \simeq G$. Since the on-shell prequantum 2-bundle 
$\nabla_{\mathrm{CS}}^1$ trivializes over these inclusions, as exhibited by
diagrams
$$
  \raisebox{20pt}{
  \xymatrix{
    Q_i
	\ar[r]
	\ar[d]
	&
	\ast
	\ar[d]
	\\
	\mathbf{B}G \ar[r]^-{\nabla_{\mathrm{CS}^2}} & \mathbf{B}^3 U(1)_{\mathrm{conn}^2}
  }
  }
  \,,
$$
the on-shell prequantum 3-bundle $\nabla_{\mathrm{CS}}^2$ extends to a diagram
of relative cocylces of the form
$$
  \raisebox{20pt}{
  \xymatrix{
    Q_0 \ar[r] \ar[d] & \ast \ar[d]
	\\
	\mathbf{B}G \ar[r]^-{\nabla_{\mathrm{CS}}^2} & \mathbf{B}(\mathbf{B}^3 U(1)_{\mathrm{conn}})
	\\
	Q_1 \ar[r] \ar[u] & \ast \ar[u]
  }
  }
  \,,
$$
hence, under forming homotopy fiber products, to the WZW-2-connection $\Omega \nabla^2_{\mathrm{CS}} : G \to \mathbf{B}^2 U(1)$ on the extended phase space $G$.

In the next section we see another example of this phenomenon.

\subsubsection{Higher prequantum 2d Poisson-Chern-Simons theory and quantum mechanics}
 \label{DeformationQuantizationInQuantumMechanics}
 \index{holography!1d mechanics/2d Poisson-CS} 
 
 We consider here the boundary prequantum theory of the 
 non-perturbative 2d Poisson-Chern-Simons theory
 and indicate how its quantization yields the quantization of 
 the corresponding Poisson manifold, regarded as a boundary condition.
 More on this quantization is below in \ref{HolographicGeometricQuantizationOfPoisson}.
 
 \medskip
 
A non-degenerate and binary invariant polynomial which induces a
pre-2-plectic structure 
on the moduli stack of a higher Chern-Simons type theory

$$
  \omega :=
  \langle F_{(-)} F_{(-)}\rangle
  :
  \xymatrix{
    \tau_2 \exp(\mathfrak{P})_{\mathrm{conn}}
    \to
    \Omega^3_{\mathrm{cl}}
  }
$$
exists precisely on Poisson Lie algebroids $\mathfrak{P}$, induced
from Poisson manifolds $(X,\pi)$.
The differential Lie integration method described above yields a 
$(\mathbf{B}(\mathbb{R}/\Gamma))$-prequantization
$$
  \raisebox{20pt}{
  \xymatrix{
	&
	\mathbf{B}^2 (\mathbb{R}/\Gamma)_{\mathrm{conn}}
	\ar[d]^{F_{(-)}}
    \\
    \tau_1 \exp(\mathfrak{P})_{\mathrm{conn}}
	\ar[ur]^{\nabla_P}
	\ar[r]_{\mathbf{\omega}}
	&
	\Omega^3_{\mathrm{cl}}
  }
  }
  \,.
$$
The action functional of this higher prequantum field theory over a
closed oriented 2-dimensional smooth manifold $\Sigma_2$ is, again 
by \cite{FiorenzaSatiSchreiberIV, frs}, the transgression
of the higher prequantum bundle to codimension 0
$$
  \exp\left(
    \int_{\Sigma_2} [\Sigma_2, \nabla_P]
  \right)
  :
  \xymatrix{
    [\Sigma_2, \tau_1\exp(\mathfrak{P})_{\mathrm{conn}}]
	\ar[r]
	&
	\mathbb{R}/\Gamma
  }
  \,.
$$
We observe now that two complementary sectors of this higher prequantum 2d 
Poisson Chern-Simons field theory
$\nabla_P$ lead a separate life of their own in the literature:
on the one hand the sector where the bundle structures and hence the nontrivial ``instanton sectors'' 
  of the field configurations are ignored and only the globally defined 
  connection differential 
  form data is retained;
   and on the other hand the complementary sector where only these bundle structures/
   instanton sectors are considered and the connection data is ignored:
\begin{enumerate}
\item
The restriction of the action functional $\exp(\int_{\Sigma_2}[\Sigma_2, \nabla_P])$ 
to the linearized theory -- hence 
along the canonical inclusion 
$\Omega(\Sigma, \mathfrak{P}) \hookrightarrow [\Sigma_2, \exp(\mathfrak{P})_{\mathrm{conn}}]$
of globally defined $\mathfrak{P}$-valued forms into all $\exp(\mathfrak{P})$-principal connections
--
is the action functional of the \emph{Poisson sigma-model}. 
\item
  The restriction of the moduli stack of fields 
  $\tau_1\exp(\mathfrak{P})_{\mathrm{conn}}$ 
  to just $\tau_1\exp(\mathfrak{P})$ obtained by forgetting the differential refinement
  (the connection data) und just remembering the underlying $\exp(\mathfrak{P})$-principal 
  bundles, yields what is known as the \emph{symplectic groupoid} of $\mathfrak{P}$.

  Precisely: while the prequantum 2-bundle $\nabla_P$
  does not descend along the forgetful map 
  $\tau_1\exp(\mathfrak{P})_{\mathrm{conN}} \to \tau_1\exp(\mathfrak{P})$
  from moduli of $\tau_1\exp(\mathfrak{P})$-principal connections to their
  underlying $\tau_1(\exp(\mathfrak{P}))$-principal bundles,
  its version $\nabla_P^1$ ``without curving'', 
  given by def. \ref{PrincipalConnectionWithoutTopDegreeForms},
  does descend (this is as for 3d Chern-Simons theory discussed above in 
  \ref{ExtendedWZW}) and so does hence its curvature $\omega^1$, which
  has  coefficients in $\mathbf{B} \Omega^2_{\mathrm{cl}}$
  instead of $\Omega^3_{\mathrm{cl}}$:
  $$
    \raisebox{20pt}{
    \xymatrix{
	  & 
	  \mathbf{B}\left( \mathbf{B}(\mathbb{R}/\Gamma)_{\mathrm{conn}} \right)
	  \ar[d]^{\mathbf{B}F_{(-)}}
	  \\
	  \tau_1 \exp(\mathfrak{P})
	  \ar[ur]^{\nabla_P^1}
	  \ar[r]^{\omega^1}
	  &
	  \mathbf{B} \Omega^2_{\mathrm{cl}}
	}
	}
	\,.
  $$
  If here the smooth groupoid $\tau_1\exp(\mathfrak{P}) \in \mathrm{Smooth}\infty\mathrm{Grpd}$
  happens to have a presentation by a Lie groupoid 
  under the canonical inclusion of Lie groupoids into smooth $\infty$-groupoids
  (this is an integrability condition
  on $\mathfrak{P}$) then equipped with the de Rham hypercohomology 3-cocycle $\omega^1$
  it is called in the literature a \emph{pre-quasi-symplectic groupoid} \cite{LGXu}.
  If moreover the de Rham hypercohomology 3-cocycle $\omega^1$ --
  which in general is given by 3-form data and 2-form data on a {\v C}ech simplicial presheaf
  that resolves $\tau_1\exp(\mathfrak{P})$  --
  happens to be represented by
  just a globally defined 2-form on the manifold of morphisms of the Lie groupoid
  (which is then necessarily closed and ``multiplicative''),
  then this local data is called a (pre-)\emph{symplectic groupoid},
  see \cite{Eli} for a review and further pointers to the literature. 
\end{enumerate}

So in the case that the descended (pre-)2-plectic form
$\omega^1 : \tau_1 \exp(\mathfrak{P}) \to \mathbf{B}\Omega^2_{\mathrm{cl}}$ 
of the higher prequantum 2d Poisson Chern-Simons theory
is represented by a multiplicative symplectic 2-form on the manifold of morphisms
of the Lie groupoid $\tau_1 \exp(\mathfrak{P}) $, then one is faced with a situation
that looks like ordinary symplectic geometry subject to a 
kind of equivariance condition. This is the perspective from which 
symplectic groupoids were originally introduced and from which they are mostly studied
in the literature (with the exception at least of \cite{LGXu}, where the 
higher geometric nature of the setup is made explict): as a means to 
re-code Poisson geometry in terms of ordinary symplectic geometry.
The goal of finding a sensible geometric quantization of symplectic groupoids
(and hence in some sense of Poisson manifolds, this we come back to below)
was finally achieved in \cite{Eli}. 

\medskip

In order to further understand the conceptual role of the prequantum 
2-bundle $\nabla^1_{\mathfrak{P}}$,
notice that by the discussion in \ref{HigherCSFromLieIntegration}, following 
\cite{CattaneoFelder}, we may think of 
the symplectic groupoid $\tau_1 \exp(\mathfrak{P})$ as the 
extended reduced phase space of the open string Poisson-Chern-Simons theory.
More precisely, if $\mathfrak{C}_1, \mathfrak{C}_1 \hookrightarrow \mathfrak{P}$
are two sub-Lie algebroids, then the homotopy fiber product 
$\mathbf{Phase}_{\mathfrak{C}_0, \mathfrak{C}_1}$ in 
$$
  \xymatrix@C=0pt{
    & \mathbf{Phase}_{\mathfrak{C}_0, \mathfrak{C}_1}
	\ar[dl]\ar[dr]
	\\
	\tau_1 \exp(\mathfrak{C}_0) \ar[dr] && \tau_1 \exp(\mathfrak{C}_1) \ar[dl]
	\\
	&
	\tau_1 \exp(\mathfrak{P})
  }
$$
should be the ordinary reduced phase space of open strings that stretch
between $\mathfrak{C}_0$ and $\mathfrak{C}_1$, regarded as D-branes. 
Unwinding the definitions shows that this
is precisely what is shown in \cite{CattaneoFelderCoisotropic}: for
$\mathfrak{C}_0, \mathfrak{C}_1 \hookrightarrow \mathfrak{P}$ two Lagrangian
sub-Lie algebroids (hence over coisotropic submanifolds of $X$) 
the homotopy fiber product stack $\mathbf{Phase}_{\mathfrak{C}_0, \mathfrak{C}_1}$
is the symplectic reduction of the open $\mathfrak{C}_0$-$\mathfrak{C}_1$-string phase space.

Notice that the condition that $\mathfrak{C}_i \hookrightarrow \mathfrak{P}$ 
be Lagrangian sub-Lie algebroids means that 
restricted to them the prequantum 2-bundle becomes flat, hence that we have 
commuting squares
$$
  \raisebox{20pt}{
  \xymatrix{
    \tau_1 \exp(\mathfrak{C}_i)
	\ar[r]
	\ar[d]
	&
	\flat \mathbf{B}^2 (\mathbb{R}/\Gamma)
	\ar[d]
	\\
	\tau_1 \exp(\mathfrak{P})
	\ar[r]^-{\nabla_P^1}
	&
	\mathbf{B}(\mathbf{B}(\mathbb{R}/\Gamma)_{\mathrm{conn}})
  }
  }\,.
$$
If the inclusions are even such $\nabla_P^1$ entirely trivializes
over them, hence that we have diagrams
$$
  \raisebox{20pt}{
  \xymatrix{
    \tau_1 \exp(\mathfrak{C}_i)
	\ar[r]^{\nabla_{\mathfrak{C}_i}}
	\ar[d]
	&
	\ast
	\ar[d]
	\\
	\tau_1 \exp(\mathfrak{P})
	\ar[r]^-{\nabla_P^1}
	&
	\mathbf{B}(\mathbf{B}(\mathbb{R}/\Gamma)_{\mathrm{conn}})
  }
  }
  \,,
$$
then under forming homotopy fiber products the prequantum 2-bundle
$\nabla_P^1$ induces a prequantum 1-bundle on the open string phase space
by the D-brane-relative looping of the on-shell prequantum 2-bundle:
$$
  \nabla_{\mathfrak{C}_0} \underset{\nabla_P^1}{\times}
  \nabla_{\mathfrak{C}_1}
  :
  \xymatrix{
    \mathbf{Phase}_{\mathfrak{C}_0, \mathfrak{C}_1)}
	\ar[r]
	&
	\mathbf{B} (\mathbb{R}/\Gamma)_{\mathrm{conn}}
  }
  \,.
$$

\medskip

We now review the steps in the geometric quantization of the symplectic groupoid
due to \cite{Eli}
 -- hence the full geometric quantization of the prequantization $\nabla_P^1$ --
while discussing along the way the natural re-interpretation of the
steps involved from the point of view of the higher geometric prequantum theory of
2d Poisson Chern-Simons theory.

Consider therefore $\nabla_P^1$, as above, as the $(\mathbf{B}U(1))$-prequantum 2-bundle 
of 2d Poisson Chern-Simons theory according to def. \ref{DiffModuliStack}.
If we have a genuine symplectic groupoid instead of a pre-quasi-symplectic groupoid then
it makes sense ask for this prequantization to be presented by a {\v C}ech-Deligne
3-cocycle on $\tau_1 \exp(\mathfrak{P})$ which is given just by a multiplicative
circle-bundle with connection on the space of morphisms of the symplectic groupoid, 
and otherwise trivial local data on the space of objects.
While this is unlikely to be the most general higher prequantization of the 2d Poisson Chern-Simons
theory, this is the choice that admits to think of the situation as if it were
a setup in traditional symplectic geometry equipped with 
an equivariance- or ``multiplicativity''-constraint, as opposed to a setup in higher 2-plectic geometry.
(Such a ``multiplicative circle bundle'' on the space of morphisms of a
Lie groupoid is just like the transition bundle that appears in the definition of a bundle
gerbe, only that here the underlying groupoid is not a {\v C}ech groupoid resolving
a plain manifold, but is, in general, a genuine non-trivial Lie groupoid.)

Such a multiplicative prequantum bundle is the traditional notion of prequantization of a symplectic groupoid 
and is also considered in \cite{Eli}. The central construction there is that of the  
convolution $C^\ast$-algebra 
$\mathcal{A}(\nabla^1)_{\mathrm{pq}}$ of sections 
of the multiplicative prequantum bundle on the
space of morphisms of the symplectic groupoid, and its subalgebra 
$$
  \mathcal{A}(\nabla^1_P)_{\mathrm{q}} \hookrightarrow \mathcal{A}(\nabla^1_P)_{\mathrm{pq}}
$$ 
of polarized sections, once a suitable kind of polarization has been chosen.
Observe then that convolution algebras of sections of transition bundles
of bundle gerbes have a natural interpretation in the higher geometry
of the corresponding higher prequantum bundle $\nabla^1$: 
by \cite{TXLG04} and section 5 of \cite{CareyJohnsonMurray} 
these are the algebras whose modules are the unitary bundles which are twisted 
by $\nabla^1$: the ``bundle gerbe modules''.

By \ref{StrucTwistedCohomology} and by the discussion above in \ref{ExtendedWZW},
$\nabla^1_P$-twisted unitary bundles are equivalently the (pre-)quantum 2-states of
$\nabla^1_P$ regarded as a prequantum 2-bundle. These hence form a category
$\mathcal{A}(\nabla^1_P)_q \mathrm{Mod}$ of modules, and such categories of modules
are naturally interpreted, by the discussion in the appendix of
\cite{Schreiber08} as \emph{2-modules} with \emph{2-basis} the linear category 
$\mathbf{B}\mathcal{A}(\nabla^1_P)_q$:
$$
  \left\{
    \begin{tabular}{c}
	  quantum \mbox{2}-states of \\
	  higher prequantum 2d Poisson Chern-Simons theory
	\end{tabular}
  \right\}
  \;\;\simeq\;\;
  \mathcal{A}(\nabla^1_P)_q \mathrm{Mod}
  \;\;\in \;\;
  2\mathrm{Mod}
  \,.
$$

This resolves what might be a conceptual puzzlement concerning
the construction in \cite{Eli} in view of the usual story of geometric quantization: 
ordinarily geometric quantization directly produces the
space of states of a theory, while it requires more work to obtain the algebra of quantum 
observables acting on that. In \cite{Eli} it superficially seems 
to be the other way around, an algebra
drops out as a direct result of the quantization procedure. 
However, from the point of view of higher prequantum geometry this algebra
\emph{is} (a 2-basis for) the 2-space of 2-states;
and indeed obtaining the \emph{2-algebra} or \emph{higher quantum operators}
which would act on these 2-states does require more work (and has not been discussed yet).

Of course \cite{Eli} amplifies a different perspective on the central
result obtained there: 
that $\mathcal{A}(\nabla^1_P)_q$ is also a \emph{strict $C^\ast$-deformation
quantization} of the Poisson manifold that corresponds to the Poisson Lie algebroid
$\mathfrak{P}$! From the point of view of higher prequantum theory this 
says that the higher-geometric quantized 2d Poisson Chern-Simons theory 
has a 2-space of quantum 2-states
in codimension 2 that encodes the correlators (commutators) of a 1-dimensional 
quantum mechanical system. In other words, we see that the construction in 
\cite{Eli} is implicitly a ``holographic'' (strict deformation-)quantization of a Poisson manifold
by directly higher-geometric quantizing instead a 2-dimensional QFT. 

Notice that this statement is an analogue in $C^\ast$-deformation quantization
to the seminal result on \emph{formal} deformation quantization of Poisson manifolds:
The general formula that Kontsevich had given for the formal deformation quantization 
of a Poisson manifold was found by Cattaneo-Felder to be the point-particle limit of the
3-point function of the corresponding 2d Poisson sigma-model \cite{CattaneoFelder}. 
A similar result is discussed in \cite{GukovWitten}. 
There the 2d A-model (which is a special case of the Poisson sigma-model) 
is shown to holographically encode the quantization of its target space
symplectic manifold regarded as a 1d quantum field theory. 

In summary, the following table indicates how the ``holographic'' formal deformation
quantization of Poisson manifolds by Kontsevich-Cattaneo-Felder is
analogous to the ``holographic'' strict deformation quantization of
Poisson manifolds by \cite{Eli}, when reinterpreted in higher prequantum theory
as discussed above.

\medskip

\begin{tabular}{|l||c|c|}
  \hline
  & \begin{tabular}{c} perturbative formal algebraic \\ quantization \end{tabular} & 
   \begin{tabular}{c} non-perturbative geometric \\ quantization \end{tabular}
  \\
  \hline
  \hline
  \begin{tabular}{l}
    quantization of
	\\
	Poisson manifold
  \end{tabular}  & formal deformation quantization & strict $C^\ast$-deformation quantization
  \\
  \hline
  \begin{tabular}{l}
  ``holographically'' related \\
  2d field theory
  \end{tabular}
  &
  Poisson sigma-model & 2d Poisson Chern-Simons theory
  \\
  \hline
  \begin{tabular}{l}
    moduli stack of fields
	\\
	of the 2d field theory
  \end{tabular} & Poisson Lie algebroid & symplectic groupoid
  \\
  \hline
  \begin{tabular}{l}
    quantization of
	\\
	holographically related
	\\
	2d field theory
  \end{tabular}
  &
  \begin{tabular}{l}
  perturbative quantization of \\
  Poisson sigma-model 
  \end{tabular} & 
  \begin{tabular}{l}
    higher geometric quantization \\
	of 2d Poisson Chern-Simons theory
  \end{tabular}
  \\
  \hline
  \begin{tabular}{l}
  1d observable algebra\\
  is holographically \\ identified with...
  \end{tabular}
  &
  \begin{tabular}{l}point-particle limit 
  \\
  of 3-point function
  \end{tabular}
  &
  \begin{tabular}{c}
  basis for 2-space
  \\
  of quantum 2-states
  \end{tabular} 
  \\
  \hline
\end{tabular}

\medskip

More details on this higher geometric interpretation of traditional symplectic groupoid
quantization are discussed below in \ref{MotivicQuantizationApplications}.

\subsubsection{Higher prequantum 6d WZW-type models and the smooth $\mathrm{fivebrane}$-6-group}
\label{AnExtended6dSomething}

We close the overview of examples by providing a brief outlook on higher dimensional
examples in general, and on certain higher prequantum field theories in dimensions seven and
six in particular.

To appreciate the following pattern, recall that 
in  \ref{ExtendedWZW} above we discussed how the universal 
$G$-Chern-Simons $(\mathbf{B}^2 U(1))$-principal connection $\nabla_{\mathrm{CS}}$
over $\mathbf{B}G_{\mathrm{conn}}$ transgresses to the Wess-Zumino-Witten
$\mathbf{B}U(1)$-principal connection $\nabla_{\mathrm{WZW}}$
on $G$ itself. At the level of the underlying principal $\infty$-bundles
$\nabla_{\mathrm{CS}}^0$ and $\nabla_{\mathrm{WZW}}^0$ this relation holds
very generally:

for $G \in \mathrm{Grp}(\mathbf{H})$ any $\infty$-group, and 
$A \in \mathrm{Grp}_{n+1}(\mathbf{H})$ any sufficiently highly deloopable 
$\infty$-group (def. \ref{BraidedGroups}) in any $\infty$-topos $\mathbf{H}$,
consider a class in smooth $\infty$-group cohomology, \ref{StrucRepresentations},
$$  
  c
  \in 
  H^{n+1}_{\mathrm{grp}}(G,A)
  =
  H^{n+1}(\mathbf{B}G,A)
  \,,
$$
hence a universal characteristic class for $G$-principal $\infty$-bundles,
represented by a smooth cocycle   
$$
  \nabla^0_{\mathrm{CS}} :  \xymatrix{\mathbf{B}G \ar[r] & \mathbf{B}^{n+1} A}
  \,.
$$
Along the above lines we may think of the corresponding $\mathbf{B}^n A$-principal
$\infty$-bundle over $\mathbf{B}G$ as a \emph{universal $\infty$-Chern-Simons bundle}.
By example \ref{GroupExtension} this is the delooped $\infty$-group extension 
which is classified by $\nabla^0_{\mathrm{CS}}$ regarded as an $\infty$-group cocycle.
The looping of this cocycle exists 
$$
  \nabla^0_{\mathrm{WZW}} := \Omega \nabla^0_{\mathrm{CS}}
  : 
  \xymatrix{G \ar[r] & \mathbf{B}^n A}
  \,.
$$
and modulates a $\mathbf{B}^{n-1}A$-principal
bundle over the $\infty$-group $G$ itself: the $\infty$-group extension itself
that is classified by $\nabla^0_{\mathrm{CS}}$ according to example \ref{GroupExtension}.
This is the corresponding WZW $\infty$-bundle.

For example, for the case that $G \in \mathrm{Grp}(\mathrm{Smooth}\infty\mathrm{Grpd})$
is a compact Lie group and $A = U(1)$ is the smooth circle group, then by example
\ref{SegalBrylinski} there is an essentially unique refinement of every integral
cohomology class $k \in  H^{4}(B G, \mathbb{Z})$ to such a smooth cocycle
$\nabla^0_{\mathrm{CS}} : \mathbf{B}G \to \mathbf{B}^3 U(1)$. This $k$ is the 
\emph{level} of $G$-Chern-Simons theory and $\nabla^0_{\mathrm{CS}}$ modulates
the corresponding higher prequantum bundle of 3d $G$-Chern-Simons theory as in \ref{ExtendedWZW}
above. Moreover, the looping $\nabla_{\mathrm{WZW}}^0 \simeq \Omega \nabla^0_{\mathrm{CS}}$
modulates the ``WZW gerbe'', as discussed there.

\medskip

Now restrict attention to the next higher example of such pairs of 
higher Chern-Simons/higher WZW bundles, as seen by the tower of examples induced by the 
smooth Whitehead tower of $\mathbf{B}O$, \ref{SmoothAndDifferentialWhiteheadtower}:
the universal Chern-Simons 7-bundle on the smooth $\mathrm{String}$-2 group and the 
corresponding Wess-Zumino-Witten 6-bundle on $\mathrm{String}$ itself.

To motivate this as part of a theory of physics, first consider a simpler example of a 7-dimensional 
Chern-Simons type theory, namely the cup-product $U(1)$-Chern-Simons theory
in 7 dimensions, for which the ``holographic'' relation to an interesting 6d
theory is fairly well understood.
This is the theory whose de-transgression
is given  
by the higher prequantum 7-bundle on the universal moduli 3-stack $\mathbf{B}^3 U(1)_{\mathrm{conn}}$ 
of $\mathbf{B}^2 U(1)$-principal connections that is
modulated by the smooth and differential refinement of the cup product $\cup$ in ordinary differential cohomology:
$$
  \begin{tabular}{c|c}
  \xymatrix{
    \mathbf{B}^3 U(1)_{\mathrm{conn}}
	\ar[rr]^{(-)\hat \cup(-)}
	\ar[d]^{u_{\mathbf{B}^3 U(1)}}
	&&
	\mathbf{B}^7 U(1)_{\mathrm{conn}}
	\ar[d]^{u_{\mathbf{B}^7 U(1)}}
	\\
	\mathbf{B}^3 U(1)
	\ar[rr]^{(-)\cup(-)}
	\ar[d]^\int
	&&
	\mathbf{B}^7 U(1)
	\ar[d]^\int
	\\
	K(\mathbb{Z}_4)
	\ar[rr]^{(-)\cup(-)}
	&&
	K(\mathbb{Z}_8)
  }
  &
  \xymatrix{
    \nabla_{7\mathrm{AbCS}}
	\\
    \nabla_{7\mathrm{AbCS}}^0	
	\\
    \int \nabla_{7\mathrm{AbCS}}^0
  }
  \end{tabular}
  \,.
$$
(Or rather, the theory to consider for the full holographic relation
is a quadratic refinement of this cup pairing. The higher geometric refinement 
of this we discuss in \ref{supergravityCField}, 
but in the present discussion we will suppress this, for simplicity).

While precise and reliable statements are getting scarce as one proceeds with the
physics literature into the study of these systems, the following four
seminal physics articles seem to represent the present understanding
of the story by which this 7d theory is related to a 6d theory in higher
generalization of how 3d Chern-Simons theory is related to the 2d WZW model.

\begin{enumerate}
\item In \cite{Witten} it was argued that the space of states that 
the (ordinary) geometric
quantization of $\nabla_{7\mathrm{AbCS}}$ assigns to a closed 6d manifold 
$\Sigma$ is naturally identified with the space of conformal blocks of 
a self-dual 2-form higher gauge theory on $\Sigma$. 
Moreover, this 6d theory is part of the worldvolume theory of 
a single M5-brane and the above 7d Chern-Simons theory is 
the abelian Chern-Simons sector of the 11-dimensional supergravity Lagrangian
compactified to a 7-manifold whose boundary is the 6d M5-brane worldvolume.

\item Then in \cite{Maldacena} a more general relation between 
the 6d theory and 11-dimensional supergravity compactified on a 4-sphere
to an asymptotically anti-de Sitter space was argued for. This is what is
today called $\mathrm{AdS}_7/\mathrm{CFT}_6$-duality, a sibling of the
$\mathrm{AdS}_5/\mathrm{CFT}_4$-duality which has received a large amount 
of attention since then.

\item As a kind of synthesis of the previous two items, in \cite{Witten98} it is 
argued for both $\mathrm{AdS}_5/\mathrm{CFT}_4$ and 
$\mathrm{AdS}_7/\mathrm{CFT}_6$ the conformal blocks on the CFT-side
are obtained already by keeping on the supergravity side \emph{only} the Chern-Simons terms
inside the full supergravity action.

\item  At the same time it is known that the abelian Chern-Simons term
in the 11-dimensional supergravity action relevant for 
$\mathrm{AdS}_7/\mathrm{CFT}_6$ is not in general just
the abelian Chern-Simons term $\nabla_{7\mathrm{AbCS}}$ considered in the above references:
more accurately it receives Green-Schwarz-type quantum corrections
that make it a \emph{nonabelian} Chern-Simons term \cite{DLM}. 
\end{enumerate}

In \cite{FiorenzaSatiSchreiberI} we observed that
these items together, taken at face value, imply that more generally it must be the quantum-corrected nonabelian 
7d Chern-Simons Lagrangian
inside 11-dimensional supergravity which is relevant for the holographic 
description of the 2-form sector of the 6d worldvolume theory of M5-branes.
(See \cite{Freed} for comments on this 6d theory as an extended QFT related to extended 7d
Chern-Simons theory.)
Moreover, in \ref{supergravityCField} we observe that the natural lift of the
``flux quantization condition'' \cite{Witten} 
-- which is an \emph{equation} between cohomology classes of fields
in 11d-supergravity -- to moduli stacks of fields (hence to higher prequantum geometry)
is given by the corresponding \emph{homotopy pullback} of these moduli fields, 
as usual in homotopy theory. 
We showed that this homotopy pullback is the smooth moduli 2-stack 
$\mathbf{B}\mathrm{String}^{2\mathbf{a}}_{\mathrm{conn}}$ of twisted $\mathrm{String}$-principal 
2-connections, unifying the Spin-connection (the field of gravity) and the 
3-form $C$-field into a single higher gauge field in higher prequantum geometry.

The nonabelian 7-dimensional Chern-Simons-type Lagrangian on String-2-connections
obtained this way in \cite{FiorenzaSatiSchreiberI} 
is the sum of some cup product terms and one indecomposable term.
Moreover, the refinement specifically of the indecomposable term to higher prequantum geometry
is the stacky and differential refinement $\tfrac{1}{6}\hat{\mathbf{p}}_2$ of the universal fractional second
Pontryagin class $\tfrac{1}{2}p_2$, which was constructed in \cite{FSS} as reviewed in 
\ref{HigherCSFromLieIntegration} above:
$$
  \begin{tabular}{c|c}
  \xymatrix{
    \mathbf{B}\mathrm{String}_{\mathrm{conn}}
	\ar[rr]^-{\tfrac{1}{6}\hat {\mathbf{p}}_2}
	\ar[d]^{u_{\mathbf{B}\mathrm{String}}}
	&&
	\mathbf{B}^7 U(1)_{\mathrm{conn}}
	\ar[d]^{p_{\mathbf{B}^7 U(1)}}
	\\
    \mathbf{B}\mathrm{String}
	\ar[rr]^-{\tfrac{1}{6}{\mathbf{p}}_2}
	\ar[d]^\int
	&&
	\mathbf{B}^7 U(1)
	\ar[d]^\int
	\\
	B O \langle 8\rangle
	\ar[rr]^{\tfrac{1}{6}p_2}
	&&
	K(\mathbb{Z}, 8)
  }
  &
  \xymatrix{
    \nabla_{7\mathrm{CS}}
    \\
    \nabla_{7\mathrm{CS}}^0
    \\
    \int \nabla_{7\mathrm{CS}}^0
  }
  \end{tabular}
  \,.
$$
Quite  independently of whatever role this extended 7d Chern-Simons theory has as a sector in  
$\mathrm{AdS}_7/\mathrm{CFT}_6$ duality,
this is the natural next example in higher prequantum theory after that of
3d $\mathrm{Spin}$-Chern-Simons theory.

In \cite{FSS} it was shown that the prequantum 7-bundle of this nonabelian 7d Chern-Simons
theory over the moduli stack of its instanton sectors, hence over $\mathbf{B}\mathrm{String}$,
is the delooping of a smooth refinement of the \emph{Fivebrane group}, 
\ref{NS5 and twisted Fivebrane Structures},
to the smooth Fivebrane 6-group, \ref{FivebraneSixGroup}: 
$$
  \raisebox{20pt}{
  \xymatrix{
    \mathbf{B}\mathrm{Fivebrane}
	\ar[d]
	\\
	\mathbf{B}\mathrm{String} \ar[rr]^{\nabla^0_{7\mathrm{CS}}}
	&&
	\mathbf{B}^7 U(1)
	\,.
  }
  }
$$
Moreover, by the above general discussion this induces a WZW-type 6-bundle over the
smooth String 2-group itself, whose total space is the Fivebrane group itself
$$
  \raisebox{20pt}{
  \xymatrix{
    \mathrm{Fivebrane}
	\ar[d]
	\\
	\mathrm{String}
	\ar[rr]^{\nabla_{6\mathrm{WZW}}^0 }
	&&
	\mathbf{B}^6 U(1)
  }}
  \,.
$$
Therefore, in view of the discussion in \ref{ExtendedWZW}, it is natural to expect 
a 6-dimensional higher analog of traditional 2d WZW theory whose underlying 
higher prequantum 6-bundle is $\nabla_{6\mathrm{WZW}}$. However, the 
lift of this discussion from just instanton sectors to the full moduli stack of 
fields is more subtle than in the 3d/2d case and deserves 
a separate discussion elsewhere. (This is ongoing joint work with Hisham Sati.)

\newpage
\subsection{Higher Chern-Simons field theory}
\label{ChernSimonsFunctional}
\index{Chern-Simons functionals}

We consider the realization of the general abstract
\emph{$\infty$-Chern-Simons functionals} from \ref{StrucChern-SimonsTheory}
in the context of smooth, synthetic-differential and super-cohesion.
We discuss general aspects of the class of quantum field theories defined this way
and then identify a list of special cases of interest.
This section builds on \cite{hgp} and \cite{FS}.

\medskip

\begin{itemize}
  \item \ref{InfinityChernSimonsFieldTheory} -- {Higher extended $\infty$-Chern-Simons theory}
  \begin{itemize}
    \item \ref{FibInt} -- {Fiber integration and extended Chern-Simons functionals}
	\item \ref{ConstructionExtendedChern-SimonsFromL} -- {Construction from $L_\infty$-cocycles}
  \end{itemize}
  \item \ref{HigherCupProductCSTheories} -- {Higher cup-product Chern-Simons theories}
  \item {Examples}
  \begin{itemize}
  \item \ref{VolumeHolonomy} -- Volume holonomy
  \item \ref{1dCSTheories} -- 1d Chern-Simons functionals
  \item \ref{3dCSTheories} -- 3d Chern-Simons functionals
  \begin{itemize}
    \item \ref{InfinCSOrdinaryCS} -- Ordinary Chern-Simons theory
    \item \ref{InfinCSDW} -- Ordinary Dijkgraaf-Witten theory
  \end{itemize}
  \item \ref{4dCSFunctionals} -- 4d Chern-Simons functionals
  \begin{itemize}
    \item \ref{InfinCSBF} -- 4d BF theory and topological Yang-Mills theory
	\item \ref{4dYetterModel} -- 4d Yetter model
  \end{itemize}
  \item \ref{GaugeCouplingOfBranes} -- Abelian gauge coupling of branes
  \item \ref{HigherAbelianChernSimons} -- Higher abelian Chern-Simons functionals
  \begin{itemize}
    \item \ref{HigherAbelianCS} -- $(4k+3)$d $U(1)$-Chern-Simons functionals;
    \item \ref{HigherElectric} -- Higher electric coupling and higher gauge anomalies.
  \end{itemize} 
  \item \ref{InfinCS7d} -- 7d Chern-Simons functionals
  \begin{itemize}
    \item \ref{CupProductTheoryOfTwo3DCSTheories}
      -- The cup product of a 3d CS theory with itself;
    \item \ref{InfinCS7CSOnString2} -- 
      7d CS theory on string 2-connection fields;
    \item \ref{7dCSInSugraOnAdS7} --
      7d CS theory in 11d supergravity on $\mathrm{AdS}_7$.
  \end{itemize}
  \item \ref{HigherElectric} -- Higher electric coupling and higher gauge anomalies
  \item \ref{CSFTAction} -- Action of closed string field theory type
  \item \ref{InfinCSAKSZ} -- AKSZ $\sigma$-models
  \begin{itemize}
  \item \ref{OrdinaryChernSimonsTheoryAsAKSZ} -- Ordinary Chern-Simons as AKSZ theory
  \item \ref{section.PoissonSigmaModel} -- Poisson $\sigma$-model
  \item \ref{section.CourantSigmaModel} -- Courant $\sigma$-model
  \item \ref{section.HigherAbelianCSTheory} -- Higher abelian Chern-Simons theory in dimension $4k+3$
  \end{itemize}
  \end{itemize}
\end{itemize}

\newpage

\subsubsection{$\infty$-Chern-Simons field theory}
\label{InfinityChernSimonsFieldTheory}
\index{Chern-Simons functionals!field theory}

By prop. \ref{FirstFractionalDifferentialPontrjagin} 
the action functional of ordinary Chern-Simons theory \cite{FreedCS} 
for a simple Lie group $G$ may be 
understood as being the volume holonomy, \ref{SmoothStrucChernSimons}, 
of the Chern-Simons circle $3$-bundle with connection that 
the refined Chern-Weil homomorphism assigns to any connection on a $G$-principal bundle.

We may observe that all the ingredients of this statement have their general abstract analogs 
in any cohesive $\infty$-topos $\mathbf{H}$: for any 
cohesive $\infty$-group $G$ and any representatative 
$\mathbf{c} : \mathbf{B}G \to \mathbf{B}^n A$ of a characteristic class for $G$ 
there is canonically the induced $\infty$-Chern-Weil homomorphism, \ref{StrucChern-WeilHomomorphism} 
$$
  L_{\mathbf{c}} : \mathbf{H}_{\mathrm{conn}}(-,\mathbf{B}G) \to \mathbf{H}_{\mathrm{diff}}^n(-)
$$ 
that sends intrinsic $G$-connections to cocycles in 
intrinsic differential cohomology with coefficients in $A$. 
This may be thought of as the \emph{Lagrangian} of the $\infty$-Chern-Simons theory induced by $\mathbf{c}$. 

In the cohesive $\infty$-topos $\mathrm{Smooth}\infty\mathrm{Grpd}$ of smooth $\infty$-groupoids,
\ref{SmoothInfgrpds}, we deduced in \ref{SmoothStrucChernSimons} a natural general abstract procedure 
for integration of 
$L_{\mathbf{c}}$ over an $n$-dimensional parameter space $\Sigma \in \mathbf{H}$
by a realization of the general abstract construction described in 
\ref{StrucChern-SimonsTheory}.
The resulting smooth function 
$$
  \exp(S_{\mathbf{c}}) : [\Sigma,\mathbf{B}G_{\mathrm{conn}}] \to U(1)
$$ 
is the exponentiated action functional of $\infty$-Chern-Simons theory on the 
smooth $\infty$-groupoid of field configurations.
It may be regarded itself as a degree-0 characteristic class on the space of field configurations. 
As such, its differential refinement
$d \exp(S_{\mathbf{c}}) : [\Sigma, \mathbf{B}G_{\mathrm{conn}}] \to \mathbf{\flat}_{\mathrm{\mathrm{dR}}} \mathbf{B}U(1)$ 
is the Euler-Lagrange equation of the theory. 

We show that this construction subsumes the action functional of ordinary Chern-Simons theory, 
of Dijkgraaf-Witten theory,
of BF-theory coupled to topological Yang-Mills theory, 
of all versions of AKSZ theory including the Poisson sigma-model and the Courant sigma model in lowest 
degree, as well as of higher Chern-Simons supergravity.

\paragraph{Fiber integration and extended Chern-Simons functionals}
\label{FibInt}

We discuss fiber integration in ordinary differential cohomology refined to
smooth higher stacks and how this turns every differential characteristic maps
into a tower of extended higher Chern-Simons action functionals in all
codimensions.

This section draws from \cite{FiorenzaSatiSchreiberIV}.

\medskip

One of the basic properties of $\infty$-toposes is that they are
\emph{cartesian closed}. This means that:
\begin{fact}
For every two objects $X, A \in \mathbf{H}$
-- hence for every two smooth higher stacks -- there is another object denoted 
$[X,A] \in \mathbf{H}$ that behaves like the ``space of smooth maps from $X$ to $A$.''
in that for every further $Y \in \mathbf{H}$ there is a natural
equivalence of cocycle $\infty$-groupoids of the form
$$
  \mathbf{H}(X \times Y, A) \simeq \mathbf{H}(Y, [X,A])
  \,,
$$
saying that cocycles with coefficients in $[X,A]$ on $Y$ are 
naturally equivalent to 
$A$-cocycles on the product $X \times Y$.
\end{fact}
\begin{remark}
The object $[X,A]$ is in category theory known as the \emph{internal hom} object,
but in applications to physics and stacks it is often better known as the
``families version'' of $A$-cocycles on $Y$, in that for each smooth parameter space 
$U \in \mathrm{SmthMfd}$, the elements of $[X,A](U)$ are 
``$U$-parameterized families of $A$-cocycles on $X$'', namely $A$-cocycles on $X \times U$.
This follows from the above characterizing formula and the Yoneda lemma:
$$
  \xymatrix{
    [X,A](U) 
	\ar[rr]_-{\mathrm{Yoneda}}^-\simeq
	&&
	\mathbf{H}(U, [X,A])
	\ar[r]^-\simeq
	&
	\mathbf{H}(X\times U, A)
	\,.
  }
$$
\end{remark}
Notably for $G$ a smooth $\infty$-group and $A = \mathbf{B}G_{\mathrm{conn}}$
a moduli $\infty$-stack of smooth $G$-principal $\infty$-bundles with connection
the object
$$
  [\Sigma_k, \mathbf{B}G_{\mathrm{conn}}] \in \mathbf{H}
$$
is the smooth higher moduli stack of $G$-connection \emph{on $\Sigma_k$}. 
It assigns to a test manifold $U$ the $\infty$-groupoid of $U$-parameterized
families of $G$-$\infty$-connections, namely of $G$-$\infty$-connections on 
$X \times U$.
This is the 
smooth higher stack incarnation of the configuration space of higher $G$-gauge theory
on $\Sigma_k$.

\begin{example}
  In the discussion of anomaly polynomials in heterotic string theory over a 
  10-dimensional spacetime $X$ one encounters degree-12 differential forms
  $I_4 \wedge I_8$, where $I_i$ is a degree $i$ polynomial in characteristic forms. 
  Clearly these cannot live on $X$, as every 12-form on $X$, 
  given by an element in the hom-$\infty$-groupoid 
  $$
    \xymatrix{
      \mathbf{H}(X, \Omega^{12}(-))
	  \ar[rr]_-{\mathrm{Yoneda}}^-\simeq
	  &&
	  \Omega^{12}(X)
	}
  $$
  is trivial.
  Instead, these differential forms are elements in the internal hom $[X, \Omega^{12}(-)]$, which means
  that for every choice of smooth parameter space $U$ there is a smooth 12-form
  on $X \times U$, such that this system of forms transforms naturally in $U$.
  
  \medskip
  Below  we discuss how such anomaly forms appear from
  morphisms of higher moduli stacks 
  $$
    \mathbf{c}_{\mathrm{conn}} : \mathbf{B}G_{\mathrm{conn}} \to \mathbf{B}^{11}U(1)_{\mathrm{conn}} 
  $$
  for $\mathbf{B}G_{\mathrm{conn}}$ the higher moduli stack of supergravity field
  configurations by sending
  the families of moduli of field configurations on spacetime $X$ to their
  anomaly form:
  $$
	\xymatrix{
      [X, \mathbf{B}G_{\mathrm{conn}}]
	  \ar[rr]^-{[X, \mathbf{c}_{\mathrm{conn}}]}
	  &&
	  [X, \mathbf{B}^{11} U(1)_{\mathrm{conn}}]
	  \ar[rr]^-{[X, \mathrm{curv}]}
	  &&
	 [X, \Omega^{12}(-)]
	 }
	 \,.
  $$
\end{example}
We now discuss how such families of $n$-cocycles on some $X$ can be integrated over
$X$ to yield $(n-\mathrm{dim}(X))$-cocycles. Recall from \ref{FiberIntegrationOfOrdinaryDifferentialCocycles}:
\begin{proposition}
  Let $\Sigma_k$ be a closed (= compact and without boundary) oriented smooth
  manifold of dimension $k$. 
  Then for every $n \geq k$ there is a natural morphism of smooth higher stacks
  $$
    \exp(2 \pi i \int_{\Sigma_k} (-))
	:
	[\Sigma_k, \mathbf{B}^n U(1)_{\mathrm{conn}}]
	\to
	\mathbf{B}^{n-k}U(1)_{\mathrm{conn}}
  $$
  from the moduli $n$-stack of circle $n$-bundles with connection on $\Sigma_k$
  to the moduli $(n-k)$-stack of smooth circle $(n-k)$-bundles with connection
  such that
  \begin{enumerate}
    \item 
	     for $k = n$ this yields a $U(1)$-valued gauge invariant smooth function
  $$
      \exp(2 \pi i \int_{\Sigma_k} (-))
	  :
	  [\Sigma_n, \mathbf{B}^n U(1)_{\mathrm{conn}}]
	  \to
	  U(1)\;,
  $$
  which is the \emph{$n$-volume holonomy} of a circle $n$-connection over the
  ``$n$-dimensional Wilson volume'' $\Sigma_n$;
  \item
    for $k_1, k_2 \in \mathbb{N}$ with $k_1+k_2 \leq n$ we have
	$$
	  \exp(2 \pi i \int_{\Sigma_{k_1}}(-)) \circ 
	  \exp(2 \pi i \int_{\Sigma_{k_2}}(-))
      \simeq
	  \exp(2 \pi i \int_{\Sigma_{k_1 } \times \Sigma_{k_2}}(-))
	  \,.
	  $$
  \end{enumerate}
  \label{FiberInte}
\end{proposition}
\proof
  Since $\mathbf{B}^n U(1)_{\mathrm{conn}}$ is fibrant in 
  the projective local model structure
  $[\mathrm{CartSp}^{\mathrm{op}}, \mathrm{sSet}]_{\mathrm{proj}, \mathrm{loc}}$
  (since every circle $n$-bundle with connection on a Cartesian space is trivializable)
  the mapping stack $[\Sigma_k, \mathbf{B}^n U(1)_{\mathrm{conn}}]$
  is presented for any choice of good open
  cover $\{U_i \to \Sigma_k\}$ by the simplicial presheaf
  $$
    U \mapsto [\mathrm{CartSp}^{\mathrm{op}}, \mathrm{sSet}](\check{C}(\mathcal{U}) \times U, 
	\mathbf{B}^n U(1)_{\mathrm{conn}})
	\,,
  $$
  where $\check{C}(\mathcal{U})$ is the \v{C}ech nerve of the open
  cover $\{U_i \to \Sigma_k\}$.
  Therefore a morphism as claimed is given by natural fiber integration 
  of Deligne hypercohomology along product bundles
  $\Sigma_k \times U \to U$ for closed $\Sigma_k$. This has been constructed for instance
  in \cite{GomiTerashima}.
\endofproof

\begin{definition}
Let
$
  \mathbf{c}_{\mathrm{conn}}
  :
  \mathbf{B}G_{\mathrm{conn}}
  \to
  \mathbf{B}^n U(1)_{\mathrm{conn}}
$
be a differential characteristic map. Then for $\Sigma_k$
a closed smooth manifold of dimension $k \leq n$, we call
$$
  \exp(2 \pi i \int_{\Sigma_k} [\Sigma_k, \mathbf{c}_{\mathrm{conn}}] )
  :
  \xymatrix{
    [\Sigma_k, \mathbf{B}G_{\mathrm{conn}}]
	\ar[rr]^-{[\Sigma_k, \mathbf{c}_{\mathrm{conn}}]}
	&&
	[\Sigma_k,\mathbf{B}^n U(1)_{\mathrm{conn}}]
	\ar[rrr]^{\exp(2 \pi i \int_{\Sigma_k}(-))}
	&&&
	\mathbf{B}^{n-k}U(1)_{\mathrm{conn}}
  }
$$
the \emph{off-shell prequantum $(n-k)$-bundle of extended 
$\mathbf{c}_{\mathrm{conn}}$-$\infty$-Chern-Simons theory}. 
For $n = k$ we have a \emph{circle 0-bundle} 
$$
  \exp(2 \pi i \int_{\Sigma_n} [\Sigma_n, \mathbf{c}_{\mathrm{conn}}] )
  :
  \xymatrix{
    [\Sigma_n, \mathbf{B}G_{\mathrm{conn}}]
	\ar[rr]^-{[\Sigma_n, \mathbf{c}_{\mathrm{conn}}]}
	&&
	[\Sigma_n,\mathbf{B}^n U(1)_{\mathrm{conn}}]
	\ar[rrr]^{~~~~~~~\exp(2 \pi i \int_{\Sigma_n}(-))}
	&&&
	U(1)  }\;,
$$
which we call the \emph{action functional} of the theory.
\label{FiberIntegration}
\end{definition}
\noindent This construction subsumes several fundamental aspects of Chern-Simons theory:
\begin{enumerate}
  \item gauge invariance and smoothness of the (extended) action functionals, remark \ref{gauge invariance and smoothness};
  \item inclusion of instanton sectors (nontrivial gauge $\infty$-bundles), remark \ref{action functional on instanton sectors};
  \item level quantization, remark \ref{level quantization};
  \item definition on non-bounding manifolds and 
    relation to (higher) topological Yang-Mills on bounding manifolds, 
	 remark \ref{definition on non-bounding manifolds}.
\end{enumerate}
We discuss these in more detail in the following remarks, as indicated.

\begin{remark}[Gauge invariance and smoothness]
 \label{gauge invariance and smoothness}
  Since $U(1) \in \mathbf{H}$ is an ordinary manifold (after forgetting the group structure),
  a 0-stack with no non-trivial morphisms 
  (no gauge transformation),
  the action functional $\exp(2 \pi i \int_{\Sigma_n} [\Sigma_n, \mathbf{c}_{\mathrm{conn}}] )$
  takes every morphism in the moduli stack of field configurations to the identity.
  But these morphisms are the \emph{gauge transformations}, and so this says that
  $\exp(2 \pi i \int_{\Sigma_n} [\Sigma_n, \mathbf{c}_{\mathrm{conn}}] )$
  is \emph{gauge invariant}, as befits a gauge theory action functional.
  To make this more explicit, notice that
  $$
    \mathbf{H}(\Sigma_n, \mathbf{B} G_{\mathrm{conn}})
	\simeq
	[\Sigma_n, \mathbf{B}G_{\mathrm{conn}}](*)
  $$
  is the evaluation of the moduli stack on the point, hence the $\infty$-groupoid of smooth families
  of field configurations which are trivially parameterized.
  Moreover
  $$
    H^1_{\mathrm{conn}}(\Sigma_n, G) := \pi_0
    \mathbf{H}(\Sigma_n, \mathbf{B} G_{\mathrm{conn}})
  $$
  is the set of gauge equivalent such field configurations. Then the statement that
  the action functional is both gauge invariant and smooth is the statement that it
  can be extended from $H^1_{\mathrm{conn}}(\Sigma_n, G)$ 
  (supposing that it were given there as a function $\exp(i S(-))$ by other means) via
  $\mathbf{H}(\Sigma_n, \mathbf{B} G_{\mathrm{conn}})$ to $[\Sigma_n, \mathbf{B}G_{\mathrm{conn}}]$
  $$
    \xymatrix{
	  H^1_{\mathrm{conn}}(\Sigma_n, G) 
	  \ar[d] 
	  \ar[rr]^-{\exp(i S(-))} && U(1) 
	  \\
	  \mathbf{H}(\Sigma_n, \mathbf{B} G_{\mathrm{conn}})
	  \ar[d]
	  && & \mbox{gauge invariance}
	  \\
	  [\Sigma_n, \mathbf{B}G_{\mathrm{conn}}]
	  \ar[uurr]_{\exp(2 \pi i \int_{\Sigma_n} [\Sigma_n,\mathbf{c}_{\mathrm{conn}}] ) }
	  &&&
	  \mbox{smoothness}~.
}
  $$
  
\end{remark}

\vspace{3mm}
\begin{remark}[Definition on instanton sectors] 
 \label{action functional on instanton sectors}
Ordinary 3-dimensional Chern-Simons theory is
often discussed for the special case only when the gauge group $G$ is connected 
and simply connected. This
yields a drastic simplification compared to the general case; 
since for every Lie group the second homotopy group $\pi_2(G)$ is trivial, 
and since the homotopy groups of the classifying
space $BG$ are those of $G$ shifted up in degree by one, 
this implies that $BG$ is 3-connected and hence that
every continuous map $\Sigma_3 \to B G$ out of a 3-manifold is homotopic to the trivial map.
This implies that every $G$-principal bundle over $\Sigma_3$ is trivializable. 
As a result, the moduli stack of $G$-gauge fields on $\Sigma_3$, which
a priori is $[\Sigma_3, \mathbf{B}G_{\mathrm{conn}}]$, becomes in this case equivalent 
to just the moduli stack of trivial $G$-bundles with
(non-trivial) connection on $\Sigma_3$, which is identified with the groupoid of 
just $\mathfrak{g}$-valued 1-forms on $\Sigma_3$, and
gauge transformations between these, which is indeed the familiar configurations 
space for 3-dimensional
$G$-Chern-Simons theory.

\vspace{3mm}
One should compare this to the case of 4-dimensional $G$-gauge theory on a 
4-dimensional manifold $\Sigma_4$,
such as $G$-Yang-Mills theory. By the same argument as before, in this case G-principal bundles may be nontrivial,
but are classified enirely by the second Chern class (or first Pontrjagin class) 
$[c_2] \in H^4(\Sigma_4, \pi(G))$.
In Yang-Mills theory with $G = SU(n)$, this class is known as the 
\emph{instanton number} of the gauge field.

\vspace{3mm}
The simplest case where non-trivial classes occur already in dimension 3 
is the non-simply connected
gauge group $G = U(1)$, discussed in section \ref{3dU1CS} below. 
Here the moduli stack of fields 
$[\Sigma_3, \mathbf{B}U(1)_{\mathrm{conn}}]$ contains configurations which are not
given by globally defined 1-forms, but by connections on non-trivial circle bundles. 
By analogy with the case
of $SU(n)$-Yang-Mills theory, we will loosely refer to such field configurations 
as instanton field congurations,
too. In this case it is the first Chern class $[c_1] \in H^2(X,\mathbb{Z})$ 
that measures the non-triviality of the bundle.
If the first Chern-class of a $U(1)$-gauge field configurations happens to vanish, 
then the gauge field is again given by just a 1-form $A \in \Omega^1(\Sigma_3)$, 
the familiar gauge potential of electromagnetism. The value of the
3d Chern-Simons action functional on such a non-instanton configuration 
is simply the familiar expression 
$$
  \exp(i S(A)) = \exp(2 \pi i \int_{\Sigma_3} A \wedge d_{\mathrm{dR}} A)
  \,,
$$
where on the right we have the ordinary integration of the 3-form 
$A \wedge d A $ over $\Sigma_3$.

\vspace{3mm}
In the general case, however, when the configuration 
in $[\Sigma_3, \mathbf{B}U(1)_{\mathrm{conn}}]$ has non-trivial first Chern class,
the expression for the value of the action functional on this configuration is 
more complicated. If we pick a
good open cover $\{U_i \to \Sigma_3\}$, then we can arrange that locally on 
each patch $U_i$ the gauge field is given by
a 1-form $A_i$ and the contribution of the action functional over $U_i$ 
by $\exp(2 \pi i \int_{\Sigma_3} A_i \wedge d A_i)$ as above. But
in such a decomposition there are further terms to be included to get the correct action functional. This is
what the construction in Prop. \ref{FiberIntegration} achieves.
\end{remark}

\begin{remark}[Level quantization]
  \label{level quantization}
  Traditionally, Chern-Simons theory in 3-dimensions
  with gauge group a connected and simply connected group $G$ 
  comes in a family parameterized by a \emph{level} $k \in \mathbb{Z}$.
  This level is secretly the cohomology class of the differential characteristic
  map 
  $$
    \mathbf{c}_{\mathrm{conn}} : \mathbf{B}G_{\mathrm{conn}} \to \mathbf{B}^3 U(1)_{\mathrm{conn}}
  $$
  (constructed in \cite{FSS}) in 
  $$
    H_{\mathrm{smooth}}^3(B G, U(1)) \simeq H^4(B G, \mathbb{Z}) \simeq \mathbb{Z}
	\,.
  $$
  So the traditional level is a cohomological shadow of the differential characteristic map 
  that we interpret as the off-shell prequantum $n$-bundle in full codimension $n$
  (down on the point). Notice that for a general smooth $\infty$-group $G$ the cohomology
  group $H^{n+1}(B G , \mathbb{Z})$ need not be equivalent to $\mathbb{Z}$ and so 
  in general the level need not be an integer.
  For for every smooth $\infty$-group $G$, and given 
  a morphism of moduli stacks 
  $\mathbf{c}_{\mathrm{conn}} : \mathbf{B}G_{\mathrm{conn}} \to \mathbf{B}^n U(1)_{\mathrm{conn}}$,
  also every integral multiple $k \mathbf{c}_{\mathrm{conn}}$ gives an
  $n$-dimensional Chern-Simons theory, ``at $k$-fold level''. The converse is in 
  general hard to establish: whether a given $\mathbf{c}_{\mathrm{conn}}$ can be divided
  by an integer. 
  For instance for 3-dimensional Chern-Simons theory division by 2 may be possible
  for Spin-structure. For 7-dimensional Chern-Simons theory division by 6 may be possible
  in the presence of String-structure \cite{FiorenzaSatiSchreiberI}.
\end{remark}

\begin{remark}
  \label{definition on non-bounding manifolds}
  Ordinary 3-dimensional Chern-Simons theory is often defined on 
  bounding 3-manifolds $\Sigma_3$ by
  $$
    \exp(i S(\nabla)) = \exp(2 \pi i k \int_{\Sigma_4} 
	\langle F_{\widehat \nabla} \wedge F_{\widehat \nabla}\rangle)
	\,,
  $$
  where $\Sigma_4$ is any 4-manifold with $\Sigma_3 = \partial \Sigma_4$ and where
  $\widehat \nabla$ is any extension of the gauge field configuration from 
  $\Sigma_3$ to $\Sigma_4$. Similar expressions exist for higher dimensional Chern-Simons theories.
  If one takes these expressions to be the actual definition of Chern-Simons
  action functional, then one needs extra discussion for which manifolds 
  (with desired structure) are bounding, hence which vanish in the respective
  cobordism ring, and, more seriously, one needs to include those which are not
  bounding from the discussion.
  For example, in type IIB string theory one encounters the cobordism group 
$\Omega_{11}^{\rm Spin}(K(\Z, 6))$ \cite{Witten96}, which is 
proven to vanish in \cite{KS2}, meaning that all the desired manifolds happen
to be bounding.

\vspace{3mm}
We emphasize that our formula in Prop. \ref{FiberIntegration} applies generally,
 whether or not a manifold
 is bounding. Moreover, it is guaranteed that 
 \emph{if} $\Sigma_n$ happens to be bounding after all, then the action
 functional is equivalently given by integrating a higher curvature invariant over
 a bounding $(n+1)$-dimensional manifold.
 At the level of differential cohomology classes $H^n_{\mathrm{conn}}(-, U(1))$ this is 
 the well-known property 
 (a review and further pointers are given in \cite{HopkinsSinger})
 which is an explicit axiom in the equivalent
formulation by Cheeger-Simons differential characters:
 a Cheeger-Simons differential character of degree $(n+1)$ 
 is by definition a group homomorphism from closed $n$-manifolds to $U(1)$ such 
 that whenever the $n$-manifold happens to be bounding, the value in $U(1)$
 is given by the exponentiated integral of a smooth $(n+1)$-form over any bounding manifold.
 
 \vspace{3mm}
 With reference to such differential characters Chern-Simons action 
 functions have been formulated for instance in \cite{Witten96,Witten98}.
 The sheaf hypercohomology classes of the Deligne complex that we are concerned with  
 here are well known to be equivalent
 to these differential characters, and {\v C}ech-Deligne cohoomology has the 
 advantage that with results
 such as \cite{GomiTerashima} invoked in 
 Prop. \ref{FiberInte} above, it yields explict formulas for the action functional
 on non-bounding manifolds in terms of local differential form data.
\end{remark}

\paragraph{Construction from $L_\infty$-cocycles}
\label{ConstructionExtendedChern-SimonsFromL}

We discuss the construction of $\infty$-Chern-Simons functionals from 
differential refinements of $L_\infty$-algebra cocycles.

This section draws from \cite{FiorenzaSatiSchreiber}. 

\medskip

Recall for the following the construction of the $\infty$-Chern-Weil homomorphism
by Lie integration of Chern-Simons elements, \ref{SmoothStrucInfChernWeil}, for 
$L_\infty$-algebroids, \ref{StrucSynthLie}.

A Chern-Simons element $\mathrm{cs}$ witnessing the transgression from 
an invariant polynomial $\langle - \rangle$ to a cocycle $\mu$ is equivalently a commuting diagram of the form
$$
  \xymatrix{
    \mathrm{CE}(\mathfrak{a})
    \ar@{<-}[r]^{\mu} &
    \mathrm{CE}(b^{n}\mathbb{R})
    & \mbox{cocycle}
    \\
    \mathrm{W}(\mathfrak{a}) \ar[u]
    \ar@{<-}[r]^{cs}&
    \mathrm{W}(b^n \mathbb{R}) \ar[u]
    &
    \mbox{Chern-Simons element}
    \\
    \mathrm{inv}(\mathfrak{a}) \ar[u]
    \ar@{<-}[r]^{\langle-\rangle} &
    \mathrm{inv}(b^n \mathbb{R}) \ar[u]
    &
    \mbox{invariant polynomial}
  }
$$
in $\mathrm{dgAlg}_{\mathbb{R}}$.
 On the other hand, an $n$-connection with values in a Lie $n$-algebroid $\mathfrak{a}$ 
is a span of simplicial presheaves
\[
  \xymatrix{
    {\hat \Sigma} \ar[d]^{\simeq} \ar[r]^<<<<{\nabla} & \mathbf{cosk}\exp(\mathfrak{a})_{\mathrm{conn}}
    \\
    \Sigma
  }
\]
with coefficients in the simplicial presheaf 
$\mathbf{cosk}_{n+1} \exp(\mathfrak{a})_{conn}$, def. \ref{ChW}, 
that sends $U \in \mathrm{CartSp}$  
to the $(n+1)$-coskeleton, def. \ref{coskeleton}, 
of the simplicial set, which in degree $k$ is the set of commuting diagrams
$$
    \xymatrix{
       \Omega^\bullet_{\mathrm{\mathrm{vert}}}(U \times \Delta^k)
         \ar@{<-}[r]^{A_{\mathrm{\mathrm{vert}}}} &
       \mathrm{CE}(\mathfrak{a})
       & \mbox{transition function}
       \\
       \Omega^\bullet(U \times \Delta^k) \ar[u]
         \ar@{<-}[r]^{A} &
       \mathrm{W}(\mathfrak{a}) \ar[u]
       & \mbox{connection forms}
       \\
       \Omega^\bullet(U) \ar[u]
         \ar@{<-}[r]^{\langle F_A \rangle} &
       \mathrm{inv}(\mathfrak{a}) \ar[u]
       & \mbox{curvature characteristic forms}
    }
    \,,  
$$
such that the curvature forms $F_A$ 
of the $\infty$-Lie algebroid valued differential forms $A$ on $U \times \Delta^k
$ with values in $\mathfrak{a}$ in the middle
are horizontal.  

If $\mu$ is an $\infty$-Lie algebroid cocycle of degree $n$,  
then the $\infty$-Chern-Weil homomorphism operates by sending an $\infty$-connection 
given by a {\v C}ech cocycle with values in simplicial sets of such commuting diagrams to 
the obvious pasting composite
$$
  \xymatrix{
    \Omega^\bullet_{\mathrm{\mathrm{vert}}}(U \times \Delta^k)
     \ar@{<-}[r]^{A_{\mathrm{\mathrm{vert}}}}&
    \mathrm{CE}(\mathfrak{a})
    \ar@{<-}[r]^{\mu}&
    \mathrm{CE}(b^{n}\mathbb{R})
    &
    : \mu(A_{\mathrm{\mathrm{vert}}})
    &
    \\
    \Omega^\bullet(U \times \Delta^k) \ar[u]
     \ar@{<-}[r]^{A}&
    \mathrm{W}(\mathfrak{a}) \ar[u]
    \ar@{<-}[r]^{cs}&
    \mathrm{W}(b^n \mathbb{R}) \ar[u]
    &
    : \mathrm{cs}(A)
    &
    \mbox{Chern-Simons form}
    \\
    \Omega^\bullet(U) \ar[u]
     \ar@{<-}[r]^{\langle F_A \rangle} &
    \mathrm{inv}(\mathfrak{a}) \ar[u]
    \ar@{<-}[r]^{\langle-\rangle}&
    \mathrm{inv}(b^n \mathbb{R}) \ar[u]
    &
    : \langle F_A \rangle
    &
     \mbox{curvature}
  }
  \,.
$$
Under the map to the coskeleton the group of such cocycles for line $n$-bundle 
with connection is quotiented by the discrete group $\Gamma$ of periods of $\mu$, 
such that the $\infty$-Chern-Weil homomorphism is given by sending the 
$\infty$-connection $\nabla$ to
$$
  \xymatrix{
    {\hat \Sigma} 
    \ar[d]^\simeq
    \ar[r]^<<<<{\nabla}     
     & \mathbf{cosk}_n\exp(\mathfrak{a})_{\mathrm{conn}}
     \ar[r]^{\exp(\mathrm{cs})}
     &
     \mathbf{B}^n (\mathbb{R}/\Gamma)_{\mathrm{conn}}
    \\
    \Sigma
  }
  \,.
$$
This presents a circle $n$-bundle with connection, \ref{SmoothStrucDifferentialCohomology}, 
whose connection $n$-form is locally given by the Chern-Simons form $\mathrm{cs}(A)$. 
This is the Lagrangian of the $\infty$-Chern-Simons theory defined by 
$(\mathfrak{a},\langle - \rangle)$ and evaluated on the given $\infty$-connection. 
If $\Sigma$ is a smooth manifold of dimension $n$, then the higher holonomy, \ref{SmoothStrucChernSimons}, 
of this circle $n$-bundle over $\Sigma$ is the value of the Chern-Simons action. 
After a suitable gauge transformation this is given by the integral
$$
  \exp(i S(A)) = \exp(i \int_\Sigma \mathrm{cs}(A))
  \,,
$$
the value of the $\infty$-Chern-Simons action functional on the $\infty$-connection $A$.
\begin{proposition}
  \index{Chern-Simons functionals!equations of motion}
  Let $\mathfrak{g}$ be an $L_\infty$-algebra and $\langle -, \cdots, -\rangle$
  an invariant polynomial on $\mathfrak{g}$.
Then the $\infty$-connections $A$ with values in $\mathfrak{g}$ that satisfy the 
equations of motion of the corresponding $\infty$-Chern-Simons theory are precisely those for which
$$
  \langle -, F_A \wedge F_A \wedge \cdots F_A \rangle = 0
  \,,
$$
as a morphism $\mathfrak{g} \to \Omega^\bullet(\Sigma)$, where $F_A$ denotes the (in general inhomogeneous) curvature form of $A$.

In particular for binary and non-degenerate invariant polynomials the equations of motion are
$$
  F_A = 0
  \,.
$$
\end{proposition}
\proof
Let $A  \in \Omega(\Sigma \times I, \mathfrak{g})$ be a 1-parameter variation of $A(t = 0)$, 
that vanishes on the boundary $\partial \Sigma$. Here 
we write $t : [0,1] \to \mathbb{R}$ for the canonical coordinate on the interval.

$A(0)$ is critical if
$$
  \left(\frac{d}{d t} \int_{\Sigma} \mathrm{cs}(A)\right)_{t = 0} = 0
$$
for all extensions $A$ of $A(0)$. Using Cartan's magic formula 
and the Stokes theorem the left hand expression is 
$$
  \begin{aligned}
    \left(\frac{d}{d t}\int_{\Sigma} \mathrm{cs}(A)\right)_{t = 0}
      & =
     \left(\int_{\Sigma} \frac{d}{d t}\mathrm{cs}(A)\right)_{t = 0}
      \\
     & =
     \left(
     \int_{\Sigma} d \iota_{\partial t} \mathrm{cs}(A)
     + 
     \int_{\Sigma} \iota_{\partial_t} d \mathrm{cs}(A)
     \right)_{t = 0}
     \\
     & = 
     \left(
     \int_{\Sigma} d_\Sigma ( \iota_{\partial t} \mathrm{cs}(A))
     + 
     \int_{\Sigma} \iota_{\partial_t} \langle F_A \wedge \cdots F_A \rangle
     \right)_{t = 0}
     \\
     & =
     \left( \int_{\partial \Sigma} \iota_{\partial t} \mathrm{cs}(A)
     +
     n \int_{\Sigma}  \langle (\frac{d}{d t}A) \wedge \cdots F_A \rangle
     \right)_{t = 0}
     \\
     & =
     \left(
     n \int_{\Sigma}  \langle (\frac{d}{d t}A) \wedge \cdots F_A 
\rangle
     \right)_{t = 0}
  \end{aligned}
  \,.
$$
Here we used that $\iota_{\partial_t} F_A = \frac{d}{d t}A$ and that by assumption 
this vanishes on $\partial \Sigma$. Since $\frac{d}{d t}A $ can have arbitrary values, the
claim follows.
\endofproof

\subsubsection{Higher cup-product Chern-Simons theories}
\label{HigherCupProductCSTheories}
\label{CupProductCS}

We discuss a class of $\infty$-Chern-Simons functionals induced
from a smooth differential refinement of the \emph{cup-product}
on integral cohomology.

\medskip

This section draws from \cite{FiorenzaSatiSchreiberIV}.

\paragraph{General construction}

A crucial property of the Dold-Kan map, as discussed in \ref{SheafAndNonabelianDoldKan}, 
is the following.
\begin{proposition}
\label{cup-products}
Let $A, B$ and $C$ be presheaves of chain complexes concentrated in non-negative degrees, and let $\cup: A\otimes B\to C$ be a morphism  of presheaves of chain complexes. Then the Dold-Kan map induces a natural morphism of simplicial preseheaves $\cup_{\mathrm{DK}}: \mathrm{DK}(A)\times \mathrm{DK}(B)\to \mathrm{DK}(C)$
\end{proposition}
\begin{proof}
Both the categories $\mathrm{Ch}_\bullet^+$ and $\mathrm{sAb}$ are monoidal categories under the respective standard tensor products
(on $\mathrm{Ch}_\bullet^+$ this is given by direct sums of tensor products of abelian groups with fixed total degree 
and on $\mathrm{sAb}$ by the degreewise tensor product of abelian groups),
and the functor $\Gamma$ is lax monoidal with respect to these structures, 
i.e., for any $V, W \in \mathrm{Ch}_\bullet^+$ we have natural 
weak equivalences
$$
  \nabla_{V,W} : \Gamma(V) \otimes \Gamma(W) \to \Gamma(V \otimes W)
  \,.
$$
These are not isomorphisms, as they would be for a \emph{strong} monoidal functor,
but they are weak equivalences.
 The forgetful functor $F$ is the right adjoint to the functor forming degreewise the free abelian group on a set, therefore it preserves products and hence  there are natural isomorphisms
$$
   F(V \times W) \xrightarrow{\simeq} F(V) \times F(W)
   \,,
$$
for all $V,W \in \mathrm{sAb}$.
Finally, by the definition of tensor product, there are universal natural 
quotient maps $V, W \in \mathrm{sAb}$
$$
  p_{V,W} : V \times W \to V \otimes W
  \,.
$$
The morphism $\cup_{\mathrm{DK}}$ is then defined as the composition indicated in the following diagram:
$$
  \xymatrix{
    \mathrm{DK}(A) \times \mathrm{DK}(B)
    	 \ar@{=}[d]
	 \ar[rrrr]^{\mathbf{\cup}_{\mathrm{DK}}}
	 &&&&
	 \mathrm{DK}(C)
	 \ar@{=}[dd]
	 \\
	 F(\Gamma(A)) \times F(\Gamma(B))
	 \ar[d]^\simeq 
	 \\
     F(\Gamma(A) \times \Gamma(B))
	 \ar[r]^{F(p)}
	 &
     F(\Gamma(A) \otimes \Gamma(B))
	 \ar[r]^-{\hspace{-2mm}F(\nabla)}
	 &
     F(\Gamma(A\otimes B))
	 \ar[rr]^-{F(\Gamma(\cup))}
	 &&
	 F(\Gamma(C))\;.
  }
$$
\end{proof}

Given the presentation $\mathbf{H} \simeq L_W [\mathcal{C}^{\mathrm{op}}, \mathrm{sSet}]$,
for every presheaf of chain complexes $A$ on $\mathcal{C}$ we obtain a corresponding
$\infty$-stack, the $\infty$-stackification of the image of $A$
under the Dold-Kan map, which we will denote by the same symbol: $\mathrm{DK}(A) \in \mathbf{H}$.

\begin{definition}
 \label{DKGivesStack}
For $A \in [\mathcal{C}^{\mathrm{op}}, \mathrm{Ab}]$ a sheaf of abelian groups, we write 
$A[n] \in [\mathcal{C}^{\mathrm{op}}, \mathrm{Ch}_\bullet^+]$ for the corresponding presheaf
of chain complexes concentrated on $A$ in degree $n$, and
$$
  \mathbf{B}^n A \simeq \mathrm{DK}(A[n]) \in \mathbf{H}
$$
for the corresponding $\infty$-stack.
\end{definition}

In this case the corresponding cohomology
$$
  H^n(X,A) = \pi_0 \mathbf{H}(X, \mathbf{B}^n A)
$$
is the traditional \emph{sheaf cohomology} of $X$ with coefficients in $A$.
More generally, if $A \in [\mathcal{C}^{\mathrm{op}}, \mathrm{Ch}_\bullet^+]$ is a sheaf
of chain complexes not necessarily concentrated in one degree, then 
$$
  H^0(X, A) := \pi_0 \mathbf{H}(X, A)
$$
is what traditionally is called the \emph{sheaf hypercohomology} of 
$X$ with coefficients in $A$.
The central coefficient object in which we are interested here is the 
sheaf of chain complexes called the \emph{Deligne complex}, to which we 
now turn.

The \emph{Beilinson-Deligne cup product} is an explicit presentation of the cup product in ordinary differential cohomology for the case that the latter is modeled by the \v{C}ech-Deligne cohomology. 

\begin{definition}
The Beilinson-Deligne cup product is the morphism of sheaves of chain complexes 
$$
\cup_{\mathrm{BD}}: \mathbb{Z}[p+1]^\infty_D\otimes  \mathbb{Z}[q+1]^\infty_D ~\longrightarrow~ \mathbb{Z}[(p+1)+(q+1)]^\infty_D,
$$
given on homogeneous elements $\alpha$, $\beta$ as follows:

$$
  \alpha \cup_{\mathrm{BD}} \beta :=
  \left.
    \begin{cases}
    \alpha \wedge \beta  = \alpha \beta  &\text{ if } \mathrm{deg}(\alpha) = p+1\;.
    \\
    \alpha \wedge d_{\mathrm{dR}}\beta &\text{ if }  \mathrm{deg}(\alpha) \leq p 
    \text{ and } \mathrm{deg}(\beta) = 0
    \\
    0  &\text{ otherwise}\;.
    \end{cases}
  \right.
  \,.
$$
\end{definition}
\begin{remark}
 When restricted to the diagonal in the case that $p= q$, 
 this means that the cup product sends a $p$-form $\alpha$ to the $(2p+1)$-form 
 $\alpha \wedge d_{\mathrm{dR}} \alpha$. This is of course the local Lagrangian for 
 cup product Chern-Simons theory of $p$-forms. We discuss this case in detail in
 section \ref{4k+3}.
\end{remark}
The Beilinson-Deligne cup product is associative and commutative up to homotopy, so it induces an associative and commutative cup product on ordinary differential cohomology.
A survey of this can be found in \cite{Brylinski} (around Prop. 1.5.8 there).

\begin{definition}
For $p,q \in \mathbb{N}$
the morphism of simplicial presheaves
$$
  \mathbf{\cup}_{\mathrm{conn}} : 
    \mathbf{B}^{p}U(1)_{\mathrm{conn}} \times
   \mathbf{B}^{q}U(1)_{\mathrm{conn}}
   \to
   \mathbf{B}^{p+q+1} U(1)_{\mathrm{conn}}
$$
is the morphism associated to the Beilinson-Deligne cup product $\cup_{\mathrm{BD}}: \mathbb{Z}[p+1]^\infty_D\otimes  \mathbb{Z}[q+1]^\infty_D ~\longrightarrow~ \mathbb{Z}[p+q+2]^\infty_D$ by Proposition \ref{cup-products}.
 \label{CupOnStacks}
 \label{ExtendedDifferentialCup}
\end{definition}

\vspace{3mm}
\noindent Since
the Beilinson-Deligne cup product is associative up to homotopy, this induces a well defined morphism
$$
\mathbf{B}^{n_1}U(1)_{\mathrm{conn}}\times \mathbf{B}^{n_2}U(1)_{\mathrm{conn}}\times \cdots \times \mathbf{B}^{n_{k+1}}U(1)_{\mathrm{conn}}\to \mathbf{B}^{n_1+\cdots+n_{k+1}+k}U(1)_{\mathrm{conn}}.
$$
In particular, if $n_1=\cdots=n_{k+1}=3$, we find
$$
\left(\mathbf{B}^{3}U(1)_{\mathrm{conn}}\right)^{k+1}\to \mathbf{B}^{4k+3}U(1)_{\mathrm{conn}}.
$$
Furthermore, we see from the explicit expression of the Beilinson-Deligne cup product that, on a local chart $U$, if the 3-form datum of a connection on a $U(1)$-3-bundle is the 3-form $C$, then the $4k+3$-form local datum for the corresponding connection on the associated $U(1)$-$(4k+3)$-bundle is
\(
C\wedge \underbrace{dC\wedge \cdots \wedge{dC}}_{k\text{ times}}.
\label{Eq ktimes}
\)

\subsubsection{Higher differential Dixmier-Douady class and higher dimensional $U(1)$-holonomy}
\label{VolumeHolonomy}

The \emph{degenerate} or rather \emph{tautological} 
case of extended $\infty$-Chern-Simons theories nevertheless
deserves special attention, since it appears universally in all other examples:
that where the extended action functional is the \emph{identity} morphism
$$
  (\mathbf{DD}_n)_{\mathrm{conn}} 
    : 
   \xymatrix{
	\mathbf{B}^n U(1)_{\mathrm{conn}} 
     \ar[r]^{\mathrm{id}}
	 &
	 \mathbf{B}^n U(1)_{\mathrm{conn}}
    }\;,
$$
for some $n \in \mathbb{N}$.
Trivial as this may seem, this is the differential refinement of what 
is called the (higher) \emph{universal Dixmier-Douady class} 
the higher universal first Chern class -- of circle $n$-bundles
/ bundle $(n-1)$-gerbes, which on the topological classifying space $B^n U(1)$
is the weak homotopy equivalence
$$
  \mathrm{DD}_n
  :
  \xymatrix{
    B^n U(1)
	\ar[r]^{\hspace{-3mm}\simeq}
	&
	K(\mathbb{Z}, n+1)
  }
  \,.
$$
Therefore, we are entitled to consider
$(\mathbf{DD}_n)_{\mathrm{conn}}$ as the extended action functional
of an $n$-dimensional $\infty$-Chern-Simons theory. Over an $n$-dimensional
manifold $\Sigma_n$ the moduli $n$-stack of field configurations is 
that of circle $n$-bundles with connection on $\Sigma_n$.
In generalization to how a circle 1-bundle with connection
has a \emph{holonomy} over closed 1-dimensional manifolds, we note
 that
 a circle $n$-connection
has a \emph{$n$-volume holonomy} over the $n$-dimensional manifold $\Sigma_n$.
This is the ordinary (codimension-0) action functional associated to $(\mathbf{DD}_n)_{\mathrm{conn}}$
regarded as an extended action functional:
$$
  \mathrm{hol} := \exp(2 \pi i \int_{\Sigma_n} [\Sigma_n, (\mathbf{DD}_n)_{\mathrm{conn}}])
  :
  [\Sigma_n, \mathbf{B}^n U(1)_{\mathrm{conn}}]
  \to 
  U(1)
  \,.
$$
This formulation makes it manifest that, for $G$ any smooth $\infty$-group and
$\mathbf{c}_{\mathrm{conn}} : \mathbf{B}G_{\mathrm{conn}} \to \mathbf{B}^n U(1)_{\mathrm{conn}}$
any extended $\infty$-Chern-Simons action functional in codimension $n$, the 
induced action functional is indeed the $n$-volume holonomy of a family of 
``Chern-Simons circle $n$-connections'', in that we have
$$
  \exp(2 \pi i \int_{\Sigma_n} [\Sigma_n, \mathbf{c}_{\mathrm{conn}}])
  \simeq
  \mathrm{hol}_{\mathbf{c}_{\mathrm{conn}}}
  \,.
$$
This is most familiar in the case where the moduli $\infty$-stack $\mathbf{B}G_{\mathrm{conn}}$
is replaced with an ordinary smooth oriented manifold $X$ 
(of any dimension and not necessarily
compact). In this case $\mathbf{c}_{\mathrm{conn}} : X \to \mathbf{B}^n U(1)_{\mathrm{conn}}$
modulates a circle $n$-bundle with connection $\nabla$ on this smooth manifold. 
Now regarding
this as an extended Chern-Simons action function in codimension $n$ means to 
\begin{enumerate}
  \item
    take the moduli stack of fields over a given closed oriented
	manifold $\Sigma_n$ to be $[\Sigma_n,X]$, which is simply the
	space of maps between these manifolds, equipped with its natural 
	(``diffeological'') smooth structure (for instance the smooth loop space
	$L X$ when $n = 1$ and $\Sigma_n = S^1$);
  \item 
	take the value of the action functional on a field configuration
	$\phi : \Sigma_n \to X$ to be the $n$-volume holonomy
	of $\nabla$
	$$
	  \mathrm{hol}_\nabla(\phi)
	  =
	  \exp(2 \pi i \int_{\Sigma_n} [\Sigma_n, \mathbf{c}_{\mathrm{conn}}] )
	  :
	  \xymatrix{
	    [\Sigma_n, X]
		\ar[rr]^-{[\Sigma_n,\mathbf{c}_{\mathrm{conn}}]}
		&&
		[\Sigma_n, \mathbf{B}^n U(1)_{\mathrm{conn}}]
		\ar[rrr]^-{\exp(2 \pi \int_{\Sigma_n}(-))}
		&&&
		U(1)
	  }
	  \,.
	$$
\end{enumerate}
Using the proof of Prop. \ref{FiberInte} to unwind this in terms of local differential
form data, this reproduces the familiar formulas for (higher) $U(1)$-holonomy.

\subsubsection{1d Chern-Simons functionals}
\label{1dCSTheories}

We discuss examples of the intrinsic notion of 
$\infty$-Chern-Simons action functionals, \ref{SmoothStrucChernSimons},
over 1-dimensional base spaces.

\begin{example}
  For some $n \in \mathbb{N}$ let
  $$
    \mathrm{tr} : \mathfrak{u}(n) \to \mathfrak{u}(1) \simeq \mathbb{R}
  $$
  be the trace function, with respect to the canonical identification of
  $\mathfrak{u}(n)$ with the Lie algebra of skew-Hermitean complex
  matrices.
  
  This is both a 1-cocycle as well as an invariant polynomial on $\mathfrak{u}(n)$, 
  the former corresponding to a degree-1 element in the Chevalley-Eilenberg 
  algebra $\mathrm{CE}(\mathfrak{u}(n))$ and 
  the latter corresponding to 
  an element 
  $d_{\mathrm{W}} c \in \mathrm{W}(\mathfrak{u}(n))$ of degree 2
  in the Weil algebra. Hence $c$ is also the corresponding
  Chern-Simons element, def. \ref{TransgressionAndCSElements}.
  By prop. \ref{PresentationOfDifferentialFirstChernClass} this 
  controls the universal differential first Chern class.
  
  The corresponding Chern-Simons action functional is defined on the
  groupoid of $\mathfrak{u}(n)$-valued differential 1-forms on 
  a line segment $\Sigma$ and given by
  $$
    A \mapsto \int_\Sigma \mathrm{tr}(A)
	\,.
  $$
  Any choice of coordinates $\Sigma \hookrightarrow \mathbb{R}$
  canonically identifies $A \in \Omega^1(\Sigma, \mathfrak{u}(n))$
  with a $\mathfrak{u}(n)$-valued function $\phi$. We may think of 
  $\bar \phi := \int_\Sigma A = \int_\Sigma \phi d t$ as the average 
  of this function. In terms of this the action functional is 
  simply the trace function itself
  $$
    \bar \phi \mapsto \mathrm{tr}(\bar \phi)
	\,.
  $$  
Degenerate as this case is, it is sometimes useful to regard the trace as
an example of 1-dimensional Chern-Simons theory, for instance in the
context of large-$N$ compactified gauge theory as discussed in
\cite{Nair}.  
\end{example}

\begin{example}
  Below in \ref{InfinCSAKSZ} we discuss in detail how
  (derived) $L_\infty$-algebroids equipped with 
  non-degenerate binary invariant polynomials
  of \emph{grade} 0 (hence total degree 2)
  give rise to 1-dimensional Chern-Simons theories.
\end{example}

\subsubsection{3d Chern-Simons functionals}
\label{3dCSTheories}

We discuss examples of the intrinsic notion of 
$\infty$-Chern-Simons action functionals, \ref{SmoothStrucChernSimons},
over 3-dimensional base spaces. This includes the archetypical
example of ordinary 3-dimensional Chern-Simons theory,
but also its discrete analog, Dijkgraaf-Witten theory.

\begin{itemize}
  \item \ref{InfinCSOrdinaryCS} -- Ordinary Chern-Simons theory;
  \item \ref{InfinCSDW} -- Ordinary Dijkgraaf-Witten theory.
\end{itemize}

\paragraph{Ordinary Chern-Simons theory for simply connected simple gauge group}
 \label{InfinCSOrdinaryCS}
 \index{Chern-Simons functionals!ordinary Chern-Simons theory}

We discuss the action functional of ordinary 3-dimensional 
Chern-Simons theory (see \cite{FreedCS} for a survey) from the 
point of view of intrinsic Chern-Simons action functionals in
$\mathrm{Smooth}\infty\mathrm{Grpd}$.

\subparagraph{Extended Lagrangian and action functional}

\begin{theorem}
  Let $G$ be a simply connected compact simple Lie group. 
  For 
  $$
    [c] \in H^4(B G, \mathbb{Z}) \simeq \mathbb{Z}
  $$
  a universal characteristic class that generates the degree-4
  integral cohomology of the classifying space $B G$, there is
  an essentially unique smooth lift $\mathbf{c}$ of 
  the characteristic map $c$ of the form
  $$
    \mathbf{c} : \mathbf{B}G \to \mathbf{B}^3 U(1)
	\;\;\;\;\;
	\in
	\mathrm{Smooth}\infty\mathrm{Grpd}
  $$
  on the smooth moduli stack $\mathbf{B}G$ of smooth $G$-principal
  bundles with values in the smooth moduli 3-stack of smooth circle 3-bundles.
  The differential refinement
  $$
    \mathbf{L} : \mathbf{B}G_{\mathrm{conn}} \to 
	 \mathbf{B}^3 U(1)_{\mathrm{conn}}
	\;\;\;\;\;
	\in
	\mathrm{Smooth}\infty\mathrm{Grpd}
  $$
  to the moduli stacks of the corresponding $n$-bundles with $n$-connections
  induces over any any closed oriented 3-dimensional smooth manifold $\Sigma$
  a smooth functional
  $$
    \exp(i S_{\mathrm{CS}}(-))
	:=
	 \exp(2 \pi i \int_{\Sigma} [\Sigma, \mathbf{L}])
	:
    \xymatrix{
      [\Sigma, \mathbf{B}G_{\mathrm{conn}}]
	  \ar[r]^{\hat {\mathbf{c}}}
	  &
	  [\Sigma, \mathbf{B}^3 U(1)_{\mathrm{conn}}]
	  \ar[rr]^-{\exp(2\pi i\int_\Sigma(-))}
	  &&
	  U(1)
	}
  $$
  on the moduli stack of $G$-principal connections on $\Sigma$,
  which on objects $A \in \Omega^1(\Sigma,\mathfrak{g})$ is 
  the exponentiated Chern-Simons action functional
  $$
    \exp(i S_{\mathrm{CS}}(A))
	=
	\exp(i \int_\Sigma \langle A \wedge d_{\mathrm{dR}} A\rangle
	+ 
	\frac{1}{6} \langle A \wedge [A \wedge A]\rangle
	)
	\,.
  $$
\end{theorem}
\proof  
  This is theorem \ref{FirstFractionalDifferentialPontrjagin}
  combined with \ref{IntrinsicIntegrationInSmooth}.
\endofproof
For more computational details that go into this see also
\ref{OrdinaryChernSimonsTheoryAsAKSZ} below

\subparagraph{The extended phase spaces}

Let $G$ be a connected and simply connected Lie group. 
We discuss the nature of the moduli of $G$-principal connections $G \mathbf{Conn}(\Sigma)$
according to \ref{SmoothStructDifferentialModuli} for various choices of $\Sigma$.

\begin{proposition}
 There is an equivalence
 $$
   \mathrm{hol}
   :
   \xymatrix{
     G\mathbf{Conn}(S^1)
	 \ar[r]^-\simeq
	 &
     G /\!/_{\mathrm{Ad}}G
   }
 $$
 in $\mathrm{Smooth}\infty\mathrm{Grpd}$ between the moduli stack of $G$-principal connections
 on the circle, def. \ref{GLieGroupBGConnXByFiberProductOfSharpn}, and the quotient groupoid
 of the adjoint action of $G$ on itself. 
 This is given by sending $G$-principal connections to their holonomy 
 (for any chosen basepoint on the circle).
\end{proposition}
\proof
  We show that for each $U \in \mathrm{CartSp}$ the 
  morphism of groupoids $\mathrm{hol}_U$ is an equivalence of groupoids. 
  
  For $f : U \to G$ a smooth function, since $G$ is connected and $U$ is topologically
  contractible, we may find a smooth homotopy
  $$
    \eta : [0,1] \times U  \to G
  $$
  with $\eta(0)$ constant on the neutral element in a neighbourhood of
  $\{0\}\times U$ and with $\eta(1) = f$ in a neighbourhood of $\{1\}\times U$.
  Let then $\eta \mathbf{d}_{[0,1]} \eta^{-1} \in \Omega^1(U\times S^1, \mathfrak{g})$.
  This is a connection 1-form on $U \times S^1 $ whose holonomy is $f$. Hence 
  $\mathrm{hol}_U$ is essentially surjective.
  
  Next, consider $A, A' \in \Omega^1(U \times S^1, \mathfrak{g})$ two connection
  1-forms (legs along $S^1$). Observe that for each point $u \in U$ a gauge transformation
  $g_u : A_u \to A'_u$ is fixed already by its value at the basepoint of $S^1$
  and moreover it has to satisfy
  $$
    \mathrm{hol}(A_u) = g_u \mathrm{hol}(A'_u) g_u^{-1}
	\,.
  $$
  This is because for every $t \in [0,1]$ the gauge transformation needs to satisfy the 
  parallel transprt naturality condition
  $$
    \raisebox{20pt}{
    \xymatrix{
	  {*} \ar[r]^{g_u(t)} & {*}
	  \\
	  {*} \ar[r]_{g_u(0)} \ar[u]^{\mathrm{tra}_{A_u}(0,t)} & {*} \ar[u]_{\mathrm{tra}_{A'_u}(0,t)}	  
	}
	}
	\;\;\;\;\;\;
	\in */\!/G
	\,,
  $$
  where $\mathrm{tra}_{A_u}(0,t)$ is the parallel transport of the connection $A_u$
  along $[0,t]$.
  
  This says that $\mathrm{hol}_{U}$ is also full and faithful. Hence it is an 
  equivalence.
\endofproof

\begin{remark}
  We have a dashed lift in 
  $$
    \raisebox{20pt}{
    \xymatrix{
	 & [S^1, \mathbf{B}G_{\mathrm{conn}}]
	 \ar[d]^{\mathrm{conc}}
	 \\
	 & G\mathbf{Conn}(S^1)
	 \ar[d]_\simeq^{\mathrm{hol}}
	 \\
	 G\ar[r] \ar@/^.6pc/@{-->}[uur] & G/\!/_{\mathrm{Ad}}G
	}
	}
	\,,
  $$
  where the top right morphism is the canonical projection of remark 
  \ref{ProjectionFromMappingSpaceIntoBGconnToDifferentialModuli}, 
  and where the bottom horizontal morphism is the canonical projection map. 
\end{remark}

\begin{proposition}
  There is an equivalence
  $$
    G\mathbf{Conn}(*) \simeq \mathbf{B}G
	\,.
  $$
\end{proposition}

\paragraph{Ordinary 3d $U(1)$-Chern-Simons theory and generalized $B_n$-geometry}
  \label{3dU1CS}

Ordinary 3-dimensional $U(1)$-Chern-Simons theory 
on a closed oriented manifold $\Sigma_3$ 
contains field configurations which are given by globally defined 1-forms
$A \in \Omega^1(\Sigma_3)$ and on which the action functional is given by
the familiar expression
$$
  \exp(i S(A)) = \exp(2 \pi i k \int_{\Sigma_3} A \wedge d_{\mathrm{dR}} A)
  \,.
$$
More generally, though, a field configuration of the theory is a connection 
$\nabla$ on 
a $U(1)$-principal bundle $P \to \Sigma_3$ and this simple formula
is modified, from being the exponential of the ordinary integral of 
the wedge product of two differential forms, to the fiber integration
in differential cohomology, Def. \ref{FiberInte}, of the differential cup-product,
Def. \ref{CupOnStacks}:
$$
  \exp(i S(\nabla)) = \exp(2 \pi i k \int_{\Sigma_3} \nabla \cup_{\mathrm{conn}} \nabla)
  \,.
$$
This defines the action functional on the set 
$H^1_{\mathrm{conn}}(\Sigma_3, U(1))$ of equivalence classes of $U(1)$-principal bundles
with connection
$$
  \exp(i S(-)) : H^1_{\mathrm{conn}}(\Sigma_3) \to U(1)
  \,.
$$
That the action functional is gauge invariant means that it extends from
a function on gauge equivalence classes to a functor on the groupoid 
$\mathbf{H}^1_{\mathrm{conn}}(\Sigma_3, U(1))$, whose objects are actual
$U(1)$-principal connections, and whose morphsims are smooth gauge transformations
between these:
$$
  \exp(i S(-)) : \mathbf{H}^1_{\mathrm{conn}}(\Sigma_3) \to U(1)
  \,.
$$
Finally, that the action functional depends \emph{smoothly} on the connections
means that it extends further to the moduli stack of fields to a morphism of stacks
$$
  \exp(i S(-)) : [\Sigma_3, \mathbf{B} U(1)_{\mathrm{conn}}] \to U(1)
  \,.
$$

\medskip
The fully extended prequantum circle 3-bundle of this extended 3d Chern-Simons theory
is that of the two-species theory restricted along the 
diagonal $\Delta : \mathbf{B}U(1)_{\mathrm{conn}} 
\to \mathbf{B}U(1)_{\mathrm{conn}}\times \mathbf{B}U(1)_{\mathrm{conn}}$.
This is the homotopy fiber of the smooth cup square in these degrees.

\medskip
According to \cite{Hitchin} 
aspects of the differential geometry of 
the homotopy fiber of a differential refinement of this 
cup square are captured by the ``generalized geometry of $B_n$-type''
that was suggested in section 2.4 of \cite{Baraglia}.
In view of the relation of the same structure to differential T-duality discussed above
one is led to expect that ``generalized geometry of $B_n$-type''
captures aspects of the differential cohomology
on fiber products of torus bundles that exhibit auto T-duality on differential K-theory. 
Indeed, such a relation is pointed out in \cite{Bouwknegt}\footnote{
Thanks, once more, to Alexander Kahle, for discussion of this point, at \emph{String-Math 2012}.}. 

\paragraph{Ordinary Dijkgraaf-Witten theory}
 \label{InfinCSDW}
  \index{Chern-Simons functionals!Dijkgraaf-Witten theory}

Dijkgraaf-Witten theory (see \cite{FreedQuinn} for a survey) is 
commonly understood as
the analog of Chern-Simons theory 
for discrete structure groups. We show that this becomes a precise 
and systematic statement in 
$\mathrm{Smooth} \infty \mathrm{Grpd}$: the Dijkgraaf-Witten action functional is that induced from 
applying the $\infty$-Chern-Simons homomorphism to a characteristic class of the form 
$\mathrm{Disc} B G \to \mathbf{B}^3 U(1)$, for 
$\mathrm{Disc} : \infty \mathrm{Grpd} \to \mathrm{Smooth} \infty \mathrm{Grpd}$ 
the canonical embeedding of discrete $\infty$-groupoids, 
\ref{DiscreteInfGroupoids}, into all smooth $\infty$-groupoids.

\medskip

Let $G \in \mathrm{Grp} \to \infty \mathrm{Grpd} \stackrel{\mathrm{Disc}}{\to} \mathrm{Smooth} \infty \mathrm{Grpd}$
be a discrete group regarded as an $\infty$-group object in 
discrete $\infty$-groupoids and hence as a smooth $\infty$-groupoid with discrete smooth cohesion. 
Write $B G = K(G,1) \in \infty \mathrm{Grpd}$ for its delooping in $\infty\mathrm{Grpd}$ and 
$\mathbf{B}G = \mathrm{Disc} B G$ for its delooping in $\mathrm{Smooth}\infty \mathrm{Grpd}$. 

We also write $\Gamma \mathbf{B}^n U(1) \simeq K(U(1), n)$. 
Notice that this is different from $B^n U(1) \simeq \Pi \mathbf{B}U(1)$, 
reflecting the fact that $U(1)$ has non-discrete smooth structure.
\begin{proposition}
For $G$ a discrete group, morphisms $\mathbf{B}G \to \mathbf{B}^n U(1)$ correspond precisely 
to cocycles in the ordinary group cohomology of $G$ with coefficients in the discrete 
group underlying the circle group
$$
  \pi_0 \mathrm{Smooth}\infty \mathrm{Grpd}(\mathbf{B}G, \mathbf{B}^n U(1))
  \simeq
   H^n_{\mathrm{Grp}}(G,U(1))
  \,.
$$
\end{proposition}
\proof
By the $(\mathrm{Disc} \dashv \Gamma)$-adjunction we have
$$
  \mathrm{Smooth}\infty \mathrm{Grpd}(\mathbf{B}G, \mathbf{B}^n U(1))
  \simeq
  \infty \mathrm{Grpd}(B G, K(U(1),n))
  \,.
$$
\endofproof
\begin{proposition}
For $G$ discrete 
\begin{itemize}
\item the intrinsic de Rham cohomology of $\mathbf{B}G$ is trivial
  $$
    \mathrm{Smooth} \infty \mathrm{Grpd}(\mathbf{B}G, \mathbf{\flat}_{\mathrm{\mathrm{dR}}}\mathbf{B}^n U((1))
    \simeq *
    ;
  $$
\item all $G$-principal bundles have a unique flat connection
  $$
    \mathrm{Smooth}\infty\mathrm{Grpd}(X, \mathbf{B}G)
     \simeq
    \mathrm{Smooth}\infty \mathrm{Grpd}(\Pi(X), \mathbf{B}G)
    \,.
  $$
\end{itemize}
\end{proposition}
\proof
By the $(\mathrm{Disc} \dashv \Gamma)$-adjunction and using that 
$\Gamma \circ \mathbf{\flat}_{\mathrm{\mathrm{dR}}} K \simeq {*}$ for all $K$.
\endofproof
It follows that for $G$ discrete 
\begin{itemize}
\item any characteristic class 
$\mathbf{c} : \mathbf{B}G \to \mathbf{B}^n U(1)$ is a group cocycle;
\item 
  the $\infty$-Chern-Weil homomorphism coincides with postcomposition with this class
  $$
    \mathbf{H}(\Sigma, \mathbf{B}G) \to \mathbf{H}(\Sigma, \mathbf{B}^n U(1))
    \,.
  $$
\end{itemize}
\begin{proposition}
For $G$ discrete and $\mathbf{c} : \mathbf{B}G \to \mathbf{B}^3 U(1)$ 
any group 3-cocycle, the $\infty$-Chern-Simons theory action functional on a 3-dimensional 
manifold $\Sigma$
$$
  \mathrm{Smooth}\infty \mathrm{Grpd}(\Sigma, \mathbf{B}G)
  \to 
  U(1)
$$
is the action functional of Dijkgraaf-Witten theory.
\end{proposition}
\proof
By proposition \ref{IntrinsicIntegrationInSmooth} the morphism is given
by evaluation of the pullback of the cocycle $\alpha : B G \to B^3 U(1)$
along a given $\nabla : \Pi(\Sigma) \to B G$, on the fundamental 
homology class of $\Sigma$. This is the definition of the 
Dijkgraaf-Witten action (for instance equation (1.2) in \cite{FreedQuinn}).

\endofproof

\subsubsection{4d Chern-Simons functionals}
\label{4dCSFunctionals}

We discuss some 4-dimensional Chern-Simons functionals
\begin{itemize}
  \item \ref{InfinCSBF} -- 4d BF theory and topological Yang-Mills;
  \item \ref{4dYetterModel} -- 4d Yetter model.
\end{itemize}

\paragraph{BF theory and topological Yang-Mills theory}
 \label{InfinCSBF}
   \index{Chern-Simons functionals!BF-theory}
  \index{Chern-Simons functionals!topological Yang-Mills theory}

We discuss how the action functional of nonabelian
\emph{BF-theory} \cite{Horowitz} in 4-dimensions 
with a ``cosmological constant'' and coupled to topological 
Yang-Mills theory is a higher Chern-Simons theory. 

\medskip

Let $\mathfrak{g} = (\mathfrak{g}_2 \stackrel{\partial}{\to} \mathfrak{g}_1)$ be a strict Lie 2-algebra, coming from a differential crossed module,
def. \ref{DifferentialCrossedModule}, as indicated.
Let $\exp(\mathfrak{g})$ be the universal Lie integration, 
according to def. \ref{ExponentiatedLInftyAlgbra}.
Field configurations with values in $\exp(\mathfrak{g})$ are locally
Lie 2-algebra valued forms $(A \in \Omega^1(\Sigma, \mathfrak{g}_0))$
and $B \in \Omega^2(\Sigma, \mathfrak{g}_1)$ as in prop. 
\ref{2GroupoidOfLie2AlgebraValuedForms}.

The following observation is due to \cite{SSSI}.
\begin{proposition}
We have
\begin{enumerate} 
\item every invariant polynomial 
$\langle -\rangle_{\mathfrak{g}_1} \in \mathrm{inv}(\mathfrak{g}_1)$ 
on $\mathfrak{g}_1$ gives rise, 
under the canonical inclusion $\mathrm{inv}(\mathfrak{g}_1) \hookrightarrow
\mathrm{W}(\mathfrak{g})$, not to an invariant polynomial,
but to a Chern-Simons element on $\mathfrak{g}$, 
exhibiting the transgression to a trivial $L_\infty$-algebra cocycle;

\item for $\mathfrak{g}_1$ a semisimple Lie algebra and $\langle - \rangle_{\mathfrak{g}_1}$ 
the Killing form, $\Sigma$ a 4-dimensional compact manifold,
the corresponding Chern-Simons action functional 
$$
  \exp(i S_{\langle -\rangle_{\mathfrak{g}_1}})
  : 
  [\Sigma, \exp(\mathfrak{g})_{\mathrm{conn}}]
  \to 
  \mathbf{B}^4 \mathbb{R}_{\mathrm{conn}}
$$
on Lie 2-algebra valued forms is
   $$
     \xymatrix{
       \Omega^\bullet(X) \ar@{<-}[r]^{(A,B)}
        &
     \mathrm{W}(\mathfrak{g}_2 \to \mathfrak{g}_1)
      \ar@{<-}[rr]^{(\langle - \rangle_{\mathfrak{g}_1}, d_W \langle - \rangle_{\mathfrak{g}_1} )}
      &&
     \mathrm{W}(b^{n-1} \mathbb{R})
     }
   $$
   the sum of the action functionals of topological Yang-Mills theory with 
  BF-theory with cosmological constant:
   $$
     \mathrm{cs}_{\langle-\rangle_{\mathfrak{g}_1}}(A,B) 
     = 
     \langle F_A \wedge F_A\rangle_{\mathfrak{g}_1}
     - 
     2\langle F_A \wedge \partial B\rangle_{\mathfrak{g}_1}
     +
     2\langle \partial B \wedge \partial B\rangle_{\mathfrak{g}_1}   
     \,,
   $$
   where $F_A$ is the ordinary curvature 2-form of $A$.
\end{enumerate}
\end{proposition}
\proof
For $\{t_a\}$ a basis of $\mathfrak{g}_1$ and $\{b_i\}$ a basis of $\mathfrak{g}_2$ we have
$$
  d_{\mathrm{W}(\mathfrak{g})} : \mathbf{d} t^a \mapsto 
  d_{\mathrm{W}(\mathfrak{g}_1)}
  + 
  \partial^a{}_i \mathbf{d} b^i
  \,.
$$
Therefore with $\langle -\rangle_{\mathfrak{g}_1} 
  = P_{a_1 \cdots a_n} \mathbf{d} r^{a_1} \wedge \cdots \mathbf{d} t^{a_n}$ we have
$$
  d_{\mathrm{W}(\mathfrak{g})} \langle - \rangle_{\mathfrak{g}_1}
  = 
  n P_{a_1 \cdots a_n}\partial^{a_1}{}_i  
    \mathbf{d} b^{i} \wedge \cdots \mathbf{d} t^{a_n}
  \,.
$$
The right hand is a polynomial in the shifted generators of 
$\mathrm{W}(\mathfrak{g})$, and hence an invariant polynomial on 
$\mathfrak{g}$. Therefore $\langle - \rangle_{\mathfrak{g}_1}$ is a 
Chern-Simons element for it. 

Now for $(A,B) \in \Omega^1(U \times \Delta^k, \mathfrak{g})$ an $L_\infty$-algebra-valued form, 
we have that the 2-form curvature is
$$
  F_{(A,B)}^1 = F_A - \partial B
  \,.
$$
Therefore
$$
  \begin{aligned}
    \mathrm{cs}_{\langle -\rangle_{\mathfrak{g}_1}}(A,B)
    & = \langle 
	     F_{(A,B)}^1
		 \wedge
	     F_{(A,B)}^1
	  \rangle_{\mathfrak{g}_1}
    \\
    & = 
    \langle F_A \wedge F_A\rangle_{\mathfrak{g}_1}
    - 
    2\langle F_A \wedge \partial B\rangle_{\mathfrak{g}_1}
    +
    2\langle \partial B \wedge \partial B\rangle_{\mathfrak{g}_1}   
  \end{aligned}
  \,.
$$
\endofproof

\paragraph{4d Yetter model}
 \label{4dYetterModel}
  \index{Chern-Simons functionals!4d Yetter model}

The discussion of 3-dimensional Dijkgraaf-Witten theory as in 
\ref{InfinCSDW} goes through verbatim for discrete groups 
generalized to discrete $\infty$-groups $G$, \ref{DiscStrucCohesiveGroups},
and cocycles $\alpha : \mathbf{B}G \to \mathbf{B}^n U(1)$
of any degree $n$. A field configurations over an $n$-dimensional
manifold $\Sigma$ is a $G$-principal $\infty$-bundle,
\ref{DiscStrucPrincipalInfinityBundles}, necessarily flat, and
the induced action functional
$$
  \exp(i S_\alpha) : \mathbf{H}(\Sigma, \mathbf{B}G)
  \to 
  U(1)
$$
sends a $G$-principal $\infty$-bundle classified by a cocycle
$g : \Sigma \to \mathbf{B}G$ to the canonical pairing of the
singular cocycle corresponging to 
$\alpha(g) : \Sigma \to \mathbf{B}G \stackrel{\alpha}{\to} \mathbf{B}^n U(1)$
with the fundamental class of $\Sigma$.

For $n = 4$ such action functionals sometimes go by the name 
``Yetter model'' \cite{Mackaay}\cite{MartinsPorter}, 
in honor of \cite{Yetter}, which however did non consider a 
nontrivial 4-cocycle.

\subsubsection{Abelian gauge coupling of branes}
\label{GaugeCouplingOfBranes}
\index{Chern-Simons functionals!gauge coupling of branes}

The gauge coupling term in the action of an $(n-1)$-brane charged under
an abelian $n$-form background gauge field (electromagnetism, $B$-field, $C$-field,
etc.) is an example of an $\infty$-Chern-Simons functional.
We spell this out in a moment. 
Here one typically considers the target space of the $(n-1)$-brane to be a smooth
manifold or at most an orbifold. The formal structure, however, allows
to consider target spaces that are arbitrary  
smooth $\infty$-groupoids / smooth $\infty$-stacks. 
When generalized to this class of target spaces,
the class of brane gauge coupling functionals in fact coincides with that
of \emph{all} $\infty$-Chern-Simons functionals. 
Conversely, every $\infty$-Chern-Simons theory in dimension $n$
may be regarded as the field theory of a ``topological $(n-1)$-brane'' 
whose target space is the higher moduli stack of field configurations
of the given $\infty$-Chern-Simons theory. 

\medskip

For $X$ a smooth manifold, let $c \in H^{n+1}(X, \mathbb{Z})$
be a class in integral cohomology, to be called the 
higher \emph{background magnetic charge}. A smooth refinement of this 
class to a morphism 
$$
  \mathbf{c} : X \to \mathbf{B}^n U(1)
$$
is a circle $n$-bundle on $X$, whose topological class is $c$
$$
  \hat {\mathbf{c}} : X \to \mathbf{B}^n U(1)_{\mathrm{conn}}
$$
A differential refinement of this is a choice of refinement to 
a circle $n$-bundle with connection $\nabla$. 

Now let $\Sigma$ the compact $n$-dimensional worldvolume of an 
$(n-1)$-brane. Then $[\Sigma, X]$ is the diffeological space 
(def. \ref{DiffeologicalSpace}) of smooth maps $\phi : \Sigma \to X$.
The induced $\infty$-Chern-Simons functional
$$
  \xymatrix{
    \exp(i S_{\hat {\mathbf{c}}})
	:
    [\Sigma, X]
	\ar[rr]^{[\hat {\mathbf{c}}, \Sigma]}
	&&
	[\Sigma, \mathbf{B}^n U(1)_{\mathrm{conn}}]
	\ar[rr]^{\int_\Sigma}
	&&
	U(1)
  }
$$
is the ordinary $n$-volume holonomy of $\nabla$ over trajectories 
$\phi : \Sigma \to X$.

\subsubsection{Higher abelian Chern-Simons functionals}
\label{HigherAbelianChernSimons}
\index{Chern-Simons functionals!higher abelian CS functionals}

We discuss higher Chern-Simons functionals on higher abelian gauge fields,
notably on circle $n$-bundles with connection.

\begin{itemize}
  \item \ref{HigherAbelianCS} -- $(4k+3)$d $U(1)$-Chern-Simons functionals;
  \item \ref{HigherElectric} -- Higher electric coupling and higher gauge anomalies.
\end{itemize}

 \paragraph{$(4k+3)$d $U(1)$-Chern-Simons functionals}
 \label{HigherAbelianCS}
 \label{4k+3}
 \index{Chern-Simons functionals!$(4k+3)$d functional}

We discuss higher dimensional abelian Chern-Simons theories in dimension
$4 k +3$.

\medskip

The basic ideas can be found in \cite{HopkinsSinger}. We refine
the discussion there from differential cohomology classes to higher moduli stacks 
of differential cocycles. 
The case in dimension 3 ($k = 0$) is
discussed for instance in \cite{GuadagniniThuillier}.
The case in dimension 7 ($k = 1$)  is the higher Chern-Simons theory whose
holographic boundary theory encodes the self-dual
2-form gauge theory on the single 5-brane \cite{Witten}.
Generally, for every $k$ the $(4k+3)$-dimensional 
abelian Chern-Simons theory induces a self-dual higher
gauge theory holographically on its boundary, see \cite{BelovMoore}.

\medskip

\begin{proposition}
  \label{SmoothDifferentialRefinementOfCupProduct}
  The cup product in integral cohomology
  $$
    (-)\cup(-)
	: 
	H^{k+1}(-,\mathbb{Z}) \times H^{l+1}(-,\mathbb{Z})
	\to 
	H^{k+l+2}(-, \mathbb{Z})
  $$
  has a smooth and differential refinement to the moduli 
  $\infty$-stacks $\mathbf{B}^n U(1)_{\mathrm{conn}}$,
  prop. \ref{BnU1conn}, for circle $n$-bundles with connection
  $$
    (-) \hat{\mathbf{\cup}} (-)
	:
	\mathbf{B}^k U(1)_{\mathrm{conn}} 
	  \times
	\mathbf{B}^l U(1)_{\mathrm{conn}} 
	\to
	\mathbf{B}^{k+l+1} U(1)_{\mathrm{conn}}
	\,.
  $$
\end{proposition}
\proof
  By the discussion in \ref{SmoothStrucDifferentialCohomology}
  we have that $\mathbf{B}^k U(1)_{\mathrm{conn}}$ is presented
  by the simplicial presheaf
  $$
    \Xi \mathbb{Z}_D^\infty[k+1]
	\in
	[\mathrm{CartSp}^{\mathrm{op}}, \mathrm{sSet}].
	\,,
  $$
  which is the image of the Deligne-Beilinson complex, 
  def. \ref{DeligneComplex},  under the 
  Dold-Kan correspondence, prop. \ref{SheafAndNonabelianDoldKan}.
  A lift of the cup product to the Deligne complex is 
  given by the \emph{Deligne-Beilinson cup product}
  \cite{Deligne}\cite{Beilinson}. Since the Dold-Kan functor
  $\Xi : [\mathrm{CartSp}^{\mathrm{op}}, \mathrm{Ch}_{\bullet}]
  \to [\mathrm{CartSp}^{\mathrm{op}}, \mathrm{sSet}]$
  is right adjoint, it preserves products and hence this 
cup product.  
\endofproof

\begin{definition}
  \label{ActionFunctionalForHigherAbelianCS}
  Let $\Sigma$ be a compact manifold of dimension $4 k + 3$ for $k \in \mathbb{N}$. 
  Consider the moduli stack $[\Sigma, \mathbf{B}^{k} U(1)_{\mathrm{conn}}]$
  of circle $(2k+1)$-bundles with connection on $\Sigma$. 
  
  On this space, the 
  action functional of higher abelian Chern-Simons theory is defined to be the 
  composite
  $$
    \exp(i S(-))
	: 
	\xymatrix{
	  [\Sigma, \mathbf{B}^{2k+1}U(1)_{\mathrm{conn}}]
	  \ar[rr]^{(-) \hat{\mathbf{\cup}} (-)}
	  &&
	  [\Sigma, \mathbf{B}^{4k+3} U(1)_{\mathrm{conn}}]
	  \ar[rr]^{\int_\Sigma}
	  &&
	  U(1)
	}
	\,.
  $$
\end{definition}
\begin{observation}
  When restricted to differential $(2k+1)$-forms, regarded as connections
  on trivial circle $(2k+1)$-bundles 
  $$
    \Omega^{2k+1}(\Sigma) \hookrightarrow
  [\Sigma, \mathbf{B}^{2k+1} U(1)_{\mathrm{conn}}]
  $$
  this action functional sends a $(2k+1)$-form $C$ to 
  $$
    \exp(i S(C)) = \exp(i \int_\Sigma C \wedge d_{\mathrm{dR}} C)
	\,.
  $$
  
  From this expression one sees directly why the corresponding functional
is not interesting in the remaining dimensions, because for even degree forms
we have $C \wedge d C = \frac{1}{2} d (C \wedge C)$ and hence for these the 
above functional would be constant.
\end{observation}

\paragraph{Higher electric coupling and higher gauge anomalies}
 \label{HigherElectric}
  \index{Chern-Simons functionals!higher electric coupling terms}

The action functional of ordinary Maxwell electromagnetism in the presence of
an electric background current involves a differential cup-product term
similar to that in def. \ref{ActionFunctionalForHigherAbelianCS}. This has
a direct generalization to higher electromagnetic fields and the corresponding
higher electric currents. If, moreover, a background \emph{magnetic} current is
present, then this action functional is, in general, anomalous. The 
``higher gauge anomalies'' in higher dimensional supergravity theories arise this
way. This is discussed in \cite{Freed}.

Here we refine this discussion from differential cohomology classes to higher
moduli stacks of differential cocycles. 

\medskip

\begin{definition}
  Let $\Sigma$ be a compact smooth manifold of dimension $d$.
  
  By prop. \ref{SmoothDifferentialRefinementOfCupProduct} the universal cup product class 
  $$
    (-)\cup (-) : B^n U(1) \times B^{d-n-1} U(1) \to B^{d} U(1)
  $$
  for any $0 \leq n \leq d$ has a smooth and differential refinement $\hat {\mathbf{\cup}}$.
  We write  
  $$ 
    \exp(i S_{\cup})
	:
	\xymatrix{
	  [\Sigma, \mathbf{B}^{n}U(1)_{\mathrm{conn}} \times \mathbf{B}^{d-n-1}U(1)_{\mathrm{conn}}]
	  \ar[rr]^<<<<<<<<{(-) \hat{\mathbf{\cup}} (-)}
	  &&
	  [\Sigma, \mathbf{B}^{d} U(1)_{\mathrm{conn}}]
	  \ar[rr]^{\int_\Sigma}
	  &&
	  U(1)
	}
  $$
  for the corresponding higher Chern-Simons action functional
  on the higher moduli stack of \emph{pairs} consisting of an $n$-connection and an 
  $(d-n-1)$-connection on $\Sigma$.
\end{definition}
\begin{remark}
  When restricted to pairs of differential forms 
  $$
    (B_1, B_2) \in \Omega^{n}(\Sigma) \times \Omega^{d-n-1}(\Sigma)
	\hookrightarrow
	[\Sigma, \mathbf{B}^{n}U(1)_{\mathrm{conn}} \times \mathbf{B}^{d-n-1}U(1)_{\mathrm{conn}}]
  $$
  this functional sends
  $$
    (B_1, B_2) \mapsto \exp(i \int_\Sigma B_1 \wedge d B_2)
	\,.
  $$
\end{remark}
The higher Chern-Simons functional of def. \ref{HigherAbelianCS}
is the \emph{diagonal} of this functional, where $B_1 = B_2$. We now
consider another variant, where only $B_1$ is taken to vary, but
$B_2$ is regarded as fixed.

Let $X$ be an $d$-dimensional manifold. The configuration space of
higher electromagnetic fields of degree $n$ on $X$ is the moduli stack
of circle $n$-bundles with connection
$[X, \mathbf{B}^n U(1)_{\mathrm{conn}}]$ on $X$. 
\begin{definition}
An \emph{electric background current} on $X$ for degree $p$ electromagnetism
is a circle $(d-n-1)$-bundle with connection 
$\hat j_{\mathrm{el}} : X \to \mathbf{B}^{d-n-1}U(1)_{\mathrm{conn}}$.

The \emph{electric coupling action functional} of the higher electromagnetic
field in the presence of the background electric current is
$$
  \exp(i S_{\mathrm{el}})
   : 
  \xymatrix{
    [X, \mathbf{B}^n U(1)_{\mathrm{conn}}]
    \ar[rr]^{(-)\hat{\mathbf{\cup}} \hat j_{\mathrm{el}}}
	&&
	[X, \mathbf{B}^d U(1)_{\mathrm{conn}}]
	\ar[rr]^{\int_X}
	&&
	U(1)
  }
  \,,
$$
where the first morphism is the differentially refined cup
product from prop. \ref{SmoothDifferentialRefinementOfCupProduct}.
\end{definition}
\begin{remark}
  For the case of ordinary Maxwell theory, 
  with $n = 1$ and $d = 4$, the electric current is a circle 2-bundle with connection.
  Its curvature 3-form is traditionally denoted $j_{\mathrm{el}}$.
  If $X$ is equipped with Lorentzian structure, then its integral over a (compact) 
  spatial slice is the background \emph{electric charge}. Integrality of this
  value, following from the nature of differential cohomology, 
  is the \emph{Dirac charge quantization} that makes electric charge appear in
  integral multiples of a fixed unit charge. 
  
  For $A \in \Omega^1(X) \to [X, \mathbf{B}U(1)_{\mathrm{conn}}]$
  a globally defined connection 1-form, the above action functional is given by
  $$
    A \mapsto \exp(i \int_X A \wedge j_{\mathrm{el}})
	\,.
  $$  
  In the limiting case that the background electric charge is that carried by
  a charged point particle, $j_{\mathrm{el}}$ is the 
  current which is Poincar{\'e}-dual to the trajectory $\gamma : S^1 \to X$
  of the particle. In this case the above goes to
  $$
    \cdots \to \exp(i \int_\Sigma A)
	\,,
  $$
  hence the line holonomy of $A$ along the trajectory of the background charge.
\end{remark}

(...)

\subsubsection{7d Chern-Simons functionals}
 \label{InfinCS7d}
  \index{Chern-Simons functionals!7-dimensional CS theory }

We discuss some higher Chern-Simons functionals over 7-dimensional
parameter spaces.
\begin{itemize}
  \item \ref{CupProductTheoryOfTwo3DCSTheories}
    -- The cup product of a 3d CS theory with itself;
  \item \ref{InfinCS7CSOnString2} --
    7d CS theory on string 2-connection fields;
  \item \ref{7dCSInSugraOnAdS7} --
    7d CS theory in 11d supergravity on $\mathrm{AdS}_7$.
\end{itemize}

This section draws from \cite{FiorenzaSatiSchreiberI}.

\medskip

\paragraph{The cup product of a 3d CS theory with itself}
  \label{CupProductTheoryOfTwo3DCSTheories}
  \index{Chern-Simons functionals!7d CS theory!cup product of two 3d CS theories}

Let $G$ be a compact and simply connected simple Lie group and
consider from \ref{InfinCSOrdinaryCS} the canonical differential characteristic map
for the induced 3d Chern-Simons theory 
$$
  \hat {\mathbf{c}}
   :
  \mathbf{B}G_{\mathrm{conn}}
  \to 
  \mathbf{B}^3 U(1)_{\mathrm{conn}}
  \,.
$$
We consider the differentially refined \emph{cup product}, 
prop. \ref{SmoothDifferentialRefinementOfCupProduct}, of this differential
characteristic map with itself.
\begin{observation}
  \label{DifferentialCupProductOfP1WithItself}
  The topological degree-8 class
  $$
    c \cup c
	:
	\xymatrix{
	  B G 
	  \ar[r]^<<<<{(c,c)}
	  &
	  K(\mathbb{Z},4)
      \times
	  K(\mathbb{Z},4)
	  \ar[r]^<<<<\cup
	  &
	  K(\mathbb{Z}, 8)
	  }
  $$
  has a smooth and differential refinement of the form
  $$
    \hat {\mathbf{c}} \hat {\mathbf{\cup}} \hat {\mathbf{c}}
	: 
	\xymatrix{
	  \mathbf{B}G_{\mathrm{conn}}
	  \ar[r]^<<<<<<<{\hat {\mathbf{c}}}
	  &
	  \mathbf{B}^3 U(1)_{\mathrm{conn}}
	  \times
	  \mathbf{B}^3 U(1)_{\mathrm{conn}}
	  \ar[r]^<<<<<{\hat {\mathbf{\cup}}}
	  &
	  \mathbf{B}^7 U(1)_{\mathrm{conn}}
	}
	\,.
  $$
\end{observation}
\proof  
  By the discussion in \ref{HigherAbelianCS}.
\endofproof
\begin{definition}
  Let $\Sigma$ be a compact smooth manifold of dimension 7.
  The higher Chern-Simons functional
  $$  
    \exp(i S_{\mathrm{CS}}(-))
	:
    \xymatrix{	
      [\Sigma,\mathbf{B}G_{\mathrm{conn}}]
	  \ar[r]^<<<<{\hat {\mathbf{c}} \hat {\mathbf{\cup}} \hat {\mathbf{c}}}
	  &
	  [\Sigma,\mathbf{B}^7 U(1)_{\mathrm{conn}}]
	  \ar[r]^<<<<{\int_\Sigma}
	  &
	  U(1)
	}
  $$
  defines the \emph{cup product Chern-Simons theory} induced by
  $\mathbf{c}$.
\end{definition}
\begin{remark}
  For ordinary Chern-Simons theory, \ref{InfinCSOrdinaryCS},
  the assumption that $G$ is simply connected implies that 
  $B G$ is 3-connected, hence that every $G$-principal bundle
  on a 3-dimensional $\Sigma$ is trivializable, so that 
  $G$-principal connections on $\Sigma$ can be identified with
  $\mathfrak{g}$-valued differential forms on $\Sigma$. This
  is no longer in general the case over a 7-dimensional
  $\Sigma$.
\end{remark}
\begin{proposition}  
  If a field configuration $A \in [\Sigma, \mathbf{B}G_{\mathrm{conn}}]$
happens to have trivial underlying bundle, then the value of the 
cup product CS theory action functional is given by
$$
  \exp(i S_{\mathrm{CS}}(A))
   = 
   \int_\Sigma
    \mathrm{CS}(A) \wedge \langle F_A \wedge F_A\rangle
	\,,
$$
  where $\mathrm{CS}(-)$ is the Lagrangian of ordinary
  Chern-Simons theory, \ref{InfinCSOrdinaryCS}.
\end{proposition}

\paragraph{7d CS theory on string 2-connection fields}
 \label{InfinCS7CSOnString2}
  \index{Chern-Simons functionals!7d CS theory!on string 2-connection fields}

By theorem \ref{SecondFractionalDifferentialPontryagin}
we have a canonical differential characteristic map
$$
  \frac{1}{6}
  \hat{\mathbf{p}}_2
  : 
  \mathbf{B}\mathrm{String}_{\mathrm{conn}}
  \to 
  \mathbf{B}^7 U(1)_{\mathrm{conn}}
$$
from the smooth moduli 2-stack of $\mathrm{String}$-2-connections,
\ref{String2ConnectionsFromLieIntegration}, with values in 
the smooth moduli 7-stack of circle 7-bundles (bundle 6-gerbes)
with connection. 
This induces a 7-dimensional Chern-Simons theory.
\begin{definition} 
  For $\Sigma$ a compact 7-dimensional smooth manifold,
  define $\exp(i S_{\frac{1}{6}p_2}(-))$ to be the 
  Chern-Simons action functional induced by the 
  universal differential second fractional Pontryagin class,
  theorem \ref{SecondFractionalDifferentialPontryagin},
  $$
    \exp(i S_{\frac{1}{6}p_2}(-))
	:
	\xymatrix{
  	  [\Sigma, \mathbf{B} \mathrm{String}_{\mathrm{conn}}]
	  \ar[r]^{\frac{1}{6}\hat{\mathbf{p}_2}}
	  &
	  [\Sigma, \mathbf{B}^7 U(1)_{\mathrm{conn}}]
	  \ar[r]^<<<<{\int_\Sigma}
	  &
	  U(1)
	}
	\,.
  $$
\end{definition}
Recall from \ref{String2ConnectionsFromLieIntegration}
the different incarnations of the local differential form data
for string 2-connections.
\begin{proposition}
  Over a 7-dimensional $\Sigma$ every field configuration
  $(A,B) \in [\Sigma, \mathbf{B}\mathrm{String}_{\mathrm{conn}}]$
  is a string 2-connection whose underlying $\mathrm{String}$-principal
  2-bundle is trivial.
  \begin{itemize}
  \item In terms of the strict $\mathfrak{string}$ Lie 2-algebra
  from def. \ref{StrictStringLie2Algebra} this is presented by
  a pair of nonabelian differential forms 
  $A \in \Omega^1(\Sigma, P_* \mathfrak{so})$,
  $B \in \Omega^2(\Sigma, \hat \Omega_* \mathfrak{so})$.
  The above action functional takes this to
  $$
    \begin{aligned}
    \exp(i S_{\frac{1}{6}p_2}(A,B))
	& =
	\int_{\Sigma} \mathrm{CS}_7(A(1))
	\\
	&=  \int_{\Sigma} (
  \langle A_e \wedge d A_e \wedge dA_e \wedge dA_e \rangle
  +
  k_1\langle A_e \wedge [A_e \wedge A_e] \wedge dA_e \wedge dA_e \rangle
  \\
  & + 
  k_2\langle A_e \wedge [A_e \wedge A_e] \wedge [A_e \wedge A_e] \wedge dA_e \rangle
  +
  k_3\langle A_e \wedge [A_e \wedge A_e] \wedge [A_e \wedge A_e] 
  \wedge [A_e \wedge A_e] \rangle
	)
  \end{aligned}
	\,,
  $$
  where $A_e \in \Omega^1(\Sigma, \mathfrak{so})$ is the 1-form of endpoint
  values of $A$ in the path Lie algebra, and where the integrand is
  the degree-7 Chern-Simons element of the quaternary invariant polynomial
  on $\mathfrak{so}$.
  \item
In terms of the skeletal $\mathfrak{string}$ Lie 2-algebra
  from def. \ref{SkeletalStringLie2Algebra} this is presented by
  a pair of differential forms 
  $A \in \Omega^1(\Sigma, \mathfrak{so})$,
  $B \in \Omega^2(\Sigma, \mathbb{R})$.
  The above action functional takes this to
  $$
    \exp(i S_{\frac{1}{6}p_2}(A,B))
	=
	\int_{\Sigma} \mathrm{CS}_7(A)
	\,.
  $$
  \end{itemize}
\end{proposition}

\paragraph{7d CS theory in 11d supergravity on $\mathrm{AdS}_7$}
 \label{7dCSInSugraOnAdS7}
  \index{Chern-Simons functionals!7d CS theory!in 11d supergravity}

The two 7-dimensional Chern-Simons theories
from \ref{CupProductTheoryOfTwo3DCSTheories} and
\ref{InfinCS7d} can be merged to a 7d theory defined
on field configurations that are 2-connections
with values in the $\mathrm{String}$-2-group
from def. \ref{smoothStringC}. We define and dicuss this
higher Chern-Simons theory below in \ref{AdS7CSTheoryDefAndProperties}.
In \ref{AdS7CSTheoryMotivation} we argue that this 
7d Chern-Simons theory plays a role in 
$\mathrm{AdS}_7/\mathrm{CFT}_6$-duality \cite{AGMOO}.

\subparagraph{Motivation from $\mathrm{AdS}_7 / \mathrm{CFT}_6$-holography}
\label{AdS7CSTheoryMotivation}
\index{holography!$\mathrm{AdS}_7/\mathrm{CFT}_6$}

We give here an argument that 
the 7-dimensional nonabelian gauge theory discussed in
section \ref{AdS7CSTheoryDefAndProperties} is the Chern-Simons part of 
11-dimensional supergravity on $\mathrm{AdS}_7 \times S^4$
with 4-form flux on the $S^4$-factor and 
with quantum anomaly cancellation conditions taken into account.
We moreover argue that this implies that the 
states of this 7-dimensional CS theory over a 7-dimensional
manifold encode the conformal blocks 
of the 6-dimensional worldvolume theory of coincident M5-branes. 
The argument is based on the available but incomplete
knowledge about $\mathrm{AdS}/\mathrm{CFT}$-duality,
such as reviewed in \cite{AGMOO}, and cohomological
effects in M-theory as reviewed and discussed in \cite{Sati10}.

\medskip

There are two, seemingly different, realizations of the
\emph{holographic principle} in quantum field theory.  
On the one hand, Chern-Simons theories in dimension $4k+3$
have spaces of states that can be identified with spaces
of correlators of $(4k+2)$-dimensional conformal field theories
(spaces of ``conformal blocks'') on their boundary. 
For the case $k = 0$ this was discussed in \cite{WittenCS}, 
for the case $k = 1$ in \cite{Witten96}. On the other hand, 
$\mathrm{AdS}/\mathrm{CFT}$ duality (see \cite{AGMOO} for a review)  
identifies
correlators of $d$-dimensional CFTs with states of compatifications
of string theory, or M-theory, on asymptotically anti-de Sitter
spacetimes of dimension $d+1$ (see \cite{Witten98a}).

In \cite{Witten98} it was pointed out that these two mechanisms are 
in fact closely related. A detailed analysis of the
$\mathrm{AdS}_5/\mathrm{SYM}_4$-duality shows that the
spaces of correlators of the 4-dimensional theory can be identified
with the spaces of states obtained by geometric quantization
just of the Chern-Simons term 
in the effective action of 
type II string theory on $\mathrm{AdS}_5$, which locally
reads
$$
  (B_{\mathrm{NS}}, B_{\mathrm{RR}})
  \mapsto 
  N \int_{\mathrm{AdS}_5} B_{\mathrm{NS}} \wedge d B_{\mathrm{RR}}
  \,,
$$
where $B_{\mathrm{NS}}$ is the local Neveu-Schwarz 2-form field, 
$B_{\mathrm{RR}}$ is the local RR 2-form field, and where  
$N$ is the RR 5-form flux picked up from integration over the $S^5$ factor.

As briefly indicated there, the similar form of the Chern-Simons term of
11-dimensional supergravity (M-theory) on $\mathrm{AdS}_7$
suggests that an analogous argument shows that under 
$\mathrm{AdS}_7$/$\mathrm{CFT}_6$-duality the conformal blocks
of the $(2,0)$-superconformal theory are identified with the 
geometric quantization of a 7-dimensional Chern-Simons theory.
In \cite{Witten98} that Chern-Simons action is taken, 
locally on $\mathrm{AdS}_7$, to be
$$
\begin{aligned}
  C_3 &\mapsto \int_{\mathrm{AdS}_7 \times S^4}
    C_3 \wedge G_4 \wedge G_4
	& =
	N \int_{\mathrm{AdS}_7} C_3 \wedge d C_3
\end{aligned}
\,,
$$
where now $C_3$ is the local incarnation of the 
supergravity $C$-field, \ref{SupergravityCFieldInSupergravity}, 
where $G_4$ is its curvature 4-form locally equal to $d C_3$, 
and where
$$
  N := \int_{S^4} G_4
$$
is the $C$-field flux on the 4-spehere factor.

This is the $(4 \cdot 1 + 3 = 7) $-dimensional abelian Chern-Simons theory,
\ref{section.HigherAbelianCSTheory},
shown in \cite{Witten96} to induce on its 6-dimensional boundary the self-dual
2-form -- in the \emph{abelian} case.

In order to generalize this to the nonabelian case of interest,
we notice that there is a term missing in the above
Lagrangian. The quantum anomaly cancellation in 11-dimensional supergravity
is known from \cite{DLM}(3.14) to require a corrected Lagrangian whose Chern-Simons
term locally reads
$$
\begin{aligned}
  (\omega, C_3) &\mapsto \int_{\mathrm{AdS}_7 \times S^4}
    C_3 \wedge \left( G_4 \wedge G_4 - I_8^{\mathrm{dR}}(\omega)\right)
	\,,
\end{aligned}
$$
where $\omega$ is the spin connection form, locally, and where
$8 I_8^{\mathrm{dR}}(\omega)$ is a de Rham representative of the integral 
cohomology class
\(
  8 I_8 
    = 
	\frac{1}{6}p_2 
    - 
    8 (\frac{1}{2}p_1) \cup (\frac{1}{2}p_1)
  \,,
  \label{I8}
\)
with $\frac{1}{2}p_1$ and $\frac{1}{6}p_2$  the first and second fractional Pontrjagin classes,
prop. \ref{FirstFracPontryagin}, prop. \ref{SecondFracPontryagin}, 
respectively, of the given $\mathrm{Spin}$ bundle over 11-dimensional
spacetime $X$.

This means that after passing to the effective theory on 
$\mathrm{AdS}_7$, this corrected Lagrangian picks up another
7-dimensional Chern-Simons term, now one depending on 
\emph{nonablian} fields (with values in $\mathrm{Spin}$ and $E_8$). Locally this reads
\(
  \begin{aligned}
     S_{7d\mathrm{CS}}
	 :
     (\omega, C_3) & \mapsto 
	 N\int_{\mathrm{AdS}_7} C_3 \wedge d C_3
	 -
	 \frac{N}{8}\int_{\mathrm{Ads}_7} \mathrm{CS}_{8 I_8}(\omega)
  \end{aligned}
  \label{The7dLagrangianInMotivation}
  \,.
\)
where $\mathrm{CS}_{8 I_8}(\omega)$ is a Chern-Simons form for 
$8 I_8^{\mathrm{dR}}(\omega)$, defined locally by
$$
  d \mathrm{CS}_{8 I_8}(\omega) = 8 I_8^{\mathrm{dR}}(\omega)
  \,.
$$

But this action functional, which is locally a functional of a
3-form and a $\mathrm{Spin}$-connection, cannot globally be of this form, 
already because the field that looks locally like a $\mathrm{Spin}$ connection 
cannot globally be a $\mathrm{Spin}$ connection.
To see this, notice from the discussion of the 
$C$-field in \ref{supergravityCField}, that there
is a quantization condition on the supergravity fields on the 
11-dimensional $X$
\cite{WittenFluxQuantization}, which in cohomology requires the
identity
$$
  2[G_4] = \frac{1}{2}p_1 +  2 a \;\;\;
  \in H^4(X,\mathbb{Z})
  \,,
$$
where on the right we have the canonical characteristic 4-class 
$a$, prop. \ref{HomotopyGroupsOfBE8},  of an `auxiliary' $E_8$
bundle on 11-dimensional spacetime.
Moreover, we expect that when restricted to the vicinity of 
the asymptotic boundary of $\mathrm{AdS}_7$,  
\begin{itemize}
\item the class of $G_4$  vanishes;
\item the $E_8$-bundle becomes equipped with a connection, too
 (the $E_8$-field ``becomes dynamical'');
\end{itemize}
in analogy to what happens at the boundary for the Ho{\v r}ava-Witten
compactification of the 11-dimensional theory \cite{HoravaWitten},
as discussed in \ref{CFieldRestrictionToTheBoundary}.
Since, moreover, the states of the topological TFT that we are
after are obtained already from geometric quantization, 
\ref{StrucGeometricPrequantization},  of the
theory in the vicinity $\Sigma \times I$ of a boundary $\Sigma$,
we find the field configurations of the 7-dimensional theory
are to satisfy the constraint in cohomology
\(
 \frac{1}{2}p_1 + 2 a=0
\label{cond}
 \,.
\)
\vspace{3mm}
Imposing this condition has two effects. 
\begin{enumerate}
\item 
The first is that, according to 
\ref{TwistedDifferentialStructures}, what locally looks like
a spin-connection is globally instead a 
\emph{twisted differential String structure}, \ref{HigherSpinStructure},
or equivalently a 
\emph{2-connection on a twisted 
$\mathrm{String}$-principal 2-bundle}, where the twist is given by
the class $2 a$.  
By \ref{Principal3BundlesAndTwisted2Bundles} 
the total space of such a principal 2-bundle may be
identified  with a (twisted) \emph{nonabelian bundle gerbe}.
Therefore the configuration space of fields of the effective 7-dimensional
nonabelian Chern-Simons action above should not involve just Spin connection forms,
but \emph{$\mathrm{String}$-2-connection} form data. 
By \ref{String2ConnectionsFromLieIntegration} there is a gauge in which this
is locally given by nonabelian 2-form field data with values in 
the loop group of $\mathrm{Spin}$. 

\item 
The second effect is that on the space of twisted String-2-connections,
the differential 4-form $\mathrm{tr}(F_\omega \wedge F_\omega)$,
that under the Chern-Weil homomorphism represents the image of 
$\frac{1}{2}p_1$ in de Rham cohomology, according to \ref{CocyclesForDifferentialStringStructures},
locally satisfies
$$
  d H_3 = \langle F_\omega \wedge F_\omega \rangle - 2 \langle F_A \wedge F_A \rangle
  \,,
$$
where $H_3$ is the 3-form curvature component of the $\mathrm{String}$-2-connection,
and where $F_A$ is the curvature of 
a connection on the $E_8$ bundle, locally given by an 
$\mathfrak{e}_8$-valued 1-form $A$.
Therefore with the quantization condition of the $C$-field
taken into account, the 7-dimensional Chern-Simons action (\ref{The7dLagrangianInMotivation}) 
becomes
\(
  S_{7d\mathrm{CS}} 
   =
	 N
	 \int_{\mathrm{AdS}_7}
	 \left(
	  C_3 \wedge d C_3
	 -
	 \frac{1}{8}
	  H_3 \wedge d H _3
	 -
	 \frac{1}{4}
     (H_3 + 2 \mathrm{CS}_{a}(A) \wedge \mathrm{tr}(F_\omega \wedge F_\omega)
	+
	\frac{1}{8}
    \mathrm{CS}_{\frac{1}{6}\hat {\mathbf{p}}_2}(\omega)
	\right)
  \,.
  \label{AdS7CS7ActionInMotivation}
\)
Here the first two terms are 7-dimensional abelian Chern-Simons actions as before,
for fields that are both locally abelian three forms (but have
very different global nature). The second two terms, however,
are action functionals for \emph{nonabelian} Chern-Simons theories.
The third term involves the familiar Chern-Simons 3-form 
of the $E_8$-connection familiar from 3-dimensional Chern-Simons theory
$$
  \mathrm{CS}_{a}(A) = 
  \mathrm{tr}(A \wedge d A) 
  +
  \frac{2}{3}\mathrm{tr}(A \wedge A \wedge A)
  \,.
$$
Finally the fourth term is the 
Chern-Simons 7-form that is locally
induced, under the Chern-Weil homomorphism, from the quartic invariant
polynomial $\langle -,-,-,-\rangle : \mathfrak{so}^{\otimes 4} \to \mathbb{R}$ 
on the special orthogonal 
Lie algebra $\mathfrak{so}$, in direct analogy to how 
standard 3-dimensional Chern-Simons theory is induced under
Chern-Weil theory from the quadratic invariant polynomial (the Killing form)
$\langle -,-\rangle : \mathfrak{so} \otimes \mathfrak{so} \to \mathbb{R}$:
$$
\begin{aligned}
  \mathrm{CS}_7(\omega)
  = & 
  \langle \omega \wedge d \omega \wedge d\omega \wedge d\omega \rangle
  +
  k_1\langle \omega \wedge [\omega \wedge \omega] \wedge d\omega \wedge d\omega \rangle
  \\
  & + 
  k_2\langle \omega \wedge [\omega \wedge \omega] \wedge [\omega \wedge \omega] \wedge d\omega \rangle
  +
  k_3\langle \omega \wedge [\omega \wedge \omega] \wedge [\omega \wedge \omega] 
  \wedge [\omega \wedge \omega] \rangle
 \end{aligned}
  \,.
$$
\end{enumerate}

This line of arguments suggests that the Chern-Simons term
that governs 11-dimensional supergravity on $\mathrm{AdS}_7 \times S^4$
is an action functional on 
fields that are twisted $\mathrm{String}$-2-connections
such that the action functional is locally given by
(\ref{AdS7CS7ActionInMotivation}). In \ref{AdS7CSTheoryDefAndProperties}
we show that a Chern-Simons theory satisfying these properties
naturally arises from the differential characteristic maps
discussed above in \ref{CupProductTheoryOfTwo3DCSTheories} 
and \ref{InfinCS7d}.

\subparagraph{Definition and properties}
\label{AdS7CSTheoryDefAndProperties}

We discuss now a twisted combination of the two 7-dimensional
Chern-Simons action functionals from \ref{CupProductTheoryOfTwo3DCSTheories} and 
\ref{InfinCS7d} which naturally lives on the moduli 2-stack
$C \mathrm{Field}(-)^{\mathrm{bdr}}$ of boundary $C$-field
configurations from \ref{CFieldBdr}. We show that on 
$\infty$-connection field configurations whose underlying $\infty$-bundles
are trivial, this functional reduces to that given in 
equation (\ref{AdS7CS7ActionInMotivation}).

\medskip

It is instructive to first consider the simple special case where 
the $E_8$ is trivial. In this case the boundary moduli stack 
$C \mathrm{Field}^{\mathrm{bdr}'}$ 
from observation \ref{BoundaryCFieldInTermsOfStringE8}
restricts to just that of string 2-connections, 
$\mathbf{B}\mathrm{String}_{\mathrm{conn}}$.

\begin{definition}
  Write $8 \hat {\mathbf{I}}_8$ for the smooth universal 
  differential characteristic cocycle
  $$
    8 \hat {\mathbf{I}}_8
	:
	\xymatrix{
  	 \mathbf{B}\mathrm{String}_{\mathrm{conn}}
	 \ar[rrr]^{
	   (\frac{1}{6}\hat {\mathbf{p}}_2) 
	   - 
	 8 (\frac{1}{2}\hat {\mathbf{p}_1} \hat{\mathbf{\cup}} \frac{1}{2}\hat {\mathbf{p}_1})
	 }
	 &&&
	 \mathbf{B}^7 U(1)_{\mathrm{conn}}
	}
	\,,
  $$
  where $\frac{1}{6}\hat{\mathbf{p}}_2$ is 
  the differential second fractional Pontryagin class 
  from theorem \ref{SecondFractionalDifferentialPontryagin}
  and where $\frac{1}{2}\hat {\mathbf{p}_1} \hat{\mathbf{\cup}} \frac{1}{2}\hat {\mathbf{p}_1}$
  is the differential cup product class from 
  observation \ref{DifferentialCupProductOfP1WithItself}.
\end{definition}
\begin{definition}
  \label{The7DAction}
  For $\Sigma$ a compact smooth manifold of dimension 7,
  the canonically induced action functional $\exp(i S_{8 I_8}(-))$  
  from def. \ref{CanonicalChernSimonsFunctional},
  on the moduli 2-stack of $\mathrm{String}$-2-connections
  is the composite
  $$
    \exp(i S_{8 I_8}(-))
	:
	\xymatrix{
	 [\Sigma, \mathbf{B}\mathrm{String}_{\mathrm{conn}}]
	 \ar[r]^{8 \hat{ \mathbf{I}}_8}
	 &
	 [\Sigma,\mathbf{B}^7 U(1)_{\mathrm{conn}}]
	 \ar[r]^<<<<{\int_\Sigma}
	 &
	 U(1)
	}
	\,.
  $$
\end{definition}
We give now an explicit description of the field configurations
in $[\Sigma, \mathbf{B}\mathrm{String}_{\mathrm{conn}}]$
and of the value of $\exp(i S_{8 I_8}(-))$ on these 
in terms of differential form data.
\begin{proposition}
  \label{48I8FieldConfigurationAsCechCocycleData}
  A field configuration in 
  $[\Sigma, \mathbf{B}\mathrm{String}_{\mathrm{conn}}]
  \in \mathrm{Smooth}\infty\mathrm{Grpd}$
  is presented in the model category
  $[\mathrm{CartSp}^{\mathrm{op}}, \mathrm{sSet}]_{\mathrm{proj}, \mathrm{loc}}$,
  \ref{SmoothInfgrpds},  
  by a correspondence of simplicial presheaves
  $$
    \xymatrix{
	 C(\{U_i\}) 
	  \ar[r]^<<<<\phi
	  \ar[d]^\simeq
	  &
	 \mathbf{cosk}_3 
	 \exp(b\mathbb{R} \to
	  \mathfrak{so}_{\mu})_{\tilde {\mathrm{conn}}}
	  \\
	  \Sigma
	}
	\,,
  $$
  where $\mathfrak{so}_\mu$ is the skeletal String Lie 2-algebra,
  def. \ref{SkeletalStringLie2Algebra},
  and where on the right we have the adapted differential coefficient object 
  from prop. \ref{PresentationByFibration};
  such that the projection 
  $$
    \xymatrix{
	 C(\{U_i\}) 
	  \ar[r]^<<<<\phi
	 &
	 \mathbf{cosk}_3 \exp(b\mathbb{R} \to \mathfrak{so}_{\mu})_{\tilde {\mathrm{conn}}}
      \ar[r]
	  &
      \mathbf{B}^3 U(1)_{\mathrm{conn}}	  
	}
  $$  
  has a class.

  The underlying nonabelian cohomology class of such a cocycle
  is that of a $\mathrm{String}$-principal 2-bundle.
    
  The local connection and 
  curvature differential form data over a patch $U_i$ is
  $$
      \begin{array}{ll}
      F_\omega =& d \omega + \frac{1}{2}[\omega\wedge \omega]
      \\
      H_3 =& \nabla B := d B + \mathrm{CS}(\omega)
      \\
      d F_\omega =& - [\omega \wedge F_\omega]
      \\
      d H_3 =& 
		 \langle F_\omega \wedge F_\omega \rangle
    \end{array}
  $$
\end{proposition}
\proof
  Without the constraint on the $C$-field this is 
  the description of twisted 
  $\mathrm{String}$-2-connections
  of observation \ref{LocalDataOfTwistedDifferentialStringStruc}
  where the twist is the $C$-field.
  The condition above picks out the untwisted case, where the
  $C$-field is trivialized. What remains is an 
  untwisted $\mathrm{String}$-principal 2-bundle.
  
  The local differential form data is found from the modified
  Weil algebra of 
  $(b\mathbb{R} \to
	  (\mathfrak{so})_{\mu_{\mathfrak{so}}})
	  $
indicated on the right of the following diagram
  $$
  \left(
    \begin{array}{ll}
      F_\omega =& d \omega + \frac{1}{2}[\omega\wedge \omega]
      \\
      H_3 =& \nabla B := d B + CS(\omega) - C_3 
      \\
      \mathcal{G}_4 =& d C_3
      \\
      d F_\omega =& - [\omega \wedge F_\omega]
      \\
      d H_3 =& 
		 \langle F_\omega \wedge F_\omega \rangle
		 -
		 \mathcal{G}_4 
      \\
      d \mathcal{G}_4 =& 0
    \end{array}
  \right)_i
  \;\;\;\;
  \stackrel{
    \begin{array}{ll}
       t_{\mathfrak{so}}^a & \mapsto \omega^a
       \\
       r_{\mathfrak{so}}^a & \mapsto F_\omega
       \\
       b & \mapsto B
       \\
       c & \mapsto C_3
       \\
       h & \mapsto H_3
       \\
       g & \mapsto \mathcal{G}_4
    \end{array}
  }{\xymatrix{\ar@{<-|}[rrr]&&&}}
  \;\;\;\;
  \left(
    \begin{array}{ll}
       r_{\mathfrak{so}}^a  =& d t_{\mathfrak{so}}^a 
	     + 
		 \frac{1}{2}C_{\mathfrak{so}}^a{}_{b c} t_{\mathfrak{so}}^b \wedge t_{\mathfrak{so}}^c  
       \\
       h = & d b + \mathrm{cs}_{\mathfrak{so}}  - c     
       \\
       g =& d c
       \\
       d r_{\mathfrak{so}}^a  =&  - C^a{}_{b c} t_{\mathfrak{so}}^b \wedge r_{\mathfrak{so}}^a
       \\
       d h =& \langle -,-\rangle - g
       \\
       d g =& 0
    \end{array}
  \right)
  \,.
$$  
\endofproof
\begin{remark}
  While the 2-form $B$ in the presentation
  used in the above proof is abelian, the
  total collection of forms is still connection data with coefficients
  in the nonabelian Lie 2-algebra $\mathfrak{string}$. 
  We explained in 
  remark \ref{TwoPointsOfViewOnStringConnectionData},
  that there is a choice of local
  gauge in which the nonabelianness of the 2-form becomes manifest.
  For the discussion of the above proposition,
  however, this gauge is not the most convenient
  one, and it is more convenient to exhibit the local cocycle data in the
  above form, which corresponds to the second gauge of
  remark
  \ref{TwoPointsOfViewOnStringConnectionData}.
  
  This is an example of a general principle in 
  higher nonabelian gauge theory (``higher gerbe theory'').
  Due to the higher gauge invariances, 
  the local component presentation of a given structure does
  not usually manifestly exhibit the gauge-invariant information
  in an obvious way.
\end{remark}
\begin{proposition}
  Let $\phi \in [\Sigma, \mathbf{B}\mathrm{String}_{\mathrm{conn}}]$
  be a field configuration which, in the presentation of 
  prop. \ref{48I8FieldConfigurationAsCechCocycleData}, is defined
  over a single patch $U = \Sigma$. 
  
  Then the action functional of def. \ref{The7DAction}
  sends this to 
  $$
    \exp(i S_{8 I_8}(\omega, H_3))
	=
	\exp\left( i 
	\int_\Sigma
	\left(
	 -
	 8  H_3 \wedge d H _3
	+
    \mathrm{CS}_{\frac{1}{6}\hat {\mathbf{p}}_2}(\omega)
	\right)
	\right)
	\,.
  $$
\end{proposition}
\proof   
   The first term is that of the 
   cup product theory, \ref{CupProductTheoryOfTwo3DCSTheories},
   after using the identity 
   $\mathrm{tr}(F_\omega \wedge F_\omega) = d H_3$ which holds
   on the configuration space of String-2-connections 
   by prop. \ref{48I8FieldConfigurationAsCechCocycleData}. 
   The second term is that of the $\frac{1}{6}p_2$-Chern-Simons
   theory from \ref{InfinCS7d}.
\endofproof
\begin{remark}
  Therefore comparison with equation (\ref{AdS7CS7ActionInMotivation})
  shows that the action functional $S_{8 I_8}$
  has all the properties that in \ref{AdS7CSTheoryMotivation}
  we argued that the effective 7-dimensional Chern-Simons 
  theory inside 11-dimensional supergravity compactified
  on $S^4$ should have, in the following special case:
  \begin{itemize}
    \item the $C$-field flux on $S^4$ is $N = 8$;
  \end{itemize}
  and
  \begin{itemize}  
    \item the $E_8$-field is trivial;
	\item the $C$-field on $\Sigma$ is trivial.
  \end{itemize}
  \end{remark}

By choosing any multiple of $8 \hat {\mathbf{I}}_8$ one can obtain
$C$-field flux of arbitrary multiples of 8. In order to obtain $C$-field
flux that is not a multiple of 8 one needs to discuss further divisibility of
$8 \hat {\mathbf{I}}_8$.

\medskip

We discuss now a refinements of $S_{8 I_8}$
that generalize away from the last two of these special conditions to
obtain the full form of (\ref{AdS7CS7ActionInMotivation}).

Recall from def. \ref{CFieldBdr}
the higher moduli stack $C \mathrm{Field}^{\mathrm{bdr}}$
of supergravity $C$-field configurations, which 
by remark. \ref{BoundaryCFieldInTermsOfStringE8} 
is the moduli 3-stack of twisted $\mathrm{String}^{2\mathbf{a}}$-connections. 
We consider now an action functional on this 
configuration stack.
 
Following remark \ref{FieldContentOfStringaConnection} we write 
a corresponding field configuration, $\phi \in C\mathrm{Field}^{\mathrm{bdr}}(\Sigma)$, 
whose underlying topological class
is trivial as a tuple of forms
$$
  (\omega, A, B_2, C_3) \in 
  \Omega^1(\Sigma, \mathfrak{so})
  \times
  \Omega^1(\Sigma, \mathfrak{e}_8)
  \times 
  \Omega^2(\Sigma)
  \times 
  \Omega^3(\Sigma)
$$
and set
$$
  H_3 := d B_2 + \mathrm{cs}(\omega) - \mathrm{cs}(A)
  \,.
$$
 Recall that by prop. \ref{String2aByLieIntegration} this object has a 
  presentation by Lie integration as \ref{CocyclesForDifferentialStringStructures}
  as a sub-simplicial set
  $$
	\mathbf{cosk}_3
	\exp( 
       (\mathbb{R} \to \mathfrak{so} \oplus \mathfrak{e}_8)_{
	      \mu^{\mathfrak{so}}_3 - 2 \mu_3^{\mathfrak{e}_8}
	   }	
	)_{\mathrm{conn}}
	\,.
  $$
  In terms of this presentation we have an evident differential 
characteristic class given by the Lie integration of the Chern-Simons
element $\mathrm{cs}_{\frac{1}{6}p_2} - 8\mathrm{cs}_{\frac{1}{2}o_1 \cup \frac{1}{2}p_1}$. 
\begin{definition}   
  \label{TheFull7DAction}
  Write $\hat {\mathbf{I}_8}$ for the smooth universal 
  characteristic map given by the composite
  $$
    \xymatrix{
	  \mathbf{B}\mathrm{String}^{2\mathbf{a}}
	  \ar[rrr]^{
	    \exp(\mathrm{cs}_{\frac{1}{6}p_2} - 8\mathrm{cs}_{\frac{1}{2}p_1 \cup \frac{1}{2}p_1})
	  }
	  &&&
	  [\Sigma,\mathbf{B}^7 (\mathbb{R}/K)_{\mathrm{conn}}]
	}
	\,,
  $$
  where the second morphism is the 
  $\infty$-Chern-Weil homomorphism of $I_8$, according to 
  \ref{SmoothStrucInfChernWeil}, with $K \subset \mathbb{R}$
  the given sublattice of periods.
  
  Write
  $$
    \exp( i S_{I_8}(-)) : 
	\xymatrix{
  	  \mathbf{B}\mathrm{String}^{2\mathbf{a}}_{\mathrm{conn}}
	  \ar[r]^<<<<{\hat {\mathbf{I}}_8}
	  &
	  [\Sigma, \mathbf{B}^7 (\mathbb{R}/K)_{\mathrm{conn}}]
	  \ar[r]^<<<<{\int_\Sigma}
	  &
	  \mathbb{R}/K
	}
  $$
  for the corresponding action functional.
\end{definition}
Finally we obtain the refinement of the 7-dimensional Chern-Simons
action (\ref{AdS7CS7ActionInMotivation}) to the full higher moduli stack
of boundary $C$-field configurations.
\begin{proposition}
  Let $\phi \in C\mathrm{Field}^{\mathrm{bdr}}(\Sigma))$
  be a boundary $C$-field  configuration according to 
  remark. \ref{BoundaryCFieldInTermsOfStringE8}, 
  whose underlying $\mathrm{String}^{2\mathbf{a}}$-principal
 2-bundle is trivial,  which is hence 
 a quadruple of forms 
 $$
   \phi = (\omega, A, B_2, C_3) 
     \in \Omega^1(\Sigma, \mathfrak{so}) \times \Omega^1(\Sigma, \mathfrak{e}_8)
	 \times \Omega^2(\Sigma) \times \Omega^3(\Sigma)
	 \,.
 $$  
  The combination of the action functional of def. \ref{ActionFunctionalForHigherAbelianCS}
  and the 
  action functional of def. \ref{TheFull7DAction}
  sends this to 
  $$
    \exp(i S(C_3))\exp(i S_{8 I_8}(\omega, A, B_2))
	=
	\int_\Sigma
	C_3 \wedge d C_3
	+
	8
	\left(
	 H_3 \wedge d H_3 + (H_3 + \mathrm{cs}(A)) \wedge \langle F_\omega \wedge F_\omega \rangle 
	 + 
    \frac{1}{8}\mathrm{cs}_{\frac{1}{6}p_2}(\omega)	 
	\right)
	\;
	\mathrm{mod} K
	\,,
  $$
  where $H_3 = d B + \mathrm{cs}(\omega) - 2\mathrm{cs}(A)$.
\end{proposition}
\proof
By the nature of the $\exp(-)$-construction we have
$$
  \begin{aligned}
    \exp(i S_{8 I_8}(\omega, A, B))
	&=
	\int_\Sigma
	\left(
    8 \mathrm{cs}(\omega) \wedge d \mathrm{cs}(\omega)
	+
    \mathrm{cs}_{\frac{1}{6}p_2}(\omega)
	\right)
 \end{aligned}
 \,.
$$
Inserting here the equation for $H_3$ satisfied by the 
$\mathrm{String}^{2\mathbf{a}}$-connections yields
$$
\begin{aligned}
	\cdots & = 
	\int_\Sigma
	\left(
    8 (H_3 + 2\mathrm{cs}(A) - d B) \wedge d (H_3 + 2\mathrm{cs}(A) - d B)
	+
    \mathrm{cs}_{\frac{1}{6}p_2}(\omega)
	\right)
	\\
	& = 
	\int_\Sigma
	\left(
    8 (H_3 + 2\mathrm{cs}(A)) \wedge d (H_3 + 2\mathrm{cs}(A))
	+
    \mathrm{cs}_{\frac{1}{6}p_2}(\omega)
	\right)
	\\
	& =
	\int_\Sigma
	8
	\left(
	 H_3 \wedge d H_3 + (H_3 + 2\mathrm{cs}(A)) \wedge \langle F_\omega \wedge F_\omega \rangle 
	 + 
    \frac{1}{8}\mathrm{cs}_{\frac{1}{6}p_2}(\omega)	 
	\right)
  \end{aligned}
  \,.
$$
\endofproof

\subsubsection{Action of closed string field theory type}
 \label{CSFTAction}
  \index{Chern-Simons functionals!string field theory}

We discuss the form of $\infty$-Chern-Simons Lagrangians, \ref{InfinityChernSimonsFieldTheory}, on 
general $L_\infty$-algebras equipped with a quadratic invariant
polynomial. The resulting action functionals have the form
of that of closed string field theory \cite{Zwiebach}.

\medskip

\begin{proposition}
Let $\mathfrak{g}$ be any $L_\infty$-algebra equipped with 
a quadratic invariant polynomial $\langle -,-\rangle$. 

  The $\infty$-Chern-Simons functional associated with this
  data is
$$
  S : A \mapsto 
    \int_\Sigma
	   \left(
	     \langle A \wedge d_{\mathrm{dR}} A\rangle
		 +
		 \sum_{k = 1}^\infty 
		 \frac{2}{(k+1)!}
		 \langle A \wedge [A\wedge \cdots A]_k\rangle
    	\right)
		\,,
$$
where 
$$
  [-,\cdots,-] : \mathfrak{g}^{\otimes k} \to \mathfrak{g}
$$
is the $k$-ary bracket of $\mathfrak{g}$ (prop. \ref{LInfinityAlgebraFromDGAlgebra}).
\end{proposition}
\proof
  There is a canonical contracting homotopy operator
  $$
    \tau : \mathrm{W}(\mathfrak{g}) \to \mathrm{W}(\mathfrak{g})
  $$
  such that $[d_{\mathrm{W}}, \tau] = \mathrm{Id}_{\mathrm{W}(\mathfrak{g})}$.
  Accordingly a Chern-Simons element, def. \ref{TransgressionAndCSElements}, for
  $\langle -,-\rangle$ is given by
  $$
    \mathrm{cs} := \tau \langle -,-\rangle
	\,.
  $$
  We claim that this is indeed the Lagrangian for the above action functional.

  To see this, first choose a basis $\{t_a\}$ and write
$$
  P_{a b} := \langle t_a , t_b\rangle
$$
for the components of the invariant polynomial in that basis and
$$
  C^a_{a_1, \cdots, a_k} := [t_{a_1}, \cdots, t_{a_k}]_k^a
$$
as well as
$$
  C_{a_0 , a_1, \cdots, a_k} := P_{a_0 a} C^a_{a_1, \cdots, a_k}
$$
for the structure constant of the $k$-ary brackets.

  In terms of this we need to show that
  $$
    \mathrm{cs} 
	  = 
	P_{a b} t^a \wedge d_{\mathrm{W}} t^b
    +
    \sum_{k = 1}^\infty
    \frac{2}{(k+1)!} C_{a_0, \cdots, a_k} t^{a_0} \wedge \cdots \wedge t^{a_k}
    \,. 	
  $$
  
  The computation is best understood via the free dg-algebra
  $F(\mathfrak{g})$ on the graded vector space $\mathfrak{g}^*$,
  which in the above basis we may take to be generated by
  elements $\{t^a , \mathbf{d}t^a\}$. There is a dg-algebra 
  isomorphism
  $$
    F(\mathfrak{g}) \stackrel{\simeq}{\to} \mathrm{W}(\mathfrak{g})
  $$
  given by sending $t^a \mapsto t^a$ and 
  $\mathbf{d}t^a \mapsto d_{\mathrm{CE}(\mathfrak{g})} + r^a$.
  
  On $F(\mathfrak{g})$ the contracting homotopy is evidently given by the 
  map $\frac{1}{L} h$, where $L$ is the word length operator in the above basis and $h$ the graded derivation which sends $t^a \mapsto 0$ and $\mathbf{d}t^a \mapsto t^a$. Therefore $\tau$ is given by
  $$
    \xymatrix{
	  \mathrm{W}(\mathfrak{g}) \ar[d]^\simeq
	  \ar[r]^\tau & \mathrm{W}(\mathfrak{g}) 
	  \\
	  F(\mathfrak{g})
	  \ar[r]^{\frac{1}{L} h}
	  &
	  F(\mathfrak{g})
	  \ar[u]_\simeq
	}
	\,.
  $$
  With this we obtain
  $$
    \begin{aligned}
	  \mathrm{cs} & := \tau \langle -,-\rangle
	  \\
	  & = \tau
	  P_{a b}
	  \left(
	    d_{\mathrm{W}}t^a + \sum_{k = 1}^\infty C^a_{a_1, \cdots, a_k}
		t^{a_1}\wedge \cdots \wedge t^{a_k}
	  \right)
	  \wedge
	  \left(
	    d_{\mathrm{W}}t^b + \sum_{k = 1}^\infty C^b_{b_1, \cdots, b_k}
		t^{b_1}\wedge \cdots \wedge t^{b_k}
	  \right)
	  \\
	  & = 
	  P_{a b } t^a \wedge d_{\mathrm{W}}t^b
	  +
      \sum_{k = 1}^\infty	  
	  \frac{2}{k! (k+1)}
	  P_{a b} C^b{}_{b_1, \cdots, b_k} t^a \wedge t^{b_1} \wedge \cdots \wedge t^{b_k} 
	\end{aligned}
	\,.
  $$
\endofproof
\begin{remark}
  \index{string field theory}
  If here $\Sigma$ is a completely odd-graded dg-manifold, such 
  as $\Sigma = \mathbb{R}^{0|3}$, then this is the kind of action functional that appears in closed string field theory \cite{Zwiebach}\cite{KajiuraStasheff}.
  In this case the underlying space of the (super-)$L_\infty$-algebra $\mathfrak{g}$ is the BRST complex of the closed (super-)string and $[-,\cdots,-]_k$ is the string's tree-level $(k+1)$-point function. 
\end{remark}

\subsubsection{Non-perturbative AKSZ theory}
 \label{InfinCSAKSZ}
  \index{Chern-Simons functionals!AKSZ $\sigma$-models}
  \index{$\sigma$-model!AKSZ theory}

We now consider \emph{symplectic Lie $n$-algebroids $\mathfrak{P}$}.
These carry canonical invariant polynomials $\omega$. 
We show that the $\infty$-Chern-Simons action functional
associated to such $\omega$ is the locally the action functional of the 
\emph{AKSZ $\sigma$-model quantum field theory} with target space $\mathfrak{P}$
(due to \cite{AKSZ}, usefully reviewed in \cite{RoytenbergAKSZ}).
Globally it is a non-perturbative refinement of the AKSZ $\sigma$-model
with possibly non-trivial instanton sectors of fields.

This section is based on \cite{frs}. 

\begin{itemize}
  \item AKSZ $\sigma$-models --
    \ref{section.AKSZSigmaModels};
  \item \ref{section.AKSZ_theory} -- The AKSZ action as a Chern-Simons functional ;
  \item \ref{OrdinaryChernSimonsTheoryAsAKSZ} -- Ordinary Chern-Simons theory;
  \item \ref{section.PoissonSigmaModel} -- Poisson $\sigma$-model;
  \item \ref{section.CourantSigmaModel} -- Courant $\sigma$-model;
  \item \ref{section.HigherAbelianCSTheory} -- Higher abelian Chern-Simons theory.
\end{itemize}

\paragraph{AKSZ $\sigma$-Models}
\label{section.AKSZSigmaModels}
\index{AKSZ $\sigma$-models}

The class of topological field theories
known as \emph{AKSZ $\sigma$-models}\cite{AKSZ} 
contains in dimension 3 ordinary Chern-Simons theory 
(see \cite{FreedCS} for a comprehensive review)
as well as its Lie algebroid generalization 
(the \emph{Courant $\sigma$-model} \cite{Ikeda}), and
in dimension 2 the Poisson $\sigma$-model 
(see \cite{CattaneoFelder} for a review). It is 
therefore clear that the AKSZ construction 
is \emph{some} sort of generalized Chern-Simons theory.
Here we demonstrate that this statement is true also in 
a useful precise sense.

Our discussion proceeds from the observation that 
the standard Chern-Simons action
functional has a systematic origin in 
Chern-Weil theory (see for instance \cite{GHV} for 
a classical textbook treatment and \cite{HopkinsSinger} for 
the refinement to differential cohomology that we need here):
 
The refined Chern-Weil homomorphism assigns to any invariant 
polynomial $\langle -\rangle : \mathfrak{g}^{\otimes_n} \to \mathbb{R}$ 
on a Lie algebra $\mathfrak{g}$ 
of compact type a map that sends $\mathfrak{g}$-connections $\nabla$ 
on a smooth manifold $X$ to  cocycles 
$[\hat {\mathbf{p}}_{\langle-\rangle}(\nabla)] \in
H^{n+1}_{\mathrm{diff}}(X)$
in \emph{ordinary differential cohomology}.  
These differential cocycles 
refine the \emph{curvature characteristic class} 
$[\langle F_\nabla \rangle] \in H_{dR}^{n+1}(X)$ in de Rham cohomology
to a fully fledged \emph{line $n$-bundle with connection}, also known as
a \emph{bundle $(n-1)$-gerbe with connection}. 
And just as an ordinary line bundle (a ``line 1-bundle'') with connection
assigns holonomy to curves, so a line $n$-bundle with connection
assigns holonomy 
$\mathrm{hol}_{\hat {\mathbf{p}}}(\Sigma)$ 
to $n$-dimensional trajectories $\Sigma \to X$. 
For the special case where $\langle -\rangle$ 
is the Killing form polynomial 
and $X = \Sigma$ with $\dim\Sigma = 3$ one finds that this 
volume holonomy map
$\nabla \mapsto \mathrm{hol}_{\hat {\mathrm{p}}_{\langle -\rangle}(\nabla)}(\Sigma)$
is precisely the standard Chern-Simons action functional.
Similarly, for $\langle -\rangle$ any higher invariant polynomial
this holonomy action functional has as Lagrangian the 
corresponding higher Chern-Simons form. 
In summary, this means that Chern-Simons-type action functionals
on Lie algebra-valued connections are the images of the 
refined Chern-Weil homomorphism.
\par
In 
\ref{StrucChern-WeilHomomorphism}
a generalization of the 
Chern-Weil homomorphism to \emph{higher} (``derived'')
differential geometry has been established. In this context
smooth manifolds are generalized 
first to orbifolds, then to general Lie groupoids, 
to Lie 2-groupoids and finally to smooth $\infty$-groupoids 
(smooth $\infty$-stacks), while Lie algebras are generalized to
Lie $2$-algebras etc., up to $L_\infty$-algebras 
and more generally to Lie $n$-algebroids and 
finally to $L_\infty$-algebroids. 

In this context one has for $\mathfrak{a}$ 
any $L_\infty$-algebroid a natural notion of $\mathfrak{a}$-valued 
$\infty$-connections on $\exp(\mathfrak{a})$-principal 
smooth $\infty$-bundles
(where $\exp(\mathfrak{a})$ is a smooth $\infty$-groupoid obtained by
Lie integration from $\mathfrak{a}$).
By analyzing the abstractly defined higher Chern-Weil homomorphism
in this context one finds a direct higher analog of the above
situation: there is a notion of invariant polynomials $\langle-\rangle$ on 
an $L_\infty$-algebroid $\mathfrak{a}$ 
and these induce maps from $\mathfrak{a}$-valued $\infty$-connections
to line $n$-bundles with connections as before . 
\par
This construction drastically simplifies when one restricts attention to
\emph{trivial} $\infty$-bundles with (nontrivial) 
$\mathfrak{a}$-connections. Over a smooth manifold $\Sigma$
these are simply given by dg-algebra homomorphisms 
\[
  A : \mathrm{W}(\mathfrak{a}) \to \Omega^\bullet(\Sigma)
  \,,
\]
where $\mathrm{W}(\mathfrak{a})$ is the \emph{Weil algebra} of the 
$L_\infty$-algebroid $\mathfrak{a}$ \cite{SSSI}, 
and $\Omega^\bullet(\Sigma)$ is the de Rham algebra of $\Sigma$ 
(which is indeed the Weil algebra of $\Sigma$ thought of as an 
$L_\infty$-algebroid concentrated in degree 0). 
Then for $\langle-\rangle \in \mathrm{W}(\mathfrak{a})$ an invariant
polynomial, the corresponding $\infty$-Chern-Weil homomorphism
is presented by a choice of ``Chern-Simons element''
$\mathrm{cs} \in \mathrm{W}(\mathfrak{a})$, which exhibits the
\emph{transgression} of $\langle-\rangle$ to an $L_\infty$-cocycle
(the higher analog of a cocycle in Lie algebra cohomology):
the dg-morphism $A$ naturally maps the Chern-Simons element 
$\mathrm{cs}$ of $A$ to a differential form $\mathrm{cs}(A) \in \Omega^\bullet(\Sigma)$  and its integral is the corresponding 
$\infty$-Chern-Simons action functional $S_{\langle -\rangle}$
$$
  S_{\langle-\rangle}
  :
  A \mapsto \mathrm{hol}_{\hat {\mathbf{p}_{\langle-\rangle}}}(\Sigma)
    = 
    \int_\Sigma \mathrm{cs}_{\langle-\rangle}(A)
  \,.
$$
\par
Even though trivial $\infty$-bundles with $\mathfrak{a}$-connections
are a very particular subcase of the general $\infty$-Chern-Weil
theory, they are rich enough to contain AKSZ theory. Namely, here
we show that a symplectic dg-manifold of grade $n$
-- which is  the geometrical datum of 
the target space defining an AKSZ $\sigma$-model --
is naturally equivalently an $L_\infty$-algebroid 
$\mathfrak{P}$ endowed with a
quadratic and non-degenerate invariant polynomial $\omega$
of grade $n$. 
Moreover, under this identification the canonical Hamiltonian
$\pi$ on the symplectic target dg-manifold is identified 
as an $L_\infty$-cocycle on $\mathfrak{P}$. Finally, the
invariant polynomial $\omega$ is naturally in transgression with the
cocycle $\pi$ via a Chern-Simons element $\mathrm{cs}_\omega$ that turns 
out to be the Lagrangian of the AKSZ $\sigma$-model:
$$
  \int_\Sigma L_{\mathrm{AKSZ}}(-) 
   = 
  \int_\Sigma \mathrm{cs}_{\omega}(-)
  \,.
$$
(An explicit description of $L_{\mathrm{AKSZ}}$ is given below
in def. \ref{TheAKSZAction})

In summary this means that we find the following
dictionary of concepts:\\
\begin{center}
\begin{tabular}{c|c|c}
  {\bf Chern-Weil theory} && {\bf AKSZ theory}
  \\
  \hline
  &&
  \\
  cocycle & $\pi$ &  Hamiltonian
  \\
   &&
  \\
  transgression element & $\mathrm{cs}$ & Lagrangian
  \\
    &&
  \\
  invariant polynomial & $\omega$ & symplectic structure
  \\
  &&
  \\
  \hline
\end{tabular}
\end{center}
More precisely, we (explain and then) prove here the following theorem:
\begin{theorem}
  \label{AKSZIsCW}
  For $(\mathfrak{P}, \omega)$ an $L_\infty$-algebroid with
  a quadratic non-degenerate invariant polynomial, the
  corresponding $\infty$-Chern-Weil homomorphism
  $$
    \nabla \mapsto \mathrm{hol}_{\hat {\mathbf{p}_\omega}}(\Sigma)
  $$
  sends $\mathfrak{P}$-valued $\infty$-connections $\nabla$
  to their corresponding exponentiated AKSZ action
  $$
    \cdots = \exp(i \int_\Sigma L_{\mathrm{AKSZ}}(\nabla))
    \,.
  $$
  \,.
\end{theorem}
The local differential form data involved in this statement 
is at the focus of attention in this 
section
here and 
contained in prop. \ref{TheAKSZActionFromCS} below. 

We consider, in definition \ref{TheAKSZAction} below,
for any symplectic dg-manifold $(X,\omega)$ a functional
$S_{\mathrm{AKSZ}}$ on spaces of maps $\mathfrak{T}\Sigma \to X$
of smooth graded manifolds. While only this precise definition
is referred to in the remainder of the section, we begin by
indicating informally the original motivation of $S_{\mathrm{AKSZ}}$. 
The reader uncomfortable with these somewhat vague considerations
can take note of def. \ref{TheAKSZAction} and then skip to the
next section.

Generally, a \emph{$\sigma$-model field theory} is, roughly, one 
\begin{enumerate}
\item whose fields over a space $\Sigma$ are 
maps $\phi : \Sigma \to X$ 
to some space $X$;
\item
 whose action functional is, apart from a kinetic term,
 the transgression of some kind of cocycle  on $X$
 to the mapping space $\mathrm{Map}(\Sigma,X)$.
\end{enumerate}
Here the terms ``space'', ``maps'' and ``cocycles''
are to be made precise in a suitable context.
One says that $\Sigma$ is the \emph{worldvolume}, 
$X$ is the \emph{target space} and the cocycle is
the \emph{background gauge field}.

For instance, an ordinary charged particle 
(such as an electron) is described by a $\sigma$-model 
where $\Sigma = (0,t) \subset \mathbb{R}$ is the abstract
\emph{worldline}, where $X$ is a (pseudo-)Riemannian smooth manifold
(for instance our spacetime), and where the background cocycle
is a line bundle with connection on $X$ (a degree-2 cocycle
in ordinary differential cohomology of $X$, representing a
background \emph{electromagnetic field}). Up to a kinetic term, 
the action functional is the holonomy of the connection over 
a given curve $\phi : \Sigma \to X$. 
A textbook discussion of these
standard kinds of $\sigma$-models is, for instance, in 
\cite{DeligneMorgan}.

The $\sigma$-models which we consider here are 
\emph{higher} generalizations of this example,
where the background gauge field is a cocycle of higher degree
(a higher bundle with connection) and where the worldvolume is
accordingly higher dimensional. In addition, $X$ is allowed to be 
not just a manifold, but an approximation to a 
\emph{higher orbifold} (a smooth $\infty$-groupoid).

More precisely, here we take the category of spaces
to be $\mathrm{SmoothDgMfd}$ from def. \ref{DGManifolds}.
We take target space to be a symplectic dg-manifold $(X,\omega)$
and the worldvolume to be the shifted tangent bundle 
$\mathfrak{T}\Sigma$ of a compact smooth manifold $\Sigma$.
Following \cite{AKSZ}, one may imagine that we can form a 
smooth $\mathbb{Z}$-graded mapping space
$\mathrm{Maps}(\mathfrak{T}\Sigma,X)$ 
of smooth graded manifolds. On this space the canonical
vector fields $v_\Sigma$ and $v_X$ naturally have commuting 
actions from the left
and from the right, respectively,
so that their sum $v_\Sigma + v_X$ 
equips $\mathrm{Maps}(\mathfrak{T}\Sigma,X)$ itself with the structure of 
a differential graded smooth manifold.

Next we take the ``cocycle'' on $X$ (to be made precise
in the next section) to be 
the Hamiltonian $\pi$ (def. \ref{Hamiltonians}) 
of $v_X$ with respect to the symplectic
structure $\omega$, according to def. \ref{SymplecticDgManifold}.
One wants to assume that there is a kind of 
Riemannian structure on $\mathfrak{T}\Sigma$
that allows us to form the transgression
$$
  \int_{\mathfrak{T}\Sigma} \mathrm{ev}^* \omega
  :=
  p_! \mathrm{ev}^* \omega
$$
by pull-push through the canonical correspondence
$$
  \xymatrix{
     \mathrm{Maps}(\mathfrak{T}\Sigma,X)
     \ar@{<-}[r]^{p}
     &
     \mathrm{Maps}(\mathfrak{T}\Sigma,X) \times \mathfrak{T}\Sigma
     \ar[r]^<<<<{\mathrm{ev}}
     &
     X
  }
  \,.
$$
When one succeeds in making this precise, one expects to 
find that $\int_{\mathfrak{T}\Sigma} \mathrm{ev}^* \omega$ is in 
turn a symplectic structure on the mapping space. 

This implies that the vector field
$v_\Sigma + v_X$ on mapping space has a Hamiltonian 
$$
  \mathbf{S} \in C^\infty(\mathrm{Maps}(\mathfrak{T}\Sigma,X))
  \,,\;\;\mbox{s.t.}\;\;
  \mathbf{d}\mathbf{S} = \iota_{v_\Sigma + v_x} 
  \int_{\mathfrak{T}\Sigma} \mathrm{ev}^* \omega
  \,.
$$
The grade-0 component
$$
  S_{\mathrm{AKSZ}}
  :=
  \mathbf{S}|_{\mathrm{Maps}(\mathfrak{T}\Sigma,X)_0}
$$ 
constitutes a functional on the space
of morphisms of graded manifolds $\phi : \mathfrak{T}\Sigma \to X$. This is
the \emph{AKSZ action functional} defining the AKSZ $\sigma$-model
with target space $X$ and background field/cocycle $\omega$.

In \cite{AKSZ}, this procedure is indicated only somewhat vaguely. 
The focus of attention there is on a discussion, from this perspective,
of the action functionals
of the 2-dimensional $\sigma$-models called the 
\emph{A-model} and the \emph{B-model}.
In \cite{RoytenbergAKSZ} a more detailed discussion of
the general construction is given, including an explicit formula
for $\mathbf{S}$, and
hence for $S_{\mathrm{AKSZ}}$. 
That formula is the following:
\begin{definition}
 \label{TheAKSZAction}
 \index{$\sigma$-model!AKSZ action functional}
For $(X,\omega)$ a symplectic dg-manifold of grade $n$
with global Darboux coordinates $\{x^a\}$, 
$\Sigma$ a smooth compact manifold of dimension $(n+1)$
and $k \in \mathbb{R}$, the
\emph{AKSZ action functional}
$$
  S_{\mathrm{AKSZ}} : 
   \mathrm{SmoothGrMfd}(\mathfrak{T}\Sigma, X)
  \to
  \mathbb{R}
$$
is
$$
  S_{\mathrm{AKSZ}} 
   :
  \phi
  \mapsto
  \int_{\Sigma} 
  \left(
    \frac{1}{2}\omega_{ab} \phi^a \wedge d_{\mathrm{dR}}\phi^b
    -
    \phi^* \pi
  \right)
  \,,
$$
where $\pi$ is the Hamiltonian for $v_X$ with respect to $\omega$
and where on the
right we are interpreting fields as forms on $\Sigma$ according to
prop. \ref{FormsFromGrManifoldMaps}.
\end{definition}
This formula hence defines an infinite class of $\sigma$-models
depending on the target space structure $(X, \omega)$. 
(One can also consider arbitrary relative factors between the first and
the second term, but below we shall find that the above choice is
singled out).
In \cite{AKSZ}, it was
already noticed that ordinary Chern-Simons theory is a special
case of this for $\omega$ of grade 2,  as is the Poisson $\sigma$-model
for $\omega$ of grade 1 (and hence, as shown there, also the A-model
and the B-model). 
The main example in \cite{RoytenbergAKSZ} spells out the
general case for $\omega$ of grade 2, which is called the
\emph{Courant $\sigma$-model} there. 
(We review and re-derive
all these examples in detail 
below.)

One nice aspect of this construction is that it follows immediately
that the full Hamiltonian $\mathbf{S}$ on the mapping space satisfies
$\{\mathbf{S}, \mathbf{S}\} = 0$. Moreover, using the standard formula
for the internal hom of chain complexes, one finds that the cohomology
of $(\mathrm{Maps}(\mathfrak{T}\Sigma,X), v_\Sigma + v_X)$ in degree 0
is the space of functions on those fields that satisfy the Euler-Lagrange
equations of $S_{\mathrm{AKSZ}}$. Taken together, these facts imply that 
$\mathbf{S}$ is a solution of the ``master equation''
of a BV-BRST complex for the quantum field theory
defined by $S_{\mathrm{AKSZ}}$. This is a crucial ingredient for the 
quantization of the model, and this is what the AKSZ construction
is mostly used for in the literature (for instance \cite{CattaneoFelder}).

Here we want to focus on another nice aspect of the AKSZ-construction:
it hints at a deeper reason for \emph{why} the $\sigma$-models of 
this type are special. It is indeed one of the very few
proposals for what a general abstract mechanism might be that picks 
out among the vast space of all possible local action functionals 
those that seem to be of relevance ``in nature''. 

We now proceed to show that the class of action functionals
$S_{\mathrm{AKSZ}}$ are precisely those that higher Chern-Weil theory
canonically associates to target data $(X,\omega)$. 
Since higher Chern-Weil theory in turn is canonically given 
on very general abstract grounds, 
this in 
a sense amounts to a derivation of $S_{\mathrm{AKSZ}}$ from 
``first principles'', and it shows that a wealth of very general 
theory applies to these systems. 

\paragraph{The AKSZ action as an $\infty$-Chern-Simons functional}
\label{section.AKSZ_theory}
\index{AKSZ $\sigma$-models!nonperturbative}

We now show how an $L_\infty$-algebroid $\mathfrak{a}$ endowed with a triple $(\pi,\mathrm{cs},\omega)$ consisting of a Chern-Simons element transgressing an invariant polynomial $\omega$ to a cocycle $\pi$ defines an AKSZ-type $\sigma$-model action. The starting point is to take as target space the tangent Lie $\infty$-algebroid $\mathfrak{T}\mathfrak{a}$, i.e., to consider as \emph{space of fields} of the theory the space of maps $\mathrm{Maps}(\mathfrak{T}\Sigma,\mathfrak{T}\mathfrak{a})$ from the worldsheet $\Sigma$ to $\mathfrak{T}\mathfrak{a}$. Dually, this is the space of morphisms of dgcas from
$\mathrm{W}(\mathfrak{a})$ to $\Omega^\bullet(\Sigma)$, i.e., the space of degree 1 $\mathfrak{a}$-valued differential forms on $\Sigma$ from definition \ref{definition.a-valued-differential-form}.

\begin{remark}
As we noticed in the introduction, in the context of the AKSZ $\sigma$-model a degree 1 $\mathfrak{a}$-valued differential form on $\Sigma$ should be thought of as the datum of a (notrivial) $\mathfrak{a}$-valued connection on a trivial principal $\infty$-bundle on $\Sigma$. 
\end{remark}
Now that we have defined the space of fields, we have to define the action. 
We have seen in definition \ref{definition.chern-simons-form} that a degree 1 $\mathfrak{a}$-valued differential form $A$ on $\Sigma$ maps the Chern-Simons element $\mathrm{cs}\in \mathrm{W}(\mathfrak{a})$ to a differential form $\mathrm{cs}(A)$ on $\Sigma$. Integrating this differential form on $\Sigma$ will therefore give an AKSZ-type action which
is naturally interpreted as an higher Chern-Simons action functional:
$$
\begin{aligned}
\mathrm{Maps}(\mathfrak{T}\Sigma,\mathfrak{T}\mathfrak{a}) &\to \mathbb{R}\\
A&\mapsto \int_\Sigma \mathrm{cs}(A).
\end{aligned}
$$

Theorem \ref{AKSZIsCW} then reduces to showing that, when $\{\mathfrak{a}, (\pi,\mathrm{cs},\omega)\}$ is the set of $L_\infty$-algebroid data arising from a symplectic Lie $n$-algebroid $(\mathfrak{P}, \omega)$, the AKSZ-type action dscribed above is precisely the AKSZ action for $(\mathfrak{P}, \omega)$. More precisely, this is stated as follows.
\begin{proposition}
  \label{TheAKSZActionFromCS}
  For $(\mathfrak{P}, \omega)$ a symplectic Lie $n$-algebroid 
  coming by proposition \ref{SymplDgSpaceAsLAlgd}
  from a symplectic dg-manifold of positive grade $n$ with 
  global Darboux chart, the action functional induced by 
  the canonical Chern-Simons element 
  $$
     \mathrm{cs} \in \mathrm{W}(\mathfrak{P})
  $$ 
  from proposition \ref{TheCSElement}
  is the AKSZ action from definition \ref{TheAKSZAction}:
  $$
    \int_\Sigma \mathrm{cs}
     =
    \int_\Sigma L_{\mathrm{AKSZ}}
    \,.
  $$
  In fact the two Lagrangians differ at most by an exact term
  $$
    \mathrm{cs} \sim L_{\mathrm{AKSZ}}
    \,.
  $$
\end{proposition}
\proof
We have seen in remark \ref{remark.local_cs} that in Darboux coordinates $\{x^a\}$ where
$$
  \omega = \frac{1}{2}\omega_{a b} \mathbf{d}x^a \wedge \mathbf{d}x^b
$$ 
the Chern-Simons element from proposition \ref{TheCSElement} is
given by
$$
  \mathrm{cs} 
    = \frac{1}{n}
  \left(
    \mathrm{deg}(x^a) \,
    \omega_{a b} x^a \wedge d_{\mathrm{W}(\mathfrak{P})} x^b
    -
    n \pi
  \right)\,.
  $$
This means that for $\Sigma$ an $(n+1)$-dimensional manifold and
$$
  \Omega^\bullet(\Sigma) 
    \leftarrow 
  \mathrm{W}(\mathfrak{P})
   :
  \phi
$$
a (degree 1) $\mathfrak{P}$-valued differential form on $\Sigma$
we have 
$$
  \begin{aligned}
      \int_\Sigma \mathrm{cs}(\phi)
      &= 
      \frac{1}{n}
      \int_\Sigma
      \left(
       \sum_{a,b} 
       \mathrm{deg}(x^a)\,\omega_{a b} \phi^a \wedge  d_{\mathrm{dR}} \phi^b 
       -
       n \pi(\phi) 
     \right)
  \end{aligned}
  \,,
$$
where we used $\phi(d_{\mathrm{W}(\mathfrak{P})} x^b)=d_{\mathrm{dR}} \phi^b$, as in  remark \ref{ImagesOfdAnddWUnderForms}.
Here the asymmetry in the coefficients of the first term is only
apparent. Using integration by parts on a closed $\Sigma$ 
we have
$$
  \begin{aligned}
    \int_\Sigma
    \sum_{a,b} \mathrm{deg}(x^a)\,\omega_{a b} \phi^a \wedge  d_{\mathrm{dR}} \phi^b 
    & =
    \int_\Sigma
     \sum_{a,b} (-1)^{1+\mathrm{deg}(x^a)}\mathrm{deg}(x^a)\,\omega_{a b} (d_{\mathrm{dR}} \phi^a) \wedge   \phi^b 
    \\
    & =
    \int_\Sigma
    \sum_{a,b} (-1)^{(1+\mathrm{deg}(x^a))(1+\mathrm{deg}(x^b))}\mathrm{deg}(x^a)\,\omega_{a b} 
  \phi^b \wedge (d_{\mathrm{dR}} \phi^a)     
    \\
    & =
    \int_\Sigma \sum_{a,b} \mathrm{deg}(x^b)\,\omega_{a b} 
      \phi^a \wedge (d_{\mathrm{dR}} \phi^b)     
  \end{aligned}
  \,,
$$
where in the last step we switched the indices on $\omega$ and used that
$\omega_{ab} = (-1)^{(1+\mathrm{deg}(x^a))(1+\mathrm{deg}(x^b))} \omega_{b a}$.
Therefore
$$
  \begin{aligned}
    \int_\Sigma
    \sum_{a,b} \mathrm{deg}(x^a)\,\omega_{a b} \phi^a \wedge  d_{\mathrm{dR}} \phi^b 
    & =
    \frac{1}{2}
    \int_\Sigma
    \sum_{a,b} \mathrm{deg}(x^a)\,\omega_{a b} \phi^a \wedge  d_{\mathrm{dR}} \phi^b 
    +
    \frac{1}{2}
    \int_\Sigma
    \sum_{a,b} \mathrm{deg}(x^b)\,\omega_{a b} \phi^a \wedge  d_{\mathrm{dR}} \phi^b 
    \\
    &=
    \frac{n}{2}
    \int_\Sigma
      \omega_{a b} \phi^a \wedge  d_{\mathrm{dR}} \phi^b 
    \,.
  \end{aligned}
  \,.
$$
Using this in the above expression for the action yields 
$$
  \int_\Sigma \mathrm{cs}(\phi)
  = 
  \int_\Sigma
   \left(
   \frac{1}{2}\omega_{ab} \phi^a \wedge d_{\mathrm{dR}} \phi^b
   -
   \pi(\phi)
   \right)
   \,,
$$
which is the formula for the action functional from definition \ref{TheAKSZAction}.
\endofproof

We now unwind the general statement of proposition \ref{TheAKSZActionFromCS}
and its ingredients
in the central examples of interest, from proposition \ref{SymplecticPoissonCourant}:
the ordinary Chern-Simons action functional, the Poisson $\sigma$-model Lagrangian, and the Courant $\sigma$-model Lagrangian.
(The ordinary Chern-Simons model is the special 
case of the Courant $\sigma$-model for $\mathfrak{P}$ having as
base manifold the point. But since it is the archetype of
all models considered here, it deserves its own discussion.)

By the very content of proposition \ref{TheAKSZActionFromCS}
there are no surprises here and the following essentially
amounts to a review of the standard formulas for these 
examples. But it may be helpful to see our general 
$\infty$-Lie theoretic derivation of these formulas spelled out 
in concrete cases, if only to carefully track the various signs
and prefactors.

\paragraph{Ordinary Chern-Simons theory}
\label{OrdinaryChernSimonsTheoryAsAKSZ}
\index{Chern-Simons functionals!ordinary Chern-Simons theory}
\index{$\sigma$-model!ordinary Chern-Simons theory}

Let $\mathfrak{P} = b\mathfrak{g}$
be a semisimple Lie algebra regarded as an $L_\infty$-algebroid 
with base space the point
and let $\omega := \langle -,-\rangle\in \mathrm{W}(b\mathfrak{g})$ be 
its Killing form invariant polynomial. Then 
$(b \mathfrak{g}, \langle -,-\rangle)$ is a symplectic Lie 2-algebroid.

For $\{t^a\}$ a dual basis
for $\mathfrak{g}$, being generators of grade 1 in 
$\mathrm{W}(\mathfrak{g})$ we have
$$
  d_{\mathrm{W}} t^a = - \frac{1}{2}C^a{}_{b c} t^a \wedge t^b + 
   \mathbf{d}t^a
$$
where $C^a{}_{b c} := t^a([t_b,t_c])$ and
$$
  \omega = \frac{1}{2} P_{a b} \mathbf{d}t^a \wedge \mathbf{d}t^b
  \,,
$$
where $P_{ab} := \langle t_a, t_b \rangle$.
The Hamiltonian cocycle $\pi$ from prop. \ref{CanonicalCocycleOfSymplecticLieAlgebroid}
is
$$
  \begin{aligned}
    \pi
     & =
     \frac{1}{2+1}\iota_v \iota_\epsilon \omega
     \\
     &= \frac{1}{3} \iota_v P_{ab} t^a \wedge \mathbf{d}t^b
     \\
     & =-\frac{1}{6}P_{ab}C^b_{cd}t^a\wedge t^c\wedge t^d
     \\
     & =: -\frac{1}{6}C_{abc}t^a\wedge t^b\wedge t^c.
  \end{aligned}
$$
Therefore the Chern-Simons element from 
prop. \ref{TheCSElement} is found to be
$$
\begin{aligned}
 \mathrm{cs}
  &=
  \frac{1}{2}\left(P_{ab}t^a\wedge\mathbf{d}t^b
  -
  \frac{1}{6}C_{abc}t^a\wedge t^b\wedge t^c\right)
   \\
   &=
   \frac{1}{2}\left(P_{ab}t^a\wedge d_\mathrm{W}t^b
   +
   \frac{1}{3}C_{abc}t^a\wedge t^b\wedge t^c\right).
\end{aligned}
$$
This is indeed, up to an overall factor $1/2$, the familiar standard choice of Chern-Simons element on a Lie algebra. To see this more explicitly,
notice that evaluated on a
$\mathfrak{g}$-valued connection form
$$
  \Omega^\bullet(\Sigma) \leftarrow \mathrm{W}(b\mathfrak{g}) : A
$$
this is
$$
  2 \mathrm{cs}(A) 
   = 
  \langle A \wedge F_A\rangle 
   - 
  \frac{1}{6}\langle A \wedge [A, A]\rangle
   = 
  \langle A \wedge d_{dR}A\rangle 
   + 
  \frac{1}{3}\langle A \wedge [A, A]\rangle
  \,.
$$
If $\mathfrak{g}$ is a matrix Lie algebra then the Killing form is 
proportional to the trace of the matrix product: $\langle t_a,t_b\rangle = \mathrm{tr}(t_a t_b)$. In this case we have
$$
  \begin{aligned}
    \langle A \wedge [A, A]\rangle
    &=
    A^a  \wedge A^b \wedge A^c \,\mathrm{tr}(t_a (t_b t_c - t_c t_b))
    \\
    &=
    2 A^a \wedge A^b \wedge A^c \,\mathrm{tr}(t_a t_b t_c )
    \\
    &= 
    2 \,\mathrm{tr}(A \wedge A \wedge A)
  \end{aligned}
$$
and hence
$$
  2 \mathrm{cs}(A) 
   = \mathrm{tr}\left(
A \wedge F_A
   - 
  \frac{1}{3} A \wedge A \wedge A\right)
   = 
\mathrm{tr}\left( A \wedge d_{dR}A 
   + 
  \frac{2}{3} A \wedge  A \wedge A\right)
  \,.
$$

\paragraph{Poisson $\sigma$-model}
\label{section.PoissonSigmaModel}
\index{Chern-Simons functionals!Poisson $\sigma$-model}
\index{$\sigma$-model!Poisson $\sigma$-model}

Let  $(M, \{-,-\})$ be a Poisson manifold and 
let $\mathfrak{P}$ be the corresponding 
Poisson Lie algebroid. This is a symplectic Lie 1-algebroid.
Over a chart for the shifted cotangent bundle $T^*[-1]X$
with coordinates $\{x^i\}$ of degree 0
and  $\{\partial_i\}$ of degree 1, respectively, we have
$$
  d_{\mathrm{W}} x^i = -\pi^{i j}\mathbf{\partial}_j + \mathbf{d}x^i;
$$
where $\pi^{i j} := \{x^i , x^j\}$ and
$$
  \omega = \mathbf{d}x^i \wedge \mathbf{d}\partial_i
  \,.
$$
The Hamiltonian cocycle from prop. \ref{CanonicalCocycleOfSymplecticLieAlgebroid}
is
$$
\pi= \frac{1}{2}\iota_v \iota_\epsilon  \omega=- \frac{1}{2} \pi^{ij} \partial_i \wedge \partial_j
$$
and the Chern-Simons element from prop. \ref{TheCSElement} is
$$
  \begin{aligned}
  \mathrm{cs}
  &= \iota_\epsilon \omega + \pi
  \\
  &= \partial_i \wedge \mathbf{d}x^i 
     - \frac{1}{2}\pi^{ij}\partial_i\wedge\partial_j
  \end{aligned}
  \,.
$$
In terms of $d_{\mathrm{W}}$ instead of $\mathbf{d}$ 
this is 
$$
  \begin{aligned}
    \mathrm{cs} & = \partial_i \wedge d_{\mathrm{W}}x^i -  \pi
    \\
    &= 
    \partial_i \wedge d_{\mathrm{W}}x^i + \frac{1}{2}\pi^{ij}\partial_i \partial_j\,.
      \end{aligned}
$$
So for $\Sigma$ a 2-manifold and
$$
  \Omega^\bullet(\Sigma) \leftarrow \mathrm{W}(\mathfrak{P}) : (X,\eta)
$$
a Poisson-Lie algebroid valued differential form on $\Sigma$
-- which in components is a function 
$X: \Sigma \to M$ and a 1-form 
$\eta \in \Omega^1(\Sigma, X^* T^* M)$ -- 
the corresponding AKSZ action is
$$
 \int_\Sigma \mathrm{cs}(X,\eta) 
   = \int_\Sigma
  \eta \wedge d_{\mathrm{dR}}X    
    + 
   \frac{1}{2}\pi^{ij}(X)\eta_i \wedge \eta_j
  \,.
$$
This is the Lagrangian of the Poisson $\sigma$-model 
\cite{CattaneoFelder}.

\paragraph{Courant $\sigma$-model}
\label{section.CourantSigmaModel}
\index{Chern-Simons functionals!Courant $\sigma$-model}
\index{$\sigma$-model!Courant $\sigma$-model}

A Courant algebroid is a symplectic Lie 2-algebroid. 
By the previous example this is a higher analog of a 
Poisson manifold. Expressed in components in the language of ordinary 
differential geometry, a Courant algebroid is a vector bundle $E$ 
over a manifold $M_{0}$, equipped with: a non-degenerate bilinear form
$\langle \cdot,\cdot \rangle$ on the fibers, a bilinear bracket
$[\cdot,\cdot]$ on sections $\Gamma(E)$, and a bundle map (called the
anchor) $\rho \colon E \to TM$, satisfying several compatibility
conditions. The bracket $[\cdot,\cdot]$ may be required to be
skew-symmetric (Def.\ 2.3.2 in \cite{RoytenbergCourant}), in which case it gives
rise to a Lie 2-algebra structure, or, alternatively, it may be required to
satisfy a Jacobi-like identity (Def.\ 2.6.1 in \cite{RoytenbergCourant}), in
which case it gives a Leibniz algebra structure.

It was shown in \cite{RoytenbergCourant} that Courant algebroids $E \to M_{0}$ 
in this component form
are in 1-1 correspondance with (non-negatively graded) grade 2
symplectic dg-manifolds $(M,v)$. Via this correspondance, $M$ is obtained
as a particular symplectic submanifold of $T^{\ast}[2]E[1]$ equipped
with its canonical symplectic structure.

Let $(M,v)$ be a Courant algebroid as above. In Darboux coordinates, the
symplectic structure is
\[
\omega = \mathbf{d}p_{i} \wedge \mathbf{d}q^{i} + \frac{1}{2}g_{ab} 
\mathbf{d}\xi^{a} \wedge  \mathbf{d}\xi^{b},
\]
with
\[
\deg{q^{i}}=0, ~ \deg{\xi^{a}}=1, ~ \deg{p_{i}}=2,
\]
and $g_{ab}$ are constants. The Chevalley-Eilenberg differential
corresponds to the vector field:
\[
v = P^{i}_{a} \xi^{a} \frac{\partial}{\partial q^{i}}
+ g^{a b} \bigl( P^{i}_{b}p_{i} - \frac{1}{2} T_{bcd}
\xi^{c} \xi^{d} \bigr) \frac{\partial}{\partial \xi^{a}}
+\left (-\frac{\partial P^{j}_{a}}{\partial q^{i}} \xi^{a}p_{j}
+ \frac{1}{6} \frac{\partial T_{abc}}{\partial q^{i}} \xi^{a} \xi^{b}
\xi^{c} \right)\frac{\partial}{\partial p_{i}}.
\]
Here $P^{i}_{a}=P^{i}_{a}(q)$ and $T_{abc}=T_{abc}(q)$ are particular degree zero
functions encoding the Courant algebroid structure.
Hence, the differential on the Weil algebra is:
\begin{align*}
d_{W} q^{i} &= P^{i}_{a} \xi^{a} + \mathbf{d} q^{i} \\
d_{W} \xi^{a} &= g^{a b} \bigl( P^{i}_{b}p_{i} - \frac{1}{2} T_{bcd}
\xi^{c} \xi^{d} \bigr) + \mathbf{d} \xi^{a} \\
d_{W} p_{i} &= -\frac{\partial P^{j}_{a}}{\partial q^{i}} \xi^{a}p_{j}
+ \frac{1}{6} \frac{\partial T_{abc}}{\partial q^{i}} \xi^{a} \xi^{b} \xi^{c}
+ \mathbf{d} p_{i}.
\end{align*}

Following remark. \ref{remark.local_hamiltonian}, we construct the
corresponding Hamiltonian cocycle from prop. 
\ref{CanonicalCocycleOfSymplecticLieAlgebroid}:
\begin{align*}
\pi &= \frac{1}{n+1}  \omega_{ab}\deg(x^a) x^a \wedge v^b\\
&= \frac{1}{3}\bigl( 2 p_{i} \wedge v(q^{i}) + g_{a b} \xi^{a}
\wedge v(\xi^{b}) \bigr)\\
&= \frac{1}{3}\bigl( 2 p_{i} P^{i}_{a} \xi^{a} + 
\xi^{a} P^{i}_{a}p_{i} - \frac{1}{2} T_{abc} \xi^{a}\xi^{b} \xi^{c} \bigr)\\
&= P^{i}_{a} \xi^{a} p_{i} - \frac{1}{6} T_{abc} \xi^{a}\xi^{b} \xi^{c}.
\end{align*}

The Chern-Simons element 
from prop. \ref{TheCSElement} is:
\begin{align*}
  \mathrm{cs} &= 
  \frac{1}{2}
  \left( 
    \sum_{a b} \deg(x^a) \,\omega_{a b} x^a \wedge d_{W}x^b  -  2 \pi
  \right)
   \\
   &= 
   p_{i} d_{W} q^{i} + \frac{1}{2}g_{ab} \xi^{a} d_{W} \xi^{b} -  \pi\\
   &=
   p_{i} d_{W} q^{i} + \frac{1}{2}g_{ab} \xi^{a} d_{W} \xi^{b}-P^{i}_{a} \xi^{a} p_{i} + \frac{1}{6} T_{abc} \xi^{a}\xi^{b} \xi^{c}.
\end{align*}
So for a map   
\[
\Omega^\bullet(\Sigma) \leftarrow \mathrm{W}(\mathfrak{P}) : (X,A,F)
\]
where $\Sigma$ is a closed 3-manifold, we have
\[
  \int_\Sigma\mathrm{cs}(X,A,F)
  = 
 \int_\Sigma F_{i} \wedge d_{\mathrm{dR}} X^{i} 
  +
  \frac{1}{2}g_{ab} A^{a} \wedge
  d_{\mathrm{dR}} A^{b}
  -
  P^{i}_{a} A^{a} \wedge F_{i}  
  + 
  \frac{1}{6} T_{abc} A^{a} \wedge A^{b} \wedge A^{c}. 
\]
This is the AKSZ action for the Courant algebroid $\sigma$-model
from \cite{Ikeda} \cite{RoytenbergCourant}\cite{RoytenbergAKSZ}.

\paragraph{Higher abelian Chern-Simons theory in $d = 4k+3$}
 \label{section.HigherAbelianCSTheory}
\index{$\sigma$-model!higher dimensional abelian Chern-Simons theory}
\index{Chern-Simons functionals!higher dimensional abelian Chern-Simons theory}

We discuss higher abelian Chern-Simons theory, \ref{HigherAbelianCS},
from the point of view of AKSZ theory.

\medskip

For $k \in \mathbb{N}$, let $\mathfrak{a}$ be the 
delooping of the line Lie $2k$-algebra, def. \ref{LineLieNAlgebra}: 
$\mathfrak{a} = b^{2k+1}\mathbb{R}$.
By observation \ref{InvariantPolynomialsOfLineLienAlgebra} there is, up to scale,
a unique binary invariant polynomial on $b^{2k+1}\mathbb{R}$,
and this is the wedge product of the unique generating 
unary invariant polynomial $\gamma$ in degree $2k+2$ with itself:
$$
  \omega
  := 
  \gamma \wedge \gamma
  \in 
  \mathrm{W}(b^{4k+4}\mathbb{R})
  \,.
$$
This invariant polynomial is clearly non-degenerate: 
for $c$ the canonical generator of $\mathrm{CE}(b^{2k+1}\mathbb{R})$
we have
$$
  \omega = \mathbf{d}c \wedge \mathbf{d}c
  \,.
$$ 
Therefore $(b^{2k+1}\mathbb{R}, \omega)$ induces an 
$\infty$-Chern-Simons theory of AKSZ $\sigma$-model type
in dimension $n+1 = 4k+3$.
(On the other hand, on $b^{2k}\mathbb{R}$ there is only the 0 
binary invariant polynomial, so that no AKSZ-$\sigma$-models
are induced from $b^{2k}\mathbb{R}$.)

The Hamiltonian cocycle from 
prop. \ref{CanonicalCocycleOfSymplecticLieAlgebroid} vanishes
$$
  \pi = 0
$$ 
because the differential $d_{\mathrm{CE}(b^{2k+1}\mathbb{R})}$ is trivial.
The Chern-Simons element from prop. \ref{TheCSElement} is
$$
  \mathrm{cs} = c \wedge \mathbf{d}c
  \,.
$$
A field configuration, def. \ref{definition.a-valued-differential-form}, 
of this $\sigma$-model over a $(2k+3)$-dimensional manifold
$$
  \Omega^\bullet(\Sigma) \leftarrow
  \mathrm{W}(b^{2k+1})
  : C
$$
is simply a $(2k+1)$-form. The AKSZ action functional in this case
is
$$
  S_{AKSZ} : C \mapsto \int_\Sigma C \wedge d_{dR}C
  \,.
$$
The simplicity of this discussion is deceptive. It 
results from the fact that here we are looking at
$\infty$-Chern-Simons theory for universal Lie integrations and
for topologically trivial $\infty$-bundles. 
More generally the $\infty$-Chern-Simons theory for $\mathfrak{a} = b^{2k+1}\mathbb{R}$
is  nontrivial and rich, as discussed in \ref{HigherAbelianCS}.
Its configuration space is that of 
\emph{circle $(2k+1)$-bundles with connection} (\ref{SmoothStrucDifferentialCohomology}) 
on $\Sigma$, classified
by ordinary differential cohomology in degree $2k+2$, and the 
action functional is given by the fiber integration in differential 
cohomology to the point
over the Beilinson-Deligne cup product, which is locally
given by the above formula, but contains global twists.

\newpage
\subsection{Higher Wess-Zumino-Witten field theory}
 \label{WZWApplications}
 \label{HigherWZWLocalPrequantumFieldTheory}
 \index{Wess-Zumino-Witten functionals!examples}
 \index{$\sigma$-model!Wess-Zumino-Witten type}

We discuss examples of higher WZW functionals, def. \ref{StrucWZWFunctional}.

This section draws from \cite{InfinityWZW}.

\medskip

\begin{itemize}
  \item \ref{TraditionalWZW} --  Introduction: Traditional WZW and the need for higher WZW
  \item \ref{WZWLienLalgebraicFormulation} -- Lie $n$-algebraic formulation
  \item \ref{WZWBoundaryConditionsAndBraneIntersectionLaws} -- Boundary conditions and branes
  \item \ref{SuperBranesAndTheirIntersectionLaws} -- Super $p$-branes and the brane bouquet
\end{itemize}

\subsubsection{Introduction: Traditional WZW and the need for higher WZW}
\label{TraditionalWZW}
\index{Wess-Zumino-Witten functionals!traditional}

For $G$ be a simple Lie group, write $\mathfrak{g}$ for its semisimple Lie algebra.
The Killing form invariant polynomial $\langle -,-\rangle : \mathrm{Sym}^2 \mathfrak{g} \to \mathbb{R}$ induces the canonical Lie algebra 3-cocycle
$$
  \mu := \langle-,[-,-]\rangle : \mathrm{Alt}^3(\mathfrak{g}) \to \mathbb{R}
$$
which by left-translation along the group defines the canonical 
closed and left-invariant 3-form
$$
  \langle \theta \wedge [\theta\wedge \theta]\rangle
  \in 
  \Omega^3_{\mathrm{cl}, \mathrm{L}}(G)
  \,,
$$
where $\theta \in \Omega^1_{\mathrm{flat}}(G,\mathfrak{g})$ is the 
canonical \emph{Maurer-Cartan form} on  $G$.
What is called the \emph{Wess-Zumino-Witten sigma-model} induced by this data 
(see for instance \cite{Gawedzki}  for a decent review) is
the prequantum field theory given by an action functional, which 
to a smooth map $\Sigma_2 \to G$ out of a 
closed oriented smooth 2-manifold assigns the product of the standard 
exponentiated kinetic action with an exponentiated ``surface holonomy''
of a 2-form connection whose curvature 3-form is 
$\langle \theta \wedge [\theta\wedge \theta]\rangle$.

\medskip
In the special case that $\phi : \Sigma_2 \to G$ happens to factor through a contractible open subset $U$ of $G$ --
notably in the \emph{perturbative expansion} about maps constant on a point --
the Poincar{\'e} lemma implies that one can find a potential 2-form $B \in \Omega^2(U)$
with $d B = \langle \theta \wedge [\theta\wedge \theta]\rangle|_{U}$ and with this
perturbative perspective understood 
one may take the action functional to be simply of the naive form that is often considered
in the literature:
$$
 \exp(i S_{\mathrm{WZW}})
 :=
 \exp\left(i \int_{\Sigma^2} \mathcal{L}_{\mathrm{WZW}}\right)
  \;:\; 
  \phi \mapsto \exp\left(2\pi i \int_{\Sigma_2} \phi^\ast B\right)
  \,.
$$

\medskip
There are plenty of hints and some known examples which point to 
the fact that this construction
of the standard WZW model is just one in a large class of examples of 
higher dimensional boundary local (pre-)quantum field theories,
\ref{LocalPrequantumFieldTheories}, 
 which generalize 
traditional WZW theory in two ways:
\begin{enumerate}
  \item the cocycle $\mu$ is allowed to be of arbitrary degree;
  \item the Lie algebra $\mathfrak{g}$ is allowed to be a 
   (super-)\emph{Lie $n$-algebra} for $n \geq 1$ ($L_\infty$-algebra). 
\end{enumerate}
One famous class of examples of the first point are the Green-Schwarz
type action functionals for the super $p$-branes of string/M-theory \cite{AETW}.
These are the higher dimensional analog of the action functional for the
superstring that was first given in \cite{GreenSchwarz} and then recognized as 
a super WZW-model in \cite{HM}, induced from an exceptional 3-cocycle
on super-Minkowski spacetime of bosonic dimension 10,
regarded a super-translation Lie algebra.
Thess higher dimensional Green-Schwarz type $\sigma$-model
action functionals are accordingly induced by higher 
exceptional super-Lie algebra cocycles on super-Minkowski
spacetime, regarded as a super-translation Lie algebra. 
Remarkably, while ordinary Minkowski spacetime is cohomologically 
fairly uninteresting, super-Minkowski spacetime has a 
finite number of \emph{exceptional} super-cohomology classes. 
The higher dimensional WZW models induced by the corresponding 
higher exceptional cocycles
account precisely for the $\sigma$-models of those super-$p$-branes in 
string/M-theory which are pure $\sigma$-models, in that they do not
carry (higher) gauge fields (``tensor multiplets'') on their worldvolume, 
a fact known as ``the old brane scan'' \cite{AETW}. This includes, for instance, the
heterotic superstring and the M2-brane, but excludes the D-branes and the M5-brane.

\medskip
However, as we discuss below in 
section \ref{SuperBranesAndTheirIntersectionLaws}, this restriction 
to pure $\sigma$-model branes without ``tensor multiplet'' fields on
their worldvolume is due
to the restriction to ordinary super Lie algebras, hence to super Lie $n$-algebras
for just $n = 1$. If one allows genuinely higher WZW models which are given
by higher cocycles on Lie $n$-algebras for higher $n$, then \emph{all} the
fbranes of string/M-theory are described by higher WZW $\sigma$-models.
This is an incarnation of the general fact that in higher differential geometry, 
the distinction between $\sigma$-models and (higher) gauge theory
disappears, as (higher) gauge theories are equivalently $\sigma$-models
whose target space is a smooth higher moduli \emph{stack}, infinitesimally
approximated by a Lie $n$-algebra for higher $n$.

\medskip
This general phenomenon is particularly interesting for the M5-brane
(see for instance the Introduction of \cite{FiorenzaSatiSchreiberI} for plenty of pointers to the
literature on this). According to 
the higher Chern-Simons-theoretic formulation of $\mathrm{AdS}_7/\mathrm{CFT}_6$ in 
\cite{Witten}, the 6-dimensional $(2,0)$-superconformal worldvolume
theory of the M5-brane is related to the 7-dimensional Chern-Simons term
in 11-dimensional supergravity compactified on a 4-sphere in direct analogy
to the famous relation 
of 2d WZW theory to the 3d-Chern-Simons theory
controled by the cocycle $\mu$ (see \cite{Gawedzki} for a review). 
In \ref{InfinCS7d} and \ref{supergravityCField} we have discussed the 
bosonic nonabelian (quantum corrected) component of this 7d Chern-Simons theory
as a higher gauge local prequantum field theory; the discussion here
provides the fermionic terms and the formalization
of the 6d WZW-type theory induced from a (flat) 7-dimensional Chern-Simons theory.

\medskip
Up to the last section in this paper we discuss general aspects 
and examples of higher WZW-type 
sigma-models in the rational/perturbative approximation, 
where only the curvature $n$-form matters
while its lift to a genuine cocycle in differential cohomology is ignored.
However, in order to define already the traditional WZW action 
functional in a sensible way on \emph{all}
maps to $G$, one needs a more global description of the WZW term $\mathcal{L}_{\mathrm{WZW}}$.
Since \cite{Ga, FreedWitten}, 
this is understood to be a circle 2-connection/bundle gerbe/Deligne 3-cocycle whose curvature 3-form is 
$\langle \theta \wedge [\theta\wedge \theta]\rangle$, hence a
\emph{higher prequantization} \cite{hgp} of the curvature 3-form, 
which we write as a lift of maps 
of smooth higher stacks
$$
  \xymatrix{
    && \mathbf{B}^2 U(1)_{\mathrm{conn}}
	\ar[d]^{H_{(-)}}
    \\
    G \ar@{-->}[urr]^-{\mathcal{L}_{\mathrm{WZW}}}
	\ar[rr]_{\langle \theta \wedge [\theta\wedge \theta]\rangle}
	&&
	\Omega^3_{\mathrm{cl}} \;,
  }
$$
where $\mathbf{B}^2 U(1)_{\mathrm{conn}}$ denotes the smooth 2-stack of 
smooth circle 2-connections.
Then for $\phi : \Sigma_2 \to G$ a smooth map from a closed 
oriented 2-manifold to $G$, 
the globally defined value of the action functional is 
the corresponding \emph{surface holonomy} expressed as the composite
$$
  \exp(i S_{\mathrm{WZW}})
  \;:=\;
  \exp\left(
    2 \pi i \int_{\Sigma_2} [(-),\mathcal{L}_{\mathrm{WZW}}]
  \right)
  \;:\;
  \xymatrix{
    [\Sigma,G]
	\ar[rr]^-{[\Sigma, \mathcal{L}_{\mathrm{WZW}}]}
	&&
	[\Sigma,\mathbf{B}^2 U(1)_{\mathrm{conn}}]
	\ar[rrr]^-{\exp(2 \pi i \int_{\Sigma_2})(-) }
	&&&
	U(1)
  }
  \,,
$$
of the functorial mapping stack construction followed by a stacky refinement
of fiber integration in differential cohomology, \ref{FiberIntegrationOfOrdinaryDifferentialCocycles}.

\medskip
Recall that by the discussion in \ref{StrucWZWFunctional} we have a 
general universal construction of 
such non-perturbative refinements of all the local higher WZW terms considered here,
and that these are in precise 
sense boundary local prequantum field theories, \ref{BoundaryFieldTheory}, for flat higher
Chern-Simons type local prequantum field theories
(which is in line with the Chern-Simons theoretic holography in \cite{Witten}).
Therefore we know in principle how to 
quantize them non-perturbatively in generalized cohomology, 
discussed below in \ref{MotivicQuantizationApplications}.

\subsubsection{Lie $n$-algebraic formulation of perturbative higher WZW}
\label{WZWLienLalgebraicFormulation}
\index{Wess-Zumino-Witten functionals!Lie $n$-algebraic formulation}

We start with the traditional WZW model and show how in 
this example we may usefully reformulate its rationalized/perturbative aspects 
in terms of Lie $n$-algebraic structures. Then we  
naturally and seamlessly generalize to a definition of higher WZW-type $\sigma$-models.

\medskip
We briefly recall the notion of $L_\infty$-algebra valued 
differential forms/connections
from to establish notation in the present context. All the actual $L_\infty$-homotopy 
theory that we need can be found discussed or referenced in \cite{LocalObservables}.
Just for simplicity of exposition and since it is sufficient for the present discussion,
here we take all $L_\infty$-algebras to be of finite type, hence degreewise finite dimensional; see \cite{Pridham}
for the general discussion in terms of pro-objects.

\medskip
A (super-)\emph{Lie $n$-algebra}, def. \ref{SuperLInfinityAlgebra}, is a (super-)$L_\infty$-algebra
concentrated in the lowest $n$ degrees. 
Given a \mbox{(super-)}$L_\infty$-algebra $\mathfrak{g}$, 
we write 
$\mathrm{CE}(\mathfrak{g})$ for its \emph{Chevalley-Eilenberg algebra}; which is
a $(\mathbb{Z}\times \mathbb{Z}_2)$-graded commutative dg-algebra 
with the property that the underlying graded 
super-algebra is the free graded commutative super-algebra on the dual graded super vector space
$\mathfrak{g}[1]^\ast$. These are the dg-algebras which in parts of the
supergravity literature are referred to as ``FDA''s,
a term introduced in \cite{Nieuwenhuizen} and 
then picked up in \cite{AFR, DF, CDF} and followups. 
Precisely all the (super-)dg-algebras of this \emph{semi-free} form arise as 
Chevalley-Eilenberg algebras of (super-)$L_\infty$-algebras this way, 
and a homomorphism of $L_\infty$-algebras $f : \mathfrak{g} \to \mathfrak{h}$
is equivalently a homomorphism of dg-algebras of the form
$f^\ast : \mathrm{CE}(\mathfrak{h}) \to \mathrm{CE}(\mathfrak{g})$. 
See \cite{LocalObservables} for a review in the context of the higher prequantum geometry
of relevance here and for further pointers to the literature on 
$L_\infty$-algebras and their homotopy theory.

\begin{definition}
  For $\mathfrak{g}$ a Lie $n$-algebra, and $X$ a smooth manifold,
  a \emph{flat $\mathfrak{g}$-valued differential form} on $X$ 
  (of total degree 1, with $\mathfrak{g}$ regarded as cohomologically graded) is equivalently
  a morphism of dg-algebras $A^\ast : \mathrm{CE}(\mathfrak{g}) \to \Omega^\bullet_{\mathrm{dR}}(X)$
  to the de Rham complex.
  Dually we write this as\footnote{The reader familiar with 
  $L_\infty$-algebr\emph{oids} should take this as shorthand for the 
  $L_\infty$-algebroid homomorphism from the 
  tangent Lie algebroid of $X$ to the delooping of the $L_\infty$-algebra $\frak{g}$.} 
  $A : X \to \mathfrak{g}$.
  These differential forms naturally pull back along maps of smooth manifolds, and we write
$
  \Omega_{\mathrm{flat}}^1(-,\mathfrak{g}) 
$
for the sheaf, on smooth manifolds, of flat $\mathfrak{g}$-valued differential forms
of total degree 1. 
\label{gValuedDifferentialForms}
\end{definition}
Notice that, in general, these forms of total degree 1 involve differential forms
of higher degree with coefficients in higher degree elements of the $L_\infty$-algebra:
\begin{example}
  For $n \in \mathbb{N}$ write 
  $\R[n]$ for the abelian Lie $n$-algebra concentrated
on $\mathbb{R}$ in degree $-n$. Its Chevalley Eilenberg algebra is the dg-algebra
which is genuinely free on a single generator in degree $n+1$. A
flat 
$\R[n]$-valued differential form is equivalently just an ordinary 
closed differential $(n+1)$-form
$$
  \Omega^1_{\mathrm{flat}}(-, \R[n])
  \simeq
  \Omega^{n+1}_{\mathrm{cl}}
  \,.
$$
\end{example}
\begin{definition}
A \emph{$(p+2)$-cocycle} $\mu$ on a Lie $n$-algebra $\mathfrak{g}$ is a degree $p+2$
closed element
in the corresponding Chevalley-Eilenberg algebra $\mu \in \mathrm{CE}(\mathfrak{g})$.
\label{cocycle}
\end{definition}
\begin{remark}
A $(p+2)$-cocycle on $\mathfrak{g}$ is equivalently a map of dg-algebras
$\mathrm{CE}(
\R[p+1])
\to \mathrm{CE}(\mathfrak{g})$ and
hence, equivalently, a map of $L_\infty$-algebras of the form
$\mu : \mathfrak{g} \to \R[p+1]$.
So, if $\{t_a\}$ is a basis for the graded vector space underlying $\mathfrak{g}$, 
then the differential $d_{\mathrm{CE}}$ is given in components by
$$
 d_{\mathrm{CE}}\, t^a = \sum_{i \in \mathbb{N}} C^a{}_{a_1 \cdots a_{i}}
  t^{a_1} \wedge \cdots t^{a_{i}}
  \,,
$$
where $\{C^a{}_{a_1 \cdots a_i}\}$ are the structure constants of the $i$-ary
bracket of $\mathfrak{g}$. Consequently, a degree $p+2$ cocycle is a degree $(p+2)$-element
$$
  \mu = \sum_{i} \mu_{a_1 \dots a_i} t^{a_1} \wedge \cdots t^{a_i}
$$
such that
$d_{\mathrm{CE}}\, \mu = 0$.
\end{remark}
\begin{example}
  For $\{t_a\}$ a basis as above and
  $
    \omega \in \Omega^1_{\mathrm{flat}}(X,\mathfrak{g})$
    a $\frak{g}$-valued 1-form on $X$, 
  the pullback of the cocycle is the closed differential $(p+2)$-form
  which in  components  reads
  $$
    \mu(\omega) 
	  = 
	  \sum_{i} \mu_{a_1 \cdots a_i} \omega^{a_1} \wedge \cdots \wedge \omega^{a_i}
	\,,
  $$
  where $\omega^{a} = \omega(t^a)$.
\end{example}
\begin{remark} 
Composition $\omega \mapsto (\xymatrix{X \ar[r]^\omega & \mathfrak{g} \ar[r]^-\mu & \mathbb{R}[p+1]})$
of $\mathfrak{g}$-valued differential forms $\omega$ with an $L_\infty$-cocycle 
$\mu$ yields a 
homomorphism of sheaves
$$
  \Omega^1_{\mathrm{flat}}(-,\mu) 
   : 
   \xymatrix{
     \Omega_{\mathrm{flat}}(-, \mathfrak{g}) 
     \ar[r]
	 &
    \Omega^{p+2}_{\mathrm{cl}}
  }
  \,.
$$
This is the sheaf incarnation of $\mu$ regarded as a universal differential form
on the space of all flat $\mathfrak{g}$-valued differential forms. 
\end{remark}
\begin{example}
By the Yoneda lemma, for $X$ a smooth manifold, 
morphisms\footnote{of sheaves, by thinking of $X$ as the sheaf $C^\infty(-, X)$.}
 $X \to \Omega^1_{\mathrm{flat}}(-, \mathfrak{g})$ are equivalently 
just flat $\mathfrak{g}$-valued differential forms on $X$. 
Specifically, for $G$ an ordinary Lie group, its Maurer-Cartan form
is equivalently a map
$$
  \theta : 
  \xymatrix{
    G \ar[r] & \Omega^1_{\mathrm{flat}}(-, \mathfrak{g})
  }
  \,.
$$
Therefore, given a field configuration $\phi : \Sigma_2 \to G$
of the traditional WZW model, 
postcomposition with $\theta$ turns this into 
$$
  \phi^\ast \theta 
    : 
  \xymatrix{
    \Sigma 
	   \ar[r]^-{\phi}
	&
	G
	\ar[r]^-\theta
	&
	\Omega^1_{\mathrm{flat}}(-,\mathfrak{g})
  }
  \,.
$$
Here if $\mathfrak{g}$ is represented as a matrix Lie algebra then this is the popular
expression $\phi^\ast \theta = \phi^{-1} d \phi$
\label{Yoneda}
\end{example}
\begin{definition}
 Given an $L_\infty$-algebra $\mathfrak{g}$ equipped with a cocycle
 $\mu : \mathfrak{g} \to \mathbb{R}[p+1]$ of degree $p+2$, 
 a \emph{perturbative $\sigma$-model datum} for $(\mathfrak{g},\mu)$ is 
 a triple consisting of 
 \begin{itemize}
   \item a space $X$;
   \item equipped with a flat $\mathfrak{g}$-valued differential form 
   $\theta_{\mathrm{global}} : X \to \Omega^1_{\mathrm{flat}}(-,\mathfrak{g})$ 
    (a ``global Maurer-Cartan form'');
     \item and equipped with a factorization $\mathcal{L}_{\mathrm{WZW}}$
 through $d_{\mathrm{dR}}$ of $\mu(\theta_{\mathrm{global}})$, 
 as expressed in the following diagram
$$
  \xymatrix{ 
	 &
	 X
	 \ar[r]^-{\theta_{\mathrm{global}}}
	 \ar[dr]_{\mathcal{L}_{\mathrm{WZW}}}
	 &
	 \Omega_{\mathrm{flat}}(-, \mathfrak{g})
	 \ar[rr]^-{\mu} 
	 &&
	 \Omega^{p+2}_{\mathrm{cl}}\;.
	 \\
	 && \Omega^{p+1}
	 \ar[urr]_{d_{\mathrm{dR}}}
  }
$$
\end{itemize}
The \emph{action functional} associated with this data is the functional
$$
  S_{\mathrm{WZW}}
   :
  \xymatrix{
    [\Sigma,X]
	\ar[r]
	&
	\mathbb{R}
  }
$$
given by
$$
  \phi \mapsto \int_{\Sigma} \mathcal{L}_{\mathrm{WZW}}(\phi)
  \,,
$$
where the integrand is the differential form  
$$
  \mathcal{L}_{\mathrm{WZW}}(\phi)
  : 
  \xymatrix{
    \Sigma \ar[r]^\phi & X \ar[rr]^-{\mathcal{L}_{\mathrm{WZW}}}
	&&
	\Omega^{p+1}_{\mathrm{cl}}
  }
  \,.
$$
\label{rationaldatum}
\end{definition}
\begin{remark}
Here $X$ actually need not be a (super-)manifold but may be a 
smooth higher \mbox{(super-)} stack, \ref{SuperInfinityGroupoids}.
\end{remark}
\begin{remark}
 The notation $\theta_{\rm global}$ serves to stress the fact that 
we are considering globally defined one-forms on $X$ as opposed to cocycles
in hypercohomology, which is where the higher Maurer-Cartan forms
on \emph{higher} (super-)Lie groups take values, due to presence of 
nontrivial higher gauge transformations. See \ref{StrucWZWFunctional}.
\end{remark}
\begin{remark}
The diagram in Def. \ref{rationaldatum} manifestly captures a local 
description, when  $X$ is a contractible manifold.
An immediate global version is captured by the following diagram
$$
  \xymatrix{
     \Sigma \ar[r]^{\eta} 
	 &
	 X
	 \ar[r]^-{\theta_{\mathrm{global}}}
	 \ar[dr]_{\mathcal{L}_{\mathrm{WZW}}}
	 &
	 \Omega_{\mathrm{flat}}(-, \mathfrak{g})
	 \ar[rr]^-{\mu} 
	 &&
	 \Omega^{p+2}_{\mathrm{cl}}\;,
	 \\
	 && \mathbf{B}^{p+1} U(1)_{\mathrm{conn}}
	 \ar[urr]_{F_{(-)}}
  }
$$
where  $\mathbf{B}^{p+1} U(1)_{\mathrm{conn}}$ is the stack of 
$U(1)$-$(p+1)$-bundles with connections,
and $F_{(-)}$ is the curvature morphism; see, for instance, \cite{FSS}.   
This globalization is what one sees, for example, in the ordinary 
WZW model. 
\end{remark}

Finally, we notice for discussion in the examples
one aspect of the higher symmetries of such perturbative higher WZW models:
\begin{definition}
  Given a (super-) $L_\infty$-algebra $\mathfrak{g}$,
  its \emph{graded Lie algebra of infinitesimal automorphisms}
  is the Lie algebra whose elements are graded derivations 
  $v \in \mathrm{Der}(\mathrm{Sym}^\bullet \mathfrak{g}[1]^\ast)$
  on the graded algebra underlying its Chevalley-Eilenberg algebra
  $\mathrm{CE}(\mathfrak{g})$, 
 acting as the corresponding Lie derivatives.
  \label{LieAlgebraOfSymmetries}
\end{definition}

\subsubsection{Boundary conditions and brane intersection laws}
\label{WZWBoundaryConditionsAndBraneIntersectionLaws}
\index{Wess-Zumino-Witten functionals!brane intersection laws}

In the context of fully extended (i.e. local) topological prequantum field theories, 
\ref{LocalPrequantumFieldTheories}
one has the following general notion of boundary condition, see
\ref{BoundaryFieldTheory}.
\begin{definition}
A \emph{prequantum boundary condition for an open brane}  
(hence a ``background brane'' on which the given brane may end) is given 
by boundary gauge trivializations $\phi_{\mathrm{bdr}}$ of the Lagrangian restricted to
the boundary fields, hence by diagrams of the form
$$
	  \raisebox{23pt}{
	  \xymatrix{
	    & {\rm Boundary~ Field}
		  \ar[dl]
		  \ar[dr]_{\ }="s"
		\\
		\ast 
		\ar[dr]_0^{\ }="t"
		&& {\rm Bulk~ Fields} \ar[dl]^{\rm~~ Lagrangian}
		\\
		& {\rm Phases}\;,
		\ar@{=>}^{\phi_{\mathrm{bdr}}}_\simeq "s"; "t"
	  }}
	$$
	where ``Phases" denotes generally the space where the Lagrangian takes values.
	\label{BoundaryConditions}
\end{definition}
Specializing this general principle to our current situation, we have the following
\begin{definition}
  A \emph{boundary condition} for a rational $\sigma$-model datum,
  $(X,\mathfrak{g}, \mu)$ of Def. \ref{rationaldatum}, is 
  \begin{enumerate}
    \item an $L_\infty$-algebra $Q$ and a homomorphism
	  $Q \longrightarrow  \mathfrak{g}$,
    \item equipped with a  homotopy $\phi_{\mathrm{brd}}$ of $L_\infty$-algebras
    morphisms 
		$
	  \raisebox{23pt}{
	  \xymatrix{
	    & Q
		  \ar[dl]
		  \ar[dr]_{\ }="s"
		\\
		\ast 
		\ar[dr]_0^{\ }="t"
		&& \mathfrak{g}\;. \ar[dl]^{\mu}
		\\
		& \mathbb{R}[p+1]
		\ar@{=>}^{\phi_{\mathrm{bdr}}} "s"; "t"
	  }}
	$
  \end{enumerate}
  \label{BoundaryCondition}
\end{definition}
\begin{remark}[Background branes]
 Since $\mathfrak{g}$ is to be thought of as the \emph{spacetime target} for a $\sigma$-model,
 we are to think of $Q \to \mathfrak{g}$ in Def. \ref{BoundaryCondition} 
 as a \emph{background brane} ``inside'' spacetime. 
 For instance, as demonstrated below in Section \ref{SuperBranesAndTheirIntersectionLaws},
 it may be a D-brane in 10-dimensional super-Minkowski space on which the 
 open superstring ends, or it may be the M5-brane in 11-dimensional
 super-Minowski spacetime on which the open M2-brane ends.
 To say then that the $p$-brane described
 by the $\sigma$-model may end on this background brane $Q$
 means to consider worldvolume manifolds $\Sigma_{n}$ with boundaries
 $\partial \Sigma_{p+1} \hookrightarrow \Sigma_{p+1}$
 and \emph{boundary field configurations} $(\phi, \phi|_{\partial})$ 
 making the left square in the following diagram commute:
$$
  \xymatrix@R=6pt{
    \partial \Sigma_{p+1} 
	  \ar[rr]^{\phi|_{\partial \Sigma}} 
	\ar[dd]
	&& Q
	 \ar[rr]_{\ }="s" \ar[dd]^{\ }="t"
	&& \ast
	\ar[dd]
    \\
	\\
    \Sigma_{p+1}
	\ar[rr]^-{\phi}
	&&
	\mathfrak{g}
	\ar[rr]_-{\mu}
	&&
	\mathbb{R}[p+1]\;.
	\ar@{=>}^{\mathrm{\phi}_{\mathrm{bdr}}} "s"; "t"
  }
$$
The commutativity of the diagram on the left encodes precisely that 
the boundary of the $p$-brane
is to sit inside the background brane $Q$.
But now -- by the defining universal property of the homotopy pullback of
super $L_\infty$-algebras -- this means, equivalently, that the 
background brane embedding map
$Q \to \mathfrak{g}$ factors through 
the \emph{homotopy fiber}
of the cocycle $\mu$.
If we denote this homotopy fiber by  $\widehat{\mathfrak{g}}$,
then  we have an essentially unique factorization as follows
$$
  \raisebox{20pt}{
  \xymatrix{
    \partial \Sigma_{p+1} \ar[rr]^{\phi|_{\partial \Sigma}} \ar[d]
	&& 
	Q  \ar[d]^{\ }
	\ar@{-->}[rr]
	&&
	\widehat{\mathfrak{g}}
	\ar[rr]_{\ }="s"
	\ar[d]^{\ }="t"
	&& \ast
	\ar[d]
    \\
    \Sigma_{p+1}
	\ar[rr]^-{\phi}
	&&
	\mathfrak{g}
	\ar@{=}[rr]
	&&
	\mathfrak{g}
	\ar[rr]_-{\mu}
	&&
	\mathbb{R}[p+1]\;,
	\ar@{=>}^{\phi_{\mathrm{bdr}}^{\mathrm{univ.}}} "s"; "t"
  }
  }
$$
where now $\widehat{\mathfrak{g}} \to \mathfrak{g}$ is the
\emph{homotopy fiber} $\widehat{\mathfrak{g}}$
of the cocycle $\mu$. Notice that here in homotopy theory \emph{all} diagrams
appearing are understood to be filled by homotopies/gauge transformations, but only
if we want to label them explicitly do we display them.
\label{HowToThinkOfExtensionsAsBackgroundBranes}
\end{remark}
The crucial implication to emphasize is that what used to be regarded as a 
background brane $Q$ on which the $\sigma$-model brane $\Sigma_n$ may end
is itself characterized by a $\sigma$-model map
$Q \to \widehat{\mathfrak{g}}$, not to the original 
target space $\mathfrak{g}$, but to the \emph{extended target space}
$\widehat{\mathfrak{g}}$.  In the class of examples discussed below in 
Section \ref{SuperBranesAndTheirIntersectionLaws}, this 
extended target space is precisely the
\emph{extended superspace} in the sense of \cite{CdAIP}.

\begin{remark}
  The $L_\infty$-algebra $\widehat{\mathfrak{g}} \to \mathfrak{g}$
  is the \emph{extension} of $\mathfrak{g}$ classified by the cocycle
  $\mu$, in generalization to the traditional extension of Lie algebras
 classified by 2-cocycles.
 If $\mu$ is an $(n_2+1)$-cocycle on an $n_1$-Lie algebra $\mathfrak{g}$
 for $n_1 \leq n_2$, then the extended $L_\infty$-algebra
 $\widehat{\mathfrak{g}}$ is an Lie $n_2$-algebra.
 See \cite{LocalObservables} for more details on this.
 \label{extension}
\end{remark}
\begin{proposition}
The Chevalley-Eilenberg algebra 
${\rm CE}(\widehat{\frak{g}})$
of the extension $\widehat{\frak{g}}$ of $\mathfrak{g}$ by a cocycle
$\mu$
 admits, up to equivalence, a very simple description;
namely, it is the differential graded algebra obtained from ${\rm CE}(\frak{g})$
by adding a single generator $c_n$ in degree $n$ subject to the 
relation 
$$
  d_{{\rm CE}(\widehat{\frak{g}})} \, c_n=\mu
  \,.
$$
Here we are viewing $\mu$ as a degree 
$n+1$ element in ${\rm CE}(\frak{g})$, and hence also in 
${\rm CE}(\widehat{\frak{g}})$.
\label{CEAlgebrasOfExtensions}
\end{proposition}
\proof
  First observe that we have a commuting diagram of (super-)dg-algebras
  of the form
  $$
    \raisebox{20pt}{
    \xymatrix{
	  \mathrm{CE}\left(\widehat{\mathfrak{g}}\right)
	  &
	  \mathrm{CE}\left(\left(\mathbb{R} \stackrel{\mathrm{id}}{\to} \mathbb{R}\right)[n-1] \right)
	  \ar[l]
	  \\
	  \mathrm{CE}\left(\mathfrak{g}\right)
	  \ar[u]
	  &
	  \mathrm{CE}\left(\mathbb{R}[n]\right)
	  \ar[u]
	  \ar[l]
	}}
	\,.
  $$ 
  Here the top left dg-algebra is the dg-algebra of the above statement, the
  top morphism is the one that sends the unique degree-$(n+1)$-generator 
  to $\mu$ and the unique degree-$n$ generator to $c_n$,
  the vertical morphisms are the evident inclusions, and the
  bottom morphism is the given cocycle. Consider the dual
  diagram of $L_\infty$-algebras
  $$
    \raisebox{20pt}{
    \xymatrix{
	  \widehat{\mathfrak{g}}
	  \ar[r]
	  \ar[d]
	  &
	  (\mathbb{R} \stackrel{\mathrm{id}}{\to} \mathbb{R})[n-1]
	  \ar[d]
	  \\
	  \mathfrak{g}
	  \ar[r]^{\mu}
	  &
	  \mathbb{R}[n]\;.
	}
	}
  $$
  Then observe that the underlying graded vector spaces here
  form a pullback diagram of linear maps 
  (the linear components of the $L_\infty$-morphisms).
  From this the statement follows directly with the recognition theorem for 
  $L_\infty$-homotopy fibers, 
  theorem 3.1.13 in \cite{LocalObservables}. 
\endofproof
\begin{remark}
  The construction appearing in  Prop. \ref{CEAlgebrasOfExtensions}
  is of course well familiar in the ``FDA''-technique in the
  supergravity literature \cite{CDF}, and we recall famous examples below in 
  Section   \ref{SuperBranesAndTheirIntersectionLaws}. The point to highlight 
  here is that this construction has a universal $L_\infty$-homotopy-theoretic meaning,
  in the way described above.
\end{remark}
The crucial consequence of this discussion is the following:
\begin{remark}
If the extension
$\widehat{\mathfrak{g}}$ itself carries a cocycle 
$\mu_Q : \widehat{\mathfrak{g}} \to \mathbb{R}[n]$
and  we are able to find a local potential/Lagrangian $\mathcal{L}_{\mathrm{WZW}}$ for the 
closed $(n+1)$-form $\mu_Q$ (which by \ref{StrucWZWFunctional} is always the case), 
then this exhibits the background brane $Q$ itself as a rational 
WZW $\sigma$-model, now propagating not on the original 
``target spacetime'' $\mathfrak{g}$ but on the ``extended spacetime''
$\widehat{\mathfrak{g}}$.
\end{remark}
\begin{remark}
Iterating this process gives rise to a tower of extensions and cocycles
$$
  \raisebox{20pt}{
  \xymatrix{
  \ar@{..}[d] && 
  \\
  \widehat{\widehat{\mathfrak{g}}} \ar[d]
   \ar[rr]^-{\mu_3} && \mathbb{R}[n_3]
  \\
    \widehat{\mathfrak{g}} \ar[d]
     \ar[rr]^-{\mu_2} && \mathbb{R}[n_2]
	\\
	\mathfrak{g} \ar[rr]^-{\mu_1} && \mathbb{R}[n_1]\;,
  }
  }
$$
which is like a 
Whitehead tower in rational homotopy theory, only that the cocycles
in each degree here are not required to be the lowest-degree nontrivial ones. 
In fact, the actual rational Whitehead tower is an example of this.
In the language of Sullivan's formulation of rational homotopy theory
this says that $\mathfrak{g}_n$ is exhibited by a
sequence of cell attachments as a \emph{relative Sullivan algebra}
relative to $\mathfrak{g}$.
\end{remark}
Since this is an important concept for the present purpose, we give it a name:
\begin{definition}
Given an $L_\infty$-algebra $\mathfrak{g}$, 
the {\it brane bouquet of $\frak{g}$} is  
the rooted tree consisting of, iteratively, all possible 
equivalence classes of nontrivial $\mathbb{R}[\bullet]$ extensions 
(corresponding to equivalence classes of nontrivial $\mathbb{R}[\bullet]$-cocycles)
starting with $\mathfrak{g}$ as the root.
\end{definition} 
$$
  \xymatrix{
     & \mathfrak{g}_{2,1} \ar[dr] & \cdots & \mathfrak{g}_{2,k} \ar[dl]	 
	 && \mathfrak{g}_{3,1} \ar[dl]
     \\
     && \mathfrak{g}_{1,1} \ar[dr] && \mathfrak{g}_{1,2} \ar[dl] & \mathfrak{g}_{3,2} \ar[l]
     \\
     &&& \mathfrak{g} & & \mathfrak{g}_{3,3} \ar[ul]
	 \\
	 &&& \mathfrak{g}_3 \ar[u]
     \\
	 &&& \ar@{..}[u]
  }
$$
This \emph{brane bouquet} construction in $L_\infty$-homotopy theory
that we introduced serves to organize and formalize the following two 
physical heuristics. 
\begin{remark}[Brane intersection laws]
By the discussion above in Remark \ref{HowToThinkOfExtensionsAsBackgroundBranes},
each piece of a brane bouquet of the form
 $$
  \raisebox{20pt}{
  \xymatrix{
    \mathfrak{g}_2 \ar[d]
     \ar[rr]^-{\mu_2} && \mathbb{R}[n_2]
	\\
	\mathfrak{g}_1 \ar[rr]^-{\mu_1} && \mathbb{R}[n_1]
  }
  }
$$
 may be thought of as encoding a
{\it brane intersection law},
 meaning that the WZW $\sigma$-model brane corresponding to 
 $(\frak{g}_1, \mu_1)$ can end on the WZW $\sigma$-model brane corresponding 
 to $(\frak{g}_2, \mu_2)$.
 Therefore, the brane bouquet of some $L_\infty$-algebra
 $\mathfrak{g}$ lists all the possible $\sigma$-model branes and all
 their intersection laws in the ``target spacetime'' $\mathfrak{g}$.
 \label{BraneIntersectionLaw}
\end{remark}
\begin{remark}[Brane condensates]
 To see how to think of the extensions $\widehat{\mathfrak{g}}$
 as ``extended spacetimes'', observe that by Prop. \ref{CEAlgebrasOfExtensions}
 and Def. \ref{gValuedDifferentialForms} a $\sigma$-model on 
 the extension $\widehat{\mathfrak{g}}$ of $\mathfrak{g}$ which is classified by a
 $(p+2)$-cocycle $\mu$ is equivalently a $\sigma$-model on $\mathfrak{g}$
 together with an $p$-form higher gauge field on its worldvolume, 
 one whose curvature $(p+1)$-form satisfies a twisted Bianchi identity controled by $\mu$.
 The examples discussed below in Section \ref{SuperBranesAndTheirIntersectionLaws} 
 show that this $p$-form field (``tensor field'' in the brane literature) 
 is that which is  ``sourced'' by the charged boundaries of the original $\sigma$-model 
 branes on $\mathfrak{g}$. 
 For instance for superstrings ending on D-branes it is the Chan-Paton gauge field
 sourced by the endpoints of the open string, and for M2-branes ending on 
 M5-branes it is the latter's B-field which is sourced by the self-dual strings
 at the boundary of the M2-brane.
 In conclusion, this means that we may think of the extension 
 $\widehat{\mathfrak{g}}$ as being the original spacetime $\mathfrak{g}$
 but \emph{filled with a condensate} of branes whose $\sigma$-model
 is induced by $\mu$.
 \label{BraneCondensates}
\end{remark}

\subsubsection{Example: Super $p$-branes and their intersection laws}
 \label{SuperBranesAndTheirIntersectionLaws}
 \label{SuperpBranes}
  \index{Wess-Zumino-Witten functionals!super $p$-branes}
  \index{$\sigma$-model!super $p$-branes}

 We now discuss higher rational/perturbative WZW models on super-Minkowski spacetime
 regarded as the super-translation Lie algebra over itself,
 as well as on the \emph{extended superspaces} which arise
 as exceptional super Lie $n$-algebra extensions of the super-translation
 Lie algebra. This is the local description of super $p$-brane $\sigma$-models
 propagating on a supergravity background spacetime, \ref{Supergravity}.  
 We show then that by the brane intersection laws
 of Remark \ref{BraneIntersectionLaw} this reproduces precisely the 
 super $p$-brane content of string/M-theory including the $p$-branes
 with tensor multiplet fields, notably including the D-branes and the M5-brane.
 The discussion is based on the work initiated in 
 \cite{DF} and further developed in articles including \cite{CdAIP}.
 The point here is to show that this ``FDA''-technology is 
 naturally and usefully reformulated in terms of super-$L_\infty$-homotopy
 theory, and that this serves to clarify and illuminate various points that 
 have not been seen, and are indeed hard to see, via the ``FDA''-perspective.

 \medskip
 We set up some basic notation concerning the super-translation- and 
 the super-Poincar{\'e} super Lie algebras, following \cite{DF}. For more background see
 lecture 3 of \cite{FreedLectures} and appendix B of 
 \cite{PolchinskiBook}.
 
 Write $\mathfrak{o}(d-1,1)$ for the Lie algebra of the Lorentz group
in dimension $d$. If $\{\omega_{a}{}^b\}_{a,b}$ is the canonical basis
of Lie algebra elements, then the Chevalley-Eilenberg algebra
$\mathrm{CE}(\mathfrak{o}(d-1,1))$ is generated from elements 
$\{\omega^{a}{}_b\}_{a,b}$ in degree $(1,\mathrm{even})$ with the 
differential 
given by\footnote{Here and in all of the following a summation over repeated indices is
understood.}
$
  d_{\mathrm{CE}}\, \omega^{a}{}_b := \omega^{a}{}_c\wedge \omega^{c}{}_b
 $. 
Next, write $\mathfrak{iso}(d-1,1)$ for the Poincar{\'e} Lie algebra.
Its Chevalley-Eilenberg algebra in turn is generated from the
$\{\omega^{a}{}_b\}$ 
as before together with further generators $\{e^a\}_a$ 
in degree $(1,\mathrm{even})$ with 
differential given by
$
  d_{\mathrm{CE}}\, e^a := \omega^{a}{}_b \wedge e^b
  $. 
Now for $N$ denoting a real spinor representation of $\mathfrak{o}(d-1,1)$,
also called the number of supersymmetries (see for instance part 3 of \cite{FreedLectures}), 
write $\{\Gamma^a\}$ for a representation of the Clifford algebra
in this representation and $\{\Psi_{\alpha}\}_\alpha$ for the corresponding
basis elements of the spinor representation. 
There is then an essentially unique symmetric $\mathrm{Spin}(d-1,1)$-equivariant bilinear map
from two spinors to a vector, traditionally written in components as
$$
  (\psi_1, \psi_2)^a := \tfrac{i}{2}\overline{\psi} \Gamma^a \psi 
  \,.
$$
This induces the
super Poincar{\'e} Lie algebra $\mathfrak{siso}_N(d-1,1)$ whose
Chevalley-Eilenberg super-dg-algebra is generated from the generators as above
together with generators $\{\Psi^{\alpha}\}$ in degree $(1,\mathrm{odd})$
with the differential now defined as follows
$$
  \begin{aligned}
    d_{\mathrm{CE}} \, \omega^{a}{}_b &= \omega^{a}{}_c \wedge \omega^{c}{}_b\;,
	\\
	d_{\mathrm{CE}} \, e^a & 
	  = 
	  \omega^{a}{}_b \wedge e^b + \tfrac{i}{2}\overline{\psi} \wedge \Gamma^a \psi\;,
	\\
	d_{\mathrm{CE}} \, \psi^\alpha & = \tfrac{1}{4} \omega^{a}{}_b \wedge \Gamma^{a}{}_b \Psi\;.
  \end{aligned}
$$
Here and in the following $\Gamma^{a_1 \cdots a_p}$ denotes the skew-symmetrized
product of the Clifford matrices and in the above matrix multiplication is
understood whenever the corresponding indices are not displayed.
In summary, the degrees are
$$
  \mathrm{deg}(e^a) = (1, \mathrm{even}),
  \;\;\;\;\;
  \mathrm{deg}(\omega^a) = (1, \mathrm{even}),
  \;\;\;\;\;
  \mathrm{deg}(\psi^\alpha) = (1, \mathrm{odd}),
  \;\;\;\;\;
  \mathrm{deg}(d_{\mathrm{CE}}) = (1, \mathrm{even})\;.
$$
Notice that this means that, for instance, $e^{a_1} \wedge e^{a_2} = - e^{a_1}\wedge e^{a_2}$
and $e^a \wedge \psi^\alpha = - \psi^\alpha \wedge e^a$ but 
$\psi^{\alpha_1} \wedge \psi^{\alpha_2} = + \psi^{\alpha^2} \wedge \psi^{\alpha_1}$.
\begin{example}
  For $\Sigma$ a supermanifold of dimension $(d;N)$, a flat $\mathfrak{siso}(d-1,1)$-valued
  differential form $A : \mathrm{CE}(\mathfrak{siso}(d-1,1) \to \Omega^{\bullet}_{\mathrm{dR}}(\Sigma)$, according to 
  Def. \ref{gValuedDifferentialForms} and subject to the constraint that the $\mathbb{R}^{d;N}$-component is induced
  from the tangent space of $\Sigma$ (this makes it a \emph{Cartan connection})
  is
  \begin{enumerate}
    \item a \emph{vielbein}  field $E^a := A(e^a)$,
	\item with a \emph{Levi-Civita connection} $\Omega^{a}{}_b := A(\omega^a{}_b)$ (graviton),
	\item a spinor-valued 1-form field $\psi^\alpha := A(\psi^\alpha)$ (gravitino),
  \end{enumerate}
  subject to the flatness constraints which here say that the torsion of 
  of the Levi-Civita connection is the super-torsion 
  $\tau = \overline{\Psi}\wedge \Gamma^a \Psi \wedge E_a$ and that the
  Riemann curvature vanishes. This is the gravitational field content 
  (for vanishing field strength here, one can of course also consider non-flat
  fields) 
  of supergravity on $\Sigma$, formulated in 
  first order formalism. 
  By passing to $L_\infty$-extensions of 
  $\mathfrak{siso}$ this is the fomulation 
  of supergravity fields which seamlessly generalizes to the higher gauge
  fields that higher supergravities contain, including their correct
  higher gauge transformations. This is the perspective on supergravity
  originating around the article \cite{DF} and expanded on in the textbook
  \cite{CDF}. Recognizing the ``FDA''-language used in this book
  as secretly being about Lie $n$-algebra homotopy theory 
  (the ``FDA''s are really Chevalley-Eilenberg algebras super-$L_\infty$-algebras)
  allows one to uncover some natural and 
  powerful higher gauge theory and geometric homotopy theory
  hidden in traditional supergravity literature.
\end{example}
The \emph{super translation Lie algebra} corresponding to the above is the quotient
$$
  \mathbb{R}^{d;N} := \mathfrak{siso}(d-1,1)/\mathfrak{o}(d-1,1)
$$
whose CE-algebra is as above but with the $\{\omega^{a}{}_b\}$ discarded.
We may think of the underlying super vector space of $\mathbb{R}^{d;N}$ as
$N$-super Minkowski spacetime of dimension $d$, i.e. with $N$ supersymmetries. 
Regarded as a supermanifold,
it has canonical super-coordinates $\{x^a, \vartheta^\alpha\}$
and the CE-generators $e^a$ and $\psi^\alpha$ above may be identified
under the general equivalence 
$\mathrm{CE}(\mathfrak{g}) \simeq \Omega^\bullet_{\mathrm{L}}(G)$
(for a (super-)Lie group $G$ with (super-)Lie algebra $\mathfrak{g}$)
with the corresponding canonical left-invariant differential forms on 
this supermanifold:
$$
  \begin{aligned}
    e^a & = d_{\mathrm{dR}}\, x^a 
	  + 
	  \overline{\vartheta} \Gamma^a \,d_{\mathrm{dR}}\, \vartheta\;,
	\\
	\psi^\alpha & = d_{\mathrm{dR}}\, \vartheta^\alpha\;.
  \end{aligned}
$$
This defines a morphism 
$\theta : {\rm CE}(\R^{d;N}) \to \Omega^{\bullet | \bullet}(\R^{d;N})$
to super-differential forms on super Minkowski space,
and via Def. \ref{gValuedDifferentialForms} this is the Maurer-Cartan form,
Example \ref{Yoneda},
on the super\emph{group} $\mathbb{R}^{d;N}$ of supergranslations
As such $\{e^a, \psi^\alpha\}$ is the canonical \emph{super-vielbein}
on super-Minkowski spacetime. 

\medskip
Notice that the only non-trivial piece of the above CE-differential
remaining on $\mathrm{CE}(\mathbb{R}^{d;N})$ is
$$
  d_{\mathrm{CE}(\mathbb{R}^{d;N})}\, e^a = \overline{\psi} \wedge \Gamma^a\psi
  \,.
$$
Dually this is the single non-trivial super-Lie bracket on $\mathbb{R}^{d;N}$,
the one which pairs two spinors to a vector. 
All the exceptional cocycles considered in the following exclusively
are controled by just this equation and Lorentz invariance.

\medskip

We next consider various branches of the brane bouquet,
Def. \ref{BraneBouquet}, of these super-spacetimes $\mathbb{R}^{d,N}$.

\begin{itemize}
  \item \ref{OldBraneScan} --$\sigma$-model super $p$-branes --- The old brane scan
  \item \ref{TypeIIASuperStringEndingOnDBrane} -- Type IIA superstring ending on D-branes and the D0-brane condensate
  \item \ref{TypeIIBSuperstringEndingOnDBranes} -- Type IIB superstring ending on D-branes and S-duality
  \item \ref{The5brane} -- The M-theory 5-brane and the M-theory super Lie algebra
  \item \ref{CompleteBraneBouquet} -- The complete brane bouquet of string/M-theory
\end{itemize}

\paragraph{$\Sigma$-model super $p$-branes --- The old brane scan}
\label{OldBraneScan}
\index{brane!old brane scan}

As usual, we write $N$ for a choice of number of irreducible 
real (Majorana) representations of $\mathrm{Spin}(d-1,1)$, and 
$N = (N_+, N_-)$ if there are two inequivalent chiral minimal representations.
For instance, two important cases are 
\begin{center}
\begin{tabular}{|c||c|}
  \hline
  $d = 10$ & $d = 11$
  \\
  \hline
  $N = (1,0) = \mathbf{16}$ & $N = 1 = \mathbf{32}$
  \\
  \hline
\end{tabular}
\end{center}
For $0 \leq p \leq 9$ consider the dual bispinor element
$$
  \mu_p 
    :=
  e^{a_1} \wedge \cdots \wedge e^{a_p}
  \wedge
  (\overline{\psi}
  \wedge \Gamma^{a_1 \cdots a_p} \psi)
  \;\in \mathrm{CE}(\mathbb{R}^{d;N})
  \,,
$$
where here and in the following the parentheses are just to guide the reader's eye.
Observe that the differential of this element is of the form
$$
  d_{\mathrm{CE}}\, \mu_p
  \;\propto\;
  e^{a_1} \wedge \cdots \wedge e^{a_{p-1}}
  \wedge
  (
  \overline{\psi}
  \Gamma^{a_1 \cdots a_p}\wedge \psi)
  \wedge
  (\overline{\psi} \wedge \Gamma^{a_p} \psi)
  \,.
$$
This is zero precisely if after skew-symmetrization of the indices, the 
spinorial expression
$$
  \overline{\psi}
  \Gamma^{[a_1 \cdots a_p}\wedge \psi
  \wedge
  \overline{\psi} \wedge \Gamma^{a_p]} \psi
  = 0  
$$ 
vanishes identically (on all spinor components). 
The spinorial relations which control this are the \emph{Fierz identities}.
If this expression vanishes, then $\mu_p$ is a $(p+2)$-cocycle on 
$\mathbb{R}^{d;N=1}$, Def. \ref{cocycle}, hence a homomorphism of 
super Lie $n$-algebras of the form
$$
  \mu_p 
    : 
  \xymatrix{
    \mathbb{R}^{d;N=1}
	\ar[rr]
	&&
	\mathbb{R}[p+1]
  }
  \,.
$$
If this is the case then, by Def. \ref{rationaldatum}, this defines a $\sigma$-model
$p$-brane propagating on $\mathbb{R}^{d;N=1}$.

\medskip
The combinations of $d$ and $p$ for which this is the case 
had originally been worked out in \cite{AETW}. 
The interpretation
in terms of super-Lie algebra cohomology was clearly laid out in 
\cite{AT}. See \cite{BrandtII, BrandtIII, Brandt} for a rigorous treatment and comprehensive classification for all $N$. The non-trivial
cases (those where $\mu_p$ is closed but not itself a differential) correspond 
precisely to the non-empty entries in the following table.\\

\medskip

\begin{center}
\begin{tabular}{|r||c|c|c|c|c|c|c|c|c|c|}
  \hline
     ${d}\backslash p$ &   & $1$ & $2$ & $3$ & $4$ & $5$ & $6$ & $7$ & $8$ & $9$ 
	 \\[5pt]
	 \hline \hline
	 $11$ & & & 
	  \begin{tabular}{c} (1) \\  $\mathfrak{m}2\mathfrak{brane}$ \end{tabular}  & 
	 \hspace{30pt} & \hspace{30pt} &  
	 &&& &
	 \\[5pt]
	 \hline
	 $10$ &		
	   & \hspace{-.4cm}\begin{tabular}{c} 
	       (1,0) \\  $\mathfrak{string}_{\mathrm{het}}$
		\end{tabular}\hspace{-.4cm} 
		& 
		& 
		& 
		& 
	  \hspace{-.4cm}\begin{tabular}{c} 
	    (1,0) \\ $\mathfrak{ns}5\mathfrak{brane}_{\mathrm{het}}$
	  \end{tabular}\hspace{-.4cm}
	  &
	  &
	  &
	  &
	 \\[5pt]
	 \hline
	 $9$ & & & & & (1) & & & & & 
	 \\[5pt]
	 \hline
	 $8$  & & & & (1) & & & & & & 
	 \\[5pt]
	 \hline
	 $7$  & & & (1) & & & & & & & 
	 \\[5pt]
	 \hline
	 $6$  & & \hspace{-.4cm}\begin{tabular}{c} 
	           (1,0) \\  $\mathfrak{littlestring}$
		 \end{tabular} \hspace{-.4cm}
		 & & (1,0) & & & & & &
	 \\[5pt]
	 \hline
	 $5$ & &  & (1) & & & & & & &
	 \\[5pt]
	 \hline
	 $4$  & & (1) & (1) &&&&& & &
	 \\[5pt]
	 \hline
	 $3$  & & (1) &&&&& & & &
	 \\[5pt]
	 \hline
  \end{tabular}
\end{center}

  \vspace{4mm}
This table is known as the ``old brane scan'' for string/M-theory.
Each non-empty entry corresponds to a $p$-brane WZW-type $\sigma$-model action
functional of Green-Schwarz type. 
For $(d= 10, p = 1)$ this is the original
Green-Schwarz action functional for the superstring \cite{GreenSchwarz}
and, therefore, we write $\mathfrak{string}_{\mathrm{het}}$ in the respective
entry of the table 
(similarly there are cocycles for type II strings, discussed in the following sections), 
which at the same time is to denote the 
super Lie 2-algebra extension of $\mathbb{R}^{10, N=1}$ that is
classified by $\mu_p$ in this dimension, according to Remark \ref{extension}:
$$
  \xymatrix{
    \mathfrak{string}_{\mathrm{het}}
	\ar[d]
	\\
	\mathbb{R}^{10;N=(1,0)}
	\ar[rr]^-{\mu_1}
	&&
	\mathbb{R}[2]\;.
  }
$$
This Lie 2-algebra has been highlighted in \cite{BHII}.

\medskip
Analogously we write $\mathfrak{m}2\mathfrak{brane}$ for the
super Lie 3-algebra extension of $\mathbb{R}^{11;N=1}$ classified by
the nontrivial cocycle $\mu_{2}$ in dimension 11 
(this was called the \emph{supergravity Lie 3-algebra}
$\mathfrak{sugra}_{11}$ in \cite{SSSI})
$$
  \raisebox{20pt}{
  \xymatrix{
    \mathfrak{m}2\mathfrak{brane}
	\ar[d]
	\\
	\mathbb{R}^{11;N=1}
	\ar[rr]^-{\mu_2}
	&&
	\mathbb{R}[3]\;,
  }}
$$
and so on.

\medskip
While it was a pleasant insight back then that so many of the extended
objects of string/M-theory do appear from just super-Lie algebra cohomology
this way in the above table, it was perhaps just as curious that not all of them
appeared. Later other tabulations of string/M-branes were compiled, based
on less mathematically well defined physical principles \cite{Duff}. 
These ``new brane scans'' are what make the above an ``old brane scan''.
But we will show next that if only we allow ourselves to pass from 
(super-)Lie algebra theory to (super-) Lie $n$-algebra theory, then 
the old brane scan turns out to be part of a brane bouquet that 
accurately incorporates all the information of the ``new brane scan'',
all the branes of the new brane scan, altogether with their 
intersection laws, with their tensor multiplet field content and its
correct higher gauge transformation laws.

\paragraph{Type IIA superstring ending on D-branes and the D0-brane condensate}
\label{TypeIIASuperStringEndingOnDBrane}
\index{brane!D-brane!in type IIA}
\index{brane!D0-brane condensate}
\index{Wess-Zumino-Witten functionals!super $p$-branes!D-brane type IIA}

We consider the branes in type IIA string theory and point out how
their $L_\infty$-homotopy theoretic formulation serves to 
provide a formal statement and proof of the folklore relation between
type IIA string theory with a D0-brane condensate and M-theory.

\medskip

Write $N = (1,1) = \mathbf{16} + \mathbf{16}^\prime$ 
for the Dirac representation of $\mathrm{Spin}(9,1)$
given by two 16-dimensional real irreducible representations of opposite chirality.
We write $\{\Gamma^a\}_{a = 1, \cdots, 10}$ for the corresponding
representation of the Clifford algebra and 
$\Gamma^{11} := \Gamma^1 \Gamma^2 \cdots \Gamma^{10}$ for the chirality operator.
Finally write $\mathbb{R}^{10;N=(1,1)}$ for the corresponding
super-translation Lie algebra, the super-Minkowski spacetime of type IIA
string theory.

\begin{definition}
The type IIA 3-cocycle is
$$
  \mu_{\mathfrak{string}_{\mathrm{IIA}}}
  :=
  \overline{\psi} \wedge \Gamma^a \Gamma^{11} \psi \wedge e^a
  \;\;:\;\;
  \xymatrix{
    \mathbb{R}^{10;N=(1,1)}
	\ar[r]
	&
	\mathbb{R}[2]
  }
  \,.
$$
The type IIA superstring super Lie 2-algebra is the corresponding 
super $L_\infty$-extension
$$
  \xymatrix{
    \mathfrak{string}_{\mathrm{IIA}}
	\ar[d]
	\\
	\mathbb{R}^{10; N=(1,1)}
	\ar[rr]^-{\mu_{\mathfrak{string}_{\mathrm{IIA}}}}
	&&
	\mathbb{R}[2]\;.
  }
$$
Its Chevalley-Eilenberg algebra is that of $\mathbb{R}^{10;N=(1,1)}$
with one generator $F$ in degree $(2,\mathrm{even})$ adjoined and with 
its differential being
$$
  d_{\mathrm{CE}}\, F = \mu_{\mathfrak{string}_{\mathrm{IIA}}} = 
  \overline{\psi} \wedge \Gamma^a \Gamma^{11} \psi \wedge e^a.
$$
\label{TypeIIALie2Algebra}
\end{definition}
This dg-algebra appears as equation (6.3) in \cite{CdAIP}. It can also be deduced from
 op.cit. that the IIA string Lie 2-algebra of Def. \ref{TypeIIALie2Algebra} 
carries exceptional cocycles of 
degrees $p+2 \in \{2,4,6,8,10\}$ of the form
\(
  \begin{aligned}
  \mu_{\mathfrak{d}p\mathfrak{brane}} 
    & := 
	C \wedge e^F
	\\
	& :=
    \sum_{k=0}^{(p+2)/2} c^p_k
    \left(
      e^{a_1}\wedge \cdots \wedge e^{a_{p-2k}}
    \right)
    \wedge
    \left(
      \overline{\psi}\Gamma^{a_1} \cdots \Gamma^{a_{p-2k}}\Gamma^{11}\psi
    \right)
    \underbrace{F \wedge \cdots \wedge F}_{\mbox{$k$ factors}}\;,
  \end{aligned}
  \label{DBraneCocyclesInIIA}
\)
where $\{c_k^p \in \mathbb{R}\}$ are some coefficients, and where $C$
denotes the inhomogeneous element of $\mathrm{CE}(\mathbb{R}^{10;N=(1,1)})$
defined by the second line.
For each $p \in \{0,2,4,6,8\}$ there is, up to a global rescaling, a unique choice of
the coefficients $c^p_k$ that make this a cocycle.
This is shown on p. 19 of \cite{CdAIP}.
\begin{remark}
  Here the identification with physics terminology is as follows
  \begin{itemize}
    \item $F$ is the field strength of the \emph{Chan-Paton gauge field}
	on the D-brane, a ``tensor field'' that happens to be a ``vector field'';
	\item $C = \sum_{p} k^p \overline{\psi} \underbrace{e \wedge \cdots \wedge e}_{\mbox{$p$ factors}} \psi$  is the \emph{RR-field}.
  \end{itemize}
\end{remark}
It is interesting to notice the special nature of the cocoycle for the D0-brane:
\begin{remark}
  According to (\ref{DBraneCocyclesInIIA}) for $p = 0$, the cocycle
  defining the D0-brane as a higher WZW $\sigma$-model is just
  $$
    \mu_{\mathfrak{d}0\mathfrak{brane}}
	=
	\overline{\psi}\Gamma^{11}\psi
	\;.
  $$
  Since this independent of the generator $F$, it restricts
  to a cocycle on just $\mathbb{R}^{10;N=(1,1)}$ itself.
  \label{D0cocycleRestrictsToUnextendedSpacetime}
\end{remark}
Concerning this, we highlight the following fact, 
which is mathematically elementary but physically noteworthy
(see also Section 2.1 of \cite{CdAIP}),
as it has conceptual consequences for arriving at M-theory starting from 
type IIA string theory.
\begin{proposition}
  The extension of 10-dimensional type IIA super-Minkowski spacetime
  $\mathbb{R}^{10;N=(1,1)}$ by the 
  D0-brane cocycle as in Remark \ref{D0cocycleRestrictsToUnextendedSpacetime}
  is the 11-dimensional super-Minkowski spacetime of
  11-dimensional supergravity/M-theory:
  $$
    \xymatrix{
	  \mathbb{R}^{11;N=1}
	  \ar[d]
	  \\
	  \mathbb{R}^{10;N=(1,1)}
	  \ar[rr]^-{\mu_{\mathfrak{d}0\mathfrak{brane}}}
	  &&
	  \mathbb{R}[1]\;.
	}
  $$
  \label{11dIsExtensionOf10DByD0Cocycle}
\end{proposition}
\proof
  By Prop. \ref{CEAlgebrasOfExtensions} the Chevalley-Eilenberg algebra of 
  the extension classified by 
  $\mu_{\mathfrak{d}0\mathfrak{brane}}$ is 
  that of $\mathbb{R}^{10;N=(1,1)}$ with one new generator
  $e^{11}$ in degree $(1,\mathrm{even})$ adjoined and with 
  its differential defined to be
  $$
    d_{\mathrm{CE}}\, e^{11} = \mu_{\mathfrak{d}0\mathfrak{brane}}
	= \overline{\psi}\Gamma^{11}\psi
	\,.
  $$
  An elementary basic fact of Spin representation theory 
  says that the $N = 1$-representation of 
  the Spin group $\mathrm{Spin}(10,1)$ in odd dimensions
  is the $N= (1,1)$-representation of 
  the even dimensional Spin group $\mathrm{Spin}(9,1)$
  regarded as a representation of the Clifford algebra
  $\{\Gamma^a\}_{a = 1}^{10}$ with $\Gamma^{11}$ adjoined as in
  Def. \ref{TypeIIALie2Algebra}. Using this, the above
  extended CE-algebra is exactly that of $\mathbb{R}^{11;N=1}$
  \,.
\endofproof
\begin{remark}
  In view of Remark \ref{BraneCondensates}
  the content of Prop. \ref{11dIsExtensionOf10DByD0Cocycle}
  translates to heuristic physics language as:
  \emph{A condensate of D0-branes turns the 10-dimensional
  type IIA super-spacetime into the 11-dimensional spacetime of
  11d-supergravity/M-theory.} Alternatively: 
\emph{The condensation of D0-branes makes an 11th dimension of
 spacetime appear}.

\medskip
 In this form the statement is along the lines of the 
standard folklore relation between type IIA string theory and M-theory,
which says that type IIA with $N$ D0-branes in it is M-theory
compactified on a circle whose radius scales with $N$;
see for instance \cite{BFSS, Polchinski}. See also
\cite{Kong} for similar remarks motivated from phenomena in 2-dimensional
boundary conformal field theory.
Here in the formalization via higher WZW $\sigma$-models 
a version of this statement
becomes a theorem, Prop. \ref{11dIsExtensionOf10DByD0Cocycle}.
  \label{11thDimensionAppearsFromD0Condensate}
\end{remark}
\begin{remark}
  The mechanism of remark \ref{11thDimensionAppearsFromD0Condensate} 
  appears at several places in the brane bouquet. 
  First of all, since by Prop. \ref{DBraneCocyclesInIIA} the
  D0-brane cocycle is a summand in each type IIA D-brane cocycle,
  it follows via the above translation from $L_\infty$-homotopy theory
  to physics language that:
  \emph{Any type IIA D-brane condensate extends 10-dimensional type IIA
  super-spacetime to 11-dimensional super-spacetime.}
  If we lift attention again from the special case of D-branes of type IIA string theory
  to general higher WZW-type $\sigma$-models, then this mechanism
  is seen to generalize: the 10-dimensional super-Minkowski spacetime 
  itself is an extension of the \emph{super-point} by 
  10-cocycles (one for each dimension):
  $$
    \xymatrix{
	  \mathbb{R}^{10;N=(1,1)}
	  \ar[d]
	  \\
	  \mathbb{R}^{0;N=(1,1)}
	  \ar[rrr]^{\sum_{a = 1}^{10}\overline{(-)}\Gamma^a (-)}
	  &&&
	  \mathbb{R}[1]\;.
	}
  $$  
  Here the cocycle describes 10 different 0-brane $\sigma$-models,
  each propagating on the super-point as their target super-spacetime.
  Again, by remark \ref{BraneCondensates}, this mathematical 
  fact is a formalization and proof of what in physics language is the 
  statement that  \emph{Spacetime itself emerges from the abstract dynamics of 0-branes.}
  This is close to another famous folklore statement about 
  string theory. In our context it is a theorem.
\end{remark}

\paragraph{Type IIB superstring ending on D-branes and S-duality}
\label{TypeIIBSuperstringEndingOnDBranes}
\index{brane!D-brane!in type IIB}
\index{S-duality}
\index{Wess-Zumino-Witten functionals!super $p$-branes!D-brane type IIB}

We consider the branes in type IIB string theory 
as examples of higher WZW-type $\sigma$-model field theories 
and observe how
their $L_\infty$-homotopy theoretic formulation serves to 
provide a formal statement of the prequantum S-duality
equivalence between F-strings and D-strings and their unification
as $(p,q)$-string bound states.

\medskip
Write $N = (2,0) = \mathbf{16} + \mathbf{16}$ 
for the direct sum representation of $\mathrm{Spin}(9,1)$
given by two 16-dimensional real irreducible representations of 
the same chirality.
We write $\{\Gamma^a\}_{a = 1, \cdots, 10}$ for the corresponding
representation of the Clifford algebra on one copy of $\mathbf{16}$ 
and $\Gamma^a \otimes \sigma^i$ for the linear maps 
on their direct sum representation that act as the $i$th Pauli matrix
on $\mathbb{C}^2$ with components $\Gamma^a$, under the canonical identification
$\mathbf{16}\oplus \mathbf{16} \simeq \mathbf{16}\otimes \mathbb{C}^2$.
Finally write $\mathbb{R}^{10;N=(2,0)}$ for the corresponding
super-translation Lie algebra, the super-Minkowski spacetime
of type IIB string theory.

\medskip
There is a cocycle 
$\mu_{\mathfrak{string}_{\mathrm{IIB}}} \in \mathrm{CE}(\mathbb{R}^{10;N=(2,0)})$ 
given by
$$
  \mu_{\mathfrak{string}_{\mathrm{IIB}}}
  =
  \overline{\psi}\wedge (\Gamma^a \otimes \sigma^3) \psi \wedge e^a
  \,.
$$
The corresponding WZW $\sigma$-model is the Green-Schwarz formulation of the 
fundamental type IIB string. Of course we could use 
in this formula any of the $\sigma^i$, but one fixed such choice we are 
to call the type IIB string. That the other choices are equivalent is
the statement of \emph{S-duality}, to which we come in a moment.
The corresponding $L_\infty$-algebra extension, hence 
by Remark \ref{BraneCondensates} the 
IIB spacetime ``with string condensate'' is the homotopy fiber
$$
  \xymatrix{
    \mathfrak{string}_{\mathrm{IIB}}
	\ar[d]
	\\
	\mathbb{R}^{10;N=(2,0)}
	\ar[rr]^-{\mu_{\mathfrak{string}_{\mathrm{IIB}}}}
	&&
	\mathbb{R}[2]\;.
  }
$$
As for type IIA, 
its Chevalley-Eilenberg algebra $\mathrm{CE}(\mathfrak{string}_{\mathrm{IIB}})$
is that of $\mathbb{R}^{10;N=(2,0)}$ with one generator $F$ in degree
$(2,\mathrm{even})$ adjoined. The differential of that is now given by
$$
  \begin{aligned}
    d_{\mathrm{CE}}\, F 
	& = 
	\mu_{\mathfrak{string}_{\mathrm{IIB}}}
	\\
	& = \overline{\psi} \wedge (\Gamma^a \otimes \sigma^3) \psi \wedge e_a\;.
  \end{aligned}
$$
Now this Lie 2-algebra itself carries exceptional cocycles 
of degree $(p+2)$ 
for $p \in \{1,3,5,7,9\}$ of the form
\(
  \begin{aligned}
  \mu_{\mathfrak{d}p\mathfrak{brane}} 
    & := 
	C \wedge e^F
	\\
	& :=
    \sum_{k=0}^{(p+2)/2+1} c^p_k
    \left(
      e^{a_1}\wedge \cdots \wedge e^{a_{p-2k}}
    \right)
    \wedge
    \left(
      \overline{\psi}\wedge (\Gamma^{a_1} \cdots \Gamma^{a_{p-2k}}\otimes \sigma^{1/2})\psi
    \right)
    \underbrace{F \wedge \cdots \wedge F}_{\mbox{$k$ factors}}\;,
  \end{aligned}
  \label{DBraneCocyclesInIIB}
\)
where on the right the notation $\sigma^{1/2}$
is to mean that $\sigma^1$ appears in summands with an odd number of 
generators ``$e$'', and $\sigma^2$
in the other summands.
The corresponding WZW models are those of the type IIB D-branes.

\begin{remark}
  According to expression (\ref{DBraneCocyclesInIIB}) the cocycle of the D1-brane is of 
  the form
  $$
    \mu_{\mathfrak{d}1\mathfrak{}brane}
	=
	\overline{\psi} \wedge (\Gamma^a \otimes \sigma^1) \wedge e^a
	\,,
  $$
  which is the same form as that of $\mu_{\mathfrak{string}_{\mathrm{IIB}}}$
  itself, only that $\sigma^3$ is replaced by $\sigma^1$. 
  In fact since this is the D-brane cocycle which is 
  independent of the new generator $F$, it restricts to a cocycle
  on just $\mathbb{R}^{10;N=(2,0)}$ itself.
  So the cocycle for the ``F-string'' in type IIB is on the same
  footing as that of the ``D-string''. Both differ only by a 
  ``rotation'' in an internal space.
\end{remark}
\begin{remark}
There is a circle worth of $L_\infty$-automorphisms
$$
  S(\alpha) : \mathbb{R}^{10;N=(2,0)}\to \mathbb{R}^{10;N=(2,0)}\;,
$$
hence a group homomorphism
$$  
  U(1) \to \mathrm{Aut}(\mathbb{R}^{10;N=(2,0)})\;,
$$
given dually on Chevalley-Eilenberg algebras by
$$
  \begin{aligned}
    e^a &\mapsto  e^a
	\\
	\psi &\mapsto \exp(\alpha \sigma^2) \psi\;.
  \end{aligned}  
$$
This mixes the cocycles for the F-string and for the D-string in that
for a quarter rotation it turns one into the other
$$
    S(\pi/4)^\ast (\mu_{\mathfrak{string}_{\mathrm{IIA}}}) 
     = \mu_{\mathfrak{d}1\mathfrak{brane}}\;,
$$
and for a rotation by a general angle it produces a corresponding
superposition of both.
In particular, we can form \emph{bound states} of $F$-strings and D1-branes
by adding these cocycles
$$
  \mu_{(p,q)\mathfrak{string}}
  =
  p \,\mu_{\mathfrak{string}_{\mathrm{IIB}}}
  + 
  q \,\mu_{\mathfrak{d}1\mathfrak{brane}}
  \;\;
  \in \mathrm{CE}(\mathbb{R}^{10;N=(2,0)})
  \,.
$$
These define the $(p,q)$-string bound states as WZW-type $\sigma$-models.
\end{remark}

\paragraph{The M-theory 5-brane and the M-theory super Lie algebra}
\label{The5brane}
\index{brane!M5-brane}
\index{M-theory!M5-brane}
\index{Wess-Zumino-Witten functionals!super $p$-branes!M5-brane}

We discuss here the single M5-brane as a higher WZW-type $\sigma$-model,
show that it is defined by a 7-cocycle on the M2-brane super Lie-3 algebra
and observe that this 7-cocycle is indeed the relevant fermionic
7d Chern-Simons term of 11-dimensional supergravity compactified on 
$S^4$, as required by $\mathrm{AdS}_7/\mathrm{CFT}_6$ 
in the Chern-Simons interpretation of \cite{Witten}. We see that the
truncation of the symmetry algebra of this higher 5-brane superalgebra
to degree 0 is the ``M-algebra''.

\medskip

Write $N = 1 = \mathbf{32}$ for the irreducible real representation of
$\mathrm{Spin}(10,1)$. Write  $\{\Gamma^a\}_{a = 1}^{11}$ for the 
corresponding representation of the Clifford algebra. Finally write
$\mathbb{R}^{11; N =1}$ for the corresponding super-translation Lie algebra.
According to the old brane scan in section \ref{OldBraneScan}, 
the exceptional Lorentz-invariant 
cocycle for the M2-brane is

$$
  \mu_{\mathfrak{m}2\mathfrak{brane}} 
    = 
  \overline{\psi} \wedge \Gamma^{a b} \psi \wedge e^a \wedge e^b\;.
$$
The Green-Schwarz action functional for the M2-brane is the $\sigma$-model defined
by this cocycle
$$
  \xymatrix{
    \mathbb{R}^{11;N}
	\ar[rr]^{\mu_{\mathfrak{m}2\mathfrak{brane}}}
	&&
	\mathbb{R}[3]\;.
  }
$$

By the $L_\infty$-theoretic brane intersection law
of Remark \ref{BraneIntersectionLaw}, for the M2-brane to end on another kind
of brane, that other WZW model is to have 
the extended spacetime $\mu_{\mathfrak{m}2\mathfrak{brane}}$
(the original spacetime including a condensate of M2s) as its 
target space. 
By Prop. \ref{CEAlgebrasOfExtensions},
  the Chevalley-Eilenberg algebra of the M2-brane algebra is 
  obtained from that of the super-Poincar{\'e} Lie algebra by 
  adding one more generator $c_3$ with $\mathrm{deg}(c_3) = (3,\mathrm{even})$
  with differential defined by
  $$
    \begin{aligned}
    d_{\mathrm{CE}} \,c_3
	& :=
	\mu_{\mathfrak{m}2\mathfrak{brane}}
	\\
	&=
	\overline{\psi} \wedge \Gamma^{a b} \psi \wedge e^a \wedge e^b
	\end{aligned}
	\,.
  $$
We can then define an extended spacetime Maurer-Cartan form $\hat \theta$ in 
$\Omega^1_{\rm flat}(\R^{11;N}, \mathfrak{m2brane})$,
extending the canonical Maurer-Cartan form $\theta$ in
$\Omega^1_{\rm flat}(\R^{11;N}, \R^{d;N})$,
by picking any 3-form $C_3 \in \Omega^3(\mathbb{R}^{11;N})$ such that
$d_{\mathrm{dR}} C_3 = \overline{\psi} \Gamma^{a b} \wedge \psi \wedge e^a \wedge e^b$.

\medskip
Next, for every $(n+1)$ cocycle on $\mathfrak{m}2\mathfrak{brane}$ we get 
an $n$-dimensional WZW model defined on $\mathbb{R}^{11;N}$ this way.
In particular, the next one we meet is the M5-brane cocycle.
  Indeed, there is the degree-7 cocycle
  $$
    \mu_7 
	  = 
	\overline{\psi} \Gamma^{a_1 \cdots a_5} \psi e^{a_1} \wedge \cdots e^{a_5}
	 +
	C_3 \wedge \overline{\psi} \Gamma^{a b} \psi \wedge e^a \wedge e^b
	\;:\;
	\xymatrix{
	 \mathfrak{m}2\mathfrak{brane}
	 \ar[rr]
	 &&
	 \mathbb{R}[6]
	}
	\,
  $$
that was first observed in \cite{DF}, then rediscovered several times, for instance in 
  \cite{Sezgin}, in \cite{BLNPST1} and in \cite{CdAIP}. 
  Here we identify it as an $L_\infty$ 7-cocycle on 
  the $\mathfrak{m2brane}$ super Lie 3-algebra. The 
  $L_\infty$-extension of $\mathfrak{m2brane}$ associated with the 7-cocycle is 
  a super Lie 6-algebra that we call  $\mathfrak{m5brane}$.  

\medskip
It follows from this, with remark \ref{BraneIntersectionLaw}, 
that the M2-brane may end on a M5-brane whose WZW term
$\mathcal{L}_{\mathrm{WZW}}$ locally satisfies
$$
  d 
  \mathcal{L}_{\mathrm{WZW}}
  =\mu_7=
  \overline{\psi} \Gamma^{a_1 \cdots a_5} \psi e^{a_1} \wedge \cdots e^{a_5}
	 +
  C_3 \wedge \overline{\psi} \Gamma^{a b} \psi \wedge e^a \wedge e^b
$$
This is precisely what in \cite{BLNPST1} is argued to be the 
action functional of the M5-brane (here displayed in the absence of the 
bosonic contribution of the C-field).
However, in order to get the expected structure of gauge transformations,
we need to go further. Namely, while the above  
local expression for the action functional appears to be correct on the nose, 
its  gauge transformations are not as expected for the M5: for the 
M5-brane worldvolume theory the 2-form with curvature $C_3$ is 
supposed to be a genuine higher 2-form gauge field on the worldvolume,
directly analogous to the Neveu-Schwarz B-field of 10-dimensional 
supergravity spacetime; see \cite{FiorenzaSatiSchreiberI}.
 As such, it is to have gauge transformations
parameterized by 1-forms. But in the above formulation fields are maps
$\Sigma_6 \to \mathbb{R}^{11;N}$ into spacetime itself, and as such
have no gauge transformations at all. 
We can fix this by finding a better space $\hat X$. In fact we should take that 
to be $\mathfrak{m}2\mathfrak{brane}$ itself. As indicated above, this is an extension
$$
  \xymatrix{
   \mathbb{R}[2] \ar[r] 
	 & 
	 \mathfrak{m}2\mathfrak{brane} 
	  \ar[r]
	 &
	 \mathbb{R}^{11; N}
  }\;,
$$
and, hence, a twisted product of spacetime with $\mathbb{R}[2]$,
the infinitesimal version of the moduli space of 2-form connections.
 This is the infinitesimal approximation to the WZW construction in 
 \ref{StrucWZWFunctional}.

\begin{remark}
  By $\mathrm{AdS}_7/\mathrm{CFT}_6$ duality and by \cite{Witten} the M5-brane is supposed to be
  the 6-dimensional WZW model which is holographically related to 
  the 7-dimensional Chern-Simons term inside 11-dimensional supergravity
  compactified on a 4-sphere in analogy to how the traditional 2d WZW model
  is the holographic dual of ordinary 3d Chern-Simons theory. 
  By our discussion here that 7d Chern-Simons theory ought to be the
  one given by the 7-cocycle. Indeed, we observe that this 7-cocycle
  does appear in the compactification according to D'Auria-Fre \cite{DF}.
  Back in that article these authors worked locally and discarded precisely
  this term as a global derivative, but in fact it is a topological term
  as befits a Chern-Simons term and may \emph{not} be discarded globally.
  This connects the discussion here to the holographic 
  $\mathrm{AdS}_7$/$\mathrm{CFT}_6$-description of the \emph{single} M5-brane.
  Now a coincident $N$-tuple of M5-branes is supposed to be determined by
  a semisimple Lie algebra and nonabelian higher gauge field data. 
  Since $\mathrm{AdS}_7/\mathrm{CFT}_6$ is still supposed to apply, we are
  to consider the \emph{nonabelian} contributions to the 7-dimensional
  Chern-Simons term in 11d sugra compactified to $\mathrm{AdS}_7$. 
  These follow from the 11-dimensional anomaly cancellation and charge 
  quantization. Putting this together as discussed in \ref{InfinCS7d}
  yields the corresponding 7d Chern-Simons theory. Among other terms it is controled
  by the canonical 7-cocycle $\mu_{7}^{\mathfrak{so}}$ 
  on the semisimple Lie algebra $\mathfrak{so}$. Since this extends 
  evidently to a cocycle also on the super Poincar{\'e} Lie algebra,
  we may just add it to the bispinorial cocycle that defines the single M5, 
  to get
  $$
    \xymatrix{
	  \mathbb{R}^{11;N=1} \times \mathfrak{so}(10,1)
	  \ar[rrrrr]^-{\overline{\psi}e^5\psi 
	   + 
	   \langle \omega \wedge [\omega \wedge \omega] \wedge [\omega \wedge \omega]\wedge [\omega \wedge \omega]  \rangle}
	  &&&&&
	  \mathbb{R}[6]
	}
	\,.
  $$
  By the general theory indicated here this defines a 6-dimensional 
  WZW model. By the discussion in \ref{InfinCS7d} and \ref{supergravityCField} it satisfies
  all the conditions imposed by holography. It is to be expected that this
  is part of the description of the nonabelian M5-brane.
\end{remark}

Finally it is interesting to consider the symmetries of the M5-brane
higher WZW model obtained this way.
\begin{definition}
  \label{11dPoincarePolyvectorExtension}
  The \emph{polyvector extension} \cite{ACDP} of $\mathfrak{sIso}(10,1)$
  -- called the \emph{M-theory Lie algebra} \cite{Sezgin} --
  is the super Lie algebra obtained by adjoining to $\mathfrak{sIso}(10,1)$ generators
  $\{Q_\alpha, Z^{ab}\}$ that transform as spinors with respect to the 
  existing generators, and whose non-vanishing brackets among themselves are
 \begin{eqnarray}
    \left[ Q_\alpha, Q_\beta \right] &=& i(C \Gamma^a)_{\alpha \beta} P_a
	+ (C \Gamma_{a b}) Z^{a b}\;,
	\nonumber\\
    \left[ Q_\alpha, 	Z^{ab}\right] &= &2 i (C \Gamma^{[a})_{\alpha \beta} Q^{b]\beta}
	\,.
	\nonumber
  \end{eqnarray}
\end{definition}
\begin{proposition}
  The degree-0 piece of the 
  graded Lie algebra of infinitesimal automorphisms
  of $\mathfrak{m}2\mathfrak{brane}$, Def. \ref{LieAlgebraOfSymmetries}, 
  is the ``M-theory algebra'' polyvector extension of the 11d super Poincar{\'e} algebra of
  Def. \ref{11dPoincarePolyvectorExtension}.
\end{proposition}
\proof
  We leave this as an exercise to the reader. 
  Hint: under the identification of FDA-language with ingredients
  of $L_\infty$-homotopy theory as discussed here, one can see that 
  this involves the computations displayed in \cite{Castellani}.
\endofproof

\paragraph{The complete brane bouquet of string/M-theory}
\label{CompleteBraneBouquet}
\index{brane!brane bouquet}
\index{Wess-Zumino-Witten functionals!super $p$-branes!bouquet}
 
We have discussed various higher super Lie $n$-algebras of 
super-spacetime. Here we now sum up, list all the relevant extensions
and fit them into the full brane bouquet.
To state the brane bouquet, we first need names for all the branches that
it has

\begin{definition}
The \emph{refined brane scan} is the following collection of
values of triples $(d,p,N)$.
%

\vspace{4mm}
\hspace{-1.0cm}\begin{tabular}{|r||c|c|c|c|c|c|c|c|c|c|ccccc|}
  \hline
  &&&&&&&&&&
  \\
     ${D \atop =}$ & $p = 0$ & $1$ & $2$ & $3$ & $4$ & $5$ & $6$ & $7$ & $8$ & $9$ 
	 \\
	 \hline \hline
	 $11$ & & & 
	  \hspace{-.4cm}\begin{tabular}{ll} (1) 
	    & $\mathfrak{m}2\mathfrak{brane}$ \end{tabular}\hspace{-.4cm} & 
	 \hspace{30pt} & \hspace{30pt} & 
	 \begin{tabular}{ll} (1) \hspace{-.4cm} & $\mathfrak{m}5\mathfrak{brane}$ \end{tabular} &&&&
	 \\
	 \hline
	 $10$ &		
	   \hspace{-.4cm}
	   \begin{tabular}{cc}
		  (1,1) \\ \hspace{0cm}  $\mathfrak{D}0\mathfrak{brane}$ 
		\end{tabular}
	   \hspace{-.4cm}
	 & \hspace{-.3cm}\begin{tabular}{ll} 
	       (1,0) & \hspace{-.4cm} $\mathfrak{string}_{\mathrm{het}}$
    	\\ (1,1) & \hspace{-.4cm} $\mathfrak{string}_{\mathrm{IIA}}$ 
     	\\ (2,0) & \hspace{-.4cm} $\mathfrak{string}_{\mathrm{IIB}}$ 
		\\ (2,0) & \hspace{-.4cm} $\mathfrak{D}1\mathfrak{brane}$
		\end{tabular}\hspace{-.4cm} 
		& 
		\hspace{-.4cm}\begin{tabular}{cc}
		  (1,1) \\  $\mathfrak{D}2\mathfrak{brane}$ 
		\end{tabular}\hspace{-.4cm}
		& 
		\hspace{-.4cm}\begin{tabular}{cc}
		  (2,0)\\ $\mathfrak{D}3\mathfrak{brane}$ 
		\end{tabular}\hspace{-.4cm}
		& 
		\hspace{-.4cm}\begin{tabular}{cc}
		  (1,1) \\ $\mathfrak{D}4\mathfrak{brane}$ 
		\end{tabular}\hspace{-.4cm}
		& 
	  \hspace{-.4cm}\begin{tabular}{ll} 
	    (1,0) & \hspace{-.4cm} $\mathfrak{ns}5\mathfrak{brane}_{\mathrm{het}}$
		\\
	    (1,1) & \hspace{-.4cm} $\mathfrak{ns}5\mathfrak{brane}_{\mathrm{IIA}}$
	    \\
	    (2,0) & \hspace{-.4cm} $\mathfrak{ns}5\mathfrak{brane}_{\mathrm{IIB}}$
		\\
		(2,0) & \hspace{-.4cm} $\mathfrak{D}5\mathfrak{brane}$
	  \end{tabular}\hspace{-.4cm}
	  &
		\hspace{-.4cm}\begin{tabular}{cc}
		  (1,1) \\$\mathfrak{D}6\mathfrak{brane}$ 
		\end{tabular}\hspace{-.4cm}
	  &
		\hspace{-.4cm}\begin{tabular}{cc}
		  (2,0) \\  $\mathfrak{D}7\mathfrak{brane}$ 
		\end{tabular}\hspace{-.4cm}
	  &
		\hspace{-.4cm}\begin{tabular}{cc}
		  (1,1) \\ $\mathfrak{D}8\mathfrak{brane}$ 
		\end{tabular}\hspace{-.4cm}
	  &
		\hspace{-.4cm}\begin{tabular}{cc}
		  (2,0) \\ $\mathfrak{D}9\mathfrak{brane}$ 
		\end{tabular}\hspace{-.4cm}
	 \\
	 \hline
	 $9$ & & & & & (1) & & & & & 
	 \\
	 \hline
	 $8$  & & & & (1) & & & & & &
	 \\
	 \hline
	 $7$  & & & (1) & & & & & & &
	 \\
	 \hline
	 $6$  & & \hspace{-.4cm}\begin{tabular}{ll} 
 	           (2,0) & \hspace{-.4cm} $\mathfrak{sdstring}$ 
		 \end{tabular} \hspace{-.4cm}
		 & & (2,0) & & & & & &
	 \\
	 \hline
	 $5$ & &  & (1) & & & & & &&  
	 \\
	 \hline
	 $4$  & & (1) & (1) &&&& && &
	 \\
	 \hline
	 $3$  & & (1) &&& && & &
	 \\
	 \hline
  \end{tabular}
  \label{refinedbranescan}
\end{definition}


\noindent The entries of this table denote super-$L_\infty$-algebras 
that organize themselves
as nodes in the brane bouquet 
according to the following proposition.

\begin{proposition}[The brane bouquet]
There exists a system of higher super-Lie-$n$-algebra extensions of the 
super-translation Lie algebra $\mathbb{R}^{d;N}$ 
for $(d = 11, N=1)$,  $(d = 10, N=(1,1))$, for $(d = 10, N = (2,0))$
and for $(d = 6, N = (2,0))$,
which is jointly given by the following diagram 
$$
  \hspace{-.6cm}
  \xymatrix@C=2pt{
    && && && \mathfrak{ns}5\mathfrak{brane}_{\mathrm{IIA}} 
    \\ 
    && 
	&& \fbox{$\mathfrak{D}0\mathfrak{brane}$} \ar[drr] 
	& \fbox{$\mathfrak{D}2\mathfrak{brane}$} \ar[dr]
	& \fbox{$\mathfrak{D}4\mathfrak{brane}$} \ar[d]
	& \fbox{$\mathfrak{D}6\mathfrak{brane}$} \ar[dl]
	& \fbox{$\mathfrak{D}8\mathfrak{brane}$} \ar[dll]
    \\
    & \ar[ur]^{\mathrm{KK}}& 
	& \mathfrak{sdstring} \ar[drrr]|{{d = 6} \atop {N = (2,0)}}  
	&
	&& \mathfrak{string}_{\mathrm{IIA}} \ar[d]|-{{d=10} \atop {N=(1,1)}}
	&& \mathfrak{string}_{\mathrm{het}} \ar[dll]|-{{d=10}\atop {N = 1}} 
	&& \mathfrak{littlestring}_{\mathrm{het}} \ar[dllll]|-{{d=6}\atop {N = 1}}
	&&
     \ar@{<->}[dd]^{\mbox{T}}	
	&&
    \\
    && \fbox{$\mathfrak{m}5\mathfrak{brane}$} \ar[rr]  
	&& \mathfrak{m}2\mathfrak{brane} \ar[rr]|-{{d=11} \atop {N=1}}
	&& \mathbb{R}^{d;N}
	&& 
	&& \mathfrak{ns}5\mathfrak{brane}_{\mathrm{het}} \ar[llll]|-{{d = 10}\atop {N =1}}
	&&
	\\
	&& 
	&& 
	& \mathfrak{string}_{\mathrm{IIB}} \ar[ur]|-{{d = 10}\atop {N=(2,0)}}
	\ar@{.}[r]
	& (p,q)\mathfrak{string}_{\mathrm{IIB}} \ar[u]|-{{d = 10}\atop {N=(2,0)}}
	\ar@{.}[r]
	& \mathfrak{Dstring} \ar[ul]|-{{d = 10}\atop {N=(2,0)}}
	&&
	&&
	&&
    \\
    && 
	&& \fbox{$(p,q)1\mathfrak{brane}$} \ar[urr]
	&\fbox{$(p,q)3\mathfrak{brane}$} \ar[ur]
	& \fbox{$(p,q)5\mathfrak{brane}$} \ar[u]
	& \fbox{$(p,q)7\mathfrak{brane}$} \ar[ul]
	& \fbox{$(p,q)9\mathfrak{brane}$} \ar[ull]
	\\
	&& &&  & \ar@{<->}[rr]_S &  && 
  }
$$ 

where
\begin{itemize}
  \item 
    An object in this diagram is precisely a super-Lie-$(p+1)$-algebra
	extension of the super translation algebra $\R^{d;N}$, with 
	$(d,p,N)$ as given by the entries of the same name in the
	refined brane scan, def. \ref{refinedbranescan};
  \item every morphism 
  is a super-Lie $(p+1)$-algebra extension by an exceptional $\mathbb{R}$-valued
  $\mathfrak{o}(d)$-invariant super-$L_\infty$-cocycle 
  of degree $p+2$ on the domain of the morphism;
  \item 
    the unboxed morphisms are hence super Lie $(p+1)$-algebra extensions of 
	$\mathbb{R}^{d;N}$ by a super Lie algebra $(p+2)$-cocycle, hence are homotopy 
	fibers 
	of the form
	$$
	  \xymatrix{
	     p\mathfrak{brane}
		 \ar[d]
		 \ar[rr]_<{\rfloor}
		 &&
		 \ast
		 \ar[d]
		 \\
		 \mathbb{R}^{d;N}
		 \ar[rr]^-{\mathrm{some\;cocycle}}
		 &&
		 \mathbb{R}[p+1]\;,
	  }
	$$
  \item and the boxed super-$L_\infty$-algebras are super Lie $(p+1)$-algebra
  extensions of genuine super-$L_\infty$-algebras (which are not plain super Lie algebras),
  again by $\mathbb{R}$-cocycles
	$$
	  \xymatrix{
	     p_2\mathfrak{brane}
		 \ar[d]
		 \ar[rr]_<{\rfloor}
		 &&
		 \ast
		 \ar[d]
		 \\
		 p_1\mathfrak{brane}
		 \ar[rr]^-{\mathrm{some\;cocycle}}
		 &&
		 \mathbb{R}[p_2+1]\;.
	  }
	$$
\end{itemize}
 \label{BraneBouquet}
\end{proposition}
\proof
  Using prop. \ref{CEAlgebrasOfExtensions} 
  and the dictionary that we have established above 
  between the language used in the 
  physics literature (``FDA''s) and super-$L_\infty$-algebra homotopy theory, 
  this is a translation of the following results that 
  can be found scattered in the literature (some of which were discussed in the
  previous sections).
  \begin{itemize}
  \item All $N=1$-extensions of $\mathbb{R}^{d;N=1}$ 
    are those corresponding to the 
  ``old brane scan'' \cite{AETW}. 
  Specifically the cocycle which classifies the super Lie 3-algebra extension 
  $\mathfrak{m}2\mathfrak{brane} \to \R^{11;1}$ 
  had been found earlier in the context of supergravity around
  equation
  (3.12) of \cite{DF}. These authors also explicitly write down
  the ``FDA'' that then in \cite{SSSI} was recognized as the
  Chevalley-Eilenberg algebra of the super Lie 3-algebra
  $\mathfrak{m}2\mathfrak{brane}$ (there called the ``supergravity Lie 3-algebra''). 
   Later all these cocycles appear in the 
  systematic classification of super Lie algebra cohomology
  in \cite{BrandtII, BrandtIII, Brandt}. 

  \item The 7-cocycle classifying the super-Lie-6-algebra 
   extension $\mathfrak{m}5\mathfrak{brane} \to \mathfrak{m}2 \mathfrak{brane}$
   together with that extension itself 
   can be traced back, in FDA-language, 
   to (3.26) in \cite{DF}. This is maybe still the only previous
   reference that makes explicit the Lie 6-algebra extension (as an ``FDA''),
   but the corresponding 7-cocycle itself has later been rediscovered several times,
   more or less explicitly.  For instance it appears as equations (6) and (9) in
   \cite{BLNPST1}. 
   A systematic discussion is 
   in section 8 of \cite{CdAIP}.
   
   \item The extension $\mathfrak{string}_{\mathrm{IIA}} \to \mathbb{R}^{10;N=(1,1)}$
   by a super Lie algebra 3-coycle and the cocycles for the further higher extensions
   $\mathfrak{D}(2n)\mathfrak{brane} \to \mathfrak{string}_{\mathrm{IIA}}$
can be traced back to section 6 of \cite{CdAIP}.

   \item The extension $\mathfrak{string}_{\mathrm{IIB}} \to \mathbb{R}^{10;N=(2,0)}$
   by a super Lie algebra 2-coycle and the cocycles for the further higher extensions
   $\mathfrak{D}(2n+1)\mathfrak{brane} \to \mathfrak{string}_{\mathrm{IIA}}$,
   as well as the extension 
   $\mathfrak{ns}5\mathfrak{brane}_{\mathrm{IIB}} \to \mathfrak{Dstring}$
   follow from section 2 of \cite{Sak2}.
   
   \end{itemize}
\endofproof
\begin{remark}
  \index{M-theory!brane bouquet}
  The look of the brane bouquet, Prop. \ref{BraneBouquet},
  is reminiscent of the famous cartoon that displays the 
  conjectured coupling limits of string/M-theory,
  e.g. figure 4 in \cite{cartoon}, or fig. 1 in \cite{Polchinski}.
  Contrary to that cartoon, the brane bouquet is a theorem.
  Of course that cartoon alludes to more details 
  of the nature of string/M-theory than we are
  currently discussing here, but all the more should it
  be worthwhile to have a formalism that makes precise
  at least the basic structure, so as to be able to proceed
  from solid foundations.
\end{remark}

\subsection{Local boundary and defect prequantum field theory}
\label{LocalBoundaryAndDefectPrequantumFieldTheory}

We now discuss examples and applications of the general mechanism of
higher local prequantum boundary and defect field theory, \ref{LocalPrequantumFieldTheories}. 
Our main interest here is the hierarchy of boundary and defect structures
relating higher Chern-Simons-type field theories to higher 
Wess-Zumino-Witten type field theories.

\medskip
We start in section
\begin{itemize}
  \item \ref{VacuumDefects} -- {\it Vacuum defects from spontaneous symmetry breaking}
\end{itemize}
with discussion of how the general abstract theory in Section 
\ref{LocalPrequantumFieldTheories} 
of correspondence spaces in higher
homotopy types nicely captures the traditional notions in physics phenomenology
of \emph{spontaneous symmetry breaking vacuum defects} 
called \emph{cosmic monopoles}, \emph{cosmic strings}
and \emph{cosmic domain walls}, including the traditional rules by which these may end on 
each other. This discussion uses a minimum of mathematical sophistication
(just some homotopy pullbacks) but may serve to nicely illustrate the interpretation
of the abstract formalism in actual realistic physics. Readers not interested in this interpretation
may want to skip this section.

\medskip
Our main example here is then 
\begin{itemize}
  \item \ref{HigherCSAsLocalPrequantumFT} -- 
  {\it Higher Chern-Simons local prequantum field theory}
\end{itemize}
where we observe that in the $\infty$-topos $\mathbf{H}$ of smooth stacks
 there is 
a canonical tower of topological higher local prequantum field
theories whose cascade of higher codimension defects naturally induce
higher Chern-Simons type prequantum field theories and their associated theories.

\subsubsection{Vacuum defects from spontaneous symmetry breaking}
\label{VacuumDefects}

In particle physics phenomenology and cosmology, 
there is a traditional notion of \emph{defects in the vacuum structure}
of gauge field theories which exhibit spontaneous symmetry breaking,
such as in the Higgs mechanism. 
A review of these ideas is in \cite{VilenkinShellard}. 
A discussion of how such vacuum defects
due to symmetry breaking may end on each other, and hence
form a network of defects of varying codimension, is in \cite{PreskillVilenkin}.
Here we briefly review the mechanism indicated in the latter article and then 
show how it is neatly formalized within 
the general notion of defect field theories as
in Section \ref{DefectTheory}. This is intended to serve as an illustration of 
the physical interpretation of the abstract notion of defects in field theories
and of their formalization by correspondences, particularly.
Readers not interested in physics phenomenology may want to skip this section.

\medskip

Consider an inclusion of topological groups
$
  H \hookrightarrow G
$.
Here we are to think of $G$ as the gauge group (more mathematically precise: structure group) 
of a gauge theory
and of $H \hookrightarrow G$ as the subgroup that is preserved by 
any one of its degenerate vacua (for instance in a Higgs mechanism), hence the 
gauge group that remains after spontaneous symmetry breaking. 
In this case
the quotient space (coset space) $G/H$ is the moduli space of vacuum 
configurations,
so that a vacuum configuration up to continuous deformations on a spacetime $\Sigma$ 
is given by 
the homotopy class of a map from $\Sigma$ to $G/H$.

\medskip
Traditionally a \emph{codimension-$k$ defect in the vacuum structure} of a theory
with such spontaneous symmetry breaking is a spacetime locally of the form
$\mathbb{R}^{n} - (D^{k} \times \mathbb{R}^{n-k})$ with a vacuum classified locally by 
a the homotopy class of a map 
$$ 
S^{k-1}\simeq  \mathbb{R}^{n} - (D^{k} \times \mathbb{R}^{n-k}) \to G/H
  \,,
$$
hence by an element of the $(k-1)$-st homotopy group of $G/H$.
If this element is non-trivial, one says that the vacuum has a 
\emph{codimension-$k$ defect}.
Specifically in an $(n = 4)$-dimensional spacetime $\Sigma$
\begin{itemize}
  \item for $k = 1$ this is called a \emph{domain wall};
  \item for $k = 2$ this is called a \emph{cosmic string};
  \item for $k = 3$ this is called a \emph{monopole}.
\end{itemize}

Next consider a sequence of inclusions of topological groups
$$
  H_2 \hookrightarrow H_1 \hookrightarrow H_0 = G
  \,.
$$
Along the above lines this is now to be thought of as describing the
breaking of a symmetry group $G = H_0$ first to $H_1$ at some energy scale $E_1$,
and then a further breaking down to $H_2$ at some lower energy scale $E_2$.
So at the high energy scale the moduli space of vacuum structures
is 
$G/H_1 = H_0/H_1$
as before.
But at the low energy scale the moduli space of vacuum structures is now $H_1/H_2$.
If there is a vacuum defect at low energy,
 classified by a map $S^{k-1}\to H_1/H_2$,
then if it is ``heated up'' or rather if it ``tunnels'' by a quantum fluctuation
through the energy barrier, 
it becomes instead a defect classified by a map to $H_0/H_2$,
namely by the composite
$$
S^{k-1} \to H_1/H_2 \to H_0/H_2
  \,.
$$
Here the map on the right is the fiber inclusion of the 
$H_1$-associated $H_1/H_2$-fiber bundle
$$  
  H_1/H_2 \to H_0/H_2 \to H_0/H_1
$$
naturally induced by the sequence of broken symmetry groups.
The heated defect may 
be unstable, hence given by a trivial element in the $(k-1)$-st homotopy group of $H_0/H_2$, 
even if the former is not, in which case one says that the original defect
is \emph{metastable}. In terms of diagrams, metastability of the
low energy defect means precisely that its classifying map 
$S^{k-1} \to H_1/H_2$  extends
to a homotopy commutative diagram
of the form
$$
  \raisebox{42pt}{
  \xymatrix{
    S^{k-1} \ar[r] \ar[d] &H_1/H_2 \ar[d]
	\\
	D^k \ar[r] & H_0/H_2
  }}
  \,,
$$
where the left vertical arrow is the boundary inclusion $S^{k-1}\hookrightarrow D^k$.
Now according to \cite{PreskillVilenkin}, the decay of a metastable
low-energy vacuum defect of codimension-$k$ leads to the formation of a 
stable high-energy defect of codimension-$(k+1)$ at its decaying boundary. For instance a metastable cosmic string defect in the low energy vacuum structure
is supposed to be able to end (decay) on a cosmic monopole defect in the high energy 
vacuum structure.

\medskip
We now turn to a formalization of this story.
By Def. \ref{CodimensionkDefect}, 
the discussion in \cite{PreskillVilenkin} shows that the transition from metastable 
codimension-$k$ defects in the low energy vacuum structure to stable high-energy $(k+1)$-defects 
should be represented
by a correspondence of the form
$$
  \xymatrix{
   [\Pi(S^{k-1}),\mathbf{\Pi}(H_1/H_2)]
	&
	[\Pi(S^k), \mathbf{\Pi}(H_0/H_1)]
	\ar[l]
	\ar[r]
	&
    \ast
  }
  \,,
$$
exhibiting the high energy defects as boundary data for the low
energy defects.

\medskip
To see how to obtain this in line with the phenomenological 
story, observe that the heating/tunneling process as well as the
decay process of the heated defects are naturally represented by
the maps on the left and the right of the following diagram, respectively:
$$
  \raisebox{5pt}{
  \xymatrix{
    && & \ast
	\ar[dl]^-{~~D^k \to H_0/H_2}
    \\
    [\Pi(S^{k-1}), \mathbf{\Pi}(H_1/H_2)]
	\ar[dr]_{\hspace{-2.8cm}H_1/H_2 \to H_0/H_2 }
	&& 
	 [\Pi((D^k), \mathbf{\Pi}(H_0/H_2)] \ar[dl]^{\hspace{11mm}S^{k-1}\hookrightarrow D^k}
	\\
	&  [\Pi(S^{k-1}), \mathbf{\Pi}(H_0/H_2)]
  }
  }\,.
$$
%
The left map sends a low energy defect to its high energy version, the right map
sends a high energy decay process to the field configuration which is decaying.
For a specific spatially localized defect process $D^k \to H_0/H_2$ 
we are to pick one point in the space
of defect processes, which is what the top right map reflects.
Therefore, the moduli space of decay processes of metastable low energy defects 
is precisely the homotopy fiber product of these two maps, namely
the space of pairs consisting of a low energy defect and a localized decay process of
its heated version (up to a pertinent gauge transformation that identifies
the heated defect with the field configuration which decays). By the 
above fiber sequence of quotient spaces one finds that this 
homotopy pullback is $[\Pi(S^{k-1}, \Omega\Pi(H_0/H_1))]$. Hence, in conclusion,
we find the desired correspondence as the top part of the following homotopy pullback diagram
$$
  \raisebox{84pt}{
  \xymatrix@C=0pt{
    & & [\Pi(S^{k}), \Pi(H_0/H_1)] \ar[d]
    \\
    & & [\Pi(S^{k-1}), \Omega\Pi(H_0/H_1)]
	\ar[dl] \ar[dr]
    \\
    & [S^{k-1}\to \Pi(H_1/H_2),  D^k \to \Pi(H_0/H_2)] 
	\ar[dl] \ar[dr]
	& \mbox{\tiny (pb)}
	& \ast \ar[dl]
    \\
    [\Pi(S^{k-1}), \Pi(H_1/H_2)]
	\ar[dr]_{\hspace{-3.2cm}[\Pi(S^{k-1}), \Pi(H_1/H_2) \to \Pi(H_0/H_2) ]}
	&\mbox{\tiny (pb)}& 
	[\Pi(D^k), \Pi(H_0/H_2)] \ar[dl]^{\hspace{1.6cm}[\Pi(S^{k-1}\hookrightarrow \Pi(D^k)), \Pi(H_0/H_2)]}
	\\
	& [\Pi(S^{k-1}), \Pi(H_0/H_2)]
  }
  }\,.
$$
In summary, this diagram encodes the phenomenological story of the decay of metastable defects as 
follows:
$$
{\footnotesize
  \hspace{-.5cm}
  \xymatrix{
    && 
    \fbox{\begin{tabular}{c} codimension-$(k+1)$ defect \\ at high energy \end{tabular}}
	\ar[dl]_{\hspace{-4mm}\mbox{induces}}
	\ar[dr]^{~~\mbox{induces}}
	\ar@/_4.6pc/[ddll]_-{\begin{tabular}{c} codim-$(k+1)$ defect \\ bounding \\ a codim-$k$ defect\end{tabular}}
    \\
    & 
    \fbox{\begin{tabular}{c} pairs of \\ codimension-$k$ defects \\ 
	 and their decay processes \end{tabular}}
	 \ar[dl]
	 \ar[dr]
	 &&
	 \fbox{\begin{tabular}{c} fixed localized \\ decay process\end{tabular}}
	 \ar[dl]
    \\
    \fbox{\begin{tabular}{c} codimension-$k$ defects \\ at low energy\end{tabular}}
	\ar[dr]_{\mbox{tunneling}}
	&&
    \fbox{\begin{tabular}{c} decay processes \end{tabular}}
	\ar[dl]^-{\mbox{apply to}}
	\\
	&
	\fbox{\begin{tabular}{c} codimension-$k$ defects \\ raised to high energy\end{tabular}}
  }
  }
$$

\subsubsection{Higher Chern-Simons local prequantum boundary field theory}
 \label{HigherCSAsLocalPrequantumFT}

We now turn to the class of those local prequantum field theories which deserve to be termed
of \emph{Chern-Simons type}. We show that these arise rather canonically
as the boundary data for the canonical differential cohomological structure
of Prop. \ref{ThePasting} which is exhibited by every cohesive $\infty$-topos
$\mathbf{H}$.

\paragraph{Survey: towers of boundaries, corners, ... and of circle reductions}

We discuss in the following towers/hierarchies of iterated defects
of increasing codimension of a universal topological Yang-Mills theory. 
Most of these defects, however, are best recognized after ``gluing their endpoints''
after which they equivalently become circle-reductions/transgression to loop space
of the original theory. 
Restricted to the archetypical case of 3d Chern-Simons theory,
the following discussion essentially goes through the following diagram:
$$
\raisebox{113pt}
{\xymatrix{
    \fbox{3d CS}~
	\ar@{^{(}->}[r]
	\ar[d]^-{S^1}
	&
	\fbox{4d tYM}
	\ar[d]^-{S^1}
	\\
	\fbox{2d WZW}~
	\ar@{^{(}->}[r]
	\ar[d]^-{S^1}
	&
	\fbox{3d tYM}
	\ar[d]^-{S^1}
	\\
	\fbox{1d Wilson line}~
	\ar@{^{(}->}[r]
	\ar[d]^-{S^1}
	&
	\fbox{2d tYM}
	\ar[d]^-{S^1}
	\\
	\fbox{3dCS action}~
	\ar@{^{(}->}[r]
	&
	\fbox{1d tYM}
  }
  }\;.
$$
This is  a pattern of iterated higher codimension 
corners and iterated circle reductions which had long been emphasized by
Hisham Sati 
to govern the grand structure of hierarchies of theories inside string/M-theory
\cite{SatiDuality, SatiC, F}.
For instance there should be a tower of this kind which instead of 3d Chern-Simons theory
has 11-dimensional supergravity, or ``M-theory''
as follows:

$$
\xymatrix{
& 
Z^{12}
\ar[dr]^{S^1}
 &
\\
Y^{11} 
\ar[dr]^{S^1}
\ar@{^{(}->}[ur]^\partial
&&
Y^{11'}
\\
&
X^{10}
\ar@{^{(}->}[ur]_\partial 
&
}
\qquad
\qquad
\xymatrix{
& 
\fbox{12d~ {\rm Bounding~ theory}}
\ar[dr]^{S^1}
 &
\\
\fbox{\rm M-theory} 
\ar[dr]^{S^1}
\ar@{^{(}->}[ur]^\partial
&&
\fbox{\rm Bounding~theory~ for~ IIA}
\\
&
\fbox{\rm Type~ IIA}
\ar@{^{(}->}[ur]_\partial 
&
}
$$
From this descent further such towers in string/M-theory, and for each one can have 
various extensions deeper in dimensions, via both dimensional reductions 
and boundaries. For instance, Edward Witten has
been exploring a system of reductions \cite{Witten11} which in (small) parts 
involves a system roughly as follows
$$
  \xymatrix{
    \fbox{M-theory}
    \ar[d]^{S^4}
    \\
	\fbox{7d CS = $\mathrm{KK}_{S^4}$ of 11d CS}
	\ar[d]^-{\partial}
	\\
	\fbox{(2,0) 6d QFT on M5}
	\ar[d]^-{S^1}
	\\
	\fbox{5d sYM}
	\ar[d]^-{S^1}
	\\
	\fbox{4d sYM}
	\ar[d]^-{T^2}
	\\
	\fbox{Langlands duality~for~$G$-bundles~on~curves.}
  }
$$
The second entry from the top indicates the 7-dimensional Chern-Simons theory/term
arising from the Kaluza-Klein reduction on the 4-sphere 
of the corresponding term in eleven dimensions \cite{FiorenzaSatiSchreiberI},
this we discuss below in \ref{7dCSInSugraOnAdS7}. The next is the 6-dimensional
boundary field theory of this, whose Green-Schwarz contribution we discuss
in \ref{SuperBranesAndTheirIntersectionLaws}.

\paragraph{$d = n+1$: Universal topological Yang-Mills theory}
\label{tYM}

We consider a simple theory where fields are closed differential forms and the Lagrangian 
being the integral of that form. We start with the abelian but higher case and later
we get nonabelian theories by introducing boundaries.

\medskip

First recall from the following system of higher differential moduli stacks.

\begin{proposition}
For $n \in \mathbb{N}$ we have a pasting diagram of homotopy pullback
squares in $\mathbf{H} = \mathrm{Smooth}\infty \mathrm{Grpd}$ of the form
$$
\raisebox{82pt}{
  \xymatrix{
    \mathbf{B}^n U(1)_{\mathrm{conn}}
	\ar[rr]^{\rm forget~(non-flat)}_{\rm connection}
	\ar[d]_{\mathrm{curv}}
	&&
	\mathbf{B}^n U(1)
	\ar[rr]
	\ar[d]
	&&
	\ast
	\ar[d]
	\\
	\Omega^{n+1}_{\mathrm{cl}}
	\ar[rr]^-{\rm inclusion}
	&&
	\flat_{\mathrm{dR}}\mathbf{B}^{n+1}U(1)
	\ar[rr]^{FA}
	\ar[d]
	&&
	\flat \mathbf{B}^{n+1} U(1)
	\ar[d]_{\rm forget}^{\rm flat~connection}
	\\
	&& \ast \ar[rr] 
	&& \mathbf{B}^{n+1}U(1)
  }
  }
  \,,
$$
where $FA$ is the map that takes a flat curvature form
and interprets it as a connection on the trivial bundle of one higher degree.
\label{ThePasting}
\end{proposition}

As a special case of Prop. \ref{EveryMapToAGroupObjectIsFullyDualizableInSpansInSlice} we
have:
\begin{proposition}
  For $n \in \mathbb{N}$, the morphism 
  $$
    \exp(i S_{\mathrm{tYM}}) \colon \Omega^{n+1}_{\mathrm{cl}} 
    \longrightarrow
    \flat \mathbf{B}^{n+1}U(1)
  $$ 
  in prop. \ref{ThePasting}, 
  regarded as an object
  $$
    \left[
	  \raisebox{20pt}{
	  \xymatrix{
	    \Omega^{n+1}_{\mathrm{cl}}
		\ar[d]^{\exp\left(\tfrac{i}{\hbar} S_{\mathrm{tYM}} \right)}
		\\
		\flat \mathbf{B}^{n+1}U(1)
	  }
	  }
	\right]
	\in 
	\mathrm{Corr}_n(\mathbf{H}_{/\flat \mathbf{B}^n U(1)})^\otimes
	\,,
  $$
  is fully dualizable, with dual $\exp\left(-\tfrac{i}{\hbar}S_{\mathrm{tYM}}\right)$.
  \label{localtYMActionisFullyDualizable}
\end{proposition}
\begin{definition}
  For $n \in \mathbb{N}$,
  we call the local prequantum field theory defined by 
  the fully dualizable object $S_{\mathrm{tYM}}$ 
  of Prop. \ref{localtYMActionisFullyDualizable} the
  \emph{universal topological Yang-Mills} local prequantum field theory
  $$
    \exp\left( \tfrac{i}{\hbar} S_{\mathrm{tYM}}\right)
	:
	\mathrm{Bord}_{n+1}^\otimes \to \mathrm{Corr}_{n+1}(\mathbf{H}_{/\flat \mathbf{B}^{n+1} U(1)})^\otimes
	\,.
  $$
  \label{StyM}
\end{definition}
This terminology is justified below in Remark \ref{BoundaryConditionsFortYMAndChernWeil}.
We will encounter this theory again in later sections. 

\paragraph{$d = n + 0$: Higher Chern-Simons field theories}
\label{HigherChernSimonsFieldTheoriesAsBoundariesOftYM}

We discuss now the boundary conditions of the universal topological
Yang-Mills local prequanutm field theory
\begin{remark}
  The universal boundary condition for $S_{\mathrm{tYM}}$
  according  to Def. \ref{TerminalBoundaryCondition} 
  is given 
  by the top rectangle in Prop. \ref{ThePasting}, naturally regarded
  as a correspondence in the slice:
  $$
    \raisebox{41pt}{
    \xymatrix@R=19pt@C=9pt{
	  & \mathbf{B}^n U(1)_{\mathrm{conn}}
	  \ar[dl] 
	  \ar[dr]^-{F_{(-)}}
	  \\
	  \ast \ar[dr] && \Omega^{n+1}_{\mathrm{cl}} 
	  \ar[dl]^-{\exp\left(\tfrac{i}{\hbar} S_{\mathrm{tYM}} \right)}
	  \\
	  & \flat \mathbf{B}^{n+1}U(1)
	}
	}
	\,.
  $$
  So by Prop. \ref{TerminalBoundaryConditionIsTerminal} 
  there is a natural equivalence of $\infty$-categories
  $$
    \mathrm{Bdr}\left(
      \exp\left(\tfrac{i}{\hbar} S_{\mathrm{tYM}}\right)
    \right)
	\simeq
	\mathbf{H}_{/\mathbf{B}^n U(1)_{\mathrm{conn}}}
  $$
  between the $\infty$-category of boundary conditions for 
  the universal topological Yang-Mills theory in dimension $(n+1)$ and the
  slice $\infty$-topos of $\mathbf{H}$ over the moduli stack of $U(1)$-$n$-connections.  
  \label{UniversalBoundaryCondition}
\end{remark}
\begin{corollary}
   The $(\infty,1)$-category of boundary conditions for the 
   universal topological Yang-Mills local prequantum field theory 
   $S_{\mathrm{tYM}}$ are equivalently 
   \emph{$\infty$-Chern-Simons local prequantum field theories}
   \cite{FiorenzaSatiSchreiberCS}: moduli stacks 
   $\mathbf{Fields}_{{\partial}} \in \mathbf{H}$ equipped with 
   a prequantum $n$-bundle \cite{hgp}
  $$
    \nabla_{\mathrm{CS}}
	:
    \mathbf{Fields}_{{\partial}}
	\to 
	\mathbf{B}^n U(1)_{\mathrm{conn}}
	\,.
  $$
  The automorphism $\infty$-group of a given boundary condition 
  for $S_{\mathrm{tYM}}$ is hence equivalently the
  \emph{quantomorphism} $\infty$-group of the corresponding
  Chern-Simons theory \cite{hgp}.
  \label{BoundaryConditionsFortopologicalYangMills}
\end{corollary}
\proof
  This is just a special case of Prop. \ref{TerminalBoundaryConditionIsTerminal}.
  Explicitly, 
  by the universal property of the homotopy pullback in $\mathbf{H}$, 
  given any boundary condition for $S_{\mathrm{tYM}}$, 
  hence by Remark \ref{BoundaryDatum} a diagram in $\mathbf{H}$
  of the form
  $$
    \raisebox{42pt}{
    \xymatrix{
	  & \mathbf{Fields}_{ \partial}
	  \ar[dl]
	  \ar[dr]^{~~\langle F_{(-)} \wedge \cdots \wedge F_{(-)}\rangle}_{\ }="s"
	  \\
	  \ast \ar[dr]_-0^{\ }="t" && \Omega^{n+1}_{\mathrm{cl}}\;,
	   \ar[dl]^-{\exp\left(\tfrac{i}{\hbar} S_{\mathrm{tYM}} \right)}
	  \\
	  & \flat \mathbf{B}^{n+1}U(1)
	  \ar@{=>}^\nabla "s"; "t"
	}
	}
  $$
  this is equivalent to the dashed morphism in the diagram
  $$
    \raisebox{80pt}{
    \xymatrix{
	  & \mathbf{Fields}_{ \partial}
	  \ar@/_.7pc/[ddl]
	  \ar@/^.7pc/[ddr]^{\langle F_{(-)}\wedge \cdots \wedge F_{(-)}\rangle}
	  \ar@{-->}[d]^\nabla
	  \\
	  & \mathbf{B}^n U(1)_{\mathrm{conn}}
	  \ar[dl]
	  \ar[dr]|{F_{(-)}}
	  \\
	  \ast \ar[dr]_-0 && \Omega^{n+1}_{\mathrm{cl}}\;.
	   \ar[dl]^-{\exp\left(\tfrac{i}{\hbar} S_{\mathrm{tYM}} \right)}
	  \\
	  & \flat \mathbf{B}^{n+1}U(1)
	}
	}
  $$
\endofproof
\begin{remark}
  Observe that while the space of phases of the bulk field theory is
  $\flat \mathbf{B}^{n+1}U(1)$, we may now regard $\mathbf{B}^n U(1)_{\mathrm{conn}}$
  as the space of spaces of the boundary field theory.
\end{remark}
In order to interpret this, notice the following.
\begin{remark}
  For the special case that $\mathbf{Fields}_{\partial}$
  is a moduli stack $\mathbf{B}G_{\mathrm{conn}}$ of $G$-principal $\infty$-connections
  for some $\infty$-group $G$, we may think of the morphism
  $$
    \langle F_{(-)} \wedge \cdots F_{(-)}\rangle 
	 : 
	 \mathbf{B}G_{\mathrm{conn}} \to \Omega^{n+1}_{\mathrm{cl}}
  $$
  as encoding an \emph{invariant polynomial} $\langle -,\cdots, -\rangle$ 
  on (the $\infty$-Lie algebra of) $G$
  \cite{FSS}. By extrapolation from this case
  we may also speak of invariant polynomials if 
  $\mathbf{Fields}|_{\partial\Sigma}$
  is of more general form, in which case we have invariant polynomials
  on \emph{smooth $\infty$-groupoids}.
  Restricting to the group-al case just for definiteness, 
  notice that a boundary field configuration, which by 
  Prop. \ref{RelativeBoundaryFieldConfigurations} is given by
  $$
  \raisebox{42pt}{
    \xymatrix{
	  {\partial}\Sigma \times U \ar[r]^\nabla 
	  \ar[d]
	  & \mathbf{B}G_{\mathrm{conn}}
	  \ar[d]
	  \\
	  \Sigma \times U \ar[r]^\omega & \Omega^{n+1}_{\mathrm{cl}}
	}
	},
  $$
  forces the closed $(n+1)$-form $\omega$ of the bulk theory to become
  the $\infty$-Chern-Weil form of a $G$-principal $\infty$-connection with 
  respect to the invariant polynomial $\langle -, \cdots, - \rangle$ at the boundary:
  $$
    \omega|_{{\partial}\Sigma} = 
	\langle 
	   F_\nabla \wedge \cdots \wedge F_\nabla
	\rangle
	\,.
  $$
  For $G$ an ordinary Lie group, this is known as the 
  \emph{Lagrangian for topological $G$-Yang-Mills theory}. More generally, 
  for $G$ any smooth $\infty$-group, we may hence think of this as the Lagrangian
  of a topological $\infty$-Yang-Mills theory. 
  
  \medskip
  Specifically for the \emph{universal}
  boundary condition 
  $\mathbf{Fields}_{\partial} = \mathbf{B}^n U(1)_{\mathrm{conn}}$
  of Remark \ref{UniversalBoundaryCondition} we find a field theory which assigns
  $U(1)$-$n$-connections $\nabla$ to $n$-dimensional manifolds $\Sigma_{n}$ and closed $(n+1)$-forms $\omega$
  on $(n+1)$-dimensional manifolds $\Sigma_{n+1}$, such that whenever the latter bounds the former, the exponentiated integral of $\omega$ equals the $n$-\emph{volume holonomy}
  of $\nabla$. This is just the relation between circle $n$-connections and their curvatures
  which is captured by the axioms of \emph{Cheeger-Simons differential characters}.
  Hence it makes sense to call the higher topological Yang-Mills theory which is
  induced from the universal boundary condition the \emph{Cheeger-Simons theory}
  in the given dimension.
  
  \medskip
  However, Corollary \ref{BoundaryConditionsFortopologicalYangMills} says more:
  the universality of the Cheeger-Simons theory as a boundary condition for
  topological Yang-Mills theory means that a consistent such boundary condition
  is necessarily not just an invariant polynomial, but is a lift
  of that from de Rham cocycles to differential cohomology. This means that it is a
  \emph{refined $\infty$-Chern-Weil homomorphism} in the sense of \cite{FSS}
  of the invariant polynomial in the sense of \cite{hgp}.
  Equivalently it is a \emph{higher prequantum field theory}. 
    In either case a 
  lift $\nabla$ in the diagram
  $$
    \raisebox{44pt}{
    \xymatrix{
	    && \mathbf{B}^n U(1)_{\mathrm{conn}}
	    \ar[d]
	  \\
	  \mathbf{B}G_{\mathrm{conn}}
	  \ar[urr]^{\nabla}
	  \ar[rr]_-{\langle F_{(-)}\wedge \cdots \wedge F_{(-)}\rangle}
	  &&
	  \Omega^{n+1}_{\mathrm{cl}}\;.
	}}
  $$
  \label{BoundaryConditionsFortYMAndChernWeil}
\end{remark}
\begin{example}
  For the canonical binary invariant polynomial $\langle -,-\rangle$ on a 
  simply connected semisimple Lie group $G$ such as $\mathrm{Spin}$
  or $\mathrm{SU}$ (the \emph{Killing form}) a consistent boundary
  condition, as in Remark \ref{BoundaryConditionsFortYMAndChernWeil}, is
  provided by the differential refinement of the first fractional Pontrjagin
  class $\tfrac{1}{2}p_1$ and of the second Chern class $c_2$, respectively, that have been 
constructed in \cite{FSS}:
  $$
    \raisebox{42pt}{
    \xymatrix{
	  && \mathbf{B}^3 U(1)_{\mathrm{conn}}
	  \ar[d]
	  \\
	  \mathbf{B}\mathrm{Spin}_{\mathrm{conn}}
	  \ar[rr]_-{\langle F_{(-)}\wedge F_{(-)}\rangle}
	  \ar[urr]^{\tfrac{1}{2}\widehat{\mathbf{p}_1}}
	  &&
	  \Omega^4_{\mathrm{cl}}\;,
	}}
	\qquad \qquad \qquad
    \raisebox{42pt}{
    \xymatrix{
	  && \mathbf{B}^3 U(1)_{\mathrm{conn}}
	  \ar[d]
	  \\
	  \mathbf{B}\mathrm{SU}_{\mathrm{conn}}
	  \ar[rr]_-{\langle F_{(-)}\wedge F_{(-)}\rangle}
	  \ar[urr]^{\widehat{\mathbf{c}}_2}
	  &&
	  \Omega^4_{\mathrm{cl}}\;.
	}}
  $$  
  Furthermore, for the canonical quaternary invariant polynomial on the 
  smooth String-2-group (see appendix of \cite{FiorenzaSatiSchreiberI} for a review)
  a consistent boundary
  condition as in Remark \ref{BoundaryConditionsFortYMAndChernWeil} is
  provided by the differential refinement of the second fractional Pontrjagin
  class $\tfrac{1}{6}p_2$ that has also been constructed in\cite{FSS}:  
  $$
    \raisebox{42pt}{
    \xymatrix{
	  &&& \mathbf{B}^7 U(1)_{\mathrm{conn}}
	  \ar[d]
	  \\
	  \mathbf{B}\mathrm{String}_{\mathrm{conn}}
	  \ar[rrr]_-{\langle F_{(-)}\wedge F_{(-)}\wedge F_{(-)}\wedge F_{(-)}\rangle}
	  \ar[urrr]^-{\tfrac{1}{6}\widehat{\mathbf{p}}_2}
	  &&&
	  \Omega^8_{\mathrm{cl}}\;.
	}}
  $$
  This describes a 7-dimensional Chern-Simons theory of nonabelian 2-form
  connections \cite{FiorenzaSatiSchreiberI}.
  \label{PontrjaginAndChernAsBoundaryConditions}
\end{example}

\paragraph{$d = n-1$: Topological Chern-Simons boundaries}
\label{TopologicalBoundaries}

We now consider codimension-2 corners of the universal 
topological Yang-Mills theory, hence codimension-1 boundaries
of higher Chern-Simons theories. These turn out to be related to 
Wess-Zumino-Witten like theories. 
Further below in Section \ref{HigherWZWLocalPrequantumFieldTheory}
we discuss that a natural differential variant of this 
type of theories also arises as $\infty$-Chern-Simons theories themselves.

\medskip
For characterizing the data 
assigned by a field theory to such corners, 
we will need to consider the generalization of the following traditional
situation.
\begin{example}
  For $(X,\omega)$ a symplectic manifold, $\omega \in \Omega^2_{\mathrm{cl}}(X)$,
  a submanifold $Y \to X$ is \emph{isotropic} if $\omega|_{Y} = 0$,
  and \emph{Lagrangian} if in addition $\mathrm{dim}(X) = 2 \mathrm{dim}(Y) $.
  If
  $(X,\omega)$ is equipped with a \emph{prequantum bundle}, namely a lift 
  $\nabla$ in 
  $$
    \raisebox{42pt}{
    \xymatrix{
	  & \mathbf{B}U(1)_{\mathrm{conn}}
	  \ar[d]^{F_{(-)}}
	  \\
	  X 
	  \ar[ur]^-\nabla
	  \ar[r]_\omega
	  &
	  \Omega^2_{\mathrm{cl}}\;,
	}}
  $$
  then we may ask not only for a trivialization of the symplectic form $\omega$ but even 
  of the connection $\nabla$ on $Y$:
  $
    \nabla|_{Y} \simeq 0
  $.
  If this exists, then traditionally $Y$ is called
  a \emph{Bohr-Sommerfeld leaf} of $(X,\nabla)$, at least when 
  $Y$ is one leaf of a foliation of $X$ by Lagrangian submanifolds.
\end{example}
Hence we set generally:
\begin{definition}
  Given  a space $X \in \mathbf{H}$ and a connection $\nabla : X \to \mathbf{B}^n U(1)_{\mathrm{conn}}$,
  a \emph{Bohr-Sommerfeld isotropic space} of $(X,\nabla)$ is a diagram of the form
  $$
    \raisebox{20pt}{
    \xymatrix{
	  Y \ar[r]_>{\ }="s" \ar[d] & 0 \ar[d]
	  \\
	  X \ar[r]_-\nabla^<{\ }="t" & \mathbf{B}^n U(1)_{\mathrm{conn}}
	  \ar@{=>} "s"; "t"
	}}
  $$
  in $\mathbf{H}$.
  \label{BohrSommerfeldIsotropicSpace}
\end{definition}
\begin{remark}
  The \emph{universal} Bohr-Sommerfeld isotropic space over 
  $(X,\nabla)$ is the homotopy fiber $\mathrm{fib}(\nabla) \to X$
  of $\nabla$. In a sense this is the ``maximal'' Bohr-Sommerfeld isotropic
  space over $(X,\nabla)$, as every other one factors through this, essentially
  uniquely. Below we see that these are equivalently the universal codimension-2
  corners of higher Chern-Simons theory.
  While the  property of being ``isotropic and maximally so'' is reminiscent
  of Lagrangian submanifolds, it seems unclear what the notion of 
  Lagrangian submanifold should refine to generally in higher prequantum geometry,
  if anything.
\end{remark}
\begin{proposition}
  A corner, Def. \ref{BordWithBoundaryAndCorners}, for 
  the universal topological Yang-Mills theory, Def. \ref{StyM}, 
  from a non-trivial to a trivial boundary condition,
  hence 
  a boundary condition for an $\infty$-Chern-Simons theory, 
  Corollary \ref{AssignmentToCobordismInducedByObjectInSpans},
  $\nabla : \mathbf{Fields}_{\partial} \to \mathbf{B}^n U(1)_{\mathrm{conn}}$,
  is equivalently a \emph{Bohr-Sommerfeld isotropic space} of boundary fields, Def. 
  \ref{BohrSommerfeldIsotropicSpace}, namely 
  a map
  $$
    \mathbf{Fields}_{{ \partial}{ \partial} } \to 
	\mathbf{Fields}_{\partial}
  $$
  such that $F_{\nabla|_{ \partial^2}} = 0$
  and equipped with a homotopy
  $
    \nabla|_{\mathbf{Fields}_{{ \partial}{\partial}}} \simeq 0
  $.
\end{proposition}
\proof
  The boundary condition for $\nabla$ is a correspondence-of-correspondences from
  $$
    \raisebox{20pt}{
    \xymatrix{
	  & \mathbf{Fields}_{ \partial}
	  \ar[dl]
	  \ar[dr]_{\ }="s"
	  \\
	  \ast \ar[dr]_0^{\ }="t" 
	  && 
	  \Omega^{n+1}_{\mathrm{cl}} 
	  \ar[dl]^{\exp\left(\tfrac{i}{\hbar} S_{\mathrm{tYM}} \right)}
	  \\
	  & \flat \mathbf{B}^{n+1}U(1)
	  \ar@{=>}^\nabla "s"; "t"
	}
	}
  $$
  to 
  $$
    \raisebox{42pt}{
    \xymatrix{
	  & \ast
	  \ar[dl]
	  \ar[dr]_{\ }="s"
	  \\
	  \ast \ar[dr]_0^{\ }="t" 
	  && 
	  \Omega^{n+1}_{\mathrm{cl}} \ar[dl]^{\exp\left(\tfrac{i}{\hbar} S_{\mathrm{tYM}} \right)}
	  \\
	  & \flat \mathbf{B}^{n+1}U(1)
	  \ar@{=>}^0 "s"; "t"
	}
	}
	\,.
  $$
  The tip of this correspondence-of-correspondences is a correspondence of the form
  $$
    \raisebox{40pt}{
    \xymatrix{
	  & \mathbf{Fields}_{{{ \partial}}{ \partial}}
	  \ar[dl] 
	  \ar[dr]
	  \\
	  \ast && \mathbf{Fields}_{\partial}
	}
	}
	\,,
  $$
  hence is just a map as on the right. The correspondence-of-correspondences is then filled
  with a second order homotopy between $\nabla$, regarded as a homotopy,
  and the 0-homotopy.
  Unwinding what this means in view of def. \ref{DKPresentationOfU1Coefficients},
  one sees that this homotopy is given by a {\v C}ech-Deligne cochain
  $
    (\cdots, A^{\nabla_{\mathrm{bdr}}}, 0, 0)
  $
  such that
  $$
    D (\cdots, A^{\nabla_{\mathrm{bdr}}}, 0, 0) 
	= (\cdots, A^\nabla, 0)|_{{{ \partial}}{\partial}}
	\,,
  $$
  where 
  $$
    D(\cdots, A^\nabla, 0) = (0, 0, \cdots, 0, \omega)|_{\partial}
	\,.
  $$
\endofproof
\begin{example}[Topological boundary for 3d Chern-Simons theory]
This is in accord with what is proposed as the data on codimension-1 defects
for ordinary Chern-Simons theory on p. 11 of \cite{KapustinSaulina}.
They propose (somewhat implicitly in their text) 
that the boundary connection should be such that $U$-component of
$\langle F_{\nabla}\wedge F_\nabla \rangle$ vanishes at each point of 
$\Sigma$. But for us the fields are 
$A : \Pi(\Sigma) \times U \to \mathbf{B}G_{conn}$, hence are flat along 
$\Sigma$, hence that component vanishes anyway. As a result, the 
proposal in \cite{KapustinSaulina} essentially comes down to asking that
boundary fields $\nabla$ are the maximal solution
to trivializing $\langle F_{\nabla}\wedge F_\nabla \rangle$. 
If we refine this statement from de Rham cocycles to differential cohomology, 
we arrive at the above picture.
\end{example}
\begin{remark}
  Chern-Simons theory is famously related to Wess-Zumino-Witten theory
  in codimension-1. However, WZW theory is not directly a ``topological boundary''
  of Chern-Simons theory. Below in Section 
   \ref{WZWFieldTheories} we show that 
  (the topological sector of)
  pre-quantum WZW theory is a codimension-1 \emph{defect} from $\exp(i S_{\mathrm{CS}})$ to itself,
  via $\exp(i S_{\mathrm{tYM}})$. 
\end{remark}
\begin{remark}
  So the \emph{universal} boundary condition for $\infty$-Chern-Simons 
  local prequantum field theory
  $\nabla : \mathbf{Fields} \to \mathbf{B}^n  U(1)_{\mathrm{conn}}$
  (regarded itself as a boundary condition for its topological Yang-Mills theory)
  is the homotopy fiber of $\nabla$.
\end{remark}
\begin{example}
  Let $\mathfrak{P}$ be 
  Poisson Lie algebroid and 
  $\nabla : \tau_1 \exp(\mathfrak{P}) \to \mathbf{B}^2 (\mathbb{R}/\Gamma)_{\mathrm{conn}}$
  the prequantum 2-bundle of the corresponding 2d Poisson-Chern-Simons prequantum field theory.
  A maximally isotropic sub-Lie algebroid $\mathfrak{C} \hookrightarrow \mathfrak{P}$
  is identified in \cite{CattaneoFelderCoisotropic} with a D-brane for the theory.
  See \cite{hgp} (...)
\end{example}
Further developing Example \ref{PontrjaginAndChernAsBoundaryConditions},
we have by \cite{FSS} the following. 
\begin{example}
  The universal boundary condition for ordinary $\mathrm{Spin}$ Chern-Simons theory
  regarded as a local prequantum field theory
  $\tfrac{1}{2}\widehat{\mathbf{p}}_1 : \mathbf{B}\mathrm{Spin}_{\mathrm{conn}}
  \to \mathbf{B}^3 U(1)_{\mathrm{conn}}$ is the moduli stack of 
  $\mathrm{String}$-2-connections
  $$
    \xymatrix{
      \mathbf{B}\mathrm{String}_{\mathrm{conn}} 
	   \ar[r] 
	   &
	  \mathbf{B}\mathrm{Spin}_{\mathrm{conn}}
	   \ar[r]^{\tfrac{1}{2}\widehat{\mathbf{p}}_1}
	   &
	   \mathbf{B}^3 U(1)_{\mathrm{conn}}
	 }
	 \,.
  $$
  The universal boundary condition for 
  7-dimensional $\mathrm{String}$-Chern-Simons local prequantum field theory
  \cite{FiorenzaSatiSchreiberI}
  $\tfrac{1}{6}\widehat{\mathbf{p}}_2 : \mathbf{B}\mathrm{String}_{\mathrm{conn}}
  \to \mathbf{B}^7 U(1)_{\mathrm{conn}}$ is the moduli stack of 
  $\mathrm{Fivebrane}$-6-connections
  $$
    \xymatrix{
      \mathbf{B}\mathrm{Fivebrane}_{\mathrm{conn}} 
	   \ar[r] 
	   &
	  \mathbf{B}\mathrm{String}_{\mathrm{conn}}
	   \ar[r]^{\tfrac{1}{6}\widehat{\mathbf{p}}_2}
	   &
	   \mathbf{B}^7 U(1)_{\mathrm{conn}}
	 }
	 \,.
  $$
\end{example}
\begin{examples}
  A rich variant of this class of examples of topological 
  prequantum boundary conditions turns out to be the intersection 
  laws of Green-Schwarz type super $p$-branes. We discuss this in details 
  in Section \ref{SuperpBranes} below.
\end{examples}

\paragraph{$d = n-k$: Holonomy defects}

The higher parallel transport of an $n$-connection over a $k$-dimensional
manifold with boundary takes values in sections of the transgression of the 
$n$-bundle to an $(n-k+1)$-bundle over the boundary. Here we discuss
this construction at the level of moduli stacks and then observe that 
it is naturally interpreted in terms of defects for 
higher topological Yang-Mills/higher Chern-Simons theory. 
The Wess-Zumino-Witten defects and the Wilson line/surface defects in the
following sections \ref{WZWFieldTheories} and \ref{WilsonLoopDefects} 
build on this class of examples. 

\medskip

First observe that a particularly simple 
boundary condition for topological Yang-Mills theory is 
to take the connection to be trivial on the boundary via
the following 
$$
  \raisebox{20pt}{
  \xymatrix{
    & \Omega^n \ar[dr]^{d}_{\ }="s"
	\ar[dl]
	\\
	\ast \ar[dr]_-0^{\ }="t" 
	&& 
	\Omega^{n+1}_{\mathrm{cl}} \ar[dl]^{\exp\left(\tfrac{i}{\hbar} S_{\mathrm{tYM}}^{n+1} \right)}
	\\
	& \flat \mathbf{B}^{n+1}U(1)
	\ar@{=>} "s"; "t"
  }
  }
  \;\;\;
  \simeq
  \;\;\;
  \raisebox{30pt}{
  \xymatrix{
    & \Omega^n
	\ar@/^1pc/[ddr]^{d}
	\ar@/_1pc/[ddl]
	\ar@{-->}[d]
	\\
	&\mathbf{B}^n U(1)_{\mathrm{conn}}
	\ar[dl]
	\ar[dr]|{F_{(-)}}
	\\
	\ast 
	\ar[dr]_-0
	&& \Omega^{n+1}_{\mathrm{cl}}\;,
	\ar[dl]^-{\exp\left( \tfrac{i}{\hbar} S_{\mathrm{tYM}}^{n+1} \right)}
	\\
	& \flat \mathbf{B}^{n+1}U(1)
  }}
$$
which corresponds to the inclusion 
$$
  \Omega^n \hookrightarrow \mathbf{B}^n (1)_{\mathrm{conn}}
$$
of globally defined differential $n$-forms regarded as 
connections on trivial $n$-bundles.

\paragraph{$d = n-1$: Wess-Zumino-Witten field theories}
\label{WZWFieldTheories}

We now consider codimension-1 defects for higher Chern-Simons theories,
hence codimension-2 corners for topological Yang-Mills theory.

\begin{remark}
In \cite{FuchsRunkelSchweigert} 2-dimensional (rational) conformal field theories (CFTs) of WZW type
have been constructed and classified by assigning to a punctured marked
surface $\Sigma$ a CFT $n$-point function which is induced by applying the Reshetikhin-Turaev 3d TQFT
functor (hence local quantum Chern-Simons theory) to a 3-d cobordism cobounding the
``double'' of the marked surface. In the case that $\Sigma$ is orientable 
and without boundary, this is the 3d cylinder $\Sigma \times [-1,1]$ over $\Sigma$.
In the language of extended QFTs with defects this 
construction of a 2d theory from a 3d theory may 
be formulated as a realization of 2d WZW theory as a codimension-1 defect
in 3d Chern-Simons theory. The two chiral halves of the WZW theory correspond to the
two ``phases'' of the 3d theory which are separated by the defect $\Sigma$. 
This perspective of \cite{FuchsRunkelSchweigert} has later been amplified in \cite{KapustinSaulinab}. 
\end{remark}

Now let
$$
  \exp(i S_{\mathrm{CS}}) =  
  \mathbf{c} : \mathbf{B}G_{\mathrm{conn}} \to \mathbf{B}^3 U(1)_{\mathrm{conn}}
$$
be a Chern-Simons prequantum field theory.
We have a $G$-principal bundle with connection $(P, \nabla)$ 
over the 1-disk ${D}^1$, i.e the interval, whose boundary is the 
0-sphere, i.e. composed of two points, schematically indicated in the following
diagram
$$
\xymatrix{
(P|_\partial, \nabla|_\partial) \ar[d] \ar[r] & (P, \nabla) \ar[d] 
& & 
\\
\ast \coprod \ast
~\ar@{^{(}->}[r]^-\partial
& 
{D}^1
\ar[r]^-\nabla
&
\mathbf{B}G_{\rm conn} 
\ar[rr]^-{\mathbf{c}}
&&
\mathbf{B}^3U(1)_{\rm conn}\;.
}
$$

\begin{definition}
  The \emph{Wess-Zumino-Witten defect} is the morphism
  $$
    \xymatrix{
      \exp\left( \tfrac{i}{\hbar} S_{\mathrm{CS}} \circ p_1 - iS_{\mathrm{CS}} \circ p_2\right)
	  \ar[r]
      &	  
	  \exp\left(\tfrac{i}{\hbar} S_{\mathrm{tYM}}\right)
	}
  $$
  in $\mathrm{Corr}(\mathbf{H}_{/\mathbf{B}^3 U(1)_{\mathrm{conn}}})$
  given in $\mathbf{H}$ by the 
  the transgression, Def. \ref{Transgression}, of the Chern-Simons connection 
  over the 1-disk
$$
  \raisebox{43pt}{
  \xymatrix{
    && [D^1, \mathbf{B}G_{\mathrm{conn}}]
	\ar[dr]_{\ }="s"
	\ar[dl]_{(-)|_{\partial D^1}}
	\\
	[S^0, \mathbf{B}G_{\mathrm{conn}}]
	\ar@{}[r]|\simeq
	&\mathbf{B}G_{\mathrm{conn}} 
	 \times
	\mathbf{B}G_{\mathrm{conn}}
	\ar[dr]_{\mathbf{c}\circ p_1 - \mathbf{c}\circ p_2}^{\ }="t"
	&&
	\Omega^3_{\mathrm{cl}}
	\ar[dl]
	\\
	&& \mathbf{B}^3 U(1)_{\mathrm{conn}}
	\ar@{=>}|{\exp\left( \tfrac{i}{\hbar} \int_{D^1}\left[D^1, \mathbf{c}\right] \right)} "s"; "t"
  }}
  \,.
$$
 \label{WZWDefect}
\end{definition}
\begin{remark}
\label{cap insert}
  This is a codimension-1 defect of $S^3_{\mathrm{tYM}}$ according to
  Def. \ref{CodimensionkDefect}.
  It may be visualized as a 1-dimensional ``cap'' 
$$
  \xymatrix{
    \ast
	 \ar@/^7pc/[dd]|{\large\mathbf{\ast}}
	 &&&
	\\
	&&& \fbox{\rm insertion, vertex operator, defect, $\cdots$}
	\\
	\ast &&&
  }
$$
  for a single copy of the CS-theory, whose 0-dimensional
  tip carries a tYM-theory.
  Here a closed 3-form is what is responsible for the defect, hence
  the name ``defect field". 
     By duality we may straighten this 
  structure and visualize it schematically as
  $$
    \mathrm{WZW}
	= 
    \left\{
	\raisebox{44pt}{
    \xymatrix{
	  \ar@{-}[dd] & \mathrm{CS}^l
	  \\
	  {\ast} & \mathrm{tYM}
	  \\
	  & \mathrm{CS}^r\;.
	}
	}
	\right.
  $$
\end{remark}
This defect becomes a plain boundary for the tYM-theory when the
left end is attached to a boundary that couples the left with the
right part of the CS-theory:
\begin{definition}
The \emph{Wess-Zumino-Witten codimension-2} corner in 4d topological Yang-Mills theory
is the boundary 
$$
  \xymatrix{
    \mathbb{I} \ar[r] \ar[d] & \mathbb{I} \ar[d]
    \\
    \mathbb{I}
	\ar[r]_{S^3_{\mathrm{tYM}}}
	&
	S^4_{\mathrm{tYM}}
  }
$$
of the boundary
3d tYM theory given as a diagram in $\mathbf{H}$
by the composite
$$
  \raisebox{33pt}{
  \xymatrix{
    & G
	\ar@/^1pc/[ddr]^{}
	\ar@/_1pc/[ddl]
	\ar@{-->}[d]|{\exp(i S_{\mathrm{WZW}})}
	\\
	&\mathbf{B}^{2}U(1)_{\mathrm{conn}}
	\ar[dl]
	\ar[dr]^{F_{(-)}}_{\ }="s"
	\\
	\ast \ar[dr]_-0^{\ }="t" 
	&& \Omega^{3}_{\mathrm{cl}} 
	\ar[dl]^-{ \exp\left( \tfrac{i}{\hbar} S_{\mathrm{tYM}}^3 \right)}
	\\
	& \mathbf{B}^{3}U(1)_{\mathrm{conn}}
	\ar@{=>}^{\nabla_{\mathrm{ChS}}} "s"; "t"
  }}
  \;\; \qquad:= \qquad
  \;\;
  \raisebox{30pt}{
  \xymatrix{
    && G
	\ar[dl]
	\ar[dr]
	\\
    & \ast 
	 \ar[dl]
	 \ar[dr]
	&& [D^1, \mathbf{B}G_{\mathrm{conn}}]
	\ar[dl]|{(-)|_{\partial D^1}}
	\ar[dr]_{\ }="s"
    \\
    \ast \ar[drr]_-0 && [S^0, \mathbf{B}G_{\mathrm{conn}}]
	\ar[d]|>>>>{\mathbf{c} \circ p_1 - \mathbf{c}\circ p_2}^{\ }="t"
	&&
	\Omega^3_{\mathrm{cl}}\;.
	\ar[dll]^-{ \exp\left( \tfrac{i}{\hbar} S_{\mathrm{tYM}}^3 \right)}
	\\
	&& \mathbf{B}^3 U(1)_{\mathrm{conn}}
	\ar@{=>}|<<<<<<<<<<<<<{\exp\left(\tfrac{i}{\hbar} \int_{D^1}\left(-\right)\right) } "s"; "t"
  }
  }
$$
Here the bottom right square is that of Def. \ref{WZWDefect},
the bottom left square is filled with the evident equivalence,
and the map $G \to [S^1, \mathbf{B}G_{\mathrm{conn}}]$
in the top square
is given by resolving the simply connected Lie group $G$ by
its based path space
$P_* G$, regarded as a diffeological space. Then each path uniquely 
arises as the parallel transport 
of a $G$-principal connection on the interval and two paths with the same 
endpoint have a unique gauge transformation relating them.
 \label{WZWBoundaryoftYM}
\end{definition}
\begin{remark}
It is important to highlight that 
$G$ here is the differential concretification of the pullback in the middle,
as discussed in \cite{hgp}. 
\end{remark}
\begin{proposition}
  The morphism
  $$
    \exp\left(\tfrac{i}{\hbar} S_{\mathrm{WZW}} \right) : G \to \mathbf{B}^2 U(1)_{\mathrm{conn}}
  $$
  from Def. \ref{WZWBoundaryoftYM} is the WZW-2-connection
  (the ``WZW gerbe''/``WZW B-field'').
\end{proposition}
\proof
This follows along the lines of the discussion in \cite{FiorenzaSatiSchreiberCS},
where it was found that the composite
$$
  \xymatrix{
    G \ar[r] & [S^1, \mathbf{B}G_{\mathrm{conn}}]
	\ar[r]^-{[S^1, \mathbf{c}]}
	&
	[S^1, \mathbf{B}^3 U(1)_{\mathrm{conn}}]
	\ar[rrr]^-{\exp\left(\tfrac{i}{\hbar} \int_{S^1}\left(-\right)\right)}
	&&&
	\mathbf{B}^2 U(1)_{\mathrm{conn}}
  }
$$
is the (topological part of) the localized  WZW action.
\endofproof

\paragraph{$d = n-2$: Wilson loop/Wilson surface field theories}
\label{WilsonLoopDefects}
\index{Chern-Simons functionals!Wilson loops}

In \ref{WithWilsonLoops} (\cite{FiorenzaSatiSchreiberCS}) a description of how 
Wilson loop line defects in 3d Chern-Simons theory is given by 
the following data.
Let $\lambda \in \mathfrak{g}$ be a regular weight, corresponding via
Borel-Weil-Bott to the irreducible representation which labels the Wilson loop.
Then the stabilizer subgroup $G_\lambda \hookrightarrow G$ of $\lambda$
under the adjoint action is a maximal torus $G_\lambda \simeq T \hookrightarrow G$
and $G/G_\lambda \simeq \mathcal{O}_\lambda$ is the coadjoint orbit. 
Integrality of $\lambda$ means that pairing with $\lambda$ 
constitutes a morphism of moduli stacks of the form
$$
  S_{W} :  
  \xymatrix{
    \Omega^1(-,\mathfrak{g})//T 
    \ar[rr]^-{\mathbf{\langle} \lambda,-\mathbf{\rangle}}
    && \mathbf{B}U(1)_{\mathrm{conn}}
  }
  \,.
$$
This is the local Lagrangian/the prequantum bundle of the Wilson loop theory
in that there is a diagram
$$
  \xymatrix{
    \mathcal{O}_\lambda \ar[r]^-{\mathrm{fib}(\mathbf{J})}
	\ar[d]
	&
	\Omega^1(-,\mathfrak{g})//T 
	\ar[rr]^{\mathbf{\langle} \lambda,-\mathbf{\rangle}} 
	\ar[d]^{\mathbf{J}}
	&&
	\mathbf{B} U(1)_{\mathrm{conn}}
	\\
	\ast \ar[r] & \mathbf{B}G_{\mathrm{conn}}
	\ar[rr]^-{\mathbf{c}} 
	&& \mathbf{B}^3 U(1)_{\mathrm{conn}}\;,
  }
$$
whose top composite is the Kirillov prequantum bundle on the coadjoint orbit 
and which is such that a Chern-Simons + Wilson loop field configuration $(A,g)$ is a diagram
$$
  \xymatrix{
    S^1 \ar[rr]^-{A|_{S^1}^g}_>{\ }="s" \ar@{^{(}->}[d] && \Omega^1(-,\mathfrak{g})//T
	\ar[d]^{\mathbf{J}}
	\\
	\Sigma_3
	\ar[rr]_A^<{\ }="t"
	&&
	\mathbf{B}G_{\mathrm{conn}}
	\ar@{=>}^g "s"; "t"
  }
$$
and the corresponding action functional is the product of 
$\mathbf{\langle} \lambda, -\mathbf{\rangle}$ transgressed over $S^1$ and 
$\mathbf{c}$ transgressed over $\Sigma_3$.

\medskip
We now interpret this formally as a codimension-2 defect of Chern-Simons theory
analogous to the WZW defect,
hence as a codimension-3 structure in the ambient tYM theory.

\begin{definition}
Let $\phi : D^2 \to S^2$ be a smooth function which on the interior 
  of $S^2$ is a diffeomorphism on $S^2 -\{\ast\}$.
  The \emph{universal Wilson line/Wilson surface defect} is, as a diagram in 
  $\mathbf{H}$, the transgression diagram, Def. \ref{Transgression}
$$
  \xymatrix{
    & [S^2, \mathbf{B}G_{\mathrm{conn}}]
	\ar[dl]
	\ar[dr]^{~~[\phi, \mathbf{B}G_{\mathrm{conn}}]}
	\ar[dd]|{[\phi|_{S^1}, \mathbf{B}G_{\mathrm{conn}}]}
    \\
    [\Pi(S^1), \mathbf{B}G_{\mathrm{conn}}]
	\ar[dr]
    & & [D^2, \mathbf{B}G_{\mathrm{conn}}]
	\ar[dl]|{(-)|_{\partial D^2}}
	\ar[dr]_{\ }="s"
    \\
    & [S^1, \mathbf{B}G_{\mathrm{conn}}]
	\ar[dr]_{\hspace{-10mm}\exp\left(\tfrac{i}{\hbar} \int_{S^1}\left[S^1, \mathbf{c}\right]\right)}^{\ }="t"
	&&
	\Omega^2_{\mathrm{cl}}\;.
	\ar[dl]
	\\
	& & \mathbf{B}^2 U(1)_{\mathrm{conn}}
	\ar@{=>}|{\exp\left(\tfrac{i}{\hbar} \int_{D^2} \left[D^2, \mathbf{c}\right]\right))} "s"; "t"
  }
$$
 \label{TerminalWilsonDefect}
\end{definition}
\begin{remark}
 This is a codimension-2 defect according to Def. \ref{CodimensionkDefect}. 
 It may be visualized as a 2-dimensional CS theory cap
$$
  \xymatrix{
    \ast
	 \ar@{-}@/^2.4pc/[dd]
	 \ar@{--}@/_2.1pc/[dd]
	 \ar@/^7pc/[dd]|{\large\mathbf{\ast}}
	\\
	\\
	\ast
  }
$$
with a tYM-theory sitting at the very tip. By duality there is the
corresponding straightened picture
$$
  \xymatrix@R=4pt@C=2pt{
    && && \ast \ar@{--}[dddd]
    \\
	\\
    \ast \ar@{-}[rrrruu] \ar@{-}[dddd]
	\\
	&& {\ast}
	\\
	&& && \ast
	\\
	\\
	\ast \ar@{-}[rrrruu]
  }
  \qquad  \qquad
  \xymatrix@R=4pt@C=2pt{
    && && \mathrm{CS} 
    \\
	\\
    \mathrm{CS} 
	\\
	&& \mathrm{tYM}
	\\
	&& && \mathrm{CS}
	\\
	\\
	\mathrm{CS} 
  }  
$$
\end{remark}
We now define a defect that factors through the universal Wilson defect 
of Def. \ref{TerminalWilsonDefect} and reproduces the 
traditional Wilson line action functional. 
To that end, let $\nabla_{S^2} : S^2 \to \mathbf{B}T_{\mathrm{conn}}$
be a $T$-principal connection on the 2-sphere, where $T$ is the 
maximal torus of $G$. We may identify the integral of its curvature 2-form
over the sphere with the weight $\lambda$,
$$
  \lambda = \int_{S^2} F_{\nabla_{S^2}}\;.
$$
Then consider the morphism
$$
 {p_1^\ast \nabla_{S^1} +  p_2^\ast(-)}
  : 
  \xymatrix{
    \Omega^1(-,\mathfrak{g})//T
    \ar[r]
	&
    [S^2, \mathbf{B}G_{\mathrm{conn}}]
  }
$$
in $\mathbf{H}$
which over a test manifold $U \in \mathrm{CartSp}$ sends a connection 1-form
$A \in \Omega^1(U,\mathfrak{g})$ to
$$
    p_1^\ast \nabla_{S^2} + p_2^\ast A \in \mathbf{H}(S^2 \times U, \mathbf{B}G_{\mathrm{conn}})
	\,.
$$
This is indeed a homomorphism since $T$ is abelian.
\begin{proposition}
We have the following
$$ 
  \raisebox{30pt}{
  \xymatrix{
    & \Omega^1(-,\mathfrak{g})//T
	\ar@/^1pc/[ddr]
	\ar@/_1pc/[ddl]
	\ar@{-->}[d]|{\langle \lambda, -\rangle}
	\\
	&\mathbf{B}U(1)_{\mathrm{conn}}
	\ar[dr]|{F_{(-)}}_{\ }="s"
	\ar[dl]
	\\
	\ast \ar[dr]_0^{\ }="t" && \Omega^2_{\mathrm{cl}} 
	\ar[dl]^{\exp\left(\tfrac{i}{\hbar} S_{\mathrm{tYM}}^2 \right)}
	\\
	& \mathbf{B}^2 U(1)_{\mathrm{conn}}
	\ar@{=>}^{\nabla_{\mathrm{CS}}} "s"; "t"
  }}
  \;\; \quad
  \simeq \quad
  \;\;
  \raisebox{30pt}{
  \xymatrix{
    & \Omega^1(-,\mathfrak{g})//T
	\ar[dr]^{~~~p_1^\ast \nabla_{S^2} +  p_2^\ast(-)}
	\ar[dl]
    \\
    \ast \ar[dr]
	& & [S^2, \mathbf{B}G_{\mathrm{conn}}]
	\ar[dl]|{(-)|_{\partial S^2}}
	\ar[dr]_{\ }="s"
    \\
    &[\emptyset, \mathbf{B}G_{\mathrm{conn}}]
	\ar[dr]_{
	  \hspace{-10mm}\exp\left(
	    \tfrac{i}{\hbar} \int_{\emptyset}\left[\emptyset, \mathbf{c}\right]
	   \right)
	}^{\ }="t"
	&&
	\Omega^2_{\mathrm{cl}}\;.
	\ar[dl]
	\\
	& 	& \mathbf{B}^2 U(1)_{\mathrm{conn}}
	\ar@{=>}|{
	    \exp\left( \tfrac{i}{\hbar} \int_{S^2} \left[S^2, -\right]\right)
	 } "s"; "t"
  }
  }
$$
\end{proposition}
\proof
  By construction and since $T$ is abelian, the component of the 
  Chern-Simons form of $p_1^\ast \nabla_{S^2} + p_2^\ast A$ with 
  two legs along $S^2$ is proportional to $\langle F_{\nabla_{S^2}} \wedge A\rangle$.
  Hence its fiber integral over $S^2 \times U \to U$ is 
  $$
    \int_{S^2} \langle F_{\nabla_{S^2}} \wedge A\rangle
	= 
	\langle \lambda, A\rangle
	\,.
  $$
\endofproof
Therefore in conclusion we find that we can axiomatize Wilson loops 
in 3d Chern-Simons theory be the following
defect structure.
\begin{definition}
  The {\it Wilson line defect} is
$$
  \xymatrix{
    & \Omega^1(-,\mathfrak{g})/\!/T
	\ar[d]^{p_1^\ast \nabla_{S^2} + p_2^\ast(-)}
    \\
    & [S^2, \mathbf{B}G_{\mathrm{conn}}]
	\ar[dl]
	\ar[dr]^{~~[\phi, \mathbf{B}G_{\mathrm{conn}}]}
    \\
    [\Pi(S^1), \mathbf{B}G_{\mathrm{conn}}]
	\ar[dr]
    & & [D^2, \mathbf{B}G_{\mathrm{conn}}]
	\ar[dl]|	{(-)|_{\partial D^2}}
	\ar[dr]_{\ }="s"
    \\
    & [S^1, \mathbf{B}G_{\mathrm{conn}}]
	\ar[dr]_{\hspace{-12mm}\exp\left(\tfrac{i}{\hbar} \int_{S^1}\left[S^1, \mathbf{c}\right]\right)}^{\ }="t"
	&&
	\Omega^2_{\mathrm{cl}}\;.
	\ar[dl]
	\\
	& & \mathbf{B}^2 U(1)_{\mathrm{conn}}
	\ar@{=>}|{\exp\left(\tfrac{i}{\hbar} \int_{D^2} \left[D^2, \mathbf{c}\right]\right)} "s"; "t"
  }
$$\end{definition}

\newpage

\section{Outlook: Motivic quantization of local prequantum field theory}
\label{MotivicQuantizationApplications} 
\index{quantization}
\index{prequantum field theory!quantization}

Above in \ref{LocalPrequantumFieldTheories} we discussed how local Lagrangian topological
prequantum field theory with boundaries and defects is axiomatized in terms of 
local action functionals/higher prequantum bundles 
$\exp\left( \tfrac{i}{\hbar} S\right )$ on moduli stacks $\mathbf{Fields}$ 
corresponding to diagrams of symmetric monoidal $(\infty,n)$-categories of the form
$$
  \xymatrix{
      {\mathrm{Bord}^{\mathrm{sing}}_n}^{\otimes}
      \ar[rr]^-{\exp\left( \tfrac{i}{\hbar} S\right)}
      \ar[drr]_{\mathbf{Fields}}
      &&
      \mathrm{Corr}_n(\mathbf{H}_{/\mathbf{Phases}})^\otimes
      \ar[d]
      \\
      && \mathrm{Corr}_n(\mathbf{H})^\otimes
   }
   \,,
$$
where $\mathbf{H}$ is the ambient cohesive $\infty$-topos.

By folk lore, a prequantum field theory 
with action functional $\exp\left( \frac{i}{\hbar} S \right)$
on $\mathbf{Fields}$
is supposed to induce a genuine quantum field theory given by 
a monoidal functor to some $E$-\emph{linear} $(\infty,n)$-category $E \mathrm{Mod}_n$
(for some ground ring $E$)
by a process that integrates the contributions of 
$\exp\left(\tfrac{i}{S}(\phi)\right) \in \mathbf{Phases}$ over all 
$\phi \in \mathbf{Fields}$ as $E$-modules, after a specified embedding
$\mathbf{Phases} \hookrightarrow E\mathrm{Mod}$. This \emph{path integral}
\cite{FeynmanHibbs, ZJ}
is traditionally denoted by the symbols on the left of
$$
  \underset{\phi \in \mathbf{Fields}}{\int}
   \exp\left(
      \tfrac{i}{\hbar}
      S(\phi)
   \right)
   D\phi
   \;:\;
   {\mathrm{Bord}_n^{\mathrm{sing}}}^\otimes
   \longrightarrow
   E \mathrm{Mod}_n^\otimes
   \,,
$$
where $D \phi$ is meant to suggest an integration measure on $\mathbf{Fields}$.

We now indicate how to naturally and usefully make formal sense of this 
inside the tangent cohesive $\infty$-topos $T \mathbf{H}$, \ref{DiagramsOfCohesiveGroupoids}
by fiber integration in twisted stable cohomology, \ref{TangentCohesionCohomology}.
Then we indicate how in examples this general process reproduces traditional
geometric quantization, \ref{GeometricQuantizationInIntroduction}, 
as the boundary quantum field theory of the 
non-perturbative local prequantum 
2d Poisson-Chern-Simons theory of \ref{DeformationQuantizationInQuantumMechanics}.
We also indicate how higher analogs of this serve to quantize higher 
WZW-type $p$-brane $\sigma$-models as in \ref{WZWApplications}
as boundary theories of the higher prequantum Chern-Simons theories in \ref{ChernSimonsFunctional}.

The material outlined here is due to \cite{Nuiten, MotQuant}. 

\medskip

\begin{itemize}
  \item \ref{CohomologicalQuantizationOfCorrespondences} --
    Cohomological quantization of correspondences in a cohesive slice
  \item \ref{QuantumParticleAtEndOfTheString} -- The quantum particle at the boundary of the string
  \item \ref{QuantumStringAtBoundaryOfMembrane} -- The quantum string at the boundary of the membrane
\end{itemize}

\subsubsection{Cohomological quantization of correspondences in a cohesive slice}
\label{CohomologicalQuantizationOfCorrespondences}

To begin the discussion at the absolute fundamentals, 
notice that the two hallmarks of quantum theory are (e.g. \cite{Dirac})
\begin{enumerate}
  \item the superposition principle
  \item quantum interference
  \,.
\end{enumerate}
These say that 1. quantum phases may be additively combined such that 
2. there are relations that make combinations sum to zero.
This informal statement is formally captured by the passage from the circle group $U(1)$ of phases
in which action functionals $\exp\left(\tfrac{i}{\hbar}S\right)$ traditionally take 
values, as discussed in \ref{ActionFunctionalsAndCorrespondences},
first to the group ring $\mathbb{Z}[U(1)]$,
which is the universal context for adding phases, and then second to 
the quotient ring defined by the relations that regard addition of phases
inside the complex numbers:
$$
  \xymatrix{
    U(1)
    \ar@{^{(}->}[rr]^-{\mbox{\scriptsize superposition}}
    &&
    \mathbb{Z}[U(1)]
    \ar@{->>}[rr]^-{\mbox{\scriptsize interference}}
    &&
    \mathbb{C}
  }
  \,.
$$
What is supposed to be integrated by the path integral is
the thus linearized action functional $\rho \exp\left(\tfrac{i}{\hbar}S\right)$.

Observe now that the free group ring construction $\mathbb{Z}[-]$ is the
left adjoint in an adjunction
$$
  \xymatrix{
    \mathrm{CRing}
    \ar@{<-}@<+3pt>[rr]^{\mathbb{Z}[-]}
    \ar@<-3pt>[rr]_{\mathrm{GL}_1}
    &&
    \mathrm{AbGrp}
  }
$$
between commutative rings and abelian groups, whose right adjoint is the
functor that sends a ring to its group of units. 
The corresponding adjunct of the map $\mbox{interference} \circ \mbox{superposition}$
above is the group homomorphims
$$
  \rho
  :
  U(1)
  \longrightarrow
  \mathrm{GL}_1(\mathbb{C})
  \,,
$$
which canonically embeds $U(1)$ into the group of units of the 
complex numbers.
Being a Lie group homomorphism, this deloops to a morphism of moduli stacks
$$
  \mathbf{B}\rho
  :
  \xymatrix{
    \mathbf{B}U(1) \ar[r] & \mathbf{B}GL_1(\mathbb{C})
    \ar[r]^\simeq &  \mathbb{C}\mathbf{Line}
    \ar@^{^{(}->}[r]
    &
    \mathbb{C}\mathbf{Mod}
  }
  \,,
$$
where on the right we observe that $\mathbf{B}\mathrm{GL}_1(\mathbb{C})$
is equivalently the moduli stack of smooth complex line bundles which
sits inside the moduli stack of smooth complex vector bundles.

Using this, we can canonically turn a action functional as in 
\ref{ActionFunctionalsAndCorrespondences}
into the corresponding integral kernel, 
$$
  \xymatrix{
    & \mathbf{Fields}_{\mathrm{traj}}
    \ar[dl]
    \ar[dr]_{\ }="s"
    \\
    \mathbf{Fields}_{\mathrm{in}} \ar[dr]|{\mathbf{L}_{\mathrm{in}}}^{\ }="t" 
    \ar@/_1pc/[dddr]_{\chi_{\mathrm{in}}}
    && 
    \mathbf{Fields}_{\mathrm{out}} \ar[dl]|{\mathbf{L}_{\mathrm{out}}}
    \ar@/^1pc/[dddl]^{\chi_{\mathrm{out}}}
    \\
    & \mathbf{B}U(1)_{\mathrm{conn}}
    \ar[d]^{\mathbf{B}\rho}
    \\
    & \mathbf{B}\mathrm{GL}_1(\mathbb{C})
    \ar@{^{(}->}[d]
    \\
    & \mathbb{C}\mathrm{Mod}
    \ar@{=>}|{\exp\left(\tfrac{i}{\hbar} S_{\mathrm{traj}} \right)} "s"; "t"
  }
$$
regarded as a gauge transformation between an 
incoming prequantum bundle $\chi_{\mathrm{in}}$ and an outgoing prequantum bundle
$\chi_{\mathrm{out}}$. Write then
$\mathbb{C}^{\chi_{\mathrm{in}}}(\mathbf{Fields}_{\mathrm{in}})$
and
$\mathbb{C}^{\chi_{\mathrm{out}}}(\mathbf{Fields}_{\mathrm{out}})$
for the space of sections of these prequantum line bundles,
then the path integral over 
$\exp\left(\tfrac{i}{\hbar} S_{\mathrm{traj}} \right)$ is supposed to produce 
a linear map
$$
  \mathbb{C}^{\chi_{\mathrm{in}}}(\mathbf{Fields}_{\mathrm{in}})
  \longrightarrow
  \mathbb{C}^{\chi_{\mathrm{out}}}(\mathbf{Fields}_{\mathrm{out}})
  \,,
$$
which would be the \emph{quantum propagator} that takes incoming 
quantum states/wave functions to outgoing quantum states.

This storty is quantum \emph{mechanics}, hence 1-dimensional quantum field theory.
By the discussion of higher dimensional 
local prequantum field theory in  \ref{LocalPrequantumFieldTheories},
for an $n$-dimensional local prequantum field theory the local 
action functional in full codimension is given by a 
Lagrangian $\mathbf{L}$ which is a map of the form
$$
  \mathbf{L}
  :
  \mathbf{Fields} \longrightarrow \mathbf{B}^n U(1)_{\mathrm{conn}}
  \,.
$$
By direct analogy with the above discussion, we are therefore led to consider
the following linearization of such local Lagrangians.

First observe, based on \cite{AndoBlumbergGepner}, that the above adjunction 
$(\mathbb{Z}[-] \dashv \mathrm{GL}_1)$ generalizes to an $\infty$-adjunction
between
$E_\infty$-rings in the $\infty$-topos $\mathbf{\mathbf{H}}$
(see \cite{LurieAlgebra})
and abelian $\infty$-group objects in $\mathbf{H}$, def. \ref{AbelianInfinityGroup}:
$$
  \xymatrix{
    \mathrm{CRing}(\mathbf{H})
    \ar@{<-}@<+3pt>[rr]^{\mathbb{S}[-]}
    \ar@<-3pt>[rr]_{\mathrm{GL}_1}
    &&
    \mathrm{AbGrp}(\mathbf{H})
  }
  \,.
$$
Therefore for an $n$-dimensional field theory we are to look for 
a quotient $E_\infty$-ring $E$ of the $\infty$-group
$E_\infty$-ring $\mathbb{S}[\mathbf{B}^{n-1}U(1)]$ on the circle $n$-group
(def. \ref{Circlengroup})
$$
  \xymatrix{
    \mathbf{B}^{n-1}U(1)
    \ar@{^{(}->}[rr]^-{\mbox{\scriptsize superposition}}
    &&
    \mathbb{S}[\mathbf{B}U(1)]
    \ar@{->>}[rr]^-{\mbox{\scriptsize interference}}
    &&
    \mathbf{E}
  }
  \,.
$$
For instance for $n = 2$ there is a natural choice of quotient:
by a smooth refinement of Snaith's theorem \cite{Snaith} 
we can take\footnote{I am grateful to Thomas Nikolaus for discussion of this point.}
$$
  \mathbf{KU} := \mathbb{S}[\mathbf{B}U(1)][\mathbf{\beta}^{-1}]
$$
to be the localization of the smooth group $\infty$-ring of the
circle 2-group at the smooth Hopf bundle Bott element. Since
these are $\infty$-colimit constructions, they are preserved by 
geometric realization, \ref{SmoothStrucHomotopy}
and Snaith's theorem then says that this is the traditional complex
K-theory spectrum
$$
  \Pi(\mathbf{KU}) \simeq \mathbb{S}[K(\mathbb{Z},2)][\beta^{-1}]
  \simeq
  \mathrm{KU}
  \,.
$$
We learn from this that 2-dimensional local pre-quantum field theory
should have a natural quantization in K-theory, and indeed it does,
as we describe in \ref{QuantumParticleAtEndOfTheString} below.

In this fashion, for suitable choices of \emph{higher superposition principles}
$$
  \rho : \mathbf{B}^{n-1}U(1) \longrightarrow \mathrm{GL}_1(E)
$$
we may $E$-linearize local action functionals to higher integral kernels given
by pasting composites of the form
$$
  \xymatrix{
    & \mathbf{Fields}_{\mathrm{traj}}
    \ar[dl]_{i_{\mathrm{in}}}
    \ar[dr]^{i_{\mathrm{out}}}_{\ }="s"
    \\
    \mathbf{Fields}_{\mathrm{in}} \ar[dr]|{\mathbf{L}_{\mathrm{in}}}^{\ }="t" 
    \ar@/_1pc/[dddr]_{\chi_{\mathrm{in}}}
    && 
    \mathbf{Fields}_{\mathrm{out}} \ar[dl]|{\mathbf{L}_{\mathrm{out}}}
    \ar@/^1pc/[dddl]^{\chi_{\mathrm{out}}}
    \\
    & \mathbf{B}^n U(1)_{\mathrm{conn}}
    \ar[d]^{\mathbf{B}\rho}
    \\
    & \mathbf{B}\mathrm{GL}_1(E)
    \ar@{^{(}->}[d]
    \\
    & E\mathrm{Mod}
    \ar@{=>}|{\exp\left(\tfrac{i}{\hbar} S_{\mathrm{traj}} \right)} "s"; "t"
  }
  \,.
$$
Notice that such a diagram is a correspondence in $\mathbf{H}$ whose
correspondence space $\mathbf{Fields}_{\mathrm{traj}}$ is equipped with a
cocycle in $(\chi_{\mathrm{in}}, \chi_{\mathrm{out}})$-twisted
bivariant $E$-cohomology theory. This is the broad structure of 
representatives of \emph{pure motives} (see e.g. 
\cite{Sujatha}), here generalized to cohesive 
higher moduli stacks and twisted bivariant generalized cohomology.

Forming the $E$-modules of sections of these higher prequantum $E$-line bundles precisely means
forming $\chi$-twisted $E$-cohomology spectra as in \ref{TangentCohesionCohomology}.
Then a correspondence as above yields a co-correspondence of $E$-module spectra
of the form
$$
  E^{\chi_{\mathrm{in}}}(\mathbf{Fields}_{\mathrm{in}})
  \stackrel{i_{\mathrm{in}}^\ast}{\longrightarrow}
  E^{i_{\mathrm{in}}^\ast \chi_{\mathrm{in}}}(\mathbf{Fields}_{\mathrm{traj}})
  \simeq
  E^{i_{\mathrm{out}}^\ast \chi_{\mathrm{out}}}(\mathbf{Fields}_{\mathrm{traj}})
  \stackrel{i_{\mathrm{out}}^\ast }{\longleftarrow}(\mathbf{Fields}_{\mathrm{out}})
  E^{\chi_{\mathrm{out}}}(\mathbf{Fields}_{\mathrm{out}})  
  \,.
$$
Now \emph{integration} in this context means to dualize the morphism on the
right (with respect to the tensor monoidal structure on $E\mathrm{Mod}$) 
and to choose an equivalence of the $E$-modules with their (fiberwise) duals,
this is the choice of \emph{orientation in twisted $E$-cohomology}.
The obstructions for such an orienation to exist by itself and
moreover to exist in a consistent way that is compatible composition 
of correspondence (as indicated in \cite{FHMConsistent}) are the 
\emph{quantum anomalies}.
\index{anomaly cancellation!in cohomological quantization}
If the orientation exists and has been consistently 
chosen, then we can form the \emph{twisted Umkehr map}
\cite{ABGUmkehr} $(i_{\mathrm{out}})_!$, see \cite{Nuiten} for a review.
This finally turns the above integral kernel into the 
\emph{higher quantum propagator}
$$
  \xymatrix{
    E^{\chi_{\mathrm{in}}}(\mathbf{Fields}_{\mathrm{in}})
     \ar[rrr]^-{ (i_{\mathrm{out}})_! \circ i_{\mathrm{in}}^\ast }
    &&&
    E^{\chi_{\mathrm{out}}}(\mathbf{Fields}_{\mathrm{out}})  
   }
  \,.
$$

In our first example in \ref{QuantumParticleAtEndOfTheString} below,
in the case of $n = 2$ with $E = \mathrm{KU}$ as above, this
cohomological quantization is presented over moduli stacks which 
are Lie groupoids by the KK-theoretic constructions in 
\cite{BMRS}, via \cite{TXLG04}. 
In \cite{KKMotive} it is shown that this indeed canonically
maps into noncommutative motives (see e.g \cite{BGT}).

\medskip

In summary, the quantization of of pre-quantum correspondences in 
the slice of a cohesive $\infty$-topos via fiber integration in 
twisted stable cohomology corresponds to lifts of the original 
pre-quantum field theory as shown in the following diagram:
$$
  \xymatrix{
      && && \mathrm{Corr}^{\mathrm{or}}_n(\mathbf{H}_{/\mathbf{B}\mathrm{GL}_1(\mathbf{E})})^\otimes
        \ar[d]
      \ar[rr]|-{\underset{(-)}{\int} (-)}
      &&
      E \mathrm{Mod}_n
      \\
      {\mathrm{Bord}^{\mathrm{sing}}_n}^{\otimes}
      \ar[rr]|-{\exp\left( \tfrac{i}{\hbar} S\right)}
      \ar[drr]_{\mathbf{Fields}}
      \ar[urrrr]|-{\exp\left( \tfrac{i}{\hbar} S\right) D(-)}
      \ar@/^3.5pc/[urrrrrr]^{\underset{\phi \in \mathbf{Fields}}{\int} 
         \exp\left(\tfrac{i}{\hbar} S(\phi)\right) D\phi}
      &&
      \mathrm{Corr}_n(\mathbf{H}_{/\mathbf{Phases}})^\otimes
      \ar[d]
      \ar[rr]|{\rho}
      &&
      \mathrm{Corr}_n(\mathbf{H}_{/\mathbf{B}\mathrm{GL}_1(\mathbf{E})})^\otimes      
      \\
      && \mathrm{Corr}_n(\mathbf{H})^\otimes
   }
   \,,
$$
Here
\begin{itemize}
\item $\mathbf{Fields}$ is the higher moduli stack of pre-quantum fields;
\item $\exp\left(\tfrac{i}{\hbar}S \right)$ is the specified local action functional 
on $\mathbf{Fields}$, defining the given pre-quantum field theory;
\item $\rho$ is the chosen higher superposition principle, linearizing in $E$-cohomology;
\item $\exp\left(\tfrac{i}{\hbar}S \right) D(-)$ is a lift of the local action functional
 to consistently twisted $E$-oriented correspondences, hence is a choice of 
 \emph{cohomological path integral measure} on $\mathbf{Fields}$;
\item $\underset{\phi \in \mathbf{Fields}}{\int} \exp\left(\tfrac{i}{\hbar}S(\phi) \right) D(\phi)$
 is the composition of the latter the previous item with the pull-push operation,
 this is the cohomological realization of the \emph{path integral}.
\end{itemize}

This quantization process is particularly interesting for the boundary prequantum field theories
discussed in \ref{BoundaryFieldTheory}, where it yields quantization of 
\emph{geometric} (non-topological) field theories as the ``holographic'' 
boundaries of topological field theories in one dimension higher.

\subsubsection{The quantum particle at the boundary of the string}
\label{QuantumParticleAtEndOfTheString}

We indicate how traditional geometric quantization of symplectic
manifolds (e.g. \cite{BrylinskiLoop}) arises as the motivic
quantization of the canonical boundary field theory to the 
non-perturbative 2d Poisson-Chern-Simons field theory of 
\ref{DeformationQuantizationInQuantumMechanics}. 
Then we observe that the same process more generally yields
a notion of geometric quantization of Poisson manifolds.
Given a compact Lie group $G$ we find a pre-quantum defect
between two Poisson-Chern-Simons boundary field theories
whose cohomological quantization yields Kirillov's orbit method
in the K-theoretic incarnation given by Freed-Hopkins-Teleman.

\paragraph{Holographic geometric quantization of Poisson manifolds}
\label{HolographicGeometricQuantizationOfPoisson}
\index{holography!1d mechanics/2d Poisson-CS}

Given a Poisson manifold $(X,\pi)$, 
by the general construction of \ref{InfinCSAKSZ},
as described in \ref{DeformationQuantizationInQuantumMechanics}, there
is a 2d Chern-Simons field theory induced by it, whose moduli stack of 
instanton sector of fields is the \emph{symplectic groupoid}
$\mathrm{SymplGrpd}(X,\pi)$, on which the local action functional
induces a $\mathbf{B}U(1)$-principal 2-bundle
$$
  \chi : \mathrm{SymplGrpd}(X,\pi) \longrightarrow \mathbf{B}^2 U(1)
  \,.
$$
Now one observes that this is such that it trivializes when restricted along the
canonical inclusion $X \longrightarrow \mathrm{SymplGrpd}(X,\pi)$
(which is the canonical atlas,  def. \ref{atlas}). 
By the discussion in \ref{BoundaryFieldTheory} this induces a
boundary field theory for the 2d Poisson-Chern-Simons theory,
exhibited by the correspondence
$$
  \raisebox{20pt}{
  \xymatrix{
    & X  \ar[dr]^i_{\ }="s"
    \ar[dl]
    \\
    \ast 
    \ar[dr]^{\ }="t"
    && \mathrm{SymplGrpd}(X,\pi)
    \ar[dl]^\chi
    \\
    &
    \mathbf{B}^2 U(1)
    \ar@{=>}^\xi "s"; "t"
  }
  }
  \,.
$$
Here the trivialization $\xi$ is a \emph{prequantum bundle} on the
Poisson manifold $X$.

Now since this is a correspondence with higher phases in $\mathbf{B}^2 U(1)$,
by the above discussion this situation is naturally quantized in 
twisted complex $K$-theory
$$
  \raisebox{20pt}{
  \xymatrix{
    & X  \ar[dr]^i_{\ }="s"
    \ar[dl]
    \\
    \ast 
    \ar[dr]^{\ }="t"
    && \mathrm{SymplGrpd}(X,\pi)
    \ar[dl]^\chi
    \\
    &
    \mathbf{B}^2 U(1)
    \ar[d]^\rho
    \\
    & 
    \mathbf{B} \mathrm{GL}_1(\mathrm{KU})
    \ar@{=>}^\xi "s"; "t"
  }
  }
  \,.
$$
If the morphism $i : X \longrightarrow \mathrm{SymplGrpd}(X,\pi)$
can be and is equipped with an orientation in $\chi$-twisted K-theory
(which usually involves in particular that we assume that $X$ is compact), then 
pull-push yields a map of $\mathrm{KU}$-module spectra of the form
$$
  i_! \xi
  : 
  \mathrm{KU}
  \stackrel{}{\longrightarrow}
  \mathrm{KU}^\chi(\mathrm{SymplGrpd(X,\pi)})
  \,.
$$
This is equivalently an element in the twisted K-cohomology
of the symplectic groupoid
$$
  i_! \xi \in \mathrm{KU}^\chi(\mathrm{SymplGrpd(X,\pi)})
$$
and this is to be regarded as the quantization of the Poisson manifold
$(X,\pi)$.

Notice that the symplectic groupoid $\mathrm{SymplGrpd}(X,\pi)$ is a resolution of the 
space of symplectic leaves of $(X,\pi)$. Therefore a class in 
$\mathrm{KU}^\chi(\mathrm{SymplGrpd(X,\pi)})$ may be thought of as providing
one (virtual) vector space for each symplectic leaf, to be thought of as the
space of quantum states.
Specifically, when $(X,\pi) = (X, \omega^{-1})$ happens to be a symplectic
manifold, then $\mathrm{SymplGrpd}(X,\pi) \simeq \ast$ (as smooth stacks)
and so the above yields an element of the K-theory of the point. 
One checks that this coincides with the K-theoretic description of
traditional geometric quantization. 

If there is a compact group $G$ of Hamiltonians acting on $(X,\omega^{-1})$
by Hamiltonian actions, then by the discussion in 
\ref{TheClassicalActionFunctionalPrequantizesLagrangianCorrespondences}, \ref{QuantomorphismGroup} 
we obtain a $G$-equivariant version of this situation exhibited by 
a correspondence of quotient stacks
$$
  \raisebox{20pt}{
  \xymatrix{
    & X/\!/G  \ar[dr]_{\ }="s"
    \ar[dl]
    \\
    \ast/\!/G
    \ar[dr]^{\ }="t"
    && \ast /\!/G
    \ar[dl]
    \\
    &
    \mathbf{B}^2 U(1)
    \ar[d]^\rho
    \\
    & 
    \mathbf{B} \mathrm{GL}_1(\mathrm{KU})
    \ar@{=>}^\xi "s"; "t"
  }
  }
  \,.
$$
This now has a cohomological quantization if the $G$-action preserves the
choice of K-orientation. If that comes from a K{\"a}hler polarization then this is the
familiar condition on quantizability of prequantum operators. Now the pull-push
quantization acts as
$$
  i_! : 
  \xymatrix{
    K(X/\!/G)
    \simeq 
    K_G(X)
    \longrightarrow
    K_G(\ast)
    \simeq
    K(\ast /\!/G)
    \simeq
    \mathrm{Rep}(G)
  }
$$
and refines 
$i_! \chi$ to a cocycle in $G$-equivariant K-theory. 
This is represented by the Hilbert space of quantum states equipped with its
$G$-action by quantum observables. See section 5.2.1 in \cite{Nuiten}.

Notice that this cohomological/geometric quantization of Poisson manifolds 
as the 1d boundary field theory of the non-perturbative 2d Poisson-Chern-Simons theory
is conceptually analogous to the perturbative formal deformation quantization 
of Poisson manifolds given in \cite{Kontsevich}, which in \cite{CaFe} was found to be induced
by the perturbative Poisson sigma-model. Another similar holographic
quantization of 1d mechanics by a 2d string sigma-model was given in \cite{GukovWitten}.

\medskip 

More generally, regard the dual vector space $\mathfrak{g}^\ast$
of the Lie algebra of $G$ as a Poisson manifold with its Lie-Poisson structure.
The corresponding symplectic groupoid is the adjoint action groupoid
$$
  \mathrm{SymplGrpd}(\mathfrak{g}^\ast, [-,-])
  \simeq
  \mathfrak{g}^\ast /\!/G
  \,.
$$
Then for $O_\lambda$ any regular coadjoint orbit, we get the following
correspondence of correspondences in $\mathbf{H}$:
$$
  \xymatrix{
    \ast & \ast \ar[l] \ar[r] & \ast & \mbox{trivial theory}
    \\
    O_\lambda/\!/G \ar[u]\ar[d] & O_\lambda \ar[l] \ar@{^{(}->}[r] \ar[u] \ar[d] & 
    \mathfrak{g}^\ast \ar[u] \ar[d] & \mbox{boundary}
    \\
    \ast/\!/G &  O_\lambda/\!/G \ar[l] \ar[r] & \mathfrak{g}^\ast/\!/G & \mbox{2dCS}
    \\
    \mbox{2dCS} & \mbox{defect} & \mbox{2dCS}
  }
  \,.
$$
\index{defect field theory!universal orbit method}
Here on the left we have the symplectic manifold $O_\lambda$ regarded as a $G$-equivariant
boundary field theory of its 2d Poisson-Chern-Simons theory as just discussed. On the right
we have the Poisson manifold $\mathfrak{g}^\ast$ regarded as the boundary of
its 2d Poisson-Chern-Simons theory as above. 
The whole diagram is equipped with maps down to $\mathbf{B}^2 U(1)$ 
as in prop. \ref{prop mesh corner}, which we will not display here.
As such the diagram exhibits a defect at which these two boundary conditions meet.
The quantization of the left boundary theory yields a space of states realized as an element
in the representation ring of $G$, which is the K-theoretic formulation
of Kirillov's orbit method. The quantization of the defect yields
a quantum operator which produces such representations from prequantum bundles
over $\mathfrak{g}^\ast$. Analysis of the details shows that this defect operator
is effectively the ``inverse universal orbit method'' of 
theorem 1.28 in \cite{FHTii}. See \ref{WithWilsonLoops} above for discussion of the
orbit method and its physical interpretation. 
See section 5.2.2 of \cite{Nuiten} for more details on this defect quantization.
\index{orbit method!universal}

\paragraph{D-brane charge}
\label{DBraneCharge}
\index{D-brane!charge in K-theory}

While by the above the endpoints of the Poisson-Chern-Simons string 
serve to exhibit every possible mechanical system, we can also 
consider endpoints of the (topological sector) of the type II superstring.
This is the $\sigma$-model on a target manifold $X$ equipped with a
background B-field whose instanton sector we denote by
$\chi_B : X \longrightarrow \mathbf{B}^2 U(1)$. 
A boundary condition for this which is given by a submanifold $Q \hookrightarrow X$ is 
precisely a choice of trivialization $\xi$ of $\chi_B$ on that submanifold
$$
  \raisebox{20pt}{
  \xymatrix{
    & Q  \ar[dr]^i_{\ }="s"
    \ar[dl]
    \\
    \ast 
    \ar[dr]^{\ }="t"
    && X
    \ar[dl]^{\chi_B}
    \\
    &
    \mathbf{B}^2 U(1)
    \ar@{=>}^\xi "s"; "t"
  }
  }
  \,.
$$
This $\xi$ now plays the role of the Chan-Paton gauge field on the \emph{D-brane}
$Q$. The cohomological quantization of this correspondence in K-theory
as above exists if $Q \hookrightarrow X$  is $\chi$-twisted K-orientable,
which is precisely the Freed-Witten anomaly cancellation condition,
see \ref{OpenStringGaugeCoupling}. In this case the quantization yields the class
$$
  i_! \xi \in K^{\chi_B}(X)
$$
in the B-field twisted K-theory of spacetime. This is what is known as the 
\emph{D-brane charge} of $(Q,\xi)$ \cite{BMRS}. See section 5.2.4 of
\cite{Nuiten} for more details.

\subsubsection{The quantum string at the boundary of the membrane}
\label{QuantumStringAtBoundaryOfMembrane}
\index{M-theory!O9-plane charge}
\index{M-theory!M5-brane charge}

We consider the above holographic quantization of the particle at the
boundary of the string in one dimension higher, and indicate 
the cohomological quantization of the string at the boundary of the membrane.

To that end, let $\chi_C : Y/\!/\mathbb{Z}_2 \longrightarrow \mathbf{B}^3 U(1)$
be (the 3-bundle underlying) the supergravity $C$-field
on an 11-dimensional Ho{\v r}ava-Witten spacetime, as discussed above in \ref{supergravityCField}.
By the discussion there, this class is of the form
$\tfrac{1}{2}p_1 + 2 a$ and on the higher orientifold plane 10-dimensional boundary
$X \hookrightarrow Y$ this class trivializes. By the discussion in \ref{HigherSpinStructure}
the choice of trivialization $\chi_B$ is the twisted B-field of the
heterotic string theory on $X$, representing a $2a$-twisted string structure.
$$
  \raisebox{20pt}{
  \xymatrix{
    & X  \ar[dr]^i_{\ }="s"
    \ar[dl]
    \\
    \ast 
    \ar[dr]^{\ }="t"
    && Y/\!/\mathbb{Z}_2
    \ar[dl]^{\chi_C}
    \\
    &
    \mathbf{B}^3 U(1)
    \ar@{=>}^{\chi_B} "s"; "t"
  }
  }
  \,.
$$
Now since $\chi_C$ is the background field for the topological sector
of the M2-brane, \ref{OldBraneScan},
by the general discussion of \ref{BoundaryFieldTheory} 
we see that this diagram exhibits a boundary condition for the M2-brane
ending on the O9-plane $X$. This boundary string of the M2 on the O9
is the heterotic string. By comparison with \ref{QuantumParticleAtEndOfTheString}
we see that therefore the cohomological quantization of this correspondence
should yield the partition function of the string.

To see this, notice that by the general discussion in 
\ref{CohomologicalQuantizationOfCorrespondences} we are to choose 
a higher superposition principle by finding an $E_\infty$-ring $E$
such that there is a natural higher group homomorphism
$$
  \mathbf{B}^2 U(1) \longrightarrow \mathbf{\Pi}\mathbf{B}^2 U(1)
  \simeq K(\mathbb{Z},3) \longrightarrow \mathrm{GL}_1(E)
  \,.
$$
There is one famous such choice \cite{AndoBlumbergGepner}, namely $E = \mathrm{tmf}$, the 
``universal elliptic cohomology theory''. 
$$
  \raisebox{20pt}{
  \xymatrix{
    & X  \ar[dr]^i_{\ }="s"
    \ar[dl]
    \\
    \ast 
    \ar[dr]^{\ }="t"
    && Y/\!/\mathbb{Z}_2
    \ar[dl]^{\chi_C}
    \\
    &
    \mathbf{B}^3 U(1)
    \ar[d]^\rho
    \\
    & \mathbf{B}\mathrm{GL}_1(\mathrm{tmf})
    \ar@{=>}^{\chi_B} "s"; "t"
  }
  }
  \,.
$$
A cohomological quantization of this exists now if we have an 
orientation in twisted $\mathrm{tmf}$, hence by twisted string structures.\footnote{
This point in \cite{AndoBlumbergGepner} was originally amplified by Hisham Sati
in work that led to the discussion in \ref{HeteroticStringAndTwistedStringStructures} above.}
If it exists, push-forward in $\mathrm{tmf}$ yields 
\cite{AHR} the (twisted \cite{CHZ})
\emph{Witten genus}. By the seminal result of \cite{WittenElliptic}
this is the partition function of the heterotic string. 

By analogy with \ref{DBraneCharge} we may also regard this 
push-forward in $\mathrm{tmf}$ as the ``O9-plane charge''. 
An analogous story should apply to the cohomological quantization of the
M2-brane ending not on the O9-plane,
but on the M5-brane, by \ref{The5brane}. In this case the result of the
quantization would be an M5-brane charge in $\mathrm{tmf}$, induced
not by the heterotic, but by the self-dual string on the M5. 
That this is the case is the statement of proposal 6.13 in \cite{Sati10}.

\newpage

\newpage

\printindex
	
\end{document}